\documentclass[
aps,
onecolumn,
12pt,
tightenlines,
nofootinbib,
superscriptaddress,
floatfix,
prd
]{revtex4-2}
\usepackage{subcaption}
\usepackage{gensymb}
\usepackage[utf8]{inputenc}
\usepackage[printonlyused]{acronym}
\usepackage{amssymb}
\usepackage{mwe}
\usepackage{comment} 
\usepackage{acronym}
\usepackage{amsmath} % or simply amstext
\usepackage[utf8]{inputenc}
\usepackage[OT1]{fontenc}
\usepackage{siunitx}
\usepackage[T1]{fontenc}

\newcommand{\Chandra}{\emph{Chandra }}
\newcommand{\Einstein}{\emph{Einstein }}
\newcommand{\XMM}{\emph{XMM }}
\newcommand{\ROSAT}{\emph{ROSAT }}
\newcommand{\ASCA}{\emph{ASCA }}
\newcommand{\RXTE}{\emph{RXTE }}
\newcommand{\Uhuru}{\emph{Uhuru }}
\newcommand{\LIGO}{\emph{LIGO }}
\newcommand{\LISA}{\emph{LISA }}

\usepackage{siunitx}
\usepackage{physics}
\usepackage{braket}
\usepackage{amsmath}	% AMS Math Package
\usepackage{amsthm}		% Theorem Formatting
\usepackage{amssymb}	% Math symbols such as \mathbb

\usepackage{mathrsfs, mathtools}

\usepackage{letltxmacro} % closed roots

\usepackage{color}
\usepackage{bm}

\usepackage{tgpagella}

\usepackage{graphicx}

\usepackage{fancyhdr}

\usepackage{lastpage}

\usepackage{titlesec}

\usepackage{rotating}
\usepackage{setspace}

\usepackage[printonlyused]{acronym}
\usepackage{aas_macros}
\usepackage{xspace}

\usepackage{datetime}

\usepackage{lettrine}

\usepackage{anyfontsize}

\usepackage[colorlinks=true
,urlcolor=DARKBLUE
,anchorcolor=DARKBLUE
,citecolor=DARKBLUE
,filecolor=DARKBLUE
,linkcolor=DARKBLUE
,menucolor=DARKBLUE
]{hyperref}
\titleformat{\section}
       {\centering\normalfont\fontsize{14}{17}\bfseries}{\thesection}{1em}{}

\numberwithin{equation}{section}

\makeatletter 
    \renewcommand{\thefigure}{{\bf\@arabic\c@figure}}
\makeatother

\usepackage{tcolorbox}
\definecolor{grey}{rgb}{0.4,0.4,0.4}
\definecolor{dullmagenta}{rgb}{0.4,0,0.4}
\definecolor{darkblue}{rgb}{0,0,0.4}
\definecolor{midblue}{rgb}{0,0,0.7}
\definecolor{midred}{rgb}{0.5,0,0}
\definecolor{orange}{rgb}{1,0.5,0}
\definecolor{lightbrown}{rgb}{0.75,0.5,0.25}
\definecolor{tan}{cmyk}{0.14,0.42,0.56,0}
\definecolor{djunglegreen}{cmyk}{0.99,0,0.52,0}
\definecolor{lightgreen}{rgb}{0,1,0}
\definecolor{olivegreen}{cmyk}{0.64,0,0.95,0.40}
\definecolor{midgreen}{rgb}{0.0,0.675,0.0}
\definecolor{darkgreen}{rgb}{0,0.5,0}
\definecolor{pink}{rgb}{1,0.078,0.57}

\definecolor{MONZA}{HTML}{CF000F}
\definecolor{DARKBLUE}{HTML}{00008b}
\definecolor{DARKMAGENTA}{HTML}{8b008b}

\newcommand{\vs}{\vspace}

\makeatletter
\let\oldr@@t\r@@t
\def\r@@t#1#2{%
\setbox0=\hbox{$\oldr@@t#1{#2\,}$}\dimen0=\ht0
\advance\dimen0-0.2\ht0
\setbox2=\hbox{\vrule height\ht0 depth -\dimen0}%
{\box0\lower0.4pt\box2}}
\LetLtxMacro{\oldsqrt}{\sqrt}
\renewcommand*{\sqrt}[2][\ ]{\oldsqrt[#1]{#2}}
\makeatother

\newcommand{\la}{\ensuremath{\leftarrow}}

\newcommand{\be}{\begin{equation}}
\newcommand{\ee}{\end{equation}}
\newcommand{\ba}{\begin{eqnarray}}
\newcommand{\ea}{\end{eqnarray}}

\smallskipamount
\settimeformat{ampmtime}

\def\ga{\mathrel{\raise.3ex\hbox{$>$\kern-.75em\lower1ex\hbox{$\sim$}}}}
\def\la{\mathrel{\raise.3ex\hbox{$<$\kern-.75em\lower1ex\hbox{$\sim$}}}}

\newcommand{\Msun}{\ensuremath{\,\rm{M}_{\odot}}\xspace}
\newcommand{\Ms}{\Msun}

\newcommand{\Rsun}{\ensuremath{\,\rm{R}_{\odot}}\xspace}
\newcommand{\Rs}{\Rsun}
\newcommand{\Zs}{{\rm Z}_\odot}

\newcommand{\Mzams}{{\rm M}_{\rm ZAMS}}

\newcommand{\Moneint}{\ensuremath{{M}_{\rm{1}}}\xspace}
\newcommand{\Mtwoint}{\ensuremath{{M}_{\rm{2}}}\xspace}
\newcommand{\aint}{\ensuremath{{a}_{\rm{0}}}\xspace}

\newcommand*\diff{\mathop{}\!\mathrm{d}}

\newcommand{\Gyr}{\ensuremath{\,\mathrm{Gyr}}\xspace}

\newcommand{\GpcminThree}{\ensuremath{\,\rm{Gpc}^{-3}}\xspace}

\acrodef{BH}{black hole}
\acrodefplural{BH}[BHs]{black holes}

\acrodef{BBH}{Binary black hole}
\acrodefplural{BBH}[BBHs]{binary black holes}

\acrodef{IMBH}{intermediate-mass black hole}
\acrodefplural{IMBH}[IMBHs]{intermediate-mass black holes}

\acrodef{SMBH}{supermassive black hole}
\acrodefplural{SMBH}[SMBHs]{supermassive black holes}

\acrodef{PBH}{primordial black hole}
\acrodefplural{PBH}[PBHs]{primordial black holes}

\acrodef{BHXB}{black hole X-ray binary}
\acrodefplural{BHXB}{black hole X-ray binaries}

\acrodef{GC}{globular cluster}
\acrodefplural{GC}[GCs]{globular clusters}

\acrodef{NC}{nuclear cluster}
\acrodefplural{NC}[NCs]{nuclear clusters}

\acrodef{OC}{open cluster}
\acrodefplural{OC}[OCs]{open clusters}

\acrodef{YMC}{young massive cluster}
\acrodefplural{YMC}[YMCs]{young massive clusters}

\acrodef{MLT}{mixing-length theory}
\acrodef{HR}{Hertzsprung-Russell}
\acrodef{MS}{main sequence}
\acrodef{ZAMS}{zero-age MS}
\acrodef{CHeB}{core helium burning}
\acrodef{RGB}{red giant branch}
\acrodef{RG}{red giant}
\acrodef{RSG}{red super-giant}
\acrodef{HB}{horizontal branch}
\acrodef{AGB}{asymptotic giant branch}
\acrodef{TP-AGB}{thermally-pulsating asymptotic giant branch}
\acrodef{LBV}{luminous blue variable}
\acrodef{WR}{Wolf-Rayet}

\acrodef{WD}{white dwarf}
\acrodefplural{WDs}{white dwarfs}
\acrodef{NS}{neutron star}
\acrodefplural{NSs}{neutron stars}
\acrodef{CO}{compact object}
\acrodefplural{COs}{compact objects}

\acrodef{CC}{core collapse}
\acrodef{ECSN}{electron-capture supernova}
\acrodef{PI}{pair instability}
\acrodef{PISN}{pair instability supernova}
\acrodef{PPISN}{pulsational pair instability supernova}
\acrodef{compas}{\texttt{COMPAS}}
\acrodef{mobse}{\texttt{MOBSE}}
\acrodef{cBH}{Black hole}
\acrodefplural{cBH}[BHs]{Black holes}

\acrodef{sBH}{stellar-mass black hole}
\acrodefplural{sBH}[sBHs]{Stellar-mass black holes}

\acrodef{IMBH}{intermediate-mass black hole}
\acrodefplural{IMBH}[IMBHs]{intermediate-mass black holes}

\acrodef{SMBH}{supermassive black hole}
\acrodefplural{SMBH}[SMBHs]{Supermassive black holes}

\acrodef{IMRI}{intermediate-mass ratio inspiral}
\acrodefplural{IMRI}[IMRIs]{intermediate-mass ratio inspirals}

\acrodef{BBH}{binary black hole}
\acrodefplural{BBH}{binary black holes}

\acrodef{GC}{globular cluster}
\acrodefplural{GC}[GCs]{Globular clusters}

\acrodef{NSC}{nuclear star cluster}
\acrodefplural{NSC}[NSCs]{Nuclear star clusters}

\acrodef{OC}{open cluster}
\acrodefplural{OC}[OCs]{open clusters}

\acrodef{YMC}{young massive cluster}
\acrodefplural{YMC}[YMCs]{Young massive clusters}

\acrodef{AGN}[AGN]{active galactic nucleus}
\acrodefplural{AGN}[AGNs]{active galactic nuclei}

\acrodef{IMF}{initial mass function}

\acrodef{GW}{gravitational wave}
\acrodefplural{GW}{gravitational waves}
\acrodef{LVK}{LIGO-Virgo-KAGRA collaboration}
\acrodef{LISA}{Laser Interferometer Space Antenna}

\acrodef{VMS}{very massive star}
\acrodefplural{VMSs}[VMS]{very massive stars}

\acrodef{ZAMS}{zero-age main sequence}
\acrodef{MS}{main sequence}
\acrodef{WD}{white dwarf}
\acrodefplural{WDs}{white dwarfs}
\acrodef{NS}{neutron star}
\acrodefplural{NSs}{neutron stars}
\acrodef{CO}{compact object}
\acrodefplural{COs}{compact objects}
\acrodef{DCO}{double compact object}
\acrodefplural{DCOs}{double compact objects}
\acrodef{RLOF}{Roche Lobe Overflow}

\acrodef{CE}{Common Envelope}
\acrodef{MW}{Milky Way}

\acrodef{SSE}{\texttt{SSE}}
\acrodef{BSE}{\texttt{BSE}}
\acrodef{MOCCA}{\texttt{MOCCA}}

\makeatletter
\def\l@subsubsection#1#2{}
\makeatother

\headwidth=\textwidth

\begin{document}

%%%%%%%%%%%%%%%%%%%%%%%%%%%%%%%%%%%%%%%%%%%
\title{\Large {\huge S}tellar {\huge B}lack {\huge H}oles and {\huge C}ompact {\huge S}tellar {\huge R}emnants}

%%%%%%%%%%%%%%%%%%%%%%%%%%%%%%%%%%%%%%%%%%%

\author{Guglielmo Costa}
\email{guglielmo.costa.astro@gmail.com}
\affiliation{Univ Lyon, Univ Lyon1, Ens de Lyon, CNRS,Centre de Recherche Astrophysique de Lyon UMR5574,  F-69230 Saint-Genis-Laval, France}
\author{Martyna Chru{\'s}li{\'n}ska}
\email{mchruslinska@mpa-garching.mpg.de}
\affiliation{Max Planck Institute for Astrophysics, Karl Schwarzchild Strasse 1, D-85748 Garching, Germany}
\author{Jakub Klencki}
\email{jakub.klencki@eso.org}
\affiliation{European Southern Observatory, Karl-Schwarzschild-Strasse 2, 85748 Garching, Germany}
\author{Floor S. Broekgaarden}
\email{fsb2127@columbia.edu}
\affiliation{Department of Astronomy and Columbia Astrophysics Laboratory, Columbia University, 550 W 120th St, New York, NY 10027, USA}
\affiliation{Simons Society of Fellows, Simons Foundation, New York, NY 10010, USA}
\affiliation{William H. Miller III Department of Physics and Astronomy, Johns Hopkins University, Baltimore, Maryland 21218, USA}
\author{Carl L.~Rodriguez}
\email{carl.rodriguez@unc.edu}
\affiliation{University of North Carolina at Chapel Hill, 120 E.~Cameron Ave, Chapel Hill, NC 27599}
\author{Tana D. Joseph}
\email{t.d.joseph@uva.nl}
\affiliation{Anton Pannekoek Institute for Astronomy, University of Amsterdam, Science Park 904, 1098 XH Amsterdam, The Netherlands}
\author{Sara Saracino}
\email{s.saracino@ljmu.ac.uk}
\affiliation{Astrophysics Research Institute, Liverpool John Moores University, 146 Brownlow Hill, Liverpool L3 5RF, UK}
    
%%%%%%%%%%%%%%%%%%%%%%%%%%%%%%%%%%%%%%%%%%%
\date{\formatdate{\day}{\month}{\year}, \currenttime}

%%%%%%%%%%%%%%%%%%%%%%%%%%%%%%%%%%%%%%%%%%%
\begin{abstract}
\vs{2mm}
\begin{tcolorbox}
The recent observations of \acp{GW} by the \ac{LVK} have provided a new opportunity for studying our Universe. By detecting several merging events of \acp{BH}, \ac{LVK} has spurred the astronomical community to improve theoretical models of single, binary, and multiple star evolution in order to better understand the formation of \ac{BBH} systems and interpret their observed properties. The final \ac{BBH} system configuration before the merger depends on several processes, including those related to the evolution of the inner stellar structure and those due to the interaction with the companion and the environment (such as in stellar clusters). This Chapter provides a summary of the formation scenarios of stellar \acp{BH} in single, binary, and multiple systems. We review all the important physical processes that affect the formation of \acp{BH} and discuss the methodologies used to detect these elusive objects and constrain their properties.
\end{tcolorbox}
\end{abstract}

%%%%%%%%%%%%%%%%%%%%%%%%%%%%%%%%%%%%%%%%%%%
\begin{center}
    \phantom{\fontsize{50}{50}\selectfont I}\\
    {
    \fontsize{30}{10}\selectfont 
    {\fontsize{30}{20}\selectfont S}tellar-{\fontsize{30}{20}\selectfont B}lack 
    {\fontsize{30}{20}\selectfont H}oles and
    {\fontsize{30}{20}\selectfont C}ompact
    {\fontsize{30}{20}\selectfont S}tellar
    {\fontsize{30}{20}\selectfont R}emnants}
    \\[90mm]
    {\noindent\makebox[\linewidth]{\resizebox{0.3333\linewidth}{1pt}{$\bullet$}}\bigskip}\\[2mm]
    {\fontsize{18}{5}\selectfont Guglielmo Costa$*$}\\[3mm]
    {\fontsize{18}{5}\selectfont Martyna Chru{\'s}li{\'n}ska$*$}\\[3mm]
    {\fontsize{18}{5}\selectfont Jakub Klencki$*$}\\[3mm]
    {\fontsize{18}{5}\selectfont Floor Broekgaarden$*$}\\[3mm]
    {\fontsize{18}{5}\selectfont Carl L.~Rodriguez$*$}\\[3mm]
    {\fontsize{18}{5}\selectfont Tana D. Joseph$*$}\\[3mm]
    {\fontsize{18}{5}\selectfont Sara Saracino$*$}\\[3mm]
    {$*$all authors have equally contributed to this chapter}
\end{center}
\newpage
\maketitle

%%%%%%%%%%%%%%%%%%%%%%%%%%%%%%%%%%%%%%%%%%%
\newpage
\tableofcontents
\newpage

%%%%%%%%%%%%%%%%%%%%%%%%%%%%%%%%%%%%%%%%%%%
%%%%%%%%%%%%%%%%%%%%%%%%%%%%%%%%%%%%%%%%%%%

%%%% COSTA %%%%
\part{Stellar evolution: single stars \\ \Large{Guglielmo Costa}}
\label{part:stars}
\section{Introduction}

\acp{BH} are the densest and most extreme objects in the Universe. Their existence was hypothesised in 1916 by Schwarzschild \citep{1916AbhKP1916..189S} as a solution to Einstein's general relativity equations of space and time.
Theoretically, they can have any mass, and commonly, they are grouped into three families: stellar-mass \acp{BH}, with a mass up to $100~\Ms$, \acp{IMBH} with mass in the range between $100$ and $10^6~\Ms$, and \acp{SMBH} with mass above $10^6~\Ms$.  
The recent detection of the first \ac{GW} by coalescing stellar mass \acp{BH} \citep{2016PhRvL.116f1102A, 2016ApJ...818L..22A} confirmed (once again) Einstein's theory of relativity and the existence of \acp{BH}. Moreover, it opens a new field of study, which uses \acp{GW} for observing and studying the Universe.
During the last few years, the interferometers of the \ac{LVK} have observed about 90 merger events \citep{2021PhRvX..11b1053A, 2021arXiv211103606T}, all with progenitor \acp{BH} masses in the stellar mass range. 
The formation scenarios of such kind of \acp{BH} and \acp{BBH} is still debated. 
Although the general idea of single stellar mass \acp{BH}' formation is commonly accepted -- that a \ac{BH} could be created from the gravitational collapse of a massive star after the end of all nuclear-burning stages as postulated in 1939 by Oppenheimer, Volkoff, and Snyder \cite{1939PhRv...56..455O, 1939PhRv...55..374O} -- the details of the formation are still discussed.
Uncertainties in many physical processes may significantly change the standard picture of massive stars evolution, ultimately changing the predicted final outcome.
On top of that, the coupling of two stars (or \acp{BH}) is needed to form a \ac{BBH} that will merge due to \ac{GW} emission. This adds a new layer of complexity to the problem. 
Nowadays, the two main studied formation scenarios of \acp{BBH} are the \emph{isolated channel} -- in which two stars are born together and evolve through all the binary evolutionary processes until they form a \ac{BBH} system -- and the \emph{dynamical channel} -- in which a population of stars born in a cluster, interact multiple times during the evolution leading to the formation of \acp{BBH}. 
Both scenarios help to interpret and understand the data obtained by the \ac{GW} detectors. 

In this first Part, we will focus on single stars and briefly review the evolution of massive stars, which can be quite complex.
After a short presentation of the main concepts of the stellar theory and stellar classification, we will introduce the main physical processes that affect the evolution of massive stars and their outcome, namely, stellar winds, mixing processes, and final explosive (or not) phases.

%-------------------------------------------------------------------

\section{Overview on stellar evolution}
\label{sec:overview}

Stars are self-gravitating objects composed of (almost) fully ionised gas in hydrostatic and thermal equilibrium. 
Their evolution is regulated by the thermonuclear reactions that act in their cores. 
The energy provided by the endothermic nuclear reactions is the fundamental ingredient in keeping them up against gravity and shining over time.
At the beginning of their life, stars are mainly composed of hydrogen ($68 - 75 \%$) and helium ($24-30 \%$), with a smaller part of heavier chemical elements ($\sim 0 - 2 \%$), which astronomers commonly refer to as ``metals''. 
The metal content of stars in mass fraction, Z, is also called metallicity.
Usually, stellar models adopt a chemical composition scaled with the solar one, but its exact composition remains a matter of debate \citep{1998SSRv...85..161G, 2009ARA&A..47..481A, 2011SoPh..268..255C, 2021A&A...653A.141A}.
Although stellar evolution is quite a complex process, the nuclear evolution of the central regions could be simply described as a series of burning cycles\footnote{Here, we are using the astronomers' terminology. Although we are talking about nuclear fusion processes, we refer to them as burning or combustion processes.}.
Each cycle can be divided into four steps, which are: (i) core contraction, (ii) core heating, (iii) nuclear burning, and (iv) fuel exhaustion \citep{2013sse..book.....K}. At the end of each cycle, the star must readjust to a new equilibrium, and a new cycle (or evolutionary phase) begins. 
During their lifetime, stars progressively burn chemical elements from the lighter ones (i.e. hydrogen) to the heavier ones.
Stars spend most of their life, $85 - 90 \%$, burning hydrogen in their cores. This phase is called \ac{MS}, which is the locus in the observed luminosity versus surface temperature diagram in which stars are more abundant (see Section~\ref{subsec:HR}). 
The second longest phase is \ac{CHeB}, in which stars spend $10 - 15 \%$ of their lifetime \citep{2009pfer.book.....M}. 
Depending on their initial mass (see Section~\ref{subsec:mass_ranges}), stars may ignite and burn elements heavier than helium.
The advanced phases are much faster than the other two main burning phases, and the evolution from carbon-burning ignition to the final explosion (or collapse) is only $\sim 0.001$ of the total lifetime. The stellar lifetime mainly depends on the star's mass (see Section~\ref{subsec:mass_ranges}), but also on the metallicity and initial angular momentum.

\subsection{Historical background}
The theory of stellar structure and evolution has been developed since the beginning of the twentieth century.
Pioneering work has been done by \citet{1918ApJ....48..205E}, which set the ground for the theoretical understanding of stars. 
Later, after the discovery of the tunnel effect by Gamov in 1928, \citet{1929ZPhy...54..656A} first postulated that thermonuclear reactions in stellar cores could be an energy source for stars.
At the same time, the breakthrough works of \citet{1925PhDT.........1P} and \citet{1929ApJ....70...11R} suggested that hydrogen is the most abundant element in the Sun and other stars.
Approximately ten years later, \citet{1938PhRv...54..248B} and \citet{1939PhRv...55..434B} described the \textit{proton-proton} chain and the carbon-nitrogen-oxygen (\textit{CNO}) cycle, which are the two main channels for the formation of helium from hydrogen.
The proton-proton chain consists of a series of consecutive proton captures. 
The first part leads to the formation of $^3$He; then, the chain can continue through three different branches (namely \textit{pp1}, \textit{pp2}, \textit{pp3}). pp1 is initially the most frequent chain, transforming six protons into one helium plus two protons. Later, at higher temperatures, pp2 and pp3 increase their relative importance. 
Both convert one $^7$Be plus a proton into two $^4$He atoms.
The CNO cycle becomes dominant over the proton-proton chain for temperatures above $1.5 \times 10^7$ K. It requires the presence of some C, N, or O elements that act as catalysts for the formation of helium.

In 1952, \citet{1952ApJ...115..326S} described the main reactions of \ac{CHeB}. During this phase, stars convert helium into carbon through the \textit{triple-$\alpha$} reaction\footnote{An $\alpha$ particle is a $^4$He nucleus. The triple-$\alpha$ reaction requires temperatures above $10^8$ K and proceeds in two steps, since the triple encounter is very improbable. The first is $^{4}$He + $^4$He $\rightarrow$ $^{8}$Be, and the second is $^{8}$Be + $^4$He $\rightarrow$ $^{12}$C + $\gamma$.}, then carbon to oxygen with $^{12}$C + $^4$He $\rightarrow$ $^{16}$O + $\gamma$ and finally oxygen to neon with $^{16}$O + $^4$He $\rightarrow$ $^{20}$Ne + $\gamma$.
For a detailed description of nuclear reaction rates in stars, we refer the reader to \citet{2007nps..book.....I}. 

The stellar theory received a great boost in the 1950s and 1960s, thanks to the arrival of the first computers and the development of computational sciences. 
In the late 1950s, the first models of the Sun appeared in the literature \citep{1957ApJ...125..233S}. 
During the second half of the 1960s, innovative numerical methods allowed astronomers to solve stellar equations in a fast and reliable fashion, and the first simulations of the evolution of stars during the main evolutionary phases appeared \citep{1964ApJ...139..306H, 1967MComP...7..129K}. 
More details on stellar structure and numerical methodologies can be found in \citet{2009pfer.book.....M, 2013sse..book.....K}.

After those first important steps, stellar evolution theory has been improved thanks to more detailed treatments and careful calibrations of stellar processes, such as stellar winds, internal mixing, advanced burning stages, rotation, and magnetic fields, to name a few.

Nowadays, it is possible to model stars with a wide range of initial masses (between $\sim 0.08~\Ms$ and $\sim 1000~\Ms$, where $\Ms = 1.989\times 10^{33}\,$g is the mass of the Sun) and compositions (from almost metal-free stars to supra-solar abundances, see Section~\ref{sec:phys_processes}).
Stellar evolution theory is useful not only for studying stars themselves but also because it is the backbone of many other astrophysical fields \citep{1983psen.book.....C}.
For instance, stellar models are used to study the ages and properties of star clusters \citep{2018ApJ...863...67G, 2018ApJ...864L...3G, 2019A&A...631A.128C, 2019ApJ...871...20S, 2022Univ....8..359M, 2023MNRAS.522.2429M}, to understand galaxies and Milky Way formation and evolution \citep{2017MNRAS.466.1903G, 2018MNRAS.479..994R, 2019MNRAS.485.1384V, 2022ApJS..262...22D}, supernova, gamma-ray burst progenitors and chemical yields \citep{2006A&A...460..199Y, 2016ApJ...821...38S, 2018ApJS..237...13L, 2020ApJ...901..114A}, obtain host-star parameters for exo-planets \citep{2017A&A...597A..14G, 2019A&A...624A..94M, 2020A&A...644A..68M}, and derive the \ac{CO} mass spectrum of single and binary populations \citep{2017MNRAS.470.4739S, 2018MNRAS.481.1908K, 2019ApJ...882..121S, 2020ApJ...888...76M, 2021ApJ...912L..31W, 2023MNRAS.524..426I, 2023ApJS..264...45F} to tackle the open problems raised by the new observed gravitational waves events \citep{2016PhRvL.116f1102A, 2020ApJ...900L..13A, 2020PhRvL.125j1102A}.

\subsection{Mass ranges and stellar remnants}
\label{subsec:mass_ranges}

From a purely theoretical point of view, the evolution of a star could be uniquely determined by just three parameters: 
the initial mass, the chemical composition, and the angular momentum. 
Without worrying about metallicity and rotation, stars' evolution and final fate are mainly determined by their initial mass at the first order of approximation.  

\begin{figure*}   
\includegraphics[width=\columnwidth]{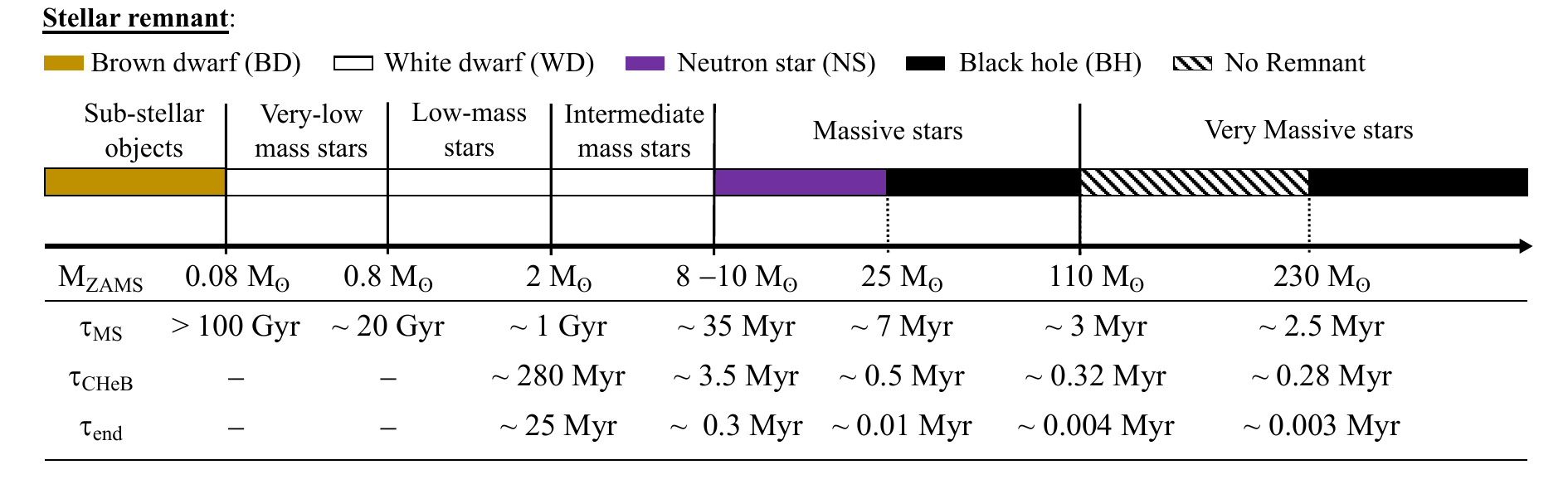}
\caption{This figure shows the stars' classification depending on the initial mass ($\Mzams$).
Different colours indicate the final remnant of the evolution.
$\tau_\mathrm{MS}$ and $\tau_\mathrm{CHeB}$ are the lifetimes of the \ac{MS} and \ac{CHeB}, respectively. $\tau_\mathrm{end}$ is the timescale of the last phases, i.e. from the end of the \ac{CHeB} to the formation of the \ac{CO}.
The transition masses and the timescales can change for different metallicities. The limits for very low-mass, low-mass, and intermediate-mass stars are taken from \citet{2022A&A...665A.126N}. Transition mass between the formation of \acp{NS} and \acp{BH} is taken from \citet{2012ApJ...749...91F}. Limits of the no remnant zone are taken from \citet{2017MNRAS.470.4739S}, with $Z = 2 \times 10^{-4}$.}
    \label{fig:mass_range}
\end{figure*}

At the low end of the mass range ($\lesssim 0.07 - 0.08~\Ms$), there are Brown dwarfs, i.e. sub-stellar objects that do not reach the conditions for burning hydrogen stably. Such objects burn deuterium in the early stages and, later, are maintained by the internal heat content \citep{2000ARA&A..38..337C}.
Stars with initial masses from $\sim 0.08~\Ms$ to $\sim 0.8~\Ms$ are called very low-mass stars. 
They are massive enough to reach the thermal equilibrium and stably burn hydrogen. 
The \ac{MS} lifetime of these stars is longer than the age of the universe. 
After \ac{MS}, the helium core contracts until gravity is balanced by the degeneracy pressure of the electrons. Finally, they die as helium \acp{WD} after the ejection of the envelope. 
In the mass range from $\sim 0.8~\Ms$ to $\sim 2~\Ms$, there are the low-mass stars.
They are massive enough to ignite helium under strongly degenerate conditions (usually referred to as He-flash). 
After the end of \ac{CHeB}, the carbon-oxygen core begins to contract, and the electrons become degenerate. Ultimately, these stars die as carbon-oxygen \acp{WD} \citep{2015A&A...578A.117C}. 
Very low-mass and low-mass stars (with an initial mass $\lesssim 2~\Ms$) are the most common stars in the universe \citep{1955ApJ...121..161S, 2001MNRAS.322..231K}.
These stars may live from a few to hundreds of Gyr in the \ac{MS} phase. 
Low-mass stars can be divided into sub-groups depending on their structure. Stars with a mass below $\sim 1$ $\Ms$ have radiative cores and convective envelopes (like our Sun). On the other hand, low-mass stars with a mass greater than $\sim 1$ $\Ms$ already develop a convective core and a radiative envelope \citep{2022A&A...665A.126N}. 

Stars in the mass range between $\sim 2$ $\Ms$ and $\sim 8$ $\Ms$ are called intermediate-mass stars. They are massive enough to ignite helium in non-degenerate cores (i.e. avoiding the He-flash).
These stars may live from about 0.1 to a few Gyrs. During the hydrogen-burning phase, they have a well-formed convective core and a radiative envelope. 
After the \ac{CHeB}, both low-mass and intermediate-mass stars evolve by contracting their cores. The contraction continues until the pressure by electron degeneracy is high enough to sustain the core. These stars are not massive enough to ignite carbon in their cores and, over time, winds completely peel off the stellar envelope, exposing the degenerate core. Such stars produce planetary nebulae and end their lives as \acp{WD}.

Between about 8~$\Ms$ and 10~$\Ms$, stars are massive enough to ignite carbon in their cores. The final fate of such stars depends on the mass of the neon-oxygen-magnesium core after the carbon-burning phase. If the core mass remains below the critical limit of $\sim 1.37 ~\Ms$ \citep{1984ApJ...277..791N}, stars will end their life as neon-oxygen \acp{WD}.
Electron capture reactions lead to core collapse if the core mass grows to $1.37 ~\Ms$ (due to the burning of the helium in the shell above the core). 
Such stars end their life with an electron-capture supernova, and a low mass \ac{NS} is the outcome of the evolution (see Section~\ref{subsec:fate} for more details). 
If the star builds a core more massive than $\sim 1.37 ~ \Ms$, it ignites neon and proceeds through the advanced burning phases, like massive stars (see below). 
Such stars will end their lives as \acp{NS}. 
These limits for intermediate-mass stars depend on the stellar properties (metallicity, rotation, extra mixing, etc.) and on the numerical details of the models \citep{2007A&A...476..893S, 2008ApJ...675..614P, 2013sse..book.....K}.

Stars with masses greater than about 10~$\Ms$ are called massive stars. 
They are the most luminous and hottest stars in the Universe and the shortest-lived (from hundreds to a few million years). 
After the core carbon burning, these stars sequentially proceed to advanced burning stages in non-degenerate conditions (neon-, oxygen- and silicon-burning) up to the formation of iron-group elements. 
Their final configuration (usually referred to as the onion skin model) consists of a series of shells composed of heavier elements going from the surface toward the centre, in which there is the iron core.
Massive stars end their life with a \ac{CC} supernova event or with a direct collapse without supernova explosion (see Section~\ref{subsec:fate} for more details).
In the first approximation, massive stars with an initial mass below $\sim$20 - 25~$\Ms$ end their life as \acp{NS}, while those above $\sim 25~\Ms$ form \acp{BH}. 
Very massive stars, with an initial mass greater than about 110~$\Ms$, could leave no remnant if they enter the \ac{PI} regime. \ac{PI} could trigger an explosive ignition of oxygen that can completely destroy the star, ultimately resulting in a powerful explosion known as \ac{PISN} \citep[][]{2017ApJ...836..244W}. 
From the theoretical point of view, the upper mass that a star can have is still a debated problem. However, observations of the R136 stellar cluster in the Large Magellanic Cloud have found stars with a mass of $\sim300~\Ms$ \citep{2010MNRAS.408..731C}.

Figure~\ref{fig:mass_range} shows a summary of the stars' classification presented so far. It is worth stressing again that such mass limits do not depend solely on the initial mass. Other factors, such as stellar metallicity and rotation, can play a role. 
Moreover, the theoretical uncertainties on the stellar processes, such as internal mixing and stellar winds, still make these limits quite uncertain, particularly for massive stars, where small changes in the input physics can significantly impact their stellar properties, evolution, and final fate.
Multiple studies have demonstrated that different models of massive stars, based on varying physical processes like mixing, mass loss rates, and rotation, can lead to very different evolutions and outcomes \citep[see][]{2013A&A...560A..16M, 2015MNRAS.447.3115J, 2020MNRAS.497.4549A}.

\subsection{The HR diagram}
\label{subsec:HR}

The \ac{HR} diagram is one of the main tools astronomers use to characterise stars. 
It shows the relation between the stellar effective temperature ($T_{eff}$, i.e. the temperature at the stellar photosphere) and the luminosity\footnote{The luminosity is the total amount of radiation emitted per unit of time.} ($L$) of stars. 
The effective temperature is strictly related to the luminosity and the radius ($R$) of the star through the following relation:
\begin{equation}
    L = 4\pi R^2 \sigma T_{eff}^4,
    \label{eq:teff}
\end{equation}
where $\sigma$ is the Stefan-Boltzmann constant of radiation. 

The diagram is named after \citet{1911POPot..22A...1H} and \citet{1914PA.....22..275R}, who discovered that stars do not randomly scatter along this diagram but fall into several groups.
This remarkable discovery strongly affected the development of modern astronomy and astrophysics. 
The \ac{HR} diagram is also referred to as the ``colour-magnitude diagram'' since the stellar colour (or spectral type) is related to $T_{eff}$ (stars look blue for $T_{eff} > 10^4$ K, yellow for $10^4 < T_{eff} < 5\times 10^3$ K or red for $T_{eff} < 5 \times 10^3$ K), and the absolute magnitude is related to the luminosity.
In this Part, we call the \ac{HR} diagram the plot of luminosity versus effective temperature. 

The most striking feature of the \ac{HR} diagram is that the position of stars (i.e. stars' luminosity and effective temperature) depends on their evolutionary stage.
Most of the stars that we observe lie on the \ac{MS}, a narrow diagonal band in the diagram in which they are converting hydrogen into helium. The locus on the diagram where stars ignite thermonuclear reactions and enter \ac{MS} is called \ac{ZAMS}. 
This is assumed as the beginning of evolution, and the mass in this stage ($\Mzams$) is the initial mass. 

\begin{figure*}
	\includegraphics[width=\columnwidth]{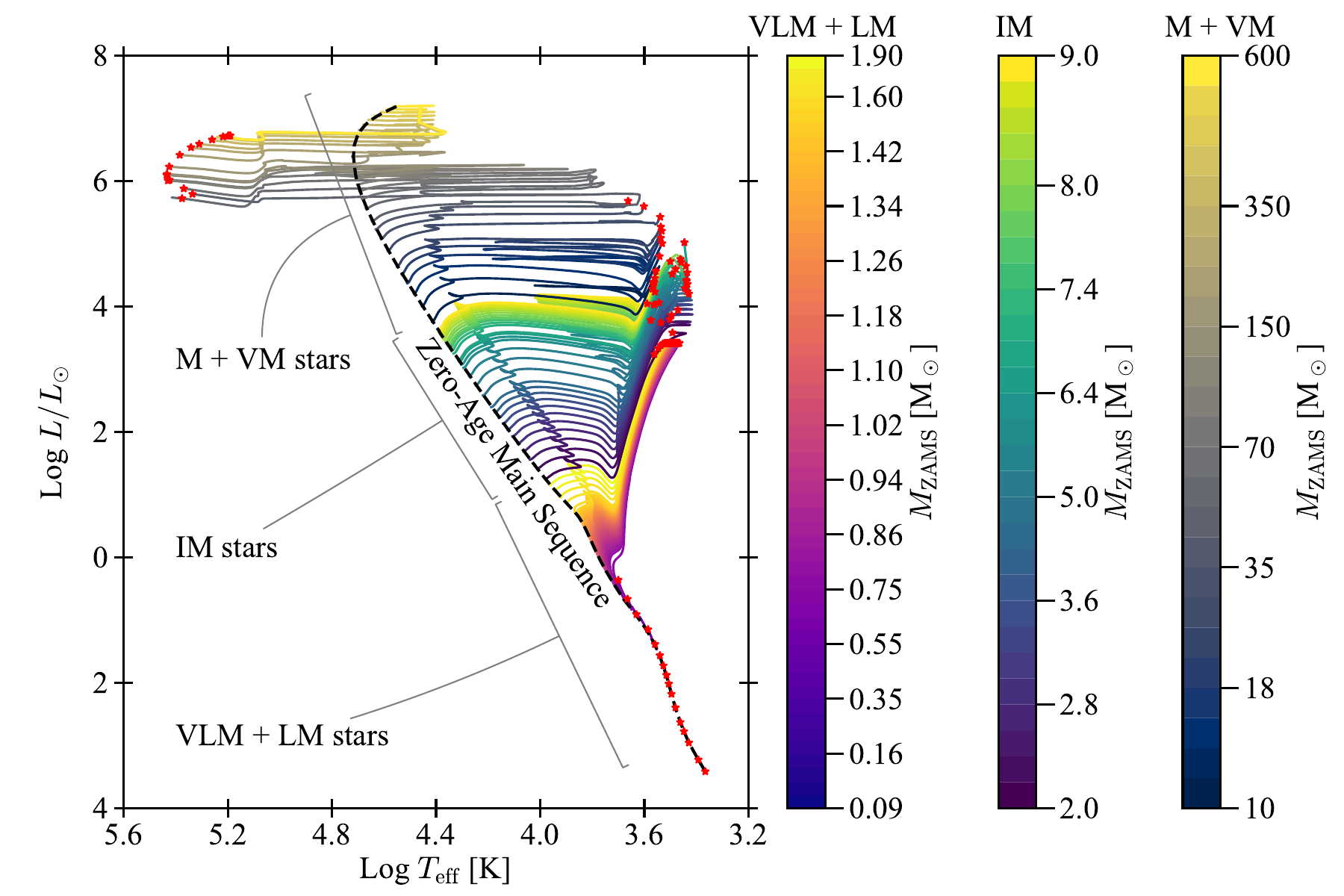}
    \caption{The plot shows the \ac{HR} diagram of a set of tracks with $Z=0.014$ computed with \textsc{parsec} stellar evolutionary code \citep{2021MNRAS.501.4514C, 2022A&A...665A.126N}. The "plasma" colour map indicates the $\Mzams$ of very low- and low-mass (VLM + LM) stars. The "viridis" map indicates the intermediate-mass (IM) stars, while the "cividis" map indicates the massive and very massive (M + VM) stars. The dashed diagonal black line shows the star's location at the \ac{ZAMS}. The small red stars show the final position of the star at the end of the evolution (for stars with $\Mzams > 2~\Ms$), at the tip of the \ac{RGB} (for stars with $ 0.7 < \Mzams/\Ms < 2$), or after 30~Gyr (for stars with $\Mzams < 0.70~\Ms$). The \ac{CHeB} (or horizontal-branch) phase of stars with $ 0.7 < \Mzams/\Ms < 2$ is not shown for clarity purposes.
    }
    \label{fig:HR}
\end{figure*}

Figure~\ref{fig:HR} shows the position of the \ac{ZAMS} in the diagram for stellar models computed with different initial masses. At the \ac{ZAMS}, stars are almost totally homogeneous, and their luminosity is proportional to their mass ($L \propto \Mzams^{3.5}$).

Stars with a mass $\Mzams \lesssim 0.75~\Ms$ barely evolve in the \ac{HR} diagram, and they remain near the \ac{ZAMS} for a time larger than the age of the Universe (the Hubble time).
Stars with $\Mzams \gtrsim 0.8~\Ms$ are massive enough to finish the \ac{MS} phase before one Hubble time. 
When hydrogen is depleted, the core contracts and the star's envelope expands at almost constant luminosity. In the \ac{HR} diagram, the star moves toward lower temperatures (because of the relation shown in Equation~\ref{eq:teff}) and becomes a \ac{RG}. 
The locus where \ac{RG} stars lie in the \ac{HR} diagram is called \ac{RGB}. In this phase, stars burn hydrogen in a shell above the helium core and start becoming more and more luminous (ascending the branch). 
During the \ac{RGB} phase, stars develop a deep convective envelope which dredges up the hydrogen-burning products (mainly nitrogen, but also isotopes of carbon and oxygen), which enrich the stellar surface (this process is called \emph{first dredge-up}).

After the \ac{RGB} phase, stars massive enough ignite helium and enter the \ac{CHeB} phase. 
The new energy source provided by the nuclear burning of helium leads the core to expand and the envelope to contract. The effective temperature increases again, and the star moves to the bluer part of the \ac{HR} diagram.
During the \ac{CHeB} phase, intermediate-mass and massive stars may draw a loop in the \ac{HR} diagram, usually referred to as the \textit{blue loop}\footnote{The blue loop may cross an instability strip, in which stars begin to pulsate as classical Cepheids \citep[see][]{1993ApJS...86..541C}. These stars are important cosmic distance indicators due to the correlation between pulsation period and luminosity \citep[see][]{2001ApJ...553...47F, 2021ApJ...908L...6R}.}. 
After \ac{CHeB}, the core contracts, and the envelope expands again and moves toward the red part of the \ac{HR} diagram again. 

Low-mass and intermediate-mass stars move to the \ac{AGB} phase. 
\ac{AGB} stars move to the reddest and most luminous part of the diagram, running near and almost parallel to the Hayashi line\footnote{The Hayashi line is the locus of fully convective stars in the \ac{HR} diagram \citep{2013sse..book.....K}. It is located at very low temperatures and defines a forbidden region (on its right) in the \ac{HR} diagram for the stars.} \citep{1961PASJ...13..450H}.
In the early-\ac{AGB} phase, stars burn helium in a shell above the carbon-oxygen core and undergo another dredge-up episode (called \emph{second dredge-up}).
After this stage, stars enter the \ac{AGB} phase, which is characterised by two burning shells above the carbon-oxygen core (H- and He- shell burning at the same time).
The double shell burning phase is thermally unstable and leads to a cyclic series of thermal pulses, which increase and decrease the star's radius, luminosity and effective temperature. 
This phase is called \ac{TP-AGB}. 
After each pulse, the convective envelope extends deeper into the star and eventually extracts products of the He-shell burning, which enrich the stellar surface (a process known as \emph{third dredge-up}).
\ac{AGB} stars suffer from powerful winds driven by dust and are one of the polluters which enrich the interstellar medium with heavy elements.

Stars with $8~\Ms \lesssim \Mzams \lesssim 10~\Ms$ evolve similarly to the most massive intermediate-mass stars. In the advanced phases, they move to the \ac{AGB}, but being more luminous, are labelled as super-\ac{AGB} stars \citep{1996ApJ...460..489R}.
They can ignite carbon in their core and die as \acp{WD} or \acp{NS} depending on the oxygen-neon-magnesium core mass (see Section~\ref{subsec:mass_ranges} and \ref{subsec:fate}).

Massive and very massive stars (with $\Mzams \gtrsim 10~\Ms$) may have a very different evolution and fate, depending on their initial properties, and the physics adopted \citep{1998A&ARv...9...63V, 2012ARA&A..50..107L, 2015A&A...575A..60M}. 
Stars with a mass $\Mzams \lesssim 35~\Ms$ evolve similarly to most massive intermediate-mass stars, with the difference that after the \ac{CHeB}, they move to lower temperatures, becoming \ac{RSG}. 
After carbon ignition, the evolution becomes so fast that stars no longer move in the \ac{HR} diagram and explode as \acp{RSG}.
Although the core evolution is similar, stars with $\Mzams \gtrsim 35~\Ms$ may end their lives differently depending on the metallicity. 
This is because the main driver of the evolution of such stars is the mass loss via stellar winds. Winds may be strong due to the high luminosity of such stars, but they also are very sensitive to the stellar metal content (see Section~\ref{subsec:winds}).
Metal-poor massive stars (Z~$\leq 0.001$) do not experience strong winds during their life (they retain most of their envelope), and after the \ac{CHeB} phase, they evolve and explode as \ac{RSG} stars. 
On the other hand, massive stars with a higher metal content (Z~$> 0.005$) may lose the largest part of their H-rich envelope during the \ac{MS} and \ac{CHeB} phases. 
Such stripped stars move toward the hotter (blue) part of the \ac{HR} diagram, becoming \ac{WR} stars\footnote{\ac{WR} are named after \citet{1867CRAS...65..292W}, who discovered a few stars in the Cygnus constellation with the peculiar broad emission lines.}. 
\acp{WR} are stars with stellar surfaces impoverished in hydrogen. 
From the observational point of view, they are divided into several groups depending on their spectrum. 
The two main groups are the WN stars which show the H-burning products (mainly helium and nitrogen) at their surface, and the WC stars, which show the He-burning products (carbon and oxygen). 
\ac{WR} stars are considered a possible final evolutionary stage of massive stars, i.e. the Conti scenario \citep{1975MSRSL...9..193C, 1987ARA&A..25..113A, 1994ARA&A..32..227M, 2007ARA&A..45..177C,  2022ApJ...931..157A}.
These stars are very luminous, hot and compact in radius. Their relatively small radius could play a role in the binary evolution (\acp{WR} could avoid episodes of mass transfer at variance with the very expanded \acp{RSG}), but also in the final \ac{CC} explosion \citep[see Section~\ref{subsec:fate},][]{2018MNRAS.476.2366F}. Figure~\ref{fig:HR} shows the position of massive stars at the end of the oxygen-burning phase (red stars) for models with an initial metallicity of Z~=~0.014 and no rotation.

Observations of Galactic massive stars revealed the existence of a `forbidden' region in which few stars are observed in the \ac{HR} diagram, the so-called Humphreys-Davidson limit \citep{1994PASP..106.1025H}. 
Theoretically, stars near this region are expected to approach the Eddington luminosity\footnote{The Eddington limit is the luminosity at which radiation pressure balances the star's surface gravity of the star \citep{1926ics..book.....E}. 
Generally, the ratio of radiative luminosity to the Eddington luminosity, i.e. the Eddington factor, is defined as $\Gamma_{E} = L \kappa_\mathrm{e} / 4 \pi c G M$, where $L$ is the stellar luminosity, $\kappa_\mathrm{e}$ is the electron scattering opacity, $c$ is the speed of light, $G$ is the gravity constant, and $M$ is the stellar mass. When the star is near the limit, $\Gamma_{E}$ approaches 1.} and experiences powerful winds (see Section~\ref{subsec:winds} for more details).

% ------------------------------------------------------------------------------------

\section{Main physical processes that affect life and death of massive stars}
\label{sec:phys_processes}

Massive stars play a pivotal role in the history of the Universe.
They are principal actors in the galaxies' chemical and spectral evolution due to their strong radiation and chemical nucleosynthesis footprint. 
They are the progenitors of powerful supernov{\ae}, \acp{NS} and \acp{BH}. 
Moreover, massive stars enrich and perturb the interstellar medium with their strong winds and supernova ejections, possibly driving (or quenching) new star formation episodes.

The life of massive stars is shorter than that of low-mass and intermediate-mass stars (as seen in Section~\ref{subsec:mass_ranges}). 
Therefore, they are much rarer, and stellar models become even more important to study and understand the physics at play.
Although state-of-the-art models of massive stars have reached a very high level of detail, there is still a significant degree of uncertainty regarding their evolutionary path and final fate. 
In particular, attention must be paid to every proposed relation between the initial mass ($\Mzams$) and the final \ac{CO} mass  ($M_\mathrm{rem}$), because tiny changes in the adopted physics may lead to big differences at the end of the evolution.
The main processes that affect massive stars' evolution are convection, stellar winds, rotation, magnetic fields, and the physics of the final explosion. These processes are reviewed in the following Section. 

The stellar initial composition (i.e. the metallicity) also plays a role, but it is mostly related to the stellar winds (see Section~\ref{subsec:winds}). 
Generally, at \ac{ZAMS}, stars with a low metal content are more compact than their metal-rich counterparts (mainly because of their lower opacity in the envelope), and they build slightly larger cores at the end of the \ac{MS} and \ac{CHeB} phases. 
The first stars in the Universe (namely Population III or Pop III) are particularly interesting. They are almost metal-free stars ($Z\sim0$), enriched only by the Big Bang nucleosynthesis. 
At variance with more metal-rich stars, simulations suggest that Pop III stars form with a top-heavy initial mass function, predicting that massive Pop III stars were not rare in the early Universe \citep{2013RPPh...76k2901B}.
The predicted high number of massive stars in the early Universe, in addition to their properties -- during the \ac{MS}, they are strong emitters of UV radiation, because very hot and luminous -- elect them between the major actors in the reionisation of the Universe \citep{2002A&A...382...28S, 2013RPPh...76k2901B, 2023ARA&A..61...65K}. 
Moreover, various studies suggest that Pop III stars may be the progenitors of the \ac{BBH} merger events that we observe today \citep{2020ApJ...903L..40L, 2021MNRAS.501.4514C, 2021MNRAS.502L..40F, 2021MNRAS.505.2170T, 2021MNRAS.504L..28K, 2023MNRAS.525.2891C, 2023MNRAS.524..307S}.

It is worth recalling that most massive stars live in binary \citep{2012Sci...337..444S} or even in multiple systems \citep{2017ApJS..230...15M}. Binary evolutionary processes may have a strong impact on the lives of massive stars and their final fate \citep{2009A&A...497..243D, 2012ARA&A..50..107L, 2016A&A...588A..50M, 2017PASA...34...58E, 2022ARA&A..60..455E}. Therefore, treating them as individual stars might give a partial picture. 
Binary processes, such as Roche lobe overflow, common envelope or even stellar collision/merger, could form peculiar stellar objects, such as stripped stars or ``puffy" stars. 
The former has a higher core-to-envelope mass ratio than standard stars with the same initial conditions and at the same evolutionary phases. The latter has a core-to-envelope mass ratio lower than standard stars (i.e., smaller cores and oversized envelopes). 
Such peculiar stars have a different evolutionary path and final fate than their standard counterparts \citep{2020A&A...638A..55K, 2020ApJ...904L..13R, 2021A&A...656A..58L, 2022MNRAS.tmp.3494B, 2022MNRAS.516.1072C}. Binary processes and their impact on the final \acp{BH} masses are reviewed in Part \ref{part:binaries} of this book.

\subsection{Convection and overshooting}

Convection is a physical process that transports the excess energy from an inner (hotter) region to an outer (cooler) region in stars. 
When a region becomes thermally unstable\footnote{A region of a star is considered thermally stable if a perturbed element (eddy or bubble) of material experiences a restoring force that brings it back to the starting position. On the contrary, a region is said to be unstable if the perturbed element experiences a force which drives it away from the starting position.}, convection arises, and large eddies of material rise and sink, mixing chemical elements and efficiently transporting energy. 
In the cores of massive stars, convection provides fresh fuel to the active burning layers and brings the ashes of nuclear reactions outward. 
It is a key process that regulates the lifetimes of each burning phase, affecting the final pre-supernova structure and explosion likelihood 
\citep[see][and references therein]{2012ARA&A..50..107L, 2014ApJ...783...10S, 2019A&A...625A.132S, 2020MNRAS.496.1967K}.
Convection is still one of the open problems of stellar physics, and its treatment in stellar simulations is still under debate \citep{1987A&A...188...49R, 1991A&A...252..179Z, 2012ARA&A..50..107L, 2017RSOS....470192S, 2020MNRAS.496.1967K, 2022A&A...659A.193A}. 

In stellar models, it is difficult to achieve a comprehensive treatment of convection. It is a very complex physical process that requires full 3-D hydrodynamical simulations to be addressed. 
In the past decades, several steps forward have been done with 2-D and 3-D hydrodynamical simulations \citep[e.g.][]{1996A&A...313..497F, 2006ApJ...642.1057H, 2007ApJ...667..448M, 2013A&A...557A..26M, 2018arXiv181004659A, 2019ApJ...882...18A, 2022A&A...659A.193A, 2022MNRAS.515.4013R}, but such models are still limited in the time and space domains to be consistently applied in stellar evolutionary models. 
Typical timescales of stellar evolution are several orders of magnitude larger than those resolved by 3-D simulations \citep[e.g.][]{2017MNRAS.471..279C}.

The most widely used theory to treat convective zones in 1-D stellar evolutionary codes is the \ac{MLT} \citep{1958ZA.....46..108B}.
\ac{MLT} provides an approximate methodology for treating convection locally and in 1-D, assuming hydrostatic equilibrium, constant energy flux, and local thermodynamic equilibrium. There are more refined theories for treating convection \citep[such as][]{1978A&A....65..281R, 1991ApJ...370..295C}, but they are less common in stellar computations.
The \ac{MLT} framework is characterised by the mixing length distance, $l_\mathrm{MLT}$, which is the distance covered by the convective bubble before it dissolves into its surroundings. 
It is defined as $l_\mathrm{MLT} = \alpha_\mathrm{MLT} H_\mathrm{P}$, where $\alpha_\mathrm{MLT}$ is the efficiency parameter, and $H_\mathrm{P}$ is the pressure scale height. 
In stellar models, the $\alpha_\mathrm{MLT}$ parameter is usually calibrated to match the Sun's luminosity, radius, and age, and its value may range between 1 and 2.5 \citep{2008Ap&SS.316...43E, 2012MNRAS.427..127B, 2015A&A...573A..89M}. 
Generally, once calibrated, $\alpha_\mathrm{MLT}$ is assumed to be constant for all masses and not dependent on other quantities, such as metallicity.
Whether $\alpha_\mathrm{MLT}$ may vary with stellar parameters is still an open debate; some authors suggest that it may depend on the effective temperature or metallicity \citep[e.g.][]{2015A&A...573A..89M, 2018ApJ...856...10J, 2018ApJ...858...28V}. 
However, its calibration from stars other than Sun could be biased since the $\alpha_\mathrm{MLT}$ uncertainties are also correlated to uncertainties from other sources, such as the stellar surface boundary condition \citep[see][]{2018ApJ...860..131C}.

The success of \ac{MLT} in describing the unstable regions is due to its simple implementation and good empirical agreement with the observations once properly calibrated.
In 1-D stellar models, \ac{MLT} must be coupled with a criterion to define the convectively unstable borders and a prescription to include the convective boundary mixing, also known as overshooting. We discuss both in the following.
For further details on convection and convection treatment in stellar models, we suggest \citet{2004cgps.book.....W}, \citet{2009pfer.book.....M}, and \citet{2013sse..book.....K}.

\subsubsection{Defining the convective borders}
\label{subsubsec:conv}

Convection may arise in different stellar regions and at different evolutionary phases, depending on the local physical condition of the matter. 
The two most used prescriptions to define the boundaries of the unstable region are the \textit{Schwarzschild} \citep{1958ses..book.....S} and \textit{Ledoux} \citep{1947ApJ...105..305L} criteria.
The former criterion states that the region is stable if
\begin{equation}
    \nabla_\mathrm{rad} < \nabla_\mathrm{ad},
    \label{eq:schw_crit}
\end{equation}
where $\nabla_\mathrm{rad}$ is the temperature gradient in the case of stable radiative energy transport, while $\nabla_\mathrm{ad}$ is the adiabatic temperature gradient for the unstable case.

The \textit{Ledoux} criterion states that matter is stable when
\begin{equation}
    \nabla_\mathrm{rad} < \nabla_\mathrm{ad} + \frac{\varphi}{\delta}\nabla_\mu,
    \label{eq:led_crit}
\end{equation}
where $\varphi =\left(\partial\ln\rho / \partial\ln \mu \right)_{P,T}$ and $\delta = - \left(\partial\ln\rho / \partial\ln T \right)_{P,\mu}$
are thermodynamical derivatives and $\nabla_\mu = \partial\ln\mu / \partial\ln P$ is the molecular weight gradient. For a perfect gas, $\varphi = \delta = 1$.
In the case of a chemical homogeneous region, $\nabla_\mu = 0$, the two criteria are identical.
The two temperature gradients are defined as follows
\begin{align}
    \nabla_\mathrm{rad} &= \frac{3}{16 \pi a c G}\frac{\kappa L P}{M T^4} \\
    \nabla_\mathrm{ad} &= \frac{P \delta}{C_P \rho T}, 
    \label{eq:grad_def}
\end{align}
where $a$ is the radiation constant, $c$ is the speed of light, $G$ the gravitational constant. The other variables are the local thermodynamical quantities of a shell which encloses a mass $M$, where $\kappa$ is the opacity, $L$ is the luminosity, $P$ the pressure, $T$ the temperature, $\rho$ the density, $C_P$ is the specific heat at constant pressure.
\begin{figure}
	\includegraphics[width=0.5\columnwidth]{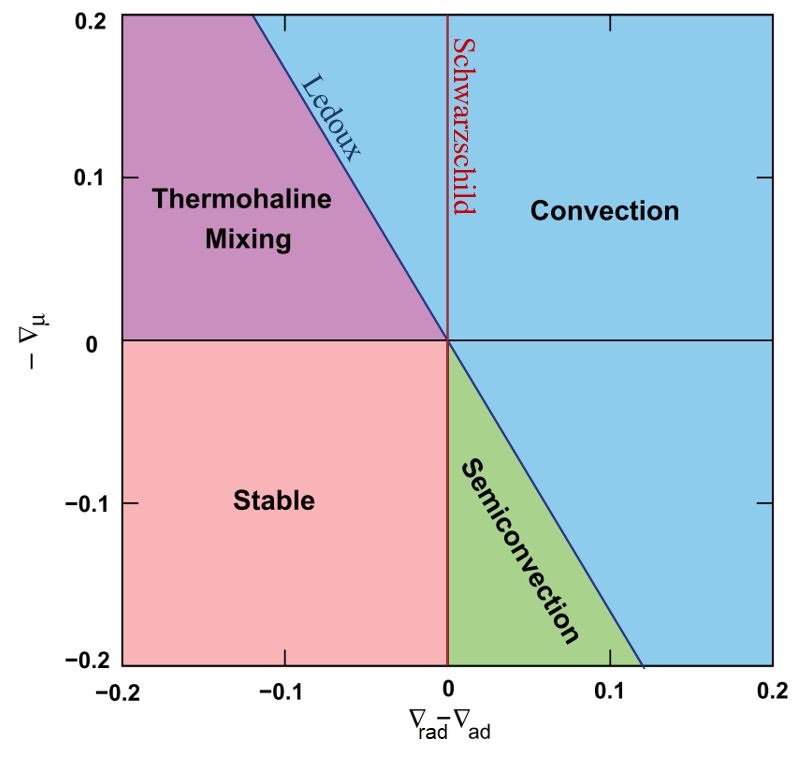}
    \caption{This plot shows the main stability criteria for stellar models (assuming a perfect gas).
    The red vertical line indicates the \textit{Schwarzschild} one, while the blue line indicates the \textit{Ledoux} one. 
    Coloured areas indicate the expected mixing processes. The pink area indicates the region where the material is stable, and no mixing occurs. The blue area indicates the region where the material is unstable, and convection occurs.
    The green area indicates where semiconvection occurs. The magenta indicates where thermohaline mixing can occur.
    % \textbf{
    Figure adapted from \citep{2014PASA...31...30K} 
    % (THIS FIGURE NEEDS COPYRIGHTS!)}
    .
    }
    \label{fig:conv}
\end{figure}

Figure~\ref{fig:conv} shows a way to visualise various mixing processes that may occur in stars. 
The regions of the star in which $\nabla_\mathrm{rad} > \nabla_\mathrm{ad} + \frac{\varphi}{\delta}\nabla_\mu$, are fully convective (upper right part of the figure), while regions which satisfy the \textit{Schwarzschild} criterion (i.e., $\nabla_\mathrm{rad} < \nabla_\mathrm{ad}$) are stable, and the energy is transported radiatively (lower left part of the figure).
When matter is stable according to the \textit{Ledoux} criterion but unstable according to the \textit{Schwarzschild} criterion (i.e., $\nabla_\mathrm{ad} < \nabla_\mathrm{rad} < \nabla_\mathrm{ad} + \frac{\varphi}{\delta}\nabla_\mu$), semi-convection instability occurs (green lower middle part of the figure). 
Semi-convection may occur during the \ac{MS} of massive stars, in which the core progressively retreats, leaving a chemical profile above it. 
In such regions, the radiative gradient increases and becomes larger than the adiabatic one, but the positive molecular gradient ($\nabla_{\mu} > 0$, i.e. $\mu$ increases toward the centre) stabilises the layer. 
That layer becomes vibrationally unstable, and the plasma starts to oscillate around an equilibrium point, producing slow mixing \citep{1958ApJ...128..348S, 1966PASJ...18..374K,1991A&A...252..669L, 2018AnRFM..50..275G, 2022ApJ...928L..10A}.
In 1-D stellar models, semi-convection can be treated as a diffusive process \citep{1985A&A...145..179L}, with a typical timescale similar to the thermal one, which is longer than the convection timescale but shorter than the nuclear one. 
In long phases such as \ac{MS}, mixing by semiconvection does not affect the evolution much, and models computed with the \textit{Ledoux} criterion are similar to those computed with the \textit{Schwarzschild} one \citep{1985A&A...145..179L}. 
Studies of convection with 3-D simulations show that the semi-convection oscillations produce a mixing which moves the unstable boundary to the one defined by the \textit{Schwarzschild} criterion \citep{1966PASJ...18..374K, 2019ApJ...882...18A, 2022ApJ...928L..10A}.
On the other hand, semi-convection could play a role in post-\ac{MS} phases, affecting, for instance, the helium core size at the end of \ac{CHeB} \citep[see][and references therein]{1986ARA&A..24..329C, 2012ARA&A..50..107L}. However, the possible interplay of semi-convection with other processes, such as overshooting, remains a matter of discussion \citep{2013sse..book.....K}.

In regions where matter is stable according to the \textit{Schwarzschild} criterion, but there is a $\mu$ inversion (when $\nabla_{\mu} < 0$), thermohaline mixing occurs (upper left part of the figure).
\citet{1972ApJ...172..165U} developed a theory of thermohaline mixing, which was then generalised by \citet{1980A&A....91..175K}. Later, \citet{2006Sci...314.1580E} stressed the importance of thermohaline mixing in RGB, discovering that the hydrogen-burning shell can generate a $\mu$ inversion that can trigger the mixing \citep[see also][for further details]{2007A&A...467L..15C, 2014PASA...31...30K}. 
Thermohaline mixing could be an important process to be included in binary evolutionary models if the accretor star gains helium-rich matter from the donor star \citep{2001A&A...369..939W}.

To conclude, there is still an open debate in the stellar community on what criterion should be used.
In the literature, there exist sets of stellar models computed either with the \textit{Schwarzschild} criterion \citep[e.g.][]{2012A&A...537A.146E, 2018MNRAS.480..538R, 2018ApJ...856..125H, 2019A&A...631A..77A, 2022A&A...665A.126N}, or with the \textit{Ledoux} one \citep[e.g.][]{2000ApJ...528..368H, 2011A&A...530A.115B, 2016ApJ...823..102C, 2018ApJS..237...13L, 2022A&A...658A.125S}.

\subsubsection{Convective boundary mixing or overshooting}
\label{subsubsec:overshoot}

Several studies have found that the mixing provided only by the convectively unstable regions was insufficient to reproduce stellar observations, and an additional source of mixing was needed beyond the convective limit \citep[e.g.][]{1975A&A....40..303M, 1981A&A...102...25B, 1991A&A...252..179Z, 2017ApJ...841...69R, 2019ApJ...876..134C, 2021RvMP...93a5001A}. This led to the concept of overshooting \citep[see][]{1991A&A...252..179Z}. 
3-D simulations show that this type of extra mixing, called convective boundary mixing \citep{2013ApJ...772...37D}, is generated from the interplay of different processes \citep[e.g.][]{2019ApJ...882...18A}. 

In the 1-D framework of the \ac{MLT}, only the radial velocity is considered, and the main idea behind the overshooting process is the following. When a rising (sinking) bubble reaches the convective border, the acceleration is zero, but the velocity is not. 
Therefore, the bubble overshoots the convective border due to its inertia and enters the stable radiative region, in which the restoring forces brake and stop the bubble. 
This process mixes matter between the convective and radiative regions.
The core overshooting process affects several properties of stars, such as lifetime, luminosity, and core mass. It affects not only the \ac{MS} but also post-\ac{MS} burning phases \citep{2007ApJ...667..448M, 2013ApJ...772...37D, 2017MNRAS.471..279C, 2022MNRAS.515.4013R}.

In 1-D stellar evolutionary models, overshooting is implemented in different ways, and the most common are described below.  
The first method is the so-called \emph{penetrative convection} scheme \citep{1965ApJ...142.1468S, 1973ApJ...184..191S, 1991A&A...252..179Z}, in which the convective border is moved by an overshooting distance, 
$d_\mathrm{ov} = \lambda_\mathrm{ov} H_\mathrm{P}$, where $\lambda_\mathrm{ov}$ is the overshooting parameter. The overshooting region is then treated as unstable, and the chemical elements are fully mixed. 
Some authors compute $d_\mathrm{ov}$ by integrating the path of convective bubbles across the unstable border \citep[called the ballistic approach, ][]{1975A&A....40..303M, 1981A&A...102...25B, 1991A&A...252..179Z}.

A second methodology is the exponential decaying overshooting in the diffusive scheme \citep{1997A&A...324L..81H}. Extra mixing is applied in the region above the convective border, with a diffusion coefficient ($D_\mathrm{ov}$) that exponentially decays. It is defined as
\begin{equation}
    D_\mathrm{ov} = D_\mathrm{0} \exp{ \left( 2\frac{r - r_0}{f_\mathrm{ov} H_\mathrm{P}}\right)}, 
\end{equation}
where $D_\mathrm{0}$ is the diffusion coefficient at the border and is calculated with the typical convective element velocities of \ac{MLT}, $r_0$ is the position of the convective border, $f_\mathrm{ov}$ is the parameter used to calibrate the mixing efficiency \citep{1997A&A...324L..81H}. This approach is based on hydrodynamical simulations by \citet{1996A&A...313..497F}.

A third approach assumes an overshooting which does not affect the thermal structure (i.e. the layer remains radiative), but the chemical elements are mixed instantaneously \citep{2004ApJ...612..168P, 2008ApJS..178...89D}. Even in this case, a parameterised overshooting distance is adopted, as in the first method.

In all methods, the free parameters, e.g. $\lambda_\mathrm{ov}$ and $f_\mathrm{ov}$, must be calibrated on observations. 
A good calibration of the internal stellar mixing in stellar models is fundamental because it has implications not only on the stellar properties during the \ac{MS} but also on the post-\ac{MS} evolution \citep{2020MNRAS.496.1967K, 2021A&A...655A..29J}.
For this purpose, several works have used different approaches, such as the position of the turn-off in stellar clusters, the stellar properties of wide eclipsing binaries, and asteroseismology measurements \citep[e.g.][]{2017ApJ...841...69R, 2019ApJ...876..134C, 2021A&A...655A..29J}.

Studies on wide binaries have shown a dependence of the overshooting efficiency with the stellar mass. In particular, there is a smooth transition between stars with radiative cores ($\Mzams \leq 1~\Ms$) that do not have core overshooting and stars with well-developed convective cores ($\Mzams \geq 1.5~\Ms$). 
For higher masses, it is usually assumed that the overshooting efficiency is constant with mass \citep[][]{2019ApJ...876..134C, 2019MNRAS.485.4641C}. 
However, the investigation is still ongoing, and other studies suggest an increasing efficiency of the overshooting for more massive stars \citep{2014A&A...570L..13C, 2020A&A...637A..60T, 2021A&A...648A.126M, 2021MNRAS.503.4208S}.

New interesting results have been found with asteroseismology, which is the study of stellar interiors through stellar oscillation \citep{2021RvMP...93a5001A}. 
Asteroseismic measurements allow a better constraining of the stellar internal mixing \citep{2021RvMP...93a5001A, 2021NatAs...5..715P, 2021A&A...655A..29J}, since stellar oscillations are sensitive to the inner properties of stars, such as the angular momentum distribution. 
However, there are still not many asteroseismic observations of massive stars, and the most recent studies include stars with $\Mzams < 25 ~\Ms$ \citep{2022MNRAS.513.3191S}.

On the modelling side, 3-D hydrodynamical simulations are taking several steps forward to improve the treatment of convective boundary mixing in different stellar phases, which could then be included in 1-D codes \citep[e.g.,][]{2021MNRAS.503.4208S}. 
Moreover, 3-D hydrodynamical simulations revealed that convective boundary mixing might excite internal gravity waves, which could propagate and dissipate in the radiative regions transporting angular momentum and contributing to mixing \citep{2017ApJ...848L...1R, 2021RvMP...93a5001A}.

Theoretically, overshooting should be applied to all convective regions. For instance, many works \citep[e.g.][]{1991A&A...244...95A, 2014MNRAS.445.4287T} have been considering the overshoot at the bottom of the convective envelopes of \ac{RGB} stars, which is called the convective envelope overshooting (or undershooting). 
Recently, it has been found that undershooting could play a role during \ac{CHeB} phases of metal-poor and very massive stars \citep{2021MNRAS.501.4514C, 2023ApJ...944...40V}. 
The undershooting may lead to dredge-up episodes, which extract mass from the helium or carbon-oxygen core, affecting the stellar global stability against the \ac{PI} and, thus, the likelihood of exploding as a \ac{PISN} (Section~\ref{subsec:fate}).

For reviews on convection in stellar interiors, we refer to \citet{2017RSOS....470192S, 2018AnRFM..50..275G, 2019ARA&A..57...35A, 2021FrASS...8...53E} and references therein.

% ------------------------------------------------------------------------------------

\subsection{Mass loss through stellar winds}
\label{subsec:winds}

Stellar winds are one of the key processes in the evolution of massive stars. 
They have an impact on the stellar structure, affecting the evolution and determine the final fate of stars \citep[e.g. ][]{1986ARA&A..24..329C, 1998A&ARv...9...63V, 2012ARA&A..50..107L, 2015A&A...575A..60M, 2017A&A...603A.118R} - which depends on their final pre-supernova internal structure configuration \citep{1999ApJ...522..413F, 2001ApJ...550..372F, 2003ApJ...591..288H, 2011ApJ...730...70O, 2012ApJ...749...91F, 2013ApJ...762..126O}. 
Mass loss due to stellar winds not only affects the compact remnants mass spectrum of single stars \citep{2010ApJ...714.1217B, 2012ApJ...749...91F, 2015MNRAS.451.4086S, 2020MNRAS.497.4549A}, 
but also plays a fundamental role in shaping the mass spectrum of binary populations that, in turn, affects the predicted populations of merging binary sources of \acp{GW} \citep{2013MNRAS.429.2298M, 2017MNRAS.470.4739S, 2018MNRAS.474.2959G, 2019MNRAS.485..889S, 2021hgwa.bookE..16M, 2022Galax..10...76S, 2023MNRAS.524..426I}.
Therefore, studying and improving our understanding of stellar winds is fundamental.
Winds also affect the position of the star in the \ac{HR} diagram, therefore, influencing the UV ionizing photon emission of such stars during their life. 
Moreover, winds influence the stellar chemical yields, which define how much different types of stars contribute to the metal enrichment of galaxies \citep[e.g.][]{2021A&A...650A.203G}. 

Stellar winds are a very efficient mass loss mechanism for massive stars with initial mass $\Mzams > 25 ~ \Ms$. This is related to their very high luminosity. 
Usually, stellar winds are not modelled on-the-fly in 1-D stellar evolutionary codes but are included with recipes obtained from observational results or theoretical studies. 
In the following, the most common wind treatments are briefly described. For a more comprehensive review of stellar winds of massive stars, we suggest the following references \citet{2008A&ARv..16..209P, 2017RSPTA.37560269V, 2021arXiv210908164V}.

The idea that radiation pressure could exert a net force against gravity on the outer layers of stars was established about a century ago by \citep{1918JRASC..12..357K, 1919ApJ....50..220S}. But in the late 1960s, with spectroscopic observations, it was discovered by \citet{1967ApJ...147.1017M} that massive stars suffer mass loss driven by line absorption winds. From such observations, they estimated that OB stars might suffer from winds between $10^{-4}$ and $10^{-6}~\Ms$/yr, which peel off the H-rich envelope during \ac{MS}. 
In the early 1970s, the first line-driven wind theory was developed by \citet{1970ApJ...159..879L} and later improved by \citet[][ the so-called CAK theory]{1975ApJ...195..157C}. 
% The main idea behind such stellar wind theories was that the interaction between radiation and opaque matter generates a net force against gravity in the outer skirt of stars. 
Within these theories, the rate of mass loss is uniquely determined by the luminosity, mass, and radius of the star. 
The CAK theory consists of a method to compute the resulting acceleration from spectral line opacities as a multiplier of the electron scattering acceleration. The force multiplier depends on the local opacity of the matter in which absorption occurs.
The difficult part was to compute all the spectral line contributions to calculate the mass loss rate. 
During the 1980s, more and more spectral lines were added to the CAK theory \citep{1982ApJ...259..282A, 1987A&A...173..293K}, and a mass loss metallicity dependence  -- $\dot{\mathrm{M}}(Z) \propto Z^{\alpha}$, with $0.5 < \alpha < 0.9$ -- was established. Stars with lower metallicity retain more mass during their life compared to their high metallicity counterparts and may form more massive \ac{CO} remnants.

Another approach was introduced by \citet{1985ApJ...288..679A}, which consists of tracking the path and the (multi-)scattering of photons through the stellar atmosphere with Monte Carlo techniques. This method was used by \citet{2000A&A...362..295V, 2001A&A...369..574V} to compute the mass loss of massive OB stars for a wide range of stellar parameters and metallicities. 
They derived a mass loss recipe that includes the so-called \emph{bi-stability jump} in $T_\mathrm{eff}$. Such a jump depends on the opacity bump resulting from the recombination of Fe IV to Fe III in the stellar atmospheres, at about 20,000 - 25,000 K, which can significantly increase the mass loss. 
Furthermore, \citet{2001A&A...369..574V} found that the metallicity exponent ($\alpha$) of the mass loss rate is constant in a wide range of metallicities $1/30 \leq Z/\Zs \leq 3$, with $\Zs = 0.019$ \citep{1989GeCoA..53..197A}. 
The general dependence of the mass loss rate on the metallicity is $\dot{M} \propto Z^{0.69}$ for $T_\mathrm{eff} \geq 25000$ K and $\dot{M} \propto Z^{0.64}$ for $T_\mathrm{eff} \leq 25000$ K. 
This mass loss recipe is only valid for hot-star winds and is adopted as a standard by many stellar evolutionary codes. 
The mass loss estimated by \citet{2001A&A...369..574V} is no longer valid for \acp{RSG} or when the star approaches the Eddington limit. 

For \ac{RSG} stars, the mass loss is less constrained than that of hot stars, and stellar modellers usually rely on empirical recipes. The most adopted prescription is given by \citet{1988A&AS...72..259D}, which only shows a dependence on the star's luminosity.
However, this recipe has been shown to overestimate mass loss in post-MS phases, and more updated empirical prescriptions -- calibrated on more recent observations -- can be found in the literature \citep[e.g.,][]{2020MNRAS.492.5994B}. 
Some studies suggest that the mass loss in \acp{RSG} could be driven by the radiation pressure on dust, analogously to \ac{AGB} stars \citep{2005A&A...438..273V}.
There is yet no well-validated theory yet to treat \acp{RSG} winds, recently \citet{2021A&A...646A.180K} proposed a model in which strong turbulence in the stellar atmosphere may trigger a mass loss able to reproduce the observed data without the inclusion of dust opacities.  

When massive stars become very luminous, they can approach the Eddington limit. 
Theoretical studies of radiative wind models have shown that stellar winds increase strongly when stars approach the Eddington limit \citep{2008A&A...482..945G, 2011A&A...531A.132V}. This is the case for \acp{LBV} and \acp{WR}. 

\acp{LBV} are objects that change their spectral type on time scales of years or decades \citep{1994PASP..106.1025H}. 
The reason for such changes of spectral type (i.e. in $T_\mathrm{eff}$) has not yet been fully understood. 
Some studies suggest that it could be related to the envelope inflation caused by the proximity to the Eddington limit \citep{2012A&A...538A..40G, 2017RSPTA.37560269V, 2021A&A...646A.180K, 2022A&A...668A..90A}. 
Recently, \citet{2021A&A...647A..99G} studied stellar wind mass loss feedback on the envelope structure in \ac{LBV} stars near the Eddington limit. They found that stars perform cycles of hot and cool envelope configurations. Such studies open the door to a consistent theoretical treatment of the \ac{LBV} phase in stellar evolution.

If the stellar winds are strong enough to remove large amounts of the envelope, the star may become a \ac{WR} \citep[Conti scenario,][]{1975MSRSL...9..193C, 1987ARA&A..25..113A, 1994ARA&A..32..227M, 2007ARA&A..45..177C,  2022ApJ...931..157A}.
\acp{WR} experience a high mass loss due to their high luminosity and to the proximity of the Fe-opacity bump (at a temperature of about 10$^5$ K) near the surface. 
When stellar winds are very effective, the H-rich envelope could be completely peeled off, and the star becomes a naked pure He star.
For \ac{WR} stars, the empirical mass loss prescription of \citet{2000A&A...360..227N} is widely adopted in stellar evolutionary codes. 
Recently, new physically consistent models of stellar winds have been computed by \citet{2020MNRAS.499..873S}. They found a complex luminosity-to-mass and Z-dependent breakdown of WR-type, predicting weaker mass loss winds for \acp{WR} and pure-He stars of few $\Ms$ than older mass loss prescriptions. 

In the case of very low metallicity, line-driven stellar winds are suppressed. Thus, it is usually assumed that Pop III very massive stars arrive at the final pre-supernova stage with about the initial mass, $\Mzams$.
However, it is known that very massive stars suffer possible mass loss driven by radial pulsations caused by nuclear reactions (the so-called $\epsilon$-mechanism) or by opacity ($\kappa$-mechanism). For more details, see \citet[][]{2001ApJ...550..890B}. 
In recent work, \citet{2020ApJ...902...81N} and later \citet{2023ApJ...944...40V} investigate the effect of pulsation driven winds on primordial stars, finding that the pulsation-driven mass loss could be substantial even for zero-metallicity stars.

Although there have been several recent improvements in stellar wind models -- thanks to improved methodologies and better calibration on new observations -- there are still uncertainties of a factor of 2 - 3 in mass loss recipes \citep{2021arXiv210908164V}, and the scaling with metallicity remains still debated \citep[e.g.][]{2022A&A...665A.133G, 2022MNRAS.511.5104M}.
These uncertainties are mainly due to wind clumpiness (i.e. the homogeneity of winds), the treatment of radiative transfer, and wind hydrodynamics in stellar atmospheres \citep[see][]{2021arXiv210908164V}. 
Such uncertainties may completely change the outcome of massive stars' evolution with $\Mzams > 60~\Ms$, which may end as \acp{BH} or be completely peeled off, ending as \acp{NS}. 

From the stellar modellers' point of view, the most widely adopted formalism for treating mass loss is a collection of prescriptions that uses \citet{2001A&A...369..574V} for OB stars, \citet{2000A&A...360..227N} for WR stars and \citet{1988A&AS...72..259D} for \acp{RSG} and all other stellar types \citep{2017PASA...34...58E, 2022ARA&A..60..455E}. 
Some authors slightly modified the list of prescriptions to include the Eddington factor dependency by \citet{2011A&A...531A.132V} \citep[e.g.,][]{2014MNRAS.445.4287T, 2015MNRAS.452.1068C}, or more recent \acp{WR} mass loss by \citet{2019A&A...621A..92S, 2020MNRAS.499..873S} \citep[e.g.,][]{2021MNRAS.501.4514C, 2021MNRAS.505.4874H}.

\begin{figure}
    \centering
	\includegraphics[width=0.7\columnwidth]{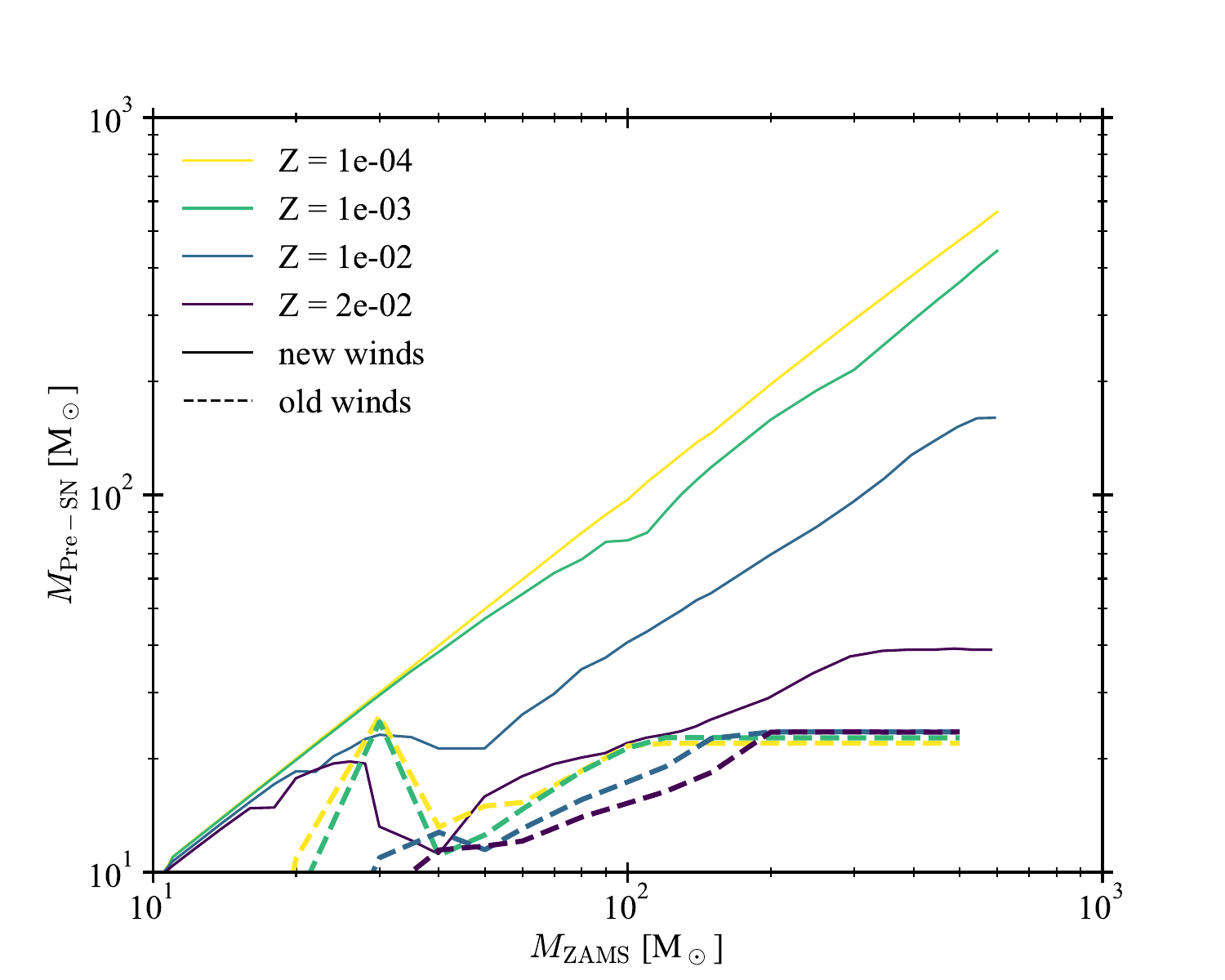}
        \caption{The plot shows the pre-supernova mass ($M_\mathrm{Pre-SN}$) vs the initial mass ($\Mzams$) for stars computed with different initial metal content (Z), in different colours, and with two different wind prescriptions. The continuous lines indicate the final masses of \textsc{parsec} tracks computed with new recipes for stellar winds \citep{2021MNRAS.501.4514C}. The dashed lines show pre-supernova masses computed with \textsc{sse} code \citep{2000MNRAS.315..543H}, which adopt older winds prescriptions for OB stars \citep{1990A&A...231..134N}, instead of \citet{2001A&A...369..574V}. A similar plot can be found in \citet{2010ApJ...714.1217B} and \citet{2021arXiv210908164V}.
    }
    \label{fig:winds}
\end{figure}

Figure \ref{fig:winds} shows a comparison between the final pre-supernova masses versus the initial mass of stars computed with the \textsc{sse/bse} populations-synthesis code \citep{2000MNRAS.315..543H} which adopts the \citet{1990A&A...231..134N} winds prescriptions for OB stars, and the \textsc{parsec} stellar evolutionary code \citep[see][ and references therein]{2015MNRAS.452.1068C, 2021MNRAS.501.4514C} which adopts prescriptions by \citet{2001A&A...369..574V} for stellar winds. 
The figure clearly shows the differences between the two wind prescriptions and how much the predicted pre-supernova mass could change depending on the stellar metallicity. In particular, stars computed with \textsc{sse} show an almost unique trend for all stars computed with different initial metallicity. On the other hand, pre-supernova models computed with \textsc{parsec} show a clear trend with metallicity. 
Stars with a low metal content retain almost all their initial mass, while models with higher metallicity have lower pre-supernova masses.

% ------------------------------------------------------------------------------------

\subsection{Rotation and magnetic fields}
\label{subsec:rotation}

\subsubsection{Stellar rotation}
\label{subsubsec:rot}
Rotation is a ubiquitous property of all stars (and objects) in the Universe. It has been found in stars of all spectral types \citep{2009LNP...765..207R}. Interestingly, a dichotomy between low-mass stars and more massive stars has been found. The former are generally slow rotators \citep{2014ApJS..211...24M}, while the latter show a wide spread of possible rotational velocities and could be fast rotators \citep{2007A&A...466..277H, 2013A&A...550A.109D, 2013A&A...560A..29R, 2020A&A...634A..51B, 2022A&A...667A.100S}.

As mentioned above, rotation is a fundamental parameter affecting the stellar evolutionary path. 
It affects stars during all phases and, especially in massive stars, rotation-induced instabilities combined with stellar winds can significantly change their evolution and death \citep{2000ApJ...528..368H, 2000A&A...361..159M, 2004A&A...425..649H, 2006A&A...460..199Y, 2006ApJ...637..914W, 2012A&A...542A..29G, 2012ARA&A..50..107L, 2013ApJ...764...21C, 2015A&A...573A..71K, 2016ApJ...823..102C, 2021FrASS...8...53E}. 
Stellar lifetimes are generally increased by rotation, from 25\% (for a 9~$\Ms$ star) to 15\% (for a 120~$\Ms$ star)\citep{2009pfer.book.....M}.
Rotation impacts how massive stars interact with the environment, affecting the number of ionizing photons emitted and the final yields \citep[e.g.,][]{2018ApJS..237...13L, 2022A&A...663A...1G}.

Rotation is an inherently 3-D phenomenon, and considering its main effects on the 1-D stellar evolutionary models is not easy. Unfortunately, the adoption of 1-D codes is necessary to follow stars throughout their evolution, from the \ac{ZAMS} to the final phases. Three-dimensional codes are computationally costly, and usually, the simulations are restricted to limited portions of stars and for very short timescales \citep[e.g.,][]{2003ASPC..293..147K, 2017MNRAS.471..279C, 2022MNRAS.509..818M}. There are also two-dimensional stellar evolution codes \citep{2016JCoPh.318..277R}, which can simulate the whole star taking into account rotational effects self-consistently, but they are few and slower than 1-D codes. Hence, there is still a lack of big sets of 2-D models which span the different stellar phases.

During the last century, much effort has been made to improve our theoretical understanding of rotating stars. Several pioneering works have opened the path to 1-D modelling of rotation \citep{1924MNRAS..84..665V, 1925Obs....48...73E, 1950MNRAS.110..548S, 1969ARA&A...7..665S, 1970stro.coll...20K,  1976ApJ...210..184E, 1978trs..book.....T, 1992A&A...265..115Z, 1992A&A...253..173C, 1997A&A...321..465M}.
Of course, this dimensional reduction has some drawbacks. Several approximations must be adopted (such as the \emph{Roche} model to describe the star's surface), and some parameters (related to the rotational mixing efficiency) must be calibrated to produce stellar models capable of nicely reproducing observations.
In the following, we briefly review how rotation affects the evolution of stars. We refer the reader to the enlightening book by \citet{2009pfer.book.....M} for a more detailed overview of stellar rotation.

Rotation induces two main effects that influence the star.
The first one is the reduced gravity at any point of the star (except on the rotation axis) due to the centrifugal force. 
This changes the shape of the star, which is no longer spherical but oblate, with a polar radius smaller than the equatorial one. 
In rotating stars, the stellar surface is no longer defined by an equipotential surface because the rotational potential must be considered. 
The Roche methodology is usually adopted to describe the stellar surface, at least as a first-level approximation \citep{1970stro.coll...20K, 1976ApJ...210..184E}. 
Using spherical coordinates and assuming the symmetry along the rotation axes, the stellar surface can be described as follows:
%%%%
\begin{equation}
    - \frac{GM}{R(\theta)} - \frac{1}{2}\Omega^2 R(\theta)^2 \sin^2 \theta = - \frac{GM}{R_\mathrm{pole}}, 
    \label{eq:roche}
\end{equation}
%%%%
where $M$ is the mass of the star, $R(\theta)$ is the surface radius as a function of the co-latitude $\theta$, which is 0 at the pole and 90 at the equator, $\Omega$ is the angular velocity (assuming a constant rotation along the co-latitude), and $R_\mathrm{pole}$ is the polar radius. This equality comes from the assumption that rotation does not modify the polar radius. More detailed models predict a decrease in the polar radius as the rotation increases \citep{1970A&A.....8...76S, 2008A&A...478..467E, 2011ApJ...735...69D}. 
The rotation rate is a convenient quantity to describe the amount of rotation, and it is defined as:
\begin{equation}
    \omega = \frac{\Omega}{\Omega_\mathrm{crit}}, 
    \label{eq:rot_rate}
\end{equation}
where $\Omega_\mathrm{crit} = (2/3)^{3/2}\sqrt{GM/R^3_\mathrm{pole}}$ is the critical break-up velocity. 
By rearranging and solving Eq.~\ref{eq:roche}, it can be found that the stellar shape is just a function of the rotation rate $\omega$, parameter, which can go from 0, in case of no rotation, to 1, in case of break-up velocity. 
When the star reaches the critical velocity, the centrifugal forces balance gravity, the outer layers become unbound, and the latter can be lost through mechanical mass loss \citep[see ][]{2013A&A...553A..24G}. 
To include the geometrical distortion caused by rotation in 1-D stellar evolutionary codes, the \emph{shellular rotation law} is adopted \cite{1997A&A...321..465M}.
This assumption is valid when the horizontal turbulence is stronger than the radial one and homogenizes chemical composition and velocity gradients along isobars \citep{1992A&A...265..115Z, 1997A&A...321..465M}.
In practice, the equations of the stellar structure are modified, and two form factors, namely $f_T$ and $f_P$, are included in the momentum equation and energy transport equation. These factors are a function of the rotation rate $\omega$ and depend on the shape, surface, and volume of the stellar shell (which are computed from Eq.~\ref{eq:roche}). For more details on the methodologies adopted to include rotation in stellar evolutionary codes, we refer the readers to \citet[][]{1997A&A...321..465M, 2009pfer.book.....M, 2021FrASS...8...53E} and references therein.

The departure from the spherical shape leads to the so-called \emph{Von Zeipel effect} \citep{1924MNRAS..84..665V, 1933MNRAS..93..539C}. Von Zeipel has shown that in a rotating star, the local surface brightness is proportional to the local effective gravity. This drives the effective temperature to be no longer constant along the stellar surface. 
Rotating stars are hotter at the poles and colder at the equator.
This effect is also known as \emph{gravity darkening} \citep[][]{1999A&A...347..185M, 2006ApJ...645..664A, 2006ApJ...643..460L, 2011A&A...533A..43E, 2013A&A...552A..35E, 2014A&A...566A..21G, 2016A&A...588A..15C, 2019MNRAS.488..696G, 2023A&A...669L..11A}.
The dependence of the effective temperature on the co-latitude ($\theta$) could be expressed as $T_\mathrm{eff}(\theta) \propto g_\mathrm{eff}(\omega,\theta)^\beta$, where $\beta=1/4$ is the Von Zeipel parameter. 
Several studies predicted lower values for this parameter \citep[e.g., 0.08 from][]{1967ZA.....65...89L}. Recently, a different analytical description of gravity darkening has been given by \citet{2011A&A...533A..43E}, in which the dependence on the $\beta$ parameter is removed. Interferometric observations of rapidly rotating stars by \citet{2014A&A...569A..10D} found good agreement with the theoretical description by \citet{2011A&A...533A..43E}.

Rotation also enhances the mass loss rates by stellar winds. This is due to the reduced effective gravity along the stellar surface. 
Usually, the mass loss enhancement is included in stellar models by including a multiplicative factor (or function) to the adopted wind mass loss prescription, which depends on the rotation rate $\omega$. One of the most popular prescriptions is given by \citet{1986ApJ...311..701F}, which comes from fits on numerical results and reads:
\begin{equation}
    \dot{M}(\omega) = \dot{M}(\omega = 0) \left ( 1 - \frac{v}{v_{crit}} \right)^\xi,
    \label{eq:M_loss_enhanc}
\end{equation}
where $\xi$ is generally assumed to be $0.43$ \citep[][]{1993ApJ...409..429B, 1998A&A...329..551L}, and $v_{crit}$ is the tangential equatorial break-up velocity, defined as $v_{crit}^2 = GM (1-\Gamma_e)/R$, and $\Gamma_e = L/L_E$ is the Eddington factor. This prescription is adopted by several stellar evolution codes that treat rotation \citep[e.g.,][]{2000ApJ...528..368H, 2011A&A...530A.115B, 2013ApJ...764...21C, 2013ApJS..208....4P, 2019A&A...631A.128C}. 
However, this formulation does not include the effect of gravity darkening, namely the dependence of $\Gamma_e$ on the rotation rate and co-latitude. 
More detailed treatments that consider this effect predict anisotropic winds stronger at the poles than at the equator \citep[][]{1999A&A...347..185M, 2000A&A...361..159M, 2011A&A...527A..52G}.

The second main effect induced by rotation is an extra mixing that acts in the stars' radiative regions.
Several mixing processes are induced by rotation \citep[see][]{2000ApJ...528..368H, 2009pfer.book.....M}, but the main two are meridional circulation and shear mixing. 
The first process is related to the Von Zeipel effect: the thermal imbalance between the poles and the equatorial regions induces large-scale currents that transport material and angular momentum within the star \citep[also known as \emph{Eddington-Sweet circulation},][]{1925Obs....48...73E, 1950MNRAS.110..548S}. 
The second process (shear mixing) occurs if the star has a non-flat rotation profile (i.e. does not rotate as a solid body), and it is due to the friction between two nearby shells, which rotate with different angular velocities \citep[e.g.][]{1974IAUS...59..185Z}.
These two processes induce extra-mixing in the radiative regions of stars, transporting processed nuclear products (like nitrogen) from the stellar core up to the envelope and fresh, unprocessed material from the envelope into the core. 
The extra-mixing enhances the stellar lifetimes, makes the star more luminous and enriches the stellar surface with non-pristine chemical elements.
\citet{1992A&A...265..115Z} and later \citet{1998A&A...334.1000M} showed that the transport of angular momentum through such processes could be described in a 1-D framework using an advective-diffusive equation. 
The radial component of the meridional circulation contributes to the advective transport, while the shear friction contributes to the diffusive one. 
\citet{1992A&A...253..173C} showed that the transport of chemical elements could be treated with a diffusion equation, in which the meridional circulation is also included as a diffusive process.
Several evolutionary codes include rotation-induced mixing with the scheme described above \citep{2003A&A...399..603P, 2008Ap&SS.316...43E,  2012MNRAS.419..748P, 2013ApJ...764...21C, 2013A&A...549A..74M}. 
However, it has been shown that suitable matches with observations could also be obtained with the much simpler diffusive approach to treat angular momentum transport \citep{2000ApJ...528..368H, 2011A&A...530A.115B, 2013ApJS..208....4P, 2019MNRAS.485.4641C}. 

The actual rotation implementation in stellar models requires adopting two parameters to regulate the rotational mixing efficiency, $f_c$ and $f_\mu$. The first parameter directly multiplies the diffusion coefficient in the chemical diffusion equation. The second parameter multiplies the mean molecular weight gradient, which tends to inhibit the transport and has a stabilizing effect. 
Since the direct measurements of projected rotational velocities ($v \sin i$, where $i$ is the inclination angle between the stellar rotation axes and the line of sight) give only hints of the surface rotational velocities, other observable must be used to probe the internal rotational mixing efficiency.
A typical methodology to calibrate the rotational mixing efficiency is to compare models with data obtained with the VLT-FLAMES survey, which provided the nitrogen surface abundances, and projected rotational velocities for a large sample of massive stars \citep{2005Msngr.122...36E, 2006A&A...456..623E, 2007A&A...466..277H}. Assuming that rotational mixing is the primary process that drives the N-enrichment of the stellar surface, the comparison could be a good tracer of the mixing efficiency. 
This calibration is adopted by several stellar codes that treat rotation \citep[see][]{2007A&A...466..277H, 2011A&A...530A.115B, 2012ARA&A..50..107L, 2013ApJ...764...21C}. 
However, other processes could compete with rotational mixing (like overshooting, internal gravity waves, and magnetic fields). Therefore, the calibration of the rotational mixing efficiency remains an open problem \citep[see][ and references therein]{2021FrASS...8...53E}. 
Asteroseismology promises to improve our understanding of stellar interiors since it can trace the angular momentum transport in the stars and give information on their rotation profiles \citep{2019ARA&A..57...35A, 2022ApJ...940...49P, 2021RvMP...93a5001A, 2022ARA&A..60...31K}. 
Unfortunately, not many massive stars are analyzed with asteroseismology techniques, but new observations with the TESS space mission are coming \citep{2022ARA&A..60...31K}.

\begin{figure*}
    \includegraphics[width=0.75\textwidth]{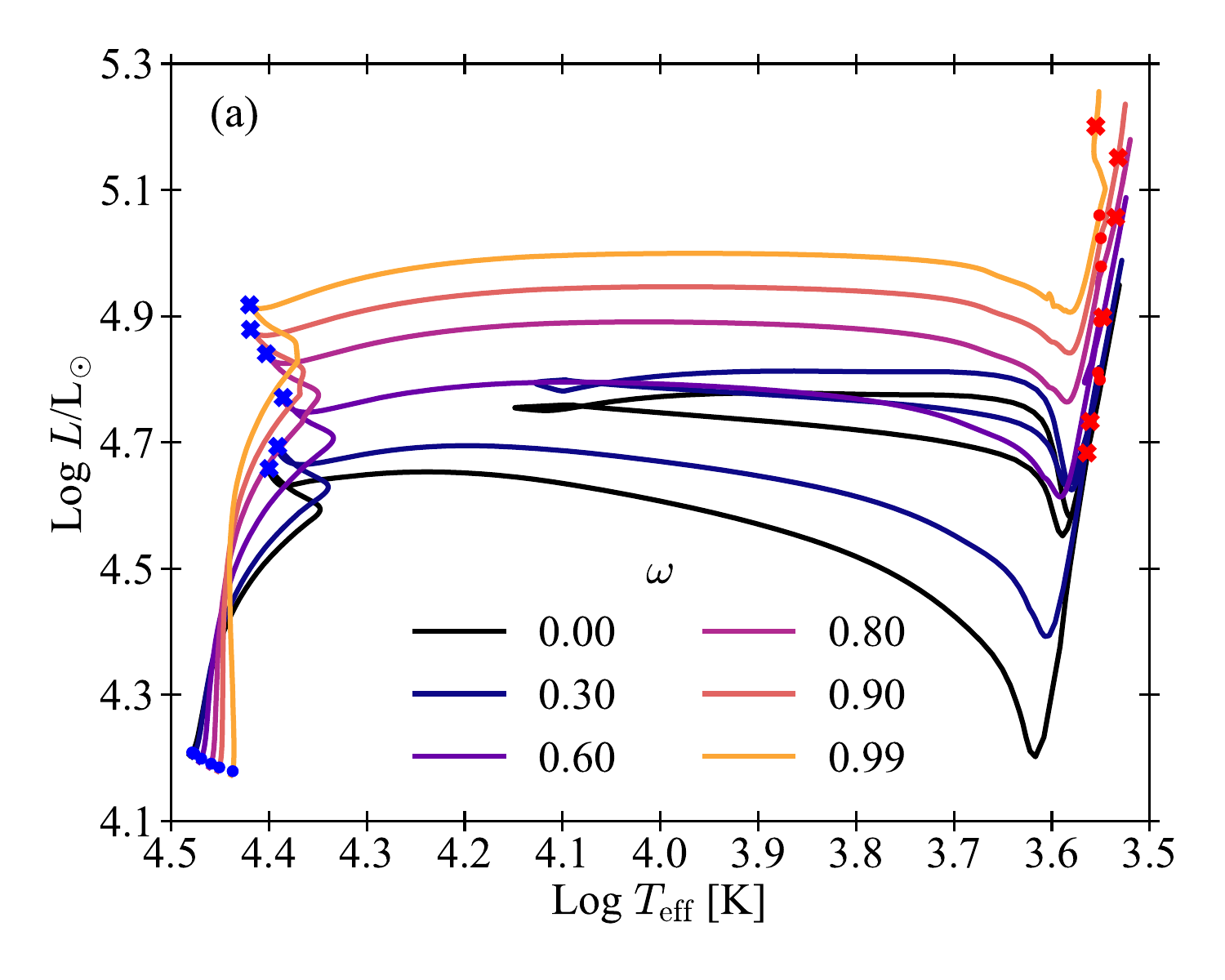} \\          \includegraphics[width=0.78\textwidth]{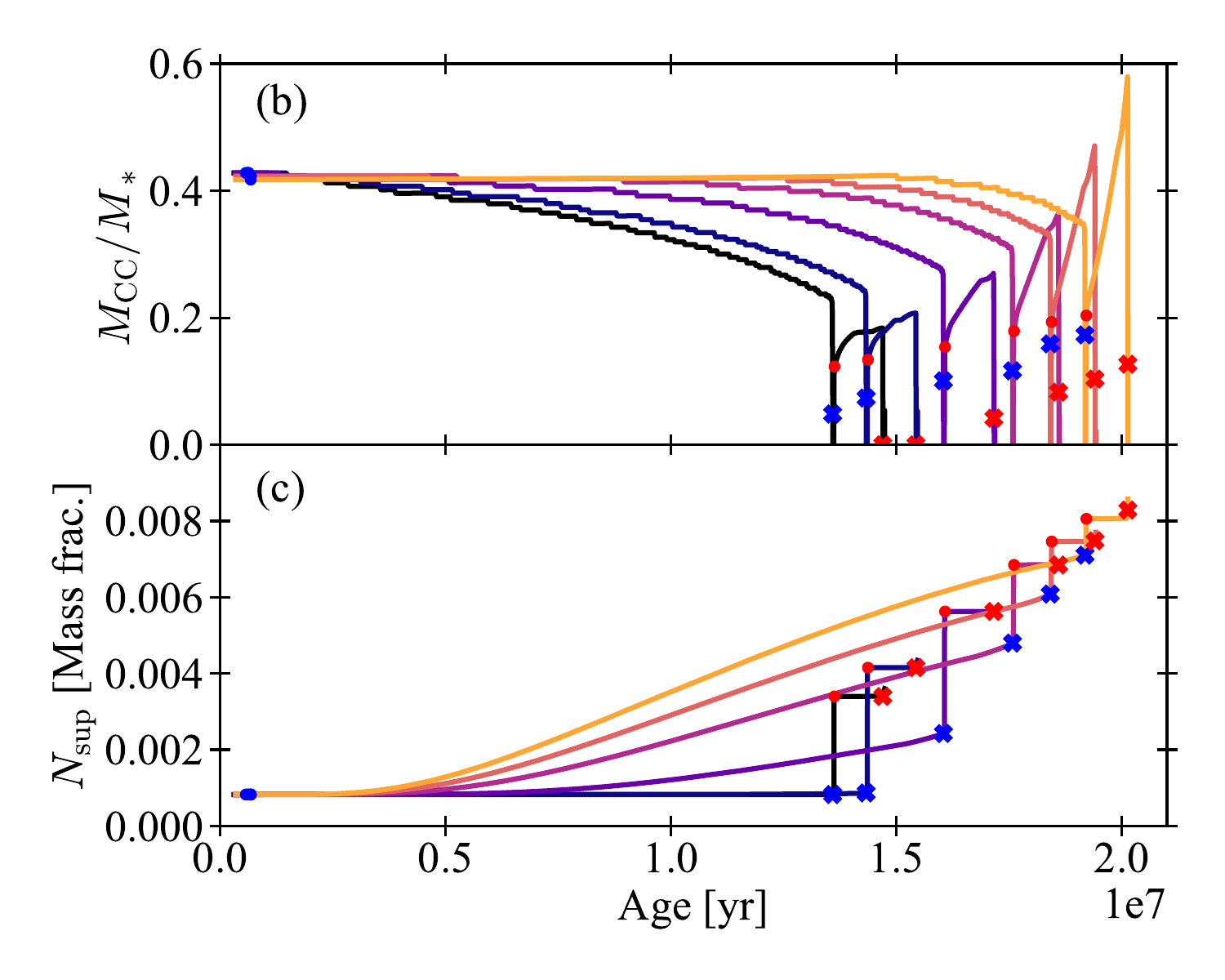}
    \caption{The figure shows the properties of tracks with $14~\Ms$ and $Z = 0.014$ with different initial rotation rates ($\omega$). Panel (a): the \ac{HR} diagram. The blue circles (crosses) indicate the beginning (end) of the \ac{MS} phase. The red circles (crosses) indicate the beginning (end) of the \ac{CHeB} phase. Panel (b): convective core in mass fraction versus time. Panel (c): evolution of the surface nitrogen abundances. Markers indicate the different phases as in panel (a). All the tracks are taken from the online public repository \url{http://stev.oapd.inaf.it/PARSEC}, and a detailed description is given by \citet{2022A&A...665A.126N}.}
    \label{fig:rot}
\end{figure*}

Figure~\ref{fig:rot} shows examples of massive stars' evolution with different initial rotation rates ($\omega$).
In the \ac{HR} diagram, at the \ac{ZAMS}, rotating stars look redder and slightly less luminous than the non-rotating ones. In this phase, only geometrical distortion is at play. 
Later, during \ac{MS}, the mixing processes begin to provide fresh fuel to the core and extract the processed material. 
The stellar surface starts to be enriched by nitrogen. The amount of enrichment depends on the rotation rate. 
Rotating stars live longer in the \ac{MS} and, at the end of the H-burning, they are more luminous and have bigger helium cores \citep{2000A&A...361..159M, 2012A&A...537A.146E, 2016ApJ...823..102C}. 
After \ac{MS}, the envelope expands and the surface angular velocity drops (due to angular momentum conservation). 
The first dredge-up causes a very fast increase in the nitrogen surface abundance. 
Nonetheless, the enrichment due to rotation is still appreciable.
During \ac{CHeB}, rotation may also affect the occurrence of the blue loops.

It is important to note that the gravity darkening influences the star's position in the \ac{HR} diagram and the colour-magnitude diagram, depending on the inclination angle, $i$, between the stellar rotation axes and the line of sight. 
Ignoring rotation and gravity darkening could lead to systematic errors in many properties inferred from data-model comparisons, such as cluster age estimates \citep{2015ApJ...807...58B, 2015ApJ...807...24B, 2018ApJ...863...67G}, and overestimation of the mixing needed to reproduce the luminosity of evolved stars \citep{2019MNRAS.485.4641C}.

As anticipated above, rotation may strongly affect the final phases of stars. 
Mass loss enhancement in rotating stars leads to smaller pre-supernovae masses than the non-rotating counterparts, possibly leading to smaller remnant masses. 
However, rotating stars build bigger He- and CO-cores than non-rotating stars (which affects the explosion likelihood), therefore the final result is not easy to infer \citep{2018ApJS..237...13L, 2019A&A...627A..24G, 2020A&A...640L..18M, 2020ApJ...888...76M, 2021ApJ...912L..31W}. More details on the final fates are given in Section~\ref{subsec:fate}).

In the case of a high rotational mixing efficiency, a very peculiar evolution is predicted for fast-rotating stellar models: the so-called \emph{quasi-chemically homogeneous evolution}. 
In this type of evolution, the star is almost entirely homogenised and remains more compact, evolving to higher luminosities and becoming bluer \citep{1987A&A...178..159M, 2005A&A...443..643Y}. 
The quasi-chemically homogeneous evolution could be an important evolutionary scenario for some supernova, gamma-ray burst events or \acp{GW} progenitors \citep[e.g.][]{ 2012A&A...542A..29G, 2013A&A...554A..23M, 2016MNRAS.460.3545D, 2021MNRAS.505..663R}.

\subsubsection{Magnetic fields}
\label{subsubsec:mag_field}
Magnetic fields may affect stars' evolution in several ways \citep{2005ApJ...626..350H, 2005A&A...440.1041M, 2012SSRv..166..145W, 2017MNRAS.466.1052P, 2020ApJ...900...98G, 2021A&A...646A..19T}. 
There are two different kinds of magnetic fields included in stellar models \citep[see][for a review]{2017RSOS....460271B}.
The first is a magnetic field created and maintained by some dynamo processes in the internal regions of stars,
e.g. generated by differential rotation \citep[Tayler-Spruit dynamo mechanism,][]{1973MNRAS.161..365T, 2002A&A...381..923S}. 
In this case, magnetic fields contribute to the transport of angular momentum and chemical elements in the radiative part of the star \citep{2003A&A...411..543M, 2004A&A...422..225M, 2005A&A...440.1041M, 2005ApJ...626..350H, 2012MNRAS.424.2358P}. 
The other kind is the surface magnetic fossil field. 
It could be the remnant field of the stars' formation process, surviving in a stable configuration. 
The fossil magnetic fields may contribute to the magnetic braking (an increase of angular momentum loss) and the mass loss quenching \citep[due to the coupling between winds and magnetic field, ][]{2011A&A...525L..11M, 2016A&A...587A.105A, 2017A&A...599L...5G, 2017MNRAS.466.1052P, 2019MNRAS.485.5843K, 2022MNRAS.517.2028K}.
The magnetic braking leads to the removal of an additional amount of angular momentum.
It is treated as a torque acting at the stellar surface and is described with an expression obtained from the 3-D simulations results by \citet{2002ApJ...576..413U, 2008MNRAS.385...97U}. 
\citet{2017MNRAS.466.1052P} and \citet{2017A&A...599L...5G} found that massive magnetic stars at solar metallicity could be progenitors of heavy stellar mass black holes and pair-instability supernovae if the magnetic braking process is at play.

How much the magnetic field affects the evolution of the internal angular momentum is still uncertain. Depending on its strength, the magnetic field could force the solid-body rotation, thus suppressing the rotation mixing. 
Otherwise, it could lead to quasi-chemically homogeneous evolution. 
New constraints on the internal angular momentum transport could be achieved with the combination of observational measurements and detailed models, but the debate is still ongoing. 
Results have been obtained by \citet{2019MNRAS.485.3661F} from the comparison of new models (which include a modified version of the \emph{Tayler-Spruit dynamo}) with asteroseismology measurements of low-mass stellar cores rotation rates in the giant phase and with \ac{WD} spins. 
New models by \citet{2019ApJ...881L...1F} predict very low spins for \acp{BH} due to the efficient dissipation of the angular momentum acting during the evolution of the progenitor stars.
However, \citet{2019A&A...626L...1E} found that the formalism proposed by \citet{2019MNRAS.485.3661F} does not match the constraints on the internal rotation rates of sub- and red-giants. 
Recently, new calibrations of the Tayler-Spruit dynamo have been found by \citet{2022A&A...664L..16E}, and also other processes like the magneto-rotational instability are investigated  \citep[see][]{2022A&A...665A.147G}.

% ------------------------------------------------------------------------------------

\subsection{Final phases: Supernov{\ae}, core collapse, and pair instability}
\label{subsec:fate}

Will it explode or not\footnote{This is a simplified version of the question raised by \citet{2012ARNPS..62..407J} at the beginning of his noteworthy review. It reads: \textit{When,  why,  and  how  can  the  catastrophic  in-fall  of  the  core  of  a  massive  star  be  reversed  to trigger the powerful ejection of the stellar mantle and envelope in a supernova explosion?}.}? 
This is the question that every modeller ask themselves when they launch a new simulation of a collapsing stellar core. 
Trying to answer this question is not an easy task because the physics of such phases, particularly the mechanism which transforms the collapse into an explosion, is remarkably complex and a matter of active research \cite[see][]{2015PASA...32....9F, 2016ARNPS..66..341J, 2020LRCA....6....4M, 2023IAUS..362..215M}. 
Moreover, multiple studies have shown that the likelihood of a star to explode (known as \emph{explodability}) is highly sensitive to its pre-supernova structure configuration \citep{2012ApJ...757...69U, 2016ApJ...818..124E, 2016MNRAS.460..742M, 2016ApJ...821...38S, 2018ApJ...860...93S, 2020ApJ...890...43C}. This, in turn, depends on the star's initial properties and the physical processes that impact its evolution. As a result, any uncertainties in these processes can affect the ultimate fate of the star.

\subsubsection{Core collapse and supernova engines}
\label{subsubsec:CCSN}

As already seen in \ref{subsec:mass_ranges}, massive and very massive stars could die in several ways, leaving \acp{NS}, \acp{BH}, or nothing.
The death of a massive star ($\sim 10 - 110~\Ms$) begins when the iron core is formed. 
After iron, forming heavier nuclei through nuclear fusion is an endothermic process that requires energy. Therefore, nuclear reactions can no longer support the inner core against gravity. When the mass of the iron core reaches $\sim 1.4~\Ms$ \citep[the Chandrasekhar mass,][]{1931MNRAS..91..456C}, the pressure support of the degenerate relativistic electrons is not sufficient to prevent collapse.
Within typical free-falling timescales (i.e. a few milliseconds), the core collapses from a radius of about 1-2 thousand km to a few tens of km, dramatically increasing temperature and density.
When the central temperature reaches about $10^{10}$ K, the \emph{photo-disintegration} of the Fe-group nuclei into $\alpha$ particles becomes the dominant process. 
This process requires energy and causes an acceleration of the collapse. 
Moreover, the \emph{electron-capture} process removes free electrons and enriches the core of neutrons and neutrinos.
When nuclear densities of about $2.7\times 10^{14}$ g cm$^{-3}$ are reached \citep{2012ARNPS..62..407J}, the neutron degeneracy pressure is high enough to sustain the in-falling core. 
The collapse overshoots the equilibrium point, and the in-fall continues until the repulsive nuclear forces make the core bounce back, creating a shock wave which propagates outward. 
This is the so-called \emph{bounce-shock mechanism} scenario, first proposed by \citet{1966ApJ...143..626C}. 
In principle, the shock wave can propagate to the external layers and generate a supernova. 
In practice, it gradually loses energy as it impacts the in-falling material until it stalls (a few hundred kilometres from the centre).
At this point, a new energy source is required to unlock and revive the stalled shock and obtain the final supernova explosion. 
Many engines have been proposed to revive the shock \citep[see][]{2012ARNPS..62..407J}, and the most promising framework is the \emph{neutrino-heating} mechanism \citep{1985ApJ...295...14B}.
Several instabilities (such as \emph{Rayleigh-Taylor} and the \emph{standing accretion shock instability}, SASI) could develop in the region between the proto-\ac{NS} (i.e., where the shock is launched) and the radius at which the shock stalls. 
These instabilities drive convection (sustained by the neutrinos' energy), which can transport heat to the outer stellar layers.
If the energy of the convective region is higher than the ram pressure of the infalling layers, a successful supernova explosion occurs \citep[this is referred to as \emph{convection-enhanced neutrino-driven supernova explosion}, ][]{1996A&A...306..167J, 2012ApJ...749...91F, 2012ARNPS..62..407J, 2023IAUS..362..215M}.

State-of-the-art 2-D and 3-D simulations found that the neutrino transport processes drive a symmetric explosion only in the low mass regime ($\Mzams < 10 ~\Ms$) while generating asymmetric explosions for more massive stars. 
The asymmetric explosion can give a kick to the newborn \ac{CO} (called \emph{natal kick}), which could go from $\sim 10~\mathrm{km\,s^{-1}}$ to $\sim 1000~\mathrm{km\,s^{-1}}$, and depends on the structural configuration of the progenitor \citep[see][and references therein]{2012ARNPS..62..407J, 2013A&A...552A.126W}. 
It is expected that \acp{NS} receive stronger kicks than \acp{BH} \citep{2012ApJ...749...91F, 2012MNRAS.425.2799R, 2021hgwa.bookE..16M, 2022Galax..10...76S}.
Natal kicks may play an important role in the creation of \ac{CO} binaries, both in isolated and in cluster environments, and can affect the spin of the newborn \acp{CO} \citep{2012ApJ...749...91F, 2018MNRAS.474.2959G, 2019MNRAS.482.2234G, 2020ApJ...891..141G, 2020MNRAS.499.3214M, 2021ApJ...920..157C}. The effects on binary evolution are further discussed in Part~\ref{part:binaries}.

Multi-dimensional simulations of stellar collapse are computationally very expensive. 
Usually, they resolve only the stellar core (to catch the detailed physics) and for timescales that rarely exceed a few seconds \citep{2012ARNPS..62..407J, 2016PASA...33...48M, 2023IAUS..362..215M}. 
Despite the success of the 2-D and 3-D simulation of exploding stars, it is still difficult to connect the core-collapse supernova explosions to first principles. 
It is not easy to study the process systematically and statistically, exploring all the possible pre-supernova configurations. 
Therefore, we are still relegated to simplified analytic or parameterized numerical 1-D models to explore the progenitor's parameter space (such as metallicity, mixing, winds, rotation, etc.) and investigate the supernova nucleosynthesis of the elements.
A popular approach to model the initial shock wave is the so-called ``piston'' model \citep{1995ApJS..101..181W, 2008ApJ...679..639Z}. Another one consists of inducing artificial supernova explosions through the injection of energy in the pre-supernova model, such as a ``thermal'' bomb \citep{1991ApJ...370..630A, 1996ApJ...460..408T} or a ``kinetic'' bomb \citep{2003ApJ...592..404L}. 
In all three cases, the energy source is placed just above the proto-\ac{NS} and is parameterized with a few free parameters, usually calibrated on the supernova SN 1987A \citep{2012ApJ...757...69U, 2017hsn..book..513L}. 
Another strategy is based on parameterized 1-D simulations, which use analytical formulae to model the heating of the engine \citep{2011ApJ...730...70O, 2012ApJ...757...69U, 2015ApJ...806..275P, 2016ApJ...818..124E, 2016ApJ...821...38S}, or on simple ordinary differential equations for the properties of the explosion and the remnant \citep{2016MNRAS.460..742M}.
These approaches permit investigating the progenitor-explosion connection, i.e., determining the stars' explodability.
From the population synthesis point of view, several useful prescriptions have been proposed to predict the stars' explodability from their pre-supernova configurations. Some examples are given in the following. 

\citet{2012ApJ...749...91F} proposed an analytic methodology to estimate the explosion properties based on the stellar structural configuration at the onset of the collapse \citep[see also][]{1999ApJ...522..413F}.
They suggest that the final mass of the \ac{CO} depends mostly on the carbon-oxygen core mass and the star's total mass. The first determines the explodability, and the second the amount of fallback material on the proto-\ac{NS}. 
In their work, \citet{2012ApJ...749...91F}, identified two possible supernova mechanisms, the \textit{rapid} one, in which the explosion happens within $\sim 250$ ms and the \textit{delayed} one in which the explosion happens after $\sim 250$ ms. The two models have a strong impact on the fallback material, thus on the final \ac{CO} mass \citep[e.g. see][]{2022Galax..10...76S}.

\begin{figure*}
       \includegraphics[width=\textwidth]{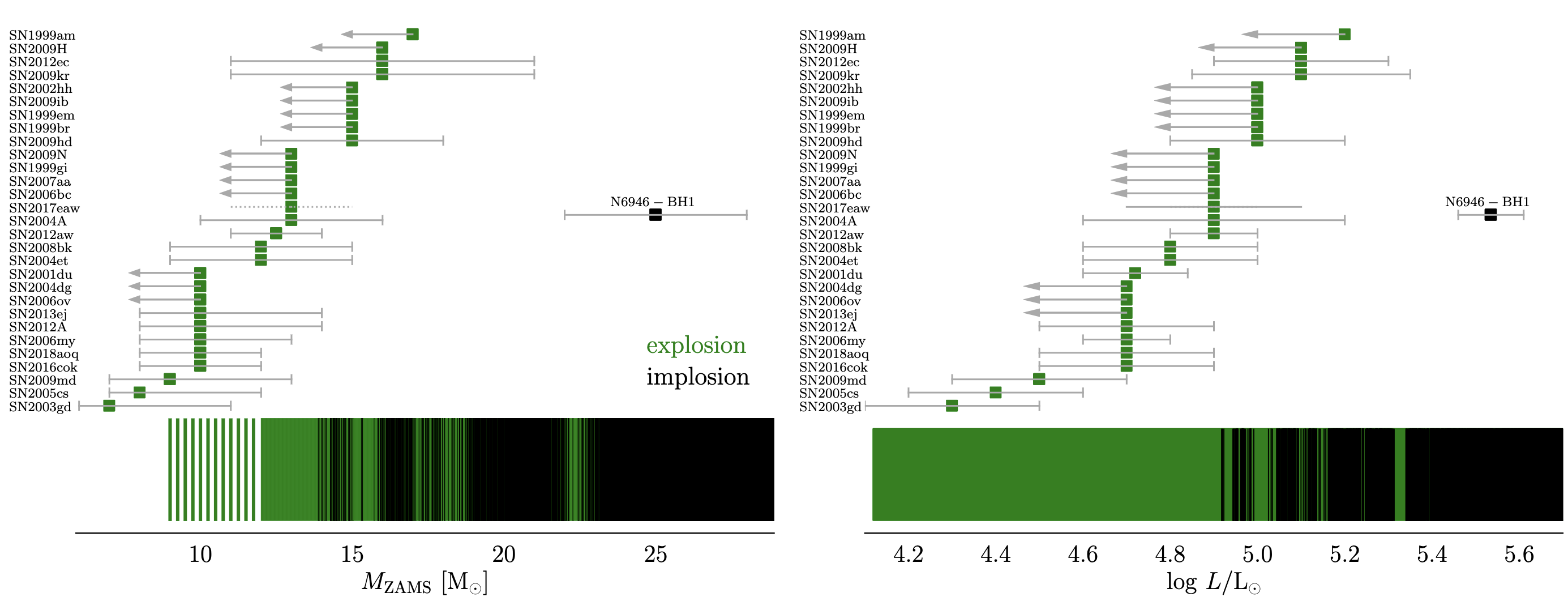}
    \caption{The Figure shows a compilation of Type II supernova progenitors' initial mass and luminosity estimates from direct imaging studies (green squares on the top). The only implosion progenitor candidate is shown in black. The vertical bars, on the bottom, show the final outcomes for non-rotating solar metallicity progenitors \citep{2016ApJ...821...38S, 2018ApJ...860...93S} based on a sample calibrated neutrino-driven explosion model \citep{2016ApJ...818..124E}. The comparison is made on the initial mass (left), and on the luminosity (right). Figure adapted from \citet{2020MNRAS.492.2578S}. 
    % \textbf{NEED PERMISSION!}
    }
    \label{fig:explosion}
\end{figure*}

\citet{2011ApJ...730...70O} identified a useful parameter called \emph{compactness}, $\xi_{2.5}$,
considered to be indicative of the star's likelihood of exploding. The compactness is defined as follows:
\begin{equation}
    \xi_{2.5} = \frac{M/\Ms}{R(M=2.5\Ms)/1000~\mathrm{km}}
    \label{eq:compactness}
\end{equation}
where $R$ is the radial coordinate that encloses the mass $M$ at the time of the core bounce. \citet{2011ApJ...730...70O} suggested that stars with a larger compactness value (above 0.45) are more likely to collapse to a \ac{BH}, avoiding the supernova explosion. 
Later, \citet{2012ApJ...757...69U} 
proposed a lower threshold for the compactness between successful and unsuccessful supernovae (between 0.15 and 0.30). 
Such studies found that the star's resulting explodability is not simply connected to the compactness parameter.
Identifying a precise transition between stars that explode or not with the compactness is difficult and could be model dependent. Rather, compactness may suggest if a star ``tends" to explode or not \citep{2012ApJ...757...69U, 2014ApJ...783...10S, 2016ApJ...821...38S}. 

\citet{2016ApJ...818..124E} found a more refined approach to determine the final fates by using a two-parameters model. 
The parameters that drive the explodability are $M_4$, the mass enclosed at the location where the dimensionless entropy per baryon is $s = 4$, and $\mu_4$, the radial derivative of mass at that location. 
They showed that successful explosions could be distinguished in the $\mu_4$ vs $M_4\mu_4$ plane. They also found almost no fallback in successful supernovae.

To summarize, models with $\Mzams < 15~\Ms$ explode easily, and those with $\Mzams > 40~\Ms$ mostly collapse to \acp{BH}. The fate of stars in the range $15 - 40~\Ms$ is harder to predict. 
There are islands of explodability, and the fate is sensitive to the final pre-supernova structure of the star and to the carbon-oxygen core composition \citep{2020MNRAS.499.2803P}. Figure~\ref{fig:explosion} shows an example of comparison between data obtained from observations and models of supernova explosions.
For the interested reader on the comparison between various approaches and their impact on population synthesis simulation, we suggest the recent paper by \citet{2022MNRAS.511..903P}.

With the prescriptions for determining the explodability described so far, it is possible to derive the final mass of the \ac{CO} from the pre-supernova star. However, they do not include all the processes which could affect the final mass in case of collapse to a \ac{BH}.
For instance, recent works \citep[][]{2018MNRAS.476.2366F, 2021MNRAS.503.2108P, 2022MNRAS.512.4503R} showed that possible mass loss ejections could occur even in the case of a failed core-collapse supernova.
These authors found that after the core bounce, neutrinos ejection or shock revival could generate shock waves which propagate outwards, driving the ejection of the external layers of the stars. The amount of mass lost via this mechanism, and therefore, the final \ac{BH} mass, depends on the final stellar configuration. \citet{2018MNRAS.476.2366F} found that \ac{RSG} stars could lose up to about 5 $\Ms$ during the final collapse, while the more compact \ac{WR} stars just $\sim 0.01~\Ms$.

\subsubsection{Electron-capture and pair-instability supernov{\ae}}
\label{subsubsec:EC_PI}
So far, we have described the standard core-collapse supernovae scenario, but other peculiar ending fates could generate supernova explosions. 

At the low-mass end of massive stars \citep[$8-10~\Ms$,][]{2008ApJ...675..614P}, stars are massive enough to ignite carbon off-centre and form an Oxygen-Neon-Magnesium core sustained by electron degeneracy. 
At this stage, electron capture on magnesium and then on neon nuclei begins, removing part of the pressure support against gravity and driving the core collapse \citep[see][and references therein]{1980PASJ...32..303M, 1982ApJ...253..798N, 2013ApJ...772..150J}. 
Like in the iron-core collapse case, the collapse proceeds until a proto-\ac{NS} forms and the strong nuclear forces make the core bounce back. In this case, at variance with the iron core-collapse case, the density profile of the star above the core is much steeper.
Therefore, the bounce-back shock does not stall and proceeds through the external layers, producing a successful \ac{ECSN} \citep{2019A&A...622A..74J}. \acp{ECSN} are expected to give low kicks to the newborn \ac{NS} \citep{2018ApJ...865...61G}. 

In the initial mass range above $\sim 110~\Ms$, theoretical models predict that stars undergo the \ac{PI} \citep{2017ApJ...836..244W}. 
This dynamical instability is triggered by the creation of the electron-positron pair when temperatures of $\sim 0.7 - 3 \times 10^9$ K and densities of $10^3 - 10^6$ g cm$^{-3}$ are reached. 
Generally, such conditions are typical when massive carbon-oxygen cores are formed \citep{1967PhRvL..18..379B, 1984ApJ...280..825B}.
The pair creation process converts photon energy into the newly created particles' rest mass, decreasing the pressure support to the star and destabilising the hydrostatic equilibrium. Such instability may trigger the collapse of the core and a successive explosive ignition of the oxygen. 
If the oxygen explosion is powerful enough, it can cause the star's total disruption, creating a \ac{PISN} and leaving no remnant.
If the oxygen burns more gently, several ignitions of oxygen could generate pulses, which may extract tens of solar masses before the star becomes stable again. Then, the star performs \acf{CC}, leaving a \ac{BH}. This is the \ac{PPISN}.
If the oxygen explosion does not provide enough energy to lead the star to the explosion, the collapse continues until the formation of a \ac{BH}.

From results of 1-D hydrodynamical simulations of non-rotating stars, \citet{2017ApJ...836..244W} found that stars with helium cores between $\sim 34~\Ms$ and $\sim 64~\Ms$ explode as \ac{PPISN}; stars with helium cores between $\sim 64~\Ms$ and $\sim 130~\Ms$ are totally disrupted by the oxygen flash and thus explode as \ac{PISN}; finally, stars with cores more massive than $\sim 130~\Ms$ perform a direct collapse.
Such theoretical models predict the carving of a gap in the \acp{BH} mass spectrum (known as the \emph{PI mass gap}), which goes from $\sim 50-60~\Ms$ to $\sim 130~\Ms$ \citep{2016A&A...594A..97B, 2017MNRAS.470.4739S, 2019ApJ...882..121S}.
Such limits have been put under question after the detection of the GW190521 merger event \citep{2020ApJ...900L..13A, 2020PhRvL.125j1102A}, in which the mass of the primary \ac{BH} ($\sim 85~\Ms$) has been found to lie in the ``forbidden'' PI mass gap.

Several studies demonstrated the sensitivity of the mass gap edges to uncertainties in the massive star's evolution \citep[see][]{2019ApJ...887...53F, 2020ApJ...902L..36F, 2020ApJ...888...76M, 2020A&A...640L..18M, 2021MNRAS.501.4514C, 2021ApJ...912L..31W, 2021MNRAS.505.2170T, 2021MNRAS.502L..40F, 2021MNRAS.504..146V}, leaving the possibility that the primary \ac{BH} of GW190521 formed via single stellar evolution. 
Among the various uncertainties proposed, the main ones are the $^{12}$C($\alpha$, $\gamma$)$^{16}$O nuclear reaction rate, the fate of the envelope in collapse, and the role of convection in advanced phases.
$^{12}$C($\alpha$, $\gamma$)$^{16}$O is one of the most uncertain reaction rates because it is difficult to measure \citep{2017RvMP...89c5007D}. It directly affects the C/O at the end of the \ac{CHeB} phase. 
\citet{2019ApJ...887...53F, 2020ApJ...902L..36F} investigated the effect of the $^{12}$C($\alpha$, $\gamma$)$^{16}$O uncertainties on the final fate of pure-He stars, finding that the lower edge of the mass gap may change more than $\sim 15~\Ms$.
\citet{2020ApJ...888...76M} found that the lower edge could increase by $\sim 25~\Ms$ if it is assumed that the stellar envelope falls into the \ac{BH} during the \ac{CC}. 
Later, \citet{2021MNRAS.501.4514C} investigated the role of $^{12}$C($\alpha$, $\gamma$)$^{16}$O, the fate of the envelope and convection (convective undershooting) altogether. They concluded that the mass gap shrinks assuming lower $^{12}$C($\alpha$, $\gamma$)$^{16}$O, and it could even disappear when adopting $^{12}$C($\alpha$, $\gamma$)$^{16}$O $- 3 \sigma$ (where $\sigma$ is the 68\% confidence error). 
In particular, stars with $\Mzams > 110~\Ms$ may experience deep dredge-up episodes in the \ac{CHeB} phase, which extracts matter from the core, lowering their mass. Consequently, such stars may avoid the \ac{PI} and produce a \acp{BH} in the gap.

% The relation between the $\Mzams$ and the final \ac{BH} mass is complex because it is affected by the stellar evolution uncertainties.
% Frequently, the mass of the helium core or the carbon-oxygen core at the end of the \ac{CHeB} phase is used as a proxy for the final fate of stars that enter the \ac{PI} regime \citep[see][and references therein]{2012ApJ...749...91F, 2017MNRAS.470.4739S, 2022arXiv221111774I}.

\section{Summary}
\label{sec:Summary}
In the last eight years, the \ac{GW} detectors of the LIGO-Virgo-KAGRA collaboration gathered a continuously growing collection of merging events mainly involving stellar mass \acp{BH}. 
These recent observations have deeply impacted the astrophysical community, not only from the observational point of view but also from the theoretical one. 
Understanding the formation channels of \acp{BBH} mergers and studying the progenitors' evolution is a fundamental step to interpreting the observed \ac{GW} events.
These formation scenarios of \acp{BBH} involve evolutionary processes in single stars, binary stars, and multiple systems.

This contribution reviews the main aspects of the evolution of single-massive stars. 
We briefly summarized how the adoption of computational models greatly improved our knowledge due to the high degree of complexity involved.
State-of-the-art 1-D stellar models have reached unprecedented detail in the description of stellar processes, and they can simulate the life of stars from their formation to the final explosion.
These tools allow us to study physics at play in stellar interiors and link a star's initial properties to a final compact remnant.
However, the final picture, which links a newborn star to its final outcome, is blurred by the still big uncertainties of several processes that occur in massive stars. Important processes such as stellar convection, stellar winds, rotation and magnetic fields may influence the evolution of stars and totally change the final stellar configuration before the supernova explosion. 
Convection is one of the open problems in stellar astrophysics. It is a fully 3-D hydro-dynamical process, and its inclusion in 1-D stellar evolutionary codes implies the adoption of free parameters that must be calibrated. For this purpose, large-scale stellar surveys, new asteroseismic measurements and multi-D models are fundamental ingredients to improve the evolutionary models.
Stellar winds are the main processes which characterize the life of massive stars. They are mainly linked to stellar metallicity and play a role in the evolution of the stellar radius, which can enormously affect the outcome of the binary evolution (see Part \ref{part:binaries}).
They also affect the pre-supernova mass of the stars. 
Advanced stellar winds models promise to tackle the biggest uncertainties, which regard wind clumpiness, the radiative transfer treatment and the hydro-dynamics in stellar atmospheres.
Rotation and magnetic fields are two important processes which can impact several stellar properties during and at the end of the evolution. They can affect the stellar geometry, the internal mixing, the mass loss and the angular momentum distribution over time. Properties of the \acp{BH}, like the mass and the spin, are deeply linked to these processes. 
The final stage of massive stars is an important phase which links the pre-supernova properties to the final \ac{CO} remnant.
The general idea that \acp{BH} with stellar masses can originate from the core collapse of massive stars is quite accepted by the astrophysical community.
However, the details of the collapse are still debated and are a matter of active research. Several prescriptions are given in the literature to predict the explodability of a star from its pre-supernova configuration. Yet, intrinsic uncertainties due to the processes' complexity are still big, and different models predict a different fate for identical initial pre-supernova configurations.
On top of that, all the uncertainties on the processes of stellar evolution propagate to the final collapse. This new layer of complexity adds more uncertainties to the relation between the initial and final stellar mass.

Finally, massive stars are observed to be mostly part of binary or multiple systems; therefore, studying them as single objects do not give the complete picture. Binary processes, like mass transfer or common envelope phases, could drastically change the evolution of a binary and, under certain conditions, could lead to the formation of merging \acp{BBH}, which we could observe with \ac{GW} interferometers. These processes will be discussed in the following Parts.

%%%%%%%%%%%%%%Chruslinska%%%%%%%%%%%
\newpage
\part{Stellar evolution: binaries \\ \Large{Martyna Chru{\'s}li{\'n}ska and Jakub Klencki}}
\label{part:binaries}
\section{Introduction}

\begin{figure}
  \centering
    \includegraphics[width=0.5\textwidth]{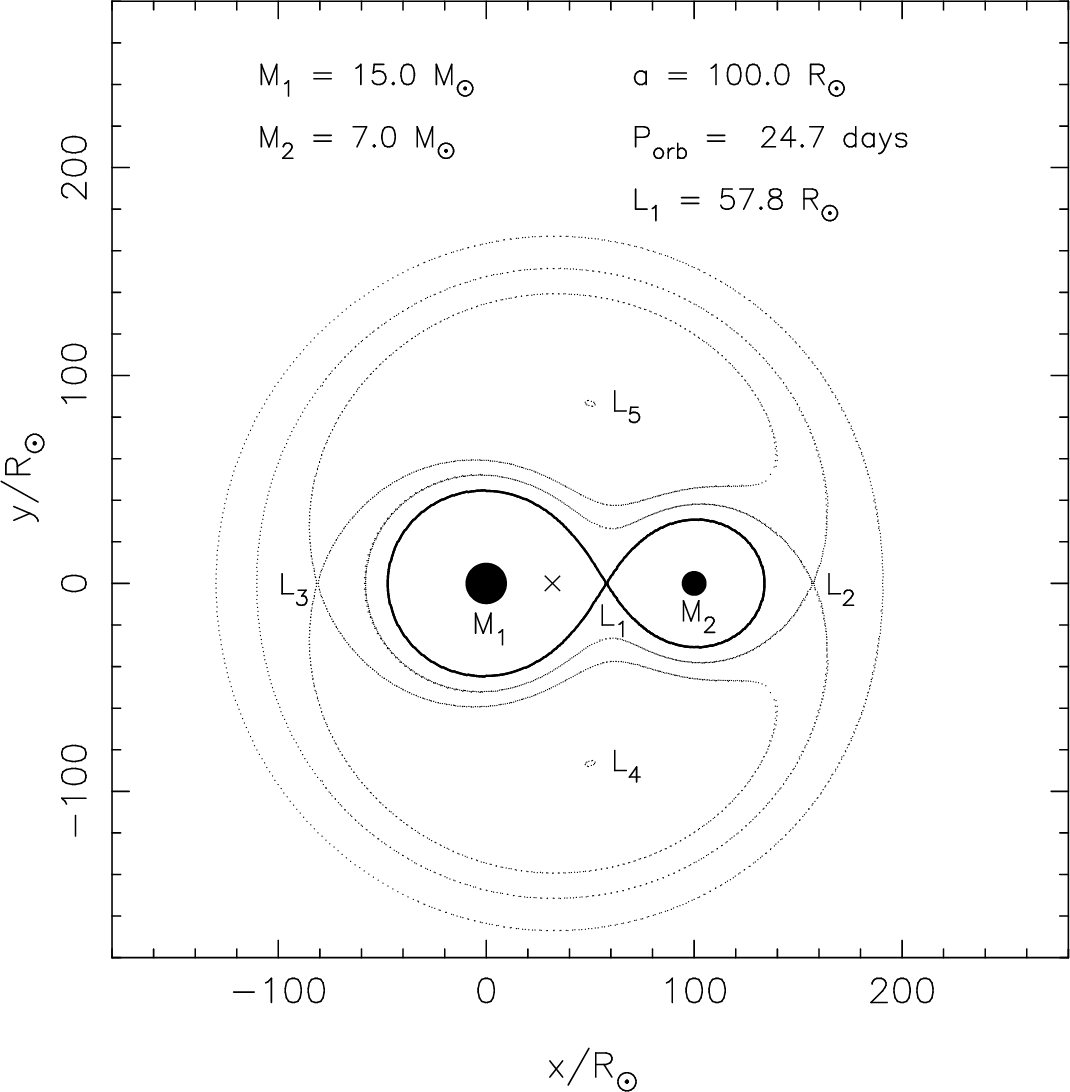}
    \caption{A cross-section in the orbital plane of a binary system. Lines indicate equipotential surfaces in the co-rotating frame of a circular orbit. Masses are marked as ``M1'' and ``M2'', Lagrangian points as `L1' through to `L5'. The center of mass is indicated with a cross. The thick curve defines the Roche lobes of the two components. Figure adapted from \citet{2006csxs.book..623T}.
    % \textbf{NEED PERMISSION}
    }
    \label{fig:roche_lobes}
\end{figure}

It is evident that the vast majority of massive stars are in binaries or in higher-order multiples \cite{1998AJ....115..821M,2007ApJ...670..747K,2012Sci...337..444S,2013ARA&A..51..269D,2014ApJS..215...15S,2017ApJS..230...15M}, i.e. they form a gravitationally bound system with at least one other nearby star. 
The fraction of stars formed in binaries increases with mass, exceeding 85\% for O-type stars (\ac{BH} progenitors). 
The key difference between the life of a single star (to large extent determined by its initial mass and metallicity) and a star in a binary is the opportunity to engage in mass transfer interaction(s) with the companion. 
If a star expands in a binary with a sufficiently narrow orbit, its outer layers may reach and go beyond the point where the forces towards both binary components are equal.
Specific locations where the gravitational forces from the two stars acting on a point mass co-rotating with the system cancel out are called Lagrangian points (see Figure~\ref{fig:roche_lobes}).
When the matter of an expanding star reaches the inner Lagrangian point, a star is said to fill its Roche lobe.
The Roche-lobe overflowing material is then lost from the star and enters the potential well of its companion, which can accrete some fraction of that material \cite{1985ibs..book.....P}.
Mass exchange in a binary can severely affect observable characteristics of both stars and make their future evolution significantly different from that of single stars with the same birth properties \cite{1971ARA&A...9..183P,1977ARA&A..15..127T,1991A&A...252..159V, 1998A&ARv...9...63V,2012ARA&A..50..107L,2017PASA...34....1D}. 
Mass transfer episodes modify the properties of the binary (orbital separation, eccentricity, masses and spins of the binary components), in some cases enabling the formation of various interesting classes of systems (e.g. low/high mass X-ray binaries, ultraluminous X-ray sources) and transients (e.g. \ac{DCO} mergers, type Ia supernovae and hydrogen/helium poor core collapse supernovae).
This Part focuses on the evolution of massive binaries and introduces the mass transfer (Section \ref{sec:M_exchange}) and other processes acting in such systems (Section \ref{sec: other processes}) that are relevant for the formation and properties of stellar origin \acp{BH}.
\\
Massive stars (\ac{NS}/\ac{BH} progenitors) are often born in binaries with separations that are comparable to the sizes that these stars reach during their evolution.
Consequently, the majority ($\gtrsim$70\%) is expected to exchange mass with their companions at some point during their life \cite{2012Sci...337..444S,2017ApJS..230...15M}.
In other words, the formation of stellar origin \acp{BH} is typically preceded by at least one phase of mass transfer interaction.
Single stellar evolution approximation may not allow for an accurate description of their properties and for the correct interpretation of observations of the systems/transients related to massive stars and stellar \acp{BH}.\\

\section{Mass exchange}
\label{sec:M_exchange}
One of the key factors that determines how mass exchange affects the evolution of a binary is
the evolutionary phase of the two stars at the onset of the interaction.
% when the interactions takes place,
% i.e. at what evolutionary phase of the binary components one of the stars starts to overfill its Roche lobe.
We discuss this in Section \ref{sec: radius evolution}, where we describe how massive stars change radius during their evolution and highlight the relevant uncertainties in the current stellar models.
Another important characteristic of a mass exchange is its stability: depending on whether the mass transfer proceeds in a dynamically stable way or leads to the common-envelope evolution, it can produce vastly different binary evolution outcomes.
We discuss this aspect in Section \ref{sec: stability}.
Finally, we consider the effects of a mass exchange on the binary orbit (Sections \ref{sec: orbit stable MT} and \ref{sec: orbit CE}), the properties of the donor star (mass loser, Section \ref{sec: impact on donor}) and those of the accretor (mass gainer, Sections \ref{sec: stellar accretor} and \ref{sec: BH accretor}).
Despite mass transfer in a binary has been a subject of intense research spanning many years, a vast array of open questions remains.
Throughout this section we highlight some of those questions and unknowns that are relevant for the formation and evolution of \ac{BH} binaries.
\\
\subsection{When: evolution of stellar radii}\label{sec: radius evolution}

\begin{figure}[h!]
  \begin{minipage}[c]{0.7\textwidth}
    \includegraphics[width=1.\textwidth]{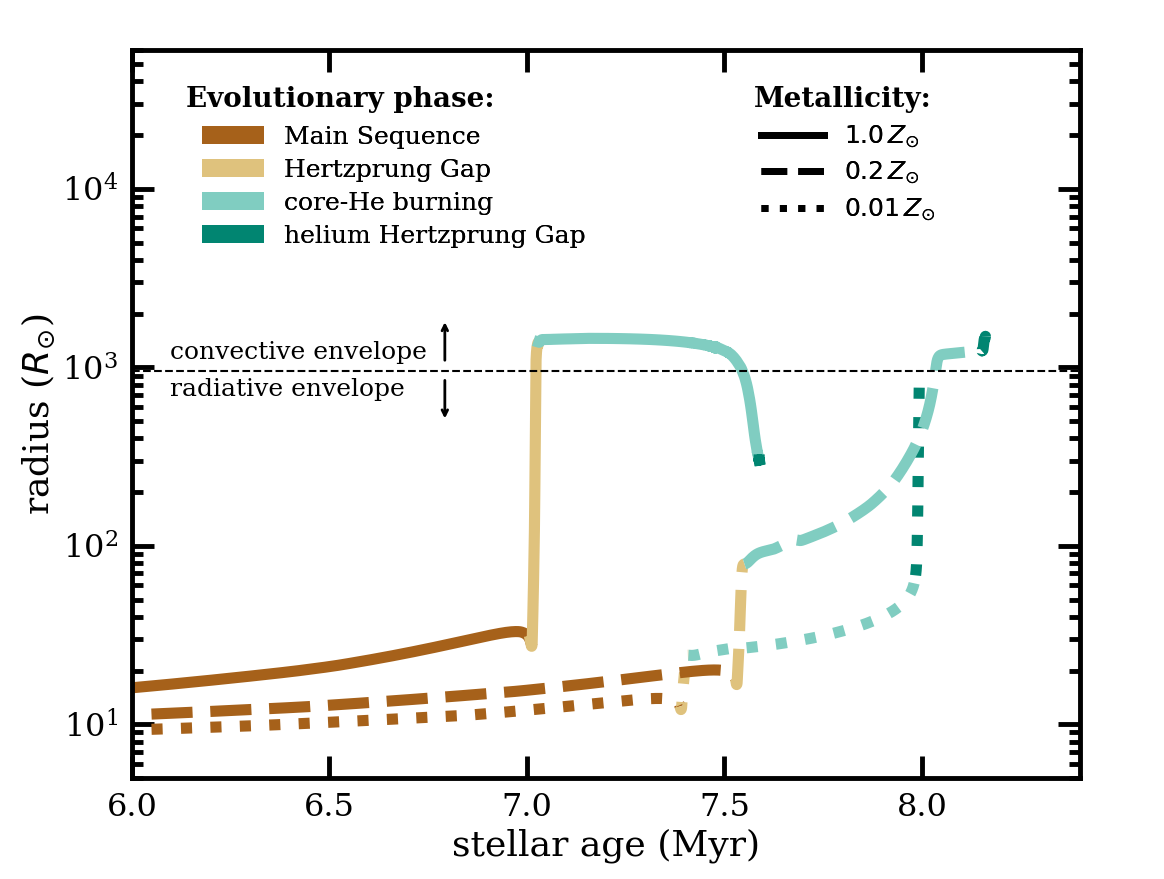}
  \end{minipage}\hfill
  \begin{minipage}[c]{0.3\textwidth}
    \caption{
Radius evolution of a 25 $\Ms$ single star model at three metallicities (representative of the Milky Way like galaxies, star forming dwarf galaxies such as the Small Magellanic Cloud and young galaxies in the early Universe). Different colours correspond to different evolutionary phases.
The horizontal dashed line roughly separates the radii above/below which the model stars have convective/radiative envelopes.
Modified figure from \cite{2020A&A...638A..55K}. Here solar metallicity $Z_{\odot}$=0.017
    }             
    \label{fig: radius evolution}
  \end{minipage}
\end{figure}

The size of a star changes substantially during its evolution. Massive stars tend to slowly expand during core hydrogen- and \ac{CHeB} stages. On top of that, when hydrogen or helium in the center becomes exhausted, the energy production drops down and the core begins to contract. In response, the structure of the entire star needs to adjust and the outer layers (the envelope) rapidly expand in the so-called "mirror effect".
The rate at which a star burns its nuclear fuel ($1/\tau_{\rm nuc}$, where $\tau_{\rm nuc}$ is the nuclear timescale\footnote{
Computed as a ratio of the energy produced in the nuclear reactions $E_{nuc}$ to the luminosity of the star $L$: 
$\tau_{nuc} = \frac{E_{nuc}}{L}\approx 10^{10} \frac{M}{\Ms} \frac{L_{\odot}}{L}$ yr  on the \ac{MS} (and $>$10 times shorter at later burning stages). Using the approximate mass-luminosity relation for the massive main sequence stars $L\propto M^{3.5}$ leads to $\tau_{nuc}\propto M^{-2.5}$.
This last scaling is appropriate for stars with masses roughly in the range $2\Ms \lesssim M \lesssim 50 \Ms$; more massive stars have a much weaker dependence ($L\propto M$, otherwise the estimated $\tau_{nuc}$ would be unrealistically short) \cite{1990sse..book.....K}.
}) increases with mass. Therefore, the initially more massive star in a binary (primary) is the first to expand, fill its Roche lobe, and become a donor in a mass transfer.
The mass exchange is usually initiated
during those evolutionary phases that are accompanied by a significant radial expansion. 
What are those phases and when the star of a given mass reaches its maximum size depends on its birth metallicity (see Figure \ref{fig: radius evolution}).

Three cases of mass transfer are conventionally distinguished in the literature (case A, B and C respectively, \cite{1967ZA.....65..251K}). 
In case A, the donor star fills its Roche-lobe as a result of its slow expansion on the \ac{MS}, i.e, during core-hydrogen burning. However, different definitions of case B/C mass transfer are used in the literature - case B either refers to the mass transfer initiated after the end of core-hydrogen burning but before core helium ignition (case C then corresponds to the interaction during helium shell burning or later) or encompasses interactions initiated between the end of core-hydrogen burning to the end of the \ac{CHeB} (case C then corresponds to the mass transfer happening after the end of the \ac{CHeB}).
The parameter space for case A type of interaction is relatively small as it is limited to systems with short pre-interaction periods of about 10 days or less, depending on mass and metallicity. The more massive and the more metal-rich the star, the larger it gets during the \ac{MS}, which increases the chance for a case A interaction \cite{2011A&A...530A.115B, 2017A&A...597A..71S,2020A&A...638A..55K}.
In particular, the envelopes of some of the most massive and metal rich main sequence stars ($\gtrsim 40 \rm M_{\odot}$ at solar metallicity) in which the gravity and radiation pressure at the surface are comparable (i.e. approaching the Eddington limit) might become inflated \cite{1926ics..book.....E, 
2015wrs..conf..241L,2015A&A...580A..20S,2017A&A...597A..71S,2022A&A...668A..90A}. 
% might become inflated
% when the radiation pressure at the surface starts to balance gravitational force
% due to their proximity to the Eddington limit (i.e. when the gravity is equal to the radiation pressure at the surface; \cite{1926ics..book.....E, 
% 2015wrs..conf..241L,2015A&A...580A..20S,2017A&A...597A..71S,2021arXiv211202801A}). 
Such inflated envelopes could reach the sizes of red supergiants ($\sim 1000 \rm \ R_{\odot}$).
As a consequence, most binary systems with such stars would initiate mass transfer already during the \ac{MS} \cite{2020A&A...638A..55K}. Yet, little effort has been put so far into understanding of mass transfer from an inflated-envelope star \cite[though see][]{2006A&A...450..219P}. 
1D stellar evolution codes (the most common and computationally feasible tool) were shown to overpredict the phenomenon of envelope inflation compared to more accurate 3D hydrodynamic simulations of massive star envelopes \cite{2015ApJ...813...74J,2020ApJ...902...67S}.

Following the stage of core-hydrogen burning, massive stars expand rapidly (on the thermal timescale) until regaining equilibrium as \ac{CHeB} stars.
During this brief phase (the Hertzsprung gap) the radii of solar-like metallicity stars can increase by even 2 orders of magnitude, making it the most significant phase of expansion and the most common starting point for a mass transfer interaction in metal-rich environments. The maximum size is reached when the star becomes a \ac{RSG} and has an outer convective envelope.
\\
However, as illustrated in Figure~\ref{fig: radius evolution}, the degree of Hertzprung gap expansion depends on metallicity \cite{1982ApJS...49..447B,1991A&A...245..548B}. Low-metallicity massive stars may remain compact for longer, with the bulk of the expansion occurring only after the onset of \ac{CHeB}.  
\\
Very low metallicity massive stars (the $0.01 Z_{\odot}$ example in Figure~\ref{fig: radius evolution}) may experience a phase of significant radial expansion even later in their evolution, i.e. following the core-He depletion (during the so-called helium Hertzprung gap phase) \cite{2019A&A...627A..24G,2001A&A...371..152M}.
Therefore, such stars are likely to initiate mass transfer in the very late stages of their evolution ($\lesssim$10$^{3-4}$ yr before the core collapse in the case of \ac{BH} progenitors) \cite{2020A&A...638A..55K}.
\\
\newline
The metallicity trend in stellar radius evolution illustrated in Figure \ref{fig: radius evolution} results from a combination of various metallicity dependent effects. 
Birth composition affects factors such as the nuclear burning efficiency, radiative opacity, and mean molecular weight (entering the equation of state), non-trivially affecting stellar evolution \cite{2022arXiv220611316X}.
The increase of the radius of \ac{MS} stars with metallicity is found to be a robust prediction of stellar models. In contrast, the extent of rapid post-\ac{MS} expansion of massive stars was shown to be sensitive to the degree of chemical mixing above the convective core through processes such as overshooting, semiconvection, or rotationally-induced mixing \cite{1982ApJS...49..447B,1992ApJ...390..136S,1995A&A...295..685L,2019A&A...625A.132S,2020A&A...638A..55K,2020MNRAS.497.4549A,2020A&A...635A.175H,2021MNRAS.506.4473S,2022MNRAS.512.4116F}.
Efficiency of those mixing processes is largely unknown 
(particularly in the case of massive stars)
and its observational calibration is an area of active research \cite{2008MNRAS.384.1109E,2011A&A...530A.116B,2016ApJ...823..102C,2018A&A...611A..75S,2019ApJ...885..143B,2020MNRAS.496.1967K,2020FrASS...7...70B,2021A&A...655A..29J,2021MNRAS.503.4208S,2022ApJ...926..169A,2022ApJ...926..169A}. 
Therefore, while the general trend with metallicity (i.e. lower metallicity stars expanding later in their evolution) is to be expected and is broadly consistent with the local observations, the exact degree of post-\ac{MS} expansion at any given mass and metallicity is very uncertain.
One particular consequence of this trend is that a binary with the same initial parameters formed at different metallicity is likely to enter mass transfer at different ages and evolutionary stages of the donor.
As we discuss in the next section, this may affect the course and outcome of the mass transfer interaction itself \cite{2022A&A...662A..56K}.
\\
Besides the evolutionary stage, 
another characteristic of the donor star that is decisive for how the mass transfer proceeds is the type of its outer envelope.
As further discussed in Section \ref{sec: stability}, whether the outer envelope it is dominated by radiative or convective energy transport is crucial for the mass transfer stability.
The boundary between those two regimes has been roughly indicated in Figure \ref{fig: radius evolution}.
% i.e. whether it is dominated by radiative or convective energy transport. 
Unlike low- and intermediate-mass stars,
massive stars show relatively little expansion after their envelopes become convective \cite{2020A&A...638A..55K}. This means that mass transfer from a radiative-envelope donor is much more common in the massive star regime. 
% As further discussed in Section \ref{sec: stability}, the type of the outer envelope is a crucial factor in the determination of mass transfer stability.
\\
The expansion of massive stars may be limited as a result of very high mass-loss rates in winds or exceptionally strong internal chemical mixing. 
The empirical Humphreys-Davidson limit in the \ac{HR} diagram
 may be an indication of such a suppressed expansion in the high mass regime.
% \footnote{The  Humphreys-Davidson limit appears for stars with the initial masses of $\gtrsim 40 M_{\odot}$.}
% The limit sets a boundary above which
Almost no stars in the local Universe are found in this high luminosity-low temperature part of the \ac{HR} diagram that is expected to be occupied by expanded stars with initial masses of roughly $\gtrsim 40 M_{\odot}$ \cite{1979ApJ...232..409H,1994PASP..106.1025H,1998ApJ...504..200U}.
% at luminosities exceeding this empirical limit \cite{1979ApJ...232..409H,1994PASP..106.1025H,1998ApJ...504..200U}, that should be occupied by stars with initial masses of  $\gtrsim 40 M_{\odot}$.
Both extensive mass-loss (e.g., \ac{LBV} winds, see also Section~\ref{subsec:winds}) \cite{2004ApJ...616..525O,2008A&A...482..945G,2011A&A...535A..56G,2015MNRAS.452.1068C,2016MNRAS.458.1214Q,2021A&A...647A..99G} and abnormally high convective-core overshooting during the \ac{MS} \cite{2021MNRAS.503.1884G,2021MNRAS.506.4473S,2022MNRAS.512.5717A,2022A&A...668A..90A} were invoked as possible factors restricting the expansion of stars beyond the Humphreys-Davidson limit. %(see also Chapter ??).
% The empirical Humphreys-Davidson limit in the HR diagram\footnote{The  Humphreys-Davidson limit appears for stars with the initial masses of $\gtrsim 40 M_{\odot}$.}, above which almost no stars are found to exist in the Local Universe \cite{1979ApJ...232..409H,1994PASP..106.1025H,1998ApJ...504..200U}, has been suggested to be the result of extensive mass-loss (such as LBV winds), possibly related to the proximity of the Eddington limit \cite{2004ApJ...616..525O,2011A&A...535A..56G,2015MNRAS.452.1068C,2016MNRAS.458.1214Q,2021A&A...647A..99G}. The suppressed expansion above the Humphreys-Davidson limit might also be the result of abnormally high convective-core overshooting during the Main Sequence \cite{2021MNRAS.503.1884G,2021MNRAS.506.4473S,2022MNRAS.512.5717A,2022A&A...668A..90A}.
In case of very massive ($\gtrsim 100 M_{\odot}$) high metallicity stars, mass loss during the \ac{MS} may already be strong enough for them to avoid any post-\ac{MS} expansion. Instead, such stars may shrink to become hot and compact helium-rich \ac{WR} stars \cite{2013MNRAS.433.1114Y,2015A&A...573A..71K}.
\\
Finally, in presence of strong rotationally-induced mixing, massive metal-poor stars may become chemically homogeneous (see Section \ref{subsec:rotation} and Section \ref{sec: CHE}, and references therein). Such stars completely avoid expanding into classical giants. The maximum sizes that they reach throughout their evolution are of the order of $\sim 10 \ \Rs$, which significantly limits the parameter space for mass transfer interactions.
\\
All these factors that influence the evolution of radii of massive stars strongly affect the parameter space for binary evolution leading to the formation of systems containing stellar \acp{BH} \cite{2020MNRAS.497.4549A,2020A&A...638A..55K,2022ApJ...925...69B,2022arXiv221115800R}.

\subsection{How: stability of mass transfer}\label{sec: stability}
The nature and stability of the mass exchange is crucial for the future fate and evolution of the binary.
During the interaction both the size of the donor star and that of its Roche lobe change in response to mass and angular momentum loss/transfer.
The companion star (the accretor) also reacts to the gained mass, either by expansion or contraction, depending on whether its envelope is radiative or convective \cite{1976ApJ...206..509U,1983Ap&SS..96...37H,1994A&A...290..129V}, see also Section~\ref{sec:accretors}. Expansion of radiative-envelope accretors may lead to the formation of contact binaries and additional angular momentum loss from outer Langrangian points \cite{1994A&A...290..119P,1998A&ARv...9...63V,2001A&A...369..939W,2008MNRAS.390.1577G,2020ApJS..249....9G}. This complicated process is not yet well understood and is usually neglected in considerations of mass transfer stability, which are focused on the behavior of the donor star.
\\
To first order, the stability of a mass transfer interaction depends on whether the radius of the donor ($R_{\rm donor}$) expands or contracts with respect to the donor's Roche lobe radius ($R_{L}$) as a result of mass exchange.\footnote{The Roche lobe radius is defined as the radius of a sphere enclosing the same volume as the volume of the Roche lobe.} 
The stability analysis is thus commonly done by comparing the mass-radius response exponents: $\zeta_{RL}=d\text{log}R_{L}/d\text{log}M_{\rm donor}$ and $\zeta_{Rd}=d\text{log}R_{\rm donor}/d\text{log}M_{\rm donor}$, where $M_{\rm donor}$ is the donor mass and $d M_{\rm donor}<0$ during mass transfer.
The condition for mass transfer stability is then $\zeta_{Rd}\geq \zeta_{RL}$, 
 and the challenge that one needs to face boils down the correct determination of those coefficients.
This is by no means an easy task - a great portion of our knowledge of stellar evolution is condensed in $\zeta_{Rd}$ and its accurate determination is limited by the related uncertainties (see Section \ref{When: Rd}). 
In turn, all the complexity of the mass and angular momentum loss and exchange experienced by the binary system is hidden within the single parameter ${\zeta_{RL}}$ (see Section \ref{When: RL}).

\subsubsection{Roche lobe radius}\label{When: RL}

$R_{L}$ of the donor can be conveniently approximated with the Eggleton's formula \cite{1983ApJ...268..368E}:
\begin{equation}\label{eq: RL}
    \frac{R_{L}}{a} = \frac{0.49 \ q^{2/3}}{0.6 \ q^{2/3} + \text{ln}(1+q^{1/3})}
\end{equation}
and depends only on the orbital separation $a$ and the donor to accretor mass ratio $q=\frac{M_{\rm donor}}{M_{\rm accretor}}$.
However, both quantities change (either increase or decrease) as a result of the mass transfer.
The evolution of those parameters depends on what fraction of mass lost by the donor is accreted by the companion (i.e. on the conservativeness of the mass transfer $\beta$, defined as $\rm \dot{M}_{\rm accretor} = -\beta \ \dot{M}_{\rm donor}$) and on the amount of angular momentum that is carried away from the system by the non-accreted matter.
Therefore, the rate at which $R_{L}$ changes as the donor loses mass crucially depends on the characteristics of the mass transfer.
It is also affected by all the other processes 
that change the orbital separation and masses of either of the binary components (see Section \ref{sec: other processes}).
Many of such processes may happen simultaneously with the mass transfer (e.g. tidal interactions, mass and angular momentum loss through stellar winds, angular momentum loss through \ac{GW} emission from a binary). Their effect can be neglected in the mass transfer stability analysis if the associated mass/angular momentum loss is much smaller or happens on a much longer timescale than the binary interaction.
\\
The simplest case of binary interaction is a fully conservative mass transfer, in which all the mass lost by the donor is accreted by the companion and the angular momentum of the binary is conserved, i.e. the quantity:
\begin{equation}\label{eq: Jorb}
J_{orb}=M_{\rm donor} M_{\rm accretor} \left( \frac{G a (1-e^{2})}{M_{\rm donor} + M_{\rm accretor}}\right)^{1/2}
\end{equation}
$J_{orb}$=const. in the conservative case.
Here $G$ is the gravitational constant and $e$ eccentricity of the orbit.
Tidal forces are thought to efficiently circularize the binary orbit before the onset of the interaction and if that is the case  $e\approx0$ (but see Section \ref{sec: tides}).
Using equations \ref{eq: RL} and \ref{eq: Jorb} one can then easily show that ${\zeta_{RL}}=2.13 q - 1.67$ depends solely by the mass ratio.
That is why the critical mass ratio $q_{crit}$ is often invoked to discuss the stability of the interaction:  
for any ${\zeta_{Rd}}$ one can determine $q_{crit}$ such that if $q>q_{crit}$  the mass transfer is unstable.
In a non-conservative case, increasing mass loss from the system (lowering $\beta$) generally tends to decrease ${\zeta_{RL}}$ (and increase $q_{crit}$), therefore having a stabilising effect on the mass transfer.
However, ${\zeta_{RL}}$ determination (and its behaviour as a function of $q$; see e.g. Figure 4 in \cite{1997A&A...327..620S}) heavily depends on how the angular momentum is removed from the orbit,
in particular on the amount of specific (i.e. per unit mass) angular momentum $j_{loss}$ carried away by the escaping matter.
During the mass transfer $j_{loss}$ is often assumed to be the same as the specific angular momentum of the matter at the accretor's orbit around the center of mass.
This corresponds to the idealized scenario in which the mass is ejected from the accretor's surface layers/inner part of the accretion disk with high velocity (so that it does not exchange any angular momentum with the system; this mode of mass loss is often called ``isotropic re-emission'').
In practice, $j_{loss}$ may be greater than that, for instance, if the accretor's Roche lobe is filled with mass which (or some fraction of it) then starts to escape the system from the outer Lagrangian point.
In general, increasing $j_{loss}$ works to decrease the orbital separation (and therefore to decrease $R_{L}$), increasing the chances for mass transfer instability.
Both $\beta$ and $j_{loss}$ are uncertain and often treated as free parameters when modelling binary interactions.

\subsubsection{Donor radius 
and its response to mass loss}\label{When: Rd}

In an idealized case, any removal of mass from a star drives it out of both the hydrostatic and thermal equilibrium \cite{1985ibs..book...39W,1987ApJ...318..794H,1997A&A...327..620S}.
% In an attempt to regain equilibrium, it can either expand or shrink.
A star responds to mass loss on two different timescales and there is no single $\zeta_{\rm Rd}$ value.
The initial response is to regain hydrostatic equilibrium, which happens on the dynamical timescale ($\sim$ days to years, $t_{\rm dyn} \propto R^{3/2} M^{-1/2}$)\cite{1996ApJ...458..301K}.
On the longer thermal timescale, the star regains thermal equilibrium, i.e. the balance between energy production, transport, and release from the surface 
($t_{\rm th} \propto R^{-1} M^{2} L^{-1}$, which is typically $\sim 10^{3} - 10^{5}$ yr for massive stars)\cite{1996ApJ...458..301K}.
\\
In the classical approach, the dynamical response of the star to mass loss, characterized by $\zeta^{\rm dyn}_{Rd}$, has been studied in an adiabatic approximation, using stellar models with a fixed entropy profile \cite{1987ApJ...318..794H,2010ApJ...717..724G}. It was shown that the adiabatic response of a star to mass loss is dictated by the entropy profile at the surface layers, which in turn depends on whether the star has a radiative or a convective outer envelope. 
A convective-envelope star, with a flat entropy profile, tends to expand
($\zeta^{\rm dyn}_{Rd} < 0$), whereas a radiative-envelope star, with a steep entropy gradient, shrinks in respond to mass loss ($\zeta^{\rm dyn}_{Rd} > 0$).
As a results, mass transfer from convective-envelope donors is much more prone to become dynamically unstable.
This general rule of thumb can be demonstrated analytically \cite{1965AcA....15...89P,1997A&A...327..620S} and was confirmed in some of the early full binary evolution calculations \cite{1973PASP...85..769P,1977ApJ...211..486W}. 
\\
The use of the adiabatic approximation is justified so long as the dynamical timescale is too short for any significant heat transfer within a star \cite{1987ApJ...318..794H,2010ApJ...717..724G,2015ApJ...812...40G}. This, however, was shown to not be true 
in the outer envelopes of (massive) giant stars where thermal relaxation is fast and cannot be neglected even on the dynamical timescale \cite{2011ApJ...739L..48W, 2015MNRAS.449.4415P}. The way to circumvent this issue is to either to allow for some thermal relaxation to take place \cite{2020ApJ...899..132G} or, ideally, to compute full binary evolution models without the adiabatic assumption \cite{2017MNRAS.465.2092P,2022arXiv220912707T}.
While the importance of this effect is yet to be quantified for the entire parameter space of donor stars, current results yield $\zeta^{\rm dyn}_{Rd}$ values that are lower(higher) for stars with convective(radiative) envelopes compared to the adiabatic models.

Predicting the stellar response to mass loss on a longer, thermal timescale $\zeta^{\rm th}_{Rd}$ is more complex. This response depends on the entire internal structure, which can  vary for different stars depending on their evolutionary stage and previous binary interactions \cite{2019A&A...628A..19Q,2022A&A...662A..56K}. In general, (at least initially) most stars tend to thermally expand \cite{1960ApJ...132..146M,1989PhDT.........7H,2020ApJS..249....9G}.
As we discuss in the following sections,
this leads to a phase of thermal-timescale mass transfer which can either (a) strip the entire envelope at the rate $\dot{M_{\rm th}} \approx M_{\rm donor} / t_{\rm th}$, (b) slow down to the nuclear timescale once a partially-stripped donor is able to regain thermal equilibrium within its Roche lobe, or (c) runaway into instability $\dot{M} >> \dot{M_{\rm th}}$ if the donor keeps expanding with respect to its Roche lobe or a delayed-dynamical instability with $\zeta^{\rm dyn}_{Rd} < \zeta_{RL}$ occurs at a later stage of the interaction. 

\subsubsection{Combining the two: stability criteria and the mass transfer timescales}
Based on the dynamical and thermal responses of the donor star to mass loss, one can deduce whether the mass transfer is stable as well as the timescale, on which the mass is exchanged. If both $\zeta_{RL}<\zeta^{\rm dyn}_{Rd}$ and $\zeta_{RL}<\zeta^{\rm th}_{Rd}$, the donor star contracts with respect to its Roche lobe on both the dynamical and the thermal timescale. The mass transfer rate is then dictated by the nuclear evolution of the donor, combined with the orbital evolution of the system due to other processes such as tides, magnetic braking, gravitational radiation, wind mass loss, (see Section~\ref{sec: other processes}).
Those other processes typically play a minor role in the orbital evolution during other, rapid mass transfer cases discussed below.
The timescale of nuclear-timescale mass transfer can still span orders of magnitude, depending mostly on the mass and evolutionary stage of the donor star. For instance, in low mass X-ray binaries, with typical donor masses of $\lesssim 1 \Ms$, the timescale of the (nuclear) mass transfer is billions of years ($\tau_{nuc}\gtrsim 10^{10}$ yr). Meanwhile, a \ac{BH} high-mass X-ray binary (donor mass $\gtrsim 10 \Ms$) evolving on a nuclear timescale would have a lifetime of a few millions of years ($\tau_{nuc}\sim 10^{7}$ yr).
\\
If $\zeta_{RL}<\zeta^{\rm dyn}_{Rd}$ but $\zeta_{RL}>\zeta^{\rm th}_{Rd}$, the donor star is able to regain hydrostatic equilibrium (and maintain it during mass transfer) but is unable to regain thermal equilibrium. In a futile attempt to do so, the star keeps expanding with respect to its Roche lobe, driving a thermal-timescale mass transfer.
Stable thermal-timescale mass transfer in massive binaries typically leads to mass transfer rates of the order of $\sim 10^{-3} \rm M_{\odot} / yr$, with the entire phase lasting for about $\lesssim 10^4$ yr.
The short timescale means that one is unlikely to observe a system during this phase.
\\
Finally, in the case when $\zeta_{RL}>\zeta^{\rm dyn}_{Rd}$ and $\zeta_{RL}>\zeta^{\rm th}_{Rd}$ the donor star is unable to regain hydrostatic equilibrium without continuously expanding with respect to its Roche lobe on the dynamical timescale. This runaway process is thought to lead to a \ac{CE} phase \cite{1975PhDT.......165W,1976IAUS...73...75P,1976IAUS...73...35V,2013A&ARv..21...59I,2020cee..book.....I}.
In a simple picture of \ac{CE} evolution, the envelope expands so much that it engulfs the entire binary system \cite{2020ApJ...893..106M}.
At some point during this process, the envelope loses co-rotation with respect to the binary orbit\footnote{Assuming the rotation of the donor was synchronized due to strong tides in the phase leading up to \ac{CE} evolution}. 
The companion, now moving through a gaseous medium of the expanded envelope, experiences increased drag force and starts to spiral in deeper into the shared envelope. If there is no way to stop this process, the companion sinks into the core of the giant donor and the binary components merge.
To avoid this scenario, the envelope needs to be ejected from the system, which requires that some amount of energy (at least equal to its binding energy $E_{bind}$) is deposited into the envelope.
The commonly considered source of energy that can be (with some efficiency $\alpha_{CE}$) dissipated into the envelope is that of the shrinking binary orbit. However, many different processes, with various efficiencies, may contribute to the unbinding of the envelope.
Both sides of the energy balance equation that one effectively needs to consider:
\begin{equation} \label{eq: CE}
E_{bind}=\alpha_{CE}\Delta E_{orb} \ \text{(+other energy sources)}
\end{equation}
 are very uncertain and the treatment of the \ac{CE} evolution is one of the biggest sources of uncertainty in binary evolution, inherited by the population models of binary-related systems and transients.
Dynamically unstable interaction is an extremely short-lived phase and it is even less likely to be directly observed than thermal-timescale mass transfer.
In both cases, observational constraints on such binary interactions have to rely mostly on the information carried by post interaction products.
\\
The mass transfer may be dynamically unstable from the beginning, but it may also gradually evolve into instability after a phase of stable mass transfer interaction. This was demonstrated with some of the early detailed binary evolution calculations \cite{1977ApJ...211..486W}
and also found in the adiabatic mass-loss model \cite{1987ApJ...318..794H}. 
Those results show that only after the substantial fraction of the outer envelope 
has been stripped in a stable way, $\zeta^{\rm dyn}_{Rd}$ decreases enough to 
lead to instability.
This phenomenon is referred to as ``delayed'' dynamical instability. A delayed instability may also arise as a result of a thermal-timescale mass transfer.
The system in which the donor star is gradually increasing in size with respect to its Roche lobe may lose a non-negligible amount of mass through the outer Langrangian points (L2/L3).
The matter escaping through outer Lagrangian points carries a lot of angular momentum away from the binary, driving an increased shrinkage of the orbit \cite{2017ApJ...850...59P}.
Understanding the conditions leading to such a delayed instability is an area of active research \cite{2017MNRAS.465.2092P,2020ApJ...899..132G,2021A&A...650A.107M,2022arXiv220912707T,2023ApJ...945....7G}. 
The relatively long duration on which the instability arises (hundreds of years) calls for the use of detailed stellar evolution codes. However, most of such codes are limited to 1D and an interacting binary (with stars highly distorted from their spherical shapes) is clearly not a 1D problem.
While simplifications are unavoidable to make the task computationally feasible, it is certainly worth the effort, given that most cases of unstable mass transfer evolution in the massive-star regime are likely the result of such an instability \cite{2017MNRAS.465.2092P}. This stems from the fact that most massive donors have outer radiative envelopes \cite{2020A&A...638A..55K}, for which an immediate (adiabatic) instability would only occur in cases of extreme mass ratios \cite{2015ApJ...812...40G,2020ApJ...899..132G}. One consequence of a delayed dynamical instability is that at the onset of the \ac{CE} phase the binary is already surrounded by a torus of non-accreted matter, which may affect the observational signatures of optical transient associated with \ac{CE} evolution known as luminous red novae \cite{2011A&A...528A.114T,2016MNRAS.461.2527P,2017MNRAS.471.3200M,2017ApJ...835..282M,2021A&A...653A.134B}.
\\
The above considerations still neglect the potential response of the accreting star to the added mass. 
If the companion is donated mass on a timescale that is shorter than its thermal readjustment, it is driven out of thermal equilibrium.
That is the case, for instance, if the binary enters a thermal timescale mass transfer from a more massive donor to a less massive stellar companion.
Stellar response to mass gain is the reverse to that expected due to mass loss: accreting star with a convective envelope shrinks, while that with a radiative envelope expands and may fill its own Roche lobe leading to a contact binary and possibly affecting the stability of the interaction \cite{1977PASJ...29..249N,1983Ap&SS..96...37H,1984Ap&SS.104...83H,1994A&A...290..129V,1995A&A...297..483B,2021ApJ...923..277R}.
We caution that this is often not accounted for in the binary mass transfer models which treat the accretor as a point mass or are limited to non-conservative mass transfer.
\\
\newline
In summary, the mass transfer is likely to be stable if the donor is radiative ($\zeta^{\rm dyn}_{Rd}>0$) and the binary mass ratio $q$ is not too extreme.
The exact value of the critical mass ratio for massive radiative donors may be a strong function of the orbital separation \cite{2020ApJ...899..132G}.
Conversely, if the donor has a deep outer convective envelope (characterized by $\zeta^{\rm dyn}_{Rd}<0$) then in most cases the mass transfer will become dynamically unstable.
Which stars, and when during their evolution, develop deep convective envelopes is a strong function of both birth mass and metallicity (see Sec.~\ref{sec: radius evolution} and references therein).
In general, the higher the mass of a star and the lower the metallicity, the smaller the parameter space for it to initiate mass transfer at a convective-envelope state.
In the context of the evolution of \ac{BH} binary progenitors this means that:
i) high mass, low metallicity \ac{BH} progenitors evolving in binaries are more likely to avoid the \ac{CE} evolution and only interact through stable mass transfer;
ii) low mass, high metallicity \ac{BH} progenitors evolving in binaries are more likely to engage in \ac{CE} evolution \cite{2021A&A...645A..54K,2021A&A...650A.107M}.
This dichotomy may lead to systematic differences in \ac{BH} merger times and properties of \ac{BH} mergers that originate from low and high metallicity binary progenitors \cite{2021ApJ...922..110G,2021A&A...647A.153B,2022ApJ...931...17V}.
\\
The major sources of uncertainty entering the determination of the stability of binary interaction are related to the amount of mass and angular momentum escaping the system with the non-accreted matter, the stellar (envelope) structure of the donor (setting its response to mass loss), and the response of the accretor star to the rapid gain of mass and angular momentum.
\\
\newline

\subsection{Mass transfer: effect on the evolution}

\subsubsection{Impact on the orbit: stable mass transfer case} \label{sec: orbit stable MT}

The orbital evolution of a binary that undergoes stable mass transfer evolution can be described by investigating the change in its orbital angular momentum $J_{orb}$. Using equation \ref{eq: Jorb} one can derive the general formula:
\begin{equation} 
\rm 2 \frac{\dot{J}_{orb}}{J_{orb}} = 2 \frac{ \dot{M}_{\rm donor}}{M_{\rm donor}} + 2\frac{\dot{M}_{\rm accretor}}{ M_{\rm accretor}} -  \frac{\dot{M}_{\rm donor} + \dot{M}_{\rm accretor} }{M_{\rm donor} +  M_{\rm accretor}} + \frac{\dot{a}}{a} + \frac{(-2 e \dot{e} )}{(1-e^2)}
\end{equation}
It is often assumed that at the onset of the \ac{RLOF} the orbit is fully circularised by tides, i.e. $e\approx0$ therefore allowing to neglect the last term (see e.g. \cite{1997A&A...327..620S,2005MNRAS.358..544R,2007ApJ...667.1170S,2009ApJ...702.1387S, 2010ApJ...724..546S,2016ApJ...825...70D,2016ApJ...825...71D, 2019ApJ...872..119H} for the more general case with non-zero $e$).
The specific angular momentum of the ejected matter, $\rm j_{loss}$, is often expressed in the units of the specific angular momentum of the binary by introducing a parameter $\gamma$:
\begin{equation}
\begin{split}
\label{eq: hloss}
\rm j_{loss} = \gamma \frac{J_{orb}}{M_{\rm donor} + M_{\rm accretor} } =  \frac{\dot{J}_{orb}}{\dot{M}_{\rm donor} + \dot{M}_{\rm accretor} } , \\
\rm \rightarrow \frac{\dot{J}_{orb}}{J_{orb} } =  \frac{\gamma (1 - \beta)\dot{M}_{\rm donor}}{M_{\rm donor} + M_{\rm accretor} }.
\end{split}
\end{equation}
The effect on the orbit then depends on the
conservativeness $\beta = - \dot{M}_{\rm accretor}/\dot{M}_{\rm donor}$ and $\gamma$ parameters:
\begin{equation}
\label{eq: sep evol}
\rm \frac{\dot{a}}{a} =  -2 \frac{\dot{M}_{\rm donor}}{M_{\rm donor}} \left( 1 - \beta \frac{M_{\rm donor}}{M_{\rm accretor}} -(\gamma + \frac{1}{2})(1 - \beta) \frac{M_{\rm donor}}{M_{\rm donor} + M_{\rm accretor} }   \right)
\end{equation}

\begin{figure}[h!]
  \begin{minipage}[c]{0.6\textwidth}
    \includegraphics[width=1.\textwidth]{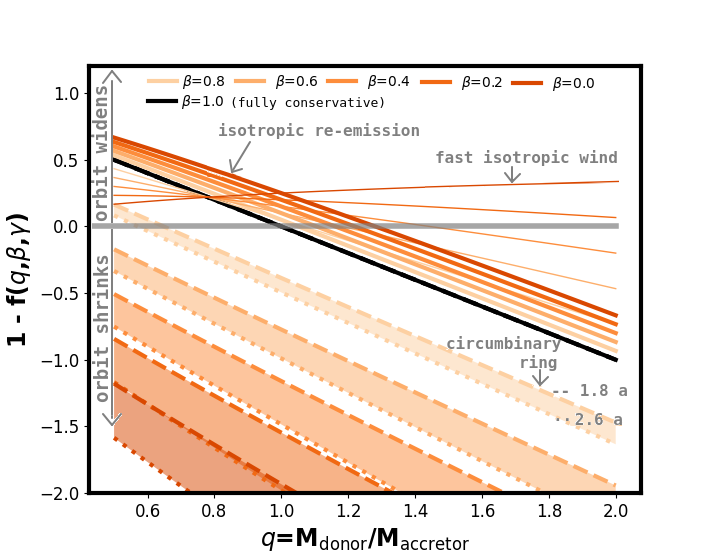}
  \end{minipage}\hfill
  \begin{minipage}[c]{0.4\textwidth}
    \caption{
    Visual representation of the expression
    1 - $f(q, \beta, \gamma) = 1 - q \beta  -(\gamma + \frac{1}{2})(1 - \beta) \frac{q}{1 + q }$ from equation \ref{eq: sep evol} that, for a given set of parameters $\beta$ and $\gamma$ determines the evolution of the orbital separation of the binary with mass ratio $q$.
    If this expression is negative, the orbital separation decreases as the donor loses mass.
    Black line corresponds to the conservative mass transfer case.
    Colors indicate different values of $\beta$, different line styles indicate different modes of angular momentum loss.
    Thick solid lines correspond to isotropic re-emission ($\gamma = q$), thin solid lines to fast isotropic wind ($\gamma = 1/q$),
    colored ranges span the range of $\gamma=\gamma_{\rm ring}$ of the circumbinary ring with $a_{\rm ring}$ between 1.8$a$ (dashed) and 2.6$a$ (dotted).
    }             
    \label{fig: orbital evolution}
  \end{minipage}
\end{figure}
Because $\rm \dot{M}_{\rm donor} < 0$, the orbit shrinks ($\rm \dot{a} < 0$) when the expression in the parenthesis in equation \ref{eq: sep evol} is negative (see Figure \ref{fig: orbital evolution} for the illustration). This is, for instance, the case for conservative mass transfer ($\beta$=1, $\gamma$=0) as long as $q>1$. As soon as the mass ratio reverses (and $q$ drops below $1$), the orbit begins to expand.
The parameters $\beta$ and $\gamma$, and the way how they depend on the properties of particular stars and binaries, are usually poorly known.
The value of $\gamma$ strongly depends on where the non-accreted matter is lost from the system.
The most commonly considered, idealized scenarios include the ``fast (Jeans) isotropic emission'' from the vicinity of the donor (typical assumption for mass loss through stellar wind), in which case $\gamma = \frac{M_{\rm accretor}}{M_{\rm donor}}$ or the
``isotropic re-emission'' from the proximity of the accretor, leading to $\gamma = \gamma_{\rm iso}  = \frac{M_{\rm donor}}{M_{\rm accretor}}$ (typical assumption for mass loss during non-conservative mass transfer).
The latter case is also though to be a good approximation for the mass ejected through jets or fast disk-winds launched in the proximity of a compact accretor in X-ray binaries. 
However, if (part of) the ejecta is slow-moving, the non-accreted mass may be escaping with notably higher specific angular momentum than in both the above cases. Such ejecta continues to exert torques on the binary orbit and gain angular momentum while moving outwards. This may happen, for instance, when part of the non-accreted mass forms a curcumbinary disk or torus, as seems to be the case in the SS433 system, an X-ray binary with a stellar \ac{BH} accretor that currently evolves through a face of stable and largely non-conservative mass transfer \cite{2004ASPRv..12....1F,2006MNRAS.370..399B,2020NewAR..8901542C}. A formation of a circumbinary disk may also occur when the Roche lobe of the accretor becomes overfilled with mass, such that it reaches the critical potential surface corresponding to the outer Lagrangian point $L_{2}$. The matter will then flow through $L_{2}$ and, not having enough energy to escape, starts orbiting the system on a Keplerian orbit at a certain distance $a_{\rm ring}$ from its center of mass.
In ejection through a circumbinary disk scenario, the specific angular momentum of the escaping mass $\gamma$=$\gamma_{\rm ring}$ depends on the $a_{\rm ring}/a$ ratio \cite{1994ApJ...421..651A}:
\begin{equation}
\rm \gamma_{\rm ring} = \frac{(M_{\rm donor} + M_{\rm accretor})^{2}}{M_{\rm donor}M_{\rm accretor}} \sqrt{\frac{a_{\rm ring}}{a}} = \frac{(q+1)^{2}}{q} \sqrt{\frac{a_{\rm ring}}{a}} 
\end{equation}{}
The three modes of angular momentum loss introduced above are compared in Figure \ref{fig: orbital evolution}. For the illustration of the circumbinary ring angular momentum loss scheme, we follow \cite{1994ApJ...421..651A} and consider
$a_{\rm ring}/a$ between $1.8$ and $2.6$ as  plausible values.\\
It is worth noticing that the wind-type of mass loss typically leads to the widening of the orbit (especially in the lack of wind accretion $\beta$=0). Wind mass loss rate of massive stars $\dot{M}_{\rm donor; \ wind}$ is expected to heavily depend on the metallicity (the higher the metallicity, the higher $\dot{M}_{\rm donor; \ wind}$ \cite{1987A&A...173..293K,2000A&A...362..295V,2001A&A...369..574V,2007A&A...473..603M,2021MNRAS.504.2051V}). This means that when similar types of systems are compared, one can expect higher metallicity binaries to be wider.
One can also notice that if the non-accreted mass carries sufficiently high specific angular momentum (as in the example of the circumbinary ring), the orbit can shrink for a wide range of mass ratios and conservativeness parameters.\\
Understanding the extent to which the orbital separation can decrease as a result of stable mass transfer is crucial in the context of the formation of merging \ac{BBH} systems.
For the \ac{BBH} system to merge within the time shorter than the age of the Universe,
its separation needs to be smaller than a few 10 $\Rs$ (depending on the masses, see Section \ref{sec: GW}).
Typical \ac{BH} progenitors reach radii $\gtrsim 100 \ \Rs$ during their lifetime (see Figure \ref{fig: radius evolution}), so the formation of a \ac{BBH} merger through isolated  evolution typically requires a mass transfer interaction that significantly shrinks the orbital separation (see also Section \ref{sec:Formation_pathGW}).
Whether this is feasible with stable mass transfer evolution is a matter of an ongoing debate in the literature \cite{2017MNRAS.471.4256V,2021A&A...645A..54K,2021A&A...650A.107M,2021A&A...647A.153B,2021ApJ...922..110G,2022ApJ...940..184V}.
An alternative, that was long considered the dominant mechanism responsible for the drastic orbital tightening of the \ac{BBH} merger progenitors formed in isolated binaries, is the \ac{CE} evolution.

\subsubsection{Impact on the orbit: unstable (\ac{CE}) case} \label{sec: orbit CE}

Once a binary is deemed to undergo an unstable mass transfer and \ac{CE} evolution, the most important question that one
has to answer is whether the system is going to survive this interaction.
If the answer is negative, the binary components merge and a single object is formed, possibly with some unique or exotic properties. Stellar mergers are hypothesized to lead to the formation of, for instance,  massive stars with strong magnetic fields \cite{2019Natur.574..211S}, blue stragglers and blue \ac{MS} in clusters \cite{1996ApJ...468..797L, 2013MNRAS.434.3497G,2022NatAs...6..480W}, Thorne-Zytkov objects \cite{1977ApJ...212..832T,1995MNRAS.274..485P,2022MNRAS.513.4802A}, isolated binary-origin \ac{BH} with mass in the pair-instability gap \cite{2020MNRAS.497.1043D,2020ApJ...904L..13R,2022MNRAS.516.1072C}.
The details of the merging process as well as the fate and properties of the merger product are uncertain.
If the binary can avoid the merger during the CE evolution, one needs to determine the post-interaction properties of the system - in particular the final separation.
\ac{CE} is a complex problem that is inherently challenging to model, as it requires accounting for physical effects happening on very different spatial and temporal scales \cite{1993PASP..105.1373I,2013A&ARv..21...59I,2000ARA&A..38..113T,2020cee..book.....I}.
In the big picture, the outcome of \ac{CE} evolution crucially depends on the binding energy of the envelope as well as on the efficiency with which energy from various sources is deposited in the envelope (and can ultimately lead to its ejection).
That is the essence of the  so-called energy formalism \cite{1984ApJ...277..355W}, which allows one to obtain a crude estimate of the final (post-\ac{CE} ejection) orbital separation of the binary and is commonly used e.g. in population synthesis studies \cite{1996A&A...310..489L,1996A&A...309..179P,2008ApJS..174..223B,2014A&A...564A.134M,2018MNRAS.481.1908K,2018MNRAS.479.4391M,2019MNRAS.485..889S,2020ApJ...898...71B,2022ApJS..258...34R} (see also Section \ref{sec:Intro_GWpaleo}).
In this framework, the difference in orbital energy between the \ac{CE} onset and \ac{CE} ejection 
multiplied by an efficiency parameter $\alpha_{CE}$ is compared against the energy needed to unbind the envelope (sometimes additional energy sources, contributing with different efficiencies are considered as in equation \ref{eq: CE}).
Another parameter,  $\lambda_{CE}$ is used to characterise the binding energy of the envelope of the donor as follows:
\begin{equation}
    E_{bind} = \frac{G M_{\rm donor} M_{\rm env}}{\lambda_{CE} R_{\rm donor}}
\end{equation}
here $M_{\rm env}$ is the mass of donor's the envelope.
Equation \ref{eq: CE} then takes the form:
\begin{equation}\label{eq: CE energy balance}
    \frac{G M_{\rm donor; ini} M_{\rm env}}{\lambda_{CE} R_{\rm donor}} = \alpha_{CE} \left( 
    \frac{G M_{\rm donor; fin} M_{\rm accretor; fin}}{2 a_{fin}} - \frac{G M_{\rm donor; ini} M_{\rm accretor; ini}}{2 a_{ini}}
    \right) \ (+ \text{other energy sources})
\end{equation}
where the indices \textit{`ini'} and \textit{`fin'} refer to the values of masses and separations at the CE onset and after the CE ejection, respectively, and $M_{\rm donor; fin}= M_{\rm donor; ini}-M_{\rm env}$.
In the simplest formulation, if we neglect the other energy sources term in Equation~\ref{eq: CE energy balance}\footnote{In practice, if energy sources other than the orbital energy term are included, then usually it is done in an implicit way, e.g. by increasing the value of $\alpha_{CE}$ or decreasing the $\lambda_{CE}$ parameter.}, the final separation becomes:
\begin{equation}
\label{eq. postCEsep}
a_{fin} = M_{\rm donor; fin} M_{\rm accretor; fin}
\left(\frac{ M_{\rm donor; ini} M_{\rm accretor; ini}}{a_{ini}} + \frac{2 \alpha_{CE}}{\lambda_{CE}} \frac{M_{\rm donor} M_{\rm env}}{R_{\rm donor}}\right)^{-1}
\end{equation}
Cases when $a_{fin}$ is so small that the helium core of the donor star would overflow its Roche lobe are interpreted as \ac{CE} events leading to stellar mergers. In the energy formalism, all the complexity of the \ac{CE} process is then hidden within the $\alpha_{CE}$ and $\lambda_{CE}$ parameters.
\\
It can be seen that the higher the $\lambda_{CE}$, the more loosely bound the envelope.
The value of $\lambda_{CE}$ can be obtained from 1D stellar models.
It was shown that $\lambda_{CE}$ can take values spanning orders of magnitude, depending on the envelope type (radiative or convective) and the evolutionary stage and internal structure of the \ac{CE} donor \cite{2010ApJ...716..114X,2016A&A...596A..58K}.
Detailed calculations show that $\lambda_{CE}$ tends to increase with the increasing fraction of mass contained in the outer convective envelope and that this increase may be quite drastic - $E_{bind}$ can drop by more than a factor of 10 as the envelope of a massive giant becomes convective \cite{2021A&A...645A..54K}.

In order to compute $E_{bind}$ from a stellar model one needs to assume the exact location of the boundary between the envelope (that is to be ejected) and the core of the giant CE donor. Because most of the binding energy is located in deep envelope layers near the core, the assumed core-envelope boundary can have a substantial effect on the value of $E_{bind}$ (even more than an order of magnitude difference in the case of convective-envelope donors \cite{2000A&A...360.1043D,2001A&A...369..170T,2016A&A...596A..58K}). Multiple criteria for the core-envelope boundary for CE evolution have been proposed in the literature but no consensus is reached \cite{2013A&ARv..21...59I,2020cee..book.....I}.  Notably, depending on the amount of hydrogen-rich material left on top of the helium core after the envelope is ejected, a CE phase may be immediately followed by another phase of (stable) mass transfer \cite{2011ApJ...730...76I,2019ApJ...883L..45F,2021A&A...645A..54K,2021arXiv211112112M,2022MNRAS.511.2326V,2022ApJ...937L..42H}.
This would continue to affect the orbital separation of the system and is not captured by the simple parametrisation in Equation~\ref{eq. postCEsep}.
\\
The second parameter in Eqn.~\ref{eq. postCEsep}, $\alpha_{CE}$,  represents the efficiency with which orbital energy is deposited inside the envelope during the inspiral. In principle, its value should thus be limited to $\alpha_{CE} < 1$. 
However, any result pointing to $\alpha_{CE} > 1$ is not always unphysical as it may indicate that additional energy sources are present in the system and contribute in ejecting the envelope (or that the $E_{bind}$ term has been overestimated). The most likely agents for additional energy input are the internal energy of the envelope (gas and radiation) and the energy released upon recombination of hydrogen and helium \cite{2015MNRAS.450L..39N,2015MNRAS.447.2181I}. These energy sources may not act immediately as they require the envelope to expand, decompress, cool, and recombine.
Some of the released energy is then reprocessed into kinetic energy (and may help to remove the envelope) while the rest is likely radiated away and lost from the system \cite{2018ApJ...863L..14S,2018MNRAS.478.1818G}.
The efficiency of these processes is unclear and remains an area of active research, with some of the key aspects being the timescale on which the internal energy is released and where in the envelope it occurs \cite{2016MNRAS.460.3992N,2022MNRAS.512.5462L,2022MNRAS.516.4669L}.
It is important to realize that the potential contribution of internal energy is often already incorporated in the $\lambda_{CE}$ values reported in the literature (rather than in the energy term and $\alpha_{CE}$), where it is assumed to lower the binding energy of the envelope with maximum eficiency (i.e. neglecting radiative loses) \cite{2010ApJ...716..114X,2014A&A...563A..83C,2016RAA....16..126W,2016A&A...596A..58K,2021A&A...645A..54K}.
There is no consistency in the literature in this respect and when combining different $\lambda_{CE}$ and $\alpha_{CE}$ estimates one should make sure that no potential energy sources (in particular the internal energy contribution) are counted twice.
\\
Constraining the value of $\alpha_{CE}$ has been at the forefront of \ac{CE} research for a long time. On the theoretical side, 3D hydrodynamic simulations of the \ac{CE} inspiral suggest $\alpha_{CE} < 1$ (e.g. $\alpha_{CE}$ = 0.046-0.25 in \citealt{2017MNRAS.470.1788C}, $\alpha_{CE} \lesssim 0.5$ based on \citealt{2015MNRAS.450L..39N,2016MNRAS.460.3992N}, $\alpha_{CE} \lesssim 0.4$ found by \citealt{2020arXiv201106630L}, and $\alpha_{CE}$ = 0.3-0.7 across the simulations of \citealt{2022A&A...660L...8O,2020A&A...644A..60S,2022MNRAS.512.5462L} when combined with $\lambda_{CE}$ that includes internal energy). A notable result from simulations are cases when the inspiral slows down or ceases without ejecting the entire envelope (e.g. just the convective part), which lowers the amount of binding energy that needs to be overcome and, at the face value, increases the effective $\alpha_{CE}$ value to even $\gtrsim 1-2$ \cite{2019ApJ...883L..45F,2022MNRAS.512.5462L,2022A&A...667A..72M}.
It is important to realize that such a variant of \ac{CE} evolution, most likely only realistic in the case of convective-envelope donors with a steep density gradient at the bottom of the convective zone, is likely to be followed by another phase of mass transfer that would strip the remaining H-rich layers \cite{2011ApJ...730...76I,2020A&A...637A...6L,2022MNRAS.511.2326V}.
 The final orbital separation (and the effective value of $\alpha_{CE}$) may further be affected by tidal torques from the recently ejected material \cite{2023arXiv230200691G} and is likely to depend on the mass ratio \cite{2022A&A...667A..72M}.
Finally, it is worth noting that the vast majority of 3D simulations of the \ac{CE} phase to date has been limited to convective-envelope donors, which are the typical case in the low- to intermediate-mass regime, but may not be the most common case of donors in the massive star regime \cite{2020A&A...638A..55K}.
\\
Observationally, $\alpha_{CE}$ can be constrained by studying post-\ac{CE} systems (e.g. double white-dwarf binaries) and reconstructing their pre-\ac{CE} orbits \cite{2000A&A...360.1011N,2010A&A...520A..86Z,2011MNRAS.411.2277D,2011A&A...536A..42Z,2012MNRAS.419..287D,2022arXiv221102036S}. While this approach suffers from uncertainties in the reconstruction of pre-\ac{CE} parameters, it generally points out to $\alpha_{CE} < 1$ values, in line with most theoretical results. A notable exception are \ac{BH} low mass X-ray binary systems, the formation of which through a \ac{CE} evolution channel would possibly require $\alpha_{CE} > 1$ efficiencies \cite{1999ApJ...521..723K,2003MNRAS.341..385P}, although see \cite{2021A&A...645A..54K}. 

In the context of \acp{BH}, the energy budget for \ac{CE} evolution leading to the formation of merging binary \ac{BH} systems has been discussed extensively by \cite{2014A&A...564A.134M,2016A&A...596A..58K,2021A&A...645A..54K,2021A&A...650A.107M}. These authors conclude that in the case of massive giants (\ac{BH} progenitors) the envelope binding energies are generally too large to allow for a successful \ac{CE} ejection for $\alpha_{CE} \leq 1$, unless the \ac{CE} phase is initiated by a convective-envelope star: a red supergiant (but see also \cite{2021A&A...649A.114G}). This might be a severe limitation for evolutionary channels involving a \ac{CE} phase:  whereas in the low-mass regime mass transfer from a convective donor is expected to be common, in the high-mass regime of \ac{BH} progenitors it is limited to a narrow range of orbital periods \cite{2008AIPC..990..230D,2020A&A...638A..55K} and may be completely impossible above a certain mass ($\sim 40-50 \rm M_{\odot}$) corresponding to the the Humphreys-Davidson limit \cite{1979ApJ...232..409H,1994PASP..106.1025H,2018MNRAS.478.3138D}. 

Importantly, the conclusion that \ac{CE} events from radiative-envelope donors result in mergers are based on 1D hydrostatic stellar models and ad hoc assumptions regarding the onset of the \ac{CE} phase as well as the final state following the inspiral \cite{2014A&A...564A.134M,2016A&A...596A..58K,2021A&A...645A..54K,2021A&A...650A.107M}.
All these factors are uncertain. 
Firstly, the internal envelope structure of a \ac{CE} donor cannot always be robustly predicted from 1D stellar models. For example, \citet{2021A&A...645A..54K} showed that two red supergiant models of a similar mass, radius, and luminosity can be characterized by significantly different envelope density profiles and $E_{bind}$ values, an effect arising from differences in their internal chemical abundance profiles. Recently, \citet{2022arXiv220615338R} demonstrated that the envelope structure of past accretors shows systematic structural differences compared to envelopes of single stars, leading to lower $E_{bind}$ values. This is important in the context of binary \ac{BH} merger formation. In the classical formation scenario of such systems, the secondary star becomes a \ac{CE} donor when the primary is already a \ac{BH} and after evolving through a mass transfer episode initiated by the primary.

Secondly, the onset of the \ac{CE} phase is not immediate but is preceded by a phase of pre-\ac{CE} mass transfer before the interaction becomes dynamically unstable \cite{2020ApJ...893..106M}. In the case of radiative-envelope donors, this phase can last for hundreds of years, with a significant (but poorly constrained) fraction of the donor's envelope becoming stripped even before the \ac{CE} inspiral begins \cite{2013A&ARv..21...59I,2020ApJ...895...29M,2021A&A...653A.134B,2021A&A...645A..54K}. 
On the far end of the \ac{CE} phase, it is unclear where exactly in the envelope may the inspiral slow down to the point where the further orbital evolution is governed by secular processes such as stable mass exchange (stripping the final envelope layers) and tidal torques between the orbit and all the matter ejected in the process, forming a rich and complex circumbinary medium.  

Last but not least, an energy source that is rarely considered is the feedback from accretion onto the compact object as it spirals in through the envelope. While Eddington-limited accretion luminosity (see Section \ref{sec: BH accretor}) would not necessarily be a 
significant addition to the energy budget \cite{2016A&A...596A..58K,2021A&A...645A..54K}, the accretion rate may exceed this limit due to neutrino cooling or photon trapping \cite{2004NewAR..48..843E,2015ApJ...798L..19M,2020ApJ...897..130D}. If an accretion disk forms (which is not necessarily the case \cite{2017ApJ...845..173M}) then disk winds and jets could transfer some of the super-Eddington power into kinetic energy, thus having a massive impact in the envelope ejection process \cite{2000ApJ...532..540A,2015ApJ...800..114S}.

\subsubsection{Impact on the donor star}\label{sec: impact on donor}

As a result of the mass transfer, the donor star loses some fraction or even the entire envelope and becomes a stripped star \cite{1969A&A.....3...83K,1967AcA....17..355P,1992ApJ...391..246P,1996ASPC...96..111H,2002PASA...19..233P,2012ARA&A..50..107L}.
The basic observable properties (e.g. radius, effective temperature and surface composition) of such a star are vastly different compared to a non-stripped star of the same mass.
If the stripping process is efficient, hot inner layers of the star are exposed and the mass-loser appears as a compact helium-rich stripped star.
This further leads to an increase in the wind mass loss that it experiences and the amount of ionising radiation that it produces \cite{2002PASA...19..233P,2003A&A...400...63V,2007ARA&A..45..177C,2009MNRAS.400.1019E,2017A&A...608A..11G,2019ApJ...878...49W}.
Especially if the interaction happens early enough during the evolution of a star, it can limit the growth of the core and reduce its compactness, thus affecting the outcome of the core collapse and the formation of a compact remant with respect to single star evolution predictions \cite{1992ApJ...391..246P,2019ApJ...878...49W,2020MNRAS.499.2803P,2021A&A...656A..58L,2021A&A...645A...5S,2021ApJ...916L...5V}.
In this context, early mass transfer is the one that can have an impact on the properties of the helium core of the star.
Helium core mass and and the inner carbon-oxygen core (formed as a result of core-helium burning) to a large extent determine the subsequent evolution and the final fate of a star \cite{1978ApJ...219.1008A,1980SSRv...25..155S}. Whatever happens to the outer layers of a star once the carbon-oxygen core mass is established is found to have relatively little impact on the core-collapse \cite{2004ApJ...612.1044P,2014ApJ...783...10S,2020MNRAS.499.2803P}.
However, it may still affect the observable characteristics of the potential supernova (in particular their classification as being envelope stripped, based on the lack of H or He lines in the supernova spectrum \cite{2017hsn..book..195G}).
Furthermore, the mass remaining in the envelope sets the limit on the total amount of mass that can be ejected or accreted during the formation of the compact object.
% potentially also affecting the magnitude of the natal kick velocity that it gains at formation \cite{2017ApJ...837...84J}.
% Both mass ejecta and the natal kick leave imprint on the orbital parameters of the binary, and may even lead to its disruption \cite{1961BAN....15..265B,1998A&A...330.1047T}.
% The latter is particularly important in the context of binary evolution - mass ejection from the collapsing star can significantly affect the orbital parameters, or even unbind the system \cite{1961BAN....15..265B,1998A&A...330.1047T}, regardless of whether the newborn compact object gains velocity (natal kick) at its formation or not.

\paragraph{Early mass transfer maximizes the differences with respect to single star evolution:}

In view of the example shown in Figure \ref{fig: radius evolution}, early and efficient stripping can be expected for high metallicity \ac{BH} progenitors, that are likely to become donors when rapidly expanding on the Hertzsprung gap and get quickly stripped of their hydrogen rich envelopes.
Because the envelope loss takes place shortly after the onset of hydrogen-shell burning, there is not enough time for the helium core to grow in mass any further. 
The exposed helium core of a massive star (a Wolf-Rayet star) experiences strong stellar winds that instead reduce the core mass \cite{1987ARA&A..25..113A,1996ASPC...96..111H,2000A&A...360..227N,2007ARA&A..45..177C}.
If the collapse of its core eventually leads to an explosion, it is likely to be observed as a stripped-envelope supernova (i.e. classified as a hydrogen or possibly even helium poor supernova type Ib/c).
Recent models suggest that stars that encounter this kind of binary interactions tend to form lower mass compact objects \cite{2021A&A...645A...5S,2021ApJ...916L...5V}.
If that is the case, higher initial mass is required to produce a stellar \ac{BH}, which acts to reduce the expected number of \ac{BH} forming per unit of star formation \cite{2021A&A...645A...5S}.
\\
In contrast, if a star becomes a donor during the core helium burning phase or even later, its helium core may still grow to some extent before the hydrogen envelope is stripped. 
More gradual expansion during this phase allows the mass transfer to last much longer (nuclear timescale) than in the case of mass transfer from a Hertzsprung gap donor (thermal timescale) \cite{2022A&A...662A..56K}.
If the Wolf-Rayet star is formed as a result of such a mass transfer, its lifetime (and the associated strong wind mass loss) is thus shorter. 
If the mass-loser eventually explodes as a supernova, the observable event may still show signs of stripping but the He lines are likely to be present (i.e. it might be classified as hydrogen-deficient supernova type IIb or Ib).
\\
Onset of mass transfer at an even more advanced evolutionary stage (after core-helium depletion) further reduces the
importance of the interaction for the remaining evolution of the donor. Unless the mass transfer is dynamically unstable, there might not be enough time to strip the entire hydrogen envelope of the star before the core-collapse.
The helium core is already established at this point of evolution.

\paragraph{Fully or partially stripped (and why it matters):}

The amount of residual H-rich envelope left atop the helium core after mass transfer was shown to depend on metallicity, with metal-poor stars
retaining more hydrogen after the interaction \cite{2017MNRAS.470.3970Y,2017A&A...608A..11G,2020A&A...637A...6L,2022A&A...662A..56K}. The process of envelope stripping may become even less efficient in the case of massive core-helium burning donors (common in metal-poor populations), which were found to interact on the long nuclear timescale and/or lead to partially stripped stars with $\gtrsim 5-30\%$ of the initial envelope left unstripped until the end of core-helium burning, depending on binary parameters \cite{2022A&A...662A..56K}. Such stars are predicted to be much larger and cooler than normal stripped (helium) stars, producing far less ionizing radiation. The possibility of partial envelope stripping is usually neglected in studies that approximate binary evolution effects by considering naked helium stars as the starting point. Similarly, it is often assumed that as a result of a successful \ac{CE} evolution
is the ejection of the entire hydrogen-rich envelope, leaving behind a helium-star remnant. This is likely not strictly the case, as the \ac{CE} inspiral may stall before all the hydrogen-rich layers are ejected \cite{2011ApJ...730...76I,2021arXiv211112112M,2022MNRAS.511.2326V,2022ApJ...937L..42H}.
\\
Understanding the exact degree of envelope stripping experienced by donors in various types of binary interactions is important not only because of the effect it has on their observable characteristics, but also because it may significantly affect the subsequent evolution of the system as a whole.
Surface properties of a star determine its wind mass loss and consequently the orbital evolution of a post-interaction binary (and likely the final separation before the core collapse). Additionally, stars that retain even a tiny hydrogen-rich envelope can significantly re-expand later during their evolution, in contrast to their fully stripped counterparts that remain compact \cite{2020A&A...637A...6L}, leading to another phase of \ac{RLOF}. 
\\
Finally, the mass remaining in the envelope of the star at the end of its life may affect the properties of the binary that remains after the core collapse.
In general, the formation of a compact object is thought to be accompanied by some amount of mass ejection from the outer layers of its immediate progenitor. 
If significant, such mass ejection from the collapsing star can considerably affect the orbital parameters, or even unbind the system
 \cite{1961BAN....15..265B,1998A&A...330.1047T}.
Furthermore, a newly born compact object is thought to gain velocity (natal kick) due to asymmetries (either in mass loss or neutrino loss \cite{2006ApJS..163..335F,2013MNRAS.434.1355J,2017ApJ...837...84J}) involved in the process of its formation. While in the case of \acp{NS} there is ample observational evidence that at least some fraction of them forms with substantial natal kick ($\gtrsim$a few hundred km/s) \cite{1970ApJ...160..979G,
1994Natur.369..127L,
1997MNRAS.291..569H,
2002ApJ...568..289A,
2002ApJ...573..283P,
2005MNRAS.360..974H,
2017A&A...608A..57V,
2020MNRAS.494.3663I,
2021MNRAS.508.3345I}, this is currently much less clear in the case of \ac{BH} formation  \cite{2003Sci...300.1119M,
2004MNRAS.354..355J,
2007ApJ...668..430D,
2009ApJ...697.1057F,
2012MNRAS.425.2799R,
2016MNRAS.456..578M,
2017MNRAS.467..298R,
2022arXiv221102158K,
2022NatAs...6.1085S}.
In some scenarios, natal kick velocity has been proposed to scale with the mass of the ejected envelope (the lower the mass of the ejecta, the lower the kick velocity \cite{2017ApJ...837...84J}). 
If that is the case, lower mass of the envelope remaining at the core collapse can increase the chances that the binary remains bound after the compact object formation.
It has been suggested that for sufficiently massive progenitors, \ac{BH} formation may occur via the collapse of the entire star and the mass loss can be completely avoided \cite{2011ApJ...730...70O}. In such a scenario, higher envelope mass allows for the formation of a more massive compact object.

\subsubsection{Impact on the accretor: stellar accretors}\label{sec: stellar accretor}

\label{sec:accretors}

Stars spend most of their life on the main sequence and the less massive component of a binary is likely to still be at this stage of evolution at the onset of the first mass transfer episode.
During a mass exchange, some fraction of the matter lost by the donor can reach and may be accreted by the companion (thereby called the accretor).
This not only affects its observable characteristics (e.g. increased luminosity, fast rotation, envelope enriched with the CNO-processed material from the donor), but may also significantly affect its subsequent evolution \cite{1977PASJ...29..249N,1983Ap&SS..96...37H,1984Ap&SS.104...83H,1994A&A...290..129V,1995A&A...297..483B,2021ApJ...923..277R}.

\paragraph{Rejuvenation and core growth (or lack thereof):}

The accretor has to adjust its size and structure to accommodate additional material.
As noted earlier, in some cases this may lead the accretor to filling its own Roche lobe, possibly affecting the stability and characteristics of the mass transfer.
Furthermore, as the accretor responds to the new mass, fresh hydrogen-rich fuel from the outer layers can be mixed into its convective core, `rejuvenating' the star - its further evolution resembles that of a younger star with a higher initial mass
 \cite{1983Ap&SS..96...37H,1984Ap&SS.104...83H,2021ApJ...923..277R}.
Such response to accretion affects the structure of the (helium-rich) core-envelope boundary of the star. That layer is where the steep rise in the density profile happens, and hence it stores most of the binding energy of the stellar envelope. 
\cite{2022arXiv220615338R} show that stars that evolved through a mass transfer phase as accretors may have systematically lower envelope binding energies  compared to single stars of the same  radius and mass.
If such stars subsequently initiate a \ac{CE}, the envelope may be easier to eject and leave behind a wider binary compared to the case with \ac{CE} donors whose core-envelope boundary was not affected by earlier accretion.
Finally (inversely to the donor's case), early enough accretion leading to rejuvenation allows for the formation of a more massive He core and consequently, \ac{NS}/\ac{BH} formation from initially less massive stars compared to single stellar evolution predictions.
Rejuvenation is sensitive, for instance, to the composition of the accreted matter \cite{2007MNRAS.376...61D} and to uncertain stellar evolution model assumptions - in particular, it may be completely avoided if internal mixing (semiconvection) is inefficient \cite{1995A&A...297..483B}.
Unless they are fully rejuvenated, accretors are bound to have different core to envelope mass ratios than single stars (likely having undermassive helium cores and overmassive envelopes for their mass). This was shown to affect the radius evolution of a star \cite{2013MNRAS.434.3497G,2014ApJ...796..121J} and is a potentially important (but poorly explored) effect for future binary interactions.
 \\
At the moment, the impact of mass transfer on the accretor star briefly discussed above is not taken into account in most current binary evolution population models, in which it is customary to approximate the subsequent evolution of the accretor with a single (possibly fast-rotating) stellar model.

\paragraph{Spin-up (followed by the likely spin-down):}

The matter that is transferred to the accretor carries angular momentum and spins up the star \cite{1981A&A...102...17P,1991A&A...241..419P,1991ApJ...370..604P}.
The transferred angular momentum can be particularly significant if the accretion occurs through a disc formed around the star, in which case the accretor may reach its critical/break-up rotation ($\omega_{crit}=(GM_{*}/R_{*}^{3})^{1/2}$ i.e. the limiting rotational velocity at which the centrifugal force at the equator is equal to the gravitational force) after accreting only a small fraction of its initial mass \cite{1981A&A...102...17P}.
This can possibly be avoided in very tight binaries (with orbital periods $\lesssim$a few days), where efficient tidal interactions can prevent the star from reaching critical rotation by locking its spin period to the orbital period.
However, if the star reaches its critical rotation, it cannot accomodate any more angular momentum.
This possibly halts further mass accretion \cite{2005A&A...435.1013P,2007AIPC..948..321D}, either making the mass transfer highly non-conservative or forcing the mass to accumulate in the disk (from which it might eventually still be accreted if the stellar rotation decreases).
On the other hand, \cite{1991ApJ...370..597P,1991ApJ...370..604P} argue that the critically rotating star can continue accreting mass from an accretion disk without accreting angular momentum as a result of viscous coupling.\\
Increased rotation increases the wind mass loss rate \cite{1986ApJ...311..701F,2000A&A...361..159M}, may significantly enhance the mixing inside the star (\cite{1974IAUS...59..185Z,1987A&A...178..159M,2000ApJ...528..368H}; see Section \ref{sec: CHE} for further discussion of this point) and overall, can have a profound impact on stellar evolution \cite{2000ARA&A..38..143M}.
However, it is unclear whether the star can retain high rotation rate throughout its evolution.
In the context of \ac{BH} properties, it is particularly important to understand whether the accretion-enhanced rotation can be transported all the way to the core and retained until the core-collapse, so that it can potentially lead to the formation of a highly spinning compact object, and possibly a long gamma ray burst progenitor \cite{2005A&A...443..643Y,2007A&A...465L..29C}.
To this end, one needs to understand i) how much angular momentum is lost by the star during its subsequent evolution and ii) how is the angular momentum transported and distributed within the star (the internal rotation does not need to be uniform, rotation of the core may decouple from the envelope).
Apart from binary effects (mass transfer and tidal interactions) that obviously need to be considered to answer i), one has to account for stellar winds.
At low metallicity, wind mass loss is smaller and the associated removal of angular momentum during evolution much weaker, in principle allowing to retain higher rotation \cite{2001A&A...373..555M}.
As long as the rotation of the core and the envelope remain coupled, the loss of angular momentum from the envelope also slows down the rotation of the core (and vice versa).
In fact, some mechanism to extract the angular momentum from the core appears necessary to avoid overpredicting the rotation rates of young \ac{NS} and \ac{WD} \cite{2005ApJ...626..350H,2008A&A...481L..87S}.
Such mechanism can be provided by magnetic torques and the Spruit-Tayler dynamo model \cite{2002A&A...381..923S}
% that mostly removes angular momentum from cores of main-sequence stars 
appears sufficient to address the above issue \cite{2005ApJ...626..350H,2008A&A...481L..87S}.
However, this mechanism alone cannot explain the asteroseismologically observed angular momentum loss of the core of red giants and helium core-burning low mass stars \cite{2014ApJ...788...93C,2019MNRAS.485.3661F,2018A&A...616A..24G}.
This led to the conclusion that more efficient angular momentum transport may be required, as in the re-formulation of the  Spruit-Tayler dynamo model proposed by \cite{2019MNRAS.485.3661F}.
If those arguments apply to more massive \ac{BH} progenitors, the model of \cite{2019MNRAS.485.3661F} predicts that the magnetic torques remove most of the angular momentum from the helium core just after the main sequence and consequently, \ac{BH} natal spins are expected to be small \cite{2019ApJ...881L...1F}.
Possible exceptions from this rule involve \ac{BH} progenitors spun up through tidal interactions in close binaries either at very low metallicity or at the WR stage in WR+BH binaries (see Section \ref{sec: tides}).
However, high observational spin estimates of \ac{BH} in high mass X-ray binaries \cite{2008ApJ...679L..37L,2014ApJ...790...29G,2021ARA&A..59..117R}, if correct, appear challenging to reproduce with the very efficient angular momentum transport in \ac{BH} progenitor interiors (e.g. \cite{2019ApJ...870L..18Q} but see \cite{2017ApJ...846L..15B}).
Alternative, more moderate angular momentum transport schemes are also explored in the massive star models \cite{2008Ap&SS.316...43E,2012A&A...537A.146E,2020A&A...636A.104B}.
\\
\newline
In summary, during the mass transfer the accretor gains (possibly a very significant amount of) mass and angular momentum.
However, this does not necessarily mean that its further evolution can be accurately described as that of a highly-spinning star of it's new mass - predominantly because the accretor's core might not fully experience the mass and angular momentum gained by its outer layers.
Its post-interaction structure (in particular core-envelope boundary and core to envelope mass ratio) is likely different than predicted by any single star models, with possible (and largely unexplored) consequences for the subsequent phases of binary interactions.

\subsubsection{Impact on the accretor: BH accretors}\label{sec: BH accretor}

\ac{BH} in a binary can accrete mass and angular momentum from its companion when the star evolves to fill its Roche lobe.
This can obviously affect its natal properties, however, to what degree is currently uncertain. We briefly discuss the related problems and commonly made assumptions below.

\paragraph{How much mass can a \ac{BH} accrete?}

Accretion onto the compact object is often assumed to be limited by the Eddington accretion rate ($\dot{M}_{Edd}$), defined as the maximum accretion rate at which gravity can overcome radiation pressure of the emission driven by spherical accretion:
\begin{equation}\label{eq: Eddington accretion}
    \dot{M}_{Edd} = \frac{4 \pi G M_{BH}}{\eta \kappa c} \approx 2.6 \times 10^{-8} \Ms\text{yr}^{-1} \left(\frac{M_{BH}}{\Ms} \right) \left(\frac{\eta}{0.1} \right)^{-1} \left(\frac{1+X}{1+0.7} \right)^{-1}
\end{equation}
where $M_{BH}$ is the \ac{BH} mass, $c$ is the speed of light, $\eta \sim 0.1$  is the efficiency at which the \ac{BH} is converting the accreted rest mass into radiative energy (\cite{1970Natur.226...64B,2003MNRAS.341..385P}; the luminosity driven by accretion of mass at a rate $\dot{M}$ is $L_{BH} = \eta \dot{M}_{acc} c$), and $\kappa=0.2 (1+X) \ \text{cm}^{2}/\text{g}$ 
is the radiative opacity due to pure electron scattering (a reasonable assumption for massive stars) for composition with hydrogen mass fraction X ($\sim$0.7 for H-rich donor stars and 0 for H-deficient donor stars).
% $\sigma_{T}/m_{p}$
% where $\sigma_{T}$ is the Thompson scattering cross section and $m_{p}$ proton mass.
\\
If $\dot{M}_{Edd}$ is exceeded, the excess mass is lost from the system and is typically assumed to carry away the specific angular momentum of the \ac{BH} in its orbital motion (i.e. $\gamma_{iso}$).
It can be seen that $\dot{M}_{Edd}$ imposes a very strong limit on the \ac{BH} accretion in the most massive stellar binaries:
even if a 10 $\Ms$ \ac{BH} was to accrete at the Eddington rate from its companion during its entire lifetime (i.e. $\approx$hydrogen burning nuclear timescale $\lesssim$ 10 Myr for \ac{BH} progenitors), the accreted mass would be  $\sim$2.6 $\Ms$.

\paragraph{Is Eddington accretion the limit?}

Whether accreting \acp{BH} can significantly violate the Eddington limit, and under which conditions, is debated.
Clear cases where the equation \ref{eq: Eddington accretion} does not hold are the lack of spherical symmetry  (e.g. in cases with strongly collimated radiation), or highly non-steady accretion.
The spherical symmetry condition is not satisfied in many models of \ac{BH} accretion. 
Various existing disk accretion models allow for moderately super-Eddington (e.g. 'slim' disks \cite{1988ApJ...332..646A,2009ApJS..183..171S})
to highly super-Eddington accretion rates (e.g. 'Polish doughnuts'-like models \cite{1978A&A....63..221A,1980A&A....88...23P}), with different rates reported with modern numerical models (see e.g. reviews  \cite{2007A&ARv..15....1D,2013LRR....16....1A}).
Furthermore, super-Eddington accretion has been suggested as the likely explanation for a wide range of observed phenomena,
in particular ultraluminous X-ray
sources (\cite{2015NatPh..11..551F,2017ARA&A..55..303K}; at least some of which are known to contain accreting remnants of massive star evolution) and the galactic binary system SS433 \cite{2004ASPRv..12....1F}.
Highly super-Eddington accretion could in principle allow for the \ac{BH} formation within the \ac{PISN} mass gap in the course of regular binary evolution \cite{2020ApJ...897..100V}.

\paragraph{How much can a \ac{BH} spin up?}

Accretion of mass and of the associated angular momentum  also affects the natal spin of a \ac{BH}.
To discuss the \ac{BH} spin the dimensionless spin parameter $a_{spin}$ is often used, which is defined as follows:
\begin{equation}\label{eq: BH spin}
    a_{spin} = \frac{c J_{BH}}{G M_{BH}^{2}}
\end{equation}
Here $J_{BH}$ is the \ac{BH} angular momentum.
In the idealised case of accretion happening from the last stable circular orbit of a thin disk around a \ac{BH}, one can describe the $a_{spin}$ of initially non-spinning \ac{BH} as follows \cite{1970Natur.226...64B,1974ApJ...191..507T}
\begin{equation}\label{eq: BH spin2}
    a_{spin} = \left( \frac{2}{3} \right)^{1/2} \frac{M_{BH; ini}}{M_{BH; fin}}  \left(  4 - \sqrt{18\frac{M_{BH; ini}}{M_{BH; fin}}^{2} -2} \right)
\end{equation}
which holds for $M_{BH; fin}<\sqrt{6}M_{BH; ini}$.
\\
If the disk geometry is more complex, the disk is thick or if the accretion does not happen from the disk at all (e.g. if the specific angular momentum of the infalling matter is not sufficient to form a disk or if the disk is destroyed), \ac{BH} spin up is likely to be less efficient than what would be estimated with equation \ref{eq: BH spin2} \cite{1974ApJ...191..507T}.
Furthermore, it has been shown that the presence of magnetic fields within the inner part of the accretion disk may substantially limit the spin up or even slow down the \ac{BH} rotation \cite{2010MNRAS.403L..74K,2015MNRAS.446.1829C}.
Even within the validity of the framework used to derive equation \ref{eq: BH spin2}, to spin up a non-rotating \ac{BH} from $a_{spin}=0$ to nearly maximum rotation $a_{spin}\sim1$, the \ac{BH} must accrete at least $\Delta M$ = 1.85$M_{BH; ini}$ \cite{1999MNRAS.305..654K}. 
As discussed above, if the accretion rate is Eddington limited, in typical binaries the \ac{BH} can only accrete a small fraction of that mass \cite{1999MNRAS.305..654K}.
On the other hand, if the accretion exceeds the Eddington limit, it is unlikely to occur through a thin disk, in which case \ac{BH} spin obtained with eq. \ref{eq: BH spin2} only yields an upper limit. 

\paragraph{Can \acp{BH} accrete non-negligible mass (and spin up) during the \ac{CE}?}

A somewhat separate question is how much mass and angular momentum can be realistically accreted by the binary components inspiralling during the \ac{CE}.
In this case the geometry and the rate of mass transfer is distinctly different from the stable \ac{RLOF}, possibly allowing for a non-negligible accretion despite the extremely short timescale.
In particular, in case of compact object accretors, both high $\dot{M}$ and non-spherical geometry could in principle allow to surpass the Eddington limit.
However, evolution within the \ac{CE} is very challenging to simulate and the complete self-consistent model including accretion is not yet available (see extensive discussion of the related challenges in \cite{2020cee..book.....I}). To approximate the accretion during this phase, it is common to refer to the idealized case of the capture of matter by a point mass moving supersonically with respect to a uniform medium (the Bondi–Hoyle–Lyttleton accretion, as described in \cite{1944MNRAS.104..273B,1952MNRAS.112..195B, 2004NewAR..48..843E}) and assume some small fraction ($\sim$a few percent) of the resulting accretion rate.
This reduction initially came from the realisation that the value expected from Bondi–Hoyle–Lyttleton accretion would lead to unrealistic level of accretion during the \ac{CE}.
In particular, the highly super-Eddington accretion onto compact objects in \ac{CE} resulting from this scheme would notoriously cause \ac{NS} to accrete sufficient mass to collapse to low-mass \ac{BH}, leading to discrepancies with observations \cite{2007ApJ...662..504B}.
The accretion rate is also theoretically expected to be much smaller than in the idealized case, as many of the assumptions of the Bondi–Hoyle–Lyttleton scheme are not satisfied during the \ac{CE} (e.g. the flow is non-axisymmetric and nonuniform; see also discussion in \cite{2020cee..book.....I}).
This is further supported by the results of 3D hydrodynamic simulations \cite{2008ApJ...672L..41R,2015ApJ...803...41M,2017ApJ...838...56M}, that suggest accretion rates an order of magnitude below the Bondi–Hoyle–Lyttleton prediction and possibly below the Eddington-limited rate.

\section{Other relevant binary processes}\label{sec: other processes}

\subsection{Tidal forces: effect on spins and orbit} \label{sec: tides}

Gravitational pull by a binary companion causes deformations of the (non-degenerate) star.
These deformations are time-varying,
for instance, when the spin periods of the binary components are not synchronised or if the orbit is eccentric.
This leads to dissipation of energy through tidal friction  and allows for the exchange of angular momentum and kinetic energy between the two stars and their orbit
\cite{1977A&A....57..383Z,1981A&A....99..126H}.
The strength of the tidal interactions strongly increases with the ratio of the stellar radius to the orbital separation \cite{1977A&A....57..383Z,1981A&A....99..126H}.
Therefore, they become an important factor in the evolution of close binaries, with component(s) filling significant fraction of their Roche lobe \cite{2002MNRAS.329..897H}.
\\
Tidal interactions act to bring the system towards circularisation, to synchronise the binary component spins with the orbital period (tidal locking) and to align the spins perpendicular to the orbital plane.
The timescale at which this equilibrium state can be achieved depends on the efficiency of the processes that are responsible for the energy dissipation (see e.g. discussion in \cite{2002MNRAS.329..897H,2009A&A...497..243D}).
Regardless of that mechanism, in stellar binaries with orbital periods below a few days, the tides are  expected to be so strong that the stars are forced to rotate synchronously with the orbital motion  \cite{2009A&A...497..243D}.
The timescale for circularisation is typically expected to be short enough to bring the eccentricity to $e\approx0$ before either of the binary components overfills its Roche-lobe  \cite{2002MNRAS.329..897H}.
However, some observational results seem to challenge this conclusion; e.g. \cite{2008A&A...480..797B,2014A&A...564A...1B}. This does not necessarily mean that tides are inefficient, but might indicate the presence of other processes that are counteracting tidal circularisation and driving eccentricity in those cases \cite{2008A&A...480..797B,2015A&A...579A..49V}. 
Nevertheless, one should remain cautious when assuming that the mass transfer always happens when the orbit is circularized.
Even in the absence of eccentricity-pumping processes, for highly eccentric systems the mass exchange may proceed for some time on a non-circular orbit, which greatly complicates the description of the orbital evolution \cite{
2007ApJ...667.1170S,2009ApJ...702.1387S,
2010ApJ...724..546S,2016ApJ...825...70D,2016ApJ...825...71D}. In such cases the mass transfer rate may be strongly boosted near the periastron (such behaviour would lead to, for instance, increasing luminosity of the accretion-powered emitters such as X-ray binaries near periastron).
In binary population synthesis simulations,
tidal circularisation is typically assumed to happen instantaneously once certain conditions are met.
Importantly, together with this simplifying assumption comes the one about the separation of the circularised orbit $a_{circ}$.
The common choice is to assume that the angular momentum is conserved throughout this process (only the energy is lost from the binary as it circularizes), 
leading to  $a_{circ}=a_{pre;circ}(1-e^{2})$  \cite{1996A&A...309..179P,2002ApJ...572..407B}.
Other choices are also used, for instance, some authors assume $a_{circ}$ equals the periastron distance of the eccentric orbit at which the circularisation condition was met, i.e. $a_{circ}=a_{pre;circ}(1-e)$, e.g. \cite{2023ApJS..264...45F}.
\\
\newline
Two possibilities offered by strong tidal interactions in close binaries are particularly interesting in the context of \acp{BH} and the formation of \ac{BH} mergers: i) tidally induced chemically homogeneous evolution of the binary ii) tidal spin up of the immediate \ac{BH} progenitor. We briefly discuss them below.

\subsubsection{Chemically homogeneous evolution}
\label{sec: CHE}

As noted in Section \ref{sec: stellar accretor}, rapid rotation induces internal mixing. Such mixing can operate also in the regions where convection (providing the conventional way of re-distributing material within stars) is inefficient.
Therefore it can reduce, or - in extreme cases, completely remove the composition gradients between the central and outer layers of the star \cite{1987A&A...178..159M,1992A&A...265..115Z,1997A&A...321..465M}.
In the latter case, a helium star with no hydrogen envelope is formed at the end of the main sequence.
In contrast to the conventional radial evolution shown in Figure \ref{fig: radius evolution}, such `chemically homogeneous' stars avoid the strong radial expansion of their envelope during the phases of core contraction \cite{1987A&A...178..159M,1997A&A...321..465M}.
Those stars can therefore evolve much closer to each other in a binary and avoid merging with their stellar companion. This observation led to the proposal of chemically homogeneous evolution as a possible binary route to form \ac{BH} mergers, unique in its lack of need for any processes that can drastically tighten the birth orbit \cite{2016A&A...588A..50M,2016MNRAS.458.2634M,2016MNRAS.460.3545D}.
Furthermore, the extreme mixing allows for the formation of more massive stellar cores (as all available fuel is used in nuclear reactions), and therefore more massive compact objects.
Chemically homogeneous evolution is believed to be restricted to stars more massive than $\gtrsim$ 30 M$_{\odot}$ (i.e. \ac{BH} progenitors) that are rapidly rotating already at the onset of the core hydrogen burning (i.e. before the composition gradient at the core-envelope boundary develops, as its presence may inhibit efficient rotational mixing \cite{1997A&A...321..465M}).
\\
One way to facilitate chemically homogeneous evolution is to form two massive stars sufficiently close to each other so that they are forced to rotate rapidly through strong tidal interactions.
In such a case, both binary components undergo chemically homogeneous evolution.
Maintaining the tidal locking is then crucial, which means that chemically homogeneous evolution is expected to preferentially happen at low metallicity: strong wind mass loss experienced by massive, high metallicity stars tends to increase the binary separation. This can remove the tidal locking and allow the stars to significantly spin down, inhibiting chemically homogeneous evolution \cite{2009A&A...497..243D}.
If tidal locking can be kept throughout the evolution, such a scenario can potentially lead to the formation of rapidly spinning \acp{BH}.\\
Alternatively, accretion during the mass transfer can efficiently spin up the secondary star (Section \ref{sec: stellar accretor}). If this happens early enough during its evolution, it may allow for the chemically homogeneous evolution of the secondary \cite{2007A&A...465L..29C}.
Also in this scenario low metallicity is favored, as stellar winds extract the angular momentum from the star and spin it down.

\subsubsection{Tidal spin up of the immediate BH progenitor}

As discussed in Section \ref{sec: stellar accretor}, if the angular momentum is efficiently transported within the star and can be exchanged between the core and the envelope, the natal spins of \ac{BH} are likely to be very small \cite{2018A&A...616A..28Q,2019ApJ...881L...1F}.
Even if the \ac{BH} progenitor was born rapidly spinning, during its post-main sequence expansion the initial angular momentum is mostly transported to its outer layers.
Those layers are then likely lost when it initiates the mass transfer in a binary and/or (excluding the lowest metallicity stars) due to wind mass loss.
\\
However, in very close binaries, there is a window of opportunity for the secondary to gain rotation at the very end of its life and therefore maintain it until the collapse to a \ac{BH}.
This window is opened up by tides.
Tidal locking of a tight system consisting of a compact object and a helium/\ac{WR} star can lead to a strong spinup of the latter 
\cite{2012MNRAS.425..470C,2016MNRAS.462..844K,2017ApJ...842..111H,2018MNRAS.473.4174Z,2018A&A...616A..28Q,2020A&A...636A.104B,2020A&A...635A..97B,2021ApJ...921L...2O}. 
At the same time, mass lost by the helium star through winds widens the orbit and may eventually lead to tidal decoupling and spin down the star.
Therefore, to understand the net effect it is important to properly account for the coupled effect of tides and stellar winds on the evolution of the system.
In general, such tidal spin up i) may only be effective for a fraction of the (already very short) lifetime of a WR+BH/NS system and ii) for the binary with otherwise the same properties, can operate for longer at lower metallicity.
The first point means that this scenario likely requires that the progenitor system evolves through a \ac{CE}, which allows to reach the small binary separation at which the spinup of the secondary is efficient \cite{2021A&A...647A.153B}.
Such tight WR+BH binaries are the progenitors of merging BH+BH systems and are expected to constitute some (uncertain) fraction of the observed gravitational wave events \cite{2020A&A...636A.104B,2020A&A...635A..97B,2021ApJ...921L..15G,2021ApJ...921L...2O}.
The second point is somewhat counteracted by the fact that massive low metallicity stars in binaries are likely to fill their Roche-lobes later during their evolution than their high metallicity counterparts (see Section \ref{sec: radius evolution}).
This means that if a low metallicity binary evolves into a tight BH + WR, the remaining lifetime of the \ac{WR} is likely to be shorter than for the analogous high metallicity system.
Nonetheless, the results discussed in \cite{2020A&A...635A..97B} suggest that overall, this scenario produces a higher fraction of highly spinning immediate \ac{BH} progenitors at low metallicity.
\\
In principle, the mechanism discussed above may produce highly spinning \acp{BH} (even maximally spinning) from secondaries evolving in isolated binaries even if the angular momentum transport in stellar interiors is very efficient.
However, the link between the spin of the immediate \ac{BH} progenitor and the natal \ac{BH} spin (both its magnitude and orientation) may be highly non-straightforward, as it depends on the details of the core-collapse and the interaction of the proto-compact object with the ejected/falling back material, e.g. \cite{2017ApJ...846L..15B,2019arXiv190404835B,2022arXiv220502541T}.

\subsection{Orbital evolution due to gravitational wave emission}\label{sec: GW}

Massive, accelerating objects whose motion leads to variations in the overall mass distribution of the system (specifically, leads to changing quadrupole moment) emit gravitational waves \citep{1916AnP...354..769E}.
Binary systems are a well known example of astrophysical sources of gravitational waves.
The emitted gravitational waves carry energy and angular momentum away from the system, leading to a gradual decrease of the orbital separation.
The accurate measurement of such orbital decay of the binary pulsar PSR 1913+16 allowed for the first indirect detection of gravitational waves \citep{1975ApJ...195L..51H,1982ApJ...253..908T}.
In stellar binaries, other processes typically dominate the orbital evolution over gravitational waves emission and the latter can be neglected.
However, for sufficiently tight systems, in which only compact objects and low mass/stripped stars can fit, the timescale for the angular momentum loss in gravitational waves can be high enough to significantly affect their orbital evolution or even drive the mass transfer \cite{1962ApJ...136..312K}.
Some of such systems located in our Galaxy will be detectable as gravitational waves sources with the proposed space-based gravitational wave detectors \cite{2022arXiv220306016A}.
Finally, evolution of a double compact object (binary composed of \acp{BH} and/or \acp{NS}) that is not affected by interactions with any other objects is fully determined by the gravitational wave emission.
gravitational wave-emission induced inspiral eventually leads to its merger - the emitted signal is the strongest around this phase and that is when it can be directly measured by current detectors.
For a binary whose orbital evolution is affected solely by gravitational wave emission and whose component masses $m_{1}$ and $m_{2}$ are constant,
the time until the merger can be estimated by solving the set of equations 
describing the related rate of change of the orbital separation ($a$) and eccentricity ($e$) 
in the lowest (quadrupolar) order \citep{1964PhRv..136.1224P}:
\begin{equation}\label{eq: GW separation decay}
 <\frac{da}{dt}> = -\frac{64}{5}\frac{G^{3} \mu M^{2}}{c^{5}a^{3} (1-e^{2})^{7/2}} \left(1 + \frac{73}{24}e^{2} + \frac{37}{96}e^{4}\right)
\end{equation}
\begin{equation} \label{eq: GW e decay}
 <\frac{de}{dt}> = -\frac{304}{15}e \frac{G^{3} \mu M^{2}}{c^{5}a^{4} (1-e^{2})^{5/2}} \left(1 + \frac{121}{304}e^{2}\right)
\end{equation}
where $M=m_{1}+m_{2}$ is the total mass of the binary and $\mu=m_{1}m_{2}/M$ is the reduced mass.
\begin{figure}[h!]
  \begin{minipage}[c]{0.6\textwidth}
    \includegraphics[width=1.\textwidth]{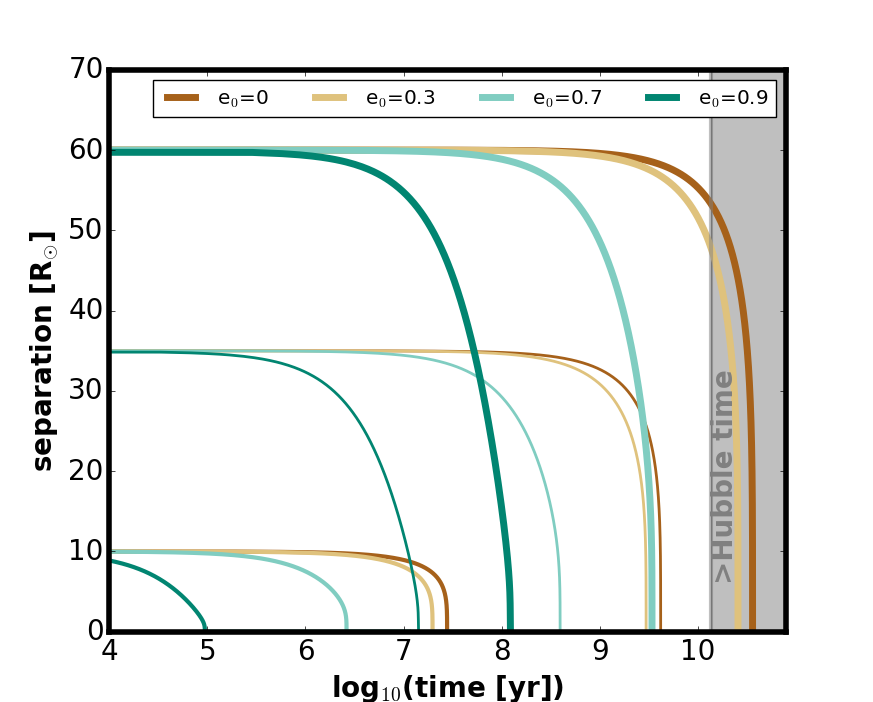}
  \end{minipage}\hfill
  \begin{minipage}[c]{0.4\textwidth}
    \caption{
    Orbital separation decay of 30 $\Ms$ + 30 $\Ms$ binary for different initial separations and eccentricities $e_{0}$ driven purely by gravitational wave emission (obtained by solving equations \ref{eq: GW separation decay} and \ref{eq: GW e decay}).
    If the lines fall within the gray band, the merger timescale of the corresponding binary exceeds the Hubble time.
    }             
    \label{fig: Peters formulae}
  \end{minipage}
\end{figure}
Example solutions to the above equations are shown in Figure \ref{fig: Peters formulae}.
It can be seen that the merger timescale $\tau_{\rm mr}$ strongly depends on both the starting separation $a_{0}$ and eccentricity $e_{0}$.
In case of circular orbits, this timescale $\tau_{\rm mr; circ}$ can be obtained directly from equation \ref{eq: GW separation decay}:
\begin{equation}\label{eq: Tmr}
\tau_{\rm mr; circ} = \frac{5 c^{5}}{256 G^{3}} \frac{a_{0}^{4}}{m_{1} m_{2} M}
\end{equation}
while the following analytical fit allows to obtain the merger time in case of eccentric orbits
\cite{2021RNAAS...5..223M}:
\begin{equation}
    \tau_{\rm mr} \approx \tau_{\rm mr; circ} \left(1 + 0.27 e_{0}^{10} + 0.33 e_{0}^{20} + 0.2 e_{0}^{1000} \right)\left(1-e_{0}^{2} \right)^{7/2}
    \label{eq:tinsp}
\end{equation}
An equal-mass m$_{1}$=m$_{2}$=30$\Ms$, circular binary with $a_{0}$=10$\Rs$ will merge within only $\tau_{\rm mr; circ}\approx$28 Myr, while for the same binary with $a_{0}$=30$\Rs$ the merger timescale becomes $\tau_{\rm mr; circ}\approx$2 Gyr.
A binary on a highly eccentric orbit with e$_{0}$=0.7 will merge in only about 10\% of the 
time that would be estimated for the same masses and $a_{0}$ with eq. \ref{eq: Tmr}. 
\\
From eq. \ref{eq: Tmr} one can estimate the maximum separation that a (circular) binary can have at its formation to merge within the time shorter than the age of the Universe ($\tau_{0}\sim 14$Gyr) due to gravitational wave emission:
\begin{equation}
    a_{\rm merging}< \left(\frac{1}{0.15} \ \frac{m_{1} m_{2} (m_{1}+m_{2})}{\Ms^{3}}\right)^{1/4} \left(\frac{\tau_{0}}{\rm Gyr}\right)^{1/4} \ \Rs
\end{equation}
The above equation yields $a_{\rm merging}\lesssim 50 \ \Rs$ for the considered m$_{1}$=m$_{2}$=30$\Ms$ binary.
This is at least 10 times smaller than the radii that typical massive stars reach during their evolution (see Section \ref{sec: radius evolution}).
To end up as such a close BH+BH system, an isolated massive star binary must either i) evolve through mass transfer phase(s) that bring the stars close together during their evolution or ii) prevent stars in initially close obits from expanding and merging prematurely (as in the chemically homogeneous evolution scenario discussed in Section \ref{sec: CHE}).

\subsection{Magnetic braking}

    Magnetic braking is the loss of angular momentum through stellar wind of a magnetic star \cite{1967ApJ...148..217W,1972ApJ...171..565S,1981A&A...100L...7V,1987MNRAS.226...57M,2011ApJS..194...28K}.
This mode of angular momentum loss
concerns primarily low mass (0.3 - 1.5 $\Ms$), main sequence stars.
Such objects have radiative cores and convective envelopes in which magnetic fields can be amplified and sustained by the action of magnetic dynamo \cite{1982A&A...106...58S,1983ApJ...275..713R}.
Even though the amount of mass such stars lose through winds is often negligible, the associated angular momentum loss may be significant because the magnetic field
of the star forces the wind matter to corotate to a large distance from the stellar surface.
Magnetic braking can completely slow down a star during its main sequence evolution, unless it is tidally synchronised with a binary companion.
In such case, tidal torques replenish the rotational angular momentum of the star and extract it from the orbit. 
Therefore, while counteracting the effect of magnetic braking, tides lead to orbital shrinkage and possibly drive the binary to \ac{RLOF}.
This process is important, for instance, in low mass X-ray binaries (and, in fact, appears necessary to explain the observed high mass transfer rates of some of those systems), where the coupled effects due to tides and magnetic braking need to be considered \cite{1981A&A...100L...7V,1993ARA&A..31...93V,2014MNRAS.444..542R}.
The fraction of magnetic stars appears to decrease with increasing mass: \cite{2009ARA&A..47..333D} estimate it to be $<$15\% for intermediate/high-mass stars.
However, their wind mass-loss is no longer negligible and if such stars experience magnetic braking, the angular momentum loss is amplified accordingly.

\section{Summary}

The progenitors of stellar-origin \acp{BH} and \acp{NS}
are very likely to exchange mass and angular momentum with another star,
and that likelihood appears to increase with the initial stellar mass.
Those exchanges happen through stable and/or unstable mass transfer phase(s) and tides.
Such interactions lead to \acp{CO} with properties
that may substantially differ from those predicted within the single star evolution theory.
The properties of \acp{CO} (masses, spins) that remain in binaries after their formation can be further altered by accretion from the companion.
In particular, isolated binary progenitors of
\ac{BH}/\ac{NS} mergers (currently detectable \ac{GW} sources)
must form with relatively narrow initial separations and
likely interact multiple times during their evolution.
Binary processes are therefore expected to leave a strong imprint
on their observable characteristics (see Part \ref{part:observations}).
\\
Good quantitative understanding of the effects of those processes is currently hampered by uncertainties pertaining to the physics of binary interactions, in particular criteria for mass transfer stability and the amount of mass and angular momentum lost during various types of interactions.
Models of binary evolution also inherit uncertainties present in the detailed models of single stars. Of particular importance in the context of binaries are those affecting the evolution of stellar radii and the envelope structure (crucial when determining the characteristics of the mass transfer), angular momentum transport in stellar interiors (particularly important to understand the effect of mass transfer on the evolution of the accretor),
core-collapse physics 
(determining the properties of the binary right after the formation of the \ac{CO}, in particular, deciding whether the system remains bound or not) and  stellar winds, introducing a metallicity-dependent effect on the evolution of the orbit, stellar masses and spins (cfr. Section \ref{sec:phys_processes}).
Finally, the structure of post-interaction stars produced in binaries may be different than predicted by any single star evolution model. The use of the latter models to approximate the evolution of individual binary components (in particular that of the secondary) is a common practice.
However, it means that some relevant effects may be missed by current models. This adds uncertainty to the current binary evolution predictions that is largely unexplored.
\\
It becomes evident 
that birth metallicity plays an important role 
in the evolution of \ac{BH} progenitors in binaries  
- probably much more so than in the case of systems with lower mass stars.
There are multiple observational clues that systems involving compact remnants of massive stars may become more common and/or more extreme at low metallicity (
\cite{2006ApJ...638L..63L,2006AcA....56..333S,2009MNRAS.395L..71M,2016Natur.534..512B,2017MNRAS.470.3566C,2017ApJ...834..170G,2018MNRAS.473.1258S,2020MNRAS.498.4790K,2021ApJ...907...17L}, (see also Part \ref{part:observations}). As highlighted throughout this section, such link to low metallicity is also present in the current models. In the latter case, it stems mostly from two effects:
i) stellar wind mass loss rate decreasing with metallicity and ii) later radial expansion of lower metallicity massive stars.
The first allows to retain more mass and angular momentum within the \ac{BH} progenitor (and so within the binary, limiting the orbital expansion due to winds).
This allows for the formation of more massive and potentially - highly spinning \acp{BH} in the course of isolated binary evolution, and opens up the possibility of chemically homogeneous evolution.
The second leads to the expectation that low metallicity \ac{BH} progenitors in binaries are more likely to initiate mass transfer during the \ac{CHeB} phase of their evolution rather than on the Hertzsprung Gap.
Such interaction is much more likely to be stable, to last longer (increasing the chances of observing interacting \ac{BH} binaries at low metallicity), and to lead to wider post-interaction orbits than the common envelope evolution (likely encountered by \ac{BH} progenitor binaries at high metallicity).
Both of those effects are uncertain
and need to be better quantified in future studies.
Understanding and characterising differences in the evolution and fates of massive stars at different (in particular low) metallicities becomes increasingly more important:
with \acp{GW} we are already detecting mergers of afterlives of massive stars that formed anywhere in the past (in a wide range of environments and metallicities) \cite{2022arXiv220610622C},
and with the electromagnetic observations we are beginning to uncover young galaxies in the early Universe. 
Their emission is dominated by the most massive stars that are likely to be very metal-poor and affected by binary interactions \cite{2022ARA&A..60..455E}.

%%%%%%%%%%%Broekgaarden%%%%%%%%%%%%%%%%%%%%%
\newpage
\part{Black Holes in binaries: formation channels - isolated \\ \Large{Floor Broekgaarden}}
\label{part:GWpaleo}
\section{Introduction: The new frontier of \ac{GW} paleontology}
\label{sec:Intro_GWpaleo}
%

%%%%
\begin{figure*}
    \centering
\includegraphics[width=0.9\textwidth]{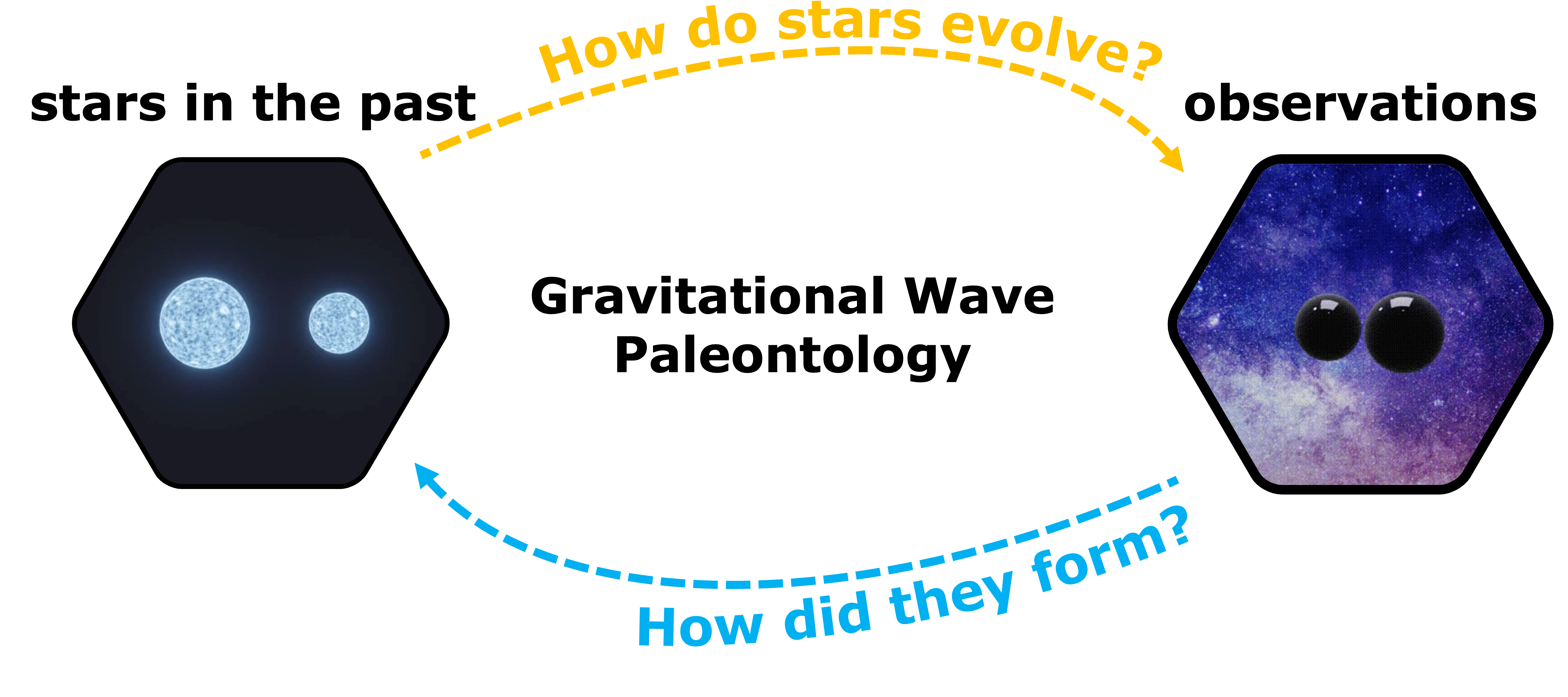} 
    \caption[Gravitational wave paleontology schematic]{Depiction of \ac{GW} paleontology. Observations of colliding \acp{BH} and \acp{NS} are used as ``fossils'' to study how massive binary stars lived in the past. Similarly to paleontology, this consists of learning about how the collisions formed (backward modelling) as well as modelling theories of how stars live to find what mergers they would leave behind (forward modelling). \acp{BH} are represented with marbles for visualization purposes.}
    \label{fig-ch0:gw-paleontology-plot}
\end{figure*}
%%%%

\begin{figure*}
    \centering
\includegraphics[width=0.9\textwidth]{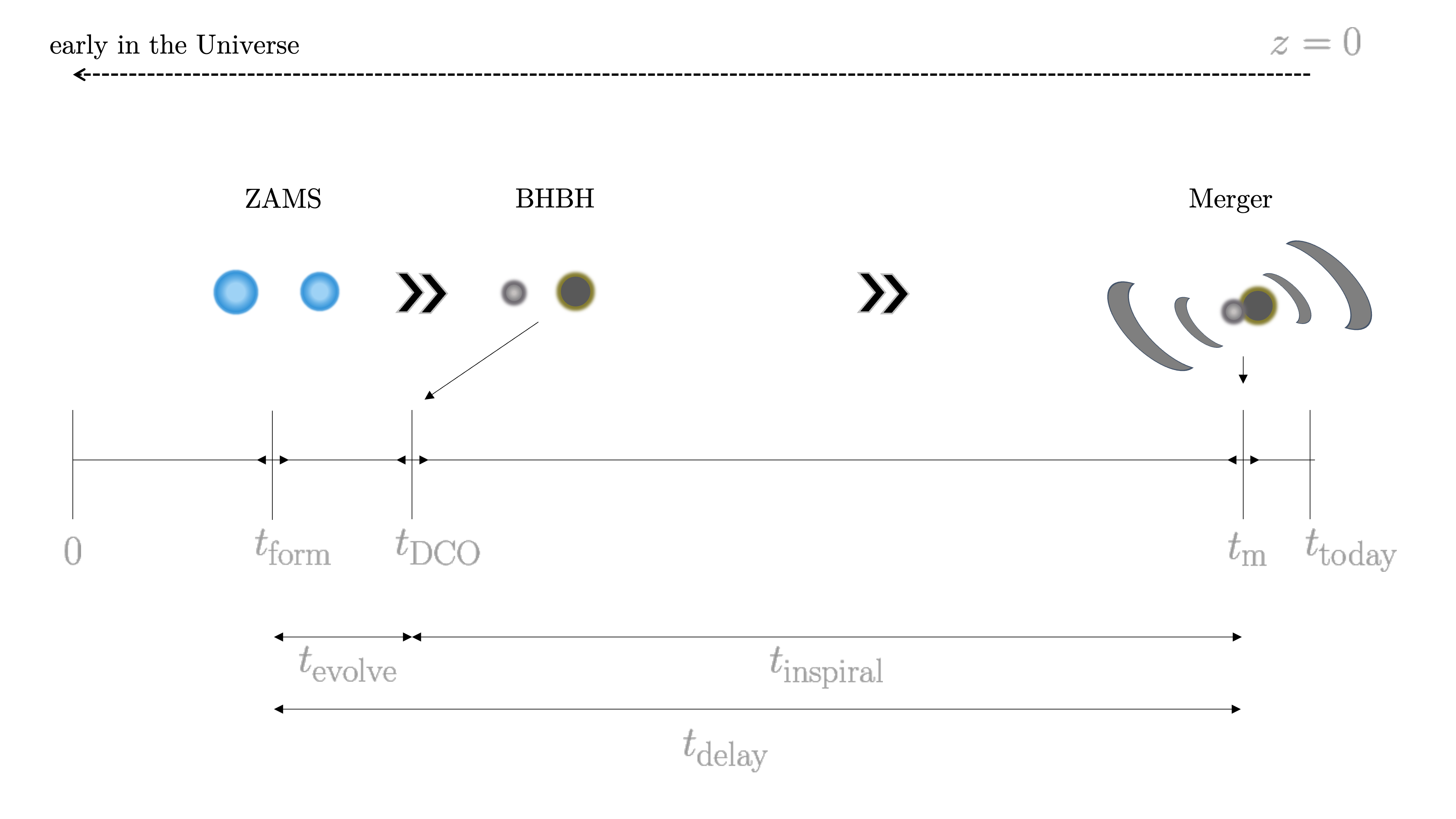} 
    \caption{Schematic display of the different times in the formation and
evolution of a binary system that impact the time $t_{\rm{m}}$ at which a \ac{BBH}
system will merge in the Universe. The inspiral times can span many Myr to Gyr such that \ac{BH} and \ac{NS} mergers observed with \ac{GW} detectors today probe massive stars that formed at different cosmic times, in different environments, and with different properties. }
    \label{fig-delay-times}
\end{figure*}

Probing millions of massive stars in different environments across our vast Universe could become reality with a new emerging frontier in astrophysics called \textit{gravitational-wave paleontology}: studying the formation, lives, and deaths of massive stars and the \acp{BH} they leave behind using the \ac{GW} observations of \ac{DCO} mergers such as binary \acp{BH}  as ``massive star fossils''. This is particularly exciting as \acp{GW} probe massive stars and \ac{BH} physics over cosmic scales. 
Namely, although massive stars evolve into \acp{BH} and \acp{NS} within just a few Myr, the time it takes for the \ac{DCO} to inspiral, i.e., the time between the formation of the \ac{DCO} and the \ac{DCO} merger (see Figure~\ref{fig-delay-times}), can span a wide range of durations from Myr to many Gyr \citep[e.g.,][]{2016A&A...589A..64M, 2022ApJ...940L..18Z}. 
This is because the \ac{DCO} orbital angular momentum loss by \ac{GW} radiation is really small for most systems. For a binary consisting of two \acp{CO} with masses \Moneint, \Mtwoint, and semi-major axis $a$ the change in separation due to \ac{GW} radiation is given by \citet[see Equation \ref{eq: GW separation decay}][]{1964peters} 
% as
% \begin{equation}
%     \left(\frac{\diff a}{\diff t}\right) = - \frac{64}{5} \frac{G^3}{c^5} \frac{\Moneint \Mtwoint}{a^3} \frac{(\Moneint + \Mtwoint)}{(1-e^2)^{7/2} } (1 + \frac{73}{24}e^2 + \frac{37}{96}e^4), 
% \end{equation}
% with $G$ the gravitational constant, $c$ the speed of light, and $e$ the eccentricity. 
For the case of a circular orbit consisting of two \acp{CO} formed with an initial separation \aint the timescale of inspiral can be derived from Equation \ref{eq:tinsp}
%
% \begin{equation}\label{eq:ch0-delay-times}
%     t_{\rm{insp}} = \frac{5}{256} \frac{c^5}{G^3} \frac{\aint^4}{\Moneint \, \Mtwoint (\Moneint + \Mtwoint)} \approx 0.15  \Gyr \left( \frac{\aint}{\Rsun}\right)^4 \left( \frac{\Moneint \Mtwoint (\Moneint+\Mtwoint)}{\Msun} \right)^{-1}.    
% \end{equation}
%

Importantly, this equation reveals that the inspiral time is strongly correlated with the initial separation $\aint$ of the \ac{DCO} system, leading to a vast range of inspiral times even for a small range of separations, and making typical delay times of order \Gyr common for systems with $\aint \gtrsim 2\Rsun$. This means that the observed population of \ac{DCO} mergers, even the current ``local'' detections observed by LIGO, Virgo, and KAGRA, probe a mixture of events with progenitor systems that could have formed throughout the entire cosmic history. This makes \acp{GW} a potential unique probe of massive stars through cosmic history. At the same time, this mixture also makes it challenging to unravel the formation histories of \ac{GW} sources as the \ac{GW} detections do not directly measure the properties of their progenitor stars.

%%%%
%%%%

\section{Formation pathways to \ac{GW} sources}
\label{sec:Formation_pathGW}

\subsection{The separation challenge}

 Historically, it has been challenged whether \ac{DCO} systems that merge within the age of our Universe could exist at all.  This is because all formation channels have to overcome the so-called ``separation challenge''. Using 14\Gyr as a proxy for the age of the Universe, Equation~\ref{eq:tinsp} provides an estimate for the maximum possible separation for a \ac{DCO} system such that the inspiral can have taken place by today. Setting $t_{\rm{insp}} \lesssim 14\Gyr$ gives for a binary consisting of two 1\Msun \acp{CO} that the minimum separation required is $\aint \lesssim 3.7\Rsun$ (see also \citep{MandelBroekgaarden:2021}). This is extremely small! To give some intuition, imagine scaling down everything in the Universe by a factor of $10^{10}$. The diameter of the sun, $2\Rsun \approx 14 \cdot 10^8\,\rm{m}$, then becomes $14\,\rm{cm}$, or about half the size of a cabbage. The diameters of massive stars are slightly larger: about $\sim 4 \Rsun$ for $8\Msun$ stars\footnote{On the \ac{ZAMS}.}, or about one cabbage, and even larger for more massive stars (with the radii approximately scaling as $M^{0.8}$). In this scaled-down Universe, two \acp{CO} in a binary system need to be a few cabbage distances apart in order to merge within the age of the Universe. Intuitively one might thus start out with two massive stars that are a $\sim 4 \Rsun $ (two cabbages) apart that then eventually form the required \ac{DCO} system with a separation of about $\sim 4\Rsun$ that can merge within the age of the Universe. However, massive stars typically expand during their lives to $100-1000$ times their initial size, e.g., during the Hertzsprung Gap phase \citep{1982ApJS...49..447B, MandelFarmer:2018}. Such an expanding star would completely engulf the other star if the stars are only a few solar radii apart, and the system would merge as stars, never making a \ac{DCO} system \citep[e.g.,][]{Schneider:2015}.  Importantly, all formation pathways thus need to find a way to overcome or avoid this separation challenge, such that the system can form a tight \ac{DCO} system that can merge within the age of our Universe and be observed with \ac{GW} observation in the current day.

\begin{figure*}
    \centering
\includegraphics[width=0.9\textwidth]{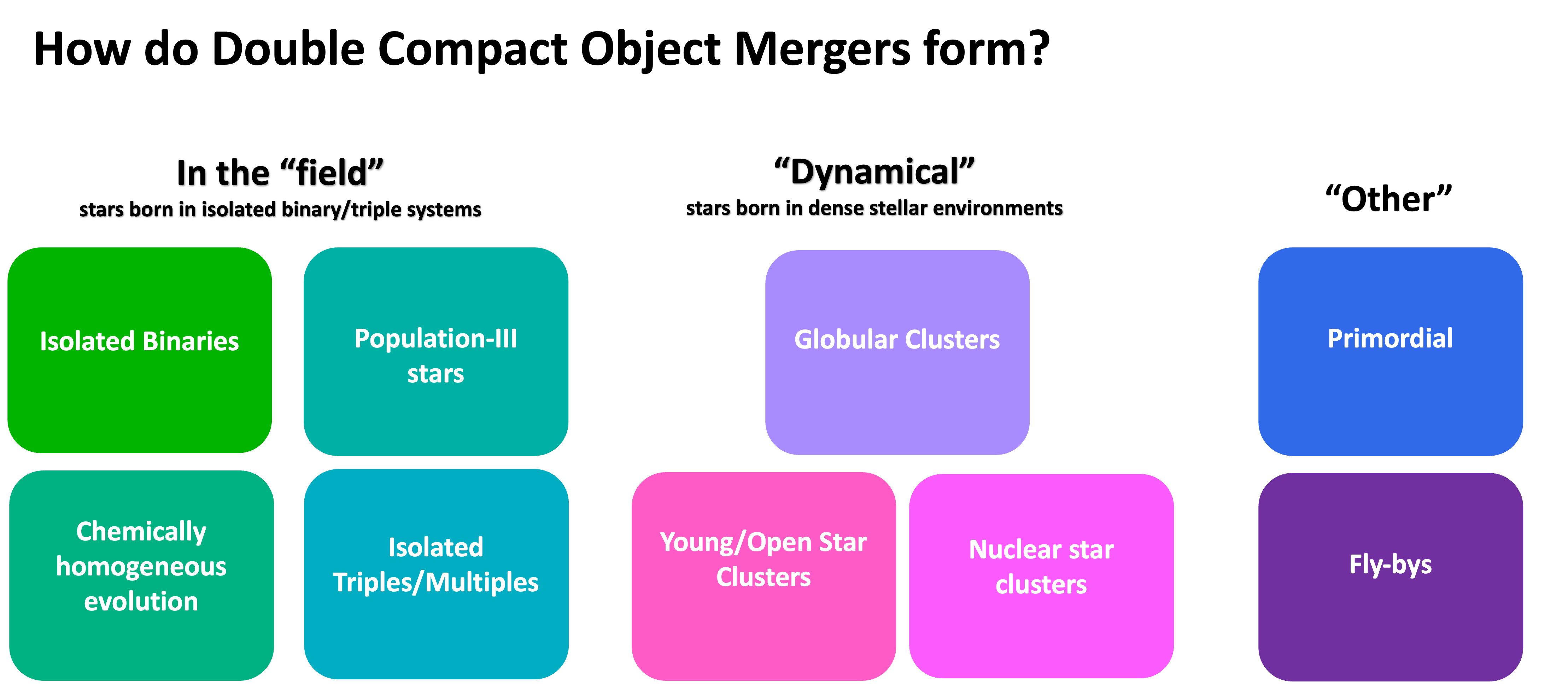} 
    \caption{Summary of proposed formation channels to form \ac{DCO} mergers}
    \label{fig-ch0:formation-channels-review}
\end{figure*}

\subsection{Formation Channels}
The main formation scenario leading to \ac{DCO} coalescences is still under debate.  Several theories have been proposed. Figure~\ref{fig-ch0:formation-channels-review} gives an overview of the most commonly discussed channels that we discuss in more detail below. The formation channels are classically divided into two categories: stars born in the ``field'', which evolve isolated as a binary, triple, or higher order multiples, and dynamical channels, where the stars are born in dense stellar environments such as globular clusters and young open star clusters and they merge due to gravitational dynamical interactions. This Section focuses on binaries from the isolated channel, which we will discuss in great detail below.  The other channels are also  briefly mentioned below and we refer the reader for more details to the reviews by \citep{ 2021hgwa.bookE..16M, MandelBroekgaarden:2021, GerosaFishbach:2021, MandelFarmer:2018, ArcaSedda:2023review, 2022Galax..10...76S}, and references therein, and to Part \ref{part:FCdynamical}.

\begin{figure*}
\includegraphics[width=1\textwidth]{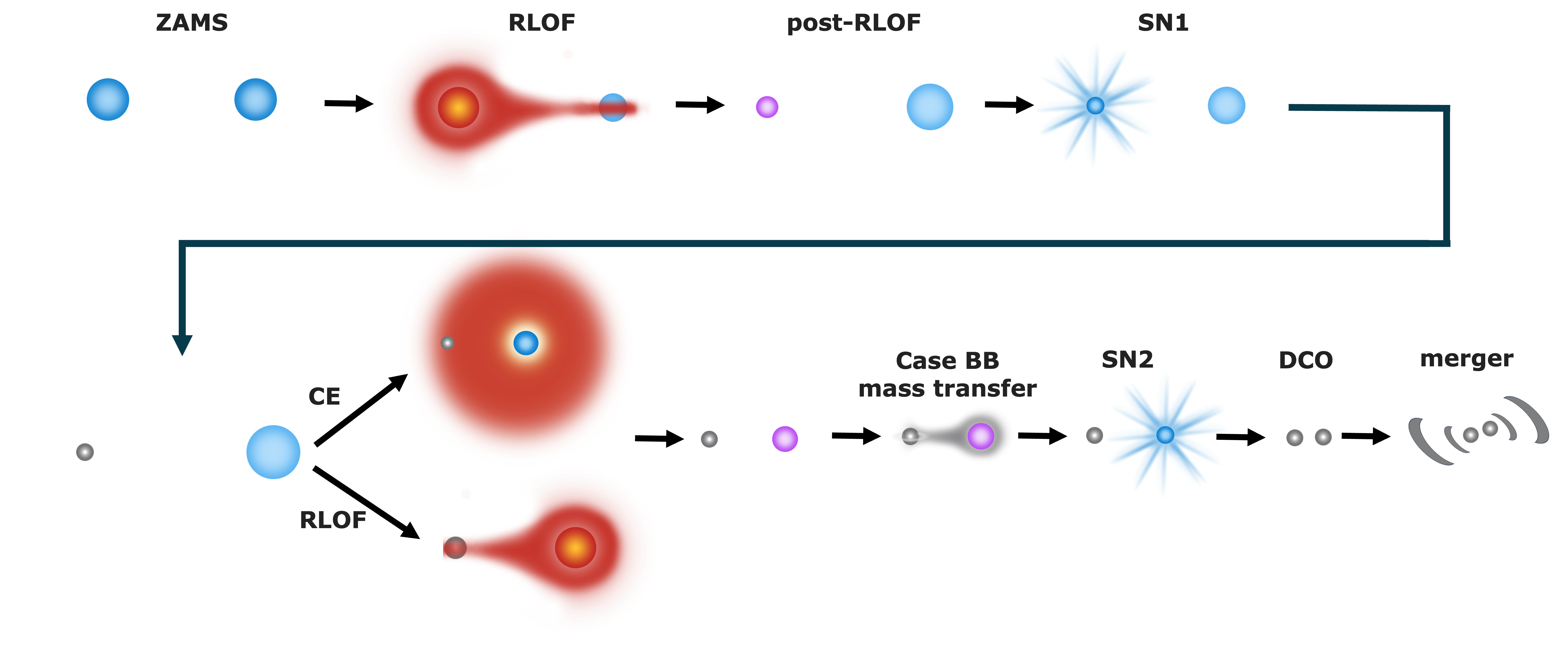}
  \caption{Schematic depiction of the isolated formation channel. The acronyms stand for zero-age main sequence (ZAMS), Roche lobe overflow (RLOF), supernova (SN), common-envelope phase (CE), and double compact object merger (DCO). This figure is created using icons from \cite{ZenvignaGomezImages}.}
  \label{fig:formation-channels-sketch-ch-1}
\end{figure*}

\subsubsection{Isolated Binary Evolution formation channel ($N=2$)}
The most commonly studied channel is perhaps the isolated binary evolution formation channel. This formation pathway starts with two ($N=2$, with N the number of stars) massive stars formed on the \ac{ZAMS} in an isolated binary with a relatively wide orbit and overcomes the separation challenge by drastically shrinking the orbit later on in the evolution of the binary stars. 
We describe a typical example of the evolution below, which is shown schematically in Figure~\ref{fig:formation-channels-sketch-ch-1}. 

\begin{enumerate}
    \item First, two massive stars are formed on the \ac{ZAMS} with a large separation (i.e., with typical orbital separations of  $a \gtrsim 100$--$1000$\Rsun). 
    \item The most massive star in the binary, the primary, evolves first off the \ac{ZAMS} when the hydrogen in the stellar core is almost completely depleted from fusing hydrogen into helium, stopping nuclear fusion in the core. This lowers the pressure outwards causing the core to contract,  creating a (further) density gradient, and releasing gravitational binding energy that initiates fusion to continue in a shell surrounding the core (hydrogen shell burning). This leads the hydrogen envelope to expand to radii of $\sim100$--$1000\Rsun$ for these massive stars, depending on the birth metallicity\footnote{Where the metallicity $Z$ is here the fraction of elements heavier than hydrogen and helium in the star.} of the stars\citep[e.g.,][]{MandelFarmer:2018}. 
    The star is now in the Hertzsprung Gap phase. For a sufficiently close binary, the star's radius will exceed its so-called Roche Lobe radius, which defines the equipotential surface within which material is bound to the star. From observations, this is expected to happen for $\sim 70\%$ of all massive star binaries \citep{2012Sci...337..444S}\footnote{But not necessarily all as a Hertzsprung Gap star.}. As a result, a mass transfer phase is initiated during which mass can flow from the donor star to its companion, the secondary. Under the right conditions, this mass transfer phase can be long-lasting (thermal timescale of the donor star) and proceed stably \citep[see, e.g.,][and references therein]{Schneider:2015} called stable \ac{RLOF}. During this mass transfer the companion star can potentially accrete a significant fraction of the transferred mass from the donor star, depending on the secondary's thermal timescale and response to the mass transfer.  
    \item Eventually, the mass transfer stops, leaving behind a stripped-envelope primary star that consists of a helium core and likely a small hydrogen envelope layer \citep{Laplace:2020}, which can be observed as a \ac{WR} star \citep[e.g.,][]{2007ARA&A..45..177C, 2018A&A...615A..78G}. 
    \item Eventually, the primary star will end its life and likely undergo a supernova and leave behind the first \ac{CO}; a \ac{NS} or \ac{BH}. During the collapse into the \ac{CO}, the system might receive a significant kick of hundreds of $\rm{km}/\rm{s}$, that disrupts the majority of binaries \citep[e.g.,][]{DeDonder:1997, Eldridge:2011, 2017A&A...603A.118R}. 
    \item In case the binary stays bound, which might be the case for less than $\lesssim 5\%$ of the systems \citep[e.g.,][]{2017A&A...603A.118R, VignaGomez:2018},  where disruption depends on the magnitude and orientation of the supernova kick, the separation of the binary and the amount of ejected mass \citep[e.g.][]{1975A&A....39...61F,1998A&A...330.1047T}. The end result is a binary with a \ac{CO} and a star. These might be observed as dormant \acp{BH} in Gaia \citep{Tanikawa:2022GaiaBH, El-Badry:2023GaiaBHone, El-Badry:2023GaiaBHtwo} or as stellar wind-fed X-ray binaries \citep[e.g.,][]{RemillardMcClintock:2006, Corral-Santana:2016, Kretschmar:2019}. 
    \item The secondary will eventually also evolve off the \ac{ZAMS} and expand, initiating a mass transfer phase. This mass transfer is thought to likely be non-conservative because the \ac{CO} companion can only accrete a very limited fraction of the donated mass (typically assumed as the Eddington limit). This leads the binary to rapidly shrink the orbital separation. Consequently, this leads to a runaway process as the secondary's Roche Lobe shrinks rapidly with decreasing separation, leading the star to even more overflow its Roche Lobe. This leads to a dynamically unstable mass transfer phase during which the \ac{CO} is engulfed in the envelope of the secondary, a common-envelope phase \citep{Paczynski:1976, ivanova2013common}. The \ac{CO} is thought to rapidly spiral in due to the drag forces it experiences from the surrounding gas envelope. In some cases the orbital energy can be deposited into the envelope, allowing it to heat up and expand and eventually be successfully ejected, halting the inspiral. If this happens, the resulting binary consists of the \ac{CO} in a tight orbit with the stripped secondary star, an example of a possible observed system in this phase is Cyg X-3 \citep{2013ApJ...764...96B, 2013MNRAS.429L.104Z}. If the envelope ejection fails, the system results in a merger of the star with the \ac{CO}, which can possibly form a  Thorne--{\.Z}ytkow object \citep{1977ApJ...212..832T} or lead to peculiar supernovae \citep[e.g.][]{2012ApJ...752L...2C,2016MNRAS.455.4351P,2020ApJ...892...13S}.
    
    Alternatively, the reversed mass transfer phase can proceed stably under certain circumstances and ranges of orbital periods and mass ratios for the binary \citep[e.g.,][]{WoodsIvanova:2011, Ge:2015, Inayoshi:2017, Pavlovskii:2017, vandenHeuvel:2017,  Gallegos-Garcia:2021hti, Marchant:2021}.  Based on this, several authors have shown that a formation pathway with only stable mass transfer (avoiding a common-envelope event) could even dominate the merger rate for \ac{BBH} mergers \citep{Neijssel:2019, Olejak:2021CE, Shao:2021} and that the role of common-envelope phases in the formation of \ac{DCO} sources might be overestimated by population models \citep[e.g.,][]{Marchant:2021, Klencki:2020convective, Klencki:2021}. 
    
    \item At the end of this mass transfer phase the system now consists of a binary with a \ac{CO} and the stripped helium secondary star in a tight orbit. 
    \item In some cases, a second mass transfer phase can take place when the helium star expands. This is sometimes referred to in the literature as case-BB mass transfer from the helium star onto the primary \ac{CO} \citep[cf.][]{2003MNRAS.344..629D}. This typically occurs when the secondary is a relatively low mass helium star as they expand to larger radii compared to more massive helium stars  \citep[e.g.,][]{2000MNRAS.315..543H, 2003MNRAS.344..629D}
    This is thought to be especially important and frequent in the formation of binary neutron star systems \citep{Tauris:2015, tauris2017formation}. 
    \item Eventually the secondary also undergoes a supernova and forms the second \ac{CO}. The tight orbit leads the system to likely stay bound, but a decent fraction of binaries might still disrupt during this second supernova. 
    \item The \ac{DCO} system will slowly inspiral due to the loss of orbital energy through \acp{GW} and eventually merge after millions to many billions of years \citep{Peters:1964}. If the binary has a tight enough orbit, as described by Equation~\ref{eq:tinsp}, it can merge within the age of the Universe and be observed as a \ac{GW} source by current ground-base \acp{GW} detectors. 

\end{enumerate}

As described in the formation channel above, there are many steps in the evolutionary pathway that a massive binary star system needs to survive in order to form a \ac{DCO} that merges in the age of our Universe. In most cases, however, the system will instead disrupt during one of the supernovae and/or merge during a mass transfer event. The formation of a \ac{DCO} merger is a rare outcome, and it can thus be challenging to find what exact properties of the pairs of stars make it survive all steps and form a \ac{DCO} system that survives eventually billions of years until it collides as a \ac{DCO} system and forms a \ac{GW} source.  
We can write down an analytically approximate equation to calculate the \ac{DCO} merger rate that this channel can produce. For \ac{BBH} mergers, this comes down to an equation along the lines of:
\begin{align}
\mathcal{R}_\textrm{BBH} &= f_\textrm{init sep} \times f_\textrm{primary} \times f_\textrm{q} \times f_\textrm{survive SN1} \times f_\textrm{CE} \times f_\textrm{survive SN2} \nonumber \times f_\textrm{merge} \\ 
  &\sim \ 0.25 \ \times \ 0.0007 \ \times \ 0.7   \times 0.05 \times 0.2 \times 0.5 \times 0.5 \approx 3 \times 10^{-7}. \label{DrakeIsolatedFloor}
\end{align}

 First, we need the binary system to be close enough to initiate a mass transfer phase during its lifetime but not too close that it undergoes a stellar merger at ZAMS. From studies including \citet{Schneider:2015} it appears that typically among the four orders of magnitude in separation, about one order of magnitude gives the binaries the right separation for it to undergo mass transfer without a stellar merger. This gives a factor $f_{\rm{init sep}} = 0.25$, given the uniform logarithm distribution of stars in separation \citep{Opik:1924}.  Note that this separation condition also takes into account the fraction of massive stars that are in binaries, as binary systems with too wide orbits are the effectively single stars in \citet{Sana:2012}, giving a binary fraction of $\sim 100\%$ when this separation range is corrected for. 
Second, we multiply by $f_\textrm{primary}$, the fraction of stars where the primary is massive enough to form a \ac{BH}. This is approximately for stars with masses $20 \lesssim \Moneint \lesssim 200$\Msun \citep[see][for more details]{Heger:2003}. And integrating this over a \citet{Kroupa:2001} initial mass function, gives approximately  $f_\textrm{primary} = (\int_{8\Msun}^{20\Msun} \Moneint^{-2.3} \diff \Moneint) / (\int_{0.08\Msun}^{200\Msun} \Moneint^{-2.3} \diff \Moneint)) \approx 0.0007$, where we choose 200\Msun as the maximum mass of massive stars \citep[e.g.,][]{Kalari:2022}, and 0.08\Msun is chosen as the approximate minimum mass for a star below which a brown dwarf is formed instead \citep[e.g.,][]{vonBoetticher:2017, Dieterich:2018}. 
Third, the secondary also needs to be massive enough to form the other \ac{CO} (\ac{NS}). Moreover, the first mass transfer phase in the binary is likely unstable if the secondary is significantly less massive than the primary at the onset of the mass transfer leading to a stellar merger \citep{Schneider:2015}. For this simple calculation, we assume a mass ratio of $q = \Moneint/\Mtwoint \gtrsim 0.3$ at the \ac{ZAMS} is required for stable mass transfer \citep[e.g.,][]{Ulrich:1976, Kippenhahn:1977, Neo:1977, Wellstein:2001, deMink:2007SMC, Schneider:2015}. From observations the mass ratio distribution of stars is found to be roughly consistent with a uniform distribution \citep[cf.][]{1991MNRAS.250..701T,1992ApJ...401..265M, 1994A&A...282..801G, 2007ApJ...670..747K, Sana:2012, 2014ApJS..213...34K}. This gives a factor $f_{\rm{q}} \sim 0.7$. 
Fourth, the binary system typically has a wide orbit by the time of the first supernova and likely disrupts by the significant natal kicks of hundreds of $\rm{km}/\rm{s}$ the system can receive when the \ac{NS} forms \citep[e.g.,][]{Hobbs:2005, Verbunt:2017}. Studies have found that this likely disrupts $\gtrsim 95\%$ of the systems at this point giving a $f_\textrm{SN,1} \approx 0.05$ \citep{VignaGomez:2018, Renzo:2019}. 
Fifth, the common-envelope phase might well be one of the most uncertain aspects of the binary's life. What stars and what fraction of stars can successfully eject the envelope and form a tight system is uncertain. An example from the literature for NSNS mergers is a fraction of $f_{\rm{CE}} \approx 0.2$ \citep[e.g.][]{VignaGomez:2018}. 
Sixth, the system also needs to survive the second supernova, this is more likely since the system has a much tighter orbit after the common envelope phase. We estimate the factor with a factor $f_{\rm{SN,2}}\approx0.5$ \citep[cf.][]{VignaGomez:2018}. 
Seventh, once the NSNS binary has formed it needs to have a tight enough orbit to merge within the age of the Universe so that it can be observable today. This likely removes another half of the systems. Giving a factor $f_{\rm{merge}}\approx0.5$. 

Combined this gives that a fraction of about $3 \times 10^{-7}$ of all stars forms a NSNS merger, making this a rare outcome among stellar populations.  
Assuming a local star formation rate of $\approx 1.5 \times 10^7\, M_{\odot}$ Gpc$^{-3}$ yr$^{-1}$ \citep[cf. Equation 15 in][]{MadauDickinson:2014} with the average mass of stars of  $\sim 0.3 M_{\odot}$ (from the initial mass function; \citealt{Kroupa:2001}), the yield of $f_\textrm{NSNS} = 9 \times 10^{-7}$ gives a NSNS merger rate of
\begin{align}
    \mathcal{R}_{\rm{NSNS}} \approx  \frac{1.5 \times 10^7\, \Msun \rm{Gpc}^{-3} \rm{yr}^{-1}}{0.3\Msun} \times 3 \times 10^{-7} \approx 17 \GpcminThree,
\end{align}
which falls inside the merger rates inferred from gravitational-wave observations, as well as typical simulation results \citep{MandelBroekgaarden:2021}.

In practice, there are many uncertainties in the physical processes underlying the formation, evolution, and deaths of massive stars and how these are modeled when investigating the formation pathways of  \ac{DCO} systems. Examples include uncertainties in the initial conditions \citep{deMinkBelczynski:2015, 2017ApJS..230...15M, 2018ApJ...855...20G, 2018A&A...619A..77K, 2018Sci...359...69S, Klencki:2018}, the metallicity-dependent star formation history \citep{Lamberts:2016, 2017MNRAS.472.2422M, Chruslinska:2019obsSFRD, Tang:2020, duBuisson2020,Briel:2021, Chu:2021, Chruslinska:2021, 2021ApJ...907..110B}, stellar winds \citep{Belczynski:2010, 2000A&A...362..295V,2001A&A...369..574V, Vink:2011, Renzo:2017, Vink:2017}, stellar rotation \citep[e.g.,][]{deMink:2007, Mapelli:2020}, the evolutionary tracks \citep{Kruckow:2018, 2020MNRAS.497.4549A, 2020arXiv201016333B, 2020A&A...637A...6L},  the expansion of the envelopes \citep[e.g.,][]{Laplace:2020, Romagnolo:2022}, common envelope phase (and its modeling) \citep{Webbink:1984, deKool:1990,  2012ApJ...759...52D,2016A&A...596A..58K,Kruckow:2018,2021arXiv210205649O, 2020A&A...638A..55K, Klencki:2021, 2020PASA...37...38V, 2021arXiv210205649O, Ivanova:2020book,2021A&A...650A.107M, Olejak:2021CE}, mass transfer processes \citep[e.g.,][]{KippenhahnMeyerHofmeister:1977, Soberman:1997,  Brown:2001, Sepinsky:2010,  claeys2014theoretical,  ge2015adiabatic,  Schneider:2015, Dosopoulou:2016,  Menon:2021, 2021MNRAS.505.3873B, Vinciguerra:2020, Schneider:2020, Belczynski:2021-uncertainStellarEvolution, Cehula:2023}, supernovae kicks \citep{1975Natur.253..698K, 1994Natur.369..127L, 2002ApJ...568..289A,   2002ApJ...571..906B, 2002ApJ...571L..37P, 2004ApJ...612.1044P, 2005MNRAS.360..974H, 2010ApJ...719..722S, 2015MNRAS.454.3073S,  2016MNRAS.456.4089B, 2017ApJ...846..170T, Verbunt:2017, Verbunt:2017, 2018ApJ...865...61G, 2018MNRAS.479.3675M, Igoshev:2020}, \\
supernova remnant masses \citep{1996ApJ...470L..61K, 2012ApJ...749...91F,janka2012explosion, 2015ApJ...812...24F, 2015ApJ...808..186L,  2016MNRAS.460..742M, 2017ApJ...850L..19M, 2018MNRAS.478.1377A, 2020arXiv200106102S, 2020CQGra..37d5006A, Abbott:2020gw190425,   2020MNRAS.491.2715B, 2020MNRAS.495.2786E,  RomanGarza:2020, 2020MNRAS.499.3214M, 2021arXiv210612381V,  Dabrowny:2021,  Mandel:2021MNRAS}, black hole spin up processes \citep{Spruit:2001tz, Spruit:2002, Belczynski:2017gds, Fuller:2019MNRAS,  Gerosa:2018wbw, Zevin:2020gxf, Qin:2018vaa,Fuller:2019sxi, ZevinBavera:2022, 2022ApJ...930...26S, Olejak:2021iux}, whether there is a mass gap between black holes and neutron stars \citep[e.g.,][]{2014ApJ...785...28K,2015ApJ...801...90P, 1995ApJS..101..181W,2001ApJ...554..548F, 2012ApJ...757...36K, 2016MNRAS.458.3012W, 2020A&A...636A..20W, Zevin:2020-lower-mass-gap, 2020ApJ...890...51E, 2020MNRAS.495.3751C} and the existence of a black hole mass gap due to pair-instability supernova \citep{1964ApJS....9..201F, 1967PhRvL..18..379B, 2017ApJ...836..244W, 2019ApJ...887...53F, 2017ApJ...851L..25F,2018ApJ...856..173T,Abbott:2021GWTC1, 2019PhRvD.100d3012W, 2019MNRAS.484.4216R,2020PhRvD.102h3026G, Woosley:2002,2019ApJ...878...49W,2021arXiv210307933W}. These uncertainties can all impact the outcome of stellar populations, making it challenging to model them all or rely on a limited set of models. On the other hand, it also means that observations of \ac{DCO} mergers with ground-based gravitational-wave detectors, as well as other observations, have the potential to test and constrain the physics of these evolutionary phases.

\subsection{Other formation channels}
Many formation channels other than the isolated binary evolution channel have been proposed. All of these also overcome the separation channel. We describe them briefly and refer the interested readers to the reviews from \citet{Mapelli:2021review, GerosaFishbach:2021, MandelFarmer:2018}, \citet{ArcaSedda:2023review}, and \citet{2022Galax..10...76S}, and references therein, and to Part \ref{part:FCdynamical} for more details. 

\begin{itemize}

    \item \textbf{Population-III stars:} in this scenario, the binary consists of two massive metal-free (or extremely metal-poor) stars in a tight orbit. Such metal-poor stars are expected to exist in the early Universe as the first generation of stars that have not yet been exposed to enrichment processes that would lead to higher metallicities. These stars are thought to avoid extensive radial expansion of the envelope and avoid the separation challenge. Stellar winds and radial expansion are driven by the presence of heavier metals, causing these stars to experience a drastically reduced radial expansion and self-stripping through winds \citep{Vink:2001, Marigo:2001, Lovegrove:2013, Fernandez:2018, Kinugawa:2021}. Moreover, the lack of metals might also favor the formation of higher mass stars (a flatter initial mass function), further boosting the potential contribution from this channel \citep[see][and references therein]{Belczynski:2004popIII, Kinugawa:2014, Inayoshi:2016, Hartwig:2016, Belczynski:2017, Hijikawa:2021, Kinugawa:2020, LiuBromm:2021, Tanikawa:2021}.  
    
    \item \textbf{Chemically Homogeneous Evolution:} in this scenario, a pair of massive metal-poor stars starts in a close enough orbit such that the stars are tidally locked. The low metallicity again suppresses strong winds and reduces the radial expansion of the stellar envelopes. The tidal locking leads the stars to spin up to a rotation close to the stellar break-up frequency that leads to efficient mixing of the helium out of the stellar core into the envelope and fresh hydrogen from the envelope into the stellar core. This allows the stars to evolve chemically homogeneously and use almost the entire hydrogen for fusion until eventually, the entire star has formed a naked helium star \citep{Eddington:1925, Sweet:1950, EndalSofia:1978, Heger:2000, MaederMeynet:2000, Yoon:2006}. The heavy helium stars do not expand much and can avoid mass transfer phases during the rest of their lives, avoiding the separation challenge. 
    The stars will eventually collapse into \acp{BH} and form a \ac{BBH} system with two \acp{BH} of similar masses ($\sim 30\Msun$) that are likely highly spinning and can contribute to the \ac{DCO} population observed today \citep{MandelDeMink:2016, deMinkMandel:2016, Marchant:2016, duBuisson2020, Riley:2020}. In rare cases, this formation channel can also form BHNS mergers \citep{Marchant:2017}.

    \item \textbf{Isolated triples or higher order multiples:} massive stars are also commonly formed in triples or higher order multiples \citep[e.g.][]{MoeDiStefano:2017}. In such systems, the stars might experience similar evolutionary processes as described in the isolated binary evolution channel, but additionally, the additional star(s) can impact the system both dynamically and through binary-type interactions (supernovae, mass transfer, etc.). Several studies have extensively investigated the evolution of triples and higher order multiples and found that they can significantly contribute to the formation of \ac{DCO} systems \citep[e.g.,][]{SilsbeeTremaine:2017, Antonini:2017, RodriguezAntonini:2018,  FragioneLoeb:2019a,  Martinez:2020, HamersThompson:2019, Trani:2021, VynatheyaHamers:2021, Stegmann:2022}. Another proposed channel comes from field binaries that might be perturbed by flybys from other stars that can increase the eccentricity and lead to higher merger rates; a channel that lies in between isolated and dynamical formation like triples and higher order multiples \citep{Raveh:2022, MichaelyNaoz:2022}.

    \item \textbf{Dynamical channels:} instead, \ac{DCO} mergers might dominantly form from stars in dense stellar environments such as globular clusters, young star clusters, and nuclear star clusters in the centers of galaxies.. In all these channels the dynamical interactions play a dominant role in the formation of the \ac{DCO} system\footnote{Where we note that isolated triples and higher multiples also belong to this category}, other examples for this are the formation of \ac{DCO} in hierarchical 3-body systems with a \ac{SMBH}, and in AGN disks (see also Part \ref{part:FCdynamical}). 

    \item \textbf{Exotic channels:} other, more exotic formation channels have also been suggested. This includes the formation of \ac{BBH} mergers from primordial black holes, or from other physics beyond the standard model \citep[e.g.,][]{Bird:2016, AliHaimoud:2017, Sakstein:2020, Croker:2021}. These are also discussed in Part \ref{part:FCdynamical}. 
    
\end{itemize}

Each of these channels comes with its own model uncertainties, contributing to the complexity of gravitational-wave paleontology. On the other hand, the rapidly increasing population of \ac{BBH} mergers is a probe of all these physical processes that are underlying the formation and lives of \acp{BH} and their massive star progenitors.

\newpage
\part{Black Holes in binaries: formation channels - dynamical \\ \Large{Carl L.~Rodriguez}}
\label{part:FCdynamical}

\section{$N=3$; Triples and Three-body Encounters}
\label{s:triples}

The first extension beyond binary physics is to consider the presence of a third body.  But whether this triple system is stable or not changes the long-term dynamical evolution of the system, as well as the properties of the BBHs it produces.  Configurations of three gravitational point particles which are known to be stable for many orbital periods are often referred to as hierarchical triples; these can be thought of as ``a binary within a binary'' (that is, there exists an inner binary with semi-major axis $a_1$, and an outer tertiary that orbits the binary's center of mass at a wider separation $a_2$).  Due to their separation in scales, these systems typically do not exchange energy between their inner and outer orbits (e.g.~the Earth-Moon-Sun system).  Decreasing the outer orbital separation, however, also decreases the stability of the system; while there does not exist a good first-principles derivation of this stability criterion, numerical experiments \citep{Mardling2001} building off the analytic arguments of \citep{Eggleton1995} have shown that triples with"

\begin{equation}
\frac{a_2}{a_1} > \frac{2.8}{1-e_2} \left[\left(1+\frac{M_t}{M_b}\right)\left(\frac{1+e_2}{\sqrt{1-e_2}}\right)\right]^{2/5} \left(1-0.3 \frac{I}{\pi}\right)
\label{eqn:mardling}
\end{equation}

\noindent for a system with inner-binary mass $M_b$, tertiary mass $M_t$, inner and outer eccentricities of $e_1$ and $e_2$, and a mutual inclination of $I$, tend to be long-lived and stable (though see \cite{Zhang2023} for an updated study focusing on the extreme-mass ratio limit).  We will refer to systems where Eqn.~\ref{eqn:mardling} is not met as democratic triples (in contrast to stable hierarchical triples).  

\subsection{Hierarchical Triples}

While hierarchical triples (in the absence of tides or other dissapative forces) do not transfer energy between the inner and outer orbits (and therefore do not 
alter their semi-major axes),  the key feature of these systems is their exchange of angular momentum over many orbital periods.  These 
secular processes, as they are often called, have been known for many decades, going back to the original papers by Lidov \citep{Lidov1962} and Kozai 
\citep{Kozai1962} for whom the Lidov-Kozai (LK, though often shortened to ``Kozai'') mechanism is named.  These hierarchical triples are of interest for a wide range of systems in 
astrophysics, from planetary dynamics, to the inclination of asteroids, to GW source production. 

To lowest (quadrupolar) order, the LK mechanism involves an exchange of angular momentum between the mutual orbit of the two binaries and the inner binary.  If 
we assume a system in which both the inner and outer binaries are initially circular, and one of the inner masses is zero (i.e.~the test-mass approximation), 
one can show that the maximum eccentricity the inner binary can achieve during these oscillations is:

\begin{equation}
	e_{\rm{max}} = \sqrt{1-\frac{5}{3}\cos^2 i_0}
\end{equation}

\noindent where $i_0$ is the initial inclination of the inner binary with respect to the outer binary.   See \cite{Naoz2016} for a detailed review.  In this regime, the inner binary can reach arbitrarily large eccentricities which, since the timescale for binaries to merge due to GW emission scales as $T_{\rm GW} \propto (1-e_0^2)^7/2$ \cite[][c.f.~Equation 5.14, ]{Peters1964}, can rapidly drive even wide binaries to merge within a Hubble time.

The first studies of this effect in the context of stellar-mass BHs were performed in the early 2000s by \cite{Miller2002} in the context of BBHs formed through four-body interactions in GCs.  It was further noted by \cite{Wen2003} that these systems could potentially maintain their eccentricity while passing into the LIGO sensitivity band, providing a mechanism for distinguishing dynamically-formed binaries from those formed from isolated evolution (although subsequent studies have shown the percentage of mergers with $e > 0.1$ in the LIGO band to be substantially lower than the reported 30\% from \cite{Wen2003}).   We note that there exists significant additional complications one can consider to this process, such as the presence of higher-order terms in the Kozai-Lidov expansion (which give rise to the so-called eccentric Kozai-Lidov mechanism, e.g.,\cite{Ford2000,Blaes2002,Naoz2013,Liu2015,Liu2015b}), complicated multi-scale cross-terms between the three-body and post-Newtonian expansions \cite[e.g.,][]{Naoz2013a,Will2014,Lim2020,Liu2020,Will2021,Kuntz2023}, and even the breakdown of the Kozai-Lidov orbit approximation entirely \cite{Antonini2014,Antognini2014} (which are responsible for producing the most eccentric of the triple-induced mergers).

Since then, many authors have explored the potential contribution of triples formed through in a variety of astrophysical environments.  Observational evidence suggests that massive O-stars above 16 $M_{\odot}$ (the progenitors of BHs) have an average of 2.1 companions \citep[][Table 13]{moe2017}, making CO triples a natural product of star formation.  These triples can undergo Kozai oscillations that drive their inner binaries to merge \cite[e.g.,][]{Silsbee2017,Antonini2017,Rodriguez2018a,Stegmann2022,Dorozsmai2023}, with further studies showing that such triple-mediated mergers exhibit unique mass ratios \cite[e.g.,][]{Su2021,Martinez2022} \& spin-orbit alignments \cite[e.g.,][]{Antonini2018,Rodriguez2018a,Liu2018,Yu2020,Fragione2020a} that may distinguish them from either field-formed binaries or binaries formed through other dynamical channels (although eccentricity remains the most reliable metric for distinguishing triple-induced mergers, despite it's degeneracy with other dynamical channels).  

In addition to massive stellar triples,  CO triples are often formed or found in other dynamical environments.  In stellar clusters, such as OCs and GCs, triples are a common byproduct of binary-binary encounters, in which two binaries dynamically interact and (typically) eject one of the components of the wider binary, leaving a hierarchical triple and a single star/CO \cite{Miller2002,Antonini2016,Kimpson2016,Banerjee2018,Martinez2020,Britt2021,ArcaSedda2021,Trani2022}.  Note that this process becomes more efficient in smaller clusters such as OCs; more massive clusters such as GCs and galactic nuclei form triples with similar masses (with the rate of such mergers in GCs being two orders of magnitude lower than those from standard binary hardening that we describe in Section \ref{s:gc}, e.g. \cite{Antonini2016,Martinez2020}).  

The one notable exception to this is nuclear star clusters containing a super-massive BH.  In these clusters, stellar binaries can form hierarchical triples where the SMBH acts as the tertiary driving the stellar-mass inner binary to merger \cite[e.g.,][]{Antonini2012,VanLandingham2016,Stephan2016,Hoang2018,Hamers2018,ArcaSedda2020,Wang2021}.  We note that, unlike triples with similar mass tertiary companions, stellar-mass binary/SMBH triples may be directly confirmed by the measurement of Doppler shift and other effects from the SMBH with space-based GW detectors \cite[e.g.,][]{Meiron2017,Torres-Orjuela2019,Wong2019}.

\subsection{Democratic Triples}

\begin{figure}[h]
    \centering
    %\vspace{-0.3cm}
    \includegraphics[width=1.0\textwidth]{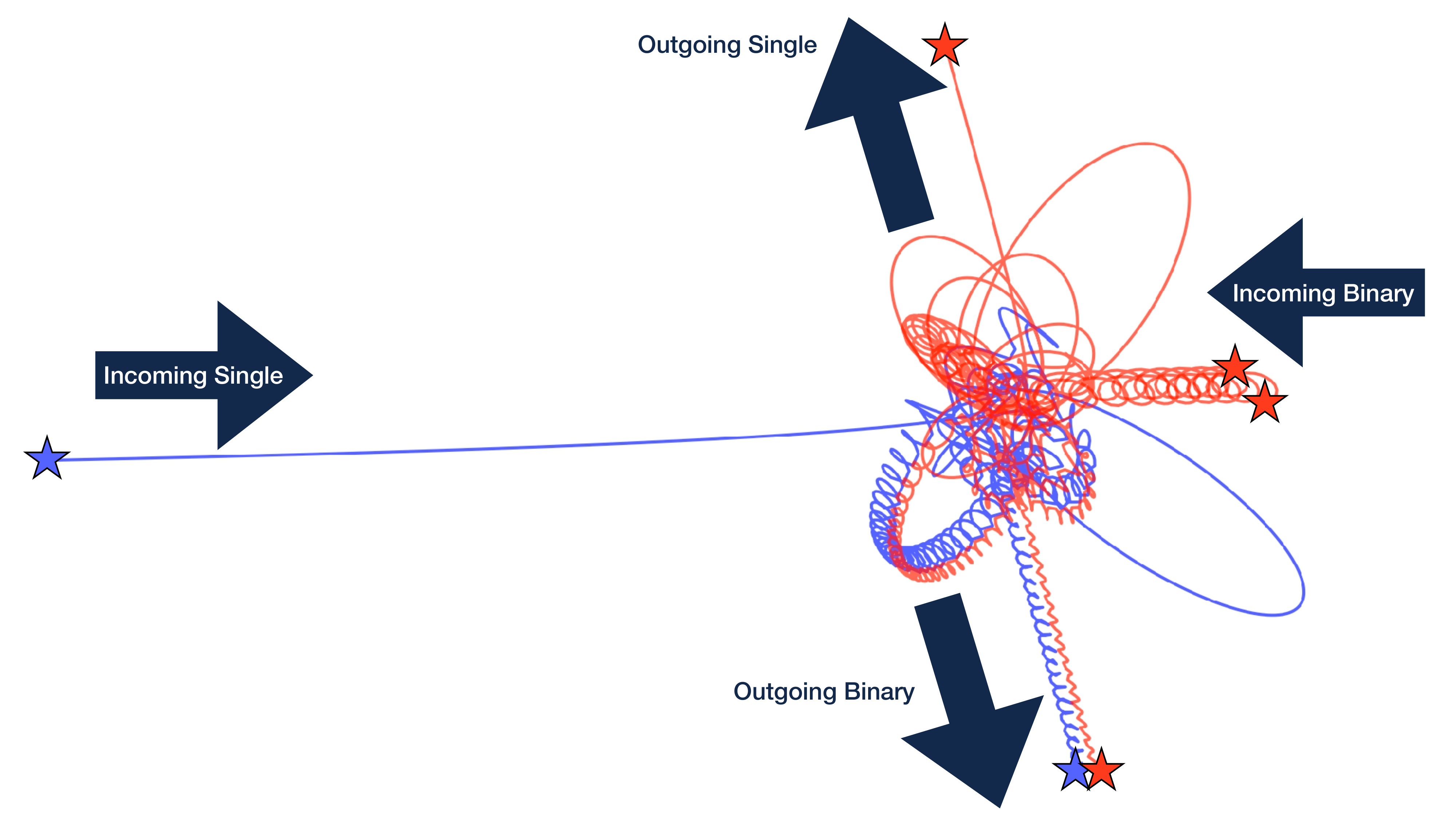}
    \caption{An example strong encounter between a binary and a single object in which the incoming particle exchanges into the binary, ejecting one of the original binary components out to infinity.  These encounters occur frequently between massive stars and/or BHs in dense star clusters such as OCs, GCs, and GN, and are critical for the production of BBH mergers.  This scattering was performed with the \textit{Fewbody} code \citet{Fregeau2004}.}
    \label{fig:scattering}
\end{figure}

While hierarchical triples are of significant interest due to their potential long-term stability, there exists a long and relevant literature on short-lived democractic triples, typically formed by the strong interaction of a single star or CO and a binary.  See, e.g., Figure \ref{fig:scattering}.  The fate of triples formed via these "scattering encounters" was first studied numerically 40 years ago by Hut and Bahcall \cite{Hut1983b,Hut1983,Hut1983a}, and has since spawned an entire industry dedicated to understanding the outcome of these encounters.  

Because of the high densities required for such encounters to take place, they are typically found only in the cores of dense star clusters such as OCs, GCs, and NSCs.  The one notable exception is the possibility of encounters between stars and wide binaries in galaxies, where many of the same interactions that occur in clusters can be replicated due to the large interaction cross sections of ultra-wide binaries.  See \cite{Michaely2019,Michaely2020}.  The effect of these encounters on BBH mergers depends on whether such encounters primarily occur among massive stars before they form BHs (as often happens in lower-density OCs) or among BHs and BBHs in the cores of dense clusters such a GCs and GN.  We will now describe these various pathways.

\section{$N=\mathcal{O}(10^3-10^5)$; Open Clusters}
\label{s:oc}

As we add more particles to the systems we consideration, the number of possible interactions increases as well.  Despite the previous sections where have assumed that most stars form as either isolated binaries or triples, we in fact know that most star formation occurs in clusters, whether they be unbound associations or YMCs such as OCs \cite{Lada2003}.  And while the majority of massive stars are born with at least one companion \cite{moe2017}, most of these binaries and triples should actually seen as living (at least at first) within dynamically active environments, where encounters with neighboring stars can modify the evolutionary path the binary would have followed had it evolved in isolation.  These modifications can take many forms, such as exchange encounters (like the one pictured in Figure \ref{fig:scattering}) in which one component is replaced with another, or resonant encounters in which the orbital period and eccentricity of the binary are dynamically scrambled to a state that would be inaccessible to a system evolving in isolation.  And while the small escape speeds of OCs mean that a single encounter is often enough to eject the stellar binary from the cluster, if many, or even the majority, of binaries formed through normal star formation undergoes a single dynamical encounter, it can have a profound effect on the population of merging BBH.

Over the past two decades, many authors have used direct $N$-body techniques to explore the formation, evolution, and destruction of these clusters and the the BBHs they produce \cite[e.g.,][]{PortegiesZwart2000,Banerjee2010,Ziosi2014,Arca-Sedda2016,Kimpson2016,Banerjee2017,Banerjee2018,Banerjee2018b,Rastello2019,DiCarlo2019,DiCarlo2020}.  By assuming that the majority of star formation occurs in these environments, they can reproduce a significant fraction (or even potentially all) of the BBH mergers observed by LIGO \cite[e.g.,][]{Kumamoto2020,DiCarlo2020}.  The interactions, and particularly the dynamical exchange encounters, that occur among massive stars in these clusters create binaries that can move along unique evolutionary pathways, producing binaries with significantly larger masses than those found through isolated binary evolution.   Dynamical encounters can also alter the alignment between the spin and orbital angular momenta of the binaries \cite{Trani2021,Banerjee2021}, another key observable of dynamical formation (though see \S \ref{ss:observing} where we discuss spin-orbit alignment in detail).

However, one of the most unique outcomes of stellar encounters prior to BH formation is the potential to produce individual BHs more massive than those that can be formed from stellar collapse alone.  In single stars, it is largely assumed that BHs cannot form with masses between $\sim 50M_{\odot}$ and $120M_{\odot}$, due to the presence of pulsational pair instabilities and pair-instability supernova \cite[e.g.,][]{Woosley2019,Farmer2019}.    But when two stars merge in OCs, their merger products can potentially bypass this limit by developing unique core-to-envelope mass ratios that cannot be realized through single or binary stellar evolution \cite[e.g.,][]{DiCarlo2020a,Banerjee2021}.  These stars, upon their formation, can then exchange into other stellar binaries, eventually forming BBHs with masses that cannot be the result of isolated stellar evolution, a key indicator of dynamical formation (though as we will see below, not a unique one, as repeated BH mergers can also create BHs with masses in the pair-instability mass gap).  Of course, stellar mergers produce merger products with inflated radii, that then in turn increases the possibility of successive stellar mergers.  These potential ``runaway'' collisions have been studied for nearly two decades \cite{PortegiesZwart2004,Freitag2006a,Freitag2006,Gaburov2008,Mapelli2016,DiCarlo2021}, and represents both an important channel for the production of both the heaviest LIGO sources as well as the potential seeds of intermediate-mass BHs and SMBHs.

\section{$N=\mathcal{O}(10^4-10^7)$; Globular Clusters}
\label{s:gc}

As we consider systems with more particles (and correspondingly deeper gravitational potentials), the primary mechanism of BBH formation shifts.  While smaller clusters largely rely on a natal population of massive binaries to create their BBHs, the stellar binaries that eventually become merging BBHs in OCs only undergo about one strong encounter before they are ejected from the cluster.  But as we move to larger systems, such as GCs these primordial binaries become less relevant to the formation of BBHs for two reasons: first, the higher mass (and correspondingly higher escape speed) of the clusters means that significant fractions of BHs can be retained by the cluster after formation \citep{Morscher2015}.  This means that, unlike OCs, BBHs actually remain (and form dynamically!) in the clusters that create them.  And second, because of their deeper central potentials, these binaries undergo many repeated encounters before they are ejected from the cluster.  These formally chaotic three-body encounters erase any information about the initial state of the binary, yielding a population of BBHs whose origin is purely a function of gravitational dynamics.

While their formation is still an area of active research, it is easiest to consider GCs as forming from a single burst of star formation $\sim10-14$ Gyr ago.  Within 10 Myr, the massive stars in the cluster collapse, creating a population of stellar-mass BHs; assuming the BHs receive low or no natal kicks due to increased supernova fallback \citep{Belczynski2006,Fryer2012} or direct-collapse \citep{Fryer2001} scenarios, a significant fraction \citep[anywhere from 65-90\%,][]{Morscher2013,Morscher2015} can be initially retained by the cluster.  The BHs experience a drag force arising from dynamical friction \cite[e.g.,][\S 8.1]{Binney2008}, causing their orbits to decay, and driving them into the cluster core where they form a subsystem of BHs.  %The timescale for this mass segregration goes roughly as $T_{\rm MS} \approx \frac{m_{\rm BH}}{\left<m\right>} T_{\rm rlx}$, where $m_{\rm BH}$ is the mass of the BH, $\left<m\right>$ is the average stellar mass, and $T_{\rm rlx}$ is the half-mass relaxation time of the cluster \citep[e.g.,][]{spitzer1987}.  Assuming a $30M_{\odot}$ BH surrounded by $0.5M_{\odot}$ stars and using GC properties from \citep{baumgardt2018}, 90\% of the GCs in the MW will have completed BH mass segregation within $\sim 200$ Myr.  Once this process is complete, the central regions of a GC can contain 100s to 1000s of BHs at spatial densities of $\gtrsim10^4 \rm{~pc}^{-3}$ \citep[assuming a typical GC model from the catalog described in][CARL DOUBLE CHECK NUMBER]{kremer2019b}.
These BH subsystems are simultaneously key to understanding the overall evolution of GCs as well as the BBHs they produce.  As the cluster collapses due to relaxation \citep[e.g.,][]{Binney2008}, the density of BHs in the core increases.  Eventually, the density becomes high enough to facilitate three-body interactions, where BBHs are either formed through encounters of three single BHs \citep[][]{Sigurdsson1993,Kulkarni1993,Ivanova2005,OLeary2006,Morscher2013} or three-body encounters between pre-existing binaries and single BHs (such as those described in \S \ref{s:triples}).  Once a BBH has been formed it will continue to participate in encounters with other BHs each of which will, on statistical average, shrink the BBH's orbital period \cite{Heggie1975}.  This process continues until either the BBH merges due to GW emission \citep[typically in eccentric binaries, e.g.,][]{Rodriguez2018c,samsing2018}, or the recoil of the binary from the three-body encounter is sufficiently large to eject it from the cluster entirely, where it may or may not merge later due to GW emission.  As a result, it is largely the central escape speed of the cluster that sets how compact a BBH will be \citep{Moody2009,Rodriguez2016a}.  GCs typically process their BHs from most to least massive, forming binaries with the heaviest BHs first before moving on to the lower-mass objects.   This, combined with their age and low stellar metallicity, makes GCs typically more efficient at producing heavier BBH mergers in the local Universe; for a $30M_{\odot}+30M_{\odot}$ BBH like GW150914, the optimal cluster to produce a merger in the local universe has a present-day mass of $\sim3-6\times10^5M_{\odot}$ \citep{Rodriguez2016b}, corresponding roughly to the upper quartile of GC masses in the MW \citep{baumgardt2018}. 

%First, the central regions of the cluster, where the majority of the BBH formation and hardening occurs, are dominated by the most massive BHs in the cluster (due to mass segregation).  These BHs must be forged into binaries and ejected before the lighter BHs can move into the center, meaning that GCs process their BH subsystem from most-to-least massive \cite{Morscher2015}.  At the same time, nearly 40 years worth of detailed $N$-body scattering experiments \citep{Hut,Hut1983b,Hut1983a,Heggie1993,Sigurdsson1993a} have shown that the hardening three-body encounters preferentially produce binaries with near equal masses, with lower-mass companions being more likely to be ejected from a binary in favor of a more-massive BH.  In other words, GCs process through their BH subsystems in order of decreasing mass, with a strong preference for producing equal-mass binaries.

%Despite the complexity of their dynamical evolution and formation these two facts---the linear relationship between galaxy mass and number of GCs, and the orbital period of the BBHs it produces---makes estimating both the BBH production rate and their properties relatively straight forward.  
Over the past three decades, many authors have used either direct-summation or Monte Carlo-based $N$-body simulations \cite[e.g.,][]{Tanikawa2013,PortegiesZwart2000,Morawski2018,Hong2018,Downing2010,Downing2011,Chatterjee2017,Chatterjee2016a,Bae2014,Askar2016,Aarseth2012,Rodriguez2015a,Rodriguez2016a,Rodriguez2018,Chattopadhyay2022,Fujii2017} or semi-analytic techniques \cite[e.g.,][]{OLeary2006,Moody2009,Fragione2018,Rodriguez2018b,choksi2019,Antonini2020,Antonini2020a,Antonini2022,Kritos2022} to study BBH formation from GCs, with most estimates of the BBH merger rate in the local universe \citep[going back to][]{Sigurdsson1993} hovering around $\sim 10\rm{~Gpc}^{-3}$.  While this number is potentially sufficient to explain the full population of BBHs detected so far \citep{Rodriguez2021}, more sophisticated analyses suggest that they are one of several channels contributing to the LVK GW detections \cite[e.g.,][]{zevin2021}.  

While there are still uncertainties associated with the GC channel (mostly related to GC formation and the evolution of massive stars and compact-object formation), the gas-free dynamics of these environments does allow several robust predictions to be made for the BBH population, and in many ways, GCs represent the first environments (in terms of increasing numbers of particles), where BBH formation is purely dynamical, and does \emph{not} depend on the properties of the initial binary population.  This transition marks the introduction of several observable features that are relevant to GCs, NSCs, and AGN, which we will now describe.    %The first of these is that because the central regions of the cluster, where the majority of the BBH formation and hardening occurs, are dominated by the most massive BHs in the cluster (due to mass segregation).  These BHs must be forged into binaries and ejected before the lighter BHs can move into the center, meaning that GCs process their BH subsystem from most-to-least massive \cite{Morscher2015}.  At the same time, nearly 40 years worth of detailed $N$-body scattering experiments \citep{Hut,Hut1983b,Hut1983a,Heggie1993,Sigurdsson1993a} have shown that the hardening three-body encounters preferentially produce binaries with near equal masses, with lower-mass companions being more likely to be ejected from a binary in favor of a more-massive BH.  In other words, GCs process through their BH subsystems in order of decreasing mass, with a strong preference for producing equal-mass binaries.

\subsection{Observational Signatures of Purely Dynamical Formation}
\label{ss:observing}

First, because BBHs in GCs are primarily created and driven to merger through chaotic three-body encounters, the spins and orbital angular momenta of the BBHs are essentially randomized after a single strong encounter, causing them to be isotropically distributed.  GW detectors are primarily sensitive to the projection of the BH spins along the direction of the binary angular momenta, making the presence of spin-orbit anti-alignment in the BBH population a key observable \citep[e.g.,][]{Abbott2021}.  Isolated binaries are expected to form BBHs with spins largely aligned with their orbit, while both hierarchical triples and OCs tend to push the distribution away from aligned and towards spin-orbit misalignment (while still maintaining a preference for spin alignment).  In effect, as the number of particles increases, the isotropy of the distrubtion increases, from nearly aligned when $N=2$ to isotropic once $N\gtrsim 10^5$.   

Second, the strong encounters that produce BBHs also facilitate eccentric mergers through GW captures during democratic triples.  Simulations of both isolated scatterings and full Monte Carlo $N$-body models have suggested that $\sim 20\%$ of BBH mergers that occur inside GCs \cite{Rodriguez2018c,Samsing2018a} are the result of GW captures during strong encounters where the emission of GWs during close hyperbolic encounters is sufficient to create bound binaries from previously unbound BHs.  Roughly half of these captures will reach a GW frequency of 10Hz with $e\gtrsim0.1$ and be detectable by LIGO.  As was described in \S \ref{s:triples}, eccentricity remains a key observable of dynamical formation for BBHs, and while the ratio between in-cluster mergers and ejected mergers can change depending on the central escape speed (from roughly half-and-half for typical GCs to nearly 90\% or more for very massive clusters, e.g., \citep{Antonini2019,rodriguez2020}), the fraction of in-cluster mergers with eccentricity above 0.1 remains largely constant.

Finally, if two BHs merge inside a cluster, there is nothing to stop the new BH they form from participating in further mergers. It is well known from numerical and analytic relativity that merging BBHs receive a recoil kick due to the asymmetric emission of GWs \cite[e.g.,][]{Merritt2004,Campanelli2007,Gonzalez2007,Lousto2008,Gualandris2008}, with the kicks of highly-spinning, non-equal-mass BH mergers reaching as high as $5000~\rm{km}/\rm{s}$ \cite{Lousto2011}.  However, if BHs are born with low spins from stellar collapse (as suggested by some studies of angular momentum transport in massive stars, e.g., \cite{Fuller2019}), then the merger products of near-equal mass BHs can sometimes be retained by their host clusters, where they can potentially find new BH companions through three-body encounters and merge again.  The number of previous merger events that a BH has undergone is referred to as its ``generation'', with BHs born from stellar collapse being ``first-generation'', BH formed from a single previous BH merger being ``second-generation'', and so on (following \cite{Gerosa2017}).  GCs, with their relatively low central escape speeds ($\lesssim 100~\rm{km}/\rm{s}$ for the most massive clusters, though potentially higher at formation \cite{rodriguez2020}), can easily produce BBH mergers with second-generation BH components when the BH spins from stellar collapse are small \cite[e.g.,][]{Rodriguez2018,Rodriguez2019}, but they rarely retain BBH third or higher generation mergers.  While OCs rely exclusively on stellar mergers to create BHs in the pair-instability mass gap, GCs represent the first environment where BH growth is facilitated by repeated BH mergers.  But to produce multiple successive generations of BH mergers, or even intermediate mass BHs, we must consider even deeper potential wells.

\section{$N=\mathcal{O}(10^5-10^8)$; Nuclear Star Clusters}

The final frontier of dense star clusters, both in terms of the number of particles, and computational difficulty, remains the NSCs that populate the centers of many galaxies.  These environments, which can span from $N=10^5$ (similar to low-mass GCs) to $N\gtrsim10^8$ numbers of stars and binaries, represent the highest density stellar regions known, with some NSCs (such as the nuclei in M32 and the MW) exceeding $\sim 10^7M_{\odot}/\rm{pc}^3$ (on scales of the central $\sim 0.1\rm{pc}$ and $\sim 0.01\rm{pc}$ respectively, \cite{Lauer1998,Schodel2018}).  Observations of nearby NSCs \cite{Pechetti2020} showing a clear trend between the mass of the host galaxy and the mass and density profile of their NSCs with more massive galaxies hosting more massive NSCs with flatter profiles.  See \cite{neumayerNuclearStarClusters2020} for a detailed review of our current observational understanding of NSCs, their hosts, and their potential evolutionary paths.

Unlike GCs, which are relatively simple stellar environments (typically spherical and forming in a single burst of star formation lasting $\lesssim 10\rm{Myr}$), NSC sadmit many complicated structures (e.g.~disks, triaxiality), formation histories (possibly forming from infalling GCs \cite[e.g.,][]{Tremaine1975}, \textit{in situ} star formation \cite[e.g.,][]{Loose1982}, or a combination of the two; see \S 7 of \cite{neumayerNuclearStarClusters2020}), with the formation history and other properties being closely related to the evolutionary history and morphology of their host galaxies.  Their internal dynamics is also complicated by the presence of SMBHs, which cause a significant increase in stellar density and velocity dispersion within the central regions of the NSC (via the formation of a ``Bahcall-Wolf Cusp'' \cite{Bahcall1976,Bahcall1977}).  

Given the extreme computational difficulties of modeling NSCs, most studies of BBH formation in these environments have relied upon analytic models of varying degrees of sophistication, with assumptions that change dramatically depending on whether or not an SMBH is present.  We now explore both of these scenarios.

\subsection{GN without SMBHs}
While the actual occupation fraction of SMBHs inNSCsis highly uncertain, there are GN, such as the nucleus of M33 \cite{Gebhardt2001}, that are known to lack SMBHs. In many ways,NSCswithout central-massive BHs can be modeled (to lowest order) as ultra-massive GCs with more complicated star-formation histories.  In these systems, the predominant mechanism for forming and merging BHs is three-body binary formation and dynamical hardening, and many of their observable properties, such as the BH spin orientations, are similar to that of GCs (see \S \ref{ss:observing}).  But, because of their substantially higher central densities (and correspondingly higher central escape velocities), they can both retain lighter-mass BHs following their supernova \cite[e.g.][]{Mapelli2021} and more easily retain multiple generations of BH merger products, easily producing BHs in the pair-instability mass gap or in the intermediate-mass regime \cite[e.g.,][]{Antonini2019,Fragione2020,Mapelli2021}.

Because all but the least massive NSCs remain beyond the capabilities of direct $N$-body integrators, most studies of BH mergers in nuclei have relied on a combination of three-body scatterings \cite[e.g.,][]{Miller2009,Leigh2018,Codazzo2023} and semi-analytic techniques \cite[e.g.,][]{Antonini2016a,Fragione2020,Mapelli2021,Fragione2022,Fragione2023,Chattopadhyay2023}.  Because of their higher escape speeds, a greater fraction of NSC BH mergers occur within the cluster (as compared to GCs), which in turn produces roughly four times as many highly eccentric mergers in the LIGO band (with $\sim 20\%$ passing 10Hz with $e>0.1$, \cite{Chattopadhyay2023}).

\subsection{GN with SMBHs}

In NSCs with SMBHs, the central regions of the NSC become completely dominated by the Keplerian potential of the central BH which, in combination with two-body relaxation puhsing the system towards dynamical equilibrium, drives the system towards a high density, high velocity dispersion configuration known as a Bahcall-Wolf Cusp \cite{Bahcall1976,Bahcall1977}.  As with GCs, BHs tend to mass segregate into the centers of SMBH cusps, increasing their density towards the central regions of the GN.  However, the increased velocity dispersion in these central regions also tends to suppress binary formation via three-body encounters, where the rate of binary formation scales as $1/v^9$, \cite[][]{Hut1985}.  Instead, the formation of CO mergers in SMBH cusps is largely powered by two-body GW capture, similar to eccentric BBH production in GCs.  This idea has been invoked for decades \cite[e.g.,][]{Giersz1985,Quinlan1987,Quinlan1989} to explain formation of hard binaries in dense stellar systems and the long-term evolution of GN.  

Because the velocity distribution and mass segregation of BHs in cusps can be well understood from collisional theory \cite[e.g.,][]{Hopman2006,Alexander2009}, the rates, mass, and eccentricity distributions of BHs merging in NSC cusps can be predicted analytically (for a given SMBH mass and stellar/BH population).  The first detailed treatment of this came from \cite{OLeary2009}, which showed that the production of mergers is dominated by BHs closest to the SMBH.  More recent work by \cite{Gondan2018,Gondan2023} has shown that this channel produces a strong correlation between the mass of the merging BHs and their eccentricity (as more massive BHs segregate deeper into the cusp where the Keplerian velocities are higher, requiring a closer encounter pericenter for GW capture to be effective).  While easily analytically tractable, the overall merger rate of BBHs in NSC cusps is likely lower than that induced by the Kozai-Lidov process about the SMBH in the absence of gas.

In the presence of gas, however, the dynamics changes considerably.  In particular, in active galactic nuclei (AGN), the presence of a gas disk can significantly alter the dynamics of the NSC.  Once a disk forms in an NSC, BHs find themselves either ground down into the disk from isotropically-distributed orbits in the NSC \cite[e.g.][]{Artymowicz1993,Karas2001,Bartos2017}, with the gas drag from repeated plunges through the disk causes the inclination of stellar orbits to align with the disk, or they form \textit{in situ} in the fragmenting disk itself \cite[e.g.,][]{Stone2017}.  Once embedded within the disk, the BHs begin to experience torques due to both gas drag and gravitational torques from the disk itself, causing the BHs to migrate throughout the disk (in a fashion very similar to planetary migration within proto-planetary disks, \cite{McKernan2012,McKernan2014}).  

While the rates of BH mergers in AGN disks are highly uncertain (spanning several orders of magnitude \cite[e.g.,][]{McKernan2018}), the effective reduction of the dynamics into a 2-dimensional disk alters several of the key features of dynamical formation in BBH mergers.  First, by focusing three-body encounters into a co-planar environment, it has been shown that the number of highly-eccentric mergers increases dramatically compared to 3-dimensional environments, with $\gtrsim 30\%$ of mergers retaining significant eccentricity into the LIGO band \cite{Tagawa2021,Samsing2022}, while interactions with the gaseous disk itself may also produce eccentric mergers through evection resonances \cite{Munoz2022} (though we note that 3D hydrodynamical simulations of BBHs in AGN disks have come to less definitive results \cite{Dempsey2022}).  At the same time, the proximity of these mergers to the central-massive BH means that any AGN-produced BH mergers occur in environments with extremely large escape speeds, potentially facilitating many generations of hierarchical BH mergers \cite{Yang2019,Secunda2019,Tagawa2020}.  While such an efficiency of hierarchical BH production may in fact \emph{overproduce} massive BHs compared to the LIGO/Virgo population \cite[e.g.,][]{Zevin2022}, the finite lifetime of AGN disks places a natural upper-limit on the production efficiency of such high-mass mergers.

Of course, the most tantalizing potential observable of the AGN channel is the potential production of a bright optical transient associated with the merger of a BH embedded in the disk \cite[e.g.,][]{McKernan2019,Tagawa2023}.  A joint observation of both an electro-magnetic (EM) transient and GWs from a BH merger would not only definitively confirm the AGN channel as a potential source of BBH mergers, but would allow for a detailed probe of the interior of AGN environments \cite{McKernan2019}, and even independent standard-siren measurements of the Hubble constant \cite{Gayathri2021,Chen2022} (at redshifts significantly higher than those of binary neutron star mergers).  While there were initial claims that the high-mass BBH GW190521 may have had an associated EM flair detected by the Zwicky Transient Factory \cite{Graham2020}, to date no EM counterpart has been confidently associated with a GW event \cite{Ashton2021,Palmese2021}, and current surveys suggest the contribution from the brightest AGN may be limited \cite{Veronesi2023}.  However, in spite of the significant uncertainties and difficulties modeling these environments, the potential of producing an EM counterpart makes the AGN channel one of the most compelling dynamical environments and an area of significant active research.

\label{s:nsc}

%%%%%%%%%%%%%%%%%Saracino - Dale%%%%%%%%%%%%%%%%%%%%%%%%
\newpage
\part{Black holes in binaries through observations: X-ray, detached, lensing, LIGO \\ \Large{Tana D. Joseph and Sara Saracino}}
\label{part:observations}

{\it Please note: This is a rapidly evolving field, especially as regards observations, so we apologize if some references are missing, especially very recent ones. We have included all the studies that, to the best of our knowledge, have provided the most significant impact to the field.}

\section{Introduction}
\label{sec:Introduction}

Previous Chapters have extensively discussed the importance of stellar-mass black holes (BHs), their formation channels and their abundance in the Universe. The main goal of this Chapter is instead to provide the reader with clear tools to understand how to reveal these elusive objects and how their properties can be constrained. We start this journey by dividing stellar-mass BHs into three main categories: 1) isolated BHs, 2) BHs in binary systems with luminous companions (stars) and 3) BHs in binaries with compact objects, such as another BH or a neutron star (NS). Each of these categories must be treated separately, as the techniques and observational instruments we use to detect them are mostly different. For the sake of simplicity, in describing the different methodologies we will not consider triples or multiple systems.

\section{Isolated Black Holes}
\label{isolated}
Being the product of the evolution of very massive stars (M $>$ 20$M_{\odot}$), BHs are expected to constitute 0.1\% of all objects populating the Universe \cite{1983bhwd.book.....S,1992A&A...262...97V,1994ApJ...423..659B,1998ApJ...496..155S} and once produced, we are unaware of any mechanism that can destroy them. They can have very different properties depending on how they form and what kind of interactions they experience during their lives. A fraction of these objects are isolated BHs, meaning that they are not part of a binary system, hence they cannot be detected thanks to interactions with their companions or any motions around them (see the following Sections for a detailed description of BHs in binaries). There are three main reasons why a large fraction of isolated BHs exists: 1) Not all massive stars are born in binaries, a fraction of them are born and evolve as single stars \cite{2017ApJS..230...15M}. 2) If two stars are in a tight binary orbit, they can experience a common envelope phase that can bring the two stars to merge and form a massive star which would then explode as supernovae, leaving behind an isolated BH. 3) In a wide binary composed of two stars, one of which is quite massive, it could happen that the BH produced after the supernovae explosion receive enough “natal kick” to become unbound \cite{1961BAN....15..265B}.\\

If a BH lives isolated in the Universe, it has very little chance of interacting with any other object so that the effect of this interaction would allow its detection. In addition, the accretion rate of material from the surrounding is expected to be so low that it does not emit any detectable high-frequency radiation (for example X-rays). Thus, discovering an isolated BH is very challenging and there is only one specific phenomenon that appears to be promising in this regard: the {\it gravitational microlensing}.\\

\begin{figure}
    \centering
	\includegraphics[width=0.7\textwidth]{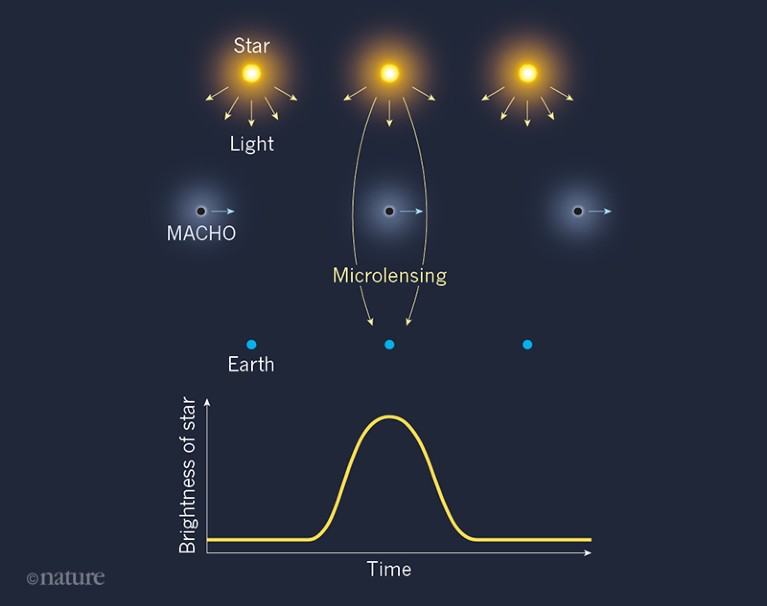}
    \caption{from Pietrzy{\'n}ski et al. \cite{2018Natur.562..349P}: it illustrates the gravitational microlensing produced by a BH. The deflection of the light of a background star when it passes in front of a BH produces two images of the same star, temporarily amplifying its brightness. These two images are too close to being spatially resolved but the difference in brightness between them produces a shift in the source's position that can be measured using astrometry and can provide strong constraints on the nature of the lens.}
    \label{fig:nasarelease}
\end{figure}

A microlensing event occurs when either a star or a compact object (called lens hereafter) passes almost exactly in front of a background star (called source hereafter). As predicted by general relativity \cite{1936Sci....84..506E}, the lens magnifies the image of the source, producing an apparent amplification of its brightness. Moreover, it slightly shifts the apparent position of the source \cite{1995AJ....110.1427M,1995A&A...294..287H,1995ApJ...453...37W}. Figure \ref{fig:nasarelease} visually illustrates the concept of gravitational microlensing, when the lens is a BH. When the BH passes nearly in front of a background star, it bends light from the star towards Earth, which temporarily amplifies the star’s observed brightness. BHs are able to warp space-time so much that it noticeably alters the source’s apparent location in the sky. The two images caused by lensing are very close to each other to be spatially resolved, but the change in brightness of the two images produces a shift in the position of the centroid of the source that can be detected via high-precision astrometry. By exploiting {\it photometric microlensing} to study how the luminosity of the source varies over time and {\it astrometric microlensing} to measure the amount of relativistic deflection of the apparent position of the source as the lens passes in front of it, the mass of the lens can be measured directly and its real nature inferred. Having both information, in fact, is very important as it allows us to break the degeneracy that exists between the mass of the lens and the relative source-lens proper motion. Unfortunately this is not the case for most detectable microlensing events, for which we only have the photometric information.\\

The mathematical formalism that physicists use to describe a microlensing event is rather complicated but a brief description of the main parameters involved can help better understand the phenomenon and their intrinsic limitations. The perfect alignment of the source, the lens and the observer, which tends to occur very rarely, produces a special type of microlensing event, where the reflected image is not observed as a single point, but as a ring of light around the lens. This ring is called the Einstein ring with an angular radius $\theta_{E}$ known as the angular Einstein radius. The latter is an important quantity used to describe the characteristic scale of a gravitational microlensing event, regardless of the exact observer/lens/source configuration, and it is defined by the equation:
\begin{equation} \label{e:eq1_lensing}
 \theta_{E} = \sqrt{\frac{4GM_{L}}{c^2}\frac{\pi_{LS}}{1 au}} 
\end{equation}
where $M_{L}$ is the mass of the lens and $\pi_{LS}$ is the relative lens-source parallax. The latter can also be written as:
\begin{equation} \label{e:eq3_lensing}
 \pi_{LS} = \pi_{L} - \pi_{S} = 1au \left(\frac{1}{D_{L}} -\frac{1}{D_{S}}\right)
\end{equation}
with $D_{L}$ and $D_{S}$ as distances between observer and lens and between observer and source, respectively. 
The Einstein radius $\theta_{E}$ cannot be obtained directly from the study of the magnification light curve of the microlensing event\footnote{In the rare case of perfect alignment observer/lens/source it is possible to infer the angular Einstein radius $\theta_{E}$ by using photometric information alone.}, hence other quantities need to be defined. The only quantity that contains some information about the physical properties of the event, especially about the lens itself, and can be directly inferred from the light curve is instead the timescale $t_{E}$ of the event, i.e. the time it takes the source to traverse an angular distance of $\theta_{E}$ and can be written as follows:
\begin{equation}
 t_{E} = \frac{\theta_{E}}{\mu_{LS}}
\end{equation}
where $\mu_{LS}$ is the relative proper motion between lens and source. In a microlensing event, $t_{E}$ can vary from a few days up to several hundreds days. In long-duration microlensing events, the annual parallax tends to lead to prominent departures in the photometric signature \cite{1992ApJ...396..104G,1995ApJ...454L.125A}, so that we can also infer the microlensing parallax parameter $\pi_{E}$:
\begin{equation} \label{e:eq2_lensing}
 \pi_{E} = \frac{\pi_{LS}}{\theta_{E}}
\end{equation}
From Eq. \ref{e:eq1_lensing} we know that the more massive the lens, the longer the duration of the event itself. In this context, microlensing events involving massive and compact objects such as BHs as lenses are the most likely to have sufficient duration to detect the microlensing parallax effect using photometry.\\

To obtain the mass of the lens, however, the photometric information is unfortunately not sufficient, even when the parallax $\pi_{E}$ is known, as $M_{L}$ and $\pi_{LS}$ are degenerate. The only way to disentangle these two contributions and get an accurate estimate of the mass of the lens is by monitoring and measuring the small astrometric shift of the position of the source induced by the presence of the lens. This can be done using high-precision astrometry from space, e.g. with HST (see an example later in the text). Indeed, the astrometric shift ${\bf \delta(u)}$ is a two-dimensional position angle which is proportional to the angular Einstein radius $\theta_{E}$ according to the expression:
\begin{equation}
    \delta({\bf u}) = \frac{{\bf u}}{u^2 +2}\theta_{E}
\end{equation}
where the vector ${\bf u}$ defines the angular separation between lens and source and it changes with time. By substituting Eq. \ref{e:eq1_lensing} in Eq. \ref{e:eq2_lensing}, we can derive an expression for the lens mass:
\begin{equation}
    M_{L} = \frac{\theta_{E}}{k\pi_{E}}
\end{equation}
where $\theta_{E}$ comes from precise, multi-epoch astrometry and $\pi_{E}$ comes from photometry.
If the distance of the source is known, for example because the microlensing event is detected in a specific region of the Galaxy (e.g. the Galactic bulge, whose distance is known), then we can use this information to constrain the distance of the lens as well, from Eq. \ref{e:eq3_lensing}. If this is not the case, we can still overcome the limitation if some spectroscopic data of the source is available. \\

Searching for isolated stellar-mass BHs via gravitational microlensing is a promising method as massive objects such as BHs are capable of producing greater space-time distortions than any other medium-mass star, so they can be distinguished. In addition, detailed Milky Way microlensing simulation codes have recently demonstrated that when plotting different sources (stars, white dwarfs, NS or BHs) in the $\pi_{E}$ - $t_{E}$ plane, BHs tend to segregate in a specific region of this parameter space, corresponding to long Einstein crossing times and relatively low microlensing parallaxes (see Fig. 13 in \cite{2020ApJ...889...31L}). This means that selecting microlensing events with such properties is an ideal way to maximize the chances of detecting events involving BH lenses.\\   

Over the past twenty years there has been a growing interest in microlensing events and several microlensing survey programs have been designed for this purpose, some of which are currently still in operation. The Optical Gravitational Lensing Experiment (OGLE-IV; \cite{1992AcA....42..253U,2015AcA....65....1U}), the Expérience pour la Recherche d'Objets Sombres (EROS; \cite{1993Natur.365..623A}), the Microlensing Observations in Astrophysics collaboration (MOA-II; \cite{2001MNRAS.327..868B,2008ExA....22...51S}), and the Korea Microlensing Telescope Network (KMTNet; \cite{2016JKAS...49...37K}) are the best examples. These surveys carry out photometric monitoring of several fields, both in the Milky Way (mostly the Galactic bulge) and in Galactic satellites (for example the Magellanic Clouds), allowing to construct light curves for an unlimited amount of objects. Thanks to their precious work over many years, these surveys have provided the scientific community not only with more than 30,000 microlensing events, but also with a detailed and long list of classified variables, from pulsators (e.g., RR-Lyrae or Cepheids) to eclipsing or ellipsoidal variables.\\

Unfortunately, it is more difficult to have accurate astrometric measurements and it is only recently that people have started to have access to this type of data, thanks to important international collaborations (e.g., HSTPROMO \cite{2014ApJ...797..115B} which exploits high-resolution HST observations of globular clusters across multiple epochs) or space missions (e.g. GAIA \cite{2016A&A...595A...1G}, which instead scans the sky multiple times to measure stars' positions and proper motions\footnote{We note here that the astrometric power of GAIA is very limited in crowded fields like in the direction of the Galactic bulge.}), opening to the concrete possibility of carrying out a search for isolated BHs through astrometric microlensing in the Universe.\\

To date, only a few isolated BHs have claimed to be detected via this technique, and interestingly, all of them have been found in the direction of the Galactic bulge, where there is a higher rate of microlensing events due to the high density of stars. It is interesting to note that, historically, all the searches for isolated BHs have focused primarily on microlensing events with $\pi_{E}$ $\sim$ 0.1 or higher, essentially excluding all those events we now know are most likely produced by BHs. This could be one of the reasons for the lack of isolated BH detections, despite the enormous observational effort of the last few decades.\\

MACHO-96-BLG-5 and MACHO-98-BLG-6 are the first two microlensing events that scientists have proposed to be linked to the presence of isolated BHs, with a mass of the order of 6 $M_{\odot}$ \cite{2006ApJ...651.1092N}. These events were first detected and observed using long-term photometric monitoring \cite{2002ApJ...579..639B} as part of the MACHO project, which used microlensing events to specifically search for MAssive Compact Halo Objects \cite{1993Natur.365..621A}. However, since these estimates were derived by fitting only the light curves (an approach that may lead to degeneracies), they were strongly influenced by the imposed Galactic model prior. Abdurrahman et al. \cite{2021ApJ...912..146A} recently re-analyzed the data of these two sources, complementing the light curves with new high-resolution near infrared data. Their work strengthens the case for an isolated BH to be the lens in both microlensing events, for relative lens-source proper motions above 0.81 mas/yr for MACHO-96-BLG-5, and 2.48 mas/yr for MACHO-98-BLG-6. A future astrometric/spectroscopic examination of the sources will be crucial: 1) to obtain an independent measurement of the main parameters of the sources (i.e. lens-source parallax, lens mass); and 2) to search for every possible contribution from a second object such as a non-BH compact object or a brown dwarf lens. This will provide a definitive answer on the nature of these objects.\\

\begin{figure}
    \centering
	\includegraphics[width=0.9\textwidth]{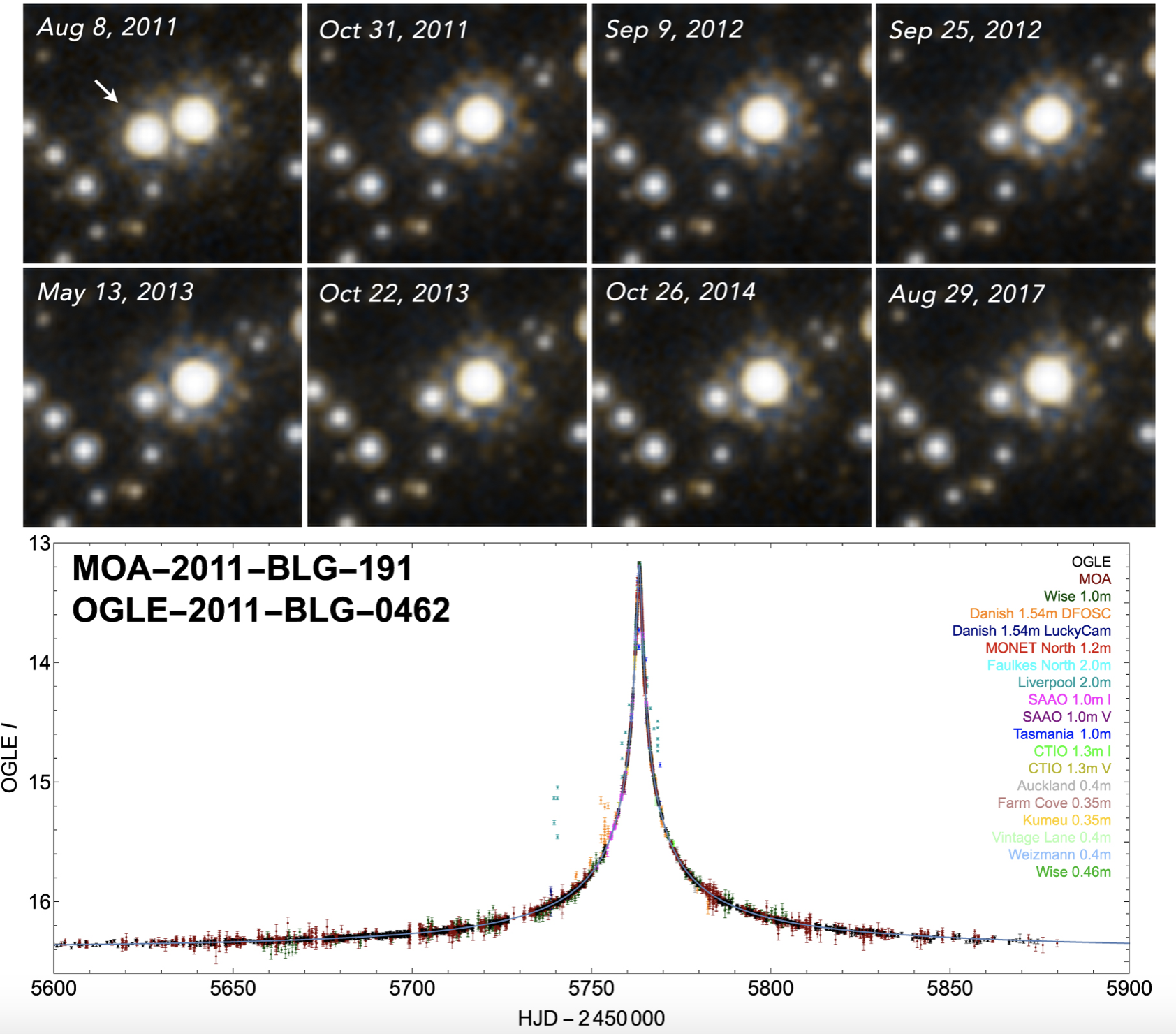}
    \caption{combined from Sahu et al. \cite{2022ApJ...933...83S}: {\it Top panel:} The source of the microlensing event MOA-11-191/OGLE-11-462 as observed in eight epochs of HST observations (from 2011 to 2017). The source is highlighted with a white arrow in the leftmost panel. All stars in the field have similar luminosity over the epochs while the magnification of MOA-11-191/OGLE-11-462 in Epochs 1 and 2 (produced by the lens passage) is evident compared to the remaining epochs. Epochs 2 and 3 are instead those where a maximum astrometric shift is measured. {\it Bottom panel:} The photometric light curve of the microlensing event MOA-11-191/OGLE-11-462 built by combining ground-based observational data collected from different facilities is shown along with the best-fitting model light curve. The instruments are labeled directly in the figure with different colors. The shape of the magnification curve is characteristic of a microlensing event.}
    \label{fig:microlensing}
\end{figure}

The first and only unambiguous detection of an isolated stellar-mass BH in the Galactic bulge, but more generally thus far in the Universe, has been recently reported by two different groups, Lam et al. \cite{2022ApJ...933L..23L} and Sahu et al. \cite{2022ApJ...933...83S}, almost at the same time. The microlensing event has been detected by MOA and OGLE, has a long duration ($t_{E}$ $\approx$ 270 d) and is labeled as MOA-2011-BLG-191 or OGLE-2011-BLG-0462. The magnification of the source caused by the passage of the lens is shown in Figure \ref{fig:microlensing} in two different ways: as a sample of HST observations acquired from 2011 to 2017 where both the apparent change in luminosity and the astrometric deflection can be visualized ({\it top panel}) and as the photometric light curve of the source drawn over time by multiple facilities ({\it bottom panel}). The two teams have used almost the same set of multi-epoch observations from Hubble to obtain precise astrometry of this magnified source and derived both the mass of the lens and its distance. They both agree that the lens must be a compact source rather than a luminous object, but they disagree on its current mass, thus on the interpretation of its nature. Lam and colleagues \cite{2022ApJ...933L..23L} estimated the source to be relatively nearby, at a distance of 0.70-1.92 kpc and derived a mass of 1.6-4.4 $M_{\odot}$, consistent with either a NS or a BH, while Sahu and collaborators \cite{2022ApJ...933...83S} measured a distance of 1.58 $\pm$ 0.18 kpc, implying an inferred lens mass of 7.1 $\pm$ 1.3 $M_{\odot}$, only compatible with it being an isolated BH.\\

Both groups noted an inconsistency between the best fit solution preferred by the photometric data and the one from the astrometric data but they adopted slightly different approaches to try and investigate this aspect further. Lam et al. \cite{2022ApJ...933L..23L} in particular used two different weights to analyze the results of the microlensing models: a default weight (DW) which gave the same weight to each measurement and uncertainty, and an equal weight (EW) where the two data sets were weighted in the same way. Of the two, the DW solution, that better agrees with the photometric data, is the one closer to the solution found in the study by Sahu et al. \cite{2022ApJ...933...83S}, although some differences still remain. The EW solution, instead, is more in agreement with the astrometric data but predicts a slightly lower distance for the lens, as well as a lower mass, falling in the NS mass regime.\\

Shortly after this detection, Mereghetti et al. \cite{2022ApJ...934...62M} have started an observational campaign with the Chandra, XMM-Newton and INTEGRAL satellites to search for any X-ray emission coming from this object, with the aim of independently binding its nature. They found that the inferred X-ray luminosity of this object is consistent with the small radiative efficiency expected for a BH more than for a NS. Upon reanalysis of the available photometric and astrometric data, {Mr{\'o}z} et al. \cite{2022ApJ...937L..24M} were able to say the last word on this story: according to them, the lens is indeed an isolated BH (with mass M = 7.88 $\pm$ 0.82 $M_{\odot}$ at a distance of $D_{L}$= 1.49 $\pm$ 0.12 kpc). They investigated the presence of systematic errors both in the photometric and/or astrometric data but they also evaluated the impact of the presence of a bright star located $\approx$ 0.4" away from the microlensing source (see Fig. \ref{fig:microlensing}) in the estimate of the astrometric shift of the source. They came to the conclusion that, although in general the presence of a blending source might produce erroneous results and need to be properly taken into account, in this case systematic errors in the astrometric solution were the primary cause of the low mass measured by the Lam et al. team.\\

The blind search for isolated BHs is still in an embryonic state, so any new detections coming in the future will represent an important step forward in the field. The third data release (DR3) of the GAIA satellite (more details about the mission can be found in Section \ref{astrometry}), for example, already contains 363 new microlensing events \cite{2022arXiv220606121W} that no other surveys have detected before and the prospect of combining both the light curves and the astrometric information that GAIA will continue to provide until the end of the mission looks extremely promising. The development of new dedicated tools for searching for transient astrometric lensing events caused by dark compact objects in GAIA DR4 will also greatly help in this regard \cite{2023arXiv230100822C}.\\

The detection of isolated BHs through gravitational microlensing will receive a decisive boost in the future thanks to the advent of a new and advanced facility, the Nancy Grace Roman Space Telescope and its Galactic Bulge Time Domain Survey (GBTDS). With its wide field of view (0.28 square degrees), super resolution (0.11 arcsec/pixel) and ability to simultaneously obtain precise photometry and astrometry, Roman is expected to detect and allow the characterization of several thousand microlensing events \cite{2015arXiv150303757S,2019ApJS..241....3P}. It will observe the Galactic bulge and other fields multiple times over the years, allowing us to collect a rich sample of photometric and astrometric measurements of known and unknown microlensing events. For such a reason, Roman will enable the GBTDS to expand the catalog of isolated BHs by order(s) of magnitude, yielding exquisite determinations of their masses, distances, and proper motions, as well as opening up the opportunity to finally perform BH population studies. \\

A recent work by Saradjan \& Sahu \cite{2023AJ....165...96S} have simulated a numerous sample of microlensing events as seen by Roman, in order to effectively quantify the role that this instrument will have in the future research for isolated BHs. Since there is not yet an accurate mass function for isolated BHs based on observations, they performed the simulations by considering several mass functions (e.g. $dN/dM \propto M^{-0.5}, \propto M^{-1}~ \text{and} \propto M^{-2}$). The conclusion of the study is that, considering the entire lifetime of the mission, the Roman telescope will be able to detect 3-4, 15-17 and 22-24 isolated BHs through astrometric microlensing (depending of the mass function adopted) and will estimate their physical properties with reasonable accuracy. In particular, they estimate that the relative errors on physical parameters such as mass, lens-observer distance and proper motion will be respectively less than 1, 5, 10\%. Advances in this field will provide important and independent insights into our understanding of the formation and evolution of BHs.

\section{Black Holes in binary systems with luminous companions}
The second family of stellar-mass BHs consists of all those objects that orbit a star in a binary system. The physics governing binary systems is rather complicated and still not fully understood but for the purpose of understanding this Section, it is important to mention some basic concepts. In a binary system, each of the two components is surrounded by a teardrop-shaped region, called {\it Roche Lobe}, where self-gravity dominates over the gravitational pull of the companion. If the star is inside of this region, the material is considered bound. Usually approximated as a sphere of the same volume, the Roche Lobe have size which depends on the mass difference between the two stars and on the orbital separation of the system, therefore on how far apart the two components are from each other. According to \cite{1983ApJ...268..368E}, the size of a Roche Lobe can be approximated to:
\begin{equation}
    \frac{r_1}{A} = \frac{0.49q^{2/3}}{0.6q^{2/3}+ ln(1+q^{1/3})}
\end{equation}
where $r_1$ is the radius of the Roche Lobe of the primary component (the more massive by convention), A is the angular separation of the system, while $q$ is the mass ratio defined here as $M_1/M_2$. To obtain the Roche Lobe size for the secondary component, it is sufficient to replace label 1 with label 2 in the previous formula.\\ 

During the evolution of a binary system, both the radius of the stars and the size of the Roche Lobes can vary, for several reasons. The first happens as the stars evolve to different stages of their evolution (for example expanding when evolving from the main sequence to the red giant branch), while the latter occurs when the orbit of the binary shrinks or widens due to dynamical interactions and close passages of third stars (especially in crowded fields) or the mutual influence of the two stars. As soon as the size of the most massive object (the one that evolves faster as predicted by stellar evolution theory) exceeds that of its Roche Lobe, the material ``located outside" the Roche Lobe becomes unbound, triggering three possible scenarios. This material can indeed: 1) escape the system completely, reducing the total mass of the system; 2) orbit both stars; 3) fall onto the companion's Roche Lobe, with consequent mass transfer and accretion. Each of these events significantly modifies the configuration of the binary. For those interested in the details, a comprehensive review on binary systems can be found in Chapter 1.\\

Depending on the status of the system, it can be classified as either non-interacting or interacting binary, with pretty significant differences in how they can be observed. The simplest and most secure way to detect a BH (or a NS) in a binary system with a star is if the two objects are actually interacting, as the accretion of material from the star onto the compact object can trigger the emission of energy at high-frequency (for example X-rays or gamma-rays) but also at low-frequency (in radio) which can be detected through tailored observations in these bands.
Unfortunately this is not the case if the BH is in a non-interacting system, also called a detached configuration. As the word itself explains, this means that the companion has not yet filled its Roche Lobe and has not yet begun to transfer material to the BH, therefore they are in a quiescent state and their presence can only be revealed using alternative techniques. These two configurations will be discussed in more detail in two dedicated sections later.

\subsection{Black Holes in interacting systems}

In this Section we will discuss interacting black hole binary systems, their observational history and their connection to gravitational wave physics. \\

%\begin{comment}
X-ray binaries (XRBs) consist of a compact object (black hole or neutron star) accreting matter from a gravitationally bound companion (or donor star). These sources are multiwavelength objects and are most luminous in the X-ray regime, with $\rm{L_X}>10^{36}$\,erg/s. The relativistic accretion of material from the companion star to the compact object is what powers this enormous energy output. XRBs are the only class of binary system that can be electromagnetically detected outside the Local Group, a small ($\sim$1 Mpc) group of 50 galaxies that includes the Milky Way \cite[see e.g.][]{2000glg..book.....V,2007ApJ...656L..13I}, making these sources very useful in the study of binary stellar systems. \\

X-ray binaries were some of the first X-ray sources to be detected and have been actively studied for several decades \cite[see e.g.][]{2004astro.ph.10536P}. There are over 400 known XRBs in the Galaxy and the Magellanic Clouds \cite{2006A&A...455.1165L,2007A&A...469..807L}. \\

The study of X-ray binaries allow us to probe physics in extreme conditions; from regions of high magnetic field strength to relativistic accretion and strong gravity \cite[see e.g.][]{2004astro.ph.10536P}. They also help us to gain insight into other astrophysical systems. \\

\begin{figure}
    \centering
    \includegraphics[width=0.7\textwidth]{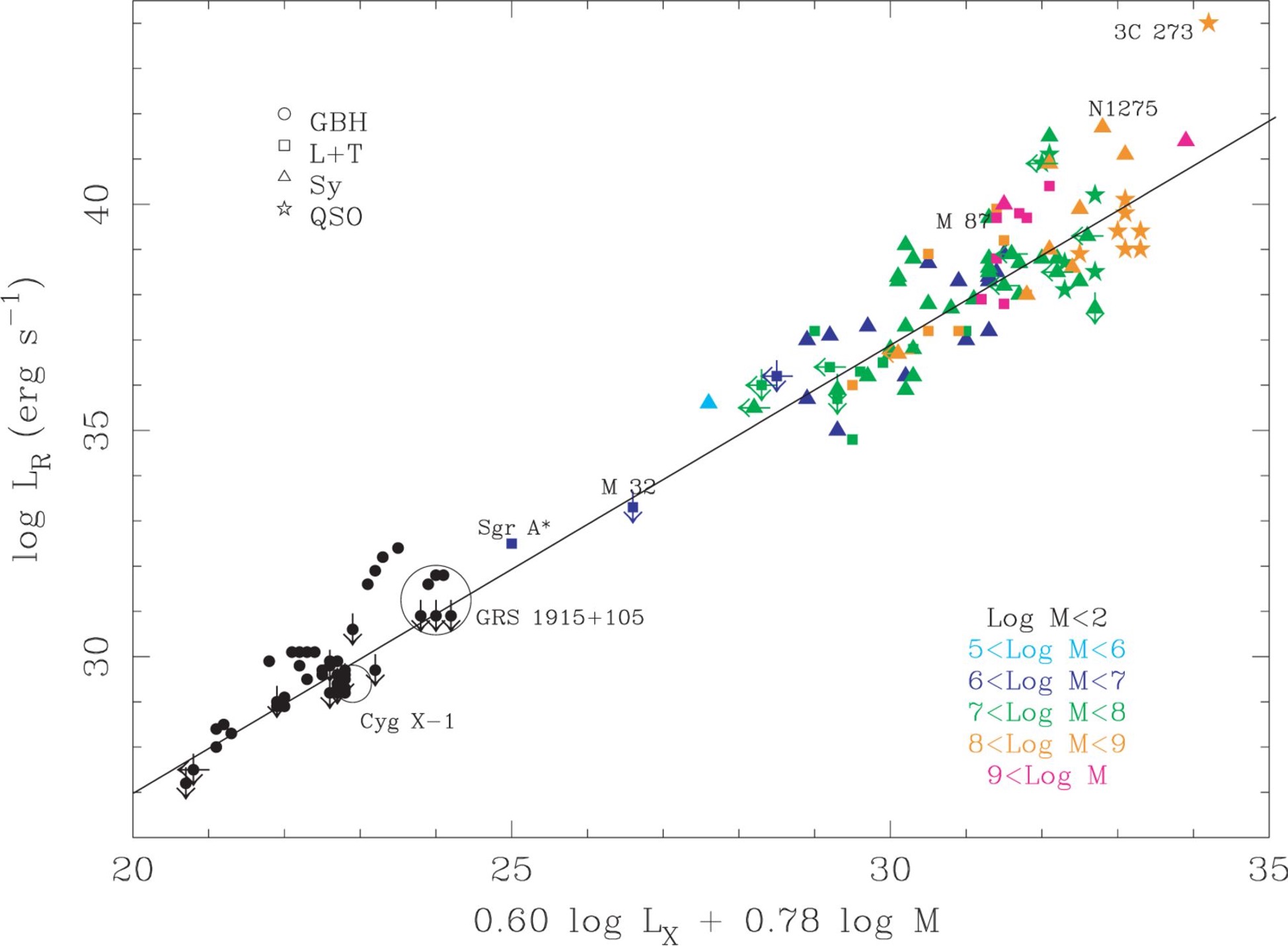}
    \caption{Taken from \cite{2003MNRAS.345.1057M}. This is an edge-on view of the Fundamental Plane of black hole activity, across a black hole mass range spanning 9 orders of magnitude.}
    \label{FP}
\end{figure}

For instance, black hole X-ray binaries (BHBs) can allow us to explore the physics of black holes on much shorter time scales than exhibited by supermassive black holes. A study of BHBs with low accretion rates uncovered a power law relation between radio ($L_R$) and X-ray luminosity ($L_X$), with $L_R \sim L_X^{0.7}$ \cite{1998A&A...337..460H,2003MNRAS.344...60G}. Extending this work to include supermassive black holes in active galactic nuclei, \cite{2003MNRAS.345.1057M,2004A&A...414..895F} revealed a relation between X-ray luminosity ($L_X$), radio luminosity ($L_R$) and black hole mass $M$, often called the Fundamental Plane of black hole activity (see Fig.\,\ref{FP}), with 

\begin{equation}
\rm{log}\,\rm{L_R} \sim 0.6\,log\,\rm{L_X} + log\,M    
\end{equation}

This relation suggests that for black holes with low accretion rates, the physical processes governing how accreted material is converted into into radiative energy could be universal over the entire black hole mass range, from stellar mass black holes to supermassive black holes \cite[see ][and refernces therein]{Gultekin_2019}.\\

XRBs are predicted to be possible progenitors of some of the type of gravitational wave sources that are already being detected \cite[][and references therein]{2017A&A...604A..55M, 2017JApA...38...45V,2018A&A...619A..77K}. They are also predicted to be sources of detectable gravitational waves themselves \cite[see e.g.][and references therein]{2023arXiv230106243Q}. In light of this, there has been a renewed interest in XRBs from many branches of astronomy.\\

In this work, we will be exclusively focusing on XRBs with a black hole accretor (black hole X-ray binaries; BHBs). In the following sections, we will go over the past, present and future of X-ray detectors and observatories. We will discuss the types of X-ray binaries and  their links to gravitational wave physics, and give an overview of the observational signatures of black hole binaries. \\

\subsubsection{Development of electromagnetic detection methods for X-ray binaries}

The Earth's atmosphere is opaque to X-rays from space. Thus, in order to detect this high energy emission from astronomical objects, X-ray telescopes must be placed above the atmosphere using balloons, rockets or satellites. The first 
X-ray detection outside of the Earth's atmosphere was from the our Sun's corona, in 1949, using Geiger counters launched on a sounding rocket \cite{burnight1949soft}.
To date, over 60 X-ray instruments and telescopes have been used by astronomers\footnote{A list of all satellite and rocketborne X-ray missions can be found at \url{https://heasarc.gsfc.nasa.gov/docs/heasarc/missions/}}. \\

\begin{figure}
    \centering
    \includegraphics[width=0.7\textwidth]{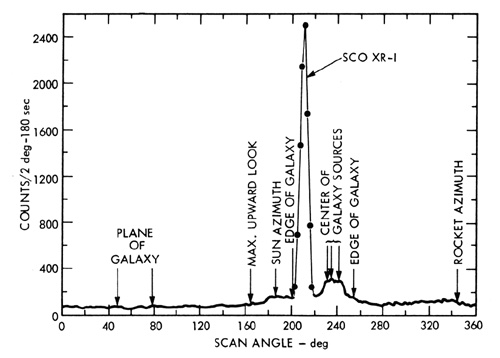}
    \caption{Taken from \cite{2010exru.book.....S}. This data shows the X-ray photon count rate from a three minute long  observation taken with a rocket-borne detector in 1967 in the direction of Scorpius X-1 and including the Galactic centre. We see clearly that Scorpius X-1 is a very bright X-ray  source.}
    \label{ScoX1}
\end{figure}

The first X-ray binary, and indeed the first extrasolar X-ray source, to be discovered was Scorpius X-1 \cite{1962PhRvL...9..439G}. In a mission lasting 350\,s, an Aerobee rocket equipped with proportional counter detectors observed what is still the brightest apparent source of X-rays in the sky, after the Sun. Figure\,\ref{ScoX1} shows some early observational X-ray data of Scorpius X-1 taken in 1967 with detectors on a sounding rocket. After considering the X-ray and optical data for Scorpius X-1 it was concluded that it must be a neutron star  accreting from a binary companion \cite{1967ApJ...148L...1S}. \\

In 1964, another rocket equipped with Geiger counters observed a strong Galactic X-ray source in the Cygnus constellation \cite{1965Sci...147..394B}. This source, called Cygnus X-1, would later be found to be a XRB containing the first compact object widely accepted to be black hole. \\

\begin{figure}
    \centering
    \includegraphics[width=0.7\textwidth]{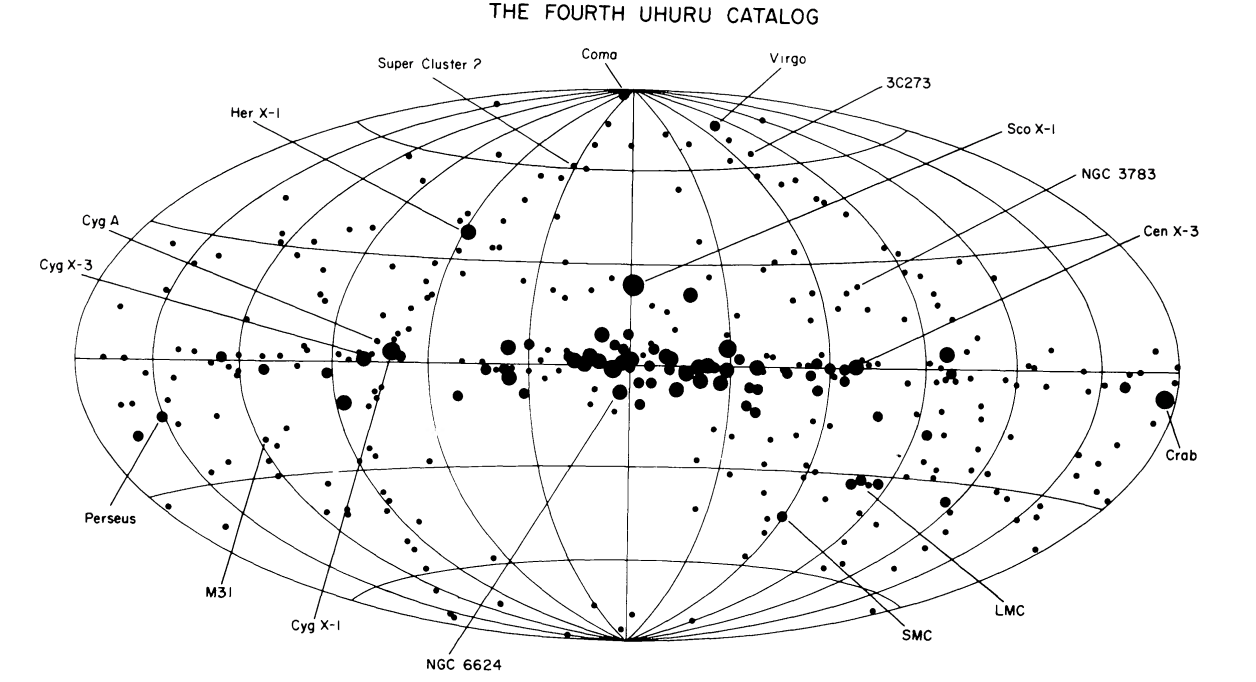}
    \caption{Taken from \cite{1978ApJS...38..357F}. The X-ray sources from the fourth and final \Uhuru source catalogue displayed in Galactic coordinates. The size of the symbols is proportional to the logarithm of the peak source intensity.}
    \label{4U}
\end{figure}

The first dedicated X-ray satellite, \emph{Uhuru} was launched in 1970 \cite{1971ApJ...165L..27G}. The mission lasted for just over two years and a total of 339 X-ray sources were detected \cite{1978ApJS...38..357F}, including galaxy clusters, supernova remnants and binary star systems (see Fig.\,\ref{4U}). \emph{Uhuru} detections allowed us to understand that most the X-ray stars that had been observed were in fact black holes or neutron stars accreting from a binary companion. Indeed it was observations from\Uhuru that helped to determine that Cygnus X-1 contained a black hole \cite{1971ApJ...166L...1O,1971ApJ...170L..21S}. \\

The systematic study of XRBs became possible with the launch of the \emph{Einstein} X-ray Observatory \cite[\emph{Einstein};][]{1979ApJ...230..540G}. A space-based instrument launched in November 1978, it was the first fully imaging X-ray satellite telescope. \emph{Einstein} boasted an angular resolution of tens of arcseconds (compared to the few arcminute resolution of the rocket borne telescopes at the time \cite[see e.g.][]{1975ITNS...22..616G}), a field of view of a few arcminutes and a sensitivity nearly 1000 times better than previous instruments and operated in the 0.3–4 keV range. \\

Prior to \Einstein's launch, only six extragalactic X-ray binaries had been detected (excluding the $\sim$60 XRBs in the Milky Way); five in the Large Magellanic Cloud and one in the Small Magellanic Cloud \cite[][and references therein]{1984PASP...96..913H}. With \Einstein, over 100 XRBs or XRB candidates were found in 10 Local Group galaxies, making it possible to compare XRB populations in different environments. For instance, it was found that high mass X-ray binaries (see section \ref{HMXBs}) were found preferentially in later type galaxies whereas low mass X-ray binaries (see section \ref{LMXBs}) were more likely to be found in earlier types. This finding was not thought to be only due to the increasing number of massive stars in later type galaxies. \Einstein also revealed that a significant number of XRBs were associated with globular clusters and that XRB host clusters were brighter than average \cite[ e.g.][see also sections \ref{HMXBs} and \ref{LMXBs}]{1975ApJ...199L.143C}.\\

\begin{figure}
    \centering
    \includegraphics[width=0.8\textwidth]{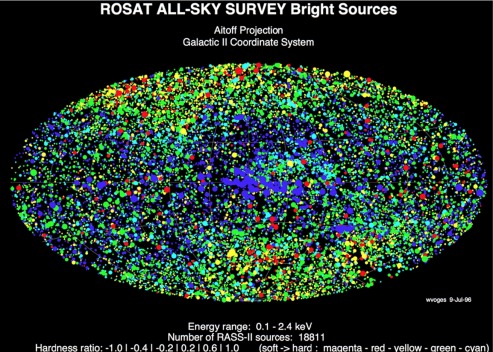}
    \caption{Taken from \url{https://www.mpe.mpg.de/971847/RH_all_sky_survey}. This image shows the 18\,811 bright point sources detected by \ROSAT in the all-sky survey. A total of 150\,000 sources were detected by \ROSAT in this survey, 25 times more than in all previous X-ray satellites combined.}
    \label{4U2}
\end{figure}

The \emph{Einstein Observatory} (also called HEAO-2)) ushered in a new and exciting age in X-ray astronomy. The three large X-ray satellite telescopes that followed, the \emph{R\"{o}ntgen} Satellite \cite[\ROSAT; launched in June 1990][]{1982AdSpR...2d.241T} , the \emph{Advanced Satellite for Cosmology and Astrophysics} \cite[\ASCA; launched in Febraury 1993][]{1994PASJ...46L..37T} and the \emph{Rossi X-ray Timing Explorer} \cite[\RXTE; launched in December 1995][]{1996SPIE.2808...59J} had instruments that provided the next big improvements on Einstein. Each of these three telescopes were optimized for one aspect of X-ray observations that complimented the other two. \ASCA, with it's focus on X-ray spectroscopy, had a large energy range and good energy resolution. It was also the first X-ray telescope to make use of large CCD detectors. \ROSAT was designed for X-ray imaging, therefore it had a large field of view (2\degree) and very good spatial resolution. \RXTE was a timing instrument and offered microsecond time resolution. Figure \ref{4U2} shows a selection of sources detected by \ROSAT during the all-sky survey (see Table\,\ref{telescope_compare} for a comparison of various telescope characteristics.)\\

\begin{figure}
    \centering
    \includegraphics[width=0.7\textwidth]{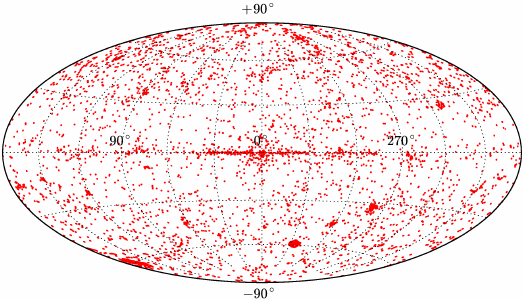}
    \caption{Taken from \cite{2016A&A...590A...1R}. This image shows the nearly 400\,000 X-ray sources detected by \XMM using public observations up to 2013 December 31.}
    \label{4U3}
\end{figure}

The next significant advancement in X-ray astronomy came with the launching of the \emph{Chandra} X-ray Observatory and the \emph{XMM-Newton} Observatory.
The \emph{Chandra} X-ray Observatory was launched in 1999 \cite{2000SPIE.4012....2W}. This telescope has subarcsecond spatial resolution and an energy range of 0.1–10 keV. The \emph{XMM-Newton} Observatory, launched later that same year, operates in a similar energy range (0.1–15 keV). \XMM has a much coarser angular resolution ($\sim$\,$15"$) than \Chandra, but has an effective area many times larger (4500\,cm$^2$ at 1 keV) \cite{2012OptEn..51a1009L}. Figure \ref{4U3} shows over 40,000 \XMM sources detected until the end of 2013. Using these telescopes, it is now possible to carry detailed studies of X-ray binaries, and other X-ray sources, in galaxies well outside the Local Group \cite[see e.g. ][]{2016MNRAS.460.4513P,2022A&A...664A..41R}. \\

With their complimentary specifications and their greatly improved capabilities compared to predecessors, \Chandra and \XMM have been the main work horse instruments for the study of X-ray binaries for over two decades. Table\,\ref{telescope_compare} shows the comparison between the technical characteristics of \Chandra, \XMM and previous X-ray satellites. \\

\begin{table}[!htb]
    \caption{A comparison of X-ray telescope characteristics. This table has been adapted from Table\,28 of the \emph{XMM-Newton} Users Handbook (\url{https://xmm-tools.cosmos.esa.int/external/xmm\_user\_support/documentation/uhb/xmmcomp.html}). Data are taken from this table unless otherwise indicated. See also \url{https://heasarc.gsfc.nasa.gov/docs/heasarc/missions/comparison.html} for comparisons of \Chandra, \XMM and other X-ray telescopes.}
    %\centering
    \resizebox{\textwidth}{!}{
    \begin{tabular}{|l|c|c|c|c|c|c|c|}
    \hline
        Telescope & Energy & Effective & Angular & Energy & Time & Orbital \\
                  & range & area & resolution & resolution & resolution & target \\
         &  (keV) & at 1\,keV (cm$^2$) & (arcsec) & at 1\,keV (eV)  & (s) & visibility (hour) \\
    \hline
        \Chandra & 0.1$-$10 & 555 (ACIS-S)\footnote{For more information on \Chandra 's ACIS instrument see \url{https://cxc.harvard.edu/proposer/POG/html/chap6.html\#tth_chAp6}.} & 0.5 & 1 (HETG)\footnote{For  more information on \Chandra 's ACIS instrument see \url{https://cxc.harvard.edu/proposer/POG/html/chap8.html}.}& 3.2\footnote{taken from \url{https://cxc.harvard.edu/cal/Acis/index.html}.} (ACIS) & 44.4 \\
        \XMM & $0.15-12$ & 4650 & 15 & 4 (RGS)\footnote{Fore more infomation on \XMM's RGS instrument see \url{https://www.cosmos.esa.int/web/xmm-newton/technical-details-rgs}.} & 0.07\footnote{Taken from \url{https://www.cosmos.esa.int/web/xmm-newton/technical-details-epic}.} (pn)\footnote{For more information on \XMM's pn instrument see \url{https://www.cosmos.esa.int/web/xmm-newton/technical-details-epic\#5.1.2}.} & 36.7\\
        \ASCA & $0.5-10$ & 350 & 174 & 100 &  $6\times 10^{-5}$ (GIS)\footnote{Taken from \cite{10.1093/pasj/48.2.171} and see same source for more information of \ASCA's GIS instrument.} & 0.9 \\
        \ROSAT & $0.1-2.4$ & 400 & 3.5 & 500 & $6\times 10^{-5}$ (HRI)\footnote{Taken from \url{https://heasarc.gsfc.nasa.gov/docs/rosat/hri.html} and see same link for more information on \ROSAT's HRI instrument.} & 1.3 \\
        \RXTE & $2-250 $ & \emph{n.a.} & \emph{n.a.} & \emph{n.a.} & $10^{-6}$ & 1 \\
        \Einstein\footnote{Data taken from Table\,1 in\cite{1979ApJ...230..540G}, see same source for more information on \Einstein instruments} & $0.3-4$ & 300 & 2 & $10-100$ (FPCS) & 8$\times10^{-6}$ (HRI) & $0.3-0.6$\footnote{Taken from \url{https://heasarc.gsfc.nasa.gov/docs/einstein/HEAO-B_Guest_Observers_Guide.pdf}} \\
    \hline    
    \end{tabular}}
    \label{telescope_compare}
\end{table}

There are currently about a dozen X-ray satellite observatories operating, including \Chandra and \XMM\footnote{see \url{https://heasarc.gsfc.nasa.gov/docs/heasarc/missions/active.html} for a list of active X-ray observatories}. This is a  global effort with China, Europe, India, Japan, Russia and the US all involved in launching and running missions, many of them jointly \cite{2005SSRv..120..165B,2009A&A...502..995T,2009PASJ...61..999M,2011ICRC....6..351T,2012SPIE.8443E..13G,2013ApJ...770..103H,2014SPIE.9144E..1SS,WEISSKOPF20161179,10.1117/12.2312053,Zhang_2020,2021A&A...647A...1P}. A list of planned upcoming and future X-ray missions can be found at \url{https://heasarc.gsfc.nasa.gov/docs/heasarc/missions/upcoming.html}. \\

\subsubsection{Types of X-ray binaries}

\begin{figure}
    \centering
    \includegraphics[width=0.7\textwidth]{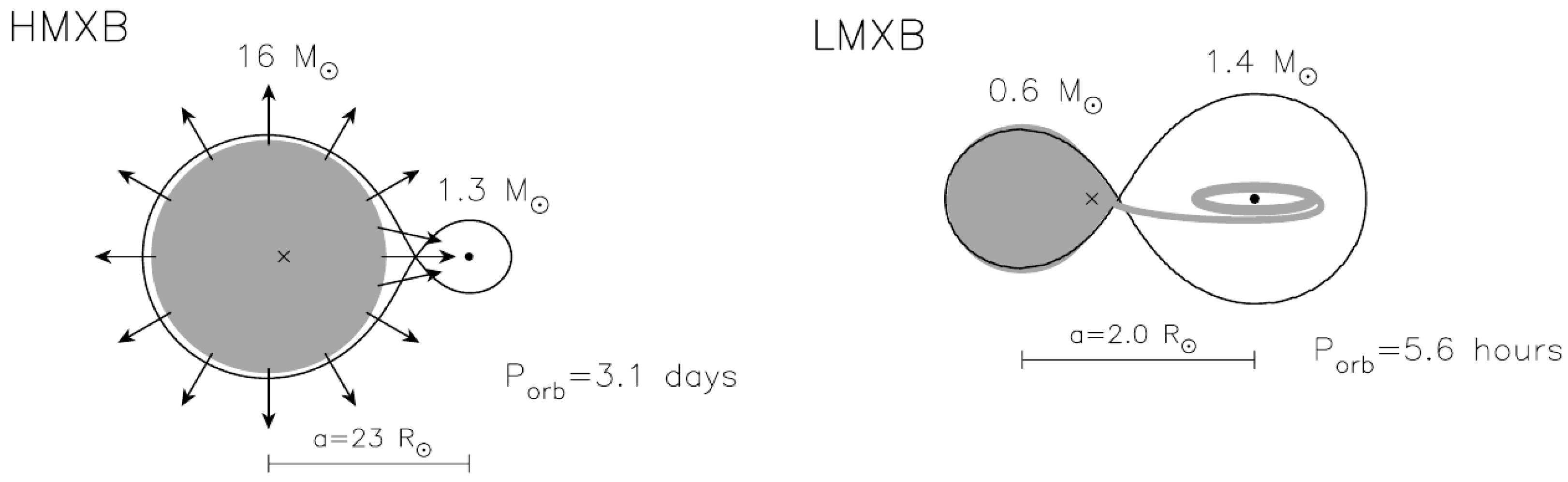}
    \caption{Taken from \cite{2006csxs.book..623T}. The typical configurations for a high mass (left) and low mass (right) X-ray binary (this sketch depicts neutron star binaries, but black hole binaries have the same composition). The Roche lobe overflow accretion process is depicted here for the LMXB and windfed accretion is shown for the HMXB, although Roche lobe overflow can take place in HMXBs as well.  Note the difference in size and mass for two types of binary.}
    \label{XRBs}
\end{figure}

X-ray binaries are generally classified by the mass of the companion star, as the nature of this companion determines the time scales and accretion mode of the binary. Thus we have high mass X-ray binaries, low mass X-ray binaries (see Fig.\,\ref{XRBs}), intermediate mass X-ray binaries, depending on the mass of the companion. XRBs can also be classified by their X-ray emission profiles, such as ultraluminous X-ray sources and super soft X-ray sources (the latter are sources with a white dwarf accretor and so will not be discussed further in this). \\

\paragraph{\textbf{High Mass X-ray Binaries}} \label{HMXBs} 
High mass X-ray binaries (HMXBs) are XRBs with a giant or subgiant donor star with mass $>10$\,M$_{\odot}$, typically O or B main sequence stars \cite[see e.g.][]{2006csxs.book..623T} and the optical luminosity is greater than the X-ray. HMXBs are associated with young stellar populations and they are a good tracer of recent star formation \cite[see e.g.][]{1974ASSL...43.....G,2003ChJAS...3..257G,2012MNRAS.419.2095M}. \\

There are two categories of HMXBs: standard HMXBs and Be star XRBs \cite[][ and references therein]{2006csxs.book..623T}. The standard HMXBs have supergiant O or B type companion stars, are persistent X-ray sources, have low eccentricity ($e <0.1$) and short orbital periods ($<10$ days). The mass transfer in these sources proceeds via one of two processes:
\begin{itemize}
    \item Roche lobe overflow. Roche lobe overflow (RLOF) occurs in a binary when the companion star expands due to its evolution, filling its Roche lobe, the tear-shaped surface of gravitational equipotential. Matter from the companion then flows through the inner Lagrangian point (the point between the binary components where their respective Roche lobes intersect, i.e. their respective gravitational potentials are equal; see Fig.\,\ref{XRBs}) onto the compact object \cite[see e.g.][]{1994inbi.conf.....S}. 
    \item Wind-fed accretion. The O and B companion stars in HMXBs have dense, strong, high-velocity stellar winds. The orbit of the compact object is inside this stellar wind and the compact object accretes material directly from the companion's wind \cite{2003astro.ph..8020C,2013AdSpR..52.2132C}.  \\
\end{itemize}

The second class of HMXBs are Be X-ray binaries (BeXRBs), so called because they have B-emission star (Be-star) companions, rapidly rotating stars with a variable disk of circumstellar materal around its equator. These systems have higher eccentricity ($ \lesssim 0.2- 0.5$) and wider orbits (P$\sim 10 - 35$\,days) than the standard HMXBs \cite{1976Natur.259..292M}. In these systems, accretion occurs when the compact object's orbit intersects with the circumstellar disk of the Be-star; rare instances of transient RLOF have also been observed. Due to the high eccentricity of the compact object orbit and the variable nature of the circumstellar disk, BeXRBs are transient and highly variable X-ray sources \cite{2003astro.ph..8020C,2006csxs.book..623T}.\\ 

%Make a table?? see Laycock et al. 2015, "Chandra and XMM monitoring of the black hole X-ray binary IC 10 X-1", Table A1

Only 10 BH HXMBs or BH HMXB candidates are currently known,  with 7 of these being extragalactic sources \cite{1972Natur.235..271B,1973ApJ...182..549H,2000ApJ...541..308H,2001MNRAS.320..316N,2007ApJ...669L..21P,2015MNRAS.452L..31L,2007Natur.449..872O,2007A&A...461L...9C,2010MNRAS.403L..41C,2009ApJ...697..573O,2012MNRAS.421.1103C,2013MNRAS.436.3380E,2019ApJ...877...57Q}. These sources are of particular interest since HMXBs are thought to be progenitors of some of the gravitational wave sources we observe with instruments such as the advanced Laser Interferometer Gravitational-Wave Observatory (a\LIGO\footnote{See \url{https://www.ligo.org/} for details on this instrument.}), such as BBHs and BH-NS binaries \cite[see e.g.][and references therein. See also section \ref{BH+compact_objects}]{2016Natur.534..512B,2016MNRAS.460.3545D,2016A&A...588A..50M,2017JApA...38...45V,2017MNRAS.471.4256V}. \\

However, the current models for the evolution of HMXBs to double degenerate binaries that merge within a Hubble timescale depends on several poorly constrained aspects of stellar and binary evolution. These uncertainties include metallicity, the common envelop phase of binary evolution, star-formation rates and massive star evolution \cite[e.g.][]{2014ApJ...789..120B,2018A&A...619A..77K}. \\

\paragraph{\textbf{Low Mass X-ray Binaries}} \label{LMXBs}

Low mass X-ray binaries (LMXBs) consist of an accreting compact object with a low mass ($\lesssim$1\,M$_{\odot}$) companion \cite{2006csxs.book..623T}. The companion stars are faint and rarely observed, with $\rm{L_{opt}}/\rm{L_X} \ll 0.1$ \cite{1995xrbi.nasa....1W}. Mass transfer in LMXBs occurs via RLOF. \\

These systems have orbital periods of between 11 minutes and about 20 days. LMXBs are associated with old stellar populations like those found in the Galactic bulge, elliptical galaxies and globular clusters \cite{2006csxs.book..623T}.\\

LMXB population size scales with the stellar mass and age of the host galaxy \cite[e.g. ][and references therein]{2012A&A...546A..36Z}. It was also found that  younger stellar populations have fewer faint LMXBs and more bright ones per unit mass than older populations. LMXB populations can also be used to probe the nature and distribution of dark matter in galaxies \cite[see e.g.][]{10.1111/j.1745-3933.2005.00132.x,Chan_2023}.\\

\emph{LMXBs in dense environments}

Globular clusters (GCs) are old ($ \gtrsim 1$\,Gyr), gravitationally bound stellar clusters that orbit galaxies. These clusters are compact systems with half light radii of a few parsecs, masses in the range
$\sim 10^4 - 10^6$\,M$_{\odot}$, optical luminosities of M$_V$ = -5 to -10 and increasing stellar density towards the cluster centre \cite[e.g.][]{1992PASP..104..981H,2006ARA&A..44..193B}. \\

Globular clusters have a high rate of dynamical stellar interactions, making these clusters very efficient at forming LMXBs. Binaries are formed in GCs via such processes as tidal capture, exchange interactions and direct collisions \cite[][and references therein]{1975ApJ...199L.143C,1975MNRAS.172P..15F,1976MNRAS.175P...1H,2005ApJ...621L.109I,2012ApJ...760L..24I}. \\ 

Ten percent of the LMXBs associated with the Milky Way reside in its GCs \cite[][and references therein]{2007A&A...469..807L}. Milky Way LMXBs are therefore more than two orders of magnitude more likely to be found in a GC than in the field of the Galaxy per unit stellar mass \cite[see e.g.][]{2002ASPC..265..289V}. For elliptical galaxies,  $20\% - 70\%$ of LMXBs are found in GCs \cite[see e.g.][]{2001ApJ...557L..35A, 2001ApJ...556..533S,2003AAS...203.3104K, 2009ApJ...703..829K}.\\

LMXBs with hydrogen-poor donors and orbital periods of less than an hour are called ultracompact X-ray binaries (UCXBs). Currently, there are about 30 UCXBs and UCXB candidates known in the Milky Way \cite{2007A&A...465..953I}. UCXBs are fives time more common in globular clusters than in the field of the Milky Way. Only one black hole UCXB in a GC has been discovered to date \cite{2017MNRAS.467.2199B}. \\

Future gravitational wave observatories, such as the Laser Interferometer Space Antenna (\LISA) will be able to detect binary black holes and black hole UCXBs in Milky Way globular clusters and perhaps even in GCs as far out at the Virgo Cluster \cite[see e.g.][]{2005MNRAS.363L..56H, PhysRevLett.120.191103, Chen_2020}. These gravitational wave detections will provide robust mass estimates for the binary components, thereby constraining GC black hole formation and evolutionary models.\\

\paragraph{\textbf{IMXBs}}	
For completeness, we note that X-ray binaries with intermediate mass (IMXB) companion stars  (1–10\,M$_{\odot}$) do exist \citep[see e.g.][]{1975ApJ...198L.109V}. Accretion onto the compact object in an IMXB proceeds via Roche lobe overflow. Due to the relatively high mass ratio between the compact object and donor star, this accretion phase proceeds very rapidly, lasting only a few thousand years \cite[see][and references therein]{2006csxs.book..623T}. These short time scales make these systems difficult to observe and only a small number of these systems are known. Additionally, the evolution of IMXBs is thought to be a  formation channel for low mass X-ray binaries \cite[see e.g.][and references therein]{2001ASSL..264..355P}. Thus IMXBs are not discussed further in this work. 	\\

\paragraph{\textbf{ULXs}} \label{ULXs}
Ultraluminous X-ray sources (ULXs) are extragalactic point-like X-ray sources located outside of the centre of a galaxy, with an apparent peak X-ray luminosity $> 10^{39}$\,erg/s, the Eddington luminosity of an 8 M$_{\odot}$ black hole \cite[see][for a review]{2017ARA&A..55..303K} (see also \ref{acc_disk_props}). With their high X-ray luminosities, ULXs are important objects in the study of relativistic accretion as they challenge our understanding of black hole formation and relativistic accretion \cite[][and references therein]{2015MNRAS.447.3243M,2017ARA&A..55..303K,2019MNRAS.482L..24M,2019MNRAS.489..282M,2021MNRAS.507.2777D}. \\

\begin{figure}
    \centering
	\includegraphics[width=0.5\textwidth]{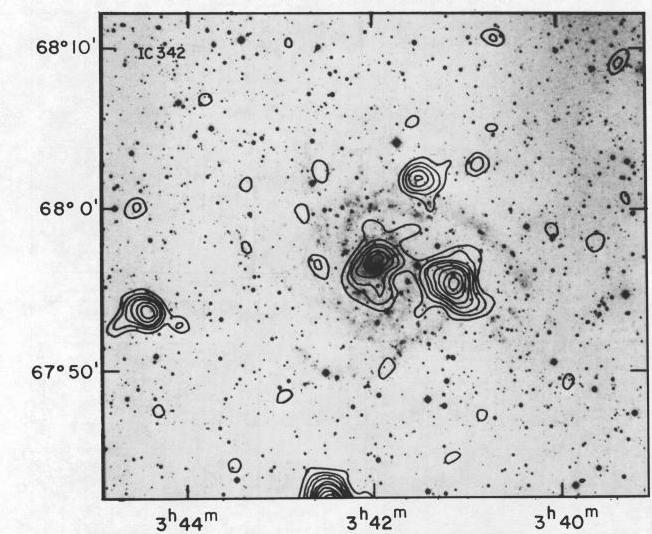}
    \caption{Figure\,2 from \cite{1987ApJ...315...46F}, showing the X-ray contour map of the spiral galaxy IC\,342 overlaid on the Palomar plate image of the galaxy. The image shows two bright point sources to the NE and SE of the central source with estimated X-ray luminosities of $\gtrsim 2 \times 10^{39}$\,erg/s. }
    \label{Einstein_ULXs}
\end{figure}
 
ULXs were first detected in the 1970s by the \emph{Einstein} Observatory \cite{1983adsx.conf..141L, 1989ARA&A..27...87F} (see Fig.\,\ref{Einstein_ULXs}). These sources were initially thought to be intermediate mass black holes (IMBHs) with mass range $100-10^5$\,M$_{\odot}$, since the Eddington luminosity, $L_{\rm X}$ scales with accretor mass, $M$,  as 

\begin{equation}
\rm{L_X} \approx 1.3 \times 10^{38} (M/M_{\odot})\, erg/s     
\end{equation} 

\cite[][and references therein]{2004IJMPD..13....1M} (see also \ref{acc_disk_props}). IMBHs are important objects from a cosmological standpoint, as they are thought to be the seed black holes from which the observed supermassive black holes in the universe grow \cite{2009MNRAS.400.1911V,2010A&ARv..18..279V,2017IJMPD..2630021M}. Subsequent studies showed that the predicted formation rates of accreting  IMBHs cannot explain the entirety of the observed population of ULXs \cite{2004MNRAS.347L..18K,2006ApJ...640..918M}. \\

%Comparison of images for ULXs from Einstein, ROSAT and Chandra (see Miller and Colbert, 2004, fig. 2) \\

\begin{figure}
    \centering
	\includegraphics[width=0.5\textwidth]{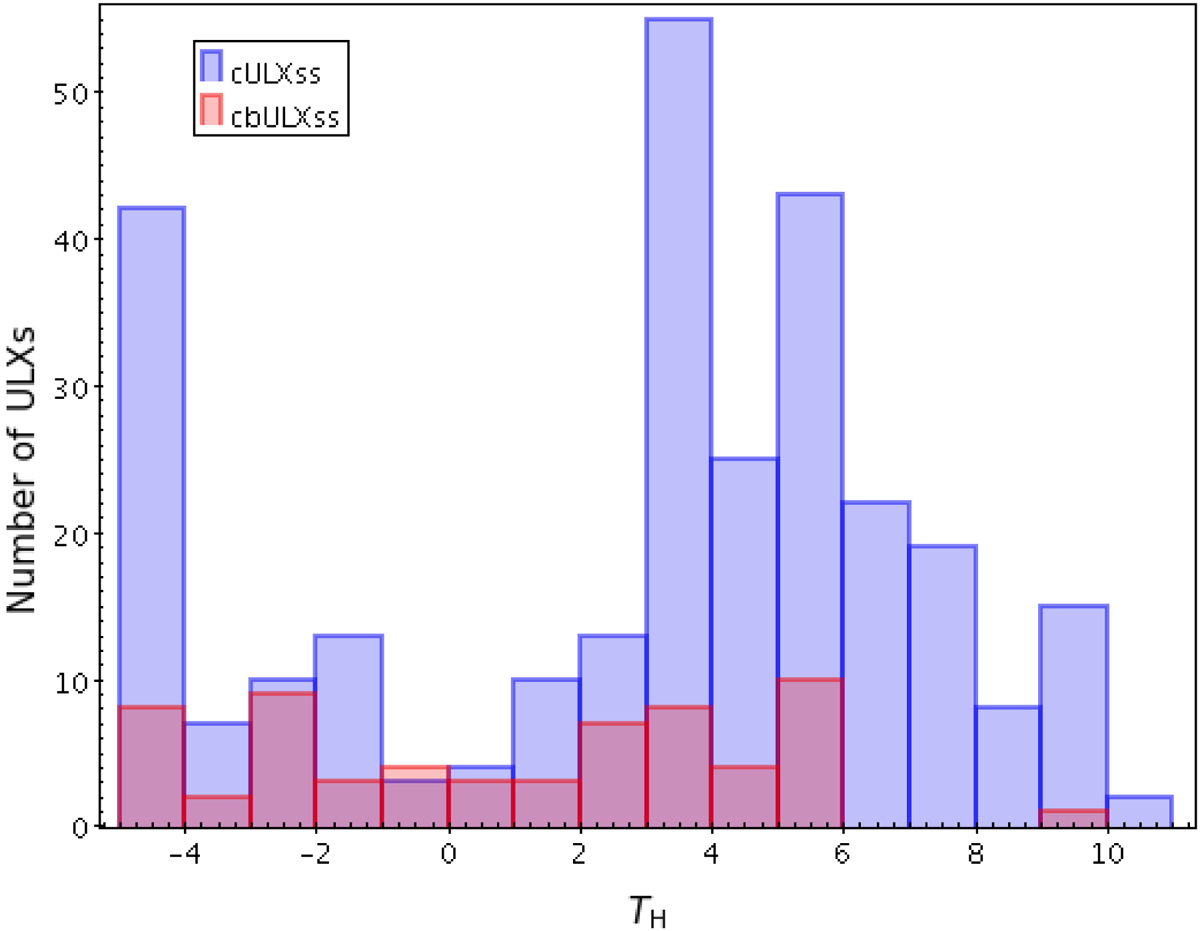}
    \caption{Figure\,9 from \cite{2022A&A...659A.188B}, showing the ULX population size by host galaxy Hubble type ($T_{H}$). The vast majority of ULXs in the catalogue are found in spiral galaxies ($0<T_{H}<10$) with a spike in elliptical galaxies ($T_{H}<-3$). The blue histogram represents the distance and luminosity complete sample and the red represents the bright ($\rm{L_X}$ $>5\times10^{40}$\,erg/s) subsample.}
    \label{fig:ULXhist}
\end{figure}

\emph{ULXs in low-density density environments} \\
ULXs are more likely to be found in star-forming galaxies \cite{2011ApJ...741...49S} and younger ellipticals with recent star formation \cite{2004ApJ...611..846K,2010ApJ...721.1523K}, suggesting that ULXs are an extension of the high luminosity X-ray binary population, which scales with both star-formation rate and stellar mass of the host galaxy \cite{2003MNRAS.339..793G,2004NuPhS.132..369G,2012MNRAS.419.2095M}. Using observational data from the \emph{Chandra} and \emph{XMM-Newton} X-ray telescopes and cross-matching with optical catalogues, \cite{2022A&A...659A.188B} compiled a catalogue of over 700 ULX sources and candidates. They found that ULXs are preferentially located in spiral galaxies and areas with active star formation, consistent with previous work (see Fig. \ref{fig:ULXhist}). \\

Today, ULXs are mostly thought to be the result of super-Eddington accretion onto stellar mass black holes and neutron stars in X-ray binaries \cite{2014Natur.514..202B,2016ApJ...831L..14F,2016MNRAS.462.4371I,2017MNRAS.466L..48I,2018MNRAS.476L..45C}. There is a subset of extremely high luminosity sources, so called hyperluminous X-ray sources (HLXs; $\rm{L_X}>10^{41}$\,erg/s), that are better explained by accretion onto higher mass black holes \cite{2012Sci...337..554W,2015MNRAS.448.1893M,2016AN....337..349B}. \\

\cite{2019ApJ...882..181B} compiled a catalogue of over 100 HLX candidates out to a redshift of 0.9 by combining X-ray and optical observations. This sample of HLXs is a useful starting point to search for IMBHs electromagnetically and they identified 22 sources as IMBH candidates. \\

\paragraph{\textbf{ULXs in high-density environments}} 
There is a small population of about 20 ULXs that resides in extragalactic globular clusters (GCs) \cite[][and references therein]{2021MNRAS.504.1545D}. Unlike ULXs in the field of galaxies, GC ULXS form via dynamical interactions, due to the high stellar density of GCs. The donor stars for GC ULXs are also old, low mass and hydrogen poor, compared to the high mass stars we find in star forming regions in the field of galaxies. The presence of ULXs in GCs means these X-ray sources can be easily separated from background AGN. An analysis of this population of X-ray sources revealed that ULXs are preferentially found in brighter GCs, but that the presence of a ULX in a GC is not correlated with the optical colour of the cluster \cite{10.1093/mnras/stz479,10.1093/mnras/staa1963}.\\

\begin{figure}
    \centering
	\includegraphics[width=0.7\textwidth]{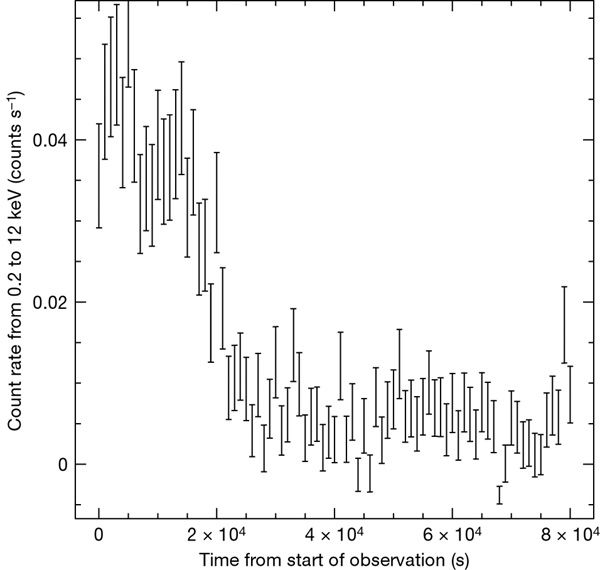}
    \caption{Taken from \cite{2007Natur.445..183M}, showing the X-ray light curve of XMMU\,J122939.7+075333, the first ULX detected in a globular cluster (GC). The source is located with in the GC RZ2019 of the giant elliptical galaxy NGC\,44772. The light curve shows an estimated peak X-ray luminosity of $\sim4 \times 10^{39}$\,erg/s.}
    \label{RZ2019}
\end{figure}

The first globular cluster ULX source was discovered by \cite[][see Fig.\,\ref{RZ2019}]{2007Natur.445..183M} using \XMM data. The source flux decreased by a factor of seven from an estimated peak flux of $\sim 4\times10^{39}$\,erg/s over the course of a few hours. This rapid variability does not support the scenario that the X-ray flux is from a superposition of neutrons stars in the GC. Indeed, rapid X-ray flux variability is one of the ways we can distinguish a black hole X-ray binary (BHB) from other X-ray sources in GCs \cite{2004ApJ...601L.171K}.\\

BHBs are the most likely sources of GC ULXs. Thus the study of ULXs in GCs can help shed light on the formation, evolution and merger history of black holes in GCs \cite[see e.g][]{2017A&A...604A..55M}. However, it is important to note given the extragalactic nature of these sources, we are unable to determine mass estimates for the components of these X-ray binaries electromagnetically. Fortunately, we can turn to gravitational wave detections to help solve this problem.\\

Initially, it was thought that any black holes formed in GCs would either cause the cluster to be destroyed \cite{1969ApJ...158L.139S} or the BHs would be ejected via dynamical interactions fairly early on in the cluster lifetime. \cite{1993Natur.364..421K,1993Natur.364..423S}. Globular clusters are now thought to be very efficient at forming binary black hole systems (BBHs) and other binaries due to the high density of stars, especially in the cluster core \cite[see e.g.][]{2016ApJ...833L...1A,PhysRevD.100.043027}. Indeed, GCs are now thought to be likely hosts of some of the gravitational waves sources observed by a\LIGO \cite[see e.g.][]{2017ApJ...836L..26C}. As with other LMXBs in GCs, the study of GC ULXs will be greatly enhanced with future gravitational wave observations \cite[see e.g. ][]{2021hgwa.bookE..24G}.\\

\paragraph{\textbf{Finding black hole X-ray binaries}} \label{MW_BHBs}

\begin{figure}
    \centering
	\includegraphics[width=0.7\textwidth]{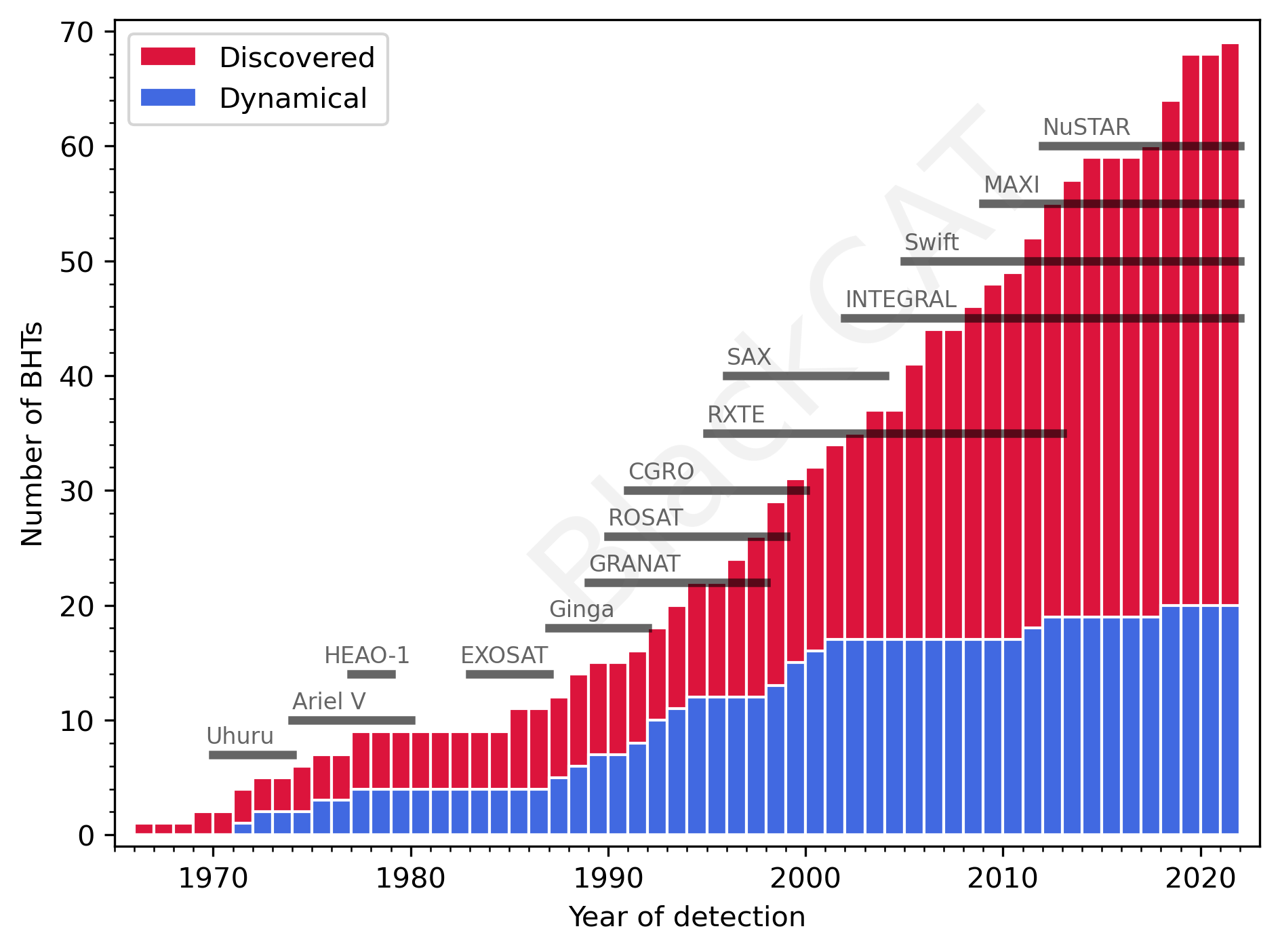}
    \caption{Taken from \cite{2016A&A...587A..61C}, last updated in December 2021. The cumulative histogram of stellar mass black hole candidates in transient XRBs. The red histogram shows the BH candidates and the blue bars represent dynamically confirmed BHs. The horizontal black lines indicate the lifetimes of the X-ray observatories that have discovered SXTs during an X-ray outburst.}
    \label{BH_hist}
\end{figure} 

As of the end of 2022, there are over 60 BHs and BH candidates known in our Galaxy \cite{2016A&A...587A..61C}. The vast majority are LMXBs detected during an outburst, where the X-ray luminosty increases by at least two orders of magnitude for weeks to months before returning to a quiescent state for between one and 50 years, with L$_X \sim 10^{32}$\,erg/s \cite{2006csxs.book..623T}. These types of LMXBs are called soft X-ray transients (SXTs). The first black hole SXT, A0620-00, was identified by \cite{1986ApJ...308..110M} in the 1980s. The source was initially discovered in August 1975 when it  went into outburst, becoming the brightest extra-solar X-ray source ever detected at the time \cite{1975Natur.257..656E}. A0620-00 remained bright for several weeks before returning to quiescence. \\

With most LMXBs, the optical flux from the low mass donor star is not detected due to the very bright optical emission produced from the X-ray heated gas around the compact object. In the case of soft X-ray transients, when the X-ray source is in a quiescent X-ray emission state ($< 10^{33}$\,erg/s; see next section), the X-ray heating and subsequent optical emission are diminished to the point where the optical emission from the donor star becomes detectable. The donor star optical spectrum can then be used to determine the orbital parameters of the SXT, including the masses of the binary components \cite[see e.g.][and references therein]{1986ApJ...308..110M,2006csxs.book..623T}, (see also \ref{dynamical} later in this chapter). \\

%-  See also de Sá, et al. An Overview of Compact Star Populations and Some of Its Open Problems. Galaxies 2023, 11, 19. https://doi.org/10.3390/galaxies11010019 \\

In the next sections, we discuss the accretion disk physics that leads to observed X-ray outbursts as well as the X-ray emission states of black hole binaries. \\

\subsubsection{Accretion physics in X-ray binaries}\label{acc_disk_props}

\paragraph{\textbf{The accretion disk}} 

As mentioned before, the vast energy output of XRBs is the result of accretion of material from the donor star onto the compact object. In the case of RLOF matter moving from the donor to towards the black hole object will have too much angular momentum to fall directly onto the compact object \cite{1973A&A....24..337S}. Instead the material flows into circular orbit around the black hole, since this is the lowest energy orbit for a given amount of angular momentum \cite[e.g.][]{2003astro.ph..1118K}. Viscous processes within the accreting material then cause the angular momentum and energy to be transferred outwards. The material spreads outwards into a series of concentric annuli moving in circular, Keplarian orbits, called an accretion disk.
\\

The viscosity in accretion disks is not the typical molecular viscosity due to interactions between particles. It widely thought that the viscosity is due to the interactions between the weak magnetic fields within the accretion disk and the accretion disk material; this is called the magneto-rotational instability model \cite{velikhov1959stability,chandrasekhar1960stability,balbus1991powerful}.\\

As the accretion material loses angular momentum and kinetic energy, it moves further down the gravitational potential well of the black hole, releasing gravitational potential energy. The amount of rest mass energy released as radiation depends on the type of compact object. For black holes, up to 40\% of of the accreted material rest mass energy can be released as radiation \cite{1981ARA&A..19..137P} .\\

The conditions to form an accretion disk are not necessarily present for wind-fed HMXBs. The stellar wind of O and B star companions do not carry as much angular momentum as the material coming from a RLOF scenario. Only one wind-fed HMXB has shown evidence of having formed an accretion disk \cite[][and references therein]{2023Galax..11...19D,10.1093/mnras/staa162}. \\

\paragraph{\textbf{Properties of the accretion disk}} %\label{acc_disk_props} 

The structure and radiation emission of the accretion disk is largely determined by the rate at which material flows from the companion star to the accretion disk and towards the black hole \cite{1973A&A....24..337S}. This accretion rate ($\dot{M}$) is related to the disk luminosity by

\begin{equation}
L_{disk} = \eta \dot{M} c^2  \,erg/s,
\end{equation} 

where $\eta$ is the efficiency with which gravitational potential energy released as radiation and $c$ is the speed of light. \\

At a critical value of the accretion rate $\dot{M}_{crit}$, the resultant disk luminosity produces enough of radiation pressure on the disk electrons to balance the compact object's gravitational force on the disk protons. For a compact object of mass $M$, 

\begin{equation}
\dot{M}_{crit} = 3 \times 10^8 \eta M/year.
\end{equation} 

The resultant disk luminosity, called the Eddington luminosity ($L_{Edd}$) is then,

\begin{equation}
L_{Edd} = \frac{4 \pi G M m_P c}{\sigma_T}  \,erg/s,
\end{equation} 

where $G$ is the gravitational constant, $m_P$ is the mass of a proton, and $\sigma_T$ is the Thompson scattering cross section of an electron \cite[see e.g.][]{1973A&A....24..337S}. \\

We note that the Eddington luminosity is calculated under the assumptions that the accretion disk is composed entirely of pure hydrogen that is fully ionised and that the radiation is isotropic. Despite this, $L_{Edd}$ is used widely to parameterise the luminosity of XRBs. \\

When $\dot{M} < \dot{M}_{crit}$, radiative processes cool the disk enough for it to be geometrically thin, but optically thick, such that the disk scale height, $H \ll R$, where $R$ is the disk radius \cite[see e.g.][]{1973A&A....24..337S,1981ARA&A..19..137P}. The thin accretion disk temperature profile is then

\begin{equation}
T(R) = \biggl\{ \frac{3GM\dot{M}}{8 \pi \sigma R^3}  \left[ 1-\beta \left( \frac{R_{in}}{R} \right ) \right] \biggr\}^{1/4} \,K,
\end{equation} 

where $R_{in}$ is the inner radius if the accretion disk, $\beta$ is a dimensionless parameter determined by the boundary condition at $R_{in}$ and $\sigma$ is the Stefan-Boltzmann constant. \\

The smallest, stable, Keplerian circular orbit that the accretion disk can extend to depends on the type of compact object \cite{1973A&A....24..337S}. For a BH, this is the innermost stable circular orbit
(ISCO), equal to 3 times the Schwarzschild radius ($R_{Schwarzschild} = 2\frac{GM}{c^2}$). The general relativistic effects in the strong gravitational field of the BH prevent stable orbits smaller than the ISCO. \\

Because the thin accretion disk rotates at roughly Keplerian velocities, half the disk energy is released as radiation and the other half is changed to kinetic energy. Black holes have no solid surface, instead they are surrounded by an intangible boundary called an event horizon that encompasses the region of spacetime that can no longer communicate with the universe outside it. Thus the kinetic energy within the accretion disk is lost once it crosses the event horizon \cite{2006csxs.book..157M}. \\

\paragraph{\textbf{The multicolour disk model}} 

The geometrically thin, optically thick accretion disk described above is most commonly modeled as a superposition of black body spectra, with a maximum temperature at the innermost radius ($T_{in}$), the so-called "multi-colour disk" (MCD) model \cite{1984PASJ...36..741M}. This a non-relativistic model that does not take into account the effects of strong gravity near the inner disk radius ($R_{in}$). Yet this MCD approximation is widely used to parameterise XRB spectra and is able to fit the spectral data well. The disk luminosity ($L_{disk}$) is then obtained by integrating over the radial distance of the disk and we have 

\begin{equation}
L_{disk} = 4 \pi \sigma R_{in}^2 T^4_{in}  \,erg/s.
\end{equation} 

For black holes, we have

\begin{equation}
\rm{M} \propto T^{-2}_{in}L^{1/2}_{disk},
\end{equation} 

therefore more massive black holes have a lower characteristic disk temperature \cite[e.g.][]{2011NewAR..55..166F}. \\

Merloni et al. (2000) \cite{10.1046/j.1365-8711.2000.03226.x} do caution that MCD model is an oversimplification of the accretion disk than can have significant impact on disk parameters obtained when the model is applied. Taking into account such processes as Doppler blurring and gravitational redshift, they showed that the MCD model systematically underestimates the value of $R_{in}$.\\

\paragraph{\textbf{Deviation from the standard accretion disk model}}

When $\dot{M}> \dot{M}_{crit}$, the accretion disk structure changes; it becomes both optically and geometrically thick, a so-called 'slim' disk. The radial temperature profile then goes from $T(R) \propto R^{-3/4}$ in the thin disk regime, to $T(R) \propto R^{-1/2}$ \cite[and references therein][]{1999ApJ...520..298Q}. \\

When the accretion rate is very low ($< 0.01 M_{crit}$), the energy released from viscous processes is not radiated away as efficiently as in the standard thin disk scenario and the energy is instead stored in the disk material. This causes the inner disk to swell and become geometrically thick, optically thin and become hot, with the electron temperatures of about 100\,keV \cite{1996ApJ...462..136N,1999ApJ...520..298Q}. As a result, the inner accretion disk moves away from the compact object and the disk becomes truncated. The  hot, optically thin accretion material and the energy inside it are then advected towards the black hole. Only a small fraction of the energy is released before the accretion material reaches the event horizon and is lost. This regime is called the advection dominated accretion flow \cite[ADAF;][]{1997ApJ...489..865E} and the radiative efficiency is less than 1\%. \\

\paragraph{\textbf{The disk instability model}} 

For an accretion disk to have a stable flow of matter through it, the accretion material must be fully ionised throughout the disk. This stable flow will result in a persistently bright X-ray source. As mentioned earlier, XRBs can display strong flux variability in the form out outbursts on timescales of weeks to months. The most widely accepted explanation for this variability is the disk instability model, whereby instabilities in the accretion disk are caused by changing ionisation states of the accretion disk material \cite{2001NewAR..45..449L,2002apa..book.....F}. \\

Over a range of accretion rates, the accretion disk is thought to be in either a cool, low ionisation, low viscosity state or a hot, high ionisation, high viscosity state. The accretion disk opacity is depends very sensitively on the ionisation level of the accretion disk material and therefore on the disk temperature. Thus small changes in disk temperature cause rapid changes between the hot and cool states. \\

To ensure that the hydrogen in the accretion disk is fully ionised, the disk temperature must be greater than 6500\,K, the ionisation temperature for hydrogen \cite[see e.g.][]{2003ASPC..308..121K}. The disk temperature declines with increasing disk radius, therefore for a stable accretion disk, the outer disk temperature must be greater than 6500\,K. \\

If the outer accretion disk temperature is lower than 6500\,K, the disk will contain areas of neutral hydrogen and these will quickly spread throughout the disk. Cool, neutral accretion material does not flow as as easily as ionised matter, thus the accretion flow towards the black hole decreases. This reduction in accretion rate then leads to a reduction in luminosity and also changes the structure of the accretion disk. \\

The changes in $\dot{M}$ described by the disk instability model, and the subsequent changes in the nature of the accretion disk, lead to very distinct radiation emission states. In the next section, we discuss the salient XRB properties that arise due to the variations in the accretion rate. \\

\subsubsection{Electromagnetic observational signatures of black hole binaries} 

\paragraph{\textbf{Black hole binary X-ray emission states}} 
Black hole binaries are known to transition between quasi-stable emission states, where the X-ray spectrum is dominated by either thermal or non-thermal emission. These emission states are a result of the disk instability model described above \cite[see ][for a full review of black hole emission states]{2006ARA&A..44...49R}. In the following sections I will describe the canonical BHB X-ray emission states as well as the accretion flow and physical accretion disk features related to them.\\

\emph{The quiescent state} \\
As described in previous sections, BH LMXBs spend most of their lifetime in a quiescent X-ray state, where $10^{30.5} < L_X < 10^{33}$\,erg/s. The accretion rate is very low in this state, with $\dot{M} < 0.01M_{crit}$ \cite[see e.g.][]{1997ApJ...489..865E,2006csxs.book..157M}. In the quiescent state, the accretion disk does not extend to the ISCO. Instead, observations of BHB show that the disk is truncated at a large inner disk radius \cite{1996ApJ...462..136N,1997ApJ...489..865E,2001ApJ...555..477M,2003ApJ...593..435M}. The ADAF model is invoked to
explain the low luminosity of the quiescent state as discussed in the previous section. \\

The leading model for the origin of the X-ray emission in the quiescent state is as follows: A corona of hot ($100 – 300$\,keV) particles surrounds the accretion disk and the ADAF \cite[e.g.][and references therein]{1997ApJ...489..865E,2006csxs.book..623T}. The X-ray emission in the quiescent state comes from soft (low energy) photons originating in the accretion disk being Compton up-scattered by the corona, giving rise to to non-thermal, hard emission with energies up to 100\,keV. The energy spectrum of this emission is well fit by a power law (PL) spectral model, with

\begin{equation}
F(E) \propto E^{-\Gamma},
\end{equation} 

where F is the flux, E in the energy and $\Gamma$ is the photon specrtal index. For the quiescent state $\Gamma$ is in the range $1.7-2.1$ \cite[e.g.][]{2006csxs.book..157M}. \\

\emph{The thermal dominant state} 

This emission state is characterised by a large ($>75\%$) thermal emission contribution component and the accretion rate is estimated be $0.1 - 0.5 $\,$\dot{M}/\dot{M}_{crit}$. The spectrum can be modelled by a MCD, with inner disk temperature of $0.7 - 1.5$\,keV \cite{1997ApJ...489..865E,2006csxs.book..157M}. In this state, there is also sometimes a hard, non-thermal component to the spectrum, modelled by steep power law, $2.1 < \Gamma < 4.8$. The flux variability in the thermal dominant state is weak and the accretion disk is thought to be in the classic thin disk configuration of \cite{1973A&A....24..337S}.\\

\emph{The hard state}

The hard emission state is defined by hard, non-thermal emission, where $> 80\%$ of the emission in the $2 - 20$\,keV range can be well modelled by a power law with $1.7 < \Gamma < 2.1$ \cite{2006csxs.book..157M}. A soft X-ray contribution is also present, which can be modelled as a cooler disk emission component of $T_{in} = 0.1-0.2$\,keV. For some BHBs, there is also an excess of hard emission ($20 - 100$\,keV), thought to arise from the reflection of the PL component emission off the inner accretion disk. \\

The hard state is sometimes regarded as a less extreme version of the quiescent state and the accretion rate is thought lie between those of the quiescent and thermal dominant state. As with the quiescent state, the inner accretion disk in the hard state is truncated to a larger radius than the ISCO \cite{1997ApJ...489..865E,2006csxs.book..157M}.\\

The nature of the accretion flow within the inner disk radius is still a matter of debate. Several versions of the ADAF model have been put forward, but no one set of models can entirely predict or replicate the observed X-ray behaviour \cite[see][for reviews]{2014ARA&A..52..529Y,2020ApJ...889L..18Y}.\\

The hard emission state in BHBs also is associated with synchrotron radio emission from the a quasi-steady, compact jet \cite[][and references therein]{2006csxs.book..381F}. The jets emission disappears when the BHB transitions from the hard state to the thermal dominant state and there is a strong correlation between the jet launching mechanisms and the nature of the accretion disk. \\

\emph{The steep power law state}

The steep power law state (SPL; formerly called the very high state) is characterised by non-thermal emission which is well modelled by a power law model with a spectral index, $\Gamma > 2.4$. At least 20\% of the overall flux is from the SPL component. In this state, $L_X$ is typically $>0.2L_{Edd}$ and the accretion rate is $\gtrsim 0.5\dot{M}_{crit}$ \cite[see][and references therein]{2006csxs.book..157M}.\\

The exact physical processes that give rise to the SPL are still being debated. \cite{2000ApJ...542..703Z} and \cite{2002ApJ...567.1057T} present a model where the steep power law emission is due to Compton upscattering of the lower energy accretion disk photons by the high energy corona electrons. Later, \cite{2011RAA....11..631Y} put forward the following scenario: high energy ($>$\,MeV energies) photons produced via synchrotron emission from relativistic electrons in highly magnetised areas near the ISCO. These high energy seed photons are then down-scattered by electrons in the corona to produce the SPL. Another model uses the expected nature of the inner accretion disk at high luminosities to produce the steep power law emission \cite{2014MNRAS.438.3352D}. At high luminosites, the inner accretion disk is radiation dominated, resulting in optically thin material producing significant flux. As a consequence, the gas temperature increases rapidly inwards, leading to  locally saturated Compton spectra with rapidly increasing peak energies, the sum of which results in a steep power law spectral component. \\

\paragraph{\textbf{Black hole binary timing signatures}} X-ray binaries display a wide range of flux variability across the electromagnetic spectrum and on various timescales. We have already mentioned flux variations on timescales of years to decades (quiescent state), weeks to months (X-ray outbursts) and days to minutes (orbital periods). In the X-ray regime, flux variability can also occur on millisecond timescales. This aperiodic, fast variability is an important tool in the study of the strong gravity near compact objects \cite[see][]{2004astro.ph.10551V}. \\

The spectral emission states and flux variability in XRBs are correlated, with both arising due to the physical processes occurring in the inner accretion disk. \cite{1971SvA....15..377S} first postulated that accretion onto stellar mass compact objects will result in millisecond timescale variability. This fast, quasi-periodic variability was thought to be the result of clumps of accretion material orbiting in the inner disk near the compact object \cite{syunyaev1973variability}. Observational evidence of this short timescale variability was provided by the \RXTE mission, which had millisecond time resolution. \\

In BHBs, this variability is made up of longer (noise) and shorter timescale features. The short timescale features are called quasi-periodic oscillations (QPOs) with frequencies in the range  0.01 to 450\,Hz. QPOs are modelled as Lorentzian functions and have coherence parameters, $Q = \nu/FWHM \gtrsim 2$, where $\nu$ is the centroid frequency  \cite[e.g.][]{2000MNRAS.318..361N,2004astro.ph.10551V}. QPOs are associated with X-ray spectral state transitions as well as non-thermal emission states. Quasi-periodic oscillations can be divided into two subgroups: low frequency QPOs (LFGQPOs; $\sim 0.1–30$\,Hz) and high frequency QPOs (HFQPOs; $40–450$\,Hz). \\

LFQPOs have been observed in BHBs in the SPL, hard and SPL to hard transition states. The frequencies and amplitudes of LFQPOs have been found to be correlated to both the PL and thermal components of these spectral states  \cite[see e.g.][]{1999ApJ...527..321M,2000ApJ...531..537S,2000MNRAS.312..151R,2003A&A...397..729V}. Thus studying LFQPOs can help us understand the PL spectral component and LFQPOs data can be used to constrain models of non-thermal emission in BHBs \cite[see][]{2006ARA&A..44...49R}.\\

HFQPOs are transient and have lower amplitudes than LFQPOs and they have been observed to occur when the BHB is in the steep power law state. Their frequencies correspond to that expected for accretion material orbiting near the ISCO of a roughly 10\,M$_{\odot}$ BH. It is thought that HFQPOs are related to the mass and spin of the BH. Combining HFQPO data with BH mass estimates are a good way to constrain the spin of the BH \cite[see \S 8.2.4 in][]{2006ARA&A..44...49R}.

\subsection{Black Holes in non-interacting systems}\label{noninteractingBHs}
In this Section we will describe what are the main observational strategies that can be used to unveil stellar-mass BHs which are part of non-interacting (or quiescent\footnote{In this Chapter we give the words "non-interacting" and "quiescent" the same meaning. In the literature you can find these words referring to slightly different configurations or evolutionary phases.}) binary systems.\\

Not being able to rely on X-ray, gamma-ray or radio emission from these sources, the only way we can unveil a BH in a quiescent binary is by measuring the dynamics of the companion luminous star. There are two main indirect techniques that can be used to unveil a BH-star binary: 1) {\it spectroscopically} by measuring period changes on the spectral lines of the luminous companion or 2) {\it astrometrically} (for example with GAIA), by tracking the orbital motion of a luminous star as a NS or BH pulls it around. We will explore these two approaches by highlighting the observational studies that have been conducted so far in the field.

\subsubsection{Spectroscopic detection of quiescent Black Holes}
\label{dynamical}
\begin{figure}
    \centering
	\includegraphics[width=0.55\textwidth]{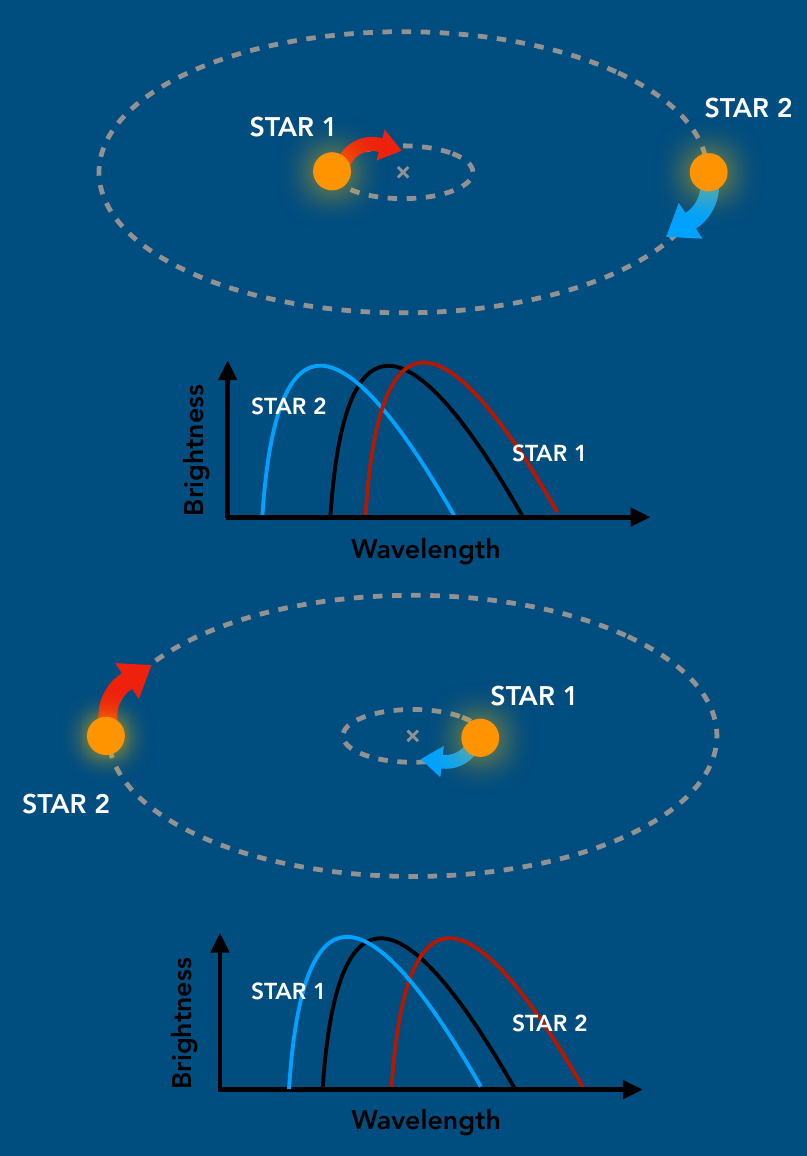}
    \caption{Schematic representation of the methodology used to detect spectroscopic binaries. This is the case of a stellar binary. Depending on the position of a star (star 1 or 2) relative to the binary's center of mass it may appear to move towards or away from the observer. In the first case the light emitted by this star appears blueshifted, however in the second it appears redshifted. By monitoring these shifts over time it is possible to construct the radial velocity curve of both star 1 and star 2, independently.}
    \label{fig:radialcurve}
\end{figure}

In a binary system made of two stars, the reciprocal motion of the two companions with respect to their mutual center of mass causes the line-of-sight components of their velocities to vary in antiphase with each other (also referred to as reflex motion) as a function of time, and this produces a periodic signal, whose periodicity depends on the orbital period of the system under investigation, directly linked to the distance between the two stars in the system. The amplitude of the signal is instead related to the masses involved in the binary. By measuring the change in wavelengths of a few lines of interest and/or the full spectrum (i.e., Doppler shift) for the two stars over a period of time, it is possible to construct a radial velocity curve. Figure \ref{fig:radialcurve} illustrates the case of a stellar binary. Depending on their positions relative to the binary's center of mass, star 1 and star 2 (as labeled in the figure) appear to move towards and away from the observer. This motion makes the light from the stars appear slightly bluer when they are moving towards the observer, and slightly redder when moving away. This is what is called Doppler effect.\\

Binaries detected using this method (also called radial velocity method) are named {\it spectroscopic binaries} and can be divided in two groups: 1) Binaries where the two stars are almost equally bright, called {\it double-lined spectroscopic binaries} (or SB2) ; 2) Binaries where one component is significantly brighter than the other, called {\it single-lined spectroscopic binaries} (or SB1).
While for SB2 binaries we can determine the radial velocities of the two components ($v_{1}$ and $v_{2}$), for SB1 binaries we can instead only measure the radial velocities of the brightest component (usually called $v_{1}$).\\ 
The radial velocity or Doppler shift method has an important limitation. It does not allow to determine the position of the binaries' orbits in space, thus the observed velocities $v$ are not the true velocities $v_0$ of the stars but their velocities projected into the line of sight. These two quantities are linked to each other via the following relationship: 
\begin{equation}
     v = v_{0}~sin~i
\end{equation}
where the inclination $i$ is the angle between the line of sight and the normal of the orbital plane. This directly affects the measurement of the radial velocity semi-amplitude of the binary $K$, which is also defined as $K = K_{0}sin~i$, where $K_0$ and $K$ represent the true and observed amplitudes, respectively.
By using a simple mathematical formulation we can understand what are the main properties of a binary obtainable with the radial velocity method, and what differentiates SB1 from SB2 binaries.
If we consider a binary system where the two stars move in a circular orbit around their common center of mass, and we assume that the radii of the orbits of the two components are $a_1$ and $a_2$, respectively, by using the equation of the center of mass $m_{1}a_{1}$ = $m_{2}a_{2}$ and considering $a = a_{1} + a_{2}$, we can rewrite the orbital radius of the first component as a function of the masses involved in the system as follows:
\begin{equation} \label{e:eq1_BH} 
    a_1 = \frac{a~m_2}{m_1+m_2}
\end{equation}
The true semi-amplitude velocity $K_0$ can instead be rewritten as: 
\begin{equation}
    K_{0,1} = \frac{2\pi a_{1}}{P},
\end{equation}
where $P$ is the orbital period of the binary. By substituting $K_{0}$ from the above equation, we obtain: 
\begin{equation} \label{e:eq2_BH} 
    K_{1} = \frac{2\pi a_{1}~sin~i}{P}
\end{equation}
Substituting Eq. \ref{e:eq1_BH} in Eq. \ref{e:eq2_BH}, we then obtain: 
\begin{equation} 
K_{1} = \frac{2\pi a}{P}\frac{m_{2}sin~i}{m_{1} + m_{2}}
\end{equation} \label{e:eq5_BH}
By solving the equation for the semi-major axis $a$ of the binary and by applying the Third Kepler's law, we derive the {\it mass function} of the binary system:
\begin{equation} \label{e:eq3_BH} 
   f(m) = \frac{m_2^3~sin~i^3}{(m_1+m_2)^2} = \frac{K_1^3 P}{2\pi G}
\end{equation}
This formula, which only applies to binaries in circular orbits, can be generalized to take into account the effect of eccentricity $e$ as follows:
\begin{equation} \label{e:eq4_BH}
   f(m) = \frac{m_2^3~sin~i^3}{(m_1+m_2)^2} = \frac{K_1^3 P}{2\pi G}(1-e^2)^{3/2}
\end{equation}
The right hand side of Eq. \ref{e:eq3_BH} (or Eq. \ref{e:eq4_BH}) can be fully inferred from radial velocity data in a spectroscopic binary, as $K_1$, $P$ and $e$ are all measurable quantities.
In an SB2 binary, where the spectral lines of both components can be observed, we measure not only $K_1$ but also $K_2$ and an expression similar to Eq. \ref{e:eq3_BH} (or Eq. \ref{e:eq4_BH} in the general case) can be written, with the labels reversed. By using Eq. \ref{e:eq5_BH} and the definition of the center of mass, we obtain the following:
\begin{equation}
    \frac{K_1}{K_2}=\frac{a_1}{a_2}=\frac{m_2}{m_1}
\end{equation}
This means that in an SB2 binary we can derive the mass ratio of the two components. However, since we do not have any direct information about $sin~i$, we cannot obtain the true masses $m_{1}$ and $m_{2}$. Only if the true mass of one of the two stars is determined via alternative methods (e.g. using photometric data to derive the spectral energy distribution of the star, hence its mass), then the mass of the companion is immediately deduced.\\

The situation becomes significantly more complicated in SB1 binaries, where only the spectral lines of the brightest component can be observed. In this case the mass ratio of the visible vs unseen object in the binary is unknown and the only information that can be retrieved is the system mass function in Eq. \ref{e:eq4_BH}. Even if we know the mass of the bright component (that we call here primary), we are not able to determine the true mass $m_{0,2}$ of the secondary unseen star, but only $m_{2} = m_{0,2}~sin^3~i$. This is the so-called {\it $m\;sin~i$ degeneracy}, i.e. the same value of $m_{2}$ can be produced either by a low inclination $i$ (close to a face-on orbit, $i=0^{\circ}$) and a high mass or by a high $i$ (close to an edge-on orbit, $i=90^{\circ}$) and a low mass for the star. There are a few cases in which this degeneracy between mass and inclination can eventually be broken and we will discuss them briefly in the following. In all other cases scientists determine a {\it minimum mass} for the secondary unseen star, by assuming that the binary orbit is perfectly edge-on, thus by assuming that it has an orbital inclination of $90^{\circ}$. This way $sin^3~(i=90^{\circ})$=1 and this term can be removed. Every other inclination value in the range between $0^{\circ}$ and $90^{\circ}$ provides a larger value for the true mass of the secondary unseen companion.\\

As anticipated, there is a way to break the degeneracy between mass and inclination in spectroscopic binaries, and that happens when the light modulation of the system as a function of time (light curve) is observed together with its radial velocity curve. More specifically:
\begin{itemize}
    \item In eclipsing binaries, i.e. when the system inclination is such that one star passes in front of the other and partially obscures the light coming from it. The shape and depth of the eclipses can be uniquely solved to give the inclination and hence the true masses of the individual stars.
    \item In non-eclipsing close binary systems, or when one component is not seen, then the ellipsoidal modulation of the seen component depends on the mass ratio and the inclination. Together with the radial velocity curve, this can then give unique masses for the components.
\end{itemize}

Promising BH-stellar binary candidates are those that meet two criteria: i) the periodic signal due to the motion of the star can be clearly detected but the reflex motion due to a visible companion is not observed; and ii) the analysis of the radial velocity curve suggests the unseen component to be more massive than the observed one. However, before misinterpreting the data, some caution is required as there may be several reasons why a visible companion, although present, does not appear in the composite spectra (see the discussion in the next two Sections for details).\\

Spectroscopic surveys are one of the main strategy for dynamically detecting non-interacting binaries. Any spectrograph can be used to perform these studies, with the most useful instruments for this purpose being either multi-object spectrographs (e.g. GIRAFFE at the Very Large Telescope - VLT) or spectrographs equipped with an integral field unit (IFU) such as the VLT/MUSE. Both can be used to obtain simultaneous spectra of many objects, hence performing blind but efficient searches. By observing the same field for a number of times, and extracting a spectrum for a significant number of sources in the field of view, it is possible to identify those objects showing radial velocity variations and among them, those orbiting massive (and likely) dark companions. Unfortunately, the only surveys that have multi-epoch spectroscopy that is rich enough for a time-series analysis are APOGEE (Apache Point Observatory Galactic Evolution Experiment \cite{2017AJ....154...94M}), LAMOST (Large Sky Area Multi- Object Fiber Spectroscopic Telescope \cite{2012RAA....12.1197C}), RAVE (RAdial Velocity Experiment \cite{2006AJ....132.1645S}) and Gaia-ESO \cite{2012Msngr.147...25G}, which came into play only in recent years. This limitation has severely hampered the detection of BH-stellar binaries in the past via this method. More surveys (WEAVE \cite{2014SPIE.9147E..0LD} and 4MOST \cite{2019Msngr.175....3D}) will be in operation in the coming years, surely providing a boost in the field, thanks to a significant increase in both the number and depth of the observations.\\

As mentioned above, an important caveat to keep in mind in this type of study is that, when only radial velocity measurements are available, the binary system cannot be fully characterized and only a minimum companion mass can be inferred for the unseen source. This is due to the fact that the inclination of the system is unconstrained. When radial velocities are instead complemented with additional pieces of information, for example, data showing the photometric variability of the system as a function of time (i.e. light curves), then these degeneracies can be removed and the current mass of the two components can be finally estimated.
Unlike spectroscopy, there are many photometric surveys, some even quite dated, which scan the sky both in the direction of the Milky Way and of Galactic satellites (e.g. the Large and Small Magellanic Clouds). Most of them have been designed to study variable stars in different regions of the Universe and for different purposes. Among the most popular we find VVV \cite{2010NewA...15..433M}, VMC \cite{2011A&A...527A.116C}, MACHO \cite{2000ApJ...542..281A}, EROS-2 \cite{1998A&A...332....1P}, ASAS-SN \cite{2017PASP..129j4502K} and TESS \cite{2015JATIS...1a4003R}. As already mentioned in Section \ref{isolated}, OGLE also has to be added to the list, which over the last twenty years has been an extraordinary resource of interesting objects.\\

When talking about dynamically detecting quiescent BHs in binaries, it is important to make a distinction between two different types of environment where they can be found. They can be located in high density fields, like the central regions of massive star clusters (dense collections of thousands to million of gravitationally bound stars) or in low density fields, which are the most abundant in the Universe.
Various mechanisms are at play in these two environments, so the BH population they host can be very different. Below we will describe the low and high density fields separately, highlighting the main results on the detection of quiescent BHs in both.
\\
\paragraph{\textbf{Quiescent Black Holes in low-density fields}}\label{a}
Our Universe is largely very low-density, populated by {\it collisionless} systems, such as galaxies, where stars are mainly moving in the collective gravitational field, and interactions or collisions between individual stars are extremely rare if at all absent. When a stellar-mass BH is detected in this type of environments, it is very likely that it might be the remnant of the death of a massive star that formed either isolated or in a "primordial" binary that exploded as a supernova, rather than the product of a more complex formation mechanism (e.g., hierarchical assembly of low-mass BHs). The population of BHs that can be sampled by randomly observing the field is as close as possible to an ``undisturbed" distribution but does not necessarily resemble what we can find in other types of environments.\\

Many studies in the literature have postulated that millions of BHs should exist in our Galaxy \cite{1966SvA....10..251G,1969ApJ...156.1013T,1983bhwd.book.....S,1992A&A...262...97V,1994ApJ...423..659B,1996ApJ...457..834T,1998ApJ...496..155S}, and detecting a good fraction of them would be extremely important not only to understand the physical processes that trigger their formation but also to characterize their properties (mass distribution, spin), to date completely unknown.
Unfortunately, until very recently, no quiescent non-interacting BH-stellar binaries have been found in radial velocity searches. As previously anticipated, one of the reasons is the lack of dedicated spectroscopic surveys capable of scanning the sky a sufficient number of times to capture a significant number of sources to analyze. Another, even more critical, is the difficulty of interpreting the results coming from the few available.\\

Casares et al. \cite{2014Natur.505..378C} reported the first detection of an X-ray quiescent stellar-mass BH of 3-6 $M_{\odot}$ in the Milky Way, in a binary system with a classical Be star (i.e., fast rotating B star that displays strong emission lines) in a mass range between 10 and 16 $M_{\odot}$. The system, MWC 656, has been identified as the candidate counterpart of a potential $\gamma$-ray flare, hence it does not represent the best example of a fully dynamically detected system. Subsequent radio and X-ray emissions have been found to be associated with the system, potentially strengthening the case for this to be a BH. A re-analysis of the spectral variability properties of the source, however, has recently questioned this interpretation, suggesting a hot subdwarf as the most likely companion \cite{2022arXiv220812315R}. Another interesting object, AS 386, has been studied by Khokhlov et al. \cite{2018ApJ...856..158K}, showing H$\alpha$ emission and near-infrared excess. Thanks to low and high-resolution spectroscopy and multi-band optical and near-infrared photometry the authors have suggested that the system is as follows: a binary of a B-type star surrounded by circumstellar gas and dust, exhibiting the B[e] phenomenon and a high mass ($\sim$7 $M_{\odot}$) invisible companion which could not be anything other than a BH. This system is one of the very few to date that has not been questioned.\\

\begin{figure}%[htb]
    \centering
        \includegraphics[width=0.475\linewidth]{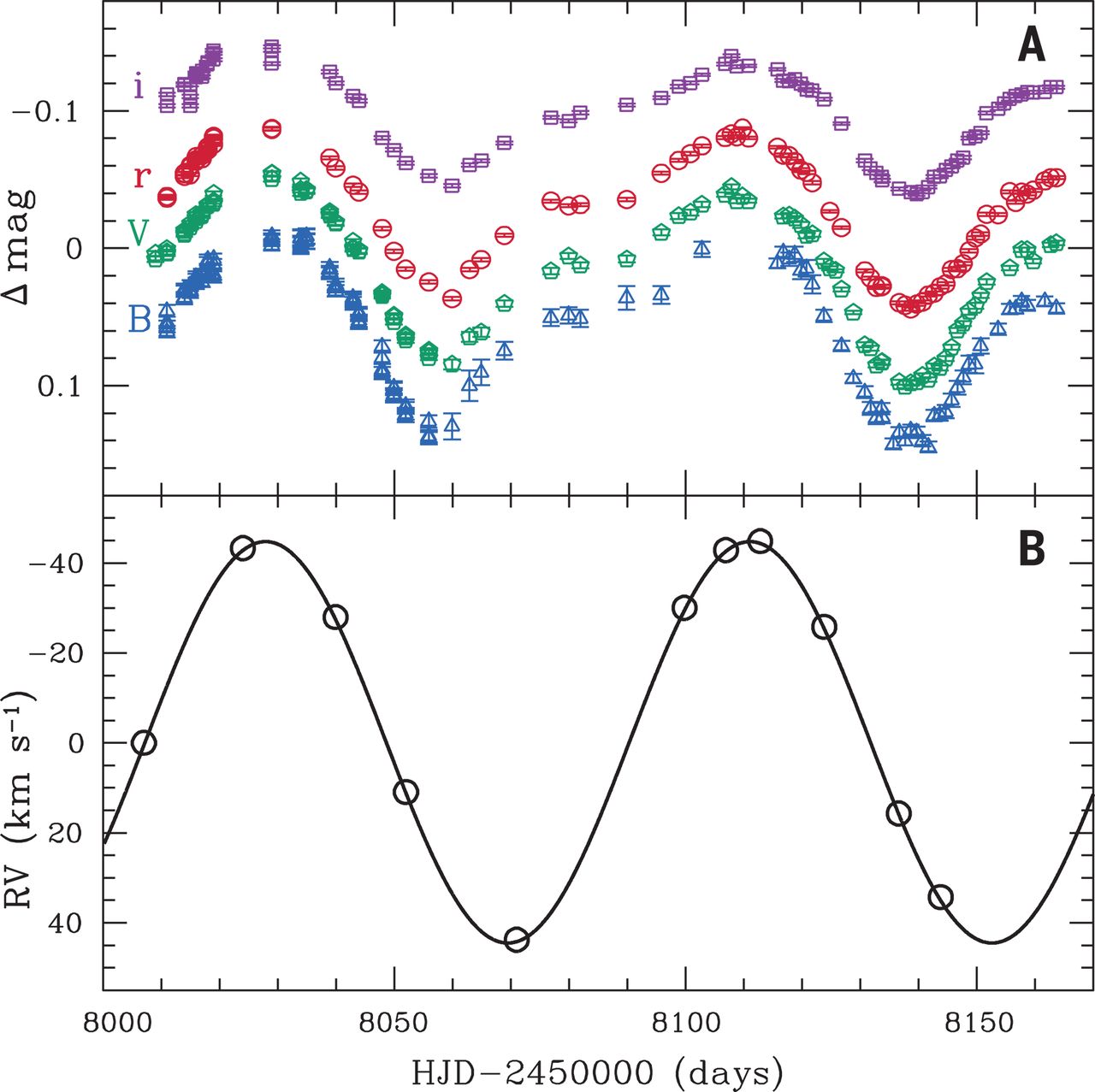}
        \includegraphics[width=0.475\linewidth]{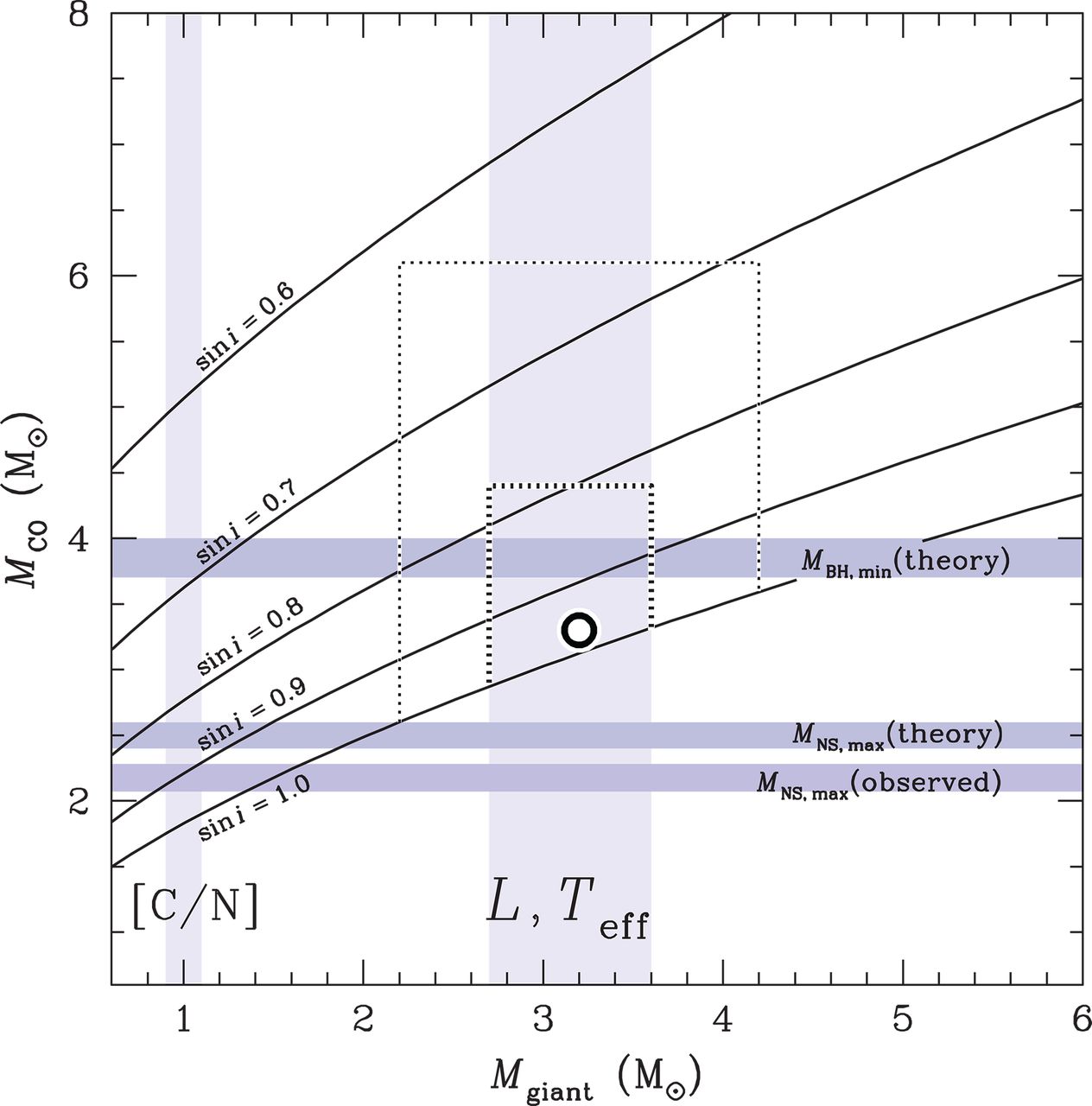}
    \caption{from Thompson et al. \cite{2019Sci...366..637T}: {\it Left panel:} The ASAS-SN light curves of the binary in 4 different filters, and the TRES radial velocity curve used to constrain the orbital properties of the system. {\it Right panel:} The mass of the unseen companion (the compact object) as a function of the mass of the giant. The blue vertical shaded areas define the allowable mass range for the giant given its T\textsubscript{eff} and L, while the solid black curves correspond to different inclinations of the orbits. The black and white dots identify the more likely masses for both components in the binary.}
    \label{fig:thompson}
\end{figure}

The first detection of a Galactic non-interacting stellar-mass BH in a binary system with a sensibly smaller luminous companion was reported by Thompson et al. \cite{2019Sci...366..637T}. Indeed, the invisible companion has an estimated mass of $\sim$3.3 $M_{\odot}$ and it is in a binary system with a giant star (2MASS J05215658+4359220). The system has an orbital period of $\sim$83 days, a semi-amplitude velocity K$\sim$44.6 km/s, near-zero eccentricity and an inclination close to 90 degrees. The detection has been possible thanks to the analysis of the radial velocity curve of the system provided by APOGEE and TRES, complemented by multi-band ASAS-SN and TESS light curves. In Figure \ref{fig:thompson}, left panel, both radial velocity and light curves of the binary system are shown. Given the uncertainties in the estimated mass of the invisible companion, it is compatible with both a BH or a NS, falling in the so-called mass gap between these two types of objects \cite{2011ApJ...741..103F}. Right panel of Figure \ref{fig:thompson} shows the constraints on the mass of the compact object based on the properties of the visible star (L, T\textsubscript{eff}) and on the predicted inclination of the system itself based on the observed photometric variability.
Van den Heuvel \& Tauris \cite{2020Sci...368.3282V} analyzed the same set of data available for the system but came to a rather different conclusion. Apparently, if the estimated mass of the giant star is lower by a factor of 3 from what suggested by \cite{2019Sci...366..637T} ($\sim$ 1 $M_{\odot}$), the system would likely be a tertiary instead of a binary system. In this case, the unseen component would be compatible with a close binary system, made of two low mass main sequence stars, which would produce no detectable signals in the spectral energy distribution, as well as providing a simple explanation why no X-ray emission is detected. To date, none of the two scenarios have been investigated further, hence it is still unclear whether this system contains or not a BH. If confirmed, it would be an extremely important discovery as it would help building statistics on these systems.\\

Starting in 2019, a number of non-interacting BH candidates have been reported in the literature, most of which with associated massive companions (i.e. O-type and B-type stars). For most of those systems, alternative explanations have been proposed that do not require the presence of a BH in them. While the alternative interpretation has been confirmed by further analysis/data for a number of these objects, additional investigations need to be performed for others, in order to provide a firm conclusion on the nature of these systems.
We briefly discuss some of these individual systems here as they represent the most recent claims of BHs found in the field, and even when it was later proved that a BH did not exist in the systems, they are still very useful examples that help us understand 1) how challenging these types of studies are; 2) what are the main caveats that scientists need to be aware of when dealing with these systems; and 3) what are the main tests to be performed to be sure that a stellar-mass BH actually resides in the system.\\

-  LB-1 is a Galactic binary system. According to Liu et al. \cite{2019Natur.575..618L}, it is made of a massive BH of $\sim$68 $M_{\odot}$ orbiting around a B-type star with period P = 78.9 $\pm$ 0.3 d, semi-amplitude velocity K = 52.8 $\pm$ 0.7 km/s, and eccentricity e = 0.03 $\pm$ 0.01. The radial velocity curve of the system was measured with a sample of 26 LAMOST, 21 GTC and 7 Keck observations obtained over two years. By studying the motion of the B-type star and the accompanying H$\alpha$ emission line, that they interpreted linked to an accretion disk orbiting around the invisible companion, the authors concluded that the invisible companion should be relatively massive, hence it could only be a BH.
When published, this finding drew much attention from the scientific community as it was the most massive stellar BH ever reported (even more massive than theoretical predictions), with strong implications for our interpretations of gravitational wave (GW) signals.
A number of groups independently analyzed the system from different perspectives, using alternative techniques, and the final answer came shortly after: LB-1 does not contain a BH. Instead, it is a rare binary system of a stripped star (the former mass donor) and a Be star rotating nearly at its critical velocity (the former mass accretor) \cite{2021A&A...649A.167L,2020MNRAS.495.2786E,2020A&A...639L...6S,2020MNRAS.493L..22E}. The high rotation of the B-type star creates a rotationally supported equatorial disk responsible for the H$\alpha$ emission that Liu and collaborators wrongly attributed to the presence of a BH in the system. Moreover, when a star rotates very fast, its spectral lines widen due to this effect, so that the lines of a presumed second bright companion can be easily overlooked, making it impossible to detect and separate the two contributions. The use of the shift-and-add spectral disentangling technique (\cite{1998PASP..110.1416M,2006A&A...448..283G}) was fundamental to separate the flux contribution of the two sources from the composite spectrum and guide to the correct interpretation of the data.\\

-  HR 6819 is a naked-eye Galactic binary system. Using FEROS echelle spectra and TESS light curves, Rivinius et al. \cite{2020A&A...637L...3R} suggested that HR 6819 is a hierarchical triple made of a classical Be star in a wide orbit with an unconstrained period around an inner 40 d binary consisting of a B3 III star and a BH in a circular orbit. They measured a radial-velocity semi-amplitude K of 61.3 km/s for the inner star and by assuming a minimum mass of $\sim$5 $M_{\odot}$ for it, they deduced a mass for the unseen object $>$ 4.2 $M_{\odot}$, indeed compatible with a BH.
Similarly to what happened for LB-1, also HR 6819 triggered lots of investigations within the community which ended up with different teams independently suggesting that the system is not a triple, but instead a binary system consisting of a stripped B-type primary and a rapidly-rotating Be star formed from a previous mass transfer (\cite{2020A&A...641A..43B,2021MNRAS.502.3436E}). This is basically the same kind of system proposed for LB-1, although the interpretation of the two systems was initially very different, and the presence of the companion in the spectra has been overlooked for exactly the same reason.
The final word on whether or not HR 6819 contains a BH came very recently, when new high-angular resolution data from the VLTI/GRAVITY revealed the presence of a stellar companion with an angular resolution of $\sim$1.2 mas from the primary, that moves on the plane of the sky over a timescale compatible with the known spectroscopic period of 40 d \cite{2022A&A...659L...3F}.
Thanks to further data and more sophisticated tools, it has now been established that neither LB-1 nor HR 6819 contain a BH.\\

In the past couple of years, two more claims of quiescent BH detections in the Milky Way have been reported in the literature: V723 Mon (“the Unicorn”, \cite{2021MNRAS.504.2577J}) and 2M04123153+6738486 (“the Giraffe”, \cite{2022MNRAS.516.5945J}). Both of them have been proposed to be binary systems of giant stars with low-mass/faint companions, which they interpreted as BHs of a few solar masses, around 2-3 $M_{\odot}$. Masuda and collaborators \cite{2021ApJ...910L..17M} also presented the results of further investigations in support of the dark nature of the companion in “the Unicorn”.
El-Badry and colleagues \cite{2022MNRAS.512.5620E}, on the other hand, have recently proposed an alternative explanation for both systems, namely that they are stripped giants with subgiant companions.
The authors used a combination of spectral disentangling and fitting of binary spectral models to investigate the presence of any luminous companion in both systems. They applied the spectral disentangling technique to the Keck/HIRES spectra available for “the Unicorn” and “the Giraffe” and were able to actually identify the presence of an additional stellar source in both systems. In “the Giraffe”, the companion was also detected in the near-infrared APOGEE spectra. They found the secondaries to contribute about half the light in the optical, while they dominate at the bluest observed wavelengths.\\

HD 96670 in the Carina OB2 association is one of the last binary systems claimed to contain a quiescent stellar-mass BH \cite{2021ApJ...913...48G}. It is a binary system of a $\sim$22 $M_{\odot}$ O-type star orbiting around a $\sim$7 $M_{\odot}$ BH with an orbital period of $\sim$5 d and possibly a third component. Similarly to the case of MWC 656, an emission at high-frequency (in X-rays) was found to be associated with the system triggering further photometric and spectroscopic analyses, hence the detection, if later confirmed from further analyses, would not come from a blind dynamical search.\\

Very recently, a new X-ray quiet BH (VFTS 243) has been discovered in the field of the Large Magellanic Cloud (LMC), in the direction of the Tarantula nebula, to be the companion of a massive (M$\sim$25 $M_{\odot}$) O-type star \cite{2022NatAs...6.1085S}. The authors have used both radial velocity curves from the Fibre Large Array Multi Element Spectrograph (FLAMES) at the VLT and OGLE light curves to constrain the nature of the binary system. In particular, it has an orbital period P$\sim$10.05 d, an almost circular orbit (e$\sim$0.017) and a semi-amplitude velocity K$\sim$81 km/s. Shenar and collaborators have applied a spectral disentangling technique to set an upper limit to the flux contribution of any putative luminous companion and found that any luminous companion would exceed this limit significantly. From the analysis of the light curves and the very small ellipsoidal variations they estimated the possible inclination of the system to be larger than 40 degrees and the mass of the unseen companion to be of at least 9 $M_{\odot}$, which can only be a BH. Under the same observational campaign targeting the Tarantula nebula in the LMC, Shenar et al. \cite{2022A&A...665A.148S} have also looked at other SB1 binaries and they have identified two other binaries hosting as many putative BH candidates. Their nature, however, is much more uncertain and a proper study of these systems is required.\\

Given the number of binaries with quiescent BHs that have been questioned in the recent years, Mahy et al. \cite{2022A&A...664A.159M} have suggested a method to overcome this problem in the future, which is based on the combination of 1) high-resolution spectroscopy, 2) high-precision space-based photometry, and 3) state-of-the-art spectral disentangling. These three ingredients together can allow to actually constrain the nature of unseen companions in systems classified as SB1 in the literature. To show the power of this approach, they selected a sample of 32 Galactic binaries hosting an O-type or an early B-type star. This sample contained also two systems that, from X-rays observations, are already known to host a BH (Cyg X-1 \cite{2011ApJ...742...84O,2021Sci...371.1046M}) and a NS (HD 74194 \cite{2015A&A...583L...4G}), respectively. Their analysis demonstrates that, when good data are available, even very faint companions can be detected. They found a secondary luminous companion in 17 out of 32 binaries in their sample, while in the remaining systems they were not able to detect any secondary star, either because it was too faint, or because it was dark. The two most promising candidates hosting a BH were actually Cyg X-1 and HD 130298, where the estimated mass of the unseen companion was more than 7 $M_{\odot}$. We know that Cyg X-1 hosts a BH since many years now, while HD 130298 was presented as another system likely hosting a quiescent BH candidate. Further multi-wavelength observations will confirm the nature of the latter but thus far it represents one of the few promising candidates in the literature. Two more binaries have been recently identified by Fu et al. \cite{2022ApJ...940..126F} as potential hosts of as many compact objects, by inspecting the GAIA DR3 catalog and modelling TESS light curves of the targets, but even in this case there is no certainty about the nature of the companions. Follow-up observations will establish this.\\

To conclude, currently no quiescent BHs in stellar binaries have been unambiguously unveiled in the Milky Way field through a purely dynamical approach based on radial velocity searches, other than that reported by Khokhlov et al. \cite{2018ApJ...856..158K} and by Mahy et al. \cite{2022A&A...664A.159M} (high-mass companions). In HD 96670, the only other system not questioned so far, a counterpart at high frequency has triggered the dynamical search, so technically we cannot classify it as a binary with a dormant BH. The only extra-galactic quiescent BH detected and confirmed to date is that recently reported in the LMC field by Shenar et al. \cite{2022NatAs...6.1085S}.
The near lack of BHs detected so far is interesting but not unexpected, given that they are all individual detections based on relatively small samples. It is even more justified if we consider how challenging these studies are.\\

It is interesting to note, however, that most of the binary systems so far proposed to host a dormant BH (confirmed or later disputed) have relatively massive O- and B-type star companions. Due to the large masses involved, these systems are by far the most difficult to model, thus being the most prone to interpretation errors. On the other side, they are the ones with the highest expected fraction of BH companions. Indeed, recent theoretical calculations predict that about 3\% of massive O or early-B stars in binary systems should have a dormant BH as a companion \cite{2019ApJ...885..151S,2020A&A...638A..39L}. If this theory is correct, large numbers of these binaries are actually hiding out in the Universe and it may be only a matter of time before they are detected. However, it is important to underline that O and B stars are very massive and short-lived objects. The reason they are largely undetected may simply be an observational limitation: their lifetime are short enough to make their observations unlikely. An intriguing aspect to mention is also that many of these systems, at some point in their evolution, may effectively become double degenerate binaries (i.e., binaries with two compact objects such as BH-NS or BH-BH binaries), thus the possible progenitors of GW sources.\\

In light of these theoretical studies \cite{2019ApJ...885..151S,2020A&A...638A..39L}, the small fraction of quiescent BHs detected so far in the field, the increasing number of interacting BHs in binaries detected through X-ray or radio observations, and the large number of GW signals produced by BH-NS or BH-BH binaries revealed by the LIGO-VIRGO-KAGRA consortium since 2015 \cite{2020CQGra..37e5002A} can probably be reconciled. The main reasons for the apparent discrepancy we see in their numbers may indeed be the following: 1) the latter detection methods are much more efficient in identifying stellar-mass BHs as their signature is clear and cannot be misinterpreted; 2) in comparison, radio and X-ray detectors tend to scan a much larger area of the sky and for GWs facilities it is even more the case. The study of these complementary detections (see the related sections of this Chapter) is very instructive, since it is expected that the local density of BHs in the field is made up of BHs randomly but almost equally distributed in these three different groups: 1) non-interacting BHs; 2) interacting BHs; 3) BH-BH binaries, hence there is real hope that future investigations may finally reveal the conspicuous number of these quiescent BHs that are now missing.
\\
\paragraph{\textbf{Quiescent Black Holes in high-density fields}}\label{b}
In a Universe predominantly dominated by low-density structures, high-density fields such as star clusters are very special environments, where specific mechanisms can occur, triggering interesting phenomena or the formation of exotic objects (such as cataclysmic variables, blue straggler stars or millisecond pulsars \cite{2005Sci...307..892R,2006ApJ...646L.143P,2009Natur.462.1028F}) that are not observed elsewhere. Star clusters are a class of cosmic objects made up of $10^{3} - 10^{6}$ gravitationally bound stars, which share nearly the same metallicity and age among them, since they are born during the same burst of star formation (at least as a first approximation, but see the review by \cite{2018ARA&A..56...83B} for a recent discussion about the multiple population phenomenon in star clusters).\\ %This picture has drastically changed in the last two decades after the discovery of the multiple population phenomenon \cite{2018ARA&A..56...83B} in star clusters. This phenomenon is not relevant for the purposes of this Chapter, so it can be ignored.

Due to the very high number of stars located in a very limited volume, star clusters are {\it collisional} systems, meaning that all stars during their lifetimes experience several interactions with other cluster members, both stellar or compact objects. These close interactions are the main engines of their dynamical evolution and may lead to changes in the properties of the clusters themselves with time but also trigger the formation and destruction of binary systems, which are present in significant numbers within them. In addition, there is observational evidence that more massive stars (e.g., main-sequence OB stars) are found more preferentially in binaries or higher order multiples than lighter ones (see \cite{2007ApJ...670..747K,2012Sci...337..444S,2014ApJS..215...15S}). Initially made up of many massive stars, star clusters are the ideal places to look at for finding stellar-mass BHs, both interacting and dormant. In this regard, massive star clusters have recently been placed under the magnifying glass, after the GW190521 event was detected \cite{2020PhRvL.125j1102A,2020ApJ...900L..13A}. This was the first intermediate mass BH discovered ($\sim$150 $M_{\odot}$), which formed from the merger of two massive BHs (60 $M_{\odot}$ and 85 $M_{\odot}$, respectively). Interestingly, these progenitor BHs might be themselves the product of a previous binary BH merger or stellar collision (as single star evolution struggles to produce such massive BHs; refer to Chapter 1, Section 1a, for an exhaustive discussion of the topic). Two more GW events of this type (GW190426 and GW200220) have been detected in recent years that involved very massive BH candidates, which merged to produce final BHs of more than 100 $M_{\odot}$, or even close to 200 $M_{\odot}$. BHs in star clusters have the real chance to pair up dynamically, thus high-density environments can actually be considered potential nurseries of many GW sources.\\

Until very recently, the innermost regions of star clusters, both in the Milky Way and in external galaxies, were completely inaccessible to perform spectroscopic studies, due to observational limitations, so until 2018 no dynamical detections of dormant stellar-mass BHs in clusters were claimed. Thanks to spectrographs equipped with IFUs, such as MUSE at the VLT (the best currently available in terms of field coverage and both spatial and spectral resolution), scientists have started to observe the innermost regions of clusters, trying to resolve and extract a spectra for every single source detected. The central regions are the most interesting for this type of science, as this is where massive and compact objects are most likely to be found, due to a mechanism at play in star clusters called mass segregation, i.e. more massive objects sink into the center while lighter sources tend to lie on the outskirts.\\

Giesers et al. \cite{2018MNRAS.475L..15G} used MUSE multi-epoch spectroscopic data of the old ($\sim$ 12 Gyr) Galactic globular cluster NGC 3201 spanning a time baseline of a few years to start a blind systematic survey of binaries and, among them, those with likely dark companions. Performing a blind search is very powerful for two main reasons: 1) we know that the highest chance to detect stellar-mass BHs is in the center compared to the outskirts, but we do not know exactly where they are in the cluster (at which direction / what distance from the center); 2) we are not biased toward detecting only a specific sample of objects, which always happens when handling targeted observations.\\

\begin{figure}
    \centering
	\includegraphics[width=0.8\textwidth]{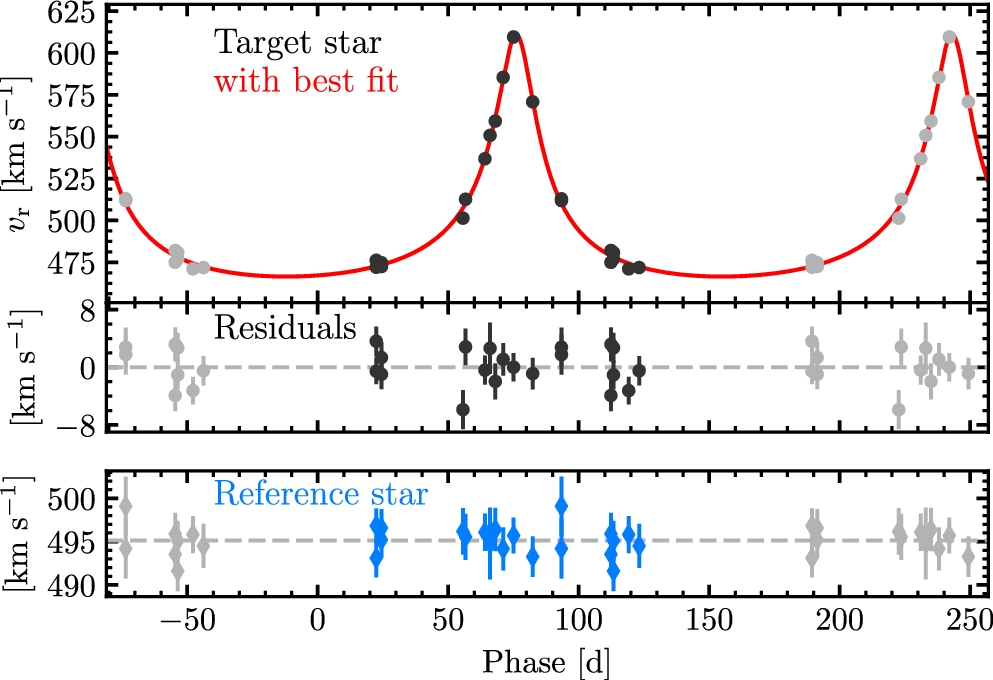}
    \caption{from Giesers et al. \cite{2018MNRAS.475L..15G}: The radial velocity curve of the first BH candidate detected in NGC 3201, from MUSE observations. The red solid line represents the best fit solution for the system, thanks to a comparison with a large library of possible orbits. The radial velocities of a nearby cluster member showing no signs of variability (i.e. a single star) are presented in blue for comparison.}
    \label{fig:giesers}
\end{figure}

Using this novel technique, they were able to detect a stellar-mass BH in the cluster, in an orbit rather eccentric (e$\sim$0.6) with a period of $\sim$167 d and a semi-amplitude velocity K$\sim$70km/s with a subgiant star companion. Since the luminous star was part of the cluster, they were able to exploit the advantage of knowing exactly the age, the metallicity, the distance and the extinction in the direction of the cluster, to infer the mass of the luminous star through a comparison with stellar evolutionary models (MIST, \cite{2016ApJ...823..102C}). They found a mass of $\sim$0.8 $M_{\odot}$ for the star, and thanks to the orbital properties of the binary system, they derived a minimum mass of 4.36 $M_{\odot}$ for the unseen companion, which can be confidently claimed to be a BH (see Figure \ref{fig:giesers}). After analyzing the entire sample of binaries in NGC 3201, they identified another BH candidate and a binary whose orbital solutions were compatible with a BH or NS companion. These solutions have been published along with the properties of several other stellar binaries in \cite{2019A&A...632A...3G}.\\

This was a very important result for two reasons: 1) it represented the first detection of non-interacting stellar-mass BHs in star clusters; 2) given the old age of NGC 3201, these detections were a further confirmation that BHs can actually reside in star clusters for a Hubble time or more. The latter is in contrast with the general understanding of star clusters in the early 1990s. In fact, it was believed that, due to interactions and dynamical processes, any BH formed within them would be expelled, leaving star clusters essentially devoid of these objects in a very short timescale (t $<$ 1 Gyr, \cite{1993Natur.364..421K}). On the other side, the results by Giesers and colleagues \cite{2018MNRAS.475L..15G,2019A&A...632A...3G} are in good agreement with what more advanced and sophisticated cluster models have actually predicted, i.e. star clusters are able to retain their BHs for longer times, even longer than a Hubble time \cite{2015ApJ...800....9M,2018ApJ...864...13W,2018MNRAS.478.1844A,2018MNRAS.479.4652A,2020ApJ...898..162W}.\\

Unlike in the field, uncovering the BH population of star clusters can actually help to answer some crucial long-standing questions about stellar astrophysics, such as: 1) how strong are natal kicks in clusters? 2) does a merger event always produce a natal kick? 3) how likely BHs are to be ejected from globular clusters through dynamical interactions? Answering these questions would be important to understand, for example, if globular clusters are the main factories of GWs, but also to investigate how likely it is that BHs with masses larger than 100 $M_{\odot}$ (like GW190521) are formed (via hierarchical merging of lighter stellar-mass BHs) and are held within globular clusters, or whether denser environments with deeper gravitational potential are required for this, such as nuclear star clusters \cite{2020A&ARv..28....4N}.\\

Until very recently, no other attempts have been made to reveal further dormant BHs in star clusters using the radial velocity method. Saracino et al. \cite{2022MNRAS.511.2914S} exploited the approach developed by Giesers and collaborators to look for any evidence of stellar-mass BHs in a massive ($\sim10^5$ $M_{\odot}$) cluster of $\sim$100 Myr located in the LMC. They chose a young cluster, where dynamical processes have not yet occurred, to study the initial distribution of BHs in clusters, which is still completely unknown.
In this work, the authors claimed to have detected a binary system consisting of a BH of $\sim$11 $M_{\odot}$ orbiting almost circularly a main-sequence B-type star with a period of 5.04 d and a semi-amplitude of K $\sim$ 140.5$\pm3$ km/s, which they named NGC1850 BH1. This finding was the result of a combined analysis of the radial velocity curve constructed from 17 multi-epoch observations with MUSE and the I and V light curves provided by OGLE-IV. Since the binary is a member of the cluster and its position in the CMD is consistent with a main sequence turn-off star, the authors derived a mass for the luminous object of $\sim$4.9 $M_{\odot}$ from the comparison with stellar isochrones of age, metallicity, extinction and distance of NGC 1850.\\

Right after the detection, an alternative explanation has been proposed for the binary system NGC1850 BH1 \cite{2022MNRAS.511L..24E,2022MNRAS.511L..77S}, which does not require the presence of a BH. Both groups suggested that this system is a post-mass transfer binary in a rare state of its evolution and it consists of a bloated-stripped star and a more massive (possibly rapidly rotating) B-type star. They argued that the stripped star is about 5 times less massive than what Saracino and collaborators assumed, but that its luminosity is still consistent with that of the observed star. Furthermore, they predicted that the massive B-type star would be expected to contribute 10 to 100\% to the total flux in the MUSE spectra and would be overlooked due to the rapid rotation of the star itself.\\
%The true nature of NGC1850 BH1 is still much debated and additional observations (e.g., high-resolution spectroscopy with a wide wavelength coverage) and modelling are required to provide a definitive answer about the nature of this system. 

After a re-analysis of the MUSE spectra, the original authors \cite{2023MNRAS.521.3162S} confirmed that the visible star has a lower mass compared to what initially thought but they also made the following findings: 1) a larger radial velocity semi-amplitude ($K_{\rm 2,revised} = 176\pm3$ km/s) than in the discovery paper, which translates in a mass function $f_{\rm revised}$=2.83 $M_{\odot}$ for the system; 2) the unseen source contributes less than 10\% to the total light in the visual; 3) an accretion disk seems to be present in NGC1850 BH1.
These new observational constraints indicate that, regardless of the mass of the visible star, the unseen companion must be rather massive. Its nature, however, is still unconstrained. This could either be a BH with an accretion disk or a massive luminous star enshrouded in an optically think disk that almost completely blocks its light. Further data and modelling are needed to provide a definitive answer as to whether or not NGC1850 BH1 contains a BH.\\

A recent addition to the list of dormant BHs detected in star clusters is NGC 2004-115 \cite{2022A&A...665A.180L}. It lies on the periphery of the young cluster NGC 2004, on the northern edge of the LMC. The authors used VLT/FLAMES and SALT-HRS spectroscopic data covering a baseline of about 20 years to shed light on the nature of the system: a triple system consisting of an inner binary with period P=2.92 d, semi-amplitude velocity K$\sim$62.4 km/s and mass function f(m)=0.07 $M_{\odot}$. By assuming the visible star is a B-type main sequence of 8.6 $M_{\odot}$, they inferred a mass for the invisible companion greater than 25 $M_{\odot}$. An upper limit on the X-ray luminosity coming from the binary has been set by XMM-Newton and it is consistent with having a quiescent BH in the system. The presence of a massive BH in NGC 2004-115 was later questioned by \cite{2022MNRAS.511.3089E} as such a high mass results from an inferred orbital inclination of only 9 degrees for the system, which is not consistent with the modelling of both the OGLE and the MACHO light curves of the system. A triple system in which the 25 $M_{\odot}$ BH is replaced by a 2-3 $M_{\odot}$ luminous companion, and with an outer Be star appears to be the only scenario capable of reproducing the observed light curves and radial velocity curves of the system simultaneously.\\

As part of a multi-epoch VLT/FLAMES spectroscopic campaign aimed at constraining the binary fraction of OB stars in high density environments, Banyard et al. \cite{2022A&A...658A..69B,2022arXiv221007149B} have recently provided an important contribution to the search for quiescent BHs in clusters. Indeed, they studied the population of B stars in the young (7 Myrs) open cluster NGC 6231 in the Milky Way, measuring an observed binary fraction of 33 $\pm$ 5\% (which increases to 52 $\pm$ 8\% if corrected for the observational biases) and characterised the multiplicity properties of the identified binary systems. By integrating medium-resolution optical spectroscopy, ground-based photometry and high-cadence space-based photometry, the authors found that 7 of the 15 SB1 systems in the cluster are indeed SB2 systems with mass ratios as low as q = 0.1. Of the 8 remaining systems, they ruled out 4 ambiguous systems and identified a sample of 4 binaries with a high probability of harboring a compact object. They calculated the minimum mass for the unseen companions, finding that two have masses between 1 and 3.5 $M_{\odot}$ and fall within the NS mass range, while the other two have mass ranges between 2.5 to 8 $M_{\odot}$ and 1.6 to 26 $M_{\odot}$, respectively, and so are possible BH candidates. The most massive unseen source is part of the CD-41 11038 system. However, as the authors themselves mention in the paper, the detection limits of their methodology do not exclude that some or all of these candidates have lower-mass main sequence companions, stripped star companions or are part of triple systems. Follow up photometric and/or interferometric observations are therefore needed to answer this question.\\

To recap, only the two (possibly three) BH candidates in NGC 3201 are currently recognized as robust non-interacting BH detections in star clusters within the scientific community, but NGC 6231 appears to be another promising source of compact objects, again underscoring the fact that crowded fields such as clusters are ideal places to look for these degenerate sources. The methods that Giesers and Banyard have developed together with their collaborators are extremely promising and have the potential to lead to the discovery of many more of these objects in other clusters if such a systematic search is performed.
The work by Giesers et al. is especially important in light of the new instruments that will be available to the community in the coming years, starting from HARMONI at the ELT in 2027 \cite{2010SPIE.7735E..2IT}. This IFU spectrograph will have a superb spatial resolution (up to 0.004 arcsec, compared to 0.2 arcsec of MUSE-WFM) and thanks to the extremely large collecting area will be able to observe and resolve the central regions of star clusters more distant than the Magellanic Clouds into individual stars, a science impossible to accomplish with current facilities. By overcoming the crowding issue in nearby star clusters, scientists will be able to look for non-interacting stellar-mass BHs in binaries where the luminous stars are much fainter on the main sequence than has been possible hitherto.\\

To conclude and improve the readability of Section \ref{dynamical} (a, b), we present in Table \ref{summary_BHs} a summary of all the systems discussed, where those hosting quiescent BHs are included, together with those subsequently questioned and proven to be BH imposters.
\begin{table}[]
\caption{A summary of all systems claimed to date to host a quiescent BH is presented below, including only those systems detected via spectroscopic investigation. Both low-density and high-density environments are included, and the systems are classified and catalogued based on the estimated mass of the luminous companions. For completeness, confirmed BHs and BH imposters are listed.}
\centering
\resizebox{\textwidth}{!}{
\begin{tabular}{l|l|l|l|l|l}
\hline
Name            & Z   & Envr.         & Proposed    & Alternative(s)         & References                                                   \\ \hline
{\it High - mass}   &     & &    &                     &                                                              \\
{\it companions}   &     & &    &                     &                                                              \\
LB-1            & MW  & Field                  & B + BH        & Be + stripped (str) & {\footnotesize Liu+\cite{2019Natur.575..618L}; Shenar+\cite{2020A&A...639L...6S}; El-Badry\cite{2020MNRAS.493L..22E}}\\               
MWC 656         & MW  & Field                  & Be + BH       & Be + sdO ?          & {\footnotesize Casares+\cite{2014Natur.505..378C}; Rivinius+\cite{2022arXiv220812315R}}                                  \\
NGC 1850 BH1    & LMC & Cluster                & B + BH        & B + str ?           & {\footnotesize Saracino+\cite{2022MNRAS.511.2914S,2023MNRAS.521.3162S}; El-Badry+\cite{2022MNRAS.511L..24E};} \\ 
&     & &    &                     & {\footnotesize Stevance+\cite{2022MNRAS.511L..77S}}           \\
HR 6819         & MW  & Field                  & (B+BH)+Be   & Be + str            & {\footnotesize Rivinius+\cite{2020A&A...637L...3R}; Bodensteiner+\cite{2020A&A...641A..43B};} \\
&     & &    &                     &  {\footnotesize El-Badry+\cite{2021MNRAS.502.3436E}; Frost+\cite{2022A&A...659L...3F}} \\
NGC 2004 \# 115 & LMC & Cluster                & (B+BH)+Be   & (B+A) + Be ?        & {\footnotesize Lennon+\cite{2022A&A...665A.180L}; El-Badry+\cite{2022MNRAS.511.3089E}}                                  \\
HD 96670        & MW  & Field                  & (O+BH) + ?    & - - -                   & {\footnotesize Gomez+\cite{2021ApJ...913...48G}}                                                   \\
AS 386          & MW  & Field                  & B{[}e{]} + BH & - - -                   & {\footnotesize Khokhlov+\cite{2018ApJ...856..158K}}                                                \\
VFTS 243        & LMC & Field                  & O + BH        & - - -       & {\footnotesize Shenar+\cite{2022NatAs...6.1085S}}                                                  \\
HD 130298       & MW  & Field                  & O + BH        & - - -       & {\footnotesize Mahy+\cite{2022A&A...664A.159M}}                                                    \\ \hline
{\it Low - mass}   &     & &    &                     &                                                              \\
{\it companions}   &     & &    &                     &                                                              \\
NGC3201 \#12560 & MW  & Cluster                & MS + BH       & - - -       & {\footnotesize Giesers+\cite{2018MNRAS.475L..15G}}                                                 \\
NGC3201 \#21859 & MW  & Cluster                & MS + BH       & - - -       & {\footnotesize Giesers+\cite{2019A&A...632A...3G}}                                                 \\
NGC3201 \#5132  & MW  & Cluster                & MS + BH/NS       & - - -             & {\footnotesize Giesers+\cite{2019A&A...632A...3G}}                                                 \\
Unicorn (V723 Mon)         & MW  & Field                  & RG + BH       & RG + str subgiant ? & {\footnotesize Jayasinghe+\cite{2021MNRAS.504.2577J}; Masuda+\cite{2021ApJ...910L..17M};} \\  
&     & &    &                     & {\footnotesize El-Badry+\cite{2022MNRAS.512.5620E}}                 \\
Giraffe (2M0412)       & MW  & Field                  & RG + BH  &  RG + str subgiant ?                   &      {\footnotesize Jayasinghe+\cite{2022MNRAS.516.5945J}; El-Badry+\cite{2022MNRAS.512.5620E}} \\ 
2MASS  & MW & Field & RG + BH/NS      & RG + (MS+MS) ? & {\footnotesize Thompson+\cite{2019Sci...366..637T}; Van den Heuvel+} \\
J05215658+4359220   &     & &    &     &  {\footnotesize
\cite{2020Sci...368.3282V}} \\ \hline
\end{tabular}}
\label{summary_BHs}
\end{table}

\subsubsection{Astrometric detection of quiescent Black Holes}
\label{astrometry}
As already explained in the previous Sections, the dynamical detection of a stellar-BH binary totally relies on the possibility to obtain multi-epoch spectroscopic data for the source of interest or for a very large sample of objects in order to conduct a blind search. When they are not available, alternative methods can instead be used to detect a quiescent BH orbiting around a luminous companion. Indeed, binary systems can also be detected “astrometrically”, by measuring the perturbations in the motion of the stars by their unseen companions. This technique has been used previously to identify “ordinary" binary systems and exoplanets, but can also be extended to discover any kind of binaries and it is particularly useful in identifying non-interacting compact binaries.
Briefly, {\it astrometric} binaries are systems made of two sources visible in astronomical observations as single sources. Three are the main reasons why the companion source is not observed: it is either too faint to be seen or too close to the primary to be resolved as two stars. Alternatively, it is a compact remnant (a BH or a NS) and cannot be observed due to its dark nature. What is observed in these cases is that the primary has a telltale wavy motion across the sky inconsistent with it being a single star. It instead wobbles about an unseen centre of mass, clearly indicating that another object exists in the system and gravitationally affects the motion of the bright, observed star. By analysing the motion of this star over time it is possible to place important constraints on the orbital properties of the system and on the masses involved.\\

\begin{figure}
    \centering
	\includegraphics[width=\textwidth]{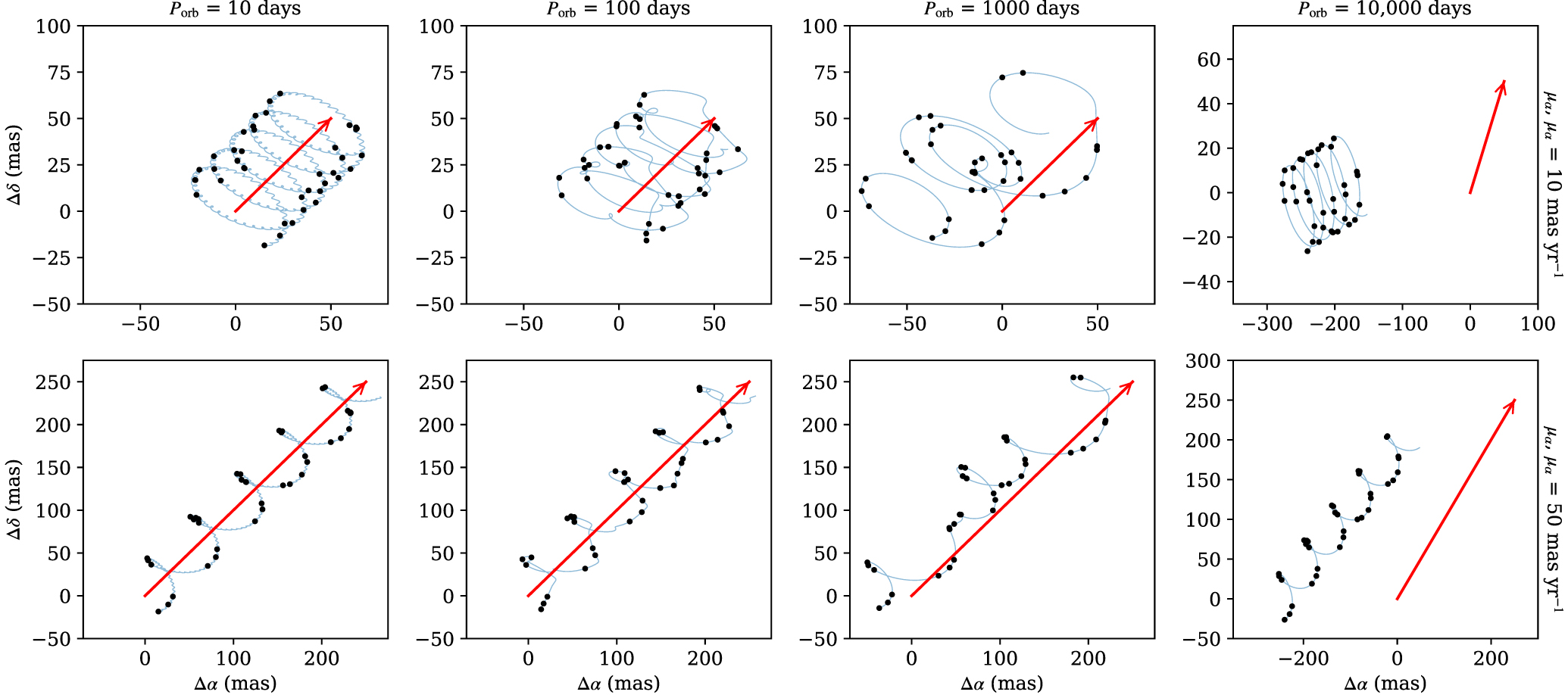}
    \caption{from Andrews et al. \cite{2019ApJ...886...68A}: The astrometric motion of a luminous star of 1 $M_{\odot}$ orbiting around an unseen companion of 1.4 $M_{\odot}$, for different orbital periods (from 10 d to 10000 d) and two different proper motions (10 mas/yr and 50 mas/yr). The binary is placed at a relatively close distance of 35 pc to magnify any astrometric effects and the black dots show the positions of the star over many GAIA epochs. For orbital periods much shorter than a year, the orbital motion and parallax can be disentangled. However, for P = 100-1000 d, the positions become extremely complex.}
    \label{fig:astrometric}
\end{figure}

As discussed in Section \ref{dynamical}, the dynamical detection of binaries through radial velocity measurements allows for a robust estimate of the mass function of the system, but unfortunately does not provide any constraints on its inclination. This limitation is effectively exceeded by astrometry, as it also allows measuring the binary’s orientation in space, including its inclination angle. This means that if the mass of the luminous component can be derived from independent analyses/data, this is sufficient to break the degeneracy between the different parameters and have a complete characterization of the system \cite{2019ApJ...886...68A}.
In order to detect astrometric binaries, it is necessary to measure the position on the sky of many objects and to trace their motion over time. Moreover, this has to be done with extremely high precision in order to pick up even very small displacements. In this respect, Figure \ref{fig:astrometric} shows the astrometric motion of a 1 $M_{\odot}$ visible star orbiting around an unseen companion of 1.4 $M_{\odot}$ for different orbital periods (increasing from left to right) and proper motion (increasing from top to bottom), from Andrews et al. \cite{2019ApJ...886...68A}. The astrometric effects are amplified as the binary is placed at a distance of only 35 pc.\\

The Hipparcos space astrometry mission led by the European Space Agency (ESA) was a pioneering project in this respect and it was able to pinpoint the positions of more than 100,000 stars with high precision, opening up the concrete possibility of detecting new binaries using this method. Several astrometric binaries have been detected thanks to this mission in the literature \cite{1997A&AS..122..571M,1998A&AS..133..149M,1998A&A...330..585M} but there have been no claims of binaries containing a non-interacting BH companion.
Now that Hipparcos has been replaced by the new ESA space astrometry mission GAIA, the ability to detect quiescent BHs by analyzing the ``wobble" in their proper motions over time is more real than ever (see \cite{2017ApJ...850L..13B} and \cite{2017MNRAS.470.2611M}). The GAIA satellite was launched at the end of 2013 with the aim of providing 3D motion, luminosity, temperature and composition of almost a billion stars in our Galaxy and beyond, and so far it has been a revolution in many astronomical fields (e.g. galactic archaeology).\\

Although the first GAIA astrometric orbital solutions for binary sources have been published only in its third release in June 2022, many scientists have started to analyze the data searching for a way to uncover astrometric BH binaries well before this date. As an example, \cite{2022A&A...658A.129J} focused on the number of OB + BH systems (i.e. BHs in binary systems with massive companions, $>$ 8 $M_{\odot}$) that GAIA would have been detected as binaries in the DR3 and DR4. They took as a reference the 2\textsuperscript{nd} Alma Luminous Star Catalogue (ALS II) containing a sample of massive OB-type stars in GAIA and estimated that 77\% of the OB + BH binaries in ALS II would be detected as binaries in DR3, of which 89\% would be unambiguously identifiable as OB + BH binaries. These systems would cover an uncharted parameter space of long-period binaries (10 d $<$ P $<$ 1000 d). They predicted that these numbers would increase to 85\% and 82\%, respectively, by the end of the mission, as they would also detect systems with both shorter and longer periods than those predicted in DR3. It is interesting to note that, after an accurate cleaning procedure performed by the GAIA team, most of these sources have been removed from the DR3 sample, hence we do not have any information on these sources yet. However, GAIA DR3 still represents the game changer in this field, as scientists from around the world have finally access to orbital solutions (periods, eccentricities, etc.) for a sample of over $10^5$ astrometric binaries \cite{2021A&A...649A...1G,2022arXiv220605726H,2022arXiv220605439H}. The analysis is still ongoing but this huge data set will grow and refine year by year, with the concrete possibility of starting to identify which of these astrometric binaries may actually contain massive invisible companions, i.e. compact objects such as NSs or BHs. \\

Andrews et al. \cite{2022arXiv220700680A} performed a first attempt in this direction: they analyzed the whole sample of GAIA astrometric binaries under the conservative assumption that the companions were dark and therefore the photocenters of the systems followed the observed stars. They assumed the luminous stars to be of 1 $M_{\odot}$ and selected only those systems having a companion more massive than 1.4 $M_{\odot}$, this to exclude any contamination from white dwarf-like objects. After removing binaries with suspiciously long orbital periods relative to the duration of the mission to date and applying further cuts, they ended up with a sample of 24 binaries with companion masses ranging from 1.35 to 2.7 $M_{\odot}$, compatible with both NS or low-mass BH companions. These systems show interesting characteristics, like periods of a few years and orbital velocities of the order of $\sim$20 km/s. For at least 8 of these objects a subsequent spectroscopic follow-up was performed and the inspection of their spectra has confirmed that there is no evidence for a second luminous component. \\

On the same line but using a different approach, Shahaf et al. \cite{2023MNRAS.518.2991S} also looked at the sample of GAIA astrometric binaries searching for possible compact object companions. Their triage technique, already proposed in \cite{2019MNRAS.487.5610S}, relies on a good knowledge of the mass-luminosity relation (for main-sequence stars only), in order to discard all the single stars that for one reason or another are contaminating the sample. They compiled a sample of 177 systems with highly-probable non-luminous massive companions, only 20\% of the total sample provided by GAIA. Among them, they identified 8 BH candidates with masses larger than 2.4 $M_{\odot}$. The most promising candidate among them was soon after the subject of two independent spectroscopic follow-ups which recently confirmed its dark and compact nature, estimating a mass for the BH of approximately 10 $M_{\odot}$ \cite{2023MNRAS.518.1057E,2022arXiv221005003C}. As in \cite{2023MNRAS.518.1057E}, we refer to the source as GAIA BH1.\\

\begin{figure}
    \centering
	\includegraphics[width=\textwidth]{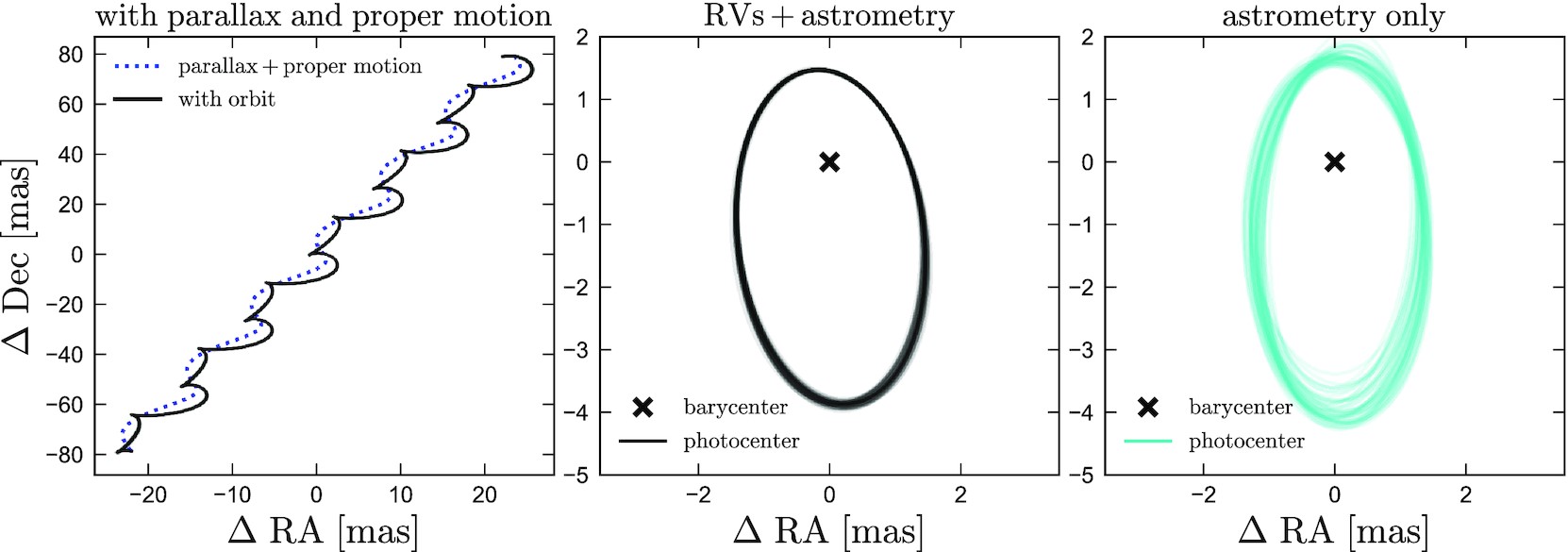}
    \caption{from El-Badry et al. \cite{2023MNRAS.518.1057E}: {\it Left:} The predicted astrometric orbit of the photocentre of GAIA BH1 on the sky over a period of 6 years. The black line indicates the total motion due to proper motion, parallax, and orbital motion while the dotted blue line shows the contribution due to parallax and proper motion alone. The predicted motion of the source, with parallax and proper motion removed is instead shown in the {\it middle panel} based on the joint astrometry and RV solution and in the {\it right panel} by using the pure astrometry solution. The joint analysis provides the most stringent constraints on the system.}
    \label{fig:gaiaBH}
\end{figure}

GAIA BH1 represents the first BH identified via its astrometric orbital solution from GAIA as well as the nearest BH ever detected (in the solar neighborhood, d = 474 pc). Although a slight discrepancy in the predicted mass of the BH between the two groups, they both confirm that it is in a binary system with a slowly-rotating G dwarf (T\textsubscript{eff} = 5850 K, log g = 4.5, and M = 0.93 $M_{\odot}$) and the system has an orbital period of 185.6 d. The predicted motion of GAIA BH1’s photocentre on the sky over a 6-yr window and its predicted orbit are shown in Figure \ref{fig:gaiaBH}, from \cite{2023MNRAS.518.1057E}. This detection is pretty robust, as consistent results have been obtained by using 1) GAIA astrometric data alone; 2) spectroscopic data alone (acquired as a follow-up program to the astrometric detection); or 3) a joint analysis of both measurements. The latter obviously provided the most stringent constraints on the properties of the system.\\

A very similar claim has been made almost at the same time from another group \cite{2023ApJ...946...79T}. They analyzed a sample of $\sim$65,000 binaries with both time-resolved spectroscopic and astrometric data in Gaia DR3 and identified a peculiar system made of a red giant star (Gaia DR3 5870569352746779008) and an unseen companion of a minimum mass $>$ 5.25 $M_{\odot}$. Although the mass of the visible star was rather uncertain, they argued that the unseen source is a BH candidate at a 99\% confidence level, based on the orbital parameters of the system. Their conclusions were entirely dependent on the reliability of the Gaia solution.\\

El-Badry and collaborators \cite{2023MNRAS.521.4323E} obtained additional spectra for this specific target, in the course of a broader program to spectroscopically follow-up Gaia DR3 binaries suspected to contain compact objects. These data confirmed the compact and dark nature of the invisible companion, which they called Gaia BH2. They were also able to better constrain the properties of the system itself. A bright (G = 12.3), nearby (d = 1.16 kpc) star on the lower red giant branch with mass M = 1.07 $\pm$ 0.19 $M_{\odot}$ orbiting a BH of mass $M_{BH}$=8.9 $\pm$ 0.3 $M_{\odot}$ with a moderate eccentricity e=0.52 and a relative low inclination angle i = 35\degree. Interestingly, this system has the longest period (1277 $\pm$ 1 days) among those discovered so far, slightly lower than what predicted by the Gaia solution (1352.25 $\pm$ 45.50 days) and reported by Tanikawa and colleagues \cite{2023ApJ...946...79T}.\\

\begin{figure}
    \centering
	\includegraphics[width=\textwidth]{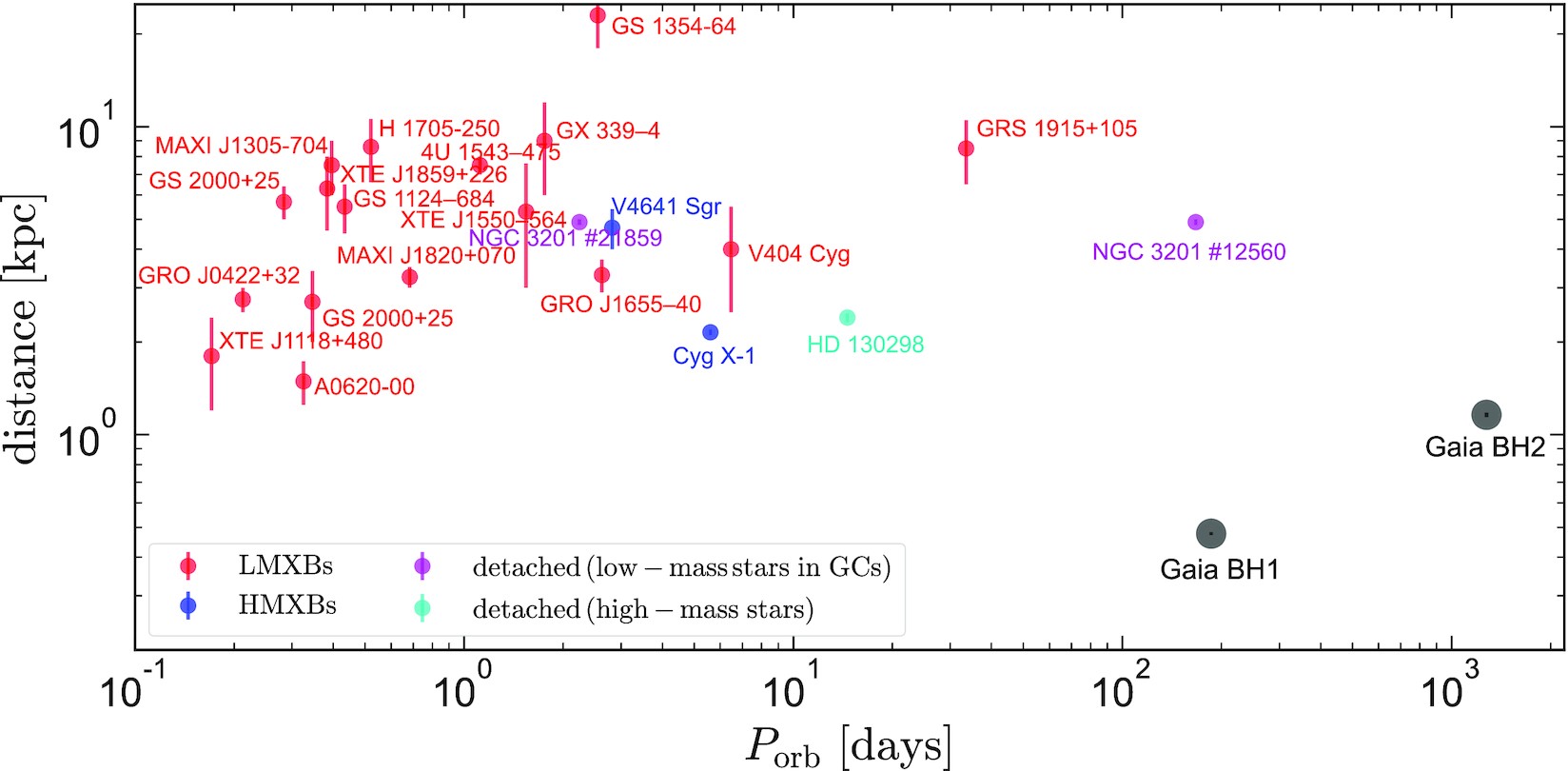}
    \caption{from El-Badry et al. \cite{2023MNRAS.521.4323E}: Gaia BH1 and BH2 (black points) are compared to Galactic BHs detected via accretion (in X-rays) in the reference plane of distance and orbital period. As indicated in the legend, the red and blue dots represent LMXBs and HMXBs respectively, while the cyan and purple dots identify detached BHs detected so far either in the field or in star clusters. As can be seen, the systems discovered by GAIA are in a part of this parameter space unexplored by other known BHs, i.e. with longer periods and closer to Earth than the rest of BHs.}
    \label{fig:gaiaBH2}
\end{figure}

Despite extensive research, to date GAIA BH1 and BH2 appear to be the only two quiescent BHs enclosed in the Gaia DR3 sample of binaries that can be detected. Figure \ref{fig:gaiaBH2} shows where the population of known accreting BHs detected via X-ray observations are located as a function of period and distance from us. When Gaia BH1 and Gaia BH2 are included (large black points) they occupy a very different parameter space based on their properties. They have significantly longer periods and closer to Earth than the rest. We do not have enough statistics to make any claims, but it appears that Gaia BH1 and BH2 may be the tip of the iceberg of the hitherto totally unknown binary population of wide-orbit black holes.\\

Gaia BH1 and Gaia BH2 unfortunately will not be progenitors of any GW signals as the luminous companion of the compact object is a low-mass star and there is no chance that the system may become a BH-NS binary at any point in time. However, the study of these systems remains extremely important as it allows us to build a more complete view of the actual population(s) of NSs and BHs existing in the Universe and if (at all) their properties change as a function of the environment.\\

The release of Gaia DR4 data in late 2025 as well as further investigations of GAIA DR3 will probably reveal many other interesting systems (e.g. massive white dwarfs and NSs). For what concerns BHs, based on recent theoretical predictions, Chawla et al. \cite{2022ApJ...931..107C}) argued that the extended Gaia mission can astrometrically resolve $\approx$ 30-300 detached BH-stellar binaries in the Milky Way, with a high probability of finding even some compatible with being the possible progenitors of GW sources.\\ 

While this is all very exciting, it is also important to mention that the observations obtained with GAIA are not custom tailored, so it is very likely that its next data releases will be affected by similar systematic uncertainties, resulting in a complex selection function \cite{2023arXiv230317738C}. This unfortunately makes a range of population studies difficult with GAIA. This limitation will surely be overcome by the GBTD Survey, which is designed to be performed with the future Roman Space Telescope. The huge grasp and direct imaging capabilities of this new facility will enable the desired astrometric accuracy to unravel binaries hosting compact objects, either BHs or NSs. There is no doubt that these are very exciting times to be an astronomer.

\section{Black Holes in binaries with compact objects} \label{BH+compact_objects}
With the advent of gravitational wave detections, we are now able to directly observe and study binary stellar mass black holes as well as black holes with neutron star companions and the remnants of their mergers. Follow-up electromagnetic observations allow us to further investigate the host environments, ejecta and merger remnants of these double compact object binaries. In this section we will discuss the so-called multi-messenger approach, i.e. combining traditional electromagnetic observations with gravitational wave and astrophysical particle observations, for detecting double compact object binaries. This includes binary black holes (BBHs), black hole neutron-star binaries (BH-NS) and black hole-white dwarf (BH-WD) binaries.

\subsection{Gravitational waves}

\subsubsection{What are gravitational waves?}
Gravitational waves (GWs) are distortions of space-time itself that move at the speed of light, carrying gravitational energy. The existence of gravitational waves was first proposed by Heaviside in 1893. After studying Maxwell’s equations for electromagnetic waves, Heaviside put forward that gravitational waves could propagate in a similar way \cite{Heaviside}. \\

In 1905, Poincare argued, using Lorentz transformations, that gravitation generates waves, similar to electromagnetism \cite{1906RCMP...21..129P}. His calculations showed that these waves of gravitation should move at the speed of light. \\

\begin{figure}
    \centering
    \includegraphics[width=0.7\textwidth]{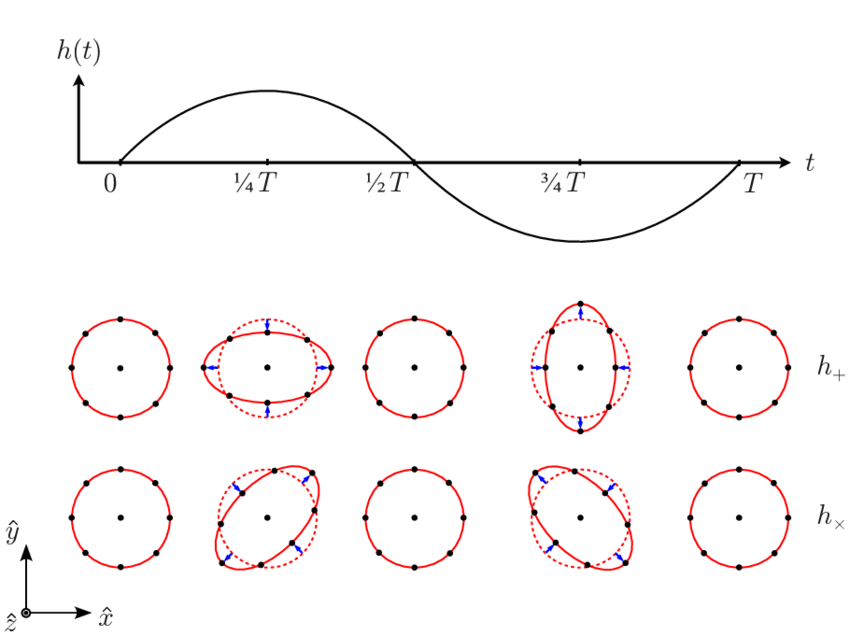}
    \caption{Taken from \cite{2017ogw..book....1L}. A gravitational wave of period $T$ and amplitude $h$, propagating along the $\hat z$ direction (top panel). The bottom panel shows the effects of the $+$ and $\times$ polarizations on a ring of free falling test particles, in an inertial frame.}
    \label{gravwaves}
\end{figure}

In 1916, Einstein published his theory of general relativity, which showed that accelerating massive objects cause space-time to be distorted and that these distortions would propagate outward, carrying gravitational energy \cite{1916SPAW.......688E, 1918SPAW.......154E}. Linearizing the field equations from the full theory of general relativity allowed Einstein to solve to them and show that these distortions, which Einstein called gravitational waves, are transverse planes waves that move at the speed of light. The energy carried by these waves is related to the third time derivative of the quadrupole moment of the objects creating the waves. Gravitational waves have two polarizations, $+$ and $\times$. The quadrupole nature of gravitational waves, with its two polarizations, is illustrated in Figure\,\ref{gravwaves}.\\

As the distance from the origin of the waves increases, the linear wave solutions coincide with those of the full theory. A full calculation of the general relativity field equations is beyond the scope of this work. For a full treatment see \cite{Christensen_2018} and references therein.\\

\subsubsection{Types of gravitational waves}
\begin{figure}
    \centering
        \centering
        \includegraphics[width=0.8\linewidth]{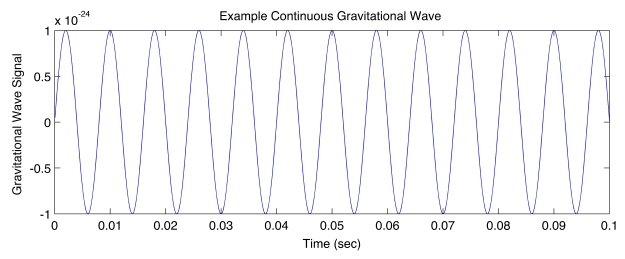}%
        \hfill
        \includegraphics[width=0.8\linewidth]{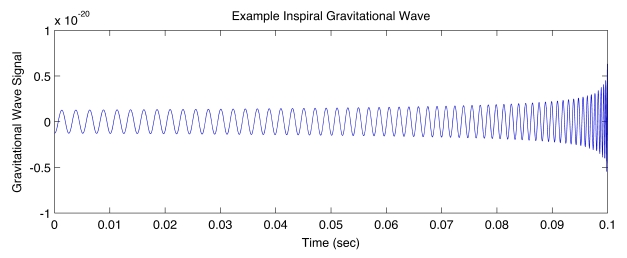}
         \hfill
        \includegraphics[width=0.8\linewidth]{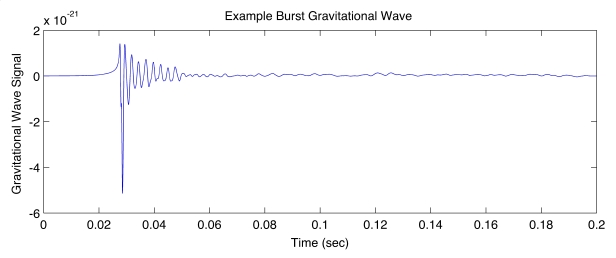}
         \hfill
        \includegraphics[width=0.8\linewidth]{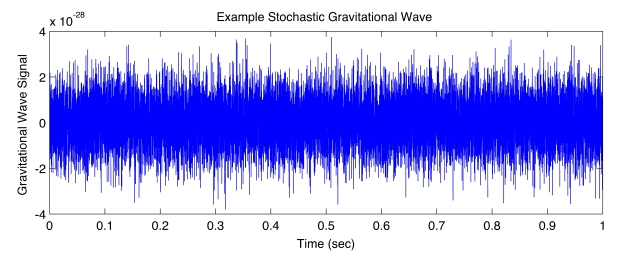}
    \caption{Examples of the four gravitational wave forms showing their shape and evolution in the time domain. Image credit: A. Stuver/LIGO.}
    \label{GW_forms}
\end{figure}

There are four main types of gravitational wave forms (see figure\,\ref{GW_forms} and also \url{https://www.ligo.org/science/GW-Sources.php}), depending on the motion and the mass distributions of the objects causing them, namely:
\begin{itemize}
    \item Continuous gravitational waves are long duration gravitational emissions with fairly stable frequencies produced by a single object with an asymmetrical mass distribution or a pair of objects in a wide, well defined orbit \citep[see e.g.][and references therein]{universe7120474}. In the astrophysical scenario, rapidly rotating neutron stars with surface deformations and binary compact objects in wide orbits, with negligible gravitational energy loss, are sources of continuous GWs.
    \item Inspiral gravitational waves are caused by binary systems with shrinking orbits \cite[see e.g.][]{2014LRR....17....2B}. As the binary components orbit each other, gravitational energy is lost over time, very slowing at first (in the form of continuous GWs). Interactions with matter surrounding the binary can also lead to increased loss of orbital angular momentum. Over time, this loss of orbital energy eventually causes the orbital distance to start to shrink. The orbital period increases and the orbital speeds increase, leading to an increase in the frequency of the emitted GWs. The frequency reaches a maximum at the moment when the binary components merge. In astrophysics, binaries consisting of compact objects are strong sources of inspiral GWs. 
    After the binary merges, the remnant object oscillates, resulting in a "ring down" GW component. This is the final part of the merger process and the GW is characterised by sharply decreasing amplitude and increasing frequency.
    \item Burst gravitational waves are short-duration ($\ll$1\,s), transient events of unknown origin. Potential astrophysical sources for bursts include certain neutron star oscillations or instabilities, core collapse of massive stars and sources related to gamma-ray bursts, but we anticipate many as-yet unknown sources of burst GWs as well \cite[see e.g.][and references therein]{Burst_GWs}.
    \item The stochastic (primordial) gravitational wave background is a result of the incoherent superposition of unresolved gravitational emission. In the astrophysical sense, this stochastic GW background is the result of physical processes in the early universe, such as inflation; analogous to the cosmic microwave background radiation \cite{Christensen_2018}.
\end{itemize} 

\subsubsection{Gravitational wave sources}
The frequency of GW emission depends on the nature of the massive objects producing the waves. Table\,\ref{GW_table} lists the GW frequency ranges and the types of astrophysical sources that emit GWs in that range \cite[see e.g.][]{1995pnac.conf..160T}. \\

\begin{table}
    \centering
    \resizebox{\textwidth}{!}{
    \begin{tabular}{|c|l|}
    \hline 
        Gravitational wave& Gravitational wave source  \\
        period (Hz) &  \\
         \hline
        1--10$^4$ & core collapse supernovae; Galactic neutron stars; extragalactic compact binaries (BBHs, BH-NS);  \\
         & sources of stochastic background GWs, i.e. early universe processes, like the big bang, \\
         & phase transitions in the early universe and vibrating cosmic loops and strings \\
        \hline        
       10$^{-4}$ --1 & Galactic compact binaries (e.g. binary WDs, BHBs, NS binaries); extragalactic compact objects  \\
       & spiralling into supermassive black holes (extreme mass ratio inspiral); inspiral and merging of \\
       &supermassive BHs, sources of stochastic background GWs, i.e. early universe processes\\
       \hline 
        10$^{-9}$ -- 10$^{-4}$  & Binary supermassive black holes;  sources of stochastic background GWs, i.e. early universe processes\\
        \hline
        $< 10^{-9}$ & sources of stochastic background GWs, i.e. early universe processes, like the big bang, \\
         & phase transitions in the early universe and vibrating cosmic loops and strings \\
        \hline
    \end{tabular}}
    \caption{Lists of astrophysical gravitational wave sources by gravitational wave frequency range. Adapted from \cite{1995pnac.conf..160T}. }
    \label{GW_table}
\end{table}

\subsection{Gravitational waves detectors}

\begin{figure}
    \centering
    \includegraphics[width=0.75\linewidth]{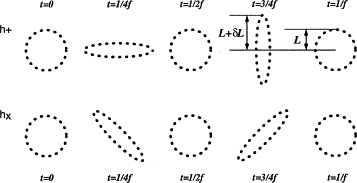}
    \caption{Taken from \cite{BRADY200333}. The relative distortion of an object of size $L$ and matter caused by a passing gravitational wave of frequency, $f$. The gravitational wave causes distortions of the object of size $\delta L$.}
    \label{strain}
\end{figure}

So, how do we detect gravitational waves? We do so by measuring the way a passing GW distorts space and matter. As illustrated in figure\,\ref{strain}, a passing GW causes an object of size $L$ to contract one direction and expand in the perpendicular direction and then spacetime expands in the first direction while contracting in the perpendicular direction. The relative size of these distortions ($\delta L/L$) of spacetime, called strain (h), is what gravitational wave detectors measure \cite[see e.g.][]{1995pnac.conf..160T, BRADY200333}. \\

All accelerating massive objects in the universe generate gravitational waves. However, only sufficiently massive objects, like BHs and NSs, produce detectable GWs. For example, a binary NS, each with mass 1.4\,$M_{\odot}$, at a distance of 15\,Mpc, in a circular orbit of radius 20\,km and orbital frequency of 1000\,Hz will produce a strain here on Earth of $h \sim 10^{-21}$ \cite{2016arXiv160500761C}. This means a detector on Earth of size 1\,km, would experience a distortion of $10^{-18}$\,m due to this this GW. For reference, the average size of an atomic nucleus is on the order of $10^{-17}$\,m. We therefore require extremely sensitive and innovative methods to detect these subtle ripples in spacetime.\\

\subsubsection{Indirect detection methods}\label{indirect}
Gravitational waves can be indirectly detected by observing their physical effects on astrophysical objects, like neutron stars or even other gravitational waves. The main types of indirect gravitational wave detection methods are:

\begin{itemize}
    \item Pulsar timing arrays. Pulsars are extremely accurate standard clocks. Passing gravitational waves will cause the time signatures of the pulsars to change as the distance between the observing telescopes and the pulsars themselves are changing. Pulsar timing arrays (PTAs) are a set of pulsars that are regularly monitored by ground based radio telescopes to track and analyzing their radio emission pulse arrival times. Pulsar timing arrays  are sensitive to GWs with nanohertz frequencies and below, i.e. GWs generated by binary supermassive BHs and stochastic GWs \cite[see e.g.][]{2021Symm...13.2418M, 2021hgwa.bookE...4V}. The main PTAs currently operating are the European Pulsar Timing Array (EPTA) \cite{2016MNRAS.455.1665B}, the Indian Pulsar Timing Array (InPTA) \cite{2018JApA...39...51J}, the North American Nanohertz Observatory for Gravitational Waves (NANOGrav) \cite{2021ApJS..252....5A}, the Parkes Pulsar Timing Array \cite{Kerr_2020}. These PTAs also work together to form the International Pulsar Timing Array (IPTA) \cite{Hobbs_2010,10.1093/mnras/stw347}
    \item Fast radio bursts (FRBs) can be used in a similar way to PTAs. For strongly lensed, repeating FRBs, their pulse arrival times can also be monitored, studied to infer the presence of GWs \cite{PhysRevD.103.063017}.
    \item Astrometric measurements of distant objects can identify the deflection of photons due to nanohertz frequency GWs \cite{1990NCimB.105.1141B,1996ApJ...465..566P,PhysRevD.83.024024}.
    \item The polarization of the cosmic microwave background (CMB) can give us insight into stochastic GWs from the beginning of the universe. These so-called primordial GWs, with frequencies in the $10^{-18}$\,Hz range, cause CMB photons to become polarized and this polarization is the focus of future CMB monitoring campaigns \cite{PhysRevLett.78.2058,PhysRevLett.78.2054}.
    \item Gravitational wave timing arrays will operate in a similar way to PTAs and FRBs to search for stochastic GWs. The continuous, high frequency Galactic GW background will experience phase modulations from the lower frequency stochastic waves and by studying these phase distortions, we can gain insight into the nature of the stochastic GWs.
 \cite{PhysRevD.105.044005}.
\end{itemize}

\subsubsection{Direct detection methods}\label{direct}

\paragraph{\textbf{Resonant mass detectors}}

The first experiments to search for direct evidence of gravitational waves started in the 1960s using resonant mass detectors were \cite{PhysRevLett.17.1228,PhysRevLett.18.498}. A resonant mass detector (or antenna) is a large ($\sim 1$\,m), solid body, usually metallic, that is isolated form surrounding vibrations. The strains from a passing GW will excite the body's resonant frequency and thereby be amplified to detectable levels. In the late 1960s and early 1970s, there were claims of detected gravitational wave emissions, but none of these results were ever replicated \cite{2004PhP.....6...42L}. \\

Second generation resonant mass antennas developed in the 1908s and 1990s used cryogenic methods to increase sensitivity \cite{Francesco_Ronga_2006,2011RAA....11....1A}. The third generation of detectors were spherical cryogenic antennas built in the 2000s \cite{2003CQGra..20S.143D,2002CQGra..19.1949A}. So far, no detections of gravitational waves have been made with resonant mass detectors.

\begin{figure}
    \centering
    \includegraphics[width=0.75\linewidth]{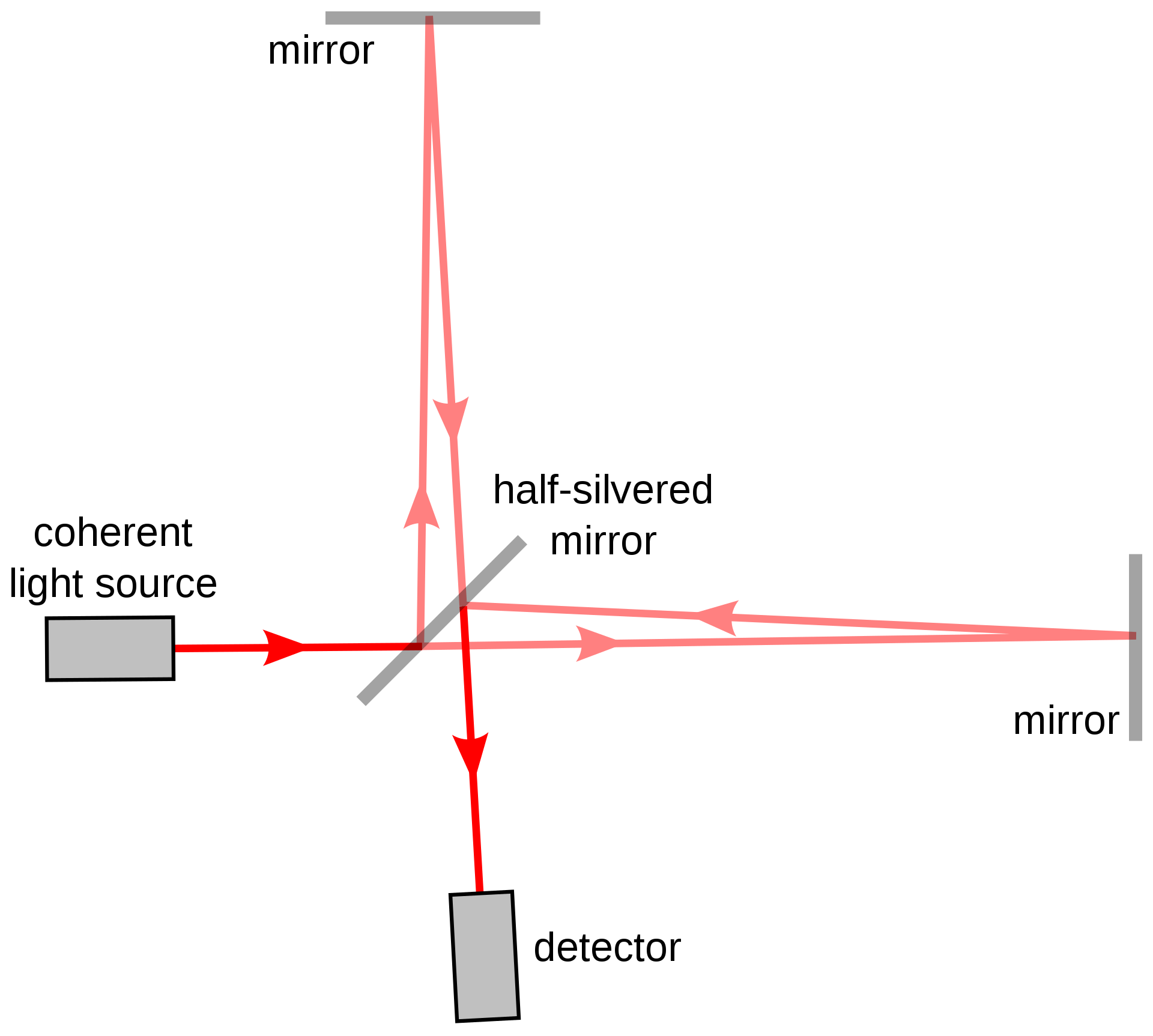}
    \caption{A schematic diagram of a Michelson interferometer. A laser beam is fired towards an angled mirror, which splits the beam into two orthogonal sub-beams. The sub-beams travel to mirrors situated at the end of each orthogonal arm, bounce back and then return to recombine in the detector instrument. Credit:Stannered, via Wikimedia Commons}
    \label{interferometer}
\end{figure}

\paragraph{\textbf{Laser interferometers}} 
The development of laser interferometers as gravitational wave detectors started in the 1960s \cite{Hill2017}.
An interferometer works in the following way (see figure\,\ref{interferometer}): a beam of coherent light (in the case of GW detectors, a laser) is fired towards an angled, partially silvered mirror that splits the light into two sub-beams. The sub-beams travel down two orthogonal arms of the interferometer that have mirrors at the end. The sub-beams bounce off the mirrors, travel back up the arms and recombine in a detector. \\

If the length of each arm is the same (i.e the light travel time is the same), the sub-beams will recombine constructively. However, if a gravitational wave has passed through the interferometer, the strain will change the arm lengths and the light travels will not be equal. The light will then not combine constructively and instead interference occurs. These inference patterns provides information on the GW that caused them.  \\

\begin{figure}
    \centering
    \includegraphics[width=0.9\linewidth]{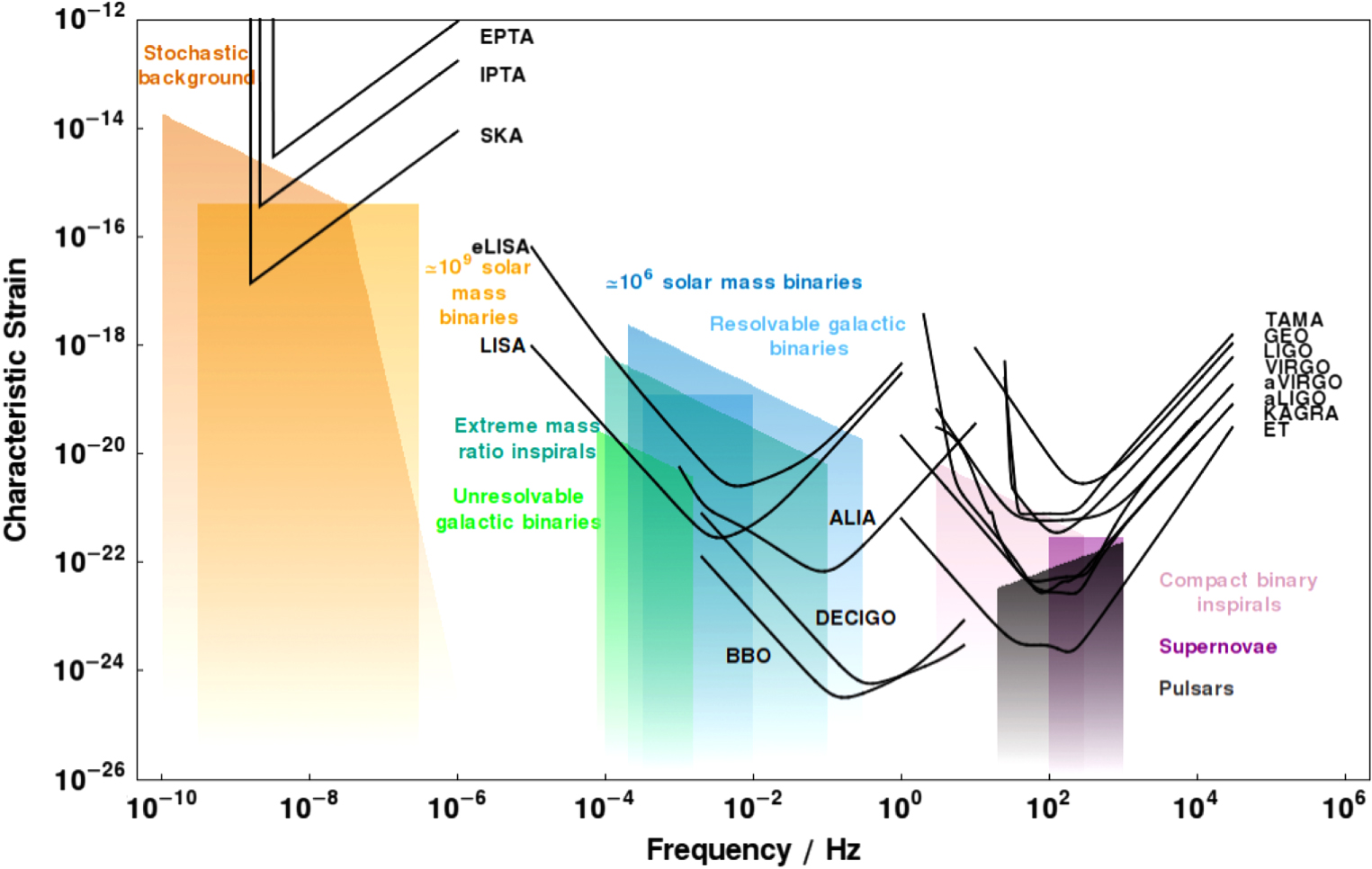}
    \caption{Taken from \cite{Moore_2015}. A comparison of past, present and future gravitational wave laser interferometer detectors, showing their respective strain sensitivities as a function of their frequency range.}
    \label{GW_laser_detectors}
\end{figure}

The earliest GW laser interferometer detectors detectors were operational in the 1990s and early 2000s. These first generation instruments were ground based detectors and served as valuable prototypes for many of the currently operating GW detectors. These include the Cryogenic Laser Interferometer Observatory \cite[CLIO;][]{Takashi_Uchiyama_2006,Yamamoto_2008}, GEO600 \cite{B_Willke_2002,S_Gossler_2002}, the Laser Interferometer Gravitational-Wave Observatory \cite[\LIGO;][]{1992Sci...256..325A}, TAMA\,300 \cite{2003CQGra..20S.593T} and Virgo \cite{CARON1996107}. Some of these are still under development and in use. Many of these instruments also carried out joint observations as part of their science and development strategies \cite{CARON1996107,Rejean_J_Dupuis_2006,2006JPhCS..39...36S,PhysRevD.76.042001,Abbott_2008,PhysRevD.81.102001,PhysRevD.89.122004}. Running joint observations increases sensitivity and improves sky localisation of the source. For a comparison of the frequent ranges and strain sensitivities of these instruments see \cite{Moore_2015} and figure\,\ref{GW_laser_detectors}.\\

Second generation laser interferometer GW detectors started operations or construction in the 2010s. Some were building on existing instruments by implementing more advanced technology such as cryogenics and quantum light squeezing to reduce noise \cite[see e.g.][]{2012mgm..conf..628H}. Thus GEO600 became GEO High Frequency \cite{2016CQGra..33g5009D}, \LIGO became advanced \LIGO \cite[aLIGO;][]{LIGO_2015} and Virgo became advanced Virgo \cite[aVirgo;][]{Acernese_2014}. \\

A new instrument, the Kamioka Gravitational Wave Detector (KAGRA) in Japan resides underground \cite{galaxies10030063}. The Indian Initiative in Gravitational-wave Observations \cite[IndIGO;][]{2013IJMPD..2241010U} is currently under construction. \\

The improvements made on the second generation of detectors led to the first direct detection of GWs in 2015 (see section \ref{direct_obs}). See \cite{Moore_2015} and figure\,\ref{GW_laser_detectors} for a comparison of of first and second generation instruments. \\

There are number third generation GW laser interferometer currently in the design study and development phases (see \cite{Moore_2015} and figure\,\ref{GW_laser_detectors} for a comparison of of first, second and third generation instruments). Most notably, Einstein Telescope (not be confused with the \Einstein X-ray Observatory!), a ground-based detector whose location is yet to be determined \cite[see e.g.][]{The_Einstein_Telescope}. Another proposed ground-based GW detector is the Cosmic Explorer \cite{reitze2019cosmic}. Both of these instruments are expected to start operations in the 2030s.\\

There are also several space-based GW detectors being planned. These include the DECi-hertz Interferometer Gravitational wave Observatory (DECIGO) \cite{10.1093_ptep_ptab019}, the  Laser Interferometer Space Antenna \cite[\LISA; ][]{Lisa_2017}, the Taiji Program in Space \cite{ruan2020taiji} and the TianQin Project \cite{2016CQGra..33c5010L}. All of these missions have projected launch dates in the 2030s. \\

\paragraph{\textbf{Other detector designs}} 

We remind the reader that gravitational wave detector design is a rapidly growing field and we can at this stage only offer a snapshot of current and future developments. There are several novel gravitational wave detector designs currently in development. \\

These innovative detectors propose to cover GW frequencies from $10^{-7}$\,Hz to up to tens of gigahertz and strain sensitivities down to $10^{-31}$. The detector designs include atom interferometers, torsion-bar antennas, coupled superconductors and microwave beam polarization measurements \cite[see e.g.][]{R_Ballantini_2003,Cruise_2005,Canuel_2018,Li-Baker_HFRGW_Detector}\\

\subsection{Gravitational wave detections}

\subsubsection{Indirect detections }\label{indirect_obs}
In 1974, the first binary pulsar system was discovered \cite{1975ApJ...195L..51H}. For many years after the discovery, the pulse arrival times of the binary were monitored. The data showed that the binary orbit was shrinking precisely as predicted by general relativity if gravitational emission was responsible for the loss of orbital energy \cite{1982ApJ...253..908T,1989ApJ...345..434T,2010ApJ...722.1030W}. This provided unambiguous, though indirect, proof of the existence of gravitational waves as predicted by general relativity. This work and initial discovery of the source won the 1993 Nobel Prize in Physics\footnote{\url{https://www.nobelprize.org/prizes/physics/1993/press-release/}}. \\

\subsubsection{Direct detections}\label{direct_obs}

\begin{figure}
    \centering 
   
        \includegraphics[width=0.8\linewidth]{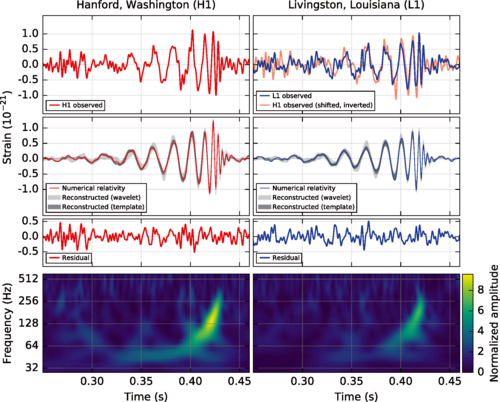}
    \caption{Taken from \cite{2016PhRvL.116f1102A}. The wave forms of the first BBH generated GW ever detected, GW150914. The top six panels show the strain as a function of time for GW150914, with the \LIGO Hanford arm on the right and the Livingstone arm's data on the left. The bottom two panels show the frequency and amplitude of the GW as a function of time.}
    \label{LIGO_first_detection}
\end{figure}

The first direct detection of gravitational waves occurred in September 2015, during the very first observing run of a\LIGO \cite{2016PhRvL.116f1102A}. The GWs originated from a coalescing binary BH system at a distance of $410^{+160}_{-180}$\,Mpc, with component masses of $36^{+5}_{-4}$\,M$_{\odot}$ and $29^{+4}_{-4}$\,M$_{\odot}$ and the merger remnant mass of $62^{+4}_{-4}$\,M$_{\odot}$. $3.0^{+0.5}_{-0.5}$\,M$_{\odot}$c$^2$ was radiated away in gravitational waves (see figures\,\ref{LIGO_first_detection}). The detected signal is called GW150914. This long awaited result and the \LIGO work that preceded it won the 2017 Nobel Prize in Physics\footnote{\url{https://www.nobelprize.org/prizes/physics/1993/press-release/}}. \\

\begin{figure}
    \centering   
        \includegraphics[width=0.8\linewidth]{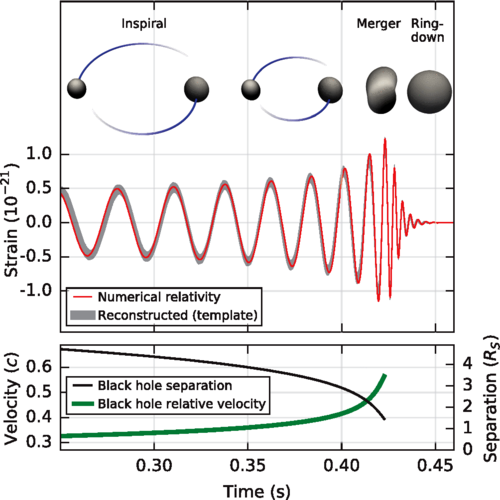}
    \caption{Taken from \cite{2016PhRvL.116f1102A}. The evolution of the binary components, GW strain, orbital velocity and orbital separation as a function of time for GW150914. }
    \label{LIGO_first_BBH}
\end{figure}

Subsequently, first a\LIGO and aVirgo, followed by KAGRA and GEO600, have operated their observing runs jointly, as shown in table\,\ref{LVK_observing}. \\

\begin{table}
    \centering
    \begin{tabular}{|c|l|l|}
        \hline
        Observing run  & Date & Detectors  \\
        \hline
       O1  & Sept 2015 -- Jan 2016 & a\LIGO  \\
        \hline
       O2  & Nov 2016 -- August 2017 & a\LIGO, and aVirgo for the final month  \\
        \hline
       O3  & April 2019 -- March 2020 & a\LIGO, aVirgo\\
        \hline
       O3GK  & April 2020 & GEO and KAGRA \\
        \hline
       O4 & May 2023 -- January 2025 & a\LIGO, aVirgo, KAGRA\\
        \hline
    \end{tabular}
    \caption{Table showing the dates of the first observing runs for current laser interferometer GW detectors. Taken from \url{https://dcc.ligo.org/public/0183/M2200155/006/LVK_citation.pdf} and \url{https://observing.docs.ligo.org/plan/}}.
    \label{LVK_observing}
\end{table}

\begin{figure}
    \centering
    \includegraphics[width=0.9\linewidth]{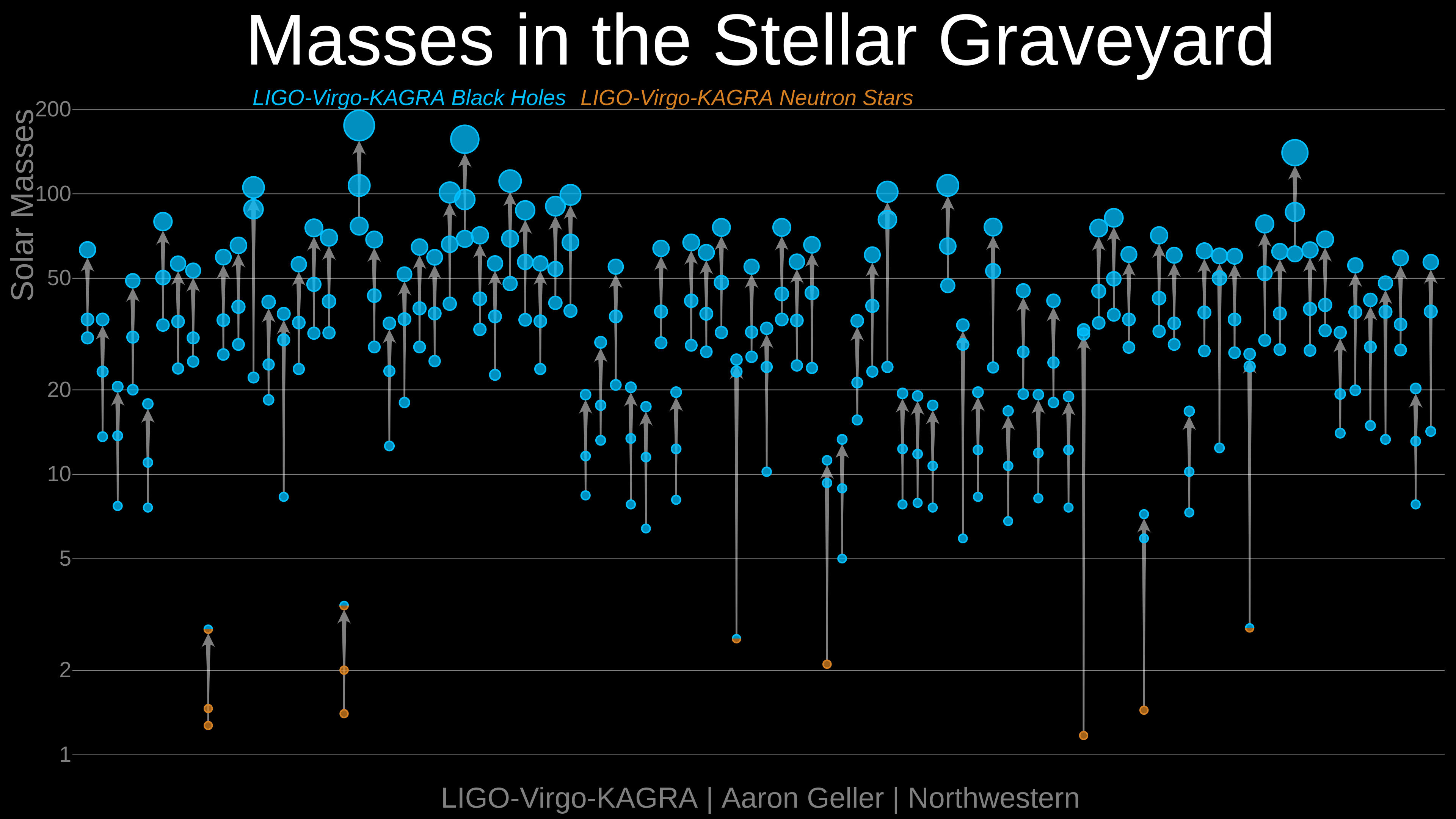}
    \caption{A plot showing the masses of all compact objects detected by LIGO/Virgo, with black holes in blue and neutron stars in orange, arranged in order of their discovery date. Credit: LIGO-Virgo-KAGRA/Aaron Geller/Northwestern.}
    \label{GWTC_detections}
\end{figure}

Over the course of the first two observing runs, 10 BBHs were detected \cite{2019ApJ...882L..24A}. The first two BH-NS binaries were detected in January 2020, during O2 \citep{Abbott:2021-first-NSBH}. 
Altogether, GWs from 90 compact object binaries have been detected, with most of them being BBHs \cite{2021PhRvX..11b1053A,2021arXiv211103606T}. Figure\,\ref{GWTC_detections} shows these detections in order of their discovery date.

\section{Acknowledgements}
\vs{-3mm}
Guglielmo Costa (GC) and Sara Saracino (SS) thank the editors of the book for giving them the opportunity to contribute to this chapter. 
SS acknowledges funding from the UK Science and Technology Facilities Council (STFC) under the grant no. R276234. GC acknowledges Alessandro Ballone, Thanh C. Nguyen, Guglielmo Volpato, Kendall G. Shepherd, Francesco Addari and the editors for reading the contribution and providing comments that helped to improve the manuscript. GC acknowledges support from the Agence Nationale de la Recherche grant POPSYCLE number ANR-19-CE31-0022. FSB acknowledges support for this work through the NASA FINESST scholarship 80NSSC22K1601 and from the Simons Foundation as part of the Simons Foundation Society of Fellows under award number 1141468. TDJ is grateful for financial support from the Nederlandse Organisatie voor Wetenschappelijk Onderzoek (NWO) through the Projectruimte and VIDI grants (Nissanke). JK acknowledges support from an ESO Fellowship.

%%%%%%%%%%%%%%%%%%%%%%%%%%%%%%%%%%%%%%%%%%%
%%%%%%%%%%%%%%%%%%%%%%%%%%%%%%%%%%%%%%%%%%%
%\interlinepenalty=10000 % to avoid splitting bibliography entries
\interlinepenalty=10000 % to avoid splitting bibliography entries
\section{Bibliography}
\bibliography{final}

%apsrev4-2.bst 2019-01-14 (MD) hand-edited version of apsrev4-1.bst
%Control: key (0)
%Control: author (8) initials jnrlst
%Control: editor formatted (1) identically to author
%Control: production of article title (0) allowed
%Control: page (0) single
%Control: year (1) truncated
%Control: production of eprint (0) enabled
\providecommand{\noopsort}[1]{}
\begin{thebibliography}{1389}%
\makeatletter
\providecommand \@ifxundefined [1]{%
 \@ifx{#1\undefined}
}%
\providecommand \@ifnum [1]{%
 \ifnum #1\expandafter \@firstoftwo
 \else \expandafter \@secondoftwo
 \fi
}%
\providecommand \@ifx [1]{%
 \ifx #1\expandafter \@firstoftwo
 \else \expandafter \@secondoftwo
 \fi
}%
\providecommand \natexlab [1]{#1}%
\providecommand \enquote  [1]{``#1''}%
\providecommand \bibnamefont  [1]{#1}%
\providecommand \bibfnamefont [1]{#1}%
\providecommand \citenamefont [1]{#1}%
\providecommand \href@noop [0]{\@secondoftwo}%
\providecommand \href [0]{\begingroup \@sanitize@url \@href}%
\providecommand \@href[1]{\@@startlink{#1}\@@href}%
\providecommand \@@href[1]{\endgroup#1\@@endlink}%
\providecommand \@sanitize@url [0]{\catcode `\\12\catcode `\$12\catcode
  `\&12\catcode `\#12\catcode `\^12\catcode `\_12\catcode `\%12\relax}%
\providecommand \@@startlink[1]{}%
\providecommand \@@endlink[0]{}%
\providecommand \url  [0]{\begingroup\@sanitize@url \@url }%
\providecommand \@url [1]{\endgroup\@href {#1}{\urlprefix }}%
\providecommand \urlprefix  [0]{URL }%
\providecommand \Eprint [0]{\href }%
\providecommand \doibase [0]{https://doi.org/}%
\providecommand \selectlanguage [0]{\@gobble}%
\providecommand \bibinfo  [0]{\@secondoftwo}%
\providecommand \bibfield  [0]{\@secondoftwo}%
\providecommand \translation [1]{[#1]}%
\providecommand \BibitemOpen [0]{}%
\providecommand \bibitemStop [0]{}%
\providecommand \bibitemNoStop [0]{.\EOS\space}%
\providecommand \EOS [0]{\spacefactor3000\relax}%
\providecommand \BibitemShut  [1]{\csname bibitem#1\endcsname}%
\let\auto@bib@innerbib\@empty
%</preamble>
\bibitem [{\citenamefont {{Schwarzschild}}(1916)}]{1916AbhKP1916..189S}%
  \BibitemOpen
  \bibfield  {author} {\bibinfo {author} {\bibfnamefont {K.}~\bibnamefont
  {{Schwarzschild}}},\ }\bibfield  {title} {\bibinfo {title} {{On the
  Gravitational Field of a Mass Point According to Einstein's Theory}},\
  }\href@noop {} {\bibfield  {journal} {\bibinfo  {journal} {Abh. Konigl.
  Preuss. Akad. Wissenschaften Jahre 1906,92, Berlin,1907}\ }\textbf {\bibinfo
  {volume} {1916}},\ \bibinfo {pages} {189} (\bibinfo {year}
  {1916})}\BibitemShut {NoStop}%
\bibitem [{\citenamefont {{Abbott}}\ \emph
  {et~al.}(2016{\natexlab{a}})\citenamefont {{Abbott}}, \citenamefont
  {{Abbott}}, \citenamefont {{Abbott}}, \citenamefont {{Abernathy}},
  \citenamefont {{Acernese}}, \citenamefont {{Ackley}}, \citenamefont
  {{Adams}}, \citenamefont {{Adams}}, \citenamefont {{Addesso}}, \citenamefont
  {{Adhikari}}, \citenamefont {{Adya}}, \citenamefont {{Affeldt}},
  \citenamefont {{Agathos}}, \citenamefont {{Agatsuma}}, \citenamefont
  {{Aggarwal}}, \citenamefont {{Aguiar}}, \citenamefont {{Aiello}},
  \citenamefont {{Ain}}, \citenamefont {{LIGO Scientific Collaboration}},\ and\
  \citenamefont {{Virgo Collaboration}}}]{2016PhRvL.116f1102A}%
  \BibitemOpen
  \bibfield  {author} {\bibinfo {author} {\bibfnamefont {B.~P.}\ \bibnamefont
  {{Abbott}}}, \bibinfo {author} {\bibfnamefont {R.}~\bibnamefont {{Abbott}}},
  \bibinfo {author} {\bibfnamefont {T.~D.}\ \bibnamefont {{Abbott}}}, \bibinfo
  {author} {\bibfnamefont {M.~R.}\ \bibnamefont {{Abernathy}}}, \bibinfo
  {author} {\bibfnamefont {F.}~\bibnamefont {{Acernese}}}, \bibinfo {author}
  {\bibfnamefont {K.}~\bibnamefont {{Ackley}}}, \bibinfo {author}
  {\bibfnamefont {C.}~\bibnamefont {{Adams}}}, \bibinfo {author} {\bibfnamefont
  {T.}~\bibnamefont {{Adams}}}, \bibinfo {author} {\bibfnamefont
  {P.}~\bibnamefont {{Addesso}}}, \bibinfo {author} {\bibfnamefont {R.~X.}\
  \bibnamefont {{Adhikari}}}, \bibinfo {author} {\bibfnamefont {V.~B.}\
  \bibnamefont {{Adya}}}, \bibinfo {author} {\bibfnamefont {C.}~\bibnamefont
  {{Affeldt}}}, \bibinfo {author} {\bibfnamefont {M.}~\bibnamefont
  {{Agathos}}}, \bibinfo {author} {\bibfnamefont {K.}~\bibnamefont
  {{Agatsuma}}}, \bibinfo {author} {\bibfnamefont {N.}~\bibnamefont
  {{Aggarwal}}}, \bibinfo {author} {\bibfnamefont {O.~D.}\ \bibnamefont
  {{Aguiar}}}, \bibinfo {author} {\bibfnamefont {L.}~\bibnamefont {{Aiello}}},
  \bibinfo {author} {\bibnamefont {{Ain}}}, \bibinfo {author} {\bibnamefont
  {{LIGO Scientific Collaboration}}},\ and\ \bibinfo {author} {\bibnamefont
  {{Virgo Collaboration}}},\ }\bibfield  {title} {\bibinfo {title}
  {{Observation of Gravitational Waves from a Binary Black Hole Merger}},\
  }\href {https://doi.org/10.1103/PhysRevLett.116.061102} {\bibfield  {journal}
  {\bibinfo  {journal} {\prl}\ }\textbf {\bibinfo {volume} {116}},\ \bibinfo
  {eid} {061102} (\bibinfo {year} {2016}{\natexlab{a}})},\ \Eprint
  {https://arxiv.org/abs/1602.03837} {arXiv:1602.03837 [gr-qc]} \BibitemShut
  {NoStop}%
\bibitem [{\citenamefont {{Abbott}}\ \emph
  {et~al.}(2016{\natexlab{b}})\citenamefont {{Abbott}}, \citenamefont
  {{Abbott}}, \citenamefont {{Abbott}}, \citenamefont {{Abernathy}},
  \citenamefont {{Acernese}}, \citenamefont {{Ackley}}, \citenamefont
  {{Adams}}, \citenamefont {{Adams}}, \citenamefont {{Addesso}}, \citenamefont
  {{Adhikari}}, \citenamefont {{Adya}}, \citenamefont {{Affeldt}},
  \citenamefont {{Agathos}}, \citenamefont {{Agatsuma}}, \citenamefont
  {{Aggarwal}}, \citenamefont {{Aguiar}}, \citenamefont {{Aiello}},
  \citenamefont {{Ain}}, \citenamefont {{Ajith}}, \citenamefont {{Allen}},
  \citenamefont {{Allocca}}, \citenamefont {{Altin}}, \citenamefont
  {{Anderson}}, \citenamefont {{Anderson}}, \citenamefont {{Arai}},
  \citenamefont {{Araya}}, \citenamefont {{Arceneaux}}, \citenamefont
  {{Areeda}}, \citenamefont {{Arnaud}}, \citenamefont {{Arun}}, \citenamefont
  {{Ascenzi}}, \citenamefont {{Ashton}}, \citenamefont {{Ast}}, \citenamefont
  {{Aston}}, \citenamefont {{Astone}}, \citenamefont {{Aufmuth}}, \citenamefont
  {{Aulbert}}, \citenamefont {{LIGO Scientific Collaboration}},\ and\
  \citenamefont {{Virgo Collaboration}}}]{2016ApJ...818L..22A}%
  \BibitemOpen
  \bibfield  {author} {\bibinfo {author} {\bibfnamefont {B.~P.}\ \bibnamefont
  {{Abbott}}}, \bibinfo {author} {\bibfnamefont {R.}~\bibnamefont {{Abbott}}},
  \bibinfo {author} {\bibfnamefont {T.~D.}\ \bibnamefont {{Abbott}}}, \bibinfo
  {author} {\bibfnamefont {M.~R.}\ \bibnamefont {{Abernathy}}}, \bibinfo
  {author} {\bibfnamefont {F.}~\bibnamefont {{Acernese}}}, \bibinfo {author}
  {\bibfnamefont {K.}~\bibnamefont {{Ackley}}}, \bibinfo {author}
  {\bibfnamefont {C.}~\bibnamefont {{Adams}}}, \bibinfo {author} {\bibfnamefont
  {T.}~\bibnamefont {{Adams}}}, \bibinfo {author} {\bibfnamefont
  {P.}~\bibnamefont {{Addesso}}}, \bibinfo {author} {\bibfnamefont {R.~X.}\
  \bibnamefont {{Adhikari}}}, \bibinfo {author} {\bibfnamefont {V.~B.}\
  \bibnamefont {{Adya}}}, \bibinfo {author} {\bibfnamefont {C.}~\bibnamefont
  {{Affeldt}}}, \bibinfo {author} {\bibfnamefont {M.}~\bibnamefont
  {{Agathos}}}, \bibinfo {author} {\bibfnamefont {K.}~\bibnamefont
  {{Agatsuma}}}, \bibinfo {author} {\bibfnamefont {N.}~\bibnamefont
  {{Aggarwal}}}, \bibinfo {author} {\bibfnamefont {O.~D.}\ \bibnamefont
  {{Aguiar}}}, \bibinfo {author} {\bibfnamefont {L.}~\bibnamefont {{Aiello}}},
  \bibinfo {author} {\bibfnamefont {A.}~\bibnamefont {{Ain}}}, \bibinfo
  {author} {\bibfnamefont {P.}~\bibnamefont {{Ajith}}}, \bibinfo {author}
  {\bibfnamefont {B.}~\bibnamefont {{Allen}}}, \bibinfo {author} {\bibfnamefont
  {A.}~\bibnamefont {{Allocca}}}, \bibinfo {author} {\bibfnamefont {P.~A.}\
  \bibnamefont {{Altin}}}, \bibinfo {author} {\bibfnamefont {S.~B.}\
  \bibnamefont {{Anderson}}}, \bibinfo {author} {\bibfnamefont {W.~G.}\
  \bibnamefont {{Anderson}}}, \bibinfo {author} {\bibfnamefont
  {K.}~\bibnamefont {{Arai}}}, \bibinfo {author} {\bibfnamefont {M.~C.}\
  \bibnamefont {{Araya}}}, \bibinfo {author} {\bibfnamefont {C.~C.}\
  \bibnamefont {{Arceneaux}}}, \bibinfo {author} {\bibfnamefont {J.~S.}\
  \bibnamefont {{Areeda}}}, \bibinfo {author} {\bibfnamefont {N.}~\bibnamefont
  {{Arnaud}}}, \bibinfo {author} {\bibfnamefont {K.~G.}\ \bibnamefont
  {{Arun}}}, \bibinfo {author} {\bibfnamefont {S.}~\bibnamefont {{Ascenzi}}},
  \bibinfo {author} {\bibfnamefont {G.}~\bibnamefont {{Ashton}}}, \bibinfo
  {author} {\bibfnamefont {M.}~\bibnamefont {{Ast}}}, \bibinfo {author}
  {\bibfnamefont {S.~M.}\ \bibnamefont {{Aston}}}, \bibinfo {author}
  {\bibfnamefont {P.}~\bibnamefont {{Astone}}}, \bibinfo {author}
  {\bibfnamefont {P.}~\bibnamefont {{Aufmuth}}}, \bibinfo {author}
  {\bibfnamefont {J.}~\bibnamefont {{Aulbert}}}, \bibinfo {author}
  {\bibnamefont {{LIGO Scientific Collaboration}}},\ and\ \bibinfo {author}
  {\bibnamefont {{Virgo Collaboration}}},\ }\bibfield  {title} {\bibinfo
  {title} {{Astrophysical Implications of the Binary Black-hole Merger
  GW150914}},\ }\href {https://doi.org/10.3847/2041-8205/818/2/L22} {\bibfield
  {journal} {\bibinfo  {journal} {\apjl}\ }\textbf {\bibinfo {volume} {818}},\
  \bibinfo {eid} {L22} (\bibinfo {year} {2016}{\natexlab{b}})},\ \Eprint
  {https://arxiv.org/abs/1602.03846} {arXiv:1602.03846 [astro-ph.HE]}
  \BibitemShut {NoStop}%
\bibitem [{\citenamefont {{Abbott}}\ \emph
  {et~al.}(2021{\natexlab{a}})\citenamefont {{Abbott}}, \citenamefont
  {{Abbott}}, \citenamefont {{Abraham}}, \citenamefont {{Acernese}},
  \citenamefont {{Ackley}}, \citenamefont {{Adams}}, \citenamefont {{Adams}},
  \citenamefont {{Adhikari}}, \citenamefont {{Adya}}, \citenamefont
  {{Affeldt}}, \citenamefont {{Agathos}}, \citenamefont {{LIGO Scientific
  Collaboration}},\ and\ \citenamefont {{Virgo
  Collaboration}}}]{2021PhRvX..11b1053A}%
  \BibitemOpen
  \bibfield  {author} {\bibinfo {author} {\bibfnamefont {R.}~\bibnamefont
  {{Abbott}}}, \bibinfo {author} {\bibfnamefont {T.~D.}\ \bibnamefont
  {{Abbott}}}, \bibinfo {author} {\bibfnamefont {S.}~\bibnamefont {{Abraham}}},
  \bibinfo {author} {\bibfnamefont {F.}~\bibnamefont {{Acernese}}}, \bibinfo
  {author} {\bibfnamefont {K.}~\bibnamefont {{Ackley}}}, \bibinfo {author}
  {\bibfnamefont {A.}~\bibnamefont {{Adams}}}, \bibinfo {author} {\bibfnamefont
  {C.}~\bibnamefont {{Adams}}}, \bibinfo {author} {\bibfnamefont {R.~X.}\
  \bibnamefont {{Adhikari}}}, \bibinfo {author} {\bibfnamefont {V.~B.}\
  \bibnamefont {{Adya}}}, \bibinfo {author} {\bibfnamefont {C.}~\bibnamefont
  {{Affeldt}}}, \bibinfo {author} {\bibfnamefont {M.}~\bibnamefont
  {{Agathos}}}, \bibinfo {author} {\bibnamefont {{LIGO Scientific
  Collaboration}}},\ and\ \bibinfo {author} {\bibnamefont {{Virgo
  Collaboration}}},\ }\bibfield  {title} {\bibinfo {title} {{GWTC-2: Compact
  Binary Coalescences Observed by LIGO and Virgo during the First Half of the
  Third Observing Run}},\ }\href {https://doi.org/10.1103/PhysRevX.11.021053}
  {\bibfield  {journal} {\bibinfo  {journal} {Physical Review X}\ }\textbf
  {\bibinfo {volume} {11}},\ \bibinfo {eid} {021053} (\bibinfo {year}
  {2021}{\natexlab{a}})},\ \Eprint {https://arxiv.org/abs/2010.14527}
  {arXiv:2010.14527 [gr-qc]} \BibitemShut {NoStop}%
\bibitem [{\citenamefont {{The LIGO Scientific Collaboration}}\ \emph
  {et~al.}(2021)\citenamefont {{The LIGO Scientific Collaboration}},
  \citenamefont {{the Virgo Collaboration}}, \citenamefont {{the KAGRA
  Collaboration}}, \citenamefont {{Abbott}}, \citenamefont {{Abbott}},
  \citenamefont {{Acernese}}, \citenamefont {{Ackley}}, \citenamefont
  {{Adams}}, \citenamefont {{Adhikari}}, \citenamefont {{Adhikari}},
  \citenamefont {{Adya}}, \citenamefont {{Affeldt}}, \citenamefont
  {{Agarwal}},\ and\ \citenamefont {{Agathos}}}]{2021arXiv211103606T}%
  \BibitemOpen
  \bibfield  {author} {\bibinfo {author} {\bibnamefont {{The LIGO Scientific
  Collaboration}}}, \bibinfo {author} {\bibnamefont {{the Virgo
  Collaboration}}}, \bibinfo {author} {\bibnamefont {{the KAGRA
  Collaboration}}}, \bibinfo {author} {\bibfnamefont {R.}~\bibnamefont
  {{Abbott}}}, \bibinfo {author} {\bibfnamefont {T.~D.}\ \bibnamefont
  {{Abbott}}}, \bibinfo {author} {\bibfnamefont {F.}~\bibnamefont
  {{Acernese}}}, \bibinfo {author} {\bibfnamefont {K.}~\bibnamefont
  {{Ackley}}}, \bibinfo {author} {\bibfnamefont {C.}~\bibnamefont {{Adams}}},
  \bibinfo {author} {\bibfnamefont {N.}~\bibnamefont {{Adhikari}}}, \bibinfo
  {author} {\bibfnamefont {R.~X.}\ \bibnamefont {{Adhikari}}}, \bibinfo
  {author} {\bibfnamefont {V.~B.}\ \bibnamefont {{Adya}}}, \bibinfo {author}
  {\bibfnamefont {C.}~\bibnamefont {{Affeldt}}}, \bibinfo {author}
  {\bibfnamefont {D.}~\bibnamefont {{Agarwal}}},\ and\ \bibinfo {author}
  {\bibfnamefont {M.}~\bibnamefont {{Agathos}}},\ }\bibfield  {title} {\bibinfo
  {title} {{GWTC-3: Compact Binary Coalescences Observed by LIGO and Virgo
  During the Second Part of the Third Observing Run}},\ }\href
  {https://doi.org/10.48550/arXiv.2111.03606} {\bibfield  {journal} {\bibinfo
  {journal} {arXiv e-prints}\ ,\ \bibinfo {eid} {arXiv:2111.03606}} (\bibinfo
  {year} {2021})},\ \Eprint {https://arxiv.org/abs/2111.03606}
  {arXiv:2111.03606 [gr-qc]} \BibitemShut {NoStop}%
\bibitem [{\citenamefont {{Oppenheimer}}\ and\ \citenamefont
  {{Snyder}}(1939)}]{1939PhRv...56..455O}%
  \BibitemOpen
  \bibfield  {author} {\bibinfo {author} {\bibfnamefont {J.~R.}\ \bibnamefont
  {{Oppenheimer}}}\ and\ \bibinfo {author} {\bibfnamefont {H.}~\bibnamefont
  {{Snyder}}},\ }\bibfield  {title} {\bibinfo {title} {{On Continued
  Gravitational Contraction}},\ }\href {https://doi.org/10.1103/PhysRev.56.455}
  {\bibfield  {journal} {\bibinfo  {journal} {Physical Review}\ }\textbf
  {\bibinfo {volume} {56}},\ \bibinfo {pages} {455} (\bibinfo {year}
  {1939})}\BibitemShut {NoStop}%
\bibitem [{\citenamefont {{Oppenheimer}}\ and\ \citenamefont
  {{Volkoff}}(1939)}]{1939PhRv...55..374O}%
  \BibitemOpen
  \bibfield  {author} {\bibinfo {author} {\bibfnamefont {J.~R.}\ \bibnamefont
  {{Oppenheimer}}}\ and\ \bibinfo {author} {\bibfnamefont {G.~M.}\ \bibnamefont
  {{Volkoff}}},\ }\bibfield  {title} {\bibinfo {title} {{On Massive Neutron
  Cores}},\ }\href {https://doi.org/10.1103/PhysRev.55.374} {\bibfield
  {journal} {\bibinfo  {journal} {Physical Review}\ }\textbf {\bibinfo {volume}
  {55}},\ \bibinfo {pages} {374} (\bibinfo {year} {1939})}\BibitemShut
  {NoStop}%
\bibitem [{\citenamefont {{Grevesse}}\ and\ \citenamefont
  {{Sauval}}(1998)}]{1998SSRv...85..161G}%
  \BibitemOpen
  \bibfield  {author} {\bibinfo {author} {\bibfnamefont {N.}~\bibnamefont
  {{Grevesse}}}\ and\ \bibinfo {author} {\bibfnamefont {A.~J.}\ \bibnamefont
  {{Sauval}}},\ }\bibfield  {title} {\bibinfo {title} {{Standard Solar
  Composition}},\ }\href {https://doi.org/10.1023/A:1005161325181} {\bibfield
  {journal} {\bibinfo  {journal} {\ssr}\ }\textbf {\bibinfo {volume} {85}},\
  \bibinfo {pages} {161} (\bibinfo {year} {1998})}\BibitemShut {NoStop}%
\bibitem [{\citenamefont {{Asplund}}\ \emph {et~al.}(2009)\citenamefont
  {{Asplund}}, \citenamefont {{Grevesse}}, \citenamefont {{Sauval}},\ and\
  \citenamefont {{Scott}}}]{2009ARA&A..47..481A}%
  \BibitemOpen
  \bibfield  {author} {\bibinfo {author} {\bibfnamefont {M.}~\bibnamefont
  {{Asplund}}}, \bibinfo {author} {\bibfnamefont {N.}~\bibnamefont
  {{Grevesse}}}, \bibinfo {author} {\bibfnamefont {A.~J.}\ \bibnamefont
  {{Sauval}}},\ and\ \bibinfo {author} {\bibfnamefont {P.}~\bibnamefont
  {{Scott}}},\ }\bibfield  {title} {\bibinfo {title} {{The Chemical Composition
  of the Sun}},\ }\href
  {https://doi.org/10.1146/annurev.astro.46.060407.145222} {\bibfield
  {journal} {\bibinfo  {journal} {\araa}\ }\textbf {\bibinfo {volume} {47}},\
  \bibinfo {pages} {481} (\bibinfo {year} {2009})},\ \Eprint
  {https://arxiv.org/abs/0909.0948} {arXiv:0909.0948 [astro-ph.SR]}
  \BibitemShut {NoStop}%
\bibitem [{\citenamefont {{Caffau}}\ \emph {et~al.}(2011)\citenamefont
  {{Caffau}}, \citenamefont {{Ludwig}}, \citenamefont {{Steffen}},
  \citenamefont {{Freytag}},\ and\ \citenamefont
  {{Bonifacio}}}]{2011SoPh..268..255C}%
  \BibitemOpen
  \bibfield  {author} {\bibinfo {author} {\bibfnamefont {E.}~\bibnamefont
  {{Caffau}}}, \bibinfo {author} {\bibfnamefont {H.~G.}\ \bibnamefont
  {{Ludwig}}}, \bibinfo {author} {\bibfnamefont {M.}~\bibnamefont {{Steffen}}},
  \bibinfo {author} {\bibfnamefont {B.}~\bibnamefont {{Freytag}}},\ and\
  \bibinfo {author} {\bibfnamefont {P.}~\bibnamefont {{Bonifacio}}},\
  }\bibfield  {title} {\bibinfo {title} {{Solar Chemical Abundances Determined
  with a CO5BOLD 3D Model Atmosphere}},\ }\href
  {https://doi.org/10.1007/s11207-010-9541-4} {\bibfield  {journal} {\bibinfo
  {journal} {\solphys}\ }\textbf {\bibinfo {volume} {268}},\ \bibinfo {pages}
  {255} (\bibinfo {year} {2011})},\ \Eprint {https://arxiv.org/abs/1003.1190}
  {arXiv:1003.1190 [astro-ph.SR]} \BibitemShut {NoStop}%
\bibitem [{\citenamefont {{Asplund}}\ \emph {et~al.}(2021)\citenamefont
  {{Asplund}}, \citenamefont {{Amarsi}},\ and\ \citenamefont
  {{Grevesse}}}]{2021A&A...653A.141A}%
  \BibitemOpen
  \bibfield  {author} {\bibinfo {author} {\bibfnamefont {M.}~\bibnamefont
  {{Asplund}}}, \bibinfo {author} {\bibfnamefont {A.~M.}\ \bibnamefont
  {{Amarsi}}},\ and\ \bibinfo {author} {\bibfnamefont {N.}~\bibnamefont
  {{Grevesse}}},\ }\bibfield  {title} {\bibinfo {title} {{The chemical make-up
  of the Sun: A 2020 vision}},\ }\href
  {https://doi.org/10.1051/0004-6361/202140445} {\bibfield  {journal} {\bibinfo
   {journal} {\aap}\ }\textbf {\bibinfo {volume} {653}},\ \bibinfo {eid} {A141}
  (\bibinfo {year} {2021})},\ \Eprint {https://arxiv.org/abs/2105.01661}
  {arXiv:2105.01661 [astro-ph.SR]} \BibitemShut {NoStop}%
\bibitem [{\citenamefont {{Kippenhahn}}\ \emph {et~al.}(2013)\citenamefont
  {{Kippenhahn}}, \citenamefont {{Weigert}},\ and\ \citenamefont
  {{Weiss}}}]{2013sse..book.....K}%
  \BibitemOpen
  \bibfield  {author} {\bibinfo {author} {\bibfnamefont {R.}~\bibnamefont
  {{Kippenhahn}}}, \bibinfo {author} {\bibfnamefont {A.}~\bibnamefont
  {{Weigert}}},\ and\ \bibinfo {author} {\bibfnamefont {A.}~\bibnamefont
  {{Weiss}}},\ }\href {https://doi.org/10.1007/978-3-642-30304-3} {\emph
  {\bibinfo {title} {{Stellar Structure and Evolution}}}}\ (\bibinfo {year}
  {2013})\BibitemShut {NoStop}%
\bibitem [{\citenamefont {{Maeder}}(2009)}]{2009pfer.book.....M}%
  \BibitemOpen
  \bibfield  {author} {\bibinfo {author} {\bibfnamefont {A.}~\bibnamefont
  {{Maeder}}},\ }\href {https://doi.org/10.1007/978-3-540-76949-1} {\emph
  {\bibinfo {title} {{Physics, Formation and Evolution of Rotating Stars}}}}\
  (\bibinfo {year} {2009})\BibitemShut {NoStop}%
\bibitem [{\citenamefont {{Eddington}}(1918)}]{1918ApJ....48..205E}%
  \BibitemOpen
  \bibfield  {author} {\bibinfo {author} {\bibfnamefont {A.~S.}\ \bibnamefont
  {{Eddington}}},\ }\bibfield  {title} {\bibinfo {title} {{On the Conditions in
  the Interior of a Star}},\ }\href {https://doi.org/10.1086/142427} {\bibfield
   {journal} {\bibinfo  {journal} {\apj}\ }\textbf {\bibinfo {volume} {48}},\
  \bibinfo {pages} {205} (\bibinfo {year} {1918})}\BibitemShut {NoStop}%
\bibitem [{\citenamefont {{Atkinson}}\ and\ \citenamefont
  {{Houtermans}}(1929)}]{1929ZPhy...54..656A}%
  \BibitemOpen
  \bibfield  {author} {\bibinfo {author} {\bibfnamefont {R.~D.~E.}\
  \bibnamefont {{Atkinson}}}\ and\ \bibinfo {author} {\bibfnamefont {F.~G.}\
  \bibnamefont {{Houtermans}}},\ }\bibfield  {title} {\bibinfo {title} {{Zur
  Frage der Aufbaum{\"o}glichkeit der Elemente in Sternen}},\ }\href
  {https://doi.org/10.1007/BF01341595} {\bibfield  {journal} {\bibinfo
  {journal} {Zeitschrift fur Physik}\ }\textbf {\bibinfo {volume} {54}},\
  \bibinfo {pages} {656} (\bibinfo {year} {1929})}\BibitemShut {NoStop}%
\bibitem [{\citenamefont {{Payne}}(1925)}]{1925PhDT.........1P}%
  \BibitemOpen
  \bibfield  {author} {\bibinfo {author} {\bibfnamefont {C.~H.}\ \bibnamefont
  {{Payne}}},\ }\emph {\bibinfo {title} {{Stellar Atmospheres; a Contribution
  to the Observational Study of High Temperature in the Reversing Layers of
  Stars.}}},\ \href@noop {} {Ph.D. thesis},\ \bibinfo  {school} {RADCLIFFE
  COLLEGE.} (\bibinfo {year} {1925})\BibitemShut {NoStop}%
\bibitem [{\citenamefont {{Russell}}(1929)}]{1929ApJ....70...11R}%
  \BibitemOpen
  \bibfield  {author} {\bibinfo {author} {\bibfnamefont {H.~N.}\ \bibnamefont
  {{Russell}}},\ }\bibfield  {title} {\bibinfo {title} {{On the Composition of
  the Sun's Atmosphere}},\ }\href {https://doi.org/10.1086/143197} {\bibfield
  {journal} {\bibinfo  {journal} {\apj}\ }\textbf {\bibinfo {volume} {70}},\
  \bibinfo {pages} {11} (\bibinfo {year} {1929})}\BibitemShut {NoStop}%
\bibitem [{\citenamefont {{Bethe}}\ and\ \citenamefont
  {{Critchfield}}(1938)}]{1938PhRv...54..248B}%
  \BibitemOpen
  \bibfield  {author} {\bibinfo {author} {\bibfnamefont {H.~A.}\ \bibnamefont
  {{Bethe}}}\ and\ \bibinfo {author} {\bibfnamefont {C.~L.}\ \bibnamefont
  {{Critchfield}}},\ }\bibfield  {title} {\bibinfo {title} {{The Formation of
  Deuterons by Proton Combination}},\ }\href
  {https://doi.org/10.1103/PhysRev.54.248} {\bibfield  {journal} {\bibinfo
  {journal} {Physical Review}\ }\textbf {\bibinfo {volume} {54}},\ \bibinfo
  {pages} {248} (\bibinfo {year} {1938})}\BibitemShut {NoStop}%
\bibitem [{\citenamefont {{Bethe}}(1939)}]{1939PhRv...55..434B}%
  \BibitemOpen
  \bibfield  {author} {\bibinfo {author} {\bibfnamefont {H.~A.}\ \bibnamefont
  {{Bethe}}},\ }\bibfield  {title} {\bibinfo {title} {{Energy Production in
  Stars}},\ }\href {https://doi.org/10.1103/PhysRev.55.434} {\bibfield
  {journal} {\bibinfo  {journal} {Physical Review}\ }\textbf {\bibinfo {volume}
  {55}},\ \bibinfo {pages} {434} (\bibinfo {year} {1939})}\BibitemShut
  {NoStop}%
\bibitem [{\citenamefont {{Salpeter}}(1952)}]{1952ApJ...115..326S}%
  \BibitemOpen
  \bibfield  {author} {\bibinfo {author} {\bibfnamefont {E.~E.}\ \bibnamefont
  {{Salpeter}}},\ }\bibfield  {title} {\bibinfo {title} {{Nuclear Reactions in
  Stars Without Hydrogen.}},\ }\href {https://doi.org/10.1086/145546}
  {\bibfield  {journal} {\bibinfo  {journal} {\apj}\ }\textbf {\bibinfo
  {volume} {115}},\ \bibinfo {pages} {326} (\bibinfo {year}
  {1952})}\BibitemShut {NoStop}%
\bibitem [{\citenamefont {{Iliadis}}(2007)}]{2007nps..book.....I}%
  \BibitemOpen
  \bibfield  {author} {\bibinfo {author} {\bibfnamefont {C.}~\bibnamefont
  {{Iliadis}}},\ }\href@noop {} {\emph {\bibinfo {title} {{Nuclear Physics of
  Stars}}}}\ (\bibinfo {year} {2007})\BibitemShut {NoStop}%
\bibitem [{\citenamefont {{Schwarzschild}}\ \emph {et~al.}(1957)\citenamefont
  {{Schwarzschild}}, \citenamefont {{Howard}},\ and\ \citenamefont
  {{H{\"a}rm}}}]{1957ApJ...125..233S}%
  \BibitemOpen
  \bibfield  {author} {\bibinfo {author} {\bibfnamefont {M.}~\bibnamefont
  {{Schwarzschild}}}, \bibinfo {author} {\bibfnamefont {R.}~\bibnamefont
  {{Howard}}},\ and\ \bibinfo {author} {\bibfnamefont {R.}~\bibnamefont
  {{H{\"a}rm}}},\ }\bibfield  {title} {\bibinfo {title} {{Inhomogeneous Stellar
  Models. V. a. Solar Model with Convective Envelope and Inhomogeneous
  Interior.}},\ }\href {https://doi.org/10.1086/146297} {\bibfield  {journal}
  {\bibinfo  {journal} {\apj}\ }\textbf {\bibinfo {volume} {125}},\ \bibinfo
  {pages} {233} (\bibinfo {year} {1957})}\BibitemShut {NoStop}%
\bibitem [{\citenamefont {{Henyey}}\ \emph {et~al.}(1964)\citenamefont
  {{Henyey}}, \citenamefont {{Forbes}},\ and\ \citenamefont
  {{Gould}}}]{1964ApJ...139..306H}%
  \BibitemOpen
  \bibfield  {author} {\bibinfo {author} {\bibfnamefont {L.~G.}\ \bibnamefont
  {{Henyey}}}, \bibinfo {author} {\bibfnamefont {J.~E.}\ \bibnamefont
  {{Forbes}}},\ and\ \bibinfo {author} {\bibfnamefont {N.~L.}\ \bibnamefont
  {{Gould}}},\ }\bibfield  {title} {\bibinfo {title} {{A New Method of
  Automatic Computation of Stellar Evolution.}},\ }\href
  {https://doi.org/10.1086/147754} {\bibfield  {journal} {\bibinfo  {journal}
  {\apj}\ }\textbf {\bibinfo {volume} {139}},\ \bibinfo {pages} {306} (\bibinfo
  {year} {1964})}\BibitemShut {NoStop}%
\bibitem [{\citenamefont {{Kippenhahn}}\ \emph {et~al.}(1967)\citenamefont
  {{Kippenhahn}}, \citenamefont {{Weigert}},\ and\ \citenamefont
  {{Hofmeister}}}]{1967MComP...7..129K}%
  \BibitemOpen
  \bibfield  {author} {\bibinfo {author} {\bibfnamefont {R.}~\bibnamefont
  {{Kippenhahn}}}, \bibinfo {author} {\bibfnamefont {A.}~\bibnamefont
  {{Weigert}}},\ and\ \bibinfo {author} {\bibfnamefont {E.}~\bibnamefont
  {{Hofmeister}}},\ }\bibfield  {title} {\bibinfo {title} {{Methods for
  Calculating Stellar Evolution}},\ }\href@noop {} {\bibfield  {journal}
  {\bibinfo  {journal} {Methods in Computational Physics}\ }\textbf {\bibinfo
  {volume} {7}},\ \bibinfo {pages} {129} (\bibinfo {year} {1967})}\BibitemShut
  {NoStop}%
\bibitem [{\citenamefont {{Clayton}}(1983)}]{1983psen.book.....C}%
  \BibitemOpen
  \bibfield  {author} {\bibinfo {author} {\bibfnamefont {D.~D.}\ \bibnamefont
  {{Clayton}}},\ }\href@noop {} {\emph {\bibinfo {title} {{Principles of
  stellar evolution and nucleosynthesis}}}}\ (\bibinfo {year}
  {1983})\BibitemShut {NoStop}%
\bibitem [{\citenamefont {{Gossage}}\ \emph {et~al.}(2018)\citenamefont
  {{Gossage}}, \citenamefont {{Conroy}}, \citenamefont {{Dotter}},
  \citenamefont {{Choi}}, \citenamefont {{Rosenfield}}, \citenamefont
  {{Cargile}},\ and\ \citenamefont {{Dolphin}}}]{2018ApJ...863...67G}%
  \BibitemOpen
  \bibfield  {author} {\bibinfo {author} {\bibfnamefont {S.}~\bibnamefont
  {{Gossage}}}, \bibinfo {author} {\bibfnamefont {C.}~\bibnamefont {{Conroy}}},
  \bibinfo {author} {\bibfnamefont {A.}~\bibnamefont {{Dotter}}}, \bibinfo
  {author} {\bibfnamefont {J.}~\bibnamefont {{Choi}}}, \bibinfo {author}
  {\bibfnamefont {P.}~\bibnamefont {{Rosenfield}}}, \bibinfo {author}
  {\bibfnamefont {P.}~\bibnamefont {{Cargile}}},\ and\ \bibinfo {author}
  {\bibfnamefont {A.}~\bibnamefont {{Dolphin}}},\ }\bibfield  {title} {\bibinfo
  {title} {{Age Determinations of the Hyades, Praesepe, and Pleiades via MESA
  Models with Rotation}},\ }\href {https://doi.org/10.3847/1538-4357/aad0a0}
  {\bibfield  {journal} {\bibinfo  {journal} {\apj}\ }\textbf {\bibinfo
  {volume} {863}},\ \bibinfo {eid} {67} (\bibinfo {year} {2018})},\ \Eprint
  {https://arxiv.org/abs/1804.06441} {arXiv:1804.06441 [astro-ph.SR]}
  \BibitemShut {NoStop}%
\bibitem [{\citenamefont {{Goudfrooij}}\ \emph {et~al.}(2018)\citenamefont
  {{Goudfrooij}}, \citenamefont {{Girardi}}, \citenamefont {{Bellini}},
  \citenamefont {{Bressan}}, \citenamefont {{Correnti}},\ and\ \citenamefont
  {{Costa}}}]{2018ApJ...864L...3G}%
  \BibitemOpen
  \bibfield  {author} {\bibinfo {author} {\bibfnamefont {P.}~\bibnamefont
  {{Goudfrooij}}}, \bibinfo {author} {\bibfnamefont {L.}~\bibnamefont
  {{Girardi}}}, \bibinfo {author} {\bibfnamefont {A.}~\bibnamefont
  {{Bellini}}}, \bibinfo {author} {\bibfnamefont {A.}~\bibnamefont
  {{Bressan}}}, \bibinfo {author} {\bibfnamefont {M.}~\bibnamefont
  {{Correnti}}},\ and\ \bibinfo {author} {\bibfnamefont {G.}~\bibnamefont
  {{Costa}}},\ }\bibfield  {title} {\bibinfo {title} {{The Minimum Mass of
  Rotating Main-sequence Stars and its Impact on the Nature of Extended
  Main-sequence Turnoffs in Intermediate-age Star Clusters in the Magellanic
  Clouds}},\ }\href {https://doi.org/10.3847/2041-8213/aada0f} {\bibfield
  {journal} {\bibinfo  {journal} {\apjl}\ }\textbf {\bibinfo {volume} {864}},\
  \bibinfo {eid} {L3} (\bibinfo {year} {2018})},\ \Eprint
  {https://arxiv.org/abs/1807.04737} {arXiv:1807.04737 [astro-ph.SR]}
  \BibitemShut {NoStop}%
\bibitem [{\citenamefont {{Costa}}\ \emph
  {et~al.}(2019{\natexlab{a}})\citenamefont {{Costa}}, \citenamefont
  {{Girardi}}, \citenamefont {{Bressan}}, \citenamefont {{Chen}}, \citenamefont
  {{Goudfrooij}}, \citenamefont {{Marigo}}, \citenamefont {{Rodrigues}},\ and\
  \citenamefont {{Lanza}}}]{2019A&A...631A.128C}%
  \BibitemOpen
  \bibfield  {author} {\bibinfo {author} {\bibfnamefont {G.}~\bibnamefont
  {{Costa}}}, \bibinfo {author} {\bibfnamefont {L.}~\bibnamefont {{Girardi}}},
  \bibinfo {author} {\bibfnamefont {A.}~\bibnamefont {{Bressan}}}, \bibinfo
  {author} {\bibfnamefont {Y.}~\bibnamefont {{Chen}}}, \bibinfo {author}
  {\bibfnamefont {P.}~\bibnamefont {{Goudfrooij}}}, \bibinfo {author}
  {\bibfnamefont {P.}~\bibnamefont {{Marigo}}}, \bibinfo {author}
  {\bibfnamefont {T.~S.}\ \bibnamefont {{Rodrigues}}},\ and\ \bibinfo {author}
  {\bibfnamefont {A.}~\bibnamefont {{Lanza}}},\ }\bibfield  {title} {\bibinfo
  {title} {{Multiple stellar populations in NGC 1866. New clues from Cepheids
  and colour-magnitude diagram}},\ }\href
  {https://doi.org/10.1051/0004-6361/201936409} {\bibfield  {journal} {\bibinfo
   {journal} {\aap}\ }\textbf {\bibinfo {volume} {631}},\ \bibinfo {eid} {A128}
  (\bibinfo {year} {2019}{\natexlab{a}})},\ \Eprint
  {https://arxiv.org/abs/1909.01907} {arXiv:1909.01907 [astro-ph.SR]}
  \BibitemShut {NoStop}%
\bibitem [{\citenamefont {{Sz{\'e}csi}}\ and\ \citenamefont
  {{W{\"u}nsch}}(2019)}]{2019ApJ...871...20S}%
  \BibitemOpen
  \bibfield  {author} {\bibinfo {author} {\bibfnamefont {D.}~\bibnamefont
  {{Sz{\'e}csi}}}\ and\ \bibinfo {author} {\bibfnamefont {R.}~\bibnamefont
  {{W{\"u}nsch}}},\ }\bibfield  {title} {\bibinfo {title} {{Role of Supergiants
  in the Formation of Globular Clusters}},\ }\href
  {https://doi.org/10.3847/1538-4357/aaf4be} {\bibfield  {journal} {\bibinfo
  {journal} {\apj}\ }\textbf {\bibinfo {volume} {871}},\ \bibinfo {eid} {20}
  (\bibinfo {year} {2019})},\ \Eprint {https://arxiv.org/abs/1809.01395}
  {arXiv:1809.01395 [astro-ph.GA]} \BibitemShut {NoStop}%
\bibitem [{\citenamefont {{Milone}}\ and\ \citenamefont
  {{Marino}}(2022)}]{2022Univ....8..359M}%
  \BibitemOpen
  \bibfield  {author} {\bibinfo {author} {\bibfnamefont {A.~P.}\ \bibnamefont
  {{Milone}}}\ and\ \bibinfo {author} {\bibfnamefont {A.~F.}\ \bibnamefont
  {{Marino}}},\ }\bibfield  {title} {\bibinfo {title} {{Multiple Populations in
  Star Clusters}},\ }\href {https://doi.org/10.3390/universe8070359} {\bibfield
   {journal} {\bibinfo  {journal} {Universe}\ }\textbf {\bibinfo {volume}
  {8}},\ \bibinfo {pages} {359} (\bibinfo {year} {2022})},\ \Eprint
  {https://arxiv.org/abs/2206.10564} {arXiv:2206.10564 [astro-ph.GA]}
  \BibitemShut {NoStop}%
\bibitem [{\citenamefont {{Milone}}\ \emph {et~al.}(2023)\citenamefont
  {{Milone}}, \citenamefont {{Marino}}, \citenamefont {{Dotter}}, \citenamefont
  {{Ziliotto}}, \citenamefont {{Dondoglio}}, \citenamefont {{Cordoni}},
  \citenamefont {{Jang}}, \citenamefont {{Lagioia}}, \citenamefont
  {{Legnardi}}, \citenamefont {{Mohandasan}}, \citenamefont {{Tailo}},
  \citenamefont {{Yong}}, \citenamefont {{Baimukhametova}},\ and\ \citenamefont
  {{Carlos}}}]{2023MNRAS.522.2429M}%
  \BibitemOpen
  \bibfield  {author} {\bibinfo {author} {\bibfnamefont {A.~P.}\ \bibnamefont
  {{Milone}}}, \bibinfo {author} {\bibfnamefont {A.~F.}\ \bibnamefont
  {{Marino}}}, \bibinfo {author} {\bibfnamefont {A.}~\bibnamefont {{Dotter}}},
  \bibinfo {author} {\bibfnamefont {T.}~\bibnamefont {{Ziliotto}}}, \bibinfo
  {author} {\bibfnamefont {E.}~\bibnamefont {{Dondoglio}}}, \bibinfo {author}
  {\bibfnamefont {G.}~\bibnamefont {{Cordoni}}}, \bibinfo {author}
  {\bibfnamefont {S.}~\bibnamefont {{Jang}}}, \bibinfo {author} {\bibfnamefont
  {E.~P.}\ \bibnamefont {{Lagioia}}}, \bibinfo {author} {\bibfnamefont {M.~V.}\
  \bibnamefont {{Legnardi}}}, \bibinfo {author} {\bibfnamefont
  {A.}~\bibnamefont {{Mohandasan}}}, \bibinfo {author} {\bibfnamefont
  {M.}~\bibnamefont {{Tailo}}}, \bibinfo {author} {\bibfnamefont
  {D.}~\bibnamefont {{Yong}}}, \bibinfo {author} {\bibfnamefont
  {S.}~\bibnamefont {{Baimukhametova}}},\ and\ \bibinfo {author} {\bibfnamefont
  {M.}~\bibnamefont {{Carlos}}},\ }\bibfield  {title} {\bibinfo {title}
  {{Multiple stellar populations in globular clusters with JWST: an NIRCam view
  of 47 Tucanae}},\ }\href {https://doi.org/10.1093/mnras/stad1041} {\bibfield
  {journal} {\bibinfo  {journal} {\mnras}\ }\textbf {\bibinfo {volume} {522}},\
  \bibinfo {pages} {2429} (\bibinfo {year} {2023})},\ \Eprint
  {https://arxiv.org/abs/2301.10889} {arXiv:2301.10889 [astro-ph.SR]}
  \BibitemShut {NoStop}%
\bibitem [{\citenamefont {{Gatto}}\ \emph {et~al.}(2017)\citenamefont
  {{Gatto}}, \citenamefont {{Walch}}, \citenamefont {{Naab}}, \citenamefont
  {{Girichidis}}, \citenamefont {{W{\"u}nsch}}, \citenamefont {{Glover}},
  \citenamefont {{Klessen}}, \citenamefont {{Clark}}, \citenamefont {{Peters}},
  \citenamefont {{Derigs}}, \citenamefont {{Baczynski}},\ and\ \citenamefont
  {{Puls}}}]{2017MNRAS.466.1903G}%
  \BibitemOpen
  \bibfield  {author} {\bibinfo {author} {\bibfnamefont {A.}~\bibnamefont
  {{Gatto}}}, \bibinfo {author} {\bibfnamefont {S.}~\bibnamefont {{Walch}}},
  \bibinfo {author} {\bibfnamefont {T.}~\bibnamefont {{Naab}}}, \bibinfo
  {author} {\bibfnamefont {P.}~\bibnamefont {{Girichidis}}}, \bibinfo {author}
  {\bibfnamefont {R.}~\bibnamefont {{W{\"u}nsch}}}, \bibinfo {author}
  {\bibfnamefont {S.~C.~O.}\ \bibnamefont {{Glover}}}, \bibinfo {author}
  {\bibfnamefont {R.~S.}\ \bibnamefont {{Klessen}}}, \bibinfo {author}
  {\bibfnamefont {P.~C.}\ \bibnamefont {{Clark}}}, \bibinfo {author}
  {\bibfnamefont {T.}~\bibnamefont {{Peters}}}, \bibinfo {author}
  {\bibfnamefont {D.}~\bibnamefont {{Derigs}}}, \bibinfo {author}
  {\bibfnamefont {C.}~\bibnamefont {{Baczynski}}},\ and\ \bibinfo {author}
  {\bibfnamefont {J.}~\bibnamefont {{Puls}}},\ }\bibfield  {title} {\bibinfo
  {title} {{The SILCC project - III. Regulation of star formation and outflows
  by stellar winds and supernovae}},\ }\href
  {https://doi.org/10.1093/mnras/stw3209} {\bibfield  {journal} {\bibinfo
  {journal} {\mnras}\ }\textbf {\bibinfo {volume} {466}},\ \bibinfo {pages}
  {1903} (\bibinfo {year} {2017})},\ \Eprint {https://arxiv.org/abs/1606.05346}
  {arXiv:1606.05346 [astro-ph.GA]} \BibitemShut {NoStop}%
\bibitem [{\citenamefont {{Rosdahl}}\ \emph {et~al.}(2018)\citenamefont
  {{Rosdahl}}, \citenamefont {{Katz}}, \citenamefont {{Blaizot}}, \citenamefont
  {{Kimm}}, \citenamefont {{Michel-Dansac}}, \citenamefont {{Garel}},
  \citenamefont {{Haehnelt}}, \citenamefont {{Ocvirk}},\ and\ \citenamefont
  {{Teyssier}}}]{2018MNRAS.479..994R}%
  \BibitemOpen
  \bibfield  {author} {\bibinfo {author} {\bibfnamefont {J.}~\bibnamefont
  {{Rosdahl}}}, \bibinfo {author} {\bibfnamefont {H.}~\bibnamefont {{Katz}}},
  \bibinfo {author} {\bibfnamefont {J.}~\bibnamefont {{Blaizot}}}, \bibinfo
  {author} {\bibfnamefont {T.}~\bibnamefont {{Kimm}}}, \bibinfo {author}
  {\bibfnamefont {L.}~\bibnamefont {{Michel-Dansac}}}, \bibinfo {author}
  {\bibfnamefont {T.}~\bibnamefont {{Garel}}}, \bibinfo {author} {\bibfnamefont
  {M.}~\bibnamefont {{Haehnelt}}}, \bibinfo {author} {\bibfnamefont
  {P.}~\bibnamefont {{Ocvirk}}},\ and\ \bibinfo {author} {\bibfnamefont
  {R.}~\bibnamefont {{Teyssier}}},\ }\bibfield  {title} {\bibinfo {title} {{The
  SPHINX cosmological simulations of the first billion years: the impact of
  binary stars on reionization}},\ }\href
  {https://doi.org/10.1093/mnras/sty1655} {\bibfield  {journal} {\bibinfo
  {journal} {\mnras}\ }\textbf {\bibinfo {volume} {479}},\ \bibinfo {pages}
  {994} (\bibinfo {year} {2018})},\ \Eprint {https://arxiv.org/abs/1801.07259}
  {arXiv:1801.07259 [astro-ph.GA]} \BibitemShut {NoStop}%
\bibitem [{\citenamefont {{Valentini}}\ \emph {et~al.}(2019)\citenamefont
  {{Valentini}}, \citenamefont {{Borgani}}, \citenamefont {{Bressan}},
  \citenamefont {{Murante}}, \citenamefont {{Tornatore}},\ and\ \citenamefont
  {{Monaco}}}]{2019MNRAS.485.1384V}%
  \BibitemOpen
  \bibfield  {author} {\bibinfo {author} {\bibfnamefont {M.}~\bibnamefont
  {{Valentini}}}, \bibinfo {author} {\bibfnamefont {S.}~\bibnamefont
  {{Borgani}}}, \bibinfo {author} {\bibfnamefont {A.}~\bibnamefont
  {{Bressan}}}, \bibinfo {author} {\bibfnamefont {G.}~\bibnamefont
  {{Murante}}}, \bibinfo {author} {\bibfnamefont {L.}~\bibnamefont
  {{Tornatore}}},\ and\ \bibinfo {author} {\bibfnamefont {P.}~\bibnamefont
  {{Monaco}}},\ }\bibfield  {title} {\bibinfo {title} {{Chemical evolution of
  disc galaxies from cosmological simulations}},\ }\href
  {https://doi.org/10.1093/mnras/stz492} {\bibfield  {journal} {\bibinfo
  {journal} {\mnras}\ }\textbf {\bibinfo {volume} {485}},\ \bibinfo {pages}
  {1384} (\bibinfo {year} {2019})},\ \Eprint {https://arxiv.org/abs/1902.05955}
  {arXiv:1902.05955 [astro-ph.GA]} \BibitemShut {NoStop}%
\bibitem [{\citenamefont {{Dal Tio}}\ \emph {et~al.}(2022)\citenamefont {{Dal
  Tio}}, \citenamefont {{Pastorelli}}, \citenamefont {{Mazzi}}, \citenamefont
  {{Trabucchi}}, \citenamefont {{Costa}}, \citenamefont {{Jacques}},
  \citenamefont {{Pieres}}, \citenamefont {{Girardi}}, \citenamefont {{Chen}},
  \citenamefont {{Olsen}}, \citenamefont {{Juric}}, \citenamefont
  {{Ivezi{\'c}}}, \citenamefont {{Yoachim}}, \citenamefont {{Clarkson}},
  \citenamefont {{Marigo}}, \citenamefont {{Rodrigues}}, \citenamefont
  {{Zaggia}}, \citenamefont {{Barbieri}}, \citenamefont {{Momany}},
  \citenamefont {{Bressan}}, \citenamefont {{Nikutta}},\ and\ \citenamefont
  {{da Costa}}}]{2022ApJS..262...22D}%
  \BibitemOpen
  \bibfield  {author} {\bibinfo {author} {\bibfnamefont {P.}~\bibnamefont {{Dal
  Tio}}}, \bibinfo {author} {\bibfnamefont {G.}~\bibnamefont {{Pastorelli}}},
  \bibinfo {author} {\bibfnamefont {A.}~\bibnamefont {{Mazzi}}}, \bibinfo
  {author} {\bibfnamefont {M.}~\bibnamefont {{Trabucchi}}}, \bibinfo {author}
  {\bibfnamefont {G.}~\bibnamefont {{Costa}}}, \bibinfo {author} {\bibfnamefont
  {A.}~\bibnamefont {{Jacques}}}, \bibinfo {author} {\bibfnamefont
  {A.}~\bibnamefont {{Pieres}}}, \bibinfo {author} {\bibfnamefont
  {L.}~\bibnamefont {{Girardi}}}, \bibinfo {author} {\bibfnamefont
  {Y.}~\bibnamefont {{Chen}}}, \bibinfo {author} {\bibfnamefont {K.~A.~G.}\
  \bibnamefont {{Olsen}}}, \bibinfo {author} {\bibfnamefont {M.}~\bibnamefont
  {{Juric}}}, \bibinfo {author} {\bibfnamefont {{\v{Z}}.}~\bibnamefont
  {{Ivezi{\'c}}}}, \bibinfo {author} {\bibfnamefont {P.}~\bibnamefont
  {{Yoachim}}}, \bibinfo {author} {\bibfnamefont {W.~I.}\ \bibnamefont
  {{Clarkson}}}, \bibinfo {author} {\bibfnamefont {P.}~\bibnamefont
  {{Marigo}}}, \bibinfo {author} {\bibfnamefont {T.~S.}\ \bibnamefont
  {{Rodrigues}}}, \bibinfo {author} {\bibfnamefont {S.}~\bibnamefont
  {{Zaggia}}}, \bibinfo {author} {\bibfnamefont {M.}~\bibnamefont
  {{Barbieri}}}, \bibinfo {author} {\bibfnamefont {Y.}~\bibnamefont
  {{Momany}}}, \bibinfo {author} {\bibfnamefont {A.}~\bibnamefont {{Bressan}}},
  \bibinfo {author} {\bibfnamefont {R.}~\bibnamefont {{Nikutta}}},\ and\
  \bibinfo {author} {\bibfnamefont {L.~N.}\ \bibnamefont {{da Costa}}},\
  }\bibfield  {title} {\bibinfo {title} {{Simulating the Legacy Survey of Space
  and Time Stellar Content with TRILEGAL}},\ }\href
  {https://doi.org/10.3847/1538-4365/ac7be6} {\bibfield  {journal} {\bibinfo
  {journal} {\apjs}\ }\textbf {\bibinfo {volume} {262}},\ \bibinfo {eid} {22}
  (\bibinfo {year} {2022})},\ \Eprint {https://arxiv.org/abs/2208.00829}
  {arXiv:2208.00829 [astro-ph.GA]} \BibitemShut {NoStop}%
\bibitem [{\citenamefont {{Yoon}}\ \emph
  {et~al.}(2006{\natexlab{a}})\citenamefont {{Yoon}}, \citenamefont
  {{Langer}},\ and\ \citenamefont {{Norman}}}]{2006A&A...460..199Y}%
  \BibitemOpen
  \bibfield  {author} {\bibinfo {author} {\bibfnamefont {S.~C.}\ \bibnamefont
  {{Yoon}}}, \bibinfo {author} {\bibfnamefont {N.}~\bibnamefont {{Langer}}},\
  and\ \bibinfo {author} {\bibfnamefont {C.}~\bibnamefont {{Norman}}},\
  }\bibfield  {title} {\bibinfo {title} {{Single star progenitors of long
  gamma-ray bursts. I. Model grids and redshift dependent GRB rate}},\ }\href
  {https://doi.org/10.1051/0004-6361:20065912} {\bibfield  {journal} {\bibinfo
  {journal} {\aap}\ }\textbf {\bibinfo {volume} {460}},\ \bibinfo {pages} {199}
  (\bibinfo {year} {2006}{\natexlab{a}})},\ \Eprint
  {https://arxiv.org/abs/astro-ph/0606637} {arXiv:astro-ph/0606637 [astro-ph]}
  \BibitemShut {NoStop}%
\bibitem [{\citenamefont {{Sukhbold}}\ \emph {et~al.}(2016)\citenamefont
  {{Sukhbold}}, \citenamefont {{Ertl}}, \citenamefont {{Woosley}},
  \citenamefont {{Brown}},\ and\ \citenamefont
  {{Janka}}}]{2016ApJ...821...38S}%
  \BibitemOpen
  \bibfield  {author} {\bibinfo {author} {\bibfnamefont {T.}~\bibnamefont
  {{Sukhbold}}}, \bibinfo {author} {\bibfnamefont {T.}~\bibnamefont {{Ertl}}},
  \bibinfo {author} {\bibfnamefont {S.~E.}\ \bibnamefont {{Woosley}}}, \bibinfo
  {author} {\bibfnamefont {J.~M.}\ \bibnamefont {{Brown}}},\ and\ \bibinfo
  {author} {\bibfnamefont {H.~T.}\ \bibnamefont {{Janka}}},\ }\bibfield
  {title} {\bibinfo {title} {{Core-collapse Supernovae from 9 to 120 Solar
  Masses Based on Neutrino-powered Explosions}},\ }\href
  {https://doi.org/10.3847/0004-637X/821/1/38} {\bibfield  {journal} {\bibinfo
  {journal} {\apj}\ }\textbf {\bibinfo {volume} {821}},\ \bibinfo {eid} {38}
  (\bibinfo {year} {2016})},\ \Eprint {https://arxiv.org/abs/1510.04643}
  {arXiv:1510.04643 [astro-ph.HE]} \BibitemShut {NoStop}%
\bibitem [{\citenamefont {{Limongi}}\ and\ \citenamefont
  {{Chieffi}}(2018)}]{2018ApJS..237...13L}%
  \BibitemOpen
  \bibfield  {author} {\bibinfo {author} {\bibfnamefont {M.}~\bibnamefont
  {{Limongi}}}\ and\ \bibinfo {author} {\bibfnamefont {A.}~\bibnamefont
  {{Chieffi}}},\ }\bibfield  {title} {\bibinfo {title} {{Presupernova Evolution
  and Explosive Nucleosynthesis of Rotating Massive Stars in the Metallicity
  Range -3 {\ensuremath{\leq}} [Fe/H] {\ensuremath{\leq}} 0}},\ }\href
  {https://doi.org/10.3847/1538-4365/aacb24} {\bibfield  {journal} {\bibinfo
  {journal} {\apjs}\ }\textbf {\bibinfo {volume} {237}},\ \bibinfo {eid} {13}
  (\bibinfo {year} {2018})},\ \Eprint {https://arxiv.org/abs/1805.09640}
  {arXiv:1805.09640 [astro-ph.SR]} \BibitemShut {NoStop}%
\bibitem [{\citenamefont {{Aguilera-Dena}}\ \emph {et~al.}(2020)\citenamefont
  {{Aguilera-Dena}}, \citenamefont {{Langer}}, \citenamefont {{Antoniadis}},\
  and\ \citenamefont {{M{\"u}ller}}}]{2020ApJ...901..114A}%
  \BibitemOpen
  \bibfield  {author} {\bibinfo {author} {\bibfnamefont {D.~R.}\ \bibnamefont
  {{Aguilera-Dena}}}, \bibinfo {author} {\bibfnamefont {N.}~\bibnamefont
  {{Langer}}}, \bibinfo {author} {\bibfnamefont {J.}~\bibnamefont
  {{Antoniadis}}},\ and\ \bibinfo {author} {\bibfnamefont {B.}~\bibnamefont
  {{M{\"u}ller}}},\ }\bibfield  {title} {\bibinfo {title} {{Precollapse
  Properties of Superluminous Supernovae and Long Gamma-Ray Burst Progenitor
  Models}},\ }\href {https://doi.org/10.3847/1538-4357/abb138} {\bibfield
  {journal} {\bibinfo  {journal} {\apj}\ }\textbf {\bibinfo {volume} {901}},\
  \bibinfo {eid} {114} (\bibinfo {year} {2020})},\ \Eprint
  {https://arxiv.org/abs/2008.09132} {arXiv:2008.09132 [astro-ph.SR]}
  \BibitemShut {NoStop}%
\bibitem [{\citenamefont {{Gallet}}\ \emph {et~al.}(2017)\citenamefont
  {{Gallet}}, \citenamefont {{Charbonnel}}, \citenamefont {{Amard}},
  \citenamefont {{Brun}}, \citenamefont {{Palacios}},\ and\ \citenamefont
  {{Mathis}}}]{2017A&A...597A..14G}%
  \BibitemOpen
  \bibfield  {author} {\bibinfo {author} {\bibfnamefont {F.}~\bibnamefont
  {{Gallet}}}, \bibinfo {author} {\bibfnamefont {C.}~\bibnamefont
  {{Charbonnel}}}, \bibinfo {author} {\bibfnamefont {L.}~\bibnamefont
  {{Amard}}}, \bibinfo {author} {\bibfnamefont {S.}~\bibnamefont {{Brun}}},
  \bibinfo {author} {\bibfnamefont {A.}~\bibnamefont {{Palacios}}},\ and\
  \bibinfo {author} {\bibfnamefont {S.}~\bibnamefont {{Mathis}}},\ }\bibfield
  {title} {\bibinfo {title} {{Impacts of stellar evolution and dynamics on the
  habitable zone: The role of rotation and magnetic activity}},\ }\href
  {https://doi.org/10.1051/0004-6361/201629034} {\bibfield  {journal} {\bibinfo
   {journal} {\aap}\ }\textbf {\bibinfo {volume} {597}},\ \bibinfo {eid} {A14}
  (\bibinfo {year} {2017})},\ \Eprint {https://arxiv.org/abs/1608.06772}
  {arXiv:1608.06772 [astro-ph.EP]} \BibitemShut {NoStop}%
\bibitem [{\citenamefont {{Maldonado}}\ \emph {et~al.}(2019)\citenamefont
  {{Maldonado}}, \citenamefont {{Villaver}}, \citenamefont {{Eiroa}},\ and\
  \citenamefont {{Micela}}}]{2019A&A...624A..94M}%
  \BibitemOpen
  \bibfield  {author} {\bibinfo {author} {\bibfnamefont {J.}~\bibnamefont
  {{Maldonado}}}, \bibinfo {author} {\bibfnamefont {E.}~\bibnamefont
  {{Villaver}}}, \bibinfo {author} {\bibfnamefont {C.}~\bibnamefont
  {{Eiroa}}},\ and\ \bibinfo {author} {\bibfnamefont {G.}~\bibnamefont
  {{Micela}}},\ }\bibfield  {title} {\bibinfo {title} {{Connecting substellar
  and stellar formation: the role of the host star's metallicity}},\ }\href
  {https://doi.org/10.1051/0004-6361/201833827} {\bibfield  {journal} {\bibinfo
   {journal} {\aap}\ }\textbf {\bibinfo {volume} {624}},\ \bibinfo {eid} {A94}
  (\bibinfo {year} {2019})},\ \Eprint {https://arxiv.org/abs/1903.01141}
  {arXiv:1903.01141 [astro-ph.SR]} \BibitemShut {NoStop}%
\bibitem [{\citenamefont {{Maldonado}}\ \emph {et~al.}(2020)\citenamefont
  {{Maldonado}}, \citenamefont {{Micela}}, \citenamefont {{Baratella}},
  \citenamefont {{D'Orazi}}, \citenamefont {{Affer}}, \citenamefont {{Biazzo}},
  \citenamefont {{Lanza}}, \citenamefont {{Maggio}}, \citenamefont
  {{Gonz{\'a}lez Hern{\'a}ndez}}, \citenamefont {{Perger}}, \citenamefont
  {{Pinamonti}}, \citenamefont {{Scandariato}}, \citenamefont {{Sozzetti}},
  \citenamefont {{Locci}}, \citenamefont {{Di Maio}}, \citenamefont
  {{Bignamini}}, \citenamefont {{Claudi}}, \citenamefont {{Molinari}},
  \citenamefont {{Rebolo}}, \citenamefont {{Ribas}}, \citenamefont
  {{Toledo-Padr{\'o}n}}, \citenamefont {{Covino}}, \citenamefont {{Desidera}},
  \citenamefont {{Herrero}}, \citenamefont {{Morales}}, \citenamefont
  {{Su{\'a}rez-Mascare{\~n}o}}, \citenamefont {{Pagano}}, \citenamefont
  {{Petralia}}, \citenamefont {{Piotto}},\ and\ \citenamefont
  {{Poretti}}}]{2020A&A...644A..68M}%
  \BibitemOpen
  \bibfield  {author} {\bibinfo {author} {\bibfnamefont {J.}~\bibnamefont
  {{Maldonado}}}, \bibinfo {author} {\bibfnamefont {G.}~\bibnamefont
  {{Micela}}}, \bibinfo {author} {\bibfnamefont {M.}~\bibnamefont
  {{Baratella}}}, \bibinfo {author} {\bibfnamefont {V.}~\bibnamefont
  {{D'Orazi}}}, \bibinfo {author} {\bibfnamefont {L.}~\bibnamefont {{Affer}}},
  \bibinfo {author} {\bibfnamefont {K.}~\bibnamefont {{Biazzo}}}, \bibinfo
  {author} {\bibfnamefont {A.~F.}\ \bibnamefont {{Lanza}}}, \bibinfo {author}
  {\bibfnamefont {A.}~\bibnamefont {{Maggio}}}, \bibinfo {author}
  {\bibfnamefont {J.~I.}\ \bibnamefont {{Gonz{\'a}lez Hern{\'a}ndez}}},
  \bibinfo {author} {\bibfnamefont {M.}~\bibnamefont {{Perger}}}, \bibinfo
  {author} {\bibfnamefont {M.}~\bibnamefont {{Pinamonti}}}, \bibinfo {author}
  {\bibfnamefont {G.}~\bibnamefont {{Scandariato}}}, \bibinfo {author}
  {\bibfnamefont {A.}~\bibnamefont {{Sozzetti}}}, \bibinfo {author}
  {\bibfnamefont {D.}~\bibnamefont {{Locci}}}, \bibinfo {author} {\bibfnamefont
  {C.}~\bibnamefont {{Di Maio}}}, \bibinfo {author} {\bibfnamefont
  {A.}~\bibnamefont {{Bignamini}}}, \bibinfo {author} {\bibfnamefont
  {R.}~\bibnamefont {{Claudi}}}, \bibinfo {author} {\bibfnamefont
  {E.}~\bibnamefont {{Molinari}}}, \bibinfo {author} {\bibfnamefont
  {R.}~\bibnamefont {{Rebolo}}}, \bibinfo {author} {\bibfnamefont
  {I.}~\bibnamefont {{Ribas}}}, \bibinfo {author} {\bibfnamefont
  {B.}~\bibnamefont {{Toledo-Padr{\'o}n}}}, \bibinfo {author} {\bibfnamefont
  {E.}~\bibnamefont {{Covino}}}, \bibinfo {author} {\bibfnamefont
  {S.}~\bibnamefont {{Desidera}}}, \bibinfo {author} {\bibfnamefont
  {E.}~\bibnamefont {{Herrero}}}, \bibinfo {author} {\bibfnamefont {J.~C.}\
  \bibnamefont {{Morales}}}, \bibinfo {author} {\bibfnamefont {A.}~\bibnamefont
  {{Su{\'a}rez-Mascare{\~n}o}}}, \bibinfo {author} {\bibfnamefont
  {I.}~\bibnamefont {{Pagano}}}, \bibinfo {author} {\bibfnamefont
  {A.}~\bibnamefont {{Petralia}}}, \bibinfo {author} {\bibfnamefont
  {G.}~\bibnamefont {{Piotto}}},\ and\ \bibinfo {author} {\bibfnamefont
  {E.}~\bibnamefont {{Poretti}}},\ }\bibfield  {title} {\bibinfo {title}
  {{HADES RV programme with HARPS-N at TNG. XII. The abundance signature of M
  dwarf stars with planets}},\ }\href
  {https://doi.org/10.1051/0004-6361/202039478} {\bibfield  {journal} {\bibinfo
   {journal} {\aap}\ }\textbf {\bibinfo {volume} {644}},\ \bibinfo {eid} {A68}
  (\bibinfo {year} {2020})},\ \Eprint {https://arxiv.org/abs/2010.14867}
  {arXiv:2010.14867 [astro-ph.SR]} \BibitemShut {NoStop}%
\bibitem [{\citenamefont {{Spera}}\ and\ \citenamefont
  {{Mapelli}}(2017)}]{2017MNRAS.470.4739S}%
  \BibitemOpen
  \bibfield  {author} {\bibinfo {author} {\bibfnamefont {M.}~\bibnamefont
  {{Spera}}}\ and\ \bibinfo {author} {\bibfnamefont {M.}~\bibnamefont
  {{Mapelli}}},\ }\bibfield  {title} {\bibinfo {title} {{Very massive stars,
  pair-instability supernovae and intermediate-mass black holes with the sevn
  code}},\ }\href {https://doi.org/10.1093/mnras/stx1576} {\bibfield  {journal}
  {\bibinfo  {journal} {\mnras}\ }\textbf {\bibinfo {volume} {470}},\ \bibinfo
  {pages} {4739} (\bibinfo {year} {2017})},\ \Eprint
  {https://arxiv.org/abs/1706.06109} {arXiv:1706.06109 [astro-ph.SR]}
  \BibitemShut {NoStop}%
\bibitem [{\citenamefont {{Kruckow}}\ \emph
  {et~al.}(2018{\natexlab{a}})\citenamefont {{Kruckow}}, \citenamefont
  {{Tauris}}, \citenamefont {{Langer}}, \citenamefont {{Kramer}},\ and\
  \citenamefont {{Izzard}}}]{2018MNRAS.481.1908K}%
  \BibitemOpen
  \bibfield  {author} {\bibinfo {author} {\bibfnamefont {M.~U.}\ \bibnamefont
  {{Kruckow}}}, \bibinfo {author} {\bibfnamefont {T.~M.}\ \bibnamefont
  {{Tauris}}}, \bibinfo {author} {\bibfnamefont {N.}~\bibnamefont {{Langer}}},
  \bibinfo {author} {\bibfnamefont {M.}~\bibnamefont {{Kramer}}},\ and\
  \bibinfo {author} {\bibfnamefont {R.~G.}\ \bibnamefont {{Izzard}}},\
  }\bibfield  {title} {\bibinfo {title} {{Progenitors of gravitational wave
  mergers: binary evolution with the stellar grid-based code COMBINE}},\ }\href
  {https://doi.org/10.1093/mnras/sty2190} {\bibfield  {journal} {\bibinfo
  {journal} {\mnras}\ }\textbf {\bibinfo {volume} {481}},\ \bibinfo {pages}
  {1908} (\bibinfo {year} {2018}{\natexlab{a}})},\ \Eprint
  {https://arxiv.org/abs/1801.05433} {arXiv:1801.05433 [astro-ph.SR]}
  \BibitemShut {NoStop}%
\bibitem [{\citenamefont {{Stevenson}}\ \emph {et~al.}(2019)\citenamefont
  {{Stevenson}}, \citenamefont {{Sampson}}, \citenamefont {{Powell}},
  \citenamefont {{Vigna-G{\'o}mez}}, \citenamefont {{Neijssel}}, \citenamefont
  {{Sz{\'e}csi}},\ and\ \citenamefont {{Mandel}}}]{2019ApJ...882..121S}%
  \BibitemOpen
  \bibfield  {author} {\bibinfo {author} {\bibfnamefont {S.}~\bibnamefont
  {{Stevenson}}}, \bibinfo {author} {\bibfnamefont {M.}~\bibnamefont
  {{Sampson}}}, \bibinfo {author} {\bibfnamefont {J.}~\bibnamefont {{Powell}}},
  \bibinfo {author} {\bibfnamefont {A.}~\bibnamefont {{Vigna-G{\'o}mez}}},
  \bibinfo {author} {\bibfnamefont {C.~J.}\ \bibnamefont {{Neijssel}}},
  \bibinfo {author} {\bibfnamefont {D.}~\bibnamefont {{Sz{\'e}csi}}},\ and\
  \bibinfo {author} {\bibfnamefont {I.}~\bibnamefont {{Mandel}}},\ }\bibfield
  {title} {\bibinfo {title} {{The Impact of Pair-instability Mass Loss on the
  Binary Black Hole Mass Distribution}},\ }\href
  {https://doi.org/10.3847/1538-4357/ab3981} {\bibfield  {journal} {\bibinfo
  {journal} {\apj}\ }\textbf {\bibinfo {volume} {882}},\ \bibinfo {eid} {121}
  (\bibinfo {year} {2019})},\ \Eprint {https://arxiv.org/abs/1904.02821}
  {arXiv:1904.02821 [astro-ph.HE]} \BibitemShut {NoStop}%
\bibitem [{\citenamefont {{Mapelli}}\ \emph
  {et~al.}(2020{\natexlab{a}})\citenamefont {{Mapelli}}, \citenamefont
  {{Spera}}, \citenamefont {{Montanari}}, \citenamefont {{Limongi}},
  \citenamefont {{Chieffi}}, \citenamefont {{Giacobbo}}, \citenamefont
  {{Bressan}},\ and\ \citenamefont {{Bouffanais}}}]{2020ApJ...888...76M}%
  \BibitemOpen
  \bibfield  {author} {\bibinfo {author} {\bibfnamefont {M.}~\bibnamefont
  {{Mapelli}}}, \bibinfo {author} {\bibfnamefont {M.}~\bibnamefont {{Spera}}},
  \bibinfo {author} {\bibfnamefont {E.}~\bibnamefont {{Montanari}}}, \bibinfo
  {author} {\bibfnamefont {M.}~\bibnamefont {{Limongi}}}, \bibinfo {author}
  {\bibfnamefont {A.}~\bibnamefont {{Chieffi}}}, \bibinfo {author}
  {\bibfnamefont {N.}~\bibnamefont {{Giacobbo}}}, \bibinfo {author}
  {\bibfnamefont {A.}~\bibnamefont {{Bressan}}},\ and\ \bibinfo {author}
  {\bibfnamefont {Y.}~\bibnamefont {{Bouffanais}}},\ }\bibfield  {title}
  {\bibinfo {title} {{Impact of the Rotation and Compactness of Progenitors on
  the Mass of Black Holes}},\ }\href {https://doi.org/10.3847/1538-4357/ab584d}
  {\bibfield  {journal} {\bibinfo  {journal} {\apj}\ }\textbf {\bibinfo
  {volume} {888}},\ \bibinfo {eid} {76} (\bibinfo {year}
  {2020}{\natexlab{a}})},\ \Eprint {https://arxiv.org/abs/1909.01371}
  {arXiv:1909.01371 [astro-ph.HE]} \BibitemShut {NoStop}%
\bibitem [{\citenamefont {{Woosley}}\ and\ \citenamefont
  {{Heger}}(2021{\natexlab{a}})}]{2021ApJ...912L..31W}%
  \BibitemOpen
  \bibfield  {author} {\bibinfo {author} {\bibfnamefont {S.~E.}\ \bibnamefont
  {{Woosley}}}\ and\ \bibinfo {author} {\bibfnamefont {A.}~\bibnamefont
  {{Heger}}},\ }\bibfield  {title} {\bibinfo {title} {{The Pair-instability
  Mass Gap for Black Holes}},\ }\href
  {https://doi.org/10.3847/2041-8213/abf2c4} {\bibfield  {journal} {\bibinfo
  {journal} {\apjl}\ }\textbf {\bibinfo {volume} {912}},\ \bibinfo {eid} {L31}
  (\bibinfo {year} {2021}{\natexlab{a}})},\ \Eprint
  {https://arxiv.org/abs/2103.07933} {arXiv:2103.07933 [astro-ph.SR]}
  \BibitemShut {NoStop}%
\bibitem [{\citenamefont {{Iorio}}\ \emph {et~al.}(2023)\citenamefont
  {{Iorio}}, \citenamefont {{Mapelli}}, \citenamefont {{Costa}}, \citenamefont
  {{Spera}}, \citenamefont {{Escobar}}, \citenamefont {{Sgalletta}},
  \citenamefont {{Trani}}, \citenamefont {{Korb}}, \citenamefont
  {{Santoliquido}}, \citenamefont {{Dall'Amico}}, \citenamefont {{Gaspari}},\
  and\ \citenamefont {{Bressan}}}]{2023MNRAS.524..426I}%
  \BibitemOpen
  \bibfield  {author} {\bibinfo {author} {\bibfnamefont {G.}~\bibnamefont
  {{Iorio}}}, \bibinfo {author} {\bibfnamefont {M.}~\bibnamefont {{Mapelli}}},
  \bibinfo {author} {\bibfnamefont {G.}~\bibnamefont {{Costa}}}, \bibinfo
  {author} {\bibfnamefont {M.}~\bibnamefont {{Spera}}}, \bibinfo {author}
  {\bibfnamefont {G.~J.}\ \bibnamefont {{Escobar}}}, \bibinfo {author}
  {\bibfnamefont {C.}~\bibnamefont {{Sgalletta}}}, \bibinfo {author}
  {\bibfnamefont {A.~A.}\ \bibnamefont {{Trani}}}, \bibinfo {author}
  {\bibfnamefont {E.}~\bibnamefont {{Korb}}}, \bibinfo {author} {\bibfnamefont
  {F.}~\bibnamefont {{Santoliquido}}}, \bibinfo {author} {\bibfnamefont
  {M.}~\bibnamefont {{Dall'Amico}}}, \bibinfo {author} {\bibfnamefont
  {N.}~\bibnamefont {{Gaspari}}},\ and\ \bibinfo {author} {\bibfnamefont
  {A.}~\bibnamefont {{Bressan}}},\ }\bibfield  {title} {\bibinfo {title}
  {{Compact object mergers: exploring uncertainties from stellar and binary
  evolution with SEVN}},\ }\href {https://doi.org/10.1093/mnras/stad1630}
  {\bibfield  {journal} {\bibinfo  {journal} {\mnras}\ }\textbf {\bibinfo
  {volume} {524}},\ \bibinfo {pages} {426} (\bibinfo {year} {2023})},\ \Eprint
  {https://arxiv.org/abs/2211.11774} {arXiv:2211.11774 [astro-ph.HE]}
  \BibitemShut {NoStop}%
\bibitem [{\citenamefont {{Fragos}}\ \emph {et~al.}(2023)\citenamefont
  {{Fragos}}, \citenamefont {{Andrews}}, \citenamefont {{Bavera}},
  \citenamefont {{Berry}}, \citenamefont {{Coughlin}}, \citenamefont
  {{Dotter}}, \citenamefont {{Giri}}, \citenamefont {{Kalogera}}, \citenamefont
  {{Katsaggelos}}, \citenamefont {{Kovlakas}}, \citenamefont {{Lalvani}},
  \citenamefont {{Misra}}, \citenamefont {{Srivastava}}, \citenamefont {{Qin}},
  \citenamefont {{Rocha}}, \citenamefont {{Rom{\'a}n-Garza}}, \citenamefont
  {{Serra}}, \citenamefont {{Stahle}}, \citenamefont {{Sun}}, \citenamefont
  {{Teng}}, \citenamefont {{Trajcevski}}, \citenamefont {{Tran}}, \citenamefont
  {{Xing}}, \citenamefont {{Zapartas}},\ and\ \citenamefont
  {{Zevin}}}]{2023ApJS..264...45F}%
  \BibitemOpen
  \bibfield  {author} {\bibinfo {author} {\bibfnamefont {T.}~\bibnamefont
  {{Fragos}}}, \bibinfo {author} {\bibfnamefont {J.~J.}\ \bibnamefont
  {{Andrews}}}, \bibinfo {author} {\bibfnamefont {S.~S.}\ \bibnamefont
  {{Bavera}}}, \bibinfo {author} {\bibfnamefont {C.~P.~L.}\ \bibnamefont
  {{Berry}}}, \bibinfo {author} {\bibfnamefont {S.}~\bibnamefont {{Coughlin}}},
  \bibinfo {author} {\bibfnamefont {A.}~\bibnamefont {{Dotter}}}, \bibinfo
  {author} {\bibfnamefont {P.}~\bibnamefont {{Giri}}}, \bibinfo {author}
  {\bibfnamefont {V.}~\bibnamefont {{Kalogera}}}, \bibinfo {author}
  {\bibfnamefont {A.}~\bibnamefont {{Katsaggelos}}}, \bibinfo {author}
  {\bibfnamefont {K.}~\bibnamefont {{Kovlakas}}}, \bibinfo {author}
  {\bibfnamefont {S.}~\bibnamefont {{Lalvani}}}, \bibinfo {author}
  {\bibfnamefont {D.}~\bibnamefont {{Misra}}}, \bibinfo {author} {\bibfnamefont
  {P.~M.}\ \bibnamefont {{Srivastava}}}, \bibinfo {author} {\bibfnamefont
  {Y.}~\bibnamefont {{Qin}}}, \bibinfo {author} {\bibfnamefont {K.~A.}\
  \bibnamefont {{Rocha}}}, \bibinfo {author} {\bibfnamefont {J.}~\bibnamefont
  {{Rom{\'a}n-Garza}}}, \bibinfo {author} {\bibfnamefont {J.~G.}\ \bibnamefont
  {{Serra}}}, \bibinfo {author} {\bibfnamefont {P.}~\bibnamefont {{Stahle}}},
  \bibinfo {author} {\bibfnamefont {M.}~\bibnamefont {{Sun}}}, \bibinfo
  {author} {\bibfnamefont {X.}~\bibnamefont {{Teng}}}, \bibinfo {author}
  {\bibfnamefont {G.}~\bibnamefont {{Trajcevski}}}, \bibinfo {author}
  {\bibfnamefont {N.~H.}\ \bibnamefont {{Tran}}}, \bibinfo {author}
  {\bibfnamefont {Z.}~\bibnamefont {{Xing}}}, \bibinfo {author} {\bibfnamefont
  {E.}~\bibnamefont {{Zapartas}}},\ and\ \bibinfo {author} {\bibfnamefont
  {M.}~\bibnamefont {{Zevin}}},\ }\bibfield  {title} {\bibinfo {title}
  {{POSYDON: A General-purpose Population Synthesis Code with Detailed
  Binary-evolution Simulations}},\ }\href
  {https://doi.org/10.3847/1538-4365/ac90c1} {\bibfield  {journal} {\bibinfo
  {journal} {\apjs}\ }\textbf {\bibinfo {volume} {264}},\ \bibinfo {eid} {45}
  (\bibinfo {year} {2023})},\ \Eprint {https://arxiv.org/abs/2202.05892}
  {arXiv:2202.05892 [astro-ph.SR]} \BibitemShut {NoStop}%
\bibitem [{\citenamefont {{Abbott}}\ \emph
  {et~al.}(2020{\natexlab{a}})\citenamefont {{Abbott}}, \citenamefont
  {{Abbott}}, \citenamefont {{Abraham}}, \citenamefont {{Acernese}},
  \citenamefont {{Ackley}}, \citenamefont {{Adams}}, \citenamefont
  {{Adhikari}}, \citenamefont {{Adya}}, \citenamefont {{Affeldt}},
  \citenamefont {{Agathos}}, \citenamefont {{Agatsuma}}, \citenamefont
  {{Aggarwal}}, \citenamefont {{Aguiar}}, \citenamefont {{Aich}}, \citenamefont
  {{Aiello}}, \citenamefont {{Ain}}, \citenamefont {{LIGO Scientific
  Collaboration}},\ and\ \citenamefont {{Virgo
  Collaboration}}}]{2020ApJ...900L..13A}%
  \BibitemOpen
  \bibfield  {author} {\bibinfo {author} {\bibfnamefont {R.}~\bibnamefont
  {{Abbott}}}, \bibinfo {author} {\bibfnamefont {T.~D.}\ \bibnamefont
  {{Abbott}}}, \bibinfo {author} {\bibfnamefont {S.}~\bibnamefont {{Abraham}}},
  \bibinfo {author} {\bibfnamefont {F.}~\bibnamefont {{Acernese}}}, \bibinfo
  {author} {\bibfnamefont {K.}~\bibnamefont {{Ackley}}}, \bibinfo {author}
  {\bibfnamefont {C.}~\bibnamefont {{Adams}}}, \bibinfo {author} {\bibfnamefont
  {R.~X.}\ \bibnamefont {{Adhikari}}}, \bibinfo {author} {\bibfnamefont
  {V.~B.}\ \bibnamefont {{Adya}}}, \bibinfo {author} {\bibfnamefont
  {C.}~\bibnamefont {{Affeldt}}}, \bibinfo {author} {\bibfnamefont
  {M.}~\bibnamefont {{Agathos}}}, \bibinfo {author} {\bibfnamefont
  {K.}~\bibnamefont {{Agatsuma}}}, \bibinfo {author} {\bibfnamefont
  {N.}~\bibnamefont {{Aggarwal}}}, \bibinfo {author} {\bibfnamefont {O.~D.}\
  \bibnamefont {{Aguiar}}}, \bibinfo {author} {\bibfnamefont {A.}~\bibnamefont
  {{Aich}}}, \bibinfo {author} {\bibfnamefont {L.}~\bibnamefont {{Aiello}}},
  \bibinfo {author} {\bibfnamefont {A.}~\bibnamefont {{Ain}}}, \bibinfo
  {author} {\bibnamefont {{LIGO Scientific Collaboration}}},\ and\ \bibinfo
  {author} {\bibnamefont {{Virgo Collaboration}}},\ }\bibfield  {title}
  {\bibinfo {title} {{Properties and Astrophysical Implications of the 150
  M$_{{\ensuremath{\odot}}}$ Binary Black Hole Merger GW190521}},\ }\href
  {https://doi.org/10.3847/2041-8213/aba493} {\bibfield  {journal} {\bibinfo
  {journal} {\apjl}\ }\textbf {\bibinfo {volume} {900}},\ \bibinfo {eid} {L13}
  (\bibinfo {year} {2020}{\natexlab{a}})},\ \Eprint
  {https://arxiv.org/abs/2009.01190} {arXiv:2009.01190 [astro-ph.HE]}
  \BibitemShut {NoStop}%
\bibitem [{\citenamefont {{Abbott}}\ \emph
  {et~al.}(2020{\natexlab{b}})\citenamefont {{Abbott}}, \citenamefont
  {{Abbott}}, \citenamefont {{Abraham}}, \citenamefont {{Acernese}},
  \citenamefont {{Ackley}}, \citenamefont {{Adams}}, \citenamefont
  {{Adhikari}}, \citenamefont {{Adya}}, \citenamefont {{Affeldt}},
  \citenamefont {{Agathos}}, \citenamefont {{Agatsuma}}, \citenamefont
  {{Aggarwal}}, \citenamefont {{Aguiar}}, \citenamefont {{Aich}}, \citenamefont
  {{Aiello}}, \citenamefont {{Ain}}, \citenamefont {{Ajith}}, \citenamefont
  {{Akcay}}, \citenamefont {{Allen}}, \citenamefont {{Allocca}}, \citenamefont
  {{Altin}}, \citenamefont {{Amato}}, \citenamefont {{Anand}}, \citenamefont
  {{LIGO Scientific Collaboration}},\ and\ \citenamefont {{Virgo
  Collaboration}}}]{2020PhRvL.125j1102A}%
  \BibitemOpen
  \bibfield  {author} {\bibinfo {author} {\bibfnamefont {R.}~\bibnamefont
  {{Abbott}}}, \bibinfo {author} {\bibfnamefont {T.~D.}\ \bibnamefont
  {{Abbott}}}, \bibinfo {author} {\bibfnamefont {S.}~\bibnamefont {{Abraham}}},
  \bibinfo {author} {\bibfnamefont {F.}~\bibnamefont {{Acernese}}}, \bibinfo
  {author} {\bibfnamefont {K.}~\bibnamefont {{Ackley}}}, \bibinfo {author}
  {\bibfnamefont {C.}~\bibnamefont {{Adams}}}, \bibinfo {author} {\bibfnamefont
  {R.~X.}\ \bibnamefont {{Adhikari}}}, \bibinfo {author} {\bibfnamefont
  {V.~B.}\ \bibnamefont {{Adya}}}, \bibinfo {author} {\bibfnamefont
  {C.}~\bibnamefont {{Affeldt}}}, \bibinfo {author} {\bibfnamefont
  {M.}~\bibnamefont {{Agathos}}}, \bibinfo {author} {\bibfnamefont
  {K.}~\bibnamefont {{Agatsuma}}}, \bibinfo {author} {\bibfnamefont
  {N.}~\bibnamefont {{Aggarwal}}}, \bibinfo {author} {\bibfnamefont {O.~D.}\
  \bibnamefont {{Aguiar}}}, \bibinfo {author} {\bibfnamefont {A.}~\bibnamefont
  {{Aich}}}, \bibinfo {author} {\bibfnamefont {L.}~\bibnamefont {{Aiello}}},
  \bibinfo {author} {\bibfnamefont {A.}~\bibnamefont {{Ain}}}, \bibinfo
  {author} {\bibfnamefont {P.}~\bibnamefont {{Ajith}}}, \bibinfo {author}
  {\bibfnamefont {S.}~\bibnamefont {{Akcay}}}, \bibinfo {author} {\bibfnamefont
  {G.}~\bibnamefont {{Allen}}}, \bibinfo {author} {\bibfnamefont
  {A.}~\bibnamefont {{Allocca}}}, \bibinfo {author} {\bibfnamefont {P.~A.}\
  \bibnamefont {{Altin}}}, \bibinfo {author} {\bibfnamefont {A.}~\bibnamefont
  {{Amato}}}, \bibinfo {author} {\bibfnamefont {S.}~\bibnamefont {{Anand}}},
  \bibinfo {author} {\bibnamefont {{LIGO Scientific Collaboration}}},\ and\
  \bibinfo {author} {\bibnamefont {{Virgo Collaboration}}},\ }\bibfield
  {title} {\bibinfo {title} {{GW190521: A Binary Black Hole Merger with a Total
  Mass of 150 M$_{{\ensuremath{\odot}}}$}},\ }\href
  {https://doi.org/10.1103/PhysRevLett.125.101102} {\bibfield  {journal}
  {\bibinfo  {journal} {\prl}\ }\textbf {\bibinfo {volume} {125}},\ \bibinfo
  {eid} {101102} (\bibinfo {year} {2020}{\natexlab{b}})},\ \Eprint
  {https://arxiv.org/abs/2009.01075} {arXiv:2009.01075 [gr-qc]} \BibitemShut
  {NoStop}%
\bibitem [{\citenamefont {{Nguyen}}\ \emph {et~al.}(2022)\citenamefont
  {{Nguyen}}, \citenamefont {{Costa}}, \citenamefont {{Girardi}}, \citenamefont
  {{Volpato}}, \citenamefont {{Bressan}}, \citenamefont {{Chen}}, \citenamefont
  {{Marigo}}, \citenamefont {{Fu}},\ and\ \citenamefont
  {{Goudfrooij}}}]{2022A&A...665A.126N}%
  \BibitemOpen
  \bibfield  {author} {\bibinfo {author} {\bibfnamefont {C.~T.}\ \bibnamefont
  {{Nguyen}}}, \bibinfo {author} {\bibfnamefont {G.}~\bibnamefont {{Costa}}},
  \bibinfo {author} {\bibfnamefont {L.}~\bibnamefont {{Girardi}}}, \bibinfo
  {author} {\bibfnamefont {G.}~\bibnamefont {{Volpato}}}, \bibinfo {author}
  {\bibfnamefont {A.}~\bibnamefont {{Bressan}}}, \bibinfo {author}
  {\bibfnamefont {Y.}~\bibnamefont {{Chen}}}, \bibinfo {author} {\bibfnamefont
  {P.}~\bibnamefont {{Marigo}}}, \bibinfo {author} {\bibfnamefont
  {X.}~\bibnamefont {{Fu}}},\ and\ \bibinfo {author} {\bibfnamefont
  {P.}~\bibnamefont {{Goudfrooij}}},\ }\bibfield  {title} {\bibinfo {title}
  {{PARSEC V2.0: Stellar tracks and isochrones of low- and intermediate-mass
  stars with rotation}},\ }\href {https://doi.org/10.1051/0004-6361/202244166}
  {\bibfield  {journal} {\bibinfo  {journal} {\aap}\ }\textbf {\bibinfo
  {volume} {665}},\ \bibinfo {eid} {A126} (\bibinfo {year} {2022})},\ \Eprint
  {https://arxiv.org/abs/2207.08642} {arXiv:2207.08642 [astro-ph.SR]}
  \BibitemShut {NoStop}%
\bibitem [{\citenamefont {{Fryer}}\ \emph {et~al.}(2012)\citenamefont
  {{Fryer}}, \citenamefont {{Belczynski}}, \citenamefont {{Wiktorowicz}},
  \citenamefont {{Dominik}}, \citenamefont {{Kalogera}},\ and\ \citenamefont
  {{Holz}}}]{2012ApJ...749...91F}%
  \BibitemOpen
  \bibfield  {author} {\bibinfo {author} {\bibfnamefont {C.~L.}\ \bibnamefont
  {{Fryer}}}, \bibinfo {author} {\bibfnamefont {K.}~\bibnamefont
  {{Belczynski}}}, \bibinfo {author} {\bibfnamefont {G.}~\bibnamefont
  {{Wiktorowicz}}}, \bibinfo {author} {\bibfnamefont {M.}~\bibnamefont
  {{Dominik}}}, \bibinfo {author} {\bibfnamefont {V.}~\bibnamefont
  {{Kalogera}}},\ and\ \bibinfo {author} {\bibfnamefont {D.~E.}\ \bibnamefont
  {{Holz}}},\ }\bibfield  {title} {\bibinfo {title} {{Compact Remnant Mass
  Function: Dependence on the Explosion Mechanism and Metallicity}},\ }\href
  {https://doi.org/10.1088/0004-637X/749/1/91} {\bibfield  {journal} {\bibinfo
  {journal} {\apj}\ }\textbf {\bibinfo {volume} {749}},\ \bibinfo {eid} {91}
  (\bibinfo {year} {2012})},\ \Eprint {https://arxiv.org/abs/1110.1726}
  {arXiv:1110.1726 [astro-ph.SR]} \BibitemShut {NoStop}%
\bibitem [{\citenamefont {{Chabrier}}\ and\ \citenamefont
  {{Baraffe}}(2000)}]{2000ARA&A..38..337C}%
  \BibitemOpen
  \bibfield  {author} {\bibinfo {author} {\bibfnamefont {G.}~\bibnamefont
  {{Chabrier}}}\ and\ \bibinfo {author} {\bibfnamefont {I.}~\bibnamefont
  {{Baraffe}}},\ }\bibfield  {title} {\bibinfo {title} {{Theory of Low-Mass
  Stars and Substellar Objects}},\ }\href
  {https://doi.org/10.1146/annurev.astro.38.1.337} {\bibfield  {journal}
  {\bibinfo  {journal} {\araa}\ }\textbf {\bibinfo {volume} {38}},\ \bibinfo
  {pages} {337} (\bibinfo {year} {2000})},\ \Eprint
  {https://arxiv.org/abs/astro-ph/0006383} {arXiv:astro-ph/0006383 [astro-ph]}
  \BibitemShut {NoStop}%
\bibitem [{\citenamefont {{Chantereau}}\ \emph {et~al.}(2015)\citenamefont
  {{Chantereau}}, \citenamefont {{Charbonnel}},\ and\ \citenamefont
  {{Decressin}}}]{2015A&A...578A.117C}%
  \BibitemOpen
  \bibfield  {author} {\bibinfo {author} {\bibfnamefont {W.}~\bibnamefont
  {{Chantereau}}}, \bibinfo {author} {\bibfnamefont {C.}~\bibnamefont
  {{Charbonnel}}},\ and\ \bibinfo {author} {\bibfnamefont {T.}~\bibnamefont
  {{Decressin}}},\ }\bibfield  {title} {\bibinfo {title} {{Evolution of
  long-lived globular cluster stars. I. Grid of stellar models with helium
  enhancement at [Fe/H] = -1.75}},\ }\href
  {https://doi.org/10.1051/0004-6361/201525929} {\bibfield  {journal} {\bibinfo
   {journal} {\aap}\ }\textbf {\bibinfo {volume} {578}},\ \bibinfo {eid} {A117}
  (\bibinfo {year} {2015})},\ \Eprint {https://arxiv.org/abs/1504.01878}
  {arXiv:1504.01878 [astro-ph.SR]} \BibitemShut {NoStop}%
\bibitem [{\citenamefont {{Salpeter}}(1955)}]{1955ApJ...121..161S}%
  \BibitemOpen
  \bibfield  {author} {\bibinfo {author} {\bibfnamefont {E.~E.}\ \bibnamefont
  {{Salpeter}}},\ }\bibfield  {title} {\bibinfo {title} {{The Luminosity
  Function and Stellar Evolution.}},\ }\href {https://doi.org/10.1086/145971}
  {\bibfield  {journal} {\bibinfo  {journal} {\apj}\ }\textbf {\bibinfo
  {volume} {121}},\ \bibinfo {pages} {161} (\bibinfo {year}
  {1955})}\BibitemShut {NoStop}%
\bibitem [{\citenamefont {{Kroupa}}(2001{\natexlab{a}})}]{2001MNRAS.322..231K}%
  \BibitemOpen
  \bibfield  {author} {\bibinfo {author} {\bibfnamefont {P.}~\bibnamefont
  {{Kroupa}}},\ }\bibfield  {title} {\bibinfo {title} {{On the variation of the
  initial mass function}},\ }\href
  {https://doi.org/10.1046/j.1365-8711.2001.04022.x} {\bibfield  {journal}
  {\bibinfo  {journal} {\mnras}\ }\textbf {\bibinfo {volume} {322}},\ \bibinfo
  {pages} {231} (\bibinfo {year} {2001}{\natexlab{a}})},\ \Eprint
  {https://arxiv.org/abs/astro-ph/0009005} {arXiv:astro-ph/0009005 [astro-ph]}
  \BibitemShut {NoStop}%
\bibitem [{\citenamefont {{Nomoto}}(1984)}]{1984ApJ...277..791N}%
  \BibitemOpen
  \bibfield  {author} {\bibinfo {author} {\bibfnamefont {K.}~\bibnamefont
  {{Nomoto}}},\ }\bibfield  {title} {\bibinfo {title} {{Evolution of 8-10 solar
  mass stars toward electron capture supernovae. I - Formation of
  electron-degenerate O + NE + MG cores.}},\ }\href
  {https://doi.org/10.1086/161749} {\bibfield  {journal} {\bibinfo  {journal}
  {\apj}\ }\textbf {\bibinfo {volume} {277}},\ \bibinfo {pages} {791} (\bibinfo
  {year} {1984})}\BibitemShut {NoStop}%
\bibitem [{\citenamefont {{Siess}}(2007)}]{2007A&A...476..893S}%
  \BibitemOpen
  \bibfield  {author} {\bibinfo {author} {\bibfnamefont {L.}~\bibnamefont
  {{Siess}}},\ }\bibfield  {title} {\bibinfo {title} {{Evolution of massive AGB
  stars. II. model properties at non-solar metallicity and the fate of
  Super-AGB stars}},\ }\href {https://doi.org/10.1051/0004-6361:20078132}
  {\bibfield  {journal} {\bibinfo  {journal} {\aap}\ }\textbf {\bibinfo
  {volume} {476}},\ \bibinfo {pages} {893} (\bibinfo {year}
  {2007})}\BibitemShut {NoStop}%
\bibitem [{\citenamefont {{Poelarends}}\ \emph {et~al.}(2008)\citenamefont
  {{Poelarends}}, \citenamefont {{Herwig}}, \citenamefont {{Langer}},\ and\
  \citenamefont {{Heger}}}]{2008ApJ...675..614P}%
  \BibitemOpen
  \bibfield  {author} {\bibinfo {author} {\bibfnamefont {A.~J.~T.}\
  \bibnamefont {{Poelarends}}}, \bibinfo {author} {\bibfnamefont
  {F.}~\bibnamefont {{Herwig}}}, \bibinfo {author} {\bibfnamefont
  {N.}~\bibnamefont {{Langer}}},\ and\ \bibinfo {author} {\bibfnamefont
  {A.}~\bibnamefont {{Heger}}},\ }\bibfield  {title} {\bibinfo {title} {{The
  Supernova Channel of Super-AGB Stars}},\ }\href
  {https://doi.org/10.1086/520872} {\bibfield  {journal} {\bibinfo  {journal}
  {\apj}\ }\textbf {\bibinfo {volume} {675}},\ \bibinfo {pages} {614} (\bibinfo
  {year} {2008})},\ \Eprint {https://arxiv.org/abs/0705.4643} {arXiv:0705.4643
  [astro-ph]} \BibitemShut {NoStop}%
\bibitem [{\citenamefont {{Woosley}}(2017)}]{2017ApJ...836..244W}%
  \BibitemOpen
  \bibfield  {author} {\bibinfo {author} {\bibfnamefont {S.~E.}\ \bibnamefont
  {{Woosley}}},\ }\bibfield  {title} {\bibinfo {title} {{Pulsational
  Pair-instability Supernovae}},\ }\href
  {https://doi.org/10.3847/1538-4357/836/2/244} {\bibfield  {journal} {\bibinfo
   {journal} {\apj}\ }\textbf {\bibinfo {volume} {836}},\ \bibinfo {eid} {244}
  (\bibinfo {year} {2017})},\ \Eprint {https://arxiv.org/abs/1608.08939}
  {arXiv:1608.08939 [astro-ph.HE]} \BibitemShut {NoStop}%
\bibitem [{\citenamefont {{Crowther}}\ \emph
  {et~al.}(2010{\natexlab{a}})\citenamefont {{Crowther}}, \citenamefont
  {{Schnurr}}, \citenamefont {{Hirschi}}, \citenamefont {{Yusof}},
  \citenamefont {{Parker}}, \citenamefont {{Goodwin}},\ and\ \citenamefont
  {{Kassim}}}]{2010MNRAS.408..731C}%
  \BibitemOpen
  \bibfield  {author} {\bibinfo {author} {\bibfnamefont {P.~A.}\ \bibnamefont
  {{Crowther}}}, \bibinfo {author} {\bibfnamefont {O.}~\bibnamefont
  {{Schnurr}}}, \bibinfo {author} {\bibfnamefont {R.}~\bibnamefont
  {{Hirschi}}}, \bibinfo {author} {\bibfnamefont {N.}~\bibnamefont {{Yusof}}},
  \bibinfo {author} {\bibfnamefont {R.~J.}\ \bibnamefont {{Parker}}}, \bibinfo
  {author} {\bibfnamefont {S.~P.}\ \bibnamefont {{Goodwin}}},\ and\ \bibinfo
  {author} {\bibfnamefont {H.~A.}\ \bibnamefont {{Kassim}}},\ }\bibfield
  {title} {\bibinfo {title} {{The R136 star cluster hosts several stars whose
  individual masses greatly exceed the accepted 150M$_{solar}$ stellar mass
  limit}},\ }\href {https://doi.org/10.1111/j.1365-2966.2010.17167.x}
  {\bibfield  {journal} {\bibinfo  {journal} {\mnras}\ }\textbf {\bibinfo
  {volume} {408}},\ \bibinfo {pages} {731} (\bibinfo {year}
  {2010}{\natexlab{a}})},\ \Eprint {https://arxiv.org/abs/1007.3284}
  {arXiv:1007.3284 [astro-ph.SR]} \BibitemShut {NoStop}%
\bibitem [{\citenamefont {{Martins}}\ and\ \citenamefont
  {{Palacios}}(2013)}]{2013A&A...560A..16M}%
  \BibitemOpen
  \bibfield  {author} {\bibinfo {author} {\bibfnamefont {F.}~\bibnamefont
  {{Martins}}}\ and\ \bibinfo {author} {\bibfnamefont {A.}~\bibnamefont
  {{Palacios}}},\ }\bibfield  {title} {\bibinfo {title} {{A comparison of
  evolutionary tracks for single Galactic massive stars}},\ }\href
  {https://doi.org/10.1051/0004-6361/201322480} {\bibfield  {journal} {\bibinfo
   {journal} {\aap}\ }\textbf {\bibinfo {volume} {560}},\ \bibinfo {eid} {A16}
  (\bibinfo {year} {2013})},\ \Eprint {https://arxiv.org/abs/1310.7218}
  {arXiv:1310.7218 [astro-ph.SR]} \BibitemShut {NoStop}%
\bibitem [{\citenamefont {{Jones}}\ \emph {et~al.}(2015)\citenamefont
  {{Jones}}, \citenamefont {{Hirschi}}, \citenamefont {{Pignatari}},
  \citenamefont {{Heger}}, \citenamefont {{Georgy}}, \citenamefont
  {{Nishimura}}, \citenamefont {{Fryer}},\ and\ \citenamefont
  {{Herwig}}}]{2015MNRAS.447.3115J}%
  \BibitemOpen
  \bibfield  {author} {\bibinfo {author} {\bibfnamefont {S.}~\bibnamefont
  {{Jones}}}, \bibinfo {author} {\bibfnamefont {R.}~\bibnamefont {{Hirschi}}},
  \bibinfo {author} {\bibfnamefont {M.}~\bibnamefont {{Pignatari}}}, \bibinfo
  {author} {\bibfnamefont {A.}~\bibnamefont {{Heger}}}, \bibinfo {author}
  {\bibfnamefont {C.}~\bibnamefont {{Georgy}}}, \bibinfo {author}
  {\bibfnamefont {N.}~\bibnamefont {{Nishimura}}}, \bibinfo {author}
  {\bibfnamefont {C.}~\bibnamefont {{Fryer}}},\ and\ \bibinfo {author}
  {\bibfnamefont {F.}~\bibnamefont {{Herwig}}},\ }\bibfield  {title} {\bibinfo
  {title} {{Code dependencies of pre-supernova evolution and nucleosynthesis in
  massive stars: evolution to the end of core helium burning}},\ }\href
  {https://doi.org/10.1093/mnras/stu2657} {\bibfield  {journal} {\bibinfo
  {journal} {\mnras}\ }\textbf {\bibinfo {volume} {447}},\ \bibinfo {pages}
  {3115} (\bibinfo {year} {2015})},\ \Eprint {https://arxiv.org/abs/1412.6518}
  {arXiv:1412.6518 [astro-ph.SR]} \BibitemShut {NoStop}%
\bibitem [{\citenamefont {{Agrawal}}\ \emph {et~al.}(2020)\citenamefont
  {{Agrawal}}, \citenamefont {{Hurley}}, \citenamefont {{Stevenson}},
  \citenamefont {{Sz{\'e}csi}},\ and\ \citenamefont
  {{Flynn}}}]{2020MNRAS.497.4549A}%
  \BibitemOpen
  \bibfield  {author} {\bibinfo {author} {\bibfnamefont {P.}~\bibnamefont
  {{Agrawal}}}, \bibinfo {author} {\bibfnamefont {J.}~\bibnamefont {{Hurley}}},
  \bibinfo {author} {\bibfnamefont {S.}~\bibnamefont {{Stevenson}}}, \bibinfo
  {author} {\bibfnamefont {D.}~\bibnamefont {{Sz{\'e}csi}}},\ and\ \bibinfo
  {author} {\bibfnamefont {C.}~\bibnamefont {{Flynn}}},\ }\bibfield  {title}
  {\bibinfo {title} {{The fates of massive stars: exploring uncertainties in
  stellar evolution with METISSE}},\ }\href
  {https://doi.org/10.1093/mnras/staa2264} {\bibfield  {journal} {\bibinfo
  {journal} {\mnras}\ }\textbf {\bibinfo {volume} {497}},\ \bibinfo {pages}
  {4549} (\bibinfo {year} {2020})},\ \Eprint {https://arxiv.org/abs/2005.13177}
  {arXiv:2005.13177 [astro-ph.SR]} \BibitemShut {NoStop}%
\bibitem [{\citenamefont {{Hertzsprung}}(1911)}]{1911POPot..22A...1H}%
  \BibitemOpen
  \bibfield  {author} {\bibinfo {author} {\bibfnamefont {E.}~\bibnamefont
  {{Hertzsprung}}},\ }\bibfield  {title} {\bibinfo {title} {{Number 63.
  Zweiundzwanzigsten Bandes Erstes Stuck. Uber die verwendung photographischer
  effecktiver wellenlangen zur bestimmung von farbenaquivalenten}},\
  }\href@noop {} {\bibfield  {journal} {\bibinfo  {journal} {Publikationen des
  Astrophysikalischen Observatoriums zu Potsdam}\ }\textbf {\bibinfo {volume}
  {22}},\ \bibinfo {pages} {A1} (\bibinfo {year} {1911})}\BibitemShut {NoStop}%
\bibitem [{\citenamefont {{Russell}}(1914)}]{1914PA.....22..275R}%
  \BibitemOpen
  \bibfield  {author} {\bibinfo {author} {\bibfnamefont {H.~N.}\ \bibnamefont
  {{Russell}}},\ }\bibfield  {title} {\bibinfo {title} {{Relations Between the
  Spectra and Other Characteristics of the Stars}},\ }\href@noop {} {\bibfield
  {journal} {\bibinfo  {journal} {Popular Astronomy}\ }\textbf {\bibinfo
  {volume} {22}},\ \bibinfo {pages} {275} (\bibinfo {year} {1914})}\BibitemShut
  {NoStop}%
\bibitem [{\citenamefont {{Costa}}\ \emph {et~al.}(2021)\citenamefont
  {{Costa}}, \citenamefont {{Bressan}}, \citenamefont {{Mapelli}},
  \citenamefont {{Marigo}}, \citenamefont {{Iorio}},\ and\ \citenamefont
  {{Spera}}}]{2021MNRAS.501.4514C}%
  \BibitemOpen
  \bibfield  {author} {\bibinfo {author} {\bibfnamefont {G.}~\bibnamefont
  {{Costa}}}, \bibinfo {author} {\bibfnamefont {A.}~\bibnamefont {{Bressan}}},
  \bibinfo {author} {\bibfnamefont {M.}~\bibnamefont {{Mapelli}}}, \bibinfo
  {author} {\bibfnamefont {P.}~\bibnamefont {{Marigo}}}, \bibinfo {author}
  {\bibfnamefont {G.}~\bibnamefont {{Iorio}}},\ and\ \bibinfo {author}
  {\bibfnamefont {M.}~\bibnamefont {{Spera}}},\ }\bibfield  {title} {\bibinfo
  {title} {{Formation of GW190521 from stellar evolution: the impact of the
  hydrogen-rich envelope, dredge-up, and $^{12}$C({\ensuremath{\alpha}},
  {\ensuremath{\gamma}})$^{16}$O rate on the pair-instability black hole mass
  gap}},\ }\href {https://doi.org/10.1093/mnras/staa3916} {\bibfield  {journal}
  {\bibinfo  {journal} {\mnras}\ }\textbf {\bibinfo {volume} {501}},\ \bibinfo
  {pages} {4514} (\bibinfo {year} {2021})},\ \Eprint
  {https://arxiv.org/abs/2010.02242} {arXiv:2010.02242 [astro-ph.SR]}
  \BibitemShut {NoStop}%
\bibitem [{\citenamefont {{Chiosi}}\ \emph {et~al.}(1993)\citenamefont
  {{Chiosi}}, \citenamefont {{Wood}},\ and\ \citenamefont
  {{Capitanio}}}]{1993ApJS...86..541C}%
  \BibitemOpen
  \bibfield  {author} {\bibinfo {author} {\bibfnamefont {C.}~\bibnamefont
  {{Chiosi}}}, \bibinfo {author} {\bibfnamefont {P.~R.}\ \bibnamefont
  {{Wood}}},\ and\ \bibinfo {author} {\bibfnamefont {N.}~\bibnamefont
  {{Capitanio}}},\ }\bibfield  {title} {\bibinfo {title} {{Theoretical Models
  of Cepheid Variables and Their BVI C Colors and Magnitudes}},\ }\href
  {https://doi.org/10.1086/191790} {\bibfield  {journal} {\bibinfo  {journal}
  {\apjs}\ }\textbf {\bibinfo {volume} {86}},\ \bibinfo {pages} {541} (\bibinfo
  {year} {1993})}\BibitemShut {NoStop}%
\bibitem [{\citenamefont {{Freedman}}\ \emph {et~al.}(2001)\citenamefont
  {{Freedman}}, \citenamefont {{Madore}}, \citenamefont {{Gibson}},
  \citenamefont {{Ferrarese}}, \citenamefont {{Kelson}}, \citenamefont
  {{Sakai}}, \citenamefont {{Mould}}, \citenamefont {{Kennicutt}},
  \citenamefont {{Ford}}, \citenamefont {{Graham}}, \citenamefont {{Huchra}},
  \citenamefont {{Hughes}}, \citenamefont {{Illingworth}}, \citenamefont
  {{Macri}},\ and\ \citenamefont {{Stetson}}}]{2001ApJ...553...47F}%
  \BibitemOpen
  \bibfield  {author} {\bibinfo {author} {\bibfnamefont {W.~L.}\ \bibnamefont
  {{Freedman}}}, \bibinfo {author} {\bibfnamefont {B.~F.}\ \bibnamefont
  {{Madore}}}, \bibinfo {author} {\bibfnamefont {B.~K.}\ \bibnamefont
  {{Gibson}}}, \bibinfo {author} {\bibfnamefont {L.}~\bibnamefont
  {{Ferrarese}}}, \bibinfo {author} {\bibfnamefont {D.~D.}\ \bibnamefont
  {{Kelson}}}, \bibinfo {author} {\bibfnamefont {S.}~\bibnamefont {{Sakai}}},
  \bibinfo {author} {\bibfnamefont {J.~R.}\ \bibnamefont {{Mould}}}, \bibinfo
  {author} {\bibfnamefont {J.}~\bibnamefont {{Kennicutt}}, \bibfnamefont
  {Robert~C.}}, \bibinfo {author} {\bibfnamefont {H.~C.}\ \bibnamefont
  {{Ford}}}, \bibinfo {author} {\bibfnamefont {J.~A.}\ \bibnamefont
  {{Graham}}}, \bibinfo {author} {\bibfnamefont {J.~P.}\ \bibnamefont
  {{Huchra}}}, \bibinfo {author} {\bibfnamefont {S.~M.~G.}\ \bibnamefont
  {{Hughes}}}, \bibinfo {author} {\bibfnamefont {G.~D.}\ \bibnamefont
  {{Illingworth}}}, \bibinfo {author} {\bibfnamefont {L.~M.}\ \bibnamefont
  {{Macri}}},\ and\ \bibinfo {author} {\bibfnamefont {P.~B.}\ \bibnamefont
  {{Stetson}}},\ }\bibfield  {title} {\bibinfo {title} {{Final Results from the
  Hubble Space Telescope Key Project to Measure the Hubble Constant}},\ }\href
  {https://doi.org/10.1086/320638} {\bibfield  {journal} {\bibinfo  {journal}
  {\apj}\ }\textbf {\bibinfo {volume} {553}},\ \bibinfo {pages} {47} (\bibinfo
  {year} {2001})},\ \Eprint {https://arxiv.org/abs/astro-ph/0012376}
  {arXiv:astro-ph/0012376 [astro-ph]} \BibitemShut {NoStop}%
\bibitem [{\citenamefont {{Riess}}\ \emph {et~al.}(2021)\citenamefont
  {{Riess}}, \citenamefont {{Casertano}}, \citenamefont {{Yuan}}, \citenamefont
  {{Bowers}}, \citenamefont {{Macri}}, \citenamefont {{Zinn}},\ and\
  \citenamefont {{Scolnic}}}]{2021ApJ...908L...6R}%
  \BibitemOpen
  \bibfield  {author} {\bibinfo {author} {\bibfnamefont {A.~G.}\ \bibnamefont
  {{Riess}}}, \bibinfo {author} {\bibfnamefont {S.}~\bibnamefont
  {{Casertano}}}, \bibinfo {author} {\bibfnamefont {W.}~\bibnamefont {{Yuan}}},
  \bibinfo {author} {\bibfnamefont {J.~B.}\ \bibnamefont {{Bowers}}}, \bibinfo
  {author} {\bibfnamefont {L.}~\bibnamefont {{Macri}}}, \bibinfo {author}
  {\bibfnamefont {J.~C.}\ \bibnamefont {{Zinn}}},\ and\ \bibinfo {author}
  {\bibfnamefont {D.}~\bibnamefont {{Scolnic}}},\ }\bibfield  {title} {\bibinfo
  {title} {{Cosmic Distances Calibrated to 1\% Precision with Gaia EDR3
  Parallaxes and Hubble Space Telescope Photometry of 75 Milky Way Cepheids
  Confirm Tension with {\ensuremath{\Lambda}}CDM}},\ }\href
  {https://doi.org/10.3847/2041-8213/abdbaf} {\bibfield  {journal} {\bibinfo
  {journal} {\apjl}\ }\textbf {\bibinfo {volume} {908}},\ \bibinfo {eid} {L6}
  (\bibinfo {year} {2021})},\ \Eprint {https://arxiv.org/abs/2012.08534}
  {arXiv:2012.08534 [astro-ph.CO]} \BibitemShut {NoStop}%
\bibitem [{\citenamefont {{Hayashi}}(1961)}]{1961PASJ...13..450H}%
  \BibitemOpen
  \bibfield  {author} {\bibinfo {author} {\bibfnamefont {C.}~\bibnamefont
  {{Hayashi}}},\ }\bibfield  {title} {\bibinfo {title} {{Stellar evolution in
  early phases of gravitational contraction.}},\ }\href@noop {} {\bibfield
  {journal} {\bibinfo  {journal} {\pasj}\ }\textbf {\bibinfo {volume} {13}},\
  \bibinfo {pages} {450} (\bibinfo {year} {1961})}\BibitemShut {NoStop}%
\bibitem [{\citenamefont {{Ritossa}}\ \emph {et~al.}(1996)\citenamefont
  {{Ritossa}}, \citenamefont {{Garcia-Berro}},\ and\ \citenamefont
  {{Iben}}}]{1996ApJ...460..489R}%
  \BibitemOpen
  \bibfield  {author} {\bibinfo {author} {\bibfnamefont {C.}~\bibnamefont
  {{Ritossa}}}, \bibinfo {author} {\bibfnamefont {E.}~\bibnamefont
  {{Garcia-Berro}}},\ and\ \bibinfo {author} {\bibfnamefont {J.}~\bibnamefont
  {{Iben}}, \bibfnamefont {Icko}},\ }\bibfield  {title} {\bibinfo {title} {{On
  the Evolution of Stars That Form Electron-degenerate Cores Processed by
  Carbon Burning. II. Isotope Abundances and Thermal Pulses in a 10 M$_{sun}$
  Model with an ONe Core and Applications to Long-Period Variables, Classical
  Novae, and Accretion-induced Collapse}},\ }\href
  {https://doi.org/10.1086/176987} {\bibfield  {journal} {\bibinfo  {journal}
  {\apj}\ }\textbf {\bibinfo {volume} {460}},\ \bibinfo {pages} {489} (\bibinfo
  {year} {1996})}\BibitemShut {NoStop}%
\bibitem [{\citenamefont {{Vanbeveren}}\ \emph {et~al.}(1998)\citenamefont
  {{Vanbeveren}}, \citenamefont {{De Loore}},\ and\ \citenamefont {{Van
  Rensbergen}}}]{1998A&ARv...9...63V}%
  \BibitemOpen
  \bibfield  {author} {\bibinfo {author} {\bibfnamefont {D.}~\bibnamefont
  {{Vanbeveren}}}, \bibinfo {author} {\bibfnamefont {C.}~\bibnamefont {{De
  Loore}}},\ and\ \bibinfo {author} {\bibfnamefont {W.}~\bibnamefont {{Van
  Rensbergen}}},\ }\bibfield  {title} {\bibinfo {title} {{Massive stars}},\
  }\href {https://doi.org/10.1007/s001590050015} {\bibfield  {journal}
  {\bibinfo  {journal} {\aapr}\ }\textbf {\bibinfo {volume} {9}},\ \bibinfo
  {pages} {63} (\bibinfo {year} {1998})}\BibitemShut {NoStop}%
\bibitem [{\citenamefont {{Langer}}(2012)}]{2012ARA&A..50..107L}%
  \BibitemOpen
  \bibfield  {author} {\bibinfo {author} {\bibfnamefont {N.}~\bibnamefont
  {{Langer}}},\ }\bibfield  {title} {\bibinfo {title} {{Presupernova Evolution
  of Massive Single and Binary Stars}},\ }\href
  {https://doi.org/10.1146/annurev-astro-081811-125534} {\bibfield  {journal}
  {\bibinfo  {journal} {\araa}\ }\textbf {\bibinfo {volume} {50}},\ \bibinfo
  {pages} {107} (\bibinfo {year} {2012})},\ \Eprint
  {https://arxiv.org/abs/1206.5443} {arXiv:1206.5443 [astro-ph.SR]}
  \BibitemShut {NoStop}%
\bibitem [{\citenamefont {{Meynet}}\ \emph {et~al.}(2015)\citenamefont
  {{Meynet}}, \citenamefont {{Chomienne}}, \citenamefont {{Ekstr{\"o}m}},
  \citenamefont {{Georgy}}, \citenamefont {{Granada}}, \citenamefont {{Groh}},
  \citenamefont {{Maeder}}, \citenamefont {{Eggenberger}}, \citenamefont
  {{Levesque}},\ and\ \citenamefont {{Massey}}}]{2015A&A...575A..60M}%
  \BibitemOpen
  \bibfield  {author} {\bibinfo {author} {\bibfnamefont {G.}~\bibnamefont
  {{Meynet}}}, \bibinfo {author} {\bibfnamefont {V.}~\bibnamefont
  {{Chomienne}}}, \bibinfo {author} {\bibfnamefont {S.}~\bibnamefont
  {{Ekstr{\"o}m}}}, \bibinfo {author} {\bibfnamefont {C.}~\bibnamefont
  {{Georgy}}}, \bibinfo {author} {\bibfnamefont {A.}~\bibnamefont {{Granada}}},
  \bibinfo {author} {\bibfnamefont {J.}~\bibnamefont {{Groh}}}, \bibinfo
  {author} {\bibfnamefont {A.}~\bibnamefont {{Maeder}}}, \bibinfo {author}
  {\bibfnamefont {P.}~\bibnamefont {{Eggenberger}}}, \bibinfo {author}
  {\bibfnamefont {E.}~\bibnamefont {{Levesque}}},\ and\ \bibinfo {author}
  {\bibfnamefont {P.}~\bibnamefont {{Massey}}},\ }\bibfield  {title} {\bibinfo
  {title} {{Impact of mass-loss on the evolution and pre-supernova properties
  of red supergiants}},\ }\href {https://doi.org/10.1051/0004-6361/201424671}
  {\bibfield  {journal} {\bibinfo  {journal} {\aap}\ }\textbf {\bibinfo
  {volume} {575}},\ \bibinfo {eid} {A60} (\bibinfo {year} {2015})},\ \Eprint
  {https://arxiv.org/abs/1410.8721} {arXiv:1410.8721 [astro-ph.SR]}
  \BibitemShut {NoStop}%
\bibitem [{\citenamefont {{Wolf}}\ and\ \citenamefont
  {{Rayet}}(1867)}]{1867CRAS...65..292W}%
  \BibitemOpen
  \bibfield  {author} {\bibinfo {author} {\bibfnamefont {C.~J.~E.}\
  \bibnamefont {{Wolf}}}\ and\ \bibinfo {author} {\bibfnamefont
  {G.}~\bibnamefont {{Rayet}}},\ }\bibfield  {title} {\bibinfo {title}
  {{Spectroscopie stellaire}},\ }\href@noop {} {\bibfield  {journal} {\bibinfo
  {journal} {Academie des Sciences Paris Comptes Rendus}\ }\textbf {\bibinfo
  {volume} {65}},\ \bibinfo {pages} {292} (\bibinfo {year} {1867})}\BibitemShut
  {NoStop}%
\bibitem [{\citenamefont {{Conti}}(1975)}]{1975MSRSL...9..193C}%
  \BibitemOpen
  \bibfield  {author} {\bibinfo {author} {\bibfnamefont {P.~S.}\ \bibnamefont
  {{Conti}}},\ }\bibfield  {title} {\bibinfo {title} {{On the relationship
  between Of and WR stars.}},\ }\href@noop {} {\bibfield  {journal} {\bibinfo
  {journal} {Memoires of the Societe Royale des Sciences de Liege}\ }\textbf
  {\bibinfo {volume} {9}},\ \bibinfo {pages} {193} (\bibinfo {year}
  {1975})}\BibitemShut {NoStop}%
\bibitem [{\citenamefont {{Abbott}}\ and\ \citenamefont
  {{Conti}}(1987)}]{1987ARA&A..25..113A}%
  \BibitemOpen
  \bibfield  {author} {\bibinfo {author} {\bibfnamefont {D.~C.}\ \bibnamefont
  {{Abbott}}}\ and\ \bibinfo {author} {\bibfnamefont {P.~S.}\ \bibnamefont
  {{Conti}}},\ }\bibfield  {title} {\bibinfo {title} {{Wolf-rayet stars.}},\
  }\href {https://doi.org/10.1146/annurev.aa.25.090187.000553} {\bibfield
  {journal} {\bibinfo  {journal} {\araa}\ }\textbf {\bibinfo {volume} {25}},\
  \bibinfo {pages} {113} (\bibinfo {year} {1987})}\BibitemShut {NoStop}%
\bibitem [{\citenamefont {{Maeder}}\ and\ \citenamefont
  {{Conti}}(1994)}]{1994ARA&A..32..227M}%
  \BibitemOpen
  \bibfield  {author} {\bibinfo {author} {\bibfnamefont {A.}~\bibnamefont
  {{Maeder}}}\ and\ \bibinfo {author} {\bibfnamefont {P.~S.}\ \bibnamefont
  {{Conti}}},\ }\bibfield  {title} {\bibinfo {title} {{Massive Star Populations
  in Nearby Galaxies}},\ }\href
  {https://doi.org/10.1146/annurev.astro.32.1.227} {\bibfield  {journal}
  {\bibinfo  {journal} {\araa}\ }\textbf {\bibinfo {volume} {32}},\ \bibinfo
  {pages} {227} (\bibinfo {year} {1994})}\BibitemShut {NoStop}%
\bibitem [{\citenamefont {{Crowther}}(2007)}]{2007ARA&A..45..177C}%
  \BibitemOpen
  \bibfield  {author} {\bibinfo {author} {\bibfnamefont {P.~A.}\ \bibnamefont
  {{Crowther}}},\ }\bibfield  {title} {\bibinfo {title} {{Physical Properties
  of Wolf-Rayet Stars}},\ }\href
  {https://doi.org/10.1146/annurev.astro.45.051806.110615} {\bibfield
  {journal} {\bibinfo  {journal} {\araa}\ }\textbf {\bibinfo {volume} {45}},\
  \bibinfo {pages} {177} (\bibinfo {year} {2007})},\ \Eprint
  {https://arxiv.org/abs/astro-ph/0610356} {arXiv:astro-ph/0610356 [astro-ph]}
  \BibitemShut {NoStop}%
\bibitem [{\citenamefont {{Aadland}}\ \emph {et~al.}(2022)\citenamefont
  {{Aadland}}, \citenamefont {{Massey}}, \citenamefont {{Hillier}},
  \citenamefont {{Morrell}}, \citenamefont {{Neugent}},\ and\ \citenamefont
  {{Eldridge}}}]{2022ApJ...931..157A}%
  \BibitemOpen
  \bibfield  {author} {\bibinfo {author} {\bibfnamefont {E.}~\bibnamefont
  {{Aadland}}}, \bibinfo {author} {\bibfnamefont {P.}~\bibnamefont {{Massey}}},
  \bibinfo {author} {\bibfnamefont {D.~J.}\ \bibnamefont {{Hillier}}}, \bibinfo
  {author} {\bibfnamefont {N.~I.}\ \bibnamefont {{Morrell}}}, \bibinfo {author}
  {\bibfnamefont {K.~F.}\ \bibnamefont {{Neugent}}},\ and\ \bibinfo {author}
  {\bibfnamefont {J.~J.}\ \bibnamefont {{Eldridge}}},\ }\bibfield  {title}
  {\bibinfo {title} {{WO-type Wolf-Rayet Stars: The Last Hurrah of Massive Star
  Evolution}},\ }\href {https://doi.org/10.3847/1538-4357/ac66e7} {\bibfield
  {journal} {\bibinfo  {journal} {\apj}\ }\textbf {\bibinfo {volume} {931}},\
  \bibinfo {eid} {157} (\bibinfo {year} {2022})},\ \Eprint
  {https://arxiv.org/abs/2204.04258} {arXiv:2204.04258 [astro-ph.SR]}
  \BibitemShut {NoStop}%
\bibitem [{\citenamefont {{Fern{\'a}ndez}}\ \emph
  {et~al.}(2018{\natexlab{a}})\citenamefont {{Fern{\'a}ndez}}, \citenamefont
  {{Quataert}}, \citenamefont {{Kashiyama}},\ and\ \citenamefont
  {{Coughlin}}}]{2018MNRAS.476.2366F}%
  \BibitemOpen
  \bibfield  {author} {\bibinfo {author} {\bibfnamefont {R.}~\bibnamefont
  {{Fern{\'a}ndez}}}, \bibinfo {author} {\bibfnamefont {E.}~\bibnamefont
  {{Quataert}}}, \bibinfo {author} {\bibfnamefont {K.}~\bibnamefont
  {{Kashiyama}}},\ and\ \bibinfo {author} {\bibfnamefont {E.~R.}\ \bibnamefont
  {{Coughlin}}},\ }\bibfield  {title} {\bibinfo {title} {{Mass ejection in
  failed supernovae: variation with stellar progenitor}},\ }\href
  {https://doi.org/10.1093/mnras/sty306} {\bibfield  {journal} {\bibinfo
  {journal} {\mnras}\ }\textbf {\bibinfo {volume} {476}},\ \bibinfo {pages}
  {2366} (\bibinfo {year} {2018}{\natexlab{a}})},\ \Eprint
  {https://arxiv.org/abs/1710.01735} {arXiv:1710.01735 [astro-ph.HE]}
  \BibitemShut {NoStop}%
\bibitem [{\citenamefont {{Humphreys}}\ and\ \citenamefont
  {{Davidson}}(1994)}]{1994PASP..106.1025H}%
  \BibitemOpen
  \bibfield  {author} {\bibinfo {author} {\bibfnamefont {R.~M.}\ \bibnamefont
  {{Humphreys}}}\ and\ \bibinfo {author} {\bibfnamefont {K.}~\bibnamefont
  {{Davidson}}},\ }\bibfield  {title} {\bibinfo {title} {{The Luminous Blue
  Variables: Astrophysical Geysers}},\ }\href {https://doi.org/10.1086/133478}
  {\bibfield  {journal} {\bibinfo  {journal} {\pasp}\ }\textbf {\bibinfo
  {volume} {106}},\ \bibinfo {pages} {1025} (\bibinfo {year}
  {1994})}\BibitemShut {NoStop}%
\bibitem [{\citenamefont {{Eddington}}(1926)}]{1926ics..book.....E}%
  \BibitemOpen
  \bibfield  {author} {\bibinfo {author} {\bibfnamefont {A.~S.}\ \bibnamefont
  {{Eddington}}},\ }\href@noop {} {\emph {\bibinfo {title} {{The Internal
  Constitution of the Stars}}}}\ (\bibinfo {year} {1926})\BibitemShut {NoStop}%
\bibitem [{\citenamefont {{Bromm}}(2013)}]{2013RPPh...76k2901B}%
  \BibitemOpen
  \bibfield  {author} {\bibinfo {author} {\bibfnamefont {V.}~\bibnamefont
  {{Bromm}}},\ }\bibfield  {title} {\bibinfo {title} {{Formation of the first
  stars}},\ }\href {https://doi.org/10.1088/0034-4885/76/11/112901} {\bibfield
  {journal} {\bibinfo  {journal} {Reports on Progress in Physics}\ }\textbf
  {\bibinfo {volume} {76}},\ \bibinfo {eid} {112901} (\bibinfo {year}
  {2013})},\ \Eprint {https://arxiv.org/abs/1305.5178} {arXiv:1305.5178
  [astro-ph.CO]} \BibitemShut {NoStop}%
\bibitem [{\citenamefont {{Schaerer}}(2002)}]{2002A&A...382...28S}%
  \BibitemOpen
  \bibfield  {author} {\bibinfo {author} {\bibfnamefont {D.}~\bibnamefont
  {{Schaerer}}},\ }\bibfield  {title} {\bibinfo {title} {{On the properties of
  massive Population III stars and metal-free stellar populations}},\ }\href
  {https://doi.org/10.1051/0004-6361:20011619} {\bibfield  {journal} {\bibinfo
  {journal} {\aap}\ }\textbf {\bibinfo {volume} {382}},\ \bibinfo {pages} {28}
  (\bibinfo {year} {2002})},\ \Eprint {https://arxiv.org/abs/astro-ph/0110697}
  {arXiv:astro-ph/0110697 [astro-ph]} \BibitemShut {NoStop}%
\bibitem [{\citenamefont {{Klessen}}\ and\ \citenamefont
  {{Glover}}(2023)}]{2023ARA&A..61...65K}%
  \BibitemOpen
  \bibfield  {author} {\bibinfo {author} {\bibfnamefont {R.~S.}\ \bibnamefont
  {{Klessen}}}\ and\ \bibinfo {author} {\bibfnamefont {S.~C.~O.}\ \bibnamefont
  {{Glover}}},\ }\bibfield  {title} {\bibinfo {title} {{The First Stars:
  Formation, Properties, and Impact}},\ }\href
  {https://doi.org/10.1146/annurev-astro-071221-053453} {\bibfield  {journal}
  {\bibinfo  {journal} {\araa}\ }\textbf {\bibinfo {volume} {61}},\ \bibinfo
  {pages} {65} (\bibinfo {year} {2023})},\ \Eprint
  {https://arxiv.org/abs/2303.12500} {arXiv:2303.12500 [astro-ph.CO]}
  \BibitemShut {NoStop}%
\bibitem [{\citenamefont {{Liu}}\ and\ \citenamefont
  {{Bromm}}(2020)}]{2020ApJ...903L..40L}%
  \BibitemOpen
  \bibfield  {author} {\bibinfo {author} {\bibfnamefont {B.}~\bibnamefont
  {{Liu}}}\ and\ \bibinfo {author} {\bibfnamefont {V.}~\bibnamefont
  {{Bromm}}},\ }\bibfield  {title} {\bibinfo {title} {{The Population III
  Origin of GW190521}},\ }\href {https://doi.org/10.3847/2041-8213/abc552}
  {\bibfield  {journal} {\bibinfo  {journal} {\apjl}\ }\textbf {\bibinfo
  {volume} {903}},\ \bibinfo {eid} {L40} (\bibinfo {year} {2020})},\ \Eprint
  {https://arxiv.org/abs/2009.11447} {arXiv:2009.11447 [astro-ph.GA]}
  \BibitemShut {NoStop}%
\bibitem [{\citenamefont {{Farrell}}\ \emph {et~al.}(2021)\citenamefont
  {{Farrell}}, \citenamefont {{Groh}}, \citenamefont {{Hirschi}}, \citenamefont
  {{Murphy}}, \citenamefont {{Kaiser}}, \citenamefont {{Ekstr{\"o}m}},
  \citenamefont {{Georgy}},\ and\ \citenamefont
  {{Meynet}}}]{2021MNRAS.502L..40F}%
  \BibitemOpen
  \bibfield  {author} {\bibinfo {author} {\bibfnamefont {E.}~\bibnamefont
  {{Farrell}}}, \bibinfo {author} {\bibfnamefont {J.~H.}\ \bibnamefont
  {{Groh}}}, \bibinfo {author} {\bibfnamefont {R.}~\bibnamefont {{Hirschi}}},
  \bibinfo {author} {\bibfnamefont {L.}~\bibnamefont {{Murphy}}}, \bibinfo
  {author} {\bibfnamefont {E.}~\bibnamefont {{Kaiser}}}, \bibinfo {author}
  {\bibfnamefont {S.}~\bibnamefont {{Ekstr{\"o}m}}}, \bibinfo {author}
  {\bibfnamefont {C.}~\bibnamefont {{Georgy}}},\ and\ \bibinfo {author}
  {\bibfnamefont {G.}~\bibnamefont {{Meynet}}},\ }\bibfield  {title} {\bibinfo
  {title} {{Is GW190521 the merger of black holes from the first stellar
  generations?}},\ }\href {https://doi.org/10.1093/mnrasl/slaa196} {\bibfield
  {journal} {\bibinfo  {journal} {\mnras}\ }\textbf {\bibinfo {volume} {502}},\
  \bibinfo {pages} {L40} (\bibinfo {year} {2021})},\ \Eprint
  {https://arxiv.org/abs/2009.06585} {arXiv:2009.06585 [astro-ph.SR]}
  \BibitemShut {NoStop}%
\bibitem [{\citenamefont {{Tanikawa}}\ \emph
  {et~al.}(2021{\natexlab{a}})\citenamefont {{Tanikawa}}, \citenamefont
  {{Kinugawa}}, \citenamefont {{Yoshida}}, \citenamefont {{Hijikawa}},\ and\
  \citenamefont {{Umeda}}}]{2021MNRAS.505.2170T}%
  \BibitemOpen
  \bibfield  {author} {\bibinfo {author} {\bibfnamefont {A.}~\bibnamefont
  {{Tanikawa}}}, \bibinfo {author} {\bibfnamefont {T.}~\bibnamefont
  {{Kinugawa}}}, \bibinfo {author} {\bibfnamefont {T.}~\bibnamefont
  {{Yoshida}}}, \bibinfo {author} {\bibfnamefont {K.}~\bibnamefont
  {{Hijikawa}}},\ and\ \bibinfo {author} {\bibfnamefont {H.}~\bibnamefont
  {{Umeda}}},\ }\bibfield  {title} {\bibinfo {title} {{Population III binary
  black holes: effects of convective overshooting on formation of GW190521}},\
  }\href {https://doi.org/10.1093/mnras/stab1421} {\bibfield  {journal}
  {\bibinfo  {journal} {\mnras}\ }\textbf {\bibinfo {volume} {505}},\ \bibinfo
  {pages} {2170} (\bibinfo {year} {2021}{\natexlab{a}})},\ \Eprint
  {https://arxiv.org/abs/2010.07616} {arXiv:2010.07616 [astro-ph.HE]}
  \BibitemShut {NoStop}%
\bibitem [{\citenamefont {{Kinugawa}}\ \emph
  {et~al.}(2021{\natexlab{a}})\citenamefont {{Kinugawa}}, \citenamefont
  {{Nakamura}},\ and\ \citenamefont {{Nakano}}}]{2021MNRAS.504L..28K}%
  \BibitemOpen
  \bibfield  {author} {\bibinfo {author} {\bibfnamefont {T.}~\bibnamefont
  {{Kinugawa}}}, \bibinfo {author} {\bibfnamefont {T.}~\bibnamefont
  {{Nakamura}}},\ and\ \bibinfo {author} {\bibfnamefont {H.}~\bibnamefont
  {{Nakano}}},\ }\bibfield  {title} {\bibinfo {title} {{Gravitational waves
  from Population III binary black holes are consistent with LIGO/Virgo O3a
  data for the chirp mass larger than {\ensuremath{\sim}}20
  M$_{{\ensuremath{\odot}}}$}},\ }\href
  {https://doi.org/10.1093/mnrasl/slab032} {\bibfield  {journal} {\bibinfo
  {journal} {\mnras}\ }\textbf {\bibinfo {volume} {504}},\ \bibinfo {pages}
  {L28} (\bibinfo {year} {2021}{\natexlab{a}})},\ \Eprint
  {https://arxiv.org/abs/2103.00797} {arXiv:2103.00797 [astro-ph.HE]}
  \BibitemShut {NoStop}%
\bibitem [{\citenamefont {{Costa}}\ \emph {et~al.}(2023)\citenamefont
  {{Costa}}, \citenamefont {{Mapelli}}, \citenamefont {{Iorio}}, \citenamefont
  {{Santoliquido}}, \citenamefont {{Escobar}}, \citenamefont {{Klessen}},\ and\
  \citenamefont {{Bressan}}}]{2023MNRAS.525.2891C}%
  \BibitemOpen
  \bibfield  {author} {\bibinfo {author} {\bibfnamefont {G.}~\bibnamefont
  {{Costa}}}, \bibinfo {author} {\bibfnamefont {M.}~\bibnamefont {{Mapelli}}},
  \bibinfo {author} {\bibfnamefont {G.}~\bibnamefont {{Iorio}}}, \bibinfo
  {author} {\bibfnamefont {F.}~\bibnamefont {{Santoliquido}}}, \bibinfo
  {author} {\bibfnamefont {G.~J.}\ \bibnamefont {{Escobar}}}, \bibinfo {author}
  {\bibfnamefont {R.~S.}\ \bibnamefont {{Klessen}}},\ and\ \bibinfo {author}
  {\bibfnamefont {A.}~\bibnamefont {{Bressan}}},\ }\bibfield  {title} {\bibinfo
  {title} {{Massive binary black holes from Population II and III stars}},\
  }\href {https://doi.org/10.1093/mnras/stad2443} {\bibfield  {journal}
  {\bibinfo  {journal} {\mnras}\ }\textbf {\bibinfo {volume} {525}},\ \bibinfo
  {pages} {2891} (\bibinfo {year} {2023})},\ \Eprint
  {https://arxiv.org/abs/2303.15511} {arXiv:2303.15511 [astro-ph.GA]}
  \BibitemShut {NoStop}%
\bibitem [{\citenamefont {{Santoliquido}}\ \emph {et~al.}(2023)\citenamefont
  {{Santoliquido}}, \citenamefont {{Mapelli}}, \citenamefont {{Iorio}},
  \citenamefont {{Costa}}, \citenamefont {{Glover}}, \citenamefont {{Hartwig}},
  \citenamefont {{Klessen}},\ and\ \citenamefont
  {{Merli}}}]{2023MNRAS.524..307S}%
  \BibitemOpen
  \bibfield  {author} {\bibinfo {author} {\bibfnamefont {F.}~\bibnamefont
  {{Santoliquido}}}, \bibinfo {author} {\bibfnamefont {M.}~\bibnamefont
  {{Mapelli}}}, \bibinfo {author} {\bibfnamefont {G.}~\bibnamefont {{Iorio}}},
  \bibinfo {author} {\bibfnamefont {G.}~\bibnamefont {{Costa}}}, \bibinfo
  {author} {\bibfnamefont {S.~C.~O.}\ \bibnamefont {{Glover}}}, \bibinfo
  {author} {\bibfnamefont {T.}~\bibnamefont {{Hartwig}}}, \bibinfo {author}
  {\bibfnamefont {R.~S.}\ \bibnamefont {{Klessen}}},\ and\ \bibinfo {author}
  {\bibfnamefont {L.}~\bibnamefont {{Merli}}},\ }\bibfield  {title} {\bibinfo
  {title} {{Binary black hole mergers from population III stars: uncertainties
  from star formation and binary star properties}},\ }\href
  {https://doi.org/10.1093/mnras/stad1860} {\bibfield  {journal} {\bibinfo
  {journal} {\mnras}\ }\textbf {\bibinfo {volume} {524}},\ \bibinfo {pages}
  {307} (\bibinfo {year} {2023})},\ \Eprint {https://arxiv.org/abs/2303.15515}
  {arXiv:2303.15515 [astro-ph.GA]} \BibitemShut {NoStop}%
\bibitem [{\citenamefont {{Sana}}\ \emph
  {et~al.}(2012{\natexlab{a}})\citenamefont {{Sana}}, \citenamefont {{de
  Mink}}, \citenamefont {{de Koter}}, \citenamefont {{Langer}}, \citenamefont
  {{Evans}}, \citenamefont {{Gieles}}, \citenamefont {{Gosset}}, \citenamefont
  {{Izzard}}, \citenamefont {{Le Bouquin}},\ and\ \citenamefont
  {{Schneider}}}]{2012Sci...337..444S}%
  \BibitemOpen
  \bibfield  {author} {\bibinfo {author} {\bibfnamefont {H.}~\bibnamefont
  {{Sana}}}, \bibinfo {author} {\bibfnamefont {S.~E.}\ \bibnamefont {{de
  Mink}}}, \bibinfo {author} {\bibfnamefont {A.}~\bibnamefont {{de Koter}}},
  \bibinfo {author} {\bibfnamefont {N.}~\bibnamefont {{Langer}}}, \bibinfo
  {author} {\bibfnamefont {C.~J.}\ \bibnamefont {{Evans}}}, \bibinfo {author}
  {\bibfnamefont {M.}~\bibnamefont {{Gieles}}}, \bibinfo {author}
  {\bibfnamefont {E.}~\bibnamefont {{Gosset}}}, \bibinfo {author}
  {\bibfnamefont {R.~G.}\ \bibnamefont {{Izzard}}}, \bibinfo {author}
  {\bibfnamefont {J.~B.}\ \bibnamefont {{Le Bouquin}}},\ and\ \bibinfo {author}
  {\bibfnamefont {F.~R.~N.}\ \bibnamefont {{Schneider}}},\ }\bibfield  {title}
  {\bibinfo {title} {{Binary Interaction Dominates the Evolution of Massive
  Stars}},\ }\href {https://doi.org/10.1126/science.1223344} {\bibfield
  {journal} {\bibinfo  {journal} {Science}\ }\textbf {\bibinfo {volume}
  {337}},\ \bibinfo {pages} {444} (\bibinfo {year} {2012}{\natexlab{a}})},\
  \Eprint {https://arxiv.org/abs/1207.6397} {arXiv:1207.6397 [astro-ph.SR]}
  \BibitemShut {NoStop}%
\bibitem [{\citenamefont {{Moe}}\ and\ \citenamefont {{Di
  Stefano}}(2017{\natexlab{a}})}]{2017ApJS..230...15M}%
  \BibitemOpen
  \bibfield  {author} {\bibinfo {author} {\bibfnamefont {M.}~\bibnamefont
  {{Moe}}}\ and\ \bibinfo {author} {\bibfnamefont {R.}~\bibnamefont {{Di
  Stefano}}},\ }\bibfield  {title} {\bibinfo {title} {{Mind Your Ps and Qs: The
  Interrelation between Period (P) and Mass-ratio (Q) Distributions of Binary
  Stars}},\ }\href {https://doi.org/10.3847/1538-4365/aa6fb6} {\bibfield
  {journal} {\bibinfo  {journal} {\apjs}\ }\textbf {\bibinfo {volume} {230}},\
  \bibinfo {eid} {15} (\bibinfo {year} {2017}{\natexlab{a}})},\ \Eprint
  {https://arxiv.org/abs/1606.05347} {arXiv:1606.05347 [astro-ph.SR]}
  \BibitemShut {NoStop}%
\bibitem [{\citenamefont {{de Mink}}\ \emph {et~al.}(2009)\citenamefont {{de
  Mink}}, \citenamefont {{Cantiello}}, \citenamefont {{Langer}}, \citenamefont
  {{Pols}}, \citenamefont {{Brott}},\ and\ \citenamefont
  {{Yoon}}}]{2009A&A...497..243D}%
  \BibitemOpen
  \bibfield  {author} {\bibinfo {author} {\bibfnamefont {S.~E.}\ \bibnamefont
  {{de Mink}}}, \bibinfo {author} {\bibfnamefont {M.}~\bibnamefont
  {{Cantiello}}}, \bibinfo {author} {\bibfnamefont {N.}~\bibnamefont
  {{Langer}}}, \bibinfo {author} {\bibfnamefont {O.~R.}\ \bibnamefont
  {{Pols}}}, \bibinfo {author} {\bibfnamefont {I.}~\bibnamefont {{Brott}}},\
  and\ \bibinfo {author} {\bibfnamefont {S.~C.}\ \bibnamefont {{Yoon}}},\
  }\bibfield  {title} {\bibinfo {title} {{Rotational mixing in massive
  binaries. Detached short-period systems}},\ }\href
  {https://doi.org/10.1051/0004-6361/200811439} {\bibfield  {journal} {\bibinfo
   {journal} {\aap}\ }\textbf {\bibinfo {volume} {497}},\ \bibinfo {pages}
  {243} (\bibinfo {year} {2009})},\ \Eprint {https://arxiv.org/abs/0902.1751}
  {arXiv:0902.1751 [astro-ph.SR]} \BibitemShut {NoStop}%
\bibitem [{\citenamefont {{Marchant}}\ \emph
  {et~al.}(2016{\natexlab{a}})\citenamefont {{Marchant}}, \citenamefont
  {{Langer}}, \citenamefont {{Podsiadlowski}}, \citenamefont {{Tauris}},\ and\
  \citenamefont {{Moriya}}}]{2016A&A...588A..50M}%
  \BibitemOpen
  \bibfield  {author} {\bibinfo {author} {\bibfnamefont {P.}~\bibnamefont
  {{Marchant}}}, \bibinfo {author} {\bibfnamefont {N.}~\bibnamefont
  {{Langer}}}, \bibinfo {author} {\bibfnamefont {P.}~\bibnamefont
  {{Podsiadlowski}}}, \bibinfo {author} {\bibfnamefont {T.~M.}\ \bibnamefont
  {{Tauris}}},\ and\ \bibinfo {author} {\bibfnamefont {T.~J.}\ \bibnamefont
  {{Moriya}}},\ }\bibfield  {title} {\bibinfo {title} {{A new route towards
  merging massive black holes}},\ }\href
  {https://doi.org/10.1051/0004-6361/201628133} {\bibfield  {journal} {\bibinfo
   {journal} {\aap}\ }\textbf {\bibinfo {volume} {588}},\ \bibinfo {eid} {A50}
  (\bibinfo {year} {2016}{\natexlab{a}})},\ \Eprint
  {https://arxiv.org/abs/1601.03718} {arXiv:1601.03718 [astro-ph.SR]}
  \BibitemShut {NoStop}%
\bibitem [{\citenamefont {{Eldridge}}\ \emph {et~al.}(2017)\citenamefont
  {{Eldridge}}, \citenamefont {{Stanway}}, \citenamefont {{Xiao}},
  \citenamefont {{McClelland}}, \citenamefont {{Taylor}}, \citenamefont {{Ng}},
  \citenamefont {{Greis}},\ and\ \citenamefont {{Bray}}}]{2017PASA...34...58E}%
  \BibitemOpen
  \bibfield  {author} {\bibinfo {author} {\bibfnamefont {J.~J.}\ \bibnamefont
  {{Eldridge}}}, \bibinfo {author} {\bibfnamefont {E.~R.}\ \bibnamefont
  {{Stanway}}}, \bibinfo {author} {\bibfnamefont {L.}~\bibnamefont {{Xiao}}},
  \bibinfo {author} {\bibfnamefont {L.~A.~S.}\ \bibnamefont {{McClelland}}},
  \bibinfo {author} {\bibfnamefont {G.}~\bibnamefont {{Taylor}}}, \bibinfo
  {author} {\bibfnamefont {M.}~\bibnamefont {{Ng}}}, \bibinfo {author}
  {\bibfnamefont {S.~M.~L.}\ \bibnamefont {{Greis}}},\ and\ \bibinfo {author}
  {\bibfnamefont {J.~C.}\ \bibnamefont {{Bray}}},\ }\bibfield  {title}
  {\bibinfo {title} {{Binary Population and Spectral Synthesis Version 2.1:
  Construction, Observational Verification, and New Results}},\ }\href
  {https://doi.org/10.1017/pasa.2017.51} {\bibfield  {journal} {\bibinfo
  {journal} {\pasa}\ }\textbf {\bibinfo {volume} {34}},\ \bibinfo {eid} {e058}
  (\bibinfo {year} {2017})},\ \Eprint {https://arxiv.org/abs/1710.02154}
  {arXiv:1710.02154 [astro-ph.SR]} \BibitemShut {NoStop}%
\bibitem [{\citenamefont {{Eldridge}}\ and\ \citenamefont
  {{Stanway}}(2022)}]{2022ARA&A..60..455E}%
  \BibitemOpen
  \bibfield  {author} {\bibinfo {author} {\bibfnamefont {J.~J.}\ \bibnamefont
  {{Eldridge}}}\ and\ \bibinfo {author} {\bibfnamefont {E.~R.}\ \bibnamefont
  {{Stanway}}},\ }\bibfield  {title} {\bibinfo {title} {{New Insights into the
  Evolution of Massive Stars and Their Effects on Our Understanding of Early
  Galaxies}},\ }\href {https://doi.org/10.1146/annurev-astro-052920-100646}
  {\bibfield  {journal} {\bibinfo  {journal} {\araa}\ }\textbf {\bibinfo
  {volume} {60}},\ \bibinfo {pages} {455} (\bibinfo {year} {2022})},\ \Eprint
  {https://arxiv.org/abs/2202.01413} {arXiv:2202.01413 [astro-ph.GA]}
  \BibitemShut {NoStop}%
\bibitem [{\citenamefont {{Klencki}}\ \emph {et~al.}(2020)\citenamefont
  {{Klencki}}, \citenamefont {{Nelemans}}, \citenamefont {{Istrate}},\ and\
  \citenamefont {{Pols}}}]{2020A&A...638A..55K}%
  \BibitemOpen
  \bibfield  {author} {\bibinfo {author} {\bibfnamefont {J.}~\bibnamefont
  {{Klencki}}}, \bibinfo {author} {\bibfnamefont {G.}~\bibnamefont
  {{Nelemans}}}, \bibinfo {author} {\bibfnamefont {A.~G.}\ \bibnamefont
  {{Istrate}}},\ and\ \bibinfo {author} {\bibfnamefont {O.}~\bibnamefont
  {{Pols}}},\ }\bibfield  {title} {\bibinfo {title} {{Massive donors in
  interacting binaries: effect of metallicity}},\ }\href
  {https://doi.org/10.1051/0004-6361/202037694} {\bibfield  {journal} {\bibinfo
   {journal} {\aap}\ }\textbf {\bibinfo {volume} {638}},\ \bibinfo {eid} {A55}
  (\bibinfo {year} {2020})},\ \Eprint {https://arxiv.org/abs/2004.00628}
  {arXiv:2004.00628 [astro-ph.SR]} \BibitemShut {NoStop}%
\bibitem [{\citenamefont {{Renzo}}\ \emph {et~al.}(2020)\citenamefont
  {{Renzo}}, \citenamefont {{Cantiello}}, \citenamefont {{Metzger}},\ and\
  \citenamefont {{Jiang}}}]{2020ApJ...904L..13R}%
  \BibitemOpen
  \bibfield  {author} {\bibinfo {author} {\bibfnamefont {M.}~\bibnamefont
  {{Renzo}}}, \bibinfo {author} {\bibfnamefont {M.}~\bibnamefont
  {{Cantiello}}}, \bibinfo {author} {\bibfnamefont {B.~D.}\ \bibnamefont
  {{Metzger}}},\ and\ \bibinfo {author} {\bibfnamefont {Y.~F.}\ \bibnamefont
  {{Jiang}}},\ }\bibfield  {title} {\bibinfo {title} {{The Stellar Merger
  Scenario for Black Holes in the Pair-instability Gap}},\ }\href
  {https://doi.org/10.3847/2041-8213/abc6a6} {\bibfield  {journal} {\bibinfo
  {journal} {\apjl}\ }\textbf {\bibinfo {volume} {904}},\ \bibinfo {eid} {L13}
  (\bibinfo {year} {2020})},\ \Eprint {https://arxiv.org/abs/2010.00705}
  {arXiv:2010.00705 [astro-ph.SR]} \BibitemShut {NoStop}%
\bibitem [{\citenamefont {{Laplace}}\ \emph {et~al.}(2021)\citenamefont
  {{Laplace}}, \citenamefont {{Justham}}, \citenamefont {{Renzo}},
  \citenamefont {{G{\"o}tberg}}, \citenamefont {{Farmer}}, \citenamefont
  {{Vartanyan}},\ and\ \citenamefont {{de Mink}}}]{2021A&A...656A..58L}%
  \BibitemOpen
  \bibfield  {author} {\bibinfo {author} {\bibfnamefont {E.}~\bibnamefont
  {{Laplace}}}, \bibinfo {author} {\bibfnamefont {S.}~\bibnamefont
  {{Justham}}}, \bibinfo {author} {\bibfnamefont {M.}~\bibnamefont {{Renzo}}},
  \bibinfo {author} {\bibfnamefont {Y.}~\bibnamefont {{G{\"o}tberg}}}, \bibinfo
  {author} {\bibfnamefont {R.}~\bibnamefont {{Farmer}}}, \bibinfo {author}
  {\bibfnamefont {D.}~\bibnamefont {{Vartanyan}}},\ and\ \bibinfo {author}
  {\bibfnamefont {S.~E.}\ \bibnamefont {{de Mink}}},\ }\bibfield  {title}
  {\bibinfo {title} {{Different to the core: The pre-supernova structures of
  massive single and binary-stripped stars}},\ }\href
  {https://doi.org/10.1051/0004-6361/202140506} {\bibfield  {journal} {\bibinfo
   {journal} {\aap}\ }\textbf {\bibinfo {volume} {656}},\ \bibinfo {eid} {A58}
  (\bibinfo {year} {2021})},\ \Eprint {https://arxiv.org/abs/2102.05036}
  {arXiv:2102.05036 [astro-ph.SR]} \BibitemShut {NoStop}%
\bibitem [{\citenamefont {{Ballone}}\ \emph {et~al.}(2022)\citenamefont
  {{Ballone}}, \citenamefont {{Costa}}, \citenamefont {{Mapelli}},
  \citenamefont {{MacLeod}}, \citenamefont {{Torniamenti}},\ and\ \citenamefont
  {{Pacheco-Arias}}}]{2022MNRAS.tmp.3494B}%
  \BibitemOpen
  \bibfield  {author} {\bibinfo {author} {\bibfnamefont {A.}~\bibnamefont
  {{Ballone}}}, \bibinfo {author} {\bibfnamefont {G.}~\bibnamefont {{Costa}}},
  \bibinfo {author} {\bibfnamefont {M.}~\bibnamefont {{Mapelli}}}, \bibinfo
  {author} {\bibfnamefont {M.}~\bibnamefont {{MacLeod}}}, \bibinfo {author}
  {\bibfnamefont {S.}~\bibnamefont {{Torniamenti}}},\ and\ \bibinfo {author}
  {\bibfnamefont {J.~M.}\ \bibnamefont {{Pacheco-Arias}}},\ }\bibfield  {title}
  {\bibinfo {title} {{Formation of black holes in the pair-instability mass
  gap: Hydrodynamical simulations of a head-on massive star collision}},\
  }\bibfield  {journal} {\bibinfo  {journal} {\mnras}\ }\href
  {https://doi.org/10.1093/mnras/stac3752} {10.1093/mnras/stac3752} (\bibinfo
  {year} {2022}),\ \Eprint {https://arxiv.org/abs/2204.03493} {arXiv:2204.03493
  [astro-ph.SR]} \BibitemShut {NoStop}%
\bibitem [{\citenamefont {{Costa}}\ \emph {et~al.}(2022)\citenamefont
  {{Costa}}, \citenamefont {{Ballone}}, \citenamefont {{Mapelli}},\ and\
  \citenamefont {{Bressan}}}]{2022MNRAS.516.1072C}%
  \BibitemOpen
  \bibfield  {author} {\bibinfo {author} {\bibfnamefont {G.}~\bibnamefont
  {{Costa}}}, \bibinfo {author} {\bibfnamefont {A.}~\bibnamefont {{Ballone}}},
  \bibinfo {author} {\bibfnamefont {M.}~\bibnamefont {{Mapelli}}},\ and\
  \bibinfo {author} {\bibfnamefont {A.}~\bibnamefont {{Bressan}}},\ }\bibfield
  {title} {\bibinfo {title} {{Formation of black holes in the pair-instability
  mass gap: Evolution of a post-collision star}},\ }\href
  {https://doi.org/10.1093/mnras/stac2222} {\bibfield  {journal} {\bibinfo
  {journal} {\mnras}\ }\textbf {\bibinfo {volume} {516}},\ \bibinfo {pages}
  {1072} (\bibinfo {year} {2022})},\ \Eprint {https://arxiv.org/abs/2204.03492}
  {arXiv:2204.03492 [astro-ph.SR]} \BibitemShut {NoStop}%
\bibitem [{\citenamefont {{Sukhbold}}\ and\ \citenamefont
  {{Woosley}}(2014)}]{2014ApJ...783...10S}%
  \BibitemOpen
  \bibfield  {author} {\bibinfo {author} {\bibfnamefont {T.}~\bibnamefont
  {{Sukhbold}}}\ and\ \bibinfo {author} {\bibfnamefont {S.~E.}\ \bibnamefont
  {{Woosley}}},\ }\bibfield  {title} {\bibinfo {title} {{The Compactness of
  Presupernova Stellar Cores}},\ }\href
  {https://doi.org/10.1088/0004-637X/783/1/10} {\bibfield  {journal} {\bibinfo
  {journal} {\apj}\ }\textbf {\bibinfo {volume} {783}},\ \bibinfo {eid} {10}
  (\bibinfo {year} {2014})},\ \Eprint {https://arxiv.org/abs/1311.6546}
  {arXiv:1311.6546 [astro-ph.SR]} \BibitemShut {NoStop}%
\bibitem [{\citenamefont {{Schootemeijer}}\ \emph {et~al.}(2019)\citenamefont
  {{Schootemeijer}}, \citenamefont {{Langer}}, \citenamefont {{Grin}},\ and\
  \citenamefont {{Wang}}}]{2019A&A...625A.132S}%
  \BibitemOpen
  \bibfield  {author} {\bibinfo {author} {\bibfnamefont {A.}~\bibnamefont
  {{Schootemeijer}}}, \bibinfo {author} {\bibfnamefont {N.}~\bibnamefont
  {{Langer}}}, \bibinfo {author} {\bibfnamefont {N.~J.}\ \bibnamefont
  {{Grin}}},\ and\ \bibinfo {author} {\bibfnamefont {C.}~\bibnamefont
  {{Wang}}},\ }\bibfield  {title} {\bibinfo {title} {{Constraining mixing in
  massive stars in the Small Magellanic Cloud}},\ }\href
  {https://doi.org/10.1051/0004-6361/201935046} {\bibfield  {journal} {\bibinfo
   {journal} {\aap}\ }\textbf {\bibinfo {volume} {625}},\ \bibinfo {eid} {A132}
  (\bibinfo {year} {2019})},\ \Eprint {https://arxiv.org/abs/1903.10423}
  {arXiv:1903.10423 [astro-ph.SR]} \BibitemShut {NoStop}%
\bibitem [{\citenamefont {{Kaiser}}\ \emph {et~al.}(2020)\citenamefont
  {{Kaiser}}, \citenamefont {{Hirschi}}, \citenamefont {{Arnett}},
  \citenamefont {{Georgy}}, \citenamefont {{Scott}},\ and\ \citenamefont
  {{Cristini}}}]{2020MNRAS.496.1967K}%
  \BibitemOpen
  \bibfield  {author} {\bibinfo {author} {\bibfnamefont {E.~A.}\ \bibnamefont
  {{Kaiser}}}, \bibinfo {author} {\bibfnamefont {R.}~\bibnamefont {{Hirschi}}},
  \bibinfo {author} {\bibfnamefont {W.~D.}\ \bibnamefont {{Arnett}}}, \bibinfo
  {author} {\bibfnamefont {C.}~\bibnamefont {{Georgy}}}, \bibinfo {author}
  {\bibfnamefont {L.~J.~A.}\ \bibnamefont {{Scott}}},\ and\ \bibinfo {author}
  {\bibfnamefont {A.}~\bibnamefont {{Cristini}}},\ }\bibfield  {title}
  {\bibinfo {title} {{Relative importance of convective uncertainties in
  massive stars}},\ }\href {https://doi.org/10.1093/mnras/staa1595} {\bibfield
  {journal} {\bibinfo  {journal} {\mnras}\ }\textbf {\bibinfo {volume} {496}},\
  \bibinfo {pages} {1967} (\bibinfo {year} {2020})},\ \Eprint
  {https://arxiv.org/abs/2006.01877} {arXiv:2006.01877 [astro-ph.SR]}
  \BibitemShut {NoStop}%
\bibitem [{\citenamefont {{Renzini}}(1987)}]{1987A&A...188...49R}%
  \BibitemOpen
  \bibfield  {author} {\bibinfo {author} {\bibfnamefont {A.}~\bibnamefont
  {{Renzini}}},\ }\bibfield  {title} {\bibinfo {title} {{Some embarrassments in
  current treatments of convective overshooting}},\ }\href@noop {} {\bibfield
  {journal} {\bibinfo  {journal} {\aap}\ }\textbf {\bibinfo {volume} {188}},\
  \bibinfo {pages} {49} (\bibinfo {year} {1987})}\BibitemShut {NoStop}%
\bibitem [{\citenamefont {{Zahn}}(1991)}]{1991A&A...252..179Z}%
  \BibitemOpen
  \bibfield  {author} {\bibinfo {author} {\bibfnamefont {J.~P.}\ \bibnamefont
  {{Zahn}}},\ }\bibfield  {title} {\bibinfo {title} {{Convective penetration in
  stellar interiors.}},\ }\href@noop {} {\bibfield  {journal} {\bibinfo
  {journal} {\aap}\ }\textbf {\bibinfo {volume} {252}},\ \bibinfo {pages} {179}
  (\bibinfo {year} {1991})}\BibitemShut {NoStop}%
\bibitem [{\citenamefont {{Salaris}}\ and\ \citenamefont
  {{Cassisi}}(2017)}]{2017RSOS....470192S}%
  \BibitemOpen
  \bibfield  {author} {\bibinfo {author} {\bibfnamefont {M.}~\bibnamefont
  {{Salaris}}}\ and\ \bibinfo {author} {\bibfnamefont {S.}~\bibnamefont
  {{Cassisi}}},\ }\bibfield  {title} {\bibinfo {title} {{Chemical element
  transport in stellar evolution models}},\ }\href
  {https://doi.org/10.1098/rsos.170192} {\bibfield  {journal} {\bibinfo
  {journal} {Royal Society Open Science}\ }\textbf {\bibinfo {volume} {4}},\
  \bibinfo {eid} {170192} (\bibinfo {year} {2017})},\ \Eprint
  {https://arxiv.org/abs/1707.07454} {arXiv:1707.07454 [astro-ph.SR]}
  \BibitemShut {NoStop}%
\bibitem [{\citenamefont {{Andrassy}}\ \emph {et~al.}(2022)\citenamefont
  {{Andrassy}}, \citenamefont {{Higl}}, \citenamefont {{Mao}}, \citenamefont
  {{Moc{\'a}k}}, \citenamefont {{Vlaykov}}, \citenamefont {{Arnett}},
  \citenamefont {{Baraffe}}, \citenamefont {{Campbell}}, \citenamefont
  {{Constantino}}, \citenamefont {{Edelmann}}, \citenamefont {{Goffrey}},
  \citenamefont {{Guillet}}, \citenamefont {{Herwig}}, \citenamefont
  {{Hirschi}}, \citenamefont {{Horst}}, \citenamefont {{Leidi}}, \citenamefont
  {{Meakin}}, \citenamefont {{Pratt}}, \citenamefont {{Rizzuti}}, \citenamefont
  {{R{\"o}pke}},\ and\ \citenamefont {{Woodward}}}]{2022A&A...659A.193A}%
  \BibitemOpen
  \bibfield  {author} {\bibinfo {author} {\bibfnamefont {R.}~\bibnamefont
  {{Andrassy}}}, \bibinfo {author} {\bibfnamefont {J.}~\bibnamefont {{Higl}}},
  \bibinfo {author} {\bibfnamefont {H.}~\bibnamefont {{Mao}}}, \bibinfo
  {author} {\bibfnamefont {M.}~\bibnamefont {{Moc{\'a}k}}}, \bibinfo {author}
  {\bibfnamefont {D.~G.}\ \bibnamefont {{Vlaykov}}}, \bibinfo {author}
  {\bibfnamefont {W.~D.}\ \bibnamefont {{Arnett}}}, \bibinfo {author}
  {\bibfnamefont {I.}~\bibnamefont {{Baraffe}}}, \bibinfo {author}
  {\bibfnamefont {S.~W.}\ \bibnamefont {{Campbell}}}, \bibinfo {author}
  {\bibfnamefont {T.}~\bibnamefont {{Constantino}}}, \bibinfo {author}
  {\bibfnamefont {P.~V.~F.}\ \bibnamefont {{Edelmann}}}, \bibinfo {author}
  {\bibfnamefont {T.}~\bibnamefont {{Goffrey}}}, \bibinfo {author}
  {\bibfnamefont {T.}~\bibnamefont {{Guillet}}}, \bibinfo {author}
  {\bibfnamefont {F.}~\bibnamefont {{Herwig}}}, \bibinfo {author}
  {\bibfnamefont {R.}~\bibnamefont {{Hirschi}}}, \bibinfo {author}
  {\bibfnamefont {L.}~\bibnamefont {{Horst}}}, \bibinfo {author} {\bibfnamefont
  {G.}~\bibnamefont {{Leidi}}}, \bibinfo {author} {\bibfnamefont
  {C.}~\bibnamefont {{Meakin}}}, \bibinfo {author} {\bibfnamefont
  {J.}~\bibnamefont {{Pratt}}}, \bibinfo {author} {\bibfnamefont
  {F.}~\bibnamefont {{Rizzuti}}}, \bibinfo {author} {\bibfnamefont {F.~K.}\
  \bibnamefont {{R{\"o}pke}}},\ and\ \bibinfo {author} {\bibfnamefont
  {P.}~\bibnamefont {{Woodward}}},\ }\bibfield  {title} {\bibinfo {title}
  {{Dynamics in a stellar convective layer and at its boundary: Comparison of
  five 3D hydrodynamics codes}},\ }\href
  {https://doi.org/10.1051/0004-6361/202142557} {\bibfield  {journal} {\bibinfo
   {journal} {\aap}\ }\textbf {\bibinfo {volume} {659}},\ \bibinfo {eid} {A193}
  (\bibinfo {year} {2022})},\ \Eprint {https://arxiv.org/abs/2111.01165}
  {arXiv:2111.01165 [astro-ph.SR]} \BibitemShut {NoStop}%
\bibitem [{\citenamefont {{Freytag}}\ \emph {et~al.}(1996)\citenamefont
  {{Freytag}}, \citenamefont {{Ludwig}},\ and\ \citenamefont
  {{Steffen}}}]{1996A&A...313..497F}%
  \BibitemOpen
  \bibfield  {author} {\bibinfo {author} {\bibfnamefont {B.}~\bibnamefont
  {{Freytag}}}, \bibinfo {author} {\bibfnamefont {H.~G.}\ \bibnamefont
  {{Ludwig}}},\ and\ \bibinfo {author} {\bibfnamefont {M.}~\bibnamefont
  {{Steffen}}},\ }\bibfield  {title} {\bibinfo {title} {{Hydrodynamical models
  of stellar convection. The role of overshoot in DA white dwarfs, A-type
  stars, and the Sun.}},\ }\href@noop {} {\bibfield  {journal} {\bibinfo
  {journal} {\aap}\ }\textbf {\bibinfo {volume} {313}},\ \bibinfo {pages} {497}
  (\bibinfo {year} {1996})}\BibitemShut {NoStop}%
\bibitem [{\citenamefont {{Herwig}}\ \emph {et~al.}(2006)\citenamefont
  {{Herwig}}, \citenamefont {{Freytag}}, \citenamefont {{Hueckstaedt}},\ and\
  \citenamefont {{Timmes}}}]{2006ApJ...642.1057H}%
  \BibitemOpen
  \bibfield  {author} {\bibinfo {author} {\bibfnamefont {F.}~\bibnamefont
  {{Herwig}}}, \bibinfo {author} {\bibfnamefont {B.}~\bibnamefont {{Freytag}}},
  \bibinfo {author} {\bibfnamefont {R.~M.}\ \bibnamefont {{Hueckstaedt}}},\
  and\ \bibinfo {author} {\bibfnamefont {F.~X.}\ \bibnamefont {{Timmes}}},\
  }\bibfield  {title} {\bibinfo {title} {{Hydrodynamic Simulations of He Shell
  Flash Convection}},\ }\href {https://doi.org/10.1086/501119} {\bibfield
  {journal} {\bibinfo  {journal} {\apj}\ }\textbf {\bibinfo {volume} {642}},\
  \bibinfo {pages} {1057} (\bibinfo {year} {2006})},\ \Eprint
  {https://arxiv.org/abs/astro-ph/0601164} {arXiv:astro-ph/0601164 [astro-ph]}
  \BibitemShut {NoStop}%
\bibitem [{\citenamefont {{Meakin}}\ and\ \citenamefont
  {{Arnett}}(2007)}]{2007ApJ...667..448M}%
  \BibitemOpen
  \bibfield  {author} {\bibinfo {author} {\bibfnamefont {C.~A.}\ \bibnamefont
  {{Meakin}}}\ and\ \bibinfo {author} {\bibfnamefont {D.}~\bibnamefont
  {{Arnett}}},\ }\bibfield  {title} {\bibinfo {title} {{Turbulent Convection in
  Stellar Interiors. I. Hydrodynamic Simulation}},\ }\href
  {https://doi.org/10.1086/520318} {\bibfield  {journal} {\bibinfo  {journal}
  {\apj}\ }\textbf {\bibinfo {volume} {667}},\ \bibinfo {pages} {448} (\bibinfo
  {year} {2007})},\ \Eprint {https://arxiv.org/abs/astro-ph/0611315}
  {arXiv:astro-ph/0611315 [astro-ph]} \BibitemShut {NoStop}%
\bibitem [{\citenamefont {{Magic}}\ \emph {et~al.}(2013)\citenamefont
  {{Magic}}, \citenamefont {{Collet}}, \citenamefont {{Asplund}}, \citenamefont
  {{Trampedach}}, \citenamefont {{Hayek}}, \citenamefont {{Chiavassa}},
  \citenamefont {{Stein}},\ and\ \citenamefont
  {{Nordlund}}}]{2013A&A...557A..26M}%
  \BibitemOpen
  \bibfield  {author} {\bibinfo {author} {\bibfnamefont {Z.}~\bibnamefont
  {{Magic}}}, \bibinfo {author} {\bibfnamefont {R.}~\bibnamefont {{Collet}}},
  \bibinfo {author} {\bibfnamefont {M.}~\bibnamefont {{Asplund}}}, \bibinfo
  {author} {\bibfnamefont {R.}~\bibnamefont {{Trampedach}}}, \bibinfo {author}
  {\bibfnamefont {W.}~\bibnamefont {{Hayek}}}, \bibinfo {author} {\bibfnamefont
  {A.}~\bibnamefont {{Chiavassa}}}, \bibinfo {author} {\bibfnamefont {R.~F.}\
  \bibnamefont {{Stein}}},\ and\ \bibinfo {author} {\bibfnamefont
  {A.}~\bibnamefont {{Nordlund}}},\ }\bibfield  {title} {\bibinfo {title} {{The
  Stagger-grid: A grid of 3D stellar atmosphere models. I. Methods and general
  properties}},\ }\href {https://doi.org/10.1051/0004-6361/201321274}
  {\bibfield  {journal} {\bibinfo  {journal} {\aap}\ }\textbf {\bibinfo
  {volume} {557}},\ \bibinfo {eid} {A26} (\bibinfo {year} {2013})},\ \Eprint
  {https://arxiv.org/abs/1302.2621} {arXiv:1302.2621 [astro-ph.SR]}
  \BibitemShut {NoStop}%
\bibitem [{\citenamefont {{Arnett}}\ \emph {et~al.}(2018)\citenamefont
  {{Arnett}}, \citenamefont {{Hirschi}}, \citenamefont {{Campbell}},
  \citenamefont {{Moc{\'a}k}}, \citenamefont {{Georgy}}, \citenamefont
  {{Meakin}}, \citenamefont {{Cristini}}, \citenamefont {{Scott}},
  \citenamefont {{Kaiser}},\ and\ \citenamefont
  {{Viallet}}}]{2018arXiv181004659A}%
  \BibitemOpen
  \bibfield  {author} {\bibinfo {author} {\bibfnamefont {W.~D.}\ \bibnamefont
  {{Arnett}}}, \bibinfo {author} {\bibfnamefont {R.}~\bibnamefont {{Hirschi}}},
  \bibinfo {author} {\bibfnamefont {S.~W.}\ \bibnamefont {{Campbell}}},
  \bibinfo {author} {\bibfnamefont {M.}~\bibnamefont {{Moc{\'a}k}}}, \bibinfo
  {author} {\bibfnamefont {C.}~\bibnamefont {{Georgy}}}, \bibinfo {author}
  {\bibfnamefont {C.}~\bibnamefont {{Meakin}}}, \bibinfo {author}
  {\bibfnamefont {A.}~\bibnamefont {{Cristini}}}, \bibinfo {author}
  {\bibfnamefont {L.~J.~A.}\ \bibnamefont {{Scott}}}, \bibinfo {author}
  {\bibfnamefont {E.~A.}\ \bibnamefont {{Kaiser}}},\ and\ \bibinfo {author}
  {\bibfnamefont {M.}~\bibnamefont {{Viallet}}},\ }\bibfield  {title} {\bibinfo
  {title} {{3D Simulations and MLT: II. Onsager's Ideal Turbulence}},\
  }\href@noop {} {\bibfield  {journal} {\bibinfo  {journal} {arXiv e-prints}\
  ,\ \bibinfo {eid} {arXiv:1810.04659}} (\bibinfo {year} {2018})},\ \Eprint
  {https://arxiv.org/abs/1810.04659} {arXiv:1810.04659 [astro-ph.SR]}
  \BibitemShut {NoStop}%
\bibitem [{\citenamefont {{Arnett}}\ \emph {et~al.}(2019)\citenamefont
  {{Arnett}}, \citenamefont {{Meakin}}, \citenamefont {{Hirschi}},
  \citenamefont {{Cristini}}, \citenamefont {{Georgy}}, \citenamefont
  {{Campbell}}, \citenamefont {{Scott}}, \citenamefont {{Kaiser}},
  \citenamefont {{Viallet}},\ and\ \citenamefont
  {{Moc{\'a}k}}}]{2019ApJ...882...18A}%
  \BibitemOpen
  \bibfield  {author} {\bibinfo {author} {\bibfnamefont {W.~D.}\ \bibnamefont
  {{Arnett}}}, \bibinfo {author} {\bibfnamefont {C.}~\bibnamefont {{Meakin}}},
  \bibinfo {author} {\bibfnamefont {R.}~\bibnamefont {{Hirschi}}}, \bibinfo
  {author} {\bibfnamefont {A.}~\bibnamefont {{Cristini}}}, \bibinfo {author}
  {\bibfnamefont {C.}~\bibnamefont {{Georgy}}}, \bibinfo {author}
  {\bibfnamefont {S.}~\bibnamefont {{Campbell}}}, \bibinfo {author}
  {\bibfnamefont {L.~J.~A.}\ \bibnamefont {{Scott}}}, \bibinfo {author}
  {\bibfnamefont {E.~A.}\ \bibnamefont {{Kaiser}}}, \bibinfo {author}
  {\bibfnamefont {M.}~\bibnamefont {{Viallet}}},\ and\ \bibinfo {author}
  {\bibfnamefont {M.}~\bibnamefont {{Moc{\'a}k}}},\ }\bibfield  {title}
  {\bibinfo {title} {{3D Simulations and MLT. I. Renzini{\textquoteright}s
  Critique}},\ }\href {https://doi.org/10.3847/1538-4357/ab21d9} {\bibfield
  {journal} {\bibinfo  {journal} {\apj}\ }\textbf {\bibinfo {volume} {882}},\
  \bibinfo {eid} {18} (\bibinfo {year} {2019})},\ \Eprint
  {https://arxiv.org/abs/1810.04653} {arXiv:1810.04653 [astro-ph.SR]}
  \BibitemShut {NoStop}%
\bibitem [{\citenamefont {{Rizzuti}}\ \emph {et~al.}(2022)\citenamefont
  {{Rizzuti}}, \citenamefont {{Hirschi}}, \citenamefont {{Georgy}},
  \citenamefont {{Arnett}}, \citenamefont {{Meakin}},\ and\ \citenamefont
  {{Murphy}}}]{2022MNRAS.515.4013R}%
  \BibitemOpen
  \bibfield  {author} {\bibinfo {author} {\bibfnamefont {F.}~\bibnamefont
  {{Rizzuti}}}, \bibinfo {author} {\bibfnamefont {R.}~\bibnamefont
  {{Hirschi}}}, \bibinfo {author} {\bibfnamefont {C.}~\bibnamefont {{Georgy}}},
  \bibinfo {author} {\bibfnamefont {W.~D.}\ \bibnamefont {{Arnett}}}, \bibinfo
  {author} {\bibfnamefont {C.}~\bibnamefont {{Meakin}}},\ and\ \bibinfo
  {author} {\bibfnamefont {A.~S.}\ \bibnamefont {{Murphy}}},\ }\bibfield
  {title} {\bibinfo {title} {{Realistic 3D hydrodynamics simulations find
  significant turbulent entrainment in massive stars}},\ }\href
  {https://doi.org/10.1093/mnras/stac1981} {\bibfield  {journal} {\bibinfo
  {journal} {\mnras}\ }\textbf {\bibinfo {volume} {515}},\ \bibinfo {pages}
  {4013} (\bibinfo {year} {2022})},\ \Eprint {https://arxiv.org/abs/2207.03223}
  {arXiv:2207.03223 [astro-ph.SR]} \BibitemShut {NoStop}%
\bibitem [{\citenamefont {{Cristini}}\ \emph {et~al.}(2017)\citenamefont
  {{Cristini}}, \citenamefont {{Meakin}}, \citenamefont {{Hirschi}},
  \citenamefont {{Arnett}}, \citenamefont {{Georgy}}, \citenamefont
  {{Viallet}},\ and\ \citenamefont {{Walkington}}}]{2017MNRAS.471..279C}%
  \BibitemOpen
  \bibfield  {author} {\bibinfo {author} {\bibfnamefont {A.}~\bibnamefont
  {{Cristini}}}, \bibinfo {author} {\bibfnamefont {C.}~\bibnamefont
  {{Meakin}}}, \bibinfo {author} {\bibfnamefont {R.}~\bibnamefont {{Hirschi}}},
  \bibinfo {author} {\bibfnamefont {D.}~\bibnamefont {{Arnett}}}, \bibinfo
  {author} {\bibfnamefont {C.}~\bibnamefont {{Georgy}}}, \bibinfo {author}
  {\bibfnamefont {M.}~\bibnamefont {{Viallet}}},\ and\ \bibinfo {author}
  {\bibfnamefont {I.}~\bibnamefont {{Walkington}}},\ }\bibfield  {title}
  {\bibinfo {title} {{3D hydrodynamic simulations of carbon burning in massive
  stars}},\ }\href {https://doi.org/10.1093/mnras/stx1535} {\bibfield
  {journal} {\bibinfo  {journal} {\mnras}\ }\textbf {\bibinfo {volume} {471}},\
  \bibinfo {pages} {279} (\bibinfo {year} {2017})},\ \Eprint
  {https://arxiv.org/abs/1610.05173} {arXiv:1610.05173 [astro-ph.SR]}
  \BibitemShut {NoStop}%
\bibitem [{\citenamefont {{B{\"o}hm-Vitense}}(1958)}]{1958ZA.....46..108B}%
  \BibitemOpen
  \bibfield  {author} {\bibinfo {author} {\bibfnamefont {E.}~\bibnamefont
  {{B{\"o}hm-Vitense}}},\ }\bibfield  {title} {\bibinfo {title} {{{\"U}ber die
  Wasserstoffkonvektionszone in Sternen verschiedener Effektivtemperaturen und
  Leuchtkr{\"a}fte. Mit 5 Textabbildungen}},\ }\href@noop {} {\bibfield
  {journal} {\bibinfo  {journal} {\zap}\ }\textbf {\bibinfo {volume} {46}},\
  \bibinfo {pages} {108} (\bibinfo {year} {1958})}\BibitemShut {NoStop}%
\bibitem [{\citenamefont {{Roxburgh}}(1978)}]{1978A&A....65..281R}%
  \BibitemOpen
  \bibfield  {author} {\bibinfo {author} {\bibfnamefont {I.~W.}\ \bibnamefont
  {{Roxburgh}}},\ }\bibfield  {title} {\bibinfo {title} {{Convection and
  stellar structure.}},\ }\href@noop {} {\bibfield  {journal} {\bibinfo
  {journal} {\aap}\ }\textbf {\bibinfo {volume} {65}},\ \bibinfo {pages} {281}
  (\bibinfo {year} {1978})}\BibitemShut {NoStop}%
\bibitem [{\citenamefont {{Canuto}}\ and\ \citenamefont
  {{Mazzitelli}}(1991)}]{1991ApJ...370..295C}%
  \BibitemOpen
  \bibfield  {author} {\bibinfo {author} {\bibfnamefont {V.~M.}\ \bibnamefont
  {{Canuto}}}\ and\ \bibinfo {author} {\bibfnamefont {I.}~\bibnamefont
  {{Mazzitelli}}},\ }\bibfield  {title} {\bibinfo {title} {{Stellar Turbulent
  Convection: A New Model and Applications}},\ }\href
  {https://doi.org/10.1086/169815} {\bibfield  {journal} {\bibinfo  {journal}
  {\apj}\ }\textbf {\bibinfo {volume} {370}},\ \bibinfo {pages} {295} (\bibinfo
  {year} {1991})}\BibitemShut {NoStop}%
\bibitem [{\citenamefont {{Eggenberger}}\ \emph {et~al.}(2008)\citenamefont
  {{Eggenberger}}, \citenamefont {{Meynet}}, \citenamefont {{Maeder}},
  \citenamefont {{Hirschi}}, \citenamefont {{Charbonnel}}, \citenamefont
  {{Talon}},\ and\ \citenamefont {{Ekstr{\"o}m}}}]{2008Ap&SS.316...43E}%
  \BibitemOpen
  \bibfield  {author} {\bibinfo {author} {\bibfnamefont {P.}~\bibnamefont
  {{Eggenberger}}}, \bibinfo {author} {\bibfnamefont {G.}~\bibnamefont
  {{Meynet}}}, \bibinfo {author} {\bibfnamefont {A.}~\bibnamefont {{Maeder}}},
  \bibinfo {author} {\bibfnamefont {R.}~\bibnamefont {{Hirschi}}}, \bibinfo
  {author} {\bibfnamefont {C.}~\bibnamefont {{Charbonnel}}}, \bibinfo {author}
  {\bibfnamefont {S.}~\bibnamefont {{Talon}}},\ and\ \bibinfo {author}
  {\bibfnamefont {S.}~\bibnamefont {{Ekstr{\"o}m}}},\ }\bibfield  {title}
  {\bibinfo {title} {{The Geneva stellar evolution code}},\ }\href
  {https://doi.org/10.1007/s10509-007-9511-y} {\bibfield  {journal} {\bibinfo
  {journal} {\apss}\ }\textbf {\bibinfo {volume} {316}},\ \bibinfo {pages} {43}
  (\bibinfo {year} {2008})}\BibitemShut {NoStop}%
\bibitem [{\citenamefont {{Bressan}}\ \emph {et~al.}(2012)\citenamefont
  {{Bressan}}, \citenamefont {{Marigo}}, \citenamefont {{Girardi}},
  \citenamefont {{Salasnich}}, \citenamefont {{Dal Cero}}, \citenamefont
  {{Rubele}},\ and\ \citenamefont {{Nanni}}}]{2012MNRAS.427..127B}%
  \BibitemOpen
  \bibfield  {author} {\bibinfo {author} {\bibfnamefont {A.}~\bibnamefont
  {{Bressan}}}, \bibinfo {author} {\bibfnamefont {P.}~\bibnamefont {{Marigo}}},
  \bibinfo {author} {\bibfnamefont {L.}~\bibnamefont {{Girardi}}}, \bibinfo
  {author} {\bibfnamefont {B.}~\bibnamefont {{Salasnich}}}, \bibinfo {author}
  {\bibfnamefont {C.}~\bibnamefont {{Dal Cero}}}, \bibinfo {author}
  {\bibfnamefont {S.}~\bibnamefont {{Rubele}}},\ and\ \bibinfo {author}
  {\bibfnamefont {A.}~\bibnamefont {{Nanni}}},\ }\bibfield  {title} {\bibinfo
  {title} {{PARSEC: stellar tracks and isochrones with the PAdova and TRieste
  Stellar Evolution Code}},\ }\href
  {https://doi.org/10.1111/j.1365-2966.2012.21948.x} {\bibfield  {journal}
  {\bibinfo  {journal} {\mnras}\ }\textbf {\bibinfo {volume} {427}},\ \bibinfo
  {pages} {127} (\bibinfo {year} {2012})},\ \Eprint
  {https://arxiv.org/abs/1208.4498} {arXiv:1208.4498 [astro-ph.SR]}
  \BibitemShut {NoStop}%
\bibitem [{\citenamefont {{Magic}}\ \emph {et~al.}(2015)\citenamefont
  {{Magic}}, \citenamefont {{Weiss}},\ and\ \citenamefont
  {{Asplund}}}]{2015A&A...573A..89M}%
  \BibitemOpen
  \bibfield  {author} {\bibinfo {author} {\bibfnamefont {Z.}~\bibnamefont
  {{Magic}}}, \bibinfo {author} {\bibfnamefont {A.}~\bibnamefont {{Weiss}}},\
  and\ \bibinfo {author} {\bibfnamefont {M.}~\bibnamefont {{Asplund}}},\
  }\bibfield  {title} {\bibinfo {title} {{The Stagger-grid: A grid of 3D
  stellar atmosphere models. III. The relation to mixing length convection
  theory}},\ }\href {https://doi.org/10.1051/0004-6361/201423760} {\bibfield
  {journal} {\bibinfo  {journal} {\aap}\ }\textbf {\bibinfo {volume} {573}},\
  \bibinfo {eid} {A89} (\bibinfo {year} {2015})},\ \Eprint
  {https://arxiv.org/abs/1403.1062} {arXiv:1403.1062 [astro-ph.SR]}
  \BibitemShut {NoStop}%
\bibitem [{\citenamefont {{Joyce}}\ and\ \citenamefont
  {{Chaboyer}}(2018)}]{2018ApJ...856...10J}%
  \BibitemOpen
  \bibfield  {author} {\bibinfo {author} {\bibfnamefont {M.}~\bibnamefont
  {{Joyce}}}\ and\ \bibinfo {author} {\bibfnamefont {B.}~\bibnamefont
  {{Chaboyer}}},\ }\bibfield  {title} {\bibinfo {title} {{Not All Stars Are the
  Sun: Empirical Calibration of the Mixing Length for Metal-poor Stars Using
  One-dimensional Stellar Evolution Models}},\ }\href
  {https://doi.org/10.3847/1538-4357/aab200} {\bibfield  {journal} {\bibinfo
  {journal} {\apj}\ }\textbf {\bibinfo {volume} {856}},\ \bibinfo {eid} {10}
  (\bibinfo {year} {2018})},\ \Eprint {https://arxiv.org/abs/1712.05082}
  {arXiv:1712.05082 [astro-ph.SR]} \BibitemShut {NoStop}%
\bibitem [{\citenamefont {{Viani}}\ \emph {et~al.}(2018)\citenamefont
  {{Viani}}, \citenamefont {{Basu}}, \citenamefont {{Ong J.}}, \citenamefont
  {{Bonaca}},\ and\ \citenamefont {{Chaplin}}}]{2018ApJ...858...28V}%
  \BibitemOpen
  \bibfield  {author} {\bibinfo {author} {\bibfnamefont {L.~S.}\ \bibnamefont
  {{Viani}}}, \bibinfo {author} {\bibfnamefont {S.}~\bibnamefont {{Basu}}},
  \bibinfo {author} {\bibfnamefont {M.~J.}\ \bibnamefont {{Ong J.}}}, \bibinfo
  {author} {\bibfnamefont {A.}~\bibnamefont {{Bonaca}}},\ and\ \bibinfo
  {author} {\bibfnamefont {W.~J.}\ \bibnamefont {{Chaplin}}},\ }\bibfield
  {title} {\bibinfo {title} {{Investigating the Metallicity-Mixing-length
  Relation}},\ }\href {https://doi.org/10.3847/1538-4357/aab7eb} {\bibfield
  {journal} {\bibinfo  {journal} {\apj}\ }\textbf {\bibinfo {volume} {858}},\
  \bibinfo {eid} {28} (\bibinfo {year} {2018})},\ \Eprint
  {https://arxiv.org/abs/1803.05924} {arXiv:1803.05924 [astro-ph.SR]}
  \BibitemShut {NoStop}%
\bibitem [{\citenamefont {{Choi}}\ \emph {et~al.}(2018)\citenamefont {{Choi}},
  \citenamefont {{Dotter}}, \citenamefont {{Conroy}},\ and\ \citenamefont
  {{Ting}}}]{2018ApJ...860..131C}%
  \BibitemOpen
  \bibfield  {author} {\bibinfo {author} {\bibfnamefont {J.}~\bibnamefont
  {{Choi}}}, \bibinfo {author} {\bibfnamefont {A.}~\bibnamefont {{Dotter}}},
  \bibinfo {author} {\bibfnamefont {C.}~\bibnamefont {{Conroy}}},\ and\
  \bibinfo {author} {\bibfnamefont {Y.-S.}\ \bibnamefont {{Ting}}},\ }\bibfield
   {title} {\bibinfo {title} {{On the Red Giant Branch: Ambiguity in the
  Surface Boundary Condition Leads to {\ensuremath{\approx}}100 K Uncertainty
  in Model Effective Temperatures}},\ }\href
  {https://doi.org/10.3847/1538-4357/aac435} {\bibfield  {journal} {\bibinfo
  {journal} {\apj}\ }\textbf {\bibinfo {volume} {860}},\ \bibinfo {eid} {131}
  (\bibinfo {year} {2018})},\ \Eprint {https://arxiv.org/abs/1805.04112}
  {arXiv:1805.04112 [astro-ph.SR]} \BibitemShut {NoStop}%
\bibitem [{\citenamefont {{Weiss}}\ \emph {et~al.}(2004)\citenamefont
  {{Weiss}}, \citenamefont {{Hillebrandt}}, \citenamefont {{Thomas}},\ and\
  \citenamefont {{Ritter}}}]{2004cgps.book.....W}%
  \BibitemOpen
  \bibfield  {author} {\bibinfo {author} {\bibfnamefont {A.}~\bibnamefont
  {{Weiss}}}, \bibinfo {author} {\bibfnamefont {W.}~\bibnamefont
  {{Hillebrandt}}}, \bibinfo {author} {\bibfnamefont {H.~C.}\ \bibnamefont
  {{Thomas}}},\ and\ \bibinfo {author} {\bibfnamefont {H.}~\bibnamefont
  {{Ritter}}},\ }\href@noop {} {\emph {\bibinfo {title} {{Cox and Giuli's
  Principles of Stellar Structure}}}}\ (\bibinfo {year} {2004})\BibitemShut
  {NoStop}%
\bibitem [{\citenamefont {{Schwarzschild}}(1958)}]{1958ses..book.....S}%
  \BibitemOpen
  \bibfield  {author} {\bibinfo {author} {\bibfnamefont {M.}~\bibnamefont
  {{Schwarzschild}}},\ }\href@noop {} {\emph {\bibinfo {title} {{Structure and
  evolution of the stars.}}}}\ (\bibinfo {year} {1958})\BibitemShut {NoStop}%
\bibitem [{\citenamefont {{Ledoux}}(1947)}]{1947ApJ...105..305L}%
  \BibitemOpen
  \bibfield  {author} {\bibinfo {author} {\bibfnamefont {P.}~\bibnamefont
  {{Ledoux}}},\ }\bibfield  {title} {\bibinfo {title} {{Stellar Models with
  Convection and with Discontinuity of the Mean Molecular Weight}},\ }\href
  {https://doi.org/10.1086/144905} {\bibfield  {journal} {\bibinfo  {journal}
  {\apj}\ }\textbf {\bibinfo {volume} {105}},\ \bibinfo {pages} {305} (\bibinfo
  {year} {1947})}\BibitemShut {NoStop}%
\bibitem [{\citenamefont {{Karakas}}\ and\ \citenamefont
  {{Lattanzio}}(2014)}]{2014PASA...31...30K}%
  \BibitemOpen
  \bibfield  {author} {\bibinfo {author} {\bibfnamefont {A.~I.}\ \bibnamefont
  {{Karakas}}}\ and\ \bibinfo {author} {\bibfnamefont {J.~C.}\ \bibnamefont
  {{Lattanzio}}},\ }\bibfield  {title} {\bibinfo {title} {{The Dawes Review 2:
  Nucleosynthesis and Stellar Yields of Low- and Intermediate-Mass Single
  Stars}},\ }\href {https://doi.org/10.1017/pasa.2014.21} {\bibfield  {journal}
  {\bibinfo  {journal} {\pasa}\ }\textbf {\bibinfo {volume} {31}},\ \bibinfo
  {eid} {e030} (\bibinfo {year} {2014})},\ \Eprint
  {https://arxiv.org/abs/1405.0062} {arXiv:1405.0062 [astro-ph.SR]}
  \BibitemShut {NoStop}%
\bibitem [{\citenamefont {{Schwarzschild}}\ and\ \citenamefont
  {{H{\"a}rm}}(1958)}]{1958ApJ...128..348S}%
  \BibitemOpen
  \bibfield  {author} {\bibinfo {author} {\bibfnamefont {M.}~\bibnamefont
  {{Schwarzschild}}}\ and\ \bibinfo {author} {\bibfnamefont {R.}~\bibnamefont
  {{H{\"a}rm}}},\ }\bibfield  {title} {\bibinfo {title} {{Evolution of Very
  Massive Stars.}},\ }\href {https://doi.org/10.1086/146548} {\bibfield
  {journal} {\bibinfo  {journal} {\apj}\ }\textbf {\bibinfo {volume} {128}},\
  \bibinfo {pages} {348} (\bibinfo {year} {1958})}\BibitemShut {NoStop}%
\bibitem [{\citenamefont {{Kato}}(1966)}]{1966PASJ...18..374K}%
  \BibitemOpen
  \bibfield  {author} {\bibinfo {author} {\bibfnamefont {S.}~\bibnamefont
  {{Kato}}},\ }\bibfield  {title} {\bibinfo {title} {{Overstable Convection in
  a Medium Stratified in Mean Molecular Weight}},\ }\href@noop {} {\bibfield
  {journal} {\bibinfo  {journal} {\pasj}\ }\textbf {\bibinfo {volume} {18}},\
  \bibinfo {pages} {374} (\bibinfo {year} {1966})}\BibitemShut {NoStop}%
\bibitem [{\citenamefont {{Langer}}(1991)}]{1991A&A...252..669L}%
  \BibitemOpen
  \bibfield  {author} {\bibinfo {author} {\bibfnamefont {N.}~\bibnamefont
  {{Langer}}},\ }\bibfield  {title} {\bibinfo {title} {{Evolution of massive
  stars in the Large Magellanic Cloud : models with semiconvection.}},\
  }\href@noop {} {\bibfield  {journal} {\bibinfo  {journal} {\aap}\ }\textbf
  {\bibinfo {volume} {252}},\ \bibinfo {pages} {669} (\bibinfo {year}
  {1991})}\BibitemShut {NoStop}%
\bibitem [{\citenamefont {{Garaud}}(2018)}]{2018AnRFM..50..275G}%
  \BibitemOpen
  \bibfield  {author} {\bibinfo {author} {\bibfnamefont {P.}~\bibnamefont
  {{Garaud}}},\ }\bibfield  {title} {\bibinfo {title} {{Double-Diffusive
  Convection at Low Prandtl Number}},\ }\href
  {https://doi.org/10.1146/annurev-fluid-122316-045234} {\bibfield  {journal}
  {\bibinfo  {journal} {Annual Review of Fluid Mechanics}\ }\textbf {\bibinfo
  {volume} {50}},\ \bibinfo {pages} {275} (\bibinfo {year} {2018})}\BibitemShut
  {NoStop}%
\bibitem [{\citenamefont {{Anders}}\ \emph
  {et~al.}(2022{\natexlab{a}})\citenamefont {{Anders}}, \citenamefont
  {{Jermyn}}, \citenamefont {{Lecoanet}}, \citenamefont {{Fraser}},
  \citenamefont {{Cresswell}}, \citenamefont {{Joyce}},\ and\ \citenamefont
  {{Fuentes}}}]{2022ApJ...928L..10A}%
  \BibitemOpen
  \bibfield  {author} {\bibinfo {author} {\bibfnamefont {E.~H.}\ \bibnamefont
  {{Anders}}}, \bibinfo {author} {\bibfnamefont {A.~S.}\ \bibnamefont
  {{Jermyn}}}, \bibinfo {author} {\bibfnamefont {D.}~\bibnamefont
  {{Lecoanet}}}, \bibinfo {author} {\bibfnamefont {A.~E.}\ \bibnamefont
  {{Fraser}}}, \bibinfo {author} {\bibfnamefont {I.~G.}\ \bibnamefont
  {{Cresswell}}}, \bibinfo {author} {\bibfnamefont {M.}~\bibnamefont
  {{Joyce}}},\ and\ \bibinfo {author} {\bibfnamefont {J.~R.}\ \bibnamefont
  {{Fuentes}}},\ }\bibfield  {title} {\bibinfo {title} {{Schwarzschild and
  Ledoux are Equivalent on Evolutionary Timescales}},\ }\href
  {https://doi.org/10.3847/2041-8213/ac5cb5} {\bibfield  {journal} {\bibinfo
  {journal} {\apjl}\ }\textbf {\bibinfo {volume} {928}},\ \bibinfo {eid} {L10}
  (\bibinfo {year} {2022}{\natexlab{a}})},\ \Eprint
  {https://arxiv.org/abs/2203.06186} {arXiv:2203.06186 [astro-ph.SR]}
  \BibitemShut {NoStop}%
\bibitem [{\citenamefont {{Langer}}\ \emph {et~al.}(1985)\citenamefont
  {{Langer}}, \citenamefont {{El Eid}},\ and\ \citenamefont
  {{Fricke}}}]{1985A&A...145..179L}%
  \BibitemOpen
  \bibfield  {author} {\bibinfo {author} {\bibfnamefont {N.}~\bibnamefont
  {{Langer}}}, \bibinfo {author} {\bibfnamefont {M.~F.}\ \bibnamefont {{El
  Eid}}},\ and\ \bibinfo {author} {\bibfnamefont {K.~J.}\ \bibnamefont
  {{Fricke}}},\ }\bibfield  {title} {\bibinfo {title} {{Evolution of massive
  stars with semiconvective diffusion}},\ }\href@noop {} {\bibfield  {journal}
  {\bibinfo  {journal} {\aap}\ }\textbf {\bibinfo {volume} {145}},\ \bibinfo
  {pages} {179} (\bibinfo {year} {1985})}\BibitemShut {NoStop}%
\bibitem [{\citenamefont {{Chiosi}}\ and\ \citenamefont
  {{Maeder}}(1986)}]{1986ARA&A..24..329C}%
  \BibitemOpen
  \bibfield  {author} {\bibinfo {author} {\bibfnamefont {C.}~\bibnamefont
  {{Chiosi}}}\ and\ \bibinfo {author} {\bibfnamefont {A.}~\bibnamefont
  {{Maeder}}},\ }\bibfield  {title} {\bibinfo {title} {{The evolution of
  massive stars with mass loss.}},\ }\href
  {https://doi.org/10.1146/annurev.aa.24.090186.001553} {\bibfield  {journal}
  {\bibinfo  {journal} {\araa}\ }\textbf {\bibinfo {volume} {24}},\ \bibinfo
  {pages} {329} (\bibinfo {year} {1986})}\BibitemShut {NoStop}%
\bibitem [{\citenamefont {{Ulrich}}(1972)}]{1972ApJ...172..165U}%
  \BibitemOpen
  \bibfield  {author} {\bibinfo {author} {\bibfnamefont {R.~K.}\ \bibnamefont
  {{Ulrich}}},\ }\bibfield  {title} {\bibinfo {title} {{Thermohaline Convection
  in Stellar Interiors.}},\ }\href {https://doi.org/10.1086/151336} {\bibfield
  {journal} {\bibinfo  {journal} {\apj}\ }\textbf {\bibinfo {volume} {172}},\
  \bibinfo {pages} {165} (\bibinfo {year} {1972})}\BibitemShut {NoStop}%
\bibitem [{\citenamefont {{Kippenhahn}}\ \emph {et~al.}(1980)\citenamefont
  {{Kippenhahn}}, \citenamefont {{Ruschenplatt}},\ and\ \citenamefont
  {{Thomas}}}]{1980A&A....91..175K}%
  \BibitemOpen
  \bibfield  {author} {\bibinfo {author} {\bibfnamefont {R.}~\bibnamefont
  {{Kippenhahn}}}, \bibinfo {author} {\bibfnamefont {G.}~\bibnamefont
  {{Ruschenplatt}}},\ and\ \bibinfo {author} {\bibfnamefont {H.~C.}\
  \bibnamefont {{Thomas}}},\ }\bibfield  {title} {\bibinfo {title} {{The time
  scale of thermohaline mixing in stars}},\ }\href@noop {} {\bibfield
  {journal} {\bibinfo  {journal} {\aap}\ }\textbf {\bibinfo {volume} {91}},\
  \bibinfo {pages} {175} (\bibinfo {year} {1980})}\BibitemShut {NoStop}%
\bibitem [{\citenamefont {{Eggleton}}\ \emph {et~al.}(2006)\citenamefont
  {{Eggleton}}, \citenamefont {{Dearborn}},\ and\ \citenamefont
  {{Lattanzio}}}]{2006Sci...314.1580E}%
  \BibitemOpen
  \bibfield  {author} {\bibinfo {author} {\bibfnamefont {P.~P.}\ \bibnamefont
  {{Eggleton}}}, \bibinfo {author} {\bibfnamefont {D.~S.~P.}\ \bibnamefont
  {{Dearborn}}},\ and\ \bibinfo {author} {\bibfnamefont {J.~C.}\ \bibnamefont
  {{Lattanzio}}},\ }\bibfield  {title} {\bibinfo {title} {{Deep Mixing of
  $^{3}$He: Reconciling Big Bang and Stellar Nucleosynthesis}},\ }\href
  {https://doi.org/10.1126/science.1133065} {\bibfield  {journal} {\bibinfo
  {journal} {Science}\ }\textbf {\bibinfo {volume} {314}},\ \bibinfo {pages}
  {1580} (\bibinfo {year} {2006})},\ \Eprint
  {https://arxiv.org/abs/astro-ph/0611039} {arXiv:astro-ph/0611039 [astro-ph]}
  \BibitemShut {NoStop}%
\bibitem [{\citenamefont {{Charbonnel}}\ and\ \citenamefont
  {{Zahn}}(2007)}]{2007A&A...467L..15C}%
  \BibitemOpen
  \bibfield  {author} {\bibinfo {author} {\bibfnamefont {C.}~\bibnamefont
  {{Charbonnel}}}\ and\ \bibinfo {author} {\bibfnamefont {J.~P.}\ \bibnamefont
  {{Zahn}}},\ }\bibfield  {title} {\bibinfo {title} {{Thermohaline mixing: a
  physical mechanism governing the photospheric composition of low-mass
  giants}},\ }\href {https://doi.org/10.1051/0004-6361:20077274} {\bibfield
  {journal} {\bibinfo  {journal} {\aap}\ }\textbf {\bibinfo {volume} {467}},\
  \bibinfo {pages} {L15} (\bibinfo {year} {2007})},\ \Eprint
  {https://arxiv.org/abs/astro-ph/0703302} {arXiv:astro-ph/0703302 [astro-ph]}
  \BibitemShut {NoStop}%
\bibitem [{\citenamefont {{Wellstein}}\ \emph
  {et~al.}(2001{\natexlab{a}})\citenamefont {{Wellstein}}, \citenamefont
  {{Langer}},\ and\ \citenamefont {{Braun}}}]{2001A&A...369..939W}%
  \BibitemOpen
  \bibfield  {author} {\bibinfo {author} {\bibfnamefont {S.}~\bibnamefont
  {{Wellstein}}}, \bibinfo {author} {\bibfnamefont {N.}~\bibnamefont
  {{Langer}}},\ and\ \bibinfo {author} {\bibfnamefont {H.}~\bibnamefont
  {{Braun}}},\ }\bibfield  {title} {\bibinfo {title} {{Formation of contact in
  massive close binaries}},\ }\href
  {https://doi.org/10.1051/0004-6361:20010151} {\bibfield  {journal} {\bibinfo
  {journal} {\aap}\ }\textbf {\bibinfo {volume} {369}},\ \bibinfo {pages} {939}
  (\bibinfo {year} {2001}{\natexlab{a}})},\ \Eprint
  {https://arxiv.org/abs/astro-ph/0102244} {arXiv:astro-ph/0102244 [astro-ph]}
  \BibitemShut {NoStop}%
\bibitem [{\citenamefont {{Ekstr{\"o}m}}\ \emph {et~al.}(2012)\citenamefont
  {{Ekstr{\"o}m}}, \citenamefont {{Georgy}}, \citenamefont {{Eggenberger}},
  \citenamefont {{Meynet}}, \citenamefont {{Mowlavi}}, \citenamefont
  {{Wyttenbach}}, \citenamefont {{Granada}}, \citenamefont {{Decressin}},
  \citenamefont {{Hirschi}}, \citenamefont {{Frischknecht}}, \citenamefont
  {{Charbonnel}},\ and\ \citenamefont {{Maeder}}}]{2012A&A...537A.146E}%
  \BibitemOpen
  \bibfield  {author} {\bibinfo {author} {\bibfnamefont {S.}~\bibnamefont
  {{Ekstr{\"o}m}}}, \bibinfo {author} {\bibfnamefont {C.}~\bibnamefont
  {{Georgy}}}, \bibinfo {author} {\bibfnamefont {P.}~\bibnamefont
  {{Eggenberger}}}, \bibinfo {author} {\bibfnamefont {G.}~\bibnamefont
  {{Meynet}}}, \bibinfo {author} {\bibfnamefont {N.}~\bibnamefont {{Mowlavi}}},
  \bibinfo {author} {\bibfnamefont {A.}~\bibnamefont {{Wyttenbach}}}, \bibinfo
  {author} {\bibfnamefont {A.}~\bibnamefont {{Granada}}}, \bibinfo {author}
  {\bibfnamefont {T.}~\bibnamefont {{Decressin}}}, \bibinfo {author}
  {\bibfnamefont {R.}~\bibnamefont {{Hirschi}}}, \bibinfo {author}
  {\bibfnamefont {U.}~\bibnamefont {{Frischknecht}}}, \bibinfo {author}
  {\bibfnamefont {C.}~\bibnamefont {{Charbonnel}}},\ and\ \bibinfo {author}
  {\bibfnamefont {A.}~\bibnamefont {{Maeder}}},\ }\bibfield  {title} {\bibinfo
  {title} {{Grids of stellar models with rotation. I. Models from 0.8 to 120
  M$_{{\ensuremath{\odot}}}$ at solar metallicity (Z = 0.014)}},\ }\href
  {https://doi.org/10.1051/0004-6361/201117751} {\bibfield  {journal} {\bibinfo
   {journal} {\aap}\ }\textbf {\bibinfo {volume} {537}},\ \bibinfo {eid} {A146}
  (\bibinfo {year} {2012})},\ \Eprint {https://arxiv.org/abs/1110.5049}
  {arXiv:1110.5049 [astro-ph.SR]} \BibitemShut {NoStop}%
\bibitem [{\citenamefont {{Ritter}}\ \emph {et~al.}(2018)\citenamefont
  {{Ritter}}, \citenamefont {{Herwig}}, \citenamefont {{Jones}}, \citenamefont
  {{Pignatari}}, \citenamefont {{Fryer}},\ and\ \citenamefont
  {{Hirschi}}}]{2018MNRAS.480..538R}%
  \BibitemOpen
  \bibfield  {author} {\bibinfo {author} {\bibfnamefont {C.}~\bibnamefont
  {{Ritter}}}, \bibinfo {author} {\bibfnamefont {F.}~\bibnamefont {{Herwig}}},
  \bibinfo {author} {\bibfnamefont {S.}~\bibnamefont {{Jones}}}, \bibinfo
  {author} {\bibfnamefont {M.}~\bibnamefont {{Pignatari}}}, \bibinfo {author}
  {\bibfnamefont {C.}~\bibnamefont {{Fryer}}},\ and\ \bibinfo {author}
  {\bibfnamefont {R.}~\bibnamefont {{Hirschi}}},\ }\bibfield  {title} {\bibinfo
  {title} {{NuGrid stellar data set - II. Stellar yields from H to Bi for
  stellar models with M$_{ZAMS}$ = 1-25 M$_{{\ensuremath{\odot}}}$ and Z =
  0.0001-0.02}},\ }\href {https://doi.org/10.1093/mnras/sty1729} {\bibfield
  {journal} {\bibinfo  {journal} {\mnras}\ }\textbf {\bibinfo {volume} {480}},\
  \bibinfo {pages} {538} (\bibinfo {year} {2018})},\ \Eprint
  {https://arxiv.org/abs/1709.08677} {arXiv:1709.08677 [astro-ph.SR]}
  \BibitemShut {NoStop}%
\bibitem [{\citenamefont {{Hidalgo}}\ \emph {et~al.}(2018)\citenamefont
  {{Hidalgo}}, \citenamefont {{Pietrinferni}}, \citenamefont {{Cassisi}},
  \citenamefont {{Salaris}}, \citenamefont {{Mucciarelli}}, \citenamefont
  {{Savino}}, \citenamefont {{Aparicio}}, \citenamefont {{Silva Aguirre}},\
  and\ \citenamefont {{Verma}}}]{2018ApJ...856..125H}%
  \BibitemOpen
  \bibfield  {author} {\bibinfo {author} {\bibfnamefont {S.~L.}\ \bibnamefont
  {{Hidalgo}}}, \bibinfo {author} {\bibfnamefont {A.}~\bibnamefont
  {{Pietrinferni}}}, \bibinfo {author} {\bibfnamefont {S.}~\bibnamefont
  {{Cassisi}}}, \bibinfo {author} {\bibfnamefont {M.}~\bibnamefont
  {{Salaris}}}, \bibinfo {author} {\bibfnamefont {A.}~\bibnamefont
  {{Mucciarelli}}}, \bibinfo {author} {\bibfnamefont {A.}~\bibnamefont
  {{Savino}}}, \bibinfo {author} {\bibfnamefont {A.}~\bibnamefont
  {{Aparicio}}}, \bibinfo {author} {\bibfnamefont {V.}~\bibnamefont {{Silva
  Aguirre}}},\ and\ \bibinfo {author} {\bibfnamefont {K.}~\bibnamefont
  {{Verma}}},\ }\bibfield  {title} {\bibinfo {title} {{The Updated BaSTI
  Stellar Evolution Models and Isochrones. I. Solar-scaled Calculations}},\
  }\href {https://doi.org/10.3847/1538-4357/aab158} {\bibfield  {journal}
  {\bibinfo  {journal} {\apj}\ }\textbf {\bibinfo {volume} {856}},\ \bibinfo
  {eid} {125} (\bibinfo {year} {2018})},\ \Eprint
  {https://arxiv.org/abs/1802.07319} {arXiv:1802.07319 [astro-ph.GA]}
  \BibitemShut {NoStop}%
\bibitem [{\citenamefont {{Amard}}\ \emph {et~al.}(2019)\citenamefont
  {{Amard}}, \citenamefont {{Palacios}}, \citenamefont {{Charbonnel}},
  \citenamefont {{Gallet}}, \citenamefont {{Georgy}}, \citenamefont
  {{Lagarde}},\ and\ \citenamefont {{Siess}}}]{2019A&A...631A..77A}%
  \BibitemOpen
  \bibfield  {author} {\bibinfo {author} {\bibfnamefont {L.}~\bibnamefont
  {{Amard}}}, \bibinfo {author} {\bibfnamefont {A.}~\bibnamefont {{Palacios}}},
  \bibinfo {author} {\bibfnamefont {C.}~\bibnamefont {{Charbonnel}}}, \bibinfo
  {author} {\bibfnamefont {F.}~\bibnamefont {{Gallet}}}, \bibinfo {author}
  {\bibfnamefont {C.}~\bibnamefont {{Georgy}}}, \bibinfo {author}
  {\bibfnamefont {N.}~\bibnamefont {{Lagarde}}},\ and\ \bibinfo {author}
  {\bibfnamefont {L.}~\bibnamefont {{Siess}}},\ }\bibfield  {title} {\bibinfo
  {title} {{First grids of low-mass stellar models and isochrones with
  self-consistent treatment of rotation. From 0.2 to 1.5
  M$_{{\ensuremath{\odot}}}$ at seven metallicities from PMS to TAMS}},\ }\href
  {https://doi.org/10.1051/0004-6361/201935160} {\bibfield  {journal} {\bibinfo
   {journal} {\aap}\ }\textbf {\bibinfo {volume} {631}},\ \bibinfo {eid} {A77}
  (\bibinfo {year} {2019})},\ \Eprint {https://arxiv.org/abs/1905.08516}
  {arXiv:1905.08516 [astro-ph.SR]} \BibitemShut {NoStop}%
\bibitem [{\citenamefont {{Heger}}\ \emph
  {et~al.}(2000{\natexlab{a}})\citenamefont {{Heger}}, \citenamefont
  {{Langer}},\ and\ \citenamefont {{Woosley}}}]{2000ApJ...528..368H}%
  \BibitemOpen
  \bibfield  {author} {\bibinfo {author} {\bibfnamefont {A.}~\bibnamefont
  {{Heger}}}, \bibinfo {author} {\bibfnamefont {N.}~\bibnamefont {{Langer}}},\
  and\ \bibinfo {author} {\bibfnamefont {S.~E.}\ \bibnamefont {{Woosley}}},\
  }\bibfield  {title} {\bibinfo {title} {{Presupernova Evolution of Rotating
  Massive Stars. I. Numerical Method and Evolution of the Internal Stellar
  Structure}},\ }\href {https://doi.org/10.1086/308158} {\bibfield  {journal}
  {\bibinfo  {journal} {\apj}\ }\textbf {\bibinfo {volume} {528}},\ \bibinfo
  {pages} {368} (\bibinfo {year} {2000}{\natexlab{a}})},\ \Eprint
  {https://arxiv.org/abs/astro-ph/9904132} {arXiv:astro-ph/9904132 [astro-ph]}
  \BibitemShut {NoStop}%
\bibitem [{\citenamefont {{Brott}}\ \emph
  {et~al.}(2011{\natexlab{a}})\citenamefont {{Brott}}, \citenamefont {{de
  Mink}}, \citenamefont {{Cantiello}}, \citenamefont {{Langer}}, \citenamefont
  {{de Koter}}, \citenamefont {{Evans}}, \citenamefont {{Hunter}},
  \citenamefont {{Trundle}},\ and\ \citenamefont
  {{Vink}}}]{2011A&A...530A.115B}%
  \BibitemOpen
  \bibfield  {author} {\bibinfo {author} {\bibfnamefont {I.}~\bibnamefont
  {{Brott}}}, \bibinfo {author} {\bibfnamefont {S.~E.}\ \bibnamefont {{de
  Mink}}}, \bibinfo {author} {\bibfnamefont {M.}~\bibnamefont {{Cantiello}}},
  \bibinfo {author} {\bibfnamefont {N.}~\bibnamefont {{Langer}}}, \bibinfo
  {author} {\bibfnamefont {A.}~\bibnamefont {{de Koter}}}, \bibinfo {author}
  {\bibfnamefont {C.~J.}\ \bibnamefont {{Evans}}}, \bibinfo {author}
  {\bibfnamefont {I.}~\bibnamefont {{Hunter}}}, \bibinfo {author}
  {\bibfnamefont {C.}~\bibnamefont {{Trundle}}},\ and\ \bibinfo {author}
  {\bibfnamefont {J.~S.}\ \bibnamefont {{Vink}}},\ }\bibfield  {title}
  {\bibinfo {title} {{Rotating massive main-sequence stars. I. Grids of
  evolutionary models and isochrones}},\ }\href
  {https://doi.org/10.1051/0004-6361/201016113} {\bibfield  {journal} {\bibinfo
   {journal} {\aap}\ }\textbf {\bibinfo {volume} {530}},\ \bibinfo {eid} {A115}
  (\bibinfo {year} {2011}{\natexlab{a}})},\ \Eprint
  {https://arxiv.org/abs/1102.0530} {arXiv:1102.0530 [astro-ph.SR]}
  \BibitemShut {NoStop}%
\bibitem [{\citenamefont {{Choi}}\ \emph {et~al.}(2016)\citenamefont {{Choi}},
  \citenamefont {{Dotter}}, \citenamefont {{Conroy}}, \citenamefont
  {{Cantiello}}, \citenamefont {{Paxton}},\ and\ \citenamefont
  {{Johnson}}}]{2016ApJ...823..102C}%
  \BibitemOpen
  \bibfield  {author} {\bibinfo {author} {\bibfnamefont {J.}~\bibnamefont
  {{Choi}}}, \bibinfo {author} {\bibfnamefont {A.}~\bibnamefont {{Dotter}}},
  \bibinfo {author} {\bibfnamefont {C.}~\bibnamefont {{Conroy}}}, \bibinfo
  {author} {\bibfnamefont {M.}~\bibnamefont {{Cantiello}}}, \bibinfo {author}
  {\bibfnamefont {B.}~\bibnamefont {{Paxton}}},\ and\ \bibinfo {author}
  {\bibfnamefont {B.~D.}\ \bibnamefont {{Johnson}}},\ }\bibfield  {title}
  {\bibinfo {title} {{Mesa Isochrones and Stellar Tracks (MIST). I.
  Solar-scaled Models}},\ }\href {https://doi.org/10.3847/0004-637X/823/2/102}
  {\bibfield  {journal} {\bibinfo  {journal} {\apj}\ }\textbf {\bibinfo
  {volume} {823}},\ \bibinfo {eid} {102} (\bibinfo {year} {2016})},\ \Eprint
  {https://arxiv.org/abs/1604.08592} {arXiv:1604.08592 [astro-ph.SR]}
  \BibitemShut {NoStop}%
\bibitem [{\citenamefont {{Sz{\'e}csi}}\ \emph {et~al.}(2022)\citenamefont
  {{Sz{\'e}csi}}, \citenamefont {{Agrawal}}, \citenamefont {{W{\"u}nsch}},\
  and\ \citenamefont {{Langer}}}]{2022A&A...658A.125S}%
  \BibitemOpen
  \bibfield  {author} {\bibinfo {author} {\bibfnamefont {D.}~\bibnamefont
  {{Sz{\'e}csi}}}, \bibinfo {author} {\bibfnamefont {P.}~\bibnamefont
  {{Agrawal}}}, \bibinfo {author} {\bibfnamefont {R.}~\bibnamefont
  {{W{\"u}nsch}}},\ and\ \bibinfo {author} {\bibfnamefont {N.}~\bibnamefont
  {{Langer}}},\ }\bibfield  {title} {\bibinfo {title} {{Bonn Optimized Stellar
  Tracks (BoOST). Simulated populations of massive and very massive stars for
  astrophysical applications}},\ }\href
  {https://doi.org/10.1051/0004-6361/202141536} {\bibfield  {journal} {\bibinfo
   {journal} {\aap}\ }\textbf {\bibinfo {volume} {658}},\ \bibinfo {eid} {A125}
  (\bibinfo {year} {2022})},\ \Eprint {https://arxiv.org/abs/2004.08203}
  {arXiv:2004.08203 [astro-ph.SR]} \BibitemShut {NoStop}%
\bibitem [{\citenamefont {{Maeder}}(1975)}]{1975A&A....40..303M}%
  \BibitemOpen
  \bibfield  {author} {\bibinfo {author} {\bibfnamefont {A.}~\bibnamefont
  {{Maeder}}},\ }\bibfield  {title} {\bibinfo {title} {{Stellar evolution III:
  the overshooting from convective cores.}},\ }\href@noop {} {\bibfield
  {journal} {\bibinfo  {journal} {\aap}\ }\textbf {\bibinfo {volume} {40}},\
  \bibinfo {pages} {303} (\bibinfo {year} {1975})}\BibitemShut {NoStop}%
\bibitem [{\citenamefont {{Bressan}}\ \emph {et~al.}(1981)\citenamefont
  {{Bressan}}, \citenamefont {{Chiosi}},\ and\ \citenamefont
  {{Bertelli}}}]{1981A&A...102...25B}%
  \BibitemOpen
  \bibfield  {author} {\bibinfo {author} {\bibfnamefont {A.~G.}\ \bibnamefont
  {{Bressan}}}, \bibinfo {author} {\bibfnamefont {C.}~\bibnamefont
  {{Chiosi}}},\ and\ \bibinfo {author} {\bibfnamefont {G.}~\bibnamefont
  {{Bertelli}}},\ }\bibfield  {title} {\bibinfo {title} {{Mass loss and
  overshooting in massive stars}},\ }\href@noop {} {\bibfield  {journal}
  {\bibinfo  {journal} {\aap}\ }\textbf {\bibinfo {volume} {102}},\ \bibinfo
  {pages} {25} (\bibinfo {year} {1981})}\BibitemShut {NoStop}%
\bibitem [{\citenamefont {{Rosenfield}}\ \emph {et~al.}(2017)\citenamefont
  {{Rosenfield}}, \citenamefont {{Girardi}}, \citenamefont {{Williams}},
  \citenamefont {{Johnson}}, \citenamefont {{Dolphin}}, \citenamefont
  {{Bressan}}, \citenamefont {{Weisz}}, \citenamefont {{Dalcanton}},
  \citenamefont {{Fouesneau}},\ and\ \citenamefont
  {{Kalirai}}}]{2017ApJ...841...69R}%
  \BibitemOpen
  \bibfield  {author} {\bibinfo {author} {\bibfnamefont {P.}~\bibnamefont
  {{Rosenfield}}}, \bibinfo {author} {\bibfnamefont {L.}~\bibnamefont
  {{Girardi}}}, \bibinfo {author} {\bibfnamefont {B.~F.}\ \bibnamefont
  {{Williams}}}, \bibinfo {author} {\bibfnamefont {L.~C.}\ \bibnamefont
  {{Johnson}}}, \bibinfo {author} {\bibfnamefont {A.}~\bibnamefont
  {{Dolphin}}}, \bibinfo {author} {\bibfnamefont {A.}~\bibnamefont
  {{Bressan}}}, \bibinfo {author} {\bibfnamefont {D.}~\bibnamefont {{Weisz}}},
  \bibinfo {author} {\bibfnamefont {J.~J.}\ \bibnamefont {{Dalcanton}}},
  \bibinfo {author} {\bibfnamefont {M.}~\bibnamefont {{Fouesneau}}},\ and\
  \bibinfo {author} {\bibfnamefont {J.}~\bibnamefont {{Kalirai}}},\ }\bibfield
  {title} {\bibinfo {title} {{A New Approach to Convective Core Overshooting:
  Probabilistic Constraints from Color-Magnitude Diagrams of LMC Clusters}},\
  }\href {https://doi.org/10.3847/1538-4357/aa70a2} {\bibfield  {journal}
  {\bibinfo  {journal} {\apj}\ }\textbf {\bibinfo {volume} {841}},\ \bibinfo
  {eid} {69} (\bibinfo {year} {2017})},\ \Eprint
  {https://arxiv.org/abs/1705.00618} {arXiv:1705.00618 [astro-ph.SR]}
  \BibitemShut {NoStop}%
\bibitem [{\citenamefont {{Claret}}\ and\ \citenamefont
  {{Torres}}(2019)}]{2019ApJ...876..134C}%
  \BibitemOpen
  \bibfield  {author} {\bibinfo {author} {\bibfnamefont {A.}~\bibnamefont
  {{Claret}}}\ and\ \bibinfo {author} {\bibfnamefont {G.}~\bibnamefont
  {{Torres}}},\ }\bibfield  {title} {\bibinfo {title} {{The Dependence of
  Convective Core Overshooting on Stellar Mass: Reality Check and Additional
  Evidence}},\ }\href {https://doi.org/10.3847/1538-4357/ab1589} {\bibfield
  {journal} {\bibinfo  {journal} {\apj}\ }\textbf {\bibinfo {volume} {876}},\
  \bibinfo {eid} {134} (\bibinfo {year} {2019})},\ \Eprint
  {https://arxiv.org/abs/1904.02714} {arXiv:1904.02714 [astro-ph.SR]}
  \BibitemShut {NoStop}%
\bibitem [{\citenamefont {{Aerts}}(2021)}]{2021RvMP...93a5001A}%
  \BibitemOpen
  \bibfield  {author} {\bibinfo {author} {\bibfnamefont {C.}~\bibnamefont
  {{Aerts}}},\ }\bibfield  {title} {\bibinfo {title} {{Probing the interior
  physics of stars through asteroseismology}},\ }\href
  {https://doi.org/10.1103/RevModPhys.93.015001} {\bibfield  {journal}
  {\bibinfo  {journal} {Reviews of Modern Physics}\ }\textbf {\bibinfo {volume}
  {93}},\ \bibinfo {eid} {015001} (\bibinfo {year} {2021})},\ \Eprint
  {https://arxiv.org/abs/1912.12300} {arXiv:1912.12300 [astro-ph.SR]}
  \BibitemShut {NoStop}%
\bibitem [{\citenamefont {{Denissenkov}}\ \emph {et~al.}(2013)\citenamefont
  {{Denissenkov}}, \citenamefont {{Herwig}}, \citenamefont {{Truran}},\ and\
  \citenamefont {{Paxton}}}]{2013ApJ...772...37D}%
  \BibitemOpen
  \bibfield  {author} {\bibinfo {author} {\bibfnamefont {P.~A.}\ \bibnamefont
  {{Denissenkov}}}, \bibinfo {author} {\bibfnamefont {F.}~\bibnamefont
  {{Herwig}}}, \bibinfo {author} {\bibfnamefont {J.~W.}\ \bibnamefont
  {{Truran}}},\ and\ \bibinfo {author} {\bibfnamefont {B.}~\bibnamefont
  {{Paxton}}},\ }\bibfield  {title} {\bibinfo {title} {{The C-flame Quenching
  by Convective Boundary Mixing in Super-AGB Stars and the Formation of Hybrid
  C/O/Ne White Dwarfs and SN Progenitors}},\ }\href
  {https://doi.org/10.1088/0004-637X/772/1/37} {\bibfield  {journal} {\bibinfo
  {journal} {\apj}\ }\textbf {\bibinfo {volume} {772}},\ \bibinfo {eid} {37}
  (\bibinfo {year} {2013})},\ \Eprint {https://arxiv.org/abs/1305.2649}
  {arXiv:1305.2649 [astro-ph.SR]} \BibitemShut {NoStop}%
\bibitem [{\citenamefont {{Saslaw}}\ and\ \citenamefont
  {{Schwarzschild}}(1965)}]{1965ApJ...142.1468S}%
  \BibitemOpen
  \bibfield  {author} {\bibinfo {author} {\bibfnamefont {W.~C.}\ \bibnamefont
  {{Saslaw}}}\ and\ \bibinfo {author} {\bibfnamefont {M.}~\bibnamefont
  {{Schwarzschild}}},\ }\bibfield  {title} {\bibinfo {title} {{Overshooting
  from Stellar Convective Cores.}},\ }\href {https://doi.org/10.1086/148430}
  {\bibfield  {journal} {\bibinfo  {journal} {\apj}\ }\textbf {\bibinfo
  {volume} {142}},\ \bibinfo {pages} {1468} (\bibinfo {year}
  {1965})}\BibitemShut {NoStop}%
\bibitem [{\citenamefont {{Shaviv}}\ and\ \citenamefont
  {{Salpeter}}(1973)}]{1973ApJ...184..191S}%
  \BibitemOpen
  \bibfield  {author} {\bibinfo {author} {\bibfnamefont {G.}~\bibnamefont
  {{Shaviv}}}\ and\ \bibinfo {author} {\bibfnamefont {E.~E.}\ \bibnamefont
  {{Salpeter}}},\ }\bibfield  {title} {\bibinfo {title} {{Convective
  Overshooting in Stellar Interior Models}},\ }\href
  {https://doi.org/10.1086/152318} {\bibfield  {journal} {\bibinfo  {journal}
  {\apj}\ }\textbf {\bibinfo {volume} {184}},\ \bibinfo {pages} {191} (\bibinfo
  {year} {1973})}\BibitemShut {NoStop}%
\bibitem [{\citenamefont {{Herwig}}\ \emph {et~al.}(1997)\citenamefont
  {{Herwig}}, \citenamefont {{Bloecker}}, \citenamefont {{Schoenberner}},\ and\
  \citenamefont {{El Eid}}}]{1997A&A...324L..81H}%
  \BibitemOpen
  \bibfield  {author} {\bibinfo {author} {\bibfnamefont {F.}~\bibnamefont
  {{Herwig}}}, \bibinfo {author} {\bibfnamefont {T.}~\bibnamefont
  {{Bloecker}}}, \bibinfo {author} {\bibfnamefont {D.}~\bibnamefont
  {{Schoenberner}}},\ and\ \bibinfo {author} {\bibfnamefont {M.}~\bibnamefont
  {{El Eid}}},\ }\bibfield  {title} {\bibinfo {title} {{Stellar evolution of
  low and intermediate-mass stars. IV. Hydrodynamically-based overshoot and
  nucleosynthesis in AGB stars.}},\ }\href@noop {} {\bibfield  {journal}
  {\bibinfo  {journal} {\aap}\ }\textbf {\bibinfo {volume} {324}},\ \bibinfo
  {pages} {L81} (\bibinfo {year} {1997})},\ \Eprint
  {https://arxiv.org/abs/astro-ph/9706122} {arXiv:astro-ph/9706122 [astro-ph]}
  \BibitemShut {NoStop}%
\bibitem [{\citenamefont {{Pietrinferni}}\ \emph {et~al.}(2004)\citenamefont
  {{Pietrinferni}}, \citenamefont {{Cassisi}}, \citenamefont {{Salaris}},\ and\
  \citenamefont {{Castelli}}}]{2004ApJ...612..168P}%
  \BibitemOpen
  \bibfield  {author} {\bibinfo {author} {\bibfnamefont {A.}~\bibnamefont
  {{Pietrinferni}}}, \bibinfo {author} {\bibfnamefont {S.}~\bibnamefont
  {{Cassisi}}}, \bibinfo {author} {\bibfnamefont {M.}~\bibnamefont
  {{Salaris}}},\ and\ \bibinfo {author} {\bibfnamefont {F.}~\bibnamefont
  {{Castelli}}},\ }\bibfield  {title} {\bibinfo {title} {{A Large Stellar
  Evolution Database for Population Synthesis Studies. I. Scaled Solar Models
  and Isochrones}},\ }\href {https://doi.org/10.1086/422498} {\bibfield
  {journal} {\bibinfo  {journal} {\apj}\ }\textbf {\bibinfo {volume} {612}},\
  \bibinfo {pages} {168} (\bibinfo {year} {2004})},\ \Eprint
  {https://arxiv.org/abs/astro-ph/0405193} {arXiv:astro-ph/0405193 [astro-ph]}
  \BibitemShut {NoStop}%
\bibitem [{\citenamefont {{Dotter}}\ \emph {et~al.}(2008)\citenamefont
  {{Dotter}}, \citenamefont {{Chaboyer}}, \citenamefont {{Jevremovi{\'c}}},
  \citenamefont {{Kostov}}, \citenamefont {{Baron}},\ and\ \citenamefont
  {{Ferguson}}}]{2008ApJS..178...89D}%
  \BibitemOpen
  \bibfield  {author} {\bibinfo {author} {\bibfnamefont {A.}~\bibnamefont
  {{Dotter}}}, \bibinfo {author} {\bibfnamefont {B.}~\bibnamefont
  {{Chaboyer}}}, \bibinfo {author} {\bibfnamefont {D.}~\bibnamefont
  {{Jevremovi{\'c}}}}, \bibinfo {author} {\bibfnamefont {V.}~\bibnamefont
  {{Kostov}}}, \bibinfo {author} {\bibfnamefont {E.}~\bibnamefont {{Baron}}},\
  and\ \bibinfo {author} {\bibfnamefont {J.~W.}\ \bibnamefont {{Ferguson}}},\
  }\bibfield  {title} {\bibinfo {title} {{The Dartmouth Stellar Evolution
  Database}},\ }\href {https://doi.org/10.1086/589654} {\bibfield  {journal}
  {\bibinfo  {journal} {\apjs}\ }\textbf {\bibinfo {volume} {178}},\ \bibinfo
  {pages} {89} (\bibinfo {year} {2008})},\ \Eprint
  {https://arxiv.org/abs/0804.4473} {arXiv:0804.4473 [astro-ph]} \BibitemShut
  {NoStop}%
\bibitem [{\citenamefont {{Johnston}}(2021)}]{2021A&A...655A..29J}%
  \BibitemOpen
  \bibfield  {author} {\bibinfo {author} {\bibfnamefont {C.}~\bibnamefont
  {{Johnston}}},\ }\bibfield  {title} {\bibinfo {title} {{One size does not fit
  all: Evidence for a range of mixing efficiencies in stellar evolution
  calculations}},\ }\href {https://doi.org/10.1051/0004-6361/202141080}
  {\bibfield  {journal} {\bibinfo  {journal} {\aap}\ }\textbf {\bibinfo
  {volume} {655}},\ \bibinfo {eid} {A29} (\bibinfo {year} {2021})},\ \Eprint
  {https://arxiv.org/abs/2107.09075} {arXiv:2107.09075 [astro-ph.SR]}
  \BibitemShut {NoStop}%
\bibitem [{\citenamefont {{Costa}}\ \emph
  {et~al.}(2019{\natexlab{b}})\citenamefont {{Costa}}, \citenamefont
  {{Girardi}}, \citenamefont {{Bressan}}, \citenamefont {{Marigo}},
  \citenamefont {{Rodrigues}}, \citenamefont {{Chen}}, \citenamefont
  {{Lanza}},\ and\ \citenamefont {{Goudfrooij}}}]{2019MNRAS.485.4641C}%
  \BibitemOpen
  \bibfield  {author} {\bibinfo {author} {\bibfnamefont {G.}~\bibnamefont
  {{Costa}}}, \bibinfo {author} {\bibfnamefont {L.}~\bibnamefont {{Girardi}}},
  \bibinfo {author} {\bibfnamefont {A.}~\bibnamefont {{Bressan}}}, \bibinfo
  {author} {\bibfnamefont {P.}~\bibnamefont {{Marigo}}}, \bibinfo {author}
  {\bibfnamefont {T.~S.}\ \bibnamefont {{Rodrigues}}}, \bibinfo {author}
  {\bibfnamefont {Y.}~\bibnamefont {{Chen}}}, \bibinfo {author} {\bibfnamefont
  {A.}~\bibnamefont {{Lanza}}},\ and\ \bibinfo {author} {\bibfnamefont
  {P.}~\bibnamefont {{Goudfrooij}}},\ }\bibfield  {title} {\bibinfo {title}
  {{Mixing by overshooting and rotation in intermediate-mass stars}},\ }\href
  {https://doi.org/10.1093/mnras/stz728} {\bibfield  {journal} {\bibinfo
  {journal} {\mnras}\ }\textbf {\bibinfo {volume} {485}},\ \bibinfo {pages}
  {4641} (\bibinfo {year} {2019}{\natexlab{b}})},\ \Eprint
  {https://arxiv.org/abs/1903.04368} {arXiv:1903.04368 [astro-ph.SR]}
  \BibitemShut {NoStop}%
\bibitem [{\citenamefont {{Castro}}\ \emph {et~al.}(2014)\citenamefont
  {{Castro}}, \citenamefont {{Fossati}}, \citenamefont {{Langer}},
  \citenamefont {{Sim{\'o}n-D{\'\i}az}}, \citenamefont {{Schneider}},\ and\
  \citenamefont {{Izzard}}}]{2014A&A...570L..13C}%
  \BibitemOpen
  \bibfield  {author} {\bibinfo {author} {\bibfnamefont {N.}~\bibnamefont
  {{Castro}}}, \bibinfo {author} {\bibfnamefont {L.}~\bibnamefont {{Fossati}}},
  \bibinfo {author} {\bibfnamefont {N.}~\bibnamefont {{Langer}}}, \bibinfo
  {author} {\bibfnamefont {S.}~\bibnamefont {{Sim{\'o}n-D{\'\i}az}}}, \bibinfo
  {author} {\bibfnamefont {F.~R.~N.}\ \bibnamefont {{Schneider}}},\ and\
  \bibinfo {author} {\bibfnamefont {R.~G.}\ \bibnamefont {{Izzard}}},\
  }\bibfield  {title} {\bibinfo {title} {{The spectroscopic Hertzsprung-Russell
  diagram of Galactic massive stars}},\ }\href
  {https://doi.org/10.1051/0004-6361/201425028} {\bibfield  {journal} {\bibinfo
   {journal} {\aap}\ }\textbf {\bibinfo {volume} {570}},\ \bibinfo {eid} {L13}
  (\bibinfo {year} {2014})},\ \Eprint {https://arxiv.org/abs/1410.3499}
  {arXiv:1410.3499 [astro-ph.SR]} \BibitemShut {NoStop}%
\bibitem [{\citenamefont {{Tkachenko}}\ \emph {et~al.}(2020)\citenamefont
  {{Tkachenko}}, \citenamefont {{Pavlovski}}, \citenamefont {{Johnston}},
  \citenamefont {{Pedersen}}, \citenamefont {{Michielsen}}, \citenamefont
  {{Bowman}}, \citenamefont {{Southworth}}, \citenamefont {{Tsymbal}},\ and\
  \citenamefont {{Aerts}}}]{2020A&A...637A..60T}%
  \BibitemOpen
  \bibfield  {author} {\bibinfo {author} {\bibfnamefont {A.}~\bibnamefont
  {{Tkachenko}}}, \bibinfo {author} {\bibfnamefont {K.}~\bibnamefont
  {{Pavlovski}}}, \bibinfo {author} {\bibfnamefont {C.}~\bibnamefont
  {{Johnston}}}, \bibinfo {author} {\bibfnamefont {M.~G.}\ \bibnamefont
  {{Pedersen}}}, \bibinfo {author} {\bibfnamefont {M.}~\bibnamefont
  {{Michielsen}}}, \bibinfo {author} {\bibfnamefont {D.~M.}\ \bibnamefont
  {{Bowman}}}, \bibinfo {author} {\bibfnamefont {J.}~\bibnamefont
  {{Southworth}}}, \bibinfo {author} {\bibfnamefont {V.}~\bibnamefont
  {{Tsymbal}}},\ and\ \bibinfo {author} {\bibfnamefont {C.}~\bibnamefont
  {{Aerts}}},\ }\bibfield  {title} {\bibinfo {title} {{The mass discrepancy in
  intermediate- and high-mass eclipsing binaries: The need for higher
  convective core masses}},\ }\href
  {https://doi.org/10.1051/0004-6361/202037452} {\bibfield  {journal} {\bibinfo
   {journal} {\aap}\ }\textbf {\bibinfo {volume} {637}},\ \bibinfo {eid} {A60}
  (\bibinfo {year} {2020})},\ \Eprint {https://arxiv.org/abs/2003.08982}
  {arXiv:2003.08982 [astro-ph.SR]} \BibitemShut {NoStop}%
\bibitem [{\citenamefont {{Martinet}}\ \emph {et~al.}(2021)\citenamefont
  {{Martinet}}, \citenamefont {{Meynet}}, \citenamefont {{Ekstr{\"o}m}},
  \citenamefont {{Sim{\'o}n-D{\'\i}az}}, \citenamefont {{Holgado}},
  \citenamefont {{Castro}}, \citenamefont {{Georgy}}, \citenamefont
  {{Eggenberger}}, \citenamefont {{Buldgen}}, \citenamefont {{Salmon}},
  \citenamefont {{Hirschi}}, \citenamefont {{Groh}}, \citenamefont
  {{Farrell}},\ and\ \citenamefont {{Murphy}}}]{2021A&A...648A.126M}%
  \BibitemOpen
  \bibfield  {author} {\bibinfo {author} {\bibfnamefont {S.}~\bibnamefont
  {{Martinet}}}, \bibinfo {author} {\bibfnamefont {G.}~\bibnamefont
  {{Meynet}}}, \bibinfo {author} {\bibfnamefont {S.}~\bibnamefont
  {{Ekstr{\"o}m}}}, \bibinfo {author} {\bibfnamefont {S.}~\bibnamefont
  {{Sim{\'o}n-D{\'\i}az}}}, \bibinfo {author} {\bibfnamefont {G.}~\bibnamefont
  {{Holgado}}}, \bibinfo {author} {\bibfnamefont {N.}~\bibnamefont {{Castro}}},
  \bibinfo {author} {\bibfnamefont {C.}~\bibnamefont {{Georgy}}}, \bibinfo
  {author} {\bibfnamefont {P.}~\bibnamefont {{Eggenberger}}}, \bibinfo {author}
  {\bibfnamefont {G.}~\bibnamefont {{Buldgen}}}, \bibinfo {author}
  {\bibfnamefont {S.}~\bibnamefont {{Salmon}}}, \bibinfo {author}
  {\bibfnamefont {R.}~\bibnamefont {{Hirschi}}}, \bibinfo {author}
  {\bibfnamefont {J.}~\bibnamefont {{Groh}}}, \bibinfo {author} {\bibfnamefont
  {E.}~\bibnamefont {{Farrell}}},\ and\ \bibinfo {author} {\bibfnamefont
  {L.}~\bibnamefont {{Murphy}}},\ }\bibfield  {title} {\bibinfo {title}
  {{Convective core sizes in rotating massive stars. I. Constraints from solar
  metallicity OB field stars}},\ }\href
  {https://doi.org/10.1051/0004-6361/202039426} {\bibfield  {journal} {\bibinfo
   {journal} {\aap}\ }\textbf {\bibinfo {volume} {648}},\ \bibinfo {eid} {A126}
  (\bibinfo {year} {2021})},\ \Eprint {https://arxiv.org/abs/2103.03672}
  {arXiv:2103.03672 [astro-ph.SR]} \BibitemShut {NoStop}%
\bibitem [{\citenamefont {{Scott}}\ \emph {et~al.}(2021)\citenamefont
  {{Scott}}, \citenamefont {{Hirschi}}, \citenamefont {{Georgy}}, \citenamefont
  {{Arnett}}, \citenamefont {{Meakin}}, \citenamefont {{Kaiser}}, \citenamefont
  {{Ekstr{\"o}m}},\ and\ \citenamefont {{Yusof}}}]{2021MNRAS.503.4208S}%
  \BibitemOpen
  \bibfield  {author} {\bibinfo {author} {\bibfnamefont {L.~J.~A.}\
  \bibnamefont {{Scott}}}, \bibinfo {author} {\bibfnamefont {R.}~\bibnamefont
  {{Hirschi}}}, \bibinfo {author} {\bibfnamefont {C.}~\bibnamefont {{Georgy}}},
  \bibinfo {author} {\bibfnamefont {W.~D.}\ \bibnamefont {{Arnett}}}, \bibinfo
  {author} {\bibfnamefont {C.}~\bibnamefont {{Meakin}}}, \bibinfo {author}
  {\bibfnamefont {E.~A.}\ \bibnamefont {{Kaiser}}}, \bibinfo {author}
  {\bibfnamefont {S.}~\bibnamefont {{Ekstr{\"o}m}}},\ and\ \bibinfo {author}
  {\bibfnamefont {N.}~\bibnamefont {{Yusof}}},\ }\bibfield  {title} {\bibinfo
  {title} {{Convective core entrainment in 1D main-sequence stellar models}},\
  }\href {https://doi.org/10.1093/mnras/stab752} {\bibfield  {journal}
  {\bibinfo  {journal} {\mnras}\ }\textbf {\bibinfo {volume} {503}},\ \bibinfo
  {pages} {4208} (\bibinfo {year} {2021})},\ \Eprint
  {https://arxiv.org/abs/2103.06196} {arXiv:2103.06196 [astro-ph.SR]}
  \BibitemShut {NoStop}%
\bibitem [{\citenamefont {{Pedersen}}\ \emph {et~al.}(2021)\citenamefont
  {{Pedersen}}, \citenamefont {{Aerts}}, \citenamefont {{P{\'a}pics}},
  \citenamefont {{Michielsen}}, \citenamefont {{Gebruers}}, \citenamefont
  {{Rogers}}, \citenamefont {{Molenberghs}}, \citenamefont {{Burssens}},
  \citenamefont {{Garcia}},\ and\ \citenamefont
  {{Bowman}}}]{2021NatAs...5..715P}%
  \BibitemOpen
  \bibfield  {author} {\bibinfo {author} {\bibfnamefont {M.~G.}\ \bibnamefont
  {{Pedersen}}}, \bibinfo {author} {\bibfnamefont {C.}~\bibnamefont {{Aerts}}},
  \bibinfo {author} {\bibfnamefont {P.~I.}\ \bibnamefont {{P{\'a}pics}}},
  \bibinfo {author} {\bibfnamefont {M.}~\bibnamefont {{Michielsen}}}, \bibinfo
  {author} {\bibfnamefont {S.}~\bibnamefont {{Gebruers}}}, \bibinfo {author}
  {\bibfnamefont {T.~M.}\ \bibnamefont {{Rogers}}}, \bibinfo {author}
  {\bibfnamefont {G.}~\bibnamefont {{Molenberghs}}}, \bibinfo {author}
  {\bibfnamefont {S.}~\bibnamefont {{Burssens}}}, \bibinfo {author}
  {\bibfnamefont {S.}~\bibnamefont {{Garcia}}},\ and\ \bibinfo {author}
  {\bibfnamefont {D.~M.}\ \bibnamefont {{Bowman}}},\ }\bibfield  {title}
  {\bibinfo {title} {{Internal mixing of rotating stars inferred from dipole
  gravity modes}},\ }\href {https://doi.org/10.1038/s41550-021-01351-x}
  {\bibfield  {journal} {\bibinfo  {journal} {Nature Astronomy}\ }\textbf
  {\bibinfo {volume} {5}},\ \bibinfo {pages} {715} (\bibinfo {year} {2021})},\
  \Eprint {https://arxiv.org/abs/2105.04533} {arXiv:2105.04533 [astro-ph.SR]}
  \BibitemShut {NoStop}%
\bibitem [{\citenamefont {{Southworth}}\ and\ \citenamefont
  {{Bowman}}(2022)}]{2022MNRAS.513.3191S}%
  \BibitemOpen
  \bibfield  {author} {\bibinfo {author} {\bibfnamefont {J.}~\bibnamefont
  {{Southworth}}}\ and\ \bibinfo {author} {\bibfnamefont {D.~M.}\ \bibnamefont
  {{Bowman}}},\ }\bibfield  {title} {\bibinfo {title} {{High-mass pulsators in
  eclipsing binaries observed using TESS}},\ }\href
  {https://doi.org/10.1093/mnras/stac875} {\bibfield  {journal} {\bibinfo
  {journal} {\mnras}\ }\textbf {\bibinfo {volume} {513}},\ \bibinfo {pages}
  {3191} (\bibinfo {year} {2022})},\ \Eprint {https://arxiv.org/abs/2203.15365}
  {arXiv:2203.15365 [astro-ph.SR]} \BibitemShut {NoStop}%
\bibitem [{\citenamefont {{Rogers}}\ and\ \citenamefont
  {{McElwaine}}(2017)}]{2017ApJ...848L...1R}%
  \BibitemOpen
  \bibfield  {author} {\bibinfo {author} {\bibfnamefont {T.~M.}\ \bibnamefont
  {{Rogers}}}\ and\ \bibinfo {author} {\bibfnamefont {J.~N.}\ \bibnamefont
  {{McElwaine}}},\ }\bibfield  {title} {\bibinfo {title} {{On the Chemical
  Mixing Induced by Internal Gravity Waves}},\ }\href
  {https://doi.org/10.3847/2041-8213/aa8d13} {\bibfield  {journal} {\bibinfo
  {journal} {\apjl}\ }\textbf {\bibinfo {volume} {848}},\ \bibinfo {eid} {L1}
  (\bibinfo {year} {2017})},\ \Eprint {https://arxiv.org/abs/1709.04920}
  {arXiv:1709.04920 [astro-ph.SR]} \BibitemShut {NoStop}%
\bibitem [{\citenamefont {{Alongi}}\ \emph {et~al.}(1991)\citenamefont
  {{Alongi}}, \citenamefont {{Bertelli}}, \citenamefont {{Bressan}},\ and\
  \citenamefont {{Chiosi}}}]{1991A&A...244...95A}%
  \BibitemOpen
  \bibfield  {author} {\bibinfo {author} {\bibfnamefont {M.}~\bibnamefont
  {{Alongi}}}, \bibinfo {author} {\bibfnamefont {G.}~\bibnamefont
  {{Bertelli}}}, \bibinfo {author} {\bibfnamefont {A.}~\bibnamefont
  {{Bressan}}},\ and\ \bibinfo {author} {\bibfnamefont {C.}~\bibnamefont
  {{Chiosi}}},\ }\bibfield  {title} {\bibinfo {title} {{Effects of envelope
  overshoot on stellar models.}},\ }\href@noop {} {\bibfield  {journal}
  {\bibinfo  {journal} {\aap}\ }\textbf {\bibinfo {volume} {244}},\ \bibinfo
  {pages} {95} (\bibinfo {year} {1991})}\BibitemShut {NoStop}%
\bibitem [{\citenamefont {{Tang}}\ \emph {et~al.}(2014)\citenamefont {{Tang}},
  \citenamefont {{Bressan}}, \citenamefont {{Rosenfield}}, \citenamefont
  {{Slemer}}, \citenamefont {{Marigo}}, \citenamefont {{Girardi}},\ and\
  \citenamefont {{Bianchi}}}]{2014MNRAS.445.4287T}%
  \BibitemOpen
  \bibfield  {author} {\bibinfo {author} {\bibfnamefont {J.}~\bibnamefont
  {{Tang}}}, \bibinfo {author} {\bibfnamefont {A.}~\bibnamefont {{Bressan}}},
  \bibinfo {author} {\bibfnamefont {P.}~\bibnamefont {{Rosenfield}}}, \bibinfo
  {author} {\bibfnamefont {A.}~\bibnamefont {{Slemer}}}, \bibinfo {author}
  {\bibfnamefont {P.}~\bibnamefont {{Marigo}}}, \bibinfo {author}
  {\bibfnamefont {L.}~\bibnamefont {{Girardi}}},\ and\ \bibinfo {author}
  {\bibfnamefont {L.}~\bibnamefont {{Bianchi}}},\ }\bibfield  {title} {\bibinfo
  {title} {{New PARSEC evolutionary tracks of massive stars at low metallicity:
  testing canonical stellar evolution in nearby star-forming dwarf galaxies}},\
  }\href {https://doi.org/10.1093/mnras/stu2029} {\bibfield  {journal}
  {\bibinfo  {journal} {\mnras}\ }\textbf {\bibinfo {volume} {445}},\ \bibinfo
  {pages} {4287} (\bibinfo {year} {2014})},\ \Eprint
  {https://arxiv.org/abs/1410.1745} {arXiv:1410.1745 [astro-ph.SR]}
  \BibitemShut {NoStop}%
\bibitem [{\citenamefont {{Volpato}}\ \emph {et~al.}(2023)\citenamefont
  {{Volpato}}, \citenamefont {{Marigo}}, \citenamefont {{Costa}}, \citenamefont
  {{Bressan}}, \citenamefont {{Trabucchi}},\ and\ \citenamefont
  {{Girardi}}}]{2023ApJ...944...40V}%
  \BibitemOpen
  \bibfield  {author} {\bibinfo {author} {\bibfnamefont {G.}~\bibnamefont
  {{Volpato}}}, \bibinfo {author} {\bibfnamefont {P.}~\bibnamefont {{Marigo}}},
  \bibinfo {author} {\bibfnamefont {G.}~\bibnamefont {{Costa}}}, \bibinfo
  {author} {\bibfnamefont {A.}~\bibnamefont {{Bressan}}}, \bibinfo {author}
  {\bibfnamefont {M.}~\bibnamefont {{Trabucchi}}},\ and\ \bibinfo {author}
  {\bibfnamefont {L.}~\bibnamefont {{Girardi}}},\ }\bibfield  {title} {\bibinfo
  {title} {{A Study of Primordial Very Massive Star Evolution}},\ }\href
  {https://doi.org/10.3847/1538-4357/acac91} {\bibfield  {journal} {\bibinfo
  {journal} {\apj}\ }\textbf {\bibinfo {volume} {944}},\ \bibinfo {eid} {40}
  (\bibinfo {year} {2023})},\ \Eprint {https://arxiv.org/abs/2212.09629}
  {arXiv:2212.09629 [astro-ph.SR]} \BibitemShut {NoStop}%
\bibitem [{\citenamefont {{Aerts}}\ \emph {et~al.}(2019)\citenamefont
  {{Aerts}}, \citenamefont {{Mathis}},\ and\ \citenamefont
  {{Rogers}}}]{2019ARA&A..57...35A}%
  \BibitemOpen
  \bibfield  {author} {\bibinfo {author} {\bibfnamefont {C.}~\bibnamefont
  {{Aerts}}}, \bibinfo {author} {\bibfnamefont {S.}~\bibnamefont {{Mathis}}},\
  and\ \bibinfo {author} {\bibfnamefont {T.~M.}\ \bibnamefont {{Rogers}}},\
  }\bibfield  {title} {\bibinfo {title} {{Angular Momentum Transport in Stellar
  Interiors}},\ }\href {https://doi.org/10.1146/annurev-astro-091918-104359}
  {\bibfield  {journal} {\bibinfo  {journal} {\araa}\ }\textbf {\bibinfo
  {volume} {57}},\ \bibinfo {pages} {35} (\bibinfo {year} {2019})},\ \Eprint
  {https://arxiv.org/abs/1809.07779} {arXiv:1809.07779 [astro-ph.SR]}
  \BibitemShut {NoStop}%
\bibitem [{\citenamefont {{Ekstr{\"o}m}}(2021)}]{2021FrASS...8...53E}%
  \BibitemOpen
  \bibfield  {author} {\bibinfo {author} {\bibfnamefont {S.}~\bibnamefont
  {{Ekstr{\"o}m}}},\ }\bibfield  {title} {\bibinfo {title} {{Massive star
  modelling and nucleosynthesis}},\ }\href
  {https://doi.org/10.3389/fspas.2021.617765} {\bibfield  {journal} {\bibinfo
  {journal} {Frontiers in Astronomy and Space Sciences}\ }\textbf {\bibinfo
  {volume} {8}},\ \bibinfo {eid} {53} (\bibinfo {year} {2021})}\BibitemShut
  {NoStop}%
\bibitem [{\citenamefont {{Renzo}}\ \emph
  {et~al.}(2017{\natexlab{a}})\citenamefont {{Renzo}}, \citenamefont {{Ott}},
  \citenamefont {{Shore}},\ and\ \citenamefont {{de
  Mink}}}]{2017A&A...603A.118R}%
  \BibitemOpen
  \bibfield  {author} {\bibinfo {author} {\bibfnamefont {M.}~\bibnamefont
  {{Renzo}}}, \bibinfo {author} {\bibfnamefont {C.~D.}\ \bibnamefont {{Ott}}},
  \bibinfo {author} {\bibfnamefont {S.~N.}\ \bibnamefont {{Shore}}},\ and\
  \bibinfo {author} {\bibfnamefont {S.~E.}\ \bibnamefont {{de Mink}}},\
  }\bibfield  {title} {\bibinfo {title} {{Systematic survey of the effects of
  wind mass loss algorithms on the evolution of single massive stars}},\ }\href
  {https://doi.org/10.1051/0004-6361/201730698} {\bibfield  {journal} {\bibinfo
   {journal} {\aap}\ }\textbf {\bibinfo {volume} {603}},\ \bibinfo {eid} {A118}
  (\bibinfo {year} {2017}{\natexlab{a}})},\ \Eprint
  {https://arxiv.org/abs/1703.09705} {arXiv:1703.09705 [astro-ph.SR]}
  \BibitemShut {NoStop}%
\bibitem [{\citenamefont {{Fryer}}(1999)}]{1999ApJ...522..413F}%
  \BibitemOpen
  \bibfield  {author} {\bibinfo {author} {\bibfnamefont {C.~L.}\ \bibnamefont
  {{Fryer}}},\ }\bibfield  {title} {\bibinfo {title} {{Mass Limits For Black
  Hole Formation}},\ }\href {https://doi.org/10.1086/307647} {\bibfield
  {journal} {\bibinfo  {journal} {\apj}\ }\textbf {\bibinfo {volume} {522}},\
  \bibinfo {pages} {413} (\bibinfo {year} {1999})},\ \Eprint
  {https://arxiv.org/abs/astro-ph/9902315} {arXiv:astro-ph/9902315 [astro-ph]}
  \BibitemShut {NoStop}%
\bibitem [{\citenamefont {{Fryer}}\ \emph {et~al.}(2001)\citenamefont
  {{Fryer}}, \citenamefont {{Woosley}},\ and\ \citenamefont
  {{Heger}}}]{2001ApJ...550..372F}%
  \BibitemOpen
  \bibfield  {author} {\bibinfo {author} {\bibfnamefont {C.~L.}\ \bibnamefont
  {{Fryer}}}, \bibinfo {author} {\bibfnamefont {S.~E.}\ \bibnamefont
  {{Woosley}}},\ and\ \bibinfo {author} {\bibfnamefont {A.}~\bibnamefont
  {{Heger}}},\ }\bibfield  {title} {\bibinfo {title} {{Pair-Instability
  Supernovae, Gravity Waves, and Gamma-Ray Transients}},\ }\href
  {https://doi.org/10.1086/319719} {\bibfield  {journal} {\bibinfo  {journal}
  {\apj}\ }\textbf {\bibinfo {volume} {550}},\ \bibinfo {pages} {372} (\bibinfo
  {year} {2001})},\ \Eprint {https://arxiv.org/abs/astro-ph/0007176}
  {arXiv:astro-ph/0007176 [astro-ph]} \BibitemShut {NoStop}%
\bibitem [{\citenamefont {{Heger}}\ \emph
  {et~al.}(2003{\natexlab{a}})\citenamefont {{Heger}}, \citenamefont {{Fryer}},
  \citenamefont {{Woosley}}, \citenamefont {{Langer}},\ and\ \citenamefont
  {{Hartmann}}}]{2003ApJ...591..288H}%
  \BibitemOpen
  \bibfield  {author} {\bibinfo {author} {\bibfnamefont {A.}~\bibnamefont
  {{Heger}}}, \bibinfo {author} {\bibfnamefont {C.~L.}\ \bibnamefont
  {{Fryer}}}, \bibinfo {author} {\bibfnamefont {S.~E.}\ \bibnamefont
  {{Woosley}}}, \bibinfo {author} {\bibfnamefont {N.}~\bibnamefont
  {{Langer}}},\ and\ \bibinfo {author} {\bibfnamefont {D.~H.}\ \bibnamefont
  {{Hartmann}}},\ }\bibfield  {title} {\bibinfo {title} {{How Massive Single
  Stars End Their Life}},\ }\href {https://doi.org/10.1086/375341} {\bibfield
  {journal} {\bibinfo  {journal} {\apj}\ }\textbf {\bibinfo {volume} {591}},\
  \bibinfo {pages} {288} (\bibinfo {year} {2003}{\natexlab{a}})},\ \Eprint
  {https://arxiv.org/abs/astro-ph/0212469} {arXiv:astro-ph/0212469 [astro-ph]}
  \BibitemShut {NoStop}%
\bibitem [{\citenamefont {{O'Connor}}\ and\ \citenamefont
  {{Ott}}(2011)}]{2011ApJ...730...70O}%
  \BibitemOpen
  \bibfield  {author} {\bibinfo {author} {\bibfnamefont {E.}~\bibnamefont
  {{O'Connor}}}\ and\ \bibinfo {author} {\bibfnamefont {C.~D.}\ \bibnamefont
  {{Ott}}},\ }\bibfield  {title} {\bibinfo {title} {{Black Hole Formation in
  Failing Core-Collapse Supernovae}},\ }\href
  {https://doi.org/10.1088/0004-637X/730/2/70} {\bibfield  {journal} {\bibinfo
  {journal} {\apj}\ }\textbf {\bibinfo {volume} {730}},\ \bibinfo {eid} {70}
  (\bibinfo {year} {2011})},\ \Eprint {https://arxiv.org/abs/1010.5550}
  {arXiv:1010.5550 [astro-ph.HE]} \BibitemShut {NoStop}%
\bibitem [{\citenamefont {{O'Connor}}\ and\ \citenamefont
  {{Ott}}(2013)}]{2013ApJ...762..126O}%
  \BibitemOpen
  \bibfield  {author} {\bibinfo {author} {\bibfnamefont {E.}~\bibnamefont
  {{O'Connor}}}\ and\ \bibinfo {author} {\bibfnamefont {C.~D.}\ \bibnamefont
  {{Ott}}},\ }\bibfield  {title} {\bibinfo {title} {{The Progenitor Dependence
  of the Pre-explosion Neutrino Emission in Core-collapse Supernovae}},\ }\href
  {https://doi.org/10.1088/0004-637X/762/2/126} {\bibfield  {journal} {\bibinfo
   {journal} {\apj}\ }\textbf {\bibinfo {volume} {762}},\ \bibinfo {eid} {126}
  (\bibinfo {year} {2013})},\ \Eprint {https://arxiv.org/abs/1207.1100}
  {arXiv:1207.1100 [astro-ph.HE]} \BibitemShut {NoStop}%
\bibitem [{\citenamefont {{Belczynski}}\ \emph
  {et~al.}(2010{\natexlab{a}})\citenamefont {{Belczynski}}, \citenamefont
  {{Bulik}}, \citenamefont {{Fryer}}, \citenamefont {{Ruiter}}, \citenamefont
  {{Valsecchi}}, \citenamefont {{Vink}},\ and\ \citenamefont
  {{Hurley}}}]{2010ApJ...714.1217B}%
  \BibitemOpen
  \bibfield  {author} {\bibinfo {author} {\bibfnamefont {K.}~\bibnamefont
  {{Belczynski}}}, \bibinfo {author} {\bibfnamefont {T.}~\bibnamefont
  {{Bulik}}}, \bibinfo {author} {\bibfnamefont {C.~L.}\ \bibnamefont
  {{Fryer}}}, \bibinfo {author} {\bibfnamefont {A.}~\bibnamefont {{Ruiter}}},
  \bibinfo {author} {\bibfnamefont {F.}~\bibnamefont {{Valsecchi}}}, \bibinfo
  {author} {\bibfnamefont {J.~S.}\ \bibnamefont {{Vink}}},\ and\ \bibinfo
  {author} {\bibfnamefont {J.~R.}\ \bibnamefont {{Hurley}}},\ }\bibfield
  {title} {\bibinfo {title} {{On the Maximum Mass of Stellar Black Holes}},\
  }\href {https://doi.org/10.1088/0004-637X/714/2/1217} {\bibfield  {journal}
  {\bibinfo  {journal} {\apj}\ }\textbf {\bibinfo {volume} {714}},\ \bibinfo
  {pages} {1217} (\bibinfo {year} {2010}{\natexlab{a}})},\ \Eprint
  {https://arxiv.org/abs/0904.2784} {arXiv:0904.2784 [astro-ph.SR]}
  \BibitemShut {NoStop}%
\bibitem [{\citenamefont {{Spera}}\ \emph {et~al.}(2015)\citenamefont
  {{Spera}}, \citenamefont {{Mapelli}},\ and\ \citenamefont
  {{Bressan}}}]{2015MNRAS.451.4086S}%
  \BibitemOpen
  \bibfield  {author} {\bibinfo {author} {\bibfnamefont {M.}~\bibnamefont
  {{Spera}}}, \bibinfo {author} {\bibfnamefont {M.}~\bibnamefont {{Mapelli}}},\
  and\ \bibinfo {author} {\bibfnamefont {A.}~\bibnamefont {{Bressan}}},\
  }\bibfield  {title} {\bibinfo {title} {{The mass spectrum of compact remnants
  from the PARSEC stellar evolution tracks}},\ }\href
  {https://doi.org/10.1093/mnras/stv1161} {\bibfield  {journal} {\bibinfo
  {journal} {\mnras}\ }\textbf {\bibinfo {volume} {451}},\ \bibinfo {pages}
  {4086} (\bibinfo {year} {2015})},\ \Eprint {https://arxiv.org/abs/1505.05201}
  {arXiv:1505.05201 [astro-ph.SR]} \BibitemShut {NoStop}%
\bibitem [{\citenamefont {{Mapelli}}\ \emph {et~al.}(2013)\citenamefont
  {{Mapelli}}, \citenamefont {{Zampieri}}, \citenamefont {{Ripamonti}},\ and\
  \citenamefont {{Bressan}}}]{2013MNRAS.429.2298M}%
  \BibitemOpen
  \bibfield  {author} {\bibinfo {author} {\bibfnamefont {M.}~\bibnamefont
  {{Mapelli}}}, \bibinfo {author} {\bibfnamefont {L.}~\bibnamefont
  {{Zampieri}}}, \bibinfo {author} {\bibfnamefont {E.}~\bibnamefont
  {{Ripamonti}}},\ and\ \bibinfo {author} {\bibfnamefont {A.}~\bibnamefont
  {{Bressan}}},\ }\bibfield  {title} {\bibinfo {title} {{Dynamics of stellar
  black holes in young star clusters with different metallicities - I.
  Implications for X-ray binaries}},\ }\href
  {https://doi.org/10.1093/mnras/sts500} {\bibfield  {journal} {\bibinfo
  {journal} {\mnras}\ }\textbf {\bibinfo {volume} {429}},\ \bibinfo {pages}
  {2298} (\bibinfo {year} {2013})},\ \Eprint {https://arxiv.org/abs/1211.6441}
  {arXiv:1211.6441 [astro-ph.HE]} \BibitemShut {NoStop}%
\bibitem [{\citenamefont {{Giacobbo}}\ \emph {et~al.}(2018)\citenamefont
  {{Giacobbo}}, \citenamefont {{Mapelli}},\ and\ \citenamefont
  {{Spera}}}]{2018MNRAS.474.2959G}%
  \BibitemOpen
  \bibfield  {author} {\bibinfo {author} {\bibfnamefont {N.}~\bibnamefont
  {{Giacobbo}}}, \bibinfo {author} {\bibfnamefont {M.}~\bibnamefont
  {{Mapelli}}},\ and\ \bibinfo {author} {\bibfnamefont {M.}~\bibnamefont
  {{Spera}}},\ }\bibfield  {title} {\bibinfo {title} {{Merging black hole
  binaries: the effects of progenitor's metallicity, mass-loss rate and
  Eddington factor}},\ }\href {https://doi.org/10.1093/mnras/stx2933}
  {\bibfield  {journal} {\bibinfo  {journal} {\mnras}\ }\textbf {\bibinfo
  {volume} {474}},\ \bibinfo {pages} {2959} (\bibinfo {year} {2018})},\ \Eprint
  {https://arxiv.org/abs/1711.03556} {arXiv:1711.03556 [astro-ph.SR]}
  \BibitemShut {NoStop}%
\bibitem [{\citenamefont {{Spera}}\ \emph {et~al.}(2019)\citenamefont
  {{Spera}}, \citenamefont {{Mapelli}}, \citenamefont {{Giacobbo}},
  \citenamefont {{Trani}}, \citenamefont {{Bressan}},\ and\ \citenamefont
  {{Costa}}}]{2019MNRAS.485..889S}%
  \BibitemOpen
  \bibfield  {author} {\bibinfo {author} {\bibfnamefont {M.}~\bibnamefont
  {{Spera}}}, \bibinfo {author} {\bibfnamefont {M.}~\bibnamefont {{Mapelli}}},
  \bibinfo {author} {\bibfnamefont {N.}~\bibnamefont {{Giacobbo}}}, \bibinfo
  {author} {\bibfnamefont {A.~A.}\ \bibnamefont {{Trani}}}, \bibinfo {author}
  {\bibfnamefont {A.}~\bibnamefont {{Bressan}}},\ and\ \bibinfo {author}
  {\bibfnamefont {G.}~\bibnamefont {{Costa}}},\ }\bibfield  {title} {\bibinfo
  {title} {{Merging black hole binaries with the SEVN code}},\ }\href
  {https://doi.org/10.1093/mnras/stz359} {\bibfield  {journal} {\bibinfo
  {journal} {\mnras}\ }\textbf {\bibinfo {volume} {485}},\ \bibinfo {pages}
  {889} (\bibinfo {year} {2019})},\ \Eprint {https://arxiv.org/abs/1809.04605}
  {arXiv:1809.04605 [astro-ph.HE]} \BibitemShut {NoStop}%
\bibitem [{\citenamefont
  {{Mapelli}}(2021{\natexlab{a}})}]{2021hgwa.bookE..16M}%
  \BibitemOpen
  \bibfield  {author} {\bibinfo {author} {\bibfnamefont {M.}~\bibnamefont
  {{Mapelli}}},\ }\bibfield  {title} {\bibinfo {title} {{Formation Channels of
  Single and Binary Stellar-Mass Black Holes}},\ }in\ \href
  {https://doi.org/10.1007/978-981-15-4702-7_16-1} {\emph {\bibinfo {booktitle}
  {Handbook of Gravitational Wave Astronomy}}}\ (\bibinfo {year} {2021})\
  p.~\bibinfo {pages} {16}\BibitemShut {NoStop}%
\bibitem [{\citenamefont {{Spera}}\ \emph {et~al.}(2022)\citenamefont
  {{Spera}}, \citenamefont {{Trani}},\ and\ \citenamefont
  {{Mencagli}}}]{2022Galax..10...76S}%
  \BibitemOpen
  \bibfield  {author} {\bibinfo {author} {\bibfnamefont {M.}~\bibnamefont
  {{Spera}}}, \bibinfo {author} {\bibfnamefont {A.~A.}\ \bibnamefont
  {{Trani}}},\ and\ \bibinfo {author} {\bibfnamefont {M.}~\bibnamefont
  {{Mencagli}}},\ }\bibfield  {title} {\bibinfo {title} {{Compact Binary
  Coalescences: Astrophysical Processes and Lessons Learned}},\ }\href
  {https://doi.org/10.3390/galaxies10040076} {\bibfield  {journal} {\bibinfo
  {journal} {Galaxies}\ }\textbf {\bibinfo {volume} {10}},\ \bibinfo {pages}
  {76} (\bibinfo {year} {2022})},\ \Eprint {https://arxiv.org/abs/2206.15392}
  {arXiv:2206.15392 [astro-ph.HE]} \BibitemShut {NoStop}%
\bibitem [{\citenamefont {{Goswami}}\ \emph {et~al.}(2021)\citenamefont
  {{Goswami}}, \citenamefont {{Slemer}}, \citenamefont {{Marigo}},
  \citenamefont {{Bressan}}, \citenamefont {{Silva}}, \citenamefont {{Spera}},
  \citenamefont {{Boco}}, \citenamefont {{Grisoni}}, \citenamefont
  {{Pantoni}},\ and\ \citenamefont {{Lapi}}}]{2021A&A...650A.203G}%
  \BibitemOpen
  \bibfield  {author} {\bibinfo {author} {\bibfnamefont {S.}~\bibnamefont
  {{Goswami}}}, \bibinfo {author} {\bibfnamefont {A.}~\bibnamefont {{Slemer}}},
  \bibinfo {author} {\bibfnamefont {P.}~\bibnamefont {{Marigo}}}, \bibinfo
  {author} {\bibfnamefont {A.}~\bibnamefont {{Bressan}}}, \bibinfo {author}
  {\bibfnamefont {L.}~\bibnamefont {{Silva}}}, \bibinfo {author} {\bibfnamefont
  {M.}~\bibnamefont {{Spera}}}, \bibinfo {author} {\bibfnamefont
  {L.}~\bibnamefont {{Boco}}}, \bibinfo {author} {\bibfnamefont
  {V.}~\bibnamefont {{Grisoni}}}, \bibinfo {author} {\bibfnamefont
  {L.}~\bibnamefont {{Pantoni}}},\ and\ \bibinfo {author} {\bibfnamefont
  {A.}~\bibnamefont {{Lapi}}},\ }\bibfield  {title} {\bibinfo {title} {{The
  effects of the initial mass function on Galactic chemical enrichment}},\
  }\href {https://doi.org/10.1051/0004-6361/202039842} {\bibfield  {journal}
  {\bibinfo  {journal} {\aap}\ }\textbf {\bibinfo {volume} {650}},\ \bibinfo
  {eid} {A203} (\bibinfo {year} {2021})},\ \Eprint
  {https://arxiv.org/abs/2104.05680} {arXiv:2104.05680 [astro-ph.GA]}
  \BibitemShut {NoStop}%
\bibitem [{\citenamefont {{Puls}}\ \emph {et~al.}(2008)\citenamefont {{Puls}},
  \citenamefont {{Vink}},\ and\ \citenamefont
  {{Najarro}}}]{2008A&ARv..16..209P}%
  \BibitemOpen
  \bibfield  {author} {\bibinfo {author} {\bibfnamefont {J.}~\bibnamefont
  {{Puls}}}, \bibinfo {author} {\bibfnamefont {J.~S.}\ \bibnamefont {{Vink}}},\
  and\ \bibinfo {author} {\bibfnamefont {F.}~\bibnamefont {{Najarro}}},\
  }\bibfield  {title} {\bibinfo {title} {{Mass loss from hot massive stars}},\
  }\href {https://doi.org/10.1007/s00159-008-0015-8} {\bibfield  {journal}
  {\bibinfo  {journal} {\aapr}\ }\textbf {\bibinfo {volume} {16}},\ \bibinfo
  {pages} {209} (\bibinfo {year} {2008})},\ \Eprint
  {https://arxiv.org/abs/0811.0487} {arXiv:0811.0487 [astro-ph]} \BibitemShut
  {NoStop}%
\bibitem [{\citenamefont {{Vink}}(2017{\natexlab{a}})}]{2017RSPTA.37560269V}%
  \BibitemOpen
  \bibfield  {author} {\bibinfo {author} {\bibfnamefont {J.~S.}\ \bibnamefont
  {{Vink}}},\ }\bibfield  {title} {\bibinfo {title} {{Mass loss and stellar
  superwinds}},\ }\href {https://doi.org/10.1098/rsta.2016.0269} {\bibfield
  {journal} {\bibinfo  {journal} {Philosophical Transactions of the Royal
  Society of London Series A}\ }\textbf {\bibinfo {volume} {375}},\ \bibinfo
  {eid} {20160269} (\bibinfo {year} {2017}{\natexlab{a}})},\ \Eprint
  {https://arxiv.org/abs/1610.00578} {arXiv:1610.00578 [astro-ph.SR]}
  \BibitemShut {NoStop}%
\bibitem [{\citenamefont {{Vink}}(2021)}]{2021arXiv210908164V}%
  \BibitemOpen
  \bibfield  {author} {\bibinfo {author} {\bibfnamefont {J.~S.}\ \bibnamefont
  {{Vink}}},\ }\bibfield  {title} {\bibinfo {title} {{Theory and Diagnostics of
  Hot Star Mass Loss}},\ }\href@noop {} {\bibfield  {journal} {\bibinfo
  {journal} {arXiv e-prints}\ ,\ \bibinfo {eid} {arXiv:2109.08164}} (\bibinfo
  {year} {2021})},\ \Eprint {https://arxiv.org/abs/2109.08164}
  {arXiv:2109.08164 [astro-ph.SR]} \BibitemShut {NoStop}%
\bibitem [{\citenamefont {{Klotz}}(1918)}]{1918JRASC..12..357K}%
  \BibitemOpen
  \bibfield  {author} {\bibinfo {author} {\bibfnamefont {O.}~\bibnamefont
  {{Klotz}}},\ }\bibfield  {title} {\bibinfo {title} {{Light Pressure}},\
  }\href@noop {} {\bibfield  {journal} {\bibinfo  {journal} {\jrasc}\ }\textbf
  {\bibinfo {volume} {12}},\ \bibinfo {pages} {357} (\bibinfo {year}
  {1918})}\BibitemShut {NoStop}%
\bibitem [{\citenamefont {{Saha}}(1919)}]{1919ApJ....50..220S}%
  \BibitemOpen
  \bibfield  {author} {\bibinfo {author} {\bibfnamefont {M.~N.}\ \bibnamefont
  {{Saha}}},\ }\bibfield  {title} {\bibinfo {title} {{On Radiation-Pressure and
  the Quantum Theory}},\ }\href {https://doi.org/10.1086/142497} {\bibfield
  {journal} {\bibinfo  {journal} {\apj}\ }\textbf {\bibinfo {volume} {50}},\
  \bibinfo {pages} {220} (\bibinfo {year} {1919})}\BibitemShut {NoStop}%
\bibitem [{\citenamefont {{Morton}}(1967)}]{1967ApJ...147.1017M}%
  \BibitemOpen
  \bibfield  {author} {\bibinfo {author} {\bibfnamefont {D.~C.}\ \bibnamefont
  {{Morton}}},\ }\bibfield  {title} {\bibinfo {title} {{The Far-Ultraviolet
  Spectra of Six Stars in Orion}},\ }\href {https://doi.org/10.1086/149091}
  {\bibfield  {journal} {\bibinfo  {journal} {\apj}\ }\textbf {\bibinfo
  {volume} {147}},\ \bibinfo {pages} {1017} (\bibinfo {year}
  {1967})}\BibitemShut {NoStop}%
\bibitem [{\citenamefont {{Lucy}}\ and\ \citenamefont
  {{Solomon}}(1970)}]{1970ApJ...159..879L}%
  \BibitemOpen
  \bibfield  {author} {\bibinfo {author} {\bibfnamefont {L.~B.}\ \bibnamefont
  {{Lucy}}}\ and\ \bibinfo {author} {\bibfnamefont {P.~M.}\ \bibnamefont
  {{Solomon}}},\ }\bibfield  {title} {\bibinfo {title} {{Mass Loss by Hot
  Stars}},\ }\href {https://doi.org/10.1086/150365} {\bibfield  {journal}
  {\bibinfo  {journal} {\apj}\ }\textbf {\bibinfo {volume} {159}},\ \bibinfo
  {pages} {879} (\bibinfo {year} {1970})}\BibitemShut {NoStop}%
\bibitem [{\citenamefont {{Castor}}\ \emph {et~al.}(1975)\citenamefont
  {{Castor}}, \citenamefont {{Abbott}},\ and\ \citenamefont
  {{Klein}}}]{1975ApJ...195..157C}%
  \BibitemOpen
  \bibfield  {author} {\bibinfo {author} {\bibfnamefont {J.~I.}\ \bibnamefont
  {{Castor}}}, \bibinfo {author} {\bibfnamefont {D.~C.}\ \bibnamefont
  {{Abbott}}},\ and\ \bibinfo {author} {\bibfnamefont {R.~I.}\ \bibnamefont
  {{Klein}}},\ }\bibfield  {title} {\bibinfo {title} {{Radiation-driven winds
  in Of stars.}},\ }\href {https://doi.org/10.1086/153315} {\bibfield
  {journal} {\bibinfo  {journal} {\apj}\ }\textbf {\bibinfo {volume} {195}},\
  \bibinfo {pages} {157} (\bibinfo {year} {1975})}\BibitemShut {NoStop}%
\bibitem [{\citenamefont {{Abbott}}(1982)}]{1982ApJ...259..282A}%
  \BibitemOpen
  \bibfield  {author} {\bibinfo {author} {\bibfnamefont {D.~C.}\ \bibnamefont
  {{Abbott}}},\ }\bibfield  {title} {\bibinfo {title} {{The theory of
  radiatively driven stellar winds. II. The line acceleration.}},\ }\href
  {https://doi.org/10.1086/160166} {\bibfield  {journal} {\bibinfo  {journal}
  {\apj}\ }\textbf {\bibinfo {volume} {259}},\ \bibinfo {pages} {282} (\bibinfo
  {year} {1982})}\BibitemShut {NoStop}%
\bibitem [{\citenamefont {{Kudritzki}}\ \emph {et~al.}(1987)\citenamefont
  {{Kudritzki}}, \citenamefont {{Pauldrach}},\ and\ \citenamefont
  {{Puls}}}]{1987A&A...173..293K}%
  \BibitemOpen
  \bibfield  {author} {\bibinfo {author} {\bibfnamefont {R.~P.}\ \bibnamefont
  {{Kudritzki}}}, \bibinfo {author} {\bibfnamefont {A.}~\bibnamefont
  {{Pauldrach}}},\ and\ \bibinfo {author} {\bibfnamefont {J.}~\bibnamefont
  {{Puls}}},\ }\bibfield  {title} {\bibinfo {title} {{Radiation driven winds of
  hot luminous stars. II. Wind models for O-stars in the Magellanic clouds.}},\
  }\href@noop {} {\bibfield  {journal} {\bibinfo  {journal} {\aap}\ }\textbf
  {\bibinfo {volume} {173}},\ \bibinfo {pages} {293} (\bibinfo {year}
  {1987})}\BibitemShut {NoStop}%
\bibitem [{\citenamefont {{Abbott}}\ and\ \citenamefont
  {{Lucy}}(1985)}]{1985ApJ...288..679A}%
  \BibitemOpen
  \bibfield  {author} {\bibinfo {author} {\bibfnamefont {D.~C.}\ \bibnamefont
  {{Abbott}}}\ and\ \bibinfo {author} {\bibfnamefont {L.~B.}\ \bibnamefont
  {{Lucy}}},\ }\bibfield  {title} {\bibinfo {title} {{Multiline transfer and
  the dynamics of stellar winds.}},\ }\href {https://doi.org/10.1086/162834}
  {\bibfield  {journal} {\bibinfo  {journal} {\apj}\ }\textbf {\bibinfo
  {volume} {288}},\ \bibinfo {pages} {679} (\bibinfo {year}
  {1985})}\BibitemShut {NoStop}%
\bibitem [{\citenamefont {{Vink}}\ \emph {et~al.}(2000)\citenamefont {{Vink}},
  \citenamefont {{de Koter}},\ and\ \citenamefont
  {{Lamers}}}]{2000A&A...362..295V}%
  \BibitemOpen
  \bibfield  {author} {\bibinfo {author} {\bibfnamefont {J.~S.}\ \bibnamefont
  {{Vink}}}, \bibinfo {author} {\bibfnamefont {A.}~\bibnamefont {{de Koter}}},\
  and\ \bibinfo {author} {\bibfnamefont {H.~J.~G.~L.~M.}\ \bibnamefont
  {{Lamers}}},\ }\bibfield  {title} {\bibinfo {title} {{New theoretical
  mass-loss rates of O and B stars}},\ }\href@noop {} {\bibfield  {journal}
  {\bibinfo  {journal} {\aap}\ }\textbf {\bibinfo {volume} {362}},\ \bibinfo
  {pages} {295} (\bibinfo {year} {2000})},\ \Eprint
  {https://arxiv.org/abs/astro-ph/0008183} {arXiv:astro-ph/0008183 [astro-ph]}
  \BibitemShut {NoStop}%
\bibitem [{\citenamefont {{Vink}}\ \emph
  {et~al.}(2001{\natexlab{a}})\citenamefont {{Vink}}, \citenamefont {{de
  Koter}},\ and\ \citenamefont {{Lamers}}}]{2001A&A...369..574V}%
  \BibitemOpen
  \bibfield  {author} {\bibinfo {author} {\bibfnamefont {J.~S.}\ \bibnamefont
  {{Vink}}}, \bibinfo {author} {\bibfnamefont {A.}~\bibnamefont {{de Koter}}},\
  and\ \bibinfo {author} {\bibfnamefont {H.~J.~G.~L.~M.}\ \bibnamefont
  {{Lamers}}},\ }\bibfield  {title} {\bibinfo {title} {{Mass-loss predictions
  for O and B stars as a function of metallicity}},\ }\href
  {https://doi.org/10.1051/0004-6361:20010127} {\bibfield  {journal} {\bibinfo
  {journal} {\aap}\ }\textbf {\bibinfo {volume} {369}},\ \bibinfo {pages} {574}
  (\bibinfo {year} {2001}{\natexlab{a}})},\ \Eprint
  {https://arxiv.org/abs/astro-ph/0101509} {arXiv:astro-ph/0101509 [astro-ph]}
  \BibitemShut {NoStop}%
\bibitem [{\citenamefont {{Anders}}\ and\ \citenamefont
  {{Grevesse}}(1989)}]{1989GeCoA..53..197A}%
  \BibitemOpen
  \bibfield  {author} {\bibinfo {author} {\bibfnamefont {E.}~\bibnamefont
  {{Anders}}}\ and\ \bibinfo {author} {\bibfnamefont {N.}~\bibnamefont
  {{Grevesse}}},\ }\bibfield  {title} {\bibinfo {title} {{Abundances of the
  elements: Meteoritic and solar}},\ }\href
  {https://doi.org/10.1016/0016-7037(89)90286-X} {\bibfield  {journal}
  {\bibinfo  {journal} {\gca}\ }\textbf {\bibinfo {volume} {53}},\ \bibinfo
  {pages} {197} (\bibinfo {year} {1989})}\BibitemShut {NoStop}%
\bibitem [{\citenamefont {{de Jager}}\ \emph {et~al.}(1988)\citenamefont {{de
  Jager}}, \citenamefont {{Nieuwenhuijzen}},\ and\ \citenamefont {{van der
  Hucht}}}]{1988A&AS...72..259D}%
  \BibitemOpen
  \bibfield  {author} {\bibinfo {author} {\bibfnamefont {C.}~\bibnamefont {{de
  Jager}}}, \bibinfo {author} {\bibfnamefont {H.}~\bibnamefont
  {{Nieuwenhuijzen}}},\ and\ \bibinfo {author} {\bibfnamefont {K.~A.}\
  \bibnamefont {{van der Hucht}}},\ }\bibfield  {title} {\bibinfo {title}
  {{Mass loss rates in the Hertzsprung-Russell diagram.}},\ }\href@noop {}
  {\bibfield  {journal} {\bibinfo  {journal} {\aaps}\ }\textbf {\bibinfo
  {volume} {72}},\ \bibinfo {pages} {259} (\bibinfo {year} {1988})}\BibitemShut
  {NoStop}%
\bibitem [{\citenamefont {{Beasor}}\ \emph {et~al.}(2020)\citenamefont
  {{Beasor}}, \citenamefont {{Davies}}, \citenamefont {{Smith}}, \citenamefont
  {{van Loon}}, \citenamefont {{Gehrz}},\ and\ \citenamefont
  {{Figer}}}]{2020MNRAS.492.5994B}%
  \BibitemOpen
  \bibfield  {author} {\bibinfo {author} {\bibfnamefont {E.~R.}\ \bibnamefont
  {{Beasor}}}, \bibinfo {author} {\bibfnamefont {B.}~\bibnamefont {{Davies}}},
  \bibinfo {author} {\bibfnamefont {N.}~\bibnamefont {{Smith}}}, \bibinfo
  {author} {\bibfnamefont {J.~T.}\ \bibnamefont {{van Loon}}}, \bibinfo
  {author} {\bibfnamefont {R.~D.}\ \bibnamefont {{Gehrz}}},\ and\ \bibinfo
  {author} {\bibfnamefont {D.~F.}\ \bibnamefont {{Figer}}},\ }\bibfield
  {title} {\bibinfo {title} {{A new mass-loss rate prescription for red
  supergiants}},\ }\href {https://doi.org/10.1093/mnras/staa255} {\bibfield
  {journal} {\bibinfo  {journal} {\mnras}\ }\textbf {\bibinfo {volume} {492}},\
  \bibinfo {pages} {5994} (\bibinfo {year} {2020})},\ \Eprint
  {https://arxiv.org/abs/2001.07222} {arXiv:2001.07222 [astro-ph.SR]}
  \BibitemShut {NoStop}%
\bibitem [{\citenamefont {{van Loon}}\ \emph {et~al.}(2005)\citenamefont {{van
  Loon}}, \citenamefont {{Cioni}}, \citenamefont {{Zijlstra}},\ and\
  \citenamefont {{Loup}}}]{2005A&A...438..273V}%
  \BibitemOpen
  \bibfield  {author} {\bibinfo {author} {\bibfnamefont {J.~T.}\ \bibnamefont
  {{van Loon}}}, \bibinfo {author} {\bibfnamefont {M.~R.~L.}\ \bibnamefont
  {{Cioni}}}, \bibinfo {author} {\bibfnamefont {A.~A.}\ \bibnamefont
  {{Zijlstra}}},\ and\ \bibinfo {author} {\bibfnamefont {C.}~\bibnamefont
  {{Loup}}},\ }\bibfield  {title} {\bibinfo {title} {{An empirical formula for
  the mass-loss rates of dust-enshrouded red supergiants and oxygen-rich
  Asymptotic Giant Branch stars}},\ }\href
  {https://doi.org/10.1051/0004-6361:20042555} {\bibfield  {journal} {\bibinfo
  {journal} {\aap}\ }\textbf {\bibinfo {volume} {438}},\ \bibinfo {pages} {273}
  (\bibinfo {year} {2005})},\ \Eprint {https://arxiv.org/abs/astro-ph/0504379}
  {arXiv:astro-ph/0504379 [astro-ph]} \BibitemShut {NoStop}%
\bibitem [{\citenamefont {{Kee}}\ \emph {et~al.}(2021)\citenamefont {{Kee}},
  \citenamefont {{Sundqvist}}, \citenamefont {{Decin}}, \citenamefont {{de
  Koter}},\ and\ \citenamefont {{Sana}}}]{2021A&A...646A.180K}%
  \BibitemOpen
  \bibfield  {author} {\bibinfo {author} {\bibfnamefont {N.~D.}\ \bibnamefont
  {{Kee}}}, \bibinfo {author} {\bibfnamefont {J.~O.}\ \bibnamefont
  {{Sundqvist}}}, \bibinfo {author} {\bibfnamefont {L.}~\bibnamefont
  {{Decin}}}, \bibinfo {author} {\bibfnamefont {A.}~\bibnamefont {{de
  Koter}}},\ and\ \bibinfo {author} {\bibfnamefont {H.}~\bibnamefont
  {{Sana}}},\ }\bibfield  {title} {\bibinfo {title} {{Analytic,
  dust-independent mass-loss rates for red supergiant winds initiated by
  turbulent pressure}},\ }\href {https://doi.org/10.1051/0004-6361/202039224}
  {\bibfield  {journal} {\bibinfo  {journal} {\aap}\ }\textbf {\bibinfo
  {volume} {646}},\ \bibinfo {eid} {A180} (\bibinfo {year} {2021})},\ \Eprint
  {https://arxiv.org/abs/2101.03070} {arXiv:2101.03070 [astro-ph.SR]}
  \BibitemShut {NoStop}%
\bibitem [{\citenamefont {{Gr{\"a}fener}}\ and\ \citenamefont
  {{Hamann}}(2008)}]{2008A&A...482..945G}%
  \BibitemOpen
  \bibfield  {author} {\bibinfo {author} {\bibfnamefont {G.}~\bibnamefont
  {{Gr{\"a}fener}}}\ and\ \bibinfo {author} {\bibfnamefont {W.~R.}\
  \bibnamefont {{Hamann}}},\ }\bibfield  {title} {\bibinfo {title} {{Mass loss
  from late-type WN stars and its Z-dependence. Very massive stars approaching
  the Eddington limit}},\ }\href {https://doi.org/10.1051/0004-6361:20066176}
  {\bibfield  {journal} {\bibinfo  {journal} {\aap}\ }\textbf {\bibinfo
  {volume} {482}},\ \bibinfo {pages} {945} (\bibinfo {year} {2008})},\ \Eprint
  {https://arxiv.org/abs/0803.0866} {arXiv:0803.0866 [astro-ph]} \BibitemShut
  {NoStop}%
\bibitem [{\citenamefont {{Vink}}\ \emph {et~al.}(2011)\citenamefont {{Vink}},
  \citenamefont {{Muijres}}, \citenamefont {{Anthonisse}}, \citenamefont {{de
  Koter}}, \citenamefont {{Gr{\"a}fener}},\ and\ \citenamefont
  {{Langer}}}]{2011A&A...531A.132V}%
  \BibitemOpen
  \bibfield  {author} {\bibinfo {author} {\bibfnamefont {J.~S.}\ \bibnamefont
  {{Vink}}}, \bibinfo {author} {\bibfnamefont {L.~E.}\ \bibnamefont
  {{Muijres}}}, \bibinfo {author} {\bibfnamefont {B.}~\bibnamefont
  {{Anthonisse}}}, \bibinfo {author} {\bibfnamefont {A.}~\bibnamefont {{de
  Koter}}}, \bibinfo {author} {\bibfnamefont {G.}~\bibnamefont
  {{Gr{\"a}fener}}},\ and\ \bibinfo {author} {\bibfnamefont {N.}~\bibnamefont
  {{Langer}}},\ }\bibfield  {title} {\bibinfo {title} {{Wind modelling of very
  massive stars up to 300 solar masses}},\ }\href
  {https://doi.org/10.1051/0004-6361/201116614} {\bibfield  {journal} {\bibinfo
   {journal} {\aap}\ }\textbf {\bibinfo {volume} {531}},\ \bibinfo {eid} {A132}
  (\bibinfo {year} {2011})},\ \Eprint {https://arxiv.org/abs/1105.0556}
  {arXiv:1105.0556 [astro-ph.SR]} \BibitemShut {NoStop}%
\bibitem [{\citenamefont {{Gr{\"a}fener}}\ \emph {et~al.}(2012)\citenamefont
  {{Gr{\"a}fener}}, \citenamefont {{Owocki}},\ and\ \citenamefont
  {{Vink}}}]{2012A&A...538A..40G}%
  \BibitemOpen
  \bibfield  {author} {\bibinfo {author} {\bibfnamefont {G.}~\bibnamefont
  {{Gr{\"a}fener}}}, \bibinfo {author} {\bibfnamefont {S.~P.}\ \bibnamefont
  {{Owocki}}},\ and\ \bibinfo {author} {\bibfnamefont {J.~S.}\ \bibnamefont
  {{Vink}}},\ }\bibfield  {title} {\bibinfo {title} {{Stellar envelope
  inflation near the Eddington limit. Implications for the radii of Wolf-Rayet
  stars and luminous blue variables}},\ }\href
  {https://doi.org/10.1051/0004-6361/201117497} {\bibfield  {journal} {\bibinfo
   {journal} {\aap}\ }\textbf {\bibinfo {volume} {538}},\ \bibinfo {eid} {A40}
  (\bibinfo {year} {2012})},\ \Eprint {https://arxiv.org/abs/1112.1910}
  {arXiv:1112.1910 [astro-ph.SR]} \BibitemShut {NoStop}%
\bibitem [{\citenamefont {{Agrawal}}\ \emph
  {et~al.}(2022{\natexlab{a}})\citenamefont {{Agrawal}}, \citenamefont
  {{Stevenson}}, \citenamefont {{Sz{\'e}csi}},\ and\ \citenamefont
  {{Hurley}}}]{2022A&A...668A..90A}%
  \BibitemOpen
  \bibfield  {author} {\bibinfo {author} {\bibfnamefont {P.}~\bibnamefont
  {{Agrawal}}}, \bibinfo {author} {\bibfnamefont {S.}~\bibnamefont
  {{Stevenson}}}, \bibinfo {author} {\bibfnamefont {D.}~\bibnamefont
  {{Sz{\'e}csi}}},\ and\ \bibinfo {author} {\bibfnamefont {J.}~\bibnamefont
  {{Hurley}}},\ }\bibfield  {title} {\bibinfo {title} {{A systematic study of
  super-Eddington layers in the envelopes of massive stars}},\ }\href
  {https://doi.org/10.1051/0004-6361/202244044} {\bibfield  {journal} {\bibinfo
   {journal} {\aap}\ }\textbf {\bibinfo {volume} {668}},\ \bibinfo {eid} {A90}
  (\bibinfo {year} {2022}{\natexlab{a}})},\ \Eprint
  {https://arxiv.org/abs/2112.02801} {arXiv:2112.02801 [astro-ph.SR]}
  \BibitemShut {NoStop}%
\bibitem [{\citenamefont {{Grassitelli}}\ \emph {et~al.}(2021)\citenamefont
  {{Grassitelli}}, \citenamefont {{Langer}}, \citenamefont {{Mackey}},
  \citenamefont {{Gr{\"a}fener}}, \citenamefont {{Grin}}, \citenamefont
  {{Sander}},\ and\ \citenamefont {{Vink}}}]{2021A&A...647A..99G}%
  \BibitemOpen
  \bibfield  {author} {\bibinfo {author} {\bibfnamefont {L.}~\bibnamefont
  {{Grassitelli}}}, \bibinfo {author} {\bibfnamefont {N.}~\bibnamefont
  {{Langer}}}, \bibinfo {author} {\bibfnamefont {J.}~\bibnamefont {{Mackey}}},
  \bibinfo {author} {\bibfnamefont {G.}~\bibnamefont {{Gr{\"a}fener}}},
  \bibinfo {author} {\bibfnamefont {N.~J.}\ \bibnamefont {{Grin}}}, \bibinfo
  {author} {\bibfnamefont {A.~A.~C.}\ \bibnamefont {{Sander}}},\ and\ \bibinfo
  {author} {\bibfnamefont {J.~S.}\ \bibnamefont {{Vink}}},\ }\bibfield  {title}
  {\bibinfo {title} {{Wind-envelope interaction as the origin of the slow
  cyclic brightness variations of luminous blue variables}},\ }\href
  {https://doi.org/10.1051/0004-6361/202038298} {\bibfield  {journal} {\bibinfo
   {journal} {\aap}\ }\textbf {\bibinfo {volume} {647}},\ \bibinfo {eid} {A99}
  (\bibinfo {year} {2021})},\ \Eprint {https://arxiv.org/abs/2012.00023}
  {arXiv:2012.00023 [astro-ph.SR]} \BibitemShut {NoStop}%
\bibitem [{\citenamefont {{Nugis}}\ and\ \citenamefont
  {{Lamers}}(2000)}]{2000A&A...360..227N}%
  \BibitemOpen
  \bibfield  {author} {\bibinfo {author} {\bibfnamefont {T.}~\bibnamefont
  {{Nugis}}}\ and\ \bibinfo {author} {\bibfnamefont {H.~J.~G.~L.~M.}\
  \bibnamefont {{Lamers}}},\ }\bibfield  {title} {\bibinfo {title} {{Mass-loss
  rates of Wolf-Rayet stars as a function of stellar parameters}},\ }\href@noop
  {} {\bibfield  {journal} {\bibinfo  {journal} {\aap}\ }\textbf {\bibinfo
  {volume} {360}},\ \bibinfo {pages} {227} (\bibinfo {year}
  {2000})}\BibitemShut {NoStop}%
\bibitem [{\citenamefont {{Sander}}\ and\ \citenamefont
  {{Vink}}(2020)}]{2020MNRAS.499..873S}%
  \BibitemOpen
  \bibfield  {author} {\bibinfo {author} {\bibfnamefont {A.~A.~C.}\
  \bibnamefont {{Sander}}}\ and\ \bibinfo {author} {\bibfnamefont {J.~S.}\
  \bibnamefont {{Vink}}},\ }\bibfield  {title} {\bibinfo {title} {{On the
  nature of massive helium star winds and Wolf-Rayet-type mass-loss}},\ }\href
  {https://doi.org/10.1093/mnras/staa2712} {\bibfield  {journal} {\bibinfo
  {journal} {\mnras}\ }\textbf {\bibinfo {volume} {499}},\ \bibinfo {pages}
  {873} (\bibinfo {year} {2020})},\ \Eprint {https://arxiv.org/abs/2009.01849}
  {arXiv:2009.01849 [astro-ph.SR]} \BibitemShut {NoStop}%
\bibitem [{\citenamefont {{Baraffe}}\ \emph {et~al.}(2001)\citenamefont
  {{Baraffe}}, \citenamefont {{Heger}},\ and\ \citenamefont
  {{Woosley}}}]{2001ApJ...550..890B}%
  \BibitemOpen
  \bibfield  {author} {\bibinfo {author} {\bibfnamefont {I.}~\bibnamefont
  {{Baraffe}}}, \bibinfo {author} {\bibfnamefont {A.}~\bibnamefont {{Heger}}},\
  and\ \bibinfo {author} {\bibfnamefont {S.~E.}\ \bibnamefont {{Woosley}}},\
  }\bibfield  {title} {\bibinfo {title} {{On the Stability of Very Massive
  Primordial Stars}},\ }\href {https://doi.org/10.1086/319808} {\bibfield
  {journal} {\bibinfo  {journal} {\apj}\ }\textbf {\bibinfo {volume} {550}},\
  \bibinfo {pages} {890} (\bibinfo {year} {2001})},\ \Eprint
  {https://arxiv.org/abs/astro-ph/0009410} {arXiv:astro-ph/0009410 [astro-ph]}
  \BibitemShut {NoStop}%
\bibitem [{\citenamefont {{Nakauchi}}\ \emph {et~al.}(2020)\citenamefont
  {{Nakauchi}}, \citenamefont {{Inayoshi}},\ and\ \citenamefont
  {{Omukai}}}]{2020ApJ...902...81N}%
  \BibitemOpen
  \bibfield  {author} {\bibinfo {author} {\bibfnamefont {D.}~\bibnamefont
  {{Nakauchi}}}, \bibinfo {author} {\bibfnamefont {K.}~\bibnamefont
  {{Inayoshi}}},\ and\ \bibinfo {author} {\bibfnamefont {K.}~\bibnamefont
  {{Omukai}}},\ }\bibfield  {title} {\bibinfo {title} {{Pulsation-driven Mass
  Loss from Massive Stars behind Stellar Mergers in Metal-poor Dense
  Clusters}},\ }\href {https://doi.org/10.3847/1538-4357/abb463} {\bibfield
  {journal} {\bibinfo  {journal} {\apj}\ }\textbf {\bibinfo {volume} {902}},\
  \bibinfo {eid} {81} (\bibinfo {year} {2020})},\ \Eprint
  {https://arxiv.org/abs/2008.13647} {arXiv:2008.13647 [astro-ph.SR]}
  \BibitemShut {NoStop}%
\bibitem [{\citenamefont {{Gormaz-Matamala}}\ \emph {et~al.}(2022)\citenamefont
  {{Gormaz-Matamala}}, \citenamefont {{Cur{\'e}}}, \citenamefont {{Meynet}},
  \citenamefont {{Cuadra}}, \citenamefont {{Groh}},\ and\ \citenamefont
  {{Murphy}}}]{2022A&A...665A.133G}%
  \BibitemOpen
  \bibfield  {author} {\bibinfo {author} {\bibfnamefont {A.~C.}\ \bibnamefont
  {{Gormaz-Matamala}}}, \bibinfo {author} {\bibfnamefont {M.}~\bibnamefont
  {{Cur{\'e}}}}, \bibinfo {author} {\bibfnamefont {G.}~\bibnamefont
  {{Meynet}}}, \bibinfo {author} {\bibfnamefont {J.}~\bibnamefont {{Cuadra}}},
  \bibinfo {author} {\bibfnamefont {J.~H.}\ \bibnamefont {{Groh}}},\ and\
  \bibinfo {author} {\bibfnamefont {L.~J.}\ \bibnamefont {{Murphy}}},\
  }\bibfield  {title} {\bibinfo {title} {{Evolution of massive stars with new
  hydrodynamic wind models}},\ }\href
  {https://doi.org/10.1051/0004-6361/202243959} {\bibfield  {journal} {\bibinfo
   {journal} {\aap}\ }\textbf {\bibinfo {volume} {665}},\ \bibinfo {eid} {A133}
  (\bibinfo {year} {2022})},\ \Eprint {https://arxiv.org/abs/2207.04786}
  {arXiv:2207.04786 [astro-ph.SR]} \BibitemShut {NoStop}%
\bibitem [{\citenamefont {{Marcolino}}\ \emph {et~al.}(2022)\citenamefont
  {{Marcolino}}, \citenamefont {{Bouret}}, \citenamefont {{Rocha-Pinto}},
  \citenamefont {{Bernini-Peron}},\ and\ \citenamefont
  {{Vink}}}]{2022MNRAS.511.5104M}%
  \BibitemOpen
  \bibfield  {author} {\bibinfo {author} {\bibfnamefont {W.~L.~F.}\
  \bibnamefont {{Marcolino}}}, \bibinfo {author} {\bibfnamefont {J.~C.}\
  \bibnamefont {{Bouret}}}, \bibinfo {author} {\bibfnamefont {H.~J.}\
  \bibnamefont {{Rocha-Pinto}}}, \bibinfo {author} {\bibfnamefont
  {M.}~\bibnamefont {{Bernini-Peron}}},\ and\ \bibinfo {author} {\bibfnamefont
  {J.~S.}\ \bibnamefont {{Vink}}},\ }\bibfield  {title} {\bibinfo {title}
  {{Wind properties of Milky Way and SMC massive stars: empirical Z dependence
  from CMFGEN models}},\ }\href {https://doi.org/10.1093/mnras/stac452}
  {\bibfield  {journal} {\bibinfo  {journal} {\mnras}\ }\textbf {\bibinfo
  {volume} {511}},\ \bibinfo {pages} {5104} (\bibinfo {year} {2022})},\ \Eprint
  {https://arxiv.org/abs/2202.07811} {arXiv:2202.07811 [astro-ph.GA]}
  \BibitemShut {NoStop}%
\bibitem [{\citenamefont {{Chen}}\ \emph {et~al.}(2015)\citenamefont {{Chen}},
  \citenamefont {{Bressan}}, \citenamefont {{Girardi}}, \citenamefont
  {{Marigo}}, \citenamefont {{Kong}},\ and\ \citenamefont
  {{Lanza}}}]{2015MNRAS.452.1068C}%
  \BibitemOpen
  \bibfield  {author} {\bibinfo {author} {\bibfnamefont {Y.}~\bibnamefont
  {{Chen}}}, \bibinfo {author} {\bibfnamefont {A.}~\bibnamefont {{Bressan}}},
  \bibinfo {author} {\bibfnamefont {L.}~\bibnamefont {{Girardi}}}, \bibinfo
  {author} {\bibfnamefont {P.}~\bibnamefont {{Marigo}}}, \bibinfo {author}
  {\bibfnamefont {X.}~\bibnamefont {{Kong}}},\ and\ \bibinfo {author}
  {\bibfnamefont {A.}~\bibnamefont {{Lanza}}},\ }\bibfield  {title} {\bibinfo
  {title} {{PARSEC evolutionary tracks of massive stars up to 350
  M$_{{\ensuremath{\odot}}}$ at metallicities 0.0001 {\ensuremath{\leq}} Z
  {\ensuremath{\leq}} 0.04}},\ }\href {https://doi.org/10.1093/mnras/stv1281}
  {\bibfield  {journal} {\bibinfo  {journal} {\mnras}\ }\textbf {\bibinfo
  {volume} {452}},\ \bibinfo {pages} {1068} (\bibinfo {year} {2015})},\ \Eprint
  {https://arxiv.org/abs/1506.01681} {arXiv:1506.01681 [astro-ph.SR]}
  \BibitemShut {NoStop}%
\bibitem [{\citenamefont {{Sander}}\ \emph {et~al.}(2019)\citenamefont
  {{Sander}}, \citenamefont {{Hamann}}, \citenamefont {{Todt}}, \citenamefont
  {{Hainich}}, \citenamefont {{Shenar}}, \citenamefont {{Ramachandran}},\ and\
  \citenamefont {{Oskinova}}}]{2019A&A...621A..92S}%
  \BibitemOpen
  \bibfield  {author} {\bibinfo {author} {\bibfnamefont {A.~A.~C.}\
  \bibnamefont {{Sander}}}, \bibinfo {author} {\bibfnamefont {W.~R.}\
  \bibnamefont {{Hamann}}}, \bibinfo {author} {\bibfnamefont {H.}~\bibnamefont
  {{Todt}}}, \bibinfo {author} {\bibfnamefont {R.}~\bibnamefont {{Hainich}}},
  \bibinfo {author} {\bibfnamefont {T.}~\bibnamefont {{Shenar}}}, \bibinfo
  {author} {\bibfnamefont {V.}~\bibnamefont {{Ramachandran}}},\ and\ \bibinfo
  {author} {\bibfnamefont {L.~M.}\ \bibnamefont {{Oskinova}}},\ }\bibfield
  {title} {\bibinfo {title} {{The Galactic WC and WO stars. The impact of
  revised distances from Gaia DR2 and their role as massive black hole
  progenitors}},\ }\href {https://doi.org/10.1051/0004-6361/201833712}
  {\bibfield  {journal} {\bibinfo  {journal} {\aap}\ }\textbf {\bibinfo
  {volume} {621}},\ \bibinfo {eid} {A92} (\bibinfo {year} {2019})},\ \Eprint
  {https://arxiv.org/abs/1807.04293} {arXiv:1807.04293 [astro-ph.GA]}
  \BibitemShut {NoStop}%
\bibitem [{\citenamefont {{Higgins}}\ \emph {et~al.}(2021)\citenamefont
  {{Higgins}}, \citenamefont {{Sander}}, \citenamefont {{Vink}},\ and\
  \citenamefont {{Hirschi}}}]{2021MNRAS.505.4874H}%
  \BibitemOpen
  \bibfield  {author} {\bibinfo {author} {\bibfnamefont {E.~R.}\ \bibnamefont
  {{Higgins}}}, \bibinfo {author} {\bibfnamefont {A.~A.~C.}\ \bibnamefont
  {{Sander}}}, \bibinfo {author} {\bibfnamefont {J.~S.}\ \bibnamefont
  {{Vink}}},\ and\ \bibinfo {author} {\bibfnamefont {R.}~\bibnamefont
  {{Hirschi}}},\ }\bibfield  {title} {\bibinfo {title} {{Evolution of
  Wolf-Rayet stars as black hole progenitors}},\ }\href
  {https://doi.org/10.1093/mnras/stab1548} {\bibfield  {journal} {\bibinfo
  {journal} {\mnras}\ }\textbf {\bibinfo {volume} {505}},\ \bibinfo {pages}
  {4874} (\bibinfo {year} {2021})},\ \Eprint {https://arxiv.org/abs/2105.12139}
  {arXiv:2105.12139 [astro-ph.SR]} \BibitemShut {NoStop}%
\bibitem [{\citenamefont {{Hurley}}\ \emph {et~al.}(2000)\citenamefont
  {{Hurley}}, \citenamefont {{Pols}},\ and\ \citenamefont
  {{Tout}}}]{2000MNRAS.315..543H}%
  \BibitemOpen
  \bibfield  {author} {\bibinfo {author} {\bibfnamefont {J.~R.}\ \bibnamefont
  {{Hurley}}}, \bibinfo {author} {\bibfnamefont {O.~R.}\ \bibnamefont
  {{Pols}}},\ and\ \bibinfo {author} {\bibfnamefont {C.~A.}\ \bibnamefont
  {{Tout}}},\ }\bibfield  {title} {\bibinfo {title} {{Comprehensive analytic
  formulae for stellar evolution as a function of mass and metallicity}},\
  }\href {https://doi.org/10.1046/j.1365-8711.2000.03426.x} {\bibfield
  {journal} {\bibinfo  {journal} {\mnras}\ }\textbf {\bibinfo {volume} {315}},\
  \bibinfo {pages} {543} (\bibinfo {year} {2000})},\ \Eprint
  {https://arxiv.org/abs/astro-ph/0001295} {arXiv:astro-ph/0001295 [astro-ph]}
  \BibitemShut {NoStop}%
\bibitem [{\citenamefont {{Nieuwenhuijzen}}\ and\ \citenamefont {{de
  Jager}}(1990)}]{1990A&A...231..134N}%
  \BibitemOpen
  \bibfield  {author} {\bibinfo {author} {\bibfnamefont {H.}~\bibnamefont
  {{Nieuwenhuijzen}}}\ and\ \bibinfo {author} {\bibfnamefont {C.}~\bibnamefont
  {{de Jager}}},\ }\bibfield  {title} {\bibinfo {title} {{Parametrization of
  stellar rates of mass loss as functions of the fundamental stellar parameters
  M, L, and R.}},\ }\href@noop {} {\bibfield  {journal} {\bibinfo  {journal}
  {\aap}\ }\textbf {\bibinfo {volume} {231}},\ \bibinfo {pages} {134} (\bibinfo
  {year} {1990})}\BibitemShut {NoStop}%
\bibitem [{\citenamefont {{Royer}}(2009)}]{2009LNP...765..207R}%
  \BibitemOpen
  \bibfield  {author} {\bibinfo {author} {\bibfnamefont {F.}~\bibnamefont
  {{Royer}}},\ }\bibfield  {title} {\bibinfo {title} {{On the Rotation of
  A-Type Stars}},\ }in\ \href {https://doi.org/10.1007/978-3-540-87831-5_9}
  {\emph {\bibinfo {booktitle} {The Rotation of Sun and Stars}}},\ Vol.\
  \bibinfo {volume} {765}\ (\bibinfo {year} {2009})\ pp.\ \bibinfo {pages}
  {207--230}\BibitemShut {NoStop}%
\bibitem [{\citenamefont {{McQuillan}}\ \emph {et~al.}(2014)\citenamefont
  {{McQuillan}}, \citenamefont {{Mazeh}},\ and\ \citenamefont
  {{Aigrain}}}]{2014ApJS..211...24M}%
  \BibitemOpen
  \bibfield  {author} {\bibinfo {author} {\bibfnamefont {A.}~\bibnamefont
  {{McQuillan}}}, \bibinfo {author} {\bibfnamefont {T.}~\bibnamefont
  {{Mazeh}}},\ and\ \bibinfo {author} {\bibfnamefont {S.}~\bibnamefont
  {{Aigrain}}},\ }\bibfield  {title} {\bibinfo {title} {{Rotation Periods of
  34,030 Kepler Main-sequence Stars: The Full Autocorrelation Sample}},\ }\href
  {https://doi.org/10.1088/0067-0049/211/2/24} {\bibfield  {journal} {\bibinfo
  {journal} {\apjs}\ }\textbf {\bibinfo {volume} {211}},\ \bibinfo {eid} {24}
  (\bibinfo {year} {2014})},\ \Eprint {https://arxiv.org/abs/1402.5694}
  {arXiv:1402.5694 [astro-ph.SR]} \BibitemShut {NoStop}%
\bibitem [{\citenamefont {{Hunter}}\ \emph {et~al.}(2007)\citenamefont
  {{Hunter}}, \citenamefont {{Dufton}}, \citenamefont {{Smartt}}, \citenamefont
  {{Ryans}}, \citenamefont {{Evans}}, \citenamefont {{Lennon}}, \citenamefont
  {{Trundle}}, \citenamefont {{Hubeny}},\ and\ \citenamefont
  {{Lanz}}}]{2007A&A...466..277H}%
  \BibitemOpen
  \bibfield  {author} {\bibinfo {author} {\bibfnamefont {I.}~\bibnamefont
  {{Hunter}}}, \bibinfo {author} {\bibfnamefont {P.~L.}\ \bibnamefont
  {{Dufton}}}, \bibinfo {author} {\bibfnamefont {S.~J.}\ \bibnamefont
  {{Smartt}}}, \bibinfo {author} {\bibfnamefont {R.~S.~I.}\ \bibnamefont
  {{Ryans}}}, \bibinfo {author} {\bibfnamefont {C.~J.}\ \bibnamefont
  {{Evans}}}, \bibinfo {author} {\bibfnamefont {D.~J.}\ \bibnamefont
  {{Lennon}}}, \bibinfo {author} {\bibfnamefont {C.}~\bibnamefont {{Trundle}}},
  \bibinfo {author} {\bibfnamefont {I.}~\bibnamefont {{Hubeny}}},\ and\
  \bibinfo {author} {\bibfnamefont {T.}~\bibnamefont {{Lanz}}},\ }\bibfield
  {title} {\bibinfo {title} {{The VLT-FLAMES survey of massive stars: surface
  chemical compositions of B-type stars in the Magellanic Clouds}},\ }\href
  {https://doi.org/10.1051/0004-6361:20066148} {\bibfield  {journal} {\bibinfo
  {journal} {\aap}\ }\textbf {\bibinfo {volume} {466}},\ \bibinfo {pages} {277}
  (\bibinfo {year} {2007})},\ \Eprint {https://arxiv.org/abs/astro-ph/0609710}
  {arXiv:astro-ph/0609710 [astro-ph]} \BibitemShut {NoStop}%
\bibitem [{\citenamefont {{Dufton}}\ \emph {et~al.}(2013)\citenamefont
  {{Dufton}}, \citenamefont {{Langer}}, \citenamefont {{Dunstall}},
  \citenamefont {{Evans}}, \citenamefont {{Brott}}, \citenamefont {{de Mink}},
  \citenamefont {{Howarth}}, \citenamefont {{Kennedy}}, \citenamefont
  {{McEvoy}}, \citenamefont {{Potter}}, \citenamefont {{Ram{\'\i}rez-Agudelo}},
  \citenamefont {{Sana}}, \citenamefont {{Sim{\'o}n-D{\'\i}az}}, \citenamefont
  {{Taylor}},\ and\ \citenamefont {{Vink}}}]{2013A&A...550A.109D}%
  \BibitemOpen
  \bibfield  {author} {\bibinfo {author} {\bibfnamefont {P.~L.}\ \bibnamefont
  {{Dufton}}}, \bibinfo {author} {\bibfnamefont {N.}~\bibnamefont {{Langer}}},
  \bibinfo {author} {\bibfnamefont {P.~R.}\ \bibnamefont {{Dunstall}}},
  \bibinfo {author} {\bibfnamefont {C.~J.}\ \bibnamefont {{Evans}}}, \bibinfo
  {author} {\bibfnamefont {I.}~\bibnamefont {{Brott}}}, \bibinfo {author}
  {\bibfnamefont {S.~E.}\ \bibnamefont {{de Mink}}}, \bibinfo {author}
  {\bibfnamefont {I.~D.}\ \bibnamefont {{Howarth}}}, \bibinfo {author}
  {\bibfnamefont {M.}~\bibnamefont {{Kennedy}}}, \bibinfo {author}
  {\bibfnamefont {C.}~\bibnamefont {{McEvoy}}}, \bibinfo {author}
  {\bibfnamefont {A.~T.}\ \bibnamefont {{Potter}}}, \bibinfo {author}
  {\bibfnamefont {O.~H.}\ \bibnamefont {{Ram{\'\i}rez-Agudelo}}}, \bibinfo
  {author} {\bibfnamefont {H.}~\bibnamefont {{Sana}}}, \bibinfo {author}
  {\bibfnamefont {S.}~\bibnamefont {{Sim{\'o}n-D{\'\i}az}}}, \bibinfo {author}
  {\bibfnamefont {W.}~\bibnamefont {{Taylor}}},\ and\ \bibinfo {author}
  {\bibfnamefont {J.~S.}\ \bibnamefont {{Vink}}},\ }\bibfield  {title}
  {\bibinfo {title} {{The VLT-FLAMES Tarantula Survey. X. Evidence for a
  bimodal distribution of rotational velocities for the single early B-type
  stars}},\ }\href
  {https://doi.org/10.1051/0004-6361/20122027310.48550/arXiv.1212.2424}
  {\bibfield  {journal} {\bibinfo  {journal} {\aap}\ }\textbf {\bibinfo
  {volume} {550}},\ \bibinfo {eid} {A109} (\bibinfo {year} {2013})},\ \Eprint
  {https://arxiv.org/abs/1212.2424} {arXiv:1212.2424 [astro-ph.SR]}
  \BibitemShut {NoStop}%
\bibitem [{\citenamefont {{Ram{\'\i}rez-Agudelo}}\ \emph
  {et~al.}(2013)\citenamefont {{Ram{\'\i}rez-Agudelo}}, \citenamefont
  {{Sim{\'o}n-D{\'\i}az}}, \citenamefont {{Sana}}, \citenamefont {{de Koter}},
  \citenamefont {{Sab{\'\i}n-Sanjul{\'\i}an}}, \citenamefont {{de Mink}},
  \citenamefont {{Dufton}}, \citenamefont {{Gr{\"a}fener}}, \citenamefont
  {{Evans}}, \citenamefont {{Herrero}}, \citenamefont {{Langer}}, \citenamefont
  {{Lennon}}, \citenamefont {{Ma{\'\i}z Apell{\'a}niz}}, \citenamefont
  {{Markova}}, \citenamefont {{Najarro}}, \citenamefont {{Puls}}, \citenamefont
  {{Taylor}},\ and\ \citenamefont {{Vink}}}]{2013A&A...560A..29R}%
  \BibitemOpen
  \bibfield  {author} {\bibinfo {author} {\bibfnamefont {O.~H.}\ \bibnamefont
  {{Ram{\'\i}rez-Agudelo}}}, \bibinfo {author} {\bibfnamefont {S.}~\bibnamefont
  {{Sim{\'o}n-D{\'\i}az}}}, \bibinfo {author} {\bibfnamefont {H.}~\bibnamefont
  {{Sana}}}, \bibinfo {author} {\bibfnamefont {A.}~\bibnamefont {{de Koter}}},
  \bibinfo {author} {\bibfnamefont {C.}~\bibnamefont
  {{Sab{\'\i}n-Sanjul{\'\i}an}}}, \bibinfo {author} {\bibfnamefont {S.~E.}\
  \bibnamefont {{de Mink}}}, \bibinfo {author} {\bibfnamefont {P.~L.}\
  \bibnamefont {{Dufton}}}, \bibinfo {author} {\bibfnamefont {G.}~\bibnamefont
  {{Gr{\"a}fener}}}, \bibinfo {author} {\bibfnamefont {C.~J.}\ \bibnamefont
  {{Evans}}}, \bibinfo {author} {\bibfnamefont {A.}~\bibnamefont {{Herrero}}},
  \bibinfo {author} {\bibfnamefont {N.}~\bibnamefont {{Langer}}}, \bibinfo
  {author} {\bibfnamefont {D.~J.}\ \bibnamefont {{Lennon}}}, \bibinfo {author}
  {\bibfnamefont {J.}~\bibnamefont {{Ma{\'\i}z Apell{\'a}niz}}}, \bibinfo
  {author} {\bibfnamefont {N.}~\bibnamefont {{Markova}}}, \bibinfo {author}
  {\bibfnamefont {F.}~\bibnamefont {{Najarro}}}, \bibinfo {author}
  {\bibfnamefont {J.}~\bibnamefont {{Puls}}}, \bibinfo {author} {\bibfnamefont
  {W.~D.}\ \bibnamefont {{Taylor}}},\ and\ \bibinfo {author} {\bibfnamefont
  {J.~S.}\ \bibnamefont {{Vink}}},\ }\bibfield  {title} {\bibinfo {title} {{The
  VLT-FLAMES Tarantula Survey. XII. Rotational velocities of the single O-type
  stars}},\ }\href {https://doi.org/10.1051/0004-6361/201321986} {\bibfield
  {journal} {\bibinfo  {journal} {\aap}\ }\textbf {\bibinfo {volume} {560}},\
  \bibinfo {eid} {A29} (\bibinfo {year} {2013})},\ \Eprint
  {https://arxiv.org/abs/1309.2929} {arXiv:1309.2929 [astro-ph.SR]}
  \BibitemShut {NoStop}%
\bibitem [{\citenamefont {{Bodensteiner}}\ \emph
  {et~al.}(2020{\natexlab{a}})\citenamefont {{Bodensteiner}}, \citenamefont
  {{Sana}}, \citenamefont {{Mahy}}, \citenamefont {{Patrick}}, \citenamefont
  {{de Koter}}, \citenamefont {{de Mink}}, \citenamefont {{Evans}},
  \citenamefont {{G{\"o}tberg}}, \citenamefont {{Langer}}, \citenamefont
  {{Lennon}}, \citenamefont {{Schneider}},\ and\ \citenamefont
  {{Tramper}}}]{2020A&A...634A..51B}%
  \BibitemOpen
  \bibfield  {author} {\bibinfo {author} {\bibfnamefont {J.}~\bibnamefont
  {{Bodensteiner}}}, \bibinfo {author} {\bibfnamefont {H.}~\bibnamefont
  {{Sana}}}, \bibinfo {author} {\bibfnamefont {L.}~\bibnamefont {{Mahy}}},
  \bibinfo {author} {\bibfnamefont {L.~R.}\ \bibnamefont {{Patrick}}}, \bibinfo
  {author} {\bibfnamefont {A.}~\bibnamefont {{de Koter}}}, \bibinfo {author}
  {\bibfnamefont {S.~E.}\ \bibnamefont {{de Mink}}}, \bibinfo {author}
  {\bibfnamefont {C.~J.}\ \bibnamefont {{Evans}}}, \bibinfo {author}
  {\bibfnamefont {Y.}~\bibnamefont {{G{\"o}tberg}}}, \bibinfo {author}
  {\bibfnamefont {N.}~\bibnamefont {{Langer}}}, \bibinfo {author}
  {\bibfnamefont {D.~J.}\ \bibnamefont {{Lennon}}}, \bibinfo {author}
  {\bibfnamefont {F.~R.~N.}\ \bibnamefont {{Schneider}}},\ and\ \bibinfo
  {author} {\bibfnamefont {F.}~\bibnamefont {{Tramper}}},\ }\bibfield  {title}
  {\bibinfo {title} {{The young massive SMC cluster NGC 330 seen by MUSE. I.
  Observations and stellar content}},\ }\href
  {https://doi.org/10.1051/0004-6361/201936743} {\bibfield  {journal} {\bibinfo
   {journal} {\aap}\ }\textbf {\bibinfo {volume} {634}},\ \bibinfo {eid} {A51}
  (\bibinfo {year} {2020}{\natexlab{a}})},\ \Eprint
  {https://arxiv.org/abs/1911.03477} {arXiv:1911.03477 [astro-ph.SR]}
  \BibitemShut {NoStop}%
\bibitem [{\citenamefont {{Schootemeijer}}\ \emph {et~al.}(2022)\citenamefont
  {{Schootemeijer}}, \citenamefont {{Lennon}}, \citenamefont {{Garcia}},
  \citenamefont {{Langer}}, \citenamefont {{Hastings}},\ and\ \citenamefont
  {{Sch{\"u}rmann}}}]{2022A&A...667A.100S}%
  \BibitemOpen
  \bibfield  {author} {\bibinfo {author} {\bibfnamefont {A.}~\bibnamefont
  {{Schootemeijer}}}, \bibinfo {author} {\bibfnamefont {D.~J.}\ \bibnamefont
  {{Lennon}}}, \bibinfo {author} {\bibfnamefont {M.}~\bibnamefont {{Garcia}}},
  \bibinfo {author} {\bibfnamefont {N.}~\bibnamefont {{Langer}}}, \bibinfo
  {author} {\bibfnamefont {B.}~\bibnamefont {{Hastings}}},\ and\ \bibinfo
  {author} {\bibfnamefont {C.}~\bibnamefont {{Sch{\"u}rmann}}},\ }\bibfield
  {title} {\bibinfo {title} {{A census of OBe stars in nearby metal-poor dwarf
  galaxies reveals a high fraction of extreme rotators}},\ }\href
  {https://doi.org/10.1051/0004-6361/202244730} {\bibfield  {journal} {\bibinfo
   {journal} {\aap}\ }\textbf {\bibinfo {volume} {667}},\ \bibinfo {eid} {A100}
  (\bibinfo {year} {2022})},\ \Eprint {https://arxiv.org/abs/2209.04943}
  {arXiv:2209.04943 [astro-ph.GA]} \BibitemShut {NoStop}%
\bibitem [{\citenamefont {{Maeder}}\ and\ \citenamefont
  {{Meynet}}(2000{\natexlab{a}})}]{2000A&A...361..159M}%
  \BibitemOpen
  \bibfield  {author} {\bibinfo {author} {\bibfnamefont {A.}~\bibnamefont
  {{Maeder}}}\ and\ \bibinfo {author} {\bibfnamefont {G.}~\bibnamefont
  {{Meynet}}},\ }\bibfield  {title} {\bibinfo {title} {{Stellar evolution with
  rotation. VI. The Eddington and Omega -limits, the rotational mass loss for
  OB and LBV stars}},\ }\href@noop {} {\bibfield  {journal} {\bibinfo
  {journal} {\aap}\ }\textbf {\bibinfo {volume} {361}},\ \bibinfo {pages} {159}
  (\bibinfo {year} {2000}{\natexlab{a}})},\ \Eprint
  {https://arxiv.org/abs/astro-ph/0006405} {arXiv:astro-ph/0006405 [astro-ph]}
  \BibitemShut {NoStop}%
\bibitem [{\citenamefont {{Hirschi}}\ \emph {et~al.}(2004)\citenamefont
  {{Hirschi}}, \citenamefont {{Meynet}},\ and\ \citenamefont
  {{Maeder}}}]{2004A&A...425..649H}%
  \BibitemOpen
  \bibfield  {author} {\bibinfo {author} {\bibfnamefont {R.}~\bibnamefont
  {{Hirschi}}}, \bibinfo {author} {\bibfnamefont {G.}~\bibnamefont
  {{Meynet}}},\ and\ \bibinfo {author} {\bibfnamefont {A.}~\bibnamefont
  {{Maeder}}},\ }\bibfield  {title} {\bibinfo {title} {{Stellar evolution with
  rotation. XII. Pre-supernova models}},\ }\href
  {https://doi.org/10.1051/0004-6361:20041095} {\bibfield  {journal} {\bibinfo
  {journal} {\aap}\ }\textbf {\bibinfo {volume} {425}},\ \bibinfo {pages} {649}
  (\bibinfo {year} {2004})},\ \Eprint {https://arxiv.org/abs/astro-ph/0406552}
  {arXiv:astro-ph/0406552 [astro-ph]} \BibitemShut {NoStop}%
\bibitem [{\citenamefont {{Woosley}}\ and\ \citenamefont
  {{Heger}}(2006)}]{2006ApJ...637..914W}%
  \BibitemOpen
  \bibfield  {author} {\bibinfo {author} {\bibfnamefont {S.~E.}\ \bibnamefont
  {{Woosley}}}\ and\ \bibinfo {author} {\bibfnamefont {A.}~\bibnamefont
  {{Heger}}},\ }\bibfield  {title} {\bibinfo {title} {{The Progenitor Stars of
  Gamma-Ray Bursts}},\ }\href {https://doi.org/10.1086/498500} {\bibfield
  {journal} {\bibinfo  {journal} {\apj}\ }\textbf {\bibinfo {volume} {637}},\
  \bibinfo {pages} {914} (\bibinfo {year} {2006})},\ \Eprint
  {https://arxiv.org/abs/astro-ph/0508175} {arXiv:astro-ph/0508175 [astro-ph]}
  \BibitemShut {NoStop}%
\bibitem [{\citenamefont {{Georgy}}\ \emph {et~al.}(2012)\citenamefont
  {{Georgy}}, \citenamefont {{Ekstr{\"o}m}}, \citenamefont {{Meynet}},
  \citenamefont {{Massey}}, \citenamefont {{Levesque}}, \citenamefont
  {{Hirschi}}, \citenamefont {{Eggenberger}},\ and\ \citenamefont
  {{Maeder}}}]{2012A&A...542A..29G}%
  \BibitemOpen
  \bibfield  {author} {\bibinfo {author} {\bibfnamefont {C.}~\bibnamefont
  {{Georgy}}}, \bibinfo {author} {\bibfnamefont {S.}~\bibnamefont
  {{Ekstr{\"o}m}}}, \bibinfo {author} {\bibfnamefont {G.}~\bibnamefont
  {{Meynet}}}, \bibinfo {author} {\bibfnamefont {P.}~\bibnamefont {{Massey}}},
  \bibinfo {author} {\bibfnamefont {E.~M.}\ \bibnamefont {{Levesque}}},
  \bibinfo {author} {\bibfnamefont {R.}~\bibnamefont {{Hirschi}}}, \bibinfo
  {author} {\bibfnamefont {P.}~\bibnamefont {{Eggenberger}}},\ and\ \bibinfo
  {author} {\bibfnamefont {A.}~\bibnamefont {{Maeder}}},\ }\bibfield  {title}
  {\bibinfo {title} {{Grids of stellar models with rotation. II. WR populations
  and supernovae/GRB progenitors at Z = 0.014}},\ }\href
  {https://doi.org/10.1051/0004-6361/201118340} {\bibfield  {journal} {\bibinfo
   {journal} {\aap}\ }\textbf {\bibinfo {volume} {542}},\ \bibinfo {eid} {A29}
  (\bibinfo {year} {2012})},\ \Eprint {https://arxiv.org/abs/1203.5243}
  {arXiv:1203.5243 [astro-ph.SR]} \BibitemShut {NoStop}%
\bibitem [{\citenamefont {{Chieffi}}\ and\ \citenamefont
  {{Limongi}}(2013)}]{2013ApJ...764...21C}%
  \BibitemOpen
  \bibfield  {author} {\bibinfo {author} {\bibfnamefont {A.}~\bibnamefont
  {{Chieffi}}}\ and\ \bibinfo {author} {\bibfnamefont {M.}~\bibnamefont
  {{Limongi}}},\ }\bibfield  {title} {\bibinfo {title} {{Pre-supernova
  Evolution of Rotating Solar Metallicity Stars in the Mass Range 13-120 M
  $_{{\ensuremath{\odot}}}$ and their Explosive Yields}},\ }\href
  {https://doi.org/10.1088/0004-637X/764/1/21} {\bibfield  {journal} {\bibinfo
  {journal} {\apj}\ }\textbf {\bibinfo {volume} {764}},\ \bibinfo {eid} {21}
  (\bibinfo {year} {2013})}\BibitemShut {NoStop}%
\bibitem [{\citenamefont {{K{\"o}hler}}\ \emph {et~al.}(2015)\citenamefont
  {{K{\"o}hler}}, \citenamefont {{Langer}}, \citenamefont {{de Koter}},
  \citenamefont {{de Mink}}, \citenamefont {{Crowther}}, \citenamefont
  {{Evans}}, \citenamefont {{Gr{\"a}fener}}, \citenamefont {{Sana}},
  \citenamefont {{Sanyal}}, \citenamefont {{Schneider}},\ and\ \citenamefont
  {{Vink}}}]{2015A&A...573A..71K}%
  \BibitemOpen
  \bibfield  {author} {\bibinfo {author} {\bibfnamefont {K.}~\bibnamefont
  {{K{\"o}hler}}}, \bibinfo {author} {\bibfnamefont {N.}~\bibnamefont
  {{Langer}}}, \bibinfo {author} {\bibfnamefont {A.}~\bibnamefont {{de
  Koter}}}, \bibinfo {author} {\bibfnamefont {S.~E.}\ \bibnamefont {{de
  Mink}}}, \bibinfo {author} {\bibfnamefont {P.~A.}\ \bibnamefont
  {{Crowther}}}, \bibinfo {author} {\bibfnamefont {C.~J.}\ \bibnamefont
  {{Evans}}}, \bibinfo {author} {\bibfnamefont {G.}~\bibnamefont
  {{Gr{\"a}fener}}}, \bibinfo {author} {\bibfnamefont {H.}~\bibnamefont
  {{Sana}}}, \bibinfo {author} {\bibfnamefont {D.}~\bibnamefont {{Sanyal}}},
  \bibinfo {author} {\bibfnamefont {F.~R.~N.}\ \bibnamefont {{Schneider}}},\
  and\ \bibinfo {author} {\bibfnamefont {J.~S.}\ \bibnamefont {{Vink}}},\
  }\bibfield  {title} {\bibinfo {title} {{The evolution of rotating very
  massive stars with LMC composition}},\ }\href
  {https://doi.org/10.1051/0004-6361/201424356} {\bibfield  {journal} {\bibinfo
   {journal} {\aap}\ }\textbf {\bibinfo {volume} {573}},\ \bibinfo {eid} {A71}
  (\bibinfo {year} {2015})},\ \Eprint {https://arxiv.org/abs/1501.03794}
  {arXiv:1501.03794 [astro-ph.SR]} \BibitemShut {NoStop}%
\bibitem [{\citenamefont {{Goswami}}\ \emph {et~al.}(2022)\citenamefont
  {{Goswami}}, \citenamefont {{Silva}}, \citenamefont {{Bressan}},
  \citenamefont {{Grisoni}}, \citenamefont {{Costa}}, \citenamefont {{Marigo}},
  \citenamefont {{Granato}}, \citenamefont {{Lapi}},\ and\ \citenamefont
  {{Spera}}}]{2022A&A...663A...1G}%
  \BibitemOpen
  \bibfield  {author} {\bibinfo {author} {\bibfnamefont {S.}~\bibnamefont
  {{Goswami}}}, \bibinfo {author} {\bibfnamefont {L.}~\bibnamefont {{Silva}}},
  \bibinfo {author} {\bibfnamefont {A.}~\bibnamefont {{Bressan}}}, \bibinfo
  {author} {\bibfnamefont {V.}~\bibnamefont {{Grisoni}}}, \bibinfo {author}
  {\bibfnamefont {G.}~\bibnamefont {{Costa}}}, \bibinfo {author} {\bibfnamefont
  {P.}~\bibnamefont {{Marigo}}}, \bibinfo {author} {\bibfnamefont {G.~L.}\
  \bibnamefont {{Granato}}}, \bibinfo {author} {\bibfnamefont {A.}~\bibnamefont
  {{Lapi}}},\ and\ \bibinfo {author} {\bibfnamefont {M.}~\bibnamefont
  {{Spera}}},\ }\bibfield  {title} {\bibinfo {title} {{Impact of very massive
  stars on the chemical evolution of extremely metal-poor galaxies}},\ }\href
  {https://doi.org/10.1051/0004-6361/202142031} {\bibfield  {journal} {\bibinfo
   {journal} {\aap}\ }\textbf {\bibinfo {volume} {663}},\ \bibinfo {eid} {A1}
  (\bibinfo {year} {2022})},\ \Eprint {https://arxiv.org/abs/2205.03402}
  {arXiv:2205.03402 [astro-ph.GA]} \BibitemShut {NoStop}%
\bibitem [{\citenamefont {{Kuhlen}}\ \emph {et~al.}(2003)\citenamefont
  {{Kuhlen}}, \citenamefont {{Woosley}},\ and\ \citenamefont
  {{Glatzmaier}}}]{2003ASPC..293..147K}%
  \BibitemOpen
  \bibfield  {author} {\bibinfo {author} {\bibfnamefont {M.}~\bibnamefont
  {{Kuhlen}}}, \bibinfo {author} {\bibfnamefont {W.~E.}\ \bibnamefont
  {{Woosley}}},\ and\ \bibinfo {author} {\bibfnamefont {G.~A.}\ \bibnamefont
  {{Glatzmaier}}},\ }\bibfield  {title} {\bibinfo {title} {{3D Anelastic
  Simulations of Convection in Massive Stars}},\ }in\ \href@noop {} {\emph
  {\bibinfo {booktitle} {3D Stellar Evolution}}},\ \bibinfo {series}
  {Astronomical Society of the Pacific Conference Series}, Vol.\ \bibinfo
  {volume} {293},\ \bibinfo {editor} {edited by\ \bibinfo {editor}
  {\bibfnamefont {S.}~\bibnamefont {{Turcotte}}}, \bibinfo {editor}
  {\bibfnamefont {S.~C.}\ \bibnamefont {{Keller}}},\ and\ \bibinfo {editor}
  {\bibfnamefont {R.~M.}\ \bibnamefont {{Cavallo}}}}\ (\bibinfo {year} {2003})\
  p.\ \bibinfo {pages} {147},\ \Eprint {https://arxiv.org/abs/astro-ph/0210557}
  {arXiv:astro-ph/0210557 [astro-ph]} \BibitemShut {NoStop}%
\bibitem [{\citenamefont {{McNeill}}\ and\ \citenamefont
  {{M{\"u}ller}}(2022)}]{2022MNRAS.509..818M}%
  \BibitemOpen
  \bibfield  {author} {\bibinfo {author} {\bibfnamefont {L.~O.}\ \bibnamefont
  {{McNeill}}}\ and\ \bibinfo {author} {\bibfnamefont {B.}~\bibnamefont
  {{M{\"u}ller}}},\ }\bibfield  {title} {\bibinfo {title} {{Differential
  rotation in a 3D simulation of oxygen shell burning}},\ }\href
  {https://doi.org/10.1093/mnras/stab3076} {\bibfield  {journal} {\bibinfo
  {journal} {\mnras}\ }\textbf {\bibinfo {volume} {509}},\ \bibinfo {pages}
  {818} (\bibinfo {year} {2022})},\ \Eprint {https://arxiv.org/abs/2107.00173}
  {arXiv:2107.00173 [astro-ph.SR]} \BibitemShut {NoStop}%
\bibitem [{\citenamefont {{Rieutord}}\ \emph {et~al.}(2016)\citenamefont
  {{Rieutord}}, \citenamefont {{Espinosa Lara}},\ and\ \citenamefont
  {{Putigny}}}]{2016JCoPh.318..277R}%
  \BibitemOpen
  \bibfield  {author} {\bibinfo {author} {\bibfnamefont {M.}~\bibnamefont
  {{Rieutord}}}, \bibinfo {author} {\bibfnamefont {F.}~\bibnamefont {{Espinosa
  Lara}}},\ and\ \bibinfo {author} {\bibfnamefont {B.}~\bibnamefont
  {{Putigny}}},\ }\bibfield  {title} {\bibinfo {title} {{An algorithm for
  computing the 2D structure of fast rotating stars}},\ }\href
  {https://doi.org/10.1016/j.jcp.2016.05.011} {\bibfield  {journal} {\bibinfo
  {journal} {Journal of Computational Physics}\ }\textbf {\bibinfo {volume}
  {318}},\ \bibinfo {pages} {277} (\bibinfo {year} {2016})},\ \Eprint
  {https://arxiv.org/abs/1605.02359} {arXiv:1605.02359 [astro-ph.SR]}
  \BibitemShut {NoStop}%
\bibitem [{\citenamefont {{von Zeipel}}(1924)}]{1924MNRAS..84..665V}%
  \BibitemOpen
  \bibfield  {author} {\bibinfo {author} {\bibfnamefont {H.}~\bibnamefont {{von
  Zeipel}}},\ }\bibfield  {title} {\bibinfo {title} {{The radiative equilibrium
  of a rotating system of gaseous masses}},\ }\href
  {https://doi.org/10.1093/mnras/84.9.665} {\bibfield  {journal} {\bibinfo
  {journal} {\mnras}\ }\textbf {\bibinfo {volume} {84}},\ \bibinfo {pages}
  {665} (\bibinfo {year} {1924})}\BibitemShut {NoStop}%
\bibitem [{\citenamefont
  {{Eddington}}(1925{\natexlab{a}})}]{1925Obs....48...73E}%
  \BibitemOpen
  \bibfield  {author} {\bibinfo {author} {\bibfnamefont {A.~S.}\ \bibnamefont
  {{Eddington}}},\ }\bibfield  {title} {\bibinfo {title} {{Circulating currents
  in rotating stars}},\ }\href@noop {} {\bibfield  {journal} {\bibinfo
  {journal} {The Observatory}\ }\textbf {\bibinfo {volume} {48}},\ \bibinfo
  {pages} {73} (\bibinfo {year} {1925}{\natexlab{a}})}\BibitemShut {NoStop}%
\bibitem [{\citenamefont {{Sweet}}(1950{\natexlab{a}})}]{1950MNRAS.110..548S}%
  \BibitemOpen
  \bibfield  {author} {\bibinfo {author} {\bibfnamefont {P.~A.}\ \bibnamefont
  {{Sweet}}},\ }\bibfield  {title} {\bibinfo {title} {{The importance of
  rotation in stellar evolution}},\ }\href
  {https://doi.org/10.1093/mnras/110.6.548} {\bibfield  {journal} {\bibinfo
  {journal} {\mnras}\ }\textbf {\bibinfo {volume} {110}},\ \bibinfo {pages}
  {548} (\bibinfo {year} {1950}{\natexlab{a}})}\BibitemShut {NoStop}%
\bibitem [{\citenamefont {{Strittmatter}}(1969)}]{1969ARA&A...7..665S}%
  \BibitemOpen
  \bibfield  {author} {\bibinfo {author} {\bibfnamefont {P.~A.}\ \bibnamefont
  {{Strittmatter}}},\ }\bibfield  {title} {\bibinfo {title} {{Stellar
  Rotation}},\ }\href {https://doi.org/10.1146/annurev.aa.07.090169.003313}
  {\bibfield  {journal} {\bibinfo  {journal} {\araa}\ }\textbf {\bibinfo
  {volume} {7}},\ \bibinfo {pages} {665} (\bibinfo {year} {1969})}\BibitemShut
  {NoStop}%
\bibitem [{\citenamefont {{Kippenhahn}}\ and\ \citenamefont
  {{Thomas}}(1970)}]{1970stro.coll...20K}%
  \BibitemOpen
  \bibfield  {author} {\bibinfo {author} {\bibfnamefont {R.}~\bibnamefont
  {{Kippenhahn}}}\ and\ \bibinfo {author} {\bibfnamefont {H.~C.}\ \bibnamefont
  {{Thomas}}},\ }\bibfield  {title} {\bibinfo {title} {{A Simple Method for the
  Solution of the Stellar Structure Equations Including Rotation and Tidal
  Forces}},\ }in\ \href@noop {} {\emph {\bibinfo {booktitle} {IAU Colloq. 4:
  Stellar Rotation}}},\ \bibinfo {editor} {edited by\ \bibinfo {editor}
  {\bibfnamefont {A.}~\bibnamefont {{Slettebak}}}}\ (\bibinfo {year} {1970})\
  p.~\bibinfo {pages} {20}\BibitemShut {NoStop}%
\bibitem [{\citenamefont {{Endal}}\ and\ \citenamefont
  {{Sofia}}(1976)}]{1976ApJ...210..184E}%
  \BibitemOpen
  \bibfield  {author} {\bibinfo {author} {\bibfnamefont {A.~S.}\ \bibnamefont
  {{Endal}}}\ and\ \bibinfo {author} {\bibfnamefont {S.}~\bibnamefont
  {{Sofia}}},\ }\bibfield  {title} {\bibinfo {title} {{The evolution of
  rotating stars. I. Method and exploratory calculations for a 7 M sun
  star.}},\ }\href {https://doi.org/10.1086/154817} {\bibfield  {journal}
  {\bibinfo  {journal} {\apj}\ }\textbf {\bibinfo {volume} {210}},\ \bibinfo
  {pages} {184} (\bibinfo {year} {1976})}\BibitemShut {NoStop}%
\bibitem [{\citenamefont {{Tassoul}}(1978)}]{1978trs..book.....T}%
  \BibitemOpen
  \bibfield  {author} {\bibinfo {author} {\bibfnamefont {J.-L.}\ \bibnamefont
  {{Tassoul}}},\ }\href@noop {} {\emph {\bibinfo {title} {{Theory of rotating
  stars}}}}\ (\bibinfo {year} {1978})\BibitemShut {NoStop}%
\bibitem [{\citenamefont {{Zahn}}(1992)}]{1992A&A...265..115Z}%
  \BibitemOpen
  \bibfield  {author} {\bibinfo {author} {\bibfnamefont {J.~P.}\ \bibnamefont
  {{Zahn}}},\ }\bibfield  {title} {\bibinfo {title} {{Circulation and
  turbulence in rotating stars.}},\ }\href@noop {} {\bibfield  {journal}
  {\bibinfo  {journal} {\aap}\ }\textbf {\bibinfo {volume} {265}},\ \bibinfo
  {pages} {115} (\bibinfo {year} {1992})}\BibitemShut {NoStop}%
\bibitem [{\citenamefont {{Chaboyer}}\ and\ \citenamefont
  {{Zahn}}(1992)}]{1992A&A...253..173C}%
  \BibitemOpen
  \bibfield  {author} {\bibinfo {author} {\bibfnamefont {B.}~\bibnamefont
  {{Chaboyer}}}\ and\ \bibinfo {author} {\bibfnamefont {J.~P.}\ \bibnamefont
  {{Zahn}}},\ }\bibfield  {title} {\bibinfo {title} {{Effect of horizontal
  turbulent diffusion on transport by meridional circulation.}},\ }\href@noop
  {} {\bibfield  {journal} {\bibinfo  {journal} {\aap}\ }\textbf {\bibinfo
  {volume} {253}},\ \bibinfo {pages} {173} (\bibinfo {year}
  {1992})}\BibitemShut {NoStop}%
\bibitem [{\citenamefont {{Meynet}}\ and\ \citenamefont
  {{Maeder}}(1997)}]{1997A&A...321..465M}%
  \BibitemOpen
  \bibfield  {author} {\bibinfo {author} {\bibfnamefont {G.}~\bibnamefont
  {{Meynet}}}\ and\ \bibinfo {author} {\bibfnamefont {A.}~\bibnamefont
  {{Maeder}}},\ }\bibfield  {title} {\bibinfo {title} {{Stellar evolution with
  rotation. I. The computational method and the inhibiting effect of the
  {\ensuremath{\mu}}-gradient.}},\ }\href@noop {} {\bibfield  {journal}
  {\bibinfo  {journal} {\aap}\ }\textbf {\bibinfo {volume} {321}},\ \bibinfo
  {pages} {465} (\bibinfo {year} {1997})}\BibitemShut {NoStop}%
\bibitem [{\citenamefont {{Sackmann}}(1970)}]{1970A&A.....8...76S}%
  \BibitemOpen
  \bibfield  {author} {\bibinfo {author} {\bibfnamefont {I.~J.}\ \bibnamefont
  {{Sackmann}}},\ }\bibfield  {title} {\bibinfo {title} {{Rapid Uniform
  Rotation Along the Main Sequence II}},\ }\href@noop {} {\bibfield  {journal}
  {\bibinfo  {journal} {\aap}\ }\textbf {\bibinfo {volume} {8}},\ \bibinfo
  {pages} {76} (\bibinfo {year} {1970})}\BibitemShut {NoStop}%
\bibitem [{\citenamefont {{Ekstr{\"o}m}}\ \emph {et~al.}(2008)\citenamefont
  {{Ekstr{\"o}m}}, \citenamefont {{Meynet}}, \citenamefont {{Maeder}},\ and\
  \citenamefont {{Barblan}}}]{2008A&A...478..467E}%
  \BibitemOpen
  \bibfield  {author} {\bibinfo {author} {\bibfnamefont {S.}~\bibnamefont
  {{Ekstr{\"o}m}}}, \bibinfo {author} {\bibfnamefont {G.}~\bibnamefont
  {{Meynet}}}, \bibinfo {author} {\bibfnamefont {A.}~\bibnamefont {{Maeder}}},\
  and\ \bibinfo {author} {\bibfnamefont {F.}~\bibnamefont {{Barblan}}},\
  }\bibfield  {title} {\bibinfo {title} {{Evolution towards the critical limit
  and the origin of Be stars}},\ }\href
  {https://doi.org/10.1051/0004-6361:20078095} {\bibfield  {journal} {\bibinfo
  {journal} {\aap}\ }\textbf {\bibinfo {volume} {478}},\ \bibinfo {pages} {467}
  (\bibinfo {year} {2008})},\ \Eprint {https://arxiv.org/abs/0711.1735}
  {arXiv:0711.1735 [astro-ph]} \BibitemShut {NoStop}%
\bibitem [{\citenamefont {{Deupree}}(2011)}]{2011ApJ...735...69D}%
  \BibitemOpen
  \bibfield  {author} {\bibinfo {author} {\bibfnamefont {R.~G.}\ \bibnamefont
  {{Deupree}}},\ }\bibfield  {title} {\bibinfo {title} {{Structure of Uniformly
  Rotating Stars}},\ }\href {https://doi.org/10.1088/0004-637X/735/2/69}
  {\bibfield  {journal} {\bibinfo  {journal} {\apj}\ }\textbf {\bibinfo
  {volume} {735}},\ \bibinfo {eid} {69} (\bibinfo {year} {2011})},\ \Eprint
  {https://arxiv.org/abs/1105.0588} {arXiv:1105.0588 [astro-ph.SR]}
  \BibitemShut {NoStop}%
\bibitem [{\citenamefont {{Georgy}}\ \emph {et~al.}(2013)\citenamefont
  {{Georgy}}, \citenamefont {{Ekstr{\"o}m}}, \citenamefont {{Granada}},
  \citenamefont {{Meynet}}, \citenamefont {{Mowlavi}}, \citenamefont
  {{Eggenberger}},\ and\ \citenamefont {{Maeder}}}]{2013A&A...553A..24G}%
  \BibitemOpen
  \bibfield  {author} {\bibinfo {author} {\bibfnamefont {C.}~\bibnamefont
  {{Georgy}}}, \bibinfo {author} {\bibfnamefont {S.}~\bibnamefont
  {{Ekstr{\"o}m}}}, \bibinfo {author} {\bibfnamefont {A.}~\bibnamefont
  {{Granada}}}, \bibinfo {author} {\bibfnamefont {G.}~\bibnamefont {{Meynet}}},
  \bibinfo {author} {\bibfnamefont {N.}~\bibnamefont {{Mowlavi}}}, \bibinfo
  {author} {\bibfnamefont {P.}~\bibnamefont {{Eggenberger}}},\ and\ \bibinfo
  {author} {\bibfnamefont {A.}~\bibnamefont {{Maeder}}},\ }\bibfield  {title}
  {\bibinfo {title} {{Populations of rotating stars. I. Models from 1.7 to 15
  M$_{{\ensuremath{\odot}}}$ at Z = 0.014, 0.006, and 0.002 with
  {\ensuremath{\Omega}}/{\ensuremath{\Omega}}$_{crit}$ between 0 and 1}},\
  }\href {https://doi.org/10.1051/0004-6361/201220558} {\bibfield  {journal}
  {\bibinfo  {journal} {\aap}\ }\textbf {\bibinfo {volume} {553}},\ \bibinfo
  {eid} {A24} (\bibinfo {year} {2013})},\ \Eprint
  {https://arxiv.org/abs/1303.2321} {arXiv:1303.2321 [astro-ph.SR]}
  \BibitemShut {NoStop}%
\bibitem [{\citenamefont {{Chandrasekhar}}(1933)}]{1933MNRAS..93..539C}%
  \BibitemOpen
  \bibfield  {author} {\bibinfo {author} {\bibfnamefont {S.}~\bibnamefont
  {{Chandrasekhar}}},\ }\bibfield  {title} {\bibinfo {title} {{The equilibrium
  of distorted polytropes. IV. the rotational and the tidal distortions as
  functions of the density distribution}},\ }\href
  {https://doi.org/10.1093/mnras/93.8.539} {\bibfield  {journal} {\bibinfo
  {journal} {\mnras}\ }\textbf {\bibinfo {volume} {93}},\ \bibinfo {pages}
  {539} (\bibinfo {year} {1933})}\BibitemShut {NoStop}%
\bibitem [{\citenamefont {{Maeder}}(1999)}]{1999A&A...347..185M}%
  \BibitemOpen
  \bibfield  {author} {\bibinfo {author} {\bibfnamefont {A.}~\bibnamefont
  {{Maeder}}},\ }\bibfield  {title} {\bibinfo {title} {{Stellar evolution with
  rotation IV: von Zeipel's theorem and anisotropic losses of mass and angular
  momentum}},\ }\href@noop {} {\bibfield  {journal} {\bibinfo  {journal}
  {\aap}\ }\textbf {\bibinfo {volume} {347}},\ \bibinfo {pages} {185} (\bibinfo
  {year} {1999})}\BibitemShut {NoStop}%
\bibitem [{\citenamefont {{Aufdenberg}}\ \emph {et~al.}(2006)\citenamefont
  {{Aufdenberg}}, \citenamefont {{M{\'e}rand}}, \citenamefont {{Coud{\'e} du
  Foresto}}, \citenamefont {{Absil}}, \citenamefont {{Di Folco}}, \citenamefont
  {{Kervella}}, \citenamefont {{Ridgway}}, \citenamefont {{Berger}},
  \citenamefont {{ten Brummelaar}}, \citenamefont {{McAlister}}, \citenamefont
  {{Sturmann}}, \citenamefont {{Sturmann}},\ and\ \citenamefont
  {{Turner}}}]{2006ApJ...645..664A}%
  \BibitemOpen
  \bibfield  {author} {\bibinfo {author} {\bibfnamefont {J.~P.}\ \bibnamefont
  {{Aufdenberg}}}, \bibinfo {author} {\bibfnamefont {A.}~\bibnamefont
  {{M{\'e}rand}}}, \bibinfo {author} {\bibfnamefont {V.}~\bibnamefont
  {{Coud{\'e} du Foresto}}}, \bibinfo {author} {\bibfnamefont {O.}~\bibnamefont
  {{Absil}}}, \bibinfo {author} {\bibfnamefont {E.}~\bibnamefont {{Di Folco}}},
  \bibinfo {author} {\bibfnamefont {P.}~\bibnamefont {{Kervella}}}, \bibinfo
  {author} {\bibfnamefont {S.~T.}\ \bibnamefont {{Ridgway}}}, \bibinfo {author}
  {\bibfnamefont {D.~H.}\ \bibnamefont {{Berger}}}, \bibinfo {author}
  {\bibfnamefont {T.~A.}\ \bibnamefont {{ten Brummelaar}}}, \bibinfo {author}
  {\bibfnamefont {H.~A.}\ \bibnamefont {{McAlister}}}, \bibinfo {author}
  {\bibfnamefont {J.}~\bibnamefont {{Sturmann}}}, \bibinfo {author}
  {\bibfnamefont {L.}~\bibnamefont {{Sturmann}}},\ and\ \bibinfo {author}
  {\bibfnamefont {N.~H.}\ \bibnamefont {{Turner}}},\ }\bibfield  {title}
  {\bibinfo {title} {{First Results from the CHARA Array. VII. Long-Baseline
  Interferometric Measurements of Vega Consistent with a Pole-On, Rapidly
  Rotating Star}},\ }\href {https://doi.org/10.1086/504149} {\bibfield
  {journal} {\bibinfo  {journal} {\apj}\ }\textbf {\bibinfo {volume} {645}},\
  \bibinfo {pages} {664} (\bibinfo {year} {2006})},\ \Eprint
  {https://arxiv.org/abs/astro-ph/0603327} {arXiv:astro-ph/0603327 [astro-ph]}
  \BibitemShut {NoStop}%
\bibitem [{\citenamefont {{Lovekin}}\ \emph {et~al.}(2006)\citenamefont
  {{Lovekin}}, \citenamefont {{Deupree}},\ and\ \citenamefont
  {{Short}}}]{2006ApJ...643..460L}%
  \BibitemOpen
  \bibfield  {author} {\bibinfo {author} {\bibfnamefont {C.~C.}\ \bibnamefont
  {{Lovekin}}}, \bibinfo {author} {\bibfnamefont {R.~G.}\ \bibnamefont
  {{Deupree}}},\ and\ \bibinfo {author} {\bibfnamefont {C.~I.}\ \bibnamefont
  {{Short}}},\ }\bibfield  {title} {\bibinfo {title} {{Surface Temperature and
  Synthetic Spectral Energy Distributions for Rotationally Deformed Stars}},\
  }\href {https://doi.org/10.1086/501492} {\bibfield  {journal} {\bibinfo
  {journal} {\apj}\ }\textbf {\bibinfo {volume} {643}},\ \bibinfo {pages} {460}
  (\bibinfo {year} {2006})},\ \Eprint {https://arxiv.org/abs/astro-ph/0602084}
  {arXiv:astro-ph/0602084 [astro-ph]} \BibitemShut {NoStop}%
\bibitem [{\citenamefont {{Espinosa Lara}}\ and\ \citenamefont
  {{Rieutord}}(2011)}]{2011A&A...533A..43E}%
  \BibitemOpen
  \bibfield  {author} {\bibinfo {author} {\bibfnamefont {F.}~\bibnamefont
  {{Espinosa Lara}}}\ and\ \bibinfo {author} {\bibfnamefont {M.}~\bibnamefont
  {{Rieutord}}},\ }\bibfield  {title} {\bibinfo {title} {{Gravity darkening in
  rotating stars}},\ }\href {https://doi.org/10.1051/0004-6361/201117252}
  {\bibfield  {journal} {\bibinfo  {journal} {\aap}\ }\textbf {\bibinfo
  {volume} {533}},\ \bibinfo {eid} {A43} (\bibinfo {year} {2011})},\ \Eprint
  {https://arxiv.org/abs/1109.3038} {arXiv:1109.3038 [astro-ph.SR]}
  \BibitemShut {NoStop}%
\bibitem [{\citenamefont {{Espinosa Lara}}\ and\ \citenamefont
  {{Rieutord}}(2013)}]{2013A&A...552A..35E}%
  \BibitemOpen
  \bibfield  {author} {\bibinfo {author} {\bibfnamefont {F.}~\bibnamefont
  {{Espinosa Lara}}}\ and\ \bibinfo {author} {\bibfnamefont {M.}~\bibnamefont
  {{Rieutord}}},\ }\bibfield  {title} {\bibinfo {title} {{Self-consistent 2D
  models of fast-rotating early-type stars}},\ }\href
  {https://doi.org/10.1051/0004-6361/201220844} {\bibfield  {journal} {\bibinfo
   {journal} {\aap}\ }\textbf {\bibinfo {volume} {552}},\ \bibinfo {eid} {A35}
  (\bibinfo {year} {2013})},\ \Eprint {https://arxiv.org/abs/1212.0778}
  {arXiv:1212.0778 [astro-ph.SR]} \BibitemShut {NoStop}%
\bibitem [{\citenamefont {{Georgy}}\ \emph {et~al.}(2014)\citenamefont
  {{Georgy}}, \citenamefont {{Granada}}, \citenamefont {{Ekstr{\"o}m}},
  \citenamefont {{Meynet}}, \citenamefont {{Anderson}}, \citenamefont
  {{Wyttenbach}}, \citenamefont {{Eggenberger}},\ and\ \citenamefont
  {{Maeder}}}]{2014A&A...566A..21G}%
  \BibitemOpen
  \bibfield  {author} {\bibinfo {author} {\bibfnamefont {C.}~\bibnamefont
  {{Georgy}}}, \bibinfo {author} {\bibfnamefont {A.}~\bibnamefont {{Granada}}},
  \bibinfo {author} {\bibfnamefont {S.}~\bibnamefont {{Ekstr{\"o}m}}}, \bibinfo
  {author} {\bibfnamefont {G.}~\bibnamefont {{Meynet}}}, \bibinfo {author}
  {\bibfnamefont {R.~I.}\ \bibnamefont {{Anderson}}}, \bibinfo {author}
  {\bibfnamefont {A.}~\bibnamefont {{Wyttenbach}}}, \bibinfo {author}
  {\bibfnamefont {P.}~\bibnamefont {{Eggenberger}}},\ and\ \bibinfo {author}
  {\bibfnamefont {A.}~\bibnamefont {{Maeder}}},\ }\bibfield  {title} {\bibinfo
  {title} {{Populations of rotating stars. III. SYCLIST, the new Geneva
  population synthesis code}},\ }\href
  {https://doi.org/10.1051/0004-6361/201423881} {\bibfield  {journal} {\bibinfo
   {journal} {\aap}\ }\textbf {\bibinfo {volume} {566}},\ \bibinfo {eid} {A21}
  (\bibinfo {year} {2014})},\ \Eprint {https://arxiv.org/abs/1404.6952}
  {arXiv:1404.6952 [astro-ph.SR]} \BibitemShut {NoStop}%
\bibitem [{\citenamefont {{Claret}}(2016)}]{2016A&A...588A..15C}%
  \BibitemOpen
  \bibfield  {author} {\bibinfo {author} {\bibfnamefont {A.}~\bibnamefont
  {{Claret}}},\ }\bibfield  {title} {\bibinfo {title} {{Theoretical gravity
  darkening as a function of optical depth. A first approach to fast rotating
  stars}},\ }\href {https://doi.org/10.1051/0004-6361/201527336} {\bibfield
  {journal} {\bibinfo  {journal} {\aap}\ }\textbf {\bibinfo {volume} {588}},\
  \bibinfo {eid} {A15} (\bibinfo {year} {2016})},\ \Eprint
  {https://arxiv.org/abs/1606.00834} {arXiv:1606.00834 [astro-ph.SR]}
  \BibitemShut {NoStop}%
\bibitem [{\citenamefont {{Girardi}}\ \emph {et~al.}(2019)\citenamefont
  {{Girardi}}, \citenamefont {{Costa}}, \citenamefont {{Chen}}, \citenamefont
  {{Goudfrooij}}, \citenamefont {{Bressan}}, \citenamefont {{Marigo}},\ and\
  \citenamefont {{Bellini}}}]{2019MNRAS.488..696G}%
  \BibitemOpen
  \bibfield  {author} {\bibinfo {author} {\bibfnamefont {L.}~\bibnamefont
  {{Girardi}}}, \bibinfo {author} {\bibfnamefont {G.}~\bibnamefont {{Costa}}},
  \bibinfo {author} {\bibfnamefont {Y.}~\bibnamefont {{Chen}}}, \bibinfo
  {author} {\bibfnamefont {P.}~\bibnamefont {{Goudfrooij}}}, \bibinfo {author}
  {\bibfnamefont {A.}~\bibnamefont {{Bressan}}}, \bibinfo {author}
  {\bibfnamefont {P.}~\bibnamefont {{Marigo}}},\ and\ \bibinfo {author}
  {\bibfnamefont {A.}~\bibnamefont {{Bellini}}},\ }\bibfield  {title} {\bibinfo
  {title} {{On the photometric signature of fast rotators}},\ }\href
  {https://doi.org/10.1093/mnras/stz1767} {\bibfield  {journal} {\bibinfo
  {journal} {\mnras}\ }\textbf {\bibinfo {volume} {488}},\ \bibinfo {pages}
  {696} (\bibinfo {year} {2019})},\ \Eprint {https://arxiv.org/abs/1907.00688}
  {arXiv:1907.00688 [astro-ph.SR]} \BibitemShut {NoStop}%
\bibitem [{\citenamefont {{Abdul-Masih}}(2023)}]{2023A&A...669L..11A}%
  \BibitemOpen
  \bibfield  {author} {\bibinfo {author} {\bibfnamefont {M.}~\bibnamefont
  {{Abdul-Masih}}},\ }\bibfield  {title} {\bibinfo {title} {{Effects of
  rotation on the spectroscopic observables of massive stars}},\ }\href
  {https://doi.org/10.1051/0004-6361/202245653} {\bibfield  {journal} {\bibinfo
   {journal} {\aap}\ }\textbf {\bibinfo {volume} {669}},\ \bibinfo {eid} {L11}
  (\bibinfo {year} {2023})},\ \Eprint {https://arxiv.org/abs/2212.10485}
  {arXiv:2212.10485 [astro-ph.SR]} \BibitemShut {NoStop}%
\bibitem [{\citenamefont {{Lucy}}(1967)}]{1967ZA.....65...89L}%
  \BibitemOpen
  \bibfield  {author} {\bibinfo {author} {\bibfnamefont {L.~B.}\ \bibnamefont
  {{Lucy}}},\ }\bibfield  {title} {\bibinfo {title} {{Gravity-Darkening for
  Stars with Convective Envelopes}},\ }\href@noop {} {\bibfield  {journal}
  {\bibinfo  {journal} {\zap}\ }\textbf {\bibinfo {volume} {65}},\ \bibinfo
  {pages} {89} (\bibinfo {year} {1967})}\BibitemShut {NoStop}%
\bibitem [{\citenamefont {{Domiciano de Souza}}\ \emph
  {et~al.}(2014)\citenamefont {{Domiciano de Souza}}, \citenamefont
  {{Kervella}}, \citenamefont {{Moser Faes}}, \citenamefont {{Dalla Vedova}},
  \citenamefont {{M{\'e}rand}}, \citenamefont {{Le Bouquin}}, \citenamefont
  {{Espinosa Lara}}, \citenamefont {{Rieutord}}, \citenamefont {{Bendjoya}},
  \citenamefont {{Carciofi}}, \citenamefont {{Hadjara}}, \citenamefont
  {{Millour}},\ and\ \citenamefont {{Vakili}}}]{2014A&A...569A..10D}%
  \BibitemOpen
  \bibfield  {author} {\bibinfo {author} {\bibfnamefont {A.}~\bibnamefont
  {{Domiciano de Souza}}}, \bibinfo {author} {\bibfnamefont {P.}~\bibnamefont
  {{Kervella}}}, \bibinfo {author} {\bibfnamefont {D.}~\bibnamefont {{Moser
  Faes}}}, \bibinfo {author} {\bibfnamefont {G.}~\bibnamefont {{Dalla
  Vedova}}}, \bibinfo {author} {\bibfnamefont {A.}~\bibnamefont
  {{M{\'e}rand}}}, \bibinfo {author} {\bibfnamefont {J.~B.}\ \bibnamefont {{Le
  Bouquin}}}, \bibinfo {author} {\bibfnamefont {F.}~\bibnamefont {{Espinosa
  Lara}}}, \bibinfo {author} {\bibfnamefont {M.}~\bibnamefont {{Rieutord}}},
  \bibinfo {author} {\bibfnamefont {P.}~\bibnamefont {{Bendjoya}}}, \bibinfo
  {author} {\bibfnamefont {A.~C.}\ \bibnamefont {{Carciofi}}}, \bibinfo
  {author} {\bibfnamefont {M.}~\bibnamefont {{Hadjara}}}, \bibinfo {author}
  {\bibfnamefont {F.}~\bibnamefont {{Millour}}},\ and\ \bibinfo {author}
  {\bibfnamefont {F.}~\bibnamefont {{Vakili}}},\ }\bibfield  {title} {\bibinfo
  {title} {{The environment of the fast rotating star Achernar. III.
  Photospheric parameters revealed by the VLTI}},\ }\href
  {https://doi.org/10.1051/0004-6361/201424144} {\bibfield  {journal} {\bibinfo
   {journal} {\aap}\ }\textbf {\bibinfo {volume} {569}},\ \bibinfo {eid} {A10}
  (\bibinfo {year} {2014})}\BibitemShut {NoStop}%
\bibitem [{\citenamefont {{Friend}}\ and\ \citenamefont
  {{Abbott}}(1986)}]{1986ApJ...311..701F}%
  \BibitemOpen
  \bibfield  {author} {\bibinfo {author} {\bibfnamefont {D.~B.}\ \bibnamefont
  {{Friend}}}\ and\ \bibinfo {author} {\bibfnamefont {D.~C.}\ \bibnamefont
  {{Abbott}}},\ }\bibfield  {title} {\bibinfo {title} {{The Theory of
  Radiatively Driven Stellar Winds. III. Wind Models with Finite Disk
  Correction and Rotation}},\ }\href {https://doi.org/10.1086/164809}
  {\bibfield  {journal} {\bibinfo  {journal} {\apj}\ }\textbf {\bibinfo
  {volume} {311}},\ \bibinfo {pages} {701} (\bibinfo {year}
  {1986})}\BibitemShut {NoStop}%
\bibitem [{\citenamefont {{Bjorkman}}\ and\ \citenamefont
  {{Cassinelli}}(1993)}]{1993ApJ...409..429B}%
  \BibitemOpen
  \bibfield  {author} {\bibinfo {author} {\bibfnamefont {J.~E.}\ \bibnamefont
  {{Bjorkman}}}\ and\ \bibinfo {author} {\bibfnamefont {J.~P.}\ \bibnamefont
  {{Cassinelli}}},\ }\bibfield  {title} {\bibinfo {title} {{Equatorial Disk
  Formation around Rotating Stars Due to Ram Pressure Confinement by the
  Stellar Wind}},\ }\href {https://doi.org/10.1086/172676} {\bibfield
  {journal} {\bibinfo  {journal} {\apj}\ }\textbf {\bibinfo {volume} {409}},\
  \bibinfo {pages} {429} (\bibinfo {year} {1993})}\BibitemShut {NoStop}%
\bibitem [{\citenamefont {{Langer}}(1998)}]{1998A&A...329..551L}%
  \BibitemOpen
  \bibfield  {author} {\bibinfo {author} {\bibfnamefont {N.}~\bibnamefont
  {{Langer}}},\ }\bibfield  {title} {\bibinfo {title} {{Coupled mass and
  angular momentum loss of massive main sequence stars}},\ }\href@noop {}
  {\bibfield  {journal} {\bibinfo  {journal} {\aap}\ }\textbf {\bibinfo
  {volume} {329}},\ \bibinfo {pages} {551} (\bibinfo {year}
  {1998})}\BibitemShut {NoStop}%
\bibitem [{\citenamefont {{Paxton}}\ \emph {et~al.}(2013)\citenamefont
  {{Paxton}}, \citenamefont {{Cantiello}}, \citenamefont {{Arras}},
  \citenamefont {{Bildsten}}, \citenamefont {{Brown}}, \citenamefont
  {{Dotter}}, \citenamefont {{Mankovich}}, \citenamefont {{Montgomery}},
  \citenamefont {{Stello}}, \citenamefont {{Timmes}},\ and\ \citenamefont
  {{Townsend}}}]{2013ApJS..208....4P}%
  \BibitemOpen
  \bibfield  {author} {\bibinfo {author} {\bibfnamefont {B.}~\bibnamefont
  {{Paxton}}}, \bibinfo {author} {\bibfnamefont {M.}~\bibnamefont
  {{Cantiello}}}, \bibinfo {author} {\bibfnamefont {P.}~\bibnamefont
  {{Arras}}}, \bibinfo {author} {\bibfnamefont {L.}~\bibnamefont {{Bildsten}}},
  \bibinfo {author} {\bibfnamefont {E.~F.}\ \bibnamefont {{Brown}}}, \bibinfo
  {author} {\bibfnamefont {A.}~\bibnamefont {{Dotter}}}, \bibinfo {author}
  {\bibfnamefont {C.}~\bibnamefont {{Mankovich}}}, \bibinfo {author}
  {\bibfnamefont {M.~H.}\ \bibnamefont {{Montgomery}}}, \bibinfo {author}
  {\bibfnamefont {D.}~\bibnamefont {{Stello}}}, \bibinfo {author}
  {\bibfnamefont {F.~X.}\ \bibnamefont {{Timmes}}},\ and\ \bibinfo {author}
  {\bibfnamefont {R.}~\bibnamefont {{Townsend}}},\ }\bibfield  {title}
  {\bibinfo {title} {{Modules for Experiments in Stellar Astrophysics (MESA):
  Planets, Oscillations, Rotation, and Massive Stars}},\ }\href
  {https://doi.org/10.1088/0067-0049/208/1/4} {\bibfield  {journal} {\bibinfo
  {journal} {\apjs}\ }\textbf {\bibinfo {volume} {208}},\ \bibinfo {eid} {4}
  (\bibinfo {year} {2013})},\ \Eprint {https://arxiv.org/abs/1301.0319}
  {arXiv:1301.0319 [astro-ph.SR]} \BibitemShut {NoStop}%
\bibitem [{\citenamefont {{Georgy}}\ \emph {et~al.}(2011)\citenamefont
  {{Georgy}}, \citenamefont {{Meynet}},\ and\ \citenamefont
  {{Maeder}}}]{2011A&A...527A..52G}%
  \BibitemOpen
  \bibfield  {author} {\bibinfo {author} {\bibfnamefont {C.}~\bibnamefont
  {{Georgy}}}, \bibinfo {author} {\bibfnamefont {G.}~\bibnamefont {{Meynet}}},\
  and\ \bibinfo {author} {\bibfnamefont {A.}~\bibnamefont {{Maeder}}},\
  }\bibfield  {title} {\bibinfo {title} {{Effects of anisotropic winds on
  massive star evolution}},\ }\href
  {https://doi.org/10.1051/0004-6361/200913797} {\bibfield  {journal} {\bibinfo
   {journal} {\aap}\ }\textbf {\bibinfo {volume} {527}},\ \bibinfo {eid} {A52}
  (\bibinfo {year} {2011})},\ \Eprint {https://arxiv.org/abs/1011.6581}
  {arXiv:1011.6581 [astro-ph.SR]} \BibitemShut {NoStop}%
\bibitem [{\citenamefont {{Zahn}}(1974)}]{1974IAUS...59..185Z}%
  \BibitemOpen
  \bibfield  {author} {\bibinfo {author} {\bibfnamefont {J.~P.}\ \bibnamefont
  {{Zahn}}},\ }\bibfield  {title} {\bibinfo {title} {{Rotational Instabilities
  and Stellar Evolution}},\ }in\ \href@noop {} {\emph {\bibinfo {booktitle}
  {Stellar Instability and Evolution}}},\ Vol.~\bibinfo {volume} {59},\
  \bibinfo {editor} {edited by\ \bibinfo {editor} {\bibfnamefont
  {P.}~\bibnamefont {{Ledoux}}}, \bibinfo {editor} {\bibfnamefont
  {A.}~\bibnamefont {{Noels}}},\ and\ \bibinfo {editor} {\bibfnamefont {A.~W.}\
  \bibnamefont {{Rodgers}}}}\ (\bibinfo {year} {1974})\ p.\ \bibinfo {pages}
  {185}\BibitemShut {NoStop}%
\bibitem [{\citenamefont {{Maeder}}\ and\ \citenamefont
  {{Zahn}}(1998)}]{1998A&A...334.1000M}%
  \BibitemOpen
  \bibfield  {author} {\bibinfo {author} {\bibfnamefont {A.}~\bibnamefont
  {{Maeder}}}\ and\ \bibinfo {author} {\bibfnamefont {J.-P.}\ \bibnamefont
  {{Zahn}}},\ }\bibfield  {title} {\bibinfo {title} {{Stellar evolution with
  rotation. III. Meridional circulation with MU -gradients and
  non-stationarity}},\ }\href@noop {} {\bibfield  {journal} {\bibinfo
  {journal} {\aap}\ }\textbf {\bibinfo {volume} {334}},\ \bibinfo {pages}
  {1000} (\bibinfo {year} {1998})}\BibitemShut {NoStop}%
\bibitem [{\citenamefont {{Palacios}}\ \emph {et~al.}(2003)\citenamefont
  {{Palacios}}, \citenamefont {{Talon}}, \citenamefont {{Charbonnel}},\ and\
  \citenamefont {{Forestini}}}]{2003A&A...399..603P}%
  \BibitemOpen
  \bibfield  {author} {\bibinfo {author} {\bibfnamefont {A.}~\bibnamefont
  {{Palacios}}}, \bibinfo {author} {\bibfnamefont {S.}~\bibnamefont {{Talon}}},
  \bibinfo {author} {\bibfnamefont {C.}~\bibnamefont {{Charbonnel}}},\ and\
  \bibinfo {author} {\bibfnamefont {M.}~\bibnamefont {{Forestini}}},\
  }\bibfield  {title} {\bibinfo {title} {{Rotational mixing in low-mass stars.
  I Effect of the mu-gradients in main sequence and subgiant Pop I stars}},\
  }\href {https://doi.org/10.1051/0004-6361:20021759} {\bibfield  {journal}
  {\bibinfo  {journal} {\aap}\ }\textbf {\bibinfo {volume} {399}},\ \bibinfo
  {pages} {603} (\bibinfo {year} {2003})},\ \Eprint
  {https://arxiv.org/abs/astro-ph/0210516} {arXiv:astro-ph/0210516 [astro-ph]}
  \BibitemShut {NoStop}%
\bibitem [{\citenamefont {{Potter}}\ \emph
  {et~al.}(2012{\natexlab{a}})\citenamefont {{Potter}}, \citenamefont
  {{Tout}},\ and\ \citenamefont {{Eldridge}}}]{2012MNRAS.419..748P}%
  \BibitemOpen
  \bibfield  {author} {\bibinfo {author} {\bibfnamefont {A.~T.}\ \bibnamefont
  {{Potter}}}, \bibinfo {author} {\bibfnamefont {C.~A.}\ \bibnamefont
  {{Tout}}},\ and\ \bibinfo {author} {\bibfnamefont {J.~J.}\ \bibnamefont
  {{Eldridge}}},\ }\bibfield  {title} {\bibinfo {title} {{Towards a unified
  model of stellar rotation}},\ }\href
  {https://doi.org/10.1111/j.1365-2966.2011.19737.x} {\bibfield  {journal}
  {\bibinfo  {journal} {\mnras}\ }\textbf {\bibinfo {volume} {419}},\ \bibinfo
  {pages} {748} (\bibinfo {year} {2012}{\natexlab{a}})},\ \Eprint
  {https://arxiv.org/abs/1109.0993} {arXiv:1109.0993 [astro-ph.SR]}
  \BibitemShut {NoStop}%
\bibitem [{\citenamefont {{Marques}}\ \emph {et~al.}(2013)\citenamefont
  {{Marques}}, \citenamefont {{Goupil}}, \citenamefont {{Lebreton}},
  \citenamefont {{Talon}}, \citenamefont {{Palacios}}, \citenamefont
  {{Belkacem}}, \citenamefont {{Ouazzani}}, \citenamefont {{Mosser}},
  \citenamefont {{Moya}}, \citenamefont {{Morel}}, \citenamefont {{Pichon}},
  \citenamefont {{Mathis}}, \citenamefont {{Zahn}}, \citenamefont
  {{Turck-Chi{\`e}ze}},\ and\ \citenamefont {{Nghiem}}}]{2013A&A...549A..74M}%
  \BibitemOpen
  \bibfield  {author} {\bibinfo {author} {\bibfnamefont {J.~P.}\ \bibnamefont
  {{Marques}}}, \bibinfo {author} {\bibfnamefont {M.~J.}\ \bibnamefont
  {{Goupil}}}, \bibinfo {author} {\bibfnamefont {Y.}~\bibnamefont
  {{Lebreton}}}, \bibinfo {author} {\bibfnamefont {S.}~\bibnamefont {{Talon}}},
  \bibinfo {author} {\bibfnamefont {A.}~\bibnamefont {{Palacios}}}, \bibinfo
  {author} {\bibfnamefont {K.}~\bibnamefont {{Belkacem}}}, \bibinfo {author}
  {\bibfnamefont {R.~M.}\ \bibnamefont {{Ouazzani}}}, \bibinfo {author}
  {\bibfnamefont {B.}~\bibnamefont {{Mosser}}}, \bibinfo {author}
  {\bibfnamefont {A.}~\bibnamefont {{Moya}}}, \bibinfo {author} {\bibfnamefont
  {P.}~\bibnamefont {{Morel}}}, \bibinfo {author} {\bibfnamefont
  {B.}~\bibnamefont {{Pichon}}}, \bibinfo {author} {\bibfnamefont
  {S.}~\bibnamefont {{Mathis}}}, \bibinfo {author} {\bibfnamefont {J.~P.}\
  \bibnamefont {{Zahn}}}, \bibinfo {author} {\bibfnamefont {S.}~\bibnamefont
  {{Turck-Chi{\`e}ze}}},\ and\ \bibinfo {author} {\bibfnamefont {P.~A.~P.}\
  \bibnamefont {{Nghiem}}},\ }\bibfield  {title} {\bibinfo {title} {{Seismic
  diagnostics for transport of angular momentum in stars. I. Rotational
  splittings from the pre-main sequence to the red-giant branch}},\ }\href
  {https://doi.org/10.1051/0004-6361/201220211} {\bibfield  {journal} {\bibinfo
   {journal} {\aap}\ }\textbf {\bibinfo {volume} {549}},\ \bibinfo {eid} {A74}
  (\bibinfo {year} {2013})},\ \Eprint {https://arxiv.org/abs/1211.1271}
  {arXiv:1211.1271 [astro-ph.SR]} \BibitemShut {NoStop}%
\bibitem [{\citenamefont {{Evans}}\ \emph {et~al.}(2005)\citenamefont
  {{Evans}}, \citenamefont {{Smartt}}, \citenamefont {{Lennon}}, \citenamefont
  {{Dufton}}, \citenamefont {{Hunter}}, \citenamefont {{Mokiem}}, \citenamefont
  {{de Koter}},\ and\ \citenamefont {{Irwin}}}]{2005Msngr.122...36E}%
  \BibitemOpen
  \bibfield  {author} {\bibinfo {author} {\bibfnamefont {C.}~\bibnamefont
  {{Evans}}}, \bibinfo {author} {\bibfnamefont {S.}~\bibnamefont {{Smartt}}},
  \bibinfo {author} {\bibfnamefont {D.}~\bibnamefont {{Lennon}}}, \bibinfo
  {author} {\bibfnamefont {P.}~\bibnamefont {{Dufton}}}, \bibinfo {author}
  {\bibfnamefont {I.}~\bibnamefont {{Hunter}}}, \bibinfo {author}
  {\bibfnamefont {R.}~\bibnamefont {{Mokiem}}}, \bibinfo {author}
  {\bibfnamefont {A.}~\bibnamefont {{de Koter}}},\ and\ \bibinfo {author}
  {\bibfnamefont {M.}~\bibnamefont {{Irwin}}},\ }\bibfield  {title} {\bibinfo
  {title} {{The VLT-FLAMES Survey of Massive Stars}},\ }\href@noop {}
  {\bibfield  {journal} {\bibinfo  {journal} {The Messenger}\ }\textbf
  {\bibinfo {volume} {122}},\ \bibinfo {pages} {36} (\bibinfo {year}
  {2005})}\BibitemShut {NoStop}%
\bibitem [{\citenamefont {{Evans}}\ \emph {et~al.}(2006)\citenamefont
  {{Evans}}, \citenamefont {{Lennon}}, \citenamefont {{Smartt}},\ and\
  \citenamefont {{Trundle}}}]{2006A&A...456..623E}%
  \BibitemOpen
  \bibfield  {author} {\bibinfo {author} {\bibfnamefont {C.~J.}\ \bibnamefont
  {{Evans}}}, \bibinfo {author} {\bibfnamefont {D.~J.}\ \bibnamefont
  {{Lennon}}}, \bibinfo {author} {\bibfnamefont {S.~J.}\ \bibnamefont
  {{Smartt}}},\ and\ \bibinfo {author} {\bibfnamefont {C.}~\bibnamefont
  {{Trundle}}},\ }\bibfield  {title} {\bibinfo {title} {{The VLT-FLAMES survey
  of massive stars: observations centered on the Magellanic Cloud clusters NGC
  330, NGC 346, NGC 2004, and the N11 region}},\ }\href
  {https://doi.org/10.1051/0004-6361:20064988} {\bibfield  {journal} {\bibinfo
  {journal} {\aap}\ }\textbf {\bibinfo {volume} {456}},\ \bibinfo {pages} {623}
  (\bibinfo {year} {2006})},\ \Eprint {https://arxiv.org/abs/astro-ph/0606405}
  {arXiv:astro-ph/0606405 [astro-ph]} \BibitemShut {NoStop}%
\bibitem [{\citenamefont {{Pedersen}}(2022)}]{2022ApJ...940...49P}%
  \BibitemOpen
  \bibfield  {author} {\bibinfo {author} {\bibfnamefont {M.~G.}\ \bibnamefont
  {{Pedersen}}},\ }\bibfield  {title} {\bibinfo {title} {{Internal Rotation and
  Inclinations of Slowly Pulsating B Stars: Evidence of Interior Angular
  Momentum Transport}},\ }\href {https://doi.org/10.3847/1538-4357/ac947f}
  {\bibfield  {journal} {\bibinfo  {journal} {\apj}\ }\textbf {\bibinfo
  {volume} {940}},\ \bibinfo {eid} {49} (\bibinfo {year} {2022})},\ \Eprint
  {https://arxiv.org/abs/2208.14497} {arXiv:2208.14497 [astro-ph.SR]}
  \BibitemShut {NoStop}%
\bibitem [{\citenamefont {{Kurtz}}(2022)}]{2022ARA&A..60...31K}%
  \BibitemOpen
  \bibfield  {author} {\bibinfo {author} {\bibfnamefont {D.~W.}\ \bibnamefont
  {{Kurtz}}},\ }\bibfield  {title} {\bibinfo {title} {{Asteroseismology Across
  the Hertzsprung-Russell Diagram}},\ }\href
  {https://doi.org/10.1146/annurev-astro-052920-094232} {\bibfield  {journal}
  {\bibinfo  {journal} {\araa}\ }\textbf {\bibinfo {volume} {60}},\ \bibinfo
  {pages} {31} (\bibinfo {year} {2022})}\BibitemShut {NoStop}%
\bibitem [{\citenamefont {{Brandt}}\ and\ \citenamefont
  {{Huang}}(2015{\natexlab{a}})}]{2015ApJ...807...58B}%
  \BibitemOpen
  \bibfield  {author} {\bibinfo {author} {\bibfnamefont {T.~D.}\ \bibnamefont
  {{Brandt}}}\ and\ \bibinfo {author} {\bibfnamefont {C.~X.}\ \bibnamefont
  {{Huang}}},\ }\bibfield  {title} {\bibinfo {title} {{Bayesian Ages for
  Early-type Stars from Isochrones Including Rotation, and a Possible Old Age
  for the Hyades}},\ }\href {https://doi.org/10.1088/0004-637X/807/1/58}
  {\bibfield  {journal} {\bibinfo  {journal} {\apj}\ }\textbf {\bibinfo
  {volume} {807}},\ \bibinfo {eid} {58} (\bibinfo {year}
  {2015}{\natexlab{a}})},\ \Eprint {https://arxiv.org/abs/1501.04404}
  {arXiv:1501.04404 [astro-ph.SR]} \BibitemShut {NoStop}%
\bibitem [{\citenamefont {{Brandt}}\ and\ \citenamefont
  {{Huang}}(2015{\natexlab{b}})}]{2015ApJ...807...24B}%
  \BibitemOpen
  \bibfield  {author} {\bibinfo {author} {\bibfnamefont {T.~D.}\ \bibnamefont
  {{Brandt}}}\ and\ \bibinfo {author} {\bibfnamefont {C.~X.}\ \bibnamefont
  {{Huang}}},\ }\bibfield  {title} {\bibinfo {title} {{The Age and Age Spread
  of the Praesepe and Hyades Clusters: a Consistent,
  \raisebox{-0.5ex}\textasciitilde800 Myr Picture from Rotating Stellar
  Models}},\ }\href {https://doi.org/10.1088/0004-637X/807/1/24} {\bibfield
  {journal} {\bibinfo  {journal} {\apj}\ }\textbf {\bibinfo {volume} {807}},\
  \bibinfo {eid} {24} (\bibinfo {year} {2015}{\natexlab{b}})},\ \Eprint
  {https://arxiv.org/abs/1504.00004} {arXiv:1504.00004 [astro-ph.SR]}
  \BibitemShut {NoStop}%
\bibitem [{\citenamefont {{Groh}}\ \emph {et~al.}(2019)\citenamefont {{Groh}},
  \citenamefont {{Ekstr{\"o}m}}, \citenamefont {{Georgy}}, \citenamefont
  {{Meynet}}, \citenamefont {{Choplin}}, \citenamefont {{Eggenberger}},
  \citenamefont {{Hirschi}}, \citenamefont {{Maeder}}, \citenamefont
  {{Murphy}}, \citenamefont {{Boian}},\ and\ \citenamefont
  {{Farrell}}}]{2019A&A...627A..24G}%
  \BibitemOpen
  \bibfield  {author} {\bibinfo {author} {\bibfnamefont {J.~H.}\ \bibnamefont
  {{Groh}}}, \bibinfo {author} {\bibfnamefont {S.}~\bibnamefont
  {{Ekstr{\"o}m}}}, \bibinfo {author} {\bibfnamefont {C.}~\bibnamefont
  {{Georgy}}}, \bibinfo {author} {\bibfnamefont {G.}~\bibnamefont {{Meynet}}},
  \bibinfo {author} {\bibfnamefont {A.}~\bibnamefont {{Choplin}}}, \bibinfo
  {author} {\bibfnamefont {P.}~\bibnamefont {{Eggenberger}}}, \bibinfo {author}
  {\bibfnamefont {R.}~\bibnamefont {{Hirschi}}}, \bibinfo {author}
  {\bibfnamefont {A.}~\bibnamefont {{Maeder}}}, \bibinfo {author}
  {\bibfnamefont {L.~J.}\ \bibnamefont {{Murphy}}}, \bibinfo {author}
  {\bibfnamefont {I.}~\bibnamefont {{Boian}}},\ and\ \bibinfo {author}
  {\bibfnamefont {E.~J.}\ \bibnamefont {{Farrell}}},\ }\bibfield  {title}
  {\bibinfo {title} {{Grids of stellar models with rotation. IV. Models from
  1.7 to 120 M$_{{\ensuremath{\odot}}}$ at a metallicity Z = 0.0004}},\ }\href
  {https://doi.org/10.1051/0004-6361/201833720} {\bibfield  {journal} {\bibinfo
   {journal} {\aap}\ }\textbf {\bibinfo {volume} {627}},\ \bibinfo {eid} {A24}
  (\bibinfo {year} {2019})},\ \Eprint {https://arxiv.org/abs/1904.04009}
  {arXiv:1904.04009 [astro-ph.SR]} \BibitemShut {NoStop}%
\bibitem [{\citenamefont {{Marchant}}\ and\ \citenamefont
  {{Moriya}}(2020)}]{2020A&A...640L..18M}%
  \BibitemOpen
  \bibfield  {author} {\bibinfo {author} {\bibfnamefont {P.}~\bibnamefont
  {{Marchant}}}\ and\ \bibinfo {author} {\bibfnamefont {T.~J.}\ \bibnamefont
  {{Moriya}}},\ }\bibfield  {title} {\bibinfo {title} {{The impact of stellar
  rotation on the black hole mass-gap from pair-instability supernovae}},\
  }\href {https://doi.org/10.1051/0004-6361/202038902} {\bibfield  {journal}
  {\bibinfo  {journal} {\aap}\ }\textbf {\bibinfo {volume} {640}},\ \bibinfo
  {eid} {L18} (\bibinfo {year} {2020})},\ \Eprint
  {https://arxiv.org/abs/2007.06220} {arXiv:2007.06220 [astro-ph.HE]}
  \BibitemShut {NoStop}%
\bibitem [{\citenamefont {{Maeder}}(1987)}]{1987A&A...178..159M}%
  \BibitemOpen
  \bibfield  {author} {\bibinfo {author} {\bibfnamefont {A.}~\bibnamefont
  {{Maeder}}},\ }\bibfield  {title} {\bibinfo {title} {{Evidences for a
  bifurcation in massive star evolution. The ON-blue stragglers.}},\
  }\href@noop {} {\bibfield  {journal} {\bibinfo  {journal} {\aap}\ }\textbf
  {\bibinfo {volume} {178}},\ \bibinfo {pages} {159} (\bibinfo {year}
  {1987})}\BibitemShut {NoStop}%
\bibitem [{\citenamefont {{Yoon}}\ and\ \citenamefont
  {{Langer}}(2005)}]{2005A&A...443..643Y}%
  \BibitemOpen
  \bibfield  {author} {\bibinfo {author} {\bibfnamefont {S.~C.}\ \bibnamefont
  {{Yoon}}}\ and\ \bibinfo {author} {\bibfnamefont {N.}~\bibnamefont
  {{Langer}}},\ }\bibfield  {title} {\bibinfo {title} {{Evolution of rapidly
  rotating metal-poor massive stars towards gamma-ray bursts}},\ }\href
  {https://doi.org/10.1051/0004-6361:20054030} {\bibfield  {journal} {\bibinfo
  {journal} {\aap}\ }\textbf {\bibinfo {volume} {443}},\ \bibinfo {pages} {643}
  (\bibinfo {year} {2005})},\ \Eprint {https://arxiv.org/abs/astro-ph/0508242}
  {arXiv:astro-ph/0508242 [astro-ph]} \BibitemShut {NoStop}%
\bibitem [{\citenamefont {{Martins}}\ \emph {et~al.}(2013)\citenamefont
  {{Martins}}, \citenamefont {{Depagne}}, \citenamefont {{Russeil}},\ and\
  \citenamefont {{Mahy}}}]{2013A&A...554A..23M}%
  \BibitemOpen
  \bibfield  {author} {\bibinfo {author} {\bibfnamefont {F.}~\bibnamefont
  {{Martins}}}, \bibinfo {author} {\bibfnamefont {E.}~\bibnamefont
  {{Depagne}}}, \bibinfo {author} {\bibfnamefont {D.}~\bibnamefont
  {{Russeil}}},\ and\ \bibinfo {author} {\bibfnamefont {L.}~\bibnamefont
  {{Mahy}}},\ }\bibfield  {title} {\bibinfo {title} {{Evidence of
  quasi-chemically homogeneous evolution of massive stars up to solar
  metallicity}},\ }\href {https://doi.org/10.1051/0004-6361/201321282}
  {\bibfield  {journal} {\bibinfo  {journal} {\aap}\ }\textbf {\bibinfo
  {volume} {554}},\ \bibinfo {eid} {A23} (\bibinfo {year} {2013})},\ \Eprint
  {https://arxiv.org/abs/1304.3337} {arXiv:1304.3337 [astro-ph.SR]}
  \BibitemShut {NoStop}%
\bibitem [{\citenamefont {{de Mink}}\ and\ \citenamefont
  {{Mandel}}(2016)}]{2016MNRAS.460.3545D}%
  \BibitemOpen
  \bibfield  {author} {\bibinfo {author} {\bibfnamefont {S.~E.}\ \bibnamefont
  {{de Mink}}}\ and\ \bibinfo {author} {\bibfnamefont {I.}~\bibnamefont
  {{Mandel}}},\ }\bibfield  {title} {\bibinfo {title} {{The chemically
  homogeneous evolutionary channel for binary black hole mergers: rates and
  properties of gravitational-wave events detectable by advanced LIGO}},\
  }\href {https://doi.org/10.1093/mnras/stw1219} {\bibfield  {journal}
  {\bibinfo  {journal} {\mnras}\ }\textbf {\bibinfo {volume} {460}},\ \bibinfo
  {pages} {3545} (\bibinfo {year} {2016})},\ \Eprint
  {https://arxiv.org/abs/1603.02291} {arXiv:1603.02291 [astro-ph.HE]}
  \BibitemShut {NoStop}%
\bibitem [{\citenamefont {{Riley}}\ \emph
  {et~al.}(2021{\natexlab{a}})\citenamefont {{Riley}}, \citenamefont
  {{Mandel}}, \citenamefont {{Marchant}}, \citenamefont {{Butler}},
  \citenamefont {{Nathaniel}}, \citenamefont {{Neijssel}}, \citenamefont
  {{Shortt}},\ and\ \citenamefont {{Vigna-G{\'o}mez}}}]{2021MNRAS.505..663R}%
  \BibitemOpen
  \bibfield  {author} {\bibinfo {author} {\bibfnamefont {J.}~\bibnamefont
  {{Riley}}}, \bibinfo {author} {\bibfnamefont {I.}~\bibnamefont {{Mandel}}},
  \bibinfo {author} {\bibfnamefont {P.}~\bibnamefont {{Marchant}}}, \bibinfo
  {author} {\bibfnamefont {E.}~\bibnamefont {{Butler}}}, \bibinfo {author}
  {\bibfnamefont {K.}~\bibnamefont {{Nathaniel}}}, \bibinfo {author}
  {\bibfnamefont {C.}~\bibnamefont {{Neijssel}}}, \bibinfo {author}
  {\bibfnamefont {S.}~\bibnamefont {{Shortt}}},\ and\ \bibinfo {author}
  {\bibfnamefont {A.}~\bibnamefont {{Vigna-G{\'o}mez}}},\ }\bibfield  {title}
  {\bibinfo {title} {{Chemically homogeneous evolution: a rapid population
  synthesis approach}},\ }\href {https://doi.org/10.1093/mnras/stab1291}
  {\bibfield  {journal} {\bibinfo  {journal} {\mnras}\ }\textbf {\bibinfo
  {volume} {505}},\ \bibinfo {pages} {663} (\bibinfo {year}
  {2021}{\natexlab{a}})},\ \Eprint {https://arxiv.org/abs/2010.00002}
  {arXiv:2010.00002 [astro-ph.SR]} \BibitemShut {NoStop}%
\bibitem [{\citenamefont {{Heger}}\ \emph {et~al.}(2005)\citenamefont
  {{Heger}}, \citenamefont {{Woosley}},\ and\ \citenamefont
  {{Spruit}}}]{2005ApJ...626..350H}%
  \BibitemOpen
  \bibfield  {author} {\bibinfo {author} {\bibfnamefont {A.}~\bibnamefont
  {{Heger}}}, \bibinfo {author} {\bibfnamefont {S.~E.}\ \bibnamefont
  {{Woosley}}},\ and\ \bibinfo {author} {\bibfnamefont {H.~C.}\ \bibnamefont
  {{Spruit}}},\ }\bibfield  {title} {\bibinfo {title} {{Presupernova Evolution
  of Differentially Rotating Massive Stars Including Magnetic Fields}},\ }\href
  {https://doi.org/10.1086/429868} {\bibfield  {journal} {\bibinfo  {journal}
  {\apj}\ }\textbf {\bibinfo {volume} {626}},\ \bibinfo {pages} {350} (\bibinfo
  {year} {2005})},\ \Eprint {https://arxiv.org/abs/astro-ph/0409422}
  {arXiv:astro-ph/0409422 [astro-ph]} \BibitemShut {NoStop}%
\bibitem [{\citenamefont {{Maeder}}\ and\ \citenamefont
  {{Meynet}}(2005)}]{2005A&A...440.1041M}%
  \BibitemOpen
  \bibfield  {author} {\bibinfo {author} {\bibfnamefont {A.}~\bibnamefont
  {{Maeder}}}\ and\ \bibinfo {author} {\bibfnamefont {G.}~\bibnamefont
  {{Meynet}}},\ }\bibfield  {title} {\bibinfo {title} {{Stellar evolution with
  rotation and magnetic fields. III. The interplay of circulation and
  dynamo}},\ }\href {https://doi.org/10.1051/0004-6361:20053261} {\bibfield
  {journal} {\bibinfo  {journal} {\aap}\ }\textbf {\bibinfo {volume} {440}},\
  \bibinfo {pages} {1041} (\bibinfo {year} {2005})},\ \Eprint
  {https://arxiv.org/abs/astro-ph/0506347} {arXiv:astro-ph/0506347 [astro-ph]}
  \BibitemShut {NoStop}%
\bibitem [{\citenamefont {{Walder}}\ \emph {et~al.}(2012)\citenamefont
  {{Walder}}, \citenamefont {{Folini}},\ and\ \citenamefont
  {{Meynet}}}]{2012SSRv..166..145W}%
  \BibitemOpen
  \bibfield  {author} {\bibinfo {author} {\bibfnamefont {R.}~\bibnamefont
  {{Walder}}}, \bibinfo {author} {\bibfnamefont {D.}~\bibnamefont {{Folini}}},\
  and\ \bibinfo {author} {\bibfnamefont {G.}~\bibnamefont {{Meynet}}},\
  }\bibfield  {title} {\bibinfo {title} {{Magnetic Fields in Massive Stars,
  Their Winds, and Their Nebulae}},\ }\href
  {https://doi.org/10.1007/s11214-011-9771-2} {\bibfield  {journal} {\bibinfo
  {journal} {\ssr}\ }\textbf {\bibinfo {volume} {166}},\ \bibinfo {pages} {145}
  (\bibinfo {year} {2012})},\ \Eprint {https://arxiv.org/abs/1103.3777}
  {arXiv:1103.3777 [astro-ph.SR]} \BibitemShut {NoStop}%
\bibitem [{\citenamefont {{Petit}}\ \emph {et~al.}(2017)\citenamefont
  {{Petit}}, \citenamefont {{Keszthelyi}}, \citenamefont {{MacInnis}},
  \citenamefont {{Cohen}}, \citenamefont {{Townsend}}, \citenamefont {{Wade}},
  \citenamefont {{Thomas}}, \citenamefont {{Owocki}}, \citenamefont {{Puls}},\
  and\ \citenamefont {{ud-Doula}}}]{2017MNRAS.466.1052P}%
  \BibitemOpen
  \bibfield  {author} {\bibinfo {author} {\bibfnamefont {V.}~\bibnamefont
  {{Petit}}}, \bibinfo {author} {\bibfnamefont {Z.}~\bibnamefont
  {{Keszthelyi}}}, \bibinfo {author} {\bibfnamefont {R.}~\bibnamefont
  {{MacInnis}}}, \bibinfo {author} {\bibfnamefont {D.~H.}\ \bibnamefont
  {{Cohen}}}, \bibinfo {author} {\bibfnamefont {R.~H.~D.}\ \bibnamefont
  {{Townsend}}}, \bibinfo {author} {\bibfnamefont {G.~A.}\ \bibnamefont
  {{Wade}}}, \bibinfo {author} {\bibfnamefont {S.~L.}\ \bibnamefont
  {{Thomas}}}, \bibinfo {author} {\bibfnamefont {S.~P.}\ \bibnamefont
  {{Owocki}}}, \bibinfo {author} {\bibfnamefont {J.}~\bibnamefont {{Puls}}},\
  and\ \bibinfo {author} {\bibfnamefont {A.}~\bibnamefont {{ud-Doula}}},\
  }\bibfield  {title} {\bibinfo {title} {{Magnetic massive stars as progenitors
  of `heavy' stellar-mass black holes}},\ }\href
  {https://doi.org/10.1093/mnras/stw3126} {\bibfield  {journal} {\bibinfo
  {journal} {\mnras}\ }\textbf {\bibinfo {volume} {466}},\ \bibinfo {pages}
  {1052} (\bibinfo {year} {2017})},\ \Eprint {https://arxiv.org/abs/1611.08964}
  {arXiv:1611.08964 [astro-ph.SR]} \BibitemShut {NoStop}%
\bibitem [{\citenamefont {{Groh}}\ \emph {et~al.}(2020)\citenamefont {{Groh}},
  \citenamefont {{Farrell}}, \citenamefont {{Meynet}}, \citenamefont {{Smith}},
  \citenamefont {{Murphy}}, \citenamefont {{Allan}}, \citenamefont {{Georgy}},\
  and\ \citenamefont {{Ekstroem}}}]{2020ApJ...900...98G}%
  \BibitemOpen
  \bibfield  {author} {\bibinfo {author} {\bibfnamefont {J.~H.}\ \bibnamefont
  {{Groh}}}, \bibinfo {author} {\bibfnamefont {E.~J.}\ \bibnamefont
  {{Farrell}}}, \bibinfo {author} {\bibfnamefont {G.}~\bibnamefont {{Meynet}}},
  \bibinfo {author} {\bibfnamefont {N.}~\bibnamefont {{Smith}}}, \bibinfo
  {author} {\bibfnamefont {L.}~\bibnamefont {{Murphy}}}, \bibinfo {author}
  {\bibfnamefont {A.~P.}\ \bibnamefont {{Allan}}}, \bibinfo {author}
  {\bibfnamefont {C.}~\bibnamefont {{Georgy}}},\ and\ \bibinfo {author}
  {\bibfnamefont {S.}~\bibnamefont {{Ekstroem}}},\ }\bibfield  {title}
  {\bibinfo {title} {{Massive Black Holes Regulated by Luminous Blue Variable
  Mass Loss and Magnetic Fields}},\ }\href
  {https://doi.org/10.3847/1538-4357/aba2c8} {\bibfield  {journal} {\bibinfo
  {journal} {\apj}\ }\textbf {\bibinfo {volume} {900}},\ \bibinfo {eid} {98}
  (\bibinfo {year} {2020})},\ \Eprint {https://arxiv.org/abs/1912.00994}
  {arXiv:1912.00994 [astro-ph.SR]} \BibitemShut {NoStop}%
\bibitem [{\citenamefont {{Takahashi}}\ and\ \citenamefont
  {{Langer}}(2021)}]{2021A&A...646A..19T}%
  \BibitemOpen
  \bibfield  {author} {\bibinfo {author} {\bibfnamefont {K.}~\bibnamefont
  {{Takahashi}}}\ and\ \bibinfo {author} {\bibfnamefont {N.}~\bibnamefont
  {{Langer}}},\ }\bibfield  {title} {\bibinfo {title} {{Modeling of
  magneto-rotational stellar evolution. I. Method and first applications}},\
  }\href {https://doi.org/10.1051/0004-6361/202039253} {\bibfield  {journal}
  {\bibinfo  {journal} {\aap}\ }\textbf {\bibinfo {volume} {646}},\ \bibinfo
  {eid} {A19} (\bibinfo {year} {2021})},\ \Eprint
  {https://arxiv.org/abs/2010.13909} {arXiv:2010.13909 [astro-ph.SR]}
  \BibitemShut {NoStop}%
\bibitem [{\citenamefont {{Braithwaite}}\ and\ \citenamefont
  {{Spruit}}(2017)}]{2017RSOS....460271B}%
  \BibitemOpen
  \bibfield  {author} {\bibinfo {author} {\bibfnamefont {J.}~\bibnamefont
  {{Braithwaite}}}\ and\ \bibinfo {author} {\bibfnamefont {H.~C.}\ \bibnamefont
  {{Spruit}}},\ }\bibfield  {title} {\bibinfo {title} {{Magnetic fields in
  non-convective regions of stars}},\ }\href
  {https://doi.org/10.1098/rsos.160271} {\bibfield  {journal} {\bibinfo
  {journal} {Royal Society Open Science}\ }\textbf {\bibinfo {volume} {4}},\
  \bibinfo {eid} {160271} (\bibinfo {year} {2017})},\ \Eprint
  {https://arxiv.org/abs/1510.03198} {arXiv:1510.03198 [astro-ph.SR]}
  \BibitemShut {NoStop}%
\bibitem [{\citenamefont {{Tayler}}(1973)}]{1973MNRAS.161..365T}%
  \BibitemOpen
  \bibfield  {author} {\bibinfo {author} {\bibfnamefont {R.~J.}\ \bibnamefont
  {{Tayler}}},\ }\bibfield  {title} {\bibinfo {title} {{The adiabatic stability
  of stars containing magnetic fields-I.Toroidal fields}},\ }\href
  {https://doi.org/10.1093/mnras/161.4.365} {\bibfield  {journal} {\bibinfo
  {journal} {\mnras}\ }\textbf {\bibinfo {volume} {161}},\ \bibinfo {pages}
  {365} (\bibinfo {year} {1973})}\BibitemShut {NoStop}%
\bibitem [{\citenamefont {{Spruit}}(2002)}]{2002A&A...381..923S}%
  \BibitemOpen
  \bibfield  {author} {\bibinfo {author} {\bibfnamefont {H.~C.}\ \bibnamefont
  {{Spruit}}},\ }\bibfield  {title} {\bibinfo {title} {{Dynamo action by
  differential rotation in a stably stratified stellar interior}},\ }\href
  {https://doi.org/10.1051/0004-6361:20011465} {\bibfield  {journal} {\bibinfo
  {journal} {\aap}\ }\textbf {\bibinfo {volume} {381}},\ \bibinfo {pages} {923}
  (\bibinfo {year} {2002})},\ \Eprint {https://arxiv.org/abs/astro-ph/0108207}
  {arXiv:astro-ph/0108207 [astro-ph]} \BibitemShut {NoStop}%
\bibitem [{\citenamefont {{Maeder}}\ and\ \citenamefont
  {{Meynet}}(2003)}]{2003A&A...411..543M}%
  \BibitemOpen
  \bibfield  {author} {\bibinfo {author} {\bibfnamefont {A.}~\bibnamefont
  {{Maeder}}}\ and\ \bibinfo {author} {\bibfnamefont {G.}~\bibnamefont
  {{Meynet}}},\ }\bibfield  {title} {\bibinfo {title} {{Stellar evolution with
  rotation and magnetic fields. I. The relative importance of rotational and
  magnetic effects}},\ }\href {https://doi.org/10.1051/0004-6361:20031491}
  {\bibfield  {journal} {\bibinfo  {journal} {\aap}\ }\textbf {\bibinfo
  {volume} {411}},\ \bibinfo {pages} {543} (\bibinfo {year} {2003})},\ \Eprint
  {https://arxiv.org/abs/astro-ph/0309672} {arXiv:astro-ph/0309672 [astro-ph]}
  \BibitemShut {NoStop}%
\bibitem [{\citenamefont {{Maeder}}\ and\ \citenamefont
  {{Meynet}}(2004)}]{2004A&A...422..225M}%
  \BibitemOpen
  \bibfield  {author} {\bibinfo {author} {\bibfnamefont {A.}~\bibnamefont
  {{Maeder}}}\ and\ \bibinfo {author} {\bibfnamefont {G.}~\bibnamefont
  {{Meynet}}},\ }\bibfield  {title} {\bibinfo {title} {{Stellar evolution with
  rotation and magnetic fields. II. General equations for the transport by
  Tayler-Spruit dynamo}},\ }\href {https://doi.org/10.1051/0004-6361:20034583}
  {\bibfield  {journal} {\bibinfo  {journal} {\aap}\ }\textbf {\bibinfo
  {volume} {422}},\ \bibinfo {pages} {225} (\bibinfo {year} {2004})},\ \Eprint
  {https://arxiv.org/abs/astro-ph/0404417} {arXiv:astro-ph/0404417 [astro-ph]}
  \BibitemShut {NoStop}%
\bibitem [{\citenamefont {{Potter}}\ \emph
  {et~al.}(2012{\natexlab{b}})\citenamefont {{Potter}}, \citenamefont
  {{Chitre}},\ and\ \citenamefont {{Tout}}}]{2012MNRAS.424.2358P}%
  \BibitemOpen
  \bibfield  {author} {\bibinfo {author} {\bibfnamefont {A.~T.}\ \bibnamefont
  {{Potter}}}, \bibinfo {author} {\bibfnamefont {S.~M.}\ \bibnamefont
  {{Chitre}}},\ and\ \bibinfo {author} {\bibfnamefont {C.~A.}\ \bibnamefont
  {{Tout}}},\ }\bibfield  {title} {\bibinfo {title} {{Stellar evolution of
  massive stars with a radiative {\ensuremath{\alpha}}-{\ensuremath{\Omega}}
  dynamo}},\ }\href {https://doi.org/10.1111/j.1365-2966.2012.21409.x}
  {\bibfield  {journal} {\bibinfo  {journal} {\mnras}\ }\textbf {\bibinfo
  {volume} {424}},\ \bibinfo {pages} {2358} (\bibinfo {year}
  {2012}{\natexlab{b}})},\ \Eprint {https://arxiv.org/abs/1205.6477}
  {arXiv:1205.6477 [astro-ph.SR]} \BibitemShut {NoStop}%
\bibitem [{\citenamefont {{Meynet}}\ \emph {et~al.}(2011)\citenamefont
  {{Meynet}}, \citenamefont {{Eggenberger}},\ and\ \citenamefont
  {{Maeder}}}]{2011A&A...525L..11M}%
  \BibitemOpen
  \bibfield  {author} {\bibinfo {author} {\bibfnamefont {G.}~\bibnamefont
  {{Meynet}}}, \bibinfo {author} {\bibfnamefont {P.}~\bibnamefont
  {{Eggenberger}}},\ and\ \bibinfo {author} {\bibfnamefont {A.}~\bibnamefont
  {{Maeder}}},\ }\bibfield  {title} {\bibinfo {title} {{Massive star models
  with magnetic braking}},\ }\href
  {https://doi.org/10.1051/0004-6361/201016017} {\bibfield  {journal} {\bibinfo
   {journal} {\aap}\ }\textbf {\bibinfo {volume} {525}},\ \bibinfo {eid} {L11}
  (\bibinfo {year} {2011})},\ \Eprint {https://arxiv.org/abs/1011.5795}
  {arXiv:1011.5795 [astro-ph.SR]} \BibitemShut {NoStop}%
\bibitem [{\citenamefont {{Amard}}\ \emph {et~al.}(2016)\citenamefont
  {{Amard}}, \citenamefont {{Palacios}}, \citenamefont {{Charbonnel}},
  \citenamefont {{Gallet}},\ and\ \citenamefont
  {{Bouvier}}}]{2016A&A...587A.105A}%
  \BibitemOpen
  \bibfield  {author} {\bibinfo {author} {\bibfnamefont {L.}~\bibnamefont
  {{Amard}}}, \bibinfo {author} {\bibfnamefont {A.}~\bibnamefont {{Palacios}}},
  \bibinfo {author} {\bibfnamefont {C.}~\bibnamefont {{Charbonnel}}}, \bibinfo
  {author} {\bibfnamefont {F.}~\bibnamefont {{Gallet}}},\ and\ \bibinfo
  {author} {\bibfnamefont {J.}~\bibnamefont {{Bouvier}}},\ }\bibfield  {title}
  {\bibinfo {title} {{Rotating models of young solar-type stars. Exploring
  braking laws and angular momentum transport processes}},\ }\href
  {https://doi.org/10.1051/0004-6361/20152734910.48550/arXiv.1601.01904}
  {\bibfield  {journal} {\bibinfo  {journal} {\aap}\ }\textbf {\bibinfo
  {volume} {587}},\ \bibinfo {eid} {A105} (\bibinfo {year} {2016})},\ \Eprint
  {https://arxiv.org/abs/1601.01904} {arXiv:1601.01904 [astro-ph.SR]}
  \BibitemShut {NoStop}%
\bibitem [{\citenamefont {{Georgy}}\ \emph {et~al.}(2017)\citenamefont
  {{Georgy}}, \citenamefont {{Meynet}}, \citenamefont {{Ekstr{\"o}m}},
  \citenamefont {{Wade}}, \citenamefont {{Petit}}, \citenamefont
  {{Keszthelyi}},\ and\ \citenamefont {{Hirschi}}}]{2017A&A...599L...5G}%
  \BibitemOpen
  \bibfield  {author} {\bibinfo {author} {\bibfnamefont {C.}~\bibnamefont
  {{Georgy}}}, \bibinfo {author} {\bibfnamefont {G.}~\bibnamefont {{Meynet}}},
  \bibinfo {author} {\bibfnamefont {S.}~\bibnamefont {{Ekstr{\"o}m}}}, \bibinfo
  {author} {\bibfnamefont {G.~A.}\ \bibnamefont {{Wade}}}, \bibinfo {author}
  {\bibfnamefont {V.}~\bibnamefont {{Petit}}}, \bibinfo {author} {\bibfnamefont
  {Z.}~\bibnamefont {{Keszthelyi}}},\ and\ \bibinfo {author} {\bibfnamefont
  {R.}~\bibnamefont {{Hirschi}}},\ }\bibfield  {title} {\bibinfo {title}
  {{Possible pair-instability supernovae at solar metallicity from magnetic
  stellar progenitors}},\ }\href {https://doi.org/10.1051/0004-6361/201730401}
  {\bibfield  {journal} {\bibinfo  {journal} {\aap}\ }\textbf {\bibinfo
  {volume} {599}},\ \bibinfo {eid} {L5} (\bibinfo {year} {2017})},\ \Eprint
  {https://arxiv.org/abs/1702.02340} {arXiv:1702.02340 [astro-ph.SR]}
  \BibitemShut {NoStop}%
\bibitem [{\citenamefont {{Keszthelyi}}\ \emph {et~al.}(2019)\citenamefont
  {{Keszthelyi}}, \citenamefont {{Meynet}}, \citenamefont {{Georgy}},
  \citenamefont {{Wade}}, \citenamefont {{Petit}},\ and\ \citenamefont
  {{David-Uraz}}}]{2019MNRAS.485.5843K}%
  \BibitemOpen
  \bibfield  {author} {\bibinfo {author} {\bibfnamefont {Z.}~\bibnamefont
  {{Keszthelyi}}}, \bibinfo {author} {\bibfnamefont {G.}~\bibnamefont
  {{Meynet}}}, \bibinfo {author} {\bibfnamefont {C.}~\bibnamefont {{Georgy}}},
  \bibinfo {author} {\bibfnamefont {G.~A.}\ \bibnamefont {{Wade}}}, \bibinfo
  {author} {\bibfnamefont {V.}~\bibnamefont {{Petit}}},\ and\ \bibinfo {author}
  {\bibfnamefont {A.}~\bibnamefont {{David-Uraz}}},\ }\bibfield  {title}
  {\bibinfo {title} {{The effects of surface fossil magnetic fields on massive
  star evolution: I. Magnetic field evolution, mass-loss quenching, and
  magnetic braking}},\ }\href {https://doi.org/10.1093/mnras/stz772} {\bibfield
   {journal} {\bibinfo  {journal} {\mnras}\ }\textbf {\bibinfo {volume}
  {485}},\ \bibinfo {pages} {5843} (\bibinfo {year} {2019})},\ \Eprint
  {https://arxiv.org/abs/1902.09333} {arXiv:1902.09333 [astro-ph.SR]}
  \BibitemShut {NoStop}%
\bibitem [{\citenamefont {{Keszthelyi}}\ \emph {et~al.}(2022)\citenamefont
  {{Keszthelyi}}, \citenamefont {{de Koter}}, \citenamefont {{G{\"o}tberg}},
  \citenamefont {{Meynet}}, \citenamefont {{Brands}}, \citenamefont {{Petit}},
  \citenamefont {{Carrington}}, \citenamefont {{David-Uraz}}, \citenamefont
  {{Geen}}, \citenamefont {{Georgy}}, \citenamefont {{Hirschi}}, \citenamefont
  {{Puls}}, \citenamefont {{Ramalatswa}}, \citenamefont {{Shultz}},\ and\
  \citenamefont {{ud-Doula}}}]{2022MNRAS.517.2028K}%
  \BibitemOpen
  \bibfield  {author} {\bibinfo {author} {\bibfnamefont {Z.}~\bibnamefont
  {{Keszthelyi}}}, \bibinfo {author} {\bibfnamefont {A.}~\bibnamefont {{de
  Koter}}}, \bibinfo {author} {\bibfnamefont {Y.}~\bibnamefont
  {{G{\"o}tberg}}}, \bibinfo {author} {\bibfnamefont {G.}~\bibnamefont
  {{Meynet}}}, \bibinfo {author} {\bibfnamefont {S.~A.}\ \bibnamefont
  {{Brands}}}, \bibinfo {author} {\bibfnamefont {V.}~\bibnamefont {{Petit}}},
  \bibinfo {author} {\bibfnamefont {M.}~\bibnamefont {{Carrington}}}, \bibinfo
  {author} {\bibfnamefont {A.}~\bibnamefont {{David-Uraz}}}, \bibinfo {author}
  {\bibfnamefont {S.~T.}\ \bibnamefont {{Geen}}}, \bibinfo {author}
  {\bibfnamefont {C.}~\bibnamefont {{Georgy}}}, \bibinfo {author}
  {\bibfnamefont {R.}~\bibnamefont {{Hirschi}}}, \bibinfo {author}
  {\bibfnamefont {J.}~\bibnamefont {{Puls}}}, \bibinfo {author} {\bibfnamefont
  {K.~J.}\ \bibnamefont {{Ramalatswa}}}, \bibinfo {author} {\bibfnamefont
  {M.~E.}\ \bibnamefont {{Shultz}}},\ and\ \bibinfo {author} {\bibfnamefont
  {A.}~\bibnamefont {{ud-Doula}}},\ }\bibfield  {title} {\bibinfo {title} {{The
  effects of surface fossil magnetic fields on massive star evolution: IV.
  Grids of models at Solar, LMC, and SMC metallicities}},\ }\href
  {https://doi.org/10.1093/mnras/stac2598} {\bibfield  {journal} {\bibinfo
  {journal} {\mnras}\ }\textbf {\bibinfo {volume} {517}},\ \bibinfo {pages}
  {2028} (\bibinfo {year} {2022})},\ \Eprint {https://arxiv.org/abs/2209.06350}
  {arXiv:2209.06350 [astro-ph.SR]} \BibitemShut {NoStop}%
\bibitem [{\citenamefont {{ud-Doula}}\ and\ \citenamefont
  {{Owocki}}(2002)}]{2002ApJ...576..413U}%
  \BibitemOpen
  \bibfield  {author} {\bibinfo {author} {\bibfnamefont {A.}~\bibnamefont
  {{ud-Doula}}}\ and\ \bibinfo {author} {\bibfnamefont {S.~P.}\ \bibnamefont
  {{Owocki}}},\ }\bibfield  {title} {\bibinfo {title} {{Dynamical Simulations
  of Magnetically Channeled Line-driven Stellar Winds. I. Isothermal,
  Nonrotating, Radially Driven Flow}},\ }\href {https://doi.org/10.1086/341543}
  {\bibfield  {journal} {\bibinfo  {journal} {\apj}\ }\textbf {\bibinfo
  {volume} {576}},\ \bibinfo {pages} {413} (\bibinfo {year} {2002})},\ \Eprint
  {https://arxiv.org/abs/astro-ph/0201195} {arXiv:astro-ph/0201195 [astro-ph]}
  \BibitemShut {NoStop}%
\bibitem [{\citenamefont {{Ud-Doula}}\ \emph {et~al.}(2008)\citenamefont
  {{Ud-Doula}}, \citenamefont {{Owocki}},\ and\ \citenamefont
  {{Townsend}}}]{2008MNRAS.385...97U}%
  \BibitemOpen
  \bibfield  {author} {\bibinfo {author} {\bibfnamefont {A.}~\bibnamefont
  {{Ud-Doula}}}, \bibinfo {author} {\bibfnamefont {S.~P.}\ \bibnamefont
  {{Owocki}}},\ and\ \bibinfo {author} {\bibfnamefont {R.~H.~D.}\ \bibnamefont
  {{Townsend}}},\ }\bibfield  {title} {\bibinfo {title} {{Dynamical simulations
  of magnetically channelled line-driven stellar winds - II. The effects of
  field-aligned rotation}},\ }\href
  {https://doi.org/10.1111/j.1365-2966.2008.12840.x} {\bibfield  {journal}
  {\bibinfo  {journal} {\mnras}\ }\textbf {\bibinfo {volume} {385}},\ \bibinfo
  {pages} {97} (\bibinfo {year} {2008})},\ \Eprint
  {https://arxiv.org/abs/0712.2780} {arXiv:0712.2780 [astro-ph]} \BibitemShut
  {NoStop}%
\bibitem [{\citenamefont {{Fuller}}\ \emph
  {et~al.}(2019{\natexlab{a}})\citenamefont {{Fuller}}, \citenamefont
  {{Piro}},\ and\ \citenamefont {{Jermyn}}}]{2019MNRAS.485.3661F}%
  \BibitemOpen
  \bibfield  {author} {\bibinfo {author} {\bibfnamefont {J.}~\bibnamefont
  {{Fuller}}}, \bibinfo {author} {\bibfnamefont {A.~L.}\ \bibnamefont
  {{Piro}}},\ and\ \bibinfo {author} {\bibfnamefont {A.~S.}\ \bibnamefont
  {{Jermyn}}},\ }\bibfield  {title} {\bibinfo {title} {{Slowing the spins of
  stellar cores}},\ }\href {https://doi.org/10.1093/mnras/stz514} {\bibfield
  {journal} {\bibinfo  {journal} {\mnras}\ }\textbf {\bibinfo {volume} {485}},\
  \bibinfo {pages} {3661} (\bibinfo {year} {2019}{\natexlab{a}})},\ \Eprint
  {https://arxiv.org/abs/1902.08227} {arXiv:1902.08227 [astro-ph.SR]}
  \BibitemShut {NoStop}%
\bibitem [{\citenamefont {{Fuller}}\ and\ \citenamefont
  {{Ma}}(2019)}]{2019ApJ...881L...1F}%
  \BibitemOpen
  \bibfield  {author} {\bibinfo {author} {\bibfnamefont {J.}~\bibnamefont
  {{Fuller}}}\ and\ \bibinfo {author} {\bibfnamefont {L.}~\bibnamefont
  {{Ma}}},\ }\bibfield  {title} {\bibinfo {title} {{Most Black Holes Are Born
  Very Slowly Rotating}},\ }\href {https://doi.org/10.3847/2041-8213/ab339b}
  {\bibfield  {journal} {\bibinfo  {journal} {\apjl}\ }\textbf {\bibinfo
  {volume} {881}},\ \bibinfo {eid} {L1} (\bibinfo {year} {2019})},\ \Eprint
  {https://arxiv.org/abs/1907.03714} {arXiv:1907.03714 [astro-ph.SR]}
  \BibitemShut {NoStop}%
\bibitem [{\citenamefont {{Eggenberger}}\ \emph {et~al.}(2019)\citenamefont
  {{Eggenberger}}, \citenamefont {{Buldgen}},\ and\ \citenamefont
  {{Salmon}}}]{2019A&A...626L...1E}%
  \BibitemOpen
  \bibfield  {author} {\bibinfo {author} {\bibfnamefont {P.}~\bibnamefont
  {{Eggenberger}}}, \bibinfo {author} {\bibfnamefont {G.}~\bibnamefont
  {{Buldgen}}},\ and\ \bibinfo {author} {\bibfnamefont {S.~J.~A.~J.}\
  \bibnamefont {{Salmon}}},\ }\bibfield  {title} {\bibinfo {title} {{Rotation
  rate of the solar core as a key constraint to magnetic angular momentum
  transport in stellar interiors}},\ }\href
  {https://doi.org/10.1051/0004-6361/201935509} {\bibfield  {journal} {\bibinfo
   {journal} {\aap}\ }\textbf {\bibinfo {volume} {626}},\ \bibinfo {eid} {L1}
  (\bibinfo {year} {2019})},\ \Eprint {https://arxiv.org/abs/1911.06343}
  {arXiv:1911.06343 [astro-ph.SR]} \BibitemShut {NoStop}%
\bibitem [{\citenamefont {{Eggenberger}}\ \emph {et~al.}(2022)\citenamefont
  {{Eggenberger}}, \citenamefont {{Moyano}},\ and\ \citenamefont {{den
  Hartogh}}}]{2022A&A...664L..16E}%
  \BibitemOpen
  \bibfield  {author} {\bibinfo {author} {\bibfnamefont {P.}~\bibnamefont
  {{Eggenberger}}}, \bibinfo {author} {\bibfnamefont {F.~D.}\ \bibnamefont
  {{Moyano}}},\ and\ \bibinfo {author} {\bibfnamefont {J.~W.}\ \bibnamefont
  {{den Hartogh}}},\ }\bibfield  {title} {\bibinfo {title} {{Rotation in
  stellar interiors: General formulation and an asteroseismic-calibrated
  transport by the Tayler instability}},\ }\href
  {https://doi.org/10.1051/0004-6361/202243781} {\bibfield  {journal} {\bibinfo
   {journal} {\aap}\ }\textbf {\bibinfo {volume} {664}},\ \bibinfo {eid} {L16}
  (\bibinfo {year} {2022})}\BibitemShut {NoStop}%
\bibitem [{\citenamefont {{Griffiths}}\ \emph {et~al.}(2022)\citenamefont
  {{Griffiths}}, \citenamefont {{Eggenberger}}, \citenamefont {{Meynet}},
  \citenamefont {{Moyano}},\ and\ \citenamefont
  {{Aloy}}}]{2022A&A...665A.147G}%
  \BibitemOpen
  \bibfield  {author} {\bibinfo {author} {\bibfnamefont {A.}~\bibnamefont
  {{Griffiths}}}, \bibinfo {author} {\bibfnamefont {P.}~\bibnamefont
  {{Eggenberger}}}, \bibinfo {author} {\bibfnamefont {G.}~\bibnamefont
  {{Meynet}}}, \bibinfo {author} {\bibfnamefont {F.}~\bibnamefont {{Moyano}}},\
  and\ \bibinfo {author} {\bibfnamefont {M.-{\'A}.}\ \bibnamefont {{Aloy}}},\
  }\bibfield  {title} {\bibinfo {title} {{The magneto-rotational instability in
  massive stars}},\ }\href {https://doi.org/10.1051/0004-6361/202243599}
  {\bibfield  {journal} {\bibinfo  {journal} {\aap}\ }\textbf {\bibinfo
  {volume} {665}},\ \bibinfo {eid} {A147} (\bibinfo {year} {2022})},\ \Eprint
  {https://arxiv.org/abs/2204.00016} {arXiv:2204.00016 [astro-ph.SR]}
  \BibitemShut {NoStop}%
\bibitem [{\citenamefont {{Janka}}(2012)}]{2012ARNPS..62..407J}%
  \BibitemOpen
  \bibfield  {author} {\bibinfo {author} {\bibfnamefont {H.-T.}\ \bibnamefont
  {{Janka}}},\ }\bibfield  {title} {\bibinfo {title} {{Explosion Mechanisms of
  Core-Collapse Supernovae}},\ }\href
  {https://doi.org/10.1146/annurev-nucl-102711-09490110.48550/arXiv.1206.2503}
  {\bibfield  {journal} {\bibinfo  {journal} {Annual Review of Nuclear and
  Particle Science}\ }\textbf {\bibinfo {volume} {62}},\ \bibinfo {pages} {407}
  (\bibinfo {year} {2012})},\ \Eprint {https://arxiv.org/abs/1206.2503}
  {arXiv:1206.2503 [astro-ph.SR]} \BibitemShut {NoStop}%
\bibitem [{\citenamefont {{Foglizzo}}\ \emph {et~al.}(2015)\citenamefont
  {{Foglizzo}}, \citenamefont {{Kazeroni}}, \citenamefont {{Guilet}},
  \citenamefont {{Masset}}, \citenamefont {{Gonz{\'a}lez}}, \citenamefont
  {{Krueger}}, \citenamefont {{Novak}}, \citenamefont {{Oertel}}, \citenamefont
  {{Margueron}}, \citenamefont {{Faure}}, \citenamefont {{Martin}},
  \citenamefont {{Blottiau}}, \citenamefont {{Peres}},\ and\ \citenamefont
  {{Durand}}}]{2015PASA...32....9F}%
  \BibitemOpen
  \bibfield  {author} {\bibinfo {author} {\bibfnamefont {T.}~\bibnamefont
  {{Foglizzo}}}, \bibinfo {author} {\bibfnamefont {R.}~\bibnamefont
  {{Kazeroni}}}, \bibinfo {author} {\bibfnamefont {J.}~\bibnamefont
  {{Guilet}}}, \bibinfo {author} {\bibfnamefont {F.}~\bibnamefont {{Masset}}},
  \bibinfo {author} {\bibfnamefont {M.}~\bibnamefont {{Gonz{\'a}lez}}},
  \bibinfo {author} {\bibfnamefont {B.~K.}\ \bibnamefont {{Krueger}}}, \bibinfo
  {author} {\bibfnamefont {J.}~\bibnamefont {{Novak}}}, \bibinfo {author}
  {\bibfnamefont {M.}~\bibnamefont {{Oertel}}}, \bibinfo {author}
  {\bibfnamefont {J.}~\bibnamefont {{Margueron}}}, \bibinfo {author}
  {\bibfnamefont {J.}~\bibnamefont {{Faure}}}, \bibinfo {author} {\bibfnamefont
  {N.}~\bibnamefont {{Martin}}}, \bibinfo {author} {\bibfnamefont
  {P.}~\bibnamefont {{Blottiau}}}, \bibinfo {author} {\bibfnamefont
  {B.}~\bibnamefont {{Peres}}},\ and\ \bibinfo {author} {\bibfnamefont
  {G.}~\bibnamefont {{Durand}}},\ }\bibfield  {title} {\bibinfo {title} {{The
  Explosion Mechanism of Core-Collapse Supernovae: Progress in Supernova Theory
  and Experiments}},\ }\href
  {https://doi.org/10.1017/pasa.2015.910.48550/arXiv.1501.01334} {\bibfield
  {journal} {\bibinfo  {journal} {\pasa}\ }\textbf {\bibinfo {volume} {32}},\
  \bibinfo {eid} {e009} (\bibinfo {year} {2015})},\ \Eprint
  {https://arxiv.org/abs/1501.01334} {arXiv:1501.01334 [astro-ph.HE]}
  \BibitemShut {NoStop}%
\bibitem [{\citenamefont {{Janka}}\ \emph {et~al.}(2016)\citenamefont
  {{Janka}}, \citenamefont {{Melson}},\ and\ \citenamefont
  {{Summa}}}]{2016ARNPS..66..341J}%
  \BibitemOpen
  \bibfield  {author} {\bibinfo {author} {\bibfnamefont {H.-T.}\ \bibnamefont
  {{Janka}}}, \bibinfo {author} {\bibfnamefont {T.}~\bibnamefont {{Melson}}},\
  and\ \bibinfo {author} {\bibfnamefont {A.}~\bibnamefont {{Summa}}},\
  }\bibfield  {title} {\bibinfo {title} {{Physics of Core-Collapse Supernovae
  in Three Dimensions: A Sneak Preview}},\ }\href
  {https://doi.org/10.1146/annurev-nucl-102115-04474710.48550/arXiv.1602.05576}
  {\bibfield  {journal} {\bibinfo  {journal} {Annual Review of Nuclear and
  Particle Science}\ }\textbf {\bibinfo {volume} {66}},\ \bibinfo {pages} {341}
  (\bibinfo {year} {2016})},\ \Eprint {https://arxiv.org/abs/1602.05576}
  {arXiv:1602.05576 [astro-ph.SR]} \BibitemShut {NoStop}%
\bibitem [{\citenamefont {{Mezzacappa}}\ \emph {et~al.}(2020)\citenamefont
  {{Mezzacappa}}, \citenamefont {{Endeve}}, \citenamefont {{Messer}},\ and\
  \citenamefont {{Bruenn}}}]{2020LRCA....6....4M}%
  \BibitemOpen
  \bibfield  {author} {\bibinfo {author} {\bibfnamefont {A.}~\bibnamefont
  {{Mezzacappa}}}, \bibinfo {author} {\bibfnamefont {E.}~\bibnamefont
  {{Endeve}}}, \bibinfo {author} {\bibfnamefont {O.~E.~B.}\ \bibnamefont
  {{Messer}}},\ and\ \bibinfo {author} {\bibfnamefont {S.~W.}\ \bibnamefont
  {{Bruenn}}},\ }\bibfield  {title} {\bibinfo {title} {{Physical, numerical,
  and computational challenges of modeling neutrino transport in core-collapse
  supernovae}},\ }\href
  {https://doi.org/10.1007/s41115-020-00010-810.48550/arXiv.2010.09013}
  {\bibfield  {journal} {\bibinfo  {journal} {Living Reviews in Computational
  Astrophysics}\ }\textbf {\bibinfo {volume} {6}},\ \bibinfo {eid} {4}
  (\bibinfo {year} {2020})},\ \Eprint {https://arxiv.org/abs/2010.09013}
  {arXiv:2010.09013 [astro-ph.HE]} \BibitemShut {NoStop}%
\bibitem [{\citenamefont {{Mezzacappa}}(2023)}]{2023IAUS..362..215M}%
  \BibitemOpen
  \bibfield  {author} {\bibinfo {author} {\bibfnamefont {A.}~\bibnamefont
  {{Mezzacappa}}},\ }\bibfield  {title} {\bibinfo {title} {{Toward Realistic
  Models of Core Collapse Supernovae: A Brief Review}},\ }\href
  {https://doi.org/10.1017/S174392132200183110.48550/arXiv.2205.13438}
  {\bibfield  {journal} {\bibinfo  {journal} {IAU Symposium}\ }\textbf
  {\bibinfo {volume} {362}},\ \bibinfo {pages} {215} (\bibinfo {year}
  {2023})},\ \Eprint {https://arxiv.org/abs/2205.13438} {arXiv:2205.13438
  [astro-ph.SR]} \BibitemShut {NoStop}%
\bibitem [{\citenamefont {{Ugliano}}\ \emph {et~al.}(2012)\citenamefont
  {{Ugliano}}, \citenamefont {{Janka}}, \citenamefont {{Marek}},\ and\
  \citenamefont {{Arcones}}}]{2012ApJ...757...69U}%
  \BibitemOpen
  \bibfield  {author} {\bibinfo {author} {\bibfnamefont {M.}~\bibnamefont
  {{Ugliano}}}, \bibinfo {author} {\bibfnamefont {H.-T.}\ \bibnamefont
  {{Janka}}}, \bibinfo {author} {\bibfnamefont {A.}~\bibnamefont {{Marek}}},\
  and\ \bibinfo {author} {\bibfnamefont {A.}~\bibnamefont {{Arcones}}},\
  }\bibfield  {title} {\bibinfo {title} {{Progenitor-explosion Connection and
  Remnant Birth Masses for Neutrino-driven Supernovae of Iron-core
  Progenitors}},\ }\href {https://doi.org/10.1088/0004-637X/757/1/69}
  {\bibfield  {journal} {\bibinfo  {journal} {\apj}\ }\textbf {\bibinfo
  {volume} {757}},\ \bibinfo {eid} {69} (\bibinfo {year} {2012})},\ \Eprint
  {https://arxiv.org/abs/1205.3657} {arXiv:1205.3657 [astro-ph.SR]}
  \BibitemShut {NoStop}%
\bibitem [{\citenamefont {{Ertl}}\ \emph {et~al.}(2016)\citenamefont {{Ertl}},
  \citenamefont {{Janka}}, \citenamefont {{Woosley}}, \citenamefont
  {{Sukhbold}},\ and\ \citenamefont {{Ugliano}}}]{2016ApJ...818..124E}%
  \BibitemOpen
  \bibfield  {author} {\bibinfo {author} {\bibfnamefont {T.}~\bibnamefont
  {{Ertl}}}, \bibinfo {author} {\bibfnamefont {H.~T.}\ \bibnamefont {{Janka}}},
  \bibinfo {author} {\bibfnamefont {S.~E.}\ \bibnamefont {{Woosley}}}, \bibinfo
  {author} {\bibfnamefont {T.}~\bibnamefont {{Sukhbold}}},\ and\ \bibinfo
  {author} {\bibfnamefont {M.}~\bibnamefont {{Ugliano}}},\ }\bibfield  {title}
  {\bibinfo {title} {{A Two-parameter Criterion for Classifying the
  Explodability of Massive Stars by the Neutrino-driven Mechanism}},\ }\href
  {https://doi.org/10.3847/0004-637X/818/2/124} {\bibfield  {journal} {\bibinfo
   {journal} {\apj}\ }\textbf {\bibinfo {volume} {818}},\ \bibinfo {eid} {124}
  (\bibinfo {year} {2016})},\ \Eprint {https://arxiv.org/abs/1503.07522}
  {arXiv:1503.07522 [astro-ph.SR]} \BibitemShut {NoStop}%
\bibitem [{\citenamefont {{M{\"u}ller}}\ \emph {et~al.}(2016)\citenamefont
  {{M{\"u}ller}}, \citenamefont {{Heger}}, \citenamefont {{Liptai}},\ and\
  \citenamefont {{Cameron}}}]{2016MNRAS.460..742M}%
  \BibitemOpen
  \bibfield  {author} {\bibinfo {author} {\bibfnamefont {B.}~\bibnamefont
  {{M{\"u}ller}}}, \bibinfo {author} {\bibfnamefont {A.}~\bibnamefont
  {{Heger}}}, \bibinfo {author} {\bibfnamefont {D.}~\bibnamefont {{Liptai}}},\
  and\ \bibinfo {author} {\bibfnamefont {J.~B.}\ \bibnamefont {{Cameron}}},\
  }\bibfield  {title} {\bibinfo {title} {{A simple approach to the supernova
  progenitor-explosion connection}},\ }\href
  {https://doi.org/10.1093/mnras/stw1083} {\bibfield  {journal} {\bibinfo
  {journal} {\mnras}\ }\textbf {\bibinfo {volume} {460}},\ \bibinfo {pages}
  {742} (\bibinfo {year} {2016})},\ \Eprint {https://arxiv.org/abs/1602.05956}
  {arXiv:1602.05956 [astro-ph.SR]} \BibitemShut {NoStop}%
\bibitem [{\citenamefont {{Sukhbold}}\ \emph {et~al.}(2018)\citenamefont
  {{Sukhbold}}, \citenamefont {{Woosley}},\ and\ \citenamefont
  {{Heger}}}]{2018ApJ...860...93S}%
  \BibitemOpen
  \bibfield  {author} {\bibinfo {author} {\bibfnamefont {T.}~\bibnamefont
  {{Sukhbold}}}, \bibinfo {author} {\bibfnamefont {S.~E.}\ \bibnamefont
  {{Woosley}}},\ and\ \bibinfo {author} {\bibfnamefont {A.}~\bibnamefont
  {{Heger}}},\ }\bibfield  {title} {\bibinfo {title} {{A High-resolution Study
  of Presupernova Core Structure}},\ }\href
  {https://doi.org/10.3847/1538-4357/aac2da} {\bibfield  {journal} {\bibinfo
  {journal} {\apj}\ }\textbf {\bibinfo {volume} {860}},\ \bibinfo {eid} {93}
  (\bibinfo {year} {2018})},\ \Eprint {https://arxiv.org/abs/1710.03243}
  {arXiv:1710.03243 [astro-ph.HE]} \BibitemShut {NoStop}%
\bibitem [{\citenamefont {{Chieffi}}\ and\ \citenamefont
  {{Limongi}}(2020)}]{2020ApJ...890...43C}%
  \BibitemOpen
  \bibfield  {author} {\bibinfo {author} {\bibfnamefont {A.}~\bibnamefont
  {{Chieffi}}}\ and\ \bibinfo {author} {\bibfnamefont {M.}~\bibnamefont
  {{Limongi}}},\ }\bibfield  {title} {\bibinfo {title} {{The Presupernova Core
  Mass-Radius Relation of Massive Stars: Understanding Its Formation and
  Evolution}},\ }\href {https://doi.org/10.3847/1538-4357/ab6739} {\bibfield
  {journal} {\bibinfo  {journal} {\apj}\ }\textbf {\bibinfo {volume} {890}},\
  \bibinfo {eid} {43} (\bibinfo {year} {2020})},\ \Eprint
  {https://arxiv.org/abs/1911.08988} {arXiv:1911.08988 [astro-ph.SR]}
  \BibitemShut {NoStop}%
\bibitem [{\citenamefont {{Chandrasekhar}}(1931)}]{1931MNRAS..91..456C}%
  \BibitemOpen
  \bibfield  {author} {\bibinfo {author} {\bibfnamefont {S.}~\bibnamefont
  {{Chandrasekhar}}},\ }\bibfield  {title} {\bibinfo {title} {{The highly
  collapsed configurations of a stellar mass}},\ }\href
  {https://doi.org/10.1093/mnras/91.5.456} {\bibfield  {journal} {\bibinfo
  {journal} {\mnras}\ }\textbf {\bibinfo {volume} {91}},\ \bibinfo {pages}
  {456} (\bibinfo {year} {1931})}\BibitemShut {NoStop}%
\bibitem [{\citenamefont {{Colgate}}\ and\ \citenamefont
  {{White}}(1966)}]{1966ApJ...143..626C}%
  \BibitemOpen
  \bibfield  {author} {\bibinfo {author} {\bibfnamefont {S.~A.}\ \bibnamefont
  {{Colgate}}}\ and\ \bibinfo {author} {\bibfnamefont {R.~H.}\ \bibnamefont
  {{White}}},\ }\bibfield  {title} {\bibinfo {title} {{The Hydrodynamic
  Behavior of Supernovae Explosions}},\ }\href {https://doi.org/10.1086/148549}
  {\bibfield  {journal} {\bibinfo  {journal} {\apj}\ }\textbf {\bibinfo
  {volume} {143}},\ \bibinfo {pages} {626} (\bibinfo {year}
  {1966})}\BibitemShut {NoStop}%
\bibitem [{\citenamefont {{Bethe}}\ and\ \citenamefont
  {{Wilson}}(1985)}]{1985ApJ...295...14B}%
  \BibitemOpen
  \bibfield  {author} {\bibinfo {author} {\bibfnamefont {H.~A.}\ \bibnamefont
  {{Bethe}}}\ and\ \bibinfo {author} {\bibfnamefont {J.~R.}\ \bibnamefont
  {{Wilson}}},\ }\bibfield  {title} {\bibinfo {title} {{Revival of a stalled
  supernova shock by neutrino heating}},\ }\href
  {https://doi.org/10.1086/163343} {\bibfield  {journal} {\bibinfo  {journal}
  {\apj}\ }\textbf {\bibinfo {volume} {295}},\ \bibinfo {pages} {14} (\bibinfo
  {year} {1985})}\BibitemShut {NoStop}%
\bibitem [{\citenamefont {{Janka}}\ and\ \citenamefont
  {{Mueller}}(1996)}]{1996A&A...306..167J}%
  \BibitemOpen
  \bibfield  {author} {\bibinfo {author} {\bibfnamefont {H.~T.}\ \bibnamefont
  {{Janka}}}\ and\ \bibinfo {author} {\bibfnamefont {E.}~\bibnamefont
  {{Mueller}}},\ }\bibfield  {title} {\bibinfo {title} {{Neutrino heating,
  convection, and the mechanism of Type-II supernova explosions.}},\
  }\href@noop {} {\bibfield  {journal} {\bibinfo  {journal} {\aap}\ }\textbf
  {\bibinfo {volume} {306}},\ \bibinfo {pages} {167} (\bibinfo {year}
  {1996})}\BibitemShut {NoStop}%
\bibitem [{\citenamefont {{Wongwathanarat}}\ \emph {et~al.}(2013)\citenamefont
  {{Wongwathanarat}}, \citenamefont {{Janka}},\ and\ \citenamefont
  {{M{\"u}ller}}}]{2013A&A...552A.126W}%
  \BibitemOpen
  \bibfield  {author} {\bibinfo {author} {\bibfnamefont {A.}~\bibnamefont
  {{Wongwathanarat}}}, \bibinfo {author} {\bibfnamefont {H.~T.}\ \bibnamefont
  {{Janka}}},\ and\ \bibinfo {author} {\bibfnamefont {E.}~\bibnamefont
  {{M{\"u}ller}}},\ }\bibfield  {title} {\bibinfo {title} {{Three-dimensional
  neutrino-driven supernovae: Neutron star kicks, spins, and asymmetric
  ejection of nucleosynthesis products}},\ }\href
  {https://doi.org/10.1051/0004-6361/201220636} {\bibfield  {journal} {\bibinfo
   {journal} {\aap}\ }\textbf {\bibinfo {volume} {552}},\ \bibinfo {eid} {A126}
  (\bibinfo {year} {2013})},\ \Eprint {https://arxiv.org/abs/1210.8148}
  {arXiv:1210.8148 [astro-ph.HE]} \BibitemShut {NoStop}%
\bibitem [{\citenamefont {{Repetto}}\ \emph {et~al.}(2012)\citenamefont
  {{Repetto}}, \citenamefont {{Davies}},\ and\ \citenamefont
  {{Sigurdsson}}}]{2012MNRAS.425.2799R}%
  \BibitemOpen
  \bibfield  {author} {\bibinfo {author} {\bibfnamefont {S.}~\bibnamefont
  {{Repetto}}}, \bibinfo {author} {\bibfnamefont {M.~B.}\ \bibnamefont
  {{Davies}}},\ and\ \bibinfo {author} {\bibfnamefont {S.}~\bibnamefont
  {{Sigurdsson}}},\ }\bibfield  {title} {\bibinfo {title} {{Investigating
  stellar-mass black hole kicks}},\ }\href
  {https://doi.org/10.1111/j.1365-2966.2012.21549.x} {\bibfield  {journal}
  {\bibinfo  {journal} {\mnras}\ }\textbf {\bibinfo {volume} {425}},\ \bibinfo
  {pages} {2799} (\bibinfo {year} {2012})},\ \Eprint
  {https://arxiv.org/abs/1203.3077} {arXiv:1203.3077 [astro-ph.GA]}
  \BibitemShut {NoStop}%
\bibitem [{\citenamefont {{Giacobbo}}\ and\ \citenamefont
  {{Mapelli}}(2019)}]{2019MNRAS.482.2234G}%
  \BibitemOpen
  \bibfield  {author} {\bibinfo {author} {\bibfnamefont {N.}~\bibnamefont
  {{Giacobbo}}}\ and\ \bibinfo {author} {\bibfnamefont {M.}~\bibnamefont
  {{Mapelli}}},\ }\bibfield  {title} {\bibinfo {title} {{The impact of
  electron-capture supernovae on merging double neutron stars}},\ }\href
  {https://doi.org/10.1093/mnras/sty2848} {\bibfield  {journal} {\bibinfo
  {journal} {\mnras}\ }\textbf {\bibinfo {volume} {482}},\ \bibinfo {pages}
  {2234} (\bibinfo {year} {2019})},\ \Eprint {https://arxiv.org/abs/1805.11100}
  {arXiv:1805.11100 [astro-ph.SR]} \BibitemShut {NoStop}%
\bibitem [{\citenamefont {{Giacobbo}}\ and\ \citenamefont
  {{Mapelli}}(2020)}]{2020ApJ...891..141G}%
  \BibitemOpen
  \bibfield  {author} {\bibinfo {author} {\bibfnamefont {N.}~\bibnamefont
  {{Giacobbo}}}\ and\ \bibinfo {author} {\bibfnamefont {M.}~\bibnamefont
  {{Mapelli}}},\ }\bibfield  {title} {\bibinfo {title} {{Revising Natal Kick
  Prescriptions in Population Synthesis Simulations}},\ }\href
  {https://doi.org/10.3847/1538-4357/ab7335} {\bibfield  {journal} {\bibinfo
  {journal} {\apj}\ }\textbf {\bibinfo {volume} {891}},\ \bibinfo {eid} {141}
  (\bibinfo {year} {2020})},\ \Eprint {https://arxiv.org/abs/1909.06385}
  {arXiv:1909.06385 [astro-ph.HE]} \BibitemShut {NoStop}%
\bibitem [{\citenamefont {{Mandel}}\ and\ \citenamefont
  {{M{\"u}ller}}(2020)}]{2020MNRAS.499.3214M}%
  \BibitemOpen
  \bibfield  {author} {\bibinfo {author} {\bibfnamefont {I.}~\bibnamefont
  {{Mandel}}}\ and\ \bibinfo {author} {\bibfnamefont {B.}~\bibnamefont
  {{M{\"u}ller}}},\ }\bibfield  {title} {\bibinfo {title} {{Simple recipes for
  compact remnant masses and natal kicks}},\ }\href
  {https://doi.org/10.1093/mnras/staa3043} {\bibfield  {journal} {\bibinfo
  {journal} {\mnras}\ }\textbf {\bibinfo {volume} {499}},\ \bibinfo {pages}
  {3214} (\bibinfo {year} {2020})},\ \Eprint {https://arxiv.org/abs/2006.08360}
  {arXiv:2006.08360 [astro-ph.HE]} \BibitemShut {NoStop}%
\bibitem [{\citenamefont {{Callister}}\ \emph {et~al.}(2021)\citenamefont
  {{Callister}}, \citenamefont {{Farr}},\ and\ \citenamefont
  {{Renzo}}}]{2021ApJ...920..157C}%
  \BibitemOpen
  \bibfield  {author} {\bibinfo {author} {\bibfnamefont {T.~A.}\ \bibnamefont
  {{Callister}}}, \bibinfo {author} {\bibfnamefont {W.~M.}\ \bibnamefont
  {{Farr}}},\ and\ \bibinfo {author} {\bibfnamefont {M.}~\bibnamefont
  {{Renzo}}},\ }\bibfield  {title} {\bibinfo {title} {{State of the Field:
  Binary Black Hole Natal Kicks and Prospects for Isolated Field Formation
  after GWTC-2}},\ }\href {https://doi.org/10.3847/1538-4357/ac1347} {\bibfield
   {journal} {\bibinfo  {journal} {\apj}\ }\textbf {\bibinfo {volume} {920}},\
  \bibinfo {eid} {157} (\bibinfo {year} {2021})},\ \Eprint
  {https://arxiv.org/abs/2011.09570} {arXiv:2011.09570 [astro-ph.HE]}
  \BibitemShut {NoStop}%
\bibitem [{\citenamefont {{M{\"u}ller}}(2016)}]{2016PASA...33...48M}%
  \BibitemOpen
  \bibfield  {author} {\bibinfo {author} {\bibfnamefont {B.}~\bibnamefont
  {{M{\"u}ller}}},\ }\bibfield  {title} {\bibinfo {title} {{The Status of
  Multi-Dimensional Core-Collapse Supernova Models}},\ }\href
  {https://doi.org/10.1017/pasa.2016.40} {\bibfield  {journal} {\bibinfo
  {journal} {\pasa}\ }\textbf {\bibinfo {volume} {33}},\ \bibinfo {eid} {e048}
  (\bibinfo {year} {2016})},\ \Eprint {https://arxiv.org/abs/1608.03274}
  {arXiv:1608.03274 [astro-ph.SR]} \BibitemShut {NoStop}%
\bibitem [{\citenamefont {{Woosley}}\ and\ \citenamefont
  {{Weaver}}(1995)}]{1995ApJS..101..181W}%
  \BibitemOpen
  \bibfield  {author} {\bibinfo {author} {\bibfnamefont {S.~E.}\ \bibnamefont
  {{Woosley}}}\ and\ \bibinfo {author} {\bibfnamefont {T.~A.}\ \bibnamefont
  {{Weaver}}},\ }\bibfield  {title} {\bibinfo {title} {{The Evolution and
  Explosion of Massive Stars. II. Explosive Hydrodynamics and
  Nucleosynthesis}},\ }\href {https://doi.org/10.1086/192237} {\bibfield
  {journal} {\bibinfo  {journal} {\apjs}\ }\textbf {\bibinfo {volume} {101}},\
  \bibinfo {pages} {181} (\bibinfo {year} {1995})}\BibitemShut {NoStop}%
\bibitem [{\citenamefont {{Zhang}}\ \emph {et~al.}(2008)\citenamefont
  {{Zhang}}, \citenamefont {{Woosley}},\ and\ \citenamefont
  {{Heger}}}]{2008ApJ...679..639Z}%
  \BibitemOpen
  \bibfield  {author} {\bibinfo {author} {\bibfnamefont {W.}~\bibnamefont
  {{Zhang}}}, \bibinfo {author} {\bibfnamefont {S.~E.}\ \bibnamefont
  {{Woosley}}},\ and\ \bibinfo {author} {\bibfnamefont {A.}~\bibnamefont
  {{Heger}}},\ }\bibfield  {title} {\bibinfo {title} {{Fallback and Black Hole
  Production in Massive Stars}},\ }\href {https://doi.org/10.1086/526404}
  {\bibfield  {journal} {\bibinfo  {journal} {\apj}\ }\textbf {\bibinfo
  {volume} {679}},\ \bibinfo {pages} {639} (\bibinfo {year} {2008})},\ \Eprint
  {https://arxiv.org/abs/astro-ph/0701083} {arXiv:astro-ph/0701083 [astro-ph]}
  \BibitemShut {NoStop}%
\bibitem [{\citenamefont {{Aufderheide}}\ \emph {et~al.}(1991)\citenamefont
  {{Aufderheide}}, \citenamefont {{Baron}},\ and\ \citenamefont
  {{Thielemann}}}]{1991ApJ...370..630A}%
  \BibitemOpen
  \bibfield  {author} {\bibinfo {author} {\bibfnamefont {M.~B.}\ \bibnamefont
  {{Aufderheide}}}, \bibinfo {author} {\bibfnamefont {E.}~\bibnamefont
  {{Baron}}},\ and\ \bibinfo {author} {\bibfnamefont {F.~K.}\ \bibnamefont
  {{Thielemann}}},\ }\bibfield  {title} {\bibinfo {title} {{Shock Waves and
  Nucleosynthesis in Type II Supernovae}},\ }\href
  {https://doi.org/10.1086/169849} {\bibfield  {journal} {\bibinfo  {journal}
  {\apj}\ }\textbf {\bibinfo {volume} {370}},\ \bibinfo {pages} {630} (\bibinfo
  {year} {1991})}\BibitemShut {NoStop}%
\bibitem [{\citenamefont {{Thielemann}}\ \emph {et~al.}(1996)\citenamefont
  {{Thielemann}}, \citenamefont {{Nomoto}},\ and\ \citenamefont
  {{Hashimoto}}}]{1996ApJ...460..408T}%
  \BibitemOpen
  \bibfield  {author} {\bibinfo {author} {\bibfnamefont {F.-K.}\ \bibnamefont
  {{Thielemann}}}, \bibinfo {author} {\bibfnamefont {K.}~\bibnamefont
  {{Nomoto}}},\ and\ \bibinfo {author} {\bibfnamefont {M.-A.}\ \bibnamefont
  {{Hashimoto}}},\ }\bibfield  {title} {\bibinfo {title} {{Core-Collapse
  Supernovae and Their Ejecta}},\ }\href {https://doi.org/10.1086/176980}
  {\bibfield  {journal} {\bibinfo  {journal} {\apj}\ }\textbf {\bibinfo
  {volume} {460}},\ \bibinfo {pages} {408} (\bibinfo {year}
  {1996})}\BibitemShut {NoStop}%
\bibitem [{\citenamefont {{Limongi}}\ and\ \citenamefont
  {{Chieffi}}(2003)}]{2003ApJ...592..404L}%
  \BibitemOpen
  \bibfield  {author} {\bibinfo {author} {\bibfnamefont {M.}~\bibnamefont
  {{Limongi}}}\ and\ \bibinfo {author} {\bibfnamefont {A.}~\bibnamefont
  {{Chieffi}}},\ }\bibfield  {title} {\bibinfo {title} {{Evolution, Explosion,
  and Nucleosynthesis of Core-Collapse Supernovae}},\ }\href
  {https://doi.org/10.1086/375703} {\bibfield  {journal} {\bibinfo  {journal}
  {\apj}\ }\textbf {\bibinfo {volume} {592}},\ \bibinfo {pages} {404} (\bibinfo
  {year} {2003})},\ \Eprint {https://arxiv.org/abs/astro-ph/0304185}
  {arXiv:astro-ph/0304185 [astro-ph]} \BibitemShut {NoStop}%
\bibitem [{\citenamefont {{Limongi}}(2017)}]{2017hsn..book..513L}%
  \BibitemOpen
  \bibfield  {author} {\bibinfo {author} {\bibfnamefont {M.}~\bibnamefont
  {{Limongi}}},\ }\bibfield  {title} {\bibinfo {title} {{Supernovae from
  Massive Stars}},\ }in\ \href {https://doi.org/10.1007/978-3-319-21846-5_119}
  {\emph {\bibinfo {booktitle} {Handbook of Supernovae}}},\ \bibinfo {editor}
  {edited by\ \bibinfo {editor} {\bibfnamefont {A.~W.}\ \bibnamefont
  {{Alsabti}}}\ and\ \bibinfo {editor} {\bibfnamefont {P.}~\bibnamefont
  {{Murdin}}}}\ (\bibinfo {year} {2017})\ p.\ \bibinfo {pages}
  {513}\BibitemShut {NoStop}%
\bibitem [{\citenamefont {{Perego}}\ \emph {et~al.}(2015)\citenamefont
  {{Perego}}, \citenamefont {{Hempel}}, \citenamefont {{Fr{\"o}hlich}},
  \citenamefont {{Ebinger}}, \citenamefont {{Eichler}}, \citenamefont
  {{Casanova}}, \citenamefont {{Liebend{\"o}rfer}},\ and\ \citenamefont
  {{Thielemann}}}]{2015ApJ...806..275P}%
  \BibitemOpen
  \bibfield  {author} {\bibinfo {author} {\bibfnamefont {A.}~\bibnamefont
  {{Perego}}}, \bibinfo {author} {\bibfnamefont {M.}~\bibnamefont {{Hempel}}},
  \bibinfo {author} {\bibfnamefont {C.}~\bibnamefont {{Fr{\"o}hlich}}},
  \bibinfo {author} {\bibfnamefont {K.}~\bibnamefont {{Ebinger}}}, \bibinfo
  {author} {\bibfnamefont {M.}~\bibnamefont {{Eichler}}}, \bibinfo {author}
  {\bibfnamefont {J.}~\bibnamefont {{Casanova}}}, \bibinfo {author}
  {\bibfnamefont {M.}~\bibnamefont {{Liebend{\"o}rfer}}},\ and\ \bibinfo
  {author} {\bibfnamefont {F.~K.}\ \bibnamefont {{Thielemann}}},\ }\bibfield
  {title} {\bibinfo {title} {{PUSHing Core-collapse Supernovae to Explosions in
  Spherical Symmetry I: the Model and the Case of SN 1987A}},\ }\href
  {https://doi.org/10.1088/0004-637X/806/2/275} {\bibfield  {journal} {\bibinfo
   {journal} {\apj}\ }\textbf {\bibinfo {volume} {806}},\ \bibinfo {eid} {275}
  (\bibinfo {year} {2015})},\ \Eprint {https://arxiv.org/abs/1501.02845}
  {arXiv:1501.02845 [astro-ph.SR]} \BibitemShut {NoStop}%
\bibitem [{\citenamefont {{Sukhbold}}\ and\ \citenamefont
  {{Adams}}(2020)}]{2020MNRAS.492.2578S}%
  \BibitemOpen
  \bibfield  {author} {\bibinfo {author} {\bibfnamefont {T.}~\bibnamefont
  {{Sukhbold}}}\ and\ \bibinfo {author} {\bibfnamefont {S.}~\bibnamefont
  {{Adams}}},\ }\bibfield  {title} {\bibinfo {title} {{Missing red supergiants
  and carbon burning}},\ }\href {https://doi.org/10.1093/mnras/staa059}
  {\bibfield  {journal} {\bibinfo  {journal} {\mnras}\ }\textbf {\bibinfo
  {volume} {492}},\ \bibinfo {pages} {2578} (\bibinfo {year} {2020})},\ \Eprint
  {https://arxiv.org/abs/1905.00474} {arXiv:1905.00474 [astro-ph.HE]}
  \BibitemShut {NoStop}%
\bibitem [{\citenamefont {{Patton}}\ and\ \citenamefont
  {{Sukhbold}}(2020)}]{2020MNRAS.499.2803P}%
  \BibitemOpen
  \bibfield  {author} {\bibinfo {author} {\bibfnamefont {R.~A.}\ \bibnamefont
  {{Patton}}}\ and\ \bibinfo {author} {\bibfnamefont {T.}~\bibnamefont
  {{Sukhbold}}},\ }\bibfield  {title} {\bibinfo {title} {{Towards a realistic
  explosion landscape for binary population synthesis}},\ }\href
  {https://doi.org/10.1093/mnras/staa3029} {\bibfield  {journal} {\bibinfo
  {journal} {\mnras}\ }\textbf {\bibinfo {volume} {499}},\ \bibinfo {pages}
  {2803} (\bibinfo {year} {2020})},\ \Eprint {https://arxiv.org/abs/2005.03055}
  {arXiv:2005.03055 [astro-ph.SR]} \BibitemShut {NoStop}%
\bibitem [{\citenamefont {{Patton}}\ \emph {et~al.}(2022)\citenamefont
  {{Patton}}, \citenamefont {{Sukhbold}},\ and\ \citenamefont
  {{Eldridge}}}]{2022MNRAS.511..903P}%
  \BibitemOpen
  \bibfield  {author} {\bibinfo {author} {\bibfnamefont {R.~A.}\ \bibnamefont
  {{Patton}}}, \bibinfo {author} {\bibfnamefont {T.}~\bibnamefont
  {{Sukhbold}}},\ and\ \bibinfo {author} {\bibfnamefont {J.~J.}\ \bibnamefont
  {{Eldridge}}},\ }\bibfield  {title} {\bibinfo {title} {{Comparing compact
  object distributions from mass- and presupernova core structure-based
  prescriptions}},\ }\href {https://doi.org/10.1093/mnras/stab3797} {\bibfield
  {journal} {\bibinfo  {journal} {\mnras}\ }\textbf {\bibinfo {volume} {511}},\
  \bibinfo {pages} {903} (\bibinfo {year} {2022})},\ \Eprint
  {https://arxiv.org/abs/2106.05978} {arXiv:2106.05978 [astro-ph.HE]}
  \BibitemShut {NoStop}%
\bibitem [{\citenamefont {{Powell}}\ \emph {et~al.}(2021)\citenamefont
  {{Powell}}, \citenamefont {{M{\"u}ller}},\ and\ \citenamefont
  {{Heger}}}]{2021MNRAS.503.2108P}%
  \BibitemOpen
  \bibfield  {author} {\bibinfo {author} {\bibfnamefont {J.}~\bibnamefont
  {{Powell}}}, \bibinfo {author} {\bibfnamefont {B.}~\bibnamefont
  {{M{\"u}ller}}},\ and\ \bibinfo {author} {\bibfnamefont {A.}~\bibnamefont
  {{Heger}}},\ }\bibfield  {title} {\bibinfo {title} {{The final core collapse
  of pulsational pair instability supernovae}},\ }\href
  {https://doi.org/10.1093/mnras/stab614} {\bibfield  {journal} {\bibinfo
  {journal} {\mnras}\ }\textbf {\bibinfo {volume} {503}},\ \bibinfo {pages}
  {2108} (\bibinfo {year} {2021})},\ \Eprint {https://arxiv.org/abs/2101.06889}
  {arXiv:2101.06889 [astro-ph.HE]} \BibitemShut {NoStop}%
\bibitem [{\citenamefont {{Rahman}}\ \emph {et~al.}(2022)\citenamefont
  {{Rahman}}, \citenamefont {{Janka}}, \citenamefont {{Stockinger}},\ and\
  \citenamefont {{Woosley}}}]{2022MNRAS.512.4503R}%
  \BibitemOpen
  \bibfield  {author} {\bibinfo {author} {\bibfnamefont {N.}~\bibnamefont
  {{Rahman}}}, \bibinfo {author} {\bibfnamefont {H.~T.}\ \bibnamefont
  {{Janka}}}, \bibinfo {author} {\bibfnamefont {G.}~\bibnamefont
  {{Stockinger}}},\ and\ \bibinfo {author} {\bibfnamefont {S.~E.}\ \bibnamefont
  {{Woosley}}},\ }\bibfield  {title} {\bibinfo {title} {{Pulsational
  pair-instability supernovae: gravitational collapse, black hole formation,
  and beyond}},\ }\href {https://doi.org/10.1093/mnras/stac758} {\bibfield
  {journal} {\bibinfo  {journal} {\mnras}\ }\textbf {\bibinfo {volume} {512}},\
  \bibinfo {pages} {4503} (\bibinfo {year} {2022})},\ \Eprint
  {https://arxiv.org/abs/2112.09707} {arXiv:2112.09707 [astro-ph.HE]}
  \BibitemShut {NoStop}%
\bibitem [{\citenamefont {{Miyaji}}\ \emph {et~al.}(1980)\citenamefont
  {{Miyaji}}, \citenamefont {{Nomoto}}, \citenamefont {{Yokoi}},\ and\
  \citenamefont {{Sugimoto}}}]{1980PASJ...32..303M}%
  \BibitemOpen
  \bibfield  {author} {\bibinfo {author} {\bibfnamefont {S.}~\bibnamefont
  {{Miyaji}}}, \bibinfo {author} {\bibfnamefont {K.}~\bibnamefont {{Nomoto}}},
  \bibinfo {author} {\bibfnamefont {K.}~\bibnamefont {{Yokoi}}},\ and\ \bibinfo
  {author} {\bibfnamefont {D.}~\bibnamefont {{Sugimoto}}},\ }\bibfield  {title}
  {\bibinfo {title} {{Supernova triggered by electron captures.}},\ }\href@noop
  {} {\bibfield  {journal} {\bibinfo  {journal} {\pasj}\ }\textbf {\bibinfo
  {volume} {32}},\ \bibinfo {pages} {303} (\bibinfo {year} {1980})}\BibitemShut
  {NoStop}%
\bibitem [{\citenamefont {{Nomoto}}(1982)}]{1982ApJ...253..798N}%
  \BibitemOpen
  \bibfield  {author} {\bibinfo {author} {\bibfnamefont {K.}~\bibnamefont
  {{Nomoto}}},\ }\bibfield  {title} {\bibinfo {title} {{Accreting white dwarf
  models for type I supernovae. I - Presupernova evolution and triggering
  mechanisms}},\ }\href {https://doi.org/10.1086/159682} {\bibfield  {journal}
  {\bibinfo  {journal} {\apj}\ }\textbf {\bibinfo {volume} {253}},\ \bibinfo
  {pages} {798} (\bibinfo {year} {1982})}\BibitemShut {NoStop}%
\bibitem [{\citenamefont {{Jones}}\ \emph {et~al.}(2013)\citenamefont
  {{Jones}}, \citenamefont {{Hirschi}}, \citenamefont {{Nomoto}}, \citenamefont
  {{Fischer}}, \citenamefont {{Timmes}}, \citenamefont {{Herwig}},
  \citenamefont {{Paxton}}, \citenamefont {{Toki}}, \citenamefont {{Suzuki}},
  \citenamefont {{Mart{\'\i}nez-Pinedo}}, \citenamefont {{Lam}},\ and\
  \citenamefont {{Bertolli}}}]{2013ApJ...772..150J}%
  \BibitemOpen
  \bibfield  {author} {\bibinfo {author} {\bibfnamefont {S.}~\bibnamefont
  {{Jones}}}, \bibinfo {author} {\bibfnamefont {R.}~\bibnamefont {{Hirschi}}},
  \bibinfo {author} {\bibfnamefont {K.}~\bibnamefont {{Nomoto}}}, \bibinfo
  {author} {\bibfnamefont {T.}~\bibnamefont {{Fischer}}}, \bibinfo {author}
  {\bibfnamefont {F.~X.}\ \bibnamefont {{Timmes}}}, \bibinfo {author}
  {\bibfnamefont {F.}~\bibnamefont {{Herwig}}}, \bibinfo {author}
  {\bibfnamefont {B.}~\bibnamefont {{Paxton}}}, \bibinfo {author}
  {\bibfnamefont {H.}~\bibnamefont {{Toki}}}, \bibinfo {author} {\bibfnamefont
  {T.}~\bibnamefont {{Suzuki}}}, \bibinfo {author} {\bibfnamefont
  {G.}~\bibnamefont {{Mart{\'\i}nez-Pinedo}}}, \bibinfo {author} {\bibfnamefont
  {Y.~H.}\ \bibnamefont {{Lam}}},\ and\ \bibinfo {author} {\bibfnamefont
  {M.~G.}\ \bibnamefont {{Bertolli}}},\ }\bibfield  {title} {\bibinfo {title}
  {{Advanced Burning Stages and Fate of 8-10 M $_{{\ensuremath{\odot}}}$
  Stars}},\ }\href {https://doi.org/10.1088/0004-637X/772/2/150} {\bibfield
  {journal} {\bibinfo  {journal} {\apj}\ }\textbf {\bibinfo {volume} {772}},\
  \bibinfo {eid} {150} (\bibinfo {year} {2013})},\ \Eprint
  {https://arxiv.org/abs/1306.2030} {arXiv:1306.2030 [astro-ph.SR]}
  \BibitemShut {NoStop}%
\bibitem [{\citenamefont {{Jones}}\ \emph {et~al.}(2019)\citenamefont
  {{Jones}}, \citenamefont {{R{\"o}pke}}, \citenamefont {{Fryer}},
  \citenamefont {{Ruiter}}, \citenamefont {{Seitenzahl}}, \citenamefont
  {{Nittler}}, \citenamefont {{Ohlmann}}, \citenamefont {{Reifarth}},
  \citenamefont {{Pignatari}},\ and\ \citenamefont
  {{Belczynski}}}]{2019A&A...622A..74J}%
  \BibitemOpen
  \bibfield  {author} {\bibinfo {author} {\bibfnamefont {S.}~\bibnamefont
  {{Jones}}}, \bibinfo {author} {\bibfnamefont {F.~K.}\ \bibnamefont
  {{R{\"o}pke}}}, \bibinfo {author} {\bibfnamefont {C.}~\bibnamefont
  {{Fryer}}}, \bibinfo {author} {\bibfnamefont {A.~J.}\ \bibnamefont
  {{Ruiter}}}, \bibinfo {author} {\bibfnamefont {I.~R.}\ \bibnamefont
  {{Seitenzahl}}}, \bibinfo {author} {\bibfnamefont {L.~R.}\ \bibnamefont
  {{Nittler}}}, \bibinfo {author} {\bibfnamefont {S.~T.}\ \bibnamefont
  {{Ohlmann}}}, \bibinfo {author} {\bibfnamefont {R.}~\bibnamefont
  {{Reifarth}}}, \bibinfo {author} {\bibfnamefont {M.}~\bibnamefont
  {{Pignatari}}},\ and\ \bibinfo {author} {\bibfnamefont {K.}~\bibnamefont
  {{Belczynski}}},\ }\bibfield  {title} {\bibinfo {title} {{Remnants and ejecta
  of thermonuclear electron-capture supernovae. Constraining oxygen-neon
  deflagrations in high-density white dwarfs}},\ }\href
  {https://doi.org/10.1051/0004-6361/201834381} {\bibfield  {journal} {\bibinfo
   {journal} {\aap}\ }\textbf {\bibinfo {volume} {622}},\ \bibinfo {eid} {A74}
  (\bibinfo {year} {2019})},\ \Eprint {https://arxiv.org/abs/1812.08230}
  {arXiv:1812.08230 [astro-ph.SR]} \BibitemShut {NoStop}%
\bibitem [{\citenamefont {{Gessner}}\ and\ \citenamefont
  {{Janka}}(2018)}]{2018ApJ...865...61G}%
  \BibitemOpen
  \bibfield  {author} {\bibinfo {author} {\bibfnamefont {A.}~\bibnamefont
  {{Gessner}}}\ and\ \bibinfo {author} {\bibfnamefont {H.-T.}\ \bibnamefont
  {{Janka}}},\ }\bibfield  {title} {\bibinfo {title} {{Hydrodynamical
  Neutron-star Kicks in Electron-capture Supernovae and Implications for the
  CRAB Supernova}},\ }\href {https://doi.org/10.3847/1538-4357/aadbae}
  {\bibfield  {journal} {\bibinfo  {journal} {\apj}\ }\textbf {\bibinfo
  {volume} {865}},\ \bibinfo {eid} {61} (\bibinfo {year} {2018})},\ \Eprint
  {https://arxiv.org/abs/1802.05274} {arXiv:1802.05274 [astro-ph.HE]}
  \BibitemShut {NoStop}%
\bibitem [{\citenamefont {{Barkat}}\ \emph {et~al.}(1967)\citenamefont
  {{Barkat}}, \citenamefont {{Rakavy}},\ and\ \citenamefont
  {{Sack}}}]{1967PhRvL..18..379B}%
  \BibitemOpen
  \bibfield  {author} {\bibinfo {author} {\bibfnamefont {Z.}~\bibnamefont
  {{Barkat}}}, \bibinfo {author} {\bibfnamefont {G.}~\bibnamefont {{Rakavy}}},\
  and\ \bibinfo {author} {\bibfnamefont {N.}~\bibnamefont {{Sack}}},\
  }\bibfield  {title} {\bibinfo {title} {{Dynamics of Supernova Explosion
  Resulting from Pair Formation}},\ }\href
  {https://doi.org/10.1103/PhysRevLett.18.379} {\bibfield  {journal} {\bibinfo
  {journal} {\prl}\ }\textbf {\bibinfo {volume} {18}},\ \bibinfo {pages} {379}
  (\bibinfo {year} {1967})}\BibitemShut {NoStop}%
\bibitem [{\citenamefont {{Bond}}\ \emph {et~al.}(1984)\citenamefont {{Bond}},
  \citenamefont {{Arnett}},\ and\ \citenamefont
  {{Carr}}}]{1984ApJ...280..825B}%
  \BibitemOpen
  \bibfield  {author} {\bibinfo {author} {\bibfnamefont {J.~R.}\ \bibnamefont
  {{Bond}}}, \bibinfo {author} {\bibfnamefont {W.~D.}\ \bibnamefont
  {{Arnett}}},\ and\ \bibinfo {author} {\bibfnamefont {B.~J.}\ \bibnamefont
  {{Carr}}},\ }\bibfield  {title} {\bibinfo {title} {{The evolution and fate of
  Very Massive Objects}},\ }\href {https://doi.org/10.1086/162057} {\bibfield
  {journal} {\bibinfo  {journal} {\apj}\ }\textbf {\bibinfo {volume} {280}},\
  \bibinfo {pages} {825} (\bibinfo {year} {1984})}\BibitemShut {NoStop}%
\bibitem [{\citenamefont {{Belczynski}}\ \emph
  {et~al.}(2016{\natexlab{a}})\citenamefont {{Belczynski}}, \citenamefont
  {{Heger}}, \citenamefont {{Gladysz}}, \citenamefont {{Ruiter}}, \citenamefont
  {{Woosley}}, \citenamefont {{Wiktorowicz}}, \citenamefont {{Chen}},
  \citenamefont {{Bulik}}, \citenamefont {{O'Shaughnessy}}, \citenamefont
  {{Holz}}, \citenamefont {{Fryer}},\ and\ \citenamefont
  {{Berti}}}]{2016A&A...594A..97B}%
  \BibitemOpen
  \bibfield  {author} {\bibinfo {author} {\bibfnamefont {K.}~\bibnamefont
  {{Belczynski}}}, \bibinfo {author} {\bibfnamefont {A.}~\bibnamefont
  {{Heger}}}, \bibinfo {author} {\bibfnamefont {W.}~\bibnamefont {{Gladysz}}},
  \bibinfo {author} {\bibfnamefont {A.~J.}\ \bibnamefont {{Ruiter}}}, \bibinfo
  {author} {\bibfnamefont {S.}~\bibnamefont {{Woosley}}}, \bibinfo {author}
  {\bibfnamefont {G.}~\bibnamefont {{Wiktorowicz}}}, \bibinfo {author}
  {\bibfnamefont {H.~Y.}\ \bibnamefont {{Chen}}}, \bibinfo {author}
  {\bibfnamefont {T.}~\bibnamefont {{Bulik}}}, \bibinfo {author} {\bibfnamefont
  {R.}~\bibnamefont {{O'Shaughnessy}}}, \bibinfo {author} {\bibfnamefont
  {D.~E.}\ \bibnamefont {{Holz}}}, \bibinfo {author} {\bibfnamefont {C.~L.}\
  \bibnamefont {{Fryer}}},\ and\ \bibinfo {author} {\bibfnamefont
  {E.}~\bibnamefont {{Berti}}},\ }\bibfield  {title} {\bibinfo {title} {{The
  effect of pair-instability mass loss on black-hole mergers}},\ }\href
  {https://doi.org/10.1051/0004-6361/201628980} {\bibfield  {journal} {\bibinfo
   {journal} {\aap}\ }\textbf {\bibinfo {volume} {594}},\ \bibinfo {eid} {A97}
  (\bibinfo {year} {2016}{\natexlab{a}})},\ \Eprint
  {https://arxiv.org/abs/1607.03116} {arXiv:1607.03116 [astro-ph.HE]}
  \BibitemShut {NoStop}%
\bibitem [{\citenamefont {{Farmer}}\ \emph
  {et~al.}(2019{\natexlab{a}})\citenamefont {{Farmer}}, \citenamefont
  {{Renzo}}, \citenamefont {{de Mink}}, \citenamefont {{Marchant}},\ and\
  \citenamefont {{Justham}}}]{2019ApJ...887...53F}%
  \BibitemOpen
  \bibfield  {author} {\bibinfo {author} {\bibfnamefont {R.}~\bibnamefont
  {{Farmer}}}, \bibinfo {author} {\bibfnamefont {M.}~\bibnamefont {{Renzo}}},
  \bibinfo {author} {\bibfnamefont {S.~E.}\ \bibnamefont {{de Mink}}}, \bibinfo
  {author} {\bibfnamefont {P.}~\bibnamefont {{Marchant}}},\ and\ \bibinfo
  {author} {\bibfnamefont {S.}~\bibnamefont {{Justham}}},\ }\bibfield  {title}
  {\bibinfo {title} {{Mind the Gap: The Location of the Lower Edge of the
  Pair-instability Supernova Black Hole Mass Gap}},\ }\href
  {https://doi.org/10.3847/1538-4357/ab518b} {\bibfield  {journal} {\bibinfo
  {journal} {\apj}\ }\textbf {\bibinfo {volume} {887}},\ \bibinfo {eid} {53}
  (\bibinfo {year} {2019}{\natexlab{a}})},\ \Eprint
  {https://arxiv.org/abs/1910.12874} {arXiv:1910.12874 [astro-ph.SR]}
  \BibitemShut {NoStop}%
\bibitem [{\citenamefont {{Farmer}}\ \emph {et~al.}(2020)\citenamefont
  {{Farmer}}, \citenamefont {{Renzo}}, \citenamefont {{de Mink}}, \citenamefont
  {{Fishbach}},\ and\ \citenamefont {{Justham}}}]{2020ApJ...902L..36F}%
  \BibitemOpen
  \bibfield  {author} {\bibinfo {author} {\bibfnamefont {R.}~\bibnamefont
  {{Farmer}}}, \bibinfo {author} {\bibfnamefont {M.}~\bibnamefont {{Renzo}}},
  \bibinfo {author} {\bibfnamefont {S.~E.}\ \bibnamefont {{de Mink}}}, \bibinfo
  {author} {\bibfnamefont {M.}~\bibnamefont {{Fishbach}}},\ and\ \bibinfo
  {author} {\bibfnamefont {S.}~\bibnamefont {{Justham}}},\ }\bibfield  {title}
  {\bibinfo {title} {{Constraints from Gravitational-wave Detections of Binary
  Black Hole Mergers on the $^{12}$C({\ensuremath{\alpha}},
  {\ensuremath{\gamma}})$^{16}$O Rate}},\ }\href
  {https://doi.org/10.3847/2041-8213/abbadd} {\bibfield  {journal} {\bibinfo
  {journal} {\apjl}\ }\textbf {\bibinfo {volume} {902}},\ \bibinfo {eid} {L36}
  (\bibinfo {year} {2020})},\ \Eprint {https://arxiv.org/abs/2006.06678}
  {arXiv:2006.06678 [astro-ph.HE]} \BibitemShut {NoStop}%
\bibitem [{\citenamefont {{Vink}}\ \emph {et~al.}(2021)\citenamefont {{Vink}},
  \citenamefont {{Higgins}}, \citenamefont {{Sander}},\ and\ \citenamefont
  {{Sabhahit}}}]{2021MNRAS.504..146V}%
  \BibitemOpen
  \bibfield  {author} {\bibinfo {author} {\bibfnamefont {J.~S.}\ \bibnamefont
  {{Vink}}}, \bibinfo {author} {\bibfnamefont {E.~R.}\ \bibnamefont
  {{Higgins}}}, \bibinfo {author} {\bibfnamefont {A.~A.~C.}\ \bibnamefont
  {{Sander}}},\ and\ \bibinfo {author} {\bibfnamefont {G.~N.}\ \bibnamefont
  {{Sabhahit}}},\ }\bibfield  {title} {\bibinfo {title} {{Maximum black hole
  mass across cosmic time}},\ }\href {https://doi.org/10.1093/mnras/stab842}
  {\bibfield  {journal} {\bibinfo  {journal} {\mnras}\ }\textbf {\bibinfo
  {volume} {504}},\ \bibinfo {pages} {146} (\bibinfo {year} {2021})},\ \Eprint
  {https://arxiv.org/abs/2010.11730} {arXiv:2010.11730 [astro-ph.HE]}
  \BibitemShut {NoStop}%
\bibitem [{\citenamefont {{deBoer}}\ \emph {et~al.}(2017)\citenamefont
  {{deBoer}}, \citenamefont {{G{\"o}rres}}, \citenamefont {{Wiescher}},
  \citenamefont {{Azuma}}, \citenamefont {{Best}}, \citenamefont {{Brune}},
  \citenamefont {{Fields}}, \citenamefont {{Jones}}, \citenamefont
  {{Pignatari}}, \citenamefont {{Sayre}}, \citenamefont {{Smith}},
  \citenamefont {{Timmes}},\ and\ \citenamefont
  {{Uberseder}}}]{2017RvMP...89c5007D}%
  \BibitemOpen
  \bibfield  {author} {\bibinfo {author} {\bibfnamefont {R.~J.}\ \bibnamefont
  {{deBoer}}}, \bibinfo {author} {\bibfnamefont {J.}~\bibnamefont
  {{G{\"o}rres}}}, \bibinfo {author} {\bibfnamefont {M.}~\bibnamefont
  {{Wiescher}}}, \bibinfo {author} {\bibfnamefont {R.~E.}\ \bibnamefont
  {{Azuma}}}, \bibinfo {author} {\bibfnamefont {A.}~\bibnamefont {{Best}}},
  \bibinfo {author} {\bibfnamefont {C.~R.}\ \bibnamefont {{Brune}}}, \bibinfo
  {author} {\bibfnamefont {C.~E.}\ \bibnamefont {{Fields}}}, \bibinfo {author}
  {\bibfnamefont {S.}~\bibnamefont {{Jones}}}, \bibinfo {author} {\bibfnamefont
  {M.}~\bibnamefont {{Pignatari}}}, \bibinfo {author} {\bibfnamefont
  {D.}~\bibnamefont {{Sayre}}}, \bibinfo {author} {\bibfnamefont
  {K.}~\bibnamefont {{Smith}}}, \bibinfo {author} {\bibfnamefont {F.~X.}\
  \bibnamefont {{Timmes}}},\ and\ \bibinfo {author} {\bibfnamefont
  {E.}~\bibnamefont {{Uberseder}}},\ }\bibfield  {title} {\bibinfo {title}
  {{The $^{12}$C({\ensuremath{\alpha}} ,{\ensuremath{\gamma}} )$^{16}$O
  reaction and its implications for stellar helium burning}},\ }\href
  {https://doi.org/10.1103/RevModPhys.89.035007} {\bibfield  {journal}
  {\bibinfo  {journal} {Reviews of Modern Physics}\ }\textbf {\bibinfo {volume}
  {89}},\ \bibinfo {eid} {035007} (\bibinfo {year} {2017})},\ \Eprint
  {https://arxiv.org/abs/1709.03144} {arXiv:1709.03144 [nucl-ex]} \BibitemShut
  {NoStop}%
\bibitem [{\citenamefont {{Mason}}\ \emph {et~al.}(1998)\citenamefont
  {{Mason}}, \citenamefont {{Gies}}, \citenamefont {{Hartkopf}}, \citenamefont
  {{Bagnuolo}}, \citenamefont {{ten Brummelaar}},\ and\ \citenamefont
  {{McAlister}}}]{1998AJ....115..821M}%
  \BibitemOpen
  \bibfield  {author} {\bibinfo {author} {\bibfnamefont {B.~D.}\ \bibnamefont
  {{Mason}}}, \bibinfo {author} {\bibfnamefont {D.~R.}\ \bibnamefont {{Gies}}},
  \bibinfo {author} {\bibfnamefont {W.~I.}\ \bibnamefont {{Hartkopf}}},
  \bibinfo {author} {\bibfnamefont {J.}~\bibnamefont {{Bagnuolo}},
  \bibfnamefont {William~G.}}, \bibinfo {author} {\bibfnamefont
  {T.}~\bibnamefont {{ten Brummelaar}}},\ and\ \bibinfo {author} {\bibfnamefont
  {H.~A.}\ \bibnamefont {{McAlister}}},\ }\bibfield  {title} {\bibinfo {title}
  {{ICCD Speckle Observations of Binary Stars. XIX. an
  Astrometric/Spectroscopic Survey of O Stars}},\ }\href
  {https://doi.org/10.1086/300234} {\bibfield  {journal} {\bibinfo  {journal}
  {\aj}\ }\textbf {\bibinfo {volume} {115}},\ \bibinfo {pages} {821} (\bibinfo
  {year} {1998})}\BibitemShut {NoStop}%
\bibitem [{\citenamefont {{Kobulnicky}}\ and\ \citenamefont
  {{Fryer}}(2007)}]{2007ApJ...670..747K}%
  \BibitemOpen
  \bibfield  {author} {\bibinfo {author} {\bibfnamefont {H.~A.}\ \bibnamefont
  {{Kobulnicky}}}\ and\ \bibinfo {author} {\bibfnamefont {C.~L.}\ \bibnamefont
  {{Fryer}}},\ }\bibfield  {title} {\bibinfo {title} {{A New Look at the Binary
  Characteristics of Massive Stars}},\ }\href {https://doi.org/10.1086/522073}
  {\bibfield  {journal} {\bibinfo  {journal} {\apj}\ }\textbf {\bibinfo
  {volume} {670}},\ \bibinfo {pages} {747} (\bibinfo {year}
  {2007})}\BibitemShut {NoStop}%
\bibitem [{\citenamefont {{Duch{\^e}ne}}\ and\ \citenamefont
  {{Kraus}}(2013)}]{2013ARA&A..51..269D}%
  \BibitemOpen
  \bibfield  {author} {\bibinfo {author} {\bibfnamefont {G.}~\bibnamefont
  {{Duch{\^e}ne}}}\ and\ \bibinfo {author} {\bibfnamefont {A.}~\bibnamefont
  {{Kraus}}},\ }\bibfield  {title} {\bibinfo {title} {{Stellar Multiplicity}},\
  }\href {https://doi.org/10.1146/annurev-astro-081710-102602} {\bibfield
  {journal} {\bibinfo  {journal} {\araa}\ }\textbf {\bibinfo {volume} {51}},\
  \bibinfo {pages} {269} (\bibinfo {year} {2013})},\ \Eprint
  {https://arxiv.org/abs/1303.3028} {arXiv:1303.3028 [astro-ph.SR]}
  \BibitemShut {NoStop}%
\bibitem [{\citenamefont {{Sana}}\ \emph {et~al.}(2014)\citenamefont {{Sana}},
  \citenamefont {{Le Bouquin}}, \citenamefont {{Lacour}}, \citenamefont
  {{Berger}}, \citenamefont {{Duvert}}, \citenamefont {{Gauchet}},
  \citenamefont {{Norris}}, \citenamefont {{Olofsson}}, \citenamefont
  {{Pickel}}, \citenamefont {{Zins}}, \citenamefont {{Absil}}, \citenamefont
  {{de Koter}}, \citenamefont {{Kratter}}, \citenamefont {{Schnurr}},\ and\
  \citenamefont {{Zinnecker}}}]{2014ApJS..215...15S}%
  \BibitemOpen
  \bibfield  {author} {\bibinfo {author} {\bibfnamefont {H.}~\bibnamefont
  {{Sana}}}, \bibinfo {author} {\bibfnamefont {J.~B.}\ \bibnamefont {{Le
  Bouquin}}}, \bibinfo {author} {\bibfnamefont {S.}~\bibnamefont {{Lacour}}},
  \bibinfo {author} {\bibfnamefont {J.~P.}\ \bibnamefont {{Berger}}}, \bibinfo
  {author} {\bibfnamefont {G.}~\bibnamefont {{Duvert}}}, \bibinfo {author}
  {\bibfnamefont {L.}~\bibnamefont {{Gauchet}}}, \bibinfo {author}
  {\bibfnamefont {B.}~\bibnamefont {{Norris}}}, \bibinfo {author}
  {\bibfnamefont {J.}~\bibnamefont {{Olofsson}}}, \bibinfo {author}
  {\bibfnamefont {D.}~\bibnamefont {{Pickel}}}, \bibinfo {author}
  {\bibfnamefont {G.}~\bibnamefont {{Zins}}}, \bibinfo {author} {\bibfnamefont
  {O.}~\bibnamefont {{Absil}}}, \bibinfo {author} {\bibfnamefont
  {A.}~\bibnamefont {{de Koter}}}, \bibinfo {author} {\bibfnamefont
  {K.}~\bibnamefont {{Kratter}}}, \bibinfo {author} {\bibfnamefont
  {O.}~\bibnamefont {{Schnurr}}},\ and\ \bibinfo {author} {\bibfnamefont
  {H.}~\bibnamefont {{Zinnecker}}},\ }\bibfield  {title} {\bibinfo {title}
  {{Southern Massive Stars at High Angular Resolution: Observational Campaign
  and Companion Detection}},\ }\href
  {https://doi.org/10.1088/0067-0049/215/1/15} {\bibfield  {journal} {\bibinfo
  {journal} {\apjs}\ }\textbf {\bibinfo {volume} {215}},\ \bibinfo {eid} {15}
  (\bibinfo {year} {2014})},\ \Eprint {https://arxiv.org/abs/1409.6304}
  {arXiv:1409.6304 [astro-ph.SR]} \BibitemShut {NoStop}%
\bibitem [{\citenamefont {{Pringle}}\ and\ \citenamefont
  {{Wade}}(1985)}]{1985ibs..book.....P}%
  \BibitemOpen
  \bibfield  {author} {\bibinfo {author} {\bibfnamefont {J.~E.}\ \bibnamefont
  {{Pringle}}}\ and\ \bibinfo {author} {\bibfnamefont {R.~A.}\ \bibnamefont
  {{Wade}}},\ }\href@noop {} {\emph {\bibinfo {title} {{Interacting binary
  stars}}}}\ (\bibinfo {year} {1985})\BibitemShut {NoStop}%
\bibitem [{\citenamefont {{Paczy{\'n}ski}}(1971)}]{1971ARA&A...9..183P}%
  \BibitemOpen
  \bibfield  {author} {\bibinfo {author} {\bibfnamefont {B.}~\bibnamefont
  {{Paczy{\'n}ski}}},\ }\bibfield  {title} {\bibinfo {title} {{Evolutionary
  Processes in Close Binary Systems}},\ }\href
  {https://doi.org/10.1146/annurev.aa.09.090171.001151} {\bibfield  {journal}
  {\bibinfo  {journal} {\araa}\ }\textbf {\bibinfo {volume} {9}},\ \bibinfo
  {pages} {183} (\bibinfo {year} {1971})}\BibitemShut {NoStop}%
\bibitem [{\citenamefont {{Thomas}}(1977)}]{1977ARA&A..15..127T}%
  \BibitemOpen
  \bibfield  {author} {\bibinfo {author} {\bibfnamefont {H.~C.}\ \bibnamefont
  {{Thomas}}},\ }\bibfield  {title} {\bibinfo {title} {{Consequences of mass
  transfer in close binary systems.}},\ }\href
  {https://doi.org/10.1146/annurev.aa.15.090177.001015} {\bibfield  {journal}
  {\bibinfo  {journal} {\araa}\ }\textbf {\bibinfo {volume} {15}},\ \bibinfo
  {pages} {127} (\bibinfo {year} {1977})}\BibitemShut {NoStop}%
\bibitem [{\citenamefont {{Vanbeveren}}(1991)}]{1991A&A...252..159V}%
  \BibitemOpen
  \bibfield  {author} {\bibinfo {author} {\bibfnamefont {D.}~\bibnamefont
  {{Vanbeveren}}},\ }\bibfield  {title} {\bibinfo {title} {{The evolution of
  massive close binaries revised.}},\ }\href@noop {} {\bibfield  {journal}
  {\bibinfo  {journal} {\aap}\ }\textbf {\bibinfo {volume} {252}},\ \bibinfo
  {pages} {159} (\bibinfo {year} {1991})}\BibitemShut {NoStop}%
\bibitem [{\citenamefont {{De Marco}}\ and\ \citenamefont
  {{Izzard}}(2017)}]{2017PASA...34....1D}%
  \BibitemOpen
  \bibfield  {author} {\bibinfo {author} {\bibfnamefont {O.}~\bibnamefont {{De
  Marco}}}\ and\ \bibinfo {author} {\bibfnamefont {R.~G.}\ \bibnamefont
  {{Izzard}}},\ }\bibfield  {title} {\bibinfo {title} {{Dawes Review 6: The
  Impact of Companions on Stellar Evolution}},\ }\href
  {https://doi.org/10.1017/pasa.2016.52} {\bibfield  {journal} {\bibinfo
  {journal} {\pasa}\ }\textbf {\bibinfo {volume} {34}},\ \bibinfo {eid} {e001}
  (\bibinfo {year} {2017})},\ \Eprint {https://arxiv.org/abs/1611.03542}
  {arXiv:1611.03542 [astro-ph.SR]} \BibitemShut {NoStop}%
\bibitem [{\citenamefont {{Tauris}}\ and\ \citenamefont {{van den
  Heuvel}}(2006)}]{2006csxs.book..623T}%
  \BibitemOpen
  \bibfield  {author} {\bibinfo {author} {\bibfnamefont {T.~M.}\ \bibnamefont
  {{Tauris}}}\ and\ \bibinfo {author} {\bibfnamefont {E.~P.~J.}\ \bibnamefont
  {{van den Heuvel}}},\ }\bibinfo {title} {{Formation and evolution of compact
  stellar X-ray sources}},\ in\ \href@noop {} {\emph {\bibinfo {booktitle} {In:
  Compact stellar X-ray sources. Edited by Walter Lewin \&amp; Michiel van der
  Klis. Cambridge Astrophysics Series, No. 39. Cambridge, UK: Cambridge
  University Press, ISBN 978-0-521-82659-4, ISBN 0-521-82659-4, DOI:
  10.2277/0521826594, 2006, p. 623 - 665}}},\ Vol.~\bibinfo {volume} {39}\
  (\bibinfo {year} {2006})\ pp.\ \bibinfo {pages} {623--665}\BibitemShut
  {NoStop}%
\bibitem [{\citenamefont {{Kippenhahn}}\ and\ \citenamefont
  {{Weigert}}(1990)}]{1990sse..book.....K}%
  \BibitemOpen
  \bibfield  {author} {\bibinfo {author} {\bibfnamefont {R.}~\bibnamefont
  {{Kippenhahn}}}\ and\ \bibinfo {author} {\bibfnamefont {A.}~\bibnamefont
  {{Weigert}}},\ }\href@noop {} {\emph {\bibinfo {title} {{Stellar Structure
  and Evolution}}}}\ (\bibinfo {year} {1990})\BibitemShut {NoStop}%
\bibitem [{\citenamefont {{Kippenhahn}}\ and\ \citenamefont
  {{Weigert}}(1967)}]{1967ZA.....65..251K}%
  \BibitemOpen
  \bibfield  {author} {\bibinfo {author} {\bibfnamefont {R.}~\bibnamefont
  {{Kippenhahn}}}\ and\ \bibinfo {author} {\bibfnamefont {A.}~\bibnamefont
  {{Weigert}}},\ }\bibfield  {title} {\bibinfo {title} {{Entwicklung in engen
  Doppelsternsystemen I. Massenaustausch vor und nach Beendigung des zentralen
  Wasserstoff-Brennens}},\ }\href@noop {} {\bibfield  {journal} {\bibinfo
  {journal} {\zap}\ }\textbf {\bibinfo {volume} {65}},\ \bibinfo {pages} {251}
  (\bibinfo {year} {1967})}\BibitemShut {NoStop}%
\bibitem [{\citenamefont {{Sanyal}}\ \emph {et~al.}(2017)\citenamefont
  {{Sanyal}}, \citenamefont {{Langer}}, \citenamefont {{Sz{\'e}csi}},
  \citenamefont {{-C Yoon}},\ and\ \citenamefont
  {{Grassitelli}}}]{2017A&A...597A..71S}%
  \BibitemOpen
  \bibfield  {author} {\bibinfo {author} {\bibfnamefont {D.}~\bibnamefont
  {{Sanyal}}}, \bibinfo {author} {\bibfnamefont {N.}~\bibnamefont {{Langer}}},
  \bibinfo {author} {\bibfnamefont {D.}~\bibnamefont {{Sz{\'e}csi}}}, \bibinfo
  {author} {\bibfnamefont {S.}~\bibnamefont {{-C Yoon}}},\ and\ \bibinfo
  {author} {\bibfnamefont {L.}~\bibnamefont {{Grassitelli}}},\ }\bibfield
  {title} {\bibinfo {title} {{Metallicity dependence of envelope inflation in
  massive stars}},\ }\href {https://doi.org/10.1051/0004-6361/201629612}
  {\bibfield  {journal} {\bibinfo  {journal} {\aap}\ }\textbf {\bibinfo
  {volume} {597}},\ \bibinfo {eid} {A71} (\bibinfo {year} {2017})},\ \Eprint
  {https://arxiv.org/abs/1611.07280} {arXiv:1611.07280 [astro-ph.SR]}
  \BibitemShut {NoStop}%
\bibitem [{\citenamefont {{Langer}}\ \emph {et~al.}(2015)\citenamefont
  {{Langer}}, \citenamefont {{Sanyal}}, \citenamefont {{Grassitelli}},\ and\
  \citenamefont {{Sz{\'e}sci}}}]{2015wrs..conf..241L}%
  \BibitemOpen
  \bibfield  {author} {\bibinfo {author} {\bibfnamefont {N.}~\bibnamefont
  {{Langer}}}, \bibinfo {author} {\bibfnamefont {D.}~\bibnamefont {{Sanyal}}},
  \bibinfo {author} {\bibfnamefont {L.}~\bibnamefont {{Grassitelli}}},\ and\
  \bibinfo {author} {\bibfnamefont {D.}~\bibnamefont {{Sz{\'e}sci}}},\
  }\bibfield  {title} {\bibinfo {title} {{The stellar Eddington limit}},\ }in\
  \href@noop {} {\emph {\bibinfo {booktitle} {Wolf-Rayet Stars}}},\ \bibinfo
  {editor} {edited by\ \bibinfo {editor} {\bibfnamefont {W.-R.}\ \bibnamefont
  {{Hamann}}}, \bibinfo {editor} {\bibfnamefont {A.}~\bibnamefont {{Sander}}},\
  and\ \bibinfo {editor} {\bibfnamefont {H.}~\bibnamefont {{Todt}}}}\ (\bibinfo
  {year} {2015})\ pp.\ \bibinfo {pages} {241--244}\BibitemShut {NoStop}%
\bibitem [{\citenamefont {{Sanyal}}\ \emph {et~al.}(2015)\citenamefont
  {{Sanyal}}, \citenamefont {{Grassitelli}}, \citenamefont {{Langer}},\ and\
  \citenamefont {{Bestenlehner}}}]{2015A&A...580A..20S}%
  \BibitemOpen
  \bibfield  {author} {\bibinfo {author} {\bibfnamefont {D.}~\bibnamefont
  {{Sanyal}}}, \bibinfo {author} {\bibfnamefont {L.}~\bibnamefont
  {{Grassitelli}}}, \bibinfo {author} {\bibfnamefont {N.}~\bibnamefont
  {{Langer}}},\ and\ \bibinfo {author} {\bibfnamefont {J.~M.}\ \bibnamefont
  {{Bestenlehner}}},\ }\bibfield  {title} {\bibinfo {title} {{Massive
  main-sequence stars evolving at the Eddington limit}},\ }\href
  {https://doi.org/10.1051/0004-6361/201525945} {\bibfield  {journal} {\bibinfo
   {journal} {\aap}\ }\textbf {\bibinfo {volume} {580}},\ \bibinfo {eid} {A20}
  (\bibinfo {year} {2015})},\ \Eprint {https://arxiv.org/abs/1506.02997}
  {arXiv:1506.02997 [astro-ph.SR]} \BibitemShut {NoStop}%
\bibitem [{\citenamefont {{Petrovic}}\ \emph {et~al.}(2006)\citenamefont
  {{Petrovic}}, \citenamefont {{Pols}},\ and\ \citenamefont
  {{Langer}}}]{2006A&A...450..219P}%
  \BibitemOpen
  \bibfield  {author} {\bibinfo {author} {\bibfnamefont {J.}~\bibnamefont
  {{Petrovic}}}, \bibinfo {author} {\bibfnamefont {O.}~\bibnamefont {{Pols}}},\
  and\ \bibinfo {author} {\bibfnamefont {N.}~\bibnamefont {{Langer}}},\
  }\bibfield  {title} {\bibinfo {title} {{Are luminous and metal-rich
  Wolf-Rayet stars inflated?}},\ }\href
  {https://doi.org/10.1051/0004-6361:20035837} {\bibfield  {journal} {\bibinfo
  {journal} {\aap}\ }\textbf {\bibinfo {volume} {450}},\ \bibinfo {pages} {219}
  (\bibinfo {year} {2006})}\BibitemShut {NoStop}%
\bibitem [{\citenamefont {{Jiang}}\ \emph {et~al.}(2015)\citenamefont
  {{Jiang}}, \citenamefont {{Cantiello}}, \citenamefont {{Bildsten}},
  \citenamefont {{Quataert}},\ and\ \citenamefont
  {{Blaes}}}]{2015ApJ...813...74J}%
  \BibitemOpen
  \bibfield  {author} {\bibinfo {author} {\bibfnamefont {Y.-F.}\ \bibnamefont
  {{Jiang}}}, \bibinfo {author} {\bibfnamefont {M.}~\bibnamefont
  {{Cantiello}}}, \bibinfo {author} {\bibfnamefont {L.}~\bibnamefont
  {{Bildsten}}}, \bibinfo {author} {\bibfnamefont {E.}~\bibnamefont
  {{Quataert}}},\ and\ \bibinfo {author} {\bibfnamefont {O.}~\bibnamefont
  {{Blaes}}},\ }\bibfield  {title} {\bibinfo {title} {{Local Radiation
  Hydrodynamic Simulations of Massive Star Envelopes at the Iron Opacity
  Peak}},\ }\href {https://doi.org/10.1088/0004-637X/813/1/74} {\bibfield
  {journal} {\bibinfo  {journal} {\apj}\ }\textbf {\bibinfo {volume} {813}},\
  \bibinfo {eid} {74} (\bibinfo {year} {2015})},\ \Eprint
  {https://arxiv.org/abs/1509.05417} {arXiv:1509.05417 [astro-ph.SR]}
  \BibitemShut {NoStop}%
\bibitem [{\citenamefont {{Schultz}}\ \emph {et~al.}(2020)\citenamefont
  {{Schultz}}, \citenamefont {{Bildsten}},\ and\ \citenamefont
  {{Jiang}}}]{2020ApJ...902...67S}%
  \BibitemOpen
  \bibfield  {author} {\bibinfo {author} {\bibfnamefont {W.~C.}\ \bibnamefont
  {{Schultz}}}, \bibinfo {author} {\bibfnamefont {L.}~\bibnamefont
  {{Bildsten}}},\ and\ \bibinfo {author} {\bibfnamefont {Y.-F.}\ \bibnamefont
  {{Jiang}}},\ }\bibfield  {title} {\bibinfo {title} {{Convectively Driven 3D
  Turbulence in Massive Star Envelopes. I. A 1D Implementation of Diffusive
  Radiative Transport}},\ }\href {https://doi.org/10.3847/1538-4357/abb405}
  {\bibfield  {journal} {\bibinfo  {journal} {\apj}\ }\textbf {\bibinfo
  {volume} {902}},\ \bibinfo {eid} {67} (\bibinfo {year} {2020})},\ \Eprint
  {https://arxiv.org/abs/2009.01238} {arXiv:2009.01238 [astro-ph.SR]}
  \BibitemShut {NoStop}%
\bibitem [{\citenamefont {{Brunish}}\ and\ \citenamefont
  {{Truran}}(1982)}]{1982ApJS...49..447B}%
  \BibitemOpen
  \bibfield  {author} {\bibinfo {author} {\bibfnamefont {W.~M.}\ \bibnamefont
  {{Brunish}}}\ and\ \bibinfo {author} {\bibfnamefont {J.~W.}\ \bibnamefont
  {{Truran}}},\ }\bibfield  {title} {\bibinfo {title} {{The evolution of
  massive stars. II. The influence of initial composition and mass loss.}},\
  }\href {https://doi.org/10.1086/190806} {\bibfield  {journal} {\bibinfo
  {journal} {\apjs}\ }\textbf {\bibinfo {volume} {49}},\ \bibinfo {pages} {447}
  (\bibinfo {year} {1982})}\BibitemShut {NoStop}%
\bibitem [{\citenamefont {{Baraffe}}\ and\ \citenamefont {{El
  Eid}}(1991)}]{1991A&A...245..548B}%
  \BibitemOpen
  \bibfield  {author} {\bibinfo {author} {\bibfnamefont {I.}~\bibnamefont
  {{Baraffe}}}\ and\ \bibinfo {author} {\bibfnamefont {M.~F.}\ \bibnamefont
  {{El Eid}}},\ }\bibfield  {title} {\bibinfo {title} {{Evolution of massive
  stars with variable initial compositions}},\ }\href@noop {} {\bibfield
  {journal} {\bibinfo  {journal} {\aap}\ }\textbf {\bibinfo {volume} {245}},\
  \bibinfo {pages} {548} (\bibinfo {year} {1991})}\BibitemShut {NoStop}%
\bibitem [{\citenamefont {{Marigo}}\ \emph
  {et~al.}(2001{\natexlab{a}})\citenamefont {{Marigo}}, \citenamefont
  {{Girardi}}, \citenamefont {{Chiosi}},\ and\ \citenamefont
  {{Wood}}}]{2001A&A...371..152M}%
  \BibitemOpen
  \bibfield  {author} {\bibinfo {author} {\bibfnamefont {P.}~\bibnamefont
  {{Marigo}}}, \bibinfo {author} {\bibfnamefont {L.}~\bibnamefont {{Girardi}}},
  \bibinfo {author} {\bibfnamefont {C.}~\bibnamefont {{Chiosi}}},\ and\
  \bibinfo {author} {\bibfnamefont {P.~R.}\ \bibnamefont {{Wood}}},\ }\bibfield
   {title} {\bibinfo {title} {{Zero-metallicity stars. I. Evolution at constant
  mass}},\ }\href {https://doi.org/10.1051/0004-6361:20010309} {\bibfield
  {journal} {\bibinfo  {journal} {\aap}\ }\textbf {\bibinfo {volume} {371}},\
  \bibinfo {pages} {152} (\bibinfo {year} {2001}{\natexlab{a}})},\ \Eprint
  {https://arxiv.org/abs/astro-ph/0102253} {arXiv:astro-ph/0102253 [astro-ph]}
  \BibitemShut {NoStop}%
\bibitem [{\citenamefont {{Xin}}\ \emph {et~al.}(2022)\citenamefont {{Xin}},
  \citenamefont {{Renzo}},\ and\ \citenamefont
  {{Metzger}}}]{2022arXiv220611316X}%
  \BibitemOpen
  \bibfield  {author} {\bibinfo {author} {\bibfnamefont {C.}~\bibnamefont
  {{Xin}}}, \bibinfo {author} {\bibfnamefont {M.}~\bibnamefont {{Renzo}}},\
  and\ \bibinfo {author} {\bibfnamefont {B.~D.}\ \bibnamefont {{Metzger}}},\
  }\bibfield  {title} {\bibinfo {title} {{Dissecting the microphysics behind
  the metallicity-dependence of massive stars radii}},\ }\href@noop {}
  {\bibfield  {journal} {\bibinfo  {journal} {arXiv e-prints}\ ,\ \bibinfo
  {eid} {arXiv:2206.11316}} (\bibinfo {year} {2022})},\ \Eprint
  {https://arxiv.org/abs/2206.11316} {arXiv:2206.11316 [astro-ph.SR]}
  \BibitemShut {NoStop}%
\bibitem [{\citenamefont {{Stothers}}\ and\ \citenamefont
  {{Chin}}(1992)}]{1992ApJ...390..136S}%
  \BibitemOpen
  \bibfield  {author} {\bibinfo {author} {\bibfnamefont {R.~B.}\ \bibnamefont
  {{Stothers}}}\ and\ \bibinfo {author} {\bibfnamefont {C.-W.}\ \bibnamefont
  {{Chin}}},\ }\bibfield  {title} {\bibinfo {title} {{Stellar Evolution in Blue
  Populous Clusters of the Small Magellanic Cloud and the Problems of Envelope
  Semiconvection and Convective Core Overshooting}},\ }\href
  {https://doi.org/10.1086/171266} {\bibfield  {journal} {\bibinfo  {journal}
  {\apj}\ }\textbf {\bibinfo {volume} {390}},\ \bibinfo {pages} {136} (\bibinfo
  {year} {1992})}\BibitemShut {NoStop}%
\bibitem [{\citenamefont {{Langer}}\ and\ \citenamefont
  {{Maeder}}(1995)}]{1995A&A...295..685L}%
  \BibitemOpen
  \bibfield  {author} {\bibinfo {author} {\bibfnamefont {N.}~\bibnamefont
  {{Langer}}}\ and\ \bibinfo {author} {\bibfnamefont {A.}~\bibnamefont
  {{Maeder}}},\ }\bibfield  {title} {\bibinfo {title} {{The problem of the
  blue-to-red supergiant ratio in galaxies.}},\ }\href@noop {} {\bibfield
  {journal} {\bibinfo  {journal} {\aap}\ }\textbf {\bibinfo {volume} {295}},\
  \bibinfo {pages} {685} (\bibinfo {year} {1995})}\BibitemShut {NoStop}%
\bibitem [{\citenamefont {{Higgins}}\ and\ \citenamefont
  {{Vink}}(2020)}]{2020A&A...635A.175H}%
  \BibitemOpen
  \bibfield  {author} {\bibinfo {author} {\bibfnamefont {E.~R.}\ \bibnamefont
  {{Higgins}}}\ and\ \bibinfo {author} {\bibfnamefont {J.~S.}\ \bibnamefont
  {{Vink}}},\ }\bibfield  {title} {\bibinfo {title} {{Theoretical investigation
  of the Humphreys-Davidson limit at high and low metallicity}},\ }\href
  {https://doi.org/10.1051/0004-6361/201937374} {\bibfield  {journal} {\bibinfo
   {journal} {\aap}\ }\textbf {\bibinfo {volume} {635}},\ \bibinfo {eid} {A175}
  (\bibinfo {year} {2020})},\ \Eprint {https://arxiv.org/abs/2002.07204}
  {arXiv:2002.07204 [astro-ph.SR]} \BibitemShut {NoStop}%
\bibitem [{\citenamefont {{Sabhahit}}\ \emph {et~al.}(2021)\citenamefont
  {{Sabhahit}}, \citenamefont {{Vink}}, \citenamefont {{Higgins}},\ and\
  \citenamefont {{Sander}}}]{2021MNRAS.506.4473S}%
  \BibitemOpen
  \bibfield  {author} {\bibinfo {author} {\bibfnamefont {G.~N.}\ \bibnamefont
  {{Sabhahit}}}, \bibinfo {author} {\bibfnamefont {J.~S.}\ \bibnamefont
  {{Vink}}}, \bibinfo {author} {\bibfnamefont {E.~R.}\ \bibnamefont
  {{Higgins}}},\ and\ \bibinfo {author} {\bibfnamefont {A.~A.~C.}\ \bibnamefont
  {{Sander}}},\ }\bibfield  {title} {\bibinfo {title} {{Superadiabaticity and
  the metallicity independence of the Humphreys-Davidson limit}},\ }\href
  {https://doi.org/10.1093/mnras/stab1948} {\bibfield  {journal} {\bibinfo
  {journal} {\mnras}\ }\textbf {\bibinfo {volume} {506}},\ \bibinfo {pages}
  {4473} (\bibinfo {year} {2021})},\ \Eprint {https://arxiv.org/abs/2107.02183}
  {arXiv:2107.02183 [astro-ph.SR]} \BibitemShut {NoStop}%
\bibitem [{\citenamefont {{Farrell}}\ \emph {et~al.}(2022)\citenamefont
  {{Farrell}}, \citenamefont {{Groh}}, \citenamefont {{Meynet}},\ and\
  \citenamefont {{Eldridge}}}]{2022MNRAS.512.4116F}%
  \BibitemOpen
  \bibfield  {author} {\bibinfo {author} {\bibfnamefont {E.}~\bibnamefont
  {{Farrell}}}, \bibinfo {author} {\bibfnamefont {J.~H.}\ \bibnamefont
  {{Groh}}}, \bibinfo {author} {\bibfnamefont {G.}~\bibnamefont {{Meynet}}},\
  and\ \bibinfo {author} {\bibfnamefont {J.~J.}\ \bibnamefont {{Eldridge}}},\
  }\bibfield  {title} {\bibinfo {title} {{Numerical experiments to help
  understand cause and effect in massive star evolution}},\ }\href
  {https://doi.org/10.1093/mnras/stac538} {\bibfield  {journal} {\bibinfo
  {journal} {\mnras}\ }\textbf {\bibinfo {volume} {512}},\ \bibinfo {pages}
  {4116} (\bibinfo {year} {2022})},\ \Eprint {https://arxiv.org/abs/2109.02488}
  {arXiv:2109.02488 [astro-ph.SR]} \BibitemShut {NoStop}%
\bibitem [{\citenamefont {{Eldridge}}\ \emph {et~al.}(2008)\citenamefont
  {{Eldridge}}, \citenamefont {{Izzard}},\ and\ \citenamefont
  {{Tout}}}]{2008MNRAS.384.1109E}%
  \BibitemOpen
  \bibfield  {author} {\bibinfo {author} {\bibfnamefont {J.~J.}\ \bibnamefont
  {{Eldridge}}}, \bibinfo {author} {\bibfnamefont {R.~G.}\ \bibnamefont
  {{Izzard}}},\ and\ \bibinfo {author} {\bibfnamefont {C.~A.}\ \bibnamefont
  {{Tout}}},\ }\bibfield  {title} {\bibinfo {title} {{The effect of massive
  binaries on stellar populations and supernova progenitors}},\ }\href
  {https://doi.org/10.1111/j.1365-2966.2007.12738.x} {\bibfield  {journal}
  {\bibinfo  {journal} {\mnras}\ }\textbf {\bibinfo {volume} {384}},\ \bibinfo
  {pages} {1109} (\bibinfo {year} {2008})},\ \Eprint
  {https://arxiv.org/abs/0711.3079} {arXiv:0711.3079 [astro-ph]} \BibitemShut
  {NoStop}%
\bibitem [{\citenamefont {{Brott}}\ \emph
  {et~al.}(2011{\natexlab{b}})\citenamefont {{Brott}}, \citenamefont {{Evans}},
  \citenamefont {{Hunter}}, \citenamefont {{de Koter}}, \citenamefont
  {{Langer}}, \citenamefont {{Dufton}}, \citenamefont {{Cantiello}},
  \citenamefont {{Trundle}}, \citenamefont {{Lennon}}, \citenamefont {{de
  Mink}}, \citenamefont {{Yoon}},\ and\ \citenamefont
  {{Anders}}}]{2011A&A...530A.116B}%
  \BibitemOpen
  \bibfield  {author} {\bibinfo {author} {\bibfnamefont {I.}~\bibnamefont
  {{Brott}}}, \bibinfo {author} {\bibfnamefont {C.~J.}\ \bibnamefont
  {{Evans}}}, \bibinfo {author} {\bibfnamefont {I.}~\bibnamefont {{Hunter}}},
  \bibinfo {author} {\bibfnamefont {A.}~\bibnamefont {{de Koter}}}, \bibinfo
  {author} {\bibfnamefont {N.}~\bibnamefont {{Langer}}}, \bibinfo {author}
  {\bibfnamefont {P.~L.}\ \bibnamefont {{Dufton}}}, \bibinfo {author}
  {\bibfnamefont {M.}~\bibnamefont {{Cantiello}}}, \bibinfo {author}
  {\bibfnamefont {C.}~\bibnamefont {{Trundle}}}, \bibinfo {author}
  {\bibfnamefont {D.~J.}\ \bibnamefont {{Lennon}}}, \bibinfo {author}
  {\bibfnamefont {S.~E.}\ \bibnamefont {{de Mink}}}, \bibinfo {author}
  {\bibfnamefont {S.~C.}\ \bibnamefont {{Yoon}}},\ and\ \bibinfo {author}
  {\bibfnamefont {P.}~\bibnamefont {{Anders}}},\ }\bibfield  {title} {\bibinfo
  {title} {{Rotating massive main-sequence stars. II. Simulating a population
  of LMC early B-type stars as a test of rotational mixing}},\ }\href
  {https://doi.org/10.1051/0004-6361/201016114} {\bibfield  {journal} {\bibinfo
   {journal} {\aap}\ }\textbf {\bibinfo {volume} {530}},\ \bibinfo {eid} {A116}
  (\bibinfo {year} {2011}{\natexlab{b}})},\ \Eprint
  {https://arxiv.org/abs/1102.0766} {arXiv:1102.0766 [astro-ph.SR]}
  \BibitemShut {NoStop}%
\bibitem [{\citenamefont {{Schootemeijer}}\ and\ \citenamefont
  {{Langer}}(2018)}]{2018A&A...611A..75S}%
  \BibitemOpen
  \bibfield  {author} {\bibinfo {author} {\bibfnamefont {A.}~\bibnamefont
  {{Schootemeijer}}}\ and\ \bibinfo {author} {\bibfnamefont {N.}~\bibnamefont
  {{Langer}}},\ }\bibfield  {title} {\bibinfo {title} {{Wolf-Rayet stars in the
  Small Magellanic Cloud as testbed for massive star evolution}},\ }\href
  {https://doi.org/10.1051/0004-6361/201731895} {\bibfield  {journal} {\bibinfo
   {journal} {\aap}\ }\textbf {\bibinfo {volume} {611}},\ \bibinfo {eid} {A75}
  (\bibinfo {year} {2018})},\ \Eprint {https://arxiv.org/abs/1709.08727}
  {arXiv:1709.08727 [astro-ph.SR]} \BibitemShut {NoStop}%
\bibitem [{\citenamefont {{Bellinger}}\ \emph {et~al.}(2019)\citenamefont
  {{Bellinger}}, \citenamefont {{Basu}}, \citenamefont {{Hekker}},\ and\
  \citenamefont {{Christensen-Dalsgaard}}}]{2019ApJ...885..143B}%
  \BibitemOpen
  \bibfield  {author} {\bibinfo {author} {\bibfnamefont {E.~P.}\ \bibnamefont
  {{Bellinger}}}, \bibinfo {author} {\bibfnamefont {S.}~\bibnamefont {{Basu}}},
  \bibinfo {author} {\bibfnamefont {S.}~\bibnamefont {{Hekker}}},\ and\
  \bibinfo {author} {\bibfnamefont {J.}~\bibnamefont
  {{Christensen-Dalsgaard}}},\ }\bibfield  {title} {\bibinfo {title} {{Testing
  Stellar Evolution with Asteroseismic Inversions of a Main-sequence Star
  Harboring a Small Convective Core}},\ }\href
  {https://doi.org/10.3847/1538-4357/ab4a0d} {\bibfield  {journal} {\bibinfo
  {journal} {\apj}\ }\textbf {\bibinfo {volume} {885}},\ \bibinfo {eid} {143}
  (\bibinfo {year} {2019})},\ \Eprint {https://arxiv.org/abs/1910.00603}
  {arXiv:1910.00603 [astro-ph.SR]} \BibitemShut {NoStop}%
\bibitem [{\citenamefont {{Bowman}}(2020)}]{2020FrASS...7...70B}%
  \BibitemOpen
  \bibfield  {author} {\bibinfo {author} {\bibfnamefont {D.~M.}\ \bibnamefont
  {{Bowman}}},\ }\bibfield  {title} {\bibinfo {title} {{Asteroseismology of
  high-mass stars: new insights of stellar interiors with space telescopes}},\
  }\href {https://doi.org/10.3389/fspas.2020.578584} {\bibfield  {journal}
  {\bibinfo  {journal} {Frontiers in Astronomy and Space Sciences}\ }\textbf
  {\bibinfo {volume} {7}},\ \bibinfo {eid} {70} (\bibinfo {year} {2020})},\
  \Eprint {https://arxiv.org/abs/2008.11162} {arXiv:2008.11162 [astro-ph.SR]}
  \BibitemShut {NoStop}%
\bibitem [{\citenamefont {{Anders}}\ \emph
  {et~al.}(2022{\natexlab{b}})\citenamefont {{Anders}}, \citenamefont
  {{Jermyn}}, \citenamefont {{Lecoanet}},\ and\ \citenamefont
  {{Brown}}}]{2022ApJ...926..169A}%
  \BibitemOpen
  \bibfield  {author} {\bibinfo {author} {\bibfnamefont {E.~H.}\ \bibnamefont
  {{Anders}}}, \bibinfo {author} {\bibfnamefont {A.~S.}\ \bibnamefont
  {{Jermyn}}}, \bibinfo {author} {\bibfnamefont {D.}~\bibnamefont
  {{Lecoanet}}},\ and\ \bibinfo {author} {\bibfnamefont {B.~P.}\ \bibnamefont
  {{Brown}}},\ }\bibfield  {title} {\bibinfo {title} {{Stellar Convective
  Penetration: Parameterized Theory and Dynamical Simulations}},\ }\href
  {https://doi.org/10.3847/1538-4357/ac408d} {\bibfield  {journal} {\bibinfo
  {journal} {\apj}\ }\textbf {\bibinfo {volume} {926}},\ \bibinfo {eid} {169}
  (\bibinfo {year} {2022}{\natexlab{b}})},\ \Eprint
  {https://arxiv.org/abs/2110.11356} {arXiv:2110.11356 [astro-ph.SR]}
  \BibitemShut {NoStop}%
\bibitem [{\citenamefont {{Klencki}}\ \emph {et~al.}(2022)\citenamefont
  {{Klencki}}, \citenamefont {{Istrate}}, \citenamefont {{Nelemans}},\ and\
  \citenamefont {{Pols}}}]{2022A&A...662A..56K}%
  \BibitemOpen
  \bibfield  {author} {\bibinfo {author} {\bibfnamefont {J.}~\bibnamefont
  {{Klencki}}}, \bibinfo {author} {\bibfnamefont {A.}~\bibnamefont
  {{Istrate}}}, \bibinfo {author} {\bibfnamefont {G.}~\bibnamefont
  {{Nelemans}}},\ and\ \bibinfo {author} {\bibfnamefont {O.}~\bibnamefont
  {{Pols}}},\ }\bibfield  {title} {\bibinfo {title} {{Partial-envelope
  stripping and nuclear-timescale mass transfer from evolved supergiants at low
  metallicity}},\ }\href {https://doi.org/10.1051/0004-6361/202142701}
  {\bibfield  {journal} {\bibinfo  {journal} {\aap}\ }\textbf {\bibinfo
  {volume} {662}},\ \bibinfo {eid} {A56} (\bibinfo {year} {2022})},\ \Eprint
  {https://arxiv.org/abs/2111.10271} {arXiv:2111.10271 [astro-ph.SR]}
  \BibitemShut {NoStop}%
\bibitem [{\citenamefont {{Humphreys}}\ and\ \citenamefont
  {{Davidson}}(1979)}]{1979ApJ...232..409H}%
  \BibitemOpen
  \bibfield  {author} {\bibinfo {author} {\bibfnamefont {R.~M.}\ \bibnamefont
  {{Humphreys}}}\ and\ \bibinfo {author} {\bibfnamefont {K.}~\bibnamefont
  {{Davidson}}},\ }\bibfield  {title} {\bibinfo {title} {{Studies of luminous
  stars in nearby galaxies. III. Comments on the evolution of the most massive
  stars in the Milky Way and the Large Magellanic Cloud.}},\ }\href
  {https://doi.org/10.1086/157301} {\bibfield  {journal} {\bibinfo  {journal}
  {\apj}\ }\textbf {\bibinfo {volume} {232}},\ \bibinfo {pages} {409} (\bibinfo
  {year} {1979})}\BibitemShut {NoStop}%
\bibitem [{\citenamefont {{Ulmer}}\ and\ \citenamefont
  {{Fitzpatrick}}(1998)}]{1998ApJ...504..200U}%
  \BibitemOpen
  \bibfield  {author} {\bibinfo {author} {\bibfnamefont {A.}~\bibnamefont
  {{Ulmer}}}\ and\ \bibinfo {author} {\bibfnamefont {E.~L.}\ \bibnamefont
  {{Fitzpatrick}}},\ }\bibfield  {title} {\bibinfo {title} {{Revisiting the
  Modified Eddington Limit for Massive Stars}},\ }\href
  {https://doi.org/10.1086/306048} {\bibfield  {journal} {\bibinfo  {journal}
  {\apj}\ }\textbf {\bibinfo {volume} {504}},\ \bibinfo {pages} {200} (\bibinfo
  {year} {1998})},\ \Eprint {https://arxiv.org/abs/astro-ph/9708264}
  {arXiv:astro-ph/9708264 [astro-ph]} \BibitemShut {NoStop}%
\bibitem [{\citenamefont {{Owocki}}\ \emph {et~al.}(2004)\citenamefont
  {{Owocki}}, \citenamefont {{Gayley}},\ and\ \citenamefont
  {{Shaviv}}}]{2004ApJ...616..525O}%
  \BibitemOpen
  \bibfield  {author} {\bibinfo {author} {\bibfnamefont {S.~P.}\ \bibnamefont
  {{Owocki}}}, \bibinfo {author} {\bibfnamefont {K.~G.}\ \bibnamefont
  {{Gayley}}},\ and\ \bibinfo {author} {\bibfnamefont {N.~J.}\ \bibnamefont
  {{Shaviv}}},\ }\bibfield  {title} {\bibinfo {title} {{A Porosity-Length
  Formalism for Photon-Tiring-limited Mass Loss from Stars above the Eddington
  Limit}},\ }\href {https://doi.org/10.1086/424910} {\bibfield  {journal}
  {\bibinfo  {journal} {\apj}\ }\textbf {\bibinfo {volume} {616}},\ \bibinfo
  {pages} {525} (\bibinfo {year} {2004})},\ \Eprint
  {https://arxiv.org/abs/astro-ph/0409573} {arXiv:astro-ph/0409573 [astro-ph]}
  \BibitemShut {NoStop}%
\bibitem [{\citenamefont {{Gr{\"a}fener}}\ \emph {et~al.}(2011)\citenamefont
  {{Gr{\"a}fener}}, \citenamefont {{Vink}}, \citenamefont {{de Koter}},\ and\
  \citenamefont {{Langer}}}]{2011A&A...535A..56G}%
  \BibitemOpen
  \bibfield  {author} {\bibinfo {author} {\bibfnamefont {G.}~\bibnamefont
  {{Gr{\"a}fener}}}, \bibinfo {author} {\bibfnamefont {J.~S.}\ \bibnamefont
  {{Vink}}}, \bibinfo {author} {\bibfnamefont {A.}~\bibnamefont {{de Koter}}},\
  and\ \bibinfo {author} {\bibfnamefont {N.}~\bibnamefont {{Langer}}},\
  }\bibfield  {title} {\bibinfo {title} {{The Eddington factor as the key to
  understand the winds of the most massive stars. Evidence for a
  {\ensuremath{\Gamma}}-dependence of Wolf-Rayet type mass loss}},\ }\href
  {https://doi.org/10.1051/0004-6361/201116701} {\bibfield  {journal} {\bibinfo
   {journal} {\aap}\ }\textbf {\bibinfo {volume} {535}},\ \bibinfo {eid} {A56}
  (\bibinfo {year} {2011})},\ \Eprint {https://arxiv.org/abs/1106.5361}
  {arXiv:1106.5361 [astro-ph.SR]} \BibitemShut {NoStop}%
\bibitem [{\citenamefont {{Quataert}}\ \emph {et~al.}(2016)\citenamefont
  {{Quataert}}, \citenamefont {{Fern{\'a}ndez}}, \citenamefont {{Kasen}},
  \citenamefont {{Klion}},\ and\ \citenamefont
  {{Paxton}}}]{2016MNRAS.458.1214Q}%
  \BibitemOpen
  \bibfield  {author} {\bibinfo {author} {\bibfnamefont {E.}~\bibnamefont
  {{Quataert}}}, \bibinfo {author} {\bibfnamefont {R.}~\bibnamefont
  {{Fern{\'a}ndez}}}, \bibinfo {author} {\bibfnamefont {D.}~\bibnamefont
  {{Kasen}}}, \bibinfo {author} {\bibfnamefont {H.}~\bibnamefont {{Klion}}},\
  and\ \bibinfo {author} {\bibfnamefont {B.}~\bibnamefont {{Paxton}}},\
  }\bibfield  {title} {\bibinfo {title} {{Super-Eddington stellar winds driven
  by near-surface energy deposition}},\ }\href
  {https://doi.org/10.1093/mnras/stw365} {\bibfield  {journal} {\bibinfo
  {journal} {\mnras}\ }\textbf {\bibinfo {volume} {458}},\ \bibinfo {pages}
  {1214} (\bibinfo {year} {2016})},\ \Eprint {https://arxiv.org/abs/1509.06370}
  {arXiv:1509.06370 [astro-ph.SR]} \BibitemShut {NoStop}%
\bibitem [{\citenamefont {{Gilkis}}\ \emph {et~al.}(2021)\citenamefont
  {{Gilkis}}, \citenamefont {{Shenar}}, \citenamefont {{Ramachandran}},
  \citenamefont {{Jermyn}}, \citenamefont {{Mahy}}, \citenamefont {{Oskinova}},
  \citenamefont {{Arcavi}},\ and\ \citenamefont
  {{Sana}}}]{2021MNRAS.503.1884G}%
  \BibitemOpen
  \bibfield  {author} {\bibinfo {author} {\bibfnamefont {A.}~\bibnamefont
  {{Gilkis}}}, \bibinfo {author} {\bibfnamefont {T.}~\bibnamefont {{Shenar}}},
  \bibinfo {author} {\bibfnamefont {V.}~\bibnamefont {{Ramachandran}}},
  \bibinfo {author} {\bibfnamefont {A.~S.}\ \bibnamefont {{Jermyn}}}, \bibinfo
  {author} {\bibfnamefont {L.}~\bibnamefont {{Mahy}}}, \bibinfo {author}
  {\bibfnamefont {L.~M.}\ \bibnamefont {{Oskinova}}}, \bibinfo {author}
  {\bibfnamefont {I.}~\bibnamefont {{Arcavi}}},\ and\ \bibinfo {author}
  {\bibfnamefont {H.}~\bibnamefont {{Sana}}},\ }\bibfield  {title} {\bibinfo
  {title} {{The excess of cool supergiants from contemporary stellar evolution
  models defies the metallicity-independent Humphreys-Davidson limit}},\ }\href
  {https://doi.org/10.1093/mnras/stab383} {\bibfield  {journal} {\bibinfo
  {journal} {\mnras}\ }\textbf {\bibinfo {volume} {503}},\ \bibinfo {pages}
  {1884} (\bibinfo {year} {2021})},\ \Eprint {https://arxiv.org/abs/2102.03102}
  {arXiv:2102.03102 [astro-ph.SR]} \BibitemShut {NoStop}%
\bibitem [{\citenamefont {{Agrawal}}\ \emph
  {et~al.}(2022{\natexlab{b}})\citenamefont {{Agrawal}}, \citenamefont
  {{Sz{\'e}csi}}, \citenamefont {{Stevenson}}, \citenamefont {{Eldridge}},\
  and\ \citenamefont {{Hurley}}}]{2022MNRAS.512.5717A}%
  \BibitemOpen
  \bibfield  {author} {\bibinfo {author} {\bibfnamefont {P.}~\bibnamefont
  {{Agrawal}}}, \bibinfo {author} {\bibfnamefont {D.}~\bibnamefont
  {{Sz{\'e}csi}}}, \bibinfo {author} {\bibfnamefont {S.}~\bibnamefont
  {{Stevenson}}}, \bibinfo {author} {\bibfnamefont {J.~J.}\ \bibnamefont
  {{Eldridge}}},\ and\ \bibinfo {author} {\bibfnamefont {J.}~\bibnamefont
  {{Hurley}}},\ }\bibfield  {title} {\bibinfo {title} {{Explaining the
  differences in massive star models from various simulations}},\ }\href
  {https://doi.org/10.1093/mnras/stac930} {\bibfield  {journal} {\bibinfo
  {journal} {\mnras}\ }\textbf {\bibinfo {volume} {512}},\ \bibinfo {pages}
  {5717} (\bibinfo {year} {2022}{\natexlab{b}})},\ \Eprint
  {https://arxiv.org/abs/2112.02800} {arXiv:2112.02800 [astro-ph.SR]}
  \BibitemShut {NoStop}%
\bibitem [{\citenamefont {{Yusof}}\ \emph {et~al.}(2013)\citenamefont
  {{Yusof}}, \citenamefont {{Hirschi}}, \citenamefont {{Meynet}}, \citenamefont
  {{Crowther}}, \citenamefont {{Ekstr{\"o}m}}, \citenamefont {{Frischknecht}},
  \citenamefont {{Georgy}}, \citenamefont {{Abu Kassim}},\ and\ \citenamefont
  {{Schnurr}}}]{2013MNRAS.433.1114Y}%
  \BibitemOpen
  \bibfield  {author} {\bibinfo {author} {\bibfnamefont {N.}~\bibnamefont
  {{Yusof}}}, \bibinfo {author} {\bibfnamefont {R.}~\bibnamefont {{Hirschi}}},
  \bibinfo {author} {\bibfnamefont {G.}~\bibnamefont {{Meynet}}}, \bibinfo
  {author} {\bibfnamefont {P.~A.}\ \bibnamefont {{Crowther}}}, \bibinfo
  {author} {\bibfnamefont {S.}~\bibnamefont {{Ekstr{\"o}m}}}, \bibinfo {author}
  {\bibfnamefont {U.}~\bibnamefont {{Frischknecht}}}, \bibinfo {author}
  {\bibfnamefont {C.}~\bibnamefont {{Georgy}}}, \bibinfo {author}
  {\bibfnamefont {H.}~\bibnamefont {{Abu Kassim}}},\ and\ \bibinfo {author}
  {\bibfnamefont {O.}~\bibnamefont {{Schnurr}}},\ }\bibfield  {title} {\bibinfo
  {title} {{Evolution and fate of very massive stars}},\ }\href
  {https://doi.org/10.1093/mnras/stt794} {\bibfield  {journal} {\bibinfo
  {journal} {\mnras}\ }\textbf {\bibinfo {volume} {433}},\ \bibinfo {pages}
  {1114} (\bibinfo {year} {2013})},\ \Eprint {https://arxiv.org/abs/1305.2099}
  {arXiv:1305.2099 [astro-ph.SR]} \BibitemShut {NoStop}%
\bibitem [{\citenamefont {{Belczynski}}\ \emph {et~al.}(2022)\citenamefont
  {{Belczynski}}, \citenamefont {{Romagnolo}}, \citenamefont {{Olejak}},
  \citenamefont {{Klencki}}, \citenamefont {{Chattopadhyay}}, \citenamefont
  {{Stevenson}}, \citenamefont {{Coleman Miller}}, \citenamefont {{Lasota}},\
  and\ \citenamefont {{Crowther}}}]{2022ApJ...925...69B}%
  \BibitemOpen
  \bibfield  {author} {\bibinfo {author} {\bibfnamefont {K.}~\bibnamefont
  {{Belczynski}}}, \bibinfo {author} {\bibfnamefont {A.}~\bibnamefont
  {{Romagnolo}}}, \bibinfo {author} {\bibfnamefont {A.}~\bibnamefont
  {{Olejak}}}, \bibinfo {author} {\bibfnamefont {J.}~\bibnamefont {{Klencki}}},
  \bibinfo {author} {\bibfnamefont {D.}~\bibnamefont {{Chattopadhyay}}},
  \bibinfo {author} {\bibfnamefont {S.}~\bibnamefont {{Stevenson}}}, \bibinfo
  {author} {\bibfnamefont {M.}~\bibnamefont {{Coleman Miller}}}, \bibinfo
  {author} {\bibfnamefont {J.~P.}\ \bibnamefont {{Lasota}}},\ and\ \bibinfo
  {author} {\bibfnamefont {P.~A.}\ \bibnamefont {{Crowther}}},\ }\bibfield
  {title} {\bibinfo {title} {{The Uncertain Future of Massive Binaries Obscures
  the Origin of LIGO/Virgo Sources}},\ }\href
  {https://doi.org/10.3847/1538-4357/ac375a} {\bibfield  {journal} {\bibinfo
  {journal} {\apj}\ }\textbf {\bibinfo {volume} {925}},\ \bibinfo {eid} {69}
  (\bibinfo {year} {2022})},\ \Eprint {https://arxiv.org/abs/2108.10885}
  {arXiv:2108.10885 [astro-ph.HE]} \BibitemShut {NoStop}%
\bibitem [{\citenamefont {{Romagnolo}}\ \emph
  {et~al.}(2022{\natexlab{a}})\citenamefont {{Romagnolo}}, \citenamefont
  {{Belczynski}}, \citenamefont {{Klencki}}, \citenamefont {{Agrawal}},
  \citenamefont {{Shenar}},\ and\ \citenamefont
  {{Sz{\'e}csi}}}]{2022arXiv221115800R}%
  \BibitemOpen
  \bibfield  {author} {\bibinfo {author} {\bibfnamefont {A.}~\bibnamefont
  {{Romagnolo}}}, \bibinfo {author} {\bibfnamefont {K.}~\bibnamefont
  {{Belczynski}}}, \bibinfo {author} {\bibfnamefont {J.}~\bibnamefont
  {{Klencki}}}, \bibinfo {author} {\bibfnamefont {P.}~\bibnamefont
  {{Agrawal}}}, \bibinfo {author} {\bibfnamefont {T.}~\bibnamefont
  {{Shenar}}},\ and\ \bibinfo {author} {\bibfnamefont {D.}~\bibnamefont
  {{Sz{\'e}csi}}},\ }\bibfield  {title} {\bibinfo {title} {{The role of stellar
  expansion on the formation of gravitational wave sources}},\ }\href
  {https://doi.org/10.48550/arXiv.2211.15800} {\bibfield  {journal} {\bibinfo
  {journal} {arXiv e-prints}\ ,\ \bibinfo {eid} {arXiv:2211.15800}} (\bibinfo
  {year} {2022}{\natexlab{a}})},\ \Eprint {https://arxiv.org/abs/2211.15800}
  {arXiv:2211.15800 [astro-ph.HE]} \BibitemShut {NoStop}%
\bibitem [{\citenamefont {{Ulrich}}\ and\ \citenamefont
  {{Burger}}(1976{\natexlab{a}})}]{1976ApJ...206..509U}%
  \BibitemOpen
  \bibfield  {author} {\bibinfo {author} {\bibfnamefont {R.~K.}\ \bibnamefont
  {{Ulrich}}}\ and\ \bibinfo {author} {\bibfnamefont {H.~L.}\ \bibnamefont
  {{Burger}}},\ }\bibfield  {title} {\bibinfo {title} {{The accreting component
  of mass-exchange binaries.}},\ }\href {https://doi.org/10.1086/154406}
  {\bibfield  {journal} {\bibinfo  {journal} {\apj}\ }\textbf {\bibinfo
  {volume} {206}},\ \bibinfo {pages} {509} (\bibinfo {year}
  {1976}{\natexlab{a}})}\BibitemShut {NoStop}%
\bibitem [{\citenamefont {{Hellings}}(1983)}]{1983Ap&SS..96...37H}%
  \BibitemOpen
  \bibfield  {author} {\bibinfo {author} {\bibfnamefont {P.}~\bibnamefont
  {{Hellings}}},\ }\bibfield  {title} {\bibinfo {title} {{Phenomenological
  Study of Massive Accretion Stars}},\ }\href
  {https://doi.org/10.1007/BF00661941} {\bibfield  {journal} {\bibinfo
  {journal} {\apss}\ }\textbf {\bibinfo {volume} {96}},\ \bibinfo {pages} {37}
  (\bibinfo {year} {1983})}\BibitemShut {NoStop}%
\bibitem [{\citenamefont {{Vanbeveren}}\ and\ \citenamefont {{De
  Loore}}(1994)}]{1994A&A...290..129V}%
  \BibitemOpen
  \bibfield  {author} {\bibinfo {author} {\bibfnamefont {D.}~\bibnamefont
  {{Vanbeveren}}}\ and\ \bibinfo {author} {\bibfnamefont {C.}~\bibnamefont {{De
  Loore}}},\ }\bibfield  {title} {\bibinfo {title} {{The evolution of the mass
  gainer in massive close binaries.}},\ }\href@noop {} {\bibfield  {journal}
  {\bibinfo  {journal} {\aap}\ }\textbf {\bibinfo {volume} {290}},\ \bibinfo
  {pages} {129} (\bibinfo {year} {1994})}\BibitemShut {NoStop}%
\bibitem [{\citenamefont {{Pols}}(1994)}]{1994A&A...290..119P}%
  \BibitemOpen
  \bibfield  {author} {\bibinfo {author} {\bibfnamefont {O.~R.}\ \bibnamefont
  {{Pols}}},\ }\bibfield  {title} {\bibinfo {title} {{Case A evolution of
  massive close binaries: formation of contact systems and possible reversal of
  the supernova order}},\ }\href@noop {} {\bibfield  {journal} {\bibinfo
  {journal} {\aap}\ }\textbf {\bibinfo {volume} {290}},\ \bibinfo {pages} {119}
  (\bibinfo {year} {1994})}\BibitemShut {NoStop}%
\bibitem [{\citenamefont {{Gazeas}}\ and\ \citenamefont
  {{St{\c{e}}pie{\'n}}}(2008)}]{2008MNRAS.390.1577G}%
  \BibitemOpen
  \bibfield  {author} {\bibinfo {author} {\bibfnamefont {K.}~\bibnamefont
  {{Gazeas}}}\ and\ \bibinfo {author} {\bibfnamefont {K.}~\bibnamefont
  {{St{\c{e}}pie{\'n}}}},\ }\bibfield  {title} {\bibinfo {title} {{Angular
  momentum and mass evolution of contact binaries}},\ }\href
  {https://doi.org/10.1111/j.1365-2966.2008.13844.x} {\bibfield  {journal}
  {\bibinfo  {journal} {\mnras}\ }\textbf {\bibinfo {volume} {390}},\ \bibinfo
  {pages} {1577} (\bibinfo {year} {2008})},\ \Eprint
  {https://arxiv.org/abs/0803.0212} {arXiv:0803.0212 [astro-ph]} \BibitemShut
  {NoStop}%
\bibitem [{\citenamefont {{Ge}}\ \emph
  {et~al.}(2020{\natexlab{a}})\citenamefont {{Ge}}, \citenamefont {{Webbink}},\
  and\ \citenamefont {{Han}}}]{2020ApJS..249....9G}%
  \BibitemOpen
  \bibfield  {author} {\bibinfo {author} {\bibfnamefont {H.}~\bibnamefont
  {{Ge}}}, \bibinfo {author} {\bibfnamefont {R.~F.}\ \bibnamefont
  {{Webbink}}},\ and\ \bibinfo {author} {\bibfnamefont {Z.}~\bibnamefont
  {{Han}}},\ }\bibfield  {title} {\bibinfo {title} {{The Thermal Equilibrium
  Mass-loss Model and Its Applications in Binary Evolution}},\ }\href
  {https://doi.org/10.3847/1538-4365/ab98f6} {\bibfield  {journal} {\bibinfo
  {journal} {\apjs}\ }\textbf {\bibinfo {volume} {249}},\ \bibinfo {eid} {9}
  (\bibinfo {year} {2020}{\natexlab{a}})},\ \Eprint
  {https://arxiv.org/abs/2006.00774} {arXiv:2006.00774 [astro-ph.SR]}
  \BibitemShut {NoStop}%
\bibitem [{\citenamefont {{Eggleton}}(1983)}]{1983ApJ...268..368E}%
  \BibitemOpen
  \bibfield  {author} {\bibinfo {author} {\bibfnamefont {P.~P.}\ \bibnamefont
  {{Eggleton}}},\ }\bibfield  {title} {\bibinfo {title} {{Aproximations to the
  radii of Roche lobes.}},\ }\href {https://doi.org/10.1086/160960} {\bibfield
  {journal} {\bibinfo  {journal} {\apj}\ }\textbf {\bibinfo {volume} {268}},\
  \bibinfo {pages} {368} (\bibinfo {year} {1983})}\BibitemShut {NoStop}%
\bibitem [{\citenamefont {{Soberman}}\ \emph
  {et~al.}(1997{\natexlab{a}})\citenamefont {{Soberman}}, \citenamefont
  {{Phinney}},\ and\ \citenamefont {{van den Heuvel}}}]{1997A&A...327..620S}%
  \BibitemOpen
  \bibfield  {author} {\bibinfo {author} {\bibfnamefont {G.~E.}\ \bibnamefont
  {{Soberman}}}, \bibinfo {author} {\bibfnamefont {E.~S.}\ \bibnamefont
  {{Phinney}}},\ and\ \bibinfo {author} {\bibfnamefont {E.~P.~J.}\ \bibnamefont
  {{van den Heuvel}}},\ }\bibfield  {title} {\bibinfo {title} {{Stability
  criteria for mass transfer in binary stellar evolution.}},\ }\href@noop {}
  {\bibfield  {journal} {\bibinfo  {journal} {\aap}\ }\textbf {\bibinfo
  {volume} {327}},\ \bibinfo {pages} {620} (\bibinfo {year}
  {1997}{\natexlab{a}})},\ \Eprint {https://arxiv.org/abs/astro-ph/9703016}
  {astro-ph/9703016} \BibitemShut {NoStop}%
\bibitem [{\citenamefont {{Webbink}}(1985)}]{1985ibs..book...39W}%
  \BibitemOpen
  \bibfield  {author} {\bibinfo {author} {\bibfnamefont {R.~F.}\ \bibnamefont
  {{Webbink}}},\ }\bibfield  {title} {\bibinfo {title} {{Stellar evolution and
  binaries}},\ }in\ \href@noop {} {\emph {\bibinfo {booktitle} {Interacting
  Binary Stars}}},\ \bibinfo {editor} {edited by\ \bibinfo {editor}
  {\bibfnamefont {J.~E.}\ \bibnamefont {{Pringle}}}\ and\ \bibinfo {editor}
  {\bibfnamefont {R.~A.}\ \bibnamefont {{Wade}}}}\ (\bibinfo {year} {1985})\
  p.~\bibinfo {pages} {39}\BibitemShut {NoStop}%
\bibitem [{\citenamefont {{Hjellming}}\ and\ \citenamefont
  {{Webbink}}(1987)}]{1987ApJ...318..794H}%
  \BibitemOpen
  \bibfield  {author} {\bibinfo {author} {\bibfnamefont {M.~S.}\ \bibnamefont
  {{Hjellming}}}\ and\ \bibinfo {author} {\bibfnamefont {R.~F.}\ \bibnamefont
  {{Webbink}}},\ }\bibfield  {title} {\bibinfo {title} {{Thresholds for Rapid
  Mass Transfer in Binary System. I. Polytropic Models}},\ }\href
  {https://doi.org/10.1086/165412} {\bibfield  {journal} {\bibinfo  {journal}
  {\apj}\ }\textbf {\bibinfo {volume} {318}},\ \bibinfo {pages} {794} (\bibinfo
  {year} {1987})}\BibitemShut {NoStop}%
\bibitem [{\citenamefont {{Kalogera}}\ and\ \citenamefont
  {{Webbink}}(1996)}]{1996ApJ...458..301K}%
  \BibitemOpen
  \bibfield  {author} {\bibinfo {author} {\bibfnamefont {V.}~\bibnamefont
  {{Kalogera}}}\ and\ \bibinfo {author} {\bibfnamefont {R.~F.}\ \bibnamefont
  {{Webbink}}},\ }\bibfield  {title} {\bibinfo {title} {{Formation of Low-Mass
  X-Ray Binaries. I. Constraints on Hydrogen-rich Donors at the Onset of the
  X-Ray Phase}},\ }\href {https://doi.org/10.1086/176813} {\bibfield  {journal}
  {\bibinfo  {journal} {\apj}\ }\textbf {\bibinfo {volume} {458}},\ \bibinfo
  {pages} {301} (\bibinfo {year} {1996})},\ \Eprint
  {https://arxiv.org/abs/astro-ph/9508072} {arXiv:astro-ph/9508072 [astro-ph]}
  \BibitemShut {NoStop}%
\bibitem [{\citenamefont {{Ge}}\ \emph {et~al.}(2010)\citenamefont {{Ge}},
  \citenamefont {{Hjellming}}, \citenamefont {{Webbink}}, \citenamefont
  {{Chen}},\ and\ \citenamefont {{Han}}}]{2010ApJ...717..724G}%
  \BibitemOpen
  \bibfield  {author} {\bibinfo {author} {\bibfnamefont {H.}~\bibnamefont
  {{Ge}}}, \bibinfo {author} {\bibfnamefont {M.~S.}\ \bibnamefont
  {{Hjellming}}}, \bibinfo {author} {\bibfnamefont {R.~F.}\ \bibnamefont
  {{Webbink}}}, \bibinfo {author} {\bibfnamefont {X.}~\bibnamefont {{Chen}}},\
  and\ \bibinfo {author} {\bibfnamefont {Z.}~\bibnamefont {{Han}}},\ }\bibfield
   {title} {\bibinfo {title} {{Adiabatic Mass Loss in Binary Stars. I.
  Computational Method}},\ }\href {https://doi.org/10.1088/0004-637X/717/2/724}
  {\bibfield  {journal} {\bibinfo  {journal} {\apj}\ }\textbf {\bibinfo
  {volume} {717}},\ \bibinfo {pages} {724} (\bibinfo {year} {2010})},\ \Eprint
  {https://arxiv.org/abs/1005.3099} {arXiv:1005.3099 [astro-ph.SR]}
  \BibitemShut {NoStop}%
\bibitem [{\citenamefont {{Paczy{\'n}ski}}(1965)}]{1965AcA....15...89P}%
  \BibitemOpen
  \bibfield  {author} {\bibinfo {author} {\bibfnamefont {B.}~\bibnamefont
  {{Paczy{\'n}ski}}},\ }\bibfield  {title} {\bibinfo {title} {{Cataclysmic
  Variables among Binary Stars I. U Geminorum Stars}},\ }\href@noop {}
  {\bibfield  {journal} {\bibinfo  {journal} {\actaa}\ }\textbf {\bibinfo
  {volume} {15}},\ \bibinfo {pages} {89} (\bibinfo {year} {1965})}\BibitemShut
  {NoStop}%
\bibitem [{\citenamefont {{Plavec}}\ \emph {et~al.}(1973)\citenamefont
  {{Plavec}}, \citenamefont {{Ulrich}},\ and\ \citenamefont
  {{Polidan}}}]{1973PASP...85..769P}%
  \BibitemOpen
  \bibfield  {author} {\bibinfo {author} {\bibfnamefont {M.}~\bibnamefont
  {{Plavec}}}, \bibinfo {author} {\bibfnamefont {R.~K.}\ \bibnamefont
  {{Ulrich}}},\ and\ \bibinfo {author} {\bibfnamefont {R.~S.}\ \bibnamefont
  {{Polidan}}},\ }\bibfield  {title} {\bibinfo {title} {{Mass Loss from
  Convective Envelopes of Giant Components of Close Binary Systems}},\ }\href
  {https://doi.org/10.1086/129546} {\bibfield  {journal} {\bibinfo  {journal}
  {\pasp}\ }\textbf {\bibinfo {volume} {85}},\ \bibinfo {pages} {769} (\bibinfo
  {year} {1973})}\BibitemShut {NoStop}%
\bibitem [{\citenamefont {{Webbink}}(1977)}]{1977ApJ...211..486W}%
  \BibitemOpen
  \bibfield  {author} {\bibinfo {author} {\bibfnamefont {R.~F.}\ \bibnamefont
  {{Webbink}}},\ }\bibfield  {title} {\bibinfo {title} {{The evolution of
  low-mass close binary systems. III. 1.50 M sun: unsteady mass loss and
  shrinking secondaries.}},\ }\href {https://doi.org/10.1086/154956} {\bibfield
   {journal} {\bibinfo  {journal} {\apj}\ }\textbf {\bibinfo {volume} {211}},\
  \bibinfo {pages} {486} (\bibinfo {year} {1977})}\BibitemShut {NoStop}%
\bibitem [{\citenamefont {{Ge}}\ \emph
  {et~al.}(2015{\natexlab{a}})\citenamefont {{Ge}}, \citenamefont {{Webbink}},
  \citenamefont {{Chen}},\ and\ \citenamefont {{Han}}}]{2015ApJ...812...40G}%
  \BibitemOpen
  \bibfield  {author} {\bibinfo {author} {\bibfnamefont {H.}~\bibnamefont
  {{Ge}}}, \bibinfo {author} {\bibfnamefont {R.~F.}\ \bibnamefont {{Webbink}}},
  \bibinfo {author} {\bibfnamefont {X.}~\bibnamefont {{Chen}}},\ and\ \bibinfo
  {author} {\bibfnamefont {Z.}~\bibnamefont {{Han}}},\ }\bibfield  {title}
  {\bibinfo {title} {{Adiabatic Mass Loss in Binary Stars. II. From Zero-age
  Main Sequence to the Base of the Giant Branch}},\ }\href
  {https://doi.org/10.1088/0004-637X/812/1/40} {\bibfield  {journal} {\bibinfo
  {journal} {\apj}\ }\textbf {\bibinfo {volume} {812}},\ \bibinfo {eid} {40}
  (\bibinfo {year} {2015}{\natexlab{a}})},\ \Eprint
  {https://arxiv.org/abs/1507.04843} {arXiv:1507.04843 [astro-ph.SR]}
  \BibitemShut {NoStop}%
\bibitem [{\citenamefont {{Woods}}\ and\ \citenamefont
  {{Ivanova}}(2011{\natexlab{a}})}]{2011ApJ...739L..48W}%
  \BibitemOpen
  \bibfield  {author} {\bibinfo {author} {\bibfnamefont {T.~E.}\ \bibnamefont
  {{Woods}}}\ and\ \bibinfo {author} {\bibfnamefont {N.}~\bibnamefont
  {{Ivanova}}},\ }\bibfield  {title} {\bibinfo {title} {{Can We Trust Models
  for Adiabatic Mass Loss?}},\ }\href
  {https://doi.org/10.1088/2041-8205/739/2/L48} {\bibfield  {journal} {\bibinfo
   {journal} {\apjl}\ }\textbf {\bibinfo {volume} {739}},\ \bibinfo {eid} {L48}
  (\bibinfo {year} {2011}{\natexlab{a}})},\ \Eprint
  {https://arxiv.org/abs/1108.2752} {arXiv:1108.2752 [astro-ph.SR]}
  \BibitemShut {NoStop}%
\bibitem [{\citenamefont {{Pavlovskii}}\ and\ \citenamefont
  {{Ivanova}}(2015)}]{2015MNRAS.449.4415P}%
  \BibitemOpen
  \bibfield  {author} {\bibinfo {author} {\bibfnamefont {K.}~\bibnamefont
  {{Pavlovskii}}}\ and\ \bibinfo {author} {\bibfnamefont {N.}~\bibnamefont
  {{Ivanova}}},\ }\bibfield  {title} {\bibinfo {title} {{Mass transfer from
  giant donors}},\ }\href {https://doi.org/10.1093/mnras/stv619} {\bibfield
  {journal} {\bibinfo  {journal} {\mnras}\ }\textbf {\bibinfo {volume} {449}},\
  \bibinfo {pages} {4415} (\bibinfo {year} {2015})},\ \Eprint
  {https://arxiv.org/abs/1410.5109} {arXiv:1410.5109 [astro-ph.SR]}
  \BibitemShut {NoStop}%
\bibitem [{\citenamefont {{Ge}}\ \emph
  {et~al.}(2020{\natexlab{b}})\citenamefont {{Ge}}, \citenamefont {{Webbink}},
  \citenamefont {{Chen}},\ and\ \citenamefont {{Han}}}]{2020ApJ...899..132G}%
  \BibitemOpen
  \bibfield  {author} {\bibinfo {author} {\bibfnamefont {H.}~\bibnamefont
  {{Ge}}}, \bibinfo {author} {\bibfnamefont {R.~F.}\ \bibnamefont {{Webbink}}},
  \bibinfo {author} {\bibfnamefont {X.}~\bibnamefont {{Chen}}},\ and\ \bibinfo
  {author} {\bibfnamefont {Z.}~\bibnamefont {{Han}}},\ }\bibfield  {title}
  {\bibinfo {title} {{Adiabatic Mass Loss in Binary Stars. III. From the Base
  of the Red Giant Branch to the Tip of the Asymptotic Giant Branch}},\ }\href
  {https://doi.org/10.3847/1538-4357/aba7b7} {\bibfield  {journal} {\bibinfo
  {journal} {\apj}\ }\textbf {\bibinfo {volume} {899}},\ \bibinfo {eid} {132}
  (\bibinfo {year} {2020}{\natexlab{b}})},\ \Eprint
  {https://arxiv.org/abs/2007.09848} {arXiv:2007.09848 [astro-ph.SR]}
  \BibitemShut {NoStop}%
\bibitem [{\citenamefont {{Pavlovskii}}\ \emph
  {et~al.}(2017{\natexlab{a}})\citenamefont {{Pavlovskii}}, \citenamefont
  {{Ivanova}}, \citenamefont {{Belczynski}},\ and\ \citenamefont
  {{Van}}}]{2017MNRAS.465.2092P}%
  \BibitemOpen
  \bibfield  {author} {\bibinfo {author} {\bibfnamefont {K.}~\bibnamefont
  {{Pavlovskii}}}, \bibinfo {author} {\bibfnamefont {N.}~\bibnamefont
  {{Ivanova}}}, \bibinfo {author} {\bibfnamefont {K.}~\bibnamefont
  {{Belczynski}}},\ and\ \bibinfo {author} {\bibfnamefont {K.~X.}\ \bibnamefont
  {{Van}}},\ }\bibfield  {title} {\bibinfo {title} {{Stability of mass transfer
  from massive giants: double black hole binary formation and ultraluminous
  X-ray sources}},\ }\href {https://doi.org/10.1093/mnras/stw2786} {\bibfield
  {journal} {\bibinfo  {journal} {\mnras}\ }\textbf {\bibinfo {volume} {465}},\
  \bibinfo {pages} {2092} (\bibinfo {year} {2017}{\natexlab{a}})},\ \Eprint
  {https://arxiv.org/abs/1606.04921} {arXiv:1606.04921 [astro-ph.HE]}
  \BibitemShut {NoStop}%
\bibitem [{\citenamefont {{Temmink}}\ \emph {et~al.}(2023)\citenamefont
  {{Temmink}}, \citenamefont {{Pols}}, \citenamefont {{Justham}}, \citenamefont
  {{Istrate}},\ and\ \citenamefont {{Toonen}}}]{2022arXiv220912707T}%
  \BibitemOpen
  \bibfield  {author} {\bibinfo {author} {\bibfnamefont {K.~D.}\ \bibnamefont
  {{Temmink}}}, \bibinfo {author} {\bibfnamefont {O.~R.}\ \bibnamefont
  {{Pols}}}, \bibinfo {author} {\bibfnamefont {S.}~\bibnamefont {{Justham}}},
  \bibinfo {author} {\bibfnamefont {A.~G.}\ \bibnamefont {{Istrate}}},\ and\
  \bibinfo {author} {\bibfnamefont {S.}~\bibnamefont {{Toonen}}},\ }\bibfield
  {title} {\bibinfo {title} {{Coping with loss. Stability of mass transfer from
  post-main-sequence donor stars}},\ }\href
  {https://doi.org/10.1051/0004-6361/202244137} {\bibfield  {journal} {\bibinfo
   {journal} {\aap}\ }\textbf {\bibinfo {volume} {669}},\ \bibinfo {eid} {A45}
  (\bibinfo {year} {2023})},\ \Eprint {https://arxiv.org/abs/2209.12707}
  {arXiv:2209.12707 [astro-ph.SR]} \BibitemShut {NoStop}%
\bibitem [{\citenamefont {{Quast}}\ \emph {et~al.}(2019)\citenamefont
  {{Quast}}, \citenamefont {{Langer}},\ and\ \citenamefont
  {{Tauris}}}]{2019A&A...628A..19Q}%
  \BibitemOpen
  \bibfield  {author} {\bibinfo {author} {\bibfnamefont {M.}~\bibnamefont
  {{Quast}}}, \bibinfo {author} {\bibfnamefont {N.}~\bibnamefont {{Langer}}},\
  and\ \bibinfo {author} {\bibfnamefont {T.~M.}\ \bibnamefont {{Tauris}}},\
  }\bibfield  {title} {\bibinfo {title} {{Mass transfer on a nuclear timescale
  in models of supergiant and ultra-luminous X-ray binaries}},\ }\href
  {https://doi.org/10.1051/0004-6361/201935453} {\bibfield  {journal} {\bibinfo
   {journal} {\aap}\ }\textbf {\bibinfo {volume} {628}},\ \bibinfo {eid} {A19}
  (\bibinfo {year} {2019})},\ \Eprint {https://arxiv.org/abs/1903.04995}
  {arXiv:1903.04995 [astro-ph.SR]} \BibitemShut {NoStop}%
\bibitem [{\citenamefont {{Morton}}(1960)}]{1960ApJ...132..146M}%
  \BibitemOpen
  \bibfield  {author} {\bibinfo {author} {\bibfnamefont {D.~C.}\ \bibnamefont
  {{Morton}}},\ }\bibfield  {title} {\bibinfo {title} {{Evolutionary Mass
  Exchange in Close Binary Systems.}},\ }\href {https://doi.org/10.1086/146908}
  {\bibfield  {journal} {\bibinfo  {journal} {\apj}\ }\textbf {\bibinfo
  {volume} {132}},\ \bibinfo {pages} {146} (\bibinfo {year}
  {1960})}\BibitemShut {NoStop}%
\bibitem [{\citenamefont {{Hjellming}}(1989)}]{1989PhDT.........7H}%
  \BibitemOpen
  \bibfield  {author} {\bibinfo {author} {\bibfnamefont {M.~S.}\ \bibnamefont
  {{Hjellming}}},\ }\emph {\bibinfo {title} {{Rapid Mass Transfer in Binary
  Systems.}}},\ \href@noop {} {Ph.D. thesis},\ \bibinfo  {school} {University
  of Illinois, Urbana-Champaign} (\bibinfo {year} {1989})\BibitemShut {NoStop}%
\bibitem [{\citenamefont {{Webbink}}(1975)}]{1975PhDT.......165W}%
  \BibitemOpen
  \bibfield  {author} {\bibinfo {author} {\bibfnamefont {R.~F.}\ \bibnamefont
  {{Webbink}}},\ }\emph {\bibinfo {title} {{The evolution of low-mass close
  binary systems}}},\ \href@noop {} {Ph.D. thesis},\ \bibinfo  {school}
  {University of Cambridge, UK} (\bibinfo {year} {1975})\BibitemShut {NoStop}%
\bibitem [{\citenamefont {{Paczynski}}(1976)}]{1976IAUS...73...75P}%
  \BibitemOpen
  \bibfield  {author} {\bibinfo {author} {\bibfnamefont {B.}~\bibnamefont
  {{Paczynski}}},\ }\bibfield  {title} {\bibinfo {title} {{Common Envelope
  Binaries}},\ }in\ \href@noop {} {\emph {\bibinfo {booktitle} {Structure and
  Evolution of Close Binary Systems}}},\ \bibinfo {series} {IAU Symposium},
  Vol.~\bibinfo {volume} {73},\ \bibinfo {editor} {edited by\ \bibinfo {editor}
  {\bibfnamefont {P.}~\bibnamefont {{Eggleton}}}, \bibinfo {editor}
  {\bibfnamefont {S.}~\bibnamefont {{Mitton}}},\ and\ \bibinfo {editor}
  {\bibfnamefont {J.}~\bibnamefont {{Whelan}}}}\ (\bibinfo {year} {1976})\
  p.~\bibinfo {pages} {75}\BibitemShut {NoStop}%
\bibitem [{\citenamefont {{{\noopsort{Van den Heuvel}}{van den
  Heuvel}}}(1976)}]{1976IAUS...73...35V}%
  \BibitemOpen
  \bibfield  {author} {\bibinfo {author} {\bibfnamefont {E.~P.~J.}\
  \bibnamefont {{{\noopsort{Van den Heuvel}}{van den Heuvel}}}},\ }\bibfield
  {title} {\bibinfo {title} {{Late Stages of Close Binary Systems}},\ }in\
  \href@noop {} {\emph {\bibinfo {booktitle} {Structure and Evolution of Close
  Binary Systems}}},\ \bibinfo {series} {IAU Symposium}, Vol.~\bibinfo {volume}
  {73},\ \bibinfo {editor} {edited by\ \bibinfo {editor} {\bibfnamefont
  {P.}~\bibnamefont {{Eggleton}}}, \bibinfo {editor} {\bibfnamefont
  {S.}~\bibnamefont {{Mitton}}},\ and\ \bibinfo {editor} {\bibfnamefont
  {J.}~\bibnamefont {{Whelan}}}}\ (\bibinfo {year} {1976})\ p.~\bibinfo {pages}
  {35}\BibitemShut {NoStop}%
\bibitem [{\citenamefont {{Ivanova}}\ \emph {et~al.}(2013)\citenamefont
  {{Ivanova}}, \citenamefont {{Justham}}, \citenamefont {{Chen}}, \citenamefont
  {{De Marco}}, \citenamefont {{Fryer}}, \citenamefont {{Gaburov}},
  \citenamefont {{Ge}}, \citenamefont {{Glebbeek}}, \citenamefont {{Han}},
  \citenamefont {{Li}}, \citenamefont {{Lu}}, \citenamefont {{Marsh}},
  \citenamefont {{Podsiadlowski}}, \citenamefont {{Potter}}, \citenamefont
  {{Soker}}, \citenamefont {{Taam}}, \citenamefont {{Tauris}}, \citenamefont
  {{van den Heuvel}},\ and\ \citenamefont {{Webbink}}}]{2013A&ARv..21...59I}%
  \BibitemOpen
  \bibfield  {author} {\bibinfo {author} {\bibfnamefont {N.}~\bibnamefont
  {{Ivanova}}}, \bibinfo {author} {\bibfnamefont {S.}~\bibnamefont
  {{Justham}}}, \bibinfo {author} {\bibfnamefont {X.}~\bibnamefont {{Chen}}},
  \bibinfo {author} {\bibfnamefont {O.}~\bibnamefont {{De Marco}}}, \bibinfo
  {author} {\bibfnamefont {C.~L.}\ \bibnamefont {{Fryer}}}, \bibinfo {author}
  {\bibfnamefont {E.}~\bibnamefont {{Gaburov}}}, \bibinfo {author}
  {\bibfnamefont {H.}~\bibnamefont {{Ge}}}, \bibinfo {author} {\bibfnamefont
  {E.}~\bibnamefont {{Glebbeek}}}, \bibinfo {author} {\bibfnamefont
  {Z.}~\bibnamefont {{Han}}}, \bibinfo {author} {\bibfnamefont {X.~D.}\
  \bibnamefont {{Li}}}, \bibinfo {author} {\bibfnamefont {G.}~\bibnamefont
  {{Lu}}}, \bibinfo {author} {\bibfnamefont {T.}~\bibnamefont {{Marsh}}},
  \bibinfo {author} {\bibfnamefont {P.}~\bibnamefont {{Podsiadlowski}}},
  \bibinfo {author} {\bibfnamefont {A.}~\bibnamefont {{Potter}}}, \bibinfo
  {author} {\bibfnamefont {N.}~\bibnamefont {{Soker}}}, \bibinfo {author}
  {\bibfnamefont {R.}~\bibnamefont {{Taam}}}, \bibinfo {author} {\bibfnamefont
  {T.~M.}\ \bibnamefont {{Tauris}}}, \bibinfo {author} {\bibfnamefont
  {E.~P.~J.}\ \bibnamefont {{van den Heuvel}}},\ and\ \bibinfo {author}
  {\bibfnamefont {R.~F.}\ \bibnamefont {{Webbink}}},\ }\bibfield  {title}
  {\bibinfo {title} {{Common envelope evolution: where we stand and how we can
  move forward}},\ }\href {https://doi.org/10.1007/s00159-013-0059-2}
  {\bibfield  {journal} {\bibinfo  {journal} {\aapr}\ }\textbf {\bibinfo
  {volume} {21}},\ \bibinfo {eid} {59} (\bibinfo {year} {2013})},\ \Eprint
  {https://arxiv.org/abs/1209.4302} {arXiv:1209.4302 [astro-ph.HE]}
  \BibitemShut {NoStop}%
\bibitem [{\citenamefont {{Ivanova}}\ \emph
  {et~al.}(2020{\natexlab{a}})\citenamefont {{Ivanova}}, \citenamefont
  {{Justham}},\ and\ \citenamefont {{Ricker}}}]{2020cee..book.....I}%
  \BibitemOpen
  \bibfield  {author} {\bibinfo {author} {\bibfnamefont {N.}~\bibnamefont
  {{Ivanova}}}, \bibinfo {author} {\bibfnamefont {S.}~\bibnamefont
  {{Justham}}},\ and\ \bibinfo {author} {\bibfnamefont {P.}~\bibnamefont
  {{Ricker}}},\ }\href {https://doi.org/10.1088/2514-3433/abb6f0} {\emph
  {\bibinfo {title} {{Common Envelope Evolution}}}}\ (\bibinfo {year}
  {2020})\BibitemShut {NoStop}%
\bibitem [{\citenamefont {{MacLeod}}\ and\ \citenamefont
  {{Loeb}}(2020{\natexlab{a}})}]{2020ApJ...893..106M}%
  \BibitemOpen
  \bibfield  {author} {\bibinfo {author} {\bibfnamefont {M.}~\bibnamefont
  {{MacLeod}}}\ and\ \bibinfo {author} {\bibfnamefont {A.}~\bibnamefont
  {{Loeb}}},\ }\bibfield  {title} {\bibinfo {title} {{Runaway Coalescence of
  Pre-common-envelope Stellar Binaries}},\ }\href
  {https://doi.org/10.3847/1538-4357/ab822e} {\bibfield  {journal} {\bibinfo
  {journal} {\apj}\ }\textbf {\bibinfo {volume} {893}},\ \bibinfo {eid} {106}
  (\bibinfo {year} {2020}{\natexlab{a}})},\ \Eprint
  {https://arxiv.org/abs/1912.05545} {arXiv:1912.05545 [astro-ph.SR]}
  \BibitemShut {NoStop}%
\bibitem [{\citenamefont {{Pejcha}}\ \emph {et~al.}(2017)\citenamefont
  {{Pejcha}}, \citenamefont {{Metzger}}, \citenamefont {{Tyles}},\ and\
  \citenamefont {{Tomida}}}]{2017ApJ...850...59P}%
  \BibitemOpen
  \bibfield  {author} {\bibinfo {author} {\bibfnamefont {O.}~\bibnamefont
  {{Pejcha}}}, \bibinfo {author} {\bibfnamefont {B.~D.}\ \bibnamefont
  {{Metzger}}}, \bibinfo {author} {\bibfnamefont {J.~G.}\ \bibnamefont
  {{Tyles}}},\ and\ \bibinfo {author} {\bibfnamefont {K.}~\bibnamefont
  {{Tomida}}},\ }\bibfield  {title} {\bibinfo {title} {{Pre-explosion Spiral
  Mass Loss of a Binary Star Merger}},\ }\href
  {https://doi.org/10.3847/1538-4357/aa95b9} {\bibfield  {journal} {\bibinfo
  {journal} {\apj}\ }\textbf {\bibinfo {volume} {850}},\ \bibinfo {eid} {59}
  (\bibinfo {year} {2017})},\ \Eprint {https://arxiv.org/abs/1710.02533}
  {arXiv:1710.02533 [astro-ph.SR]} \BibitemShut {NoStop}%
\bibitem [{\citenamefont {{Marchant}}\ \emph
  {et~al.}(2021{\natexlab{a}})\citenamefont {{Marchant}}, \citenamefont
  {{Pappas}}, \citenamefont {{Gallegos-Garcia}}, \citenamefont {{Berry}},
  \citenamefont {{Taam}}, \citenamefont {{Kalogera}},\ and\ \citenamefont
  {{Podsiadlowski}}}]{2021A&A...650A.107M}%
  \BibitemOpen
  \bibfield  {author} {\bibinfo {author} {\bibfnamefont {P.}~\bibnamefont
  {{Marchant}}}, \bibinfo {author} {\bibfnamefont {K.~M.~W.}\ \bibnamefont
  {{Pappas}}}, \bibinfo {author} {\bibfnamefont {M.}~\bibnamefont
  {{Gallegos-Garcia}}}, \bibinfo {author} {\bibfnamefont {C.~P.~L.}\
  \bibnamefont {{Berry}}}, \bibinfo {author} {\bibfnamefont {R.~E.}\
  \bibnamefont {{Taam}}}, \bibinfo {author} {\bibfnamefont {V.}~\bibnamefont
  {{Kalogera}}},\ and\ \bibinfo {author} {\bibfnamefont {P.}~\bibnamefont
  {{Podsiadlowski}}},\ }\bibfield  {title} {\bibinfo {title} {{The role of mass
  transfer and common envelope evolution in the formation of merging binary
  black holes}},\ }\href {https://doi.org/10.1051/0004-6361/202039992}
  {\bibfield  {journal} {\bibinfo  {journal} {\aap}\ }\textbf {\bibinfo
  {volume} {650}},\ \bibinfo {eid} {A107} (\bibinfo {year}
  {2021}{\natexlab{a}})},\ \Eprint {https://arxiv.org/abs/2103.09243}
  {arXiv:2103.09243 [astro-ph.SR]} \BibitemShut {NoStop}%
\bibitem [{\citenamefont {{Ge}}\ \emph {et~al.}(2023)\citenamefont {{Ge}},
  \citenamefont {{Tout}}, \citenamefont {{Chen}}, \citenamefont {{Sarkar}},
  \citenamefont {{Walton}},\ and\ \citenamefont {{Han}}}]{2023ApJ...945....7G}%
  \BibitemOpen
  \bibfield  {author} {\bibinfo {author} {\bibfnamefont {H.}~\bibnamefont
  {{Ge}}}, \bibinfo {author} {\bibfnamefont {C.~A.}\ \bibnamefont {{Tout}}},
  \bibinfo {author} {\bibfnamefont {X.}~\bibnamefont {{Chen}}}, \bibinfo
  {author} {\bibfnamefont {A.}~\bibnamefont {{Sarkar}}}, \bibinfo {author}
  {\bibfnamefont {D.~J.}\ \bibnamefont {{Walton}}},\ and\ \bibinfo {author}
  {\bibfnamefont {Z.}~\bibnamefont {{Han}}},\ }\bibfield  {title} {\bibinfo
  {title} {{Criteria for Dynamical Timescale Mass Transfer of Metal-poor
  Intermediate-mass Stars}},\ }\href {https://doi.org/10.3847/1538-4357/acb7e9}
  {\bibfield  {journal} {\bibinfo  {journal} {\apj}\ }\textbf {\bibinfo
  {volume} {945}},\ \bibinfo {eid} {7} (\bibinfo {year} {2023})},\ \Eprint
  {https://arxiv.org/abs/2302.00183} {arXiv:2302.00183 [astro-ph.SR]}
  \BibitemShut {NoStop}%
\bibitem [{\citenamefont {{Tylenda}}\ \emph {et~al.}(2011)\citenamefont
  {{Tylenda}}, \citenamefont {{Hajduk}}, \citenamefont {{Kami{\'n}ski}},
  \citenamefont {{Udalski}}, \citenamefont {{Soszy{\'n}ski}}, \citenamefont
  {{Szyma{\'n}ski}}, \citenamefont {{Kubiak}}, \citenamefont
  {{Pietrzy{\'n}ski}}, \citenamefont {{Poleski}}, \citenamefont
  {{Wyrzykowski}},\ and\ \citenamefont {{Ulaczyk}}}]{2011A&A...528A.114T}%
  \BibitemOpen
  \bibfield  {author} {\bibinfo {author} {\bibfnamefont {R.}~\bibnamefont
  {{Tylenda}}}, \bibinfo {author} {\bibfnamefont {M.}~\bibnamefont {{Hajduk}}},
  \bibinfo {author} {\bibfnamefont {T.}~\bibnamefont {{Kami{\'n}ski}}},
  \bibinfo {author} {\bibfnamefont {A.}~\bibnamefont {{Udalski}}}, \bibinfo
  {author} {\bibfnamefont {I.}~\bibnamefont {{Soszy{\'n}ski}}}, \bibinfo
  {author} {\bibfnamefont {M.~K.}\ \bibnamefont {{Szyma{\'n}ski}}}, \bibinfo
  {author} {\bibfnamefont {M.}~\bibnamefont {{Kubiak}}}, \bibinfo {author}
  {\bibfnamefont {G.}~\bibnamefont {{Pietrzy{\'n}ski}}}, \bibinfo {author}
  {\bibfnamefont {R.}~\bibnamefont {{Poleski}}}, \bibinfo {author}
  {\bibfnamefont {{\L}.}~\bibnamefont {{Wyrzykowski}}},\ and\ \bibinfo {author}
  {\bibfnamefont {K.}~\bibnamefont {{Ulaczyk}}},\ }\bibfield  {title} {\bibinfo
  {title} {{V1309 Scorpii: merger of a contact binary}},\ }\href
  {https://doi.org/10.1051/0004-6361/201016221} {\bibfield  {journal} {\bibinfo
   {journal} {\aap}\ }\textbf {\bibinfo {volume} {528}},\ \bibinfo {eid} {A114}
  (\bibinfo {year} {2011})},\ \Eprint {https://arxiv.org/abs/1012.0163}
  {arXiv:1012.0163 [astro-ph.SR]} \BibitemShut {NoStop}%
\bibitem [{\citenamefont {{Pejcha}}\ \emph
  {et~al.}(2016{\natexlab{a}})\citenamefont {{Pejcha}}, \citenamefont
  {{Metzger}},\ and\ \citenamefont {{Tomida}}}]{2016MNRAS.461.2527P}%
  \BibitemOpen
  \bibfield  {author} {\bibinfo {author} {\bibfnamefont {O.}~\bibnamefont
  {{Pejcha}}}, \bibinfo {author} {\bibfnamefont {B.~D.}\ \bibnamefont
  {{Metzger}}},\ and\ \bibinfo {author} {\bibfnamefont {K.}~\bibnamefont
  {{Tomida}}},\ }\bibfield  {title} {\bibinfo {title} {{Binary stellar mergers
  with marginally bound ejecta: excretion discs, inflated envelopes, outflows,
  and their luminous transients}},\ }\href
  {https://doi.org/10.1093/mnras/stw1481} {\bibfield  {journal} {\bibinfo
  {journal} {\mnras}\ }\textbf {\bibinfo {volume} {461}},\ \bibinfo {pages}
  {2527} (\bibinfo {year} {2016}{\natexlab{a}})},\ \Eprint
  {https://arxiv.org/abs/1604.07414} {arXiv:1604.07414 [astro-ph.SR]}
  \BibitemShut {NoStop}%
\bibitem [{\citenamefont {{Metzger}}\ and\ \citenamefont
  {{Pejcha}}(2017)}]{2017MNRAS.471.3200M}%
  \BibitemOpen
  \bibfield  {author} {\bibinfo {author} {\bibfnamefont {B.~D.}\ \bibnamefont
  {{Metzger}}}\ and\ \bibinfo {author} {\bibfnamefont {O.}~\bibnamefont
  {{Pejcha}}},\ }\bibfield  {title} {\bibinfo {title} {{Shock-powered light
  curves of luminous red novae as signatures of pre-dynamical mass-loss in
  stellar mergers}},\ }\href {https://doi.org/10.1093/mnras/stx1768} {\bibfield
   {journal} {\bibinfo  {journal} {\mnras}\ }\textbf {\bibinfo {volume}
  {471}},\ \bibinfo {pages} {3200} (\bibinfo {year} {2017})},\ \Eprint
  {https://arxiv.org/abs/1705.03895} {arXiv:1705.03895 [astro-ph.HE]}
  \BibitemShut {NoStop}%
\bibitem [{\citenamefont {{MacLeod}}\ \emph
  {et~al.}(2017{\natexlab{a}})\citenamefont {{MacLeod}}, \citenamefont
  {{Macias}}, \citenamefont {{Ramirez-Ruiz}}, \citenamefont {{Grindlay}},
  \citenamefont {{Batta}},\ and\ \citenamefont
  {{Montes}}}]{2017ApJ...835..282M}%
  \BibitemOpen
  \bibfield  {author} {\bibinfo {author} {\bibfnamefont {M.}~\bibnamefont
  {{MacLeod}}}, \bibinfo {author} {\bibfnamefont {P.}~\bibnamefont {{Macias}}},
  \bibinfo {author} {\bibfnamefont {E.}~\bibnamefont {{Ramirez-Ruiz}}},
  \bibinfo {author} {\bibfnamefont {J.}~\bibnamefont {{Grindlay}}}, \bibinfo
  {author} {\bibfnamefont {A.}~\bibnamefont {{Batta}}},\ and\ \bibinfo {author}
  {\bibfnamefont {G.}~\bibnamefont {{Montes}}},\ }\bibfield  {title} {\bibinfo
  {title} {{Lessons from the Onset of a Common Envelope Episode: the Remarkable
  M31 2015 Luminous Red Nova Outburst}},\ }\href
  {https://doi.org/10.3847/1538-4357/835/2/282} {\bibfield  {journal} {\bibinfo
   {journal} {\apj}\ }\textbf {\bibinfo {volume} {835}},\ \bibinfo {eid} {282}
  (\bibinfo {year} {2017}{\natexlab{a}})},\ \Eprint
  {https://arxiv.org/abs/1605.01493} {arXiv:1605.01493 [astro-ph.SR]}
  \BibitemShut {NoStop}%
\bibitem [{\citenamefont {{Blagorodnova}}\ \emph {et~al.}(2021)\citenamefont
  {{Blagorodnova}}, \citenamefont {{Klencki}}, \citenamefont {{Pejcha}},
  \citenamefont {{Vreeswijk}}, \citenamefont {{Bond}}, \citenamefont
  {{Burdge}}, \citenamefont {{De}}, \citenamefont {{Fremling}}, \citenamefont
  {{Gehrz}}, \citenamefont {{Jencson}}, \citenamefont {{Kasliwal}},
  \citenamefont {{Kupfer}}, \citenamefont {{Lau}}, \citenamefont {{Masci}},\
  and\ \citenamefont {{Rich}}}]{2021A&A...653A.134B}%
  \BibitemOpen
  \bibfield  {author} {\bibinfo {author} {\bibfnamefont {N.}~\bibnamefont
  {{Blagorodnova}}}, \bibinfo {author} {\bibfnamefont {J.}~\bibnamefont
  {{Klencki}}}, \bibinfo {author} {\bibfnamefont {O.}~\bibnamefont {{Pejcha}}},
  \bibinfo {author} {\bibfnamefont {P.~M.}\ \bibnamefont {{Vreeswijk}}},
  \bibinfo {author} {\bibfnamefont {H.~E.}\ \bibnamefont {{Bond}}}, \bibinfo
  {author} {\bibfnamefont {K.~B.}\ \bibnamefont {{Burdge}}}, \bibinfo {author}
  {\bibfnamefont {K.}~\bibnamefont {{De}}}, \bibinfo {author} {\bibfnamefont
  {C.}~\bibnamefont {{Fremling}}}, \bibinfo {author} {\bibfnamefont {R.~D.}\
  \bibnamefont {{Gehrz}}}, \bibinfo {author} {\bibfnamefont {J.~E.}\
  \bibnamefont {{Jencson}}}, \bibinfo {author} {\bibfnamefont {M.~M.}\
  \bibnamefont {{Kasliwal}}}, \bibinfo {author} {\bibfnamefont
  {T.}~\bibnamefont {{Kupfer}}}, \bibinfo {author} {\bibfnamefont {R.~M.}\
  \bibnamefont {{Lau}}}, \bibinfo {author} {\bibfnamefont {F.~J.}\ \bibnamefont
  {{Masci}}},\ and\ \bibinfo {author} {\bibfnamefont {M.~R.}\ \bibnamefont
  {{Rich}}},\ }\bibfield  {title} {\bibinfo {title} {{The luminous red nova AT
  2018bwo in NGC 45 and its binary yellow supergiant progenitor}},\ }\href
  {https://doi.org/10.1051/0004-6361/202140525} {\bibfield  {journal} {\bibinfo
   {journal} {\aap}\ }\textbf {\bibinfo {volume} {653}},\ \bibinfo {eid} {A134}
  (\bibinfo {year} {2021})},\ \Eprint {https://arxiv.org/abs/2102.05662}
  {arXiv:2102.05662 [astro-ph.SR]} \BibitemShut {NoStop}%
\bibitem [{\citenamefont {{Neo}}\ \emph
  {et~al.}(1977{\natexlab{a}})\citenamefont {{Neo}}, \citenamefont {{Miyaji}},
  \citenamefont {{Nomoto}},\ and\ \citenamefont
  {{Sugimoto}}}]{1977PASJ...29..249N}%
  \BibitemOpen
  \bibfield  {author} {\bibinfo {author} {\bibfnamefont {S.}~\bibnamefont
  {{Neo}}}, \bibinfo {author} {\bibfnamefont {S.}~\bibnamefont {{Miyaji}}},
  \bibinfo {author} {\bibfnamefont {K.}~\bibnamefont {{Nomoto}}},\ and\
  \bibinfo {author} {\bibfnamefont {D.}~\bibnamefont {{Sugimoto}}},\ }\bibfield
   {title} {\bibinfo {title} {{Effect of Rapid Mass Accretion onto the
  Main-Sequence Stars}},\ }\href@noop {} {\bibfield  {journal} {\bibinfo
  {journal} {\pasj}\ }\textbf {\bibinfo {volume} {29}},\ \bibinfo {pages} {249}
  (\bibinfo {year} {1977}{\natexlab{a}})}\BibitemShut {NoStop}%
\bibitem [{\citenamefont {{Hellings}}(1984)}]{1984Ap&SS.104...83H}%
  \BibitemOpen
  \bibfield  {author} {\bibinfo {author} {\bibfnamefont {P.}~\bibnamefont
  {{Hellings}}},\ }\bibfield  {title} {\bibinfo {title} {{The post-RLOF
  structure of the secondary components in close binary systems, with an
  application to masses of Wolf-Rayet stars}},\ }\href
  {https://doi.org/10.1007/BF00653994} {\bibfield  {journal} {\bibinfo
  {journal} {\apss}\ }\textbf {\bibinfo {volume} {104}},\ \bibinfo {pages} {83}
  (\bibinfo {year} {1984})}\BibitemShut {NoStop}%
\bibitem [{\citenamefont {{Braun}}\ and\ \citenamefont
  {{Langer}}(1995)}]{1995A&A...297..483B}%
  \BibitemOpen
  \bibfield  {author} {\bibinfo {author} {\bibfnamefont {H.}~\bibnamefont
  {{Braun}}}\ and\ \bibinfo {author} {\bibfnamefont {N.}~\bibnamefont
  {{Langer}}},\ }\bibfield  {title} {\bibinfo {title} {{Effects of accretion
  onto massive main sequence stars.}},\ }\href@noop {} {\bibfield  {journal}
  {\bibinfo  {journal} {\aap}\ }\textbf {\bibinfo {volume} {297}},\ \bibinfo
  {pages} {483} (\bibinfo {year} {1995})}\BibitemShut {NoStop}%
\bibitem [{\citenamefont {{Renzo}}\ and\ \citenamefont
  {{G{\"o}tberg}}(2021)}]{2021ApJ...923..277R}%
  \BibitemOpen
  \bibfield  {author} {\bibinfo {author} {\bibfnamefont {M.}~\bibnamefont
  {{Renzo}}}\ and\ \bibinfo {author} {\bibfnamefont {Y.}~\bibnamefont
  {{G{\"o}tberg}}},\ }\bibfield  {title} {\bibinfo {title} {{Evolution of
  Accretor Stars in Massive Binaries: Broader Implications from Modeling
  {\ensuremath{\zeta}} Ophiuchi}},\ }\href
  {https://doi.org/10.3847/1538-4357/ac29c5} {\bibfield  {journal} {\bibinfo
  {journal} {\apj}\ }\textbf {\bibinfo {volume} {923}},\ \bibinfo {eid} {277}
  (\bibinfo {year} {2021})},\ \Eprint {https://arxiv.org/abs/2107.10933}
  {arXiv:2107.10933 [astro-ph.SR]} \BibitemShut {NoStop}%
\bibitem [{\citenamefont {{Klencki}}\ \emph
  {et~al.}(2021{\natexlab{a}})\citenamefont {{Klencki}}, \citenamefont
  {{Nelemans}}, \citenamefont {{Istrate}},\ and\ \citenamefont
  {{Chruslinska}}}]{2021A&A...645A..54K}%
  \BibitemOpen
  \bibfield  {author} {\bibinfo {author} {\bibfnamefont {J.}~\bibnamefont
  {{Klencki}}}, \bibinfo {author} {\bibfnamefont {G.}~\bibnamefont
  {{Nelemans}}}, \bibinfo {author} {\bibfnamefont {A.~G.}\ \bibnamefont
  {{Istrate}}},\ and\ \bibinfo {author} {\bibfnamefont {M.}~\bibnamefont
  {{Chruslinska}}},\ }\bibfield  {title} {\bibinfo {title} {{It has to be cool:
  Supergiant progenitors of binary black hole mergers from common-envelope
  evolution}},\ }\href {https://doi.org/10.1051/0004-6361/202038707} {\bibfield
   {journal} {\bibinfo  {journal} {\aap}\ }\textbf {\bibinfo {volume} {645}},\
  \bibinfo {eid} {A54} (\bibinfo {year} {2021}{\natexlab{a}})},\ \Eprint
  {https://arxiv.org/abs/2006.11286} {arXiv:2006.11286 [astro-ph.SR]}
  \BibitemShut {NoStop}%
\bibitem [{\citenamefont {{Gallegos-Garcia}}\ \emph
  {et~al.}(2021{\natexlab{a}})\citenamefont {{Gallegos-Garcia}}, \citenamefont
  {{Berry}}, \citenamefont {{Marchant}},\ and\ \citenamefont
  {{Kalogera}}}]{2021ApJ...922..110G}%
  \BibitemOpen
  \bibfield  {author} {\bibinfo {author} {\bibfnamefont {M.}~\bibnamefont
  {{Gallegos-Garcia}}}, \bibinfo {author} {\bibfnamefont {C.~P.~L.}\
  \bibnamefont {{Berry}}}, \bibinfo {author} {\bibfnamefont {P.}~\bibnamefont
  {{Marchant}}},\ and\ \bibinfo {author} {\bibfnamefont {V.}~\bibnamefont
  {{Kalogera}}},\ }\bibfield  {title} {\bibinfo {title} {{Binary Black Hole
  Formation with Detailed Modeling: Stable Mass Transfer Leads to Lower Merger
  Rates}},\ }\href {https://doi.org/10.3847/1538-4357/ac2610} {\bibfield
  {journal} {\bibinfo  {journal} {\apj}\ }\textbf {\bibinfo {volume} {922}},\
  \bibinfo {eid} {110} (\bibinfo {year} {2021}{\natexlab{a}})},\ \Eprint
  {https://arxiv.org/abs/2107.05702} {arXiv:2107.05702 [astro-ph.HE]}
  \BibitemShut {NoStop}%
\bibitem [{\citenamefont {{Bavera}}\ \emph
  {et~al.}(2021{\natexlab{a}})\citenamefont {{Bavera}}, \citenamefont
  {{Fragos}}, \citenamefont {{Zevin}}, \citenamefont {{Berry}}, \citenamefont
  {{Marchant}}, \citenamefont {{Andrews}}, \citenamefont {{Coughlin}},
  \citenamefont {{Dotter}}, \citenamefont {{Kovlakas}}, \citenamefont
  {{Misra}}, \citenamefont {{Serra-Perez}}, \citenamefont {{Qin}},
  \citenamefont {{Rocha}}, \citenamefont {{Rom{\'a}n-Garza}}, \citenamefont
  {{Tran}},\ and\ \citenamefont {{Zapartas}}}]{2021A&A...647A.153B}%
  \BibitemOpen
  \bibfield  {author} {\bibinfo {author} {\bibfnamefont {S.~S.}\ \bibnamefont
  {{Bavera}}}, \bibinfo {author} {\bibfnamefont {T.}~\bibnamefont {{Fragos}}},
  \bibinfo {author} {\bibfnamefont {M.}~\bibnamefont {{Zevin}}}, \bibinfo
  {author} {\bibfnamefont {C.~P.~L.}\ \bibnamefont {{Berry}}}, \bibinfo
  {author} {\bibfnamefont {P.}~\bibnamefont {{Marchant}}}, \bibinfo {author}
  {\bibfnamefont {J.~J.}\ \bibnamefont {{Andrews}}}, \bibinfo {author}
  {\bibfnamefont {S.}~\bibnamefont {{Coughlin}}}, \bibinfo {author}
  {\bibfnamefont {A.}~\bibnamefont {{Dotter}}}, \bibinfo {author}
  {\bibfnamefont {K.}~\bibnamefont {{Kovlakas}}}, \bibinfo {author}
  {\bibfnamefont {D.}~\bibnamefont {{Misra}}}, \bibinfo {author} {\bibfnamefont
  {J.~G.}\ \bibnamefont {{Serra-Perez}}}, \bibinfo {author} {\bibfnamefont
  {Y.}~\bibnamefont {{Qin}}}, \bibinfo {author} {\bibfnamefont {K.~A.}\
  \bibnamefont {{Rocha}}}, \bibinfo {author} {\bibfnamefont {J.}~\bibnamefont
  {{Rom{\'a}n-Garza}}}, \bibinfo {author} {\bibfnamefont {N.~H.}\ \bibnamefont
  {{Tran}}},\ and\ \bibinfo {author} {\bibfnamefont {E.}~\bibnamefont
  {{Zapartas}}},\ }\bibfield  {title} {\bibinfo {title} {{The impact of
  mass-transfer physics on the observable properties of field binary black hole
  populations}},\ }\href {https://doi.org/10.1051/0004-6361/202039804}
  {\bibfield  {journal} {\bibinfo  {journal} {\aap}\ }\textbf {\bibinfo
  {volume} {647}},\ \bibinfo {eid} {A153} (\bibinfo {year}
  {2021}{\natexlab{a}})},\ \Eprint {https://arxiv.org/abs/2010.16333}
  {arXiv:2010.16333 [astro-ph.HE]} \BibitemShut {NoStop}%
\bibitem [{\citenamefont {{van Son}}\ \emph
  {et~al.}(2022{\natexlab{a}})\citenamefont {{van Son}}, \citenamefont {{de
  Mink}}, \citenamefont {{Callister}}, \citenamefont {{Justham}}, \citenamefont
  {{Renzo}}, \citenamefont {{Wagg}}, \citenamefont {{Broekgaarden}},
  \citenamefont {{Kummer}}, \citenamefont {{Pakmor}},\ and\ \citenamefont
  {{Mandel}}}]{2022ApJ...931...17V}%
  \BibitemOpen
  \bibfield  {author} {\bibinfo {author} {\bibfnamefont {L.~A.~C.}\
  \bibnamefont {{van Son}}}, \bibinfo {author} {\bibfnamefont {S.~E.}\
  \bibnamefont {{de Mink}}}, \bibinfo {author} {\bibfnamefont {T.}~\bibnamefont
  {{Callister}}}, \bibinfo {author} {\bibfnamefont {S.}~\bibnamefont
  {{Justham}}}, \bibinfo {author} {\bibfnamefont {M.}~\bibnamefont {{Renzo}}},
  \bibinfo {author} {\bibfnamefont {T.}~\bibnamefont {{Wagg}}}, \bibinfo
  {author} {\bibfnamefont {F.~S.}\ \bibnamefont {{Broekgaarden}}}, \bibinfo
  {author} {\bibfnamefont {F.}~\bibnamefont {{Kummer}}}, \bibinfo {author}
  {\bibfnamefont {R.}~\bibnamefont {{Pakmor}}},\ and\ \bibinfo {author}
  {\bibfnamefont {I.}~\bibnamefont {{Mandel}}},\ }\bibfield  {title} {\bibinfo
  {title} {{The Redshift Evolution of the Binary Black Hole Merger Rate: A
  Weighty Matter}},\ }\href {https://doi.org/10.3847/1538-4357/ac64a3}
  {\bibfield  {journal} {\bibinfo  {journal} {\apj}\ }\textbf {\bibinfo
  {volume} {931}},\ \bibinfo {eid} {17} (\bibinfo {year}
  {2022}{\natexlab{a}})},\ \Eprint {https://arxiv.org/abs/2110.01634}
  {arXiv:2110.01634 [astro-ph.HE]} \BibitemShut {NoStop}%
\bibitem [{\citenamefont {{Reg{\"o}s}}\ \emph {et~al.}(2005)\citenamefont
  {{Reg{\"o}s}}, \citenamefont {{Bailey}},\ and\ \citenamefont
  {{Mardling}}}]{2005MNRAS.358..544R}%
  \BibitemOpen
  \bibfield  {author} {\bibinfo {author} {\bibfnamefont {E.}~\bibnamefont
  {{Reg{\"o}s}}}, \bibinfo {author} {\bibfnamefont {V.~C.}\ \bibnamefont
  {{Bailey}}},\ and\ \bibinfo {author} {\bibfnamefont {R.}~\bibnamefont
  {{Mardling}}},\ }\bibfield  {title} {\bibinfo {title} {{Mass transfer in
  eccentric binary stars}},\ }\href
  {https://doi.org/10.1111/j.1365-2966.2005.08813.x} {\bibfield  {journal}
  {\bibinfo  {journal} {\mnras}\ }\textbf {\bibinfo {volume} {358}},\ \bibinfo
  {pages} {544} (\bibinfo {year} {2005})}\BibitemShut {NoStop}%
\bibitem [{\citenamefont {{Sepinsky}}\ \emph {et~al.}(2007)\citenamefont
  {{Sepinsky}}, \citenamefont {{Willems}}, \citenamefont {{Kalogera}},\ and\
  \citenamefont {{Rasio}}}]{2007ApJ...667.1170S}%
  \BibitemOpen
  \bibfield  {author} {\bibinfo {author} {\bibfnamefont {J.~F.}\ \bibnamefont
  {{Sepinsky}}}, \bibinfo {author} {\bibfnamefont {B.}~\bibnamefont
  {{Willems}}}, \bibinfo {author} {\bibfnamefont {V.}~\bibnamefont
  {{Kalogera}}},\ and\ \bibinfo {author} {\bibfnamefont {F.~A.}\ \bibnamefont
  {{Rasio}}},\ }\bibfield  {title} {\bibinfo {title} {{Interacting Binaries
  with Eccentric Orbits: Secular Orbital Evolution Due to Conservative Mass
  Transfer}},\ }\href {https://doi.org/10.1086/520911} {\bibfield  {journal}
  {\bibinfo  {journal} {\apj}\ }\textbf {\bibinfo {volume} {667}},\ \bibinfo
  {pages} {1170} (\bibinfo {year} {2007})},\ \Eprint
  {https://arxiv.org/abs/0706.4312} {arXiv:0706.4312 [astro-ph]} \BibitemShut
  {NoStop}%
\bibitem [{\citenamefont {{Sepinsky}}\ \emph {et~al.}(2009)\citenamefont
  {{Sepinsky}}, \citenamefont {{Willems}}, \citenamefont {{Kalogera}},\ and\
  \citenamefont {{Rasio}}}]{2009ApJ...702.1387S}%
  \BibitemOpen
  \bibfield  {author} {\bibinfo {author} {\bibfnamefont {J.~F.}\ \bibnamefont
  {{Sepinsky}}}, \bibinfo {author} {\bibfnamefont {B.}~\bibnamefont
  {{Willems}}}, \bibinfo {author} {\bibfnamefont {V.}~\bibnamefont
  {{Kalogera}}},\ and\ \bibinfo {author} {\bibfnamefont {F.~A.}\ \bibnamefont
  {{Rasio}}},\ }\bibfield  {title} {\bibinfo {title} {{Interacting Binaries
  with Eccentric Orbits. II. Secular Orbital Evolution due to Non-conservative
  Mass Transfer}},\ }\href {https://doi.org/10.1088/0004-637X/702/2/1387}
  {\bibfield  {journal} {\bibinfo  {journal} {\apj}\ }\textbf {\bibinfo
  {volume} {702}},\ \bibinfo {pages} {1387} (\bibinfo {year} {2009})},\ \Eprint
  {https://arxiv.org/abs/0903.0621} {arXiv:0903.0621 [astro-ph.SR]}
  \BibitemShut {NoStop}%
\bibitem [{\citenamefont {{Sepinsky}}\ \emph
  {et~al.}(2010{\natexlab{a}})\citenamefont {{Sepinsky}}, \citenamefont
  {{Willems}}, \citenamefont {{Kalogera}},\ and\ \citenamefont
  {{Rasio}}}]{2010ApJ...724..546S}%
  \BibitemOpen
  \bibfield  {author} {\bibinfo {author} {\bibfnamefont {J.~F.}\ \bibnamefont
  {{Sepinsky}}}, \bibinfo {author} {\bibfnamefont {B.}~\bibnamefont
  {{Willems}}}, \bibinfo {author} {\bibfnamefont {V.}~\bibnamefont
  {{Kalogera}}},\ and\ \bibinfo {author} {\bibfnamefont {F.~A.}\ \bibnamefont
  {{Rasio}}},\ }\bibfield  {title} {\bibinfo {title} {{Interacting Binaries
  with Eccentric Orbits. III. Orbital Evolution due to Direct Impact and
  Self-Accretion}},\ }\href {https://doi.org/10.1088/0004-637X/724/1/546}
  {\bibfield  {journal} {\bibinfo  {journal} {\apj}\ }\textbf {\bibinfo
  {volume} {724}},\ \bibinfo {pages} {546} (\bibinfo {year}
  {2010}{\natexlab{a}})},\ \Eprint {https://arxiv.org/abs/1005.0625}
  {arXiv:1005.0625 [astro-ph.SR]} \BibitemShut {NoStop}%
\bibitem [{\citenamefont {{Dosopoulou}}\ and\ \citenamefont
  {{Kalogera}}(2016{\natexlab{a}})}]{2016ApJ...825...70D}%
  \BibitemOpen
  \bibfield  {author} {\bibinfo {author} {\bibfnamefont {F.}~\bibnamefont
  {{Dosopoulou}}}\ and\ \bibinfo {author} {\bibfnamefont {V.}~\bibnamefont
  {{Kalogera}}},\ }\bibfield  {title} {\bibinfo {title} {{Orbital Evolution of
  Mass-transferring Eccentric Binary Systems. I. Phase-dependent Evolution}},\
  }\href {https://doi.org/10.3847/0004-637X/825/1/70} {\bibfield  {journal}
  {\bibinfo  {journal} {\apj}\ }\textbf {\bibinfo {volume} {825}},\ \bibinfo
  {eid} {70} (\bibinfo {year} {2016}{\natexlab{a}})},\ \Eprint
  {https://arxiv.org/abs/1603.06592} {arXiv:1603.06592 [astro-ph.SR]}
  \BibitemShut {NoStop}%
\bibitem [{\citenamefont {{Dosopoulou}}\ and\ \citenamefont
  {{Kalogera}}(2016{\natexlab{b}})}]{2016ApJ...825...71D}%
  \BibitemOpen
  \bibfield  {author} {\bibinfo {author} {\bibfnamefont {F.}~\bibnamefont
  {{Dosopoulou}}}\ and\ \bibinfo {author} {\bibfnamefont {V.}~\bibnamefont
  {{Kalogera}}},\ }\bibfield  {title} {\bibinfo {title} {{Orbital Evolution of
  Mass-transferring Eccentric Binary Systems. II. Secular Evolution}},\ }\href
  {https://doi.org/10.3847/0004-637X/825/1/71} {\bibfield  {journal} {\bibinfo
  {journal} {\apj}\ }\textbf {\bibinfo {volume} {825}},\ \bibinfo {eid} {71}
  (\bibinfo {year} {2016}{\natexlab{b}})},\ \Eprint
  {https://arxiv.org/abs/1603.06593} {arXiv:1603.06593 [astro-ph.SR]}
  \BibitemShut {NoStop}%
\bibitem [{\citenamefont {{Hamers}}\ and\ \citenamefont
  {{Dosopoulou}}(2019)}]{2019ApJ...872..119H}%
  \BibitemOpen
  \bibfield  {author} {\bibinfo {author} {\bibfnamefont {A.~S.}\ \bibnamefont
  {{Hamers}}}\ and\ \bibinfo {author} {\bibfnamefont {F.}~\bibnamefont
  {{Dosopoulou}}},\ }\bibfield  {title} {\bibinfo {title} {{An Analytic Model
  for Mass Transfer in Binaries with Arbitrary Eccentricity, with Applications
  to Triple-star Systems}},\ }\href {https://doi.org/10.3847/1538-4357/ab001d}
  {\bibfield  {journal} {\bibinfo  {journal} {\apj}\ }\textbf {\bibinfo
  {volume} {872}},\ \bibinfo {eid} {119} (\bibinfo {year} {2019})},\ \Eprint
  {https://arxiv.org/abs/1812.05624} {arXiv:1812.05624 [astro-ph.SR]}
  \BibitemShut {NoStop}%
\bibitem [{\citenamefont {{Fabrika}}(2004)}]{2004ASPRv..12....1F}%
  \BibitemOpen
  \bibfield  {author} {\bibinfo {author} {\bibfnamefont {S.}~\bibnamefont
  {{Fabrika}}},\ }\bibfield  {title} {\bibinfo {title} {{The jets and
  supercritical accretion disk in SS433}},\ }\href@noop {} {\bibfield
  {journal} {\bibinfo  {journal} {\apspr}\ }\textbf {\bibinfo {volume} {12}},\
  \bibinfo {pages} {1} (\bibinfo {year} {2004})},\ \Eprint
  {https://arxiv.org/abs/astro-ph/0603390} {arXiv:astro-ph/0603390 [astro-ph]}
  \BibitemShut {NoStop}%
\bibitem [{\citenamefont {{Begelman}}\ \emph {et~al.}(2006)\citenamefont
  {{Begelman}}, \citenamefont {{King}},\ and\ \citenamefont
  {{Pringle}}}]{2006MNRAS.370..399B}%
  \BibitemOpen
  \bibfield  {author} {\bibinfo {author} {\bibfnamefont {M.~C.}\ \bibnamefont
  {{Begelman}}}, \bibinfo {author} {\bibfnamefont {A.~R.}\ \bibnamefont
  {{King}}},\ and\ \bibinfo {author} {\bibfnamefont {J.~E.}\ \bibnamefont
  {{Pringle}}},\ }\bibfield  {title} {\bibinfo {title} {{The nature of SS433
  and the ultraluminous X-ray sources}},\ }\href
  {https://doi.org/10.1111/j.1365-2966.2006.10469.x} {\bibfield  {journal}
  {\bibinfo  {journal} {\mnras}\ }\textbf {\bibinfo {volume} {370}},\ \bibinfo
  {pages} {399} (\bibinfo {year} {2006})},\ \Eprint
  {https://arxiv.org/abs/astro-ph/0604497} {arXiv:astro-ph/0604497 [astro-ph]}
  \BibitemShut {NoStop}%
\bibitem [{\citenamefont {{Cherepashchuk}}\ \emph {et~al.}(2020)\citenamefont
  {{Cherepashchuk}}, \citenamefont {{Postnov}}, \citenamefont {{Molkov}},
  \citenamefont {{Antokhina}},\ and\ \citenamefont
  {{Belinski}}}]{2020NewAR..8901542C}%
  \BibitemOpen
  \bibfield  {author} {\bibinfo {author} {\bibfnamefont {A.}~\bibnamefont
  {{Cherepashchuk}}}, \bibinfo {author} {\bibfnamefont {K.}~\bibnamefont
  {{Postnov}}}, \bibinfo {author} {\bibfnamefont {S.}~\bibnamefont {{Molkov}}},
  \bibinfo {author} {\bibfnamefont {E.}~\bibnamefont {{Antokhina}}},\ and\
  \bibinfo {author} {\bibfnamefont {A.}~\bibnamefont {{Belinski}}},\ }\bibfield
   {title} {\bibinfo {title} {{SS433: A massive X-ray binary in an advanced
  evolutionary stage}},\ }\href {https://doi.org/10.1016/j.newar.2020.101542}
  {\bibfield  {journal} {\bibinfo  {journal} {\nar}\ }\textbf {\bibinfo
  {volume} {89}},\ \bibinfo {eid} {101542} (\bibinfo {year} {2020})},\ \Eprint
  {https://arxiv.org/abs/1905.02938} {arXiv:1905.02938 [astro-ph.HE]}
  \BibitemShut {NoStop}%
\bibitem [{\citenamefont {{Artymowicz}}\ and\ \citenamefont
  {{Lubow}}(1994)}]{1994ApJ...421..651A}%
  \BibitemOpen
  \bibfield  {author} {\bibinfo {author} {\bibfnamefont {P.}~\bibnamefont
  {{Artymowicz}}}\ and\ \bibinfo {author} {\bibfnamefont {S.~H.}\ \bibnamefont
  {{Lubow}}},\ }\bibfield  {title} {\bibinfo {title} {{Dynamics of Binary-Disk
  Interaction. I. Resonances and Disk Gap Sizes}},\ }\href
  {https://doi.org/10.1086/173679} {\bibfield  {journal} {\bibinfo  {journal}
  {\apj}\ }\textbf {\bibinfo {volume} {421}},\ \bibinfo {pages} {651} (\bibinfo
  {year} {1994})}\BibitemShut {NoStop}%
\bibitem [{\citenamefont {{Mokiem}}\ \emph {et~al.}(2007)\citenamefont
  {{Mokiem}}, \citenamefont {{de Koter}}, \citenamefont {{Vink}}, \citenamefont
  {{Puls}}, \citenamefont {{Evans}}, \citenamefont {{Smartt}}, \citenamefont
  {{Crowther}}, \citenamefont {{Herrero}}, \citenamefont {{Langer}},
  \citenamefont {{Lennon}}, \citenamefont {{Najarro}},\ and\ \citenamefont
  {{Villamariz}}}]{2007A&A...473..603M}%
  \BibitemOpen
  \bibfield  {author} {\bibinfo {author} {\bibfnamefont {M.~R.}\ \bibnamefont
  {{Mokiem}}}, \bibinfo {author} {\bibfnamefont {A.}~\bibnamefont {{de
  Koter}}}, \bibinfo {author} {\bibfnamefont {J.~S.}\ \bibnamefont {{Vink}}},
  \bibinfo {author} {\bibfnamefont {J.}~\bibnamefont {{Puls}}}, \bibinfo
  {author} {\bibfnamefont {C.~J.}\ \bibnamefont {{Evans}}}, \bibinfo {author}
  {\bibfnamefont {S.~J.}\ \bibnamefont {{Smartt}}}, \bibinfo {author}
  {\bibfnamefont {P.~A.}\ \bibnamefont {{Crowther}}}, \bibinfo {author}
  {\bibfnamefont {A.}~\bibnamefont {{Herrero}}}, \bibinfo {author}
  {\bibfnamefont {N.}~\bibnamefont {{Langer}}}, \bibinfo {author}
  {\bibfnamefont {D.~J.}\ \bibnamefont {{Lennon}}}, \bibinfo {author}
  {\bibfnamefont {F.}~\bibnamefont {{Najarro}}},\ and\ \bibinfo {author}
  {\bibfnamefont {M.~R.}\ \bibnamefont {{Villamariz}}},\ }\bibfield  {title}
  {\bibinfo {title} {{The empirical metallicity dependence of the mass-loss
  rate of O- and early B-type stars}},\ }\href
  {https://doi.org/10.1051/0004-6361:20077545} {\bibfield  {journal} {\bibinfo
  {journal} {\aap}\ }\textbf {\bibinfo {volume} {473}},\ \bibinfo {pages} {603}
  (\bibinfo {year} {2007})},\ \Eprint {https://arxiv.org/abs/0708.2042}
  {arXiv:0708.2042 [astro-ph]} \BibitemShut {NoStop}%
\bibitem [{\citenamefont {{Vink}}\ and\ \citenamefont
  {{Sander}}(2021)}]{2021MNRAS.504.2051V}%
  \BibitemOpen
  \bibfield  {author} {\bibinfo {author} {\bibfnamefont {J.~S.}\ \bibnamefont
  {{Vink}}}\ and\ \bibinfo {author} {\bibfnamefont {A.~A.~C.}\ \bibnamefont
  {{Sander}}},\ }\bibfield  {title} {\bibinfo {title} {{Metallicity-dependent
  wind parameter predictions for OB stars}},\ }\href
  {https://doi.org/10.1093/mnras/stab90210.48550/arXiv.2103.12736} {\bibfield
  {journal} {\bibinfo  {journal} {\mnras}\ }\textbf {\bibinfo {volume} {504}},\
  \bibinfo {pages} {2051} (\bibinfo {year} {2021})},\ \Eprint
  {https://arxiv.org/abs/2103.12736} {arXiv:2103.12736 [astro-ph.SR]}
  \BibitemShut {NoStop}%
\bibitem [{\citenamefont {{van den Heuvel}}\ \emph
  {et~al.}(2017{\natexlab{a}})\citenamefont {{van den Heuvel}}, \citenamefont
  {{Portegies Zwart}},\ and\ \citenamefont {{de Mink}}}]{2017MNRAS.471.4256V}%
  \BibitemOpen
  \bibfield  {author} {\bibinfo {author} {\bibfnamefont {E.~P.~J.}\
  \bibnamefont {{van den Heuvel}}}, \bibinfo {author} {\bibfnamefont {S.~F.}\
  \bibnamefont {{Portegies Zwart}}},\ and\ \bibinfo {author} {\bibfnamefont
  {S.~E.}\ \bibnamefont {{de Mink}}},\ }\bibfield  {title} {\bibinfo {title}
  {{Forming short-period Wolf-Rayet X-ray binaries and double black holes
  through stable mass transfer}},\ }\href
  {https://doi.org/10.1093/mnras/stx1430} {\bibfield  {journal} {\bibinfo
  {journal} {\mnras}\ }\textbf {\bibinfo {volume} {471}},\ \bibinfo {pages}
  {4256} (\bibinfo {year} {2017}{\natexlab{a}})},\ \Eprint
  {https://arxiv.org/abs/1701.02355} {arXiv:1701.02355 [astro-ph.SR]}
  \BibitemShut {NoStop}%
\bibitem [{\citenamefont {{van Son}}\ \emph
  {et~al.}(2022{\natexlab{b}})\citenamefont {{van Son}}, \citenamefont {{de
  Mink}}, \citenamefont {{Renzo}}, \citenamefont {{Justham}}, \citenamefont
  {{Zapartas}}, \citenamefont {{Breivik}}, \citenamefont {{Callister}},
  \citenamefont {{Farr}},\ and\ \citenamefont
  {{Conroy}}}]{2022ApJ...940..184V}%
  \BibitemOpen
  \bibfield  {author} {\bibinfo {author} {\bibfnamefont {L.~A.~C.}\
  \bibnamefont {{van Son}}}, \bibinfo {author} {\bibfnamefont {S.~E.}\
  \bibnamefont {{de Mink}}}, \bibinfo {author} {\bibfnamefont {M.}~\bibnamefont
  {{Renzo}}}, \bibinfo {author} {\bibfnamefont {S.}~\bibnamefont {{Justham}}},
  \bibinfo {author} {\bibfnamefont {E.}~\bibnamefont {{Zapartas}}}, \bibinfo
  {author} {\bibfnamefont {K.}~\bibnamefont {{Breivik}}}, \bibinfo {author}
  {\bibfnamefont {T.}~\bibnamefont {{Callister}}}, \bibinfo {author}
  {\bibfnamefont {W.~M.}\ \bibnamefont {{Farr}}},\ and\ \bibinfo {author}
  {\bibfnamefont {C.}~\bibnamefont {{Conroy}}},\ }\bibfield  {title} {\bibinfo
  {title} {{No Peaks without Valleys: The Stable Mass Transfer Channel for
  Gravitational-wave Sources in Light of the Neutron Star-Black Hole Mass
  Gap}},\ }\href {https://doi.org/10.3847/1538-4357/ac9b0a} {\bibfield
  {journal} {\bibinfo  {journal} {\apj}\ }\textbf {\bibinfo {volume} {940}},\
  \bibinfo {eid} {184} (\bibinfo {year} {2022}{\natexlab{b}})},\ \Eprint
  {https://arxiv.org/abs/2209.13609} {arXiv:2209.13609 [astro-ph.HE]}
  \BibitemShut {NoStop}%
\bibitem [{\citenamefont {{Schneider}}\ \emph {et~al.}(2019)\citenamefont
  {{Schneider}}, \citenamefont {{Ohlmann}}, \citenamefont {{Podsiadlowski}},
  \citenamefont {{R{\"o}pke}}, \citenamefont {{Balbus}}, \citenamefont
  {{Pakmor}},\ and\ \citenamefont {{Springel}}}]{2019Natur.574..211S}%
  \BibitemOpen
  \bibfield  {author} {\bibinfo {author} {\bibfnamefont {F.~R.~N.}\
  \bibnamefont {{Schneider}}}, \bibinfo {author} {\bibfnamefont {S.~T.}\
  \bibnamefont {{Ohlmann}}}, \bibinfo {author} {\bibfnamefont {P.}~\bibnamefont
  {{Podsiadlowski}}}, \bibinfo {author} {\bibfnamefont {F.~K.}\ \bibnamefont
  {{R{\"o}pke}}}, \bibinfo {author} {\bibfnamefont {S.~A.}\ \bibnamefont
  {{Balbus}}}, \bibinfo {author} {\bibfnamefont {R.}~\bibnamefont {{Pakmor}}},\
  and\ \bibinfo {author} {\bibfnamefont {V.}~\bibnamefont {{Springel}}},\
  }\bibfield  {title} {\bibinfo {title} {{Stellar mergers as the origin of
  magnetic massive stars}},\ }\href {https://doi.org/10.1038/s41586-019-1621-5}
  {\bibfield  {journal} {\bibinfo  {journal} {\nat}\ }\textbf {\bibinfo
  {volume} {574}},\ \bibinfo {pages} {211} (\bibinfo {year} {2019})},\ \Eprint
  {https://arxiv.org/abs/1910.14058} {arXiv:1910.14058 [astro-ph.SR]}
  \BibitemShut {NoStop}%
\bibitem [{\citenamefont {{Lombardi}}\ \emph {et~al.}(1996)\citenamefont
  {{Lombardi}}, \citenamefont {{Rasio}},\ and\ \citenamefont
  {{Shapiro}}}]{1996ApJ...468..797L}%
  \BibitemOpen
  \bibfield  {author} {\bibinfo {author} {\bibfnamefont {J.}~\bibnamefont
  {{Lombardi}}, \bibfnamefont {James~C.}}, \bibinfo {author} {\bibfnamefont
  {F.~A.}\ \bibnamefont {{Rasio}}},\ and\ \bibinfo {author} {\bibfnamefont
  {S.~L.}\ \bibnamefont {{Shapiro}}},\ }\bibfield  {title} {\bibinfo {title}
  {{Collisions of Main-Sequence Stars and the Formation of Blue Stragglers in
  Globular Clusters}},\ }\href {https://doi.org/10.1086/177736} {\bibfield
  {journal} {\bibinfo  {journal} {\apj}\ }\textbf {\bibinfo {volume} {468}},\
  \bibinfo {pages} {797} (\bibinfo {year} {1996})},\ \Eprint
  {https://arxiv.org/abs/astro-ph/9511074} {arXiv:astro-ph/9511074 [astro-ph]}
  \BibitemShut {NoStop}%
\bibitem [{\citenamefont {{Glebbeek}}\ \emph {et~al.}(2013)\citenamefont
  {{Glebbeek}}, \citenamefont {{Gaburov}}, \citenamefont {{Portegies Zwart}},\
  and\ \citenamefont {{Pols}}}]{2013MNRAS.434.3497G}%
  \BibitemOpen
  \bibfield  {author} {\bibinfo {author} {\bibfnamefont {E.}~\bibnamefont
  {{Glebbeek}}}, \bibinfo {author} {\bibfnamefont {E.}~\bibnamefont
  {{Gaburov}}}, \bibinfo {author} {\bibfnamefont {S.}~\bibnamefont {{Portegies
  Zwart}}},\ and\ \bibinfo {author} {\bibfnamefont {O.~R.}\ \bibnamefont
  {{Pols}}},\ }\bibfield  {title} {\bibinfo {title} {{Structure and evolution
  of high-mass stellar mergers}},\ }\href
  {https://doi.org/10.1093/mnras/stt1268} {\bibfield  {journal} {\bibinfo
  {journal} {\mnras}\ }\textbf {\bibinfo {volume} {434}},\ \bibinfo {pages}
  {3497} (\bibinfo {year} {2013})},\ \Eprint {https://arxiv.org/abs/1307.2445}
  {arXiv:1307.2445 [astro-ph.SR]} \BibitemShut {NoStop}%
\bibitem [{\citenamefont {{Wang}}\ \emph {et~al.}(2022)\citenamefont {{Wang}},
  \citenamefont {{Langer}}, \citenamefont {{Schootemeijer}}, \citenamefont
  {{Milone}}, \citenamefont {{Hastings}}, \citenamefont {{Xu}}, \citenamefont
  {{Bodensteiner}}, \citenamefont {{Sana}}, \citenamefont {{Castro}},
  \citenamefont {{Lennon}}, \citenamefont {{Marchant}}, \citenamefont {{de
  Koter}},\ and\ \citenamefont {{de Mink}}}]{2022NatAs...6..480W}%
  \BibitemOpen
  \bibfield  {author} {\bibinfo {author} {\bibfnamefont {C.}~\bibnamefont
  {{Wang}}}, \bibinfo {author} {\bibfnamefont {N.}~\bibnamefont {{Langer}}},
  \bibinfo {author} {\bibfnamefont {A.}~\bibnamefont {{Schootemeijer}}},
  \bibinfo {author} {\bibfnamefont {A.}~\bibnamefont {{Milone}}}, \bibinfo
  {author} {\bibfnamefont {B.}~\bibnamefont {{Hastings}}}, \bibinfo {author}
  {\bibfnamefont {X.-T.}\ \bibnamefont {{Xu}}}, \bibinfo {author}
  {\bibfnamefont {J.}~\bibnamefont {{Bodensteiner}}}, \bibinfo {author}
  {\bibfnamefont {H.}~\bibnamefont {{Sana}}}, \bibinfo {author} {\bibfnamefont
  {N.}~\bibnamefont {{Castro}}}, \bibinfo {author} {\bibfnamefont {D.~J.}\
  \bibnamefont {{Lennon}}}, \bibinfo {author} {\bibfnamefont {P.}~\bibnamefont
  {{Marchant}}}, \bibinfo {author} {\bibfnamefont {A.}~\bibnamefont {{de
  Koter}}},\ and\ \bibinfo {author} {\bibfnamefont {S.~E.}\ \bibnamefont {{de
  Mink}}},\ }\bibfield  {title} {\bibinfo {title} {{Stellar mergers as the
  origin of the blue main-sequence band in young star clusters}},\ }\href
  {https://doi.org/10.1038/s41550-021-01597-5} {\bibfield  {journal} {\bibinfo
  {journal} {Nature Astronomy}\ }\textbf {\bibinfo {volume} {6}},\ \bibinfo
  {pages} {480} (\bibinfo {year} {2022})},\ \Eprint
  {https://arxiv.org/abs/2202.05552} {arXiv:2202.05552 [astro-ph.SR]}
  \BibitemShut {NoStop}%
\bibitem [{\citenamefont {{Thorne}}\ and\ \citenamefont
  {{Zytkow}}(1977)}]{1977ApJ...212..832T}%
  \BibitemOpen
  \bibfield  {author} {\bibinfo {author} {\bibfnamefont {K.~S.}\ \bibnamefont
  {{Thorne}}}\ and\ \bibinfo {author} {\bibfnamefont {A.~N.}\ \bibnamefont
  {{Zytkow}}},\ }\bibfield  {title} {\bibinfo {title} {{Stars with degenerate
  neutron cores. I. Structure of equilibrium models.}},\ }\href
  {https://doi.org/10.1086/155109} {\bibfield  {journal} {\bibinfo  {journal}
  {\apj}\ }\textbf {\bibinfo {volume} {212}},\ \bibinfo {pages} {832} (\bibinfo
  {year} {1977})}\BibitemShut {NoStop}%
\bibitem [{\citenamefont {{Podsiadlowski}}\ \emph {et~al.}(1995)\citenamefont
  {{Podsiadlowski}}, \citenamefont {{Cannon}},\ and\ \citenamefont
  {{Rees}}}]{1995MNRAS.274..485P}%
  \BibitemOpen
  \bibfield  {author} {\bibinfo {author} {\bibfnamefont {P.}~\bibnamefont
  {{Podsiadlowski}}}, \bibinfo {author} {\bibfnamefont {R.~C.}\ \bibnamefont
  {{Cannon}}},\ and\ \bibinfo {author} {\bibfnamefont {M.~J.}\ \bibnamefont
  {{Rees}}},\ }\bibfield  {title} {\bibinfo {title} {{The evolution and final
  fate of massive Thorne-Zytkow objects}},\ }\href
  {https://doi.org/10.1093/mnras/274.2.485} {\bibfield  {journal} {\bibinfo
  {journal} {\mnras}\ }\textbf {\bibinfo {volume} {274}},\ \bibinfo {pages}
  {485} (\bibinfo {year} {1995})}\BibitemShut {NoStop}%
\bibitem [{\citenamefont {{Ablimit}}\ \emph {et~al.}(2022)\citenamefont
  {{Ablimit}}, \citenamefont {{Podsiadlowski}}, \citenamefont {{Hirai}},\ and\
  \citenamefont {{Wicker}}}]{2022MNRAS.513.4802A}%
  \BibitemOpen
  \bibfield  {author} {\bibinfo {author} {\bibfnamefont {I.}~\bibnamefont
  {{Ablimit}}}, \bibinfo {author} {\bibfnamefont {P.}~\bibnamefont
  {{Podsiadlowski}}}, \bibinfo {author} {\bibfnamefont {R.}~\bibnamefont
  {{Hirai}}},\ and\ \bibinfo {author} {\bibfnamefont {J.}~\bibnamefont
  {{Wicker}}},\ }\bibfield  {title} {\bibinfo {title} {{Stellar
  core-merger-induced collapse: new formation pathways for black holes,
  Thorne-{\.Z}ytkow objects, magnetars, and superluminous supernovae}},\ }\href
  {https://doi.org/10.1093/mnras/stac631} {\bibfield  {journal} {\bibinfo
  {journal} {\mnras}\ }\textbf {\bibinfo {volume} {513}},\ \bibinfo {pages}
  {4802} (\bibinfo {year} {2022})},\ \Eprint {https://arxiv.org/abs/2108.08430}
  {arXiv:2108.08430 [astro-ph.HE]} \BibitemShut {NoStop}%
\bibitem [{\citenamefont {{Di Carlo}}\ \emph
  {et~al.}(2020{\natexlab{a}})\citenamefont {{Di Carlo}}, \citenamefont
  {{Mapelli}}, \citenamefont {{Bouffanais}}, \citenamefont {{Giacobbo}},
  \citenamefont {{Santoliquido}}, \citenamefont {{Bressan}}, \citenamefont
  {{Spera}},\ and\ \citenamefont {{Haardt}}}]{2020MNRAS.497.1043D}%
  \BibitemOpen
  \bibfield  {author} {\bibinfo {author} {\bibfnamefont {U.~N.}\ \bibnamefont
  {{Di Carlo}}}, \bibinfo {author} {\bibfnamefont {M.}~\bibnamefont
  {{Mapelli}}}, \bibinfo {author} {\bibfnamefont {Y.}~\bibnamefont
  {{Bouffanais}}}, \bibinfo {author} {\bibfnamefont {N.}~\bibnamefont
  {{Giacobbo}}}, \bibinfo {author} {\bibfnamefont {F.}~\bibnamefont
  {{Santoliquido}}}, \bibinfo {author} {\bibfnamefont {A.}~\bibnamefont
  {{Bressan}}}, \bibinfo {author} {\bibfnamefont {M.}~\bibnamefont {{Spera}}},\
  and\ \bibinfo {author} {\bibfnamefont {F.}~\bibnamefont {{Haardt}}},\
  }\bibfield  {title} {\bibinfo {title} {{Binary black holes in the pair
  instability mass gap}},\ }\href {https://doi.org/10.1093/mnras/staa1997}
  {\bibfield  {journal} {\bibinfo  {journal} {\mnras}\ }\textbf {\bibinfo
  {volume} {497}},\ \bibinfo {pages} {1043} (\bibinfo {year}
  {2020}{\natexlab{a}})},\ \Eprint {https://arxiv.org/abs/1911.01434}
  {arXiv:1911.01434 [astro-ph.HE]} \BibitemShut {NoStop}%
\bibitem [{\citenamefont {{Iben}}\ and\ \citenamefont
  {{Livio}}(1993)}]{1993PASP..105.1373I}%
  \BibitemOpen
  \bibfield  {author} {\bibinfo {author} {\bibfnamefont {J.}~\bibnamefont
  {{Iben}}, \bibfnamefont {Icko}}\ and\ \bibinfo {author} {\bibfnamefont
  {M.}~\bibnamefont {{Livio}}},\ }\bibfield  {title} {\bibinfo {title} {{Common
  Envelopes in Binary Star Evolution}},\ }\href
  {https://doi.org/10.1086/133321} {\bibfield  {journal} {\bibinfo  {journal}
  {\pasp}\ }\textbf {\bibinfo {volume} {105}},\ \bibinfo {pages} {1373}
  (\bibinfo {year} {1993})}\BibitemShut {NoStop}%
\bibitem [{\citenamefont {{Taam}}\ and\ \citenamefont
  {{Sandquist}}(2000)}]{2000ARA&A..38..113T}%
  \BibitemOpen
  \bibfield  {author} {\bibinfo {author} {\bibfnamefont {R.~E.}\ \bibnamefont
  {{Taam}}}\ and\ \bibinfo {author} {\bibfnamefont {E.~L.}\ \bibnamefont
  {{Sandquist}}},\ }\bibfield  {title} {\bibinfo {title} {{Common Envelope
  Evolution of Massive Binary Stars}},\ }\href
  {https://doi.org/10.1146/annurev.astro.38.1.113} {\bibfield  {journal}
  {\bibinfo  {journal} {\araa}\ }\textbf {\bibinfo {volume} {38}},\ \bibinfo
  {pages} {113} (\bibinfo {year} {2000})}\BibitemShut {NoStop}%
\bibitem [{\citenamefont
  {{Webbink}}(1984{\natexlab{a}})}]{1984ApJ...277..355W}%
  \BibitemOpen
  \bibfield  {author} {\bibinfo {author} {\bibfnamefont {R.~F.}\ \bibnamefont
  {{Webbink}}},\ }\bibfield  {title} {\bibinfo {title} {{Double white dwarfs as
  progenitors of R Coronae Borealis stars and type I supernovae.}},\ }\href
  {https://doi.org/10.1086/161701} {\bibfield  {journal} {\bibinfo  {journal}
  {\apj}\ }\textbf {\bibinfo {volume} {277}},\ \bibinfo {pages} {355} (\bibinfo
  {year} {1984}{\natexlab{a}})}\BibitemShut {NoStop}%
\bibitem [{\citenamefont {{Lipunov}}\ \emph {et~al.}(1996)\citenamefont
  {{Lipunov}}, \citenamefont {{Postnov}},\ and\ \citenamefont
  {{Prokhorov}}}]{1996A&A...310..489L}%
  \BibitemOpen
  \bibfield  {author} {\bibinfo {author} {\bibfnamefont {V.~M.}\ \bibnamefont
  {{Lipunov}}}, \bibinfo {author} {\bibfnamefont {K.~A.}\ \bibnamefont
  {{Postnov}}},\ and\ \bibinfo {author} {\bibfnamefont {M.~E.}\ \bibnamefont
  {{Prokhorov}}},\ }\bibfield  {title} {\bibinfo {title} {{The Scenario
  Machine: restrictions on key parameters of binary evolution.}},\ }\href@noop
  {} {\bibfield  {journal} {\bibinfo  {journal} {\aap}\ }\textbf {\bibinfo
  {volume} {310}},\ \bibinfo {pages} {489} (\bibinfo {year}
  {1996})}\BibitemShut {NoStop}%
\bibitem [{\citenamefont {{Portegies Zwart}}\ and\ \citenamefont
  {{Verbunt}}(1996)}]{1996A&A...309..179P}%
  \BibitemOpen
  \bibfield  {author} {\bibinfo {author} {\bibfnamefont {S.~F.}\ \bibnamefont
  {{Portegies Zwart}}}\ and\ \bibinfo {author} {\bibfnamefont {F.}~\bibnamefont
  {{Verbunt}}},\ }\bibfield  {title} {\bibinfo {title} {{Population synthesis
  of high-mass binaries.}},\ }\href@noop {} {\bibfield  {journal} {\bibinfo
  {journal} {\aap}\ }\textbf {\bibinfo {volume} {309}},\ \bibinfo {pages} {179}
  (\bibinfo {year} {1996})}\BibitemShut {NoStop}%
\bibitem [{\citenamefont {{Belczynski}}\ \emph {et~al.}(2008)\citenamefont
  {{Belczynski}}, \citenamefont {{Kalogera}}, \citenamefont {{Rasio}},
  \citenamefont {{Taam}}, \citenamefont {{Zezas}}, \citenamefont {{Bulik}},
  \citenamefont {{Maccarone}},\ and\ \citenamefont
  {{Ivanova}}}]{2008ApJS..174..223B}%
  \BibitemOpen
  \bibfield  {author} {\bibinfo {author} {\bibfnamefont {K.}~\bibnamefont
  {{Belczynski}}}, \bibinfo {author} {\bibfnamefont {V.}~\bibnamefont
  {{Kalogera}}}, \bibinfo {author} {\bibfnamefont {F.~A.}\ \bibnamefont
  {{Rasio}}}, \bibinfo {author} {\bibfnamefont {R.~E.}\ \bibnamefont {{Taam}}},
  \bibinfo {author} {\bibfnamefont {A.}~\bibnamefont {{Zezas}}}, \bibinfo
  {author} {\bibfnamefont {T.}~\bibnamefont {{Bulik}}}, \bibinfo {author}
  {\bibfnamefont {T.~J.}\ \bibnamefont {{Maccarone}}},\ and\ \bibinfo {author}
  {\bibfnamefont {N.}~\bibnamefont {{Ivanova}}},\ }\bibfield  {title} {\bibinfo
  {title} {{Compact Object Modeling with the StarTrack Population Synthesis
  Code}},\ }\href {https://doi.org/10.1086/521026} {\bibfield  {journal}
  {\bibinfo  {journal} {\apjs}\ }\textbf {\bibinfo {volume} {174}},\ \bibinfo
  {pages} {223} (\bibinfo {year} {2008})},\ \Eprint
  {https://arxiv.org/abs/astro-ph/0511811} {arXiv:astro-ph/0511811 [astro-ph]}
  \BibitemShut {NoStop}%
\bibitem [{\citenamefont {{Mennekens}}\ and\ \citenamefont
  {{Vanbeveren}}(2014)}]{2014A&A...564A.134M}%
  \BibitemOpen
  \bibfield  {author} {\bibinfo {author} {\bibfnamefont {N.}~\bibnamefont
  {{Mennekens}}}\ and\ \bibinfo {author} {\bibfnamefont {D.}~\bibnamefont
  {{Vanbeveren}}},\ }\bibfield  {title} {\bibinfo {title} {{Massive double
  compact object mergers: gravitational wave sources and r-process element
  production sites}},\ }\href {https://doi.org/10.1051/0004-6361/201322198}
  {\bibfield  {journal} {\bibinfo  {journal} {\aap}\ }\textbf {\bibinfo
  {volume} {564}},\ \bibinfo {eid} {A134} (\bibinfo {year} {2014})},\ \Eprint
  {https://arxiv.org/abs/1307.0959} {arXiv:1307.0959} \BibitemShut {NoStop}%
\bibitem [{\citenamefont {{Mapelli}}\ and\ \citenamefont
  {{Giacobbo}}(2018)}]{2018MNRAS.479.4391M}%
  \BibitemOpen
  \bibfield  {author} {\bibinfo {author} {\bibfnamefont {M.}~\bibnamefont
  {{Mapelli}}}\ and\ \bibinfo {author} {\bibfnamefont {N.}~\bibnamefont
  {{Giacobbo}}},\ }\bibfield  {title} {\bibinfo {title} {{The cosmic merger
  rate of neutron stars and black holes}},\ }\href
  {https://doi.org/10.1093/mnras/sty1613} {\bibfield  {journal} {\bibinfo
  {journal} {\mnras}\ }\textbf {\bibinfo {volume} {479}},\ \bibinfo {pages}
  {4391} (\bibinfo {year} {2018})},\ \Eprint {https://arxiv.org/abs/1806.04866}
  {arXiv:1806.04866 [astro-ph.HE]} \BibitemShut {NoStop}%
\bibitem [{\citenamefont {{Breivik}}\ \emph {et~al.}(2020)\citenamefont
  {{Breivik}}, \citenamefont {{Coughlin}}, \citenamefont {{Zevin}},
  \citenamefont {{Rodriguez}}, \citenamefont {{Kremer}}, \citenamefont {{Ye}},
  \citenamefont {{Andrews}}, \citenamefont {{Kurkowski}}, \citenamefont
  {{Digman}}, \citenamefont {{Larson}},\ and\ \citenamefont
  {et~al.}}]{2020ApJ...898...71B}%
  \BibitemOpen
  \bibfield  {author} {\bibinfo {author} {\bibfnamefont {K.}~\bibnamefont
  {{Breivik}}}, \bibinfo {author} {\bibfnamefont {S.}~\bibnamefont
  {{Coughlin}}}, \bibinfo {author} {\bibfnamefont {M.}~\bibnamefont {{Zevin}}},
  \bibinfo {author} {\bibfnamefont {C.~L.}\ \bibnamefont {{Rodriguez}}},
  \bibinfo {author} {\bibfnamefont {K.}~\bibnamefont {{Kremer}}}, \bibinfo
  {author} {\bibfnamefont {C.~S.}\ \bibnamefont {{Ye}}}, \bibinfo {author}
  {\bibfnamefont {J.~J.}\ \bibnamefont {{Andrews}}}, \bibinfo {author}
  {\bibfnamefont {M.}~\bibnamefont {{Kurkowski}}}, \bibinfo {author}
  {\bibfnamefont {M.~C.}\ \bibnamefont {{Digman}}}, \bibinfo {author}
  {\bibfnamefont {S.~L.}\ \bibnamefont {{Larson}}},\ and\ \bibinfo {author}
  {\bibnamefont {et~al.}},\ }\bibfield  {title} {\bibinfo {title} {{COSMIC
  Variance in Binary Population Synthesis}},\ }\href
  {https://doi.org/10.3847/1538-4357/ab9d85} {\bibfield  {journal} {\bibinfo
  {journal} {\apj}\ }\textbf {\bibinfo {volume} {898}},\ \bibinfo {eid} {71}
  (\bibinfo {year} {2020})},\ \Eprint {https://arxiv.org/abs/1911.00903}
  {arXiv:1911.00903 [astro-ph.HE]} \BibitemShut {NoStop}%
\bibitem [{\citenamefont {{Riley}}\ \emph {et~al.}(2022)\citenamefont
  {{Riley}}, \citenamefont {{Agrawal}}, \citenamefont {{Barrett}},
  \citenamefont {{Boyett}}, \citenamefont {{Broekgaarden}}, \citenamefont
  {{Chattopadhyay}}, \citenamefont {{Gaebel}}, \citenamefont {{Gittins}},
  \citenamefont {{Hirai}}, \citenamefont {{Howitt}}, \citenamefont {{Justham}},
  \citenamefont {{Khandelwal}}, \citenamefont {{Kummer}}, \citenamefont
  {{Lau}}, \citenamefont {{Mandel}}, \citenamefont {{de Mink}}, \citenamefont
  {{Neijssel}}, \citenamefont {{Riley}}, \citenamefont {{van Son}},
  \citenamefont {{Stevenson}}, \citenamefont {{Vigna-G{\'o}mez}}, \citenamefont
  {{Vinciguerra}}, \citenamefont {{Wagg}}, \citenamefont {{Willcox}},\ and\
  \citenamefont {{Team Compas}}}]{2022ApJS..258...34R}%
  \BibitemOpen
  \bibfield  {author} {\bibinfo {author} {\bibfnamefont {J.}~\bibnamefont
  {{Riley}}}, \bibinfo {author} {\bibfnamefont {P.}~\bibnamefont {{Agrawal}}},
  \bibinfo {author} {\bibfnamefont {J.~W.}\ \bibnamefont {{Barrett}}}, \bibinfo
  {author} {\bibfnamefont {K.~N.~K.}\ \bibnamefont {{Boyett}}}, \bibinfo
  {author} {\bibfnamefont {F.~S.}\ \bibnamefont {{Broekgaarden}}}, \bibinfo
  {author} {\bibfnamefont {D.}~\bibnamefont {{Chattopadhyay}}}, \bibinfo
  {author} {\bibfnamefont {S.~M.}\ \bibnamefont {{Gaebel}}}, \bibinfo {author}
  {\bibfnamefont {F.}~\bibnamefont {{Gittins}}}, \bibinfo {author}
  {\bibfnamefont {R.}~\bibnamefont {{Hirai}}}, \bibinfo {author} {\bibfnamefont
  {G.}~\bibnamefont {{Howitt}}}, \bibinfo {author} {\bibfnamefont
  {S.}~\bibnamefont {{Justham}}}, \bibinfo {author} {\bibfnamefont
  {L.}~\bibnamefont {{Khandelwal}}}, \bibinfo {author} {\bibfnamefont
  {F.}~\bibnamefont {{Kummer}}}, \bibinfo {author} {\bibfnamefont {M.~Y.~M.}\
  \bibnamefont {{Lau}}}, \bibinfo {author} {\bibfnamefont {I.}~\bibnamefont
  {{Mandel}}}, \bibinfo {author} {\bibfnamefont {S.~E.}\ \bibnamefont {{de
  Mink}}}, \bibinfo {author} {\bibfnamefont {C.}~\bibnamefont {{Neijssel}}},
  \bibinfo {author} {\bibfnamefont {T.}~\bibnamefont {{Riley}}}, \bibinfo
  {author} {\bibfnamefont {L.}~\bibnamefont {{van Son}}}, \bibinfo {author}
  {\bibfnamefont {S.}~\bibnamefont {{Stevenson}}}, \bibinfo {author}
  {\bibfnamefont {A.}~\bibnamefont {{Vigna-G{\'o}mez}}}, \bibinfo {author}
  {\bibfnamefont {S.}~\bibnamefont {{Vinciguerra}}}, \bibinfo {author}
  {\bibfnamefont {T.}~\bibnamefont {{Wagg}}}, \bibinfo {author} {\bibfnamefont
  {R.}~\bibnamefont {{Willcox}}},\ and\ \bibinfo {author} {\bibnamefont {{Team
  Compas}}},\ }\bibfield  {title} {\bibinfo {title} {{Rapid Stellar and Binary
  Population Synthesis with COMPAS}},\ }\href
  {https://doi.org/10.3847/1538-4365/ac416c} {\bibfield  {journal} {\bibinfo
  {journal} {\apjs}\ }\textbf {\bibinfo {volume} {258}},\ \bibinfo {eid} {34}
  (\bibinfo {year} {2022})},\ \Eprint {https://arxiv.org/abs/2109.10352}
  {arXiv:2109.10352 [astro-ph.IM]} \BibitemShut {NoStop}%
\bibitem [{\citenamefont {{Xu}}\ and\ \citenamefont
  {{Li}}(2010)}]{2010ApJ...716..114X}%
  \BibitemOpen
  \bibfield  {author} {\bibinfo {author} {\bibfnamefont {X.-J.}\ \bibnamefont
  {{Xu}}}\ and\ \bibinfo {author} {\bibfnamefont {X.-D.}\ \bibnamefont
  {{Li}}},\ }\bibfield  {title} {\bibinfo {title} {{On the Binding Energy
  Parameter {\ensuremath{\lambda}} of Common Envelope Evolution}},\ }\href
  {https://doi.org/10.1088/0004-637X/716/1/114} {\bibfield  {journal} {\bibinfo
   {journal} {\apj}\ }\textbf {\bibinfo {volume} {716}},\ \bibinfo {pages}
  {114} (\bibinfo {year} {2010})},\ \Eprint {https://arxiv.org/abs/1004.4957}
  {arXiv:1004.4957 [astro-ph.SR]} \BibitemShut {NoStop}%
\bibitem [{\citenamefont {{Kruckow}}\ \emph {et~al.}(2016)\citenamefont
  {{Kruckow}}, \citenamefont {{Tauris}}, \citenamefont {{Langer}},
  \citenamefont {{Sz{\'e}csi}}, \citenamefont {{Marchant}},\ and\ \citenamefont
  {{Podsiadlowski}}}]{2016A&A...596A..58K}%
  \BibitemOpen
  \bibfield  {author} {\bibinfo {author} {\bibfnamefont {M.~U.}\ \bibnamefont
  {{Kruckow}}}, \bibinfo {author} {\bibfnamefont {T.~M.}\ \bibnamefont
  {{Tauris}}}, \bibinfo {author} {\bibfnamefont {N.}~\bibnamefont {{Langer}}},
  \bibinfo {author} {\bibfnamefont {D.}~\bibnamefont {{Sz{\'e}csi}}}, \bibinfo
  {author} {\bibfnamefont {P.}~\bibnamefont {{Marchant}}},\ and\ \bibinfo
  {author} {\bibfnamefont {P.}~\bibnamefont {{Podsiadlowski}}},\ }\bibfield
  {title} {\bibinfo {title} {{Common-envelope ejection in massive binary stars.
  Implications for the progenitors of GW150914 and GW151226}},\ }\href
  {https://doi.org/10.1051/0004-6361/201629420} {\bibfield  {journal} {\bibinfo
   {journal} {\aap}\ }\textbf {\bibinfo {volume} {596}},\ \bibinfo {eid} {A58}
  (\bibinfo {year} {2016})},\ \Eprint {https://arxiv.org/abs/1610.04417}
  {arXiv:1610.04417 [astro-ph.SR]} \BibitemShut {NoStop}%
\bibitem [{\citenamefont {{Dewi}}\ and\ \citenamefont
  {{Tauris}}(2000)}]{2000A&A...360.1043D}%
  \BibitemOpen
  \bibfield  {author} {\bibinfo {author} {\bibfnamefont {J.~D.~M.}\
  \bibnamefont {{Dewi}}}\ and\ \bibinfo {author} {\bibfnamefont {T.~M.}\
  \bibnamefont {{Tauris}}},\ }\bibfield  {title} {\bibinfo {title} {{On the
  energy equation and efficiency parameter of the common envelope evolution}},\
  }\href@noop {} {\bibfield  {journal} {\bibinfo  {journal} {\aap}\ }\textbf
  {\bibinfo {volume} {360}},\ \bibinfo {pages} {1043} (\bibinfo {year}
  {2000})},\ \Eprint {https://arxiv.org/abs/astro-ph/0007034}
  {arXiv:astro-ph/0007034 [astro-ph]} \BibitemShut {NoStop}%
\bibitem [{\citenamefont {{Tauris}}\ and\ \citenamefont
  {{Dewi}}(2001)}]{2001A&A...369..170T}%
  \BibitemOpen
  \bibfield  {author} {\bibinfo {author} {\bibfnamefont {T.~M.}\ \bibnamefont
  {{Tauris}}}\ and\ \bibinfo {author} {\bibfnamefont {J.~D.~M.}\ \bibnamefont
  {{Dewi}}},\ }\bibfield  {title} {\bibinfo {title} {{Research Note On the
  binding energy parameter of common envelope evolution. Dependency on the
  definition of the stellar core boundary during spiral-in}},\ }\href
  {https://doi.org/10.1051/0004-6361:20010099} {\bibfield  {journal} {\bibinfo
  {journal} {\aap}\ }\textbf {\bibinfo {volume} {369}},\ \bibinfo {pages} {170}
  (\bibinfo {year} {2001})},\ \Eprint {https://arxiv.org/abs/astro-ph/0101530}
  {arXiv:astro-ph/0101530 [astro-ph]} \BibitemShut {NoStop}%
\bibitem [{\citenamefont {{Ivanova}}(2011)}]{2011ApJ...730...76I}%
  \BibitemOpen
  \bibfield  {author} {\bibinfo {author} {\bibfnamefont {N.}~\bibnamefont
  {{Ivanova}}},\ }\bibfield  {title} {\bibinfo {title} {{Common Envelope: On
  the Mass and the Fate of the Remnant}},\ }\href
  {https://doi.org/10.1088/0004-637X/730/2/76} {\bibfield  {journal} {\bibinfo
  {journal} {\apj}\ }\textbf {\bibinfo {volume} {730}},\ \bibinfo {eid} {76}
  (\bibinfo {year} {2011})},\ \Eprint {https://arxiv.org/abs/1101.2863}
  {arXiv:1101.2863 [astro-ph.SR]} \BibitemShut {NoStop}%
\bibitem [{\citenamefont {{Fragos}}\ \emph {et~al.}(2019)\citenamefont
  {{Fragos}}, \citenamefont {{Andrews}}, \citenamefont {{Ramirez-Ruiz}},
  \citenamefont {{Meynet}}, \citenamefont {{Kalogera}}, \citenamefont
  {{Taam}},\ and\ \citenamefont {{Zezas}}}]{2019ApJ...883L..45F}%
  \BibitemOpen
  \bibfield  {author} {\bibinfo {author} {\bibfnamefont {T.}~\bibnamefont
  {{Fragos}}}, \bibinfo {author} {\bibfnamefont {J.~J.}\ \bibnamefont
  {{Andrews}}}, \bibinfo {author} {\bibfnamefont {E.}~\bibnamefont
  {{Ramirez-Ruiz}}}, \bibinfo {author} {\bibfnamefont {G.}~\bibnamefont
  {{Meynet}}}, \bibinfo {author} {\bibfnamefont {V.}~\bibnamefont
  {{Kalogera}}}, \bibinfo {author} {\bibfnamefont {R.~E.}\ \bibnamefont
  {{Taam}}},\ and\ \bibinfo {author} {\bibfnamefont {A.}~\bibnamefont
  {{Zezas}}},\ }\bibfield  {title} {\bibinfo {title} {{The Complete Evolution
  of a Neutron-star Binary through a Common Envelope Phase Using 1D
  Hydrodynamic Simulations}},\ }\href
  {https://doi.org/10.3847/2041-8213/ab40d1} {\bibfield  {journal} {\bibinfo
  {journal} {\apjl}\ }\textbf {\bibinfo {volume} {883}},\ \bibinfo {eid} {L45}
  (\bibinfo {year} {2019})},\ \Eprint {https://arxiv.org/abs/1907.12573}
  {arXiv:1907.12573 [astro-ph.HE]} \BibitemShut {NoStop}%
\bibitem [{\citenamefont {{Moreno}}\ \emph {et~al.}(2021)\citenamefont
  {{Moreno}}, \citenamefont {{Schneider}}, \citenamefont {{Roepke}},
  \citenamefont {{Ohlmann}}, \citenamefont {{Pakmor}}, \citenamefont
  {{Podsiadlowski}},\ and\ \citenamefont {{Sand}}}]{2021arXiv211112112M}%
  \BibitemOpen
  \bibfield  {author} {\bibinfo {author} {\bibfnamefont {M.~M.}\ \bibnamefont
  {{Moreno}}}, \bibinfo {author} {\bibfnamefont {F.~R.~N.}\ \bibnamefont
  {{Schneider}}}, \bibinfo {author} {\bibfnamefont {F.~K.}\ \bibnamefont
  {{Roepke}}}, \bibinfo {author} {\bibfnamefont {S.~T.}\ \bibnamefont
  {{Ohlmann}}}, \bibinfo {author} {\bibfnamefont {R.}~\bibnamefont {{Pakmor}}},
  \bibinfo {author} {\bibfnamefont {P.}~\bibnamefont {{Podsiadlowski}}},\ and\
  \bibinfo {author} {\bibfnamefont {C.}~\bibnamefont {{Sand}}},\ }\bibfield
  {title} {\bibinfo {title} {{From 3D hydrodynamic simulations of
  common-envelope interaction to gravitational-wave mergers}},\ }\href@noop {}
  {\bibfield  {journal} {\bibinfo  {journal} {arXiv e-prints}\ ,\ \bibinfo
  {eid} {arXiv:2111.12112}} (\bibinfo {year} {2021})},\ \Eprint
  {https://arxiv.org/abs/2111.12112} {arXiv:2111.12112 [astro-ph.SR]}
  \BibitemShut {NoStop}%
\bibitem [{\citenamefont {{Vigna-G{\'o}mez}}\ \emph {et~al.}(2022)\citenamefont
  {{Vigna-G{\'o}mez}}, \citenamefont {{Wassink}}, \citenamefont {{Klencki}},
  \citenamefont {{Istrate}}, \citenamefont {{Nelemans}},\ and\ \citenamefont
  {{Mandel}}}]{2022MNRAS.511.2326V}%
  \BibitemOpen
  \bibfield  {author} {\bibinfo {author} {\bibfnamefont {A.}~\bibnamefont
  {{Vigna-G{\'o}mez}}}, \bibinfo {author} {\bibfnamefont {M.}~\bibnamefont
  {{Wassink}}}, \bibinfo {author} {\bibfnamefont {J.}~\bibnamefont
  {{Klencki}}}, \bibinfo {author} {\bibfnamefont {A.}~\bibnamefont
  {{Istrate}}}, \bibinfo {author} {\bibfnamefont {G.}~\bibnamefont
  {{Nelemans}}},\ and\ \bibinfo {author} {\bibfnamefont {I.}~\bibnamefont
  {{Mandel}}},\ }\bibfield  {title} {\bibinfo {title} {{Stellar response after
  stripping as a model for common-envelope outcomes}},\ }\href
  {https://doi.org/10.1093/mnras/stac237} {\bibfield  {journal} {\bibinfo
  {journal} {\mnras}\ }\textbf {\bibinfo {volume} {511}},\ \bibinfo {pages}
  {2326} (\bibinfo {year} {2022})},\ \Eprint {https://arxiv.org/abs/2107.14526}
  {arXiv:2107.14526 [astro-ph.HE]} \BibitemShut {NoStop}%
\bibitem [{\citenamefont {{Hirai}}\ and\ \citenamefont
  {{Mandel}}(2022)}]{2022ApJ...937L..42H}%
  \BibitemOpen
  \bibfield  {author} {\bibinfo {author} {\bibfnamefont {R.}~\bibnamefont
  {{Hirai}}}\ and\ \bibinfo {author} {\bibfnamefont {I.}~\bibnamefont
  {{Mandel}}},\ }\bibfield  {title} {\bibinfo {title} {{A Two-stage Formalism
  for Common-envelope Phases of Massive Stars}},\ }\href
  {https://doi.org/10.3847/2041-8213/ac9519} {\bibfield  {journal} {\bibinfo
  {journal} {\apjl}\ }\textbf {\bibinfo {volume} {937}},\ \bibinfo {eid} {L42}
  (\bibinfo {year} {2022})},\ \Eprint {https://arxiv.org/abs/2209.05328}
  {arXiv:2209.05328 [astro-ph.SR]} \BibitemShut {NoStop}%
\bibitem [{\citenamefont {{Nandez}}\ \emph {et~al.}(2015)\citenamefont
  {{Nandez}}, \citenamefont {{Ivanova}},\ and\ \citenamefont
  {{Lombardi}}}]{2015MNRAS.450L..39N}%
  \BibitemOpen
  \bibfield  {author} {\bibinfo {author} {\bibfnamefont {J.~L.~A.}\
  \bibnamefont {{Nandez}}}, \bibinfo {author} {\bibfnamefont {N.}~\bibnamefont
  {{Ivanova}}},\ and\ \bibinfo {author} {\bibfnamefont {J.~C.~J.}\ \bibnamefont
  {{Lombardi}}},\ }\bibfield  {title} {\bibinfo {title} {{Recombination energy
  in double white dwarf formation.}},\ }\href
  {https://doi.org/10.1093/mnrasl/slv043} {\bibfield  {journal} {\bibinfo
  {journal} {\mnras}\ }\textbf {\bibinfo {volume} {450}},\ \bibinfo {pages}
  {L39} (\bibinfo {year} {2015})},\ \Eprint {https://arxiv.org/abs/1503.02750}
  {arXiv:1503.02750 [astro-ph.SR]} \BibitemShut {NoStop}%
\bibitem [{\citenamefont {{Ivanova}}\ \emph {et~al.}(2015)\citenamefont
  {{Ivanova}}, \citenamefont {{Justham}},\ and\ \citenamefont
  {{Podsiadlowski}}}]{2015MNRAS.447.2181I}%
  \BibitemOpen
  \bibfield  {author} {\bibinfo {author} {\bibfnamefont {N.}~\bibnamefont
  {{Ivanova}}}, \bibinfo {author} {\bibfnamefont {S.}~\bibnamefont
  {{Justham}}},\ and\ \bibinfo {author} {\bibfnamefont {P.}~\bibnamefont
  {{Podsiadlowski}}},\ }\bibfield  {title} {\bibinfo {title} {{On the role of
  recombination in common-envelope ejections}},\ }\href
  {https://doi.org/10.1093/mnras/stu2582} {\bibfield  {journal} {\bibinfo
  {journal} {\mnras}\ }\textbf {\bibinfo {volume} {447}},\ \bibinfo {pages}
  {2181} (\bibinfo {year} {2015})},\ \Eprint {https://arxiv.org/abs/1409.3260}
  {arXiv:1409.3260 [astro-ph.SR]} \BibitemShut {NoStop}%
\bibitem [{\citenamefont {{Soker}}\ \emph {et~al.}(2018)\citenamefont
  {{Soker}}, \citenamefont {{Grichener}},\ and\ \citenamefont
  {{Sabach}}}]{2018ApJ...863L..14S}%
  \BibitemOpen
  \bibfield  {author} {\bibinfo {author} {\bibfnamefont {N.}~\bibnamefont
  {{Soker}}}, \bibinfo {author} {\bibfnamefont {A.}~\bibnamefont
  {{Grichener}}},\ and\ \bibinfo {author} {\bibfnamefont {E.}~\bibnamefont
  {{Sabach}}},\ }\bibfield  {title} {\bibinfo {title} {{Radiating the Hydrogen
  Recombination Energy during Common Envelope Evolution}},\ }\href
  {https://doi.org/10.3847/2041-8213/aad736} {\bibfield  {journal} {\bibinfo
  {journal} {\apjl}\ }\textbf {\bibinfo {volume} {863}},\ \bibinfo {eid} {L14}
  (\bibinfo {year} {2018})},\ \Eprint {https://arxiv.org/abs/1805.08543}
  {arXiv:1805.08543 [astro-ph.SR]} \BibitemShut {NoStop}%
\bibitem [{\citenamefont {{Grichener}}\ \emph {et~al.}(2018)\citenamefont
  {{Grichener}}, \citenamefont {{Sabach}},\ and\ \citenamefont
  {{Soker}}}]{2018MNRAS.478.1818G}%
  \BibitemOpen
  \bibfield  {author} {\bibinfo {author} {\bibfnamefont {A.}~\bibnamefont
  {{Grichener}}}, \bibinfo {author} {\bibfnamefont {E.}~\bibnamefont
  {{Sabach}}},\ and\ \bibinfo {author} {\bibfnamefont {N.}~\bibnamefont
  {{Soker}}},\ }\bibfield  {title} {\bibinfo {title} {{The limited role of
  recombination energy in common envelope removal}},\ }\href
  {https://doi.org/10.1093/mnras/sty1178} {\bibfield  {journal} {\bibinfo
  {journal} {\mnras}\ }\textbf {\bibinfo {volume} {478}},\ \bibinfo {pages}
  {1818} (\bibinfo {year} {2018})},\ \Eprint {https://arxiv.org/abs/1803.05864}
  {arXiv:1803.05864 [astro-ph.SR]} \BibitemShut {NoStop}%
\bibitem [{\citenamefont {{Nandez}}\ and\ \citenamefont
  {{Ivanova}}(2016)}]{2016MNRAS.460.3992N}%
  \BibitemOpen
  \bibfield  {author} {\bibinfo {author} {\bibfnamefont {J.~L.~A.}\
  \bibnamefont {{Nandez}}}\ and\ \bibinfo {author} {\bibfnamefont
  {N.}~\bibnamefont {{Ivanova}}},\ }\bibfield  {title} {\bibinfo {title}
  {{Common envelope events with low-mass giants: understanding the energy
  budget}},\ }\href {https://doi.org/10.1093/mnras/stw1266} {\bibfield
  {journal} {\bibinfo  {journal} {\mnras}\ }\textbf {\bibinfo {volume} {460}},\
  \bibinfo {pages} {3992} (\bibinfo {year} {2016})},\ \Eprint
  {https://arxiv.org/abs/1606.04922} {arXiv:1606.04922 [astro-ph.SR]}
  \BibitemShut {NoStop}%
\bibitem [{\citenamefont {{Lau}}\ \emph
  {et~al.}(2022{\natexlab{a}})\citenamefont {{Lau}}, \citenamefont {{Hirai}},
  \citenamefont {{Gonz{\'a}lez-Bol{\'\i}var}}, \citenamefont {{Price}},
  \citenamefont {{De Marco}},\ and\ \citenamefont
  {{Mandel}}}]{2022MNRAS.512.5462L}%
  \BibitemOpen
  \bibfield  {author} {\bibinfo {author} {\bibfnamefont {M.~Y.~M.}\
  \bibnamefont {{Lau}}}, \bibinfo {author} {\bibfnamefont {R.}~\bibnamefont
  {{Hirai}}}, \bibinfo {author} {\bibfnamefont {M.}~\bibnamefont
  {{Gonz{\'a}lez-Bol{\'\i}var}}}, \bibinfo {author} {\bibfnamefont {D.~J.}\
  \bibnamefont {{Price}}}, \bibinfo {author} {\bibfnamefont {O.}~\bibnamefont
  {{De Marco}}},\ and\ \bibinfo {author} {\bibfnamefont {I.}~\bibnamefont
  {{Mandel}}},\ }\bibfield  {title} {\bibinfo {title} {{Common envelopes in
  massive stars: towards the role of radiation pressure and recombination
  energy in ejecting red supergiant envelopes}},\ }\href
  {https://doi.org/10.1093/mnras/stac049} {\bibfield  {journal} {\bibinfo
  {journal} {\mnras}\ }\textbf {\bibinfo {volume} {512}},\ \bibinfo {pages}
  {5462} (\bibinfo {year} {2022}{\natexlab{a}})},\ \Eprint
  {https://arxiv.org/abs/2111.00923} {arXiv:2111.00923 [astro-ph.SR]}
  \BibitemShut {NoStop}%
\bibitem [{\citenamefont {{Lau}}\ \emph
  {et~al.}(2022{\natexlab{b}})\citenamefont {{Lau}}, \citenamefont {{Hirai}},
  \citenamefont {{Price}},\ and\ \citenamefont
  {{Mandel}}}]{2022MNRAS.516.4669L}%
  \BibitemOpen
  \bibfield  {author} {\bibinfo {author} {\bibfnamefont {M.~Y.~M.}\
  \bibnamefont {{Lau}}}, \bibinfo {author} {\bibfnamefont {R.}~\bibnamefont
  {{Hirai}}}, \bibinfo {author} {\bibfnamefont {D.~J.}\ \bibnamefont
  {{Price}}},\ and\ \bibinfo {author} {\bibfnamefont {I.}~\bibnamefont
  {{Mandel}}},\ }\bibfield  {title} {\bibinfo {title} {{Common envelopes in
  massive stars II: The distinct roles of hydrogen and helium recombination}},\
  }\href {https://doi.org/10.1093/mnras/stac2490} {\bibfield  {journal}
  {\bibinfo  {journal} {\mnras}\ }\textbf {\bibinfo {volume} {516}},\ \bibinfo
  {pages} {4669} (\bibinfo {year} {2022}{\natexlab{b}})},\ \Eprint
  {https://arxiv.org/abs/2206.06411} {arXiv:2206.06411 [astro-ph.SR]}
  \BibitemShut {NoStop}%
\bibitem [{\citenamefont {{Claeys}}\ \emph {et~al.}(2014)\citenamefont
  {{Claeys}}, \citenamefont {{Pols}}, \citenamefont {{Izzard}}, \citenamefont
  {{Vink}},\ and\ \citenamefont {{Verbunt}}}]{2014A&A...563A..83C}%
  \BibitemOpen
  \bibfield  {author} {\bibinfo {author} {\bibfnamefont {J.~S.~W.}\
  \bibnamefont {{Claeys}}}, \bibinfo {author} {\bibfnamefont {O.~R.}\
  \bibnamefont {{Pols}}}, \bibinfo {author} {\bibfnamefont {R.~G.}\
  \bibnamefont {{Izzard}}}, \bibinfo {author} {\bibfnamefont {J.}~\bibnamefont
  {{Vink}}},\ and\ \bibinfo {author} {\bibfnamefont {F.~W.~M.}\ \bibnamefont
  {{Verbunt}}},\ }\bibfield  {title} {\bibinfo {title} {{Theoretical
  uncertainties of the Type Ia supernova rate}},\ }\href
  {https://doi.org/10.1051/0004-6361/201322714} {\bibfield  {journal} {\bibinfo
   {journal} {\aap}\ }\textbf {\bibinfo {volume} {563}},\ \bibinfo {eid} {A83}
  (\bibinfo {year} {2014})},\ \Eprint {https://arxiv.org/abs/1401.2895}
  {arXiv:1401.2895 [astro-ph.SR]} \BibitemShut {NoStop}%
\bibitem [{\citenamefont {{Wang}}\ \emph {et~al.}(2016)\citenamefont {{Wang}},
  \citenamefont {{Jia}},\ and\ \citenamefont {{Li}}}]{2016RAA....16..126W}%
  \BibitemOpen
  \bibfield  {author} {\bibinfo {author} {\bibfnamefont {C.}~\bibnamefont
  {{Wang}}}, \bibinfo {author} {\bibfnamefont {K.}~\bibnamefont {{Jia}}},\ and\
  \bibinfo {author} {\bibfnamefont {X.-D.}\ \bibnamefont {{Li}}},\ }\bibfield
  {title} {\bibinfo {title} {{The binding energy parameter for common envelope
  evolution}},\ }\href {https://doi.org/10.1088/1674-4527/16/8/126} {\bibfield
  {journal} {\bibinfo  {journal} {Research in Astronomy and Astrophysics}\
  }\textbf {\bibinfo {volume} {16}},\ \bibinfo {eid} {126} (\bibinfo {year}
  {2016})},\ \Eprint {https://arxiv.org/abs/1605.03668} {arXiv:1605.03668
  [astro-ph.SR]} \BibitemShut {NoStop}%
\bibitem [{\citenamefont {{Clayton}}\ \emph {et~al.}(2017)\citenamefont
  {{Clayton}}, \citenamefont {{Podsiadlowski}}, \citenamefont {{Ivanova}},\
  and\ \citenamefont {{Justham}}}]{2017MNRAS.470.1788C}%
  \BibitemOpen
  \bibfield  {author} {\bibinfo {author} {\bibfnamefont {M.}~\bibnamefont
  {{Clayton}}}, \bibinfo {author} {\bibfnamefont {P.}~\bibnamefont
  {{Podsiadlowski}}}, \bibinfo {author} {\bibfnamefont {N.}~\bibnamefont
  {{Ivanova}}},\ and\ \bibinfo {author} {\bibfnamefont {S.}~\bibnamefont
  {{Justham}}},\ }\bibfield  {title} {\bibinfo {title} {{Episodic mass
  ejections from common-envelope objects}},\ }\href
  {https://doi.org/10.1093/mnras/stx1290} {\bibfield  {journal} {\bibinfo
  {journal} {\mnras}\ }\textbf {\bibinfo {volume} {470}},\ \bibinfo {pages}
  {1788} (\bibinfo {year} {2017})},\ \Eprint {https://arxiv.org/abs/1705.08457}
  {arXiv:1705.08457 [astro-ph.SR]} \BibitemShut {NoStop}%
\bibitem [{\citenamefont {{Law-Smith}}\ \emph {et~al.}(2020)\citenamefont
  {{Law-Smith}}, \citenamefont {{Everson}}, \citenamefont {{Ramirez-Ruiz}},
  \citenamefont {{de Mink}}, \citenamefont {{van Son}}, \citenamefont
  {{G{\"o}tberg}}, \citenamefont {{Zellmann}}, \citenamefont
  {{Vigna-G{\'o}mez}}, \citenamefont {{Renzo}}, \citenamefont {{Wu}},
  \citenamefont {{Schr{\o}der}}, \citenamefont {{Foley}},\ and\ \citenamefont
  {{Hutchinson-Smith}}}]{2020arXiv201106630L}%
  \BibitemOpen
  \bibfield  {author} {\bibinfo {author} {\bibfnamefont {J.~A.~P.}\
  \bibnamefont {{Law-Smith}}}, \bibinfo {author} {\bibfnamefont {R.~W.}\
  \bibnamefont {{Everson}}}, \bibinfo {author} {\bibfnamefont {E.}~\bibnamefont
  {{Ramirez-Ruiz}}}, \bibinfo {author} {\bibfnamefont {S.~E.}\ \bibnamefont
  {{de Mink}}}, \bibinfo {author} {\bibfnamefont {L.~A.~C.}\ \bibnamefont {{van
  Son}}}, \bibinfo {author} {\bibfnamefont {Y.}~\bibnamefont {{G{\"o}tberg}}},
  \bibinfo {author} {\bibfnamefont {S.}~\bibnamefont {{Zellmann}}}, \bibinfo
  {author} {\bibfnamefont {A.}~\bibnamefont {{Vigna-G{\'o}mez}}}, \bibinfo
  {author} {\bibfnamefont {M.}~\bibnamefont {{Renzo}}}, \bibinfo {author}
  {\bibfnamefont {S.}~\bibnamefont {{Wu}}}, \bibinfo {author} {\bibfnamefont
  {S.~L.}\ \bibnamefont {{Schr{\o}der}}}, \bibinfo {author} {\bibfnamefont
  {R.~J.}\ \bibnamefont {{Foley}}},\ and\ \bibinfo {author} {\bibfnamefont
  {T.}~\bibnamefont {{Hutchinson-Smith}}},\ }\bibfield  {title} {\bibinfo
  {title} {{Successful Common Envelope Ejection and Binary Neutron Star
  Formation in 3D Hydrodynamics}},\ }\href
  {https://doi.org/10.48550/arXiv.2011.06630} {\bibfield  {journal} {\bibinfo
  {journal} {arXiv e-prints}\ ,\ \bibinfo {eid} {arXiv:2011.06630}} (\bibinfo
  {year} {2020})},\ \Eprint {https://arxiv.org/abs/2011.06630}
  {arXiv:2011.06630 [astro-ph.HE]} \BibitemShut {NoStop}%
\bibitem [{\citenamefont {{Ondratschek}}\ \emph {et~al.}(2022)\citenamefont
  {{Ondratschek}}, \citenamefont {{R{\"o}pke}}, \citenamefont {{Schneider}},
  \citenamefont {{Fendt}}, \citenamefont {{Sand}}, \citenamefont {{Ohlmann}},
  \citenamefont {{Pakmor}},\ and\ \citenamefont
  {{Springel}}}]{2022A&A...660L...8O}%
  \BibitemOpen
  \bibfield  {author} {\bibinfo {author} {\bibfnamefont {P.~A.}\ \bibnamefont
  {{Ondratschek}}}, \bibinfo {author} {\bibfnamefont {F.~K.}\ \bibnamefont
  {{R{\"o}pke}}}, \bibinfo {author} {\bibfnamefont {F.~R.~N.}\ \bibnamefont
  {{Schneider}}}, \bibinfo {author} {\bibfnamefont {C.}~\bibnamefont
  {{Fendt}}}, \bibinfo {author} {\bibfnamefont {C.}~\bibnamefont {{Sand}}},
  \bibinfo {author} {\bibfnamefont {S.~T.}\ \bibnamefont {{Ohlmann}}}, \bibinfo
  {author} {\bibfnamefont {R.}~\bibnamefont {{Pakmor}}},\ and\ \bibinfo
  {author} {\bibfnamefont {V.}~\bibnamefont {{Springel}}},\ }\bibfield  {title}
  {\bibinfo {title} {{Bipolar planetary nebulae from common-envelope evolution
  of binary stars}},\ }\href {https://doi.org/10.1051/0004-6361/202142478}
  {\bibfield  {journal} {\bibinfo  {journal} {\aap}\ }\textbf {\bibinfo
  {volume} {660}},\ \bibinfo {eid} {L8} (\bibinfo {year} {2022})},\ \Eprint
  {https://arxiv.org/abs/2110.13177} {arXiv:2110.13177 [astro-ph.SR]}
  \BibitemShut {NoStop}%
\bibitem [{\citenamefont {{Sand}}\ \emph {et~al.}(2020)\citenamefont {{Sand}},
  \citenamefont {{Ohlmann}}, \citenamefont {{Schneider}}, \citenamefont
  {{Pakmor}},\ and\ \citenamefont {{R{\"o}pke}}}]{2020A&A...644A..60S}%
  \BibitemOpen
  \bibfield  {author} {\bibinfo {author} {\bibfnamefont {C.}~\bibnamefont
  {{Sand}}}, \bibinfo {author} {\bibfnamefont {S.~T.}\ \bibnamefont
  {{Ohlmann}}}, \bibinfo {author} {\bibfnamefont {F.~R.~N.}\ \bibnamefont
  {{Schneider}}}, \bibinfo {author} {\bibfnamefont {R.}~\bibnamefont
  {{Pakmor}}},\ and\ \bibinfo {author} {\bibfnamefont {F.~K.}\ \bibnamefont
  {{R{\"o}pke}}},\ }\bibfield  {title} {\bibinfo {title} {{Common-envelope
  evolution with an asymptotic giant branch star}},\ }\href
  {https://doi.org/10.1051/0004-6361/202038992} {\bibfield  {journal} {\bibinfo
   {journal} {\aap}\ }\textbf {\bibinfo {volume} {644}},\ \bibinfo {eid} {A60}
  (\bibinfo {year} {2020})},\ \Eprint {https://arxiv.org/abs/2007.11000}
  {arXiv:2007.11000 [astro-ph.SR]} \BibitemShut {NoStop}%
\bibitem [{\citenamefont {{Moreno}}\ \emph {et~al.}(2022)\citenamefont
  {{Moreno}}, \citenamefont {{Schneider}}, \citenamefont {{R{\"o}pke}},
  \citenamefont {{Ohlmann}}, \citenamefont {{Pakmor}}, \citenamefont
  {{Podsiadlowski}},\ and\ \citenamefont {{Sand}}}]{2022A&A...667A..72M}%
  \BibitemOpen
  \bibfield  {author} {\bibinfo {author} {\bibfnamefont {M.~M.}\ \bibnamefont
  {{Moreno}}}, \bibinfo {author} {\bibfnamefont {F.~R.~N.}\ \bibnamefont
  {{Schneider}}}, \bibinfo {author} {\bibfnamefont {F.~K.}\ \bibnamefont
  {{R{\"o}pke}}}, \bibinfo {author} {\bibfnamefont {S.~T.}\ \bibnamefont
  {{Ohlmann}}}, \bibinfo {author} {\bibfnamefont {R.}~\bibnamefont {{Pakmor}}},
  \bibinfo {author} {\bibfnamefont {P.}~\bibnamefont {{Podsiadlowski}}},\ and\
  \bibinfo {author} {\bibfnamefont {C.}~\bibnamefont {{Sand}}},\ }\bibfield
  {title} {\bibinfo {title} {{From 3D hydrodynamic simulations of
  common-envelope interaction to gravitational-wave mergers}},\ }\href
  {https://doi.org/10.1051/0004-6361/202142731} {\bibfield  {journal} {\bibinfo
   {journal} {\aap}\ }\textbf {\bibinfo {volume} {667}},\ \bibinfo {eid} {A72}
  (\bibinfo {year} {2022})},\ \Eprint {https://arxiv.org/abs/2111.12112}
  {arXiv:2111.12112 [astro-ph.SR]} \BibitemShut {NoStop}%
\bibitem [{\citenamefont {{Laplace}}\ \emph
  {et~al.}(2020{\natexlab{a}})\citenamefont {{Laplace}}, \citenamefont
  {{G{\"o}tberg}}, \citenamefont {{de Mink}}, \citenamefont {{Justham}},\ and\
  \citenamefont {{Farmer}}}]{2020A&A...637A...6L}%
  \BibitemOpen
  \bibfield  {author} {\bibinfo {author} {\bibfnamefont {E.}~\bibnamefont
  {{Laplace}}}, \bibinfo {author} {\bibfnamefont {Y.}~\bibnamefont
  {{G{\"o}tberg}}}, \bibinfo {author} {\bibfnamefont {S.~E.}\ \bibnamefont {{de
  Mink}}}, \bibinfo {author} {\bibfnamefont {S.}~\bibnamefont {{Justham}}},\
  and\ \bibinfo {author} {\bibfnamefont {R.}~\bibnamefont {{Farmer}}},\
  }\bibfield  {title} {\bibinfo {title} {{The expansion of stripped-envelope
  stars: Consequences for supernovae and gravitational-wave progenitors}},\
  }\href {https://doi.org/10.1051/0004-6361/201937300} {\bibfield  {journal}
  {\bibinfo  {journal} {\aap}\ }\textbf {\bibinfo {volume} {637}},\ \bibinfo
  {eid} {A6} (\bibinfo {year} {2020}{\natexlab{a}})},\ \Eprint
  {https://arxiv.org/abs/2003.01120} {arXiv:2003.01120 [astro-ph.SR]}
  \BibitemShut {NoStop}%
\bibitem [{\citenamefont {{Gagnier}}\ and\ \citenamefont
  {{Pejcha}}(2023)}]{2023arXiv230200691G}%
  \BibitemOpen
  \bibfield  {author} {\bibinfo {author} {\bibfnamefont {D.}~\bibnamefont
  {{Gagnier}}}\ and\ \bibinfo {author} {\bibfnamefont {O.}~\bibnamefont
  {{Pejcha}}},\ }\bibfield  {title} {\bibinfo {title} {{Post-dynamical inspiral
  phase of common envelope evolution: Binary orbit evolution and angular
  momentum transport}},\ }\href {https://doi.org/10.48550/arXiv.2302.00691}
  {\bibfield  {journal} {\bibinfo  {journal} {arXiv e-prints}\ ,\ \bibinfo
  {eid} {arXiv:2302.00691}} (\bibinfo {year} {2023})},\ \Eprint
  {https://arxiv.org/abs/2302.00691} {arXiv:2302.00691 [astro-ph.SR]}
  \BibitemShut {NoStop}%
\bibitem [{\citenamefont {{Nelemans}}\ \emph {et~al.}(2000)\citenamefont
  {{Nelemans}}, \citenamefont {{Verbunt}}, \citenamefont {{Yungelson}},\ and\
  \citenamefont {{Portegies Zwart}}}]{2000A&A...360.1011N}%
  \BibitemOpen
  \bibfield  {author} {\bibinfo {author} {\bibfnamefont {G.}~\bibnamefont
  {{Nelemans}}}, \bibinfo {author} {\bibfnamefont {F.}~\bibnamefont
  {{Verbunt}}}, \bibinfo {author} {\bibfnamefont {L.~R.}\ \bibnamefont
  {{Yungelson}}},\ and\ \bibinfo {author} {\bibfnamefont {S.~F.}\ \bibnamefont
  {{Portegies Zwart}}},\ }\bibfield  {title} {\bibinfo {title} {{Reconstructing
  the evolution of double helium white dwarfs: envelope loss without
  spiral-in}},\ }\href@noop {} {\bibfield  {journal} {\bibinfo  {journal}
  {\aap}\ }\textbf {\bibinfo {volume} {360}},\ \bibinfo {pages} {1011}
  (\bibinfo {year} {2000})},\ \Eprint {https://arxiv.org/abs/astro-ph/0006216}
  {arXiv:astro-ph/0006216 [astro-ph]} \BibitemShut {NoStop}%
\bibitem [{\citenamefont {{Zorotovic}}\ \emph {et~al.}(2010)\citenamefont
  {{Zorotovic}}, \citenamefont {{Schreiber}}, \citenamefont {{G{\"a}nsicke}},\
  and\ \citenamefont {{Nebot G{\'o}mez-Mor{\'a}n}}}]{2010A&A...520A..86Z}%
  \BibitemOpen
  \bibfield  {author} {\bibinfo {author} {\bibfnamefont {M.}~\bibnamefont
  {{Zorotovic}}}, \bibinfo {author} {\bibfnamefont {M.~R.}\ \bibnamefont
  {{Schreiber}}}, \bibinfo {author} {\bibfnamefont {B.~T.}\ \bibnamefont
  {{G{\"a}nsicke}}},\ and\ \bibinfo {author} {\bibfnamefont {A.}~\bibnamefont
  {{Nebot G{\'o}mez-Mor{\'a}n}}},\ }\bibfield  {title} {\bibinfo {title}
  {{Post-common-envelope binaries from SDSS. IX: Constraining the
  common-envelope efficiency}},\ }\href
  {https://doi.org/10.1051/0004-6361/200913658} {\bibfield  {journal} {\bibinfo
   {journal} {\aap}\ }\textbf {\bibinfo {volume} {520}},\ \bibinfo {eid} {A86}
  (\bibinfo {year} {2010})},\ \Eprint {https://arxiv.org/abs/1006.1621}
  {arXiv:1006.1621 [astro-ph.SR]} \BibitemShut {NoStop}%
\bibitem [{\citenamefont {{De Marco}}\ \emph {et~al.}(2011)\citenamefont {{De
  Marco}}, \citenamefont {{Passy}}, \citenamefont {{Moe}}, \citenamefont
  {{Herwig}}, \citenamefont {{Mac Low}},\ and\ \citenamefont
  {{Paxton}}}]{2011MNRAS.411.2277D}%
  \BibitemOpen
  \bibfield  {author} {\bibinfo {author} {\bibfnamefont {O.}~\bibnamefont {{De
  Marco}}}, \bibinfo {author} {\bibfnamefont {J.-C.}\ \bibnamefont {{Passy}}},
  \bibinfo {author} {\bibfnamefont {M.}~\bibnamefont {{Moe}}}, \bibinfo
  {author} {\bibfnamefont {F.}~\bibnamefont {{Herwig}}}, \bibinfo {author}
  {\bibfnamefont {M.-M.}\ \bibnamefont {{Mac Low}}},\ and\ \bibinfo {author}
  {\bibfnamefont {B.}~\bibnamefont {{Paxton}}},\ }\bibfield  {title} {\bibinfo
  {title} {{On the {\ensuremath{\alpha}} formalism for the common envelope
  interaction}},\ }\href {https://doi.org/10.1111/j.1365-2966.2010.17891.x}
  {\bibfield  {journal} {\bibinfo  {journal} {\mnras}\ }\textbf {\bibinfo
  {volume} {411}},\ \bibinfo {pages} {2277} (\bibinfo {year} {2011})},\ \Eprint
  {https://arxiv.org/abs/1010.4374} {arXiv:1010.4374 [astro-ph.SR]}
  \BibitemShut {NoStop}%
\bibitem [{\citenamefont {{Zorotovic}}\ \emph {et~al.}(2011)\citenamefont
  {{Zorotovic}}, \citenamefont {{Schreiber}},\ and\ \citenamefont
  {{G{\"a}nsicke}}}]{2011A&A...536A..42Z}%
  \BibitemOpen
  \bibfield  {author} {\bibinfo {author} {\bibfnamefont {M.}~\bibnamefont
  {{Zorotovic}}}, \bibinfo {author} {\bibfnamefont {M.~R.}\ \bibnamefont
  {{Schreiber}}},\ and\ \bibinfo {author} {\bibfnamefont {B.~T.}\ \bibnamefont
  {{G{\"a}nsicke}}},\ }\bibfield  {title} {\bibinfo {title} {{Post common
  envelope binaries from SDSS. XI. The white dwarf mass distributions of CVs
  and pre-CVs}},\ }\href {https://doi.org/10.1051/0004-6361/201116626}
  {\bibfield  {journal} {\bibinfo  {journal} {\aap}\ }\textbf {\bibinfo
  {volume} {536}},\ \bibinfo {eid} {A42} (\bibinfo {year} {2011})},\ \Eprint
  {https://arxiv.org/abs/1108.4600} {arXiv:1108.4600 [astro-ph.SR]}
  \BibitemShut {NoStop}%
\bibitem [{\citenamefont {{Davis}}\ \emph {et~al.}(2012)\citenamefont
  {{Davis}}, \citenamefont {{Kolb}},\ and\ \citenamefont
  {{Knigge}}}]{2012MNRAS.419..287D}%
  \BibitemOpen
  \bibfield  {author} {\bibinfo {author} {\bibfnamefont {P.~J.}\ \bibnamefont
  {{Davis}}}, \bibinfo {author} {\bibfnamefont {U.}~\bibnamefont {{Kolb}}},\
  and\ \bibinfo {author} {\bibfnamefont {C.}~\bibnamefont {{Knigge}}},\
  }\bibfield  {title} {\bibinfo {title} {{Is the common envelope ejection
  efficiency a function of the binary parameters?}},\ }\href
  {https://doi.org/10.1111/j.1365-2966.2011.19690.x} {\bibfield  {journal}
  {\bibinfo  {journal} {\mnras}\ }\textbf {\bibinfo {volume} {419}},\ \bibinfo
  {pages} {287} (\bibinfo {year} {2012})},\ \Eprint
  {https://arxiv.org/abs/1106.4741} {arXiv:1106.4741 [astro-ph.SR]}
  \BibitemShut {NoStop}%
\bibitem [{\citenamefont {{Scherbak}}\ and\ \citenamefont
  {{Fuller}}(2022)}]{2022arXiv221102036S}%
  \BibitemOpen
  \bibfield  {author} {\bibinfo {author} {\bibfnamefont {P.}~\bibnamefont
  {{Scherbak}}}\ and\ \bibinfo {author} {\bibfnamefont {J.}~\bibnamefont
  {{Fuller}}},\ }\bibfield  {title} {\bibinfo {title} {{White dwarf binaries
  suggest a common envelope efficiency $\alpha \sim 1/3$}},\ }\href@noop {}
  {\bibfield  {journal} {\bibinfo  {journal} {arXiv e-prints}\ ,\ \bibinfo
  {eid} {arXiv:2211.02036}} (\bibinfo {year} {2022})},\ \Eprint
  {https://arxiv.org/abs/2211.02036} {arXiv:2211.02036 [astro-ph.SR]}
  \BibitemShut {NoStop}%
\bibitem [{\citenamefont {{Kalogera}}(1999)}]{1999ApJ...521..723K}%
  \BibitemOpen
  \bibfield  {author} {\bibinfo {author} {\bibfnamefont {V.}~\bibnamefont
  {{Kalogera}}},\ }\bibfield  {title} {\bibinfo {title} {{Donor Stars in Black
  Hole X-Ray Binaries}},\ }\href {https://doi.org/10.1086/307562} {\bibfield
  {journal} {\bibinfo  {journal} {\apj}\ }\textbf {\bibinfo {volume} {521}},\
  \bibinfo {pages} {723} (\bibinfo {year} {1999})},\ \Eprint
  {https://arxiv.org/abs/astro-ph/9903417} {arXiv:astro-ph/9903417 [astro-ph]}
  \BibitemShut {NoStop}%
\bibitem [{\citenamefont {{Podsiadlowski}}\ \emph {et~al.}(2003)\citenamefont
  {{Podsiadlowski}}, \citenamefont {{Rappaport}},\ and\ \citenamefont
  {{Han}}}]{2003MNRAS.341..385P}%
  \BibitemOpen
  \bibfield  {author} {\bibinfo {author} {\bibfnamefont {P.}~\bibnamefont
  {{Podsiadlowski}}}, \bibinfo {author} {\bibfnamefont {S.}~\bibnamefont
  {{Rappaport}}},\ and\ \bibinfo {author} {\bibfnamefont {Z.}~\bibnamefont
  {{Han}}},\ }\bibfield  {title} {\bibinfo {title} {{On the formation and
  evolution of black hole binaries}},\ }\href
  {https://doi.org/10.1046/j.1365-8711.2003.06464.x} {\bibfield  {journal}
  {\bibinfo  {journal} {\mnras}\ }\textbf {\bibinfo {volume} {341}},\ \bibinfo
  {pages} {385} (\bibinfo {year} {2003})},\ \Eprint
  {https://arxiv.org/abs/astro-ph/0207153} {arXiv:astro-ph/0207153 [astro-ph]}
  \BibitemShut {NoStop}%
\bibitem [{\citenamefont {{Garc{\'\i}a}}\ \emph {et~al.}(2021)\citenamefont
  {{Garc{\'\i}a}}, \citenamefont {{Simaz Bunzel}}, \citenamefont {{Chaty}},
  \citenamefont {{Porter}},\ and\ \citenamefont
  {{Chassande-Mottin}}}]{2021A&A...649A.114G}%
  \BibitemOpen
  \bibfield  {author} {\bibinfo {author} {\bibfnamefont {F.}~\bibnamefont
  {{Garc{\'\i}a}}}, \bibinfo {author} {\bibfnamefont {A.}~\bibnamefont {{Simaz
  Bunzel}}}, \bibinfo {author} {\bibfnamefont {S.}~\bibnamefont {{Chaty}}},
  \bibinfo {author} {\bibfnamefont {E.}~\bibnamefont {{Porter}}},\ and\
  \bibinfo {author} {\bibfnamefont {E.}~\bibnamefont {{Chassande-Mottin}}},\
  }\bibfield  {title} {\bibinfo {title} {{Progenitors of low-mass binary
  black-hole mergers in the isolated binary evolution scenario}},\ }\href
  {https://doi.org/10.1051/0004-6361/202038357} {\bibfield  {journal} {\bibinfo
   {journal} {\aap}\ }\textbf {\bibinfo {volume} {649}},\ \bibinfo {eid} {A114}
  (\bibinfo {year} {2021})},\ \Eprint {https://arxiv.org/abs/2103.03161}
  {arXiv:2103.03161 [astro-ph.HE]} \BibitemShut {NoStop}%
\bibitem [{\citenamefont {{de Mink}}\ \emph
  {et~al.}(2008{\natexlab{a}})\citenamefont {{de Mink}}, \citenamefont
  {{Pols}},\ and\ \citenamefont {{Yoon}}}]{2008AIPC..990..230D}%
  \BibitemOpen
  \bibfield  {author} {\bibinfo {author} {\bibfnamefont {S.~E.}\ \bibnamefont
  {{de Mink}}}, \bibinfo {author} {\bibfnamefont {O.~R.}\ \bibnamefont
  {{Pols}}},\ and\ \bibinfo {author} {\bibfnamefont {S.~C.}\ \bibnamefont
  {{Yoon}}},\ }\bibfield  {title} {\bibinfo {title} {{Binaries at Low
  Metallicity: Ranges For Case A, B and C Mass Transfer}},\ }in\ \href
  {https://doi.org/10.1063/1.2905549} {\emph {\bibinfo {booktitle} {First Stars
  III}}},\ \bibinfo {series} {American Institute of Physics Conference Series},
  Vol.\ \bibinfo {volume} {990},\ \bibinfo {editor} {edited by\ \bibinfo
  {editor} {\bibfnamefont {B.~W.}\ \bibnamefont {{O'Shea}}}\ and\ \bibinfo
  {editor} {\bibfnamefont {A.}~\bibnamefont {{Heger}}}}\ (\bibinfo {year}
  {2008})\ pp.\ \bibinfo {pages} {230--232},\ \Eprint
  {https://arxiv.org/abs/0710.1010} {arXiv:0710.1010 [astro-ph]} \BibitemShut
  {NoStop}%
\bibitem [{\citenamefont {{Davies}}\ \emph {et~al.}(2018)\citenamefont
  {{Davies}}, \citenamefont {{Crowther}},\ and\ \citenamefont
  {{Beasor}}}]{2018MNRAS.478.3138D}%
  \BibitemOpen
  \bibfield  {author} {\bibinfo {author} {\bibfnamefont {B.}~\bibnamefont
  {{Davies}}}, \bibinfo {author} {\bibfnamefont {P.~A.}\ \bibnamefont
  {{Crowther}}},\ and\ \bibinfo {author} {\bibfnamefont {E.~R.}\ \bibnamefont
  {{Beasor}}},\ }\bibfield  {title} {\bibinfo {title} {{The luminosities of
  cool supergiants in the Magellanic Clouds, and the Humphreys-Davidson limit
  revisited}},\ }\href {https://doi.org/10.1093/mnras/sty1302} {\bibfield
  {journal} {\bibinfo  {journal} {\mnras}\ }\textbf {\bibinfo {volume} {478}},\
  \bibinfo {pages} {3138} (\bibinfo {year} {2018})},\ \Eprint
  {https://arxiv.org/abs/1804.06417} {arXiv:1804.06417 [astro-ph.SR]}
  \BibitemShut {NoStop}%
\bibitem [{\citenamefont {{Renzo}}\ \emph {et~al.}(2022)\citenamefont
  {{Renzo}}, \citenamefont {{Zapartas}}, \citenamefont {{Justham}},
  \citenamefont {{Breivik}}, \citenamefont {{Lau}}, \citenamefont {{Farmer}},
  \citenamefont {{Cantiello}},\ and\ \citenamefont
  {{Metzger}}}]{2022arXiv220615338R}%
  \BibitemOpen
  \bibfield  {author} {\bibinfo {author} {\bibfnamefont {M.}~\bibnamefont
  {{Renzo}}}, \bibinfo {author} {\bibfnamefont {E.}~\bibnamefont {{Zapartas}}},
  \bibinfo {author} {\bibfnamefont {S.}~\bibnamefont {{Justham}}}, \bibinfo
  {author} {\bibfnamefont {K.}~\bibnamefont {{Breivik}}}, \bibinfo {author}
  {\bibfnamefont {M.}~\bibnamefont {{Lau}}}, \bibinfo {author} {\bibfnamefont
  {R.}~\bibnamefont {{Farmer}}}, \bibinfo {author} {\bibfnamefont
  {M.}~\bibnamefont {{Cantiello}}},\ and\ \bibinfo {author} {\bibfnamefont
  {B.~D.}\ \bibnamefont {{Metzger}}},\ }\bibfield  {title} {\bibinfo {title}
  {{Rejuvenated accretors have less bound envelopes: Impact of Roche lobe
  overflow on subsequent common envelope events}},\ }\href@noop {} {\bibfield
  {journal} {\bibinfo  {journal} {arXiv e-prints}\ ,\ \bibinfo {eid}
  {arXiv:2206.15338}} (\bibinfo {year} {2022})},\ \Eprint
  {https://arxiv.org/abs/2206.15338} {arXiv:2206.15338 [astro-ph.SR]}
  \BibitemShut {NoStop}%
\bibitem [{\citenamefont {{MacLeod}}\ and\ \citenamefont
  {{Loeb}}(2020{\natexlab{b}})}]{2020ApJ...895...29M}%
  \BibitemOpen
  \bibfield  {author} {\bibinfo {author} {\bibfnamefont {M.}~\bibnamefont
  {{MacLeod}}}\ and\ \bibinfo {author} {\bibfnamefont {A.}~\bibnamefont
  {{Loeb}}},\ }\bibfield  {title} {\bibinfo {title} {{Pre-common-envelope Mass
  Loss from Coalescing Binary Systems}},\ }\href
  {https://doi.org/10.3847/1538-4357/ab89b6} {\bibfield  {journal} {\bibinfo
  {journal} {\apj}\ }\textbf {\bibinfo {volume} {895}},\ \bibinfo {eid} {29}
  (\bibinfo {year} {2020}{\natexlab{b}})},\ \Eprint
  {https://arxiv.org/abs/2003.01123} {arXiv:2003.01123 [astro-ph.SR]}
  \BibitemShut {NoStop}%
\bibitem [{\citenamefont {{Edgar}}(2004)}]{2004NewAR..48..843E}%
  \BibitemOpen
  \bibfield  {author} {\bibinfo {author} {\bibfnamefont {R.}~\bibnamefont
  {{Edgar}}},\ }\bibfield  {title} {\bibinfo {title} {{A review of
  Bondi-Hoyle-Lyttleton accretion}},\ }\href
  {https://doi.org/10.1016/j.newar.2004.06.001} {\bibfield  {journal} {\bibinfo
   {journal} {\nar}\ }\textbf {\bibinfo {volume} {48}},\ \bibinfo {pages} {843}
  (\bibinfo {year} {2004})},\ \Eprint {https://arxiv.org/abs/astro-ph/0406166}
  {arXiv:astro-ph/0406166 [astro-ph]} \BibitemShut {NoStop}%
\bibitem [{\citenamefont {{MacLeod}}\ and\ \citenamefont
  {{Ramirez-Ruiz}}(2015{\natexlab{a}})}]{2015ApJ...798L..19M}%
  \BibitemOpen
  \bibfield  {author} {\bibinfo {author} {\bibfnamefont {M.}~\bibnamefont
  {{MacLeod}}}\ and\ \bibinfo {author} {\bibfnamefont {E.}~\bibnamefont
  {{Ramirez-Ruiz}}},\ }\bibfield  {title} {\bibinfo {title} {{On the
  Accretion-fed Growth of Neutron Stars during Common Envelope}},\ }\href
  {https://doi.org/10.1088/2041-8205/798/1/L19} {\bibfield  {journal} {\bibinfo
   {journal} {\apjl}\ }\textbf {\bibinfo {volume} {798}},\ \bibinfo {eid} {L19}
  (\bibinfo {year} {2015}{\natexlab{a}})},\ \Eprint
  {https://arxiv.org/abs/1410.5421} {arXiv:1410.5421 [astro-ph.SR]}
  \BibitemShut {NoStop}%
\bibitem [{\citenamefont {{De}}\ \emph {et~al.}(2020)\citenamefont {{De}},
  \citenamefont {{MacLeod}}, \citenamefont {{Everson}}, \citenamefont
  {{Antoni}}, \citenamefont {{Mandel}},\ and\ \citenamefont
  {{Ramirez-Ruiz}}}]{2020ApJ...897..130D}%
  \BibitemOpen
  \bibfield  {author} {\bibinfo {author} {\bibfnamefont {S.}~\bibnamefont
  {{De}}}, \bibinfo {author} {\bibfnamefont {M.}~\bibnamefont {{MacLeod}}},
  \bibinfo {author} {\bibfnamefont {R.~W.}\ \bibnamefont {{Everson}}}, \bibinfo
  {author} {\bibfnamefont {A.}~\bibnamefont {{Antoni}}}, \bibinfo {author}
  {\bibfnamefont {I.}~\bibnamefont {{Mandel}}},\ and\ \bibinfo {author}
  {\bibfnamefont {E.}~\bibnamefont {{Ramirez-Ruiz}}},\ }\bibfield  {title}
  {\bibinfo {title} {{Common Envelope Wind Tunnel: The Effects of Binary Mass
  Ratio and Implications for the Accretion-driven Growth of LIGO Binary Black
  Holes}},\ }\href {https://doi.org/10.3847/1538-4357/ab9ac6} {\bibfield
  {journal} {\bibinfo  {journal} {\apj}\ }\textbf {\bibinfo {volume} {897}},\
  \bibinfo {eid} {130} (\bibinfo {year} {2020})},\ \Eprint
  {https://arxiv.org/abs/1910.13333} {arXiv:1910.13333 [astro-ph.SR]}
  \BibitemShut {NoStop}%
\bibitem [{\citenamefont {{Murguia-Berthier}}\ \emph
  {et~al.}(2017)\citenamefont {{Murguia-Berthier}}, \citenamefont {{MacLeod}},
  \citenamefont {{Ramirez-Ruiz}}, \citenamefont {{Antoni}},\ and\ \citenamefont
  {{Macias}}}]{2017ApJ...845..173M}%
  \BibitemOpen
  \bibfield  {author} {\bibinfo {author} {\bibfnamefont {A.}~\bibnamefont
  {{Murguia-Berthier}}}, \bibinfo {author} {\bibfnamefont {M.}~\bibnamefont
  {{MacLeod}}}, \bibinfo {author} {\bibfnamefont {E.}~\bibnamefont
  {{Ramirez-Ruiz}}}, \bibinfo {author} {\bibfnamefont {A.}~\bibnamefont
  {{Antoni}}},\ and\ \bibinfo {author} {\bibfnamefont {P.}~\bibnamefont
  {{Macias}}},\ }\bibfield  {title} {\bibinfo {title} {{Accretion Disk Assembly
  During Common Envelope Evolution: Implications for Feedback and LIGO Binary
  Black Hole Formation}},\ }\href {https://doi.org/10.3847/1538-4357/aa8140}
  {\bibfield  {journal} {\bibinfo  {journal} {\apj}\ }\textbf {\bibinfo
  {volume} {845}},\ \bibinfo {eid} {173} (\bibinfo {year} {2017})},\ \Eprint
  {https://arxiv.org/abs/1705.04698} {arXiv:1705.04698 [astro-ph.SR]}
  \BibitemShut {NoStop}%
\bibitem [{\citenamefont {{Armitage}}\ and\ \citenamefont
  {{Livio}}(2000)}]{2000ApJ...532..540A}%
  \BibitemOpen
  \bibfield  {author} {\bibinfo {author} {\bibfnamefont {P.~J.}\ \bibnamefont
  {{Armitage}}}\ and\ \bibinfo {author} {\bibfnamefont {M.}~\bibnamefont
  {{Livio}}},\ }\bibfield  {title} {\bibinfo {title} {{Black Hole Formation via
  Hypercritical Accretion during Common-Envelope Evolution}},\ }\href
  {https://doi.org/10.1086/308548} {\bibfield  {journal} {\bibinfo  {journal}
  {\apj}\ }\textbf {\bibinfo {volume} {532}},\ \bibinfo {pages} {540} (\bibinfo
  {year} {2000})},\ \Eprint {https://arxiv.org/abs/astro-ph/9906028}
  {arXiv:astro-ph/9906028 [astro-ph]} \BibitemShut {NoStop}%
\bibitem [{\citenamefont {{Soker}}(2015)}]{2015ApJ...800..114S}%
  \BibitemOpen
  \bibfield  {author} {\bibinfo {author} {\bibfnamefont {N.}~\bibnamefont
  {{Soker}}},\ }\bibfield  {title} {\bibinfo {title} {{Close Stellar Binary
  Systems by Grazing Envelope Evolution}},\ }\href
  {https://doi.org/10.1088/0004-637X/800/2/114} {\bibfield  {journal} {\bibinfo
   {journal} {\apj}\ }\textbf {\bibinfo {volume} {800}},\ \bibinfo {eid} {114}
  (\bibinfo {year} {2015})},\ \Eprint {https://arxiv.org/abs/1410.5363}
  {arXiv:1410.5363 [astro-ph.SR]} \BibitemShut {NoStop}%
\bibitem [{\citenamefont {{Kippenhahn}}(1969)}]{1969A&A.....3...83K}%
  \BibitemOpen
  \bibfield  {author} {\bibinfo {author} {\bibfnamefont {R.}~\bibnamefont
  {{Kippenhahn}}},\ }\bibfield  {title} {\bibinfo {title} {{Mass Exchange in a
  Massive Close Binary System}},\ }\href@noop {} {\bibfield  {journal}
  {\bibinfo  {journal} {\aap}\ }\textbf {\bibinfo {volume} {3}},\ \bibinfo
  {pages} {83} (\bibinfo {year} {1969})}\BibitemShut {NoStop}%
\bibitem [{\citenamefont {{Paczy{\'n}ski}}(1967)}]{1967AcA....17..355P}%
  \BibitemOpen
  \bibfield  {author} {\bibinfo {author} {\bibfnamefont {B.}~\bibnamefont
  {{Paczy{\'n}ski}}},\ }\bibfield  {title} {\bibinfo {title} {{Evolution of
  Close Binaries. V. The Evolution of Massive Binaries and the Formation of the
  Wolf-Rayet Stars}},\ }\href@noop {} {\bibfield  {journal} {\bibinfo
  {journal} {\actaa}\ }\textbf {\bibinfo {volume} {17}},\ \bibinfo {pages}
  {355} (\bibinfo {year} {1967})}\BibitemShut {NoStop}%
\bibitem [{\citenamefont {{Podsiadlowski}}\ \emph {et~al.}(1992)\citenamefont
  {{Podsiadlowski}}, \citenamefont {{Joss}},\ and\ \citenamefont
  {{Hsu}}}]{1992ApJ...391..246P}%
  \BibitemOpen
  \bibfield  {author} {\bibinfo {author} {\bibfnamefont {P.}~\bibnamefont
  {{Podsiadlowski}}}, \bibinfo {author} {\bibfnamefont {P.~C.}\ \bibnamefont
  {{Joss}}},\ and\ \bibinfo {author} {\bibfnamefont {J.~J.~L.}\ \bibnamefont
  {{Hsu}}},\ }\bibfield  {title} {\bibinfo {title} {{Presupernova Evolution in
  Massive Interacting Binaries}},\ }\href {https://doi.org/10.1086/171341}
  {\bibfield  {journal} {\bibinfo  {journal} {\apj}\ }\textbf {\bibinfo
  {volume} {391}},\ \bibinfo {pages} {246} (\bibinfo {year}
  {1992})}\BibitemShut {NoStop}%
\bibitem [{\citenamefont {{Hillier}}(1996)}]{1996ASPC...96..111H}%
  \BibitemOpen
  \bibfield  {author} {\bibinfo {author} {\bibfnamefont {D.~J.}\ \bibnamefont
  {{Hillier}}},\ }\bibfield  {title} {\bibinfo {title} {{Wolf-Rayet stars and
  stellar winds}},\ }in\ \href@noop {} {\emph {\bibinfo {booktitle} {Hydrogen
  Deficient Stars}}},\ \bibinfo {series} {Astronomical Society of the Pacific
  Conference Series}, Vol.~\bibinfo {volume} {96},\ \bibinfo {editor} {edited
  by\ \bibinfo {editor} {\bibfnamefont {C.~S.}\ \bibnamefont {{Jeffery}}}\ and\
  \bibinfo {editor} {\bibfnamefont {U.}~\bibnamefont {{Heber}}}}\ (\bibinfo
  {year} {1996})\ p.\ \bibinfo {pages} {111}\BibitemShut {NoStop}%
\bibitem [{\citenamefont {{Pols}}\ and\ \citenamefont
  {{Dewi}}(2002)}]{2002PASA...19..233P}%
  \BibitemOpen
  \bibfield  {author} {\bibinfo {author} {\bibfnamefont {O.~R.}\ \bibnamefont
  {{Pols}}}\ and\ \bibinfo {author} {\bibfnamefont {J.~D.~M.}\ \bibnamefont
  {{Dewi}}},\ }\bibfield  {title} {\bibinfo {title} {{Helium-star Mass Loss and
  Its Implications for Black Hole Formation and Supernova Progenitors}},\
  }\href {https://doi.org/10.1071/AS01121} {\bibfield  {journal} {\bibinfo
  {journal} {\pasa}\ }\textbf {\bibinfo {volume} {19}},\ \bibinfo {pages} {233}
  (\bibinfo {year} {2002})},\ \Eprint {https://arxiv.org/abs/astro-ph/0203308}
  {arXiv:astro-ph/0203308 [astro-ph]} \BibitemShut {NoStop}%
\bibitem [{\citenamefont {{Van Bever}}\ and\ \citenamefont
  {{Vanbeveren}}(2003)}]{2003A&A...400...63V}%
  \BibitemOpen
  \bibfield  {author} {\bibinfo {author} {\bibfnamefont {J.}~\bibnamefont {{Van
  Bever}}}\ and\ \bibinfo {author} {\bibfnamefont {D.}~\bibnamefont
  {{Vanbeveren}}},\ }\bibfield  {title} {\bibinfo {title} {{The effects of
  binaries on the evolution of Wolf-Rayet type spectral features in
  starbursts}},\ }\href {https://doi.org/10.1051/0004-6361:20021884} {\bibfield
   {journal} {\bibinfo  {journal} {\aap}\ }\textbf {\bibinfo {volume} {400}},\
  \bibinfo {pages} {63} (\bibinfo {year} {2003})}\BibitemShut {NoStop}%
\bibitem [{\citenamefont {{Eldridge}}\ and\ \citenamefont
  {{Stanway}}(2009)}]{2009MNRAS.400.1019E}%
  \BibitemOpen
  \bibfield  {author} {\bibinfo {author} {\bibfnamefont {J.~J.}\ \bibnamefont
  {{Eldridge}}}\ and\ \bibinfo {author} {\bibfnamefont {E.~R.}\ \bibnamefont
  {{Stanway}}},\ }\bibfield  {title} {\bibinfo {title} {{Spectral population
  synthesis including massive binaries}},\ }\href
  {https://doi.org/10.1111/j.1365-2966.2009.15514.x} {\bibfield  {journal}
  {\bibinfo  {journal} {\mnras}\ }\textbf {\bibinfo {volume} {400}},\ \bibinfo
  {pages} {1019} (\bibinfo {year} {2009})},\ \Eprint
  {https://arxiv.org/abs/0908.1386} {arXiv:0908.1386 [astro-ph.CO]}
  \BibitemShut {NoStop}%
\bibitem [{\citenamefont {{G{\"o}tberg}}\ \emph {et~al.}(2017)\citenamefont
  {{G{\"o}tberg}}, \citenamefont {{de Mink}},\ and\ \citenamefont
  {{Groh}}}]{2017A&A...608A..11G}%
  \BibitemOpen
  \bibfield  {author} {\bibinfo {author} {\bibfnamefont {Y.}~\bibnamefont
  {{G{\"o}tberg}}}, \bibinfo {author} {\bibfnamefont {S.~E.}\ \bibnamefont {{de
  Mink}}},\ and\ \bibinfo {author} {\bibfnamefont {J.~H.}\ \bibnamefont
  {{Groh}}},\ }\bibfield  {title} {\bibinfo {title} {{Ionizing spectra of stars
  that lose their envelope through interaction with a binary companion: role of
  metallicity}},\ }\href {https://doi.org/10.1051/0004-6361/201730472}
  {\bibfield  {journal} {\bibinfo  {journal} {\aap}\ }\textbf {\bibinfo
  {volume} {608}},\ \bibinfo {eid} {A11} (\bibinfo {year} {2017})},\ \Eprint
  {https://arxiv.org/abs/1701.07439} {arXiv:1701.07439 [astro-ph.SR]}
  \BibitemShut {NoStop}%
\bibitem [{\citenamefont
  {{Woosley}}(2019{\natexlab{a}})}]{2019ApJ...878...49W}%
  \BibitemOpen
  \bibfield  {author} {\bibinfo {author} {\bibfnamefont {S.~E.}\ \bibnamefont
  {{Woosley}}},\ }\bibfield  {title} {\bibinfo {title} {{The Evolution of
  Massive Helium Stars, Including Mass Loss}},\ }\href
  {https://doi.org/10.3847/1538-4357/ab1b41} {\bibfield  {journal} {\bibinfo
  {journal} {\apj}\ }\textbf {\bibinfo {volume} {878}},\ \bibinfo {eid} {49}
  (\bibinfo {year} {2019}{\natexlab{a}})},\ \Eprint
  {https://arxiv.org/abs/1901.00215} {arXiv:1901.00215 [astro-ph.SR]}
  \BibitemShut {NoStop}%
\bibitem [{\citenamefont {{Schneider}}\ \emph
  {et~al.}(2021{\natexlab{a}})\citenamefont {{Schneider}}, \citenamefont
  {{Podsiadlowski}},\ and\ \citenamefont {{M{\"u}ller}}}]{2021A&A...645A...5S}%
  \BibitemOpen
  \bibfield  {author} {\bibinfo {author} {\bibfnamefont {F.~R.~N.}\
  \bibnamefont {{Schneider}}}, \bibinfo {author} {\bibfnamefont
  {P.}~\bibnamefont {{Podsiadlowski}}},\ and\ \bibinfo {author} {\bibfnamefont
  {B.}~\bibnamefont {{M{\"u}ller}}},\ }\bibfield  {title} {\bibinfo {title}
  {{Pre-supernova evolution, compact-object masses, and explosion properties of
  stripped binary stars}},\ }\href
  {https://doi.org/10.1051/0004-6361/202039219} {\bibfield  {journal} {\bibinfo
   {journal} {\aap}\ }\textbf {\bibinfo {volume} {645}},\ \bibinfo {eid} {A5}
  (\bibinfo {year} {2021}{\natexlab{a}})},\ \Eprint
  {https://arxiv.org/abs/2008.08599} {arXiv:2008.08599 [astro-ph.SR]}
  \BibitemShut {NoStop}%
\bibitem [{\citenamefont {{Vartanyan}}\ \emph {et~al.}(2021)\citenamefont
  {{Vartanyan}}, \citenamefont {{Laplace}}, \citenamefont {{Renzo}},
  \citenamefont {{G{\"o}tberg}}, \citenamefont {{Burrows}},\ and\ \citenamefont
  {{de Mink}}}]{2021ApJ...916L...5V}%
  \BibitemOpen
  \bibfield  {author} {\bibinfo {author} {\bibfnamefont {D.}~\bibnamefont
  {{Vartanyan}}}, \bibinfo {author} {\bibfnamefont {E.}~\bibnamefont
  {{Laplace}}}, \bibinfo {author} {\bibfnamefont {M.}~\bibnamefont {{Renzo}}},
  \bibinfo {author} {\bibfnamefont {Y.}~\bibnamefont {{G{\"o}tberg}}}, \bibinfo
  {author} {\bibfnamefont {A.}~\bibnamefont {{Burrows}}},\ and\ \bibinfo
  {author} {\bibfnamefont {S.~E.}\ \bibnamefont {{de Mink}}},\ }\bibfield
  {title} {\bibinfo {title} {{Binary-stripped Stars as Core-collapse Supernovae
  Progenitors}},\ }\href {https://doi.org/10.3847/2041-8213/ac0b42} {\bibfield
  {journal} {\bibinfo  {journal} {\apjl}\ }\textbf {\bibinfo {volume} {916}},\
  \bibinfo {eid} {L5} (\bibinfo {year} {2021})},\ \Eprint
  {https://arxiv.org/abs/2104.03317} {arXiv:2104.03317 [astro-ph.SR]}
  \BibitemShut {NoStop}%
\bibitem [{\citenamefont {{Arnett}}(1978)}]{1978ApJ...219.1008A}%
  \BibitemOpen
  \bibfield  {author} {\bibinfo {author} {\bibfnamefont {W.~D.}\ \bibnamefont
  {{Arnett}}},\ }\bibfield  {title} {\bibinfo {title} {{On the bulk yields of
  nucleosynthesis from massive stars.}},\ }\href
  {https://doi.org/10.1086/155865} {\bibfield  {journal} {\bibinfo  {journal}
  {\apj}\ }\textbf {\bibinfo {volume} {219}},\ \bibinfo {pages} {1008}
  (\bibinfo {year} {1978})}\BibitemShut {NoStop}%
\bibitem [{\citenamefont {{Sugimoto}}\ and\ \citenamefont
  {{Nomoto}}(1980)}]{1980SSRv...25..155S}%
  \BibitemOpen
  \bibfield  {author} {\bibinfo {author} {\bibfnamefont {D.}~\bibnamefont
  {{Sugimoto}}}\ and\ \bibinfo {author} {\bibfnamefont {K.}~\bibnamefont
  {{Nomoto}}},\ }\bibfield  {title} {\bibinfo {title} {{Pre Supernova Models
  and Supernovae}},\ }\href {https://doi.org/10.1007/BF00212318} {\bibfield
  {journal} {\bibinfo  {journal} {\ssr}\ }\textbf {\bibinfo {volume} {25}},\
  \bibinfo {pages} {155} (\bibinfo {year} {1980})}\BibitemShut {NoStop}%
\bibitem [{\citenamefont {{Podsiadlowski}}\ \emph {et~al.}(2004)\citenamefont
  {{Podsiadlowski}}, \citenamefont {{Langer}}, \citenamefont {{Poelarends}},
  \citenamefont {{Rappaport}}, \citenamefont {{Heger}},\ and\ \citenamefont
  {{Pfahl}}}]{2004ApJ...612.1044P}%
  \BibitemOpen
  \bibfield  {author} {\bibinfo {author} {\bibfnamefont {P.}~\bibnamefont
  {{Podsiadlowski}}}, \bibinfo {author} {\bibfnamefont {N.}~\bibnamefont
  {{Langer}}}, \bibinfo {author} {\bibfnamefont {A.~J.~T.}\ \bibnamefont
  {{Poelarends}}}, \bibinfo {author} {\bibfnamefont {S.}~\bibnamefont
  {{Rappaport}}}, \bibinfo {author} {\bibfnamefont {A.}~\bibnamefont
  {{Heger}}},\ and\ \bibinfo {author} {\bibfnamefont {E.}~\bibnamefont
  {{Pfahl}}},\ }\bibfield  {title} {\bibinfo {title} {{The Effects of Binary
  Evolution on the Dynamics of Core Collapse and Neutron Star Kicks}},\ }\href
  {https://doi.org/10.1086/421713} {\bibfield  {journal} {\bibinfo  {journal}
  {\apj}\ }\textbf {\bibinfo {volume} {612}},\ \bibinfo {pages} {1044}
  (\bibinfo {year} {2004})},\ \Eprint {https://arxiv.org/abs/astro-ph/0309588}
  {arXiv:astro-ph/0309588 [astro-ph]} \BibitemShut {NoStop}%
\bibitem [{\citenamefont {{Gal-Yam}}(2017)}]{2017hsn..book..195G}%
  \BibitemOpen
  \bibfield  {author} {\bibinfo {author} {\bibfnamefont {A.}~\bibnamefont
  {{Gal-Yam}}},\ }\bibfield  {title} {\bibinfo {title} {{Observational and
  Physical Classification of Supernovae}},\ }in\ \href
  {https://doi.org/10.1007/978-3-319-21846-5_35} {\emph {\bibinfo {booktitle}
  {Handbook of Supernovae}}},\ \bibinfo {editor} {edited by\ \bibinfo {editor}
  {\bibfnamefont {A.~W.}\ \bibnamefont {{Alsabti}}}\ and\ \bibinfo {editor}
  {\bibfnamefont {P.}~\bibnamefont {{Murdin}}}}\ (\bibinfo {year} {2017})\ p.\
  \bibinfo {pages} {195}\BibitemShut {NoStop}%
\bibitem [{\citenamefont {{Yoon}}(2017)}]{2017MNRAS.470.3970Y}%
  \BibitemOpen
  \bibfield  {author} {\bibinfo {author} {\bibfnamefont {S.-C.}\ \bibnamefont
  {{Yoon}}},\ }\bibfield  {title} {\bibinfo {title} {{Towards a better
  understanding of the evolution of Wolf-Rayet stars and Type Ib/Ic supernova
  progenitors}},\ }\href {https://doi.org/10.1093/mnras/stx1496} {\bibfield
  {journal} {\bibinfo  {journal} {\mnras}\ }\textbf {\bibinfo {volume} {470}},\
  \bibinfo {pages} {3970} (\bibinfo {year} {2017})},\ \Eprint
  {https://arxiv.org/abs/1706.04716} {arXiv:1706.04716 [astro-ph.SR]}
  \BibitemShut {NoStop}%
\bibitem [{\citenamefont {{Blaauw}}(1961)}]{1961BAN....15..265B}%
  \BibitemOpen
  \bibfield  {author} {\bibinfo {author} {\bibfnamefont {A.}~\bibnamefont
  {{Blaauw}}},\ }\bibfield  {title} {\bibinfo {title} {{On the origin of the O-
  and B-type stars with high velocities (the ``run-away'' stars), and some
  related problems}},\ }\href@noop {} {\bibfield  {journal} {\bibinfo
  {journal} {\bain}\ }\textbf {\bibinfo {volume} {15}},\ \bibinfo {pages} {265}
  (\bibinfo {year} {1961})}\BibitemShut {NoStop}%
\bibitem [{\citenamefont {{Tauris}}\ and\ \citenamefont
  {{Takens}}(1998)}]{1998A&A...330.1047T}%
  \BibitemOpen
  \bibfield  {author} {\bibinfo {author} {\bibfnamefont {T.~M.}\ \bibnamefont
  {{Tauris}}}\ and\ \bibinfo {author} {\bibfnamefont {R.~J.}\ \bibnamefont
  {{Takens}}},\ }\bibfield  {title} {\bibinfo {title} {{Runaway velocities of
  stellar components originating from disrupted binaries via asymmetric
  supernova explosions}},\ }\href@noop {} {\bibfield  {journal} {\bibinfo
  {journal} {\aap}\ }\textbf {\bibinfo {volume} {330}},\ \bibinfo {pages}
  {1047} (\bibinfo {year} {1998})}\BibitemShut {NoStop}%
\bibitem [{\citenamefont {{Fryer}}\ and\ \citenamefont
  {{Kusenko}}(2006)}]{2006ApJS..163..335F}%
  \BibitemOpen
  \bibfield  {author} {\bibinfo {author} {\bibfnamefont {C.~L.}\ \bibnamefont
  {{Fryer}}}\ and\ \bibinfo {author} {\bibfnamefont {A.}~\bibnamefont
  {{Kusenko}}},\ }\bibfield  {title} {\bibinfo {title} {{Effects of
  Neutrino-driven Kicks on the Supernova Explosion Mechanism}},\ }\href
  {https://doi.org/10.1086/500933} {\bibfield  {journal} {\bibinfo  {journal}
  {\apjs}\ }\textbf {\bibinfo {volume} {163}},\ \bibinfo {pages} {335}
  (\bibinfo {year} {2006})},\ \Eprint {https://arxiv.org/abs/astro-ph/0512033}
  {arXiv:astro-ph/0512033 [astro-ph]} \BibitemShut {NoStop}%
\bibitem [{\citenamefont {{Janka}}(2013)}]{2013MNRAS.434.1355J}%
  \BibitemOpen
  \bibfield  {author} {\bibinfo {author} {\bibfnamefont {H.-T.}\ \bibnamefont
  {{Janka}}},\ }\bibfield  {title} {\bibinfo {title} {{Natal kicks of stellar
  mass black holes by asymmetric mass ejection in fallback supernovae}},\
  }\href {https://doi.org/10.1093/mnras/stt1106} {\bibfield  {journal}
  {\bibinfo  {journal} {\mnras}\ }\textbf {\bibinfo {volume} {434}},\ \bibinfo
  {pages} {1355} (\bibinfo {year} {2013})},\ \Eprint
  {https://arxiv.org/abs/1306.0007} {arXiv:1306.0007 [astro-ph.SR]}
  \BibitemShut {NoStop}%
\bibitem [{\citenamefont {{Janka}}(2017)}]{2017ApJ...837...84J}%
  \BibitemOpen
  \bibfield  {author} {\bibinfo {author} {\bibfnamefont {H.-T.}\ \bibnamefont
  {{Janka}}},\ }\bibfield  {title} {\bibinfo {title} {{Neutron Star Kicks by
  the Gravitational Tug-boat Mechanism in Asymmetric Supernova Explosions:
  Progenitor and Explosion Dependence}},\ }\href
  {https://doi.org/10.3847/1538-4357/aa618e} {\bibfield  {journal} {\bibinfo
  {journal} {\apj}\ }\textbf {\bibinfo {volume} {837}},\ \bibinfo {eid} {84}
  (\bibinfo {year} {2017})},\ \Eprint {https://arxiv.org/abs/1611.07562}
  {arXiv:1611.07562 [astro-ph.HE]} \BibitemShut {NoStop}%
\bibitem [{\citenamefont {{Gunn}}\ and\ \citenamefont
  {{Ostriker}}(1970)}]{1970ApJ...160..979G}%
  \BibitemOpen
  \bibfield  {author} {\bibinfo {author} {\bibfnamefont {J.~E.}\ \bibnamefont
  {{Gunn}}}\ and\ \bibinfo {author} {\bibfnamefont {J.~P.}\ \bibnamefont
  {{Ostriker}}},\ }\bibfield  {title} {\bibinfo {title} {{On the Nature of
  Pulsars. III. Analysis of Observations}},\ }\href
  {https://doi.org/10.1086/150487} {\bibfield  {journal} {\bibinfo  {journal}
  {\apj}\ }\textbf {\bibinfo {volume} {160}},\ \bibinfo {pages} {979} (\bibinfo
  {year} {1970})}\BibitemShut {NoStop}%
\bibitem [{\citenamefont {{Lyne}}\ and\ \citenamefont
  {{Lorimer}}(1994)}]{1994Natur.369..127L}%
  \BibitemOpen
  \bibfield  {author} {\bibinfo {author} {\bibfnamefont {A.~G.}\ \bibnamefont
  {{Lyne}}}\ and\ \bibinfo {author} {\bibfnamefont {D.~R.}\ \bibnamefont
  {{Lorimer}}},\ }\bibfield  {title} {\bibinfo {title} {{High birth velocities
  of radio pulsars}},\ }\href {https://doi.org/10.1038/369127a0} {\bibfield
  {journal} {\bibinfo  {journal} {\nat}\ }\textbf {\bibinfo {volume} {369}},\
  \bibinfo {pages} {127} (\bibinfo {year} {1994})}\BibitemShut {NoStop}%
\bibitem [{\citenamefont {{Hansen}}\ and\ \citenamefont
  {{Phinney}}(1997)}]{1997MNRAS.291..569H}%
  \BibitemOpen
  \bibfield  {author} {\bibinfo {author} {\bibfnamefont {B.~M.~S.}\
  \bibnamefont {{Hansen}}}\ and\ \bibinfo {author} {\bibfnamefont {E.~S.}\
  \bibnamefont {{Phinney}}},\ }\bibfield  {title} {\bibinfo {title} {{The
  pulsar kick velocity distribution}},\ }\href
  {https://doi.org/10.1093/mnras/291.3.569} {\bibfield  {journal} {\bibinfo
  {journal} {\mnras}\ }\textbf {\bibinfo {volume} {291}},\ \bibinfo {pages}
  {569} (\bibinfo {year} {1997})},\ \Eprint
  {https://arxiv.org/abs/astro-ph/9708071} {arXiv:astro-ph/9708071 [astro-ph]}
  \BibitemShut {NoStop}%
\bibitem [{\citenamefont {{Arzoumanian}}\ \emph {et~al.}(2002)\citenamefont
  {{Arzoumanian}}, \citenamefont {{Chernoff}},\ and\ \citenamefont
  {{Cordes}}}]{2002ApJ...568..289A}%
  \BibitemOpen
  \bibfield  {author} {\bibinfo {author} {\bibfnamefont {Z.}~\bibnamefont
  {{Arzoumanian}}}, \bibinfo {author} {\bibfnamefont {D.~F.}\ \bibnamefont
  {{Chernoff}}},\ and\ \bibinfo {author} {\bibfnamefont {J.~M.}\ \bibnamefont
  {{Cordes}}},\ }\bibfield  {title} {\bibinfo {title} {{The Velocity
  Distribution of Isolated Radio Pulsars}},\ }\href
  {https://doi.org/10.1086/338805} {\bibfield  {journal} {\bibinfo  {journal}
  {\apj}\ }\textbf {\bibinfo {volume} {568}},\ \bibinfo {pages} {289} (\bibinfo
  {year} {2002})},\ \Eprint {https://arxiv.org/abs/astro-ph/0106159}
  {arXiv:astro-ph/0106159 [astro-ph]} \BibitemShut {NoStop}%
\bibitem [{\citenamefont {{Pfahl}}\ \emph
  {et~al.}(2002{\natexlab{a}})\citenamefont {{Pfahl}}, \citenamefont
  {{Rappaport}},\ and\ \citenamefont {{Podsiadlowski}}}]{2002ApJ...573..283P}%
  \BibitemOpen
  \bibfield  {author} {\bibinfo {author} {\bibfnamefont {E.}~\bibnamefont
  {{Pfahl}}}, \bibinfo {author} {\bibfnamefont {S.}~\bibnamefont
  {{Rappaport}}},\ and\ \bibinfo {author} {\bibfnamefont {P.}~\bibnamefont
  {{Podsiadlowski}}},\ }\bibfield  {title} {\bibinfo {title} {{A Comprehensive
  Study of Neutron Star Retention in Globular Clusters}},\ }\href
  {https://doi.org/10.1086/340494} {\bibfield  {journal} {\bibinfo  {journal}
  {\apj}\ }\textbf {\bibinfo {volume} {573}},\ \bibinfo {pages} {283} (\bibinfo
  {year} {2002}{\natexlab{a}})},\ \Eprint
  {https://arxiv.org/abs/astro-ph/0106141} {arXiv:astro-ph/0106141 [astro-ph]}
  \BibitemShut {NoStop}%
\bibitem [{\citenamefont {{Hobbs}}\ \emph
  {et~al.}(2005{\natexlab{a}})\citenamefont {{Hobbs}}, \citenamefont
  {{Lorimer}}, \citenamefont {{Lyne}},\ and\ \citenamefont
  {{Kramer}}}]{2005MNRAS.360..974H}%
  \BibitemOpen
  \bibfield  {author} {\bibinfo {author} {\bibfnamefont {G.}~\bibnamefont
  {{Hobbs}}}, \bibinfo {author} {\bibfnamefont {D.~R.}\ \bibnamefont
  {{Lorimer}}}, \bibinfo {author} {\bibfnamefont {A.~G.}\ \bibnamefont
  {{Lyne}}},\ and\ \bibinfo {author} {\bibfnamefont {M.}~\bibnamefont
  {{Kramer}}},\ }\bibfield  {title} {\bibinfo {title} {{A statistical study of
  233 pulsar proper motions}},\ }\href
  {https://doi.org/10.1111/j.1365-2966.2005.09087.x} {\bibfield  {journal}
  {\bibinfo  {journal} {\mnras}\ }\textbf {\bibinfo {volume} {360}},\ \bibinfo
  {pages} {974} (\bibinfo {year} {2005}{\natexlab{a}})},\ \Eprint
  {https://arxiv.org/abs/astro-ph/0504584} {astro-ph/0504584} \BibitemShut
  {NoStop}%
\bibitem [{\citenamefont {{Verbunt}}\ \emph
  {et~al.}(2017{\natexlab{a}})\citenamefont {{Verbunt}}, \citenamefont
  {{Igoshev}},\ and\ \citenamefont {{Cator}}}]{2017A&A...608A..57V}%
  \BibitemOpen
  \bibfield  {author} {\bibinfo {author} {\bibfnamefont {F.}~\bibnamefont
  {{Verbunt}}}, \bibinfo {author} {\bibfnamefont {A.}~\bibnamefont
  {{Igoshev}}},\ and\ \bibinfo {author} {\bibfnamefont {E.}~\bibnamefont
  {{Cator}}},\ }\bibfield  {title} {\bibinfo {title} {{The observed velocity
  distribution of young pulsars}},\ }\href
  {https://doi.org/10.1051/0004-6361/201731518} {\bibfield  {journal} {\bibinfo
   {journal} {\aap}\ }\textbf {\bibinfo {volume} {608}},\ \bibinfo {eid} {A57}
  (\bibinfo {year} {2017}{\natexlab{a}})},\ \Eprint
  {https://arxiv.org/abs/1708.08281} {arXiv:1708.08281 [astro-ph.HE]}
  \BibitemShut {NoStop}%
\bibitem [{\citenamefont
  {{Igoshev}}(2020{\natexlab{a}})}]{2020MNRAS.494.3663I}%
  \BibitemOpen
  \bibfield  {author} {\bibinfo {author} {\bibfnamefont {A.~P.}\ \bibnamefont
  {{Igoshev}}},\ }\bibfield  {title} {\bibinfo {title} {{The observed velocity
  distribution of young pulsars - II. Analysis of complete
  PSR{\ensuremath{\pi}}}},\ }\href {https://doi.org/10.1093/mnras/staa958}
  {\bibfield  {journal} {\bibinfo  {journal} {\mnras}\ }\textbf {\bibinfo
  {volume} {494}},\ \bibinfo {pages} {3663} (\bibinfo {year}
  {2020}{\natexlab{a}})},\ \Eprint {https://arxiv.org/abs/2002.01367}
  {arXiv:2002.01367 [astro-ph.HE]} \BibitemShut {NoStop}%
\bibitem [{\citenamefont {{Igoshev}}\ \emph {et~al.}(2021)\citenamefont
  {{Igoshev}}, \citenamefont {{Chruslinska}}, \citenamefont {{Dorozsmai}},\
  and\ \citenamefont {{Toonen}}}]{2021MNRAS.508.3345I}%
  \BibitemOpen
  \bibfield  {author} {\bibinfo {author} {\bibfnamefont {A.~P.}\ \bibnamefont
  {{Igoshev}}}, \bibinfo {author} {\bibfnamefont {M.}~\bibnamefont
  {{Chruslinska}}}, \bibinfo {author} {\bibfnamefont {A.}~\bibnamefont
  {{Dorozsmai}}},\ and\ \bibinfo {author} {\bibfnamefont {S.}~\bibnamefont
  {{Toonen}}},\ }\bibfield  {title} {\bibinfo {title} {{Combined analysis of
  neutron star natal kicks using proper motions and parallax measurements for
  radio pulsars and Be X-ray binaries}},\ }\href
  {https://doi.org/10.1093/mnras/stab2734} {\bibfield  {journal} {\bibinfo
  {journal} {\mnras}\ }\textbf {\bibinfo {volume} {508}},\ \bibinfo {pages}
  {3345} (\bibinfo {year} {2021})},\ \Eprint {https://arxiv.org/abs/2109.10362}
  {arXiv:2109.10362 [astro-ph.HE]} \BibitemShut {NoStop}%
\bibitem [{\citenamefont {{Mirabel}}\ and\ \citenamefont
  {{Rodrigues}}(2003)}]{2003Sci...300.1119M}%
  \BibitemOpen
  \bibfield  {author} {\bibinfo {author} {\bibfnamefont {I.~F.}\ \bibnamefont
  {{Mirabel}}}\ and\ \bibinfo {author} {\bibfnamefont {I.}~\bibnamefont
  {{Rodrigues}}},\ }\bibfield  {title} {\bibinfo {title} {{Formation of a Black
  Hole in the Dark}},\ }\href {https://doi.org/10.1126/science.1083451}
  {\bibfield  {journal} {\bibinfo  {journal} {Science}\ }\textbf {\bibinfo
  {volume} {300}},\ \bibinfo {pages} {1119} (\bibinfo {year} {2003})},\ \Eprint
  {https://arxiv.org/abs/astro-ph/0305205} {arXiv:astro-ph/0305205 [astro-ph]}
  \BibitemShut {NoStop}%
\bibitem [{\citenamefont {{Jonker}}\ and\ \citenamefont
  {{Nelemans}}(2004)}]{2004MNRAS.354..355J}%
  \BibitemOpen
  \bibfield  {author} {\bibinfo {author} {\bibfnamefont {P.~G.}\ \bibnamefont
  {{Jonker}}}\ and\ \bibinfo {author} {\bibfnamefont {G.}~\bibnamefont
  {{Nelemans}}},\ }\bibfield  {title} {\bibinfo {title} {{The distances to
  Galactic low-mass X-ray binaries: consequences for black hole luminosities
  and kicks}},\ }\href {https://doi.org/10.1111/j.1365-2966.2004.08193.x}
  {\bibfield  {journal} {\bibinfo  {journal} {\mnras}\ }\textbf {\bibinfo
  {volume} {354}},\ \bibinfo {pages} {355} (\bibinfo {year} {2004})},\ \Eprint
  {https://arxiv.org/abs/astro-ph/0407168} {arXiv:astro-ph/0407168 [astro-ph]}
  \BibitemShut {NoStop}%
\bibitem [{\citenamefont {{Dhawan}}\ \emph {et~al.}(2007)\citenamefont
  {{Dhawan}}, \citenamefont {{Mirabel}}, \citenamefont {{Rib{\'o}}},\ and\
  \citenamefont {{Rodrigues}}}]{2007ApJ...668..430D}%
  \BibitemOpen
  \bibfield  {author} {\bibinfo {author} {\bibfnamefont {V.}~\bibnamefont
  {{Dhawan}}}, \bibinfo {author} {\bibfnamefont {I.~F.}\ \bibnamefont
  {{Mirabel}}}, \bibinfo {author} {\bibfnamefont {M.}~\bibnamefont
  {{Rib{\'o}}}},\ and\ \bibinfo {author} {\bibfnamefont {I.}~\bibnamefont
  {{Rodrigues}}},\ }\bibfield  {title} {\bibinfo {title} {{Kinematics of Black
  Hole X-Ray Binary GRS 1915+105}},\ }\href {https://doi.org/10.1086/520111}
  {\bibfield  {journal} {\bibinfo  {journal} {\apj}\ }\textbf {\bibinfo
  {volume} {668}},\ \bibinfo {pages} {430} (\bibinfo {year} {2007})},\ \Eprint
  {https://arxiv.org/abs/0705.1800} {arXiv:0705.1800 [astro-ph]} \BibitemShut
  {NoStop}%
\bibitem [{\citenamefont {{Fragos}}\ \emph {et~al.}(2009)\citenamefont
  {{Fragos}}, \citenamefont {{Willems}}, \citenamefont {{Kalogera}},
  \citenamefont {{Ivanova}}, \citenamefont {{Rockefeller}}, \citenamefont
  {{Fryer}},\ and\ \citenamefont {{Young}}}]{2009ApJ...697.1057F}%
  \BibitemOpen
  \bibfield  {author} {\bibinfo {author} {\bibfnamefont {T.}~\bibnamefont
  {{Fragos}}}, \bibinfo {author} {\bibfnamefont {B.}~\bibnamefont {{Willems}}},
  \bibinfo {author} {\bibfnamefont {V.}~\bibnamefont {{Kalogera}}}, \bibinfo
  {author} {\bibfnamefont {N.}~\bibnamefont {{Ivanova}}}, \bibinfo {author}
  {\bibfnamefont {G.}~\bibnamefont {{Rockefeller}}}, \bibinfo {author}
  {\bibfnamefont {C.~L.}\ \bibnamefont {{Fryer}}},\ and\ \bibinfo {author}
  {\bibfnamefont {P.~A.}\ \bibnamefont {{Young}}},\ }\bibfield  {title}
  {\bibinfo {title} {{Understanding Compact Object Formation and Natal Kicks.
  II. The Case of XTE J1118 + 480}},\ }\href
  {https://doi.org/10.1088/0004-637X/697/2/1057} {\bibfield  {journal}
  {\bibinfo  {journal} {\apj}\ }\textbf {\bibinfo {volume} {697}},\ \bibinfo
  {pages} {1057} (\bibinfo {year} {2009})},\ \Eprint
  {https://arxiv.org/abs/0809.1588} {arXiv:0809.1588} \BibitemShut {NoStop}%
\bibitem [{\citenamefont {{Mandel}}(2016)}]{2016MNRAS.456..578M}%
  \BibitemOpen
  \bibfield  {author} {\bibinfo {author} {\bibfnamefont {I.}~\bibnamefont
  {{Mandel}}},\ }\bibfield  {title} {\bibinfo {title} {{Estimates of black hole
  natal kick velocities from observations of low-mass X-ray binaries}},\ }\href
  {https://doi.org/10.1093/mnras/stv2733} {\bibfield  {journal} {\bibinfo
  {journal} {\mnras}\ }\textbf {\bibinfo {volume} {456}},\ \bibinfo {pages}
  {578} (\bibinfo {year} {2016})},\ \Eprint {https://arxiv.org/abs/1510.03871}
  {arXiv:1510.03871 [astro-ph.HE]} \BibitemShut {NoStop}%
\bibitem [{\citenamefont {{Repetto}}\ \emph {et~al.}(2017)\citenamefont
  {{Repetto}}, \citenamefont {{Igoshev}},\ and\ \citenamefont
  {{Nelemans}}}]{2017MNRAS.467..298R}%
  \BibitemOpen
  \bibfield  {author} {\bibinfo {author} {\bibfnamefont {S.}~\bibnamefont
  {{Repetto}}}, \bibinfo {author} {\bibfnamefont {A.~P.}\ \bibnamefont
  {{Igoshev}}},\ and\ \bibinfo {author} {\bibfnamefont {G.}~\bibnamefont
  {{Nelemans}}},\ }\bibfield  {title} {\bibinfo {title} {{The Galactic
  distribution of X-ray binaries and its implications for compact object
  formation and natal kicks}},\ }\href {https://doi.org/10.1093/mnras/stx027}
  {\bibfield  {journal} {\bibinfo  {journal} {\mnras}\ }\textbf {\bibinfo
  {volume} {467}},\ \bibinfo {pages} {298} (\bibinfo {year} {2017})},\ \Eprint
  {https://arxiv.org/abs/1701.01347} {arXiv:1701.01347 [astro-ph.HE]}
  \BibitemShut {NoStop}%
\bibitem [{\citenamefont {{Kimball}}\ \emph {et~al.}(2022)\citenamefont
  {{Kimball}}, \citenamefont {{Imperato}}, \citenamefont {{Kalogera}},
  \citenamefont {{Rocha}}, \citenamefont {{Doctor}}, \citenamefont {{Andrews}},
  \citenamefont {{Dotter}}, \citenamefont {{Zapartas}}, \citenamefont
  {{Bavera}}, \citenamefont {{Kovlakas}}, \citenamefont {{Fragos}},
  \citenamefont {{Srivastava}}, \citenamefont {{Misra}}, \citenamefont
  {{Sun}},\ and\ \citenamefont {{Xing}}}]{2022arXiv221102158K}%
  \BibitemOpen
  \bibfield  {author} {\bibinfo {author} {\bibfnamefont {C.}~\bibnamefont
  {{Kimball}}}, \bibinfo {author} {\bibfnamefont {S.}~\bibnamefont
  {{Imperato}}}, \bibinfo {author} {\bibfnamefont {V.}~\bibnamefont
  {{Kalogera}}}, \bibinfo {author} {\bibfnamefont {K.~A.}\ \bibnamefont
  {{Rocha}}}, \bibinfo {author} {\bibfnamefont {Z.}~\bibnamefont {{Doctor}}},
  \bibinfo {author} {\bibfnamefont {J.~J.}\ \bibnamefont {{Andrews}}}, \bibinfo
  {author} {\bibfnamefont {A.}~\bibnamefont {{Dotter}}}, \bibinfo {author}
  {\bibfnamefont {E.}~\bibnamefont {{Zapartas}}}, \bibinfo {author}
  {\bibfnamefont {S.~S.}\ \bibnamefont {{Bavera}}}, \bibinfo {author}
  {\bibfnamefont {K.}~\bibnamefont {{Kovlakas}}}, \bibinfo {author}
  {\bibfnamefont {T.}~\bibnamefont {{Fragos}}}, \bibinfo {author}
  {\bibfnamefont {P.~M.}\ \bibnamefont {{Srivastava}}}, \bibinfo {author}
  {\bibfnamefont {D.}~\bibnamefont {{Misra}}}, \bibinfo {author} {\bibfnamefont
  {M.}~\bibnamefont {{Sun}}},\ and\ \bibinfo {author} {\bibfnamefont
  {Z.}~\bibnamefont {{Xing}}},\ }\bibfield  {title} {\bibinfo {title} {{A Black
  Hole Kicked At Birth: MAXI J1305-704}},\ }\href@noop {} {\bibfield  {journal}
  {\bibinfo  {journal} {arXiv e-prints}\ ,\ \bibinfo {eid} {arXiv:2211.02158}}
  (\bibinfo {year} {2022})},\ \Eprint {https://arxiv.org/abs/2211.02158}
  {arXiv:2211.02158 [astro-ph.HE]} \BibitemShut {NoStop}%
\bibitem [{\citenamefont {{Shenar}}\ \emph
  {et~al.}(2022{\natexlab{a}})\citenamefont {{Shenar}}, \citenamefont {{Sana}},
  \citenamefont {{Mahy}}, \citenamefont {{El-Badry}}, \citenamefont
  {{Marchant}}, \citenamefont {{Langer}}, \citenamefont {{Hawcroft}},
  \citenamefont {{Fabry}}, \citenamefont {{Sen}}, \citenamefont {{Almeida}},
  \citenamefont {{Abdul-Masih}}, \citenamefont {{Bodensteiner}}, \citenamefont
  {{Crowther}}, \citenamefont {{Gieles}}, \citenamefont {{Gromadzki}},
  \citenamefont {{H{\'e}nault-Brunet}}, \citenamefont {{Herrero}},
  \citenamefont {{Koter}}, \citenamefont {{Iwanek}}, \citenamefont
  {{Koz{\l}owski}}, \citenamefont {{Lennon}}, \citenamefont {{Apell{\'a}niz}},
  \citenamefont {{Mr{\'o}z}}, \citenamefont {{Moffat}}, \citenamefont
  {{Picco}}, \citenamefont {{Pietrukowicz}}, \citenamefont {{Poleski}},
  \citenamefont {{Rybicki}}, \citenamefont {{Schneider}}, \citenamefont
  {{Skowron}}, \citenamefont {{Skowron}}, \citenamefont {{Soszy{\'n}ski}},
  \citenamefont {{Szyma{\'n}ski}}, \citenamefont {{Toonen}}, \citenamefont
  {{Udalski}}, \citenamefont {{Ulaczyk}}, \citenamefont {{Vink}},\ and\
  \citenamefont {{Wrona}}}]{2022NatAs...6.1085S}%
  \BibitemOpen
  \bibfield  {author} {\bibinfo {author} {\bibfnamefont {T.}~\bibnamefont
  {{Shenar}}}, \bibinfo {author} {\bibfnamefont {H.}~\bibnamefont {{Sana}}},
  \bibinfo {author} {\bibfnamefont {L.}~\bibnamefont {{Mahy}}}, \bibinfo
  {author} {\bibfnamefont {K.}~\bibnamefont {{El-Badry}}}, \bibinfo {author}
  {\bibfnamefont {P.}~\bibnamefont {{Marchant}}}, \bibinfo {author}
  {\bibfnamefont {N.}~\bibnamefont {{Langer}}}, \bibinfo {author}
  {\bibfnamefont {C.}~\bibnamefont {{Hawcroft}}}, \bibinfo {author}
  {\bibfnamefont {M.}~\bibnamefont {{Fabry}}}, \bibinfo {author} {\bibfnamefont
  {K.}~\bibnamefont {{Sen}}}, \bibinfo {author} {\bibfnamefont {L.~A.}\
  \bibnamefont {{Almeida}}}, \bibinfo {author} {\bibfnamefont {M.}~\bibnamefont
  {{Abdul-Masih}}}, \bibinfo {author} {\bibfnamefont {J.}~\bibnamefont
  {{Bodensteiner}}}, \bibinfo {author} {\bibfnamefont {P.~A.}\ \bibnamefont
  {{Crowther}}}, \bibinfo {author} {\bibfnamefont {M.}~\bibnamefont
  {{Gieles}}}, \bibinfo {author} {\bibfnamefont {M.}~\bibnamefont
  {{Gromadzki}}}, \bibinfo {author} {\bibfnamefont {V.}~\bibnamefont
  {{H{\'e}nault-Brunet}}}, \bibinfo {author} {\bibfnamefont {A.}~\bibnamefont
  {{Herrero}}}, \bibinfo {author} {\bibfnamefont {A.~d.}\ \bibnamefont
  {{Koter}}}, \bibinfo {author} {\bibfnamefont {P.}~\bibnamefont {{Iwanek}}},
  \bibinfo {author} {\bibfnamefont {S.}~\bibnamefont {{Koz{\l}owski}}},
  \bibinfo {author} {\bibfnamefont {D.~J.}\ \bibnamefont {{Lennon}}}, \bibinfo
  {author} {\bibfnamefont {J.~M.}\ \bibnamefont {{Apell{\'a}niz}}}, \bibinfo
  {author} {\bibfnamefont {P.}~\bibnamefont {{Mr{\'o}z}}}, \bibinfo {author}
  {\bibfnamefont {A.~F.~J.}\ \bibnamefont {{Moffat}}}, \bibinfo {author}
  {\bibfnamefont {A.}~\bibnamefont {{Picco}}}, \bibinfo {author} {\bibfnamefont
  {P.}~\bibnamefont {{Pietrukowicz}}}, \bibinfo {author} {\bibfnamefont
  {R.}~\bibnamefont {{Poleski}}}, \bibinfo {author} {\bibfnamefont
  {K.}~\bibnamefont {{Rybicki}}}, \bibinfo {author} {\bibfnamefont {F.~R.~N.}\
  \bibnamefont {{Schneider}}}, \bibinfo {author} {\bibfnamefont {D.~M.}\
  \bibnamefont {{Skowron}}}, \bibinfo {author} {\bibfnamefont {J.}~\bibnamefont
  {{Skowron}}}, \bibinfo {author} {\bibfnamefont {I.}~\bibnamefont
  {{Soszy{\'n}ski}}}, \bibinfo {author} {\bibfnamefont {M.~K.}\ \bibnamefont
  {{Szyma{\'n}ski}}}, \bibinfo {author} {\bibfnamefont {S.}~\bibnamefont
  {{Toonen}}}, \bibinfo {author} {\bibfnamefont {A.}~\bibnamefont {{Udalski}}},
  \bibinfo {author} {\bibfnamefont {K.}~\bibnamefont {{Ulaczyk}}}, \bibinfo
  {author} {\bibfnamefont {J.~S.}\ \bibnamefont {{Vink}}},\ and\ \bibinfo
  {author} {\bibfnamefont {M.}~\bibnamefont {{Wrona}}},\ }\bibfield  {title}
  {\bibinfo {title} {{An X-ray-quiet black hole born with a negligible kick in
  a massive binary within the Large Magellanic Cloud}},\ }\href
  {https://doi.org/10.1038/s41550-022-01730-y} {\bibfield  {journal} {\bibinfo
  {journal} {Nature Astronomy}\ }\textbf {\bibinfo {volume} {6}},\ \bibinfo
  {pages} {1085} (\bibinfo {year} {2022}{\natexlab{a}})},\ \Eprint
  {https://arxiv.org/abs/2207.07675} {arXiv:2207.07675 [astro-ph.HE]}
  \BibitemShut {NoStop}%
\bibitem [{\citenamefont {{Dray}}\ and\ \citenamefont
  {{Tout}}(2007)}]{2007MNRAS.376...61D}%
  \BibitemOpen
  \bibfield  {author} {\bibinfo {author} {\bibfnamefont {L.~M.}\ \bibnamefont
  {{Dray}}}\ and\ \bibinfo {author} {\bibfnamefont {C.~A.}\ \bibnamefont
  {{Tout}}},\ }\bibfield  {title} {\bibinfo {title} {{On rejuvenation in
  massive binary systems}},\ }\href
  {https://doi.org/10.1111/j.1365-2966.2007.11431.x} {\bibfield  {journal}
  {\bibinfo  {journal} {\mnras}\ }\textbf {\bibinfo {volume} {376}},\ \bibinfo
  {pages} {61} (\bibinfo {year} {2007})},\ \Eprint
  {https://arxiv.org/abs/astro-ph/0612539} {arXiv:astro-ph/0612539 [astro-ph]}
  \BibitemShut {NoStop}%
\bibitem [{\citenamefont {{Justham}}\ \emph {et~al.}(2014)\citenamefont
  {{Justham}}, \citenamefont {{Podsiadlowski}},\ and\ \citenamefont
  {{Vink}}}]{2014ApJ...796..121J}%
  \BibitemOpen
  \bibfield  {author} {\bibinfo {author} {\bibfnamefont {S.}~\bibnamefont
  {{Justham}}}, \bibinfo {author} {\bibfnamefont {P.}~\bibnamefont
  {{Podsiadlowski}}},\ and\ \bibinfo {author} {\bibfnamefont {J.~S.}\
  \bibnamefont {{Vink}}},\ }\bibfield  {title} {\bibinfo {title} {{Luminous
  Blue Variables and Superluminous Supernovae from Binary Mergers}},\ }\href
  {https://doi.org/10.1088/0004-637X/796/2/121} {\bibfield  {journal} {\bibinfo
   {journal} {\apj}\ }\textbf {\bibinfo {volume} {796}},\ \bibinfo {eid} {121}
  (\bibinfo {year} {2014})},\ \Eprint {https://arxiv.org/abs/1410.2426}
  {arXiv:1410.2426 [astro-ph.SR]} \BibitemShut {NoStop}%
\bibitem [{\citenamefont {{Packet}}(1981)}]{1981A&A...102...17P}%
  \BibitemOpen
  \bibfield  {author} {\bibinfo {author} {\bibfnamefont {W.}~\bibnamefont
  {{Packet}}},\ }\bibfield  {title} {\bibinfo {title} {{On the spin-up of the
  mass accreting component in a close binary system}},\ }\href@noop {}
  {\bibfield  {journal} {\bibinfo  {journal} {\aap}\ }\textbf {\bibinfo
  {volume} {102}},\ \bibinfo {pages} {17} (\bibinfo {year} {1981})}\BibitemShut
  {NoStop}%
\bibitem [{\citenamefont {{Pols}}\ \emph {et~al.}(1991)\citenamefont {{Pols}},
  \citenamefont {{Cote}}, \citenamefont {{Waters}},\ and\ \citenamefont
  {{Heise}}}]{1991A&A...241..419P}%
  \BibitemOpen
  \bibfield  {author} {\bibinfo {author} {\bibfnamefont {O.~R.}\ \bibnamefont
  {{Pols}}}, \bibinfo {author} {\bibfnamefont {J.}~\bibnamefont {{Cote}}},
  \bibinfo {author} {\bibfnamefont {L.~B.~F.~M.}\ \bibnamefont {{Waters}}},\
  and\ \bibinfo {author} {\bibfnamefont {J.}~\bibnamefont {{Heise}}},\
  }\bibfield  {title} {\bibinfo {title} {{The formation of Be stars through
  close binary evolution.}},\ }\href@noop {} {\bibfield  {journal} {\bibinfo
  {journal} {\aap}\ }\textbf {\bibinfo {volume} {241}},\ \bibinfo {pages} {419}
  (\bibinfo {year} {1991})}\BibitemShut {NoStop}%
\bibitem [{\citenamefont {{Popham}}\ and\ \citenamefont
  {{Narayan}}(1991)}]{1991ApJ...370..604P}%
  \BibitemOpen
  \bibfield  {author} {\bibinfo {author} {\bibfnamefont {R.}~\bibnamefont
  {{Popham}}}\ and\ \bibinfo {author} {\bibfnamefont {R.}~\bibnamefont
  {{Narayan}}},\ }\bibfield  {title} {\bibinfo {title} {{Does Accretion Cease
  When a Star Approaches Breakup?}},\ }\href {https://doi.org/10.1086/169847}
  {\bibfield  {journal} {\bibinfo  {journal} {\apj}\ }\textbf {\bibinfo
  {volume} {370}},\ \bibinfo {pages} {604} (\bibinfo {year}
  {1991})}\BibitemShut {NoStop}%
\bibitem [{\citenamefont {{Petrovic}}\ \emph {et~al.}(2005)\citenamefont
  {{Petrovic}}, \citenamefont {{Langer}},\ and\ \citenamefont {{van der
  Hucht}}}]{2005A&A...435.1013P}%
  \BibitemOpen
  \bibfield  {author} {\bibinfo {author} {\bibfnamefont {J.}~\bibnamefont
  {{Petrovic}}}, \bibinfo {author} {\bibfnamefont {N.}~\bibnamefont
  {{Langer}}},\ and\ \bibinfo {author} {\bibfnamefont {K.~A.}\ \bibnamefont
  {{van der Hucht}}},\ }\bibfield  {title} {\bibinfo {title} {{Constraining the
  mass transfer in massive binaries through progenitor evolution models of
  Wolf-Rayet+O binaries}},\ }\href {https://doi.org/10.1051/0004-6361:20042368}
  {\bibfield  {journal} {\bibinfo  {journal} {\aap}\ }\textbf {\bibinfo
  {volume} {435}},\ \bibinfo {pages} {1013} (\bibinfo {year} {2005})},\ \Eprint
  {https://arxiv.org/abs/astro-ph/0504242} {arXiv:astro-ph/0504242 [astro-ph]}
  \BibitemShut {NoStop}%
\bibitem [{\citenamefont {{de Mink}}\ \emph
  {et~al.}(2007{\natexlab{a}})\citenamefont {{de Mink}}, \citenamefont
  {{Pols}},\ and\ \citenamefont {{Glebbeek}}}]{2007AIPC..948..321D}%
  \BibitemOpen
  \bibfield  {author} {\bibinfo {author} {\bibfnamefont {S.~E.}\ \bibnamefont
  {{de Mink}}}, \bibinfo {author} {\bibfnamefont {O.~R.}\ \bibnamefont
  {{Pols}}},\ and\ \bibinfo {author} {\bibfnamefont {E.}~\bibnamefont
  {{Glebbeek}}},\ }\bibfield  {title} {\bibinfo {title} {{Critically-rotating
  Stars in Binaries-An Unsolved Problem}},\ }in\ \href
  {https://doi.org/10.1063/1.2818989} {\emph {\bibinfo {booktitle} {Unsolved
  Problems in Stellar Physics: A Conference in Honor of Douglas Gough}}},\
  \bibinfo {series} {American Institute of Physics Conference Series}, Vol.\
  \bibinfo {volume} {948},\ \bibinfo {editor} {edited by\ \bibinfo {editor}
  {\bibfnamefont {R.~J.}\ \bibnamefont {{Stancliffe}}}, \bibinfo {editor}
  {\bibfnamefont {G.}~\bibnamefont {{Houdek}}}, \bibinfo {editor}
  {\bibfnamefont {R.~G.}\ \bibnamefont {{Martin}}},\ and\ \bibinfo {editor}
  {\bibfnamefont {C.~A.}\ \bibnamefont {{Tout}}}}\ (\bibinfo {year} {2007})\
  pp.\ \bibinfo {pages} {321--325},\ \Eprint {https://arxiv.org/abs/0709.2285}
  {arXiv:0709.2285 [astro-ph]} \BibitemShut {NoStop}%
\bibitem [{\citenamefont {{Paczynski}}(1991)}]{1991ApJ...370..597P}%
  \BibitemOpen
  \bibfield  {author} {\bibinfo {author} {\bibfnamefont {B.}~\bibnamefont
  {{Paczynski}}},\ }\bibfield  {title} {\bibinfo {title} {{A Polytropic Model
  of an Accretion Disk, a Boundary Layer, and a Star}},\ }\href
  {https://doi.org/10.1086/169846} {\bibfield  {journal} {\bibinfo  {journal}
  {\apj}\ }\textbf {\bibinfo {volume} {370}},\ \bibinfo {pages} {597} (\bibinfo
  {year} {1991})}\BibitemShut {NoStop}%
\bibitem [{\citenamefont {{Maeder}}\ and\ \citenamefont
  {{Meynet}}(2000{\natexlab{b}})}]{2000ARA&A..38..143M}%
  \BibitemOpen
  \bibfield  {author} {\bibinfo {author} {\bibfnamefont {A.}~\bibnamefont
  {{Maeder}}}\ and\ \bibinfo {author} {\bibfnamefont {G.}~\bibnamefont
  {{Meynet}}},\ }\bibfield  {title} {\bibinfo {title} {{The Evolution of
  Rotating Stars}},\ }\href {https://doi.org/10.1146/annurev.astro.38.1.143}
  {\bibfield  {journal} {\bibinfo  {journal} {\araa}\ }\textbf {\bibinfo
  {volume} {38}},\ \bibinfo {pages} {143} (\bibinfo {year}
  {2000}{\natexlab{b}})},\ \Eprint {https://arxiv.org/abs/astro-ph/0004204}
  {arXiv:astro-ph/0004204 [astro-ph]} \BibitemShut {NoStop}%
\bibitem [{\citenamefont {{Cantiello}}\ \emph {et~al.}(2007)\citenamefont
  {{Cantiello}}, \citenamefont {{Yoon}}, \citenamefont {{Langer}},\ and\
  \citenamefont {{Livio}}}]{2007A&A...465L..29C}%
  \BibitemOpen
  \bibfield  {author} {\bibinfo {author} {\bibfnamefont {M.}~\bibnamefont
  {{Cantiello}}}, \bibinfo {author} {\bibfnamefont {S.~C.}\ \bibnamefont
  {{Yoon}}}, \bibinfo {author} {\bibfnamefont {N.}~\bibnamefont {{Langer}}},\
  and\ \bibinfo {author} {\bibfnamefont {M.}~\bibnamefont {{Livio}}},\
  }\bibfield  {title} {\bibinfo {title} {{Binary star progenitors of long
  gamma-ray bursts}},\ }\href {https://doi.org/10.1051/0004-6361:20077115}
  {\bibfield  {journal} {\bibinfo  {journal} {\aap}\ }\textbf {\bibinfo
  {volume} {465}},\ \bibinfo {pages} {L29} (\bibinfo {year} {2007})},\ \Eprint
  {https://arxiv.org/abs/astro-ph/0702540} {arXiv:astro-ph/0702540 [astro-ph]}
  \BibitemShut {NoStop}%
\bibitem [{\citenamefont {{Maeder}}\ and\ \citenamefont
  {{Meynet}}(2001)}]{2001A&A...373..555M}%
  \BibitemOpen
  \bibfield  {author} {\bibinfo {author} {\bibfnamefont {A.}~\bibnamefont
  {{Maeder}}}\ and\ \bibinfo {author} {\bibfnamefont {G.}~\bibnamefont
  {{Meynet}}},\ }\bibfield  {title} {\bibinfo {title} {{Stellar evolution with
  rotation. VII. . Low metallicity models and the blue to red supergiant ratio
  in the SMC}},\ }\href {https://doi.org/10.1051/0004-6361:20010596} {\bibfield
   {journal} {\bibinfo  {journal} {\aap}\ }\textbf {\bibinfo {volume} {373}},\
  \bibinfo {pages} {555} (\bibinfo {year} {2001})},\ \Eprint
  {https://arxiv.org/abs/astro-ph/0105051} {arXiv:astro-ph/0105051 [astro-ph]}
  \BibitemShut {NoStop}%
\bibitem [{\citenamefont {{Suijs}}\ \emph {et~al.}(2008)\citenamefont
  {{Suijs}}, \citenamefont {{Langer}}, \citenamefont {{Poelarends}},
  \citenamefont {{Yoon}}, \citenamefont {{Heger}},\ and\ \citenamefont
  {{Herwig}}}]{2008A&A...481L..87S}%
  \BibitemOpen
  \bibfield  {author} {\bibinfo {author} {\bibfnamefont {M.~P.~L.}\
  \bibnamefont {{Suijs}}}, \bibinfo {author} {\bibfnamefont {N.}~\bibnamefont
  {{Langer}}}, \bibinfo {author} {\bibfnamefont {A.~J.}\ \bibnamefont
  {{Poelarends}}}, \bibinfo {author} {\bibfnamefont {S.~C.}\ \bibnamefont
  {{Yoon}}}, \bibinfo {author} {\bibfnamefont {A.}~\bibnamefont {{Heger}}},\
  and\ \bibinfo {author} {\bibfnamefont {F.}~\bibnamefont {{Herwig}}},\
  }\bibfield  {title} {\bibinfo {title} {{White dwarf spins from low-mass
  stellar evolution models}},\ }\href
  {https://doi.org/10.1051/0004-6361:200809411} {\bibfield  {journal} {\bibinfo
   {journal} {\aap}\ }\textbf {\bibinfo {volume} {481}},\ \bibinfo {pages}
  {L87} (\bibinfo {year} {2008})},\ \Eprint {https://arxiv.org/abs/0802.3286}
  {arXiv:0802.3286 [astro-ph]} \BibitemShut {NoStop}%
\bibitem [{\citenamefont {{Cantiello}}\ \emph {et~al.}(2014)\citenamefont
  {{Cantiello}}, \citenamefont {{Mankovich}}, \citenamefont {{Bildsten}},
  \citenamefont {{Christensen-Dalsgaard}},\ and\ \citenamefont
  {{Paxton}}}]{2014ApJ...788...93C}%
  \BibitemOpen
  \bibfield  {author} {\bibinfo {author} {\bibfnamefont {M.}~\bibnamefont
  {{Cantiello}}}, \bibinfo {author} {\bibfnamefont {C.}~\bibnamefont
  {{Mankovich}}}, \bibinfo {author} {\bibfnamefont {L.}~\bibnamefont
  {{Bildsten}}}, \bibinfo {author} {\bibfnamefont {J.}~\bibnamefont
  {{Christensen-Dalsgaard}}},\ and\ \bibinfo {author} {\bibfnamefont
  {B.}~\bibnamefont {{Paxton}}},\ }\bibfield  {title} {\bibinfo {title}
  {{Angular Momentum Transport within Evolved Low-mass Stars}},\ }\href
  {https://doi.org/10.1088/0004-637X/788/1/93} {\bibfield  {journal} {\bibinfo
  {journal} {\apj}\ }\textbf {\bibinfo {volume} {788}},\ \bibinfo {eid} {93}
  (\bibinfo {year} {2014})},\ \Eprint {https://arxiv.org/abs/1405.1419}
  {arXiv:1405.1419 [astro-ph.SR]} \BibitemShut {NoStop}%
\bibitem [{\citenamefont {{Gehan}}\ \emph {et~al.}(2018)\citenamefont
  {{Gehan}}, \citenamefont {{Mosser}}, \citenamefont {{Michel}}, \citenamefont
  {{Samadi}},\ and\ \citenamefont {{Kallinger}}}]{2018A&A...616A..24G}%
  \BibitemOpen
  \bibfield  {author} {\bibinfo {author} {\bibfnamefont {C.}~\bibnamefont
  {{Gehan}}}, \bibinfo {author} {\bibfnamefont {B.}~\bibnamefont {{Mosser}}},
  \bibinfo {author} {\bibfnamefont {E.}~\bibnamefont {{Michel}}}, \bibinfo
  {author} {\bibfnamefont {R.}~\bibnamefont {{Samadi}}},\ and\ \bibinfo
  {author} {\bibfnamefont {T.}~\bibnamefont {{Kallinger}}},\ }\bibfield
  {title} {\bibinfo {title} {{Core rotation braking on the red giant branch for
  various mass ranges}},\ }\href {https://doi.org/10.1051/0004-6361/201832822}
  {\bibfield  {journal} {\bibinfo  {journal} {\aap}\ }\textbf {\bibinfo
  {volume} {616}},\ \bibinfo {eid} {A24} (\bibinfo {year} {2018})},\ \Eprint
  {https://arxiv.org/abs/1802.04558} {arXiv:1802.04558 [astro-ph.SR]}
  \BibitemShut {NoStop}%
\bibitem [{\citenamefont {{Liu}}\ \emph {et~al.}(2008)\citenamefont {{Liu}},
  \citenamefont {{McClintock}}, \citenamefont {{Narayan}}, \citenamefont
  {{Davis}},\ and\ \citenamefont {{Orosz}}}]{2008ApJ...679L..37L}%
  \BibitemOpen
  \bibfield  {author} {\bibinfo {author} {\bibfnamefont {J.}~\bibnamefont
  {{Liu}}}, \bibinfo {author} {\bibfnamefont {J.~E.}\ \bibnamefont
  {{McClintock}}}, \bibinfo {author} {\bibfnamefont {R.}~\bibnamefont
  {{Narayan}}}, \bibinfo {author} {\bibfnamefont {S.~W.}\ \bibnamefont
  {{Davis}}},\ and\ \bibinfo {author} {\bibfnamefont {J.~A.}\ \bibnamefont
  {{Orosz}}},\ }\bibfield  {title} {\bibinfo {title} {{Precise Measurement of
  the Spin Parameter of the Stellar-Mass Black Hole M33 X-7}},\ }\href
  {https://doi.org/10.1086/588840} {\bibfield  {journal} {\bibinfo  {journal}
  {\apjl}\ }\textbf {\bibinfo {volume} {679}},\ \bibinfo {pages} {L37}
  (\bibinfo {year} {2008})},\ \Eprint {https://arxiv.org/abs/0803.1834}
  {arXiv:0803.1834 [astro-ph]} \BibitemShut {NoStop}%
\bibitem [{\citenamefont {{Gou}}\ \emph {et~al.}(2014)\citenamefont {{Gou}},
  \citenamefont {{McClintock}}, \citenamefont {{Remillard}}, \citenamefont
  {{Steiner}}, \citenamefont {{Reid}}, \citenamefont {{Orosz}}, \citenamefont
  {{Narayan}}, \citenamefont {{Hanke}},\ and\ \citenamefont
  {{Garc{\'\i}a}}}]{2014ApJ...790...29G}%
  \BibitemOpen
  \bibfield  {author} {\bibinfo {author} {\bibfnamefont {L.}~\bibnamefont
  {{Gou}}}, \bibinfo {author} {\bibfnamefont {J.~E.}\ \bibnamefont
  {{McClintock}}}, \bibinfo {author} {\bibfnamefont {R.~A.}\ \bibnamefont
  {{Remillard}}}, \bibinfo {author} {\bibfnamefont {J.~F.}\ \bibnamefont
  {{Steiner}}}, \bibinfo {author} {\bibfnamefont {M.~J.}\ \bibnamefont
  {{Reid}}}, \bibinfo {author} {\bibfnamefont {J.~A.}\ \bibnamefont {{Orosz}}},
  \bibinfo {author} {\bibfnamefont {R.}~\bibnamefont {{Narayan}}}, \bibinfo
  {author} {\bibfnamefont {M.}~\bibnamefont {{Hanke}}},\ and\ \bibinfo {author}
  {\bibfnamefont {J.}~\bibnamefont {{Garc{\'\i}a}}},\ }\bibfield  {title}
  {\bibinfo {title} {{Confirmation via the Continuum-fitting Method that the
  Spin of the Black Hole in Cygnus X-1 Is Extreme}},\ }\href
  {https://doi.org/10.1088/0004-637X/790/1/29} {\bibfield  {journal} {\bibinfo
  {journal} {\apj}\ }\textbf {\bibinfo {volume} {790}},\ \bibinfo {eid} {29}
  (\bibinfo {year} {2014})},\ \Eprint {https://arxiv.org/abs/1308.4760}
  {arXiv:1308.4760 [astro-ph.HE]} \BibitemShut {NoStop}%
\bibitem [{\citenamefont {{Reynolds}}(2021)}]{2021ARA&A..59..117R}%
  \BibitemOpen
  \bibfield  {author} {\bibinfo {author} {\bibfnamefont {C.~S.}\ \bibnamefont
  {{Reynolds}}},\ }\bibfield  {title} {\bibinfo {title} {{Observational
  Constraints on Black Hole Spin}},\ }\href
  {https://doi.org/10.1146/annurev-astro-112420-035022} {\bibfield  {journal}
  {\bibinfo  {journal} {\araa}\ }\textbf {\bibinfo {volume} {59}},\ \bibinfo
  {pages} {117} (\bibinfo {year} {2021})},\ \Eprint
  {https://arxiv.org/abs/2011.08948} {arXiv:2011.08948 [astro-ph.HE]}
  \BibitemShut {NoStop}%
\bibitem [{\citenamefont {{Qin}}\ \emph {et~al.}(2019)\citenamefont {{Qin}},
  \citenamefont {{Marchant}}, \citenamefont {{Fragos}}, \citenamefont
  {{Meynet}},\ and\ \citenamefont {{Kalogera}}}]{2019ApJ...870L..18Q}%
  \BibitemOpen
  \bibfield  {author} {\bibinfo {author} {\bibfnamefont {Y.}~\bibnamefont
  {{Qin}}}, \bibinfo {author} {\bibfnamefont {P.}~\bibnamefont {{Marchant}}},
  \bibinfo {author} {\bibfnamefont {T.}~\bibnamefont {{Fragos}}}, \bibinfo
  {author} {\bibfnamefont {G.}~\bibnamefont {{Meynet}}},\ and\ \bibinfo
  {author} {\bibfnamefont {V.}~\bibnamefont {{Kalogera}}},\ }\bibfield  {title}
  {\bibinfo {title} {{On the Origin of Black Hole Spin in High-mass X-Ray
  Binaries}},\ }\href {https://doi.org/10.3847/2041-8213/aaf97b} {\bibfield
  {journal} {\bibinfo  {journal} {\apjl}\ }\textbf {\bibinfo {volume} {870}},\
  \bibinfo {eid} {L18} (\bibinfo {year} {2019})},\ \Eprint
  {https://arxiv.org/abs/1810.13016} {arXiv:1810.13016 [astro-ph.SR]}
  \BibitemShut {NoStop}%
\bibitem [{\citenamefont {{Batta}}\ \emph {et~al.}(2017)\citenamefont
  {{Batta}}, \citenamefont {{Ramirez-Ruiz}},\ and\ \citenamefont
  {{Fryer}}}]{2017ApJ...846L..15B}%
  \BibitemOpen
  \bibfield  {author} {\bibinfo {author} {\bibfnamefont {A.}~\bibnamefont
  {{Batta}}}, \bibinfo {author} {\bibfnamefont {E.}~\bibnamefont
  {{Ramirez-Ruiz}}},\ and\ \bibinfo {author} {\bibfnamefont {C.}~\bibnamefont
  {{Fryer}}},\ }\bibfield  {title} {\bibinfo {title} {{The Formation of Rapidly
  Rotating Black Holes in High-mass X-Ray Binaries}},\ }\href
  {https://doi.org/10.3847/2041-8213/aa8506} {\bibfield  {journal} {\bibinfo
  {journal} {\apjl}\ }\textbf {\bibinfo {volume} {846}},\ \bibinfo {eid} {L15}
  (\bibinfo {year} {2017})},\ \Eprint {https://arxiv.org/abs/1708.00570}
  {arXiv:1708.00570 [astro-ph.HE]} \BibitemShut {NoStop}%
\bibitem [{\citenamefont {{Belczynski}}\ \emph {et~al.}(2020)\citenamefont
  {{Belczynski}}, \citenamefont {{Klencki}}, \citenamefont {{Fields}},
  \citenamefont {{Olejak}}, \citenamefont {{Berti}}, \citenamefont {{Meynet}},
  \citenamefont {{Fryer}}, \citenamefont {{Holz}}, \citenamefont
  {{O'Shaughnessy}}, \citenamefont {{Brown}},\ and\ \citenamefont
  {et~al.}}]{2020A&A...636A.104B}%
  \BibitemOpen
  \bibfield  {author} {\bibinfo {author} {\bibfnamefont {K.}~\bibnamefont
  {{Belczynski}}}, \bibinfo {author} {\bibfnamefont {J.}~\bibnamefont
  {{Klencki}}}, \bibinfo {author} {\bibfnamefont {C.~E.}\ \bibnamefont
  {{Fields}}}, \bibinfo {author} {\bibfnamefont {A.}~\bibnamefont {{Olejak}}},
  \bibinfo {author} {\bibfnamefont {E.}~\bibnamefont {{Berti}}}, \bibinfo
  {author} {\bibfnamefont {G.}~\bibnamefont {{Meynet}}}, \bibinfo {author}
  {\bibfnamefont {C.~L.}\ \bibnamefont {{Fryer}}}, \bibinfo {author}
  {\bibfnamefont {D.~E.}\ \bibnamefont {{Holz}}}, \bibinfo {author}
  {\bibfnamefont {R.}~\bibnamefont {{O'Shaughnessy}}}, \bibinfo {author}
  {\bibfnamefont {D.~A.}\ \bibnamefont {{Brown}}},\ and\ \bibinfo {author}
  {\bibnamefont {et~al.}},\ }\bibfield  {title} {\bibinfo {title}
  {{Evolutionary roads leading to low effective spins, high black hole masses,
  and O1/O2 rates for LIGO/Virgo binary black holes}},\ }\href
  {https://doi.org/10.1051/0004-6361/201936528} {\bibfield  {journal} {\bibinfo
   {journal} {\aap}\ }\textbf {\bibinfo {volume} {636}},\ \bibinfo {eid} {A104}
  (\bibinfo {year} {2020})},\ \Eprint {https://arxiv.org/abs/1706.07053}
  {arXiv:1706.07053 [astro-ph.HE]} \BibitemShut {NoStop}%
\bibitem [{\citenamefont {{Bardeen}}(1970)}]{1970Natur.226...64B}%
  \BibitemOpen
  \bibfield  {author} {\bibinfo {author} {\bibfnamefont {J.~M.}\ \bibnamefont
  {{Bardeen}}},\ }\bibfield  {title} {\bibinfo {title} {{Kerr Metric Black
  Holes}},\ }\href {https://doi.org/10.1038/226064a0} {\bibfield  {journal}
  {\bibinfo  {journal} {\nat}\ }\textbf {\bibinfo {volume} {226}},\ \bibinfo
  {pages} {64} (\bibinfo {year} {1970})}\BibitemShut {NoStop}%
\bibitem [{\citenamefont {{Abramowicz}}\ \emph {et~al.}(1988)\citenamefont
  {{Abramowicz}}, \citenamefont {{Czerny}}, \citenamefont {{Lasota}},\ and\
  \citenamefont {{Szuszkiewicz}}}]{1988ApJ...332..646A}%
  \BibitemOpen
  \bibfield  {author} {\bibinfo {author} {\bibfnamefont {M.~A.}\ \bibnamefont
  {{Abramowicz}}}, \bibinfo {author} {\bibfnamefont {B.}~\bibnamefont
  {{Czerny}}}, \bibinfo {author} {\bibfnamefont {J.~P.}\ \bibnamefont
  {{Lasota}}},\ and\ \bibinfo {author} {\bibfnamefont {E.}~\bibnamefont
  {{Szuszkiewicz}}},\ }\bibfield  {title} {\bibinfo {title} {{Slim Accretion
  Disks}},\ }\href {https://doi.org/10.1086/166683} {\bibfield  {journal}
  {\bibinfo  {journal} {\apj}\ }\textbf {\bibinfo {volume} {332}},\ \bibinfo
  {pages} {646} (\bibinfo {year} {1988})}\BibitemShut {NoStop}%
\bibitem [{\citenamefont {{S{\k{a}}dowski}}(2009)}]{2009ApJS..183..171S}%
  \BibitemOpen
  \bibfield  {author} {\bibinfo {author} {\bibfnamefont {A.}~\bibnamefont
  {{S{\k{a}}dowski}}},\ }\bibfield  {title} {\bibinfo {title} {{Slim Disks
  Around Kerr Black Holes Revisited}},\ }\href
  {https://doi.org/10.1088/0067-0049/183/2/171} {\bibfield  {journal} {\bibinfo
   {journal} {\apjs}\ }\textbf {\bibinfo {volume} {183}},\ \bibinfo {pages}
  {171} (\bibinfo {year} {2009})},\ \Eprint {https://arxiv.org/abs/0906.0355}
  {arXiv:0906.0355 [astro-ph.HE]} \BibitemShut {NoStop}%
\bibitem [{\citenamefont {{Abramowicz}}\ \emph {et~al.}(1978)\citenamefont
  {{Abramowicz}}, \citenamefont {{Jaroszynski}},\ and\ \citenamefont
  {{Sikora}}}]{1978A&A....63..221A}%
  \BibitemOpen
  \bibfield  {author} {\bibinfo {author} {\bibfnamefont {M.}~\bibnamefont
  {{Abramowicz}}}, \bibinfo {author} {\bibfnamefont {M.}~\bibnamefont
  {{Jaroszynski}}},\ and\ \bibinfo {author} {\bibfnamefont {M.}~\bibnamefont
  {{Sikora}}},\ }\bibfield  {title} {\bibinfo {title} {{Relativistic, accreting
  disks.}},\ }\href@noop {} {\bibfield  {journal} {\bibinfo  {journal} {\aap}\
  }\textbf {\bibinfo {volume} {63}},\ \bibinfo {pages} {221} (\bibinfo {year}
  {1978})}\BibitemShut {NoStop}%
\bibitem [{\citenamefont {{Paczy{\'n}sky}}\ and\ \citenamefont
  {{Wiita}}(1980)}]{1980A&A....88...23P}%
  \BibitemOpen
  \bibfield  {author} {\bibinfo {author} {\bibfnamefont {B.}~\bibnamefont
  {{Paczy{\'n}sky}}}\ and\ \bibinfo {author} {\bibfnamefont {P.~J.}\
  \bibnamefont {{Wiita}}},\ }\bibfield  {title} {\bibinfo {title} {{Thick
  Accretion Disks and Supercritical Luminosities}},\ }\href@noop {} {\bibfield
  {journal} {\bibinfo  {journal} {\aap}\ }\textbf {\bibinfo {volume} {88}},\
  \bibinfo {pages} {23} (\bibinfo {year} {1980})}\BibitemShut {NoStop}%
\bibitem [{\citenamefont {{Done}}\ \emph {et~al.}(2007)\citenamefont {{Done}},
  \citenamefont {{Gierli{\'n}ski}},\ and\ \citenamefont
  {{Kubota}}}]{2007A&ARv..15....1D}%
  \BibitemOpen
  \bibfield  {author} {\bibinfo {author} {\bibfnamefont {C.}~\bibnamefont
  {{Done}}}, \bibinfo {author} {\bibfnamefont {M.}~\bibnamefont
  {{Gierli{\'n}ski}}},\ and\ \bibinfo {author} {\bibfnamefont {A.}~\bibnamefont
  {{Kubota}}},\ }\bibfield  {title} {\bibinfo {title} {{Modelling the behaviour
  of accretion flows in X-ray binaries. Everything you always wanted to know
  about accretion but were afraid to ask}},\ }\href
  {https://doi.org/10.1007/s00159-007-0006-1} {\bibfield  {journal} {\bibinfo
  {journal} {\aapr}\ }\textbf {\bibinfo {volume} {15}},\ \bibinfo {pages} {1}
  (\bibinfo {year} {2007})},\ \Eprint {https://arxiv.org/abs/0708.0148}
  {arXiv:0708.0148 [astro-ph]} \BibitemShut {NoStop}%
\bibitem [{\citenamefont {{Abramowicz}}\ and\ \citenamefont
  {{Fragile}}(2013)}]{2013LRR....16....1A}%
  \BibitemOpen
  \bibfield  {author} {\bibinfo {author} {\bibfnamefont {M.~A.}\ \bibnamefont
  {{Abramowicz}}}\ and\ \bibinfo {author} {\bibfnamefont {P.~C.}\ \bibnamefont
  {{Fragile}}},\ }\bibfield  {title} {\bibinfo {title} {{Foundations of Black
  Hole Accretion Disk Theory}},\ }\href {https://doi.org/10.12942/lrr-2013-1}
  {\bibfield  {journal} {\bibinfo  {journal} {Living Reviews in Relativity}\
  }\textbf {\bibinfo {volume} {16}},\ \bibinfo {eid} {1} (\bibinfo {year}
  {2013})},\ \Eprint {https://arxiv.org/abs/1104.5499} {arXiv:1104.5499
  [astro-ph.HE]} \BibitemShut {NoStop}%
\bibitem [{\citenamefont {{Fabrika}}\ \emph {et~al.}(2015)\citenamefont
  {{Fabrika}}, \citenamefont {{Ueda}}, \citenamefont {{Vinokurov}},
  \citenamefont {{Sholukhova}},\ and\ \citenamefont
  {{Shidatsu}}}]{2015NatPh..11..551F}%
  \BibitemOpen
  \bibfield  {author} {\bibinfo {author} {\bibfnamefont {S.}~\bibnamefont
  {{Fabrika}}}, \bibinfo {author} {\bibfnamefont {Y.}~\bibnamefont {{Ueda}}},
  \bibinfo {author} {\bibfnamefont {A.}~\bibnamefont {{Vinokurov}}}, \bibinfo
  {author} {\bibfnamefont {O.}~\bibnamefont {{Sholukhova}}},\ and\ \bibinfo
  {author} {\bibfnamefont {M.}~\bibnamefont {{Shidatsu}}},\ }\bibfield  {title}
  {\bibinfo {title} {{Supercritical accretion disks in ultraluminous X-ray
  sources and SS 433}},\ }\href {https://doi.org/10.1038/nphys3348} {\bibfield
  {journal} {\bibinfo  {journal} {Nature Physics}\ }\textbf {\bibinfo {volume}
  {11}},\ \bibinfo {pages} {551} (\bibinfo {year} {2015})},\ \Eprint
  {https://arxiv.org/abs/1512.00435} {arXiv:1512.00435 [astro-ph.HE]}
  \BibitemShut {NoStop}%
\bibitem [{\citenamefont {{Kaaret}}\ \emph {et~al.}(2017)\citenamefont
  {{Kaaret}}, \citenamefont {{Feng}},\ and\ \citenamefont
  {{Roberts}}}]{2017ARA&A..55..303K}%
  \BibitemOpen
  \bibfield  {author} {\bibinfo {author} {\bibfnamefont {P.}~\bibnamefont
  {{Kaaret}}}, \bibinfo {author} {\bibfnamefont {H.}~\bibnamefont {{Feng}}},\
  and\ \bibinfo {author} {\bibfnamefont {T.~P.}\ \bibnamefont {{Roberts}}},\
  }\bibfield  {title} {\bibinfo {title} {{Ultraluminous X-Ray Sources}},\
  }\href {https://doi.org/10.1146/annurev-astro-091916-055259} {\bibfield
  {journal} {\bibinfo  {journal} {\araa}\ }\textbf {\bibinfo {volume} {55}},\
  \bibinfo {pages} {303} (\bibinfo {year} {2017})},\ \Eprint
  {https://arxiv.org/abs/1703.10728} {arXiv:1703.10728 [astro-ph.HE]}
  \BibitemShut {NoStop}%
\bibitem [{\citenamefont {{van Son}}\ \emph {et~al.}(2020)\citenamefont {{van
  Son}}, \citenamefont {{De Mink}}, \citenamefont {{Broekgaarden}},
  \citenamefont {{Renzo}}, \citenamefont {{Justham}}, \citenamefont
  {{Laplace}}, \citenamefont {{Mor{\'a}n-Fraile}}, \citenamefont {{Hendriks}},\
  and\ \citenamefont {{Farmer}}}]{2020ApJ...897..100V}%
  \BibitemOpen
  \bibfield  {author} {\bibinfo {author} {\bibfnamefont {L.~A.~C.}\
  \bibnamefont {{van Son}}}, \bibinfo {author} {\bibfnamefont {S.~E.}\
  \bibnamefont {{De Mink}}}, \bibinfo {author} {\bibfnamefont {F.~S.}\
  \bibnamefont {{Broekgaarden}}}, \bibinfo {author} {\bibfnamefont
  {M.}~\bibnamefont {{Renzo}}}, \bibinfo {author} {\bibfnamefont
  {S.}~\bibnamefont {{Justham}}}, \bibinfo {author} {\bibfnamefont
  {E.}~\bibnamefont {{Laplace}}}, \bibinfo {author} {\bibfnamefont
  {J.}~\bibnamefont {{Mor{\'a}n-Fraile}}}, \bibinfo {author} {\bibfnamefont
  {D.~D.}\ \bibnamefont {{Hendriks}}},\ and\ \bibinfo {author} {\bibfnamefont
  {R.}~\bibnamefont {{Farmer}}},\ }\bibfield  {title} {\bibinfo {title}
  {{Polluting the Pair-instability Mass Gap for Binary Black Holes through
  Super-Eddington Accretion in Isolated Binaries}},\ }\href
  {https://doi.org/10.3847/1538-4357/ab9809} {\bibfield  {journal} {\bibinfo
  {journal} {\apj}\ }\textbf {\bibinfo {volume} {897}},\ \bibinfo {eid} {100}
  (\bibinfo {year} {2020})},\ \Eprint {https://arxiv.org/abs/2004.05187}
  {arXiv:2004.05187 [astro-ph.HE]} \BibitemShut {NoStop}%
\bibitem [{\citenamefont {{Thorne}}(1974)}]{1974ApJ...191..507T}%
  \BibitemOpen
  \bibfield  {author} {\bibinfo {author} {\bibfnamefont {K.~S.}\ \bibnamefont
  {{Thorne}}},\ }\bibfield  {title} {\bibinfo {title} {{Disk-Accretion onto a
  Black Hole. II. Evolution of the Hole}},\ }\href
  {https://doi.org/10.1086/152991} {\bibfield  {journal} {\bibinfo  {journal}
  {\apj}\ }\textbf {\bibinfo {volume} {191}},\ \bibinfo {pages} {507} (\bibinfo
  {year} {1974})}\BibitemShut {NoStop}%
\bibitem [{\citenamefont {{Kato}}\ \emph {et~al.}(2010)\citenamefont {{Kato}},
  \citenamefont {{Miyoshi}}, \citenamefont {{Takahashi}}, \citenamefont
  {{Negoro}},\ and\ \citenamefont {{Matsumoto}}}]{2010MNRAS.403L..74K}%
  \BibitemOpen
  \bibfield  {author} {\bibinfo {author} {\bibfnamefont {Y.}~\bibnamefont
  {{Kato}}}, \bibinfo {author} {\bibfnamefont {M.}~\bibnamefont {{Miyoshi}}},
  \bibinfo {author} {\bibfnamefont {R.}~\bibnamefont {{Takahashi}}}, \bibinfo
  {author} {\bibfnamefont {H.}~\bibnamefont {{Negoro}}},\ and\ \bibinfo
  {author} {\bibfnamefont {R.}~\bibnamefont {{Matsumoto}}},\ }\bibfield
  {title} {\bibinfo {title} {{Measuring spin of a supermassive black hole at
  the Galactic centre - implications for a unique spin}},\ }\href
  {https://doi.org/10.1111/j.1745-3933.2010.00818.x} {\bibfield  {journal}
  {\bibinfo  {journal} {\mnras}\ }\textbf {\bibinfo {volume} {403}},\ \bibinfo
  {pages} {L74} (\bibinfo {year} {2010})},\ \Eprint
  {https://arxiv.org/abs/0906.5423} {arXiv:0906.5423 [astro-ph.GA]}
  \BibitemShut {NoStop}%
\bibitem [{\citenamefont {{Chashkina}}\ and\ \citenamefont
  {{Abolmasov}}(2015)}]{2015MNRAS.446.1829C}%
  \BibitemOpen
  \bibfield  {author} {\bibinfo {author} {\bibfnamefont {A.}~\bibnamefont
  {{Chashkina}}}\ and\ \bibinfo {author} {\bibfnamefont {P.}~\bibnamefont
  {{Abolmasov}}},\ }\bibfield  {title} {\bibinfo {title} {{Black hole spin
  evolution affected by magnetic field decay}},\ }\href
  {https://doi.org/10.1093/mnras/stu2078} {\bibfield  {journal} {\bibinfo
  {journal} {\mnras}\ }\textbf {\bibinfo {volume} {446}},\ \bibinfo {pages}
  {1829} (\bibinfo {year} {2015})},\ \Eprint {https://arxiv.org/abs/1410.1203}
  {arXiv:1410.1203 [astro-ph.HE]} \BibitemShut {NoStop}%
\bibitem [{\citenamefont {{King}}\ and\ \citenamefont
  {{Kolb}}(1999)}]{1999MNRAS.305..654K}%
  \BibitemOpen
  \bibfield  {author} {\bibinfo {author} {\bibfnamefont {A.~R.}\ \bibnamefont
  {{King}}}\ and\ \bibinfo {author} {\bibfnamefont {U.}~\bibnamefont
  {{Kolb}}},\ }\bibfield  {title} {\bibinfo {title} {{The evolution of black
  hole mass and angular momentum}},\ }\href
  {https://doi.org/10.1046/j.1365-8711.1999.02482.x} {\bibfield  {journal}
  {\bibinfo  {journal} {\mnras}\ }\textbf {\bibinfo {volume} {305}},\ \bibinfo
  {pages} {654} (\bibinfo {year} {1999})},\ \Eprint
  {https://arxiv.org/abs/astro-ph/9901296} {arXiv:astro-ph/9901296 [astro-ph]}
  \BibitemShut {NoStop}%
\bibitem [{\citenamefont {{Bondi}}\ and\ \citenamefont
  {{Hoyle}}(1944)}]{1944MNRAS.104..273B}%
  \BibitemOpen
  \bibfield  {author} {\bibinfo {author} {\bibfnamefont {H.}~\bibnamefont
  {{Bondi}}}\ and\ \bibinfo {author} {\bibfnamefont {F.}~\bibnamefont
  {{Hoyle}}},\ }\bibfield  {title} {\bibinfo {title} {{On the mechanism of
  accretion by stars}},\ }\href {https://doi.org/10.1093/mnras/104.5.273}
  {\bibfield  {journal} {\bibinfo  {journal} {\mnras}\ }\textbf {\bibinfo
  {volume} {104}},\ \bibinfo {pages} {273} (\bibinfo {year}
  {1944})}\BibitemShut {NoStop}%
\bibitem [{\citenamefont {{Bondi}}(1952)}]{1952MNRAS.112..195B}%
  \BibitemOpen
  \bibfield  {author} {\bibinfo {author} {\bibfnamefont {H.}~\bibnamefont
  {{Bondi}}},\ }\bibfield  {title} {\bibinfo {title} {{On spherically
  symmetrical accretion}},\ }\href {https://doi.org/10.1093/mnras/112.2.195}
  {\bibfield  {journal} {\bibinfo  {journal} {\mnras}\ }\textbf {\bibinfo
  {volume} {112}},\ \bibinfo {pages} {195} (\bibinfo {year}
  {1952})}\BibitemShut {NoStop}%
\bibitem [{\citenamefont {{Belczynski}}\ \emph {et~al.}(2007)\citenamefont
  {{Belczynski}}, \citenamefont {{Taam}}, \citenamefont {{Kalogera}},
  \citenamefont {{Rasio}},\ and\ \citenamefont
  {{Bulik}}}]{2007ApJ...662..504B}%
  \BibitemOpen
  \bibfield  {author} {\bibinfo {author} {\bibfnamefont {K.}~\bibnamefont
  {{Belczynski}}}, \bibinfo {author} {\bibfnamefont {R.~E.}\ \bibnamefont
  {{Taam}}}, \bibinfo {author} {\bibfnamefont {V.}~\bibnamefont {{Kalogera}}},
  \bibinfo {author} {\bibfnamefont {F.~A.}\ \bibnamefont {{Rasio}}},\ and\
  \bibinfo {author} {\bibfnamefont {T.}~\bibnamefont {{Bulik}}},\ }\bibfield
  {title} {\bibinfo {title} {{On the Rarity of Double Black Hole Binaries:
  Consequences for Gravitational Wave Detection}},\ }\href
  {https://doi.org/10.1086/513562} {\bibfield  {journal} {\bibinfo  {journal}
  {\apj}\ }\textbf {\bibinfo {volume} {662}},\ \bibinfo {pages} {504} (\bibinfo
  {year} {2007})},\ \Eprint {https://arxiv.org/abs/astro-ph/0612032}
  {arXiv:astro-ph/0612032 [astro-ph]} \BibitemShut {NoStop}%
\bibitem [{\citenamefont {{Ricker}}\ and\ \citenamefont
  {{Taam}}(2008)}]{2008ApJ...672L..41R}%
  \BibitemOpen
  \bibfield  {author} {\bibinfo {author} {\bibfnamefont {P.~M.}\ \bibnamefont
  {{Ricker}}}\ and\ \bibinfo {author} {\bibfnamefont {R.~E.}\ \bibnamefont
  {{Taam}}},\ }\bibfield  {title} {\bibinfo {title} {{The Interaction of
  Stellar Objects within a Common Envelope}},\ }\href
  {https://doi.org/10.1086/526343} {\bibfield  {journal} {\bibinfo  {journal}
  {\apjl}\ }\textbf {\bibinfo {volume} {672}},\ \bibinfo {pages} {L41}
  (\bibinfo {year} {2008})},\ \Eprint {https://arxiv.org/abs/0710.3631}
  {arXiv:0710.3631 [astro-ph]} \BibitemShut {NoStop}%
\bibitem [{\citenamefont {{MacLeod}}\ and\ \citenamefont
  {{Ramirez-Ruiz}}(2015{\natexlab{b}})}]{2015ApJ...803...41M}%
  \BibitemOpen
  \bibfield  {author} {\bibinfo {author} {\bibfnamefont {M.}~\bibnamefont
  {{MacLeod}}}\ and\ \bibinfo {author} {\bibfnamefont {E.}~\bibnamefont
  {{Ramirez-Ruiz}}},\ }\bibfield  {title} {\bibinfo {title} {{Asymmetric
  Accretion Flows within a Common Envelope}},\ }\href
  {https://doi.org/10.1088/0004-637X/803/1/41} {\bibfield  {journal} {\bibinfo
  {journal} {\apj}\ }\textbf {\bibinfo {volume} {803}},\ \bibinfo {eid} {41}
  (\bibinfo {year} {2015}{\natexlab{b}})},\ \Eprint
  {https://arxiv.org/abs/1410.3823} {arXiv:1410.3823 [astro-ph.SR]}
  \BibitemShut {NoStop}%
\bibitem [{\citenamefont {{MacLeod}}\ \emph
  {et~al.}(2017{\natexlab{b}})\citenamefont {{MacLeod}}, \citenamefont
  {{Antoni}}, \citenamefont {{Murguia-Berthier}}, \citenamefont {{Macias}},\
  and\ \citenamefont {{Ramirez-Ruiz}}}]{2017ApJ...838...56M}%
  \BibitemOpen
  \bibfield  {author} {\bibinfo {author} {\bibfnamefont {M.}~\bibnamefont
  {{MacLeod}}}, \bibinfo {author} {\bibfnamefont {A.}~\bibnamefont {{Antoni}}},
  \bibinfo {author} {\bibfnamefont {A.}~\bibnamefont {{Murguia-Berthier}}},
  \bibinfo {author} {\bibfnamefont {P.}~\bibnamefont {{Macias}}},\ and\
  \bibinfo {author} {\bibfnamefont {E.}~\bibnamefont {{Ramirez-Ruiz}}},\
  }\bibfield  {title} {\bibinfo {title} {{Common Envelope Wind Tunnel:
  Coefficients of Drag and Accretion in a Simplified Context for Studying Flows
  around Objects Embedded within Stellar Envelopes}},\ }\href
  {https://doi.org/10.3847/1538-4357/aa6117} {\bibfield  {journal} {\bibinfo
  {journal} {\apj}\ }\textbf {\bibinfo {volume} {838}},\ \bibinfo {eid} {56}
  (\bibinfo {year} {2017}{\natexlab{b}})},\ \Eprint
  {https://arxiv.org/abs/1704.02372} {arXiv:1704.02372 [astro-ph.SR]}
  \BibitemShut {NoStop}%
\bibitem [{\citenamefont {{Zahn}}(1977)}]{1977A&A....57..383Z}%
  \BibitemOpen
  \bibfield  {author} {\bibinfo {author} {\bibfnamefont {J.~P.}\ \bibnamefont
  {{Zahn}}},\ }\bibfield  {title} {\bibinfo {title} {{Reprint of
  1977A\&amp;A....57..383Z. Tidal friction in close binary stars.}},\
  }\href@noop {} {\bibfield  {journal} {\bibinfo  {journal} {\aap}\ }\textbf
  {\bibinfo {volume} {500}},\ \bibinfo {pages} {121} (\bibinfo {year}
  {1977})}\BibitemShut {NoStop}%
\bibitem [{\citenamefont {{Hut}}(1981)}]{1981A&A....99..126H}%
  \BibitemOpen
  \bibfield  {author} {\bibinfo {author} {\bibfnamefont {P.}~\bibnamefont
  {{Hut}}},\ }\bibfield  {title} {\bibinfo {title} {{Tidal evolution in close
  binary systems.}},\ }\href@noop {} {\bibfield  {journal} {\bibinfo  {journal}
  {\aap}\ }\textbf {\bibinfo {volume} {99}},\ \bibinfo {pages} {126} (\bibinfo
  {year} {1981})}\BibitemShut {NoStop}%
\bibitem [{\citenamefont {{Hurley}}\ \emph {et~al.}(2002)\citenamefont
  {{Hurley}}, \citenamefont {{Tout}},\ and\ \citenamefont
  {{Pols}}}]{2002MNRAS.329..897H}%
  \BibitemOpen
  \bibfield  {author} {\bibinfo {author} {\bibfnamefont {J.~R.}\ \bibnamefont
  {{Hurley}}}, \bibinfo {author} {\bibfnamefont {C.~A.}\ \bibnamefont
  {{Tout}}},\ and\ \bibinfo {author} {\bibfnamefont {O.~R.}\ \bibnamefont
  {{Pols}}},\ }\bibfield  {title} {\bibinfo {title} {{Evolution of binary stars
  and the effect of tides on binary populations}},\ }\href
  {https://doi.org/10.1046/j.1365-8711.2002.05038.x} {\bibfield  {journal}
  {\bibinfo  {journal} {\mnras}\ }\textbf {\bibinfo {volume} {329}},\ \bibinfo
  {pages} {897} (\bibinfo {year} {2002})},\ \Eprint
  {https://arxiv.org/abs/astro-ph/0201220} {arXiv:astro-ph/0201220 [astro-ph]}
  \BibitemShut {NoStop}%
\bibitem [{\citenamefont {{Bona{\v{c}}i{\'c} Marinovi{\'c}}}\ \emph
  {et~al.}(2008)\citenamefont {{Bona{\v{c}}i{\'c} Marinovi{\'c}}},
  \citenamefont {{Glebbeek}},\ and\ \citenamefont
  {{Pols}}}]{2008A&A...480..797B}%
  \BibitemOpen
  \bibfield  {author} {\bibinfo {author} {\bibfnamefont {A.~A.}\ \bibnamefont
  {{Bona{\v{c}}i{\'c} Marinovi{\'c}}}}, \bibinfo {author} {\bibfnamefont
  {E.}~\bibnamefont {{Glebbeek}}},\ and\ \bibinfo {author} {\bibfnamefont
  {O.~R.}\ \bibnamefont {{Pols}}},\ }\bibfield  {title} {\bibinfo {title}
  {{Orbital eccentricities of binary systems with a former AGB star}},\ }\href
  {https://doi.org/10.1051/0004-6361:20078297} {\bibfield  {journal} {\bibinfo
  {journal} {\aap}\ }\textbf {\bibinfo {volume} {480}},\ \bibinfo {pages} {797}
  (\bibinfo {year} {2008})},\ \Eprint {https://arxiv.org/abs/0710.4859}
  {arXiv:0710.4859 [astro-ph]} \BibitemShut {NoStop}%
\bibitem [{\citenamefont {{Boffin}}\ \emph {et~al.}(2014)\citenamefont
  {{Boffin}}, \citenamefont {{Hillen}}, \citenamefont {{Berger}}, \citenamefont
  {{Jorissen}}, \citenamefont {{Blind}}, \citenamefont {{Le Bouquin}},
  \citenamefont {{Miko{\l}ajewska}},\ and\ \citenamefont
  {{Lazareff}}}]{2014A&A...564A...1B}%
  \BibitemOpen
  \bibfield  {author} {\bibinfo {author} {\bibfnamefont {H.~M.~J.}\
  \bibnamefont {{Boffin}}}, \bibinfo {author} {\bibfnamefont {M.}~\bibnamefont
  {{Hillen}}}, \bibinfo {author} {\bibfnamefont {J.~P.}\ \bibnamefont
  {{Berger}}}, \bibinfo {author} {\bibfnamefont {A.}~\bibnamefont
  {{Jorissen}}}, \bibinfo {author} {\bibfnamefont {N.}~\bibnamefont {{Blind}}},
  \bibinfo {author} {\bibfnamefont {J.~B.}\ \bibnamefont {{Le Bouquin}}},
  \bibinfo {author} {\bibfnamefont {J.}~\bibnamefont {{Miko{\l}ajewska}}},\
  and\ \bibinfo {author} {\bibfnamefont {B.}~\bibnamefont {{Lazareff}}},\
  }\bibfield  {title} {\bibinfo {title} {{Roche-lobe filling factor of
  mass-transferring red giants: the PIONIER view}},\ }\href
  {https://doi.org/10.1051/0004-6361/201323194} {\bibfield  {journal} {\bibinfo
   {journal} {\aap}\ }\textbf {\bibinfo {volume} {564}},\ \bibinfo {eid} {A1}
  (\bibinfo {year} {2014})},\ \Eprint {https://arxiv.org/abs/1402.1798}
  {arXiv:1402.1798 [astro-ph.SR]} \BibitemShut {NoStop}%
\bibitem [{\citenamefont {{Vos}}\ \emph {et~al.}(2015)\citenamefont {{Vos}},
  \citenamefont {{{\O}stensen}}, \citenamefont {{Marchant}},\ and\
  \citenamefont {{Van Winckel}}}]{2015A&A...579A..49V}%
  \BibitemOpen
  \bibfield  {author} {\bibinfo {author} {\bibfnamefont {J.}~\bibnamefont
  {{Vos}}}, \bibinfo {author} {\bibfnamefont {R.~H.}\ \bibnamefont
  {{{\O}stensen}}}, \bibinfo {author} {\bibfnamefont {P.}~\bibnamefont
  {{Marchant}}},\ and\ \bibinfo {author} {\bibfnamefont {H.}~\bibnamefont {{Van
  Winckel}}},\ }\bibfield  {title} {\bibinfo {title} {{Testing eccentricity
  pumping mechanisms to model eccentric long-period sdB binaries with MESA}},\
  }\href {https://doi.org/10.1051/0004-6361/201526019} {\bibfield  {journal}
  {\bibinfo  {journal} {\aap}\ }\textbf {\bibinfo {volume} {579}},\ \bibinfo
  {eid} {A49} (\bibinfo {year} {2015})},\ \Eprint
  {https://arxiv.org/abs/1505.03293} {arXiv:1505.03293 [astro-ph.SR]}
  \BibitemShut {NoStop}%
\bibitem [{\citenamefont {{Belczynski}}\ \emph {et~al.}(2002)\citenamefont
  {{Belczynski}}, \citenamefont {{Kalogera}},\ and\ \citenamefont
  {{Bulik}}}]{2002ApJ...572..407B}%
  \BibitemOpen
  \bibfield  {author} {\bibinfo {author} {\bibfnamefont {K.}~\bibnamefont
  {{Belczynski}}}, \bibinfo {author} {\bibfnamefont {V.}~\bibnamefont
  {{Kalogera}}},\ and\ \bibinfo {author} {\bibfnamefont {T.}~\bibnamefont
  {{Bulik}}},\ }\bibfield  {title} {\bibinfo {title} {{A Comprehensive Study of
  Binary Compact Objects as Gravitational Wave Sources: Evolutionary Channels,
  Rates, and Physical Properties}},\ }\href {https://doi.org/10.1086/340304}
  {\bibfield  {journal} {\bibinfo  {journal} {\apj}\ }\textbf {\bibinfo
  {volume} {572}},\ \bibinfo {pages} {407} (\bibinfo {year} {2002})},\ \Eprint
  {https://arxiv.org/abs/astro-ph/0111452} {astro-ph/0111452} \BibitemShut
  {NoStop}%
\bibitem [{\citenamefont {{Mandel}}\ and\ \citenamefont {{de
  Mink}}(2016{\natexlab{a}})}]{2016MNRAS.458.2634M}%
  \BibitemOpen
  \bibfield  {author} {\bibinfo {author} {\bibfnamefont {I.}~\bibnamefont
  {{Mandel}}}\ and\ \bibinfo {author} {\bibfnamefont {S.~E.}\ \bibnamefont {{de
  Mink}}},\ }\bibfield  {title} {\bibinfo {title} {{Merging binary black holes
  formed through chemically homogeneous evolution in short-period stellar
  binaries}},\ }\href {https://doi.org/10.1093/mnras/stw379} {\bibfield
  {journal} {\bibinfo  {journal} {\mnras}\ }\textbf {\bibinfo {volume} {458}},\
  \bibinfo {pages} {2634} (\bibinfo {year} {2016}{\natexlab{a}})},\ \Eprint
  {https://arxiv.org/abs/1601.00007} {arXiv:1601.00007 [astro-ph.HE]}
  \BibitemShut {NoStop}%
\bibitem [{\citenamefont {{Qin}}\ \emph {et~al.}(2018)\citenamefont {{Qin}},
  \citenamefont {{Fragos}}, \citenamefont {{Meynet}}, \citenamefont
  {{Andrews}}, \citenamefont {{S{\o}rensen}},\ and\ \citenamefont
  {{Song}}}]{2018A&A...616A..28Q}%
  \BibitemOpen
  \bibfield  {author} {\bibinfo {author} {\bibfnamefont {Y.}~\bibnamefont
  {{Qin}}}, \bibinfo {author} {\bibfnamefont {T.}~\bibnamefont {{Fragos}}},
  \bibinfo {author} {\bibfnamefont {G.}~\bibnamefont {{Meynet}}}, \bibinfo
  {author} {\bibfnamefont {J.}~\bibnamefont {{Andrews}}}, \bibinfo {author}
  {\bibfnamefont {M.}~\bibnamefont {{S{\o}rensen}}},\ and\ \bibinfo {author}
  {\bibfnamefont {H.~F.}\ \bibnamefont {{Song}}},\ }\bibfield  {title}
  {\bibinfo {title} {{The spin of the second-born black hole in coalescing
  binary black holes}},\ }\href {https://doi.org/10.1051/0004-6361/201832839}
  {\bibfield  {journal} {\bibinfo  {journal} {\aap}\ }\textbf {\bibinfo
  {volume} {616}},\ \bibinfo {eid} {A28} (\bibinfo {year} {2018})},\ \Eprint
  {https://arxiv.org/abs/1802.05738} {arXiv:1802.05738 [astro-ph.SR]}
  \BibitemShut {NoStop}%
\bibitem [{\citenamefont {{Church}}\ \emph {et~al.}(2012)\citenamefont
  {{Church}}, \citenamefont {{Kim}}, \citenamefont {{Levan}},\ and\
  \citenamefont {{Davies}}}]{2012MNRAS.425..470C}%
  \BibitemOpen
  \bibfield  {author} {\bibinfo {author} {\bibfnamefont {R.~P.}\ \bibnamefont
  {{Church}}}, \bibinfo {author} {\bibfnamefont {C.}~\bibnamefont {{Kim}}},
  \bibinfo {author} {\bibfnamefont {A.~J.}\ \bibnamefont {{Levan}}},\ and\
  \bibinfo {author} {\bibfnamefont {M.~B.}\ \bibnamefont {{Davies}}},\
  }\bibfield  {title} {\bibinfo {title} {{The properties of long gamma-ray
  bursts in massive compact binaries}},\ }\href
  {https://doi.org/10.1111/j.1365-2966.2012.21480.x} {\bibfield  {journal}
  {\bibinfo  {journal} {\mnras}\ }\textbf {\bibinfo {volume} {425}},\ \bibinfo
  {pages} {470} (\bibinfo {year} {2012})},\ \Eprint
  {https://arxiv.org/abs/1205.0552} {arXiv:1205.0552 [astro-ph.HE]}
  \BibitemShut {NoStop}%
\bibitem [{\citenamefont {{Kushnir}}\ \emph {et~al.}(2016)\citenamefont
  {{Kushnir}}, \citenamefont {{Zaldarriaga}}, \citenamefont {{Kollmeier}},\
  and\ \citenamefont {{Waldman}}}]{2016MNRAS.462..844K}%
  \BibitemOpen
  \bibfield  {author} {\bibinfo {author} {\bibfnamefont {D.}~\bibnamefont
  {{Kushnir}}}, \bibinfo {author} {\bibfnamefont {M.}~\bibnamefont
  {{Zaldarriaga}}}, \bibinfo {author} {\bibfnamefont {J.~A.}\ \bibnamefont
  {{Kollmeier}}},\ and\ \bibinfo {author} {\bibfnamefont {R.}~\bibnamefont
  {{Waldman}}},\ }\bibfield  {title} {\bibinfo {title} {{GW150914: spin-based
  constraints on the merger time of the progenitor system}},\ }\href
  {https://doi.org/10.1093/mnras/stw1684} {\bibfield  {journal} {\bibinfo
  {journal} {\mnras}\ }\textbf {\bibinfo {volume} {462}},\ \bibinfo {pages}
  {844} (\bibinfo {year} {2016})},\ \Eprint {https://arxiv.org/abs/1605.03839}
  {arXiv:1605.03839 [astro-ph.HE]} \BibitemShut {NoStop}%
\bibitem [{\citenamefont {{Hotokezaka}}\ and\ \citenamefont
  {{Piran}}(2017)}]{2017ApJ...842..111H}%
  \BibitemOpen
  \bibfield  {author} {\bibinfo {author} {\bibfnamefont {K.}~\bibnamefont
  {{Hotokezaka}}}\ and\ \bibinfo {author} {\bibfnamefont {T.}~\bibnamefont
  {{Piran}}},\ }\bibfield  {title} {\bibinfo {title} {{Implications of the Low
  Binary Black Hole Aligned Spins Observed by LIGO}},\ }\href
  {https://doi.org/10.3847/1538-4357/aa6f61} {\bibfield  {journal} {\bibinfo
  {journal} {\apj}\ }\textbf {\bibinfo {volume} {842}},\ \bibinfo {eid} {111}
  (\bibinfo {year} {2017})},\ \Eprint {https://arxiv.org/abs/1702.03952}
  {arXiv:1702.03952 [astro-ph.HE]} \BibitemShut {NoStop}%
\bibitem [{\citenamefont {{Zaldarriaga}}\ \emph {et~al.}(2018)\citenamefont
  {{Zaldarriaga}}, \citenamefont {{Kushnir}},\ and\ \citenamefont
  {{Kollmeier}}}]{2018MNRAS.473.4174Z}%
  \BibitemOpen
  \bibfield  {author} {\bibinfo {author} {\bibfnamefont {M.}~\bibnamefont
  {{Zaldarriaga}}}, \bibinfo {author} {\bibfnamefont {D.}~\bibnamefont
  {{Kushnir}}},\ and\ \bibinfo {author} {\bibfnamefont {J.~A.}\ \bibnamefont
  {{Kollmeier}}},\ }\bibfield  {title} {\bibinfo {title} {{The expected spins
  of gravitational wave sources with isolated field binary progenitors}},\
  }\href {https://doi.org/10.1093/mnras/stx2577} {\bibfield  {journal}
  {\bibinfo  {journal} {\mnras}\ }\textbf {\bibinfo {volume} {473}},\ \bibinfo
  {pages} {4174} (\bibinfo {year} {2018})},\ \Eprint
  {https://arxiv.org/abs/1702.00885} {arXiv:1702.00885 [astro-ph.HE]}
  \BibitemShut {NoStop}%
\bibitem [{\citenamefont {{Bavera}}\ \emph {et~al.}(2020)\citenamefont
  {{Bavera}}, \citenamefont {{Fragos}}, \citenamefont {{Qin}}, \citenamefont
  {{Zapartas}}, \citenamefont {{Neijssel}}, \citenamefont {{Mandel}},
  \citenamefont {{Batta}}, \citenamefont {{Gaebel}}, \citenamefont
  {{Kimball}},\ and\ \citenamefont {{Stevenson}}}]{2020A&A...635A..97B}%
  \BibitemOpen
  \bibfield  {author} {\bibinfo {author} {\bibfnamefont {S.~S.}\ \bibnamefont
  {{Bavera}}}, \bibinfo {author} {\bibfnamefont {T.}~\bibnamefont {{Fragos}}},
  \bibinfo {author} {\bibfnamefont {Y.}~\bibnamefont {{Qin}}}, \bibinfo
  {author} {\bibfnamefont {E.}~\bibnamefont {{Zapartas}}}, \bibinfo {author}
  {\bibfnamefont {C.~J.}\ \bibnamefont {{Neijssel}}}, \bibinfo {author}
  {\bibfnamefont {I.}~\bibnamefont {{Mandel}}}, \bibinfo {author}
  {\bibfnamefont {A.}~\bibnamefont {{Batta}}}, \bibinfo {author} {\bibfnamefont
  {S.~M.}\ \bibnamefont {{Gaebel}}}, \bibinfo {author} {\bibfnamefont
  {C.}~\bibnamefont {{Kimball}}},\ and\ \bibinfo {author} {\bibfnamefont
  {S.}~\bibnamefont {{Stevenson}}},\ }\bibfield  {title} {\bibinfo {title}
  {{The origin of spin in binary black holes. Predicting the distributions of
  the main observables of Advanced LIGO}},\ }\href
  {https://doi.org/10.1051/0004-6361/201936204} {\bibfield  {journal} {\bibinfo
   {journal} {\aap}\ }\textbf {\bibinfo {volume} {635}},\ \bibinfo {eid} {A97}
  (\bibinfo {year} {2020})},\ \Eprint {https://arxiv.org/abs/1906.12257}
  {arXiv:1906.12257 [astro-ph.HE]} \BibitemShut {NoStop}%
\bibitem [{\citenamefont {{Olejak}}\ and\ \citenamefont
  {{Belczynski}}(2021)}]{2021ApJ...921L...2O}%
  \BibitemOpen
  \bibfield  {author} {\bibinfo {author} {\bibfnamefont {A.}~\bibnamefont
  {{Olejak}}}\ and\ \bibinfo {author} {\bibfnamefont {K.}~\bibnamefont
  {{Belczynski}}},\ }\bibfield  {title} {\bibinfo {title} {{The Implications of
  High Black Hole Spins for the Origin of Binary Black Hole Mergers}},\ }\href
  {https://doi.org/10.3847/2041-8213/ac2f48} {\bibfield  {journal} {\bibinfo
  {journal} {\apjl}\ }\textbf {\bibinfo {volume} {921}},\ \bibinfo {eid} {L2}
  (\bibinfo {year} {2021})},\ \Eprint {https://arxiv.org/abs/2109.06872}
  {arXiv:2109.06872 [astro-ph.HE]} \BibitemShut {NoStop}%
\bibitem [{\citenamefont {{Galaudage}}\ \emph {et~al.}(2021)\citenamefont
  {{Galaudage}}, \citenamefont {{Talbot}}, \citenamefont {{Nagar}},
  \citenamefont {{Jain}}, \citenamefont {{Thrane}},\ and\ \citenamefont
  {{Mandel}}}]{2021ApJ...921L..15G}%
  \BibitemOpen
  \bibfield  {author} {\bibinfo {author} {\bibfnamefont {S.}~\bibnamefont
  {{Galaudage}}}, \bibinfo {author} {\bibfnamefont {C.}~\bibnamefont
  {{Talbot}}}, \bibinfo {author} {\bibfnamefont {T.}~\bibnamefont {{Nagar}}},
  \bibinfo {author} {\bibfnamefont {D.}~\bibnamefont {{Jain}}}, \bibinfo
  {author} {\bibfnamefont {E.}~\bibnamefont {{Thrane}}},\ and\ \bibinfo
  {author} {\bibfnamefont {I.}~\bibnamefont {{Mandel}}},\ }\bibfield  {title}
  {\bibinfo {title} {{Building Better Spin Models for Merging Binary Black
  Holes: Evidence for Nonspinning and Rapidly Spinning Nearly Aligned
  Subpopulations}},\ }\href {https://doi.org/10.3847/2041-8213/ac2f3c}
  {\bibfield  {journal} {\bibinfo  {journal} {\apjl}\ }\textbf {\bibinfo
  {volume} {921}},\ \bibinfo {eid} {L15} (\bibinfo {year} {2021})},\ \Eprint
  {https://arxiv.org/abs/2109.02424} {arXiv:2109.02424 [gr-qc]} \BibitemShut
  {NoStop}%
\bibitem [{\citenamefont {{Batta}}\ and\ \citenamefont
  {{Ramirez-Ruiz}}(2019)}]{2019arXiv190404835B}%
  \BibitemOpen
  \bibfield  {author} {\bibinfo {author} {\bibfnamefont {A.}~\bibnamefont
  {{Batta}}}\ and\ \bibinfo {author} {\bibfnamefont {E.}~\bibnamefont
  {{Ramirez-Ruiz}}},\ }\bibfield  {title} {\bibinfo {title} {{Accretion
  Feedback from newly-formed black holes and its implications for LIGO
  Sources}},\ }\href@noop {} {\bibfield  {journal} {\bibinfo  {journal} {arXiv
  e-prints}\ ,\ \bibinfo {eid} {arXiv:1904.04835}} (\bibinfo {year} {2019})},\
  \Eprint {https://arxiv.org/abs/1904.04835} {arXiv:1904.04835 [astro-ph.HE]}
  \BibitemShut {NoStop}%
\bibitem [{\citenamefont {{Tauris}}(2022)}]{2022arXiv220502541T}%
  \BibitemOpen
  \bibfield  {author} {\bibinfo {author} {\bibfnamefont {T.~M.}\ \bibnamefont
  {{Tauris}}},\ }\bibfield  {title} {\bibinfo {title} {{Tossing Black Hole Spin
  Axes}},\ }\href@noop {} {\bibfield  {journal} {\bibinfo  {journal} {arXiv
  e-prints}\ ,\ \bibinfo {eid} {arXiv:2205.02541}} (\bibinfo {year} {2022})},\
  \Eprint {https://arxiv.org/abs/2205.02541} {arXiv:2205.02541 [astro-ph.HE]}
  \BibitemShut {NoStop}%
\bibitem [{\citenamefont
  {{Einstein}}(1916{\natexlab{a}})}]{1916AnP...354..769E}%
  \BibitemOpen
  \bibfield  {author} {\bibinfo {author} {\bibfnamefont {A.}~\bibnamefont
  {{Einstein}}},\ }\bibfield  {title} {\bibinfo {title} {{Die Grundlage der
  allgemeinen Relativit{\"a}tstheorie}},\ }\href
  {https://doi.org/10.1002/andp.19163540702} {\bibfield  {journal} {\bibinfo
  {journal} {Annalen der Physik}\ }\textbf {\bibinfo {volume} {354}},\ \bibinfo
  {pages} {769} (\bibinfo {year} {1916}{\natexlab{a}})}\BibitemShut {NoStop}%
\bibitem [{\citenamefont {{Hulse}}\ and\ \citenamefont
  {{Taylor}}(1975)}]{1975ApJ...195L..51H}%
  \BibitemOpen
  \bibfield  {author} {\bibinfo {author} {\bibfnamefont {R.~A.}\ \bibnamefont
  {{Hulse}}}\ and\ \bibinfo {author} {\bibfnamefont {J.~H.}\ \bibnamefont
  {{Taylor}}},\ }\bibfield  {title} {\bibinfo {title} {{Discovery of a pulsar
  in a binary system.}},\ }\href {https://doi.org/10.1086/181708} {\bibfield
  {journal} {\bibinfo  {journal} {\apjl}\ }\textbf {\bibinfo {volume} {195}},\
  \bibinfo {pages} {L51} (\bibinfo {year} {1975})}\BibitemShut {NoStop}%
\bibitem [{\citenamefont {{Taylor}}\ and\ \citenamefont
  {{Weisberg}}(1982)}]{1982ApJ...253..908T}%
  \BibitemOpen
  \bibfield  {author} {\bibinfo {author} {\bibfnamefont {J.~H.}\ \bibnamefont
  {{Taylor}}}\ and\ \bibinfo {author} {\bibfnamefont {J.~M.}\ \bibnamefont
  {{Weisberg}}},\ }\bibfield  {title} {\bibinfo {title} {{A new test of general
  relativity - Gravitational radiation and the binary pulsar PSR 1913+16}},\
  }\href {https://doi.org/10.1086/159690} {\bibfield  {journal} {\bibinfo
  {journal} {\apj}\ }\textbf {\bibinfo {volume} {253}},\ \bibinfo {pages} {908}
  (\bibinfo {year} {1982})}\BibitemShut {NoStop}%
\bibitem [{\citenamefont {{Kraft}}\ \emph {et~al.}(1962)\citenamefont
  {{Kraft}}, \citenamefont {{Mathews}},\ and\ \citenamefont
  {{Greenstein}}}]{1962ApJ...136..312K}%
  \BibitemOpen
  \bibfield  {author} {\bibinfo {author} {\bibfnamefont {R.~P.}\ \bibnamefont
  {{Kraft}}}, \bibinfo {author} {\bibfnamefont {J.}~\bibnamefont {{Mathews}}},\
  and\ \bibinfo {author} {\bibfnamefont {J.~L.}\ \bibnamefont {{Greenstein}}},\
  }\bibfield  {title} {\bibinfo {title} {{Binary Stars among Cataclysmic
  Variables. II. Nova WZ Sagittae: a Possible Radiator of Gravitational
  Waves.}},\ }\href {https://doi.org/10.1086/147381} {\bibfield  {journal}
  {\bibinfo  {journal} {\apj}\ }\textbf {\bibinfo {volume} {136}},\ \bibinfo
  {pages} {312} (\bibinfo {year} {1962})}\BibitemShut {NoStop}%
\bibitem [{\citenamefont {{Amaro-Seoane}}\ \emph {et~al.}(2022)\citenamefont
  {{Amaro-Seoane}}, \citenamefont {{Andrews}}, \citenamefont {{Arca Sedda}},
  \citenamefont {{Askar}}, \citenamefont {{Balasov}}, \citenamefont {{Bartos}},
  \citenamefont {{Bavera}}, \citenamefont {{Bellovary}}, \citenamefont
  {{Berry}}, \citenamefont {{Berti}}, \citenamefont {{Bianchi}}, \citenamefont
  {{Blecha}}, \citenamefont {{Blondin}}, \citenamefont {{Bogdanovi{\'c}}},
  \citenamefont {{Boissier}}, \citenamefont {{Bonetti}}, \citenamefont
  {{Bonoli}}, \citenamefont {{Bortolas}}, \citenamefont {{Breivik}},
  \citenamefont {{Capelo}}, \citenamefont {{Caramete}}, \citenamefont
  {{Catorini}}, \citenamefont {{Charisi}}, \citenamefont {{Chaty}},
  \citenamefont {{Chen}}, \citenamefont {{Chru{\'s}li{\'n}ska}}, \citenamefont
  {{Chua}}, \citenamefont {{Church}}, \citenamefont {{Colpi}}, \citenamefont
  {{D'Orazio}}, \citenamefont {{Danielski}}, \citenamefont {{Davies}},
  \citenamefont {{Dayal}}, \citenamefont {{De Rosa}}, \citenamefont
  {{Derdzinski}}, \citenamefont {{Destounis}}, \citenamefont {{Dotti}},
  \citenamefont {{Du{\c{t}}an}}, \citenamefont {{Dvorkin}}, \citenamefont
  {{Fabj}}, \citenamefont {{Foglizzo}}, \citenamefont {{Ford}}, \citenamefont
  {{Fouvry}}, \citenamefont {{Fragkos}}, \citenamefont {{Fryer}}, \citenamefont
  {{Gaspari}}, \citenamefont {{Gerosa}}, \citenamefont {{Graziani}},
  \citenamefont {{Groot}}, \citenamefont {{Habouzit}}, \citenamefont
  {{Haggard}}, \citenamefont {{Haiman}}, \citenamefont {{Han}}, \citenamefont
  {{Istrate}}, \citenamefont {{Johansson}}, \citenamefont {{Khan}},
  \citenamefont {{Kimpson}}, \citenamefont {{Kokkotas}}, \citenamefont
  {{Kong}}, \citenamefont {{Korol}}, \citenamefont {{Kremer}}, \citenamefont
  {{Kupfer}}, \citenamefont {{Lamberts}}, \citenamefont {{Larson}},
  \citenamefont {{Lau}}, \citenamefont {{Liu}}, \citenamefont
  {{Lloyd-Ronning}}, \citenamefont {{Lodato}}, \citenamefont {{Lupi}},
  \citenamefont {{Ma}}, \citenamefont {{Maccarone}}, \citenamefont {{Mandel}},
  \citenamefont {{Mangiagli}}, \citenamefont {{Mapelli}}, \citenamefont
  {{Mathis}}, \citenamefont {{Mayer}}, \citenamefont {{McGee}}, \citenamefont
  {{McKernan}}, \citenamefont {{Miller}}, \citenamefont {{Mota}}, \citenamefont
  {{Mumpower}}, \citenamefont {{Nasim}}, \citenamefont {{Nelemans}},
  \citenamefont {{Noble}}, \citenamefont {{Pacucci}}, \citenamefont
  {{Panessa}}, \citenamefont {{Paschalidis}}, \citenamefont {{Pfister}},
  \citenamefont {{Porquet}}, \citenamefont {{Quenby}}, \citenamefont
  {{R{\"o}pke}}, \citenamefont {{Regan}}, \citenamefont {{Rosswog}},
  \citenamefont {{Ruiter}}, \citenamefont {{Ruiz}}, \citenamefont {{Runnoe}},
  \citenamefont {{Schneider}}, \citenamefont {{Schnittman}}, \citenamefont
  {{Secunda}}, \citenamefont {{Sesana}}, \citenamefont {{Seto}}, \citenamefont
  {{Shao}}, \citenamefont {{Shapiro}}, \citenamefont {{Sopuerta}},
  \citenamefont {{Stone}}, \citenamefont {{Suvorov}}, \citenamefont
  {{Tamanini}}, \citenamefont {{Tamfal}}, \citenamefont {{Tauris}},
  \citenamefont {{Temmink}}, \citenamefont {{Tomsick}}, \citenamefont
  {{Toonen}}, \citenamefont {{Torres-Orjuela}}, \citenamefont {{Toscani}},
  \citenamefont {{Tsokaros}}, \citenamefont {{Unal}}, \citenamefont
  {{V{\'a}zquez-Aceves}}, \citenamefont {{Valiante}}, \citenamefont {{van
  Putten}}, \citenamefont {{van Roestel}}, \citenamefont {{Vignali}},
  \citenamefont {{Volonteri}}, \citenamefont {{Wu}}, \citenamefont {{Younsi}},
  \citenamefont {{Yu}}, \citenamefont {{Zane}}, \citenamefont {{Zwick}},
  \citenamefont {{Antonini}}, \citenamefont {{Baibhav}}, \citenamefont
  {{Barausse}}, \citenamefont {{Bonilla Rivera}}, \citenamefont {{Branchesi}},
  \citenamefont {{Branduardi-Raymont}}, \citenamefont {{Burdge}}, \citenamefont
  {{Chakraborty}}, \citenamefont {{Cuadra}}, \citenamefont {{Dage}},
  \citenamefont {{Davis}}, \citenamefont {{de Mink}}, \citenamefont
  {{Decarli}}, \citenamefont {{Doneva}}, \citenamefont {{Escoffier}},
  \citenamefont {{Gandhi}}, \citenamefont {{Haardt}}, \citenamefont {{Lousto}},
  \citenamefont {{Nissanke}}, \citenamefont {{Nordhaus}}, \citenamefont
  {{O'Shaughnessy}}, \citenamefont {{Portegies Zwart}}, \citenamefont
  {{Pound}}, \citenamefont {{Schussler}}, \citenamefont {{Sergijenko}},
  \citenamefont {{Spallicci}}, \citenamefont {{Vernieri}},\ and\ \citenamefont
  {{Vigna-G{\'o}mez}}}]{2022arXiv220306016A}%
  \BibitemOpen
  \bibfield  {author} {\bibinfo {author} {\bibfnamefont {P.}~\bibnamefont
  {{Amaro-Seoane}}}, \bibinfo {author} {\bibfnamefont {J.}~\bibnamefont
  {{Andrews}}}, \bibinfo {author} {\bibfnamefont {M.}~\bibnamefont {{Arca
  Sedda}}}, \bibinfo {author} {\bibfnamefont {A.}~\bibnamefont {{Askar}}},
  \bibinfo {author} {\bibfnamefont {R.}~\bibnamefont {{Balasov}}}, \bibinfo
  {author} {\bibfnamefont {I.}~\bibnamefont {{Bartos}}}, \bibinfo {author}
  {\bibfnamefont {S.~S.}\ \bibnamefont {{Bavera}}}, \bibinfo {author}
  {\bibfnamefont {J.}~\bibnamefont {{Bellovary}}}, \bibinfo {author}
  {\bibfnamefont {C.~P.~L.}\ \bibnamefont {{Berry}}}, \bibinfo {author}
  {\bibfnamefont {E.}~\bibnamefont {{Berti}}}, \bibinfo {author} {\bibfnamefont
  {S.}~\bibnamefont {{Bianchi}}}, \bibinfo {author} {\bibfnamefont
  {L.}~\bibnamefont {{Blecha}}}, \bibinfo {author} {\bibfnamefont
  {S.}~\bibnamefont {{Blondin}}}, \bibinfo {author} {\bibfnamefont
  {T.}~\bibnamefont {{Bogdanovi{\'c}}}}, \bibinfo {author} {\bibfnamefont
  {S.}~\bibnamefont {{Boissier}}}, \bibinfo {author} {\bibfnamefont
  {M.}~\bibnamefont {{Bonetti}}}, \bibinfo {author} {\bibfnamefont
  {S.}~\bibnamefont {{Bonoli}}}, \bibinfo {author} {\bibfnamefont
  {E.}~\bibnamefont {{Bortolas}}}, \bibinfo {author} {\bibfnamefont
  {K.}~\bibnamefont {{Breivik}}}, \bibinfo {author} {\bibfnamefont {P.~R.}\
  \bibnamefont {{Capelo}}}, \bibinfo {author} {\bibfnamefont {L.}~\bibnamefont
  {{Caramete}}}, \bibinfo {author} {\bibfnamefont {F.}~\bibnamefont
  {{Catorini}}}, \bibinfo {author} {\bibfnamefont {M.}~\bibnamefont
  {{Charisi}}}, \bibinfo {author} {\bibfnamefont {S.}~\bibnamefont {{Chaty}}},
  \bibinfo {author} {\bibfnamefont {X.}~\bibnamefont {{Chen}}}, \bibinfo
  {author} {\bibfnamefont {M.}~\bibnamefont {{Chru{\'s}li{\'n}ska}}}, \bibinfo
  {author} {\bibfnamefont {A.~J.~K.}\ \bibnamefont {{Chua}}}, \bibinfo {author}
  {\bibfnamefont {R.}~\bibnamefont {{Church}}}, \bibinfo {author}
  {\bibfnamefont {M.}~\bibnamefont {{Colpi}}}, \bibinfo {author} {\bibfnamefont
  {D.}~\bibnamefont {{D'Orazio}}}, \bibinfo {author} {\bibfnamefont
  {C.}~\bibnamefont {{Danielski}}}, \bibinfo {author} {\bibfnamefont {M.~B.}\
  \bibnamefont {{Davies}}}, \bibinfo {author} {\bibfnamefont {P.}~\bibnamefont
  {{Dayal}}}, \bibinfo {author} {\bibfnamefont {A.}~\bibnamefont {{De Rosa}}},
  \bibinfo {author} {\bibfnamefont {A.}~\bibnamefont {{Derdzinski}}}, \bibinfo
  {author} {\bibfnamefont {K.}~\bibnamefont {{Destounis}}}, \bibinfo {author}
  {\bibfnamefont {M.}~\bibnamefont {{Dotti}}}, \bibinfo {author} {\bibfnamefont
  {I.}~\bibnamefont {{Du{\c{t}}an}}}, \bibinfo {author} {\bibfnamefont
  {I.}~\bibnamefont {{Dvorkin}}}, \bibinfo {author} {\bibfnamefont
  {G.}~\bibnamefont {{Fabj}}}, \bibinfo {author} {\bibfnamefont
  {T.}~\bibnamefont {{Foglizzo}}}, \bibinfo {author} {\bibfnamefont
  {S.}~\bibnamefont {{Ford}}}, \bibinfo {author} {\bibfnamefont {J.-B.}\
  \bibnamefont {{Fouvry}}}, \bibinfo {author} {\bibfnamefont {T.}~\bibnamefont
  {{Fragkos}}}, \bibinfo {author} {\bibfnamefont {C.}~\bibnamefont {{Fryer}}},
  \bibinfo {author} {\bibfnamefont {M.}~\bibnamefont {{Gaspari}}}, \bibinfo
  {author} {\bibfnamefont {D.}~\bibnamefont {{Gerosa}}}, \bibinfo {author}
  {\bibfnamefont {L.}~\bibnamefont {{Graziani}}}, \bibinfo {author}
  {\bibfnamefont {P.~J.}\ \bibnamefont {{Groot}}}, \bibinfo {author}
  {\bibfnamefont {M.}~\bibnamefont {{Habouzit}}}, \bibinfo {author}
  {\bibfnamefont {D.}~\bibnamefont {{Haggard}}}, \bibinfo {author}
  {\bibfnamefont {Z.}~\bibnamefont {{Haiman}}}, \bibinfo {author}
  {\bibfnamefont {W.-B.}\ \bibnamefont {{Han}}}, \bibinfo {author}
  {\bibfnamefont {A.}~\bibnamefont {{Istrate}}}, \bibinfo {author}
  {\bibfnamefont {P.~H.}\ \bibnamefont {{Johansson}}}, \bibinfo {author}
  {\bibfnamefont {F.~M.}\ \bibnamefont {{Khan}}}, \bibinfo {author}
  {\bibfnamefont {T.}~\bibnamefont {{Kimpson}}}, \bibinfo {author}
  {\bibfnamefont {K.}~\bibnamefont {{Kokkotas}}}, \bibinfo {author}
  {\bibfnamefont {A.}~\bibnamefont {{Kong}}}, \bibinfo {author} {\bibfnamefont
  {V.}~\bibnamefont {{Korol}}}, \bibinfo {author} {\bibfnamefont
  {K.}~\bibnamefont {{Kremer}}}, \bibinfo {author} {\bibfnamefont
  {T.}~\bibnamefont {{Kupfer}}}, \bibinfo {author} {\bibfnamefont
  {A.}~\bibnamefont {{Lamberts}}}, \bibinfo {author} {\bibfnamefont
  {S.}~\bibnamefont {{Larson}}}, \bibinfo {author} {\bibfnamefont
  {M.}~\bibnamefont {{Lau}}}, \bibinfo {author} {\bibfnamefont
  {D.}~\bibnamefont {{Liu}}}, \bibinfo {author} {\bibfnamefont
  {N.}~\bibnamefont {{Lloyd-Ronning}}}, \bibinfo {author} {\bibfnamefont
  {G.}~\bibnamefont {{Lodato}}}, \bibinfo {author} {\bibfnamefont
  {A.}~\bibnamefont {{Lupi}}}, \bibinfo {author} {\bibfnamefont {C.-P.}\
  \bibnamefont {{Ma}}}, \bibinfo {author} {\bibfnamefont {T.}~\bibnamefont
  {{Maccarone}}}, \bibinfo {author} {\bibfnamefont {I.}~\bibnamefont
  {{Mandel}}}, \bibinfo {author} {\bibfnamefont {A.}~\bibnamefont
  {{Mangiagli}}}, \bibinfo {author} {\bibfnamefont {M.}~\bibnamefont
  {{Mapelli}}}, \bibinfo {author} {\bibfnamefont {S.}~\bibnamefont {{Mathis}}},
  \bibinfo {author} {\bibfnamefont {L.}~\bibnamefont {{Mayer}}}, \bibinfo
  {author} {\bibfnamefont {S.}~\bibnamefont {{McGee}}}, \bibinfo {author}
  {\bibfnamefont {B.}~\bibnamefont {{McKernan}}}, \bibinfo {author}
  {\bibfnamefont {M.~C.}\ \bibnamefont {{Miller}}}, \bibinfo {author}
  {\bibfnamefont {D.~F.}\ \bibnamefont {{Mota}}}, \bibinfo {author}
  {\bibfnamefont {M.}~\bibnamefont {{Mumpower}}}, \bibinfo {author}
  {\bibfnamefont {S.~S.}\ \bibnamefont {{Nasim}}}, \bibinfo {author}
  {\bibfnamefont {G.}~\bibnamefont {{Nelemans}}}, \bibinfo {author}
  {\bibfnamefont {S.}~\bibnamefont {{Noble}}}, \bibinfo {author} {\bibfnamefont
  {F.}~\bibnamefont {{Pacucci}}}, \bibinfo {author} {\bibfnamefont
  {F.}~\bibnamefont {{Panessa}}}, \bibinfo {author} {\bibfnamefont
  {V.}~\bibnamefont {{Paschalidis}}}, \bibinfo {author} {\bibfnamefont
  {H.}~\bibnamefont {{Pfister}}}, \bibinfo {author} {\bibfnamefont
  {D.}~\bibnamefont {{Porquet}}}, \bibinfo {author} {\bibfnamefont
  {J.}~\bibnamefont {{Quenby}}}, \bibinfo {author} {\bibfnamefont
  {F.}~\bibnamefont {{R{\"o}pke}}}, \bibinfo {author} {\bibfnamefont
  {J.}~\bibnamefont {{Regan}}}, \bibinfo {author} {\bibfnamefont
  {S.}~\bibnamefont {{Rosswog}}}, \bibinfo {author} {\bibfnamefont
  {A.}~\bibnamefont {{Ruiter}}}, \bibinfo {author} {\bibfnamefont
  {M.}~\bibnamefont {{Ruiz}}}, \bibinfo {author} {\bibfnamefont
  {J.}~\bibnamefont {{Runnoe}}}, \bibinfo {author} {\bibfnamefont
  {R.}~\bibnamefont {{Schneider}}}, \bibinfo {author} {\bibfnamefont
  {J.}~\bibnamefont {{Schnittman}}}, \bibinfo {author} {\bibfnamefont
  {A.}~\bibnamefont {{Secunda}}}, \bibinfo {author} {\bibfnamefont
  {A.}~\bibnamefont {{Sesana}}}, \bibinfo {author} {\bibfnamefont
  {N.}~\bibnamefont {{Seto}}}, \bibinfo {author} {\bibfnamefont
  {L.}~\bibnamefont {{Shao}}}, \bibinfo {author} {\bibfnamefont
  {S.}~\bibnamefont {{Shapiro}}}, \bibinfo {author} {\bibfnamefont
  {C.}~\bibnamefont {{Sopuerta}}}, \bibinfo {author} {\bibfnamefont
  {N.}~\bibnamefont {{Stone}}}, \bibinfo {author} {\bibfnamefont
  {A.}~\bibnamefont {{Suvorov}}}, \bibinfo {author} {\bibfnamefont
  {N.}~\bibnamefont {{Tamanini}}}, \bibinfo {author} {\bibfnamefont
  {T.}~\bibnamefont {{Tamfal}}}, \bibinfo {author} {\bibfnamefont
  {T.}~\bibnamefont {{Tauris}}}, \bibinfo {author} {\bibfnamefont
  {K.}~\bibnamefont {{Temmink}}}, \bibinfo {author} {\bibfnamefont
  {J.}~\bibnamefont {{Tomsick}}}, \bibinfo {author} {\bibfnamefont
  {S.}~\bibnamefont {{Toonen}}}, \bibinfo {author} {\bibfnamefont
  {A.}~\bibnamefont {{Torres-Orjuela}}}, \bibinfo {author} {\bibfnamefont
  {M.}~\bibnamefont {{Toscani}}}, \bibinfo {author} {\bibfnamefont
  {A.}~\bibnamefont {{Tsokaros}}}, \bibinfo {author} {\bibfnamefont
  {C.}~\bibnamefont {{Unal}}}, \bibinfo {author} {\bibfnamefont
  {V.}~\bibnamefont {{V{\'a}zquez-Aceves}}}, \bibinfo {author} {\bibfnamefont
  {R.}~\bibnamefont {{Valiante}}}, \bibinfo {author} {\bibfnamefont
  {M.}~\bibnamefont {{van Putten}}}, \bibinfo {author} {\bibfnamefont
  {J.}~\bibnamefont {{van Roestel}}}, \bibinfo {author} {\bibfnamefont
  {C.}~\bibnamefont {{Vignali}}}, \bibinfo {author} {\bibfnamefont
  {M.}~\bibnamefont {{Volonteri}}}, \bibinfo {author} {\bibfnamefont
  {K.}~\bibnamefont {{Wu}}}, \bibinfo {author} {\bibfnamefont {Z.}~\bibnamefont
  {{Younsi}}}, \bibinfo {author} {\bibfnamefont {S.}~\bibnamefont {{Yu}}},
  \bibinfo {author} {\bibfnamefont {S.}~\bibnamefont {{Zane}}}, \bibinfo
  {author} {\bibfnamefont {L.}~\bibnamefont {{Zwick}}}, \bibinfo {author}
  {\bibfnamefont {F.}~\bibnamefont {{Antonini}}}, \bibinfo {author}
  {\bibfnamefont {V.}~\bibnamefont {{Baibhav}}}, \bibinfo {author}
  {\bibfnamefont {E.}~\bibnamefont {{Barausse}}}, \bibinfo {author}
  {\bibfnamefont {A.}~\bibnamefont {{Bonilla Rivera}}}, \bibinfo {author}
  {\bibfnamefont {M.}~\bibnamefont {{Branchesi}}}, \bibinfo {author}
  {\bibfnamefont {G.}~\bibnamefont {{Branduardi-Raymont}}}, \bibinfo {author}
  {\bibfnamefont {K.}~\bibnamefont {{Burdge}}}, \bibinfo {author}
  {\bibfnamefont {S.}~\bibnamefont {{Chakraborty}}}, \bibinfo {author}
  {\bibfnamefont {J.}~\bibnamefont {{Cuadra}}}, \bibinfo {author}
  {\bibfnamefont {K.}~\bibnamefont {{Dage}}}, \bibinfo {author} {\bibfnamefont
  {B.}~\bibnamefont {{Davis}}}, \bibinfo {author} {\bibfnamefont {S.~E.}\
  \bibnamefont {{de Mink}}}, \bibinfo {author} {\bibfnamefont {R.}~\bibnamefont
  {{Decarli}}}, \bibinfo {author} {\bibfnamefont {D.}~\bibnamefont {{Doneva}}},
  \bibinfo {author} {\bibfnamefont {S.}~\bibnamefont {{Escoffier}}}, \bibinfo
  {author} {\bibfnamefont {P.}~\bibnamefont {{Gandhi}}}, \bibinfo {author}
  {\bibfnamefont {F.}~\bibnamefont {{Haardt}}}, \bibinfo {author}
  {\bibfnamefont {C.~O.}\ \bibnamefont {{Lousto}}}, \bibinfo {author}
  {\bibfnamefont {S.}~\bibnamefont {{Nissanke}}}, \bibinfo {author}
  {\bibfnamefont {J.}~\bibnamefont {{Nordhaus}}}, \bibinfo {author}
  {\bibfnamefont {R.}~\bibnamefont {{O'Shaughnessy}}}, \bibinfo {author}
  {\bibfnamefont {S.}~\bibnamefont {{Portegies Zwart}}}, \bibinfo {author}
  {\bibfnamefont {A.}~\bibnamefont {{Pound}}}, \bibinfo {author} {\bibfnamefont
  {F.}~\bibnamefont {{Schussler}}}, \bibinfo {author} {\bibfnamefont
  {O.}~\bibnamefont {{Sergijenko}}}, \bibinfo {author} {\bibfnamefont
  {A.}~\bibnamefont {{Spallicci}}}, \bibinfo {author} {\bibfnamefont
  {D.}~\bibnamefont {{Vernieri}}},\ and\ \bibinfo {author} {\bibfnamefont
  {A.}~\bibnamefont {{Vigna-G{\'o}mez}}},\ }\bibfield  {title} {\bibinfo
  {title} {{Astrophysics with the Laser Interferometer Space Antenna}},\
  }\href@noop {} {\bibfield  {journal} {\bibinfo  {journal} {arXiv e-prints}\
  ,\ \bibinfo {eid} {arXiv:2203.06016}} (\bibinfo {year} {2022})},\ \Eprint
  {https://arxiv.org/abs/2203.06016} {arXiv:2203.06016 [gr-qc]} \BibitemShut
  {NoStop}%
\bibitem [{\citenamefont {{Peters}}(1964{\natexlab{a}})}]{1964PhRv..136.1224P}%
  \BibitemOpen
  \bibfield  {author} {\bibinfo {author} {\bibfnamefont {P.~C.}\ \bibnamefont
  {{Peters}}},\ }\bibfield  {title} {\bibinfo {title} {{Gravitational Radiation
  and the Motion of Two Point Masses}},\ }\href
  {https://doi.org/10.1103/PhysRev.136.B1224} {\bibfield  {journal} {\bibinfo
  {journal} {Physical Review}\ }\textbf {\bibinfo {volume} {136}},\ \bibinfo
  {pages} {1224} (\bibinfo {year} {1964}{\natexlab{a}})}\BibitemShut {NoStop}%
\bibitem [{\citenamefont {{Mandel}}(2021)}]{2021RNAAS...5..223M}%
  \BibitemOpen
  \bibfield  {author} {\bibinfo {author} {\bibfnamefont {I.}~\bibnamefont
  {{Mandel}}},\ }\bibfield  {title} {\bibinfo {title} {{An Accurate Analytical
  Fit to the Gravitational-wave Inspiral Duration for Eccentric Binaries}},\
  }\href {https://doi.org/10.3847/2515-5172/ac2d35} {\bibfield  {journal}
  {\bibinfo  {journal} {Research Notes of the American Astronomical Society}\
  }\textbf {\bibinfo {volume} {5}},\ \bibinfo {eid} {223} (\bibinfo {year}
  {2021})},\ \Eprint {https://arxiv.org/abs/2110.09254} {arXiv:2110.09254
  [astro-ph.HE]} \BibitemShut {NoStop}%
\bibitem [{\citenamefont {{Weber}}\ and\ \citenamefont
  {{Davis}}(1967)}]{1967ApJ...148..217W}%
  \BibitemOpen
  \bibfield  {author} {\bibinfo {author} {\bibfnamefont {E.~J.}\ \bibnamefont
  {{Weber}}}\ and\ \bibinfo {author} {\bibfnamefont {J.}~\bibnamefont
  {{Davis}}, \bibfnamefont {Leverett}},\ }\bibfield  {title} {\bibinfo {title}
  {{The Angular Momentum of the Solar Wind}},\ }\href
  {https://doi.org/10.1086/149138} {\bibfield  {journal} {\bibinfo  {journal}
  {\apj}\ }\textbf {\bibinfo {volume} {148}},\ \bibinfo {pages} {217} (\bibinfo
  {year} {1967})}\BibitemShut {NoStop}%
\bibitem [{\citenamefont {{Skumanich}}(1972)}]{1972ApJ...171..565S}%
  \BibitemOpen
  \bibfield  {author} {\bibinfo {author} {\bibfnamefont {A.}~\bibnamefont
  {{Skumanich}}},\ }\bibfield  {title} {\bibinfo {title} {{Time Scales for Ca
  II Emission Decay, Rotational Braking, and Lithium Depletion}},\ }\href
  {https://doi.org/10.1086/151310} {\bibfield  {journal} {\bibinfo  {journal}
  {\apj}\ }\textbf {\bibinfo {volume} {171}},\ \bibinfo {pages} {565} (\bibinfo
  {year} {1972})}\BibitemShut {NoStop}%
\bibitem [{\citenamefont {{Verbunt}}\ and\ \citenamefont
  {{Zwaan}}(1981)}]{1981A&A...100L...7V}%
  \BibitemOpen
  \bibfield  {author} {\bibinfo {author} {\bibfnamefont {F.}~\bibnamefont
  {{Verbunt}}}\ and\ \bibinfo {author} {\bibfnamefont {C.}~\bibnamefont
  {{Zwaan}}},\ }\bibfield  {title} {\bibinfo {title} {{Magnetic braking in
  low-mass X-ray binaries.}},\ }\href@noop {} {\bibfield  {journal} {\bibinfo
  {journal} {\aap}\ }\textbf {\bibinfo {volume} {100}},\ \bibinfo {pages} {L7}
  (\bibinfo {year} {1981})}\BibitemShut {NoStop}%
\bibitem [{\citenamefont {{Mestel}}\ and\ \citenamefont
  {{Spruit}}(1987)}]{1987MNRAS.226...57M}%
  \BibitemOpen
  \bibfield  {author} {\bibinfo {author} {\bibfnamefont {L.}~\bibnamefont
  {{Mestel}}}\ and\ \bibinfo {author} {\bibfnamefont {H.~C.}\ \bibnamefont
  {{Spruit}}},\ }\bibfield  {title} {\bibinfo {title} {{On magnetic braking of
  late-type stars}},\ }\href {https://doi.org/10.1093/mnras/226.1.57}
  {\bibfield  {journal} {\bibinfo  {journal} {\mnras}\ }\textbf {\bibinfo
  {volume} {226}},\ \bibinfo {pages} {57} (\bibinfo {year} {1987})}\BibitemShut
  {NoStop}%
\bibitem [{\citenamefont {{Knigge}}\ \emph {et~al.}(2011)\citenamefont
  {{Knigge}}, \citenamefont {{Baraffe}},\ and\ \citenamefont
  {{Patterson}}}]{2011ApJS..194...28K}%
  \BibitemOpen
  \bibfield  {author} {\bibinfo {author} {\bibfnamefont {C.}~\bibnamefont
  {{Knigge}}}, \bibinfo {author} {\bibfnamefont {I.}~\bibnamefont
  {{Baraffe}}},\ and\ \bibinfo {author} {\bibfnamefont {J.}~\bibnamefont
  {{Patterson}}},\ }\bibfield  {title} {\bibinfo {title} {{The Evolution of
  Cataclysmic Variables as Revealed by Their Donor Stars}},\ }\href
  {https://doi.org/10.1088/0067-0049/194/2/28} {\bibfield  {journal} {\bibinfo
  {journal} {\apjs}\ }\textbf {\bibinfo {volume} {194}},\ \bibinfo {eid} {28}
  (\bibinfo {year} {2011})},\ \Eprint {https://arxiv.org/abs/1102.2440}
  {arXiv:1102.2440 [astro-ph.SR]} \BibitemShut {NoStop}%
\bibitem [{\citenamefont {{Spruit}}\ and\ \citenamefont {{van
  Ballegooijen}}(1982)}]{1982A&A...106...58S}%
  \BibitemOpen
  \bibfield  {author} {\bibinfo {author} {\bibfnamefont {H.~C.}\ \bibnamefont
  {{Spruit}}}\ and\ \bibinfo {author} {\bibfnamefont {A.~A.}\ \bibnamefont
  {{van Ballegooijen}}},\ }\bibfield  {title} {\bibinfo {title} {{Stability of
  toroidal flux tubes in stars}},\ }\href@noop {} {\bibfield  {journal}
  {\bibinfo  {journal} {\aap}\ }\textbf {\bibinfo {volume} {106}},\ \bibinfo
  {pages} {58} (\bibinfo {year} {1982})}\BibitemShut {NoStop}%
\bibitem [{\citenamefont {{Rappaport}}\ \emph {et~al.}(1983)\citenamefont
  {{Rappaport}}, \citenamefont {{Verbunt}},\ and\ \citenamefont
  {{Joss}}}]{1983ApJ...275..713R}%
  \BibitemOpen
  \bibfield  {author} {\bibinfo {author} {\bibfnamefont {S.}~\bibnamefont
  {{Rappaport}}}, \bibinfo {author} {\bibfnamefont {F.}~\bibnamefont
  {{Verbunt}}},\ and\ \bibinfo {author} {\bibfnamefont {P.~C.}\ \bibnamefont
  {{Joss}}},\ }\bibfield  {title} {\bibinfo {title} {{A new technique for
  calculations of binary stellar evolution application to magnetic braking.}},\
  }\href {https://doi.org/10.1086/161569} {\bibfield  {journal} {\bibinfo
  {journal} {\apj}\ }\textbf {\bibinfo {volume} {275}},\ \bibinfo {pages} {713}
  (\bibinfo {year} {1983})}\BibitemShut {NoStop}%
\bibitem [{\citenamefont {{Verbunt}}(1993)}]{1993ARA&A..31...93V}%
  \BibitemOpen
  \bibfield  {author} {\bibinfo {author} {\bibfnamefont {F.}~\bibnamefont
  {{Verbunt}}},\ }\bibfield  {title} {\bibinfo {title} {{Origin and evolution
  of X-ray binaries and binary radio pulsars.}},\ }\href
  {https://doi.org/10.1146/annurev.aa.31.090193.000521} {\bibfield  {journal}
  {\bibinfo  {journal} {\araa}\ }\textbf {\bibinfo {volume} {31}},\ \bibinfo
  {pages} {93} (\bibinfo {year} {1993})}\BibitemShut {NoStop}%
\bibitem [{\citenamefont {{Repetto}}\ and\ \citenamefont
  {{Nelemans}}(2014)}]{2014MNRAS.444..542R}%
  \BibitemOpen
  \bibfield  {author} {\bibinfo {author} {\bibfnamefont {S.}~\bibnamefont
  {{Repetto}}}\ and\ \bibinfo {author} {\bibfnamefont {G.}~\bibnamefont
  {{Nelemans}}},\ }\bibfield  {title} {\bibinfo {title} {{The coupled effect of
  tides and stellar winds on the evolution of compact binaries}},\ }\href
  {https://doi.org/10.1093/mnras/stu1454} {\bibfield  {journal} {\bibinfo
  {journal} {\mnras}\ }\textbf {\bibinfo {volume} {444}},\ \bibinfo {pages}
  {542} (\bibinfo {year} {2014})},\ \Eprint {https://arxiv.org/abs/1407.5182}
  {arXiv:1407.5182 [astro-ph.SR]} \BibitemShut {NoStop}%
\bibitem [{\citenamefont {{Donati}}\ and\ \citenamefont
  {{Landstreet}}(2009)}]{2009ARA&A..47..333D}%
  \BibitemOpen
  \bibfield  {author} {\bibinfo {author} {\bibfnamefont {J.~F.}\ \bibnamefont
  {{Donati}}}\ and\ \bibinfo {author} {\bibfnamefont {J.~D.}\ \bibnamefont
  {{Landstreet}}},\ }\bibfield  {title} {\bibinfo {title} {{Magnetic Fields of
  Nondegenerate Stars}},\ }\href
  {https://doi.org/10.1146/annurev-astro-082708-101833} {\bibfield  {journal}
  {\bibinfo  {journal} {\araa}\ }\textbf {\bibinfo {volume} {47}},\ \bibinfo
  {pages} {333} (\bibinfo {year} {2009})},\ \Eprint
  {https://arxiv.org/abs/0904.1938} {arXiv:0904.1938 [astro-ph.SR]}
  \BibitemShut {NoStop}%
\bibitem [{\citenamefont {{Langer}}\ and\ \citenamefont
  {{Norman}}(2006)}]{2006ApJ...638L..63L}%
  \BibitemOpen
  \bibfield  {author} {\bibinfo {author} {\bibfnamefont {N.}~\bibnamefont
  {{Langer}}}\ and\ \bibinfo {author} {\bibfnamefont {C.~A.}\ \bibnamefont
  {{Norman}}},\ }\bibfield  {title} {\bibinfo {title} {{On the Collapsar Model
  of Long Gamma-Ray Bursts:Constraints from Cosmic Metallicity Evolution}},\
  }\href {https://doi.org/10.1086/500363} {\bibfield  {journal} {\bibinfo
  {journal} {\apjl}\ }\textbf {\bibinfo {volume} {638}},\ \bibinfo {pages}
  {L63} (\bibinfo {year} {2006})},\ \Eprint
  {https://arxiv.org/abs/astro-ph/0512271} {arXiv:astro-ph/0512271 [astro-ph]}
  \BibitemShut {NoStop}%
\bibitem [{\citenamefont {{Stanek}}\ \emph {et~al.}(2006)\citenamefont
  {{Stanek}}, \citenamefont {{Gnedin}}, \citenamefont {{Beacom}}, \citenamefont
  {{Gould}}, \citenamefont {{Johnson}}, \citenamefont {{Kollmeier}},
  \citenamefont {{Modjaz}}, \citenamefont {{Pinsonneault}}, \citenamefont
  {{Pogge}},\ and\ \citenamefont {{Weinberg}}}]{2006AcA....56..333S}%
  \BibitemOpen
  \bibfield  {author} {\bibinfo {author} {\bibfnamefont {K.~Z.}\ \bibnamefont
  {{Stanek}}}, \bibinfo {author} {\bibfnamefont {O.~Y.}\ \bibnamefont
  {{Gnedin}}}, \bibinfo {author} {\bibfnamefont {J.~F.}\ \bibnamefont
  {{Beacom}}}, \bibinfo {author} {\bibfnamefont {A.~P.}\ \bibnamefont
  {{Gould}}}, \bibinfo {author} {\bibfnamefont {J.~A.}\ \bibnamefont
  {{Johnson}}}, \bibinfo {author} {\bibfnamefont {J.~A.}\ \bibnamefont
  {{Kollmeier}}}, \bibinfo {author} {\bibfnamefont {M.}~\bibnamefont
  {{Modjaz}}}, \bibinfo {author} {\bibfnamefont {M.~H.}\ \bibnamefont
  {{Pinsonneault}}}, \bibinfo {author} {\bibfnamefont {R.}~\bibnamefont
  {{Pogge}}},\ and\ \bibinfo {author} {\bibfnamefont {D.~H.}\ \bibnamefont
  {{Weinberg}}},\ }\bibfield  {title} {\bibinfo {title} {{Protecting Life in
  the Milky Way: Metals Keep the GRBs Away}},\ }\href@noop {} {\bibfield
  {journal} {\bibinfo  {journal} {\actaa}\ }\textbf {\bibinfo {volume} {56}},\
  \bibinfo {pages} {333} (\bibinfo {year} {2006})},\ \Eprint
  {https://arxiv.org/abs/astro-ph/0604113} {arXiv:astro-ph/0604113 [astro-ph]}
  \BibitemShut {NoStop}%
\bibitem [{\citenamefont {{Mapelli}}\ \emph {et~al.}(2009)\citenamefont
  {{Mapelli}}, \citenamefont {{Colpi}},\ and\ \citenamefont
  {{Zampieri}}}]{2009MNRAS.395L..71M}%
  \BibitemOpen
  \bibfield  {author} {\bibinfo {author} {\bibfnamefont {M.}~\bibnamefont
  {{Mapelli}}}, \bibinfo {author} {\bibfnamefont {M.}~\bibnamefont {{Colpi}}},\
  and\ \bibinfo {author} {\bibfnamefont {L.}~\bibnamefont {{Zampieri}}},\
  }\bibfield  {title} {\bibinfo {title} {{Low metallicity and ultra-luminous
  X-ray sources in the Cartwheel galaxy}},\ }\href
  {https://doi.org/10.1111/j.1745-3933.2009.00645.x} {\bibfield  {journal}
  {\bibinfo  {journal} {\mnras}\ }\textbf {\bibinfo {volume} {395}},\ \bibinfo
  {pages} {L71} (\bibinfo {year} {2009})},\ \Eprint
  {https://arxiv.org/abs/0902.3540} {arXiv:0902.3540 [astro-ph.HE]}
  \BibitemShut {NoStop}%
\bibitem [{\citenamefont {{Belczynski}}\ \emph
  {et~al.}(2016{\natexlab{b}})\citenamefont {{Belczynski}}, \citenamefont
  {{Holz}}, \citenamefont {{Bulik}},\ and\ \citenamefont
  {{O'Shaughnessy}}}]{2016Natur.534..512B}%
  \BibitemOpen
  \bibfield  {author} {\bibinfo {author} {\bibfnamefont {K.}~\bibnamefont
  {{Belczynski}}}, \bibinfo {author} {\bibfnamefont {D.~E.}\ \bibnamefont
  {{Holz}}}, \bibinfo {author} {\bibfnamefont {T.}~\bibnamefont {{Bulik}}},\
  and\ \bibinfo {author} {\bibfnamefont {R.}~\bibnamefont {{O'Shaughnessy}}},\
  }\bibfield  {title} {\bibinfo {title} {{The first gravitational-wave source
  from the isolated evolution of two stars in the 40-100 solar mass range}},\
  }\href {https://doi.org/10.1038/nature18322} {\bibfield  {journal} {\bibinfo
  {journal} {\nat}\ }\textbf {\bibinfo {volume} {534}},\ \bibinfo {pages} {512}
  (\bibinfo {year} {2016}{\natexlab{b}})},\ \Eprint
  {https://arxiv.org/abs/1602.04531} {arXiv:1602.04531 [astro-ph.HE]}
  \BibitemShut {NoStop}%
\bibitem [{\citenamefont {{Chen}}\ \emph {et~al.}(2017)\citenamefont {{Chen}},
  \citenamefont {{Smartt}}, \citenamefont {{Yates}}, \citenamefont {{Nicholl}},
  \citenamefont {{Kr{\"u}hler}}, \citenamefont {{Schady}}, \citenamefont
  {{Dennefeld}},\ and\ \citenamefont {{Inserra}}}]{2017MNRAS.470.3566C}%
  \BibitemOpen
  \bibfield  {author} {\bibinfo {author} {\bibfnamefont {T.-W.}\ \bibnamefont
  {{Chen}}}, \bibinfo {author} {\bibfnamefont {S.~J.}\ \bibnamefont
  {{Smartt}}}, \bibinfo {author} {\bibfnamefont {R.~M.}\ \bibnamefont
  {{Yates}}}, \bibinfo {author} {\bibfnamefont {M.}~\bibnamefont {{Nicholl}}},
  \bibinfo {author} {\bibfnamefont {T.}~\bibnamefont {{Kr{\"u}hler}}}, \bibinfo
  {author} {\bibfnamefont {P.}~\bibnamefont {{Schady}}}, \bibinfo {author}
  {\bibfnamefont {M.}~\bibnamefont {{Dennefeld}}},\ and\ \bibinfo {author}
  {\bibfnamefont {C.}~\bibnamefont {{Inserra}}},\ }\bibfield  {title} {\bibinfo
  {title} {{Superluminous supernova progenitors have a half-solar metallicity
  threshold}},\ }\href {https://doi.org/10.1093/mnras/stx1428} {\bibfield
  {journal} {\bibinfo  {journal} {\mnras}\ }\textbf {\bibinfo {volume} {470}},\
  \bibinfo {pages} {3566} (\bibinfo {year} {2017})},\ \Eprint
  {https://arxiv.org/abs/1605.04925} {arXiv:1605.04925 [astro-ph.GA]}
  \BibitemShut {NoStop}%
\bibitem [{\citenamefont {{Graham}}\ and\ \citenamefont
  {{Fruchter}}(2017)}]{2017ApJ...834..170G}%
  \BibitemOpen
  \bibfield  {author} {\bibinfo {author} {\bibfnamefont {J.~F.}\ \bibnamefont
  {{Graham}}}\ and\ \bibinfo {author} {\bibfnamefont {A.~S.}\ \bibnamefont
  {{Fruchter}}},\ }\bibfield  {title} {\bibinfo {title} {{The Relative Rate of
  LGRB Formation as a Function of Metallicity}},\ }\href
  {https://doi.org/10.3847/1538-4357/834/2/170} {\bibfield  {journal} {\bibinfo
   {journal} {\apj}\ }\textbf {\bibinfo {volume} {834}},\ \bibinfo {eid} {170}
  (\bibinfo {year} {2017})},\ \Eprint {https://arxiv.org/abs/1511.01079}
  {arXiv:1511.01079 [astro-ph.HE]} \BibitemShut {NoStop}%
\bibitem [{\citenamefont {{Schulze}}\ \emph {et~al.}(2018)\citenamefont
  {{Schulze}}, \citenamefont {{Kr{\"u}hler}}, \citenamefont {{Leloudas}},
  \citenamefont {{Gorosabel}}, \citenamefont {{Mehner}}, \citenamefont
  {{Buchner}}, \citenamefont {{Kim}}, \citenamefont {{Ibar}}, \citenamefont
  {{Amor{\'\i}n}}, \citenamefont {{Herrero-Illana}}, \citenamefont
  {{Anderson}}, \citenamefont {{Bauer}}, \citenamefont {{Christensen}},
  \citenamefont {{de Pasquale}}, \citenamefont {{de Ugarte Postigo}},
  \citenamefont {{Gallazzi}}, \citenamefont {{Hjorth}}, \citenamefont
  {{Morrell}}, \citenamefont {{Malesani}}, \citenamefont {{Sparre}},
  \citenamefont {{Stalder}}, \citenamefont {{Stark}}, \citenamefont
  {{Th{\"o}ne}},\ and\ \citenamefont {{Wheeler}}}]{2018MNRAS.473.1258S}%
  \BibitemOpen
  \bibfield  {author} {\bibinfo {author} {\bibfnamefont {S.}~\bibnamefont
  {{Schulze}}}, \bibinfo {author} {\bibfnamefont {T.}~\bibnamefont
  {{Kr{\"u}hler}}}, \bibinfo {author} {\bibfnamefont {G.}~\bibnamefont
  {{Leloudas}}}, \bibinfo {author} {\bibfnamefont {J.}~\bibnamefont
  {{Gorosabel}}}, \bibinfo {author} {\bibfnamefont {A.}~\bibnamefont
  {{Mehner}}}, \bibinfo {author} {\bibfnamefont {J.}~\bibnamefont {{Buchner}}},
  \bibinfo {author} {\bibfnamefont {S.}~\bibnamefont {{Kim}}}, \bibinfo
  {author} {\bibfnamefont {E.}~\bibnamefont {{Ibar}}}, \bibinfo {author}
  {\bibfnamefont {R.}~\bibnamefont {{Amor{\'\i}n}}}, \bibinfo {author}
  {\bibfnamefont {R.}~\bibnamefont {{Herrero-Illana}}}, \bibinfo {author}
  {\bibfnamefont {J.~P.}\ \bibnamefont {{Anderson}}}, \bibinfo {author}
  {\bibfnamefont {F.~E.}\ \bibnamefont {{Bauer}}}, \bibinfo {author}
  {\bibfnamefont {L.}~\bibnamefont {{Christensen}}}, \bibinfo {author}
  {\bibfnamefont {M.}~\bibnamefont {{de Pasquale}}}, \bibinfo {author}
  {\bibfnamefont {A.}~\bibnamefont {{de Ugarte Postigo}}}, \bibinfo {author}
  {\bibfnamefont {A.}~\bibnamefont {{Gallazzi}}}, \bibinfo {author}
  {\bibfnamefont {J.}~\bibnamefont {{Hjorth}}}, \bibinfo {author}
  {\bibfnamefont {N.}~\bibnamefont {{Morrell}}}, \bibinfo {author}
  {\bibfnamefont {D.}~\bibnamefont {{Malesani}}}, \bibinfo {author}
  {\bibfnamefont {M.}~\bibnamefont {{Sparre}}}, \bibinfo {author}
  {\bibfnamefont {B.}~\bibnamefont {{Stalder}}}, \bibinfo {author}
  {\bibfnamefont {A.~A.}\ \bibnamefont {{Stark}}}, \bibinfo {author}
  {\bibfnamefont {C.~C.}\ \bibnamefont {{Th{\"o}ne}}},\ and\ \bibinfo {author}
  {\bibfnamefont {J.~C.}\ \bibnamefont {{Wheeler}}},\ }\bibfield  {title}
  {\bibinfo {title} {{Cosmic evolution and metal aversion in superluminous
  supernova host galaxies}},\ }\href {https://doi.org/10.1093/mnras/stx2352}
  {\bibfield  {journal} {\bibinfo  {journal} {\mnras}\ }\textbf {\bibinfo
  {volume} {473}},\ \bibinfo {pages} {1258} (\bibinfo {year} {2018})},\ \Eprint
  {https://arxiv.org/abs/1612.05978} {arXiv:1612.05978 [astro-ph.GA]}
  \BibitemShut {NoStop}%
\bibitem [{\citenamefont {{Kovlakas}}\ \emph {et~al.}(2020)\citenamefont
  {{Kovlakas}}, \citenamefont {{Zezas}}, \citenamefont {{Andrews}},
  \citenamefont {{Basu-Zych}}, \citenamefont {{Fragos}}, \citenamefont
  {{Hornschemeier}}, \citenamefont {{Lehmer}},\ and\ \citenamefont
  {{Ptak}}}]{2020MNRAS.498.4790K}%
  \BibitemOpen
  \bibfield  {author} {\bibinfo {author} {\bibfnamefont {K.}~\bibnamefont
  {{Kovlakas}}}, \bibinfo {author} {\bibfnamefont {A.}~\bibnamefont {{Zezas}}},
  \bibinfo {author} {\bibfnamefont {J.~J.}\ \bibnamefont {{Andrews}}}, \bibinfo
  {author} {\bibfnamefont {A.}~\bibnamefont {{Basu-Zych}}}, \bibinfo {author}
  {\bibfnamefont {T.}~\bibnamefont {{Fragos}}}, \bibinfo {author}
  {\bibfnamefont {A.}~\bibnamefont {{Hornschemeier}}}, \bibinfo {author}
  {\bibfnamefont {B.}~\bibnamefont {{Lehmer}}},\ and\ \bibinfo {author}
  {\bibfnamefont {A.}~\bibnamefont {{Ptak}}},\ }\bibfield  {title} {\bibinfo
  {title} {{A census of ultraluminous X-ray sources in the local Universe}},\
  }\href {https://doi.org/10.1093/mnras/staa2481} {\bibfield  {journal}
  {\bibinfo  {journal} {\mnras}\ }\textbf {\bibinfo {volume} {498}},\ \bibinfo
  {pages} {4790} (\bibinfo {year} {2020})},\ \Eprint
  {https://arxiv.org/abs/2008.10572} {arXiv:2008.10572 [astro-ph.GA]}
  \BibitemShut {NoStop}%
\bibitem [{\citenamefont {{Lehmer}}\ \emph {et~al.}(2021)\citenamefont
  {{Lehmer}}, \citenamefont {{Eufrasio}}, \citenamefont {{Basu-Zych}},
  \citenamefont {{Doore}}, \citenamefont {{Fragos}}, \citenamefont
  {{Garofali}}, \citenamefont {{Kovlakas}}, \citenamefont {{Williams}},
  \citenamefont {{Zezas}},\ and\ \citenamefont
  {{Santana-Silva}}}]{2021ApJ...907...17L}%
  \BibitemOpen
  \bibfield  {author} {\bibinfo {author} {\bibfnamefont {B.~D.}\ \bibnamefont
  {{Lehmer}}}, \bibinfo {author} {\bibfnamefont {R.~T.}\ \bibnamefont
  {{Eufrasio}}}, \bibinfo {author} {\bibfnamefont {A.}~\bibnamefont
  {{Basu-Zych}}}, \bibinfo {author} {\bibfnamefont {K.}~\bibnamefont
  {{Doore}}}, \bibinfo {author} {\bibfnamefont {T.}~\bibnamefont {{Fragos}}},
  \bibinfo {author} {\bibfnamefont {K.}~\bibnamefont {{Garofali}}}, \bibinfo
  {author} {\bibfnamefont {K.}~\bibnamefont {{Kovlakas}}}, \bibinfo {author}
  {\bibfnamefont {B.~F.}\ \bibnamefont {{Williams}}}, \bibinfo {author}
  {\bibfnamefont {A.}~\bibnamefont {{Zezas}}},\ and\ \bibinfo {author}
  {\bibfnamefont {L.}~\bibnamefont {{Santana-Silva}}},\ }\bibfield  {title}
  {\bibinfo {title} {{The Metallicity Dependence of the High-mass X-Ray Binary
  Luminosity Function}},\ }\href {https://doi.org/10.3847/1538-4357/abcec1}
  {\bibfield  {journal} {\bibinfo  {journal} {\apj}\ }\textbf {\bibinfo
  {volume} {907}},\ \bibinfo {eid} {17} (\bibinfo {year} {2021})},\ \Eprint
  {https://arxiv.org/abs/2011.09476} {arXiv:2011.09476 [astro-ph.GA]}
  \BibitemShut {NoStop}%
\bibitem [{\citenamefont {{Chru{\'s}li{\'n}ska}}(2022)}]{2022arXiv220610622C}%
  \BibitemOpen
  \bibfield  {author} {\bibinfo {author} {\bibfnamefont {M.}~\bibnamefont
  {{Chru{\'s}li{\'n}ska}}},\ }\bibfield  {title} {\bibinfo {title} {{Chemical
  evolution of the Universe and its consequences for gravitational-wave
  astrophysics}},\ }\href@noop {} {\bibfield  {journal} {\bibinfo  {journal}
  {arXiv e-prints}\ ,\ \bibinfo {eid} {arXiv:2206.10622}} (\bibinfo {year}
  {2022})},\ \Eprint {https://arxiv.org/abs/2206.10622} {arXiv:2206.10622
  [astro-ph.GA]} \BibitemShut {NoStop}%
\bibitem [{\citenamefont {{Mennekens}}\ and\ \citenamefont
  {{Vanbeveren}}(2016)}]{2016A&A...589A..64M}%
  \BibitemOpen
  \bibfield  {author} {\bibinfo {author} {\bibfnamefont {N.}~\bibnamefont
  {{Mennekens}}}\ and\ \bibinfo {author} {\bibfnamefont {D.}~\bibnamefont
  {{Vanbeveren}}},\ }\bibfield  {title} {\bibinfo {title} {{The delay time
  distribution of massive double compact star mergers}},\ }\href
  {https://doi.org/10.1051/0004-6361/201628193} {\bibfield  {journal} {\bibinfo
   {journal} {\aap}\ }\textbf {\bibinfo {volume} {589}},\ \bibinfo {eid} {A64}
  (\bibinfo {year} {2016})},\ \Eprint {https://arxiv.org/abs/1601.06966}
  {arXiv:1601.06966 [astro-ph.SR]} \BibitemShut {NoStop}%
\bibitem [{\citenamefont {{Zevin}}\ \emph {et~al.}(2022)\citenamefont
  {{Zevin}}, \citenamefont {{Nugent}}, \citenamefont {{Adhikari}},
  \citenamefont {{Fong}}, \citenamefont {{Holz}},\ and\ \citenamefont
  {{Kelley}}}]{2022ApJ...940L..18Z}%
  \BibitemOpen
  \bibfield  {author} {\bibinfo {author} {\bibfnamefont {M.}~\bibnamefont
  {{Zevin}}}, \bibinfo {author} {\bibfnamefont {A.~E.}\ \bibnamefont
  {{Nugent}}}, \bibinfo {author} {\bibfnamefont {S.}~\bibnamefont
  {{Adhikari}}}, \bibinfo {author} {\bibfnamefont {W.-f.}\ \bibnamefont
  {{Fong}}}, \bibinfo {author} {\bibfnamefont {D.~E.}\ \bibnamefont {{Holz}}},\
  and\ \bibinfo {author} {\bibfnamefont {L.~Z.}\ \bibnamefont {{Kelley}}},\
  }\bibfield  {title} {\bibinfo {title} {{Observational Inference on the Delay
  Time Distribution of Short Gamma-Ray Bursts}},\ }\href
  {https://doi.org/10.3847/2041-8213/ac91cd} {\bibfield  {journal} {\bibinfo
  {journal} {\apjl}\ }\textbf {\bibinfo {volume} {940}},\ \bibinfo {eid} {L18}
  (\bibinfo {year} {2022})},\ \Eprint {https://arxiv.org/abs/2206.02814}
  {arXiv:2206.02814 [astro-ph.HE]} \BibitemShut {NoStop}%
\bibitem [{\citenamefont {{Peters}}(1964{\natexlab{b}})}]{1964peters}%
  \BibitemOpen
  \bibfield  {author} {\bibinfo {author} {\bibfnamefont {P.~C.}\ \bibnamefont
  {{Peters}}},\ }\bibfield  {title} {\bibinfo {title} {{Gravitational Radiation
  and the Motion of Two Point Masses}},\ }\href
  {https://doi.org/10.1103/PhysRev.136.B1224} {\bibfield  {journal} {\bibinfo
  {journal} {Physical Review}\ }\textbf {\bibinfo {volume} {136}},\ \bibinfo
  {pages} {1224} (\bibinfo {year} {1964}{\natexlab{b}})}\BibitemShut {NoStop}%
\bibitem [{\citenamefont {{Mandel}}\ and\ \citenamefont
  {{Broekgaarden}}(2022)}]{MandelBroekgaarden:2021}%
  \BibitemOpen
  \bibfield  {author} {\bibinfo {author} {\bibfnamefont {I.}~\bibnamefont
  {{Mandel}}}\ and\ \bibinfo {author} {\bibfnamefont {F.~S.}\ \bibnamefont
  {{Broekgaarden}}},\ }\bibfield  {title} {\bibinfo {title} {{Rates of compact
  object coalescences}},\ }\href {https://doi.org/10.1007/s41114-021-00034-3}
  {\bibfield  {journal} {\bibinfo  {journal} {Living Reviews in Relativity}\
  }\textbf {\bibinfo {volume} {25}},\ \bibinfo {eid} {1} (\bibinfo {year}
  {2022})},\ \Eprint {https://arxiv.org/abs/2107.14239} {arXiv:2107.14239
  [astro-ph.HE]} \BibitemShut {NoStop}%
\bibitem [{\citenamefont {{Mandel}}\ and\ \citenamefont
  {{Farmer}}(2022)}]{MandelFarmer:2018}%
  \BibitemOpen
  \bibfield  {author} {\bibinfo {author} {\bibfnamefont {I.}~\bibnamefont
  {{Mandel}}}\ and\ \bibinfo {author} {\bibfnamefont {A.}~\bibnamefont
  {{Farmer}}},\ }\bibfield  {title} {\bibinfo {title} {{Merging stellar-mass
  binary black holes}},\ }\href {https://doi.org/10.1016/j.physrep.2022.01.003}
  {\bibfield  {journal} {\bibinfo  {journal} {\physrep}\ }\textbf {\bibinfo
  {volume} {955}},\ \bibinfo {pages} {1} (\bibinfo {year} {2022})},\ \Eprint
  {https://arxiv.org/abs/1806.05820} {arXiv:1806.05820 [astro-ph.HE]}
  \BibitemShut {NoStop}%
\bibitem [{\citenamefont {{Schneider}}\ \emph {et~al.}(2015)\citenamefont
  {{Schneider}}, \citenamefont {{Izzard}}, \citenamefont {{Langer}},\ and\
  \citenamefont {{de Mink}}}]{Schneider:2015}%
  \BibitemOpen
  \bibfield  {author} {\bibinfo {author} {\bibfnamefont {F.~R.~N.}\
  \bibnamefont {{Schneider}}}, \bibinfo {author} {\bibfnamefont {R.~G.}\
  \bibnamefont {{Izzard}}}, \bibinfo {author} {\bibfnamefont {N.}~\bibnamefont
  {{Langer}}},\ and\ \bibinfo {author} {\bibfnamefont {S.~E.}\ \bibnamefont
  {{de Mink}}},\ }\bibfield  {title} {\bibinfo {title} {{Evolution of Mass
  Functions of Coeval Stars through Wind Mass Loss and Binary Interactions}},\
  }\href {https://doi.org/10.1088/0004-637X/805/1/20} {\bibfield  {journal}
  {\bibinfo  {journal} {\apj}\ }\textbf {\bibinfo {volume} {805}},\ \bibinfo
  {eid} {20} (\bibinfo {year} {2015})},\ \Eprint
  {https://arxiv.org/abs/1504.01735} {arXiv:1504.01735 [astro-ph.SR]}
  \BibitemShut {NoStop}%
\bibitem [{\citenamefont {{Gerosa}}\ and\ \citenamefont
  {{Fishbach}}(2021)}]{GerosaFishbach:2021}%
  \BibitemOpen
  \bibfield  {author} {\bibinfo {author} {\bibfnamefont {D.}~\bibnamefont
  {{Gerosa}}}\ and\ \bibinfo {author} {\bibfnamefont {M.}~\bibnamefont
  {{Fishbach}}},\ }\bibfield  {title} {\bibinfo {title} {{Hierarchical mergers
  of stellar-mass black holes and their gravitational-wave signatures}},\
  }\href {https://doi.org/10.1038/s41550-021-01398-w} {\bibfield  {journal}
  {\bibinfo  {journal} {Nature Astronomy}\ }\textbf {\bibinfo {volume} {5}},\
  \bibinfo {pages} {749} (\bibinfo {year} {2021})},\ \Eprint
  {https://arxiv.org/abs/2105.03439} {arXiv:2105.03439 [astro-ph.HE]}
  \BibitemShut {NoStop}%
\bibitem [{\citenamefont {{Arca Sedda}}\ \emph {et~al.}(2023)\citenamefont
  {{Arca Sedda}}, \citenamefont {{Naoz}},\ and\ \citenamefont
  {{Kocsis}}}]{ArcaSedda:2023review}%
  \BibitemOpen
  \bibfield  {author} {\bibinfo {author} {\bibfnamefont {M.}~\bibnamefont
  {{Arca Sedda}}}, \bibinfo {author} {\bibfnamefont {S.}~\bibnamefont
  {{Naoz}}},\ and\ \bibinfo {author} {\bibfnamefont {B.}~\bibnamefont
  {{Kocsis}}},\ }\bibfield  {title} {\bibinfo {title} {{Quiescent and Active
  Galactic Nuclei as Factories of Merging Compact Objects in the Era of
  Gravitational Wave Astronomy}},\ }\href
  {https://doi.org/10.3390/universe9030138} {\bibfield  {journal} {\bibinfo
  {journal} {Universe}\ }\textbf {\bibinfo {volume} {9}},\ \bibinfo {pages}
  {138} (\bibinfo {year} {2023})},\ \Eprint {https://arxiv.org/abs/2302.14071}
  {arXiv:2302.14071 [astro-ph.GA]} \BibitemShut {NoStop}%
\bibitem [{\citenamefont {Vigna-Gómez}\ and\ \citenamefont
  {Rebagliato}(2020)}]{ZenvignaGomezImages}%
  \BibitemOpen
  \bibfield  {author} {\bibinfo {author} {\bibfnamefont {A.}~\bibnamefont
  {Vigna-Gómez}}\ and\ \bibinfo {author} {\bibfnamefont {T.}~\bibnamefont
  {Rebagliato}},\ }\bibfield  {title} {\bibinfo {title} {{Graphics:
  Evolutionary pathways leading to Double Neutron Star formation}}\ }\href
  {https://doi.org/10.5281/zenodo.3634498} {10.5281/zenodo.3634498} (\bibinfo
  {year} {2020})\BibitemShut {NoStop}%
\bibitem [{\citenamefont {{Laplace}}\ \emph
  {et~al.}(2020{\natexlab{b}})\citenamefont {{Laplace}}, \citenamefont
  {{G{\"o}tberg}}, \citenamefont {{de Mink}}, \citenamefont {{Justham}},\ and\
  \citenamefont {{Farmer}}}]{Laplace:2020}%
  \BibitemOpen
  \bibfield  {author} {\bibinfo {author} {\bibfnamefont {E.}~\bibnamefont
  {{Laplace}}}, \bibinfo {author} {\bibfnamefont {Y.}~\bibnamefont
  {{G{\"o}tberg}}}, \bibinfo {author} {\bibfnamefont {S.~E.}\ \bibnamefont {{de
  Mink}}}, \bibinfo {author} {\bibfnamefont {S.}~\bibnamefont {{Justham}}},\
  and\ \bibinfo {author} {\bibfnamefont {R.}~\bibnamefont {{Farmer}}},\
  }\bibfield  {title} {\bibinfo {title} {{The expansion of stripped-envelope
  stars: Consequences for supernovae and gravitational-wave progenitors}},\
  }\href {https://doi.org/10.1051/0004-6361/201937300} {\bibfield  {journal}
  {\bibinfo  {journal} {\aap}\ }\textbf {\bibinfo {volume} {637}},\ \bibinfo
  {eid} {A6} (\bibinfo {year} {2020}{\natexlab{b}})},\ \Eprint
  {https://arxiv.org/abs/2003.01120} {arXiv:2003.01120 [astro-ph.SR]}
  \BibitemShut {NoStop}%
\bibitem [{\citenamefont {{G{\"o}tberg}}\ \emph {et~al.}(2018)\citenamefont
  {{G{\"o}tberg}}, \citenamefont {{de Mink}}, \citenamefont {{Groh}},
  \citenamefont {{Kupfer}}, \citenamefont {{Crowther}}, \citenamefont
  {{Zapartas}},\ and\ \citenamefont {{Renzo}}}]{2018A&A...615A..78G}%
  \BibitemOpen
  \bibfield  {author} {\bibinfo {author} {\bibfnamefont {Y.}~\bibnamefont
  {{G{\"o}tberg}}}, \bibinfo {author} {\bibfnamefont {S.~E.}\ \bibnamefont {{de
  Mink}}}, \bibinfo {author} {\bibfnamefont {J.~H.}\ \bibnamefont {{Groh}}},
  \bibinfo {author} {\bibfnamefont {T.}~\bibnamefont {{Kupfer}}}, \bibinfo
  {author} {\bibfnamefont {P.~A.}\ \bibnamefont {{Crowther}}}, \bibinfo
  {author} {\bibfnamefont {E.}~\bibnamefont {{Zapartas}}},\ and\ \bibinfo
  {author} {\bibfnamefont {M.}~\bibnamefont {{Renzo}}},\ }\bibfield  {title}
  {\bibinfo {title} {{Spectral models for binary products: Unifying subdwarfs
  and Wolf-Rayet stars as a sequence of stripped-envelope stars}},\ }\href
  {https://doi.org/10.1051/0004-6361/201732274} {\bibfield  {journal} {\bibinfo
   {journal} {\aap}\ }\textbf {\bibinfo {volume} {615}},\ \bibinfo {eid} {A78}
  (\bibinfo {year} {2018})},\ \Eprint {https://arxiv.org/abs/1802.03018}
  {arXiv:1802.03018 [astro-ph.SR]} \BibitemShut {NoStop}%
\bibitem [{\citenamefont {{De Donder}}\ \emph {et~al.}(1997)\citenamefont {{De
  Donder}}, \citenamefont {{Vanbeveren}},\ and\ \citenamefont {{van
  Bever}}}]{DeDonder:1997}%
  \BibitemOpen
  \bibfield  {author} {\bibinfo {author} {\bibfnamefont {E.}~\bibnamefont {{De
  Donder}}}, \bibinfo {author} {\bibfnamefont {D.}~\bibnamefont
  {{Vanbeveren}}},\ and\ \bibinfo {author} {\bibfnamefont {J.}~\bibnamefont
  {{van Bever}}},\ }\bibfield  {title} {\bibinfo {title} {{The number of O-type
  runaways, the number of O and Wolf-Rayet stars with a compact companion and
  the formation rate of double pulsars predicted by massive close binary
  evolution.}},\ }\href@noop {} {\bibfield  {journal} {\bibinfo  {journal}
  {\aap}\ }\textbf {\bibinfo {volume} {318}},\ \bibinfo {pages} {812} (\bibinfo
  {year} {1997})}\BibitemShut {NoStop}%
\bibitem [{\citenamefont {{Eldridge}}\ \emph {et~al.}(2011)\citenamefont
  {{Eldridge}}, \citenamefont {{Langer}},\ and\ \citenamefont
  {{Tout}}}]{Eldridge:2011}%
  \BibitemOpen
  \bibfield  {author} {\bibinfo {author} {\bibfnamefont {J.~J.}\ \bibnamefont
  {{Eldridge}}}, \bibinfo {author} {\bibfnamefont {N.}~\bibnamefont
  {{Langer}}},\ and\ \bibinfo {author} {\bibfnamefont {C.~A.}\ \bibnamefont
  {{Tout}}},\ }\bibfield  {title} {\bibinfo {title} {{Runaway stars as
  progenitors of supernovae and gamma-ray bursts}},\ }\href
  {https://doi.org/10.1111/j.1365-2966.2011.18650.x} {\bibfield  {journal}
  {\bibinfo  {journal} {\mnras}\ }\textbf {\bibinfo {volume} {414}},\ \bibinfo
  {pages} {3501} (\bibinfo {year} {2011})},\ \Eprint
  {https://arxiv.org/abs/1103.1877} {arXiv:1103.1877 [astro-ph.SR]}
  \BibitemShut {NoStop}%
\bibitem [{\citenamefont {{Vigna-G{\'o}mez}}\ \emph {et~al.}(2018)\citenamefont
  {{Vigna-G{\'o}mez}}, \citenamefont {{Neijssel}}, \citenamefont {{Stevenson}},
  \citenamefont {{Barrett}}, \citenamefont {{Belczynski}}, \citenamefont
  {{Justham}}, \citenamefont {{de Mink}}, \citenamefont {{M{\"u}ller}},
  \citenamefont {{Podsiadlowski}}, \citenamefont {{Renzo}}, \citenamefont
  {{Sz{\'e}csi}},\ and\ \citenamefont {{Mandel}}}]{VignaGomez:2018}%
  \BibitemOpen
  \bibfield  {author} {\bibinfo {author} {\bibfnamefont {A.}~\bibnamefont
  {{Vigna-G{\'o}mez}}}, \bibinfo {author} {\bibfnamefont {C.~J.}\ \bibnamefont
  {{Neijssel}}}, \bibinfo {author} {\bibfnamefont {S.}~\bibnamefont
  {{Stevenson}}}, \bibinfo {author} {\bibfnamefont {J.~W.}\ \bibnamefont
  {{Barrett}}}, \bibinfo {author} {\bibfnamefont {K.}~\bibnamefont
  {{Belczynski}}}, \bibinfo {author} {\bibfnamefont {S.}~\bibnamefont
  {{Justham}}}, \bibinfo {author} {\bibfnamefont {S.~E.}\ \bibnamefont {{de
  Mink}}}, \bibinfo {author} {\bibfnamefont {B.}~\bibnamefont {{M{\"u}ller}}},
  \bibinfo {author} {\bibfnamefont {P.}~\bibnamefont {{Podsiadlowski}}},
  \bibinfo {author} {\bibfnamefont {M.}~\bibnamefont {{Renzo}}}, \bibinfo
  {author} {\bibfnamefont {D.}~\bibnamefont {{Sz{\'e}csi}}},\ and\ \bibinfo
  {author} {\bibfnamefont {I.}~\bibnamefont {{Mandel}}},\ }\bibfield  {title}
  {\bibinfo {title} {{On the formation history of Galactic double neutron
  stars}},\ }\href {https://doi.org/10.1093/mnras/sty2463} {\bibfield
  {journal} {\bibinfo  {journal} {\mnras}\ }\textbf {\bibinfo {volume} {481}},\
  \bibinfo {pages} {4009} (\bibinfo {year} {2018})},\ \Eprint
  {https://arxiv.org/abs/1805.07974} {arXiv:1805.07974 [astro-ph.SR]}
  \BibitemShut {NoStop}%
\bibitem [{\citenamefont {{Flannery}}\ and\ \citenamefont {{van den
  Heuvel}}(1975)}]{1975A&A....39...61F}%
  \BibitemOpen
  \bibfield  {author} {\bibinfo {author} {\bibfnamefont {B.~P.}\ \bibnamefont
  {{Flannery}}}\ and\ \bibinfo {author} {\bibfnamefont {E.~P.~J.}\ \bibnamefont
  {{van den Heuvel}}},\ }\bibfield  {title} {\bibinfo {title} {{On the origin
  of the binary pulsar PSR 1913+16.}},\ }\href@noop {} {\bibfield  {journal}
  {\bibinfo  {journal} {\aap}\ }\textbf {\bibinfo {volume} {39}},\ \bibinfo
  {pages} {61} (\bibinfo {year} {1975})}\BibitemShut {NoStop}%
\bibitem [{\citenamefont {{Tanikawa}}\ \emph {et~al.}(2022)\citenamefont
  {{Tanikawa}}, \citenamefont {{Hattori}}, \citenamefont {{Kawanaka}},
  \citenamefont {{Kinugawa}}, \citenamefont {{Shikauchi}},\ and\ \citenamefont
  {{Tsuna}}}]{Tanikawa:2022GaiaBH}%
  \BibitemOpen
  \bibfield  {author} {\bibinfo {author} {\bibfnamefont {A.}~\bibnamefont
  {{Tanikawa}}}, \bibinfo {author} {\bibfnamefont {K.}~\bibnamefont
  {{Hattori}}}, \bibinfo {author} {\bibfnamefont {N.}~\bibnamefont
  {{Kawanaka}}}, \bibinfo {author} {\bibfnamefont {T.}~\bibnamefont
  {{Kinugawa}}}, \bibinfo {author} {\bibfnamefont {M.}~\bibnamefont
  {{Shikauchi}}},\ and\ \bibinfo {author} {\bibfnamefont {D.}~\bibnamefont
  {{Tsuna}}},\ }\bibfield  {title} {\bibinfo {title} {{Search for a Black Hole
  Binary in Gaia DR3 Astrometric Binary Stars with Spectroscopic Data}},\
  }\href {https://doi.org/10.48550/arXiv.2209.05632} {\bibfield  {journal}
  {\bibinfo  {journal} {arXiv e-prints}\ ,\ \bibinfo {eid} {arXiv:2209.05632}}
  (\bibinfo {year} {2022})},\ \Eprint {https://arxiv.org/abs/2209.05632}
  {arXiv:2209.05632 [astro-ph.SR]} \BibitemShut {NoStop}%
\bibitem [{\citenamefont {{El-Badry}}\ \emph
  {et~al.}(2023{\natexlab{a}})\citenamefont {{El-Badry}}, \citenamefont
  {{Rix}}, \citenamefont {{Quataert}}, \citenamefont {{Howard}}, \citenamefont
  {{Isaacson}}, \citenamefont {{Fuller}}, \citenamefont {{Hawkins}},
  \citenamefont {{Breivik}}, \citenamefont {{Wong}}, \citenamefont
  {{Rodriguez}}, \citenamefont {{Conroy}}, \citenamefont {{Shahaf}},
  \citenamefont {{Mazeh}}, \citenamefont {{Arenou}}, \citenamefont {{Burdge}},
  \citenamefont {{Bashi}}, \citenamefont {{Faigler}}, \citenamefont {{Weisz}},
  \citenamefont {{Seeburger}}, \citenamefont {{Almada Monter}},\ and\
  \citenamefont {{Wojno}}}]{El-Badry:2023GaiaBHone}%
  \BibitemOpen
  \bibfield  {author} {\bibinfo {author} {\bibfnamefont {K.}~\bibnamefont
  {{El-Badry}}}, \bibinfo {author} {\bibfnamefont {H.-W.}\ \bibnamefont
  {{Rix}}}, \bibinfo {author} {\bibfnamefont {E.}~\bibnamefont {{Quataert}}},
  \bibinfo {author} {\bibfnamefont {A.~W.}\ \bibnamefont {{Howard}}}, \bibinfo
  {author} {\bibfnamefont {H.}~\bibnamefont {{Isaacson}}}, \bibinfo {author}
  {\bibfnamefont {J.}~\bibnamefont {{Fuller}}}, \bibinfo {author}
  {\bibfnamefont {K.}~\bibnamefont {{Hawkins}}}, \bibinfo {author}
  {\bibfnamefont {K.}~\bibnamefont {{Breivik}}}, \bibinfo {author}
  {\bibfnamefont {K.~W.~K.}\ \bibnamefont {{Wong}}}, \bibinfo {author}
  {\bibfnamefont {A.~C.}\ \bibnamefont {{Rodriguez}}}, \bibinfo {author}
  {\bibfnamefont {C.}~\bibnamefont {{Conroy}}}, \bibinfo {author}
  {\bibfnamefont {S.}~\bibnamefont {{Shahaf}}}, \bibinfo {author}
  {\bibfnamefont {T.}~\bibnamefont {{Mazeh}}}, \bibinfo {author} {\bibfnamefont
  {F.}~\bibnamefont {{Arenou}}}, \bibinfo {author} {\bibfnamefont {K.~B.}\
  \bibnamefont {{Burdge}}}, \bibinfo {author} {\bibfnamefont {D.}~\bibnamefont
  {{Bashi}}}, \bibinfo {author} {\bibfnamefont {S.}~\bibnamefont {{Faigler}}},
  \bibinfo {author} {\bibfnamefont {D.~R.}\ \bibnamefont {{Weisz}}}, \bibinfo
  {author} {\bibfnamefont {R.}~\bibnamefont {{Seeburger}}}, \bibinfo {author}
  {\bibfnamefont {S.}~\bibnamefont {{Almada Monter}}},\ and\ \bibinfo {author}
  {\bibfnamefont {J.}~\bibnamefont {{Wojno}}},\ }\bibfield  {title} {\bibinfo
  {title} {{A Sun-like star orbiting a black hole}},\ }\href
  {https://doi.org/10.1093/mnras/stac3140} {\bibfield  {journal} {\bibinfo
  {journal} {\mnras}\ }\textbf {\bibinfo {volume} {518}},\ \bibinfo {pages}
  {1057} (\bibinfo {year} {2023}{\natexlab{a}})},\ \Eprint
  {https://arxiv.org/abs/2209.06833} {arXiv:2209.06833 [astro-ph.SR]}
  \BibitemShut {NoStop}%
\bibitem [{\citenamefont {{El-Badry}}\ \emph
  {et~al.}(2023{\natexlab{b}})\citenamefont {{El-Badry}}, \citenamefont
  {{Rix}}, \citenamefont {{Cendes}}, \citenamefont {{Rodriguez}}, \citenamefont
  {{Conroy}}, \citenamefont {{Quataert}}, \citenamefont {{Hawkins}},
  \citenamefont {{Zari}}, \citenamefont {{Hobson}}, \citenamefont {{Breivik}},
  \citenamefont {{Rau}}, \citenamefont {{Berger}}, \citenamefont {{Shahaf}},
  \citenamefont {{Seeburger}}, \citenamefont {{Burdge}}, \citenamefont
  {{Latham}}, \citenamefont {{Buchhave}}, \citenamefont {{Bieryla}},
  \citenamefont {{Bashi}}, \citenamefont {{Mazeh}},\ and\ \citenamefont
  {{Faigler}}}]{El-Badry:2023GaiaBHtwo}%
  \BibitemOpen
  \bibfield  {author} {\bibinfo {author} {\bibfnamefont {K.}~\bibnamefont
  {{El-Badry}}}, \bibinfo {author} {\bibfnamefont {H.-W.}\ \bibnamefont
  {{Rix}}}, \bibinfo {author} {\bibfnamefont {Y.}~\bibnamefont {{Cendes}}},
  \bibinfo {author} {\bibfnamefont {A.~C.}\ \bibnamefont {{Rodriguez}}},
  \bibinfo {author} {\bibfnamefont {C.}~\bibnamefont {{Conroy}}}, \bibinfo
  {author} {\bibfnamefont {E.}~\bibnamefont {{Quataert}}}, \bibinfo {author}
  {\bibfnamefont {K.}~\bibnamefont {{Hawkins}}}, \bibinfo {author}
  {\bibfnamefont {E.}~\bibnamefont {{Zari}}}, \bibinfo {author} {\bibfnamefont
  {M.}~\bibnamefont {{Hobson}}}, \bibinfo {author} {\bibfnamefont
  {K.}~\bibnamefont {{Breivik}}}, \bibinfo {author} {\bibfnamefont
  {A.}~\bibnamefont {{Rau}}}, \bibinfo {author} {\bibfnamefont
  {E.}~\bibnamefont {{Berger}}}, \bibinfo {author} {\bibfnamefont
  {S.}~\bibnamefont {{Shahaf}}}, \bibinfo {author} {\bibfnamefont
  {R.}~\bibnamefont {{Seeburger}}}, \bibinfo {author} {\bibfnamefont {K.~B.}\
  \bibnamefont {{Burdge}}}, \bibinfo {author} {\bibfnamefont {D.~W.}\
  \bibnamefont {{Latham}}}, \bibinfo {author} {\bibfnamefont {L.~A.}\
  \bibnamefont {{Buchhave}}}, \bibinfo {author} {\bibfnamefont
  {A.}~\bibnamefont {{Bieryla}}}, \bibinfo {author} {\bibfnamefont
  {D.}~\bibnamefont {{Bashi}}}, \bibinfo {author} {\bibfnamefont
  {T.}~\bibnamefont {{Mazeh}}},\ and\ \bibinfo {author} {\bibfnamefont
  {S.}~\bibnamefont {{Faigler}}},\ }\bibfield  {title} {\bibinfo {title} {{A
  red giant orbiting a black hole}},\ }\href
  {https://doi.org/10.48550/arXiv.2302.07880} {\bibfield  {journal} {\bibinfo
  {journal} {arXiv e-prints}\ ,\ \bibinfo {eid} {arXiv:2302.07880}} (\bibinfo
  {year} {2023}{\natexlab{b}})},\ \Eprint {https://arxiv.org/abs/2302.07880}
  {arXiv:2302.07880 [astro-ph.SR]} \BibitemShut {NoStop}%
\bibitem [{\citenamefont {{Remillard}}\ and\ \citenamefont
  {{McClintock}}(2006{\natexlab{a}})}]{RemillardMcClintock:2006}%
  \BibitemOpen
  \bibfield  {author} {\bibinfo {author} {\bibfnamefont {R.~A.}\ \bibnamefont
  {{Remillard}}}\ and\ \bibinfo {author} {\bibfnamefont {J.~E.}\ \bibnamefont
  {{McClintock}}},\ }\bibfield  {title} {\bibinfo {title} {{X-Ray Properties of
  Black-Hole Binaries}},\ }\href
  {https://doi.org/10.1146/annurev.astro.44.051905.092532} {\bibfield
  {journal} {\bibinfo  {journal} {\araa}\ }\textbf {\bibinfo {volume} {44}},\
  \bibinfo {pages} {49} (\bibinfo {year} {2006}{\natexlab{a}})},\ \Eprint
  {https://arxiv.org/abs/arXiv:astro-ph/0606352} {arXiv:astro-ph/0606352}
  \BibitemShut {NoStop}%
\bibitem [{\citenamefont {{Corral-Santana}}\ \emph
  {et~al.}(2016{\natexlab{a}})\citenamefont {{Corral-Santana}}, \citenamefont
  {{Casares}}, \citenamefont {{Mu{\~n}oz-Darias}}, \citenamefont {{Bauer}},
  \citenamefont {{Mart{\'\i}nez-Pais}},\ and\ \citenamefont
  {{Russell}}}]{Corral-Santana:2016}%
  \BibitemOpen
  \bibfield  {author} {\bibinfo {author} {\bibfnamefont {J.~M.}\ \bibnamefont
  {{Corral-Santana}}}, \bibinfo {author} {\bibfnamefont {J.}~\bibnamefont
  {{Casares}}}, \bibinfo {author} {\bibfnamefont {T.}~\bibnamefont
  {{Mu{\~n}oz-Darias}}}, \bibinfo {author} {\bibfnamefont {F.~E.}\ \bibnamefont
  {{Bauer}}}, \bibinfo {author} {\bibfnamefont {I.~G.}\ \bibnamefont
  {{Mart{\'\i}nez-Pais}}},\ and\ \bibinfo {author} {\bibfnamefont {D.~M.}\
  \bibnamefont {{Russell}}},\ }\bibfield  {title} {\bibinfo {title} {{BlackCAT:
  A catalogue of stellar-mass black holes in X-ray transients}},\ }\href
  {https://doi.org/10.1051/0004-6361/201527130} {\bibfield  {journal} {\bibinfo
   {journal} {\aap}\ }\textbf {\bibinfo {volume} {587}},\ \bibinfo {eid} {A61}
  (\bibinfo {year} {2016}{\natexlab{a}})},\ \Eprint
  {https://arxiv.org/abs/1510.08869} {arXiv:1510.08869 [astro-ph.HE]}
  \BibitemShut {NoStop}%
\bibitem [{\citenamefont {{Kretschmar}}\ \emph {et~al.}(2019)\citenamefont
  {{Kretschmar}}, \citenamefont {{F{\"u}rst}}, \citenamefont {{Sidoli}},
  \citenamefont {{Bozzo}}, \citenamefont {{Alfonso-Garz{\'o}n}}, \citenamefont
  {{Bodaghee}}, \citenamefont {{Chaty}}, \citenamefont {{Chernyakova}},
  \citenamefont {{Ferrigno}}, \citenamefont {{Manousakis}}, \citenamefont
  {{Negueruela}}, \citenamefont {{Postnov}}, \citenamefont {{Paizis}},
  \citenamefont {{Reig}}, \citenamefont {{Rodes-Roca}}, \citenamefont
  {{Tsygankov}}, \citenamefont {{Bird}}, \citenamefont {{Bissinger n{\'e}
  K{\"u}hnel}}, \citenamefont {{Blay}}, \citenamefont {{Caballero}},
  \citenamefont {{Coe}}, \citenamefont {{Domingo}}, \citenamefont
  {{Doroshenko}}, \citenamefont {{Ducci}}, \citenamefont {{Falanga}},
  \citenamefont {{Grebenev}}, \citenamefont {{Grinberg}}, \citenamefont
  {{Hemphill}}, \citenamefont {{Kreykenbohm}}, \citenamefont {{Kreykenbohm
  n{\'e} Fritz}}, \citenamefont {{Li}}, \citenamefont {{Lutovinov}},
  \citenamefont {{Mart{\'\i}nez-N{\'u}{\~n}ez}}, \citenamefont {{Mas-Hesse}},
  \citenamefont {{Masetti}}, \citenamefont {{McBride}}, \citenamefont
  {{Neronov}}, \citenamefont {{Pottschmidt}}, \citenamefont {{Rodriguez}},
  \citenamefont {{Romano}}, \citenamefont {{Rothschild}}, \citenamefont
  {{Santangelo}}, \citenamefont {{Sguera}}, \citenamefont {{Staubert}},
  \citenamefont {{Tomsick}}, \citenamefont {{Torrej{\'o}n}}, \citenamefont
  {{Torres}}, \citenamefont {{Walter}}, \citenamefont {{Wilms}}, \citenamefont
  {{Wilson-Hodge}},\ and\ \citenamefont {{Zhang}}}]{Kretschmar:2019}%
  \BibitemOpen
  \bibfield  {author} {\bibinfo {author} {\bibfnamefont {P.}~\bibnamefont
  {{Kretschmar}}}, \bibinfo {author} {\bibfnamefont {F.}~\bibnamefont
  {{F{\"u}rst}}}, \bibinfo {author} {\bibfnamefont {L.}~\bibnamefont
  {{Sidoli}}}, \bibinfo {author} {\bibfnamefont {E.}~\bibnamefont {{Bozzo}}},
  \bibinfo {author} {\bibfnamefont {J.}~\bibnamefont {{Alfonso-Garz{\'o}n}}},
  \bibinfo {author} {\bibfnamefont {A.}~\bibnamefont {{Bodaghee}}}, \bibinfo
  {author} {\bibfnamefont {S.}~\bibnamefont {{Chaty}}}, \bibinfo {author}
  {\bibfnamefont {M.}~\bibnamefont {{Chernyakova}}}, \bibinfo {author}
  {\bibfnamefont {C.}~\bibnamefont {{Ferrigno}}}, \bibinfo {author}
  {\bibfnamefont {A.}~\bibnamefont {{Manousakis}}}, \bibinfo {author}
  {\bibfnamefont {I.}~\bibnamefont {{Negueruela}}}, \bibinfo {author}
  {\bibfnamefont {K.}~\bibnamefont {{Postnov}}}, \bibinfo {author}
  {\bibfnamefont {A.}~\bibnamefont {{Paizis}}}, \bibinfo {author}
  {\bibfnamefont {P.}~\bibnamefont {{Reig}}}, \bibinfo {author} {\bibfnamefont
  {J.~J.}\ \bibnamefont {{Rodes-Roca}}}, \bibinfo {author} {\bibfnamefont
  {S.}~\bibnamefont {{Tsygankov}}}, \bibinfo {author} {\bibfnamefont {A.~J.}\
  \bibnamefont {{Bird}}}, \bibinfo {author} {\bibfnamefont {M.}~\bibnamefont
  {{Bissinger n{\'e} K{\"u}hnel}}}, \bibinfo {author} {\bibfnamefont
  {P.}~\bibnamefont {{Blay}}}, \bibinfo {author} {\bibfnamefont
  {I.}~\bibnamefont {{Caballero}}}, \bibinfo {author} {\bibfnamefont {M.~J.}\
  \bibnamefont {{Coe}}}, \bibinfo {author} {\bibfnamefont {A.}~\bibnamefont
  {{Domingo}}}, \bibinfo {author} {\bibfnamefont {V.}~\bibnamefont
  {{Doroshenko}}}, \bibinfo {author} {\bibfnamefont {L.}~\bibnamefont
  {{Ducci}}}, \bibinfo {author} {\bibfnamefont {M.}~\bibnamefont {{Falanga}}},
  \bibinfo {author} {\bibfnamefont {S.~A.}\ \bibnamefont {{Grebenev}}},
  \bibinfo {author} {\bibfnamefont {V.}~\bibnamefont {{Grinberg}}}, \bibinfo
  {author} {\bibfnamefont {P.}~\bibnamefont {{Hemphill}}}, \bibinfo {author}
  {\bibfnamefont {I.}~\bibnamefont {{Kreykenbohm}}}, \bibinfo {author}
  {\bibfnamefont {S.}~\bibnamefont {{Kreykenbohm n{\'e} Fritz}}}, \bibinfo
  {author} {\bibfnamefont {J.}~\bibnamefont {{Li}}}, \bibinfo {author}
  {\bibfnamefont {A.~A.}\ \bibnamefont {{Lutovinov}}}, \bibinfo {author}
  {\bibfnamefont {S.}~\bibnamefont {{Mart{\'\i}nez-N{\'u}{\~n}ez}}}, \bibinfo
  {author} {\bibfnamefont {J.~M.}\ \bibnamefont {{Mas-Hesse}}}, \bibinfo
  {author} {\bibfnamefont {N.}~\bibnamefont {{Masetti}}}, \bibinfo {author}
  {\bibfnamefont {V.~A.}\ \bibnamefont {{McBride}}}, \bibinfo {author}
  {\bibfnamefont {A.}~\bibnamefont {{Neronov}}}, \bibinfo {author}
  {\bibfnamefont {K.}~\bibnamefont {{Pottschmidt}}}, \bibinfo {author}
  {\bibfnamefont {J.}~\bibnamefont {{Rodriguez}}}, \bibinfo {author}
  {\bibfnamefont {P.}~\bibnamefont {{Romano}}}, \bibinfo {author}
  {\bibfnamefont {R.~E.}\ \bibnamefont {{Rothschild}}}, \bibinfo {author}
  {\bibfnamefont {A.}~\bibnamefont {{Santangelo}}}, \bibinfo {author}
  {\bibfnamefont {V.}~\bibnamefont {{Sguera}}}, \bibinfo {author}
  {\bibfnamefont {R.}~\bibnamefont {{Staubert}}}, \bibinfo {author}
  {\bibfnamefont {J.~A.}\ \bibnamefont {{Tomsick}}}, \bibinfo {author}
  {\bibfnamefont {J.~M.}\ \bibnamefont {{Torrej{\'o}n}}}, \bibinfo {author}
  {\bibfnamefont {D.~F.}\ \bibnamefont {{Torres}}}, \bibinfo {author}
  {\bibfnamefont {R.}~\bibnamefont {{Walter}}}, \bibinfo {author}
  {\bibfnamefont {J.}~\bibnamefont {{Wilms}}}, \bibinfo {author} {\bibfnamefont
  {C.~A.}\ \bibnamefont {{Wilson-Hodge}}},\ and\ \bibinfo {author}
  {\bibfnamefont {S.}~\bibnamefont {{Zhang}}},\ }\bibfield  {title} {\bibinfo
  {title} {{Advances in Understanding High-Mass X-ray Binaries with INTEGRALand
  Future Directions}},\ }\href {https://doi.org/10.1016/j.newar.2020.101546}
  {\bibfield  {journal} {\bibinfo  {journal} {\nar}\ }\textbf {\bibinfo
  {volume} {86}},\ \bibinfo {eid} {101546} (\bibinfo {year} {2019})},\ \Eprint
  {https://arxiv.org/abs/2009.03244} {arXiv:2009.03244 [astro-ph.HE]}
  \BibitemShut {NoStop}%
\bibitem [{\citenamefont {{Paczy{\'n}ski}}(1976)}]{Paczynski:1976}%
  \BibitemOpen
  \bibfield  {author} {\bibinfo {author} {\bibfnamefont {B.}~\bibnamefont
  {{Paczy{\'n}ski}}},\ }\bibfield  {title} {\bibinfo {title} {{Common Envelope
  Binaries}},\ }in\ \href@noop {} {\emph {\bibinfo {booktitle} {Structure and
  Evolution of Close Binary Systems}}},\ \bibinfo {series} {IAU Symposium},
  Vol.~\bibinfo {volume} {73},\ \bibinfo {editor} {edited by\ \bibinfo {editor}
  {\bibfnamefont {P.}~\bibnamefont {{Eggleton}}}, \bibinfo {editor}
  {\bibfnamefont {S.}~\bibnamefont {{Mitton}}},\ and\ \bibinfo {editor}
  {\bibfnamefont {J.}~\bibnamefont {{Whelan}}}}\ (\bibinfo {year} {1976})\
  p.~\bibinfo {pages} {75}\BibitemShut {NoStop}%
\bibitem [{\citenamefont {Ivanova}\ \emph {et~al.}(2013)\citenamefont
  {Ivanova}, \citenamefont {Justham}, \citenamefont {Chen}, \citenamefont
  {De~Marco}, \citenamefont {Fryer}, \citenamefont {Gaburov}, \citenamefont
  {Ge}, \citenamefont {Glebbeek}, \citenamefont {Han}, \citenamefont {Li} \emph
  {et~al.}}]{ivanova2013common}%
  \BibitemOpen
  \bibfield  {author} {\bibinfo {author} {\bibfnamefont {N.}~\bibnamefont
  {Ivanova}}, \bibinfo {author} {\bibfnamefont {S.}~\bibnamefont {Justham}},
  \bibinfo {author} {\bibfnamefont {X.}~\bibnamefont {Chen}}, \bibinfo {author}
  {\bibfnamefont {O.}~\bibnamefont {De~Marco}}, \bibinfo {author}
  {\bibfnamefont {C.}~\bibnamefont {Fryer}}, \bibinfo {author} {\bibfnamefont
  {E.}~\bibnamefont {Gaburov}}, \bibinfo {author} {\bibfnamefont
  {H.}~\bibnamefont {Ge}}, \bibinfo {author} {\bibfnamefont {E.}~\bibnamefont
  {Glebbeek}}, \bibinfo {author} {\bibfnamefont {Z.}~\bibnamefont {Han}},
  \bibinfo {author} {\bibfnamefont {X.-D.}\ \bibnamefont {Li}}, \emph
  {et~al.},\ }\bibfield  {title} {\bibinfo {title} {Common envelope evolution:
  where we stand and how we can move forward},\ }\href@noop {} {\bibfield
  {journal} {\bibinfo  {journal} {The Astronomy \& Astrophysics Review}\
  }\textbf {\bibinfo {volume} {21}},\ \bibinfo {pages} {59} (\bibinfo {year}
  {2013})}\BibitemShut {NoStop}%
\bibitem [{\citenamefont {{Belczynski}}\ \emph {et~al.}(2013)\citenamefont
  {{Belczynski}}, \citenamefont {{Bulik}}, \citenamefont {{Mandel}},
  \citenamefont {{Sathyaprakash}}, \citenamefont {{Zdziarski}},\ and\
  \citenamefont {{Miko{\l}ajewska}}}]{2013ApJ...764...96B}%
  \BibitemOpen
  \bibfield  {author} {\bibinfo {author} {\bibfnamefont {K.}~\bibnamefont
  {{Belczynski}}}, \bibinfo {author} {\bibfnamefont {T.}~\bibnamefont
  {{Bulik}}}, \bibinfo {author} {\bibfnamefont {I.}~\bibnamefont {{Mandel}}},
  \bibinfo {author} {\bibfnamefont {B.~S.}\ \bibnamefont {{Sathyaprakash}}},
  \bibinfo {author} {\bibfnamefont {A.~A.}\ \bibnamefont {{Zdziarski}}},\ and\
  \bibinfo {author} {\bibfnamefont {J.}~\bibnamefont {{Miko{\l}ajewska}}},\
  }\bibfield  {title} {\bibinfo {title} {{Cyg X-3: A Galactic Double Black Hole
  or Black-hole-Neutron-star Progenitor}},\ }\href
  {https://doi.org/10.1088/0004-637X/764/1/96} {\bibfield  {journal} {\bibinfo
  {journal} {\apj}\ }\textbf {\bibinfo {volume} {764}},\ \bibinfo {eid} {96}
  (\bibinfo {year} {2013})},\ \Eprint {https://arxiv.org/abs/1209.2658}
  {arXiv:1209.2658 [astro-ph.HE]} \BibitemShut {NoStop}%
\bibitem [{\citenamefont {{Zdziarski}}\ \emph {et~al.}(2013)\citenamefont
  {{Zdziarski}}, \citenamefont {{Mikolajewska}},\ and\ \citenamefont
  {{Belczynski}}}]{2013MNRAS.429L.104Z}%
  \BibitemOpen
  \bibfield  {author} {\bibinfo {author} {\bibfnamefont {A.~A.}\ \bibnamefont
  {{Zdziarski}}}, \bibinfo {author} {\bibfnamefont {J.}~\bibnamefont
  {{Mikolajewska}}},\ and\ \bibinfo {author} {\bibfnamefont {K.}~\bibnamefont
  {{Belczynski}}},\ }\bibfield  {title} {\bibinfo {title} {{Cyg X-3: a low-mass
  black hole or a neutron star.}},\ }\href
  {https://doi.org/10.1093/mnrasl/sls035} {\bibfield  {journal} {\bibinfo
  {journal} {\mnras}\ }\textbf {\bibinfo {volume} {429}},\ \bibinfo {pages}
  {L104} (\bibinfo {year} {2013})},\ \Eprint {https://arxiv.org/abs/1208.5455}
  {arXiv:1208.5455 [astro-ph.HE]} \BibitemShut {NoStop}%
\bibitem [{\citenamefont {{Chevalier}}(2012)}]{2012ApJ...752L...2C}%
  \BibitemOpen
  \bibfield  {author} {\bibinfo {author} {\bibfnamefont {R.~A.}\ \bibnamefont
  {{Chevalier}}},\ }\bibfield  {title} {\bibinfo {title} {{Common Envelope
  Evolution Leading to Supernovae with Dense Interaction}},\ }\href
  {https://doi.org/10.1088/2041-8205/752/1/L2} {\bibfield  {journal} {\bibinfo
  {journal} {\apjl}\ }\textbf {\bibinfo {volume} {752}},\ \bibinfo {eid} {L2}
  (\bibinfo {year} {2012})},\ \Eprint {https://arxiv.org/abs/1204.3300}
  {arXiv:1204.3300 [astro-ph.HE]} \BibitemShut {NoStop}%
\bibitem [{\citenamefont {{Pejcha}}\ \emph
  {et~al.}(2016{\natexlab{b}})\citenamefont {{Pejcha}}, \citenamefont
  {{Metzger}},\ and\ \citenamefont {{Tomida}}}]{2016MNRAS.455.4351P}%
  \BibitemOpen
  \bibfield  {author} {\bibinfo {author} {\bibfnamefont {O.}~\bibnamefont
  {{Pejcha}}}, \bibinfo {author} {\bibfnamefont {B.~D.}\ \bibnamefont
  {{Metzger}}},\ and\ \bibinfo {author} {\bibfnamefont {K.}~\bibnamefont
  {{Tomida}}},\ }\bibfield  {title} {\bibinfo {title} {{Cool and luminous
  transients from mass-losing binary stars}},\ }\href
  {https://doi.org/10.1093/mnras/stv2592} {\bibfield  {journal} {\bibinfo
  {journal} {\mnras}\ }\textbf {\bibinfo {volume} {455}},\ \bibinfo {pages}
  {4351} (\bibinfo {year} {2016}{\natexlab{b}})},\ \Eprint
  {https://arxiv.org/abs/1509.02531} {arXiv:1509.02531 [astro-ph.SR]}
  \BibitemShut {NoStop}%
\bibitem [{\citenamefont {{Schr{\o}der}}\ \emph {et~al.}(2020)\citenamefont
  {{Schr{\o}der}}, \citenamefont {{MacLeod}}, \citenamefont {{Loeb}},
  \citenamefont {{Vigna-G{\'o}mez}},\ and\ \citenamefont
  {{Mandel}}}]{2020ApJ...892...13S}%
  \BibitemOpen
  \bibfield  {author} {\bibinfo {author} {\bibfnamefont {S.~L.}\ \bibnamefont
  {{Schr{\o}der}}}, \bibinfo {author} {\bibfnamefont {M.}~\bibnamefont
  {{MacLeod}}}, \bibinfo {author} {\bibfnamefont {A.}~\bibnamefont {{Loeb}}},
  \bibinfo {author} {\bibfnamefont {A.}~\bibnamefont {{Vigna-G{\'o}mez}}},\
  and\ \bibinfo {author} {\bibfnamefont {I.}~\bibnamefont {{Mandel}}},\
  }\bibfield  {title} {\bibinfo {title} {{Explosions Driven by the Coalescence
  of a Compact Object with the Core of a Massive-star Companion inside a Common
  Envelope: Circumstellar Properties, Light Curves, and Population
  Statistics}},\ }\href {https://doi.org/10.3847/1538-4357/ab7014} {\bibfield
  {journal} {\bibinfo  {journal} {\apj}\ }\textbf {\bibinfo {volume} {892}},\
  \bibinfo {eid} {13} (\bibinfo {year} {2020})},\ \Eprint
  {https://arxiv.org/abs/1906.04189} {arXiv:1906.04189 [astro-ph.HE]}
  \BibitemShut {NoStop}%
\bibitem [{\citenamefont {{Woods}}\ and\ \citenamefont
  {{Ivanova}}(2011{\natexlab{b}})}]{WoodsIvanova:2011}%
  \BibitemOpen
  \bibfield  {author} {\bibinfo {author} {\bibfnamefont {T.~E.}\ \bibnamefont
  {{Woods}}}\ and\ \bibinfo {author} {\bibfnamefont {N.}~\bibnamefont
  {{Ivanova}}},\ }\bibfield  {title} {\bibinfo {title} {{Can We Trust Models
  for Adiabatic Mass Loss?}},\ }\href
  {https://doi.org/10.1088/2041-8205/739/2/L48} {\bibfield  {journal} {\bibinfo
   {journal} {\apjl}\ }\textbf {\bibinfo {volume} {739}},\ \bibinfo {eid} {L48}
  (\bibinfo {year} {2011}{\natexlab{b}})},\ \Eprint
  {https://arxiv.org/abs/1108.2752} {arXiv:1108.2752 [astro-ph.SR]}
  \BibitemShut {NoStop}%
\bibitem [{\citenamefont {{Ge}}\ \emph
  {et~al.}(2015{\natexlab{b}})\citenamefont {{Ge}}, \citenamefont {{Webbink}},
  \citenamefont {{Chen}},\ and\ \citenamefont {{Han}}}]{Ge:2015}%
  \BibitemOpen
  \bibfield  {author} {\bibinfo {author} {\bibfnamefont {H.}~\bibnamefont
  {{Ge}}}, \bibinfo {author} {\bibfnamefont {R.~F.}\ \bibnamefont {{Webbink}}},
  \bibinfo {author} {\bibfnamefont {X.}~\bibnamefont {{Chen}}},\ and\ \bibinfo
  {author} {\bibfnamefont {Z.}~\bibnamefont {{Han}}},\ }\bibfield  {title}
  {\bibinfo {title} {{Adiabatic Mass Loss in Binary Stars. II. From Zero-age
  Main Sequence to the Base of the Giant Branch}},\ }\href
  {https://doi.org/10.1088/0004-637X/812/1/40} {\bibfield  {journal} {\bibinfo
  {journal} {\apj}\ }\textbf {\bibinfo {volume} {812}},\ \bibinfo {eid} {40}
  (\bibinfo {year} {2015}{\natexlab{b}})},\ \Eprint
  {https://arxiv.org/abs/1507.04843} {arXiv:1507.04843 [astro-ph.SR]}
  \BibitemShut {NoStop}%
\bibitem [{\citenamefont {{Inayoshi}}\ \emph {et~al.}(2017)\citenamefont
  {{Inayoshi}}, \citenamefont {{Hirai}}, \citenamefont {{Kinugawa}},\ and\
  \citenamefont {{Hotokezaka}}}]{Inayoshi:2017}%
  \BibitemOpen
  \bibfield  {author} {\bibinfo {author} {\bibfnamefont {K.}~\bibnamefont
  {{Inayoshi}}}, \bibinfo {author} {\bibfnamefont {R.}~\bibnamefont {{Hirai}}},
  \bibinfo {author} {\bibfnamefont {T.}~\bibnamefont {{Kinugawa}}},\ and\
  \bibinfo {author} {\bibfnamefont {K.}~\bibnamefont {{Hotokezaka}}},\
  }\bibfield  {title} {\bibinfo {title} {{Formation pathway of Population III
  coalescing binary black holes through stable mass transfer}},\ }\href
  {https://doi.org/10.1093/mnras/stx757} {\bibfield  {journal} {\bibinfo
  {journal} {\mnras}\ }\textbf {\bibinfo {volume} {468}},\ \bibinfo {pages}
  {5020} (\bibinfo {year} {2017})},\ \Eprint {https://arxiv.org/abs/1701.04823}
  {arXiv:1701.04823 [astro-ph.HE]} \BibitemShut {NoStop}%
\bibitem [{\citenamefont {{Pavlovskii}}\ \emph
  {et~al.}(2017{\natexlab{b}})\citenamefont {{Pavlovskii}}, \citenamefont
  {{Ivanova}}, \citenamefont {{Belczynski}},\ and\ \citenamefont
  {{Van}}}]{Pavlovskii:2017}%
  \BibitemOpen
  \bibfield  {author} {\bibinfo {author} {\bibfnamefont {K.}~\bibnamefont
  {{Pavlovskii}}}, \bibinfo {author} {\bibfnamefont {N.}~\bibnamefont
  {{Ivanova}}}, \bibinfo {author} {\bibfnamefont {K.}~\bibnamefont
  {{Belczynski}}},\ and\ \bibinfo {author} {\bibfnamefont {K.~X.}\ \bibnamefont
  {{Van}}},\ }\bibfield  {title} {\bibinfo {title} {{Stability of mass transfer
  from massive giants: double black hole binary formation and ultraluminous
  X-ray sources}},\ }\href {https://doi.org/10.1093/mnras/stw2786} {\bibfield
  {journal} {\bibinfo  {journal} {\mnras}\ }\textbf {\bibinfo {volume} {465}},\
  \bibinfo {pages} {2092} (\bibinfo {year} {2017}{\natexlab{b}})},\ \Eprint
  {https://arxiv.org/abs/1606.04921} {arXiv:1606.04921 [astro-ph.HE]}
  \BibitemShut {NoStop}%
\bibitem [{\citenamefont {{van den Heuvel}}\ \emph
  {et~al.}(2017{\natexlab{b}})\citenamefont {{van den Heuvel}}, \citenamefont
  {{Portegies Zwart}},\ and\ \citenamefont {{de Mink}}}]{vandenHeuvel:2017}%
  \BibitemOpen
  \bibfield  {author} {\bibinfo {author} {\bibfnamefont {E.~P.~J.}\
  \bibnamefont {{van den Heuvel}}}, \bibinfo {author} {\bibfnamefont {S.~F.}\
  \bibnamefont {{Portegies Zwart}}},\ and\ \bibinfo {author} {\bibfnamefont
  {S.~E.}\ \bibnamefont {{de Mink}}},\ }\bibfield  {title} {\bibinfo {title}
  {{Forming short-period Wolf-Rayet X-ray binaries and double black holes
  through stable mass transfer}},\ }\href
  {https://doi.org/10.1093/mnras/stx1430} {\bibfield  {journal} {\bibinfo
  {journal} {\mnras}\ }\textbf {\bibinfo {volume} {471}},\ \bibinfo {pages}
  {4256} (\bibinfo {year} {2017}{\natexlab{b}})},\ \Eprint
  {https://arxiv.org/abs/1701.02355} {arXiv:1701.02355 [astro-ph.SR]}
  \BibitemShut {NoStop}%
\bibitem [{\citenamefont {{Gallegos-Garcia}}\ \emph
  {et~al.}(2021{\natexlab{b}})\citenamefont {{Gallegos-Garcia}}, \citenamefont
  {{Berry}}, \citenamefont {{Marchant}},\ and\ \citenamefont
  {{Kalogera}}}]{Gallegos-Garcia:2021hti}%
  \BibitemOpen
  \bibfield  {author} {\bibinfo {author} {\bibfnamefont {M.}~\bibnamefont
  {{Gallegos-Garcia}}}, \bibinfo {author} {\bibfnamefont {C.~P.~L.}\
  \bibnamefont {{Berry}}}, \bibinfo {author} {\bibfnamefont {P.}~\bibnamefont
  {{Marchant}}},\ and\ \bibinfo {author} {\bibfnamefont {V.}~\bibnamefont
  {{Kalogera}}},\ }\bibfield  {title} {\bibinfo {title} {{Binary Black Hole
  Formation with Detailed Modeling: Stable Mass Transfer Leads to Lower Merger
  Rates}},\ }\href {https://doi.org/10.3847/1538-4357/ac2610} {\bibfield
  {journal} {\bibinfo  {journal} {\apj}\ }\textbf {\bibinfo {volume} {922}},\
  \bibinfo {eid} {110} (\bibinfo {year} {2021}{\natexlab{b}})},\ \Eprint
  {https://arxiv.org/abs/2107.05702} {arXiv:2107.05702 [astro-ph.HE]}
  \BibitemShut {NoStop}%
\bibitem [{\citenamefont {{Marchant}}\ \emph
  {et~al.}(2021{\natexlab{b}})\citenamefont {{Marchant}}, \citenamefont
  {{Pappas}}, \citenamefont {{Gallegos-Garcia}}, \citenamefont {{Berry}},
  \citenamefont {{Taam}}, \citenamefont {{Kalogera}},\ and\ \citenamefont
  {{Podsiadlowski}}}]{Marchant:2021}%
  \BibitemOpen
  \bibfield  {author} {\bibinfo {author} {\bibfnamefont {P.}~\bibnamefont
  {{Marchant}}}, \bibinfo {author} {\bibfnamefont {K.~M.~W.}\ \bibnamefont
  {{Pappas}}}, \bibinfo {author} {\bibfnamefont {M.}~\bibnamefont
  {{Gallegos-Garcia}}}, \bibinfo {author} {\bibfnamefont {C.~P.~L.}\
  \bibnamefont {{Berry}}}, \bibinfo {author} {\bibfnamefont {R.~E.}\
  \bibnamefont {{Taam}}}, \bibinfo {author} {\bibfnamefont {V.}~\bibnamefont
  {{Kalogera}}},\ and\ \bibinfo {author} {\bibfnamefont {P.}~\bibnamefont
  {{Podsiadlowski}}},\ }\bibfield  {title} {\bibinfo {title} {{The role of mass
  transfer and common envelope evolution in the formation of merging binary
  black holes}},\ }\href {https://doi.org/10.1051/0004-6361/202039992}
  {\bibfield  {journal} {\bibinfo  {journal} {\aap}\ }\textbf {\bibinfo
  {volume} {650}},\ \bibinfo {eid} {A107} (\bibinfo {year}
  {2021}{\natexlab{b}})},\ \Eprint {https://arxiv.org/abs/2103.09243}
  {arXiv:2103.09243 [astro-ph.SR]} \BibitemShut {NoStop}%
\bibitem [{\citenamefont {{Neijssel}}\ \emph {et~al.}(2019)\citenamefont
  {{Neijssel}}, \citenamefont {{Vigna-G{\'o}mez}}, \citenamefont {{Stevenson}},
  \citenamefont {{Barrett}}, \citenamefont {{Gaebel}}, \citenamefont
  {{Broekgaarden}}, \citenamefont {{de Mink}}, \citenamefont {{Sz{\'e}csi}},
  \citenamefont {{Vinciguerra}},\ and\ \citenamefont
  {{Mandel}}}]{Neijssel:2019}%
  \BibitemOpen
  \bibfield  {author} {\bibinfo {author} {\bibfnamefont {C.~J.}\ \bibnamefont
  {{Neijssel}}}, \bibinfo {author} {\bibfnamefont {A.}~\bibnamefont
  {{Vigna-G{\'o}mez}}}, \bibinfo {author} {\bibfnamefont {S.}~\bibnamefont
  {{Stevenson}}}, \bibinfo {author} {\bibfnamefont {J.~W.}\ \bibnamefont
  {{Barrett}}}, \bibinfo {author} {\bibfnamefont {S.~M.}\ \bibnamefont
  {{Gaebel}}}, \bibinfo {author} {\bibfnamefont {F.~S.}\ \bibnamefont
  {{Broekgaarden}}}, \bibinfo {author} {\bibfnamefont {S.~E.}\ \bibnamefont
  {{de Mink}}}, \bibinfo {author} {\bibfnamefont {D.}~\bibnamefont
  {{Sz{\'e}csi}}}, \bibinfo {author} {\bibfnamefont {S.}~\bibnamefont
  {{Vinciguerra}}},\ and\ \bibinfo {author} {\bibfnamefont {I.}~\bibnamefont
  {{Mandel}}},\ }\bibfield  {title} {\bibinfo {title} {{The effect of the
  metallicity-specific star formation history on double compact object
  mergers}},\ }\href {https://doi.org/10.1093/mnras/stz2840} {\bibfield
  {journal} {\bibinfo  {journal} {\mnras}\ }\textbf {\bibinfo {volume} {490}},\
  \bibinfo {pages} {3740} (\bibinfo {year} {2019})},\ \Eprint
  {https://arxiv.org/abs/1906.08136} {arXiv:1906.08136 [astro-ph.SR]}
  \BibitemShut {NoStop}%
\bibitem [{\citenamefont {{Olejak}}\ \emph
  {et~al.}(2021{\natexlab{a}})\citenamefont {{Olejak}}, \citenamefont
  {{Belczynski}},\ and\ \citenamefont {{Ivanova}}}]{Olejak:2021CE}%
  \BibitemOpen
  \bibfield  {author} {\bibinfo {author} {\bibfnamefont {A.}~\bibnamefont
  {{Olejak}}}, \bibinfo {author} {\bibfnamefont {K.}~\bibnamefont
  {{Belczynski}}},\ and\ \bibinfo {author} {\bibfnamefont {N.}~\bibnamefont
  {{Ivanova}}},\ }\bibfield  {title} {\bibinfo {title} {{Impact of common
  envelope development criteria on the formation of LIGO/Virgo sources}},\
  }\href {https://doi.org/10.1051/0004-6361/202140520} {\bibfield  {journal}
  {\bibinfo  {journal} {\aap}\ }\textbf {\bibinfo {volume} {651}},\ \bibinfo
  {eid} {A100} (\bibinfo {year} {2021}{\natexlab{a}})},\ \Eprint
  {https://arxiv.org/abs/2102.05649} {arXiv:2102.05649 [astro-ph.HE]}
  \BibitemShut {NoStop}%
\bibitem [{\citenamefont {{Shao}}\ and\ \citenamefont
  {{Li}}(2021)}]{Shao:2021}%
  \BibitemOpen
  \bibfield  {author} {\bibinfo {author} {\bibfnamefont {Y.}~\bibnamefont
  {{Shao}}}\ and\ \bibinfo {author} {\bibfnamefont {X.-D.}\ \bibnamefont
  {{Li}}},\ }\bibfield  {title} {\bibinfo {title} {{Population Synthesis of
  Black Hole Binaries with Compact Star Companions}},\ }\href@noop {}
  {\bibfield  {journal} {\bibinfo  {journal} {arXiv e-prints}\ } (\bibinfo
  {year} {2021})},\ \Eprint {https://arxiv.org/abs/2107.03565}
  {arXiv:2107.03565 [astro-ph.HE]} \BibitemShut {NoStop}%
\bibitem [{\citenamefont {{Klencki}}\ \emph
  {et~al.}(2021{\natexlab{b}})\citenamefont {{Klencki}}, \citenamefont
  {{Nelemans}}, \citenamefont {{Istrate}},\ and\ \citenamefont
  {{Chruslinska}}}]{Klencki:2020convective}%
  \BibitemOpen
  \bibfield  {author} {\bibinfo {author} {\bibfnamefont {J.}~\bibnamefont
  {{Klencki}}}, \bibinfo {author} {\bibfnamefont {G.}~\bibnamefont
  {{Nelemans}}}, \bibinfo {author} {\bibfnamefont {A.~G.}\ \bibnamefont
  {{Istrate}}},\ and\ \bibinfo {author} {\bibfnamefont {M.}~\bibnamefont
  {{Chruslinska}}},\ }\bibfield  {title} {\bibinfo {title} {{It has to be cool:
  Supergiant progenitors of binary black hole mergers from common-envelope
  evolution}},\ }\href {https://doi.org/10.1051/0004-6361/202038707} {\bibfield
   {journal} {\bibinfo  {journal} {\aap}\ }\textbf {\bibinfo {volume} {645}},\
  \bibinfo {eid} {A54} (\bibinfo {year} {2021}{\natexlab{b}})},\ \Eprint
  {https://arxiv.org/abs/2006.11286} {arXiv:2006.11286 [astro-ph.SR]}
  \BibitemShut {NoStop}%
\bibitem [{\citenamefont {{Klencki}}\ \emph
  {et~al.}(2021{\natexlab{c}})\citenamefont {{Klencki}}, \citenamefont
  {{Nelemans}}, \citenamefont {{Istrate}},\ and\ \citenamefont
  {{Chruslinska}}}]{Klencki:2021}%
  \BibitemOpen
  \bibfield  {author} {\bibinfo {author} {\bibfnamefont {J.}~\bibnamefont
  {{Klencki}}}, \bibinfo {author} {\bibfnamefont {G.}~\bibnamefont
  {{Nelemans}}}, \bibinfo {author} {\bibfnamefont {A.~G.}\ \bibnamefont
  {{Istrate}}},\ and\ \bibinfo {author} {\bibfnamefont {M.}~\bibnamefont
  {{Chruslinska}}},\ }\bibfield  {title} {\bibinfo {title} {{It has to be cool:
  Supergiant progenitors of binary black hole mergers from common-envelope
  evolution}},\ }\href {https://doi.org/10.1051/0004-6361/202038707} {\bibfield
   {journal} {\bibinfo  {journal} {\aap}\ }\textbf {\bibinfo {volume} {645}},\
  \bibinfo {eid} {A54} (\bibinfo {year} {2021}{\natexlab{c}})},\ \Eprint
  {https://arxiv.org/abs/2006.11286} {arXiv:2006.11286 [astro-ph.SR]}
  \BibitemShut {NoStop}%
\bibitem [{\citenamefont {{Dewi}}\ and\ \citenamefont
  {{Pols}}(2003)}]{2003MNRAS.344..629D}%
  \BibitemOpen
  \bibfield  {author} {\bibinfo {author} {\bibfnamefont {J.~D.~M.}\
  \bibnamefont {{Dewi}}}\ and\ \bibinfo {author} {\bibfnamefont {O.~R.}\
  \bibnamefont {{Pols}}},\ }\bibfield  {title} {\bibinfo {title} {{The late
  stages of evolution of helium star-neutron star binaries and the formation of
  double neutron star systems}},\ }\href
  {https://doi.org/10.1046/j.1365-8711.2003.06844.x} {\bibfield  {journal}
  {\bibinfo  {journal} {\mnras}\ }\textbf {\bibinfo {volume} {344}},\ \bibinfo
  {pages} {629} (\bibinfo {year} {2003})},\ \Eprint
  {https://arxiv.org/abs/astro-ph/0306066} {arXiv:astro-ph/0306066 [astro-ph]}
  \BibitemShut {NoStop}%
\bibitem [{\citenamefont {{Tauris}}\ \emph {et~al.}(2015)\citenamefont
  {{Tauris}}, \citenamefont {{Langer}},\ and\ \citenamefont
  {{Podsiadlowski}}}]{Tauris:2015}%
  \BibitemOpen
  \bibfield  {author} {\bibinfo {author} {\bibfnamefont {T.~M.}\ \bibnamefont
  {{Tauris}}}, \bibinfo {author} {\bibfnamefont {N.}~\bibnamefont {{Langer}}},\
  and\ \bibinfo {author} {\bibfnamefont {P.}~\bibnamefont {{Podsiadlowski}}},\
  }\bibfield  {title} {\bibinfo {title} {{Ultra-stripped supernovae:
  progenitors and fate}},\ }\href {https://doi.org/10.1093/mnras/stv990}
  {\bibfield  {journal} {\bibinfo  {journal} {\mnras}\ }\textbf {\bibinfo
  {volume} {451}},\ \bibinfo {pages} {2123} (\bibinfo {year} {2015})},\ \Eprint
  {https://arxiv.org/abs/1505.00270} {arXiv:1505.00270 [astro-ph.SR]}
  \BibitemShut {NoStop}%
\bibitem [{\citenamefont {Tauris}\ \emph {et~al.}(2017)\citenamefont {Tauris},
  \citenamefont {Kramer}, \citenamefont {Freire}, \citenamefont {Wex},
  \citenamefont {Janka}, \citenamefont {Langer}, \citenamefont {Podsiadlowski},
  \citenamefont {Bozzo}, \citenamefont {Chaty}, \citenamefont {Kruckow} \emph
  {et~al.}}]{tauris2017formation}%
  \BibitemOpen
  \bibfield  {author} {\bibinfo {author} {\bibfnamefont {T.}~\bibnamefont
  {Tauris}}, \bibinfo {author} {\bibfnamefont {M.}~\bibnamefont {Kramer}},
  \bibinfo {author} {\bibfnamefont {P.}~\bibnamefont {Freire}}, \bibinfo
  {author} {\bibfnamefont {N.}~\bibnamefont {Wex}}, \bibinfo {author}
  {\bibfnamefont {H.-T.}\ \bibnamefont {Janka}}, \bibinfo {author}
  {\bibfnamefont {N.}~\bibnamefont {Langer}}, \bibinfo {author} {\bibfnamefont
  {P.}~\bibnamefont {Podsiadlowski}}, \bibinfo {author} {\bibfnamefont
  {E.}~\bibnamefont {Bozzo}}, \bibinfo {author} {\bibfnamefont
  {S.}~\bibnamefont {Chaty}}, \bibinfo {author} {\bibfnamefont
  {M.}~\bibnamefont {Kruckow}}, \emph {et~al.},\ }\bibfield  {title} {\bibinfo
  {title} {Formation of double neutron star systems},\ }\href@noop {}
  {\bibfield  {journal} {\bibinfo  {journal} {The Astrophysical Journal}\
  }\textbf {\bibinfo {volume} {846}},\ \bibinfo {pages} {170} (\bibinfo {year}
  {2017})}\BibitemShut {NoStop}%
\bibitem [{\citenamefont {{Peters}}(1964{\natexlab{c}})}]{Peters:1964}%
  \BibitemOpen
  \bibfield  {author} {\bibinfo {author} {\bibfnamefont {P.~C.}\ \bibnamefont
  {{Peters}}},\ }\bibfield  {title} {\bibinfo {title} {{Gravitational Radiation
  and the Motion of Two Point Masses}},\ }\href
  {https://doi.org/10.1103/PhysRev.136.B1224} {\bibfield  {journal} {\bibinfo
  {journal} {Physical Review}\ }\textbf {\bibinfo {volume} {136}},\ \bibinfo
  {pages} {1224} (\bibinfo {year} {1964}{\natexlab{c}})}\BibitemShut {NoStop}%
\bibitem [{\citenamefont {{{\"O}pik}}(1924)}]{Opik:1924}%
  \BibitemOpen
  \bibfield  {author} {\bibinfo {author} {\bibfnamefont {E.}~\bibnamefont
  {{{\"O}pik}}},\ }\bibfield  {title} {\bibinfo {title} {{Statistical Studies
  of Double Stars: On the Distribution of Relative Luminosities and Distances
  of Double Stars in the Harvard Revised Photometry North of Declination -31
  degrees}},\ }\href {http://muuseum.to.ee/Main/HTML/vol__xxv_-_1924.htm}
  {\bibfield  {journal} {\bibinfo  {journal} {Publ. Obs. Astron. Univ. Tartu}\
  }\textbf {\bibinfo {volume} {25}},\ \bibinfo {pages} {1} (\bibinfo {year}
  {1924})}\BibitemShut {NoStop}%
\bibitem [{\citenamefont {{Sana}}\ \emph
  {et~al.}(2012{\natexlab{b}})\citenamefont {{Sana}}, \citenamefont {{de
  Mink}}, \citenamefont {{de Koter}}, \citenamefont {{Langer}}, \citenamefont
  {{Evans}}, \citenamefont {{Gieles}}, \citenamefont {{Gosset}}, \citenamefont
  {{Izzard}}, \citenamefont {{Le Bouquin}},\ and\ \citenamefont
  {{Schneider}}}]{Sana:2012}%
  \BibitemOpen
  \bibfield  {author} {\bibinfo {author} {\bibfnamefont {H.}~\bibnamefont
  {{Sana}}}, \bibinfo {author} {\bibfnamefont {S.~E.}\ \bibnamefont {{de
  Mink}}}, \bibinfo {author} {\bibfnamefont {A.}~\bibnamefont {{de Koter}}},
  \bibinfo {author} {\bibfnamefont {N.}~\bibnamefont {{Langer}}}, \bibinfo
  {author} {\bibfnamefont {C.~J.}\ \bibnamefont {{Evans}}}, \bibinfo {author}
  {\bibfnamefont {M.}~\bibnamefont {{Gieles}}}, \bibinfo {author}
  {\bibfnamefont {E.}~\bibnamefont {{Gosset}}}, \bibinfo {author}
  {\bibfnamefont {R.~G.}\ \bibnamefont {{Izzard}}}, \bibinfo {author}
  {\bibfnamefont {J.-B.}\ \bibnamefont {{Le Bouquin}}},\ and\ \bibinfo {author}
  {\bibfnamefont {F.~R.~N.}\ \bibnamefont {{Schneider}}},\ }\bibfield  {title}
  {\bibinfo {title} {{Binary Interaction Dominates the Evolution of Massive
  Stars}},\ }\href {https://doi.org/10.1126/science.1223344} {\bibfield
  {journal} {\bibinfo  {journal} {Science}\ }\textbf {\bibinfo {volume}
  {337}},\ \bibinfo {pages} {444} (\bibinfo {year} {2012}{\natexlab{b}})},\
  \Eprint {https://arxiv.org/abs/1207.6397} {arXiv:1207.6397 [astro-ph.SR]}
  \BibitemShut {NoStop}%
\bibitem [{\citenamefont {{Heger}}\ \emph
  {et~al.}(2003{\natexlab{b}})\citenamefont {{Heger}}, \citenamefont {{Fryer}},
  \citenamefont {{Woosley}}, \citenamefont {{Langer}},\ and\ \citenamefont
  {{Hartmann}}}]{Heger:2003}%
  \BibitemOpen
  \bibfield  {author} {\bibinfo {author} {\bibfnamefont {A.}~\bibnamefont
  {{Heger}}}, \bibinfo {author} {\bibfnamefont {C.~L.}\ \bibnamefont
  {{Fryer}}}, \bibinfo {author} {\bibfnamefont {S.~E.}\ \bibnamefont
  {{Woosley}}}, \bibinfo {author} {\bibfnamefont {N.}~\bibnamefont
  {{Langer}}},\ and\ \bibinfo {author} {\bibfnamefont {D.~H.}\ \bibnamefont
  {{Hartmann}}},\ }\bibfield  {title} {\bibinfo {title} {{How Massive Single
  Stars End Their Life}},\ }\href {https://doi.org/10.1086/375341} {\bibfield
  {journal} {\bibinfo  {journal} {\apj}\ }\textbf {\bibinfo {volume} {591}},\
  \bibinfo {pages} {288} (\bibinfo {year} {2003}{\natexlab{b}})},\ \Eprint
  {https://arxiv.org/abs/astro-ph/0212469} {arXiv:astro-ph/0212469 [astro-ph]}
  \BibitemShut {NoStop}%
\bibitem [{\citenamefont {{Kroupa}}(2001{\natexlab{b}})}]{Kroupa:2001}%
  \BibitemOpen
  \bibfield  {author} {\bibinfo {author} {\bibfnamefont {P.}~\bibnamefont
  {{Kroupa}}},\ }\bibfield  {title} {\bibinfo {title} {{On the variation of the
  initial mass function}},\ }\href
  {https://doi.org/10.1046/j.1365-8711.2001.04022.x} {\bibfield  {journal}
  {\bibinfo  {journal} {\mnras}\ }\textbf {\bibinfo {volume} {322}},\ \bibinfo
  {pages} {231} (\bibinfo {year} {2001}{\natexlab{b}})},\ \Eprint
  {https://arxiv.org/abs/astro-ph/0009005} {astro-ph/0009005} \BibitemShut
  {NoStop}%
\bibitem [{\citenamefont {{Kalari}}\ \emph {et~al.}(2022)\citenamefont
  {{Kalari}}, \citenamefont {{Horch}}, \citenamefont {{Salinas}}, \citenamefont
  {{Vink}}, \citenamefont {{Andersen}}, \citenamefont {{Bestenlehner}},\ and\
  \citenamefont {{Rubio}}}]{Kalari:2022}%
  \BibitemOpen
  \bibfield  {author} {\bibinfo {author} {\bibfnamefont {V.~M.}\ \bibnamefont
  {{Kalari}}}, \bibinfo {author} {\bibfnamefont {E.~P.}\ \bibnamefont
  {{Horch}}}, \bibinfo {author} {\bibfnamefont {R.}~\bibnamefont {{Salinas}}},
  \bibinfo {author} {\bibfnamefont {J.~S.}\ \bibnamefont {{Vink}}}, \bibinfo
  {author} {\bibfnamefont {M.}~\bibnamefont {{Andersen}}}, \bibinfo {author}
  {\bibfnamefont {J.~M.}\ \bibnamefont {{Bestenlehner}}},\ and\ \bibinfo
  {author} {\bibfnamefont {M.}~\bibnamefont {{Rubio}}},\ }\bibfield  {title}
  {\bibinfo {title} {{Resolving the Core of R136 in the Optical}},\ }\href
  {https://doi.org/10.3847/1538-4357/ac8424} {\bibfield  {journal} {\bibinfo
  {journal} {\apj}\ }\textbf {\bibinfo {volume} {935}},\ \bibinfo {eid} {162}
  (\bibinfo {year} {2022})},\ \Eprint {https://arxiv.org/abs/2207.13078}
  {arXiv:2207.13078 [astro-ph.SR]} \BibitemShut {NoStop}%
\bibitem [{\citenamefont {{von Boetticher}}\ \emph {et~al.}(2017)\citenamefont
  {{von Boetticher}}, \citenamefont {{Triaud}}, \citenamefont {{Queloz}},
  \citenamefont {{Gill}}, \citenamefont {{Lendl}}, \citenamefont {{Delrez}},
  \citenamefont {{Anderson}}, \citenamefont {{Collier Cameron}}, \citenamefont
  {{Faedi}}, \citenamefont {{Gillon}}, \citenamefont {{G{\'o}mez Maqueo Chew}},
  \citenamefont {{Hebb}}, \citenamefont {{Hellier}}, \citenamefont {{Jehin}},
  \citenamefont {{Maxted}}, \citenamefont {{Martin}}, \citenamefont {{Pepe}},
  \citenamefont {{Pollacco}}, \citenamefont {{S{\'e}gransan}}, \citenamefont
  {{Smalley}}, \citenamefont {{Udry}},\ and\ \citenamefont
  {{West}}}]{vonBoetticher:2017}%
  \BibitemOpen
  \bibfield  {author} {\bibinfo {author} {\bibfnamefont {A.}~\bibnamefont {{von
  Boetticher}}}, \bibinfo {author} {\bibfnamefont {A.~H.~M.~J.}\ \bibnamefont
  {{Triaud}}}, \bibinfo {author} {\bibfnamefont {D.}~\bibnamefont {{Queloz}}},
  \bibinfo {author} {\bibfnamefont {S.}~\bibnamefont {{Gill}}}, \bibinfo
  {author} {\bibfnamefont {M.}~\bibnamefont {{Lendl}}}, \bibinfo {author}
  {\bibfnamefont {L.}~\bibnamefont {{Delrez}}}, \bibinfo {author}
  {\bibfnamefont {D.~R.}\ \bibnamefont {{Anderson}}}, \bibinfo {author}
  {\bibfnamefont {A.}~\bibnamefont {{Collier Cameron}}}, \bibinfo {author}
  {\bibfnamefont {F.}~\bibnamefont {{Faedi}}}, \bibinfo {author} {\bibfnamefont
  {M.}~\bibnamefont {{Gillon}}}, \bibinfo {author} {\bibfnamefont
  {Y.}~\bibnamefont {{G{\'o}mez Maqueo Chew}}}, \bibinfo {author}
  {\bibfnamefont {L.}~\bibnamefont {{Hebb}}}, \bibinfo {author} {\bibfnamefont
  {C.}~\bibnamefont {{Hellier}}}, \bibinfo {author} {\bibfnamefont
  {E.}~\bibnamefont {{Jehin}}}, \bibinfo {author} {\bibfnamefont {P.~F.~L.}\
  \bibnamefont {{Maxted}}}, \bibinfo {author} {\bibfnamefont {D.~V.}\
  \bibnamefont {{Martin}}}, \bibinfo {author} {\bibfnamefont {F.}~\bibnamefont
  {{Pepe}}}, \bibinfo {author} {\bibfnamefont {D.}~\bibnamefont {{Pollacco}}},
  \bibinfo {author} {\bibfnamefont {D.}~\bibnamefont {{S{\'e}gransan}}},
  \bibinfo {author} {\bibfnamefont {B.}~\bibnamefont {{Smalley}}}, \bibinfo
  {author} {\bibfnamefont {S.}~\bibnamefont {{Udry}}},\ and\ \bibinfo {author}
  {\bibfnamefont {R.}~\bibnamefont {{West}}},\ }\bibfield  {title} {\bibinfo
  {title} {{The EBLM project. III. A Saturn-size low-mass star at the
  hydrogen-burning limit}},\ }\href
  {https://doi.org/10.1051/0004-6361/201731107} {\bibfield  {journal} {\bibinfo
   {journal} {\aap}\ }\textbf {\bibinfo {volume} {604}},\ \bibinfo {eid} {L6}
  (\bibinfo {year} {2017})},\ \Eprint {https://arxiv.org/abs/1706.08781}
  {arXiv:1706.08781 [astro-ph.SR]} \BibitemShut {NoStop}%
\bibitem [{\citenamefont {{Dieterich}}\ \emph {et~al.}(2018)\citenamefont
  {{Dieterich}}, \citenamefont {{Weinberger}}, \citenamefont {{Boss}},
  \citenamefont {{Henry}}, \citenamefont {{Jao}}, \citenamefont {{Gagn{\'e}}},
  \citenamefont {{Astraatmadja}}, \citenamefont {{Thompson}},\ and\
  \citenamefont {{Anglada-Escud{\'e}}}}]{Dieterich:2018}%
  \BibitemOpen
  \bibfield  {author} {\bibinfo {author} {\bibfnamefont {S.~B.}\ \bibnamefont
  {{Dieterich}}}, \bibinfo {author} {\bibfnamefont {A.~J.}\ \bibnamefont
  {{Weinberger}}}, \bibinfo {author} {\bibfnamefont {A.~P.}\ \bibnamefont
  {{Boss}}}, \bibinfo {author} {\bibfnamefont {T.~J.}\ \bibnamefont {{Henry}}},
  \bibinfo {author} {\bibfnamefont {W.-C.}\ \bibnamefont {{Jao}}}, \bibinfo
  {author} {\bibfnamefont {J.}~\bibnamefont {{Gagn{\'e}}}}, \bibinfo {author}
  {\bibfnamefont {T.~L.}\ \bibnamefont {{Astraatmadja}}}, \bibinfo {author}
  {\bibfnamefont {M.~A.}\ \bibnamefont {{Thompson}}},\ and\ \bibinfo {author}
  {\bibfnamefont {G.}~\bibnamefont {{Anglada-Escud{\'e}}}},\ }\bibfield
  {title} {\bibinfo {title} {{Dynamical Masses of $\eta$ Indi B and C: Two
  Massive Brown Dwarfs at the Edge of the Stellar-substellar Boundary}},\
  }\href {https://doi.org/10.3847/1538-4357/aadadc} {\bibfield  {journal}
  {\bibinfo  {journal} {\apj}\ }\textbf {\bibinfo {volume} {865}},\ \bibinfo
  {eid} {28} (\bibinfo {year} {2018})},\ \Eprint
  {https://arxiv.org/abs/1807.09880} {arXiv:1807.09880 [astro-ph.SR]}
  \BibitemShut {NoStop}%
\bibitem [{\citenamefont {{Ulrich}}\ and\ \citenamefont
  {{Burger}}(1976{\natexlab{b}})}]{Ulrich:1976}%
  \BibitemOpen
  \bibfield  {author} {\bibinfo {author} {\bibfnamefont {R.~K.}\ \bibnamefont
  {{Ulrich}}}\ and\ \bibinfo {author} {\bibfnamefont {H.~L.}\ \bibnamefont
  {{Burger}}},\ }\bibfield  {title} {\bibinfo {title} {{The accreting component
  of mass-exchange binaries.}},\ }\href {https://doi.org/10.1086/154406}
  {\bibfield  {journal} {\bibinfo  {journal} {\apj}\ }\textbf {\bibinfo
  {volume} {206}},\ \bibinfo {pages} {509} (\bibinfo {year}
  {1976}{\natexlab{b}})}\BibitemShut {NoStop}%
\bibitem [{\citenamefont {{Kippenhahn}}\ and\ \citenamefont
  {{Meyer-Hofmeister}}(1977{\natexlab{a}})}]{Kippenhahn:1977}%
  \BibitemOpen
  \bibfield  {author} {\bibinfo {author} {\bibfnamefont {R.}~\bibnamefont
  {{Kippenhahn}}}\ and\ \bibinfo {author} {\bibfnamefont {E.}~\bibnamefont
  {{Meyer-Hofmeister}}},\ }\bibfield  {title} {\bibinfo {title} {{On the radii
  of accreting main sequence stars.}},\ }\href@noop {} {\bibfield  {journal}
  {\bibinfo  {journal} {\aap}\ }\textbf {\bibinfo {volume} {54}},\ \bibinfo
  {pages} {539} (\bibinfo {year} {1977}{\natexlab{a}})}\BibitemShut {NoStop}%
\bibitem [{\citenamefont {{Neo}}\ \emph
  {et~al.}(1977{\natexlab{b}})\citenamefont {{Neo}}, \citenamefont {{Miyaji}},
  \citenamefont {{Nomoto}},\ and\ \citenamefont {{Sugimoto}}}]{Neo:1977}%
  \BibitemOpen
  \bibfield  {author} {\bibinfo {author} {\bibfnamefont {S.}~\bibnamefont
  {{Neo}}}, \bibinfo {author} {\bibfnamefont {S.}~\bibnamefont {{Miyaji}}},
  \bibinfo {author} {\bibfnamefont {K.}~\bibnamefont {{Nomoto}}},\ and\
  \bibinfo {author} {\bibfnamefont {D.}~\bibnamefont {{Sugimoto}}},\ }\bibfield
   {title} {\bibinfo {title} {{Effect of Rapid Mass Accretion onto the
  Main-Sequence Stars}},\ }\href@noop {} {\bibfield  {journal} {\bibinfo
  {journal} {\pasj}\ }\textbf {\bibinfo {volume} {29}},\ \bibinfo {pages} {249}
  (\bibinfo {year} {1977}{\natexlab{b}})}\BibitemShut {NoStop}%
\bibitem [{\citenamefont {{Wellstein}}\ \emph
  {et~al.}(2001{\natexlab{b}})\citenamefont {{Wellstein}}, \citenamefont
  {{Langer}},\ and\ \citenamefont {{Braun}}}]{Wellstein:2001}%
  \BibitemOpen
  \bibfield  {author} {\bibinfo {author} {\bibfnamefont {S.}~\bibnamefont
  {{Wellstein}}}, \bibinfo {author} {\bibfnamefont {N.}~\bibnamefont
  {{Langer}}},\ and\ \bibinfo {author} {\bibfnamefont {H.}~\bibnamefont
  {{Braun}}},\ }\bibfield  {title} {\bibinfo {title} {{Formation of contact in
  massive close binaries}},\ }\href
  {https://doi.org/10.1051/0004-6361:20010151} {\bibfield  {journal} {\bibinfo
  {journal} {\aap}\ }\textbf {\bibinfo {volume} {369}},\ \bibinfo {pages} {939}
  (\bibinfo {year} {2001}{\natexlab{b}})},\ \Eprint
  {https://arxiv.org/abs/astro-ph/0102244} {arXiv:astro-ph/0102244 [astro-ph]}
  \BibitemShut {NoStop}%
\bibitem [{\citenamefont {{de Mink}}\ \emph
  {et~al.}(2007{\natexlab{b}})\citenamefont {{de Mink}}, \citenamefont
  {{Pols}},\ and\ \citenamefont {{Hilditch}}}]{deMink:2007SMC}%
  \BibitemOpen
  \bibfield  {author} {\bibinfo {author} {\bibfnamefont {S.~E.}\ \bibnamefont
  {{de Mink}}}, \bibinfo {author} {\bibfnamefont {O.~R.}\ \bibnamefont
  {{Pols}}},\ and\ \bibinfo {author} {\bibfnamefont {R.~W.}\ \bibnamefont
  {{Hilditch}}},\ }\bibfield  {title} {\bibinfo {title} {{Efficiency of mass
  transfer in massive close binaries. Tests from double-lined eclipsing
  binaries in the SMC}},\ }\href {https://doi.org/10.1051/0004-6361:20067007}
  {\bibfield  {journal} {\bibinfo  {journal} {\aap}\ }\textbf {\bibinfo
  {volume} {467}},\ \bibinfo {pages} {1181} (\bibinfo {year}
  {2007}{\natexlab{b}})},\ \Eprint {https://arxiv.org/abs/astro-ph/0703480}
  {astro-ph/0703480} \BibitemShut {NoStop}%
\bibitem [{\citenamefont {{Tout}}(1991)}]{1991MNRAS.250..701T}%
  \BibitemOpen
  \bibfield  {author} {\bibinfo {author} {\bibfnamefont {C.~A.}\ \bibnamefont
  {{Tout}}},\ }\bibfield  {title} {\bibinfo {title} {{On the relation between
  the mass-ratio distribution in binary stars and the mass function for single
  stars.}},\ }\href {https://doi.org/10.1093/mnras/250.4.701} {\bibfield
  {journal} {\bibinfo  {journal} {\mnras}\ }\textbf {\bibinfo {volume} {250}},\
  \bibinfo {pages} {701} (\bibinfo {year} {1991})}\BibitemShut {NoStop}%
\bibitem [{\citenamefont {{Mazeh}}\ \emph {et~al.}(1992)\citenamefont
  {{Mazeh}}, \citenamefont {{Goldberg}}, \citenamefont {{Duquennoy}},\ and\
  \citenamefont {{Mayor}}}]{1992ApJ...401..265M}%
  \BibitemOpen
  \bibfield  {author} {\bibinfo {author} {\bibfnamefont {T.}~\bibnamefont
  {{Mazeh}}}, \bibinfo {author} {\bibfnamefont {D.}~\bibnamefont {{Goldberg}}},
  \bibinfo {author} {\bibfnamefont {A.}~\bibnamefont {{Duquennoy}}},\ and\
  \bibinfo {author} {\bibfnamefont {M.}~\bibnamefont {{Mayor}}},\ }\bibfield
  {title} {\bibinfo {title} {{On the Mass-Ratio Distribution of Spectroscopic
  Binaries with Solar-Type Primaries}},\ }\href
  {https://doi.org/10.1086/172058} {\bibfield  {journal} {\bibinfo  {journal}
  {\apj}\ }\textbf {\bibinfo {volume} {401}},\ \bibinfo {pages} {265} (\bibinfo
  {year} {1992})}\BibitemShut {NoStop}%
\bibitem [{\citenamefont {{Goldberg}}\ and\ \citenamefont
  {{Mazeh}}(1994)}]{1994A&A...282..801G}%
  \BibitemOpen
  \bibfield  {author} {\bibinfo {author} {\bibfnamefont {D.}~\bibnamefont
  {{Goldberg}}}\ and\ \bibinfo {author} {\bibfnamefont {T.}~\bibnamefont
  {{Mazeh}}},\ }\bibfield  {title} {\bibinfo {title} {{The mass-ratio
  distribution of the spectroscopic binaries in the Pleiades.}},\ }\href@noop
  {} {\bibfield  {journal} {\bibinfo  {journal} {\aap}\ }\textbf {\bibinfo
  {volume} {282}},\ \bibinfo {pages} {801} (\bibinfo {year}
  {1994})}\BibitemShut {NoStop}%
\bibitem [{\citenamefont {{Kobulnicky}}\ \emph {et~al.}(2014)\citenamefont
  {{Kobulnicky}}, \citenamefont {{Kiminki}}, \citenamefont {{Lundquist}},
  \citenamefont {{Burke}}, \citenamefont {{Chapman}}, \citenamefont {{Keller}},
  \citenamefont {{Lester}}, \citenamefont {{Rolen}}, \citenamefont {{Topel}},
  \citenamefont {{Bhattacharjee}}, \citenamefont {{Smullen}}, \citenamefont
  {{Vargas {\'A}lvarez}}, \citenamefont {{Runnoe}}, \citenamefont {{Dale}},\
  and\ \citenamefont {{Brotherton}}}]{2014ApJS..213...34K}%
  \BibitemOpen
  \bibfield  {author} {\bibinfo {author} {\bibfnamefont {H.~A.}\ \bibnamefont
  {{Kobulnicky}}}, \bibinfo {author} {\bibfnamefont {D.~C.}\ \bibnamefont
  {{Kiminki}}}, \bibinfo {author} {\bibfnamefont {M.~J.}\ \bibnamefont
  {{Lundquist}}}, \bibinfo {author} {\bibfnamefont {J.}~\bibnamefont
  {{Burke}}}, \bibinfo {author} {\bibfnamefont {J.}~\bibnamefont {{Chapman}}},
  \bibinfo {author} {\bibfnamefont {E.}~\bibnamefont {{Keller}}}, \bibinfo
  {author} {\bibfnamefont {K.}~\bibnamefont {{Lester}}}, \bibinfo {author}
  {\bibfnamefont {E.~K.}\ \bibnamefont {{Rolen}}}, \bibinfo {author}
  {\bibfnamefont {E.}~\bibnamefont {{Topel}}}, \bibinfo {author} {\bibfnamefont
  {A.}~\bibnamefont {{Bhattacharjee}}}, \bibinfo {author} {\bibfnamefont
  {R.~A.}\ \bibnamefont {{Smullen}}}, \bibinfo {author} {\bibfnamefont {C.~A.}\
  \bibnamefont {{Vargas {\'A}lvarez}}}, \bibinfo {author} {\bibfnamefont
  {J.~C.}\ \bibnamefont {{Runnoe}}}, \bibinfo {author} {\bibfnamefont {D.~A.}\
  \bibnamefont {{Dale}}},\ and\ \bibinfo {author} {\bibfnamefont {M.~M.}\
  \bibnamefont {{Brotherton}}},\ }\bibfield  {title} {\bibinfo {title} {{Toward
  Complete Statistics of Massive Binary Stars: Penultimate Results from the
  Cygnus OB2 Radial Velocity Survey}},\ }\href
  {https://doi.org/10.1088/0067-0049/213/2/34} {\bibfield  {journal} {\bibinfo
  {journal} {\apjs}\ }\textbf {\bibinfo {volume} {213}},\ \bibinfo {eid} {34}
  (\bibinfo {year} {2014})},\ \Eprint {https://arxiv.org/abs/1406.6655}
  {arXiv:1406.6655 [astro-ph.SR]} \BibitemShut {NoStop}%
\bibitem [{\citenamefont {{Hobbs}}\ \emph
  {et~al.}(2005{\natexlab{b}})\citenamefont {{Hobbs}}, \citenamefont
  {{Lorimer}}, \citenamefont {{Lyne}},\ and\ \citenamefont
  {{Kramer}}}]{Hobbs:2005}%
  \BibitemOpen
  \bibfield  {author} {\bibinfo {author} {\bibfnamefont {G.}~\bibnamefont
  {{Hobbs}}}, \bibinfo {author} {\bibfnamefont {D.~R.}\ \bibnamefont
  {{Lorimer}}}, \bibinfo {author} {\bibfnamefont {A.~G.}\ \bibnamefont
  {{Lyne}}},\ and\ \bibinfo {author} {\bibfnamefont {M.}~\bibnamefont
  {{Kramer}}},\ }\bibfield  {title} {\bibinfo {title} {{A statistical study of
  233 pulsar proper motions}},\ }\href
  {https://doi.org/10.1111/j.1365-2966.2005.09087.x} {\bibfield  {journal}
  {\bibinfo  {journal} {\mnras}\ }\textbf {\bibinfo {volume} {360}},\ \bibinfo
  {pages} {974} (\bibinfo {year} {2005}{\natexlab{b}})},\ \Eprint
  {https://arxiv.org/abs/astro-ph/0504584} {astro-ph/0504584} \BibitemShut
  {NoStop}%
\bibitem [{\citenamefont {{Verbunt}}\ \emph
  {et~al.}(2017{\natexlab{b}})\citenamefont {{Verbunt}}, \citenamefont
  {{Igoshev}},\ and\ \citenamefont {{Cator}}}]{Verbunt:2017}%
  \BibitemOpen
  \bibfield  {author} {\bibinfo {author} {\bibfnamefont {F.}~\bibnamefont
  {{Verbunt}}}, \bibinfo {author} {\bibfnamefont {A.}~\bibnamefont
  {{Igoshev}}},\ and\ \bibinfo {author} {\bibfnamefont {E.}~\bibnamefont
  {{Cator}}},\ }\bibfield  {title} {\bibinfo {title} {{The observed velocity
  distribution of young pulsars}},\ }\href
  {https://doi.org/10.1051/0004-6361/201731518} {\bibfield  {journal} {\bibinfo
   {journal} {\aap}\ }\textbf {\bibinfo {volume} {608}},\ \bibinfo {eid} {A57}
  (\bibinfo {year} {2017}{\natexlab{b}})},\ \Eprint
  {https://arxiv.org/abs/1708.08281} {arXiv:1708.08281 [astro-ph.HE]}
  \BibitemShut {NoStop}%
\bibitem [{\citenamefont {{Renzo}}\ \emph {et~al.}(2019)\citenamefont
  {{Renzo}}, \citenamefont {{Zapartas}}, \citenamefont {{de Mink}},
  \citenamefont {{G{\"o}tberg}}, \citenamefont {{Justham}}, \citenamefont
  {{Farmer}}, \citenamefont {{Izzard}}, \citenamefont {{Toonen}},\ and\
  \citenamefont {{Sana}}}]{Renzo:2019}%
  \BibitemOpen
  \bibfield  {author} {\bibinfo {author} {\bibfnamefont {M.}~\bibnamefont
  {{Renzo}}}, \bibinfo {author} {\bibfnamefont {E.}~\bibnamefont {{Zapartas}}},
  \bibinfo {author} {\bibfnamefont {S.~E.}\ \bibnamefont {{de Mink}}}, \bibinfo
  {author} {\bibfnamefont {Y.}~\bibnamefont {{G{\"o}tberg}}}, \bibinfo {author}
  {\bibfnamefont {S.}~\bibnamefont {{Justham}}}, \bibinfo {author}
  {\bibfnamefont {R.~J.}\ \bibnamefont {{Farmer}}}, \bibinfo {author}
  {\bibfnamefont {R.~G.}\ \bibnamefont {{Izzard}}}, \bibinfo {author}
  {\bibfnamefont {S.}~\bibnamefont {{Toonen}}},\ and\ \bibinfo {author}
  {\bibfnamefont {H.}~\bibnamefont {{Sana}}},\ }\bibfield  {title} {\bibinfo
  {title} {{Massive runaway and walkaway stars. A study of the kinematical
  imprints of the physical processes governing the evolution and explosion of
  their binary progenitors}},\ }\href
  {https://doi.org/10.1051/0004-6361/201833297} {\bibfield  {journal} {\bibinfo
   {journal} {\aap}\ }\textbf {\bibinfo {volume} {624}},\ \bibinfo {eid} {A66}
  (\bibinfo {year} {2019})},\ \Eprint {https://arxiv.org/abs/1804.09164}
  {arXiv:1804.09164 [astro-ph.SR]} \BibitemShut {NoStop}%
\bibitem [{\citenamefont {{Madau}}\ and\ \citenamefont
  {{Dickinson}}(2014)}]{MadauDickinson:2014}%
  \BibitemOpen
  \bibfield  {author} {\bibinfo {author} {\bibfnamefont {P.}~\bibnamefont
  {{Madau}}}\ and\ \bibinfo {author} {\bibfnamefont {M.}~\bibnamefont
  {{Dickinson}}},\ }\bibfield  {title} {\bibinfo {title} {{Cosmic
  Star-Formation History}},\ }\href
  {https://doi.org/10.1146/annurev-astro-081811-125615} {\bibfield  {journal}
  {\bibinfo  {journal} {\araa}\ }\textbf {\bibinfo {volume} {52}},\ \bibinfo
  {pages} {415} (\bibinfo {year} {2014})},\ \Eprint
  {https://arxiv.org/abs/1403.0007} {arXiv:1403.0007} \BibitemShut {NoStop}%
\bibitem [{\citenamefont {{\noopsort{De Mink}}{de Mink}}\ and\ \citenamefont
  {{Belczynski}}(2015)}]{deMinkBelczynski:2015}%
  \BibitemOpen
  \bibfield  {author} {\bibinfo {author} {\bibfnamefont {S.~E.}\ \bibnamefont
  {{\noopsort{De Mink}}{de Mink}}}\ and\ \bibinfo {author} {\bibfnamefont
  {K.}~\bibnamefont {{Belczynski}}},\ }\bibfield  {title} {\bibinfo {title}
  {{Merger Rates of Double Neutron Stars and Stellar Origin Black Holes: The
  Impact of Initial Conditions on Binary Evolution Predictions}},\ }\href
  {https://doi.org/10.1088/0004-637X/814/1/58} {\bibfield  {journal} {\bibinfo
  {journal} {\apj}\ }\textbf {\bibinfo {volume} {814}},\ \bibinfo {eid} {58}
  (\bibinfo {year} {2015})},\ \Eprint {https://arxiv.org/abs/1506.03573}
  {arXiv:1506.03573 [astro-ph.HE]} \BibitemShut {NoStop}%
\bibitem [{\citenamefont {{Gennaro}}\ \emph {et~al.}(2018)\citenamefont
  {{Gennaro}}, \citenamefont {{Tchernyshyov}}, \citenamefont {{Brown}},
  \citenamefont {{Geha}}, \citenamefont {{Avila}}, \citenamefont
  {{Guhathakurta}}, \citenamefont {{Kalirai}}, \citenamefont {{Kirby}},
  \citenamefont {{Renzini}}, \citenamefont {{Simon}}, \citenamefont
  {{Tumlinson}},\ and\ \citenamefont {{Vargas}}}]{2018ApJ...855...20G}%
  \BibitemOpen
  \bibfield  {author} {\bibinfo {author} {\bibfnamefont {M.}~\bibnamefont
  {{Gennaro}}}, \bibinfo {author} {\bibfnamefont {K.}~\bibnamefont
  {{Tchernyshyov}}}, \bibinfo {author} {\bibfnamefont {T.~M.}\ \bibnamefont
  {{Brown}}}, \bibinfo {author} {\bibfnamefont {M.}~\bibnamefont {{Geha}}},
  \bibinfo {author} {\bibfnamefont {R.~J.}\ \bibnamefont {{Avila}}}, \bibinfo
  {author} {\bibfnamefont {P.}~\bibnamefont {{Guhathakurta}}}, \bibinfo
  {author} {\bibfnamefont {J.~S.}\ \bibnamefont {{Kalirai}}}, \bibinfo {author}
  {\bibfnamefont {E.~N.}\ \bibnamefont {{Kirby}}}, \bibinfo {author}
  {\bibfnamefont {A.}~\bibnamefont {{Renzini}}}, \bibinfo {author}
  {\bibfnamefont {J.~D.}\ \bibnamefont {{Simon}}}, \bibinfo {author}
  {\bibfnamefont {J.}~\bibnamefont {{Tumlinson}}},\ and\ \bibinfo {author}
  {\bibfnamefont {L.~C.}\ \bibnamefont {{Vargas}}},\ }\bibfield  {title}
  {\bibinfo {title} {{Evidence of a Non-universal Stellar Initial Mass
  Function. Insights from HST Optical Imaging of Six Ultra-faint Dwarf Milky
  Way Satellites}},\ }\href {https://doi.org/10.3847/1538-4357/aaa973}
  {\bibfield  {journal} {\bibinfo  {journal} {\apj}\ }\textbf {\bibinfo
  {volume} {855}},\ \bibinfo {eid} {20} (\bibinfo {year} {2018})},\ \Eprint
  {https://arxiv.org/abs/1801.06195} {arXiv:1801.06195 [astro-ph.GA]}
  \BibitemShut {NoStop}%
\bibitem [{\citenamefont {{Klencki}}\ \emph
  {et~al.}(2018{\natexlab{a}})\citenamefont {{Klencki}}, \citenamefont {{Moe}},
  \citenamefont {{Gladysz}}, \citenamefont {{Chruslinska}}, \citenamefont
  {{Holz}},\ and\ \citenamefont {{Belczynski}}}]{2018A&A...619A..77K}%
  \BibitemOpen
  \bibfield  {author} {\bibinfo {author} {\bibfnamefont {J.}~\bibnamefont
  {{Klencki}}}, \bibinfo {author} {\bibfnamefont {M.}~\bibnamefont {{Moe}}},
  \bibinfo {author} {\bibfnamefont {W.}~\bibnamefont {{Gladysz}}}, \bibinfo
  {author} {\bibfnamefont {M.}~\bibnamefont {{Chruslinska}}}, \bibinfo {author}
  {\bibfnamefont {D.~E.}\ \bibnamefont {{Holz}}},\ and\ \bibinfo {author}
  {\bibfnamefont {K.}~\bibnamefont {{Belczynski}}},\ }\bibfield  {title}
  {\bibinfo {title} {{Impact of inter-correlated initial binary parameters on
  double black hole and neutron star mergers}},\ }\href
  {https://doi.org/10.1051/0004-6361/201833025} {\bibfield  {journal} {\bibinfo
   {journal} {\aap}\ }\textbf {\bibinfo {volume} {619}},\ \bibinfo {eid} {A77}
  (\bibinfo {year} {2018}{\natexlab{a}})},\ \Eprint
  {https://arxiv.org/abs/1808.07889} {arXiv:1808.07889 [astro-ph.HE]}
  \BibitemShut {NoStop}%
\bibitem [{\citenamefont {{Schneider}}\ \emph {et~al.}(2018)\citenamefont
  {{Schneider}}, \citenamefont {{Sana}}, \citenamefont {{Evans}}, \citenamefont
  {{Bestenlehner}}, \citenamefont {{Castro}}, \citenamefont {{Fossati}},
  \citenamefont {{Gr{\"a}fener}}, \citenamefont {{Langer}}, \citenamefont
  {{Ram{\'\i}rez-Agudelo}}, \citenamefont {{Sab{\'\i}n-Sanjuli{\'a}n}},
  \citenamefont {{Sim{\'o}n-D{\'\i}az}}, \citenamefont {{Tramper}},
  \citenamefont {{Crowther}}, \citenamefont {{de Koter}}, \citenamefont {{de
  Mink}}, \citenamefont {{Dufton}}, \citenamefont {{Garcia}}, \citenamefont
  {{Gieles}}, \citenamefont {{H{\'e}nault-Brunet}}, \citenamefont {{Herrero}},
  \citenamefont {{Izzard}}, \citenamefont {{Kalari}}, \citenamefont {{Lennon}},
  \citenamefont {{Ma{\'\i}z Apell{\'a}niz}}, \citenamefont {{Markova}},
  \citenamefont {{Najarro}}, \citenamefont {{Podsiadlowski}}, \citenamefont
  {{Puls}}, \citenamefont {{Taylor}}, \citenamefont {{van Loon}}, \citenamefont
  {{Vink}},\ and\ \citenamefont {{Norman}}}]{2018Sci...359...69S}%
  \BibitemOpen
  \bibfield  {author} {\bibinfo {author} {\bibfnamefont {F.~R.~N.}\
  \bibnamefont {{Schneider}}}, \bibinfo {author} {\bibfnamefont
  {H.}~\bibnamefont {{Sana}}}, \bibinfo {author} {\bibfnamefont {C.~J.}\
  \bibnamefont {{Evans}}}, \bibinfo {author} {\bibfnamefont {J.~M.}\
  \bibnamefont {{Bestenlehner}}}, \bibinfo {author} {\bibfnamefont
  {N.}~\bibnamefont {{Castro}}}, \bibinfo {author} {\bibfnamefont
  {L.}~\bibnamefont {{Fossati}}}, \bibinfo {author} {\bibfnamefont
  {G.}~\bibnamefont {{Gr{\"a}fener}}}, \bibinfo {author} {\bibfnamefont
  {N.}~\bibnamefont {{Langer}}}, \bibinfo {author} {\bibfnamefont {O.~H.}\
  \bibnamefont {{Ram{\'\i}rez-Agudelo}}}, \bibinfo {author} {\bibfnamefont
  {C.}~\bibnamefont {{Sab{\'\i}n-Sanjuli{\'a}n}}}, \bibinfo {author}
  {\bibfnamefont {S.}~\bibnamefont {{Sim{\'o}n-D{\'\i}az}}}, \bibinfo {author}
  {\bibfnamefont {F.}~\bibnamefont {{Tramper}}}, \bibinfo {author}
  {\bibfnamefont {P.~A.}\ \bibnamefont {{Crowther}}}, \bibinfo {author}
  {\bibfnamefont {A.}~\bibnamefont {{de Koter}}}, \bibinfo {author}
  {\bibfnamefont {S.~E.}\ \bibnamefont {{de Mink}}}, \bibinfo {author}
  {\bibfnamefont {P.~L.}\ \bibnamefont {{Dufton}}}, \bibinfo {author}
  {\bibfnamefont {M.}~\bibnamefont {{Garcia}}}, \bibinfo {author}
  {\bibfnamefont {M.}~\bibnamefont {{Gieles}}}, \bibinfo {author}
  {\bibfnamefont {V.}~\bibnamefont {{H{\'e}nault-Brunet}}}, \bibinfo {author}
  {\bibfnamefont {A.}~\bibnamefont {{Herrero}}}, \bibinfo {author}
  {\bibfnamefont {R.~G.}\ \bibnamefont {{Izzard}}}, \bibinfo {author}
  {\bibfnamefont {V.}~\bibnamefont {{Kalari}}}, \bibinfo {author}
  {\bibfnamefont {D.~J.}\ \bibnamefont {{Lennon}}}, \bibinfo {author}
  {\bibfnamefont {J.}~\bibnamefont {{Ma{\'\i}z Apell{\'a}niz}}}, \bibinfo
  {author} {\bibfnamefont {N.}~\bibnamefont {{Markova}}}, \bibinfo {author}
  {\bibfnamefont {F.}~\bibnamefont {{Najarro}}}, \bibinfo {author}
  {\bibfnamefont {P.}~\bibnamefont {{Podsiadlowski}}}, \bibinfo {author}
  {\bibfnamefont {J.}~\bibnamefont {{Puls}}}, \bibinfo {author} {\bibfnamefont
  {W.~D.}\ \bibnamefont {{Taylor}}}, \bibinfo {author} {\bibfnamefont {J.~T.}\
  \bibnamefont {{van Loon}}}, \bibinfo {author} {\bibfnamefont {J.~S.}\
  \bibnamefont {{Vink}}},\ and\ \bibinfo {author} {\bibfnamefont
  {C.}~\bibnamefont {{Norman}}},\ }\bibfield  {title} {\bibinfo {title} {{An
  excess of massive stars in the local 30 Doradus starburst}},\ }\href
  {https://doi.org/10.1126/science.aan0106} {\bibfield  {journal} {\bibinfo
  {journal} {Science}\ }\textbf {\bibinfo {volume} {359}},\ \bibinfo {pages}
  {69} (\bibinfo {year} {2018})},\ \Eprint {https://arxiv.org/abs/1801.03107}
  {arXiv:1801.03107 [astro-ph.SR]} \BibitemShut {NoStop}%
\bibitem [{\citenamefont {{Klencki}}\ \emph
  {et~al.}(2018{\natexlab{b}})\citenamefont {{Klencki}}, \citenamefont {{Moe}},
  \citenamefont {{Gladysz}}, \citenamefont {{Chruslinska}}, \citenamefont
  {{Holz}},\ and\ \citenamefont {{Belczynski}}}]{Klencki:2018}%
  \BibitemOpen
  \bibfield  {author} {\bibinfo {author} {\bibfnamefont {J.}~\bibnamefont
  {{Klencki}}}, \bibinfo {author} {\bibfnamefont {M.}~\bibnamefont {{Moe}}},
  \bibinfo {author} {\bibfnamefont {W.}~\bibnamefont {{Gladysz}}}, \bibinfo
  {author} {\bibfnamefont {M.}~\bibnamefont {{Chruslinska}}}, \bibinfo {author}
  {\bibfnamefont {D.~E.}\ \bibnamefont {{Holz}}},\ and\ \bibinfo {author}
  {\bibfnamefont {K.}~\bibnamefont {{Belczynski}}},\ }\bibfield  {title}
  {\bibinfo {title} {{Impact of inter-correlated initial binary parameters on
  double black hole and neutron star mergers}},\ }\href
  {https://doi.org/10.1051/0004-6361/201833025} {\bibfield  {journal} {\bibinfo
   {journal} {\aap}\ }\textbf {\bibinfo {volume} {619}},\ \bibinfo {eid} {A77}
  (\bibinfo {year} {2018}{\natexlab{b}})},\ \Eprint
  {https://arxiv.org/abs/1808.07889} {arXiv:1808.07889 [astro-ph.HE]}
  \BibitemShut {NoStop}%
\bibitem [{\citenamefont {{Lamberts}}\ \emph {et~al.}(2016)\citenamefont
  {{Lamberts}}, \citenamefont {{Garrison-Kimmel}}, \citenamefont {{Clausen}},\
  and\ \citenamefont {{Hopkins}}}]{Lamberts:2016}%
  \BibitemOpen
  \bibfield  {author} {\bibinfo {author} {\bibfnamefont {A.}~\bibnamefont
  {{Lamberts}}}, \bibinfo {author} {\bibfnamefont {S.}~\bibnamefont
  {{Garrison-Kimmel}}}, \bibinfo {author} {\bibfnamefont {D.~R.}\ \bibnamefont
  {{Clausen}}},\ and\ \bibinfo {author} {\bibfnamefont {P.~F.}\ \bibnamefont
  {{Hopkins}}},\ }\bibfield  {title} {\bibinfo {title} {{When and where did
  GW150914 form?}},\ }\href {https://doi.org/10.1093/mnrasl/slw152} {\bibfield
  {journal} {\bibinfo  {journal} {\mnras}\ }\textbf {\bibinfo {volume} {463}},\
  \bibinfo {pages} {L31} (\bibinfo {year} {2016})},\ \Eprint
  {https://arxiv.org/abs/1605.08783} {arXiv:1605.08783 [astro-ph.HE]}
  \BibitemShut {NoStop}%
\bibitem [{\citenamefont {{Mapelli}}\ \emph {et~al.}(2017)\citenamefont
  {{Mapelli}}, \citenamefont {{Giacobbo}}, \citenamefont {{Ripamonti}},\ and\
  \citenamefont {{Spera}}}]{2017MNRAS.472.2422M}%
  \BibitemOpen
  \bibfield  {author} {\bibinfo {author} {\bibfnamefont {M.}~\bibnamefont
  {{Mapelli}}}, \bibinfo {author} {\bibfnamefont {N.}~\bibnamefont
  {{Giacobbo}}}, \bibinfo {author} {\bibfnamefont {E.}~\bibnamefont
  {{Ripamonti}}},\ and\ \bibinfo {author} {\bibfnamefont {M.}~\bibnamefont
  {{Spera}}},\ }\bibfield  {title} {\bibinfo {title} {{The cosmic merger rate
  of stellar black hole binaries from the Illustris simulation}},\ }\href
  {https://doi.org/10.1093/mnras/stx2123} {\bibfield  {journal} {\bibinfo
  {journal} {\mnras}\ }\textbf {\bibinfo {volume} {472}},\ \bibinfo {pages}
  {2422} (\bibinfo {year} {2017})},\ \Eprint {https://arxiv.org/abs/1708.05722}
  {arXiv:1708.05722 [astro-ph.GA]} \BibitemShut {NoStop}%
\bibitem [{\citenamefont {{Chru{\'s}li{\'n}ska}}\ and\ \citenamefont
  {{Nelemans}}(2019)}]{Chruslinska:2019obsSFRD}%
  \BibitemOpen
  \bibfield  {author} {\bibinfo {author} {\bibfnamefont {M.}~\bibnamefont
  {{Chru{\'s}li{\'n}ska}}}\ and\ \bibinfo {author} {\bibfnamefont
  {G.}~\bibnamefont {{Nelemans}}},\ }\bibfield  {title} {\bibinfo {title}
  {{Metallicity of stars formed throughout the cosmic history based on the
  observational properties of star-forming galaxies}},\ }\href
  {https://doi.org/10.1093/mnras/stz2057} {\bibfield  {journal} {\bibinfo
  {journal} {\mnras}\ }\textbf {\bibinfo {volume} {488}},\ \bibinfo {pages}
  {5300} (\bibinfo {year} {2019})},\ \Eprint {https://arxiv.org/abs/1907.11243}
  {arXiv:1907.11243 [astro-ph.GA]} \BibitemShut {NoStop}%
\bibitem [{\citenamefont {{Tang}}\ \emph {et~al.}(2020)\citenamefont {{Tang}},
  \citenamefont {{Eldridge}}, \citenamefont {{Stanway}},\ and\ \citenamefont
  {{Bray}}}]{Tang:2020}%
  \BibitemOpen
  \bibfield  {author} {\bibinfo {author} {\bibfnamefont {P.~N.}\ \bibnamefont
  {{Tang}}}, \bibinfo {author} {\bibfnamefont {J.~J.}\ \bibnamefont
  {{Eldridge}}}, \bibinfo {author} {\bibfnamefont {E.~R.}\ \bibnamefont
  {{Stanway}}},\ and\ \bibinfo {author} {\bibfnamefont {J.~C.}\ \bibnamefont
  {{Bray}}},\ }\bibfield  {title} {\bibinfo {title} {{Dependence of
  gravitational wave transient rates on cosmic star formation and metallicity
  evolution history}},\ }\href {https://doi.org/10.1093/mnrasl/slz183}
  {\bibfield  {journal} {\bibinfo  {journal} {\mnras}\ }\textbf {\bibinfo
  {volume} {493}},\ \bibinfo {pages} {L6} (\bibinfo {year} {2020})},\ \Eprint
  {https://arxiv.org/abs/1912.04474} {arXiv:1912.04474 [astro-ph.GA]}
  \BibitemShut {NoStop}%
\bibitem [{\citenamefont {{\noopsort{Du Buisson}}{du Buisson}}\ \emph
  {et~al.}(2020)\citenamefont {{\noopsort{Du Buisson}}{du Buisson}},
  \citenamefont {{Marchant}}, \citenamefont {{Podsiadlowski}}, \citenamefont
  {{Kobayashi}}, \citenamefont {{Abdalla}}, \citenamefont {{Taylor}},
  \citenamefont {{Mandel}}, \citenamefont {{de Mink}}, \citenamefont
  {{Moriya}},\ and\ \citenamefont {{Langer}}}]{duBuisson2020}%
  \BibitemOpen
  \bibfield  {author} {\bibinfo {author} {\bibfnamefont {L.}~\bibnamefont
  {{\noopsort{Du Buisson}}{du Buisson}}}, \bibinfo {author} {\bibfnamefont
  {P.}~\bibnamefont {{Marchant}}}, \bibinfo {author} {\bibfnamefont
  {P.}~\bibnamefont {{Podsiadlowski}}}, \bibinfo {author} {\bibfnamefont
  {C.}~\bibnamefont {{Kobayashi}}}, \bibinfo {author} {\bibfnamefont {F.~B.}\
  \bibnamefont {{Abdalla}}}, \bibinfo {author} {\bibfnamefont {P.}~\bibnamefont
  {{Taylor}}}, \bibinfo {author} {\bibfnamefont {I.}~\bibnamefont {{Mandel}}},
  \bibinfo {author} {\bibfnamefont {S.~E.}\ \bibnamefont {{de Mink}}}, \bibinfo
  {author} {\bibfnamefont {T.~J.}\ \bibnamefont {{Moriya}}},\ and\ \bibinfo
  {author} {\bibfnamefont {N.}~\bibnamefont {{Langer}}},\ }\bibfield  {title}
  {\bibinfo {title} {{Cosmic rates of black hole mergers and pair-instability
  supernovae from chemically homogeneous binary evolution}},\ }\href
  {https://doi.org/10.1093/mnras/staa3225} {\bibfield  {journal} {\bibinfo
  {journal} {\mnras}\ }\textbf {\bibinfo {volume} {499}},\ \bibinfo {pages}
  {5941} (\bibinfo {year} {2020})},\ \Eprint {https://arxiv.org/abs/2002.11630}
  {arXiv:2002.11630 [astro-ph.HE]} \BibitemShut {NoStop}%
\bibitem [{\citenamefont {{Briel}}\ \emph {et~al.}(2021)\citenamefont
  {{Briel}}, \citenamefont {{Eldridge}}, \citenamefont {{Stanway}},
  \citenamefont {{Stevance}},\ and\ \citenamefont {{Chrimes}}}]{Briel:2021}%
  \BibitemOpen
  \bibfield  {author} {\bibinfo {author} {\bibfnamefont {M.~M.}\ \bibnamefont
  {{Briel}}}, \bibinfo {author} {\bibfnamefont {J.~J.}\ \bibnamefont
  {{Eldridge}}}, \bibinfo {author} {\bibfnamefont {E.~R.}\ \bibnamefont
  {{Stanway}}}, \bibinfo {author} {\bibfnamefont {H.~F.}\ \bibnamefont
  {{Stevance}}},\ and\ \bibinfo {author} {\bibfnamefont {A.~A.}\ \bibnamefont
  {{Chrimes}}},\ }\bibfield  {title} {\bibinfo {title} {{Estimating Transient
  Rates from Cosmological Simulations and BPASS}},\ }\href@noop {} {\bibfield
  {journal} {\bibinfo  {journal} {arXiv e-prints}\ ,\ \bibinfo {eid}
  {arXiv:2111.08124}} (\bibinfo {year} {2021})},\ \Eprint
  {https://arxiv.org/abs/2111.08124} {arXiv:2111.08124 [astro-ph.CO]}
  \BibitemShut {NoStop}%
\bibitem [{\citenamefont {{Chu}}\ \emph {et~al.}(2021)\citenamefont {{Chu}},
  \citenamefont {{Yu}},\ and\ \citenamefont {{Lu}}}]{Chu:2021}%
  \BibitemOpen
  \bibfield  {author} {\bibinfo {author} {\bibfnamefont {Q.}~\bibnamefont
  {{Chu}}}, \bibinfo {author} {\bibfnamefont {S.}~\bibnamefont {{Yu}}},\ and\
  \bibinfo {author} {\bibfnamefont {Y.}~\bibnamefont {{Lu}}},\ }\bibfield
  {title} {\bibinfo {title} {{Formation and Evolution of Binary Neutron Stars:
  Mergers and Their Host Galaxies}},\ }\bibfield  {journal} {\bibinfo
  {journal} {\mnras}\ }\href {https://doi.org/10.1093/mnras/stab2882}
  {10.1093/mnras/stab2882} (\bibinfo {year} {2021}),\ \Eprint
  {https://arxiv.org/abs/2110.04687} {arXiv:2110.04687 [astro-ph.GA]}
  \BibitemShut {NoStop}%
\bibitem [{\citenamefont {{Chru{\'s}li{\'n}ska}}\ \emph
  {et~al.}(2021)\citenamefont {{Chru{\'s}li{\'n}ska}}, \citenamefont
  {{Nelemans}}, \citenamefont {{Boco}},\ and\ \citenamefont
  {{Lapi}}}]{Chruslinska:2021}%
  \BibitemOpen
  \bibfield  {author} {\bibinfo {author} {\bibfnamefont {M.}~\bibnamefont
  {{Chru{\'s}li{\'n}ska}}}, \bibinfo {author} {\bibfnamefont {G.}~\bibnamefont
  {{Nelemans}}}, \bibinfo {author} {\bibfnamefont {L.}~\bibnamefont {{Boco}}},\
  and\ \bibinfo {author} {\bibfnamefont {A.}~\bibnamefont {{Lapi}}},\
  }\bibfield  {title} {\bibinfo {title} {{The impact of the FMR and starburst
  galaxies on the (low-metallicity) cosmic star formation history}},\
  }\bibfield  {journal} {\bibinfo  {journal} {\mnras}\ }\href
  {https://doi.org/10.1093/mnras/stab2690} {10.1093/mnras/stab2690} (\bibinfo
  {year} {2021}),\ \Eprint {https://arxiv.org/abs/2109.06187} {arXiv:2109.06187
  [astro-ph.GA]} \BibitemShut {NoStop}%
\bibitem [{\citenamefont {{Boco}}\ \emph {et~al.}(2021)\citenamefont {{Boco}},
  \citenamefont {{Lapi}}, \citenamefont {{Chruslinska}}, \citenamefont
  {{Donevski}}, \citenamefont {{Sicilia}},\ and\ \citenamefont
  {{Danese}}}]{2021ApJ...907..110B}%
  \BibitemOpen
  \bibfield  {author} {\bibinfo {author} {\bibfnamefont {L.}~\bibnamefont
  {{Boco}}}, \bibinfo {author} {\bibfnamefont {A.}~\bibnamefont {{Lapi}}},
  \bibinfo {author} {\bibfnamefont {M.}~\bibnamefont {{Chruslinska}}}, \bibinfo
  {author} {\bibfnamefont {D.}~\bibnamefont {{Donevski}}}, \bibinfo {author}
  {\bibfnamefont {A.}~\bibnamefont {{Sicilia}}},\ and\ \bibinfo {author}
  {\bibfnamefont {L.}~\bibnamefont {{Danese}}},\ }\bibfield  {title} {\bibinfo
  {title} {{Evolution of Galaxy Star Formation and Metallicity: Impact on
  Double Compact Object Mergers}},\ }\href
  {https://doi.org/10.3847/1538-4357/abd3a0} {\bibfield  {journal} {\bibinfo
  {journal} {\apj}\ }\textbf {\bibinfo {volume} {907}},\ \bibinfo {eid} {110}
  (\bibinfo {year} {2021})},\ \Eprint {https://arxiv.org/abs/2012.02800}
  {arXiv:2012.02800 [astro-ph.GA]} \BibitemShut {NoStop}%
\bibitem [{\citenamefont {{Belczynski}}\ \emph
  {et~al.}(2010{\natexlab{b}})\citenamefont {{Belczynski}}, \citenamefont
  {{Dominik}}, \citenamefont {{Bulik}}, \citenamefont {{O'Shaughnessy}},
  \citenamefont {{Fryer}},\ and\ \citenamefont {{Holz}}}]{Belczynski:2010}%
  \BibitemOpen
  \bibfield  {author} {\bibinfo {author} {\bibfnamefont {K.}~\bibnamefont
  {{Belczynski}}}, \bibinfo {author} {\bibfnamefont {M.}~\bibnamefont
  {{Dominik}}}, \bibinfo {author} {\bibfnamefont {T.}~\bibnamefont {{Bulik}}},
  \bibinfo {author} {\bibfnamefont {R.}~\bibnamefont {{O'Shaughnessy}}},
  \bibinfo {author} {\bibfnamefont {C.}~\bibnamefont {{Fryer}}},\ and\ \bibinfo
  {author} {\bibfnamefont {D.~E.}\ \bibnamefont {{Holz}}},\ }\bibfield  {title}
  {\bibinfo {title} {{The Effect of Metallicity on the Detection Prospects for
  Gravitational Waves}},\ }\href {https://doi.org/10.1088/2041-8205/715/2/L138}
  {\bibfield  {journal} {\bibinfo  {journal} {\apjl}\ }\textbf {\bibinfo
  {volume} {715}},\ \bibinfo {pages} {L138} (\bibinfo {year}
  {2010}{\natexlab{b}})},\ \Eprint {https://arxiv.org/abs/1004.0386}
  {arXiv:1004.0386 [astro-ph.HE]} \BibitemShut {NoStop}%
\bibitem [{\citenamefont {{Vink}}(2011)}]{Vink:2011}%
  \BibitemOpen
  \bibfield  {author} {\bibinfo {author} {\bibfnamefont {J.~S.}\ \bibnamefont
  {{Vink}}},\ }\bibfield  {title} {\bibinfo {title} {{The theory of stellar
  winds}},\ }\href {https://doi.org/10.1007/s10509-011-0636-7} {\bibfield
  {journal} {\bibinfo  {journal} {\apss}\ }\textbf {\bibinfo {volume} {336}},\
  \bibinfo {pages} {163} (\bibinfo {year} {2011})},\ \Eprint
  {https://arxiv.org/abs/1112.0952} {arXiv:1112.0952 [astro-ph.SR]}
  \BibitemShut {NoStop}%
\bibitem [{\citenamefont {{Renzo}}\ \emph
  {et~al.}(2017{\natexlab{b}})\citenamefont {{Renzo}}, \citenamefont {{Ott}},
  \citenamefont {{Shore}},\ and\ \citenamefont {{de Mink}}}]{Renzo:2017}%
  \BibitemOpen
  \bibfield  {author} {\bibinfo {author} {\bibfnamefont {M.}~\bibnamefont
  {{Renzo}}}, \bibinfo {author} {\bibfnamefont {C.~D.}\ \bibnamefont {{Ott}}},
  \bibinfo {author} {\bibfnamefont {S.~N.}\ \bibnamefont {{Shore}}},\ and\
  \bibinfo {author} {\bibfnamefont {S.~E.}\ \bibnamefont {{de Mink}}},\
  }\bibfield  {title} {\bibinfo {title} {{Systematic survey of the effects of
  wind mass loss algorithms on the evolution of single massive stars}},\ }\href
  {https://doi.org/10.1051/0004-6361/201730698} {\bibfield  {journal} {\bibinfo
   {journal} {\aap}\ }\textbf {\bibinfo {volume} {603}},\ \bibinfo {eid} {A118}
  (\bibinfo {year} {2017}{\natexlab{b}})},\ \Eprint
  {https://arxiv.org/abs/1703.09705} {arXiv:1703.09705 [astro-ph.SR]}
  \BibitemShut {NoStop}%
\bibitem [{\citenamefont {{Vink}}(2017{\natexlab{b}})}]{Vink:2017}%
  \BibitemOpen
  \bibfield  {author} {\bibinfo {author} {\bibfnamefont {J.~S.}\ \bibnamefont
  {{Vink}}},\ }\bibfield  {title} {\bibinfo {title} {{Winds from stripped
  low-mass helium stars and Wolf-Rayet stars}},\ }\href
  {https://doi.org/10.1051/0004-6361/201731902} {\bibfield  {journal} {\bibinfo
   {journal} {\aap}\ }\textbf {\bibinfo {volume} {607}},\ \bibinfo {eid} {L8}
  (\bibinfo {year} {2017}{\natexlab{b}})},\ \Eprint
  {https://arxiv.org/abs/1710.02010} {arXiv:1710.02010 [astro-ph.SR]}
  \BibitemShut {NoStop}%
\bibitem [{\citenamefont {{de Mink}}\ \emph
  {et~al.}(2008{\natexlab{b}})\citenamefont {{de Mink}}, \citenamefont
  {{Pols}},\ and\ \citenamefont {{Yoon}}}]{deMink:2007}%
  \BibitemOpen
  \bibfield  {author} {\bibinfo {author} {\bibfnamefont {S.~E.}\ \bibnamefont
  {{de Mink}}}, \bibinfo {author} {\bibfnamefont {O.~R.}\ \bibnamefont
  {{Pols}}},\ and\ \bibinfo {author} {\bibfnamefont {S.~C.}\ \bibnamefont
  {{Yoon}}},\ }\bibfield  {title} {\bibinfo {title} {{Binaries at Low
  Metallicity: Ranges For Case A, B and C Mass Transfer}},\ }in\ \href
  {https://doi.org/10.1063/1.2905549} {\emph {\bibinfo {booktitle} {First Stars
  III}}},\ \bibinfo {series} {American Institute of Physics Conference Series},
  Vol.\ \bibinfo {volume} {990},\ \bibinfo {editor} {edited by\ \bibinfo
  {editor} {\bibfnamefont {B.~W.}\ \bibnamefont {{O'Shea}}}\ and\ \bibinfo
  {editor} {\bibfnamefont {A.}~\bibnamefont {{Heger}}}}\ (\bibinfo {year}
  {2008})\ pp.\ \bibinfo {pages} {230--232},\ \Eprint
  {https://arxiv.org/abs/0710.1010} {arXiv:0710.1010 [astro-ph]} \BibitemShut
  {NoStop}%
\bibitem [{\citenamefont {{Mapelli}}\ \emph
  {et~al.}(2020{\natexlab{b}})\citenamefont {{Mapelli}}, \citenamefont
  {{Spera}}, \citenamefont {{Montanari}}, \citenamefont {{Limongi}},
  \citenamefont {{Chieffi}}, \citenamefont {{Giacobbo}}, \citenamefont
  {{Bressan}},\ and\ \citenamefont {{Bouffanais}}}]{Mapelli:2020}%
  \BibitemOpen
  \bibfield  {author} {\bibinfo {author} {\bibfnamefont {M.}~\bibnamefont
  {{Mapelli}}}, \bibinfo {author} {\bibfnamefont {M.}~\bibnamefont {{Spera}}},
  \bibinfo {author} {\bibfnamefont {E.}~\bibnamefont {{Montanari}}}, \bibinfo
  {author} {\bibfnamefont {M.}~\bibnamefont {{Limongi}}}, \bibinfo {author}
  {\bibfnamefont {A.}~\bibnamefont {{Chieffi}}}, \bibinfo {author}
  {\bibfnamefont {N.}~\bibnamefont {{Giacobbo}}}, \bibinfo {author}
  {\bibfnamefont {A.}~\bibnamefont {{Bressan}}},\ and\ \bibinfo {author}
  {\bibfnamefont {Y.}~\bibnamefont {{Bouffanais}}},\ }\bibfield  {title}
  {\bibinfo {title} {{Impact of the Rotation and Compactness of Progenitors on
  the Mass of Black Holes}},\ }\href {https://doi.org/10.3847/1538-4357/ab584d}
  {\bibfield  {journal} {\bibinfo  {journal} {\apj}\ }\textbf {\bibinfo
  {volume} {888}},\ \bibinfo {eid} {76} (\bibinfo {year}
  {2020}{\natexlab{b}})},\ \Eprint {https://arxiv.org/abs/1909.01371}
  {arXiv:1909.01371 [astro-ph.HE]} \BibitemShut {NoStop}%
\bibitem [{\citenamefont {{Kruckow}}\ \emph
  {et~al.}(2018{\natexlab{b}})\citenamefont {{Kruckow}}, \citenamefont
  {{Tauris}}, \citenamefont {{Langer}}, \citenamefont {{Kramer}},\ and\
  \citenamefont {{Izzard}}}]{Kruckow:2018}%
  \BibitemOpen
  \bibfield  {author} {\bibinfo {author} {\bibfnamefont {M.~U.}\ \bibnamefont
  {{Kruckow}}}, \bibinfo {author} {\bibfnamefont {T.~M.}\ \bibnamefont
  {{Tauris}}}, \bibinfo {author} {\bibfnamefont {N.}~\bibnamefont {{Langer}}},
  \bibinfo {author} {\bibfnamefont {M.}~\bibnamefont {{Kramer}}},\ and\
  \bibinfo {author} {\bibfnamefont {R.~G.}\ \bibnamefont {{Izzard}}},\
  }\bibfield  {title} {\bibinfo {title} {{Progenitors of gravitational wave
  mergers: binary evolution with the stellar grid-based code COMBINE}},\ }\href
  {https://doi.org/10.1093/mnras/sty2190} {\bibfield  {journal} {\bibinfo
  {journal} {\mnras}\ }\textbf {\bibinfo {volume} {481}},\ \bibinfo {pages}
  {1908} (\bibinfo {year} {2018}{\natexlab{b}})},\ \Eprint
  {https://arxiv.org/abs/1801.05433} {arXiv:1801.05433 [astro-ph.SR]}
  \BibitemShut {NoStop}%
\bibitem [{\citenamefont {{Bavera}}\ \emph
  {et~al.}(2021{\natexlab{b}})\citenamefont {{Bavera}}, \citenamefont
  {{Fragos}}, \citenamefont {{Zevin}}, \citenamefont {{Berry}}, \citenamefont
  {{Marchant}}, \citenamefont {{Andrews}}, \citenamefont {{Coughlin}},
  \citenamefont {{Dotter}}, \citenamefont {{Kovlakas}}, \citenamefont
  {{Misra}}, \citenamefont {{Serra-Perez}}, \citenamefont {{Qin}},
  \citenamefont {{Rocha}}, \citenamefont {{Rom{\'a}n-Garza}}, \citenamefont
  {{Tran}},\ and\ \citenamefont {{Zapartas}}}]{2020arXiv201016333B}%
  \BibitemOpen
  \bibfield  {author} {\bibinfo {author} {\bibfnamefont {S.~S.}\ \bibnamefont
  {{Bavera}}}, \bibinfo {author} {\bibfnamefont {T.}~\bibnamefont {{Fragos}}},
  \bibinfo {author} {\bibfnamefont {M.}~\bibnamefont {{Zevin}}}, \bibinfo
  {author} {\bibfnamefont {C.~P.~L.}\ \bibnamefont {{Berry}}}, \bibinfo
  {author} {\bibfnamefont {P.}~\bibnamefont {{Marchant}}}, \bibinfo {author}
  {\bibfnamefont {J.~J.}\ \bibnamefont {{Andrews}}}, \bibinfo {author}
  {\bibfnamefont {S.}~\bibnamefont {{Coughlin}}}, \bibinfo {author}
  {\bibfnamefont {A.}~\bibnamefont {{Dotter}}}, \bibinfo {author}
  {\bibfnamefont {K.}~\bibnamefont {{Kovlakas}}}, \bibinfo {author}
  {\bibfnamefont {D.}~\bibnamefont {{Misra}}}, \bibinfo {author} {\bibfnamefont
  {J.~G.}\ \bibnamefont {{Serra-Perez}}}, \bibinfo {author} {\bibfnamefont
  {Y.}~\bibnamefont {{Qin}}}, \bibinfo {author} {\bibfnamefont {K.~A.}\
  \bibnamefont {{Rocha}}}, \bibinfo {author} {\bibfnamefont {J.}~\bibnamefont
  {{Rom{\'a}n-Garza}}}, \bibinfo {author} {\bibfnamefont {N.~H.}\ \bibnamefont
  {{Tran}}},\ and\ \bibinfo {author} {\bibfnamefont {E.}~\bibnamefont
  {{Zapartas}}},\ }\bibfield  {title} {\bibinfo {title} {{The impact of
  mass-transfer physics on the observable properties of field binary black hole
  populations}},\ }\href {https://doi.org/10.1051/0004-6361/202039804}
  {\bibfield  {journal} {\bibinfo  {journal} {\aap}\ }\textbf {\bibinfo
  {volume} {647}},\ \bibinfo {eid} {A153} (\bibinfo {year}
  {2021}{\natexlab{b}})},\ \Eprint {https://arxiv.org/abs/2010.16333}
  {arXiv:2010.16333 [astro-ph.HE]} \BibitemShut {NoStop}%
\bibitem [{\citenamefont {{Romagnolo}}\ \emph
  {et~al.}(2022{\natexlab{b}})\citenamefont {{Romagnolo}}, \citenamefont
  {{Belczynski}}, \citenamefont {{Klencki}}, \citenamefont {{Agrawal}},
  \citenamefont {{Shenar}},\ and\ \citenamefont
  {{Sz{\'e}csi}}}]{Romagnolo:2022}%
  \BibitemOpen
  \bibfield  {author} {\bibinfo {author} {\bibfnamefont {A.}~\bibnamefont
  {{Romagnolo}}}, \bibinfo {author} {\bibfnamefont {K.}~\bibnamefont
  {{Belczynski}}}, \bibinfo {author} {\bibfnamefont {J.}~\bibnamefont
  {{Klencki}}}, \bibinfo {author} {\bibfnamefont {P.}~\bibnamefont
  {{Agrawal}}}, \bibinfo {author} {\bibfnamefont {T.}~\bibnamefont
  {{Shenar}}},\ and\ \bibinfo {author} {\bibfnamefont {D.}~\bibnamefont
  {{Sz{\'e}csi}}},\ }\bibfield  {title} {\bibinfo {title} {{The role of stellar
  expansion on the formation of gravitational wave sources}},\ }\href
  {https://doi.org/10.48550/arXiv.2211.15800} {\bibfield  {journal} {\bibinfo
  {journal} {arXiv e-prints}\ ,\ \bibinfo {eid} {arXiv:2211.15800}} (\bibinfo
  {year} {2022}{\natexlab{b}})},\ \Eprint {https://arxiv.org/abs/2211.15800}
  {arXiv:2211.15800 [astro-ph.HE]} \BibitemShut {NoStop}%
\bibitem [{\citenamefont {{Webbink}}(1984{\natexlab{b}})}]{Webbink:1984}%
  \BibitemOpen
  \bibfield  {author} {\bibinfo {author} {\bibfnamefont {R.~F.}\ \bibnamefont
  {{Webbink}}},\ }\bibfield  {title} {\bibinfo {title} {{Double white dwarfs as
  progenitors of R Coronae Borealis stars and Type I supernovae}},\ }\href
  {https://doi.org/10.1086/161701} {\bibfield  {journal} {\bibinfo  {journal}
  {\apj}\ }\textbf {\bibinfo {volume} {277}},\ \bibinfo {pages} {355} (\bibinfo
  {year} {1984}{\natexlab{b}})}\BibitemShut {NoStop}%
\bibitem [{\citenamefont {{de Kool}}(1990)}]{deKool:1990}%
  \BibitemOpen
  \bibfield  {author} {\bibinfo {author} {\bibfnamefont {M.}~\bibnamefont {{de
  Kool}}},\ }\bibfield  {title} {\bibinfo {title} {{Common Envelope Evolution
  and Double Cores of Planetary Nebulae}},\ }\href
  {https://doi.org/10.1086/168974} {\bibfield  {journal} {\bibinfo  {journal}
  {\apj}\ }\textbf {\bibinfo {volume} {358}},\ \bibinfo {pages} {189} (\bibinfo
  {year} {1990})}\BibitemShut {NoStop}%
\bibitem [{\citenamefont {{Dominik}}\ \emph {et~al.}(2012)\citenamefont
  {{Dominik}}, \citenamefont {{Belczynski}}, \citenamefont {{Fryer}},
  \citenamefont {{Holz}}, \citenamefont {{Berti}}, \citenamefont {{Bulik}},
  \citenamefont {{Mand el}},\ and\ \citenamefont
  {{O'Shaughnessy}}}]{2012ApJ...759...52D}%
  \BibitemOpen
  \bibfield  {author} {\bibinfo {author} {\bibfnamefont {M.}~\bibnamefont
  {{Dominik}}}, \bibinfo {author} {\bibfnamefont {K.}~\bibnamefont
  {{Belczynski}}}, \bibinfo {author} {\bibfnamefont {C.}~\bibnamefont
  {{Fryer}}}, \bibinfo {author} {\bibfnamefont {D.~E.}\ \bibnamefont {{Holz}}},
  \bibinfo {author} {\bibfnamefont {E.}~\bibnamefont {{Berti}}}, \bibinfo
  {author} {\bibfnamefont {T.}~\bibnamefont {{Bulik}}}, \bibinfo {author}
  {\bibfnamefont {I.}~\bibnamefont {{Mand el}}},\ and\ \bibinfo {author}
  {\bibfnamefont {R.}~\bibnamefont {{O'Shaughnessy}}},\ }\bibfield  {title}
  {\bibinfo {title} {{Double Compact Objects. I. The Significance of the Common
  Envelope on Merger Rates}},\ }\href
  {https://doi.org/10.1088/0004-637X/759/1/52} {\bibfield  {journal} {\bibinfo
  {journal} {\apj}\ }\textbf {\bibinfo {volume} {759}},\ \bibinfo {eid} {52}
  (\bibinfo {year} {2012})},\ \Eprint {https://arxiv.org/abs/1202.4901}
  {arXiv:1202.4901 [astro-ph.HE]} \BibitemShut {NoStop}%
\bibitem [{\citenamefont {{Olejak}}\ \emph
  {et~al.}(2021{\natexlab{b}})\citenamefont {{Olejak}}, \citenamefont
  {{Belczynski}},\ and\ \citenamefont {{Ivanova}}}]{2021arXiv210205649O}%
  \BibitemOpen
  \bibfield  {author} {\bibinfo {author} {\bibfnamefont {A.}~\bibnamefont
  {{Olejak}}}, \bibinfo {author} {\bibfnamefont {K.}~\bibnamefont
  {{Belczynski}}},\ and\ \bibinfo {author} {\bibfnamefont {N.}~\bibnamefont
  {{Ivanova}}},\ }\bibfield  {title} {\bibinfo {title} {{Impact of common
  envelope development criteria on the formation of LIGO/Virgo sources}},\
  }\href {https://doi.org/10.1051/0004-6361/202140520} {\bibfield  {journal}
  {\bibinfo  {journal} {\aap}\ }\textbf {\bibinfo {volume} {651}},\ \bibinfo
  {eid} {A100} (\bibinfo {year} {2021}{\natexlab{b}})},\ \Eprint
  {https://arxiv.org/abs/2102.05649} {arXiv:2102.05649 [astro-ph.HE]}
  \BibitemShut {NoStop}%
\bibitem [{\citenamefont {{Vigna-G{\'o}mez}}\ \emph {et~al.}(2020)\citenamefont
  {{Vigna-G{\'o}mez}}, \citenamefont {{MacLeod}}, \citenamefont {{Neijssel}},
  \citenamefont {{Broekgaarden}}, \citenamefont {{Justham}}, \citenamefont
  {{Howitt}}, \citenamefont {{de Mink}}, \citenamefont {{Vinciguerra}},\ and\
  \citenamefont {{Mandel}}}]{2020PASA...37...38V}%
  \BibitemOpen
  \bibfield  {author} {\bibinfo {author} {\bibfnamefont {A.}~\bibnamefont
  {{Vigna-G{\'o}mez}}}, \bibinfo {author} {\bibfnamefont {M.}~\bibnamefont
  {{MacLeod}}}, \bibinfo {author} {\bibfnamefont {C.~J.}\ \bibnamefont
  {{Neijssel}}}, \bibinfo {author} {\bibfnamefont {F.~S.}\ \bibnamefont
  {{Broekgaarden}}}, \bibinfo {author} {\bibfnamefont {S.}~\bibnamefont
  {{Justham}}}, \bibinfo {author} {\bibfnamefont {G.}~\bibnamefont {{Howitt}}},
  \bibinfo {author} {\bibfnamefont {S.~E.}\ \bibnamefont {{de Mink}}}, \bibinfo
  {author} {\bibfnamefont {S.}~\bibnamefont {{Vinciguerra}}},\ and\ \bibinfo
  {author} {\bibfnamefont {I.}~\bibnamefont {{Mandel}}},\ }\bibfield  {title}
  {\bibinfo {title} {{Common envelope episodes that lead to double neutron star
  formation}},\ }\href {https://doi.org/10.1017/pasa.2020.31} {\bibfield
  {journal} {\bibinfo  {journal} {\pasa}\ }\textbf {\bibinfo {volume} {37}},\
  \bibinfo {eid} {e038} (\bibinfo {year} {2020})},\ \Eprint
  {https://arxiv.org/abs/2001.09829} {arXiv:2001.09829 [astro-ph.SR]}
  \BibitemShut {NoStop}%
\bibitem [{\citenamefont {{Ivanova}}\ \emph
  {et~al.}(2020{\natexlab{b}})\citenamefont {{Ivanova}}, \citenamefont
  {{Justham}},\ and\ \citenamefont {{Ricker}}}]{Ivanova:2020book}%
  \BibitemOpen
  \bibfield  {author} {\bibinfo {author} {\bibfnamefont {N.}~\bibnamefont
  {{Ivanova}}}, \bibinfo {author} {\bibfnamefont {S.}~\bibnamefont
  {{Justham}}},\ and\ \bibinfo {author} {\bibfnamefont {P.}~\bibnamefont
  {{Ricker}}},\ }\href {https://doi.org/10.1088/2514-3433/abb6f0} {\emph
  {\bibinfo {title} {{Common Envelope Evolution}}}}\ (\bibinfo {year}
  {2020})\BibitemShut {NoStop}%
\bibitem [{\citenamefont {{Kippenhahn}}\ and\ \citenamefont
  {{Meyer-Hofmeister}}(1977{\natexlab{b}})}]{KippenhahnMeyerHofmeister:1977}%
  \BibitemOpen
  \bibfield  {author} {\bibinfo {author} {\bibfnamefont {R.}~\bibnamefont
  {{Kippenhahn}}}\ and\ \bibinfo {author} {\bibfnamefont {E.}~\bibnamefont
  {{Meyer-Hofmeister}}},\ }\bibfield  {title} {\bibinfo {title} {{On the radii
  of accreting main sequence stars.}},\ }\href@noop {} {\bibfield  {journal}
  {\bibinfo  {journal} {\aap}\ }\textbf {\bibinfo {volume} {54}},\ \bibinfo
  {pages} {539} (\bibinfo {year} {1977}{\natexlab{b}})}\BibitemShut {NoStop}%
\bibitem [{\citenamefont {{Soberman}}\ \emph
  {et~al.}(1997{\natexlab{b}})\citenamefont {{Soberman}}, \citenamefont
  {{Phinney}},\ and\ \citenamefont {{van den Heuvel}}}]{Soberman:1997}%
  \BibitemOpen
  \bibfield  {author} {\bibinfo {author} {\bibfnamefont {G.~E.}\ \bibnamefont
  {{Soberman}}}, \bibinfo {author} {\bibfnamefont {E.~S.}\ \bibnamefont
  {{Phinney}}},\ and\ \bibinfo {author} {\bibfnamefont {E.~P.~J.}\ \bibnamefont
  {{van den Heuvel}}},\ }\bibfield  {title} {\bibinfo {title} {{Stability
  criteria for mass transfer in binary stellar evolution.}},\ }\href@noop {}
  {\bibfield  {journal} {\bibinfo  {journal} {\aap}\ }\textbf {\bibinfo
  {volume} {327}},\ \bibinfo {pages} {620} (\bibinfo {year}
  {1997}{\natexlab{b}})},\ \Eprint {https://arxiv.org/abs/astro-ph/9703016}
  {arXiv:astro-ph/9703016 [astro-ph]} \BibitemShut {NoStop}%
\bibitem [{\citenamefont {{Brown}}\ \emph {et~al.}(2001)\citenamefont
  {{Brown}}, \citenamefont {{Heger}}, \citenamefont {{Langer}}, \citenamefont
  {{Lee}}, \citenamefont {{Wellstein}},\ and\ \citenamefont
  {{Bethe}}}]{Brown:2001}%
  \BibitemOpen
  \bibfield  {author} {\bibinfo {author} {\bibfnamefont {G.~E.}\ \bibnamefont
  {{Brown}}}, \bibinfo {author} {\bibfnamefont {A.}~\bibnamefont {{Heger}}},
  \bibinfo {author} {\bibfnamefont {N.}~\bibnamefont {{Langer}}}, \bibinfo
  {author} {\bibfnamefont {C.~H.}\ \bibnamefont {{Lee}}}, \bibinfo {author}
  {\bibfnamefont {S.}~\bibnamefont {{Wellstein}}},\ and\ \bibinfo {author}
  {\bibfnamefont {H.~A.}\ \bibnamefont {{Bethe}}},\ }\bibfield  {title}
  {\bibinfo {title} {{Formation of high mass X-ray black hole binaries}},\
  }\href {https://doi.org/10.1016/S1384-1076(01)00077-X} {\bibfield  {journal}
  {\bibinfo  {journal} {\na}\ }\textbf {\bibinfo {volume} {6}},\ \bibinfo
  {pages} {457} (\bibinfo {year} {2001})},\ \Eprint
  {https://arxiv.org/abs/astro-ph/0102379} {arXiv:astro-ph/0102379 [astro-ph]}
  \BibitemShut {NoStop}%
\bibitem [{\citenamefont {{Sepinsky}}\ \emph
  {et~al.}(2010{\natexlab{b}})\citenamefont {{Sepinsky}}, \citenamefont
  {{Willems}}, \citenamefont {{Kalogera}},\ and\ \citenamefont
  {{Rasio}}}]{Sepinsky:2010}%
  \BibitemOpen
  \bibfield  {author} {\bibinfo {author} {\bibfnamefont {J.~F.}\ \bibnamefont
  {{Sepinsky}}}, \bibinfo {author} {\bibfnamefont {B.}~\bibnamefont
  {{Willems}}}, \bibinfo {author} {\bibfnamefont {V.}~\bibnamefont
  {{Kalogera}}},\ and\ \bibinfo {author} {\bibfnamefont {F.~A.}\ \bibnamefont
  {{Rasio}}},\ }\bibfield  {title} {\bibinfo {title} {{Interacting Binaries
  with Eccentric Orbits. III. Orbital Evolution due to Direct Impact and
  Self-Accretion}},\ }\href {https://doi.org/10.1088/0004-637X/724/1/546}
  {\bibfield  {journal} {\bibinfo  {journal} {\apj}\ }\textbf {\bibinfo
  {volume} {724}},\ \bibinfo {pages} {546} (\bibinfo {year}
  {2010}{\natexlab{b}})},\ \Eprint {https://arxiv.org/abs/1005.0625}
  {arXiv:1005.0625 [astro-ph.SR]} \BibitemShut {NoStop}%
\bibitem [{\citenamefont {Claeys}\ \emph {et~al.}(2014)\citenamefont {Claeys},
  \citenamefont {Pols}, \citenamefont {Izzard}, \citenamefont {Vink},\ and\
  \citenamefont {Verbunt}}]{claeys2014theoretical}%
  \BibitemOpen
  \bibfield  {author} {\bibinfo {author} {\bibfnamefont {J.}~\bibnamefont
  {Claeys}}, \bibinfo {author} {\bibfnamefont {O.}~\bibnamefont {Pols}},
  \bibinfo {author} {\bibfnamefont {R.}~\bibnamefont {Izzard}}, \bibinfo
  {author} {\bibfnamefont {J.}~\bibnamefont {Vink}},\ and\ \bibinfo {author}
  {\bibfnamefont {F.}~\bibnamefont {Verbunt}},\ }\bibfield  {title} {\bibinfo
  {title} {Theoretical uncertainties of the type ia supernova rate},\
  }\href@noop {} {\bibfield  {journal} {\bibinfo  {journal} {\aap}\ }\textbf
  {\bibinfo {volume} {563}},\ \bibinfo {pages} {A83} (\bibinfo {year}
  {2014})}\BibitemShut {NoStop}%
\bibitem [{\citenamefont {Ge}\ \emph {et~al.}(2015)\citenamefont {Ge},
  \citenamefont {Webbink}, \citenamefont {Chen},\ and\ \citenamefont
  {Han}}]{ge2015adiabatic}%
  \BibitemOpen
  \bibfield  {author} {\bibinfo {author} {\bibfnamefont {H.}~\bibnamefont
  {Ge}}, \bibinfo {author} {\bibfnamefont {R.~F.}\ \bibnamefont {Webbink}},
  \bibinfo {author} {\bibfnamefont {X.}~\bibnamefont {Chen}},\ and\ \bibinfo
  {author} {\bibfnamefont {Z.}~\bibnamefont {Han}},\ }\bibfield  {title}
  {\bibinfo {title} {Adiabatic mass loss in binary stars. ii. from zero-age
  main sequence to the base of the giant branch},\ }\href@noop {} {\bibfield
  {journal} {\bibinfo  {journal} {The Astrophysical Journal}\ }\textbf
  {\bibinfo {volume} {812}},\ \bibinfo {pages} {40} (\bibinfo {year}
  {2015})}\BibitemShut {NoStop}%
\bibitem [{\citenamefont {{Dosopoulou}}\ and\ \citenamefont
  {{Kalogera}}(2016{\natexlab{c}})}]{Dosopoulou:2016}%
  \BibitemOpen
  \bibfield  {author} {\bibinfo {author} {\bibfnamefont {F.}~\bibnamefont
  {{Dosopoulou}}}\ and\ \bibinfo {author} {\bibfnamefont {V.}~\bibnamefont
  {{Kalogera}}},\ }\bibfield  {title} {\bibinfo {title} {{Orbital Evolution of
  Mass-transferring Eccentric Binary Systems. II. Secular Evolution}},\ }\href
  {https://doi.org/10.3847/0004-637X/825/1/71} {\bibfield  {journal} {\bibinfo
  {journal} {\apj}\ }\textbf {\bibinfo {volume} {825}},\ \bibinfo {eid} {71}
  (\bibinfo {year} {2016}{\natexlab{c}})},\ \Eprint
  {https://arxiv.org/abs/1603.06593} {arXiv:1603.06593 [astro-ph.SR]}
  \BibitemShut {NoStop}%
\bibitem [{\citenamefont {{Menon}}\ \emph {et~al.}(2021)\citenamefont
  {{Menon}}, \citenamefont {{Langer}}, \citenamefont {{de Mink}}, \citenamefont
  {{Justham}}, \citenamefont {{Sen}}, \citenamefont {{Sz{\'e}csi}},
  \citenamefont {{de Koter}}, \citenamefont {{Abdul-Masih}}, \citenamefont
  {{Sana}}, \citenamefont {{Mahy}},\ and\ \citenamefont
  {{Marchant}}}]{Menon:2021}%
  \BibitemOpen
  \bibfield  {author} {\bibinfo {author} {\bibfnamefont {A.}~\bibnamefont
  {{Menon}}}, \bibinfo {author} {\bibfnamefont {N.}~\bibnamefont {{Langer}}},
  \bibinfo {author} {\bibfnamefont {S.~E.}\ \bibnamefont {{de Mink}}}, \bibinfo
  {author} {\bibfnamefont {S.}~\bibnamefont {{Justham}}}, \bibinfo {author}
  {\bibfnamefont {K.}~\bibnamefont {{Sen}}}, \bibinfo {author} {\bibfnamefont
  {D.}~\bibnamefont {{Sz{\'e}csi}}}, \bibinfo {author} {\bibfnamefont
  {A.}~\bibnamefont {{de Koter}}}, \bibinfo {author} {\bibfnamefont
  {M.}~\bibnamefont {{Abdul-Masih}}}, \bibinfo {author} {\bibfnamefont
  {H.}~\bibnamefont {{Sana}}}, \bibinfo {author} {\bibfnamefont
  {L.}~\bibnamefont {{Mahy}}},\ and\ \bibinfo {author} {\bibfnamefont
  {P.}~\bibnamefont {{Marchant}}},\ }\bibfield  {title} {\bibinfo {title}
  {{Detailed evolutionary models of massive contact binaries - I. Model grids
  and synthetic populations for the Magellanic Clouds}},\ }\href
  {https://doi.org/10.1093/mnras/stab2276} {\bibfield  {journal} {\bibinfo
  {journal} {\mnras}\ }\textbf {\bibinfo {volume} {507}},\ \bibinfo {pages}
  {5013} (\bibinfo {year} {2021})},\ \Eprint {https://arxiv.org/abs/2011.13459}
  {arXiv:2011.13459 [astro-ph.SR]} \BibitemShut {NoStop}%
\bibitem [{\citenamefont {{Bouffanais}}\ \emph {et~al.}(2021)\citenamefont
  {{Bouffanais}}, \citenamefont {{Mapelli}}, \citenamefont {{Santoliquido}},
  \citenamefont {{Giacobbo}}, \citenamefont {{Iorio}},\ and\ \citenamefont
  {{Costa}}}]{2021MNRAS.505.3873B}%
  \BibitemOpen
  \bibfield  {author} {\bibinfo {author} {\bibfnamefont {Y.}~\bibnamefont
  {{Bouffanais}}}, \bibinfo {author} {\bibfnamefont {M.}~\bibnamefont
  {{Mapelli}}}, \bibinfo {author} {\bibfnamefont {F.}~\bibnamefont
  {{Santoliquido}}}, \bibinfo {author} {\bibfnamefont {N.}~\bibnamefont
  {{Giacobbo}}}, \bibinfo {author} {\bibfnamefont {G.}~\bibnamefont
  {{Iorio}}},\ and\ \bibinfo {author} {\bibfnamefont {G.}~\bibnamefont
  {{Costa}}},\ }\bibfield  {title} {\bibinfo {title} {{Constraining accretion
  efficiency in massive binary stars with LIGO -Virgo black holes}},\ }\href
  {https://doi.org/10.1093/mnras/stab1589} {\bibfield  {journal} {\bibinfo
  {journal} {\mnras}\ }\textbf {\bibinfo {volume} {505}},\ \bibinfo {pages}
  {3873} (\bibinfo {year} {2021})},\ \Eprint {https://arxiv.org/abs/2010.11220}
  {arXiv:2010.11220 [astro-ph.HE]} \BibitemShut {NoStop}%
\bibitem [{\citenamefont {{Vinciguerra}}\ \emph {et~al.}(2020)\citenamefont
  {{Vinciguerra}}, \citenamefont {{Neijssel}}, \citenamefont
  {{Vigna-G{\'o}mez}}, \citenamefont {{Mandel}}, \citenamefont
  {{Podsiadlowski}}, \citenamefont {{Maccarone}}, \citenamefont {{Nicholl}},
  \citenamefont {{Kingdon}}, \citenamefont {{Perry}},\ and\ \citenamefont
  {{Salemi}}}]{Vinciguerra:2020}%
  \BibitemOpen
  \bibfield  {author} {\bibinfo {author} {\bibfnamefont {S.}~\bibnamefont
  {{Vinciguerra}}}, \bibinfo {author} {\bibfnamefont {C.~J.}\ \bibnamefont
  {{Neijssel}}}, \bibinfo {author} {\bibfnamefont {A.}~\bibnamefont
  {{Vigna-G{\'o}mez}}}, \bibinfo {author} {\bibfnamefont {I.}~\bibnamefont
  {{Mandel}}}, \bibinfo {author} {\bibfnamefont {P.}~\bibnamefont
  {{Podsiadlowski}}}, \bibinfo {author} {\bibfnamefont {T.~J.}\ \bibnamefont
  {{Maccarone}}}, \bibinfo {author} {\bibfnamefont {M.}~\bibnamefont
  {{Nicholl}}}, \bibinfo {author} {\bibfnamefont {S.}~\bibnamefont
  {{Kingdon}}}, \bibinfo {author} {\bibfnamefont {A.}~\bibnamefont {{Perry}}},\
  and\ \bibinfo {author} {\bibfnamefont {F.}~\bibnamefont {{Salemi}}},\
  }\bibfield  {title} {\bibinfo {title} {{Be X-ray binaries in the SMC as
  indicators of mass-transfer efficiency}},\ }\href
  {https://doi.org/10.1093/mnras/staa2177} {\bibfield  {journal} {\bibinfo
  {journal} {\mnras}\ }\textbf {\bibinfo {volume} {498}},\ \bibinfo {pages}
  {4705} (\bibinfo {year} {2020})},\ \Eprint {https://arxiv.org/abs/2003.00195}
  {arXiv:2003.00195 [astro-ph.HE]} \BibitemShut {NoStop}%
\bibitem [{\citenamefont {{Schneider}}\ \emph
  {et~al.}(2021{\natexlab{b}})\citenamefont {{Schneider}}, \citenamefont
  {{Podsiadlowski}},\ and\ \citenamefont {{M{\"u}ller}}}]{Schneider:2020}%
  \BibitemOpen
  \bibfield  {author} {\bibinfo {author} {\bibfnamefont {F.~R.~N.}\
  \bibnamefont {{Schneider}}}, \bibinfo {author} {\bibfnamefont
  {P.}~\bibnamefont {{Podsiadlowski}}},\ and\ \bibinfo {author} {\bibfnamefont
  {B.}~\bibnamefont {{M{\"u}ller}}},\ }\bibfield  {title} {\bibinfo {title}
  {{Pre-supernova evolution, compact-object masses, and explosion properties of
  stripped binary stars}},\ }\href
  {https://doi.org/10.1051/0004-6361/202039219} {\bibfield  {journal} {\bibinfo
   {journal} {\aap}\ }\textbf {\bibinfo {volume} {645}},\ \bibinfo {eid} {A5}
  (\bibinfo {year} {2021}{\natexlab{b}})},\ \Eprint
  {https://arxiv.org/abs/2008.08599} {arXiv:2008.08599 [astro-ph.SR]}
  \BibitemShut {NoStop}%
\bibitem [{\citenamefont {{Belczynski}}\ \emph {et~al.}(2021)\citenamefont
  {{Belczynski}}, \citenamefont {{Romagnolo}}, \citenamefont {{Olejak}},
  \citenamefont {{Klencki}}, \citenamefont {{Chattopadhyay}}, \citenamefont
  {{Stevenson}}, \citenamefont {{Miller}}, \citenamefont {{Lasota}},\ and\
  \citenamefont {{Crowther}}}]{Belczynski:2021-uncertainStellarEvolution}%
  \BibitemOpen
  \bibfield  {author} {\bibinfo {author} {\bibfnamefont {K.}~\bibnamefont
  {{Belczynski}}}, \bibinfo {author} {\bibfnamefont {A.}~\bibnamefont
  {{Romagnolo}}}, \bibinfo {author} {\bibfnamefont {A.}~\bibnamefont
  {{Olejak}}}, \bibinfo {author} {\bibfnamefont {J.}~\bibnamefont {{Klencki}}},
  \bibinfo {author} {\bibfnamefont {D.}~\bibnamefont {{Chattopadhyay}}},
  \bibinfo {author} {\bibfnamefont {S.}~\bibnamefont {{Stevenson}}}, \bibinfo
  {author} {\bibfnamefont {M.~C.}\ \bibnamefont {{Miller}}}, \bibinfo {author}
  {\bibfnamefont {J.~P.}\ \bibnamefont {{Lasota}}},\ and\ \bibinfo {author}
  {\bibfnamefont {P.~A.}\ \bibnamefont {{Crowther}}},\ }\bibfield  {title}
  {\bibinfo {title} {{The Uncertain Future of Massive Binaries Obscures the
  Origin of LIGO/Virgo Sources}},\ }\href@noop {} {\bibfield  {journal}
  {\bibinfo  {journal} {arXiv e-prints}\ ,\ \bibinfo {eid} {arXiv:2108.10885}}
  (\bibinfo {year} {2021})},\ \Eprint {https://arxiv.org/abs/2108.10885}
  {arXiv:2108.10885 [astro-ph.HE]} \BibitemShut {NoStop}%
\bibitem [{\citenamefont {{Cehula}}\ and\ \citenamefont
  {{Pejcha}}(2023)}]{Cehula:2023}%
  \BibitemOpen
  \bibfield  {author} {\bibinfo {author} {\bibfnamefont {J.}~\bibnamefont
  {{Cehula}}}\ and\ \bibinfo {author} {\bibfnamefont {O.}~\bibnamefont
  {{Pejcha}}},\ }\bibfield  {title} {\bibinfo {title} {{A theory of mass
  transfer in binary stars}},\ }\href
  {https://doi.org/10.48550/arXiv.2303.05526} {\bibfield  {journal} {\bibinfo
  {journal} {arXiv e-prints}\ ,\ \bibinfo {eid} {arXiv:2303.05526}} (\bibinfo
  {year} {2023})},\ \Eprint {https://arxiv.org/abs/2303.05526}
  {arXiv:2303.05526 [astro-ph.SR]} \BibitemShut {NoStop}%
\bibitem [{\citenamefont {{Katz}}(1975)}]{1975Natur.253..698K}%
  \BibitemOpen
  \bibfield  {author} {\bibinfo {author} {\bibfnamefont {J.~I.}\ \bibnamefont
  {{Katz}}},\ }\bibfield  {title} {\bibinfo {title} {{Two kinds of stellar
  collapse}},\ }\href {https://doi.org/10.1038/253698a0} {\bibfield  {journal}
  {\bibinfo  {journal} {\nat}\ }\textbf {\bibinfo {volume} {253}},\ \bibinfo
  {pages} {698} (\bibinfo {year} {1975})}\BibitemShut {NoStop}%
\bibitem [{\citenamefont {{Brisken}}\ \emph {et~al.}(2002)\citenamefont
  {{Brisken}}, \citenamefont {{Benson}}, \citenamefont {{Goss}},\ and\
  \citenamefont {{Thorsett}}}]{2002ApJ...571..906B}%
  \BibitemOpen
  \bibfield  {author} {\bibinfo {author} {\bibfnamefont {W.~F.}\ \bibnamefont
  {{Brisken}}}, \bibinfo {author} {\bibfnamefont {J.~M.}\ \bibnamefont
  {{Benson}}}, \bibinfo {author} {\bibfnamefont {W.~M.}\ \bibnamefont
  {{Goss}}},\ and\ \bibinfo {author} {\bibfnamefont {S.~E.}\ \bibnamefont
  {{Thorsett}}},\ }\bibfield  {title} {\bibinfo {title} {{Very Long Baseline
  Array Measurement of Nine Pulsar Parallaxes}},\ }\href
  {https://doi.org/10.1086/340098} {\bibfield  {journal} {\bibinfo  {journal}
  {\apj}\ }\textbf {\bibinfo {volume} {571}},\ \bibinfo {pages} {906} (\bibinfo
  {year} {2002})},\ \Eprint {https://arxiv.org/abs/astro-ph/0204105}
  {arXiv:astro-ph/0204105 [astro-ph]} \BibitemShut {NoStop}%
\bibitem [{\citenamefont {{Pfahl}}\ \emph
  {et~al.}(2002{\natexlab{b}})\citenamefont {{Pfahl}}, \citenamefont
  {{Rappaport}},\ and\ \citenamefont {{Podsiadlowski}}}]{2002ApJ...571L..37P}%
  \BibitemOpen
  \bibfield  {author} {\bibinfo {author} {\bibfnamefont {E.}~\bibnamefont
  {{Pfahl}}}, \bibinfo {author} {\bibfnamefont {S.}~\bibnamefont
  {{Rappaport}}},\ and\ \bibinfo {author} {\bibfnamefont {P.}~\bibnamefont
  {{Podsiadlowski}}},\ }\bibfield  {title} {\bibinfo {title} {{On the
  Population of Wind-accreting Neutron Stars in the Galaxy}},\ }\href
  {https://doi.org/10.1086/341197} {\bibfield  {journal} {\bibinfo  {journal}
  {\apjl}\ }\textbf {\bibinfo {volume} {571}},\ \bibinfo {pages} {L37}
  (\bibinfo {year} {2002}{\natexlab{b}})},\ \Eprint
  {https://arxiv.org/abs/astro-ph/0203266} {arXiv:astro-ph/0203266 [astro-ph]}
  \BibitemShut {NoStop}%
\bibitem [{\citenamefont {{Schwab}}\ \emph {et~al.}(2010)\citenamefont
  {{Schwab}}, \citenamefont {{Podsiadlowski}},\ and\ \citenamefont
  {{Rappaport}}}]{2010ApJ...719..722S}%
  \BibitemOpen
  \bibfield  {author} {\bibinfo {author} {\bibfnamefont {J.}~\bibnamefont
  {{Schwab}}}, \bibinfo {author} {\bibfnamefont {P.}~\bibnamefont
  {{Podsiadlowski}}},\ and\ \bibinfo {author} {\bibfnamefont {S.}~\bibnamefont
  {{Rappaport}}},\ }\bibfield  {title} {\bibinfo {title} {{Further Evidence for
  the Bimodal Distribution of Neutron-star Masses}},\ }\href
  {https://doi.org/10.1088/0004-637X/719/1/722} {\bibfield  {journal} {\bibinfo
   {journal} {\apj}\ }\textbf {\bibinfo {volume} {719}},\ \bibinfo {pages}
  {722} (\bibinfo {year} {2010})},\ \Eprint {https://arxiv.org/abs/1006.4584}
  {arXiv:1006.4584 [astro-ph.HE]} \BibitemShut {NoStop}%
\bibitem [{\citenamefont {{Suwa}}\ \emph {et~al.}(2015)\citenamefont {{Suwa}},
  \citenamefont {{Yoshida}}, \citenamefont {{Shibata}}, \citenamefont
  {{Umeda}},\ and\ \citenamefont {{Takahashi}}}]{2015MNRAS.454.3073S}%
  \BibitemOpen
  \bibfield  {author} {\bibinfo {author} {\bibfnamefont {Y.}~\bibnamefont
  {{Suwa}}}, \bibinfo {author} {\bibfnamefont {T.}~\bibnamefont {{Yoshida}}},
  \bibinfo {author} {\bibfnamefont {M.}~\bibnamefont {{Shibata}}}, \bibinfo
  {author} {\bibfnamefont {H.}~\bibnamefont {{Umeda}}},\ and\ \bibinfo {author}
  {\bibfnamefont {K.}~\bibnamefont {{Takahashi}}},\ }\bibfield  {title}
  {\bibinfo {title} {{Neutrino-driven explosions of ultra-stripped Type Ic
  supernovae generating binary neutron stars}},\ }\href
  {https://doi.org/10.1093/mnras/stv2195} {\bibfield  {journal} {\bibinfo
  {journal} {\mnras}\ }\textbf {\bibinfo {volume} {454}},\ \bibinfo {pages}
  {3073} (\bibinfo {year} {2015})},\ \Eprint {https://arxiv.org/abs/1506.08827}
  {arXiv:1506.08827 [astro-ph.HE]} \BibitemShut {NoStop}%
\bibitem [{\citenamefont {{Beniamini}}\ and\ \citenamefont
  {{Piran}}(2016)}]{2016MNRAS.456.4089B}%
  \BibitemOpen
  \bibfield  {author} {\bibinfo {author} {\bibfnamefont {P.}~\bibnamefont
  {{Beniamini}}}\ and\ \bibinfo {author} {\bibfnamefont {T.}~\bibnamefont
  {{Piran}}},\ }\bibfield  {title} {\bibinfo {title} {{Formation of double
  neutron star systems as implied by observations}},\ }\href
  {https://doi.org/10.1093/mnras/stv2903} {\bibfield  {journal} {\bibinfo
  {journal} {\mnras}\ }\textbf {\bibinfo {volume} {456}},\ \bibinfo {pages}
  {4089} (\bibinfo {year} {2016})},\ \Eprint {https://arxiv.org/abs/1510.03111}
  {arXiv:1510.03111 [astro-ph.HE]} \BibitemShut {NoStop}%
\bibitem [{\citenamefont {{Tauris}}\ \emph {et~al.}(2017)\citenamefont
  {{Tauris}}, \citenamefont {{Kramer}}, \citenamefont {{Freire}}, \citenamefont
  {{Wex}}, \citenamefont {{Janka}}, \citenamefont {{Langer}}, \citenamefont
  {{Podsiadlowski}}, \citenamefont {{Bozzo}}, \citenamefont {{Chaty}},
  \citenamefont {{Kruckow}}, \citenamefont {{van den Heuvel}}, \citenamefont
  {{Antoniadis}}, \citenamefont {{Breton}},\ and\ \citenamefont
  {{Champion}}}]{2017ApJ...846..170T}%
  \BibitemOpen
  \bibfield  {author} {\bibinfo {author} {\bibfnamefont {T.~M.}\ \bibnamefont
  {{Tauris}}}, \bibinfo {author} {\bibfnamefont {M.}~\bibnamefont {{Kramer}}},
  \bibinfo {author} {\bibfnamefont {P.~C.~C.}\ \bibnamefont {{Freire}}},
  \bibinfo {author} {\bibfnamefont {N.}~\bibnamefont {{Wex}}}, \bibinfo
  {author} {\bibfnamefont {H.~T.}\ \bibnamefont {{Janka}}}, \bibinfo {author}
  {\bibfnamefont {N.}~\bibnamefont {{Langer}}}, \bibinfo {author}
  {\bibfnamefont {P.}~\bibnamefont {{Podsiadlowski}}}, \bibinfo {author}
  {\bibfnamefont {E.}~\bibnamefont {{Bozzo}}}, \bibinfo {author} {\bibfnamefont
  {S.}~\bibnamefont {{Chaty}}}, \bibinfo {author} {\bibfnamefont {M.~U.}\
  \bibnamefont {{Kruckow}}}, \bibinfo {author} {\bibfnamefont {E.~P.~J.}\
  \bibnamefont {{van den Heuvel}}}, \bibinfo {author} {\bibfnamefont
  {J.}~\bibnamefont {{Antoniadis}}}, \bibinfo {author} {\bibfnamefont {R.~P.}\
  \bibnamefont {{Breton}}},\ and\ \bibinfo {author} {\bibfnamefont {D.~J.}\
  \bibnamefont {{Champion}}},\ }\bibfield  {title} {\bibinfo {title}
  {{Formation of Double Neutron Star Systems}},\ }\href
  {https://doi.org/10.3847/1538-4357/aa7e89} {\bibfield  {journal} {\bibinfo
  {journal} {\apj}\ }\textbf {\bibinfo {volume} {846}},\ \bibinfo {eid} {170}
  (\bibinfo {year} {2017})},\ \Eprint {https://arxiv.org/abs/1706.09438}
  {arXiv:1706.09438 [astro-ph.HE]} \BibitemShut {NoStop}%
\bibitem [{\citenamefont {{M{\"u}ller}}\ \emph {et~al.}(2018)\citenamefont
  {{M{\"u}ller}}, \citenamefont {{Gay}}, \citenamefont {{Heger}}, \citenamefont
  {{Tauris}},\ and\ \citenamefont {{Sim}}}]{2018MNRAS.479.3675M}%
  \BibitemOpen
  \bibfield  {author} {\bibinfo {author} {\bibfnamefont {B.}~\bibnamefont
  {{M{\"u}ller}}}, \bibinfo {author} {\bibfnamefont {D.~W.}\ \bibnamefont
  {{Gay}}}, \bibinfo {author} {\bibfnamefont {A.}~\bibnamefont {{Heger}}},
  \bibinfo {author} {\bibfnamefont {T.~M.}\ \bibnamefont {{Tauris}}},\ and\
  \bibinfo {author} {\bibfnamefont {S.~A.}\ \bibnamefont {{Sim}}},\ }\bibfield
  {title} {\bibinfo {title} {{Multidimensional simulations of ultrastripped
  supernovae to shock breakout}},\ }\href
  {https://doi.org/10.1093/mnras/sty1683} {\bibfield  {journal} {\bibinfo
  {journal} {\mnras}\ }\textbf {\bibinfo {volume} {479}},\ \bibinfo {pages}
  {3675} (\bibinfo {year} {2018})},\ \Eprint {https://arxiv.org/abs/1803.03388}
  {arXiv:1803.03388 [astro-ph.SR]} \BibitemShut {NoStop}%
\bibitem [{\citenamefont {{Igoshev}}(2020{\natexlab{b}})}]{Igoshev:2020}%
  \BibitemOpen
  \bibfield  {author} {\bibinfo {author} {\bibfnamefont {A.~P.}\ \bibnamefont
  {{Igoshev}}},\ }\bibfield  {title} {\bibinfo {title} {{The observed velocity
  distribution of young pulsars - II. Analysis of complete
  PSR{\ensuremath{\pi}}}},\ }\href {https://doi.org/10.1093/mnras/staa958}
  {\bibfield  {journal} {\bibinfo  {journal} {\mnras}\ }\textbf {\bibinfo
  {volume} {494}},\ \bibinfo {pages} {3663} (\bibinfo {year}
  {2020}{\natexlab{b}})},\ \Eprint {https://arxiv.org/abs/2002.01367}
  {arXiv:2002.01367 [astro-ph.HE]} \BibitemShut {NoStop}%
\bibitem [{\citenamefont {{Kalogera}}\ and\ \citenamefont
  {{Baym}}(1996)}]{1996ApJ...470L..61K}%
  \BibitemOpen
  \bibfield  {author} {\bibinfo {author} {\bibfnamefont {V.}~\bibnamefont
  {{Kalogera}}}\ and\ \bibinfo {author} {\bibfnamefont {G.}~\bibnamefont
  {{Baym}}},\ }\bibfield  {title} {\bibinfo {title} {{The Maximum Mass of a
  Neutron Star}},\ }\href {https://doi.org/10.1086/310296} {\bibfield
  {journal} {\bibinfo  {journal} {\apjl}\ }\textbf {\bibinfo {volume} {470}},\
  \bibinfo {pages} {L61} (\bibinfo {year} {1996})},\ \Eprint
  {https://arxiv.org/abs/astro-ph/9608059} {arXiv:astro-ph/9608059 [astro-ph]}
  \BibitemShut {NoStop}%
\bibitem [{\citenamefont {Janka}(2012)}]{janka2012explosion}%
  \BibitemOpen
  \bibfield  {author} {\bibinfo {author} {\bibfnamefont {H.-T.}\ \bibnamefont
  {Janka}},\ }\bibfield  {title} {\bibinfo {title} {Explosion mechanisms of
  core-collapse supernovae},\ }\href@noop {} {\bibfield  {journal} {\bibinfo
  {journal} {Annual Review of Nuclear and Particle Science}\ }\textbf {\bibinfo
  {volume} {62}},\ \bibinfo {pages} {407} (\bibinfo {year} {2012})}\BibitemShut
  {NoStop}%
\bibitem [{\citenamefont {{Fryer}}\ \emph {et~al.}(2015)\citenamefont
  {{Fryer}}, \citenamefont {{Belczynski}}, \citenamefont {{Ramirez-Ruiz}},
  \citenamefont {{Rosswog}}, \citenamefont {{Shen}},\ and\ \citenamefont
  {{Steiner}}}]{2015ApJ...812...24F}%
  \BibitemOpen
  \bibfield  {author} {\bibinfo {author} {\bibfnamefont {C.~L.}\ \bibnamefont
  {{Fryer}}}, \bibinfo {author} {\bibfnamefont {K.}~\bibnamefont
  {{Belczynski}}}, \bibinfo {author} {\bibfnamefont {E.}~\bibnamefont
  {{Ramirez-Ruiz}}}, \bibinfo {author} {\bibfnamefont {S.}~\bibnamefont
  {{Rosswog}}}, \bibinfo {author} {\bibfnamefont {G.}~\bibnamefont {{Shen}}},\
  and\ \bibinfo {author} {\bibfnamefont {A.~W.}\ \bibnamefont {{Steiner}}},\
  }\bibfield  {title} {\bibinfo {title} {{The Fate of the Compact Remnant in
  Neutron Star Mergers}},\ }\href {https://doi.org/10.1088/0004-637X/812/1/24}
  {\bibfield  {journal} {\bibinfo  {journal} {\apj}\ }\textbf {\bibinfo
  {volume} {812}},\ \bibinfo {eid} {24} (\bibinfo {year} {2015})},\ \Eprint
  {https://arxiv.org/abs/1504.07605} {arXiv:1504.07605 [astro-ph.HE]}
  \BibitemShut {NoStop}%
\bibitem [{\citenamefont {{Lawrence}}\ \emph {et~al.}(2015)\citenamefont
  {{Lawrence}}, \citenamefont {{Tervala}}, \citenamefont {{Bedaque}},\ and\
  \citenamefont {{Miller}}}]{2015ApJ...808..186L}%
  \BibitemOpen
  \bibfield  {author} {\bibinfo {author} {\bibfnamefont {S.}~\bibnamefont
  {{Lawrence}}}, \bibinfo {author} {\bibfnamefont {J.~G.}\ \bibnamefont
  {{Tervala}}}, \bibinfo {author} {\bibfnamefont {P.~F.}\ \bibnamefont
  {{Bedaque}}},\ and\ \bibinfo {author} {\bibfnamefont {M.~C.}\ \bibnamefont
  {{Miller}}},\ }\bibfield  {title} {\bibinfo {title} {{An Upper Bound on
  Neutron Star Masses from Models of Short Gamma-Ray Bursts}},\ }\href
  {https://doi.org/10.1088/0004-637X/808/2/186} {\bibfield  {journal} {\bibinfo
   {journal} {\apj}\ }\textbf {\bibinfo {volume} {808}},\ \bibinfo {eid} {186}
  (\bibinfo {year} {2015})},\ \Eprint {https://arxiv.org/abs/1505.00231}
  {arXiv:1505.00231 [astro-ph.HE]} \BibitemShut {NoStop}%
\bibitem [{\citenamefont {{Margalit}}\ and\ \citenamefont
  {{Metzger}}(2017)}]{2017ApJ...850L..19M}%
  \BibitemOpen
  \bibfield  {author} {\bibinfo {author} {\bibfnamefont {B.}~\bibnamefont
  {{Margalit}}}\ and\ \bibinfo {author} {\bibfnamefont {B.~D.}\ \bibnamefont
  {{Metzger}}},\ }\bibfield  {title} {\bibinfo {title} {{Constraining the
  Maximum Mass of Neutron Stars from Multi-messenger Observations of
  GW170817}},\ }\href {https://doi.org/10.3847/2041-8213/aa991c} {\bibfield
  {journal} {\bibinfo  {journal} {\apjl}\ }\textbf {\bibinfo {volume} {850}},\
  \bibinfo {eid} {L19} (\bibinfo {year} {2017})},\ \Eprint
  {https://arxiv.org/abs/1710.05938} {arXiv:1710.05938 [astro-ph.HE]}
  \BibitemShut {NoStop}%
\bibitem [{\citenamefont {{Alsing}}\ \emph {et~al.}(2018)\citenamefont
  {{Alsing}}, \citenamefont {{Silva}},\ and\ \citenamefont
  {{Berti}}}]{2018MNRAS.478.1377A}%
  \BibitemOpen
  \bibfield  {author} {\bibinfo {author} {\bibfnamefont {J.}~\bibnamefont
  {{Alsing}}}, \bibinfo {author} {\bibfnamefont {H.~O.}\ \bibnamefont
  {{Silva}}},\ and\ \bibinfo {author} {\bibfnamefont {E.}~\bibnamefont
  {{Berti}}},\ }\bibfield  {title} {\bibinfo {title} {{Evidence for a maximum
  mass cut-off in the neutron star mass distribution and constraints on the
  equation of state}},\ }\href {https://doi.org/10.1093/mnras/sty1065}
  {\bibfield  {journal} {\bibinfo  {journal} {\mnras}\ }\textbf {\bibinfo
  {volume} {478}},\ \bibinfo {pages} {1377} (\bibinfo {year} {2018})},\ \Eprint
  {https://arxiv.org/abs/1709.07889} {arXiv:1709.07889 [astro-ph.HE]}
  \BibitemShut {NoStop}%
\bibitem [{\citenamefont {{Sarin}}\ \emph {et~al.}(2020)\citenamefont
  {{Sarin}}, \citenamefont {{Lasky}},\ and\ \citenamefont
  {{Ashton}}}]{2020arXiv200106102S}%
  \BibitemOpen
  \bibfield  {author} {\bibinfo {author} {\bibfnamefont {N.}~\bibnamefont
  {{Sarin}}}, \bibinfo {author} {\bibfnamefont {P.~D.}\ \bibnamefont
  {{Lasky}}},\ and\ \bibinfo {author} {\bibfnamefont {G.}~\bibnamefont
  {{Ashton}}},\ }\bibfield  {title} {\bibinfo {title} {{Gravitational waves or
  deconfined quarks: What causes the premature collapse of neutron stars born
  in short gamma-ray bursts?}},\ }\href
  {https://doi.org/10.1103/PhysRevD.101.063021} {\bibfield  {journal} {\bibinfo
   {journal} {\prd}\ }\textbf {\bibinfo {volume} {101}},\ \bibinfo {eid}
  {063021} (\bibinfo {year} {2020})},\ \Eprint
  {https://arxiv.org/abs/2001.06102} {arXiv:2001.06102 [astro-ph.HE]}
  \BibitemShut {NoStop}%
\bibitem [{\citenamefont {{Abbott}}\ \emph
  {et~al.}(2020{\natexlab{c}})\citenamefont {{Abbott}}, \citenamefont
  {{Abbott}}, \citenamefont {{Abbott}}, \citenamefont {{Abraham}},
  \citenamefont {{Acernese}}, \citenamefont {{Ackley}}, \citenamefont
  {{Adams}}, \citenamefont {{Adya}}, \citenamefont {{Affeldt}}, \citenamefont
  {{Agathos}}, \citenamefont {{Agatsuma}}, \citenamefont {{Aggarwal}},
  \citenamefont {{Aguiar}}, \citenamefont {{Aiello}}, \citenamefont {{Ain}},
  \citenamefont {{Ajith}}, \citenamefont {{Allen}}, \citenamefont {{Allocca}},
  \citenamefont {{Aloy}}, \citenamefont {{Altin}}, \citenamefont {{Amato}},
  \citenamefont {{Anand}}, \citenamefont {{Ananyeva}}, \citenamefont
  {{Anderson}}, \citenamefont {{Zelenova}}, \citenamefont {{Zendri}},
  \citenamefont {{Zevin}}, \citenamefont {{Zhang}}, \citenamefont {{Zhang}},
  \citenamefont {{Zhang}}, \citenamefont {{Zhao}}, \citenamefont {{Zhao}},
  \citenamefont {{Zhou}}, \citenamefont {{Zhou}}, \citenamefont {{Zhu}},
  \citenamefont {{Zimmerman}}, \citenamefont {{Zlochower}}, \citenamefont
  {{Zucker}}, \citenamefont {{Zweizig}}, \citenamefont {{(The LIGO Scientific
  Collaboration}},\ and\ \citenamefont {{Virgo
  Collaboration)}}}]{2020CQGra..37d5006A}%
  \BibitemOpen
  \bibfield  {author} {\bibinfo {author} {\bibfnamefont {B.~P.}\ \bibnamefont
  {{Abbott}}}, \bibinfo {author} {\bibfnamefont {R.}~\bibnamefont {{Abbott}}},
  \bibinfo {author} {\bibfnamefont {T.~D.}\ \bibnamefont {{Abbott}}}, \bibinfo
  {author} {\bibfnamefont {S.}~\bibnamefont {{Abraham}}}, \bibinfo {author}
  {\bibfnamefont {F.}~\bibnamefont {{Acernese}}}, \bibinfo {author}
  {\bibfnamefont {K.}~\bibnamefont {{Ackley}}}, \bibinfo {author}
  {\bibfnamefont {C.}~\bibnamefont {{Adams}}}, \bibinfo {author} {\bibfnamefont
  {V.~B.}\ \bibnamefont {{Adya}}}, \bibinfo {author} {\bibfnamefont
  {C.}~\bibnamefont {{Affeldt}}}, \bibinfo {author} {\bibfnamefont
  {M.}~\bibnamefont {{Agathos}}}, \bibinfo {author} {\bibfnamefont
  {K.}~\bibnamefont {{Agatsuma}}}, \bibinfo {author} {\bibfnamefont
  {N.}~\bibnamefont {{Aggarwal}}}, \bibinfo {author} {\bibfnamefont {O.~D.}\
  \bibnamefont {{Aguiar}}}, \bibinfo {author} {\bibfnamefont {L.}~\bibnamefont
  {{Aiello}}}, \bibinfo {author} {\bibfnamefont {A.}~\bibnamefont {{Ain}}},
  \bibinfo {author} {\bibfnamefont {P.}~\bibnamefont {{Ajith}}}, \bibinfo
  {author} {\bibfnamefont {G.}~\bibnamefont {{Allen}}}, \bibinfo {author}
  {\bibfnamefont {A.}~\bibnamefont {{Allocca}}}, \bibinfo {author}
  {\bibfnamefont {M.~A.}\ \bibnamefont {{Aloy}}}, \bibinfo {author}
  {\bibfnamefont {P.~A.}\ \bibnamefont {{Altin}}}, \bibinfo {author}
  {\bibfnamefont {A.}~\bibnamefont {{Amato}}}, \bibinfo {author} {\bibfnamefont
  {S.}~\bibnamefont {{Anand}}}, \bibinfo {author} {\bibfnamefont
  {A.}~\bibnamefont {{Ananyeva}}}, \bibinfo {author} {\bibfnamefont {S.~B.}\
  \bibnamefont {{Anderson}}}, \bibinfo {author} {\bibfnamefont
  {T.}~\bibnamefont {{Zelenova}}}, \bibinfo {author} {\bibfnamefont {J.~P.}\
  \bibnamefont {{Zendri}}}, \bibinfo {author} {\bibfnamefont {M.}~\bibnamefont
  {{Zevin}}}, \bibinfo {author} {\bibfnamefont {J.}~\bibnamefont {{Zhang}}},
  \bibinfo {author} {\bibfnamefont {L.}~\bibnamefont {{Zhang}}}, \bibinfo
  {author} {\bibfnamefont {T.}~\bibnamefont {{Zhang}}}, \bibinfo {author}
  {\bibfnamefont {C.}~\bibnamefont {{Zhao}}}, \bibinfo {author} {\bibfnamefont
  {G.}~\bibnamefont {{Zhao}}}, \bibinfo {author} {\bibfnamefont
  {M.}~\bibnamefont {{Zhou}}}, \bibinfo {author} {\bibfnamefont
  {Z.}~\bibnamefont {{Zhou}}}, \bibinfo {author} {\bibfnamefont {X.~J.}\
  \bibnamefont {{Zhu}}}, \bibinfo {author} {\bibfnamefont {A.~B.}\ \bibnamefont
  {{Zimmerman}}}, \bibinfo {author} {\bibfnamefont {Y.}~\bibnamefont
  {{Zlochower}}}, \bibinfo {author} {\bibfnamefont {M.~E.}\ \bibnamefont
  {{Zucker}}}, \bibinfo {author} {\bibfnamefont {J.}~\bibnamefont {{Zweizig}}},
  \bibinfo {author} {\bibnamefont {{(The LIGO Scientific Collaboration}}},\
  and\ \bibinfo {author} {\bibnamefont {{Virgo Collaboration)}}},\ }\bibfield
  {title} {\bibinfo {title} {{Model comparison from LIGO-Virgo data on
  GW170817's binary components and consequences for the merger remnant}},\
  }\href {https://doi.org/10.1088/1361-6382/ab5f7c} {\bibfield  {journal}
  {\bibinfo  {journal} {Classical and Quantum Gravity}\ }\textbf {\bibinfo
  {volume} {37}},\ \bibinfo {eid} {045006} (\bibinfo {year}
  {2020}{\natexlab{c}})},\ \Eprint {https://arxiv.org/abs/1908.01012}
  {arXiv:1908.01012 [gr-qc]} \BibitemShut {NoStop}%
\bibitem [{\citenamefont {{Abbott}}\ \emph
  {et~al.}(2020{\natexlab{d}})\citenamefont {{Abbott}}, \citenamefont
  {{Abbott}}, \citenamefont {{Abbott}}, \citenamefont {{Abraham}},
  \citenamefont {{Acernese}}, \citenamefont {{Ackley}}, \citenamefont
  {{Adams}}, \citenamefont {{Adhikari}}, \citenamefont {{Adya}}, \citenamefont
  {{Affeldt}}, \citenamefont {{Agathos}}, \citenamefont {{Agatsuma}},
  \citenamefont {{Aggarwal}}, \citenamefont {{Aguiar}}, \citenamefont
  {{Aiello}}, \citenamefont {{Ain}}, \citenamefont {{Ajith}}, \citenamefont
  {{Allen}}, \citenamefont {{Allocca}}, \citenamefont {{Aloy}}, \citenamefont
  {{Altin}}, \citenamefont {{Amato}}, \citenamefont {{Anand}}, \citenamefont
  {{Ananyeva}}, \citenamefont {{Anderson}}, \citenamefont {{Anderson}},
  \citenamefont {{Angelova}}, \citenamefont {{Antier}}, \citenamefont
  {{Appert}}, \citenamefont {{Arai}}, \citenamefont {{Araya}}, \citenamefont
  {{Areeda}}, \citenamefont {{Ar{\`e}ne}}, \citenamefont {{Arnaud}},
  \citenamefont {{Aronson}}, \citenamefont {{Arun}}, \citenamefont {{Ascenzi}},
  \citenamefont {{Ashton}}, \citenamefont {{Aston}}, \citenamefont {{Astone}},
  \citenamefont {{Aubin}}, \citenamefont {{Aufmuth}}, \citenamefont
  {{AultONeal}}, \citenamefont {{Austin}}, \citenamefont {{Avendano}},
  \citenamefont {{Avila-Alvarez}}, \citenamefont {{Babak}}, \citenamefont
  {{Bacon}}, \citenamefont {{Badaracco}}, \citenamefont {{Bader}},
  \citenamefont {{Bae}}, \citenamefont {{Baird}}, \citenamefont {{Baker}},
  \citenamefont {{Baldaccini}}, \citenamefont {{Ballardin}}, \citenamefont
  {{Ballmer}}, \citenamefont {{Bals}}, \citenamefont {{Banagiri}},
  \citenamefont {{Barayoga}}, \citenamefont {{Barbieri}}, \citenamefont
  {{Barclay}}, \citenamefont {{Barish}}, \citenamefont {{Barker}},
  \citenamefont {{Barkett}}, \citenamefont {{Barnum}}, \citenamefont
  {{Barone}}, \citenamefont {{Barr}}, \citenamefont {{Barsotti}}, \citenamefont
  {{Barsuglia}}, \citenamefont {{Williams}}, \citenamefont {{Williamson}},
  \citenamefont {{Willis}}, \citenamefont {{Willke}}, \citenamefont
  {{Winkler}}, \citenamefont {{Wipf}}, \citenamefont {{Wittel}}, \citenamefont
  {{Woan}}, \citenamefont {{Woehler}}, \citenamefont {{Wofford}}, \citenamefont
  {{Wright}}, \citenamefont {{Wu}}, \citenamefont {{Wysocki}}, \citenamefont
  {{Xiao}}, \citenamefont {{Xu}}, \citenamefont {{Yamamoto}}, \citenamefont
  {{Yancey}}, \citenamefont {{Yang}}, \citenamefont {{Yang}}, \citenamefont
  {{Yang}}, \citenamefont {{Yap}}, \citenamefont {{Yazback}}, \citenamefont
  {{Yeeles}}, \citenamefont {{Yu}}, \citenamefont {{Yu}}, \citenamefont
  {{Yuen}}, \citenamefont {{Zadro{\.z}ny}}, \citenamefont {{Zadro{\.z}ny}},
  \citenamefont {{Zanolin}}, \citenamefont {{Zelenova}}, \citenamefont
  {{Zendri}}, \citenamefont {{Zevin}}, \citenamefont {{Zhang}}, \citenamefont
  {{Zhang}}, \citenamefont {{Zhang}}, \citenamefont {{Zhao}}, \citenamefont
  {{Zhao}}, \citenamefont {{Zhou}}, \citenamefont {{Zhou}}, \citenamefont
  {{Zhu}}, \citenamefont {{Zimmerman}}, \citenamefont {{Zucker}},\ and\
  \citenamefont {{Zweizig}}}]{Abbott:2020gw190425}%
  \BibitemOpen
  \bibfield  {author} {\bibinfo {author} {\bibfnamefont {B.~P.}\ \bibnamefont
  {{Abbott}}}, \bibinfo {author} {\bibfnamefont {R.}~\bibnamefont {{Abbott}}},
  \bibinfo {author} {\bibfnamefont {T.~D.}\ \bibnamefont {{Abbott}}}, \bibinfo
  {author} {\bibfnamefont {S.}~\bibnamefont {{Abraham}}}, \bibinfo {author}
  {\bibfnamefont {F.}~\bibnamefont {{Acernese}}}, \bibinfo {author}
  {\bibfnamefont {K.}~\bibnamefont {{Ackley}}}, \bibinfo {author}
  {\bibfnamefont {C.}~\bibnamefont {{Adams}}}, \bibinfo {author} {\bibfnamefont
  {R.~X.}\ \bibnamefont {{Adhikari}}}, \bibinfo {author} {\bibfnamefont
  {V.~B.}\ \bibnamefont {{Adya}}}, \bibinfo {author} {\bibfnamefont
  {C.}~\bibnamefont {{Affeldt}}}, \bibinfo {author} {\bibfnamefont
  {M.}~\bibnamefont {{Agathos}}}, \bibinfo {author} {\bibfnamefont
  {K.}~\bibnamefont {{Agatsuma}}}, \bibinfo {author} {\bibfnamefont
  {N.}~\bibnamefont {{Aggarwal}}}, \bibinfo {author} {\bibfnamefont {O.~D.}\
  \bibnamefont {{Aguiar}}}, \bibinfo {author} {\bibfnamefont {L.}~\bibnamefont
  {{Aiello}}}, \bibinfo {author} {\bibfnamefont {A.}~\bibnamefont {{Ain}}},
  \bibinfo {author} {\bibfnamefont {P.}~\bibnamefont {{Ajith}}}, \bibinfo
  {author} {\bibfnamefont {G.}~\bibnamefont {{Allen}}}, \bibinfo {author}
  {\bibfnamefont {A.}~\bibnamefont {{Allocca}}}, \bibinfo {author}
  {\bibfnamefont {M.~A.}\ \bibnamefont {{Aloy}}}, \bibinfo {author}
  {\bibfnamefont {P.~A.}\ \bibnamefont {{Altin}}}, \bibinfo {author}
  {\bibfnamefont {A.}~\bibnamefont {{Amato}}}, \bibinfo {author} {\bibfnamefont
  {S.}~\bibnamefont {{Anand}}}, \bibinfo {author} {\bibfnamefont
  {A.}~\bibnamefont {{Ananyeva}}}, \bibinfo {author} {\bibfnamefont {S.~B.}\
  \bibnamefont {{Anderson}}}, \bibinfo {author} {\bibfnamefont {W.~G.}\
  \bibnamefont {{Anderson}}}, \bibinfo {author} {\bibfnamefont {S.~V.}\
  \bibnamefont {{Angelova}}}, \bibinfo {author} {\bibfnamefont
  {S.}~\bibnamefont {{Antier}}}, \bibinfo {author} {\bibfnamefont
  {S.}~\bibnamefont {{Appert}}}, \bibinfo {author} {\bibfnamefont
  {K.}~\bibnamefont {{Arai}}}, \bibinfo {author} {\bibfnamefont {M.~C.}\
  \bibnamefont {{Araya}}}, \bibinfo {author} {\bibfnamefont {J.~S.}\
  \bibnamefont {{Areeda}}}, \bibinfo {author} {\bibfnamefont {M.}~\bibnamefont
  {{Ar{\`e}ne}}}, \bibinfo {author} {\bibfnamefont {N.}~\bibnamefont
  {{Arnaud}}}, \bibinfo {author} {\bibfnamefont {S.~M.}\ \bibnamefont
  {{Aronson}}}, \bibinfo {author} {\bibfnamefont {K.~G.}\ \bibnamefont
  {{Arun}}}, \bibinfo {author} {\bibfnamefont {S.}~\bibnamefont {{Ascenzi}}},
  \bibinfo {author} {\bibfnamefont {G.}~\bibnamefont {{Ashton}}}, \bibinfo
  {author} {\bibfnamefont {S.~M.}\ \bibnamefont {{Aston}}}, \bibinfo {author}
  {\bibfnamefont {P.}~\bibnamefont {{Astone}}}, \bibinfo {author}
  {\bibfnamefont {F.}~\bibnamefont {{Aubin}}}, \bibinfo {author} {\bibfnamefont
  {P.}~\bibnamefont {{Aufmuth}}}, \bibinfo {author} {\bibfnamefont
  {K.}~\bibnamefont {{AultONeal}}}, \bibinfo {author} {\bibfnamefont
  {C.}~\bibnamefont {{Austin}}}, \bibinfo {author} {\bibfnamefont
  {V.}~\bibnamefont {{Avendano}}}, \bibinfo {author} {\bibfnamefont
  {A.}~\bibnamefont {{Avila-Alvarez}}}, \bibinfo {author} {\bibfnamefont
  {S.}~\bibnamefont {{Babak}}}, \bibinfo {author} {\bibfnamefont
  {P.}~\bibnamefont {{Bacon}}}, \bibinfo {author} {\bibfnamefont
  {F.}~\bibnamefont {{Badaracco}}}, \bibinfo {author} {\bibfnamefont
  {M.~K.~M.}\ \bibnamefont {{Bader}}}, \bibinfo {author} {\bibfnamefont
  {S.}~\bibnamefont {{Bae}}}, \bibinfo {author} {\bibfnamefont
  {J.}~\bibnamefont {{Baird}}}, \bibinfo {author} {\bibfnamefont {P.~T.}\
  \bibnamefont {{Baker}}}, \bibinfo {author} {\bibfnamefont {F.}~\bibnamefont
  {{Baldaccini}}}, \bibinfo {author} {\bibfnamefont {G.}~\bibnamefont
  {{Ballardin}}}, \bibinfo {author} {\bibfnamefont {S.~W.}\ \bibnamefont
  {{Ballmer}}}, \bibinfo {author} {\bibfnamefont {A.}~\bibnamefont {{Bals}}},
  \bibinfo {author} {\bibfnamefont {S.}~\bibnamefont {{Banagiri}}}, \bibinfo
  {author} {\bibfnamefont {J.~C.}\ \bibnamefont {{Barayoga}}}, \bibinfo
  {author} {\bibfnamefont {C.}~\bibnamefont {{Barbieri}}}, \bibinfo {author}
  {\bibfnamefont {S.~E.}\ \bibnamefont {{Barclay}}}, \bibinfo {author}
  {\bibfnamefont {B.~C.}\ \bibnamefont {{Barish}}}, \bibinfo {author}
  {\bibfnamefont {D.}~\bibnamefont {{Barker}}}, \bibinfo {author}
  {\bibfnamefont {K.}~\bibnamefont {{Barkett}}}, \bibinfo {author}
  {\bibfnamefont {S.}~\bibnamefont {{Barnum}}}, \bibinfo {author}
  {\bibfnamefont {F.}~\bibnamefont {{Barone}}}, \bibinfo {author}
  {\bibfnamefont {B.}~\bibnamefont {{Barr}}}, \bibinfo {author} {\bibfnamefont
  {L.}~\bibnamefont {{Barsotti}}}, \bibinfo {author} {\bibfnamefont
  {M.}~\bibnamefont {{Barsuglia}}}, \bibinfo {author} {\bibfnamefont
  {D.}~\bibnamefont {{Williams}}}, \bibinfo {author} {\bibfnamefont {A.~R.}\
  \bibnamefont {{Williamson}}}, \bibinfo {author} {\bibfnamefont {J.~L.}\
  \bibnamefont {{Willis}}}, \bibinfo {author} {\bibfnamefont {B.}~\bibnamefont
  {{Willke}}}, \bibinfo {author} {\bibfnamefont {W.}~\bibnamefont {{Winkler}}},
  \bibinfo {author} {\bibfnamefont {C.~C.}\ \bibnamefont {{Wipf}}}, \bibinfo
  {author} {\bibfnamefont {H.}~\bibnamefont {{Wittel}}}, \bibinfo {author}
  {\bibfnamefont {G.}~\bibnamefont {{Woan}}}, \bibinfo {author} {\bibfnamefont
  {J.}~\bibnamefont {{Woehler}}}, \bibinfo {author} {\bibfnamefont {J.~K.}\
  \bibnamefont {{Wofford}}}, \bibinfo {author} {\bibfnamefont {J.~L.}\
  \bibnamefont {{Wright}}}, \bibinfo {author} {\bibfnamefont {D.~S.}\
  \bibnamefont {{Wu}}}, \bibinfo {author} {\bibfnamefont {D.~M.}\ \bibnamefont
  {{Wysocki}}}, \bibinfo {author} {\bibfnamefont {S.}~\bibnamefont {{Xiao}}},
  \bibinfo {author} {\bibfnamefont {R.}~\bibnamefont {{Xu}}}, \bibinfo {author}
  {\bibfnamefont {H.}~\bibnamefont {{Yamamoto}}}, \bibinfo {author}
  {\bibfnamefont {C.~C.}\ \bibnamefont {{Yancey}}}, \bibinfo {author}
  {\bibfnamefont {L.}~\bibnamefont {{Yang}}}, \bibinfo {author} {\bibfnamefont
  {Y.}~\bibnamefont {{Yang}}}, \bibinfo {author} {\bibfnamefont
  {Z.}~\bibnamefont {{Yang}}}, \bibinfo {author} {\bibfnamefont {M.~J.}\
  \bibnamefont {{Yap}}}, \bibinfo {author} {\bibfnamefont {M.}~\bibnamefont
  {{Yazback}}}, \bibinfo {author} {\bibfnamefont {D.~W.}\ \bibnamefont
  {{Yeeles}}}, \bibinfo {author} {\bibfnamefont {H.}~\bibnamefont {{Yu}}},
  \bibinfo {author} {\bibfnamefont {H.}~\bibnamefont {{Yu}}}, \bibinfo {author}
  {\bibfnamefont {S.~H.~R.}\ \bibnamefont {{Yuen}}}, \bibinfo {author}
  {\bibfnamefont {A.~K.}\ \bibnamefont {{Zadro{\.z}ny}}}, \bibinfo {author}
  {\bibfnamefont {A.}~\bibnamefont {{Zadro{\.z}ny}}}, \bibinfo {author}
  {\bibfnamefont {M.}~\bibnamefont {{Zanolin}}}, \bibinfo {author}
  {\bibfnamefont {T.}~\bibnamefont {{Zelenova}}}, \bibinfo {author}
  {\bibfnamefont {J.~P.}\ \bibnamefont {{Zendri}}}, \bibinfo {author}
  {\bibfnamefont {M.}~\bibnamefont {{Zevin}}}, \bibinfo {author} {\bibfnamefont
  {J.}~\bibnamefont {{Zhang}}}, \bibinfo {author} {\bibfnamefont
  {L.}~\bibnamefont {{Zhang}}}, \bibinfo {author} {\bibfnamefont
  {T.}~\bibnamefont {{Zhang}}}, \bibinfo {author} {\bibfnamefont
  {C.}~\bibnamefont {{Zhao}}}, \bibinfo {author} {\bibfnamefont
  {G.}~\bibnamefont {{Zhao}}}, \bibinfo {author} {\bibfnamefont
  {M.}~\bibnamefont {{Zhou}}}, \bibinfo {author} {\bibfnamefont
  {Z.}~\bibnamefont {{Zhou}}}, \bibinfo {author} {\bibfnamefont {X.~J.}\
  \bibnamefont {{Zhu}}}, \bibinfo {author} {\bibfnamefont {A.~B.}\ \bibnamefont
  {{Zimmerman}}}, \bibinfo {author} {\bibfnamefont {M.~E.}\ \bibnamefont
  {{Zucker}}},\ and\ \bibinfo {author} {\bibfnamefont {J.}~\bibnamefont
  {{Zweizig}}},\ }\bibfield  {title} {\bibinfo {title} {{GW190425: Observation
  of a Compact Binary Coalescence with Total Mass {\ensuremath{\sim}} 3.4
  M$_{\odot}$}},\ }\href {https://doi.org/10.3847/2041-8213/ab75f5} {\bibfield
  {journal} {\bibinfo  {journal} {\apjl}\ }\textbf {\bibinfo {volume} {892}},\
  \bibinfo {eid} {L3} (\bibinfo {year} {2020}{\natexlab{d}})},\ \Eprint
  {https://arxiv.org/abs/2001.01761} {arXiv:2001.01761 [astro-ph.HE]}
  \BibitemShut {NoStop}%
\bibitem [{\citenamefont {{Burrows}}\ \emph {et~al.}(2020)\citenamefont
  {{Burrows}}, \citenamefont {{Radice}}, \citenamefont {{Vartanyan}},
  \citenamefont {{Nagakura}}, \citenamefont {{Skinner}},\ and\ \citenamefont
  {{Dolence}}}]{2020MNRAS.491.2715B}%
  \BibitemOpen
  \bibfield  {author} {\bibinfo {author} {\bibfnamefont {A.}~\bibnamefont
  {{Burrows}}}, \bibinfo {author} {\bibfnamefont {D.}~\bibnamefont {{Radice}}},
  \bibinfo {author} {\bibfnamefont {D.}~\bibnamefont {{Vartanyan}}}, \bibinfo
  {author} {\bibfnamefont {H.}~\bibnamefont {{Nagakura}}}, \bibinfo {author}
  {\bibfnamefont {M.~A.}\ \bibnamefont {{Skinner}}},\ and\ \bibinfo {author}
  {\bibfnamefont {J.~C.}\ \bibnamefont {{Dolence}}},\ }\bibfield  {title}
  {\bibinfo {title} {{The overarching framework of core-collapse supernova
  explosions as revealed by 3D FORNAX simulations}},\ }\href
  {https://doi.org/10.1093/mnras/stz3223} {\bibfield  {journal} {\bibinfo
  {journal} {\mnras}\ }\textbf {\bibinfo {volume} {491}},\ \bibinfo {pages}
  {2715} (\bibinfo {year} {2020})},\ \Eprint {https://arxiv.org/abs/1909.04152}
  {arXiv:1909.04152 [astro-ph.HE]} \BibitemShut {NoStop}%
\bibitem [{\citenamefont {{Eldridge}}\ \emph {et~al.}(2020)\citenamefont
  {{Eldridge}}, \citenamefont {{Stanway}}, \citenamefont {{Breivik}},
  \citenamefont {{Casey}}, \citenamefont {{Steeghs}},\ and\ \citenamefont
  {{Stevance}}}]{2020MNRAS.495.2786E}%
  \BibitemOpen
  \bibfield  {author} {\bibinfo {author} {\bibfnamefont {J.~J.}\ \bibnamefont
  {{Eldridge}}}, \bibinfo {author} {\bibfnamefont {E.~R.}\ \bibnamefont
  {{Stanway}}}, \bibinfo {author} {\bibfnamefont {K.}~\bibnamefont
  {{Breivik}}}, \bibinfo {author} {\bibfnamefont {A.~R.}\ \bibnamefont
  {{Casey}}}, \bibinfo {author} {\bibfnamefont {D.~T.~H.}\ \bibnamefont
  {{Steeghs}}},\ and\ \bibinfo {author} {\bibfnamefont {H.~F.}\ \bibnamefont
  {{Stevance}}},\ }\bibfield  {title} {\bibinfo {title} {{Weighing in on black
  hole binaries with BPASS: LB-1 does not contain a 70 M$_{\odot}$ black
  hole}},\ }\href {https://doi.org/10.1093/mnras/staa1324} {\bibfield
  {journal} {\bibinfo  {journal} {\mnras}\ }\textbf {\bibinfo {volume} {495}},\
  \bibinfo {pages} {2786} (\bibinfo {year} {2020})},\ \Eprint
  {https://arxiv.org/abs/1912.03599} {arXiv:1912.03599 [astro-ph.SR]}
  \BibitemShut {NoStop}%
\bibitem [{\citenamefont {{Rom{\'a}n-Garza}}\ \emph {et~al.}(2021)\citenamefont
  {{Rom{\'a}n-Garza}}, \citenamefont {{Bavera}}, \citenamefont {{Fragos}},
  \citenamefont {{Zapartas}}, \citenamefont {{Misra}}, \citenamefont
  {{Andrews}}, \citenamefont {{Coughlin}}, \citenamefont {{Dotter}},
  \citenamefont {{Kovlakas}}, \citenamefont {{Serra}}, \citenamefont {{Qin}},
  \citenamefont {{Rocha}},\ and\ \citenamefont {{Tran}}}]{RomanGarza:2020}%
  \BibitemOpen
  \bibfield  {author} {\bibinfo {author} {\bibfnamefont {J.}~\bibnamefont
  {{Rom{\'a}n-Garza}}}, \bibinfo {author} {\bibfnamefont {S.~S.}\ \bibnamefont
  {{Bavera}}}, \bibinfo {author} {\bibfnamefont {T.}~\bibnamefont {{Fragos}}},
  \bibinfo {author} {\bibfnamefont {E.}~\bibnamefont {{Zapartas}}}, \bibinfo
  {author} {\bibfnamefont {D.}~\bibnamefont {{Misra}}}, \bibinfo {author}
  {\bibfnamefont {J.}~\bibnamefont {{Andrews}}}, \bibinfo {author}
  {\bibfnamefont {S.}~\bibnamefont {{Coughlin}}}, \bibinfo {author}
  {\bibfnamefont {A.}~\bibnamefont {{Dotter}}}, \bibinfo {author}
  {\bibfnamefont {K.}~\bibnamefont {{Kovlakas}}}, \bibinfo {author}
  {\bibfnamefont {J.~G.}\ \bibnamefont {{Serra}}}, \bibinfo {author}
  {\bibfnamefont {Y.}~\bibnamefont {{Qin}}}, \bibinfo {author} {\bibfnamefont
  {K.~A.}\ \bibnamefont {{Rocha}}},\ and\ \bibinfo {author} {\bibfnamefont
  {N.~H.}\ \bibnamefont {{Tran}}},\ }\bibfield  {title} {\bibinfo {title} {{The
  Role of Core-collapse Physics in the Observability of Black Hole Neutron Star
  Mergers as Multimessenger Sources}},\ }\href
  {https://doi.org/10.3847/2041-8213/abf42c} {\bibfield  {journal} {\bibinfo
  {journal} {\apjl}\ }\textbf {\bibinfo {volume} {912}},\ \bibinfo {eid} {L23}
  (\bibinfo {year} {2021})},\ \Eprint {https://arxiv.org/abs/2012.02274}
  {arXiv:2012.02274 [astro-ph.HE]} \BibitemShut {NoStop}%
\bibitem [{\citenamefont {{Vigna-G{\'o}mez}}\ \emph {et~al.}(2021)\citenamefont
  {{Vigna-G{\'o}mez}}, \citenamefont {{Schr{\o}der}}, \citenamefont
  {{Ramirez-Ruiz}}, \citenamefont {{Aguilera-Dena}}, \citenamefont {{Batta}},
  \citenamefont {{Langer}},\ and\ \citenamefont
  {{Wilcox}}}]{2021arXiv210612381V}%
  \BibitemOpen
  \bibfield  {author} {\bibinfo {author} {\bibfnamefont {A.}~\bibnamefont
  {{Vigna-G{\'o}mez}}}, \bibinfo {author} {\bibfnamefont {S.~L.}\ \bibnamefont
  {{Schr{\o}der}}}, \bibinfo {author} {\bibfnamefont {E.}~\bibnamefont
  {{Ramirez-Ruiz}}}, \bibinfo {author} {\bibfnamefont {D.~R.}\ \bibnamefont
  {{Aguilera-Dena}}}, \bibinfo {author} {\bibfnamefont {A.}~\bibnamefont
  {{Batta}}}, \bibinfo {author} {\bibfnamefont {N.}~\bibnamefont {{Langer}}},\
  and\ \bibinfo {author} {\bibfnamefont {R.}~\bibnamefont {{Wilcox}}},\
  }\bibfield  {title} {\bibinfo {title} {{The formation of heavy, radio-quiet
  neutron star binaries and the origin of GW190425}},\ }\href@noop {}
  {\bibfield  {journal} {\bibinfo  {journal} {arXiv e-prints}\ ,\ \bibinfo
  {eid} {arXiv:2106.12381}} (\bibinfo {year} {2021})},\ \Eprint
  {https://arxiv.org/abs/2106.12381} {arXiv:2106.12381 [astro-ph.HE]}
  \BibitemShut {NoStop}%
\bibitem [{\citenamefont {{Dabrowny}}\ \emph {et~al.}(2021)\citenamefont
  {{Dabrowny}}, \citenamefont {{Giacobbo}},\ and\ \citenamefont
  {{Gerosa}}}]{Dabrowny:2021}%
  \BibitemOpen
  \bibfield  {author} {\bibinfo {author} {\bibfnamefont {M.}~\bibnamefont
  {{Dabrowny}}}, \bibinfo {author} {\bibfnamefont {N.}~\bibnamefont
  {{Giacobbo}}},\ and\ \bibinfo {author} {\bibfnamefont {D.}~\bibnamefont
  {{Gerosa}}},\ }\bibfield  {title} {\bibinfo {title} {{Modeling the outcome of
  supernova explosions in binary population synthesis using the stellar
  compactness}},\ }\bibfield  {journal} {\bibinfo  {journal} {Rendiconti
  Lincei. Scienze Fisiche e Naturali}\ }\href
  {https://doi.org/10.1007/s12210-021-01019-8} {10.1007/s12210-021-01019-8}
  (\bibinfo {year} {2021}),\ \Eprint {https://arxiv.org/abs/2106.12541}
  {arXiv:2106.12541 [astro-ph.HE]} \BibitemShut {NoStop}%
\bibitem [{\citenamefont {{Mandel}}\ \emph {et~al.}(2021)\citenamefont
  {{Mandel}}, \citenamefont {{M{\"u}ller}}, \citenamefont {{Riley}},
  \citenamefont {{de Mink}}, \citenamefont {{Vigna-G{\'o}mez}},\ and\
  \citenamefont {{Chattopadhyay}}}]{Mandel:2021MNRAS}%
  \BibitemOpen
  \bibfield  {author} {\bibinfo {author} {\bibfnamefont {I.}~\bibnamefont
  {{Mandel}}}, \bibinfo {author} {\bibfnamefont {B.}~\bibnamefont
  {{M{\"u}ller}}}, \bibinfo {author} {\bibfnamefont {J.}~\bibnamefont
  {{Riley}}}, \bibinfo {author} {\bibfnamefont {S.~E.}\ \bibnamefont {{de
  Mink}}}, \bibinfo {author} {\bibfnamefont {A.}~\bibnamefont
  {{Vigna-G{\'o}mez}}},\ and\ \bibinfo {author} {\bibfnamefont
  {D.}~\bibnamefont {{Chattopadhyay}}},\ }\bibfield  {title} {\bibinfo {title}
  {{Binary population synthesis with probabilistic remnant mass and kick
  prescriptions}},\ }\href {https://doi.org/10.1093/mnras/staa3390} {\bibfield
  {journal} {\bibinfo  {journal} {\mnras}\ }\textbf {\bibinfo {volume} {500}},\
  \bibinfo {pages} {1380} (\bibinfo {year} {2021})},\ \Eprint
  {https://arxiv.org/abs/2007.03890} {arXiv:2007.03890 [astro-ph.HE]}
  \BibitemShut {NoStop}%
\bibitem [{\citenamefont {Spruit}(2002)}]{Spruit:2001tz}%
  \BibitemOpen
  \bibfield  {author} {\bibinfo {author} {\bibfnamefont {H.~C.}\ \bibnamefont
  {Spruit}},\ }\bibfield  {title} {\bibinfo {title} {{Dynamo action by
  differential rotation in a stably stratified stellar interior}},\ }\href
  {https://doi.org/10.1051/0004-6361:20011465} {\bibfield  {journal} {\bibinfo
  {journal} {Astron. Astrophys.}\ }\textbf {\bibinfo {volume} {381}},\ \bibinfo
  {pages} {923} (\bibinfo {year} {2002})},\ \Eprint
  {https://arxiv.org/abs/astro-ph/0108207} {arXiv:astro-ph/0108207}
  \BibitemShut {NoStop}%
\bibitem [{\citenamefont {{Spruit}}(2002)}]{Spruit:2002}%
  \BibitemOpen
  \bibfield  {author} {\bibinfo {author} {\bibfnamefont {H.~C.}\ \bibnamefont
  {{Spruit}}},\ }\bibfield  {title} {\bibinfo {title} {{Dynamo action by
  differential rotation in a stably stratified stellar interior}},\ }\href
  {https://doi.org/10.1051/0004-6361:20011465} {\bibfield  {journal} {\bibinfo
  {journal} {\aap}\ }\textbf {\bibinfo {volume} {381}},\ \bibinfo {pages} {923}
  (\bibinfo {year} {2002})},\ \Eprint {https://arxiv.org/abs/astro-ph/0108207}
  {arXiv:astro-ph/0108207 [astro-ph]} \BibitemShut {NoStop}%
\bibitem [{\citenamefont {Belczynski}\ \emph {et~al.}(2020)\citenamefont
  {Belczynski} \emph {et~al.}}]{Belczynski:2017gds}%
  \BibitemOpen
  \bibfield  {author} {\bibinfo {author} {\bibfnamefont {K.}~\bibnamefont
  {Belczynski}} \emph {et~al.},\ }\bibfield  {title} {\bibinfo {title}
  {{Evolutionary roads leading to low effective spins, high black hole masses,
  and O1/O2 rates for LIGO/Virgo binary black holes}},\ }\href
  {https://doi.org/10.1051/0004-6361/201936528} {\bibfield  {journal} {\bibinfo
   {journal} {Astron. Astrophys.}\ }\textbf {\bibinfo {volume} {636}},\
  \bibinfo {pages} {A104} (\bibinfo {year} {2020})},\ \Eprint
  {https://arxiv.org/abs/1706.07053} {arXiv:1706.07053 [astro-ph.HE]}
  \BibitemShut {NoStop}%
\bibitem [{\citenamefont {{Fuller}}\ \emph
  {et~al.}(2019{\natexlab{b}})\citenamefont {{Fuller}}, \citenamefont
  {{Piro}},\ and\ \citenamefont {{Jermyn}}}]{Fuller:2019MNRAS}%
  \BibitemOpen
  \bibfield  {author} {\bibinfo {author} {\bibfnamefont {J.}~\bibnamefont
  {{Fuller}}}, \bibinfo {author} {\bibfnamefont {A.~L.}\ \bibnamefont
  {{Piro}}},\ and\ \bibinfo {author} {\bibfnamefont {A.~S.}\ \bibnamefont
  {{Jermyn}}},\ }\bibfield  {title} {\bibinfo {title} {{Slowing the spins of
  stellar cores}},\ }\href {https://doi.org/10.1093/mnras/stz514} {\bibfield
  {journal} {\bibinfo  {journal} {\mnras}\ }\textbf {\bibinfo {volume} {485}},\
  \bibinfo {pages} {3661} (\bibinfo {year} {2019}{\natexlab{b}})},\ \Eprint
  {https://arxiv.org/abs/1902.08227} {arXiv:1902.08227 [astro-ph.SR]}
  \BibitemShut {NoStop}%
\bibitem [{\citenamefont {Gerosa}\ \emph {et~al.}(2018)\citenamefont {Gerosa},
  \citenamefont {Berti}, \citenamefont {O'Shaughnessy}, \citenamefont
  {Belczynski}, \citenamefont {Kesden}, \citenamefont {Wysocki},\ and\
  \citenamefont {Gladysz}}]{Gerosa:2018wbw}%
  \BibitemOpen
  \bibfield  {author} {\bibinfo {author} {\bibfnamefont {D.}~\bibnamefont
  {Gerosa}}, \bibinfo {author} {\bibfnamefont {E.}~\bibnamefont {Berti}},
  \bibinfo {author} {\bibfnamefont {R.}~\bibnamefont {O'Shaughnessy}}, \bibinfo
  {author} {\bibfnamefont {K.}~\bibnamefont {Belczynski}}, \bibinfo {author}
  {\bibfnamefont {M.}~\bibnamefont {Kesden}}, \bibinfo {author} {\bibfnamefont
  {D.}~\bibnamefont {Wysocki}},\ and\ \bibinfo {author} {\bibfnamefont
  {W.}~\bibnamefont {Gladysz}},\ }\bibfield  {title} {\bibinfo {title} {{Spin
  orientations of merging black holes formed from the evolution of stellar
  binaries}},\ }\href {https://doi.org/10.1103/PhysRevD.98.084036} {\bibfield
  {journal} {\bibinfo  {journal} {Phys. Rev. D}\ }\textbf {\bibinfo {volume}
  {98}},\ \bibinfo {pages} {084036} (\bibinfo {year} {2018})},\ \Eprint
  {https://arxiv.org/abs/1808.02491} {arXiv:1808.02491 [astro-ph.HE]}
  \BibitemShut {NoStop}%
\bibitem [{\citenamefont {Zevin}\ \emph {et~al.}(2020)\citenamefont {Zevin},
  \citenamefont {Berry}, \citenamefont {Coughlin}, \citenamefont
  {Chatziioannou},\ and\ \citenamefont {Vitale}}]{Zevin:2020gxf}%
  \BibitemOpen
  \bibfield  {author} {\bibinfo {author} {\bibfnamefont {M.}~\bibnamefont
  {Zevin}}, \bibinfo {author} {\bibfnamefont {C.~P.~L.}\ \bibnamefont {Berry}},
  \bibinfo {author} {\bibfnamefont {S.}~\bibnamefont {Coughlin}}, \bibinfo
  {author} {\bibfnamefont {K.}~\bibnamefont {Chatziioannou}},\ and\ \bibinfo
  {author} {\bibfnamefont {S.}~\bibnamefont {Vitale}},\ }\bibfield  {title}
  {\bibinfo {title} {{You Can\textquoteright{}t Always Get What You Want: The
  Impact of Prior Assumptions on Interpreting GW190412}},\ }\href
  {https://doi.org/10.3847/2041-8213/aba8ef} {\bibfield  {journal} {\bibinfo
  {journal} {Astrophys. J. Lett.}\ }\textbf {\bibinfo {volume} {899}},\
  \bibinfo {pages} {L17} (\bibinfo {year} {2020})},\ \Eprint
  {https://arxiv.org/abs/2006.11293} {arXiv:2006.11293 [astro-ph.HE]}
  \BibitemShut {NoStop}%
\bibitem [{\citenamefont {Qin}\ \emph {et~al.}(2018)\citenamefont {Qin},
  \citenamefont {Fragos}, \citenamefont {Meynet}, \citenamefont {Andrews},
  \citenamefont {S\o{}rensen},\ and\ \citenamefont {Song}}]{Qin:2018vaa}%
  \BibitemOpen
  \bibfield  {author} {\bibinfo {author} {\bibfnamefont {Y.}~\bibnamefont
  {Qin}}, \bibinfo {author} {\bibfnamefont {T.}~\bibnamefont {Fragos}},
  \bibinfo {author} {\bibfnamefont {G.}~\bibnamefont {Meynet}}, \bibinfo
  {author} {\bibfnamefont {J.}~\bibnamefont {Andrews}}, \bibinfo {author}
  {\bibfnamefont {M.}~\bibnamefont {S\o{}rensen}},\ and\ \bibinfo {author}
  {\bibfnamefont {H.~F.}\ \bibnamefont {Song}},\ }\bibfield  {title} {\bibinfo
  {title} {{The spin of the second-born black hole in coalescing binary black
  holes}},\ }\href {https://doi.org/10.1051/0004-6361/201832839} {\bibfield
  {journal} {\bibinfo  {journal} {Astron. Astrophys.}\ }\textbf {\bibinfo
  {volume} {616}},\ \bibinfo {pages} {A28} (\bibinfo {year} {2018})},\ \Eprint
  {https://arxiv.org/abs/1802.05738} {arXiv:1802.05738 [astro-ph.SR]}
  \BibitemShut {NoStop}%
\bibitem [{\citenamefont {Fuller}\ and\ \citenamefont
  {Ma}(2019)}]{Fuller:2019sxi}%
  \BibitemOpen
  \bibfield  {author} {\bibinfo {author} {\bibfnamefont {J.}~\bibnamefont
  {Fuller}}\ and\ \bibinfo {author} {\bibfnamefont {L.}~\bibnamefont {Ma}},\
  }\bibfield  {title} {\bibinfo {title} {{Most Black Holes are Born Very Slowly
  Rotating}},\ }\href {https://doi.org/10.3847/2041-8213/ab339b} {\bibfield
  {journal} {\bibinfo  {journal} {Astrophys. J. Lett.}\ }\textbf {\bibinfo
  {volume} {881}},\ \bibinfo {pages} {L1} (\bibinfo {year} {2019})},\ \Eprint
  {https://arxiv.org/abs/1907.03714} {arXiv:1907.03714 [astro-ph.SR]}
  \BibitemShut {NoStop}%
\bibitem [{\citenamefont {{Zevin}}\ and\ \citenamefont
  {{Bavera}}(2022)}]{ZevinBavera:2022}%
  \BibitemOpen
  \bibfield  {author} {\bibinfo {author} {\bibfnamefont {M.}~\bibnamefont
  {{Zevin}}}\ and\ \bibinfo {author} {\bibfnamefont {S.~S.}\ \bibnamefont
  {{Bavera}}},\ }\bibfield  {title} {\bibinfo {title} {{Suspicious Siblings:
  The Distribution of Mass and Spin across Component Black Holes in Isolated
  Binary Evolution}},\ }\href {https://doi.org/10.3847/1538-4357/ac6f5d}
  {\bibfield  {journal} {\bibinfo  {journal} {\apj}\ }\textbf {\bibinfo
  {volume} {933}},\ \bibinfo {eid} {86} (\bibinfo {year} {2022})},\ \Eprint
  {https://arxiv.org/abs/2203.02515} {arXiv:2203.02515 [astro-ph.HE]}
  \BibitemShut {NoStop}%
\bibitem [{\citenamefont {{Shao}}\ and\ \citenamefont
  {{Li}}(2022)}]{2022ApJ...930...26S}%
  \BibitemOpen
  \bibfield  {author} {\bibinfo {author} {\bibfnamefont {Y.}~\bibnamefont
  {{Shao}}}\ and\ \bibinfo {author} {\bibfnamefont {X.-D.}\ \bibnamefont
  {{Li}}},\ }\bibfield  {title} {\bibinfo {title} {{Stable Mass Transfer Can
  Explain Massive Binary Black Hole Mergers with a High-spin Component}},\
  }\href {https://doi.org/10.3847/1538-4357/ac61da} {\bibfield  {journal}
  {\bibinfo  {journal} {\apj}\ }\textbf {\bibinfo {volume} {930}},\ \bibinfo
  {eid} {26} (\bibinfo {year} {2022})},\ \Eprint
  {https://arxiv.org/abs/2203.14529} {arXiv:2203.14529 [astro-ph.HE]}
  \BibitemShut {NoStop}%
\bibitem [{\citenamefont {Olejak}\ and\ \citenamefont
  {Belczynski}(2021)}]{Olejak:2021iux}%
  \BibitemOpen
  \bibfield  {author} {\bibinfo {author} {\bibfnamefont {A.}~\bibnamefont
  {Olejak}}\ and\ \bibinfo {author} {\bibfnamefont {K.}~\bibnamefont
  {Belczynski}},\ }\bibfield  {title} {\bibinfo {title} {{The Implications of
  High BH Spins on the Origin of BH\textendash{}BH Mergers}},\ }\href
  {https://doi.org/10.3847/2041-8213/ac2f48} {\bibfield  {journal} {\bibinfo
  {journal} {Astrophys. J. Lett.}\ }\textbf {\bibinfo {volume} {921}},\
  \bibinfo {pages} {L2} (\bibinfo {year} {2021})},\ \Eprint
  {https://arxiv.org/abs/2109.06872} {arXiv:2109.06872 [astro-ph.HE]}
  \BibitemShut {NoStop}%
\bibitem [{\citenamefont {{Kochanek}}(2014)}]{2014ApJ...785...28K}%
  \BibitemOpen
  \bibfield  {author} {\bibinfo {author} {\bibfnamefont {C.~S.}\ \bibnamefont
  {{Kochanek}}},\ }\bibfield  {title} {\bibinfo {title} {{Failed Supernovae
  Explain the Compact Remnant Mass Function}},\ }\href
  {https://doi.org/10.1088/0004-637X/785/1/28} {\bibfield  {journal} {\bibinfo
  {journal} {\apj}\ }\textbf {\bibinfo {volume} {785}},\ \bibinfo {eid} {28}
  (\bibinfo {year} {2014})},\ \Eprint {https://arxiv.org/abs/1308.0013}
  {arXiv:1308.0013 [astro-ph.HE]} \BibitemShut {NoStop}%
\bibitem [{\citenamefont {{Pejcha}}\ and\ \citenamefont
  {{Thompson}}(2015)}]{2015ApJ...801...90P}%
  \BibitemOpen
  \bibfield  {author} {\bibinfo {author} {\bibfnamefont {O.}~\bibnamefont
  {{Pejcha}}}\ and\ \bibinfo {author} {\bibfnamefont {T.~A.}\ \bibnamefont
  {{Thompson}}},\ }\bibfield  {title} {\bibinfo {title} {{The Landscape of the
  Neutrino Mechanism of Core-collapse Supernovae: Neutron Star and Black Hole
  Mass Functions, Explosion Energies, and Nickel Yields}},\ }\href
  {https://doi.org/10.1088/0004-637X/801/2/90} {\bibfield  {journal} {\bibinfo
  {journal} {\apj}\ }\textbf {\bibinfo {volume} {801}},\ \bibinfo {eid} {90}
  (\bibinfo {year} {2015})},\ \Eprint {https://arxiv.org/abs/1409.0540}
  {arXiv:1409.0540 [astro-ph.HE]} \BibitemShut {NoStop}%
\bibitem [{\citenamefont {{Fryer}}\ and\ \citenamefont
  {{Kalogera}}(2001)}]{2001ApJ...554..548F}%
  \BibitemOpen
  \bibfield  {author} {\bibinfo {author} {\bibfnamefont {C.~L.}\ \bibnamefont
  {{Fryer}}}\ and\ \bibinfo {author} {\bibfnamefont {V.}~\bibnamefont
  {{Kalogera}}},\ }\bibfield  {title} {\bibinfo {title} {{Theoretical Black
  Hole Mass Distributions}},\ }\href {https://doi.org/10.1086/321359}
  {\bibfield  {journal} {\bibinfo  {journal} {\apj}\ }\textbf {\bibinfo
  {volume} {554}},\ \bibinfo {pages} {548} (\bibinfo {year} {2001})},\ \Eprint
  {https://arxiv.org/abs/astro-ph/9911312} {arXiv:astro-ph/9911312 [astro-ph]}
  \BibitemShut {NoStop}%
\bibitem [{\citenamefont {{Kreidberg}}\ \emph {et~al.}(2012)\citenamefont
  {{Kreidberg}}, \citenamefont {{Bailyn}}, \citenamefont {{Farr}},\ and\
  \citenamefont {{Kalogera}}}]{2012ApJ...757...36K}%
  \BibitemOpen
  \bibfield  {author} {\bibinfo {author} {\bibfnamefont {L.}~\bibnamefont
  {{Kreidberg}}}, \bibinfo {author} {\bibfnamefont {C.~D.}\ \bibnamefont
  {{Bailyn}}}, \bibinfo {author} {\bibfnamefont {W.~M.}\ \bibnamefont
  {{Farr}}},\ and\ \bibinfo {author} {\bibfnamefont {V.}~\bibnamefont
  {{Kalogera}}},\ }\bibfield  {title} {\bibinfo {title} {{Mass Measurements of
  Black Holes in X-Ray Transients: Is There a Mass Gap?}},\ }\href
  {https://doi.org/10.1088/0004-637X/757/1/36} {\bibfield  {journal} {\bibinfo
  {journal} {\apj}\ }\textbf {\bibinfo {volume} {757}},\ \bibinfo {eid} {36}
  (\bibinfo {year} {2012})},\ \Eprint {https://arxiv.org/abs/1205.1805}
  {arXiv:1205.1805 [astro-ph.HE]} \BibitemShut {NoStop}%
\bibitem [{\citenamefont {{Wyrzykowski}}\ \emph {et~al.}(2016)\citenamefont
  {{Wyrzykowski}}, \citenamefont {{Kostrzewa-Rutkowska}}, \citenamefont
  {{Skowron}}, \citenamefont {{Rybicki}}, \citenamefont {{Mr{\'o}z}},
  \citenamefont {{Koz{\l}owski}}, \citenamefont {{Udalski}}, \citenamefont
  {{Szyma{\'n}ski}}, \citenamefont {{Pietrzy{\'n}ski}}, \citenamefont
  {{Soszy{\'n}ski}}, \citenamefont {{Ulaczyk}}, \citenamefont {{Pietrukowicz}},
  \citenamefont {{Poleski}}, \citenamefont {{Pawlak}}, \citenamefont
  {{I{\l}kiewicz}},\ and\ \citenamefont {{Rattenbury}}}]{2016MNRAS.458.3012W}%
  \BibitemOpen
  \bibfield  {author} {\bibinfo {author} {\bibfnamefont {{\L}.}~\bibnamefont
  {{Wyrzykowski}}}, \bibinfo {author} {\bibfnamefont {Z.}~\bibnamefont
  {{Kostrzewa-Rutkowska}}}, \bibinfo {author} {\bibfnamefont {J.}~\bibnamefont
  {{Skowron}}}, \bibinfo {author} {\bibfnamefont {K.~A.}\ \bibnamefont
  {{Rybicki}}}, \bibinfo {author} {\bibfnamefont {P.}~\bibnamefont
  {{Mr{\'o}z}}}, \bibinfo {author} {\bibfnamefont {S.}~\bibnamefont
  {{Koz{\l}owski}}}, \bibinfo {author} {\bibfnamefont {A.}~\bibnamefont
  {{Udalski}}}, \bibinfo {author} {\bibfnamefont {M.~K.}\ \bibnamefont
  {{Szyma{\'n}ski}}}, \bibinfo {author} {\bibfnamefont {G.}~\bibnamefont
  {{Pietrzy{\'n}ski}}}, \bibinfo {author} {\bibfnamefont {I.}~\bibnamefont
  {{Soszy{\'n}ski}}}, \bibinfo {author} {\bibfnamefont {K.}~\bibnamefont
  {{Ulaczyk}}}, \bibinfo {author} {\bibfnamefont {P.}~\bibnamefont
  {{Pietrukowicz}}}, \bibinfo {author} {\bibfnamefont {R.}~\bibnamefont
  {{Poleski}}}, \bibinfo {author} {\bibfnamefont {M.}~\bibnamefont {{Pawlak}}},
  \bibinfo {author} {\bibfnamefont {K.}~\bibnamefont {{I{\l}kiewicz}}},\ and\
  \bibinfo {author} {\bibfnamefont {N.~J.}\ \bibnamefont {{Rattenbury}}},\
  }\bibfield  {title} {\bibinfo {title} {{Black hole, neutron star and white
  dwarf candidates from microlensing with OGLE-III}},\ }\href
  {https://doi.org/10.1093/mnras/stw426} {\bibfield  {journal} {\bibinfo
  {journal} {\mnras}\ }\textbf {\bibinfo {volume} {458}},\ \bibinfo {pages}
  {3012} (\bibinfo {year} {2016})},\ \Eprint {https://arxiv.org/abs/1509.04899}
  {arXiv:1509.04899 [astro-ph.SR]} \BibitemShut {NoStop}%
\bibitem [{\citenamefont {{Wyrzykowski}}\ and\ \citenamefont
  {{Mandel}}(2020)}]{2020A&A...636A..20W}%
  \BibitemOpen
  \bibfield  {author} {\bibinfo {author} {\bibfnamefont {{\L}.}~\bibnamefont
  {{Wyrzykowski}}}\ and\ \bibinfo {author} {\bibfnamefont {I.}~\bibnamefont
  {{Mandel}}},\ }\bibfield  {title} {\bibinfo {title} {{Constraining the masses
  of microlensing black holes and the mass gap with Gaia DR2}},\ }\href
  {https://doi.org/10.1051/0004-6361/201935842} {\bibfield  {journal} {\bibinfo
   {journal} {\aap}\ }\textbf {\bibinfo {volume} {636}},\ \bibinfo {eid} {A20}
  (\bibinfo {year} {2020})},\ \Eprint {https://arxiv.org/abs/1904.07789}
  {arXiv:1904.07789 [astro-ph.SR]} \BibitemShut {NoStop}%
\bibitem [{\citenamefont {{Zevin}}\ \emph {et~al.}(2020)\citenamefont
  {{Zevin}}, \citenamefont {{Spera}}, \citenamefont {{Berry}},\ and\
  \citenamefont {{Kalogera}}}]{Zevin:2020-lower-mass-gap}%
  \BibitemOpen
  \bibfield  {author} {\bibinfo {author} {\bibfnamefont {M.}~\bibnamefont
  {{Zevin}}}, \bibinfo {author} {\bibfnamefont {M.}~\bibnamefont {{Spera}}},
  \bibinfo {author} {\bibfnamefont {C.~P.~L.}\ \bibnamefont {{Berry}}},\ and\
  \bibinfo {author} {\bibfnamefont {V.}~\bibnamefont {{Kalogera}}},\ }\bibfield
   {title} {\bibinfo {title} {{Exploring the Lower Mass Gap and Unequal Mass
  Regime in Compact Binary Evolution}},\ }\href
  {https://doi.org/10.3847/2041-8213/aba74e} {\bibfield  {journal} {\bibinfo
  {journal} {\apjl}\ }\textbf {\bibinfo {volume} {899}},\ \bibinfo {eid} {L1}
  (\bibinfo {year} {2020})},\ \Eprint {https://arxiv.org/abs/2006.14573}
  {arXiv:2006.14573 [astro-ph.HE]} \BibitemShut {NoStop}%
\bibitem [{\citenamefont {{Ertl}}\ \emph {et~al.}(2020)\citenamefont {{Ertl}},
  \citenamefont {{Woosley}}, \citenamefont {{Sukhbold}},\ and\ \citenamefont
  {{Janka}}}]{2020ApJ...890...51E}%
  \BibitemOpen
  \bibfield  {author} {\bibinfo {author} {\bibfnamefont {T.}~\bibnamefont
  {{Ertl}}}, \bibinfo {author} {\bibfnamefont {S.~E.}\ \bibnamefont
  {{Woosley}}}, \bibinfo {author} {\bibfnamefont {T.}~\bibnamefont
  {{Sukhbold}}},\ and\ \bibinfo {author} {\bibfnamefont {H.~T.}\ \bibnamefont
  {{Janka}}},\ }\bibfield  {title} {\bibinfo {title} {{The Explosion of Helium
  Stars Evolved with Mass Loss}},\ }\href
  {https://doi.org/10.3847/1538-4357/ab6458} {\bibfield  {journal} {\bibinfo
  {journal} {\apj}\ }\textbf {\bibinfo {volume} {890}},\ \bibinfo {eid} {51}
  (\bibinfo {year} {2020})},\ \Eprint {https://arxiv.org/abs/1910.01641}
  {arXiv:1910.01641 [astro-ph.HE]} \BibitemShut {NoStop}%
\bibitem [{\citenamefont {{Chan}}\ \emph {et~al.}(2020)\citenamefont {{Chan}},
  \citenamefont {{M{\"u}ller}},\ and\ \citenamefont
  {{Heger}}}]{2020MNRAS.495.3751C}%
  \BibitemOpen
  \bibfield  {author} {\bibinfo {author} {\bibfnamefont {C.}~\bibnamefont
  {{Chan}}}, \bibinfo {author} {\bibfnamefont {B.}~\bibnamefont
  {{M{\"u}ller}}},\ and\ \bibinfo {author} {\bibfnamefont {A.}~\bibnamefont
  {{Heger}}},\ }\bibfield  {title} {\bibinfo {title} {{The impact of fallback
  on the compact remnants and chemical yields of core-collapse supernovae}},\
  }\href {https://doi.org/10.1093/mnras/staa1431} {\bibfield  {journal}
  {\bibinfo  {journal} {\mnras}\ }\textbf {\bibinfo {volume} {495}},\ \bibinfo
  {pages} {3751} (\bibinfo {year} {2020})},\ \Eprint
  {https://arxiv.org/abs/2003.04320} {arXiv:2003.04320 [astro-ph.SR]}
  \BibitemShut {NoStop}%
\bibitem [{\citenamefont {{Fowler}}\ and\ \citenamefont
  {{Hoyle}}(1964)}]{1964ApJS....9..201F}%
  \BibitemOpen
  \bibfield  {author} {\bibinfo {author} {\bibfnamefont {W.~A.}\ \bibnamefont
  {{Fowler}}}\ and\ \bibinfo {author} {\bibfnamefont {F.}~\bibnamefont
  {{Hoyle}}},\ }\bibfield  {title} {\bibinfo {title} {{Neutrino Processes and
  Pair Formation in Massive Stars and Supernovae.}},\ }\href
  {https://doi.org/10.1086/190103} {\bibfield  {journal} {\bibinfo  {journal}
  {\apjs}\ }\textbf {\bibinfo {volume} {9}},\ \bibinfo {pages} {201} (\bibinfo
  {year} {1964})}\BibitemShut {NoStop}%
\bibitem [{\citenamefont {{Fishbach}}\ and\ \citenamefont
  {{Holz}}(2017)}]{2017ApJ...851L..25F}%
  \BibitemOpen
  \bibfield  {author} {\bibinfo {author} {\bibfnamefont {M.}~\bibnamefont
  {{Fishbach}}}\ and\ \bibinfo {author} {\bibfnamefont {D.~E.}\ \bibnamefont
  {{Holz}}},\ }\bibfield  {title} {\bibinfo {title} {{Where Are
  LIGO{\textquoteright}s Big Black Holes?}},\ }\href
  {https://doi.org/10.3847/2041-8213/aa9bf6} {\bibfield  {journal} {\bibinfo
  {journal} {\apjl}\ }\textbf {\bibinfo {volume} {851}},\ \bibinfo {eid} {L25}
  (\bibinfo {year} {2017})},\ \Eprint {https://arxiv.org/abs/1709.08584}
  {arXiv:1709.08584 [astro-ph.HE]} \BibitemShut {NoStop}%
\bibitem [{\citenamefont {{Talbot}}\ and\ \citenamefont
  {{Thrane}}(2018)}]{2018ApJ...856..173T}%
  \BibitemOpen
  \bibfield  {author} {\bibinfo {author} {\bibfnamefont {C.}~\bibnamefont
  {{Talbot}}}\ and\ \bibinfo {author} {\bibfnamefont {E.}~\bibnamefont
  {{Thrane}}},\ }\bibfield  {title} {\bibinfo {title} {{Measuring the Binary
  Black Hole Mass Spectrum with an Astrophysically Motivated
  Parameterization}},\ }\href {https://doi.org/10.3847/1538-4357/aab34c}
  {\bibfield  {journal} {\bibinfo  {journal} {\apj}\ }\textbf {\bibinfo
  {volume} {856}},\ \bibinfo {eid} {173} (\bibinfo {year} {2018})},\ \Eprint
  {https://arxiv.org/abs/1801.02699} {arXiv:1801.02699 [astro-ph.HE]}
  \BibitemShut {NoStop}%
\bibitem [{\citenamefont {{Abbott}}\ \emph
  {et~al.}(2019{\natexlab{a}})\citenamefont {{Abbott}} \emph
  {et~al.}}]{Abbott:2021GWTC1}%
  \BibitemOpen
  \bibfield  {author} {\bibinfo {author} {\bibfnamefont {B.~P.}\ \bibnamefont
  {{Abbott}}} \emph {et~al.},\ }\bibfield  {title} {\bibinfo {title} {{GWTC-1:
  A Gravitational-Wave Transient Catalog of Compact Binary Mergers Observed by
  LIGO and Virgo during the First and Second Observing Runs}},\ }\href
  {https://doi.org/10.1103/PhysRevX.9.031040} {\bibfield  {journal} {\bibinfo
  {journal} {Physical Review X}\ }\textbf {\bibinfo {volume} {9}},\ \bibinfo
  {eid} {031040} (\bibinfo {year} {2019}{\natexlab{a}})},\ \Eprint
  {https://arxiv.org/abs/1811.12907} {arXiv:1811.12907 [astro-ph.HE]}
  \BibitemShut {NoStop}%
\bibitem [{\citenamefont {{Wysocki}}\ \emph {et~al.}(2019)\citenamefont
  {{Wysocki}}, \citenamefont {{Lange}},\ and\ \citenamefont
  {{O'Shaughnessy}}}]{2019PhRvD.100d3012W}%
  \BibitemOpen
  \bibfield  {author} {\bibinfo {author} {\bibfnamefont {D.}~\bibnamefont
  {{Wysocki}}}, \bibinfo {author} {\bibfnamefont {J.}~\bibnamefont {{Lange}}},\
  and\ \bibinfo {author} {\bibfnamefont {R.}~\bibnamefont {{O'Shaughnessy}}},\
  }\bibfield  {title} {\bibinfo {title} {{Reconstructing phenomenological
  distributions of compact binaries via gravitational wave observations}},\
  }\href {https://doi.org/10.1103/PhysRevD.100.043012} {\bibfield  {journal}
  {\bibinfo  {journal} {\prd}\ }\textbf {\bibinfo {volume} {100}},\ \bibinfo
  {eid} {043012} (\bibinfo {year} {2019})},\ \Eprint
  {https://arxiv.org/abs/1805.06442} {arXiv:1805.06442 [gr-qc]} \BibitemShut
  {NoStop}%
\bibitem [{\citenamefont {{Roulet}}\ and\ \citenamefont
  {{Zaldarriaga}}(2019)}]{2019MNRAS.484.4216R}%
  \BibitemOpen
  \bibfield  {author} {\bibinfo {author} {\bibfnamefont {J.}~\bibnamefont
  {{Roulet}}}\ and\ \bibinfo {author} {\bibfnamefont {M.}~\bibnamefont
  {{Zaldarriaga}}},\ }\bibfield  {title} {\bibinfo {title} {{Constraints on
  binary black hole populations from LIGO-Virgo detections}},\ }\href
  {https://doi.org/10.1093/mnras/stz226} {\bibfield  {journal} {\bibinfo
  {journal} {\mnras}\ }\textbf {\bibinfo {volume} {484}},\ \bibinfo {pages}
  {4216} (\bibinfo {year} {2019})},\ \Eprint {https://arxiv.org/abs/1806.10610}
  {arXiv:1806.10610 [astro-ph.HE]} \BibitemShut {NoStop}%
\bibitem [{\citenamefont {{Galaudage}}\ \emph {et~al.}(2020)\citenamefont
  {{Galaudage}}, \citenamefont {{Talbot}},\ and\ \citenamefont
  {{Thrane}}}]{2020PhRvD.102h3026G}%
  \BibitemOpen
  \bibfield  {author} {\bibinfo {author} {\bibfnamefont {S.}~\bibnamefont
  {{Galaudage}}}, \bibinfo {author} {\bibfnamefont {C.}~\bibnamefont
  {{Talbot}}},\ and\ \bibinfo {author} {\bibfnamefont {E.}~\bibnamefont
  {{Thrane}}},\ }\bibfield  {title} {\bibinfo {title} {{Gravitational-wave
  inference in the catalog era: Evolving priors and marginal events}},\ }\href
  {https://doi.org/10.1103/PhysRevD.102.083026} {\bibfield  {journal} {\bibinfo
   {journal} {\prd}\ }\textbf {\bibinfo {volume} {102}},\ \bibinfo {eid}
  {083026} (\bibinfo {year} {2020})},\ \Eprint
  {https://arxiv.org/abs/1912.09708} {arXiv:1912.09708 [astro-ph.HE]}
  \BibitemShut {NoStop}%
\bibitem [{\citenamefont {{Woosley}}\ \emph {et~al.}(2002)\citenamefont
  {{Woosley}}, \citenamefont {{Heger}},\ and\ \citenamefont
  {{Weaver}}}]{Woosley:2002}%
  \BibitemOpen
  \bibfield  {author} {\bibinfo {author} {\bibfnamefont {S.~E.}\ \bibnamefont
  {{Woosley}}}, \bibinfo {author} {\bibfnamefont {A.}~\bibnamefont {{Heger}}},\
  and\ \bibinfo {author} {\bibfnamefont {T.~A.}\ \bibnamefont {{Weaver}}},\
  }\bibfield  {title} {\bibinfo {title} {{The evolution and explosion of
  massive stars}},\ }\href {https://doi.org/10.1103/RevModPhys.74.1015}
  {\bibfield  {journal} {\bibinfo  {journal} {Reviews of Modern Physics}\
  }\textbf {\bibinfo {volume} {74}},\ \bibinfo {pages} {1015} (\bibinfo {year}
  {2002})}\BibitemShut {NoStop}%
\bibitem [{\citenamefont {{Woosley}}\ and\ \citenamefont
  {{Heger}}(2021{\natexlab{b}})}]{2021arXiv210307933W}%
  \BibitemOpen
  \bibfield  {author} {\bibinfo {author} {\bibfnamefont {S.~E.}\ \bibnamefont
  {{Woosley}}}\ and\ \bibinfo {author} {\bibfnamefont {A.}~\bibnamefont
  {{Heger}}},\ }\bibfield  {title} {\bibinfo {title} {{The Pair-instability
  Mass Gap for Black Holes}},\ }\href
  {https://doi.org/10.3847/2041-8213/abf2c4} {\bibfield  {journal} {\bibinfo
  {journal} {\apjl}\ }\textbf {\bibinfo {volume} {912}},\ \bibinfo {eid} {L31}
  (\bibinfo {year} {2021}{\natexlab{b}})},\ \Eprint
  {https://arxiv.org/abs/2103.07933} {arXiv:2103.07933 [astro-ph.SR]}
  \BibitemShut {NoStop}%
\bibitem [{\citenamefont {{Mapelli}}(2021{\natexlab{b}})}]{Mapelli:2021review}%
  \BibitemOpen
  \bibfield  {author} {\bibinfo {author} {\bibfnamefont {M.}~\bibnamefont
  {{Mapelli}}},\ }\bibinfo {title} {{Formation Channels of Single and Binary
  Stellar-Mass Black Holes}},\ in\ \href
  {https://doi.org/10.1007/978-981-15-4702-7\_16-1} {\emph {\bibinfo
  {booktitle} {Handbook of Gravitational Wave Astronomy. Edited by C. Bambi}}}\
  (\bibinfo {year} {2021})\ p.~\bibinfo {pages} {4}\BibitemShut {NoStop}%
\bibitem [{\citenamefont {{Vink}}\ \emph
  {et~al.}(2001{\natexlab{b}})\citenamefont {{Vink}}, \citenamefont {{de
  Koter}},\ and\ \citenamefont {{Lamers}}}]{Vink:2001}%
  \BibitemOpen
  \bibfield  {author} {\bibinfo {author} {\bibfnamefont {J.~S.}\ \bibnamefont
  {{Vink}}}, \bibinfo {author} {\bibfnamefont {A.}~\bibnamefont {{de Koter}}},\
  and\ \bibinfo {author} {\bibfnamefont {H.~J.~G.~L.~M.}\ \bibnamefont
  {{Lamers}}},\ }\bibfield  {title} {\bibinfo {title} {{Mass-loss predictions
  for O and B stars as a function of metallicity}},\ }\href
  {https://doi.org/10.1051/0004-6361:20010127} {\bibfield  {journal} {\bibinfo
  {journal} {\aap}\ }\textbf {\bibinfo {volume} {369}},\ \bibinfo {pages} {574}
  (\bibinfo {year} {2001}{\natexlab{b}})},\ \Eprint
  {https://arxiv.org/abs/astro-ph/0101509} {astro-ph/0101509} \BibitemShut
  {NoStop}%
\bibitem [{\citenamefont {{Marigo}}\ \emph
  {et~al.}(2001{\natexlab{b}})\citenamefont {{Marigo}}, \citenamefont
  {{Girardi}}, \citenamefont {{Chiosi}},\ and\ \citenamefont
  {{Wood}}}]{Marigo:2001}%
  \BibitemOpen
  \bibfield  {author} {\bibinfo {author} {\bibfnamefont {P.}~\bibnamefont
  {{Marigo}}}, \bibinfo {author} {\bibfnamefont {L.}~\bibnamefont {{Girardi}}},
  \bibinfo {author} {\bibfnamefont {C.}~\bibnamefont {{Chiosi}}},\ and\
  \bibinfo {author} {\bibfnamefont {P.~R.}\ \bibnamefont {{Wood}}},\ }\bibfield
   {title} {\bibinfo {title} {{Zero-metallicity stars. I. Evolution at constant
  mass}},\ }\href {https://doi.org/10.1051/0004-6361:20010309} {\bibfield
  {journal} {\bibinfo  {journal} {\aap}\ }\textbf {\bibinfo {volume} {371}},\
  \bibinfo {pages} {152} (\bibinfo {year} {2001}{\natexlab{b}})},\ \Eprint
  {https://arxiv.org/abs/astro-ph/0102253} {arXiv:astro-ph/0102253 [astro-ph]}
  \BibitemShut {NoStop}%
\bibitem [{\citenamefont {{Lovegrove}}\ and\ \citenamefont
  {{Woosley}}(2013)}]{Lovegrove:2013}%
  \BibitemOpen
  \bibfield  {author} {\bibinfo {author} {\bibfnamefont {E.}~\bibnamefont
  {{Lovegrove}}}\ and\ \bibinfo {author} {\bibfnamefont {S.~E.}\ \bibnamefont
  {{Woosley}}},\ }\bibfield  {title} {\bibinfo {title} {{Very Low Energy
  Supernovae from Neutrino Mass Loss}},\ }\href
  {https://doi.org/10.1088/0004-637X/769/2/109} {\bibfield  {journal} {\bibinfo
   {journal} {\apj}\ }\textbf {\bibinfo {volume} {769}},\ \bibinfo {eid} {109}
  (\bibinfo {year} {2013})},\ \Eprint {https://arxiv.org/abs/1303.5055}
  {arXiv:1303.5055 [astro-ph.HE]} \BibitemShut {NoStop}%
\bibitem [{\citenamefont {{Fern{\'a}ndez}}\ \emph
  {et~al.}(2018{\natexlab{b}})\citenamefont {{Fern{\'a}ndez}}, \citenamefont
  {{Quataert}}, \citenamefont {{Kashiyama}},\ and\ \citenamefont
  {{Coughlin}}}]{Fernandez:2018}%
  \BibitemOpen
  \bibfield  {author} {\bibinfo {author} {\bibfnamefont {R.}~\bibnamefont
  {{Fern{\'a}ndez}}}, \bibinfo {author} {\bibfnamefont {E.}~\bibnamefont
  {{Quataert}}}, \bibinfo {author} {\bibfnamefont {K.}~\bibnamefont
  {{Kashiyama}}},\ and\ \bibinfo {author} {\bibfnamefont {E.~R.}\ \bibnamefont
  {{Coughlin}}},\ }\bibfield  {title} {\bibinfo {title} {{Mass ejection in
  failed supernovae: variation with stellar progenitor}},\ }\href
  {https://doi.org/10.1093/mnras/sty306} {\bibfield  {journal} {\bibinfo
  {journal} {\mnras}\ }\textbf {\bibinfo {volume} {476}},\ \bibinfo {pages}
  {2366} (\bibinfo {year} {2018}{\natexlab{b}})},\ \Eprint
  {https://arxiv.org/abs/1710.01735} {arXiv:1710.01735 [astro-ph.HE]}
  \BibitemShut {NoStop}%
\bibitem [{\citenamefont {{Kinugawa}}\ \emph
  {et~al.}(2021{\natexlab{b}})\citenamefont {{Kinugawa}}, \citenamefont
  {{Nakamura}},\ and\ \citenamefont {{Nakano}}}]{Kinugawa:2021}%
  \BibitemOpen
  \bibfield  {author} {\bibinfo {author} {\bibfnamefont {T.}~\bibnamefont
  {{Kinugawa}}}, \bibinfo {author} {\bibfnamefont {T.}~\bibnamefont
  {{Nakamura}}},\ and\ \bibinfo {author} {\bibfnamefont {H.}~\bibnamefont
  {{Nakano}}},\ }\bibfield  {title} {\bibinfo {title} {{Formation of binary
  black holes similar to GW190521 with a total mass of 150 M solar masses from
  Population III binary star evolution}},\ }\href
  {https://doi.org/10.1093/mnrasl/slaa191} {\bibfield  {journal} {\bibinfo
  {journal} {\mnras}\ }\textbf {\bibinfo {volume} {501}},\ \bibinfo {pages}
  {L49} (\bibinfo {year} {2021}{\natexlab{b}})},\ \Eprint
  {https://arxiv.org/abs/2009.06922} {arXiv:2009.06922 [astro-ph.HE]}
  \BibitemShut {NoStop}%
\bibitem [{\citenamefont {{Belczynski}}\ \emph {et~al.}(2004)\citenamefont
  {{Belczynski}}, \citenamefont {{Bulik}},\ and\ \citenamefont
  {{Rudak}}}]{Belczynski:2004popIII}%
  \BibitemOpen
  \bibfield  {author} {\bibinfo {author} {\bibfnamefont {K.}~\bibnamefont
  {{Belczynski}}}, \bibinfo {author} {\bibfnamefont {T.}~\bibnamefont
  {{Bulik}}},\ and\ \bibinfo {author} {\bibfnamefont {B.}~\bibnamefont
  {{Rudak}}},\ }\bibfield  {title} {\bibinfo {title} {{The First Stellar Binary
  Black Holes: The Strongest Gravitational Wave Burst Sources}},\ }\href
  {https://doi.org/10.1086/422172} {\bibfield  {journal} {\bibinfo  {journal}
  {\apjl}\ }\textbf {\bibinfo {volume} {608}},\ \bibinfo {pages} {L45}
  (\bibinfo {year} {2004})},\ \Eprint
  {https://arxiv.org/abs/arXiv:astro-ph/0403361} {arXiv:astro-ph/0403361}
  \BibitemShut {NoStop}%
\bibitem [{\citenamefont {{Kinugawa}}\ \emph {et~al.}(2014)\citenamefont
  {{Kinugawa}}, \citenamefont {{Inayoshi}}, \citenamefont {{Hotokezaka}},
  \citenamefont {{Nakauchi}},\ and\ \citenamefont
  {{Nakamura}}}]{Kinugawa:2014}%
  \BibitemOpen
  \bibfield  {author} {\bibinfo {author} {\bibfnamefont {T.}~\bibnamefont
  {{Kinugawa}}}, \bibinfo {author} {\bibfnamefont {K.}~\bibnamefont
  {{Inayoshi}}}, \bibinfo {author} {\bibfnamefont {K.}~\bibnamefont
  {{Hotokezaka}}}, \bibinfo {author} {\bibfnamefont {D.}~\bibnamefont
  {{Nakauchi}}},\ and\ \bibinfo {author} {\bibfnamefont {T.}~\bibnamefont
  {{Nakamura}}},\ }\bibfield  {title} {\bibinfo {title} {{Possible indirect
  confirmation of the existence of Pop III massive stars by gravitational
  wave}},\ }\href {https://doi.org/10.1093/mnras/stu1022} {\bibfield  {journal}
  {\bibinfo  {journal} {\mnras}\ }\textbf {\bibinfo {volume} {442}},\ \bibinfo
  {pages} {2963} (\bibinfo {year} {2014})},\ \Eprint
  {https://arxiv.org/abs/1402.6672} {arXiv:1402.6672 [astro-ph.HE]}
  \BibitemShut {NoStop}%
\bibitem [{\citenamefont {{Inayoshi}}\ \emph {et~al.}(2016)\citenamefont
  {{Inayoshi}}, \citenamefont {{Kashiyama}}, \citenamefont {{Visbal}},\ and\
  \citenamefont {{Haiman}}}]{Inayoshi:2016}%
  \BibitemOpen
  \bibfield  {author} {\bibinfo {author} {\bibfnamefont {K.}~\bibnamefont
  {{Inayoshi}}}, \bibinfo {author} {\bibfnamefont {K.}~\bibnamefont
  {{Kashiyama}}}, \bibinfo {author} {\bibfnamefont {E.}~\bibnamefont
  {{Visbal}}},\ and\ \bibinfo {author} {\bibfnamefont {Z.}~\bibnamefont
  {{Haiman}}},\ }\bibfield  {title} {\bibinfo {title} {{Gravitational wave
  background from Population III binary black holes consistent with cosmic
  reionization}},\ }\href {https://doi.org/10.1093/mnras/stw1431} {\bibfield
  {journal} {\bibinfo  {journal} {\mnras}\ }\textbf {\bibinfo {volume} {461}},\
  \bibinfo {pages} {2722} (\bibinfo {year} {2016})},\ \Eprint
  {https://arxiv.org/abs/1603.06921} {arXiv:1603.06921 [astro-ph.GA]}
  \BibitemShut {NoStop}%
\bibitem [{\citenamefont {{Hartwig}}\ \emph {et~al.}(2016)\citenamefont
  {{Hartwig}}, \citenamefont {{Volonteri}}, \citenamefont {{Bromm}},
  \citenamefont {{Klessen}}, \citenamefont {{Barausse}}, \citenamefont
  {{Magg}},\ and\ \citenamefont {{Stacy}}}]{Hartwig:2016}%
  \BibitemOpen
  \bibfield  {author} {\bibinfo {author} {\bibfnamefont {T.}~\bibnamefont
  {{Hartwig}}}, \bibinfo {author} {\bibfnamefont {M.}~\bibnamefont
  {{Volonteri}}}, \bibinfo {author} {\bibfnamefont {V.}~\bibnamefont
  {{Bromm}}}, \bibinfo {author} {\bibfnamefont {R.~S.}\ \bibnamefont
  {{Klessen}}}, \bibinfo {author} {\bibfnamefont {E.}~\bibnamefont
  {{Barausse}}}, \bibinfo {author} {\bibfnamefont {M.}~\bibnamefont {{Magg}}},\
  and\ \bibinfo {author} {\bibfnamefont {A.}~\bibnamefont {{Stacy}}},\
  }\bibfield  {title} {\bibinfo {title} {{Gravitational waves from the remnants
  of the first stars}},\ }\href {https://doi.org/10.1093/mnrasl/slw074}
  {\bibfield  {journal} {\bibinfo  {journal} {\mnras}\ }\textbf {\bibinfo
  {volume} {460}},\ \bibinfo {pages} {L74} (\bibinfo {year} {2016})},\ \Eprint
  {https://arxiv.org/abs/1603.05655} {arXiv:1603.05655 [astro-ph.GA]}
  \BibitemShut {NoStop}%
\bibitem [{\citenamefont {{Belczynski}}\ \emph {et~al.}(2018)\citenamefont
  {{Belczynski}}, \citenamefont {{Askar}}, \citenamefont {{Arca-Sedda}},
  \citenamefont {{Chruslinska}}, \citenamefont {{Donnari}}, \citenamefont
  {{Giersz}}, \citenamefont {{Benacquista}}, \citenamefont {{Spurzem}},
  \citenamefont {{Jin}}, \citenamefont {{Wiktorowicz}},\ and\ \citenamefont
  {{Belloni}}}]{Belczynski:2017}%
  \BibitemOpen
  \bibfield  {author} {\bibinfo {author} {\bibfnamefont {K.}~\bibnamefont
  {{Belczynski}}}, \bibinfo {author} {\bibfnamefont {A.}~\bibnamefont
  {{Askar}}}, \bibinfo {author} {\bibfnamefont {M.}~\bibnamefont
  {{Arca-Sedda}}}, \bibinfo {author} {\bibfnamefont {M.}~\bibnamefont
  {{Chruslinska}}}, \bibinfo {author} {\bibfnamefont {M.}~\bibnamefont
  {{Donnari}}}, \bibinfo {author} {\bibfnamefont {M.}~\bibnamefont {{Giersz}}},
  \bibinfo {author} {\bibfnamefont {M.}~\bibnamefont {{Benacquista}}}, \bibinfo
  {author} {\bibfnamefont {R.}~\bibnamefont {{Spurzem}}}, \bibinfo {author}
  {\bibfnamefont {D.}~\bibnamefont {{Jin}}}, \bibinfo {author} {\bibfnamefont
  {G.}~\bibnamefont {{Wiktorowicz}}},\ and\ \bibinfo {author} {\bibfnamefont
  {D.}~\bibnamefont {{Belloni}}},\ }\bibfield  {title} {\bibinfo {title} {{The
  origin of the first neutron star - neutron star merger}},\ }\href
  {https://doi.org/10.1051/0004-6361/201732428} {\bibfield  {journal} {\bibinfo
   {journal} {\aap}\ }\textbf {\bibinfo {volume} {615}},\ \bibinfo {eid} {A91}
  (\bibinfo {year} {2018})},\ \Eprint {https://arxiv.org/abs/1712.00632}
  {arXiv:1712.00632 [astro-ph.HE]} \BibitemShut {NoStop}%
\bibitem [{\citenamefont {{Hijikawa}}\ \emph {et~al.}(2021)\citenamefont
  {{Hijikawa}}, \citenamefont {{Tanikawa}}, \citenamefont {{Kinugawa}},
  \citenamefont {{Yoshida}},\ and\ \citenamefont {{Umeda}}}]{Hijikawa:2021}%
  \BibitemOpen
  \bibfield  {author} {\bibinfo {author} {\bibfnamefont {K.}~\bibnamefont
  {{Hijikawa}}}, \bibinfo {author} {\bibfnamefont {A.}~\bibnamefont
  {{Tanikawa}}}, \bibinfo {author} {\bibfnamefont {T.}~\bibnamefont
  {{Kinugawa}}}, \bibinfo {author} {\bibfnamefont {T.}~\bibnamefont
  {{Yoshida}}},\ and\ \bibinfo {author} {\bibfnamefont {H.}~\bibnamefont
  {{Umeda}}},\ }\bibfield  {title} {\bibinfo {title} {{On the population III
  binary black hole mergers beyond the pair-instability mass gap}},\ }\bibfield
   {journal} {\bibinfo  {journal} {\mnras}\ }\href
  {https://doi.org/10.1093/mnrasl/slab052} {10.1093/mnrasl/slab052} (\bibinfo
  {year} {2021}),\ \Eprint {https://arxiv.org/abs/2104.13384} {arXiv:2104.13384
  [astro-ph.HE]} \BibitemShut {NoStop}%
\bibitem [{\citenamefont {{Kinugawa}}\ \emph {et~al.}(2020)\citenamefont
  {{Kinugawa}}, \citenamefont {{Nakamura}},\ and\ \citenamefont
  {{Nakano}}}]{Kinugawa:2020}%
  \BibitemOpen
  \bibfield  {author} {\bibinfo {author} {\bibfnamefont {T.}~\bibnamefont
  {{Kinugawa}}}, \bibinfo {author} {\bibfnamefont {T.}~\bibnamefont
  {{Nakamura}}},\ and\ \bibinfo {author} {\bibfnamefont {H.}~\bibnamefont
  {{Nakano}}},\ }\bibfield  {title} {\bibinfo {title} {{Chirp mass and spin of
  binary black holes from first star remnants}},\ }\href
  {https://doi.org/10.1093/mnras/staa2511} {\bibfield  {journal} {\bibinfo
  {journal} {\mnras}\ }\textbf {\bibinfo {volume} {498}},\ \bibinfo {pages}
  {3946} (\bibinfo {year} {2020})},\ \Eprint {https://arxiv.org/abs/2005.09795}
  {arXiv:2005.09795 [astro-ph.HE]} \BibitemShut {NoStop}%
\bibitem [{\citenamefont {{Liu}}\ and\ \citenamefont
  {{Bromm}}(2021)}]{LiuBromm:2021}%
  \BibitemOpen
  \bibfield  {author} {\bibinfo {author} {\bibfnamefont {B.}~\bibnamefont
  {{Liu}}}\ and\ \bibinfo {author} {\bibfnamefont {V.}~\bibnamefont
  {{Bromm}}},\ }\bibfield  {title} {\bibinfo {title} {{Gravitational waves from
  the remnants of the first stars in nuclear star clusters}},\ }\href@noop {}
  {\bibfield  {journal} {\bibinfo  {journal} {arXiv e-prints}\ } (\bibinfo
  {year} {2021})},\ \Eprint {https://arxiv.org/abs/2106.02244}
  {arXiv:2106.02244 [astro-ph.GA]} \BibitemShut {NoStop}%
\bibitem [{\citenamefont {{Tanikawa}}\ \emph
  {et~al.}(2021{\natexlab{b}})\citenamefont {{Tanikawa}}, \citenamefont
  {{Yoshida}}, \citenamefont {{Kinugawa}}, \citenamefont {{Trani}},
  \citenamefont {{Hosokawa}}, \citenamefont {{Susa}},\ and\ \citenamefont
  {{Omukai}}}]{Tanikawa:2021}%
  \BibitemOpen
  \bibfield  {author} {\bibinfo {author} {\bibfnamefont {A.}~\bibnamefont
  {{Tanikawa}}}, \bibinfo {author} {\bibfnamefont {T.}~\bibnamefont
  {{Yoshida}}}, \bibinfo {author} {\bibfnamefont {T.}~\bibnamefont
  {{Kinugawa}}}, \bibinfo {author} {\bibfnamefont {A.~A.}\ \bibnamefont
  {{Trani}}}, \bibinfo {author} {\bibfnamefont {T.}~\bibnamefont {{Hosokawa}}},
  \bibinfo {author} {\bibfnamefont {H.}~\bibnamefont {{Susa}}},\ and\ \bibinfo
  {author} {\bibfnamefont {K.}~\bibnamefont {{Omukai}}},\ }\bibfield  {title}
  {\bibinfo {title} {{Merger rate density of binary black holes through
  isolated Population I, II, and III binary star evolution}},\ }\href@noop {}
  {\bibfield  {journal} {\bibinfo  {journal} {arXiv e-prints}\ ,\ \bibinfo
  {eid} {arXiv:2110.10846}} (\bibinfo {year} {2021}{\natexlab{b}})},\ \Eprint
  {https://arxiv.org/abs/2110.10846} {arXiv:2110.10846 [astro-ph.HE]}
  \BibitemShut {NoStop}%
\bibitem [{\citenamefont {{Eddington}}(1925{\natexlab{b}})}]{Eddington:1925}%
  \BibitemOpen
  \bibfield  {author} {\bibinfo {author} {\bibfnamefont {A.~S.}\ \bibnamefont
  {{Eddington}}},\ }\bibfield  {title} {\bibinfo {title} {{Circulating currents
  in rotating stars}},\ }\href@noop {} {\bibfield  {journal} {\bibinfo
  {journal} {The Observatory}\ }\textbf {\bibinfo {volume} {48}},\ \bibinfo
  {pages} {73} (\bibinfo {year} {1925}{\natexlab{b}})}\BibitemShut {NoStop}%
\bibitem [{\citenamefont {{Sweet}}(1950{\natexlab{b}})}]{Sweet:1950}%
  \BibitemOpen
  \bibfield  {author} {\bibinfo {author} {\bibfnamefont {P.~A.}\ \bibnamefont
  {{Sweet}}},\ }\bibfield  {title} {\bibinfo {title} {{The importance of
  rotation in stellar evolution}},\ }\href
  {https://doi.org/10.1093/mnras/110.6.548} {\bibfield  {journal} {\bibinfo
  {journal} {\mnras}\ }\textbf {\bibinfo {volume} {110}},\ \bibinfo {pages}
  {548} (\bibinfo {year} {1950}{\natexlab{b}})}\BibitemShut {NoStop}%
\bibitem [{\citenamefont {{Endal}}\ and\ \citenamefont
  {{Sofia}}(1978)}]{EndalSofia:1978}%
  \BibitemOpen
  \bibfield  {author} {\bibinfo {author} {\bibfnamefont {A.~S.}\ \bibnamefont
  {{Endal}}}\ and\ \bibinfo {author} {\bibfnamefont {S.}~\bibnamefont
  {{Sofia}}},\ }\bibfield  {title} {\bibinfo {title} {{The evolution of
  rotating stars. II - Calculations with time-dependent redistribution of
  angular momentum for 7- and 10-solar-mass stars}},\ }\href
  {https://doi.org/10.1086/155904} {\bibfield  {journal} {\bibinfo  {journal}
  {\apj}\ }\textbf {\bibinfo {volume} {220}},\ \bibinfo {pages} {279} (\bibinfo
  {year} {1978})}\BibitemShut {NoStop}%
\bibitem [{\citenamefont {{Heger}}\ \emph
  {et~al.}(2000{\natexlab{b}})\citenamefont {{Heger}}, \citenamefont
  {{Langer}},\ and\ \citenamefont {{Woosley}}}]{Heger:2000}%
  \BibitemOpen
  \bibfield  {author} {\bibinfo {author} {\bibfnamefont {A.}~\bibnamefont
  {{Heger}}}, \bibinfo {author} {\bibfnamefont {N.}~\bibnamefont {{Langer}}},\
  and\ \bibinfo {author} {\bibfnamefont {S.~E.}\ \bibnamefont {{Woosley}}},\
  }\bibfield  {title} {\bibinfo {title} {{Presupernova Evolution of Rotating
  Massive Stars. I. Numerical Method and Evolution of the Internal Stellar
  Structure}},\ }\href {https://doi.org/10.1086/308158} {\bibfield  {journal}
  {\bibinfo  {journal} {\apj}\ }\textbf {\bibinfo {volume} {528}},\ \bibinfo
  {pages} {368} (\bibinfo {year} {2000}{\natexlab{b}})},\ \Eprint
  {https://arxiv.org/abs/astro-ph/9904132} {astro-ph/9904132} \BibitemShut
  {NoStop}%
\bibitem [{\citenamefont {{Maeder}}\ and\ \citenamefont
  {{Meynet}}(2000{\natexlab{c}})}]{MaederMeynet:2000}%
  \BibitemOpen
  \bibfield  {author} {\bibinfo {author} {\bibfnamefont {A.}~\bibnamefont
  {{Maeder}}}\ and\ \bibinfo {author} {\bibfnamefont {G.}~\bibnamefont
  {{Meynet}}},\ }\bibfield  {title} {\bibinfo {title} {{The Evolution of
  Rotating Stars}},\ }\href {https://doi.org/10.1146/annurev.astro.38.1.143}
  {\bibfield  {journal} {\bibinfo  {journal} {\araa}\ }\textbf {\bibinfo
  {volume} {38}},\ \bibinfo {pages} {143} (\bibinfo {year}
  {2000}{\natexlab{c}})},\ \Eprint {https://arxiv.org/abs/astro-ph/0004204}
  {astro-ph/0004204} \BibitemShut {NoStop}%
\bibitem [{\citenamefont {{Yoon}}\ \emph
  {et~al.}(2006{\natexlab{b}})\citenamefont {{Yoon}}, \citenamefont
  {{Langer}},\ and\ \citenamefont {{Norman}}}]{Yoon:2006}%
  \BibitemOpen
  \bibfield  {author} {\bibinfo {author} {\bibfnamefont {S.-C.}\ \bibnamefont
  {{Yoon}}}, \bibinfo {author} {\bibfnamefont {N.}~\bibnamefont {{Langer}}},\
  and\ \bibinfo {author} {\bibfnamefont {C.}~\bibnamefont {{Norman}}},\
  }\bibfield  {title} {\bibinfo {title} {{Single star progenitors of long
  gamma-ray bursts. I. Model grids and redshift dependent GRB rate}},\ }\href
  {https://doi.org/10.1051/0004-6361:20065912} {\bibfield  {journal} {\bibinfo
  {journal} {\aap}\ }\textbf {\bibinfo {volume} {460}},\ \bibinfo {pages} {199}
  (\bibinfo {year} {2006}{\natexlab{b}})},\ \Eprint
  {https://arxiv.org/abs/astro-ph/0606637} {astro-ph/0606637} \BibitemShut
  {NoStop}%
\bibitem [{\citenamefont {{Mandel}}\ and\ \citenamefont {{de
  Mink}}(2016{\natexlab{b}})}]{MandelDeMink:2016}%
  \BibitemOpen
  \bibfield  {author} {\bibinfo {author} {\bibfnamefont {I.}~\bibnamefont
  {{Mandel}}}\ and\ \bibinfo {author} {\bibfnamefont {S.~E.}\ \bibnamefont {{de
  Mink}}},\ }\bibfield  {title} {\bibinfo {title} {{Merging binary black holes
  formed through chemically homogeneous evolution in short-period stellar
  binaries}},\ }\href {https://doi.org/10.1093/mnras/stw379} {\bibfield
  {journal} {\bibinfo  {journal} {\mnras}\ }\textbf {\bibinfo {volume} {458}},\
  \bibinfo {pages} {2634} (\bibinfo {year} {2016}{\natexlab{b}})},\ \Eprint
  {https://arxiv.org/abs/1601.00007} {arXiv:1601.00007 [astro-ph.HE]}
  \BibitemShut {NoStop}%
\bibitem [{\citenamefont {{\noopsort{De Mink}}{de Mink}}\ and\ \citenamefont
  {{Mandel}}(2016)}]{deMinkMandel:2016}%
  \BibitemOpen
  \bibfield  {author} {\bibinfo {author} {\bibfnamefont {S.~E.}\ \bibnamefont
  {{\noopsort{De Mink}}{de Mink}}}\ and\ \bibinfo {author} {\bibfnamefont
  {I.}~\bibnamefont {{Mandel}}},\ }\bibfield  {title} {\bibinfo {title} {{The
  chemically homogeneous evolutionary channel for binary black hole mergers:
  rates and properties of gravitational-wave events detectable by advanced
  LIGO}},\ }\href {https://doi.org/10.1093/mnras/stw1219} {\bibfield  {journal}
  {\bibinfo  {journal} {\mnras}\ }\textbf {\bibinfo {volume} {460}},\ \bibinfo
  {pages} {3545} (\bibinfo {year} {2016})},\ \Eprint
  {https://arxiv.org/abs/1603.02291} {arXiv:1603.02291 [astro-ph.HE]}
  \BibitemShut {NoStop}%
\bibitem [{\citenamefont {{Marchant}}\ \emph
  {et~al.}(2016{\natexlab{b}})\citenamefont {{Marchant}}, \citenamefont
  {{Langer}}, \citenamefont {{Podsiadlowski}}, \citenamefont {{Tauris}},\ and\
  \citenamefont {{Moriya}}}]{Marchant:2016}%
  \BibitemOpen
  \bibfield  {author} {\bibinfo {author} {\bibfnamefont {P.}~\bibnamefont
  {{Marchant}}}, \bibinfo {author} {\bibfnamefont {N.}~\bibnamefont
  {{Langer}}}, \bibinfo {author} {\bibfnamefont {P.}~\bibnamefont
  {{Podsiadlowski}}}, \bibinfo {author} {\bibfnamefont {T.~M.}\ \bibnamefont
  {{Tauris}}},\ and\ \bibinfo {author} {\bibfnamefont {T.~J.}\ \bibnamefont
  {{Moriya}}},\ }\bibfield  {title} {\bibinfo {title} {{A new route towards
  merging massive black holes}},\ }\href
  {https://doi.org/10.1051/0004-6361/201628133} {\bibfield  {journal} {\bibinfo
   {journal} {\aap}\ }\textbf {\bibinfo {volume} {588}},\ \bibinfo {eid} {A50}
  (\bibinfo {year} {2016}{\natexlab{b}})},\ \Eprint
  {https://arxiv.org/abs/1601.03718} {arXiv:1601.03718 [astro-ph.SR]}
  \BibitemShut {NoStop}%
\bibitem [{\citenamefont {{Riley}}\ \emph
  {et~al.}(2021{\natexlab{b}})\citenamefont {{Riley}}, \citenamefont
  {{Mandel}}, \citenamefont {{Marchant}}, \citenamefont {{Butler}},
  \citenamefont {{Nathaniel}}, \citenamefont {{Neijssel}}, \citenamefont
  {{Shortt}},\ and\ \citenamefont {{Vigna-G{\'o}mez}}}]{Riley:2020}%
  \BibitemOpen
  \bibfield  {author} {\bibinfo {author} {\bibfnamefont {J.}~\bibnamefont
  {{Riley}}}, \bibinfo {author} {\bibfnamefont {I.}~\bibnamefont {{Mandel}}},
  \bibinfo {author} {\bibfnamefont {P.}~\bibnamefont {{Marchant}}}, \bibinfo
  {author} {\bibfnamefont {E.}~\bibnamefont {{Butler}}}, \bibinfo {author}
  {\bibfnamefont {K.}~\bibnamefont {{Nathaniel}}}, \bibinfo {author}
  {\bibfnamefont {C.}~\bibnamefont {{Neijssel}}}, \bibinfo {author}
  {\bibfnamefont {S.}~\bibnamefont {{Shortt}}},\ and\ \bibinfo {author}
  {\bibfnamefont {A.}~\bibnamefont {{Vigna-G{\'o}mez}}},\ }\bibfield  {title}
  {\bibinfo {title} {{Chemically homogeneous evolution: a rapid population
  synthesis approach}},\ }\href {https://doi.org/10.1093/mnras/stab1291}
  {\bibfield  {journal} {\bibinfo  {journal} {\mnras}\ }\textbf {\bibinfo
  {volume} {505}},\ \bibinfo {pages} {663} (\bibinfo {year}
  {2021}{\natexlab{b}})},\ \Eprint {https://arxiv.org/abs/2010.00002}
  {arXiv:2010.00002 [astro-ph.SR]} \BibitemShut {NoStop}%
\bibitem [{\citenamefont {{Marchant}}\ \emph
  {et~al.}(2017{\natexlab{a}})\citenamefont {{Marchant}}, \citenamefont
  {{Langer}}, \citenamefont {{Podsiadlowski}}, \citenamefont {{Tauris}},
  \citenamefont {{de Mink}}, \citenamefont {{Mandel}},\ and\ \citenamefont
  {{Moriya}}}]{Marchant:2017}%
  \BibitemOpen
  \bibfield  {author} {\bibinfo {author} {\bibfnamefont {P.}~\bibnamefont
  {{Marchant}}}, \bibinfo {author} {\bibfnamefont {N.}~\bibnamefont
  {{Langer}}}, \bibinfo {author} {\bibfnamefont {P.}~\bibnamefont
  {{Podsiadlowski}}}, \bibinfo {author} {\bibfnamefont {T.~M.}\ \bibnamefont
  {{Tauris}}}, \bibinfo {author} {\bibfnamefont {S.}~\bibnamefont {{de Mink}}},
  \bibinfo {author} {\bibfnamefont {I.}~\bibnamefont {{Mandel}}},\ and\
  \bibinfo {author} {\bibfnamefont {T.~J.}\ \bibnamefont {{Moriya}}},\
  }\bibfield  {title} {\bibinfo {title} {{Ultra-luminous X-ray sources and
  neutron-star-black-hole mergers from very massive close binaries at low
  metallicity}},\ }\href {https://doi.org/10.1051/0004-6361/201630188}
  {\bibfield  {journal} {\bibinfo  {journal} {\aap}\ }\textbf {\bibinfo
  {volume} {604}},\ \bibinfo {eid} {A55} (\bibinfo {year}
  {2017}{\natexlab{a}})},\ \Eprint {https://arxiv.org/abs/1705.04734}
  {arXiv:1705.04734 [astro-ph.HE]} \BibitemShut {NoStop}%
\bibitem [{\citenamefont {{Moe}}\ and\ \citenamefont {{Di
  Stefano}}(2017{\natexlab{b}})}]{MoeDiStefano:2017}%
  \BibitemOpen
  \bibfield  {author} {\bibinfo {author} {\bibfnamefont {M.}~\bibnamefont
  {{Moe}}}\ and\ \bibinfo {author} {\bibfnamefont {R.}~\bibnamefont {{Di
  Stefano}}},\ }\bibfield  {title} {\bibinfo {title} {{Mind Your Ps and Qs: The
  Interrelation between Period (P) and Mass-ratio (Q) Distributions of Binary
  Stars}},\ }\href {https://doi.org/10.3847/1538-4365/aa6fb6} {\bibfield
  {journal} {\bibinfo  {journal} {\apjs}\ }\textbf {\bibinfo {volume} {230}},\
  \bibinfo {eid} {15} (\bibinfo {year} {2017}{\natexlab{b}})},\ \Eprint
  {https://arxiv.org/abs/1606.05347} {arXiv:1606.05347 [astro-ph.SR]}
  \BibitemShut {NoStop}%
\bibitem [{\citenamefont {{Silsbee}}\ and\ \citenamefont
  {{Tremaine}}(2017{\natexlab{a}})}]{SilsbeeTremaine:2017}%
  \BibitemOpen
  \bibfield  {author} {\bibinfo {author} {\bibfnamefont {K.}~\bibnamefont
  {{Silsbee}}}\ and\ \bibinfo {author} {\bibfnamefont {S.}~\bibnamefont
  {{Tremaine}}},\ }\bibfield  {title} {\bibinfo {title} {{Lidov-Kozai Cycles
  with Gravitational Radiation: Merging Black Holes in Isolated Triple
  Systems}},\ }\href {https://doi.org/10.3847/1538-4357/aa5729} {\bibfield
  {journal} {\bibinfo  {journal} {\apj}\ }\textbf {\bibinfo {volume} {836}},\
  \bibinfo {eid} {39} (\bibinfo {year} {2017}{\natexlab{a}})},\ \Eprint
  {https://arxiv.org/abs/1608.07642} {arXiv:1608.07642 [astro-ph.HE]}
  \BibitemShut {NoStop}%
\bibitem [{\citenamefont {{Antonini}}\ \emph {et~al.}(2017)\citenamefont
  {{Antonini}}, \citenamefont {{Toonen}},\ and\ \citenamefont
  {{Hamers}}}]{Antonini:2017}%
  \BibitemOpen
  \bibfield  {author} {\bibinfo {author} {\bibfnamefont {F.}~\bibnamefont
  {{Antonini}}}, \bibinfo {author} {\bibfnamefont {S.}~\bibnamefont
  {{Toonen}}},\ and\ \bibinfo {author} {\bibfnamefont {A.~S.}\ \bibnamefont
  {{Hamers}}},\ }\bibfield  {title} {\bibinfo {title} {{Binary Black Hole
  Mergers from Field Triples: Properties, Rates, and the Impact of Stellar
  Evolution}},\ }\href {https://doi.org/10.3847/1538-4357/aa6f5e} {\bibfield
  {journal} {\bibinfo  {journal} {\apj}\ }\textbf {\bibinfo {volume} {841}},\
  \bibinfo {eid} {77} (\bibinfo {year} {2017})},\ \Eprint
  {https://arxiv.org/abs/1703.06614} {arXiv:1703.06614 [astro-ph.GA]}
  \BibitemShut {NoStop}%
\bibitem [{\citenamefont {{Rodriguez}}\ and\ \citenamefont
  {{Antonini}}(2018)}]{RodriguezAntonini:2018}%
  \BibitemOpen
  \bibfield  {author} {\bibinfo {author} {\bibfnamefont {C.~L.}\ \bibnamefont
  {{Rodriguez}}}\ and\ \bibinfo {author} {\bibfnamefont {F.}~\bibnamefont
  {{Antonini}}},\ }\bibfield  {title} {\bibinfo {title} {{A Triple Origin for
  the Heavy and Low-spin Binary Black Holes Detected by {LIGO}/VIRGO}},\ }\href
  {https://doi.org/10.3847/1538-4357/aacea4} {\bibfield  {journal} {\bibinfo
  {journal} {\apj}\ }\textbf {\bibinfo {volume} {863}},\ \bibinfo {eid} {7}
  (\bibinfo {year} {2018})},\ \Eprint {https://arxiv.org/abs/1805.08212}
  {arXiv:1805.08212 [astro-ph.HE]} \BibitemShut {NoStop}%
\bibitem [{\citenamefont {{Fragione}}\ and\ \citenamefont
  {{Loeb}}(2019)}]{FragioneLoeb:2019a}%
  \BibitemOpen
  \bibfield  {author} {\bibinfo {author} {\bibfnamefont {G.}~\bibnamefont
  {{Fragione}}}\ and\ \bibinfo {author} {\bibfnamefont {A.}~\bibnamefont
  {{Loeb}}},\ }\bibfield  {title} {\bibinfo {title} {{Black hole-neutron star
  mergers from triples}},\ }\href {https://doi.org/10.1093/mnras/stz1131}
  {\bibfield  {journal} {\bibinfo  {journal} {\mnras}\ }\textbf {\bibinfo
  {volume} {486}},\ \bibinfo {pages} {4443} (\bibinfo {year} {2019})},\ \Eprint
  {https://arxiv.org/abs/1903.10511} {arXiv:1903.10511 [astro-ph.GA]}
  \BibitemShut {NoStop}%
\bibitem [{\citenamefont {{Martinez}}\ \emph
  {et~al.}(2020{\natexlab{a}})\citenamefont {{Martinez}}, \citenamefont
  {{Fragione}}, \citenamefont {{Kremer}}, \citenamefont {{Chatterjee}},
  \citenamefont {{Rodriguez}}, \citenamefont {{Samsing}}, \citenamefont {{Ye}},
  \citenamefont {{Weatherford}}, \citenamefont {{Zevin}}, \citenamefont
  {{Naoz}},\ and\ \citenamefont {{Rasio}}}]{Martinez:2020}%
  \BibitemOpen
  \bibfield  {author} {\bibinfo {author} {\bibfnamefont {M.~A.~S.}\
  \bibnamefont {{Martinez}}}, \bibinfo {author} {\bibfnamefont
  {G.}~\bibnamefont {{Fragione}}}, \bibinfo {author} {\bibfnamefont
  {K.}~\bibnamefont {{Kremer}}}, \bibinfo {author} {\bibfnamefont
  {S.}~\bibnamefont {{Chatterjee}}}, \bibinfo {author} {\bibfnamefont {C.~L.}\
  \bibnamefont {{Rodriguez}}}, \bibinfo {author} {\bibfnamefont
  {J.}~\bibnamefont {{Samsing}}}, \bibinfo {author} {\bibfnamefont {C.~S.}\
  \bibnamefont {{Ye}}}, \bibinfo {author} {\bibfnamefont {N.~C.}\ \bibnamefont
  {{Weatherford}}}, \bibinfo {author} {\bibfnamefont {M.}~\bibnamefont
  {{Zevin}}}, \bibinfo {author} {\bibfnamefont {S.}~\bibnamefont {{Naoz}}},\
  and\ \bibinfo {author} {\bibfnamefont {F.~A.}\ \bibnamefont {{Rasio}}},\
  }\bibfield  {title} {\bibinfo {title} {{Black Hole Mergers from Hierarchical
  Triples in Dense Star Clusters}},\ }\href
  {https://doi.org/10.3847/1538-4357/abba25} {\bibfield  {journal} {\bibinfo
  {journal} {\apj}\ }\textbf {\bibinfo {volume} {903}},\ \bibinfo {eid} {67}
  (\bibinfo {year} {2020}{\natexlab{a}})},\ \Eprint
  {https://arxiv.org/abs/2009.08468} {arXiv:2009.08468 [astro-ph.GA]}
  \BibitemShut {NoStop}%
\bibitem [{\citenamefont {{Hamers}}\ and\ \citenamefont
  {{Thompson}}(2019)}]{HamersThompson:2019}%
  \BibitemOpen
  \bibfield  {author} {\bibinfo {author} {\bibfnamefont {A.~S.}\ \bibnamefont
  {{Hamers}}}\ and\ \bibinfo {author} {\bibfnamefont {T.~A.}\ \bibnamefont
  {{Thompson}}},\ }\bibfield  {title} {\bibinfo {title} {{Double Neutron Star
  Mergers from Hierarchical Triple-star Systems}},\ }\href
  {https://doi.org/10.3847/1538-4357/ab3b06} {\bibfield  {journal} {\bibinfo
  {journal} {\apj}\ }\textbf {\bibinfo {volume} {883}},\ \bibinfo {eid} {23}
  (\bibinfo {year} {2019})},\ \Eprint {https://arxiv.org/abs/1907.08297}
  {arXiv:1907.08297 [astro-ph.HE]} \BibitemShut {NoStop}%
\bibitem [{\citenamefont {{Trani}}\ \emph
  {et~al.}(2021{\natexlab{a}})\citenamefont {{Trani}}, \citenamefont
  {{Rastello}}, \citenamefont {{Di Carlo}}, \citenamefont {{Santoliquido}},
  \citenamefont {{Tanikawa}},\ and\ \citenamefont {{Mapelli}}}]{Trani:2021}%
  \BibitemOpen
  \bibfield  {author} {\bibinfo {author} {\bibfnamefont {A.~A.}\ \bibnamefont
  {{Trani}}}, \bibinfo {author} {\bibfnamefont {S.}~\bibnamefont {{Rastello}}},
  \bibinfo {author} {\bibfnamefont {U.~N.}\ \bibnamefont {{Di Carlo}}},
  \bibinfo {author} {\bibfnamefont {F.}~\bibnamefont {{Santoliquido}}},
  \bibinfo {author} {\bibfnamefont {A.}~\bibnamefont {{Tanikawa}}},\ and\
  \bibinfo {author} {\bibfnamefont {M.}~\bibnamefont {{Mapelli}}},\ }\bibfield
  {title} {\bibinfo {title} {{Compact Object Mergers in Hierarchical Triples
  from Low-Mass Young Star Clusters}},\ }\href@noop {} {\bibfield  {journal}
  {\bibinfo  {journal} {arXiv e-prints}\ } (\bibinfo {year}
  {2021}{\natexlab{a}})},\ \Eprint {https://arxiv.org/abs/2111.06388}
  {arXiv:2111.06388 [astro-ph.HE]} \BibitemShut {NoStop}%
\bibitem [{\citenamefont {{Vynatheya}}\ and\ \citenamefont
  {{Hamers}}(2021)}]{VynatheyaHamers:2021}%
  \BibitemOpen
  \bibfield  {author} {\bibinfo {author} {\bibfnamefont {P.}~\bibnamefont
  {{Vynatheya}}}\ and\ \bibinfo {author} {\bibfnamefont {A.~S.}\ \bibnamefont
  {{Hamers}}},\ }\bibfield  {title} {\bibinfo {title} {{How important is
  secular evolution for black hole and neutron star mergers in 2+2 and 3+1
  quadruple-star systems?}},\ }\href@noop {} {\bibfield  {journal} {\bibinfo
  {journal} {arXiv e-prints}\ } (\bibinfo {year} {2021})},\ \Eprint
  {https://arxiv.org/abs/2110.14680} {arXiv:2110.14680 [astro-ph.HE]}
  \BibitemShut {NoStop}%
\bibitem [{\citenamefont {{Stegmann}}\ \emph
  {et~al.}(2022{\natexlab{a}})\citenamefont {{Stegmann}}, \citenamefont
  {{Antonini}}, \citenamefont {{Schneider}}, \citenamefont {{Tiwari}},\ and\
  \citenamefont {{Chattopadhyay}}}]{Stegmann:2022}%
  \BibitemOpen
  \bibfield  {author} {\bibinfo {author} {\bibfnamefont {J.}~\bibnamefont
  {{Stegmann}}}, \bibinfo {author} {\bibfnamefont {F.}~\bibnamefont
  {{Antonini}}}, \bibinfo {author} {\bibfnamefont {F.~R.~N.}\ \bibnamefont
  {{Schneider}}}, \bibinfo {author} {\bibfnamefont {V.}~\bibnamefont
  {{Tiwari}}},\ and\ \bibinfo {author} {\bibfnamefont {D.}~\bibnamefont
  {{Chattopadhyay}}},\ }\bibfield  {title} {\bibinfo {title} {{Binary black
  hole mergers from merged stars in the Galactic field}},\ }\href
  {https://doi.org/10.1103/PhysRevD.106.023014} {\bibfield  {journal} {\bibinfo
   {journal} {\prd}\ }\textbf {\bibinfo {volume} {106}},\ \bibinfo {eid}
  {023014} (\bibinfo {year} {2022}{\natexlab{a}})},\ \Eprint
  {https://arxiv.org/abs/2203.16544} {arXiv:2203.16544 [astro-ph.GA]}
  \BibitemShut {NoStop}%
\bibitem [{\citenamefont {{Raveh}}\ \emph {et~al.}(2022)\citenamefont
  {{Raveh}}, \citenamefont {{Michaely}},\ and\ \citenamefont
  {{Perets}}}]{Raveh:2022}%
  \BibitemOpen
  \bibfield  {author} {\bibinfo {author} {\bibfnamefont {Y.}~\bibnamefont
  {{Raveh}}}, \bibinfo {author} {\bibfnamefont {E.}~\bibnamefont
  {{Michaely}}},\ and\ \bibinfo {author} {\bibfnamefont {H.~B.}\ \bibnamefont
  {{Perets}}},\ }\bibfield  {title} {\bibinfo {title} {{Detailed properties of
  gravitational-wave mergers from flyby perturbations of wide binary black
  holes in the field}},\ }\href {https://doi.org/10.1093/mnras/stac1605}
  {\bibfield  {journal} {\bibinfo  {journal} {\mnras}\ }\textbf {\bibinfo
  {volume} {514}},\ \bibinfo {pages} {4246} (\bibinfo {year} {2022})},\ \Eprint
  {https://arxiv.org/abs/2204.12506} {arXiv:2204.12506 [astro-ph.HE]}
  \BibitemShut {NoStop}%
\bibitem [{\citenamefont {{Michaely}}\ and\ \citenamefont
  {{Naoz}}(2022)}]{MichaelyNaoz:2022}%
  \BibitemOpen
  \bibfield  {author} {\bibinfo {author} {\bibfnamefont {E.}~\bibnamefont
  {{Michaely}}}\ and\ \bibinfo {author} {\bibfnamefont {S.}~\bibnamefont
  {{Naoz}}},\ }\bibfield  {title} {\bibinfo {title} {{Ultra wide black-hole -
  neutron star binaries as a possible source for gravitational waves and short
  gamma ray bursts}},\ }\href@noop {} {\bibfield  {journal} {\bibinfo
  {journal} {arXiv e-prints}\ ,\ \bibinfo {eid} {arXiv:2205.15040}} (\bibinfo
  {year} {2022})},\ \Eprint {https://arxiv.org/abs/2205.15040}
  {arXiv:2205.15040 [astro-ph.HE]} \BibitemShut {NoStop}%
\bibitem [{\citenamefont {{Bird}}\ \emph {et~al.}(2016)\citenamefont {{Bird}},
  \citenamefont {{Cholis}}, \citenamefont {{Mu{\~n}oz}}, \citenamefont
  {{Ali-Ha{\"i}moud}}, \citenamefont {{Kamionkowski}}, \citenamefont
  {{Kovetz}}, \citenamefont {{Raccanelli}},\ and\ \citenamefont
  {{Riess}}}]{Bird:2016}%
  \BibitemOpen
  \bibfield  {author} {\bibinfo {author} {\bibfnamefont {S.}~\bibnamefont
  {{Bird}}}, \bibinfo {author} {\bibfnamefont {I.}~\bibnamefont {{Cholis}}},
  \bibinfo {author} {\bibfnamefont {J.~B.}\ \bibnamefont {{Mu{\~n}oz}}},
  \bibinfo {author} {\bibfnamefont {Y.}~\bibnamefont {{Ali-Ha{\"i}moud}}},
  \bibinfo {author} {\bibfnamefont {M.}~\bibnamefont {{Kamionkowski}}},
  \bibinfo {author} {\bibfnamefont {E.~D.}\ \bibnamefont {{Kovetz}}}, \bibinfo
  {author} {\bibfnamefont {A.}~\bibnamefont {{Raccanelli}}},\ and\ \bibinfo
  {author} {\bibfnamefont {A.~G.}\ \bibnamefont {{Riess}}},\ }\bibfield
  {title} {\bibinfo {title} {{Did LIGO Detect Dark Matter?}},\ }\href
  {https://doi.org/10.1103/PhysRevLett.116.201301} {\bibfield  {journal}
  {\bibinfo  {journal} {\prl}\ }\textbf {\bibinfo {volume} {116}},\ \bibinfo
  {eid} {201301} (\bibinfo {year} {2016})},\ \Eprint
  {https://arxiv.org/abs/1603.00464} {arXiv:1603.00464} \BibitemShut {NoStop}%
\bibitem [{\citenamefont {{Ali-Ha{\"\i}moud}}\ \emph
  {et~al.}(2017)\citenamefont {{Ali-Ha{\"\i}moud}}, \citenamefont {{Kovetz}},\
  and\ \citenamefont {{Kamionkowski}}}]{AliHaimoud:2017}%
  \BibitemOpen
  \bibfield  {author} {\bibinfo {author} {\bibfnamefont {Y.}~\bibnamefont
  {{Ali-Ha{\"\i}moud}}}, \bibinfo {author} {\bibfnamefont {E.~D.}\ \bibnamefont
  {{Kovetz}}},\ and\ \bibinfo {author} {\bibfnamefont {M.}~\bibnamefont
  {{Kamionkowski}}},\ }\bibfield  {title} {\bibinfo {title} {{Merger rate of
  primordial black-hole binaries}},\ }\href
  {https://doi.org/10.1103/PhysRevD.96.123523} {\bibfield  {journal} {\bibinfo
  {journal} {\prd}\ }\textbf {\bibinfo {volume} {96}},\ \bibinfo {eid} {123523}
  (\bibinfo {year} {2017})},\ \Eprint {https://arxiv.org/abs/1709.06576}
  {arXiv:1709.06576 [astro-ph.CO]} \BibitemShut {NoStop}%
\bibitem [{\citenamefont {{Sakstein}}\ \emph {et~al.}(2020)\citenamefont
  {{Sakstein}}, \citenamefont {{Croon}}, \citenamefont {{McDermott}},
  \citenamefont {{Straight}},\ and\ \citenamefont {{Baxter}}}]{Sakstein:2020}%
  \BibitemOpen
  \bibfield  {author} {\bibinfo {author} {\bibfnamefont {J.}~\bibnamefont
  {{Sakstein}}}, \bibinfo {author} {\bibfnamefont {D.}~\bibnamefont {{Croon}}},
  \bibinfo {author} {\bibfnamefont {S.~D.}\ \bibnamefont {{McDermott}}},
  \bibinfo {author} {\bibfnamefont {M.~C.}\ \bibnamefont {{Straight}}},\ and\
  \bibinfo {author} {\bibfnamefont {E.~J.}\ \bibnamefont {{Baxter}}},\
  }\bibfield  {title} {\bibinfo {title} {{Beyond the Standard Model
  Explanations of GW190521}},\ }\href
  {https://doi.org/10.1103/PhysRevLett.125.261105} {\bibfield  {journal}
  {\bibinfo  {journal} {\prl}\ }\textbf {\bibinfo {volume} {125}},\ \bibinfo
  {eid} {261105} (\bibinfo {year} {2020})},\ \Eprint
  {https://arxiv.org/abs/2009.01213} {arXiv:2009.01213 [gr-qc]} \BibitemShut
  {NoStop}%
\bibitem [{\citenamefont {{Croker}}\ \emph {et~al.}(2021)\citenamefont
  {{Croker}}, \citenamefont {{Zevin}}, \citenamefont {{Farrah}}, \citenamefont
  {{Nishimura}},\ and\ \citenamefont {{Tarl{\'e}}}}]{Croker:2021}%
  \BibitemOpen
  \bibfield  {author} {\bibinfo {author} {\bibfnamefont {K.~S.}\ \bibnamefont
  {{Croker}}}, \bibinfo {author} {\bibfnamefont {M.}~\bibnamefont {{Zevin}}},
  \bibinfo {author} {\bibfnamefont {D.}~\bibnamefont {{Farrah}}}, \bibinfo
  {author} {\bibfnamefont {K.~A.}\ \bibnamefont {{Nishimura}}},\ and\ \bibinfo
  {author} {\bibfnamefont {G.}~\bibnamefont {{Tarl{\'e}}}},\ }\bibfield
  {title} {\bibinfo {title} {{Cosmologically Coupled Compact Objects: A
  Single-parameter Model for {LIGO}-{Virgo} Mass and Redshift Distributions}},\
  }\href {https://doi.org/10.3847/2041-8213/ac2fad} {\bibfield  {journal}
  {\bibinfo  {journal} {\apjl}\ }\textbf {\bibinfo {volume} {921}},\ \bibinfo
  {eid} {L22} (\bibinfo {year} {2021})},\ \Eprint
  {https://arxiv.org/abs/2109.08146} {arXiv:2109.08146 [gr-qc]} \BibitemShut
  {NoStop}%
\bibitem [{\citenamefont {{Mardling}}\ and\ \citenamefont
  {{Aarseth}}(2001)}]{Mardling2001}%
  \BibitemOpen
  \bibfield  {author} {\bibinfo {author} {\bibfnamefont {R.~A.}\ \bibnamefont
  {{Mardling}}}\ and\ \bibinfo {author} {\bibfnamefont {S.~J.}\ \bibnamefont
  {{Aarseth}}},\ }\bibfield  {title} {\bibinfo {title} {{Tidal interactions in
  star cluster simulations}},\ }\href
  {https://doi.org/10.1046/j.1365-8711.2001.03974.x} {\bibfield  {journal}
  {\bibinfo  {journal} {\mnras}\ }\textbf {\bibinfo {volume} {321}},\ \bibinfo
  {pages} {398} (\bibinfo {year} {2001})}\BibitemShut {NoStop}%
\bibitem [{\citenamefont {{Eggleton}}\ and\ \citenamefont
  {{Kiseleva}}(1995)}]{Eggleton1995}%
  \BibitemOpen
  \bibfield  {author} {\bibinfo {author} {\bibfnamefont {P.}~\bibnamefont
  {{Eggleton}}}\ and\ \bibinfo {author} {\bibfnamefont {L.}~\bibnamefont
  {{Kiseleva}}},\ }\bibfield  {title} {\bibinfo {title} {{An Empirical
  Condition for Stability of Hierarchical Triple Systems}},\ }\href
  {https://doi.org/10.1086/176611} {\bibfield  {journal} {\bibinfo  {journal}
  {\apj}\ }\textbf {\bibinfo {volume} {455}},\ \bibinfo {pages} {640} (\bibinfo
  {year} {1995})}\BibitemShut {NoStop}%
\bibitem [{\citenamefont {{Zhang}}\ \emph {et~al.}(2023)\citenamefont
  {{Zhang}}, \citenamefont {{Naoz}},\ and\ \citenamefont {{Will}}}]{Zhang2023}%
  \BibitemOpen
  \bibfield  {author} {\bibinfo {author} {\bibfnamefont {E.}~\bibnamefont
  {{Zhang}}}, \bibinfo {author} {\bibfnamefont {S.}~\bibnamefont {{Naoz}}},\
  and\ \bibinfo {author} {\bibfnamefont {C.~M.}\ \bibnamefont {{Will}}},\
  }\bibfield  {title} {\bibinfo {title} {{A Stability Timescale for
  Nonhierarchical Three-body Systems}},\ }\href
  {https://doi.org/10.3847/1538-4357/acd782} {\bibfield  {journal} {\bibinfo
  {journal} {\apj}\ }\textbf {\bibinfo {volume} {952}},\ \bibinfo {eid} {103}
  (\bibinfo {year} {2023})},\ \Eprint {https://arxiv.org/abs/2301.08271}
  {arXiv:2301.08271 [astro-ph.GA]} \BibitemShut {NoStop}%
\bibitem [{\citenamefont {Lidov}(1962)}]{Lidov1962}%
  \BibitemOpen
  \bibfield  {author} {\bibinfo {author} {\bibfnamefont {M.}~\bibnamefont
  {Lidov}},\ }\bibfield  {title} {\bibinfo {title} {The evolution of orbits of
  artificial satellites of planets under the action of gravitational
  perturbations of external bodies},\ }\href
  {https://doi.org/10.1016/0032-0633(62)90129-0} {\bibfield  {journal}
  {\bibinfo  {journal} {Planetary and Space Science}\ }\textbf {\bibinfo
  {volume} {9}},\ \bibinfo {pages} {719} (\bibinfo {year} {1962})}\BibitemShut
  {NoStop}%
\bibitem [{\citenamefont {Kozai}(1962)}]{Kozai1962}%
  \BibitemOpen
  \bibfield  {author} {\bibinfo {author} {\bibfnamefont {Y.}~\bibnamefont
  {Kozai}},\ }\bibfield  {title} {\bibinfo {title} {Secular perturbations of
  asteroids with high inclination and eccentricity},\ }\href
  {https://doi.org/10.1086/108790} {\bibfield  {journal} {\bibinfo  {journal}
  {The Astronomical Journal}\ }\textbf {\bibinfo {volume} {67}},\ \bibinfo
  {pages} {591} (\bibinfo {year} {1962})}\BibitemShut {NoStop}%
\bibitem [{\citenamefont {Naoz}(2016)}]{Naoz2016}%
  \BibitemOpen
  \bibfield  {author} {\bibinfo {author} {\bibfnamefont {S.}~\bibnamefont
  {Naoz}},\ }\bibfield  {title} {\bibinfo {title} {The {{Eccentric Kozai-Lidov
  Effect}} and {{Its Applications}}},\ }\href
  {https://doi.org/10.1146/annurev-astro-081915-023315} {\bibfield  {journal}
  {\bibinfo  {journal} {Annual Review of Astronomy and Astrophysics}\ }\textbf
  {\bibinfo {volume} {54}},\ \bibinfo {pages} {441} (\bibinfo {year} {2016})},\
  \Eprint {https://arxiv.org/abs/1601.07175} {arxiv:1601.07175} \BibitemShut
  {NoStop}%
\bibitem [{\citenamefont {Peters}(1964)}]{Peters1964}%
  \BibitemOpen
  \bibfield  {author} {\bibinfo {author} {\bibfnamefont {P.}~\bibnamefont
  {Peters}},\ }\bibfield  {title} {\bibinfo {title} {Gravitational
  {{Radiation}} and the {{Motion}} of {{Two Point Masses}}},\ }\href
  {https://doi.org/10.1103/PhysRev.136.B1224} {\bibfield  {journal} {\bibinfo
  {journal} {Physical Review}\ }\textbf {\bibinfo {volume} {136}},\ \bibinfo
  {pages} {B1224} (\bibinfo {year} {1964})}\BibitemShut {NoStop}%
\bibitem [{\citenamefont {{Miller}}\ and\ \citenamefont
  {{Hamilton}}(2002)}]{Miller2002}%
  \BibitemOpen
  \bibfield  {author} {\bibinfo {author} {\bibfnamefont {M.~C.}\ \bibnamefont
  {{Miller}}}\ and\ \bibinfo {author} {\bibfnamefont {D.~P.}\ \bibnamefont
  {{Hamilton}}},\ }\bibfield  {title} {\bibinfo {title} {{Four-Body Effects in
  Globular Cluster Black Hole Coalescence}},\ }\href
  {https://doi.org/10.1086/341788} {\bibfield  {journal} {\bibinfo  {journal}
  {\apj}\ }\textbf {\bibinfo {volume} {576}},\ \bibinfo {pages} {894} (\bibinfo
  {year} {2002})},\ \Eprint {https://arxiv.org/abs/astro-ph/0202298}
  {arXiv:astro-ph/0202298 [astro-ph]} \BibitemShut {NoStop}%
\bibitem [{\citenamefont {Wen}(2003)}]{Wen2003}%
  \BibitemOpen
  \bibfield  {author} {\bibinfo {author} {\bibfnamefont {L.}~\bibnamefont
  {Wen}},\ }\bibfield  {title} {\bibinfo {title} {On the {{Eccentricity
  Distribution}} of {{Coalescing Black Hole Binaries Driven}} by the {{Kozai
  Mechanism}} in {{Globular Clusters}}},\ }\href
  {https://doi.org/10.1086/378794} {\bibfield  {journal} {\bibinfo  {journal}
  {The Astrophysical Journal}\ }\textbf {\bibinfo {volume} {598}},\ \bibinfo
  {pages} {419} (\bibinfo {year} {2003})}\BibitemShut {NoStop}%
\bibitem [{\citenamefont {Ford}\ \emph {et~al.}(2000)\citenamefont {Ford},
  \citenamefont {Kozinsky},\ and\ \citenamefont {Rasio}}]{Ford2000}%
  \BibitemOpen
  \bibfield  {author} {\bibinfo {author} {\bibfnamefont {E.~B.}\ \bibnamefont
  {Ford}}, \bibinfo {author} {\bibfnamefont {B.}~\bibnamefont {Kozinsky}},\
  and\ \bibinfo {author} {\bibfnamefont {F.~A.}\ \bibnamefont {Rasio}},\
  }\bibfield  {title} {\bibinfo {title} {Secular {{Evolution}} of
  {{Hierarchical Triple Star Systems}}},\ }\href
  {https://doi.org/10.1086/308815} {\bibfield  {journal} {\bibinfo  {journal}
  {The Astrophysical Journal}\ }\textbf {\bibinfo {volume} {535}},\ \bibinfo
  {pages} {385} (\bibinfo {year} {2000})},\ \Eprint
  {https://arxiv.org/abs/astro-ph/9905348} {arxiv:astro-ph/9905348}
  \BibitemShut {NoStop}%
\bibitem [{\citenamefont {Blaes}\ \emph {et~al.}(2002)\citenamefont {Blaes},
  \citenamefont {Lee},\ and\ \citenamefont {Socrates}}]{Blaes2002}%
  \BibitemOpen
  \bibfield  {author} {\bibinfo {author} {\bibfnamefont {O.}~\bibnamefont
  {Blaes}}, \bibinfo {author} {\bibfnamefont {M.~H.}\ \bibnamefont {Lee}},\
  and\ \bibinfo {author} {\bibfnamefont {A.}~\bibnamefont {Socrates}},\
  }\bibfield  {title} {\bibinfo {title} {The {{Kozai Mechanism}} and the
  {{Evolution}} of {{Binary Supermassive Black Holes}}},\ }\href
  {https://doi.org/10.1086/342655} {\bibfield  {journal} {\bibinfo  {journal}
  {The Astrophysical Journal}\ }\textbf {\bibinfo {volume} {578}},\ \bibinfo
  {pages} {775} (\bibinfo {year} {2002})}\BibitemShut {NoStop}%
\bibitem [{\citenamefont {Naoz}\ \emph {et~al.}(2013)\citenamefont {Naoz},
  \citenamefont {Farr}, \citenamefont {Lithwick}, \citenamefont {Rasio},\ and\
  \citenamefont {Teyssandier}}]{Naoz2013}%
  \BibitemOpen
  \bibfield  {author} {\bibinfo {author} {\bibfnamefont {S.}~\bibnamefont
  {Naoz}}, \bibinfo {author} {\bibfnamefont {W.~M.}\ \bibnamefont {Farr}},
  \bibinfo {author} {\bibfnamefont {Y.}~\bibnamefont {Lithwick}}, \bibinfo
  {author} {\bibfnamefont {F.~A.}\ \bibnamefont {Rasio}},\ and\ \bibinfo
  {author} {\bibfnamefont {J.}~\bibnamefont {Teyssandier}},\ }\bibfield
  {title} {\bibinfo {title} {Secular dynamics in hierarchical three-body
  systems},\ }\href {https://doi.org/10.1093/mnras/stt302} {\bibfield
  {journal} {\bibinfo  {journal} {Monthly Notices of the Royal Astronomical
  Society}\ }\textbf {\bibinfo {volume} {431}},\ \bibinfo {pages} {2155}
  (\bibinfo {year} {2013})},\ \Eprint {https://arxiv.org/abs/1107.2414}
  {arxiv:1107.2414} \BibitemShut {NoStop}%
\bibitem [{\citenamefont {Liu}\ \emph {et~al.}(2015)\citenamefont {Liu},
  \citenamefont {Mu{\~n}oz},\ and\ \citenamefont {Lai}}]{Liu2015}%
  \BibitemOpen
  \bibfield  {author} {\bibinfo {author} {\bibfnamefont {B.}~\bibnamefont
  {Liu}}, \bibinfo {author} {\bibfnamefont {D.~J.}\ \bibnamefont {Mu{\~n}oz}},\
  and\ \bibinfo {author} {\bibfnamefont {D.}~\bibnamefont {Lai}},\ }\bibfield
  {title} {\bibinfo {title} {Suppression of extreme orbital evolution in triple
  systems with short-range forces},\ }\href
  {https://doi.org/10.1093/mnras/stu2396} {\bibfield  {journal} {\bibinfo
  {journal} {Monthly Notices of the Royal Astronomical Society}\ }\textbf
  {\bibinfo {volume} {447}},\ \bibinfo {pages} {747} (\bibinfo {year}
  {2015})}\BibitemShut {NoStop}%
\bibitem [{\citenamefont {{Liu}}\ \emph {et~al.}(2015)\citenamefont {{Liu}},
  \citenamefont {{Lai}},\ and\ \citenamefont {{Yuan}}}]{Liu2015b}%
  \BibitemOpen
  \bibfield  {author} {\bibinfo {author} {\bibfnamefont {B.}~\bibnamefont
  {{Liu}}}, \bibinfo {author} {\bibfnamefont {D.}~\bibnamefont {{Lai}}},\ and\
  \bibinfo {author} {\bibfnamefont {Y.-F.}\ \bibnamefont {{Yuan}}},\ }\bibfield
   {title} {\bibinfo {title} {{Merging compact binaries in hierarchical triple
  systems: Resonant excitation of binary eccentricity}},\ }\href
  {https://doi.org/10.1103/PhysRevD.92.124048} {\bibfield  {journal} {\bibinfo
  {journal} {\prd}\ }\textbf {\bibinfo {volume} {92}},\ \bibinfo {eid} {124048}
  (\bibinfo {year} {2015})},\ \Eprint {https://arxiv.org/abs/1511.07365}
  {arXiv:1511.07365 [astro-ph.HE]} \BibitemShut {NoStop}%
\bibitem [{\citenamefont {{Naoz}}\ \emph {et~al.}(2013)\citenamefont {{Naoz}},
  \citenamefont {{Kocsis}}, \citenamefont {{Loeb}},\ and\ \citenamefont
  {{Yunes}}}]{Naoz2013a}%
  \BibitemOpen
  \bibfield  {author} {\bibinfo {author} {\bibfnamefont {S.}~\bibnamefont
  {{Naoz}}}, \bibinfo {author} {\bibfnamefont {B.}~\bibnamefont {{Kocsis}}},
  \bibinfo {author} {\bibfnamefont {A.}~\bibnamefont {{Loeb}}},\ and\ \bibinfo
  {author} {\bibfnamefont {N.}~\bibnamefont {{Yunes}}},\ }\bibfield  {title}
  {\bibinfo {title} {{Resonant Post-Newtonian Eccentricity Excitation in
  Hierarchical Three-body Systems}},\ }\href
  {https://doi.org/10.1088/0004-637X/773/2/187} {\bibfield  {journal} {\bibinfo
   {journal} {\apj}\ }\textbf {\bibinfo {volume} {773}},\ \bibinfo {eid} {187}
  (\bibinfo {year} {2013})},\ \Eprint {https://arxiv.org/abs/1206.4316}
  {arXiv:1206.4316 [astro-ph.SR]} \BibitemShut {NoStop}%
\bibitem [{\citenamefont {{Will}}(2014)}]{Will2014}%
  \BibitemOpen
  \bibfield  {author} {\bibinfo {author} {\bibfnamefont {C.~M.}\ \bibnamefont
  {{Will}}},\ }\bibfield  {title} {\bibinfo {title} {{Incorporating
  post-Newtonian effects in N-body dynamics}},\ }\href
  {https://doi.org/10.1103/PhysRevD.89.044043} {\bibfield  {journal} {\bibinfo
  {journal} {\prd}\ }\textbf {\bibinfo {volume} {89}},\ \bibinfo {eid} {044043}
  (\bibinfo {year} {2014})},\ \Eprint {https://arxiv.org/abs/1312.1289}
  {arXiv:1312.1289 [astro-ph.GA]} \BibitemShut {NoStop}%
\bibitem [{\citenamefont {{Lim}}\ and\ \citenamefont
  {{Rodriguez}}(2020)}]{Lim2020}%
  \BibitemOpen
  \bibfield  {author} {\bibinfo {author} {\bibfnamefont {H.}~\bibnamefont
  {{Lim}}}\ and\ \bibinfo {author} {\bibfnamefont {C.~L.}\ \bibnamefont
  {{Rodriguez}}},\ }\bibfield  {title} {\bibinfo {title} {{Relativistic
  three-body effects in hierarchical triples}},\ }\href
  {https://doi.org/10.1103/PhysRevD.102.064033} {\bibfield  {journal} {\bibinfo
   {journal} {\prd}\ }\textbf {\bibinfo {volume} {102}},\ \bibinfo {eid}
  {064033} (\bibinfo {year} {2020})},\ \Eprint
  {https://arxiv.org/abs/2001.03654} {arXiv:2001.03654 [astro-ph.HE]}
  \BibitemShut {NoStop}%
\bibitem [{\citenamefont {{Liu}}\ and\ \citenamefont {{Lai}}(2020)}]{Liu2020}%
  \BibitemOpen
  \bibfield  {author} {\bibinfo {author} {\bibfnamefont {B.}~\bibnamefont
  {{Liu}}}\ and\ \bibinfo {author} {\bibfnamefont {D.}~\bibnamefont {{Lai}}},\
  }\bibfield  {title} {\bibinfo {title} {{Merging compact binaries near a
  rotating supermassive black hole: Eccentricity excitation due to apsidal
  precession resonance}},\ }\href {https://doi.org/10.1103/PhysRevD.102.023020}
  {\bibfield  {journal} {\bibinfo  {journal} {\prd}\ }\textbf {\bibinfo
  {volume} {102}},\ \bibinfo {eid} {023020} (\bibinfo {year} {2020})},\ \Eprint
  {https://arxiv.org/abs/2004.10205} {arXiv:2004.10205 [astro-ph.HE]}
  \BibitemShut {NoStop}%
\bibitem [{\citenamefont {{Will}}(2021)}]{Will2021}%
  \BibitemOpen
  \bibfield  {author} {\bibinfo {author} {\bibfnamefont {C.~M.}\ \bibnamefont
  {{Will}}},\ }\bibfield  {title} {\bibinfo {title} {{Higher-order effects in
  the dynamics of hierarchical triple systems: Quadrupole-squared terms}},\
  }\href {https://doi.org/10.1103/PhysRevD.103.063003} {\bibfield  {journal}
  {\bibinfo  {journal} {\prd}\ }\textbf {\bibinfo {volume} {103}},\ \bibinfo
  {eid} {063003} (\bibinfo {year} {2021})},\ \Eprint
  {https://arxiv.org/abs/2011.13286} {arXiv:2011.13286 [astro-ph.EP]}
  \BibitemShut {NoStop}%
\bibitem [{\citenamefont {{Kuntz}}\ \emph {et~al.}(2023)\citenamefont
  {{Kuntz}}, \citenamefont {{Serra}},\ and\ \citenamefont
  {{Trincherini}}}]{Kuntz2023}%
  \BibitemOpen
  \bibfield  {author} {\bibinfo {author} {\bibfnamefont {A.}~\bibnamefont
  {{Kuntz}}}, \bibinfo {author} {\bibfnamefont {F.}~\bibnamefont {{Serra}}},\
  and\ \bibinfo {author} {\bibfnamefont {E.}~\bibnamefont {{Trincherini}}},\
  }\bibfield  {title} {\bibinfo {title} {{Effective two-body approach to the
  hierarchical three-body problem: Quadrupole to 1PN}},\ }\href
  {https://doi.org/10.1103/PhysRevD.107.044011} {\bibfield  {journal} {\bibinfo
   {journal} {\prd}\ }\textbf {\bibinfo {volume} {107}},\ \bibinfo {eid}
  {044011} (\bibinfo {year} {2023})},\ \Eprint
  {https://arxiv.org/abs/2210.13493} {arXiv:2210.13493 [gr-qc]} \BibitemShut
  {NoStop}%
\bibitem [{\citenamefont {Antonini}\ \emph {et~al.}(2014)\citenamefont
  {Antonini}, \citenamefont {Murray},\ and\ \citenamefont
  {Mikkola}}]{Antonini2014}%
  \BibitemOpen
  \bibfield  {author} {\bibinfo {author} {\bibfnamefont {F.}~\bibnamefont
  {Antonini}}, \bibinfo {author} {\bibfnamefont {N.}~\bibnamefont {Murray}},\
  and\ \bibinfo {author} {\bibfnamefont {S.}~\bibnamefont {Mikkola}},\
  }\bibfield  {title} {\bibinfo {title} {{{BLACK HOLE TRIPLE DYNAMICS}}: {{A
  BREAKDOWN OF THE ORBIT AVERAGE APPROXIMATION AND IMPLICATIONS FOR
  GRAVITATIONAL WAVE DETECTIONS}}},\ }\href
  {https://doi.org/10.1088/0004-637X/781/1/45} {\bibfield  {journal} {\bibinfo
  {journal} {The Astrophysical Journal}\ }\textbf {\bibinfo {volume} {781}},\
  \bibinfo {pages} {45} (\bibinfo {year} {2014})}\BibitemShut {NoStop}%
\bibitem [{\citenamefont {Antognini}\ \emph {et~al.}(2014)\citenamefont
  {Antognini}, \citenamefont {Shappee}, \citenamefont {Thompson},\ and\
  \citenamefont {{Amaro-Seoane}}}]{Antognini2014}%
  \BibitemOpen
  \bibfield  {author} {\bibinfo {author} {\bibfnamefont {J.~M.}\ \bibnamefont
  {Antognini}}, \bibinfo {author} {\bibfnamefont {B.~J.}\ \bibnamefont
  {Shappee}}, \bibinfo {author} {\bibfnamefont {T.~A.}\ \bibnamefont
  {Thompson}},\ and\ \bibinfo {author} {\bibfnamefont {P.}~\bibnamefont
  {{Amaro-Seoane}}},\ }\bibfield  {title} {\bibinfo {title} {Rapid eccentricity
  oscillations and the mergers of compact objects in hierarchical triples},\
  }\href {https://doi.org/10.1093/mnras/stu039} {\bibfield  {journal} {\bibinfo
   {journal} {Monthly Notices of the Royal Astronomical Society}\ }\textbf
  {\bibinfo {volume} {439}},\ \bibinfo {pages} {1079} (\bibinfo {year}
  {2014})}\BibitemShut {NoStop}%
\bibitem [{\citenamefont {Moe}\ and\ \citenamefont
  {Di~Stefano}(2017)}]{moe2017}%
  \BibitemOpen
  \bibfield  {author} {\bibinfo {author} {\bibfnamefont {M.}~\bibnamefont
  {Moe}}\ and\ \bibinfo {author} {\bibfnamefont {R.}~\bibnamefont
  {Di~Stefano}},\ }\bibfield  {title} {\bibinfo {title} {Mind {{Your Ps}} and
  {{Qs}}: {{The Interrelation}} between {{Period}} ( {{P}} ) and {{Mass-ratio}}
  ( {{Q}} ) {{Distributions}} of {{Binary Stars}}},\ }\href
  {https://doi.org/10.3847/1538-4365/aa6fb6} {\bibfield  {journal} {\bibinfo
  {journal} {The Astrophysical Journal Supplement Series}\ }\textbf {\bibinfo
  {volume} {230}},\ \bibinfo {pages} {15} (\bibinfo {year} {2017})},\ \Eprint
  {https://arxiv.org/abs/1606.05347} {arXiv:1606.05347} \BibitemShut {NoStop}%
\bibitem [{\citenamefont {{Silsbee}}\ and\ \citenamefont
  {{Tremaine}}(2017{\natexlab{b}})}]{Silsbee2017}%
  \BibitemOpen
  \bibfield  {author} {\bibinfo {author} {\bibfnamefont {K.}~\bibnamefont
  {{Silsbee}}}\ and\ \bibinfo {author} {\bibfnamefont {S.}~\bibnamefont
  {{Tremaine}}},\ }\bibfield  {title} {\bibinfo {title} {{Lidov-Kozai Cycles
  with Gravitational Radiation: Merging Black Holes in Isolated Triple
  Systems}},\ }\href {https://doi.org/10.3847/1538-4357/aa5729} {\bibfield
  {journal} {\bibinfo  {journal} {\apj}\ }\textbf {\bibinfo {volume} {836}},\
  \bibinfo {eid} {39} (\bibinfo {year} {2017}{\natexlab{b}})},\ \Eprint
  {https://arxiv.org/abs/1608.07642} {arXiv:1608.07642 [astro-ph.HE]}
  \BibitemShut {NoStop}%
\bibitem [{\citenamefont {Antonini}\ \emph {et~al.}(2017)\citenamefont
  {Antonini}, \citenamefont {Toonen},\ and\ \citenamefont
  {Hamers}}]{Antonini2017}%
  \BibitemOpen
  \bibfield  {author} {\bibinfo {author} {\bibfnamefont {F.}~\bibnamefont
  {Antonini}}, \bibinfo {author} {\bibfnamefont {S.}~\bibnamefont {Toonen}},\
  and\ \bibinfo {author} {\bibfnamefont {A.~S.}\ \bibnamefont {Hamers}},\
  }\bibfield  {title} {\bibinfo {title} {Binary {{Black Hole Mergers}} from
  {{Field Triples}}: {{Properties}}, {{Rates}}, and the {{Impact}} of {{Stellar
  Evolution}}},\ }\href {https://doi.org/10.3847/1538-4357/aa6f5e} {\bibfield
  {journal} {\bibinfo  {journal} {The Astrophysical Journal}\ }\textbf
  {\bibinfo {volume} {841}},\ \bibinfo {pages} {77} (\bibinfo {year} {2017})},\
  \Eprint {https://arxiv.org/abs/1703.06614} {arxiv:1703.06614} \BibitemShut
  {NoStop}%
\bibitem [{\citenamefont {Rodriguez}\ and\ \citenamefont
  {Antonini}(2018)}]{Rodriguez2018a}%
  \BibitemOpen
  \bibfield  {author} {\bibinfo {author} {\bibfnamefont {C.~L.}\ \bibnamefont
  {Rodriguez}}\ and\ \bibinfo {author} {\bibfnamefont {F.}~\bibnamefont
  {Antonini}},\ }\bibfield  {title} {\bibinfo {title} {A {{Triple Origin}} for
  the {{Heavy}} and {{Low-Spin Binary Black Holes Detected}} by
  {{LIGO}}/{{Virgo}}},\ }\href {https://doi.org/10.3847/1538-4357/aacea4}
  {\bibfield  {journal} {\bibinfo  {journal} {The Astrophysical Journal}\
  }\textbf {\bibinfo {volume} {863}},\ \bibinfo {pages} {7} (\bibinfo {year}
  {2018})},\ \Eprint {https://arxiv.org/abs/1805.08212} {arxiv:1805.08212}
  \BibitemShut {NoStop}%
\bibitem [{\citenamefont {{Stegmann}}\ \emph
  {et~al.}(2022{\natexlab{b}})\citenamefont {{Stegmann}}, \citenamefont
  {{Antonini}},\ and\ \citenamefont {{Moe}}}]{Stegmann2022}%
  \BibitemOpen
  \bibfield  {author} {\bibinfo {author} {\bibfnamefont {J.}~\bibnamefont
  {{Stegmann}}}, \bibinfo {author} {\bibfnamefont {F.}~\bibnamefont
  {{Antonini}}},\ and\ \bibinfo {author} {\bibfnamefont {M.}~\bibnamefont
  {{Moe}}},\ }\bibfield  {title} {\bibinfo {title} {{Evolution of massive
  stellar triples and implications for compact object binary formation}},\
  }\href {https://doi.org/10.1093/mnras/stac2192} {\bibfield  {journal}
  {\bibinfo  {journal} {\mnras}\ }\textbf {\bibinfo {volume} {516}},\ \bibinfo
  {pages} {1406} (\bibinfo {year} {2022}{\natexlab{b}})},\ \Eprint
  {https://arxiv.org/abs/2112.10786} {arXiv:2112.10786 [astro-ph.SR]}
  \BibitemShut {NoStop}%
\bibitem [{\citenamefont {{Dorozsmai}}\ \emph {et~al.}(2023)\citenamefont
  {{Dorozsmai}}, \citenamefont {{Toonen}}, \citenamefont {{Vigna-G{\'o}mez}},
  \citenamefont {{de Mink}},\ and\ \citenamefont {{Kummer}}}]{Dorozsmai2023}%
  \BibitemOpen
  \bibfield  {author} {\bibinfo {author} {\bibfnamefont {A.}~\bibnamefont
  {{Dorozsmai}}}, \bibinfo {author} {\bibfnamefont {S.}~\bibnamefont
  {{Toonen}}}, \bibinfo {author} {\bibfnamefont {A.}~\bibnamefont
  {{Vigna-G{\'o}mez}}}, \bibinfo {author} {\bibfnamefont {S.~E.}\ \bibnamefont
  {{de Mink}}},\ and\ \bibinfo {author} {\bibfnamefont {F.}~\bibnamefont
  {{Kummer}}},\ }\bibfield  {title} {\bibinfo {title} {{Stellar triples with
  chemically homogeneously evolving inner binaries}},\ }\href
  {https://doi.org/10.48550/arXiv.2307.04793} {\bibfield  {journal} {\bibinfo
  {journal} {arXiv e-prints}\ ,\ \bibinfo {eid} {arXiv:2307.04793}} (\bibinfo
  {year} {2023})},\ \Eprint {https://arxiv.org/abs/2307.04793}
  {arXiv:2307.04793 [astro-ph.SR]} \BibitemShut {NoStop}%
\bibitem [{\citenamefont {{Su}}\ \emph {et~al.}(2021)\citenamefont {{Su}},
  \citenamefont {{Liu}},\ and\ \citenamefont {{Lai}}}]{Su2021}%
  \BibitemOpen
  \bibfield  {author} {\bibinfo {author} {\bibfnamefont {Y.}~\bibnamefont
  {{Su}}}, \bibinfo {author} {\bibfnamefont {B.}~\bibnamefont {{Liu}}},\ and\
  \bibinfo {author} {\bibfnamefont {D.}~\bibnamefont {{Lai}}},\ }\bibfield
  {title} {\bibinfo {title} {{The mass-ratio distribution of tertiary-induced
  binary black hole mergers}},\ }\href {https://doi.org/10.1093/mnras/stab1617}
  {\bibfield  {journal} {\bibinfo  {journal} {\mnras}\ }\textbf {\bibinfo
  {volume} {505}},\ \bibinfo {pages} {3681} (\bibinfo {year} {2021})},\ \Eprint
  {https://arxiv.org/abs/2103.01963} {arXiv:2103.01963 [astro-ph.HE]}
  \BibitemShut {NoStop}%
\bibitem [{\citenamefont {{Martinez}}\ \emph {et~al.}(2022)\citenamefont
  {{Martinez}}, \citenamefont {{Rodriguez}},\ and\ \citenamefont
  {{Fragione}}}]{Martinez2022}%
  \BibitemOpen
  \bibfield  {author} {\bibinfo {author} {\bibfnamefont {M.~A.~S.}\
  \bibnamefont {{Martinez}}}, \bibinfo {author} {\bibfnamefont {C.~L.}\
  \bibnamefont {{Rodriguez}}},\ and\ \bibinfo {author} {\bibfnamefont
  {G.}~\bibnamefont {{Fragione}}},\ }\bibfield  {title} {\bibinfo {title} {{On
  the Mass Ratio Distribution of Black Hole Mergers in Triple Systems}},\
  }\href {https://doi.org/10.3847/1538-4357/ac8d55} {\bibfield  {journal}
  {\bibinfo  {journal} {\apj}\ }\textbf {\bibinfo {volume} {937}},\ \bibinfo
  {eid} {78} (\bibinfo {year} {2022})},\ \Eprint
  {https://arxiv.org/abs/2105.01671} {arXiv:2105.01671 [astro-ph.SR]}
  \BibitemShut {NoStop}%
\bibitem [{\citenamefont {{Antonini}}\ \emph {et~al.}(2018)\citenamefont
  {{Antonini}}, \citenamefont {{Rodriguez}}, \citenamefont {{Petrovich}},\ and\
  \citenamefont {{Fischer}}}]{Antonini2018}%
  \BibitemOpen
  \bibfield  {author} {\bibinfo {author} {\bibfnamefont {F.}~\bibnamefont
  {{Antonini}}}, \bibinfo {author} {\bibfnamefont {C.~L.}\ \bibnamefont
  {{Rodriguez}}}, \bibinfo {author} {\bibfnamefont {C.}~\bibnamefont
  {{Petrovich}}},\ and\ \bibinfo {author} {\bibfnamefont {C.~L.}\ \bibnamefont
  {{Fischer}}},\ }\bibfield  {title} {\bibinfo {title} {{Precessional dynamics
  of black hole triples: binary mergers with near-zero effective spin}},\
  }\href {https://doi.org/10.1093/mnrasl/sly126} {\bibfield  {journal}
  {\bibinfo  {journal} {\mnras}\ }\textbf {\bibinfo {volume} {480}},\ \bibinfo
  {pages} {L58} (\bibinfo {year} {2018})},\ \Eprint
  {https://arxiv.org/abs/1711.07142} {arXiv:1711.07142 [astro-ph.HE]}
  \BibitemShut {NoStop}%
\bibitem [{\citenamefont {{Liu}}\ and\ \citenamefont {{Lai}}(2018)}]{Liu2018}%
  \BibitemOpen
  \bibfield  {author} {\bibinfo {author} {\bibfnamefont {B.}~\bibnamefont
  {{Liu}}}\ and\ \bibinfo {author} {\bibfnamefont {D.}~\bibnamefont {{Lai}}},\
  }\bibfield  {title} {\bibinfo {title} {{Black Hole and Neutron Star Binary
  Mergers in Triple Systems: Merger Fraction and Spin-Orbit Misalignment}},\
  }\href {https://doi.org/10.3847/1538-4357/aad09f} {\bibfield  {journal}
  {\bibinfo  {journal} {\apj}\ }\textbf {\bibinfo {volume} {863}},\ \bibinfo
  {eid} {68} (\bibinfo {year} {2018})},\ \Eprint
  {https://arxiv.org/abs/1805.03202} {arXiv:1805.03202 [astro-ph.HE]}
  \BibitemShut {NoStop}%
\bibitem [{\citenamefont {{Yu}}\ \emph {et~al.}(2020)\citenamefont {{Yu}},
  \citenamefont {{Ma}}, \citenamefont {{Giesler}},\ and\ \citenamefont
  {{Chen}}}]{Yu2020}%
  \BibitemOpen
  \bibfield  {author} {\bibinfo {author} {\bibfnamefont {H.}~\bibnamefont
  {{Yu}}}, \bibinfo {author} {\bibfnamefont {S.}~\bibnamefont {{Ma}}}, \bibinfo
  {author} {\bibfnamefont {M.}~\bibnamefont {{Giesler}}},\ and\ \bibinfo
  {author} {\bibfnamefont {Y.}~\bibnamefont {{Chen}}},\ }\bibfield  {title}
  {\bibinfo {title} {{Spin and eccentricity evolution in triple systems: From
  the Lidov-Kozai interaction to the final merger of the inner binary}},\
  }\href {https://doi.org/10.1103/PhysRevD.102.123009} {\bibfield  {journal}
  {\bibinfo  {journal} {\prd}\ }\textbf {\bibinfo {volume} {102}},\ \bibinfo
  {eid} {123009} (\bibinfo {year} {2020})},\ \Eprint
  {https://arxiv.org/abs/2007.12978} {arXiv:2007.12978 [gr-qc]} \BibitemShut
  {NoStop}%
\bibitem [{\citenamefont {{Fragione}}\ and\ \citenamefont
  {{Kocsis}}(2020)}]{Fragione2020a}%
  \BibitemOpen
  \bibfield  {author} {\bibinfo {author} {\bibfnamefont {G.}~\bibnamefont
  {{Fragione}}}\ and\ \bibinfo {author} {\bibfnamefont {B.}~\bibnamefont
  {{Kocsis}}},\ }\bibfield  {title} {\bibinfo {title} {{Effective spin
  distribution of black hole mergers in triples}},\ }\href
  {https://doi.org/10.1093/mnras/staa443} {\bibfield  {journal} {\bibinfo
  {journal} {\mnras}\ }\textbf {\bibinfo {volume} {493}},\ \bibinfo {pages}
  {3920} (\bibinfo {year} {2020})},\ \Eprint {https://arxiv.org/abs/1910.00407}
  {arXiv:1910.00407 [astro-ph.GA]} \BibitemShut {NoStop}%
\bibitem [{\citenamefont {{Antonini}}\ \emph {et~al.}(2016)\citenamefont
  {{Antonini}}, \citenamefont {{Chatterjee}}, \citenamefont {{Rodriguez}},
  \citenamefont {{Morscher}}, \citenamefont {{Pattabiraman}}, \citenamefont
  {{Kalogera}},\ and\ \citenamefont {{Rasio}}}]{Antonini2016}%
  \BibitemOpen
  \bibfield  {author} {\bibinfo {author} {\bibfnamefont {F.}~\bibnamefont
  {{Antonini}}}, \bibinfo {author} {\bibfnamefont {S.}~\bibnamefont
  {{Chatterjee}}}, \bibinfo {author} {\bibfnamefont {C.~L.}\ \bibnamefont
  {{Rodriguez}}}, \bibinfo {author} {\bibfnamefont {M.}~\bibnamefont
  {{Morscher}}}, \bibinfo {author} {\bibfnamefont {B.}~\bibnamefont
  {{Pattabiraman}}}, \bibinfo {author} {\bibfnamefont {V.}~\bibnamefont
  {{Kalogera}}},\ and\ \bibinfo {author} {\bibfnamefont {F.~A.}\ \bibnamefont
  {{Rasio}}},\ }\bibfield  {title} {\bibinfo {title} {{Black Hole Mergers and
  Blue Stragglers from Hierarchical Triples Formed in Globular Clusters}},\
  }\href {https://doi.org/10.3847/0004-637X/816/2/65} {\bibfield  {journal}
  {\bibinfo  {journal} {\apj}\ }\textbf {\bibinfo {volume} {816}},\ \bibinfo
  {eid} {65} (\bibinfo {year} {2016})},\ \Eprint
  {https://arxiv.org/abs/1509.05080} {arXiv:1509.05080 [astro-ph.GA]}
  \BibitemShut {NoStop}%
\bibitem [{\citenamefont {{Kimpson}}\ \emph {et~al.}(2016)\citenamefont
  {{Kimpson}}, \citenamefont {{Spera}}, \citenamefont {{Mapelli}},\ and\
  \citenamefont {{Ziosi}}}]{Kimpson2016}%
  \BibitemOpen
  \bibfield  {author} {\bibinfo {author} {\bibfnamefont {T.~O.}\ \bibnamefont
  {{Kimpson}}}, \bibinfo {author} {\bibfnamefont {M.}~\bibnamefont {{Spera}}},
  \bibinfo {author} {\bibfnamefont {M.}~\bibnamefont {{Mapelli}}},\ and\
  \bibinfo {author} {\bibfnamefont {B.~M.}\ \bibnamefont {{Ziosi}}},\
  }\bibfield  {title} {\bibinfo {title} {{Hierarchical black hole triples in
  young star clusters: impact of Kozai-Lidov resonance on mergers}},\ }\href
  {https://doi.org/10.1093/mnras/stw2085} {\bibfield  {journal} {\bibinfo
  {journal} {\mnras}\ }\textbf {\bibinfo {volume} {463}},\ \bibinfo {pages}
  {2443} (\bibinfo {year} {2016})},\ \Eprint {https://arxiv.org/abs/1608.05422}
  {arXiv:1608.05422 [astro-ph.GA]} \BibitemShut {NoStop}%
\bibitem [{\citenamefont {{Banerjee}}(2018{\natexlab{a}})}]{Banerjee2018}%
  \BibitemOpen
  \bibfield  {author} {\bibinfo {author} {\bibfnamefont {S.}~\bibnamefont
  {{Banerjee}}},\ }\bibfield  {title} {\bibinfo {title} {{Stellar-mass black
  holes in young massive and open stellar clusters and their role in
  gravitational-wave generation - II}},\ }\href
  {https://doi.org/10.1093/mnras/stx2347} {\bibfield  {journal} {\bibinfo
  {journal} {\mnras}\ }\textbf {\bibinfo {volume} {473}},\ \bibinfo {pages}
  {909} (\bibinfo {year} {2018}{\natexlab{a}})},\ \Eprint
  {https://arxiv.org/abs/1707.00922} {arXiv:1707.00922 [astro-ph.HE]}
  \BibitemShut {NoStop}%
\bibitem [{\citenamefont {{Martinez}}\ \emph
  {et~al.}(2020{\natexlab{b}})\citenamefont {{Martinez}}, \citenamefont
  {{Fragione}}, \citenamefont {{Kremer}}, \citenamefont {{Chatterjee}},
  \citenamefont {{Rodriguez}}, \citenamefont {{Samsing}}, \citenamefont {{Ye}},
  \citenamefont {{Weatherford}}, \citenamefont {{Zevin}}, \citenamefont
  {{Naoz}},\ and\ \citenamefont {{Rasio}}}]{Martinez2020}%
  \BibitemOpen
  \bibfield  {author} {\bibinfo {author} {\bibfnamefont {M.~A.~S.}\
  \bibnamefont {{Martinez}}}, \bibinfo {author} {\bibfnamefont
  {G.}~\bibnamefont {{Fragione}}}, \bibinfo {author} {\bibfnamefont
  {K.}~\bibnamefont {{Kremer}}}, \bibinfo {author} {\bibfnamefont
  {S.}~\bibnamefont {{Chatterjee}}}, \bibinfo {author} {\bibfnamefont {C.~L.}\
  \bibnamefont {{Rodriguez}}}, \bibinfo {author} {\bibfnamefont
  {J.}~\bibnamefont {{Samsing}}}, \bibinfo {author} {\bibfnamefont {C.~S.}\
  \bibnamefont {{Ye}}}, \bibinfo {author} {\bibfnamefont {N.~C.}\ \bibnamefont
  {{Weatherford}}}, \bibinfo {author} {\bibfnamefont {M.}~\bibnamefont
  {{Zevin}}}, \bibinfo {author} {\bibfnamefont {S.}~\bibnamefont {{Naoz}}},\
  and\ \bibinfo {author} {\bibfnamefont {F.~A.}\ \bibnamefont {{Rasio}}},\
  }\bibfield  {title} {\bibinfo {title} {{Black Hole Mergers from Hierarchical
  Triples in Dense Star Clusters}},\ }\href
  {https://doi.org/10.3847/1538-4357/abba25} {\bibfield  {journal} {\bibinfo
  {journal} {\apj}\ }\textbf {\bibinfo {volume} {903}},\ \bibinfo {eid} {67}
  (\bibinfo {year} {2020}{\natexlab{b}})},\ \Eprint
  {https://arxiv.org/abs/2009.08468} {arXiv:2009.08468 [astro-ph.GA]}
  \BibitemShut {NoStop}%
\bibitem [{\citenamefont {{Britt}}\ \emph {et~al.}(2021)\citenamefont
  {{Britt}}, \citenamefont {{Johanson}}, \citenamefont {{Wood}}, \citenamefont
  {{Miller}},\ and\ \citenamefont {{Michaely}}}]{Britt2021}%
  \BibitemOpen
  \bibfield  {author} {\bibinfo {author} {\bibfnamefont {D.}~\bibnamefont
  {{Britt}}}, \bibinfo {author} {\bibfnamefont {B.}~\bibnamefont {{Johanson}}},
  \bibinfo {author} {\bibfnamefont {L.}~\bibnamefont {{Wood}}}, \bibinfo
  {author} {\bibfnamefont {M.~C.}\ \bibnamefont {{Miller}}},\ and\ \bibinfo
  {author} {\bibfnamefont {E.}~\bibnamefont {{Michaely}}},\ }\bibfield  {title}
  {\bibinfo {title} {{Binary black hole mergers from hierarchical triples in
  open clusters}},\ }\href {https://doi.org/10.1093/mnras/stab1570} {\bibfield
  {journal} {\bibinfo  {journal} {\mnras}\ }\textbf {\bibinfo {volume} {505}},\
  \bibinfo {pages} {3844} (\bibinfo {year} {2021})},\ \Eprint
  {https://arxiv.org/abs/2103.14706} {arXiv:2103.14706 [astro-ph.HE]}
  \BibitemShut {NoStop}%
\bibitem [{\citenamefont {{Arca Sedda}}\ \emph {et~al.}(2021)\citenamefont
  {{Arca Sedda}}, \citenamefont {{Li}},\ and\ \citenamefont
  {{Kocsis}}}]{ArcaSedda2021}%
  \BibitemOpen
  \bibfield  {author} {\bibinfo {author} {\bibfnamefont {M.}~\bibnamefont
  {{Arca Sedda}}}, \bibinfo {author} {\bibfnamefont {G.}~\bibnamefont {{Li}}},\
  and\ \bibinfo {author} {\bibfnamefont {B.}~\bibnamefont {{Kocsis}}},\
  }\bibfield  {title} {\bibinfo {title} {{Order in the chaos. Eccentric black
  hole binary mergers in triples formed via strong binary-binary
  scatterings}},\ }\href {https://doi.org/10.1051/0004-6361/202038795}
  {\bibfield  {journal} {\bibinfo  {journal} {\aap}\ }\textbf {\bibinfo
  {volume} {650}},\ \bibinfo {eid} {A189} (\bibinfo {year} {2021})},\ \Eprint
  {https://arxiv.org/abs/1805.06458} {arXiv:1805.06458 [astro-ph.HE]}
  \BibitemShut {NoStop}%
\bibitem [{\citenamefont {{Trani}}\ \emph {et~al.}(2022)\citenamefont
  {{Trani}}, \citenamefont {{Rastello}}, \citenamefont {{Di Carlo}},
  \citenamefont {{Santoliquido}}, \citenamefont {{Tanikawa}},\ and\
  \citenamefont {{Mapelli}}}]{Trani2022}%
  \BibitemOpen
  \bibfield  {author} {\bibinfo {author} {\bibfnamefont {A.~A.}\ \bibnamefont
  {{Trani}}}, \bibinfo {author} {\bibfnamefont {S.}~\bibnamefont {{Rastello}}},
  \bibinfo {author} {\bibfnamefont {U.~N.}\ \bibnamefont {{Di Carlo}}},
  \bibinfo {author} {\bibfnamefont {F.}~\bibnamefont {{Santoliquido}}},
  \bibinfo {author} {\bibfnamefont {A.}~\bibnamefont {{Tanikawa}}},\ and\
  \bibinfo {author} {\bibfnamefont {M.}~\bibnamefont {{Mapelli}}},\ }\bibfield
  {title} {\bibinfo {title} {{Compact object mergers in hierarchical triples
  from low-mass young star clusters}},\ }\href
  {https://doi.org/10.1093/mnras/stac122} {\bibfield  {journal} {\bibinfo
  {journal} {\mnras}\ }\textbf {\bibinfo {volume} {511}},\ \bibinfo {pages}
  {1362} (\bibinfo {year} {2022})},\ \Eprint {https://arxiv.org/abs/2111.06388}
  {arXiv:2111.06388 [astro-ph.HE]} \BibitemShut {NoStop}%
\bibitem [{\citenamefont {Antonini}\ and\ \citenamefont
  {Perets}(2012)}]{Antonini2012}%
  \BibitemOpen
  \bibfield  {author} {\bibinfo {author} {\bibfnamefont {F.}~\bibnamefont
  {Antonini}}\ and\ \bibinfo {author} {\bibfnamefont {H.~B.}\ \bibnamefont
  {Perets}},\ }\bibfield  {title} {\bibinfo {title} {{{SECULAR EVOLUTION OF
  COMPACT BINARIES NEAR MASSIVE BLACK HOLES}}: {{GRAVITATIONAL WAVE SOURCES AND
  OTHER EXOTICA}}},\ }\href {https://doi.org/10.1088/0004-637X/757/1/27}
  {\bibfield  {journal} {\bibinfo  {journal} {The Astrophysical Journal}\
  }\textbf {\bibinfo {volume} {757}},\ \bibinfo {pages} {27} (\bibinfo {year}
  {2012})}\BibitemShut {NoStop}%
\bibitem [{\citenamefont {VanLandingham}\ \emph {et~al.}(2016)\citenamefont
  {VanLandingham}, \citenamefont {Miller}, \citenamefont {Hamilton},\ and\
  \citenamefont {Richardson}}]{VanLandingham2016}%
  \BibitemOpen
  \bibfield  {author} {\bibinfo {author} {\bibfnamefont {J.~H.}\ \bibnamefont
  {VanLandingham}}, \bibinfo {author} {\bibfnamefont {M.~C.}\ \bibnamefont
  {Miller}}, \bibinfo {author} {\bibfnamefont {D.~P.}\ \bibnamefont
  {Hamilton}},\ and\ \bibinfo {author} {\bibfnamefont {D.~C.}\ \bibnamefont
  {Richardson}},\ }\bibfield  {title} {\bibinfo {title} {{{THE ROLE OF THE
  KOZAI}}\textendash{{LIDOV MECHANISM IN BLACK HOLE BINARY MERGERS IN GALACTIC
  CENTERS}}},\ }\href {https://doi.org/10.3847/0004-637X/828/2/77} {\bibfield
  {journal} {\bibinfo  {journal} {The Astrophysical Journal}\ }\textbf
  {\bibinfo {volume} {828}},\ \bibinfo {pages} {77} (\bibinfo {year} {2016})},\
  \Eprint {https://arxiv.org/abs/1604.04948} {arxiv:1604.04948} \BibitemShut
  {NoStop}%
\bibitem [{\citenamefont {Stephan}\ \emph {et~al.}(2016)\citenamefont
  {Stephan}, \citenamefont {Naoz}, \citenamefont {Ghez}, \citenamefont
  {Witzel}, \citenamefont {Sitarski}, \citenamefont {Do},\ and\ \citenamefont
  {Kocsis}}]{Stephan2016}%
  \BibitemOpen
  \bibfield  {author} {\bibinfo {author} {\bibfnamefont {A.~P.}\ \bibnamefont
  {Stephan}}, \bibinfo {author} {\bibfnamefont {S.}~\bibnamefont {Naoz}},
  \bibinfo {author} {\bibfnamefont {A.~M.}\ \bibnamefont {Ghez}}, \bibinfo
  {author} {\bibfnamefont {G.}~\bibnamefont {Witzel}}, \bibinfo {author}
  {\bibfnamefont {B.~N.}\ \bibnamefont {Sitarski}}, \bibinfo {author}
  {\bibfnamefont {T.}~\bibnamefont {Do}},\ and\ \bibinfo {author}
  {\bibfnamefont {B.}~\bibnamefont {Kocsis}},\ }\bibfield  {title} {\bibinfo
  {title} {Merging binaries in the {{Galactic Center}}: The eccentric
  {{Kozai}}\textendash{{Lidov}} mechanism with stellar evolution},\ }\href
  {https://doi.org/10.1093/mnras/stw1220} {\bibfield  {journal} {\bibinfo
  {journal} {Monthly Notices of the Royal Astronomical Society}\ }\textbf
  {\bibinfo {volume} {460}},\ \bibinfo {pages} {3494} (\bibinfo {year}
  {2016})},\ \Eprint {https://arxiv.org/abs/1603.02709} {arxiv:1603.02709}
  \BibitemShut {NoStop}%
\bibitem [{\citenamefont {Hoang}\ \emph {et~al.}(2018)\citenamefont {Hoang},
  \citenamefont {Naoz}, \citenamefont {Kocsis}, \citenamefont {Rasio},\ and\
  \citenamefont {Dosopoulou}}]{Hoang2018}%
  \BibitemOpen
  \bibfield  {author} {\bibinfo {author} {\bibfnamefont {B.-M.}\ \bibnamefont
  {Hoang}}, \bibinfo {author} {\bibfnamefont {S.}~\bibnamefont {Naoz}},
  \bibinfo {author} {\bibfnamefont {B.}~\bibnamefont {Kocsis}}, \bibinfo
  {author} {\bibfnamefont {F.~A.}\ \bibnamefont {Rasio}},\ and\ \bibinfo
  {author} {\bibfnamefont {F.}~\bibnamefont {Dosopoulou}},\ }\bibfield  {title}
  {\bibinfo {title} {Black {{Hole Mergers}} in {{Galactic Nuclei Induced}} by
  the {{Eccentric Kozai}}\textendash{{Lidov Effect}}},\ }\href
  {https://doi.org/10.3847/1538-4357/aaafce} {\bibfield  {journal} {\bibinfo
  {journal} {The Astrophysical Journal}\ }\textbf {\bibinfo {volume} {856}},\
  \bibinfo {pages} {140} (\bibinfo {year} {2018})},\ \Eprint
  {https://arxiv.org/abs/1706.09896} {arxiv:1706.09896} \BibitemShut {NoStop}%
\bibitem [{\citenamefont {{Hamers}}\ \emph {et~al.}(2018)\citenamefont
  {{Hamers}}, \citenamefont {{Bar-Or}}, \citenamefont {{Petrovich}},\ and\
  \citenamefont {{Antonini}}}]{Hamers2018}%
  \BibitemOpen
  \bibfield  {author} {\bibinfo {author} {\bibfnamefont {A.~S.}\ \bibnamefont
  {{Hamers}}}, \bibinfo {author} {\bibfnamefont {B.}~\bibnamefont {{Bar-Or}}},
  \bibinfo {author} {\bibfnamefont {C.}~\bibnamefont {{Petrovich}}},\ and\
  \bibinfo {author} {\bibfnamefont {F.}~\bibnamefont {{Antonini}}},\ }\bibfield
   {title} {\bibinfo {title} {{The Impact of Vector Resonant Relaxation on the
  Evolution of Binaries near a Massive Black Hole: Implications for
  Gravitational-wave Sources}},\ }\href
  {https://doi.org/10.3847/1538-4357/aadae2} {\bibfield  {journal} {\bibinfo
  {journal} {\apj}\ }\textbf {\bibinfo {volume} {865}},\ \bibinfo {eid} {2}
  (\bibinfo {year} {2018})},\ \Eprint {https://arxiv.org/abs/1805.10313}
  {arXiv:1805.10313 [astro-ph.HE]} \BibitemShut {NoStop}%
\bibitem [{\citenamefont {{Arca Sedda}}(2020)}]{ArcaSedda2020}%
  \BibitemOpen
  \bibfield  {author} {\bibinfo {author} {\bibfnamefont {M.}~\bibnamefont
  {{Arca Sedda}}},\ }\bibfield  {title} {\bibinfo {title} {{Birth, Life, and
  Death of Black Hole Binaries around Supermassive Black Holes: Dynamical
  Evolution of Gravitational Wave Sources}},\ }\href
  {https://doi.org/10.3847/1538-4357/ab723b} {\bibfield  {journal} {\bibinfo
  {journal} {\apj}\ }\textbf {\bibinfo {volume} {891}},\ \bibinfo {eid} {47}
  (\bibinfo {year} {2020})},\ \Eprint {https://arxiv.org/abs/2002.04037}
  {arXiv:2002.04037 [astro-ph.GA]} \BibitemShut {NoStop}%
\bibitem [{\citenamefont {{Wang}}\ \emph {et~al.}(2021)\citenamefont {{Wang}},
  \citenamefont {{Stephan}}, \citenamefont {{Naoz}}, \citenamefont {{Hoang}},\
  and\ \citenamefont {{Breivik}}}]{Wang2021}%
  \BibitemOpen
  \bibfield  {author} {\bibinfo {author} {\bibfnamefont {H.}~\bibnamefont
  {{Wang}}}, \bibinfo {author} {\bibfnamefont {A.~P.}\ \bibnamefont
  {{Stephan}}}, \bibinfo {author} {\bibfnamefont {S.}~\bibnamefont {{Naoz}}},
  \bibinfo {author} {\bibfnamefont {B.-M.}\ \bibnamefont {{Hoang}}},\ and\
  \bibinfo {author} {\bibfnamefont {K.}~\bibnamefont {{Breivik}}},\ }\bibfield
  {title} {\bibinfo {title} {{Gravitational-wave Signatures from Compact Object
  Binaries in the Galactic Center}},\ }\href
  {https://doi.org/10.3847/1538-4357/ac088d} {\bibfield  {journal} {\bibinfo
  {journal} {\apj}\ }\textbf {\bibinfo {volume} {917}},\ \bibinfo {eid} {76}
  (\bibinfo {year} {2021})},\ \Eprint {https://arxiv.org/abs/2010.15841}
  {arXiv:2010.15841 [astro-ph.HE]} \BibitemShut {NoStop}%
\bibitem [{\citenamefont {{Meiron}}\ \emph {et~al.}(2017)\citenamefont
  {{Meiron}}, \citenamefont {{Kocsis}},\ and\ \citenamefont
  {{Loeb}}}]{Meiron2017}%
  \BibitemOpen
  \bibfield  {author} {\bibinfo {author} {\bibfnamefont {Y.}~\bibnamefont
  {{Meiron}}}, \bibinfo {author} {\bibfnamefont {B.}~\bibnamefont {{Kocsis}}},\
  and\ \bibinfo {author} {\bibfnamefont {A.}~\bibnamefont {{Loeb}}},\
  }\bibfield  {title} {\bibinfo {title} {{Detecting Triple Systems with
  Gravitational Wave Observations}},\ }\href
  {https://doi.org/10.3847/1538-4357/834/2/200} {\bibfield  {journal} {\bibinfo
   {journal} {\apj}\ }\textbf {\bibinfo {volume} {834}},\ \bibinfo {eid} {200}
  (\bibinfo {year} {2017})},\ \Eprint {https://arxiv.org/abs/1604.02148}
  {arXiv:1604.02148 [astro-ph.HE]} \BibitemShut {NoStop}%
\bibitem [{\citenamefont {{Torres-Orjuela}}\ \emph {et~al.}(2019)\citenamefont
  {{Torres-Orjuela}}, \citenamefont {{Chen}}, \citenamefont {{Cao}},
  \citenamefont {{Amaro-Seoane}},\ and\ \citenamefont
  {{Peng}}}]{Torres-Orjuela2019}%
  \BibitemOpen
  \bibfield  {author} {\bibinfo {author} {\bibfnamefont {A.}~\bibnamefont
  {{Torres-Orjuela}}}, \bibinfo {author} {\bibfnamefont {X.}~\bibnamefont
  {{Chen}}}, \bibinfo {author} {\bibfnamefont {Z.}~\bibnamefont {{Cao}}},
  \bibinfo {author} {\bibfnamefont {P.}~\bibnamefont {{Amaro-Seoane}}},\ and\
  \bibinfo {author} {\bibfnamefont {P.}~\bibnamefont {{Peng}}},\ }\bibfield
  {title} {\bibinfo {title} {{Detecting the beaming effect of gravitational
  waves}},\ }\href {https://doi.org/10.1103/PhysRevD.100.063012} {\bibfield
  {journal} {\bibinfo  {journal} {\prd}\ }\textbf {\bibinfo {volume} {100}},\
  \bibinfo {eid} {063012} (\bibinfo {year} {2019})},\ \Eprint
  {https://arxiv.org/abs/1806.09857} {arXiv:1806.09857 [astro-ph.HE]}
  \BibitemShut {NoStop}%
\bibitem [{\citenamefont {{Wong}}\ \emph {et~al.}(2019)\citenamefont {{Wong}},
  \citenamefont {{Baibhav}},\ and\ \citenamefont {{Berti}}}]{Wong2019}%
  \BibitemOpen
  \bibfield  {author} {\bibinfo {author} {\bibfnamefont {K.~W.~K.}\
  \bibnamefont {{Wong}}}, \bibinfo {author} {\bibfnamefont {V.}~\bibnamefont
  {{Baibhav}}},\ and\ \bibinfo {author} {\bibfnamefont {E.}~\bibnamefont
  {{Berti}}},\ }\bibfield  {title} {\bibinfo {title} {{Binary radial velocity
  measurements with space-based gravitational-wave detectors}},\ }\href
  {https://doi.org/10.1093/mnras/stz2077} {\bibfield  {journal} {\bibinfo
  {journal} {\mnras}\ }\textbf {\bibinfo {volume} {488}},\ \bibinfo {pages}
  {5665} (\bibinfo {year} {2019})},\ \Eprint {https://arxiv.org/abs/1902.01402}
  {arXiv:1902.01402 [astro-ph.HE]} \BibitemShut {NoStop}%
\bibitem [{\citenamefont {Hut}\ and\ \citenamefont {Bahcall}(1983)}]{Hut1983b}%
  \BibitemOpen
  \bibfield  {author} {\bibinfo {author} {\bibfnamefont {P.}~\bibnamefont
  {Hut}}\ and\ \bibinfo {author} {\bibfnamefont {J.~N.}\ \bibnamefont
  {Bahcall}},\ }\bibfield  {title} {\bibinfo {title} {Binary-single star
  scattering. {{I}} - {{Numerical}} experiments for equal masses},\ }\href
  {https://doi.org/10.1086/160956} {\bibfield  {journal} {\bibinfo  {journal}
  {The Astrophysical Journal}\ }\textbf {\bibinfo {volume} {268}},\ \bibinfo
  {pages} {319} (\bibinfo {year} {1983})}\BibitemShut {NoStop}%
\bibitem [{\citenamefont {Hut}(1983{\natexlab{a}})}]{Hut1983}%
  \BibitemOpen
  \bibfield  {author} {\bibinfo {author} {\bibfnamefont {P.}~\bibnamefont
  {Hut}},\ }\bibfield  {title} {\bibinfo {title} {Binaries as a heat source in
  stellar dynamics - {{Release}} of binding energy},\ }\href
  {https://doi.org/10.1086/184111} {\bibfield  {journal} {\bibinfo  {journal}
  {The Astrophysical Journal Letters}\ }\textbf {\bibinfo {volume} {272}},\
  \bibinfo {pages} {L29} (\bibinfo {year} {1983}{\natexlab{a}})}\BibitemShut
  {NoStop}%
\bibitem [{\citenamefont {Hut}(1983{\natexlab{b}})}]{Hut1983a}%
  \BibitemOpen
  \bibfield  {author} {\bibinfo {author} {\bibfnamefont {P.}~\bibnamefont
  {Hut}},\ }\bibfield  {title} {\bibinfo {title} {Binary-single star
  scattering. {{II}} - {{Analytic}} approximations for high velocity},\ }\href
  {https://doi.org/10.1086/160957} {\bibfield  {journal} {\bibinfo  {journal}
  {Astrophysical Journal}\ }\textbf {\bibinfo {volume} {268}},\ \bibinfo
  {pages} {342} (\bibinfo {year} {1983}{\natexlab{b}})}\BibitemShut {NoStop}%
\bibitem [{\citenamefont {Fregeau}\ \emph {et~al.}(2004)\citenamefont
  {Fregeau}, \citenamefont {Cheung}, \citenamefont {Portegies~Zwart},\ and\
  \citenamefont {Rasio}}]{Fregeau2004}%
  \BibitemOpen
  \bibfield  {author} {\bibinfo {author} {\bibfnamefont {J.~M.}\ \bibnamefont
  {Fregeau}}, \bibinfo {author} {\bibfnamefont {P.}~\bibnamefont {Cheung}},
  \bibinfo {author} {\bibfnamefont {S.~F.}\ \bibnamefont {Portegies~Zwart}},\
  and\ \bibinfo {author} {\bibfnamefont {F.~A.}\ \bibnamefont {Rasio}},\
  }\bibfield  {title} {\bibinfo {title} {Stellar collisions during
  binary-binary and binary-single star interactions},\ }\href
  {https://doi.org/10.1111/j.1365-2966.2004.07914.x} {\bibfield  {journal}
  {\bibinfo  {journal} {Monthly Notices of the Royal Astronomical Society}\
  }\textbf {\bibinfo {volume} {352}},\ \bibinfo {pages} {1} (\bibinfo {year}
  {2004})}\BibitemShut {NoStop}%
\bibitem [{\citenamefont {{Michaely}}\ and\ \citenamefont
  {{Perets}}(2019)}]{Michaely2019}%
  \BibitemOpen
  \bibfield  {author} {\bibinfo {author} {\bibfnamefont {E.}~\bibnamefont
  {{Michaely}}}\ and\ \bibinfo {author} {\bibfnamefont {H.~B.}\ \bibnamefont
  {{Perets}}},\ }\bibfield  {title} {\bibinfo {title} {{Gravitational-wave
  Sources from Mergers of Binary Black Holes Catalyzed by Flyby Interactions in
  the Field}},\ }\href {https://doi.org/10.3847/2041-8213/ab5b9b} {\bibfield
  {journal} {\bibinfo  {journal} {\apjl}\ }\textbf {\bibinfo {volume} {887}},\
  \bibinfo {eid} {L36} (\bibinfo {year} {2019})},\ \Eprint
  {https://arxiv.org/abs/1902.01864} {arXiv:1902.01864 [astro-ph.SR]}
  \BibitemShut {NoStop}%
\bibitem [{\citenamefont {{Michaely}}\ and\ \citenamefont
  {{Perets}}(2020)}]{Michaely2020}%
  \BibitemOpen
  \bibfield  {author} {\bibinfo {author} {\bibfnamefont {E.}~\bibnamefont
  {{Michaely}}}\ and\ \bibinfo {author} {\bibfnamefont {H.~B.}\ \bibnamefont
  {{Perets}}},\ }\bibfield  {title} {\bibinfo {title} {{High rate of
  gravitational waves mergers from flyby perturbations of wide black hole
  triples in the field}},\ }\href {https://doi.org/10.1093/mnras/staa2720}
  {\bibfield  {journal} {\bibinfo  {journal} {\mnras}\ }\textbf {\bibinfo
  {volume} {498}},\ \bibinfo {pages} {4924} (\bibinfo {year} {2020})},\ \Eprint
  {https://arxiv.org/abs/2008.01094} {arXiv:2008.01094 [astro-ph.HE]}
  \BibitemShut {NoStop}%
\bibitem [{\citenamefont {Lada}\ and\ \citenamefont {Lada}(2003)}]{Lada2003}%
  \BibitemOpen
  \bibfield  {author} {\bibinfo {author} {\bibfnamefont {C.~J.}\ \bibnamefont
  {Lada}}\ and\ \bibinfo {author} {\bibfnamefont {E.~A.}\ \bibnamefont
  {Lada}},\ }\bibfield  {title} {\bibinfo {title} {Embedded {{Clusters}} in
  {{Molecular Clouds}}},\ }\href
  {https://doi.org/10.1146/annurev.astro.41.011802.094844} {\bibfield
  {journal} {\bibinfo  {journal} {Annual Review of Astronomy and Astrophysics}\
  }\textbf {\bibinfo {volume} {41}},\ \bibinfo {pages} {57} (\bibinfo {year}
  {2003})}\BibitemShut {NoStop}%
\bibitem [{\citenamefont {Portegies~Zwart}\ and\ \citenamefont
  {Mcmillan}(2000)}]{PortegiesZwart2000}%
  \BibitemOpen
  \bibfield  {author} {\bibinfo {author} {\bibfnamefont {S.~F.}\ \bibnamefont
  {Portegies~Zwart}}\ and\ \bibinfo {author} {\bibfnamefont {S.~L.~W.}\
  \bibnamefont {Mcmillan}},\ }\bibfield  {title} {\bibinfo {title} {{{BLACK
  HOLE MERGERS IN THE UNIVERSE}}},\ }\href@noop {} {\bibfield  {journal}
  {\bibinfo  {journal} {The Astrophysical Journal}\ }\textbf {\bibinfo {volume}
  {528}},\ \bibinfo {pages} {17} (\bibinfo {year} {2000})}\BibitemShut
  {NoStop}%
\bibitem [{\citenamefont {{Banerjee}}\ \emph {et~al.}(2010)\citenamefont
  {{Banerjee}}, \citenamefont {{Baumgardt}},\ and\ \citenamefont
  {{Kroupa}}}]{Banerjee2010}%
  \BibitemOpen
  \bibfield  {author} {\bibinfo {author} {\bibfnamefont {S.}~\bibnamefont
  {{Banerjee}}}, \bibinfo {author} {\bibfnamefont {H.}~\bibnamefont
  {{Baumgardt}}},\ and\ \bibinfo {author} {\bibfnamefont {P.}~\bibnamefont
  {{Kroupa}}},\ }\bibfield  {title} {\bibinfo {title} {{Stellar-mass black
  holes in star clusters: implications for gravitational wave radiation}},\
  }\href {https://doi.org/10.1111/j.1365-2966.2009.15880.x} {\bibfield
  {journal} {\bibinfo  {journal} {\mnras}\ }\textbf {\bibinfo {volume} {402}},\
  \bibinfo {pages} {371} (\bibinfo {year} {2010})},\ \Eprint
  {https://arxiv.org/abs/0910.3954} {arXiv:0910.3954 [astro-ph.SR]}
  \BibitemShut {NoStop}%
\bibitem [{\citenamefont {Ziosi}\ \emph {et~al.}(2014)\citenamefont {Ziosi},
  \citenamefont {Mapelli}, \citenamefont {Branchesi},\ and\ \citenamefont
  {Tormen}}]{Ziosi2014}%
  \BibitemOpen
  \bibfield  {author} {\bibinfo {author} {\bibfnamefont {B.~M.}\ \bibnamefont
  {Ziosi}}, \bibinfo {author} {\bibfnamefont {M.}~\bibnamefont {Mapelli}},
  \bibinfo {author} {\bibfnamefont {M.}~\bibnamefont {Branchesi}},\ and\
  \bibinfo {author} {\bibfnamefont {G.}~\bibnamefont {Tormen}},\ }\bibfield
  {title} {\bibinfo {title} {Dynamics of stellar black holes in young star
  clusters with different metallicities - {{II}}. {{Black}} hole-black hole
  binaries},\ }\href {https://doi.org/10.1093/mnras/stu824} {\bibfield
  {journal} {\bibinfo  {journal} {Monthly Notices of the Royal Astronomical
  Society}\ }\textbf {\bibinfo {volume} {441}},\ \bibinfo {pages} {3703}
  (\bibinfo {year} {2014})}\BibitemShut {NoStop}%
\bibitem [{\citenamefont {{Arca-Sedda}}(2016)}]{Arca-Sedda2016}%
  \BibitemOpen
  \bibfield  {author} {\bibinfo {author} {\bibfnamefont {M.}~\bibnamefont
  {{Arca-Sedda}}},\ }\bibfield  {title} {\bibinfo {title} {{On the formation of
  compact, massive subsystems in stellar clusters and its relation with
  intermediate-mass black holes}},\ }\href
  {https://doi.org/10.1093/mnras/stv2265} {\bibfield  {journal} {\bibinfo
  {journal} {\mnras}\ }\textbf {\bibinfo {volume} {455}},\ \bibinfo {pages}
  {35} (\bibinfo {year} {2016})},\ \Eprint {https://arxiv.org/abs/1502.01242}
  {arXiv:1502.01242 [astro-ph.GA]} \BibitemShut {NoStop}%
\bibitem [{\citenamefont {{Banerjee}}(2017)}]{Banerjee2017}%
  \BibitemOpen
  \bibfield  {author} {\bibinfo {author} {\bibfnamefont {S.}~\bibnamefont
  {{Banerjee}}},\ }\bibfield  {title} {\bibinfo {title} {{Stellar-mass black
  holes in young massive and open stellar clusters and their role in
  gravitational-wave generation}},\ }\href
  {https://doi.org/10.1093/mnras/stw3392} {\bibfield  {journal} {\bibinfo
  {journal} {\mnras}\ }\textbf {\bibinfo {volume} {467}},\ \bibinfo {pages}
  {524} (\bibinfo {year} {2017})},\ \Eprint {https://arxiv.org/abs/1611.09357}
  {arXiv:1611.09357 [astro-ph.HE]} \BibitemShut {NoStop}%
\bibitem [{\citenamefont {{Banerjee}}(2018{\natexlab{b}})}]{Banerjee2018b}%
  \BibitemOpen
  \bibfield  {author} {\bibinfo {author} {\bibfnamefont {S.}~\bibnamefont
  {{Banerjee}}},\ }\bibfield  {title} {\bibinfo {title} {{Stellar-mass black
  holes in young massive and open stellar clusters and their role in
  gravitational-wave generation III: dissecting black hole dynamics}},\ }\href
  {https://doi.org/10.1093/mnras/sty2608} {\bibfield  {journal} {\bibinfo
  {journal} {\mnras}\ }\textbf {\bibinfo {volume} {481}},\ \bibinfo {pages}
  {5123} (\bibinfo {year} {2018}{\natexlab{b}})},\ \Eprint
  {https://arxiv.org/abs/1805.06466} {arXiv:1805.06466 [astro-ph.HE]}
  \BibitemShut {NoStop}%
\bibitem [{\citenamefont {{Rastello}}\ \emph {et~al.}(2019)\citenamefont
  {{Rastello}}, \citenamefont {{Amaro-Seoane}}, \citenamefont {{Arca-Sedda}},
  \citenamefont {{Capuzzo-Dolcetta}}, \citenamefont {{Fragione}},\ and\
  \citenamefont {{Tosta e Melo}}}]{Rastello2019}%
  \BibitemOpen
  \bibfield  {author} {\bibinfo {author} {\bibfnamefont {S.}~\bibnamefont
  {{Rastello}}}, \bibinfo {author} {\bibfnamefont {P.}~\bibnamefont
  {{Amaro-Seoane}}}, \bibinfo {author} {\bibfnamefont {M.}~\bibnamefont
  {{Arca-Sedda}}}, \bibinfo {author} {\bibfnamefont {R.}~\bibnamefont
  {{Capuzzo-Dolcetta}}}, \bibinfo {author} {\bibfnamefont {G.}~\bibnamefont
  {{Fragione}}},\ and\ \bibinfo {author} {\bibfnamefont {I.}~\bibnamefont
  {{Tosta e Melo}}},\ }\bibfield  {title} {\bibinfo {title} {{Stellar black
  hole binary mergers in open clusters}},\ }\href
  {https://doi.org/10.1093/mnras/sty3193} {\bibfield  {journal} {\bibinfo
  {journal} {\mnras}\ }\textbf {\bibinfo {volume} {483}},\ \bibinfo {pages}
  {1233} (\bibinfo {year} {2019})},\ \Eprint {https://arxiv.org/abs/1811.10628}
  {arXiv:1811.10628 [astro-ph.GA]} \BibitemShut {NoStop}%
\bibitem [{\citenamefont {{Di Carlo}}\ \emph {et~al.}(2019)\citenamefont {{Di
  Carlo}}, \citenamefont {{Giacobbo}}, \citenamefont {{Mapelli}}, \citenamefont
  {{Pasquato}}, \citenamefont {{Spera}}, \citenamefont {{Wang}},\ and\
  \citenamefont {{Haardt}}}]{DiCarlo2019}%
  \BibitemOpen
  \bibfield  {author} {\bibinfo {author} {\bibfnamefont {U.~N.}\ \bibnamefont
  {{Di Carlo}}}, \bibinfo {author} {\bibfnamefont {N.}~\bibnamefont
  {{Giacobbo}}}, \bibinfo {author} {\bibfnamefont {M.}~\bibnamefont
  {{Mapelli}}}, \bibinfo {author} {\bibfnamefont {M.}~\bibnamefont
  {{Pasquato}}}, \bibinfo {author} {\bibfnamefont {M.}~\bibnamefont {{Spera}}},
  \bibinfo {author} {\bibfnamefont {L.}~\bibnamefont {{Wang}}},\ and\ \bibinfo
  {author} {\bibfnamefont {F.}~\bibnamefont {{Haardt}}},\ }\bibfield  {title}
  {\bibinfo {title} {{Merging black holes in young star clusters}},\ }\href
  {https://doi.org/10.1093/mnras/stz1453} {\bibfield  {journal} {\bibinfo
  {journal} {\mnras}\ }\textbf {\bibinfo {volume} {487}},\ \bibinfo {pages}
  {2947} (\bibinfo {year} {2019})},\ \Eprint {https://arxiv.org/abs/1901.00863}
  {arXiv:1901.00863 [astro-ph.HE]} \BibitemShut {NoStop}%
\bibitem [{\citenamefont {{Di Carlo}}\ \emph
  {et~al.}(2020{\natexlab{b}})\citenamefont {{Di Carlo}}, \citenamefont
  {{Mapelli}}, \citenamefont {{Giacobbo}}, \citenamefont {{Spera}},
  \citenamefont {{Bouffanais}}, \citenamefont {{Rastello}}, \citenamefont
  {{Santoliquido}}, \citenamefont {{Pasquato}}, \citenamefont {{Ballone}},
  \citenamefont {{Trani}}, \citenamefont {{Torniamenti}},\ and\ \citenamefont
  {{Haardt}}}]{DiCarlo2020}%
  \BibitemOpen
  \bibfield  {author} {\bibinfo {author} {\bibfnamefont {U.~N.}\ \bibnamefont
  {{Di Carlo}}}, \bibinfo {author} {\bibfnamefont {M.}~\bibnamefont
  {{Mapelli}}}, \bibinfo {author} {\bibfnamefont {N.}~\bibnamefont
  {{Giacobbo}}}, \bibinfo {author} {\bibfnamefont {M.}~\bibnamefont {{Spera}}},
  \bibinfo {author} {\bibfnamefont {Y.}~\bibnamefont {{Bouffanais}}}, \bibinfo
  {author} {\bibfnamefont {S.}~\bibnamefont {{Rastello}}}, \bibinfo {author}
  {\bibfnamefont {F.}~\bibnamefont {{Santoliquido}}}, \bibinfo {author}
  {\bibfnamefont {M.}~\bibnamefont {{Pasquato}}}, \bibinfo {author}
  {\bibfnamefont {A.}~\bibnamefont {{Ballone}}}, \bibinfo {author}
  {\bibfnamefont {A.~A.}\ \bibnamefont {{Trani}}}, \bibinfo {author}
  {\bibfnamefont {S.}~\bibnamefont {{Torniamenti}}},\ and\ \bibinfo {author}
  {\bibfnamefont {F.}~\bibnamefont {{Haardt}}},\ }\bibfield  {title} {\bibinfo
  {title} {{Binary black holes in young star clusters: the impact of
  metallicity}},\ }\href {https://doi.org/10.1093/mnras/staa2286} {\bibfield
  {journal} {\bibinfo  {journal} {\mnras}\ }\textbf {\bibinfo {volume} {498}},\
  \bibinfo {pages} {495} (\bibinfo {year} {2020}{\natexlab{b}})},\ \Eprint
  {https://arxiv.org/abs/2004.09525} {arXiv:2004.09525 [astro-ph.HE]}
  \BibitemShut {NoStop}%
\bibitem [{\citenamefont {{Kumamoto}}\ \emph {et~al.}(2020)\citenamefont
  {{Kumamoto}}, \citenamefont {{Fujii}},\ and\ \citenamefont
  {{Tanikawa}}}]{Kumamoto2020}%
  \BibitemOpen
  \bibfield  {author} {\bibinfo {author} {\bibfnamefont {J.}~\bibnamefont
  {{Kumamoto}}}, \bibinfo {author} {\bibfnamefont {M.~S.}\ \bibnamefont
  {{Fujii}}},\ and\ \bibinfo {author} {\bibfnamefont {A.}~\bibnamefont
  {{Tanikawa}}},\ }\bibfield  {title} {\bibinfo {title} {{Merger rate density
  of binary black holes formed in open clusters}},\ }\href
  {https://doi.org/10.1093/mnras/staa1440} {\bibfield  {journal} {\bibinfo
  {journal} {\mnras}\ }\textbf {\bibinfo {volume} {495}},\ \bibinfo {pages}
  {4268} (\bibinfo {year} {2020})},\ \Eprint {https://arxiv.org/abs/2001.10690}
  {arXiv:2001.10690 [astro-ph.HE]} \BibitemShut {NoStop}%
\bibitem [{\citenamefont {{Trani}}\ \emph
  {et~al.}(2021{\natexlab{b}})\citenamefont {{Trani}}, \citenamefont
  {{Tanikawa}}, \citenamefont {{Fujii}}, \citenamefont {{Leigh}},\ and\
  \citenamefont {{Kumamoto}}}]{Trani2021}%
  \BibitemOpen
  \bibfield  {author} {\bibinfo {author} {\bibfnamefont {A.~A.}\ \bibnamefont
  {{Trani}}}, \bibinfo {author} {\bibfnamefont {A.}~\bibnamefont {{Tanikawa}}},
  \bibinfo {author} {\bibfnamefont {M.~S.}\ \bibnamefont {{Fujii}}}, \bibinfo
  {author} {\bibfnamefont {N.~W.~C.}\ \bibnamefont {{Leigh}}},\ and\ \bibinfo
  {author} {\bibfnamefont {J.}~\bibnamefont {{Kumamoto}}},\ }\bibfield  {title}
  {\bibinfo {title} {{Spin misalignment of black hole binaries from young star
  clusters: implications for the origin of gravitational waves events}},\
  }\href {https://doi.org/10.1093/mnras/stab967} {\bibfield  {journal}
  {\bibinfo  {journal} {\mnras}\ }\textbf {\bibinfo {volume} {504}},\ \bibinfo
  {pages} {910} (\bibinfo {year} {2021}{\natexlab{b}})},\ \Eprint
  {https://arxiv.org/abs/2102.01689} {arXiv:2102.01689 [astro-ph.HE]}
  \BibitemShut {NoStop}%
\bibitem [{\citenamefont {{Banerjee}}(2021)}]{Banerjee2021}%
  \BibitemOpen
  \bibfield  {author} {\bibinfo {author} {\bibfnamefont {S.}~\bibnamefont
  {{Banerjee}}},\ }\bibfield  {title} {\bibinfo {title} {{Stellar-mass black
  holes in young massive and open stellar clusters - IV. Updated
  stellar-evolutionary and black hole spin models and comparisons with the
  LIGO-Virgo O1/O2 merger-event data}},\ }\href
  {https://doi.org/10.1093/mnras/staa2392} {\bibfield  {journal} {\bibinfo
  {journal} {\mnras}\ }\textbf {\bibinfo {volume} {500}},\ \bibinfo {pages}
  {3002} (\bibinfo {year} {2021})},\ \Eprint {https://arxiv.org/abs/2004.07382}
  {arXiv:2004.07382 [astro-ph.HE]} \BibitemShut {NoStop}%
\bibitem [{\citenamefont {{Woosley}}(2019{\natexlab{b}})}]{Woosley2019}%
  \BibitemOpen
  \bibfield  {author} {\bibinfo {author} {\bibfnamefont {S.~E.}\ \bibnamefont
  {{Woosley}}},\ }\bibfield  {title} {\bibinfo {title} {{The Evolution of
  Massive Helium Stars, Including Mass Loss}},\ }\href
  {https://doi.org/10.3847/1538-4357/ab1b41} {\bibfield  {journal} {\bibinfo
  {journal} {\apj}\ }\textbf {\bibinfo {volume} {878}},\ \bibinfo {eid} {49}
  (\bibinfo {year} {2019}{\natexlab{b}})},\ \Eprint
  {https://arxiv.org/abs/1901.00215} {arXiv:1901.00215 [astro-ph.SR]}
  \BibitemShut {NoStop}%
\bibitem [{\citenamefont {{Farmer}}\ \emph
  {et~al.}(2019{\natexlab{b}})\citenamefont {{Farmer}}, \citenamefont
  {{Renzo}}, \citenamefont {{de Mink}}, \citenamefont {{Marchant}},\ and\
  \citenamefont {{Justham}}}]{Farmer2019}%
  \BibitemOpen
  \bibfield  {author} {\bibinfo {author} {\bibfnamefont {R.}~\bibnamefont
  {{Farmer}}}, \bibinfo {author} {\bibfnamefont {M.}~\bibnamefont {{Renzo}}},
  \bibinfo {author} {\bibfnamefont {S.~E.}\ \bibnamefont {{de Mink}}}, \bibinfo
  {author} {\bibfnamefont {P.}~\bibnamefont {{Marchant}}},\ and\ \bibinfo
  {author} {\bibfnamefont {S.}~\bibnamefont {{Justham}}},\ }\bibfield  {title}
  {\bibinfo {title} {{Mind the Gap: The Location of the Lower Edge of the
  Pair-instability Supernova Black Hole Mass Gap}},\ }\href
  {https://doi.org/10.3847/1538-4357/ab518b} {\bibfield  {journal} {\bibinfo
  {journal} {\apj}\ }\textbf {\bibinfo {volume} {887}},\ \bibinfo {eid} {53}
  (\bibinfo {year} {2019}{\natexlab{b}})},\ \Eprint
  {https://arxiv.org/abs/1910.12874} {arXiv:1910.12874 [astro-ph.SR]}
  \BibitemShut {NoStop}%
\bibitem [{\citenamefont {{Di Carlo}}\ \emph
  {et~al.}(2020{\natexlab{c}})\citenamefont {{Di Carlo}}, \citenamefont
  {{Mapelli}}, \citenamefont {{Bouffanais}}, \citenamefont {{Giacobbo}},
  \citenamefont {{Santoliquido}}, \citenamefont {{Bressan}}, \citenamefont
  {{Spera}},\ and\ \citenamefont {{Haardt}}}]{DiCarlo2020a}%
  \BibitemOpen
  \bibfield  {author} {\bibinfo {author} {\bibfnamefont {U.~N.}\ \bibnamefont
  {{Di Carlo}}}, \bibinfo {author} {\bibfnamefont {M.}~\bibnamefont
  {{Mapelli}}}, \bibinfo {author} {\bibfnamefont {Y.}~\bibnamefont
  {{Bouffanais}}}, \bibinfo {author} {\bibfnamefont {N.}~\bibnamefont
  {{Giacobbo}}}, \bibinfo {author} {\bibfnamefont {F.}~\bibnamefont
  {{Santoliquido}}}, \bibinfo {author} {\bibfnamefont {A.}~\bibnamefont
  {{Bressan}}}, \bibinfo {author} {\bibfnamefont {M.}~\bibnamefont {{Spera}}},\
  and\ \bibinfo {author} {\bibfnamefont {F.}~\bibnamefont {{Haardt}}},\
  }\bibfield  {title} {\bibinfo {title} {{Binary black holes in the pair
  instability mass gap}},\ }\href {https://doi.org/10.1093/mnras/staa1997}
  {\bibfield  {journal} {\bibinfo  {journal} {\mnras}\ }\textbf {\bibinfo
  {volume} {497}},\ \bibinfo {pages} {1043} (\bibinfo {year}
  {2020}{\natexlab{c}})},\ \Eprint {https://arxiv.org/abs/1911.01434}
  {arXiv:1911.01434 [astro-ph.HE]} \BibitemShut {NoStop}%
\bibitem [{\citenamefont {{Portegies Zwart}}\ \emph {et~al.}(2004)\citenamefont
  {{Portegies Zwart}}, \citenamefont {{Baumgardt}}, \citenamefont {{Hut}},
  \citenamefont {{Makino}},\ and\ \citenamefont
  {{McMillan}}}]{PortegiesZwart2004}%
  \BibitemOpen
  \bibfield  {author} {\bibinfo {author} {\bibfnamefont {S.~F.}\ \bibnamefont
  {{Portegies Zwart}}}, \bibinfo {author} {\bibfnamefont {H.}~\bibnamefont
  {{Baumgardt}}}, \bibinfo {author} {\bibfnamefont {P.}~\bibnamefont {{Hut}}},
  \bibinfo {author} {\bibfnamefont {J.}~\bibnamefont {{Makino}}},\ and\
  \bibinfo {author} {\bibfnamefont {S.~L.~W.}\ \bibnamefont {{McMillan}}},\
  }\bibfield  {title} {\bibinfo {title} {{Formation of massive black holes
  through runaway collisions in dense young star clusters}},\ }\href
  {https://doi.org/10.1038/nature02448} {\bibfield  {journal} {\bibinfo
  {journal} {\nat}\ }\textbf {\bibinfo {volume} {428}},\ \bibinfo {pages} {724}
  (\bibinfo {year} {2004})},\ \Eprint {https://arxiv.org/abs/astro-ph/0402622}
  {arXiv:astro-ph/0402622 [astro-ph]} \BibitemShut {NoStop}%
\bibitem [{\citenamefont {{Freitag}}\ \emph
  {et~al.}(2006{\natexlab{a}})\citenamefont {{Freitag}}, \citenamefont
  {{Rasio}},\ and\ \citenamefont {{Baumgardt}}}]{Freitag2006a}%
  \BibitemOpen
  \bibfield  {author} {\bibinfo {author} {\bibfnamefont {M.}~\bibnamefont
  {{Freitag}}}, \bibinfo {author} {\bibfnamefont {F.~A.}\ \bibnamefont
  {{Rasio}}},\ and\ \bibinfo {author} {\bibfnamefont {H.}~\bibnamefont
  {{Baumgardt}}},\ }\bibfield  {title} {\bibinfo {title} {{Runaway collisions
  in young star clusters - I. Methods and tests}},\ }\href
  {https://doi.org/10.1111/j.1365-2966.2006.10095.x} {\bibfield  {journal}
  {\bibinfo  {journal} {\mnras}\ }\textbf {\bibinfo {volume} {368}},\ \bibinfo
  {pages} {121} (\bibinfo {year} {2006}{\natexlab{a}})},\ \Eprint
  {https://arxiv.org/abs/astro-ph/0503129} {arXiv:astro-ph/0503129 [astro-ph]}
  \BibitemShut {NoStop}%
\bibitem [{\citenamefont {{Freitag}}\ \emph
  {et~al.}(2006{\natexlab{b}})\citenamefont {{Freitag}}, \citenamefont
  {{G{\"u}rkan}},\ and\ \citenamefont {{Rasio}}}]{Freitag2006}%
  \BibitemOpen
  \bibfield  {author} {\bibinfo {author} {\bibfnamefont {M.}~\bibnamefont
  {{Freitag}}}, \bibinfo {author} {\bibfnamefont {M.~A.}\ \bibnamefont
  {{G{\"u}rkan}}},\ and\ \bibinfo {author} {\bibfnamefont {F.~A.}\ \bibnamefont
  {{Rasio}}},\ }\bibfield  {title} {\bibinfo {title} {{Runaway collisions in
  young star clusters - II. Numerical results}},\ }\href
  {https://doi.org/10.1111/j.1365-2966.2006.10096.x} {\bibfield  {journal}
  {\bibinfo  {journal} {\mnras}\ }\textbf {\bibinfo {volume} {368}},\ \bibinfo
  {pages} {141} (\bibinfo {year} {2006}{\natexlab{b}})},\ \Eprint
  {https://arxiv.org/abs/astro-ph/0503130} {arXiv:astro-ph/0503130 [astro-ph]}
  \BibitemShut {NoStop}%
\bibitem [{\citenamefont {{Gaburov}}\ \emph {et~al.}(2008)\citenamefont
  {{Gaburov}}, \citenamefont {{Gualandris}},\ and\ \citenamefont {{Portegies
  Zwart}}}]{Gaburov2008}%
  \BibitemOpen
  \bibfield  {author} {\bibinfo {author} {\bibfnamefont {E.}~\bibnamefont
  {{Gaburov}}}, \bibinfo {author} {\bibfnamefont {A.}~\bibnamefont
  {{Gualandris}}},\ and\ \bibinfo {author} {\bibfnamefont {S.}~\bibnamefont
  {{Portegies Zwart}}},\ }\bibfield  {title} {\bibinfo {title} {{On the onset
  of runaway stellar collisions in dense star clusters - I. Dynamics of the
  first collision}},\ }\href {https://doi.org/10.1111/j.1365-2966.2007.12731.x}
  {\bibfield  {journal} {\bibinfo  {journal} {\mnras}\ }\textbf {\bibinfo
  {volume} {384}},\ \bibinfo {pages} {376} (\bibinfo {year} {2008})},\ \Eprint
  {https://arxiv.org/abs/0707.0406} {arXiv:0707.0406 [astro-ph]} \BibitemShut
  {NoStop}%
\bibitem [{\citenamefont {{Mapelli}}(2016)}]{Mapelli2016}%
  \BibitemOpen
  \bibfield  {author} {\bibinfo {author} {\bibfnamefont {M.}~\bibnamefont
  {{Mapelli}}},\ }\bibfield  {title} {\bibinfo {title} {{Massive black hole
  binaries from runaway collisions: the impact of metallicity}},\ }\href
  {https://doi.org/10.1093/mnras/stw869} {\bibfield  {journal} {\bibinfo
  {journal} {\mnras}\ }\textbf {\bibinfo {volume} {459}},\ \bibinfo {pages}
  {3432} (\bibinfo {year} {2016})},\ \Eprint {https://arxiv.org/abs/1604.03559}
  {arXiv:1604.03559 [astro-ph.GA]} \BibitemShut {NoStop}%
\bibitem [{\citenamefont {{Di Carlo}}\ \emph {et~al.}(2021)\citenamefont {{Di
  Carlo}}, \citenamefont {{Mapelli}}, \citenamefont {{Pasquato}}, \citenamefont
  {{Rastello}}, \citenamefont {{Ballone}}, \citenamefont {{Dall'Amico}},
  \citenamefont {{Giacobbo}}, \citenamefont {{Iorio}}, \citenamefont {{Spera}},
  \citenamefont {{Torniamenti}},\ and\ \citenamefont {{Haardt}}}]{DiCarlo2021}%
  \BibitemOpen
  \bibfield  {author} {\bibinfo {author} {\bibfnamefont {U.~N.}\ \bibnamefont
  {{Di Carlo}}}, \bibinfo {author} {\bibfnamefont {M.}~\bibnamefont
  {{Mapelli}}}, \bibinfo {author} {\bibfnamefont {M.}~\bibnamefont
  {{Pasquato}}}, \bibinfo {author} {\bibfnamefont {S.}~\bibnamefont
  {{Rastello}}}, \bibinfo {author} {\bibfnamefont {A.}~\bibnamefont
  {{Ballone}}}, \bibinfo {author} {\bibfnamefont {M.}~\bibnamefont
  {{Dall'Amico}}}, \bibinfo {author} {\bibfnamefont {N.}~\bibnamefont
  {{Giacobbo}}}, \bibinfo {author} {\bibfnamefont {G.}~\bibnamefont {{Iorio}}},
  \bibinfo {author} {\bibfnamefont {M.}~\bibnamefont {{Spera}}}, \bibinfo
  {author} {\bibfnamefont {S.}~\bibnamefont {{Torniamenti}}},\ and\ \bibinfo
  {author} {\bibfnamefont {F.}~\bibnamefont {{Haardt}}},\ }\bibfield  {title}
  {\bibinfo {title} {{Intermediate-mass black holes from stellar mergers in
  young star clusters}},\ }\href {https://doi.org/10.1093/mnras/stab2390}
  {\bibfield  {journal} {\bibinfo  {journal} {\mnras}\ }\textbf {\bibinfo
  {volume} {507}},\ \bibinfo {pages} {5132} (\bibinfo {year} {2021})},\ \Eprint
  {https://arxiv.org/abs/2105.01085} {arXiv:2105.01085 [astro-ph.GA]}
  \BibitemShut {NoStop}%
\bibitem [{\citenamefont {Morscher}\ \emph {et~al.}(2015)\citenamefont
  {Morscher}, \citenamefont {Pattabiraman}, \citenamefont {Rodriguez},
  \citenamefont {Rasio},\ and\ \citenamefont {Umbreit}}]{Morscher2015}%
  \BibitemOpen
  \bibfield  {author} {\bibinfo {author} {\bibfnamefont {M.}~\bibnamefont
  {Morscher}}, \bibinfo {author} {\bibfnamefont {B.}~\bibnamefont
  {Pattabiraman}}, \bibinfo {author} {\bibfnamefont {C.}~\bibnamefont
  {Rodriguez}}, \bibinfo {author} {\bibfnamefont {F.~A.}\ \bibnamefont
  {Rasio}},\ and\ \bibinfo {author} {\bibfnamefont {S.}~\bibnamefont
  {Umbreit}},\ }\bibfield  {title} {\bibinfo {title} {The {{Dynamical
  Evolution}} of {{Stellar Black Holes}} in {{Globular Clusters}}},\ }\href
  {https://doi.org/10.1088/0004-637X/800/1/9} {\bibfield  {journal} {\bibinfo
  {journal} {The Astrophysical Journal}\ }\textbf {\bibinfo {volume} {800}},\
  \bibinfo {pages} {9} (\bibinfo {year} {2015})},\ \Eprint
  {https://arxiv.org/abs/1409.0866} {arXiv:1409.0866} \BibitemShut {NoStop}%
\bibitem [{\citenamefont {Belczynski}\ \emph {et~al.}(2006)\citenamefont
  {Belczynski}, \citenamefont {Sadowski}, \citenamefont {Rasio},\ and\
  \citenamefont {Bulik}}]{Belczynski2006}%
  \BibitemOpen
  \bibfield  {author} {\bibinfo {author} {\bibfnamefont {K.}~\bibnamefont
  {Belczynski}}, \bibinfo {author} {\bibfnamefont {A.}~\bibnamefont
  {Sadowski}}, \bibinfo {author} {\bibfnamefont {F.~A.}\ \bibnamefont
  {Rasio}},\ and\ \bibinfo {author} {\bibfnamefont {T.}~\bibnamefont {Bulik}},\
  }\bibfield  {title} {\bibinfo {title} {Initial {{Populations}} of {{Black
  Holes}} in {{Star Clusters}}},\ }\href@noop {} {\bibfield  {journal}
  {\bibinfo  {journal} {The Astrophysical Journal}\ }\textbf {\bibinfo {volume}
  {650}},\ \bibinfo {pages} {303} (\bibinfo {year} {2006})}\BibitemShut
  {NoStop}%
\bibitem [{\citenamefont {Fryer}\ \emph {et~al.}(2012)\citenamefont {Fryer},
  \citenamefont {Belczynski}, \citenamefont {Wiktorowicz}, \citenamefont
  {Dominik}, \citenamefont {Kalogera},\ and\ \citenamefont {Holz}}]{Fryer2012}%
  \BibitemOpen
  \bibfield  {author} {\bibinfo {author} {\bibfnamefont {C.~L.}\ \bibnamefont
  {Fryer}}, \bibinfo {author} {\bibfnamefont {K.}~\bibnamefont {Belczynski}},
  \bibinfo {author} {\bibfnamefont {G.}~\bibnamefont {Wiktorowicz}}, \bibinfo
  {author} {\bibfnamefont {M.}~\bibnamefont {Dominik}}, \bibinfo {author}
  {\bibfnamefont {V.}~\bibnamefont {Kalogera}},\ and\ \bibinfo {author}
  {\bibfnamefont {D.~E.}\ \bibnamefont {Holz}},\ }\bibfield  {title} {\bibinfo
  {title} {{{COMPACT REMNANT MASS FUNCTION}}: {{DEPENDENCE ON THE EXPLOSION
  MECHANISM AND METALLICITY}}},\ }\href
  {https://doi.org/10.1088/0004-637X/749/1/91} {\bibfield  {journal} {\bibinfo
  {journal} {The Astrophysical Journal}\ }\textbf {\bibinfo {volume} {749}},\
  \bibinfo {pages} {91} (\bibinfo {year} {2012})}\BibitemShut {NoStop}%
\bibitem [{\citenamefont {Fryer}\ and\ \citenamefont
  {Kalogera}(2001)}]{Fryer2001}%
  \BibitemOpen
  \bibfield  {author} {\bibinfo {author} {\bibfnamefont {C.~L.}\ \bibnamefont
  {Fryer}}\ and\ \bibinfo {author} {\bibfnamefont {V.}~\bibnamefont
  {Kalogera}},\ }\bibfield  {title} {\bibinfo {title} {Theoretical {{Black Hole
  Mass Distributions}}},\ }\href {https://doi.org/10.1086/321359} {\bibfield
  {journal} {\bibinfo  {journal} {The Astrophysical Journal}\ }\textbf
  {\bibinfo {volume} {554}},\ \bibinfo {pages} {548} (\bibinfo {year}
  {2001})}\BibitemShut {NoStop}%
\bibitem [{\citenamefont {Morscher}\ \emph {et~al.}(2013)\citenamefont
  {Morscher}, \citenamefont {Umbreit}, \citenamefont {Farr},\ and\
  \citenamefont {Rasio}}]{Morscher2013}%
  \BibitemOpen
  \bibfield  {author} {\bibinfo {author} {\bibfnamefont {M.}~\bibnamefont
  {Morscher}}, \bibinfo {author} {\bibfnamefont {S.}~\bibnamefont {Umbreit}},
  \bibinfo {author} {\bibfnamefont {W.~M.}\ \bibnamefont {Farr}},\ and\
  \bibinfo {author} {\bibfnamefont {F.~A.}\ \bibnamefont {Rasio}},\ }\bibfield
  {title} {\bibinfo {title} {{{RETENTION OF STELLAR-MASS BLACK HOLES IN
  GLOBULAR CLUSTERS}}},\ }\href {https://doi.org/10.1088/2041-8205/763/1/L15}
  {\bibfield  {journal} {\bibinfo  {journal} {The Astrophysical Journal}\
  }\textbf {\bibinfo {volume} {763}},\ \bibinfo {pages} {L15} (\bibinfo {year}
  {2013})},\ \Eprint {https://arxiv.org/abs/1211.3372} {arXiv:1211.3372}
  \BibitemShut {NoStop}%
\bibitem [{\citenamefont {{Binney}}\ and\ \citenamefont
  {{Tremaine}}(2008)}]{Binney2008}%
  \BibitemOpen
  \bibfield  {author} {\bibinfo {author} {\bibfnamefont {J.}~\bibnamefont
  {{Binney}}}\ and\ \bibinfo {author} {\bibfnamefont {S.}~\bibnamefont
  {{Tremaine}}},\ }\href@noop {} {\emph {\bibinfo {title} {{Galactic Dynamics:
  Second Edition}}}}\ (\bibinfo {year} {2008})\BibitemShut {NoStop}%
\bibitem [{\citenamefont {{Sigurdsson}}\ and\ \citenamefont
  {{Phinney}}(1993)}]{Sigurdsson1993}%
  \BibitemOpen
  \bibfield  {author} {\bibinfo {author} {\bibfnamefont {S.}~\bibnamefont
  {{Sigurdsson}}}\ and\ \bibinfo {author} {\bibfnamefont {E.~S.}\ \bibnamefont
  {{Phinney}}},\ }\bibfield  {title} {\bibinfo {title} {{Binary--Single Star
  Interactions in Globular Clusters}},\ }\href {https://doi.org/10.1086/173190}
  {\bibfield  {journal} {\bibinfo  {journal} {\apj}\ }\textbf {\bibinfo
  {volume} {415}},\ \bibinfo {pages} {631} (\bibinfo {year}
  {1993})}\BibitemShut {NoStop}%
\bibitem [{\citenamefont {Kulkarni}\ \emph {et~al.}(1993)\citenamefont
  {Kulkarni}, \citenamefont {Hut},\ and\ \citenamefont
  {McMillan}}]{Kulkarni1993}%
  \BibitemOpen
  \bibfield  {author} {\bibinfo {author} {\bibfnamefont {S.~R.}\ \bibnamefont
  {Kulkarni}}, \bibinfo {author} {\bibfnamefont {P.}~\bibnamefont {Hut}},\ and\
  \bibinfo {author} {\bibfnamefont {S.~J.}\ \bibnamefont {McMillan}},\
  }\bibfield  {title} {\bibinfo {title} {Stellar black holes in globular
  clusters},\ }\href@noop {} {\bibfield  {journal} {\bibinfo  {journal}
  {Nature}\ }\textbf {\bibinfo {volume} {364}},\ \bibinfo {pages} {421}
  (\bibinfo {year} {1993})}\BibitemShut {NoStop}%
\bibitem [{\citenamefont {Ivanova}\ \emph {et~al.}(2005)\citenamefont
  {Ivanova}, \citenamefont {Belczynski}, \citenamefont {Fregeau},\ and\
  \citenamefont {Rasio}}]{Ivanova2005}%
  \BibitemOpen
  \bibfield  {author} {\bibinfo {author} {\bibfnamefont {N.}~\bibnamefont
  {Ivanova}}, \bibinfo {author} {\bibfnamefont {K.}~\bibnamefont {Belczynski}},
  \bibinfo {author} {\bibfnamefont {J.~M.}\ \bibnamefont {Fregeau}},\ and\
  \bibinfo {author} {\bibfnamefont {F.~A.}\ \bibnamefont {Rasio}},\ }\bibfield
  {title} {\bibinfo {title} {The evolution of binary fractions in globular
  clusters},\ }\href {https://doi.org/10.1111/j.1365-2966.2005.08804.x}
  {\bibfield  {journal} {\bibinfo  {journal} {Monthly Notices of the Royal
  Astronomical Society}\ }\textbf {\bibinfo {volume} {358}},\ \bibinfo {pages}
  {572} (\bibinfo {year} {2005})},\ \Eprint
  {https://arxiv.org/abs/astro-ph/0501131} {arXiv:astro-ph/0501131}
  \BibitemShut {NoStop}%
\bibitem [{\citenamefont {O'Leary}\ \emph {et~al.}(2 31)\citenamefont
  {O'Leary}, \citenamefont {Rasio}, \citenamefont {Fregeau}, \citenamefont
  {Ivanova},\ and\ \citenamefont {O'Shaughnessy}}]{OLeary2006}%
  \BibitemOpen
  \bibfield  {author} {\bibinfo {author} {\bibfnamefont {R.~M.}\ \bibnamefont
  {O'Leary}}, \bibinfo {author} {\bibfnamefont {F.~A.}\ \bibnamefont {Rasio}},
  \bibinfo {author} {\bibfnamefont {J.~M.}\ \bibnamefont {Fregeau}}, \bibinfo
  {author} {\bibfnamefont {N.}~\bibnamefont {Ivanova}},\ and\ \bibinfo {author}
  {\bibfnamefont {R.}~\bibnamefont {O'Shaughnessy}},\ }\bibfield  {title}
  {\bibinfo {title} {Binary {{Mergers}} and {{Growth}} of {{Black Holes}} in
  {{Dense Star Clusters}}},\ }\href {https://doi.org/10.1086/498446} {\bibfield
   {journal} {\bibinfo  {journal} {The Astrophysical Journal}\ }\textbf
  {\bibinfo {volume} {637}},\ \bibinfo {pages} {937} (\bibinfo {year}
  {2006-02-31})}\BibitemShut {NoStop}%
\bibitem [{\citenamefont {Heggie}(1975)}]{Heggie1975}%
  \BibitemOpen
  \bibfield  {author} {\bibinfo {author} {\bibfnamefont {D.~C.}\ \bibnamefont
  {Heggie}},\ }\bibfield  {title} {\bibinfo {title} {Binary evolution in
  stellar dynamics},\ }\href@noop {} {\bibfield  {journal} {\bibinfo  {journal}
  {Monthly Notices of the Royal Astronomical Society}\ }\textbf {\bibinfo
  {volume} {173}},\ \bibinfo {pages} {729} (\bibinfo {year}
  {1975})}\BibitemShut {NoStop}%
\bibitem [{\citenamefont {Rodriguez}\ \emph
  {et~al.}(2018{\natexlab{a}})\citenamefont {Rodriguez}, \citenamefont
  {{Amaro-Seoane}}, \citenamefont {Chatterjee}, \citenamefont {Kremer},
  \citenamefont {Rasio}, \citenamefont {Samsing}, \citenamefont {Ye},\ and\
  \citenamefont {Zevin}}]{Rodriguez2018c}%
  \BibitemOpen
  \bibfield  {author} {\bibinfo {author} {\bibfnamefont {C.~L.}\ \bibnamefont
  {Rodriguez}}, \bibinfo {author} {\bibfnamefont {P.}~\bibnamefont
  {{Amaro-Seoane}}}, \bibinfo {author} {\bibfnamefont {S.}~\bibnamefont
  {Chatterjee}}, \bibinfo {author} {\bibfnamefont {K.}~\bibnamefont {Kremer}},
  \bibinfo {author} {\bibfnamefont {F.~A.}\ \bibnamefont {Rasio}}, \bibinfo
  {author} {\bibfnamefont {J.}~\bibnamefont {Samsing}}, \bibinfo {author}
  {\bibfnamefont {C.~S.}\ \bibnamefont {Ye}},\ and\ \bibinfo {author}
  {\bibfnamefont {M.}~\bibnamefont {Zevin}},\ }\bibfield  {title} {\bibinfo
  {title} {Post-{{Newtonian}} dynamics in dense star clusters: {{Formation}},
  masses, and merger rates of highly-eccentric black hole binaries},\ }\href
  {https://doi.org/10.1103/PhysRevD.98.123005} {\bibfield  {journal} {\bibinfo
  {journal} {Physical Review D}\ }\textbf {\bibinfo {volume} {98}},\ \bibinfo
  {pages} {123005} (\bibinfo {year} {2018}{\natexlab{a}})}\BibitemShut
  {NoStop}%
\bibitem [{\citenamefont {Samsing}\ and\ \citenamefont
  {D'Orazio}(2018)}]{samsing2018}%
  \BibitemOpen
  \bibfield  {author} {\bibinfo {author} {\bibfnamefont {J.}~\bibnamefont
  {Samsing}}\ and\ \bibinfo {author} {\bibfnamefont {D.~J.}\ \bibnamefont
  {D'Orazio}},\ }\bibfield  {title} {\bibinfo {title} {Black {{Hole Mergers
  From Globular Clusters Observable}} by {{LISA I}}: {{Eccentric Sources
  Originating From Relativistic N-body Dynamics}}},\ }\href
  {https://doi.org/10.1093/mnras/sty2334} {\bibfield  {journal} {\bibinfo
  {journal} {Monthly Notices of the Royal Astronomical Society}\ }\textbf
  {\bibinfo {volume} {481}},\ \bibinfo {pages} {5445} (\bibinfo {year}
  {2018})}\BibitemShut {NoStop}%
\bibitem [{\citenamefont {Moody}\ and\ \citenamefont
  {Sigurdsson}(2009)}]{Moody2009}%
  \BibitemOpen
  \bibfield  {author} {\bibinfo {author} {\bibfnamefont {K.}~\bibnamefont
  {Moody}}\ and\ \bibinfo {author} {\bibfnamefont {S.}~\bibnamefont
  {Sigurdsson}},\ }\bibfield  {title} {\bibinfo {title} {{{MODELING THE
  RETENTION PROBABILITY OF BLACK HOLES IN GLOBULAR CLUSTERS}}: {{KICKS AND
  RATES}}},\ }\href {https://doi.org/10.1088/0004-637X/690/2/1370} {\bibfield
  {journal} {\bibinfo  {journal} {The Astrophysical Journal}\ }\textbf
  {\bibinfo {volume} {690}},\ \bibinfo {pages} {1370} (\bibinfo {year}
  {2009})}\BibitemShut {NoStop}%
\bibitem [{\citenamefont {Rodriguez}\ \emph
  {et~al.}(2016{\natexlab{a}})\citenamefont {Rodriguez}, \citenamefont
  {Chatterjee},\ and\ \citenamefont {Rasio}}]{Rodriguez2016a}%
  \BibitemOpen
  \bibfield  {author} {\bibinfo {author} {\bibfnamefont {C.~L.}\ \bibnamefont
  {Rodriguez}}, \bibinfo {author} {\bibfnamefont {S.}~\bibnamefont
  {Chatterjee}},\ and\ \bibinfo {author} {\bibfnamefont {F.~A.}\ \bibnamefont
  {Rasio}},\ }\bibfield  {title} {\bibinfo {title} {Binary black hole mergers
  from globular clusters: {{Masses}}, merger rates, and the impact of stellar
  evolution},\ }\href {https://doi.org/10.1103/PhysRevD.93.084029} {\bibfield
  {journal} {\bibinfo  {journal} {Physical Review D}\ }\textbf {\bibinfo
  {volume} {93}},\ \bibinfo {pages} {084029} (\bibinfo {year}
  {2016}{\natexlab{a}})},\ \Eprint {https://arxiv.org/abs/1602.02444}
  {arXiv:1602.02444} \BibitemShut {NoStop}%
\bibitem [{\citenamefont {Rodriguez}\ \emph
  {et~al.}(2016{\natexlab{b}})\citenamefont {Rodriguez}, \citenamefont
  {Haster}, \citenamefont {Chatterjee}, \citenamefont {Kalogera},\ and\
  \citenamefont {Rasio}}]{Rodriguez2016b}%
  \BibitemOpen
  \bibfield  {author} {\bibinfo {author} {\bibfnamefont {C.~L.}\ \bibnamefont
  {Rodriguez}}, \bibinfo {author} {\bibfnamefont {C.-J.}\ \bibnamefont
  {Haster}}, \bibinfo {author} {\bibfnamefont {S.}~\bibnamefont {Chatterjee}},
  \bibinfo {author} {\bibfnamefont {V.}~\bibnamefont {Kalogera}},\ and\
  \bibinfo {author} {\bibfnamefont {F.~A.}\ \bibnamefont {Rasio}},\ }\bibfield
  {title} {\bibinfo {title} {{{DYNAMICAL FORMATION OF THE GW150914 BINARY BLACK
  HOLE}}},\ }\href {https://doi.org/10.3847/2041-8205/824/1/L8} {\bibfield
  {journal} {\bibinfo  {journal} {The Astrophysical Journal}\ }\textbf
  {\bibinfo {volume} {824}},\ \bibinfo {pages} {L8} (\bibinfo {year}
  {2016}{\natexlab{b}})}\BibitemShut {NoStop}%
\bibitem [{\citenamefont {Baumgardt}\ and\ \citenamefont
  {Hilker}(2018)}]{baumgardt2018}%
  \BibitemOpen
  \bibfield  {author} {\bibinfo {author} {\bibfnamefont {H.}~\bibnamefont
  {Baumgardt}}\ and\ \bibinfo {author} {\bibfnamefont {M.}~\bibnamefont
  {Hilker}},\ }\bibfield  {title} {\bibinfo {title} {A catalogue of masses,
  structural parameters, and velocity dispersion profiles of 112 {{Milky Way}}
  globular clusters},\ }\href {https://doi.org/10.1093/mnras/sty1057}
  {\bibfield  {journal} {\bibinfo  {journal} {Monthly Notices of the Royal
  Astronomical Society}\ }\textbf {\bibinfo {volume} {478}},\ \bibinfo {pages}
  {1520} (\bibinfo {year} {2018})}\BibitemShut {NoStop}%
\bibitem [{\citenamefont {Tanikawa}(2013)}]{Tanikawa2013}%
  \BibitemOpen
  \bibfield  {author} {\bibinfo {author} {\bibfnamefont {A.}~\bibnamefont
  {Tanikawa}},\ }\bibfield  {title} {\bibinfo {title} {Dynamical evolution of
  stellar mass black holes in dense stellar clusters: Estimate for merger rate
  of binary black holes originating from globular clusters},\ }\href
  {https://doi.org/10.1093/mnras/stt1380} {\bibfield  {journal} {\bibinfo
  {journal} {Monthly Notices of the Royal Astronomical Society}\ }\textbf
  {\bibinfo {volume} {435}},\ \bibinfo {pages} {1358} (\bibinfo {year}
  {2013})}\BibitemShut {NoStop}%
\bibitem [{\citenamefont {Morawski}\ \emph {et~al.}(2018)\citenamefont
  {Morawski}, \citenamefont {Giersz}, \citenamefont {Askar},\ and\
  \citenamefont {Belczynski}}]{Morawski2018}%
  \BibitemOpen
  \bibfield  {author} {\bibinfo {author} {\bibfnamefont {J.}~\bibnamefont
  {Morawski}}, \bibinfo {author} {\bibfnamefont {M.}~\bibnamefont {Giersz}},
  \bibinfo {author} {\bibfnamefont {A.}~\bibnamefont {Askar}},\ and\ \bibinfo
  {author} {\bibfnamefont {K.}~\bibnamefont {Belczynski}},\ }\bibfield  {title}
  {\bibinfo {title} {{{MOCCA-SURVEY Database I}}: {{Assessing GW}} kick
  retention fractions for {{BH-BH}} mergers in globular clusters},\ }\href
  {https://doi.org/10.1093/MNRAS/STY2401} {\bibfield  {journal} {\bibinfo
  {journal} {Monthly Notices of the Royal Astronomical Society}\ }\textbf
  {\bibinfo {volume} {481}},\ \bibinfo {pages} {2168} (\bibinfo {year}
  {2018})}\BibitemShut {NoStop}%
\bibitem [{\citenamefont {Hong}\ \emph {et~al.}(2018)\citenamefont {Hong},
  \citenamefont {Vesperini}, \citenamefont {Askar}, \citenamefont {Giersz},
  \citenamefont {Szkudlarek},\ and\ \citenamefont {Bulik}}]{Hong2018}%
  \BibitemOpen
  \bibfield  {author} {\bibinfo {author} {\bibfnamefont {J.}~\bibnamefont
  {Hong}}, \bibinfo {author} {\bibfnamefont {E.}~\bibnamefont {Vesperini}},
  \bibinfo {author} {\bibfnamefont {A.}~\bibnamefont {Askar}}, \bibinfo
  {author} {\bibfnamefont {M.}~\bibnamefont {Giersz}}, \bibinfo {author}
  {\bibfnamefont {M.}~\bibnamefont {Szkudlarek}},\ and\ \bibinfo {author}
  {\bibfnamefont {T.}~\bibnamefont {Bulik}},\ }\bibfield  {title} {\bibinfo
  {title} {Binary black hole mergers from globular clusters: The impact of
  globular cluster properties},\ }\href {https://doi.org/10.1093/mnras/sty2211}
  {\bibfield  {journal} {\bibinfo  {journal} {Monthly Notices of the Royal
  Astronomical Society}\ }\textbf {\bibinfo {volume} {480}},\ \bibinfo {pages}
  {5645} (\bibinfo {year} {2018})},\ \Eprint {https://arxiv.org/abs/1808.04514}
  {arXiv:1808.04514} \BibitemShut {NoStop}%
\bibitem [{\citenamefont {Downing}\ \emph {et~al.}(2010)\citenamefont
  {Downing}, \citenamefont {Benacquista}, \citenamefont {Giersz},\ and\
  \citenamefont {Spurzem}}]{Downing2010}%
  \BibitemOpen
  \bibfield  {author} {\bibinfo {author} {\bibfnamefont {J.~M.~B.}\
  \bibnamefont {Downing}}, \bibinfo {author} {\bibfnamefont {M.~J.}\
  \bibnamefont {Benacquista}}, \bibinfo {author} {\bibfnamefont
  {M.}~\bibnamefont {Giersz}},\ and\ \bibinfo {author} {\bibfnamefont
  {R.}~\bibnamefont {Spurzem}},\ }\bibfield  {title} {\bibinfo {title} {Compact
  binaries in star clusters - {{I}}. {{Black}} hole binaries inside globular
  clusters},\ }\href {https://doi.org/10.1111/j.1365-2966.2010.17040.x}
  {\bibfield  {journal} {\bibinfo  {journal} {Monthly Notices of the Royal
  Astronomical Society}\ }\textbf {\bibinfo {volume} {407}},\ \bibinfo {pages}
  {1946} (\bibinfo {year} {2010})}\BibitemShut {NoStop}%
\bibitem [{\citenamefont {Downing}\ \emph {et~al.}(2011)\citenamefont
  {Downing}, \citenamefont {Benacquista}, \citenamefont {Giersz},\ and\
  \citenamefont {Spurzem}}]{Downing2011}%
  \BibitemOpen
  \bibfield  {author} {\bibinfo {author} {\bibfnamefont {J.~M.~B.}\
  \bibnamefont {Downing}}, \bibinfo {author} {\bibfnamefont {M.~J.}\
  \bibnamefont {Benacquista}}, \bibinfo {author} {\bibfnamefont
  {M.}~\bibnamefont {Giersz}},\ and\ \bibinfo {author} {\bibfnamefont
  {R.}~\bibnamefont {Spurzem}},\ }\bibfield  {title} {\bibinfo {title} {Compact
  binaries in star clusters - {{II}}. {{Escapers}} and detection rates},\
  }\href {https://doi.org/10.1111/j.1365-2966.2011.19023.x} {\bibfield
  {journal} {\bibinfo  {journal} {Monthly Notices of the Royal Astronomical
  Society}\ }\textbf {\bibinfo {volume} {416}},\ \bibinfo {pages} {no}
  (\bibinfo {year} {2011})},\ \Eprint {https://arxiv.org/abs/1008.5060}
  {arXiv:1008.5060} \BibitemShut {NoStop}%
\bibitem [{\citenamefont {Chatterjee}\ \emph {et~al.}(2016)\citenamefont
  {Chatterjee}, \citenamefont {Rodriguez}, \citenamefont {Kalogera},\ and\
  \citenamefont {Rasio}}]{Chatterjee2017}%
  \BibitemOpen
  \bibfield  {author} {\bibinfo {author} {\bibfnamefont {S.}~\bibnamefont
  {Chatterjee}}, \bibinfo {author} {\bibfnamefont {C.~L.}\ \bibnamefont
  {Rodriguez}}, \bibinfo {author} {\bibfnamefont {V.}~\bibnamefont
  {Kalogera}},\ and\ \bibinfo {author} {\bibfnamefont {F.~A.}\ \bibnamefont
  {Rasio}},\ }\bibfield  {title} {\bibinfo {title} {Dynamical {{Formation}} of
  {{Low-Mass Merging Black Hole Binaries}} like {{GW151226}}},\ }\href
  {https://doi.org/10.3847/2041-8213/aa5caa} {\bibfield  {journal} {\bibinfo
  {journal} {The Astrophysical Journal Letters}\ }\textbf {\bibinfo {volume}
  {836}},\ \bibinfo {pages} {1} (\bibinfo {year} {2016})},\ \Eprint
  {https://arxiv.org/abs/1609.06689} {arXiv:1609.06689} \BibitemShut {NoStop}%
\bibitem [{\citenamefont {Chatterjee}\ \emph {et~al.}(2017)\citenamefont
  {Chatterjee}, \citenamefont {Rodriguez},\ and\ \citenamefont
  {Rasio}}]{Chatterjee2016a}%
  \BibitemOpen
  \bibfield  {author} {\bibinfo {author} {\bibfnamefont {S.}~\bibnamefont
  {Chatterjee}}, \bibinfo {author} {\bibfnamefont {C.~L.}\ \bibnamefont
  {Rodriguez}},\ and\ \bibinfo {author} {\bibfnamefont {F.~A.}\ \bibnamefont
  {Rasio}},\ }\bibfield  {title} {\bibinfo {title} {{{BINARY BLACK HOLES IN
  DENSE STAR CLUSTERS}}: {{EXPLORING THE THEORETICAL UNCERTAINTIES}}},\ }\href
  {https://doi.org/10.3847/1538-4357/834/1/68} {\bibfield  {journal} {\bibinfo
  {journal} {The Astrophysical Journal}\ }\textbf {\bibinfo {volume} {834}},\
  \bibinfo {pages} {68} (\bibinfo {year} {2017})}\BibitemShut {NoStop}%
\bibitem [{\citenamefont {Bae}\ \emph {et~al.}(2014)\citenamefont {Bae},
  \citenamefont {Kim},\ and\ \citenamefont {Lee}}]{Bae2014}%
  \BibitemOpen
  \bibfield  {author} {\bibinfo {author} {\bibfnamefont {Y.-B.}\ \bibnamefont
  {Bae}}, \bibinfo {author} {\bibfnamefont {C.}~\bibnamefont {Kim}},\ and\
  \bibinfo {author} {\bibfnamefont {H.~M.}\ \bibnamefont {Lee}},\ }\bibfield
  {title} {\bibinfo {title} {Compact binaries ejected from globular clusters as
  gravitational wave sources},\ }\href {https://doi.org/10.1093/mnras/stu381}
  {\bibfield  {journal} {\bibinfo  {journal} {Monthly Notices of the Royal
  Astronomical Society}\ }\textbf {\bibinfo {volume} {440}},\ \bibinfo {pages}
  {2714} (\bibinfo {year} {2014})}\BibitemShut {NoStop}%
\bibitem [{\citenamefont {Askar}\ \emph {et~al.}(2016)\citenamefont {Askar},
  \citenamefont {Szkudlarek}, \citenamefont {{Gondek-Rosi{\'n}ska}},
  \citenamefont {Giersz},\ and\ \citenamefont {Bulik}}]{Askar2016}%
  \BibitemOpen
  \bibfield  {author} {\bibinfo {author} {\bibfnamefont {A.}~\bibnamefont
  {Askar}}, \bibinfo {author} {\bibfnamefont {M.}~\bibnamefont {Szkudlarek}},
  \bibinfo {author} {\bibfnamefont {D.}~\bibnamefont {{Gondek-Rosi{\'n}ska}}},
  \bibinfo {author} {\bibfnamefont {M.}~\bibnamefont {Giersz}},\ and\ \bibinfo
  {author} {\bibfnamefont {T.}~\bibnamefont {Bulik}},\ }\bibfield  {title}
  {\bibinfo {title} {{{MOCCA-SURVEY Database I}}: {{Coalescing Binary Black
  Holes Originating From Globular Clusters}}},\ }\href
  {https://doi.org/10.1093/mnrasl/slw177} {\bibfield  {journal} {\bibinfo
  {journal} {Monthly Notices of the Royal Astronomical Society: Letters}\
  }\textbf {\bibinfo {volume} {464}},\ \bibinfo {pages} {L36} (\bibinfo {year}
  {2016})},\ \Eprint {https://arxiv.org/abs/1608.02520} {arXiv:1608.02520}
  \BibitemShut {NoStop}%
\bibitem [{\citenamefont {Aarseth}(2012)}]{Aarseth2012}%
  \BibitemOpen
  \bibfield  {author} {\bibinfo {author} {\bibfnamefont {S.~J.}\ \bibnamefont
  {Aarseth}},\ }\bibfield  {title} {\bibinfo {title} {Mergers and ejections of
  black holes in globular clusters},\ }\href
  {https://doi.org/10.1111/j.1365-2966.2012.20666.x} {\bibfield  {journal}
  {\bibinfo  {journal} {Monthly Notices of the Royal Astronomical Society}\
  }\textbf {\bibinfo {volume} {422}},\ \bibinfo {pages} {841} (\bibinfo {year}
  {2012})}\BibitemShut {NoStop}%
\bibitem [{\citenamefont {Rodriguez}\ \emph {et~al.}(2015)\citenamefont
  {Rodriguez}, \citenamefont {Morscher}, \citenamefont {Pattabiraman},
  \citenamefont {Chatterjee}, \citenamefont {Haster},\ and\ \citenamefont
  {Rasio}}]{Rodriguez2015a}%
  \BibitemOpen
  \bibfield  {author} {\bibinfo {author} {\bibfnamefont {C.~L.}\ \bibnamefont
  {Rodriguez}}, \bibinfo {author} {\bibfnamefont {M.}~\bibnamefont {Morscher}},
  \bibinfo {author} {\bibfnamefont {B.}~\bibnamefont {Pattabiraman}}, \bibinfo
  {author} {\bibfnamefont {S.}~\bibnamefont {Chatterjee}}, \bibinfo {author}
  {\bibfnamefont {C.-J.}\ \bibnamefont {Haster}},\ and\ \bibinfo {author}
  {\bibfnamefont {F.~A.}\ \bibnamefont {Rasio}},\ }\bibfield  {title} {\bibinfo
  {title} {Binary {{Black Hole Mergers}} from {{Globular Clusters}}:
  {{Implications}} for {{Advanced LIGO}}},\ }\href
  {https://doi.org/10.1103/PhysRevLett.115.051101} {\bibfield  {journal}
  {\bibinfo  {journal} {Physical Review Letters}\ }\textbf {\bibinfo {volume}
  {115}},\ \bibinfo {pages} {051101} (\bibinfo {year} {2015})}\BibitemShut
  {NoStop}%
\bibitem [{\citenamefont {Rodriguez}\ \emph
  {et~al.}(2018{\natexlab{b}})\citenamefont {Rodriguez}, \citenamefont
  {{Amaro-Seoane}}, \citenamefont {Chatterjee},\ and\ \citenamefont
  {Rasio}}]{Rodriguez2018}%
  \BibitemOpen
  \bibfield  {author} {\bibinfo {author} {\bibfnamefont {C.~L.}\ \bibnamefont
  {Rodriguez}}, \bibinfo {author} {\bibfnamefont {P.}~\bibnamefont
  {{Amaro-Seoane}}}, \bibinfo {author} {\bibfnamefont {S.}~\bibnamefont
  {Chatterjee}},\ and\ \bibinfo {author} {\bibfnamefont {F.~A.}\ \bibnamefont
  {Rasio}},\ }\bibfield  {title} {\bibinfo {title} {Post-{{Newtonian Dynamics}}
  in {{Dense Star Clusters}}: {{Highly Eccentric}}, {{Highly Spinning}}, and
  {{Repeated Binary Black Hole Mergers}}},\ }\href
  {https://doi.org/10.1103/PhysRevLett.120.151101} {\bibfield  {journal}
  {\bibinfo  {journal} {Physical Review Letters}\ }\textbf {\bibinfo {volume}
  {120}},\ \bibinfo {pages} {151101} (\bibinfo {year} {2018}{\natexlab{b}})},\
  \Eprint {https://arxiv.org/abs/1712.04937} {arXiv:1712.04937} \BibitemShut
  {NoStop}%
\bibitem [{\citenamefont {{Chattopadhyay}}\ \emph {et~al.}(2022)\citenamefont
  {{Chattopadhyay}}, \citenamefont {{Hurley}}, \citenamefont {{Stevenson}},\
  and\ \citenamefont {{Raidani}}}]{Chattopadhyay2022}%
  \BibitemOpen
  \bibfield  {author} {\bibinfo {author} {\bibfnamefont {D.}~\bibnamefont
  {{Chattopadhyay}}}, \bibinfo {author} {\bibfnamefont {J.}~\bibnamefont
  {{Hurley}}}, \bibinfo {author} {\bibfnamefont {S.}~\bibnamefont
  {{Stevenson}}},\ and\ \bibinfo {author} {\bibfnamefont {A.}~\bibnamefont
  {{Raidani}}},\ }\bibfield  {title} {\bibinfo {title} {{Dynamical double black
  holes and their host cluster properties}},\ }\href
  {https://doi.org/10.1093/mnras/stac1163} {\bibfield  {journal} {\bibinfo
  {journal} {\mnras}\ }\textbf {\bibinfo {volume} {513}},\ \bibinfo {pages}
  {4527} (\bibinfo {year} {2022})},\ \Eprint {https://arxiv.org/abs/2202.08924}
  {arXiv:2202.08924 [astro-ph.GA]} \BibitemShut {NoStop}%
\bibitem [{\citenamefont {{Fujii}}\ \emph {et~al.}(2017)\citenamefont
  {{Fujii}}, \citenamefont {{Tanikawa}},\ and\ \citenamefont
  {{Makino}}}]{Fujii2017}%
  \BibitemOpen
  \bibfield  {author} {\bibinfo {author} {\bibfnamefont {M.~S.}\ \bibnamefont
  {{Fujii}}}, \bibinfo {author} {\bibfnamefont {A.}~\bibnamefont
  {{Tanikawa}}},\ and\ \bibinfo {author} {\bibfnamefont {J.}~\bibnamefont
  {{Makino}}},\ }\bibfield  {title} {\bibinfo {title} {{The detection rates of
  merging binary black holes originating from star clusters and their mass
  function}},\ }\href {https://doi.org/10.1093/pasj/psx108} {\bibfield
  {journal} {\bibinfo  {journal} {\pasj}\ }\textbf {\bibinfo {volume} {69}},\
  \bibinfo {eid} {94} (\bibinfo {year} {2017})},\ \Eprint
  {https://arxiv.org/abs/1709.02058} {arXiv:1709.02058 [astro-ph.HE]}
  \BibitemShut {NoStop}%
\bibitem [{\citenamefont {Fragione}\ \emph {et~al.}(2018)\citenamefont
  {Fragione}, \citenamefont {Ginsburg},\ and\ \citenamefont
  {Kocsis}}]{Fragione2018}%
  \BibitemOpen
  \bibfield  {author} {\bibinfo {author} {\bibfnamefont {G.}~\bibnamefont
  {Fragione}}, \bibinfo {author} {\bibfnamefont {I.}~\bibnamefont {Ginsburg}},\
  and\ \bibinfo {author} {\bibfnamefont {B.}~\bibnamefont {Kocsis}},\
  }\bibfield  {title} {\bibinfo {title} {Gravitational {{Waves}} and
  {{Intermediate-mass Black Hole Retention}} in {{Globular Clusters}}},\ }\href
  {https://doi.org/10.3847/1538-4357/aab368} {\bibfield  {journal} {\bibinfo
  {journal} {The Astrophysical Journal}\ }\textbf {\bibinfo {volume} {856}},\
  \bibinfo {pages} {92} (\bibinfo {year} {2018})},\ \Eprint
  {https://arxiv.org/abs/1711.00483} {arXiv:1711.00483} \BibitemShut {NoStop}%
\bibitem [{\citenamefont {Rodriguez}\ and\ \citenamefont
  {Loeb}(2018)}]{Rodriguez2018b}%
  \BibitemOpen
  \bibfield  {author} {\bibinfo {author} {\bibfnamefont {C.~L.}\ \bibnamefont
  {Rodriguez}}\ and\ \bibinfo {author} {\bibfnamefont {A.}~\bibnamefont
  {Loeb}},\ }\bibfield  {title} {\bibinfo {title} {Redshift {{Evolution}} of
  the {{Black Hole Merger Rate}} from {{Globular Clusters}}},\ }\href
  {https://doi.org/10.3847/2041-8213/aae377} {\bibfield  {journal} {\bibinfo
  {journal} {The Astrophysical Journal}\ }\textbf {\bibinfo {volume} {866}},\
  \bibinfo {pages} {L5} (\bibinfo {year} {2018})},\ \Eprint
  {https://arxiv.org/abs/1809.01152} {arXiv:1809.01152} \BibitemShut {NoStop}%
\bibitem [{\citenamefont {Choksi}\ \emph {et~al.}(2019)\citenamefont {Choksi},
  \citenamefont {Volonteri}, \citenamefont {Colpi}, \citenamefont {Gnedin},\
  and\ \citenamefont {Li}}]{choksi2019}%
  \BibitemOpen
  \bibfield  {author} {\bibinfo {author} {\bibfnamefont {N.}~\bibnamefont
  {Choksi}}, \bibinfo {author} {\bibfnamefont {M.}~\bibnamefont {Volonteri}},
  \bibinfo {author} {\bibfnamefont {M.}~\bibnamefont {Colpi}}, \bibinfo
  {author} {\bibfnamefont {O.~Y.}\ \bibnamefont {Gnedin}},\ and\ \bibinfo
  {author} {\bibfnamefont {H.}~\bibnamefont {Li}},\ }\bibfield  {title}
  {\bibinfo {title} {The {{Star Clusters That Make Black Hole Binaries}} across
  {{Cosmic Time}}},\ }\href {https://doi.org/10.3847/1538-4357/aaffde}
  {\bibfield  {journal} {\bibinfo  {journal} {The Astrophysical Journal}\
  }\textbf {\bibinfo {volume} {873}},\ \bibinfo {pages} {100} (\bibinfo {year}
  {2019})},\ \Eprint {https://arxiv.org/abs/1809.01164} {arXiv:1809.01164}
  \BibitemShut {NoStop}%
\bibitem [{\citenamefont {{Antonini}}\ and\ \citenamefont
  {{Gieles}}(2020{\natexlab{a}})}]{Antonini2020}%
  \BibitemOpen
  \bibfield  {author} {\bibinfo {author} {\bibfnamefont {F.}~\bibnamefont
  {{Antonini}}}\ and\ \bibinfo {author} {\bibfnamefont {M.}~\bibnamefont
  {{Gieles}}},\ }\bibfield  {title} {\bibinfo {title} {{Population synthesis of
  black hole binary mergers from star clusters}},\ }\href
  {https://doi.org/10.1093/mnras/stz3584} {\bibfield  {journal} {\bibinfo
  {journal} {\mnras}\ }\textbf {\bibinfo {volume} {492}},\ \bibinfo {pages}
  {2936} (\bibinfo {year} {2020}{\natexlab{a}})},\ \Eprint
  {https://arxiv.org/abs/1906.11855} {arXiv:1906.11855 [astro-ph.HE]}
  \BibitemShut {NoStop}%
\bibitem [{\citenamefont {{Antonini}}\ and\ \citenamefont
  {{Gieles}}(2020{\natexlab{b}})}]{Antonini2020a}%
  \BibitemOpen
  \bibfield  {author} {\bibinfo {author} {\bibfnamefont {F.}~\bibnamefont
  {{Antonini}}}\ and\ \bibinfo {author} {\bibfnamefont {M.}~\bibnamefont
  {{Gieles}}},\ }\bibfield  {title} {\bibinfo {title} {{Merger rate of black
  hole binaries from globular clusters: Theoretical error bars and comparison
  to gravitational wave data from GWTC-2}},\ }\href
  {https://doi.org/10.1103/PhysRevD.102.123016} {\bibfield  {journal} {\bibinfo
   {journal} {\prd}\ }\textbf {\bibinfo {volume} {102}},\ \bibinfo {eid}
  {123016} (\bibinfo {year} {2020}{\natexlab{b}})},\ \Eprint
  {https://arxiv.org/abs/2009.01861} {arXiv:2009.01861 [astro-ph.HE]}
  \BibitemShut {NoStop}%
\bibitem [{\citenamefont {{Antonini}}\ \emph {et~al.}(2022)\citenamefont
  {{Antonini}}, \citenamefont {{Gieles}}, \citenamefont {{Dosopoulou}},\ and\
  \citenamefont {{Chattopadhyay}}}]{Antonini2022}%
  \BibitemOpen
  \bibfield  {author} {\bibinfo {author} {\bibfnamefont {F.}~\bibnamefont
  {{Antonini}}}, \bibinfo {author} {\bibfnamefont {M.}~\bibnamefont
  {{Gieles}}}, \bibinfo {author} {\bibfnamefont {F.}~\bibnamefont
  {{Dosopoulou}}},\ and\ \bibinfo {author} {\bibfnamefont {D.}~\bibnamefont
  {{Chattopadhyay}}},\ }\bibfield  {title} {\bibinfo {title} {{Coalescing black
  hole binaries from globular clusters: mass distributions and comparison to
  gravitational wave data from GWTC-3}},\ }\href
  {https://doi.org/10.48550/arXiv.2208.01081} {\bibfield  {journal} {\bibinfo
  {journal} {arXiv e-prints}\ ,\ \bibinfo {eid} {arXiv:2208.01081}} (\bibinfo
  {year} {2022})},\ \Eprint {https://arxiv.org/abs/2208.01081}
  {arXiv:2208.01081 [astro-ph.HE]} \BibitemShut {NoStop}%
\bibitem [{\citenamefont {{Kritos}}\ \emph {et~al.}(2022)\citenamefont
  {{Kritos}}, \citenamefont {{Strokov}}, \citenamefont {{Baibhav}},\ and\
  \citenamefont {{Berti}}}]{Kritos2022}%
  \BibitemOpen
  \bibfield  {author} {\bibinfo {author} {\bibfnamefont {K.}~\bibnamefont
  {{Kritos}}}, \bibinfo {author} {\bibfnamefont {V.}~\bibnamefont {{Strokov}}},
  \bibinfo {author} {\bibfnamefont {V.}~\bibnamefont {{Baibhav}}},\ and\
  \bibinfo {author} {\bibfnamefont {E.}~\bibnamefont {{Berti}}},\ }\bibfield
  {title} {\bibinfo {title} {{Rapster: a fast code for dynamical formation of
  black-hole binaries in dense star clusters}},\ }\href
  {https://doi.org/10.48550/arXiv.2210.10055} {\bibfield  {journal} {\bibinfo
  {journal} {arXiv e-prints}\ ,\ \bibinfo {eid} {arXiv:2210.10055}} (\bibinfo
  {year} {2022})},\ \Eprint {https://arxiv.org/abs/2210.10055}
  {arXiv:2210.10055 [astro-ph.HE]} \BibitemShut {NoStop}%
\bibitem [{\citenamefont {{Rodriguez}}\ \emph {et~al.}(2021)\citenamefont
  {{Rodriguez}}, \citenamefont {{Kremer}}, \citenamefont {{Chatterjee}},
  \citenamefont {{Fragione}}, \citenamefont {{Loeb}}, \citenamefont {{Rasio}},
  \citenamefont {{Weatherford}},\ and\ \citenamefont {{Ye}}}]{Rodriguez2021}%
  \BibitemOpen
  \bibfield  {author} {\bibinfo {author} {\bibfnamefont {C.~L.}\ \bibnamefont
  {{Rodriguez}}}, \bibinfo {author} {\bibfnamefont {K.}~\bibnamefont
  {{Kremer}}}, \bibinfo {author} {\bibfnamefont {S.}~\bibnamefont
  {{Chatterjee}}}, \bibinfo {author} {\bibfnamefont {G.}~\bibnamefont
  {{Fragione}}}, \bibinfo {author} {\bibfnamefont {A.}~\bibnamefont {{Loeb}}},
  \bibinfo {author} {\bibfnamefont {F.~A.}\ \bibnamefont {{Rasio}}}, \bibinfo
  {author} {\bibfnamefont {N.~C.}\ \bibnamefont {{Weatherford}}},\ and\
  \bibinfo {author} {\bibfnamefont {C.~S.}\ \bibnamefont {{Ye}}},\ }\bibfield
  {title} {\bibinfo {title} {{The Observed Rate of Binary Black Hole Mergers
  can be Entirely Explained by Globular Clusters}},\ }\href
  {https://doi.org/10.3847/2515-5172/abdf54} {\bibfield  {journal} {\bibinfo
  {journal} {Research Notes of the American Astronomical Society}\ }\textbf
  {\bibinfo {volume} {5}},\ \bibinfo {eid} {19} (\bibinfo {year} {2021})},\
  \Eprint {https://arxiv.org/abs/2101.07793} {arXiv:2101.07793 [astro-ph.HE]}
  \BibitemShut {NoStop}%
\bibitem [{\citenamefont {Zevin}\ \emph {et~al.}(2021)\citenamefont {Zevin},
  \citenamefont {Bavera}, \citenamefont {Berry}, \citenamefont {Kalogera},
  \citenamefont {Fragos}, \citenamefont {Marchant}, \citenamefont {Rodriguez},
  \citenamefont {Antonini}, \citenamefont {Holz},\ and\ \citenamefont
  {Pankow}}]{zevin2021}%
  \BibitemOpen
  \bibfield  {author} {\bibinfo {author} {\bibfnamefont {M.}~\bibnamefont
  {Zevin}}, \bibinfo {author} {\bibfnamefont {S.~S.}\ \bibnamefont {Bavera}},
  \bibinfo {author} {\bibfnamefont {C.~P.~L.}\ \bibnamefont {Berry}}, \bibinfo
  {author} {\bibfnamefont {V.}~\bibnamefont {Kalogera}}, \bibinfo {author}
  {\bibfnamefont {T.}~\bibnamefont {Fragos}}, \bibinfo {author} {\bibfnamefont
  {P.}~\bibnamefont {Marchant}}, \bibinfo {author} {\bibfnamefont {C.~L.}\
  \bibnamefont {Rodriguez}}, \bibinfo {author} {\bibfnamefont {F.}~\bibnamefont
  {Antonini}}, \bibinfo {author} {\bibfnamefont {D.~E.}\ \bibnamefont {Holz}},\
  and\ \bibinfo {author} {\bibfnamefont {C.}~\bibnamefont {Pankow}},\
  }\bibfield  {title} {\bibinfo {title} {One {{Channel}} to {{Rule Them All}}?
  {{Constraining}} the {{Origins}} of {{Binary Black Holes Using Multiple
  Formation Pathways}}},\ }\href {https://doi.org/10.3847/1538-4357/abe40e}
  {\bibfield  {journal} {\bibinfo  {journal} {The Astrophysical Journal}\
  }\textbf {\bibinfo {volume} {910}},\ \bibinfo {pages} {152} (\bibinfo {year}
  {2021})}\BibitemShut {NoStop}%
\bibitem [{\citenamefont {{Abbott}}\ \emph
  {et~al.}(2021{\natexlab{b}})\citenamefont {{Abbott}}, \citenamefont
  {{Abbott}}, \citenamefont {{Abraham}}, \citenamefont {{Acernese}},
  \citenamefont {{Ackley}}, \citenamefont {{Adams}}, \citenamefont {{Adams}},
  \citenamefont {{Adhikari}}, \citenamefont {{Adya}}, \citenamefont
  {{Affeldt}},\ and\ \citenamefont {et~al.}}]{Abbott2021}%
  \BibitemOpen
  \bibfield  {author} {\bibinfo {author} {\bibfnamefont {R.}~\bibnamefont
  {{Abbott}}}, \bibinfo {author} {\bibfnamefont {T.~D.}\ \bibnamefont
  {{Abbott}}}, \bibinfo {author} {\bibfnamefont {S.}~\bibnamefont {{Abraham}}},
  \bibinfo {author} {\bibfnamefont {F.}~\bibnamefont {{Acernese}}}, \bibinfo
  {author} {\bibfnamefont {K.}~\bibnamefont {{Ackley}}}, \bibinfo {author}
  {\bibfnamefont {A.}~\bibnamefont {{Adams}}}, \bibinfo {author} {\bibfnamefont
  {C.}~\bibnamefont {{Adams}}}, \bibinfo {author} {\bibfnamefont {R.~X.}\
  \bibnamefont {{Adhikari}}}, \bibinfo {author} {\bibfnamefont {V.~B.}\
  \bibnamefont {{Adya}}}, \bibinfo {author} {\bibfnamefont {C.}~\bibnamefont
  {{Affeldt}}},\ and\ \bibinfo {author} {\bibnamefont {et~al.}},\ }\bibfield
  {title} {\bibinfo {title} {{Population Properties of Compact Objects from the
  Second LIGO-Virgo Gravitational-Wave Transient Catalog}},\ }\href
  {https://doi.org/10.3847/2041-8213/abe949} {\bibfield  {journal} {\bibinfo
  {journal} {\apjl}\ }\textbf {\bibinfo {volume} {913}},\ \bibinfo {eid} {L7}
  (\bibinfo {year} {2021}{\natexlab{b}})},\ \Eprint
  {https://arxiv.org/abs/2010.14533} {arXiv:2010.14533 [astro-ph.HE]}
  \BibitemShut {NoStop}%
\bibitem [{\citenamefont {{Samsing}}\ \emph {et~al.}(2018)\citenamefont
  {{Samsing}}, \citenamefont {{Askar}},\ and\ \citenamefont
  {{Giersz}}}]{Samsing2018a}%
  \BibitemOpen
  \bibfield  {author} {\bibinfo {author} {\bibfnamefont {J.}~\bibnamefont
  {{Samsing}}}, \bibinfo {author} {\bibfnamefont {A.}~\bibnamefont {{Askar}}},\
  and\ \bibinfo {author} {\bibfnamefont {M.}~\bibnamefont {{Giersz}}},\
  }\bibfield  {title} {\bibinfo {title} {{MOCCA-SURVEY Database. I. Eccentric
  Black Hole Mergers during Binary-Single Interactions in Globular Clusters}},\
  }\href {https://doi.org/10.3847/1538-4357/aaab52} {\bibfield  {journal}
  {\bibinfo  {journal} {\apj}\ }\textbf {\bibinfo {volume} {855}},\ \bibinfo
  {eid} {124} (\bibinfo {year} {2018})},\ \Eprint
  {https://arxiv.org/abs/1712.06186} {arXiv:1712.06186 [astro-ph.HE]}
  \BibitemShut {NoStop}%
\bibitem [{\citenamefont {{Antonini}}\ \emph {et~al.}(2019)\citenamefont
  {{Antonini}}, \citenamefont {{Gieles}},\ and\ \citenamefont
  {{Gualandris}}}]{Antonini2019}%
  \BibitemOpen
  \bibfield  {author} {\bibinfo {author} {\bibfnamefont {F.}~\bibnamefont
  {{Antonini}}}, \bibinfo {author} {\bibfnamefont {M.}~\bibnamefont
  {{Gieles}}},\ and\ \bibinfo {author} {\bibfnamefont {A.}~\bibnamefont
  {{Gualandris}}},\ }\bibfield  {title} {\bibinfo {title} {{Black hole growth
  through hierarchical black hole mergers in dense star clusters: implications
  for gravitational wave detections}},\ }\href
  {https://doi.org/10.1093/mnras/stz1149} {\bibfield  {journal} {\bibinfo
  {journal} {\mnras}\ }\textbf {\bibinfo {volume} {486}},\ \bibinfo {pages}
  {5008} (\bibinfo {year} {2019})},\ \Eprint {https://arxiv.org/abs/1811.03640}
  {arXiv:1811.03640 [astro-ph.HE]} \BibitemShut {NoStop}%
\bibitem [{\citenamefont {Rodriguez}\ \emph {et~al.}(2020)\citenamefont
  {Rodriguez}, \citenamefont {Kremer}, \citenamefont {Grudi{\'c}},
  \citenamefont {Hafen}, \citenamefont {Chatterjee}, \citenamefont {Fragione},
  \citenamefont {Lamberts}, \citenamefont {Martinez}, \citenamefont {Rasio},
  \citenamefont {Weatherford},\ and\ \citenamefont {Ye}}]{rodriguez2020}%
  \BibitemOpen
  \bibfield  {author} {\bibinfo {author} {\bibfnamefont {C.~L.}\ \bibnamefont
  {Rodriguez}}, \bibinfo {author} {\bibfnamefont {K.}~\bibnamefont {Kremer}},
  \bibinfo {author} {\bibfnamefont {M.~Y.}\ \bibnamefont {Grudi{\'c}}},
  \bibinfo {author} {\bibfnamefont {Z.}~\bibnamefont {Hafen}}, \bibinfo
  {author} {\bibfnamefont {S.}~\bibnamefont {Chatterjee}}, \bibinfo {author}
  {\bibfnamefont {G.}~\bibnamefont {Fragione}}, \bibinfo {author}
  {\bibfnamefont {A.}~\bibnamefont {Lamberts}}, \bibinfo {author}
  {\bibfnamefont {M.~A.~S.}\ \bibnamefont {Martinez}}, \bibinfo {author}
  {\bibfnamefont {F.~A.}\ \bibnamefont {Rasio}}, \bibinfo {author}
  {\bibfnamefont {N.}~\bibnamefont {Weatherford}},\ and\ \bibinfo {author}
  {\bibfnamefont {C.~S.}\ \bibnamefont {Ye}},\ }\bibfield  {title} {\bibinfo
  {title} {{{GW190412}} as a {{Third-generation Black Hole Merger}} from a
  {{Super Star Cluster}}},\ }\href {https://doi.org/10.3847/2041-8213/ab961d}
  {\bibfield  {journal} {\bibinfo  {journal} {The Astrophysical Journal}\
  }\textbf {\bibinfo {volume} {896}},\ \bibinfo {pages} {L10} (\bibinfo {year}
  {2020})}\BibitemShut {NoStop}%
\bibitem [{\citenamefont {Merritt}\ \emph {et~al.}(2004)\citenamefont
  {Merritt}, \citenamefont {Milosavljevi}, \citenamefont {Favata},
  \citenamefont {Hughes},\ and\ \citenamefont {Holz}}]{Merritt2004}%
  \BibitemOpen
  \bibfield  {author} {\bibinfo {author} {\bibfnamefont {D.}~\bibnamefont
  {Merritt}}, \bibinfo {author} {\bibfnamefont {M.}~\bibnamefont
  {Milosavljevi}}, \bibinfo {author} {\bibfnamefont {M.}~\bibnamefont
  {Favata}}, \bibinfo {author} {\bibfnamefont {S.~A.}\ \bibnamefont {Hughes}},\
  and\ \bibinfo {author} {\bibfnamefont {D.~E.}\ \bibnamefont {Holz}},\
  }\bibfield  {title} {\bibinfo {title} {Consequences of {{Gravitational
  Radiation Recoil}}},\ }\href {https://doi.org/10.1086/421551} {\bibfield
  {journal} {\bibinfo  {journal} {The Astrophysical Journal}\ }\textbf
  {\bibinfo {volume} {607}},\ \bibinfo {pages} {L9} (\bibinfo {year}
  {2004})}\BibitemShut {NoStop}%
\bibitem [{\citenamefont {Campanelli}\ \emph {et~al.}(2007)\citenamefont
  {Campanelli}, \citenamefont {Lousto}, \citenamefont {Zlochower},\ and\
  \citenamefont {Merritt}}]{Campanelli2007}%
  \BibitemOpen
  \bibfield  {author} {\bibinfo {author} {\bibfnamefont {M.}~\bibnamefont
  {Campanelli}}, \bibinfo {author} {\bibfnamefont {C.}~\bibnamefont {Lousto}},
  \bibinfo {author} {\bibfnamefont {Y.}~\bibnamefont {Zlochower}},\ and\
  \bibinfo {author} {\bibfnamefont {D.}~\bibnamefont {Merritt}},\ }\bibfield
  {title} {\bibinfo {title} {Large {{Merger Recoils}} and {{Spin Flips}} from
  {{Generic Black Hole Binaries}}},\ }\href {https://doi.org/10.1086/516712}
  {\bibfield  {journal} {\bibinfo  {journal} {The Astrophysical Journal}\
  }\textbf {\bibinfo {volume} {659}},\ \bibinfo {pages} {L5} (\bibinfo {year}
  {2007})},\ \Eprint {https://arxiv.org/abs/gr-qc/0701164}
  {arxiv:gr-qc/0701164} \BibitemShut {NoStop}%
\bibitem [{\citenamefont {Gonz{\'a}lez}\ \emph {et~al.}(2007)\citenamefont
  {Gonz{\'a}lez}, \citenamefont {Sperhake}, \citenamefont {Br{\"u}gmann},
  \citenamefont {Hannam},\ and\ \citenamefont {Husa}}]{Gonzalez2007}%
  \BibitemOpen
  \bibfield  {author} {\bibinfo {author} {\bibfnamefont {J.~A.}\ \bibnamefont
  {Gonz{\'a}lez}}, \bibinfo {author} {\bibfnamefont {U.}~\bibnamefont
  {Sperhake}}, \bibinfo {author} {\bibfnamefont {B.}~\bibnamefont
  {Br{\"u}gmann}}, \bibinfo {author} {\bibfnamefont {M.}~\bibnamefont
  {Hannam}},\ and\ \bibinfo {author} {\bibfnamefont {S.}~\bibnamefont {Husa}},\
  }\bibfield  {title} {\bibinfo {title} {Maximum {{Kick}} from {{Nonspinning
  Black-Hole Binary Inspiral}}},\ }\href
  {https://doi.org/10.1103/PhysRevLett.98.091101} {\bibfield  {journal}
  {\bibinfo  {journal} {Physical Review Letters}\ }\textbf {\bibinfo {volume}
  {98}},\ \bibinfo {pages} {091101} (\bibinfo {year} {2007})}\BibitemShut
  {NoStop}%
\bibitem [{\citenamefont {Lousto}\ and\ \citenamefont
  {Zlochower}(2008)}]{Lousto2008}%
  \BibitemOpen
  \bibfield  {author} {\bibinfo {author} {\bibfnamefont {C.~O.}\ \bibnamefont
  {Lousto}}\ and\ \bibinfo {author} {\bibfnamefont {Y.}~\bibnamefont
  {Zlochower}},\ }\bibfield  {title} {\bibinfo {title} {Further insight into
  gravitational recoil},\ }\href {https://doi.org/10.1103/PhysRevD.77.044028}
  {\bibfield  {journal} {\bibinfo  {journal} {Physical Review D}\ }\textbf
  {\bibinfo {volume} {77}},\ \bibinfo {pages} {044028} (\bibinfo {year}
  {2008})}\BibitemShut {NoStop}%
\bibitem [{\citenamefont {Gualandris}\ and\ \citenamefont
  {Merritt}(2008)}]{Gualandris2008}%
  \BibitemOpen
  \bibfield  {author} {\bibinfo {author} {\bibfnamefont {A.}~\bibnamefont
  {Gualandris}}\ and\ \bibinfo {author} {\bibfnamefont {D.}~\bibnamefont
  {Merritt}},\ }\bibfield  {title} {\bibinfo {title} {Ejection of
  {{Supermassive Black Holes}} from {{Galaxy Cores}}},\ }\href
  {https://doi.org/10.1086/586877} {\bibfield  {journal} {\bibinfo  {journal}
  {The Astrophysical Journal}\ }\textbf {\bibinfo {volume} {678}},\ \bibinfo
  {pages} {780} (\bibinfo {year} {2008})},\ \Eprint
  {https://arxiv.org/abs/0708.0771} {arxiv:0708.0771} \BibitemShut {NoStop}%
\bibitem [{\citenamefont {{Lousto}}\ and\ \citenamefont
  {{Zlochower}}(2011)}]{Lousto2011}%
  \BibitemOpen
  \bibfield  {author} {\bibinfo {author} {\bibfnamefont {C.~O.}\ \bibnamefont
  {{Lousto}}}\ and\ \bibinfo {author} {\bibfnamefont {Y.}~\bibnamefont
  {{Zlochower}}},\ }\bibfield  {title} {\bibinfo {title} {{Hangup Kicks: Still
  Larger Recoils by Partial Spin-Orbit Alignment of Black-Hole Binaries}},\
  }\href {https://doi.org/10.1103/PhysRevLett.107.231102} {\bibfield  {journal}
  {\bibinfo  {journal} {\prl}\ }\textbf {\bibinfo {volume} {107}},\ \bibinfo
  {eid} {231102} (\bibinfo {year} {2011})},\ \Eprint
  {https://arxiv.org/abs/1108.2009} {arXiv:1108.2009 [gr-qc]} \BibitemShut
  {NoStop}%
\bibitem [{\citenamefont {{Fuller}}\ and\ \citenamefont
  {{Ma}}(2019)}]{Fuller2019}%
  \BibitemOpen
  \bibfield  {author} {\bibinfo {author} {\bibfnamefont {J.}~\bibnamefont
  {{Fuller}}}\ and\ \bibinfo {author} {\bibfnamefont {L.}~\bibnamefont
  {{Ma}}},\ }\bibfield  {title} {\bibinfo {title} {{Most Black Holes Are Born
  Very Slowly Rotating}},\ }\href {https://doi.org/10.3847/2041-8213/ab339b}
  {\bibfield  {journal} {\bibinfo  {journal} {\apjl}\ }\textbf {\bibinfo
  {volume} {881}},\ \bibinfo {eid} {L1} (\bibinfo {year} {2019})},\ \Eprint
  {https://arxiv.org/abs/1907.03714} {arXiv:1907.03714 [astro-ph.SR]}
  \BibitemShut {NoStop}%
\bibitem [{\citenamefont {Gerosa}\ and\ \citenamefont
  {Berti}(2017)}]{Gerosa2017}%
  \BibitemOpen
  \bibfield  {author} {\bibinfo {author} {\bibfnamefont {D.}~\bibnamefont
  {Gerosa}}\ and\ \bibinfo {author} {\bibfnamefont {E.}~\bibnamefont {Berti}},\
  }\bibfield  {title} {\bibinfo {title} {Are merging black holes born from
  stellar collapse or previous mergers?},\ }\href
  {https://doi.org/10.1103/PhysRevD.95.124046} {\bibfield  {journal} {\bibinfo
  {journal} {Physical Review D}\ }\textbf {\bibinfo {volume} {95}},\ \bibinfo
  {pages} {124046} (\bibinfo {year} {2017})},\ \Eprint
  {https://arxiv.org/abs/1703.06223} {arxiv:1703.06223} \BibitemShut {NoStop}%
\bibitem [{\citenamefont {{Rodriguez}}\ \emph {et~al.}(2019)\citenamefont
  {{Rodriguez}}, \citenamefont {{Zevin}}, \citenamefont {{Amaro-Seoane}},
  \citenamefont {{Chatterjee}}, \citenamefont {{Kremer}}, \citenamefont
  {{Rasio}},\ and\ \citenamefont {{Ye}}}]{Rodriguez2019}%
  \BibitemOpen
  \bibfield  {author} {\bibinfo {author} {\bibfnamefont {C.~L.}\ \bibnamefont
  {{Rodriguez}}}, \bibinfo {author} {\bibfnamefont {M.}~\bibnamefont
  {{Zevin}}}, \bibinfo {author} {\bibfnamefont {P.}~\bibnamefont
  {{Amaro-Seoane}}}, \bibinfo {author} {\bibfnamefont {S.}~\bibnamefont
  {{Chatterjee}}}, \bibinfo {author} {\bibfnamefont {K.}~\bibnamefont
  {{Kremer}}}, \bibinfo {author} {\bibfnamefont {F.~A.}\ \bibnamefont
  {{Rasio}}},\ and\ \bibinfo {author} {\bibfnamefont {C.~S.}\ \bibnamefont
  {{Ye}}},\ }\bibfield  {title} {\bibinfo {title} {{Black holes: The next
  generation{\textemdash}repeated mergers in dense star clusters and their
  gravitational-wave properties}},\ }\href
  {https://doi.org/10.1103/PhysRevD.100.043027} {\bibfield  {journal} {\bibinfo
   {journal} {\prd}\ }\textbf {\bibinfo {volume} {100}},\ \bibinfo {eid}
  {043027} (\bibinfo {year} {2019})},\ \Eprint
  {https://arxiv.org/abs/1906.10260} {arXiv:1906.10260 [astro-ph.HE]}
  \BibitemShut {NoStop}%
\bibitem [{\citenamefont {{Lauer}}\ \emph {et~al.}(1998)\citenamefont
  {{Lauer}}, \citenamefont {{Faber}}, \citenamefont {{Ajhar}}, \citenamefont
  {{Grillmair}},\ and\ \citenamefont {{Scowen}}}]{Lauer1998}%
  \BibitemOpen
  \bibfield  {author} {\bibinfo {author} {\bibfnamefont {T.~R.}\ \bibnamefont
  {{Lauer}}}, \bibinfo {author} {\bibfnamefont {S.~M.}\ \bibnamefont
  {{Faber}}}, \bibinfo {author} {\bibfnamefont {E.~A.}\ \bibnamefont
  {{Ajhar}}}, \bibinfo {author} {\bibfnamefont {C.~J.}\ \bibnamefont
  {{Grillmair}}},\ and\ \bibinfo {author} {\bibfnamefont {P.~A.}\ \bibnamefont
  {{Scowen}}},\ }\bibfield  {title} {\bibinfo {title} {{M32 +/- 1}},\ }\href
  {https://doi.org/10.1086/300617} {\bibfield  {journal} {\bibinfo  {journal}
  {\aj}\ }\textbf {\bibinfo {volume} {116}},\ \bibinfo {pages} {2263} (\bibinfo
  {year} {1998})},\ \Eprint {https://arxiv.org/abs/astro-ph/9806277}
  {arXiv:astro-ph/9806277 [astro-ph]} \BibitemShut {NoStop}%
\bibitem [{\citenamefont {{Sch{\"o}del}}\ \emph {et~al.}(2018)\citenamefont
  {{Sch{\"o}del}}, \citenamefont {{Gallego-Cano}}, \citenamefont {{Dong}},
  \citenamefont {{Nogueras-Lara}}, \citenamefont {{Gallego-Calvente}},
  \citenamefont {{Amaro-Seoane}},\ and\ \citenamefont
  {{Baumgardt}}}]{Schodel2018}%
  \BibitemOpen
  \bibfield  {author} {\bibinfo {author} {\bibfnamefont {R.}~\bibnamefont
  {{Sch{\"o}del}}}, \bibinfo {author} {\bibfnamefont {E.}~\bibnamefont
  {{Gallego-Cano}}}, \bibinfo {author} {\bibfnamefont {H.}~\bibnamefont
  {{Dong}}}, \bibinfo {author} {\bibfnamefont {F.}~\bibnamefont
  {{Nogueras-Lara}}}, \bibinfo {author} {\bibfnamefont {A.~T.}\ \bibnamefont
  {{Gallego-Calvente}}}, \bibinfo {author} {\bibfnamefont {P.}~\bibnamefont
  {{Amaro-Seoane}}},\ and\ \bibinfo {author} {\bibfnamefont {H.}~\bibnamefont
  {{Baumgardt}}},\ }\bibfield  {title} {\bibinfo {title} {{The distribution of
  stars around the Milky Way's central black hole. II. Diffuse light from
  sub-giants and dwarfs}},\ }\href
  {https://doi.org/10.1051/0004-6361/201730452} {\bibfield  {journal} {\bibinfo
   {journal} {\aap}\ }\textbf {\bibinfo {volume} {609}},\ \bibinfo {eid} {A27}
  (\bibinfo {year} {2018})},\ \Eprint {https://arxiv.org/abs/1701.03817}
  {arXiv:1701.03817 [astro-ph.GA]} \BibitemShut {NoStop}%
\bibitem [{\citenamefont {{Pechetti}}\ \emph {et~al.}(2020)\citenamefont
  {{Pechetti}}, \citenamefont {{Seth}}, \citenamefont {{Neumayer}},
  \citenamefont {{Georgiev}}, \citenamefont {{Kacharov}},\ and\ \citenamefont
  {{den Brok}}}]{Pechetti2020}%
  \BibitemOpen
  \bibfield  {author} {\bibinfo {author} {\bibfnamefont {R.}~\bibnamefont
  {{Pechetti}}}, \bibinfo {author} {\bibfnamefont {A.}~\bibnamefont {{Seth}}},
  \bibinfo {author} {\bibfnamefont {N.}~\bibnamefont {{Neumayer}}}, \bibinfo
  {author} {\bibfnamefont {I.}~\bibnamefont {{Georgiev}}}, \bibinfo {author}
  {\bibfnamefont {N.}~\bibnamefont {{Kacharov}}},\ and\ \bibinfo {author}
  {\bibfnamefont {M.}~\bibnamefont {{den Brok}}},\ }\bibfield  {title}
  {\bibinfo {title} {{Luminosity Models and Density Profiles for Nuclear Star
  Clusters for a Nearby Volume-limited Sample of 29 Galaxies}},\ }\href
  {https://doi.org/10.3847/1538-4357/abaaa7} {\bibfield  {journal} {\bibinfo
  {journal} {\apj}\ }\textbf {\bibinfo {volume} {900}},\ \bibinfo {eid} {32}
  (\bibinfo {year} {2020})},\ \Eprint {https://arxiv.org/abs/1911.09686}
  {arXiv:1911.09686 [astro-ph.GA]} \BibitemShut {NoStop}%
\bibitem [{\citenamefont {Neumayer}\ \emph {et~al.}(2020)\citenamefont
  {Neumayer}, \citenamefont {Seth},\ and\ \citenamefont
  {Boeker}}]{neumayerNuclearStarClusters2020}%
  \BibitemOpen
  \bibfield  {author} {\bibinfo {author} {\bibfnamefont {N.}~\bibnamefont
  {Neumayer}}, \bibinfo {author} {\bibfnamefont {A.}~\bibnamefont {Seth}},\
  and\ \bibinfo {author} {\bibfnamefont {T.}~\bibnamefont {Boeker}},\
  }\bibfield  {title} {\bibinfo {title} {Nuclear {{Star Clusters}}},\ }\href
  {https://doi.org/10.1007/s00159-020-00125-0} {\bibfield  {journal} {\bibinfo
  {journal} {The Astronomy and Astrophysics Review}\ }\textbf {\bibinfo
  {volume} {28}},\ \bibinfo {pages} {4} (\bibinfo {year} {2020})},\ \Eprint
  {https://arxiv.org/abs/2001.03626} {arxiv:2001.03626} \BibitemShut {NoStop}%
\bibitem [{\citenamefont {{Tremaine}}\ \emph {et~al.}(1975)\citenamefont
  {{Tremaine}}, \citenamefont {{Ostriker}},\ and\ \citenamefont
  {{Spitzer}}}]{Tremaine1975}%
  \BibitemOpen
  \bibfield  {author} {\bibinfo {author} {\bibfnamefont {S.~D.}\ \bibnamefont
  {{Tremaine}}}, \bibinfo {author} {\bibfnamefont {J.~P.}\ \bibnamefont
  {{Ostriker}}},\ and\ \bibinfo {author} {\bibfnamefont {J.}~\bibnamefont
  {{Spitzer}}, \bibfnamefont {L.}},\ }\bibfield  {title} {\bibinfo {title}
  {{The formation of the nuclei of galaxies. I. M31.}},\ }\href
  {https://doi.org/10.1086/153422} {\bibfield  {journal} {\bibinfo  {journal}
  {\apj}\ }\textbf {\bibinfo {volume} {196}},\ \bibinfo {pages} {407} (\bibinfo
  {year} {1975})}\BibitemShut {NoStop}%
\bibitem [{\citenamefont {{Loose}}\ \emph {et~al.}(1982)\citenamefont
  {{Loose}}, \citenamefont {{Kruegel}},\ and\ \citenamefont
  {{Tutukov}}}]{Loose1982}%
  \BibitemOpen
  \bibfield  {author} {\bibinfo {author} {\bibfnamefont {H.~H.}\ \bibnamefont
  {{Loose}}}, \bibinfo {author} {\bibfnamefont {E.}~\bibnamefont {{Kruegel}}},\
  and\ \bibinfo {author} {\bibfnamefont {A.}~\bibnamefont {{Tutukov}}},\
  }\bibfield  {title} {\bibinfo {title} {{Bursts of star formation in the
  galactic centre}},\ }\href@noop {} {\bibfield  {journal} {\bibinfo  {journal}
  {\aap}\ }\textbf {\bibinfo {volume} {105}},\ \bibinfo {pages} {342} (\bibinfo
  {year} {1982})}\BibitemShut {NoStop}%
\bibitem [{\citenamefont {Bahcall}\ and\ \citenamefont
  {Wolf}(1976)}]{Bahcall1976}%
  \BibitemOpen
  \bibfield  {author} {\bibinfo {author} {\bibfnamefont {J.~N.}\ \bibnamefont
  {Bahcall}}\ and\ \bibinfo {author} {\bibfnamefont {R.~A.}\ \bibnamefont
  {Wolf}},\ }\bibfield  {title} {\bibinfo {title} {Star distribution around a
  massive black hole in a globular cluster},\ }\href
  {https://doi.org/10.1086/154711} {\bibfield  {journal} {\bibinfo  {journal}
  {The Astrophysical Journal}\ }\textbf {\bibinfo {volume} {209}},\ \bibinfo
  {pages} {214} (\bibinfo {year} {1976})}\BibitemShut {NoStop}%
\bibitem [{\citenamefont {Bahcall}\ and\ \citenamefont
  {Wolf}(1977)}]{Bahcall1977}%
  \BibitemOpen
  \bibfield  {author} {\bibinfo {author} {\bibfnamefont {J.~N.}\ \bibnamefont
  {Bahcall}}\ and\ \bibinfo {author} {\bibfnamefont {R.~A.}\ \bibnamefont
  {Wolf}},\ }\bibfield  {title} {\bibinfo {title} {The star distribution around
  a massive black hole in a globular cluster. {{II Unequal}} star masses},\
  }\href {https://doi.org/10.1086/155534} {\bibfield  {journal} {\bibinfo
  {journal} {The Astrophysical Journal}\ }\textbf {\bibinfo {volume} {216}},\
  \bibinfo {pages} {883} (\bibinfo {year} {1977})}\BibitemShut {NoStop}%
\bibitem [{\citenamefont {{Gebhardt}}\ \emph {et~al.}(2001)\citenamefont
  {{Gebhardt}}, \citenamefont {{Lauer}}, \citenamefont {{Kormendy}},
  \citenamefont {{Pinkney}}, \citenamefont {{Bower}}, \citenamefont {{Green}},
  \citenamefont {{Gull}}, \citenamefont {{Hutchings}}, \citenamefont
  {{Kaiser}}, \citenamefont {{Nelson}}, \citenamefont {{Richstone}},\ and\
  \citenamefont {{Weistrop}}}]{Gebhardt2001}%
  \BibitemOpen
  \bibfield  {author} {\bibinfo {author} {\bibfnamefont {K.}~\bibnamefont
  {{Gebhardt}}}, \bibinfo {author} {\bibfnamefont {T.~R.}\ \bibnamefont
  {{Lauer}}}, \bibinfo {author} {\bibfnamefont {J.}~\bibnamefont {{Kormendy}}},
  \bibinfo {author} {\bibfnamefont {J.}~\bibnamefont {{Pinkney}}}, \bibinfo
  {author} {\bibfnamefont {G.~A.}\ \bibnamefont {{Bower}}}, \bibinfo {author}
  {\bibfnamefont {R.}~\bibnamefont {{Green}}}, \bibinfo {author} {\bibfnamefont
  {T.}~\bibnamefont {{Gull}}}, \bibinfo {author} {\bibfnamefont {J.~B.}\
  \bibnamefont {{Hutchings}}}, \bibinfo {author} {\bibfnamefont {M.~E.}\
  \bibnamefont {{Kaiser}}}, \bibinfo {author} {\bibfnamefont {C.~H.}\
  \bibnamefont {{Nelson}}}, \bibinfo {author} {\bibfnamefont {D.}~\bibnamefont
  {{Richstone}}},\ and\ \bibinfo {author} {\bibfnamefont {D.}~\bibnamefont
  {{Weistrop}}},\ }\bibfield  {title} {\bibinfo {title} {{M33: A Galaxy with No
  Supermassive Black Hole}},\ }\href {https://doi.org/10.1086/323481}
  {\bibfield  {journal} {\bibinfo  {journal} {\aj}\ }\textbf {\bibinfo {volume}
  {122}},\ \bibinfo {pages} {2469} (\bibinfo {year} {2001})},\ \Eprint
  {https://arxiv.org/abs/astro-ph/0107135} {arXiv:astro-ph/0107135 [astro-ph]}
  \BibitemShut {NoStop}%
\bibitem [{\citenamefont {{Mapelli}}\ \emph {et~al.}(2021)\citenamefont
  {{Mapelli}}, \citenamefont {{Dall'Amico}}, \citenamefont {{Bouffanais}},
  \citenamefont {{Giacobbo}}, \citenamefont {{Arca Sedda}}, \citenamefont
  {{Artale}}, \citenamefont {{Ballone}}, \citenamefont {{Di Carlo}},
  \citenamefont {{Iorio}}, \citenamefont {{Santoliquido}},\ and\ \citenamefont
  {{Torniamenti}}}]{Mapelli2021}%
  \BibitemOpen
  \bibfield  {author} {\bibinfo {author} {\bibfnamefont {M.}~\bibnamefont
  {{Mapelli}}}, \bibinfo {author} {\bibfnamefont {M.}~\bibnamefont
  {{Dall'Amico}}}, \bibinfo {author} {\bibfnamefont {Y.}~\bibnamefont
  {{Bouffanais}}}, \bibinfo {author} {\bibfnamefont {N.}~\bibnamefont
  {{Giacobbo}}}, \bibinfo {author} {\bibfnamefont {M.}~\bibnamefont {{Arca
  Sedda}}}, \bibinfo {author} {\bibfnamefont {M.~C.}\ \bibnamefont {{Artale}}},
  \bibinfo {author} {\bibfnamefont {A.}~\bibnamefont {{Ballone}}}, \bibinfo
  {author} {\bibfnamefont {U.~N.}\ \bibnamefont {{Di Carlo}}}, \bibinfo
  {author} {\bibfnamefont {G.}~\bibnamefont {{Iorio}}}, \bibinfo {author}
  {\bibfnamefont {F.}~\bibnamefont {{Santoliquido}}},\ and\ \bibinfo {author}
  {\bibfnamefont {S.}~\bibnamefont {{Torniamenti}}},\ }\bibfield  {title}
  {\bibinfo {title} {{Hierarchical black hole mergers in young, globular and
  nuclear star clusters: the effect of metallicity, spin and cluster
  properties}},\ }\href {https://doi.org/10.1093/mnras/stab1334} {\bibfield
  {journal} {\bibinfo  {journal} {\mnras}\ }\textbf {\bibinfo {volume} {505}},\
  \bibinfo {pages} {339} (\bibinfo {year} {2021})},\ \Eprint
  {https://arxiv.org/abs/2103.05016} {arXiv:2103.05016 [astro-ph.HE]}
  \BibitemShut {NoStop}%
\bibitem [{\citenamefont {{Fragione}}\ and\ \citenamefont
  {{Silk}}(2020)}]{Fragione2020}%
  \BibitemOpen
  \bibfield  {author} {\bibinfo {author} {\bibfnamefont {G.}~\bibnamefont
  {{Fragione}}}\ and\ \bibinfo {author} {\bibfnamefont {J.}~\bibnamefont
  {{Silk}}},\ }\bibfield  {title} {\bibinfo {title} {{Repeated mergers and
  ejection of black holes within nuclear star clusters}},\ }\href
  {https://doi.org/10.1093/mnras/staa2629} {\bibfield  {journal} {\bibinfo
  {journal} {\mnras}\ }\textbf {\bibinfo {volume} {498}},\ \bibinfo {pages}
  {4591} (\bibinfo {year} {2020})},\ \Eprint {https://arxiv.org/abs/2006.01867}
  {arXiv:2006.01867 [astro-ph.GA]} \BibitemShut {NoStop}%
\bibitem [{\citenamefont {{Miller}}\ and\ \citenamefont
  {{Lauburg}}(2009)}]{Miller2009}%
  \BibitemOpen
  \bibfield  {author} {\bibinfo {author} {\bibfnamefont {M.~C.}\ \bibnamefont
  {{Miller}}}\ and\ \bibinfo {author} {\bibfnamefont {V.~M.}\ \bibnamefont
  {{Lauburg}}},\ }\bibfield  {title} {\bibinfo {title} {{Mergers of
  Stellar-Mass Black Holes in Nuclear Star Clusters}},\ }\href
  {https://doi.org/10.1088/0004-637X/692/1/917} {\bibfield  {journal} {\bibinfo
   {journal} {\apj}\ }\textbf {\bibinfo {volume} {692}},\ \bibinfo {pages}
  {917} (\bibinfo {year} {2009})},\ \Eprint {https://arxiv.org/abs/0804.2783}
  {arXiv:0804.2783 [astro-ph]} \BibitemShut {NoStop}%
\bibitem [{\citenamefont {{Leigh}}\ \emph {et~al.}(2018)\citenamefont
  {{Leigh}}, \citenamefont {{Geller}}, \citenamefont {{McKernan}},
  \citenamefont {{Ford}}, \citenamefont {{Mac Low}}, \citenamefont
  {{Bellovary}}, \citenamefont {{Haiman}}, \citenamefont {{Lyra}},
  \citenamefont {{Samsing}}, \citenamefont {{O'Dowd}}, \citenamefont
  {{Kocsis}},\ and\ \citenamefont {{Endlich}}}]{Leigh2018}%
  \BibitemOpen
  \bibfield  {author} {\bibinfo {author} {\bibfnamefont {N.~W.~C.}\
  \bibnamefont {{Leigh}}}, \bibinfo {author} {\bibfnamefont {A.~M.}\
  \bibnamefont {{Geller}}}, \bibinfo {author} {\bibfnamefont {B.}~\bibnamefont
  {{McKernan}}}, \bibinfo {author} {\bibfnamefont {K.~E.~S.}\ \bibnamefont
  {{Ford}}}, \bibinfo {author} {\bibfnamefont {M.~M.}\ \bibnamefont {{Mac
  Low}}}, \bibinfo {author} {\bibfnamefont {J.}~\bibnamefont {{Bellovary}}},
  \bibinfo {author} {\bibfnamefont {Z.}~\bibnamefont {{Haiman}}}, \bibinfo
  {author} {\bibfnamefont {W.}~\bibnamefont {{Lyra}}}, \bibinfo {author}
  {\bibfnamefont {J.}~\bibnamefont {{Samsing}}}, \bibinfo {author}
  {\bibfnamefont {M.}~\bibnamefont {{O'Dowd}}}, \bibinfo {author}
  {\bibfnamefont {B.}~\bibnamefont {{Kocsis}}},\ and\ \bibinfo {author}
  {\bibfnamefont {S.}~\bibnamefont {{Endlich}}},\ }\bibfield  {title} {\bibinfo
  {title} {{On the rate of black hole binary mergers in galactic nuclei due to
  dynamical hardening}},\ }\href {https://doi.org/10.1093/mnras/stx3134}
  {\bibfield  {journal} {\bibinfo  {journal} {\mnras}\ }\textbf {\bibinfo
  {volume} {474}},\ \bibinfo {pages} {5672} (\bibinfo {year} {2018})},\ \Eprint
  {https://arxiv.org/abs/1711.10494} {arXiv:1711.10494 [astro-ph.GA]}
  \BibitemShut {NoStop}%
\bibitem [{\citenamefont {{Codazzo}}\ \emph {et~al.}(2023)\citenamefont
  {{Codazzo}}, \citenamefont {{Di Giovanni}}, \citenamefont {{Harms}},
  \citenamefont {{Dall'Amico}},\ and\ \citenamefont {{Mapelli}}}]{Codazzo2023}%
  \BibitemOpen
  \bibfield  {author} {\bibinfo {author} {\bibfnamefont {E.}~\bibnamefont
  {{Codazzo}}}, \bibinfo {author} {\bibfnamefont {M.}~\bibnamefont {{Di
  Giovanni}}}, \bibinfo {author} {\bibfnamefont {J.}~\bibnamefont {{Harms}}},
  \bibinfo {author} {\bibfnamefont {M.}~\bibnamefont {{Dall'Amico}}},\ and\
  \bibinfo {author} {\bibfnamefont {M.}~\bibnamefont {{Mapelli}}},\ }\bibfield
  {title} {\bibinfo {title} {{Study on the detectability of gravitational
  radiation from single-binary encounters between black holes in nuclear star
  clusters: The case of hyperbolic flybys}},\ }\href
  {https://doi.org/10.1103/PhysRevD.107.023023} {\bibfield  {journal} {\bibinfo
   {journal} {\prd}\ }\textbf {\bibinfo {volume} {107}},\ \bibinfo {eid}
  {023023} (\bibinfo {year} {2023})},\ \Eprint
  {https://arxiv.org/abs/2207.01326} {arXiv:2207.01326 [astro-ph.HE]}
  \BibitemShut {NoStop}%
\bibitem [{\citenamefont {{Antonini}}\ and\ \citenamefont
  {{Rasio}}(2016)}]{Antonini2016a}%
  \BibitemOpen
  \bibfield  {author} {\bibinfo {author} {\bibfnamefont {F.}~\bibnamefont
  {{Antonini}}}\ and\ \bibinfo {author} {\bibfnamefont {F.~A.}\ \bibnamefont
  {{Rasio}}},\ }\bibfield  {title} {\bibinfo {title} {{Merging Black Hole
  Binaries in Galactic Nuclei: Implications for Advanced-LIGO Detections}},\
  }\href {https://doi.org/10.3847/0004-637X/831/2/187} {\bibfield  {journal}
  {\bibinfo  {journal} {\apj}\ }\textbf {\bibinfo {volume} {831}},\ \bibinfo
  {eid} {187} (\bibinfo {year} {2016})},\ \Eprint
  {https://arxiv.org/abs/1606.04889} {arXiv:1606.04889 [astro-ph.HE]}
  \BibitemShut {NoStop}%
\bibitem [{\citenamefont {{Fragione}}\ \emph {et~al.}(2022)\citenamefont
  {{Fragione}}, \citenamefont {{Loeb}}, \citenamefont {{Kocsis}},\ and\
  \citenamefont {{Rasio}}}]{Fragione2022}%
  \BibitemOpen
  \bibfield  {author} {\bibinfo {author} {\bibfnamefont {G.}~\bibnamefont
  {{Fragione}}}, \bibinfo {author} {\bibfnamefont {A.}~\bibnamefont {{Loeb}}},
  \bibinfo {author} {\bibfnamefont {B.}~\bibnamefont {{Kocsis}}},\ and\
  \bibinfo {author} {\bibfnamefont {F.~A.}\ \bibnamefont {{Rasio}}},\
  }\bibfield  {title} {\bibinfo {title} {{Merger Rates of Intermediate-mass
  Black Hole Binaries in Nuclear Star Clusters}},\ }\href
  {https://doi.org/10.3847/1538-4357/ac75d0} {\bibfield  {journal} {\bibinfo
  {journal} {\apj}\ }\textbf {\bibinfo {volume} {933}},\ \bibinfo {eid} {170}
  (\bibinfo {year} {2022})},\ \Eprint {https://arxiv.org/abs/2204.03745}
  {arXiv:2204.03745 [astro-ph.HE]} \BibitemShut {NoStop}%
\bibitem [{\citenamefont {{Fragione}}\ and\ \citenamefont
  {{Rasio}}(2023)}]{Fragione2023}%
  \BibitemOpen
  \bibfield  {author} {\bibinfo {author} {\bibfnamefont {G.}~\bibnamefont
  {{Fragione}}}\ and\ \bibinfo {author} {\bibfnamefont {F.~A.}\ \bibnamefont
  {{Rasio}}},\ }\bibfield  {title} {\bibinfo {title} {{Demographics of
  Hierarchical Black Hole Mergers in Dense Star Clusters}},\ }\href
  {https://doi.org/10.3847/1538-4357/acd9c9} {\bibfield  {journal} {\bibinfo
  {journal} {\apj}\ }\textbf {\bibinfo {volume} {951}},\ \bibinfo {eid} {129}
  (\bibinfo {year} {2023})},\ \Eprint {https://arxiv.org/abs/2302.11613}
  {arXiv:2302.11613 [astro-ph.GA]} \BibitemShut {NoStop}%
\bibitem [{\citenamefont {{Chattopadhyay}}\ \emph {et~al.}(2023)\citenamefont
  {{Chattopadhyay}}, \citenamefont {{Stegmann}}, \citenamefont {{Antonini}},
  \citenamefont {{Barber}},\ and\ \citenamefont
  {{Romero-Shaw}}}]{Chattopadhyay2023}%
  \BibitemOpen
  \bibfield  {author} {\bibinfo {author} {\bibfnamefont {D.}~\bibnamefont
  {{Chattopadhyay}}}, \bibinfo {author} {\bibfnamefont {J.}~\bibnamefont
  {{Stegmann}}}, \bibinfo {author} {\bibfnamefont {F.}~\bibnamefont
  {{Antonini}}}, \bibinfo {author} {\bibfnamefont {J.}~\bibnamefont
  {{Barber}}},\ and\ \bibinfo {author} {\bibfnamefont {I.~M.}\ \bibnamefont
  {{Romero-Shaw}}},\ }\bibfield  {title} {\bibinfo {title} {{Binary black hole
  mergers in nuclear star clusters: eccentricities, spins, masses, and the
  growth of massive seeds}},\ }\href
  {https://doi.org/10.48550/arXiv.2308.10884} {\bibfield  {journal} {\bibinfo
  {journal} {arXiv e-prints}\ ,\ \bibinfo {eid} {arXiv:2308.10884}} (\bibinfo
  {year} {2023})},\ \Eprint {https://arxiv.org/abs/2308.10884}
  {arXiv:2308.10884 [astro-ph.HE]} \BibitemShut {NoStop}%
\bibitem [{\citenamefont {{Hut}}(1985)}]{Hut1985}%
  \BibitemOpen
  \bibfield  {author} {\bibinfo {author} {\bibfnamefont {P.}~\bibnamefont
  {{Hut}}},\ }\bibfield  {title} {\bibinfo {title} {{Binary formation and
  interactions with field stars.}},\ }in\ \href@noop {} {\emph {\bibinfo
  {booktitle} {Dynamics of Star Clusters}}},\ Vol.\ \bibinfo {volume} {113},\
  \bibinfo {editor} {edited by\ \bibinfo {editor} {\bibfnamefont
  {J.}~\bibnamefont {{Goodman}}}\ and\ \bibinfo {editor} {\bibfnamefont
  {P.}~\bibnamefont {{Hut}}}}\ (\bibinfo {year} {1985})\ pp.\ \bibinfo {pages}
  {231--247}\BibitemShut {NoStop}%
\bibitem [{\citenamefont {{Giersz}}(1985)}]{Giersz1985}%
  \BibitemOpen
  \bibfield  {author} {\bibinfo {author} {\bibfnamefont {M.}~\bibnamefont
  {{Giersz}}},\ }\bibfield  {title} {\bibinfo {title} {{Two-body capture in
  large N body systems. I}},\ }\href@noop {} {\bibfield  {journal} {\bibinfo
  {journal} {\actaa}\ }\textbf {\bibinfo {volume} {35}},\ \bibinfo {pages}
  {119} (\bibinfo {year} {1985})}\BibitemShut {NoStop}%
\bibitem [{\citenamefont {{Quinlan}}\ and\ \citenamefont
  {{Shapiro}}(1987)}]{Quinlan1987}%
  \BibitemOpen
  \bibfield  {author} {\bibinfo {author} {\bibfnamefont {G.~D.}\ \bibnamefont
  {{Quinlan}}}\ and\ \bibinfo {author} {\bibfnamefont {S.~L.}\ \bibnamefont
  {{Shapiro}}},\ }\bibfield  {title} {\bibinfo {title} {{The Collapse of Dense
  Star Clusters to Supermassive Black Holes: Binaries and Gravitational
  Radiation}},\ }\href {https://doi.org/10.1086/165624} {\bibfield  {journal}
  {\bibinfo  {journal} {\apj}\ }\textbf {\bibinfo {volume} {321}},\ \bibinfo
  {pages} {199} (\bibinfo {year} {1987})}\BibitemShut {NoStop}%
\bibitem [{\citenamefont {{Quinlan}}\ and\ \citenamefont
  {{Shapiro}}(1989)}]{Quinlan1989}%
  \BibitemOpen
  \bibfield  {author} {\bibinfo {author} {\bibfnamefont {G.~D.}\ \bibnamefont
  {{Quinlan}}}\ and\ \bibinfo {author} {\bibfnamefont {S.~L.}\ \bibnamefont
  {{Shapiro}}},\ }\bibfield  {title} {\bibinfo {title} {{Dynamical Evolution of
  Dense Clusters of Compact Stars}},\ }\href {https://doi.org/10.1086/167745}
  {\bibfield  {journal} {\bibinfo  {journal} {\apj}\ }\textbf {\bibinfo
  {volume} {343}},\ \bibinfo {pages} {725} (\bibinfo {year}
  {1989})}\BibitemShut {NoStop}%
\bibitem [{\citenamefont {{Hopman}}\ and\ \citenamefont
  {{Alexander}}(2006)}]{Hopman2006}%
  \BibitemOpen
  \bibfield  {author} {\bibinfo {author} {\bibfnamefont {C.}~\bibnamefont
  {{Hopman}}}\ and\ \bibinfo {author} {\bibfnamefont {T.}~\bibnamefont
  {{Alexander}}},\ }\bibfield  {title} {\bibinfo {title} {{The Effect of Mass
  Segregation on Gravitational Wave Sources near Massive Black Holes}},\ }\href
  {https://doi.org/10.1086/506273} {\bibfield  {journal} {\bibinfo  {journal}
  {\apjl}\ }\textbf {\bibinfo {volume} {645}},\ \bibinfo {pages} {L133}
  (\bibinfo {year} {2006})},\ \Eprint {https://arxiv.org/abs/astro-ph/0603324}
  {arXiv:astro-ph/0603324 [astro-ph]} \BibitemShut {NoStop}%
\bibitem [{\citenamefont {{Alexander}}\ and\ \citenamefont
  {{Hopman}}(2009)}]{Alexander2009}%
  \BibitemOpen
  \bibfield  {author} {\bibinfo {author} {\bibfnamefont {T.}~\bibnamefont
  {{Alexander}}}\ and\ \bibinfo {author} {\bibfnamefont {C.}~\bibnamefont
  {{Hopman}}},\ }\bibfield  {title} {\bibinfo {title} {{Strong Mass Segregation
  Around a Massive Black Hole}},\ }\href
  {https://doi.org/10.1088/0004-637X/697/2/1861} {\bibfield  {journal}
  {\bibinfo  {journal} {\apj}\ }\textbf {\bibinfo {volume} {697}},\ \bibinfo
  {pages} {1861} (\bibinfo {year} {2009})},\ \Eprint
  {https://arxiv.org/abs/0808.3150} {arXiv:0808.3150 [astro-ph]} \BibitemShut
  {NoStop}%
\bibitem [{\citenamefont {{O'Leary}}\ \emph {et~al.}(2009)\citenamefont
  {{O'Leary}}, \citenamefont {{Kocsis}},\ and\ \citenamefont
  {{Loeb}}}]{OLeary2009}%
  \BibitemOpen
  \bibfield  {author} {\bibinfo {author} {\bibfnamefont {R.~M.}\ \bibnamefont
  {{O'Leary}}}, \bibinfo {author} {\bibfnamefont {B.}~\bibnamefont
  {{Kocsis}}},\ and\ \bibinfo {author} {\bibfnamefont {A.}~\bibnamefont
  {{Loeb}}},\ }\bibfield  {title} {\bibinfo {title} {{Gravitational waves from
  scattering of stellar-mass black holes in galactic nuclei}},\ }\href
  {https://doi.org/10.1111/j.1365-2966.2009.14653.x} {\bibfield  {journal}
  {\bibinfo  {journal} {\mnras}\ }\textbf {\bibinfo {volume} {395}},\ \bibinfo
  {pages} {2127} (\bibinfo {year} {2009})},\ \Eprint
  {https://arxiv.org/abs/0807.2638} {arXiv:0807.2638 [astro-ph]} \BibitemShut
  {NoStop}%
\bibitem [{\citenamefont {{Gond{\'a}n}}\ \emph {et~al.}(2018)\citenamefont
  {{Gond{\'a}n}}, \citenamefont {{Kocsis}}, \citenamefont {{Raffai}},\ and\
  \citenamefont {{Frei}}}]{Gondan2018}%
  \BibitemOpen
  \bibfield  {author} {\bibinfo {author} {\bibfnamefont {L.}~\bibnamefont
  {{Gond{\'a}n}}}, \bibinfo {author} {\bibfnamefont {B.}~\bibnamefont
  {{Kocsis}}}, \bibinfo {author} {\bibfnamefont {P.}~\bibnamefont {{Raffai}}},\
  and\ \bibinfo {author} {\bibfnamefont {Z.}~\bibnamefont {{Frei}}},\
  }\bibfield  {title} {\bibinfo {title} {{Eccentric Black Hole
  Gravitational-wave Capture Sources in Galactic Nuclei: Distribution of Binary
  Parameters}},\ }\href {https://doi.org/10.3847/1538-4357/aabfee} {\bibfield
  {journal} {\bibinfo  {journal} {\apj}\ }\textbf {\bibinfo {volume} {860}},\
  \bibinfo {eid} {5} (\bibinfo {year} {2018})},\ \Eprint
  {https://arxiv.org/abs/1711.09989} {arXiv:1711.09989 [astro-ph.HE]}
  \BibitemShut {NoStop}%
\bibitem [{\citenamefont {{Gond{\'a}n}}(2023)}]{Gondan2023}%
  \BibitemOpen
  \bibfield  {author} {\bibinfo {author} {\bibfnamefont {L.}~\bibnamefont
  {{Gond{\'a}n}}},\ }\bibfield  {title} {\bibinfo {title} {{Parameter
  distributions of binary black hole mergers near supermassive black holes as
  seen by advanced gravitational wave detectors}},\ }\href
  {https://doi.org/10.1093/mnras/stac3612} {\bibfield  {journal} {\bibinfo
  {journal} {\mnras}\ }\textbf {\bibinfo {volume} {519}},\ \bibinfo {pages}
  {1856} (\bibinfo {year} {2023})},\ \Eprint {https://arxiv.org/abs/2210.02975}
  {arXiv:2210.02975 [astro-ph.HE]} \BibitemShut {NoStop}%
\bibitem [{\citenamefont {{Artymowicz}}\ \emph {et~al.}(1993)\citenamefont
  {{Artymowicz}}, \citenamefont {{Lin}},\ and\ \citenamefont
  {{Wampler}}}]{Artymowicz1993}%
  \BibitemOpen
  \bibfield  {author} {\bibinfo {author} {\bibfnamefont {P.}~\bibnamefont
  {{Artymowicz}}}, \bibinfo {author} {\bibfnamefont {D.~N.~C.}\ \bibnamefont
  {{Lin}}},\ and\ \bibinfo {author} {\bibfnamefont {E.~J.}\ \bibnamefont
  {{Wampler}}},\ }\bibfield  {title} {\bibinfo {title} {{Star Trapping and
  Metallicity Enrichment in Quasars and Active Galactic Nuclei}},\ }\href
  {https://doi.org/10.1086/172690} {\bibfield  {journal} {\bibinfo  {journal}
  {\apj}\ }\textbf {\bibinfo {volume} {409}},\ \bibinfo {pages} {592} (\bibinfo
  {year} {1993})}\BibitemShut {NoStop}%
\bibitem [{\citenamefont {{Karas}}\ and\ \citenamefont
  {{{\v{S}}ubr}}(2001)}]{Karas2001}%
  \BibitemOpen
  \bibfield  {author} {\bibinfo {author} {\bibfnamefont {V.}~\bibnamefont
  {{Karas}}}\ and\ \bibinfo {author} {\bibfnamefont {L.}~\bibnamefont
  {{{\v{S}}ubr}}},\ }\bibfield  {title} {\bibinfo {title} {{Orbital decay of
  satellites crossing an accretion disc}},\ }\href
  {https://doi.org/10.1051/0004-6361:20011009} {\bibfield  {journal} {\bibinfo
  {journal} {\aap}\ }\textbf {\bibinfo {volume} {376}},\ \bibinfo {pages} {686}
  (\bibinfo {year} {2001})},\ \Eprint {https://arxiv.org/abs/astro-ph/0107232}
  {arXiv:astro-ph/0107232 [astro-ph]} \BibitemShut {NoStop}%
\bibitem [{\citenamefont {{Bartos}}\ \emph {et~al.}(2017)\citenamefont
  {{Bartos}}, \citenamefont {{Kocsis}}, \citenamefont {{Haiman}},\ and\
  \citenamefont {{M{\'a}rka}}}]{Bartos2017}%
  \BibitemOpen
  \bibfield  {author} {\bibinfo {author} {\bibfnamefont {I.}~\bibnamefont
  {{Bartos}}}, \bibinfo {author} {\bibfnamefont {B.}~\bibnamefont {{Kocsis}}},
  \bibinfo {author} {\bibfnamefont {Z.}~\bibnamefont {{Haiman}}},\ and\
  \bibinfo {author} {\bibfnamefont {S.}~\bibnamefont {{M{\'a}rka}}},\
  }\bibfield  {title} {\bibinfo {title} {{Rapid and Bright Stellar-mass Binary
  Black Hole Mergers in Active Galactic Nuclei}},\ }\href
  {https://doi.org/10.3847/1538-4357/835/2/165} {\bibfield  {journal} {\bibinfo
   {journal} {\apj}\ }\textbf {\bibinfo {volume} {835}},\ \bibinfo {eid} {165}
  (\bibinfo {year} {2017})},\ \Eprint {https://arxiv.org/abs/1602.03831}
  {arXiv:1602.03831 [astro-ph.HE]} \BibitemShut {NoStop}%
\bibitem [{\citenamefont {{Stone}}\ \emph {et~al.}(2017)\citenamefont
  {{Stone}}, \citenamefont {{Metzger}},\ and\ \citenamefont
  {{Haiman}}}]{Stone2017}%
  \BibitemOpen
  \bibfield  {author} {\bibinfo {author} {\bibfnamefont {N.~C.}\ \bibnamefont
  {{Stone}}}, \bibinfo {author} {\bibfnamefont {B.~D.}\ \bibnamefont
  {{Metzger}}},\ and\ \bibinfo {author} {\bibfnamefont {Z.}~\bibnamefont
  {{Haiman}}},\ }\bibfield  {title} {\bibinfo {title} {{Assisted inspirals of
  stellar mass black holes embedded in AGN discs: solving the `final au
  problem'}},\ }\href {https://doi.org/10.1093/mnras/stw2260} {\bibfield
  {journal} {\bibinfo  {journal} {\mnras}\ }\textbf {\bibinfo {volume} {464}},\
  \bibinfo {pages} {946} (\bibinfo {year} {2017})},\ \Eprint
  {https://arxiv.org/abs/1602.04226} {arXiv:1602.04226 [astro-ph.GA]}
  \BibitemShut {NoStop}%
\bibitem [{\citenamefont {{McKernan}}\ \emph {et~al.}(2012)\citenamefont
  {{McKernan}}, \citenamefont {{Ford}}, \citenamefont {{Lyra}},\ and\
  \citenamefont {{Perets}}}]{McKernan2012}%
  \BibitemOpen
  \bibfield  {author} {\bibinfo {author} {\bibfnamefont {B.}~\bibnamefont
  {{McKernan}}}, \bibinfo {author} {\bibfnamefont {K.~E.~S.}\ \bibnamefont
  {{Ford}}}, \bibinfo {author} {\bibfnamefont {W.}~\bibnamefont {{Lyra}}},\
  and\ \bibinfo {author} {\bibfnamefont {H.~B.}\ \bibnamefont {{Perets}}},\
  }\bibfield  {title} {\bibinfo {title} {{Intermediate mass black holes in AGN
  discs - I. Production and growth}},\ }\href
  {https://doi.org/10.1111/j.1365-2966.2012.21486.x} {\bibfield  {journal}
  {\bibinfo  {journal} {\mnras}\ }\textbf {\bibinfo {volume} {425}},\ \bibinfo
  {pages} {460} (\bibinfo {year} {2012})},\ \Eprint
  {https://arxiv.org/abs/1206.2309} {arXiv:1206.2309 [astro-ph.GA]}
  \BibitemShut {NoStop}%
\bibitem [{\citenamefont {{McKernan}}\ \emph {et~al.}(2014)\citenamefont
  {{McKernan}}, \citenamefont {{Ford}}, \citenamefont {{Kocsis}}, \citenamefont
  {{Lyra}},\ and\ \citenamefont {{Winter}}}]{McKernan2014}%
  \BibitemOpen
  \bibfield  {author} {\bibinfo {author} {\bibfnamefont {B.}~\bibnamefont
  {{McKernan}}}, \bibinfo {author} {\bibfnamefont {K.~E.~S.}\ \bibnamefont
  {{Ford}}}, \bibinfo {author} {\bibfnamefont {B.}~\bibnamefont {{Kocsis}}},
  \bibinfo {author} {\bibfnamefont {W.}~\bibnamefont {{Lyra}}},\ and\ \bibinfo
  {author} {\bibfnamefont {L.~M.}\ \bibnamefont {{Winter}}},\ }\bibfield
  {title} {\bibinfo {title} {{Intermediate-mass black holes in AGN discs - II.
  Model predictions and observational constraints}},\ }\href
  {https://doi.org/10.1093/mnras/stu553} {\bibfield  {journal} {\bibinfo
  {journal} {\mnras}\ }\textbf {\bibinfo {volume} {441}},\ \bibinfo {pages}
  {900} (\bibinfo {year} {2014})},\ \Eprint {https://arxiv.org/abs/1403.6433}
  {arXiv:1403.6433 [astro-ph.GA]} \BibitemShut {NoStop}%
\bibitem [{\citenamefont {{McKernan}}\ \emph {et~al.}(2018)\citenamefont
  {{McKernan}}, \citenamefont {{Ford}}, \citenamefont {{Bellovary}},
  \citenamefont {{Leigh}}, \citenamefont {{Haiman}}, \citenamefont {{Kocsis}},
  \citenamefont {{Lyra}}, \citenamefont {{Mac Low}}, \citenamefont {{Metzger}},
  \citenamefont {{O'Dowd}}, \citenamefont {{Endlich}},\ and\ \citenamefont
  {{Rosen}}}]{McKernan2018}%
  \BibitemOpen
  \bibfield  {author} {\bibinfo {author} {\bibfnamefont {B.}~\bibnamefont
  {{McKernan}}}, \bibinfo {author} {\bibfnamefont {K.~E.~S.}\ \bibnamefont
  {{Ford}}}, \bibinfo {author} {\bibfnamefont {J.}~\bibnamefont {{Bellovary}}},
  \bibinfo {author} {\bibfnamefont {N.~W.~C.}\ \bibnamefont {{Leigh}}},
  \bibinfo {author} {\bibfnamefont {Z.}~\bibnamefont {{Haiman}}}, \bibinfo
  {author} {\bibfnamefont {B.}~\bibnamefont {{Kocsis}}}, \bibinfo {author}
  {\bibfnamefont {W.}~\bibnamefont {{Lyra}}}, \bibinfo {author} {\bibfnamefont
  {M.~M.}\ \bibnamefont {{Mac Low}}}, \bibinfo {author} {\bibfnamefont
  {B.}~\bibnamefont {{Metzger}}}, \bibinfo {author} {\bibfnamefont
  {M.}~\bibnamefont {{O'Dowd}}}, \bibinfo {author} {\bibfnamefont
  {S.}~\bibnamefont {{Endlich}}},\ and\ \bibinfo {author} {\bibfnamefont
  {D.~J.}\ \bibnamefont {{Rosen}}},\ }\bibfield  {title} {\bibinfo {title}
  {{Constraining Stellar-mass Black Hole Mergers in AGN Disks Detectable with
  LIGO}},\ }\href {https://doi.org/10.3847/1538-4357/aadae5} {\bibfield
  {journal} {\bibinfo  {journal} {\apj}\ }\textbf {\bibinfo {volume} {866}},\
  \bibinfo {eid} {66} (\bibinfo {year} {2018})},\ \Eprint
  {https://arxiv.org/abs/1702.07818} {arXiv:1702.07818 [astro-ph.HE]}
  \BibitemShut {NoStop}%
\bibitem [{\citenamefont {{Tagawa}}\ \emph {et~al.}(2021)\citenamefont
  {{Tagawa}}, \citenamefont {{Kocsis}}, \citenamefont {{Haiman}}, \citenamefont
  {{Bartos}}, \citenamefont {{Omukai}},\ and\ \citenamefont
  {{Samsing}}}]{Tagawa2021}%
  \BibitemOpen
  \bibfield  {author} {\bibinfo {author} {\bibfnamefont {H.}~\bibnamefont
  {{Tagawa}}}, \bibinfo {author} {\bibfnamefont {B.}~\bibnamefont {{Kocsis}}},
  \bibinfo {author} {\bibfnamefont {Z.}~\bibnamefont {{Haiman}}}, \bibinfo
  {author} {\bibfnamefont {I.}~\bibnamefont {{Bartos}}}, \bibinfo {author}
  {\bibfnamefont {K.}~\bibnamefont {{Omukai}}},\ and\ \bibinfo {author}
  {\bibfnamefont {J.}~\bibnamefont {{Samsing}}},\ }\bibfield  {title} {\bibinfo
  {title} {{Eccentric Black Hole Mergers in Active Galactic Nuclei}},\ }\href
  {https://doi.org/10.3847/2041-8213/abd4d3} {\bibfield  {journal} {\bibinfo
  {journal} {\apjl}\ }\textbf {\bibinfo {volume} {907}},\ \bibinfo {eid} {L20}
  (\bibinfo {year} {2021})},\ \Eprint {https://arxiv.org/abs/2010.10526}
  {arXiv:2010.10526 [astro-ph.HE]} \BibitemShut {NoStop}%
\bibitem [{\citenamefont {{Samsing}}\ \emph {et~al.}(2022)\citenamefont
  {{Samsing}}, \citenamefont {{Bartos}}, \citenamefont {{D'Orazio}},
  \citenamefont {{Haiman}}, \citenamefont {{Kocsis}}, \citenamefont {{Leigh}},
  \citenamefont {{Liu}}, \citenamefont {{Pessah}},\ and\ \citenamefont
  {{Tagawa}}}]{Samsing2022}%
  \BibitemOpen
  \bibfield  {author} {\bibinfo {author} {\bibfnamefont {J.}~\bibnamefont
  {{Samsing}}}, \bibinfo {author} {\bibfnamefont {I.}~\bibnamefont {{Bartos}}},
  \bibinfo {author} {\bibfnamefont {D.~J.}\ \bibnamefont {{D'Orazio}}},
  \bibinfo {author} {\bibfnamefont {Z.}~\bibnamefont {{Haiman}}}, \bibinfo
  {author} {\bibfnamefont {B.}~\bibnamefont {{Kocsis}}}, \bibinfo {author}
  {\bibfnamefont {N.~W.~C.}\ \bibnamefont {{Leigh}}}, \bibinfo {author}
  {\bibfnamefont {B.}~\bibnamefont {{Liu}}}, \bibinfo {author} {\bibfnamefont
  {M.~E.}\ \bibnamefont {{Pessah}}},\ and\ \bibinfo {author} {\bibfnamefont
  {H.}~\bibnamefont {{Tagawa}}},\ }\bibfield  {title} {\bibinfo {title} {{AGN
  as potential factories for eccentric black hole mergers}},\ }\href
  {https://doi.org/10.1038/s41586-021-04333-1} {\bibfield  {journal} {\bibinfo
  {journal} {\nat}\ }\textbf {\bibinfo {volume} {603}},\ \bibinfo {pages} {237}
  (\bibinfo {year} {2022})},\ \Eprint {https://arxiv.org/abs/2010.09765}
  {arXiv:2010.09765 [astro-ph.HE]} \BibitemShut {NoStop}%
\bibitem [{\citenamefont {{Mu{\~n}oz}}\ \emph {et~al.}(2022)\citenamefont
  {{Mu{\~n}oz}}, \citenamefont {{Stone}}, \citenamefont {{Petrovich}},\ and\
  \citenamefont {{Rasio}}}]{Munoz2022}%
  \BibitemOpen
  \bibfield  {author} {\bibinfo {author} {\bibfnamefont {D.~J.}\ \bibnamefont
  {{Mu{\~n}oz}}}, \bibinfo {author} {\bibfnamefont {N.~C.}\ \bibnamefont
  {{Stone}}}, \bibinfo {author} {\bibfnamefont {C.}~\bibnamefont
  {{Petrovich}}},\ and\ \bibinfo {author} {\bibfnamefont {F.~A.}\ \bibnamefont
  {{Rasio}}},\ }\bibfield  {title} {\bibinfo {title} {{Eccentric Mergers of
  Intermediate-Mass Black Holes from Evection Resonances in AGN Disks}},\
  }\href {https://doi.org/10.48550/arXiv.2204.06002} {\bibfield  {journal}
  {\bibinfo  {journal} {arXiv e-prints}\ ,\ \bibinfo {eid} {arXiv:2204.06002}}
  (\bibinfo {year} {2022})},\ \Eprint {https://arxiv.org/abs/2204.06002}
  {arXiv:2204.06002 [astro-ph.HE]} \BibitemShut {NoStop}%
\bibitem [{\citenamefont {{Dempsey}}\ \emph {et~al.}(2022)\citenamefont
  {{Dempsey}}, \citenamefont {{Li}}, \citenamefont {{Mishra}},\ and\
  \citenamefont {{Li}}}]{Dempsey2022}%
  \BibitemOpen
  \bibfield  {author} {\bibinfo {author} {\bibfnamefont {A.~M.}\ \bibnamefont
  {{Dempsey}}}, \bibinfo {author} {\bibfnamefont {H.}~\bibnamefont {{Li}}},
  \bibinfo {author} {\bibfnamefont {B.}~\bibnamefont {{Mishra}}},\ and\
  \bibinfo {author} {\bibfnamefont {S.}~\bibnamefont {{Li}}},\ }\bibfield
  {title} {\bibinfo {title} {{Contracting and Expanding Binary Black Holes in
  3D Low-mass AGN Disks: The Importance of Separation}},\ }\href
  {https://doi.org/10.3847/1538-4357/ac9d92} {\bibfield  {journal} {\bibinfo
  {journal} {\apj}\ }\textbf {\bibinfo {volume} {940}},\ \bibinfo {eid} {155}
  (\bibinfo {year} {2022})},\ \Eprint {https://arxiv.org/abs/2203.06534}
  {arXiv:2203.06534 [astro-ph.HE]} \BibitemShut {NoStop}%
\bibitem [{\citenamefont {{Yang}}\ \emph {et~al.}(2019)\citenamefont {{Yang}},
  \citenamefont {{Bartos}}, \citenamefont {{Gayathri}}, \citenamefont {{Ford}},
  \citenamefont {{Haiman}}, \citenamefont {{Klimenko}}, \citenamefont
  {{Kocsis}}, \citenamefont {{M{\'a}rka}}, \citenamefont {{M{\'a}rka}},
  \citenamefont {{McKernan}},\ and\ \citenamefont
  {{O'Shaughnessy}}}]{Yang2019}%
  \BibitemOpen
  \bibfield  {author} {\bibinfo {author} {\bibfnamefont {Y.}~\bibnamefont
  {{Yang}}}, \bibinfo {author} {\bibfnamefont {I.}~\bibnamefont {{Bartos}}},
  \bibinfo {author} {\bibfnamefont {V.}~\bibnamefont {{Gayathri}}}, \bibinfo
  {author} {\bibfnamefont {K.~E.~S.}\ \bibnamefont {{Ford}}}, \bibinfo {author}
  {\bibfnamefont {Z.}~\bibnamefont {{Haiman}}}, \bibinfo {author}
  {\bibfnamefont {S.}~\bibnamefont {{Klimenko}}}, \bibinfo {author}
  {\bibfnamefont {B.}~\bibnamefont {{Kocsis}}}, \bibinfo {author}
  {\bibfnamefont {S.}~\bibnamefont {{M{\'a}rka}}}, \bibinfo {author}
  {\bibfnamefont {Z.}~\bibnamefont {{M{\'a}rka}}}, \bibinfo {author}
  {\bibfnamefont {B.}~\bibnamefont {{McKernan}}},\ and\ \bibinfo {author}
  {\bibfnamefont {R.}~\bibnamefont {{O'Shaughnessy}}},\ }\bibfield  {title}
  {\bibinfo {title} {{Hierarchical Black Hole Mergers in Active Galactic
  Nuclei}},\ }\href {https://doi.org/10.1103/PhysRevLett.123.181101} {\bibfield
   {journal} {\bibinfo  {journal} {\prl}\ }\textbf {\bibinfo {volume} {123}},\
  \bibinfo {eid} {181101} (\bibinfo {year} {2019})},\ \Eprint
  {https://arxiv.org/abs/1906.09281} {arXiv:1906.09281 [astro-ph.HE]}
  \BibitemShut {NoStop}%
\bibitem [{\citenamefont {{Secunda}}\ \emph {et~al.}(2019)\citenamefont
  {{Secunda}}, \citenamefont {{Bellovary}}, \citenamefont {{Mac Low}},
  \citenamefont {{Ford}}, \citenamefont {{McKernan}}, \citenamefont {{Leigh}},
  \citenamefont {{Lyra}},\ and\ \citenamefont {{S{\'a}ndor}}}]{Secunda2019}%
  \BibitemOpen
  \bibfield  {author} {\bibinfo {author} {\bibfnamefont {A.}~\bibnamefont
  {{Secunda}}}, \bibinfo {author} {\bibfnamefont {J.}~\bibnamefont
  {{Bellovary}}}, \bibinfo {author} {\bibfnamefont {M.-M.}\ \bibnamefont {{Mac
  Low}}}, \bibinfo {author} {\bibfnamefont {K.~E.~S.}\ \bibnamefont {{Ford}}},
  \bibinfo {author} {\bibfnamefont {B.}~\bibnamefont {{McKernan}}}, \bibinfo
  {author} {\bibfnamefont {N.~W.~C.}\ \bibnamefont {{Leigh}}}, \bibinfo
  {author} {\bibfnamefont {W.}~\bibnamefont {{Lyra}}},\ and\ \bibinfo {author}
  {\bibfnamefont {Z.}~\bibnamefont {{S{\'a}ndor}}},\ }\bibfield  {title}
  {\bibinfo {title} {{Orbital Migration of Interacting Stellar Mass Black Holes
  in Disks around Supermassive Black Holes}},\ }\href
  {https://doi.org/10.3847/1538-4357/ab20ca} {\bibfield  {journal} {\bibinfo
  {journal} {\apj}\ }\textbf {\bibinfo {volume} {878}},\ \bibinfo {eid} {85}
  (\bibinfo {year} {2019})},\ \Eprint {https://arxiv.org/abs/1807.02859}
  {arXiv:1807.02859 [astro-ph.HE]} \BibitemShut {NoStop}%
\bibitem [{\citenamefont {{Tagawa}}\ \emph {et~al.}(2020)\citenamefont
  {{Tagawa}}, \citenamefont {{Haiman}},\ and\ \citenamefont
  {{Kocsis}}}]{Tagawa2020}%
  \BibitemOpen
  \bibfield  {author} {\bibinfo {author} {\bibfnamefont {H.}~\bibnamefont
  {{Tagawa}}}, \bibinfo {author} {\bibfnamefont {Z.}~\bibnamefont {{Haiman}}},\
  and\ \bibinfo {author} {\bibfnamefont {B.}~\bibnamefont {{Kocsis}}},\
  }\bibfield  {title} {\bibinfo {title} {{Making a Supermassive Star by Stellar
  Bombardment}},\ }\href {https://doi.org/10.3847/1538-4357/ab7922} {\bibfield
  {journal} {\bibinfo  {journal} {\apj}\ }\textbf {\bibinfo {volume} {892}},\
  \bibinfo {eid} {36} (\bibinfo {year} {2020})},\ \Eprint
  {https://arxiv.org/abs/1909.10517} {arXiv:1909.10517 [astro-ph.GA]}
  \BibitemShut {NoStop}%
\bibitem [{\citenamefont {{Zevin}}\ and\ \citenamefont
  {{Holz}}(2022)}]{Zevin2022}%
  \BibitemOpen
  \bibfield  {author} {\bibinfo {author} {\bibfnamefont {M.}~\bibnamefont
  {{Zevin}}}\ and\ \bibinfo {author} {\bibfnamefont {D.~E.}\ \bibnamefont
  {{Holz}}},\ }\bibfield  {title} {\bibinfo {title} {{Avoiding a Cluster
  Catastrophe: Retention Efficiency and the Binary Black Hole Mass Spectrum}},\
  }\href {https://doi.org/10.3847/2041-8213/ac853d} {\bibfield  {journal}
  {\bibinfo  {journal} {\apjl}\ }\textbf {\bibinfo {volume} {935}},\ \bibinfo
  {eid} {L20} (\bibinfo {year} {2022})},\ \Eprint
  {https://arxiv.org/abs/2205.08549} {arXiv:2205.08549 [astro-ph.HE]}
  \BibitemShut {NoStop}%
\bibitem [{\citenamefont {{McKernan}}\ \emph {et~al.}(2019)\citenamefont
  {{McKernan}}, \citenamefont {{Ford}}, \citenamefont {{Bartos}}, \citenamefont
  {{Graham}}, \citenamefont {{Lyra}}, \citenamefont {{Marka}}, \citenamefont
  {{Marka}}, \citenamefont {{Ross}}, \citenamefont {{Stern}},\ and\
  \citenamefont {{Yang}}}]{McKernan2019}%
  \BibitemOpen
  \bibfield  {author} {\bibinfo {author} {\bibfnamefont {B.}~\bibnamefont
  {{McKernan}}}, \bibinfo {author} {\bibfnamefont {K.~E.~S.}\ \bibnamefont
  {{Ford}}}, \bibinfo {author} {\bibfnamefont {I.}~\bibnamefont {{Bartos}}},
  \bibinfo {author} {\bibfnamefont {M.~J.}\ \bibnamefont {{Graham}}}, \bibinfo
  {author} {\bibfnamefont {W.}~\bibnamefont {{Lyra}}}, \bibinfo {author}
  {\bibfnamefont {S.}~\bibnamefont {{Marka}}}, \bibinfo {author} {\bibfnamefont
  {Z.}~\bibnamefont {{Marka}}}, \bibinfo {author} {\bibfnamefont {N.~P.}\
  \bibnamefont {{Ross}}}, \bibinfo {author} {\bibfnamefont {D.}~\bibnamefont
  {{Stern}}},\ and\ \bibinfo {author} {\bibfnamefont {Y.}~\bibnamefont
  {{Yang}}},\ }\bibfield  {title} {\bibinfo {title} {{Ram-pressure Stripping of
  a Kicked Hill Sphere: Prompt Electromagnetic Emission from the Merger of
  Stellar Mass Black Holes in an AGN Accretion Disk}},\ }\href
  {https://doi.org/10.3847/2041-8213/ab4886} {\bibfield  {journal} {\bibinfo
  {journal} {\apjl}\ }\textbf {\bibinfo {volume} {884}},\ \bibinfo {eid} {L50}
  (\bibinfo {year} {2019})},\ \Eprint {https://arxiv.org/abs/1907.03746}
  {arXiv:1907.03746 [astro-ph.HE]} \BibitemShut {NoStop}%
\bibitem [{\citenamefont {{Tagawa}}\ \emph {et~al.}(2023)\citenamefont
  {{Tagawa}}, \citenamefont {{Kimura}}, \citenamefont {{Haiman}}, \citenamefont
  {{Perna}},\ and\ \citenamefont {{Bartos}}}]{Tagawa2023}%
  \BibitemOpen
  \bibfield  {author} {\bibinfo {author} {\bibfnamefont {H.}~\bibnamefont
  {{Tagawa}}}, \bibinfo {author} {\bibfnamefont {S.~S.}\ \bibnamefont
  {{Kimura}}}, \bibinfo {author} {\bibfnamefont {Z.}~\bibnamefont {{Haiman}}},
  \bibinfo {author} {\bibfnamefont {R.}~\bibnamefont {{Perna}}},\ and\ \bibinfo
  {author} {\bibfnamefont {I.}~\bibnamefont {{Bartos}}},\ }\bibfield  {title}
  {\bibinfo {title} {{Observable Signature of Merging Stellar-mass Black Holes
  in Active Galactic Nuclei}},\ }\href
  {https://doi.org/10.3847/1538-4357/acc4bb} {\bibfield  {journal} {\bibinfo
  {journal} {\apj}\ }\textbf {\bibinfo {volume} {950}},\ \bibinfo {eid} {13}
  (\bibinfo {year} {2023})},\ \Eprint {https://arxiv.org/abs/2301.07111}
  {arXiv:2301.07111 [astro-ph.HE]} \BibitemShut {NoStop}%
\bibitem [{\citenamefont {{Gayathri}}\ \emph {et~al.}(2021)\citenamefont
  {{Gayathri}}, \citenamefont {{Healy}}, \citenamefont {{Lange}}, \citenamefont
  {{O'Brien}}, \citenamefont {{Szczepanczyk}}, \citenamefont {{Bartos}},
  \citenamefont {{Campanelli}}, \citenamefont {{Klimenko}}, \citenamefont
  {{Lousto}},\ and\ \citenamefont {{O'Shaughnessy}}}]{Gayathri2021}%
  \BibitemOpen
  \bibfield  {author} {\bibinfo {author} {\bibfnamefont {V.}~\bibnamefont
  {{Gayathri}}}, \bibinfo {author} {\bibfnamefont {J.}~\bibnamefont {{Healy}}},
  \bibinfo {author} {\bibfnamefont {J.}~\bibnamefont {{Lange}}}, \bibinfo
  {author} {\bibfnamefont {B.}~\bibnamefont {{O'Brien}}}, \bibinfo {author}
  {\bibfnamefont {M.}~\bibnamefont {{Szczepanczyk}}}, \bibinfo {author}
  {\bibfnamefont {I.}~\bibnamefont {{Bartos}}}, \bibinfo {author}
  {\bibfnamefont {M.}~\bibnamefont {{Campanelli}}}, \bibinfo {author}
  {\bibfnamefont {S.}~\bibnamefont {{Klimenko}}}, \bibinfo {author}
  {\bibfnamefont {C.~O.}\ \bibnamefont {{Lousto}}},\ and\ \bibinfo {author}
  {\bibfnamefont {R.}~\bibnamefont {{O'Shaughnessy}}},\ }\bibfield  {title}
  {\bibinfo {title} {{Measuring the Hubble Constant with GW190521 as an
  Eccentric black hole Merger and Its Potential Electromagnetic Counterpart}},\
  }\href {https://doi.org/10.3847/2041-8213/abe388} {\bibfield  {journal}
  {\bibinfo  {journal} {\apjl}\ }\textbf {\bibinfo {volume} {908}},\ \bibinfo
  {eid} {L34} (\bibinfo {year} {2021})}\BibitemShut {NoStop}%
\bibitem [{\citenamefont {{Chen}}\ \emph {et~al.}(2022)\citenamefont {{Chen}},
  \citenamefont {{Haster}}, \citenamefont {{Vitale}}, \citenamefont {{Farr}},\
  and\ \citenamefont {{Isi}}}]{Chen2022}%
  \BibitemOpen
  \bibfield  {author} {\bibinfo {author} {\bibfnamefont {H.-Y.}\ \bibnamefont
  {{Chen}}}, \bibinfo {author} {\bibfnamefont {C.-J.}\ \bibnamefont
  {{Haster}}}, \bibinfo {author} {\bibfnamefont {S.}~\bibnamefont {{Vitale}}},
  \bibinfo {author} {\bibfnamefont {W.~M.}\ \bibnamefont {{Farr}}},\ and\
  \bibinfo {author} {\bibfnamefont {M.}~\bibnamefont {{Isi}}},\ }\bibfield
  {title} {\bibinfo {title} {{A standard siren cosmological measurement from
  the potential GW190521 electromagnetic counterpart ZTF19abanrhr}},\ }\href
  {https://doi.org/10.1093/mnras/stac989} {\bibfield  {journal} {\bibinfo
  {journal} {\mnras}\ }\textbf {\bibinfo {volume} {513}},\ \bibinfo {pages}
  {2152} (\bibinfo {year} {2022})},\ \Eprint {https://arxiv.org/abs/2009.14057}
  {arXiv:2009.14057 [astro-ph.CO]} \BibitemShut {NoStop}%
\bibitem [{\citenamefont {{Graham}}\ \emph {et~al.}(2020)\citenamefont
  {{Graham}}, \citenamefont {{Ford}}, \citenamefont {{McKernan}}, \citenamefont
  {{Ross}}, \citenamefont {{Stern}}, \citenamefont {{Burdge}}, \citenamefont
  {{Coughlin}}, \citenamefont {{Djorgovski}}, \citenamefont {{Drake}},
  \citenamefont {{Duev}}, \citenamefont {{Kasliwal}}, \citenamefont
  {{Mahabal}}, \citenamefont {{van Velzen}}, \citenamefont {{Belecki}},
  \citenamefont {{Bellm}}, \citenamefont {{Burruss}}, \citenamefont {{Cenko}},
  \citenamefont {{Cunningham}}, \citenamefont {{Helou}}, \citenamefont
  {{Kulkarni}}, \citenamefont {{Masci}}, \citenamefont {{Prince}},
  \citenamefont {{Reiley}}, \citenamefont {{Rodriguez}}, \citenamefont
  {{Rusholme}}, \citenamefont {{Smith}},\ and\ \citenamefont
  {{Soumagnac}}}]{Graham2020}%
  \BibitemOpen
  \bibfield  {author} {\bibinfo {author} {\bibfnamefont {M.~J.}\ \bibnamefont
  {{Graham}}}, \bibinfo {author} {\bibfnamefont {K.~E.~S.}\ \bibnamefont
  {{Ford}}}, \bibinfo {author} {\bibfnamefont {B.}~\bibnamefont {{McKernan}}},
  \bibinfo {author} {\bibfnamefont {N.~P.}\ \bibnamefont {{Ross}}}, \bibinfo
  {author} {\bibfnamefont {D.}~\bibnamefont {{Stern}}}, \bibinfo {author}
  {\bibfnamefont {K.}~\bibnamefont {{Burdge}}}, \bibinfo {author}
  {\bibfnamefont {M.}~\bibnamefont {{Coughlin}}}, \bibinfo {author}
  {\bibfnamefont {S.~G.}\ \bibnamefont {{Djorgovski}}}, \bibinfo {author}
  {\bibfnamefont {A.~J.}\ \bibnamefont {{Drake}}}, \bibinfo {author}
  {\bibfnamefont {D.}~\bibnamefont {{Duev}}}, \bibinfo {author} {\bibfnamefont
  {M.}~\bibnamefont {{Kasliwal}}}, \bibinfo {author} {\bibfnamefont {A.~A.}\
  \bibnamefont {{Mahabal}}}, \bibinfo {author} {\bibfnamefont {S.}~\bibnamefont
  {{van Velzen}}}, \bibinfo {author} {\bibfnamefont {J.}~\bibnamefont
  {{Belecki}}}, \bibinfo {author} {\bibfnamefont {E.~C.}\ \bibnamefont
  {{Bellm}}}, \bibinfo {author} {\bibfnamefont {R.}~\bibnamefont {{Burruss}}},
  \bibinfo {author} {\bibfnamefont {S.~B.}\ \bibnamefont {{Cenko}}}, \bibinfo
  {author} {\bibfnamefont {V.}~\bibnamefont {{Cunningham}}}, \bibinfo {author}
  {\bibfnamefont {G.}~\bibnamefont {{Helou}}}, \bibinfo {author} {\bibfnamefont
  {S.~R.}\ \bibnamefont {{Kulkarni}}}, \bibinfo {author} {\bibfnamefont
  {F.~J.}\ \bibnamefont {{Masci}}}, \bibinfo {author} {\bibfnamefont
  {T.}~\bibnamefont {{Prince}}}, \bibinfo {author} {\bibfnamefont
  {D.}~\bibnamefont {{Reiley}}}, \bibinfo {author} {\bibfnamefont
  {H.}~\bibnamefont {{Rodriguez}}}, \bibinfo {author} {\bibfnamefont
  {B.}~\bibnamefont {{Rusholme}}}, \bibinfo {author} {\bibfnamefont {R.~M.}\
  \bibnamefont {{Smith}}},\ and\ \bibinfo {author} {\bibfnamefont {M.~T.}\
  \bibnamefont {{Soumagnac}}},\ }\bibfield  {title} {\bibinfo {title}
  {{Candidate Electromagnetic Counterpart to the Binary Black Hole Merger
  Gravitational-Wave Event S190521g$^{*}$}},\ }\href
  {https://doi.org/10.1103/PhysRevLett.124.251102} {\bibfield  {journal}
  {\bibinfo  {journal} {\prl}\ }\textbf {\bibinfo {volume} {124}},\ \bibinfo
  {eid} {251102} (\bibinfo {year} {2020})},\ \Eprint
  {https://arxiv.org/abs/2006.14122} {arXiv:2006.14122 [astro-ph.HE]}
  \BibitemShut {NoStop}%
\bibitem [{\citenamefont {{Ashton}}\ \emph {et~al.}(2021)\citenamefont
  {{Ashton}}, \citenamefont {{Ackley}}, \citenamefont {{Hernandez}},\ and\
  \citenamefont {{Piotrzkowski}}}]{Ashton2021}%
  \BibitemOpen
  \bibfield  {author} {\bibinfo {author} {\bibfnamefont {G.}~\bibnamefont
  {{Ashton}}}, \bibinfo {author} {\bibfnamefont {K.}~\bibnamefont {{Ackley}}},
  \bibinfo {author} {\bibfnamefont {I.~M.}\ \bibnamefont {{Hernandez}}},\ and\
  \bibinfo {author} {\bibfnamefont {B.}~\bibnamefont {{Piotrzkowski}}},\
  }\bibfield  {title} {\bibinfo {title} {{Current observations are insufficient
  to confidently associate the binary black hole merger GW190521 with AGN
  J124942.3 + 344929}},\ }\href {https://doi.org/10.1088/1361-6382/ac33bb}
  {\bibfield  {journal} {\bibinfo  {journal} {Classical and Quantum Gravity}\
  }\textbf {\bibinfo {volume} {38}},\ \bibinfo {eid} {235004} (\bibinfo {year}
  {2021})},\ \Eprint {https://arxiv.org/abs/2009.12346} {arXiv:2009.12346
  [astro-ph.HE]} \BibitemShut {NoStop}%
\bibitem [{\citenamefont {{Palmese}}\ \emph {et~al.}(2021)\citenamefont
  {{Palmese}}, \citenamefont {{Fishbach}}, \citenamefont {{Burke}},
  \citenamefont {{Annis}},\ and\ \citenamefont {{Liu}}}]{Palmese2021}%
  \BibitemOpen
  \bibfield  {author} {\bibinfo {author} {\bibfnamefont {A.}~\bibnamefont
  {{Palmese}}}, \bibinfo {author} {\bibfnamefont {M.}~\bibnamefont
  {{Fishbach}}}, \bibinfo {author} {\bibfnamefont {C.~J.}\ \bibnamefont
  {{Burke}}}, \bibinfo {author} {\bibfnamefont {J.}~\bibnamefont {{Annis}}},\
  and\ \bibinfo {author} {\bibfnamefont {X.}~\bibnamefont {{Liu}}},\ }\bibfield
   {title} {\bibinfo {title} {{Do LIGO/Virgo Black Hole Mergers Produce AGN
  Flares? The Case of GW190521 and Prospects for Reaching a Confident
  Association}},\ }\href {https://doi.org/10.3847/2041-8213/ac0883} {\bibfield
  {journal} {\bibinfo  {journal} {\apjl}\ }\textbf {\bibinfo {volume} {914}},\
  \bibinfo {eid} {L34} (\bibinfo {year} {2021})},\ \Eprint
  {https://arxiv.org/abs/2103.16069} {arXiv:2103.16069 [astro-ph.HE]}
  \BibitemShut {NoStop}%
\bibitem [{\citenamefont {{Veronesi}}\ \emph {et~al.}(2023)\citenamefont
  {{Veronesi}}, \citenamefont {{Rossi}},\ and\ \citenamefont {{van
  Velzen}}}]{Veronesi2023}%
  \BibitemOpen
  \bibfield  {author} {\bibinfo {author} {\bibfnamefont {N.}~\bibnamefont
  {{Veronesi}}}, \bibinfo {author} {\bibfnamefont {E.~M.}\ \bibnamefont
  {{Rossi}}},\ and\ \bibinfo {author} {\bibfnamefont {S.}~\bibnamefont {{van
  Velzen}}},\ }\bibfield  {title} {\bibinfo {title} {{The most luminous AGN do
  not produce the majority of the detected stellar-mass black hole binary
  mergers in the local Universe}},\ }\href
  {https://doi.org/10.48550/arXiv.2306.09415} {\bibfield  {journal} {\bibinfo
  {journal} {arXiv e-prints}\ ,\ \bibinfo {eid} {arXiv:2306.09415}} (\bibinfo
  {year} {2023})},\ \Eprint {https://arxiv.org/abs/2306.09415}
  {arXiv:2306.09415 [astro-ph.HE]} \BibitemShut {NoStop}%
\bibitem [{\citenamefont {{Shapiro}}\ and\ \citenamefont
  {{Teukolsky}}(1983)}]{1983bhwd.book.....S}%
  \BibitemOpen
  \bibfield  {author} {\bibinfo {author} {\bibfnamefont {S.~L.}\ \bibnamefont
  {{Shapiro}}}\ and\ \bibinfo {author} {\bibfnamefont {S.~A.}\ \bibnamefont
  {{Teukolsky}}},\ }\href@noop {} {\emph {\bibinfo {title} {{Black holes, white
  dwarfs, and neutron stars : the physics of compact objects}}}}\ (\bibinfo
  {year} {1983})\BibitemShut {NoStop}%
\bibitem [{\citenamefont {{van den Heuvel}}\ \emph {et~al.}(1992)\citenamefont
  {{van den Heuvel}}, \citenamefont {{Bhattacharya}}, \citenamefont
  {{Nomoto}},\ and\ \citenamefont {{Rappaport}}}]{1992A&A...262...97V}%
  \BibitemOpen
  \bibfield  {author} {\bibinfo {author} {\bibfnamefont {E.~P.~J.}\
  \bibnamefont {{van den Heuvel}}}, \bibinfo {author} {\bibfnamefont
  {D.}~\bibnamefont {{Bhattacharya}}}, \bibinfo {author} {\bibfnamefont
  {K.}~\bibnamefont {{Nomoto}}},\ and\ \bibinfo {author} {\bibfnamefont
  {S.~A.}\ \bibnamefont {{Rappaport}}},\ }\bibfield  {title} {\bibinfo {title}
  {{Accreting white dwarf models for CAL 83, CAL 87 and other ultrasoft X-ray
  sources in the LMC.}},\ }\href@noop {} {\bibfield  {journal} {\bibinfo
  {journal} {\aap}\ }\textbf {\bibinfo {volume} {262}},\ \bibinfo {pages} {97}
  (\bibinfo {year} {1992})}\BibitemShut {NoStop}%
\bibitem [{\citenamefont {{Brown}}\ and\ \citenamefont
  {{Bethe}}(1994)}]{1994ApJ...423..659B}%
  \BibitemOpen
  \bibfield  {author} {\bibinfo {author} {\bibfnamefont {G.~E.}\ \bibnamefont
  {{Brown}}}\ and\ \bibinfo {author} {\bibfnamefont {H.~A.}\ \bibnamefont
  {{Bethe}}},\ }\bibfield  {title} {\bibinfo {title} {{A Scenario for a Large
  Number of Low-Mass Black Holes in the Galaxy}},\ }\href
  {https://doi.org/10.1086/173844} {\bibfield  {journal} {\bibinfo  {journal}
  {\apj}\ }\textbf {\bibinfo {volume} {423}},\ \bibinfo {pages} {659} (\bibinfo
  {year} {1994})}\BibitemShut {NoStop}%
\bibitem [{\citenamefont {{Samland}}(1998)}]{1998ApJ...496..155S}%
  \BibitemOpen
  \bibfield  {author} {\bibinfo {author} {\bibfnamefont {M.}~\bibnamefont
  {{Samland}}},\ }\bibfield  {title} {\bibinfo {title} {{Modeling the Evolution
  of Disk Galaxies. II. Yields of Massive Stars}},\ }\href
  {https://doi.org/10.1086/305368} {\bibfield  {journal} {\bibinfo  {journal}
  {\apj}\ }\textbf {\bibinfo {volume} {496}},\ \bibinfo {pages} {155} (\bibinfo
  {year} {1998})}\BibitemShut {NoStop}%
\bibitem [{\citenamefont {{Einstein}}(1936)}]{1936Sci....84..506E}%
  \BibitemOpen
  \bibfield  {author} {\bibinfo {author} {\bibfnamefont {A.}~\bibnamefont
  {{Einstein}}},\ }\bibfield  {title} {\bibinfo {title} {{Lens-Like Action of a
  Star by the Deviation of Light in the Gravitational Field}},\ }\href
  {https://doi.org/10.1126/science.84.2188.506} {\bibfield  {journal} {\bibinfo
   {journal} {Science}\ }\textbf {\bibinfo {volume} {84}},\ \bibinfo {pages}
  {506} (\bibinfo {year} {1936})}\BibitemShut {NoStop}%
\bibitem [{\citenamefont {{Miyamoto}}\ and\ \citenamefont
  {{Yoshii}}(1995)}]{1995AJ....110.1427M}%
  \BibitemOpen
  \bibfield  {author} {\bibinfo {author} {\bibfnamefont {M.}~\bibnamefont
  {{Miyamoto}}}\ and\ \bibinfo {author} {\bibfnamefont {Y.}~\bibnamefont
  {{Yoshii}}},\ }\bibfield  {title} {\bibinfo {title} {{Astrometry for
  Determining the MACHO Mass and Trajectory}},\ }\href
  {https://doi.org/10.1086/117616} {\bibfield  {journal} {\bibinfo  {journal}
  {\aj}\ }\textbf {\bibinfo {volume} {110}},\ \bibinfo {pages} {1427} (\bibinfo
  {year} {1995})}\BibitemShut {NoStop}%
\bibitem [{\citenamefont {{Hog}}\ \emph {et~al.}(1995)\citenamefont {{Hog}},
  \citenamefont {{Novikov}},\ and\ \citenamefont
  {{Polnarev}}}]{1995A&A...294..287H}%
  \BibitemOpen
  \bibfield  {author} {\bibinfo {author} {\bibfnamefont {E.}~\bibnamefont
  {{Hog}}}, \bibinfo {author} {\bibfnamefont {I.~D.}\ \bibnamefont
  {{Novikov}}},\ and\ \bibinfo {author} {\bibfnamefont {A.~G.}\ \bibnamefont
  {{Polnarev}}},\ }\bibfield  {title} {\bibinfo {title} {{MACHO photometry and
  astrometry.}},\ }\href@noop {} {\bibfield  {journal} {\bibinfo  {journal}
  {\aap}\ }\textbf {\bibinfo {volume} {294}},\ \bibinfo {pages} {287} (\bibinfo
  {year} {1995})}\BibitemShut {NoStop}%
\bibitem [{\citenamefont {{Walker}}(1995)}]{1995ApJ...453...37W}%
  \BibitemOpen
  \bibfield  {author} {\bibinfo {author} {\bibfnamefont {M.~A.}\ \bibnamefont
  {{Walker}}},\ }\bibfield  {title} {\bibinfo {title} {{Microlensed Image
  Motions}},\ }\href {https://doi.org/10.1086/176367} {\bibfield  {journal}
  {\bibinfo  {journal} {\apj}\ }\textbf {\bibinfo {volume} {453}},\ \bibinfo
  {pages} {37} (\bibinfo {year} {1995})}\BibitemShut {NoStop}%
\bibitem [{\citenamefont {{Gould}}\ and\ \citenamefont
  {{Loeb}}(1992)}]{1992ApJ...396..104G}%
  \BibitemOpen
  \bibfield  {author} {\bibinfo {author} {\bibfnamefont {A.}~\bibnamefont
  {{Gould}}}\ and\ \bibinfo {author} {\bibfnamefont {A.}~\bibnamefont
  {{Loeb}}},\ }\bibfield  {title} {\bibinfo {title} {{Discovering Planetary
  Systems through Gravitational Microlenses}},\ }\href
  {https://doi.org/10.1086/171700} {\bibfield  {journal} {\bibinfo  {journal}
  {\apj}\ }\textbf {\bibinfo {volume} {396}},\ \bibinfo {pages} {104} (\bibinfo
  {year} {1992})}\BibitemShut {NoStop}%
\bibitem [{\citenamefont {{Alcock}}\ \emph {et~al.}(1995)\citenamefont
  {{Alcock}}, \citenamefont {{Allsman}}, \citenamefont {{Alves}}, \citenamefont
  {{Axelrod}}, \citenamefont {{Bennett}}, \citenamefont {{Cook}}, \citenamefont
  {{Freeman}}, \citenamefont {{Griest}}, \citenamefont {{Guern}}, \citenamefont
  {{Lehner}}, \citenamefont {{Marshall}}, \citenamefont {{Peterson}},
  \citenamefont {{Pratt}}, \citenamefont {{Quinn}}, \citenamefont {{Rodgers}},
  \citenamefont {{Stubbs}},\ and\ \citenamefont
  {{Sutherland}}}]{1995ApJ...454L.125A}%
  \BibitemOpen
  \bibfield  {author} {\bibinfo {author} {\bibfnamefont {C.}~\bibnamefont
  {{Alcock}}}, \bibinfo {author} {\bibfnamefont {R.~A.}\ \bibnamefont
  {{Allsman}}}, \bibinfo {author} {\bibfnamefont {D.}~\bibnamefont {{Alves}}},
  \bibinfo {author} {\bibfnamefont {T.~S.}\ \bibnamefont {{Axelrod}}}, \bibinfo
  {author} {\bibfnamefont {D.~P.}\ \bibnamefont {{Bennett}}}, \bibinfo {author}
  {\bibfnamefont {K.~H.}\ \bibnamefont {{Cook}}}, \bibinfo {author}
  {\bibfnamefont {K.~C.}\ \bibnamefont {{Freeman}}}, \bibinfo {author}
  {\bibfnamefont {K.}~\bibnamefont {{Griest}}}, \bibinfo {author}
  {\bibfnamefont {J.}~\bibnamefont {{Guern}}}, \bibinfo {author} {\bibfnamefont
  {M.~J.}\ \bibnamefont {{Lehner}}}, \bibinfo {author} {\bibfnamefont {S.~L.}\
  \bibnamefont {{Marshall}}}, \bibinfo {author} {\bibfnamefont {B.~A.}\
  \bibnamefont {{Peterson}}}, \bibinfo {author} {\bibfnamefont {M.~R.}\
  \bibnamefont {{Pratt}}}, \bibinfo {author} {\bibfnamefont {P.~J.}\
  \bibnamefont {{Quinn}}}, \bibinfo {author} {\bibfnamefont {A.~W.}\
  \bibnamefont {{Rodgers}}}, \bibinfo {author} {\bibfnamefont {C.~W.}\
  \bibnamefont {{Stubbs}}},\ and\ \bibinfo {author} {\bibfnamefont
  {W.}~\bibnamefont {{Sutherland}}},\ }\bibfield  {title} {\bibinfo {title}
  {{First Observation of Parallax in a Gravitational Microlensing Event}},\
  }\href {https://doi.org/10.1086/309783} {\bibfield  {journal} {\bibinfo
  {journal} {\apjl}\ }\textbf {\bibinfo {volume} {454}},\ \bibinfo {pages}
  {L125} (\bibinfo {year} {1995})},\ \Eprint
  {https://arxiv.org/abs/astro-ph/9506114} {arXiv:astro-ph/9506114 [astro-ph]}
  \BibitemShut {NoStop}%
\bibitem [{\citenamefont {{Lam}}\ \emph {et~al.}(2020)\citenamefont {{Lam}},
  \citenamefont {{Lu}}, \citenamefont {{Hosek}}, \citenamefont {{Dawson}},\
  and\ \citenamefont {{Golovich}}}]{2020ApJ...889...31L}%
  \BibitemOpen
  \bibfield  {author} {\bibinfo {author} {\bibfnamefont {C.~Y.}\ \bibnamefont
  {{Lam}}}, \bibinfo {author} {\bibfnamefont {J.~R.}\ \bibnamefont {{Lu}}},
  \bibinfo {author} {\bibfnamefont {J.}~\bibnamefont {{Hosek}}, \bibfnamefont
  {Matthew~W.}}, \bibinfo {author} {\bibfnamefont {W.~A.}\ \bibnamefont
  {{Dawson}}},\ and\ \bibinfo {author} {\bibfnamefont {N.~R.}\ \bibnamefont
  {{Golovich}}},\ }\bibfield  {title} {\bibinfo {title} {{PopSyCLE: A New
  Population Synthesis Code for Compact Object Microlensing Events}},\ }\href
  {https://doi.org/10.3847/1538-4357/ab5fd3} {\bibfield  {journal} {\bibinfo
  {journal} {\apj}\ }\textbf {\bibinfo {volume} {889}},\ \bibinfo {eid} {31}
  (\bibinfo {year} {2020})},\ \Eprint {https://arxiv.org/abs/1912.04510}
  {arXiv:1912.04510 [astro-ph.SR]} \BibitemShut {NoStop}%
\bibitem [{\citenamefont {{Pietrzy{\'n}ski}}(2018)}]{2018Natur.562..349P}%
  \BibitemOpen
  \bibfield  {author} {\bibinfo {author} {\bibfnamefont {G.}~\bibnamefont
  {{Pietrzy{\'n}ski}}},\ }\bibfield  {title} {\bibinfo {title} {{Twenty-five
  years of using microlensing to study dark matter}},\ }\href
  {https://doi.org/10.1038/d41586-018-07006-8} {\bibfield  {journal} {\bibinfo
  {journal} {\nat}\ }\textbf {\bibinfo {volume} {562}},\ \bibinfo {pages} {349}
  (\bibinfo {year} {2018})}\BibitemShut {NoStop}%
\bibitem [{\citenamefont {{Udalski}}\ \emph {et~al.}(1992)\citenamefont
  {{Udalski}}, \citenamefont {{Szymanski}}, \citenamefont {{Kaluzny}},
  \citenamefont {{Kubiak}},\ and\ \citenamefont
  {{Mateo}}}]{1992AcA....42..253U}%
  \BibitemOpen
  \bibfield  {author} {\bibinfo {author} {\bibfnamefont {A.}~\bibnamefont
  {{Udalski}}}, \bibinfo {author} {\bibfnamefont {M.}~\bibnamefont
  {{Szymanski}}}, \bibinfo {author} {\bibfnamefont {J.}~\bibnamefont
  {{Kaluzny}}}, \bibinfo {author} {\bibfnamefont {M.}~\bibnamefont
  {{Kubiak}}},\ and\ \bibinfo {author} {\bibfnamefont {M.}~\bibnamefont
  {{Mateo}}},\ }\bibfield  {title} {\bibinfo {title} {{The Optical
  Gravitational Lensing Experiment}},\ }\href@noop {} {\bibfield  {journal}
  {\bibinfo  {journal} {\actaa}\ }\textbf {\bibinfo {volume} {42}},\ \bibinfo
  {pages} {253} (\bibinfo {year} {1992})}\BibitemShut {NoStop}%
\bibitem [{\citenamefont {{Udalski}}\ \emph {et~al.}(2015)\citenamefont
  {{Udalski}}, \citenamefont {{Szyma{\'n}ski}},\ and\ \citenamefont
  {{Szyma{\'n}ski}}}]{2015AcA....65....1U}%
  \BibitemOpen
  \bibfield  {author} {\bibinfo {author} {\bibfnamefont {A.}~\bibnamefont
  {{Udalski}}}, \bibinfo {author} {\bibfnamefont {M.~K.}\ \bibnamefont
  {{Szyma{\'n}ski}}},\ and\ \bibinfo {author} {\bibfnamefont {G.}~\bibnamefont
  {{Szyma{\'n}ski}}},\ }\bibfield  {title} {\bibinfo {title} {{OGLE-IV: Fourth
  Phase of the Optical Gravitational Lensing Experiment}},\ }\href@noop {}
  {\bibfield  {journal} {\bibinfo  {journal} {\actaa}\ }\textbf {\bibinfo
  {volume} {65}},\ \bibinfo {pages} {1} (\bibinfo {year} {2015})},\ \Eprint
  {https://arxiv.org/abs/1504.05966} {arXiv:1504.05966 [astro-ph.SR]}
  \BibitemShut {NoStop}%
\bibitem [{\citenamefont {{Aubourg}}\ \emph {et~al.}(1993)\citenamefont
  {{Aubourg}}, \citenamefont {{Bareyre}}, \citenamefont {{Br{\'e}hin}},
  \citenamefont {{Gros}}, \citenamefont {{Lachi{\`e}ze-Rey}}, \citenamefont
  {{Laurent}}, \citenamefont {{Lesquoy}}, \citenamefont {{Magneville}},
  \citenamefont {{Milsztajn}}, \citenamefont {{Moscoso}}, \citenamefont
  {{Queinnec}}, \citenamefont {{Rich}}, \citenamefont {{Spiro}}, \citenamefont
  {{Vigroux}}, \citenamefont {{Zylberajch}}, \citenamefont {{Ansari}},
  \citenamefont {{Cavalier}}, \citenamefont {{Moniez}}, \citenamefont
  {{Beaulieu}}, \citenamefont {{Ferlet}}, \citenamefont {{Grison}},
  \citenamefont {{Vidal-Madjar}}, \citenamefont {{Guibert}}, \citenamefont
  {{Moreau}}, \citenamefont {{Tajahmady}}, \citenamefont {{Maurice}},
  \citenamefont {{Pr{\'e}v{\^o}t}},\ and\ \citenamefont
  {{Gry}}}]{1993Natur.365..623A}%
  \BibitemOpen
  \bibfield  {author} {\bibinfo {author} {\bibfnamefont {E.}~\bibnamefont
  {{Aubourg}}}, \bibinfo {author} {\bibfnamefont {P.}~\bibnamefont
  {{Bareyre}}}, \bibinfo {author} {\bibfnamefont {S.}~\bibnamefont
  {{Br{\'e}hin}}}, \bibinfo {author} {\bibfnamefont {M.}~\bibnamefont
  {{Gros}}}, \bibinfo {author} {\bibfnamefont {M.}~\bibnamefont
  {{Lachi{\`e}ze-Rey}}}, \bibinfo {author} {\bibfnamefont {B.}~\bibnamefont
  {{Laurent}}}, \bibinfo {author} {\bibfnamefont {E.}~\bibnamefont
  {{Lesquoy}}}, \bibinfo {author} {\bibfnamefont {C.}~\bibnamefont
  {{Magneville}}}, \bibinfo {author} {\bibfnamefont {A.}~\bibnamefont
  {{Milsztajn}}}, \bibinfo {author} {\bibfnamefont {L.}~\bibnamefont
  {{Moscoso}}}, \bibinfo {author} {\bibfnamefont {F.}~\bibnamefont
  {{Queinnec}}}, \bibinfo {author} {\bibfnamefont {J.}~\bibnamefont {{Rich}}},
  \bibinfo {author} {\bibfnamefont {M.}~\bibnamefont {{Spiro}}}, \bibinfo
  {author} {\bibfnamefont {L.}~\bibnamefont {{Vigroux}}}, \bibinfo {author}
  {\bibfnamefont {S.}~\bibnamefont {{Zylberajch}}}, \bibinfo {author}
  {\bibfnamefont {R.}~\bibnamefont {{Ansari}}}, \bibinfo {author}
  {\bibfnamefont {F.}~\bibnamefont {{Cavalier}}}, \bibinfo {author}
  {\bibfnamefont {M.}~\bibnamefont {{Moniez}}}, \bibinfo {author}
  {\bibfnamefont {J.~P.}\ \bibnamefont {{Beaulieu}}}, \bibinfo {author}
  {\bibfnamefont {R.}~\bibnamefont {{Ferlet}}}, \bibinfo {author}
  {\bibfnamefont {P.}~\bibnamefont {{Grison}}}, \bibinfo {author}
  {\bibfnamefont {A.}~\bibnamefont {{Vidal-Madjar}}}, \bibinfo {author}
  {\bibfnamefont {J.}~\bibnamefont {{Guibert}}}, \bibinfo {author}
  {\bibfnamefont {O.}~\bibnamefont {{Moreau}}}, \bibinfo {author}
  {\bibfnamefont {F.}~\bibnamefont {{Tajahmady}}}, \bibinfo {author}
  {\bibfnamefont {E.}~\bibnamefont {{Maurice}}}, \bibinfo {author}
  {\bibfnamefont {L.}~\bibnamefont {{Pr{\'e}v{\^o}t}}},\ and\ \bibinfo {author}
  {\bibfnamefont {C.}~\bibnamefont {{Gry}}},\ }\bibfield  {title} {\bibinfo
  {title} {{Evidence for gravitational microlensing by dark objects in the
  Galactic halo}},\ }\href {https://doi.org/10.1038/365623a0} {\bibfield
  {journal} {\bibinfo  {journal} {\nat}\ }\textbf {\bibinfo {volume} {365}},\
  \bibinfo {pages} {623} (\bibinfo {year} {1993})}\BibitemShut {NoStop}%
\bibitem [{\citenamefont {{Bond}}\ \emph {et~al.}(2001)\citenamefont {{Bond}},
  \citenamefont {{Abe}}, \citenamefont {{Dodd}}, \citenamefont {{Hearnshaw}},
  \citenamefont {{Honda}}, \citenamefont {{Jugaku}}, \citenamefont
  {{Kilmartin}}, \citenamefont {{Marles}}, \citenamefont {{Masuda}},
  \citenamefont {{Matsubara}}, \citenamefont {{Muraki}}, \citenamefont
  {{Nakamura}}, \citenamefont {{Nankivell}}, \citenamefont {{Noda}},
  \citenamefont {{Noguchi}}, \citenamefont {{Ohnishi}}, \citenamefont
  {{Rattenbury}}, \citenamefont {{Reid}}, \citenamefont {{Saito}},
  \citenamefont {{Sato}}, \citenamefont {{Sekiguchi}}, \citenamefont
  {{Skuljan}}, \citenamefont {{Sullivan}}, \citenamefont {{Sumi}},
  \citenamefont {{Takeuti}}, \citenamefont {{Watase}}, \citenamefont
  {{Wilkinson}}, \citenamefont {{Yamada}}, \citenamefont {{Yanagisawa}},\ and\
  \citenamefont {{Yock}}}]{2001MNRAS.327..868B}%
  \BibitemOpen
  \bibfield  {author} {\bibinfo {author} {\bibfnamefont {I.~A.}\ \bibnamefont
  {{Bond}}}, \bibinfo {author} {\bibfnamefont {F.}~\bibnamefont {{Abe}}},
  \bibinfo {author} {\bibfnamefont {R.~J.}\ \bibnamefont {{Dodd}}}, \bibinfo
  {author} {\bibfnamefont {J.~B.}\ \bibnamefont {{Hearnshaw}}}, \bibinfo
  {author} {\bibfnamefont {M.}~\bibnamefont {{Honda}}}, \bibinfo {author}
  {\bibfnamefont {J.}~\bibnamefont {{Jugaku}}}, \bibinfo {author}
  {\bibfnamefont {P.~M.}\ \bibnamefont {{Kilmartin}}}, \bibinfo {author}
  {\bibfnamefont {A.}~\bibnamefont {{Marles}}}, \bibinfo {author}
  {\bibfnamefont {K.}~\bibnamefont {{Masuda}}}, \bibinfo {author}
  {\bibfnamefont {Y.}~\bibnamefont {{Matsubara}}}, \bibinfo {author}
  {\bibfnamefont {Y.}~\bibnamefont {{Muraki}}}, \bibinfo {author}
  {\bibfnamefont {T.}~\bibnamefont {{Nakamura}}}, \bibinfo {author}
  {\bibfnamefont {G.}~\bibnamefont {{Nankivell}}}, \bibinfo {author}
  {\bibfnamefont {S.}~\bibnamefont {{Noda}}}, \bibinfo {author} {\bibfnamefont
  {C.}~\bibnamefont {{Noguchi}}}, \bibinfo {author} {\bibfnamefont
  {K.}~\bibnamefont {{Ohnishi}}}, \bibinfo {author} {\bibfnamefont {N.~J.}\
  \bibnamefont {{Rattenbury}}}, \bibinfo {author} {\bibfnamefont
  {M.}~\bibnamefont {{Reid}}}, \bibinfo {author} {\bibfnamefont
  {T.}~\bibnamefont {{Saito}}}, \bibinfo {author} {\bibfnamefont
  {H.}~\bibnamefont {{Sato}}}, \bibinfo {author} {\bibfnamefont
  {M.}~\bibnamefont {{Sekiguchi}}}, \bibinfo {author} {\bibfnamefont
  {J.}~\bibnamefont {{Skuljan}}}, \bibinfo {author} {\bibfnamefont {D.~J.}\
  \bibnamefont {{Sullivan}}}, \bibinfo {author} {\bibfnamefont
  {T.}~\bibnamefont {{Sumi}}}, \bibinfo {author} {\bibfnamefont
  {M.}~\bibnamefont {{Takeuti}}}, \bibinfo {author} {\bibfnamefont
  {Y.}~\bibnamefont {{Watase}}}, \bibinfo {author} {\bibfnamefont
  {S.}~\bibnamefont {{Wilkinson}}}, \bibinfo {author} {\bibfnamefont
  {R.}~\bibnamefont {{Yamada}}}, \bibinfo {author} {\bibfnamefont
  {T.}~\bibnamefont {{Yanagisawa}}},\ and\ \bibinfo {author} {\bibfnamefont
  {P.~C.~M.}\ \bibnamefont {{Yock}}},\ }\bibfield  {title} {\bibinfo {title}
  {{Real-time difference imaging analysis of MOA Galactic bulge observations
  during 2000}},\ }\href {https://doi.org/10.1046/j.1365-8711.2001.04776.x}
  {\bibfield  {journal} {\bibinfo  {journal} {\mnras}\ }\textbf {\bibinfo
  {volume} {327}},\ \bibinfo {pages} {868} (\bibinfo {year} {2001})},\ \Eprint
  {https://arxiv.org/abs/astro-ph/0102181} {arXiv:astro-ph/0102181 [astro-ph]}
  \BibitemShut {NoStop}%
\bibitem [{\citenamefont {{Sako}}\ \emph {et~al.}(2008)\citenamefont {{Sako}},
  \citenamefont {{Sekiguchi}}, \citenamefont {{Sasaki}}, \citenamefont
  {{Okajima}}, \citenamefont {{Abe}}, \citenamefont {{Bond}}, \citenamefont
  {{Hearnshaw}}, \citenamefont {{Itow}}, \citenamefont {{Kamiya}},
  \citenamefont {{Kilmartin}}, \citenamefont {{Masuda}}, \citenamefont
  {{Matsubara}}, \citenamefont {{Muraki}}, \citenamefont {{Rattenbury}},
  \citenamefont {{Sullivan}}, \citenamefont {{Sumi}}, \citenamefont
  {{Tristram}}, \citenamefont {{Yanagisawa}},\ and\ \citenamefont
  {{Yock}}}]{2008ExA....22...51S}%
  \BibitemOpen
  \bibfield  {author} {\bibinfo {author} {\bibfnamefont {T.}~\bibnamefont
  {{Sako}}}, \bibinfo {author} {\bibfnamefont {T.}~\bibnamefont {{Sekiguchi}}},
  \bibinfo {author} {\bibfnamefont {M.}~\bibnamefont {{Sasaki}}}, \bibinfo
  {author} {\bibfnamefont {K.}~\bibnamefont {{Okajima}}}, \bibinfo {author}
  {\bibfnamefont {F.}~\bibnamefont {{Abe}}}, \bibinfo {author} {\bibfnamefont
  {I.~A.}\ \bibnamefont {{Bond}}}, \bibinfo {author} {\bibfnamefont {J.~B.}\
  \bibnamefont {{Hearnshaw}}}, \bibinfo {author} {\bibfnamefont
  {Y.}~\bibnamefont {{Itow}}}, \bibinfo {author} {\bibfnamefont
  {K.}~\bibnamefont {{Kamiya}}}, \bibinfo {author} {\bibfnamefont {P.~M.}\
  \bibnamefont {{Kilmartin}}}, \bibinfo {author} {\bibfnamefont
  {K.}~\bibnamefont {{Masuda}}}, \bibinfo {author} {\bibfnamefont
  {Y.}~\bibnamefont {{Matsubara}}}, \bibinfo {author} {\bibfnamefont
  {Y.}~\bibnamefont {{Muraki}}}, \bibinfo {author} {\bibfnamefont {N.~J.}\
  \bibnamefont {{Rattenbury}}}, \bibinfo {author} {\bibfnamefont {D.~J.}\
  \bibnamefont {{Sullivan}}}, \bibinfo {author} {\bibfnamefont
  {T.}~\bibnamefont {{Sumi}}}, \bibinfo {author} {\bibfnamefont
  {P.}~\bibnamefont {{Tristram}}}, \bibinfo {author} {\bibfnamefont
  {T.}~\bibnamefont {{Yanagisawa}}},\ and\ \bibinfo {author} {\bibfnamefont
  {P.~C.~M.}\ \bibnamefont {{Yock}}},\ }\bibfield  {title} {\bibinfo {title}
  {{MOA-cam3: a wide-field mosaic CCD camera for a gravitational microlensing
  survey in New Zealand}},\ }\href {https://doi.org/10.1007/s10686-007-9082-5}
  {\bibfield  {journal} {\bibinfo  {journal} {Experimental Astronomy}\ }\textbf
  {\bibinfo {volume} {22}},\ \bibinfo {pages} {51} (\bibinfo {year} {2008})},\
  \Eprint {https://arxiv.org/abs/0804.0653} {arXiv:0804.0653 [astro-ph]}
  \BibitemShut {NoStop}%
\bibitem [{\citenamefont {{Kim}}\ \emph {et~al.}(2016)\citenamefont {{Kim}},
  \citenamefont {{Lee}}, \citenamefont {{Park}}, \citenamefont {{Kim}},
  \citenamefont {{Cha}}, \citenamefont {{Lee}}, \citenamefont {{Han}},
  \citenamefont {{Chun}},\ and\ \citenamefont {{Yuk}}}]{2016JKAS...49...37K}%
  \BibitemOpen
  \bibfield  {author} {\bibinfo {author} {\bibfnamefont {S.-L.}\ \bibnamefont
  {{Kim}}}, \bibinfo {author} {\bibfnamefont {C.-U.}\ \bibnamefont {{Lee}}},
  \bibinfo {author} {\bibfnamefont {B.-G.}\ \bibnamefont {{Park}}}, \bibinfo
  {author} {\bibfnamefont {D.-J.}\ \bibnamefont {{Kim}}}, \bibinfo {author}
  {\bibfnamefont {S.-M.}\ \bibnamefont {{Cha}}}, \bibinfo {author}
  {\bibfnamefont {Y.}~\bibnamefont {{Lee}}}, \bibinfo {author} {\bibfnamefont
  {C.}~\bibnamefont {{Han}}}, \bibinfo {author} {\bibfnamefont {M.-Y.}\
  \bibnamefont {{Chun}}},\ and\ \bibinfo {author} {\bibfnamefont
  {I.}~\bibnamefont {{Yuk}}},\ }\bibfield  {title} {\bibinfo {title} {{KMTNET:
  A Network of 1.6 m Wide-Field Optical Telescopes Installed at Three Southern
  Observatories}},\ }\href {https://doi.org/10.5303/JKAS.2016.49.1.37}
  {\bibfield  {journal} {\bibinfo  {journal} {Journal of Korean Astronomical
  Society}\ }\textbf {\bibinfo {volume} {49}},\ \bibinfo {pages} {37} (\bibinfo
  {year} {2016})}\BibitemShut {NoStop}%
\bibitem [{\citenamefont {{Bellini}}\ \emph {et~al.}(2014)\citenamefont
  {{Bellini}}, \citenamefont {{Anderson}}, \citenamefont {{van der Marel}},
  \citenamefont {{Watkins}}, \citenamefont {{King}}, \citenamefont
  {{Bianchini}}, \citenamefont {{Chanam{\'e}}}, \citenamefont {{Chandar}},
  \citenamefont {{Cool}}, \citenamefont {{Ferraro}}, \citenamefont {{Ford}},\
  and\ \citenamefont {{Massari}}}]{2014ApJ...797..115B}%
  \BibitemOpen
  \bibfield  {author} {\bibinfo {author} {\bibfnamefont {A.}~\bibnamefont
  {{Bellini}}}, \bibinfo {author} {\bibfnamefont {J.}~\bibnamefont
  {{Anderson}}}, \bibinfo {author} {\bibfnamefont {R.~P.}\ \bibnamefont {{van
  der Marel}}}, \bibinfo {author} {\bibfnamefont {L.~L.}\ \bibnamefont
  {{Watkins}}}, \bibinfo {author} {\bibfnamefont {I.~R.}\ \bibnamefont
  {{King}}}, \bibinfo {author} {\bibfnamefont {P.}~\bibnamefont {{Bianchini}}},
  \bibinfo {author} {\bibfnamefont {J.}~\bibnamefont {{Chanam{\'e}}}}, \bibinfo
  {author} {\bibfnamefont {R.}~\bibnamefont {{Chandar}}}, \bibinfo {author}
  {\bibfnamefont {A.~M.}\ \bibnamefont {{Cool}}}, \bibinfo {author}
  {\bibfnamefont {F.~R.}\ \bibnamefont {{Ferraro}}}, \bibinfo {author}
  {\bibfnamefont {H.}~\bibnamefont {{Ford}}},\ and\ \bibinfo {author}
  {\bibfnamefont {D.}~\bibnamefont {{Massari}}},\ }\bibfield  {title} {\bibinfo
  {title} {{Hubble Space Telescope Proper Motion (HSTPROMO) Catalogs of
  Galactic Globular Clusters. I. Sample Selection, Data Reduction, and NGC 7078
  Results}},\ }\href {https://doi.org/10.1088/0004-637X/797/2/115} {\bibfield
  {journal} {\bibinfo  {journal} {\apj}\ }\textbf {\bibinfo {volume} {797}},\
  \bibinfo {eid} {115} (\bibinfo {year} {2014})},\ \Eprint
  {https://arxiv.org/abs/1410.5820} {arXiv:1410.5820 [astro-ph.SR]}
  \BibitemShut {NoStop}%
\bibitem [{\citenamefont {{Gaia Collaboration}}\ \emph
  {et~al.}(2016)\citenamefont {{Gaia Collaboration}}, \citenamefont {{Prusti}},
  \citenamefont {{de Bruijne}}, \citenamefont {{Brown}}, \citenamefont
  {{Vallenari}}, \citenamefont {{Babusiaux}}, \citenamefont {{Bailer-Jones}},
  \citenamefont {{Bastian}}, \citenamefont {{Biermann}}, \citenamefont
  {{Evans}}, \citenamefont {{Eyer}}, \citenamefont {{Jansen}}, \citenamefont
  {{Jordi}}, \citenamefont {{Klioner}}, \citenamefont {{Lammers}},
  \citenamefont {{Lindegren}}, \citenamefont {{Luri}}, \citenamefont
  {{Mignard}}, \citenamefont {{Milligan}}, \citenamefont {{Panem}},
  \citenamefont {{Poinsignon}}, \citenamefont {{Pourbaix}}, \citenamefont
  {{Randich}}, \citenamefont {{Sarri}}, \citenamefont {{Sartoretti}},
  \citenamefont {{Siddiqui}}, \citenamefont {{Soubiran}}, \citenamefont
  {{Valette}}, \citenamefont {{van Leeuwen}}, \citenamefont {{Walton}},
  \citenamefont {{Aerts}}, \citenamefont {{Arenou}}, \citenamefont {{Cropper}},
  \citenamefont {{Drimmel}}, \citenamefont {{H{\o}g}}, \citenamefont {{Katz}},
  \citenamefont {{Lattanzi}}, \citenamefont {{O'Mullane}}, \citenamefont
  {{Grebel}}, \citenamefont {{Holland}}, \citenamefont {{Huc}}, \citenamefont
  {{Passot}}, \citenamefont {{Bramante}}, \citenamefont {{Cacciari}},
  \citenamefont {{Casta{\~n}eda}}, \citenamefont {{Chaoul}}, \citenamefont
  {{Cheek}}, \citenamefont {{De Angeli}}, \citenamefont {{Fabricius}},
  \citenamefont {{Guerra}}, \citenamefont {{Hern{\'a}ndez}}, \citenamefont
  {{Jean-Antoine-Piccolo}}, \citenamefont {{Masana}}, \citenamefont
  {{Messineo}}, \citenamefont {{Mowlavi}}, \citenamefont {{Nienartowicz}},
  \citenamefont {{Ord{\'o}{\~n}ez-Blanco}}, \citenamefont {{Panuzzo}},
  \citenamefont {{Portell}}, \citenamefont {{Richards}}, \citenamefont
  {{Riello}}, \citenamefont {{Seabroke}}, \citenamefont {{Tanga}},
  \citenamefont {{Th{\'e}venin}}, \citenamefont {{Torra}}, \citenamefont
  {{Els}}, \citenamefont {{Gracia-Abril}}, \citenamefont {{Comoretto}},
  \citenamefont {{Garcia-Reinaldos}}, \citenamefont {{Lock}}, \citenamefont
  {{Mercier}}, \citenamefont {{Altmann}}, \citenamefont {{Andrae}},
  \citenamefont {{Astraatmadja}}, \citenamefont {{Bellas-Velidis}},
  \citenamefont {{Benson}}, \citenamefont {{Berthier}}, \citenamefont
  {{Blomme}}, \citenamefont {{Busso}}, \citenamefont {{Carry}}, \citenamefont
  {{Cellino}}, \citenamefont {{Clementini}}, \citenamefont {{Cowell}},
  \citenamefont {{Creevey}}, \citenamefont {{Cuypers}}, \citenamefont
  {{Davidson}}, \citenamefont {{De Ridder}}, \citenamefont {{de Torres}},
  \citenamefont {{Delchambre}}, \citenamefont {{Dell'Oro}}, \citenamefont
  {{Ducourant}}, \citenamefont {{Fr{\'e}mat}}, \citenamefont
  {{Garc{\'\i}a-Torres}}, \citenamefont {{Gosset}}, \citenamefont
  {{Halbwachs}}, \citenamefont {{Hambly}}, \citenamefont {{Harrison}},
  \citenamefont {{Hauser}}, \citenamefont {{Hestroffer}}, \citenamefont
  {{Hodgkin}}, \citenamefont {{Huckle}}, \citenamefont {{Hutton}},
  \citenamefont {{Jasniewicz}}, \citenamefont {{Jordan}}, \citenamefont
  {{Kontizas}}, \citenamefont {{Korn}}, \citenamefont {{Lanzafame}},
  \citenamefont {{Manteiga}}, \citenamefont {{Moitinho}}, \citenamefont
  {{Muinonen}}, \citenamefont {{Osinde}}, \citenamefont {{Pancino}},
  \citenamefont {{Pauwels}}, \citenamefont {{Petit}}, \citenamefont
  {{Recio-Blanco}}, \citenamefont {{Robin}}, \citenamefont {{Sarro}},
  \citenamefont {{Siopis}}, \citenamefont {{Smith}}, \citenamefont {{Smith}},
  \citenamefont {{Sozzetti}}, \citenamefont {{Thuillot}}, \citenamefont {{van
  Reeven}}, \citenamefont {{Viala}}, \citenamefont {{Abbas}}, \citenamefont
  {{Abreu Aramburu}}, \citenamefont {{Accart}}, \citenamefont {{Aguado}},
  \citenamefont {{Allan}}, \citenamefont {{Allasia}}, \citenamefont
  {{Altavilla}}, \citenamefont {{{\'A}lvarez}}, \citenamefont {{Alves}},
  \citenamefont {{Anderson}}, \citenamefont {{Andrei}}, \citenamefont {{Anglada
  Varela}}, \citenamefont {{Antiche}}, \citenamefont {{Antoja}}, \citenamefont
  {{Ant{\'o}n}}, \citenamefont {{Arcay}}, \citenamefont {{Atzei}},
  \citenamefont {{Ayache}}, \citenamefont {{Bach}}, \citenamefont {{Baker}},
  \citenamefont {{Balaguer-N{\'u}{\~n}ez}}, \citenamefont {{Barache}},
  \citenamefont {{Barata}}, \citenamefont {{Barbier}}, \citenamefont
  {{Barblan}}, \citenamefont {{Baroni}}, \citenamefont {{Barrado y
  Navascu{\'e}s}}, \citenamefont {{Barros}}, \citenamefont {{Barstow}},
  \citenamefont {{Becciani}}, \citenamefont {{Bellazzini}}, \citenamefont
  {{Bellei}}, \citenamefont {{Bello Garc{\'\i}a}}, \citenamefont {{Belokurov}},
  \citenamefont {{Bendjoya}}, \citenamefont {{Berihuete}}, \citenamefont
  {{Bianchi}}, \citenamefont {{Bienaym{\'e}}}, \citenamefont {{Billebaud}},
  \citenamefont {{Blagorodnova}}, \citenamefont {{Blanco-Cuaresma}},
  \citenamefont {{Boch}}, \citenamefont {{Bombrun}}, \citenamefont
  {{Borrachero}}, \citenamefont {{Bouquillon}}, \citenamefont {{Bourda}},
  \citenamefont {{Bouy}}, \citenamefont {{Bragaglia}}, \citenamefont
  {{Breddels}}, \citenamefont {{Brouillet}}, \citenamefont
  {{Br{\"u}semeister}}, \citenamefont {{Bucciarelli}}, \citenamefont
  {{Budnik}}, \citenamefont {{Burgess}}, \citenamefont {{Burgon}},
  \citenamefont {{Burlacu}}, \citenamefont {{Busonero}}, \citenamefont
  {{Buzzi}}, \citenamefont {{Caffau}}, \citenamefont {{Cambras}}, \citenamefont
  {{Campbell}}, \citenamefont {{Cancelliere}}, \citenamefont {{Cantat-Gaudin}},
  \citenamefont {{Carlucci}}, \citenamefont {{Carrasco}}, \citenamefont
  {{Castellani}}, \citenamefont {{Charlot}}, \citenamefont {{Charnas}},
  \citenamefont {{Charvet}}, \citenamefont {{Chassat}}, \citenamefont
  {{Chiavassa}}, \citenamefont {{Clotet}}, \citenamefont {{Cocozza}},
  \citenamefont {{Collins}}, \citenamefont {{Collins}}, \citenamefont
  {{Costigan}}, \citenamefont {{Crifo}}, \citenamefont {{Cross}}, \citenamefont
  {{Crosta}}, \citenamefont {{Crowley}}, \citenamefont {{Dafonte}},
  \citenamefont {{Damerdji}}, \citenamefont {{Dapergolas}}, \citenamefont
  {{David}}, \citenamefont {{David}}, \citenamefont {{De Cat}}, \citenamefont
  {{de Felice}}, \citenamefont {{de Laverny}}, \citenamefont {{De Luise}},
  \citenamefont {{De March}}, \citenamefont {{de Martino}}, \citenamefont {{de
  Souza}}, \citenamefont {{Debosscher}}, \citenamefont {{del Pozo}},
  \citenamefont {{Delbo}}, \citenamefont {{Delgado}}, \citenamefont
  {{Delgado}}, \citenamefont {{di Marco}}, \citenamefont {{Di Matteo}},
  \citenamefont {{Diakite}}, \citenamefont {{Distefano}}, \citenamefont
  {{Dolding}}, \citenamefont {{Dos Anjos}}, \citenamefont {{Drazinos}},
  \citenamefont {{Dur{\'a}n}}, \citenamefont {{Dzigan}}, \citenamefont
  {{Ecale}}, \citenamefont {{Edvardsson}}, \citenamefont {{Enke}},
  \citenamefont {{Erdmann}}, \citenamefont {{Escolar}}, \citenamefont
  {{Espina}}, \citenamefont {{Evans}}, \citenamefont {{Eynard Bontemps}},
  \citenamefont {{Fabre}}, \citenamefont {{Fabrizio}}, \citenamefont
  {{Faigler}}, \citenamefont {{Falc{\~a}o}}, \citenamefont {{Farr{\`a}s
  Casas}}, \citenamefont {{Faye}}, \citenamefont {{Federici}}, \citenamefont
  {{Fedorets}}, \citenamefont {{Fern{\'a}ndez-Hern{\'a}ndez}}, \citenamefont
  {{Fernique}}, \citenamefont {{Fienga}}, \citenamefont {{Figueras}},
  \citenamefont {{Filippi}}, \citenamefont {{Findeisen}}, \citenamefont
  {{Fonti}}, \citenamefont {{Fouesneau}}, \citenamefont {{Fraile}},
  \citenamefont {{Fraser}}, \citenamefont {{Fuchs}}, \citenamefont {{Furnell}},
  \citenamefont {{Gai}}, \citenamefont {{Galleti}}, \citenamefont
  {{Galluccio}}, \citenamefont {{Garabato}}, \citenamefont
  {{Garc{\'\i}a-Sedano}}, \citenamefont {{Gar{\'e}}}, \citenamefont
  {{Garofalo}}, \citenamefont {{Garralda}}, \citenamefont {{Gavras}},
  \citenamefont {{Gerssen}}, \citenamefont {{Geyer}}, \citenamefont
  {{Gilmore}}, \citenamefont {{Girona}}, \citenamefont {{Giuffrida}},
  \citenamefont {{Gomes}}, \citenamefont {{Gonz{\'a}lez-Marcos}}, \citenamefont
  {{Gonz{\'a}lez-N{\'u}{\~n}ez}}, \citenamefont {{Gonz{\'a}lez-Vidal}},
  \citenamefont {{Granvik}}, \citenamefont {{Guerrier}}, \citenamefont
  {{Guillout}}, \citenamefont {{Guiraud}}, \citenamefont {{G{\'u}rpide}},
  \citenamefont {{Guti{\'e}rrez-S{\'a}nchez}}, \citenamefont {{Guy}},
  \citenamefont {{Haigron}}, \citenamefont {{Hatzidimitriou}}, \citenamefont
  {{Haywood}}, \citenamefont {{Heiter}}, \citenamefont {{Helmi}}, \citenamefont
  {{Hobbs}}, \citenamefont {{Hofmann}}, \citenamefont {{Holl}}, \citenamefont
  {{Holland}}, \citenamefont {{Hunt}}, \citenamefont {{Hypki}}, \citenamefont
  {{Icardi}}, \citenamefont {{Irwin}}, \citenamefont {{Jevardat de Fombelle}},
  \citenamefont {{Jofr{\'e}}}, \citenamefont {{Jonker}}, \citenamefont
  {{Jorissen}}, \citenamefont {{Julbe}}, \citenamefont {{Karampelas}},
  \citenamefont {{Kochoska}}, \citenamefont {{Kohley}}, \citenamefont
  {{Kolenberg}}, \citenamefont {{Kontizas}}, \citenamefont {{Koposov}},
  \citenamefont {{Kordopatis}}, \citenamefont {{Koubsky}}, \citenamefont
  {{Kowalczyk}}, \citenamefont {{Krone-Martins}}, \citenamefont
  {{Kudryashova}}, \citenamefont {{Kull}}, \citenamefont {{Bachchan}},
  \citenamefont {{Lacoste-Seris}}, \citenamefont {{Lanza}}, \citenamefont
  {{Lavigne}}, \citenamefont {{Le Poncin-Lafitte}}, \citenamefont {{Lebreton}},
  \citenamefont {{Lebzelter}}, \citenamefont {{Leccia}}, \citenamefont
  {{Leclerc}}, \citenamefont {{Lecoeur-Taibi}}, \citenamefont {{Lemaitre}},
  \citenamefont {{Lenhardt}}, \citenamefont {{Leroux}}, \citenamefont {{Liao}},
  \citenamefont {{Licata}}, \citenamefont {{Lindstr{\o}m}}, \citenamefont
  {{Lister}}, \citenamefont {{Livanou}}, \citenamefont {{Lobel}}, \citenamefont
  {{L{\"o}ffler}}, \citenamefont {{L{\'o}pez}}, \citenamefont {{Lopez-Lozano}},
  \citenamefont {{Lorenz}}, \citenamefont {{Loureiro}}, \citenamefont
  {{MacDonald}}, \citenamefont {{Magalh{\~a}es Fernandes}}, \citenamefont
  {{Managau}}, \citenamefont {{Mann}}, \citenamefont {{Mantelet}},
  \citenamefont {{Marchal}}, \citenamefont {{Marchant}}, \citenamefont
  {{Marconi}}, \citenamefont {{Marie}}, \citenamefont {{Marinoni}},
  \citenamefont {{Marrese}}, \citenamefont {{Marschalk{\'o}}}, \citenamefont
  {{Marshall}}, \citenamefont {{Mart{\'\i}n-Fleitas}}, \citenamefont
  {{Martino}}, \citenamefont {{Mary}}, \citenamefont {{Matijevi{\v{c}}}},
  \citenamefont {{Mazeh}}, \citenamefont {{McMillan}}, \citenamefont
  {{Messina}}, \citenamefont {{Mestre}}, \citenamefont {{Michalik}},
  \citenamefont {{Millar}}, \citenamefont {{Miranda}}, \citenamefont
  {{Molina}}, \citenamefont {{Molinaro}}, \citenamefont {{Molinaro}},
  \citenamefont {{Moln{\'a}r}}, \citenamefont {{Moniez}}, \citenamefont
  {{Montegriffo}}, \citenamefont {{Monteiro}}, \citenamefont {{Mor}},
  \citenamefont {{Mora}}, \citenamefont {{Morbidelli}}, \citenamefont
  {{Morel}}, \citenamefont {{Morgenthaler}}, \citenamefont {{Morley}},
  \citenamefont {{Morris}}, \citenamefont {{Mulone}}, \citenamefont
  {{Muraveva}}, \citenamefont {{Musella}}, \citenamefont {{Narbonne}},
  \citenamefont {{Nelemans}}, \citenamefont {{Nicastro}}, \citenamefont
  {{Noval}}, \citenamefont {{Ord{\'e}novic}}, \citenamefont
  {{Ordieres-Mer{\'e}}}, \citenamefont {{Osborne}}, \citenamefont {{Pagani}},
  \citenamefont {{Pagano}}, \citenamefont {{Pailler}}, \citenamefont
  {{Palacin}}, \citenamefont {{Palaversa}}, \citenamefont {{Parsons}},
  \citenamefont {{Paulsen}}, \citenamefont {{Pecoraro}}, \citenamefont
  {{Pedrosa}}, \citenamefont {{Pentik{\"a}inen}}, \citenamefont {{Pereira}},
  \citenamefont {{Pichon}}, \citenamefont {{Piersimoni}}, \citenamefont
  {{Pineau}}, \citenamefont {{Plachy}}, \citenamefont {{Plum}}, \citenamefont
  {{Poujoulet}}, \citenamefont {{Pr{\v{s}}a}}, \citenamefont {{Pulone}},
  \citenamefont {{Ragaini}}, \citenamefont {{Rago}}, \citenamefont {{Rambaux}},
  \citenamefont {{Ramos-Lerate}}, \citenamefont {{Ranalli}}, \citenamefont
  {{Rauw}}, \citenamefont {{Read}}, \citenamefont {{Regibo}}, \citenamefont
  {{Renk}}, \citenamefont {{Reyl{\'e}}}, \citenamefont {{Ribeiro}},
  \citenamefont {{Rimoldini}}, \citenamefont {{Ripepi}}, \citenamefont
  {{Riva}}, \citenamefont {{Rixon}}, \citenamefont {{Roelens}}, \citenamefont
  {{Romero-G{\'o}mez}}, \citenamefont {{Rowell}}, \citenamefont {{Royer}},
  \citenamefont {{Rudolph}}, \citenamefont {{Ruiz-Dern}}, \citenamefont
  {{Sadowski}}, \citenamefont {{Sagrist{\`a} Sell{\'e}s}}, \citenamefont
  {{Sahlmann}}, \citenamefont {{Salgado}}, \citenamefont {{Salguero}},
  \citenamefont {{Sarasso}}, \citenamefont {{Savietto}}, \citenamefont
  {{Schnorhk}}, \citenamefont {{Schultheis}}, \citenamefont {{Sciacca}},
  \citenamefont {{Segol}}, \citenamefont {{Segovia}}, \citenamefont
  {{Segransan}}, \citenamefont {{Serpell}}, \citenamefont {{Shih}},
  \citenamefont {{Smareglia}}, \citenamefont {{Smart}}, \citenamefont
  {{Smith}}, \citenamefont {{Solano}}, \citenamefont {{Solitro}}, \citenamefont
  {{Sordo}}, \citenamefont {{Soria Nieto}}, \citenamefont {{Souchay}},
  \citenamefont {{Spagna}}, \citenamefont {{Spoto}}, \citenamefont {{Stampa}},
  \citenamefont {{Steele}}, \citenamefont {{Steidelm{\"u}ller}}, \citenamefont
  {{Stephenson}}, \citenamefont {{Stoev}}, \citenamefont {{Suess}},
  \citenamefont {{S{\"u}veges}}, \citenamefont {{Surdej}}, \citenamefont
  {{Szabados}}, \citenamefont {{Szegedi-Elek}}, \citenamefont {{Tapiador}},
  \citenamefont {{Taris}}, \citenamefont {{Tauran}}, \citenamefont {{Taylor}},
  \citenamefont {{Teixeira}}, \citenamefont {{Terrett}}, \citenamefont
  {{Tingley}}, \citenamefont {{Trager}}, \citenamefont {{Turon}}, \citenamefont
  {{Ulla}}, \citenamefont {{Utrilla}}, \citenamefont {{Valentini}},
  \citenamefont {{van Elteren}}, \citenamefont {{Van Hemelryck}}, \citenamefont
  {{van Leeuwen}}, \citenamefont {{Varadi}}, \citenamefont {{Vecchiato}},
  \citenamefont {{Veljanoski}}, \citenamefont {{Via}}, \citenamefont
  {{Vicente}}, \citenamefont {{Vogt}}, \citenamefont {{Voss}}, \citenamefont
  {{Votruba}}, \citenamefont {{Voutsinas}}, \citenamefont {{Walmsley}},
  \citenamefont {{Weiler}}, \citenamefont {{Weingrill}}, \citenamefont
  {{Werner}}, \citenamefont {{Wevers}}, \citenamefont {{Whitehead}},
  \citenamefont {{Wyrzykowski}}, \citenamefont {{Yoldas}}, \citenamefont
  {{{\v{Z}}erjal}}, \citenamefont {{Zucker}}, \citenamefont {{Zurbach}},
  \citenamefont {{Zwitter}}, \citenamefont {{Alecu}}, \citenamefont {{Allen}},
  \citenamefont {{Allende Prieto}}, \citenamefont {{Amorim}}, \citenamefont
  {{Anglada-Escud{\'e}}}, \citenamefont {{Arsenijevic}}, \citenamefont
  {{Azaz}}, \citenamefont {{Balm}}, \citenamefont {{Beck}}, \citenamefont
  {{Bernstein}}, \citenamefont {{Bigot}}, \citenamefont {{Bijaoui}},
  \citenamefont {{Blasco}}, \citenamefont {{Bonfigli}}, \citenamefont {{Bono}},
  \citenamefont {{Boudreault}}, \citenamefont {{Bressan}}, \citenamefont
  {{Brown}}, \citenamefont {{Brunet}}, \citenamefont {{Bunclark}},
  \citenamefont {{Buonanno}}, \citenamefont {{Butkevich}}, \citenamefont
  {{Carret}}, \citenamefont {{Carrion}}, \citenamefont {{Chemin}},
  \citenamefont {{Ch{\'e}reau}}, \citenamefont {{Corcione}}, \citenamefont
  {{Darmigny}}, \citenamefont {{de Boer}}, \citenamefont {{de Teodoro}},
  \citenamefont {{de Zeeuw}}, \citenamefont {{Delle Luche}}, \citenamefont
  {{Domingues}}, \citenamefont {{Dubath}}, \citenamefont {{Fodor}},
  \citenamefont {{Fr{\'e}zouls}}, \citenamefont {{Fries}}, \citenamefont
  {{Fustes}}, \citenamefont {{Fyfe}}, \citenamefont {{Gallardo}}, \citenamefont
  {{Gallegos}}, \citenamefont {{Gardiol}}, \citenamefont {{Gebran}},
  \citenamefont {{Gomboc}}, \citenamefont {{G{\'o}mez}}, \citenamefont
  {{Grux}}, \citenamefont {{Gueguen}}, \citenamefont {{Heyrovsky}},
  \citenamefont {{Hoar}}, \citenamefont {{Iannicola}}, \citenamefont {{Isasi
  Parache}}, \citenamefont {{Janotto}}, \citenamefont {{Joliet}}, \citenamefont
  {{Jonckheere}}, \citenamefont {{Keil}}, \citenamefont {{Kim}}, \citenamefont
  {{Klagyivik}}, \citenamefont {{Klar}}, \citenamefont {{Knude}}, \citenamefont
  {{Kochukhov}}, \citenamefont {{Kolka}}, \citenamefont {{Kos}}, \citenamefont
  {{Kutka}}, \citenamefont {{Lainey}}, \citenamefont {{LeBouquin}},
  \citenamefont {{Liu}}, \citenamefont {{Loreggia}}, \citenamefont {{Makarov}},
  \citenamefont {{Marseille}}, \citenamefont {{Martayan}}, \citenamefont
  {{Martinez-Rubi}}, \citenamefont {{Massart}}, \citenamefont {{Meynadier}},
  \citenamefont {{Mignot}}, \citenamefont {{Munari}}, \citenamefont {{Nguyen}},
  \citenamefont {{Nordlander}}, \citenamefont {{Ocvirk}}, \citenamefont
  {{O'Flaherty}}, \citenamefont {{Olias Sanz}}, \citenamefont {{Ortiz}},
  \citenamefont {{Osorio}}, \citenamefont {{Oszkiewicz}}, \citenamefont
  {{Ouzounis}}, \citenamefont {{Palmer}}, \citenamefont {{Park}}, \citenamefont
  {{Pasquato}}, \citenamefont {{Peltzer}}, \citenamefont {{Peralta}},
  \citenamefont {{P{\'e}turaud}}, \citenamefont {{Pieniluoma}}, \citenamefont
  {{Pigozzi}}, \citenamefont {{Poels}}, \citenamefont {{Prat}}, \citenamefont
  {{Prod'homme}}, \citenamefont {{Raison}}, \citenamefont {{Rebordao}},
  \citenamefont {{Risquez}}, \citenamefont {{Rocca-Volmerange}}, \citenamefont
  {{Rosen}}, \citenamefont {{Ruiz-Fuertes}}, \citenamefont {{Russo}},
  \citenamefont {{Sembay}}, \citenamefont {{Serraller Vizcaino}}, \citenamefont
  {{Short}}, \citenamefont {{Siebert}}, \citenamefont {{Silva}}, \citenamefont
  {{Sinachopoulos}}, \citenamefont {{Slezak}}, \citenamefont {{Soffel}},
  \citenamefont {{Sosnowska}}, \citenamefont {{Strai{\v{z}}ys}}, \citenamefont
  {{ter Linden}}, \citenamefont {{Terrell}}, \citenamefont {{Theil}},
  \citenamefont {{Tiede}}, \citenamefont {{Troisi}}, \citenamefont
  {{Tsalmantza}}, \citenamefont {{Tur}}, \citenamefont {{Vaccari}},
  \citenamefont {{Vachier}}, \citenamefont {{Valles}}, \citenamefont {{Van
  Hamme}}, \citenamefont {{Veltz}}, \citenamefont {{Virtanen}}, \citenamefont
  {{Wallut}}, \citenamefont {{Wichmann}}, \citenamefont {{Wilkinson}},
  \citenamefont {{Ziaeepour}},\ and\ \citenamefont
  {{Zschocke}}}]{2016A&A...595A...1G}%
  \BibitemOpen
  \bibfield  {author} {\bibinfo {author} {\bibnamefont {{Gaia Collaboration}}},
  \bibinfo {author} {\bibfnamefont {T.}~\bibnamefont {{Prusti}}}, \bibinfo
  {author} {\bibfnamefont {J.~H.~J.}\ \bibnamefont {{de Bruijne}}}, \bibinfo
  {author} {\bibfnamefont {A.~G.~A.}\ \bibnamefont {{Brown}}}, \bibinfo
  {author} {\bibfnamefont {A.}~\bibnamefont {{Vallenari}}}, \bibinfo {author}
  {\bibfnamefont {C.}~\bibnamefont {{Babusiaux}}}, \bibinfo {author}
  {\bibfnamefont {C.~A.~L.}\ \bibnamefont {{Bailer-Jones}}}, \bibinfo {author}
  {\bibfnamefont {U.}~\bibnamefont {{Bastian}}}, \bibinfo {author}
  {\bibfnamefont {M.}~\bibnamefont {{Biermann}}}, \bibinfo {author}
  {\bibfnamefont {D.~W.}\ \bibnamefont {{Evans}}}, \bibinfo {author}
  {\bibfnamefont {L.}~\bibnamefont {{Eyer}}}, \bibinfo {author} {\bibfnamefont
  {F.}~\bibnamefont {{Jansen}}}, \bibinfo {author} {\bibfnamefont
  {C.}~\bibnamefont {{Jordi}}}, \bibinfo {author} {\bibfnamefont {S.~A.}\
  \bibnamefont {{Klioner}}}, \bibinfo {author} {\bibfnamefont {U.}~\bibnamefont
  {{Lammers}}}, \bibinfo {author} {\bibfnamefont {L.}~\bibnamefont
  {{Lindegren}}}, \bibinfo {author} {\bibfnamefont {X.}~\bibnamefont {{Luri}}},
  \bibinfo {author} {\bibfnamefont {F.}~\bibnamefont {{Mignard}}}, \bibinfo
  {author} {\bibfnamefont {D.~J.}\ \bibnamefont {{Milligan}}}, \bibinfo
  {author} {\bibfnamefont {C.}~\bibnamefont {{Panem}}}, \bibinfo {author}
  {\bibfnamefont {V.}~\bibnamefont {{Poinsignon}}}, \bibinfo {author}
  {\bibfnamefont {D.}~\bibnamefont {{Pourbaix}}}, \bibinfo {author}
  {\bibfnamefont {S.}~\bibnamefont {{Randich}}}, \bibinfo {author}
  {\bibfnamefont {G.}~\bibnamefont {{Sarri}}}, \bibinfo {author} {\bibfnamefont
  {P.}~\bibnamefont {{Sartoretti}}}, \bibinfo {author} {\bibfnamefont {H.~I.}\
  \bibnamefont {{Siddiqui}}}, \bibinfo {author} {\bibfnamefont
  {C.}~\bibnamefont {{Soubiran}}}, \bibinfo {author} {\bibfnamefont
  {V.}~\bibnamefont {{Valette}}}, \bibinfo {author} {\bibfnamefont
  {F.}~\bibnamefont {{van Leeuwen}}}, \bibinfo {author} {\bibfnamefont {N.~A.}\
  \bibnamefont {{Walton}}}, \bibinfo {author} {\bibfnamefont {C.}~\bibnamefont
  {{Aerts}}}, \bibinfo {author} {\bibfnamefont {F.}~\bibnamefont {{Arenou}}},
  \bibinfo {author} {\bibfnamefont {M.}~\bibnamefont {{Cropper}}}, \bibinfo
  {author} {\bibfnamefont {R.}~\bibnamefont {{Drimmel}}}, \bibinfo {author}
  {\bibfnamefont {E.}~\bibnamefont {{H{\o}g}}}, \bibinfo {author}
  {\bibfnamefont {D.}~\bibnamefont {{Katz}}}, \bibinfo {author} {\bibfnamefont
  {M.~G.}\ \bibnamefont {{Lattanzi}}}, \bibinfo {author} {\bibfnamefont
  {W.}~\bibnamefont {{O'Mullane}}}, \bibinfo {author} {\bibfnamefont {E.~K.}\
  \bibnamefont {{Grebel}}}, \bibinfo {author} {\bibfnamefont {A.~D.}\
  \bibnamefont {{Holland}}}, \bibinfo {author} {\bibfnamefont {C.}~\bibnamefont
  {{Huc}}}, \bibinfo {author} {\bibfnamefont {X.}~\bibnamefont {{Passot}}},
  \bibinfo {author} {\bibfnamefont {L.}~\bibnamefont {{Bramante}}}, \bibinfo
  {author} {\bibfnamefont {C.}~\bibnamefont {{Cacciari}}}, \bibinfo {author}
  {\bibfnamefont {J.}~\bibnamefont {{Casta{\~n}eda}}}, \bibinfo {author}
  {\bibfnamefont {L.}~\bibnamefont {{Chaoul}}}, \bibinfo {author}
  {\bibfnamefont {N.}~\bibnamefont {{Cheek}}}, \bibinfo {author} {\bibfnamefont
  {F.}~\bibnamefont {{De Angeli}}}, \bibinfo {author} {\bibfnamefont
  {C.}~\bibnamefont {{Fabricius}}}, \bibinfo {author} {\bibfnamefont
  {R.}~\bibnamefont {{Guerra}}}, \bibinfo {author} {\bibfnamefont
  {J.}~\bibnamefont {{Hern{\'a}ndez}}}, \bibinfo {author} {\bibfnamefont
  {A.}~\bibnamefont {{Jean-Antoine-Piccolo}}}, \bibinfo {author} {\bibfnamefont
  {E.}~\bibnamefont {{Masana}}}, \bibinfo {author} {\bibfnamefont
  {R.}~\bibnamefont {{Messineo}}}, \bibinfo {author} {\bibfnamefont
  {N.}~\bibnamefont {{Mowlavi}}}, \bibinfo {author} {\bibfnamefont
  {K.}~\bibnamefont {{Nienartowicz}}}, \bibinfo {author} {\bibfnamefont
  {D.}~\bibnamefont {{Ord{\'o}{\~n}ez-Blanco}}}, \bibinfo {author}
  {\bibfnamefont {P.}~\bibnamefont {{Panuzzo}}}, \bibinfo {author}
  {\bibfnamefont {J.}~\bibnamefont {{Portell}}}, \bibinfo {author}
  {\bibfnamefont {P.~J.}\ \bibnamefont {{Richards}}}, \bibinfo {author}
  {\bibfnamefont {M.}~\bibnamefont {{Riello}}}, \bibinfo {author}
  {\bibfnamefont {G.~M.}\ \bibnamefont {{Seabroke}}}, \bibinfo {author}
  {\bibfnamefont {P.}~\bibnamefont {{Tanga}}}, \bibinfo {author} {\bibfnamefont
  {F.}~\bibnamefont {{Th{\'e}venin}}}, \bibinfo {author} {\bibfnamefont
  {J.}~\bibnamefont {{Torra}}}, \bibinfo {author} {\bibfnamefont {S.~G.}\
  \bibnamefont {{Els}}}, \bibinfo {author} {\bibfnamefont {G.}~\bibnamefont
  {{Gracia-Abril}}}, \bibinfo {author} {\bibfnamefont {G.}~\bibnamefont
  {{Comoretto}}}, \bibinfo {author} {\bibfnamefont {M.}~\bibnamefont
  {{Garcia-Reinaldos}}}, \bibinfo {author} {\bibfnamefont {T.}~\bibnamefont
  {{Lock}}}, \bibinfo {author} {\bibfnamefont {E.}~\bibnamefont {{Mercier}}},
  \bibinfo {author} {\bibfnamefont {M.}~\bibnamefont {{Altmann}}}, \bibinfo
  {author} {\bibfnamefont {R.}~\bibnamefont {{Andrae}}}, \bibinfo {author}
  {\bibfnamefont {T.~L.}\ \bibnamefont {{Astraatmadja}}}, \bibinfo {author}
  {\bibfnamefont {I.}~\bibnamefont {{Bellas-Velidis}}}, \bibinfo {author}
  {\bibfnamefont {K.}~\bibnamefont {{Benson}}}, \bibinfo {author}
  {\bibfnamefont {J.}~\bibnamefont {{Berthier}}}, \bibinfo {author}
  {\bibfnamefont {R.}~\bibnamefont {{Blomme}}}, \bibinfo {author}
  {\bibfnamefont {G.}~\bibnamefont {{Busso}}}, \bibinfo {author} {\bibfnamefont
  {B.}~\bibnamefont {{Carry}}}, \bibinfo {author} {\bibfnamefont
  {A.}~\bibnamefont {{Cellino}}}, \bibinfo {author} {\bibfnamefont
  {G.}~\bibnamefont {{Clementini}}}, \bibinfo {author} {\bibfnamefont
  {S.}~\bibnamefont {{Cowell}}}, \bibinfo {author} {\bibfnamefont
  {O.}~\bibnamefont {{Creevey}}}, \bibinfo {author} {\bibfnamefont
  {J.}~\bibnamefont {{Cuypers}}}, \bibinfo {author} {\bibfnamefont
  {M.}~\bibnamefont {{Davidson}}}, \bibinfo {author} {\bibfnamefont
  {J.}~\bibnamefont {{De Ridder}}}, \bibinfo {author} {\bibfnamefont
  {A.}~\bibnamefont {{de Torres}}}, \bibinfo {author} {\bibfnamefont
  {L.}~\bibnamefont {{Delchambre}}}, \bibinfo {author} {\bibfnamefont
  {A.}~\bibnamefont {{Dell'Oro}}}, \bibinfo {author} {\bibfnamefont
  {C.}~\bibnamefont {{Ducourant}}}, \bibinfo {author} {\bibfnamefont
  {Y.}~\bibnamefont {{Fr{\'e}mat}}}, \bibinfo {author} {\bibfnamefont
  {M.}~\bibnamefont {{Garc{\'\i}a-Torres}}}, \bibinfo {author} {\bibfnamefont
  {E.}~\bibnamefont {{Gosset}}}, \bibinfo {author} {\bibfnamefont {J.~L.}\
  \bibnamefont {{Halbwachs}}}, \bibinfo {author} {\bibfnamefont {N.~C.}\
  \bibnamefont {{Hambly}}}, \bibinfo {author} {\bibfnamefont {D.~L.}\
  \bibnamefont {{Harrison}}}, \bibinfo {author} {\bibfnamefont
  {M.}~\bibnamefont {{Hauser}}}, \bibinfo {author} {\bibfnamefont
  {D.}~\bibnamefont {{Hestroffer}}}, \bibinfo {author} {\bibfnamefont {S.~T.}\
  \bibnamefont {{Hodgkin}}}, \bibinfo {author} {\bibfnamefont {H.~E.}\
  \bibnamefont {{Huckle}}}, \bibinfo {author} {\bibfnamefont {A.}~\bibnamefont
  {{Hutton}}}, \bibinfo {author} {\bibfnamefont {G.}~\bibnamefont
  {{Jasniewicz}}}, \bibinfo {author} {\bibfnamefont {S.}~\bibnamefont
  {{Jordan}}}, \bibinfo {author} {\bibfnamefont {M.}~\bibnamefont
  {{Kontizas}}}, \bibinfo {author} {\bibfnamefont {A.~J.}\ \bibnamefont
  {{Korn}}}, \bibinfo {author} {\bibfnamefont {A.~C.}\ \bibnamefont
  {{Lanzafame}}}, \bibinfo {author} {\bibfnamefont {M.}~\bibnamefont
  {{Manteiga}}}, \bibinfo {author} {\bibfnamefont {A.}~\bibnamefont
  {{Moitinho}}}, \bibinfo {author} {\bibfnamefont {K.}~\bibnamefont
  {{Muinonen}}}, \bibinfo {author} {\bibfnamefont {J.}~\bibnamefont
  {{Osinde}}}, \bibinfo {author} {\bibfnamefont {E.}~\bibnamefont {{Pancino}}},
  \bibinfo {author} {\bibfnamefont {T.}~\bibnamefont {{Pauwels}}}, \bibinfo
  {author} {\bibfnamefont {J.~M.}\ \bibnamefont {{Petit}}}, \bibinfo {author}
  {\bibfnamefont {A.}~\bibnamefont {{Recio-Blanco}}}, \bibinfo {author}
  {\bibfnamefont {A.~C.}\ \bibnamefont {{Robin}}}, \bibinfo {author}
  {\bibfnamefont {L.~M.}\ \bibnamefont {{Sarro}}}, \bibinfo {author}
  {\bibfnamefont {C.}~\bibnamefont {{Siopis}}}, \bibinfo {author}
  {\bibfnamefont {M.}~\bibnamefont {{Smith}}}, \bibinfo {author} {\bibfnamefont
  {K.~W.}\ \bibnamefont {{Smith}}}, \bibinfo {author} {\bibfnamefont
  {A.}~\bibnamefont {{Sozzetti}}}, \bibinfo {author} {\bibfnamefont
  {W.}~\bibnamefont {{Thuillot}}}, \bibinfo {author} {\bibfnamefont
  {W.}~\bibnamefont {{van Reeven}}}, \bibinfo {author} {\bibfnamefont
  {Y.}~\bibnamefont {{Viala}}}, \bibinfo {author} {\bibfnamefont
  {U.}~\bibnamefont {{Abbas}}}, \bibinfo {author} {\bibfnamefont
  {A.}~\bibnamefont {{Abreu Aramburu}}}, \bibinfo {author} {\bibfnamefont
  {S.}~\bibnamefont {{Accart}}}, \bibinfo {author} {\bibfnamefont {J.~J.}\
  \bibnamefont {{Aguado}}}, \bibinfo {author} {\bibfnamefont {P.~M.}\
  \bibnamefont {{Allan}}}, \bibinfo {author} {\bibfnamefont {W.}~\bibnamefont
  {{Allasia}}}, \bibinfo {author} {\bibfnamefont {G.}~\bibnamefont
  {{Altavilla}}}, \bibinfo {author} {\bibfnamefont {M.~A.}\ \bibnamefont
  {{{\'A}lvarez}}}, \bibinfo {author} {\bibfnamefont {J.}~\bibnamefont
  {{Alves}}}, \bibinfo {author} {\bibfnamefont {R.~I.}\ \bibnamefont
  {{Anderson}}}, \bibinfo {author} {\bibfnamefont {A.~H.}\ \bibnamefont
  {{Andrei}}}, \bibinfo {author} {\bibfnamefont {E.}~\bibnamefont {{Anglada
  Varela}}}, \bibinfo {author} {\bibfnamefont {E.}~\bibnamefont {{Antiche}}},
  \bibinfo {author} {\bibfnamefont {T.}~\bibnamefont {{Antoja}}}, \bibinfo
  {author} {\bibfnamefont {S.}~\bibnamefont {{Ant{\'o}n}}}, \bibinfo {author}
  {\bibfnamefont {B.}~\bibnamefont {{Arcay}}}, \bibinfo {author} {\bibfnamefont
  {A.}~\bibnamefont {{Atzei}}}, \bibinfo {author} {\bibfnamefont
  {L.}~\bibnamefont {{Ayache}}}, \bibinfo {author} {\bibfnamefont
  {N.}~\bibnamefont {{Bach}}}, \bibinfo {author} {\bibfnamefont {S.~G.}\
  \bibnamefont {{Baker}}}, \bibinfo {author} {\bibfnamefont {L.}~\bibnamefont
  {{Balaguer-N{\'u}{\~n}ez}}}, \bibinfo {author} {\bibfnamefont
  {C.}~\bibnamefont {{Barache}}}, \bibinfo {author} {\bibfnamefont
  {C.}~\bibnamefont {{Barata}}}, \bibinfo {author} {\bibfnamefont
  {A.}~\bibnamefont {{Barbier}}}, \bibinfo {author} {\bibfnamefont
  {F.}~\bibnamefont {{Barblan}}}, \bibinfo {author} {\bibfnamefont
  {M.}~\bibnamefont {{Baroni}}}, \bibinfo {author} {\bibfnamefont
  {D.}~\bibnamefont {{Barrado y Navascu{\'e}s}}}, \bibinfo {author}
  {\bibfnamefont {M.}~\bibnamefont {{Barros}}}, \bibinfo {author}
  {\bibfnamefont {M.~A.}\ \bibnamefont {{Barstow}}}, \bibinfo {author}
  {\bibfnamefont {U.}~\bibnamefont {{Becciani}}}, \bibinfo {author}
  {\bibfnamefont {M.}~\bibnamefont {{Bellazzini}}}, \bibinfo {author}
  {\bibfnamefont {G.}~\bibnamefont {{Bellei}}}, \bibinfo {author}
  {\bibfnamefont {A.}~\bibnamefont {{Bello Garc{\'\i}a}}}, \bibinfo {author}
  {\bibfnamefont {V.}~\bibnamefont {{Belokurov}}}, \bibinfo {author}
  {\bibfnamefont {P.}~\bibnamefont {{Bendjoya}}}, \bibinfo {author}
  {\bibfnamefont {A.}~\bibnamefont {{Berihuete}}}, \bibinfo {author}
  {\bibfnamefont {L.}~\bibnamefont {{Bianchi}}}, \bibinfo {author}
  {\bibfnamefont {O.}~\bibnamefont {{Bienaym{\'e}}}}, \bibinfo {author}
  {\bibfnamefont {F.}~\bibnamefont {{Billebaud}}}, \bibinfo {author}
  {\bibfnamefont {N.}~\bibnamefont {{Blagorodnova}}}, \bibinfo {author}
  {\bibfnamefont {S.}~\bibnamefont {{Blanco-Cuaresma}}}, \bibinfo {author}
  {\bibfnamefont {T.}~\bibnamefont {{Boch}}}, \bibinfo {author} {\bibfnamefont
  {A.}~\bibnamefont {{Bombrun}}}, \bibinfo {author} {\bibfnamefont
  {R.}~\bibnamefont {{Borrachero}}}, \bibinfo {author} {\bibfnamefont
  {S.}~\bibnamefont {{Bouquillon}}}, \bibinfo {author} {\bibfnamefont
  {G.}~\bibnamefont {{Bourda}}}, \bibinfo {author} {\bibfnamefont
  {H.}~\bibnamefont {{Bouy}}}, \bibinfo {author} {\bibfnamefont
  {A.}~\bibnamefont {{Bragaglia}}}, \bibinfo {author} {\bibfnamefont {M.~A.}\
  \bibnamefont {{Breddels}}}, \bibinfo {author} {\bibfnamefont
  {N.}~\bibnamefont {{Brouillet}}}, \bibinfo {author} {\bibfnamefont
  {T.}~\bibnamefont {{Br{\"u}semeister}}}, \bibinfo {author} {\bibfnamefont
  {B.}~\bibnamefont {{Bucciarelli}}}, \bibinfo {author} {\bibfnamefont
  {F.}~\bibnamefont {{Budnik}}}, \bibinfo {author} {\bibfnamefont
  {P.}~\bibnamefont {{Burgess}}}, \bibinfo {author} {\bibfnamefont
  {R.}~\bibnamefont {{Burgon}}}, \bibinfo {author} {\bibfnamefont
  {A.}~\bibnamefont {{Burlacu}}}, \bibinfo {author} {\bibfnamefont
  {D.}~\bibnamefont {{Busonero}}}, \bibinfo {author} {\bibfnamefont
  {R.}~\bibnamefont {{Buzzi}}}, \bibinfo {author} {\bibfnamefont
  {E.}~\bibnamefont {{Caffau}}}, \bibinfo {author} {\bibfnamefont
  {J.}~\bibnamefont {{Cambras}}}, \bibinfo {author} {\bibfnamefont
  {H.}~\bibnamefont {{Campbell}}}, \bibinfo {author} {\bibfnamefont
  {R.}~\bibnamefont {{Cancelliere}}}, \bibinfo {author} {\bibfnamefont
  {T.}~\bibnamefont {{Cantat-Gaudin}}}, \bibinfo {author} {\bibfnamefont
  {T.}~\bibnamefont {{Carlucci}}}, \bibinfo {author} {\bibfnamefont {J.~M.}\
  \bibnamefont {{Carrasco}}}, \bibinfo {author} {\bibfnamefont
  {M.}~\bibnamefont {{Castellani}}}, \bibinfo {author} {\bibfnamefont
  {P.}~\bibnamefont {{Charlot}}}, \bibinfo {author} {\bibfnamefont
  {J.}~\bibnamefont {{Charnas}}}, \bibinfo {author} {\bibfnamefont
  {P.}~\bibnamefont {{Charvet}}}, \bibinfo {author} {\bibfnamefont
  {F.}~\bibnamefont {{Chassat}}}, \bibinfo {author} {\bibfnamefont
  {A.}~\bibnamefont {{Chiavassa}}}, \bibinfo {author} {\bibfnamefont
  {M.}~\bibnamefont {{Clotet}}}, \bibinfo {author} {\bibfnamefont
  {G.}~\bibnamefont {{Cocozza}}}, \bibinfo {author} {\bibfnamefont {R.~S.}\
  \bibnamefont {{Collins}}}, \bibinfo {author} {\bibfnamefont {P.}~\bibnamefont
  {{Collins}}}, \bibinfo {author} {\bibfnamefont {G.}~\bibnamefont
  {{Costigan}}}, \bibinfo {author} {\bibfnamefont {F.}~\bibnamefont {{Crifo}}},
  \bibinfo {author} {\bibfnamefont {N.~J.~G.}\ \bibnamefont {{Cross}}},
  \bibinfo {author} {\bibfnamefont {M.}~\bibnamefont {{Crosta}}}, \bibinfo
  {author} {\bibfnamefont {C.}~\bibnamefont {{Crowley}}}, \bibinfo {author}
  {\bibfnamefont {C.}~\bibnamefont {{Dafonte}}}, \bibinfo {author}
  {\bibfnamefont {Y.}~\bibnamefont {{Damerdji}}}, \bibinfo {author}
  {\bibfnamefont {A.}~\bibnamefont {{Dapergolas}}}, \bibinfo {author}
  {\bibfnamefont {P.}~\bibnamefont {{David}}}, \bibinfo {author} {\bibfnamefont
  {M.}~\bibnamefont {{David}}}, \bibinfo {author} {\bibfnamefont
  {P.}~\bibnamefont {{De Cat}}}, \bibinfo {author} {\bibfnamefont
  {F.}~\bibnamefont {{de Felice}}}, \bibinfo {author} {\bibfnamefont
  {P.}~\bibnamefont {{de Laverny}}}, \bibinfo {author} {\bibfnamefont
  {F.}~\bibnamefont {{De Luise}}}, \bibinfo {author} {\bibfnamefont
  {R.}~\bibnamefont {{De March}}}, \bibinfo {author} {\bibfnamefont
  {D.}~\bibnamefont {{de Martino}}}, \bibinfo {author} {\bibfnamefont
  {R.}~\bibnamefont {{de Souza}}}, \bibinfo {author} {\bibfnamefont
  {J.}~\bibnamefont {{Debosscher}}}, \bibinfo {author} {\bibfnamefont
  {E.}~\bibnamefont {{del Pozo}}}, \bibinfo {author} {\bibfnamefont
  {M.}~\bibnamefont {{Delbo}}}, \bibinfo {author} {\bibfnamefont
  {A.}~\bibnamefont {{Delgado}}}, \bibinfo {author} {\bibfnamefont {H.~E.}\
  \bibnamefont {{Delgado}}}, \bibinfo {author} {\bibfnamefont {F.}~\bibnamefont
  {{di Marco}}}, \bibinfo {author} {\bibfnamefont {P.}~\bibnamefont {{Di
  Matteo}}}, \bibinfo {author} {\bibfnamefont {S.}~\bibnamefont {{Diakite}}},
  \bibinfo {author} {\bibfnamefont {E.}~\bibnamefont {{Distefano}}}, \bibinfo
  {author} {\bibfnamefont {C.}~\bibnamefont {{Dolding}}}, \bibinfo {author}
  {\bibfnamefont {S.}~\bibnamefont {{Dos Anjos}}}, \bibinfo {author}
  {\bibfnamefont {P.}~\bibnamefont {{Drazinos}}}, \bibinfo {author}
  {\bibfnamefont {J.}~\bibnamefont {{Dur{\'a}n}}}, \bibinfo {author}
  {\bibfnamefont {Y.}~\bibnamefont {{Dzigan}}}, \bibinfo {author}
  {\bibfnamefont {E.}~\bibnamefont {{Ecale}}}, \bibinfo {author} {\bibfnamefont
  {B.}~\bibnamefont {{Edvardsson}}}, \bibinfo {author} {\bibfnamefont
  {H.}~\bibnamefont {{Enke}}}, \bibinfo {author} {\bibfnamefont
  {M.}~\bibnamefont {{Erdmann}}}, \bibinfo {author} {\bibfnamefont
  {D.}~\bibnamefont {{Escolar}}}, \bibinfo {author} {\bibfnamefont
  {M.}~\bibnamefont {{Espina}}}, \bibinfo {author} {\bibfnamefont {N.~W.}\
  \bibnamefont {{Evans}}}, \bibinfo {author} {\bibfnamefont {G.}~\bibnamefont
  {{Eynard Bontemps}}}, \bibinfo {author} {\bibfnamefont {C.}~\bibnamefont
  {{Fabre}}}, \bibinfo {author} {\bibfnamefont {M.}~\bibnamefont {{Fabrizio}}},
  \bibinfo {author} {\bibfnamefont {S.}~\bibnamefont {{Faigler}}}, \bibinfo
  {author} {\bibfnamefont {A.~J.}\ \bibnamefont {{Falc{\~a}o}}}, \bibinfo
  {author} {\bibfnamefont {M.}~\bibnamefont {{Farr{\`a}s Casas}}}, \bibinfo
  {author} {\bibfnamefont {F.}~\bibnamefont {{Faye}}}, \bibinfo {author}
  {\bibfnamefont {L.}~\bibnamefont {{Federici}}}, \bibinfo {author}
  {\bibfnamefont {G.}~\bibnamefont {{Fedorets}}}, \bibinfo {author}
  {\bibfnamefont {J.}~\bibnamefont {{Fern{\'a}ndez-Hern{\'a}ndez}}}, \bibinfo
  {author} {\bibfnamefont {P.}~\bibnamefont {{Fernique}}}, \bibinfo {author}
  {\bibfnamefont {A.}~\bibnamefont {{Fienga}}}, \bibinfo {author}
  {\bibfnamefont {F.}~\bibnamefont {{Figueras}}}, \bibinfo {author}
  {\bibfnamefont {F.}~\bibnamefont {{Filippi}}}, \bibinfo {author}
  {\bibfnamefont {K.}~\bibnamefont {{Findeisen}}}, \bibinfo {author}
  {\bibfnamefont {A.}~\bibnamefont {{Fonti}}}, \bibinfo {author} {\bibfnamefont
  {M.}~\bibnamefont {{Fouesneau}}}, \bibinfo {author} {\bibfnamefont
  {E.}~\bibnamefont {{Fraile}}}, \bibinfo {author} {\bibfnamefont
  {M.}~\bibnamefont {{Fraser}}}, \bibinfo {author} {\bibfnamefont
  {J.}~\bibnamefont {{Fuchs}}}, \bibinfo {author} {\bibfnamefont
  {R.}~\bibnamefont {{Furnell}}}, \bibinfo {author} {\bibfnamefont
  {M.}~\bibnamefont {{Gai}}}, \bibinfo {author} {\bibfnamefont
  {S.}~\bibnamefont {{Galleti}}}, \bibinfo {author} {\bibfnamefont
  {L.}~\bibnamefont {{Galluccio}}}, \bibinfo {author} {\bibfnamefont
  {D.}~\bibnamefont {{Garabato}}}, \bibinfo {author} {\bibfnamefont
  {F.}~\bibnamefont {{Garc{\'\i}a-Sedano}}}, \bibinfo {author} {\bibfnamefont
  {P.}~\bibnamefont {{Gar{\'e}}}}, \bibinfo {author} {\bibfnamefont
  {A.}~\bibnamefont {{Garofalo}}}, \bibinfo {author} {\bibfnamefont
  {N.}~\bibnamefont {{Garralda}}}, \bibinfo {author} {\bibfnamefont
  {P.}~\bibnamefont {{Gavras}}}, \bibinfo {author} {\bibfnamefont
  {J.}~\bibnamefont {{Gerssen}}}, \bibinfo {author} {\bibfnamefont
  {R.}~\bibnamefont {{Geyer}}}, \bibinfo {author} {\bibfnamefont
  {G.}~\bibnamefont {{Gilmore}}}, \bibinfo {author} {\bibfnamefont
  {S.}~\bibnamefont {{Girona}}}, \bibinfo {author} {\bibfnamefont
  {G.}~\bibnamefont {{Giuffrida}}}, \bibinfo {author} {\bibfnamefont
  {M.}~\bibnamefont {{Gomes}}}, \bibinfo {author} {\bibfnamefont
  {A.}~\bibnamefont {{Gonz{\'a}lez-Marcos}}}, \bibinfo {author} {\bibfnamefont
  {J.}~\bibnamefont {{Gonz{\'a}lez-N{\'u}{\~n}ez}}}, \bibinfo {author}
  {\bibfnamefont {J.~J.}\ \bibnamefont {{Gonz{\'a}lez-Vidal}}}, \bibinfo
  {author} {\bibfnamefont {M.}~\bibnamefont {{Granvik}}}, \bibinfo {author}
  {\bibfnamefont {A.}~\bibnamefont {{Guerrier}}}, \bibinfo {author}
  {\bibfnamefont {P.}~\bibnamefont {{Guillout}}}, \bibinfo {author}
  {\bibfnamefont {J.}~\bibnamefont {{Guiraud}}}, \bibinfo {author}
  {\bibfnamefont {A.}~\bibnamefont {{G{\'u}rpide}}}, \bibinfo {author}
  {\bibfnamefont {R.}~\bibnamefont {{Guti{\'e}rrez-S{\'a}nchez}}}, \bibinfo
  {author} {\bibfnamefont {L.~P.}\ \bibnamefont {{Guy}}}, \bibinfo {author}
  {\bibfnamefont {R.}~\bibnamefont {{Haigron}}}, \bibinfo {author}
  {\bibfnamefont {D.}~\bibnamefont {{Hatzidimitriou}}}, \bibinfo {author}
  {\bibfnamefont {M.}~\bibnamefont {{Haywood}}}, \bibinfo {author}
  {\bibfnamefont {U.}~\bibnamefont {{Heiter}}}, \bibinfo {author}
  {\bibfnamefont {A.}~\bibnamefont {{Helmi}}}, \bibinfo {author} {\bibfnamefont
  {D.}~\bibnamefont {{Hobbs}}}, \bibinfo {author} {\bibfnamefont
  {W.}~\bibnamefont {{Hofmann}}}, \bibinfo {author} {\bibfnamefont
  {B.}~\bibnamefont {{Holl}}}, \bibinfo {author} {\bibfnamefont
  {G.}~\bibnamefont {{Holland}}}, \bibinfo {author} {\bibfnamefont {J.~A.~S.}\
  \bibnamefont {{Hunt}}}, \bibinfo {author} {\bibfnamefont {A.}~\bibnamefont
  {{Hypki}}}, \bibinfo {author} {\bibfnamefont {V.}~\bibnamefont {{Icardi}}},
  \bibinfo {author} {\bibfnamefont {M.}~\bibnamefont {{Irwin}}}, \bibinfo
  {author} {\bibfnamefont {G.}~\bibnamefont {{Jevardat de Fombelle}}}, \bibinfo
  {author} {\bibfnamefont {P.}~\bibnamefont {{Jofr{\'e}}}}, \bibinfo {author}
  {\bibfnamefont {P.~G.}\ \bibnamefont {{Jonker}}}, \bibinfo {author}
  {\bibfnamefont {A.}~\bibnamefont {{Jorissen}}}, \bibinfo {author}
  {\bibfnamefont {F.}~\bibnamefont {{Julbe}}}, \bibinfo {author} {\bibfnamefont
  {A.}~\bibnamefont {{Karampelas}}}, \bibinfo {author} {\bibfnamefont
  {A.}~\bibnamefont {{Kochoska}}}, \bibinfo {author} {\bibfnamefont
  {R.}~\bibnamefont {{Kohley}}}, \bibinfo {author} {\bibfnamefont
  {K.}~\bibnamefont {{Kolenberg}}}, \bibinfo {author} {\bibfnamefont
  {E.}~\bibnamefont {{Kontizas}}}, \bibinfo {author} {\bibfnamefont {S.~E.}\
  \bibnamefont {{Koposov}}}, \bibinfo {author} {\bibfnamefont {G.}~\bibnamefont
  {{Kordopatis}}}, \bibinfo {author} {\bibfnamefont {P.}~\bibnamefont
  {{Koubsky}}}, \bibinfo {author} {\bibfnamefont {A.}~\bibnamefont
  {{Kowalczyk}}}, \bibinfo {author} {\bibfnamefont {A.}~\bibnamefont
  {{Krone-Martins}}}, \bibinfo {author} {\bibfnamefont {M.}~\bibnamefont
  {{Kudryashova}}}, \bibinfo {author} {\bibfnamefont {I.}~\bibnamefont
  {{Kull}}}, \bibinfo {author} {\bibfnamefont {R.~K.}\ \bibnamefont
  {{Bachchan}}}, \bibinfo {author} {\bibfnamefont {F.}~\bibnamefont
  {{Lacoste-Seris}}}, \bibinfo {author} {\bibfnamefont {A.~F.}\ \bibnamefont
  {{Lanza}}}, \bibinfo {author} {\bibfnamefont {J.~B.}\ \bibnamefont
  {{Lavigne}}}, \bibinfo {author} {\bibfnamefont {C.}~\bibnamefont {{Le
  Poncin-Lafitte}}}, \bibinfo {author} {\bibfnamefont {Y.}~\bibnamefont
  {{Lebreton}}}, \bibinfo {author} {\bibfnamefont {T.}~\bibnamefont
  {{Lebzelter}}}, \bibinfo {author} {\bibfnamefont {S.}~\bibnamefont
  {{Leccia}}}, \bibinfo {author} {\bibfnamefont {N.}~\bibnamefont {{Leclerc}}},
  \bibinfo {author} {\bibfnamefont {I.}~\bibnamefont {{Lecoeur-Taibi}}},
  \bibinfo {author} {\bibfnamefont {V.}~\bibnamefont {{Lemaitre}}}, \bibinfo
  {author} {\bibfnamefont {H.}~\bibnamefont {{Lenhardt}}}, \bibinfo {author}
  {\bibfnamefont {F.}~\bibnamefont {{Leroux}}}, \bibinfo {author}
  {\bibfnamefont {S.}~\bibnamefont {{Liao}}}, \bibinfo {author} {\bibfnamefont
  {E.}~\bibnamefont {{Licata}}}, \bibinfo {author} {\bibfnamefont {H.~E.~P.}\
  \bibnamefont {{Lindstr{\o}m}}}, \bibinfo {author} {\bibfnamefont {T.~A.}\
  \bibnamefont {{Lister}}}, \bibinfo {author} {\bibfnamefont {E.}~\bibnamefont
  {{Livanou}}}, \bibinfo {author} {\bibfnamefont {A.}~\bibnamefont {{Lobel}}},
  \bibinfo {author} {\bibfnamefont {W.}~\bibnamefont {{L{\"o}ffler}}}, \bibinfo
  {author} {\bibfnamefont {M.}~\bibnamefont {{L{\'o}pez}}}, \bibinfo {author}
  {\bibfnamefont {A.}~\bibnamefont {{Lopez-Lozano}}}, \bibinfo {author}
  {\bibfnamefont {D.}~\bibnamefont {{Lorenz}}}, \bibinfo {author}
  {\bibfnamefont {T.}~\bibnamefont {{Loureiro}}}, \bibinfo {author}
  {\bibfnamefont {I.}~\bibnamefont {{MacDonald}}}, \bibinfo {author}
  {\bibfnamefont {T.}~\bibnamefont {{Magalh{\~a}es Fernandes}}}, \bibinfo
  {author} {\bibfnamefont {S.}~\bibnamefont {{Managau}}}, \bibinfo {author}
  {\bibfnamefont {R.~G.}\ \bibnamefont {{Mann}}}, \bibinfo {author}
  {\bibfnamefont {G.}~\bibnamefont {{Mantelet}}}, \bibinfo {author}
  {\bibfnamefont {O.}~\bibnamefont {{Marchal}}}, \bibinfo {author}
  {\bibfnamefont {J.~M.}\ \bibnamefont {{Marchant}}}, \bibinfo {author}
  {\bibfnamefont {M.}~\bibnamefont {{Marconi}}}, \bibinfo {author}
  {\bibfnamefont {J.}~\bibnamefont {{Marie}}}, \bibinfo {author} {\bibfnamefont
  {S.}~\bibnamefont {{Marinoni}}}, \bibinfo {author} {\bibfnamefont {P.~M.}\
  \bibnamefont {{Marrese}}}, \bibinfo {author} {\bibfnamefont {G.}~\bibnamefont
  {{Marschalk{\'o}}}}, \bibinfo {author} {\bibfnamefont {D.~J.}\ \bibnamefont
  {{Marshall}}}, \bibinfo {author} {\bibfnamefont {J.~M.}\ \bibnamefont
  {{Mart{\'\i}n-Fleitas}}}, \bibinfo {author} {\bibfnamefont {M.}~\bibnamefont
  {{Martino}}}, \bibinfo {author} {\bibfnamefont {N.}~\bibnamefont {{Mary}}},
  \bibinfo {author} {\bibfnamefont {G.}~\bibnamefont {{Matijevi{\v{c}}}}},
  \bibinfo {author} {\bibfnamefont {T.}~\bibnamefont {{Mazeh}}}, \bibinfo
  {author} {\bibfnamefont {P.~J.}\ \bibnamefont {{McMillan}}}, \bibinfo
  {author} {\bibfnamefont {S.}~\bibnamefont {{Messina}}}, \bibinfo {author}
  {\bibfnamefont {A.}~\bibnamefont {{Mestre}}}, \bibinfo {author}
  {\bibfnamefont {D.}~\bibnamefont {{Michalik}}}, \bibinfo {author}
  {\bibfnamefont {N.~R.}\ \bibnamefont {{Millar}}}, \bibinfo {author}
  {\bibfnamefont {B.~M.~H.}\ \bibnamefont {{Miranda}}}, \bibinfo {author}
  {\bibfnamefont {D.}~\bibnamefont {{Molina}}}, \bibinfo {author}
  {\bibfnamefont {R.}~\bibnamefont {{Molinaro}}}, \bibinfo {author}
  {\bibfnamefont {M.}~\bibnamefont {{Molinaro}}}, \bibinfo {author}
  {\bibfnamefont {L.}~\bibnamefont {{Moln{\'a}r}}}, \bibinfo {author}
  {\bibfnamefont {M.}~\bibnamefont {{Moniez}}}, \bibinfo {author}
  {\bibfnamefont {P.}~\bibnamefont {{Montegriffo}}}, \bibinfo {author}
  {\bibfnamefont {D.}~\bibnamefont {{Monteiro}}}, \bibinfo {author}
  {\bibfnamefont {R.}~\bibnamefont {{Mor}}}, \bibinfo {author} {\bibfnamefont
  {A.}~\bibnamefont {{Mora}}}, \bibinfo {author} {\bibfnamefont
  {R.}~\bibnamefont {{Morbidelli}}}, \bibinfo {author} {\bibfnamefont
  {T.}~\bibnamefont {{Morel}}}, \bibinfo {author} {\bibfnamefont
  {S.}~\bibnamefont {{Morgenthaler}}}, \bibinfo {author} {\bibfnamefont
  {T.}~\bibnamefont {{Morley}}}, \bibinfo {author} {\bibfnamefont
  {D.}~\bibnamefont {{Morris}}}, \bibinfo {author} {\bibfnamefont {A.~F.}\
  \bibnamefont {{Mulone}}}, \bibinfo {author} {\bibfnamefont {T.}~\bibnamefont
  {{Muraveva}}}, \bibinfo {author} {\bibfnamefont {I.}~\bibnamefont
  {{Musella}}}, \bibinfo {author} {\bibfnamefont {J.}~\bibnamefont
  {{Narbonne}}}, \bibinfo {author} {\bibfnamefont {G.}~\bibnamefont
  {{Nelemans}}}, \bibinfo {author} {\bibfnamefont {L.}~\bibnamefont
  {{Nicastro}}}, \bibinfo {author} {\bibfnamefont {L.}~\bibnamefont {{Noval}}},
  \bibinfo {author} {\bibfnamefont {C.}~\bibnamefont {{Ord{\'e}novic}}},
  \bibinfo {author} {\bibfnamefont {J.}~\bibnamefont {{Ordieres-Mer{\'e}}}},
  \bibinfo {author} {\bibfnamefont {P.}~\bibnamefont {{Osborne}}}, \bibinfo
  {author} {\bibfnamefont {C.}~\bibnamefont {{Pagani}}}, \bibinfo {author}
  {\bibfnamefont {I.}~\bibnamefont {{Pagano}}}, \bibinfo {author}
  {\bibfnamefont {F.}~\bibnamefont {{Pailler}}}, \bibinfo {author}
  {\bibfnamefont {H.}~\bibnamefont {{Palacin}}}, \bibinfo {author}
  {\bibfnamefont {L.}~\bibnamefont {{Palaversa}}}, \bibinfo {author}
  {\bibfnamefont {P.}~\bibnamefont {{Parsons}}}, \bibinfo {author}
  {\bibfnamefont {T.}~\bibnamefont {{Paulsen}}}, \bibinfo {author}
  {\bibfnamefont {M.}~\bibnamefont {{Pecoraro}}}, \bibinfo {author}
  {\bibfnamefont {R.}~\bibnamefont {{Pedrosa}}}, \bibinfo {author}
  {\bibfnamefont {H.}~\bibnamefont {{Pentik{\"a}inen}}}, \bibinfo {author}
  {\bibfnamefont {J.}~\bibnamefont {{Pereira}}}, \bibinfo {author}
  {\bibfnamefont {B.}~\bibnamefont {{Pichon}}}, \bibinfo {author}
  {\bibfnamefont {A.~M.}\ \bibnamefont {{Piersimoni}}}, \bibinfo {author}
  {\bibfnamefont {F.~X.}\ \bibnamefont {{Pineau}}}, \bibinfo {author}
  {\bibfnamefont {E.}~\bibnamefont {{Plachy}}}, \bibinfo {author}
  {\bibfnamefont {G.}~\bibnamefont {{Plum}}}, \bibinfo {author} {\bibfnamefont
  {E.}~\bibnamefont {{Poujoulet}}}, \bibinfo {author} {\bibfnamefont
  {A.}~\bibnamefont {{Pr{\v{s}}a}}}, \bibinfo {author} {\bibfnamefont
  {L.}~\bibnamefont {{Pulone}}}, \bibinfo {author} {\bibfnamefont
  {S.}~\bibnamefont {{Ragaini}}}, \bibinfo {author} {\bibfnamefont
  {S.}~\bibnamefont {{Rago}}}, \bibinfo {author} {\bibfnamefont
  {N.}~\bibnamefont {{Rambaux}}}, \bibinfo {author} {\bibfnamefont
  {M.}~\bibnamefont {{Ramos-Lerate}}}, \bibinfo {author} {\bibfnamefont
  {P.}~\bibnamefont {{Ranalli}}}, \bibinfo {author} {\bibfnamefont
  {G.}~\bibnamefont {{Rauw}}}, \bibinfo {author} {\bibfnamefont
  {A.}~\bibnamefont {{Read}}}, \bibinfo {author} {\bibfnamefont
  {S.}~\bibnamefont {{Regibo}}}, \bibinfo {author} {\bibfnamefont
  {F.}~\bibnamefont {{Renk}}}, \bibinfo {author} {\bibfnamefont
  {C.}~\bibnamefont {{Reyl{\'e}}}}, \bibinfo {author} {\bibfnamefont {R.~A.}\
  \bibnamefont {{Ribeiro}}}, \bibinfo {author} {\bibfnamefont {L.}~\bibnamefont
  {{Rimoldini}}}, \bibinfo {author} {\bibfnamefont {V.}~\bibnamefont
  {{Ripepi}}}, \bibinfo {author} {\bibfnamefont {A.}~\bibnamefont {{Riva}}},
  \bibinfo {author} {\bibfnamefont {G.}~\bibnamefont {{Rixon}}}, \bibinfo
  {author} {\bibfnamefont {M.}~\bibnamefont {{Roelens}}}, \bibinfo {author}
  {\bibfnamefont {M.}~\bibnamefont {{Romero-G{\'o}mez}}}, \bibinfo {author}
  {\bibfnamefont {N.}~\bibnamefont {{Rowell}}}, \bibinfo {author}
  {\bibfnamefont {F.}~\bibnamefont {{Royer}}}, \bibinfo {author} {\bibfnamefont
  {A.}~\bibnamefont {{Rudolph}}}, \bibinfo {author} {\bibfnamefont
  {L.}~\bibnamefont {{Ruiz-Dern}}}, \bibinfo {author} {\bibfnamefont
  {G.}~\bibnamefont {{Sadowski}}}, \bibinfo {author} {\bibfnamefont
  {T.}~\bibnamefont {{Sagrist{\`a} Sell{\'e}s}}}, \bibinfo {author}
  {\bibfnamefont {J.}~\bibnamefont {{Sahlmann}}}, \bibinfo {author}
  {\bibfnamefont {J.}~\bibnamefont {{Salgado}}}, \bibinfo {author}
  {\bibfnamefont {E.}~\bibnamefont {{Salguero}}}, \bibinfo {author}
  {\bibfnamefont {M.}~\bibnamefont {{Sarasso}}}, \bibinfo {author}
  {\bibfnamefont {H.}~\bibnamefont {{Savietto}}}, \bibinfo {author}
  {\bibfnamefont {A.}~\bibnamefont {{Schnorhk}}}, \bibinfo {author}
  {\bibfnamefont {M.}~\bibnamefont {{Schultheis}}}, \bibinfo {author}
  {\bibfnamefont {E.}~\bibnamefont {{Sciacca}}}, \bibinfo {author}
  {\bibfnamefont {M.}~\bibnamefont {{Segol}}}, \bibinfo {author} {\bibfnamefont
  {J.~C.}\ \bibnamefont {{Segovia}}}, \bibinfo {author} {\bibfnamefont
  {D.}~\bibnamefont {{Segransan}}}, \bibinfo {author} {\bibfnamefont
  {E.}~\bibnamefont {{Serpell}}}, \bibinfo {author} {\bibfnamefont {I.~C.}\
  \bibnamefont {{Shih}}}, \bibinfo {author} {\bibfnamefont {R.}~\bibnamefont
  {{Smareglia}}}, \bibinfo {author} {\bibfnamefont {R.~L.}\ \bibnamefont
  {{Smart}}}, \bibinfo {author} {\bibfnamefont {C.}~\bibnamefont {{Smith}}},
  \bibinfo {author} {\bibfnamefont {E.}~\bibnamefont {{Solano}}}, \bibinfo
  {author} {\bibfnamefont {F.}~\bibnamefont {{Solitro}}}, \bibinfo {author}
  {\bibfnamefont {R.}~\bibnamefont {{Sordo}}}, \bibinfo {author} {\bibfnamefont
  {S.}~\bibnamefont {{Soria Nieto}}}, \bibinfo {author} {\bibfnamefont
  {J.}~\bibnamefont {{Souchay}}}, \bibinfo {author} {\bibfnamefont
  {A.}~\bibnamefont {{Spagna}}}, \bibinfo {author} {\bibfnamefont
  {F.}~\bibnamefont {{Spoto}}}, \bibinfo {author} {\bibfnamefont
  {U.}~\bibnamefont {{Stampa}}}, \bibinfo {author} {\bibfnamefont {I.~A.}\
  \bibnamefont {{Steele}}}, \bibinfo {author} {\bibfnamefont {H.}~\bibnamefont
  {{Steidelm{\"u}ller}}}, \bibinfo {author} {\bibfnamefont {C.~A.}\
  \bibnamefont {{Stephenson}}}, \bibinfo {author} {\bibfnamefont
  {H.}~\bibnamefont {{Stoev}}}, \bibinfo {author} {\bibfnamefont {F.~F.}\
  \bibnamefont {{Suess}}}, \bibinfo {author} {\bibfnamefont {M.}~\bibnamefont
  {{S{\"u}veges}}}, \bibinfo {author} {\bibfnamefont {J.}~\bibnamefont
  {{Surdej}}}, \bibinfo {author} {\bibfnamefont {L.}~\bibnamefont
  {{Szabados}}}, \bibinfo {author} {\bibfnamefont {E.}~\bibnamefont
  {{Szegedi-Elek}}}, \bibinfo {author} {\bibfnamefont {D.}~\bibnamefont
  {{Tapiador}}}, \bibinfo {author} {\bibfnamefont {F.}~\bibnamefont {{Taris}}},
  \bibinfo {author} {\bibfnamefont {G.}~\bibnamefont {{Tauran}}}, \bibinfo
  {author} {\bibfnamefont {M.~B.}\ \bibnamefont {{Taylor}}}, \bibinfo {author}
  {\bibfnamefont {R.}~\bibnamefont {{Teixeira}}}, \bibinfo {author}
  {\bibfnamefont {D.}~\bibnamefont {{Terrett}}}, \bibinfo {author}
  {\bibfnamefont {B.}~\bibnamefont {{Tingley}}}, \bibinfo {author}
  {\bibfnamefont {S.~C.}\ \bibnamefont {{Trager}}}, \bibinfo {author}
  {\bibfnamefont {C.}~\bibnamefont {{Turon}}}, \bibinfo {author} {\bibfnamefont
  {A.}~\bibnamefont {{Ulla}}}, \bibinfo {author} {\bibfnamefont
  {E.}~\bibnamefont {{Utrilla}}}, \bibinfo {author} {\bibfnamefont
  {G.}~\bibnamefont {{Valentini}}}, \bibinfo {author} {\bibfnamefont
  {A.}~\bibnamefont {{van Elteren}}}, \bibinfo {author} {\bibfnamefont
  {E.}~\bibnamefont {{Van Hemelryck}}}, \bibinfo {author} {\bibfnamefont
  {M.}~\bibnamefont {{van Leeuwen}}}, \bibinfo {author} {\bibfnamefont
  {M.}~\bibnamefont {{Varadi}}}, \bibinfo {author} {\bibfnamefont
  {A.}~\bibnamefont {{Vecchiato}}}, \bibinfo {author} {\bibfnamefont
  {J.}~\bibnamefont {{Veljanoski}}}, \bibinfo {author} {\bibfnamefont
  {T.}~\bibnamefont {{Via}}}, \bibinfo {author} {\bibfnamefont
  {D.}~\bibnamefont {{Vicente}}}, \bibinfo {author} {\bibfnamefont
  {S.}~\bibnamefont {{Vogt}}}, \bibinfo {author} {\bibfnamefont
  {H.}~\bibnamefont {{Voss}}}, \bibinfo {author} {\bibfnamefont
  {V.}~\bibnamefont {{Votruba}}}, \bibinfo {author} {\bibfnamefont
  {S.}~\bibnamefont {{Voutsinas}}}, \bibinfo {author} {\bibfnamefont
  {G.}~\bibnamefont {{Walmsley}}}, \bibinfo {author} {\bibfnamefont
  {M.}~\bibnamefont {{Weiler}}}, \bibinfo {author} {\bibfnamefont
  {K.}~\bibnamefont {{Weingrill}}}, \bibinfo {author} {\bibfnamefont
  {D.}~\bibnamefont {{Werner}}}, \bibinfo {author} {\bibfnamefont
  {T.}~\bibnamefont {{Wevers}}}, \bibinfo {author} {\bibfnamefont
  {G.}~\bibnamefont {{Whitehead}}}, \bibinfo {author} {\bibfnamefont
  {{\L}.}~\bibnamefont {{Wyrzykowski}}}, \bibinfo {author} {\bibfnamefont
  {A.}~\bibnamefont {{Yoldas}}}, \bibinfo {author} {\bibfnamefont
  {M.}~\bibnamefont {{{\v{Z}}erjal}}}, \bibinfo {author} {\bibfnamefont
  {S.}~\bibnamefont {{Zucker}}}, \bibinfo {author} {\bibfnamefont
  {C.}~\bibnamefont {{Zurbach}}}, \bibinfo {author} {\bibfnamefont
  {T.}~\bibnamefont {{Zwitter}}}, \bibinfo {author} {\bibfnamefont
  {A.}~\bibnamefont {{Alecu}}}, \bibinfo {author} {\bibfnamefont
  {M.}~\bibnamefont {{Allen}}}, \bibinfo {author} {\bibfnamefont
  {C.}~\bibnamefont {{Allende Prieto}}}, \bibinfo {author} {\bibfnamefont
  {A.}~\bibnamefont {{Amorim}}}, \bibinfo {author} {\bibfnamefont
  {G.}~\bibnamefont {{Anglada-Escud{\'e}}}}, \bibinfo {author} {\bibfnamefont
  {V.}~\bibnamefont {{Arsenijevic}}}, \bibinfo {author} {\bibfnamefont
  {S.}~\bibnamefont {{Azaz}}}, \bibinfo {author} {\bibfnamefont
  {P.}~\bibnamefont {{Balm}}}, \bibinfo {author} {\bibfnamefont
  {M.}~\bibnamefont {{Beck}}}, \bibinfo {author} {\bibfnamefont {H.~H.}\
  \bibnamefont {{Bernstein}}}, \bibinfo {author} {\bibfnamefont
  {L.}~\bibnamefont {{Bigot}}}, \bibinfo {author} {\bibfnamefont
  {A.}~\bibnamefont {{Bijaoui}}}, \bibinfo {author} {\bibfnamefont
  {C.}~\bibnamefont {{Blasco}}}, \bibinfo {author} {\bibfnamefont
  {M.}~\bibnamefont {{Bonfigli}}}, \bibinfo {author} {\bibfnamefont
  {G.}~\bibnamefont {{Bono}}}, \bibinfo {author} {\bibfnamefont
  {S.}~\bibnamefont {{Boudreault}}}, \bibinfo {author} {\bibfnamefont
  {A.}~\bibnamefont {{Bressan}}}, \bibinfo {author} {\bibfnamefont
  {S.}~\bibnamefont {{Brown}}}, \bibinfo {author} {\bibfnamefont {P.~M.}\
  \bibnamefont {{Brunet}}}, \bibinfo {author} {\bibfnamefont {P.}~\bibnamefont
  {{Bunclark}}}, \bibinfo {author} {\bibfnamefont {R.}~\bibnamefont
  {{Buonanno}}}, \bibinfo {author} {\bibfnamefont {A.~G.}\ \bibnamefont
  {{Butkevich}}}, \bibinfo {author} {\bibfnamefont {C.}~\bibnamefont
  {{Carret}}}, \bibinfo {author} {\bibfnamefont {C.}~\bibnamefont {{Carrion}}},
  \bibinfo {author} {\bibfnamefont {L.}~\bibnamefont {{Chemin}}}, \bibinfo
  {author} {\bibfnamefont {F.}~\bibnamefont {{Ch{\'e}reau}}}, \bibinfo {author}
  {\bibfnamefont {L.}~\bibnamefont {{Corcione}}}, \bibinfo {author}
  {\bibfnamefont {E.}~\bibnamefont {{Darmigny}}}, \bibinfo {author}
  {\bibfnamefont {K.~S.}\ \bibnamefont {{de Boer}}}, \bibinfo {author}
  {\bibfnamefont {P.}~\bibnamefont {{de Teodoro}}}, \bibinfo {author}
  {\bibfnamefont {P.~T.}\ \bibnamefont {{de Zeeuw}}}, \bibinfo {author}
  {\bibfnamefont {C.}~\bibnamefont {{Delle Luche}}}, \bibinfo {author}
  {\bibfnamefont {C.~D.}\ \bibnamefont {{Domingues}}}, \bibinfo {author}
  {\bibfnamefont {P.}~\bibnamefont {{Dubath}}}, \bibinfo {author}
  {\bibfnamefont {F.}~\bibnamefont {{Fodor}}}, \bibinfo {author} {\bibfnamefont
  {B.}~\bibnamefont {{Fr{\'e}zouls}}}, \bibinfo {author} {\bibfnamefont
  {A.}~\bibnamefont {{Fries}}}, \bibinfo {author} {\bibfnamefont
  {D.}~\bibnamefont {{Fustes}}}, \bibinfo {author} {\bibfnamefont
  {D.}~\bibnamefont {{Fyfe}}}, \bibinfo {author} {\bibfnamefont
  {E.}~\bibnamefont {{Gallardo}}}, \bibinfo {author} {\bibfnamefont
  {J.}~\bibnamefont {{Gallegos}}}, \bibinfo {author} {\bibfnamefont
  {D.}~\bibnamefont {{Gardiol}}}, \bibinfo {author} {\bibfnamefont
  {M.}~\bibnamefont {{Gebran}}}, \bibinfo {author} {\bibfnamefont
  {A.}~\bibnamefont {{Gomboc}}}, \bibinfo {author} {\bibfnamefont
  {A.}~\bibnamefont {{G{\'o}mez}}}, \bibinfo {author} {\bibfnamefont
  {E.}~\bibnamefont {{Grux}}}, \bibinfo {author} {\bibfnamefont
  {A.}~\bibnamefont {{Gueguen}}}, \bibinfo {author} {\bibfnamefont
  {A.}~\bibnamefont {{Heyrovsky}}}, \bibinfo {author} {\bibfnamefont
  {J.}~\bibnamefont {{Hoar}}}, \bibinfo {author} {\bibfnamefont
  {G.}~\bibnamefont {{Iannicola}}}, \bibinfo {author} {\bibfnamefont
  {Y.}~\bibnamefont {{Isasi Parache}}}, \bibinfo {author} {\bibfnamefont
  {A.~M.}\ \bibnamefont {{Janotto}}}, \bibinfo {author} {\bibfnamefont
  {E.}~\bibnamefont {{Joliet}}}, \bibinfo {author} {\bibfnamefont
  {A.}~\bibnamefont {{Jonckheere}}}, \bibinfo {author} {\bibfnamefont
  {R.}~\bibnamefont {{Keil}}}, \bibinfo {author} {\bibfnamefont {D.~W.}\
  \bibnamefont {{Kim}}}, \bibinfo {author} {\bibfnamefont {P.}~\bibnamefont
  {{Klagyivik}}}, \bibinfo {author} {\bibfnamefont {J.}~\bibnamefont {{Klar}}},
  \bibinfo {author} {\bibfnamefont {J.}~\bibnamefont {{Knude}}}, \bibinfo
  {author} {\bibfnamefont {O.}~\bibnamefont {{Kochukhov}}}, \bibinfo {author}
  {\bibfnamefont {I.}~\bibnamefont {{Kolka}}}, \bibinfo {author} {\bibfnamefont
  {J.}~\bibnamefont {{Kos}}}, \bibinfo {author} {\bibfnamefont
  {A.}~\bibnamefont {{Kutka}}}, \bibinfo {author} {\bibfnamefont
  {V.}~\bibnamefont {{Lainey}}}, \bibinfo {author} {\bibfnamefont
  {D.}~\bibnamefont {{LeBouquin}}}, \bibinfo {author} {\bibfnamefont
  {C.}~\bibnamefont {{Liu}}}, \bibinfo {author} {\bibfnamefont
  {D.}~\bibnamefont {{Loreggia}}}, \bibinfo {author} {\bibfnamefont {V.~V.}\
  \bibnamefont {{Makarov}}}, \bibinfo {author} {\bibfnamefont {M.~G.}\
  \bibnamefont {{Marseille}}}, \bibinfo {author} {\bibfnamefont
  {C.}~\bibnamefont {{Martayan}}}, \bibinfo {author} {\bibfnamefont
  {O.}~\bibnamefont {{Martinez-Rubi}}}, \bibinfo {author} {\bibfnamefont
  {B.}~\bibnamefont {{Massart}}}, \bibinfo {author} {\bibfnamefont
  {F.}~\bibnamefont {{Meynadier}}}, \bibinfo {author} {\bibfnamefont
  {S.}~\bibnamefont {{Mignot}}}, \bibinfo {author} {\bibfnamefont
  {U.}~\bibnamefont {{Munari}}}, \bibinfo {author} {\bibfnamefont {A.~T.}\
  \bibnamefont {{Nguyen}}}, \bibinfo {author} {\bibfnamefont {T.}~\bibnamefont
  {{Nordlander}}}, \bibinfo {author} {\bibfnamefont {P.}~\bibnamefont
  {{Ocvirk}}}, \bibinfo {author} {\bibfnamefont {K.~S.}\ \bibnamefont
  {{O'Flaherty}}}, \bibinfo {author} {\bibfnamefont {A.}~\bibnamefont {{Olias
  Sanz}}}, \bibinfo {author} {\bibfnamefont {P.}~\bibnamefont {{Ortiz}}},
  \bibinfo {author} {\bibfnamefont {J.}~\bibnamefont {{Osorio}}}, \bibinfo
  {author} {\bibfnamefont {D.}~\bibnamefont {{Oszkiewicz}}}, \bibinfo {author}
  {\bibfnamefont {A.}~\bibnamefont {{Ouzounis}}}, \bibinfo {author}
  {\bibfnamefont {M.}~\bibnamefont {{Palmer}}}, \bibinfo {author}
  {\bibfnamefont {P.}~\bibnamefont {{Park}}}, \bibinfo {author} {\bibfnamefont
  {E.}~\bibnamefont {{Pasquato}}}, \bibinfo {author} {\bibfnamefont
  {C.}~\bibnamefont {{Peltzer}}}, \bibinfo {author} {\bibfnamefont
  {J.}~\bibnamefont {{Peralta}}}, \bibinfo {author} {\bibfnamefont
  {F.}~\bibnamefont {{P{\'e}turaud}}}, \bibinfo {author} {\bibfnamefont
  {T.}~\bibnamefont {{Pieniluoma}}}, \bibinfo {author} {\bibfnamefont
  {E.}~\bibnamefont {{Pigozzi}}}, \bibinfo {author} {\bibfnamefont
  {J.}~\bibnamefont {{Poels}}}, \bibinfo {author} {\bibfnamefont
  {G.}~\bibnamefont {{Prat}}}, \bibinfo {author} {\bibfnamefont
  {T.}~\bibnamefont {{Prod'homme}}}, \bibinfo {author} {\bibfnamefont
  {F.}~\bibnamefont {{Raison}}}, \bibinfo {author} {\bibfnamefont {J.~M.}\
  \bibnamefont {{Rebordao}}}, \bibinfo {author} {\bibfnamefont
  {D.}~\bibnamefont {{Risquez}}}, \bibinfo {author} {\bibfnamefont
  {B.}~\bibnamefont {{Rocca-Volmerange}}}, \bibinfo {author} {\bibfnamefont
  {S.}~\bibnamefont {{Rosen}}}, \bibinfo {author} {\bibfnamefont {M.~I.}\
  \bibnamefont {{Ruiz-Fuertes}}}, \bibinfo {author} {\bibfnamefont
  {F.}~\bibnamefont {{Russo}}}, \bibinfo {author} {\bibfnamefont
  {S.}~\bibnamefont {{Sembay}}}, \bibinfo {author} {\bibfnamefont
  {I.}~\bibnamefont {{Serraller Vizcaino}}}, \bibinfo {author} {\bibfnamefont
  {A.}~\bibnamefont {{Short}}}, \bibinfo {author} {\bibfnamefont
  {A.}~\bibnamefont {{Siebert}}}, \bibinfo {author} {\bibfnamefont
  {H.}~\bibnamefont {{Silva}}}, \bibinfo {author} {\bibfnamefont
  {D.}~\bibnamefont {{Sinachopoulos}}}, \bibinfo {author} {\bibfnamefont
  {E.}~\bibnamefont {{Slezak}}}, \bibinfo {author} {\bibfnamefont
  {M.}~\bibnamefont {{Soffel}}}, \bibinfo {author} {\bibfnamefont
  {D.}~\bibnamefont {{Sosnowska}}}, \bibinfo {author} {\bibfnamefont
  {V.}~\bibnamefont {{Strai{\v{z}}ys}}}, \bibinfo {author} {\bibfnamefont
  {M.}~\bibnamefont {{ter Linden}}}, \bibinfo {author} {\bibfnamefont
  {D.}~\bibnamefont {{Terrell}}}, \bibinfo {author} {\bibfnamefont
  {S.}~\bibnamefont {{Theil}}}, \bibinfo {author} {\bibfnamefont
  {C.}~\bibnamefont {{Tiede}}}, \bibinfo {author} {\bibfnamefont
  {L.}~\bibnamefont {{Troisi}}}, \bibinfo {author} {\bibfnamefont
  {P.}~\bibnamefont {{Tsalmantza}}}, \bibinfo {author} {\bibfnamefont
  {D.}~\bibnamefont {{Tur}}}, \bibinfo {author} {\bibfnamefont
  {M.}~\bibnamefont {{Vaccari}}}, \bibinfo {author} {\bibfnamefont
  {F.}~\bibnamefont {{Vachier}}}, \bibinfo {author} {\bibfnamefont
  {P.}~\bibnamefont {{Valles}}}, \bibinfo {author} {\bibfnamefont
  {W.}~\bibnamefont {{Van Hamme}}}, \bibinfo {author} {\bibfnamefont
  {L.}~\bibnamefont {{Veltz}}}, \bibinfo {author} {\bibfnamefont
  {J.}~\bibnamefont {{Virtanen}}}, \bibinfo {author} {\bibfnamefont {J.~M.}\
  \bibnamefont {{Wallut}}}, \bibinfo {author} {\bibfnamefont {R.}~\bibnamefont
  {{Wichmann}}}, \bibinfo {author} {\bibfnamefont {M.~I.}\ \bibnamefont
  {{Wilkinson}}}, \bibinfo {author} {\bibfnamefont {H.}~\bibnamefont
  {{Ziaeepour}}},\ and\ \bibinfo {author} {\bibfnamefont {S.}~\bibnamefont
  {{Zschocke}}},\ }\bibfield  {title} {\bibinfo {title} {{The Gaia mission}},\
  }\href {https://doi.org/10.1051/0004-6361/201629272} {\bibfield  {journal}
  {\bibinfo  {journal} {\aap}\ }\textbf {\bibinfo {volume} {595}},\ \bibinfo
  {eid} {A1} (\bibinfo {year} {2016})},\ \Eprint
  {https://arxiv.org/abs/1609.04153} {arXiv:1609.04153 [astro-ph.IM]}
  \BibitemShut {NoStop}%
\bibitem [{\citenamefont {{Nucita}}\ \emph {et~al.}(2006)\citenamefont
  {{Nucita}}, \citenamefont {{De Paolis}}, \citenamefont {{Ingrosso}},
  \citenamefont {{Elia}}, \citenamefont {{de Plaa}},\ and\ \citenamefont
  {{Kaastra}}}]{2006ApJ...651.1092N}%
  \BibitemOpen
  \bibfield  {author} {\bibinfo {author} {\bibfnamefont {A.~A.}\ \bibnamefont
  {{Nucita}}}, \bibinfo {author} {\bibfnamefont {F.}~\bibnamefont {{De
  Paolis}}}, \bibinfo {author} {\bibfnamefont {G.}~\bibnamefont {{Ingrosso}}},
  \bibinfo {author} {\bibfnamefont {D.}~\bibnamefont {{Elia}}}, \bibinfo
  {author} {\bibfnamefont {J.}~\bibnamefont {{de Plaa}}},\ and\ \bibinfo
  {author} {\bibfnamefont {J.~S.}\ \bibnamefont {{Kaastra}}},\ }\bibfield
  {title} {\bibinfo {title} {{An XMM-Newton Search for X-Ray Emission from the
  Microlensing Event MACHO-96-BLG-5}},\ }\href {https://doi.org/10.1086/507784}
  {\bibfield  {journal} {\bibinfo  {journal} {\apj}\ }\textbf {\bibinfo
  {volume} {651}},\ \bibinfo {pages} {1092} (\bibinfo {year}
  {2006})}\BibitemShut {NoStop}%
\bibitem [{\citenamefont {{Bennett}}\ \emph {et~al.}(2002)\citenamefont
  {{Bennett}}, \citenamefont {{Becker}}, \citenamefont {{Quinn}}, \citenamefont
  {{Tomaney}}, \citenamefont {{Alcock}}, \citenamefont {{Allsman}},
  \citenamefont {{Alves}}, \citenamefont {{Axelrod}}, \citenamefont {{Calitz}},
  \citenamefont {{Cook}}, \citenamefont {{Drake}}, \citenamefont {{Fragile}},
  \citenamefont {{Freeman}}, \citenamefont {{Geha}}, \citenamefont {{Griest}},
  \citenamefont {{Johnson}}, \citenamefont {{Keller}}, \citenamefont {{Laws}},
  \citenamefont {{Lehner}}, \citenamefont {{Marshall}}, \citenamefont
  {{Minniti}}, \citenamefont {{Nelson}}, \citenamefont {{Peterson}},
  \citenamefont {{Popowski}}, \citenamefont {{Pratt}}, \citenamefont {{Quinn}},
  \citenamefont {{Rhie}}, \citenamefont {{Stubbs}}, \citenamefont
  {{Sutherland}}, \citenamefont {{Vandehei}}, \citenamefont {{Welch}},
  \citenamefont {{MACHO Collaboration}},\ and\ \citenamefont {{MPS
  Collaboration}}}]{2002ApJ...579..639B}%
  \BibitemOpen
  \bibfield  {author} {\bibinfo {author} {\bibfnamefont {D.~P.}\ \bibnamefont
  {{Bennett}}}, \bibinfo {author} {\bibfnamefont {A.~C.}\ \bibnamefont
  {{Becker}}}, \bibinfo {author} {\bibfnamefont {J.~L.}\ \bibnamefont
  {{Quinn}}}, \bibinfo {author} {\bibfnamefont {A.~B.}\ \bibnamefont
  {{Tomaney}}}, \bibinfo {author} {\bibfnamefont {C.}~\bibnamefont {{Alcock}}},
  \bibinfo {author} {\bibfnamefont {R.~A.}\ \bibnamefont {{Allsman}}}, \bibinfo
  {author} {\bibfnamefont {D.~R.}\ \bibnamefont {{Alves}}}, \bibinfo {author}
  {\bibfnamefont {T.~S.}\ \bibnamefont {{Axelrod}}}, \bibinfo {author}
  {\bibfnamefont {J.~J.}\ \bibnamefont {{Calitz}}}, \bibinfo {author}
  {\bibfnamefont {K.~H.}\ \bibnamefont {{Cook}}}, \bibinfo {author}
  {\bibfnamefont {A.~J.}\ \bibnamefont {{Drake}}}, \bibinfo {author}
  {\bibfnamefont {P.~C.}\ \bibnamefont {{Fragile}}}, \bibinfo {author}
  {\bibfnamefont {K.~C.}\ \bibnamefont {{Freeman}}}, \bibinfo {author}
  {\bibfnamefont {M.}~\bibnamefont {{Geha}}}, \bibinfo {author} {\bibfnamefont
  {K.}~\bibnamefont {{Griest}}}, \bibinfo {author} {\bibfnamefont {B.~R.}\
  \bibnamefont {{Johnson}}}, \bibinfo {author} {\bibfnamefont {S.~C.}\
  \bibnamefont {{Keller}}}, \bibinfo {author} {\bibfnamefont {C.}~\bibnamefont
  {{Laws}}}, \bibinfo {author} {\bibfnamefont {M.~J.}\ \bibnamefont
  {{Lehner}}}, \bibinfo {author} {\bibfnamefont {S.~L.}\ \bibnamefont
  {{Marshall}}}, \bibinfo {author} {\bibfnamefont {D.}~\bibnamefont
  {{Minniti}}}, \bibinfo {author} {\bibfnamefont {C.~A.}\ \bibnamefont
  {{Nelson}}}, \bibinfo {author} {\bibfnamefont {B.~A.}\ \bibnamefont
  {{Peterson}}}, \bibinfo {author} {\bibfnamefont {P.}~\bibnamefont
  {{Popowski}}}, \bibinfo {author} {\bibfnamefont {M.~R.}\ \bibnamefont
  {{Pratt}}}, \bibinfo {author} {\bibfnamefont {P.~J.}\ \bibnamefont
  {{Quinn}}}, \bibinfo {author} {\bibfnamefont {S.~H.}\ \bibnamefont {{Rhie}}},
  \bibinfo {author} {\bibfnamefont {C.~W.}\ \bibnamefont {{Stubbs}}}, \bibinfo
  {author} {\bibfnamefont {W.}~\bibnamefont {{Sutherland}}}, \bibinfo {author}
  {\bibfnamefont {T.}~\bibnamefont {{Vandehei}}}, \bibinfo {author}
  {\bibfnamefont {D.}~\bibnamefont {{Welch}}}, \bibinfo {author} {\bibnamefont
  {{MACHO Collaboration}}},\ and\ \bibinfo {author} {\bibnamefont {{MPS
  Collaboration}}},\ }\bibfield  {title} {\bibinfo {title} {{Gravitational
  Microlensing Events Due to Stellar-Mass Black Holes}},\ }\href
  {https://doi.org/10.1086/342225} {\bibfield  {journal} {\bibinfo  {journal}
  {\apj}\ }\textbf {\bibinfo {volume} {579}},\ \bibinfo {pages} {639} (\bibinfo
  {year} {2002})},\ \Eprint {https://arxiv.org/abs/astro-ph/0109467}
  {arXiv:astro-ph/0109467 [astro-ph]} \BibitemShut {NoStop}%
\bibitem [{\citenamefont {{Alcock}}\ \emph {et~al.}(1993)\citenamefont
  {{Alcock}}, \citenamefont {{Akerlof}}, \citenamefont {{Allsman}},
  \citenamefont {{Axelrod}}, \citenamefont {{Bennett}}, \citenamefont {{Chan}},
  \citenamefont {{Cook}}, \citenamefont {{Freeman}}, \citenamefont {{Griest}},
  \citenamefont {{Marshall}}, \citenamefont {{Park}}, \citenamefont
  {{Perlmutter}}, \citenamefont {{Peterson}}, \citenamefont {{Pratt}},
  \citenamefont {{Quinn}}, \citenamefont {{Rodgers}}, \citenamefont
  {{Stubbs}},\ and\ \citenamefont {{Sutherland}}}]{1993Natur.365..621A}%
  \BibitemOpen
  \bibfield  {author} {\bibinfo {author} {\bibfnamefont {C.}~\bibnamefont
  {{Alcock}}}, \bibinfo {author} {\bibfnamefont {C.~W.}\ \bibnamefont
  {{Akerlof}}}, \bibinfo {author} {\bibfnamefont {R.~A.}\ \bibnamefont
  {{Allsman}}}, \bibinfo {author} {\bibfnamefont {T.~S.}\ \bibnamefont
  {{Axelrod}}}, \bibinfo {author} {\bibfnamefont {D.~P.}\ \bibnamefont
  {{Bennett}}}, \bibinfo {author} {\bibfnamefont {S.}~\bibnamefont {{Chan}}},
  \bibinfo {author} {\bibfnamefont {K.~H.}\ \bibnamefont {{Cook}}}, \bibinfo
  {author} {\bibfnamefont {K.~C.}\ \bibnamefont {{Freeman}}}, \bibinfo {author}
  {\bibfnamefont {K.}~\bibnamefont {{Griest}}}, \bibinfo {author}
  {\bibfnamefont {S.~L.}\ \bibnamefont {{Marshall}}}, \bibinfo {author}
  {\bibfnamefont {H.~S.}\ \bibnamefont {{Park}}}, \bibinfo {author}
  {\bibfnamefont {S.}~\bibnamefont {{Perlmutter}}}, \bibinfo {author}
  {\bibfnamefont {B.~A.}\ \bibnamefont {{Peterson}}}, \bibinfo {author}
  {\bibfnamefont {M.~R.}\ \bibnamefont {{Pratt}}}, \bibinfo {author}
  {\bibfnamefont {P.~J.}\ \bibnamefont {{Quinn}}}, \bibinfo {author}
  {\bibfnamefont {A.~W.}\ \bibnamefont {{Rodgers}}}, \bibinfo {author}
  {\bibfnamefont {C.~W.}\ \bibnamefont {{Stubbs}}},\ and\ \bibinfo {author}
  {\bibfnamefont {W.}~\bibnamefont {{Sutherland}}},\ }\bibfield  {title}
  {\bibinfo {title} {{Possible gravitational microlensing of a star in the
  Large Magellanic Cloud}},\ }\href {https://doi.org/10.1038/365621a0}
  {\bibfield  {journal} {\bibinfo  {journal} {\nat}\ }\textbf {\bibinfo
  {volume} {365}},\ \bibinfo {pages} {621} (\bibinfo {year} {1993})},\ \Eprint
  {https://arxiv.org/abs/astro-ph/9309052} {arXiv:astro-ph/9309052 [astro-ph]}
  \BibitemShut {NoStop}%
\bibitem [{\citenamefont {{Abdurrahman}}\ \emph {et~al.}(2021)\citenamefont
  {{Abdurrahman}}, \citenamefont {{Stephens}},\ and\ \citenamefont
  {{Lu}}}]{2021ApJ...912..146A}%
  \BibitemOpen
  \bibfield  {author} {\bibinfo {author} {\bibfnamefont {F.~N.}\ \bibnamefont
  {{Abdurrahman}}}, \bibinfo {author} {\bibfnamefont {H.~F.}\ \bibnamefont
  {{Stephens}}},\ and\ \bibinfo {author} {\bibfnamefont {J.~R.}\ \bibnamefont
  {{Lu}}},\ }\bibfield  {title} {\bibinfo {title} {{On the Possibility of
  Stellar Lenses in the Black Hole Candidate Microlensing Events MACHO-96-BLG-5
  and MACHO-98-BLG-6}},\ }\href {https://doi.org/10.3847/1538-4357/abee83}
  {\bibfield  {journal} {\bibinfo  {journal} {\apj}\ }\textbf {\bibinfo
  {volume} {912}},\ \bibinfo {eid} {146} (\bibinfo {year} {2021})},\ \Eprint
  {https://arxiv.org/abs/2103.09923} {arXiv:2103.09923 [astro-ph.SR]}
  \BibitemShut {NoStop}%
\bibitem [{\citenamefont {{Lam}}\ \emph {et~al.}(2022)\citenamefont {{Lam}},
  \citenamefont {{Lu}}, \citenamefont {{Udalski}}, \citenamefont {{Bond}},
  \citenamefont {{Bennett}}, \citenamefont {{Skowron}}, \citenamefont
  {{Mr{\'o}z}}, \citenamefont {{Poleski}}, \citenamefont {{Sumi}},
  \citenamefont {{Szyma{\'n}ski}}, \citenamefont {{Koz{\l}owski}},
  \citenamefont {{Pietrukowicz}}, \citenamefont {{Soszy{\'n}ski}},
  \citenamefont {{Ulaczyk}}, \citenamefont {{Wyrzykowski}}, \citenamefont
  {{Miyazaki}}, \citenamefont {{Suzuki}}, \citenamefont {{Koshimoto}},
  \citenamefont {{Rattenbury}}, \citenamefont {{Hosek}}, \citenamefont {{Abe}},
  \citenamefont {{Barry}}, \citenamefont {{Bhattacharya}}, \citenamefont
  {{Fukui}}, \citenamefont {{Fujii}}, \citenamefont {{Hirao}}, \citenamefont
  {{Itow}}, \citenamefont {{Kirikawa}}, \citenamefont {{Kondo}}, \citenamefont
  {{Matsubara}}, \citenamefont {{Matsumoto}}, \citenamefont {{Muraki}},
  \citenamefont {{Olmschenk}}, \citenamefont {{Ranc}}, \citenamefont
  {{Okamura}}, \citenamefont {{Satoh}}, \citenamefont {{Silva}}, \citenamefont
  {{Toda}}, \citenamefont {{Tristram}}, \citenamefont {{Vandorou}},
  \citenamefont {{Yama}}, \citenamefont {{Abrams}}, \citenamefont {{Agarwal}},
  \citenamefont {{Rose}},\ and\ \citenamefont {{Terry}}}]{2022ApJ...933L..23L}%
  \BibitemOpen
  \bibfield  {author} {\bibinfo {author} {\bibfnamefont {C.~Y.}\ \bibnamefont
  {{Lam}}}, \bibinfo {author} {\bibfnamefont {J.~R.}\ \bibnamefont {{Lu}}},
  \bibinfo {author} {\bibfnamefont {A.}~\bibnamefont {{Udalski}}}, \bibinfo
  {author} {\bibfnamefont {I.}~\bibnamefont {{Bond}}}, \bibinfo {author}
  {\bibfnamefont {D.~P.}\ \bibnamefont {{Bennett}}}, \bibinfo {author}
  {\bibfnamefont {J.}~\bibnamefont {{Skowron}}}, \bibinfo {author}
  {\bibfnamefont {P.}~\bibnamefont {{Mr{\'o}z}}}, \bibinfo {author}
  {\bibfnamefont {R.}~\bibnamefont {{Poleski}}}, \bibinfo {author}
  {\bibfnamefont {T.}~\bibnamefont {{Sumi}}}, \bibinfo {author} {\bibfnamefont
  {M.~K.}\ \bibnamefont {{Szyma{\'n}ski}}}, \bibinfo {author} {\bibfnamefont
  {S.}~\bibnamefont {{Koz{\l}owski}}}, \bibinfo {author} {\bibfnamefont
  {P.}~\bibnamefont {{Pietrukowicz}}}, \bibinfo {author} {\bibfnamefont
  {I.}~\bibnamefont {{Soszy{\'n}ski}}}, \bibinfo {author} {\bibfnamefont
  {K.}~\bibnamefont {{Ulaczyk}}}, \bibinfo {author} {\bibfnamefont
  {{\L}.}~\bibnamefont {{Wyrzykowski}}}, \bibinfo {author} {\bibfnamefont
  {S.}~\bibnamefont {{Miyazaki}}}, \bibinfo {author} {\bibfnamefont
  {D.}~\bibnamefont {{Suzuki}}}, \bibinfo {author} {\bibfnamefont
  {N.}~\bibnamefont {{Koshimoto}}}, \bibinfo {author} {\bibfnamefont {N.~J.}\
  \bibnamefont {{Rattenbury}}}, \bibinfo {author} {\bibfnamefont {M.~W.}\
  \bibnamefont {{Hosek}}}, \bibinfo {author} {\bibfnamefont {F.}~\bibnamefont
  {{Abe}}}, \bibinfo {author} {\bibfnamefont {R.}~\bibnamefont {{Barry}}},
  \bibinfo {author} {\bibfnamefont {A.}~\bibnamefont {{Bhattacharya}}},
  \bibinfo {author} {\bibfnamefont {A.}~\bibnamefont {{Fukui}}}, \bibinfo
  {author} {\bibfnamefont {H.}~\bibnamefont {{Fujii}}}, \bibinfo {author}
  {\bibfnamefont {Y.}~\bibnamefont {{Hirao}}}, \bibinfo {author} {\bibfnamefont
  {Y.}~\bibnamefont {{Itow}}}, \bibinfo {author} {\bibfnamefont
  {R.}~\bibnamefont {{Kirikawa}}}, \bibinfo {author} {\bibfnamefont
  {I.}~\bibnamefont {{Kondo}}}, \bibinfo {author} {\bibfnamefont
  {Y.}~\bibnamefont {{Matsubara}}}, \bibinfo {author} {\bibfnamefont
  {S.}~\bibnamefont {{Matsumoto}}}, \bibinfo {author} {\bibfnamefont
  {Y.}~\bibnamefont {{Muraki}}}, \bibinfo {author} {\bibfnamefont
  {G.}~\bibnamefont {{Olmschenk}}}, \bibinfo {author} {\bibfnamefont
  {C.}~\bibnamefont {{Ranc}}}, \bibinfo {author} {\bibfnamefont
  {A.}~\bibnamefont {{Okamura}}}, \bibinfo {author} {\bibfnamefont
  {Y.}~\bibnamefont {{Satoh}}}, \bibinfo {author} {\bibfnamefont {S.~I.}\
  \bibnamefont {{Silva}}}, \bibinfo {author} {\bibfnamefont {T.}~\bibnamefont
  {{Toda}}}, \bibinfo {author} {\bibfnamefont {P.~J.}\ \bibnamefont
  {{Tristram}}}, \bibinfo {author} {\bibfnamefont {A.}~\bibnamefont
  {{Vandorou}}}, \bibinfo {author} {\bibfnamefont {H.}~\bibnamefont {{Yama}}},
  \bibinfo {author} {\bibfnamefont {N.~S.}\ \bibnamefont {{Abrams}}}, \bibinfo
  {author} {\bibfnamefont {S.}~\bibnamefont {{Agarwal}}}, \bibinfo {author}
  {\bibfnamefont {S.}~\bibnamefont {{Rose}}},\ and\ \bibinfo {author}
  {\bibfnamefont {S.~K.}\ \bibnamefont {{Terry}}},\ }\bibfield  {title}
  {\bibinfo {title} {{An Isolated Mass-gap Black Hole or Neutron Star Detected
  with Astrometric Microlensing}},\ }\href
  {https://doi.org/10.3847/2041-8213/ac7442} {\bibfield  {journal} {\bibinfo
  {journal} {\apjl}\ }\textbf {\bibinfo {volume} {933}},\ \bibinfo {eid} {L23}
  (\bibinfo {year} {2022})},\ \Eprint {https://arxiv.org/abs/2202.01903}
  {arXiv:2202.01903 [astro-ph.GA]} \BibitemShut {NoStop}%
\bibitem [{\citenamefont {{Sahu}}\ \emph {et~al.}(2022)\citenamefont {{Sahu}},
  \citenamefont {{Anderson}}, \citenamefont {{Casertano}}, \citenamefont
  {{Bond}}, \citenamefont {{Udalski}}, \citenamefont {{Dominik}}, \citenamefont
  {{Calamida}}, \citenamefont {{Bellini}}, \citenamefont {{Brown}},
  \citenamefont {{Rejkuba}}, \citenamefont {{Bajaj}}, \citenamefont {{Kains}},
  \citenamefont {{Ferguson}}, \citenamefont {{Fryer}}, \citenamefont {{Yock}},
  \citenamefont {{Mr{\'o}z}}, \citenamefont {{Koz{\l}owski}}, \citenamefont
  {{Pietrukowicz}}, \citenamefont {{Poleski}}, \citenamefont {{Skowron}},
  \citenamefont {{Soszy{\'n}ski}}, \citenamefont {{Szyma{\'n}ski}},
  \citenamefont {{Ulaczyk}}, \citenamefont {{Wyrzykowski}}, \citenamefont
  {{Barry}}, \citenamefont {{Bennett}}, \citenamefont {{Bond}}, \citenamefont
  {{Hirao}}, \citenamefont {{Silva}}, \citenamefont {{Kondo}}, \citenamefont
  {{Koshimoto}}, \citenamefont {{Ranc}}, \citenamefont {{Rattenbury}},
  \citenamefont {{Sumi}}, \citenamefont {{Suzuki}}, \citenamefont {{Tristram}},
  \citenamefont {{Vandorou}}, \citenamefont {{Beaulieu}}, \citenamefont
  {{Marquette}}, \citenamefont {{Cole}}, \citenamefont {{Fouqu{\'e}}},
  \citenamefont {{Hill}}, \citenamefont {{Dieters}}, \citenamefont
  {{Coutures}}, \citenamefont {{Dominis-Prester}}, \citenamefont {{Bennett}},
  \citenamefont {{Bachelet}}, \citenamefont {{Menzies}}, \citenamefont
  {{Albrow}}, \citenamefont {{Pollard}}, \citenamefont {{Gould}}, \citenamefont
  {{Yee}}, \citenamefont {{Allen}}, \citenamefont {{Almeida}}, \citenamefont
  {{Christie}}, \citenamefont {{Drummond}}, \citenamefont {{Gal-Yam}},
  \citenamefont {{Gorbikov}}, \citenamefont {{Jablonski}}, \citenamefont
  {{Lee}}, \citenamefont {{Maoz}}, \citenamefont {{Manulis}}, \citenamefont
  {{McCormick}}, \citenamefont {{Natusch}}, \citenamefont {{Pogge}},
  \citenamefont {{Shvartzvald}}, \citenamefont {{J{\o}rgensen}}, \citenamefont
  {{Alsubai}}, \citenamefont {{Andersen}}, \citenamefont {{Bozza}},
  \citenamefont {{Novati}}, \citenamefont {{Burgdorf}}, \citenamefont
  {{Hinse}}, \citenamefont {{Hundertmark}}, \citenamefont {{Husser}},
  \citenamefont {{Kerins}}, \citenamefont {{Longa-Pe{\~n}a}}, \citenamefont
  {{Mancini}}, \citenamefont {{Penny}}, \citenamefont {{Rahvar}}, \citenamefont
  {{Ricci}}, \citenamefont {{Sajadian}}, \citenamefont {{Skottfelt}},
  \citenamefont {{Snodgrass}}, \citenamefont {{Southworth}}, \citenamefont
  {{Tregloan-Reed}}, \citenamefont {{Wambsganss}}, \citenamefont {{Wertz}},
  \citenamefont {{Tsapras}}, \citenamefont {{Street}}, \citenamefont
  {{Bramich}}, \citenamefont {{Horne}}, \citenamefont {{Steele}},\ and\
  \citenamefont {{RoboNet Collaboration}}}]{2022ApJ...933...83S}%
  \BibitemOpen
  \bibfield  {author} {\bibinfo {author} {\bibfnamefont {K.~C.}\ \bibnamefont
  {{Sahu}}}, \bibinfo {author} {\bibfnamefont {J.}~\bibnamefont {{Anderson}}},
  \bibinfo {author} {\bibfnamefont {S.}~\bibnamefont {{Casertano}}}, \bibinfo
  {author} {\bibfnamefont {H.~E.}\ \bibnamefont {{Bond}}}, \bibinfo {author}
  {\bibfnamefont {A.}~\bibnamefont {{Udalski}}}, \bibinfo {author}
  {\bibfnamefont {M.}~\bibnamefont {{Dominik}}}, \bibinfo {author}
  {\bibfnamefont {A.}~\bibnamefont {{Calamida}}}, \bibinfo {author}
  {\bibfnamefont {A.}~\bibnamefont {{Bellini}}}, \bibinfo {author}
  {\bibfnamefont {T.~M.}\ \bibnamefont {{Brown}}}, \bibinfo {author}
  {\bibfnamefont {M.}~\bibnamefont {{Rejkuba}}}, \bibinfo {author}
  {\bibfnamefont {V.}~\bibnamefont {{Bajaj}}}, \bibinfo {author} {\bibfnamefont
  {N.}~\bibnamefont {{Kains}}}, \bibinfo {author} {\bibfnamefont {H.~C.}\
  \bibnamefont {{Ferguson}}}, \bibinfo {author} {\bibfnamefont {C.~L.}\
  \bibnamefont {{Fryer}}}, \bibinfo {author} {\bibfnamefont {P.}~\bibnamefont
  {{Yock}}}, \bibinfo {author} {\bibfnamefont {P.}~\bibnamefont {{Mr{\'o}z}}},
  \bibinfo {author} {\bibfnamefont {S.}~\bibnamefont {{Koz{\l}owski}}},
  \bibinfo {author} {\bibfnamefont {P.}~\bibnamefont {{Pietrukowicz}}},
  \bibinfo {author} {\bibfnamefont {R.}~\bibnamefont {{Poleski}}}, \bibinfo
  {author} {\bibfnamefont {J.}~\bibnamefont {{Skowron}}}, \bibinfo {author}
  {\bibfnamefont {I.}~\bibnamefont {{Soszy{\'n}ski}}}, \bibinfo {author}
  {\bibfnamefont {M.~K.}\ \bibnamefont {{Szyma{\'n}ski}}}, \bibinfo {author}
  {\bibfnamefont {K.}~\bibnamefont {{Ulaczyk}}}, \bibinfo {author}
  {\bibfnamefont {{\L}.}~\bibnamefont {{Wyrzykowski}}}, \bibinfo {author}
  {\bibfnamefont {R.~K.}\ \bibnamefont {{Barry}}}, \bibinfo {author}
  {\bibfnamefont {D.~P.}\ \bibnamefont {{Bennett}}}, \bibinfo {author}
  {\bibfnamefont {I.~A.}\ \bibnamefont {{Bond}}}, \bibinfo {author}
  {\bibfnamefont {Y.}~\bibnamefont {{Hirao}}}, \bibinfo {author} {\bibfnamefont
  {S.~I.}\ \bibnamefont {{Silva}}}, \bibinfo {author} {\bibfnamefont
  {I.}~\bibnamefont {{Kondo}}}, \bibinfo {author} {\bibfnamefont
  {N.}~\bibnamefont {{Koshimoto}}}, \bibinfo {author} {\bibfnamefont
  {C.}~\bibnamefont {{Ranc}}}, \bibinfo {author} {\bibfnamefont {N.~J.}\
  \bibnamefont {{Rattenbury}}}, \bibinfo {author} {\bibfnamefont
  {T.}~\bibnamefont {{Sumi}}}, \bibinfo {author} {\bibfnamefont
  {D.}~\bibnamefont {{Suzuki}}}, \bibinfo {author} {\bibfnamefont {P.~J.}\
  \bibnamefont {{Tristram}}}, \bibinfo {author} {\bibfnamefont
  {A.}~\bibnamefont {{Vandorou}}}, \bibinfo {author} {\bibfnamefont {J.-P.}\
  \bibnamefont {{Beaulieu}}}, \bibinfo {author} {\bibfnamefont {J.-B.}\
  \bibnamefont {{Marquette}}}, \bibinfo {author} {\bibfnamefont
  {A.}~\bibnamefont {{Cole}}}, \bibinfo {author} {\bibfnamefont
  {P.}~\bibnamefont {{Fouqu{\'e}}}}, \bibinfo {author} {\bibfnamefont
  {K.}~\bibnamefont {{Hill}}}, \bibinfo {author} {\bibfnamefont
  {S.}~\bibnamefont {{Dieters}}}, \bibinfo {author} {\bibfnamefont
  {C.}~\bibnamefont {{Coutures}}}, \bibinfo {author} {\bibfnamefont
  {D.}~\bibnamefont {{Dominis-Prester}}}, \bibinfo {author} {\bibfnamefont
  {C.}~\bibnamefont {{Bennett}}}, \bibinfo {author} {\bibfnamefont
  {E.}~\bibnamefont {{Bachelet}}}, \bibinfo {author} {\bibfnamefont
  {J.}~\bibnamefont {{Menzies}}}, \bibinfo {author} {\bibfnamefont
  {M.}~\bibnamefont {{Albrow}}}, \bibinfo {author} {\bibfnamefont
  {K.}~\bibnamefont {{Pollard}}}, \bibinfo {author} {\bibfnamefont
  {A.}~\bibnamefont {{Gould}}}, \bibinfo {author} {\bibfnamefont {J.~C.}\
  \bibnamefont {{Yee}}}, \bibinfo {author} {\bibfnamefont {W.}~\bibnamefont
  {{Allen}}}, \bibinfo {author} {\bibfnamefont {L.~A.}\ \bibnamefont
  {{Almeida}}}, \bibinfo {author} {\bibfnamefont {G.}~\bibnamefont
  {{Christie}}}, \bibinfo {author} {\bibfnamefont {J.}~\bibnamefont
  {{Drummond}}}, \bibinfo {author} {\bibfnamefont {A.}~\bibnamefont
  {{Gal-Yam}}}, \bibinfo {author} {\bibfnamefont {E.}~\bibnamefont
  {{Gorbikov}}}, \bibinfo {author} {\bibfnamefont {F.}~\bibnamefont
  {{Jablonski}}}, \bibinfo {author} {\bibfnamefont {C.-U.}\ \bibnamefont
  {{Lee}}}, \bibinfo {author} {\bibfnamefont {D.}~\bibnamefont {{Maoz}}},
  \bibinfo {author} {\bibfnamefont {I.}~\bibnamefont {{Manulis}}}, \bibinfo
  {author} {\bibfnamefont {J.}~\bibnamefont {{McCormick}}}, \bibinfo {author}
  {\bibfnamefont {T.}~\bibnamefont {{Natusch}}}, \bibinfo {author}
  {\bibfnamefont {R.~W.}\ \bibnamefont {{Pogge}}}, \bibinfo {author}
  {\bibfnamefont {Y.}~\bibnamefont {{Shvartzvald}}}, \bibinfo {author}
  {\bibfnamefont {U.~G.}\ \bibnamefont {{J{\o}rgensen}}}, \bibinfo {author}
  {\bibfnamefont {K.~A.}\ \bibnamefont {{Alsubai}}}, \bibinfo {author}
  {\bibfnamefont {M.~I.}\ \bibnamefont {{Andersen}}}, \bibinfo {author}
  {\bibfnamefont {V.}~\bibnamefont {{Bozza}}}, \bibinfo {author} {\bibfnamefont
  {S.~C.}\ \bibnamefont {{Novati}}}, \bibinfo {author} {\bibfnamefont
  {M.}~\bibnamefont {{Burgdorf}}}, \bibinfo {author} {\bibfnamefont {T.~C.}\
  \bibnamefont {{Hinse}}}, \bibinfo {author} {\bibfnamefont {M.}~\bibnamefont
  {{Hundertmark}}}, \bibinfo {author} {\bibfnamefont {T.-O.}\ \bibnamefont
  {{Husser}}}, \bibinfo {author} {\bibfnamefont {E.}~\bibnamefont {{Kerins}}},
  \bibinfo {author} {\bibfnamefont {P.}~\bibnamefont {{Longa-Pe{\~n}a}}},
  \bibinfo {author} {\bibfnamefont {L.}~\bibnamefont {{Mancini}}}, \bibinfo
  {author} {\bibfnamefont {M.}~\bibnamefont {{Penny}}}, \bibinfo {author}
  {\bibfnamefont {S.}~\bibnamefont {{Rahvar}}}, \bibinfo {author}
  {\bibfnamefont {D.}~\bibnamefont {{Ricci}}}, \bibinfo {author} {\bibfnamefont
  {S.}~\bibnamefont {{Sajadian}}}, \bibinfo {author} {\bibfnamefont
  {J.}~\bibnamefont {{Skottfelt}}}, \bibinfo {author} {\bibfnamefont
  {C.}~\bibnamefont {{Snodgrass}}}, \bibinfo {author} {\bibfnamefont
  {J.}~\bibnamefont {{Southworth}}}, \bibinfo {author} {\bibfnamefont
  {J.}~\bibnamefont {{Tregloan-Reed}}}, \bibinfo {author} {\bibfnamefont
  {J.}~\bibnamefont {{Wambsganss}}}, \bibinfo {author} {\bibfnamefont
  {O.}~\bibnamefont {{Wertz}}}, \bibinfo {author} {\bibfnamefont
  {Y.}~\bibnamefont {{Tsapras}}}, \bibinfo {author} {\bibfnamefont {R.~A.}\
  \bibnamefont {{Street}}}, \bibinfo {author} {\bibfnamefont {D.~M.}\
  \bibnamefont {{Bramich}}}, \bibinfo {author} {\bibfnamefont {K.}~\bibnamefont
  {{Horne}}}, \bibinfo {author} {\bibfnamefont {I.~A.}\ \bibnamefont
  {{Steele}}},\ and\ \bibinfo {author} {\bibnamefont {{RoboNet
  Collaboration}}},\ }\bibfield  {title} {\bibinfo {title} {{An Isolated
  Stellar-mass Black Hole Detected through Astrometric Microlensing}},\ }\href
  {https://doi.org/10.3847/1538-4357/ac739e} {\bibfield  {journal} {\bibinfo
  {journal} {\apj}\ }\textbf {\bibinfo {volume} {933}},\ \bibinfo {eid} {83}
  (\bibinfo {year} {2022})},\ \Eprint {https://arxiv.org/abs/2201.13296}
  {arXiv:2201.13296 [astro-ph.SR]} \BibitemShut {NoStop}%
\bibitem [{\citenamefont {{Mereghetti}}\ \emph {et~al.}(2022)\citenamefont
  {{Mereghetti}}, \citenamefont {{Sidoli}}, \citenamefont {{Ponti}},\ and\
  \citenamefont {{Treves}}}]{2022ApJ...934...62M}%
  \BibitemOpen
  \bibfield  {author} {\bibinfo {author} {\bibfnamefont {S.}~\bibnamefont
  {{Mereghetti}}}, \bibinfo {author} {\bibfnamefont {L.}~\bibnamefont
  {{Sidoli}}}, \bibinfo {author} {\bibfnamefont {G.}~\bibnamefont {{Ponti}}},\
  and\ \bibinfo {author} {\bibfnamefont {A.}~\bibnamefont {{Treves}}},\
  }\bibfield  {title} {\bibinfo {title} {{X-Ray Observations of the Isolated
  Black Hole Candidate OGLE-2011-BLG-0462 and Other Collapsed Objects
  Discovered through Gravitational Microlensing}},\ }\href
  {https://doi.org/10.3847/1538-4357/ac7965} {\bibfield  {journal} {\bibinfo
  {journal} {\apj}\ }\textbf {\bibinfo {volume} {934}},\ \bibinfo {eid} {62}
  (\bibinfo {year} {2022})},\ \Eprint {https://arxiv.org/abs/2206.07480}
  {arXiv:2206.07480 [astro-ph.HE]} \BibitemShut {NoStop}%
\bibitem [{\citenamefont {{Mr{\'o}z}}\ \emph {et~al.}(2022)\citenamefont
  {{Mr{\'o}z}}, \citenamefont {{Udalski}},\ and\ \citenamefont
  {{Gould}}}]{2022ApJ...937L..24M}%
  \BibitemOpen
  \bibfield  {author} {\bibinfo {author} {\bibfnamefont {P.}~\bibnamefont
  {{Mr{\'o}z}}}, \bibinfo {author} {\bibfnamefont {A.}~\bibnamefont
  {{Udalski}}},\ and\ \bibinfo {author} {\bibfnamefont {A.}~\bibnamefont
  {{Gould}}},\ }\bibfield  {title} {\bibinfo {title} {{Systematic Errors as a
  Source of Mass Discrepancy in Black Hole Microlensing Event
  OGLE-2011-BLG-0462}},\ }\href {https://doi.org/10.3847/2041-8213/ac90bb}
  {\bibfield  {journal} {\bibinfo  {journal} {\apjl}\ }\textbf {\bibinfo
  {volume} {937}},\ \bibinfo {eid} {L24} (\bibinfo {year} {2022})},\ \Eprint
  {https://arxiv.org/abs/2207.10729} {arXiv:2207.10729 [astro-ph.SR]}
  \BibitemShut {NoStop}%
\bibitem [{\citenamefont {{Wyrzykowski}}\ \emph {et~al.}(2022)\citenamefont
  {{Wyrzykowski}}, \citenamefont {{Kruszy{\'n}ska}}, \citenamefont {{Rybicki}},
  \citenamefont {{Holl}}, \citenamefont {{Lec{\o}e ur-Ta{\"\i}bi}},
  \citenamefont {{Mowlavi}}, \citenamefont {{Nienartowicz}}, \citenamefont
  {{Jevardat de Fombelle}}, \citenamefont {{Rimoldini}}, \citenamefont
  {{Audard}}, \citenamefont {{Garcia-Lario}}, \citenamefont {{Gavras}},
  \citenamefont {{Evans}}, \citenamefont {{Hodgkin}},\ and\ \citenamefont
  {{Eyer}}}]{2022arXiv220606121W}%
  \BibitemOpen
  \bibfield  {author} {\bibinfo {author} {\bibfnamefont {{\L}.}~\bibnamefont
  {{Wyrzykowski}}}, \bibinfo {author} {\bibfnamefont {K.}~\bibnamefont
  {{Kruszy{\'n}ska}}}, \bibinfo {author} {\bibfnamefont {K.~A.}\ \bibnamefont
  {{Rybicki}}}, \bibinfo {author} {\bibfnamefont {B.}~\bibnamefont {{Holl}}},
  \bibinfo {author} {\bibfnamefont {I.}~\bibnamefont {{Lec{\o}e
  ur-Ta{\"\i}bi}}}, \bibinfo {author} {\bibfnamefont {N.}~\bibnamefont
  {{Mowlavi}}}, \bibinfo {author} {\bibfnamefont {K.}~\bibnamefont
  {{Nienartowicz}}}, \bibinfo {author} {\bibfnamefont {G.}~\bibnamefont
  {{Jevardat de Fombelle}}}, \bibinfo {author} {\bibfnamefont {L.}~\bibnamefont
  {{Rimoldini}}}, \bibinfo {author} {\bibfnamefont {M.}~\bibnamefont
  {{Audard}}}, \bibinfo {author} {\bibfnamefont {P.}~\bibnamefont
  {{Garcia-Lario}}}, \bibinfo {author} {\bibfnamefont {P.}~\bibnamefont
  {{Gavras}}}, \bibinfo {author} {\bibfnamefont {D.~W.}\ \bibnamefont
  {{Evans}}}, \bibinfo {author} {\bibfnamefont {S.~T.}\ \bibnamefont
  {{Hodgkin}}},\ and\ \bibinfo {author} {\bibfnamefont {L.}~\bibnamefont
  {{Eyer}}},\ }\bibfield  {title} {\bibinfo {title} {{Gaia Data Release 3:
  Microlensing Events from All Over the Sky}},\ }\href@noop {} {\bibfield
  {journal} {\bibinfo  {journal} {arXiv e-prints}\ ,\ \bibinfo {eid}
  {arXiv:2206.06121}} (\bibinfo {year} {2022})},\ \Eprint
  {https://arxiv.org/abs/2206.06121} {arXiv:2206.06121 [astro-ph.SR]}
  \BibitemShut {NoStop}%
\bibitem [{\citenamefont {{Chen}}\ \emph {et~al.}(2023)\citenamefont {{Chen}},
  \citenamefont {{Kongsore}},\ and\ \citenamefont {{Van
  Tilburg}}}]{2023arXiv230100822C}%
  \BibitemOpen
  \bibfield  {author} {\bibinfo {author} {\bibfnamefont {I.-K.}\ \bibnamefont
  {{Chen}}}, \bibinfo {author} {\bibfnamefont {M.}~\bibnamefont {{Kongsore}}},\
  and\ \bibinfo {author} {\bibfnamefont {K.}~\bibnamefont {{Van Tilburg}}},\
  }\bibfield  {title} {\bibinfo {title} {{Detecting Dark Compact Objects in
  Gaia DR4: A Data Analysis Pipeline for Transient Astrometric Lensing
  Searches}},\ }\href@noop {} {\bibfield  {journal} {\bibinfo  {journal} {arXiv
  e-prints}\ ,\ \bibinfo {eid} {arXiv:2301.00822}} (\bibinfo {year} {2023})},\
  \Eprint {https://arxiv.org/abs/2301.00822} {arXiv:2301.00822 [astro-ph.GA]}
  \BibitemShut {NoStop}%
\bibitem [{\citenamefont {{Spergel}}\ \emph {et~al.}(2015)\citenamefont
  {{Spergel}}, \citenamefont {{Gehrels}}, \citenamefont {{Baltay}},
  \citenamefont {{Bennett}}, \citenamefont {{Breckinridge}}, \citenamefont
  {{Donahue}}, \citenamefont {{Dressler}}, \citenamefont {{Gaudi}},
  \citenamefont {{Greene}}, \citenamefont {{Guyon}}, \citenamefont {{Hirata}},
  \citenamefont {{Kalirai}}, \citenamefont {{Kasdin}}, \citenamefont
  {{Macintosh}}, \citenamefont {{Moos}}, \citenamefont {{Perlmutter}},
  \citenamefont {{Postman}}, \citenamefont {{Rauscher}}, \citenamefont
  {{Rhodes}}, \citenamefont {{Wang}}, \citenamefont {{Weinberg}}, \citenamefont
  {{Benford}}, \citenamefont {{Hudson}}, \citenamefont {{Jeong}}, \citenamefont
  {{Mellier}}, \citenamefont {{Traub}}, \citenamefont {{Yamada}}, \citenamefont
  {{Capak}}, \citenamefont {{Colbert}}, \citenamefont {{Masters}},
  \citenamefont {{Penny}}, \citenamefont {{Savransky}}, \citenamefont
  {{Stern}}, \citenamefont {{Zimmerman}}, \citenamefont {{Barry}},
  \citenamefont {{Bartusek}}, \citenamefont {{Carpenter}}, \citenamefont
  {{Cheng}}, \citenamefont {{Content}}, \citenamefont {{Dekens}}, \citenamefont
  {{Demers}}, \citenamefont {{Grady}}, \citenamefont {{Jackson}}, \citenamefont
  {{Kuan}}, \citenamefont {{Kruk}}, \citenamefont {{Melton}}, \citenamefont
  {{Nemati}}, \citenamefont {{Parvin}}, \citenamefont {{Poberezhskiy}},
  \citenamefont {{Peddie}}, \citenamefont {{Ruffa}}, \citenamefont {{Wallace}},
  \citenamefont {{Whipple}}, \citenamefont {{Wollack}},\ and\ \citenamefont
  {{Zhao}}}]{2015arXiv150303757S}%
  \BibitemOpen
  \bibfield  {author} {\bibinfo {author} {\bibfnamefont {D.}~\bibnamefont
  {{Spergel}}}, \bibinfo {author} {\bibfnamefont {N.}~\bibnamefont
  {{Gehrels}}}, \bibinfo {author} {\bibfnamefont {C.}~\bibnamefont {{Baltay}}},
  \bibinfo {author} {\bibfnamefont {D.}~\bibnamefont {{Bennett}}}, \bibinfo
  {author} {\bibfnamefont {J.}~\bibnamefont {{Breckinridge}}}, \bibinfo
  {author} {\bibfnamefont {M.}~\bibnamefont {{Donahue}}}, \bibinfo {author}
  {\bibfnamefont {A.}~\bibnamefont {{Dressler}}}, \bibinfo {author}
  {\bibfnamefont {B.~S.}\ \bibnamefont {{Gaudi}}}, \bibinfo {author}
  {\bibfnamefont {T.}~\bibnamefont {{Greene}}}, \bibinfo {author}
  {\bibfnamefont {O.}~\bibnamefont {{Guyon}}}, \bibinfo {author} {\bibfnamefont
  {C.}~\bibnamefont {{Hirata}}}, \bibinfo {author} {\bibfnamefont
  {J.}~\bibnamefont {{Kalirai}}}, \bibinfo {author} {\bibfnamefont {N.~J.}\
  \bibnamefont {{Kasdin}}}, \bibinfo {author} {\bibfnamefont {B.}~\bibnamefont
  {{Macintosh}}}, \bibinfo {author} {\bibfnamefont {W.}~\bibnamefont {{Moos}}},
  \bibinfo {author} {\bibfnamefont {S.}~\bibnamefont {{Perlmutter}}}, \bibinfo
  {author} {\bibfnamefont {M.}~\bibnamefont {{Postman}}}, \bibinfo {author}
  {\bibfnamefont {B.}~\bibnamefont {{Rauscher}}}, \bibinfo {author}
  {\bibfnamefont {J.}~\bibnamefont {{Rhodes}}}, \bibinfo {author}
  {\bibfnamefont {Y.}~\bibnamefont {{Wang}}}, \bibinfo {author} {\bibfnamefont
  {D.}~\bibnamefont {{Weinberg}}}, \bibinfo {author} {\bibfnamefont
  {D.}~\bibnamefont {{Benford}}}, \bibinfo {author} {\bibfnamefont
  {M.}~\bibnamefont {{Hudson}}}, \bibinfo {author} {\bibfnamefont {W.~S.}\
  \bibnamefont {{Jeong}}}, \bibinfo {author} {\bibfnamefont {Y.}~\bibnamefont
  {{Mellier}}}, \bibinfo {author} {\bibfnamefont {W.}~\bibnamefont {{Traub}}},
  \bibinfo {author} {\bibfnamefont {T.}~\bibnamefont {{Yamada}}}, \bibinfo
  {author} {\bibfnamefont {P.}~\bibnamefont {{Capak}}}, \bibinfo {author}
  {\bibfnamefont {J.}~\bibnamefont {{Colbert}}}, \bibinfo {author}
  {\bibfnamefont {D.}~\bibnamefont {{Masters}}}, \bibinfo {author}
  {\bibfnamefont {M.}~\bibnamefont {{Penny}}}, \bibinfo {author} {\bibfnamefont
  {D.}~\bibnamefont {{Savransky}}}, \bibinfo {author} {\bibfnamefont
  {D.}~\bibnamefont {{Stern}}}, \bibinfo {author} {\bibfnamefont
  {N.}~\bibnamefont {{Zimmerman}}}, \bibinfo {author} {\bibfnamefont
  {R.}~\bibnamefont {{Barry}}}, \bibinfo {author} {\bibfnamefont
  {L.}~\bibnamefont {{Bartusek}}}, \bibinfo {author} {\bibfnamefont
  {K.}~\bibnamefont {{Carpenter}}}, \bibinfo {author} {\bibfnamefont
  {E.}~\bibnamefont {{Cheng}}}, \bibinfo {author} {\bibfnamefont
  {D.}~\bibnamefont {{Content}}}, \bibinfo {author} {\bibfnamefont
  {F.}~\bibnamefont {{Dekens}}}, \bibinfo {author} {\bibfnamefont
  {R.}~\bibnamefont {{Demers}}}, \bibinfo {author} {\bibfnamefont
  {K.}~\bibnamefont {{Grady}}}, \bibinfo {author} {\bibfnamefont
  {C.}~\bibnamefont {{Jackson}}}, \bibinfo {author} {\bibfnamefont
  {G.}~\bibnamefont {{Kuan}}}, \bibinfo {author} {\bibfnamefont
  {J.}~\bibnamefont {{Kruk}}}, \bibinfo {author} {\bibfnamefont
  {M.}~\bibnamefont {{Melton}}}, \bibinfo {author} {\bibfnamefont
  {B.}~\bibnamefont {{Nemati}}}, \bibinfo {author} {\bibfnamefont
  {B.}~\bibnamefont {{Parvin}}}, \bibinfo {author} {\bibfnamefont
  {I.}~\bibnamefont {{Poberezhskiy}}}, \bibinfo {author} {\bibfnamefont
  {C.}~\bibnamefont {{Peddie}}}, \bibinfo {author} {\bibfnamefont
  {J.}~\bibnamefont {{Ruffa}}}, \bibinfo {author} {\bibfnamefont {J.~K.}\
  \bibnamefont {{Wallace}}}, \bibinfo {author} {\bibfnamefont {A.}~\bibnamefont
  {{Whipple}}}, \bibinfo {author} {\bibfnamefont {E.}~\bibnamefont
  {{Wollack}}},\ and\ \bibinfo {author} {\bibfnamefont {F.}~\bibnamefont
  {{Zhao}}},\ }\bibfield  {title} {\bibinfo {title} {{Wide-Field InfrarRed
  Survey Telescope-Astrophysics Focused Telescope Assets WFIRST-AFTA 2015
  Report}},\ }\href@noop {} {\bibfield  {journal} {\bibinfo  {journal} {arXiv
  e-prints}\ ,\ \bibinfo {eid} {arXiv:1503.03757}} (\bibinfo {year} {2015})},\
  \Eprint {https://arxiv.org/abs/1503.03757} {arXiv:1503.03757 [astro-ph.IM]}
  \BibitemShut {NoStop}%
\bibitem [{\citenamefont {{Penny}}\ \emph {et~al.}(2019)\citenamefont
  {{Penny}}, \citenamefont {{Gaudi}}, \citenamefont {{Kerins}}, \citenamefont
  {{Rattenbury}}, \citenamefont {{Mao}}, \citenamefont {{Robin}},\ and\
  \citenamefont {{Calchi Novati}}}]{2019ApJS..241....3P}%
  \BibitemOpen
  \bibfield  {author} {\bibinfo {author} {\bibfnamefont {M.~T.}\ \bibnamefont
  {{Penny}}}, \bibinfo {author} {\bibfnamefont {B.~S.}\ \bibnamefont
  {{Gaudi}}}, \bibinfo {author} {\bibfnamefont {E.}~\bibnamefont {{Kerins}}},
  \bibinfo {author} {\bibfnamefont {N.~J.}\ \bibnamefont {{Rattenbury}}},
  \bibinfo {author} {\bibfnamefont {S.}~\bibnamefont {{Mao}}}, \bibinfo
  {author} {\bibfnamefont {A.~C.}\ \bibnamefont {{Robin}}},\ and\ \bibinfo
  {author} {\bibfnamefont {S.}~\bibnamefont {{Calchi Novati}}},\ }\bibfield
  {title} {\bibinfo {title} {{Predictions of the WFIRST Microlensing Survey. I.
  Bound Planet Detection Rates}},\ }\href
  {https://doi.org/10.3847/1538-4365/aafb69} {\bibfield  {journal} {\bibinfo
  {journal} {\apjs}\ }\textbf {\bibinfo {volume} {241}},\ \bibinfo {eid} {3}
  (\bibinfo {year} {2019})},\ \Eprint {https://arxiv.org/abs/1808.02490}
  {arXiv:1808.02490 [astro-ph.EP]} \BibitemShut {NoStop}%
\bibitem [{\citenamefont {{Sajadian}}\ and\ \citenamefont
  {{Sahu}}(2023)}]{2023AJ....165...96S}%
  \BibitemOpen
  \bibfield  {author} {\bibinfo {author} {\bibfnamefont {S.}~\bibnamefont
  {{Sajadian}}}\ and\ \bibinfo {author} {\bibfnamefont {K.~C.}\ \bibnamefont
  {{Sahu}}},\ }\bibfield  {title} {\bibinfo {title} {{Detecting isolated
  stellar-mass black holes by the Roman telescope}},\ }\href@noop {} {\bibfield
   {journal} {\bibinfo  {journal} {arXiv e-prints}\ ,\ \bibinfo {eid}
  {arXiv:2301.03812}} (\bibinfo {year} {2023})},\ \Eprint
  {https://arxiv.org/abs/2301.03812} {arXiv:2301.03812 [astro-ph.GA]}
  \BibitemShut {NoStop}%
\bibitem [{\citenamefont {{van den Bergh}}(2000)}]{2000glg..book.....V}%
  \BibitemOpen
  \bibfield  {author} {\bibinfo {author} {\bibfnamefont {S.}~\bibnamefont {{van
  den Bergh}}},\ }\href@noop {} {\emph {\bibinfo {title} {{The Galaxies of the
  Local Group}}}}\ (\bibinfo {year} {2000})\BibitemShut {NoStop}%
\bibitem [{\citenamefont {{Irwin}}\ \emph {et~al.}(2007)\citenamefont
  {{Irwin}}, \citenamefont {{Belokurov}}, \citenamefont {{Evans}},
  \citenamefont {{Ryan-Weber}}, \citenamefont {{de Jong}}, \citenamefont
  {{Koposov}}, \citenamefont {{Zucker}}, \citenamefont {{Hodgkin}},
  \citenamefont {{Gilmore}}, \citenamefont {{Prema}}, \citenamefont {{Hebb}},
  \citenamefont {{Begum}}, \citenamefont {{Fellhauer}}, \citenamefont
  {{Hewett}}, \citenamefont {{Kennicutt}}, \citenamefont {{Wilkinson}},
  \citenamefont {{Bramich}}, \citenamefont {{Vidrih}}, \citenamefont {{Rix}},
  \citenamefont {{Beers}}, \citenamefont {{Barentine}}, \citenamefont
  {{Brewington}}, \citenamefont {{Harvanek}}, \citenamefont {{Krzesinski}},
  \citenamefont {{Long}}, \citenamefont {{Nitta}},\ and\ \citenamefont
  {{Snedden}}}]{2007ApJ...656L..13I}%
  \BibitemOpen
  \bibfield  {author} {\bibinfo {author} {\bibfnamefont {M.~J.}\ \bibnamefont
  {{Irwin}}}, \bibinfo {author} {\bibfnamefont {V.}~\bibnamefont
  {{Belokurov}}}, \bibinfo {author} {\bibfnamefont {N.~W.}\ \bibnamefont
  {{Evans}}}, \bibinfo {author} {\bibfnamefont {E.~V.}\ \bibnamefont
  {{Ryan-Weber}}}, \bibinfo {author} {\bibfnamefont {J.~T.~A.}\ \bibnamefont
  {{de Jong}}}, \bibinfo {author} {\bibfnamefont {S.}~\bibnamefont
  {{Koposov}}}, \bibinfo {author} {\bibfnamefont {D.~B.}\ \bibnamefont
  {{Zucker}}}, \bibinfo {author} {\bibfnamefont {S.~T.}\ \bibnamefont
  {{Hodgkin}}}, \bibinfo {author} {\bibfnamefont {G.}~\bibnamefont
  {{Gilmore}}}, \bibinfo {author} {\bibfnamefont {P.}~\bibnamefont {{Prema}}},
  \bibinfo {author} {\bibfnamefont {L.}~\bibnamefont {{Hebb}}}, \bibinfo
  {author} {\bibfnamefont {A.}~\bibnamefont {{Begum}}}, \bibinfo {author}
  {\bibfnamefont {M.}~\bibnamefont {{Fellhauer}}}, \bibinfo {author}
  {\bibfnamefont {P.~C.}\ \bibnamefont {{Hewett}}}, \bibinfo {author}
  {\bibfnamefont {J.}~\bibnamefont {{Kennicutt}}, \bibfnamefont {R.~C.}},
  \bibinfo {author} {\bibfnamefont {M.~I.}\ \bibnamefont {{Wilkinson}}},
  \bibinfo {author} {\bibfnamefont {D.~M.}\ \bibnamefont {{Bramich}}}, \bibinfo
  {author} {\bibfnamefont {S.}~\bibnamefont {{Vidrih}}}, \bibinfo {author}
  {\bibfnamefont {H.~W.}\ \bibnamefont {{Rix}}}, \bibinfo {author}
  {\bibfnamefont {T.~C.}\ \bibnamefont {{Beers}}}, \bibinfo {author}
  {\bibfnamefont {J.~C.}\ \bibnamefont {{Barentine}}}, \bibinfo {author}
  {\bibfnamefont {H.}~\bibnamefont {{Brewington}}}, \bibinfo {author}
  {\bibfnamefont {M.}~\bibnamefont {{Harvanek}}}, \bibinfo {author}
  {\bibfnamefont {J.}~\bibnamefont {{Krzesinski}}}, \bibinfo {author}
  {\bibfnamefont {D.}~\bibnamefont {{Long}}}, \bibinfo {author} {\bibfnamefont
  {A.}~\bibnamefont {{Nitta}}},\ and\ \bibinfo {author} {\bibfnamefont {S.~A.}\
  \bibnamefont {{Snedden}}},\ }\bibfield  {title} {\bibinfo {title} {{Discovery
  of an Unusual Dwarf Galaxy in the Outskirts of the Milky Way}},\ }\href
  {https://doi.org/10.1086/512183} {\bibfield  {journal} {\bibinfo  {journal}
  {\apjl}\ }\textbf {\bibinfo {volume} {656}},\ \bibinfo {pages} {L13}
  (\bibinfo {year} {2007})},\ \Eprint {https://arxiv.org/abs/astro-ph/0701154}
  {arXiv:astro-ph/0701154 [astro-ph]} \BibitemShut {NoStop}%
\bibitem [{\citenamefont {{Psaltis}}(2004)}]{2004astro.ph.10536P}%
  \BibitemOpen
  \bibfield  {author} {\bibinfo {author} {\bibfnamefont {D.}~\bibnamefont
  {{Psaltis}}},\ }\bibfield  {title} {\bibinfo {title} {{Accreting Neutron
  Stars and Black Holes: A Decade of Discoveries}},\ }\href
  {https://doi.org/10.48550/arXiv.astro-ph/0410536} {\bibfield  {journal}
  {\bibinfo  {journal} {arXiv e-prints}\ ,\ \bibinfo {eid} {astro-ph/0410536}}
  (\bibinfo {year} {2004})},\ \Eprint {https://arxiv.org/abs/astro-ph/0410536}
  {arXiv:astro-ph/0410536 [astro-ph]} \BibitemShut {NoStop}%
\bibitem [{\citenamefont {{Liu}}\ \emph {et~al.}(2006)\citenamefont {{Liu}},
  \citenamefont {{van Paradijs}},\ and\ \citenamefont {{van den
  Heuvel}}}]{2006A&A...455.1165L}%
  \BibitemOpen
  \bibfield  {author} {\bibinfo {author} {\bibfnamefont {Q.~Z.}\ \bibnamefont
  {{Liu}}}, \bibinfo {author} {\bibfnamefont {J.}~\bibnamefont {{van
  Paradijs}}},\ and\ \bibinfo {author} {\bibfnamefont {E.~P.~J.}\ \bibnamefont
  {{van den Heuvel}}},\ }\bibfield  {title} {\bibinfo {title} {{Catalogue of
  high-mass X-ray binaries in the Galaxy (4th edition)}},\ }\href
  {https://doi.org/10.1051/0004-6361:20064987} {\bibfield  {journal} {\bibinfo
  {journal} {\aap}\ }\textbf {\bibinfo {volume} {455}},\ \bibinfo {pages}
  {1165} (\bibinfo {year} {2006})},\ \Eprint {https://arxiv.org/abs/0707.0549}
  {arXiv:0707.0549 [astro-ph]} \BibitemShut {NoStop}%
\bibitem [{\citenamefont {{Liu}}\ \emph {et~al.}(2007)\citenamefont {{Liu}},
  \citenamefont {{van Paradijs}},\ and\ \citenamefont {{van den
  Heuvel}}}]{2007A&A...469..807L}%
  \BibitemOpen
  \bibfield  {author} {\bibinfo {author} {\bibfnamefont {Q.~Z.}\ \bibnamefont
  {{Liu}}}, \bibinfo {author} {\bibfnamefont {J.}~\bibnamefont {{van
  Paradijs}}},\ and\ \bibinfo {author} {\bibfnamefont {E.~P.~J.}\ \bibnamefont
  {{van den Heuvel}}},\ }\bibfield  {title} {\bibinfo {title} {{A catalogue of
  low-mass X-ray binaries in the Galaxy, LMC, and SMC (Fourth edition)}},\
  }\href {https://doi.org/10.1051/0004-6361:20077303} {\bibfield  {journal}
  {\bibinfo  {journal} {\aap}\ }\textbf {\bibinfo {volume} {469}},\ \bibinfo
  {pages} {807} (\bibinfo {year} {2007})},\ \Eprint
  {https://arxiv.org/abs/0707.0544} {arXiv:0707.0544 [astro-ph]} \BibitemShut
  {NoStop}%
\bibitem [{\citenamefont {{Hannikainen}}\ \emph {et~al.}(1998)\citenamefont
  {{Hannikainen}}, \citenamefont {{Hunstead}}, \citenamefont
  {{Campbell-Wilson}},\ and\ \citenamefont {{Sood}}}]{1998A&A...337..460H}%
  \BibitemOpen
  \bibfield  {author} {\bibinfo {author} {\bibfnamefont {D.~C.}\ \bibnamefont
  {{Hannikainen}}}, \bibinfo {author} {\bibfnamefont {R.~W.}\ \bibnamefont
  {{Hunstead}}}, \bibinfo {author} {\bibfnamefont {D.}~\bibnamefont
  {{Campbell-Wilson}}},\ and\ \bibinfo {author} {\bibfnamefont {R.~K.}\
  \bibnamefont {{Sood}}},\ }\bibfield  {title} {\bibinfo {title} {{MOST radio
  monitoring of GX 339-4}},\ }\href
  {https://doi.org/10.48550/arXiv.astro-ph/9805332} {\bibfield  {journal}
  {\bibinfo  {journal} {\aap}\ }\textbf {\bibinfo {volume} {337}},\ \bibinfo
  {pages} {460} (\bibinfo {year} {1998})},\ \Eprint
  {https://arxiv.org/abs/astro-ph/9805332} {arXiv:astro-ph/9805332 [astro-ph]}
  \BibitemShut {NoStop}%
\bibitem [{\citenamefont {{Gallo}}\ \emph {et~al.}(2003)\citenamefont
  {{Gallo}}, \citenamefont {{Fender}},\ and\ \citenamefont
  {{Pooley}}}]{2003MNRAS.344...60G}%
  \BibitemOpen
  \bibfield  {author} {\bibinfo {author} {\bibfnamefont {E.}~\bibnamefont
  {{Gallo}}}, \bibinfo {author} {\bibfnamefont {R.~P.}\ \bibnamefont
  {{Fender}}},\ and\ \bibinfo {author} {\bibfnamefont {G.~G.}\ \bibnamefont
  {{Pooley}}},\ }\bibfield  {title} {\bibinfo {title} {{A universal radio-X-ray
  correlation in low/hard state black hole binaries}},\ }\href
  {https://doi.org/10.1046/j.1365-8711.2003.06791.x} {\bibfield  {journal}
  {\bibinfo  {journal} {\mnras}\ }\textbf {\bibinfo {volume} {344}},\ \bibinfo
  {pages} {60} (\bibinfo {year} {2003})},\ \Eprint
  {https://arxiv.org/abs/astro-ph/0305231} {arXiv:astro-ph/0305231 [astro-ph]}
  \BibitemShut {NoStop}%
\bibitem [{\citenamefont {{Merloni}}\ \emph {et~al.}(2003)\citenamefont
  {{Merloni}}, \citenamefont {{Heinz}},\ and\ \citenamefont {{di
  Matteo}}}]{2003MNRAS.345.1057M}%
  \BibitemOpen
  \bibfield  {author} {\bibinfo {author} {\bibfnamefont {A.}~\bibnamefont
  {{Merloni}}}, \bibinfo {author} {\bibfnamefont {S.}~\bibnamefont {{Heinz}}},\
  and\ \bibinfo {author} {\bibfnamefont {T.}~\bibnamefont {{di Matteo}}},\
  }\bibfield  {title} {\bibinfo {title} {{A Fundamental Plane of black hole
  activity}},\ }\href {https://doi.org/10.1046/j.1365-2966.2003.07017.x}
  {\bibfield  {journal} {\bibinfo  {journal} {\mnras}\ }\textbf {\bibinfo
  {volume} {345}},\ \bibinfo {pages} {1057} (\bibinfo {year} {2003})},\ \Eprint
  {https://arxiv.org/abs/astro-ph/0305261} {arXiv:astro-ph/0305261 [astro-ph]}
  \BibitemShut {NoStop}%
\bibitem [{\citenamefont {{Falcke}}\ \emph {et~al.}(2004)\citenamefont
  {{Falcke}}, \citenamefont {{K{\"o}rding}},\ and\ \citenamefont
  {{Markoff}}}]{2004A&A...414..895F}%
  \BibitemOpen
  \bibfield  {author} {\bibinfo {author} {\bibfnamefont {H.}~\bibnamefont
  {{Falcke}}}, \bibinfo {author} {\bibfnamefont {E.}~\bibnamefont
  {{K{\"o}rding}}},\ and\ \bibinfo {author} {\bibfnamefont {S.}~\bibnamefont
  {{Markoff}}},\ }\bibfield  {title} {\bibinfo {title} {{A scheme to unify
  low-power accreting black holes. Jet-dominated accretion flows and the
  radio/X-ray correlation}},\ }\href
  {https://doi.org/10.1051/0004-6361:20031683} {\bibfield  {journal} {\bibinfo
  {journal} {\aap}\ }\textbf {\bibinfo {volume} {414}},\ \bibinfo {pages} {895}
  (\bibinfo {year} {2004})},\ \Eprint {https://arxiv.org/abs/astro-ph/0305335}
  {arXiv:astro-ph/0305335 [astro-ph]} \BibitemShut {NoStop}%
\bibitem [{\citenamefont {Gültekin}\ \emph {et~al.}(2019)\citenamefont
  {Gültekin}, \citenamefont {King}, \citenamefont {Cackett}, \citenamefont
  {Nyland}, \citenamefont {Miller}, \citenamefont {Matteo}, \citenamefont
  {Markoff},\ and\ \citenamefont {Rupen}}]{Gultekin_2019}%
  \BibitemOpen
  \bibfield  {author} {\bibinfo {author} {\bibfnamefont {K.}~\bibnamefont
  {Gültekin}}, \bibinfo {author} {\bibfnamefont {A.~L.}\ \bibnamefont {King}},
  \bibinfo {author} {\bibfnamefont {E.~M.}\ \bibnamefont {Cackett}}, \bibinfo
  {author} {\bibfnamefont {K.}~\bibnamefont {Nyland}}, \bibinfo {author}
  {\bibfnamefont {J.~M.}\ \bibnamefont {Miller}}, \bibinfo {author}
  {\bibfnamefont {T.~D.}\ \bibnamefont {Matteo}}, \bibinfo {author}
  {\bibfnamefont {S.}~\bibnamefont {Markoff}},\ and\ \bibinfo {author}
  {\bibfnamefont {M.~P.}\ \bibnamefont {Rupen}},\ }\bibfield  {title} {\bibinfo
  {title} {The fundamental plane of black hole accretion and its use as a black
  hole-mass estimator},\ }\href {https://doi.org/10.3847/1538-4357/aaf6b9}
  {\bibfield  {journal} {\bibinfo  {journal} {The Astrophysical Journal}\
  }\textbf {\bibinfo {volume} {871}},\ \bibinfo {pages} {80} (\bibinfo {year}
  {2019})}\BibitemShut {NoStop}%
\bibitem [{\citenamefont {{Marchant}}\ \emph
  {et~al.}(2017{\natexlab{b}})\citenamefont {{Marchant}}, \citenamefont
  {{Langer}}, \citenamefont {{Podsiadlowski}}, \citenamefont {{Tauris}},
  \citenamefont {{de Mink}}, \citenamefont {{Mandel}},\ and\ \citenamefont
  {{Moriya}}}]{2017A&A...604A..55M}%
  \BibitemOpen
  \bibfield  {author} {\bibinfo {author} {\bibfnamefont {P.}~\bibnamefont
  {{Marchant}}}, \bibinfo {author} {\bibfnamefont {N.}~\bibnamefont
  {{Langer}}}, \bibinfo {author} {\bibfnamefont {P.}~\bibnamefont
  {{Podsiadlowski}}}, \bibinfo {author} {\bibfnamefont {T.~M.}\ \bibnamefont
  {{Tauris}}}, \bibinfo {author} {\bibfnamefont {S.}~\bibnamefont {{de Mink}}},
  \bibinfo {author} {\bibfnamefont {I.}~\bibnamefont {{Mandel}}},\ and\
  \bibinfo {author} {\bibfnamefont {T.~J.}\ \bibnamefont {{Moriya}}},\
  }\bibfield  {title} {\bibinfo {title} {{Ultra-luminous X-ray sources and
  neutron-star-black-hole mergers from very massive close binaries at low
  metallicity}},\ }\href {https://doi.org/10.1051/0004-6361/201630188}
  {\bibfield  {journal} {\bibinfo  {journal} {\aap}\ }\textbf {\bibinfo
  {volume} {604}},\ \bibinfo {eid} {A55} (\bibinfo {year}
  {2017}{\natexlab{b}})},\ \Eprint {https://arxiv.org/abs/1705.04734}
  {arXiv:1705.04734 [astro-ph.HE]} \BibitemShut {NoStop}%
\bibitem [{\citenamefont {{van den Heuvel}}(2017)}]{2017JApA...38...45V}%
  \BibitemOpen
  \bibfield  {author} {\bibinfo {author} {\bibfnamefont {E.~P.~J.}\
  \bibnamefont {{van den Heuvel}}},\ }\bibfield  {title} {\bibinfo {title}
  {{Formation of Double Neutron Stars, Millisecond Pulsars and Double Black
  Holes}},\ }\href {https://doi.org/10.1007/s12036-017-9458-5} {\bibfield
  {journal} {\bibinfo  {journal} {Journal of Astrophysics and Astronomy}\
  }\textbf {\bibinfo {volume} {38}},\ \bibinfo {eid} {45} (\bibinfo {year}
  {2017})},\ \Eprint {https://arxiv.org/abs/1709.07636} {arXiv:1709.07636
  [astro-ph.HE]} \BibitemShut {NoStop}%
\bibitem [{\citenamefont {{Qin}}\ \emph {et~al.}(2023)\citenamefont {{Qin}},
  \citenamefont {{Jiang}},\ and\ \citenamefont {{Chen}}}]{2023arXiv230106243Q}%
  \BibitemOpen
  \bibfield  {author} {\bibinfo {author} {\bibfnamefont {K.}~\bibnamefont
  {{Qin}}}, \bibinfo {author} {\bibfnamefont {L.}~\bibnamefont {{Jiang}}},\
  and\ \bibinfo {author} {\bibfnamefont {W.-C.}\ \bibnamefont {{Chen}}},\
  }\bibfield  {title} {\bibinfo {title} {{Black Hole Ultra-compact X-ray
  Binaries: Galactic Low-Frequency Gravitational Wave Sources}},\ }\href
  {https://doi.org/10.48550/arXiv.2301.06243} {\bibfield  {journal} {\bibinfo
  {journal} {arXiv e-prints}\ ,\ \bibinfo {eid} {arXiv:2301.06243}} (\bibinfo
  {year} {2023})},\ \Eprint {https://arxiv.org/abs/2301.06243}
  {arXiv:2301.06243 [astro-ph.HE]} \BibitemShut {NoStop}%
\bibitem [{\citenamefont {Burnight}(1949)}]{burnight1949soft}%
  \BibitemOpen
  \bibfield  {author} {\bibinfo {author} {\bibfnamefont {T.~R.}\ \bibnamefont
  {Burnight}},\ }\bibfield  {title} {\bibinfo {title} {Soft x-radiation in the
  upper atmosphere},\ }\href@noop {} {\bibfield  {journal} {\bibinfo  {journal}
  {Phys. Rev}\ }\textbf {\bibinfo {volume} {76}},\ \bibinfo {pages} {19}
  (\bibinfo {year} {1949})}\BibitemShut {NoStop}%
\bibitem [{\citenamefont {{Giacconi}}\ \emph {et~al.}(1962)\citenamefont
  {{Giacconi}}, \citenamefont {{Gursky}}, \citenamefont {{Paolini}},\ and\
  \citenamefont {{Rossi}}}]{1962PhRvL...9..439G}%
  \BibitemOpen
  \bibfield  {author} {\bibinfo {author} {\bibfnamefont {R.}~\bibnamefont
  {{Giacconi}}}, \bibinfo {author} {\bibfnamefont {H.}~\bibnamefont
  {{Gursky}}}, \bibinfo {author} {\bibfnamefont {F.~R.}\ \bibnamefont
  {{Paolini}}},\ and\ \bibinfo {author} {\bibfnamefont {B.~B.}\ \bibnamefont
  {{Rossi}}},\ }\bibfield  {title} {\bibinfo {title} {{Evidence for x Rays From
  Sources Outside the Solar System}},\ }\href
  {https://doi.org/10.1103/PhysRevLett.9.439} {\bibfield  {journal} {\bibinfo
  {journal} {\prl}\ }\textbf {\bibinfo {volume} {9}},\ \bibinfo {pages} {439}
  (\bibinfo {year} {1962})}\BibitemShut {NoStop}%
\bibitem [{\citenamefont {{Shklovsky}}(1967)}]{1967ApJ...148L...1S}%
  \BibitemOpen
  \bibfield  {author} {\bibinfo {author} {\bibfnamefont {I.~S.}\ \bibnamefont
  {{Shklovsky}}},\ }\bibfield  {title} {\bibinfo {title} {{On the Nature of the
  Source of X-Ray Emission of Sco XR-1.}},\ }\href
  {https://doi.org/10.1086/180001} {\bibfield  {journal} {\bibinfo  {journal}
  {\apjl}\ }\textbf {\bibinfo {volume} {148}},\ \bibinfo {pages} {L1} (\bibinfo
  {year} {1967})}\BibitemShut {NoStop}%
\bibitem [{\citenamefont {{Bowyer}}\ \emph {et~al.}(1965)\citenamefont
  {{Bowyer}}, \citenamefont {{Byram}}, \citenamefont {{Chubb}},\ and\
  \citenamefont {{Friedman}}}]{1965Sci...147..394B}%
  \BibitemOpen
  \bibfield  {author} {\bibinfo {author} {\bibfnamefont {S.}~\bibnamefont
  {{Bowyer}}}, \bibinfo {author} {\bibfnamefont {E.~T.}\ \bibnamefont
  {{Byram}}}, \bibinfo {author} {\bibfnamefont {T.~A.}\ \bibnamefont
  {{Chubb}}},\ and\ \bibinfo {author} {\bibfnamefont {H.}~\bibnamefont
  {{Friedman}}},\ }\bibfield  {title} {\bibinfo {title} {{Cosmic X-ray
  Sources}},\ }\href {https://doi.org/10.1126/science.147.3656.394} {\bibfield
  {journal} {\bibinfo  {journal} {Science}\ }\textbf {\bibinfo {volume}
  {147}},\ \bibinfo {pages} {394} (\bibinfo {year} {1965})}\BibitemShut
  {NoStop}%
\bibitem [{\citenamefont {{Seward}}\ and\ \citenamefont
  {{Charles}}(2010)}]{2010exru.book.....S}%
  \BibitemOpen
  \bibfield  {author} {\bibinfo {author} {\bibfnamefont {F.~D.}\ \bibnamefont
  {{Seward}}}\ and\ \bibinfo {author} {\bibfnamefont {P.~A.}\ \bibnamefont
  {{Charles}}},\ }\href@noop {} {\emph {\bibinfo {title} {{Exploring the X-ray
  Universe}}}}\ (\bibinfo {year} {2010})\BibitemShut {NoStop}%
\bibitem [{\citenamefont {{Giacconi}}\ \emph {et~al.}(1971)\citenamefont
  {{Giacconi}}, \citenamefont {{Kellogg}}, \citenamefont {{Gorenstein}},
  \citenamefont {{Gursky}},\ and\ \citenamefont
  {{Tananbaum}}}]{1971ApJ...165L..27G}%
  \BibitemOpen
  \bibfield  {author} {\bibinfo {author} {\bibfnamefont {R.}~\bibnamefont
  {{Giacconi}}}, \bibinfo {author} {\bibfnamefont {E.}~\bibnamefont
  {{Kellogg}}}, \bibinfo {author} {\bibfnamefont {P.}~\bibnamefont
  {{Gorenstein}}}, \bibinfo {author} {\bibfnamefont {H.}~\bibnamefont
  {{Gursky}}},\ and\ \bibinfo {author} {\bibfnamefont {H.}~\bibnamefont
  {{Tananbaum}}},\ }\bibfield  {title} {\bibinfo {title} {{An X-Ray Scan of the
  Galactic Plane from UHURU}},\ }\href {https://doi.org/10.1086/180711}
  {\bibfield  {journal} {\bibinfo  {journal} {\apjl}\ }\textbf {\bibinfo
  {volume} {165}},\ \bibinfo {pages} {L27} (\bibinfo {year}
  {1971})}\BibitemShut {NoStop}%
\bibitem [{\citenamefont {{Forman}}\ \emph {et~al.}(1978)\citenamefont
  {{Forman}}, \citenamefont {{Jones}}, \citenamefont {{Cominsky}},
  \citenamefont {{Julien}}, \citenamefont {{Murray}}, \citenamefont {{Peters}},
  \citenamefont {{Tananbaum}},\ and\ \citenamefont
  {{Giacconi}}}]{1978ApJS...38..357F}%
  \BibitemOpen
  \bibfield  {author} {\bibinfo {author} {\bibfnamefont {W.}~\bibnamefont
  {{Forman}}}, \bibinfo {author} {\bibfnamefont {C.}~\bibnamefont {{Jones}}},
  \bibinfo {author} {\bibfnamefont {L.}~\bibnamefont {{Cominsky}}}, \bibinfo
  {author} {\bibfnamefont {P.}~\bibnamefont {{Julien}}}, \bibinfo {author}
  {\bibfnamefont {S.}~\bibnamefont {{Murray}}}, \bibinfo {author}
  {\bibfnamefont {G.}~\bibnamefont {{Peters}}}, \bibinfo {author}
  {\bibfnamefont {H.}~\bibnamefont {{Tananbaum}}},\ and\ \bibinfo {author}
  {\bibfnamefont {R.}~\bibnamefont {{Giacconi}}},\ }\bibfield  {title}
  {\bibinfo {title} {{The fourth Uhuru catalog of X-ray sources.}},\ }\href
  {https://doi.org/10.1086/190561} {\bibfield  {journal} {\bibinfo  {journal}
  {\apjs}\ }\textbf {\bibinfo {volume} {38}},\ \bibinfo {pages} {357} (\bibinfo
  {year} {1978})}\BibitemShut {NoStop}%
\bibitem [{\citenamefont {{Oda}}\ \emph {et~al.}(1971)\citenamefont {{Oda}},
  \citenamefont {{Gorenstein}}, \citenamefont {{Gursky}}, \citenamefont
  {{Kellogg}}, \citenamefont {{Schreier}}, \citenamefont {{Tananbaum}},\ and\
  \citenamefont {{Giacconi}}}]{1971ApJ...166L...1O}%
  \BibitemOpen
  \bibfield  {author} {\bibinfo {author} {\bibfnamefont {M.}~\bibnamefont
  {{Oda}}}, \bibinfo {author} {\bibfnamefont {P.}~\bibnamefont {{Gorenstein}}},
  \bibinfo {author} {\bibfnamefont {H.}~\bibnamefont {{Gursky}}}, \bibinfo
  {author} {\bibfnamefont {E.}~\bibnamefont {{Kellogg}}}, \bibinfo {author}
  {\bibfnamefont {E.}~\bibnamefont {{Schreier}}}, \bibinfo {author}
  {\bibfnamefont {H.}~\bibnamefont {{Tananbaum}}},\ and\ \bibinfo {author}
  {\bibfnamefont {R.}~\bibnamefont {{Giacconi}}},\ }\bibfield  {title}
  {\bibinfo {title} {{X-Ray Pulsations from Cygnus X-1 Observed from UHURU}},\
  }\href {https://doi.org/10.1086/180726} {\bibfield  {journal} {\bibinfo
  {journal} {\apjl}\ }\textbf {\bibinfo {volume} {166}},\ \bibinfo {pages} {L1}
  (\bibinfo {year} {1971})}\BibitemShut {NoStop}%
\bibitem [{\citenamefont {{Schreier}}\ \emph {et~al.}(1971)\citenamefont
  {{Schreier}}, \citenamefont {{Gursky}}, \citenamefont {{Kellogg}},
  \citenamefont {{Tananbaum}},\ and\ \citenamefont
  {{Giacconi}}}]{1971ApJ...170L..21S}%
  \BibitemOpen
  \bibfield  {author} {\bibinfo {author} {\bibfnamefont {E.}~\bibnamefont
  {{Schreier}}}, \bibinfo {author} {\bibfnamefont {H.}~\bibnamefont
  {{Gursky}}}, \bibinfo {author} {\bibfnamefont {E.}~\bibnamefont {{Kellogg}}},
  \bibinfo {author} {\bibfnamefont {H.}~\bibnamefont {{Tananbaum}}},\ and\
  \bibinfo {author} {\bibfnamefont {R.}~\bibnamefont {{Giacconi}}},\ }\bibfield
   {title} {\bibinfo {title} {{Further Observations of the Pulsating X-Ray
  Source Cygnus X-1 from UHURU}},\ }\href {https://doi.org/10.1086/180833}
  {\bibfield  {journal} {\bibinfo  {journal} {\apjl}\ }\textbf {\bibinfo
  {volume} {170}},\ \bibinfo {pages} {L21} (\bibinfo {year}
  {1971})}\BibitemShut {NoStop}%
\bibitem [{\citenamefont {{Giacconi}}\ \emph {et~al.}(1979)\citenamefont
  {{Giacconi}}, \citenamefont {{Branduardi}}, \citenamefont {{Briel}},
  \citenamefont {{Epstein}}, \citenamefont {{Fabricant}}, \citenamefont
  {{Feigelson}}, \citenamefont {{Forman}}, \citenamefont {{Gorenstein}},
  \citenamefont {{Grindlay}}, \citenamefont {{Gursky}}, \citenamefont
  {{Harnden}}, \citenamefont {{Henry}}, \citenamefont {{Jones}}, \citenamefont
  {{Kellogg}}, \citenamefont {{Koch}}, \citenamefont {{Murray}}, \citenamefont
  {{Schreier}}, \citenamefont {{Seward}}, \citenamefont {{Tananbaum}},
  \citenamefont {{Topka}}, \citenamefont {{Van Speybroeck}}, \citenamefont
  {{Holt}}, \citenamefont {{Becker}}, \citenamefont {{Boldt}}, \citenamefont
  {{Serlemitsos}}, \citenamefont {{Clark}}, \citenamefont {{Canizares}},
  \citenamefont {{Markert}}, \citenamefont {{Novick}}, \citenamefont
  {{Helfand}},\ and\ \citenamefont {{Long}}}]{1979ApJ...230..540G}%
  \BibitemOpen
  \bibfield  {author} {\bibinfo {author} {\bibfnamefont {R.}~\bibnamefont
  {{Giacconi}}}, \bibinfo {author} {\bibfnamefont {G.}~\bibnamefont
  {{Branduardi}}}, \bibinfo {author} {\bibfnamefont {U.}~\bibnamefont
  {{Briel}}}, \bibinfo {author} {\bibfnamefont {A.}~\bibnamefont {{Epstein}}},
  \bibinfo {author} {\bibfnamefont {D.}~\bibnamefont {{Fabricant}}}, \bibinfo
  {author} {\bibfnamefont {E.}~\bibnamefont {{Feigelson}}}, \bibinfo {author}
  {\bibfnamefont {W.}~\bibnamefont {{Forman}}}, \bibinfo {author}
  {\bibfnamefont {P.}~\bibnamefont {{Gorenstein}}}, \bibinfo {author}
  {\bibfnamefont {J.}~\bibnamefont {{Grindlay}}}, \bibinfo {author}
  {\bibfnamefont {H.}~\bibnamefont {{Gursky}}}, \bibinfo {author}
  {\bibfnamefont {F.~R.}\ \bibnamefont {{Harnden}}}, \bibinfo {author}
  {\bibfnamefont {J.~P.}\ \bibnamefont {{Henry}}}, \bibinfo {author}
  {\bibfnamefont {C.}~\bibnamefont {{Jones}}}, \bibinfo {author} {\bibfnamefont
  {E.}~\bibnamefont {{Kellogg}}}, \bibinfo {author} {\bibfnamefont
  {D.}~\bibnamefont {{Koch}}}, \bibinfo {author} {\bibfnamefont
  {S.}~\bibnamefont {{Murray}}}, \bibinfo {author} {\bibfnamefont
  {E.}~\bibnamefont {{Schreier}}}, \bibinfo {author} {\bibfnamefont
  {F.}~\bibnamefont {{Seward}}}, \bibinfo {author} {\bibfnamefont
  {H.}~\bibnamefont {{Tananbaum}}}, \bibinfo {author} {\bibfnamefont
  {K.}~\bibnamefont {{Topka}}}, \bibinfo {author} {\bibfnamefont
  {L.}~\bibnamefont {{Van Speybroeck}}}, \bibinfo {author} {\bibfnamefont
  {S.~S.}\ \bibnamefont {{Holt}}}, \bibinfo {author} {\bibfnamefont {R.~H.}\
  \bibnamefont {{Becker}}}, \bibinfo {author} {\bibfnamefont {E.~A.}\
  \bibnamefont {{Boldt}}}, \bibinfo {author} {\bibfnamefont {P.~J.}\
  \bibnamefont {{Serlemitsos}}}, \bibinfo {author} {\bibfnamefont
  {G.}~\bibnamefont {{Clark}}}, \bibinfo {author} {\bibfnamefont
  {C.}~\bibnamefont {{Canizares}}}, \bibinfo {author} {\bibfnamefont
  {T.}~\bibnamefont {{Markert}}}, \bibinfo {author} {\bibfnamefont
  {R.}~\bibnamefont {{Novick}}}, \bibinfo {author} {\bibfnamefont
  {D.}~\bibnamefont {{Helfand}}},\ and\ \bibinfo {author} {\bibfnamefont
  {K.}~\bibnamefont {{Long}}},\ }\bibfield  {title} {\bibinfo {title} {{The
  Einstein (HEAO 2) X-ray Observatory.}},\ }\href
  {https://doi.org/10.1086/157110} {\bibfield  {journal} {\bibinfo  {journal}
  {\apj}\ }\textbf {\bibinfo {volume} {230}},\ \bibinfo {pages} {540} (\bibinfo
  {year} {1979})}\BibitemShut {NoStop}%
\bibitem [{\citenamefont {{Gorenstein}}\ \emph {et~al.}(1975)\citenamefont
  {{Gorenstein}}, \citenamefont {{Gursky}}, \citenamefont {{Harnden}},
  \citenamefont {{Decaprio}},\ and\ \citenamefont
  {{Bjorkholm}}}]{1975ITNS...22..616G}%
  \BibitemOpen
  \bibfield  {author} {\bibinfo {author} {\bibfnamefont {P.}~\bibnamefont
  {{Gorenstein}}}, \bibinfo {author} {\bibfnamefont {H.}~\bibnamefont
  {{Gursky}}}, \bibinfo {author} {\bibfnamefont {J.}~\bibnamefont {{Harnden}},
  \bibfnamefont {F.~R.}}, \bibinfo {author} {\bibfnamefont {A.}~\bibnamefont
  {{Decaprio}}},\ and\ \bibinfo {author} {\bibfnamefont {P.}~\bibnamefont
  {{Bjorkholm}}},\ }\bibfield  {title} {\bibinfo {title} {{Large area soft
  X-ray imaging system for cosmic X-ray studies from rockets.}},\ }\href
  {https://doi.org/10.1109/TNS.1975.4327715} {\bibfield  {journal} {\bibinfo
  {journal} {IEEE Transactions on Nuclear Science}\ }\textbf {\bibinfo {volume}
  {22}},\ \bibinfo {pages} {616} (\bibinfo {year} {1975})}\BibitemShut
  {NoStop}%
\bibitem [{\citenamefont {{Helfand}}(1984)}]{1984PASP...96..913H}%
  \BibitemOpen
  \bibfield  {author} {\bibinfo {author} {\bibfnamefont {D.~J.}\ \bibnamefont
  {{Helfand}}},\ }\bibfield  {title} {\bibinfo {title} {{Endpoints of stellar
  evolution : X-ray surveys of the Local Group.}},\ }\href
  {https://doi.org/10.1086/131455} {\bibfield  {journal} {\bibinfo  {journal}
  {\pasp}\ }\textbf {\bibinfo {volume} {96}},\ \bibinfo {pages} {913} (\bibinfo
  {year} {1984})}\BibitemShut {NoStop}%
\bibitem [{\citenamefont {{Clark}}(1975)}]{1975ApJ...199L.143C}%
  \BibitemOpen
  \bibfield  {author} {\bibinfo {author} {\bibfnamefont {G.~W.}\ \bibnamefont
  {{Clark}}},\ }\bibfield  {title} {\bibinfo {title} {{X-ray binaries in
  globular clusters.}},\ }\href {https://doi.org/10.1086/181869} {\bibfield
  {journal} {\bibinfo  {journal} {\apjl}\ }\textbf {\bibinfo {volume} {199}},\
  \bibinfo {pages} {L143} (\bibinfo {year} {1975})}\BibitemShut {NoStop}%
\bibitem [{\citenamefont {{Truemper}}(1982)}]{1982AdSpR...2d.241T}%
  \BibitemOpen
  \bibfield  {author} {\bibinfo {author} {\bibfnamefont {J.}~\bibnamefont
  {{Truemper}}},\ }\bibfield  {title} {\bibinfo {title} {{The ROSAT mission}},\
  }\href {https://doi.org/10.1016/0273-1177(82)90070-9} {\bibfield  {journal}
  {\bibinfo  {journal} {Advances in Space Research}\ }\textbf {\bibinfo
  {volume} {2}},\ \bibinfo {pages} {241} (\bibinfo {year} {1982})}\BibitemShut
  {NoStop}%
\bibitem [{\citenamefont {{Tanaka}}\ \emph {et~al.}(1994)\citenamefont
  {{Tanaka}}, \citenamefont {{Inoue}},\ and\ \citenamefont
  {{Holt}}}]{1994PASJ...46L..37T}%
  \BibitemOpen
  \bibfield  {author} {\bibinfo {author} {\bibfnamefont {Y.}~\bibnamefont
  {{Tanaka}}}, \bibinfo {author} {\bibfnamefont {H.}~\bibnamefont {{Inoue}}},\
  and\ \bibinfo {author} {\bibfnamefont {S.~S.}\ \bibnamefont {{Holt}}},\
  }\bibfield  {title} {\bibinfo {title} {{The X-Ray Astronomy Satellite
  ASCA}},\ }\href@noop {} {\bibfield  {journal} {\bibinfo  {journal} {\pasj}\
  }\textbf {\bibinfo {volume} {46}},\ \bibinfo {pages} {L37} (\bibinfo {year}
  {1994})}\BibitemShut {NoStop}%
\bibitem [{\citenamefont {{Jahoda}}\ \emph {et~al.}(1996)\citenamefont
  {{Jahoda}}, \citenamefont {{Swank}}, \citenamefont {{Giles}}, \citenamefont
  {{Stark}}, \citenamefont {{Strohmayer}}, \citenamefont {{Zhang}},\ and\
  \citenamefont {{Morgan}}}]{1996SPIE.2808...59J}%
  \BibitemOpen
  \bibfield  {author} {\bibinfo {author} {\bibfnamefont {K.}~\bibnamefont
  {{Jahoda}}}, \bibinfo {author} {\bibfnamefont {J.~H.}\ \bibnamefont
  {{Swank}}}, \bibinfo {author} {\bibfnamefont {A.~B.}\ \bibnamefont
  {{Giles}}}, \bibinfo {author} {\bibfnamefont {M.~J.}\ \bibnamefont
  {{Stark}}}, \bibinfo {author} {\bibfnamefont {T.}~\bibnamefont
  {{Strohmayer}}}, \bibinfo {author} {\bibfnamefont {W.}~\bibnamefont
  {{Zhang}}},\ and\ \bibinfo {author} {\bibfnamefont {E.~H.}\ \bibnamefont
  {{Morgan}}},\ }\bibfield  {title} {\bibinfo {title} {{In-orbit performance
  and calibration of the Rossi X-ray Timing Explorer (RXTE) Proportional
  Counter Array (PCA)}},\ }in\ \href {https://doi.org/10.1117/12.256034} {\emph
  {\bibinfo {booktitle} {EUV, X-Ray, and Gamma-Ray Instrumentation for
  Astronomy VII}}},\ \bibinfo {series} {Society of Photo-Optical
  Instrumentation Engineers (SPIE) Conference Series}, Vol.\ \bibinfo {volume}
  {2808},\ \bibinfo {editor} {edited by\ \bibinfo {editor} {\bibfnamefont
  {O.~H.}\ \bibnamefont {{Siegmund}}}\ and\ \bibinfo {editor} {\bibfnamefont
  {M.~A.}\ \bibnamefont {{Gummin}}}}\ (\bibinfo {year} {1996})\ pp.\ \bibinfo
  {pages} {59--70}\BibitemShut {NoStop}%
\bibitem [{\citenamefont {{Weisskopf}}\ \emph {et~al.}(2000)\citenamefont
  {{Weisskopf}}, \citenamefont {{Tananbaum}}, \citenamefont {{Van
  Speybroeck}},\ and\ \citenamefont {{O'Dell}}}]{2000SPIE.4012....2W}%
  \BibitemOpen
  \bibfield  {author} {\bibinfo {author} {\bibfnamefont {M.~C.}\ \bibnamefont
  {{Weisskopf}}}, \bibinfo {author} {\bibfnamefont {H.~D.}\ \bibnamefont
  {{Tananbaum}}}, \bibinfo {author} {\bibfnamefont {L.~P.}\ \bibnamefont {{Van
  Speybroeck}}},\ and\ \bibinfo {author} {\bibfnamefont {S.~L.}\ \bibnamefont
  {{O'Dell}}},\ }\bibfield  {title} {\bibinfo {title} {{Chandra X-ray
  Observatory (CXO): overview}},\ }in\ \href
  {https://doi.org/10.1117/12.391545} {\emph {\bibinfo {booktitle} {X-Ray
  Optics, Instruments, and Missions III}}},\ \bibinfo {series} {Society of
  Photo-Optical Instrumentation Engineers (SPIE) Conference Series}, Vol.\
  \bibinfo {volume} {4012},\ \bibinfo {editor} {edited by\ \bibinfo {editor}
  {\bibfnamefont {J.~E.}\ \bibnamefont {{Truemper}}}\ and\ \bibinfo {editor}
  {\bibfnamefont {B.}~\bibnamefont {{Aschenbach}}}}\ (\bibinfo {year} {2000})\
  pp.\ \bibinfo {pages} {2--16},\ \Eprint
  {https://arxiv.org/abs/astro-ph/0004127} {arXiv:astro-ph/0004127 [astro-ph]}
  \BibitemShut {NoStop}%
\bibitem [{\citenamefont {{Lumb}}\ \emph {et~al.}(2012)\citenamefont {{Lumb}},
  \citenamefont {{Schartel}},\ and\ \citenamefont
  {{Jansen}}}]{2012OptEn..51a1009L}%
  \BibitemOpen
  \bibfield  {author} {\bibinfo {author} {\bibfnamefont {D.~H.}\ \bibnamefont
  {{Lumb}}}, \bibinfo {author} {\bibfnamefont {N.}~\bibnamefont {{Schartel}}},\
  and\ \bibinfo {author} {\bibfnamefont {F.~A.}\ \bibnamefont {{Jansen}}},\
  }\bibfield  {title} {\bibinfo {title} {{X-ray Multi-mirror Mission
  (XMM-Newton) observatory}},\ }\href
  {https://doi.org/10.1117/1.OE.51.1.011009} {\bibfield  {journal} {\bibinfo
  {journal} {Optical Engineering}\ }\textbf {\bibinfo {volume} {51}},\ \bibinfo
  {eid} {011009-011009-11} (\bibinfo {year} {2012})}\BibitemShut {NoStop}%
\bibitem [{\citenamefont {{Papadopoulou}}\ \emph {et~al.}(2016)\citenamefont
  {{Papadopoulou}}, \citenamefont {{Phillipps}},\ and\ \citenamefont
  {{Young}}}]{2016MNRAS.460.4513P}%
  \BibitemOpen
  \bibfield  {author} {\bibinfo {author} {\bibfnamefont {M.}~\bibnamefont
  {{Papadopoulou}}}, \bibinfo {author} {\bibfnamefont {S.}~\bibnamefont
  {{Phillipps}}},\ and\ \bibinfo {author} {\bibfnamefont {A.~J.}\ \bibnamefont
  {{Young}}},\ }\bibfield  {title} {\bibinfo {title} {{X-ray sources in dwarf
  galaxies in the Virgo cluster and the nearby field}},\ }\href
  {https://doi.org/10.1093/mnras/stw1297} {\bibfield  {journal} {\bibinfo
  {journal} {\mnras}\ }\textbf {\bibinfo {volume} {460}},\ \bibinfo {pages}
  {4513} (\bibinfo {year} {2016})}\BibitemShut {NoStop}%
\bibitem [{\citenamefont {{Riccio}}\ \emph {et~al.}(2022)\citenamefont
  {{Riccio}}, \citenamefont {{Paolillo}}, \citenamefont {{Cantiello}},
  \citenamefont {{D'Abrusco}}, \citenamefont {{Jin}}, \citenamefont {{Li}},
  \citenamefont {{Puzia}}, \citenamefont {{Mieske}}, \citenamefont {{Prole}},
  \citenamefont {{Iodice}}, \citenamefont {{D'Ago}}, \citenamefont {{Gatto}},\
  and\ \citenamefont {{Spavone}}}]{2022A&A...664A..41R}%
  \BibitemOpen
  \bibfield  {author} {\bibinfo {author} {\bibfnamefont {G.}~\bibnamefont
  {{Riccio}}}, \bibinfo {author} {\bibfnamefont {M.}~\bibnamefont
  {{Paolillo}}}, \bibinfo {author} {\bibfnamefont {M.}~\bibnamefont
  {{Cantiello}}}, \bibinfo {author} {\bibfnamefont {R.}~\bibnamefont
  {{D'Abrusco}}}, \bibinfo {author} {\bibfnamefont {X.}~\bibnamefont {{Jin}}},
  \bibinfo {author} {\bibfnamefont {Z.}~\bibnamefont {{Li}}}, \bibinfo {author}
  {\bibfnamefont {T.}~\bibnamefont {{Puzia}}}, \bibinfo {author} {\bibfnamefont
  {S.}~\bibnamefont {{Mieske}}}, \bibinfo {author} {\bibfnamefont {D.~J.}\
  \bibnamefont {{Prole}}}, \bibinfo {author} {\bibfnamefont {E.}~\bibnamefont
  {{Iodice}}}, \bibinfo {author} {\bibfnamefont {G.}~\bibnamefont {{D'Ago}}},
  \bibinfo {author} {\bibfnamefont {M.}~\bibnamefont {{Gatto}}},\ and\ \bibinfo
  {author} {\bibfnamefont {M.}~\bibnamefont {{Spavone}}},\ }\bibfield  {title}
  {\bibinfo {title} {{Properties of intra-cluster low-mass X-ray binaries in
  Fornax globular clusters}},\ }\href
  {https://doi.org/10.1051/0004-6361/202142894} {\bibfield  {journal} {\bibinfo
   {journal} {\aap}\ }\textbf {\bibinfo {volume} {664}},\ \bibinfo {eid} {A41}
  (\bibinfo {year} {2022})},\ \Eprint {https://arxiv.org/abs/2206.14207}
  {arXiv:2206.14207 [astro-ph.HE]} \BibitemShut {NoStop}%
\bibitem [{\citenamefont {{Rosen}}\ \emph {et~al.}(2016)\citenamefont
  {{Rosen}}, \citenamefont {{Webb}}, \citenamefont {{Watson}}, \citenamefont
  {{Ballet}}, \citenamefont {{Barret}}, \citenamefont {{Braito}}, \citenamefont
  {{Carrera}}, \citenamefont {{Ceballos}}, \citenamefont {{Coriat}},
  \citenamefont {{Della Ceca}}, \citenamefont {{Denkinson}}, \citenamefont
  {{Esquej}}, \citenamefont {{Farrell}}, \citenamefont {{Freyberg}},
  \citenamefont {{Gris{\'e}}}, \citenamefont {{Guillout}}, \citenamefont
  {{Heil}}, \citenamefont {{Koliopanos}}, \citenamefont {{Law-Green}},
  \citenamefont {{Lamer}}, \citenamefont {{Lin}}, \citenamefont {{Martino}},
  \citenamefont {{Michel}}, \citenamefont {{Motch}}, \citenamefont {{Nebot
  Gomez-Moran}}, \citenamefont {{Page}}, \citenamefont {{Page}}, \citenamefont
  {{Page}}, \citenamefont {{Pakull}}, \citenamefont {{Pye}}, \citenamefont
  {{Read}}, \citenamefont {{Rodriguez}}, \citenamefont {{Sakano}},
  \citenamefont {{Saxton}}, \citenamefont {{Schwope}}, \citenamefont {{Scott}},
  \citenamefont {{Sturm}}, \citenamefont {{Traulsen}}, \citenamefont
  {{Yershov}},\ and\ \citenamefont {{Zolotukhin}}}]{2016A&A...590A...1R}%
  \BibitemOpen
  \bibfield  {author} {\bibinfo {author} {\bibfnamefont {S.~R.}\ \bibnamefont
  {{Rosen}}}, \bibinfo {author} {\bibfnamefont {N.~A.}\ \bibnamefont {{Webb}}},
  \bibinfo {author} {\bibfnamefont {M.~G.}\ \bibnamefont {{Watson}}}, \bibinfo
  {author} {\bibfnamefont {J.}~\bibnamefont {{Ballet}}}, \bibinfo {author}
  {\bibfnamefont {D.}~\bibnamefont {{Barret}}}, \bibinfo {author}
  {\bibfnamefont {V.}~\bibnamefont {{Braito}}}, \bibinfo {author}
  {\bibfnamefont {F.~J.}\ \bibnamefont {{Carrera}}}, \bibinfo {author}
  {\bibfnamefont {M.~T.}\ \bibnamefont {{Ceballos}}}, \bibinfo {author}
  {\bibfnamefont {M.}~\bibnamefont {{Coriat}}}, \bibinfo {author}
  {\bibfnamefont {R.}~\bibnamefont {{Della Ceca}}}, \bibinfo {author}
  {\bibfnamefont {G.}~\bibnamefont {{Denkinson}}}, \bibinfo {author}
  {\bibfnamefont {P.}~\bibnamefont {{Esquej}}}, \bibinfo {author}
  {\bibfnamefont {S.~A.}\ \bibnamefont {{Farrell}}}, \bibinfo {author}
  {\bibfnamefont {M.}~\bibnamefont {{Freyberg}}}, \bibinfo {author}
  {\bibfnamefont {F.}~\bibnamefont {{Gris{\'e}}}}, \bibinfo {author}
  {\bibfnamefont {P.}~\bibnamefont {{Guillout}}}, \bibinfo {author}
  {\bibfnamefont {L.}~\bibnamefont {{Heil}}}, \bibinfo {author} {\bibfnamefont
  {F.}~\bibnamefont {{Koliopanos}}}, \bibinfo {author} {\bibfnamefont
  {D.}~\bibnamefont {{Law-Green}}}, \bibinfo {author} {\bibfnamefont
  {G.}~\bibnamefont {{Lamer}}}, \bibinfo {author} {\bibfnamefont
  {D.}~\bibnamefont {{Lin}}}, \bibinfo {author} {\bibfnamefont
  {R.}~\bibnamefont {{Martino}}}, \bibinfo {author} {\bibfnamefont
  {L.}~\bibnamefont {{Michel}}}, \bibinfo {author} {\bibfnamefont
  {C.}~\bibnamefont {{Motch}}}, \bibinfo {author} {\bibfnamefont
  {A.}~\bibnamefont {{Nebot Gomez-Moran}}}, \bibinfo {author} {\bibfnamefont
  {C.~G.}\ \bibnamefont {{Page}}}, \bibinfo {author} {\bibfnamefont
  {K.}~\bibnamefont {{Page}}}, \bibinfo {author} {\bibfnamefont
  {M.}~\bibnamefont {{Page}}}, \bibinfo {author} {\bibfnamefont {M.~W.}\
  \bibnamefont {{Pakull}}}, \bibinfo {author} {\bibfnamefont {J.}~\bibnamefont
  {{Pye}}}, \bibinfo {author} {\bibfnamefont {A.}~\bibnamefont {{Read}}},
  \bibinfo {author} {\bibfnamefont {P.}~\bibnamefont {{Rodriguez}}}, \bibinfo
  {author} {\bibfnamefont {M.}~\bibnamefont {{Sakano}}}, \bibinfo {author}
  {\bibfnamefont {R.}~\bibnamefont {{Saxton}}}, \bibinfo {author}
  {\bibfnamefont {A.}~\bibnamefont {{Schwope}}}, \bibinfo {author}
  {\bibfnamefont {A.~E.}\ \bibnamefont {{Scott}}}, \bibinfo {author}
  {\bibfnamefont {R.}~\bibnamefont {{Sturm}}}, \bibinfo {author} {\bibfnamefont
  {I.}~\bibnamefont {{Traulsen}}}, \bibinfo {author} {\bibfnamefont
  {V.}~\bibnamefont {{Yershov}}},\ and\ \bibinfo {author} {\bibfnamefont
  {I.}~\bibnamefont {{Zolotukhin}}},\ }\bibfield  {title} {\bibinfo {title}
  {{The XMM-Newton serendipitous survey. VII. The third XMM-Newton
  serendipitous source catalogue}},\ }\href
  {https://doi.org/10.1051/0004-6361/201526416} {\bibfield  {journal} {\bibinfo
   {journal} {\aap}\ }\textbf {\bibinfo {volume} {590}},\ \bibinfo {eid} {A1}
  (\bibinfo {year} {2016})},\ \Eprint {https://arxiv.org/abs/1504.07051}
  {arXiv:1504.07051 [astro-ph.HE]} \BibitemShut {NoStop}%
\bibitem [{\citenamefont {Makishima}\ \emph {et~al.}(1996)\citenamefont
  {Makishima}, \citenamefont {Tashiro}, \citenamefont {Ebisawa}, \citenamefont
  {Ezawa}, \citenamefont {Fukazawa}, \citenamefont {Gunji}, \citenamefont
  {Hirayama}, \citenamefont {Idesawa}, \citenamefont {Ikebe}, \citenamefont
  {Ishida}, \citenamefont {Ishisaki}, \citenamefont {Iyomoto}, \citenamefont
  {Kamae}, \citenamefont {Kaneda}, \citenamefont {Kikuchi}, \citenamefont
  {Kohmura}, \citenamefont {Kubo}, \citenamefont {Matsushita}, \citenamefont
  {Matsuzaki}, \citenamefont {Mihara}, \citenamefont {Nakagawa}, \citenamefont
  {Ohashi}, \citenamefont {Saito}, \citenamefont {Sekimoto}, \citenamefont
  {Takahashi}, \citenamefont {Tamura}, \citenamefont {Tsuru}, \citenamefont
  {Ueda},\ and\ \citenamefont {Yamasaki}}]{10.1093/pasj/48.2.171}%
  \BibitemOpen
  \bibfield  {author} {\bibinfo {author} {\bibfnamefont {K.}~\bibnamefont
  {Makishima}}, \bibinfo {author} {\bibfnamefont {M.}~\bibnamefont {Tashiro}},
  \bibinfo {author} {\bibfnamefont {K.}~\bibnamefont {Ebisawa}}, \bibinfo
  {author} {\bibfnamefont {H.}~\bibnamefont {Ezawa}}, \bibinfo {author}
  {\bibfnamefont {Y.}~\bibnamefont {Fukazawa}}, \bibinfo {author}
  {\bibfnamefont {S.}~\bibnamefont {Gunji}}, \bibinfo {author} {\bibfnamefont
  {M.}~\bibnamefont {Hirayama}}, \bibinfo {author} {\bibfnamefont
  {E.}~\bibnamefont {Idesawa}}, \bibinfo {author} {\bibfnamefont
  {Y.}~\bibnamefont {Ikebe}}, \bibinfo {author} {\bibfnamefont
  {M.}~\bibnamefont {Ishida}}, \bibinfo {author} {\bibfnamefont
  {Y.}~\bibnamefont {Ishisaki}}, \bibinfo {author} {\bibfnamefont
  {N.}~\bibnamefont {Iyomoto}}, \bibinfo {author} {\bibfnamefont
  {T.}~\bibnamefont {Kamae}}, \bibinfo {author} {\bibfnamefont
  {H.}~\bibnamefont {Kaneda}}, \bibinfo {author} {\bibfnamefont
  {K.}~\bibnamefont {Kikuchi}}, \bibinfo {author} {\bibfnamefont
  {Y.}~\bibnamefont {Kohmura}}, \bibinfo {author} {\bibfnamefont
  {H.}~\bibnamefont {Kubo}}, \bibinfo {author} {\bibfnamefont {K.}~\bibnamefont
  {Matsushita}}, \bibinfo {author} {\bibfnamefont {K.}~\bibnamefont
  {Matsuzaki}}, \bibinfo {author} {\bibfnamefont {T.}~\bibnamefont {Mihara}},
  \bibinfo {author} {\bibfnamefont {K.}~\bibnamefont {Nakagawa}}, \bibinfo
  {author} {\bibfnamefont {T.}~\bibnamefont {Ohashi}}, \bibinfo {author}
  {\bibfnamefont {Y.}~\bibnamefont {Saito}}, \bibinfo {author} {\bibfnamefont
  {Y.}~\bibnamefont {Sekimoto}}, \bibinfo {author} {\bibfnamefont
  {T.}~\bibnamefont {Takahashi}}, \bibinfo {author} {\bibfnamefont
  {T.}~\bibnamefont {Tamura}}, \bibinfo {author} {\bibfnamefont
  {T.}~\bibnamefont {Tsuru}}, \bibinfo {author} {\bibfnamefont
  {Y.}~\bibnamefont {Ueda}},\ and\ \bibinfo {author} {\bibfnamefont {N.~Y.}\
  \bibnamefont {Yamasaki}},\ }\bibfield  {title} {\bibinfo {title} {{In-Orbit
  Performance of the Gas Imaging Spectrometer onboard ASCA}},\ }\href
  {https://doi.org/10.1093/pasj/48.2.171} {\bibfield  {journal} {\bibinfo
  {journal} {Publications of the Astronomical Society of Japan}\ }\textbf
  {\bibinfo {volume} {48}},\ \bibinfo {pages} {171} (\bibinfo {year} {1996})},\
  \Eprint
  {https://arxiv.org/abs/https://academic.oup.com/pasj/article-pdf/48/2/171/9712899/pasj48-0171.pdf}
  {https://academic.oup.com/pasj/article-pdf/48/2/171/9712899/pasj48-0171.pdf}
  \BibitemShut {NoStop}%
\bibitem [{\citenamefont {{Burrows}}\ \emph {et~al.}(2005)\citenamefont
  {{Burrows}}, \citenamefont {{Hill}}, \citenamefont {{Nousek}}, \citenamefont
  {{Kennea}}, \citenamefont {{Wells}}, \citenamefont {{Osborne}}, \citenamefont
  {{Abbey}}, \citenamefont {{Beardmore}}, \citenamefont {{Mukerjee}},
  \citenamefont {{Short}}, \citenamefont {{Chincarini}}, \citenamefont
  {{Campana}}, \citenamefont {{Citterio}}, \citenamefont {{Moretti}},
  \citenamefont {{Pagani}}, \citenamefont {{Tagliaferri}}, \citenamefont
  {{Giommi}}, \citenamefont {{Capalbi}}, \citenamefont {{Tamburelli}},
  \citenamefont {{Angelini}}, \citenamefont {{Cusumano}}, \citenamefont
  {{Br{\"a}uninger}}, \citenamefont {{Burkert}},\ and\ \citenamefont
  {{Hartner}}}]{2005SSRv..120..165B}%
  \BibitemOpen
  \bibfield  {author} {\bibinfo {author} {\bibfnamefont {D.~N.}\ \bibnamefont
  {{Burrows}}}, \bibinfo {author} {\bibfnamefont {J.~E.}\ \bibnamefont
  {{Hill}}}, \bibinfo {author} {\bibfnamefont {J.~A.}\ \bibnamefont
  {{Nousek}}}, \bibinfo {author} {\bibfnamefont {J.~A.}\ \bibnamefont
  {{Kennea}}}, \bibinfo {author} {\bibfnamefont {A.}~\bibnamefont {{Wells}}},
  \bibinfo {author} {\bibfnamefont {J.~P.}\ \bibnamefont {{Osborne}}}, \bibinfo
  {author} {\bibfnamefont {A.~F.}\ \bibnamefont {{Abbey}}}, \bibinfo {author}
  {\bibfnamefont {A.}~\bibnamefont {{Beardmore}}}, \bibinfo {author}
  {\bibfnamefont {K.}~\bibnamefont {{Mukerjee}}}, \bibinfo {author}
  {\bibfnamefont {A.~D.~T.}\ \bibnamefont {{Short}}}, \bibinfo {author}
  {\bibfnamefont {G.}~\bibnamefont {{Chincarini}}}, \bibinfo {author}
  {\bibfnamefont {S.}~\bibnamefont {{Campana}}}, \bibinfo {author}
  {\bibfnamefont {O.}~\bibnamefont {{Citterio}}}, \bibinfo {author}
  {\bibfnamefont {A.}~\bibnamefont {{Moretti}}}, \bibinfo {author}
  {\bibfnamefont {C.}~\bibnamefont {{Pagani}}}, \bibinfo {author}
  {\bibfnamefont {G.}~\bibnamefont {{Tagliaferri}}}, \bibinfo {author}
  {\bibfnamefont {P.}~\bibnamefont {{Giommi}}}, \bibinfo {author}
  {\bibfnamefont {M.}~\bibnamefont {{Capalbi}}}, \bibinfo {author}
  {\bibfnamefont {F.}~\bibnamefont {{Tamburelli}}}, \bibinfo {author}
  {\bibfnamefont {L.}~\bibnamefont {{Angelini}}}, \bibinfo {author}
  {\bibfnamefont {G.}~\bibnamefont {{Cusumano}}}, \bibinfo {author}
  {\bibfnamefont {H.~W.}\ \bibnamefont {{Br{\"a}uninger}}}, \bibinfo {author}
  {\bibfnamefont {W.}~\bibnamefont {{Burkert}}},\ and\ \bibinfo {author}
  {\bibfnamefont {G.~D.}\ \bibnamefont {{Hartner}}},\ }\bibfield  {title}
  {\bibinfo {title} {{The Swift X-Ray Telescope}},\ }\href
  {https://doi.org/10.1007/s11214-005-5097-2} {\bibfield  {journal} {\bibinfo
  {journal} {\ssr}\ }\textbf {\bibinfo {volume} {120}},\ \bibinfo {pages} {165}
  (\bibinfo {year} {2005})},\ \Eprint {https://arxiv.org/abs/astro-ph/0508071}
  {arXiv:astro-ph/0508071 [astro-ph]} \BibitemShut {NoStop}%
\bibitem [{\citenamefont {{Tavani}}\ \emph {et~al.}(2009)\citenamefont
  {{Tavani}}, \citenamefont {{Barbiellini}}, \citenamefont {{Argan}},
  \citenamefont {{Boffelli}}, \citenamefont {{Bulgarelli}}, \citenamefont
  {{Caraveo}}, \citenamefont {{Cattaneo}}, \citenamefont {{Chen}},
  \citenamefont {{Cocco}}, \citenamefont {{Costa}}, \citenamefont
  {{D'Ammando}}, \citenamefont {{Del Monte}}, \citenamefont {{de Paris}},
  \citenamefont {{Di Cocco}}, \citenamefont {{di Persio}}, \citenamefont
  {{Donnarumma}}, \citenamefont {{Evangelista}}, \citenamefont {{Feroci}},
  \citenamefont {{Ferrari}}, \citenamefont {{Fiorini}}, \citenamefont
  {{Fornari}}, \citenamefont {{Fuschino}}, \citenamefont {{Froysland}},
  \citenamefont {{Frutti}}, \citenamefont {{Galli}}, \citenamefont
  {{Gianotti}}, \citenamefont {{Giuliani}}, \citenamefont {{Labanti}},
  \citenamefont {{Lapshov}}, \citenamefont {{Lazzarotto}}, \citenamefont
  {{Liello}}, \citenamefont {{Lipari}}, \citenamefont {{Longo}}, \citenamefont
  {{Mattaini}}, \citenamefont {{Marisaldi}}, \citenamefont {{Mastropietro}},
  \citenamefont {{Mauri}}, \citenamefont {{Mauri}}, \citenamefont
  {{Mereghetti}}, \citenamefont {{Morelli}}, \citenamefont {{Morselli}},
  \citenamefont {{Pacciani}}, \citenamefont {{Pellizzoni}}, \citenamefont
  {{Perotti}}, \citenamefont {{Piano}}, \citenamefont {{Picozza}},
  \citenamefont {{Pontoni}}, \citenamefont {{Porrovecchio}}, \citenamefont
  {{Prest}}, \citenamefont {{Pucella}}, \citenamefont {{Rapisarda}},
  \citenamefont {{Rappoldi}}, \citenamefont {{Rossi}}, \citenamefont
  {{Rubini}}, \citenamefont {{Soffitta}}, \citenamefont {{Traci}},
  \citenamefont {{Trifoglio}}, \citenamefont {{Trois}}, \citenamefont
  {{Vallazza}}, \citenamefont {{Vercellone}}, \citenamefont {{Vittorini}},
  \citenamefont {{Zambra}}, \citenamefont {{Zanello}}, \citenamefont
  {{Pittori}}, \citenamefont {{Preger}}, \citenamefont {{Santolamazza}},
  \citenamefont {{Verrecchia}}, \citenamefont {{Giommi}}, \citenamefont
  {{Colafrancesco}}, \citenamefont {{Antonelli}}, \citenamefont {{Cutini}},
  \citenamefont {{Gasparrini}}, \citenamefont {{Stellato}}, \citenamefont
  {{Fanari}}, \citenamefont {{Primavera}}, \citenamefont {{Tamburelli}},
  \citenamefont {{Viola}}, \citenamefont {{Guarrera}}, \citenamefont
  {{Salotti}}, \citenamefont {{D'Amico}}, \citenamefont {{Marchetti}},
  \citenamefont {{Crisconio}}, \citenamefont {{Sabatini}}, \citenamefont
  {{Annoni}}, \citenamefont {{Alia}}, \citenamefont {{Longoni}}, \citenamefont
  {{Sanquerin}}, \citenamefont {{Battilana}}, \citenamefont {{Concari}},
  \citenamefont {{Dessimone}}, \citenamefont {{Grossi}}, \citenamefont
  {{Parise}}, \citenamefont {{Monzani}}, \citenamefont {{Artina}},
  \citenamefont {{Pavesi}}, \citenamefont {{Marseguerra}}, \citenamefont
  {{Nicolini}}, \citenamefont {{Scandelli}}, \citenamefont {{Soli}},
  \citenamefont {{Vettorello}}, \citenamefont {{Zardetto}}, \citenamefont
  {{Bonati}}, \citenamefont {{Maltecca}}, \citenamefont {{D'Alba}},
  \citenamefont {{Patan{\'e}}}, \citenamefont {{Babini}}, \citenamefont
  {{Onorati}}, \citenamefont {{Acquaroli}}, \citenamefont {{Angelucci}},
  \citenamefont {{Morelli}}, \citenamefont {{Agostara}}, \citenamefont
  {{Cerone}}, \citenamefont {{Michetti}}, \citenamefont {{Tempesta}},
  \citenamefont {{D'Eramo}}, \citenamefont {{Rocca}}, \citenamefont
  {{Giannini}}, \citenamefont {{Borghi}}, \citenamefont {{Garavelli}},
  \citenamefont {{Conte}}, \citenamefont {{Balasini}}, \citenamefont
  {{Ferrario}}, \citenamefont {{Vanotti}}, \citenamefont {{Collavo}},\ and\
  \citenamefont {{Giacomazzo}}}]{2009A&A...502..995T}%
  \BibitemOpen
  \bibfield  {author} {\bibinfo {author} {\bibfnamefont {M.}~\bibnamefont
  {{Tavani}}}, \bibinfo {author} {\bibfnamefont {G.}~\bibnamefont
  {{Barbiellini}}}, \bibinfo {author} {\bibfnamefont {A.}~\bibnamefont
  {{Argan}}}, \bibinfo {author} {\bibfnamefont {F.}~\bibnamefont {{Boffelli}}},
  \bibinfo {author} {\bibfnamefont {A.}~\bibnamefont {{Bulgarelli}}}, \bibinfo
  {author} {\bibfnamefont {P.}~\bibnamefont {{Caraveo}}}, \bibinfo {author}
  {\bibfnamefont {P.~W.}\ \bibnamefont {{Cattaneo}}}, \bibinfo {author}
  {\bibfnamefont {A.~W.}\ \bibnamefont {{Chen}}}, \bibinfo {author}
  {\bibfnamefont {V.}~\bibnamefont {{Cocco}}}, \bibinfo {author} {\bibfnamefont
  {E.}~\bibnamefont {{Costa}}}, \bibinfo {author} {\bibfnamefont
  {F.}~\bibnamefont {{D'Ammando}}}, \bibinfo {author} {\bibfnamefont
  {E.}~\bibnamefont {{Del Monte}}}, \bibinfo {author} {\bibfnamefont
  {G.}~\bibnamefont {{de Paris}}}, \bibinfo {author} {\bibfnamefont
  {G.}~\bibnamefont {{Di Cocco}}}, \bibinfo {author} {\bibfnamefont
  {G.}~\bibnamefont {{di Persio}}}, \bibinfo {author} {\bibfnamefont
  {I.}~\bibnamefont {{Donnarumma}}}, \bibinfo {author} {\bibfnamefont
  {Y.}~\bibnamefont {{Evangelista}}}, \bibinfo {author} {\bibfnamefont
  {M.}~\bibnamefont {{Feroci}}}, \bibinfo {author} {\bibfnamefont
  {A.}~\bibnamefont {{Ferrari}}}, \bibinfo {author} {\bibfnamefont
  {M.}~\bibnamefont {{Fiorini}}}, \bibinfo {author} {\bibfnamefont
  {F.}~\bibnamefont {{Fornari}}}, \bibinfo {author} {\bibfnamefont
  {F.}~\bibnamefont {{Fuschino}}}, \bibinfo {author} {\bibfnamefont
  {T.}~\bibnamefont {{Froysland}}}, \bibinfo {author} {\bibfnamefont
  {M.}~\bibnamefont {{Frutti}}}, \bibinfo {author} {\bibfnamefont
  {M.}~\bibnamefont {{Galli}}}, \bibinfo {author} {\bibfnamefont
  {F.}~\bibnamefont {{Gianotti}}}, \bibinfo {author} {\bibfnamefont
  {A.}~\bibnamefont {{Giuliani}}}, \bibinfo {author} {\bibfnamefont
  {C.}~\bibnamefont {{Labanti}}}, \bibinfo {author} {\bibfnamefont
  {I.}~\bibnamefont {{Lapshov}}}, \bibinfo {author} {\bibfnamefont
  {F.}~\bibnamefont {{Lazzarotto}}}, \bibinfo {author} {\bibfnamefont
  {F.}~\bibnamefont {{Liello}}}, \bibinfo {author} {\bibfnamefont
  {P.}~\bibnamefont {{Lipari}}}, \bibinfo {author} {\bibfnamefont
  {F.}~\bibnamefont {{Longo}}}, \bibinfo {author} {\bibfnamefont
  {E.}~\bibnamefont {{Mattaini}}}, \bibinfo {author} {\bibfnamefont
  {M.}~\bibnamefont {{Marisaldi}}}, \bibinfo {author} {\bibfnamefont
  {M.}~\bibnamefont {{Mastropietro}}}, \bibinfo {author} {\bibfnamefont
  {A.}~\bibnamefont {{Mauri}}}, \bibinfo {author} {\bibfnamefont
  {F.}~\bibnamefont {{Mauri}}}, \bibinfo {author} {\bibfnamefont
  {S.}~\bibnamefont {{Mereghetti}}}, \bibinfo {author} {\bibfnamefont
  {E.}~\bibnamefont {{Morelli}}}, \bibinfo {author} {\bibfnamefont
  {A.}~\bibnamefont {{Morselli}}}, \bibinfo {author} {\bibfnamefont
  {L.}~\bibnamefont {{Pacciani}}}, \bibinfo {author} {\bibfnamefont
  {A.}~\bibnamefont {{Pellizzoni}}}, \bibinfo {author} {\bibfnamefont
  {F.}~\bibnamefont {{Perotti}}}, \bibinfo {author} {\bibfnamefont
  {G.}~\bibnamefont {{Piano}}}, \bibinfo {author} {\bibfnamefont
  {P.}~\bibnamefont {{Picozza}}}, \bibinfo {author} {\bibfnamefont
  {C.}~\bibnamefont {{Pontoni}}}, \bibinfo {author} {\bibfnamefont
  {G.}~\bibnamefont {{Porrovecchio}}}, \bibinfo {author} {\bibfnamefont
  {M.}~\bibnamefont {{Prest}}}, \bibinfo {author} {\bibfnamefont
  {G.}~\bibnamefont {{Pucella}}}, \bibinfo {author} {\bibfnamefont
  {M.}~\bibnamefont {{Rapisarda}}}, \bibinfo {author} {\bibfnamefont
  {A.}~\bibnamefont {{Rappoldi}}}, \bibinfo {author} {\bibfnamefont
  {E.}~\bibnamefont {{Rossi}}}, \bibinfo {author} {\bibfnamefont
  {A.}~\bibnamefont {{Rubini}}}, \bibinfo {author} {\bibfnamefont
  {P.}~\bibnamefont {{Soffitta}}}, \bibinfo {author} {\bibfnamefont
  {A.}~\bibnamefont {{Traci}}}, \bibinfo {author} {\bibfnamefont
  {M.}~\bibnamefont {{Trifoglio}}}, \bibinfo {author} {\bibfnamefont
  {A.}~\bibnamefont {{Trois}}}, \bibinfo {author} {\bibfnamefont
  {E.}~\bibnamefont {{Vallazza}}}, \bibinfo {author} {\bibfnamefont
  {S.}~\bibnamefont {{Vercellone}}}, \bibinfo {author} {\bibfnamefont
  {V.}~\bibnamefont {{Vittorini}}}, \bibinfo {author} {\bibfnamefont
  {A.}~\bibnamefont {{Zambra}}}, \bibinfo {author} {\bibfnamefont
  {D.}~\bibnamefont {{Zanello}}}, \bibinfo {author} {\bibfnamefont
  {C.}~\bibnamefont {{Pittori}}}, \bibinfo {author} {\bibfnamefont
  {B.}~\bibnamefont {{Preger}}}, \bibinfo {author} {\bibfnamefont
  {P.}~\bibnamefont {{Santolamazza}}}, \bibinfo {author} {\bibfnamefont
  {F.}~\bibnamefont {{Verrecchia}}}, \bibinfo {author} {\bibfnamefont
  {P.}~\bibnamefont {{Giommi}}}, \bibinfo {author} {\bibfnamefont
  {S.}~\bibnamefont {{Colafrancesco}}}, \bibinfo {author} {\bibfnamefont
  {A.}~\bibnamefont {{Antonelli}}}, \bibinfo {author} {\bibfnamefont
  {S.}~\bibnamefont {{Cutini}}}, \bibinfo {author} {\bibfnamefont
  {D.}~\bibnamefont {{Gasparrini}}}, \bibinfo {author} {\bibfnamefont
  {S.}~\bibnamefont {{Stellato}}}, \bibinfo {author} {\bibfnamefont
  {G.}~\bibnamefont {{Fanari}}}, \bibinfo {author} {\bibfnamefont
  {R.}~\bibnamefont {{Primavera}}}, \bibinfo {author} {\bibfnamefont
  {F.}~\bibnamefont {{Tamburelli}}}, \bibinfo {author} {\bibfnamefont
  {F.}~\bibnamefont {{Viola}}}, \bibinfo {author} {\bibfnamefont
  {G.}~\bibnamefont {{Guarrera}}}, \bibinfo {author} {\bibfnamefont
  {L.}~\bibnamefont {{Salotti}}}, \bibinfo {author} {\bibfnamefont
  {F.}~\bibnamefont {{D'Amico}}}, \bibinfo {author} {\bibfnamefont
  {E.}~\bibnamefont {{Marchetti}}}, \bibinfo {author} {\bibfnamefont
  {M.}~\bibnamefont {{Crisconio}}}, \bibinfo {author} {\bibfnamefont
  {P.}~\bibnamefont {{Sabatini}}}, \bibinfo {author} {\bibfnamefont
  {G.}~\bibnamefont {{Annoni}}}, \bibinfo {author} {\bibfnamefont
  {S.}~\bibnamefont {{Alia}}}, \bibinfo {author} {\bibfnamefont
  {A.}~\bibnamefont {{Longoni}}}, \bibinfo {author} {\bibfnamefont
  {R.}~\bibnamefont {{Sanquerin}}}, \bibinfo {author} {\bibfnamefont
  {M.}~\bibnamefont {{Battilana}}}, \bibinfo {author} {\bibfnamefont
  {P.}~\bibnamefont {{Concari}}}, \bibinfo {author} {\bibfnamefont
  {E.}~\bibnamefont {{Dessimone}}}, \bibinfo {author} {\bibfnamefont
  {R.}~\bibnamefont {{Grossi}}}, \bibinfo {author} {\bibfnamefont
  {A.}~\bibnamefont {{Parise}}}, \bibinfo {author} {\bibfnamefont
  {F.}~\bibnamefont {{Monzani}}}, \bibinfo {author} {\bibfnamefont
  {E.}~\bibnamefont {{Artina}}}, \bibinfo {author} {\bibfnamefont
  {R.}~\bibnamefont {{Pavesi}}}, \bibinfo {author} {\bibfnamefont
  {G.}~\bibnamefont {{Marseguerra}}}, \bibinfo {author} {\bibfnamefont
  {L.}~\bibnamefont {{Nicolini}}}, \bibinfo {author} {\bibfnamefont
  {L.}~\bibnamefont {{Scandelli}}}, \bibinfo {author} {\bibfnamefont
  {L.}~\bibnamefont {{Soli}}}, \bibinfo {author} {\bibfnamefont
  {V.}~\bibnamefont {{Vettorello}}}, \bibinfo {author} {\bibfnamefont
  {E.}~\bibnamefont {{Zardetto}}}, \bibinfo {author} {\bibfnamefont
  {A.}~\bibnamefont {{Bonati}}}, \bibinfo {author} {\bibfnamefont
  {L.}~\bibnamefont {{Maltecca}}}, \bibinfo {author} {\bibfnamefont
  {E.}~\bibnamefont {{D'Alba}}}, \bibinfo {author} {\bibfnamefont
  {M.}~\bibnamefont {{Patan{\'e}}}}, \bibinfo {author} {\bibfnamefont
  {G.}~\bibnamefont {{Babini}}}, \bibinfo {author} {\bibfnamefont
  {F.}~\bibnamefont {{Onorati}}}, \bibinfo {author} {\bibfnamefont
  {L.}~\bibnamefont {{Acquaroli}}}, \bibinfo {author} {\bibfnamefont
  {M.}~\bibnamefont {{Angelucci}}}, \bibinfo {author} {\bibfnamefont
  {B.}~\bibnamefont {{Morelli}}}, \bibinfo {author} {\bibfnamefont
  {C.}~\bibnamefont {{Agostara}}}, \bibinfo {author} {\bibfnamefont
  {M.}~\bibnamefont {{Cerone}}}, \bibinfo {author} {\bibfnamefont
  {A.}~\bibnamefont {{Michetti}}}, \bibinfo {author} {\bibfnamefont
  {P.}~\bibnamefont {{Tempesta}}}, \bibinfo {author} {\bibfnamefont
  {S.}~\bibnamefont {{D'Eramo}}}, \bibinfo {author} {\bibfnamefont
  {F.}~\bibnamefont {{Rocca}}}, \bibinfo {author} {\bibfnamefont
  {F.}~\bibnamefont {{Giannini}}}, \bibinfo {author} {\bibfnamefont
  {G.}~\bibnamefont {{Borghi}}}, \bibinfo {author} {\bibfnamefont
  {B.}~\bibnamefont {{Garavelli}}}, \bibinfo {author} {\bibfnamefont
  {M.}~\bibnamefont {{Conte}}}, \bibinfo {author} {\bibfnamefont
  {M.}~\bibnamefont {{Balasini}}}, \bibinfo {author} {\bibfnamefont
  {I.}~\bibnamefont {{Ferrario}}}, \bibinfo {author} {\bibfnamefont
  {M.}~\bibnamefont {{Vanotti}}}, \bibinfo {author} {\bibfnamefont
  {E.}~\bibnamefont {{Collavo}}},\ and\ \bibinfo {author} {\bibfnamefont
  {M.}~\bibnamefont {{Giacomazzo}}},\ }\bibfield  {title} {\bibinfo {title}
  {{The AGILE Mission}},\ }\href {https://doi.org/10.1051/0004-6361/200810527}
  {\bibfield  {journal} {\bibinfo  {journal} {\aap}\ }\textbf {\bibinfo
  {volume} {502}},\ \bibinfo {pages} {995} (\bibinfo {year} {2009})},\ \Eprint
  {https://arxiv.org/abs/0807.4254} {arXiv:0807.4254 [astro-ph]} \BibitemShut
  {NoStop}%
\bibitem [{\citenamefont {{Matsuoka}}\ \emph {et~al.}(2009)\citenamefont
  {{Matsuoka}}, \citenamefont {{Kawasaki}}, \citenamefont {{Ueno}},
  \citenamefont {{Tomida}}, \citenamefont {{Kohama}}, \citenamefont {{Suzuki}},
  \citenamefont {{Adachi}}, \citenamefont {{Ishikawa}}, \citenamefont
  {{Mihara}}, \citenamefont {{Sugizaki}}, \citenamefont {{Isobe}},
  \citenamefont {{Nakagawa}}, \citenamefont {{Tsunemi}}, \citenamefont
  {{Miyata}}, \citenamefont {{Kawai}}, \citenamefont {{Kataoka}}, \citenamefont
  {{Morii}}, \citenamefont {{Yoshida}}, \citenamefont {{Negoro}}, \citenamefont
  {{Nakajima}}, \citenamefont {{Ueda}}, \citenamefont {{Chujo}}, \citenamefont
  {{Yamaoka}}, \citenamefont {{Yamazaki}}, \citenamefont {{Nakahira}},
  \citenamefont {{You}}, \citenamefont {{Ishiwata}}, \citenamefont {{Miyoshi}},
  \citenamefont {{Eguchi}}, \citenamefont {{Hiroi}}, \citenamefont
  {{Katayama}},\ and\ \citenamefont {{Ebisawa}}}]{2009PASJ...61..999M}%
  \BibitemOpen
  \bibfield  {author} {\bibinfo {author} {\bibfnamefont {M.}~\bibnamefont
  {{Matsuoka}}}, \bibinfo {author} {\bibfnamefont {K.}~\bibnamefont
  {{Kawasaki}}}, \bibinfo {author} {\bibfnamefont {S.}~\bibnamefont {{Ueno}}},
  \bibinfo {author} {\bibfnamefont {H.}~\bibnamefont {{Tomida}}}, \bibinfo
  {author} {\bibfnamefont {M.}~\bibnamefont {{Kohama}}}, \bibinfo {author}
  {\bibfnamefont {M.}~\bibnamefont {{Suzuki}}}, \bibinfo {author}
  {\bibfnamefont {Y.}~\bibnamefont {{Adachi}}}, \bibinfo {author}
  {\bibfnamefont {M.}~\bibnamefont {{Ishikawa}}}, \bibinfo {author}
  {\bibfnamefont {T.}~\bibnamefont {{Mihara}}}, \bibinfo {author}
  {\bibfnamefont {M.}~\bibnamefont {{Sugizaki}}}, \bibinfo {author}
  {\bibfnamefont {N.}~\bibnamefont {{Isobe}}}, \bibinfo {author} {\bibfnamefont
  {Y.}~\bibnamefont {{Nakagawa}}}, \bibinfo {author} {\bibfnamefont
  {H.}~\bibnamefont {{Tsunemi}}}, \bibinfo {author} {\bibfnamefont
  {E.}~\bibnamefont {{Miyata}}}, \bibinfo {author} {\bibfnamefont
  {N.}~\bibnamefont {{Kawai}}}, \bibinfo {author} {\bibfnamefont
  {J.}~\bibnamefont {{Kataoka}}}, \bibinfo {author} {\bibfnamefont
  {M.}~\bibnamefont {{Morii}}}, \bibinfo {author} {\bibfnamefont
  {A.}~\bibnamefont {{Yoshida}}}, \bibinfo {author} {\bibfnamefont
  {H.}~\bibnamefont {{Negoro}}}, \bibinfo {author} {\bibfnamefont
  {M.}~\bibnamefont {{Nakajima}}}, \bibinfo {author} {\bibfnamefont
  {Y.}~\bibnamefont {{Ueda}}}, \bibinfo {author} {\bibfnamefont
  {H.}~\bibnamefont {{Chujo}}}, \bibinfo {author} {\bibfnamefont
  {K.}~\bibnamefont {{Yamaoka}}}, \bibinfo {author} {\bibfnamefont
  {O.}~\bibnamefont {{Yamazaki}}}, \bibinfo {author} {\bibfnamefont
  {S.}~\bibnamefont {{Nakahira}}}, \bibinfo {author} {\bibfnamefont
  {T.}~\bibnamefont {{You}}}, \bibinfo {author} {\bibfnamefont
  {R.}~\bibnamefont {{Ishiwata}}}, \bibinfo {author} {\bibfnamefont
  {S.}~\bibnamefont {{Miyoshi}}}, \bibinfo {author} {\bibfnamefont
  {S.}~\bibnamefont {{Eguchi}}}, \bibinfo {author} {\bibfnamefont
  {K.}~\bibnamefont {{Hiroi}}}, \bibinfo {author} {\bibfnamefont
  {H.}~\bibnamefont {{Katayama}}},\ and\ \bibinfo {author} {\bibfnamefont
  {K.}~\bibnamefont {{Ebisawa}}},\ }\bibfield  {title} {\bibinfo {title} {{The
  MAXI Mission on the ISS: Science and Instruments for Monitoring All-Sky X-Ray
  Images}},\ }\href {https://doi.org/10.1093/pasj/61.5.999} {\bibfield
  {journal} {\bibinfo  {journal} {\pasj}\ }\textbf {\bibinfo {volume} {61}},\
  \bibinfo {pages} {999} (\bibinfo {year} {2009})},\ \Eprint
  {https://arxiv.org/abs/0906.0631} {arXiv:0906.0631 [astro-ph.IM]}
  \BibitemShut {NoStop}%
\bibitem [{\citenamefont {{Torii}}(2011)}]{2011ICRC....6..351T}%
  \BibitemOpen
  \bibfield  {author} {\bibinfo {author} {\bibfnamefont {S.}~\bibnamefont
  {{Torii}}},\ }\bibfield  {title} {\bibinfo {title} {{Overview of the CALET
  Mission to the ISS}},\ }in\ \href {https://doi.org/10.7529/ICRC2011/V06/0615}
  {\emph {\bibinfo {booktitle} {International Cosmic Ray Conference}}},\
  \bibinfo {series} {International Cosmic Ray Conference}, Vol.~\bibinfo
  {volume} {6}\ (\bibinfo {year} {2011})\ p.\ \bibinfo {pages}
  {351}\BibitemShut {NoStop}%
\bibitem [{\citenamefont {{Gendreau}}\ \emph {et~al.}(2012)\citenamefont
  {{Gendreau}}, \citenamefont {{Arzoumanian}},\ and\ \citenamefont
  {{Okajima}}}]{2012SPIE.8443E..13G}%
  \BibitemOpen
  \bibfield  {author} {\bibinfo {author} {\bibfnamefont {K.~C.}\ \bibnamefont
  {{Gendreau}}}, \bibinfo {author} {\bibfnamefont {Z.}~\bibnamefont
  {{Arzoumanian}}},\ and\ \bibinfo {author} {\bibfnamefont {T.}~\bibnamefont
  {{Okajima}}},\ }\bibfield  {title} {\bibinfo {title} {{The Neutron star
  Interior Composition ExploreR (NICER): an Explorer mission of opportunity for
  soft x-ray timing spectroscopy}},\ }in\ \href
  {https://doi.org/10.1117/12.926396} {\emph {\bibinfo {booktitle} {Space
  Telescopes and Instrumentation 2012: Ultraviolet to Gamma Ray}}},\ \bibinfo
  {series} {Society of Photo-Optical Instrumentation Engineers (SPIE)
  Conference Series}, Vol.\ \bibinfo {volume} {8443},\ \bibinfo {editor}
  {edited by\ \bibinfo {editor} {\bibfnamefont {T.}~\bibnamefont
  {{Takahashi}}}, \bibinfo {editor} {\bibfnamefont {S.~S.}\ \bibnamefont
  {{Murray}}},\ and\ \bibinfo {editor} {\bibfnamefont {J.-W.~A.}\ \bibnamefont
  {{den Herder}}}}\ (\bibinfo {year} {2012})\ p.\ \bibinfo {pages}
  {844313}\BibitemShut {NoStop}%
\bibitem [{\citenamefont {{Harrison}}\ \emph {et~al.}(2013)\citenamefont
  {{Harrison}}, \citenamefont {{Craig}}, \citenamefont {{Christensen}},
  \citenamefont {{Hailey}}, \citenamefont {{Zhang}}, \citenamefont {{Boggs}},
  \citenamefont {{Stern}}, \citenamefont {{Cook}}, \citenamefont {{Forster}},
  \citenamefont {{Giommi}}, \citenamefont {{Grefenstette}}, \citenamefont
  {{Kim}}, \citenamefont {{Kitaguchi}}, \citenamefont {{Koglin}}, \citenamefont
  {{Madsen}}, \citenamefont {{Mao}}, \citenamefont {{Miyasaka}}, \citenamefont
  {{Mori}}, \citenamefont {{Perri}}, \citenamefont {{Pivovaroff}},
  \citenamefont {{Puccetti}}, \citenamefont {{Rana}}, \citenamefont
  {{Westergaard}}, \citenamefont {{Willis}}, \citenamefont {{Zoglauer}},
  \citenamefont {{An}}, \citenamefont {{Bachetti}}, \citenamefont
  {{Barri{\`e}re}}, \citenamefont {{Bellm}}, \citenamefont {{Bhalerao}},
  \citenamefont {{Brejnholt}}, \citenamefont {{Fuerst}}, \citenamefont
  {{Liebe}}, \citenamefont {{Markwardt}}, \citenamefont {{Nynka}},
  \citenamefont {{Vogel}}, \citenamefont {{Walton}}, \citenamefont {{Wik}},
  \citenamefont {{Alexander}}, \citenamefont {{Cominsky}}, \citenamefont
  {{Hornschemeier}}, \citenamefont {{Hornstrup}}, \citenamefont {{Kaspi}},
  \citenamefont {{Madejski}}, \citenamefont {{Matt}}, \citenamefont
  {{Molendi}}, \citenamefont {{Smith}}, \citenamefont {{Tomsick}},
  \citenamefont {{Ajello}}, \citenamefont {{Ballantyne}}, \citenamefont
  {{Balokovi{\'c}}}, \citenamefont {{Barret}}, \citenamefont {{Bauer}},
  \citenamefont {{Blandford}}, \citenamefont {{Brandt}}, \citenamefont
  {{Brenneman}}, \citenamefont {{Chiang}}, \citenamefont {{Chakrabarty}},
  \citenamefont {{Chenevez}}, \citenamefont {{Comastri}}, \citenamefont
  {{Dufour}}, \citenamefont {{Elvis}}, \citenamefont {{Fabian}}, \citenamefont
  {{Farrah}}, \citenamefont {{Fryer}}, \citenamefont {{Gotthelf}},
  \citenamefont {{Grindlay}}, \citenamefont {{Helfand}}, \citenamefont
  {{Krivonos}}, \citenamefont {{Meier}}, \citenamefont {{Miller}},
  \citenamefont {{Natalucci}}, \citenamefont {{Ogle}}, \citenamefont {{Ofek}},
  \citenamefont {{Ptak}}, \citenamefont {{Reynolds}}, \citenamefont {{Rigby}},
  \citenamefont {{Tagliaferri}}, \citenamefont {{Thorsett}}, \citenamefont
  {{Treister}},\ and\ \citenamefont {{Urry}}}]{2013ApJ...770..103H}%
  \BibitemOpen
  \bibfield  {author} {\bibinfo {author} {\bibfnamefont {F.~A.}\ \bibnamefont
  {{Harrison}}}, \bibinfo {author} {\bibfnamefont {W.~W.}\ \bibnamefont
  {{Craig}}}, \bibinfo {author} {\bibfnamefont {F.~E.}\ \bibnamefont
  {{Christensen}}}, \bibinfo {author} {\bibfnamefont {C.~J.}\ \bibnamefont
  {{Hailey}}}, \bibinfo {author} {\bibfnamefont {W.~W.}\ \bibnamefont
  {{Zhang}}}, \bibinfo {author} {\bibfnamefont {S.~E.}\ \bibnamefont
  {{Boggs}}}, \bibinfo {author} {\bibfnamefont {D.}~\bibnamefont {{Stern}}},
  \bibinfo {author} {\bibfnamefont {W.~R.}\ \bibnamefont {{Cook}}}, \bibinfo
  {author} {\bibfnamefont {K.}~\bibnamefont {{Forster}}}, \bibinfo {author}
  {\bibfnamefont {P.}~\bibnamefont {{Giommi}}}, \bibinfo {author}
  {\bibfnamefont {B.~W.}\ \bibnamefont {{Grefenstette}}}, \bibinfo {author}
  {\bibfnamefont {Y.}~\bibnamefont {{Kim}}}, \bibinfo {author} {\bibfnamefont
  {T.}~\bibnamefont {{Kitaguchi}}}, \bibinfo {author} {\bibfnamefont {J.~E.}\
  \bibnamefont {{Koglin}}}, \bibinfo {author} {\bibfnamefont {K.~K.}\
  \bibnamefont {{Madsen}}}, \bibinfo {author} {\bibfnamefont {P.~H.}\
  \bibnamefont {{Mao}}}, \bibinfo {author} {\bibfnamefont {H.}~\bibnamefont
  {{Miyasaka}}}, \bibinfo {author} {\bibfnamefont {K.}~\bibnamefont {{Mori}}},
  \bibinfo {author} {\bibfnamefont {M.}~\bibnamefont {{Perri}}}, \bibinfo
  {author} {\bibfnamefont {M.~J.}\ \bibnamefont {{Pivovaroff}}}, \bibinfo
  {author} {\bibfnamefont {S.}~\bibnamefont {{Puccetti}}}, \bibinfo {author}
  {\bibfnamefont {V.~R.}\ \bibnamefont {{Rana}}}, \bibinfo {author}
  {\bibfnamefont {N.~J.}\ \bibnamefont {{Westergaard}}}, \bibinfo {author}
  {\bibfnamefont {J.}~\bibnamefont {{Willis}}}, \bibinfo {author}
  {\bibfnamefont {A.}~\bibnamefont {{Zoglauer}}}, \bibinfo {author}
  {\bibfnamefont {H.}~\bibnamefont {{An}}}, \bibinfo {author} {\bibfnamefont
  {M.}~\bibnamefont {{Bachetti}}}, \bibinfo {author} {\bibfnamefont {N.~M.}\
  \bibnamefont {{Barri{\`e}re}}}, \bibinfo {author} {\bibfnamefont {E.~C.}\
  \bibnamefont {{Bellm}}}, \bibinfo {author} {\bibfnamefont {V.}~\bibnamefont
  {{Bhalerao}}}, \bibinfo {author} {\bibfnamefont {N.~F.}\ \bibnamefont
  {{Brejnholt}}}, \bibinfo {author} {\bibfnamefont {F.}~\bibnamefont
  {{Fuerst}}}, \bibinfo {author} {\bibfnamefont {C.~C.}\ \bibnamefont
  {{Liebe}}}, \bibinfo {author} {\bibfnamefont {C.~B.}\ \bibnamefont
  {{Markwardt}}}, \bibinfo {author} {\bibfnamefont {M.}~\bibnamefont
  {{Nynka}}}, \bibinfo {author} {\bibfnamefont {J.~K.}\ \bibnamefont
  {{Vogel}}}, \bibinfo {author} {\bibfnamefont {D.~J.}\ \bibnamefont
  {{Walton}}}, \bibinfo {author} {\bibfnamefont {D.~R.}\ \bibnamefont {{Wik}}},
  \bibinfo {author} {\bibfnamefont {D.~M.}\ \bibnamefont {{Alexander}}},
  \bibinfo {author} {\bibfnamefont {L.~R.}\ \bibnamefont {{Cominsky}}},
  \bibinfo {author} {\bibfnamefont {A.~E.}\ \bibnamefont {{Hornschemeier}}},
  \bibinfo {author} {\bibfnamefont {A.}~\bibnamefont {{Hornstrup}}}, \bibinfo
  {author} {\bibfnamefont {V.~M.}\ \bibnamefont {{Kaspi}}}, \bibinfo {author}
  {\bibfnamefont {G.~M.}\ \bibnamefont {{Madejski}}}, \bibinfo {author}
  {\bibfnamefont {G.}~\bibnamefont {{Matt}}}, \bibinfo {author} {\bibfnamefont
  {S.}~\bibnamefont {{Molendi}}}, \bibinfo {author} {\bibfnamefont {D.~M.}\
  \bibnamefont {{Smith}}}, \bibinfo {author} {\bibfnamefont {J.~A.}\
  \bibnamefont {{Tomsick}}}, \bibinfo {author} {\bibfnamefont {M.}~\bibnamefont
  {{Ajello}}}, \bibinfo {author} {\bibfnamefont {D.~R.}\ \bibnamefont
  {{Ballantyne}}}, \bibinfo {author} {\bibfnamefont {M.}~\bibnamefont
  {{Balokovi{\'c}}}}, \bibinfo {author} {\bibfnamefont {D.}~\bibnamefont
  {{Barret}}}, \bibinfo {author} {\bibfnamefont {F.~E.}\ \bibnamefont
  {{Bauer}}}, \bibinfo {author} {\bibfnamefont {R.~D.}\ \bibnamefont
  {{Blandford}}}, \bibinfo {author} {\bibfnamefont {W.~N.}\ \bibnamefont
  {{Brandt}}}, \bibinfo {author} {\bibfnamefont {L.~W.}\ \bibnamefont
  {{Brenneman}}}, \bibinfo {author} {\bibfnamefont {J.}~\bibnamefont
  {{Chiang}}}, \bibinfo {author} {\bibfnamefont {D.}~\bibnamefont
  {{Chakrabarty}}}, \bibinfo {author} {\bibfnamefont {J.}~\bibnamefont
  {{Chenevez}}}, \bibinfo {author} {\bibfnamefont {A.}~\bibnamefont
  {{Comastri}}}, \bibinfo {author} {\bibfnamefont {F.}~\bibnamefont
  {{Dufour}}}, \bibinfo {author} {\bibfnamefont {M.}~\bibnamefont {{Elvis}}},
  \bibinfo {author} {\bibfnamefont {A.~C.}\ \bibnamefont {{Fabian}}}, \bibinfo
  {author} {\bibfnamefont {D.}~\bibnamefont {{Farrah}}}, \bibinfo {author}
  {\bibfnamefont {C.~L.}\ \bibnamefont {{Fryer}}}, \bibinfo {author}
  {\bibfnamefont {E.~V.}\ \bibnamefont {{Gotthelf}}}, \bibinfo {author}
  {\bibfnamefont {J.~E.}\ \bibnamefont {{Grindlay}}}, \bibinfo {author}
  {\bibfnamefont {D.~J.}\ \bibnamefont {{Helfand}}}, \bibinfo {author}
  {\bibfnamefont {R.}~\bibnamefont {{Krivonos}}}, \bibinfo {author}
  {\bibfnamefont {D.~L.}\ \bibnamefont {{Meier}}}, \bibinfo {author}
  {\bibfnamefont {J.~M.}\ \bibnamefont {{Miller}}}, \bibinfo {author}
  {\bibfnamefont {L.}~\bibnamefont {{Natalucci}}}, \bibinfo {author}
  {\bibfnamefont {P.}~\bibnamefont {{Ogle}}}, \bibinfo {author} {\bibfnamefont
  {E.~O.}\ \bibnamefont {{Ofek}}}, \bibinfo {author} {\bibfnamefont
  {A.}~\bibnamefont {{Ptak}}}, \bibinfo {author} {\bibfnamefont {S.~P.}\
  \bibnamefont {{Reynolds}}}, \bibinfo {author} {\bibfnamefont {J.~R.}\
  \bibnamefont {{Rigby}}}, \bibinfo {author} {\bibfnamefont {G.}~\bibnamefont
  {{Tagliaferri}}}, \bibinfo {author} {\bibfnamefont {S.~E.}\ \bibnamefont
  {{Thorsett}}}, \bibinfo {author} {\bibfnamefont {E.}~\bibnamefont
  {{Treister}}},\ and\ \bibinfo {author} {\bibfnamefont {C.~M.}\ \bibnamefont
  {{Urry}}},\ }\bibfield  {title} {\bibinfo {title} {{The Nuclear Spectroscopic
  Telescope Array (NuSTAR) High-energy X-Ray Mission}},\ }\href
  {https://doi.org/10.1088/0004-637X/770/2/103} {\bibfield  {journal} {\bibinfo
   {journal} {\apj}\ }\textbf {\bibinfo {volume} {770}},\ \bibinfo {eid} {103}
  (\bibinfo {year} {2013})},\ \Eprint {https://arxiv.org/abs/1301.7307}
  {arXiv:1301.7307 [astro-ph.IM]} \BibitemShut {NoStop}%
\bibitem [{\citenamefont {{Singh}}\ \emph {et~al.}(2014)\citenamefont
  {{Singh}}, \citenamefont {{Tandon}}, \citenamefont {{Agrawal}}, \citenamefont
  {{Antia}}, \citenamefont {{Manchanda}}, \citenamefont {{Yadav}},
  \citenamefont {{Seetha}}, \citenamefont {{Ramadevi}}, \citenamefont {{Rao}},
  \citenamefont {{Bhattacharya}}, \citenamefont {{Paul}}, \citenamefont
  {{Sreekumar}}, \citenamefont {{Bhattacharyya}}, \citenamefont {{Stewart}},
  \citenamefont {{Hutchings}}, \citenamefont {{Annapurni}}, \citenamefont
  {{Ghosh}}, \citenamefont {{Murthy}}, \citenamefont {{Pati}}, \citenamefont
  {{Rao}}, \citenamefont {{Stalin}}, \citenamefont {{Girish}}, \citenamefont
  {{Sankarasubramanian}}, \citenamefont {{Vadawale}}, \citenamefont
  {{Bhalerao}}, \citenamefont {{Dewangan}}, \citenamefont {{Dedhia}},
  \citenamefont {{Hingar}}, \citenamefont {{Katoch}}, \citenamefont
  {{Kothare}}, \citenamefont {{Mirza}}, \citenamefont {{Mukerjee}},
  \citenamefont {{Shah}}, \citenamefont {{Shah}}, \citenamefont {{Mohan}},
  \citenamefont {{Sangal}}, \citenamefont {{Nagabhusana}}, \citenamefont
  {{Sriram}}, \citenamefont {{Malkar}}, \citenamefont {{Sreekumar}},
  \citenamefont {{Abbey}}, \citenamefont {{Hansford}}, \citenamefont
  {{Beardmore}}, \citenamefont {{Sharma}}, \citenamefont {{Murthy}},
  \citenamefont {{Kulkarni}}, \citenamefont {{Meena}}, \citenamefont {{Babu}},\
  and\ \citenamefont {{Postma}}}]{2014SPIE.9144E..1SS}%
  \BibitemOpen
  \bibfield  {author} {\bibinfo {author} {\bibfnamefont {K.~P.}\ \bibnamefont
  {{Singh}}}, \bibinfo {author} {\bibfnamefont {S.~N.}\ \bibnamefont
  {{Tandon}}}, \bibinfo {author} {\bibfnamefont {P.~C.}\ \bibnamefont
  {{Agrawal}}}, \bibinfo {author} {\bibfnamefont {H.~M.}\ \bibnamefont
  {{Antia}}}, \bibinfo {author} {\bibfnamefont {R.~K.}\ \bibnamefont
  {{Manchanda}}}, \bibinfo {author} {\bibfnamefont {J.~S.}\ \bibnamefont
  {{Yadav}}}, \bibinfo {author} {\bibfnamefont {S.}~\bibnamefont {{Seetha}}},
  \bibinfo {author} {\bibfnamefont {M.~C.}\ \bibnamefont {{Ramadevi}}},
  \bibinfo {author} {\bibfnamefont {A.~R.}\ \bibnamefont {{Rao}}}, \bibinfo
  {author} {\bibfnamefont {D.}~\bibnamefont {{Bhattacharya}}}, \bibinfo
  {author} {\bibfnamefont {B.}~\bibnamefont {{Paul}}}, \bibinfo {author}
  {\bibfnamefont {P.}~\bibnamefont {{Sreekumar}}}, \bibinfo {author}
  {\bibfnamefont {S.}~\bibnamefont {{Bhattacharyya}}}, \bibinfo {author}
  {\bibfnamefont {G.~C.}\ \bibnamefont {{Stewart}}}, \bibinfo {author}
  {\bibfnamefont {J.}~\bibnamefont {{Hutchings}}}, \bibinfo {author}
  {\bibfnamefont {S.~A.}\ \bibnamefont {{Annapurni}}}, \bibinfo {author}
  {\bibfnamefont {S.~K.}\ \bibnamefont {{Ghosh}}}, \bibinfo {author}
  {\bibfnamefont {J.}~\bibnamefont {{Murthy}}}, \bibinfo {author}
  {\bibfnamefont {A.}~\bibnamefont {{Pati}}}, \bibinfo {author} {\bibfnamefont
  {N.~K.}\ \bibnamefont {{Rao}}}, \bibinfo {author} {\bibfnamefont {C.~S.}\
  \bibnamefont {{Stalin}}}, \bibinfo {author} {\bibfnamefont {V.}~\bibnamefont
  {{Girish}}}, \bibinfo {author} {\bibfnamefont {K.}~\bibnamefont
  {{Sankarasubramanian}}}, \bibinfo {author} {\bibfnamefont {S.}~\bibnamefont
  {{Vadawale}}}, \bibinfo {author} {\bibfnamefont {V.~B.}\ \bibnamefont
  {{Bhalerao}}}, \bibinfo {author} {\bibfnamefont {G.~C.}\ \bibnamefont
  {{Dewangan}}}, \bibinfo {author} {\bibfnamefont {D.~K.}\ \bibnamefont
  {{Dedhia}}}, \bibinfo {author} {\bibfnamefont {M.~K.}\ \bibnamefont
  {{Hingar}}}, \bibinfo {author} {\bibfnamefont {T.~B.}\ \bibnamefont
  {{Katoch}}}, \bibinfo {author} {\bibfnamefont {A.~T.}\ \bibnamefont
  {{Kothare}}}, \bibinfo {author} {\bibfnamefont {I.}~\bibnamefont {{Mirza}}},
  \bibinfo {author} {\bibfnamefont {K.}~\bibnamefont {{Mukerjee}}}, \bibinfo
  {author} {\bibfnamefont {H.}~\bibnamefont {{Shah}}}, \bibinfo {author}
  {\bibfnamefont {P.}~\bibnamefont {{Shah}}}, \bibinfo {author} {\bibfnamefont
  {R.}~\bibnamefont {{Mohan}}}, \bibinfo {author} {\bibfnamefont {A.~K.}\
  \bibnamefont {{Sangal}}}, \bibinfo {author} {\bibfnamefont {S.}~\bibnamefont
  {{Nagabhusana}}}, \bibinfo {author} {\bibfnamefont {S.}~\bibnamefont
  {{Sriram}}}, \bibinfo {author} {\bibfnamefont {J.~P.}\ \bibnamefont
  {{Malkar}}}, \bibinfo {author} {\bibfnamefont {S.}~\bibnamefont
  {{Sreekumar}}}, \bibinfo {author} {\bibfnamefont {A.~F.}\ \bibnamefont
  {{Abbey}}}, \bibinfo {author} {\bibfnamefont {G.~M.}\ \bibnamefont
  {{Hansford}}}, \bibinfo {author} {\bibfnamefont {A.~P.}\ \bibnamefont
  {{Beardmore}}}, \bibinfo {author} {\bibfnamefont {M.~R.}\ \bibnamefont
  {{Sharma}}}, \bibinfo {author} {\bibfnamefont {S.}~\bibnamefont {{Murthy}}},
  \bibinfo {author} {\bibfnamefont {R.}~\bibnamefont {{Kulkarni}}}, \bibinfo
  {author} {\bibfnamefont {G.}~\bibnamefont {{Meena}}}, \bibinfo {author}
  {\bibfnamefont {V.~C.}\ \bibnamefont {{Babu}}},\ and\ \bibinfo {author}
  {\bibfnamefont {J.}~\bibnamefont {{Postma}}},\ }\bibfield  {title} {\bibinfo
  {title} {{ASTROSAT mission}},\ }in\ \href
  {https://doi.org/10.1117/12.2062667} {\emph {\bibinfo {booktitle} {Space
  Telescopes and Instrumentation 2014: Ultraviolet to Gamma Ray}}},\ \bibinfo
  {series} {Society of Photo-Optical Instrumentation Engineers (SPIE)
  Conference Series}, Vol.\ \bibinfo {volume} {9144},\ \bibinfo {editor}
  {edited by\ \bibinfo {editor} {\bibfnamefont {T.}~\bibnamefont
  {{Takahashi}}}, \bibinfo {editor} {\bibfnamefont {J.-W.~A.}\ \bibnamefont
  {{den Herder}}},\ and\ \bibinfo {editor} {\bibfnamefont {M.}~\bibnamefont
  {{Bautz}}}}\ (\bibinfo {year} {2014})\ p.\ \bibinfo {pages}
  {91441S}\BibitemShut {NoStop}%
\bibitem [{\citenamefont {Weisskopf}\ \emph {et~al.}(2016)\citenamefont
  {Weisskopf}, \citenamefont {Ramsey}, \citenamefont {O’Dell}, \citenamefont
  {Tennant}, \citenamefont {Elsner}, \citenamefont {Soffita}, \citenamefont
  {Bellazzini}, \citenamefont {Costa}, \citenamefont {Kolodziejczak},
  \citenamefont {Kaspi}, \citenamefont {Mulieri}, \citenamefont {Marshall},
  \citenamefont {Matt},\ and\ \citenamefont {Romani}}]{WEISSKOPF20161179}%
  \BibitemOpen
  \bibfield  {author} {\bibinfo {author} {\bibfnamefont {M.~C.}\ \bibnamefont
  {Weisskopf}}, \bibinfo {author} {\bibfnamefont {B.}~\bibnamefont {Ramsey}},
  \bibinfo {author} {\bibfnamefont {S.~L.}\ \bibnamefont {O’Dell}}, \bibinfo
  {author} {\bibfnamefont {A.}~\bibnamefont {Tennant}}, \bibinfo {author}
  {\bibfnamefont {R.}~\bibnamefont {Elsner}}, \bibinfo {author} {\bibfnamefont
  {P.}~\bibnamefont {Soffita}}, \bibinfo {author} {\bibfnamefont
  {R.}~\bibnamefont {Bellazzini}}, \bibinfo {author} {\bibfnamefont
  {E.}~\bibnamefont {Costa}}, \bibinfo {author} {\bibfnamefont
  {J.}~\bibnamefont {Kolodziejczak}}, \bibinfo {author} {\bibfnamefont
  {V.}~\bibnamefont {Kaspi}}, \bibinfo {author} {\bibfnamefont
  {F.}~\bibnamefont {Mulieri}}, \bibinfo {author} {\bibfnamefont
  {H.}~\bibnamefont {Marshall}}, \bibinfo {author} {\bibfnamefont
  {G.}~\bibnamefont {Matt}},\ and\ \bibinfo {author} {\bibfnamefont
  {R.}~\bibnamefont {Romani}},\ }\bibfield  {title} {\bibinfo {title} {The
  imaging x-ray polarimetry explorer (ixpe)},\ }\href
  {https://doi.org/https://doi.org/10.1016/j.rinp.2016.10.021} {\bibfield
  {journal} {\bibinfo  {journal} {Results in Physics}\ }\textbf {\bibinfo
  {volume} {6}},\ \bibinfo {pages} {1179} (\bibinfo {year} {2016})}\BibitemShut
  {NoStop}%
\bibitem [{\citenamefont {Pavlinsky}\ \emph {et~al.}(2018)\citenamefont
  {Pavlinsky}, \citenamefont {Levin}, \citenamefont {Akimov}, \citenamefont
  {Krivchenko}, \citenamefont {Rotin}, \citenamefont {Kuznetsova},
  \citenamefont {Lapshov}, \citenamefont {Tkachenko}, \citenamefont {Krivonos},
  \citenamefont {Semena}, \citenamefont {Buntov}, \citenamefont {Glushenko},
  \citenamefont {Arefiev}, \citenamefont {Yaskovich}, \citenamefont {Grebenev},
  \citenamefont {Sazonov}, \citenamefont {Lutovinov}, \citenamefont {Molkov},
  \citenamefont {Serbinov}, \citenamefont {Kudelin}, \citenamefont {Drozdova},
  \citenamefont {Voronkov}, \citenamefont {Sunyaev}, \citenamefont {Churazov},
  \citenamefont {Gilfanov}, \citenamefont {Ramsey}, \citenamefont {O'Dell},
  \citenamefont {Kolodziejczak}, \citenamefont {Zavlin},\ and\ \citenamefont
  {Swartz}}]{10.1117/12.2312053}%
  \BibitemOpen
  \bibfield  {author} {\bibinfo {author} {\bibfnamefont {M.}~\bibnamefont
  {Pavlinsky}}, \bibinfo {author} {\bibfnamefont {V.}~\bibnamefont {Levin}},
  \bibinfo {author} {\bibfnamefont {V.}~\bibnamefont {Akimov}}, \bibinfo
  {author} {\bibfnamefont {A.}~\bibnamefont {Krivchenko}}, \bibinfo {author}
  {\bibfnamefont {A.}~\bibnamefont {Rotin}}, \bibinfo {author} {\bibfnamefont
  {M.}~\bibnamefont {Kuznetsova}}, \bibinfo {author} {\bibfnamefont
  {I.}~\bibnamefont {Lapshov}}, \bibinfo {author} {\bibfnamefont
  {A.}~\bibnamefont {Tkachenko}}, \bibinfo {author} {\bibfnamefont
  {R.}~\bibnamefont {Krivonos}}, \bibinfo {author} {\bibfnamefont
  {N.}~\bibnamefont {Semena}}, \bibinfo {author} {\bibfnamefont
  {M.}~\bibnamefont {Buntov}}, \bibinfo {author} {\bibfnamefont
  {A.}~\bibnamefont {Glushenko}}, \bibinfo {author} {\bibfnamefont
  {V.}~\bibnamefont {Arefiev}}, \bibinfo {author} {\bibfnamefont
  {A.}~\bibnamefont {Yaskovich}}, \bibinfo {author} {\bibfnamefont
  {S.}~\bibnamefont {Grebenev}}, \bibinfo {author} {\bibfnamefont
  {S.}~\bibnamefont {Sazonov}}, \bibinfo {author} {\bibfnamefont
  {A.}~\bibnamefont {Lutovinov}}, \bibinfo {author} {\bibfnamefont
  {S.}~\bibnamefont {Molkov}}, \bibinfo {author} {\bibfnamefont
  {D.}~\bibnamefont {Serbinov}}, \bibinfo {author} {\bibfnamefont
  {M.}~\bibnamefont {Kudelin}}, \bibinfo {author} {\bibfnamefont
  {T.}~\bibnamefont {Drozdova}}, \bibinfo {author} {\bibfnamefont
  {S.}~\bibnamefont {Voronkov}}, \bibinfo {author} {\bibfnamefont
  {R.}~\bibnamefont {Sunyaev}}, \bibinfo {author} {\bibfnamefont
  {E.}~\bibnamefont {Churazov}}, \bibinfo {author} {\bibfnamefont
  {M.}~\bibnamefont {Gilfanov}}, \bibinfo {author} {\bibfnamefont
  {B.}~\bibnamefont {Ramsey}}, \bibinfo {author} {\bibfnamefont {S.~L.}\
  \bibnamefont {O'Dell}}, \bibinfo {author} {\bibfnamefont {J.}~\bibnamefont
  {Kolodziejczak}}, \bibinfo {author} {\bibfnamefont {V.}~\bibnamefont
  {Zavlin}},\ and\ \bibinfo {author} {\bibfnamefont {D.}~\bibnamefont
  {Swartz}},\ }\bibfield  {title} {\bibinfo {title} {{ART-XC / SRG overview}},\
  }in\ \href {https://doi.org/10.1117/12.2312053} {\emph {\bibinfo {booktitle}
  {Space Telescopes and Instrumentation 2018: Ultraviolet to Gamma Ray}}},\
  Vol.\ \bibinfo {volume} {10699},\ \bibinfo {editor} {edited by\ \bibinfo
  {editor} {\bibfnamefont {J.-W.~A.}\ \bibnamefont {den Herder}}, \bibinfo
  {editor} {\bibfnamefont {S.}~\bibnamefont {Nikzad}},\ and\ \bibinfo {editor}
  {\bibfnamefont {K.}~\bibnamefont {Nakazawa}}},\ \bibinfo {organization}
  {International Society for Optics and Photonics}\ (\bibinfo  {publisher}
  {SPIE},\ \bibinfo {year} {2018})\ p.\ \bibinfo {pages} {106991Y}\BibitemShut
  {NoStop}%
\bibitem [{\citenamefont {Zhang}\ \emph {et~al.}(2020)\citenamefont {Zhang},
  \citenamefont {Li}, \citenamefont {Lu}, \citenamefont {Song}, \citenamefont
  {Xu}, \citenamefont {Liu}, \citenamefont {Chen}, \citenamefont {Cao},
  \citenamefont {Bu}, \citenamefont {Chang}, \citenamefont {Chen},
  \citenamefont {Chen}, \citenamefont {Chen}, \citenamefont {Chen},
  \citenamefont {Chen}, \citenamefont {Cui}, \citenamefont {Cui}, \citenamefont
  {Deng}, \citenamefont {Dong}, \citenamefont {Du}, \citenamefont {Fu},
  \citenamefont {Gao}, \citenamefont {Gao}, \citenamefont {Gao}, \citenamefont
  {Ge}, \citenamefont {Gu}, \citenamefont {Guan}, \citenamefont {Gungor},
  \citenamefont {Guo}, \citenamefont {Han}, \citenamefont {Hu}, \citenamefont
  {Huang}, \citenamefont {Huo}, \citenamefont {Jia}, \citenamefont {Jiang},
  \citenamefont {Jiang}, \citenamefont {Jin}, \citenamefont {Jin},
  \citenamefont {Li}, \citenamefont {Li}, \citenamefont {Li}, \citenamefont
  {Li}, \citenamefont {Li}, \citenamefont {Li}, \citenamefont {Li},
  \citenamefont {Li}, \citenamefont {Li}, \citenamefont {Li}, \citenamefont
  {Li}, \citenamefont {Liang}, \citenamefont {Liao}, \citenamefont {Liu},
  \citenamefont {Liu}, \citenamefont {Liu}, \citenamefont {Liu}, \citenamefont
  {Liu}, \citenamefont {Liu}, \citenamefont {Lu}, \citenamefont {Lu},
  \citenamefont {Luo}, \citenamefont {Ma}, \citenamefont {Meng}, \citenamefont
  {Nang}, \citenamefont {Nie}, \citenamefont {Ou}, \citenamefont {Qu},
  \citenamefont {Sai}, \citenamefont {Shang}, \citenamefont {Shen},
  \citenamefont {Sun}, \citenamefont {Tan}, \citenamefont {Tao}, \citenamefont
  {Tuo}, \citenamefont {Wang}, \citenamefont {Wang}, \citenamefont {Wang},
  \citenamefont {Wang}, \citenamefont {Wang}, \citenamefont {Wang},
  \citenamefont {Wang}, \citenamefont {Wen}, \citenamefont {Wu}, \citenamefont
  {Wu}, \citenamefont {Wu}, \citenamefont {Xiao}, \citenamefont {Xiong},
  \citenamefont {Yan}, \citenamefont {Yang}, \citenamefont {Yang},
  \citenamefont {Yang}, \citenamefont {Yi}, \citenamefont {Yuan}, \citenamefont
  {Zhang}, \citenamefont {Zhang}, \citenamefont {Zhang}, \citenamefont {Zhang},
  \citenamefont {Zhang}, \citenamefont {Zhang}, \citenamefont {Zhang},
  \citenamefont {Zhang}, \citenamefont {Zhang}, \citenamefont {Zhang},
  \citenamefont {Zhang}, \citenamefont {Zhang}, \citenamefont {Zhang},
  \citenamefont {Zhang}, \citenamefont {Zhang}, \citenamefont {Zhang},
  \citenamefont {Zhang}, \citenamefont {Zhang}, \citenamefont {Zhang},
  \citenamefont {Zhang}, \citenamefont {Zhao}, \citenamefont {Zhao},
  \citenamefont {Zheng}, \citenamefont {Zhou}, \citenamefont {Zhu},
  \citenamefont {Zhu},\ and\ \citenamefont {and}}]{Zhang_2020}%
  \BibitemOpen
  \bibfield  {author} {\bibinfo {author} {\bibfnamefont {S.-N.}\ \bibnamefont
  {Zhang}}, \bibinfo {author} {\bibfnamefont {T.}~\bibnamefont {Li}}, \bibinfo
  {author} {\bibfnamefont {F.}~\bibnamefont {Lu}}, \bibinfo {author}
  {\bibfnamefont {L.}~\bibnamefont {Song}}, \bibinfo {author} {\bibfnamefont
  {Y.}~\bibnamefont {Xu}}, \bibinfo {author} {\bibfnamefont {C.}~\bibnamefont
  {Liu}}, \bibinfo {author} {\bibfnamefont {Y.}~\bibnamefont {Chen}}, \bibinfo
  {author} {\bibfnamefont {X.}~\bibnamefont {Cao}}, \bibinfo {author}
  {\bibfnamefont {Q.}~\bibnamefont {Bu}}, \bibinfo {author} {\bibfnamefont
  {Z.}~\bibnamefont {Chang}}, \bibinfo {author} {\bibfnamefont
  {G.}~\bibnamefont {Chen}}, \bibinfo {author} {\bibfnamefont {L.}~\bibnamefont
  {Chen}}, \bibinfo {author} {\bibfnamefont {T.}~\bibnamefont {Chen}}, \bibinfo
  {author} {\bibfnamefont {Y.}~\bibnamefont {Chen}}, \bibinfo {author}
  {\bibfnamefont {Y.}~\bibnamefont {Chen}}, \bibinfo {author} {\bibfnamefont
  {W.}~\bibnamefont {Cui}}, \bibinfo {author} {\bibfnamefont {W.}~\bibnamefont
  {Cui}}, \bibinfo {author} {\bibfnamefont {J.}~\bibnamefont {Deng}}, \bibinfo
  {author} {\bibfnamefont {Y.}~\bibnamefont {Dong}}, \bibinfo {author}
  {\bibfnamefont {Y.}~\bibnamefont {Du}}, \bibinfo {author} {\bibfnamefont
  {M.}~\bibnamefont {Fu}}, \bibinfo {author} {\bibfnamefont {G.}~\bibnamefont
  {Gao}}, \bibinfo {author} {\bibfnamefont {H.}~\bibnamefont {Gao}}, \bibinfo
  {author} {\bibfnamefont {M.}~\bibnamefont {Gao}}, \bibinfo {author}
  {\bibfnamefont {M.}~\bibnamefont {Ge}}, \bibinfo {author} {\bibfnamefont
  {Y.}~\bibnamefont {Gu}}, \bibinfo {author} {\bibfnamefont {J.}~\bibnamefont
  {Guan}}, \bibinfo {author} {\bibfnamefont {C.}~\bibnamefont {Gungor}},
  \bibinfo {author} {\bibfnamefont {C.}~\bibnamefont {Guo}}, \bibinfo {author}
  {\bibfnamefont {D.}~\bibnamefont {Han}}, \bibinfo {author} {\bibfnamefont
  {W.}~\bibnamefont {Hu}}, \bibinfo {author} {\bibfnamefont {Y.}~\bibnamefont
  {Huang}}, \bibinfo {author} {\bibfnamefont {J.}~\bibnamefont {Huo}}, \bibinfo
  {author} {\bibfnamefont {S.}~\bibnamefont {Jia}}, \bibinfo {author}
  {\bibfnamefont {L.}~\bibnamefont {Jiang}}, \bibinfo {author} {\bibfnamefont
  {W.}~\bibnamefont {Jiang}}, \bibinfo {author} {\bibfnamefont
  {J.}~\bibnamefont {Jin}}, \bibinfo {author} {\bibfnamefont {Y.}~\bibnamefont
  {Jin}}, \bibinfo {author} {\bibfnamefont {B.}~\bibnamefont {Li}}, \bibinfo
  {author} {\bibfnamefont {C.}~\bibnamefont {Li}}, \bibinfo {author}
  {\bibfnamefont {G.}~\bibnamefont {Li}}, \bibinfo {author} {\bibfnamefont
  {M.}~\bibnamefont {Li}}, \bibinfo {author} {\bibfnamefont {W.}~\bibnamefont
  {Li}}, \bibinfo {author} {\bibfnamefont {X.}~\bibnamefont {Li}}, \bibinfo
  {author} {\bibfnamefont {X.}~\bibnamefont {Li}}, \bibinfo {author}
  {\bibfnamefont {X.}~\bibnamefont {Li}}, \bibinfo {author} {\bibfnamefont
  {Y.}~\bibnamefont {Li}}, \bibinfo {author} {\bibfnamefont {Z.}~\bibnamefont
  {Li}}, \bibinfo {author} {\bibfnamefont {Z.}~\bibnamefont {Li}}, \bibinfo
  {author} {\bibfnamefont {X.}~\bibnamefont {Liang}}, \bibinfo {author}
  {\bibfnamefont {J.}~\bibnamefont {Liao}}, \bibinfo {author} {\bibfnamefont
  {G.}~\bibnamefont {Liu}}, \bibinfo {author} {\bibfnamefont {H.}~\bibnamefont
  {Liu}}, \bibinfo {author} {\bibfnamefont {S.}~\bibnamefont {Liu}}, \bibinfo
  {author} {\bibfnamefont {X.}~\bibnamefont {Liu}}, \bibinfo {author}
  {\bibfnamefont {Y.}~\bibnamefont {Liu}}, \bibinfo {author} {\bibfnamefont
  {Y.}~\bibnamefont {Liu}}, \bibinfo {author} {\bibfnamefont {B.}~\bibnamefont
  {Lu}}, \bibinfo {author} {\bibfnamefont {X.}~\bibnamefont {Lu}}, \bibinfo
  {author} {\bibfnamefont {T.}~\bibnamefont {Luo}}, \bibinfo {author}
  {\bibfnamefont {X.}~\bibnamefont {Ma}}, \bibinfo {author} {\bibfnamefont
  {B.}~\bibnamefont {Meng}}, \bibinfo {author} {\bibfnamefont {Y.}~\bibnamefont
  {Nang}}, \bibinfo {author} {\bibfnamefont {J.}~\bibnamefont {Nie}}, \bibinfo
  {author} {\bibfnamefont {G.}~\bibnamefont {Ou}}, \bibinfo {author}
  {\bibfnamefont {J.}~\bibnamefont {Qu}}, \bibinfo {author} {\bibfnamefont
  {N.}~\bibnamefont {Sai}}, \bibinfo {author} {\bibfnamefont {R.}~\bibnamefont
  {Shang}}, \bibinfo {author} {\bibfnamefont {G.}~\bibnamefont {Shen}},
  \bibinfo {author} {\bibfnamefont {L.}~\bibnamefont {Sun}}, \bibinfo {author}
  {\bibfnamefont {Y.}~\bibnamefont {Tan}}, \bibinfo {author} {\bibfnamefont
  {L.}~\bibnamefont {Tao}}, \bibinfo {author} {\bibfnamefont {Y.}~\bibnamefont
  {Tuo}}, \bibinfo {author} {\bibfnamefont {C.}~\bibnamefont {Wang}}, \bibinfo
  {author} {\bibfnamefont {C.}~\bibnamefont {Wang}}, \bibinfo {author}
  {\bibfnamefont {G.}~\bibnamefont {Wang}}, \bibinfo {author} {\bibfnamefont
  {H.}~\bibnamefont {Wang}}, \bibinfo {author} {\bibfnamefont {J.}~\bibnamefont
  {Wang}}, \bibinfo {author} {\bibfnamefont {W.}~\bibnamefont {Wang}}, \bibinfo
  {author} {\bibfnamefont {Y.}~\bibnamefont {Wang}}, \bibinfo {author}
  {\bibfnamefont {X.}~\bibnamefont {Wen}}, \bibinfo {author} {\bibfnamefont
  {B.}~\bibnamefont {Wu}}, \bibinfo {author} {\bibfnamefont {B.}~\bibnamefont
  {Wu}}, \bibinfo {author} {\bibfnamefont {M.}~\bibnamefont {Wu}}, \bibinfo
  {author} {\bibfnamefont {G.}~\bibnamefont {Xiao}}, \bibinfo {author}
  {\bibfnamefont {S.}~\bibnamefont {Xiong}}, \bibinfo {author} {\bibfnamefont
  {L.}~\bibnamefont {Yan}}, \bibinfo {author} {\bibfnamefont {J.}~\bibnamefont
  {Yang}}, \bibinfo {author} {\bibfnamefont {S.}~\bibnamefont {Yang}}, \bibinfo
  {author} {\bibfnamefont {Y.}~\bibnamefont {Yang}}, \bibinfo {author}
  {\bibfnamefont {Q.}~\bibnamefont {Yi}}, \bibinfo {author} {\bibfnamefont
  {B.}~\bibnamefont {Yuan}}, \bibinfo {author} {\bibfnamefont {A.}~\bibnamefont
  {Zhang}}, \bibinfo {author} {\bibfnamefont {C.}~\bibnamefont {Zhang}},
  \bibinfo {author} {\bibfnamefont {C.}~\bibnamefont {Zhang}}, \bibinfo
  {author} {\bibfnamefont {F.}~\bibnamefont {Zhang}}, \bibinfo {author}
  {\bibfnamefont {H.}~\bibnamefont {Zhang}}, \bibinfo {author} {\bibfnamefont
  {J.}~\bibnamefont {Zhang}}, \bibinfo {author} {\bibfnamefont
  {Q.}~\bibnamefont {Zhang}}, \bibinfo {author} {\bibfnamefont
  {S.}~\bibnamefont {Zhang}}, \bibinfo {author} {\bibfnamefont
  {S.}~\bibnamefont {Zhang}}, \bibinfo {author} {\bibfnamefont
  {T.}~\bibnamefont {Zhang}}, \bibinfo {author} {\bibfnamefont
  {W.}~\bibnamefont {Zhang}}, \bibinfo {author} {\bibfnamefont
  {W.}~\bibnamefont {Zhang}}, \bibinfo {author} {\bibfnamefont
  {W.}~\bibnamefont {Zhang}}, \bibinfo {author} {\bibfnamefont
  {Y.}~\bibnamefont {Zhang}}, \bibinfo {author} {\bibfnamefont
  {Y.}~\bibnamefont {Zhang}}, \bibinfo {author} {\bibfnamefont
  {Y.}~\bibnamefont {Zhang}}, \bibinfo {author} {\bibfnamefont
  {Y.}~\bibnamefont {Zhang}}, \bibinfo {author} {\bibfnamefont
  {Z.}~\bibnamefont {Zhang}}, \bibinfo {author} {\bibfnamefont
  {Z.}~\bibnamefont {Zhang}}, \bibinfo {author} {\bibfnamefont
  {Z.}~\bibnamefont {Zhang}}, \bibinfo {author} {\bibfnamefont
  {H.}~\bibnamefont {Zhao}}, \bibinfo {author} {\bibfnamefont {X.}~\bibnamefont
  {Zhao}}, \bibinfo {author} {\bibfnamefont {S.}~\bibnamefont {Zheng}},
  \bibinfo {author} {\bibfnamefont {J.}~\bibnamefont {Zhou}}, \bibinfo {author}
  {\bibfnamefont {Y.}~\bibnamefont {Zhu}}, \bibinfo {author} {\bibfnamefont
  {Y.}~\bibnamefont {Zhu}},\ and\ \bibinfo {author} {\bibfnamefont {R.~Z.}\
  \bibnamefont {and}},\ }\bibfield  {title} {\bibinfo {title} {Overview to the
  hard x-ray modulation telescope (insight-{HXMT}) satellite},\ }\bibfield
  {journal} {\bibinfo  {journal} {Science China Physics, Mechanics \&
  Astronomy}\ }\textbf {\bibinfo {volume} {63}},\ \href
  {https://doi.org/10.1007/s11433-019-1432-6} {10.1007/s11433-019-1432-6}
  (\bibinfo {year} {2020})\BibitemShut {NoStop}%
\bibitem [{\citenamefont {{Predehl}}\ \emph {et~al.}(2021)\citenamefont
  {{Predehl}}, \citenamefont {{Andritschke}}, \citenamefont {{Arefiev}},
  \citenamefont {{Babyshkin}}, \citenamefont {{Batanov}}, \citenamefont
  {{Becker}}, \citenamefont {{B{\"o}hringer}}, \citenamefont {{Bogomolov}},
  \citenamefont {{Boller}}, \citenamefont {{Borm}}, \citenamefont
  {{Bornemann}}, \citenamefont {{Br{\"a}uninger}}, \citenamefont
  {{Br{\"u}ggen}}, \citenamefont {{Brunner}}, \citenamefont {{Brusa}},
  \citenamefont {{Bulbul}}, \citenamefont {{Buntov}}, \citenamefont
  {{Burwitz}}, \citenamefont {{Burkert}}, \citenamefont {{Clerc}},
  \citenamefont {{Churazov}}, \citenamefont {{Coutinho}}, \citenamefont
  {{Dauser}}, \citenamefont {{Dennerl}}, \citenamefont {{Doroshenko}},
  \citenamefont {{Eder}}, \citenamefont {{Emberger}}, \citenamefont
  {{Eraerds}}, \citenamefont {{Finoguenov}}, \citenamefont {{Freyberg}},
  \citenamefont {{Friedrich}}, \citenamefont {{Friedrich}}, \citenamefont
  {{F{\"u}rmetz}}, \citenamefont {{Georgakakis}}, \citenamefont {{Gilfanov}},
  \citenamefont {{Granato}}, \citenamefont {{Grossberger}}, \citenamefont
  {{Gueguen}}, \citenamefont {{Gureev}}, \citenamefont {{Haberl}},
  \citenamefont {{H{\"a}lker}}, \citenamefont {{Hartner}}, \citenamefont
  {{Hasinger}}, \citenamefont {{Huber}}, \citenamefont {{Ji}}, \citenamefont
  {{Kienlin}}, \citenamefont {{Kink}}, \citenamefont {{Korotkov}},
  \citenamefont {{Kreykenbohm}}, \citenamefont {{Lamer}}, \citenamefont
  {{Lomakin}}, \citenamefont {{Lapshov}}, \citenamefont {{Liu}}, \citenamefont
  {{Maitra}}, \citenamefont {{Meidinger}}, \citenamefont {{Menz}},
  \citenamefont {{Merloni}}, \citenamefont {{Mernik}}, \citenamefont {{Mican}},
  \citenamefont {{Mohr}}, \citenamefont {{M{\"u}ller}}, \citenamefont
  {{Nandra}}, \citenamefont {{Nazarov}}, \citenamefont {{Pacaud}},
  \citenamefont {{Pavlinsky}}, \citenamefont {{Perinati}}, \citenamefont
  {{Pfeffermann}}, \citenamefont {{Pietschner}}, \citenamefont {{Ramos-Ceja}},
  \citenamefont {{Rau}}, \citenamefont {{Reiffers}}, \citenamefont
  {{Reiprich}}, \citenamefont {{Robrade}}, \citenamefont {{Salvato}},
  \citenamefont {{Sanders}}, \citenamefont {{Santangelo}}, \citenamefont
  {{Sasaki}}, \citenamefont {{Scheuerle}}, \citenamefont {{Schmid}},
  \citenamefont {{Schmitt}}, \citenamefont {{Schwope}}, \citenamefont
  {{Shirshakov}}, \citenamefont {{Steinmetz}}, \citenamefont {{Stewart}},
  \citenamefont {{Str{\"u}der}}, \citenamefont {{Sunyaev}}, \citenamefont
  {{Tenzer}}, \citenamefont {{Tiedemann}}, \citenamefont {{Tr{\"u}mper}},
  \citenamefont {{Voron}}, \citenamefont {{Weber}}, \citenamefont {{Wilms}},\
  and\ \citenamefont {{Yaroshenko}}}]{2021A&A...647A...1P}%
  \BibitemOpen
  \bibfield  {author} {\bibinfo {author} {\bibfnamefont {P.}~\bibnamefont
  {{Predehl}}}, \bibinfo {author} {\bibfnamefont {R.}~\bibnamefont
  {{Andritschke}}}, \bibinfo {author} {\bibfnamefont {V.}~\bibnamefont
  {{Arefiev}}}, \bibinfo {author} {\bibfnamefont {V.}~\bibnamefont
  {{Babyshkin}}}, \bibinfo {author} {\bibfnamefont {O.}~\bibnamefont
  {{Batanov}}}, \bibinfo {author} {\bibfnamefont {W.}~\bibnamefont {{Becker}}},
  \bibinfo {author} {\bibfnamefont {H.}~\bibnamefont {{B{\"o}hringer}}},
  \bibinfo {author} {\bibfnamefont {A.}~\bibnamefont {{Bogomolov}}}, \bibinfo
  {author} {\bibfnamefont {T.}~\bibnamefont {{Boller}}}, \bibinfo {author}
  {\bibfnamefont {K.}~\bibnamefont {{Borm}}}, \bibinfo {author} {\bibfnamefont
  {W.}~\bibnamefont {{Bornemann}}}, \bibinfo {author} {\bibfnamefont
  {H.}~\bibnamefont {{Br{\"a}uninger}}}, \bibinfo {author} {\bibfnamefont
  {M.}~\bibnamefont {{Br{\"u}ggen}}}, \bibinfo {author} {\bibfnamefont
  {H.}~\bibnamefont {{Brunner}}}, \bibinfo {author} {\bibfnamefont
  {M.}~\bibnamefont {{Brusa}}}, \bibinfo {author} {\bibfnamefont
  {E.}~\bibnamefont {{Bulbul}}}, \bibinfo {author} {\bibfnamefont
  {M.}~\bibnamefont {{Buntov}}}, \bibinfo {author} {\bibfnamefont
  {V.}~\bibnamefont {{Burwitz}}}, \bibinfo {author} {\bibfnamefont
  {W.}~\bibnamefont {{Burkert}}}, \bibinfo {author} {\bibfnamefont
  {N.}~\bibnamefont {{Clerc}}}, \bibinfo {author} {\bibfnamefont
  {E.}~\bibnamefont {{Churazov}}}, \bibinfo {author} {\bibfnamefont
  {D.}~\bibnamefont {{Coutinho}}}, \bibinfo {author} {\bibfnamefont
  {T.}~\bibnamefont {{Dauser}}}, \bibinfo {author} {\bibfnamefont
  {K.}~\bibnamefont {{Dennerl}}}, \bibinfo {author} {\bibfnamefont
  {V.}~\bibnamefont {{Doroshenko}}}, \bibinfo {author} {\bibfnamefont
  {J.}~\bibnamefont {{Eder}}}, \bibinfo {author} {\bibfnamefont
  {V.}~\bibnamefont {{Emberger}}}, \bibinfo {author} {\bibfnamefont
  {T.}~\bibnamefont {{Eraerds}}}, \bibinfo {author} {\bibfnamefont
  {A.}~\bibnamefont {{Finoguenov}}}, \bibinfo {author} {\bibfnamefont
  {M.}~\bibnamefont {{Freyberg}}}, \bibinfo {author} {\bibfnamefont
  {P.}~\bibnamefont {{Friedrich}}}, \bibinfo {author} {\bibfnamefont
  {S.}~\bibnamefont {{Friedrich}}}, \bibinfo {author} {\bibfnamefont
  {M.}~\bibnamefont {{F{\"u}rmetz}}}, \bibinfo {author} {\bibfnamefont
  {A.}~\bibnamefont {{Georgakakis}}}, \bibinfo {author} {\bibfnamefont
  {M.}~\bibnamefont {{Gilfanov}}}, \bibinfo {author} {\bibfnamefont
  {S.}~\bibnamefont {{Granato}}}, \bibinfo {author} {\bibfnamefont
  {C.}~\bibnamefont {{Grossberger}}}, \bibinfo {author} {\bibfnamefont
  {A.}~\bibnamefont {{Gueguen}}}, \bibinfo {author} {\bibfnamefont
  {P.}~\bibnamefont {{Gureev}}}, \bibinfo {author} {\bibfnamefont
  {F.}~\bibnamefont {{Haberl}}}, \bibinfo {author} {\bibfnamefont
  {O.}~\bibnamefont {{H{\"a}lker}}}, \bibinfo {author} {\bibfnamefont
  {G.}~\bibnamefont {{Hartner}}}, \bibinfo {author} {\bibfnamefont
  {G.}~\bibnamefont {{Hasinger}}}, \bibinfo {author} {\bibfnamefont
  {H.}~\bibnamefont {{Huber}}}, \bibinfo {author} {\bibfnamefont
  {L.}~\bibnamefont {{Ji}}}, \bibinfo {author} {\bibfnamefont {A.~v.}\
  \bibnamefont {{Kienlin}}}, \bibinfo {author} {\bibfnamefont {W.}~\bibnamefont
  {{Kink}}}, \bibinfo {author} {\bibfnamefont {F.}~\bibnamefont {{Korotkov}}},
  \bibinfo {author} {\bibfnamefont {I.}~\bibnamefont {{Kreykenbohm}}}, \bibinfo
  {author} {\bibfnamefont {G.}~\bibnamefont {{Lamer}}}, \bibinfo {author}
  {\bibfnamefont {I.}~\bibnamefont {{Lomakin}}}, \bibinfo {author}
  {\bibfnamefont {I.}~\bibnamefont {{Lapshov}}}, \bibinfo {author}
  {\bibfnamefont {T.}~\bibnamefont {{Liu}}}, \bibinfo {author} {\bibfnamefont
  {C.}~\bibnamefont {{Maitra}}}, \bibinfo {author} {\bibfnamefont
  {N.}~\bibnamefont {{Meidinger}}}, \bibinfo {author} {\bibfnamefont
  {B.}~\bibnamefont {{Menz}}}, \bibinfo {author} {\bibfnamefont
  {A.}~\bibnamefont {{Merloni}}}, \bibinfo {author} {\bibfnamefont
  {T.}~\bibnamefont {{Mernik}}}, \bibinfo {author} {\bibfnamefont
  {B.}~\bibnamefont {{Mican}}}, \bibinfo {author} {\bibfnamefont
  {J.}~\bibnamefont {{Mohr}}}, \bibinfo {author} {\bibfnamefont
  {S.}~\bibnamefont {{M{\"u}ller}}}, \bibinfo {author} {\bibfnamefont
  {K.}~\bibnamefont {{Nandra}}}, \bibinfo {author} {\bibfnamefont
  {V.}~\bibnamefont {{Nazarov}}}, \bibinfo {author} {\bibfnamefont
  {F.}~\bibnamefont {{Pacaud}}}, \bibinfo {author} {\bibfnamefont
  {M.}~\bibnamefont {{Pavlinsky}}}, \bibinfo {author} {\bibfnamefont
  {E.}~\bibnamefont {{Perinati}}}, \bibinfo {author} {\bibfnamefont
  {E.}~\bibnamefont {{Pfeffermann}}}, \bibinfo {author} {\bibfnamefont
  {D.}~\bibnamefont {{Pietschner}}}, \bibinfo {author} {\bibfnamefont {M.~E.}\
  \bibnamefont {{Ramos-Ceja}}}, \bibinfo {author} {\bibfnamefont
  {A.}~\bibnamefont {{Rau}}}, \bibinfo {author} {\bibfnamefont
  {J.}~\bibnamefont {{Reiffers}}}, \bibinfo {author} {\bibfnamefont {T.~H.}\
  \bibnamefont {{Reiprich}}}, \bibinfo {author} {\bibfnamefont
  {J.}~\bibnamefont {{Robrade}}}, \bibinfo {author} {\bibfnamefont
  {M.}~\bibnamefont {{Salvato}}}, \bibinfo {author} {\bibfnamefont
  {J.}~\bibnamefont {{Sanders}}}, \bibinfo {author} {\bibfnamefont
  {A.}~\bibnamefont {{Santangelo}}}, \bibinfo {author} {\bibfnamefont
  {M.}~\bibnamefont {{Sasaki}}}, \bibinfo {author} {\bibfnamefont
  {H.}~\bibnamefont {{Scheuerle}}}, \bibinfo {author} {\bibfnamefont
  {C.}~\bibnamefont {{Schmid}}}, \bibinfo {author} {\bibfnamefont
  {J.}~\bibnamefont {{Schmitt}}}, \bibinfo {author} {\bibfnamefont
  {A.}~\bibnamefont {{Schwope}}}, \bibinfo {author} {\bibfnamefont
  {A.}~\bibnamefont {{Shirshakov}}}, \bibinfo {author} {\bibfnamefont
  {M.}~\bibnamefont {{Steinmetz}}}, \bibinfo {author} {\bibfnamefont
  {I.}~\bibnamefont {{Stewart}}}, \bibinfo {author} {\bibfnamefont
  {L.}~\bibnamefont {{Str{\"u}der}}}, \bibinfo {author} {\bibfnamefont
  {R.}~\bibnamefont {{Sunyaev}}}, \bibinfo {author} {\bibfnamefont
  {C.}~\bibnamefont {{Tenzer}}}, \bibinfo {author} {\bibfnamefont
  {L.}~\bibnamefont {{Tiedemann}}}, \bibinfo {author} {\bibfnamefont
  {J.}~\bibnamefont {{Tr{\"u}mper}}}, \bibinfo {author} {\bibfnamefont
  {V.}~\bibnamefont {{Voron}}}, \bibinfo {author} {\bibfnamefont
  {P.}~\bibnamefont {{Weber}}}, \bibinfo {author} {\bibfnamefont
  {J.}~\bibnamefont {{Wilms}}},\ and\ \bibinfo {author} {\bibfnamefont
  {V.}~\bibnamefont {{Yaroshenko}}},\ }\bibfield  {title} {\bibinfo {title}
  {{The eROSITA X-ray telescope on SRG}},\ }\href
  {https://doi.org/10.1051/0004-6361/202039313} {\bibfield  {journal} {\bibinfo
   {journal} {\aap}\ }\textbf {\bibinfo {volume} {647}},\ \bibinfo {eid} {A1}
  (\bibinfo {year} {2021})},\ \Eprint {https://arxiv.org/abs/2010.03477}
  {arXiv:2010.03477 [astro-ph.HE]} \BibitemShut {NoStop}%
\bibitem [{197(1974)}]{1974ASSL...43.....G}%
  \BibitemOpen
  \href {https://doi.org/10.1007/978-94-010-2105-0} {\emph {\bibinfo {title}
  {X-ray Astronomy}}}\ (\bibinfo {year} {1974})\BibitemShut {NoStop}%
\bibitem [{\citenamefont {{Grimm}}\ \emph
  {et~al.}(2003{\natexlab{a}})\citenamefont {{Grimm}}, \citenamefont
  {{Gilfanov}},\ and\ \citenamefont {{Sunyaev}}}]{2003ChJAS...3..257G}%
  \BibitemOpen
  \bibfield  {author} {\bibinfo {author} {\bibfnamefont {H.-J.}\ \bibnamefont
  {{Grimm}}}, \bibinfo {author} {\bibfnamefont {M.}~\bibnamefont
  {{Gilfanov}}},\ and\ \bibinfo {author} {\bibfnamefont {R.}~\bibnamefont
  {{Sunyaev}}},\ }\bibfield  {title} {\bibinfo {title} {{X-ray binaries in the
  Milky Way and other galaxies}},\ }\href
  {https://doi.org/10.1088/1009-9271/3/S1/257} {\bibfield  {journal} {\bibinfo
  {journal} {Chinese Journal of Astronomy and Astrophysics Supplement}\
  }\textbf {\bibinfo {volume} {3}},\ \bibinfo {pages} {257} (\bibinfo {year}
  {2003}{\natexlab{a}})}\BibitemShut {NoStop}%
\bibitem [{\citenamefont {{Mineo}}\ \emph {et~al.}(2012)\citenamefont
  {{Mineo}}, \citenamefont {{Gilfanov}},\ and\ \citenamefont
  {{Sunyaev}}}]{2012MNRAS.419.2095M}%
  \BibitemOpen
  \bibfield  {author} {\bibinfo {author} {\bibfnamefont {S.}~\bibnamefont
  {{Mineo}}}, \bibinfo {author} {\bibfnamefont {M.}~\bibnamefont
  {{Gilfanov}}},\ and\ \bibinfo {author} {\bibfnamefont {R.}~\bibnamefont
  {{Sunyaev}}},\ }\bibfield  {title} {\bibinfo {title} {{X-ray emission from
  star-forming galaxies - I. High-mass X-ray binaries}},\ }\href
  {https://doi.org/10.1111/j.1365-2966.2011.19862.x} {\bibfield  {journal}
  {\bibinfo  {journal} {\mnras}\ }\textbf {\bibinfo {volume} {419}},\ \bibinfo
  {pages} {2095} (\bibinfo {year} {2012})},\ \Eprint
  {https://arxiv.org/abs/1105.4610} {arXiv:1105.4610 [astro-ph.HE]}
  \BibitemShut {NoStop}%
\bibitem [{199(1994)}]{1994inbi.conf.....S}%
  \BibitemOpen
  \href@noop {} {\emph {\bibinfo {title} {Saas-Fee Advanced Course 22:
  Interacting Binaries}}}\ (\bibinfo {year} {1994})\BibitemShut {NoStop}%
\bibitem [{\citenamefont {{Charles}}\ and\ \citenamefont
  {{Coe}}(2003)}]{2003astro.ph..8020C}%
  \BibitemOpen
  \bibfield  {author} {\bibinfo {author} {\bibfnamefont {P.~A.}\ \bibnamefont
  {{Charles}}}\ and\ \bibinfo {author} {\bibfnamefont {M.~J.}\ \bibnamefont
  {{Coe}}},\ }\bibfield  {title} {\bibinfo {title} {{Optical, ultraviolet and
  infrared observations of X-ray binaries}},\ }\href
  {https://doi.org/10.48550/arXiv.astro-ph/0308020} {\bibfield  {journal}
  {\bibinfo  {journal} {arXiv e-prints}\ ,\ \bibinfo {eid} {astro-ph/0308020}}
  (\bibinfo {year} {2003})},\ \Eprint {https://arxiv.org/abs/astro-ph/0308020}
  {arXiv:astro-ph/0308020 [astro-ph]} \BibitemShut {NoStop}%
\bibitem [{\citenamefont {{Chaty}}(2013)}]{2013AdSpR..52.2132C}%
  \BibitemOpen
  \bibfield  {author} {\bibinfo {author} {\bibfnamefont {S.}~\bibnamefont
  {{Chaty}}},\ }\bibfield  {title} {\bibinfo {title} {{Optical/infrared
  observations unveiling the formation, nature and evolution of High-Mass X-ray
  Binaries}},\ }\href {https://doi.org/10.1016/j.asr.2013.09.010} {\bibfield
  {journal} {\bibinfo  {journal} {Advances in Space Research}\ }\textbf
  {\bibinfo {volume} {52}},\ \bibinfo {pages} {2132} (\bibinfo {year}
  {2013})},\ \Eprint {https://arxiv.org/abs/1403.0792} {arXiv:1403.0792
  [astro-ph.HE]} \BibitemShut {NoStop}%
\bibitem [{\citenamefont {{Maraschi}}\ \emph {et~al.}(1976)\citenamefont
  {{Maraschi}}, \citenamefont {{Treves}},\ and\ \citenamefont {{van den
  Heuvel}}}]{1976Natur.259..292M}%
  \BibitemOpen
  \bibfield  {author} {\bibinfo {author} {\bibfnamefont {L.}~\bibnamefont
  {{Maraschi}}}, \bibinfo {author} {\bibfnamefont {A.}~\bibnamefont
  {{Treves}}},\ and\ \bibinfo {author} {\bibfnamefont {E.~P.~J.}\ \bibnamefont
  {{van den Heuvel}}},\ }\bibfield  {title} {\bibinfo {title} {{B-emission
  Stars and X-ray sources.}},\ }\href {https://doi.org/10.1038/259292a0}
  {\bibfield  {journal} {\bibinfo  {journal} {\nat}\ }\textbf {\bibinfo
  {volume} {259}},\ \bibinfo {pages} {292} (\bibinfo {year}
  {1976})}\BibitemShut {NoStop}%
\bibitem [{\citenamefont {{Bolton}}(1972)}]{1972Natur.235..271B}%
  \BibitemOpen
  \bibfield  {author} {\bibinfo {author} {\bibfnamefont {C.~T.}\ \bibnamefont
  {{Bolton}}},\ }\bibfield  {title} {\bibinfo {title} {{Identification of
  Cygnus X-1 with HDE 226868}},\ }\href {https://doi.org/10.1038/235271b0}
  {\bibfield  {journal} {\bibinfo  {journal} {\nat}\ }\textbf {\bibinfo
  {volume} {235}},\ \bibinfo {pages} {271} (\bibinfo {year}
  {1972})}\BibitemShut {NoStop}%
\bibitem [{\citenamefont {{Hutchings}}\ \emph {et~al.}(1973)\citenamefont
  {{Hutchings}}, \citenamefont {{Crampton}}, \citenamefont {{Glaspey}},\ and\
  \citenamefont {{Walker}}}]{1973ApJ...182..549H}%
  \BibitemOpen
  \bibfield  {author} {\bibinfo {author} {\bibfnamefont {J.~B.}\ \bibnamefont
  {{Hutchings}}}, \bibinfo {author} {\bibfnamefont {D.}~\bibnamefont
  {{Crampton}}}, \bibinfo {author} {\bibfnamefont {J.}~\bibnamefont
  {{Glaspey}}},\ and\ \bibinfo {author} {\bibfnamefont {G.~A.~H.}\ \bibnamefont
  {{Walker}}},\ }\bibfield  {title} {\bibinfo {title} {{Optical observations
  and model for Cygnus X-1.}},\ }\href {https://doi.org/10.1086/152163}
  {\bibfield  {journal} {\bibinfo  {journal} {\apj}\ }\textbf {\bibinfo
  {volume} {182}},\ \bibinfo {pages} {549} (\bibinfo {year}
  {1973})}\BibitemShut {NoStop}%
\bibitem [{\citenamefont {{Hanson}}\ \emph {et~al.}(2000)\citenamefont
  {{Hanson}}, \citenamefont {{Still}},\ and\ \citenamefont
  {{Fender}}}]{2000ApJ...541..308H}%
  \BibitemOpen
  \bibfield  {author} {\bibinfo {author} {\bibfnamefont {M.~M.}\ \bibnamefont
  {{Hanson}}}, \bibinfo {author} {\bibfnamefont {M.~D.}\ \bibnamefont
  {{Still}}},\ and\ \bibinfo {author} {\bibfnamefont {R.~P.}\ \bibnamefont
  {{Fender}}},\ }\bibfield  {title} {\bibinfo {title} {{Orbital Dynamics of
  Cygnus X-3}},\ }\href {https://doi.org/10.1086/309419} {\bibfield  {journal}
  {\bibinfo  {journal} {\apj}\ }\textbf {\bibinfo {volume} {541}},\ \bibinfo
  {pages} {308} (\bibinfo {year} {2000})},\ \Eprint
  {https://arxiv.org/abs/astro-ph/0005032} {arXiv:astro-ph/0005032 [astro-ph]}
  \BibitemShut {NoStop}%
\bibitem [{\citenamefont {{Nowak}}\ \emph {et~al.}(2001)\citenamefont
  {{Nowak}}, \citenamefont {{Wilms}}, \citenamefont {{Heindl}}, \citenamefont
  {{Pottschmidt}}, \citenamefont {{Dove}},\ and\ \citenamefont
  {{Begelman}}}]{2001MNRAS.320..316N}%
  \BibitemOpen
  \bibfield  {author} {\bibinfo {author} {\bibfnamefont {M.~A.}\ \bibnamefont
  {{Nowak}}}, \bibinfo {author} {\bibfnamefont {J.}~\bibnamefont {{Wilms}}},
  \bibinfo {author} {\bibfnamefont {W.~A.}\ \bibnamefont {{Heindl}}}, \bibinfo
  {author} {\bibfnamefont {K.}~\bibnamefont {{Pottschmidt}}}, \bibinfo {author}
  {\bibfnamefont {J.~B.}\ \bibnamefont {{Dove}}},\ and\ \bibinfo {author}
  {\bibfnamefont {M.~C.}\ \bibnamefont {{Begelman}}},\ }\bibfield  {title}
  {\bibinfo {title} {{A good long look at the black hole candidates LMC X-1 and
  LMC X-3}},\ }\href {https://doi.org/10.1046/j.1365-8711.2001.03984.x}
  {\bibfield  {journal} {\bibinfo  {journal} {\mnras}\ }\textbf {\bibinfo
  {volume} {320}},\ \bibinfo {pages} {316} (\bibinfo {year} {2001})},\ \Eprint
  {https://arxiv.org/abs/astro-ph/0005487} {arXiv:astro-ph/0005487 [astro-ph]}
  \BibitemShut {NoStop}%
\bibitem [{\citenamefont {{Prestwich}}\ \emph {et~al.}(2007)\citenamefont
  {{Prestwich}}, \citenamefont {{Kilgard}}, \citenamefont {{Crowther}},
  \citenamefont {{Carpano}}, \citenamefont {{Pollock}}, \citenamefont
  {{Zezas}}, \citenamefont {{Saar}}, \citenamefont {{Roberts}},\ and\
  \citenamefont {{Ward}}}]{2007ApJ...669L..21P}%
  \BibitemOpen
  \bibfield  {author} {\bibinfo {author} {\bibfnamefont {A.~H.}\ \bibnamefont
  {{Prestwich}}}, \bibinfo {author} {\bibfnamefont {R.}~\bibnamefont
  {{Kilgard}}}, \bibinfo {author} {\bibfnamefont {P.~A.}\ \bibnamefont
  {{Crowther}}}, \bibinfo {author} {\bibfnamefont {S.}~\bibnamefont
  {{Carpano}}}, \bibinfo {author} {\bibfnamefont {A.~M.~T.}\ \bibnamefont
  {{Pollock}}}, \bibinfo {author} {\bibfnamefont {A.}~\bibnamefont {{Zezas}}},
  \bibinfo {author} {\bibfnamefont {S.~H.}\ \bibnamefont {{Saar}}}, \bibinfo
  {author} {\bibfnamefont {T.~P.}\ \bibnamefont {{Roberts}}},\ and\ \bibinfo
  {author} {\bibfnamefont {M.~J.}\ \bibnamefont {{Ward}}},\ }\bibfield  {title}
  {\bibinfo {title} {{The Orbital Period of the Wolf-Rayet Binary IC 10 X-1:
  Dynamic Evidence that the Compact Object Is a Black Hole}},\ }\href
  {https://doi.org/10.1086/523755} {\bibfield  {journal} {\bibinfo  {journal}
  {\apjl}\ }\textbf {\bibinfo {volume} {669}},\ \bibinfo {pages} {L21}
  (\bibinfo {year} {2007})},\ \Eprint {https://arxiv.org/abs/0709.2892}
  {arXiv:0709.2892 [astro-ph]} \BibitemShut {NoStop}%
\bibitem [{\citenamefont {{Laycock}}\ \emph {et~al.}(2015)\citenamefont
  {{Laycock}}, \citenamefont {{Maccarone}},\ and\ \citenamefont
  {{Christodoulou}}}]{2015MNRAS.452L..31L}%
  \BibitemOpen
  \bibfield  {author} {\bibinfo {author} {\bibfnamefont {S.~G.~T.}\
  \bibnamefont {{Laycock}}}, \bibinfo {author} {\bibfnamefont {T.~J.}\
  \bibnamefont {{Maccarone}}},\ and\ \bibinfo {author} {\bibfnamefont {D.~M.}\
  \bibnamefont {{Christodoulou}}},\ }\bibfield  {title} {\bibinfo {title}
  {{Revisiting the dynamical case for a massive black hole in IC10 X-1}},\
  }\href {https://doi.org/10.1093/mnrasl/slv082} {\bibfield  {journal}
  {\bibinfo  {journal} {\mnras}\ }\textbf {\bibinfo {volume} {452}},\ \bibinfo
  {pages} {L31} (\bibinfo {year} {2015})},\ \Eprint
  {https://arxiv.org/abs/1506.03882} {arXiv:1506.03882 [astro-ph.HE]}
  \BibitemShut {NoStop}%
\bibitem [{\citenamefont {{Orosz}}\ \emph {et~al.}(2007)\citenamefont
  {{Orosz}}, \citenamefont {{McClintock}}, \citenamefont {{Narayan}},
  \citenamefont {{Bailyn}}, \citenamefont {{Hartman}}, \citenamefont {{Macri}},
  \citenamefont {{Liu}}, \citenamefont {{Pietsch}}, \citenamefont
  {{Remillard}}, \citenamefont {{Shporer}},\ and\ \citenamefont
  {{Mazeh}}}]{2007Natur.449..872O}%
  \BibitemOpen
  \bibfield  {author} {\bibinfo {author} {\bibfnamefont {J.~A.}\ \bibnamefont
  {{Orosz}}}, \bibinfo {author} {\bibfnamefont {J.~E.}\ \bibnamefont
  {{McClintock}}}, \bibinfo {author} {\bibfnamefont {R.}~\bibnamefont
  {{Narayan}}}, \bibinfo {author} {\bibfnamefont {C.~D.}\ \bibnamefont
  {{Bailyn}}}, \bibinfo {author} {\bibfnamefont {J.~D.}\ \bibnamefont
  {{Hartman}}}, \bibinfo {author} {\bibfnamefont {L.}~\bibnamefont {{Macri}}},
  \bibinfo {author} {\bibfnamefont {J.}~\bibnamefont {{Liu}}}, \bibinfo
  {author} {\bibfnamefont {W.}~\bibnamefont {{Pietsch}}}, \bibinfo {author}
  {\bibfnamefont {R.~A.}\ \bibnamefont {{Remillard}}}, \bibinfo {author}
  {\bibfnamefont {A.}~\bibnamefont {{Shporer}}},\ and\ \bibinfo {author}
  {\bibfnamefont {T.}~\bibnamefont {{Mazeh}}},\ }\bibfield  {title} {\bibinfo
  {title} {{A 15.65-solar-mass black hole in an eclipsing binary in the nearby
  spiral galaxy M 33}},\ }\href {https://doi.org/10.1038/nature06218}
  {\bibfield  {journal} {\bibinfo  {journal} {\nat}\ }\textbf {\bibinfo
  {volume} {449}},\ \bibinfo {pages} {872} (\bibinfo {year} {2007})},\ \Eprint
  {https://arxiv.org/abs/0710.3165} {arXiv:0710.3165 [astro-ph]} \BibitemShut
  {NoStop}%
\bibitem [{\citenamefont {{Carpano}}\ \emph {et~al.}(2007)\citenamefont
  {{Carpano}}, \citenamefont {{Pollock}}, \citenamefont {{Wilms}},
  \citenamefont {{Ehle}},\ and\ \citenamefont
  {{Schirmer}}}]{2007A&A...461L...9C}%
  \BibitemOpen
  \bibfield  {author} {\bibinfo {author} {\bibfnamefont {S.}~\bibnamefont
  {{Carpano}}}, \bibinfo {author} {\bibfnamefont {A.~M.~T.}\ \bibnamefont
  {{Pollock}}}, \bibinfo {author} {\bibfnamefont {J.}~\bibnamefont {{Wilms}}},
  \bibinfo {author} {\bibfnamefont {M.}~\bibnamefont {{Ehle}}},\ and\ \bibinfo
  {author} {\bibfnamefont {M.}~\bibnamefont {{Schirmer}}},\ }\bibfield  {title}
  {\bibinfo {title} {{A Wolf-Rayet/black-hole X-ray binary candidate in NGC
  300}},\ }\href {https://doi.org/10.1051/0004-6361:20066527} {\bibfield
  {journal} {\bibinfo  {journal} {\aap}\ }\textbf {\bibinfo {volume} {461}},\
  \bibinfo {pages} {L9} (\bibinfo {year} {2007})},\ \Eprint
  {https://arxiv.org/abs/astro-ph/0611424} {arXiv:astro-ph/0611424 [astro-ph]}
  \BibitemShut {NoStop}%
\bibitem [{\citenamefont {{Crowther}}\ \emph
  {et~al.}(2010{\natexlab{b}})\citenamefont {{Crowther}}, \citenamefont
  {{Barnard}}, \citenamefont {{Carpano}}, \citenamefont {{Clark}},
  \citenamefont {{Dhillon}},\ and\ \citenamefont
  {{Pollock}}}]{2010MNRAS.403L..41C}%
  \BibitemOpen
  \bibfield  {author} {\bibinfo {author} {\bibfnamefont {P.~A.}\ \bibnamefont
  {{Crowther}}}, \bibinfo {author} {\bibfnamefont {R.}~\bibnamefont
  {{Barnard}}}, \bibinfo {author} {\bibfnamefont {S.}~\bibnamefont
  {{Carpano}}}, \bibinfo {author} {\bibfnamefont {J.~S.}\ \bibnamefont
  {{Clark}}}, \bibinfo {author} {\bibfnamefont {V.~S.}\ \bibnamefont
  {{Dhillon}}},\ and\ \bibinfo {author} {\bibfnamefont {A.~M.~T.}\ \bibnamefont
  {{Pollock}}},\ }\bibfield  {title} {\bibinfo {title} {{NGC 300 X-1 is a
  Wolf-Rayet/black hole binary}},\ }\href
  {https://doi.org/10.1111/j.1745-3933.2010.00811.x} {\bibfield  {journal}
  {\bibinfo  {journal} {\mnras}\ }\textbf {\bibinfo {volume} {403}},\ \bibinfo
  {pages} {L41} (\bibinfo {year} {2010}{\natexlab{b}})},\ \Eprint
  {https://arxiv.org/abs/1001.4616} {arXiv:1001.4616 [astro-ph.SR]}
  \BibitemShut {NoStop}%
\bibitem [{\citenamefont {{Orosz}}\ \emph {et~al.}(2009)\citenamefont
  {{Orosz}}, \citenamefont {{Steeghs}}, \citenamefont {{McClintock}},
  \citenamefont {{Torres}}, \citenamefont {{Bochkov}}, \citenamefont {{Gou}},
  \citenamefont {{Narayan}}, \citenamefont {{Blaschak}}, \citenamefont
  {{Levine}}, \citenamefont {{Remillard}}, \citenamefont {{Bailyn}},
  \citenamefont {{Dwyer}},\ and\ \citenamefont
  {{Buxton}}}]{2009ApJ...697..573O}%
  \BibitemOpen
  \bibfield  {author} {\bibinfo {author} {\bibfnamefont {J.~A.}\ \bibnamefont
  {{Orosz}}}, \bibinfo {author} {\bibfnamefont {D.}~\bibnamefont {{Steeghs}}},
  \bibinfo {author} {\bibfnamefont {J.~E.}\ \bibnamefont {{McClintock}}},
  \bibinfo {author} {\bibfnamefont {M.~A.~P.}\ \bibnamefont {{Torres}}},
  \bibinfo {author} {\bibfnamefont {I.}~\bibnamefont {{Bochkov}}}, \bibinfo
  {author} {\bibfnamefont {L.}~\bibnamefont {{Gou}}}, \bibinfo {author}
  {\bibfnamefont {R.}~\bibnamefont {{Narayan}}}, \bibinfo {author}
  {\bibfnamefont {M.}~\bibnamefont {{Blaschak}}}, \bibinfo {author}
  {\bibfnamefont {A.~M.}\ \bibnamefont {{Levine}}}, \bibinfo {author}
  {\bibfnamefont {R.~A.}\ \bibnamefont {{Remillard}}}, \bibinfo {author}
  {\bibfnamefont {C.~D.}\ \bibnamefont {{Bailyn}}}, \bibinfo {author}
  {\bibfnamefont {M.~M.}\ \bibnamefont {{Dwyer}}},\ and\ \bibinfo {author}
  {\bibfnamefont {M.}~\bibnamefont {{Buxton}}},\ }\bibfield  {title} {\bibinfo
  {title} {{A New Dynamical Model for the Black Hole Binary LMC X-1}},\ }\href
  {https://doi.org/10.1088/0004-637X/697/1/573} {\bibfield  {journal} {\bibinfo
   {journal} {\apj}\ }\textbf {\bibinfo {volume} {697}},\ \bibinfo {pages}
  {573} (\bibinfo {year} {2009})},\ \Eprint {https://arxiv.org/abs/0810.3447}
  {arXiv:0810.3447 [astro-ph]} \BibitemShut {NoStop}%
\bibitem [{\citenamefont {{Casares}}\ \emph {et~al.}(2012)\citenamefont
  {{Casares}}, \citenamefont {{Rib{\'o}}}, \citenamefont {{Ribas}},
  \citenamefont {{Paredes}}, \citenamefont {{Vilardell}},\ and\ \citenamefont
  {{Negueruela}}}]{2012MNRAS.421.1103C}%
  \BibitemOpen
  \bibfield  {author} {\bibinfo {author} {\bibfnamefont {J.}~\bibnamefont
  {{Casares}}}, \bibinfo {author} {\bibfnamefont {M.}~\bibnamefont
  {{Rib{\'o}}}}, \bibinfo {author} {\bibfnamefont {I.}~\bibnamefont {{Ribas}}},
  \bibinfo {author} {\bibfnamefont {J.~M.}\ \bibnamefont {{Paredes}}}, \bibinfo
  {author} {\bibfnamefont {F.}~\bibnamefont {{Vilardell}}},\ and\ \bibinfo
  {author} {\bibfnamefont {I.}~\bibnamefont {{Negueruela}}},\ }\bibfield
  {title} {\bibinfo {title} {{On the binary nature of the
  {\ensuremath{\gamma}}-ray sources AGL J2241+4454 (= MWC 656) and HESS
  J0632+057 (= MWC 148)}},\ }\href
  {https://doi.org/10.1111/j.1365-2966.2011.20368.x} {\bibfield  {journal}
  {\bibinfo  {journal} {\mnras}\ }\textbf {\bibinfo {volume} {421}},\ \bibinfo
  {pages} {1103} (\bibinfo {year} {2012})},\ \Eprint
  {https://arxiv.org/abs/1201.1726} {arXiv:1201.1726 [astro-ph.SR]}
  \BibitemShut {NoStop}%
\bibitem [{\citenamefont {{Esposito}}\ \emph {et~al.}(2013)\citenamefont
  {{Esposito}}, \citenamefont {{Israel}}, \citenamefont {{Sidoli}},
  \citenamefont {{Mapelli}}, \citenamefont {{Zampieri}},\ and\ \citenamefont
  {{Motta}}}]{2013MNRAS.436.3380E}%
  \BibitemOpen
  \bibfield  {author} {\bibinfo {author} {\bibfnamefont {P.}~\bibnamefont
  {{Esposito}}}, \bibinfo {author} {\bibfnamefont {G.~L.}\ \bibnamefont
  {{Israel}}}, \bibinfo {author} {\bibfnamefont {L.}~\bibnamefont {{Sidoli}}},
  \bibinfo {author} {\bibfnamefont {M.}~\bibnamefont {{Mapelli}}}, \bibinfo
  {author} {\bibfnamefont {L.}~\bibnamefont {{Zampieri}}},\ and\ \bibinfo
  {author} {\bibfnamefont {S.~E.}\ \bibnamefont {{Motta}}},\ }\bibfield
  {title} {\bibinfo {title} {{Discovery of a 6.4 h black hole binary in NGC
  4490}},\ }\href {https://doi.org/10.1093/mnras/stt1819} {\bibfield  {journal}
  {\bibinfo  {journal} {\mnras}\ }\textbf {\bibinfo {volume} {436}},\ \bibinfo
  {pages} {3380} (\bibinfo {year} {2013})},\ \Eprint
  {https://arxiv.org/abs/1309.6328} {arXiv:1309.6328 [astro-ph.HE]}
  \BibitemShut {NoStop}%
\bibitem [{\citenamefont {{Qiu}}\ \emph {et~al.}(2019)\citenamefont {{Qiu}},
  \citenamefont {{Soria}}, \citenamefont {{Wang}}, \citenamefont
  {{Wiktorowicz}}, \citenamefont {{Liu}}, \citenamefont {{Bai}}, \citenamefont
  {{Bogomazov}}, \citenamefont {{Di Stefano}}, \citenamefont {{Walton}},\ and\
  \citenamefont {{Xu}}}]{2019ApJ...877...57Q}%
  \BibitemOpen
  \bibfield  {author} {\bibinfo {author} {\bibfnamefont {Y.}~\bibnamefont
  {{Qiu}}}, \bibinfo {author} {\bibfnamefont {R.}~\bibnamefont {{Soria}}},
  \bibinfo {author} {\bibfnamefont {S.}~\bibnamefont {{Wang}}}, \bibinfo
  {author} {\bibfnamefont {G.}~\bibnamefont {{Wiktorowicz}}}, \bibinfo {author}
  {\bibfnamefont {J.}~\bibnamefont {{Liu}}}, \bibinfo {author} {\bibfnamefont
  {Y.}~\bibnamefont {{Bai}}}, \bibinfo {author} {\bibfnamefont
  {A.}~\bibnamefont {{Bogomazov}}}, \bibinfo {author} {\bibfnamefont
  {R.}~\bibnamefont {{Di Stefano}}}, \bibinfo {author} {\bibfnamefont {D.~J.}\
  \bibnamefont {{Walton}}},\ and\ \bibinfo {author} {\bibfnamefont
  {X.}~\bibnamefont {{Xu}}},\ }\bibfield  {title} {\bibinfo {title} {{CG X-1:
  An Eclipsing Wolf-Rayet ULX in the Circinus Galaxy}},\ }\href
  {https://doi.org/10.3847/1538-4357/ab16e7} {\bibfield  {journal} {\bibinfo
  {journal} {\apj}\ }\textbf {\bibinfo {volume} {877}},\ \bibinfo {eid} {57}
  (\bibinfo {year} {2019})},\ \Eprint {https://arxiv.org/abs/1904.01066}
  {arXiv:1904.01066 [astro-ph.HE]} \BibitemShut {NoStop}%
\bibitem [{\citenamefont {{Belczynski}}\ \emph {et~al.}(2014)\citenamefont
  {{Belczynski}}, \citenamefont {{Buonanno}}, \citenamefont {{Cantiello}},
  \citenamefont {{Fryer}}, \citenamefont {{Holz}}, \citenamefont {{Mandel}},
  \citenamefont {{Miller}},\ and\ \citenamefont
  {{Walczak}}}]{2014ApJ...789..120B}%
  \BibitemOpen
  \bibfield  {author} {\bibinfo {author} {\bibfnamefont {K.}~\bibnamefont
  {{Belczynski}}}, \bibinfo {author} {\bibfnamefont {A.}~\bibnamefont
  {{Buonanno}}}, \bibinfo {author} {\bibfnamefont {M.}~\bibnamefont
  {{Cantiello}}}, \bibinfo {author} {\bibfnamefont {C.~L.}\ \bibnamefont
  {{Fryer}}}, \bibinfo {author} {\bibfnamefont {D.~E.}\ \bibnamefont {{Holz}}},
  \bibinfo {author} {\bibfnamefont {I.}~\bibnamefont {{Mandel}}}, \bibinfo
  {author} {\bibfnamefont {M.~C.}\ \bibnamefont {{Miller}}},\ and\ \bibinfo
  {author} {\bibfnamefont {M.}~\bibnamefont {{Walczak}}},\ }\bibfield  {title}
  {\bibinfo {title} {{The Formation and Gravitational-wave Detection of Massive
  Stellar Black Hole Binaries}},\ }\href
  {https://doi.org/10.1088/0004-637X/789/2/120} {\bibfield  {journal} {\bibinfo
   {journal} {\apj}\ }\textbf {\bibinfo {volume} {789}},\ \bibinfo {eid} {120}
  (\bibinfo {year} {2014})},\ \Eprint {https://arxiv.org/abs/1403.0677}
  {arXiv:1403.0677 [astro-ph.HE]} \BibitemShut {NoStop}%
\bibitem [{\citenamefont {{White}}\ \emph {et~al.}(1995)\citenamefont
  {{White}}, \citenamefont {{Nagase}},\ and\ \citenamefont
  {{Parmar}}}]{1995xrbi.nasa....1W}%
  \BibitemOpen
  \bibfield  {author} {\bibinfo {author} {\bibfnamefont {N.~E.}\ \bibnamefont
  {{White}}}, \bibinfo {author} {\bibfnamefont {F.}~\bibnamefont {{Nagase}}},\
  and\ \bibinfo {author} {\bibfnamefont {A.~N.}\ \bibnamefont {{Parmar}}},\
  }\bibfield  {title} {\bibinfo {title} {{The properties of X-ray binaries.}},\
  }in\ \href@noop {} {\emph {\bibinfo {booktitle} {X-ray Binaries}}}\ (\bibinfo
  {year} {1995})\ pp.\ \bibinfo {pages} {1--57}\BibitemShut {NoStop}%
\bibitem [{\citenamefont {{Zhang}}\ \emph {et~al.}(2012)\citenamefont
  {{Zhang}}, \citenamefont {{Gilfanov}},\ and\ \citenamefont
  {{Bogd{\'a}n}}}]{2012A&A...546A..36Z}%
  \BibitemOpen
  \bibfield  {author} {\bibinfo {author} {\bibfnamefont {Z.}~\bibnamefont
  {{Zhang}}}, \bibinfo {author} {\bibfnamefont {M.}~\bibnamefont
  {{Gilfanov}}},\ and\ \bibinfo {author} {\bibfnamefont {{\'A}.}~\bibnamefont
  {{Bogd{\'a}n}}},\ }\bibfield  {title} {\bibinfo {title} {{Dependence of the
  low-mass X-ray binary population on stellar age}},\ }\href
  {https://doi.org/10.1051/0004-6361/201219015} {\bibfield  {journal} {\bibinfo
   {journal} {\aap}\ }\textbf {\bibinfo {volume} {546}},\ \bibinfo {eid} {A36}
  (\bibinfo {year} {2012})},\ \Eprint {https://arxiv.org/abs/1202.2331}
  {arXiv:1202.2331 [astro-ph.HE]} \BibitemShut {NoStop}%
\bibitem [{\citenamefont {Dehnen}\ and\ \citenamefont
  {King}(2006)}]{10.1111/j.1745-3933.2005.00132.x}%
  \BibitemOpen
  \bibfield  {author} {\bibinfo {author} {\bibfnamefont {W.}~\bibnamefont
  {Dehnen}}\ and\ \bibinfo {author} {\bibfnamefont {A.}~\bibnamefont {King}},\
  }\bibfield  {title} {\bibinfo {title} {{Probing dark matter with X-ray
  binaries}},\ }\href {https://doi.org/10.1111/j.1745-3933.2005.00132.x}
  {\bibfield  {journal} {\bibinfo  {journal} {Monthly Notices of the Royal
  Astronomical Society: Letters}\ }\textbf {\bibinfo {volume} {367}},\ \bibinfo
  {pages} {L29} (\bibinfo {year} {2006})},\ \Eprint
  {https://arxiv.org/abs/https://academic.oup.com/mnrasl/article-pdf/367/1/L29/4032300/367-1-L29.pdf}
  {https://academic.oup.com/mnrasl/article-pdf/367/1/L29/4032300/367-1-L29.pdf}
  \BibitemShut {NoStop}%
\bibitem [{\citenamefont {Chan}\ and\ \citenamefont {Lee}(2023)}]{Chan_2023}%
  \BibitemOpen
  \bibfield  {author} {\bibinfo {author} {\bibfnamefont {M.~H.}\ \bibnamefont
  {Chan}}\ and\ \bibinfo {author} {\bibfnamefont {C.~M.}\ \bibnamefont {Lee}},\
  }\bibfield  {title} {\bibinfo {title} {Indirect evidence for dark matter
  density spikes around stellar-mass black holes},\ }\href
  {https://doi.org/10.3847/2041-8213/acaafa} {\bibfield  {journal} {\bibinfo
  {journal} {The Astrophysical Journal Letters}\ }\textbf {\bibinfo {volume}
  {943}},\ \bibinfo {pages} {L11} (\bibinfo {year} {2023})}\BibitemShut
  {NoStop}%
\bibitem [{\citenamefont {{Hut}}\ \emph {et~al.}(1992)\citenamefont {{Hut}},
  \citenamefont {{McMillan}}, \citenamefont {{Goodman}}, \citenamefont
  {{Mateo}}, \citenamefont {{Phinney}}, \citenamefont {{Pryor}}, \citenamefont
  {{Richer}}, \citenamefont {{Verbunt}},\ and\ \citenamefont
  {{Weinberg}}}]{1992PASP..104..981H}%
  \BibitemOpen
  \bibfield  {author} {\bibinfo {author} {\bibfnamefont {P.}~\bibnamefont
  {{Hut}}}, \bibinfo {author} {\bibfnamefont {S.}~\bibnamefont {{McMillan}}},
  \bibinfo {author} {\bibfnamefont {J.}~\bibnamefont {{Goodman}}}, \bibinfo
  {author} {\bibfnamefont {M.}~\bibnamefont {{Mateo}}}, \bibinfo {author}
  {\bibfnamefont {E.~S.}\ \bibnamefont {{Phinney}}}, \bibinfo {author}
  {\bibfnamefont {C.}~\bibnamefont {{Pryor}}}, \bibinfo {author} {\bibfnamefont
  {H.~B.}\ \bibnamefont {{Richer}}}, \bibinfo {author} {\bibfnamefont
  {F.}~\bibnamefont {{Verbunt}}},\ and\ \bibinfo {author} {\bibfnamefont
  {M.}~\bibnamefont {{Weinberg}}},\ }\bibfield  {title} {\bibinfo {title}
  {{Binaries in Globular Clusters}},\ }\href {https://doi.org/10.1086/133085}
  {\bibfield  {journal} {\bibinfo  {journal} {\pasp}\ }\textbf {\bibinfo
  {volume} {104}},\ \bibinfo {pages} {981} (\bibinfo {year}
  {1992})}\BibitemShut {NoStop}%
\bibitem [{\citenamefont {{Brodie}}\ and\ \citenamefont
  {{Strader}}(2006)}]{2006ARA&A..44..193B}%
  \BibitemOpen
  \bibfield  {author} {\bibinfo {author} {\bibfnamefont {J.~P.}\ \bibnamefont
  {{Brodie}}}\ and\ \bibinfo {author} {\bibfnamefont {J.}~\bibnamefont
  {{Strader}}},\ }\bibfield  {title} {\bibinfo {title} {{Extragalactic Globular
  Clusters and Galaxy Formation}},\ }\href
  {https://doi.org/10.1146/annurev.astro.44.051905.092441} {\bibfield
  {journal} {\bibinfo  {journal} {\araa}\ }\textbf {\bibinfo {volume} {44}},\
  \bibinfo {pages} {193} (\bibinfo {year} {2006})},\ \Eprint
  {https://arxiv.org/abs/astro-ph/0602601} {arXiv:astro-ph/0602601 [astro-ph]}
  \BibitemShut {NoStop}%
\bibitem [{\citenamefont {{Fabian}}\ \emph {et~al.}(1975)\citenamefont
  {{Fabian}}, \citenamefont {{Pringle}},\ and\ \citenamefont
  {{Rees}}}]{1975MNRAS.172P..15F}%
  \BibitemOpen
  \bibfield  {author} {\bibinfo {author} {\bibfnamefont {A.~C.}\ \bibnamefont
  {{Fabian}}}, \bibinfo {author} {\bibfnamefont {J.~E.}\ \bibnamefont
  {{Pringle}}},\ and\ \bibinfo {author} {\bibfnamefont {M.~J.}\ \bibnamefont
  {{Rees}}},\ }\bibfield  {title} {\bibinfo {title} {{Tidal capture formation
  of binary systems and X-ray sources in globular clusters.}},\ }\href
  {https://doi.org/10.1093/mnras/172.1.15P} {\bibfield  {journal} {\bibinfo
  {journal} {\mnras}\ }\textbf {\bibinfo {volume} {172}},\ \bibinfo {pages}
  {15} (\bibinfo {year} {1975})}\BibitemShut {NoStop}%
\bibitem [{\citenamefont {{Hills}}(1976)}]{1976MNRAS.175P...1H}%
  \BibitemOpen
  \bibfield  {author} {\bibinfo {author} {\bibfnamefont {J.~G.}\ \bibnamefont
  {{Hills}}},\ }\bibfield  {title} {\bibinfo {title} {{The formation of
  binaries containing black holes by the exchange of companions and the X-ray
  sources in globular clusters.}},\ }\href
  {https://doi.org/10.1093/mnras/175.1.1P} {\bibfield  {journal} {\bibinfo
  {journal} {\mnras}\ }\textbf {\bibinfo {volume} {175}},\ \bibinfo {pages}
  {1P} (\bibinfo {year} {1976})}\BibitemShut {NoStop}%
\bibitem [{\citenamefont {{Ivanova}}\ \emph {et~al.}(2005)\citenamefont
  {{Ivanova}}, \citenamefont {{Rasio}}, \citenamefont {{Lombardi}},
  \citenamefont {{Dooley}},\ and\ \citenamefont
  {{Proulx}}}]{2005ApJ...621L.109I}%
  \BibitemOpen
  \bibfield  {author} {\bibinfo {author} {\bibfnamefont {N.}~\bibnamefont
  {{Ivanova}}}, \bibinfo {author} {\bibfnamefont {F.~A.}\ \bibnamefont
  {{Rasio}}}, \bibinfo {author} {\bibfnamefont {J.}~\bibnamefont {{Lombardi}},
  \bibfnamefont {J.~C.}}, \bibinfo {author} {\bibfnamefont {K.~L.}\
  \bibnamefont {{Dooley}}},\ and\ \bibinfo {author} {\bibfnamefont {Z.~F.}\
  \bibnamefont {{Proulx}}},\ }\bibfield  {title} {\bibinfo {title} {{Formation
  of Ultracompact X-Ray Binaries in Dense Star Clusters}},\ }\href
  {https://doi.org/10.1086/429220} {\bibfield  {journal} {\bibinfo  {journal}
  {\apjl}\ }\textbf {\bibinfo {volume} {621}},\ \bibinfo {pages} {L109}
  (\bibinfo {year} {2005})},\ \Eprint {https://arxiv.org/abs/astro-ph/0501617}
  {arXiv:astro-ph/0501617 [astro-ph]} \BibitemShut {NoStop}%
\bibitem [{\citenamefont {{Ivanova}}\ \emph {et~al.}(2012)\citenamefont
  {{Ivanova}}, \citenamefont {{Fragos}}, \citenamefont {{Kim}}, \citenamefont
  {{Fabbiano}}, \citenamefont {{Avendano Nandez}}, \citenamefont {{Lombardi}},
  \citenamefont {{Sivakoff}}, \citenamefont {{Voss}},\ and\ \citenamefont
  {{Jord{\'a}n}}}]{2012ApJ...760L..24I}%
  \BibitemOpen
  \bibfield  {author} {\bibinfo {author} {\bibfnamefont {N.}~\bibnamefont
  {{Ivanova}}}, \bibinfo {author} {\bibfnamefont {T.}~\bibnamefont {{Fragos}}},
  \bibinfo {author} {\bibfnamefont {D.~W.}\ \bibnamefont {{Kim}}}, \bibinfo
  {author} {\bibfnamefont {G.}~\bibnamefont {{Fabbiano}}}, \bibinfo {author}
  {\bibfnamefont {J.~L.}\ \bibnamefont {{Avendano Nandez}}}, \bibinfo {author}
  {\bibfnamefont {J.~C.}\ \bibnamefont {{Lombardi}}}, \bibinfo {author}
  {\bibfnamefont {G.~R.}\ \bibnamefont {{Sivakoff}}}, \bibinfo {author}
  {\bibfnamefont {R.}~\bibnamefont {{Voss}}},\ and\ \bibinfo {author}
  {\bibfnamefont {A.}~\bibnamefont {{Jord{\'a}n}}},\ }\bibfield  {title}
  {\bibinfo {title} {{On the Origin of the Metallicity Dependence in
  Dynamically formed Extragalactic Low-mass X-Ray Binaries}},\ }\href
  {https://doi.org/10.1088/2041-8205/760/2/L24} {\bibfield  {journal} {\bibinfo
   {journal} {\apjl}\ }\textbf {\bibinfo {volume} {760}},\ \bibinfo {eid} {L24}
  (\bibinfo {year} {2012})},\ \Eprint {https://arxiv.org/abs/1208.5972}
  {arXiv:1208.5972 [astro-ph.HE]} \BibitemShut {NoStop}%
\bibitem [{\citenamefont {{Verbunt}}(2002)}]{2002ASPC..265..289V}%
  \BibitemOpen
  \bibfield  {author} {\bibinfo {author} {\bibfnamefont {F.}~\bibnamefont
  {{Verbunt}}},\ }\bibfield  {title} {\bibinfo {title} {{X-ray Sources in
  {\ensuremath{\omega}} Centauri and Other Globular Clusters}},\ }in\ \href
  {https://doi.org/10.48550/arXiv.astro-ph/0111441} {\emph {\bibinfo
  {booktitle} {Omega Centauri, A Unique Window into Astrophysics}}},\ \bibinfo
  {series} {Astronomical Society of the Pacific Conference Series}, Vol.\
  \bibinfo {volume} {265},\ \bibinfo {editor} {edited by\ \bibinfo {editor}
  {\bibfnamefont {F.}~\bibnamefont {{van Leeuwen}}}, \bibinfo {editor}
  {\bibfnamefont {J.~D.}\ \bibnamefont {{Hughes}}},\ and\ \bibinfo {editor}
  {\bibfnamefont {G.}~\bibnamefont {{Piotto}}}}\ (\bibinfo {year} {2002})\ p.\
  \bibinfo {pages} {289},\ \Eprint {https://arxiv.org/abs/astro-ph/0111441}
  {arXiv:astro-ph/0111441 [astro-ph]} \BibitemShut {NoStop}%
\bibitem [{\citenamefont {{Angelini}}\ \emph {et~al.}(2001)\citenamefont
  {{Angelini}}, \citenamefont {{Loewenstein}},\ and\ \citenamefont
  {{Mushotzky}}}]{2001ApJ...557L..35A}%
  \BibitemOpen
  \bibfield  {author} {\bibinfo {author} {\bibfnamefont {L.}~\bibnamefont
  {{Angelini}}}, \bibinfo {author} {\bibfnamefont {M.}~\bibnamefont
  {{Loewenstein}}},\ and\ \bibinfo {author} {\bibfnamefont {R.~F.}\
  \bibnamefont {{Mushotzky}}},\ }\bibfield  {title} {\bibinfo {title} {{The
  X-Ray Globular Cluster Population in NGC 1399}},\ }\href
  {https://doi.org/10.1086/323026} {\bibfield  {journal} {\bibinfo  {journal}
  {\apjl}\ }\textbf {\bibinfo {volume} {557}},\ \bibinfo {pages} {L35}
  (\bibinfo {year} {2001})},\ \Eprint {https://arxiv.org/abs/astro-ph/0107362}
  {arXiv:astro-ph/0107362 [astro-ph]} \BibitemShut {NoStop}%
\bibitem [{\citenamefont {{Sarazin}}\ \emph {et~al.}(2001)\citenamefont
  {{Sarazin}}, \citenamefont {{Irwin}},\ and\ \citenamefont
  {{Bregman}}}]{2001ApJ...556..533S}%
  \BibitemOpen
  \bibfield  {author} {\bibinfo {author} {\bibfnamefont {C.~L.}\ \bibnamefont
  {{Sarazin}}}, \bibinfo {author} {\bibfnamefont {J.~A.}\ \bibnamefont
  {{Irwin}}},\ and\ \bibinfo {author} {\bibfnamefont {J.~N.}\ \bibnamefont
  {{Bregman}}},\ }\bibfield  {title} {\bibinfo {title} {{Chandra X-Ray
  Observations of the X-Ray Faint Elliptical Galaxy NGC 4697}},\ }\href
  {https://doi.org/10.1086/321618} {\bibfield  {journal} {\bibinfo  {journal}
  {\apj}\ }\textbf {\bibinfo {volume} {556}},\ \bibinfo {pages} {533} (\bibinfo
  {year} {2001})},\ \Eprint {https://arxiv.org/abs/astro-ph/0104070}
  {arXiv:astro-ph/0104070 [astro-ph]} \BibitemShut {NoStop}%
\bibitem [{\citenamefont {{Kundu}}\ \emph {et~al.}(2003)\citenamefont
  {{Kundu}}, \citenamefont {{Maccarone}},\ and\ \citenamefont
  {{Zepf}}}]{2003AAS...203.3104K}%
  \BibitemOpen
  \bibfield  {author} {\bibinfo {author} {\bibfnamefont {A.}~\bibnamefont
  {{Kundu}}}, \bibinfo {author} {\bibfnamefont {T.~J.}\ \bibnamefont
  {{Maccarone}}},\ and\ \bibinfo {author} {\bibfnamefont {S.~E.}\ \bibnamefont
  {{Zepf}}},\ }\bibfield  {title} {\bibinfo {title} {{Low Mass X-ray Binaries
  and the Globular Cluster Connection}},\ }in\ \href@noop {} {\emph {\bibinfo
  {booktitle} {American Astronomical Society Meeting Abstracts}}},\ \bibinfo
  {series} {American Astronomical Society Meeting Abstracts}, Vol.\ \bibinfo
  {volume} {203}\ (\bibinfo {year} {2003})\ p.\ \bibinfo {pages}
  {31.04}\BibitemShut {NoStop}%
\bibitem [{\citenamefont {{Kim}}\ \emph {et~al.}(2009)\citenamefont {{Kim}},
  \citenamefont {{Fabbiano}}, \citenamefont {{Brassington}}, \citenamefont
  {{Fragos}}, \citenamefont {{Kalogera}}, \citenamefont {{Zezas}},
  \citenamefont {{Jord{\'a}n}}, \citenamefont {{Sivakoff}}, \citenamefont
  {{Kundu}}, \citenamefont {{Zepf}}, \citenamefont {{Angelini}}, \citenamefont
  {{Davies}}, \citenamefont {{Gallagher}}, \citenamefont {{Juett}},
  \citenamefont {{King}}, \citenamefont {{Pellegrini}}, \citenamefont
  {{Sarazin}},\ and\ \citenamefont {{Trinchieri}}}]{2009ApJ...703..829K}%
  \BibitemOpen
  \bibfield  {author} {\bibinfo {author} {\bibfnamefont {D.~W.}\ \bibnamefont
  {{Kim}}}, \bibinfo {author} {\bibfnamefont {G.}~\bibnamefont {{Fabbiano}}},
  \bibinfo {author} {\bibfnamefont {N.~J.}\ \bibnamefont {{Brassington}}},
  \bibinfo {author} {\bibfnamefont {T.}~\bibnamefont {{Fragos}}}, \bibinfo
  {author} {\bibfnamefont {V.}~\bibnamefont {{Kalogera}}}, \bibinfo {author}
  {\bibfnamefont {A.}~\bibnamefont {{Zezas}}}, \bibinfo {author} {\bibfnamefont
  {A.}~\bibnamefont {{Jord{\'a}n}}}, \bibinfo {author} {\bibfnamefont {G.~R.}\
  \bibnamefont {{Sivakoff}}}, \bibinfo {author} {\bibfnamefont
  {A.}~\bibnamefont {{Kundu}}}, \bibinfo {author} {\bibfnamefont {S.~E.}\
  \bibnamefont {{Zepf}}}, \bibinfo {author} {\bibfnamefont {L.}~\bibnamefont
  {{Angelini}}}, \bibinfo {author} {\bibfnamefont {R.~L.}\ \bibnamefont
  {{Davies}}}, \bibinfo {author} {\bibfnamefont {J.~S.}\ \bibnamefont
  {{Gallagher}}}, \bibinfo {author} {\bibfnamefont {A.~M.}\ \bibnamefont
  {{Juett}}}, \bibinfo {author} {\bibfnamefont {A.~R.}\ \bibnamefont {{King}}},
  \bibinfo {author} {\bibfnamefont {S.}~\bibnamefont {{Pellegrini}}}, \bibinfo
  {author} {\bibfnamefont {C.~L.}\ \bibnamefont {{Sarazin}}},\ and\ \bibinfo
  {author} {\bibfnamefont {G.}~\bibnamefont {{Trinchieri}}},\ }\bibfield
  {title} {\bibinfo {title} {{Comparing GC and Field LMXBs in Elliptical
  Galaxies with Deep Chandra and Hubble Data}},\ }\href
  {https://doi.org/10.1088/0004-637X/703/1/829} {\bibfield  {journal} {\bibinfo
   {journal} {\apj}\ }\textbf {\bibinfo {volume} {703}},\ \bibinfo {pages}
  {829} (\bibinfo {year} {2009})},\ \Eprint {https://arxiv.org/abs/0902.2343}
  {arXiv:0902.2343 [astro-ph.CO]} \BibitemShut {NoStop}%
\bibitem [{\citenamefont {{in't Zand}}\ \emph {et~al.}(2007)\citenamefont
  {{in't Zand}}, \citenamefont {{Jonker}},\ and\ \citenamefont
  {{Markwardt}}}]{2007A&A...465..953I}%
  \BibitemOpen
  \bibfield  {author} {\bibinfo {author} {\bibfnamefont {J.~J.~M.}\
  \bibnamefont {{in't Zand}}}, \bibinfo {author} {\bibfnamefont {P.~G.}\
  \bibnamefont {{Jonker}}},\ and\ \bibinfo {author} {\bibfnamefont {C.~B.}\
  \bibnamefont {{Markwardt}}},\ }\bibfield  {title} {\bibinfo {title} {{Six new
  candidate ultracompact X-ray binaries}},\ }\href
  {https://doi.org/10.1051/0004-6361:20066678} {\bibfield  {journal} {\bibinfo
  {journal} {\aap}\ }\textbf {\bibinfo {volume} {465}},\ \bibinfo {pages} {953}
  (\bibinfo {year} {2007})},\ \Eprint {https://arxiv.org/abs/astro-ph/0701810}
  {arXiv:astro-ph/0701810 [astro-ph]} \BibitemShut {NoStop}%
\bibitem [{\citenamefont {{Bahramian}}\ \emph {et~al.}(2017)\citenamefont
  {{Bahramian}}, \citenamefont {{Heinke}}, \citenamefont {{Tudor}},
  \citenamefont {{Miller-Jones}}, \citenamefont {{Bogdanov}}, \citenamefont
  {{Maccarone}}, \citenamefont {{Knigge}}, \citenamefont {{Sivakoff}},
  \citenamefont {{Chomiuk}}, \citenamefont {{Strader}}, \citenamefont
  {{Garcia}},\ and\ \citenamefont {{Kallman}}}]{2017MNRAS.467.2199B}%
  \BibitemOpen
  \bibfield  {author} {\bibinfo {author} {\bibfnamefont {A.}~\bibnamefont
  {{Bahramian}}}, \bibinfo {author} {\bibfnamefont {C.~O.}\ \bibnamefont
  {{Heinke}}}, \bibinfo {author} {\bibfnamefont {V.}~\bibnamefont {{Tudor}}},
  \bibinfo {author} {\bibfnamefont {J.~C.~A.}\ \bibnamefont {{Miller-Jones}}},
  \bibinfo {author} {\bibfnamefont {S.}~\bibnamefont {{Bogdanov}}}, \bibinfo
  {author} {\bibfnamefont {T.~J.}\ \bibnamefont {{Maccarone}}}, \bibinfo
  {author} {\bibfnamefont {C.}~\bibnamefont {{Knigge}}}, \bibinfo {author}
  {\bibfnamefont {G.~R.}\ \bibnamefont {{Sivakoff}}}, \bibinfo {author}
  {\bibfnamefont {L.}~\bibnamefont {{Chomiuk}}}, \bibinfo {author}
  {\bibfnamefont {J.}~\bibnamefont {{Strader}}}, \bibinfo {author}
  {\bibfnamefont {J.~A.}\ \bibnamefont {{Garcia}}},\ and\ \bibinfo {author}
  {\bibfnamefont {T.}~\bibnamefont {{Kallman}}},\ }\bibfield  {title} {\bibinfo
  {title} {{The ultracompact nature of the black hole candidate X-ray binary 47
  Tuc X9}},\ }\href {https://doi.org/10.1093/mnras/stx166} {\bibfield
  {journal} {\bibinfo  {journal} {\mnras}\ }\textbf {\bibinfo {volume} {467}},\
  \bibinfo {pages} {2199} (\bibinfo {year} {2017})},\ \Eprint
  {https://arxiv.org/abs/1702.02167} {arXiv:1702.02167 [astro-ph.HE]}
  \BibitemShut {NoStop}%
\bibitem [{\citenamefont {{Hopman}}\ and\ \citenamefont {{Portegies
  Zwart}}(2005)}]{2005MNRAS.363L..56H}%
  \BibitemOpen
  \bibfield  {author} {\bibinfo {author} {\bibfnamefont {C.}~\bibnamefont
  {{Hopman}}}\ and\ \bibinfo {author} {\bibfnamefont {S.}~\bibnamefont
  {{Portegies Zwart}}},\ }\bibfield  {title} {\bibinfo {title} {{Gravitational
  waves from remnants of ultraluminous X-ray sources}},\ }\href
  {https://doi.org/10.1111/j.1745-3933.2005.00083.x} {\bibfield  {journal}
  {\bibinfo  {journal} {\mnras}\ }\textbf {\bibinfo {volume} {363}},\ \bibinfo
  {pages} {L56} (\bibinfo {year} {2005})},\ \Eprint
  {https://arxiv.org/abs/astro-ph/0506181} {arXiv:astro-ph/0506181 [astro-ph]}
  \BibitemShut {NoStop}%
\bibitem [{\citenamefont {Kremer}\ \emph {et~al.}(2018)\citenamefont {Kremer},
  \citenamefont {Chatterjee}, \citenamefont {Breivik}, \citenamefont
  {Rodriguez}, \citenamefont {Larson},\ and\ \citenamefont
  {Rasio}}]{PhysRevLett.120.191103}%
  \BibitemOpen
  \bibfield  {author} {\bibinfo {author} {\bibfnamefont {K.}~\bibnamefont
  {Kremer}}, \bibinfo {author} {\bibfnamefont {S.}~\bibnamefont {Chatterjee}},
  \bibinfo {author} {\bibfnamefont {K.}~\bibnamefont {Breivik}}, \bibinfo
  {author} {\bibfnamefont {C.~L.}\ \bibnamefont {Rodriguez}}, \bibinfo {author}
  {\bibfnamefont {S.~L.}\ \bibnamefont {Larson}},\ and\ \bibinfo {author}
  {\bibfnamefont {F.~A.}\ \bibnamefont {Rasio}},\ }\bibfield  {title} {\bibinfo
  {title} {Lisa sources in milky way globular clusters},\ }\href
  {https://doi.org/10.1103/PhysRevLett.120.191103} {\bibfield  {journal}
  {\bibinfo  {journal} {Phys. Rev. Lett.}\ }\textbf {\bibinfo {volume} {120}},\
  \bibinfo {pages} {191103} (\bibinfo {year} {2018})}\BibitemShut {NoStop}%
\bibitem [{\citenamefont {Chen}\ \emph {et~al.}(2020)\citenamefont {Chen},
  \citenamefont {Liu},\ and\ \citenamefont {Wang}}]{Chen_2020}%
  \BibitemOpen
  \bibfield  {author} {\bibinfo {author} {\bibfnamefont {W.-C.}\ \bibnamefont
  {Chen}}, \bibinfo {author} {\bibfnamefont {D.-D.}\ \bibnamefont {Liu}},\ and\
  \bibinfo {author} {\bibfnamefont {B.}~\bibnamefont {Wang}},\ }\bibfield
  {title} {\bibinfo {title} {Detectability of ultra-compact x-ray binaries as
  lisa sources},\ }\href {https://doi.org/10.3847/2041-8213/abae66} {\bibfield
  {journal} {\bibinfo  {journal} {The Astrophysical Journal Letters}\ }\textbf
  {\bibinfo {volume} {900}},\ \bibinfo {pages} {L8} (\bibinfo {year}
  {2020})}\BibitemShut {NoStop}%
\bibitem [{\citenamefont {{van den Heuvel}}(1975)}]{1975ApJ...198L.109V}%
  \BibitemOpen
  \bibfield  {author} {\bibinfo {author} {\bibfnamefont {E.~P.~J.}\
  \bibnamefont {{van den Heuvel}}},\ }\bibfield  {title} {\bibinfo {title}
  {{Modes of mass transfer and classes of binary X-ray sources.}},\ }\href
  {https://doi.org/10.1086/181824} {\bibfield  {journal} {\bibinfo  {journal}
  {\apjl}\ }\textbf {\bibinfo {volume} {198}},\ \bibinfo {pages} {L109}
  (\bibinfo {year} {1975})}\BibitemShut {NoStop}%
\bibitem [{\citenamefont {{Podsiadlowski}}\ \emph {et~al.}(2001)\citenamefont
  {{Podsiadlowski}}, \citenamefont {{Rappaport}},\ and\ \citenamefont
  {{Pfahl}}}]{2001ASSL..264..355P}%
  \BibitemOpen
  \bibfield  {author} {\bibinfo {author} {\bibfnamefont {P.}~\bibnamefont
  {{Podsiadlowski}}}, \bibinfo {author} {\bibfnamefont {S.}~\bibnamefont
  {{Rappaport}}},\ and\ \bibinfo {author} {\bibfnamefont {E.}~\bibnamefont
  {{Pfahl}}},\ }\bibfield  {title} {\bibinfo {title} {{Binary Population
  Synthesis: Low- and Intermediate- Mass X-Ray Binaries}},\ }in\ \href
  {https://doi.org/10.1007/978-94-015-9723-4_26} {\emph {\bibinfo {booktitle}
  {The Influence of Binaries on Stellar Population Studies}}},\ \bibinfo
  {series} {Astrophysics and Space Science Library}, Vol.\ \bibinfo {volume}
  {264},\ \bibinfo {editor} {edited by\ \bibinfo {editor} {\bibfnamefont
  {D.}~\bibnamefont {{Vanbeveren}}}}\ (\bibinfo {year} {2001})\ p.\ \bibinfo
  {pages} {355},\ \Eprint {https://arxiv.org/abs/astro-ph/0109386}
  {arXiv:astro-ph/0109386 [astro-ph]} \BibitemShut {NoStop}%
\bibitem [{\citenamefont {{Middleton}}\ \emph {et~al.}(2015)\citenamefont
  {{Middleton}}, \citenamefont {{Heil}}, \citenamefont {{Pintore}},
  \citenamefont {{Walton}},\ and\ \citenamefont
  {{Roberts}}}]{2015MNRAS.447.3243M}%
  \BibitemOpen
  \bibfield  {author} {\bibinfo {author} {\bibfnamefont {M.~J.}\ \bibnamefont
  {{Middleton}}}, \bibinfo {author} {\bibfnamefont {L.}~\bibnamefont {{Heil}}},
  \bibinfo {author} {\bibfnamefont {F.}~\bibnamefont {{Pintore}}}, \bibinfo
  {author} {\bibfnamefont {D.~J.}\ \bibnamefont {{Walton}}},\ and\ \bibinfo
  {author} {\bibfnamefont {T.~P.}\ \bibnamefont {{Roberts}}},\ }\bibfield
  {title} {\bibinfo {title} {{A spectral-timing model for ULXs in the
  supercritical regime}},\ }\href {https://doi.org/10.1093/mnras/stu2644}
  {\bibfield  {journal} {\bibinfo  {journal} {\mnras}\ }\textbf {\bibinfo
  {volume} {447}},\ \bibinfo {pages} {3243} (\bibinfo {year} {2015})},\ \Eprint
  {https://arxiv.org/abs/1412.4532} {arXiv:1412.4532 [astro-ph.HE]}
  \BibitemShut {NoStop}%
\bibitem [{\citenamefont {{Mondal}}\ and\ \citenamefont
  {{Mukhopadhyay}}(2019)}]{2019MNRAS.482L..24M}%
  \BibitemOpen
  \bibfield  {author} {\bibinfo {author} {\bibfnamefont {T.}~\bibnamefont
  {{Mondal}}}\ and\ \bibinfo {author} {\bibfnamefont {B.}~\bibnamefont
  {{Mukhopadhyay}}},\ }\bibfield  {title} {\bibinfo {title} {{Ultraluminous
  X-ray sources as magnetically powered sub-Eddington advective accretion flows
  around stellar mass black holes}},\ }\href
  {https://doi.org/10.1093/mnrasl/sly165} {\bibfield  {journal} {\bibinfo
  {journal} {\mnras}\ }\textbf {\bibinfo {volume} {482}},\ \bibinfo {pages}
  {L24} (\bibinfo {year} {2019})},\ \Eprint {https://arxiv.org/abs/1808.10461}
  {arXiv:1808.10461 [astro-ph.HE]} \BibitemShut {NoStop}%
\bibitem [{\citenamefont {{Middleton}}\ \emph {et~al.}(2019)\citenamefont
  {{Middleton}}, \citenamefont {{Fragile}}, \citenamefont {{Ingram}},\ and\
  \citenamefont {{Roberts}}}]{2019MNRAS.489..282M}%
  \BibitemOpen
  \bibfield  {author} {\bibinfo {author} {\bibfnamefont {M.~J.}\ \bibnamefont
  {{Middleton}}}, \bibinfo {author} {\bibfnamefont {P.~C.}\ \bibnamefont
  {{Fragile}}}, \bibinfo {author} {\bibfnamefont {A.}~\bibnamefont
  {{Ingram}}},\ and\ \bibinfo {author} {\bibfnamefont {T.~P.}\ \bibnamefont
  {{Roberts}}},\ }\bibfield  {title} {\bibinfo {title} {{The Lense-Thirring
  timing-accretion plane for ULXs}},\ }\href
  {https://doi.org/10.1093/mnras/stz2005} {\bibfield  {journal} {\bibinfo
  {journal} {\mnras}\ }\textbf {\bibinfo {volume} {489}},\ \bibinfo {pages}
  {282} (\bibinfo {year} {2019})},\ \Eprint {https://arxiv.org/abs/1905.02731}
  {arXiv:1905.02731 [astro-ph.HE]} \BibitemShut {NoStop}%
\bibitem [{\citenamefont {{Das}}\ \emph {et~al.}(2021)\citenamefont {{Das}},
  \citenamefont {{Nandi}}, \citenamefont {{Agrawal}}, \citenamefont
  {{Dihingia}},\ and\ \citenamefont {{Majumder}}}]{2021MNRAS.507.2777D}%
  \BibitemOpen
  \bibfield  {author} {\bibinfo {author} {\bibfnamefont {S.}~\bibnamefont
  {{Das}}}, \bibinfo {author} {\bibfnamefont {A.}~\bibnamefont {{Nandi}}},
  \bibinfo {author} {\bibfnamefont {V.~K.}\ \bibnamefont {{Agrawal}}}, \bibinfo
  {author} {\bibfnamefont {I.~K.}\ \bibnamefont {{Dihingia}}},\ and\ \bibinfo
  {author} {\bibfnamefont {S.}~\bibnamefont {{Majumder}}},\ }\bibfield  {title}
  {\bibinfo {title} {{Relativistic viscous accretion flow model for ULX
  sources: a case study for IC 342 X-1}},\ }\href
  {https://doi.org/10.1093/mnras/stab2307} {\bibfield  {journal} {\bibinfo
  {journal} {\mnras}\ }\textbf {\bibinfo {volume} {507}},\ \bibinfo {pages}
  {2777} (\bibinfo {year} {2021})},\ \Eprint {https://arxiv.org/abs/2108.02973}
  {arXiv:2108.02973 [astro-ph.HE]} \BibitemShut {NoStop}%
\bibitem [{\citenamefont {{Fabbiano}}\ and\ \citenamefont
  {{Trinchieri}}(1987)}]{1987ApJ...315...46F}%
  \BibitemOpen
  \bibfield  {author} {\bibinfo {author} {\bibfnamefont {G.}~\bibnamefont
  {{Fabbiano}}}\ and\ \bibinfo {author} {\bibfnamefont {G.}~\bibnamefont
  {{Trinchieri}}},\ }\bibfield  {title} {\bibinfo {title} {{X-Ray Observations
  of Spiral Galaxies. II. Images and Spectral Parameters of 13 Galaxies}},\
  }\href {https://doi.org/10.1086/165113} {\bibfield  {journal} {\bibinfo
  {journal} {\apj}\ }\textbf {\bibinfo {volume} {315}},\ \bibinfo {pages} {46}
  (\bibinfo {year} {1987})}\BibitemShut {NoStop}%
\bibitem [{\citenamefont {{Long}}\ and\ \citenamefont {{van
  Speybroeck}}(1983)}]{1983adsx.conf..141L}%
  \BibitemOpen
  \bibfield  {author} {\bibinfo {author} {\bibfnamefont {K.~S.}\ \bibnamefont
  {{Long}}}\ and\ \bibinfo {author} {\bibfnamefont {L.~P.}\ \bibnamefont {{van
  Speybroeck}}},\ }\bibfield  {title} {\bibinfo {title} {{Very Bright
  Non-Nuclear Sources}},\ }in\ \href@noop {} {\emph {\bibinfo {booktitle}
  {Accretion-Driven Stellar X-ray Sources}}},\ \bibinfo {editor} {edited by\
  \bibinfo {editor} {\bibfnamefont {W.~H.~G.}\ \bibnamefont {{Lewin}}}\ and\
  \bibinfo {editor} {\bibfnamefont {E.~P.~J.}\ \bibnamefont {{van den
  Heuvel}}}}\ (\bibinfo {year} {1983})\ p.\ \bibinfo {pages} {141}\BibitemShut
  {NoStop}%
\bibitem [{\citenamefont {{Fabbiano}}(1989)}]{1989ARA&A..27...87F}%
  \BibitemOpen
  \bibfield  {author} {\bibinfo {author} {\bibfnamefont {G.}~\bibnamefont
  {{Fabbiano}}},\ }\bibfield  {title} {\bibinfo {title} {{X-rays from normal
  galaxies.}},\ }\href {https://doi.org/10.1146/annurev.aa.27.090189.000511}
  {\bibfield  {journal} {\bibinfo  {journal} {\araa}\ }\textbf {\bibinfo
  {volume} {27}},\ \bibinfo {pages} {87} (\bibinfo {year} {1989})}\BibitemShut
  {NoStop}%
\bibitem [{\citenamefont {{Miller}}\ and\ \citenamefont
  {{Colbert}}(2004)}]{2004IJMPD..13....1M}%
  \BibitemOpen
  \bibfield  {author} {\bibinfo {author} {\bibfnamefont {M.~C.}\ \bibnamefont
  {{Miller}}}\ and\ \bibinfo {author} {\bibfnamefont {E.~J.~M.}\ \bibnamefont
  {{Colbert}}},\ }\bibfield  {title} {\bibinfo {title} {{Intermediate-Mass
  Black Holes}},\ }\href {https://doi.org/10.1142/S0218271804004426} {\bibfield
   {journal} {\bibinfo  {journal} {International Journal of Modern Physics D}\
  }\textbf {\bibinfo {volume} {13}},\ \bibinfo {pages} {1} (\bibinfo {year}
  {2004})},\ \Eprint {https://arxiv.org/abs/astro-ph/0308402}
  {arXiv:astro-ph/0308402 [astro-ph]} \BibitemShut {NoStop}%
\bibitem [{\citenamefont {{Volonteri}}\ and\ \citenamefont
  {{Natarajan}}(2009)}]{2009MNRAS.400.1911V}%
  \BibitemOpen
  \bibfield  {author} {\bibinfo {author} {\bibfnamefont {M.}~\bibnamefont
  {{Volonteri}}}\ and\ \bibinfo {author} {\bibfnamefont {P.}~\bibnamefont
  {{Natarajan}}},\ }\bibfield  {title} {\bibinfo {title} {{Journey to the
  M$_{BH}$-{\ensuremath{\sigma}} relation: the fate of low-mass black holes in
  the Universe}},\ }\href {https://doi.org/10.1111/j.1365-2966.2009.15577.x}
  {\bibfield  {journal} {\bibinfo  {journal} {\mnras}\ }\textbf {\bibinfo
  {volume} {400}},\ \bibinfo {pages} {1911} (\bibinfo {year} {2009})},\ \Eprint
  {https://arxiv.org/abs/0903.2262} {arXiv:0903.2262 [astro-ph.CO]}
  \BibitemShut {NoStop}%
\bibitem [{\citenamefont {{Volonteri}}(2010)}]{2010A&ARv..18..279V}%
  \BibitemOpen
  \bibfield  {author} {\bibinfo {author} {\bibfnamefont {M.}~\bibnamefont
  {{Volonteri}}},\ }\bibfield  {title} {\bibinfo {title} {{Formation of
  supermassive black holes}},\ }\href
  {https://doi.org/10.1007/s00159-010-0029-x} {\bibfield  {journal} {\bibinfo
  {journal} {\aapr}\ }\textbf {\bibinfo {volume} {18}},\ \bibinfo {pages} {279}
  (\bibinfo {year} {2010})},\ \Eprint {https://arxiv.org/abs/1003.4404}
  {arXiv:1003.4404 [astro-ph.CO]} \BibitemShut {NoStop}%
\bibitem [{\citenamefont {{Mezcua}}(2017)}]{2017IJMPD..2630021M}%
  \BibitemOpen
  \bibfield  {author} {\bibinfo {author} {\bibfnamefont {M.}~\bibnamefont
  {{Mezcua}}},\ }\bibfield  {title} {\bibinfo {title} {{Observational evidence
  for intermediate-mass black holes}},\ }\href
  {https://doi.org/10.1142/S021827181730021X} {\bibfield  {journal} {\bibinfo
  {journal} {International Journal of Modern Physics D}\ }\textbf {\bibinfo
  {volume} {26}},\ \bibinfo {eid} {1730021} (\bibinfo {year} {2017})},\ \Eprint
  {https://arxiv.org/abs/1705.09667} {arXiv:1705.09667 [astro-ph.GA]}
  \BibitemShut {NoStop}%
\bibitem [{\citenamefont {{King}}(2004)}]{2004MNRAS.347L..18K}%
  \BibitemOpen
  \bibfield  {author} {\bibinfo {author} {\bibfnamefont {A.~R.}\ \bibnamefont
  {{King}}},\ }\bibfield  {title} {\bibinfo {title} {{Ultraluminous X-ray
  sources and star formation}},\ }\href
  {https://doi.org/10.1111/j.1365-2966.2004.07403.x} {\bibfield  {journal}
  {\bibinfo  {journal} {\mnras}\ }\textbf {\bibinfo {volume} {347}},\ \bibinfo
  {pages} {L18} (\bibinfo {year} {2004})},\ \Eprint
  {https://arxiv.org/abs/astro-ph/0309450} {arXiv:astro-ph/0309450 [astro-ph]}
  \BibitemShut {NoStop}%
\bibitem [{\citenamefont {{Madhusudhan}}\ \emph {et~al.}(2006)\citenamefont
  {{Madhusudhan}}, \citenamefont {{Justham}}, \citenamefont {{Nelson}},
  \citenamefont {{Paxton}}, \citenamefont {{Pfahl}}, \citenamefont
  {{Podsiadlowski}},\ and\ \citenamefont {{Rappaport}}}]{2006ApJ...640..918M}%
  \BibitemOpen
  \bibfield  {author} {\bibinfo {author} {\bibfnamefont {N.}~\bibnamefont
  {{Madhusudhan}}}, \bibinfo {author} {\bibfnamefont {S.}~\bibnamefont
  {{Justham}}}, \bibinfo {author} {\bibfnamefont {L.}~\bibnamefont {{Nelson}}},
  \bibinfo {author} {\bibfnamefont {B.}~\bibnamefont {{Paxton}}}, \bibinfo
  {author} {\bibfnamefont {E.}~\bibnamefont {{Pfahl}}}, \bibinfo {author}
  {\bibfnamefont {P.}~\bibnamefont {{Podsiadlowski}}},\ and\ \bibinfo {author}
  {\bibfnamefont {S.}~\bibnamefont {{Rappaport}}},\ }\bibfield  {title}
  {\bibinfo {title} {{Models of Ultraluminous X-Ray Sources with
  Intermediate-Mass Black Holes}},\ }\href {https://doi.org/10.1086/500238}
  {\bibfield  {journal} {\bibinfo  {journal} {\apj}\ }\textbf {\bibinfo
  {volume} {640}},\ \bibinfo {pages} {918} (\bibinfo {year} {2006})},\ \Eprint
  {https://arxiv.org/abs/astro-ph/0511393} {arXiv:astro-ph/0511393 [astro-ph]}
  \BibitemShut {NoStop}%
\bibitem [{\citenamefont {{Swartz}}\ \emph {et~al.}(2011)\citenamefont
  {{Swartz}}, \citenamefont {{Soria}}, \citenamefont {{Tennant}},\ and\
  \citenamefont {{Yukita}}}]{2011ApJ...741...49S}%
  \BibitemOpen
  \bibfield  {author} {\bibinfo {author} {\bibfnamefont {D.~A.}\ \bibnamefont
  {{Swartz}}}, \bibinfo {author} {\bibfnamefont {R.}~\bibnamefont {{Soria}}},
  \bibinfo {author} {\bibfnamefont {A.~F.}\ \bibnamefont {{Tennant}}},\ and\
  \bibinfo {author} {\bibfnamefont {M.}~\bibnamefont {{Yukita}}},\ }\bibfield
  {title} {\bibinfo {title} {{A Complete Sample of Ultraluminous X-ray Source
  Host Galaxies}},\ }\href {https://doi.org/10.1088/0004-637X/741/1/49}
  {\bibfield  {journal} {\bibinfo  {journal} {\apj}\ }\textbf {\bibinfo
  {volume} {741}},\ \bibinfo {eid} {49} (\bibinfo {year} {2011})},\ \Eprint
  {https://arxiv.org/abs/1108.1372} {arXiv:1108.1372 [astro-ph.HE]}
  \BibitemShut {NoStop}%
\bibitem [{\citenamefont {{Kim}}\ and\ \citenamefont
  {{Fabbiano}}(2004)}]{2004ApJ...611..846K}%
  \BibitemOpen
  \bibfield  {author} {\bibinfo {author} {\bibfnamefont {D.-W.}\ \bibnamefont
  {{Kim}}}\ and\ \bibinfo {author} {\bibfnamefont {G.}~\bibnamefont
  {{Fabbiano}}},\ }\bibfield  {title} {\bibinfo {title} {{X-Ray Luminosity
  Function and Total Luminosity of Low-Mass X-Ray Binaries in Early-Type
  Galaxies}},\ }\href {https://doi.org/10.1086/422210} {\bibfield  {journal}
  {\bibinfo  {journal} {\apj}\ }\textbf {\bibinfo {volume} {611}},\ \bibinfo
  {pages} {846} (\bibinfo {year} {2004})},\ \Eprint
  {https://arxiv.org/abs/astro-ph/0312104} {arXiv:astro-ph/0312104 [astro-ph]}
  \BibitemShut {NoStop}%
\bibitem [{\citenamefont {{Kim}}\ and\ \citenamefont
  {{Fabbiano}}(2010)}]{2010ApJ...721.1523K}%
  \BibitemOpen
  \bibfield  {author} {\bibinfo {author} {\bibfnamefont {D.-W.}\ \bibnamefont
  {{Kim}}}\ and\ \bibinfo {author} {\bibfnamefont {G.}~\bibnamefont
  {{Fabbiano}}},\ }\bibfield  {title} {\bibinfo {title} {{X-ray Properties of
  Young Early-type Galaxies. I. X-ray Luminosity Function of Low-mass X-ray
  Binaries}},\ }\href {https://doi.org/10.1088/0004-637X/721/2/1523} {\bibfield
   {journal} {\bibinfo  {journal} {\apj}\ }\textbf {\bibinfo {volume} {721}},\
  \bibinfo {pages} {1523} (\bibinfo {year} {2010})},\ \Eprint
  {https://arxiv.org/abs/1004.2427} {arXiv:1004.2427 [astro-ph.CO]}
  \BibitemShut {NoStop}%
\bibitem [{\citenamefont {{Grimm}}\ \emph
  {et~al.}(2003{\natexlab{b}})\citenamefont {{Grimm}}, \citenamefont
  {{Gilfanov}},\ and\ \citenamefont {{Sunyaev}}}]{2003MNRAS.339..793G}%
  \BibitemOpen
  \bibfield  {author} {\bibinfo {author} {\bibfnamefont {H.~J.}\ \bibnamefont
  {{Grimm}}}, \bibinfo {author} {\bibfnamefont {M.}~\bibnamefont
  {{Gilfanov}}},\ and\ \bibinfo {author} {\bibfnamefont {R.}~\bibnamefont
  {{Sunyaev}}},\ }\bibfield  {title} {\bibinfo {title} {{High-mass X-ray
  binaries as a star formation rate indicator in distant galaxies}},\ }\href
  {https://doi.org/10.1046/j.1365-8711.2003.06224.x} {\bibfield  {journal}
  {\bibinfo  {journal} {\mnras}\ }\textbf {\bibinfo {volume} {339}},\ \bibinfo
  {pages} {793} (\bibinfo {year} {2003}{\natexlab{b}})},\ \Eprint
  {https://arxiv.org/abs/astro-ph/0205371} {arXiv:astro-ph/0205371 [astro-ph]}
  \BibitemShut {NoStop}%
\bibitem [{\citenamefont {{Gilfanov}}\ \emph {et~al.}(2004)\citenamefont
  {{Gilfanov}}, \citenamefont {{Grimm}},\ and\ \citenamefont
  {{Sunyaev}}}]{2004NuPhS.132..369G}%
  \BibitemOpen
  \bibfield  {author} {\bibinfo {author} {\bibfnamefont {M.}~\bibnamefont
  {{Gilfanov}}}, \bibinfo {author} {\bibfnamefont {H.~J.}\ \bibnamefont
  {{Grimm}}},\ and\ \bibinfo {author} {\bibfnamefont {R.}~\bibnamefont
  {{Sunyaev}}},\ }\bibfield  {title} {\bibinfo {title} {{HMXB, ULX and star
  formation}},\ }\href {https://doi.org/10.1016/j.nuclphysbps.2004.04.065}
  {\bibfield  {journal} {\bibinfo  {journal} {Nuclear Physics B Proceedings
  Supplements}\ }\textbf {\bibinfo {volume} {132}},\ \bibinfo {pages} {369}
  (\bibinfo {year} {2004})},\ \Eprint {https://arxiv.org/abs/astro-ph/0309725}
  {arXiv:astro-ph/0309725 [astro-ph]} \BibitemShut {NoStop}%
\bibitem [{\citenamefont {{Bernadich}}\ \emph {et~al.}(2022)\citenamefont
  {{Bernadich}}, \citenamefont {{Schwope}}, \citenamefont {{Kovlakas}},
  \citenamefont {{Zezas}},\ and\ \citenamefont
  {{Traulsen}}}]{2022A&A...659A.188B}%
  \BibitemOpen
  \bibfield  {author} {\bibinfo {author} {\bibfnamefont {M.~C.~i.}\
  \bibnamefont {{Bernadich}}}, \bibinfo {author} {\bibfnamefont {A.~D.}\
  \bibnamefont {{Schwope}}}, \bibinfo {author} {\bibfnamefont {K.}~\bibnamefont
  {{Kovlakas}}}, \bibinfo {author} {\bibfnamefont {A.}~\bibnamefont
  {{Zezas}}},\ and\ \bibinfo {author} {\bibfnamefont {I.}~\bibnamefont
  {{Traulsen}}},\ }\bibfield  {title} {\bibinfo {title} {{An expanded
  ultraluminous X-ray source catalogue}},\ }\href
  {https://doi.org/10.1051/0004-6361/202141560} {\bibfield  {journal} {\bibinfo
   {journal} {\aap}\ }\textbf {\bibinfo {volume} {659}},\ \bibinfo {eid} {A188}
  (\bibinfo {year} {2022})},\ \Eprint {https://arxiv.org/abs/2110.14562}
  {arXiv:2110.14562 [astro-ph.HE]} \BibitemShut {NoStop}%
\bibitem [{\citenamefont {{Bachetti}}\ \emph {et~al.}(2014)\citenamefont
  {{Bachetti}}, \citenamefont {{Harrison}}, \citenamefont {{Walton}},
  \citenamefont {{Grefenstette}}, \citenamefont {{Chakrabarty}}, \citenamefont
  {{F{\"u}rst}}, \citenamefont {{Barret}}, \citenamefont {{Beloborodov}},
  \citenamefont {{Boggs}}, \citenamefont {{Christensen}}, \citenamefont
  {{Craig}}, \citenamefont {{Fabian}}, \citenamefont {{Hailey}}, \citenamefont
  {{Hornschemeier}}, \citenamefont {{Kaspi}}, \citenamefont {{Kulkarni}},
  \citenamefont {{Maccarone}}, \citenamefont {{Miller}}, \citenamefont
  {{Rana}}, \citenamefont {{Stern}}, \citenamefont {{Tendulkar}}, \citenamefont
  {{Tomsick}}, \citenamefont {{Webb}},\ and\ \citenamefont
  {{Zhang}}}]{2014Natur.514..202B}%
  \BibitemOpen
  \bibfield  {author} {\bibinfo {author} {\bibfnamefont {M.}~\bibnamefont
  {{Bachetti}}}, \bibinfo {author} {\bibfnamefont {F.~A.}\ \bibnamefont
  {{Harrison}}}, \bibinfo {author} {\bibfnamefont {D.~J.}\ \bibnamefont
  {{Walton}}}, \bibinfo {author} {\bibfnamefont {B.~W.}\ \bibnamefont
  {{Grefenstette}}}, \bibinfo {author} {\bibfnamefont {D.}~\bibnamefont
  {{Chakrabarty}}}, \bibinfo {author} {\bibfnamefont {F.}~\bibnamefont
  {{F{\"u}rst}}}, \bibinfo {author} {\bibfnamefont {D.}~\bibnamefont
  {{Barret}}}, \bibinfo {author} {\bibfnamefont {A.}~\bibnamefont
  {{Beloborodov}}}, \bibinfo {author} {\bibfnamefont {S.~E.}\ \bibnamefont
  {{Boggs}}}, \bibinfo {author} {\bibfnamefont {F.~E.}\ \bibnamefont
  {{Christensen}}}, \bibinfo {author} {\bibfnamefont {W.~W.}\ \bibnamefont
  {{Craig}}}, \bibinfo {author} {\bibfnamefont {A.~C.}\ \bibnamefont
  {{Fabian}}}, \bibinfo {author} {\bibfnamefont {C.~J.}\ \bibnamefont
  {{Hailey}}}, \bibinfo {author} {\bibfnamefont {A.}~\bibnamefont
  {{Hornschemeier}}}, \bibinfo {author} {\bibfnamefont {V.}~\bibnamefont
  {{Kaspi}}}, \bibinfo {author} {\bibfnamefont {S.~R.}\ \bibnamefont
  {{Kulkarni}}}, \bibinfo {author} {\bibfnamefont {T.}~\bibnamefont
  {{Maccarone}}}, \bibinfo {author} {\bibfnamefont {J.~M.}\ \bibnamefont
  {{Miller}}}, \bibinfo {author} {\bibfnamefont {V.}~\bibnamefont {{Rana}}},
  \bibinfo {author} {\bibfnamefont {D.}~\bibnamefont {{Stern}}}, \bibinfo
  {author} {\bibfnamefont {S.~P.}\ \bibnamefont {{Tendulkar}}}, \bibinfo
  {author} {\bibfnamefont {J.}~\bibnamefont {{Tomsick}}}, \bibinfo {author}
  {\bibfnamefont {N.~A.}\ \bibnamefont {{Webb}}},\ and\ \bibinfo {author}
  {\bibfnamefont {W.~W.}\ \bibnamefont {{Zhang}}},\ }\bibfield  {title}
  {\bibinfo {title} {{An ultraluminous X-ray source powered by an accreting
  neutron star}},\ }\href {https://doi.org/10.1038/nature13791} {\bibfield
  {journal} {\bibinfo  {journal} {\nat}\ }\textbf {\bibinfo {volume} {514}},\
  \bibinfo {pages} {202} (\bibinfo {year} {2014})},\ \Eprint
  {https://arxiv.org/abs/1410.3590} {arXiv:1410.3590 [astro-ph.HE]}
  \BibitemShut {NoStop}%
\bibitem [{\citenamefont {{F{\"u}rst}}\ \emph {et~al.}(2016)\citenamefont
  {{F{\"u}rst}}, \citenamefont {{Walton}}, \citenamefont {{Harrison}},
  \citenamefont {{Stern}}, \citenamefont {{Barret}}, \citenamefont
  {{Brightman}}, \citenamefont {{Fabian}}, \citenamefont {{Grefenstette}},
  \citenamefont {{Madsen}}, \citenamefont {{Middleton}}, \citenamefont
  {{Miller}}, \citenamefont {{Pottschmidt}}, \citenamefont {{Ptak}},
  \citenamefont {{Rana}},\ and\ \citenamefont {{Webb}}}]{2016ApJ...831L..14F}%
  \BibitemOpen
  \bibfield  {author} {\bibinfo {author} {\bibfnamefont {F.}~\bibnamefont
  {{F{\"u}rst}}}, \bibinfo {author} {\bibfnamefont {D.~J.}\ \bibnamefont
  {{Walton}}}, \bibinfo {author} {\bibfnamefont {F.~A.}\ \bibnamefont
  {{Harrison}}}, \bibinfo {author} {\bibfnamefont {D.}~\bibnamefont {{Stern}}},
  \bibinfo {author} {\bibfnamefont {D.}~\bibnamefont {{Barret}}}, \bibinfo
  {author} {\bibfnamefont {M.}~\bibnamefont {{Brightman}}}, \bibinfo {author}
  {\bibfnamefont {A.~C.}\ \bibnamefont {{Fabian}}}, \bibinfo {author}
  {\bibfnamefont {B.}~\bibnamefont {{Grefenstette}}}, \bibinfo {author}
  {\bibfnamefont {K.~K.}\ \bibnamefont {{Madsen}}}, \bibinfo {author}
  {\bibfnamefont {M.~J.}\ \bibnamefont {{Middleton}}}, \bibinfo {author}
  {\bibfnamefont {J.~M.}\ \bibnamefont {{Miller}}}, \bibinfo {author}
  {\bibfnamefont {K.}~\bibnamefont {{Pottschmidt}}}, \bibinfo {author}
  {\bibfnamefont {A.}~\bibnamefont {{Ptak}}}, \bibinfo {author} {\bibfnamefont
  {V.}~\bibnamefont {{Rana}}},\ and\ \bibinfo {author} {\bibfnamefont
  {N.}~\bibnamefont {{Webb}}},\ }\bibfield  {title} {\bibinfo {title}
  {{Discovery of Coherent Pulsations from the Ultraluminous X-Ray Source NGC
  7793 P13}},\ }\href {https://doi.org/10.3847/2041-8205/831/2/L14} {\bibfield
  {journal} {\bibinfo  {journal} {\apjl}\ }\textbf {\bibinfo {volume} {831}},\
  \bibinfo {eid} {L14} (\bibinfo {year} {2016})},\ \Eprint
  {https://arxiv.org/abs/1609.07129} {arXiv:1609.07129 [astro-ph.HE]}
  \BibitemShut {NoStop}%
\bibitem [{\citenamefont {{Israel}}\ \emph {et~al.}(2016)\citenamefont
  {{Israel}}, \citenamefont {{Esposito}}, \citenamefont {{Rodr{\'\i}guez
  Castillo}},\ and\ \citenamefont {{Sidoli}}}]{2016MNRAS.462.4371I}%
  \BibitemOpen
  \bibfield  {author} {\bibinfo {author} {\bibfnamefont {G.~L.}\ \bibnamefont
  {{Israel}}}, \bibinfo {author} {\bibfnamefont {P.}~\bibnamefont
  {{Esposito}}}, \bibinfo {author} {\bibfnamefont {G.~A.}\ \bibnamefont
  {{Rodr{\'\i}guez Castillo}}},\ and\ \bibinfo {author} {\bibfnamefont
  {L.}~\bibnamefont {{Sidoli}}},\ }\bibfield  {title} {\bibinfo {title} {{The
  Chandra ACIS Timing Survey Project: glimpsing a sample of faint X-ray
  pulsators}},\ }\href {https://doi.org/10.1093/mnras/stw1897} {\bibfield
  {journal} {\bibinfo  {journal} {\mnras}\ }\textbf {\bibinfo {volume} {462}},\
  \bibinfo {pages} {4371} (\bibinfo {year} {2016})},\ \Eprint
  {https://arxiv.org/abs/1608.00077} {arXiv:1608.00077 [astro-ph.HE]}
  \BibitemShut {NoStop}%
\bibitem [{\citenamefont {{Israel}}\ \emph {et~al.}(2017)\citenamefont
  {{Israel}}, \citenamefont {{Papitto}}, \citenamefont {{Esposito}},
  \citenamefont {{Stella}}, \citenamefont {{Zampieri}}, \citenamefont
  {{Belfiore}}, \citenamefont {{Rodr{\'\i}guez Castillo}}, \citenamefont {{De
  Luca}}, \citenamefont {{Tiengo}}, \citenamefont {{Haberl}}, \citenamefont
  {{Greiner}}, \citenamefont {{Salvaterra}}, \citenamefont {{Sandrelli}},\ and\
  \citenamefont {{Lisini}}}]{2017MNRAS.466L..48I}%
  \BibitemOpen
  \bibfield  {author} {\bibinfo {author} {\bibfnamefont {G.~L.}\ \bibnamefont
  {{Israel}}}, \bibinfo {author} {\bibfnamefont {A.}~\bibnamefont {{Papitto}}},
  \bibinfo {author} {\bibfnamefont {P.}~\bibnamefont {{Esposito}}}, \bibinfo
  {author} {\bibfnamefont {L.}~\bibnamefont {{Stella}}}, \bibinfo {author}
  {\bibfnamefont {L.}~\bibnamefont {{Zampieri}}}, \bibinfo {author}
  {\bibfnamefont {A.}~\bibnamefont {{Belfiore}}}, \bibinfo {author}
  {\bibfnamefont {G.~A.}\ \bibnamefont {{Rodr{\'\i}guez Castillo}}}, \bibinfo
  {author} {\bibfnamefont {A.}~\bibnamefont {{De Luca}}}, \bibinfo {author}
  {\bibfnamefont {A.}~\bibnamefont {{Tiengo}}}, \bibinfo {author}
  {\bibfnamefont {F.}~\bibnamefont {{Haberl}}}, \bibinfo {author}
  {\bibfnamefont {J.}~\bibnamefont {{Greiner}}}, \bibinfo {author}
  {\bibfnamefont {R.}~\bibnamefont {{Salvaterra}}}, \bibinfo {author}
  {\bibfnamefont {S.}~\bibnamefont {{Sandrelli}}},\ and\ \bibinfo {author}
  {\bibfnamefont {G.}~\bibnamefont {{Lisini}}},\ }\bibfield  {title} {\bibinfo
  {title} {{Discovery of a 0.42-s pulsar in the ultraluminous X-ray source NGC
  7793 P13}},\ }\href {https://doi.org/10.1093/mnrasl/slw218} {\bibfield
  {journal} {\bibinfo  {journal} {\mnras}\ }\textbf {\bibinfo {volume} {466}},\
  \bibinfo {pages} {L48} (\bibinfo {year} {2017})},\ \Eprint
  {https://arxiv.org/abs/1609.06538} {arXiv:1609.06538 [astro-ph.HE]}
  \BibitemShut {NoStop}%
\bibitem [{\citenamefont {{Carpano}}\ \emph {et~al.}(2018)\citenamefont
  {{Carpano}}, \citenamefont {{Haberl}}, \citenamefont {{Maitra}},\ and\
  \citenamefont {{Vasilopoulos}}}]{2018MNRAS.476L..45C}%
  \BibitemOpen
  \bibfield  {author} {\bibinfo {author} {\bibfnamefont {S.}~\bibnamefont
  {{Carpano}}}, \bibinfo {author} {\bibfnamefont {F.}~\bibnamefont {{Haberl}}},
  \bibinfo {author} {\bibfnamefont {C.}~\bibnamefont {{Maitra}}},\ and\
  \bibinfo {author} {\bibfnamefont {G.}~\bibnamefont {{Vasilopoulos}}},\
  }\bibfield  {title} {\bibinfo {title} {{Discovery of pulsations from NGC 300
  ULX1 and its fast period evolution}},\ }\href
  {https://doi.org/10.1093/mnrasl/sly030} {\bibfield  {journal} {\bibinfo
  {journal} {\mnras}\ }\textbf {\bibinfo {volume} {476}},\ \bibinfo {pages}
  {L45} (\bibinfo {year} {2018})},\ \Eprint {https://arxiv.org/abs/1802.10341}
  {arXiv:1802.10341 [astro-ph.HE]} \BibitemShut {NoStop}%
\bibitem [{\citenamefont {{Webb}}\ \emph {et~al.}(2012)\citenamefont {{Webb}},
  \citenamefont {{Cseh}}, \citenamefont {{Lenc}}, \citenamefont {{Godet}},
  \citenamefont {{Barret}}, \citenamefont {{Corbel}}, \citenamefont
  {{Farrell}}, \citenamefont {{Fender}}, \citenamefont {{Gehrels}},\ and\
  \citenamefont {{Heywood}}}]{2012Sci...337..554W}%
  \BibitemOpen
  \bibfield  {author} {\bibinfo {author} {\bibfnamefont {N.}~\bibnamefont
  {{Webb}}}, \bibinfo {author} {\bibfnamefont {D.}~\bibnamefont {{Cseh}}},
  \bibinfo {author} {\bibfnamefont {E.}~\bibnamefont {{Lenc}}}, \bibinfo
  {author} {\bibfnamefont {O.}~\bibnamefont {{Godet}}}, \bibinfo {author}
  {\bibfnamefont {D.}~\bibnamefont {{Barret}}}, \bibinfo {author}
  {\bibfnamefont {S.}~\bibnamefont {{Corbel}}}, \bibinfo {author}
  {\bibfnamefont {S.}~\bibnamefont {{Farrell}}}, \bibinfo {author}
  {\bibfnamefont {R.}~\bibnamefont {{Fender}}}, \bibinfo {author}
  {\bibfnamefont {N.}~\bibnamefont {{Gehrels}}},\ and\ \bibinfo {author}
  {\bibfnamefont {I.}~\bibnamefont {{Heywood}}},\ }\bibfield  {title} {\bibinfo
  {title} {{Radio Detections During Two State Transitions of the
  Intermediate-Mass Black Hole HLX-1}},\ }\href
  {https://doi.org/10.1126/science.1222779} {\bibfield  {journal} {\bibinfo
  {journal} {Science}\ }\textbf {\bibinfo {volume} {337}},\ \bibinfo {pages}
  {554} (\bibinfo {year} {2012})},\ \Eprint {https://arxiv.org/abs/1311.6918}
  {arXiv:1311.6918 [astro-ph.HE]} \BibitemShut {NoStop}%
\bibitem [{\citenamefont {{Mezcua}}\ \emph {et~al.}(2015)\citenamefont
  {{Mezcua}}, \citenamefont {{Roberts}}, \citenamefont {{Lobanov}},\ and\
  \citenamefont {{Sutton}}}]{2015MNRAS.448.1893M}%
  \BibitemOpen
  \bibfield  {author} {\bibinfo {author} {\bibfnamefont {M.}~\bibnamefont
  {{Mezcua}}}, \bibinfo {author} {\bibfnamefont {T.~P.}\ \bibnamefont
  {{Roberts}}}, \bibinfo {author} {\bibfnamefont {A.~P.}\ \bibnamefont
  {{Lobanov}}},\ and\ \bibinfo {author} {\bibfnamefont {A.~D.}\ \bibnamefont
  {{Sutton}}},\ }\bibfield  {title} {\bibinfo {title} {{The powerful jet of an
  off-nuclear intermediate-mass black hole in the spiral galaxy NGC 2276}},\
  }\href {https://doi.org/10.1093/mnras/stv143} {\bibfield  {journal} {\bibinfo
   {journal} {\mnras}\ }\textbf {\bibinfo {volume} {448}},\ \bibinfo {pages}
  {1893} (\bibinfo {year} {2015})},\ \Eprint {https://arxiv.org/abs/1501.04897}
  {arXiv:1501.04897 [astro-ph.GA]} \BibitemShut {NoStop}%
\bibitem [{\citenamefont {{Bachetti}}(2016)}]{2016AN....337..349B}%
  \BibitemOpen
  \bibfield  {author} {\bibinfo {author} {\bibfnamefont {M.}~\bibnamefont
  {{Bachetti}}},\ }\bibfield  {title} {\bibinfo {title} {{Ultraluminous X-ray
  sources: Three exciting years}},\ }\href
  {https://doi.org/10.1002/asna.201612312} {\bibfield  {journal} {\bibinfo
  {journal} {Astronomische Nachrichten}\ }\textbf {\bibinfo {volume} {337}},\
  \bibinfo {pages} {349} (\bibinfo {year} {2016})},\ \Eprint
  {https://arxiv.org/abs/1510.05565} {arXiv:1510.05565 [astro-ph.HE]}
  \BibitemShut {NoStop}%
\bibitem [{\citenamefont {{Barrows}}\ \emph {et~al.}(2019)\citenamefont
  {{Barrows}}, \citenamefont {{Mezcua}},\ and\ \citenamefont
  {{Comerford}}}]{2019ApJ...882..181B}%
  \BibitemOpen
  \bibfield  {author} {\bibinfo {author} {\bibfnamefont {R.~S.}\ \bibnamefont
  {{Barrows}}}, \bibinfo {author} {\bibfnamefont {M.}~\bibnamefont
  {{Mezcua}}},\ and\ \bibinfo {author} {\bibfnamefont {J.~M.}\ \bibnamefont
  {{Comerford}}},\ }\bibfield  {title} {\bibinfo {title} {{A Catalog of
  Hyper-luminous X-Ray Sources and Intermediate-mass Black Hole Candidates out
  to High Redshifts}},\ }\href {https://doi.org/10.3847/1538-4357/ab338a}
  {\bibfield  {journal} {\bibinfo  {journal} {\apj}\ }\textbf {\bibinfo
  {volume} {882}},\ \bibinfo {eid} {181} (\bibinfo {year} {2019})},\ \Eprint
  {https://arxiv.org/abs/1907.08213} {arXiv:1907.08213 [astro-ph.GA]}
  \BibitemShut {NoStop}%
\bibitem [{\citenamefont {{Dage}}\ \emph {et~al.}(2021)\citenamefont {{Dage}},
  \citenamefont {{Kundu}}, \citenamefont {{Thygesen}}, \citenamefont
  {{Bahramian}}, \citenamefont {{Haggard}}, \citenamefont {{Irwin}},
  \citenamefont {{Maccarone}}, \citenamefont {{Nair}}, \citenamefont
  {{Peacock}}, \citenamefont {{Strader}},\ and\ \citenamefont
  {{Zepf}}}]{2021MNRAS.504.1545D}%
  \BibitemOpen
  \bibfield  {author} {\bibinfo {author} {\bibfnamefont {K.~C.}\ \bibnamefont
  {{Dage}}}, \bibinfo {author} {\bibfnamefont {A.}~\bibnamefont {{Kundu}}},
  \bibinfo {author} {\bibfnamefont {E.}~\bibnamefont {{Thygesen}}}, \bibinfo
  {author} {\bibfnamefont {A.}~\bibnamefont {{Bahramian}}}, \bibinfo {author}
  {\bibfnamefont {D.}~\bibnamefont {{Haggard}}}, \bibinfo {author}
  {\bibfnamefont {J.~A.}\ \bibnamefont {{Irwin}}}, \bibinfo {author}
  {\bibfnamefont {T.~J.}\ \bibnamefont {{Maccarone}}}, \bibinfo {author}
  {\bibfnamefont {S.}~\bibnamefont {{Nair}}}, \bibinfo {author} {\bibfnamefont
  {M.~B.}\ \bibnamefont {{Peacock}}}, \bibinfo {author} {\bibfnamefont
  {J.}~\bibnamefont {{Strader}}},\ and\ \bibinfo {author} {\bibfnamefont
  {S.~E.}\ \bibnamefont {{Zepf}}},\ }\bibfield  {title} {\bibinfo {title}
  {{Three ultraluminous X-ray sources hosted by globular clusters in NGC
  1316}},\ }\href {https://doi.org/10.1093/mnras/stab943} {\bibfield  {journal}
  {\bibinfo  {journal} {\mnras}\ }\textbf {\bibinfo {volume} {504}},\ \bibinfo
  {pages} {1545} (\bibinfo {year} {2021})},\ \Eprint
  {https://arxiv.org/abs/2103.16576} {arXiv:2103.16576 [astro-ph.HE]}
  \BibitemShut {NoStop}%
\bibitem [{\citenamefont {Dage}\ \emph {et~al.}(2019)\citenamefont {Dage},
  \citenamefont {Zepf}, \citenamefont {Peacock}, \citenamefont {Bahramian},
  \citenamefont {Noroozi}, \citenamefont {Kundu},\ and\ \citenamefont
  {Maccarone}}]{10.1093/mnras/stz479}%
  \BibitemOpen
  \bibfield  {author} {\bibinfo {author} {\bibfnamefont {K.~C.}\ \bibnamefont
  {Dage}}, \bibinfo {author} {\bibfnamefont {S.~E.}\ \bibnamefont {Zepf}},
  \bibinfo {author} {\bibfnamefont {M.~B.}\ \bibnamefont {Peacock}}, \bibinfo
  {author} {\bibfnamefont {A.}~\bibnamefont {Bahramian}}, \bibinfo {author}
  {\bibfnamefont {O.}~\bibnamefont {Noroozi}}, \bibinfo {author} {\bibfnamefont
  {A.}~\bibnamefont {Kundu}},\ and\ \bibinfo {author} {\bibfnamefont {T.~J.}\
  \bibnamefont {Maccarone}},\ }\bibfield  {title} {\bibinfo {title} {{X-ray
  spectral variability of ultraluminous X-ray sources in extragalactic globular
  clusters}},\ }\href {https://doi.org/10.1093/mnras/stz479} {\bibfield
  {journal} {\bibinfo  {journal} {Monthly Notices of the Royal Astronomical
  Society}\ }\textbf {\bibinfo {volume} {485}},\ \bibinfo {pages} {1694}
  (\bibinfo {year} {2019})},\ \Eprint
  {https://arxiv.org/abs/https://academic.oup.com/mnras/article-pdf/485/2/1694/28009751/stz479.pdf}
  {https://academic.oup.com/mnras/article-pdf/485/2/1694/28009751/stz479.pdf}
  \BibitemShut {NoStop}%
\bibitem [{\citenamefont {Dage}\ \emph {et~al.}(2020)\citenamefont {Dage},
  \citenamefont {Zepf}, \citenamefont {Thygesen}, \citenamefont {Bahramian},
  \citenamefont {Kundu}, \citenamefont {Maccarone}, \citenamefont {Peacock},\
  and\ \citenamefont {Strader}}]{10.1093/mnras/staa1963}%
  \BibitemOpen
  \bibfield  {author} {\bibinfo {author} {\bibfnamefont {K.~C.}\ \bibnamefont
  {Dage}}, \bibinfo {author} {\bibfnamefont {S.~E.}\ \bibnamefont {Zepf}},
  \bibinfo {author} {\bibfnamefont {E.}~\bibnamefont {Thygesen}}, \bibinfo
  {author} {\bibfnamefont {A.}~\bibnamefont {Bahramian}}, \bibinfo {author}
  {\bibfnamefont {A.}~\bibnamefont {Kundu}}, \bibinfo {author} {\bibfnamefont
  {T.~J.}\ \bibnamefont {Maccarone}}, \bibinfo {author} {\bibfnamefont {M.~B.}\
  \bibnamefont {Peacock}},\ and\ \bibinfo {author} {\bibfnamefont
  {J.}~\bibnamefont {Strader}},\ }\bibfield  {title} {\bibinfo {title} {{X-ray
  spectroscopy of newly identified ULXs associated with M87’s globular
  cluster population}},\ }\href {https://doi.org/10.1093/mnras/staa1963}
  {\bibfield  {journal} {\bibinfo  {journal} {Monthly Notices of the Royal
  Astronomical Society}\ }\textbf {\bibinfo {volume} {497}},\ \bibinfo {pages}
  {596} (\bibinfo {year} {2020})},\ \Eprint
  {https://arxiv.org/abs/https://academic.oup.com/mnras/article-pdf/497/1/596/33528071/staa1963.pdf}
  {https://academic.oup.com/mnras/article-pdf/497/1/596/33528071/staa1963.pdf}
  \BibitemShut {NoStop}%
\bibitem [{\citenamefont {{Maccarone}}\ \emph {et~al.}(2007)\citenamefont
  {{Maccarone}}, \citenamefont {{Kundu}}, \citenamefont {{Zepf}},\ and\
  \citenamefont {{Rhode}}}]{2007Natur.445..183M}%
  \BibitemOpen
  \bibfield  {author} {\bibinfo {author} {\bibfnamefont {T.~J.}\ \bibnamefont
  {{Maccarone}}}, \bibinfo {author} {\bibfnamefont {A.}~\bibnamefont
  {{Kundu}}}, \bibinfo {author} {\bibfnamefont {S.~E.}\ \bibnamefont
  {{Zepf}}},\ and\ \bibinfo {author} {\bibfnamefont {K.~L.}\ \bibnamefont
  {{Rhode}}},\ }\bibfield  {title} {\bibinfo {title} {{A black hole in a
  globular cluster}},\ }\href {https://doi.org/10.1038/nature05434} {\bibfield
  {journal} {\bibinfo  {journal} {\nat}\ }\textbf {\bibinfo {volume} {445}},\
  \bibinfo {pages} {183} (\bibinfo {year} {2007})},\ \Eprint
  {https://arxiv.org/abs/astro-ph/0701310} {astro-ph/0701310} \BibitemShut
  {NoStop}%
\bibitem [{\citenamefont {{Kalogera}}\ \emph {et~al.}(2004)\citenamefont
  {{Kalogera}}, \citenamefont {{King}},\ and\ \citenamefont
  {{Rasio}}}]{2004ApJ...601L.171K}%
  \BibitemOpen
  \bibfield  {author} {\bibinfo {author} {\bibfnamefont {V.}~\bibnamefont
  {{Kalogera}}}, \bibinfo {author} {\bibfnamefont {A.~R.}\ \bibnamefont
  {{King}}},\ and\ \bibinfo {author} {\bibfnamefont {F.~A.}\ \bibnamefont
  {{Rasio}}},\ }\bibfield  {title} {\bibinfo {title} {{Could Black Hole X-Ray
  Binaries Be Detected in Globular Clusters?}},\ }\href
  {https://doi.org/10.1086/382042} {\bibfield  {journal} {\bibinfo  {journal}
  {\apjl}\ }\textbf {\bibinfo {volume} {601}},\ \bibinfo {pages} {L171}
  (\bibinfo {year} {2004})},\ \Eprint {https://arxiv.org/abs/astro-ph/0308485}
  {arXiv:astro-ph/0308485 [astro-ph]} \BibitemShut {NoStop}%
\bibitem [{\citenamefont {{Spitzer}}(1969)}]{1969ApJ...158L.139S}%
  \BibitemOpen
  \bibfield  {author} {\bibinfo {author} {\bibfnamefont {J.}~\bibnamefont
  {{Spitzer}}, \bibfnamefont {Lyman}},\ }\bibfield  {title} {\bibinfo {title}
  {{Equipartition and the Formation of Compact Nuclei in Spherical Stellar
  Systems}},\ }\href {https://doi.org/10.1086/180451} {\bibfield  {journal}
  {\bibinfo  {journal} {\apjl}\ }\textbf {\bibinfo {volume} {158}},\ \bibinfo
  {pages} {L139} (\bibinfo {year} {1969})}\BibitemShut {NoStop}%
\bibitem [{\citenamefont {{Kulkarni}}\ \emph {et~al.}(1993)\citenamefont
  {{Kulkarni}}, \citenamefont {{Hut}},\ and\ \citenamefont
  {{McMillan}}}]{1993Natur.364..421K}%
  \BibitemOpen
  \bibfield  {author} {\bibinfo {author} {\bibfnamefont {S.~R.}\ \bibnamefont
  {{Kulkarni}}}, \bibinfo {author} {\bibfnamefont {P.}~\bibnamefont {{Hut}}},\
  and\ \bibinfo {author} {\bibfnamefont {S.}~\bibnamefont {{McMillan}}},\
  }\bibfield  {title} {\bibinfo {title} {{Stellar black holes in globular
  clusters}},\ }\href {https://doi.org/10.1038/364421a0} {\bibfield  {journal}
  {\bibinfo  {journal} {\nat}\ }\textbf {\bibinfo {volume} {364}},\ \bibinfo
  {pages} {421} (\bibinfo {year} {1993})}\BibitemShut {NoStop}%
\bibitem [{\citenamefont {{Sigurdsson}}\ and\ \citenamefont
  {{Hernquist}}(1993)}]{1993Natur.364..423S}%
  \BibitemOpen
  \bibfield  {author} {\bibinfo {author} {\bibfnamefont {S.}~\bibnamefont
  {{Sigurdsson}}}\ and\ \bibinfo {author} {\bibfnamefont {L.}~\bibnamefont
  {{Hernquist}}},\ }\bibfield  {title} {\bibinfo {title} {{Primordial black
  holes in globular clusters}},\ }\href {https://doi.org/10.1038/364423a0}
  {\bibfield  {journal} {\bibinfo  {journal} {\nat}\ }\textbf {\bibinfo
  {volume} {364}},\ \bibinfo {pages} {423} (\bibinfo {year}
  {1993})}\BibitemShut {NoStop}%
\bibitem [{\citenamefont {{Abbott}}\ \emph
  {et~al.}(2016{\natexlab{c}})\citenamefont {{Abbott}}, \citenamefont
  {{Abbott}}, \citenamefont {{Abbott}}, \citenamefont {{Abernathy}},
  \citenamefont {{Acernese}}, \citenamefont {{Ackley}}, \citenamefont
  {{Adams}}, \citenamefont {{Adams}}, \citenamefont {{Addesso}}, \citenamefont
  {{Adhikari}}, \citenamefont {{Adya}}, \citenamefont {{Affeldt}},
  \citenamefont {{Agathos}}, \citenamefont {{Agatsuma}}, \citenamefont
  {{Aggarwal}}, \citenamefont {{Aguiar}}, \citenamefont {{Aiello}},
  \citenamefont {{Ain}}, \citenamefont {{Ajith}}, \citenamefont {{Allen}},
  \citenamefont {{Allocca}}, \citenamefont {{Altin}}, \citenamefont
  {{Anderson}}, \citenamefont {{Anderson}}, \citenamefont {{Arai}},
  \citenamefont {{Araya}}, \citenamefont {{Arceneaux}}, \citenamefont
  {{Areeda}}, \citenamefont {{Arnaud}}, \citenamefont {{Arun}}, \citenamefont
  {{Ascenzi}}, \citenamefont {{Ashton}}, \citenamefont {{Ast}}, \citenamefont
  {{Aston}}, \citenamefont {{Astone}}, \citenamefont {{Aufmuth}}, \citenamefont
  {{Aulbert}}, \citenamefont {{Babak}}, \citenamefont {{Bacon}}, \citenamefont
  {{Bader}}, \citenamefont {{Baker}}, \citenamefont {{Baldaccini}},
  \citenamefont {{Ballardin}}, \citenamefont {{Ballmer}}, \citenamefont
  {{Barayoga}}, \citenamefont {{Barclay}}, \citenamefont {{Barish}},
  \citenamefont {{Barker}}, \citenamefont {{Barone}}, \citenamefont {{Barr}},
  \citenamefont {{Barsotti}}, \citenamefont {{Barsuglia}}, \citenamefont
  {{Barta}}, \citenamefont {{Bartlett}}, \citenamefont {{Bartos}},
  \citenamefont {{Bassiri}}, \citenamefont {{Basti}}, \citenamefont {{Batch}},
  \citenamefont {{Baune}}, \citenamefont {{Bavigadda}}, \citenamefont
  {{Bazzan}}, \citenamefont {{Behnke}}, \citenamefont {{Bejger}}, \citenamefont
  {{Bell}}, \citenamefont {{Bell}}, \citenamefont {{Berger}}, \citenamefont
  {{Bergman}}, \citenamefont {{Bergmann}}, \citenamefont {{Berry}},
  \citenamefont {{Bersanetti}}, \citenamefont {{Bertolini}}, \citenamefont
  {{Betzwieser}}, \citenamefont {{Bhagwat}}, \citenamefont {{Bhandare}},
  \citenamefont {{Bilenko}}, \citenamefont {{Billingsley}}, \citenamefont
  {{Birch}}, \citenamefont {{Birney}}, \citenamefont {{Biscans}}, \citenamefont
  {{Bisht}}, \citenamefont {{Bitossi}}, \citenamefont {{Biwer}}, \citenamefont
  {{Bizouard}}, \citenamefont {{Blackburn}}, \citenamefont {{Blair}},
  \citenamefont {{Blair}}, \citenamefont {{Blair}}, \citenamefont {{Bloemen}},
  \citenamefont {{Bock}}, \citenamefont {{Bodiya}}, \citenamefont {{Boer}},
  \citenamefont {{Bogaert}}, \citenamefont {{Bogan}}, \citenamefont {{Bohe}},
  \citenamefont {{Bojtos}}, \citenamefont {{Bond}}, \citenamefont {{Bondu}},
  \citenamefont {{Bonnand}}, \citenamefont {{Boom}}, \citenamefont {{Bork}},
  \citenamefont {{Boschi}}, \citenamefont {{Bose}}, \citenamefont
  {{Bouffanais}}, \citenamefont {{Bozzi}}, \citenamefont {{Bradaschia}},
  \citenamefont {{Brady}}, \citenamefont {{Braginsky}}, \citenamefont
  {{Branchesi}}, \citenamefont {{Brau}}, \citenamefont {{Briant}},
  \citenamefont {{Brillet}}, \citenamefont {{Brinkmann}}, \citenamefont
  {{Brisson}}, \citenamefont {{Brockill}}, \citenamefont {{Brooks}},
  \citenamefont {{Brown}}, \citenamefont {{Brown}}, \citenamefont {{Brown}},
  \citenamefont {{Buchanan}}, \citenamefont {{Buikema}}, \citenamefont
  {{Bulik}}, \citenamefont {{Bulten}}, \citenamefont {{Buonanno}},
  \citenamefont {{Buskulic}}, \citenamefont {{Buy}}, \citenamefont {{Byer}},
  \citenamefont {{Cadonati}}, \citenamefont {{Cagnoli}}, \citenamefont
  {{Cahillane}}, \citenamefont {{Calder{\'o}n Bustillo}}, \citenamefont
  {{Callister}}, \citenamefont {{Calloni}}, \citenamefont {{Camp}},
  \citenamefont {{Cannon}}, \citenamefont {{Cao}}, \citenamefont {{Capano}},
  \citenamefont {{Capocasa}}, \citenamefont {{Carbognani}}, \citenamefont
  {{Caride}}, \citenamefont {{Casanueva Diaz}}, \citenamefont {{Casentini}},
  \citenamefont {{Caudill}}, \citenamefont {{Cavagli{\`a}}}, \citenamefont
  {{Cavalier}}, \citenamefont {{Cavalieri}}, \citenamefont {{Cella}},
  \citenamefont {{Cepeda}}, \citenamefont {{Cerboni Baiardi}}, \citenamefont
  {{Cerretani}}, \citenamefont {{Cesarini}}, \citenamefont {{Chakraborty}},
  \citenamefont {{Chalermsongsak}}, \citenamefont {{Chamberlin}}, \citenamefont
  {{Chan}}, \citenamefont {{Chao}}, \citenamefont {{Charlton}}, \citenamefont
  {{Chassande-Mottin}}, \citenamefont {{Chen}}, \citenamefont {{Chen}},
  \citenamefont {{Cheng}}, \citenamefont {{Chincarini}}, \citenamefont
  {{Chiummo}}, \citenamefont {{Cho}}, \citenamefont {{Cho}}, \citenamefont
  {{Chow}}, \citenamefont {{Christensen}}, \citenamefont {{Chu}}, \citenamefont
  {{Chua}}, \citenamefont {{Chung}}, \citenamefont {{Ciani}}, \citenamefont
  {{Clara}}, \citenamefont {{Clark}}, \citenamefont {{Cleva}}, \citenamefont
  {{Coccia}}, \citenamefont {{Cohadon}}, \citenamefont {{Colla}}, \citenamefont
  {{Collette}}, \citenamefont {{Cominsky}}, \citenamefont {{Constancio}},
  \citenamefont {{Conte}}, \citenamefont {{Conti}}, \citenamefont {{Cook}},
  \citenamefont {{Corbitt}}, \citenamefont {{Cornish}}, \citenamefont
  {{Corsi}}, \citenamefont {{Cortese}}, \citenamefont {{Costa}}, \citenamefont
  {{Coughlin}}, \citenamefont {{Coughlin}}, \citenamefont {{Coulon}},
  \citenamefont {{Countryman}}, \citenamefont {{Couvares}}, \citenamefont
  {{Cowan}}, \citenamefont {{Coward}}, \citenamefont {{Cowart}}, \citenamefont
  {{Coyne}}, \citenamefont {{Coyne}}, \citenamefont {{Craig}}, \citenamefont
  {{Creighton}}, \citenamefont {{Cripe}}, \citenamefont {{Crowder}},
  \citenamefont {{Cumming}}, \citenamefont {{Cunningham}}, \citenamefont
  {{Cuoco}}, \citenamefont {{Dal Canton}}, \citenamefont {{Danilishin}},
  \citenamefont {{D'Antonio}}, \citenamefont {{Danzmann}}, \citenamefont
  {{Darman}}, \citenamefont {{Dattilo}}, \citenamefont {{Dave}}, \citenamefont
  {{Daveloza}}, \citenamefont {{Davier}}, \citenamefont {{Davies}},
  \citenamefont {{Daw}}, \citenamefont {{Day}}, \citenamefont {{De}},
  \citenamefont {{DeBra}}, \citenamefont {{Debreczeni}}, \citenamefont
  {{Degallaix}}, \citenamefont {{De Laurentis}}, \citenamefont
  {{Del{\'e}glise}}, \citenamefont {{Del Pozzo}}, \citenamefont {{Denker}},
  \citenamefont {{Dent}}, \citenamefont {{Dereli}}, \citenamefont
  {{Dergachev}}, \citenamefont {{De Rosa}}, \citenamefont {{DeRosa}},
  \citenamefont {{DeSalvo}}, \citenamefont {{Dhurandhar}}, \citenamefont
  {{D{\'\i}az}}, \citenamefont {{Di Fiore}}, \citenamefont {{Di Giovanni}},
  \citenamefont {{Di Lieto}}, \citenamefont {{Di Pace}}, \citenamefont {{Di
  Palma}}, \citenamefont {{Di Virgilio}}, \citenamefont {{Dojcinoski}},
  \citenamefont {{Dolique}}, \citenamefont {{Donovan}}, \citenamefont
  {{Dooley}}, \citenamefont {{Doravari}}, \citenamefont {{Douglas}},
  \citenamefont {{Downes}}, \citenamefont {{Drago}}, \citenamefont {{Drever}},
  \citenamefont {{Driggers}}, \citenamefont {{Du}}, \citenamefont {{Ducrot}},
  \citenamefont {{Dwyer}}, \citenamefont {{Edo}}, \citenamefont {{Edwards}},
  \citenamefont {{Effler}}, \citenamefont {{Eggenstein}}, \citenamefont
  {{Ehrens}}, \citenamefont {{Eichholz}}, \citenamefont {{Eikenberry}},
  \citenamefont {{Engels}}, \citenamefont {{Essick}}, \citenamefont {{Etzel}},
  \citenamefont {{Evans}}, \citenamefont {{Evans}}, \citenamefont {{Everett}},
  \citenamefont {{Factourovich}}, \citenamefont {{Fafone}}, \citenamefont
  {{Fair}}, \citenamefont {{Fairhurst}}, \citenamefont {{Fan}}, \citenamefont
  {{Fang}}, \citenamefont {{Farinon}}, \citenamefont {{Farr}}, \citenamefont
  {{Farr}}, \citenamefont {{Favata}}, \citenamefont {{Fays}}, \citenamefont
  {{Fehrmann}}, \citenamefont {{Fejer}}, \citenamefont {{Ferrante}},
  \citenamefont {{Ferreira}}, \citenamefont {{Ferrini}}, \citenamefont
  {{Fidecaro}}, \citenamefont {{Fiori}}, \citenamefont {{Fiorucci}},
  \citenamefont {{Fisher}}, \citenamefont {{Flaminio}}, \citenamefont
  {{Fletcher}}, \citenamefont {{Fong}}, \citenamefont {{Fournier}},
  \citenamefont {{Franco}}, \citenamefont {{Frasca}}, \citenamefont
  {{Frasconi}}, \citenamefont {{Frei}}, \citenamefont {{Freise}}, \citenamefont
  {{Frey}}, \citenamefont {{Frey}}, \citenamefont {{Fricke}}, \citenamefont
  {{Fritschel}}, \citenamefont {{Frolov}}, \citenamefont {{Fulda}},
  \citenamefont {{Fyffe}}, \citenamefont {{Gabbard}}, \citenamefont {{Gair}},
  \citenamefont {{Gammaitoni}}, \citenamefont {{Gaonkar}}, \citenamefont
  {{Garufi}}, \citenamefont {{Gatto}}, \citenamefont {{Gaur}}, \citenamefont
  {{Gehrels}}, \citenamefont {{Gemme}}, \citenamefont {{Gendre}}, \citenamefont
  {{Genin}}, \citenamefont {{Gennai}}, \citenamefont {{George}}, \citenamefont
  {{Gergely}}, \citenamefont {{Germain}}, \citenamefont {{Ghosh}},
  \citenamefont {{Ghosh}}, \citenamefont {{Giaime}}, \citenamefont
  {{Giardina}}, \citenamefont {{Giazotto}}, \citenamefont {{Gill}},
  \citenamefont {{Glaefke}}, \citenamefont {{Goetz}}, \citenamefont {{Goetz}},
  \citenamefont {{Gondan}}, \citenamefont {{Gonz{\'a}lez}}, \citenamefont
  {{Gonzalez Castro}}, \citenamefont {{Gopakumar}}, \citenamefont {{Gordon}},
  \citenamefont {{Gorodetsky}}, \citenamefont {{Gossan}}, \citenamefont
  {{Gosselin}}, \citenamefont {{Gouaty}}, \citenamefont {{Graef}},
  \citenamefont {{Graff}}, \citenamefont {{Granata}}, \citenamefont {{Grant}},
  \citenamefont {{Gras}}, \citenamefont {{Gray}}, \citenamefont {{Greco}},
  \citenamefont {{Green}}, \citenamefont {{Groot}}, \citenamefont {{Grote}},
  \citenamefont {{Grunewald}}, \citenamefont {{Guidi}}, \citenamefont {{Guo}},
  \citenamefont {{Gupta}}, \citenamefont {{Gupta}}, \citenamefont {{Gushwa}},
  \citenamefont {{Gustafson}}, \citenamefont {{Gustafson}}, \citenamefont
  {{Hacker}}, \citenamefont {{Hall}}, \citenamefont {{Hall}}, \citenamefont
  {{Hammond}}, \citenamefont {{Haney}}, \citenamefont {{Hanke}}, \citenamefont
  {{Hanks}}, \citenamefont {{Hanna}}, \citenamefont {{Hannam}}, \citenamefont
  {{Hanson}}, \citenamefont {{Hardwick}}, \citenamefont {{Harms}},
  \citenamefont {{Harry}}, \citenamefont {{Harry}}, \citenamefont {{Hart}},
  \citenamefont {{Hartman}}, \citenamefont {{Haster}}, \citenamefont
  {{Haughian}}, \citenamefont {{Heidmann}}, \citenamefont {{Heintze}},
  \citenamefont {{Heitmann}}, \citenamefont {{Hello}}, \citenamefont
  {{Hemming}}, \citenamefont {{Hendry}}, \citenamefont {{Heng}}, \citenamefont
  {{Hennig}}, \citenamefont {{Heptonstall}}, \citenamefont {{Heurs}},
  \citenamefont {{Hild}}, \citenamefont {{Hoak}}, \citenamefont {{Hodge}},
  \citenamefont {{Hofman}}, \citenamefont {{Hollitt}}, \citenamefont {{Holt}},
  \citenamefont {{Holz}}, \citenamefont {{Hopkins}}, \citenamefont {{Hosken}},
  \citenamefont {{Hough}}, \citenamefont {{Houston}}, \citenamefont {{Howell}},
  \citenamefont {{Hu}}, \citenamefont {{Huang}}, \citenamefont {{Huerta}},
  \citenamefont {{Huet}}, \citenamefont {{Hughey}}, \citenamefont {{Husa}},
  \citenamefont {{Huttner}}, \citenamefont {{Huynh-Dinh}}, \citenamefont
  {{Idrisy}}, \citenamefont {{Indik}}, \citenamefont {{Ingram}}, \citenamefont
  {{Inta}}, \citenamefont {{Isa}}, \citenamefont {{Isac}}, \citenamefont
  {{Isi}}, \citenamefont {{Islas}}, \citenamefont {{Isogai}}, \citenamefont
  {{Iyer}}, \citenamefont {{Izumi}}, \citenamefont {{Jacqmin}}, \citenamefont
  {{Jang}}, \citenamefont {{Jani}}, \citenamefont {{Jaranowski}}, \citenamefont
  {{Jawahar}}, \citenamefont {{Jim{\'e}nez-Forteza}}, \citenamefont
  {{Johnson}}, \citenamefont {{Jones}}, \citenamefont {{Jones}}, \citenamefont
  {{Jonker}}, \citenamefont {{Ju}}, \citenamefont {{K}}, \citenamefont
  {{Kalaghatgi}}, \citenamefont {{Kalogera}}, \citenamefont {{Kandhasamy}},
  \citenamefont {{Kang}}, \citenamefont {{Kanner}}, \citenamefont {{Karki}},
  \citenamefont {{Kasprzack}}, \citenamefont {{Katsavounidis}}, \citenamefont
  {{Katzman}}, \citenamefont {{Kaufer}}, \citenamefont {{Kaur}}, \citenamefont
  {{Kawabe}}, \citenamefont {{Kawazoe}}, \citenamefont {{K{\'e}f{\'e}lian}},
  \citenamefont {{Kehl}}, \citenamefont {{Keitel}}, \citenamefont {{Kelley}},
  \citenamefont {{Kells}}, \citenamefont {{Kennedy}}, \citenamefont {{Key}},
  \citenamefont {{Khalaidovski}}, \citenamefont {{Khalili}}, \citenamefont
  {{Khan}}, \citenamefont {{Khan}}, \citenamefont {{Khan}}, \citenamefont
  {{Khazanov}}, \citenamefont {{Kijbunchoo}}, \citenamefont {{Kim}},
  \citenamefont {{Kim}}, \citenamefont {{Kim}}, \citenamefont {{Kim}},
  \citenamefont {{Kim}}, \citenamefont {{Kim}}, \citenamefont {{King}},
  \citenamefont {{King}}, \citenamefont {{Kinzel}}, \citenamefont {{Kissel}},
  \citenamefont {{Kleybolte}}, \citenamefont {{Klimenko}}, \citenamefont
  {{Koehlenbeck}}, \citenamefont {{Kokeyama}}, \citenamefont {{Koley}},
  \citenamefont {{Kondrashov}}, \citenamefont {{Kontos}}, \citenamefont
  {{Korobko}}, \citenamefont {{Korth}}, \citenamefont {{Kowalska}},
  \citenamefont {{Kozak}}, \citenamefont {{Kringel}}, \citenamefont
  {{Krishnan}}, \citenamefont {{Kr{\'o}lak}}, \citenamefont {{Krueger}},
  \citenamefont {{Kuehn}}, \citenamefont {{Kumar}}, \citenamefont {{Kuo}},
  \citenamefont {{Kutynia}}, \citenamefont {{Lackey}}, \citenamefont
  {{Landry}}, \citenamefont {{Lange}}, \citenamefont {{Lantz}}, \citenamefont
  {{Lasky}}, \citenamefont {{Lazzarini}}, \citenamefont {{Lazzaro}},
  \citenamefont {{Leaci}}, \citenamefont {{Leavey}}, \citenamefont {{Lebigot}},
  \citenamefont {{Lee}}, \citenamefont {{Lee}}, \citenamefont {{Lee}},
  \citenamefont {{Lee}}, \citenamefont {{Lenon}}, \citenamefont {{Leonardi}},
  \citenamefont {{Leong}}, \citenamefont {{Leroy}}, \citenamefont {{Letendre}},
  \citenamefont {{Levin}}, \citenamefont {{Levine}}, \citenamefont {{Li}},
  \citenamefont {{Libson}}, \citenamefont {{Littenberg}}, \citenamefont
  {{Lockerbie}}, \citenamefont {{Logue}}, \citenamefont {{Lombardi}},
  \citenamefont {{Lord}}, \citenamefont {{Lorenzini}}, \citenamefont
  {{Loriette}}, \citenamefont {{Lormand}}, \citenamefont {{Losurdo}},
  \citenamefont {{Lough}}, \citenamefont {{L{\"u}ck}}, \citenamefont
  {{Lundgren}}, \citenamefont {{Luo}}, \citenamefont {{Lynch}}, \citenamefont
  {{Ma}}, \citenamefont {{MacDonald}}, \citenamefont {{Machenschalk}},
  \citenamefont {{MacInnis}}, \citenamefont {{Macleod}}, \citenamefont
  {{Maga{\~n}a-Sandoval}}, \citenamefont {{Magee}}, \citenamefont
  {{Mageswaran}}, \citenamefont {{Majorana}}, \citenamefont {{Maksimovic}},
  \citenamefont {{Malvezzi}}, \citenamefont {{Man}}, \citenamefont {{Mandel}},
  \citenamefont {{Mandic}}, \citenamefont {{Mangano}}, \citenamefont
  {{Mansell}}, \citenamefont {{Manske}}, \citenamefont {{Mantovani}},
  \citenamefont {{Marchesoni}}, \citenamefont {{Marion}}, \citenamefont
  {{M{\'a}rka}}, \citenamefont {{M{\'a}rka}}, \citenamefont {{Markosyan}},
  \citenamefont {{Maros}}, \citenamefont {{Martelli}}, \citenamefont
  {{Martellini}}, \citenamefont {{Martin}}, \citenamefont {{Martin}},
  \citenamefont {{Martynov}}, \citenamefont {{Marx}}, \citenamefont {{Mason}},
  \citenamefont {{Masserot}}, \citenamefont {{Massinger}}, \citenamefont
  {{Masso-Reid}}, \citenamefont {{Matichard}}, \citenamefont {{Matone}},
  \citenamefont {{Mavalvala}}, \citenamefont {{Mazumder}}, \citenamefont
  {{Mazzolo}}, \citenamefont {{McCarthy}}, \citenamefont {{McClelland}},
  \citenamefont {{McCormick}}, \citenamefont {{McGuire}}, \citenamefont
  {{McIntyre}}, \citenamefont {{McIver}}, \citenamefont {{McManus}},
  \citenamefont {{McWilliams}}, \citenamefont {{Meacher}}, \citenamefont
  {{Meadors}}, \citenamefont {{Meidam}}, \citenamefont {{Melatos}},
  \citenamefont {{Mendell}}, \citenamefont {{Mendoza-Gandara}}, \citenamefont
  {{Mercer}}, \citenamefont {{Merilh}}, \citenamefont {{Merzougui}},
  \citenamefont {{Meshkov}}, \citenamefont {{Messenger}}, \citenamefont
  {{Messick}}, \citenamefont {{Meyers}}, \citenamefont {{Mezzani}},
  \citenamefont {{Miao}}, \citenamefont {{Michel}}, \citenamefont
  {{Middleton}}, \citenamefont {{Mikhailov}}, \citenamefont {{Milano}},
  \citenamefont {{Miller}}, \citenamefont {{Millhouse}}, \citenamefont
  {{Minenkov}}, \citenamefont {{Ming}}, \citenamefont {{Mirshekari}},
  \citenamefont {{Mishra}}, \citenamefont {{Mitra}}, \citenamefont
  {{Mitrofanov}}, \citenamefont {{Mitselmakher}}, \citenamefont {{Mittleman}},
  \citenamefont {{Moggi}}, \citenamefont {{Mohan}}, \citenamefont
  {{Mohapatra}}, \citenamefont {{Montani}}, \citenamefont {{Moore}},
  \citenamefont {{Moore}}, \citenamefont {{Moraru}}, \citenamefont {{Moreno}},
  \citenamefont {{Morriss}}, \citenamefont {{Mossavi}}, \citenamefont
  {{Mours}}, \citenamefont {{Mow-Lowry}}, \citenamefont {{Mueller}},
  \citenamefont {{Mueller}}, \citenamefont {{Muir}}, \citenamefont
  {{Mukherjee}}, \citenamefont {{Mukherjee}}, \citenamefont {{Mukherjee}},
  \citenamefont {{Mukund}}, \citenamefont {{Mullavey}}, \citenamefont
  {{Munch}}, \citenamefont {{Murphy}}, \citenamefont {{Murray}}, \citenamefont
  {{Mytidis}}, \citenamefont {{Nardecchia}}, \citenamefont {{Naticchioni}},
  \citenamefont {{Nayak}}, \citenamefont {{Necula}}, \citenamefont {{Nedkova}},
  \citenamefont {{Nelemans}}, \citenamefont {{Neri}}, \citenamefont
  {{Neunzert}}, \citenamefont {{Newton}}, \citenamefont {{Nguyen}},
  \citenamefont {{Nielsen}}, \citenamefont {{Nissanke}}, \citenamefont
  {{Nitz}}, \citenamefont {{Nocera}}, \citenamefont {{Nolting}}, \citenamefont
  {{Normandin}}, \citenamefont {{Nuttall}}, \citenamefont {{Oberling}},
  \citenamefont {{Ochsner}}, \citenamefont {{O'Dell}}, \citenamefont
  {{Oelker}}, \citenamefont {{Ogin}}, \citenamefont {{Oh}}, \citenamefont
  {{Oh}}, \citenamefont {{Ohme}}, \citenamefont {{Oliver}}, \citenamefont
  {{Oppermann}}, \citenamefont {{Oram}}, \citenamefont {{O'Reilly}},
  \citenamefont {{O'Shaughnessy}}, \citenamefont {{Ottaway}}, \citenamefont
  {{Ottens}}, \citenamefont {{Overmier}}, \citenamefont {{Owen}}, \citenamefont
  {{Pai}}, \citenamefont {{Pai}}, \citenamefont {{Palamos}}, \citenamefont
  {{Palashov}}, \citenamefont {{Palomba}}, \citenamefont {{Pal-Singh}},
  \citenamefont {{Pan}}, \citenamefont {{Pankow}}, \citenamefont {{Pannarale}},
  \citenamefont {{Pant}}, \citenamefont {{Paoletti}}, \citenamefont {{Paoli}},
  \citenamefont {{Papa}}, \citenamefont {{Paris}}, \citenamefont {{Parker}},
  \citenamefont {{Pascucci}}, \citenamefont {{Pasqualetti}}, \citenamefont
  {{Passaquieti}}, \citenamefont {{Passuello}}, \citenamefont {{Patricelli}},
  \citenamefont {{Patrick}}, \citenamefont {{Pearlstone}}, \citenamefont
  {{Pedraza}}, \citenamefont {{Pedurand}}, \citenamefont {{Pekowsky}},
  \citenamefont {{Pele}}, \citenamefont {{Penn}}, \citenamefont {{Perreca}},
  \citenamefont {{Phelps}}, \citenamefont {{Piccinni}}, \citenamefont
  {{Pichot}}, \citenamefont {{Piergiovanni}}, \citenamefont {{Pierro}},
  \citenamefont {{Pillant}}, \citenamefont {{Pinard}}, \citenamefont {{Pinto}},
  \citenamefont {{Pitkin}}, \citenamefont {{Poggiani}}, \citenamefont
  {{Popolizio}}, \citenamefont {{Porter}}, \citenamefont {{Post}},
  \citenamefont {{Powell}}, \citenamefont {{Prasad}}, \citenamefont {{Predoi}},
  \citenamefont {{Premachandra}}, \citenamefont {{Prestegard}}, \citenamefont
  {{Price}}, \citenamefont {{Prijatelj}}, \citenamefont {{Principe}},
  \citenamefont {{Privitera}}, \citenamefont {{Prodi}}, \citenamefont
  {{Prokhorov}}, \citenamefont {{Puncken}}, \citenamefont {{Punturo}},
  \citenamefont {{Puppo}}, \citenamefont {{P{\"u}rrer}}, \citenamefont {{Qi}},
  \citenamefont {{Qin}}, \citenamefont {{Quetschke}}, \citenamefont
  {{Quintero}}, \citenamefont {{Quitzow-James}}, \citenamefont {{Raab}},
  \citenamefont {{Rabeling}}, \citenamefont {{Radkins}}, \citenamefont
  {{Raffai}}, \citenamefont {{Raja}}, \citenamefont {{Rakhmanov}},
  \citenamefont {{Rapagnani}}, \citenamefont {{Raymond}}, \citenamefont
  {{Razzano}}, \citenamefont {{Re}}, \citenamefont {{Read}}, \citenamefont
  {{Reed}}, \citenamefont {{Regimbau}}, \citenamefont {{Rei}}, \citenamefont
  {{Reid}}, \citenamefont {{Reitze}}, \citenamefont {{Rew}}, \citenamefont
  {{Reyes}}, \citenamefont {{Ricci}}, \citenamefont {{Riles}}, \citenamefont
  {{Robertson}}, \citenamefont {{Robie}}, \citenamefont {{Robinet}},
  \citenamefont {{Rocchi}}, \citenamefont {{Rolland}}, \citenamefont
  {{Rollins}}, \citenamefont {{Roma}}, \citenamefont {{Romano}}, \citenamefont
  {{Romanov}}, \citenamefont {{Romie}}, \citenamefont {{Rosi{\'n}ska}},
  \citenamefont {{Rowan}}, \citenamefont {{R{\"u}diger}}, \citenamefont
  {{Ruggi}}, \citenamefont {{Ryan}}, \citenamefont {{Sachdev}}, \citenamefont
  {{Sadecki}}, \citenamefont {{Sadeghian}}, \citenamefont {{Salconi}},
  \citenamefont {{Saleem}}, \citenamefont {{Salemi}}, \citenamefont
  {{Samajdar}}, \citenamefont {{Sammut}}, \citenamefont {{Sampson}},
  \citenamefont {{Sanchez}}, \citenamefont {{Sandberg}}, \citenamefont
  {{Sandeen}}, \citenamefont {{Sanders}}, \citenamefont {{Sassolas}},
  \citenamefont {{Sathyaprakash}}, \citenamefont {{Saulson}}, \citenamefont
  {{Sauter}}, \citenamefont {{Savage}}, \citenamefont {{Sawadsky}},
  \citenamefont {{Schale}}, \citenamefont {{Schilling}}, \citenamefont
  {{Schmidt}}, \citenamefont {{Schmidt}}, \citenamefont {{Schnabel}},
  \citenamefont {{Schofield}}, \citenamefont {{Sch{\"o}nbeck}}, \citenamefont
  {{Schreiber}}, \citenamefont {{Schuette}}, \citenamefont {{Schutz}},
  \citenamefont {{Scott}}, \citenamefont {{Scott}}, \citenamefont {{Sellers}},
  \citenamefont {{Sengupta}}, \citenamefont {{Sentenac}}, \citenamefont
  {{Sequino}}, \citenamefont {{Sergeev}}, \citenamefont {{Serna}},
  \citenamefont {{Setyawati}}, \citenamefont {{Sevigny}}, \citenamefont
  {{Shaddock}}, \citenamefont {{Shah}}, \citenamefont {{Shahriar}},
  \citenamefont {{Shaltev}}, \citenamefont {{Shao}}, \citenamefont {{Shapiro}},
  \citenamefont {{Shawhan}}, \citenamefont {{Sheperd}}, \citenamefont
  {{Shoemaker}}, \citenamefont {{Shoemaker}}, \citenamefont {{Siellez}},
  \citenamefont {{Siemens}}, \citenamefont {{Sigg}}, \citenamefont {{Silva}},
  \citenamefont {{Simakov}}, \citenamefont {{Singer}}, \citenamefont
  {{Singer}}, \citenamefont {{Singh}}, \citenamefont {{Singh}}, \citenamefont
  {{Singhal}}, \citenamefont {{Sintes}}, \citenamefont {{Slagmolen}},
  \citenamefont {{Smith}}, \citenamefont {{Smith}}, \citenamefont {{Smith}},
  \citenamefont {{Son}}, \citenamefont {{Sorazu}}, \citenamefont
  {{Sorrentino}}, \citenamefont {{Souradeep}}, \citenamefont {{Srivastava}},
  \citenamefont {{Staley}}, \citenamefont {{Steinke}}, \citenamefont
  {{Steinlechner}}, \citenamefont {{Steinlechner}}, \citenamefont
  {{Steinmeyer}}, \citenamefont {{Stephens}}, \citenamefont {{Stevenson}},
  \citenamefont {{Stone}}, \citenamefont {{Strain}}, \citenamefont
  {{Straniero}}, \citenamefont {{Stratta}}, \citenamefont {{Strauss}},
  \citenamefont {{Strigin}}, \citenamefont {{Sturani}}, \citenamefont
  {{Stuver}}, \citenamefont {{Summerscales}}, \citenamefont {{Sun}},
  \citenamefont {{Sutton}}, \citenamefont {{Swinkels}}, \citenamefont
  {{Szczepa{\'n}czyk}}, \citenamefont {{Tacca}}, \citenamefont {{Talukder}},
  \citenamefont {{Tanner}}, \citenamefont {{T{\'a}pai}}, \citenamefont
  {{Tarabrin}}, \citenamefont {{Taracchini}}, \citenamefont {{Taylor}},
  \citenamefont {{Theeg}}, \citenamefont {{Thirugnanasambandam}}, \citenamefont
  {{Thomas}}, \citenamefont {{Thomas}}, \citenamefont {{Thomas}}, \citenamefont
  {{Thorne}}, \citenamefont {{Thorne}}, \citenamefont {{Thrane}}, \citenamefont
  {{Tiwari}}, \citenamefont {{Tiwari}}, \citenamefont {{Tokmakov}},
  \citenamefont {{Tomlinson}}, \citenamefont {{Tonelli}}, \citenamefont
  {{Torres}}, \citenamefont {{Torrie}}, \citenamefont {{T{\"o}yr{\"a}}},
  \citenamefont {{Travasso}}, \citenamefont {{Traylor}}, \citenamefont
  {{Trifir{\`o}}}, \citenamefont {{Tringali}}, \citenamefont {{Trozzo}},
  \citenamefont {{Tse}}, \citenamefont {{Turconi}}, \citenamefont
  {{Tuyenbayev}}, \citenamefont {{Ugolini}}, \citenamefont {{Unnikrishnan}},
  \citenamefont {{Urban}}, \citenamefont {{Usman}}, \citenamefont
  {{Vahlbruch}}, \citenamefont {{Vajente}}, \citenamefont {{Valdes}},
  \citenamefont {{Vallisneri}}, \citenamefont {{van Bakel}}, \citenamefont
  {{van Beuzekom}}, \citenamefont {{van den Brand}}, \citenamefont {{Van Den
  Broeck}}, \citenamefont {{Vander-Hyde}}, \citenamefont {{van der Schaaf}},
  \citenamefont {{van Heijningen}}, \citenamefont {{van Veggel}}, \citenamefont
  {{Vardaro}}, \citenamefont {{Vass}}, \citenamefont {{Vas{\'u}th}},
  \citenamefont {{Vaulin}}, \citenamefont {{Vecchio}}, \citenamefont
  {{Vedovato}}, \citenamefont {{Veitch}}, \citenamefont {{Veitch}},
  \citenamefont {{Venkateswara}}, \citenamefont {{Verkindt}}, \citenamefont
  {{Vetrano}}, \citenamefont {{Vicer{\'e}}}, \citenamefont {{Vinciguerra}},
  \citenamefont {{Vine}}, \citenamefont {{Vinet}}, \citenamefont {{Vitale}},
  \citenamefont {{Vo}}, \citenamefont {{Vocca}}, \citenamefont {{Vorvick}},
  \citenamefont {{Voss}}, \citenamefont {{Vousden}}, \citenamefont
  {{Vyatchanin}}, \citenamefont {{Wade}}, \citenamefont {{Wade}}, \citenamefont
  {{Wade}}, \citenamefont {{Walker}}, \citenamefont {{Wallace}}, \citenamefont
  {{Walsh}}, \citenamefont {{Wang}}, \citenamefont {{Wang}}, \citenamefont
  {{Wang}}, \citenamefont {{Wang}}, \citenamefont {{Wang}}, \citenamefont
  {{Ward}}, \citenamefont {{Warner}}, \citenamefont {{Was}}, \citenamefont
  {{Weaver}}, \citenamefont {{Wei}}, \citenamefont {{Weinert}}, \citenamefont
  {{Weinstein}}, \citenamefont {{Weiss}}, \citenamefont {{Welborn}},
  \citenamefont {{Wen}}, \citenamefont {{We{\ss}els}}, \citenamefont
  {{Westphal}}, \citenamefont {{Wette}}, \citenamefont {{Whelan}},
  \citenamefont {{White}}, \citenamefont {{Whiting}}, \citenamefont
  {{Williams}}, \citenamefont {{Williamson}}, \citenamefont {{Willis}},
  \citenamefont {{Willke}}, \citenamefont {{Wimmer}}, \citenamefont
  {{Winkler}}, \citenamefont {{Wipf}}, \citenamefont {{Wittel}}, \citenamefont
  {{Woan}}, \citenamefont {{Worden}}, \citenamefont {{Wright}}, \citenamefont
  {{Wu}}, \citenamefont {{Yablon}}, \citenamefont {{Yam}}, \citenamefont
  {{Yamamoto}}, \citenamefont {{Yancey}}, \citenamefont {{Yap}}, \citenamefont
  {{Yu}}, \citenamefont {{Yvert}}, \citenamefont {{Zadro{\.z}ny}},
  \citenamefont {{Zangrando}}, \citenamefont {{Zanolin}}, \citenamefont
  {{Zendri}}, \citenamefont {{Zevin}}, \citenamefont {{Zhang}}, \citenamefont
  {{Zhang}}, \citenamefont {{Zhang}}, \citenamefont {{Zhang}}, \citenamefont
  {{Zhao}}, \citenamefont {{Zhou}}, \citenamefont {{Zhou}}, \citenamefont
  {{Zhu}}, \citenamefont {{Zucker}}, \citenamefont {{Zuraw}}, \citenamefont
  {{Zweizig}}, \citenamefont {{LIGO Scientific Collaboration}},\ and\
  \citenamefont {{Virgo Collaboration}}}]{2016ApJ...833L...1A}%
  \BibitemOpen
  \bibfield  {author} {\bibinfo {author} {\bibfnamefont {B.~P.}\ \bibnamefont
  {{Abbott}}}, \bibinfo {author} {\bibfnamefont {R.}~\bibnamefont {{Abbott}}},
  \bibinfo {author} {\bibfnamefont {T.~D.}\ \bibnamefont {{Abbott}}}, \bibinfo
  {author} {\bibfnamefont {M.~R.}\ \bibnamefont {{Abernathy}}}, \bibinfo
  {author} {\bibfnamefont {F.}~\bibnamefont {{Acernese}}}, \bibinfo {author}
  {\bibfnamefont {K.}~\bibnamefont {{Ackley}}}, \bibinfo {author}
  {\bibfnamefont {C.}~\bibnamefont {{Adams}}}, \bibinfo {author} {\bibfnamefont
  {T.}~\bibnamefont {{Adams}}}, \bibinfo {author} {\bibfnamefont
  {P.}~\bibnamefont {{Addesso}}}, \bibinfo {author} {\bibfnamefont {R.~X.}\
  \bibnamefont {{Adhikari}}}, \bibinfo {author} {\bibfnamefont {V.~B.}\
  \bibnamefont {{Adya}}}, \bibinfo {author} {\bibfnamefont {C.}~\bibnamefont
  {{Affeldt}}}, \bibinfo {author} {\bibfnamefont {M.}~\bibnamefont
  {{Agathos}}}, \bibinfo {author} {\bibfnamefont {K.}~\bibnamefont
  {{Agatsuma}}}, \bibinfo {author} {\bibfnamefont {N.}~\bibnamefont
  {{Aggarwal}}}, \bibinfo {author} {\bibfnamefont {O.~D.}\ \bibnamefont
  {{Aguiar}}}, \bibinfo {author} {\bibfnamefont {L.}~\bibnamefont {{Aiello}}},
  \bibinfo {author} {\bibfnamefont {A.}~\bibnamefont {{Ain}}}, \bibinfo
  {author} {\bibfnamefont {P.}~\bibnamefont {{Ajith}}}, \bibinfo {author}
  {\bibfnamefont {B.}~\bibnamefont {{Allen}}}, \bibinfo {author} {\bibfnamefont
  {A.}~\bibnamefont {{Allocca}}}, \bibinfo {author} {\bibfnamefont {P.~A.}\
  \bibnamefont {{Altin}}}, \bibinfo {author} {\bibfnamefont {S.~B.}\
  \bibnamefont {{Anderson}}}, \bibinfo {author} {\bibfnamefont {W.~G.}\
  \bibnamefont {{Anderson}}}, \bibinfo {author} {\bibfnamefont
  {K.}~\bibnamefont {{Arai}}}, \bibinfo {author} {\bibfnamefont {M.~C.}\
  \bibnamefont {{Araya}}}, \bibinfo {author} {\bibfnamefont {C.~C.}\
  \bibnamefont {{Arceneaux}}}, \bibinfo {author} {\bibfnamefont {J.~S.}\
  \bibnamefont {{Areeda}}}, \bibinfo {author} {\bibfnamefont {N.}~\bibnamefont
  {{Arnaud}}}, \bibinfo {author} {\bibfnamefont {K.~G.}\ \bibnamefont
  {{Arun}}}, \bibinfo {author} {\bibfnamefont {S.}~\bibnamefont {{Ascenzi}}},
  \bibinfo {author} {\bibfnamefont {G.}~\bibnamefont {{Ashton}}}, \bibinfo
  {author} {\bibfnamefont {M.}~\bibnamefont {{Ast}}}, \bibinfo {author}
  {\bibfnamefont {S.~M.}\ \bibnamefont {{Aston}}}, \bibinfo {author}
  {\bibfnamefont {P.}~\bibnamefont {{Astone}}}, \bibinfo {author}
  {\bibfnamefont {P.}~\bibnamefont {{Aufmuth}}}, \bibinfo {author}
  {\bibfnamefont {C.}~\bibnamefont {{Aulbert}}}, \bibinfo {author}
  {\bibfnamefont {S.}~\bibnamefont {{Babak}}}, \bibinfo {author} {\bibfnamefont
  {P.}~\bibnamefont {{Bacon}}}, \bibinfo {author} {\bibfnamefont {M.~K.~M.}\
  \bibnamefont {{Bader}}}, \bibinfo {author} {\bibfnamefont {P.~T.}\
  \bibnamefont {{Baker}}}, \bibinfo {author} {\bibfnamefont {F.}~\bibnamefont
  {{Baldaccini}}}, \bibinfo {author} {\bibfnamefont {G.}~\bibnamefont
  {{Ballardin}}}, \bibinfo {author} {\bibfnamefont {S.~W.}\ \bibnamefont
  {{Ballmer}}}, \bibinfo {author} {\bibfnamefont {J.~C.}\ \bibnamefont
  {{Barayoga}}}, \bibinfo {author} {\bibfnamefont {S.~E.}\ \bibnamefont
  {{Barclay}}}, \bibinfo {author} {\bibfnamefont {B.~C.}\ \bibnamefont
  {{Barish}}}, \bibinfo {author} {\bibfnamefont {D.}~\bibnamefont {{Barker}}},
  \bibinfo {author} {\bibfnamefont {F.}~\bibnamefont {{Barone}}}, \bibinfo
  {author} {\bibfnamefont {B.}~\bibnamefont {{Barr}}}, \bibinfo {author}
  {\bibfnamefont {L.}~\bibnamefont {{Barsotti}}}, \bibinfo {author}
  {\bibfnamefont {M.}~\bibnamefont {{Barsuglia}}}, \bibinfo {author}
  {\bibfnamefont {D.}~\bibnamefont {{Barta}}}, \bibinfo {author} {\bibfnamefont
  {J.}~\bibnamefont {{Bartlett}}}, \bibinfo {author} {\bibfnamefont
  {I.}~\bibnamefont {{Bartos}}}, \bibinfo {author} {\bibfnamefont
  {R.}~\bibnamefont {{Bassiri}}}, \bibinfo {author} {\bibfnamefont
  {A.}~\bibnamefont {{Basti}}}, \bibinfo {author} {\bibfnamefont {J.~C.}\
  \bibnamefont {{Batch}}}, \bibinfo {author} {\bibfnamefont {C.}~\bibnamefont
  {{Baune}}}, \bibinfo {author} {\bibfnamefont {V.}~\bibnamefont
  {{Bavigadda}}}, \bibinfo {author} {\bibfnamefont {M.}~\bibnamefont
  {{Bazzan}}}, \bibinfo {author} {\bibfnamefont {B.}~\bibnamefont {{Behnke}}},
  \bibinfo {author} {\bibfnamefont {M.}~\bibnamefont {{Bejger}}}, \bibinfo
  {author} {\bibfnamefont {A.~S.}\ \bibnamefont {{Bell}}}, \bibinfo {author}
  {\bibfnamefont {C.~J.}\ \bibnamefont {{Bell}}}, \bibinfo {author}
  {\bibfnamefont {B.~K.}\ \bibnamefont {{Berger}}}, \bibinfo {author}
  {\bibfnamefont {J.}~\bibnamefont {{Bergman}}}, \bibinfo {author}
  {\bibfnamefont {G.}~\bibnamefont {{Bergmann}}}, \bibinfo {author}
  {\bibfnamefont {C.~P.~L.}\ \bibnamefont {{Berry}}}, \bibinfo {author}
  {\bibfnamefont {D.}~\bibnamefont {{Bersanetti}}}, \bibinfo {author}
  {\bibfnamefont {A.}~\bibnamefont {{Bertolini}}}, \bibinfo {author}
  {\bibfnamefont {J.}~\bibnamefont {{Betzwieser}}}, \bibinfo {author}
  {\bibfnamefont {S.}~\bibnamefont {{Bhagwat}}}, \bibinfo {author}
  {\bibfnamefont {R.}~\bibnamefont {{Bhandare}}}, \bibinfo {author}
  {\bibfnamefont {I.~A.}\ \bibnamefont {{Bilenko}}}, \bibinfo {author}
  {\bibfnamefont {G.}~\bibnamefont {{Billingsley}}}, \bibinfo {author}
  {\bibfnamefont {J.}~\bibnamefont {{Birch}}}, \bibinfo {author} {\bibfnamefont
  {R.}~\bibnamefont {{Birney}}}, \bibinfo {author} {\bibfnamefont
  {S.}~\bibnamefont {{Biscans}}}, \bibinfo {author} {\bibfnamefont
  {A.}~\bibnamefont {{Bisht}}}, \bibinfo {author} {\bibfnamefont
  {M.}~\bibnamefont {{Bitossi}}}, \bibinfo {author} {\bibfnamefont
  {C.}~\bibnamefont {{Biwer}}}, \bibinfo {author} {\bibfnamefont {M.~A.}\
  \bibnamefont {{Bizouard}}}, \bibinfo {author} {\bibfnamefont {J.~K.}\
  \bibnamefont {{Blackburn}}}, \bibinfo {author} {\bibfnamefont {C.~D.}\
  \bibnamefont {{Blair}}}, \bibinfo {author} {\bibfnamefont {D.~G.}\
  \bibnamefont {{Blair}}}, \bibinfo {author} {\bibfnamefont {R.~M.}\
  \bibnamefont {{Blair}}}, \bibinfo {author} {\bibfnamefont {S.}~\bibnamefont
  {{Bloemen}}}, \bibinfo {author} {\bibfnamefont {O.}~\bibnamefont {{Bock}}},
  \bibinfo {author} {\bibfnamefont {T.~P.}\ \bibnamefont {{Bodiya}}}, \bibinfo
  {author} {\bibfnamefont {M.}~\bibnamefont {{Boer}}}, \bibinfo {author}
  {\bibfnamefont {G.}~\bibnamefont {{Bogaert}}}, \bibinfo {author}
  {\bibfnamefont {C.}~\bibnamefont {{Bogan}}}, \bibinfo {author} {\bibfnamefont
  {A.}~\bibnamefont {{Bohe}}}, \bibinfo {author} {\bibfnamefont
  {P.}~\bibnamefont {{Bojtos}}}, \bibinfo {author} {\bibfnamefont
  {C.}~\bibnamefont {{Bond}}}, \bibinfo {author} {\bibfnamefont
  {F.}~\bibnamefont {{Bondu}}}, \bibinfo {author} {\bibfnamefont
  {R.}~\bibnamefont {{Bonnand}}}, \bibinfo {author} {\bibfnamefont {B.~A.}\
  \bibnamefont {{Boom}}}, \bibinfo {author} {\bibfnamefont {R.}~\bibnamefont
  {{Bork}}}, \bibinfo {author} {\bibfnamefont {V.}~\bibnamefont {{Boschi}}},
  \bibinfo {author} {\bibfnamefont {S.}~\bibnamefont {{Bose}}}, \bibinfo
  {author} {\bibfnamefont {Y.}~\bibnamefont {{Bouffanais}}}, \bibinfo {author}
  {\bibfnamefont {A.}~\bibnamefont {{Bozzi}}}, \bibinfo {author} {\bibfnamefont
  {C.}~\bibnamefont {{Bradaschia}}}, \bibinfo {author} {\bibfnamefont {P.~R.}\
  \bibnamefont {{Brady}}}, \bibinfo {author} {\bibfnamefont {V.~B.}\
  \bibnamefont {{Braginsky}}}, \bibinfo {author} {\bibfnamefont
  {M.}~\bibnamefont {{Branchesi}}}, \bibinfo {author} {\bibfnamefont {J.~E.}\
  \bibnamefont {{Brau}}}, \bibinfo {author} {\bibfnamefont {T.}~\bibnamefont
  {{Briant}}}, \bibinfo {author} {\bibfnamefont {A.}~\bibnamefont {{Brillet}}},
  \bibinfo {author} {\bibfnamefont {M.}~\bibnamefont {{Brinkmann}}}, \bibinfo
  {author} {\bibfnamefont {V.}~\bibnamefont {{Brisson}}}, \bibinfo {author}
  {\bibfnamefont {P.}~\bibnamefont {{Brockill}}}, \bibinfo {author}
  {\bibfnamefont {A.~F.}\ \bibnamefont {{Brooks}}}, \bibinfo {author}
  {\bibfnamefont {D.~A.}\ \bibnamefont {{Brown}}}, \bibinfo {author}
  {\bibfnamefont {D.~D.}\ \bibnamefont {{Brown}}}, \bibinfo {author}
  {\bibfnamefont {N.~M.}\ \bibnamefont {{Brown}}}, \bibinfo {author}
  {\bibfnamefont {C.~C.}\ \bibnamefont {{Buchanan}}}, \bibinfo {author}
  {\bibfnamefont {A.}~\bibnamefont {{Buikema}}}, \bibinfo {author}
  {\bibfnamefont {T.}~\bibnamefont {{Bulik}}}, \bibinfo {author} {\bibfnamefont
  {H.~J.}\ \bibnamefont {{Bulten}}}, \bibinfo {author} {\bibfnamefont
  {A.}~\bibnamefont {{Buonanno}}}, \bibinfo {author} {\bibfnamefont
  {D.}~\bibnamefont {{Buskulic}}}, \bibinfo {author} {\bibfnamefont
  {C.}~\bibnamefont {{Buy}}}, \bibinfo {author} {\bibfnamefont {R.~L.}\
  \bibnamefont {{Byer}}}, \bibinfo {author} {\bibfnamefont {L.}~\bibnamefont
  {{Cadonati}}}, \bibinfo {author} {\bibfnamefont {G.}~\bibnamefont
  {{Cagnoli}}}, \bibinfo {author} {\bibfnamefont {C.}~\bibnamefont
  {{Cahillane}}}, \bibinfo {author} {\bibfnamefont {J.}~\bibnamefont
  {{Calder{\'o}n Bustillo}}}, \bibinfo {author} {\bibfnamefont
  {T.}~\bibnamefont {{Callister}}}, \bibinfo {author} {\bibfnamefont
  {E.}~\bibnamefont {{Calloni}}}, \bibinfo {author} {\bibfnamefont {J.~B.}\
  \bibnamefont {{Camp}}}, \bibinfo {author} {\bibfnamefont {K.~C.}\
  \bibnamefont {{Cannon}}}, \bibinfo {author} {\bibfnamefont {J.}~\bibnamefont
  {{Cao}}}, \bibinfo {author} {\bibfnamefont {C.~D.}\ \bibnamefont {{Capano}}},
  \bibinfo {author} {\bibfnamefont {E.}~\bibnamefont {{Capocasa}}}, \bibinfo
  {author} {\bibfnamefont {F.}~\bibnamefont {{Carbognani}}}, \bibinfo {author}
  {\bibfnamefont {S.}~\bibnamefont {{Caride}}}, \bibinfo {author}
  {\bibfnamefont {J.}~\bibnamefont {{Casanueva Diaz}}}, \bibinfo {author}
  {\bibfnamefont {C.}~\bibnamefont {{Casentini}}}, \bibinfo {author}
  {\bibfnamefont {S.}~\bibnamefont {{Caudill}}}, \bibinfo {author}
  {\bibfnamefont {M.}~\bibnamefont {{Cavagli{\`a}}}}, \bibinfo {author}
  {\bibfnamefont {F.}~\bibnamefont {{Cavalier}}}, \bibinfo {author}
  {\bibfnamefont {R.}~\bibnamefont {{Cavalieri}}}, \bibinfo {author}
  {\bibfnamefont {G.}~\bibnamefont {{Cella}}}, \bibinfo {author} {\bibfnamefont
  {C.~B.}\ \bibnamefont {{Cepeda}}}, \bibinfo {author} {\bibfnamefont
  {L.}~\bibnamefont {{Cerboni Baiardi}}}, \bibinfo {author} {\bibfnamefont
  {G.}~\bibnamefont {{Cerretani}}}, \bibinfo {author} {\bibfnamefont
  {E.}~\bibnamefont {{Cesarini}}}, \bibinfo {author} {\bibfnamefont
  {R.}~\bibnamefont {{Chakraborty}}}, \bibinfo {author} {\bibfnamefont
  {T.}~\bibnamefont {{Chalermsongsak}}}, \bibinfo {author} {\bibfnamefont
  {S.~J.}\ \bibnamefont {{Chamberlin}}}, \bibinfo {author} {\bibfnamefont
  {M.}~\bibnamefont {{Chan}}}, \bibinfo {author} {\bibfnamefont
  {S.}~\bibnamefont {{Chao}}}, \bibinfo {author} {\bibfnamefont
  {P.}~\bibnamefont {{Charlton}}}, \bibinfo {author} {\bibfnamefont
  {E.}~\bibnamefont {{Chassande-Mottin}}}, \bibinfo {author} {\bibfnamefont
  {H.~Y.}\ \bibnamefont {{Chen}}}, \bibinfo {author} {\bibfnamefont
  {Y.}~\bibnamefont {{Chen}}}, \bibinfo {author} {\bibfnamefont
  {C.}~\bibnamefont {{Cheng}}}, \bibinfo {author} {\bibfnamefont
  {A.}~\bibnamefont {{Chincarini}}}, \bibinfo {author} {\bibfnamefont
  {A.}~\bibnamefont {{Chiummo}}}, \bibinfo {author} {\bibfnamefont {H.~S.}\
  \bibnamefont {{Cho}}}, \bibinfo {author} {\bibfnamefont {M.}~\bibnamefont
  {{Cho}}}, \bibinfo {author} {\bibfnamefont {J.~H.}\ \bibnamefont {{Chow}}},
  \bibinfo {author} {\bibfnamefont {N.}~\bibnamefont {{Christensen}}}, \bibinfo
  {author} {\bibfnamefont {Q.}~\bibnamefont {{Chu}}}, \bibinfo {author}
  {\bibfnamefont {S.}~\bibnamefont {{Chua}}}, \bibinfo {author} {\bibfnamefont
  {S.}~\bibnamefont {{Chung}}}, \bibinfo {author} {\bibfnamefont
  {G.}~\bibnamefont {{Ciani}}}, \bibinfo {author} {\bibfnamefont
  {F.}~\bibnamefont {{Clara}}}, \bibinfo {author} {\bibfnamefont {J.~A.}\
  \bibnamefont {{Clark}}}, \bibinfo {author} {\bibfnamefont {F.}~\bibnamefont
  {{Cleva}}}, \bibinfo {author} {\bibfnamefont {E.}~\bibnamefont {{Coccia}}},
  \bibinfo {author} {\bibfnamefont {P.~F.}\ \bibnamefont {{Cohadon}}}, \bibinfo
  {author} {\bibfnamefont {A.}~\bibnamefont {{Colla}}}, \bibinfo {author}
  {\bibfnamefont {C.~G.}\ \bibnamefont {{Collette}}}, \bibinfo {author}
  {\bibfnamefont {L.}~\bibnamefont {{Cominsky}}}, \bibinfo {author}
  {\bibfnamefont {J.}~\bibnamefont {{Constancio}}, \bibfnamefont {M.}},
  \bibinfo {author} {\bibfnamefont {A.}~\bibnamefont {{Conte}}}, \bibinfo
  {author} {\bibfnamefont {L.}~\bibnamefont {{Conti}}}, \bibinfo {author}
  {\bibfnamefont {D.}~\bibnamefont {{Cook}}}, \bibinfo {author} {\bibfnamefont
  {T.~R.}\ \bibnamefont {{Corbitt}}}, \bibinfo {author} {\bibfnamefont
  {N.}~\bibnamefont {{Cornish}}}, \bibinfo {author} {\bibfnamefont
  {A.}~\bibnamefont {{Corsi}}}, \bibinfo {author} {\bibfnamefont
  {S.}~\bibnamefont {{Cortese}}}, \bibinfo {author} {\bibfnamefont {C.~A.}\
  \bibnamefont {{Costa}}}, \bibinfo {author} {\bibfnamefont {M.~W.}\
  \bibnamefont {{Coughlin}}}, \bibinfo {author} {\bibfnamefont {S.~B.}\
  \bibnamefont {{Coughlin}}}, \bibinfo {author} {\bibfnamefont {J.~P.}\
  \bibnamefont {{Coulon}}}, \bibinfo {author} {\bibfnamefont {S.~T.}\
  \bibnamefont {{Countryman}}}, \bibinfo {author} {\bibfnamefont
  {P.}~\bibnamefont {{Couvares}}}, \bibinfo {author} {\bibfnamefont {E.~E.}\
  \bibnamefont {{Cowan}}}, \bibinfo {author} {\bibfnamefont {D.~M.}\
  \bibnamefont {{Coward}}}, \bibinfo {author} {\bibfnamefont {M.~J.}\
  \bibnamefont {{Cowart}}}, \bibinfo {author} {\bibfnamefont {D.~C.}\
  \bibnamefont {{Coyne}}}, \bibinfo {author} {\bibfnamefont {R.}~\bibnamefont
  {{Coyne}}}, \bibinfo {author} {\bibfnamefont {K.}~\bibnamefont {{Craig}}},
  \bibinfo {author} {\bibfnamefont {J.~D.~E.}\ \bibnamefont {{Creighton}}},
  \bibinfo {author} {\bibfnamefont {J.}~\bibnamefont {{Cripe}}}, \bibinfo
  {author} {\bibfnamefont {S.~G.}\ \bibnamefont {{Crowder}}}, \bibinfo {author}
  {\bibfnamefont {A.}~\bibnamefont {{Cumming}}}, \bibinfo {author}
  {\bibfnamefont {L.}~\bibnamefont {{Cunningham}}}, \bibinfo {author}
  {\bibfnamefont {E.}~\bibnamefont {{Cuoco}}}, \bibinfo {author} {\bibfnamefont
  {T.}~\bibnamefont {{Dal Canton}}}, \bibinfo {author} {\bibfnamefont {S.~L.}\
  \bibnamefont {{Danilishin}}}, \bibinfo {author} {\bibfnamefont
  {S.}~\bibnamefont {{D'Antonio}}}, \bibinfo {author} {\bibfnamefont
  {K.}~\bibnamefont {{Danzmann}}}, \bibinfo {author} {\bibfnamefont {N.~S.}\
  \bibnamefont {{Darman}}}, \bibinfo {author} {\bibfnamefont {V.}~\bibnamefont
  {{Dattilo}}}, \bibinfo {author} {\bibfnamefont {I.}~\bibnamefont {{Dave}}},
  \bibinfo {author} {\bibfnamefont {H.~P.}\ \bibnamefont {{Daveloza}}},
  \bibinfo {author} {\bibfnamefont {M.}~\bibnamefont {{Davier}}}, \bibinfo
  {author} {\bibfnamefont {G.~S.}\ \bibnamefont {{Davies}}}, \bibinfo {author}
  {\bibfnamefont {E.~J.}\ \bibnamefont {{Daw}}}, \bibinfo {author}
  {\bibfnamefont {R.}~\bibnamefont {{Day}}}, \bibinfo {author} {\bibfnamefont
  {S.}~\bibnamefont {{De}}}, \bibinfo {author} {\bibfnamefont {D.}~\bibnamefont
  {{DeBra}}}, \bibinfo {author} {\bibfnamefont {G.}~\bibnamefont
  {{Debreczeni}}}, \bibinfo {author} {\bibfnamefont {J.}~\bibnamefont
  {{Degallaix}}}, \bibinfo {author} {\bibfnamefont {M.}~\bibnamefont {{De
  Laurentis}}}, \bibinfo {author} {\bibfnamefont {S.}~\bibnamefont
  {{Del{\'e}glise}}}, \bibinfo {author} {\bibfnamefont {W.}~\bibnamefont {{Del
  Pozzo}}}, \bibinfo {author} {\bibfnamefont {T.}~\bibnamefont {{Denker}}},
  \bibinfo {author} {\bibfnamefont {T.}~\bibnamefont {{Dent}}}, \bibinfo
  {author} {\bibfnamefont {H.}~\bibnamefont {{Dereli}}}, \bibinfo {author}
  {\bibfnamefont {V.}~\bibnamefont {{Dergachev}}}, \bibinfo {author}
  {\bibfnamefont {R.}~\bibnamefont {{De Rosa}}}, \bibinfo {author}
  {\bibfnamefont {R.~T.}\ \bibnamefont {{DeRosa}}}, \bibinfo {author}
  {\bibfnamefont {R.}~\bibnamefont {{DeSalvo}}}, \bibinfo {author}
  {\bibfnamefont {S.}~\bibnamefont {{Dhurandhar}}}, \bibinfo {author}
  {\bibfnamefont {M.~C.}\ \bibnamefont {{D{\'\i}az}}}, \bibinfo {author}
  {\bibfnamefont {L.}~\bibnamefont {{Di Fiore}}}, \bibinfo {author}
  {\bibfnamefont {M.}~\bibnamefont {{Di Giovanni}}}, \bibinfo {author}
  {\bibfnamefont {A.}~\bibnamefont {{Di Lieto}}}, \bibinfo {author}
  {\bibfnamefont {S.}~\bibnamefont {{Di Pace}}}, \bibinfo {author}
  {\bibfnamefont {I.}~\bibnamefont {{Di Palma}}}, \bibinfo {author}
  {\bibfnamefont {A.}~\bibnamefont {{Di Virgilio}}}, \bibinfo {author}
  {\bibfnamefont {G.}~\bibnamefont {{Dojcinoski}}}, \bibinfo {author}
  {\bibfnamefont {V.}~\bibnamefont {{Dolique}}}, \bibinfo {author}
  {\bibfnamefont {F.}~\bibnamefont {{Donovan}}}, \bibinfo {author}
  {\bibfnamefont {K.~L.}\ \bibnamefont {{Dooley}}}, \bibinfo {author}
  {\bibfnamefont {S.}~\bibnamefont {{Doravari}}}, \bibinfo {author}
  {\bibfnamefont {R.}~\bibnamefont {{Douglas}}}, \bibinfo {author}
  {\bibfnamefont {T.~P.}\ \bibnamefont {{Downes}}}, \bibinfo {author}
  {\bibfnamefont {M.}~\bibnamefont {{Drago}}}, \bibinfo {author} {\bibfnamefont
  {R.~W.~P.}\ \bibnamefont {{Drever}}}, \bibinfo {author} {\bibfnamefont
  {J.~C.}\ \bibnamefont {{Driggers}}}, \bibinfo {author} {\bibfnamefont
  {Z.}~\bibnamefont {{Du}}}, \bibinfo {author} {\bibfnamefont {M.}~\bibnamefont
  {{Ducrot}}}, \bibinfo {author} {\bibfnamefont {S.~E.}\ \bibnamefont
  {{Dwyer}}}, \bibinfo {author} {\bibfnamefont {T.~B.}\ \bibnamefont {{Edo}}},
  \bibinfo {author} {\bibfnamefont {M.~C.}\ \bibnamefont {{Edwards}}}, \bibinfo
  {author} {\bibfnamefont {A.}~\bibnamefont {{Effler}}}, \bibinfo {author}
  {\bibfnamefont {H.~B.}\ \bibnamefont {{Eggenstein}}}, \bibinfo {author}
  {\bibfnamefont {P.}~\bibnamefont {{Ehrens}}}, \bibinfo {author}
  {\bibfnamefont {J.}~\bibnamefont {{Eichholz}}}, \bibinfo {author}
  {\bibfnamefont {S.~S.}\ \bibnamefont {{Eikenberry}}}, \bibinfo {author}
  {\bibfnamefont {W.}~\bibnamefont {{Engels}}}, \bibinfo {author}
  {\bibfnamefont {R.~C.}\ \bibnamefont {{Essick}}}, \bibinfo {author}
  {\bibfnamefont {T.}~\bibnamefont {{Etzel}}}, \bibinfo {author} {\bibfnamefont
  {M.}~\bibnamefont {{Evans}}}, \bibinfo {author} {\bibfnamefont {T.~M.}\
  \bibnamefont {{Evans}}}, \bibinfo {author} {\bibfnamefont {R.}~\bibnamefont
  {{Everett}}}, \bibinfo {author} {\bibfnamefont {M.}~\bibnamefont
  {{Factourovich}}}, \bibinfo {author} {\bibfnamefont {V.}~\bibnamefont
  {{Fafone}}}, \bibinfo {author} {\bibfnamefont {H.}~\bibnamefont {{Fair}}},
  \bibinfo {author} {\bibfnamefont {S.}~\bibnamefont {{Fairhurst}}}, \bibinfo
  {author} {\bibfnamefont {X.}~\bibnamefont {{Fan}}}, \bibinfo {author}
  {\bibfnamefont {Q.}~\bibnamefont {{Fang}}}, \bibinfo {author} {\bibfnamefont
  {S.}~\bibnamefont {{Farinon}}}, \bibinfo {author} {\bibfnamefont
  {B.}~\bibnamefont {{Farr}}}, \bibinfo {author} {\bibfnamefont {W.~M.}\
  \bibnamefont {{Farr}}}, \bibinfo {author} {\bibfnamefont {M.}~\bibnamefont
  {{Favata}}}, \bibinfo {author} {\bibfnamefont {M.}~\bibnamefont {{Fays}}},
  \bibinfo {author} {\bibfnamefont {H.}~\bibnamefont {{Fehrmann}}}, \bibinfo
  {author} {\bibfnamefont {M.~M.}\ \bibnamefont {{Fejer}}}, \bibinfo {author}
  {\bibfnamefont {I.}~\bibnamefont {{Ferrante}}}, \bibinfo {author}
  {\bibfnamefont {E.~C.}\ \bibnamefont {{Ferreira}}}, \bibinfo {author}
  {\bibfnamefont {F.}~\bibnamefont {{Ferrini}}}, \bibinfo {author}
  {\bibfnamefont {F.}~\bibnamefont {{Fidecaro}}}, \bibinfo {author}
  {\bibfnamefont {I.}~\bibnamefont {{Fiori}}}, \bibinfo {author} {\bibfnamefont
  {D.}~\bibnamefont {{Fiorucci}}}, \bibinfo {author} {\bibfnamefont {R.~P.}\
  \bibnamefont {{Fisher}}}, \bibinfo {author} {\bibfnamefont {R.}~\bibnamefont
  {{Flaminio}}}, \bibinfo {author} {\bibfnamefont {M.}~\bibnamefont
  {{Fletcher}}}, \bibinfo {author} {\bibfnamefont {H.}~\bibnamefont {{Fong}}},
  \bibinfo {author} {\bibfnamefont {J.~D.}\ \bibnamefont {{Fournier}}},
  \bibinfo {author} {\bibfnamefont {S.}~\bibnamefont {{Franco}}}, \bibinfo
  {author} {\bibfnamefont {S.}~\bibnamefont {{Frasca}}}, \bibinfo {author}
  {\bibfnamefont {F.}~\bibnamefont {{Frasconi}}}, \bibinfo {author}
  {\bibfnamefont {Z.}~\bibnamefont {{Frei}}}, \bibinfo {author} {\bibfnamefont
  {A.}~\bibnamefont {{Freise}}}, \bibinfo {author} {\bibfnamefont
  {R.}~\bibnamefont {{Frey}}}, \bibinfo {author} {\bibfnamefont
  {V.}~\bibnamefont {{Frey}}}, \bibinfo {author} {\bibfnamefont {T.~T.}\
  \bibnamefont {{Fricke}}}, \bibinfo {author} {\bibfnamefont {P.}~\bibnamefont
  {{Fritschel}}}, \bibinfo {author} {\bibfnamefont {V.~V.}\ \bibnamefont
  {{Frolov}}}, \bibinfo {author} {\bibfnamefont {P.}~\bibnamefont {{Fulda}}},
  \bibinfo {author} {\bibfnamefont {M.}~\bibnamefont {{Fyffe}}}, \bibinfo
  {author} {\bibfnamefont {H.~A.~G.}\ \bibnamefont {{Gabbard}}}, \bibinfo
  {author} {\bibfnamefont {J.~R.}\ \bibnamefont {{Gair}}}, \bibinfo {author}
  {\bibfnamefont {L.}~\bibnamefont {{Gammaitoni}}}, \bibinfo {author}
  {\bibfnamefont {S.~G.}\ \bibnamefont {{Gaonkar}}}, \bibinfo {author}
  {\bibfnamefont {F.}~\bibnamefont {{Garufi}}}, \bibinfo {author}
  {\bibfnamefont {A.}~\bibnamefont {{Gatto}}}, \bibinfo {author} {\bibfnamefont
  {G.}~\bibnamefont {{Gaur}}}, \bibinfo {author} {\bibfnamefont
  {N.}~\bibnamefont {{Gehrels}}}, \bibinfo {author} {\bibfnamefont
  {G.}~\bibnamefont {{Gemme}}}, \bibinfo {author} {\bibfnamefont
  {B.}~\bibnamefont {{Gendre}}}, \bibinfo {author} {\bibfnamefont
  {E.}~\bibnamefont {{Genin}}}, \bibinfo {author} {\bibfnamefont
  {A.}~\bibnamefont {{Gennai}}}, \bibinfo {author} {\bibfnamefont
  {J.}~\bibnamefont {{George}}}, \bibinfo {author} {\bibfnamefont
  {L.}~\bibnamefont {{Gergely}}}, \bibinfo {author} {\bibfnamefont
  {V.}~\bibnamefont {{Germain}}}, \bibinfo {author} {\bibfnamefont
  {A.}~\bibnamefont {{Ghosh}}}, \bibinfo {author} {\bibfnamefont
  {S.}~\bibnamefont {{Ghosh}}}, \bibinfo {author} {\bibfnamefont {J.~A.}\
  \bibnamefont {{Giaime}}}, \bibinfo {author} {\bibfnamefont {K.~D.}\
  \bibnamefont {{Giardina}}}, \bibinfo {author} {\bibfnamefont
  {A.}~\bibnamefont {{Giazotto}}}, \bibinfo {author} {\bibfnamefont
  {K.}~\bibnamefont {{Gill}}}, \bibinfo {author} {\bibfnamefont
  {A.}~\bibnamefont {{Glaefke}}}, \bibinfo {author} {\bibfnamefont
  {E.}~\bibnamefont {{Goetz}}}, \bibinfo {author} {\bibfnamefont
  {R.}~\bibnamefont {{Goetz}}}, \bibinfo {author} {\bibfnamefont
  {L.}~\bibnamefont {{Gondan}}}, \bibinfo {author} {\bibfnamefont
  {G.}~\bibnamefont {{Gonz{\'a}lez}}}, \bibinfo {author} {\bibfnamefont
  {J.~M.}\ \bibnamefont {{Gonzalez Castro}}}, \bibinfo {author} {\bibfnamefont
  {A.}~\bibnamefont {{Gopakumar}}}, \bibinfo {author} {\bibfnamefont {N.~A.}\
  \bibnamefont {{Gordon}}}, \bibinfo {author} {\bibfnamefont {M.~L.}\
  \bibnamefont {{Gorodetsky}}}, \bibinfo {author} {\bibfnamefont {S.~E.}\
  \bibnamefont {{Gossan}}}, \bibinfo {author} {\bibfnamefont {M.}~\bibnamefont
  {{Gosselin}}}, \bibinfo {author} {\bibfnamefont {R.}~\bibnamefont
  {{Gouaty}}}, \bibinfo {author} {\bibfnamefont {C.}~\bibnamefont {{Graef}}},
  \bibinfo {author} {\bibfnamefont {P.~B.}\ \bibnamefont {{Graff}}}, \bibinfo
  {author} {\bibfnamefont {M.}~\bibnamefont {{Granata}}}, \bibinfo {author}
  {\bibfnamefont {A.}~\bibnamefont {{Grant}}}, \bibinfo {author} {\bibfnamefont
  {S.}~\bibnamefont {{Gras}}}, \bibinfo {author} {\bibfnamefont
  {C.}~\bibnamefont {{Gray}}}, \bibinfo {author} {\bibfnamefont
  {G.}~\bibnamefont {{Greco}}}, \bibinfo {author} {\bibfnamefont {A.~C.}\
  \bibnamefont {{Green}}}, \bibinfo {author} {\bibfnamefont {P.}~\bibnamefont
  {{Groot}}}, \bibinfo {author} {\bibfnamefont {H.}~\bibnamefont {{Grote}}},
  \bibinfo {author} {\bibfnamefont {S.}~\bibnamefont {{Grunewald}}}, \bibinfo
  {author} {\bibfnamefont {G.~M.}\ \bibnamefont {{Guidi}}}, \bibinfo {author}
  {\bibfnamefont {X.}~\bibnamefont {{Guo}}}, \bibinfo {author} {\bibfnamefont
  {A.}~\bibnamefont {{Gupta}}}, \bibinfo {author} {\bibfnamefont {M.~K.}\
  \bibnamefont {{Gupta}}}, \bibinfo {author} {\bibfnamefont {K.~E.}\
  \bibnamefont {{Gushwa}}}, \bibinfo {author} {\bibfnamefont {E.~K.}\
  \bibnamefont {{Gustafson}}}, \bibinfo {author} {\bibfnamefont
  {R.}~\bibnamefont {{Gustafson}}}, \bibinfo {author} {\bibfnamefont {J.~J.}\
  \bibnamefont {{Hacker}}}, \bibinfo {author} {\bibfnamefont {B.~R.}\
  \bibnamefont {{Hall}}}, \bibinfo {author} {\bibfnamefont {E.~D.}\
  \bibnamefont {{Hall}}}, \bibinfo {author} {\bibfnamefont {G.}~\bibnamefont
  {{Hammond}}}, \bibinfo {author} {\bibfnamefont {M.}~\bibnamefont {{Haney}}},
  \bibinfo {author} {\bibfnamefont {M.~M.}\ \bibnamefont {{Hanke}}}, \bibinfo
  {author} {\bibfnamefont {J.}~\bibnamefont {{Hanks}}}, \bibinfo {author}
  {\bibfnamefont {C.}~\bibnamefont {{Hanna}}}, \bibinfo {author} {\bibfnamefont
  {M.~D.}\ \bibnamefont {{Hannam}}}, \bibinfo {author} {\bibfnamefont
  {J.}~\bibnamefont {{Hanson}}}, \bibinfo {author} {\bibfnamefont
  {T.}~\bibnamefont {{Hardwick}}}, \bibinfo {author} {\bibfnamefont
  {J.}~\bibnamefont {{Harms}}}, \bibinfo {author} {\bibfnamefont {G.~M.}\
  \bibnamefont {{Harry}}}, \bibinfo {author} {\bibfnamefont {I.~W.}\
  \bibnamefont {{Harry}}}, \bibinfo {author} {\bibfnamefont {M.~J.}\
  \bibnamefont {{Hart}}}, \bibinfo {author} {\bibfnamefont {M.~T.}\
  \bibnamefont {{Hartman}}}, \bibinfo {author} {\bibfnamefont {C.~J.}\
  \bibnamefont {{Haster}}}, \bibinfo {author} {\bibfnamefont {K.}~\bibnamefont
  {{Haughian}}}, \bibinfo {author} {\bibfnamefont {A.}~\bibnamefont
  {{Heidmann}}}, \bibinfo {author} {\bibfnamefont {M.~C.}\ \bibnamefont
  {{Heintze}}}, \bibinfo {author} {\bibfnamefont {H.}~\bibnamefont
  {{Heitmann}}}, \bibinfo {author} {\bibfnamefont {P.}~\bibnamefont {{Hello}}},
  \bibinfo {author} {\bibfnamefont {G.}~\bibnamefont {{Hemming}}}, \bibinfo
  {author} {\bibfnamefont {M.}~\bibnamefont {{Hendry}}}, \bibinfo {author}
  {\bibfnamefont {I.~S.}\ \bibnamefont {{Heng}}}, \bibinfo {author}
  {\bibfnamefont {J.}~\bibnamefont {{Hennig}}}, \bibinfo {author}
  {\bibfnamefont {A.~W.}\ \bibnamefont {{Heptonstall}}}, \bibinfo {author}
  {\bibfnamefont {M.}~\bibnamefont {{Heurs}}}, \bibinfo {author} {\bibfnamefont
  {S.}~\bibnamefont {{Hild}}}, \bibinfo {author} {\bibfnamefont
  {D.}~\bibnamefont {{Hoak}}}, \bibinfo {author} {\bibfnamefont {K.~A.}\
  \bibnamefont {{Hodge}}}, \bibinfo {author} {\bibfnamefont {D.}~\bibnamefont
  {{Hofman}}}, \bibinfo {author} {\bibfnamefont {S.~E.}\ \bibnamefont
  {{Hollitt}}}, \bibinfo {author} {\bibfnamefont {K.}~\bibnamefont {{Holt}}},
  \bibinfo {author} {\bibfnamefont {D.~E.}\ \bibnamefont {{Holz}}}, \bibinfo
  {author} {\bibfnamefont {P.}~\bibnamefont {{Hopkins}}}, \bibinfo {author}
  {\bibfnamefont {D.~J.}\ \bibnamefont {{Hosken}}}, \bibinfo {author}
  {\bibfnamefont {J.}~\bibnamefont {{Hough}}}, \bibinfo {author} {\bibfnamefont
  {E.~A.}\ \bibnamefont {{Houston}}}, \bibinfo {author} {\bibfnamefont {E.~J.}\
  \bibnamefont {{Howell}}}, \bibinfo {author} {\bibfnamefont {Y.~M.}\
  \bibnamefont {{Hu}}}, \bibinfo {author} {\bibfnamefont {S.}~\bibnamefont
  {{Huang}}}, \bibinfo {author} {\bibfnamefont {E.~A.}\ \bibnamefont
  {{Huerta}}}, \bibinfo {author} {\bibfnamefont {D.}~\bibnamefont {{Huet}}},
  \bibinfo {author} {\bibfnamefont {B.}~\bibnamefont {{Hughey}}}, \bibinfo
  {author} {\bibfnamefont {S.}~\bibnamefont {{Husa}}}, \bibinfo {author}
  {\bibfnamefont {S.~H.}\ \bibnamefont {{Huttner}}}, \bibinfo {author}
  {\bibfnamefont {T.}~\bibnamefont {{Huynh-Dinh}}}, \bibinfo {author}
  {\bibfnamefont {A.}~\bibnamefont {{Idrisy}}}, \bibinfo {author}
  {\bibfnamefont {N.}~\bibnamefont {{Indik}}}, \bibinfo {author} {\bibfnamefont
  {D.~R.}\ \bibnamefont {{Ingram}}}, \bibinfo {author} {\bibfnamefont
  {R.}~\bibnamefont {{Inta}}}, \bibinfo {author} {\bibfnamefont {H.~N.}\
  \bibnamefont {{Isa}}}, \bibinfo {author} {\bibfnamefont {J.~M.}\ \bibnamefont
  {{Isac}}}, \bibinfo {author} {\bibfnamefont {M.}~\bibnamefont {{Isi}}},
  \bibinfo {author} {\bibfnamefont {G.}~\bibnamefont {{Islas}}}, \bibinfo
  {author} {\bibfnamefont {T.}~\bibnamefont {{Isogai}}}, \bibinfo {author}
  {\bibfnamefont {B.~R.}\ \bibnamefont {{Iyer}}}, \bibinfo {author}
  {\bibfnamefont {K.}~\bibnamefont {{Izumi}}}, \bibinfo {author} {\bibfnamefont
  {T.}~\bibnamefont {{Jacqmin}}}, \bibinfo {author} {\bibfnamefont
  {H.}~\bibnamefont {{Jang}}}, \bibinfo {author} {\bibfnamefont
  {K.}~\bibnamefont {{Jani}}}, \bibinfo {author} {\bibfnamefont
  {P.}~\bibnamefont {{Jaranowski}}}, \bibinfo {author} {\bibfnamefont
  {S.}~\bibnamefont {{Jawahar}}}, \bibinfo {author} {\bibfnamefont
  {F.}~\bibnamefont {{Jim{\'e}nez-Forteza}}}, \bibinfo {author} {\bibfnamefont
  {W.~W.}\ \bibnamefont {{Johnson}}}, \bibinfo {author} {\bibfnamefont {D.~I.}\
  \bibnamefont {{Jones}}}, \bibinfo {author} {\bibfnamefont {R.}~\bibnamefont
  {{Jones}}}, \bibinfo {author} {\bibfnamefont {R.~J.~G.}\ \bibnamefont
  {{Jonker}}}, \bibinfo {author} {\bibfnamefont {L.}~\bibnamefont {{Ju}}},
  \bibinfo {author} {\bibfnamefont {H.}~\bibnamefont {{K}}}, \bibinfo {author}
  {\bibfnamefont {C.~V.}\ \bibnamefont {{Kalaghatgi}}}, \bibinfo {author}
  {\bibfnamefont {V.}~\bibnamefont {{Kalogera}}}, \bibinfo {author}
  {\bibfnamefont {S.}~\bibnamefont {{Kandhasamy}}}, \bibinfo {author}
  {\bibfnamefont {G.}~\bibnamefont {{Kang}}}, \bibinfo {author} {\bibfnamefont
  {J.~B.}\ \bibnamefont {{Kanner}}}, \bibinfo {author} {\bibfnamefont
  {S.}~\bibnamefont {{Karki}}}, \bibinfo {author} {\bibfnamefont
  {M.}~\bibnamefont {{Kasprzack}}}, \bibinfo {author} {\bibfnamefont
  {E.}~\bibnamefont {{Katsavounidis}}}, \bibinfo {author} {\bibfnamefont
  {W.}~\bibnamefont {{Katzman}}}, \bibinfo {author} {\bibfnamefont
  {S.}~\bibnamefont {{Kaufer}}}, \bibinfo {author} {\bibfnamefont
  {T.}~\bibnamefont {{Kaur}}}, \bibinfo {author} {\bibfnamefont
  {K.}~\bibnamefont {{Kawabe}}}, \bibinfo {author} {\bibfnamefont
  {F.}~\bibnamefont {{Kawazoe}}}, \bibinfo {author} {\bibfnamefont
  {F.}~\bibnamefont {{K{\'e}f{\'e}lian}}}, \bibinfo {author} {\bibfnamefont
  {M.~S.}\ \bibnamefont {{Kehl}}}, \bibinfo {author} {\bibfnamefont
  {D.}~\bibnamefont {{Keitel}}}, \bibinfo {author} {\bibfnamefont {D.~B.}\
  \bibnamefont {{Kelley}}}, \bibinfo {author} {\bibfnamefont {W.}~\bibnamefont
  {{Kells}}}, \bibinfo {author} {\bibfnamefont {R.}~\bibnamefont {{Kennedy}}},
  \bibinfo {author} {\bibfnamefont {J.~S.}\ \bibnamefont {{Key}}}, \bibinfo
  {author} {\bibfnamefont {A.}~\bibnamefont {{Khalaidovski}}}, \bibinfo
  {author} {\bibfnamefont {F.~Y.}\ \bibnamefont {{Khalili}}}, \bibinfo {author}
  {\bibfnamefont {I.}~\bibnamefont {{Khan}}}, \bibinfo {author} {\bibfnamefont
  {S.}~\bibnamefont {{Khan}}}, \bibinfo {author} {\bibfnamefont
  {Z.}~\bibnamefont {{Khan}}}, \bibinfo {author} {\bibfnamefont {E.~A.}\
  \bibnamefont {{Khazanov}}}, \bibinfo {author} {\bibfnamefont
  {N.}~\bibnamefont {{Kijbunchoo}}}, \bibinfo {author} {\bibfnamefont
  {C.}~\bibnamefont {{Kim}}}, \bibinfo {author} {\bibfnamefont
  {J.}~\bibnamefont {{Kim}}}, \bibinfo {author} {\bibfnamefont
  {K.}~\bibnamefont {{Kim}}}, \bibinfo {author} {\bibfnamefont {N.-G.}\
  \bibnamefont {{Kim}}}, \bibinfo {author} {\bibfnamefont {N.}~\bibnamefont
  {{Kim}}}, \bibinfo {author} {\bibfnamefont {Y.~M.}\ \bibnamefont {{Kim}}},
  \bibinfo {author} {\bibfnamefont {E.~J.}\ \bibnamefont {{King}}}, \bibinfo
  {author} {\bibfnamefont {P.~J.}\ \bibnamefont {{King}}}, \bibinfo {author}
  {\bibfnamefont {D.~L.}\ \bibnamefont {{Kinzel}}}, \bibinfo {author}
  {\bibfnamefont {J.~S.}\ \bibnamefont {{Kissel}}}, \bibinfo {author}
  {\bibfnamefont {L.}~\bibnamefont {{Kleybolte}}}, \bibinfo {author}
  {\bibfnamefont {S.}~\bibnamefont {{Klimenko}}}, \bibinfo {author}
  {\bibfnamefont {S.~M.}\ \bibnamefont {{Koehlenbeck}}}, \bibinfo {author}
  {\bibfnamefont {K.}~\bibnamefont {{Kokeyama}}}, \bibinfo {author}
  {\bibfnamefont {S.}~\bibnamefont {{Koley}}}, \bibinfo {author} {\bibfnamefont
  {V.}~\bibnamefont {{Kondrashov}}}, \bibinfo {author} {\bibfnamefont
  {A.}~\bibnamefont {{Kontos}}}, \bibinfo {author} {\bibfnamefont
  {M.}~\bibnamefont {{Korobko}}}, \bibinfo {author} {\bibfnamefont {W.~Z.}\
  \bibnamefont {{Korth}}}, \bibinfo {author} {\bibfnamefont {I.}~\bibnamefont
  {{Kowalska}}}, \bibinfo {author} {\bibfnamefont {D.~B.}\ \bibnamefont
  {{Kozak}}}, \bibinfo {author} {\bibfnamefont {V.}~\bibnamefont {{Kringel}}},
  \bibinfo {author} {\bibfnamefont {B.}~\bibnamefont {{Krishnan}}}, \bibinfo
  {author} {\bibfnamefont {A.}~\bibnamefont {{Kr{\'o}lak}}}, \bibinfo {author}
  {\bibfnamefont {C.}~\bibnamefont {{Krueger}}}, \bibinfo {author}
  {\bibfnamefont {G.}~\bibnamefont {{Kuehn}}}, \bibinfo {author} {\bibfnamefont
  {P.}~\bibnamefont {{Kumar}}}, \bibinfo {author} {\bibfnamefont
  {L.}~\bibnamefont {{Kuo}}}, \bibinfo {author} {\bibfnamefont
  {A.}~\bibnamefont {{Kutynia}}}, \bibinfo {author} {\bibfnamefont {B.~D.}\
  \bibnamefont {{Lackey}}}, \bibinfo {author} {\bibfnamefont {M.}~\bibnamefont
  {{Landry}}}, \bibinfo {author} {\bibfnamefont {J.}~\bibnamefont {{Lange}}},
  \bibinfo {author} {\bibfnamefont {B.}~\bibnamefont {{Lantz}}}, \bibinfo
  {author} {\bibfnamefont {P.~D.}\ \bibnamefont {{Lasky}}}, \bibinfo {author}
  {\bibfnamefont {A.}~\bibnamefont {{Lazzarini}}}, \bibinfo {author}
  {\bibfnamefont {C.}~\bibnamefont {{Lazzaro}}}, \bibinfo {author}
  {\bibfnamefont {P.}~\bibnamefont {{Leaci}}}, \bibinfo {author} {\bibfnamefont
  {S.}~\bibnamefont {{Leavey}}}, \bibinfo {author} {\bibfnamefont {E.~O.}\
  \bibnamefont {{Lebigot}}}, \bibinfo {author} {\bibfnamefont {C.~H.}\
  \bibnamefont {{Lee}}}, \bibinfo {author} {\bibfnamefont {H.~K.}\ \bibnamefont
  {{Lee}}}, \bibinfo {author} {\bibfnamefont {H.~M.}\ \bibnamefont {{Lee}}},
  \bibinfo {author} {\bibfnamefont {K.}~\bibnamefont {{Lee}}}, \bibinfo
  {author} {\bibfnamefont {A.}~\bibnamefont {{Lenon}}}, \bibinfo {author}
  {\bibfnamefont {M.}~\bibnamefont {{Leonardi}}}, \bibinfo {author}
  {\bibfnamefont {J.~R.}\ \bibnamefont {{Leong}}}, \bibinfo {author}
  {\bibfnamefont {N.}~\bibnamefont {{Leroy}}}, \bibinfo {author} {\bibfnamefont
  {N.}~\bibnamefont {{Letendre}}}, \bibinfo {author} {\bibfnamefont
  {Y.}~\bibnamefont {{Levin}}}, \bibinfo {author} {\bibfnamefont {B.~M.}\
  \bibnamefont {{Levine}}}, \bibinfo {author} {\bibfnamefont {T.~G.~F.}\
  \bibnamefont {{Li}}}, \bibinfo {author} {\bibfnamefont {A.}~\bibnamefont
  {{Libson}}}, \bibinfo {author} {\bibfnamefont {T.~B.}\ \bibnamefont
  {{Littenberg}}}, \bibinfo {author} {\bibfnamefont {N.~A.}\ \bibnamefont
  {{Lockerbie}}}, \bibinfo {author} {\bibfnamefont {J.}~\bibnamefont
  {{Logue}}}, \bibinfo {author} {\bibfnamefont {A.~L.}\ \bibnamefont
  {{Lombardi}}}, \bibinfo {author} {\bibfnamefont {J.~E.}\ \bibnamefont
  {{Lord}}}, \bibinfo {author} {\bibfnamefont {M.}~\bibnamefont {{Lorenzini}}},
  \bibinfo {author} {\bibfnamefont {V.}~\bibnamefont {{Loriette}}}, \bibinfo
  {author} {\bibfnamefont {M.}~\bibnamefont {{Lormand}}}, \bibinfo {author}
  {\bibfnamefont {G.}~\bibnamefont {{Losurdo}}}, \bibinfo {author}
  {\bibfnamefont {J.~D.}\ \bibnamefont {{Lough}}}, \bibinfo {author}
  {\bibfnamefont {H.}~\bibnamefont {{L{\"u}ck}}}, \bibinfo {author}
  {\bibfnamefont {A.~P.}\ \bibnamefont {{Lundgren}}}, \bibinfo {author}
  {\bibfnamefont {J.}~\bibnamefont {{Luo}}}, \bibinfo {author} {\bibfnamefont
  {R.}~\bibnamefont {{Lynch}}}, \bibinfo {author} {\bibfnamefont
  {Y.}~\bibnamefont {{Ma}}}, \bibinfo {author} {\bibfnamefont {T.}~\bibnamefont
  {{MacDonald}}}, \bibinfo {author} {\bibfnamefont {B.}~\bibnamefont
  {{Machenschalk}}}, \bibinfo {author} {\bibfnamefont {M.}~\bibnamefont
  {{MacInnis}}}, \bibinfo {author} {\bibfnamefont {D.~M.}\ \bibnamefont
  {{Macleod}}}, \bibinfo {author} {\bibfnamefont {F.}~\bibnamefont
  {{Maga{\~n}a-Sandoval}}}, \bibinfo {author} {\bibfnamefont {R.~M.}\
  \bibnamefont {{Magee}}}, \bibinfo {author} {\bibfnamefont {M.}~\bibnamefont
  {{Mageswaran}}}, \bibinfo {author} {\bibfnamefont {E.}~\bibnamefont
  {{Majorana}}}, \bibinfo {author} {\bibfnamefont {I.}~\bibnamefont
  {{Maksimovic}}}, \bibinfo {author} {\bibfnamefont {V.}~\bibnamefont
  {{Malvezzi}}}, \bibinfo {author} {\bibfnamefont {N.}~\bibnamefont {{Man}}},
  \bibinfo {author} {\bibfnamefont {I.}~\bibnamefont {{Mandel}}}, \bibinfo
  {author} {\bibfnamefont {V.}~\bibnamefont {{Mandic}}}, \bibinfo {author}
  {\bibfnamefont {V.}~\bibnamefont {{Mangano}}}, \bibinfo {author}
  {\bibfnamefont {G.~L.}\ \bibnamefont {{Mansell}}}, \bibinfo {author}
  {\bibfnamefont {M.}~\bibnamefont {{Manske}}}, \bibinfo {author}
  {\bibfnamefont {M.}~\bibnamefont {{Mantovani}}}, \bibinfo {author}
  {\bibfnamefont {F.}~\bibnamefont {{Marchesoni}}}, \bibinfo {author}
  {\bibfnamefont {F.}~\bibnamefont {{Marion}}}, \bibinfo {author}
  {\bibfnamefont {S.}~\bibnamefont {{M{\'a}rka}}}, \bibinfo {author}
  {\bibfnamefont {Z.}~\bibnamefont {{M{\'a}rka}}}, \bibinfo {author}
  {\bibfnamefont {A.~S.}\ \bibnamefont {{Markosyan}}}, \bibinfo {author}
  {\bibfnamefont {E.}~\bibnamefont {{Maros}}}, \bibinfo {author} {\bibfnamefont
  {F.}~\bibnamefont {{Martelli}}}, \bibinfo {author} {\bibfnamefont
  {L.}~\bibnamefont {{Martellini}}}, \bibinfo {author} {\bibfnamefont {I.~W.}\
  \bibnamefont {{Martin}}}, \bibinfo {author} {\bibfnamefont {R.~M.}\
  \bibnamefont {{Martin}}}, \bibinfo {author} {\bibfnamefont {D.~V.}\
  \bibnamefont {{Martynov}}}, \bibinfo {author} {\bibfnamefont {J.~N.}\
  \bibnamefont {{Marx}}}, \bibinfo {author} {\bibfnamefont {K.}~\bibnamefont
  {{Mason}}}, \bibinfo {author} {\bibfnamefont {A.}~\bibnamefont {{Masserot}}},
  \bibinfo {author} {\bibfnamefont {T.~J.}\ \bibnamefont {{Massinger}}},
  \bibinfo {author} {\bibfnamefont {M.}~\bibnamefont {{Masso-Reid}}}, \bibinfo
  {author} {\bibfnamefont {F.}~\bibnamefont {{Matichard}}}, \bibinfo {author}
  {\bibfnamefont {L.}~\bibnamefont {{Matone}}}, \bibinfo {author}
  {\bibfnamefont {N.}~\bibnamefont {{Mavalvala}}}, \bibinfo {author}
  {\bibfnamefont {N.}~\bibnamefont {{Mazumder}}}, \bibinfo {author}
  {\bibfnamefont {G.}~\bibnamefont {{Mazzolo}}}, \bibinfo {author}
  {\bibfnamefont {R.}~\bibnamefont {{McCarthy}}}, \bibinfo {author}
  {\bibfnamefont {D.~E.}\ \bibnamefont {{McClelland}}}, \bibinfo {author}
  {\bibfnamefont {S.}~\bibnamefont {{McCormick}}}, \bibinfo {author}
  {\bibfnamefont {S.~C.}\ \bibnamefont {{McGuire}}}, \bibinfo {author}
  {\bibfnamefont {G.}~\bibnamefont {{McIntyre}}}, \bibinfo {author}
  {\bibfnamefont {J.}~\bibnamefont {{McIver}}}, \bibinfo {author}
  {\bibfnamefont {D.~J.}\ \bibnamefont {{McManus}}}, \bibinfo {author}
  {\bibfnamefont {S.~T.}\ \bibnamefont {{McWilliams}}}, \bibinfo {author}
  {\bibfnamefont {D.}~\bibnamefont {{Meacher}}}, \bibinfo {author}
  {\bibfnamefont {G.~D.}\ \bibnamefont {{Meadors}}}, \bibinfo {author}
  {\bibfnamefont {J.}~\bibnamefont {{Meidam}}}, \bibinfo {author}
  {\bibfnamefont {A.}~\bibnamefont {{Melatos}}}, \bibinfo {author}
  {\bibfnamefont {G.}~\bibnamefont {{Mendell}}}, \bibinfo {author}
  {\bibfnamefont {D.}~\bibnamefont {{Mendoza-Gandara}}}, \bibinfo {author}
  {\bibfnamefont {R.~A.}\ \bibnamefont {{Mercer}}}, \bibinfo {author}
  {\bibfnamefont {E.}~\bibnamefont {{Merilh}}}, \bibinfo {author}
  {\bibfnamefont {M.}~\bibnamefont {{Merzougui}}}, \bibinfo {author}
  {\bibfnamefont {S.}~\bibnamefont {{Meshkov}}}, \bibinfo {author}
  {\bibfnamefont {C.}~\bibnamefont {{Messenger}}}, \bibinfo {author}
  {\bibfnamefont {C.}~\bibnamefont {{Messick}}}, \bibinfo {author}
  {\bibfnamefont {P.~M.}\ \bibnamefont {{Meyers}}}, \bibinfo {author}
  {\bibfnamefont {F.}~\bibnamefont {{Mezzani}}}, \bibinfo {author}
  {\bibfnamefont {H.}~\bibnamefont {{Miao}}}, \bibinfo {author} {\bibfnamefont
  {C.}~\bibnamefont {{Michel}}}, \bibinfo {author} {\bibfnamefont
  {H.}~\bibnamefont {{Middleton}}}, \bibinfo {author} {\bibfnamefont {E.~E.}\
  \bibnamefont {{Mikhailov}}}, \bibinfo {author} {\bibfnamefont
  {L.}~\bibnamefont {{Milano}}}, \bibinfo {author} {\bibfnamefont
  {J.}~\bibnamefont {{Miller}}}, \bibinfo {author} {\bibfnamefont
  {M.}~\bibnamefont {{Millhouse}}}, \bibinfo {author} {\bibfnamefont
  {Y.}~\bibnamefont {{Minenkov}}}, \bibinfo {author} {\bibfnamefont
  {J.}~\bibnamefont {{Ming}}}, \bibinfo {author} {\bibfnamefont
  {S.}~\bibnamefont {{Mirshekari}}}, \bibinfo {author} {\bibfnamefont
  {C.}~\bibnamefont {{Mishra}}}, \bibinfo {author} {\bibfnamefont
  {S.}~\bibnamefont {{Mitra}}}, \bibinfo {author} {\bibfnamefont {V.~P.}\
  \bibnamefont {{Mitrofanov}}}, \bibinfo {author} {\bibfnamefont
  {G.}~\bibnamefont {{Mitselmakher}}}, \bibinfo {author} {\bibfnamefont
  {R.}~\bibnamefont {{Mittleman}}}, \bibinfo {author} {\bibfnamefont
  {A.}~\bibnamefont {{Moggi}}}, \bibinfo {author} {\bibfnamefont
  {M.}~\bibnamefont {{Mohan}}}, \bibinfo {author} {\bibfnamefont {S.~R.~P.}\
  \bibnamefont {{Mohapatra}}}, \bibinfo {author} {\bibfnamefont
  {M.}~\bibnamefont {{Montani}}}, \bibinfo {author} {\bibfnamefont {B.~C.}\
  \bibnamefont {{Moore}}}, \bibinfo {author} {\bibfnamefont {C.~J.}\
  \bibnamefont {{Moore}}}, \bibinfo {author} {\bibfnamefont {D.}~\bibnamefont
  {{Moraru}}}, \bibinfo {author} {\bibfnamefont {G.}~\bibnamefont {{Moreno}}},
  \bibinfo {author} {\bibfnamefont {S.~R.}\ \bibnamefont {{Morriss}}}, \bibinfo
  {author} {\bibfnamefont {K.}~\bibnamefont {{Mossavi}}}, \bibinfo {author}
  {\bibfnamefont {B.}~\bibnamefont {{Mours}}}, \bibinfo {author} {\bibfnamefont
  {C.~M.}\ \bibnamefont {{Mow-Lowry}}}, \bibinfo {author} {\bibfnamefont
  {C.~L.}\ \bibnamefont {{Mueller}}}, \bibinfo {author} {\bibfnamefont
  {G.}~\bibnamefont {{Mueller}}}, \bibinfo {author} {\bibfnamefont {A.~W.}\
  \bibnamefont {{Muir}}}, \bibinfo {author} {\bibfnamefont {A.}~\bibnamefont
  {{Mukherjee}}}, \bibinfo {author} {\bibfnamefont {D.}~\bibnamefont
  {{Mukherjee}}}, \bibinfo {author} {\bibfnamefont {S.}~\bibnamefont
  {{Mukherjee}}}, \bibinfo {author} {\bibfnamefont {N.}~\bibnamefont
  {{Mukund}}}, \bibinfo {author} {\bibfnamefont {A.}~\bibnamefont
  {{Mullavey}}}, \bibinfo {author} {\bibfnamefont {J.}~\bibnamefont {{Munch}}},
  \bibinfo {author} {\bibfnamefont {D.~J.}\ \bibnamefont {{Murphy}}}, \bibinfo
  {author} {\bibfnamefont {P.~G.}\ \bibnamefont {{Murray}}}, \bibinfo {author}
  {\bibfnamefont {A.}~\bibnamefont {{Mytidis}}}, \bibinfo {author}
  {\bibfnamefont {I.}~\bibnamefont {{Nardecchia}}}, \bibinfo {author}
  {\bibfnamefont {L.}~\bibnamefont {{Naticchioni}}}, \bibinfo {author}
  {\bibfnamefont {R.~K.}\ \bibnamefont {{Nayak}}}, \bibinfo {author}
  {\bibfnamefont {V.}~\bibnamefont {{Necula}}}, \bibinfo {author}
  {\bibfnamefont {K.}~\bibnamefont {{Nedkova}}}, \bibinfo {author}
  {\bibfnamefont {G.}~\bibnamefont {{Nelemans}}}, \bibinfo {author}
  {\bibfnamefont {M.}~\bibnamefont {{Neri}}}, \bibinfo {author} {\bibfnamefont
  {A.}~\bibnamefont {{Neunzert}}}, \bibinfo {author} {\bibfnamefont
  {G.}~\bibnamefont {{Newton}}}, \bibinfo {author} {\bibfnamefont {T.~T.}\
  \bibnamefont {{Nguyen}}}, \bibinfo {author} {\bibfnamefont {A.~B.}\
  \bibnamefont {{Nielsen}}}, \bibinfo {author} {\bibfnamefont {S.}~\bibnamefont
  {{Nissanke}}}, \bibinfo {author} {\bibfnamefont {A.}~\bibnamefont {{Nitz}}},
  \bibinfo {author} {\bibfnamefont {F.}~\bibnamefont {{Nocera}}}, \bibinfo
  {author} {\bibfnamefont {D.}~\bibnamefont {{Nolting}}}, \bibinfo {author}
  {\bibfnamefont {M.~E.}\ \bibnamefont {{Normandin}}}, \bibinfo {author}
  {\bibfnamefont {L.~K.}\ \bibnamefont {{Nuttall}}}, \bibinfo {author}
  {\bibfnamefont {J.}~\bibnamefont {{Oberling}}}, \bibinfo {author}
  {\bibfnamefont {E.}~\bibnamefont {{Ochsner}}}, \bibinfo {author}
  {\bibfnamefont {J.}~\bibnamefont {{O'Dell}}}, \bibinfo {author}
  {\bibfnamefont {E.}~\bibnamefont {{Oelker}}}, \bibinfo {author}
  {\bibfnamefont {G.~H.}\ \bibnamefont {{Ogin}}}, \bibinfo {author}
  {\bibfnamefont {J.~J.}\ \bibnamefont {{Oh}}}, \bibinfo {author}
  {\bibfnamefont {S.~H.}\ \bibnamefont {{Oh}}}, \bibinfo {author}
  {\bibfnamefont {F.}~\bibnamefont {{Ohme}}}, \bibinfo {author} {\bibfnamefont
  {M.}~\bibnamefont {{Oliver}}}, \bibinfo {author} {\bibfnamefont
  {P.}~\bibnamefont {{Oppermann}}}, \bibinfo {author} {\bibfnamefont {R.~J.}\
  \bibnamefont {{Oram}}}, \bibinfo {author} {\bibfnamefont {B.}~\bibnamefont
  {{O'Reilly}}}, \bibinfo {author} {\bibfnamefont {R.}~\bibnamefont
  {{O'Shaughnessy}}}, \bibinfo {author} {\bibfnamefont {D.~J.}\ \bibnamefont
  {{Ottaway}}}, \bibinfo {author} {\bibfnamefont {R.~S.}\ \bibnamefont
  {{Ottens}}}, \bibinfo {author} {\bibfnamefont {H.}~\bibnamefont
  {{Overmier}}}, \bibinfo {author} {\bibfnamefont {B.~J.}\ \bibnamefont
  {{Owen}}}, \bibinfo {author} {\bibfnamefont {A.}~\bibnamefont {{Pai}}},
  \bibinfo {author} {\bibfnamefont {S.~A.}\ \bibnamefont {{Pai}}}, \bibinfo
  {author} {\bibfnamefont {J.~R.}\ \bibnamefont {{Palamos}}}, \bibinfo {author}
  {\bibfnamefont {O.}~\bibnamefont {{Palashov}}}, \bibinfo {author}
  {\bibfnamefont {C.}~\bibnamefont {{Palomba}}}, \bibinfo {author}
  {\bibfnamefont {A.}~\bibnamefont {{Pal-Singh}}}, \bibinfo {author}
  {\bibfnamefont {H.}~\bibnamefont {{Pan}}}, \bibinfo {author} {\bibfnamefont
  {C.}~\bibnamefont {{Pankow}}}, \bibinfo {author} {\bibfnamefont
  {F.}~\bibnamefont {{Pannarale}}}, \bibinfo {author} {\bibfnamefont {B.~C.}\
  \bibnamefont {{Pant}}}, \bibinfo {author} {\bibfnamefont {F.}~\bibnamefont
  {{Paoletti}}}, \bibinfo {author} {\bibfnamefont {A.}~\bibnamefont {{Paoli}}},
  \bibinfo {author} {\bibfnamefont {M.~A.}\ \bibnamefont {{Papa}}}, \bibinfo
  {author} {\bibfnamefont {H.~R.}\ \bibnamefont {{Paris}}}, \bibinfo {author}
  {\bibfnamefont {W.}~\bibnamefont {{Parker}}}, \bibinfo {author}
  {\bibfnamefont {D.}~\bibnamefont {{Pascucci}}}, \bibinfo {author}
  {\bibfnamefont {A.}~\bibnamefont {{Pasqualetti}}}, \bibinfo {author}
  {\bibfnamefont {R.}~\bibnamefont {{Passaquieti}}}, \bibinfo {author}
  {\bibfnamefont {D.}~\bibnamefont {{Passuello}}}, \bibinfo {author}
  {\bibfnamefont {B.}~\bibnamefont {{Patricelli}}}, \bibinfo {author}
  {\bibfnamefont {Z.}~\bibnamefont {{Patrick}}}, \bibinfo {author}
  {\bibfnamefont {B.~L.}\ \bibnamefont {{Pearlstone}}}, \bibinfo {author}
  {\bibfnamefont {M.}~\bibnamefont {{Pedraza}}}, \bibinfo {author}
  {\bibfnamefont {R.}~\bibnamefont {{Pedurand}}}, \bibinfo {author}
  {\bibfnamefont {L.}~\bibnamefont {{Pekowsky}}}, \bibinfo {author}
  {\bibfnamefont {A.}~\bibnamefont {{Pele}}}, \bibinfo {author} {\bibfnamefont
  {S.}~\bibnamefont {{Penn}}}, \bibinfo {author} {\bibfnamefont
  {A.}~\bibnamefont {{Perreca}}}, \bibinfo {author} {\bibfnamefont
  {M.}~\bibnamefont {{Phelps}}}, \bibinfo {author} {\bibfnamefont
  {O.}~\bibnamefont {{Piccinni}}}, \bibinfo {author} {\bibfnamefont
  {M.}~\bibnamefont {{Pichot}}}, \bibinfo {author} {\bibfnamefont
  {F.}~\bibnamefont {{Piergiovanni}}}, \bibinfo {author} {\bibfnamefont
  {V.}~\bibnamefont {{Pierro}}}, \bibinfo {author} {\bibfnamefont
  {G.}~\bibnamefont {{Pillant}}}, \bibinfo {author} {\bibfnamefont
  {L.}~\bibnamefont {{Pinard}}}, \bibinfo {author} {\bibfnamefont {I.~M.}\
  \bibnamefont {{Pinto}}}, \bibinfo {author} {\bibfnamefont {M.}~\bibnamefont
  {{Pitkin}}}, \bibinfo {author} {\bibfnamefont {R.}~\bibnamefont
  {{Poggiani}}}, \bibinfo {author} {\bibfnamefont {P.}~\bibnamefont
  {{Popolizio}}}, \bibinfo {author} {\bibfnamefont {E.~K.}\ \bibnamefont
  {{Porter}}}, \bibinfo {author} {\bibfnamefont {A.}~\bibnamefont {{Post}}},
  \bibinfo {author} {\bibfnamefont {J.}~\bibnamefont {{Powell}}}, \bibinfo
  {author} {\bibfnamefont {J.}~\bibnamefont {{Prasad}}}, \bibinfo {author}
  {\bibfnamefont {V.}~\bibnamefont {{Predoi}}}, \bibinfo {author}
  {\bibfnamefont {S.~S.}\ \bibnamefont {{Premachandra}}}, \bibinfo {author}
  {\bibfnamefont {T.}~\bibnamefont {{Prestegard}}}, \bibinfo {author}
  {\bibfnamefont {L.~R.}\ \bibnamefont {{Price}}}, \bibinfo {author}
  {\bibfnamefont {M.}~\bibnamefont {{Prijatelj}}}, \bibinfo {author}
  {\bibfnamefont {M.}~\bibnamefont {{Principe}}}, \bibinfo {author}
  {\bibfnamefont {S.}~\bibnamefont {{Privitera}}}, \bibinfo {author}
  {\bibfnamefont {G.~A.}\ \bibnamefont {{Prodi}}}, \bibinfo {author}
  {\bibfnamefont {L.}~\bibnamefont {{Prokhorov}}}, \bibinfo {author}
  {\bibfnamefont {O.}~\bibnamefont {{Puncken}}}, \bibinfo {author}
  {\bibfnamefont {M.}~\bibnamefont {{Punturo}}}, \bibinfo {author}
  {\bibfnamefont {P.}~\bibnamefont {{Puppo}}}, \bibinfo {author} {\bibfnamefont
  {M.}~\bibnamefont {{P{\"u}rrer}}}, \bibinfo {author} {\bibfnamefont
  {H.}~\bibnamefont {{Qi}}}, \bibinfo {author} {\bibfnamefont {J.}~\bibnamefont
  {{Qin}}}, \bibinfo {author} {\bibfnamefont {V.}~\bibnamefont {{Quetschke}}},
  \bibinfo {author} {\bibfnamefont {E.~A.}\ \bibnamefont {{Quintero}}},
  \bibinfo {author} {\bibfnamefont {R.}~\bibnamefont {{Quitzow-James}}},
  \bibinfo {author} {\bibfnamefont {F.~J.}\ \bibnamefont {{Raab}}}, \bibinfo
  {author} {\bibfnamefont {D.~S.}\ \bibnamefont {{Rabeling}}}, \bibinfo
  {author} {\bibfnamefont {H.}~\bibnamefont {{Radkins}}}, \bibinfo {author}
  {\bibfnamefont {P.}~\bibnamefont {{Raffai}}}, \bibinfo {author}
  {\bibfnamefont {S.}~\bibnamefont {{Raja}}}, \bibinfo {author} {\bibfnamefont
  {M.}~\bibnamefont {{Rakhmanov}}}, \bibinfo {author} {\bibfnamefont
  {P.}~\bibnamefont {{Rapagnani}}}, \bibinfo {author} {\bibfnamefont
  {V.}~\bibnamefont {{Raymond}}}, \bibinfo {author} {\bibfnamefont
  {M.}~\bibnamefont {{Razzano}}}, \bibinfo {author} {\bibfnamefont
  {V.}~\bibnamefont {{Re}}}, \bibinfo {author} {\bibfnamefont {J.}~\bibnamefont
  {{Read}}}, \bibinfo {author} {\bibfnamefont {C.~M.}\ \bibnamefont {{Reed}}},
  \bibinfo {author} {\bibfnamefont {T.}~\bibnamefont {{Regimbau}}}, \bibinfo
  {author} {\bibfnamefont {L.}~\bibnamefont {{Rei}}}, \bibinfo {author}
  {\bibfnamefont {S.}~\bibnamefont {{Reid}}}, \bibinfo {author} {\bibfnamefont
  {D.~H.}\ \bibnamefont {{Reitze}}}, \bibinfo {author} {\bibfnamefont
  {H.}~\bibnamefont {{Rew}}}, \bibinfo {author} {\bibfnamefont {S.~D.}\
  \bibnamefont {{Reyes}}}, \bibinfo {author} {\bibfnamefont {F.}~\bibnamefont
  {{Ricci}}}, \bibinfo {author} {\bibfnamefont {K.}~\bibnamefont {{Riles}}},
  \bibinfo {author} {\bibfnamefont {N.~A.}\ \bibnamefont {{Robertson}}},
  \bibinfo {author} {\bibfnamefont {R.}~\bibnamefont {{Robie}}}, \bibinfo
  {author} {\bibfnamefont {F.}~\bibnamefont {{Robinet}}}, \bibinfo {author}
  {\bibfnamefont {A.}~\bibnamefont {{Rocchi}}}, \bibinfo {author}
  {\bibfnamefont {L.}~\bibnamefont {{Rolland}}}, \bibinfo {author}
  {\bibfnamefont {J.~G.}\ \bibnamefont {{Rollins}}}, \bibinfo {author}
  {\bibfnamefont {V.~J.}\ \bibnamefont {{Roma}}}, \bibinfo {author}
  {\bibfnamefont {R.}~\bibnamefont {{Romano}}}, \bibinfo {author}
  {\bibfnamefont {G.}~\bibnamefont {{Romanov}}}, \bibinfo {author}
  {\bibfnamefont {J.~H.}\ \bibnamefont {{Romie}}}, \bibinfo {author}
  {\bibfnamefont {D.}~\bibnamefont {{Rosi{\'n}ska}}}, \bibinfo {author}
  {\bibfnamefont {S.}~\bibnamefont {{Rowan}}}, \bibinfo {author} {\bibfnamefont
  {A.}~\bibnamefont {{R{\"u}diger}}}, \bibinfo {author} {\bibfnamefont
  {P.}~\bibnamefont {{Ruggi}}}, \bibinfo {author} {\bibfnamefont
  {K.}~\bibnamefont {{Ryan}}}, \bibinfo {author} {\bibfnamefont
  {S.}~\bibnamefont {{Sachdev}}}, \bibinfo {author} {\bibfnamefont
  {T.}~\bibnamefont {{Sadecki}}}, \bibinfo {author} {\bibfnamefont
  {L.}~\bibnamefont {{Sadeghian}}}, \bibinfo {author} {\bibfnamefont
  {L.}~\bibnamefont {{Salconi}}}, \bibinfo {author} {\bibfnamefont
  {M.}~\bibnamefont {{Saleem}}}, \bibinfo {author} {\bibfnamefont
  {F.}~\bibnamefont {{Salemi}}}, \bibinfo {author} {\bibfnamefont
  {A.}~\bibnamefont {{Samajdar}}}, \bibinfo {author} {\bibfnamefont
  {L.}~\bibnamefont {{Sammut}}}, \bibinfo {author} {\bibfnamefont
  {L.}~\bibnamefont {{Sampson}}}, \bibinfo {author} {\bibfnamefont {E.~J.}\
  \bibnamefont {{Sanchez}}}, \bibinfo {author} {\bibfnamefont {V.}~\bibnamefont
  {{Sandberg}}}, \bibinfo {author} {\bibfnamefont {B.}~\bibnamefont
  {{Sandeen}}}, \bibinfo {author} {\bibfnamefont {J.~R.}\ \bibnamefont
  {{Sanders}}}, \bibinfo {author} {\bibfnamefont {B.}~\bibnamefont
  {{Sassolas}}}, \bibinfo {author} {\bibfnamefont {B.~S.}\ \bibnamefont
  {{Sathyaprakash}}}, \bibinfo {author} {\bibfnamefont {P.~R.}\ \bibnamefont
  {{Saulson}}}, \bibinfo {author} {\bibfnamefont {O.}~\bibnamefont {{Sauter}}},
  \bibinfo {author} {\bibfnamefont {R.~L.}\ \bibnamefont {{Savage}}}, \bibinfo
  {author} {\bibfnamefont {A.}~\bibnamefont {{Sawadsky}}}, \bibinfo {author}
  {\bibfnamefont {P.}~\bibnamefont {{Schale}}}, \bibinfo {author}
  {\bibfnamefont {R.}~\bibnamefont {{Schilling}}}, \bibinfo {author}
  {\bibfnamefont {J.}~\bibnamefont {{Schmidt}}}, \bibinfo {author}
  {\bibfnamefont {P.}~\bibnamefont {{Schmidt}}}, \bibinfo {author}
  {\bibfnamefont {R.}~\bibnamefont {{Schnabel}}}, \bibinfo {author}
  {\bibfnamefont {R.~M.~S.}\ \bibnamefont {{Schofield}}}, \bibinfo {author}
  {\bibfnamefont {A.}~\bibnamefont {{Sch{\"o}nbeck}}}, \bibinfo {author}
  {\bibfnamefont {E.}~\bibnamefont {{Schreiber}}}, \bibinfo {author}
  {\bibfnamefont {D.}~\bibnamefont {{Schuette}}}, \bibinfo {author}
  {\bibfnamefont {B.~F.}\ \bibnamefont {{Schutz}}}, \bibinfo {author}
  {\bibfnamefont {J.}~\bibnamefont {{Scott}}}, \bibinfo {author} {\bibfnamefont
  {S.~M.}\ \bibnamefont {{Scott}}}, \bibinfo {author} {\bibfnamefont
  {D.}~\bibnamefont {{Sellers}}}, \bibinfo {author} {\bibfnamefont {A.~S.}\
  \bibnamefont {{Sengupta}}}, \bibinfo {author} {\bibfnamefont
  {D.}~\bibnamefont {{Sentenac}}}, \bibinfo {author} {\bibfnamefont
  {V.}~\bibnamefont {{Sequino}}}, \bibinfo {author} {\bibfnamefont
  {A.}~\bibnamefont {{Sergeev}}}, \bibinfo {author} {\bibfnamefont
  {G.}~\bibnamefont {{Serna}}}, \bibinfo {author} {\bibfnamefont
  {Y.}~\bibnamefont {{Setyawati}}}, \bibinfo {author} {\bibfnamefont
  {A.}~\bibnamefont {{Sevigny}}}, \bibinfo {author} {\bibfnamefont {D.~A.}\
  \bibnamefont {{Shaddock}}}, \bibinfo {author} {\bibfnamefont
  {S.}~\bibnamefont {{Shah}}}, \bibinfo {author} {\bibfnamefont {M.~S.}\
  \bibnamefont {{Shahriar}}}, \bibinfo {author} {\bibfnamefont
  {M.}~\bibnamefont {{Shaltev}}}, \bibinfo {author} {\bibfnamefont
  {Z.}~\bibnamefont {{Shao}}}, \bibinfo {author} {\bibfnamefont
  {B.}~\bibnamefont {{Shapiro}}}, \bibinfo {author} {\bibfnamefont
  {P.}~\bibnamefont {{Shawhan}}}, \bibinfo {author} {\bibfnamefont
  {A.}~\bibnamefont {{Sheperd}}}, \bibinfo {author} {\bibfnamefont {D.~H.}\
  \bibnamefont {{Shoemaker}}}, \bibinfo {author} {\bibfnamefont {D.~M.}\
  \bibnamefont {{Shoemaker}}}, \bibinfo {author} {\bibfnamefont
  {K.}~\bibnamefont {{Siellez}}}, \bibinfo {author} {\bibfnamefont
  {X.}~\bibnamefont {{Siemens}}}, \bibinfo {author} {\bibfnamefont
  {D.}~\bibnamefont {{Sigg}}}, \bibinfo {author} {\bibfnamefont {A.~D.}\
  \bibnamefont {{Silva}}}, \bibinfo {author} {\bibfnamefont {D.}~\bibnamefont
  {{Simakov}}}, \bibinfo {author} {\bibfnamefont {A.}~\bibnamefont {{Singer}}},
  \bibinfo {author} {\bibfnamefont {L.~P.}\ \bibnamefont {{Singer}}}, \bibinfo
  {author} {\bibfnamefont {A.}~\bibnamefont {{Singh}}}, \bibinfo {author}
  {\bibfnamefont {R.}~\bibnamefont {{Singh}}}, \bibinfo {author} {\bibfnamefont
  {A.}~\bibnamefont {{Singhal}}}, \bibinfo {author} {\bibfnamefont {A.~M.}\
  \bibnamefont {{Sintes}}}, \bibinfo {author} {\bibfnamefont {B.~J.~J.}\
  \bibnamefont {{Slagmolen}}}, \bibinfo {author} {\bibfnamefont {J.~R.}\
  \bibnamefont {{Smith}}}, \bibinfo {author} {\bibfnamefont {N.~D.}\
  \bibnamefont {{Smith}}}, \bibinfo {author} {\bibfnamefont {R.~J.~E.}\
  \bibnamefont {{Smith}}}, \bibinfo {author} {\bibfnamefont {E.~J.}\
  \bibnamefont {{Son}}}, \bibinfo {author} {\bibfnamefont {B.}~\bibnamefont
  {{Sorazu}}}, \bibinfo {author} {\bibfnamefont {F.}~\bibnamefont
  {{Sorrentino}}}, \bibinfo {author} {\bibfnamefont {T.}~\bibnamefont
  {{Souradeep}}}, \bibinfo {author} {\bibfnamefont {A.~K.}\ \bibnamefont
  {{Srivastava}}}, \bibinfo {author} {\bibfnamefont {A.}~\bibnamefont
  {{Staley}}}, \bibinfo {author} {\bibfnamefont {M.}~\bibnamefont {{Steinke}}},
  \bibinfo {author} {\bibfnamefont {J.}~\bibnamefont {{Steinlechner}}},
  \bibinfo {author} {\bibfnamefont {S.}~\bibnamefont {{Steinlechner}}},
  \bibinfo {author} {\bibfnamefont {D.}~\bibnamefont {{Steinmeyer}}}, \bibinfo
  {author} {\bibfnamefont {B.~C.}\ \bibnamefont {{Stephens}}}, \bibinfo
  {author} {\bibfnamefont {S.}~\bibnamefont {{Stevenson}}}, \bibinfo {author}
  {\bibfnamefont {R.}~\bibnamefont {{Stone}}}, \bibinfo {author} {\bibfnamefont
  {K.~A.}\ \bibnamefont {{Strain}}}, \bibinfo {author} {\bibfnamefont
  {N.}~\bibnamefont {{Straniero}}}, \bibinfo {author} {\bibfnamefont
  {G.}~\bibnamefont {{Stratta}}}, \bibinfo {author} {\bibfnamefont {N.~A.}\
  \bibnamefont {{Strauss}}}, \bibinfo {author} {\bibfnamefont {S.}~\bibnamefont
  {{Strigin}}}, \bibinfo {author} {\bibfnamefont {R.}~\bibnamefont
  {{Sturani}}}, \bibinfo {author} {\bibfnamefont {A.~L.}\ \bibnamefont
  {{Stuver}}}, \bibinfo {author} {\bibfnamefont {T.~Z.}\ \bibnamefont
  {{Summerscales}}}, \bibinfo {author} {\bibfnamefont {L.}~\bibnamefont
  {{Sun}}}, \bibinfo {author} {\bibfnamefont {P.~J.}\ \bibnamefont {{Sutton}}},
  \bibinfo {author} {\bibfnamefont {B.~L.}\ \bibnamefont {{Swinkels}}},
  \bibinfo {author} {\bibfnamefont {M.~J.}\ \bibnamefont {{Szczepa{\'n}czyk}}},
  \bibinfo {author} {\bibfnamefont {M.}~\bibnamefont {{Tacca}}}, \bibinfo
  {author} {\bibfnamefont {D.}~\bibnamefont {{Talukder}}}, \bibinfo {author}
  {\bibfnamefont {D.~B.}\ \bibnamefont {{Tanner}}}, \bibinfo {author}
  {\bibfnamefont {M.}~\bibnamefont {{T{\'a}pai}}}, \bibinfo {author}
  {\bibfnamefont {S.~P.}\ \bibnamefont {{Tarabrin}}}, \bibinfo {author}
  {\bibfnamefont {A.}~\bibnamefont {{Taracchini}}}, \bibinfo {author}
  {\bibfnamefont {R.}~\bibnamefont {{Taylor}}}, \bibinfo {author}
  {\bibfnamefont {T.}~\bibnamefont {{Theeg}}}, \bibinfo {author} {\bibfnamefont
  {M.~P.}\ \bibnamefont {{Thirugnanasambandam}}}, \bibinfo {author}
  {\bibfnamefont {E.~G.}\ \bibnamefont {{Thomas}}}, \bibinfo {author}
  {\bibfnamefont {M.}~\bibnamefont {{Thomas}}}, \bibinfo {author}
  {\bibfnamefont {P.}~\bibnamefont {{Thomas}}}, \bibinfo {author}
  {\bibfnamefont {K.~A.}\ \bibnamefont {{Thorne}}}, \bibinfo {author}
  {\bibfnamefont {K.~S.}\ \bibnamefont {{Thorne}}}, \bibinfo {author}
  {\bibfnamefont {E.}~\bibnamefont {{Thrane}}}, \bibinfo {author}
  {\bibfnamefont {S.}~\bibnamefont {{Tiwari}}}, \bibinfo {author}
  {\bibfnamefont {V.}~\bibnamefont {{Tiwari}}}, \bibinfo {author}
  {\bibfnamefont {K.~V.}\ \bibnamefont {{Tokmakov}}}, \bibinfo {author}
  {\bibfnamefont {C.}~\bibnamefont {{Tomlinson}}}, \bibinfo {author}
  {\bibfnamefont {M.}~\bibnamefont {{Tonelli}}}, \bibinfo {author}
  {\bibfnamefont {C.~V.}\ \bibnamefont {{Torres}}}, \bibinfo {author}
  {\bibfnamefont {C.~I.}\ \bibnamefont {{Torrie}}}, \bibinfo {author}
  {\bibfnamefont {D.}~\bibnamefont {{T{\"o}yr{\"a}}}}, \bibinfo {author}
  {\bibfnamefont {F.}~\bibnamefont {{Travasso}}}, \bibinfo {author}
  {\bibfnamefont {G.}~\bibnamefont {{Traylor}}}, \bibinfo {author}
  {\bibfnamefont {D.}~\bibnamefont {{Trifir{\`o}}}}, \bibinfo {author}
  {\bibfnamefont {M.~C.}\ \bibnamefont {{Tringali}}}, \bibinfo {author}
  {\bibfnamefont {L.}~\bibnamefont {{Trozzo}}}, \bibinfo {author}
  {\bibfnamefont {M.}~\bibnamefont {{Tse}}}, \bibinfo {author} {\bibfnamefont
  {M.}~\bibnamefont {{Turconi}}}, \bibinfo {author} {\bibfnamefont
  {D.}~\bibnamefont {{Tuyenbayev}}}, \bibinfo {author} {\bibfnamefont
  {D.}~\bibnamefont {{Ugolini}}}, \bibinfo {author} {\bibfnamefont {C.~S.}\
  \bibnamefont {{Unnikrishnan}}}, \bibinfo {author} {\bibfnamefont {A.~L.}\
  \bibnamefont {{Urban}}}, \bibinfo {author} {\bibfnamefont {S.~A.}\
  \bibnamefont {{Usman}}}, \bibinfo {author} {\bibfnamefont {H.}~\bibnamefont
  {{Vahlbruch}}}, \bibinfo {author} {\bibfnamefont {G.}~\bibnamefont
  {{Vajente}}}, \bibinfo {author} {\bibfnamefont {G.}~\bibnamefont {{Valdes}}},
  \bibinfo {author} {\bibfnamefont {M.}~\bibnamefont {{Vallisneri}}}, \bibinfo
  {author} {\bibfnamefont {N.}~\bibnamefont {{van Bakel}}}, \bibinfo {author}
  {\bibfnamefont {M.}~\bibnamefont {{van Beuzekom}}}, \bibinfo {author}
  {\bibfnamefont {J.~F.~J.}\ \bibnamefont {{van den Brand}}}, \bibinfo {author}
  {\bibfnamefont {C.}~\bibnamefont {{Van Den Broeck}}}, \bibinfo {author}
  {\bibfnamefont {D.~C.}\ \bibnamefont {{Vander-Hyde}}}, \bibinfo {author}
  {\bibfnamefont {L.}~\bibnamefont {{van der Schaaf}}}, \bibinfo {author}
  {\bibfnamefont {J.~V.}\ \bibnamefont {{van Heijningen}}}, \bibinfo {author}
  {\bibfnamefont {A.~A.}\ \bibnamefont {{van Veggel}}}, \bibinfo {author}
  {\bibfnamefont {M.}~\bibnamefont {{Vardaro}}}, \bibinfo {author}
  {\bibfnamefont {S.}~\bibnamefont {{Vass}}}, \bibinfo {author} {\bibfnamefont
  {M.}~\bibnamefont {{Vas{\'u}th}}}, \bibinfo {author} {\bibfnamefont
  {R.}~\bibnamefont {{Vaulin}}}, \bibinfo {author} {\bibfnamefont
  {A.}~\bibnamefont {{Vecchio}}}, \bibinfo {author} {\bibfnamefont
  {G.}~\bibnamefont {{Vedovato}}}, \bibinfo {author} {\bibfnamefont
  {J.}~\bibnamefont {{Veitch}}}, \bibinfo {author} {\bibfnamefont {P.~J.}\
  \bibnamefont {{Veitch}}}, \bibinfo {author} {\bibfnamefont {K.}~\bibnamefont
  {{Venkateswara}}}, \bibinfo {author} {\bibfnamefont {D.}~\bibnamefont
  {{Verkindt}}}, \bibinfo {author} {\bibfnamefont {F.}~\bibnamefont
  {{Vetrano}}}, \bibinfo {author} {\bibfnamefont {A.}~\bibnamefont
  {{Vicer{\'e}}}}, \bibinfo {author} {\bibfnamefont {S.}~\bibnamefont
  {{Vinciguerra}}}, \bibinfo {author} {\bibfnamefont {D.~J.}\ \bibnamefont
  {{Vine}}}, \bibinfo {author} {\bibfnamefont {J.~Y.}\ \bibnamefont {{Vinet}}},
  \bibinfo {author} {\bibfnamefont {S.}~\bibnamefont {{Vitale}}}, \bibinfo
  {author} {\bibfnamefont {T.}~\bibnamefont {{Vo}}}, \bibinfo {author}
  {\bibfnamefont {H.}~\bibnamefont {{Vocca}}}, \bibinfo {author} {\bibfnamefont
  {C.}~\bibnamefont {{Vorvick}}}, \bibinfo {author} {\bibfnamefont
  {D.}~\bibnamefont {{Voss}}}, \bibinfo {author} {\bibfnamefont {W.~D.}\
  \bibnamefont {{Vousden}}}, \bibinfo {author} {\bibfnamefont {S.~P.}\
  \bibnamefont {{Vyatchanin}}}, \bibinfo {author} {\bibfnamefont {A.~R.}\
  \bibnamefont {{Wade}}}, \bibinfo {author} {\bibfnamefont {L.~E.}\
  \bibnamefont {{Wade}}}, \bibinfo {author} {\bibfnamefont {M.}~\bibnamefont
  {{Wade}}}, \bibinfo {author} {\bibfnamefont {M.}~\bibnamefont {{Walker}}},
  \bibinfo {author} {\bibfnamefont {L.}~\bibnamefont {{Wallace}}}, \bibinfo
  {author} {\bibfnamefont {S.}~\bibnamefont {{Walsh}}}, \bibinfo {author}
  {\bibfnamefont {G.}~\bibnamefont {{Wang}}}, \bibinfo {author} {\bibfnamefont
  {H.}~\bibnamefont {{Wang}}}, \bibinfo {author} {\bibfnamefont
  {M.}~\bibnamefont {{Wang}}}, \bibinfo {author} {\bibfnamefont
  {X.}~\bibnamefont {{Wang}}}, \bibinfo {author} {\bibfnamefont
  {Y.}~\bibnamefont {{Wang}}}, \bibinfo {author} {\bibfnamefont {R.~L.}\
  \bibnamefont {{Ward}}}, \bibinfo {author} {\bibfnamefont {J.}~\bibnamefont
  {{Warner}}}, \bibinfo {author} {\bibfnamefont {M.}~\bibnamefont {{Was}}},
  \bibinfo {author} {\bibfnamefont {B.}~\bibnamefont {{Weaver}}}, \bibinfo
  {author} {\bibfnamefont {L.~W.}\ \bibnamefont {{Wei}}}, \bibinfo {author}
  {\bibfnamefont {M.}~\bibnamefont {{Weinert}}}, \bibinfo {author}
  {\bibfnamefont {A.~J.}\ \bibnamefont {{Weinstein}}}, \bibinfo {author}
  {\bibfnamefont {R.}~\bibnamefont {{Weiss}}}, \bibinfo {author} {\bibfnamefont
  {T.}~\bibnamefont {{Welborn}}}, \bibinfo {author} {\bibfnamefont
  {L.}~\bibnamefont {{Wen}}}, \bibinfo {author} {\bibfnamefont
  {P.}~\bibnamefont {{We{\ss}els}}}, \bibinfo {author} {\bibfnamefont
  {T.}~\bibnamefont {{Westphal}}}, \bibinfo {author} {\bibfnamefont
  {K.}~\bibnamefont {{Wette}}}, \bibinfo {author} {\bibfnamefont {J.~T.}\
  \bibnamefont {{Whelan}}}, \bibinfo {author} {\bibfnamefont {D.~J.}\
  \bibnamefont {{White}}}, \bibinfo {author} {\bibfnamefont {B.~F.}\
  \bibnamefont {{Whiting}}}, \bibinfo {author} {\bibfnamefont {R.~D.}\
  \bibnamefont {{Williams}}}, \bibinfo {author} {\bibfnamefont {A.~R.}\
  \bibnamefont {{Williamson}}}, \bibinfo {author} {\bibfnamefont {J.~L.}\
  \bibnamefont {{Willis}}}, \bibinfo {author} {\bibfnamefont {B.}~\bibnamefont
  {{Willke}}}, \bibinfo {author} {\bibfnamefont {M.~H.}\ \bibnamefont
  {{Wimmer}}}, \bibinfo {author} {\bibfnamefont {W.}~\bibnamefont {{Winkler}}},
  \bibinfo {author} {\bibfnamefont {C.~C.}\ \bibnamefont {{Wipf}}}, \bibinfo
  {author} {\bibfnamefont {H.}~\bibnamefont {{Wittel}}}, \bibinfo {author}
  {\bibfnamefont {G.}~\bibnamefont {{Woan}}}, \bibinfo {author} {\bibfnamefont
  {J.}~\bibnamefont {{Worden}}}, \bibinfo {author} {\bibfnamefont {J.~L.}\
  \bibnamefont {{Wright}}}, \bibinfo {author} {\bibfnamefont {G.}~\bibnamefont
  {{Wu}}}, \bibinfo {author} {\bibfnamefont {J.}~\bibnamefont {{Yablon}}},
  \bibinfo {author} {\bibfnamefont {W.}~\bibnamefont {{Yam}}}, \bibinfo
  {author} {\bibfnamefont {H.}~\bibnamefont {{Yamamoto}}}, \bibinfo {author}
  {\bibfnamefont {C.~C.}\ \bibnamefont {{Yancey}}}, \bibinfo {author}
  {\bibfnamefont {M.~J.}\ \bibnamefont {{Yap}}}, \bibinfo {author}
  {\bibfnamefont {H.}~\bibnamefont {{Yu}}}, \bibinfo {author} {\bibfnamefont
  {M.}~\bibnamefont {{Yvert}}}, \bibinfo {author} {\bibfnamefont
  {A.}~\bibnamefont {{Zadro{\.z}ny}}}, \bibinfo {author} {\bibfnamefont
  {L.}~\bibnamefont {{Zangrando}}}, \bibinfo {author} {\bibfnamefont
  {M.}~\bibnamefont {{Zanolin}}}, \bibinfo {author} {\bibfnamefont {J.~P.}\
  \bibnamefont {{Zendri}}}, \bibinfo {author} {\bibfnamefont {M.}~\bibnamefont
  {{Zevin}}}, \bibinfo {author} {\bibfnamefont {F.}~\bibnamefont {{Zhang}}},
  \bibinfo {author} {\bibfnamefont {L.}~\bibnamefont {{Zhang}}}, \bibinfo
  {author} {\bibfnamefont {M.}~\bibnamefont {{Zhang}}}, \bibinfo {author}
  {\bibfnamefont {Y.}~\bibnamefont {{Zhang}}}, \bibinfo {author} {\bibfnamefont
  {C.}~\bibnamefont {{Zhao}}}, \bibinfo {author} {\bibfnamefont
  {M.}~\bibnamefont {{Zhou}}}, \bibinfo {author} {\bibfnamefont
  {Z.}~\bibnamefont {{Zhou}}}, \bibinfo {author} {\bibfnamefont {X.~J.}\
  \bibnamefont {{Zhu}}}, \bibinfo {author} {\bibfnamefont {M.~E.}\ \bibnamefont
  {{Zucker}}}, \bibinfo {author} {\bibfnamefont {S.~E.}\ \bibnamefont
  {{Zuraw}}}, \bibinfo {author} {\bibfnamefont {J.}~\bibnamefont {{Zweizig}}},
  \bibinfo {author} {\bibnamefont {{LIGO Scientific Collaboration}}},\ and\
  \bibinfo {author} {\bibnamefont {{Virgo Collaboration}}},\ }\bibfield
  {title} {\bibinfo {title} {{The Rate of Binary Black Hole Mergers Inferred
  from Advanced LIGO Observations Surrounding GW150914}},\ }\href
  {https://doi.org/10.3847/2041-8205/833/1/L1} {\bibfield  {journal} {\bibinfo
  {journal} {\apjl}\ }\textbf {\bibinfo {volume} {833}},\ \bibinfo {eid} {L1}
  (\bibinfo {year} {2016}{\natexlab{c}})},\ \Eprint
  {https://arxiv.org/abs/1602.03842} {arXiv:1602.03842 [astro-ph.HE]}
  \BibitemShut {NoStop}%
\bibitem [{\citenamefont {Rodriguez}\ \emph {et~al.}(2019)\citenamefont
  {Rodriguez}, \citenamefont {Zevin}, \citenamefont {Amaro-Seoane},
  \citenamefont {Chatterjee}, \citenamefont {Kremer}, \citenamefont {Rasio},\
  and\ \citenamefont {Ye}}]{PhysRevD.100.043027}%
  \BibitemOpen
  \bibfield  {author} {\bibinfo {author} {\bibfnamefont {C.~L.}\ \bibnamefont
  {Rodriguez}}, \bibinfo {author} {\bibfnamefont {M.}~\bibnamefont {Zevin}},
  \bibinfo {author} {\bibfnamefont {P.}~\bibnamefont {Amaro-Seoane}}, \bibinfo
  {author} {\bibfnamefont {S.}~\bibnamefont {Chatterjee}}, \bibinfo {author}
  {\bibfnamefont {K.}~\bibnamefont {Kremer}}, \bibinfo {author} {\bibfnamefont
  {F.~A.}\ \bibnamefont {Rasio}},\ and\ \bibinfo {author} {\bibfnamefont
  {C.~S.}\ \bibnamefont {Ye}},\ }\bibfield  {title} {\bibinfo {title} {Black
  holes: The next generation---repeated mergers in dense star clusters and
  their gravitational-wave properties},\ }\href
  {https://doi.org/10.1103/PhysRevD.100.043027} {\bibfield  {journal} {\bibinfo
   {journal} {Phys. Rev. D}\ }\textbf {\bibinfo {volume} {100}},\ \bibinfo
  {pages} {043027} (\bibinfo {year} {2019})}\BibitemShut {NoStop}%
\bibitem [{\citenamefont {{Chatterjee}}\ \emph {et~al.}(2017)\citenamefont
  {{Chatterjee}}, \citenamefont {{Rodriguez}}, \citenamefont {{Kalogera}},\
  and\ \citenamefont {{Rasio}}}]{2017ApJ...836L..26C}%
  \BibitemOpen
  \bibfield  {author} {\bibinfo {author} {\bibfnamefont {S.}~\bibnamefont
  {{Chatterjee}}}, \bibinfo {author} {\bibfnamefont {C.~L.}\ \bibnamefont
  {{Rodriguez}}}, \bibinfo {author} {\bibfnamefont {V.}~\bibnamefont
  {{Kalogera}}},\ and\ \bibinfo {author} {\bibfnamefont {F.~A.}\ \bibnamefont
  {{Rasio}}},\ }\bibfield  {title} {\bibinfo {title} {{Dynamical Formation of
  Low-mass Merging Black Hole Binaries like GW151226}},\ }\href
  {https://doi.org/10.3847/2041-8213/aa5caa} {\bibfield  {journal} {\bibinfo
  {journal} {\apjl}\ }\textbf {\bibinfo {volume} {836}},\ \bibinfo {eid} {L26}
  (\bibinfo {year} {2017})},\ \Eprint {https://arxiv.org/abs/1609.06689}
  {arXiv:1609.06689 [astro-ph.GA]} \BibitemShut {NoStop}%
\bibitem [{\citenamefont {{Gong}}\ \emph {et~al.}(2021)\citenamefont {{Gong}},
  \citenamefont {{Xu}}, \citenamefont {{Gui}}, \citenamefont {{Huang}},\ and\
  \citenamefont {{Lau}}}]{2021hgwa.bookE..24G}%
  \BibitemOpen
  \bibfield  {author} {\bibinfo {author} {\bibfnamefont {X.}~\bibnamefont
  {{Gong}}}, \bibinfo {author} {\bibfnamefont {S.}~\bibnamefont {{Xu}}},
  \bibinfo {author} {\bibfnamefont {S.}~\bibnamefont {{Gui}}}, \bibinfo
  {author} {\bibfnamefont {S.}~\bibnamefont {{Huang}}},\ and\ \bibinfo {author}
  {\bibfnamefont {Y.-K.}\ \bibnamefont {{Lau}}},\ }\bibfield  {title} {\bibinfo
  {title} {{Mission Design for the TAIJI Mission and Structure Formation in
  Early Universe}},\ }in\ \href
  {https://doi.org/10.1007/978-981-15-4702-7_24-1} {\emph {\bibinfo {booktitle}
  {Handbook of Gravitational Wave Astronomy}}}\ (\bibinfo {year} {2021})\
  p.~\bibinfo {pages} {24}\BibitemShut {NoStop}%
\bibitem [{\citenamefont {{Corral-Santana}}\ \emph
  {et~al.}(2016{\natexlab{b}})\citenamefont {{Corral-Santana}}, \citenamefont
  {{Casares}}, \citenamefont {{Mu{\~n}oz-Darias}}, \citenamefont {{Bauer}},
  \citenamefont {{Mart{\'\i}nez-Pais}},\ and\ \citenamefont
  {{Russell}}}]{2016A&A...587A..61C}%
  \BibitemOpen
  \bibfield  {author} {\bibinfo {author} {\bibfnamefont {J.~M.}\ \bibnamefont
  {{Corral-Santana}}}, \bibinfo {author} {\bibfnamefont {J.}~\bibnamefont
  {{Casares}}}, \bibinfo {author} {\bibfnamefont {T.}~\bibnamefont
  {{Mu{\~n}oz-Darias}}}, \bibinfo {author} {\bibfnamefont {F.~E.}\ \bibnamefont
  {{Bauer}}}, \bibinfo {author} {\bibfnamefont {I.~G.}\ \bibnamefont
  {{Mart{\'\i}nez-Pais}}},\ and\ \bibinfo {author} {\bibfnamefont {D.~M.}\
  \bibnamefont {{Russell}}},\ }\bibfield  {title} {\bibinfo {title} {{BlackCAT:
  A catalogue of stellar-mass black holes in X-ray transients}},\ }\href
  {https://doi.org/10.1051/0004-6361/201527130} {\bibfield  {journal} {\bibinfo
   {journal} {\aap}\ }\textbf {\bibinfo {volume} {587}},\ \bibinfo {eid} {A61}
  (\bibinfo {year} {2016}{\natexlab{b}})},\ \Eprint
  {https://arxiv.org/abs/1510.08869} {arXiv:1510.08869 [astro-ph.HE]}
  \BibitemShut {NoStop}%
\bibitem [{\citenamefont {{McClintock}}\ and\ \citenamefont
  {{Remillard}}(1986)}]{1986ApJ...308..110M}%
  \BibitemOpen
  \bibfield  {author} {\bibinfo {author} {\bibfnamefont {J.~E.}\ \bibnamefont
  {{McClintock}}}\ and\ \bibinfo {author} {\bibfnamefont {R.~A.}\ \bibnamefont
  {{Remillard}}},\ }\bibfield  {title} {\bibinfo {title} {{The Black Hole
  Binary A0620-00}},\ }\href {https://doi.org/10.1086/164482} {\bibfield
  {journal} {\bibinfo  {journal} {\apj}\ }\textbf {\bibinfo {volume} {308}},\
  \bibinfo {pages} {110} (\bibinfo {year} {1986})}\BibitemShut {NoStop}%
\bibitem [{\citenamefont {{Elvis}}\ \emph {et~al.}(1975)\citenamefont
  {{Elvis}}, \citenamefont {{Page}}, \citenamefont {{Pounds}}, \citenamefont
  {{Ricketts}},\ and\ \citenamefont {{Turner}}}]{1975Natur.257..656E}%
  \BibitemOpen
  \bibfield  {author} {\bibinfo {author} {\bibfnamefont {M.}~\bibnamefont
  {{Elvis}}}, \bibinfo {author} {\bibfnamefont {C.~G.}\ \bibnamefont {{Page}}},
  \bibinfo {author} {\bibfnamefont {K.~A.}\ \bibnamefont {{Pounds}}}, \bibinfo
  {author} {\bibfnamefont {M.~J.}\ \bibnamefont {{Ricketts}}},\ and\ \bibinfo
  {author} {\bibfnamefont {M.~J.~L.}\ \bibnamefont {{Turner}}},\ }\bibfield
  {title} {\bibinfo {title} {{Discovery of powerful transient X-ray source
  A0620-00 with Ariel V Sky Survey Experiment}},\ }\href
  {https://doi.org/10.1038/257656a0} {\bibfield  {journal} {\bibinfo  {journal}
  {\nat}\ }\textbf {\bibinfo {volume} {257}},\ \bibinfo {pages} {656} (\bibinfo
  {year} {1975})}\BibitemShut {NoStop}%
\bibitem [{\citenamefont {{Shakura}}\ and\ \citenamefont
  {{Sunyaev}}(1973)}]{1973A&A....24..337S}%
  \BibitemOpen
  \bibfield  {author} {\bibinfo {author} {\bibfnamefont {N.~I.}\ \bibnamefont
  {{Shakura}}}\ and\ \bibinfo {author} {\bibfnamefont {R.~A.}\ \bibnamefont
  {{Sunyaev}}},\ }\bibfield  {title} {\bibinfo {title} {{Black holes in binary
  systems. Observational appearance.}},\ }\href@noop {} {\bibfield  {journal}
  {\bibinfo  {journal} {\aap}\ }\textbf {\bibinfo {volume} {24}},\ \bibinfo
  {pages} {337} (\bibinfo {year} {1973})}\BibitemShut {NoStop}%
\bibitem [{\citenamefont {{King}}(2003{\natexlab{a}})}]{2003astro.ph..1118K}%
  \BibitemOpen
  \bibfield  {author} {\bibinfo {author} {\bibfnamefont {A.~R.}\ \bibnamefont
  {{King}}},\ }\bibfield  {title} {\bibinfo {title} {{Accretion in Compact
  Binaries}},\ }\href {https://doi.org/10.48550/arXiv.astro-ph/0301118}
  {\bibfield  {journal} {\bibinfo  {journal} {arXiv e-prints}\ ,\ \bibinfo
  {eid} {astro-ph/0301118}} (\bibinfo {year} {2003}{\natexlab{a}})},\ \Eprint
  {https://arxiv.org/abs/astro-ph/0301118} {arXiv:astro-ph/0301118 [astro-ph]}
  \BibitemShut {NoStop}%
\bibitem [{\citenamefont {Velikhov}(1959)}]{velikhov1959stability}%
  \BibitemOpen
  \bibfield  {author} {\bibinfo {author} {\bibfnamefont {E.}~\bibnamefont
  {Velikhov}},\ }\bibfield  {title} {\bibinfo {title} {Stability of an ideally
  conducting liquid flowing between cylinders rotating in a magnetic field},\
  }\href@noop {} {\bibfield  {journal} {\bibinfo  {journal} {Sov. Phys. JETP}\
  }\textbf {\bibinfo {volume} {36}},\ \bibinfo {pages} {995} (\bibinfo {year}
  {1959})}\BibitemShut {NoStop}%
\bibitem [{\citenamefont {Chandrasekhar}(1960)}]{chandrasekhar1960stability}%
  \BibitemOpen
  \bibfield  {author} {\bibinfo {author} {\bibfnamefont {S.}~\bibnamefont
  {Chandrasekhar}},\ }\bibfield  {title} {\bibinfo {title} {The stability of
  non-dissipative couette flow in hydromagnetics},\ }\href@noop {} {\bibfield
  {journal} {\bibinfo  {journal} {Proceedings of the National Academy of
  Sciences}\ }\textbf {\bibinfo {volume} {46}},\ \bibinfo {pages} {253}
  (\bibinfo {year} {1960})}\BibitemShut {NoStop}%
\bibitem [{\citenamefont {Balbus}\ and\ \citenamefont
  {Hawley}(1991)}]{balbus1991powerful}%
  \BibitemOpen
  \bibfield  {author} {\bibinfo {author} {\bibfnamefont {S.~A.}\ \bibnamefont
  {Balbus}}\ and\ \bibinfo {author} {\bibfnamefont {J.~F.}\ \bibnamefont
  {Hawley}},\ }\bibfield  {title} {\bibinfo {title} {A powerful local shear
  instability in weakly magnetized disks. i-linear analysis. ii-nonlinear
  evolution},\ }\href@noop {} {\bibfield  {journal} {\bibinfo  {journal} {The
  Astrophysical Journal}\ }\textbf {\bibinfo {volume} {376}},\ \bibinfo {pages}
  {214} (\bibinfo {year} {1991})}\BibitemShut {NoStop}%
\bibitem [{\citenamefont {{Pringle}}(1981)}]{1981ARA&A..19..137P}%
  \BibitemOpen
  \bibfield  {author} {\bibinfo {author} {\bibfnamefont {J.~E.}\ \bibnamefont
  {{Pringle}}},\ }\bibfield  {title} {\bibinfo {title} {{Accretion discs in
  astrophysics}},\ }\href {https://doi.org/10.1146/annurev.aa.19.090181.001033}
  {\bibfield  {journal} {\bibinfo  {journal} {\araa}\ }\textbf {\bibinfo
  {volume} {19}},\ \bibinfo {pages} {137} (\bibinfo {year} {1981})}\BibitemShut
  {NoStop}%
\bibitem [{\citenamefont {{de S{\'a}}}\ \emph {et~al.}(2023)\citenamefont {{de
  S{\'a}}}, \citenamefont {{Bernardo}}, \citenamefont {{Bachega}},
  \citenamefont {{Rocha}}, \citenamefont {{Moraes}},\ and\ \citenamefont
  {{Horvath}}}]{2023Galax..11...19D}%
  \BibitemOpen
  \bibfield  {author} {\bibinfo {author} {\bibfnamefont {L.~M.~d.}\
  \bibnamefont {{de S{\'a}}}}, \bibinfo {author} {\bibfnamefont
  {A.}~\bibnamefont {{Bernardo}}}, \bibinfo {author} {\bibfnamefont {R.~R.~A.}\
  \bibnamefont {{Bachega}}}, \bibinfo {author} {\bibfnamefont {L.~S.}\
  \bibnamefont {{Rocha}}}, \bibinfo {author} {\bibfnamefont {P.~H.~R.~S.}\
  \bibnamefont {{Moraes}}},\ and\ \bibinfo {author} {\bibfnamefont {J.~E.}\
  \bibnamefont {{Horvath}}},\ }\bibfield  {title} {\bibinfo {title} {{An
  Overview of Compact Star Populations and Some of Its Open Problems}},\ }\href
  {https://doi.org/10.3390/galaxies11010019} {\bibfield  {journal} {\bibinfo
  {journal} {Galaxies}\ }\textbf {\bibinfo {volume} {11}},\ \bibinfo {pages}
  {19} (\bibinfo {year} {2023})},\ \Eprint {https://arxiv.org/abs/2301.07780}
  {arXiv:2301.07780 [astro-ph.HE]} \BibitemShut {NoStop}%
\bibitem [{\citenamefont {Liao}\ \emph {et~al.}(2020)\citenamefont {Liao},
  \citenamefont {Liu}, \citenamefont {Zheng},\ and\ \citenamefont
  {Gou}}]{10.1093/mnras/staa162}%
  \BibitemOpen
  \bibfield  {author} {\bibinfo {author} {\bibfnamefont {Z.}~\bibnamefont
  {Liao}}, \bibinfo {author} {\bibfnamefont {J.}~\bibnamefont {Liu}}, \bibinfo
  {author} {\bibfnamefont {X.}~\bibnamefont {Zheng}},\ and\ \bibinfo {author}
  {\bibfnamefont {L.}~\bibnamefont {Gou}},\ }\bibfield  {title} {\bibinfo
  {title} {{Spectral evidence of an accretion disc in wind-fed X-ray pulsar
  Vela X-1 during an unusual spin-up period}},\ }\href
  {https://doi.org/10.1093/mnras/staa162} {\bibfield  {journal} {\bibinfo
  {journal} {Monthly Notices of the Royal Astronomical Society}\ }\textbf
  {\bibinfo {volume} {492}},\ \bibinfo {pages} {5922} (\bibinfo {year}
  {2020})},\ \Eprint
  {https://arxiv.org/abs/https://academic.oup.com/mnras/article-pdf/492/4/5922/32444089/staa162.pdf}
  {https://academic.oup.com/mnras/article-pdf/492/4/5922/32444089/staa162.pdf}
  \BibitemShut {NoStop}%
\bibitem [{\citenamefont {{McClintock}}\ and\ \citenamefont
  {{Remillard}}(2006)}]{2006csxs.book..157M}%
  \BibitemOpen
  \bibfield  {author} {\bibinfo {author} {\bibfnamefont {J.~E.}\ \bibnamefont
  {{McClintock}}}\ and\ \bibinfo {author} {\bibfnamefont {R.~A.}\ \bibnamefont
  {{Remillard}}},\ }\bibfield  {title} {\bibinfo {title} {{Black hole
  binaries}},\ }in\ \href {https://doi.org/10.48550/arXiv.astro-ph/0306213}
  {\emph {\bibinfo {booktitle} {Compact stellar X-ray sources}}},\
  Vol.~\bibinfo {volume} {39}\ (\bibinfo {year} {2006})\ pp.\ \bibinfo {pages}
  {157--213}\BibitemShut {NoStop}%
\bibitem [{\citenamefont {{Mitsuda}}\ \emph {et~al.}(1984)\citenamefont
  {{Mitsuda}}, \citenamefont {{Inoue}}, \citenamefont {{Koyama}}, \citenamefont
  {{Makishima}}, \citenamefont {{Matsuoka}}, \citenamefont {{Ogawara}},
  \citenamefont {{Shibazaki}}, \citenamefont {{Suzuki}}, \citenamefont
  {{Tanaka}},\ and\ \citenamefont {{Hirano}}}]{1984PASJ...36..741M}%
  \BibitemOpen
  \bibfield  {author} {\bibinfo {author} {\bibfnamefont {K.}~\bibnamefont
  {{Mitsuda}}}, \bibinfo {author} {\bibfnamefont {H.}~\bibnamefont {{Inoue}}},
  \bibinfo {author} {\bibfnamefont {K.}~\bibnamefont {{Koyama}}}, \bibinfo
  {author} {\bibfnamefont {K.}~\bibnamefont {{Makishima}}}, \bibinfo {author}
  {\bibfnamefont {M.}~\bibnamefont {{Matsuoka}}}, \bibinfo {author}
  {\bibfnamefont {Y.}~\bibnamefont {{Ogawara}}}, \bibinfo {author}
  {\bibfnamefont {N.}~\bibnamefont {{Shibazaki}}}, \bibinfo {author}
  {\bibfnamefont {K.}~\bibnamefont {{Suzuki}}}, \bibinfo {author}
  {\bibfnamefont {Y.}~\bibnamefont {{Tanaka}}},\ and\ \bibinfo {author}
  {\bibfnamefont {T.}~\bibnamefont {{Hirano}}},\ }\bibfield  {title} {\bibinfo
  {title} {{Energy spectra of low-mass binary X-ray sources observed from
  Tenma.}},\ }\href@noop {} {\bibfield  {journal} {\bibinfo  {journal} {\pasj}\
  }\textbf {\bibinfo {volume} {36}},\ \bibinfo {pages} {741} (\bibinfo {year}
  {1984})}\BibitemShut {NoStop}%
\bibitem [{\citenamefont {{Feng}}\ and\ \citenamefont
  {{Soria}}(2011)}]{2011NewAR..55..166F}%
  \BibitemOpen
  \bibfield  {author} {\bibinfo {author} {\bibfnamefont {H.}~\bibnamefont
  {{Feng}}}\ and\ \bibinfo {author} {\bibfnamefont {R.}~\bibnamefont
  {{Soria}}},\ }\bibfield  {title} {\bibinfo {title} {{Ultraluminous X-ray
  sources in the Chandra and XMM-Newton era}},\ }\href
  {https://doi.org/10.1016/j.newar.2011.08.002} {\bibfield  {journal} {\bibinfo
   {journal} {\nar}\ }\textbf {\bibinfo {volume} {55}},\ \bibinfo {pages} {166}
  (\bibinfo {year} {2011})},\ \Eprint {https://arxiv.org/abs/1109.1610}
  {arXiv:1109.1610 [astro-ph.HE]} \BibitemShut {NoStop}%
\bibitem [{\citenamefont {Merloni}\ \emph {et~al.}(2000)\citenamefont
  {Merloni}, \citenamefont {Fabian},\ and\ \citenamefont
  {Ross}}]{10.1046/j.1365-8711.2000.03226.x}%
  \BibitemOpen
  \bibfield  {author} {\bibinfo {author} {\bibfnamefont {A.}~\bibnamefont
  {Merloni}}, \bibinfo {author} {\bibfnamefont {A.~C.}\ \bibnamefont
  {Fabian}},\ and\ \bibinfo {author} {\bibfnamefont {R.~R.}\ \bibnamefont
  {Ross}},\ }\bibfield  {title} {\bibinfo {title} {{On the interpretation of
  the multicolour disc model for black hole candidates}},\ }\href
  {https://doi.org/10.1046/j.1365-8711.2000.03226.x} {\bibfield  {journal}
  {\bibinfo  {journal} {Monthly Notices of the Royal Astronomical Society}\
  }\textbf {\bibinfo {volume} {313}},\ \bibinfo {pages} {193} (\bibinfo {year}
  {2000})},\ \Eprint
  {https://arxiv.org/abs/https://academic.oup.com/mnras/article-pdf/313/1/193/18411050/313-1-193.pdf}
  {https://academic.oup.com/mnras/article-pdf/313/1/193/18411050/313-1-193.pdf}
  \BibitemShut {NoStop}%
\bibitem [{\citenamefont {{Quataert}}\ and\ \citenamefont
  {{Narayan}}(1999)}]{1999ApJ...520..298Q}%
  \BibitemOpen
  \bibfield  {author} {\bibinfo {author} {\bibfnamefont {E.}~\bibnamefont
  {{Quataert}}}\ and\ \bibinfo {author} {\bibfnamefont {R.}~\bibnamefont
  {{Narayan}}},\ }\bibfield  {title} {\bibinfo {title} {{Spectral Models of
  Advection-dominated Accretion Flows with Winds}},\ }\href
  {https://doi.org/10.1086/307439} {\bibfield  {journal} {\bibinfo  {journal}
  {\apj}\ }\textbf {\bibinfo {volume} {520}},\ \bibinfo {pages} {298} (\bibinfo
  {year} {1999})},\ \Eprint {https://arxiv.org/abs/astro-ph/9810136}
  {arXiv:astro-ph/9810136 [astro-ph]} \BibitemShut {NoStop}%
\bibitem [{\citenamefont {{Narayan}}(1996)}]{1996ApJ...462..136N}%
  \BibitemOpen
  \bibfield  {author} {\bibinfo {author} {\bibfnamefont {R.}~\bibnamefont
  {{Narayan}}},\ }\bibfield  {title} {\bibinfo {title} {{Advection-dominated
  Models of Luminous Accreting Black Holes}},\ }\href
  {https://doi.org/10.1086/177136} {\bibfield  {journal} {\bibinfo  {journal}
  {\apj}\ }\textbf {\bibinfo {volume} {462}},\ \bibinfo {pages} {136} (\bibinfo
  {year} {1996})},\ \Eprint {https://arxiv.org/abs/astro-ph/9510028}
  {arXiv:astro-ph/9510028 [astro-ph]} \BibitemShut {NoStop}%
\bibitem [{\citenamefont {{Esin}}\ \emph {et~al.}(1997)\citenamefont {{Esin}},
  \citenamefont {{McClintock}},\ and\ \citenamefont
  {{Narayan}}}]{1997ApJ...489..865E}%
  \BibitemOpen
  \bibfield  {author} {\bibinfo {author} {\bibfnamefont {A.~A.}\ \bibnamefont
  {{Esin}}}, \bibinfo {author} {\bibfnamefont {J.~E.}\ \bibnamefont
  {{McClintock}}},\ and\ \bibinfo {author} {\bibfnamefont {R.}~\bibnamefont
  {{Narayan}}},\ }\bibfield  {title} {\bibinfo {title} {{Advection-Dominated
  Accretion and the Spectral States of Black Hole X-Ray Binaries: Application
  to Nova Muscae 1991}},\ }\href {https://doi.org/10.1086/304829} {\bibfield
  {journal} {\bibinfo  {journal} {\apj}\ }\textbf {\bibinfo {volume} {489}},\
  \bibinfo {pages} {865} (\bibinfo {year} {1997})},\ \Eprint
  {https://arxiv.org/abs/astro-ph/9705237} {arXiv:astro-ph/9705237 [astro-ph]}
  \BibitemShut {NoStop}%
\bibitem [{\citenamefont {{Lasota}}(2001)}]{2001NewAR..45..449L}%
  \BibitemOpen
  \bibfield  {author} {\bibinfo {author} {\bibfnamefont {J.-P.}\ \bibnamefont
  {{Lasota}}},\ }\bibfield  {title} {\bibinfo {title} {{The disc instability
  model of dwarf novae and low-mass X-ray binary transients}},\ }\href
  {https://doi.org/10.1016/S1387-6473(01)00112-9} {\bibfield  {journal}
  {\bibinfo  {journal} {\nar}\ }\textbf {\bibinfo {volume} {45}},\ \bibinfo
  {pages} {449} (\bibinfo {year} {2001})},\ \Eprint
  {https://arxiv.org/abs/astro-ph/0102072} {arXiv:astro-ph/0102072 [astro-ph]}
  \BibitemShut {NoStop}%
\bibitem [{\citenamefont {{Frank}}\ \emph {et~al.}(2002)\citenamefont
  {{Frank}}, \citenamefont {{King}},\ and\ \citenamefont
  {{Raine}}}]{2002apa..book.....F}%
  \BibitemOpen
  \bibfield  {author} {\bibinfo {author} {\bibfnamefont {J.}~\bibnamefont
  {{Frank}}}, \bibinfo {author} {\bibfnamefont {A.}~\bibnamefont {{King}}},\
  and\ \bibinfo {author} {\bibfnamefont {D.~J.}\ \bibnamefont {{Raine}}},\
  }\href@noop {} {\emph {\bibinfo {title} {{Accretion Power in Astrophysics:
  Third Edition}}}}\ (\bibinfo {year} {2002})\BibitemShut {NoStop}%
\bibitem [{\citenamefont {{King}}(2003{\natexlab{b}})}]{2003ASPC..308..121K}%
  \BibitemOpen
  \bibfield  {author} {\bibinfo {author} {\bibfnamefont {A.~R.}\ \bibnamefont
  {{King}}},\ }\bibfield  {title} {\bibinfo {title} {{Disc Instabilities in
  Soft X-ray Transients}},\ }in\ \href@noop {} {\emph {\bibinfo {booktitle}
  {From X-ray Binaries to Gamma-Ray Bursts: Jan van Paradijs Memorial
  Symposium}}},\ \bibinfo {series} {Astronomical Society of the Pacific
  Conference Series}, Vol.\ \bibinfo {volume} {308},\ \bibinfo {editor} {edited
  by\ \bibinfo {editor} {\bibfnamefont {E.~P.}\ \bibnamefont {{van den
  Heuvel}}}, \bibinfo {editor} {\bibfnamefont {L.}~\bibnamefont {{Kaper}}},
  \bibinfo {editor} {\bibfnamefont {E.}~\bibnamefont {{Rol}}},\ and\ \bibinfo
  {editor} {\bibfnamefont {R.~A.~M.~J.}\ \bibnamefont {{Wijers}}}}\ (\bibinfo
  {year} {2003})\ p.\ \bibinfo {pages} {121}\BibitemShut {NoStop}%
\bibitem [{\citenamefont {{Remillard}}\ and\ \citenamefont
  {{McClintock}}(2006{\natexlab{b}})}]{2006ARA&A..44...49R}%
  \BibitemOpen
  \bibfield  {author} {\bibinfo {author} {\bibfnamefont {R.~A.}\ \bibnamefont
  {{Remillard}}}\ and\ \bibinfo {author} {\bibfnamefont {J.~E.}\ \bibnamefont
  {{McClintock}}},\ }\bibfield  {title} {\bibinfo {title} {{X-Ray Properties of
  Black-Hole Binaries}},\ }\href
  {https://doi.org/10.1146/annurev.astro.44.051905.092532} {\bibfield
  {journal} {\bibinfo  {journal} {\araa}\ }\textbf {\bibinfo {volume} {44}},\
  \bibinfo {pages} {49} (\bibinfo {year} {2006}{\natexlab{b}})},\ \Eprint
  {https://arxiv.org/abs/astro-ph/0606352} {arXiv:astro-ph/0606352 [astro-ph]}
  \BibitemShut {NoStop}%
\bibitem [{\citenamefont {{McClintock}}\ \emph {et~al.}(2001)\citenamefont
  {{McClintock}}, \citenamefont {{Haswell}}, \citenamefont {{Garcia}},
  \citenamefont {{Drake}}, \citenamefont {{Hynes}}, \citenamefont {{Marshall}},
  \citenamefont {{Muno}}, \citenamefont {{Chaty}}, \citenamefont {{Garnavich}},
  \citenamefont {{Groot}}, \citenamefont {{Lewin}}, \citenamefont {{Mauche}},
  \citenamefont {{Miller}}, \citenamefont {{Pooley}}, \citenamefont
  {{Shrader}},\ and\ \citenamefont {{Vrtilek}}}]{2001ApJ...555..477M}%
  \BibitemOpen
  \bibfield  {author} {\bibinfo {author} {\bibfnamefont {J.~E.}\ \bibnamefont
  {{McClintock}}}, \bibinfo {author} {\bibfnamefont {C.~A.}\ \bibnamefont
  {{Haswell}}}, \bibinfo {author} {\bibfnamefont {M.~R.}\ \bibnamefont
  {{Garcia}}}, \bibinfo {author} {\bibfnamefont {J.~J.}\ \bibnamefont
  {{Drake}}}, \bibinfo {author} {\bibfnamefont {R.~I.}\ \bibnamefont
  {{Hynes}}}, \bibinfo {author} {\bibfnamefont {H.~L.}\ \bibnamefont
  {{Marshall}}}, \bibinfo {author} {\bibfnamefont {M.~P.}\ \bibnamefont
  {{Muno}}}, \bibinfo {author} {\bibfnamefont {S.}~\bibnamefont {{Chaty}}},
  \bibinfo {author} {\bibfnamefont {P.~M.}\ \bibnamefont {{Garnavich}}},
  \bibinfo {author} {\bibfnamefont {P.~J.}\ \bibnamefont {{Groot}}}, \bibinfo
  {author} {\bibfnamefont {W.~H.~G.}\ \bibnamefont {{Lewin}}}, \bibinfo
  {author} {\bibfnamefont {C.~W.}\ \bibnamefont {{Mauche}}}, \bibinfo {author}
  {\bibfnamefont {J.~M.}\ \bibnamefont {{Miller}}}, \bibinfo {author}
  {\bibfnamefont {G.~G.}\ \bibnamefont {{Pooley}}}, \bibinfo {author}
  {\bibfnamefont {C.~R.}\ \bibnamefont {{Shrader}}},\ and\ \bibinfo {author}
  {\bibfnamefont {S.~D.}\ \bibnamefont {{Vrtilek}}},\ }\bibfield  {title}
  {\bibinfo {title} {{Complete and Simultaneous Spectral Observations of the
  Black Hole X-Ray Nova XTE J1118+480}},\ }\href
  {https://doi.org/10.1086/321449} {\bibfield  {journal} {\bibinfo  {journal}
  {\apj}\ }\textbf {\bibinfo {volume} {555}},\ \bibinfo {pages} {477} (\bibinfo
  {year} {2001})},\ \Eprint {https://arxiv.org/abs/astro-ph/0103051}
  {arXiv:astro-ph/0103051 [astro-ph]} \BibitemShut {NoStop}%
\bibitem [{\citenamefont {{McClintock}}\ \emph {et~al.}(2003)\citenamefont
  {{McClintock}}, \citenamefont {{Narayan}}, \citenamefont {{Garcia}},
  \citenamefont {{Orosz}}, \citenamefont {{Remillard}},\ and\ \citenamefont
  {{Murray}}}]{2003ApJ...593..435M}%
  \BibitemOpen
  \bibfield  {author} {\bibinfo {author} {\bibfnamefont {J.~E.}\ \bibnamefont
  {{McClintock}}}, \bibinfo {author} {\bibfnamefont {R.}~\bibnamefont
  {{Narayan}}}, \bibinfo {author} {\bibfnamefont {M.~R.}\ \bibnamefont
  {{Garcia}}}, \bibinfo {author} {\bibfnamefont {J.~A.}\ \bibnamefont
  {{Orosz}}}, \bibinfo {author} {\bibfnamefont {R.~A.}\ \bibnamefont
  {{Remillard}}},\ and\ \bibinfo {author} {\bibfnamefont {S.~S.}\ \bibnamefont
  {{Murray}}},\ }\bibfield  {title} {\bibinfo {title} {{Multiwavelength
  Spectrum of the Black Hole XTE J1118+480 in Quiescence}},\ }\href
  {https://doi.org/10.1086/376406} {\bibfield  {journal} {\bibinfo  {journal}
  {\apj}\ }\textbf {\bibinfo {volume} {593}},\ \bibinfo {pages} {435} (\bibinfo
  {year} {2003})},\ \Eprint {https://arxiv.org/abs/astro-ph/0304535}
  {arXiv:astro-ph/0304535 [astro-ph]} \BibitemShut {NoStop}%
\bibitem [{\citenamefont {{Yuan}}\ and\ \citenamefont
  {{Narayan}}(2014)}]{2014ARA&A..52..529Y}%
  \BibitemOpen
  \bibfield  {author} {\bibinfo {author} {\bibfnamefont {F.}~\bibnamefont
  {{Yuan}}}\ and\ \bibinfo {author} {\bibfnamefont {R.}~\bibnamefont
  {{Narayan}}},\ }\bibfield  {title} {\bibinfo {title} {{Hot Accretion Flows
  Around Black Holes}},\ }\href
  {https://doi.org/10.1146/annurev-astro-082812-141003} {\bibfield  {journal}
  {\bibinfo  {journal} {\araa}\ }\textbf {\bibinfo {volume} {52}},\ \bibinfo
  {pages} {529} (\bibinfo {year} {2014})},\ \Eprint
  {https://arxiv.org/abs/1401.0586} {arXiv:1401.0586 [astro-ph.HE]}
  \BibitemShut {NoStop}%
\bibitem [{\citenamefont {{Yan}}\ \emph {et~al.}(2020)\citenamefont {{Yan}},
  \citenamefont {{Xie}},\ and\ \citenamefont {{Zhang}}}]{2020ApJ...889L..18Y}%
  \BibitemOpen
  \bibfield  {author} {\bibinfo {author} {\bibfnamefont {Z.}~\bibnamefont
  {{Yan}}}, \bibinfo {author} {\bibfnamefont {F.-G.}\ \bibnamefont {{Xie}}},\
  and\ \bibinfo {author} {\bibfnamefont {W.}~\bibnamefont {{Zhang}}},\
  }\bibfield  {title} {\bibinfo {title} {{Coronal Properties of Black Hole
  X-Ray Binaries in the Hard State as Seen by NuSTAR and Swift}},\ }\href
  {https://doi.org/10.3847/2041-8213/ab665e} {\bibfield  {journal} {\bibinfo
  {journal} {\apjl}\ }\textbf {\bibinfo {volume} {889}},\ \bibinfo {eid} {L18}
  (\bibinfo {year} {2020})},\ \Eprint {https://arxiv.org/abs/1912.12145}
  {arXiv:1912.12145 [astro-ph.HE]} \BibitemShut {NoStop}%
\bibitem [{\citenamefont {{Fender}}(2006)}]{2006csxs.book..381F}%
  \BibitemOpen
  \bibfield  {author} {\bibinfo {author} {\bibfnamefont {R.}~\bibnamefont
  {{Fender}}},\ }\bibfield  {title} {\bibinfo {title} {{Jets from X-ray
  binaries}},\ }in\ \href {https://doi.org/10.48550/arXiv.astro-ph/0303339}
  {\emph {\bibinfo {booktitle} {Compact stellar X-ray sources}}},\
  Vol.~\bibinfo {volume} {39}\ (\bibinfo {year} {2006})\ pp.\ \bibinfo {pages}
  {381--419}\BibitemShut {NoStop}%
\bibitem [{\citenamefont {{Zdziarski}}\ \emph {et~al.}(2000)\citenamefont
  {{Zdziarski}}, \citenamefont {{Poutanen}},\ and\ \citenamefont
  {{Johnson}}}]{2000ApJ...542..703Z}%
  \BibitemOpen
  \bibfield  {author} {\bibinfo {author} {\bibfnamefont {A.~A.}\ \bibnamefont
  {{Zdziarski}}}, \bibinfo {author} {\bibfnamefont {J.}~\bibnamefont
  {{Poutanen}}},\ and\ \bibinfo {author} {\bibfnamefont {W.~N.}\ \bibnamefont
  {{Johnson}}},\ }\bibfield  {title} {\bibinfo {title} {{Observations of
  Seyfert Galaxies by OSSE and Parameters of Their X-Ray/Gamma-Ray Sources}},\
  }\href {https://doi.org/10.1086/317046} {\bibfield  {journal} {\bibinfo
  {journal} {\apj}\ }\textbf {\bibinfo {volume} {542}},\ \bibinfo {pages} {703}
  (\bibinfo {year} {2000})},\ \Eprint {https://arxiv.org/abs/astro-ph/0006151}
  {arXiv:astro-ph/0006151 [astro-ph]} \BibitemShut {NoStop}%
\bibitem [{\citenamefont {{Titarchuk}}\ and\ \citenamefont
  {{Shrader}}(2002)}]{2002ApJ...567.1057T}%
  \BibitemOpen
  \bibfield  {author} {\bibinfo {author} {\bibfnamefont {L.}~\bibnamefont
  {{Titarchuk}}}\ and\ \bibinfo {author} {\bibfnamefont {C.~R.}\ \bibnamefont
  {{Shrader}}},\ }\bibfield  {title} {\bibinfo {title} {{Observational
  Signatures of Black Holes: Spectral and Temporal Features of XTE
  J1550-564}},\ }\href {https://doi.org/10.1086/338661} {\bibfield  {journal}
  {\bibinfo  {journal} {\apj}\ }\textbf {\bibinfo {volume} {567}},\ \bibinfo
  {pages} {1057} (\bibinfo {year} {2002})},\ \Eprint
  {https://arxiv.org/abs/astro-ph/0111272} {arXiv:astro-ph/0111272 [astro-ph]}
  \BibitemShut {NoStop}%
\bibitem [{\citenamefont {{Yan}}\ and\ \citenamefont
  {{Wang}}(2011)}]{2011RAA....11..631Y}%
  \BibitemOpen
  \bibfield  {author} {\bibinfo {author} {\bibfnamefont {L.-H.}\ \bibnamefont
  {{Yan}}}\ and\ \bibinfo {author} {\bibfnamefont {J.-C.}\ \bibnamefont
  {{Wang}}},\ }\bibfield  {title} {\bibinfo {title} {{Origin of power-law X-ray
  emission in the steep power-law state of X-ray binaries}},\ }\href
  {https://doi.org/10.1088/1674-4527/11/6/002} {\bibfield  {journal} {\bibinfo
  {journal} {Research in Astronomy and Astrophysics}\ }\textbf {\bibinfo
  {volume} {11}},\ \bibinfo {pages} {631} (\bibinfo {year} {2011})}\BibitemShut
  {NoStop}%
\bibitem [{\citenamefont {{Dexter}}\ and\ \citenamefont
  {{Blaes}}(2014)}]{2014MNRAS.438.3352D}%
  \BibitemOpen
  \bibfield  {author} {\bibinfo {author} {\bibfnamefont {J.}~\bibnamefont
  {{Dexter}}}\ and\ \bibinfo {author} {\bibfnamefont {O.}~\bibnamefont
  {{Blaes}}},\ }\bibfield  {title} {\bibinfo {title} {{A model of the steep
  power-law spectra and high-frequency quasi-periodic oscillations in luminous
  black hole X-ray binaries}},\ }\href {https://doi.org/10.1093/mnras/stu121}
  {\bibfield  {journal} {\bibinfo  {journal} {\mnras}\ }\textbf {\bibinfo
  {volume} {438}},\ \bibinfo {pages} {3352} (\bibinfo {year} {2014})},\ \Eprint
  {https://arxiv.org/abs/1312.0941} {arXiv:1312.0941 [astro-ph.HE]}
  \BibitemShut {NoStop}%
\bibitem [{\citenamefont {{van der Klis}}(2004)}]{2004astro.ph.10551V}%
  \BibitemOpen
  \bibfield  {author} {\bibinfo {author} {\bibfnamefont {M.}~\bibnamefont {{van
  der Klis}}},\ }\bibfield  {title} {\bibinfo {title} {{A review of rapid X-ray
  variability in X-ray binaries}},\ }\href
  {https://doi.org/10.48550/arXiv.astro-ph/0410551} {\bibfield  {journal}
  {\bibinfo  {journal} {arXiv e-prints}\ ,\ \bibinfo {eid} {astro-ph/0410551}}
  (\bibinfo {year} {2004})},\ \Eprint {https://arxiv.org/abs/astro-ph/0410551}
  {arXiv:astro-ph/0410551 [astro-ph]} \BibitemShut {NoStop}%
\bibitem [{\citenamefont {{Shvartsman}}(1971)}]{1971SvA....15..377S}%
  \BibitemOpen
  \bibfield  {author} {\bibinfo {author} {\bibfnamefont {V.~F.}\ \bibnamefont
  {{Shvartsman}}},\ }\bibfield  {title} {\bibinfo {title} {{Halos around
  ``Black Holes''.}},\ }\href@noop {} {\bibfield  {journal} {\bibinfo
  {journal} {\sovast}\ }\textbf {\bibinfo {volume} {15}},\ \bibinfo {pages}
  {377} (\bibinfo {year} {1971})}\BibitemShut {NoStop}%
\bibitem [{\citenamefont {Syunyaev}(1973)}]{syunyaev1973variability}%
  \BibitemOpen
  \bibfield  {author} {\bibinfo {author} {\bibfnamefont {R.}~\bibnamefont
  {Syunyaev}},\ }\bibfield  {title} {\bibinfo {title} {Variability of x rays
  from black holes with accretion disks.},\ }\href@noop {} {\bibfield
  {journal} {\bibinfo  {journal} {Soviet Astronomy}\ }\textbf {\bibinfo
  {volume} {16}},\ \bibinfo {pages} {941} (\bibinfo {year} {1973})}\BibitemShut
  {NoStop}%
\bibitem [{\citenamefont {{Nowak}}(2000)}]{2000MNRAS.318..361N}%
  \BibitemOpen
  \bibfield  {author} {\bibinfo {author} {\bibfnamefont {M.~A.}\ \bibnamefont
  {{Nowak}}},\ }\bibfield  {title} {\bibinfo {title} {{Are there three peaks in
  the power spectra of GX 339-4 and Cyg X-1?}},\ }\href
  {https://doi.org/10.1046/j.1365-8711.2000.03668.x} {\bibfield  {journal}
  {\bibinfo  {journal} {\mnras}\ }\textbf {\bibinfo {volume} {318}},\ \bibinfo
  {pages} {361} (\bibinfo {year} {2000})},\ \Eprint
  {https://arxiv.org/abs/astro-ph/0005232} {arXiv:astro-ph/0005232 [astro-ph]}
  \BibitemShut {NoStop}%
\bibitem [{\citenamefont {{Muno}}\ \emph {et~al.}(1999)\citenamefont {{Muno}},
  \citenamefont {{Morgan}},\ and\ \citenamefont
  {{Remillard}}}]{1999ApJ...527..321M}%
  \BibitemOpen
  \bibfield  {author} {\bibinfo {author} {\bibfnamefont {M.~P.}\ \bibnamefont
  {{Muno}}}, \bibinfo {author} {\bibfnamefont {E.~H.}\ \bibnamefont
  {{Morgan}}},\ and\ \bibinfo {author} {\bibfnamefont {R.~A.}\ \bibnamefont
  {{Remillard}}},\ }\bibfield  {title} {\bibinfo {title} {{Quasi-periodic
  Oscillations and Spectral States in GRS 1915+105}},\ }\href
  {https://doi.org/10.1086/308063} {\bibfield  {journal} {\bibinfo  {journal}
  {\apj}\ }\textbf {\bibinfo {volume} {527}},\ \bibinfo {pages} {321} (\bibinfo
  {year} {1999})}\BibitemShut {NoStop}%
\bibitem [{\citenamefont {{Sobczak}}\ \emph {et~al.}(2000)\citenamefont
  {{Sobczak}}, \citenamefont {{McClintock}}, \citenamefont {{Remillard}},
  \citenamefont {{Cui}}, \citenamefont {{Levine}}, \citenamefont {{Morgan}},
  \citenamefont {{Orosz}},\ and\ \citenamefont
  {{Bailyn}}}]{2000ApJ...531..537S}%
  \BibitemOpen
  \bibfield  {author} {\bibinfo {author} {\bibfnamefont {G.~J.}\ \bibnamefont
  {{Sobczak}}}, \bibinfo {author} {\bibfnamefont {J.~E.}\ \bibnamefont
  {{McClintock}}}, \bibinfo {author} {\bibfnamefont {R.~A.}\ \bibnamefont
  {{Remillard}}}, \bibinfo {author} {\bibfnamefont {W.}~\bibnamefont {{Cui}}},
  \bibinfo {author} {\bibfnamefont {A.~M.}\ \bibnamefont {{Levine}}}, \bibinfo
  {author} {\bibfnamefont {E.~H.}\ \bibnamefont {{Morgan}}}, \bibinfo {author}
  {\bibfnamefont {J.~A.}\ \bibnamefont {{Orosz}}},\ and\ \bibinfo {author}
  {\bibfnamefont {C.~D.}\ \bibnamefont {{Bailyn}}},\ }\bibfield  {title}
  {\bibinfo {title} {{Correlations between Low-Frequency Quasi-periodic
  Oscillations and Spectral Parameters in XTE J1550-564 and GRO J1655-40}},\
  }\href {https://doi.org/10.1086/308463} {\bibfield  {journal} {\bibinfo
  {journal} {\apj}\ }\textbf {\bibinfo {volume} {531}},\ \bibinfo {pages} {537}
  (\bibinfo {year} {2000})},\ \Eprint {https://arxiv.org/abs/astro-ph/9910519}
  {arXiv:astro-ph/9910519 [astro-ph]} \BibitemShut {NoStop}%
\bibitem [{\citenamefont {{Revnivtsev}}\ \emph {et~al.}(2000)\citenamefont
  {{Revnivtsev}}, \citenamefont {{Trudolyubov}},\ and\ \citenamefont
  {{Borozdin}}}]{2000MNRAS.312..151R}%
  \BibitemOpen
  \bibfield  {author} {\bibinfo {author} {\bibfnamefont {M.~G.}\ \bibnamefont
  {{Revnivtsev}}}, \bibinfo {author} {\bibfnamefont {S.~P.}\ \bibnamefont
  {{Trudolyubov}}},\ and\ \bibinfo {author} {\bibfnamefont {K.~N.}\
  \bibnamefont {{Borozdin}}},\ }\bibfield  {title} {\bibinfo {title} {{RXTE
  observations of Galactic microquasar XTE J1748-288 during its 1998
  outburst}},\ }\href {https://doi.org/10.1046/j.1365-8711.2000.03144.x}
  {\bibfield  {journal} {\bibinfo  {journal} {\mnras}\ }\textbf {\bibinfo
  {volume} {312}},\ \bibinfo {pages} {151} (\bibinfo {year}
  {2000})}\BibitemShut {NoStop}%
\bibitem [{\citenamefont {{Vignarca}}\ \emph {et~al.}(2003)\citenamefont
  {{Vignarca}}, \citenamefont {{Migliari}}, \citenamefont {{Belloni}},
  \citenamefont {{Psaltis}},\ and\ \citenamefont {{van der
  Klis}}}]{2003A&A...397..729V}%
  \BibitemOpen
  \bibfield  {author} {\bibinfo {author} {\bibfnamefont {F.}~\bibnamefont
  {{Vignarca}}}, \bibinfo {author} {\bibfnamefont {S.}~\bibnamefont
  {{Migliari}}}, \bibinfo {author} {\bibfnamefont {T.}~\bibnamefont
  {{Belloni}}}, \bibinfo {author} {\bibfnamefont {D.}~\bibnamefont
  {{Psaltis}}},\ and\ \bibinfo {author} {\bibfnamefont {M.}~\bibnamefont {{van
  der Klis}}},\ }\bibfield  {title} {\bibinfo {title} {{Tracing the power-law
  component in the energy spectrum of black hole candidates as a function of
  the QPO frequency}},\ }\href {https://doi.org/10.1051/0004-6361:20021542}
  {\bibfield  {journal} {\bibinfo  {journal} {\aap}\ }\textbf {\bibinfo
  {volume} {397}},\ \bibinfo {pages} {729} (\bibinfo {year} {2003})},\ \Eprint
  {https://arxiv.org/abs/astro-ph/0210517} {arXiv:astro-ph/0210517 [astro-ph]}
  \BibitemShut {NoStop}%
\bibitem [{\citenamefont {{Majewski}}\ \emph {et~al.}(2017)\citenamefont
  {{Majewski}}, \citenamefont {{Schiavon}}, \citenamefont {{Frinchaboy}},
  \citenamefont {{Allende Prieto}}, \citenamefont {{Barkhouser}}, \citenamefont
  {{Bizyaev}}, \citenamefont {{Blank}}, \citenamefont {{Brunner}},
  \citenamefont {{Burton}}, \citenamefont {{Carrera}}, \citenamefont
  {{Chojnowski}}, \citenamefont {{Cunha}}, \citenamefont {{Epstein}},
  \citenamefont {{Fitzgerald}}, \citenamefont {{Garc{\'\i}a P{\'e}rez}},
  \citenamefont {{Hearty}}, \citenamefont {{Henderson}}, \citenamefont
  {{Holtzman}}, \citenamefont {{Johnson}}, \citenamefont {{Lam}}, \citenamefont
  {{Lawler}}, \citenamefont {{Maseman}}, \citenamefont {{M{\'e}sz{\'a}ros}},
  \citenamefont {{Nelson}}, \citenamefont {{Nguyen}}, \citenamefont
  {{Nidever}}, \citenamefont {{Pinsonneault}}, \citenamefont {{Shetrone}},
  \citenamefont {{Smee}}, \citenamefont {{Smith}}, \citenamefont {{Stolberg}},
  \citenamefont {{Skrutskie}}, \citenamefont {{Walker}}, \citenamefont
  {{Wilson}}, \citenamefont {{Zasowski}}, \citenamefont {{Anders}},
  \citenamefont {{Basu}}, \citenamefont {{Beland}}, \citenamefont {{Blanton}},
  \citenamefont {{Bovy}}, \citenamefont {{Brownstein}}, \citenamefont
  {{Carlberg}}, \citenamefont {{Chaplin}}, \citenamefont {{Chiappini}},
  \citenamefont {{Eisenstein}}, \citenamefont {{Elsworth}}, \citenamefont
  {{Feuillet}}, \citenamefont {{Fleming}}, \citenamefont {{Galbraith-Frew}},
  \citenamefont {{Garc{\'\i}a}}, \citenamefont {{Garc{\'\i}a-Hern{\'a}ndez}},
  \citenamefont {{Gillespie}}, \citenamefont {{Girardi}}, \citenamefont
  {{Gunn}}, \citenamefont {{Hasselquist}}, \citenamefont {{Hayden}},
  \citenamefont {{Hekker}}, \citenamefont {{Ivans}}, \citenamefont
  {{Kinemuchi}}, \citenamefont {{Klaene}}, \citenamefont {{Mahadevan}},
  \citenamefont {{Mathur}}, \citenamefont {{Mosser}}, \citenamefont {{Muna}},
  \citenamefont {{Munn}}, \citenamefont {{Nichol}}, \citenamefont
  {{O'Connell}}, \citenamefont {{Parejko}}, \citenamefont {{Robin}},
  \citenamefont {{Rocha-Pinto}}, \citenamefont {{Schultheis}}, \citenamefont
  {{Serenelli}}, \citenamefont {{Shane}}, \citenamefont {{Silva Aguirre}},
  \citenamefont {{Sobeck}}, \citenamefont {{Thompson}}, \citenamefont
  {{Troup}}, \citenamefont {{Weinberg}},\ and\ \citenamefont
  {{Zamora}}}]{2017AJ....154...94M}%
  \BibitemOpen
  \bibfield  {author} {\bibinfo {author} {\bibfnamefont {S.~R.}\ \bibnamefont
  {{Majewski}}}, \bibinfo {author} {\bibfnamefont {R.~P.}\ \bibnamefont
  {{Schiavon}}}, \bibinfo {author} {\bibfnamefont {P.~M.}\ \bibnamefont
  {{Frinchaboy}}}, \bibinfo {author} {\bibfnamefont {C.}~\bibnamefont {{Allende
  Prieto}}}, \bibinfo {author} {\bibfnamefont {R.}~\bibnamefont
  {{Barkhouser}}}, \bibinfo {author} {\bibfnamefont {D.}~\bibnamefont
  {{Bizyaev}}}, \bibinfo {author} {\bibfnamefont {B.}~\bibnamefont {{Blank}}},
  \bibinfo {author} {\bibfnamefont {S.}~\bibnamefont {{Brunner}}}, \bibinfo
  {author} {\bibfnamefont {A.}~\bibnamefont {{Burton}}}, \bibinfo {author}
  {\bibfnamefont {R.}~\bibnamefont {{Carrera}}}, \bibinfo {author}
  {\bibfnamefont {S.~D.}\ \bibnamefont {{Chojnowski}}}, \bibinfo {author}
  {\bibfnamefont {K.}~\bibnamefont {{Cunha}}}, \bibinfo {author} {\bibfnamefont
  {C.}~\bibnamefont {{Epstein}}}, \bibinfo {author} {\bibfnamefont
  {G.}~\bibnamefont {{Fitzgerald}}}, \bibinfo {author} {\bibfnamefont {A.~E.}\
  \bibnamefont {{Garc{\'\i}a P{\'e}rez}}}, \bibinfo {author} {\bibfnamefont
  {F.~R.}\ \bibnamefont {{Hearty}}}, \bibinfo {author} {\bibfnamefont
  {C.}~\bibnamefont {{Henderson}}}, \bibinfo {author} {\bibfnamefont {J.~A.}\
  \bibnamefont {{Holtzman}}}, \bibinfo {author} {\bibfnamefont {J.~A.}\
  \bibnamefont {{Johnson}}}, \bibinfo {author} {\bibfnamefont {C.~R.}\
  \bibnamefont {{Lam}}}, \bibinfo {author} {\bibfnamefont {J.~E.}\ \bibnamefont
  {{Lawler}}}, \bibinfo {author} {\bibfnamefont {P.}~\bibnamefont {{Maseman}}},
  \bibinfo {author} {\bibfnamefont {S.}~\bibnamefont {{M{\'e}sz{\'a}ros}}},
  \bibinfo {author} {\bibfnamefont {M.}~\bibnamefont {{Nelson}}}, \bibinfo
  {author} {\bibfnamefont {D.~C.}\ \bibnamefont {{Nguyen}}}, \bibinfo {author}
  {\bibfnamefont {D.~L.}\ \bibnamefont {{Nidever}}}, \bibinfo {author}
  {\bibfnamefont {M.}~\bibnamefont {{Pinsonneault}}}, \bibinfo {author}
  {\bibfnamefont {M.}~\bibnamefont {{Shetrone}}}, \bibinfo {author}
  {\bibfnamefont {S.}~\bibnamefont {{Smee}}}, \bibinfo {author} {\bibfnamefont
  {V.~V.}\ \bibnamefont {{Smith}}}, \bibinfo {author} {\bibfnamefont
  {T.}~\bibnamefont {{Stolberg}}}, \bibinfo {author} {\bibfnamefont {M.~F.}\
  \bibnamefont {{Skrutskie}}}, \bibinfo {author} {\bibfnamefont
  {E.}~\bibnamefont {{Walker}}}, \bibinfo {author} {\bibfnamefont {J.~C.}\
  \bibnamefont {{Wilson}}}, \bibinfo {author} {\bibfnamefont {G.}~\bibnamefont
  {{Zasowski}}}, \bibinfo {author} {\bibfnamefont {F.}~\bibnamefont
  {{Anders}}}, \bibinfo {author} {\bibfnamefont {S.}~\bibnamefont {{Basu}}},
  \bibinfo {author} {\bibfnamefont {S.}~\bibnamefont {{Beland}}}, \bibinfo
  {author} {\bibfnamefont {M.~R.}\ \bibnamefont {{Blanton}}}, \bibinfo {author}
  {\bibfnamefont {J.}~\bibnamefont {{Bovy}}}, \bibinfo {author} {\bibfnamefont
  {J.~R.}\ \bibnamefont {{Brownstein}}}, \bibinfo {author} {\bibfnamefont
  {J.}~\bibnamefont {{Carlberg}}}, \bibinfo {author} {\bibfnamefont
  {W.}~\bibnamefont {{Chaplin}}}, \bibinfo {author} {\bibfnamefont
  {C.}~\bibnamefont {{Chiappini}}}, \bibinfo {author} {\bibfnamefont {D.~J.}\
  \bibnamefont {{Eisenstein}}}, \bibinfo {author} {\bibfnamefont
  {Y.}~\bibnamefont {{Elsworth}}}, \bibinfo {author} {\bibfnamefont
  {D.}~\bibnamefont {{Feuillet}}}, \bibinfo {author} {\bibfnamefont {S.~W.}\
  \bibnamefont {{Fleming}}}, \bibinfo {author} {\bibfnamefont {J.}~\bibnamefont
  {{Galbraith-Frew}}}, \bibinfo {author} {\bibfnamefont {R.~A.}\ \bibnamefont
  {{Garc{\'\i}a}}}, \bibinfo {author} {\bibfnamefont {D.~A.}\ \bibnamefont
  {{Garc{\'\i}a-Hern{\'a}ndez}}}, \bibinfo {author} {\bibfnamefont {B.~A.}\
  \bibnamefont {{Gillespie}}}, \bibinfo {author} {\bibfnamefont
  {L.}~\bibnamefont {{Girardi}}}, \bibinfo {author} {\bibfnamefont {J.~E.}\
  \bibnamefont {{Gunn}}}, \bibinfo {author} {\bibfnamefont {S.}~\bibnamefont
  {{Hasselquist}}}, \bibinfo {author} {\bibfnamefont {M.~R.}\ \bibnamefont
  {{Hayden}}}, \bibinfo {author} {\bibfnamefont {S.}~\bibnamefont {{Hekker}}},
  \bibinfo {author} {\bibfnamefont {I.}~\bibnamefont {{Ivans}}}, \bibinfo
  {author} {\bibfnamefont {K.}~\bibnamefont {{Kinemuchi}}}, \bibinfo {author}
  {\bibfnamefont {M.}~\bibnamefont {{Klaene}}}, \bibinfo {author}
  {\bibfnamefont {S.}~\bibnamefont {{Mahadevan}}}, \bibinfo {author}
  {\bibfnamefont {S.}~\bibnamefont {{Mathur}}}, \bibinfo {author}
  {\bibfnamefont {B.}~\bibnamefont {{Mosser}}}, \bibinfo {author}
  {\bibfnamefont {D.}~\bibnamefont {{Muna}}}, \bibinfo {author} {\bibfnamefont
  {J.~A.}\ \bibnamefont {{Munn}}}, \bibinfo {author} {\bibfnamefont {R.~C.}\
  \bibnamefont {{Nichol}}}, \bibinfo {author} {\bibfnamefont {R.~W.}\
  \bibnamefont {{O'Connell}}}, \bibinfo {author} {\bibfnamefont {J.~K.}\
  \bibnamefont {{Parejko}}}, \bibinfo {author} {\bibfnamefont {A.~C.}\
  \bibnamefont {{Robin}}}, \bibinfo {author} {\bibfnamefont {H.}~\bibnamefont
  {{Rocha-Pinto}}}, \bibinfo {author} {\bibfnamefont {M.}~\bibnamefont
  {{Schultheis}}}, \bibinfo {author} {\bibfnamefont {A.~M.}\ \bibnamefont
  {{Serenelli}}}, \bibinfo {author} {\bibfnamefont {N.}~\bibnamefont
  {{Shane}}}, \bibinfo {author} {\bibfnamefont {V.}~\bibnamefont {{Silva
  Aguirre}}}, \bibinfo {author} {\bibfnamefont {J.~S.}\ \bibnamefont
  {{Sobeck}}}, \bibinfo {author} {\bibfnamefont {B.}~\bibnamefont
  {{Thompson}}}, \bibinfo {author} {\bibfnamefont {N.~W.}\ \bibnamefont
  {{Troup}}}, \bibinfo {author} {\bibfnamefont {D.~H.}\ \bibnamefont
  {{Weinberg}}},\ and\ \bibinfo {author} {\bibfnamefont {O.}~\bibnamefont
  {{Zamora}}},\ }\bibfield  {title} {\bibinfo {title} {{The Apache Point
  Observatory Galactic Evolution Experiment (APOGEE)}},\ }\href
  {https://doi.org/10.3847/1538-3881/aa784d} {\bibfield  {journal} {\bibinfo
  {journal} {\aj}\ }\textbf {\bibinfo {volume} {154}},\ \bibinfo {eid} {94}
  (\bibinfo {year} {2017})},\ \Eprint {https://arxiv.org/abs/1509.05420}
  {arXiv:1509.05420 [astro-ph.IM]} \BibitemShut {NoStop}%
\bibitem [{\citenamefont {{Cui}}\ \emph {et~al.}(2012)\citenamefont {{Cui}},
  \citenamefont {{Zhao}}, \citenamefont {{Chu}}, \citenamefont {{Li}},
  \citenamefont {{Li}}, \citenamefont {{Zhang}}, \citenamefont {{Su}},
  \citenamefont {{Yao}}, \citenamefont {{Wang}}, \citenamefont {{Xing}},
  \citenamefont {{Li}}, \citenamefont {{Zhu}}, \citenamefont {{Wang}},
  \citenamefont {{Gu}}, \citenamefont {{Luo}}, \citenamefont {{Xu}},
  \citenamefont {{Zhang}}, \citenamefont {{Liu}}, \citenamefont {{Zhang}},
  \citenamefont {{Yang}}, \citenamefont {{Cao}}, \citenamefont {{Chen}},
  \citenamefont {{Chen}}, \citenamefont {{Chen}}, \citenamefont {{Chen}},
  \citenamefont {{Chu}}, \citenamefont {{Feng}}, \citenamefont {{Gong}},
  \citenamefont {{Hou}}, \citenamefont {{Hu}}, \citenamefont {{Hu}},
  \citenamefont {{Hu}}, \citenamefont {{Jia}}, \citenamefont {{Jiang}},
  \citenamefont {{Jiang}}, \citenamefont {{Jiang}}, \citenamefont {{Jin}},
  \citenamefont {{Li}}, \citenamefont {{Li}}, \citenamefont {{Li}},
  \citenamefont {{Liu}}, \citenamefont {{Liu}}, \citenamefont {{Lu}},
  \citenamefont {{Mao}}, \citenamefont {{Men}}, \citenamefont {{Qi}},
  \citenamefont {{Qi}}, \citenamefont {{Shi}}, \citenamefont {{Tang}},
  \citenamefont {{Tao}}, \citenamefont {{Wang}}, \citenamefont {{Wang}},
  \citenamefont {{Wang}}, \citenamefont {{Wang}}, \citenamefont {{Wang}},
  \citenamefont {{Wang}}, \citenamefont {{Wang}}, \citenamefont {{Wang}},
  \citenamefont {{Wang}}, \citenamefont {{Wang}}, \citenamefont {{Wang}},
  \citenamefont {{Wang}}, \citenamefont {{Xu}}, \citenamefont {{Xu}},
  \citenamefont {{Yang}}, \citenamefont {{Yu}}, \citenamefont {{Yuan}},
  \citenamefont {{Yuan}}, \citenamefont {{Zhai}}, \citenamefont {{Zhang}},
  \citenamefont {{Zhang}}, \citenamefont {{Zhang}}, \citenamefont {{Zhao}},
  \citenamefont {{Zhou}}, \citenamefont {{Zhou}}, \citenamefont {{Zhu}},\ and\
  \citenamefont {{Zou}}}]{2012RAA....12.1197C}%
  \BibitemOpen
  \bibfield  {author} {\bibinfo {author} {\bibfnamefont {X.-Q.}\ \bibnamefont
  {{Cui}}}, \bibinfo {author} {\bibfnamefont {Y.-H.}\ \bibnamefont {{Zhao}}},
  \bibinfo {author} {\bibfnamefont {Y.-Q.}\ \bibnamefont {{Chu}}}, \bibinfo
  {author} {\bibfnamefont {G.-P.}\ \bibnamefont {{Li}}}, \bibinfo {author}
  {\bibfnamefont {Q.}~\bibnamefont {{Li}}}, \bibinfo {author} {\bibfnamefont
  {L.-P.}\ \bibnamefont {{Zhang}}}, \bibinfo {author} {\bibfnamefont {H.-J.}\
  \bibnamefont {{Su}}}, \bibinfo {author} {\bibfnamefont {Z.-Q.}\ \bibnamefont
  {{Yao}}}, \bibinfo {author} {\bibfnamefont {Y.-N.}\ \bibnamefont {{Wang}}},
  \bibinfo {author} {\bibfnamefont {X.-Z.}\ \bibnamefont {{Xing}}}, \bibinfo
  {author} {\bibfnamefont {X.-N.}\ \bibnamefont {{Li}}}, \bibinfo {author}
  {\bibfnamefont {Y.-T.}\ \bibnamefont {{Zhu}}}, \bibinfo {author}
  {\bibfnamefont {G.}~\bibnamefont {{Wang}}}, \bibinfo {author} {\bibfnamefont
  {B.-Z.}\ \bibnamefont {{Gu}}}, \bibinfo {author} {\bibfnamefont {A.~L.}\
  \bibnamefont {{Luo}}}, \bibinfo {author} {\bibfnamefont {X.-Q.}\ \bibnamefont
  {{Xu}}}, \bibinfo {author} {\bibfnamefont {Z.-C.}\ \bibnamefont {{Zhang}}},
  \bibinfo {author} {\bibfnamefont {G.-R.}\ \bibnamefont {{Liu}}}, \bibinfo
  {author} {\bibfnamefont {H.-T.}\ \bibnamefont {{Zhang}}}, \bibinfo {author}
  {\bibfnamefont {D.-H.}\ \bibnamefont {{Yang}}}, \bibinfo {author}
  {\bibfnamefont {S.-Y.}\ \bibnamefont {{Cao}}}, \bibinfo {author}
  {\bibfnamefont {H.-Y.}\ \bibnamefont {{Chen}}}, \bibinfo {author}
  {\bibfnamefont {J.-J.}\ \bibnamefont {{Chen}}}, \bibinfo {author}
  {\bibfnamefont {K.-X.}\ \bibnamefont {{Chen}}}, \bibinfo {author}
  {\bibfnamefont {Y.}~\bibnamefont {{Chen}}}, \bibinfo {author} {\bibfnamefont
  {J.-R.}\ \bibnamefont {{Chu}}}, \bibinfo {author} {\bibfnamefont
  {L.}~\bibnamefont {{Feng}}}, \bibinfo {author} {\bibfnamefont {X.-F.}\
  \bibnamefont {{Gong}}}, \bibinfo {author} {\bibfnamefont {Y.-H.}\
  \bibnamefont {{Hou}}}, \bibinfo {author} {\bibfnamefont {H.-Z.}\ \bibnamefont
  {{Hu}}}, \bibinfo {author} {\bibfnamefont {N.-S.}\ \bibnamefont {{Hu}}},
  \bibinfo {author} {\bibfnamefont {Z.-W.}\ \bibnamefont {{Hu}}}, \bibinfo
  {author} {\bibfnamefont {L.}~\bibnamefont {{Jia}}}, \bibinfo {author}
  {\bibfnamefont {F.-H.}\ \bibnamefont {{Jiang}}}, \bibinfo {author}
  {\bibfnamefont {X.}~\bibnamefont {{Jiang}}}, \bibinfo {author} {\bibfnamefont
  {Z.-B.}\ \bibnamefont {{Jiang}}}, \bibinfo {author} {\bibfnamefont
  {G.}~\bibnamefont {{Jin}}}, \bibinfo {author} {\bibfnamefont {A.-H.}\
  \bibnamefont {{Li}}}, \bibinfo {author} {\bibfnamefont {Y.}~\bibnamefont
  {{Li}}}, \bibinfo {author} {\bibfnamefont {Y.-P.}\ \bibnamefont {{Li}}},
  \bibinfo {author} {\bibfnamefont {G.-Q.}\ \bibnamefont {{Liu}}}, \bibinfo
  {author} {\bibfnamefont {Z.-G.}\ \bibnamefont {{Liu}}}, \bibinfo {author}
  {\bibfnamefont {W.-Z.}\ \bibnamefont {{Lu}}}, \bibinfo {author}
  {\bibfnamefont {Y.-D.}\ \bibnamefont {{Mao}}}, \bibinfo {author}
  {\bibfnamefont {L.}~\bibnamefont {{Men}}}, \bibinfo {author} {\bibfnamefont
  {Y.-J.}\ \bibnamefont {{Qi}}}, \bibinfo {author} {\bibfnamefont {Z.-X.}\
  \bibnamefont {{Qi}}}, \bibinfo {author} {\bibfnamefont {H.-M.}\ \bibnamefont
  {{Shi}}}, \bibinfo {author} {\bibfnamefont {Z.-H.}\ \bibnamefont {{Tang}}},
  \bibinfo {author} {\bibfnamefont {Q.-S.}\ \bibnamefont {{Tao}}}, \bibinfo
  {author} {\bibfnamefont {D.-Q.}\ \bibnamefont {{Wang}}}, \bibinfo {author}
  {\bibfnamefont {D.}~\bibnamefont {{Wang}}}, \bibinfo {author} {\bibfnamefont
  {G.-M.}\ \bibnamefont {{Wang}}}, \bibinfo {author} {\bibfnamefont
  {H.}~\bibnamefont {{Wang}}}, \bibinfo {author} {\bibfnamefont {J.-N.}\
  \bibnamefont {{Wang}}}, \bibinfo {author} {\bibfnamefont {J.}~\bibnamefont
  {{Wang}}}, \bibinfo {author} {\bibfnamefont {J.-L.}\ \bibnamefont {{Wang}}},
  \bibinfo {author} {\bibfnamefont {J.-P.}\ \bibnamefont {{Wang}}}, \bibinfo
  {author} {\bibfnamefont {L.}~\bibnamefont {{Wang}}}, \bibinfo {author}
  {\bibfnamefont {S.-Q.}\ \bibnamefont {{Wang}}}, \bibinfo {author}
  {\bibfnamefont {Y.}~\bibnamefont {{Wang}}}, \bibinfo {author} {\bibfnamefont
  {Y.-F.}\ \bibnamefont {{Wang}}}, \bibinfo {author} {\bibfnamefont {L.-Z.}\
  \bibnamefont {{Xu}}}, \bibinfo {author} {\bibfnamefont {Y.}~\bibnamefont
  {{Xu}}}, \bibinfo {author} {\bibfnamefont {S.-H.}\ \bibnamefont {{Yang}}},
  \bibinfo {author} {\bibfnamefont {Y.}~\bibnamefont {{Yu}}}, \bibinfo {author}
  {\bibfnamefont {H.}~\bibnamefont {{Yuan}}}, \bibinfo {author} {\bibfnamefont
  {X.-Y.}\ \bibnamefont {{Yuan}}}, \bibinfo {author} {\bibfnamefont
  {C.}~\bibnamefont {{Zhai}}}, \bibinfo {author} {\bibfnamefont
  {J.}~\bibnamefont {{Zhang}}}, \bibinfo {author} {\bibfnamefont {Y.-X.}\
  \bibnamefont {{Zhang}}}, \bibinfo {author} {\bibfnamefont {Y.}~\bibnamefont
  {{Zhang}}}, \bibinfo {author} {\bibfnamefont {M.}~\bibnamefont {{Zhao}}},
  \bibinfo {author} {\bibfnamefont {F.}~\bibnamefont {{Zhou}}}, \bibinfo
  {author} {\bibfnamefont {G.-H.}\ \bibnamefont {{Zhou}}}, \bibinfo {author}
  {\bibfnamefont {J.}~\bibnamefont {{Zhu}}},\ and\ \bibinfo {author}
  {\bibfnamefont {S.-C.}\ \bibnamefont {{Zou}}},\ }\bibfield  {title} {\bibinfo
  {title} {{The Large Sky Area Multi-Object Fiber Spectroscopic Telescope
  (LAMOST)}},\ }\href {https://doi.org/10.1088/1674-4527/12/9/003} {\bibfield
  {journal} {\bibinfo  {journal} {Research in Astronomy and Astrophysics}\
  }\textbf {\bibinfo {volume} {12}},\ \bibinfo {pages} {1197} (\bibinfo {year}
  {2012})}\BibitemShut {NoStop}%
\bibitem [{\citenamefont {{Steinmetz}}\ \emph {et~al.}(2006)\citenamefont
  {{Steinmetz}}, \citenamefont {{Zwitter}}, \citenamefont {{Siebert}},
  \citenamefont {{Watson}}, \citenamefont {{Freeman}}, \citenamefont
  {{Munari}}, \citenamefont {{Campbell}}, \citenamefont {{Williams}},
  \citenamefont {{Seabroke}}, \citenamefont {{Wyse}}, \citenamefont {{Parker}},
  \citenamefont {{Bienaym{\'e}}}, \citenamefont {{Roeser}}, \citenamefont
  {{Gibson}}, \citenamefont {{Gilmore}}, \citenamefont {{Grebel}},
  \citenamefont {{Helmi}}, \citenamefont {{Navarro}}, \citenamefont {{Burton}},
  \citenamefont {{Cass}}, \citenamefont {{Dawe}}, \citenamefont {{Fiegert}},
  \citenamefont {{Hartley}}, \citenamefont {{Russell}}, \citenamefont
  {{Saunders}}, \citenamefont {{Enke}}, \citenamefont {{Bailin}}, \citenamefont
  {{Binney}}, \citenamefont {{Bland-Hawthorn}}, \citenamefont {{Boeche}},
  \citenamefont {{Dehnen}}, \citenamefont {{Eisenstein}}, \citenamefont
  {{Evans}}, \citenamefont {{Fiorucci}}, \citenamefont {{Fulbright}},
  \citenamefont {{Gerhard}}, \citenamefont {{Jauregi}}, \citenamefont {{Kelz}},
  \citenamefont {{Mijovi{\'c}}}, \citenamefont {{Minchev}}, \citenamefont
  {{Parmentier}}, \citenamefont {{Pe{\~n}arrubia}}, \citenamefont {{Quillen}},
  \citenamefont {{Read}}, \citenamefont {{Ruchti}}, \citenamefont {{Scholz}},
  \citenamefont {{Siviero}}, \citenamefont {{Smith}}, \citenamefont {{Sordo}},
  \citenamefont {{Veltz}}, \citenamefont {{Vidrih}}, \citenamefont {{von
  Berlepsch}}, \citenamefont {{Boyle}},\ and\ \citenamefont
  {{Schilbach}}}]{2006AJ....132.1645S}%
  \BibitemOpen
  \bibfield  {author} {\bibinfo {author} {\bibfnamefont {M.}~\bibnamefont
  {{Steinmetz}}}, \bibinfo {author} {\bibfnamefont {T.}~\bibnamefont
  {{Zwitter}}}, \bibinfo {author} {\bibfnamefont {A.}~\bibnamefont
  {{Siebert}}}, \bibinfo {author} {\bibfnamefont {F.~G.}\ \bibnamefont
  {{Watson}}}, \bibinfo {author} {\bibfnamefont {K.~C.}\ \bibnamefont
  {{Freeman}}}, \bibinfo {author} {\bibfnamefont {U.}~\bibnamefont {{Munari}}},
  \bibinfo {author} {\bibfnamefont {R.}~\bibnamefont {{Campbell}}}, \bibinfo
  {author} {\bibfnamefont {M.}~\bibnamefont {{Williams}}}, \bibinfo {author}
  {\bibfnamefont {G.~M.}\ \bibnamefont {{Seabroke}}}, \bibinfo {author}
  {\bibfnamefont {R.~F.~G.}\ \bibnamefont {{Wyse}}}, \bibinfo {author}
  {\bibfnamefont {Q.~A.}\ \bibnamefont {{Parker}}}, \bibinfo {author}
  {\bibfnamefont {O.}~\bibnamefont {{Bienaym{\'e}}}}, \bibinfo {author}
  {\bibfnamefont {S.}~\bibnamefont {{Roeser}}}, \bibinfo {author}
  {\bibfnamefont {B.~K.}\ \bibnamefont {{Gibson}}}, \bibinfo {author}
  {\bibfnamefont {G.}~\bibnamefont {{Gilmore}}}, \bibinfo {author}
  {\bibfnamefont {E.~K.}\ \bibnamefont {{Grebel}}}, \bibinfo {author}
  {\bibfnamefont {A.}~\bibnamefont {{Helmi}}}, \bibinfo {author} {\bibfnamefont
  {J.~F.}\ \bibnamefont {{Navarro}}}, \bibinfo {author} {\bibfnamefont
  {D.}~\bibnamefont {{Burton}}}, \bibinfo {author} {\bibfnamefont {C.~J.~P.}\
  \bibnamefont {{Cass}}}, \bibinfo {author} {\bibfnamefont {J.~A.}\
  \bibnamefont {{Dawe}}}, \bibinfo {author} {\bibfnamefont {K.}~\bibnamefont
  {{Fiegert}}}, \bibinfo {author} {\bibfnamefont {M.}~\bibnamefont
  {{Hartley}}}, \bibinfo {author} {\bibfnamefont {K.~S.}\ \bibnamefont
  {{Russell}}}, \bibinfo {author} {\bibfnamefont {W.}~\bibnamefont
  {{Saunders}}}, \bibinfo {author} {\bibfnamefont {H.}~\bibnamefont {{Enke}}},
  \bibinfo {author} {\bibfnamefont {J.}~\bibnamefont {{Bailin}}}, \bibinfo
  {author} {\bibfnamefont {J.}~\bibnamefont {{Binney}}}, \bibinfo {author}
  {\bibfnamefont {J.}~\bibnamefont {{Bland-Hawthorn}}}, \bibinfo {author}
  {\bibfnamefont {C.}~\bibnamefont {{Boeche}}}, \bibinfo {author}
  {\bibfnamefont {W.}~\bibnamefont {{Dehnen}}}, \bibinfo {author}
  {\bibfnamefont {D.~J.}\ \bibnamefont {{Eisenstein}}}, \bibinfo {author}
  {\bibfnamefont {N.~W.}\ \bibnamefont {{Evans}}}, \bibinfo {author}
  {\bibfnamefont {M.}~\bibnamefont {{Fiorucci}}}, \bibinfo {author}
  {\bibfnamefont {J.~P.}\ \bibnamefont {{Fulbright}}}, \bibinfo {author}
  {\bibfnamefont {O.}~\bibnamefont {{Gerhard}}}, \bibinfo {author}
  {\bibfnamefont {U.}~\bibnamefont {{Jauregi}}}, \bibinfo {author}
  {\bibfnamefont {A.}~\bibnamefont {{Kelz}}}, \bibinfo {author} {\bibfnamefont
  {L.}~\bibnamefont {{Mijovi{\'c}}}}, \bibinfo {author} {\bibfnamefont
  {I.}~\bibnamefont {{Minchev}}}, \bibinfo {author} {\bibfnamefont
  {G.}~\bibnamefont {{Parmentier}}}, \bibinfo {author} {\bibfnamefont
  {J.}~\bibnamefont {{Pe{\~n}arrubia}}}, \bibinfo {author} {\bibfnamefont
  {A.~C.}\ \bibnamefont {{Quillen}}}, \bibinfo {author} {\bibfnamefont {M.~A.}\
  \bibnamefont {{Read}}}, \bibinfo {author} {\bibfnamefont {G.}~\bibnamefont
  {{Ruchti}}}, \bibinfo {author} {\bibfnamefont {R.~D.}\ \bibnamefont
  {{Scholz}}}, \bibinfo {author} {\bibfnamefont {A.}~\bibnamefont {{Siviero}}},
  \bibinfo {author} {\bibfnamefont {M.~C.}\ \bibnamefont {{Smith}}}, \bibinfo
  {author} {\bibfnamefont {R.}~\bibnamefont {{Sordo}}}, \bibinfo {author}
  {\bibfnamefont {L.}~\bibnamefont {{Veltz}}}, \bibinfo {author} {\bibfnamefont
  {S.}~\bibnamefont {{Vidrih}}}, \bibinfo {author} {\bibfnamefont
  {R.}~\bibnamefont {{von Berlepsch}}}, \bibinfo {author} {\bibfnamefont
  {B.~J.}\ \bibnamefont {{Boyle}}},\ and\ \bibinfo {author} {\bibfnamefont
  {E.}~\bibnamefont {{Schilbach}}},\ }\bibfield  {title} {\bibinfo {title}
  {{The Radial Velocity Experiment (RAVE): First Data Release}},\ }\href
  {https://doi.org/10.1086/506564} {\bibfield  {journal} {\bibinfo  {journal}
  {\aj}\ }\textbf {\bibinfo {volume} {132}},\ \bibinfo {pages} {1645} (\bibinfo
  {year} {2006})},\ \Eprint {https://arxiv.org/abs/astro-ph/0606211}
  {arXiv:astro-ph/0606211 [astro-ph]} \BibitemShut {NoStop}%
\bibitem [{\citenamefont {{Gilmore}}\ \emph {et~al.}(2012)\citenamefont
  {{Gilmore}}, \citenamefont {{Randich}}, \citenamefont {{Asplund}},
  \citenamefont {{Binney}}, \citenamefont {{Bonifacio}}, \citenamefont
  {{Drew}}, \citenamefont {{Feltzing}}, \citenamefont {{Ferguson}},
  \citenamefont {{Jeffries}}, \citenamefont {{Micela}}, \citenamefont
  {{Negueruela}}, \citenamefont {{Prusti}}, \citenamefont {{Rix}},
  \citenamefont {{Vallenari}}, \citenamefont {{Alfaro}}, \citenamefont
  {{Allende-Prieto}}, \citenamefont {{Babusiaux}}, \citenamefont {{Bensby}},
  \citenamefont {{Blomme}}, \citenamefont {{Bragaglia}}, \citenamefont
  {{Flaccomio}}, \citenamefont {{Fran{\c{c}}ois}}, \citenamefont {{Irwin}},
  \citenamefont {{Koposov}}, \citenamefont {{Korn}}, \citenamefont
  {{Lanzafame}}, \citenamefont {{Pancino}}, \citenamefont {{Paunzen}},
  \citenamefont {{Recio-Blanco}}, \citenamefont {{Sacco}}, \citenamefont
  {{Smiljanic}}, \citenamefont {{Van Eck}}, \citenamefont {{Walton}},
  \citenamefont {{Aden}}, \citenamefont {{Aerts}}, \citenamefont {{Affer}},
  \citenamefont {{Alcala}}, \citenamefont {{Altavilla}}, \citenamefont
  {{Alves}}, \citenamefont {{Antoja}}, \citenamefont {{Arenou}}, \citenamefont
  {{Argiroffi}}, \citenamefont {{Asensio Ramos}}, \citenamefont
  {{Bailer-Jones}}, \citenamefont {{Balaguer-Nunez}}, \citenamefont {{Bayo}},
  \citenamefont {{Barbuy}}, \citenamefont {{Barisevicius}}, \citenamefont
  {{Barrado y Navascues}}, \citenamefont {{Battistini}}, \citenamefont {{Bellas
  Velidis}}, \citenamefont {{Bellazzini}}, \citenamefont {{Belokurov}},
  \citenamefont {{Bergemann}}, \citenamefont {{Bertelli}}, \citenamefont
  {{Biazzo}}, \citenamefont {{Bienayme}}, \citenamefont {{Bland-Hawthorn}},
  \citenamefont {{Boeche}}, \citenamefont {{Bonito}}, \citenamefont
  {{Boudreault}}, \citenamefont {{Bouvier}}, \citenamefont {{Brandao}},
  \citenamefont {{Brown}}, \citenamefont {{de Bruijne}}, \citenamefont
  {{Burleigh}}, \citenamefont {{Caballero}}, \citenamefont {{Caffau}},
  \citenamefont {{Calura}}, \citenamefont {{Capuzzo-Dolcetta}}, \citenamefont
  {{Caramazza}}, \citenamefont {{Carraro}}, \citenamefont {{Casagrande}},
  \citenamefont {{Casewell}}, \citenamefont {{Chapman}}, \citenamefont
  {{Chiappini}}, \citenamefont {{Chorniy}}, \citenamefont {{Christlieb}},
  \citenamefont {{Cignoni}}, \citenamefont {{Cocozza}}, \citenamefont
  {{Colless}}, \citenamefont {{Collet}}, \citenamefont {{Collins}},
  \citenamefont {{Correnti}}, \citenamefont {{Covino}}, \citenamefont
  {{Crnojevic}}, \citenamefont {{Cropper}}, \citenamefont {{Cunha}},
  \citenamefont {{Damiani}}, \citenamefont {{David}}, \citenamefont
  {{Delgado}}, \citenamefont {{Duffau}}, \citenamefont {{Edvardsson}},
  \citenamefont {{Eldridge}}, \citenamefont {{Enke}}, \citenamefont
  {{Eriksson}}, \citenamefont {{Evans}}, \citenamefont {{Eyer}}, \citenamefont
  {{Famaey}}, \citenamefont {{Fellhauer}}, \citenamefont {{Ferreras}},
  \citenamefont {{Figueras}}, \citenamefont {{Fiorentino}}, \citenamefont
  {{Flynn}}, \citenamefont {{Folha}}, \citenamefont {{Franciosini}},
  \citenamefont {{Frasca}}, \citenamefont {{Freeman}}, \citenamefont
  {{Fremat}}, \citenamefont {{Friel}}, \citenamefont {{Gaensicke}},
  \citenamefont {{Gameiro}}, \citenamefont {{Garzon}}, \citenamefont {{Geier}},
  \citenamefont {{Geisler}}, \citenamefont {{Gerhard}}, \citenamefont
  {{Gibson}}, \citenamefont {{Gomboc}}, \citenamefont {{Gomez}}, \citenamefont
  {{Gonzalez-Fernandez}}, \citenamefont {{Gonzalez Hernandez}}, \citenamefont
  {{Gosset}}, \citenamefont {{Grebel}}, \citenamefont {{Greimel}},
  \citenamefont {{Groenewegen}}, \citenamefont {{Grundahl}}, \citenamefont
  {{Guarcello}}, \citenamefont {{Gustafsson}}, \citenamefont {{Hadrava}},
  \citenamefont {{Hatzidimitriou}}, \citenamefont {{Hambly}}, \citenamefont
  {{Hammersley}}, \citenamefont {{Hansen}}, \citenamefont {{Haywood}},
  \citenamefont {{Heber}}, \citenamefont {{Heiter}}, \citenamefont {{Held}},
  \citenamefont {{Helmi}}, \citenamefont {{Hensler}}, \citenamefont
  {{Herrero}}, \citenamefont {{Hill}}, \citenamefont {{Hodgkin}}, \citenamefont
  {{Huelamo}}, \citenamefont {{Huxor}}, \citenamefont {{Ibata}}, \citenamefont
  {{Jackson}}, \citenamefont {{de Jong}}, \citenamefont {{Jonker}},
  \citenamefont {{Jordan}}, \citenamefont {{Jordi}}, \citenamefont
  {{Jorissen}}, \citenamefont {{Katz}}, \citenamefont {{Kawata}}, \citenamefont
  {{Keller}}, \citenamefont {{Kharchenko}}, \citenamefont {{Klement}},
  \citenamefont {{Klutsch}}, \citenamefont {{Knude}}, \citenamefont {{Koch}},
  \citenamefont {{Kochukhov}}, \citenamefont {{Kontizas}}, \citenamefont
  {{Koubsky}}, \citenamefont {{Lallement}}, \citenamefont {{de Laverny}},
  \citenamefont {{van Leeuwen}}, \citenamefont {{Lemasle}}, \citenamefont
  {{Lewis}}, \citenamefont {{Lind}}, \citenamefont {{Lindstrom}}, \citenamefont
  {{Lobel}}, \citenamefont {{Lopez Santiago}}, \citenamefont {{Lucas}},
  \citenamefont {{Ludwig}}, \citenamefont {{Lueftinger}}, \citenamefont
  {{Magrini}}, \citenamefont {{Maiz Apellaniz}}, \citenamefont {{Maldonado}},
  \citenamefont {{Marconi}}, \citenamefont {{Marino}}, \citenamefont
  {{Martayan}}, \citenamefont {{Martinez-Valpuesta}}, \citenamefont
  {{Matijevic}}, \citenamefont {{McMahon}}, \citenamefont {{Messina}},
  \citenamefont {{Meyer}}, \citenamefont {{Miglio}}, \citenamefont
  {{Mikolaitis}}, \citenamefont {{Minchev}}, \citenamefont {{Minniti}},
  \citenamefont {{Moitinho}}, \citenamefont {{Momany}}, \citenamefont
  {{Monaco}}, \citenamefont {{Montalto}}, \citenamefont {{Monteiro}},
  \citenamefont {{Monier}}, \citenamefont {{Montes}}, \citenamefont {{Mora}},
  \citenamefont {{Moraux}}, \citenamefont {{Morel}}, \citenamefont {{Mowlavi}},
  \citenamefont {{Mucciarelli}}, \citenamefont {{Munari}}, \citenamefont
  {{Napiwotzki}}, \citenamefont {{Nardetto}}, \citenamefont {{Naylor}},
  \citenamefont {{Naze}}, \citenamefont {{Nelemans}}, \citenamefont
  {{Okamoto}}, \citenamefont {{Ortolani}}, \citenamefont {{Pace}},
  \citenamefont {{Palla}}, \citenamefont {{Palous}}, \citenamefont {{Parker}},
  \citenamefont {{Penarrubia}}, \citenamefont {{Pillitteri}}, \citenamefont
  {{Piotto}}, \citenamefont {{Posbic}}, \citenamefont {{Prisinzano}},
  \citenamefont {{Puzeras}}, \citenamefont {{Quirrenbach}}, \citenamefont
  {{Ragaini}}, \citenamefont {{Read}}, \citenamefont {{Read}}, \citenamefont
  {{Reyle}}, \citenamefont {{De Ridder}}, \citenamefont {{Robichon}},
  \citenamefont {{Robin}}, \citenamefont {{Roeser}}, \citenamefont {{Romano}},
  \citenamefont {{Royer}}, \citenamefont {{Ruchti}}, \citenamefont {{Ruzicka}},
  \citenamefont {{Ryan}}, \citenamefont {{Ryde}}, \citenamefont {{Santos}},
  \citenamefont {{Sanz Forcada}}, \citenamefont {{Sarro Baro}}, \citenamefont
  {{Sbordone}}, \citenamefont {{Schilbach}}, \citenamefont {{Schmeja}},
  \citenamefont {{Schnurr}}, \citenamefont {{Schoenrich}}, \citenamefont
  {{Scholz}}, \citenamefont {{Seabroke}}, \citenamefont {{Sharma}},
  \citenamefont {{De Silva}}, \citenamefont {{Smith}}, \citenamefont
  {{Solano}}, \citenamefont {{Sordo}}, \citenamefont {{Soubiran}},
  \citenamefont {{Sousa}}, \citenamefont {{Spagna}}, \citenamefont {{Steffen}},
  \citenamefont {{Steinmetz}}, \citenamefont {{Stelzer}}, \citenamefont
  {{Stempels}}, \citenamefont {{Tabernero}}, \citenamefont {{Tautvaisiene}},
  \citenamefont {{Thevenin}}, \citenamefont {{Torra}}, \citenamefont {{Tosi}},
  \citenamefont {{Tolstoy}}, \citenamefont {{Turon}}, \citenamefont {{Walker}},
  \citenamefont {{Wambsganss}}, \citenamefont {{Worley}}, \citenamefont
  {{Venn}}, \citenamefont {{Vink}}, \citenamefont {{Wyse}}, \citenamefont
  {{Zaggia}}, \citenamefont {{Zeilinger}}, \citenamefont {{Zoccali}},
  \citenamefont {{Zorec}}, \citenamefont {{Zucker}}, \citenamefont
  {{Zwitter}},\ and\ \citenamefont {{Gaia-ESO Survey
  Team}}}]{2012Msngr.147...25G}%
  \BibitemOpen
  \bibfield  {author} {\bibinfo {author} {\bibfnamefont {G.}~\bibnamefont
  {{Gilmore}}}, \bibinfo {author} {\bibfnamefont {S.}~\bibnamefont
  {{Randich}}}, \bibinfo {author} {\bibfnamefont {M.}~\bibnamefont
  {{Asplund}}}, \bibinfo {author} {\bibfnamefont {J.}~\bibnamefont {{Binney}}},
  \bibinfo {author} {\bibfnamefont {P.}~\bibnamefont {{Bonifacio}}}, \bibinfo
  {author} {\bibfnamefont {J.}~\bibnamefont {{Drew}}}, \bibinfo {author}
  {\bibfnamefont {S.}~\bibnamefont {{Feltzing}}}, \bibinfo {author}
  {\bibfnamefont {A.}~\bibnamefont {{Ferguson}}}, \bibinfo {author}
  {\bibfnamefont {R.}~\bibnamefont {{Jeffries}}}, \bibinfo {author}
  {\bibfnamefont {G.}~\bibnamefont {{Micela}}}, \bibinfo {author}
  {\bibfnamefont {I.}~\bibnamefont {{Negueruela}}}, \bibinfo {author}
  {\bibfnamefont {T.}~\bibnamefont {{Prusti}}}, \bibinfo {author}
  {\bibfnamefont {H.~W.}\ \bibnamefont {{Rix}}}, \bibinfo {author}
  {\bibfnamefont {A.}~\bibnamefont {{Vallenari}}}, \bibinfo {author}
  {\bibfnamefont {E.}~\bibnamefont {{Alfaro}}}, \bibinfo {author}
  {\bibfnamefont {C.}~\bibnamefont {{Allende-Prieto}}}, \bibinfo {author}
  {\bibfnamefont {C.}~\bibnamefont {{Babusiaux}}}, \bibinfo {author}
  {\bibfnamefont {T.}~\bibnamefont {{Bensby}}}, \bibinfo {author}
  {\bibfnamefont {R.}~\bibnamefont {{Blomme}}}, \bibinfo {author}
  {\bibfnamefont {A.}~\bibnamefont {{Bragaglia}}}, \bibinfo {author}
  {\bibfnamefont {E.}~\bibnamefont {{Flaccomio}}}, \bibinfo {author}
  {\bibfnamefont {P.}~\bibnamefont {{Fran{\c{c}}ois}}}, \bibinfo {author}
  {\bibfnamefont {M.}~\bibnamefont {{Irwin}}}, \bibinfo {author} {\bibfnamefont
  {S.}~\bibnamefont {{Koposov}}}, \bibinfo {author} {\bibfnamefont
  {A.}~\bibnamefont {{Korn}}}, \bibinfo {author} {\bibfnamefont
  {A.}~\bibnamefont {{Lanzafame}}}, \bibinfo {author} {\bibfnamefont
  {E.}~\bibnamefont {{Pancino}}}, \bibinfo {author} {\bibfnamefont
  {E.}~\bibnamefont {{Paunzen}}}, \bibinfo {author} {\bibfnamefont
  {A.}~\bibnamefont {{Recio-Blanco}}}, \bibinfo {author} {\bibfnamefont
  {G.}~\bibnamefont {{Sacco}}}, \bibinfo {author} {\bibfnamefont
  {R.}~\bibnamefont {{Smiljanic}}}, \bibinfo {author} {\bibfnamefont
  {S.}~\bibnamefont {{Van Eck}}}, \bibinfo {author} {\bibfnamefont
  {N.}~\bibnamefont {{Walton}}}, \bibinfo {author} {\bibfnamefont
  {D.}~\bibnamefont {{Aden}}}, \bibinfo {author} {\bibfnamefont
  {C.}~\bibnamefont {{Aerts}}}, \bibinfo {author} {\bibfnamefont
  {L.}~\bibnamefont {{Affer}}}, \bibinfo {author} {\bibfnamefont {J.~M.}\
  \bibnamefont {{Alcala}}}, \bibinfo {author} {\bibfnamefont {G.}~\bibnamefont
  {{Altavilla}}}, \bibinfo {author} {\bibfnamefont {J.}~\bibnamefont
  {{Alves}}}, \bibinfo {author} {\bibfnamefont {T.}~\bibnamefont {{Antoja}}},
  \bibinfo {author} {\bibfnamefont {F.}~\bibnamefont {{Arenou}}}, \bibinfo
  {author} {\bibfnamefont {C.}~\bibnamefont {{Argiroffi}}}, \bibinfo {author}
  {\bibfnamefont {A.}~\bibnamefont {{Asensio Ramos}}}, \bibinfo {author}
  {\bibfnamefont {C.}~\bibnamefont {{Bailer-Jones}}}, \bibinfo {author}
  {\bibfnamefont {L.}~\bibnamefont {{Balaguer-Nunez}}}, \bibinfo {author}
  {\bibfnamefont {A.}~\bibnamefont {{Bayo}}}, \bibinfo {author} {\bibfnamefont
  {B.}~\bibnamefont {{Barbuy}}}, \bibinfo {author} {\bibfnamefont
  {G.}~\bibnamefont {{Barisevicius}}}, \bibinfo {author} {\bibfnamefont
  {D.}~\bibnamefont {{Barrado y Navascues}}}, \bibinfo {author} {\bibfnamefont
  {C.}~\bibnamefont {{Battistini}}}, \bibinfo {author} {\bibfnamefont
  {I.}~\bibnamefont {{Bellas Velidis}}}, \bibinfo {author} {\bibfnamefont
  {M.}~\bibnamefont {{Bellazzini}}}, \bibinfo {author} {\bibfnamefont
  {V.}~\bibnamefont {{Belokurov}}}, \bibinfo {author} {\bibfnamefont
  {M.}~\bibnamefont {{Bergemann}}}, \bibinfo {author} {\bibfnamefont
  {G.}~\bibnamefont {{Bertelli}}}, \bibinfo {author} {\bibfnamefont
  {K.}~\bibnamefont {{Biazzo}}}, \bibinfo {author} {\bibfnamefont
  {O.}~\bibnamefont {{Bienayme}}}, \bibinfo {author} {\bibfnamefont
  {J.}~\bibnamefont {{Bland-Hawthorn}}}, \bibinfo {author} {\bibfnamefont
  {C.}~\bibnamefont {{Boeche}}}, \bibinfo {author} {\bibfnamefont
  {S.}~\bibnamefont {{Bonito}}}, \bibinfo {author} {\bibfnamefont
  {S.}~\bibnamefont {{Boudreault}}}, \bibinfo {author} {\bibfnamefont
  {J.}~\bibnamefont {{Bouvier}}}, \bibinfo {author} {\bibfnamefont
  {I.}~\bibnamefont {{Brandao}}}, \bibinfo {author} {\bibfnamefont
  {A.}~\bibnamefont {{Brown}}}, \bibinfo {author} {\bibfnamefont
  {J.}~\bibnamefont {{de Bruijne}}}, \bibinfo {author} {\bibfnamefont
  {M.}~\bibnamefont {{Burleigh}}}, \bibinfo {author} {\bibfnamefont
  {J.}~\bibnamefont {{Caballero}}}, \bibinfo {author} {\bibfnamefont
  {E.}~\bibnamefont {{Caffau}}}, \bibinfo {author} {\bibfnamefont
  {F.}~\bibnamefont {{Calura}}}, \bibinfo {author} {\bibfnamefont
  {R.}~\bibnamefont {{Capuzzo-Dolcetta}}}, \bibinfo {author} {\bibfnamefont
  {M.}~\bibnamefont {{Caramazza}}}, \bibinfo {author} {\bibfnamefont
  {G.}~\bibnamefont {{Carraro}}}, \bibinfo {author} {\bibfnamefont
  {L.}~\bibnamefont {{Casagrande}}}, \bibinfo {author} {\bibfnamefont
  {S.}~\bibnamefont {{Casewell}}}, \bibinfo {author} {\bibfnamefont
  {S.}~\bibnamefont {{Chapman}}}, \bibinfo {author} {\bibfnamefont
  {C.}~\bibnamefont {{Chiappini}}}, \bibinfo {author} {\bibfnamefont
  {Y.}~\bibnamefont {{Chorniy}}}, \bibinfo {author} {\bibfnamefont
  {N.}~\bibnamefont {{Christlieb}}}, \bibinfo {author} {\bibfnamefont
  {M.}~\bibnamefont {{Cignoni}}}, \bibinfo {author} {\bibfnamefont
  {G.}~\bibnamefont {{Cocozza}}}, \bibinfo {author} {\bibfnamefont
  {M.}~\bibnamefont {{Colless}}}, \bibinfo {author} {\bibfnamefont
  {R.}~\bibnamefont {{Collet}}}, \bibinfo {author} {\bibfnamefont
  {M.}~\bibnamefont {{Collins}}}, \bibinfo {author} {\bibfnamefont
  {M.}~\bibnamefont {{Correnti}}}, \bibinfo {author} {\bibfnamefont
  {E.}~\bibnamefont {{Covino}}}, \bibinfo {author} {\bibfnamefont
  {D.}~\bibnamefont {{Crnojevic}}}, \bibinfo {author} {\bibfnamefont
  {M.}~\bibnamefont {{Cropper}}}, \bibinfo {author} {\bibfnamefont
  {M.}~\bibnamefont {{Cunha}}}, \bibinfo {author} {\bibfnamefont
  {F.}~\bibnamefont {{Damiani}}}, \bibinfo {author} {\bibfnamefont
  {M.}~\bibnamefont {{David}}}, \bibinfo {author} {\bibfnamefont
  {A.}~\bibnamefont {{Delgado}}}, \bibinfo {author} {\bibfnamefont
  {S.}~\bibnamefont {{Duffau}}}, \bibinfo {author} {\bibfnamefont
  {B.}~\bibnamefont {{Edvardsson}}}, \bibinfo {author} {\bibfnamefont
  {J.}~\bibnamefont {{Eldridge}}}, \bibinfo {author} {\bibfnamefont
  {H.}~\bibnamefont {{Enke}}}, \bibinfo {author} {\bibfnamefont
  {K.}~\bibnamefont {{Eriksson}}}, \bibinfo {author} {\bibfnamefont {N.~W.}\
  \bibnamefont {{Evans}}}, \bibinfo {author} {\bibfnamefont {L.}~\bibnamefont
  {{Eyer}}}, \bibinfo {author} {\bibfnamefont {B.}~\bibnamefont {{Famaey}}},
  \bibinfo {author} {\bibfnamefont {M.}~\bibnamefont {{Fellhauer}}}, \bibinfo
  {author} {\bibfnamefont {I.}~\bibnamefont {{Ferreras}}}, \bibinfo {author}
  {\bibfnamefont {F.}~\bibnamefont {{Figueras}}}, \bibinfo {author}
  {\bibfnamefont {G.}~\bibnamefont {{Fiorentino}}}, \bibinfo {author}
  {\bibfnamefont {C.}~\bibnamefont {{Flynn}}}, \bibinfo {author} {\bibfnamefont
  {D.}~\bibnamefont {{Folha}}}, \bibinfo {author} {\bibfnamefont
  {E.}~\bibnamefont {{Franciosini}}}, \bibinfo {author} {\bibfnamefont
  {A.}~\bibnamefont {{Frasca}}}, \bibinfo {author} {\bibfnamefont
  {K.}~\bibnamefont {{Freeman}}}, \bibinfo {author} {\bibfnamefont
  {Y.}~\bibnamefont {{Fremat}}}, \bibinfo {author} {\bibfnamefont
  {E.}~\bibnamefont {{Friel}}}, \bibinfo {author} {\bibfnamefont
  {B.}~\bibnamefont {{Gaensicke}}}, \bibinfo {author} {\bibfnamefont
  {J.}~\bibnamefont {{Gameiro}}}, \bibinfo {author} {\bibfnamefont
  {F.}~\bibnamefont {{Garzon}}}, \bibinfo {author} {\bibfnamefont
  {S.}~\bibnamefont {{Geier}}}, \bibinfo {author} {\bibfnamefont
  {D.}~\bibnamefont {{Geisler}}}, \bibinfo {author} {\bibfnamefont
  {O.}~\bibnamefont {{Gerhard}}}, \bibinfo {author} {\bibfnamefont
  {B.}~\bibnamefont {{Gibson}}}, \bibinfo {author} {\bibfnamefont
  {A.}~\bibnamefont {{Gomboc}}}, \bibinfo {author} {\bibfnamefont
  {A.}~\bibnamefont {{Gomez}}}, \bibinfo {author} {\bibfnamefont
  {C.}~\bibnamefont {{Gonzalez-Fernandez}}}, \bibinfo {author} {\bibfnamefont
  {J.}~\bibnamefont {{Gonzalez Hernandez}}}, \bibinfo {author} {\bibfnamefont
  {E.}~\bibnamefont {{Gosset}}}, \bibinfo {author} {\bibfnamefont
  {E.}~\bibnamefont {{Grebel}}}, \bibinfo {author} {\bibfnamefont
  {R.}~\bibnamefont {{Greimel}}}, \bibinfo {author} {\bibfnamefont
  {M.}~\bibnamefont {{Groenewegen}}}, \bibinfo {author} {\bibfnamefont
  {F.}~\bibnamefont {{Grundahl}}}, \bibinfo {author} {\bibfnamefont
  {M.}~\bibnamefont {{Guarcello}}}, \bibinfo {author} {\bibfnamefont
  {B.}~\bibnamefont {{Gustafsson}}}, \bibinfo {author} {\bibfnamefont
  {P.}~\bibnamefont {{Hadrava}}}, \bibinfo {author} {\bibfnamefont
  {D.}~\bibnamefont {{Hatzidimitriou}}}, \bibinfo {author} {\bibfnamefont
  {N.}~\bibnamefont {{Hambly}}}, \bibinfo {author} {\bibfnamefont
  {P.}~\bibnamefont {{Hammersley}}}, \bibinfo {author} {\bibfnamefont
  {C.}~\bibnamefont {{Hansen}}}, \bibinfo {author} {\bibfnamefont
  {M.}~\bibnamefont {{Haywood}}}, \bibinfo {author} {\bibfnamefont
  {U.}~\bibnamefont {{Heber}}}, \bibinfo {author} {\bibfnamefont
  {U.}~\bibnamefont {{Heiter}}}, \bibinfo {author} {\bibfnamefont
  {E.}~\bibnamefont {{Held}}}, \bibinfo {author} {\bibfnamefont
  {A.}~\bibnamefont {{Helmi}}}, \bibinfo {author} {\bibfnamefont
  {G.}~\bibnamefont {{Hensler}}}, \bibinfo {author} {\bibfnamefont
  {A.}~\bibnamefont {{Herrero}}}, \bibinfo {author} {\bibfnamefont
  {V.}~\bibnamefont {{Hill}}}, \bibinfo {author} {\bibfnamefont
  {S.}~\bibnamefont {{Hodgkin}}}, \bibinfo {author} {\bibfnamefont
  {N.}~\bibnamefont {{Huelamo}}}, \bibinfo {author} {\bibfnamefont
  {A.}~\bibnamefont {{Huxor}}}, \bibinfo {author} {\bibfnamefont
  {R.}~\bibnamefont {{Ibata}}}, \bibinfo {author} {\bibfnamefont
  {R.}~\bibnamefont {{Jackson}}}, \bibinfo {author} {\bibfnamefont
  {R.}~\bibnamefont {{de Jong}}}, \bibinfo {author} {\bibfnamefont
  {P.}~\bibnamefont {{Jonker}}}, \bibinfo {author} {\bibfnamefont
  {S.}~\bibnamefont {{Jordan}}}, \bibinfo {author} {\bibfnamefont
  {C.}~\bibnamefont {{Jordi}}}, \bibinfo {author} {\bibfnamefont
  {A.}~\bibnamefont {{Jorissen}}}, \bibinfo {author} {\bibfnamefont
  {D.}~\bibnamefont {{Katz}}}, \bibinfo {author} {\bibfnamefont
  {D.}~\bibnamefont {{Kawata}}}, \bibinfo {author} {\bibfnamefont
  {S.}~\bibnamefont {{Keller}}}, \bibinfo {author} {\bibfnamefont
  {N.}~\bibnamefont {{Kharchenko}}}, \bibinfo {author} {\bibfnamefont
  {R.}~\bibnamefont {{Klement}}}, \bibinfo {author} {\bibfnamefont
  {A.}~\bibnamefont {{Klutsch}}}, \bibinfo {author} {\bibfnamefont
  {J.}~\bibnamefont {{Knude}}}, \bibinfo {author} {\bibfnamefont
  {A.}~\bibnamefont {{Koch}}}, \bibinfo {author} {\bibfnamefont
  {O.}~\bibnamefont {{Kochukhov}}}, \bibinfo {author} {\bibfnamefont
  {M.}~\bibnamefont {{Kontizas}}}, \bibinfo {author} {\bibfnamefont
  {P.}~\bibnamefont {{Koubsky}}}, \bibinfo {author} {\bibfnamefont
  {R.}~\bibnamefont {{Lallement}}}, \bibinfo {author} {\bibfnamefont
  {P.}~\bibnamefont {{de Laverny}}}, \bibinfo {author} {\bibfnamefont
  {F.}~\bibnamefont {{van Leeuwen}}}, \bibinfo {author} {\bibfnamefont
  {B.}~\bibnamefont {{Lemasle}}}, \bibinfo {author} {\bibfnamefont
  {G.}~\bibnamefont {{Lewis}}}, \bibinfo {author} {\bibfnamefont
  {K.}~\bibnamefont {{Lind}}}, \bibinfo {author} {\bibfnamefont {H.~P.~E.}\
  \bibnamefont {{Lindstrom}}}, \bibinfo {author} {\bibfnamefont
  {A.}~\bibnamefont {{Lobel}}}, \bibinfo {author} {\bibfnamefont
  {J.}~\bibnamefont {{Lopez Santiago}}}, \bibinfo {author} {\bibfnamefont
  {P.}~\bibnamefont {{Lucas}}}, \bibinfo {author} {\bibfnamefont
  {H.}~\bibnamefont {{Ludwig}}}, \bibinfo {author} {\bibfnamefont
  {T.}~\bibnamefont {{Lueftinger}}}, \bibinfo {author} {\bibfnamefont
  {L.}~\bibnamefont {{Magrini}}}, \bibinfo {author} {\bibfnamefont
  {J.}~\bibnamefont {{Maiz Apellaniz}}}, \bibinfo {author} {\bibfnamefont
  {J.}~\bibnamefont {{Maldonado}}}, \bibinfo {author} {\bibfnamefont
  {G.}~\bibnamefont {{Marconi}}}, \bibinfo {author} {\bibfnamefont
  {A.}~\bibnamefont {{Marino}}}, \bibinfo {author} {\bibfnamefont
  {C.}~\bibnamefont {{Martayan}}}, \bibinfo {author} {\bibfnamefont
  {I.}~\bibnamefont {{Martinez-Valpuesta}}}, \bibinfo {author} {\bibfnamefont
  {G.}~\bibnamefont {{Matijevic}}}, \bibinfo {author} {\bibfnamefont
  {R.}~\bibnamefont {{McMahon}}}, \bibinfo {author} {\bibfnamefont
  {S.}~\bibnamefont {{Messina}}}, \bibinfo {author} {\bibfnamefont
  {M.}~\bibnamefont {{Meyer}}}, \bibinfo {author} {\bibfnamefont
  {A.}~\bibnamefont {{Miglio}}}, \bibinfo {author} {\bibfnamefont
  {S.}~\bibnamefont {{Mikolaitis}}}, \bibinfo {author} {\bibfnamefont
  {I.}~\bibnamefont {{Minchev}}}, \bibinfo {author} {\bibfnamefont
  {D.}~\bibnamefont {{Minniti}}}, \bibinfo {author} {\bibfnamefont
  {A.}~\bibnamefont {{Moitinho}}}, \bibinfo {author} {\bibfnamefont
  {Y.}~\bibnamefont {{Momany}}}, \bibinfo {author} {\bibfnamefont
  {L.}~\bibnamefont {{Monaco}}}, \bibinfo {author} {\bibfnamefont
  {M.}~\bibnamefont {{Montalto}}}, \bibinfo {author} {\bibfnamefont {M.~J.}\
  \bibnamefont {{Monteiro}}}, \bibinfo {author} {\bibfnamefont
  {R.}~\bibnamefont {{Monier}}}, \bibinfo {author} {\bibfnamefont
  {D.}~\bibnamefont {{Montes}}}, \bibinfo {author} {\bibfnamefont
  {A.}~\bibnamefont {{Mora}}}, \bibinfo {author} {\bibfnamefont
  {E.}~\bibnamefont {{Moraux}}}, \bibinfo {author} {\bibfnamefont
  {T.}~\bibnamefont {{Morel}}}, \bibinfo {author} {\bibfnamefont
  {N.}~\bibnamefont {{Mowlavi}}}, \bibinfo {author} {\bibfnamefont
  {A.}~\bibnamefont {{Mucciarelli}}}, \bibinfo {author} {\bibfnamefont
  {U.}~\bibnamefont {{Munari}}}, \bibinfo {author} {\bibfnamefont
  {R.}~\bibnamefont {{Napiwotzki}}}, \bibinfo {author} {\bibfnamefont
  {N.}~\bibnamefont {{Nardetto}}}, \bibinfo {author} {\bibfnamefont
  {T.}~\bibnamefont {{Naylor}}}, \bibinfo {author} {\bibfnamefont
  {Y.}~\bibnamefont {{Naze}}}, \bibinfo {author} {\bibfnamefont
  {G.}~\bibnamefont {{Nelemans}}}, \bibinfo {author} {\bibfnamefont
  {S.}~\bibnamefont {{Okamoto}}}, \bibinfo {author} {\bibfnamefont
  {S.}~\bibnamefont {{Ortolani}}}, \bibinfo {author} {\bibfnamefont
  {G.}~\bibnamefont {{Pace}}}, \bibinfo {author} {\bibfnamefont
  {F.}~\bibnamefont {{Palla}}}, \bibinfo {author} {\bibfnamefont
  {J.}~\bibnamefont {{Palous}}}, \bibinfo {author} {\bibfnamefont
  {R.}~\bibnamefont {{Parker}}}, \bibinfo {author} {\bibfnamefont
  {J.}~\bibnamefont {{Penarrubia}}}, \bibinfo {author} {\bibfnamefont
  {I.}~\bibnamefont {{Pillitteri}}}, \bibinfo {author} {\bibfnamefont
  {G.}~\bibnamefont {{Piotto}}}, \bibinfo {author} {\bibfnamefont
  {H.}~\bibnamefont {{Posbic}}}, \bibinfo {author} {\bibfnamefont
  {L.}~\bibnamefont {{Prisinzano}}}, \bibinfo {author} {\bibfnamefont
  {E.}~\bibnamefont {{Puzeras}}}, \bibinfo {author} {\bibfnamefont
  {A.}~\bibnamefont {{Quirrenbach}}}, \bibinfo {author} {\bibfnamefont
  {S.}~\bibnamefont {{Ragaini}}}, \bibinfo {author} {\bibfnamefont
  {J.}~\bibnamefont {{Read}}}, \bibinfo {author} {\bibfnamefont
  {M.}~\bibnamefont {{Read}}}, \bibinfo {author} {\bibfnamefont
  {C.}~\bibnamefont {{Reyle}}}, \bibinfo {author} {\bibfnamefont
  {J.}~\bibnamefont {{De Ridder}}}, \bibinfo {author} {\bibfnamefont
  {N.}~\bibnamefont {{Robichon}}}, \bibinfo {author} {\bibfnamefont
  {A.}~\bibnamefont {{Robin}}}, \bibinfo {author} {\bibfnamefont
  {S.}~\bibnamefont {{Roeser}}}, \bibinfo {author} {\bibfnamefont
  {D.}~\bibnamefont {{Romano}}}, \bibinfo {author} {\bibfnamefont
  {F.}~\bibnamefont {{Royer}}}, \bibinfo {author} {\bibfnamefont
  {G.}~\bibnamefont {{Ruchti}}}, \bibinfo {author} {\bibfnamefont
  {A.}~\bibnamefont {{Ruzicka}}}, \bibinfo {author} {\bibfnamefont
  {S.}~\bibnamefont {{Ryan}}}, \bibinfo {author} {\bibfnamefont
  {N.}~\bibnamefont {{Ryde}}}, \bibinfo {author} {\bibfnamefont
  {N.}~\bibnamefont {{Santos}}}, \bibinfo {author} {\bibfnamefont
  {J.}~\bibnamefont {{Sanz Forcada}}}, \bibinfo {author} {\bibfnamefont
  {L.~M.}\ \bibnamefont {{Sarro Baro}}}, \bibinfo {author} {\bibfnamefont
  {L.}~\bibnamefont {{Sbordone}}}, \bibinfo {author} {\bibfnamefont
  {E.}~\bibnamefont {{Schilbach}}}, \bibinfo {author} {\bibfnamefont
  {S.}~\bibnamefont {{Schmeja}}}, \bibinfo {author} {\bibfnamefont
  {O.}~\bibnamefont {{Schnurr}}}, \bibinfo {author} {\bibfnamefont
  {R.}~\bibnamefont {{Schoenrich}}}, \bibinfo {author} {\bibfnamefont {R.~D.}\
  \bibnamefont {{Scholz}}}, \bibinfo {author} {\bibfnamefont {G.}~\bibnamefont
  {{Seabroke}}}, \bibinfo {author} {\bibfnamefont {S.}~\bibnamefont
  {{Sharma}}}, \bibinfo {author} {\bibfnamefont {G.}~\bibnamefont {{De
  Silva}}}, \bibinfo {author} {\bibfnamefont {M.}~\bibnamefont {{Smith}}},
  \bibinfo {author} {\bibfnamefont {E.}~\bibnamefont {{Solano}}}, \bibinfo
  {author} {\bibfnamefont {R.}~\bibnamefont {{Sordo}}}, \bibinfo {author}
  {\bibfnamefont {C.}~\bibnamefont {{Soubiran}}}, \bibinfo {author}
  {\bibfnamefont {S.}~\bibnamefont {{Sousa}}}, \bibinfo {author} {\bibfnamefont
  {A.}~\bibnamefont {{Spagna}}}, \bibinfo {author} {\bibfnamefont
  {M.}~\bibnamefont {{Steffen}}}, \bibinfo {author} {\bibfnamefont
  {M.}~\bibnamefont {{Steinmetz}}}, \bibinfo {author} {\bibfnamefont
  {B.}~\bibnamefont {{Stelzer}}}, \bibinfo {author} {\bibfnamefont
  {E.}~\bibnamefont {{Stempels}}}, \bibinfo {author} {\bibfnamefont
  {H.}~\bibnamefont {{Tabernero}}}, \bibinfo {author} {\bibfnamefont
  {G.}~\bibnamefont {{Tautvaisiene}}}, \bibinfo {author} {\bibfnamefont
  {F.}~\bibnamefont {{Thevenin}}}, \bibinfo {author} {\bibfnamefont
  {J.}~\bibnamefont {{Torra}}}, \bibinfo {author} {\bibfnamefont
  {M.}~\bibnamefont {{Tosi}}}, \bibinfo {author} {\bibfnamefont
  {E.}~\bibnamefont {{Tolstoy}}}, \bibinfo {author} {\bibfnamefont
  {C.}~\bibnamefont {{Turon}}}, \bibinfo {author} {\bibfnamefont
  {M.}~\bibnamefont {{Walker}}}, \bibinfo {author} {\bibfnamefont
  {J.}~\bibnamefont {{Wambsganss}}}, \bibinfo {author} {\bibfnamefont
  {C.}~\bibnamefont {{Worley}}}, \bibinfo {author} {\bibfnamefont
  {K.}~\bibnamefont {{Venn}}}, \bibinfo {author} {\bibfnamefont
  {J.}~\bibnamefont {{Vink}}}, \bibinfo {author} {\bibfnamefont
  {R.}~\bibnamefont {{Wyse}}}, \bibinfo {author} {\bibfnamefont
  {S.}~\bibnamefont {{Zaggia}}}, \bibinfo {author} {\bibfnamefont
  {W.}~\bibnamefont {{Zeilinger}}}, \bibinfo {author} {\bibfnamefont
  {M.}~\bibnamefont {{Zoccali}}}, \bibinfo {author} {\bibfnamefont
  {J.}~\bibnamefont {{Zorec}}}, \bibinfo {author} {\bibfnamefont
  {D.}~\bibnamefont {{Zucker}}}, \bibinfo {author} {\bibfnamefont
  {T.}~\bibnamefont {{Zwitter}}},\ and\ \bibinfo {author} {\bibnamefont
  {{Gaia-ESO Survey Team}}},\ }\bibfield  {title} {\bibinfo {title} {{The
  Gaia-ESO Public Spectroscopic Survey}},\ }\href@noop {} {\bibfield  {journal}
  {\bibinfo  {journal} {The Messenger}\ }\textbf {\bibinfo {volume} {147}},\
  \bibinfo {pages} {25} (\bibinfo {year} {2012})}\BibitemShut {NoStop}%
\bibitem [{\citenamefont {{Dalton}}\ \emph {et~al.}(2014)\citenamefont
  {{Dalton}}, \citenamefont {{Trager}}, \citenamefont {{Abrams}}, \citenamefont
  {{Bonifacio}}, \citenamefont {{L{\'o}pez Aguerri}}, \citenamefont
  {{Middleton}}, \citenamefont {{Benn}}, \citenamefont {{Dee}}, \citenamefont
  {{Say{\`e}de}}, \citenamefont {{Lewis}}, \citenamefont {{Pragt}},
  \citenamefont {{Pico}}, \citenamefont {{Walton}}, \citenamefont {{Rey}},
  \citenamefont {{Allende Prieto}}, \citenamefont {{Pe{\~n}ate}}, \citenamefont
  {{Lhome}}, \citenamefont {{Ag{\'o}cs}}, \citenamefont {{Alonso}},
  \citenamefont {{Terrett}}, \citenamefont {{Brock}}, \citenamefont
  {{Gilbert}}, \citenamefont {{Ridings}}, \citenamefont {{Guinouard}},
  \citenamefont {{Verheijen}}, \citenamefont {{Tosh}}, \citenamefont
  {{Rogers}}, \citenamefont {{Steele}}, \citenamefont {{Stuik}}, \citenamefont
  {{Tromp}}, \citenamefont {{Jasko}}, \citenamefont {{Kragt}}, \citenamefont
  {{Lesman}}, \citenamefont {{Mottram}}, \citenamefont {{Bates}}, \citenamefont
  {{Gribbin}}, \citenamefont {{Fernando Rodriguez}}, \citenamefont {{Delgado}},
  \citenamefont {{Martin}}, \citenamefont {{Cano}}, \citenamefont {{Navarro}},
  \citenamefont {{Irwin}}, \citenamefont {{Lewis}}, \citenamefont {{Gonzalez
  Solares}}, \citenamefont {{O'Mahony}}, \citenamefont {{Bianco}},
  \citenamefont {{Zurita}}, \citenamefont {{ter Horst}}, \citenamefont
  {{Molinari}}, \citenamefont {{Lodi}}, \citenamefont {{Guerra}}, \citenamefont
  {{Vallenari}},\ and\ \citenamefont {{Baruffolo}}}]{2014SPIE.9147E..0LD}%
  \BibitemOpen
  \bibfield  {author} {\bibinfo {author} {\bibfnamefont {G.}~\bibnamefont
  {{Dalton}}}, \bibinfo {author} {\bibfnamefont {S.}~\bibnamefont {{Trager}}},
  \bibinfo {author} {\bibfnamefont {D.~C.}\ \bibnamefont {{Abrams}}}, \bibinfo
  {author} {\bibfnamefont {P.}~\bibnamefont {{Bonifacio}}}, \bibinfo {author}
  {\bibfnamefont {J.~A.}\ \bibnamefont {{L{\'o}pez Aguerri}}}, \bibinfo
  {author} {\bibfnamefont {K.}~\bibnamefont {{Middleton}}}, \bibinfo {author}
  {\bibfnamefont {C.}~\bibnamefont {{Benn}}}, \bibinfo {author} {\bibfnamefont
  {K.}~\bibnamefont {{Dee}}}, \bibinfo {author} {\bibfnamefont
  {F.}~\bibnamefont {{Say{\`e}de}}}, \bibinfo {author} {\bibfnamefont
  {I.}~\bibnamefont {{Lewis}}}, \bibinfo {author} {\bibfnamefont
  {J.}~\bibnamefont {{Pragt}}}, \bibinfo {author} {\bibfnamefont
  {S.}~\bibnamefont {{Pico}}}, \bibinfo {author} {\bibfnamefont
  {N.}~\bibnamefont {{Walton}}}, \bibinfo {author} {\bibfnamefont
  {J.}~\bibnamefont {{Rey}}}, \bibinfo {author} {\bibfnamefont
  {C.}~\bibnamefont {{Allende Prieto}}}, \bibinfo {author} {\bibfnamefont
  {J.}~\bibnamefont {{Pe{\~n}ate}}}, \bibinfo {author} {\bibfnamefont
  {E.}~\bibnamefont {{Lhome}}}, \bibinfo {author} {\bibfnamefont
  {T.}~\bibnamefont {{Ag{\'o}cs}}}, \bibinfo {author} {\bibfnamefont
  {J.}~\bibnamefont {{Alonso}}}, \bibinfo {author} {\bibfnamefont
  {D.}~\bibnamefont {{Terrett}}}, \bibinfo {author} {\bibfnamefont
  {M.}~\bibnamefont {{Brock}}}, \bibinfo {author} {\bibfnamefont
  {J.}~\bibnamefont {{Gilbert}}}, \bibinfo {author} {\bibfnamefont
  {A.}~\bibnamefont {{Ridings}}}, \bibinfo {author} {\bibfnamefont
  {I.}~\bibnamefont {{Guinouard}}}, \bibinfo {author} {\bibfnamefont
  {M.}~\bibnamefont {{Verheijen}}}, \bibinfo {author} {\bibfnamefont
  {I.}~\bibnamefont {{Tosh}}}, \bibinfo {author} {\bibfnamefont
  {K.}~\bibnamefont {{Rogers}}}, \bibinfo {author} {\bibfnamefont
  {I.}~\bibnamefont {{Steele}}}, \bibinfo {author} {\bibfnamefont
  {R.}~\bibnamefont {{Stuik}}}, \bibinfo {author} {\bibfnamefont
  {N.}~\bibnamefont {{Tromp}}}, \bibinfo {author} {\bibfnamefont
  {A.}~\bibnamefont {{Jasko}}}, \bibinfo {author} {\bibfnamefont
  {J.}~\bibnamefont {{Kragt}}}, \bibinfo {author} {\bibfnamefont
  {D.}~\bibnamefont {{Lesman}}}, \bibinfo {author} {\bibfnamefont
  {C.}~\bibnamefont {{Mottram}}}, \bibinfo {author} {\bibfnamefont
  {S.}~\bibnamefont {{Bates}}}, \bibinfo {author} {\bibfnamefont
  {F.}~\bibnamefont {{Gribbin}}}, \bibinfo {author} {\bibfnamefont
  {L.}~\bibnamefont {{Fernando Rodriguez}}}, \bibinfo {author} {\bibfnamefont
  {J.~M.}\ \bibnamefont {{Delgado}}}, \bibinfo {author} {\bibfnamefont
  {C.}~\bibnamefont {{Martin}}}, \bibinfo {author} {\bibfnamefont
  {D.}~\bibnamefont {{Cano}}}, \bibinfo {author} {\bibfnamefont
  {R.}~\bibnamefont {{Navarro}}}, \bibinfo {author} {\bibfnamefont
  {M.}~\bibnamefont {{Irwin}}}, \bibinfo {author} {\bibfnamefont
  {J.}~\bibnamefont {{Lewis}}}, \bibinfo {author} {\bibfnamefont
  {E.}~\bibnamefont {{Gonzalez Solares}}}, \bibinfo {author} {\bibfnamefont
  {N.}~\bibnamefont {{O'Mahony}}}, \bibinfo {author} {\bibfnamefont
  {A.}~\bibnamefont {{Bianco}}}, \bibinfo {author} {\bibfnamefont
  {C.}~\bibnamefont {{Zurita}}}, \bibinfo {author} {\bibfnamefont
  {R.}~\bibnamefont {{ter Horst}}}, \bibinfo {author} {\bibfnamefont
  {E.}~\bibnamefont {{Molinari}}}, \bibinfo {author} {\bibfnamefont
  {M.}~\bibnamefont {{Lodi}}}, \bibinfo {author} {\bibfnamefont
  {J.}~\bibnamefont {{Guerra}}}, \bibinfo {author} {\bibfnamefont
  {A.}~\bibnamefont {{Vallenari}}},\ and\ \bibinfo {author} {\bibfnamefont
  {A.}~\bibnamefont {{Baruffolo}}},\ }\bibfield  {title} {\bibinfo {title}
  {{Project overview and update on WEAVE: the next generation wide-field
  spectroscopy facility for the William Herschel Telescope}},\ }in\ \href
  {https://doi.org/10.1117/12.2055132} {\emph {\bibinfo {booktitle}
  {Ground-based and Airborne Instrumentation for Astronomy V}}},\ \bibinfo
  {series} {Society of Photo-Optical Instrumentation Engineers (SPIE)
  Conference Series}, Vol.\ \bibinfo {volume} {9147},\ \bibinfo {editor}
  {edited by\ \bibinfo {editor} {\bibfnamefont {S.~K.}\ \bibnamefont
  {{Ramsay}}}, \bibinfo {editor} {\bibfnamefont {I.~S.}\ \bibnamefont
  {{McLean}}},\ and\ \bibinfo {editor} {\bibfnamefont {H.}~\bibnamefont
  {{Takami}}}}\ (\bibinfo {year} {2014})\ p.\ \bibinfo {pages} {91470L},\
  \Eprint {https://arxiv.org/abs/1412.0843} {arXiv:1412.0843 [astro-ph.IM]}
  \BibitemShut {NoStop}%
\bibitem [{\citenamefont {{de Jong}}\ \emph {et~al.}(2019)\citenamefont {{de
  Jong}}, \citenamefont {{Agertz}}, \citenamefont {{Berbel}}, \citenamefont
  {{Aird}}, \citenamefont {{Alexander}}, \citenamefont {{Amarsi}},
  \citenamefont {{Anders}}, \citenamefont {{Andrae}}, \citenamefont
  {{Ansarinejad}}, \citenamefont {{Ansorge}}, \citenamefont {{Antilogus}},
  \citenamefont {{Anwand-Heerwart}}, \citenamefont {{Arentsen}}, \citenamefont
  {{Arnadottir}}, \citenamefont {{Asplund}}, \citenamefont {{Auger}},
  \citenamefont {{Azais}}, \citenamefont {{Baade}}, \citenamefont {{Baker}},
  \citenamefont {{Baker}}, \citenamefont {{Balbinot}}, \citenamefont
  {{Baldry}}, \citenamefont {{Banerji}}, \citenamefont {{Barden}},
  \citenamefont {{Barklem}}, \citenamefont {{Barth{\'e}l{\'e}my-Mazot}},
  \citenamefont {{Battistini}}, \citenamefont {{Bauer}}, \citenamefont
  {{Bell}}, \citenamefont {{Bellido-Tirado}}, \citenamefont {{Bellstedt}},
  \citenamefont {{Belokurov}}, \citenamefont {{Bensby}}, \citenamefont
  {{Bergemann}}, \citenamefont {{Bestenlehner}}, \citenamefont {{Bielby}},
  \citenamefont {{Bilicki}}, \citenamefont {{Blake}}, \citenamefont
  {{Bland-Hawthorn}}, \citenamefont {{Boeche}}, \citenamefont {{Boland}},
  \citenamefont {{Boller}}, \citenamefont {{Bongard}}, \citenamefont
  {{Bongiorno}}, \citenamefont {{Bonifacio}}, \citenamefont {{Boudon}},
  \citenamefont {{Brooks}}, \citenamefont {{Brown}}, \citenamefont {{Brown}},
  \citenamefont {{Br{\"u}ggen}}, \citenamefont {{Brynnel}}, \citenamefont
  {{Brzeski}}, \citenamefont {{Buchert}}, \citenamefont {{Buschkamp}},
  \citenamefont {{Caffau}}, \citenamefont {{Caillier}}, \citenamefont
  {{Carrick}}, \citenamefont {{Casagrande}}, \citenamefont {{Case}},
  \citenamefont {{Casey}}, \citenamefont {{Cesarini}}, \citenamefont
  {{Cescutti}}, \citenamefont {{Chapuis}}, \citenamefont {{Chiappini}},
  \citenamefont {{Childress}}, \citenamefont {{Christlieb}}, \citenamefont
  {{Church}}, \citenamefont {{Cioni}}, \citenamefont {{Cluver}}, \citenamefont
  {{Colless}}, \citenamefont {{Collett}}, \citenamefont {{Comparat}},
  \citenamefont {{Cooper}}, \citenamefont {{Couch}}, \citenamefont {{Courbin}},
  \citenamefont {{Croom}}, \citenamefont {{Croton}}, \citenamefont
  {{Daguis{\'e}}}, \citenamefont {{Dalton}}, \citenamefont {{Davies}},
  \citenamefont {{Davis}}, \citenamefont {{de Laverny}}, \citenamefont
  {{Deason}}, \citenamefont {{Dionies}}, \citenamefont {{Disseau}},
  \citenamefont {{Doel}}, \citenamefont {{D{\"o}scher}}, \citenamefont
  {{Driver}}, \citenamefont {{Dwelly}}, \citenamefont {{Eckert}}, \citenamefont
  {{Edge}}, \citenamefont {{Edvardsson}}, \citenamefont {{Youssoufi}},
  \citenamefont {{Elhaddad}}, \citenamefont {{Enke}}, \citenamefont
  {{Erfanianfar}}, \citenamefont {{Farrell}}, \citenamefont {{Fechner}},
  \citenamefont {{Feiz}}, \citenamefont {{Feltzing}}, \citenamefont
  {{Ferreras}}, \citenamefont {{Feuerstein}}, \citenamefont {{Feuillet}},
  \citenamefont {{Finoguenov}}, \citenamefont {{Ford}}, \citenamefont
  {{Fotopoulou}}, \citenamefont {{Fouesneau}}, \citenamefont {{Frenk}},
  \citenamefont {{Frey}}, \citenamefont {{Gaessler}}, \citenamefont {{Geier}},
  \citenamefont {{Gentile Fusillo}}, \citenamefont {{Gerhard}}, \citenamefont
  {{Giannantonio}}, \citenamefont {{Giannone}}, \citenamefont {{Gibson}},
  \citenamefont {{Gillingham}}, \citenamefont {{Gonz{\'a}lez-Fern{\'a}ndez}},
  \citenamefont {{Gonzalez-Solares}}, \citenamefont {{Gottloeber}},
  \citenamefont {{Gould}}, \citenamefont {{Grebel}}, \citenamefont {{Gueguen}},
  \citenamefont {{Guiglion}}, \citenamefont {{Haehnelt}}, \citenamefont
  {{Hahn}}, \citenamefont {{Hansen}}, \citenamefont {{Hartman}}, \citenamefont
  {{Hauptner}}, \citenamefont {{Hawkins}}, \citenamefont {{Haynes}},
  \citenamefont {{Haynes}}, \citenamefont {{Heiter}}, \citenamefont {{Helmi}},
  \citenamefont {{Aguayo}}, \citenamefont {{Hewett}}, \citenamefont {{Hinton}},
  \citenamefont {{Hobbs}}, \citenamefont {{Hoenig}}, \citenamefont {{Hofman}},
  \citenamefont {{Hook}}, \citenamefont {{Hopgood}}, \citenamefont {{Hopkins}},
  \citenamefont {{Hourihane}}, \citenamefont {{Howes}}, \citenamefont
  {{Howlett}}, \citenamefont {{Huet}}, \citenamefont {{Irwin}}, \citenamefont
  {{Iwert}}, \citenamefont {{Jablonka}}, \citenamefont {{Jahn}}, \citenamefont
  {{Jahnke}}, \citenamefont {{Jarno}}, \citenamefont {{Jin}}, \citenamefont
  {{Jofre}}, \citenamefont {{Johl}}, \citenamefont {{Jones}}, \citenamefont
  {{J{\"o}nsson}}, \citenamefont {{Jordan}}, \citenamefont {{Karovicova}},
  \citenamefont {{Khalatyan}}, \citenamefont {{Kelz}}, \citenamefont
  {{Kennicutt}}, \citenamefont {{King}}, \citenamefont {{Kitaura}},
  \citenamefont {{Klar}}, \citenamefont {{Klauser}}, \citenamefont {{Kneib}},
  \citenamefont {{Koch}}, \citenamefont {{Koposov}}, \citenamefont
  {{Kordopatis}}, \citenamefont {{Korn}}, \citenamefont {{Kosmalski}},
  \citenamefont {{Kotak}}, \citenamefont {{Kovalev}}, \citenamefont
  {{Kreckel}}, \citenamefont {{Kripak}}, \citenamefont {{Krumpe}},
  \citenamefont {{Kuijken}}, \citenamefont {{Kunder}}, \citenamefont
  {{Kushniruk}}, \citenamefont {{Lam}}, \citenamefont {{Lamer}}, \citenamefont
  {{Laurent}}, \citenamefont {{Lawrence}}, \citenamefont {{Lehmitz}},
  \citenamefont {{Lemasle}}, \citenamefont {{Lewis}}, \citenamefont {{Li}},
  \citenamefont {{Lidman}}, \citenamefont {{Lind}}, \citenamefont {{Liske}},
  \citenamefont {{Lizon}}, \citenamefont {{Loveday}}, \citenamefont {{Ludwig}},
  \citenamefont {{McDermid}}, \citenamefont {{Maguire}}, \citenamefont
  {{Mainieri}}, \citenamefont {{Mali}}, \citenamefont {{Mandel}}, \citenamefont
  {{Mandel}}, \citenamefont {{Mannering}}, \citenamefont {{Martell}},
  \citenamefont {{Martinez Delgado}}, \citenamefont {{Matijevic}},
  \citenamefont {{McGregor}}, \citenamefont {{McMahon}}, \citenamefont
  {{McMillan}}, \citenamefont {{Mena}}, \citenamefont {{Merloni}},
  \citenamefont {{Meyer}}, \citenamefont {{Michel}}, \citenamefont {{Micheva}},
  \citenamefont {{Migniau}}, \citenamefont {{Minchev}}, \citenamefont
  {{Monari}}, \citenamefont {{Muller}}, \citenamefont {{Murphy}}, \citenamefont
  {{Muthukrishna}}, \citenamefont {{Nandra}}, \citenamefont {{Navarro}},
  \citenamefont {{Ness}}, \citenamefont {{Nichani}}, \citenamefont {{Nichol}},
  \citenamefont {{Nicklas}}, \citenamefont {{Niederhofer}}, \citenamefont
  {{Norberg}}, \citenamefont {{Obreschkow}}, \citenamefont {{Oliver}},
  \citenamefont {{Owers}}, \citenamefont {{Pai}}, \citenamefont {{Pankratow}},
  \citenamefont {{Parkinson}}, \citenamefont {{Paschke}}, \citenamefont
  {{Paterson}}, \citenamefont {{Pecontal}}, \citenamefont {{Parry}},
  \citenamefont {{Phillips}}, \citenamefont {{Pillepich}}, \citenamefont
  {{Pinard}}, \citenamefont {{Pirard}}, \citenamefont {{Piskunov}},
  \citenamefont {{Plank}}, \citenamefont {{Pl{\"u}schke}}, \citenamefont
  {{Pons}}, \citenamefont {{Popesso}}, \citenamefont {{Power}}, \citenamefont
  {{Pragt}}, \citenamefont {{Pramskiy}}, \citenamefont {{Pryer}}, \citenamefont
  {{Quattri}}, \citenamefont {{Queiroz}}, \citenamefont {{Quirrenbach}},
  \citenamefont {{Rahurkar}}, \citenamefont {{Raichoor}}, \citenamefont
  {{Ramstedt}}, \citenamefont {{Rau}}, \citenamefont {{Recio-Blanco}},
  \citenamefont {{Reiss}}, \citenamefont {{Renaud}}, \citenamefont {{Revaz}},
  \citenamefont {{Rhode}}, \citenamefont {{Richard}}, \citenamefont
  {{Richter}}, \citenamefont {{Rix}}, \citenamefont {{Robotham}}, \citenamefont
  {{Roelfsema}}, \citenamefont {{Romaniello}}, \citenamefont {{Rosario}},
  \citenamefont {{Rothmaier}}, \citenamefont {{Roukema}}, \citenamefont
  {{Ruchti}}, \citenamefont {{Rupprecht}}, \citenamefont {{Rybizki}},
  \citenamefont {{Ryde}}, \citenamefont {{Saar}}, \citenamefont {{Sadler}},
  \citenamefont {{Sahl{\'e}n}}, \citenamefont {{Salvato}}, \citenamefont
  {{Sassolas}}, \citenamefont {{Saunders}}, \citenamefont {{Saviauk}},
  \citenamefont {{Sbordone}}, \citenamefont {{Schmidt}}, \citenamefont
  {{Schnurr}}, \citenamefont {{Scholz}}, \citenamefont {{Schwope}},
  \citenamefont {{Seifert}}, \citenamefont {{Shanks}}, \citenamefont
  {{Sheinis}}, \citenamefont {{Sivov}}, \citenamefont {{Sk{\'u}lad{\'o}ttir}},
  \citenamefont {{Smartt}}, \citenamefont {{Smedley}}, \citenamefont {{Smith}},
  \citenamefont {{Smith}}, \citenamefont {{Sorce}}, \citenamefont {{Spitler}},
  \citenamefont {{Starkenburg}}, \citenamefont {{Steinmetz}}, \citenamefont
  {{Stilz}}, \citenamefont {{Storm}}, \citenamefont {{Sullivan}}, \citenamefont
  {{Sutherland}}, \citenamefont {{Swann}}, \citenamefont {{Tamone}},
  \citenamefont {{Taylor}}, \citenamefont {{Teillon}}, \citenamefont
  {{Tempel}}, \citenamefont {{ter Horst}}, \citenamefont {{Thi}}, \citenamefont
  {{Tolstoy}}, \citenamefont {{Trager}}, \citenamefont {{Traven}},
  \citenamefont {{Tremblay}}, \citenamefont {{Tresse}}, \citenamefont
  {{Valentini}}, \citenamefont {{van de Weygaert}}, \citenamefont {{van den
  Ancker}}, \citenamefont {{Veljanoski}}, \citenamefont {{Venkatesan}},
  \citenamefont {{Wagner}}, \citenamefont {{Wagner}}, \citenamefont
  {{Walcher}}, \citenamefont {{Waller}}, \citenamefont {{Walton}},
  \citenamefont {{Wang}}, \citenamefont {{Winkler}}, \citenamefont
  {{Wisotzki}}, \citenamefont {{Worley}}, \citenamefont {{Worseck}},
  \citenamefont {{Xiang}}, \citenamefont {{Xu}}, \citenamefont {{Yong}},
  \citenamefont {{Zhao}}, \citenamefont {{Zheng}}, \citenamefont {{Zscheyge}},\
  and\ \citenamefont {{Zucker}}}]{2019Msngr.175....3D}%
  \BibitemOpen
  \bibfield  {author} {\bibinfo {author} {\bibfnamefont {R.~S.}\ \bibnamefont
  {{de Jong}}}, \bibinfo {author} {\bibfnamefont {O.}~\bibnamefont {{Agertz}}},
  \bibinfo {author} {\bibfnamefont {A.~A.}\ \bibnamefont {{Berbel}}}, \bibinfo
  {author} {\bibfnamefont {J.}~\bibnamefont {{Aird}}}, \bibinfo {author}
  {\bibfnamefont {D.~A.}\ \bibnamefont {{Alexander}}}, \bibinfo {author}
  {\bibfnamefont {A.}~\bibnamefont {{Amarsi}}}, \bibinfo {author}
  {\bibfnamefont {F.}~\bibnamefont {{Anders}}}, \bibinfo {author}
  {\bibfnamefont {R.}~\bibnamefont {{Andrae}}}, \bibinfo {author}
  {\bibfnamefont {B.}~\bibnamefont {{Ansarinejad}}}, \bibinfo {author}
  {\bibfnamefont {W.}~\bibnamefont {{Ansorge}}}, \bibinfo {author}
  {\bibfnamefont {P.}~\bibnamefont {{Antilogus}}}, \bibinfo {author}
  {\bibfnamefont {H.}~\bibnamefont {{Anwand-Heerwart}}}, \bibinfo {author}
  {\bibfnamefont {A.}~\bibnamefont {{Arentsen}}}, \bibinfo {author}
  {\bibfnamefont {A.}~\bibnamefont {{Arnadottir}}}, \bibinfo {author}
  {\bibfnamefont {M.}~\bibnamefont {{Asplund}}}, \bibinfo {author}
  {\bibfnamefont {M.}~\bibnamefont {{Auger}}}, \bibinfo {author} {\bibfnamefont
  {N.}~\bibnamefont {{Azais}}}, \bibinfo {author} {\bibfnamefont
  {D.}~\bibnamefont {{Baade}}}, \bibinfo {author} {\bibfnamefont
  {G.}~\bibnamefont {{Baker}}}, \bibinfo {author} {\bibfnamefont
  {S.}~\bibnamefont {{Baker}}}, \bibinfo {author} {\bibfnamefont
  {E.}~\bibnamefont {{Balbinot}}}, \bibinfo {author} {\bibfnamefont {I.~K.}\
  \bibnamefont {{Baldry}}}, \bibinfo {author} {\bibfnamefont {M.}~\bibnamefont
  {{Banerji}}}, \bibinfo {author} {\bibfnamefont {S.}~\bibnamefont {{Barden}}},
  \bibinfo {author} {\bibfnamefont {P.}~\bibnamefont {{Barklem}}}, \bibinfo
  {author} {\bibfnamefont {E.}~\bibnamefont {{Barth{\'e}l{\'e}my-Mazot}}},
  \bibinfo {author} {\bibfnamefont {C.}~\bibnamefont {{Battistini}}}, \bibinfo
  {author} {\bibfnamefont {S.}~\bibnamefont {{Bauer}}}, \bibinfo {author}
  {\bibfnamefont {C.~P.~M.}\ \bibnamefont {{Bell}}}, \bibinfo {author}
  {\bibfnamefont {O.}~\bibnamefont {{Bellido-Tirado}}}, \bibinfo {author}
  {\bibfnamefont {S.}~\bibnamefont {{Bellstedt}}}, \bibinfo {author}
  {\bibfnamefont {V.}~\bibnamefont {{Belokurov}}}, \bibinfo {author}
  {\bibfnamefont {T.}~\bibnamefont {{Bensby}}}, \bibinfo {author}
  {\bibfnamefont {M.}~\bibnamefont {{Bergemann}}}, \bibinfo {author}
  {\bibfnamefont {J.~M.}\ \bibnamefont {{Bestenlehner}}}, \bibinfo {author}
  {\bibfnamefont {R.}~\bibnamefont {{Bielby}}}, \bibinfo {author}
  {\bibfnamefont {M.}~\bibnamefont {{Bilicki}}}, \bibinfo {author}
  {\bibfnamefont {C.}~\bibnamefont {{Blake}}}, \bibinfo {author} {\bibfnamefont
  {J.}~\bibnamefont {{Bland-Hawthorn}}}, \bibinfo {author} {\bibfnamefont
  {C.}~\bibnamefont {{Boeche}}}, \bibinfo {author} {\bibfnamefont
  {W.}~\bibnamefont {{Boland}}}, \bibinfo {author} {\bibfnamefont
  {T.}~\bibnamefont {{Boller}}}, \bibinfo {author} {\bibfnamefont
  {S.}~\bibnamefont {{Bongard}}}, \bibinfo {author} {\bibfnamefont
  {A.}~\bibnamefont {{Bongiorno}}}, \bibinfo {author} {\bibfnamefont
  {P.}~\bibnamefont {{Bonifacio}}}, \bibinfo {author} {\bibfnamefont
  {D.}~\bibnamefont {{Boudon}}}, \bibinfo {author} {\bibfnamefont
  {D.}~\bibnamefont {{Brooks}}}, \bibinfo {author} {\bibfnamefont {M.~J.~I.}\
  \bibnamefont {{Brown}}}, \bibinfo {author} {\bibfnamefont {R.}~\bibnamefont
  {{Brown}}}, \bibinfo {author} {\bibfnamefont {M.}~\bibnamefont
  {{Br{\"u}ggen}}}, \bibinfo {author} {\bibfnamefont {J.}~\bibnamefont
  {{Brynnel}}}, \bibinfo {author} {\bibfnamefont {J.}~\bibnamefont
  {{Brzeski}}}, \bibinfo {author} {\bibfnamefont {T.}~\bibnamefont
  {{Buchert}}}, \bibinfo {author} {\bibfnamefont {P.}~\bibnamefont
  {{Buschkamp}}}, \bibinfo {author} {\bibfnamefont {E.}~\bibnamefont
  {{Caffau}}}, \bibinfo {author} {\bibfnamefont {P.}~\bibnamefont
  {{Caillier}}}, \bibinfo {author} {\bibfnamefont {J.}~\bibnamefont
  {{Carrick}}}, \bibinfo {author} {\bibfnamefont {L.}~\bibnamefont
  {{Casagrande}}}, \bibinfo {author} {\bibfnamefont {S.}~\bibnamefont
  {{Case}}}, \bibinfo {author} {\bibfnamefont {A.}~\bibnamefont {{Casey}}},
  \bibinfo {author} {\bibfnamefont {I.}~\bibnamefont {{Cesarini}}}, \bibinfo
  {author} {\bibfnamefont {G.}~\bibnamefont {{Cescutti}}}, \bibinfo {author}
  {\bibfnamefont {D.}~\bibnamefont {{Chapuis}}}, \bibinfo {author}
  {\bibfnamefont {C.}~\bibnamefont {{Chiappini}}}, \bibinfo {author}
  {\bibfnamefont {M.}~\bibnamefont {{Childress}}}, \bibinfo {author}
  {\bibfnamefont {N.}~\bibnamefont {{Christlieb}}}, \bibinfo {author}
  {\bibfnamefont {R.}~\bibnamefont {{Church}}}, \bibinfo {author}
  {\bibfnamefont {M.~R.~L.}\ \bibnamefont {{Cioni}}}, \bibinfo {author}
  {\bibfnamefont {M.}~\bibnamefont {{Cluver}}}, \bibinfo {author}
  {\bibfnamefont {M.}~\bibnamefont {{Colless}}}, \bibinfo {author}
  {\bibfnamefont {T.}~\bibnamefont {{Collett}}}, \bibinfo {author}
  {\bibfnamefont {J.}~\bibnamefont {{Comparat}}}, \bibinfo {author}
  {\bibfnamefont {A.}~\bibnamefont {{Cooper}}}, \bibinfo {author}
  {\bibfnamefont {W.}~\bibnamefont {{Couch}}}, \bibinfo {author} {\bibfnamefont
  {F.}~\bibnamefont {{Courbin}}}, \bibinfo {author} {\bibfnamefont
  {S.}~\bibnamefont {{Croom}}}, \bibinfo {author} {\bibfnamefont
  {D.}~\bibnamefont {{Croton}}}, \bibinfo {author} {\bibfnamefont
  {E.}~\bibnamefont {{Daguis{\'e}}}}, \bibinfo {author} {\bibfnamefont
  {G.}~\bibnamefont {{Dalton}}}, \bibinfo {author} {\bibfnamefont {L.~J.~M.}\
  \bibnamefont {{Davies}}}, \bibinfo {author} {\bibfnamefont {T.}~\bibnamefont
  {{Davis}}}, \bibinfo {author} {\bibfnamefont {P.}~\bibnamefont {{de
  Laverny}}}, \bibinfo {author} {\bibfnamefont {A.}~\bibnamefont {{Deason}}},
  \bibinfo {author} {\bibfnamefont {F.}~\bibnamefont {{Dionies}}}, \bibinfo
  {author} {\bibfnamefont {K.}~\bibnamefont {{Disseau}}}, \bibinfo {author}
  {\bibfnamefont {P.}~\bibnamefont {{Doel}}}, \bibinfo {author} {\bibfnamefont
  {D.}~\bibnamefont {{D{\"o}scher}}}, \bibinfo {author} {\bibfnamefont {S.~P.}\
  \bibnamefont {{Driver}}}, \bibinfo {author} {\bibfnamefont {T.}~\bibnamefont
  {{Dwelly}}}, \bibinfo {author} {\bibfnamefont {D.}~\bibnamefont {{Eckert}}},
  \bibinfo {author} {\bibfnamefont {A.}~\bibnamefont {{Edge}}}, \bibinfo
  {author} {\bibfnamefont {B.}~\bibnamefont {{Edvardsson}}}, \bibinfo {author}
  {\bibfnamefont {D.~E.}\ \bibnamefont {{Youssoufi}}}, \bibinfo {author}
  {\bibfnamefont {A.}~\bibnamefont {{Elhaddad}}}, \bibinfo {author}
  {\bibfnamefont {H.}~\bibnamefont {{Enke}}}, \bibinfo {author} {\bibfnamefont
  {G.}~\bibnamefont {{Erfanianfar}}}, \bibinfo {author} {\bibfnamefont
  {T.}~\bibnamefont {{Farrell}}}, \bibinfo {author} {\bibfnamefont
  {T.}~\bibnamefont {{Fechner}}}, \bibinfo {author} {\bibfnamefont
  {C.}~\bibnamefont {{Feiz}}}, \bibinfo {author} {\bibfnamefont
  {S.}~\bibnamefont {{Feltzing}}}, \bibinfo {author} {\bibfnamefont
  {I.}~\bibnamefont {{Ferreras}}}, \bibinfo {author} {\bibfnamefont
  {D.}~\bibnamefont {{Feuerstein}}}, \bibinfo {author} {\bibfnamefont
  {D.}~\bibnamefont {{Feuillet}}}, \bibinfo {author} {\bibfnamefont
  {A.}~\bibnamefont {{Finoguenov}}}, \bibinfo {author} {\bibfnamefont
  {D.}~\bibnamefont {{Ford}}}, \bibinfo {author} {\bibfnamefont
  {S.}~\bibnamefont {{Fotopoulou}}}, \bibinfo {author} {\bibfnamefont
  {M.}~\bibnamefont {{Fouesneau}}}, \bibinfo {author} {\bibfnamefont
  {C.}~\bibnamefont {{Frenk}}}, \bibinfo {author} {\bibfnamefont
  {S.}~\bibnamefont {{Frey}}}, \bibinfo {author} {\bibfnamefont
  {W.}~\bibnamefont {{Gaessler}}}, \bibinfo {author} {\bibfnamefont
  {S.}~\bibnamefont {{Geier}}}, \bibinfo {author} {\bibfnamefont
  {N.}~\bibnamefont {{Gentile Fusillo}}}, \bibinfo {author} {\bibfnamefont
  {O.}~\bibnamefont {{Gerhard}}}, \bibinfo {author} {\bibfnamefont
  {T.}~\bibnamefont {{Giannantonio}}}, \bibinfo {author} {\bibfnamefont
  {D.}~\bibnamefont {{Giannone}}}, \bibinfo {author} {\bibfnamefont
  {B.}~\bibnamefont {{Gibson}}}, \bibinfo {author} {\bibfnamefont
  {P.}~\bibnamefont {{Gillingham}}}, \bibinfo {author} {\bibfnamefont
  {C.}~\bibnamefont {{Gonz{\'a}lez-Fern{\'a}ndez}}}, \bibinfo {author}
  {\bibfnamefont {E.}~\bibnamefont {{Gonzalez-Solares}}}, \bibinfo {author}
  {\bibfnamefont {S.}~\bibnamefont {{Gottloeber}}}, \bibinfo {author}
  {\bibfnamefont {A.}~\bibnamefont {{Gould}}}, \bibinfo {author} {\bibfnamefont
  {E.~K.}\ \bibnamefont {{Grebel}}}, \bibinfo {author} {\bibfnamefont
  {A.}~\bibnamefont {{Gueguen}}}, \bibinfo {author} {\bibfnamefont
  {G.}~\bibnamefont {{Guiglion}}}, \bibinfo {author} {\bibfnamefont
  {M.}~\bibnamefont {{Haehnelt}}}, \bibinfo {author} {\bibfnamefont
  {T.}~\bibnamefont {{Hahn}}}, \bibinfo {author} {\bibfnamefont {C.~J.}\
  \bibnamefont {{Hansen}}}, \bibinfo {author} {\bibfnamefont {H.}~\bibnamefont
  {{Hartman}}}, \bibinfo {author} {\bibfnamefont {K.}~\bibnamefont
  {{Hauptner}}}, \bibinfo {author} {\bibfnamefont {K.}~\bibnamefont
  {{Hawkins}}}, \bibinfo {author} {\bibfnamefont {D.}~\bibnamefont {{Haynes}}},
  \bibinfo {author} {\bibfnamefont {R.}~\bibnamefont {{Haynes}}}, \bibinfo
  {author} {\bibfnamefont {U.}~\bibnamefont {{Heiter}}}, \bibinfo {author}
  {\bibfnamefont {A.}~\bibnamefont {{Helmi}}}, \bibinfo {author} {\bibfnamefont
  {C.~H.}\ \bibnamefont {{Aguayo}}}, \bibinfo {author} {\bibfnamefont
  {P.}~\bibnamefont {{Hewett}}}, \bibinfo {author} {\bibfnamefont
  {S.}~\bibnamefont {{Hinton}}}, \bibinfo {author} {\bibfnamefont
  {D.}~\bibnamefont {{Hobbs}}}, \bibinfo {author} {\bibfnamefont
  {S.}~\bibnamefont {{Hoenig}}}, \bibinfo {author} {\bibfnamefont
  {D.}~\bibnamefont {{Hofman}}}, \bibinfo {author} {\bibfnamefont
  {I.}~\bibnamefont {{Hook}}}, \bibinfo {author} {\bibfnamefont
  {J.}~\bibnamefont {{Hopgood}}}, \bibinfo {author} {\bibfnamefont
  {A.}~\bibnamefont {{Hopkins}}}, \bibinfo {author} {\bibfnamefont
  {A.}~\bibnamefont {{Hourihane}}}, \bibinfo {author} {\bibfnamefont
  {L.}~\bibnamefont {{Howes}}}, \bibinfo {author} {\bibfnamefont
  {C.}~\bibnamefont {{Howlett}}}, \bibinfo {author} {\bibfnamefont
  {T.}~\bibnamefont {{Huet}}}, \bibinfo {author} {\bibfnamefont
  {M.}~\bibnamefont {{Irwin}}}, \bibinfo {author} {\bibfnamefont
  {O.}~\bibnamefont {{Iwert}}}, \bibinfo {author} {\bibfnamefont
  {P.}~\bibnamefont {{Jablonka}}}, \bibinfo {author} {\bibfnamefont
  {T.}~\bibnamefont {{Jahn}}}, \bibinfo {author} {\bibfnamefont
  {K.}~\bibnamefont {{Jahnke}}}, \bibinfo {author} {\bibfnamefont
  {A.}~\bibnamefont {{Jarno}}}, \bibinfo {author} {\bibfnamefont
  {S.}~\bibnamefont {{Jin}}}, \bibinfo {author} {\bibfnamefont
  {P.}~\bibnamefont {{Jofre}}}, \bibinfo {author} {\bibfnamefont
  {D.}~\bibnamefont {{Johl}}}, \bibinfo {author} {\bibfnamefont
  {D.}~\bibnamefont {{Jones}}}, \bibinfo {author} {\bibfnamefont
  {H.}~\bibnamefont {{J{\"o}nsson}}}, \bibinfo {author} {\bibfnamefont
  {C.}~\bibnamefont {{Jordan}}}, \bibinfo {author} {\bibfnamefont
  {I.}~\bibnamefont {{Karovicova}}}, \bibinfo {author} {\bibfnamefont
  {A.}~\bibnamefont {{Khalatyan}}}, \bibinfo {author} {\bibfnamefont
  {A.}~\bibnamefont {{Kelz}}}, \bibinfo {author} {\bibfnamefont
  {R.}~\bibnamefont {{Kennicutt}}}, \bibinfo {author} {\bibfnamefont
  {D.}~\bibnamefont {{King}}}, \bibinfo {author} {\bibfnamefont
  {F.}~\bibnamefont {{Kitaura}}}, \bibinfo {author} {\bibfnamefont
  {J.}~\bibnamefont {{Klar}}}, \bibinfo {author} {\bibfnamefont
  {U.}~\bibnamefont {{Klauser}}}, \bibinfo {author} {\bibfnamefont {J.~P.}\
  \bibnamefont {{Kneib}}}, \bibinfo {author} {\bibfnamefont {A.}~\bibnamefont
  {{Koch}}}, \bibinfo {author} {\bibfnamefont {S.}~\bibnamefont {{Koposov}}},
  \bibinfo {author} {\bibfnamefont {G.}~\bibnamefont {{Kordopatis}}}, \bibinfo
  {author} {\bibfnamefont {A.}~\bibnamefont {{Korn}}}, \bibinfo {author}
  {\bibfnamefont {J.}~\bibnamefont {{Kosmalski}}}, \bibinfo {author}
  {\bibfnamefont {R.}~\bibnamefont {{Kotak}}}, \bibinfo {author} {\bibfnamefont
  {M.}~\bibnamefont {{Kovalev}}}, \bibinfo {author} {\bibfnamefont
  {K.}~\bibnamefont {{Kreckel}}}, \bibinfo {author} {\bibfnamefont
  {Y.}~\bibnamefont {{Kripak}}}, \bibinfo {author} {\bibfnamefont
  {M.}~\bibnamefont {{Krumpe}}}, \bibinfo {author} {\bibfnamefont
  {K.}~\bibnamefont {{Kuijken}}}, \bibinfo {author} {\bibfnamefont
  {A.}~\bibnamefont {{Kunder}}}, \bibinfo {author} {\bibfnamefont
  {I.}~\bibnamefont {{Kushniruk}}}, \bibinfo {author} {\bibfnamefont {M.~I.}\
  \bibnamefont {{Lam}}}, \bibinfo {author} {\bibfnamefont {G.}~\bibnamefont
  {{Lamer}}}, \bibinfo {author} {\bibfnamefont {F.}~\bibnamefont {{Laurent}}},
  \bibinfo {author} {\bibfnamefont {J.}~\bibnamefont {{Lawrence}}}, \bibinfo
  {author} {\bibfnamefont {M.}~\bibnamefont {{Lehmitz}}}, \bibinfo {author}
  {\bibfnamefont {B.}~\bibnamefont {{Lemasle}}}, \bibinfo {author}
  {\bibfnamefont {J.}~\bibnamefont {{Lewis}}}, \bibinfo {author} {\bibfnamefont
  {B.}~\bibnamefont {{Li}}}, \bibinfo {author} {\bibfnamefont {C.}~\bibnamefont
  {{Lidman}}}, \bibinfo {author} {\bibfnamefont {K.}~\bibnamefont {{Lind}}},
  \bibinfo {author} {\bibfnamefont {J.}~\bibnamefont {{Liske}}}, \bibinfo
  {author} {\bibfnamefont {J.~L.}\ \bibnamefont {{Lizon}}}, \bibinfo {author}
  {\bibfnamefont {J.}~\bibnamefont {{Loveday}}}, \bibinfo {author}
  {\bibfnamefont {H.~G.}\ \bibnamefont {{Ludwig}}}, \bibinfo {author}
  {\bibfnamefont {R.~M.}\ \bibnamefont {{McDermid}}}, \bibinfo {author}
  {\bibfnamefont {K.}~\bibnamefont {{Maguire}}}, \bibinfo {author}
  {\bibfnamefont {V.}~\bibnamefont {{Mainieri}}}, \bibinfo {author}
  {\bibfnamefont {S.}~\bibnamefont {{Mali}}}, \bibinfo {author} {\bibfnamefont
  {H.}~\bibnamefont {{Mandel}}}, \bibinfo {author} {\bibfnamefont
  {K.}~\bibnamefont {{Mandel}}}, \bibinfo {author} {\bibfnamefont
  {L.}~\bibnamefont {{Mannering}}}, \bibinfo {author} {\bibfnamefont
  {S.}~\bibnamefont {{Martell}}}, \bibinfo {author} {\bibfnamefont
  {D.}~\bibnamefont {{Martinez Delgado}}}, \bibinfo {author} {\bibfnamefont
  {G.}~\bibnamefont {{Matijevic}}}, \bibinfo {author} {\bibfnamefont
  {H.}~\bibnamefont {{McGregor}}}, \bibinfo {author} {\bibfnamefont
  {R.}~\bibnamefont {{McMahon}}}, \bibinfo {author} {\bibfnamefont
  {P.}~\bibnamefont {{McMillan}}}, \bibinfo {author} {\bibfnamefont
  {O.}~\bibnamefont {{Mena}}}, \bibinfo {author} {\bibfnamefont
  {A.}~\bibnamefont {{Merloni}}}, \bibinfo {author} {\bibfnamefont {M.~J.}\
  \bibnamefont {{Meyer}}}, \bibinfo {author} {\bibfnamefont {C.}~\bibnamefont
  {{Michel}}}, \bibinfo {author} {\bibfnamefont {G.}~\bibnamefont {{Micheva}}},
  \bibinfo {author} {\bibfnamefont {J.~E.}\ \bibnamefont {{Migniau}}}, \bibinfo
  {author} {\bibfnamefont {I.}~\bibnamefont {{Minchev}}}, \bibinfo {author}
  {\bibfnamefont {G.}~\bibnamefont {{Monari}}}, \bibinfo {author}
  {\bibfnamefont {R.}~\bibnamefont {{Muller}}}, \bibinfo {author}
  {\bibfnamefont {D.}~\bibnamefont {{Murphy}}}, \bibinfo {author}
  {\bibfnamefont {D.}~\bibnamefont {{Muthukrishna}}}, \bibinfo {author}
  {\bibfnamefont {K.}~\bibnamefont {{Nandra}}}, \bibinfo {author}
  {\bibfnamefont {R.}~\bibnamefont {{Navarro}}}, \bibinfo {author}
  {\bibfnamefont {M.}~\bibnamefont {{Ness}}}, \bibinfo {author} {\bibfnamefont
  {V.}~\bibnamefont {{Nichani}}}, \bibinfo {author} {\bibfnamefont
  {R.}~\bibnamefont {{Nichol}}}, \bibinfo {author} {\bibfnamefont
  {H.}~\bibnamefont {{Nicklas}}}, \bibinfo {author} {\bibfnamefont
  {F.}~\bibnamefont {{Niederhofer}}}, \bibinfo {author} {\bibfnamefont
  {P.}~\bibnamefont {{Norberg}}}, \bibinfo {author} {\bibfnamefont
  {D.}~\bibnamefont {{Obreschkow}}}, \bibinfo {author} {\bibfnamefont
  {S.}~\bibnamefont {{Oliver}}}, \bibinfo {author} {\bibfnamefont
  {M.}~\bibnamefont {{Owers}}}, \bibinfo {author} {\bibfnamefont
  {N.}~\bibnamefont {{Pai}}}, \bibinfo {author} {\bibfnamefont
  {S.}~\bibnamefont {{Pankratow}}}, \bibinfo {author} {\bibfnamefont
  {D.}~\bibnamefont {{Parkinson}}}, \bibinfo {author} {\bibfnamefont
  {J.}~\bibnamefont {{Paschke}}}, \bibinfo {author} {\bibfnamefont
  {R.}~\bibnamefont {{Paterson}}}, \bibinfo {author} {\bibfnamefont
  {A.}~\bibnamefont {{Pecontal}}}, \bibinfo {author} {\bibfnamefont
  {I.}~\bibnamefont {{Parry}}}, \bibinfo {author} {\bibfnamefont
  {D.}~\bibnamefont {{Phillips}}}, \bibinfo {author} {\bibfnamefont
  {A.}~\bibnamefont {{Pillepich}}}, \bibinfo {author} {\bibfnamefont
  {L.}~\bibnamefont {{Pinard}}}, \bibinfo {author} {\bibfnamefont
  {J.}~\bibnamefont {{Pirard}}}, \bibinfo {author} {\bibfnamefont
  {N.}~\bibnamefont {{Piskunov}}}, \bibinfo {author} {\bibfnamefont
  {V.}~\bibnamefont {{Plank}}}, \bibinfo {author} {\bibfnamefont
  {D.}~\bibnamefont {{Pl{\"u}schke}}}, \bibinfo {author} {\bibfnamefont
  {E.}~\bibnamefont {{Pons}}}, \bibinfo {author} {\bibfnamefont
  {P.}~\bibnamefont {{Popesso}}}, \bibinfo {author} {\bibfnamefont
  {C.}~\bibnamefont {{Power}}}, \bibinfo {author} {\bibfnamefont
  {J.}~\bibnamefont {{Pragt}}}, \bibinfo {author} {\bibfnamefont
  {A.}~\bibnamefont {{Pramskiy}}}, \bibinfo {author} {\bibfnamefont
  {D.}~\bibnamefont {{Pryer}}}, \bibinfo {author} {\bibfnamefont
  {M.}~\bibnamefont {{Quattri}}}, \bibinfo {author} {\bibfnamefont {A.~B.
  d.~A.}\ \bibnamefont {{Queiroz}}}, \bibinfo {author} {\bibfnamefont
  {A.}~\bibnamefont {{Quirrenbach}}}, \bibinfo {author} {\bibfnamefont
  {S.}~\bibnamefont {{Rahurkar}}}, \bibinfo {author} {\bibfnamefont
  {A.}~\bibnamefont {{Raichoor}}}, \bibinfo {author} {\bibfnamefont
  {S.}~\bibnamefont {{Ramstedt}}}, \bibinfo {author} {\bibfnamefont
  {A.}~\bibnamefont {{Rau}}}, \bibinfo {author} {\bibfnamefont
  {A.}~\bibnamefont {{Recio-Blanco}}}, \bibinfo {author} {\bibfnamefont
  {R.}~\bibnamefont {{Reiss}}}, \bibinfo {author} {\bibfnamefont
  {F.}~\bibnamefont {{Renaud}}}, \bibinfo {author} {\bibfnamefont
  {Y.}~\bibnamefont {{Revaz}}}, \bibinfo {author} {\bibfnamefont
  {P.}~\bibnamefont {{Rhode}}}, \bibinfo {author} {\bibfnamefont
  {J.}~\bibnamefont {{Richard}}}, \bibinfo {author} {\bibfnamefont {A.~D.}\
  \bibnamefont {{Richter}}}, \bibinfo {author} {\bibfnamefont {H.~W.}\
  \bibnamefont {{Rix}}}, \bibinfo {author} {\bibfnamefont {A.~S.~G.}\
  \bibnamefont {{Robotham}}}, \bibinfo {author} {\bibfnamefont
  {R.}~\bibnamefont {{Roelfsema}}}, \bibinfo {author} {\bibfnamefont
  {M.}~\bibnamefont {{Romaniello}}}, \bibinfo {author} {\bibfnamefont
  {D.}~\bibnamefont {{Rosario}}}, \bibinfo {author} {\bibfnamefont
  {F.}~\bibnamefont {{Rothmaier}}}, \bibinfo {author} {\bibfnamefont
  {B.}~\bibnamefont {{Roukema}}}, \bibinfo {author} {\bibfnamefont
  {G.}~\bibnamefont {{Ruchti}}}, \bibinfo {author} {\bibfnamefont
  {G.}~\bibnamefont {{Rupprecht}}}, \bibinfo {author} {\bibfnamefont
  {J.}~\bibnamefont {{Rybizki}}}, \bibinfo {author} {\bibfnamefont
  {N.}~\bibnamefont {{Ryde}}}, \bibinfo {author} {\bibfnamefont
  {A.}~\bibnamefont {{Saar}}}, \bibinfo {author} {\bibfnamefont
  {E.}~\bibnamefont {{Sadler}}}, \bibinfo {author} {\bibfnamefont
  {M.}~\bibnamefont {{Sahl{\'e}n}}}, \bibinfo {author} {\bibfnamefont
  {M.}~\bibnamefont {{Salvato}}}, \bibinfo {author} {\bibfnamefont
  {B.}~\bibnamefont {{Sassolas}}}, \bibinfo {author} {\bibfnamefont
  {W.}~\bibnamefont {{Saunders}}}, \bibinfo {author} {\bibfnamefont
  {A.}~\bibnamefont {{Saviauk}}}, \bibinfo {author} {\bibfnamefont
  {L.}~\bibnamefont {{Sbordone}}}, \bibinfo {author} {\bibfnamefont
  {T.}~\bibnamefont {{Schmidt}}}, \bibinfo {author} {\bibfnamefont
  {O.}~\bibnamefont {{Schnurr}}}, \bibinfo {author} {\bibfnamefont {R.~D.}\
  \bibnamefont {{Scholz}}}, \bibinfo {author} {\bibfnamefont {A.}~\bibnamefont
  {{Schwope}}}, \bibinfo {author} {\bibfnamefont {W.}~\bibnamefont
  {{Seifert}}}, \bibinfo {author} {\bibfnamefont {T.}~\bibnamefont {{Shanks}}},
  \bibinfo {author} {\bibfnamefont {A.}~\bibnamefont {{Sheinis}}}, \bibinfo
  {author} {\bibfnamefont {T.}~\bibnamefont {{Sivov}}}, \bibinfo {author}
  {\bibfnamefont {{\'A}.}~\bibnamefont {{Sk{\'u}lad{\'o}ttir}}}, \bibinfo
  {author} {\bibfnamefont {S.}~\bibnamefont {{Smartt}}}, \bibinfo {author}
  {\bibfnamefont {S.}~\bibnamefont {{Smedley}}}, \bibinfo {author}
  {\bibfnamefont {G.}~\bibnamefont {{Smith}}}, \bibinfo {author} {\bibfnamefont
  {R.}~\bibnamefont {{Smith}}}, \bibinfo {author} {\bibfnamefont
  {J.}~\bibnamefont {{Sorce}}}, \bibinfo {author} {\bibfnamefont
  {L.}~\bibnamefont {{Spitler}}}, \bibinfo {author} {\bibfnamefont
  {E.}~\bibnamefont {{Starkenburg}}}, \bibinfo {author} {\bibfnamefont
  {M.}~\bibnamefont {{Steinmetz}}}, \bibinfo {author} {\bibfnamefont
  {I.}~\bibnamefont {{Stilz}}}, \bibinfo {author} {\bibfnamefont
  {J.}~\bibnamefont {{Storm}}}, \bibinfo {author} {\bibfnamefont
  {M.}~\bibnamefont {{Sullivan}}}, \bibinfo {author} {\bibfnamefont
  {W.}~\bibnamefont {{Sutherland}}}, \bibinfo {author} {\bibfnamefont
  {E.}~\bibnamefont {{Swann}}}, \bibinfo {author} {\bibfnamefont
  {A.}~\bibnamefont {{Tamone}}}, \bibinfo {author} {\bibfnamefont {E.~N.}\
  \bibnamefont {{Taylor}}}, \bibinfo {author} {\bibfnamefont {J.}~\bibnamefont
  {{Teillon}}}, \bibinfo {author} {\bibfnamefont {E.}~\bibnamefont {{Tempel}}},
  \bibinfo {author} {\bibfnamefont {R.}~\bibnamefont {{ter Horst}}}, \bibinfo
  {author} {\bibfnamefont {W.~F.}\ \bibnamefont {{Thi}}}, \bibinfo {author}
  {\bibfnamefont {E.}~\bibnamefont {{Tolstoy}}}, \bibinfo {author}
  {\bibfnamefont {S.}~\bibnamefont {{Trager}}}, \bibinfo {author}
  {\bibfnamefont {G.}~\bibnamefont {{Traven}}}, \bibinfo {author}
  {\bibfnamefont {P.~E.}\ \bibnamefont {{Tremblay}}}, \bibinfo {author}
  {\bibfnamefont {L.}~\bibnamefont {{Tresse}}}, \bibinfo {author}
  {\bibfnamefont {M.}~\bibnamefont {{Valentini}}}, \bibinfo {author}
  {\bibfnamefont {R.}~\bibnamefont {{van de Weygaert}}}, \bibinfo {author}
  {\bibfnamefont {M.}~\bibnamefont {{van den Ancker}}}, \bibinfo {author}
  {\bibfnamefont {J.}~\bibnamefont {{Veljanoski}}}, \bibinfo {author}
  {\bibfnamefont {S.}~\bibnamefont {{Venkatesan}}}, \bibinfo {author}
  {\bibfnamefont {L.}~\bibnamefont {{Wagner}}}, \bibinfo {author}
  {\bibfnamefont {K.}~\bibnamefont {{Wagner}}}, \bibinfo {author}
  {\bibfnamefont {C.~J.}\ \bibnamefont {{Walcher}}}, \bibinfo {author}
  {\bibfnamefont {L.}~\bibnamefont {{Waller}}}, \bibinfo {author}
  {\bibfnamefont {N.}~\bibnamefont {{Walton}}}, \bibinfo {author}
  {\bibfnamefont {L.}~\bibnamefont {{Wang}}}, \bibinfo {author} {\bibfnamefont
  {R.}~\bibnamefont {{Winkler}}}, \bibinfo {author} {\bibfnamefont
  {L.}~\bibnamefont {{Wisotzki}}}, \bibinfo {author} {\bibfnamefont {C.~C.}\
  \bibnamefont {{Worley}}}, \bibinfo {author} {\bibfnamefont {G.}~\bibnamefont
  {{Worseck}}}, \bibinfo {author} {\bibfnamefont {M.}~\bibnamefont {{Xiang}}},
  \bibinfo {author} {\bibfnamefont {W.}~\bibnamefont {{Xu}}}, \bibinfo {author}
  {\bibfnamefont {D.}~\bibnamefont {{Yong}}}, \bibinfo {author} {\bibfnamefont
  {C.}~\bibnamefont {{Zhao}}}, \bibinfo {author} {\bibfnamefont
  {J.}~\bibnamefont {{Zheng}}}, \bibinfo {author} {\bibfnamefont
  {F.}~\bibnamefont {{Zscheyge}}},\ and\ \bibinfo {author} {\bibfnamefont
  {D.}~\bibnamefont {{Zucker}}},\ }\bibfield  {title} {\bibinfo {title}
  {{4MOST: Project overview and information for the First Call for
  Proposals}},\ }\href {https://doi.org/10.18727/0722-6691/5117} {\bibfield
  {journal} {\bibinfo  {journal} {The Messenger}\ }\textbf {\bibinfo {volume}
  {175}},\ \bibinfo {pages} {3} (\bibinfo {year} {2019})},\ \Eprint
  {https://arxiv.org/abs/1903.02464} {arXiv:1903.02464 [astro-ph.IM]}
  \BibitemShut {NoStop}%
\bibitem [{\citenamefont {{Minniti}}\ \emph {et~al.}(2010)\citenamefont
  {{Minniti}}, \citenamefont {{Lucas}}, \citenamefont {{Emerson}},
  \citenamefont {{Saito}}, \citenamefont {{Hempel}}, \citenamefont
  {{Pietrukowicz}}, \citenamefont {{Ahumada}}, \citenamefont {{Alonso}},
  \citenamefont {{Alonso-Garcia}}, \citenamefont {{Arias}}, \citenamefont
  {{Bandyopadhyay}}, \citenamefont {{Barb{\'a}}}, \citenamefont {{Barbuy}},
  \citenamefont {{Bedin}}, \citenamefont {{Bica}}, \citenamefont {{Borissova}},
  \citenamefont {{Bronfman}}, \citenamefont {{Carraro}}, \citenamefont
  {{Catelan}}, \citenamefont {{Clari{\'a}}}, \citenamefont {{Cross}},
  \citenamefont {{de Grijs}}, \citenamefont {{D{\'e}k{\'a}ny}}, \citenamefont
  {{Drew}}, \citenamefont {{Fari{\~n}a}}, \citenamefont {{Feinstein}},
  \citenamefont {{Fern{\'a}ndez Laj{\'u}s}}, \citenamefont {{Gamen}},
  \citenamefont {{Geisler}}, \citenamefont {{Gieren}}, \citenamefont
  {{Goldman}}, \citenamefont {{Gonzalez}}, \citenamefont {{Gunthardt}},
  \citenamefont {{Gurovich}}, \citenamefont {{Hambly}}, \citenamefont
  {{Irwin}}, \citenamefont {{Ivanov}}, \citenamefont {{Jord{\'a}n}},
  \citenamefont {{Kerins}}, \citenamefont {{Kinemuchi}}, \citenamefont
  {{Kurtev}}, \citenamefont {{L{\'o}pez-Corredoira}}, \citenamefont
  {{Maccarone}}, \citenamefont {{Masetti}}, \citenamefont {{Merlo}},
  \citenamefont {{Messineo}}, \citenamefont {{Mirabel}}, \citenamefont
  {{Monaco}}, \citenamefont {{Morelli}}, \citenamefont {{Padilla}},
  \citenamefont {{Palma}}, \citenamefont {{Parisi}}, \citenamefont {{Pignata}},
  \citenamefont {{Rejkuba}}, \citenamefont {{Roman-Lopes}}, \citenamefont
  {{Sale}}, \citenamefont {{Schreiber}}, \citenamefont {{Schr{\"o}der}},
  \citenamefont {{Smith}}, \citenamefont {{}}, \citenamefont {{Soto}},
  \citenamefont {{Tamura}}, \citenamefont {{Tappert}}, \citenamefont
  {{Thompson}}, \citenamefont {{Toledo}}, \citenamefont {{Zoccali}},\ and\
  \citenamefont {{Pietrzynski}}}]{2010NewA...15..433M}%
  \BibitemOpen
  \bibfield  {author} {\bibinfo {author} {\bibfnamefont {D.}~\bibnamefont
  {{Minniti}}}, \bibinfo {author} {\bibfnamefont {P.~W.}\ \bibnamefont
  {{Lucas}}}, \bibinfo {author} {\bibfnamefont {J.~P.}\ \bibnamefont
  {{Emerson}}}, \bibinfo {author} {\bibfnamefont {R.~K.}\ \bibnamefont
  {{Saito}}}, \bibinfo {author} {\bibfnamefont {M.}~\bibnamefont {{Hempel}}},
  \bibinfo {author} {\bibfnamefont {P.}~\bibnamefont {{Pietrukowicz}}},
  \bibinfo {author} {\bibfnamefont {A.~V.}\ \bibnamefont {{Ahumada}}}, \bibinfo
  {author} {\bibfnamefont {M.~V.}\ \bibnamefont {{Alonso}}}, \bibinfo {author}
  {\bibfnamefont {J.}~\bibnamefont {{Alonso-Garcia}}}, \bibinfo {author}
  {\bibfnamefont {J.~I.}\ \bibnamefont {{Arias}}}, \bibinfo {author}
  {\bibfnamefont {R.~M.}\ \bibnamefont {{Bandyopadhyay}}}, \bibinfo {author}
  {\bibfnamefont {R.~H.}\ \bibnamefont {{Barb{\'a}}}}, \bibinfo {author}
  {\bibfnamefont {B.}~\bibnamefont {{Barbuy}}}, \bibinfo {author}
  {\bibfnamefont {L.~R.}\ \bibnamefont {{Bedin}}}, \bibinfo {author}
  {\bibfnamefont {E.}~\bibnamefont {{Bica}}}, \bibinfo {author} {\bibfnamefont
  {J.}~\bibnamefont {{Borissova}}}, \bibinfo {author} {\bibfnamefont
  {L.}~\bibnamefont {{Bronfman}}}, \bibinfo {author} {\bibfnamefont
  {G.}~\bibnamefont {{Carraro}}}, \bibinfo {author} {\bibfnamefont
  {M.}~\bibnamefont {{Catelan}}}, \bibinfo {author} {\bibfnamefont {J.~J.}\
  \bibnamefont {{Clari{\'a}}}}, \bibinfo {author} {\bibfnamefont
  {N.}~\bibnamefont {{Cross}}}, \bibinfo {author} {\bibfnamefont
  {R.}~\bibnamefont {{de Grijs}}}, \bibinfo {author} {\bibfnamefont
  {I.}~\bibnamefont {{D{\'e}k{\'a}ny}}}, \bibinfo {author} {\bibfnamefont
  {J.~E.}\ \bibnamefont {{Drew}}}, \bibinfo {author} {\bibfnamefont
  {C.}~\bibnamefont {{Fari{\~n}a}}}, \bibinfo {author} {\bibfnamefont
  {C.}~\bibnamefont {{Feinstein}}}, \bibinfo {author} {\bibfnamefont
  {E.}~\bibnamefont {{Fern{\'a}ndez Laj{\'u}s}}}, \bibinfo {author}
  {\bibfnamefont {R.~C.}\ \bibnamefont {{Gamen}}}, \bibinfo {author}
  {\bibfnamefont {D.}~\bibnamefont {{Geisler}}}, \bibinfo {author}
  {\bibfnamefont {W.}~\bibnamefont {{Gieren}}}, \bibinfo {author}
  {\bibfnamefont {B.}~\bibnamefont {{Goldman}}}, \bibinfo {author}
  {\bibfnamefont {O.~A.}\ \bibnamefont {{Gonzalez}}}, \bibinfo {author}
  {\bibfnamefont {G.}~\bibnamefont {{Gunthardt}}}, \bibinfo {author}
  {\bibfnamefont {S.}~\bibnamefont {{Gurovich}}}, \bibinfo {author}
  {\bibfnamefont {N.~C.}\ \bibnamefont {{Hambly}}}, \bibinfo {author}
  {\bibfnamefont {M.~J.}\ \bibnamefont {{Irwin}}}, \bibinfo {author}
  {\bibfnamefont {V.~D.}\ \bibnamefont {{Ivanov}}}, \bibinfo {author}
  {\bibfnamefont {A.}~\bibnamefont {{Jord{\'a}n}}}, \bibinfo {author}
  {\bibfnamefont {E.}~\bibnamefont {{Kerins}}}, \bibinfo {author}
  {\bibfnamefont {K.}~\bibnamefont {{Kinemuchi}}}, \bibinfo {author}
  {\bibfnamefont {R.}~\bibnamefont {{Kurtev}}}, \bibinfo {author}
  {\bibfnamefont {M.}~\bibnamefont {{L{\'o}pez-Corredoira}}}, \bibinfo {author}
  {\bibfnamefont {T.}~\bibnamefont {{Maccarone}}}, \bibinfo {author}
  {\bibfnamefont {N.}~\bibnamefont {{Masetti}}}, \bibinfo {author}
  {\bibfnamefont {D.}~\bibnamefont {{Merlo}}}, \bibinfo {author} {\bibfnamefont
  {M.}~\bibnamefont {{Messineo}}}, \bibinfo {author} {\bibfnamefont {I.~F.}\
  \bibnamefont {{Mirabel}}}, \bibinfo {author} {\bibfnamefont {L.}~\bibnamefont
  {{Monaco}}}, \bibinfo {author} {\bibfnamefont {L.}~\bibnamefont {{Morelli}}},
  \bibinfo {author} {\bibfnamefont {N.}~\bibnamefont {{Padilla}}}, \bibinfo
  {author} {\bibfnamefont {T.}~\bibnamefont {{Palma}}}, \bibinfo {author}
  {\bibfnamefont {M.~C.}\ \bibnamefont {{Parisi}}}, \bibinfo {author}
  {\bibfnamefont {G.}~\bibnamefont {{Pignata}}}, \bibinfo {author}
  {\bibfnamefont {M.}~\bibnamefont {{Rejkuba}}}, \bibinfo {author}
  {\bibfnamefont {A.}~\bibnamefont {{Roman-Lopes}}}, \bibinfo {author}
  {\bibfnamefont {S.~E.}\ \bibnamefont {{Sale}}}, \bibinfo {author}
  {\bibfnamefont {M.~R.}\ \bibnamefont {{Schreiber}}}, \bibinfo {author}
  {\bibfnamefont {A.~C.}\ \bibnamefont {{Schr{\"o}der}}}, \bibinfo {author}
  {\bibfnamefont {M.}~\bibnamefont {{Smith}}}, \bibinfo {author} {\bibfnamefont
  {J.}~\bibnamefont {{}}, \bibfnamefont {L.~Sodr{\'e}}}, \bibinfo {author}
  {\bibfnamefont {M.}~\bibnamefont {{Soto}}}, \bibinfo {author} {\bibfnamefont
  {M.}~\bibnamefont {{Tamura}}}, \bibinfo {author} {\bibfnamefont
  {C.}~\bibnamefont {{Tappert}}}, \bibinfo {author} {\bibfnamefont {M.~A.}\
  \bibnamefont {{Thompson}}}, \bibinfo {author} {\bibfnamefont
  {I.}~\bibnamefont {{Toledo}}}, \bibinfo {author} {\bibfnamefont
  {M.}~\bibnamefont {{Zoccali}}},\ and\ \bibinfo {author} {\bibfnamefont
  {G.}~\bibnamefont {{Pietrzynski}}},\ }\bibfield  {title} {\bibinfo {title}
  {{VISTA Variables in the Via Lactea (VVV): The public ESO near-IR variability
  survey of the Milky Way}},\ }\href
  {https://doi.org/10.1016/j.newast.2009.12.002} {\bibfield  {journal}
  {\bibinfo  {journal} {\na}\ }\textbf {\bibinfo {volume} {15}},\ \bibinfo
  {pages} {433} (\bibinfo {year} {2010})},\ \Eprint
  {https://arxiv.org/abs/0912.1056} {arXiv:0912.1056 [astro-ph.GA]}
  \BibitemShut {NoStop}%
\bibitem [{\citenamefont {{Cioni}}\ \emph {et~al.}(2011)\citenamefont
  {{Cioni}}, \citenamefont {{Clementini}}, \citenamefont {{Girardi}},
  \citenamefont {{Guandalini}}, \citenamefont {{Gullieuszik}}, \citenamefont
  {{Miszalski}}, \citenamefont {{Moretti}}, \citenamefont {{Ripepi}},
  \citenamefont {{Rubele}}, \citenamefont {{Bagheri}}, \citenamefont {{Bekki}},
  \citenamefont {{Cross}}, \citenamefont {{de Blok}}, \citenamefont {{de
  Grijs}}, \citenamefont {{Emerson}}, \citenamefont {{Evans}}, \citenamefont
  {{Gibson}}, \citenamefont {{Gonzales-Solares}}, \citenamefont
  {{Groenewegen}}, \citenamefont {{Irwin}}, \citenamefont {{Ivanov}},
  \citenamefont {{Lewis}}, \citenamefont {{Marconi}}, \citenamefont
  {{Marquette}}, \citenamefont {{Mastropietro}}, \citenamefont {{Moore}},
  \citenamefont {{Napiwotzki}}, \citenamefont {{Naylor}}, \citenamefont
  {{Oliveira}}, \citenamefont {{Read}}, \citenamefont {{Sutorius}},
  \citenamefont {{van Loon}}, \citenamefont {{Wilkinson}},\ and\ \citenamefont
  {{Wood}}}]{2011A&A...527A.116C}%
  \BibitemOpen
  \bibfield  {author} {\bibinfo {author} {\bibfnamefont {M.~R.~L.}\
  \bibnamefont {{Cioni}}}, \bibinfo {author} {\bibfnamefont {G.}~\bibnamefont
  {{Clementini}}}, \bibinfo {author} {\bibfnamefont {L.}~\bibnamefont
  {{Girardi}}}, \bibinfo {author} {\bibfnamefont {R.}~\bibnamefont
  {{Guandalini}}}, \bibinfo {author} {\bibfnamefont {M.}~\bibnamefont
  {{Gullieuszik}}}, \bibinfo {author} {\bibfnamefont {B.}~\bibnamefont
  {{Miszalski}}}, \bibinfo {author} {\bibfnamefont {M.~I.}\ \bibnamefont
  {{Moretti}}}, \bibinfo {author} {\bibfnamefont {V.}~\bibnamefont {{Ripepi}}},
  \bibinfo {author} {\bibfnamefont {S.}~\bibnamefont {{Rubele}}}, \bibinfo
  {author} {\bibfnamefont {G.}~\bibnamefont {{Bagheri}}}, \bibinfo {author}
  {\bibfnamefont {K.}~\bibnamefont {{Bekki}}}, \bibinfo {author} {\bibfnamefont
  {N.}~\bibnamefont {{Cross}}}, \bibinfo {author} {\bibfnamefont {W.~J.~G.}\
  \bibnamefont {{de Blok}}}, \bibinfo {author} {\bibfnamefont {R.}~\bibnamefont
  {{de Grijs}}}, \bibinfo {author} {\bibfnamefont {J.~P.}\ \bibnamefont
  {{Emerson}}}, \bibinfo {author} {\bibfnamefont {C.~J.}\ \bibnamefont
  {{Evans}}}, \bibinfo {author} {\bibfnamefont {B.}~\bibnamefont {{Gibson}}},
  \bibinfo {author} {\bibfnamefont {E.}~\bibnamefont {{Gonzales-Solares}}},
  \bibinfo {author} {\bibfnamefont {M.~A.~T.}\ \bibnamefont {{Groenewegen}}},
  \bibinfo {author} {\bibfnamefont {M.}~\bibnamefont {{Irwin}}}, \bibinfo
  {author} {\bibfnamefont {V.~D.}\ \bibnamefont {{Ivanov}}}, \bibinfo {author}
  {\bibfnamefont {J.}~\bibnamefont {{Lewis}}}, \bibinfo {author} {\bibfnamefont
  {M.}~\bibnamefont {{Marconi}}}, \bibinfo {author} {\bibfnamefont {J.~B.}\
  \bibnamefont {{Marquette}}}, \bibinfo {author} {\bibfnamefont
  {C.}~\bibnamefont {{Mastropietro}}}, \bibinfo {author} {\bibfnamefont
  {B.}~\bibnamefont {{Moore}}}, \bibinfo {author} {\bibfnamefont
  {R.}~\bibnamefont {{Napiwotzki}}}, \bibinfo {author} {\bibfnamefont
  {T.}~\bibnamefont {{Naylor}}}, \bibinfo {author} {\bibfnamefont {J.~M.}\
  \bibnamefont {{Oliveira}}}, \bibinfo {author} {\bibfnamefont
  {M.}~\bibnamefont {{Read}}}, \bibinfo {author} {\bibfnamefont
  {E.}~\bibnamefont {{Sutorius}}}, \bibinfo {author} {\bibfnamefont {J.~T.}\
  \bibnamefont {{van Loon}}}, \bibinfo {author} {\bibfnamefont {M.~I.}\
  \bibnamefont {{Wilkinson}}},\ and\ \bibinfo {author} {\bibfnamefont {P.~R.}\
  \bibnamefont {{Wood}}},\ }\bibfield  {title} {\bibinfo {title} {{The VMC
  survey. I. Strategy and first data}},\ }\href
  {https://doi.org/10.1051/0004-6361/201016137} {\bibfield  {journal} {\bibinfo
   {journal} {\aap}\ }\textbf {\bibinfo {volume} {527}},\ \bibinfo {eid} {A116}
  (\bibinfo {year} {2011})},\ \Eprint {https://arxiv.org/abs/1012.5193}
  {arXiv:1012.5193 [astro-ph.CO]} \BibitemShut {NoStop}%
\bibitem [{\citenamefont {{Alcock}}\ \emph {et~al.}(2000)\citenamefont
  {{Alcock}}, \citenamefont {{Allsman}}, \citenamefont {{Alves}}, \citenamefont
  {{Axelrod}}, \citenamefont {{Becker}}, \citenamefont {{Bennett}},
  \citenamefont {{Cook}}, \citenamefont {{Dalal}}, \citenamefont {{Drake}},
  \citenamefont {{Freeman}}, \citenamefont {{Geha}}, \citenamefont {{Griest}},
  \citenamefont {{Lehner}}, \citenamefont {{Marshall}}, \citenamefont
  {{Minniti}}, \citenamefont {{Nelson}}, \citenamefont {{Peterson}},
  \citenamefont {{Popowski}}, \citenamefont {{Pratt}}, \citenamefont {{Quinn}},
  \citenamefont {{Stubbs}}, \citenamefont {{Sutherland}}, \citenamefont
  {{Tomaney}}, \citenamefont {{Vandehei}},\ and\ \citenamefont
  {{Welch}}}]{2000ApJ...542..281A}%
  \BibitemOpen
  \bibfield  {author} {\bibinfo {author} {\bibfnamefont {C.}~\bibnamefont
  {{Alcock}}}, \bibinfo {author} {\bibfnamefont {R.~A.}\ \bibnamefont
  {{Allsman}}}, \bibinfo {author} {\bibfnamefont {D.~R.}\ \bibnamefont
  {{Alves}}}, \bibinfo {author} {\bibfnamefont {T.~S.}\ \bibnamefont
  {{Axelrod}}}, \bibinfo {author} {\bibfnamefont {A.~C.}\ \bibnamefont
  {{Becker}}}, \bibinfo {author} {\bibfnamefont {D.~P.}\ \bibnamefont
  {{Bennett}}}, \bibinfo {author} {\bibfnamefont {K.~H.}\ \bibnamefont
  {{Cook}}}, \bibinfo {author} {\bibfnamefont {N.}~\bibnamefont {{Dalal}}},
  \bibinfo {author} {\bibfnamefont {A.~J.}\ \bibnamefont {{Drake}}}, \bibinfo
  {author} {\bibfnamefont {K.~C.}\ \bibnamefont {{Freeman}}}, \bibinfo {author}
  {\bibfnamefont {M.}~\bibnamefont {{Geha}}}, \bibinfo {author} {\bibfnamefont
  {K.}~\bibnamefont {{Griest}}}, \bibinfo {author} {\bibfnamefont {M.~J.}\
  \bibnamefont {{Lehner}}}, \bibinfo {author} {\bibfnamefont {S.~L.}\
  \bibnamefont {{Marshall}}}, \bibinfo {author} {\bibfnamefont
  {D.}~\bibnamefont {{Minniti}}}, \bibinfo {author} {\bibfnamefont {C.~A.}\
  \bibnamefont {{Nelson}}}, \bibinfo {author} {\bibfnamefont {B.~A.}\
  \bibnamefont {{Peterson}}}, \bibinfo {author} {\bibfnamefont
  {P.}~\bibnamefont {{Popowski}}}, \bibinfo {author} {\bibfnamefont {M.~R.}\
  \bibnamefont {{Pratt}}}, \bibinfo {author} {\bibfnamefont {P.~J.}\
  \bibnamefont {{Quinn}}}, \bibinfo {author} {\bibfnamefont {C.~W.}\
  \bibnamefont {{Stubbs}}}, \bibinfo {author} {\bibfnamefont {W.}~\bibnamefont
  {{Sutherland}}}, \bibinfo {author} {\bibfnamefont {A.~B.}\ \bibnamefont
  {{Tomaney}}}, \bibinfo {author} {\bibfnamefont {T.}~\bibnamefont
  {{Vandehei}}},\ and\ \bibinfo {author} {\bibfnamefont {D.}~\bibnamefont
  {{Welch}}},\ }\bibfield  {title} {\bibinfo {title} {{The MACHO Project:
  Microlensing Results from 5.7 Years of Large Magellanic Cloud
  Observations}},\ }\href {https://doi.org/10.1086/309512} {\bibfield
  {journal} {\bibinfo  {journal} {\apj}\ }\textbf {\bibinfo {volume} {542}},\
  \bibinfo {pages} {281} (\bibinfo {year} {2000})},\ \Eprint
  {https://arxiv.org/abs/astro-ph/0001272} {arXiv:astro-ph/0001272 [astro-ph]}
  \BibitemShut {NoStop}%
\bibitem [{\citenamefont {{Palanque-Delabrouille}}\ \emph
  {et~al.}(1998)\citenamefont {{Palanque-Delabrouille}}, \citenamefont
  {{Afonso}}, \citenamefont {{Albert}}, \citenamefont {{Andersen}},
  \citenamefont {{Ansari}}, \citenamefont {{Aubourg}}, \citenamefont
  {{Bareyre}}, \citenamefont {{Bauer}}, \citenamefont {{Beaulieu}},
  \citenamefont {{Bouquet}}, \citenamefont {{Char}}, \citenamefont {{Charlot}},
  \citenamefont {{Couchot}}, \citenamefont {{Coutures}}, \citenamefont
  {{Derue}}, \citenamefont {{Ferlet}}, \citenamefont {{Glicenstein}},
  \citenamefont {{Goldman}}, \citenamefont {{Gould}}, \citenamefont {{Graff}},
  \citenamefont {{Gros}}, \citenamefont {{Haissinski}}, \citenamefont
  {{Hamilton}}, \citenamefont {{Hardin}}, \citenamefont {{de Kat}},
  \citenamefont {{Lesquoy}}, \citenamefont {{Loup}}, \citenamefont
  {{Magneville}}, \citenamefont {{Mansoux}}, \citenamefont {{Marquette}},
  \citenamefont {{Maurice}}, \citenamefont {{Milsztajn}}, \citenamefont
  {{Moniez}}, \citenamefont {{Perdereau}}, \citenamefont {{Prevot}},
  \citenamefont {{Renault}}, \citenamefont {{Rich}}, \citenamefont {{Spiro}},
  \citenamefont {{Vidal-Madjar}}, \citenamefont {{Vigroux}}, \citenamefont
  {{Zylberajch}},\ and\ \citenamefont {{EROS
  Collaboration}}}]{1998A&A...332....1P}%
  \BibitemOpen
  \bibfield  {author} {\bibinfo {author} {\bibfnamefont {N.}~\bibnamefont
  {{Palanque-Delabrouille}}}, \bibinfo {author} {\bibfnamefont
  {C.}~\bibnamefont {{Afonso}}}, \bibinfo {author} {\bibfnamefont {J.~N.}\
  \bibnamefont {{Albert}}}, \bibinfo {author} {\bibfnamefont {J.}~\bibnamefont
  {{Andersen}}}, \bibinfo {author} {\bibfnamefont {R.}~\bibnamefont
  {{Ansari}}}, \bibinfo {author} {\bibfnamefont {E.}~\bibnamefont {{Aubourg}}},
  \bibinfo {author} {\bibfnamefont {P.}~\bibnamefont {{Bareyre}}}, \bibinfo
  {author} {\bibfnamefont {F.}~\bibnamefont {{Bauer}}}, \bibinfo {author}
  {\bibfnamefont {J.~P.}\ \bibnamefont {{Beaulieu}}}, \bibinfo {author}
  {\bibfnamefont {A.}~\bibnamefont {{Bouquet}}}, \bibinfo {author}
  {\bibfnamefont {S.}~\bibnamefont {{Char}}}, \bibinfo {author} {\bibfnamefont
  {X.}~\bibnamefont {{Charlot}}}, \bibinfo {author} {\bibfnamefont
  {F.}~\bibnamefont {{Couchot}}}, \bibinfo {author} {\bibfnamefont
  {C.}~\bibnamefont {{Coutures}}}, \bibinfo {author} {\bibfnamefont
  {F.}~\bibnamefont {{Derue}}}, \bibinfo {author} {\bibfnamefont
  {R.}~\bibnamefont {{Ferlet}}}, \bibinfo {author} {\bibfnamefont {J.~F.}\
  \bibnamefont {{Glicenstein}}}, \bibinfo {author} {\bibfnamefont
  {B.}~\bibnamefont {{Goldman}}}, \bibinfo {author} {\bibfnamefont
  {A.}~\bibnamefont {{Gould}}}, \bibinfo {author} {\bibfnamefont
  {D.}~\bibnamefont {{Graff}}}, \bibinfo {author} {\bibfnamefont
  {M.}~\bibnamefont {{Gros}}}, \bibinfo {author} {\bibfnamefont
  {J.}~\bibnamefont {{Haissinski}}}, \bibinfo {author} {\bibfnamefont {J.~C.}\
  \bibnamefont {{Hamilton}}}, \bibinfo {author} {\bibfnamefont
  {D.}~\bibnamefont {{Hardin}}}, \bibinfo {author} {\bibfnamefont
  {J.}~\bibnamefont {{de Kat}}}, \bibinfo {author} {\bibfnamefont
  {E.}~\bibnamefont {{Lesquoy}}}, \bibinfo {author} {\bibfnamefont
  {C.}~\bibnamefont {{Loup}}}, \bibinfo {author} {\bibfnamefont
  {C.}~\bibnamefont {{Magneville}}}, \bibinfo {author} {\bibfnamefont
  {B.}~\bibnamefont {{Mansoux}}}, \bibinfo {author} {\bibfnamefont {J.~B.}\
  \bibnamefont {{Marquette}}}, \bibinfo {author} {\bibfnamefont
  {E.}~\bibnamefont {{Maurice}}}, \bibinfo {author} {\bibfnamefont
  {A.}~\bibnamefont {{Milsztajn}}}, \bibinfo {author} {\bibfnamefont
  {M.}~\bibnamefont {{Moniez}}}, \bibinfo {author} {\bibfnamefont
  {O.}~\bibnamefont {{Perdereau}}}, \bibinfo {author} {\bibfnamefont
  {L.}~\bibnamefont {{Prevot}}}, \bibinfo {author} {\bibfnamefont
  {C.}~\bibnamefont {{Renault}}}, \bibinfo {author} {\bibfnamefont
  {J.}~\bibnamefont {{Rich}}}, \bibinfo {author} {\bibfnamefont
  {M.}~\bibnamefont {{Spiro}}}, \bibinfo {author} {\bibfnamefont
  {A.}~\bibnamefont {{Vidal-Madjar}}}, \bibinfo {author} {\bibfnamefont
  {L.}~\bibnamefont {{Vigroux}}}, \bibinfo {author} {\bibfnamefont
  {S.}~\bibnamefont {{Zylberajch}}},\ and\ \bibinfo {author} {\bibnamefont
  {{EROS Collaboration}}},\ }\bibfield  {title} {\bibinfo {title}
  {{Microlensing towards the Small Magellanic Cloud EROS 2 first year
  survey}},\ }\href@noop {} {\bibfield  {journal} {\bibinfo  {journal} {\aap}\
  }\textbf {\bibinfo {volume} {332}},\ \bibinfo {pages} {1} (\bibinfo {year}
  {1998})},\ \Eprint {https://arxiv.org/abs/astro-ph/9710194}
  {arXiv:astro-ph/9710194 [astro-ph]} \BibitemShut {NoStop}%
\bibitem [{\citenamefont {{Kochanek}}\ \emph {et~al.}(2017)\citenamefont
  {{Kochanek}}, \citenamefont {{Shappee}}, \citenamefont {{Stanek}},
  \citenamefont {{Holoien}}, \citenamefont {{Thompson}}, \citenamefont
  {{Prieto}}, \citenamefont {{Dong}}, \citenamefont {{Shields}}, \citenamefont
  {{Will}}, \citenamefont {{Britt}}, \citenamefont {{Perzanowski}},\ and\
  \citenamefont {{Pojma{\'n}ski}}}]{2017PASP..129j4502K}%
  \BibitemOpen
  \bibfield  {author} {\bibinfo {author} {\bibfnamefont {C.~S.}\ \bibnamefont
  {{Kochanek}}}, \bibinfo {author} {\bibfnamefont {B.~J.}\ \bibnamefont
  {{Shappee}}}, \bibinfo {author} {\bibfnamefont {K.~Z.}\ \bibnamefont
  {{Stanek}}}, \bibinfo {author} {\bibfnamefont {T.~W.~S.}\ \bibnamefont
  {{Holoien}}}, \bibinfo {author} {\bibfnamefont {T.~A.}\ \bibnamefont
  {{Thompson}}}, \bibinfo {author} {\bibfnamefont {J.~L.}\ \bibnamefont
  {{Prieto}}}, \bibinfo {author} {\bibfnamefont {S.}~\bibnamefont {{Dong}}},
  \bibinfo {author} {\bibfnamefont {J.~V.}\ \bibnamefont {{Shields}}}, \bibinfo
  {author} {\bibfnamefont {D.}~\bibnamefont {{Will}}}, \bibinfo {author}
  {\bibfnamefont {C.}~\bibnamefont {{Britt}}}, \bibinfo {author} {\bibfnamefont
  {D.}~\bibnamefont {{Perzanowski}}},\ and\ \bibinfo {author} {\bibfnamefont
  {G.}~\bibnamefont {{Pojma{\'n}ski}}},\ }\bibfield  {title} {\bibinfo {title}
  {{The All-Sky Automated Survey for Supernovae (ASAS-SN) Light Curve Server
  v1.0}},\ }\href {https://doi.org/10.1088/1538-3873/aa80d9} {\bibfield
  {journal} {\bibinfo  {journal} {\pasp}\ }\textbf {\bibinfo {volume} {129}},\
  \bibinfo {pages} {104502} (\bibinfo {year} {2017})},\ \Eprint
  {https://arxiv.org/abs/1706.07060} {arXiv:1706.07060 [astro-ph.SR]}
  \BibitemShut {NoStop}%
\bibitem [{\citenamefont {{Ricker}}\ \emph {et~al.}(2015)\citenamefont
  {{Ricker}}, \citenamefont {{Winn}}, \citenamefont {{Vanderspek}},
  \citenamefont {{Latham}}, \citenamefont {{Bakos}}, \citenamefont {{Bean}},
  \citenamefont {{Berta-Thompson}}, \citenamefont {{Brown}}, \citenamefont
  {{Buchhave}}, \citenamefont {{Butler}}, \citenamefont {{Butler}},
  \citenamefont {{Chaplin}}, \citenamefont {{Charbonneau}}, \citenamefont
  {{Christensen-Dalsgaard}}, \citenamefont {{Clampin}}, \citenamefont
  {{Deming}}, \citenamefont {{Doty}}, \citenamefont {{De Lee}}, \citenamefont
  {{Dressing}}, \citenamefont {{Dunham}}, \citenamefont {{Endl}}, \citenamefont
  {{Fressin}}, \citenamefont {{Ge}}, \citenamefont {{Henning}}, \citenamefont
  {{Holman}}, \citenamefont {{Howard}}, \citenamefont {{Ida}}, \citenamefont
  {{Jenkins}}, \citenamefont {{Jernigan}}, \citenamefont {{Johnson}},
  \citenamefont {{Kaltenegger}}, \citenamefont {{Kawai}}, \citenamefont
  {{Kjeldsen}}, \citenamefont {{Laughlin}}, \citenamefont {{Levine}},
  \citenamefont {{Lin}}, \citenamefont {{Lissauer}}, \citenamefont
  {{MacQueen}}, \citenamefont {{Marcy}}, \citenamefont {{McCullough}},
  \citenamefont {{Morton}}, \citenamefont {{Narita}}, \citenamefont
  {{Paegert}}, \citenamefont {{Palle}}, \citenamefont {{Pepe}}, \citenamefont
  {{Pepper}}, \citenamefont {{Quirrenbach}}, \citenamefont {{Rinehart}},
  \citenamefont {{Sasselov}}, \citenamefont {{Sato}}, \citenamefont {{Seager}},
  \citenamefont {{Sozzetti}}, \citenamefont {{Stassun}}, \citenamefont
  {{Sullivan}}, \citenamefont {{Szentgyorgyi}}, \citenamefont {{Torres}},
  \citenamefont {{Udry}},\ and\ \citenamefont
  {{Villasenor}}}]{2015JATIS...1a4003R}%
  \BibitemOpen
  \bibfield  {author} {\bibinfo {author} {\bibfnamefont {G.~R.}\ \bibnamefont
  {{Ricker}}}, \bibinfo {author} {\bibfnamefont {J.~N.}\ \bibnamefont
  {{Winn}}}, \bibinfo {author} {\bibfnamefont {R.}~\bibnamefont
  {{Vanderspek}}}, \bibinfo {author} {\bibfnamefont {D.~W.}\ \bibnamefont
  {{Latham}}}, \bibinfo {author} {\bibfnamefont {G.~{\'A}.}\ \bibnamefont
  {{Bakos}}}, \bibinfo {author} {\bibfnamefont {J.~L.}\ \bibnamefont {{Bean}}},
  \bibinfo {author} {\bibfnamefont {Z.~K.}\ \bibnamefont {{Berta-Thompson}}},
  \bibinfo {author} {\bibfnamefont {T.~M.}\ \bibnamefont {{Brown}}}, \bibinfo
  {author} {\bibfnamefont {L.}~\bibnamefont {{Buchhave}}}, \bibinfo {author}
  {\bibfnamefont {N.~R.}\ \bibnamefont {{Butler}}}, \bibinfo {author}
  {\bibfnamefont {R.~P.}\ \bibnamefont {{Butler}}}, \bibinfo {author}
  {\bibfnamefont {W.~J.}\ \bibnamefont {{Chaplin}}}, \bibinfo {author}
  {\bibfnamefont {D.}~\bibnamefont {{Charbonneau}}}, \bibinfo {author}
  {\bibfnamefont {J.}~\bibnamefont {{Christensen-Dalsgaard}}}, \bibinfo
  {author} {\bibfnamefont {M.}~\bibnamefont {{Clampin}}}, \bibinfo {author}
  {\bibfnamefont {D.}~\bibnamefont {{Deming}}}, \bibinfo {author}
  {\bibfnamefont {J.}~\bibnamefont {{Doty}}}, \bibinfo {author} {\bibfnamefont
  {N.}~\bibnamefont {{De Lee}}}, \bibinfo {author} {\bibfnamefont
  {C.}~\bibnamefont {{Dressing}}}, \bibinfo {author} {\bibfnamefont {E.~W.}\
  \bibnamefont {{Dunham}}}, \bibinfo {author} {\bibfnamefont {M.}~\bibnamefont
  {{Endl}}}, \bibinfo {author} {\bibfnamefont {F.}~\bibnamefont {{Fressin}}},
  \bibinfo {author} {\bibfnamefont {J.}~\bibnamefont {{Ge}}}, \bibinfo {author}
  {\bibfnamefont {T.}~\bibnamefont {{Henning}}}, \bibinfo {author}
  {\bibfnamefont {M.~J.}\ \bibnamefont {{Holman}}}, \bibinfo {author}
  {\bibfnamefont {A.~W.}\ \bibnamefont {{Howard}}}, \bibinfo {author}
  {\bibfnamefont {S.}~\bibnamefont {{Ida}}}, \bibinfo {author} {\bibfnamefont
  {J.~M.}\ \bibnamefont {{Jenkins}}}, \bibinfo {author} {\bibfnamefont
  {G.}~\bibnamefont {{Jernigan}}}, \bibinfo {author} {\bibfnamefont {J.~A.}\
  \bibnamefont {{Johnson}}}, \bibinfo {author} {\bibfnamefont {L.}~\bibnamefont
  {{Kaltenegger}}}, \bibinfo {author} {\bibfnamefont {N.}~\bibnamefont
  {{Kawai}}}, \bibinfo {author} {\bibfnamefont {H.}~\bibnamefont {{Kjeldsen}}},
  \bibinfo {author} {\bibfnamefont {G.}~\bibnamefont {{Laughlin}}}, \bibinfo
  {author} {\bibfnamefont {A.~M.}\ \bibnamefont {{Levine}}}, \bibinfo {author}
  {\bibfnamefont {D.}~\bibnamefont {{Lin}}}, \bibinfo {author} {\bibfnamefont
  {J.~J.}\ \bibnamefont {{Lissauer}}}, \bibinfo {author} {\bibfnamefont
  {P.}~\bibnamefont {{MacQueen}}}, \bibinfo {author} {\bibfnamefont
  {G.}~\bibnamefont {{Marcy}}}, \bibinfo {author} {\bibfnamefont {P.~R.}\
  \bibnamefont {{McCullough}}}, \bibinfo {author} {\bibfnamefont {T.~D.}\
  \bibnamefont {{Morton}}}, \bibinfo {author} {\bibfnamefont {N.}~\bibnamefont
  {{Narita}}}, \bibinfo {author} {\bibfnamefont {M.}~\bibnamefont {{Paegert}}},
  \bibinfo {author} {\bibfnamefont {E.}~\bibnamefont {{Palle}}}, \bibinfo
  {author} {\bibfnamefont {F.}~\bibnamefont {{Pepe}}}, \bibinfo {author}
  {\bibfnamefont {J.}~\bibnamefont {{Pepper}}}, \bibinfo {author}
  {\bibfnamefont {A.}~\bibnamefont {{Quirrenbach}}}, \bibinfo {author}
  {\bibfnamefont {S.~A.}\ \bibnamefont {{Rinehart}}}, \bibinfo {author}
  {\bibfnamefont {D.}~\bibnamefont {{Sasselov}}}, \bibinfo {author}
  {\bibfnamefont {B.}~\bibnamefont {{Sato}}}, \bibinfo {author} {\bibfnamefont
  {S.}~\bibnamefont {{Seager}}}, \bibinfo {author} {\bibfnamefont
  {A.}~\bibnamefont {{Sozzetti}}}, \bibinfo {author} {\bibfnamefont {K.~G.}\
  \bibnamefont {{Stassun}}}, \bibinfo {author} {\bibfnamefont {P.}~\bibnamefont
  {{Sullivan}}}, \bibinfo {author} {\bibfnamefont {A.}~\bibnamefont
  {{Szentgyorgyi}}}, \bibinfo {author} {\bibfnamefont {G.}~\bibnamefont
  {{Torres}}}, \bibinfo {author} {\bibfnamefont {S.}~\bibnamefont {{Udry}}},\
  and\ \bibinfo {author} {\bibfnamefont {J.}~\bibnamefont {{Villasenor}}},\
  }\bibfield  {title} {\bibinfo {title} {{Transiting Exoplanet Survey Satellite
  (TESS)}},\ }\href {https://doi.org/10.1117/1.JATIS.1.1.014003} {\bibfield
  {journal} {\bibinfo  {journal} {Journal of Astronomical Telescopes,
  Instruments, and Systems}\ }\textbf {\bibinfo {volume} {1}},\ \bibinfo {eid}
  {014003} (\bibinfo {year} {2015})}\BibitemShut {NoStop}%
\bibitem [{\citenamefont {{Guseinov}}\ and\ \citenamefont
  {{Zel'dovich}}(1966)}]{1966SvA....10..251G}%
  \BibitemOpen
  \bibfield  {author} {\bibinfo {author} {\bibfnamefont {O.~K.}\ \bibnamefont
  {{Guseinov}}}\ and\ \bibinfo {author} {\bibfnamefont {Y.~B.}\ \bibnamefont
  {{Zel'dovich}}},\ }\bibfield  {title} {\bibinfo {title} {{Collapsed Stars in
  Binary Systems}},\ }\href@noop {} {\bibfield  {journal} {\bibinfo  {journal}
  {\sovast}\ }\textbf {\bibinfo {volume} {10}},\ \bibinfo {pages} {251}
  (\bibinfo {year} {1966})}\BibitemShut {NoStop}%
\bibitem [{\citenamefont {{Trimble}}\ and\ \citenamefont
  {{Thorne}}(1969)}]{1969ApJ...156.1013T}%
  \BibitemOpen
  \bibfield  {author} {\bibinfo {author} {\bibfnamefont {V.~L.}\ \bibnamefont
  {{Trimble}}}\ and\ \bibinfo {author} {\bibfnamefont {K.~S.}\ \bibnamefont
  {{Thorne}}},\ }\bibfield  {title} {\bibinfo {title} {{Spectroscopic Binaries
  and Collapsed Stars}},\ }\href {https://doi.org/10.1086/150032} {\bibfield
  {journal} {\bibinfo  {journal} {\apj}\ }\textbf {\bibinfo {volume} {156}},\
  \bibinfo {pages} {1013} (\bibinfo {year} {1969})}\BibitemShut {NoStop}%
\bibitem [{\citenamefont {{Timmes}}\ \emph {et~al.}(1996)\citenamefont
  {{Timmes}}, \citenamefont {{Woosley}},\ and\ \citenamefont
  {{Weaver}}}]{1996ApJ...457..834T}%
  \BibitemOpen
  \bibfield  {author} {\bibinfo {author} {\bibfnamefont {F.~X.}\ \bibnamefont
  {{Timmes}}}, \bibinfo {author} {\bibfnamefont {S.~E.}\ \bibnamefont
  {{Woosley}}},\ and\ \bibinfo {author} {\bibfnamefont {T.~A.}\ \bibnamefont
  {{Weaver}}},\ }\bibfield  {title} {\bibinfo {title} {{The Neutron Star and
  Black Hole Initial Mass Function}},\ }\href {https://doi.org/10.1086/176778}
  {\bibfield  {journal} {\bibinfo  {journal} {\apj}\ }\textbf {\bibinfo
  {volume} {457}},\ \bibinfo {pages} {834} (\bibinfo {year} {1996})},\ \Eprint
  {https://arxiv.org/abs/astro-ph/9510136} {arXiv:astro-ph/9510136 [astro-ph]}
  \BibitemShut {NoStop}%
\bibitem [{\citenamefont {{Casares}}\ \emph {et~al.}(2014)\citenamefont
  {{Casares}}, \citenamefont {{Negueruela}}, \citenamefont {{Rib{\'o}}},
  \citenamefont {{Ribas}}, \citenamefont {{Paredes}}, \citenamefont
  {{Herrero}},\ and\ \citenamefont
  {{Sim{\'o}n-D{\'\i}az}}}]{2014Natur.505..378C}%
  \BibitemOpen
  \bibfield  {author} {\bibinfo {author} {\bibfnamefont {J.}~\bibnamefont
  {{Casares}}}, \bibinfo {author} {\bibfnamefont {I.}~\bibnamefont
  {{Negueruela}}}, \bibinfo {author} {\bibfnamefont {M.}~\bibnamefont
  {{Rib{\'o}}}}, \bibinfo {author} {\bibfnamefont {I.}~\bibnamefont {{Ribas}}},
  \bibinfo {author} {\bibfnamefont {J.~M.}\ \bibnamefont {{Paredes}}}, \bibinfo
  {author} {\bibfnamefont {A.}~\bibnamefont {{Herrero}}},\ and\ \bibinfo
  {author} {\bibfnamefont {S.}~\bibnamefont {{Sim{\'o}n-D{\'\i}az}}},\
  }\bibfield  {title} {\bibinfo {title} {{A Be-type star with a black-hole
  companion}},\ }\href {https://doi.org/10.1038/nature12916} {\bibfield
  {journal} {\bibinfo  {journal} {\nat}\ }\textbf {\bibinfo {volume} {505}},\
  \bibinfo {pages} {378} (\bibinfo {year} {2014})},\ \Eprint
  {https://arxiv.org/abs/1401.3711} {arXiv:1401.3711 [astro-ph.SR]}
  \BibitemShut {NoStop}%
\bibitem [{\citenamefont {{Rivinius}}\ \emph {et~al.}(2022)\citenamefont
  {{Rivinius}}, \citenamefont {{Klement}}, \citenamefont {{Chojnowski}},
  \citenamefont {{Baade}}, \citenamefont {{Shepard}},\ and\ \citenamefont
  {{Hadrava}}}]{2022arXiv220812315R}%
  \BibitemOpen
  \bibfield  {author} {\bibinfo {author} {\bibfnamefont {T.}~\bibnamefont
  {{Rivinius}}}, \bibinfo {author} {\bibfnamefont {R.}~\bibnamefont
  {{Klement}}}, \bibinfo {author} {\bibfnamefont {S.~D.}\ \bibnamefont
  {{Chojnowski}}}, \bibinfo {author} {\bibfnamefont {D.}~\bibnamefont
  {{Baade}}}, \bibinfo {author} {\bibfnamefont {K.}~\bibnamefont {{Shepard}}},\
  and\ \bibinfo {author} {\bibfnamefont {P.}~\bibnamefont {{Hadrava}}},\
  }\bibfield  {title} {\bibinfo {title} {{MWC656: A Be+BH or a Be+sdO?}},\
  }\href {https://doi.org/10.48550/arXiv.2208.12315} {\bibfield  {journal}
  {\bibinfo  {journal} {arXiv e-prints}\ ,\ \bibinfo {eid} {arXiv:2208.12315}}
  (\bibinfo {year} {2022})},\ \Eprint {https://arxiv.org/abs/2208.12315}
  {arXiv:2208.12315 [astro-ph.SR]} \BibitemShut {NoStop}%
\bibitem [{\citenamefont {{Khokhlov}}\ \emph {et~al.}(2018)\citenamefont
  {{Khokhlov}}, \citenamefont {{Miroshnichenko}}, \citenamefont {{Zharikov}},
  \citenamefont {{Manset}}, \citenamefont {{Arkharov}}, \citenamefont
  {{Efimova}}, \citenamefont {{Klimanov}}, \citenamefont {{Larionov}},
  \citenamefont {{Kusakin}}, \citenamefont {{Kokumbaeva}}, \citenamefont
  {{Omarov}}, \citenamefont {{Kuratov}}, \citenamefont {{Kuratova}},
  \citenamefont {{Rudy}}, \citenamefont {{Laag}}, \citenamefont {{Crawford}},
  \citenamefont {{Swift}}, \citenamefont {{Puetter}}, \citenamefont {{Perry}},
  \citenamefont {{Chojnowski}}, \citenamefont {{Agishev}}, \citenamefont
  {{Caton}}, \citenamefont {{Hawkins}}, \citenamefont {{Smith}}, \citenamefont
  {{Reichart}}, \citenamefont {{Kouprianov}},\ and\ \citenamefont
  {{Haislip}}}]{2018ApJ...856..158K}%
  \BibitemOpen
  \bibfield  {author} {\bibinfo {author} {\bibfnamefont {S.~A.}\ \bibnamefont
  {{Khokhlov}}}, \bibinfo {author} {\bibfnamefont {A.~S.}\ \bibnamefont
  {{Miroshnichenko}}}, \bibinfo {author} {\bibfnamefont {S.~V.}\ \bibnamefont
  {{Zharikov}}}, \bibinfo {author} {\bibfnamefont {N.}~\bibnamefont
  {{Manset}}}, \bibinfo {author} {\bibfnamefont {A.~A.}\ \bibnamefont
  {{Arkharov}}}, \bibinfo {author} {\bibfnamefont {N.}~\bibnamefont
  {{Efimova}}}, \bibinfo {author} {\bibfnamefont {S.}~\bibnamefont
  {{Klimanov}}}, \bibinfo {author} {\bibfnamefont {V.~M.}\ \bibnamefont
  {{Larionov}}}, \bibinfo {author} {\bibfnamefont {A.~V.}\ \bibnamefont
  {{Kusakin}}}, \bibinfo {author} {\bibfnamefont {R.~I.}\ \bibnamefont
  {{Kokumbaeva}}}, \bibinfo {author} {\bibfnamefont {C.~T.}\ \bibnamefont
  {{Omarov}}}, \bibinfo {author} {\bibfnamefont {K.~S.}\ \bibnamefont
  {{Kuratov}}}, \bibinfo {author} {\bibfnamefont {A.~K.}\ \bibnamefont
  {{Kuratova}}}, \bibinfo {author} {\bibfnamefont {R.~J.}\ \bibnamefont
  {{Rudy}}}, \bibinfo {author} {\bibfnamefont {E.~A.}\ \bibnamefont {{Laag}}},
  \bibinfo {author} {\bibfnamefont {K.~B.}\ \bibnamefont {{Crawford}}},
  \bibinfo {author} {\bibfnamefont {T.~K.}\ \bibnamefont {{Swift}}}, \bibinfo
  {author} {\bibfnamefont {R.~C.}\ \bibnamefont {{Puetter}}}, \bibinfo {author}
  {\bibfnamefont {R.~B.}\ \bibnamefont {{Perry}}}, \bibinfo {author}
  {\bibfnamefont {S.~D.}\ \bibnamefont {{Chojnowski}}}, \bibinfo {author}
  {\bibfnamefont {A.}~\bibnamefont {{Agishev}}}, \bibinfo {author}
  {\bibfnamefont {D.~B.}\ \bibnamefont {{Caton}}}, \bibinfo {author}
  {\bibfnamefont {R.~L.}\ \bibnamefont {{Hawkins}}}, \bibinfo {author}
  {\bibfnamefont {A.~B.}\ \bibnamefont {{Smith}}}, \bibinfo {author}
  {\bibfnamefont {D.~E.}\ \bibnamefont {{Reichart}}}, \bibinfo {author}
  {\bibfnamefont {V.~V.}\ \bibnamefont {{Kouprianov}}},\ and\ \bibinfo {author}
  {\bibfnamefont {J.~B.}\ \bibnamefont {{Haislip}}},\ }\bibfield  {title}
  {\bibinfo {title} {{Toward Understanding the B[e] Phenomenon. VII. AS 386, a
  Single-lined Binary with a Candidate Black Hole Component}},\ }\href
  {https://doi.org/10.3847/1538-4357/aab49d} {\bibfield  {journal} {\bibinfo
  {journal} {\apj}\ }\textbf {\bibinfo {volume} {856}},\ \bibinfo {eid} {158}
  (\bibinfo {year} {2018})},\ \Eprint {https://arxiv.org/abs/1803.03892}
  {arXiv:1803.03892 [astro-ph.SR]} \BibitemShut {NoStop}%
\bibitem [{\citenamefont {{Thompson}}\ \emph {et~al.}(2019)\citenamefont
  {{Thompson}}, \citenamefont {{Kochanek}}, \citenamefont {{Stanek}},
  \citenamefont {{Badenes}}, \citenamefont {{Post}}, \citenamefont
  {{Jayasinghe}}, \citenamefont {{Latham}}, \citenamefont {{Bieryla}},
  \citenamefont {{Esquerdo}}, \citenamefont {{Berlind}}, \citenamefont
  {{Calkins}}, \citenamefont {{Tayar}}, \citenamefont {{Lindegren}},
  \citenamefont {{Johnson}}, \citenamefont {{Holoien}}, \citenamefont
  {{Auchettl}},\ and\ \citenamefont {{Covey}}}]{2019Sci...366..637T}%
  \BibitemOpen
  \bibfield  {author} {\bibinfo {author} {\bibfnamefont {T.~A.}\ \bibnamefont
  {{Thompson}}}, \bibinfo {author} {\bibfnamefont {C.~S.}\ \bibnamefont
  {{Kochanek}}}, \bibinfo {author} {\bibfnamefont {K.~Z.}\ \bibnamefont
  {{Stanek}}}, \bibinfo {author} {\bibfnamefont {C.}~\bibnamefont {{Badenes}}},
  \bibinfo {author} {\bibfnamefont {R.~S.}\ \bibnamefont {{Post}}}, \bibinfo
  {author} {\bibfnamefont {T.}~\bibnamefont {{Jayasinghe}}}, \bibinfo {author}
  {\bibfnamefont {D.~W.}\ \bibnamefont {{Latham}}}, \bibinfo {author}
  {\bibfnamefont {A.}~\bibnamefont {{Bieryla}}}, \bibinfo {author}
  {\bibfnamefont {G.~A.}\ \bibnamefont {{Esquerdo}}}, \bibinfo {author}
  {\bibfnamefont {P.}~\bibnamefont {{Berlind}}}, \bibinfo {author}
  {\bibfnamefont {M.~L.}\ \bibnamefont {{Calkins}}}, \bibinfo {author}
  {\bibfnamefont {J.}~\bibnamefont {{Tayar}}}, \bibinfo {author} {\bibfnamefont
  {L.}~\bibnamefont {{Lindegren}}}, \bibinfo {author} {\bibfnamefont {J.~A.}\
  \bibnamefont {{Johnson}}}, \bibinfo {author} {\bibfnamefont {T.~W.~S.}\
  \bibnamefont {{Holoien}}}, \bibinfo {author} {\bibfnamefont {K.}~\bibnamefont
  {{Auchettl}}},\ and\ \bibinfo {author} {\bibfnamefont {K.}~\bibnamefont
  {{Covey}}},\ }\bibfield  {title} {\bibinfo {title} {{A noninteracting
  low-mass black hole-giant star binary system}},\ }\href
  {https://doi.org/10.1126/science.aau4005} {\bibfield  {journal} {\bibinfo
  {journal} {Science}\ }\textbf {\bibinfo {volume} {366}},\ \bibinfo {pages}
  {637} (\bibinfo {year} {2019})},\ \Eprint {https://arxiv.org/abs/1806.02751}
  {arXiv:1806.02751 [astro-ph.HE]} \BibitemShut {NoStop}%
\bibitem [{\citenamefont {{Farr}}\ \emph {et~al.}(2011)\citenamefont {{Farr}},
  \citenamefont {{Sravan}}, \citenamefont {{Cantrell}}, \citenamefont
  {{Kreidberg}}, \citenamefont {{Bailyn}}, \citenamefont {{Mandel}},\ and\
  \citenamefont {{Kalogera}}}]{2011ApJ...741..103F}%
  \BibitemOpen
  \bibfield  {author} {\bibinfo {author} {\bibfnamefont {W.~M.}\ \bibnamefont
  {{Farr}}}, \bibinfo {author} {\bibfnamefont {N.}~\bibnamefont {{Sravan}}},
  \bibinfo {author} {\bibfnamefont {A.}~\bibnamefont {{Cantrell}}}, \bibinfo
  {author} {\bibfnamefont {L.}~\bibnamefont {{Kreidberg}}}, \bibinfo {author}
  {\bibfnamefont {C.~D.}\ \bibnamefont {{Bailyn}}}, \bibinfo {author}
  {\bibfnamefont {I.}~\bibnamefont {{Mandel}}},\ and\ \bibinfo {author}
  {\bibfnamefont {V.}~\bibnamefont {{Kalogera}}},\ }\bibfield  {title}
  {\bibinfo {title} {{The Mass Distribution of Stellar-mass Black Holes}},\
  }\href {https://doi.org/10.1088/0004-637X/741/2/103} {\bibfield  {journal}
  {\bibinfo  {journal} {\apj}\ }\textbf {\bibinfo {volume} {741}},\ \bibinfo
  {eid} {103} (\bibinfo {year} {2011})},\ \Eprint
  {https://arxiv.org/abs/1011.1459} {arXiv:1011.1459 [astro-ph.GA]}
  \BibitemShut {NoStop}%
\bibitem [{\citenamefont {{van den Heuvel}}\ and\ \citenamefont
  {{Tauris}}(2020)}]{2020Sci...368.3282V}%
  \BibitemOpen
  \bibfield  {author} {\bibinfo {author} {\bibfnamefont {E.~P.~J.}\
  \bibnamefont {{van den Heuvel}}}\ and\ \bibinfo {author} {\bibfnamefont
  {T.~M.}\ \bibnamefont {{Tauris}}},\ }\bibfield  {title} {\bibinfo {title}
  {{Comment on {\textquotedblleft}A noninteracting low-mass black hole-giant
  star binary system{\textquotedblright}}},\ }\href
  {https://doi.org/10.1126/science.aba3282} {\bibfield  {journal} {\bibinfo
  {journal} {Science}\ }\textbf {\bibinfo {volume} {368}},\ \bibinfo {eid}
  {eaba3282} (\bibinfo {year} {2020})},\ \Eprint
  {https://arxiv.org/abs/2005.04896} {arXiv:2005.04896 [astro-ph.SR]}
  \BibitemShut {NoStop}%
\bibitem [{\citenamefont {{Liu}}\ \emph {et~al.}(2019)\citenamefont {{Liu}},
  \citenamefont {{Zhang}}, \citenamefont {{Howard}}, \citenamefont {{Bai}},
  \citenamefont {{Lu}}, \citenamefont {{Soria}}, \citenamefont {{Justham}},
  \citenamefont {{Li}}, \citenamefont {{Zheng}}, \citenamefont {{Wang}},
  \citenamefont {{Belczynski}}, \citenamefont {{Casares}}, \citenamefont
  {{Zhang}}, \citenamefont {{Yuan}}, \citenamefont {{Dong}}, \citenamefont
  {{Lei}}, \citenamefont {{Isaacson}}, \citenamefont {{Wang}}, \citenamefont
  {{Bai}}, \citenamefont {{Shao}}, \citenamefont {{Gao}}, \citenamefont
  {{Wang}}, \citenamefont {{Niu}}, \citenamefont {{Cui}}, \citenamefont
  {{Zheng}}, \citenamefont {{Mu}}, \citenamefont {{Zhang}}, \citenamefont
  {{Wang}}, \citenamefont {{Heger}}, \citenamefont {{Qi}}, \citenamefont
  {{Liao}}, \citenamefont {{Lattanzi}}, \citenamefont {{Gu}}, \citenamefont
  {{Wang}}, \citenamefont {{Wu}}, \citenamefont {{Shao}}, \citenamefont
  {{Shen}}, \citenamefont {{Wang}}, \citenamefont {{Bregman}}, \citenamefont
  {{Di Stefano}}, \citenamefont {{Liu}}, \citenamefont {{Han}}, \citenamefont
  {{Zhang}}, \citenamefont {{Wang}}, \citenamefont {{Ren}}, \citenamefont
  {{Zhang}}, \citenamefont {{Zhang}}, \citenamefont {{Wang}}, \citenamefont
  {{Cabrera-Lavers}}, \citenamefont {{Corradi}}, \citenamefont {{Rebolo}},
  \citenamefont {{Zhao}}, \citenamefont {{Zhao}}, \citenamefont {{Chu}},\ and\
  \citenamefont {{Cui}}}]{2019Natur.575..618L}%
  \BibitemOpen
  \bibfield  {author} {\bibinfo {author} {\bibfnamefont {J.}~\bibnamefont
  {{Liu}}}, \bibinfo {author} {\bibfnamefont {H.}~\bibnamefont {{Zhang}}},
  \bibinfo {author} {\bibfnamefont {A.~W.}\ \bibnamefont {{Howard}}}, \bibinfo
  {author} {\bibfnamefont {Z.}~\bibnamefont {{Bai}}}, \bibinfo {author}
  {\bibfnamefont {Y.}~\bibnamefont {{Lu}}}, \bibinfo {author} {\bibfnamefont
  {R.}~\bibnamefont {{Soria}}}, \bibinfo {author} {\bibfnamefont
  {S.}~\bibnamefont {{Justham}}}, \bibinfo {author} {\bibfnamefont
  {X.}~\bibnamefont {{Li}}}, \bibinfo {author} {\bibfnamefont {Z.}~\bibnamefont
  {{Zheng}}}, \bibinfo {author} {\bibfnamefont {T.}~\bibnamefont {{Wang}}},
  \bibinfo {author} {\bibfnamefont {K.}~\bibnamefont {{Belczynski}}}, \bibinfo
  {author} {\bibfnamefont {J.}~\bibnamefont {{Casares}}}, \bibinfo {author}
  {\bibfnamefont {W.}~\bibnamefont {{Zhang}}}, \bibinfo {author} {\bibfnamefont
  {H.}~\bibnamefont {{Yuan}}}, \bibinfo {author} {\bibfnamefont
  {Y.}~\bibnamefont {{Dong}}}, \bibinfo {author} {\bibfnamefont
  {Y.}~\bibnamefont {{Lei}}}, \bibinfo {author} {\bibfnamefont
  {H.}~\bibnamefont {{Isaacson}}}, \bibinfo {author} {\bibfnamefont
  {S.}~\bibnamefont {{Wang}}}, \bibinfo {author} {\bibfnamefont
  {Y.}~\bibnamefont {{Bai}}}, \bibinfo {author} {\bibfnamefont
  {Y.}~\bibnamefont {{Shao}}}, \bibinfo {author} {\bibfnamefont
  {Q.}~\bibnamefont {{Gao}}}, \bibinfo {author} {\bibfnamefont
  {Y.}~\bibnamefont {{Wang}}}, \bibinfo {author} {\bibfnamefont
  {Z.}~\bibnamefont {{Niu}}}, \bibinfo {author} {\bibfnamefont
  {K.}~\bibnamefont {{Cui}}}, \bibinfo {author} {\bibfnamefont
  {C.}~\bibnamefont {{Zheng}}}, \bibinfo {author} {\bibfnamefont
  {X.}~\bibnamefont {{Mu}}}, \bibinfo {author} {\bibfnamefont {L.}~\bibnamefont
  {{Zhang}}}, \bibinfo {author} {\bibfnamefont {W.}~\bibnamefont {{Wang}}},
  \bibinfo {author} {\bibfnamefont {A.}~\bibnamefont {{Heger}}}, \bibinfo
  {author} {\bibfnamefont {Z.}~\bibnamefont {{Qi}}}, \bibinfo {author}
  {\bibfnamefont {S.}~\bibnamefont {{Liao}}}, \bibinfo {author} {\bibfnamefont
  {M.}~\bibnamefont {{Lattanzi}}}, \bibinfo {author} {\bibfnamefont {W.-M.}\
  \bibnamefont {{Gu}}}, \bibinfo {author} {\bibfnamefont {J.}~\bibnamefont
  {{Wang}}}, \bibinfo {author} {\bibfnamefont {J.}~\bibnamefont {{Wu}}},
  \bibinfo {author} {\bibfnamefont {L.}~\bibnamefont {{Shao}}}, \bibinfo
  {author} {\bibfnamefont {R.}~\bibnamefont {{Shen}}}, \bibinfo {author}
  {\bibfnamefont {X.}~\bibnamefont {{Wang}}}, \bibinfo {author} {\bibfnamefont
  {J.}~\bibnamefont {{Bregman}}}, \bibinfo {author} {\bibfnamefont
  {R.}~\bibnamefont {{Di Stefano}}}, \bibinfo {author} {\bibfnamefont
  {Q.}~\bibnamefont {{Liu}}}, \bibinfo {author} {\bibfnamefont
  {Z.}~\bibnamefont {{Han}}}, \bibinfo {author} {\bibfnamefont
  {T.}~\bibnamefont {{Zhang}}}, \bibinfo {author} {\bibfnamefont
  {H.}~\bibnamefont {{Wang}}}, \bibinfo {author} {\bibfnamefont
  {J.}~\bibnamefont {{Ren}}}, \bibinfo {author} {\bibfnamefont
  {J.}~\bibnamefont {{Zhang}}}, \bibinfo {author} {\bibfnamefont
  {J.}~\bibnamefont {{Zhang}}}, \bibinfo {author} {\bibfnamefont
  {X.}~\bibnamefont {{Wang}}}, \bibinfo {author} {\bibfnamefont
  {A.}~\bibnamefont {{Cabrera-Lavers}}}, \bibinfo {author} {\bibfnamefont
  {R.}~\bibnamefont {{Corradi}}}, \bibinfo {author} {\bibfnamefont
  {R.}~\bibnamefont {{Rebolo}}}, \bibinfo {author} {\bibfnamefont
  {Y.}~\bibnamefont {{Zhao}}}, \bibinfo {author} {\bibfnamefont
  {G.}~\bibnamefont {{Zhao}}}, \bibinfo {author} {\bibfnamefont
  {Y.}~\bibnamefont {{Chu}}},\ and\ \bibinfo {author} {\bibfnamefont
  {X.}~\bibnamefont {{Cui}}},\ }\bibfield  {title} {\bibinfo {title} {{A wide
  star-black-hole binary system from radial-velocity measurements}},\ }\href
  {https://doi.org/10.1038/s41586-019-1766-2} {\bibfield  {journal} {\bibinfo
  {journal} {\nat}\ }\textbf {\bibinfo {volume} {575}},\ \bibinfo {pages} {618}
  (\bibinfo {year} {2019})},\ \Eprint {https://arxiv.org/abs/1911.11989}
  {arXiv:1911.11989 [astro-ph.SR]} \BibitemShut {NoStop}%
\bibitem [{\citenamefont {{Lennon}}\ \emph {et~al.}(2021)\citenamefont
  {{Lennon}}, \citenamefont {{Ma{\'\i}z Apell{\'a}niz}}, \citenamefont
  {{Irrgang}}, \citenamefont {{Bohlin}}, \citenamefont {{Deustua}},
  \citenamefont {{Dufton}}, \citenamefont {{Sim{\'o}n-D{\'\i}az}},
  \citenamefont {{Herrero}}, \citenamefont {{Casares}}, \citenamefont
  {{Mu{\~n}oz-Darias}}, \citenamefont {{Smartt}}, \citenamefont {{Gonz{\'a}lez
  Hern{\'a}ndez}},\ and\ \citenamefont {{de Burgos}}}]{2021A&A...649A.167L}%
  \BibitemOpen
  \bibfield  {author} {\bibinfo {author} {\bibfnamefont {D.~J.}\ \bibnamefont
  {{Lennon}}}, \bibinfo {author} {\bibfnamefont {J.}~\bibnamefont {{Ma{\'\i}z
  Apell{\'a}niz}}}, \bibinfo {author} {\bibfnamefont {A.}~\bibnamefont
  {{Irrgang}}}, \bibinfo {author} {\bibfnamefont {R.}~\bibnamefont {{Bohlin}}},
  \bibinfo {author} {\bibfnamefont {S.}~\bibnamefont {{Deustua}}}, \bibinfo
  {author} {\bibfnamefont {P.~L.}\ \bibnamefont {{Dufton}}}, \bibinfo {author}
  {\bibfnamefont {S.}~\bibnamefont {{Sim{\'o}n-D{\'\i}az}}}, \bibinfo {author}
  {\bibfnamefont {A.}~\bibnamefont {{Herrero}}}, \bibinfo {author}
  {\bibfnamefont {J.}~\bibnamefont {{Casares}}}, \bibinfo {author}
  {\bibfnamefont {T.}~\bibnamefont {{Mu{\~n}oz-Darias}}}, \bibinfo {author}
  {\bibfnamefont {S.~J.}\ \bibnamefont {{Smartt}}}, \bibinfo {author}
  {\bibfnamefont {J.~I.}\ \bibnamefont {{Gonz{\'a}lez Hern{\'a}ndez}}},\ and\
  \bibinfo {author} {\bibfnamefont {A.}~\bibnamefont {{de Burgos}}},\
  }\bibfield  {title} {\bibinfo {title} {{Hubble spectroscopy of LB-1:
  Comparison with B+black-hole and Be+stripped-star models}},\ }\href
  {https://doi.org/10.1051/0004-6361/202040253} {\bibfield  {journal} {\bibinfo
   {journal} {\aap}\ }\textbf {\bibinfo {volume} {649}},\ \bibinfo {eid} {A167}
  (\bibinfo {year} {2021})},\ \Eprint {https://arxiv.org/abs/2103.14069}
  {arXiv:2103.14069 [astro-ph.SR]} \BibitemShut {NoStop}%
\bibitem [{\citenamefont {{Shenar}}\ \emph {et~al.}(2020)\citenamefont
  {{Shenar}}, \citenamefont {{Bodensteiner}}, \citenamefont {{Abdul-Masih}},
  \citenamefont {{Fabry}}, \citenamefont {{Mahy}}, \citenamefont {{Marchant}},
  \citenamefont {{Banyard}}, \citenamefont {{Bowman}}, \citenamefont
  {{Dsilva}}, \citenamefont {{Hawcroft}}, \citenamefont {{Reggiani}},\ and\
  \citenamefont {{Sana}}}]{2020A&A...639L...6S}%
  \BibitemOpen
  \bibfield  {author} {\bibinfo {author} {\bibfnamefont {T.}~\bibnamefont
  {{Shenar}}}, \bibinfo {author} {\bibfnamefont {J.}~\bibnamefont
  {{Bodensteiner}}}, \bibinfo {author} {\bibfnamefont {M.}~\bibnamefont
  {{Abdul-Masih}}}, \bibinfo {author} {\bibfnamefont {M.}~\bibnamefont
  {{Fabry}}}, \bibinfo {author} {\bibfnamefont {L.}~\bibnamefont {{Mahy}}},
  \bibinfo {author} {\bibfnamefont {P.}~\bibnamefont {{Marchant}}}, \bibinfo
  {author} {\bibfnamefont {G.}~\bibnamefont {{Banyard}}}, \bibinfo {author}
  {\bibfnamefont {D.~M.}\ \bibnamefont {{Bowman}}}, \bibinfo {author}
  {\bibfnamefont {K.}~\bibnamefont {{Dsilva}}}, \bibinfo {author}
  {\bibfnamefont {C.}~\bibnamefont {{Hawcroft}}}, \bibinfo {author}
  {\bibfnamefont {M.}~\bibnamefont {{Reggiani}}},\ and\ \bibinfo {author}
  {\bibfnamefont {H.}~\bibnamefont {{Sana}}},\ }\bibfield  {title} {\bibinfo
  {title} {{The ``hidden'' companion in LB-1 unveiled by spectral
  disentangling}},\ }\href {https://doi.org/10.1051/0004-6361/202038275}
  {\bibfield  {journal} {\bibinfo  {journal} {\aap}\ }\textbf {\bibinfo
  {volume} {639}},\ \bibinfo {eid} {L6} (\bibinfo {year} {2020})},\ \Eprint
  {https://arxiv.org/abs/2004.12882} {arXiv:2004.12882 [astro-ph.SR]}
  \BibitemShut {NoStop}%
\bibitem [{\citenamefont {{El-Badry}}\ and\ \citenamefont
  {{Quataert}}(2020)}]{2020MNRAS.493L..22E}%
  \BibitemOpen
  \bibfield  {author} {\bibinfo {author} {\bibfnamefont {K.}~\bibnamefont
  {{El-Badry}}}\ and\ \bibinfo {author} {\bibfnamefont {E.}~\bibnamefont
  {{Quataert}}},\ }\bibfield  {title} {\bibinfo {title} {{Not so fast: LB-1 is
  unlikely to contain a 70 M$_{{\ensuremath{\odot}}}$ black hole}},\ }\href
  {https://doi.org/10.1093/mnrasl/slaa004} {\bibfield  {journal} {\bibinfo
  {journal} {\mnras}\ }\textbf {\bibinfo {volume} {493}},\ \bibinfo {pages}
  {L22} (\bibinfo {year} {2020})},\ \Eprint {https://arxiv.org/abs/1912.04185}
  {arXiv:1912.04185 [astro-ph.SR]} \BibitemShut {NoStop}%
\bibitem [{\citenamefont {{Marchenko}}\ \emph {et~al.}(1998)\citenamefont
  {{Marchenko}}, \citenamefont {{Moffat}},\ and\ \citenamefont
  {{Eenens}}}]{1998PASP..110.1416M}%
  \BibitemOpen
  \bibfield  {author} {\bibinfo {author} {\bibfnamefont {S.~V.}\ \bibnamefont
  {{Marchenko}}}, \bibinfo {author} {\bibfnamefont {A.~F.~J.}\ \bibnamefont
  {{Moffat}}},\ and\ \bibinfo {author} {\bibfnamefont {P.~R.~J.}\ \bibnamefont
  {{Eenens}}},\ }\bibfield  {title} {\bibinfo {title} {{The Wolf-Rayet Binary
  WR 141 (WN5O + O5 V-III) Revisited}},\ }\href
  {https://doi.org/10.1086/316280} {\bibfield  {journal} {\bibinfo  {journal}
  {\pasp}\ }\textbf {\bibinfo {volume} {110}},\ \bibinfo {pages} {1416}
  (\bibinfo {year} {1998})}\BibitemShut {NoStop}%
\bibitem [{\citenamefont {{Gonz{\'a}lez}}\ and\ \citenamefont
  {{Levato}}(2006)}]{2006A&A...448..283G}%
  \BibitemOpen
  \bibfield  {author} {\bibinfo {author} {\bibfnamefont {J.~F.}\ \bibnamefont
  {{Gonz{\'a}lez}}}\ and\ \bibinfo {author} {\bibfnamefont {H.}~\bibnamefont
  {{Levato}}},\ }\bibfield  {title} {\bibinfo {title} {{Separation of composite
  spectra: the spectroscopic detection of an eclipsing binary star}},\ }\href
  {https://doi.org/10.1051/0004-6361:20053177} {\bibfield  {journal} {\bibinfo
  {journal} {\aap}\ }\textbf {\bibinfo {volume} {448}},\ \bibinfo {pages} {283}
  (\bibinfo {year} {2006})}\BibitemShut {NoStop}%
\bibitem [{\citenamefont {{Rivinius}}\ \emph {et~al.}(2020)\citenamefont
  {{Rivinius}}, \citenamefont {{Baade}}, \citenamefont {{Hadrava}},
  \citenamefont {{Heida}},\ and\ \citenamefont
  {{Klement}}}]{2020A&A...637L...3R}%
  \BibitemOpen
  \bibfield  {author} {\bibinfo {author} {\bibfnamefont {T.}~\bibnamefont
  {{Rivinius}}}, \bibinfo {author} {\bibfnamefont {D.}~\bibnamefont {{Baade}}},
  \bibinfo {author} {\bibfnamefont {P.}~\bibnamefont {{Hadrava}}}, \bibinfo
  {author} {\bibfnamefont {M.}~\bibnamefont {{Heida}}},\ and\ \bibinfo {author}
  {\bibfnamefont {R.}~\bibnamefont {{Klement}}},\ }\bibfield  {title} {\bibinfo
  {title} {{A naked-eye triple system with a nonaccreting black hole in the
  inner binary}},\ }\href {https://doi.org/10.1051/0004-6361/202038020}
  {\bibfield  {journal} {\bibinfo  {journal} {\aap}\ }\textbf {\bibinfo
  {volume} {637}},\ \bibinfo {eid} {L3} (\bibinfo {year} {2020})},\ \Eprint
  {https://arxiv.org/abs/2005.02541} {arXiv:2005.02541 [astro-ph.SR]}
  \BibitemShut {NoStop}%
\bibitem [{\citenamefont {{Bodensteiner}}\ \emph
  {et~al.}(2020{\natexlab{b}})\citenamefont {{Bodensteiner}}, \citenamefont
  {{Shenar}}, \citenamefont {{Mahy}}, \citenamefont {{Fabry}}, \citenamefont
  {{Marchant}}, \citenamefont {{Abdul-Masih}}, \citenamefont {{Banyard}},
  \citenamefont {{Bowman}}, \citenamefont {{Dsilva}}, \citenamefont {{Frost}},
  \citenamefont {{Hawcroft}}, \citenamefont {{Reggiani}},\ and\ \citenamefont
  {{Sana}}}]{2020A&A...641A..43B}%
  \BibitemOpen
  \bibfield  {author} {\bibinfo {author} {\bibfnamefont {J.}~\bibnamefont
  {{Bodensteiner}}}, \bibinfo {author} {\bibfnamefont {T.}~\bibnamefont
  {{Shenar}}}, \bibinfo {author} {\bibfnamefont {L.}~\bibnamefont {{Mahy}}},
  \bibinfo {author} {\bibfnamefont {M.}~\bibnamefont {{Fabry}}}, \bibinfo
  {author} {\bibfnamefont {P.}~\bibnamefont {{Marchant}}}, \bibinfo {author}
  {\bibfnamefont {M.}~\bibnamefont {{Abdul-Masih}}}, \bibinfo {author}
  {\bibfnamefont {G.}~\bibnamefont {{Banyard}}}, \bibinfo {author}
  {\bibfnamefont {D.~M.}\ \bibnamefont {{Bowman}}}, \bibinfo {author}
  {\bibfnamefont {K.}~\bibnamefont {{Dsilva}}}, \bibinfo {author}
  {\bibfnamefont {A.~J.}\ \bibnamefont {{Frost}}}, \bibinfo {author}
  {\bibfnamefont {C.}~\bibnamefont {{Hawcroft}}}, \bibinfo {author}
  {\bibfnamefont {M.}~\bibnamefont {{Reggiani}}},\ and\ \bibinfo {author}
  {\bibfnamefont {H.}~\bibnamefont {{Sana}}},\ }\bibfield  {title} {\bibinfo
  {title} {{Is HR 6819 a triple system containing a black hole?. An alternative
  explanation}},\ }\href {https://doi.org/10.1051/0004-6361/202038682}
  {\bibfield  {journal} {\bibinfo  {journal} {\aap}\ }\textbf {\bibinfo
  {volume} {641}},\ \bibinfo {eid} {A43} (\bibinfo {year}
  {2020}{\natexlab{b}})},\ \Eprint {https://arxiv.org/abs/2006.10770}
  {arXiv:2006.10770 [astro-ph.SR]} \BibitemShut {NoStop}%
\bibitem [{\citenamefont {{El-Badry}}\ and\ \citenamefont
  {{Quataert}}(2021)}]{2021MNRAS.502.3436E}%
  \BibitemOpen
  \bibfield  {author} {\bibinfo {author} {\bibfnamefont {K.}~\bibnamefont
  {{El-Badry}}}\ and\ \bibinfo {author} {\bibfnamefont {E.}~\bibnamefont
  {{Quataert}}},\ }\bibfield  {title} {\bibinfo {title} {{A stripped-companion
  origin for Be stars: clues from the putative black holes HR 6819 and LB-1}},\
  }\href {https://doi.org/10.1093/mnras/stab285} {\bibfield  {journal}
  {\bibinfo  {journal} {\mnras}\ }\textbf {\bibinfo {volume} {502}},\ \bibinfo
  {pages} {3436} (\bibinfo {year} {2021})},\ \Eprint
  {https://arxiv.org/abs/2006.11974} {arXiv:2006.11974 [astro-ph.SR]}
  \BibitemShut {NoStop}%
\bibitem [{\citenamefont {{Frost}}\ \emph {et~al.}(2022)\citenamefont
  {{Frost}}, \citenamefont {{Bodensteiner}}, \citenamefont {{Rivinius}},
  \citenamefont {{Baade}}, \citenamefont {{Merand}}, \citenamefont {{Selman}},
  \citenamefont {{Abdul-Masih}}, \citenamefont {{Banyard}}, \citenamefont
  {{Bordier}}, \citenamefont {{Dsilva}}, \citenamefont {{Hawcroft}},
  \citenamefont {{Mahy}}, \citenamefont {{Reggiani}}, \citenamefont {{Shenar}},
  \citenamefont {{Cabezas}}, \citenamefont {{Hadrava}}, \citenamefont
  {{Heida}}, \citenamefont {{Klement}},\ and\ \citenamefont
  {{Sana}}}]{2022A&A...659L...3F}%
  \BibitemOpen
  \bibfield  {author} {\bibinfo {author} {\bibfnamefont {A.~J.}\ \bibnamefont
  {{Frost}}}, \bibinfo {author} {\bibfnamefont {J.}~\bibnamefont
  {{Bodensteiner}}}, \bibinfo {author} {\bibfnamefont {T.}~\bibnamefont
  {{Rivinius}}}, \bibinfo {author} {\bibfnamefont {D.}~\bibnamefont {{Baade}}},
  \bibinfo {author} {\bibfnamefont {A.}~\bibnamefont {{Merand}}}, \bibinfo
  {author} {\bibfnamefont {F.}~\bibnamefont {{Selman}}}, \bibinfo {author}
  {\bibfnamefont {M.}~\bibnamefont {{Abdul-Masih}}}, \bibinfo {author}
  {\bibfnamefont {G.}~\bibnamefont {{Banyard}}}, \bibinfo {author}
  {\bibfnamefont {E.}~\bibnamefont {{Bordier}}}, \bibinfo {author}
  {\bibfnamefont {K.}~\bibnamefont {{Dsilva}}}, \bibinfo {author}
  {\bibfnamefont {C.}~\bibnamefont {{Hawcroft}}}, \bibinfo {author}
  {\bibfnamefont {L.}~\bibnamefont {{Mahy}}}, \bibinfo {author} {\bibfnamefont
  {M.}~\bibnamefont {{Reggiani}}}, \bibinfo {author} {\bibfnamefont
  {T.}~\bibnamefont {{Shenar}}}, \bibinfo {author} {\bibfnamefont
  {M.}~\bibnamefont {{Cabezas}}}, \bibinfo {author} {\bibfnamefont
  {P.}~\bibnamefont {{Hadrava}}}, \bibinfo {author} {\bibfnamefont
  {M.}~\bibnamefont {{Heida}}}, \bibinfo {author} {\bibfnamefont
  {R.}~\bibnamefont {{Klement}}},\ and\ \bibinfo {author} {\bibfnamefont
  {H.}~\bibnamefont {{Sana}}},\ }\bibfield  {title} {\bibinfo {title} {{HR 6819
  is a binary system with no black hole. Revisiting the source with infrared
  interferometry and optical integral field spectroscopy}},\ }\href
  {https://doi.org/10.1051/0004-6361/202143004} {\bibfield  {journal} {\bibinfo
   {journal} {\aap}\ }\textbf {\bibinfo {volume} {659}},\ \bibinfo {eid} {L3}
  (\bibinfo {year} {2022})},\ \Eprint {https://arxiv.org/abs/2203.01359}
  {arXiv:2203.01359 [astro-ph.SR]} \BibitemShut {NoStop}%
\bibitem [{\citenamefont {{Jayasinghe}}\ \emph {et~al.}(2021)\citenamefont
  {{Jayasinghe}}, \citenamefont {{Stanek}}, \citenamefont {{Thompson}},
  \citenamefont {{Kochanek}}, \citenamefont {{Rowan}}, \citenamefont
  {{Vallely}}, \citenamefont {{Strassmeier}}, \citenamefont {{Weber}},
  \citenamefont {{Hinkle}}, \citenamefont {{Hambsch}}, \citenamefont
  {{Martin}}, \citenamefont {{Prieto}}, \citenamefont {{Pessi}}, \citenamefont
  {{Huber}}, \citenamefont {{Auchettl}}, \citenamefont {{Lopez}}, \citenamefont
  {{Ilyin}}, \citenamefont {{Badenes}}, \citenamefont {{Howard}}, \citenamefont
  {{Isaacson}},\ and\ \citenamefont {{Murphy}}}]{2021MNRAS.504.2577J}%
  \BibitemOpen
  \bibfield  {author} {\bibinfo {author} {\bibfnamefont {T.}~\bibnamefont
  {{Jayasinghe}}}, \bibinfo {author} {\bibfnamefont {K.~Z.}\ \bibnamefont
  {{Stanek}}}, \bibinfo {author} {\bibfnamefont {T.~A.}\ \bibnamefont
  {{Thompson}}}, \bibinfo {author} {\bibfnamefont {C.~S.}\ \bibnamefont
  {{Kochanek}}}, \bibinfo {author} {\bibfnamefont {D.~M.}\ \bibnamefont
  {{Rowan}}}, \bibinfo {author} {\bibfnamefont {P.~J.}\ \bibnamefont
  {{Vallely}}}, \bibinfo {author} {\bibfnamefont {K.~G.}\ \bibnamefont
  {{Strassmeier}}}, \bibinfo {author} {\bibfnamefont {M.}~\bibnamefont
  {{Weber}}}, \bibinfo {author} {\bibfnamefont {J.~T.}\ \bibnamefont
  {{Hinkle}}}, \bibinfo {author} {\bibfnamefont {F.~J.}\ \bibnamefont
  {{Hambsch}}}, \bibinfo {author} {\bibfnamefont {D.~V.}\ \bibnamefont
  {{Martin}}}, \bibinfo {author} {\bibfnamefont {J.~L.}\ \bibnamefont
  {{Prieto}}}, \bibinfo {author} {\bibfnamefont {T.}~\bibnamefont {{Pessi}}},
  \bibinfo {author} {\bibfnamefont {D.}~\bibnamefont {{Huber}}}, \bibinfo
  {author} {\bibfnamefont {K.}~\bibnamefont {{Auchettl}}}, \bibinfo {author}
  {\bibfnamefont {L.~A.}\ \bibnamefont {{Lopez}}}, \bibinfo {author}
  {\bibfnamefont {I.}~\bibnamefont {{Ilyin}}}, \bibinfo {author} {\bibfnamefont
  {C.}~\bibnamefont {{Badenes}}}, \bibinfo {author} {\bibfnamefont {A.~W.}\
  \bibnamefont {{Howard}}}, \bibinfo {author} {\bibfnamefont {H.}~\bibnamefont
  {{Isaacson}}},\ and\ \bibinfo {author} {\bibfnamefont {S.~J.}\ \bibnamefont
  {{Murphy}}},\ }\bibfield  {title} {\bibinfo {title} {{A unicorn in monoceros:
  the 3 M$_{{\ensuremath{\odot}}}$ dark companion to the bright, nearby red
  giant V723 Mon is a non-interacting, mass-gap black hole candidate}},\ }\href
  {https://doi.org/10.1093/mnras/stab907} {\bibfield  {journal} {\bibinfo
  {journal} {\mnras}\ }\textbf {\bibinfo {volume} {504}},\ \bibinfo {pages}
  {2577} (\bibinfo {year} {2021})},\ \Eprint {https://arxiv.org/abs/2101.02212}
  {arXiv:2101.02212 [astro-ph.SR]} \BibitemShut {NoStop}%
\bibitem [{\citenamefont {{Jayasinghe}}\ \emph {et~al.}(2022)\citenamefont
  {{Jayasinghe}}, \citenamefont {{Thompson}}, \citenamefont {{Kochanek}},
  \citenamefont {{Stanek}}, \citenamefont {{Rowan}}, \citenamefont {{Martin}},
  \citenamefont {{El-Badry}}, \citenamefont {{Vallely}}, \citenamefont
  {{Hinkle}}, \citenamefont {{Huber}}, \citenamefont {{Isaacson}},
  \citenamefont {{Tayar}}, \citenamefont {{Auchettl}}, \citenamefont {{Ilyin}},
  \citenamefont {{Howard}},\ and\ \citenamefont
  {{Badenes}}}]{2022MNRAS.516.5945J}%
  \BibitemOpen
  \bibfield  {author} {\bibinfo {author} {\bibfnamefont {T.}~\bibnamefont
  {{Jayasinghe}}}, \bibinfo {author} {\bibfnamefont {T.~A.}\ \bibnamefont
  {{Thompson}}}, \bibinfo {author} {\bibfnamefont {C.~S.}\ \bibnamefont
  {{Kochanek}}}, \bibinfo {author} {\bibfnamefont {K.~Z.}\ \bibnamefont
  {{Stanek}}}, \bibinfo {author} {\bibfnamefont {D.~M.}\ \bibnamefont
  {{Rowan}}}, \bibinfo {author} {\bibfnamefont {D.~V.}\ \bibnamefont
  {{Martin}}}, \bibinfo {author} {\bibfnamefont {K.}~\bibnamefont
  {{El-Badry}}}, \bibinfo {author} {\bibfnamefont {P.~J.}\ \bibnamefont
  {{Vallely}}}, \bibinfo {author} {\bibfnamefont {J.~T.}\ \bibnamefont
  {{Hinkle}}}, \bibinfo {author} {\bibfnamefont {D.}~\bibnamefont {{Huber}}},
  \bibinfo {author} {\bibfnamefont {H.}~\bibnamefont {{Isaacson}}}, \bibinfo
  {author} {\bibfnamefont {J.}~\bibnamefont {{Tayar}}}, \bibinfo {author}
  {\bibfnamefont {K.}~\bibnamefont {{Auchettl}}}, \bibinfo {author}
  {\bibfnamefont {I.}~\bibnamefont {{Ilyin}}}, \bibinfo {author} {\bibfnamefont
  {A.~W.}\ \bibnamefont {{Howard}}},\ and\ \bibinfo {author} {\bibfnamefont
  {C.}~\bibnamefont {{Badenes}}},\ }\bibfield  {title} {\bibinfo {title} {{The
  'Giraffe': discovery of a stripped red giant in an interacting binary with an
  2 M$_{{\ensuremath{\odot}}}$ lower giant}},\ }\href
  {https://doi.org/10.1093/mnras/stac2187} {\bibfield  {journal} {\bibinfo
  {journal} {\mnras}\ }\textbf {\bibinfo {volume} {516}},\ \bibinfo {pages}
  {5945} (\bibinfo {year} {2022})},\ \Eprint {https://arxiv.org/abs/2201.11131}
  {arXiv:2201.11131 [astro-ph.SR]} \BibitemShut {NoStop}%
\bibitem [{\citenamefont {{Masuda}}\ and\ \citenamefont
  {{Hirano}}(2021)}]{2021ApJ...910L..17M}%
  \BibitemOpen
  \bibfield  {author} {\bibinfo {author} {\bibfnamefont {K.}~\bibnamefont
  {{Masuda}}}\ and\ \bibinfo {author} {\bibfnamefont {T.}~\bibnamefont
  {{Hirano}}},\ }\bibfield  {title} {\bibinfo {title} {{Tidal Effects on the
  Radial Velocities of V723 Mon: Additional Evidence for a Dark 3
  M$_{{\ensuremath{\odot}}}$ Companion}},\ }\href
  {https://doi.org/10.3847/2041-8213/abecdc} {\bibfield  {journal} {\bibinfo
  {journal} {\apjl}\ }\textbf {\bibinfo {volume} {910}},\ \bibinfo {eid} {L17}
  (\bibinfo {year} {2021})},\ \Eprint {https://arxiv.org/abs/2103.05216}
  {arXiv:2103.05216 [astro-ph.SR]} \BibitemShut {NoStop}%
\bibitem [{\citenamefont {{El-Badry}}\ \emph
  {et~al.}(2022{\natexlab{a}})\citenamefont {{El-Badry}}, \citenamefont
  {{Seeburger}}, \citenamefont {{Jayasinghe}}, \citenamefont {{Rix}},
  \citenamefont {{Almada}}, \citenamefont {{Conroy}}, \citenamefont
  {{Price-Whelan}},\ and\ \citenamefont {{Burdge}}}]{2022MNRAS.512.5620E}%
  \BibitemOpen
  \bibfield  {author} {\bibinfo {author} {\bibfnamefont {K.}~\bibnamefont
  {{El-Badry}}}, \bibinfo {author} {\bibfnamefont {R.}~\bibnamefont
  {{Seeburger}}}, \bibinfo {author} {\bibfnamefont {T.}~\bibnamefont
  {{Jayasinghe}}}, \bibinfo {author} {\bibfnamefont {H.-W.}\ \bibnamefont
  {{Rix}}}, \bibinfo {author} {\bibfnamefont {S.}~\bibnamefont {{Almada}}},
  \bibinfo {author} {\bibfnamefont {C.}~\bibnamefont {{Conroy}}}, \bibinfo
  {author} {\bibfnamefont {A.~M.}\ \bibnamefont {{Price-Whelan}}},\ and\
  \bibinfo {author} {\bibfnamefont {K.}~\bibnamefont {{Burdge}}},\ }\bibfield
  {title} {\bibinfo {title} {{Unicorns and giraffes in the binary zoo: stripped
  giants with subgiant companions}},\ }\href
  {https://doi.org/10.1093/mnras/stac815} {\bibfield  {journal} {\bibinfo
  {journal} {\mnras}\ }\textbf {\bibinfo {volume} {512}},\ \bibinfo {pages}
  {5620} (\bibinfo {year} {2022}{\natexlab{a}})},\ \Eprint
  {https://arxiv.org/abs/2203.06348} {arXiv:2203.06348 [astro-ph.SR]}
  \BibitemShut {NoStop}%
\bibitem [{\citenamefont {{Gomez}}\ and\ \citenamefont
  {{Grindlay}}(2021)}]{2021ApJ...913...48G}%
  \BibitemOpen
  \bibfield  {author} {\bibinfo {author} {\bibfnamefont {S.}~\bibnamefont
  {{Gomez}}}\ and\ \bibinfo {author} {\bibfnamefont {J.~E.}\ \bibnamefont
  {{Grindlay}}},\ }\bibfield  {title} {\bibinfo {title} {{Optical Analysis and
  Modeling of HD96670, a New Black Hole X-Ray Binary Candidate}},\ }\href
  {https://doi.org/10.3847/1538-4357/abf24c} {\bibfield  {journal} {\bibinfo
  {journal} {\apj}\ }\textbf {\bibinfo {volume} {913}},\ \bibinfo {eid} {48}
  (\bibinfo {year} {2021})}\BibitemShut {NoStop}%
\bibitem [{\citenamefont {{Shenar}}\ \emph
  {et~al.}(2022{\natexlab{b}})\citenamefont {{Shenar}}, \citenamefont {{Sana}},
  \citenamefont {{Mahy}}, \citenamefont {{Ma{\'\i}z Apell{\'a}niz}},
  \citenamefont {{Crowther}}, \citenamefont {{Gromadzki}}, \citenamefont
  {{Herrero}}, \citenamefont {{Langer}}, \citenamefont {{Marchant}},
  \citenamefont {{Schneider}}, \citenamefont {{Sen}}, \citenamefont
  {{Soszy{\'n}ski}},\ and\ \citenamefont {{Toonen}}}]{2022A&A...665A.148S}%
  \BibitemOpen
  \bibfield  {author} {\bibinfo {author} {\bibfnamefont {T.}~\bibnamefont
  {{Shenar}}}, \bibinfo {author} {\bibfnamefont {H.}~\bibnamefont {{Sana}}},
  \bibinfo {author} {\bibfnamefont {L.}~\bibnamefont {{Mahy}}}, \bibinfo
  {author} {\bibfnamefont {J.}~\bibnamefont {{Ma{\'\i}z Apell{\'a}niz}}},
  \bibinfo {author} {\bibfnamefont {P.~A.}\ \bibnamefont {{Crowther}}},
  \bibinfo {author} {\bibfnamefont {M.}~\bibnamefont {{Gromadzki}}}, \bibinfo
  {author} {\bibfnamefont {A.}~\bibnamefont {{Herrero}}}, \bibinfo {author}
  {\bibfnamefont {N.}~\bibnamefont {{Langer}}}, \bibinfo {author}
  {\bibfnamefont {P.}~\bibnamefont {{Marchant}}}, \bibinfo {author}
  {\bibfnamefont {F.~R.~N.}\ \bibnamefont {{Schneider}}}, \bibinfo {author}
  {\bibfnamefont {K.}~\bibnamefont {{Sen}}}, \bibinfo {author} {\bibfnamefont
  {I.}~\bibnamefont {{Soszy{\'n}ski}}},\ and\ \bibinfo {author} {\bibfnamefont
  {S.}~\bibnamefont {{Toonen}}},\ }\bibfield  {title} {\bibinfo {title} {{The
  Tarantula Massive Binary Monitoring. VI. Characterisation of hidden
  companions in 51 single-lined O-type binaries: A flat mass-ratio distribution
  and black-hole binary candidates}},\ }\href
  {https://doi.org/10.1051/0004-6361/202244245} {\bibfield  {journal} {\bibinfo
   {journal} {\aap}\ }\textbf {\bibinfo {volume} {665}},\ \bibinfo {eid} {A148}
  (\bibinfo {year} {2022}{\natexlab{b}})},\ \Eprint
  {https://arxiv.org/abs/2207.07674} {arXiv:2207.07674 [astro-ph.SR]}
  \BibitemShut {NoStop}%
\bibitem [{\citenamefont {{Mahy}}\ \emph {et~al.}(2022)\citenamefont {{Mahy}},
  \citenamefont {{Sana}}, \citenamefont {{Shenar}}, \citenamefont {{Sen}},
  \citenamefont {{Langer}}, \citenamefont {{Marchant}}, \citenamefont
  {{Abdul-Masih}}, \citenamefont {{Banyard}}, \citenamefont {{Bodensteiner}},
  \citenamefont {{Bowman}}, \citenamefont {{Dsilva}}, \citenamefont {{Fabry}},
  \citenamefont {{Hawcroft}}, \citenamefont {{Janssens}}, \citenamefont {{Van
  Reeth}},\ and\ \citenamefont {{Eldridge}}}]{2022A&A...664A.159M}%
  \BibitemOpen
  \bibfield  {author} {\bibinfo {author} {\bibfnamefont {L.}~\bibnamefont
  {{Mahy}}}, \bibinfo {author} {\bibfnamefont {H.}~\bibnamefont {{Sana}}},
  \bibinfo {author} {\bibfnamefont {T.}~\bibnamefont {{Shenar}}}, \bibinfo
  {author} {\bibfnamefont {K.}~\bibnamefont {{Sen}}}, \bibinfo {author}
  {\bibfnamefont {N.}~\bibnamefont {{Langer}}}, \bibinfo {author}
  {\bibfnamefont {P.}~\bibnamefont {{Marchant}}}, \bibinfo {author}
  {\bibfnamefont {M.}~\bibnamefont {{Abdul-Masih}}}, \bibinfo {author}
  {\bibfnamefont {G.}~\bibnamefont {{Banyard}}}, \bibinfo {author}
  {\bibfnamefont {J.}~\bibnamefont {{Bodensteiner}}}, \bibinfo {author}
  {\bibfnamefont {D.~M.}\ \bibnamefont {{Bowman}}}, \bibinfo {author}
  {\bibfnamefont {K.}~\bibnamefont {{Dsilva}}}, \bibinfo {author}
  {\bibfnamefont {M.}~\bibnamefont {{Fabry}}}, \bibinfo {author} {\bibfnamefont
  {C.}~\bibnamefont {{Hawcroft}}}, \bibinfo {author} {\bibfnamefont
  {S.}~\bibnamefont {{Janssens}}}, \bibinfo {author} {\bibfnamefont
  {T.}~\bibnamefont {{Van Reeth}}},\ and\ \bibinfo {author} {\bibfnamefont
  {C.}~\bibnamefont {{Eldridge}}},\ }\bibfield  {title} {\bibinfo {title}
  {{Identifying quiescent compact objects in massive Galactic single-lined
  spectroscopic binaries}},\ }\href
  {https://doi.org/10.1051/0004-6361/202243147} {\bibfield  {journal} {\bibinfo
   {journal} {\aap}\ }\textbf {\bibinfo {volume} {664}},\ \bibinfo {eid} {A159}
  (\bibinfo {year} {2022})},\ \Eprint {https://arxiv.org/abs/2207.07752}
  {arXiv:2207.07752 [astro-ph.SR]} \BibitemShut {NoStop}%
\bibitem [{\citenamefont {{Orosz}}\ \emph {et~al.}(2011)\citenamefont
  {{Orosz}}, \citenamefont {{McClintock}}, \citenamefont {{Aufdenberg}},
  \citenamefont {{Remillard}}, \citenamefont {{Reid}}, \citenamefont
  {{Narayan}},\ and\ \citenamefont {{Gou}}}]{2011ApJ...742...84O}%
  \BibitemOpen
  \bibfield  {author} {\bibinfo {author} {\bibfnamefont {J.~A.}\ \bibnamefont
  {{Orosz}}}, \bibinfo {author} {\bibfnamefont {J.~E.}\ \bibnamefont
  {{McClintock}}}, \bibinfo {author} {\bibfnamefont {J.~P.}\ \bibnamefont
  {{Aufdenberg}}}, \bibinfo {author} {\bibfnamefont {R.~A.}\ \bibnamefont
  {{Remillard}}}, \bibinfo {author} {\bibfnamefont {M.~J.}\ \bibnamefont
  {{Reid}}}, \bibinfo {author} {\bibfnamefont {R.}~\bibnamefont {{Narayan}}},\
  and\ \bibinfo {author} {\bibfnamefont {L.}~\bibnamefont {{Gou}}},\ }\bibfield
   {title} {\bibinfo {title} {{The Mass of the Black Hole in Cygnus X-1}},\
  }\href {https://doi.org/10.1088/0004-637X/742/2/84} {\bibfield  {journal}
  {\bibinfo  {journal} {\apj}\ }\textbf {\bibinfo {volume} {742}},\ \bibinfo
  {eid} {84} (\bibinfo {year} {2011})},\ \Eprint
  {https://arxiv.org/abs/1106.3689} {arXiv:1106.3689 [astro-ph.HE]}
  \BibitemShut {NoStop}%
\bibitem [{\citenamefont {{Miller-Jones}}\ \emph {et~al.}(2021)\citenamefont
  {{Miller-Jones}}, \citenamefont {{Bahramian}}, \citenamefont {{Orosz}},
  \citenamefont {{Mandel}}, \citenamefont {{Gou}}, \citenamefont {{Maccarone}},
  \citenamefont {{Neijssel}}, \citenamefont {{Zhao}}, \citenamefont
  {{Zi{\'o}{\l}kowski}}, \citenamefont {{Reid}}, \citenamefont {{Uttley}},
  \citenamefont {{Zheng}}, \citenamefont {{Byun}}, \citenamefont {{Dodson}},
  \citenamefont {{Grinberg}}, \citenamefont {{Jung}}, \citenamefont {{Kim}},
  \citenamefont {{Marcote}}, \citenamefont {{Markoff}}, \citenamefont
  {{Rioja}}, \citenamefont {{Rushton}}, \citenamefont {{Russell}},
  \citenamefont {{Sivakoff}}, \citenamefont {{Tetarenko}}, \citenamefont
  {{Tudose}},\ and\ \citenamefont {{Wilms}}}]{2021Sci...371.1046M}%
  \BibitemOpen
  \bibfield  {author} {\bibinfo {author} {\bibfnamefont {J.~C.~A.}\
  \bibnamefont {{Miller-Jones}}}, \bibinfo {author} {\bibfnamefont
  {A.}~\bibnamefont {{Bahramian}}}, \bibinfo {author} {\bibfnamefont {J.~A.}\
  \bibnamefont {{Orosz}}}, \bibinfo {author} {\bibfnamefont {I.}~\bibnamefont
  {{Mandel}}}, \bibinfo {author} {\bibfnamefont {L.}~\bibnamefont {{Gou}}},
  \bibinfo {author} {\bibfnamefont {T.~J.}\ \bibnamefont {{Maccarone}}},
  \bibinfo {author} {\bibfnamefont {C.~J.}\ \bibnamefont {{Neijssel}}},
  \bibinfo {author} {\bibfnamefont {X.}~\bibnamefont {{Zhao}}}, \bibinfo
  {author} {\bibfnamefont {J.}~\bibnamefont {{Zi{\'o}{\l}kowski}}}, \bibinfo
  {author} {\bibfnamefont {M.~J.}\ \bibnamefont {{Reid}}}, \bibinfo {author}
  {\bibfnamefont {P.}~\bibnamefont {{Uttley}}}, \bibinfo {author}
  {\bibfnamefont {X.}~\bibnamefont {{Zheng}}}, \bibinfo {author} {\bibfnamefont
  {D.-Y.}\ \bibnamefont {{Byun}}}, \bibinfo {author} {\bibfnamefont
  {R.}~\bibnamefont {{Dodson}}}, \bibinfo {author} {\bibfnamefont
  {V.}~\bibnamefont {{Grinberg}}}, \bibinfo {author} {\bibfnamefont
  {T.}~\bibnamefont {{Jung}}}, \bibinfo {author} {\bibfnamefont {J.-S.}\
  \bibnamefont {{Kim}}}, \bibinfo {author} {\bibfnamefont {B.}~\bibnamefont
  {{Marcote}}}, \bibinfo {author} {\bibfnamefont {S.}~\bibnamefont
  {{Markoff}}}, \bibinfo {author} {\bibfnamefont {M.~J.}\ \bibnamefont
  {{Rioja}}}, \bibinfo {author} {\bibfnamefont {A.~P.}\ \bibnamefont
  {{Rushton}}}, \bibinfo {author} {\bibfnamefont {D.~M.}\ \bibnamefont
  {{Russell}}}, \bibinfo {author} {\bibfnamefont {G.~R.}\ \bibnamefont
  {{Sivakoff}}}, \bibinfo {author} {\bibfnamefont {A.~J.}\ \bibnamefont
  {{Tetarenko}}}, \bibinfo {author} {\bibfnamefont {V.}~\bibnamefont
  {{Tudose}}},\ and\ \bibinfo {author} {\bibfnamefont {J.}~\bibnamefont
  {{Wilms}}},\ }\bibfield  {title} {\bibinfo {title} {{Cygnus X-1 contains a
  21-solar mass black hole{\textemdash}Implications for massive star winds}},\
  }\href {https://doi.org/10.1126/science.abb3363} {\bibfield  {journal}
  {\bibinfo  {journal} {Science}\ }\textbf {\bibinfo {volume} {371}},\ \bibinfo
  {pages} {1046} (\bibinfo {year} {2021})},\ \Eprint
  {https://arxiv.org/abs/2102.09091} {arXiv:2102.09091 [astro-ph.HE]}
  \BibitemShut {NoStop}%
\bibitem [{\citenamefont {{Gamen}}\ \emph {et~al.}(2015)\citenamefont
  {{Gamen}}, \citenamefont {{Barb{\`a}}}, \citenamefont {{Walborn}},
  \citenamefont {{Morrell}}, \citenamefont {{Arias}}, \citenamefont {{Ma{\'\i}z
  Apell{\'a}niz}}, \citenamefont {{Sota}},\ and\ \citenamefont
  {{Alfaro}}}]{2015A&A...583L...4G}%
  \BibitemOpen
  \bibfield  {author} {\bibinfo {author} {\bibfnamefont {R.}~\bibnamefont
  {{Gamen}}}, \bibinfo {author} {\bibfnamefont {R.~H.}\ \bibnamefont
  {{Barb{\`a}}}}, \bibinfo {author} {\bibfnamefont {N.~R.}\ \bibnamefont
  {{Walborn}}}, \bibinfo {author} {\bibfnamefont {N.~I.}\ \bibnamefont
  {{Morrell}}}, \bibinfo {author} {\bibfnamefont {J.~I.}\ \bibnamefont
  {{Arias}}}, \bibinfo {author} {\bibfnamefont {J.}~\bibnamefont {{Ma{\'\i}z
  Apell{\'a}niz}}}, \bibinfo {author} {\bibfnamefont {A.}~\bibnamefont
  {{Sota}}},\ and\ \bibinfo {author} {\bibfnamefont {E.~J.}\ \bibnamefont
  {{Alfaro}}},\ }\bibfield  {title} {\bibinfo {title} {{The eccentric
  short-period orbit of the supergiant fast X-ray transient HD 74194 (=LM
  Vel)}},\ }\href {https://doi.org/10.1051/0004-6361/201527140} {\bibfield
  {journal} {\bibinfo  {journal} {\aap}\ }\textbf {\bibinfo {volume} {583}},\
  \bibinfo {eid} {L4} (\bibinfo {year} {2015})},\ \Eprint
  {https://arxiv.org/abs/1510.06584} {arXiv:1510.06584 [astro-ph.SR]}
  \BibitemShut {NoStop}%
\bibitem [{\citenamefont {{Fu}}\ \emph {et~al.}(2022)\citenamefont {{Fu}},
  \citenamefont {{Gu}}, \citenamefont {{Zhang}}, \citenamefont {{Yi}},
  \citenamefont {{Qi}}, \citenamefont {{Zheng}},\ and\ \citenamefont
  {{Liu}}}]{2022ApJ...940..126F}%
  \BibitemOpen
  \bibfield  {author} {\bibinfo {author} {\bibfnamefont {J.-B.}\ \bibnamefont
  {{Fu}}}, \bibinfo {author} {\bibfnamefont {W.-M.}\ \bibnamefont {{Gu}}},
  \bibinfo {author} {\bibfnamefont {Z.-X.}\ \bibnamefont {{Zhang}}}, \bibinfo
  {author} {\bibfnamefont {T.}~\bibnamefont {{Yi}}}, \bibinfo {author}
  {\bibfnamefont {S.-Y.}\ \bibnamefont {{Qi}}}, \bibinfo {author}
  {\bibfnamefont {L.-L.}\ \bibnamefont {{Zheng}}},\ and\ \bibinfo {author}
  {\bibfnamefont {J.}~\bibnamefont {{Liu}}},\ }\bibfield  {title} {\bibinfo
  {title} {{Searching for Compact Objects in Binaries with Gaia DR3}},\ }\href
  {https://doi.org/10.3847/1538-4357/ac9b4c} {\bibfield  {journal} {\bibinfo
  {journal} {\apj}\ }\textbf {\bibinfo {volume} {940}},\ \bibinfo {eid} {126}
  (\bibinfo {year} {2022})},\ \Eprint {https://arxiv.org/abs/2207.05434}
  {arXiv:2207.05434 [astro-ph.SR]} \BibitemShut {NoStop}%
\bibitem [{\citenamefont {{Shao}}\ and\ \citenamefont
  {{Li}}(2019)}]{2019ApJ...885..151S}%
  \BibitemOpen
  \bibfield  {author} {\bibinfo {author} {\bibfnamefont {Y.}~\bibnamefont
  {{Shao}}}\ and\ \bibinfo {author} {\bibfnamefont {X.-D.}\ \bibnamefont
  {{Li}}},\ }\bibfield  {title} {\bibinfo {title} {{Population Synthesis of
  Black Hole Binaries with Normal-star Companions. I. Detached Systems}},\
  }\href {https://doi.org/10.3847/1538-4357/ab4816} {\bibfield  {journal}
  {\bibinfo  {journal} {\apj}\ }\textbf {\bibinfo {volume} {885}},\ \bibinfo
  {eid} {151} (\bibinfo {year} {2019})},\ \Eprint
  {https://arxiv.org/abs/1909.11328} {arXiv:1909.11328 [astro-ph.SR]}
  \BibitemShut {NoStop}%
\bibitem [{\citenamefont {{Langer}}\ \emph {et~al.}(2020)\citenamefont
  {{Langer}}, \citenamefont {{Sch{\"u}rmann}}, \citenamefont {{Stoll}},
  \citenamefont {{Marchant}}, \citenamefont {{Lennon}}, \citenamefont {{Mahy}},
  \citenamefont {{de Mink}}, \citenamefont {{Quast}}, \citenamefont {{Riedel}},
  \citenamefont {{Sana}}, \citenamefont {{Schneider}}, \citenamefont
  {{Schootemeijer}}, \citenamefont {{Wang}}, \citenamefont {{Almeida}},
  \citenamefont {{Bestenlehner}}, \citenamefont {{Bodensteiner}}, \citenamefont
  {{Castro}}, \citenamefont {{Clark}}, \citenamefont {{Crowther}},
  \citenamefont {{Dufton}}, \citenamefont {{Evans}}, \citenamefont {{Fossati}},
  \citenamefont {{Gr{\"a}fener}}, \citenamefont {{Grassitelli}}, \citenamefont
  {{Grin}}, \citenamefont {{Hastings}}, \citenamefont {{Herrero}},
  \citenamefont {{de Koter}}, \citenamefont {{Menon}}, \citenamefont
  {{Patrick}}, \citenamefont {{Puls}}, \citenamefont {{Renzo}}, \citenamefont
  {{Sander}}, \citenamefont {{Schneider}}, \citenamefont {{Sen}}, \citenamefont
  {{Shenar}}, \citenamefont {{Sim{\'o}n-D{\'\i}as}}, \citenamefont {{Tauris}},
  \citenamefont {{Tramper}}, \citenamefont {{Vink}},\ and\ \citenamefont
  {{Xu}}}]{2020A&A...638A..39L}%
  \BibitemOpen
  \bibfield  {author} {\bibinfo {author} {\bibfnamefont {N.}~\bibnamefont
  {{Langer}}}, \bibinfo {author} {\bibfnamefont {C.}~\bibnamefont
  {{Sch{\"u}rmann}}}, \bibinfo {author} {\bibfnamefont {K.}~\bibnamefont
  {{Stoll}}}, \bibinfo {author} {\bibfnamefont {P.}~\bibnamefont {{Marchant}}},
  \bibinfo {author} {\bibfnamefont {D.~J.}\ \bibnamefont {{Lennon}}}, \bibinfo
  {author} {\bibfnamefont {L.}~\bibnamefont {{Mahy}}}, \bibinfo {author}
  {\bibfnamefont {S.~E.}\ \bibnamefont {{de Mink}}}, \bibinfo {author}
  {\bibfnamefont {M.}~\bibnamefont {{Quast}}}, \bibinfo {author} {\bibfnamefont
  {W.}~\bibnamefont {{Riedel}}}, \bibinfo {author} {\bibfnamefont
  {H.}~\bibnamefont {{Sana}}}, \bibinfo {author} {\bibfnamefont
  {P.}~\bibnamefont {{Schneider}}}, \bibinfo {author} {\bibfnamefont
  {A.}~\bibnamefont {{Schootemeijer}}}, \bibinfo {author} {\bibfnamefont
  {C.}~\bibnamefont {{Wang}}}, \bibinfo {author} {\bibfnamefont {L.~A.}\
  \bibnamefont {{Almeida}}}, \bibinfo {author} {\bibfnamefont {J.~M.}\
  \bibnamefont {{Bestenlehner}}}, \bibinfo {author} {\bibfnamefont
  {J.}~\bibnamefont {{Bodensteiner}}}, \bibinfo {author} {\bibfnamefont
  {N.}~\bibnamefont {{Castro}}}, \bibinfo {author} {\bibfnamefont
  {S.}~\bibnamefont {{Clark}}}, \bibinfo {author} {\bibfnamefont {P.~A.}\
  \bibnamefont {{Crowther}}}, \bibinfo {author} {\bibfnamefont
  {P.}~\bibnamefont {{Dufton}}}, \bibinfo {author} {\bibfnamefont {C.~J.}\
  \bibnamefont {{Evans}}}, \bibinfo {author} {\bibfnamefont {L.}~\bibnamefont
  {{Fossati}}}, \bibinfo {author} {\bibfnamefont {G.}~\bibnamefont
  {{Gr{\"a}fener}}}, \bibinfo {author} {\bibfnamefont {L.}~\bibnamefont
  {{Grassitelli}}}, \bibinfo {author} {\bibfnamefont {N.}~\bibnamefont
  {{Grin}}}, \bibinfo {author} {\bibfnamefont {B.}~\bibnamefont {{Hastings}}},
  \bibinfo {author} {\bibfnamefont {A.}~\bibnamefont {{Herrero}}}, \bibinfo
  {author} {\bibfnamefont {A.}~\bibnamefont {{de Koter}}}, \bibinfo {author}
  {\bibfnamefont {A.}~\bibnamefont {{Menon}}}, \bibinfo {author} {\bibfnamefont
  {L.}~\bibnamefont {{Patrick}}}, \bibinfo {author} {\bibfnamefont
  {J.}~\bibnamefont {{Puls}}}, \bibinfo {author} {\bibfnamefont
  {M.}~\bibnamefont {{Renzo}}}, \bibinfo {author} {\bibfnamefont {A.~A.~C.}\
  \bibnamefont {{Sander}}}, \bibinfo {author} {\bibfnamefont {F.~R.~N.}\
  \bibnamefont {{Schneider}}}, \bibinfo {author} {\bibfnamefont
  {K.}~\bibnamefont {{Sen}}}, \bibinfo {author} {\bibfnamefont
  {T.}~\bibnamefont {{Shenar}}}, \bibinfo {author} {\bibfnamefont
  {S.}~\bibnamefont {{Sim{\'o}n-D{\'\i}as}}}, \bibinfo {author} {\bibfnamefont
  {T.~M.}\ \bibnamefont {{Tauris}}}, \bibinfo {author} {\bibfnamefont
  {F.}~\bibnamefont {{Tramper}}}, \bibinfo {author} {\bibfnamefont {J.~S.}\
  \bibnamefont {{Vink}}},\ and\ \bibinfo {author} {\bibfnamefont {X.~T.}\
  \bibnamefont {{Xu}}},\ }\bibfield  {title} {\bibinfo {title} {{Properties of
  OB star-black hole systems derived from detailed binary evolution models}},\
  }\href {https://doi.org/10.1051/0004-6361/201937375} {\bibfield  {journal}
  {\bibinfo  {journal} {\aap}\ }\textbf {\bibinfo {volume} {638}},\ \bibinfo
  {eid} {A39} (\bibinfo {year} {2020})},\ \Eprint
  {https://arxiv.org/abs/1912.09826} {arXiv:1912.09826 [astro-ph.SR]}
  \BibitemShut {NoStop}%
\bibitem [{\citenamefont {{Abbott}}\ \emph
  {et~al.}(2020{\natexlab{e}})\citenamefont {{Abbott}}, \citenamefont
  {{Abbott}}, \citenamefont {{Abbott}}, \citenamefont {{Abraham}},
  \citenamefont {{Acernese}}, \citenamefont {{Ackley}}, \citenamefont
  {{Adams}}, \citenamefont {{Adya}}, \citenamefont {{Affeldt}}, \citenamefont
  {{Agathos}}, \citenamefont {{Agatsuma}}, \citenamefont {{Aggarwal}},
  \citenamefont {{Aguiar}}, \citenamefont {{Aiello}}, \citenamefont {{Ain}},
  \citenamefont {{Ajith}}, \citenamefont {{Alford}}, \citenamefont {{Allen}},
  \citenamefont {{Allocca}}, \citenamefont {{Aloy}}, \citenamefont {{Altin}},
  \citenamefont {{Amato}}, \citenamefont {{Ananyeva}}, \citenamefont
  {{Anderson}}, \citenamefont {{Anderson}}, \citenamefont {{Angelova}},
  \citenamefont {{Antier}}, \citenamefont {{Appert}}, \citenamefont {{Arai}},
  \citenamefont {{Araya}}, \citenamefont {{Areeda}}, \citenamefont
  {{Ar{\`e}ne}}, \citenamefont {{Arnaud}}, \citenamefont {{Arun}},
  \citenamefont {{Ascenzi}}, \citenamefont {{Ashton}}, \citenamefont {{Aston}},
  \citenamefont {{Astone}}, \citenamefont {{Aubin}}, \citenamefont {{Aufmuth}},
  \citenamefont {{AultONeal}}, \citenamefont {{Austin}}, \citenamefont
  {{Avendano}}, \citenamefont {{Avila-Alvarez}}, \citenamefont {{Babak}},
  \citenamefont {{Bacon}}, \citenamefont {{Badaracco}}, \citenamefont
  {{Bader}}, \citenamefont {{Bae}}, \citenamefont {{Baker}}, \citenamefont
  {{Baldaccini}}, \citenamefont {{Ballardin}}, \citenamefont {{Ballmer}},
  \citenamefont {{Banagiri}}, \citenamefont {{Barayoga}}, \citenamefont
  {{Barclay}}, \citenamefont {{Barish}}, \citenamefont {{Barker}},
  \citenamefont {{Barkett}}, \citenamefont {{Barnum}}, \citenamefont
  {{Barone}}, \citenamefont {{Barr}}, \citenamefont {{Barsotti}}, \citenamefont
  {{Barsuglia}}, \citenamefont {{Barta}}, \citenamefont {{Bartlett}},
  \citenamefont {{Bartos}}, \citenamefont {{Bassiri}}, \citenamefont {{Basti}},
  \citenamefont {{Bawaj}}, \citenamefont {{Bayley}}, \citenamefont {{Bazzan}},
  \citenamefont {{B{\'e}csy}}, \citenamefont {{Bejger}}, \citenamefont
  {{Belahcene}}, \citenamefont {{Bell}}, \citenamefont {{Beniwal}},
  \citenamefont {{Berger}}, \citenamefont {{Bergmann}}, \citenamefont
  {{Bernuzzi}}, \citenamefont {{Bero}}, \citenamefont {{Berry}}, \citenamefont
  {{Bersanetti}}, \citenamefont {{Bertolini}}, \citenamefont {{Betzwieser}},
  \citenamefont {{Bhandare}}, \citenamefont {{Bidler}}, \citenamefont
  {{Bilenko}}, \citenamefont {{Bilgili}}, \citenamefont {{Billingsley}},
  \citenamefont {{Birch}}, \citenamefont {{Birney}}, \citenamefont
  {{Birnholtz}}, \citenamefont {{Biscans}}, \citenamefont {{Biscoveanu}},
  \citenamefont {{Bisht}}, \citenamefont {{Bitossi}}, \citenamefont
  {{Bizouard}}, \citenamefont {{Blackburn}}, \citenamefont {{Blair}},
  \citenamefont {{Blair}}, \citenamefont {{Blair}}, \citenamefont {{Bloemen}},
  \citenamefont {{Bode}}, \citenamefont {{Boer}}, \citenamefont {{Boetzel}},
  \citenamefont {{Bogaert}}, \citenamefont {{Bondu}}, \citenamefont
  {{Bonilla}}, \citenamefont {{Bonnand}}, \citenamefont {{Booker}},
  \citenamefont {{Boom}}, \citenamefont {{Booth}}, \citenamefont {{Bork}},
  \citenamefont {{Boschi}}, \citenamefont {{Bose}}, \citenamefont {{Bossie}},
  \citenamefont {{Bossilkov}}, \citenamefont {{Bosveld}}, \citenamefont
  {{Bouffanais}}, \citenamefont {{Bozzi}}, \citenamefont {{Bradaschia}},
  \citenamefont {{Brady}}, \citenamefont {{Bramley}}, \citenamefont
  {{Branchesi}}, \citenamefont {{Brau}}, \citenamefont {{Briant}},
  \citenamefont {{Briggs}}, \citenamefont {{Brighenti}}, \citenamefont
  {{Brillet}}, \citenamefont {{Brinkmann}}, \citenamefont {{Brisson}},
  \citenamefont {{Brockill}}, \citenamefont {{Brooks}}, \citenamefont
  {{Brown}}, \citenamefont {{Brunett}}, \citenamefont {{Buikema}},
  \citenamefont {{Bulik}}, \citenamefont {{Bulten}}, \citenamefont
  {{Buonanno}}, \citenamefont {{Buskulic}}, \citenamefont {{Buy}},
  \citenamefont {{Byer}}, \citenamefont {{Cabero}}, \citenamefont {{Cadonati}},
  \citenamefont {{Cagnoli}}, \citenamefont {{Cahillane}}, \citenamefont
  {{Calder{\'o}n Bustillo}}, \citenamefont {{Callister}}, \citenamefont
  {{Calloni}}, \citenamefont {{Camp}}, \citenamefont {{Campbell}},
  \citenamefont {{Canepa}}, \citenamefont {{Cannon}}, \citenamefont {{Cao}},
  \citenamefont {{Cao}}, \citenamefont {{Capocasa}}, \citenamefont
  {{Carbognani}}, \citenamefont {{Caride}}, \citenamefont {{Carney}},
  \citenamefont {{Carullo}}, \citenamefont {{Casanueva Diaz}}, \citenamefont
  {{Casentini}}, \citenamefont {{Caudill}}, \citenamefont {{Cavagli{\`a}}},
  \citenamefont {{Cavalier}}, \citenamefont {{Cavalieri}}, \citenamefont
  {{Cella}}, \citenamefont {{Cerd{\'a}-Dur{\'a}n}}, \citenamefont
  {{Cerretani}}, \citenamefont {{Cesarini}}, \citenamefont {{Chaibi}},
  \citenamefont {{Chakravarti}}, \citenamefont {{Chamberlin}}, \citenamefont
  {{Chan}}, \citenamefont {{Chao}}, \citenamefont {{Charlton}}, \citenamefont
  {{Chase}}, \citenamefont {{Chassande-Mottin}}, \citenamefont {{Chatterjee}},
  \citenamefont {{Chaturvedi}}, \citenamefont {{Chatziioannou}}, \citenamefont
  {{Cheeseboro}}, \citenamefont {{Chen}}, \citenamefont {{Chen}}, \citenamefont
  {{Chen}}, \citenamefont {{Cheng}}, \citenamefont {{Cheong}}, \citenamefont
  {{Chia}}, \citenamefont {{Chincarini}}, \citenamefont {{Chiummo}},
  \citenamefont {{Cho}}, \citenamefont {{Cho}}, \citenamefont {{Cho}},
  \citenamefont {{Christensen}}, \citenamefont {{Chu}}, \citenamefont {{Chua}},
  \citenamefont {{Chung}}, \citenamefont {{Chung}}, \citenamefont {{Ciani}},
  \citenamefont {{Ciobanu}}, \citenamefont {{Ciolfi}}, \citenamefont
  {{Cipriano}}, \citenamefont {{Cirone}}, \citenamefont {{Clara}},
  \citenamefont {{Clark}}, \citenamefont {{Clearwater}}, \citenamefont
  {{Cleva}}, \citenamefont {{Cocchieri}}, \citenamefont {{Coccia}},
  \citenamefont {{Cohadon}}, \citenamefont {{Cohen}}, \citenamefont {{Colgan}},
  \citenamefont {{Colleoni}}, \citenamefont {{Collette}}, \citenamefont
  {{Collins}}, \citenamefont {{Cominsky}}, \citenamefont {{Constancio}},
  \citenamefont {{Conti}}, \citenamefont {{Cooper}}, \citenamefont {{Corban}},
  \citenamefont {{Corbitt}}, \citenamefont {{Cordero-Carri{\'o}n}},
  \citenamefont {{Corley}}, \citenamefont {{Cornish}}, \citenamefont {{Corsi}},
  \citenamefont {{Cortese}}, \citenamefont {{Costa}}, \citenamefont
  {{Cotesta}}, \citenamefont {{Coughlin}}, \citenamefont {{Coughlin}},
  \citenamefont {{Coulon}}, \citenamefont {{Countryman}}, \citenamefont
  {{Couvares}}, \citenamefont {{Covas}}, \citenamefont {{Cowan}}, \citenamefont
  {{Coward}}, \citenamefont {{Cowart}}, \citenamefont {{Coyne}}, \citenamefont
  {{Coyne}}, \citenamefont {{Creighton}}, \citenamefont {{Creighton}},
  \citenamefont {{Cripe}}, \citenamefont {{Croquette}}, \citenamefont
  {{Crowder}}, \citenamefont {{Cullen}}, \citenamefont {{Cumming}},
  \citenamefont {{Cunningham}}, \citenamefont {{Cuoco}}, \citenamefont {{Dal
  Canton}}, \citenamefont {{D{\'a}lya}}, \citenamefont {{Danilishin}},
  \citenamefont {{D'Antonio}}, \citenamefont {{Danzmann}}, \citenamefont
  {{Dasgupta}}, \citenamefont {{Da Silva Costa}}, \citenamefont {{Datrier}},
  \citenamefont {{Dattilo}}, \citenamefont {{Dave}}, \citenamefont {{Davier}},
  \citenamefont {{Davis}}, \citenamefont {{Daw}}, \citenamefont {{DeBra}},
  \citenamefont {{Deenadayalan}}, \citenamefont {{Degallaix}}, \citenamefont
  {{De Laurentis}}, \citenamefont {{Del{\'e}glise}}, \citenamefont {{Del
  Pozzo}}, \citenamefont {{DeMarchi}}, \citenamefont {{Demos}}, \citenamefont
  {{Dent}}, \citenamefont {{De Pietri}}, \citenamefont {{Derby}}, \citenamefont
  {{De Rosa}}, \citenamefont {{De Rossi}}, \citenamefont {{DeSalvo}},
  \citenamefont {{de Varona}}, \citenamefont {{Dhurandhar}}, \citenamefont
  {{D{\'\i}az}}, \citenamefont {{Dietrich}}, \citenamefont {{Di Fiore}},
  \citenamefont {{Di Giovanni}}, \citenamefont {{Di Girolamo}}, \citenamefont
  {{Di Lieto}}, \citenamefont {{Ding}}, \citenamefont {{Di Pace}},
  \citenamefont {{Di Palma}}, \citenamefont {{Di Renzo}}, \citenamefont
  {{Dmitriev}}, \citenamefont {{Doctor}}, \citenamefont {{Donovan}},
  \citenamefont {{Dooley}}, \citenamefont {{Doravari}}, \citenamefont
  {{Dorrington}}, \citenamefont {{Downes}}, \citenamefont {{Drago}},
  \citenamefont {{Driggers}}, \citenamefont {{Du}}, \citenamefont {{Ducoin}},
  \citenamefont {{Dupej}}, \citenamefont {{Dwyer}}, \citenamefont {{Easter}},
  \citenamefont {{Edo}}, \citenamefont {{Edwards}}, \citenamefont {{Effler}},
  \citenamefont {{Ehrens}}, \citenamefont {{Eichholz}}, \citenamefont
  {{Eikenberry}}, \citenamefont {{Eisenmann}}, \citenamefont {{Eisenstein}},
  \citenamefont {{Essick}}, \citenamefont {{Estelles}}, \citenamefont
  {{Estevez}}, \citenamefont {{Etienne}}, \citenamefont {{Etzel}},
  \citenamefont {{Evans}}, \citenamefont {{Evans}}, \citenamefont {{Fafone}},
  \citenamefont {{Fair}}, \citenamefont {{Fairhurst}}, \citenamefont {{Fan}},
  \citenamefont {{Farinon}}, \citenamefont {{Farr}}, \citenamefont {{Farr}},
  \citenamefont {{Fauchon-Jones}}, \citenamefont {{Favata}}, \citenamefont
  {{Fays}}, \citenamefont {{Fazio}}, \citenamefont {{Fee}}, \citenamefont
  {{Feicht}}, \citenamefont {{Fejer}}, \citenamefont {{Feng}}, \citenamefont
  {{Fernandez-Galiana}}, \citenamefont {{Ferrante}}, \citenamefont
  {{Ferreira}}, \citenamefont {{Ferreira}}, \citenamefont {{Ferrini}},
  \citenamefont {{Fidecaro}}, \citenamefont {{Fiori}}, \citenamefont
  {{Fiorucci}}, \citenamefont {{Fishbach}}, \citenamefont {{Fisher}},
  \citenamefont {{Fishner}}, \citenamefont {{Fitz-Axen}}, \citenamefont
  {{Flaminio}}, \citenamefont {{Fletcher}}, \citenamefont {{Flynn}},
  \citenamefont {{Fong}}, \citenamefont {{Font}}, \citenamefont {{Forsyth}},
  \citenamefont {{Fournier}}, \citenamefont {{Frasca}}, \citenamefont
  {{Frasconi}}, \citenamefont {{Frei}}, \citenamefont {{Freise}}, \citenamefont
  {{Frey}}, \citenamefont {{Frey}}, \citenamefont {{Fritschel}}, \citenamefont
  {{Frolov}}, \citenamefont {{Fulda}}, \citenamefont {{Fyffe}}, \citenamefont
  {{Gabbard}}, \citenamefont {{Gadre}}, \citenamefont {{Gaebel}}, \citenamefont
  {{Gair}}, \citenamefont {{Gammaitoni}}, \citenamefont {{Ganija}},
  \citenamefont {{Gaonkar}}, \citenamefont {{Garcia}}, \citenamefont
  {{Garc{\'\i}a-Quir{\'o}s}}, \citenamefont {{Garufi}}, \citenamefont
  {{Gateley}}, \citenamefont {{Gaudio}}, \citenamefont {{Gaur}}, \citenamefont
  {{Gayathri}}, \citenamefont {{Gemme}}, \citenamefont {{Genin}}, \citenamefont
  {{Gennai}}, \citenamefont {{George}}, \citenamefont {{George}}, \citenamefont
  {{Gergely}}, \citenamefont {{Germain}}, \citenamefont {{Ghonge}},
  \citenamefont {{Ghosh}}, \citenamefont {{Ghosh}}, \citenamefont {{Ghosh}},
  \citenamefont {{Giacomazzo}}, \citenamefont {{Giaime}}, \citenamefont
  {{Giardina}}, \citenamefont {{Giazotto}}, \citenamefont {{Gill}},
  \citenamefont {{Giordano}}, \citenamefont {{Glover}}, \citenamefont
  {{Godwin}}, \citenamefont {{Goetz}}, \citenamefont {{Goetz}}, \citenamefont
  {{Goncharov}}, \citenamefont {{Gonz{\'a}lez}}, \citenamefont {{Gonzalez
  Castro}}, \citenamefont {{Gopakumar}}, \citenamefont {{Gorodetsky}},
  \citenamefont {{Gossan}}, \citenamefont {{Gosselin}}, \citenamefont
  {{Gouaty}}, \citenamefont {{Grado}}, \citenamefont {{Graef}}, \citenamefont
  {{Granata}}, \citenamefont {{Grant}}, \citenamefont {{Gras}}, \citenamefont
  {{Grassia}}, \citenamefont {{Gray}}, \citenamefont {{Gray}}, \citenamefont
  {{Greco}}, \citenamefont {{Green}}, \citenamefont {{Green}}, \citenamefont
  {{Gretarsson}}, \citenamefont {{Groot}}, \citenamefont {{Grote}},
  \citenamefont {{Grunewald}}, \citenamefont {{Gruning}}, \citenamefont
  {{Guidi}}, \citenamefont {{Gulati}}, \citenamefont {{Guo}}, \citenamefont
  {{Gupta}}, \citenamefont {{Gupta}}, \citenamefont {{Gustafson}},
  \citenamefont {{Gustafson}}, \citenamefont {{Haegel}}, \citenamefont
  {{Halim}}, \citenamefont {{Hall}}, \citenamefont {{Hall}}, \citenamefont
  {{Hamilton}}, \citenamefont {{Hammond}}, \citenamefont {{Haney}},
  \citenamefont {{Hanke}}, \citenamefont {{Hanks}}, \citenamefont {{Hanna}},
  \citenamefont {{Hannam}}, \citenamefont {{Hannuksela}}, \citenamefont
  {{Hanson}}, \citenamefont {{Hardwick}}, \citenamefont {{Haris}},
  \citenamefont {{Harms}}, \citenamefont {{Harry}}, \citenamefont {{Harry}},
  \citenamefont {{Haster}}, \citenamefont {{Haughian}}, \citenamefont
  {{Hayes}}, \citenamefont {{Healy}}, \citenamefont {{Heidmann}}, \citenamefont
  {{Heintze}}, \citenamefont {{Heitmann}}, \citenamefont {{Hello}},
  \citenamefont {{Hemming}}, \citenamefont {{Hendry}}, \citenamefont {{Heng}},
  \citenamefont {{Hennig}}, \citenamefont {{Heptonstall}}, \citenamefont
  {{Vivanco}}, \citenamefont {{Heurs}}, \citenamefont {{Hild}}, \citenamefont
  {{Hinderer}}, \citenamefont {{Hoak}}, \citenamefont {{Hochheim}},
  \citenamefont {{Hofman}}, \citenamefont {{Holgado}}, \citenamefont
  {{Holland}}, \citenamefont {{Holt}}, \citenamefont {{Holz}}, \citenamefont
  {{Hopkins}}, \citenamefont {{Horst}}, \citenamefont {{Hough}}, \citenamefont
  {{Howell}}, \citenamefont {{Hoy}}, \citenamefont {{Hreibi}}, \citenamefont
  {{Huerta}}, \citenamefont {{Huet}}, \citenamefont {{Hughey}}, \citenamefont
  {{Hulko}}, \citenamefont {{Husa}}, \citenamefont {{Huttner}}, \citenamefont
  {{Huynh-Dinh}}, \citenamefont {{Idzkowski}}, \citenamefont {{Iess}},
  \citenamefont {{Ingram}}, \citenamefont {{Inta}}, \citenamefont {{Intini}},
  \citenamefont {{Irwin}}, \citenamefont {{Isa}}, \citenamefont {{Isac}},
  \citenamefont {{Isi}}, \citenamefont {{Iyer}}, \citenamefont {{Izumi}},
  \citenamefont {{Jacqmin}}, \citenamefont {{Jadhav}}, \citenamefont {{Jani}},
  \citenamefont {{Janthalur}}, \citenamefont {{Jaranowski}}, \citenamefont
  {{Jenkins}}, \citenamefont {{Jiang}}, \citenamefont {{Johnson}},
  \citenamefont {{Jones}}, \citenamefont {{Jones}}, \citenamefont {{Jones}},
  \citenamefont {{Jonker}}, \citenamefont {{Ju}}, \citenamefont {{Junker}},
  \citenamefont {{Kalaghatgi}}, \citenamefont {{Kalogera}}, \citenamefont
  {{Kamai}}, \citenamefont {{Kandhasamy}}, \citenamefont {{Kang}},
  \citenamefont {{Kanner}}, \citenamefont {{Kapadia}}, \citenamefont {{Karki}},
  \citenamefont {{Karvinen}}, \citenamefont {{Kashyap}}, \citenamefont
  {{Kasprzack}}, \citenamefont {{Katsanevas}}, \citenamefont {{Katsavounidis}},
  \citenamefont {{Katzman}}, \citenamefont {{Kaufer}}, \citenamefont
  {{Kawabe}}, \citenamefont {{Keerthana}}, \citenamefont {{K{\'e}f{\'e}lian}},
  \citenamefont {{Keitel}}, \citenamefont {{Kennedy}}, \citenamefont {{Key}},
  \citenamefont {{Khalili}}, \citenamefont {{Khan}}, \citenamefont {{Khan}},
  \citenamefont {{Khan}}, \citenamefont {{Khan}}, \citenamefont {{Khazanov}},
  \citenamefont {{Khursheed}}, \citenamefont {{Kijbunchoo}}, \citenamefont
  {{Kim}}, \citenamefont {{Kim}}, \citenamefont {{Kim}}, \citenamefont {{Kim}},
  \citenamefont {{Kim}}, \citenamefont {{Kim}}, \citenamefont {{Kim}},
  \citenamefont {{Kimball}}, \citenamefont {{King}}, \citenamefont {{King}},
  \citenamefont {{Kinley-Hanlon}}, \citenamefont {{Kirchhoff}}, \citenamefont
  {{Kissel}}, \citenamefont {{Kleybolte}}, \citenamefont {{Klika}},
  \citenamefont {{Klimenko}}, \citenamefont {{Knowles}}, \citenamefont
  {{Koch}}, \citenamefont {{Koehlenbeck}}, \citenamefont {{Koekoek}},
  \citenamefont {{Koley}}, \citenamefont {{Kondrashov}}, \citenamefont
  {{Kontos}}, \citenamefont {{Koper}}, \citenamefont {{Korobko}}, \citenamefont
  {{Korth}}, \citenamefont {{Kowalska}}, \citenamefont {{Kozak}}, \citenamefont
  {{Kringel}}, \citenamefont {{Krishnendu}}, \citenamefont {{Kr{\'o}lak}},
  \citenamefont {{Kuehn}}, \citenamefont {{Kumar}}, \citenamefont {{Kumar}},
  \citenamefont {{Kumar}}, \citenamefont {{Kumar}}, \citenamefont {{Kuo}},
  \citenamefont {{Kutynia}}, \citenamefont {{Kwang}}, \citenamefont {{Lackey}},
  \citenamefont {{Lai}}, \citenamefont {{Lam}}, \citenamefont {{Landry}},
  \citenamefont {{Lane}}, \citenamefont {{Lang}}, \citenamefont {{Lange}},
  \citenamefont {{Lantz}}, \citenamefont {{Lanza}}, \citenamefont {{Larson}},
  \citenamefont {{Lartaux-Vollard}}, \citenamefont {{Lasky}}, \citenamefont
  {{Laxen}}, \citenamefont {{Lazzarini}}, \citenamefont {{Lazzaro}},
  \citenamefont {{Leaci}}, \citenamefont {{Leavey}}, \citenamefont
  {{Lecoeuche}}, \citenamefont {{Lee}}, \citenamefont {{Lee}}, \citenamefont
  {{Lee}}, \citenamefont {{Lee}}, \citenamefont {{Lee}}, \citenamefont {{Lee}},
  \citenamefont {{Lehmann}}, \citenamefont {{Lenon}}, \citenamefont {{Leroy}},
  \citenamefont {{Letendre}}, \citenamefont {{Levin}}, \citenamefont {{Li}},
  \citenamefont {{Li}}, \citenamefont {{Li}}, \citenamefont {{Li}},
  \citenamefont {{Lin}}, \citenamefont {{Linde}}, \citenamefont {{Linker}},
  \citenamefont {{Littenberg}}, \citenamefont {{Liu}}, \citenamefont {{Liu}},
  \citenamefont {{Lo}}, \citenamefont {{Lockerbie}}, \citenamefont {{London}},
  \citenamefont {{Longo}}, \citenamefont {{Lorenzini}}, \citenamefont
  {{Loriette}}, \citenamefont {{Lormand}}, \citenamefont {{Losurdo}},
  \citenamefont {{Lough}}, \citenamefont {{Lousto}}, \citenamefont
  {{Lovelace}}, \citenamefont {{Lower}}, \citenamefont {{L{\"u}ck}},
  \citenamefont {{Lumaca}}, \citenamefont {{Lundgren}}, \citenamefont
  {{Lynch}}, \citenamefont {{Ma}}, \citenamefont {{Macas}}, \citenamefont
  {{Macfoy}}, \citenamefont {{MacInnis}}, \citenamefont {{Macleod}},
  \citenamefont {{Macquet}}, \citenamefont {{Maga{\~n}a-Sandoval}},
  \citenamefont {{Maga{\~n}a Zertuche}}, \citenamefont {{Magee}}, \citenamefont
  {{Majorana}}, \citenamefont {{Maksimovic}}, \citenamefont {{Malik}},
  \citenamefont {{Man}}, \citenamefont {{Mandic}}, \citenamefont {{Mangano}},
  \citenamefont {{Mansell}}, \citenamefont {{Manske}}, \citenamefont
  {{Mantovani}}, \citenamefont {{Marchesoni}}, \citenamefont {{Marion}},
  \citenamefont {{M{\'a}rka}}, \citenamefont {{M{\'a}rka}}, \citenamefont
  {{Markakis}}, \citenamefont {{Markosyan}}, \citenamefont {{Markowitz}},
  \citenamefont {{Maros}}, \citenamefont {{Marquina}}, \citenamefont
  {{Marsat}}, \citenamefont {{Martelli}}, \citenamefont {{Martin}},
  \citenamefont {{Martin}}, \citenamefont {{Martynov}}, \citenamefont
  {{Mason}}, \citenamefont {{Massera}}, \citenamefont {{Masserot}},
  \citenamefont {{Massinger}}, \citenamefont {{Masso-Reid}}, \citenamefont
  {{Mastrogiovanni}}, \citenamefont {{Matas}}, \citenamefont {{Matichard}},
  \citenamefont {{Matone}}, \citenamefont {{Mavalvala}}, \citenamefont
  {{Mazumder}}, \citenamefont {{McCann}}, \citenamefont {{McCarthy}},
  \citenamefont {{McClelland}}, \citenamefont {{McCormick}}, \citenamefont
  {{McCuller}}, \citenamefont {{McGuire}}, \citenamefont {{McIver}},
  \citenamefont {{McManus}}, \citenamefont {{McRae}}, \citenamefont
  {{McWilliams}}, \citenamefont {{Meacher}}, \citenamefont {{Meadors}},
  \citenamefont {{Mehmet}}, \citenamefont {{Mehta}}, \citenamefont {{Meidam}},
  \citenamefont {{Melatos}}, \citenamefont {{Mendell}}, \citenamefont
  {{Mercer}}, \citenamefont {{Mereni}}, \citenamefont {{Merilh}}, \citenamefont
  {{Merzougui}}, \citenamefont {{Meshkov}}, \citenamefont {{Messenger}},
  \citenamefont {{Messick}}, \citenamefont {{Metzdorff}}, \citenamefont
  {{Meyers}}, \citenamefont {{Miao}}, \citenamefont {{Michel}}, \citenamefont
  {{Middleton}}, \citenamefont {{Mikhailov}}, \citenamefont {{Milano}},
  \citenamefont {{Miller}}, \citenamefont {{Miller}}, \citenamefont
  {{Millhouse}}, \citenamefont {{Mills}}, \citenamefont {{Milovich-Goff}},
  \citenamefont {{Minazzoli}}, \citenamefont {{Minenkov}}, \citenamefont
  {{Mishkin}}, \citenamefont {{Mishra}}, \citenamefont {{Mistry}},
  \citenamefont {{Mitra}}, \citenamefont {{Mitrofanov}}, \citenamefont
  {{Mitselmakher}}, \citenamefont {{Mittleman}}, \citenamefont {{Mo}},
  \citenamefont {{Moffa}}, \citenamefont {{Mogushi}}, \citenamefont
  {{Mohapatra}}, \citenamefont {{Montani}}, \citenamefont {{Moore}},
  \citenamefont {{Moraru}}, \citenamefont {{Moreno}}, \citenamefont
  {{Morisaki}}, \citenamefont {{Mours}}, \citenamefont {{Mow-Lowry}},
  \citenamefont {{Mukherjee}}, \citenamefont {{Mukherjee}}, \citenamefont
  {{Mukherjee}}, \citenamefont {{Mukund}}, \citenamefont {{Mullavey}},
  \citenamefont {{Munch}}, \citenamefont {{Mu{\~n}iz}}, \citenamefont
  {{Muratore}}, \citenamefont {{Murray}}, \citenamefont {{Nagar}},
  \citenamefont {{Nardecchia}}, \citenamefont {{Naticchioni}}, \citenamefont
  {{Nayak}}, \citenamefont {{Neilson}}, \citenamefont {{Nelemans}},
  \citenamefont {{Nelson}}, \citenamefont {{Nery}}, \citenamefont {{Neunzert}},
  \citenamefont {{Ng}}, \citenamefont {{Ng}}, \citenamefont {{Nguyen}},
  \citenamefont {{Nichols}}, \citenamefont {{Nissanke}}, \citenamefont
  {{Nocera}}, \citenamefont {{North}}, \citenamefont {{Nuttall}}, \citenamefont
  {{Obergaulinger}}, \citenamefont {{Oberling}}, \citenamefont {{O'Brien}},
  \citenamefont {{O'Dea}}, \citenamefont {{Ogin}}, \citenamefont {{Oh}},
  \citenamefont {{Oh}}, \citenamefont {{Ohme}}, \citenamefont {{Ohta}},
  \citenamefont {{Okada}}, \citenamefont {{Oliver}}, \citenamefont
  {{Oppermann}}, \citenamefont {{Oram}}, \citenamefont {{O'Reilly}},
  \citenamefont {{Ormiston}}, \citenamefont {{Ortega}}, \citenamefont
  {{O'Shaughnessy}}, \citenamefont {{Ossokine}}, \citenamefont {{Ottaway}},
  \citenamefont {{Overmier}}, \citenamefont {{Owen}}, \citenamefont {{Pace}},
  \citenamefont {{Pagano}}, \citenamefont {{Page}}, \citenamefont {{Pai}},
  \citenamefont {{Pai}}, \citenamefont {{Palamos}}, \citenamefont {{Palashov}},
  \citenamefont {{Palomba}}, \citenamefont {{Pal-Singh}}, \citenamefont
  {{Pan}}, \citenamefont {{Pang}}, \citenamefont {{Pang}}, \citenamefont
  {{Pankow}}, \citenamefont {{Pannarale}}, \citenamefont {{Pant}},
  \citenamefont {{Paoletti}}, \citenamefont {{Paoli}}, \citenamefont
  {{Parida}}, \citenamefont {{Parker}}, \citenamefont {{Pascucci}},
  \citenamefont {{Pasqualetti}}, \citenamefont {{Passaquieti}}, \citenamefont
  {{Passuello}}, \citenamefont {{Patil}}, \citenamefont {{Patricelli}},
  \citenamefont {{Pearlstone}}, \citenamefont {{Pedersen}}, \citenamefont
  {{Pedraza}}, \citenamefont {{Pedurand}}, \citenamefont {{Pele}},
  \citenamefont {{Penn}}, \citenamefont {{Perez}}, \citenamefont {{Perreca}},
  \citenamefont {{Pfeiffer}}, \citenamefont {{Phelps}}, \citenamefont
  {{Phukon}}, \citenamefont {{Piccinni}}, \citenamefont {{Pichot}},
  \citenamefont {{Piergiovanni}}, \citenamefont {{Pillant}}, \citenamefont
  {{Pinard}}, \citenamefont {{Pirello}}, \citenamefont {{Pitkin}},
  \citenamefont {{Poggiani}}, \citenamefont {{Pong}}, \citenamefont
  {{Ponrathnam}}, \citenamefont {{Popolizio}}, \citenamefont {{Porter}},
  \citenamefont {{Powell}}, \citenamefont {{Prajapati}}, \citenamefont
  {{Prasad}}, \citenamefont {{Prasai}}, \citenamefont {{Prasanna}},
  \citenamefont {{Pratten}}, \citenamefont {{Prestegard}}, \citenamefont
  {{Privitera}}, \citenamefont {{Prodi}}, \citenamefont {{Prokhorov}},
  \citenamefont {{Puncken}}, \citenamefont {{Punturo}}, \citenamefont
  {{Puppo}}, \citenamefont {{P{\"u}rrer}}, \citenamefont {{Qi}}, \citenamefont
  {{Quetschke}}, \citenamefont {{Quinonez}}, \citenamefont {{Quintero}},
  \citenamefont {{Quitzow-James}}, \citenamefont {{Raab}}, \citenamefont
  {{Radkins}}, \citenamefont {{Radulescu}}, \citenamefont {{Raffai}},
  \citenamefont {{Raja}}, \citenamefont {{Rajan}}, \citenamefont
  {{Rajbhandari}}, \citenamefont {{Rakhmanov}}, \citenamefont {{Ramirez}},
  \citenamefont {{Ramos-Buades}}, \citenamefont {{Rana}}, \citenamefont
  {{Rao}}, \citenamefont {{Rapagnani}}, \citenamefont {{Raymond}},
  \citenamefont {{Razzano}}, \citenamefont {{Read}}, \citenamefont
  {{Regimbau}}, \citenamefont {{Rei}}, \citenamefont {{Reid}}, \citenamefont
  {{Reitze}}, \citenamefont {{Ren}}, \citenamefont {{Ricci}}, \citenamefont
  {{Richardson}}, \citenamefont {{Richardson}}, \citenamefont {{Ricker}},
  \citenamefont {{Riles}}, \citenamefont {{Rizzo}}, \citenamefont
  {{Robertson}}, \citenamefont {{Robie}}, \citenamefont {{Robinet}},
  \citenamefont {{Rocchi}}, \citenamefont {{Rolland}}, \citenamefont
  {{Rollins}}, \citenamefont {{Roma}}, \citenamefont {{Romanelli}},
  \citenamefont {{Romano}}, \citenamefont {{Romel}}, \citenamefont {{Romie}},
  \citenamefont {{Rose}}, \citenamefont {{Rosi{\'n}ska}}, \citenamefont
  {{Rosofsky}}, \citenamefont {{Ross}}, \citenamefont {{Rowan}}, \citenamefont
  {{R{\"u}diger}}, \citenamefont {{Ruggi}}, \citenamefont {{Rutins}},
  \citenamefont {{Ryan}}, \citenamefont {{Sachdev}}, \citenamefont {{Sadecki}},
  \citenamefont {{Sakellariadou}}, \citenamefont {{Salconi}}, \citenamefont
  {{Saleem}}, \citenamefont {{Samajdar}}, \citenamefont {{Sammut}},
  \citenamefont {{Sanchez}}, \citenamefont {{Sanchez}}, \citenamefont
  {{Sanchis-Gual}}, \citenamefont {{Sandberg}}, \citenamefont {{Sanders}},
  \citenamefont {{Santiago}}, \citenamefont {{Sarin}}, \citenamefont
  {{Sassolas}}, \citenamefont {{Sathyaprakash}}, \citenamefont {{Saulson}},
  \citenamefont {{Sauter}}, \citenamefont {{Savage}}, \citenamefont {{Schale}},
  \citenamefont {{Scheel}}, \citenamefont {{Scheuer}}, \citenamefont
  {{Schmidt}}, \citenamefont {{Schnabel}}, \citenamefont {{Schofield}},
  \citenamefont {{Sch{\"o}nbeck}}, \citenamefont {{Schreiber}}, \citenamefont
  {{Schulte}}, \citenamefont {{Schutz}}, \citenamefont {{Schwalbe}},
  \citenamefont {{Scott}}, \citenamefont {{Scott}}, \citenamefont {{Seidel}},
  \citenamefont {{Sellers}}, \citenamefont {{Sengupta}}, \citenamefont
  {{Sennett}}, \citenamefont {{Sentenac}}, \citenamefont {{Sequino}},
  \citenamefont {{Sergeev}}, \citenamefont {{Setyawati}}, \citenamefont
  {{Shaddock}}, \citenamefont {{Shaffer}}, \citenamefont {{Shahriar}},
  \citenamefont {{Shaner}}, \citenamefont {{Shao}}, \citenamefont {{Sharma}},
  \citenamefont {{Shawhan}}, \citenamefont {{Shen}}, \citenamefont {{Shink}},
  \citenamefont {{Shoemaker}}, \citenamefont {{Shoemaker}}, \citenamefont
  {{ShyamSundar}}, \citenamefont {{Siellez}}, \citenamefont {{Sieniawska}},
  \citenamefont {{Sigg}}, \citenamefont {{Silva}}, \citenamefont {{Singer}},
  \citenamefont {{Singh}}, \citenamefont {{Singhal}}, \citenamefont {{Sintes}},
  \citenamefont {{Sitmukhambetov}}, \citenamefont {{Skliris}}, \citenamefont
  {{Slagmolen}}, \citenamefont {{Slaven-Blair}}, \citenamefont {{Smith}},
  \citenamefont {{Smith}}, \citenamefont {{Somala}}, \citenamefont {{Son}},
  \citenamefont {{Sorazu}}, \citenamefont {{Sorrentino}}, \citenamefont
  {{Souradeep}}, \citenamefont {{Sowell}}, \citenamefont {{Spencer}},
  \citenamefont {{Srivastava}}, \citenamefont {{Srivastava}}, \citenamefont
  {{Staats}}, \citenamefont {{Stachie}}, \citenamefont {{Standke}},
  \citenamefont {{Steer}}, \citenamefont {{Steinke}}, \citenamefont
  {{Steinlechner}}, \citenamefont {{Steinlechner}}, \citenamefont
  {{Steinmeyer}}, \citenamefont {{Stevenson}}, \citenamefont {{Stocks}},
  \citenamefont {{Stone}}, \citenamefont {{Stops}}, \citenamefont {{Strain}},
  \citenamefont {{Stratta}}, \citenamefont {{Strigin}}, \citenamefont
  {{Strunk}}, \citenamefont {{Sturani}}, \citenamefont {{Stuver}},
  \citenamefont {{Sudhir}}, \citenamefont {{Summerscales}}, \citenamefont
  {{Sun}}, \citenamefont {{Sunil}}, \citenamefont {{Suresh}}, \citenamefont
  {{Sutton}}, \citenamefont {{Swinkels}}, \citenamefont {{Szczepa{\'n}czyk}},
  \citenamefont {{Tacca}}, \citenamefont {{Tait}}, \citenamefont {{Talbot}},
  \citenamefont {{Talukder}}, \citenamefont {{Tanner}}, \citenamefont
  {{T{\'a}pai}}, \citenamefont {{Taracchini}}, \citenamefont {{Tasson}},
  \citenamefont {{Taylor}}, \citenamefont {{Thies}}, \citenamefont {{Thomas}},
  \citenamefont {{Thomas}}, \citenamefont {{Thondapu}}, \citenamefont
  {{Thorne}}, \citenamefont {{Thrane}}, \citenamefont {{Tiwari}}, \citenamefont
  {{Tiwari}}, \citenamefont {{Tiwari}}, \citenamefont {{Toland}}, \citenamefont
  {{Tonelli}}, \citenamefont {{Tornasi}}, \citenamefont {{Torres-Forn{\'e}}},
  \citenamefont {{Torrie}}, \citenamefont {{T{\"o}yr{\"a}}}, \citenamefont
  {{Travasso}}, \citenamefont {{Traylor}}, \citenamefont {{Tringali}},
  \citenamefont {{Trovato}}, \citenamefont {{Trozzo}}, \citenamefont
  {{Trudeau}}, \citenamefont {{Tsang}}, \citenamefont {{Tse}}, \citenamefont
  {{Tso}}, \citenamefont {{Tsukada}}, \citenamefont {{Tsuna}}, \citenamefont
  {{Tuyenbayev}}, \citenamefont {{Ueno}}, \citenamefont {{Ugolini}},
  \citenamefont {{Unnikrishnan}}, \citenamefont {{Urban}}, \citenamefont
  {{Usman}}, \citenamefont {{Vahlbruch}}, \citenamefont {{Vajente}},
  \citenamefont {{Valdes}}, \citenamefont {{van Bakel}}, \citenamefont {{van
  Beuzekom}}, \citenamefont {{van den Brand}}, \citenamefont {{Van Den
  Broeck}}, \citenamefont {{Vander-Hyde}}, \citenamefont {{van Heijningen}},
  \citenamefont {{van der Schaaf}}, \citenamefont {{van Veggel}}, \citenamefont
  {{Vardaro}}, \citenamefont {{Varma}}, \citenamefont {{Vass}}, \citenamefont
  {{Vas{\'u}th}}, \citenamefont {{Vecchio}}, \citenamefont {{Vedovato}},
  \citenamefont {{Veitch}}, \citenamefont {{Veitch}}, \citenamefont
  {{Venkateswara}}, \citenamefont {{Venugopalan}}, \citenamefont {{Verkindt}},
  \citenamefont {{Vetrano}}, \citenamefont {{Vicer{\'e}}}, \citenamefont
  {{Viets}}, \citenamefont {{Vine}}, \citenamefont {{Vinet}}, \citenamefont
  {{Vitale}}, \citenamefont {{Vo}}, \citenamefont {{Vocca}}, \citenamefont
  {{Vorvick}}, \citenamefont {{Vyatchanin}}, \citenamefont {{Wade}},
  \citenamefont {{Wade}}, \citenamefont {{Wade}}, \citenamefont {{Walet}},
  \citenamefont {{Walker}}, \citenamefont {{Wallace}}, \citenamefont {{Walsh}},
  \citenamefont {{Wang}}, \citenamefont {{Wang}}, \citenamefont {{Wang}},
  \citenamefont {{Wang}}, \citenamefont {{Wang}}, \citenamefont {{Ward}},
  \citenamefont {{Warden}}, \citenamefont {{Warner}}, \citenamefont {{Was}},
  \citenamefont {{Watchi}}, \citenamefont {{Weaver}}, \citenamefont {{Wei}},
  \citenamefont {{Weinert}}, \citenamefont {{Weinstein}}, \citenamefont
  {{Weiss}}, \citenamefont {{Wellmann}}, \citenamefont {{Wen}}, \citenamefont
  {{Wessel}}, \citenamefont {{We{\ss}els}}, \citenamefont {{Westhouse}},
  \citenamefont {{Wette}}, \citenamefont {{Whelan}}, \citenamefont {{Whiting}},
  \citenamefont {{Whittle}}, \citenamefont {{Wilken}}, \citenamefont
  {{Williams}}, \citenamefont {{Williamson}}, \citenamefont {{Willis}},
  \citenamefont {{Willke}}, \citenamefont {{Wimmer}}, \citenamefont
  {{Winkler}}, \citenamefont {{Wipf}}, \citenamefont {{Wittel}}, \citenamefont
  {{Woan}}, \citenamefont {{Woehler}}, \citenamefont {{Wofford}}, \citenamefont
  {{Worden}}, \citenamefont {{Wright}}, \citenamefont {{Wu}}, \citenamefont
  {{Wysocki}}, \citenamefont {{Xiao}}, \citenamefont {{Yamamoto}},
  \citenamefont {{Yancey}}, \citenamefont {{Yang}}, \citenamefont {{Yap}},
  \citenamefont {{Yazback}}, \citenamefont {{Yeeles}}, \citenamefont {{Yu}},
  \citenamefont {{Yu}}, \citenamefont {{Yuen}}, \citenamefont {{Yvert}},
  \citenamefont {{Zadro{\.z}ny}}, \citenamefont {{Zanolin}}, \citenamefont
  {{Zelenova}}, \citenamefont {{Zendri}}, \citenamefont {{Zevin}},
  \citenamefont {{Zhang}}, \citenamefont {{Zhang}}, \citenamefont {{Zhang}},
  \citenamefont {{Zhao}}, \citenamefont {{Zhou}}, \citenamefont {{Zhou}},
  \citenamefont {{Zhu}}, \citenamefont {{Zucker}}, \citenamefont {{Zweizig}},
  \citenamefont {{LIGO Scientific Collaboration}},\ and\ \citenamefont {{Virgo
  Collaboration}}}]{2020CQGra..37e5002A}%
  \BibitemOpen
  \bibfield  {author} {\bibinfo {author} {\bibfnamefont {B.~P.}\ \bibnamefont
  {{Abbott}}}, \bibinfo {author} {\bibfnamefont {R.}~\bibnamefont {{Abbott}}},
  \bibinfo {author} {\bibfnamefont {T.~D.}\ \bibnamefont {{Abbott}}}, \bibinfo
  {author} {\bibfnamefont {S.}~\bibnamefont {{Abraham}}}, \bibinfo {author}
  {\bibfnamefont {F.}~\bibnamefont {{Acernese}}}, \bibinfo {author}
  {\bibfnamefont {K.}~\bibnamefont {{Ackley}}}, \bibinfo {author}
  {\bibfnamefont {C.}~\bibnamefont {{Adams}}}, \bibinfo {author} {\bibfnamefont
  {V.~B.}\ \bibnamefont {{Adya}}}, \bibinfo {author} {\bibfnamefont
  {C.}~\bibnamefont {{Affeldt}}}, \bibinfo {author} {\bibfnamefont
  {M.}~\bibnamefont {{Agathos}}}, \bibinfo {author} {\bibfnamefont
  {K.}~\bibnamefont {{Agatsuma}}}, \bibinfo {author} {\bibfnamefont
  {N.}~\bibnamefont {{Aggarwal}}}, \bibinfo {author} {\bibfnamefont {O.~D.}\
  \bibnamefont {{Aguiar}}}, \bibinfo {author} {\bibfnamefont {L.}~\bibnamefont
  {{Aiello}}}, \bibinfo {author} {\bibfnamefont {A.}~\bibnamefont {{Ain}}},
  \bibinfo {author} {\bibfnamefont {P.}~\bibnamefont {{Ajith}}}, \bibinfo
  {author} {\bibfnamefont {T.}~\bibnamefont {{Alford}}}, \bibinfo {author}
  {\bibfnamefont {G.}~\bibnamefont {{Allen}}}, \bibinfo {author} {\bibfnamefont
  {A.}~\bibnamefont {{Allocca}}}, \bibinfo {author} {\bibfnamefont {M.~A.}\
  \bibnamefont {{Aloy}}}, \bibinfo {author} {\bibfnamefont {P.~A.}\
  \bibnamefont {{Altin}}}, \bibinfo {author} {\bibfnamefont {A.}~\bibnamefont
  {{Amato}}}, \bibinfo {author} {\bibfnamefont {A.}~\bibnamefont {{Ananyeva}}},
  \bibinfo {author} {\bibfnamefont {S.~B.}\ \bibnamefont {{Anderson}}},
  \bibinfo {author} {\bibfnamefont {W.~G.}\ \bibnamefont {{Anderson}}},
  \bibinfo {author} {\bibfnamefont {S.~V.}\ \bibnamefont {{Angelova}}},
  \bibinfo {author} {\bibfnamefont {S.}~\bibnamefont {{Antier}}}, \bibinfo
  {author} {\bibfnamefont {S.}~\bibnamefont {{Appert}}}, \bibinfo {author}
  {\bibfnamefont {K.}~\bibnamefont {{Arai}}}, \bibinfo {author} {\bibfnamefont
  {M.~C.}\ \bibnamefont {{Araya}}}, \bibinfo {author} {\bibfnamefont {J.~S.}\
  \bibnamefont {{Areeda}}}, \bibinfo {author} {\bibfnamefont {M.}~\bibnamefont
  {{Ar{\`e}ne}}}, \bibinfo {author} {\bibfnamefont {N.}~\bibnamefont
  {{Arnaud}}}, \bibinfo {author} {\bibfnamefont {K.~G.}\ \bibnamefont
  {{Arun}}}, \bibinfo {author} {\bibfnamefont {S.}~\bibnamefont {{Ascenzi}}},
  \bibinfo {author} {\bibfnamefont {G.}~\bibnamefont {{Ashton}}}, \bibinfo
  {author} {\bibfnamefont {S.~M.}\ \bibnamefont {{Aston}}}, \bibinfo {author}
  {\bibfnamefont {P.}~\bibnamefont {{Astone}}}, \bibinfo {author}
  {\bibfnamefont {F.}~\bibnamefont {{Aubin}}}, \bibinfo {author} {\bibfnamefont
  {P.}~\bibnamefont {{Aufmuth}}}, \bibinfo {author} {\bibfnamefont
  {K.}~\bibnamefont {{AultONeal}}}, \bibinfo {author} {\bibfnamefont
  {C.}~\bibnamefont {{Austin}}}, \bibinfo {author} {\bibfnamefont
  {V.}~\bibnamefont {{Avendano}}}, \bibinfo {author} {\bibfnamefont
  {A.}~\bibnamefont {{Avila-Alvarez}}}, \bibinfo {author} {\bibfnamefont
  {S.}~\bibnamefont {{Babak}}}, \bibinfo {author} {\bibfnamefont
  {P.}~\bibnamefont {{Bacon}}}, \bibinfo {author} {\bibfnamefont
  {F.}~\bibnamefont {{Badaracco}}}, \bibinfo {author} {\bibfnamefont
  {M.~K.~M.}\ \bibnamefont {{Bader}}}, \bibinfo {author} {\bibfnamefont
  {S.}~\bibnamefont {{Bae}}}, \bibinfo {author} {\bibfnamefont {P.~T.}\
  \bibnamefont {{Baker}}}, \bibinfo {author} {\bibfnamefont {F.}~\bibnamefont
  {{Baldaccini}}}, \bibinfo {author} {\bibfnamefont {G.}~\bibnamefont
  {{Ballardin}}}, \bibinfo {author} {\bibfnamefont {S.~W.}\ \bibnamefont
  {{Ballmer}}}, \bibinfo {author} {\bibfnamefont {S.}~\bibnamefont
  {{Banagiri}}}, \bibinfo {author} {\bibfnamefont {J.~C.}\ \bibnamefont
  {{Barayoga}}}, \bibinfo {author} {\bibfnamefont {S.~E.}\ \bibnamefont
  {{Barclay}}}, \bibinfo {author} {\bibfnamefont {B.~C.}\ \bibnamefont
  {{Barish}}}, \bibinfo {author} {\bibfnamefont {D.}~\bibnamefont {{Barker}}},
  \bibinfo {author} {\bibfnamefont {K.}~\bibnamefont {{Barkett}}}, \bibinfo
  {author} {\bibfnamefont {S.}~\bibnamefont {{Barnum}}}, \bibinfo {author}
  {\bibfnamefont {F.}~\bibnamefont {{Barone}}}, \bibinfo {author}
  {\bibfnamefont {B.}~\bibnamefont {{Barr}}}, \bibinfo {author} {\bibfnamefont
  {L.}~\bibnamefont {{Barsotti}}}, \bibinfo {author} {\bibfnamefont
  {M.}~\bibnamefont {{Barsuglia}}}, \bibinfo {author} {\bibfnamefont
  {D.}~\bibnamefont {{Barta}}}, \bibinfo {author} {\bibfnamefont
  {J.}~\bibnamefont {{Bartlett}}}, \bibinfo {author} {\bibfnamefont
  {I.}~\bibnamefont {{Bartos}}}, \bibinfo {author} {\bibfnamefont
  {R.}~\bibnamefont {{Bassiri}}}, \bibinfo {author} {\bibfnamefont
  {A.}~\bibnamefont {{Basti}}}, \bibinfo {author} {\bibfnamefont
  {M.}~\bibnamefont {{Bawaj}}}, \bibinfo {author} {\bibfnamefont {J.~C.}\
  \bibnamefont {{Bayley}}}, \bibinfo {author} {\bibfnamefont {M.}~\bibnamefont
  {{Bazzan}}}, \bibinfo {author} {\bibfnamefont {B.}~\bibnamefont
  {{B{\'e}csy}}}, \bibinfo {author} {\bibfnamefont {M.}~\bibnamefont
  {{Bejger}}}, \bibinfo {author} {\bibfnamefont {I.}~\bibnamefont
  {{Belahcene}}}, \bibinfo {author} {\bibfnamefont {A.~S.}\ \bibnamefont
  {{Bell}}}, \bibinfo {author} {\bibfnamefont {D.}~\bibnamefont {{Beniwal}}},
  \bibinfo {author} {\bibfnamefont {B.~K.}\ \bibnamefont {{Berger}}}, \bibinfo
  {author} {\bibfnamefont {G.}~\bibnamefont {{Bergmann}}}, \bibinfo {author}
  {\bibfnamefont {S.}~\bibnamefont {{Bernuzzi}}}, \bibinfo {author}
  {\bibfnamefont {J.~J.}\ \bibnamefont {{Bero}}}, \bibinfo {author}
  {\bibfnamefont {C.~P.~L.}\ \bibnamefont {{Berry}}}, \bibinfo {author}
  {\bibfnamefont {D.}~\bibnamefont {{Bersanetti}}}, \bibinfo {author}
  {\bibfnamefont {A.}~\bibnamefont {{Bertolini}}}, \bibinfo {author}
  {\bibfnamefont {J.}~\bibnamefont {{Betzwieser}}}, \bibinfo {author}
  {\bibfnamefont {R.}~\bibnamefont {{Bhandare}}}, \bibinfo {author}
  {\bibfnamefont {J.}~\bibnamefont {{Bidler}}}, \bibinfo {author}
  {\bibfnamefont {I.~A.}\ \bibnamefont {{Bilenko}}}, \bibinfo {author}
  {\bibfnamefont {S.~A.}\ \bibnamefont {{Bilgili}}}, \bibinfo {author}
  {\bibfnamefont {G.}~\bibnamefont {{Billingsley}}}, \bibinfo {author}
  {\bibfnamefont {J.}~\bibnamefont {{Birch}}}, \bibinfo {author} {\bibfnamefont
  {R.}~\bibnamefont {{Birney}}}, \bibinfo {author} {\bibfnamefont
  {O.}~\bibnamefont {{Birnholtz}}}, \bibinfo {author} {\bibfnamefont
  {S.}~\bibnamefont {{Biscans}}}, \bibinfo {author} {\bibfnamefont
  {S.}~\bibnamefont {{Biscoveanu}}}, \bibinfo {author} {\bibfnamefont
  {A.}~\bibnamefont {{Bisht}}}, \bibinfo {author} {\bibfnamefont
  {M.}~\bibnamefont {{Bitossi}}}, \bibinfo {author} {\bibfnamefont {M.~A.}\
  \bibnamefont {{Bizouard}}}, \bibinfo {author} {\bibfnamefont {J.~K.}\
  \bibnamefont {{Blackburn}}}, \bibinfo {author} {\bibfnamefont {C.~D.}\
  \bibnamefont {{Blair}}}, \bibinfo {author} {\bibfnamefont {D.~G.}\
  \bibnamefont {{Blair}}}, \bibinfo {author} {\bibfnamefont {R.~M.}\
  \bibnamefont {{Blair}}}, \bibinfo {author} {\bibfnamefont {S.}~\bibnamefont
  {{Bloemen}}}, \bibinfo {author} {\bibfnamefont {N.}~\bibnamefont {{Bode}}},
  \bibinfo {author} {\bibfnamefont {M.}~\bibnamefont {{Boer}}}, \bibinfo
  {author} {\bibfnamefont {Y.}~\bibnamefont {{Boetzel}}}, \bibinfo {author}
  {\bibfnamefont {G.}~\bibnamefont {{Bogaert}}}, \bibinfo {author}
  {\bibfnamefont {F.}~\bibnamefont {{Bondu}}}, \bibinfo {author} {\bibfnamefont
  {E.}~\bibnamefont {{Bonilla}}}, \bibinfo {author} {\bibfnamefont
  {R.}~\bibnamefont {{Bonnand}}}, \bibinfo {author} {\bibfnamefont
  {P.}~\bibnamefont {{Booker}}}, \bibinfo {author} {\bibfnamefont {B.~A.}\
  \bibnamefont {{Boom}}}, \bibinfo {author} {\bibfnamefont {C.~D.}\
  \bibnamefont {{Booth}}}, \bibinfo {author} {\bibfnamefont {R.}~\bibnamefont
  {{Bork}}}, \bibinfo {author} {\bibfnamefont {V.}~\bibnamefont {{Boschi}}},
  \bibinfo {author} {\bibfnamefont {S.}~\bibnamefont {{Bose}}}, \bibinfo
  {author} {\bibfnamefont {K.}~\bibnamefont {{Bossie}}}, \bibinfo {author}
  {\bibfnamefont {V.}~\bibnamefont {{Bossilkov}}}, \bibinfo {author}
  {\bibfnamefont {J.}~\bibnamefont {{Bosveld}}}, \bibinfo {author}
  {\bibfnamefont {Y.}~\bibnamefont {{Bouffanais}}}, \bibinfo {author}
  {\bibfnamefont {A.}~\bibnamefont {{Bozzi}}}, \bibinfo {author} {\bibfnamefont
  {C.}~\bibnamefont {{Bradaschia}}}, \bibinfo {author} {\bibfnamefont {P.~R.}\
  \bibnamefont {{Brady}}}, \bibinfo {author} {\bibfnamefont {A.}~\bibnamefont
  {{Bramley}}}, \bibinfo {author} {\bibfnamefont {M.}~\bibnamefont
  {{Branchesi}}}, \bibinfo {author} {\bibfnamefont {J.~E.}\ \bibnamefont
  {{Brau}}}, \bibinfo {author} {\bibfnamefont {T.}~\bibnamefont {{Briant}}},
  \bibinfo {author} {\bibfnamefont {J.~H.}\ \bibnamefont {{Briggs}}}, \bibinfo
  {author} {\bibfnamefont {F.}~\bibnamefont {{Brighenti}}}, \bibinfo {author}
  {\bibfnamefont {A.}~\bibnamefont {{Brillet}}}, \bibinfo {author}
  {\bibfnamefont {M.}~\bibnamefont {{Brinkmann}}}, \bibinfo {author}
  {\bibfnamefont {V.}~\bibnamefont {{Brisson}}}, \bibinfo {author}
  {\bibfnamefont {P.}~\bibnamefont {{Brockill}}}, \bibinfo {author}
  {\bibfnamefont {A.~F.}\ \bibnamefont {{Brooks}}}, \bibinfo {author}
  {\bibfnamefont {D.~D.}\ \bibnamefont {{Brown}}}, \bibinfo {author}
  {\bibfnamefont {S.}~\bibnamefont {{Brunett}}}, \bibinfo {author}
  {\bibfnamefont {A.}~\bibnamefont {{Buikema}}}, \bibinfo {author}
  {\bibfnamefont {T.}~\bibnamefont {{Bulik}}}, \bibinfo {author} {\bibfnamefont
  {H.~J.}\ \bibnamefont {{Bulten}}}, \bibinfo {author} {\bibfnamefont
  {A.}~\bibnamefont {{Buonanno}}}, \bibinfo {author} {\bibfnamefont
  {D.}~\bibnamefont {{Buskulic}}}, \bibinfo {author} {\bibfnamefont
  {C.}~\bibnamefont {{Buy}}}, \bibinfo {author} {\bibfnamefont {R.~L.}\
  \bibnamefont {{Byer}}}, \bibinfo {author} {\bibfnamefont {M.}~\bibnamefont
  {{Cabero}}}, \bibinfo {author} {\bibfnamefont {L.}~\bibnamefont
  {{Cadonati}}}, \bibinfo {author} {\bibfnamefont {G.}~\bibnamefont
  {{Cagnoli}}}, \bibinfo {author} {\bibfnamefont {C.}~\bibnamefont
  {{Cahillane}}}, \bibinfo {author} {\bibfnamefont {J.}~\bibnamefont
  {{Calder{\'o}n Bustillo}}}, \bibinfo {author} {\bibfnamefont {T.~A.}\
  \bibnamefont {{Callister}}}, \bibinfo {author} {\bibfnamefont
  {E.}~\bibnamefont {{Calloni}}}, \bibinfo {author} {\bibfnamefont {J.~B.}\
  \bibnamefont {{Camp}}}, \bibinfo {author} {\bibfnamefont {W.~A.}\
  \bibnamefont {{Campbell}}}, \bibinfo {author} {\bibfnamefont
  {M.}~\bibnamefont {{Canepa}}}, \bibinfo {author} {\bibfnamefont {K.~C.}\
  \bibnamefont {{Cannon}}}, \bibinfo {author} {\bibfnamefont {H.}~\bibnamefont
  {{Cao}}}, \bibinfo {author} {\bibfnamefont {J.}~\bibnamefont {{Cao}}},
  \bibinfo {author} {\bibfnamefont {E.}~\bibnamefont {{Capocasa}}}, \bibinfo
  {author} {\bibfnamefont {F.}~\bibnamefont {{Carbognani}}}, \bibinfo {author}
  {\bibfnamefont {S.}~\bibnamefont {{Caride}}}, \bibinfo {author}
  {\bibfnamefont {M.~F.}\ \bibnamefont {{Carney}}}, \bibinfo {author}
  {\bibfnamefont {G.}~\bibnamefont {{Carullo}}}, \bibinfo {author}
  {\bibfnamefont {J.}~\bibnamefont {{Casanueva Diaz}}}, \bibinfo {author}
  {\bibfnamefont {C.}~\bibnamefont {{Casentini}}}, \bibinfo {author}
  {\bibfnamefont {S.}~\bibnamefont {{Caudill}}}, \bibinfo {author}
  {\bibfnamefont {M.}~\bibnamefont {{Cavagli{\`a}}}}, \bibinfo {author}
  {\bibfnamefont {F.}~\bibnamefont {{Cavalier}}}, \bibinfo {author}
  {\bibfnamefont {R.}~\bibnamefont {{Cavalieri}}}, \bibinfo {author}
  {\bibfnamefont {G.}~\bibnamefont {{Cella}}}, \bibinfo {author} {\bibfnamefont
  {P.}~\bibnamefont {{Cerd{\'a}-Dur{\'a}n}}}, \bibinfo {author} {\bibfnamefont
  {G.}~\bibnamefont {{Cerretani}}}, \bibinfo {author} {\bibfnamefont
  {E.}~\bibnamefont {{Cesarini}}}, \bibinfo {author} {\bibfnamefont
  {O.}~\bibnamefont {{Chaibi}}}, \bibinfo {author} {\bibfnamefont
  {K.}~\bibnamefont {{Chakravarti}}}, \bibinfo {author} {\bibfnamefont {S.~J.}\
  \bibnamefont {{Chamberlin}}}, \bibinfo {author} {\bibfnamefont
  {M.}~\bibnamefont {{Chan}}}, \bibinfo {author} {\bibfnamefont
  {S.}~\bibnamefont {{Chao}}}, \bibinfo {author} {\bibfnamefont
  {P.}~\bibnamefont {{Charlton}}}, \bibinfo {author} {\bibfnamefont {E.~A.}\
  \bibnamefont {{Chase}}}, \bibinfo {author} {\bibfnamefont {E.}~\bibnamefont
  {{Chassande-Mottin}}}, \bibinfo {author} {\bibfnamefont {D.}~\bibnamefont
  {{Chatterjee}}}, \bibinfo {author} {\bibfnamefont {M.}~\bibnamefont
  {{Chaturvedi}}}, \bibinfo {author} {\bibfnamefont {K.}~\bibnamefont
  {{Chatziioannou}}}, \bibinfo {author} {\bibfnamefont {B.~D.}\ \bibnamefont
  {{Cheeseboro}}}, \bibinfo {author} {\bibfnamefont {H.~Y.}\ \bibnamefont
  {{Chen}}}, \bibinfo {author} {\bibfnamefont {X.}~\bibnamefont {{Chen}}},
  \bibinfo {author} {\bibfnamefont {Y.}~\bibnamefont {{Chen}}}, \bibinfo
  {author} {\bibfnamefont {H.~P.}\ \bibnamefont {{Cheng}}}, \bibinfo {author}
  {\bibfnamefont {C.~K.}\ \bibnamefont {{Cheong}}}, \bibinfo {author}
  {\bibfnamefont {H.~Y.}\ \bibnamefont {{Chia}}}, \bibinfo {author}
  {\bibfnamefont {A.}~\bibnamefont {{Chincarini}}}, \bibinfo {author}
  {\bibfnamefont {A.}~\bibnamefont {{Chiummo}}}, \bibinfo {author}
  {\bibfnamefont {G.}~\bibnamefont {{Cho}}}, \bibinfo {author} {\bibfnamefont
  {H.~S.}\ \bibnamefont {{Cho}}}, \bibinfo {author} {\bibfnamefont
  {M.}~\bibnamefont {{Cho}}}, \bibinfo {author} {\bibfnamefont
  {N.}~\bibnamefont {{Christensen}}}, \bibinfo {author} {\bibfnamefont
  {Q.}~\bibnamefont {{Chu}}}, \bibinfo {author} {\bibfnamefont
  {S.}~\bibnamefont {{Chua}}}, \bibinfo {author} {\bibfnamefont {K.~W.}\
  \bibnamefont {{Chung}}}, \bibinfo {author} {\bibfnamefont {S.}~\bibnamefont
  {{Chung}}}, \bibinfo {author} {\bibfnamefont {G.}~\bibnamefont {{Ciani}}},
  \bibinfo {author} {\bibfnamefont {A.~A.}\ \bibnamefont {{Ciobanu}}}, \bibinfo
  {author} {\bibfnamefont {R.}~\bibnamefont {{Ciolfi}}}, \bibinfo {author}
  {\bibfnamefont {F.}~\bibnamefont {{Cipriano}}}, \bibinfo {author}
  {\bibfnamefont {A.}~\bibnamefont {{Cirone}}}, \bibinfo {author}
  {\bibfnamefont {F.}~\bibnamefont {{Clara}}}, \bibinfo {author} {\bibfnamefont
  {J.~A.}\ \bibnamefont {{Clark}}}, \bibinfo {author} {\bibfnamefont
  {P.}~\bibnamefont {{Clearwater}}}, \bibinfo {author} {\bibfnamefont
  {F.}~\bibnamefont {{Cleva}}}, \bibinfo {author} {\bibfnamefont
  {C.}~\bibnamefont {{Cocchieri}}}, \bibinfo {author} {\bibfnamefont
  {E.}~\bibnamefont {{Coccia}}}, \bibinfo {author} {\bibfnamefont {P.~F.}\
  \bibnamefont {{Cohadon}}}, \bibinfo {author} {\bibfnamefont {D.}~\bibnamefont
  {{Cohen}}}, \bibinfo {author} {\bibfnamefont {R.}~\bibnamefont {{Colgan}}},
  \bibinfo {author} {\bibfnamefont {M.}~\bibnamefont {{Colleoni}}}, \bibinfo
  {author} {\bibfnamefont {C.~G.}\ \bibnamefont {{Collette}}}, \bibinfo
  {author} {\bibfnamefont {C.}~\bibnamefont {{Collins}}}, \bibinfo {author}
  {\bibfnamefont {L.~R.}\ \bibnamefont {{Cominsky}}}, \bibinfo {author}
  {\bibfnamefont {J.}~\bibnamefont {{Constancio}}, \bibfnamefont {M.}},
  \bibinfo {author} {\bibfnamefont {L.}~\bibnamefont {{Conti}}}, \bibinfo
  {author} {\bibfnamefont {S.~J.}\ \bibnamefont {{Cooper}}}, \bibinfo {author}
  {\bibfnamefont {P.}~\bibnamefont {{Corban}}}, \bibinfo {author}
  {\bibfnamefont {T.~R.}\ \bibnamefont {{Corbitt}}}, \bibinfo {author}
  {\bibfnamefont {I.}~\bibnamefont {{Cordero-Carri{\'o}n}}}, \bibinfo {author}
  {\bibfnamefont {K.~R.}\ \bibnamefont {{Corley}}}, \bibinfo {author}
  {\bibfnamefont {N.}~\bibnamefont {{Cornish}}}, \bibinfo {author}
  {\bibfnamefont {A.}~\bibnamefont {{Corsi}}}, \bibinfo {author} {\bibfnamefont
  {S.}~\bibnamefont {{Cortese}}}, \bibinfo {author} {\bibfnamefont {C.~A.}\
  \bibnamefont {{Costa}}}, \bibinfo {author} {\bibfnamefont {R.}~\bibnamefont
  {{Cotesta}}}, \bibinfo {author} {\bibfnamefont {M.~W.}\ \bibnamefont
  {{Coughlin}}}, \bibinfo {author} {\bibfnamefont {S.~B.}\ \bibnamefont
  {{Coughlin}}}, \bibinfo {author} {\bibfnamefont {J.~P.}\ \bibnamefont
  {{Coulon}}}, \bibinfo {author} {\bibfnamefont {S.~T.}\ \bibnamefont
  {{Countryman}}}, \bibinfo {author} {\bibfnamefont {P.}~\bibnamefont
  {{Couvares}}}, \bibinfo {author} {\bibfnamefont {P.~B.}\ \bibnamefont
  {{Covas}}}, \bibinfo {author} {\bibfnamefont {E.~E.}\ \bibnamefont
  {{Cowan}}}, \bibinfo {author} {\bibfnamefont {D.~M.}\ \bibnamefont
  {{Coward}}}, \bibinfo {author} {\bibfnamefont {M.~J.}\ \bibnamefont
  {{Cowart}}}, \bibinfo {author} {\bibfnamefont {D.~C.}\ \bibnamefont
  {{Coyne}}}, \bibinfo {author} {\bibfnamefont {R.}~\bibnamefont {{Coyne}}},
  \bibinfo {author} {\bibfnamefont {J.~D.~E.}\ \bibnamefont {{Creighton}}},
  \bibinfo {author} {\bibfnamefont {T.~D.}\ \bibnamefont {{Creighton}}},
  \bibinfo {author} {\bibfnamefont {J.}~\bibnamefont {{Cripe}}}, \bibinfo
  {author} {\bibfnamefont {M.}~\bibnamefont {{Croquette}}}, \bibinfo {author}
  {\bibfnamefont {S.~G.}\ \bibnamefont {{Crowder}}}, \bibinfo {author}
  {\bibfnamefont {T.~J.}\ \bibnamefont {{Cullen}}}, \bibinfo {author}
  {\bibfnamefont {A.}~\bibnamefont {{Cumming}}}, \bibinfo {author}
  {\bibfnamefont {L.}~\bibnamefont {{Cunningham}}}, \bibinfo {author}
  {\bibfnamefont {E.}~\bibnamefont {{Cuoco}}}, \bibinfo {author} {\bibfnamefont
  {T.}~\bibnamefont {{Dal Canton}}}, \bibinfo {author} {\bibfnamefont
  {G.}~\bibnamefont {{D{\'a}lya}}}, \bibinfo {author} {\bibfnamefont {S.~L.}\
  \bibnamefont {{Danilishin}}}, \bibinfo {author} {\bibfnamefont
  {S.}~\bibnamefont {{D'Antonio}}}, \bibinfo {author} {\bibfnamefont
  {K.}~\bibnamefont {{Danzmann}}}, \bibinfo {author} {\bibfnamefont
  {A.}~\bibnamefont {{Dasgupta}}}, \bibinfo {author} {\bibfnamefont {C.~F.}\
  \bibnamefont {{Da Silva Costa}}}, \bibinfo {author} {\bibfnamefont
  {L.~E.~H.}\ \bibnamefont {{Datrier}}}, \bibinfo {author} {\bibfnamefont
  {V.}~\bibnamefont {{Dattilo}}}, \bibinfo {author} {\bibfnamefont
  {I.}~\bibnamefont {{Dave}}}, \bibinfo {author} {\bibfnamefont
  {M.}~\bibnamefont {{Davier}}}, \bibinfo {author} {\bibfnamefont
  {D.}~\bibnamefont {{Davis}}}, \bibinfo {author} {\bibfnamefont {E.~J.}\
  \bibnamefont {{Daw}}}, \bibinfo {author} {\bibfnamefont {D.}~\bibnamefont
  {{DeBra}}}, \bibinfo {author} {\bibfnamefont {M.}~\bibnamefont
  {{Deenadayalan}}}, \bibinfo {author} {\bibfnamefont {J.}~\bibnamefont
  {{Degallaix}}}, \bibinfo {author} {\bibfnamefont {M.}~\bibnamefont {{De
  Laurentis}}}, \bibinfo {author} {\bibfnamefont {S.}~\bibnamefont
  {{Del{\'e}glise}}}, \bibinfo {author} {\bibfnamefont {W.}~\bibnamefont {{Del
  Pozzo}}}, \bibinfo {author} {\bibfnamefont {L.~M.}\ \bibnamefont
  {{DeMarchi}}}, \bibinfo {author} {\bibfnamefont {N.}~\bibnamefont {{Demos}}},
  \bibinfo {author} {\bibfnamefont {T.}~\bibnamefont {{Dent}}}, \bibinfo
  {author} {\bibfnamefont {R.}~\bibnamefont {{De Pietri}}}, \bibinfo {author}
  {\bibfnamefont {J.}~\bibnamefont {{Derby}}}, \bibinfo {author} {\bibfnamefont
  {R.}~\bibnamefont {{De Rosa}}}, \bibinfo {author} {\bibfnamefont
  {C.}~\bibnamefont {{De Rossi}}}, \bibinfo {author} {\bibfnamefont
  {R.}~\bibnamefont {{DeSalvo}}}, \bibinfo {author} {\bibfnamefont
  {O.}~\bibnamefont {{de Varona}}}, \bibinfo {author} {\bibfnamefont
  {S.}~\bibnamefont {{Dhurandhar}}}, \bibinfo {author} {\bibfnamefont {M.~C.}\
  \bibnamefont {{D{\'\i}az}}}, \bibinfo {author} {\bibfnamefont
  {T.}~\bibnamefont {{Dietrich}}}, \bibinfo {author} {\bibfnamefont
  {L.}~\bibnamefont {{Di Fiore}}}, \bibinfo {author} {\bibfnamefont
  {M.}~\bibnamefont {{Di Giovanni}}}, \bibinfo {author} {\bibfnamefont
  {T.}~\bibnamefont {{Di Girolamo}}}, \bibinfo {author} {\bibfnamefont
  {A.}~\bibnamefont {{Di Lieto}}}, \bibinfo {author} {\bibfnamefont
  {B.}~\bibnamefont {{Ding}}}, \bibinfo {author} {\bibfnamefont
  {S.}~\bibnamefont {{Di Pace}}}, \bibinfo {author} {\bibfnamefont
  {I.}~\bibnamefont {{Di Palma}}}, \bibinfo {author} {\bibfnamefont
  {F.}~\bibnamefont {{Di Renzo}}}, \bibinfo {author} {\bibfnamefont
  {A.}~\bibnamefont {{Dmitriev}}}, \bibinfo {author} {\bibfnamefont
  {Z.}~\bibnamefont {{Doctor}}}, \bibinfo {author} {\bibfnamefont
  {F.}~\bibnamefont {{Donovan}}}, \bibinfo {author} {\bibfnamefont {K.~L.}\
  \bibnamefont {{Dooley}}}, \bibinfo {author} {\bibfnamefont {S.}~\bibnamefont
  {{Doravari}}}, \bibinfo {author} {\bibfnamefont {I.}~\bibnamefont
  {{Dorrington}}}, \bibinfo {author} {\bibfnamefont {T.~P.}\ \bibnamefont
  {{Downes}}}, \bibinfo {author} {\bibfnamefont {M.}~\bibnamefont {{Drago}}},
  \bibinfo {author} {\bibfnamefont {J.~C.}\ \bibnamefont {{Driggers}}},
  \bibinfo {author} {\bibfnamefont {Z.}~\bibnamefont {{Du}}}, \bibinfo {author}
  {\bibfnamefont {J.~G.}\ \bibnamefont {{Ducoin}}}, \bibinfo {author}
  {\bibfnamefont {P.}~\bibnamefont {{Dupej}}}, \bibinfo {author} {\bibfnamefont
  {S.~E.}\ \bibnamefont {{Dwyer}}}, \bibinfo {author} {\bibfnamefont {P.~J.}\
  \bibnamefont {{Easter}}}, \bibinfo {author} {\bibfnamefont {T.~B.}\
  \bibnamefont {{Edo}}}, \bibinfo {author} {\bibfnamefont {M.~C.}\ \bibnamefont
  {{Edwards}}}, \bibinfo {author} {\bibfnamefont {A.}~\bibnamefont {{Effler}}},
  \bibinfo {author} {\bibfnamefont {P.}~\bibnamefont {{Ehrens}}}, \bibinfo
  {author} {\bibfnamefont {J.}~\bibnamefont {{Eichholz}}}, \bibinfo {author}
  {\bibfnamefont {S.~S.}\ \bibnamefont {{Eikenberry}}}, \bibinfo {author}
  {\bibfnamefont {M.}~\bibnamefont {{Eisenmann}}}, \bibinfo {author}
  {\bibfnamefont {R.~A.}\ \bibnamefont {{Eisenstein}}}, \bibinfo {author}
  {\bibfnamefont {R.~C.}\ \bibnamefont {{Essick}}}, \bibinfo {author}
  {\bibfnamefont {H.}~\bibnamefont {{Estelles}}}, \bibinfo {author}
  {\bibfnamefont {D.}~\bibnamefont {{Estevez}}}, \bibinfo {author}
  {\bibfnamefont {Z.~B.}\ \bibnamefont {{Etienne}}}, \bibinfo {author}
  {\bibfnamefont {T.}~\bibnamefont {{Etzel}}}, \bibinfo {author} {\bibfnamefont
  {M.}~\bibnamefont {{Evans}}}, \bibinfo {author} {\bibfnamefont {T.~M.}\
  \bibnamefont {{Evans}}}, \bibinfo {author} {\bibfnamefont {V.}~\bibnamefont
  {{Fafone}}}, \bibinfo {author} {\bibfnamefont {H.}~\bibnamefont {{Fair}}},
  \bibinfo {author} {\bibfnamefont {S.}~\bibnamefont {{Fairhurst}}}, \bibinfo
  {author} {\bibfnamefont {X.}~\bibnamefont {{Fan}}}, \bibinfo {author}
  {\bibfnamefont {S.}~\bibnamefont {{Farinon}}}, \bibinfo {author}
  {\bibfnamefont {B.}~\bibnamefont {{Farr}}}, \bibinfo {author} {\bibfnamefont
  {W.~M.}\ \bibnamefont {{Farr}}}, \bibinfo {author} {\bibfnamefont {E.~J.}\
  \bibnamefont {{Fauchon-Jones}}}, \bibinfo {author} {\bibfnamefont
  {M.}~\bibnamefont {{Favata}}}, \bibinfo {author} {\bibfnamefont
  {M.}~\bibnamefont {{Fays}}}, \bibinfo {author} {\bibfnamefont
  {M.}~\bibnamefont {{Fazio}}}, \bibinfo {author} {\bibfnamefont
  {C.}~\bibnamefont {{Fee}}}, \bibinfo {author} {\bibfnamefont
  {J.}~\bibnamefont {{Feicht}}}, \bibinfo {author} {\bibfnamefont {M.~M.}\
  \bibnamefont {{Fejer}}}, \bibinfo {author} {\bibfnamefont {F.}~\bibnamefont
  {{Feng}}}, \bibinfo {author} {\bibfnamefont {A.}~\bibnamefont
  {{Fernandez-Galiana}}}, \bibinfo {author} {\bibfnamefont {I.}~\bibnamefont
  {{Ferrante}}}, \bibinfo {author} {\bibfnamefont {E.~C.}\ \bibnamefont
  {{Ferreira}}}, \bibinfo {author} {\bibfnamefont {T.~A.}\ \bibnamefont
  {{Ferreira}}}, \bibinfo {author} {\bibfnamefont {F.}~\bibnamefont
  {{Ferrini}}}, \bibinfo {author} {\bibfnamefont {F.}~\bibnamefont
  {{Fidecaro}}}, \bibinfo {author} {\bibfnamefont {I.}~\bibnamefont {{Fiori}}},
  \bibinfo {author} {\bibfnamefont {D.}~\bibnamefont {{Fiorucci}}}, \bibinfo
  {author} {\bibfnamefont {M.}~\bibnamefont {{Fishbach}}}, \bibinfo {author}
  {\bibfnamefont {R.~P.}\ \bibnamefont {{Fisher}}}, \bibinfo {author}
  {\bibfnamefont {J.~M.}\ \bibnamefont {{Fishner}}}, \bibinfo {author}
  {\bibfnamefont {M.}~\bibnamefont {{Fitz-Axen}}}, \bibinfo {author}
  {\bibfnamefont {R.}~\bibnamefont {{Flaminio}}}, \bibinfo {author}
  {\bibfnamefont {M.}~\bibnamefont {{Fletcher}}}, \bibinfo {author}
  {\bibfnamefont {E.}~\bibnamefont {{Flynn}}}, \bibinfo {author} {\bibfnamefont
  {H.}~\bibnamefont {{Fong}}}, \bibinfo {author} {\bibfnamefont {J.~A.}\
  \bibnamefont {{Font}}}, \bibinfo {author} {\bibfnamefont {P.~W.~F.}\
  \bibnamefont {{Forsyth}}}, \bibinfo {author} {\bibfnamefont {J.~D.}\
  \bibnamefont {{Fournier}}}, \bibinfo {author} {\bibfnamefont
  {S.}~\bibnamefont {{Frasca}}}, \bibinfo {author} {\bibfnamefont
  {F.}~\bibnamefont {{Frasconi}}}, \bibinfo {author} {\bibfnamefont
  {Z.}~\bibnamefont {{Frei}}}, \bibinfo {author} {\bibfnamefont
  {A.}~\bibnamefont {{Freise}}}, \bibinfo {author} {\bibfnamefont
  {R.}~\bibnamefont {{Frey}}}, \bibinfo {author} {\bibfnamefont
  {V.}~\bibnamefont {{Frey}}}, \bibinfo {author} {\bibfnamefont
  {P.}~\bibnamefont {{Fritschel}}}, \bibinfo {author} {\bibfnamefont {V.~V.}\
  \bibnamefont {{Frolov}}}, \bibinfo {author} {\bibfnamefont {P.}~\bibnamefont
  {{Fulda}}}, \bibinfo {author} {\bibfnamefont {M.}~\bibnamefont {{Fyffe}}},
  \bibinfo {author} {\bibfnamefont {H.~A.}\ \bibnamefont {{Gabbard}}}, \bibinfo
  {author} {\bibfnamefont {B.~U.}\ \bibnamefont {{Gadre}}}, \bibinfo {author}
  {\bibfnamefont {S.~M.}\ \bibnamefont {{Gaebel}}}, \bibinfo {author}
  {\bibfnamefont {J.~R.}\ \bibnamefont {{Gair}}}, \bibinfo {author}
  {\bibfnamefont {L.}~\bibnamefont {{Gammaitoni}}}, \bibinfo {author}
  {\bibfnamefont {M.~R.}\ \bibnamefont {{Ganija}}}, \bibinfo {author}
  {\bibfnamefont {S.~G.}\ \bibnamefont {{Gaonkar}}}, \bibinfo {author}
  {\bibfnamefont {A.}~\bibnamefont {{Garcia}}}, \bibinfo {author}
  {\bibfnamefont {C.}~\bibnamefont {{Garc{\'\i}a-Quir{\'o}s}}}, \bibinfo
  {author} {\bibfnamefont {F.}~\bibnamefont {{Garufi}}}, \bibinfo {author}
  {\bibfnamefont {B.}~\bibnamefont {{Gateley}}}, \bibinfo {author}
  {\bibfnamefont {S.}~\bibnamefont {{Gaudio}}}, \bibinfo {author}
  {\bibfnamefont {G.}~\bibnamefont {{Gaur}}}, \bibinfo {author} {\bibfnamefont
  {V.}~\bibnamefont {{Gayathri}}}, \bibinfo {author} {\bibfnamefont
  {G.}~\bibnamefont {{Gemme}}}, \bibinfo {author} {\bibfnamefont
  {E.}~\bibnamefont {{Genin}}}, \bibinfo {author} {\bibfnamefont
  {A.}~\bibnamefont {{Gennai}}}, \bibinfo {author} {\bibfnamefont
  {D.}~\bibnamefont {{George}}}, \bibinfo {author} {\bibfnamefont
  {J.}~\bibnamefont {{George}}}, \bibinfo {author} {\bibfnamefont
  {L.}~\bibnamefont {{Gergely}}}, \bibinfo {author} {\bibfnamefont
  {V.}~\bibnamefont {{Germain}}}, \bibinfo {author} {\bibfnamefont
  {S.}~\bibnamefont {{Ghonge}}}, \bibinfo {author} {\bibfnamefont
  {A.}~\bibnamefont {{Ghosh}}}, \bibinfo {author} {\bibfnamefont
  {A.}~\bibnamefont {{Ghosh}}}, \bibinfo {author} {\bibfnamefont
  {S.}~\bibnamefont {{Ghosh}}}, \bibinfo {author} {\bibfnamefont
  {B.}~\bibnamefont {{Giacomazzo}}}, \bibinfo {author} {\bibfnamefont {J.~A.}\
  \bibnamefont {{Giaime}}}, \bibinfo {author} {\bibfnamefont {K.~D.}\
  \bibnamefont {{Giardina}}}, \bibinfo {author} {\bibfnamefont
  {A.}~\bibnamefont {{Giazotto}}}, \bibinfo {author} {\bibfnamefont
  {K.}~\bibnamefont {{Gill}}}, \bibinfo {author} {\bibfnamefont
  {G.}~\bibnamefont {{Giordano}}}, \bibinfo {author} {\bibfnamefont
  {L.}~\bibnamefont {{Glover}}}, \bibinfo {author} {\bibfnamefont
  {P.}~\bibnamefont {{Godwin}}}, \bibinfo {author} {\bibfnamefont
  {E.}~\bibnamefont {{Goetz}}}, \bibinfo {author} {\bibfnamefont
  {R.}~\bibnamefont {{Goetz}}}, \bibinfo {author} {\bibfnamefont
  {B.}~\bibnamefont {{Goncharov}}}, \bibinfo {author} {\bibfnamefont
  {G.}~\bibnamefont {{Gonz{\'a}lez}}}, \bibinfo {author} {\bibfnamefont
  {J.~M.}\ \bibnamefont {{Gonzalez Castro}}}, \bibinfo {author} {\bibfnamefont
  {A.}~\bibnamefont {{Gopakumar}}}, \bibinfo {author} {\bibfnamefont {M.~L.}\
  \bibnamefont {{Gorodetsky}}}, \bibinfo {author} {\bibfnamefont {S.~E.}\
  \bibnamefont {{Gossan}}}, \bibinfo {author} {\bibfnamefont {M.}~\bibnamefont
  {{Gosselin}}}, \bibinfo {author} {\bibfnamefont {R.}~\bibnamefont
  {{Gouaty}}}, \bibinfo {author} {\bibfnamefont {A.}~\bibnamefont {{Grado}}},
  \bibinfo {author} {\bibfnamefont {C.}~\bibnamefont {{Graef}}}, \bibinfo
  {author} {\bibfnamefont {M.}~\bibnamefont {{Granata}}}, \bibinfo {author}
  {\bibfnamefont {A.}~\bibnamefont {{Grant}}}, \bibinfo {author} {\bibfnamefont
  {S.}~\bibnamefont {{Gras}}}, \bibinfo {author} {\bibfnamefont
  {P.}~\bibnamefont {{Grassia}}}, \bibinfo {author} {\bibfnamefont
  {C.}~\bibnamefont {{Gray}}}, \bibinfo {author} {\bibfnamefont
  {R.}~\bibnamefont {{Gray}}}, \bibinfo {author} {\bibfnamefont
  {G.}~\bibnamefont {{Greco}}}, \bibinfo {author} {\bibfnamefont {A.~C.}\
  \bibnamefont {{Green}}}, \bibinfo {author} {\bibfnamefont {R.}~\bibnamefont
  {{Green}}}, \bibinfo {author} {\bibfnamefont {E.~M.}\ \bibnamefont
  {{Gretarsson}}}, \bibinfo {author} {\bibfnamefont {P.}~\bibnamefont
  {{Groot}}}, \bibinfo {author} {\bibfnamefont {H.}~\bibnamefont {{Grote}}},
  \bibinfo {author} {\bibfnamefont {S.}~\bibnamefont {{Grunewald}}}, \bibinfo
  {author} {\bibfnamefont {P.}~\bibnamefont {{Gruning}}}, \bibinfo {author}
  {\bibfnamefont {G.~M.}\ \bibnamefont {{Guidi}}}, \bibinfo {author}
  {\bibfnamefont {H.~K.}\ \bibnamefont {{Gulati}}}, \bibinfo {author}
  {\bibfnamefont {Y.}~\bibnamefont {{Guo}}}, \bibinfo {author} {\bibfnamefont
  {A.}~\bibnamefont {{Gupta}}}, \bibinfo {author} {\bibfnamefont {M.~K.}\
  \bibnamefont {{Gupta}}}, \bibinfo {author} {\bibfnamefont {E.~K.}\
  \bibnamefont {{Gustafson}}}, \bibinfo {author} {\bibfnamefont
  {R.}~\bibnamefont {{Gustafson}}}, \bibinfo {author} {\bibfnamefont
  {L.}~\bibnamefont {{Haegel}}}, \bibinfo {author} {\bibfnamefont
  {O.}~\bibnamefont {{Halim}}}, \bibinfo {author} {\bibfnamefont {B.~R.}\
  \bibnamefont {{Hall}}}, \bibinfo {author} {\bibfnamefont {E.~D.}\
  \bibnamefont {{Hall}}}, \bibinfo {author} {\bibfnamefont {E.~Z.}\
  \bibnamefont {{Hamilton}}}, \bibinfo {author} {\bibfnamefont
  {G.}~\bibnamefont {{Hammond}}}, \bibinfo {author} {\bibfnamefont
  {M.}~\bibnamefont {{Haney}}}, \bibinfo {author} {\bibfnamefont {M.~M.}\
  \bibnamefont {{Hanke}}}, \bibinfo {author} {\bibfnamefont {J.}~\bibnamefont
  {{Hanks}}}, \bibinfo {author} {\bibfnamefont {C.}~\bibnamefont {{Hanna}}},
  \bibinfo {author} {\bibfnamefont {M.~D.}\ \bibnamefont {{Hannam}}}, \bibinfo
  {author} {\bibfnamefont {O.~A.}\ \bibnamefont {{Hannuksela}}}, \bibinfo
  {author} {\bibfnamefont {J.}~\bibnamefont {{Hanson}}}, \bibinfo {author}
  {\bibfnamefont {T.}~\bibnamefont {{Hardwick}}}, \bibinfo {author}
  {\bibfnamefont {K.}~\bibnamefont {{Haris}}}, \bibinfo {author} {\bibfnamefont
  {J.}~\bibnamefont {{Harms}}}, \bibinfo {author} {\bibfnamefont {G.~M.}\
  \bibnamefont {{Harry}}}, \bibinfo {author} {\bibfnamefont {I.~W.}\
  \bibnamefont {{Harry}}}, \bibinfo {author} {\bibfnamefont {C.~J.}\
  \bibnamefont {{Haster}}}, \bibinfo {author} {\bibfnamefont {K.}~\bibnamefont
  {{Haughian}}}, \bibinfo {author} {\bibfnamefont {F.~J.}\ \bibnamefont
  {{Hayes}}}, \bibinfo {author} {\bibfnamefont {J.}~\bibnamefont {{Healy}}},
  \bibinfo {author} {\bibfnamefont {A.}~\bibnamefont {{Heidmann}}}, \bibinfo
  {author} {\bibfnamefont {M.~C.}\ \bibnamefont {{Heintze}}}, \bibinfo {author}
  {\bibfnamefont {H.}~\bibnamefont {{Heitmann}}}, \bibinfo {author}
  {\bibfnamefont {P.}~\bibnamefont {{Hello}}}, \bibinfo {author} {\bibfnamefont
  {G.}~\bibnamefont {{Hemming}}}, \bibinfo {author} {\bibfnamefont
  {M.}~\bibnamefont {{Hendry}}}, \bibinfo {author} {\bibfnamefont {I.~S.}\
  \bibnamefont {{Heng}}}, \bibinfo {author} {\bibfnamefont {J.}~\bibnamefont
  {{Hennig}}}, \bibinfo {author} {\bibfnamefont {A.~W.}\ \bibnamefont
  {{Heptonstall}}}, \bibinfo {author} {\bibfnamefont {F.~H.}\ \bibnamefont
  {{Vivanco}}}, \bibinfo {author} {\bibfnamefont {M.}~\bibnamefont {{Heurs}}},
  \bibinfo {author} {\bibfnamefont {S.}~\bibnamefont {{Hild}}}, \bibinfo
  {author} {\bibfnamefont {T.}~\bibnamefont {{Hinderer}}}, \bibinfo {author}
  {\bibfnamefont {D.}~\bibnamefont {{Hoak}}}, \bibinfo {author} {\bibfnamefont
  {S.}~\bibnamefont {{Hochheim}}}, \bibinfo {author} {\bibfnamefont
  {D.}~\bibnamefont {{Hofman}}}, \bibinfo {author} {\bibfnamefont {A.~M.}\
  \bibnamefont {{Holgado}}}, \bibinfo {author} {\bibfnamefont {N.~A.}\
  \bibnamefont {{Holland}}}, \bibinfo {author} {\bibfnamefont {K.}~\bibnamefont
  {{Holt}}}, \bibinfo {author} {\bibfnamefont {D.~E.}\ \bibnamefont {{Holz}}},
  \bibinfo {author} {\bibfnamefont {P.}~\bibnamefont {{Hopkins}}}, \bibinfo
  {author} {\bibfnamefont {C.}~\bibnamefont {{Horst}}}, \bibinfo {author}
  {\bibfnamefont {J.}~\bibnamefont {{Hough}}}, \bibinfo {author} {\bibfnamefont
  {E.~J.}\ \bibnamefont {{Howell}}}, \bibinfo {author} {\bibfnamefont {C.~G.}\
  \bibnamefont {{Hoy}}}, \bibinfo {author} {\bibfnamefont {A.}~\bibnamefont
  {{Hreibi}}}, \bibinfo {author} {\bibfnamefont {E.~A.}\ \bibnamefont
  {{Huerta}}}, \bibinfo {author} {\bibfnamefont {D.}~\bibnamefont {{Huet}}},
  \bibinfo {author} {\bibfnamefont {B.}~\bibnamefont {{Hughey}}}, \bibinfo
  {author} {\bibfnamefont {M.}~\bibnamefont {{Hulko}}}, \bibinfo {author}
  {\bibfnamefont {S.}~\bibnamefont {{Husa}}}, \bibinfo {author} {\bibfnamefont
  {S.~H.}\ \bibnamefont {{Huttner}}}, \bibinfo {author} {\bibfnamefont
  {T.}~\bibnamefont {{Huynh-Dinh}}}, \bibinfo {author} {\bibfnamefont
  {B.}~\bibnamefont {{Idzkowski}}}, \bibinfo {author} {\bibfnamefont
  {A.}~\bibnamefont {{Iess}}}, \bibinfo {author} {\bibfnamefont
  {C.}~\bibnamefont {{Ingram}}}, \bibinfo {author} {\bibfnamefont
  {R.}~\bibnamefont {{Inta}}}, \bibinfo {author} {\bibfnamefont
  {G.}~\bibnamefont {{Intini}}}, \bibinfo {author} {\bibfnamefont
  {B.}~\bibnamefont {{Irwin}}}, \bibinfo {author} {\bibfnamefont {H.~N.}\
  \bibnamefont {{Isa}}}, \bibinfo {author} {\bibfnamefont {J.~M.}\ \bibnamefont
  {{Isac}}}, \bibinfo {author} {\bibfnamefont {M.}~\bibnamefont {{Isi}}},
  \bibinfo {author} {\bibfnamefont {B.~R.}\ \bibnamefont {{Iyer}}}, \bibinfo
  {author} {\bibfnamefont {K.}~\bibnamefont {{Izumi}}}, \bibinfo {author}
  {\bibfnamefont {T.}~\bibnamefont {{Jacqmin}}}, \bibinfo {author}
  {\bibfnamefont {S.~J.}\ \bibnamefont {{Jadhav}}}, \bibinfo {author}
  {\bibfnamefont {K.}~\bibnamefont {{Jani}}}, \bibinfo {author} {\bibfnamefont
  {N.~N.}\ \bibnamefont {{Janthalur}}}, \bibinfo {author} {\bibfnamefont
  {P.}~\bibnamefont {{Jaranowski}}}, \bibinfo {author} {\bibfnamefont {A.~C.}\
  \bibnamefont {{Jenkins}}}, \bibinfo {author} {\bibfnamefont {J.}~\bibnamefont
  {{Jiang}}}, \bibinfo {author} {\bibfnamefont {D.~S.}\ \bibnamefont
  {{Johnson}}}, \bibinfo {author} {\bibfnamefont {A.~W.}\ \bibnamefont
  {{Jones}}}, \bibinfo {author} {\bibfnamefont {D.~I.}\ \bibnamefont
  {{Jones}}}, \bibinfo {author} {\bibfnamefont {R.}~\bibnamefont {{Jones}}},
  \bibinfo {author} {\bibfnamefont {R.~J.~G.}\ \bibnamefont {{Jonker}}},
  \bibinfo {author} {\bibfnamefont {L.}~\bibnamefont {{Ju}}}, \bibinfo {author}
  {\bibfnamefont {J.}~\bibnamefont {{Junker}}}, \bibinfo {author}
  {\bibfnamefont {C.~V.}\ \bibnamefont {{Kalaghatgi}}}, \bibinfo {author}
  {\bibfnamefont {V.}~\bibnamefont {{Kalogera}}}, \bibinfo {author}
  {\bibfnamefont {B.}~\bibnamefont {{Kamai}}}, \bibinfo {author} {\bibfnamefont
  {S.}~\bibnamefont {{Kandhasamy}}}, \bibinfo {author} {\bibfnamefont
  {G.}~\bibnamefont {{Kang}}}, \bibinfo {author} {\bibfnamefont {J.~B.}\
  \bibnamefont {{Kanner}}}, \bibinfo {author} {\bibfnamefont {S.~J.}\
  \bibnamefont {{Kapadia}}}, \bibinfo {author} {\bibfnamefont {S.}~\bibnamefont
  {{Karki}}}, \bibinfo {author} {\bibfnamefont {K.~S.}\ \bibnamefont
  {{Karvinen}}}, \bibinfo {author} {\bibfnamefont {R.}~\bibnamefont
  {{Kashyap}}}, \bibinfo {author} {\bibfnamefont {M.}~\bibnamefont
  {{Kasprzack}}}, \bibinfo {author} {\bibfnamefont {S.}~\bibnamefont
  {{Katsanevas}}}, \bibinfo {author} {\bibfnamefont {E.}~\bibnamefont
  {{Katsavounidis}}}, \bibinfo {author} {\bibfnamefont {W.}~\bibnamefont
  {{Katzman}}}, \bibinfo {author} {\bibfnamefont {S.}~\bibnamefont {{Kaufer}}},
  \bibinfo {author} {\bibfnamefont {K.}~\bibnamefont {{Kawabe}}}, \bibinfo
  {author} {\bibfnamefont {N.~V.}\ \bibnamefont {{Keerthana}}}, \bibinfo
  {author} {\bibfnamefont {F.}~\bibnamefont {{K{\'e}f{\'e}lian}}}, \bibinfo
  {author} {\bibfnamefont {D.}~\bibnamefont {{Keitel}}}, \bibinfo {author}
  {\bibfnamefont {R.}~\bibnamefont {{Kennedy}}}, \bibinfo {author}
  {\bibfnamefont {J.~S.}\ \bibnamefont {{Key}}}, \bibinfo {author}
  {\bibfnamefont {F.~Y.}\ \bibnamefont {{Khalili}}}, \bibinfo {author}
  {\bibfnamefont {H.}~\bibnamefont {{Khan}}}, \bibinfo {author} {\bibfnamefont
  {I.}~\bibnamefont {{Khan}}}, \bibinfo {author} {\bibfnamefont
  {S.}~\bibnamefont {{Khan}}}, \bibinfo {author} {\bibfnamefont
  {Z.}~\bibnamefont {{Khan}}}, \bibinfo {author} {\bibfnamefont {E.~A.}\
  \bibnamefont {{Khazanov}}}, \bibinfo {author} {\bibfnamefont
  {M.}~\bibnamefont {{Khursheed}}}, \bibinfo {author} {\bibfnamefont
  {N.}~\bibnamefont {{Kijbunchoo}}}, \bibinfo {author} {\bibfnamefont {A.~X.}\
  \bibnamefont {{Kim}}}, \bibinfo {author} {\bibfnamefont {C.}~\bibnamefont
  {{Kim}}}, \bibinfo {author} {\bibfnamefont {J.~C.}\ \bibnamefont {{Kim}}},
  \bibinfo {author} {\bibfnamefont {K.}~\bibnamefont {{Kim}}}, \bibinfo
  {author} {\bibfnamefont {W.}~\bibnamefont {{Kim}}}, \bibinfo {author}
  {\bibfnamefont {W.~S.}\ \bibnamefont {{Kim}}}, \bibinfo {author}
  {\bibfnamefont {Y.~M.}\ \bibnamefont {{Kim}}}, \bibinfo {author}
  {\bibfnamefont {C.}~\bibnamefont {{Kimball}}}, \bibinfo {author}
  {\bibfnamefont {E.~J.}\ \bibnamefont {{King}}}, \bibinfo {author}
  {\bibfnamefont {P.~J.}\ \bibnamefont {{King}}}, \bibinfo {author}
  {\bibfnamefont {M.}~\bibnamefont {{Kinley-Hanlon}}}, \bibinfo {author}
  {\bibfnamefont {R.}~\bibnamefont {{Kirchhoff}}}, \bibinfo {author}
  {\bibfnamefont {J.~S.}\ \bibnamefont {{Kissel}}}, \bibinfo {author}
  {\bibfnamefont {L.}~\bibnamefont {{Kleybolte}}}, \bibinfo {author}
  {\bibfnamefont {J.~H.}\ \bibnamefont {{Klika}}}, \bibinfo {author}
  {\bibfnamefont {S.}~\bibnamefont {{Klimenko}}}, \bibinfo {author}
  {\bibfnamefont {T.~D.}\ \bibnamefont {{Knowles}}}, \bibinfo {author}
  {\bibfnamefont {P.}~\bibnamefont {{Koch}}}, \bibinfo {author} {\bibfnamefont
  {S.~M.}\ \bibnamefont {{Koehlenbeck}}}, \bibinfo {author} {\bibfnamefont
  {G.}~\bibnamefont {{Koekoek}}}, \bibinfo {author} {\bibfnamefont
  {S.}~\bibnamefont {{Koley}}}, \bibinfo {author} {\bibfnamefont
  {V.}~\bibnamefont {{Kondrashov}}}, \bibinfo {author} {\bibfnamefont
  {A.}~\bibnamefont {{Kontos}}}, \bibinfo {author} {\bibfnamefont
  {N.}~\bibnamefont {{Koper}}}, \bibinfo {author} {\bibfnamefont
  {M.}~\bibnamefont {{Korobko}}}, \bibinfo {author} {\bibfnamefont {W.~Z.}\
  \bibnamefont {{Korth}}}, \bibinfo {author} {\bibfnamefont {I.}~\bibnamefont
  {{Kowalska}}}, \bibinfo {author} {\bibfnamefont {D.~B.}\ \bibnamefont
  {{Kozak}}}, \bibinfo {author} {\bibfnamefont {V.}~\bibnamefont {{Kringel}}},
  \bibinfo {author} {\bibfnamefont {N.}~\bibnamefont {{Krishnendu}}}, \bibinfo
  {author} {\bibfnamefont {A.}~\bibnamefont {{Kr{\'o}lak}}}, \bibinfo {author}
  {\bibfnamefont {G.}~\bibnamefont {{Kuehn}}}, \bibinfo {author} {\bibfnamefont
  {A.}~\bibnamefont {{Kumar}}}, \bibinfo {author} {\bibfnamefont
  {P.}~\bibnamefont {{Kumar}}}, \bibinfo {author} {\bibfnamefont
  {R.}~\bibnamefont {{Kumar}}}, \bibinfo {author} {\bibfnamefont
  {S.}~\bibnamefont {{Kumar}}}, \bibinfo {author} {\bibfnamefont
  {L.}~\bibnamefont {{Kuo}}}, \bibinfo {author} {\bibfnamefont
  {A.}~\bibnamefont {{Kutynia}}}, \bibinfo {author} {\bibfnamefont
  {S.}~\bibnamefont {{Kwang}}}, \bibinfo {author} {\bibfnamefont {B.~D.}\
  \bibnamefont {{Lackey}}}, \bibinfo {author} {\bibfnamefont {K.~H.}\
  \bibnamefont {{Lai}}}, \bibinfo {author} {\bibfnamefont {T.~L.}\ \bibnamefont
  {{Lam}}}, \bibinfo {author} {\bibfnamefont {M.}~\bibnamefont {{Landry}}},
  \bibinfo {author} {\bibfnamefont {B.~B.}\ \bibnamefont {{Lane}}}, \bibinfo
  {author} {\bibfnamefont {R.~N.}\ \bibnamefont {{Lang}}}, \bibinfo {author}
  {\bibfnamefont {J.}~\bibnamefont {{Lange}}}, \bibinfo {author} {\bibfnamefont
  {B.}~\bibnamefont {{Lantz}}}, \bibinfo {author} {\bibfnamefont {R.~K.}\
  \bibnamefont {{Lanza}}}, \bibinfo {author} {\bibfnamefont {S.}~\bibnamefont
  {{Larson}}}, \bibinfo {author} {\bibfnamefont {A.}~\bibnamefont
  {{Lartaux-Vollard}}}, \bibinfo {author} {\bibfnamefont {P.~D.}\ \bibnamefont
  {{Lasky}}}, \bibinfo {author} {\bibfnamefont {M.}~\bibnamefont {{Laxen}}},
  \bibinfo {author} {\bibfnamefont {A.}~\bibnamefont {{Lazzarini}}}, \bibinfo
  {author} {\bibfnamefont {C.}~\bibnamefont {{Lazzaro}}}, \bibinfo {author}
  {\bibfnamefont {P.}~\bibnamefont {{Leaci}}}, \bibinfo {author} {\bibfnamefont
  {S.}~\bibnamefont {{Leavey}}}, \bibinfo {author} {\bibfnamefont {Y.~K.}\
  \bibnamefont {{Lecoeuche}}}, \bibinfo {author} {\bibfnamefont {C.~H.}\
  \bibnamefont {{Lee}}}, \bibinfo {author} {\bibfnamefont {H.~K.}\ \bibnamefont
  {{Lee}}}, \bibinfo {author} {\bibfnamefont {H.~M.}\ \bibnamefont {{Lee}}},
  \bibinfo {author} {\bibfnamefont {H.~W.}\ \bibnamefont {{Lee}}}, \bibinfo
  {author} {\bibfnamefont {J.}~\bibnamefont {{Lee}}}, \bibinfo {author}
  {\bibfnamefont {K.}~\bibnamefont {{Lee}}}, \bibinfo {author} {\bibfnamefont
  {J.}~\bibnamefont {{Lehmann}}}, \bibinfo {author} {\bibfnamefont
  {A.}~\bibnamefont {{Lenon}}}, \bibinfo {author} {\bibfnamefont
  {N.}~\bibnamefont {{Leroy}}}, \bibinfo {author} {\bibfnamefont
  {N.}~\bibnamefont {{Letendre}}}, \bibinfo {author} {\bibfnamefont
  {Y.}~\bibnamefont {{Levin}}}, \bibinfo {author} {\bibfnamefont
  {J.}~\bibnamefont {{Li}}}, \bibinfo {author} {\bibfnamefont {K.~J.~L.}\
  \bibnamefont {{Li}}}, \bibinfo {author} {\bibfnamefont {T.~G.~F.}\
  \bibnamefont {{Li}}}, \bibinfo {author} {\bibfnamefont {X.}~\bibnamefont
  {{Li}}}, \bibinfo {author} {\bibfnamefont {F.}~\bibnamefont {{Lin}}},
  \bibinfo {author} {\bibfnamefont {F.}~\bibnamefont {{Linde}}}, \bibinfo
  {author} {\bibfnamefont {S.~D.}\ \bibnamefont {{Linker}}}, \bibinfo {author}
  {\bibfnamefont {T.~B.}\ \bibnamefont {{Littenberg}}}, \bibinfo {author}
  {\bibfnamefont {J.}~\bibnamefont {{Liu}}}, \bibinfo {author} {\bibfnamefont
  {X.}~\bibnamefont {{Liu}}}, \bibinfo {author} {\bibfnamefont {R.~K.~L.}\
  \bibnamefont {{Lo}}}, \bibinfo {author} {\bibfnamefont {N.~A.}\ \bibnamefont
  {{Lockerbie}}}, \bibinfo {author} {\bibfnamefont {L.~T.}\ \bibnamefont
  {{London}}}, \bibinfo {author} {\bibfnamefont {A.}~\bibnamefont {{Longo}}},
  \bibinfo {author} {\bibfnamefont {M.}~\bibnamefont {{Lorenzini}}}, \bibinfo
  {author} {\bibfnamefont {V.}~\bibnamefont {{Loriette}}}, \bibinfo {author}
  {\bibfnamefont {M.}~\bibnamefont {{Lormand}}}, \bibinfo {author}
  {\bibfnamefont {G.}~\bibnamefont {{Losurdo}}}, \bibinfo {author}
  {\bibfnamefont {J.~D.}\ \bibnamefont {{Lough}}}, \bibinfo {author}
  {\bibfnamefont {C.~O.}\ \bibnamefont {{Lousto}}}, \bibinfo {author}
  {\bibfnamefont {G.}~\bibnamefont {{Lovelace}}}, \bibinfo {author}
  {\bibfnamefont {M.~E.}\ \bibnamefont {{Lower}}}, \bibinfo {author}
  {\bibfnamefont {H.}~\bibnamefont {{L{\"u}ck}}}, \bibinfo {author}
  {\bibfnamefont {D.}~\bibnamefont {{Lumaca}}}, \bibinfo {author}
  {\bibfnamefont {A.~P.}\ \bibnamefont {{Lundgren}}}, \bibinfo {author}
  {\bibfnamefont {R.}~\bibnamefont {{Lynch}}}, \bibinfo {author} {\bibfnamefont
  {Y.}~\bibnamefont {{Ma}}}, \bibinfo {author} {\bibfnamefont {R.}~\bibnamefont
  {{Macas}}}, \bibinfo {author} {\bibfnamefont {S.}~\bibnamefont {{Macfoy}}},
  \bibinfo {author} {\bibfnamefont {M.}~\bibnamefont {{MacInnis}}}, \bibinfo
  {author} {\bibfnamefont {D.~M.}\ \bibnamefont {{Macleod}}}, \bibinfo {author}
  {\bibfnamefont {A.}~\bibnamefont {{Macquet}}}, \bibinfo {author}
  {\bibfnamefont {F.}~\bibnamefont {{Maga{\~n}a-Sandoval}}}, \bibinfo {author}
  {\bibfnamefont {L.}~\bibnamefont {{Maga{\~n}a Zertuche}}}, \bibinfo {author}
  {\bibfnamefont {R.~M.}\ \bibnamefont {{Magee}}}, \bibinfo {author}
  {\bibfnamefont {E.}~\bibnamefont {{Majorana}}}, \bibinfo {author}
  {\bibfnamefont {I.}~\bibnamefont {{Maksimovic}}}, \bibinfo {author}
  {\bibfnamefont {A.}~\bibnamefont {{Malik}}}, \bibinfo {author} {\bibfnamefont
  {N.}~\bibnamefont {{Man}}}, \bibinfo {author} {\bibfnamefont
  {V.}~\bibnamefont {{Mandic}}}, \bibinfo {author} {\bibfnamefont
  {V.}~\bibnamefont {{Mangano}}}, \bibinfo {author} {\bibfnamefont {G.~L.}\
  \bibnamefont {{Mansell}}}, \bibinfo {author} {\bibfnamefont {M.}~\bibnamefont
  {{Manske}}}, \bibinfo {author} {\bibfnamefont {M.}~\bibnamefont
  {{Mantovani}}}, \bibinfo {author} {\bibfnamefont {F.}~\bibnamefont
  {{Marchesoni}}}, \bibinfo {author} {\bibfnamefont {F.}~\bibnamefont
  {{Marion}}}, \bibinfo {author} {\bibfnamefont {S.}~\bibnamefont
  {{M{\'a}rka}}}, \bibinfo {author} {\bibfnamefont {Z.}~\bibnamefont
  {{M{\'a}rka}}}, \bibinfo {author} {\bibfnamefont {C.}~\bibnamefont
  {{Markakis}}}, \bibinfo {author} {\bibfnamefont {A.~S.}\ \bibnamefont
  {{Markosyan}}}, \bibinfo {author} {\bibfnamefont {A.}~\bibnamefont
  {{Markowitz}}}, \bibinfo {author} {\bibfnamefont {E.}~\bibnamefont
  {{Maros}}}, \bibinfo {author} {\bibfnamefont {A.}~\bibnamefont {{Marquina}}},
  \bibinfo {author} {\bibfnamefont {S.}~\bibnamefont {{Marsat}}}, \bibinfo
  {author} {\bibfnamefont {F.}~\bibnamefont {{Martelli}}}, \bibinfo {author}
  {\bibfnamefont {I.~W.}\ \bibnamefont {{Martin}}}, \bibinfo {author}
  {\bibfnamefont {R.~M.}\ \bibnamefont {{Martin}}}, \bibinfo {author}
  {\bibfnamefont {D.~V.}\ \bibnamefont {{Martynov}}}, \bibinfo {author}
  {\bibfnamefont {K.}~\bibnamefont {{Mason}}}, \bibinfo {author} {\bibfnamefont
  {E.}~\bibnamefont {{Massera}}}, \bibinfo {author} {\bibfnamefont
  {A.}~\bibnamefont {{Masserot}}}, \bibinfo {author} {\bibfnamefont {T.~J.}\
  \bibnamefont {{Massinger}}}, \bibinfo {author} {\bibfnamefont
  {M.}~\bibnamefont {{Masso-Reid}}}, \bibinfo {author} {\bibfnamefont
  {S.}~\bibnamefont {{Mastrogiovanni}}}, \bibinfo {author} {\bibfnamefont
  {A.}~\bibnamefont {{Matas}}}, \bibinfo {author} {\bibfnamefont
  {F.}~\bibnamefont {{Matichard}}}, \bibinfo {author} {\bibfnamefont
  {L.}~\bibnamefont {{Matone}}}, \bibinfo {author} {\bibfnamefont
  {N.}~\bibnamefont {{Mavalvala}}}, \bibinfo {author} {\bibfnamefont
  {N.}~\bibnamefont {{Mazumder}}}, \bibinfo {author} {\bibfnamefont {J.~J.}\
  \bibnamefont {{McCann}}}, \bibinfo {author} {\bibfnamefont {R.}~\bibnamefont
  {{McCarthy}}}, \bibinfo {author} {\bibfnamefont {D.~E.}\ \bibnamefont
  {{McClelland}}}, \bibinfo {author} {\bibfnamefont {S.}~\bibnamefont
  {{McCormick}}}, \bibinfo {author} {\bibfnamefont {L.}~\bibnamefont
  {{McCuller}}}, \bibinfo {author} {\bibfnamefont {S.~C.}\ \bibnamefont
  {{McGuire}}}, \bibinfo {author} {\bibfnamefont {J.}~\bibnamefont {{McIver}}},
  \bibinfo {author} {\bibfnamefont {D.~J.}\ \bibnamefont {{McManus}}}, \bibinfo
  {author} {\bibfnamefont {T.}~\bibnamefont {{McRae}}}, \bibinfo {author}
  {\bibfnamefont {S.~T.}\ \bibnamefont {{McWilliams}}}, \bibinfo {author}
  {\bibfnamefont {D.}~\bibnamefont {{Meacher}}}, \bibinfo {author}
  {\bibfnamefont {G.~D.}\ \bibnamefont {{Meadors}}}, \bibinfo {author}
  {\bibfnamefont {M.}~\bibnamefont {{Mehmet}}}, \bibinfo {author}
  {\bibfnamefont {A.~K.}\ \bibnamefont {{Mehta}}}, \bibinfo {author}
  {\bibfnamefont {J.}~\bibnamefont {{Meidam}}}, \bibinfo {author}
  {\bibfnamefont {A.}~\bibnamefont {{Melatos}}}, \bibinfo {author}
  {\bibfnamefont {G.}~\bibnamefont {{Mendell}}}, \bibinfo {author}
  {\bibfnamefont {R.~A.}\ \bibnamefont {{Mercer}}}, \bibinfo {author}
  {\bibfnamefont {L.}~\bibnamefont {{Mereni}}}, \bibinfo {author}
  {\bibfnamefont {E.~L.}\ \bibnamefont {{Merilh}}}, \bibinfo {author}
  {\bibfnamefont {M.}~\bibnamefont {{Merzougui}}}, \bibinfo {author}
  {\bibfnamefont {S.}~\bibnamefont {{Meshkov}}}, \bibinfo {author}
  {\bibfnamefont {C.}~\bibnamefont {{Messenger}}}, \bibinfo {author}
  {\bibfnamefont {C.}~\bibnamefont {{Messick}}}, \bibinfo {author}
  {\bibfnamefont {R.}~\bibnamefont {{Metzdorff}}}, \bibinfo {author}
  {\bibfnamefont {P.~M.}\ \bibnamefont {{Meyers}}}, \bibinfo {author}
  {\bibfnamefont {H.}~\bibnamefont {{Miao}}}, \bibinfo {author} {\bibfnamefont
  {C.}~\bibnamefont {{Michel}}}, \bibinfo {author} {\bibfnamefont
  {H.}~\bibnamefont {{Middleton}}}, \bibinfo {author} {\bibfnamefont {E.~E.}\
  \bibnamefont {{Mikhailov}}}, \bibinfo {author} {\bibfnamefont
  {L.}~\bibnamefont {{Milano}}}, \bibinfo {author} {\bibfnamefont {A.~L.}\
  \bibnamefont {{Miller}}}, \bibinfo {author} {\bibfnamefont {A.}~\bibnamefont
  {{Miller}}}, \bibinfo {author} {\bibfnamefont {M.}~\bibnamefont
  {{Millhouse}}}, \bibinfo {author} {\bibfnamefont {J.~C.}\ \bibnamefont
  {{Mills}}}, \bibinfo {author} {\bibfnamefont {M.~C.}\ \bibnamefont
  {{Milovich-Goff}}}, \bibinfo {author} {\bibfnamefont {O.}~\bibnamefont
  {{Minazzoli}}}, \bibinfo {author} {\bibfnamefont {Y.}~\bibnamefont
  {{Minenkov}}}, \bibinfo {author} {\bibfnamefont {A.}~\bibnamefont
  {{Mishkin}}}, \bibinfo {author} {\bibfnamefont {C.}~\bibnamefont {{Mishra}}},
  \bibinfo {author} {\bibfnamefont {T.}~\bibnamefont {{Mistry}}}, \bibinfo
  {author} {\bibfnamefont {S.}~\bibnamefont {{Mitra}}}, \bibinfo {author}
  {\bibfnamefont {V.~P.}\ \bibnamefont {{Mitrofanov}}}, \bibinfo {author}
  {\bibfnamefont {G.}~\bibnamefont {{Mitselmakher}}}, \bibinfo {author}
  {\bibfnamefont {R.}~\bibnamefont {{Mittleman}}}, \bibinfo {author}
  {\bibfnamefont {G.}~\bibnamefont {{Mo}}}, \bibinfo {author} {\bibfnamefont
  {D.}~\bibnamefont {{Moffa}}}, \bibinfo {author} {\bibfnamefont
  {K.}~\bibnamefont {{Mogushi}}}, \bibinfo {author} {\bibfnamefont {S.~R.~P.}\
  \bibnamefont {{Mohapatra}}}, \bibinfo {author} {\bibfnamefont
  {M.}~\bibnamefont {{Montani}}}, \bibinfo {author} {\bibfnamefont {C.~J.}\
  \bibnamefont {{Moore}}}, \bibinfo {author} {\bibfnamefont {D.}~\bibnamefont
  {{Moraru}}}, \bibinfo {author} {\bibfnamefont {G.}~\bibnamefont {{Moreno}}},
  \bibinfo {author} {\bibfnamefont {S.}~\bibnamefont {{Morisaki}}}, \bibinfo
  {author} {\bibfnamefont {B.}~\bibnamefont {{Mours}}}, \bibinfo {author}
  {\bibfnamefont {C.~M.}\ \bibnamefont {{Mow-Lowry}}}, \bibinfo {author}
  {\bibfnamefont {A.}~\bibnamefont {{Mukherjee}}}, \bibinfo {author}
  {\bibfnamefont {D.}~\bibnamefont {{Mukherjee}}}, \bibinfo {author}
  {\bibfnamefont {S.}~\bibnamefont {{Mukherjee}}}, \bibinfo {author}
  {\bibfnamefont {N.}~\bibnamefont {{Mukund}}}, \bibinfo {author}
  {\bibfnamefont {A.}~\bibnamefont {{Mullavey}}}, \bibinfo {author}
  {\bibfnamefont {J.}~\bibnamefont {{Munch}}}, \bibinfo {author} {\bibfnamefont
  {E.~A.}\ \bibnamefont {{Mu{\~n}iz}}}, \bibinfo {author} {\bibfnamefont
  {M.}~\bibnamefont {{Muratore}}}, \bibinfo {author} {\bibfnamefont {P.~G.}\
  \bibnamefont {{Murray}}}, \bibinfo {author} {\bibfnamefont {A.}~\bibnamefont
  {{Nagar}}}, \bibinfo {author} {\bibfnamefont {I.}~\bibnamefont
  {{Nardecchia}}}, \bibinfo {author} {\bibfnamefont {L.}~\bibnamefont
  {{Naticchioni}}}, \bibinfo {author} {\bibfnamefont {R.~K.}\ \bibnamefont
  {{Nayak}}}, \bibinfo {author} {\bibfnamefont {J.}~\bibnamefont {{Neilson}}},
  \bibinfo {author} {\bibfnamefont {G.}~\bibnamefont {{Nelemans}}}, \bibinfo
  {author} {\bibfnamefont {T.~J.~N.}\ \bibnamefont {{Nelson}}}, \bibinfo
  {author} {\bibfnamefont {M.}~\bibnamefont {{Nery}}}, \bibinfo {author}
  {\bibfnamefont {A.}~\bibnamefont {{Neunzert}}}, \bibinfo {author}
  {\bibfnamefont {K.~Y.}\ \bibnamefont {{Ng}}}, \bibinfo {author}
  {\bibfnamefont {S.}~\bibnamefont {{Ng}}}, \bibinfo {author} {\bibfnamefont
  {P.}~\bibnamefont {{Nguyen}}}, \bibinfo {author} {\bibfnamefont
  {D.}~\bibnamefont {{Nichols}}}, \bibinfo {author} {\bibfnamefont
  {S.}~\bibnamefont {{Nissanke}}}, \bibinfo {author} {\bibfnamefont
  {F.}~\bibnamefont {{Nocera}}}, \bibinfo {author} {\bibfnamefont
  {C.}~\bibnamefont {{North}}}, \bibinfo {author} {\bibfnamefont {L.~K.}\
  \bibnamefont {{Nuttall}}}, \bibinfo {author} {\bibfnamefont {M.}~\bibnamefont
  {{Obergaulinger}}}, \bibinfo {author} {\bibfnamefont {J.}~\bibnamefont
  {{Oberling}}}, \bibinfo {author} {\bibfnamefont {B.~D.}\ \bibnamefont
  {{O'Brien}}}, \bibinfo {author} {\bibfnamefont {G.~D.}\ \bibnamefont
  {{O'Dea}}}, \bibinfo {author} {\bibfnamefont {G.~H.}\ \bibnamefont {{Ogin}}},
  \bibinfo {author} {\bibfnamefont {J.~J.}\ \bibnamefont {{Oh}}}, \bibinfo
  {author} {\bibfnamefont {S.~H.}\ \bibnamefont {{Oh}}}, \bibinfo {author}
  {\bibfnamefont {F.}~\bibnamefont {{Ohme}}}, \bibinfo {author} {\bibfnamefont
  {H.}~\bibnamefont {{Ohta}}}, \bibinfo {author} {\bibfnamefont {M.~A.}\
  \bibnamefont {{Okada}}}, \bibinfo {author} {\bibfnamefont {M.}~\bibnamefont
  {{Oliver}}}, \bibinfo {author} {\bibfnamefont {P.}~\bibnamefont
  {{Oppermann}}}, \bibinfo {author} {\bibfnamefont {R.~J.}\ \bibnamefont
  {{Oram}}}, \bibinfo {author} {\bibfnamefont {B.}~\bibnamefont {{O'Reilly}}},
  \bibinfo {author} {\bibfnamefont {R.~G.}\ \bibnamefont {{Ormiston}}},
  \bibinfo {author} {\bibfnamefont {L.~F.}\ \bibnamefont {{Ortega}}}, \bibinfo
  {author} {\bibfnamefont {R.}~\bibnamefont {{O'Shaughnessy}}}, \bibinfo
  {author} {\bibfnamefont {S.}~\bibnamefont {{Ossokine}}}, \bibinfo {author}
  {\bibfnamefont {D.~J.}\ \bibnamefont {{Ottaway}}}, \bibinfo {author}
  {\bibfnamefont {H.}~\bibnamefont {{Overmier}}}, \bibinfo {author}
  {\bibfnamefont {B.~J.}\ \bibnamefont {{Owen}}}, \bibinfo {author}
  {\bibfnamefont {A.~E.}\ \bibnamefont {{Pace}}}, \bibinfo {author}
  {\bibfnamefont {G.}~\bibnamefont {{Pagano}}}, \bibinfo {author}
  {\bibfnamefont {M.~A.}\ \bibnamefont {{Page}}}, \bibinfo {author}
  {\bibfnamefont {A.}~\bibnamefont {{Pai}}}, \bibinfo {author} {\bibfnamefont
  {S.~A.}\ \bibnamefont {{Pai}}}, \bibinfo {author} {\bibfnamefont {J.~R.}\
  \bibnamefont {{Palamos}}}, \bibinfo {author} {\bibfnamefont {O.}~\bibnamefont
  {{Palashov}}}, \bibinfo {author} {\bibfnamefont {C.}~\bibnamefont
  {{Palomba}}}, \bibinfo {author} {\bibfnamefont {A.}~\bibnamefont
  {{Pal-Singh}}}, \bibinfo {author} {\bibfnamefont {H.-W.}\ \bibnamefont
  {{Pan}}}, \bibinfo {author} {\bibfnamefont {B.}~\bibnamefont {{Pang}}},
  \bibinfo {author} {\bibfnamefont {P.~T.~H.}\ \bibnamefont {{Pang}}}, \bibinfo
  {author} {\bibfnamefont {C.}~\bibnamefont {{Pankow}}}, \bibinfo {author}
  {\bibfnamefont {F.}~\bibnamefont {{Pannarale}}}, \bibinfo {author}
  {\bibfnamefont {B.~C.}\ \bibnamefont {{Pant}}}, \bibinfo {author}
  {\bibfnamefont {F.}~\bibnamefont {{Paoletti}}}, \bibinfo {author}
  {\bibfnamefont {A.}~\bibnamefont {{Paoli}}}, \bibinfo {author} {\bibfnamefont
  {A.}~\bibnamefont {{Parida}}}, \bibinfo {author} {\bibfnamefont
  {W.}~\bibnamefont {{Parker}}}, \bibinfo {author} {\bibfnamefont
  {D.}~\bibnamefont {{Pascucci}}}, \bibinfo {author} {\bibfnamefont
  {A.}~\bibnamefont {{Pasqualetti}}}, \bibinfo {author} {\bibfnamefont
  {R.}~\bibnamefont {{Passaquieti}}}, \bibinfo {author} {\bibfnamefont
  {D.}~\bibnamefont {{Passuello}}}, \bibinfo {author} {\bibfnamefont
  {M.}~\bibnamefont {{Patil}}}, \bibinfo {author} {\bibfnamefont
  {B.}~\bibnamefont {{Patricelli}}}, \bibinfo {author} {\bibfnamefont {B.~L.}\
  \bibnamefont {{Pearlstone}}}, \bibinfo {author} {\bibfnamefont
  {C.}~\bibnamefont {{Pedersen}}}, \bibinfo {author} {\bibfnamefont
  {M.}~\bibnamefont {{Pedraza}}}, \bibinfo {author} {\bibfnamefont
  {R.}~\bibnamefont {{Pedurand}}}, \bibinfo {author} {\bibfnamefont
  {A.}~\bibnamefont {{Pele}}}, \bibinfo {author} {\bibfnamefont
  {S.}~\bibnamefont {{Penn}}}, \bibinfo {author} {\bibfnamefont {C.~J.}\
  \bibnamefont {{Perez}}}, \bibinfo {author} {\bibfnamefont {A.}~\bibnamefont
  {{Perreca}}}, \bibinfo {author} {\bibfnamefont {H.~P.}\ \bibnamefont
  {{Pfeiffer}}}, \bibinfo {author} {\bibfnamefont {M.}~\bibnamefont
  {{Phelps}}}, \bibinfo {author} {\bibfnamefont {K.~S.}\ \bibnamefont
  {{Phukon}}}, \bibinfo {author} {\bibfnamefont {O.~J.}\ \bibnamefont
  {{Piccinni}}}, \bibinfo {author} {\bibfnamefont {M.}~\bibnamefont
  {{Pichot}}}, \bibinfo {author} {\bibfnamefont {F.}~\bibnamefont
  {{Piergiovanni}}}, \bibinfo {author} {\bibfnamefont {G.}~\bibnamefont
  {{Pillant}}}, \bibinfo {author} {\bibfnamefont {L.}~\bibnamefont {{Pinard}}},
  \bibinfo {author} {\bibfnamefont {M.}~\bibnamefont {{Pirello}}}, \bibinfo
  {author} {\bibfnamefont {M.}~\bibnamefont {{Pitkin}}}, \bibinfo {author}
  {\bibfnamefont {R.}~\bibnamefont {{Poggiani}}}, \bibinfo {author}
  {\bibfnamefont {D.~Y.~T.}\ \bibnamefont {{Pong}}}, \bibinfo {author}
  {\bibfnamefont {S.}~\bibnamefont {{Ponrathnam}}}, \bibinfo {author}
  {\bibfnamefont {P.}~\bibnamefont {{Popolizio}}}, \bibinfo {author}
  {\bibfnamefont {E.~K.}\ \bibnamefont {{Porter}}}, \bibinfo {author}
  {\bibfnamefont {J.}~\bibnamefont {{Powell}}}, \bibinfo {author}
  {\bibfnamefont {A.~K.}\ \bibnamefont {{Prajapati}}}, \bibinfo {author}
  {\bibfnamefont {J.}~\bibnamefont {{Prasad}}}, \bibinfo {author}
  {\bibfnamefont {K.}~\bibnamefont {{Prasai}}}, \bibinfo {author}
  {\bibfnamefont {R.}~\bibnamefont {{Prasanna}}}, \bibinfo {author}
  {\bibfnamefont {G.}~\bibnamefont {{Pratten}}}, \bibinfo {author}
  {\bibfnamefont {T.}~\bibnamefont {{Prestegard}}}, \bibinfo {author}
  {\bibfnamefont {S.}~\bibnamefont {{Privitera}}}, \bibinfo {author}
  {\bibfnamefont {G.~A.}\ \bibnamefont {{Prodi}}}, \bibinfo {author}
  {\bibfnamefont {L.~G.}\ \bibnamefont {{Prokhorov}}}, \bibinfo {author}
  {\bibfnamefont {O.}~\bibnamefont {{Puncken}}}, \bibinfo {author}
  {\bibfnamefont {M.}~\bibnamefont {{Punturo}}}, \bibinfo {author}
  {\bibfnamefont {P.}~\bibnamefont {{Puppo}}}, \bibinfo {author} {\bibfnamefont
  {M.}~\bibnamefont {{P{\"u}rrer}}}, \bibinfo {author} {\bibfnamefont
  {H.}~\bibnamefont {{Qi}}}, \bibinfo {author} {\bibfnamefont {V.}~\bibnamefont
  {{Quetschke}}}, \bibinfo {author} {\bibfnamefont {P.~J.}\ \bibnamefont
  {{Quinonez}}}, \bibinfo {author} {\bibfnamefont {E.~A.}\ \bibnamefont
  {{Quintero}}}, \bibinfo {author} {\bibfnamefont {R.}~\bibnamefont
  {{Quitzow-James}}}, \bibinfo {author} {\bibfnamefont {F.~J.}\ \bibnamefont
  {{Raab}}}, \bibinfo {author} {\bibfnamefont {H.}~\bibnamefont {{Radkins}}},
  \bibinfo {author} {\bibfnamefont {N.}~\bibnamefont {{Radulescu}}}, \bibinfo
  {author} {\bibfnamefont {P.}~\bibnamefont {{Raffai}}}, \bibinfo {author}
  {\bibfnamefont {S.}~\bibnamefont {{Raja}}}, \bibinfo {author} {\bibfnamefont
  {C.}~\bibnamefont {{Rajan}}}, \bibinfo {author} {\bibfnamefont
  {B.}~\bibnamefont {{Rajbhandari}}}, \bibinfo {author} {\bibfnamefont
  {M.}~\bibnamefont {{Rakhmanov}}}, \bibinfo {author} {\bibfnamefont {K.~E.}\
  \bibnamefont {{Ramirez}}}, \bibinfo {author} {\bibfnamefont {A.}~\bibnamefont
  {{Ramos-Buades}}}, \bibinfo {author} {\bibfnamefont {J.}~\bibnamefont
  {{Rana}}}, \bibinfo {author} {\bibfnamefont {K.}~\bibnamefont {{Rao}}},
  \bibinfo {author} {\bibfnamefont {P.}~\bibnamefont {{Rapagnani}}}, \bibinfo
  {author} {\bibfnamefont {V.}~\bibnamefont {{Raymond}}}, \bibinfo {author}
  {\bibfnamefont {M.}~\bibnamefont {{Razzano}}}, \bibinfo {author}
  {\bibfnamefont {J.}~\bibnamefont {{Read}}}, \bibinfo {author} {\bibfnamefont
  {T.}~\bibnamefont {{Regimbau}}}, \bibinfo {author} {\bibfnamefont
  {L.}~\bibnamefont {{Rei}}}, \bibinfo {author} {\bibfnamefont
  {S.}~\bibnamefont {{Reid}}}, \bibinfo {author} {\bibfnamefont {D.~H.}\
  \bibnamefont {{Reitze}}}, \bibinfo {author} {\bibfnamefont {W.}~\bibnamefont
  {{Ren}}}, \bibinfo {author} {\bibfnamefont {F.}~\bibnamefont {{Ricci}}},
  \bibinfo {author} {\bibfnamefont {C.~J.}\ \bibnamefont {{Richardson}}},
  \bibinfo {author} {\bibfnamefont {J.~W.}\ \bibnamefont {{Richardson}}},
  \bibinfo {author} {\bibfnamefont {P.~M.}\ \bibnamefont {{Ricker}}}, \bibinfo
  {author} {\bibfnamefont {K.}~\bibnamefont {{Riles}}}, \bibinfo {author}
  {\bibfnamefont {M.}~\bibnamefont {{Rizzo}}}, \bibinfo {author} {\bibfnamefont
  {N.~A.}\ \bibnamefont {{Robertson}}}, \bibinfo {author} {\bibfnamefont
  {R.}~\bibnamefont {{Robie}}}, \bibinfo {author} {\bibfnamefont
  {F.}~\bibnamefont {{Robinet}}}, \bibinfo {author} {\bibfnamefont
  {A.}~\bibnamefont {{Rocchi}}}, \bibinfo {author} {\bibfnamefont
  {L.}~\bibnamefont {{Rolland}}}, \bibinfo {author} {\bibfnamefont {J.~G.}\
  \bibnamefont {{Rollins}}}, \bibinfo {author} {\bibfnamefont {V.~J.}\
  \bibnamefont {{Roma}}}, \bibinfo {author} {\bibfnamefont {M.}~\bibnamefont
  {{Romanelli}}}, \bibinfo {author} {\bibfnamefont {R.}~\bibnamefont
  {{Romano}}}, \bibinfo {author} {\bibfnamefont {C.~L.}\ \bibnamefont
  {{Romel}}}, \bibinfo {author} {\bibfnamefont {J.~H.}\ \bibnamefont
  {{Romie}}}, \bibinfo {author} {\bibfnamefont {K.}~\bibnamefont {{Rose}}},
  \bibinfo {author} {\bibfnamefont {D.}~\bibnamefont {{Rosi{\'n}ska}}},
  \bibinfo {author} {\bibfnamefont {S.~G.}\ \bibnamefont {{Rosofsky}}},
  \bibinfo {author} {\bibfnamefont {M.~P.}\ \bibnamefont {{Ross}}}, \bibinfo
  {author} {\bibfnamefont {S.}~\bibnamefont {{Rowan}}}, \bibinfo {author}
  {\bibfnamefont {A.}~\bibnamefont {{R{\"u}diger}}}, \bibinfo {author}
  {\bibfnamefont {P.}~\bibnamefont {{Ruggi}}}, \bibinfo {author} {\bibfnamefont
  {G.}~\bibnamefont {{Rutins}}}, \bibinfo {author} {\bibfnamefont
  {K.}~\bibnamefont {{Ryan}}}, \bibinfo {author} {\bibfnamefont
  {S.}~\bibnamefont {{Sachdev}}}, \bibinfo {author} {\bibfnamefont
  {T.}~\bibnamefont {{Sadecki}}}, \bibinfo {author} {\bibfnamefont
  {M.}~\bibnamefont {{Sakellariadou}}}, \bibinfo {author} {\bibfnamefont
  {L.}~\bibnamefont {{Salconi}}}, \bibinfo {author} {\bibfnamefont
  {M.}~\bibnamefont {{Saleem}}}, \bibinfo {author} {\bibfnamefont
  {A.}~\bibnamefont {{Samajdar}}}, \bibinfo {author} {\bibfnamefont
  {L.}~\bibnamefont {{Sammut}}}, \bibinfo {author} {\bibfnamefont {E.~J.}\
  \bibnamefont {{Sanchez}}}, \bibinfo {author} {\bibfnamefont {L.~E.}\
  \bibnamefont {{Sanchez}}}, \bibinfo {author} {\bibfnamefont {N.}~\bibnamefont
  {{Sanchis-Gual}}}, \bibinfo {author} {\bibfnamefont {V.}~\bibnamefont
  {{Sandberg}}}, \bibinfo {author} {\bibfnamefont {J.~R.}\ \bibnamefont
  {{Sanders}}}, \bibinfo {author} {\bibfnamefont {K.~A.}\ \bibnamefont
  {{Santiago}}}, \bibinfo {author} {\bibfnamefont {N.}~\bibnamefont {{Sarin}}},
  \bibinfo {author} {\bibfnamefont {B.}~\bibnamefont {{Sassolas}}}, \bibinfo
  {author} {\bibfnamefont {B.~S.}\ \bibnamefont {{Sathyaprakash}}}, \bibinfo
  {author} {\bibfnamefont {P.~R.}\ \bibnamefont {{Saulson}}}, \bibinfo {author}
  {\bibfnamefont {O.}~\bibnamefont {{Sauter}}}, \bibinfo {author}
  {\bibfnamefont {R.~L.}\ \bibnamefont {{Savage}}}, \bibinfo {author}
  {\bibfnamefont {P.}~\bibnamefont {{Schale}}}, \bibinfo {author}
  {\bibfnamefont {M.}~\bibnamefont {{Scheel}}}, \bibinfo {author}
  {\bibfnamefont {J.}~\bibnamefont {{Scheuer}}}, \bibinfo {author}
  {\bibfnamefont {P.}~\bibnamefont {{Schmidt}}}, \bibinfo {author}
  {\bibfnamefont {R.}~\bibnamefont {{Schnabel}}}, \bibinfo {author}
  {\bibfnamefont {R.~M.~S.}\ \bibnamefont {{Schofield}}}, \bibinfo {author}
  {\bibfnamefont {A.}~\bibnamefont {{Sch{\"o}nbeck}}}, \bibinfo {author}
  {\bibfnamefont {E.}~\bibnamefont {{Schreiber}}}, \bibinfo {author}
  {\bibfnamefont {B.~W.}\ \bibnamefont {{Schulte}}}, \bibinfo {author}
  {\bibfnamefont {B.~F.}\ \bibnamefont {{Schutz}}}, \bibinfo {author}
  {\bibfnamefont {S.~G.}\ \bibnamefont {{Schwalbe}}}, \bibinfo {author}
  {\bibfnamefont {J.}~\bibnamefont {{Scott}}}, \bibinfo {author} {\bibfnamefont
  {S.~M.}\ \bibnamefont {{Scott}}}, \bibinfo {author} {\bibfnamefont
  {E.}~\bibnamefont {{Seidel}}}, \bibinfo {author} {\bibfnamefont
  {D.}~\bibnamefont {{Sellers}}}, \bibinfo {author} {\bibfnamefont {A.~S.}\
  \bibnamefont {{Sengupta}}}, \bibinfo {author} {\bibfnamefont
  {N.}~\bibnamefont {{Sennett}}}, \bibinfo {author} {\bibfnamefont
  {D.}~\bibnamefont {{Sentenac}}}, \bibinfo {author} {\bibfnamefont
  {V.}~\bibnamefont {{Sequino}}}, \bibinfo {author} {\bibfnamefont
  {A.}~\bibnamefont {{Sergeev}}}, \bibinfo {author} {\bibfnamefont
  {Y.}~\bibnamefont {{Setyawati}}}, \bibinfo {author} {\bibfnamefont {D.~A.}\
  \bibnamefont {{Shaddock}}}, \bibinfo {author} {\bibfnamefont
  {T.}~\bibnamefont {{Shaffer}}}, \bibinfo {author} {\bibfnamefont {M.~S.}\
  \bibnamefont {{Shahriar}}}, \bibinfo {author} {\bibfnamefont {M.~B.}\
  \bibnamefont {{Shaner}}}, \bibinfo {author} {\bibfnamefont {L.}~\bibnamefont
  {{Shao}}}, \bibinfo {author} {\bibfnamefont {P.}~\bibnamefont {{Sharma}}},
  \bibinfo {author} {\bibfnamefont {P.}~\bibnamefont {{Shawhan}}}, \bibinfo
  {author} {\bibfnamefont {H.}~\bibnamefont {{Shen}}}, \bibinfo {author}
  {\bibfnamefont {R.}~\bibnamefont {{Shink}}}, \bibinfo {author} {\bibfnamefont
  {D.~H.}\ \bibnamefont {{Shoemaker}}}, \bibinfo {author} {\bibfnamefont
  {D.~M.}\ \bibnamefont {{Shoemaker}}}, \bibinfo {author} {\bibfnamefont
  {S.}~\bibnamefont {{ShyamSundar}}}, \bibinfo {author} {\bibfnamefont
  {K.}~\bibnamefont {{Siellez}}}, \bibinfo {author} {\bibfnamefont
  {M.}~\bibnamefont {{Sieniawska}}}, \bibinfo {author} {\bibfnamefont
  {D.}~\bibnamefont {{Sigg}}}, \bibinfo {author} {\bibfnamefont {A.~D.}\
  \bibnamefont {{Silva}}}, \bibinfo {author} {\bibfnamefont {L.~P.}\
  \bibnamefont {{Singer}}}, \bibinfo {author} {\bibfnamefont {N.}~\bibnamefont
  {{Singh}}}, \bibinfo {author} {\bibfnamefont {A.}~\bibnamefont {{Singhal}}},
  \bibinfo {author} {\bibfnamefont {A.~M.}\ \bibnamefont {{Sintes}}}, \bibinfo
  {author} {\bibfnamefont {S.}~\bibnamefont {{Sitmukhambetov}}}, \bibinfo
  {author} {\bibfnamefont {V.}~\bibnamefont {{Skliris}}}, \bibinfo {author}
  {\bibfnamefont {B.~J.~J.}\ \bibnamefont {{Slagmolen}}}, \bibinfo {author}
  {\bibfnamefont {T.~J.}\ \bibnamefont {{Slaven-Blair}}}, \bibinfo {author}
  {\bibfnamefont {J.~R.}\ \bibnamefont {{Smith}}}, \bibinfo {author}
  {\bibfnamefont {R.~J.~E.}\ \bibnamefont {{Smith}}}, \bibinfo {author}
  {\bibfnamefont {S.}~\bibnamefont {{Somala}}}, \bibinfo {author}
  {\bibfnamefont {E.~J.}\ \bibnamefont {{Son}}}, \bibinfo {author}
  {\bibfnamefont {B.}~\bibnamefont {{Sorazu}}}, \bibinfo {author}
  {\bibfnamefont {F.}~\bibnamefont {{Sorrentino}}}, \bibinfo {author}
  {\bibfnamefont {T.}~\bibnamefont {{Souradeep}}}, \bibinfo {author}
  {\bibfnamefont {E.}~\bibnamefont {{Sowell}}}, \bibinfo {author}
  {\bibfnamefont {A.~P.}\ \bibnamefont {{Spencer}}}, \bibinfo {author}
  {\bibfnamefont {A.~K.}\ \bibnamefont {{Srivastava}}}, \bibinfo {author}
  {\bibfnamefont {V.}~\bibnamefont {{Srivastava}}}, \bibinfo {author}
  {\bibfnamefont {K.}~\bibnamefont {{Staats}}}, \bibinfo {author}
  {\bibfnamefont {C.}~\bibnamefont {{Stachie}}}, \bibinfo {author}
  {\bibfnamefont {M.}~\bibnamefont {{Standke}}}, \bibinfo {author}
  {\bibfnamefont {D.~A.}\ \bibnamefont {{Steer}}}, \bibinfo {author}
  {\bibfnamefont {M.}~\bibnamefont {{Steinke}}}, \bibinfo {author}
  {\bibfnamefont {J.}~\bibnamefont {{Steinlechner}}}, \bibinfo {author}
  {\bibfnamefont {S.}~\bibnamefont {{Steinlechner}}}, \bibinfo {author}
  {\bibfnamefont {D.}~\bibnamefont {{Steinmeyer}}}, \bibinfo {author}
  {\bibfnamefont {S.~P.}\ \bibnamefont {{Stevenson}}}, \bibinfo {author}
  {\bibfnamefont {D.}~\bibnamefont {{Stocks}}}, \bibinfo {author}
  {\bibfnamefont {R.}~\bibnamefont {{Stone}}}, \bibinfo {author} {\bibfnamefont
  {D.~J.}\ \bibnamefont {{Stops}}}, \bibinfo {author} {\bibfnamefont {K.~A.}\
  \bibnamefont {{Strain}}}, \bibinfo {author} {\bibfnamefont {G.}~\bibnamefont
  {{Stratta}}}, \bibinfo {author} {\bibfnamefont {S.~E.}\ \bibnamefont
  {{Strigin}}}, \bibinfo {author} {\bibfnamefont {A.}~\bibnamefont {{Strunk}}},
  \bibinfo {author} {\bibfnamefont {R.}~\bibnamefont {{Sturani}}}, \bibinfo
  {author} {\bibfnamefont {A.~L.}\ \bibnamefont {{Stuver}}}, \bibinfo {author}
  {\bibfnamefont {V.}~\bibnamefont {{Sudhir}}}, \bibinfo {author}
  {\bibfnamefont {T.~Z.}\ \bibnamefont {{Summerscales}}}, \bibinfo {author}
  {\bibfnamefont {L.}~\bibnamefont {{Sun}}}, \bibinfo {author} {\bibfnamefont
  {S.}~\bibnamefont {{Sunil}}}, \bibinfo {author} {\bibfnamefont
  {J.}~\bibnamefont {{Suresh}}}, \bibinfo {author} {\bibfnamefont {P.~J.}\
  \bibnamefont {{Sutton}}}, \bibinfo {author} {\bibfnamefont {B.~L.}\
  \bibnamefont {{Swinkels}}}, \bibinfo {author} {\bibfnamefont {M.~J.}\
  \bibnamefont {{Szczepa{\'n}czyk}}}, \bibinfo {author} {\bibfnamefont
  {M.}~\bibnamefont {{Tacca}}}, \bibinfo {author} {\bibfnamefont {S.~C.}\
  \bibnamefont {{Tait}}}, \bibinfo {author} {\bibfnamefont {C.}~\bibnamefont
  {{Talbot}}}, \bibinfo {author} {\bibfnamefont {D.}~\bibnamefont
  {{Talukder}}}, \bibinfo {author} {\bibfnamefont {D.~B.}\ \bibnamefont
  {{Tanner}}}, \bibinfo {author} {\bibfnamefont {M.}~\bibnamefont
  {{T{\'a}pai}}}, \bibinfo {author} {\bibfnamefont {A.}~\bibnamefont
  {{Taracchini}}}, \bibinfo {author} {\bibfnamefont {J.~D.}\ \bibnamefont
  {{Tasson}}}, \bibinfo {author} {\bibfnamefont {R.}~\bibnamefont {{Taylor}}},
  \bibinfo {author} {\bibfnamefont {F.}~\bibnamefont {{Thies}}}, \bibinfo
  {author} {\bibfnamefont {M.}~\bibnamefont {{Thomas}}}, \bibinfo {author}
  {\bibfnamefont {P.}~\bibnamefont {{Thomas}}}, \bibinfo {author}
  {\bibfnamefont {S.~R.}\ \bibnamefont {{Thondapu}}}, \bibinfo {author}
  {\bibfnamefont {K.~A.}\ \bibnamefont {{Thorne}}}, \bibinfo {author}
  {\bibfnamefont {E.}~\bibnamefont {{Thrane}}}, \bibinfo {author}
  {\bibfnamefont {S.}~\bibnamefont {{Tiwari}}}, \bibinfo {author}
  {\bibfnamefont {S.}~\bibnamefont {{Tiwari}}}, \bibinfo {author}
  {\bibfnamefont {V.}~\bibnamefont {{Tiwari}}}, \bibinfo {author}
  {\bibfnamefont {K.}~\bibnamefont {{Toland}}}, \bibinfo {author}
  {\bibfnamefont {M.}~\bibnamefont {{Tonelli}}}, \bibinfo {author}
  {\bibfnamefont {Z.}~\bibnamefont {{Tornasi}}}, \bibinfo {author}
  {\bibfnamefont {A.}~\bibnamefont {{Torres-Forn{\'e}}}}, \bibinfo {author}
  {\bibfnamefont {C.~I.}\ \bibnamefont {{Torrie}}}, \bibinfo {author}
  {\bibfnamefont {D.}~\bibnamefont {{T{\"o}yr{\"a}}}}, \bibinfo {author}
  {\bibfnamefont {F.}~\bibnamefont {{Travasso}}}, \bibinfo {author}
  {\bibfnamefont {G.}~\bibnamefont {{Traylor}}}, \bibinfo {author}
  {\bibfnamefont {M.~C.}\ \bibnamefont {{Tringali}}}, \bibinfo {author}
  {\bibfnamefont {A.}~\bibnamefont {{Trovato}}}, \bibinfo {author}
  {\bibfnamefont {L.}~\bibnamefont {{Trozzo}}}, \bibinfo {author}
  {\bibfnamefont {R.}~\bibnamefont {{Trudeau}}}, \bibinfo {author}
  {\bibfnamefont {K.~W.}\ \bibnamefont {{Tsang}}}, \bibinfo {author}
  {\bibfnamefont {M.}~\bibnamefont {{Tse}}}, \bibinfo {author} {\bibfnamefont
  {R.}~\bibnamefont {{Tso}}}, \bibinfo {author} {\bibfnamefont
  {L.}~\bibnamefont {{Tsukada}}}, \bibinfo {author} {\bibfnamefont
  {D.}~\bibnamefont {{Tsuna}}}, \bibinfo {author} {\bibfnamefont
  {D.}~\bibnamefont {{Tuyenbayev}}}, \bibinfo {author} {\bibfnamefont
  {K.}~\bibnamefont {{Ueno}}}, \bibinfo {author} {\bibfnamefont
  {D.}~\bibnamefont {{Ugolini}}}, \bibinfo {author} {\bibfnamefont {C.~S.}\
  \bibnamefont {{Unnikrishnan}}}, \bibinfo {author} {\bibfnamefont {A.~L.}\
  \bibnamefont {{Urban}}}, \bibinfo {author} {\bibfnamefont {S.~A.}\
  \bibnamefont {{Usman}}}, \bibinfo {author} {\bibfnamefont {H.}~\bibnamefont
  {{Vahlbruch}}}, \bibinfo {author} {\bibfnamefont {G.}~\bibnamefont
  {{Vajente}}}, \bibinfo {author} {\bibfnamefont {G.}~\bibnamefont {{Valdes}}},
  \bibinfo {author} {\bibfnamefont {N.}~\bibnamefont {{van Bakel}}}, \bibinfo
  {author} {\bibfnamefont {M.}~\bibnamefont {{van Beuzekom}}}, \bibinfo
  {author} {\bibfnamefont {J.~F.~J.}\ \bibnamefont {{van den Brand}}}, \bibinfo
  {author} {\bibfnamefont {C.}~\bibnamefont {{Van Den Broeck}}}, \bibinfo
  {author} {\bibfnamefont {D.~C.}\ \bibnamefont {{Vander-Hyde}}}, \bibinfo
  {author} {\bibfnamefont {J.~V.}\ \bibnamefont {{van Heijningen}}}, \bibinfo
  {author} {\bibfnamefont {L.}~\bibnamefont {{van der Schaaf}}}, \bibinfo
  {author} {\bibfnamefont {A.~A.}\ \bibnamefont {{van Veggel}}}, \bibinfo
  {author} {\bibfnamefont {M.}~\bibnamefont {{Vardaro}}}, \bibinfo {author}
  {\bibfnamefont {V.}~\bibnamefont {{Varma}}}, \bibinfo {author} {\bibfnamefont
  {S.}~\bibnamefont {{Vass}}}, \bibinfo {author} {\bibfnamefont
  {M.}~\bibnamefont {{Vas{\'u}th}}}, \bibinfo {author} {\bibfnamefont
  {A.}~\bibnamefont {{Vecchio}}}, \bibinfo {author} {\bibfnamefont
  {G.}~\bibnamefont {{Vedovato}}}, \bibinfo {author} {\bibfnamefont
  {J.}~\bibnamefont {{Veitch}}}, \bibinfo {author} {\bibfnamefont {P.~J.}\
  \bibnamefont {{Veitch}}}, \bibinfo {author} {\bibfnamefont {K.}~\bibnamefont
  {{Venkateswara}}}, \bibinfo {author} {\bibfnamefont {G.}~\bibnamefont
  {{Venugopalan}}}, \bibinfo {author} {\bibfnamefont {D.}~\bibnamefont
  {{Verkindt}}}, \bibinfo {author} {\bibfnamefont {F.}~\bibnamefont
  {{Vetrano}}}, \bibinfo {author} {\bibfnamefont {A.}~\bibnamefont
  {{Vicer{\'e}}}}, \bibinfo {author} {\bibfnamefont {A.~D.}\ \bibnamefont
  {{Viets}}}, \bibinfo {author} {\bibfnamefont {D.~J.}\ \bibnamefont {{Vine}}},
  \bibinfo {author} {\bibfnamefont {J.~Y.}\ \bibnamefont {{Vinet}}}, \bibinfo
  {author} {\bibfnamefont {S.}~\bibnamefont {{Vitale}}}, \bibinfo {author}
  {\bibfnamefont {T.}~\bibnamefont {{Vo}}}, \bibinfo {author} {\bibfnamefont
  {H.}~\bibnamefont {{Vocca}}}, \bibinfo {author} {\bibfnamefont
  {C.}~\bibnamefont {{Vorvick}}}, \bibinfo {author} {\bibfnamefont {S.~P.}\
  \bibnamefont {{Vyatchanin}}}, \bibinfo {author} {\bibfnamefont {A.~R.}\
  \bibnamefont {{Wade}}}, \bibinfo {author} {\bibfnamefont {L.~E.}\
  \bibnamefont {{Wade}}}, \bibinfo {author} {\bibfnamefont {M.}~\bibnamefont
  {{Wade}}}, \bibinfo {author} {\bibfnamefont {R.}~\bibnamefont {{Walet}}},
  \bibinfo {author} {\bibfnamefont {M.}~\bibnamefont {{Walker}}}, \bibinfo
  {author} {\bibfnamefont {L.}~\bibnamefont {{Wallace}}}, \bibinfo {author}
  {\bibfnamefont {S.}~\bibnamefont {{Walsh}}}, \bibinfo {author} {\bibfnamefont
  {G.}~\bibnamefont {{Wang}}}, \bibinfo {author} {\bibfnamefont
  {H.}~\bibnamefont {{Wang}}}, \bibinfo {author} {\bibfnamefont {J.~Z.}\
  \bibnamefont {{Wang}}}, \bibinfo {author} {\bibfnamefont {W.~H.}\
  \bibnamefont {{Wang}}}, \bibinfo {author} {\bibfnamefont {Y.~F.}\
  \bibnamefont {{Wang}}}, \bibinfo {author} {\bibfnamefont {R.~L.}\
  \bibnamefont {{Ward}}}, \bibinfo {author} {\bibfnamefont {Z.~A.}\
  \bibnamefont {{Warden}}}, \bibinfo {author} {\bibfnamefont {J.}~\bibnamefont
  {{Warner}}}, \bibinfo {author} {\bibfnamefont {M.}~\bibnamefont {{Was}}},
  \bibinfo {author} {\bibfnamefont {J.}~\bibnamefont {{Watchi}}}, \bibinfo
  {author} {\bibfnamefont {B.}~\bibnamefont {{Weaver}}}, \bibinfo {author}
  {\bibfnamefont {L.~W.}\ \bibnamefont {{Wei}}}, \bibinfo {author}
  {\bibfnamefont {M.}~\bibnamefont {{Weinert}}}, \bibinfo {author}
  {\bibfnamefont {A.~J.}\ \bibnamefont {{Weinstein}}}, \bibinfo {author}
  {\bibfnamefont {R.}~\bibnamefont {{Weiss}}}, \bibinfo {author} {\bibfnamefont
  {F.}~\bibnamefont {{Wellmann}}}, \bibinfo {author} {\bibfnamefont
  {L.}~\bibnamefont {{Wen}}}, \bibinfo {author} {\bibfnamefont {E.~K.}\
  \bibnamefont {{Wessel}}}, \bibinfo {author} {\bibfnamefont {P.}~\bibnamefont
  {{We{\ss}els}}}, \bibinfo {author} {\bibfnamefont {J.~W.}\ \bibnamefont
  {{Westhouse}}}, \bibinfo {author} {\bibfnamefont {K.}~\bibnamefont
  {{Wette}}}, \bibinfo {author} {\bibfnamefont {J.~T.}\ \bibnamefont
  {{Whelan}}}, \bibinfo {author} {\bibfnamefont {B.~F.}\ \bibnamefont
  {{Whiting}}}, \bibinfo {author} {\bibfnamefont {C.}~\bibnamefont
  {{Whittle}}}, \bibinfo {author} {\bibfnamefont {D.~M.}\ \bibnamefont
  {{Wilken}}}, \bibinfo {author} {\bibfnamefont {D.}~\bibnamefont
  {{Williams}}}, \bibinfo {author} {\bibfnamefont {A.~R.}\ \bibnamefont
  {{Williamson}}}, \bibinfo {author} {\bibfnamefont {J.~L.}\ \bibnamefont
  {{Willis}}}, \bibinfo {author} {\bibfnamefont {B.}~\bibnamefont {{Willke}}},
  \bibinfo {author} {\bibfnamefont {M.~H.}\ \bibnamefont {{Wimmer}}}, \bibinfo
  {author} {\bibfnamefont {W.}~\bibnamefont {{Winkler}}}, \bibinfo {author}
  {\bibfnamefont {C.~C.}\ \bibnamefont {{Wipf}}}, \bibinfo {author}
  {\bibfnamefont {H.}~\bibnamefont {{Wittel}}}, \bibinfo {author}
  {\bibfnamefont {G.}~\bibnamefont {{Woan}}}, \bibinfo {author} {\bibfnamefont
  {J.}~\bibnamefont {{Woehler}}}, \bibinfo {author} {\bibfnamefont {J.~K.}\
  \bibnamefont {{Wofford}}}, \bibinfo {author} {\bibfnamefont {J.}~\bibnamefont
  {{Worden}}}, \bibinfo {author} {\bibfnamefont {J.~L.}\ \bibnamefont
  {{Wright}}}, \bibinfo {author} {\bibfnamefont {D.~S.}\ \bibnamefont {{Wu}}},
  \bibinfo {author} {\bibfnamefont {D.~M.}\ \bibnamefont {{Wysocki}}}, \bibinfo
  {author} {\bibfnamefont {L.}~\bibnamefont {{Xiao}}}, \bibinfo {author}
  {\bibfnamefont {H.}~\bibnamefont {{Yamamoto}}}, \bibinfo {author}
  {\bibfnamefont {C.~C.}\ \bibnamefont {{Yancey}}}, \bibinfo {author}
  {\bibfnamefont {L.}~\bibnamefont {{Yang}}}, \bibinfo {author} {\bibfnamefont
  {M.~J.}\ \bibnamefont {{Yap}}}, \bibinfo {author} {\bibfnamefont
  {M.}~\bibnamefont {{Yazback}}}, \bibinfo {author} {\bibfnamefont {D.~W.}\
  \bibnamefont {{Yeeles}}}, \bibinfo {author} {\bibfnamefont {H.}~\bibnamefont
  {{Yu}}}, \bibinfo {author} {\bibfnamefont {H.}~\bibnamefont {{Yu}}}, \bibinfo
  {author} {\bibfnamefont {S.~H.~R.}\ \bibnamefont {{Yuen}}}, \bibinfo {author}
  {\bibfnamefont {M.}~\bibnamefont {{Yvert}}}, \bibinfo {author} {\bibfnamefont
  {A.~K.}\ \bibnamefont {{Zadro{\.z}ny}}}, \bibinfo {author} {\bibfnamefont
  {M.}~\bibnamefont {{Zanolin}}}, \bibinfo {author} {\bibfnamefont
  {T.}~\bibnamefont {{Zelenova}}}, \bibinfo {author} {\bibfnamefont {J.~P.}\
  \bibnamefont {{Zendri}}}, \bibinfo {author} {\bibfnamefont {M.}~\bibnamefont
  {{Zevin}}}, \bibinfo {author} {\bibfnamefont {J.}~\bibnamefont {{Zhang}}},
  \bibinfo {author} {\bibfnamefont {L.}~\bibnamefont {{Zhang}}}, \bibinfo
  {author} {\bibfnamefont {T.}~\bibnamefont {{Zhang}}}, \bibinfo {author}
  {\bibfnamefont {C.}~\bibnamefont {{Zhao}}}, \bibinfo {author} {\bibfnamefont
  {M.}~\bibnamefont {{Zhou}}}, \bibinfo {author} {\bibfnamefont
  {Z.}~\bibnamefont {{Zhou}}}, \bibinfo {author} {\bibfnamefont {X.~J.}\
  \bibnamefont {{Zhu}}}, \bibinfo {author} {\bibfnamefont {M.~E.}\ \bibnamefont
  {{Zucker}}}, \bibinfo {author} {\bibfnamefont {J.}~\bibnamefont {{Zweizig}}},
  \bibinfo {author} {\bibnamefont {{LIGO Scientific Collaboration}}},\ and\
  \bibinfo {author} {\bibnamefont {{Virgo Collaboration}}},\ }\bibfield
  {title} {\bibinfo {title} {{A guide to LIGO-Virgo detector noise and
  extraction of transient gravitational-wave signals}},\ }\href
  {https://doi.org/10.1088/1361-6382/ab685e} {\bibfield  {journal} {\bibinfo
  {journal} {Classical and Quantum Gravity}\ }\textbf {\bibinfo {volume}
  {37}},\ \bibinfo {eid} {055002} (\bibinfo {year} {2020}{\natexlab{e}})},\
  \Eprint {https://arxiv.org/abs/1908.11170} {arXiv:1908.11170 [gr-qc]}
  \BibitemShut {NoStop}%
\bibitem [{\citenamefont {{Ransom}}\ \emph {et~al.}(2005)\citenamefont
  {{Ransom}}, \citenamefont {{Hessels}}, \citenamefont {{Stairs}},
  \citenamefont {{Freire}}, \citenamefont {{Camilo}}, \citenamefont {{Kaspi}},\
  and\ \citenamefont {{Kaplan}}}]{2005Sci...307..892R}%
  \BibitemOpen
  \bibfield  {author} {\bibinfo {author} {\bibfnamefont {S.~M.}\ \bibnamefont
  {{Ransom}}}, \bibinfo {author} {\bibfnamefont {J.~W.~T.}\ \bibnamefont
  {{Hessels}}}, \bibinfo {author} {\bibfnamefont {I.~H.}\ \bibnamefont
  {{Stairs}}}, \bibinfo {author} {\bibfnamefont {P.~C.~C.}\ \bibnamefont
  {{Freire}}}, \bibinfo {author} {\bibfnamefont {F.}~\bibnamefont {{Camilo}}},
  \bibinfo {author} {\bibfnamefont {V.~M.}\ \bibnamefont {{Kaspi}}},\ and\
  \bibinfo {author} {\bibfnamefont {D.~L.}\ \bibnamefont {{Kaplan}}},\
  }\bibfield  {title} {\bibinfo {title} {{Twenty-One Millisecond Pulsars in
  Terzan 5 Using the Green Bank Telescope}},\ }\href
  {https://doi.org/10.1126/science.1108632} {\bibfield  {journal} {\bibinfo
  {journal} {Science}\ }\textbf {\bibinfo {volume} {307}},\ \bibinfo {pages}
  {892} (\bibinfo {year} {2005})},\ \Eprint
  {https://arxiv.org/abs/astro-ph/0501230} {arXiv:astro-ph/0501230 [astro-ph]}
  \BibitemShut {NoStop}%
\bibitem [{\citenamefont {{Pooley}}\ and\ \citenamefont
  {{Hut}}(2006)}]{2006ApJ...646L.143P}%
  \BibitemOpen
  \bibfield  {author} {\bibinfo {author} {\bibfnamefont {D.}~\bibnamefont
  {{Pooley}}}\ and\ \bibinfo {author} {\bibfnamefont {P.}~\bibnamefont
  {{Hut}}},\ }\bibfield  {title} {\bibinfo {title} {{Dynamical Formation of
  Close Binaries in Globular Clusters: Cataclysmic Variables}},\ }\href
  {https://doi.org/10.1086/507027} {\bibfield  {journal} {\bibinfo  {journal}
  {\apjl}\ }\textbf {\bibinfo {volume} {646}},\ \bibinfo {pages} {L143}
  (\bibinfo {year} {2006})},\ \Eprint {https://arxiv.org/abs/astro-ph/0605048}
  {arXiv:astro-ph/0605048 [astro-ph]} \BibitemShut {NoStop}%
\bibitem [{\citenamefont {{Ferraro}}\ \emph {et~al.}(2009)\citenamefont
  {{Ferraro}}, \citenamefont {{Beccari}}, \citenamefont {{Dalessandro}},
  \citenamefont {{Lanzoni}}, \citenamefont {{Sills}}, \citenamefont {{Rood}},
  \citenamefont {{Pecci}}, \citenamefont {{Karakas}}, \citenamefont
  {{Miocchi}},\ and\ \citenamefont {{Bovinelli}}}]{2009Natur.462.1028F}%
  \BibitemOpen
  \bibfield  {author} {\bibinfo {author} {\bibfnamefont {F.~R.}\ \bibnamefont
  {{Ferraro}}}, \bibinfo {author} {\bibfnamefont {G.}~\bibnamefont
  {{Beccari}}}, \bibinfo {author} {\bibfnamefont {E.}~\bibnamefont
  {{Dalessandro}}}, \bibinfo {author} {\bibfnamefont {B.}~\bibnamefont
  {{Lanzoni}}}, \bibinfo {author} {\bibfnamefont {A.}~\bibnamefont {{Sills}}},
  \bibinfo {author} {\bibfnamefont {R.~T.}\ \bibnamefont {{Rood}}}, \bibinfo
  {author} {\bibfnamefont {F.~F.}\ \bibnamefont {{Pecci}}}, \bibinfo {author}
  {\bibfnamefont {A.~I.}\ \bibnamefont {{Karakas}}}, \bibinfo {author}
  {\bibfnamefont {P.}~\bibnamefont {{Miocchi}}},\ and\ \bibinfo {author}
  {\bibfnamefont {S.}~\bibnamefont {{Bovinelli}}},\ }\bibfield  {title}
  {\bibinfo {title} {{Two distinct sequences of blue straggler stars in the
  globular cluster M 30}},\ }\href {https://doi.org/10.1038/nature08607}
  {\bibfield  {journal} {\bibinfo  {journal} {\nat}\ }\textbf {\bibinfo
  {volume} {462}},\ \bibinfo {pages} {1028} (\bibinfo {year} {2009})},\ \Eprint
  {https://arxiv.org/abs/1001.1096} {arXiv:1001.1096 [astro-ph.GA]}
  \BibitemShut {NoStop}%
\bibitem [{\citenamefont {{Bastian}}\ and\ \citenamefont
  {{Lardo}}(2018)}]{2018ARA&A..56...83B}%
  \BibitemOpen
  \bibfield  {author} {\bibinfo {author} {\bibfnamefont {N.}~\bibnamefont
  {{Bastian}}}\ and\ \bibinfo {author} {\bibfnamefont {C.}~\bibnamefont
  {{Lardo}}},\ }\bibfield  {title} {\bibinfo {title} {{Multiple Stellar
  Populations in Globular Clusters}},\ }\href
  {https://doi.org/10.1146/annurev-astro-081817-051839} {\bibfield  {journal}
  {\bibinfo  {journal} {\araa}\ }\textbf {\bibinfo {volume} {56}},\ \bibinfo
  {pages} {83} (\bibinfo {year} {2018})},\ \Eprint
  {https://arxiv.org/abs/1712.01286} {arXiv:1712.01286 [astro-ph.SR]}
  \BibitemShut {NoStop}%
\bibitem [{\citenamefont {{Giesers}}\ \emph {et~al.}(2018)\citenamefont
  {{Giesers}}, \citenamefont {{Dreizler}}, \citenamefont {{Husser}},
  \citenamefont {{Kamann}}, \citenamefont {{Anglada Escud{\'e}}}, \citenamefont
  {{Brinchmann}}, \citenamefont {{Carollo}}, \citenamefont {{Roth}},
  \citenamefont {{Weilbacher}},\ and\ \citenamefont
  {{Wisotzki}}}]{2018MNRAS.475L..15G}%
  \BibitemOpen
  \bibfield  {author} {\bibinfo {author} {\bibfnamefont {B.}~\bibnamefont
  {{Giesers}}}, \bibinfo {author} {\bibfnamefont {S.}~\bibnamefont
  {{Dreizler}}}, \bibinfo {author} {\bibfnamefont {T.-O.}\ \bibnamefont
  {{Husser}}}, \bibinfo {author} {\bibfnamefont {S.}~\bibnamefont {{Kamann}}},
  \bibinfo {author} {\bibfnamefont {G.}~\bibnamefont {{Anglada Escud{\'e}}}},
  \bibinfo {author} {\bibfnamefont {J.}~\bibnamefont {{Brinchmann}}}, \bibinfo
  {author} {\bibfnamefont {C.~M.}\ \bibnamefont {{Carollo}}}, \bibinfo {author}
  {\bibfnamefont {M.~M.}\ \bibnamefont {{Roth}}}, \bibinfo {author}
  {\bibfnamefont {P.~M.}\ \bibnamefont {{Weilbacher}}},\ and\ \bibinfo {author}
  {\bibfnamefont {L.}~\bibnamefont {{Wisotzki}}},\ }\bibfield  {title}
  {\bibinfo {title} {{A detached stellar-mass black hole candidate in the
  globular cluster NGC 3201}},\ }\href {https://doi.org/10.1093/mnrasl/slx203}
  {\bibfield  {journal} {\bibinfo  {journal} {\mnras}\ }\textbf {\bibinfo
  {volume} {475}},\ \bibinfo {pages} {L15} (\bibinfo {year} {2018})},\ \Eprint
  {https://arxiv.org/abs/1801.05642} {arXiv:1801.05642 [astro-ph.SR]}
  \BibitemShut {NoStop}%
\bibitem [{\citenamefont {{Giesers}}\ \emph {et~al.}(2019)\citenamefont
  {{Giesers}}, \citenamefont {{Kamann}}, \citenamefont {{Dreizler}},
  \citenamefont {{Husser}}, \citenamefont {{Askar}}, \citenamefont
  {{G{\"o}ttgens}}, \citenamefont {{Brinchmann}}, \citenamefont {{Latour}},
  \citenamefont {{Weilbacher}}, \citenamefont {{Wendt}},\ and\ \citenamefont
  {et~al.}}]{2019A&A...632A...3G}%
  \BibitemOpen
  \bibfield  {author} {\bibinfo {author} {\bibfnamefont {B.}~\bibnamefont
  {{Giesers}}}, \bibinfo {author} {\bibfnamefont {S.}~\bibnamefont {{Kamann}}},
  \bibinfo {author} {\bibfnamefont {S.}~\bibnamefont {{Dreizler}}}, \bibinfo
  {author} {\bibfnamefont {T.-O.}\ \bibnamefont {{Husser}}}, \bibinfo {author}
  {\bibfnamefont {A.}~\bibnamefont {{Askar}}}, \bibinfo {author} {\bibfnamefont
  {F.}~\bibnamefont {{G{\"o}ttgens}}}, \bibinfo {author} {\bibfnamefont
  {J.}~\bibnamefont {{Brinchmann}}}, \bibinfo {author} {\bibfnamefont
  {M.}~\bibnamefont {{Latour}}}, \bibinfo {author} {\bibfnamefont {P.~M.}\
  \bibnamefont {{Weilbacher}}}, \bibinfo {author} {\bibfnamefont
  {M.}~\bibnamefont {{Wendt}}},\ and\ \bibinfo {author} {\bibnamefont
  {et~al.}},\ }\bibfield  {title} {\bibinfo {title} {{A stellar census in
  globular clusters with MUSE: Binaries in NGC 3201}},\ }\href
  {https://doi.org/10.1051/0004-6361/201936203} {\bibfield  {journal} {\bibinfo
   {journal} {\aap}\ }\textbf {\bibinfo {volume} {632}},\ \bibinfo {eid} {A3}
  (\bibinfo {year} {2019})},\ \Eprint {https://arxiv.org/abs/1909.04050}
  {arXiv:1909.04050 [astro-ph.SR]} \BibitemShut {NoStop}%
\bibitem [{\citenamefont {{Morscher}}\ \emph {et~al.}(2015)\citenamefont
  {{Morscher}}, \citenamefont {{Pattabiraman}}, \citenamefont {{Rodriguez}},
  \citenamefont {{Rasio}},\ and\ \citenamefont
  {{Umbreit}}}]{2015ApJ...800....9M}%
  \BibitemOpen
  \bibfield  {author} {\bibinfo {author} {\bibfnamefont {M.}~\bibnamefont
  {{Morscher}}}, \bibinfo {author} {\bibfnamefont {B.}~\bibnamefont
  {{Pattabiraman}}}, \bibinfo {author} {\bibfnamefont {C.}~\bibnamefont
  {{Rodriguez}}}, \bibinfo {author} {\bibfnamefont {F.~A.}\ \bibnamefont
  {{Rasio}}},\ and\ \bibinfo {author} {\bibfnamefont {S.}~\bibnamefont
  {{Umbreit}}},\ }\bibfield  {title} {\bibinfo {title} {{The Dynamical
  Evolution of Stellar Black Holes in Globular Clusters}},\ }\href
  {https://doi.org/10.1088/0004-637X/800/1/9} {\bibfield  {journal} {\bibinfo
  {journal} {\apj}\ }\textbf {\bibinfo {volume} {800}},\ \bibinfo {eid} {9}
  (\bibinfo {year} {2015})},\ \Eprint {https://arxiv.org/abs/1409.0866}
  {arXiv:1409.0866 [astro-ph.GA]} \BibitemShut {NoStop}%
\bibitem [{\citenamefont {{Weatherford}}\ \emph {et~al.}(2018)\citenamefont
  {{Weatherford}}, \citenamefont {{Chatterjee}}, \citenamefont {{Rodriguez}},\
  and\ \citenamefont {{Rasio}}}]{2018ApJ...864...13W}%
  \BibitemOpen
  \bibfield  {author} {\bibinfo {author} {\bibfnamefont {N.~C.}\ \bibnamefont
  {{Weatherford}}}, \bibinfo {author} {\bibfnamefont {S.}~\bibnamefont
  {{Chatterjee}}}, \bibinfo {author} {\bibfnamefont {C.~L.}\ \bibnamefont
  {{Rodriguez}}},\ and\ \bibinfo {author} {\bibfnamefont {F.~A.}\ \bibnamefont
  {{Rasio}}},\ }\bibfield  {title} {\bibinfo {title} {{Predicting Stellar-mass
  Black Hole Populations in Globular Clusters}},\ }\href
  {https://doi.org/10.3847/1538-4357/aad63d} {\bibfield  {journal} {\bibinfo
  {journal} {\apj}\ }\textbf {\bibinfo {volume} {864}},\ \bibinfo {eid} {13}
  (\bibinfo {year} {2018})},\ \Eprint {https://arxiv.org/abs/1712.03979}
  {arXiv:1712.03979 [astro-ph.SR]} \BibitemShut {NoStop}%
\bibitem [{\citenamefont {{Askar}}\ \emph {et~al.}(2018)\citenamefont
  {{Askar}}, \citenamefont {{Arca Sedda}},\ and\ \citenamefont
  {{Giersz}}}]{2018MNRAS.478.1844A}%
  \BibitemOpen
  \bibfield  {author} {\bibinfo {author} {\bibfnamefont {A.}~\bibnamefont
  {{Askar}}}, \bibinfo {author} {\bibfnamefont {M.}~\bibnamefont {{Arca
  Sedda}}},\ and\ \bibinfo {author} {\bibfnamefont {M.}~\bibnamefont
  {{Giersz}}},\ }\bibfield  {title} {\bibinfo {title} {{MOCCA-SURVEY Database
  I: Galactic globular clusters harbouring a black hole subsystem}},\ }\href
  {https://doi.org/10.1093/mnras/sty1186} {\bibfield  {journal} {\bibinfo
  {journal} {\mnras}\ }\textbf {\bibinfo {volume} {478}},\ \bibinfo {pages}
  {1844} (\bibinfo {year} {2018})},\ \Eprint {https://arxiv.org/abs/1802.05284}
  {arXiv:1802.05284 [astro-ph.GA]} \BibitemShut {NoStop}%
\bibitem [{\citenamefont {{Arca Sedda}}\ \emph {et~al.}(2018)\citenamefont
  {{Arca Sedda}}, \citenamefont {{Askar}},\ and\ \citenamefont
  {{Giersz}}}]{2018MNRAS.479.4652A}%
  \BibitemOpen
  \bibfield  {author} {\bibinfo {author} {\bibfnamefont {M.}~\bibnamefont
  {{Arca Sedda}}}, \bibinfo {author} {\bibfnamefont {A.}~\bibnamefont
  {{Askar}}},\ and\ \bibinfo {author} {\bibfnamefont {M.}~\bibnamefont
  {{Giersz}}},\ }\bibfield  {title} {\bibinfo {title} {{MOCCA-Survey Database -
  I. Unravelling black hole subsystems in globular clusters}},\ }\href
  {https://doi.org/10.1093/mnras/sty1859} {\bibfield  {journal} {\bibinfo
  {journal} {\mnras}\ }\textbf {\bibinfo {volume} {479}},\ \bibinfo {pages}
  {4652} (\bibinfo {year} {2018})},\ \Eprint {https://arxiv.org/abs/1801.00795}
  {arXiv:1801.00795 [astro-ph.GA]} \BibitemShut {NoStop}%
\bibitem [{\citenamefont {{Weatherford}}\ \emph {et~al.}(2020)\citenamefont
  {{Weatherford}}, \citenamefont {{Chatterjee}}, \citenamefont {{Kremer}},\
  and\ \citenamefont {{Rasio}}}]{2020ApJ...898..162W}%
  \BibitemOpen
  \bibfield  {author} {\bibinfo {author} {\bibfnamefont {N.~C.}\ \bibnamefont
  {{Weatherford}}}, \bibinfo {author} {\bibfnamefont {S.}~\bibnamefont
  {{Chatterjee}}}, \bibinfo {author} {\bibfnamefont {K.}~\bibnamefont
  {{Kremer}}},\ and\ \bibinfo {author} {\bibfnamefont {F.~A.}\ \bibnamefont
  {{Rasio}}},\ }\bibfield  {title} {\bibinfo {title} {{A Dynamical Survey of
  Stellar-mass Black Holes in 50 Milky Way Globular Clusters}},\ }\href
  {https://doi.org/10.3847/1538-4357/ab9f98} {\bibfield  {journal} {\bibinfo
  {journal} {\apj}\ }\textbf {\bibinfo {volume} {898}},\ \bibinfo {eid} {162}
  (\bibinfo {year} {2020})},\ \Eprint {https://arxiv.org/abs/1911.09125}
  {arXiv:1911.09125 [astro-ph.SR]} \BibitemShut {NoStop}%
\bibitem [{\citenamefont {{Neumayer}}\ \emph {et~al.}(2020)\citenamefont
  {{Neumayer}}, \citenamefont {{Seth}},\ and\ \citenamefont
  {{B{\"o}ker}}}]{2020A&ARv..28....4N}%
  \BibitemOpen
  \bibfield  {author} {\bibinfo {author} {\bibfnamefont {N.}~\bibnamefont
  {{Neumayer}}}, \bibinfo {author} {\bibfnamefont {A.}~\bibnamefont {{Seth}}},\
  and\ \bibinfo {author} {\bibfnamefont {T.}~\bibnamefont {{B{\"o}ker}}},\
  }\bibfield  {title} {\bibinfo {title} {{Nuclear star clusters}},\ }\href
  {https://doi.org/10.1007/s00159-020-00125-0} {\bibfield  {journal} {\bibinfo
  {journal} {\aapr}\ }\textbf {\bibinfo {volume} {28}},\ \bibinfo {eid} {4}
  (\bibinfo {year} {2020})},\ \Eprint {https://arxiv.org/abs/2001.03626}
  {arXiv:2001.03626 [astro-ph.GA]} \BibitemShut {NoStop}%
\bibitem [{\citenamefont {{Saracino}}\ \emph {et~al.}(2022)\citenamefont
  {{Saracino}}, \citenamefont {{Kamann}}, \citenamefont {{Guarcello}},
  \citenamefont {{Usher}}, \citenamefont {{Bastian}}, \citenamefont
  {{Cabrera-Ziri}}, \citenamefont {{Gieles}}, \citenamefont {{Dreizler}},
  \citenamefont {{Da Costa}}, \citenamefont {{Husser}},\ and\ \citenamefont
  {{H{\'e}nault-Brunet}}}]{2022MNRAS.511.2914S}%
  \BibitemOpen
  \bibfield  {author} {\bibinfo {author} {\bibfnamefont {S.}~\bibnamefont
  {{Saracino}}}, \bibinfo {author} {\bibfnamefont {S.}~\bibnamefont
  {{Kamann}}}, \bibinfo {author} {\bibfnamefont {M.~G.}\ \bibnamefont
  {{Guarcello}}}, \bibinfo {author} {\bibfnamefont {C.}~\bibnamefont
  {{Usher}}}, \bibinfo {author} {\bibfnamefont {N.}~\bibnamefont {{Bastian}}},
  \bibinfo {author} {\bibfnamefont {I.}~\bibnamefont {{Cabrera-Ziri}}},
  \bibinfo {author} {\bibfnamefont {M.}~\bibnamefont {{Gieles}}}, \bibinfo
  {author} {\bibfnamefont {S.}~\bibnamefont {{Dreizler}}}, \bibinfo {author}
  {\bibfnamefont {G.~S.}\ \bibnamefont {{Da Costa}}}, \bibinfo {author}
  {\bibfnamefont {T.~O.}\ \bibnamefont {{Husser}}},\ and\ \bibinfo {author}
  {\bibfnamefont {V.}~\bibnamefont {{H{\'e}nault-Brunet}}},\ }\bibfield
  {title} {\bibinfo {title} {{A black hole detected in the young massive LMC
  cluster NGC 1850}},\ }\href {https://doi.org/10.1093/mnras/stab3159}
  {\bibfield  {journal} {\bibinfo  {journal} {\mnras}\ }\textbf {\bibinfo
  {volume} {511}},\ \bibinfo {pages} {2914} (\bibinfo {year} {2022})},\ \Eprint
  {https://arxiv.org/abs/2111.06506} {arXiv:2111.06506 [astro-ph.GA]}
  \BibitemShut {NoStop}%
\bibitem [{\citenamefont {{El-Badry}}\ and\ \citenamefont
  {{Burdge}}(2022)}]{2022MNRAS.511L..24E}%
  \BibitemOpen
  \bibfield  {author} {\bibinfo {author} {\bibfnamefont {K.}~\bibnamefont
  {{El-Badry}}}\ and\ \bibinfo {author} {\bibfnamefont {K.~B.}\ \bibnamefont
  {{Burdge}}},\ }\bibfield  {title} {\bibinfo {title} {{NGC 1850 BH1 is another
  stripped-star binary masquerading as a black hole}},\ }\href
  {https://doi.org/10.1093/mnrasl/slab135} {\bibfield  {journal} {\bibinfo
  {journal} {\mnras}\ }\textbf {\bibinfo {volume} {511}},\ \bibinfo {pages}
  {24} (\bibinfo {year} {2022})},\ \Eprint {https://arxiv.org/abs/2111.07925}
  {arXiv:2111.07925 [astro-ph.SR]} \BibitemShut {NoStop}%
\bibitem [{\citenamefont {{Stevance}}\ \emph {et~al.}(2022)\citenamefont
  {{Stevance}}, \citenamefont {{Parsons}},\ and\ \citenamefont
  {{Eldridge}}}]{2022MNRAS.511L..77S}%
  \BibitemOpen
  \bibfield  {author} {\bibinfo {author} {\bibfnamefont {H.~F.}\ \bibnamefont
  {{Stevance}}}, \bibinfo {author} {\bibfnamefont {S.~G.}\ \bibnamefont
  {{Parsons}}},\ and\ \bibinfo {author} {\bibfnamefont {J.~J.}\ \bibnamefont
  {{Eldridge}}},\ }\bibfield  {title} {\bibinfo {title} {{To be or not to be a
  black hole: detailed binary population models as a sanity check}},\ }\href
  {https://doi.org/10.1093/mnrasl/slac001} {\bibfield  {journal} {\bibinfo
  {journal} {\mnras}\ }\textbf {\bibinfo {volume} {511}},\ \bibinfo {pages}
  {L77} (\bibinfo {year} {2022})},\ \Eprint {https://arxiv.org/abs/2112.00015}
  {arXiv:2112.00015 [astro-ph.SR]} \BibitemShut {NoStop}%
\bibitem [{\citenamefont {{Saracino}}\ \emph {et~al.}(2023)\citenamefont
  {{Saracino}}, \citenamefont {{Shenar}}, \citenamefont {{Kamann}},
  \citenamefont {{Bastian}}, \citenamefont {{Gieles}}, \citenamefont {{Usher}},
  \citenamefont {{Bodensteiner}}, \citenamefont {{Kochoska}}, \citenamefont
  {{Orosz}},\ and\ \citenamefont {{Sana}}}]{2023MNRAS.521.3162S}%
  \BibitemOpen
  \bibfield  {author} {\bibinfo {author} {\bibfnamefont {S.}~\bibnamefont
  {{Saracino}}}, \bibinfo {author} {\bibfnamefont {T.}~\bibnamefont
  {{Shenar}}}, \bibinfo {author} {\bibfnamefont {S.}~\bibnamefont {{Kamann}}},
  \bibinfo {author} {\bibfnamefont {N.}~\bibnamefont {{Bastian}}}, \bibinfo
  {author} {\bibfnamefont {M.}~\bibnamefont {{Gieles}}}, \bibinfo {author}
  {\bibfnamefont {C.}~\bibnamefont {{Usher}}}, \bibinfo {author} {\bibfnamefont
  {J.}~\bibnamefont {{Bodensteiner}}}, \bibinfo {author} {\bibfnamefont
  {A.}~\bibnamefont {{Kochoska}}}, \bibinfo {author} {\bibfnamefont {J.~A.}\
  \bibnamefont {{Orosz}}},\ and\ \bibinfo {author} {\bibfnamefont
  {H.}~\bibnamefont {{Sana}}},\ }\bibfield  {title} {\bibinfo {title} {{Updated
  radial velocities and new constraints on the nature of the unseen source in
  NGC1850 BH1}},\ }\href {https://doi.org/10.1093/mnras/stad764} {\bibfield
  {journal} {\bibinfo  {journal} {\mnras}\ }\textbf {\bibinfo {volume} {521}},\
  \bibinfo {pages} {3162} (\bibinfo {year} {2023})},\ \Eprint
  {https://arxiv.org/abs/2303.07369} {arXiv:2303.07369 [astro-ph.GA]}
  \BibitemShut {NoStop}%
\bibitem [{\citenamefont {{Lennon}}\ \emph {et~al.}(2022)\citenamefont
  {{Lennon}}, \citenamefont {{Dufton}}, \citenamefont {{Villase{\~n}or}},
  \citenamefont {{Evans}}, \citenamefont {{Langer}}, \citenamefont {{Saxton}},
  \citenamefont {{Monageng}},\ and\ \citenamefont
  {{Toonen}}}]{2022A&A...665A.180L}%
  \BibitemOpen
  \bibfield  {author} {\bibinfo {author} {\bibfnamefont {D.~J.}\ \bibnamefont
  {{Lennon}}}, \bibinfo {author} {\bibfnamefont {P.~L.}\ \bibnamefont
  {{Dufton}}}, \bibinfo {author} {\bibfnamefont {J.~I.}\ \bibnamefont
  {{Villase{\~n}or}}}, \bibinfo {author} {\bibfnamefont {C.~J.}\ \bibnamefont
  {{Evans}}}, \bibinfo {author} {\bibfnamefont {N.}~\bibnamefont {{Langer}}},
  \bibinfo {author} {\bibfnamefont {R.}~\bibnamefont {{Saxton}}}, \bibinfo
  {author} {\bibfnamefont {I.~M.}\ \bibnamefont {{Monageng}}},\ and\ \bibinfo
  {author} {\bibfnamefont {S.}~\bibnamefont {{Toonen}}},\ }\bibfield  {title}
  {\bibinfo {title} {{The VLT-FLAMES survey of massive stars. NGC 2004\#115: A
  triple system hosting a possible short period B+BH binary}},\ }\href
  {https://doi.org/10.1051/0004-6361/202142413} {\bibfield  {journal} {\bibinfo
   {journal} {\aap}\ }\textbf {\bibinfo {volume} {665}},\ \bibinfo {eid} {A180}
  (\bibinfo {year} {2022})},\ \Eprint {https://arxiv.org/abs/2111.12173}
  {arXiv:2111.12173 [astro-ph.SR]} \BibitemShut {NoStop}%
\bibitem [{\citenamefont {{El-Badry}}\ \emph
  {et~al.}(2022{\natexlab{b}})\citenamefont {{El-Badry}}, \citenamefont
  {{Burdge}},\ and\ \citenamefont {{Mr{\'o}z}}}]{2022MNRAS.511.3089E}%
  \BibitemOpen
  \bibfield  {author} {\bibinfo {author} {\bibfnamefont {K.}~\bibnamefont
  {{El-Badry}}}, \bibinfo {author} {\bibfnamefont {K.~B.}\ \bibnamefont
  {{Burdge}}},\ and\ \bibinfo {author} {\bibfnamefont {P.}~\bibnamefont
  {{Mr{\'o}z}}},\ }\bibfield  {title} {\bibinfo {title} {{NGC 2004 \#115: a
  black hole imposter containing three luminous stars}},\ }\href
  {https://doi.org/10.1093/mnras/stac274} {\bibfield  {journal} {\bibinfo
  {journal} {\mnras}\ }\textbf {\bibinfo {volume} {511}},\ \bibinfo {pages}
  {3089} (\bibinfo {year} {2022}{\natexlab{b}})},\ \Eprint
  {https://arxiv.org/abs/2112.05030} {arXiv:2112.05030 [astro-ph.SR]}
  \BibitemShut {NoStop}%
\bibitem [{\citenamefont {{Banyard}}\ \emph
  {et~al.}(2022{\natexlab{a}})\citenamefont {{Banyard}}, \citenamefont
  {{Sana}}, \citenamefont {{Mahy}}, \citenamefont {{Bodensteiner}},
  \citenamefont {{Villase{\~n}or}},\ and\ \citenamefont
  {{Evans}}}]{2022A&A...658A..69B}%
  \BibitemOpen
  \bibfield  {author} {\bibinfo {author} {\bibfnamefont {G.}~\bibnamefont
  {{Banyard}}}, \bibinfo {author} {\bibfnamefont {H.}~\bibnamefont {{Sana}}},
  \bibinfo {author} {\bibfnamefont {L.}~\bibnamefont {{Mahy}}}, \bibinfo
  {author} {\bibfnamefont {J.}~\bibnamefont {{Bodensteiner}}}, \bibinfo
  {author} {\bibfnamefont {J.~I.}\ \bibnamefont {{Villase{\~n}or}}},\ and\
  \bibinfo {author} {\bibfnamefont {C.~J.}\ \bibnamefont {{Evans}}},\
  }\bibfield  {title} {\bibinfo {title} {{The observed multiplicity properties
  of B-type stars in the Galactic young open cluster NGC 6231}},\ }\href
  {https://doi.org/10.1051/0004-6361/202141037} {\bibfield  {journal} {\bibinfo
   {journal} {\aap}\ }\textbf {\bibinfo {volume} {658}},\ \bibinfo {eid} {A69}
  (\bibinfo {year} {2022}{\natexlab{a}})},\ \Eprint
  {https://arxiv.org/abs/2108.07814} {arXiv:2108.07814 [astro-ph.SR]}
  \BibitemShut {NoStop}%
\bibitem [{\citenamefont {{Banyard}}\ \emph
  {et~al.}(2022{\natexlab{b}})\citenamefont {{Banyard}}, \citenamefont
  {{Mahy}}, \citenamefont {{Sana}}, \citenamefont {{Bodensteiner}},
  \citenamefont {{Villase{\'n}or}}, \citenamefont {{Sen}}, \citenamefont
  {{Langer}}, \citenamefont {{de Mink}}, \citenamefont {{Picco}},\ and\
  \citenamefont {{Shenar}}}]{2022arXiv221007149B}%
  \BibitemOpen
  \bibfield  {author} {\bibinfo {author} {\bibfnamefont {G.}~\bibnamefont
  {{Banyard}}}, \bibinfo {author} {\bibfnamefont {L.}~\bibnamefont {{Mahy}}},
  \bibinfo {author} {\bibfnamefont {H.}~\bibnamefont {{Sana}}}, \bibinfo
  {author} {\bibfnamefont {J.}~\bibnamefont {{Bodensteiner}}}, \bibinfo
  {author} {\bibfnamefont {J.~I.}\ \bibnamefont {{Villase{\'n}or}}}, \bibinfo
  {author} {\bibfnamefont {K.}~\bibnamefont {{Sen}}}, \bibinfo {author}
  {\bibfnamefont {N.}~\bibnamefont {{Langer}}}, \bibinfo {author}
  {\bibfnamefont {S.}~\bibnamefont {{de Mink}}}, \bibinfo {author}
  {\bibfnamefont {A.}~\bibnamefont {{Picco}}},\ and\ \bibinfo {author}
  {\bibfnamefont {T.}~\bibnamefont {{Shenar}}},\ }\bibfield  {title} {\bibinfo
  {title} {{Searching for compact objects in the single-lined spectroscopic
  binaries of the young Galactic cluster NGC 6231}},\ }\href@noop {} {\bibfield
   {journal} {\bibinfo  {journal} {arXiv e-prints}\ ,\ \bibinfo {eid}
  {arXiv:2210.07149}} (\bibinfo {year} {2022}{\natexlab{b}})},\ \Eprint
  {https://arxiv.org/abs/2210.07149} {arXiv:2210.07149 [astro-ph.SR]}
  \BibitemShut {NoStop}%
\bibitem [{\citenamefont {{Thatte}}\ \emph {et~al.}(2010)\citenamefont
  {{Thatte}}, \citenamefont {{Tecza}}, \citenamefont {{Clarke}}, \citenamefont
  {{Davies}}, \citenamefont {{Remillieux}}, \citenamefont {{Bacon}},
  \citenamefont {{Lunney}}, \citenamefont {{Arribas}}, \citenamefont
  {{Mediavilla}}, \citenamefont {{Gago}}, \citenamefont {{Bezawada}},
  \citenamefont {{Ferruit}}, \citenamefont {{Fragoso}}, \citenamefont
  {{Freeman}}, \citenamefont {{Fuentes}}, \citenamefont {{Fusco}},
  \citenamefont {{Gallie}}, \citenamefont {{Garcia}}, \citenamefont
  {{Goodsall}}, \citenamefont {{Gracia}}, \citenamefont {{Jarno}},
  \citenamefont {{Kosmalski}}, \citenamefont {{Lynn}}, \citenamefont {{McLay}},
  \citenamefont {{Montgomery}}, \citenamefont {{Pecontal}}, \citenamefont
  {{Schnetler}}, \citenamefont {{Smith}}, \citenamefont {{Sosa}}, \citenamefont
  {{Battaglia}}, \citenamefont {{Bowles}}, \citenamefont {{Colina}},
  \citenamefont {{Emsellem}}, \citenamefont {{Garcia-Perez}}, \citenamefont
  {{Gladysz}}, \citenamefont {{Hook}}, \citenamefont {{Irwin}}, \citenamefont
  {{Jarvis}}, \citenamefont {{Kennicutt}}, \citenamefont {{Levan}},
  \citenamefont {{Longmore}}, \citenamefont {{Magorrian}}, \citenamefont
  {{McCaughrean}}, \citenamefont {{Origlia}}, \citenamefont {{Rebolo}},
  \citenamefont {{Rigopoulou}}, \citenamefont {{Ryan}}, \citenamefont
  {{Swinbank}}, \citenamefont {{Tanvir}}, \citenamefont {{Tolstoy}},\ and\
  \citenamefont {{Verma}}}]{2010SPIE.7735E..2IT}%
  \BibitemOpen
  \bibfield  {author} {\bibinfo {author} {\bibfnamefont {N.}~\bibnamefont
  {{Thatte}}}, \bibinfo {author} {\bibfnamefont {M.}~\bibnamefont {{Tecza}}},
  \bibinfo {author} {\bibfnamefont {F.}~\bibnamefont {{Clarke}}}, \bibinfo
  {author} {\bibfnamefont {R.~L.}\ \bibnamefont {{Davies}}}, \bibinfo {author}
  {\bibfnamefont {A.}~\bibnamefont {{Remillieux}}}, \bibinfo {author}
  {\bibfnamefont {R.}~\bibnamefont {{Bacon}}}, \bibinfo {author} {\bibfnamefont
  {D.}~\bibnamefont {{Lunney}}}, \bibinfo {author} {\bibfnamefont
  {S.}~\bibnamefont {{Arribas}}}, \bibinfo {author} {\bibfnamefont
  {E.}~\bibnamefont {{Mediavilla}}}, \bibinfo {author} {\bibfnamefont
  {F.}~\bibnamefont {{Gago}}}, \bibinfo {author} {\bibfnamefont
  {N.}~\bibnamefont {{Bezawada}}}, \bibinfo {author} {\bibfnamefont
  {P.}~\bibnamefont {{Ferruit}}}, \bibinfo {author} {\bibfnamefont
  {A.}~\bibnamefont {{Fragoso}}}, \bibinfo {author} {\bibfnamefont
  {D.}~\bibnamefont {{Freeman}}}, \bibinfo {author} {\bibfnamefont
  {J.}~\bibnamefont {{Fuentes}}}, \bibinfo {author} {\bibfnamefont
  {T.}~\bibnamefont {{Fusco}}}, \bibinfo {author} {\bibfnamefont
  {A.}~\bibnamefont {{Gallie}}}, \bibinfo {author} {\bibfnamefont
  {A.}~\bibnamefont {{Garcia}}}, \bibinfo {author} {\bibfnamefont
  {T.}~\bibnamefont {{Goodsall}}}, \bibinfo {author} {\bibfnamefont
  {F.}~\bibnamefont {{Gracia}}}, \bibinfo {author} {\bibfnamefont
  {A.}~\bibnamefont {{Jarno}}}, \bibinfo {author} {\bibfnamefont
  {J.}~\bibnamefont {{Kosmalski}}}, \bibinfo {author} {\bibfnamefont
  {J.}~\bibnamefont {{Lynn}}}, \bibinfo {author} {\bibfnamefont
  {S.}~\bibnamefont {{McLay}}}, \bibinfo {author} {\bibfnamefont
  {D.}~\bibnamefont {{Montgomery}}}, \bibinfo {author} {\bibfnamefont
  {A.}~\bibnamefont {{Pecontal}}}, \bibinfo {author} {\bibfnamefont
  {H.}~\bibnamefont {{Schnetler}}}, \bibinfo {author} {\bibfnamefont
  {H.}~\bibnamefont {{Smith}}}, \bibinfo {author} {\bibfnamefont
  {D.}~\bibnamefont {{Sosa}}}, \bibinfo {author} {\bibfnamefont
  {G.}~\bibnamefont {{Battaglia}}}, \bibinfo {author} {\bibfnamefont
  {N.}~\bibnamefont {{Bowles}}}, \bibinfo {author} {\bibfnamefont
  {L.}~\bibnamefont {{Colina}}}, \bibinfo {author} {\bibfnamefont
  {E.}~\bibnamefont {{Emsellem}}}, \bibinfo {author} {\bibfnamefont
  {A.}~\bibnamefont {{Garcia-Perez}}}, \bibinfo {author} {\bibfnamefont
  {S.}~\bibnamefont {{Gladysz}}}, \bibinfo {author} {\bibfnamefont
  {I.}~\bibnamefont {{Hook}}}, \bibinfo {author} {\bibfnamefont
  {P.}~\bibnamefont {{Irwin}}}, \bibinfo {author} {\bibfnamefont
  {M.}~\bibnamefont {{Jarvis}}}, \bibinfo {author} {\bibfnamefont
  {R.}~\bibnamefont {{Kennicutt}}}, \bibinfo {author} {\bibfnamefont
  {A.}~\bibnamefont {{Levan}}}, \bibinfo {author} {\bibfnamefont
  {A.}~\bibnamefont {{Longmore}}}, \bibinfo {author} {\bibfnamefont
  {J.}~\bibnamefont {{Magorrian}}}, \bibinfo {author} {\bibfnamefont
  {M.}~\bibnamefont {{McCaughrean}}}, \bibinfo {author} {\bibfnamefont
  {L.}~\bibnamefont {{Origlia}}}, \bibinfo {author} {\bibfnamefont
  {R.}~\bibnamefont {{Rebolo}}}, \bibinfo {author} {\bibfnamefont
  {D.}~\bibnamefont {{Rigopoulou}}}, \bibinfo {author} {\bibfnamefont
  {S.}~\bibnamefont {{Ryan}}}, \bibinfo {author} {\bibfnamefont
  {M.}~\bibnamefont {{Swinbank}}}, \bibinfo {author} {\bibfnamefont
  {N.}~\bibnamefont {{Tanvir}}}, \bibinfo {author} {\bibfnamefont
  {E.}~\bibnamefont {{Tolstoy}}},\ and\ \bibinfo {author} {\bibfnamefont
  {A.}~\bibnamefont {{Verma}}},\ }\bibfield  {title} {\bibinfo {title}
  {{HARMONI: a single-field wide-band integral-field spectrograph for the
  European ELT}},\ }in\ \href {https://doi.org/10.1117/12.857445} {\emph
  {\bibinfo {booktitle} {Ground-based and Airborne Instrumentation for
  Astronomy III}}},\ \bibinfo {series} {Society of Photo-Optical
  Instrumentation Engineers (SPIE) Conference Series}, Vol.\ \bibinfo {volume}
  {7735},\ \bibinfo {editor} {edited by\ \bibinfo {editor} {\bibfnamefont
  {I.~S.}\ \bibnamefont {{McLean}}}, \bibinfo {editor} {\bibfnamefont {S.~K.}\
  \bibnamefont {{Ramsay}}},\ and\ \bibinfo {editor} {\bibfnamefont
  {H.}~\bibnamefont {{Takami}}}}\ (\bibinfo {year} {2010})\ p.\ \bibinfo
  {pages} {77352I}\BibitemShut {NoStop}%
\bibitem [{\citenamefont {{Andrews}}\ \emph {et~al.}(2019)\citenamefont
  {{Andrews}}, \citenamefont {{Breivik}},\ and\ \citenamefont
  {{Chatterjee}}}]{2019ApJ...886...68A}%
  \BibitemOpen
  \bibfield  {author} {\bibinfo {author} {\bibfnamefont {J.~J.}\ \bibnamefont
  {{Andrews}}}, \bibinfo {author} {\bibfnamefont {K.}~\bibnamefont
  {{Breivik}}},\ and\ \bibinfo {author} {\bibfnamefont {S.}~\bibnamefont
  {{Chatterjee}}},\ }\bibfield  {title} {\bibinfo {title} {{Weighing the
  Darkness: Astrometric Mass Measurement of Hidden Stellar Companions Using
  Gaia}},\ }\href {https://doi.org/10.3847/1538-4357/ab441f} {\bibfield
  {journal} {\bibinfo  {journal} {\apj}\ }\textbf {\bibinfo {volume} {886}},\
  \bibinfo {eid} {68} (\bibinfo {year} {2019})},\ \Eprint
  {https://arxiv.org/abs/1909.05606} {arXiv:1909.05606 [astro-ph.SR]}
  \BibitemShut {NoStop}%
\bibitem [{\citenamefont {{Martin}}\ \emph {et~al.}(1997)\citenamefont
  {{Martin}}, \citenamefont {{Mignard}},\ and\ \citenamefont
  {{Froeschle}}}]{1997A&AS..122..571M}%
  \BibitemOpen
  \bibfield  {author} {\bibinfo {author} {\bibfnamefont {C.}~\bibnamefont
  {{Martin}}}, \bibinfo {author} {\bibfnamefont {F.}~\bibnamefont
  {{Mignard}}},\ and\ \bibinfo {author} {\bibfnamefont {M.}~\bibnamefont
  {{Froeschle}}},\ }\bibfield  {title} {\bibinfo {title} {{Mass determination
  of astrometric binaries with Hipparcos. I. Theory and simulation}},\ }\href
  {https://doi.org/10.1051/aas:1997339} {\bibfield  {journal} {\bibinfo
  {journal} {\aaps}\ }\textbf {\bibinfo {volume} {122}},\ \bibinfo {pages}
  {571} (\bibinfo {year} {1997})}\BibitemShut {NoStop}%
\bibitem [{\citenamefont {{Martin}}\ \emph {et~al.}(1998)\citenamefont
  {{Martin}}, \citenamefont {{Mignard}}, \citenamefont {{Hartkopf}},\ and\
  \citenamefont {{McAlister}}}]{1998A&AS..133..149M}%
  \BibitemOpen
  \bibfield  {author} {\bibinfo {author} {\bibfnamefont {C.}~\bibnamefont
  {{Martin}}}, \bibinfo {author} {\bibfnamefont {F.}~\bibnamefont {{Mignard}}},
  \bibinfo {author} {\bibfnamefont {W.~I.}\ \bibnamefont {{Hartkopf}}},\ and\
  \bibinfo {author} {\bibfnamefont {H.~A.}\ \bibnamefont {{McAlister}}},\
  }\bibfield  {title} {\bibinfo {title} {{Mass determination of astrometric
  binaries with Hipparcos. III. New results for 28 systems}},\ }\href
  {https://doi.org/10.1051/aas:1998459} {\bibfield  {journal} {\bibinfo
  {journal} {\aaps}\ }\textbf {\bibinfo {volume} {133}},\ \bibinfo {pages}
  {149} (\bibinfo {year} {1998})}\BibitemShut {NoStop}%
\bibitem [{\citenamefont {{Martin}}\ and\ \citenamefont
  {{Mignard}}(1998)}]{1998A&A...330..585M}%
  \BibitemOpen
  \bibfield  {author} {\bibinfo {author} {\bibfnamefont {C.}~\bibnamefont
  {{Martin}}}\ and\ \bibinfo {author} {\bibfnamefont {F.}~\bibnamefont
  {{Mignard}}},\ }\bibfield  {title} {\bibinfo {title} {{Mass determination of
  astrometric binaries with Hipparcos. II. Selection of candidates and
  results}},\ }\href@noop {} {\bibfield  {journal} {\bibinfo  {journal} {\aap}\
  }\textbf {\bibinfo {volume} {330}},\ \bibinfo {pages} {585} (\bibinfo {year}
  {1998})}\BibitemShut {NoStop}%
\bibitem [{\citenamefont {{Breivik}}\ \emph {et~al.}(2017)\citenamefont
  {{Breivik}}, \citenamefont {{Chatterjee}},\ and\ \citenamefont
  {{Larson}}}]{2017ApJ...850L..13B}%
  \BibitemOpen
  \bibfield  {author} {\bibinfo {author} {\bibfnamefont {K.}~\bibnamefont
  {{Breivik}}}, \bibinfo {author} {\bibfnamefont {S.}~\bibnamefont
  {{Chatterjee}}},\ and\ \bibinfo {author} {\bibfnamefont {S.~L.}\ \bibnamefont
  {{Larson}}},\ }\bibfield  {title} {\bibinfo {title} {{Revealing Black Holes
  with Gaia}},\ }\href {https://doi.org/10.3847/2041-8213/aa97d5} {\bibfield
  {journal} {\bibinfo  {journal} {\apjl}\ }\textbf {\bibinfo {volume} {850}},\
  \bibinfo {eid} {L13} (\bibinfo {year} {2017})},\ \Eprint
  {https://arxiv.org/abs/1710.04657} {arXiv:1710.04657 [astro-ph.SR]}
  \BibitemShut {NoStop}%
\bibitem [{\citenamefont {{Mashian}}\ and\ \citenamefont
  {{Loeb}}(2017)}]{2017MNRAS.470.2611M}%
  \BibitemOpen
  \bibfield  {author} {\bibinfo {author} {\bibfnamefont {N.}~\bibnamefont
  {{Mashian}}}\ and\ \bibinfo {author} {\bibfnamefont {A.}~\bibnamefont
  {{Loeb}}},\ }\bibfield  {title} {\bibinfo {title} {{Hunting black holes with
  Gaia}},\ }\href {https://doi.org/10.1093/mnras/stx1410} {\bibfield  {journal}
  {\bibinfo  {journal} {\mnras}\ }\textbf {\bibinfo {volume} {470}},\ \bibinfo
  {pages} {2611} (\bibinfo {year} {2017})},\ \Eprint
  {https://arxiv.org/abs/1704.03455} {arXiv:1704.03455 [astro-ph.HE]}
  \BibitemShut {NoStop}%
\bibitem [{\citenamefont {{Janssens}}\ \emph {et~al.}(2022)\citenamefont
  {{Janssens}}, \citenamefont {{Shenar}}, \citenamefont {{Sana}}, \citenamefont
  {{Faigler}}, \citenamefont {{Langer}}, \citenamefont {{Marchant}},
  \citenamefont {{Mazeh}}, \citenamefont {{Sch{\"u}rmann}},\ and\ \citenamefont
  {{Shahaf}}}]{2022A&A...658A.129J}%
  \BibitemOpen
  \bibfield  {author} {\bibinfo {author} {\bibfnamefont {S.}~\bibnamefont
  {{Janssens}}}, \bibinfo {author} {\bibfnamefont {T.}~\bibnamefont
  {{Shenar}}}, \bibinfo {author} {\bibfnamefont {H.}~\bibnamefont {{Sana}}},
  \bibinfo {author} {\bibfnamefont {S.}~\bibnamefont {{Faigler}}}, \bibinfo
  {author} {\bibfnamefont {N.}~\bibnamefont {{Langer}}}, \bibinfo {author}
  {\bibfnamefont {P.}~\bibnamefont {{Marchant}}}, \bibinfo {author}
  {\bibfnamefont {T.}~\bibnamefont {{Mazeh}}}, \bibinfo {author} {\bibfnamefont
  {C.}~\bibnamefont {{Sch{\"u}rmann}}},\ and\ \bibinfo {author} {\bibfnamefont
  {S.}~\bibnamefont {{Shahaf}}},\ }\bibfield  {title} {\bibinfo {title}
  {{Uncovering astrometric black hole binaries with massive main-sequence
  companions with Gaia}},\ }\href {https://doi.org/10.1051/0004-6361/202141866}
  {\bibfield  {journal} {\bibinfo  {journal} {\aap}\ }\textbf {\bibinfo
  {volume} {658}},\ \bibinfo {eid} {A129} (\bibinfo {year} {2022})},\ \Eprint
  {https://arxiv.org/abs/2111.06427} {arXiv:2111.06427 [astro-ph.SR]}
  \BibitemShut {NoStop}%
\bibitem [{\citenamefont {{Gaia Collaboration}}\ \emph
  {et~al.}(2021)\citenamefont {{Gaia Collaboration}}, \citenamefont {{Brown}},
  \citenamefont {{Vallenari}}, \citenamefont {{Prusti}}, \citenamefont {{de
  Bruijne}}, \citenamefont {{Babusiaux}}, \citenamefont {{Biermann}},
  \citenamefont {{Creevey}}, \citenamefont {{Evans}}, \citenamefont {{Eyer}},
  \citenamefont {{Hutton}}, \citenamefont {{Jansen}}, \citenamefont {{Jordi}},
  \citenamefont {{Klioner}}, \citenamefont {{Lammers}}, \citenamefont
  {{Lindegren}}, \citenamefont {{Luri}}, \citenamefont {{Mignard}},
  \citenamefont {{Panem}}, \citenamefont {{Pourbaix}}, \citenamefont
  {{Randich}}, \citenamefont {{Sartoretti}}, \citenamefont {{Soubiran}},
  \citenamefont {{Walton}}, \citenamefont {{Arenou}}, \citenamefont
  {{Bailer-Jones}}, \citenamefont {{Bastian}}, \citenamefont {{Cropper}},
  \citenamefont {{Drimmel}}, \citenamefont {{Katz}}, \citenamefont
  {{Lattanzi}}, \citenamefont {{van Leeuwen}}, \citenamefont {{Bakker}},
  \citenamefont {{Cacciari}}, \citenamefont {{Casta{\~n}eda}}, \citenamefont
  {{De Angeli}}, \citenamefont {{Ducourant}}, \citenamefont {{Fabricius}},
  \citenamefont {{Fouesneau}}, \citenamefont {{Fr{\'e}mat}}, \citenamefont
  {{Guerra}}, \citenamefont {{Guerrier}}, \citenamefont {{Guiraud}},
  \citenamefont {{Jean-Antoine Piccolo}}, \citenamefont {{Masana}},
  \citenamefont {{Messineo}}, \citenamefont {{Mowlavi}}, \citenamefont
  {{Nicolas}}, \citenamefont {{Nienartowicz}}, \citenamefont {{Pailler}},
  \citenamefont {{Panuzzo}}, \citenamefont {{Riclet}}, \citenamefont {{Roux}},
  \citenamefont {{Seabroke}}, \citenamefont {{Sordo}}, \citenamefont {{Tanga}},
  \citenamefont {{Th{\'e}venin}}, \citenamefont {{Gracia-Abril}}, \citenamefont
  {{Portell}}, \citenamefont {{Teyssier}}, \citenamefont {{Altmann}},
  \citenamefont {{Andrae}}, \citenamefont {{Bellas-Velidis}}, \citenamefont
  {{Benson}}, \citenamefont {{Berthier}}, \citenamefont {{Blomme}},
  \citenamefont {{Brugaletta}}, \citenamefont {{Burgess}}, \citenamefont
  {{Busso}}, \citenamefont {{Carry}}, \citenamefont {{Cellino}}, \citenamefont
  {{Cheek}}, \citenamefont {{Clementini}}, \citenamefont {{Damerdji}},
  \citenamefont {{Davidson}}, \citenamefont {{Delchambre}}, \citenamefont
  {{Dell'Oro}}, \citenamefont {{Fern{\'a}ndez-Hern{\'a}ndez}}, \citenamefont
  {{Galluccio}}, \citenamefont {{Garc{\'\i}a-Lario}}, \citenamefont
  {{Garcia-Reinaldos}}, \citenamefont {{Gonz{\'a}lez-N{\'u}{\~n}ez}},
  \citenamefont {{Gosset}}, \citenamefont {{Haigron}}, \citenamefont
  {{Halbwachs}}, \citenamefont {{Hambly}}, \citenamefont {{Harrison}},
  \citenamefont {{Hatzidimitriou}}, \citenamefont {{Heiter}}, \citenamefont
  {{Hern{\'a}ndez}}, \citenamefont {{Hestroffer}}, \citenamefont {{Hodgkin}},
  \citenamefont {{Holl}}, \citenamefont {{Jan{\ss}en}}, \citenamefont
  {{Jevardat de Fombelle}}, \citenamefont {{Jordan}}, \citenamefont
  {{Krone-Martins}}, \citenamefont {{Lanzafame}}, \citenamefont
  {{L{\"o}ffler}}, \citenamefont {{Lorca}}, \citenamefont {{Manteiga}},
  \citenamefont {{Marchal}}, \citenamefont {{Marrese}}, \citenamefont
  {{Moitinho}}, \citenamefont {{Mora}}, \citenamefont {{Muinonen}},
  \citenamefont {{Osborne}}, \citenamefont {{Pancino}}, \citenamefont
  {{Pauwels}}, \citenamefont {{Petit}}, \citenamefont {{Recio-Blanco}},
  \citenamefont {{Richards}}, \citenamefont {{Riello}}, \citenamefont
  {{Rimoldini}}, \citenamefont {{Robin}}, \citenamefont {{Roegiers}},
  \citenamefont {{Rybizki}}, \citenamefont {{Sarro}}, \citenamefont {{Siopis}},
  \citenamefont {{Smith}}, \citenamefont {{Sozzetti}}, \citenamefont {{Ulla}},
  \citenamefont {{Utrilla}}, \citenamefont {{van Leeuwen}}, \citenamefont {{van
  Reeven}}, \citenamefont {{Abbas}}, \citenamefont {{Abreu Aramburu}},
  \citenamefont {{Accart}}, \citenamefont {{Aerts}}, \citenamefont {{Aguado}},
  \citenamefont {{Ajaj}}, \citenamefont {{Altavilla}}, \citenamefont
  {{{\'A}lvarez}}, \citenamefont {{{\'A}lvarez Cid-Fuentes}}, \citenamefont
  {{Alves}}, \citenamefont {{Anderson}}, \citenamefont {{Anglada Varela}},
  \citenamefont {{Antoja}}, \citenamefont {{Audard}}, \citenamefont {{Baines}},
  \citenamefont {{Baker}}, \citenamefont {{Balaguer-N{\'u}{\~n}ez}},
  \citenamefont {{Balbinot}}, \citenamefont {{Balog}}, \citenamefont
  {{Barache}}, \citenamefont {{Barbato}}, \citenamefont {{Barros}},
  \citenamefont {{Barstow}}, \citenamefont {{Bartolom{\'e}}}, \citenamefont
  {{Bassilana}}, \citenamefont {{Bauchet}}, \citenamefont {{Baudesson-Stella}},
  \citenamefont {{Becciani}}, \citenamefont {{Bellazzini}}, \citenamefont
  {{Bernet}}, \citenamefont {{Bertone}}, \citenamefont {{Bianchi}},
  \citenamefont {{Blanco-Cuaresma}}, \citenamefont {{Boch}}, \citenamefont
  {{Bombrun}}, \citenamefont {{Bossini}}, \citenamefont {{Bouquillon}},
  \citenamefont {{Bragaglia}}, \citenamefont {{Bramante}}, \citenamefont
  {{Breedt}}, \citenamefont {{Bressan}}, \citenamefont {{Brouillet}},
  \citenamefont {{Bucciarelli}}, \citenamefont {{Burlacu}}, \citenamefont
  {{Busonero}}, \citenamefont {{Butkevich}}, \citenamefont {{Buzzi}},
  \citenamefont {{Caffau}}, \citenamefont {{Cancelliere}}, \citenamefont
  {{C{\'a}novas}}, \citenamefont {{Cantat-Gaudin}}, \citenamefont {{Carballo}},
  \citenamefont {{Carlucci}}, \citenamefont {{Carnerero}}, \citenamefont
  {{Carrasco}}, \citenamefont {{Casamiquela}}, \citenamefont {{Castellani}},
  \citenamefont {{Castro-Ginard}}, \citenamefont {{Castro Sampol}},
  \citenamefont {{Chaoul}}, \citenamefont {{Charlot}}, \citenamefont
  {{Chemin}}, \citenamefont {{Chiavassa}}, \citenamefont {{Cioni}},
  \citenamefont {{Comoretto}}, \citenamefont {{Cooper}}, \citenamefont
  {{Cornez}}, \citenamefont {{Cowell}}, \citenamefont {{Crifo}}, \citenamefont
  {{Crosta}}, \citenamefont {{Crowley}}, \citenamefont {{Dafonte}},
  \citenamefont {{Dapergolas}}, \citenamefont {{David}}, \citenamefont
  {{David}}, \citenamefont {{de Laverny}}, \citenamefont {{De Luise}},
  \citenamefont {{De March}}, \citenamefont {{De Ridder}}, \citenamefont {{de
  Souza}}, \citenamefont {{de Teodoro}}, \citenamefont {{de Torres}},
  \citenamefont {{del Peloso}}, \citenamefont {{del Pozo}}, \citenamefont
  {{Delbo}}, \citenamefont {{Delgado}}, \citenamefont {{Delgado}},
  \citenamefont {{Delisle}}, \citenamefont {{Di Matteo}}, \citenamefont
  {{Diakite}}, \citenamefont {{Diener}}, \citenamefont {{Distefano}},
  \citenamefont {{Dolding}}, \citenamefont {{Eappachen}}, \citenamefont
  {{Edvardsson}}, \citenamefont {{Enke}}, \citenamefont {{Esquej}},
  \citenamefont {{Fabre}}, \citenamefont {{Fabrizio}}, \citenamefont
  {{Faigler}}, \citenamefont {{Fedorets}}, \citenamefont {{Fernique}},
  \citenamefont {{Fienga}}, \citenamefont {{Figueras}}, \citenamefont
  {{Fouron}}, \citenamefont {{Fragkoudi}}, \citenamefont {{Fraile}},
  \citenamefont {{Franke}}, \citenamefont {{Gai}}, \citenamefont {{Garabato}},
  \citenamefont {{Garcia-Gutierrez}}, \citenamefont {{Garc{\'\i}a-Torres}},
  \citenamefont {{Garofalo}}, \citenamefont {{Gavras}}, \citenamefont
  {{Gerlach}}, \citenamefont {{Geyer}}, \citenamefont {{Giacobbe}},
  \citenamefont {{Gilmore}}, \citenamefont {{Girona}}, \citenamefont
  {{Giuffrida}}, \citenamefont {{Gomel}}, \citenamefont {{Gomez}},
  \citenamefont {{Gonzalez-Santamaria}}, \citenamefont {{Gonz{\'a}lez-Vidal}},
  \citenamefont {{Granvik}}, \citenamefont {{Guti{\'e}rrez-S{\'a}nchez}},
  \citenamefont {{Guy}}, \citenamefont {{Hauser}}, \citenamefont {{Haywood}},
  \citenamefont {{Helmi}}, \citenamefont {{Hidalgo}}, \citenamefont {{Hilger}},
  \citenamefont {{H{\l}adczuk}}, \citenamefont {{Hobbs}}, \citenamefont
  {{Holland}}, \citenamefont {{Huckle}}, \citenamefont {{Jasniewicz}},
  \citenamefont {{Jonker}}, \citenamefont {{Juaristi Campillo}}, \citenamefont
  {{Julbe}}, \citenamefont {{Karbevska}}, \citenamefont {{Kervella}},
  \citenamefont {{Khanna}}, \citenamefont {{Kochoska}}, \citenamefont
  {{Kontizas}}, \citenamefont {{Kordopatis}}, \citenamefont {{Korn}},
  \citenamefont {{Kostrzewa-Rutkowska}}, \citenamefont {{Kruszy{\'n}ska}},
  \citenamefont {{Lambert}}, \citenamefont {{Lanza}}, \citenamefont {{Lasne}},
  \citenamefont {{Le Campion}}, \citenamefont {{Le Fustec}}, \citenamefont
  {{Lebreton}}, \citenamefont {{Lebzelter}}, \citenamefont {{Leccia}},
  \citenamefont {{Leclerc}}, \citenamefont {{Lecoeur-Taibi}}, \citenamefont
  {{Liao}}, \citenamefont {{Licata}}, \citenamefont {{Lindstr{\o}m}},
  \citenamefont {{Lister}}, \citenamefont {{Livanou}}, \citenamefont {{Lobel}},
  \citenamefont {{Madrero Pardo}}, \citenamefont {{Managau}}, \citenamefont
  {{Mann}}, \citenamefont {{Marchant}}, \citenamefont {{Marconi}},
  \citenamefont {{Marcos Santos}}, \citenamefont {{Marinoni}}, \citenamefont
  {{Marocco}}, \citenamefont {{Marshall}}, \citenamefont {{Martin Polo}},
  \citenamefont {{Mart{\'\i}n-Fleitas}}, \citenamefont {{Masip}}, \citenamefont
  {{Massari}}, \citenamefont {{Mastrobuono-Battisti}}, \citenamefont {{Mazeh}},
  \citenamefont {{McMillan}}, \citenamefont {{Messina}}, \citenamefont
  {{Michalik}}, \citenamefont {{Millar}}, \citenamefont {{Mints}},
  \citenamefont {{Molina}}, \citenamefont {{Molinaro}}, \citenamefont
  {{Moln{\'a}r}}, \citenamefont {{Montegriffo}}, \citenamefont {{Mor}},
  \citenamefont {{Morbidelli}}, \citenamefont {{Morel}}, \citenamefont
  {{Morris}}, \citenamefont {{Mulone}}, \citenamefont {{Munoz}}, \citenamefont
  {{Muraveva}}, \citenamefont {{Murphy}}, \citenamefont {{Musella}},
  \citenamefont {{Noval}}, \citenamefont {{Ord{\'e}novic}}, \citenamefont
  {{Orr{\`u}}}, \citenamefont {{Osinde}}, \citenamefont {{Pagani}},
  \citenamefont {{Pagano}}, \citenamefont {{Palaversa}}, \citenamefont
  {{Palicio}}, \citenamefont {{Panahi}}, \citenamefont {{Pawlak}},
  \citenamefont {{Pe{\~n}alosa Esteller}}, \citenamefont {{Penttil{\"a}}},
  \citenamefont {{Piersimoni}}, \citenamefont {{Pineau}}, \citenamefont
  {{Plachy}}, \citenamefont {{Plum}}, \citenamefont {{Poggio}}, \citenamefont
  {{Poretti}}, \citenamefont {{Poujoulet}}, \citenamefont {{Pr{\v{s}}a}},
  \citenamefont {{Pulone}}, \citenamefont {{Racero}}, \citenamefont
  {{Ragaini}}, \citenamefont {{Rainer}}, \citenamefont {{Raiteri}},
  \citenamefont {{Rambaux}}, \citenamefont {{Ramos}}, \citenamefont
  {{Ramos-Lerate}}, \citenamefont {{Re Fiorentin}}, \citenamefont {{Regibo}},
  \citenamefont {{Reyl{\'e}}}, \citenamefont {{Ripepi}}, \citenamefont
  {{Riva}}, \citenamefont {{Rixon}}, \citenamefont {{Robichon}}, \citenamefont
  {{Robin}}, \citenamefont {{Roelens}}, \citenamefont {{Rohrbasser}},
  \citenamefont {{Romero-G{\'o}mez}}, \citenamefont {{Rowell}}, \citenamefont
  {{Royer}}, \citenamefont {{Rybicki}}, \citenamefont {{Sadowski}},
  \citenamefont {{Sagrist{\`a} Sell{\'e}s}}, \citenamefont {{Sahlmann}},
  \citenamefont {{Salgado}}, \citenamefont {{Salguero}}, \citenamefont
  {{Samaras}}, \citenamefont {{Sanchez Gimenez}}, \citenamefont {{Sanna}},
  \citenamefont {{Santove{\~n}a}}, \citenamefont {{Sarasso}}, \citenamefont
  {{Schultheis}}, \citenamefont {{Sciacca}}, \citenamefont {{Segol}},
  \citenamefont {{Segovia}}, \citenamefont {{S{\'e}gransan}}, \citenamefont
  {{Semeux}}, \citenamefont {{Shahaf}}, \citenamefont {{Siddiqui}},
  \citenamefont {{Siebert}}, \citenamefont {{Siltala}}, \citenamefont
  {{Slezak}}, \citenamefont {{Smart}}, \citenamefont {{Solano}}, \citenamefont
  {{Solitro}}, \citenamefont {{Souami}}, \citenamefont {{Souchay}},
  \citenamefont {{Spagna}}, \citenamefont {{Spoto}}, \citenamefont {{Steele}},
  \citenamefont {{Steidelm{\"u}ller}}, \citenamefont {{Stephenson}},
  \citenamefont {{S{\"u}veges}}, \citenamefont {{Szabados}}, \citenamefont
  {{Szegedi-Elek}}, \citenamefont {{Taris}}, \citenamefont {{Tauran}},
  \citenamefont {{Taylor}}, \citenamefont {{Teixeira}}, \citenamefont
  {{Thuillot}}, \citenamefont {{Tonello}}, \citenamefont {{Torra}},
  \citenamefont {{Torra}}, \citenamefont {{Turon}}, \citenamefont {{Unger}},
  \citenamefont {{Vaillant}}, \citenamefont {{van Dillen}}, \citenamefont
  {{Vanel}}, \citenamefont {{Vecchiato}}, \citenamefont {{Viala}},
  \citenamefont {{Vicente}}, \citenamefont {{Voutsinas}}, \citenamefont
  {{Weiler}}, \citenamefont {{Wevers}}, \citenamefont {{Wyrzykowski}},
  \citenamefont {{Yoldas}}, \citenamefont {{Yvard}}, \citenamefont {{Zhao}},
  \citenamefont {{Zorec}}, \citenamefont {{Zucker}}, \citenamefont
  {{Zurbach}},\ and\ \citenamefont {{Zwitter}}}]{2021A&A...649A...1G}%
  \BibitemOpen
  \bibfield  {author} {\bibinfo {author} {\bibnamefont {{Gaia Collaboration}}},
  \bibinfo {author} {\bibfnamefont {A.~G.~A.}\ \bibnamefont {{Brown}}},
  \bibinfo {author} {\bibfnamefont {A.}~\bibnamefont {{Vallenari}}}, \bibinfo
  {author} {\bibfnamefont {T.}~\bibnamefont {{Prusti}}}, \bibinfo {author}
  {\bibfnamefont {J.~H.~J.}\ \bibnamefont {{de Bruijne}}}, \bibinfo {author}
  {\bibfnamefont {C.}~\bibnamefont {{Babusiaux}}}, \bibinfo {author}
  {\bibfnamefont {M.}~\bibnamefont {{Biermann}}}, \bibinfo {author}
  {\bibfnamefont {O.~L.}\ \bibnamefont {{Creevey}}}, \bibinfo {author}
  {\bibfnamefont {D.~W.}\ \bibnamefont {{Evans}}}, \bibinfo {author}
  {\bibfnamefont {L.}~\bibnamefont {{Eyer}}}, \bibinfo {author} {\bibfnamefont
  {A.}~\bibnamefont {{Hutton}}}, \bibinfo {author} {\bibfnamefont
  {F.}~\bibnamefont {{Jansen}}}, \bibinfo {author} {\bibfnamefont
  {C.}~\bibnamefont {{Jordi}}}, \bibinfo {author} {\bibfnamefont {S.~A.}\
  \bibnamefont {{Klioner}}}, \bibinfo {author} {\bibfnamefont {U.}~\bibnamefont
  {{Lammers}}}, \bibinfo {author} {\bibfnamefont {L.}~\bibnamefont
  {{Lindegren}}}, \bibinfo {author} {\bibfnamefont {X.}~\bibnamefont {{Luri}}},
  \bibinfo {author} {\bibfnamefont {F.}~\bibnamefont {{Mignard}}}, \bibinfo
  {author} {\bibfnamefont {C.}~\bibnamefont {{Panem}}}, \bibinfo {author}
  {\bibfnamefont {D.}~\bibnamefont {{Pourbaix}}}, \bibinfo {author}
  {\bibfnamefont {S.}~\bibnamefont {{Randich}}}, \bibinfo {author}
  {\bibfnamefont {P.}~\bibnamefont {{Sartoretti}}}, \bibinfo {author}
  {\bibfnamefont {C.}~\bibnamefont {{Soubiran}}}, \bibinfo {author}
  {\bibfnamefont {N.~A.}\ \bibnamefont {{Walton}}}, \bibinfo {author}
  {\bibfnamefont {F.}~\bibnamefont {{Arenou}}}, \bibinfo {author}
  {\bibfnamefont {C.~A.~L.}\ \bibnamefont {{Bailer-Jones}}}, \bibinfo {author}
  {\bibfnamefont {U.}~\bibnamefont {{Bastian}}}, \bibinfo {author}
  {\bibfnamefont {M.}~\bibnamefont {{Cropper}}}, \bibinfo {author}
  {\bibfnamefont {R.}~\bibnamefont {{Drimmel}}}, \bibinfo {author}
  {\bibfnamefont {D.}~\bibnamefont {{Katz}}}, \bibinfo {author} {\bibfnamefont
  {M.~G.}\ \bibnamefont {{Lattanzi}}}, \bibinfo {author} {\bibfnamefont
  {F.}~\bibnamefont {{van Leeuwen}}}, \bibinfo {author} {\bibfnamefont
  {J.}~\bibnamefont {{Bakker}}}, \bibinfo {author} {\bibfnamefont
  {C.}~\bibnamefont {{Cacciari}}}, \bibinfo {author} {\bibfnamefont
  {J.}~\bibnamefont {{Casta{\~n}eda}}}, \bibinfo {author} {\bibfnamefont
  {F.}~\bibnamefont {{De Angeli}}}, \bibinfo {author} {\bibfnamefont
  {C.}~\bibnamefont {{Ducourant}}}, \bibinfo {author} {\bibfnamefont
  {C.}~\bibnamefont {{Fabricius}}}, \bibinfo {author} {\bibfnamefont
  {M.}~\bibnamefont {{Fouesneau}}}, \bibinfo {author} {\bibfnamefont
  {Y.}~\bibnamefont {{Fr{\'e}mat}}}, \bibinfo {author} {\bibfnamefont
  {R.}~\bibnamefont {{Guerra}}}, \bibinfo {author} {\bibfnamefont
  {A.}~\bibnamefont {{Guerrier}}}, \bibinfo {author} {\bibfnamefont
  {J.}~\bibnamefont {{Guiraud}}}, \bibinfo {author} {\bibfnamefont
  {A.}~\bibnamefont {{Jean-Antoine Piccolo}}}, \bibinfo {author} {\bibfnamefont
  {E.}~\bibnamefont {{Masana}}}, \bibinfo {author} {\bibfnamefont
  {R.}~\bibnamefont {{Messineo}}}, \bibinfo {author} {\bibfnamefont
  {N.}~\bibnamefont {{Mowlavi}}}, \bibinfo {author} {\bibfnamefont
  {C.}~\bibnamefont {{Nicolas}}}, \bibinfo {author} {\bibfnamefont
  {K.}~\bibnamefont {{Nienartowicz}}}, \bibinfo {author} {\bibfnamefont
  {F.}~\bibnamefont {{Pailler}}}, \bibinfo {author} {\bibfnamefont
  {P.}~\bibnamefont {{Panuzzo}}}, \bibinfo {author} {\bibfnamefont
  {F.}~\bibnamefont {{Riclet}}}, \bibinfo {author} {\bibfnamefont
  {W.}~\bibnamefont {{Roux}}}, \bibinfo {author} {\bibfnamefont {G.~M.}\
  \bibnamefont {{Seabroke}}}, \bibinfo {author} {\bibfnamefont
  {R.}~\bibnamefont {{Sordo}}}, \bibinfo {author} {\bibfnamefont
  {P.}~\bibnamefont {{Tanga}}}, \bibinfo {author} {\bibfnamefont
  {F.}~\bibnamefont {{Th{\'e}venin}}}, \bibinfo {author} {\bibfnamefont
  {G.}~\bibnamefont {{Gracia-Abril}}}, \bibinfo {author} {\bibfnamefont
  {J.}~\bibnamefont {{Portell}}}, \bibinfo {author} {\bibfnamefont
  {D.}~\bibnamefont {{Teyssier}}}, \bibinfo {author} {\bibfnamefont
  {M.}~\bibnamefont {{Altmann}}}, \bibinfo {author} {\bibfnamefont
  {R.}~\bibnamefont {{Andrae}}}, \bibinfo {author} {\bibfnamefont
  {I.}~\bibnamefont {{Bellas-Velidis}}}, \bibinfo {author} {\bibfnamefont
  {K.}~\bibnamefont {{Benson}}}, \bibinfo {author} {\bibfnamefont
  {J.}~\bibnamefont {{Berthier}}}, \bibinfo {author} {\bibfnamefont
  {R.}~\bibnamefont {{Blomme}}}, \bibinfo {author} {\bibfnamefont
  {E.}~\bibnamefont {{Brugaletta}}}, \bibinfo {author} {\bibfnamefont {P.~W.}\
  \bibnamefont {{Burgess}}}, \bibinfo {author} {\bibfnamefont {G.}~\bibnamefont
  {{Busso}}}, \bibinfo {author} {\bibfnamefont {B.}~\bibnamefont {{Carry}}},
  \bibinfo {author} {\bibfnamefont {A.}~\bibnamefont {{Cellino}}}, \bibinfo
  {author} {\bibfnamefont {N.}~\bibnamefont {{Cheek}}}, \bibinfo {author}
  {\bibfnamefont {G.}~\bibnamefont {{Clementini}}}, \bibinfo {author}
  {\bibfnamefont {Y.}~\bibnamefont {{Damerdji}}}, \bibinfo {author}
  {\bibfnamefont {M.}~\bibnamefont {{Davidson}}}, \bibinfo {author}
  {\bibfnamefont {L.}~\bibnamefont {{Delchambre}}}, \bibinfo {author}
  {\bibfnamefont {A.}~\bibnamefont {{Dell'Oro}}}, \bibinfo {author}
  {\bibfnamefont {J.}~\bibnamefont {{Fern{\'a}ndez-Hern{\'a}ndez}}}, \bibinfo
  {author} {\bibfnamefont {L.}~\bibnamefont {{Galluccio}}}, \bibinfo {author}
  {\bibfnamefont {P.}~\bibnamefont {{Garc{\'\i}a-Lario}}}, \bibinfo {author}
  {\bibfnamefont {M.}~\bibnamefont {{Garcia-Reinaldos}}}, \bibinfo {author}
  {\bibfnamefont {J.}~\bibnamefont {{Gonz{\'a}lez-N{\'u}{\~n}ez}}}, \bibinfo
  {author} {\bibfnamefont {E.}~\bibnamefont {{Gosset}}}, \bibinfo {author}
  {\bibfnamefont {R.}~\bibnamefont {{Haigron}}}, \bibinfo {author}
  {\bibfnamefont {J.~L.}\ \bibnamefont {{Halbwachs}}}, \bibinfo {author}
  {\bibfnamefont {N.~C.}\ \bibnamefont {{Hambly}}}, \bibinfo {author}
  {\bibfnamefont {D.~L.}\ \bibnamefont {{Harrison}}}, \bibinfo {author}
  {\bibfnamefont {D.}~\bibnamefont {{Hatzidimitriou}}}, \bibinfo {author}
  {\bibfnamefont {U.}~\bibnamefont {{Heiter}}}, \bibinfo {author}
  {\bibfnamefont {J.}~\bibnamefont {{Hern{\'a}ndez}}}, \bibinfo {author}
  {\bibfnamefont {D.}~\bibnamefont {{Hestroffer}}}, \bibinfo {author}
  {\bibfnamefont {S.~T.}\ \bibnamefont {{Hodgkin}}}, \bibinfo {author}
  {\bibfnamefont {B.}~\bibnamefont {{Holl}}}, \bibinfo {author} {\bibfnamefont
  {K.}~\bibnamefont {{Jan{\ss}en}}}, \bibinfo {author} {\bibfnamefont
  {G.}~\bibnamefont {{Jevardat de Fombelle}}}, \bibinfo {author} {\bibfnamefont
  {S.}~\bibnamefont {{Jordan}}}, \bibinfo {author} {\bibfnamefont
  {A.}~\bibnamefont {{Krone-Martins}}}, \bibinfo {author} {\bibfnamefont
  {A.~C.}\ \bibnamefont {{Lanzafame}}}, \bibinfo {author} {\bibfnamefont
  {W.}~\bibnamefont {{L{\"o}ffler}}}, \bibinfo {author} {\bibfnamefont
  {A.}~\bibnamefont {{Lorca}}}, \bibinfo {author} {\bibfnamefont
  {M.}~\bibnamefont {{Manteiga}}}, \bibinfo {author} {\bibfnamefont
  {O.}~\bibnamefont {{Marchal}}}, \bibinfo {author} {\bibfnamefont {P.~M.}\
  \bibnamefont {{Marrese}}}, \bibinfo {author} {\bibfnamefont {A.}~\bibnamefont
  {{Moitinho}}}, \bibinfo {author} {\bibfnamefont {A.}~\bibnamefont {{Mora}}},
  \bibinfo {author} {\bibfnamefont {K.}~\bibnamefont {{Muinonen}}}, \bibinfo
  {author} {\bibfnamefont {P.}~\bibnamefont {{Osborne}}}, \bibinfo {author}
  {\bibfnamefont {E.}~\bibnamefont {{Pancino}}}, \bibinfo {author}
  {\bibfnamefont {T.}~\bibnamefont {{Pauwels}}}, \bibinfo {author}
  {\bibfnamefont {J.~M.}\ \bibnamefont {{Petit}}}, \bibinfo {author}
  {\bibfnamefont {A.}~\bibnamefont {{Recio-Blanco}}}, \bibinfo {author}
  {\bibfnamefont {P.~J.}\ \bibnamefont {{Richards}}}, \bibinfo {author}
  {\bibfnamefont {M.}~\bibnamefont {{Riello}}}, \bibinfo {author}
  {\bibfnamefont {L.}~\bibnamefont {{Rimoldini}}}, \bibinfo {author}
  {\bibfnamefont {A.~C.}\ \bibnamefont {{Robin}}}, \bibinfo {author}
  {\bibfnamefont {T.}~\bibnamefont {{Roegiers}}}, \bibinfo {author}
  {\bibfnamefont {J.}~\bibnamefont {{Rybizki}}}, \bibinfo {author}
  {\bibfnamefont {L.~M.}\ \bibnamefont {{Sarro}}}, \bibinfo {author}
  {\bibfnamefont {C.}~\bibnamefont {{Siopis}}}, \bibinfo {author}
  {\bibfnamefont {M.}~\bibnamefont {{Smith}}}, \bibinfo {author} {\bibfnamefont
  {A.}~\bibnamefont {{Sozzetti}}}, \bibinfo {author} {\bibfnamefont
  {A.}~\bibnamefont {{Ulla}}}, \bibinfo {author} {\bibfnamefont
  {E.}~\bibnamefont {{Utrilla}}}, \bibinfo {author} {\bibfnamefont
  {M.}~\bibnamefont {{van Leeuwen}}}, \bibinfo {author} {\bibfnamefont
  {W.}~\bibnamefont {{van Reeven}}}, \bibinfo {author} {\bibfnamefont
  {U.}~\bibnamefont {{Abbas}}}, \bibinfo {author} {\bibfnamefont
  {A.}~\bibnamefont {{Abreu Aramburu}}}, \bibinfo {author} {\bibfnamefont
  {S.}~\bibnamefont {{Accart}}}, \bibinfo {author} {\bibfnamefont
  {C.}~\bibnamefont {{Aerts}}}, \bibinfo {author} {\bibfnamefont {J.~J.}\
  \bibnamefont {{Aguado}}}, \bibinfo {author} {\bibfnamefont {M.}~\bibnamefont
  {{Ajaj}}}, \bibinfo {author} {\bibfnamefont {G.}~\bibnamefont {{Altavilla}}},
  \bibinfo {author} {\bibfnamefont {M.~A.}\ \bibnamefont {{{\'A}lvarez}}},
  \bibinfo {author} {\bibfnamefont {J.}~\bibnamefont {{{\'A}lvarez
  Cid-Fuentes}}}, \bibinfo {author} {\bibfnamefont {J.}~\bibnamefont
  {{Alves}}}, \bibinfo {author} {\bibfnamefont {R.~I.}\ \bibnamefont
  {{Anderson}}}, \bibinfo {author} {\bibfnamefont {E.}~\bibnamefont {{Anglada
  Varela}}}, \bibinfo {author} {\bibfnamefont {T.}~\bibnamefont {{Antoja}}},
  \bibinfo {author} {\bibfnamefont {M.}~\bibnamefont {{Audard}}}, \bibinfo
  {author} {\bibfnamefont {D.}~\bibnamefont {{Baines}}}, \bibinfo {author}
  {\bibfnamefont {S.~G.}\ \bibnamefont {{Baker}}}, \bibinfo {author}
  {\bibfnamefont {L.}~\bibnamefont {{Balaguer-N{\'u}{\~n}ez}}}, \bibinfo
  {author} {\bibfnamefont {E.}~\bibnamefont {{Balbinot}}}, \bibinfo {author}
  {\bibfnamefont {Z.}~\bibnamefont {{Balog}}}, \bibinfo {author} {\bibfnamefont
  {C.}~\bibnamefont {{Barache}}}, \bibinfo {author} {\bibfnamefont
  {D.}~\bibnamefont {{Barbato}}}, \bibinfo {author} {\bibfnamefont
  {M.}~\bibnamefont {{Barros}}}, \bibinfo {author} {\bibfnamefont {M.~A.}\
  \bibnamefont {{Barstow}}}, \bibinfo {author} {\bibfnamefont {S.}~\bibnamefont
  {{Bartolom{\'e}}}}, \bibinfo {author} {\bibfnamefont {J.~L.}\ \bibnamefont
  {{Bassilana}}}, \bibinfo {author} {\bibfnamefont {N.}~\bibnamefont
  {{Bauchet}}}, \bibinfo {author} {\bibfnamefont {A.}~\bibnamefont
  {{Baudesson-Stella}}}, \bibinfo {author} {\bibfnamefont {U.}~\bibnamefont
  {{Becciani}}}, \bibinfo {author} {\bibfnamefont {M.}~\bibnamefont
  {{Bellazzini}}}, \bibinfo {author} {\bibfnamefont {M.}~\bibnamefont
  {{Bernet}}}, \bibinfo {author} {\bibfnamefont {S.}~\bibnamefont {{Bertone}}},
  \bibinfo {author} {\bibfnamefont {L.}~\bibnamefont {{Bianchi}}}, \bibinfo
  {author} {\bibfnamefont {S.}~\bibnamefont {{Blanco-Cuaresma}}}, \bibinfo
  {author} {\bibfnamefont {T.}~\bibnamefont {{Boch}}}, \bibinfo {author}
  {\bibfnamefont {A.}~\bibnamefont {{Bombrun}}}, \bibinfo {author}
  {\bibfnamefont {D.}~\bibnamefont {{Bossini}}}, \bibinfo {author}
  {\bibfnamefont {S.}~\bibnamefont {{Bouquillon}}}, \bibinfo {author}
  {\bibfnamefont {A.}~\bibnamefont {{Bragaglia}}}, \bibinfo {author}
  {\bibfnamefont {L.}~\bibnamefont {{Bramante}}}, \bibinfo {author}
  {\bibfnamefont {E.}~\bibnamefont {{Breedt}}}, \bibinfo {author}
  {\bibfnamefont {A.}~\bibnamefont {{Bressan}}}, \bibinfo {author}
  {\bibfnamefont {N.}~\bibnamefont {{Brouillet}}}, \bibinfo {author}
  {\bibfnamefont {B.}~\bibnamefont {{Bucciarelli}}}, \bibinfo {author}
  {\bibfnamefont {A.}~\bibnamefont {{Burlacu}}}, \bibinfo {author}
  {\bibfnamefont {D.}~\bibnamefont {{Busonero}}}, \bibinfo {author}
  {\bibfnamefont {A.~G.}\ \bibnamefont {{Butkevich}}}, \bibinfo {author}
  {\bibfnamefont {R.}~\bibnamefont {{Buzzi}}}, \bibinfo {author} {\bibfnamefont
  {E.}~\bibnamefont {{Caffau}}}, \bibinfo {author} {\bibfnamefont
  {R.}~\bibnamefont {{Cancelliere}}}, \bibinfo {author} {\bibfnamefont
  {H.}~\bibnamefont {{C{\'a}novas}}}, \bibinfo {author} {\bibfnamefont
  {T.}~\bibnamefont {{Cantat-Gaudin}}}, \bibinfo {author} {\bibfnamefont
  {R.}~\bibnamefont {{Carballo}}}, \bibinfo {author} {\bibfnamefont
  {T.}~\bibnamefont {{Carlucci}}}, \bibinfo {author} {\bibfnamefont {M.~I.}\
  \bibnamefont {{Carnerero}}}, \bibinfo {author} {\bibfnamefont {J.~M.}\
  \bibnamefont {{Carrasco}}}, \bibinfo {author} {\bibfnamefont
  {L.}~\bibnamefont {{Casamiquela}}}, \bibinfo {author} {\bibfnamefont
  {M.}~\bibnamefont {{Castellani}}}, \bibinfo {author} {\bibfnamefont
  {A.}~\bibnamefont {{Castro-Ginard}}}, \bibinfo {author} {\bibfnamefont
  {P.}~\bibnamefont {{Castro Sampol}}}, \bibinfo {author} {\bibfnamefont
  {L.}~\bibnamefont {{Chaoul}}}, \bibinfo {author} {\bibfnamefont
  {P.}~\bibnamefont {{Charlot}}}, \bibinfo {author} {\bibfnamefont
  {L.}~\bibnamefont {{Chemin}}}, \bibinfo {author} {\bibfnamefont
  {A.}~\bibnamefont {{Chiavassa}}}, \bibinfo {author} {\bibfnamefont
  {M.~R.~L.}\ \bibnamefont {{Cioni}}}, \bibinfo {author} {\bibfnamefont
  {G.}~\bibnamefont {{Comoretto}}}, \bibinfo {author} {\bibfnamefont {W.~J.}\
  \bibnamefont {{Cooper}}}, \bibinfo {author} {\bibfnamefont {T.}~\bibnamefont
  {{Cornez}}}, \bibinfo {author} {\bibfnamefont {S.}~\bibnamefont {{Cowell}}},
  \bibinfo {author} {\bibfnamefont {F.}~\bibnamefont {{Crifo}}}, \bibinfo
  {author} {\bibfnamefont {M.}~\bibnamefont {{Crosta}}}, \bibinfo {author}
  {\bibfnamefont {C.}~\bibnamefont {{Crowley}}}, \bibinfo {author}
  {\bibfnamefont {C.}~\bibnamefont {{Dafonte}}}, \bibinfo {author}
  {\bibfnamefont {A.}~\bibnamefont {{Dapergolas}}}, \bibinfo {author}
  {\bibfnamefont {M.}~\bibnamefont {{David}}}, \bibinfo {author} {\bibfnamefont
  {P.}~\bibnamefont {{David}}}, \bibinfo {author} {\bibfnamefont
  {P.}~\bibnamefont {{de Laverny}}}, \bibinfo {author} {\bibfnamefont
  {F.}~\bibnamefont {{De Luise}}}, \bibinfo {author} {\bibfnamefont
  {R.}~\bibnamefont {{De March}}}, \bibinfo {author} {\bibfnamefont
  {J.}~\bibnamefont {{De Ridder}}}, \bibinfo {author} {\bibfnamefont
  {R.}~\bibnamefont {{de Souza}}}, \bibinfo {author} {\bibfnamefont
  {P.}~\bibnamefont {{de Teodoro}}}, \bibinfo {author} {\bibfnamefont
  {A.}~\bibnamefont {{de Torres}}}, \bibinfo {author} {\bibfnamefont {E.~F.}\
  \bibnamefont {{del Peloso}}}, \bibinfo {author} {\bibfnamefont
  {E.}~\bibnamefont {{del Pozo}}}, \bibinfo {author} {\bibfnamefont
  {M.}~\bibnamefont {{Delbo}}}, \bibinfo {author} {\bibfnamefont
  {A.}~\bibnamefont {{Delgado}}}, \bibinfo {author} {\bibfnamefont {H.~E.}\
  \bibnamefont {{Delgado}}}, \bibinfo {author} {\bibfnamefont {J.~B.}\
  \bibnamefont {{Delisle}}}, \bibinfo {author} {\bibfnamefont {P.}~\bibnamefont
  {{Di Matteo}}}, \bibinfo {author} {\bibfnamefont {S.}~\bibnamefont
  {{Diakite}}}, \bibinfo {author} {\bibfnamefont {C.}~\bibnamefont {{Diener}}},
  \bibinfo {author} {\bibfnamefont {E.}~\bibnamefont {{Distefano}}}, \bibinfo
  {author} {\bibfnamefont {C.}~\bibnamefont {{Dolding}}}, \bibinfo {author}
  {\bibfnamefont {D.}~\bibnamefont {{Eappachen}}}, \bibinfo {author}
  {\bibfnamefont {B.}~\bibnamefont {{Edvardsson}}}, \bibinfo {author}
  {\bibfnamefont {H.}~\bibnamefont {{Enke}}}, \bibinfo {author} {\bibfnamefont
  {P.}~\bibnamefont {{Esquej}}}, \bibinfo {author} {\bibfnamefont
  {C.}~\bibnamefont {{Fabre}}}, \bibinfo {author} {\bibfnamefont
  {M.}~\bibnamefont {{Fabrizio}}}, \bibinfo {author} {\bibfnamefont
  {S.}~\bibnamefont {{Faigler}}}, \bibinfo {author} {\bibfnamefont
  {G.}~\bibnamefont {{Fedorets}}}, \bibinfo {author} {\bibfnamefont
  {P.}~\bibnamefont {{Fernique}}}, \bibinfo {author} {\bibfnamefont
  {A.}~\bibnamefont {{Fienga}}}, \bibinfo {author} {\bibfnamefont
  {F.}~\bibnamefont {{Figueras}}}, \bibinfo {author} {\bibfnamefont
  {C.}~\bibnamefont {{Fouron}}}, \bibinfo {author} {\bibfnamefont
  {F.}~\bibnamefont {{Fragkoudi}}}, \bibinfo {author} {\bibfnamefont
  {E.}~\bibnamefont {{Fraile}}}, \bibinfo {author} {\bibfnamefont
  {F.}~\bibnamefont {{Franke}}}, \bibinfo {author} {\bibfnamefont
  {M.}~\bibnamefont {{Gai}}}, \bibinfo {author} {\bibfnamefont
  {D.}~\bibnamefont {{Garabato}}}, \bibinfo {author} {\bibfnamefont
  {A.}~\bibnamefont {{Garcia-Gutierrez}}}, \bibinfo {author} {\bibfnamefont
  {M.}~\bibnamefont {{Garc{\'\i}a-Torres}}}, \bibinfo {author} {\bibfnamefont
  {A.}~\bibnamefont {{Garofalo}}}, \bibinfo {author} {\bibfnamefont
  {P.}~\bibnamefont {{Gavras}}}, \bibinfo {author} {\bibfnamefont
  {E.}~\bibnamefont {{Gerlach}}}, \bibinfo {author} {\bibfnamefont
  {R.}~\bibnamefont {{Geyer}}}, \bibinfo {author} {\bibfnamefont
  {P.}~\bibnamefont {{Giacobbe}}}, \bibinfo {author} {\bibfnamefont
  {G.}~\bibnamefont {{Gilmore}}}, \bibinfo {author} {\bibfnamefont
  {S.}~\bibnamefont {{Girona}}}, \bibinfo {author} {\bibfnamefont
  {G.}~\bibnamefont {{Giuffrida}}}, \bibinfo {author} {\bibfnamefont
  {R.}~\bibnamefont {{Gomel}}}, \bibinfo {author} {\bibfnamefont
  {A.}~\bibnamefont {{Gomez}}}, \bibinfo {author} {\bibfnamefont
  {I.}~\bibnamefont {{Gonzalez-Santamaria}}}, \bibinfo {author} {\bibfnamefont
  {J.~J.}\ \bibnamefont {{Gonz{\'a}lez-Vidal}}}, \bibinfo {author}
  {\bibfnamefont {M.}~\bibnamefont {{Granvik}}}, \bibinfo {author}
  {\bibfnamefont {R.}~\bibnamefont {{Guti{\'e}rrez-S{\'a}nchez}}}, \bibinfo
  {author} {\bibfnamefont {L.~P.}\ \bibnamefont {{Guy}}}, \bibinfo {author}
  {\bibfnamefont {M.}~\bibnamefont {{Hauser}}}, \bibinfo {author}
  {\bibfnamefont {M.}~\bibnamefont {{Haywood}}}, \bibinfo {author}
  {\bibfnamefont {A.}~\bibnamefont {{Helmi}}}, \bibinfo {author} {\bibfnamefont
  {S.~L.}\ \bibnamefont {{Hidalgo}}}, \bibinfo {author} {\bibfnamefont
  {T.}~\bibnamefont {{Hilger}}}, \bibinfo {author} {\bibfnamefont
  {N.}~\bibnamefont {{H{\l}adczuk}}}, \bibinfo {author} {\bibfnamefont
  {D.}~\bibnamefont {{Hobbs}}}, \bibinfo {author} {\bibfnamefont
  {G.}~\bibnamefont {{Holland}}}, \bibinfo {author} {\bibfnamefont {H.~E.}\
  \bibnamefont {{Huckle}}}, \bibinfo {author} {\bibfnamefont {G.}~\bibnamefont
  {{Jasniewicz}}}, \bibinfo {author} {\bibfnamefont {P.~G.}\ \bibnamefont
  {{Jonker}}}, \bibinfo {author} {\bibfnamefont {J.}~\bibnamefont {{Juaristi
  Campillo}}}, \bibinfo {author} {\bibfnamefont {F.}~\bibnamefont {{Julbe}}},
  \bibinfo {author} {\bibfnamefont {L.}~\bibnamefont {{Karbevska}}}, \bibinfo
  {author} {\bibfnamefont {P.}~\bibnamefont {{Kervella}}}, \bibinfo {author}
  {\bibfnamefont {S.}~\bibnamefont {{Khanna}}}, \bibinfo {author}
  {\bibfnamefont {A.}~\bibnamefont {{Kochoska}}}, \bibinfo {author}
  {\bibfnamefont {M.}~\bibnamefont {{Kontizas}}}, \bibinfo {author}
  {\bibfnamefont {G.}~\bibnamefont {{Kordopatis}}}, \bibinfo {author}
  {\bibfnamefont {A.~J.}\ \bibnamefont {{Korn}}}, \bibinfo {author}
  {\bibfnamefont {Z.}~\bibnamefont {{Kostrzewa-Rutkowska}}}, \bibinfo {author}
  {\bibfnamefont {K.}~\bibnamefont {{Kruszy{\'n}ska}}}, \bibinfo {author}
  {\bibfnamefont {S.}~\bibnamefont {{Lambert}}}, \bibinfo {author}
  {\bibfnamefont {A.~F.}\ \bibnamefont {{Lanza}}}, \bibinfo {author}
  {\bibfnamefont {Y.}~\bibnamefont {{Lasne}}}, \bibinfo {author} {\bibfnamefont
  {J.~F.}\ \bibnamefont {{Le Campion}}}, \bibinfo {author} {\bibfnamefont
  {Y.}~\bibnamefont {{Le Fustec}}}, \bibinfo {author} {\bibfnamefont
  {Y.}~\bibnamefont {{Lebreton}}}, \bibinfo {author} {\bibfnamefont
  {T.}~\bibnamefont {{Lebzelter}}}, \bibinfo {author} {\bibfnamefont
  {S.}~\bibnamefont {{Leccia}}}, \bibinfo {author} {\bibfnamefont
  {N.}~\bibnamefont {{Leclerc}}}, \bibinfo {author} {\bibfnamefont
  {I.}~\bibnamefont {{Lecoeur-Taibi}}}, \bibinfo {author} {\bibfnamefont
  {S.}~\bibnamefont {{Liao}}}, \bibinfo {author} {\bibfnamefont
  {E.}~\bibnamefont {{Licata}}}, \bibinfo {author} {\bibfnamefont {E.~P.}\
  \bibnamefont {{Lindstr{\o}m}}}, \bibinfo {author} {\bibfnamefont {T.~A.}\
  \bibnamefont {{Lister}}}, \bibinfo {author} {\bibfnamefont {E.}~\bibnamefont
  {{Livanou}}}, \bibinfo {author} {\bibfnamefont {A.}~\bibnamefont {{Lobel}}},
  \bibinfo {author} {\bibfnamefont {P.}~\bibnamefont {{Madrero Pardo}}},
  \bibinfo {author} {\bibfnamefont {S.}~\bibnamefont {{Managau}}}, \bibinfo
  {author} {\bibfnamefont {R.~G.}\ \bibnamefont {{Mann}}}, \bibinfo {author}
  {\bibfnamefont {J.~M.}\ \bibnamefont {{Marchant}}}, \bibinfo {author}
  {\bibfnamefont {M.}~\bibnamefont {{Marconi}}}, \bibinfo {author}
  {\bibfnamefont {M.~M.~S.}\ \bibnamefont {{Marcos Santos}}}, \bibinfo {author}
  {\bibfnamefont {S.}~\bibnamefont {{Marinoni}}}, \bibinfo {author}
  {\bibfnamefont {F.}~\bibnamefont {{Marocco}}}, \bibinfo {author}
  {\bibfnamefont {D.~J.}\ \bibnamefont {{Marshall}}}, \bibinfo {author}
  {\bibfnamefont {L.}~\bibnamefont {{Martin Polo}}}, \bibinfo {author}
  {\bibfnamefont {J.~M.}\ \bibnamefont {{Mart{\'\i}n-Fleitas}}}, \bibinfo
  {author} {\bibfnamefont {A.}~\bibnamefont {{Masip}}}, \bibinfo {author}
  {\bibfnamefont {D.}~\bibnamefont {{Massari}}}, \bibinfo {author}
  {\bibfnamefont {A.}~\bibnamefont {{Mastrobuono-Battisti}}}, \bibinfo {author}
  {\bibfnamefont {T.}~\bibnamefont {{Mazeh}}}, \bibinfo {author} {\bibfnamefont
  {P.~J.}\ \bibnamefont {{McMillan}}}, \bibinfo {author} {\bibfnamefont
  {S.}~\bibnamefont {{Messina}}}, \bibinfo {author} {\bibfnamefont
  {D.}~\bibnamefont {{Michalik}}}, \bibinfo {author} {\bibfnamefont {N.~R.}\
  \bibnamefont {{Millar}}}, \bibinfo {author} {\bibfnamefont {A.}~\bibnamefont
  {{Mints}}}, \bibinfo {author} {\bibfnamefont {D.}~\bibnamefont {{Molina}}},
  \bibinfo {author} {\bibfnamefont {R.}~\bibnamefont {{Molinaro}}}, \bibinfo
  {author} {\bibfnamefont {L.}~\bibnamefont {{Moln{\'a}r}}}, \bibinfo {author}
  {\bibfnamefont {P.}~\bibnamefont {{Montegriffo}}}, \bibinfo {author}
  {\bibfnamefont {R.}~\bibnamefont {{Mor}}}, \bibinfo {author} {\bibfnamefont
  {R.}~\bibnamefont {{Morbidelli}}}, \bibinfo {author} {\bibfnamefont
  {T.}~\bibnamefont {{Morel}}}, \bibinfo {author} {\bibfnamefont
  {D.}~\bibnamefont {{Morris}}}, \bibinfo {author} {\bibfnamefont {A.~F.}\
  \bibnamefont {{Mulone}}}, \bibinfo {author} {\bibfnamefont {D.}~\bibnamefont
  {{Munoz}}}, \bibinfo {author} {\bibfnamefont {T.}~\bibnamefont {{Muraveva}}},
  \bibinfo {author} {\bibfnamefont {C.~P.}\ \bibnamefont {{Murphy}}}, \bibinfo
  {author} {\bibfnamefont {I.}~\bibnamefont {{Musella}}}, \bibinfo {author}
  {\bibfnamefont {L.}~\bibnamefont {{Noval}}}, \bibinfo {author} {\bibfnamefont
  {C.}~\bibnamefont {{Ord{\'e}novic}}}, \bibinfo {author} {\bibfnamefont
  {G.}~\bibnamefont {{Orr{\`u}}}}, \bibinfo {author} {\bibfnamefont
  {J.}~\bibnamefont {{Osinde}}}, \bibinfo {author} {\bibfnamefont
  {C.}~\bibnamefont {{Pagani}}}, \bibinfo {author} {\bibfnamefont
  {I.}~\bibnamefont {{Pagano}}}, \bibinfo {author} {\bibfnamefont
  {L.}~\bibnamefont {{Palaversa}}}, \bibinfo {author} {\bibfnamefont {P.~A.}\
  \bibnamefont {{Palicio}}}, \bibinfo {author} {\bibfnamefont {A.}~\bibnamefont
  {{Panahi}}}, \bibinfo {author} {\bibfnamefont {M.}~\bibnamefont {{Pawlak}}},
  \bibinfo {author} {\bibfnamefont {X.}~\bibnamefont {{Pe{\~n}alosa
  Esteller}}}, \bibinfo {author} {\bibfnamefont {A.}~\bibnamefont
  {{Penttil{\"a}}}}, \bibinfo {author} {\bibfnamefont {A.~M.}\ \bibnamefont
  {{Piersimoni}}}, \bibinfo {author} {\bibfnamefont {F.~X.}\ \bibnamefont
  {{Pineau}}}, \bibinfo {author} {\bibfnamefont {E.}~\bibnamefont {{Plachy}}},
  \bibinfo {author} {\bibfnamefont {G.}~\bibnamefont {{Plum}}}, \bibinfo
  {author} {\bibfnamefont {E.}~\bibnamefont {{Poggio}}}, \bibinfo {author}
  {\bibfnamefont {E.}~\bibnamefont {{Poretti}}}, \bibinfo {author}
  {\bibfnamefont {E.}~\bibnamefont {{Poujoulet}}}, \bibinfo {author}
  {\bibfnamefont {A.}~\bibnamefont {{Pr{\v{s}}a}}}, \bibinfo {author}
  {\bibfnamefont {L.}~\bibnamefont {{Pulone}}}, \bibinfo {author}
  {\bibfnamefont {E.}~\bibnamefont {{Racero}}}, \bibinfo {author}
  {\bibfnamefont {S.}~\bibnamefont {{Ragaini}}}, \bibinfo {author}
  {\bibfnamefont {M.}~\bibnamefont {{Rainer}}}, \bibinfo {author}
  {\bibfnamefont {C.~M.}\ \bibnamefont {{Raiteri}}}, \bibinfo {author}
  {\bibfnamefont {N.}~\bibnamefont {{Rambaux}}}, \bibinfo {author}
  {\bibfnamefont {P.}~\bibnamefont {{Ramos}}}, \bibinfo {author} {\bibfnamefont
  {M.}~\bibnamefont {{Ramos-Lerate}}}, \bibinfo {author} {\bibfnamefont
  {P.}~\bibnamefont {{Re Fiorentin}}}, \bibinfo {author} {\bibfnamefont
  {S.}~\bibnamefont {{Regibo}}}, \bibinfo {author} {\bibfnamefont
  {C.}~\bibnamefont {{Reyl{\'e}}}}, \bibinfo {author} {\bibfnamefont
  {V.}~\bibnamefont {{Ripepi}}}, \bibinfo {author} {\bibfnamefont
  {A.}~\bibnamefont {{Riva}}}, \bibinfo {author} {\bibfnamefont
  {G.}~\bibnamefont {{Rixon}}}, \bibinfo {author} {\bibfnamefont
  {N.}~\bibnamefont {{Robichon}}}, \bibinfo {author} {\bibfnamefont
  {C.}~\bibnamefont {{Robin}}}, \bibinfo {author} {\bibfnamefont
  {M.}~\bibnamefont {{Roelens}}}, \bibinfo {author} {\bibfnamefont
  {L.}~\bibnamefont {{Rohrbasser}}}, \bibinfo {author} {\bibfnamefont
  {M.}~\bibnamefont {{Romero-G{\'o}mez}}}, \bibinfo {author} {\bibfnamefont
  {N.}~\bibnamefont {{Rowell}}}, \bibinfo {author} {\bibfnamefont
  {F.}~\bibnamefont {{Royer}}}, \bibinfo {author} {\bibfnamefont {K.~A.}\
  \bibnamefont {{Rybicki}}}, \bibinfo {author} {\bibfnamefont {G.}~\bibnamefont
  {{Sadowski}}}, \bibinfo {author} {\bibfnamefont {A.}~\bibnamefont
  {{Sagrist{\`a} Sell{\'e}s}}}, \bibinfo {author} {\bibfnamefont
  {J.}~\bibnamefont {{Sahlmann}}}, \bibinfo {author} {\bibfnamefont
  {J.}~\bibnamefont {{Salgado}}}, \bibinfo {author} {\bibfnamefont
  {E.}~\bibnamefont {{Salguero}}}, \bibinfo {author} {\bibfnamefont
  {N.}~\bibnamefont {{Samaras}}}, \bibinfo {author} {\bibfnamefont
  {V.}~\bibnamefont {{Sanchez Gimenez}}}, \bibinfo {author} {\bibfnamefont
  {N.}~\bibnamefont {{Sanna}}}, \bibinfo {author} {\bibfnamefont
  {R.}~\bibnamefont {{Santove{\~n}a}}}, \bibinfo {author} {\bibfnamefont
  {M.}~\bibnamefont {{Sarasso}}}, \bibinfo {author} {\bibfnamefont
  {M.}~\bibnamefont {{Schultheis}}}, \bibinfo {author} {\bibfnamefont
  {E.}~\bibnamefont {{Sciacca}}}, \bibinfo {author} {\bibfnamefont
  {M.}~\bibnamefont {{Segol}}}, \bibinfo {author} {\bibfnamefont {J.~C.}\
  \bibnamefont {{Segovia}}}, \bibinfo {author} {\bibfnamefont {D.}~\bibnamefont
  {{S{\'e}gransan}}}, \bibinfo {author} {\bibfnamefont {D.}~\bibnamefont
  {{Semeux}}}, \bibinfo {author} {\bibfnamefont {S.}~\bibnamefont {{Shahaf}}},
  \bibinfo {author} {\bibfnamefont {H.~I.}\ \bibnamefont {{Siddiqui}}},
  \bibinfo {author} {\bibfnamefont {A.}~\bibnamefont {{Siebert}}}, \bibinfo
  {author} {\bibfnamefont {L.}~\bibnamefont {{Siltala}}}, \bibinfo {author}
  {\bibfnamefont {E.}~\bibnamefont {{Slezak}}}, \bibinfo {author}
  {\bibfnamefont {R.~L.}\ \bibnamefont {{Smart}}}, \bibinfo {author}
  {\bibfnamefont {E.}~\bibnamefont {{Solano}}}, \bibinfo {author}
  {\bibfnamefont {F.}~\bibnamefont {{Solitro}}}, \bibinfo {author}
  {\bibfnamefont {D.}~\bibnamefont {{Souami}}}, \bibinfo {author}
  {\bibfnamefont {J.}~\bibnamefont {{Souchay}}}, \bibinfo {author}
  {\bibfnamefont {A.}~\bibnamefont {{Spagna}}}, \bibinfo {author}
  {\bibfnamefont {F.}~\bibnamefont {{Spoto}}}, \bibinfo {author} {\bibfnamefont
  {I.~A.}\ \bibnamefont {{Steele}}}, \bibinfo {author} {\bibfnamefont
  {H.}~\bibnamefont {{Steidelm{\"u}ller}}}, \bibinfo {author} {\bibfnamefont
  {C.~A.}\ \bibnamefont {{Stephenson}}}, \bibinfo {author} {\bibfnamefont
  {M.}~\bibnamefont {{S{\"u}veges}}}, \bibinfo {author} {\bibfnamefont
  {L.}~\bibnamefont {{Szabados}}}, \bibinfo {author} {\bibfnamefont
  {E.}~\bibnamefont {{Szegedi-Elek}}}, \bibinfo {author} {\bibfnamefont
  {F.}~\bibnamefont {{Taris}}}, \bibinfo {author} {\bibfnamefont
  {G.}~\bibnamefont {{Tauran}}}, \bibinfo {author} {\bibfnamefont {M.~B.}\
  \bibnamefont {{Taylor}}}, \bibinfo {author} {\bibfnamefont {R.}~\bibnamefont
  {{Teixeira}}}, \bibinfo {author} {\bibfnamefont {W.}~\bibnamefont
  {{Thuillot}}}, \bibinfo {author} {\bibfnamefont {N.}~\bibnamefont
  {{Tonello}}}, \bibinfo {author} {\bibfnamefont {F.}~\bibnamefont {{Torra}}},
  \bibinfo {author} {\bibfnamefont {J.}~\bibnamefont {{Torra}}}, \bibinfo
  {author} {\bibfnamefont {C.}~\bibnamefont {{Turon}}}, \bibinfo {author}
  {\bibfnamefont {N.}~\bibnamefont {{Unger}}}, \bibinfo {author} {\bibfnamefont
  {M.}~\bibnamefont {{Vaillant}}}, \bibinfo {author} {\bibfnamefont
  {E.}~\bibnamefont {{van Dillen}}}, \bibinfo {author} {\bibfnamefont
  {O.}~\bibnamefont {{Vanel}}}, \bibinfo {author} {\bibfnamefont
  {A.}~\bibnamefont {{Vecchiato}}}, \bibinfo {author} {\bibfnamefont
  {Y.}~\bibnamefont {{Viala}}}, \bibinfo {author} {\bibfnamefont
  {D.}~\bibnamefont {{Vicente}}}, \bibinfo {author} {\bibfnamefont
  {S.}~\bibnamefont {{Voutsinas}}}, \bibinfo {author} {\bibfnamefont
  {M.}~\bibnamefont {{Weiler}}}, \bibinfo {author} {\bibfnamefont
  {T.}~\bibnamefont {{Wevers}}}, \bibinfo {author} {\bibfnamefont
  {{\L}.}~\bibnamefont {{Wyrzykowski}}}, \bibinfo {author} {\bibfnamefont
  {A.}~\bibnamefont {{Yoldas}}}, \bibinfo {author} {\bibfnamefont
  {P.}~\bibnamefont {{Yvard}}}, \bibinfo {author} {\bibfnamefont
  {H.}~\bibnamefont {{Zhao}}}, \bibinfo {author} {\bibfnamefont
  {J.}~\bibnamefont {{Zorec}}}, \bibinfo {author} {\bibfnamefont
  {S.}~\bibnamefont {{Zucker}}}, \bibinfo {author} {\bibfnamefont
  {C.}~\bibnamefont {{Zurbach}}},\ and\ \bibinfo {author} {\bibfnamefont
  {T.}~\bibnamefont {{Zwitter}}},\ }\bibfield  {title} {\bibinfo {title} {{Gaia
  Early Data Release 3. Summary of the contents and survey properties}},\
  }\href {https://doi.org/10.1051/0004-6361/202039657} {\bibfield  {journal}
  {\bibinfo  {journal} {\aap}\ }\textbf {\bibinfo {volume} {649}},\ \bibinfo
  {eid} {A1} (\bibinfo {year} {2021})},\ \Eprint
  {https://arxiv.org/abs/2012.01533} {arXiv:2012.01533 [astro-ph.GA]}
  \BibitemShut {NoStop}%
\bibitem [{\citenamefont {{Halbwachs}}\ \emph {et~al.}(2022)\citenamefont
  {{Halbwachs}}, \citenamefont {{Pourbaix}}, \citenamefont {{Arenou}},
  \citenamefont {{Galluccio}}, \citenamefont {{Guillout}}, \citenamefont
  {{Bauchet}}, \citenamefont {{Marchal}}, \citenamefont {{Sadowski}},\ and\
  \citenamefont {{Teyssier}}}]{2022arXiv220605726H}%
  \BibitemOpen
  \bibfield  {author} {\bibinfo {author} {\bibfnamefont {J.-L.}\ \bibnamefont
  {{Halbwachs}}}, \bibinfo {author} {\bibfnamefont {D.}~\bibnamefont
  {{Pourbaix}}}, \bibinfo {author} {\bibfnamefont {F.}~\bibnamefont
  {{Arenou}}}, \bibinfo {author} {\bibfnamefont {L.}~\bibnamefont
  {{Galluccio}}}, \bibinfo {author} {\bibfnamefont {P.}~\bibnamefont
  {{Guillout}}}, \bibinfo {author} {\bibfnamefont {N.}~\bibnamefont
  {{Bauchet}}}, \bibinfo {author} {\bibfnamefont {O.}~\bibnamefont
  {{Marchal}}}, \bibinfo {author} {\bibfnamefont {G.}~\bibnamefont
  {{Sadowski}}},\ and\ \bibinfo {author} {\bibfnamefont {D.}~\bibnamefont
  {{Teyssier}}},\ }\bibfield  {title} {\bibinfo {title} {{Gaia Data Release 3.
  Astrometric binary star processing}},\ }\href@noop {} {\bibfield  {journal}
  {\bibinfo  {journal} {arXiv e-prints}\ ,\ \bibinfo {eid} {arXiv:2206.05726}}
  (\bibinfo {year} {2022})},\ \Eprint {https://arxiv.org/abs/2206.05726}
  {arXiv:2206.05726 [astro-ph.SR]} \BibitemShut {NoStop}%
\bibitem [{\citenamefont {{Holl}}\ \emph {et~al.}(2022)\citenamefont {{Holl}},
  \citenamefont {{Sozzetti}}, \citenamefont {{Sahlmann}}, \citenamefont
  {{Giacobbe}}, \citenamefont {{S{\'e}gransan}}, \citenamefont {{Unger}},
  \citenamefont {{Delisle}}, \citenamefont {{Barbato}}, \citenamefont
  {{Lattanzi}}, \citenamefont {{Morbidelli}},\ and\ \citenamefont
  {{Sosnowska}}}]{2022arXiv220605439H}%
  \BibitemOpen
  \bibfield  {author} {\bibinfo {author} {\bibfnamefont {B.}~\bibnamefont
  {{Holl}}}, \bibinfo {author} {\bibfnamefont {A.}~\bibnamefont {{Sozzetti}}},
  \bibinfo {author} {\bibfnamefont {J.}~\bibnamefont {{Sahlmann}}}, \bibinfo
  {author} {\bibfnamefont {P.}~\bibnamefont {{Giacobbe}}}, \bibinfo {author}
  {\bibfnamefont {D.}~\bibnamefont {{S{\'e}gransan}}}, \bibinfo {author}
  {\bibfnamefont {N.}~\bibnamefont {{Unger}}}, \bibinfo {author} {\bibfnamefont
  {J.~B.}\ \bibnamefont {{Delisle}}}, \bibinfo {author} {\bibfnamefont
  {D.}~\bibnamefont {{Barbato}}}, \bibinfo {author} {\bibfnamefont {M.~G.}\
  \bibnamefont {{Lattanzi}}}, \bibinfo {author} {\bibfnamefont
  {R.}~\bibnamefont {{Morbidelli}}},\ and\ \bibinfo {author} {\bibfnamefont
  {D.}~\bibnamefont {{Sosnowska}}},\ }\bibfield  {title} {\bibinfo {title}
  {{Gaia DR3 astrometric orbit determination with Markov Chain Monte Carlo and
  Genetic Algorithms. Systems with stellar, substellar, and planetary mass
  companions}},\ }\href@noop {} {\bibfield  {journal} {\bibinfo  {journal}
  {arXiv e-prints}\ ,\ \bibinfo {eid} {arXiv:2206.05439}} (\bibinfo {year}
  {2022})},\ \Eprint {https://arxiv.org/abs/2206.05439} {arXiv:2206.05439
  [astro-ph.EP]} \BibitemShut {NoStop}%
\bibitem [{\citenamefont {{Andrews}}\ \emph {et~al.}(2022)\citenamefont
  {{Andrews}}, \citenamefont {{Taggart}},\ and\ \citenamefont
  {{Foley}}}]{2022arXiv220700680A}%
  \BibitemOpen
  \bibfield  {author} {\bibinfo {author} {\bibfnamefont {J.~J.}\ \bibnamefont
  {{Andrews}}}, \bibinfo {author} {\bibfnamefont {K.}~\bibnamefont
  {{Taggart}}},\ and\ \bibinfo {author} {\bibfnamefont {R.}~\bibnamefont
  {{Foley}}},\ }\bibfield  {title} {\bibinfo {title} {{A Sample of Neutron Star
  and Black Hole Binaries Detected through Gaia DR3 Astrometry}},\ }\href@noop
  {} {\bibfield  {journal} {\bibinfo  {journal} {arXiv e-prints}\ ,\ \bibinfo
  {eid} {arXiv:2207.00680}} (\bibinfo {year} {2022})},\ \Eprint
  {https://arxiv.org/abs/2207.00680} {arXiv:2207.00680 [astro-ph.SR]}
  \BibitemShut {NoStop}%
\bibitem [{\citenamefont {{Shahaf}}\ \emph {et~al.}(2023)\citenamefont
  {{Shahaf}}, \citenamefont {{Bashi}}, \citenamefont {{Mazeh}}, \citenamefont
  {{Faigler}}, \citenamefont {{Arenou}}, \citenamefont {{El-Badry}},\ and\
  \citenamefont {{Rix}}}]{2023MNRAS.518.2991S}%
  \BibitemOpen
  \bibfield  {author} {\bibinfo {author} {\bibfnamefont {S.}~\bibnamefont
  {{Shahaf}}}, \bibinfo {author} {\bibfnamefont {D.}~\bibnamefont {{Bashi}}},
  \bibinfo {author} {\bibfnamefont {T.}~\bibnamefont {{Mazeh}}}, \bibinfo
  {author} {\bibfnamefont {S.}~\bibnamefont {{Faigler}}}, \bibinfo {author}
  {\bibfnamefont {F.}~\bibnamefont {{Arenou}}}, \bibinfo {author}
  {\bibfnamefont {K.}~\bibnamefont {{El-Badry}}},\ and\ \bibinfo {author}
  {\bibfnamefont {H.~W.}\ \bibnamefont {{Rix}}},\ }\bibfield  {title} {\bibinfo
  {title} {{Triage of the Gaia DR3 astrometric orbits - I. A sample of binaries
  with probable compact companions}},\ }\href
  {https://doi.org/10.1093/mnras/stac3290} {\bibfield  {journal} {\bibinfo
  {journal} {\mnras}\ }\textbf {\bibinfo {volume} {518}},\ \bibinfo {pages}
  {2991} (\bibinfo {year} {2023})},\ \Eprint {https://arxiv.org/abs/2209.00828}
  {arXiv:2209.00828 [astro-ph.SR]} \BibitemShut {NoStop}%
\bibitem [{\citenamefont {{Shahaf}}\ \emph {et~al.}(2019)\citenamefont
  {{Shahaf}}, \citenamefont {{Mazeh}}, \citenamefont {{Faigler}},\ and\
  \citenamefont {{Holl}}}]{2019MNRAS.487.5610S}%
  \BibitemOpen
  \bibfield  {author} {\bibinfo {author} {\bibfnamefont {S.}~\bibnamefont
  {{Shahaf}}}, \bibinfo {author} {\bibfnamefont {T.}~\bibnamefont {{Mazeh}}},
  \bibinfo {author} {\bibfnamefont {S.}~\bibnamefont {{Faigler}}},\ and\
  \bibinfo {author} {\bibfnamefont {B.}~\bibnamefont {{Holl}}},\ }\bibfield
  {title} {\bibinfo {title} {{Triage of astrometric binaries - how to find
  triple systems and dormant black hole secondaries in the Gaia orbits}},\
  }\href {https://doi.org/10.1093/mnras/stz1636} {\bibfield  {journal}
  {\bibinfo  {journal} {\mnras}\ }\textbf {\bibinfo {volume} {487}},\ \bibinfo
  {pages} {5610} (\bibinfo {year} {2019})},\ \Eprint
  {https://arxiv.org/abs/1905.08542} {arXiv:1905.08542 [astro-ph.SR]}
  \BibitemShut {NoStop}%
\bibitem [{\citenamefont {{El-Badry}}\ \emph
  {et~al.}(2023{\natexlab{c}})\citenamefont {{El-Badry}}, \citenamefont
  {{Rix}}, \citenamefont {{Quataert}}, \citenamefont {{Howard}}, \citenamefont
  {{Isaacson}}, \citenamefont {{Fuller}}, \citenamefont {{Hawkins}},
  \citenamefont {{Breivik}}, \citenamefont {{Wong}}, \citenamefont
  {{Rodriguez}}, \citenamefont {{Conroy}}, \citenamefont {{Shahaf}},
  \citenamefont {{Mazeh}}, \citenamefont {{Arenou}}, \citenamefont {{Burdge}},
  \citenamefont {{Bashi}}, \citenamefont {{Faigler}}, \citenamefont {{Weisz}},
  \citenamefont {{Seeburger}}, \citenamefont {{Almada Monter}},\ and\
  \citenamefont {{Wojno}}}]{2023MNRAS.518.1057E}%
  \BibitemOpen
  \bibfield  {author} {\bibinfo {author} {\bibfnamefont {K.}~\bibnamefont
  {{El-Badry}}}, \bibinfo {author} {\bibfnamefont {H.-W.}\ \bibnamefont
  {{Rix}}}, \bibinfo {author} {\bibfnamefont {E.}~\bibnamefont {{Quataert}}},
  \bibinfo {author} {\bibfnamefont {A.~W.}\ \bibnamefont {{Howard}}}, \bibinfo
  {author} {\bibfnamefont {H.}~\bibnamefont {{Isaacson}}}, \bibinfo {author}
  {\bibfnamefont {J.}~\bibnamefont {{Fuller}}}, \bibinfo {author}
  {\bibfnamefont {K.}~\bibnamefont {{Hawkins}}}, \bibinfo {author}
  {\bibfnamefont {K.}~\bibnamefont {{Breivik}}}, \bibinfo {author}
  {\bibfnamefont {K.~W.~K.}\ \bibnamefont {{Wong}}}, \bibinfo {author}
  {\bibfnamefont {A.~C.}\ \bibnamefont {{Rodriguez}}}, \bibinfo {author}
  {\bibfnamefont {C.}~\bibnamefont {{Conroy}}}, \bibinfo {author}
  {\bibfnamefont {S.}~\bibnamefont {{Shahaf}}}, \bibinfo {author}
  {\bibfnamefont {T.}~\bibnamefont {{Mazeh}}}, \bibinfo {author} {\bibfnamefont
  {F.}~\bibnamefont {{Arenou}}}, \bibinfo {author} {\bibfnamefont {K.~B.}\
  \bibnamefont {{Burdge}}}, \bibinfo {author} {\bibfnamefont {D.}~\bibnamefont
  {{Bashi}}}, \bibinfo {author} {\bibfnamefont {S.}~\bibnamefont {{Faigler}}},
  \bibinfo {author} {\bibfnamefont {D.~R.}\ \bibnamefont {{Weisz}}}, \bibinfo
  {author} {\bibfnamefont {R.}~\bibnamefont {{Seeburger}}}, \bibinfo {author}
  {\bibfnamefont {S.}~\bibnamefont {{Almada Monter}}},\ and\ \bibinfo {author}
  {\bibfnamefont {J.}~\bibnamefont {{Wojno}}},\ }\bibfield  {title} {\bibinfo
  {title} {{A Sun-like star orbiting a black hole}},\ }\href
  {https://doi.org/10.1093/mnras/stac3140} {\bibfield  {journal} {\bibinfo
  {journal} {\mnras}\ }\textbf {\bibinfo {volume} {518}},\ \bibinfo {pages}
  {1057} (\bibinfo {year} {2023}{\natexlab{c}})},\ \Eprint
  {https://arxiv.org/abs/2209.06833} {arXiv:2209.06833 [astro-ph.SR]}
  \BibitemShut {NoStop}%
\bibitem [{\citenamefont {{Chakrabarti}}\ \emph {et~al.}(2022)\citenamefont
  {{Chakrabarti}}, \citenamefont {{Simon}}, \citenamefont {{Craig}},
  \citenamefont {{Reggiani}}, \citenamefont {{Guhathakurta}}, \citenamefont
  {{Dalba}}, \citenamefont {{Kirby}}, \citenamefont {{Chang}}, \citenamefont
  {{Hey}}, \citenamefont {{Savino}},\ and\ \citenamefont
  {{Geha}}}]{2022arXiv221005003C}%
  \BibitemOpen
  \bibfield  {author} {\bibinfo {author} {\bibfnamefont {S.}~\bibnamefont
  {{Chakrabarti}}}, \bibinfo {author} {\bibfnamefont {J.~D.}\ \bibnamefont
  {{Simon}}}, \bibinfo {author} {\bibfnamefont {P.~A.}\ \bibnamefont
  {{Craig}}}, \bibinfo {author} {\bibfnamefont {H.}~\bibnamefont {{Reggiani}}},
  \bibinfo {author} {\bibfnamefont {P.}~\bibnamefont {{Guhathakurta}}},
  \bibinfo {author} {\bibfnamefont {P.~A.}\ \bibnamefont {{Dalba}}}, \bibinfo
  {author} {\bibfnamefont {E.~N.}\ \bibnamefont {{Kirby}}}, \bibinfo {author}
  {\bibfnamefont {P.}~\bibnamefont {{Chang}}}, \bibinfo {author} {\bibfnamefont
  {D.~R.}\ \bibnamefont {{Hey}}}, \bibinfo {author} {\bibfnamefont
  {A.}~\bibnamefont {{Savino}}},\ and\ \bibinfo {author} {\bibfnamefont
  {M.}~\bibnamefont {{Geha}}},\ }\bibfield  {title} {\bibinfo {title} {{A
  non-interacting Galactic black hole candidate in a binary system with a
  main-sequence star}},\ }\href@noop {} {\bibfield  {journal} {\bibinfo
  {journal} {arXiv e-prints}\ ,\ \bibinfo {eid} {arXiv:2210.05003}} (\bibinfo
  {year} {2022})},\ \Eprint {https://arxiv.org/abs/2210.05003}
  {arXiv:2210.05003 [astro-ph.GA]} \BibitemShut {NoStop}%
\bibitem [{\citenamefont {{Tanikawa}}\ \emph {et~al.}(2023)\citenamefont
  {{Tanikawa}}, \citenamefont {{Hattori}}, \citenamefont {{Kawanaka}},
  \citenamefont {{Kinugawa}}, \citenamefont {{Shikauchi}},\ and\ \citenamefont
  {{Tsuna}}}]{2023ApJ...946...79T}%
  \BibitemOpen
  \bibfield  {author} {\bibinfo {author} {\bibfnamefont {A.}~\bibnamefont
  {{Tanikawa}}}, \bibinfo {author} {\bibfnamefont {K.}~\bibnamefont
  {{Hattori}}}, \bibinfo {author} {\bibfnamefont {N.}~\bibnamefont
  {{Kawanaka}}}, \bibinfo {author} {\bibfnamefont {T.}~\bibnamefont
  {{Kinugawa}}}, \bibinfo {author} {\bibfnamefont {M.}~\bibnamefont
  {{Shikauchi}}},\ and\ \bibinfo {author} {\bibfnamefont {D.}~\bibnamefont
  {{Tsuna}}},\ }\bibfield  {title} {\bibinfo {title} {{Search for a Black Hole
  Binary in Gaia DR3 Astrometric Binary Stars with Spectroscopic Data}},\
  }\href {https://doi.org/10.3847/1538-4357/acbf36} {\bibfield  {journal}
  {\bibinfo  {journal} {\apj}\ }\textbf {\bibinfo {volume} {946}},\ \bibinfo
  {eid} {79} (\bibinfo {year} {2023})},\ \Eprint
  {https://arxiv.org/abs/2209.05632} {arXiv:2209.05632 [astro-ph.SR]}
  \BibitemShut {NoStop}%
\bibitem [{\citenamefont {{El-Badry}}\ \emph
  {et~al.}(2023{\natexlab{d}})\citenamefont {{El-Badry}}, \citenamefont
  {{Rix}}, \citenamefont {{Cendes}}, \citenamefont {{Rodriguez}}, \citenamefont
  {{Conroy}}, \citenamefont {{Quataert}}, \citenamefont {{Hawkins}},
  \citenamefont {{Zari}}, \citenamefont {{Hobson}}, \citenamefont {{Breivik}},
  \citenamefont {{Rau}}, \citenamefont {{Berger}}, \citenamefont {{Shahaf}},
  \citenamefont {{Seeburger}}, \citenamefont {{Burdge}}, \citenamefont
  {{Latham}}, \citenamefont {{Buchhave}}, \citenamefont {{Bieryla}},
  \citenamefont {{Bashi}}, \citenamefont {{Mazeh}},\ and\ \citenamefont
  {{Faigler}}}]{2023MNRAS.521.4323E}%
  \BibitemOpen
  \bibfield  {author} {\bibinfo {author} {\bibfnamefont {K.}~\bibnamefont
  {{El-Badry}}}, \bibinfo {author} {\bibfnamefont {H.-W.}\ \bibnamefont
  {{Rix}}}, \bibinfo {author} {\bibfnamefont {Y.}~\bibnamefont {{Cendes}}},
  \bibinfo {author} {\bibfnamefont {A.~C.}\ \bibnamefont {{Rodriguez}}},
  \bibinfo {author} {\bibfnamefont {C.}~\bibnamefont {{Conroy}}}, \bibinfo
  {author} {\bibfnamefont {E.}~\bibnamefont {{Quataert}}}, \bibinfo {author}
  {\bibfnamefont {K.}~\bibnamefont {{Hawkins}}}, \bibinfo {author}
  {\bibfnamefont {E.}~\bibnamefont {{Zari}}}, \bibinfo {author} {\bibfnamefont
  {M.}~\bibnamefont {{Hobson}}}, \bibinfo {author} {\bibfnamefont
  {K.}~\bibnamefont {{Breivik}}}, \bibinfo {author} {\bibfnamefont
  {A.}~\bibnamefont {{Rau}}}, \bibinfo {author} {\bibfnamefont
  {E.}~\bibnamefont {{Berger}}}, \bibinfo {author} {\bibfnamefont
  {S.}~\bibnamefont {{Shahaf}}}, \bibinfo {author} {\bibfnamefont
  {R.}~\bibnamefont {{Seeburger}}}, \bibinfo {author} {\bibfnamefont {K.~B.}\
  \bibnamefont {{Burdge}}}, \bibinfo {author} {\bibfnamefont {D.~W.}\
  \bibnamefont {{Latham}}}, \bibinfo {author} {\bibfnamefont {L.~A.}\
  \bibnamefont {{Buchhave}}}, \bibinfo {author} {\bibfnamefont
  {A.}~\bibnamefont {{Bieryla}}}, \bibinfo {author} {\bibfnamefont
  {D.}~\bibnamefont {{Bashi}}}, \bibinfo {author} {\bibfnamefont
  {T.}~\bibnamefont {{Mazeh}}},\ and\ \bibinfo {author} {\bibfnamefont
  {S.}~\bibnamefont {{Faigler}}},\ }\bibfield  {title} {\bibinfo {title} {{A
  red giant orbiting a black hole}},\ }\href
  {https://doi.org/10.1093/mnras/stad799} {\bibfield  {journal} {\bibinfo
  {journal} {\mnras}\ }\textbf {\bibinfo {volume} {521}},\ \bibinfo {pages}
  {4323} (\bibinfo {year} {2023}{\natexlab{d}})},\ \Eprint
  {https://arxiv.org/abs/2302.07880} {arXiv:2302.07880 [astro-ph.SR]}
  \BibitemShut {NoStop}%
\bibitem [{\citenamefont {{Chawla}}\ \emph {et~al.}(2022)\citenamefont
  {{Chawla}}, \citenamefont {{Chatterjee}}, \citenamefont {{Breivik}},
  \citenamefont {{Moorthy}}, \citenamefont {{Andrews}},\ and\ \citenamefont
  {{Sanderson}}}]{2022ApJ...931..107C}%
  \BibitemOpen
  \bibfield  {author} {\bibinfo {author} {\bibfnamefont {C.}~\bibnamefont
  {{Chawla}}}, \bibinfo {author} {\bibfnamefont {S.}~\bibnamefont
  {{Chatterjee}}}, \bibinfo {author} {\bibfnamefont {K.}~\bibnamefont
  {{Breivik}}}, \bibinfo {author} {\bibfnamefont {C.~K.}\ \bibnamefont
  {{Moorthy}}}, \bibinfo {author} {\bibfnamefont {J.~J.}\ \bibnamefont
  {{Andrews}}},\ and\ \bibinfo {author} {\bibfnamefont {R.~E.}\ \bibnamefont
  {{Sanderson}}},\ }\bibfield  {title} {\bibinfo {title} {{Gaia May Detect
  Hundreds of Well-characterized Stellar Black Holes}},\ }\href
  {https://doi.org/10.3847/1538-4357/ac60a5} {\bibfield  {journal} {\bibinfo
  {journal} {\apj}\ }\textbf {\bibinfo {volume} {931}},\ \bibinfo {eid} {107}
  (\bibinfo {year} {2022})},\ \Eprint {https://arxiv.org/abs/2110.05979}
  {arXiv:2110.05979 [astro-ph.GA]} \BibitemShut {NoStop}%
\bibitem [{\citenamefont {{Castro-Ginard}}\ \emph {et~al.}(2023)\citenamefont
  {{Castro-Ginard}}, \citenamefont {{Brown}}, \citenamefont
  {{Kostrzewa-Rutkowska}}, \citenamefont {{Cantat-Gaudin}}, \citenamefont
  {{Drimmel}}, \citenamefont {{Oh}}, \citenamefont {{Belokurov}}, \citenamefont
  {{Casey}}, \citenamefont {{Fouesneau}}, \citenamefont {{Khanna}},
  \citenamefont {{Price-Whelan}},\ and\ \citenamefont
  {{Rix}}}]{2023arXiv230317738C}%
  \BibitemOpen
  \bibfield  {author} {\bibinfo {author} {\bibfnamefont {A.}~\bibnamefont
  {{Castro-Ginard}}}, \bibinfo {author} {\bibfnamefont {A.~G.~A.}\ \bibnamefont
  {{Brown}}}, \bibinfo {author} {\bibfnamefont {Z.}~\bibnamefont
  {{Kostrzewa-Rutkowska}}}, \bibinfo {author} {\bibfnamefont {T.}~\bibnamefont
  {{Cantat-Gaudin}}}, \bibinfo {author} {\bibfnamefont {R.}~\bibnamefont
  {{Drimmel}}}, \bibinfo {author} {\bibfnamefont {S.}~\bibnamefont {{Oh}}},
  \bibinfo {author} {\bibfnamefont {V.}~\bibnamefont {{Belokurov}}}, \bibinfo
  {author} {\bibfnamefont {A.~R.}\ \bibnamefont {{Casey}}}, \bibinfo {author}
  {\bibfnamefont {M.}~\bibnamefont {{Fouesneau}}}, \bibinfo {author}
  {\bibfnamefont {S.}~\bibnamefont {{Khanna}}}, \bibinfo {author}
  {\bibfnamefont {A.~M.}\ \bibnamefont {{Price-Whelan}}},\ and\ \bibinfo
  {author} {\bibfnamefont {H.~W.}\ \bibnamefont {{Rix}}},\ }\bibfield  {title}
  {\bibinfo {title} {{Estimating the selection function of Gaia DR3
  sub-samples}},\ }\href {https://doi.org/10.48550/arXiv.2303.17738} {\bibfield
   {journal} {\bibinfo  {journal} {arXiv e-prints}\ ,\ \bibinfo {eid}
  {arXiv:2303.17738}} (\bibinfo {year} {2023})},\ \Eprint
  {https://arxiv.org/abs/2303.17738} {arXiv:2303.17738 [astro-ph.GA]}
  \BibitemShut {NoStop}%
\bibitem [{\citenamefont {Heaviside}(1893)}]{Heaviside}%
  \BibitemOpen
  \bibfield  {author} {\bibinfo {author} {\bibfnamefont {O.}~\bibnamefont
  {Heaviside}},\ }\bibfield  {title} {\bibinfo {title} {{A Gravitational and
  Electromagnetic Analogy, Part I.}},\ }\href@noop {} {\bibfield  {journal}
  {\bibinfo  {journal} {The Electrician}\ }\textbf {\bibinfo {volume} {31}},\
  \bibinfo {pages} {281} (\bibinfo {year} {1893})}\BibitemShut {NoStop}%
\bibitem [{\citenamefont {{Poincar{\'e}}}(1906)}]{1906RCMP...21..129P}%
  \BibitemOpen
  \bibfield  {author} {\bibinfo {author} {\bibfnamefont {H.}~\bibnamefont
  {{Poincar{\'e}}}},\ }\bibfield  {title} {\bibinfo {title} {{Sur la dynamique
  de l'{\'e}lectron}},\ }\href {https://doi.org/10.1007/BF03013466} {\bibfield
  {journal} {\bibinfo  {journal} {Rendiconti del Circolo matematico di
  Palermo}\ }\textbf {\bibinfo {volume} {21}},\ \bibinfo {pages} {129}
  (\bibinfo {year} {1906})}\BibitemShut {NoStop}%
\bibitem [{\citenamefont
  {{Einstein}}(1916{\natexlab{b}})}]{1916SPAW.......688E}%
  \BibitemOpen
  \bibfield  {author} {\bibinfo {author} {\bibfnamefont {A.}~\bibnamefont
  {{Einstein}}},\ }\bibfield  {title} {\bibinfo {title} {{N{\"a}herungsweise
  Integration der Feldgleichungen der Gravitation}},\ }\href@noop {} {\bibfield
   {journal} {\bibinfo  {journal} {Sitzungsberichte der K\&ouml;niglich
  Preussischen Akademie der Wissenschaften}\ ,\ \bibinfo {pages} {688}}
  (\bibinfo {year} {1916}{\natexlab{b}})}\BibitemShut {NoStop}%
\bibitem [{\citenamefont {{Einstein}}(1918)}]{1918SPAW.......154E}%
  \BibitemOpen
  \bibfield  {author} {\bibinfo {author} {\bibfnamefont {A.}~\bibnamefont
  {{Einstein}}},\ }\bibfield  {title} {\bibinfo {title} {{{\"U}ber
  Gravitationswellen}},\ }\href@noop {} {\bibfield  {journal} {\bibinfo
  {journal} {Sitzungsberichte der K\&ouml;niglich Preussischen Akademie der
  Wissenschaften}\ ,\ \bibinfo {pages} {154}} (\bibinfo {year}
  {1918})}\BibitemShut {NoStop}%
\bibitem [{\citenamefont {{Le Tiec}}\ and\ \citenamefont
  {{Novak}}(2017)}]{2017ogw..book....1L}%
  \BibitemOpen
  \bibfield  {author} {\bibinfo {author} {\bibfnamefont {A.}~\bibnamefont {{Le
  Tiec}}}\ and\ \bibinfo {author} {\bibfnamefont {J.}~\bibnamefont {{Novak}}},\
  }\bibfield  {title} {\bibinfo {title} {{Theory of Gravitational Waves}},\
  }in\ \href {https://doi.org/10.1142/9789813141766_0001} {\emph {\bibinfo
  {booktitle} {An Overview of Gravitational Waves: Theory}}}\ (\bibinfo {year}
  {2017})\ pp.\ \bibinfo {pages} {1--41}\BibitemShut {NoStop}%
\bibitem [{\citenamefont {Christensen}(2018)}]{Christensen_2018}%
  \BibitemOpen
  \bibfield  {author} {\bibinfo {author} {\bibfnamefont {N.}~\bibnamefont
  {Christensen}},\ }\bibfield  {title} {\bibinfo {title} {Stochastic
  gravitational wave backgrounds},\ }\href
  {https://doi.org/10.1088/1361-6633/aae6b5} {\bibfield  {journal} {\bibinfo
  {journal} {Reports on Progress in Physics}\ }\textbf {\bibinfo {volume}
  {82}},\ \bibinfo {pages} {016903} (\bibinfo {year} {2018})}\BibitemShut
  {NoStop}%
\bibitem [{\citenamefont {Tenorio}\ \emph {et~al.}(2021)\citenamefont
  {Tenorio}, \citenamefont {Keitel},\ and\ \citenamefont
  {Sintes}}]{universe7120474}%
  \BibitemOpen
  \bibfield  {author} {\bibinfo {author} {\bibfnamefont {R.}~\bibnamefont
  {Tenorio}}, \bibinfo {author} {\bibfnamefont {D.}~\bibnamefont {Keitel}},\
  and\ \bibinfo {author} {\bibfnamefont {A.~M.}\ \bibnamefont {Sintes}},\
  }\bibfield  {title} {\bibinfo {title} {Search methods for continuous
  gravitational-wave signals from unknown sources in the advanced-detector
  era},\ }\bibfield  {journal} {\bibinfo  {journal} {Universe}\ }\textbf
  {\bibinfo {volume} {7}},\ \href {https://doi.org/10.3390/universe7120474}
  {10.3390/universe7120474} (\bibinfo {year} {2021})\BibitemShut {NoStop}%
\bibitem [{\citenamefont {{Blanchet}}(2014)}]{2014LRR....17....2B}%
  \BibitemOpen
  \bibfield  {author} {\bibinfo {author} {\bibfnamefont {L.}~\bibnamefont
  {{Blanchet}}},\ }\bibfield  {title} {\bibinfo {title} {{Gravitational
  Radiation from Post-Newtonian Sources and Inspiralling Compact Binaries}},\
  }\href {https://doi.org/10.12942/lrr-2014-2} {\bibfield  {journal} {\bibinfo
  {journal} {Living Reviews in Relativity}\ }\textbf {\bibinfo {volume} {17}},\
  \bibinfo {eid} {2} (\bibinfo {year} {2014})},\ \Eprint
  {https://arxiv.org/abs/1310.1528} {arXiv:1310.1528 [gr-qc]} \BibitemShut
  {NoStop}%
\bibitem [{\citenamefont {Beauville}\ \emph {et~al.}(2008)\citenamefont
  {Beauville}, \citenamefont {Bizouard}, \citenamefont {Blackburn},
  \citenamefont {Bosi}, \citenamefont {Brocco}, \citenamefont {Brown},
  \citenamefont {Buskulic}, \citenamefont {Cavalier}, \citenamefont
  {Chatterji}, \citenamefont {Christensen}, \citenamefont {Clapson},
  \citenamefont {Fairhurst}, \citenamefont {Grosjean}, \citenamefont {Guidi},
  \citenamefont {Hello}, \citenamefont {Heng}, \citenamefont {Hewitson},
  \citenamefont {Katsavounidis}, \citenamefont {Klimenko},\ and\ \citenamefont
  {Zanolin}}]{Burst_GWs}%
  \BibitemOpen
  \bibfield  {author} {\bibinfo {author} {\bibfnamefont {F.}~\bibnamefont
  {Beauville}}, \bibinfo {author} {\bibfnamefont {M.-A.}\ \bibnamefont
  {Bizouard}}, \bibinfo {author} {\bibfnamefont {L.}~\bibnamefont {Blackburn}},
  \bibinfo {author} {\bibfnamefont {L.}~\bibnamefont {Bosi}}, \bibinfo {author}
  {\bibfnamefont {L.}~\bibnamefont {Brocco}}, \bibinfo {author} {\bibfnamefont
  {D.}~\bibnamefont {Brown}}, \bibinfo {author} {\bibfnamefont
  {D.}~\bibnamefont {Buskulic}}, \bibinfo {author} {\bibfnamefont
  {F.}~\bibnamefont {Cavalier}}, \bibinfo {author} {\bibfnamefont
  {S.}~\bibnamefont {Chatterji}}, \bibinfo {author} {\bibfnamefont
  {N.}~\bibnamefont {Christensen}}, \bibinfo {author} {\bibfnamefont {A.-C.}\
  \bibnamefont {Clapson}}, \bibinfo {author} {\bibfnamefont {S.}~\bibnamefont
  {Fairhurst}}, \bibinfo {author} {\bibfnamefont {D.}~\bibnamefont {Grosjean}},
  \bibinfo {author} {\bibfnamefont {G.}~\bibnamefont {Guidi}}, \bibinfo
  {author} {\bibfnamefont {P.}~\bibnamefont {Hello}}, \bibinfo {author}
  {\bibfnamefont {S.}~\bibnamefont {Heng}}, \bibinfo {author} {\bibfnamefont
  {M.}~\bibnamefont {Hewitson}}, \bibinfo {author} {\bibfnamefont
  {E.}~\bibnamefont {Katsavounidis}}, \bibinfo {author} {\bibfnamefont
  {S.}~\bibnamefont {Klimenko}},\ and\ \bibinfo {author} {\bibfnamefont
  {M.}~\bibnamefont {Zanolin}},\ }\bibfield  {title} {\bibinfo {title} {A
  comparison of methods for gravitational wave burst searches from ligo and
  virgo},\ }\href {https://doi.org/10.1088/0264-9381/25/4/045002} {\bibfield
  {journal} {\bibinfo  {journal} {Classical and Quantum Gravity}\ }\textbf
  {\bibinfo {volume} {25}},\ \bibinfo {pages} {045002} (\bibinfo {year}
  {2008})}\BibitemShut {NoStop}%
\bibitem [{\citenamefont {{Thorne}}(1995)}]{1995pnac.conf..160T}%
  \BibitemOpen
  \bibfield  {author} {\bibinfo {author} {\bibfnamefont {K.~S.}\ \bibnamefont
  {{Thorne}}},\ }\bibfield  {title} {\bibinfo {title} {{Gravitational Waves}},\
  }in\ \href {https://doi.org/10.48550/arXiv.gr-qc/9506086} {\emph {\bibinfo
  {booktitle} {Particle and Nuclear Astrophysics and Cosmology in the Next
  Millenium}}},\ \bibinfo {editor} {edited by\ \bibinfo {editor} {\bibfnamefont
  {E.~W.}\ \bibnamefont {{Kolb}}}\ and\ \bibinfo {editor} {\bibfnamefont
  {R.~D.}\ \bibnamefont {{Peccei}}}}\ (\bibinfo {year} {1995})\ p.\ \bibinfo
  {pages} {160},\ \Eprint {https://arxiv.org/abs/gr-qc/9506086}
  {arXiv:gr-qc/9506086 [gr-qc]} \BibitemShut {NoStop}%
\bibitem [{\citenamefont {Brady}\ and\ \citenamefont
  {Creighton}(2003)}]{BRADY200333}%
  \BibitemOpen
  \bibfield  {author} {\bibinfo {author} {\bibfnamefont {P.~R.}\ \bibnamefont
  {Brady}}\ and\ \bibinfo {author} {\bibfnamefont {J.~D.}\ \bibnamefont
  {Creighton}},\ }\bibfield  {title} {\bibinfo {title} {Gravitational wave
  astronomy},\ }in\ \href
  {https://doi.org/https://doi.org/10.1016/B0-12-227410-5/00299-4} {\emph
  {\bibinfo {booktitle} {Encyclopedia of Physical Science and Technology (Third
  Edition)}}},\ \bibinfo {editor} {edited by\ \bibinfo {editor} {\bibfnamefont
  {R.~A.}\ \bibnamefont {Meyers}}}\ (\bibinfo  {publisher} {Academic Press},\
  \bibinfo {address} {New York},\ \bibinfo {year} {2003})\ \bibinfo {edition}
  {third edition}\ ed.,\ pp.\ \bibinfo {pages} {33--48}\BibitemShut {NoStop}%
\bibitem [{\citenamefont {{Chaudhuri}}(2016)}]{2016arXiv160500761C}%
  \BibitemOpen
  \bibfield  {author} {\bibinfo {author} {\bibfnamefont {A.~K.}\ \bibnamefont
  {{Chaudhuri}}},\ }\bibfield  {title} {\bibinfo {title} {{Gravitational Wave
  for a pedestrian}},\ }\href {https://doi.org/10.48550/arXiv.1605.00761}
  {\bibfield  {journal} {\bibinfo  {journal} {arXiv e-prints}\ ,\ \bibinfo
  {eid} {arXiv:1605.00761}} (\bibinfo {year} {2016})},\ \Eprint
  {https://arxiv.org/abs/1605.00761} {arXiv:1605.00761 [physics.pop-ph]}
  \BibitemShut {NoStop}%
\bibitem [{\citenamefont {{Maiorano}}\ \emph {et~al.}(2021)\citenamefont
  {{Maiorano}}, \citenamefont {{De Paolis}},\ and\ \citenamefont
  {{Nucita}}}]{2021Symm...13.2418M}%
  \BibitemOpen
  \bibfield  {author} {\bibinfo {author} {\bibfnamefont {M.}~\bibnamefont
  {{Maiorano}}}, \bibinfo {author} {\bibfnamefont {F.}~\bibnamefont {{De
  Paolis}}},\ and\ \bibinfo {author} {\bibfnamefont {A.~A.}\ \bibnamefont
  {{Nucita}}},\ }\bibfield  {title} {\bibinfo {title} {{Principles of
  Gravitational-Wave Detection with Pulsar Timing Arrays}},\ }\href
  {https://doi.org/10.3390/sym13122418} {\bibfield  {journal} {\bibinfo
  {journal} {Symmetry}\ }\textbf {\bibinfo {volume} {13}},\ \bibinfo {pages}
  {2418} (\bibinfo {year} {2021})},\ \Eprint {https://arxiv.org/abs/2112.08064}
  {arXiv:2112.08064 [astro-ph.GA]} \BibitemShut {NoStop}%
\bibitem [{\citenamefont {{Verbiest}}\ \emph {et~al.}(2021)\citenamefont
  {{Verbiest}}, \citenamefont {{Os{\l}owski}},\ and\ \citenamefont
  {{Burke-Spolaor}}}]{2021hgwa.bookE...4V}%
  \BibitemOpen
  \bibfield  {author} {\bibinfo {author} {\bibfnamefont {J.~P.~W.}\
  \bibnamefont {{Verbiest}}}, \bibinfo {author} {\bibfnamefont
  {S.}~\bibnamefont {{Os{\l}owski}}},\ and\ \bibinfo {author} {\bibfnamefont
  {S.}~\bibnamefont {{Burke-Spolaor}}},\ }\bibfield  {title} {\bibinfo {title}
  {{Pulsar Timing Array Experiments}},\ }in\ \href
  {https://doi.org/10.1007/978-981-15-4702-7_4-1} {\emph {\bibinfo {booktitle}
  {Handbook of Gravitational Wave Astronomy}}}\ (\bibinfo {year} {2021})\
  p.~\bibinfo {pages} {4}\BibitemShut {NoStop}%
\bibitem [{\citenamefont {{Babak}}\ \emph {et~al.}(2016)\citenamefont
  {{Babak}}, \citenamefont {{Petiteau}}, \citenamefont {{Sesana}},
  \citenamefont {{Brem}}, \citenamefont {{Rosado}}, \citenamefont {{Taylor}},
  \citenamefont {{Lassus}}, \citenamefont {{Hessels}}, \citenamefont {{Bassa}},
  \citenamefont {{Burgay}}, \citenamefont {{Caballero}}, \citenamefont
  {{Champion}}, \citenamefont {{Cognard}}, \citenamefont {{Desvignes}},
  \citenamefont {{Gair}}, \citenamefont {{Guillemot}}, \citenamefont
  {{Janssen}}, \citenamefont {{Karuppusamy}}, \citenamefont {{Kramer}},
  \citenamefont {{Lazarus}}, \citenamefont {{Lee}}, \citenamefont {{Lentati}},
  \citenamefont {{Liu}}, \citenamefont {{Mingarelli}}, \citenamefont
  {{Os{\l}owski}}, \citenamefont {{Perrodin}}, \citenamefont {{Possenti}},
  \citenamefont {{Purver}}, \citenamefont {{Sanidas}}, \citenamefont {{Smits}},
  \citenamefont {{Stappers}}, \citenamefont {{Theureau}}, \citenamefont
  {{Tiburzi}}, \citenamefont {{van Haasteren}}, \citenamefont {{Vecchio}},\
  and\ \citenamefont {{Verbiest}}}]{2016MNRAS.455.1665B}%
  \BibitemOpen
  \bibfield  {author} {\bibinfo {author} {\bibfnamefont {S.}~\bibnamefont
  {{Babak}}}, \bibinfo {author} {\bibfnamefont {A.}~\bibnamefont {{Petiteau}}},
  \bibinfo {author} {\bibfnamefont {A.}~\bibnamefont {{Sesana}}}, \bibinfo
  {author} {\bibfnamefont {P.}~\bibnamefont {{Brem}}}, \bibinfo {author}
  {\bibfnamefont {P.~A.}\ \bibnamefont {{Rosado}}}, \bibinfo {author}
  {\bibfnamefont {S.~R.}\ \bibnamefont {{Taylor}}}, \bibinfo {author}
  {\bibfnamefont {A.}~\bibnamefont {{Lassus}}}, \bibinfo {author}
  {\bibfnamefont {J.~W.~T.}\ \bibnamefont {{Hessels}}}, \bibinfo {author}
  {\bibfnamefont {C.~G.}\ \bibnamefont {{Bassa}}}, \bibinfo {author}
  {\bibfnamefont {M.}~\bibnamefont {{Burgay}}}, \bibinfo {author}
  {\bibfnamefont {R.~N.}\ \bibnamefont {{Caballero}}}, \bibinfo {author}
  {\bibfnamefont {D.~J.}\ \bibnamefont {{Champion}}}, \bibinfo {author}
  {\bibfnamefont {I.}~\bibnamefont {{Cognard}}}, \bibinfo {author}
  {\bibfnamefont {G.}~\bibnamefont {{Desvignes}}}, \bibinfo {author}
  {\bibfnamefont {J.~R.}\ \bibnamefont {{Gair}}}, \bibinfo {author}
  {\bibfnamefont {L.}~\bibnamefont {{Guillemot}}}, \bibinfo {author}
  {\bibfnamefont {G.~H.}\ \bibnamefont {{Janssen}}}, \bibinfo {author}
  {\bibfnamefont {R.}~\bibnamefont {{Karuppusamy}}}, \bibinfo {author}
  {\bibfnamefont {M.}~\bibnamefont {{Kramer}}}, \bibinfo {author}
  {\bibfnamefont {P.}~\bibnamefont {{Lazarus}}}, \bibinfo {author}
  {\bibfnamefont {K.~J.}\ \bibnamefont {{Lee}}}, \bibinfo {author}
  {\bibfnamefont {L.}~\bibnamefont {{Lentati}}}, \bibinfo {author}
  {\bibfnamefont {K.}~\bibnamefont {{Liu}}}, \bibinfo {author} {\bibfnamefont
  {C.~M.~F.}\ \bibnamefont {{Mingarelli}}}, \bibinfo {author} {\bibfnamefont
  {S.}~\bibnamefont {{Os{\l}owski}}}, \bibinfo {author} {\bibfnamefont
  {D.}~\bibnamefont {{Perrodin}}}, \bibinfo {author} {\bibfnamefont
  {A.}~\bibnamefont {{Possenti}}}, \bibinfo {author} {\bibfnamefont {M.~B.}\
  \bibnamefont {{Purver}}}, \bibinfo {author} {\bibfnamefont {S.}~\bibnamefont
  {{Sanidas}}}, \bibinfo {author} {\bibfnamefont {R.}~\bibnamefont {{Smits}}},
  \bibinfo {author} {\bibfnamefont {B.}~\bibnamefont {{Stappers}}}, \bibinfo
  {author} {\bibfnamefont {G.}~\bibnamefont {{Theureau}}}, \bibinfo {author}
  {\bibfnamefont {C.}~\bibnamefont {{Tiburzi}}}, \bibinfo {author}
  {\bibfnamefont {R.}~\bibnamefont {{van Haasteren}}}, \bibinfo {author}
  {\bibfnamefont {A.}~\bibnamefont {{Vecchio}}},\ and\ \bibinfo {author}
  {\bibfnamefont {J.~P.~W.}\ \bibnamefont {{Verbiest}}},\ }\bibfield  {title}
  {\bibinfo {title} {{European Pulsar Timing Array limits on continuous
  gravitational waves from individual supermassive black hole binaries}},\
  }\href {https://doi.org/10.1093/mnras/stv2092} {\bibfield  {journal}
  {\bibinfo  {journal} {\mnras}\ }\textbf {\bibinfo {volume} {455}},\ \bibinfo
  {pages} {1665} (\bibinfo {year} {2016})},\ \Eprint
  {https://arxiv.org/abs/1509.02165} {arXiv:1509.02165 [astro-ph.CO]}
  \BibitemShut {NoStop}%
\bibitem [{\citenamefont {{Joshi}}\ \emph {et~al.}(2018)\citenamefont
  {{Joshi}}, \citenamefont {{Arumugasamy}}, \citenamefont {{Bagchi}},
  \citenamefont {{Bandyopadhyay}}, \citenamefont {{Basu}}, \citenamefont
  {{Dhanda Batra}}, \citenamefont {{Bethapudi}}, \citenamefont {{Choudhary}},
  \citenamefont {{De}}, \citenamefont {{Dey}}, \citenamefont {{Gopakumar}},
  \citenamefont {{Gupta}}, \citenamefont {{Krishnakumar}}, \citenamefont
  {{Maan}}, \citenamefont {{Manoharan}}, \citenamefont {{Naidu}}, \citenamefont
  {{Nandi}}, \citenamefont {{Pathak}}, \citenamefont {{Surnis}},\ and\
  \citenamefont {{Susobhanan}}}]{2018JApA...39...51J}%
  \BibitemOpen
  \bibfield  {author} {\bibinfo {author} {\bibfnamefont {B.~C.}\ \bibnamefont
  {{Joshi}}}, \bibinfo {author} {\bibfnamefont {P.}~\bibnamefont
  {{Arumugasamy}}}, \bibinfo {author} {\bibfnamefont {M.}~\bibnamefont
  {{Bagchi}}}, \bibinfo {author} {\bibfnamefont {D.}~\bibnamefont
  {{Bandyopadhyay}}}, \bibinfo {author} {\bibfnamefont {A.}~\bibnamefont
  {{Basu}}}, \bibinfo {author} {\bibfnamefont {N.}~\bibnamefont {{Dhanda
  Batra}}}, \bibinfo {author} {\bibfnamefont {S.}~\bibnamefont {{Bethapudi}}},
  \bibinfo {author} {\bibfnamefont {A.}~\bibnamefont {{Choudhary}}}, \bibinfo
  {author} {\bibfnamefont {K.}~\bibnamefont {{De}}}, \bibinfo {author}
  {\bibfnamefont {L.}~\bibnamefont {{Dey}}}, \bibinfo {author} {\bibfnamefont
  {A.}~\bibnamefont {{Gopakumar}}}, \bibinfo {author} {\bibfnamefont
  {Y.}~\bibnamefont {{Gupta}}}, \bibinfo {author} {\bibfnamefont {M.~A.}\
  \bibnamefont {{Krishnakumar}}}, \bibinfo {author} {\bibfnamefont
  {Y.}~\bibnamefont {{Maan}}}, \bibinfo {author} {\bibfnamefont {P.~K.}\
  \bibnamefont {{Manoharan}}}, \bibinfo {author} {\bibfnamefont
  {A.}~\bibnamefont {{Naidu}}}, \bibinfo {author} {\bibfnamefont
  {R.}~\bibnamefont {{Nandi}}}, \bibinfo {author} {\bibfnamefont
  {D.}~\bibnamefont {{Pathak}}}, \bibinfo {author} {\bibfnamefont
  {M.}~\bibnamefont {{Surnis}}},\ and\ \bibinfo {author} {\bibfnamefont
  {A.}~\bibnamefont {{Susobhanan}}},\ }\bibfield  {title} {\bibinfo {title}
  {{Precision pulsar timing with the ORT and the GMRT and its applications in
  pulsar astrophysics}},\ }\href {https://doi.org/10.1007/s12036-018-9549-y}
  {\bibfield  {journal} {\bibinfo  {journal} {Journal of Astrophysics and
  Astronomy}\ }\textbf {\bibinfo {volume} {39}},\ \bibinfo {eid} {51} (\bibinfo
  {year} {2018})}\BibitemShut {NoStop}%
\bibitem [{\citenamefont {{Alam}}\ \emph {et~al.}(2021)\citenamefont {{Alam}},
  \citenamefont {{Arzoumanian}}, \citenamefont {{Baker}}, \citenamefont
  {{Blumer}}, \citenamefont {{Bohler}}, \citenamefont {{Brazier}},
  \citenamefont {{Brook}}, \citenamefont {{Burke-Spolaor}}, \citenamefont
  {{Caballero}}, \citenamefont {{Camuccio}}, \citenamefont {{Chamberlain}},
  \citenamefont {{Chatterjee}}, \citenamefont {{Cordes}}, \citenamefont
  {{Cornish}}, \citenamefont {{Crawford}}, \citenamefont {{Cromartie}},
  \citenamefont {{Decesar}}, \citenamefont {{Demorest}}, \citenamefont
  {{Dolch}}, \citenamefont {{Ellis}}, \citenamefont {{Ferdman}}, \citenamefont
  {{Ferrara}}, \citenamefont {{Fiore}}, \citenamefont {{Fonseca}},
  \citenamefont {{Garcia}}, \citenamefont {{Garver-Daniels}}, \citenamefont
  {{Gentile}}, \citenamefont {{Good}}, \citenamefont {{Gusdorff}},
  \citenamefont {{Halmrast}}, \citenamefont {{Hazboun}}, \citenamefont
  {{Islo}}, \citenamefont {{Jennings}}, \citenamefont {{Jessup}}, \citenamefont
  {{Jones}}, \citenamefont {{Kaiser}}, \citenamefont {{Kaplan}}, \citenamefont
  {{Kelley}}, \citenamefont {{Key}}, \citenamefont {{Lam}}, \citenamefont
  {{Lazio}}, \citenamefont {{Lorimer}}, \citenamefont {{Luo}}, \citenamefont
  {{Lynch}}, \citenamefont {{Madison}}, \citenamefont {{Maraccini}},
  \citenamefont {{McLaughlin}}, \citenamefont {{Mingarelli}}, \citenamefont
  {{Ng}}, \citenamefont {{Nguyen}}, \citenamefont {{Nice}}, \citenamefont
  {{Pennucci}}, \citenamefont {{Pol}}, \citenamefont {{Ramette}}, \citenamefont
  {{Ransom}}, \citenamefont {{Ray}}, \citenamefont {{Shapiro-Albert}},
  \citenamefont {{Siemens}}, \citenamefont {{Simon}}, \citenamefont
  {{Spiewak}}, \citenamefont {{Stairs}}, \citenamefont {{Stinebring}},
  \citenamefont {{Stovall}}, \citenamefont {{Swiggum}}, \citenamefont
  {{Taylor}}, \citenamefont {{Tripepi}}, \citenamefont {{Vallisneri}},
  \citenamefont {{Vigeland}}, \citenamefont {{Witt}}, \citenamefont {{Zhu}},\
  and\ \citenamefont {{Nanograv Collaboration}}}]{2021ApJS..252....5A}%
  \BibitemOpen
  \bibfield  {author} {\bibinfo {author} {\bibfnamefont {M.~F.}\ \bibnamefont
  {{Alam}}}, \bibinfo {author} {\bibfnamefont {Z.}~\bibnamefont
  {{Arzoumanian}}}, \bibinfo {author} {\bibfnamefont {P.~T.}\ \bibnamefont
  {{Baker}}}, \bibinfo {author} {\bibfnamefont {H.}~\bibnamefont {{Blumer}}},
  \bibinfo {author} {\bibfnamefont {K.~E.}\ \bibnamefont {{Bohler}}}, \bibinfo
  {author} {\bibfnamefont {A.}~\bibnamefont {{Brazier}}}, \bibinfo {author}
  {\bibfnamefont {P.~R.}\ \bibnamefont {{Brook}}}, \bibinfo {author}
  {\bibfnamefont {S.}~\bibnamefont {{Burke-Spolaor}}}, \bibinfo {author}
  {\bibfnamefont {K.}~\bibnamefont {{Caballero}}}, \bibinfo {author}
  {\bibfnamefont {R.~S.}\ \bibnamefont {{Camuccio}}}, \bibinfo {author}
  {\bibfnamefont {R.~L.}\ \bibnamefont {{Chamberlain}}}, \bibinfo {author}
  {\bibfnamefont {S.}~\bibnamefont {{Chatterjee}}}, \bibinfo {author}
  {\bibfnamefont {J.~M.}\ \bibnamefont {{Cordes}}}, \bibinfo {author}
  {\bibfnamefont {N.~J.}\ \bibnamefont {{Cornish}}}, \bibinfo {author}
  {\bibfnamefont {F.}~\bibnamefont {{Crawford}}}, \bibinfo {author}
  {\bibfnamefont {H.~T.}\ \bibnamefont {{Cromartie}}}, \bibinfo {author}
  {\bibfnamefont {M.~E.}\ \bibnamefont {{Decesar}}}, \bibinfo {author}
  {\bibfnamefont {P.~B.}\ \bibnamefont {{Demorest}}}, \bibinfo {author}
  {\bibfnamefont {T.}~\bibnamefont {{Dolch}}}, \bibinfo {author} {\bibfnamefont
  {J.~A.}\ \bibnamefont {{Ellis}}}, \bibinfo {author} {\bibfnamefont {R.~D.}\
  \bibnamefont {{Ferdman}}}, \bibinfo {author} {\bibfnamefont {E.~C.}\
  \bibnamefont {{Ferrara}}}, \bibinfo {author} {\bibfnamefont {W.}~\bibnamefont
  {{Fiore}}}, \bibinfo {author} {\bibfnamefont {E.}~\bibnamefont {{Fonseca}}},
  \bibinfo {author} {\bibfnamefont {Y.}~\bibnamefont {{Garcia}}}, \bibinfo
  {author} {\bibfnamefont {N.}~\bibnamefont {{Garver-Daniels}}}, \bibinfo
  {author} {\bibfnamefont {P.~A.}\ \bibnamefont {{Gentile}}}, \bibinfo {author}
  {\bibfnamefont {D.~C.}\ \bibnamefont {{Good}}}, \bibinfo {author}
  {\bibfnamefont {J.~A.}\ \bibnamefont {{Gusdorff}}}, \bibinfo {author}
  {\bibfnamefont {D.}~\bibnamefont {{Halmrast}}}, \bibinfo {author}
  {\bibfnamefont {J.~S.}\ \bibnamefont {{Hazboun}}}, \bibinfo {author}
  {\bibfnamefont {K.}~\bibnamefont {{Islo}}}, \bibinfo {author} {\bibfnamefont
  {R.~J.}\ \bibnamefont {{Jennings}}}, \bibinfo {author} {\bibfnamefont
  {C.}~\bibnamefont {{Jessup}}}, \bibinfo {author} {\bibfnamefont {M.~L.}\
  \bibnamefont {{Jones}}}, \bibinfo {author} {\bibfnamefont {A.~R.}\
  \bibnamefont {{Kaiser}}}, \bibinfo {author} {\bibfnamefont {D.~L.}\
  \bibnamefont {{Kaplan}}}, \bibinfo {author} {\bibfnamefont {L.~Z.}\
  \bibnamefont {{Kelley}}}, \bibinfo {author} {\bibfnamefont {J.~S.}\
  \bibnamefont {{Key}}}, \bibinfo {author} {\bibfnamefont {M.~T.}\ \bibnamefont
  {{Lam}}}, \bibinfo {author} {\bibfnamefont {T.~J.~W.}\ \bibnamefont
  {{Lazio}}}, \bibinfo {author} {\bibfnamefont {D.~R.}\ \bibnamefont
  {{Lorimer}}}, \bibinfo {author} {\bibfnamefont {J.}~\bibnamefont {{Luo}}},
  \bibinfo {author} {\bibfnamefont {R.~S.}\ \bibnamefont {{Lynch}}}, \bibinfo
  {author} {\bibfnamefont {D.~R.}\ \bibnamefont {{Madison}}}, \bibinfo {author}
  {\bibfnamefont {K.}~\bibnamefont {{Maraccini}}}, \bibinfo {author}
  {\bibfnamefont {M.~A.}\ \bibnamefont {{McLaughlin}}}, \bibinfo {author}
  {\bibfnamefont {C.~M.~F.}\ \bibnamefont {{Mingarelli}}}, \bibinfo {author}
  {\bibfnamefont {C.}~\bibnamefont {{Ng}}}, \bibinfo {author} {\bibfnamefont
  {B.~M.~X.}\ \bibnamefont {{Nguyen}}}, \bibinfo {author} {\bibfnamefont
  {D.~J.}\ \bibnamefont {{Nice}}}, \bibinfo {author} {\bibfnamefont {T.~T.}\
  \bibnamefont {{Pennucci}}}, \bibinfo {author} {\bibfnamefont {N.~S.}\
  \bibnamefont {{Pol}}}, \bibinfo {author} {\bibfnamefont {J.}~\bibnamefont
  {{Ramette}}}, \bibinfo {author} {\bibfnamefont {S.~M.}\ \bibnamefont
  {{Ransom}}}, \bibinfo {author} {\bibfnamefont {P.~S.}\ \bibnamefont {{Ray}}},
  \bibinfo {author} {\bibfnamefont {B.~J.}\ \bibnamefont {{Shapiro-Albert}}},
  \bibinfo {author} {\bibfnamefont {X.}~\bibnamefont {{Siemens}}}, \bibinfo
  {author} {\bibfnamefont {J.}~\bibnamefont {{Simon}}}, \bibinfo {author}
  {\bibfnamefont {R.}~\bibnamefont {{Spiewak}}}, \bibinfo {author}
  {\bibfnamefont {I.~H.}\ \bibnamefont {{Stairs}}}, \bibinfo {author}
  {\bibfnamefont {D.~R.}\ \bibnamefont {{Stinebring}}}, \bibinfo {author}
  {\bibfnamefont {K.}~\bibnamefont {{Stovall}}}, \bibinfo {author}
  {\bibfnamefont {J.~K.}\ \bibnamefont {{Swiggum}}}, \bibinfo {author}
  {\bibfnamefont {S.~R.}\ \bibnamefont {{Taylor}}}, \bibinfo {author}
  {\bibfnamefont {M.}~\bibnamefont {{Tripepi}}}, \bibinfo {author}
  {\bibfnamefont {M.}~\bibnamefont {{Vallisneri}}}, \bibinfo {author}
  {\bibfnamefont {S.~J.}\ \bibnamefont {{Vigeland}}}, \bibinfo {author}
  {\bibfnamefont {C.~A.}\ \bibnamefont {{Witt}}}, \bibinfo {author}
  {\bibfnamefont {W.}~\bibnamefont {{Zhu}}},\ and\ \bibinfo {author}
  {\bibnamefont {{Nanograv Collaboration}}},\ }\bibfield  {title} {\bibinfo
  {title} {{The NANOGrav 12.5 yr Data Set: Wideband Timing of 47 Millisecond
  Pulsars}},\ }\href {https://doi.org/10.3847/1538-4365/abc6a1} {\bibfield
  {journal} {\bibinfo  {journal} {\apjs}\ }\textbf {\bibinfo {volume} {252}},\
  \bibinfo {eid} {5} (\bibinfo {year} {2021})},\ \Eprint
  {https://arxiv.org/abs/2005.06495} {arXiv:2005.06495 [astro-ph.HE]}
  \BibitemShut {NoStop}%
\bibitem [{\citenamefont {Kerr}\ \emph {et~al.}(2020)\citenamefont {Kerr},
  \citenamefont {Reardon}, \citenamefont {Hobbs}, \citenamefont {Shannon},
  \citenamefont {Manchester}, \citenamefont {Dai}, \citenamefont {Russell},
  \citenamefont {Zhang}, \citenamefont {van Straten}, \citenamefont
  {Os{\l}owski}, \citenamefont {Parthasarathy}, \citenamefont {Spiewak},
  \citenamefont {Bailes}, \citenamefont {Bhat}, \citenamefont {Cameron},
  \citenamefont {Coles}, \citenamefont {Dempsey}, \citenamefont {Deng},
  \citenamefont {Goncharov}, \citenamefont {Kaczmarek}, \citenamefont {Keith},
  \citenamefont {Lasky}, \citenamefont {Lower}, \citenamefont {Preisig},
  \citenamefont {Sarkissian}, \citenamefont {Toomey}, \citenamefont {Wang},
  \citenamefont {Wang}, \citenamefont {Zhang},\ and\ \citenamefont
  {Zhu}}]{Kerr_2020}%
  \BibitemOpen
  \bibfield  {author} {\bibinfo {author} {\bibfnamefont {M.}~\bibnamefont
  {Kerr}}, \bibinfo {author} {\bibfnamefont {D.~J.}\ \bibnamefont {Reardon}},
  \bibinfo {author} {\bibfnamefont {G.}~\bibnamefont {Hobbs}}, \bibinfo
  {author} {\bibfnamefont {R.~M.}\ \bibnamefont {Shannon}}, \bibinfo {author}
  {\bibfnamefont {R.~N.}\ \bibnamefont {Manchester}}, \bibinfo {author}
  {\bibfnamefont {S.}~\bibnamefont {Dai}}, \bibinfo {author} {\bibfnamefont
  {C.~J.}\ \bibnamefont {Russell}}, \bibinfo {author} {\bibfnamefont
  {S.}~\bibnamefont {Zhang}}, \bibinfo {author} {\bibfnamefont
  {W.}~\bibnamefont {van Straten}}, \bibinfo {author} {\bibfnamefont
  {S.}~\bibnamefont {Os{\l}owski}}, \bibinfo {author} {\bibfnamefont
  {A.}~\bibnamefont {Parthasarathy}}, \bibinfo {author} {\bibfnamefont
  {R.}~\bibnamefont {Spiewak}}, \bibinfo {author} {\bibfnamefont
  {M.}~\bibnamefont {Bailes}}, \bibinfo {author} {\bibfnamefont {N.~D.~R.}\
  \bibnamefont {Bhat}}, \bibinfo {author} {\bibfnamefont {A.~D.}\ \bibnamefont
  {Cameron}}, \bibinfo {author} {\bibfnamefont {W.~A.}\ \bibnamefont {Coles}},
  \bibinfo {author} {\bibfnamefont {J.}~\bibnamefont {Dempsey}}, \bibinfo
  {author} {\bibfnamefont {X.}~\bibnamefont {Deng}}, \bibinfo {author}
  {\bibfnamefont {B.}~\bibnamefont {Goncharov}}, \bibinfo {author}
  {\bibfnamefont {J.~F.}\ \bibnamefont {Kaczmarek}}, \bibinfo {author}
  {\bibfnamefont {M.~J.}\ \bibnamefont {Keith}}, \bibinfo {author}
  {\bibfnamefont {P.~D.}\ \bibnamefont {Lasky}}, \bibinfo {author}
  {\bibfnamefont {M.~E.}\ \bibnamefont {Lower}}, \bibinfo {author}
  {\bibfnamefont {B.}~\bibnamefont {Preisig}}, \bibinfo {author} {\bibfnamefont
  {J.~M.}\ \bibnamefont {Sarkissian}}, \bibinfo {author} {\bibfnamefont
  {L.}~\bibnamefont {Toomey}}, \bibinfo {author} {\bibfnamefont
  {H.}~\bibnamefont {Wang}}, \bibinfo {author} {\bibfnamefont {J.}~\bibnamefont
  {Wang}}, \bibinfo {author} {\bibfnamefont {L.}~\bibnamefont {Zhang}},\ and\
  \bibinfo {author} {\bibfnamefont {X.}~\bibnamefont {Zhu}},\ }\bibfield
  {title} {\bibinfo {title} {The parkes pulsar timing array project: second
  data release},\ }\bibfield  {journal} {\bibinfo  {journal} {Publications of
  the Astronomical Society of Australia}\ }\textbf {\bibinfo {volume} {37}},\
  \href {https://doi.org/10.1017/pasa.2020.11} {10.1017/pasa.2020.11} (\bibinfo
  {year} {2020})\BibitemShut {NoStop}%
\bibitem [{\citenamefont {Hobbs}\ \emph {et~al.}(2010)\citenamefont {Hobbs},
  \citenamefont {Archibald}, \citenamefont {Arzoumanian}, \citenamefont
  {Backer}, \citenamefont {Bailes}, \citenamefont {Bhat}, \citenamefont
  {Burgay}, \citenamefont {Burke-Spolaor}, \citenamefont {Champion},
  \citenamefont {Cognard}, \citenamefont {Coles}, \citenamefont {Cordes},
  \citenamefont {Demorest}, \citenamefont {Desvignes}, \citenamefont {Ferdman},
  \citenamefont {Finn}, \citenamefont {Freire}, \citenamefont {Gonzalez},
  \citenamefont {Hessels}, \citenamefont {Hotan}, \citenamefont {Janssen},
  \citenamefont {Jenet}, \citenamefont {Jessner}, \citenamefont {Jordan},
  \citenamefont {Kaspi}, \citenamefont {Kramer}, \citenamefont {Kondratiev},
  \citenamefont {Lazio}, \citenamefont {Lazaridis}, \citenamefont {Lee},
  \citenamefont {Levin}, \citenamefont {Lommen}, \citenamefont {Lorimer},
  \citenamefont {Lynch}, \citenamefont {Lyne}, \citenamefont {Manchester},
  \citenamefont {McLaughlin}, \citenamefont {Nice}, \citenamefont {Oslowski},
  \citenamefont {Pilia}, \citenamefont {Possenti}, \citenamefont {Purver},
  \citenamefont {Ransom}, \citenamefont {Reynolds}, \citenamefont {Sanidas},
  \citenamefont {Sarkissian}, \citenamefont {Sesana}, \citenamefont {Shannon},
  \citenamefont {Siemens}, \citenamefont {Stairs}, \citenamefont {Stappers},
  \citenamefont {Stinebring}, \citenamefont {Theureau}, \citenamefont {van
  Haasteren}, \citenamefont {van Straten}, \citenamefont {Verbiest},
  \citenamefont {Yardley},\ and\ \citenamefont {You}}]{Hobbs_2010}%
  \BibitemOpen
  \bibfield  {author} {\bibinfo {author} {\bibfnamefont {G.}~\bibnamefont
  {Hobbs}}, \bibinfo {author} {\bibfnamefont {A.}~\bibnamefont {Archibald}},
  \bibinfo {author} {\bibfnamefont {Z.}~\bibnamefont {Arzoumanian}}, \bibinfo
  {author} {\bibfnamefont {D.}~\bibnamefont {Backer}}, \bibinfo {author}
  {\bibfnamefont {M.}~\bibnamefont {Bailes}}, \bibinfo {author} {\bibfnamefont
  {N.~D.~R.}\ \bibnamefont {Bhat}}, \bibinfo {author} {\bibfnamefont
  {M.}~\bibnamefont {Burgay}}, \bibinfo {author} {\bibfnamefont
  {S.}~\bibnamefont {Burke-Spolaor}}, \bibinfo {author} {\bibfnamefont
  {D.}~\bibnamefont {Champion}}, \bibinfo {author} {\bibfnamefont
  {I.}~\bibnamefont {Cognard}}, \bibinfo {author} {\bibfnamefont
  {W.}~\bibnamefont {Coles}}, \bibinfo {author} {\bibfnamefont
  {J.}~\bibnamefont {Cordes}}, \bibinfo {author} {\bibfnamefont
  {P.}~\bibnamefont {Demorest}}, \bibinfo {author} {\bibfnamefont
  {G.}~\bibnamefont {Desvignes}}, \bibinfo {author} {\bibfnamefont {R.~D.}\
  \bibnamefont {Ferdman}}, \bibinfo {author} {\bibfnamefont {L.}~\bibnamefont
  {Finn}}, \bibinfo {author} {\bibfnamefont {P.}~\bibnamefont {Freire}},
  \bibinfo {author} {\bibfnamefont {M.}~\bibnamefont {Gonzalez}}, \bibinfo
  {author} {\bibfnamefont {J.}~\bibnamefont {Hessels}}, \bibinfo {author}
  {\bibfnamefont {A.}~\bibnamefont {Hotan}}, \bibinfo {author} {\bibfnamefont
  {G.}~\bibnamefont {Janssen}}, \bibinfo {author} {\bibfnamefont
  {F.}~\bibnamefont {Jenet}}, \bibinfo {author} {\bibfnamefont
  {A.}~\bibnamefont {Jessner}}, \bibinfo {author} {\bibfnamefont
  {C.}~\bibnamefont {Jordan}}, \bibinfo {author} {\bibfnamefont
  {V.}~\bibnamefont {Kaspi}}, \bibinfo {author} {\bibfnamefont
  {M.}~\bibnamefont {Kramer}}, \bibinfo {author} {\bibfnamefont
  {V.}~\bibnamefont {Kondratiev}}, \bibinfo {author} {\bibfnamefont
  {J.}~\bibnamefont {Lazio}}, \bibinfo {author} {\bibfnamefont
  {K.}~\bibnamefont {Lazaridis}}, \bibinfo {author} {\bibfnamefont {K.~J.}\
  \bibnamefont {Lee}}, \bibinfo {author} {\bibfnamefont {Y.}~\bibnamefont
  {Levin}}, \bibinfo {author} {\bibfnamefont {A.}~\bibnamefont {Lommen}},
  \bibinfo {author} {\bibfnamefont {D.}~\bibnamefont {Lorimer}}, \bibinfo
  {author} {\bibfnamefont {R.}~\bibnamefont {Lynch}}, \bibinfo {author}
  {\bibfnamefont {A.}~\bibnamefont {Lyne}}, \bibinfo {author} {\bibfnamefont
  {R.}~\bibnamefont {Manchester}}, \bibinfo {author} {\bibfnamefont
  {M.}~\bibnamefont {McLaughlin}}, \bibinfo {author} {\bibfnamefont
  {D.}~\bibnamefont {Nice}}, \bibinfo {author} {\bibfnamefont {S.}~\bibnamefont
  {Oslowski}}, \bibinfo {author} {\bibfnamefont {M.}~\bibnamefont {Pilia}},
  \bibinfo {author} {\bibfnamefont {A.}~\bibnamefont {Possenti}}, \bibinfo
  {author} {\bibfnamefont {M.}~\bibnamefont {Purver}}, \bibinfo {author}
  {\bibfnamefont {S.}~\bibnamefont {Ransom}}, \bibinfo {author} {\bibfnamefont
  {J.}~\bibnamefont {Reynolds}}, \bibinfo {author} {\bibfnamefont
  {S.}~\bibnamefont {Sanidas}}, \bibinfo {author} {\bibfnamefont
  {J.}~\bibnamefont {Sarkissian}}, \bibinfo {author} {\bibfnamefont
  {A.}~\bibnamefont {Sesana}}, \bibinfo {author} {\bibfnamefont
  {R.}~\bibnamefont {Shannon}}, \bibinfo {author} {\bibfnamefont
  {X.}~\bibnamefont {Siemens}}, \bibinfo {author} {\bibfnamefont
  {I.}~\bibnamefont {Stairs}}, \bibinfo {author} {\bibfnamefont
  {B.}~\bibnamefont {Stappers}}, \bibinfo {author} {\bibfnamefont
  {D.}~\bibnamefont {Stinebring}}, \bibinfo {author} {\bibfnamefont
  {G.}~\bibnamefont {Theureau}}, \bibinfo {author} {\bibfnamefont
  {R.}~\bibnamefont {van Haasteren}}, \bibinfo {author} {\bibfnamefont
  {W.}~\bibnamefont {van Straten}}, \bibinfo {author} {\bibfnamefont
  {J.~P.~W.}\ \bibnamefont {Verbiest}}, \bibinfo {author} {\bibfnamefont
  {D.~R.~B.}\ \bibnamefont {Yardley}},\ and\ \bibinfo {author} {\bibfnamefont
  {X.~P.}\ \bibnamefont {You}},\ }\bibfield  {title} {\bibinfo {title} {The
  international pulsar timing array project: using pulsars as a gravitational
  wave detector},\ }\href {https://doi.org/10.1088/0264-9381/27/8/084013}
  {\bibfield  {journal} {\bibinfo  {journal} {Classical and Quantum Gravity}\
  }\textbf {\bibinfo {volume} {27}},\ \bibinfo {pages} {084013} (\bibinfo
  {year} {2010})}\BibitemShut {NoStop}%
\bibitem [{\citenamefont {Verbiest}\ \emph {et~al.}(2016)\citenamefont
  {Verbiest}, \citenamefont {Lentati}, \citenamefont {Hobbs}, \citenamefont
  {van Haasteren}, \citenamefont {Demorest}, \citenamefont {Janssen},
  \citenamefont {Wang}, \citenamefont {Desvignes}, \citenamefont {Caballero},
  \citenamefont {Keith}, \citenamefont {Champion}, \citenamefont {Arzoumanian},
  \citenamefont {Babak}, \citenamefont {Bassa}, \citenamefont {Bhat},
  \citenamefont {Brazier}, \citenamefont {Brem}, \citenamefont {Burgay},
  \citenamefont {Burke-Spolaor}, \citenamefont {Chamberlin}, \citenamefont
  {Chatterjee}, \citenamefont {Christy}, \citenamefont {Cognard}, \citenamefont
  {Cordes}, \citenamefont {Dai}, \citenamefont {Dolch}, \citenamefont {Ellis},
  \citenamefont {Ferdman}, \citenamefont {Fonseca}, \citenamefont {Gair},
  \citenamefont {Garver-Daniels}, \citenamefont {Gentile}, \citenamefont
  {Gonzalez}, \citenamefont {Graikou}, \citenamefont {Guillemot}, \citenamefont
  {Hessels}, \citenamefont {Jones}, \citenamefont {Karuppusamy}, \citenamefont
  {Kerr}, \citenamefont {Kramer}, \citenamefont {Lam}, \citenamefont {Lasky},
  \citenamefont {Lassus}, \citenamefont {Lazarus}, \citenamefont {Lazio},
  \citenamefont {Lee}, \citenamefont {Levin}, \citenamefont {Liu},
  \citenamefont {Lynch}, \citenamefont {Lyne}, \citenamefont {Mckee},
  \citenamefont {McLaughlin}, \citenamefont {McWilliams}, \citenamefont
  {Madison}, \citenamefont {Manchester}, \citenamefont {Mingarelli},
  \citenamefont {Nice}, \citenamefont {Osłowski}, \citenamefont {Palliyaguru},
  \citenamefont {Pennucci}, \citenamefont {Perera}, \citenamefont {Perrodin},
  \citenamefont {Possenti}, \citenamefont {Petiteau}, \citenamefont {Ransom},
  \citenamefont {Reardon}, \citenamefont {Rosado}, \citenamefont {Sanidas},
  \citenamefont {Sesana}, \citenamefont {Shaifullah}, \citenamefont {Shannon},
  \citenamefont {Siemens}, \citenamefont {Simon}, \citenamefont {Smits},
  \citenamefont {Spiewak}, \citenamefont {Stairs}, \citenamefont {Stappers},
  \citenamefont {Stinebring}, \citenamefont {Stovall}, \citenamefont {Swiggum},
  \citenamefont {Taylor}, \citenamefont {Theureau}, \citenamefont {Tiburzi},
  \citenamefont {Toomey}, \citenamefont {Vallisneri}, \citenamefont {van
  Straten}, \citenamefont {Vecchio}, \citenamefont {Wang}, \citenamefont {Wen},
  \citenamefont {You}, \citenamefont {Zhu},\ and\ \citenamefont
  {Zhu}}]{10.1093/mnras/stw347}%
  \BibitemOpen
  \bibfield  {author} {\bibinfo {author} {\bibfnamefont {J.~P.~W.}\
  \bibnamefont {Verbiest}}, \bibinfo {author} {\bibfnamefont {L.}~\bibnamefont
  {Lentati}}, \bibinfo {author} {\bibfnamefont {G.}~\bibnamefont {Hobbs}},
  \bibinfo {author} {\bibfnamefont {R.}~\bibnamefont {van Haasteren}}, \bibinfo
  {author} {\bibfnamefont {P.~B.}\ \bibnamefont {Demorest}}, \bibinfo {author}
  {\bibfnamefont {G.~H.}\ \bibnamefont {Janssen}}, \bibinfo {author}
  {\bibfnamefont {J.-B.}\ \bibnamefont {Wang}}, \bibinfo {author}
  {\bibfnamefont {G.}~\bibnamefont {Desvignes}}, \bibinfo {author}
  {\bibfnamefont {R.~N.}\ \bibnamefont {Caballero}}, \bibinfo {author}
  {\bibfnamefont {M.~J.}\ \bibnamefont {Keith}}, \bibinfo {author}
  {\bibfnamefont {D.~J.}\ \bibnamefont {Champion}}, \bibinfo {author}
  {\bibfnamefont {Z.}~\bibnamefont {Arzoumanian}}, \bibinfo {author}
  {\bibfnamefont {S.}~\bibnamefont {Babak}}, \bibinfo {author} {\bibfnamefont
  {C.~G.}\ \bibnamefont {Bassa}}, \bibinfo {author} {\bibfnamefont {N.~D.~R.}\
  \bibnamefont {Bhat}}, \bibinfo {author} {\bibfnamefont {A.}~\bibnamefont
  {Brazier}}, \bibinfo {author} {\bibfnamefont {P.}~\bibnamefont {Brem}},
  \bibinfo {author} {\bibfnamefont {M.}~\bibnamefont {Burgay}}, \bibinfo
  {author} {\bibfnamefont {S.}~\bibnamefont {Burke-Spolaor}}, \bibinfo {author}
  {\bibfnamefont {S.~J.}\ \bibnamefont {Chamberlin}}, \bibinfo {author}
  {\bibfnamefont {S.}~\bibnamefont {Chatterjee}}, \bibinfo {author}
  {\bibfnamefont {B.}~\bibnamefont {Christy}}, \bibinfo {author} {\bibfnamefont
  {I.}~\bibnamefont {Cognard}}, \bibinfo {author} {\bibfnamefont {J.~M.}\
  \bibnamefont {Cordes}}, \bibinfo {author} {\bibfnamefont {S.}~\bibnamefont
  {Dai}}, \bibinfo {author} {\bibfnamefont {T.}~\bibnamefont {Dolch}}, \bibinfo
  {author} {\bibfnamefont {J.~A.}\ \bibnamefont {Ellis}}, \bibinfo {author}
  {\bibfnamefont {R.~D.}\ \bibnamefont {Ferdman}}, \bibinfo {author}
  {\bibfnamefont {E.}~\bibnamefont {Fonseca}}, \bibinfo {author} {\bibfnamefont
  {J.~R.}\ \bibnamefont {Gair}}, \bibinfo {author} {\bibfnamefont {N.~E.}\
  \bibnamefont {Garver-Daniels}}, \bibinfo {author} {\bibfnamefont
  {P.}~\bibnamefont {Gentile}}, \bibinfo {author} {\bibfnamefont {M.~E.}\
  \bibnamefont {Gonzalez}}, \bibinfo {author} {\bibfnamefont {E.}~\bibnamefont
  {Graikou}}, \bibinfo {author} {\bibfnamefont {L.}~\bibnamefont {Guillemot}},
  \bibinfo {author} {\bibfnamefont {J.~W.~T.}\ \bibnamefont {Hessels}},
  \bibinfo {author} {\bibfnamefont {G.}~\bibnamefont {Jones}}, \bibinfo
  {author} {\bibfnamefont {R.}~\bibnamefont {Karuppusamy}}, \bibinfo {author}
  {\bibfnamefont {M.}~\bibnamefont {Kerr}}, \bibinfo {author} {\bibfnamefont
  {M.}~\bibnamefont {Kramer}}, \bibinfo {author} {\bibfnamefont {M.~T.}\
  \bibnamefont {Lam}}, \bibinfo {author} {\bibfnamefont {P.~D.}\ \bibnamefont
  {Lasky}}, \bibinfo {author} {\bibfnamefont {A.}~\bibnamefont {Lassus}},
  \bibinfo {author} {\bibfnamefont {P.}~\bibnamefont {Lazarus}}, \bibinfo
  {author} {\bibfnamefont {T.~J.~W.}\ \bibnamefont {Lazio}}, \bibinfo {author}
  {\bibfnamefont {K.~J.}\ \bibnamefont {Lee}}, \bibinfo {author} {\bibfnamefont
  {L.}~\bibnamefont {Levin}}, \bibinfo {author} {\bibfnamefont
  {K.}~\bibnamefont {Liu}}, \bibinfo {author} {\bibfnamefont {R.~S.}\
  \bibnamefont {Lynch}}, \bibinfo {author} {\bibfnamefont {A.~G.}\ \bibnamefont
  {Lyne}}, \bibinfo {author} {\bibfnamefont {J.}~\bibnamefont {Mckee}},
  \bibinfo {author} {\bibfnamefont {M.~A.}\ \bibnamefont {McLaughlin}},
  \bibinfo {author} {\bibfnamefont {S.~T.}\ \bibnamefont {McWilliams}},
  \bibinfo {author} {\bibfnamefont {D.~R.}\ \bibnamefont {Madison}}, \bibinfo
  {author} {\bibfnamefont {R.~N.}\ \bibnamefont {Manchester}}, \bibinfo
  {author} {\bibfnamefont {C.~M.~F.}\ \bibnamefont {Mingarelli}}, \bibinfo
  {author} {\bibfnamefont {D.~J.}\ \bibnamefont {Nice}}, \bibinfo {author}
  {\bibfnamefont {S.}~\bibnamefont {Osłowski}}, \bibinfo {author}
  {\bibfnamefont {N.~T.}\ \bibnamefont {Palliyaguru}}, \bibinfo {author}
  {\bibfnamefont {T.~T.}\ \bibnamefont {Pennucci}}, \bibinfo {author}
  {\bibfnamefont {B.~B.~P.}\ \bibnamefont {Perera}}, \bibinfo {author}
  {\bibfnamefont {D.}~\bibnamefont {Perrodin}}, \bibinfo {author}
  {\bibfnamefont {A.}~\bibnamefont {Possenti}}, \bibinfo {author}
  {\bibfnamefont {A.}~\bibnamefont {Petiteau}}, \bibinfo {author}
  {\bibfnamefont {S.~M.}\ \bibnamefont {Ransom}}, \bibinfo {author}
  {\bibfnamefont {D.}~\bibnamefont {Reardon}}, \bibinfo {author} {\bibfnamefont
  {P.~A.}\ \bibnamefont {Rosado}}, \bibinfo {author} {\bibfnamefont {S.~A.}\
  \bibnamefont {Sanidas}}, \bibinfo {author} {\bibfnamefont {A.}~\bibnamefont
  {Sesana}}, \bibinfo {author} {\bibfnamefont {G.}~\bibnamefont {Shaifullah}},
  \bibinfo {author} {\bibfnamefont {R.~M.}\ \bibnamefont {Shannon}}, \bibinfo
  {author} {\bibfnamefont {X.}~\bibnamefont {Siemens}}, \bibinfo {author}
  {\bibfnamefont {J.}~\bibnamefont {Simon}}, \bibinfo {author} {\bibfnamefont
  {R.}~\bibnamefont {Smits}}, \bibinfo {author} {\bibfnamefont
  {R.}~\bibnamefont {Spiewak}}, \bibinfo {author} {\bibfnamefont {I.~H.}\
  \bibnamefont {Stairs}}, \bibinfo {author} {\bibfnamefont {B.~W.}\
  \bibnamefont {Stappers}}, \bibinfo {author} {\bibfnamefont {D.~R.}\
  \bibnamefont {Stinebring}}, \bibinfo {author} {\bibfnamefont
  {K.}~\bibnamefont {Stovall}}, \bibinfo {author} {\bibfnamefont {J.~K.}\
  \bibnamefont {Swiggum}}, \bibinfo {author} {\bibfnamefont {S.~R.}\
  \bibnamefont {Taylor}}, \bibinfo {author} {\bibfnamefont {G.}~\bibnamefont
  {Theureau}}, \bibinfo {author} {\bibfnamefont {C.}~\bibnamefont {Tiburzi}},
  \bibinfo {author} {\bibfnamefont {L.}~\bibnamefont {Toomey}}, \bibinfo
  {author} {\bibfnamefont {M.}~\bibnamefont {Vallisneri}}, \bibinfo {author}
  {\bibfnamefont {W.}~\bibnamefont {van Straten}}, \bibinfo {author}
  {\bibfnamefont {A.}~\bibnamefont {Vecchio}}, \bibinfo {author} {\bibfnamefont
  {Y.}~\bibnamefont {Wang}}, \bibinfo {author} {\bibfnamefont {L.}~\bibnamefont
  {Wen}}, \bibinfo {author} {\bibfnamefont {X.~P.}\ \bibnamefont {You}},
  \bibinfo {author} {\bibfnamefont {W.~W.}\ \bibnamefont {Zhu}},\ and\ \bibinfo
  {author} {\bibfnamefont {X.-J.}\ \bibnamefont {Zhu}},\ }\bibfield  {title}
  {\bibinfo {title} {{The International Pulsar Timing Array: First data
  release}},\ }\href {https://doi.org/10.1093/mnras/stw347} {\bibfield
  {journal} {\bibinfo  {journal} {Monthly Notices of the Royal Astronomical
  Society}\ }\textbf {\bibinfo {volume} {458}},\ \bibinfo {pages} {1267}
  (\bibinfo {year} {2016})},\ \Eprint
  {https://arxiv.org/abs/https://academic.oup.com/mnras/article-pdf/458/2/1267/18237975/stw347.pdf}
  {https://academic.oup.com/mnras/article-pdf/458/2/1267/18237975/stw347.pdf}
  \BibitemShut {NoStop}%
\bibitem [{\citenamefont {Pearson}\ \emph {et~al.}(2021)\citenamefont
  {Pearson}, \citenamefont {Trendafilova},\ and\ \citenamefont
  {Meyers}}]{PhysRevD.103.063017}%
  \BibitemOpen
  \bibfield  {author} {\bibinfo {author} {\bibfnamefont {N.}~\bibnamefont
  {Pearson}}, \bibinfo {author} {\bibfnamefont {C.}~\bibnamefont
  {Trendafilova}},\ and\ \bibinfo {author} {\bibfnamefont {J.}~\bibnamefont
  {Meyers}},\ }\bibfield  {title} {\bibinfo {title} {Searching for
  gravitational waves with strongly lensed repeating fast radio bursts},\
  }\href {https://doi.org/10.1103/PhysRevD.103.063017} {\bibfield  {journal}
  {\bibinfo  {journal} {Phys. Rev. D}\ }\textbf {\bibinfo {volume} {103}},\
  \bibinfo {pages} {063017} (\bibinfo {year} {2021})}\BibitemShut {NoStop}%
\bibitem [{\citenamefont {{Braginsky}}\ \emph {et~al.}(1990)\citenamefont
  {{Braginsky}}, \citenamefont {{Kardashev}}, \citenamefont {{Polnarev}},\ and\
  \citenamefont {{Novikov}}}]{1990NCimB.105.1141B}%
  \BibitemOpen
  \bibfield  {author} {\bibinfo {author} {\bibfnamefont {V.~B.}\ \bibnamefont
  {{Braginsky}}}, \bibinfo {author} {\bibfnamefont {N.~S.}\ \bibnamefont
  {{Kardashev}}}, \bibinfo {author} {\bibfnamefont {A.~G.}\ \bibnamefont
  {{Polnarev}}},\ and\ \bibinfo {author} {\bibfnamefont {I.~D.}\ \bibnamefont
  {{Novikov}}},\ }\bibfield  {title} {\bibinfo {title} {{Propagation of
  electromagnetic radiation in a random field of gravitational waves and space
  radio interferometry.}},\ }\href {https://doi.org/10.1007/BF02827323}
  {\bibfield  {journal} {\bibinfo  {journal} {Nuovo Cimento B Serie}\ }\textbf
  {\bibinfo {volume} {105}},\ \bibinfo {pages} {1141} (\bibinfo {year}
  {1990})}\BibitemShut {NoStop}%
\bibitem [{\citenamefont {{Pyne}}\ \emph {et~al.}(1996)\citenamefont {{Pyne}},
  \citenamefont {{Gwinn}}, \citenamefont {{Birkinshaw}}, \citenamefont
  {{Eubanks}},\ and\ \citenamefont {{Matsakis}}}]{1996ApJ...465..566P}%
  \BibitemOpen
  \bibfield  {author} {\bibinfo {author} {\bibfnamefont {T.}~\bibnamefont
  {{Pyne}}}, \bibinfo {author} {\bibfnamefont {C.~R.}\ \bibnamefont {{Gwinn}}},
  \bibinfo {author} {\bibfnamefont {M.}~\bibnamefont {{Birkinshaw}}}, \bibinfo
  {author} {\bibfnamefont {T.~M.}\ \bibnamefont {{Eubanks}}},\ and\ \bibinfo
  {author} {\bibfnamefont {D.~N.}\ \bibnamefont {{Matsakis}}},\ }\bibfield
  {title} {\bibinfo {title} {{Gravitational Radiation and Very Long Baseline
  Interferometry}},\ }\href {https://doi.org/10.1086/177443} {\bibfield
  {journal} {\bibinfo  {journal} {\apj}\ }\textbf {\bibinfo {volume} {465}},\
  \bibinfo {pages} {566} (\bibinfo {year} {1996})},\ \Eprint
  {https://arxiv.org/abs/astro-ph/9507030} {arXiv:astro-ph/9507030 [astro-ph]}
  \BibitemShut {NoStop}%
\bibitem [{\citenamefont {Book}\ and\ \citenamefont
  {Flanagan}(2011)}]{PhysRevD.83.024024}%
  \BibitemOpen
  \bibfield  {author} {\bibinfo {author} {\bibfnamefont {L.~G.}\ \bibnamefont
  {Book}}\ and\ \bibinfo {author} {\bibfnamefont {E.~E.}\ \bibnamefont
  {Flanagan}},\ }\bibfield  {title} {\bibinfo {title} {Astrometric effects of a
  stochastic gravitational wave background},\ }\href
  {https://doi.org/10.1103/PhysRevD.83.024024} {\bibfield  {journal} {\bibinfo
  {journal} {Phys. Rev. D}\ }\textbf {\bibinfo {volume} {83}},\ \bibinfo
  {pages} {024024} (\bibinfo {year} {2011})}\BibitemShut {NoStop}%
\bibitem [{\citenamefont {Kamionkowski}\ \emph {et~al.}(1997)\citenamefont
  {Kamionkowski}, \citenamefont {Kosowsky},\ and\ \citenamefont
  {Stebbins}}]{PhysRevLett.78.2058}%
  \BibitemOpen
  \bibfield  {author} {\bibinfo {author} {\bibfnamefont {M.}~\bibnamefont
  {Kamionkowski}}, \bibinfo {author} {\bibfnamefont {A.}~\bibnamefont
  {Kosowsky}},\ and\ \bibinfo {author} {\bibfnamefont {A.}~\bibnamefont
  {Stebbins}},\ }\bibfield  {title} {\bibinfo {title} {A probe of primordial
  gravity waves and vorticity},\ }\href
  {https://doi.org/10.1103/PhysRevLett.78.2058} {\bibfield  {journal} {\bibinfo
   {journal} {Phys. Rev. Lett.}\ }\textbf {\bibinfo {volume} {78}},\ \bibinfo
  {pages} {2058} (\bibinfo {year} {1997})}\BibitemShut {NoStop}%
\bibitem [{\citenamefont {Seljak}\ and\ \citenamefont
  {Zaldarriaga}(1997)}]{PhysRevLett.78.2054}%
  \BibitemOpen
  \bibfield  {author} {\bibinfo {author} {\bibfnamefont {U.~b.~u.}\
  \bibnamefont {Seljak}}\ and\ \bibinfo {author} {\bibfnamefont
  {M.}~\bibnamefont {Zaldarriaga}},\ }\bibfield  {title} {\bibinfo {title}
  {Signature of gravity waves in the polarization of the microwave
  background},\ }\href {https://doi.org/10.1103/PhysRevLett.78.2054} {\bibfield
   {journal} {\bibinfo  {journal} {Phys. Rev. Lett.}\ }\textbf {\bibinfo
  {volume} {78}},\ \bibinfo {pages} {2054} (\bibinfo {year}
  {1997})}\BibitemShut {NoStop}%
\bibitem [{\citenamefont {Bustamante-Rosell}\ \emph {et~al.}(2022)\citenamefont
  {Bustamante-Rosell}, \citenamefont {Meyers}, \citenamefont {Pearson},
  \citenamefont {Trendafilova},\ and\ \citenamefont
  {Zimmerman}}]{PhysRevD.105.044005}%
  \BibitemOpen
  \bibfield  {author} {\bibinfo {author} {\bibfnamefont {M.~J.}\ \bibnamefont
  {Bustamante-Rosell}}, \bibinfo {author} {\bibfnamefont {J.}~\bibnamefont
  {Meyers}}, \bibinfo {author} {\bibfnamefont {N.}~\bibnamefont {Pearson}},
  \bibinfo {author} {\bibfnamefont {C.}~\bibnamefont {Trendafilova}},\ and\
  \bibinfo {author} {\bibfnamefont {A.}~\bibnamefont {Zimmerman}},\ }\bibfield
  {title} {\bibinfo {title} {Gravitational wave timing array},\ }\href
  {https://doi.org/10.1103/PhysRevD.105.044005} {\bibfield  {journal} {\bibinfo
   {journal} {Phys. Rev. D}\ }\textbf {\bibinfo {volume} {105}},\ \bibinfo
  {pages} {044005} (\bibinfo {year} {2022})}\BibitemShut {NoStop}%
\bibitem [{\citenamefont {Weber}(1966)}]{PhysRevLett.17.1228}%
  \BibitemOpen
  \bibfield  {author} {\bibinfo {author} {\bibfnamefont {J.}~\bibnamefont
  {Weber}},\ }\bibfield  {title} {\bibinfo {title} {Observation of the thermal
  fluctuations of a gravitational-wave detector},\ }\href
  {https://doi.org/10.1103/PhysRevLett.17.1228} {\bibfield  {journal} {\bibinfo
   {journal} {Phys. Rev. Lett.}\ }\textbf {\bibinfo {volume} {17}},\ \bibinfo
  {pages} {1228} (\bibinfo {year} {1966})}\BibitemShut {NoStop}%
\bibitem [{\citenamefont {Weber}(1967)}]{PhysRevLett.18.498}%
  \BibitemOpen
  \bibfield  {author} {\bibinfo {author} {\bibfnamefont {J.}~\bibnamefont
  {Weber}},\ }\bibfield  {title} {\bibinfo {title} {Gravitational radiation},\
  }\href {https://doi.org/10.1103/PhysRevLett.18.498} {\bibfield  {journal}
  {\bibinfo  {journal} {Phys. Rev. Lett.}\ }\textbf {\bibinfo {volume} {18}},\
  \bibinfo {pages} {498} (\bibinfo {year} {1967})}\BibitemShut {NoStop}%
\bibitem [{\citenamefont {{Levine}}(2004)}]{2004PhP.....6...42L}%
  \BibitemOpen
  \bibfield  {author} {\bibinfo {author} {\bibfnamefont {J.~L.}\ \bibnamefont
  {{Levine}}},\ }\bibfield  {title} {\bibinfo {title} {{Early Gravity-Wave
  Detection Experiments, 1960-1975}},\ }\href
  {https://doi.org/10.1007/s00016-003-0179-6} {\bibfield  {journal} {\bibinfo
  {journal} {Physics in Perspective}\ }\textbf {\bibinfo {volume} {6}},\
  \bibinfo {pages} {42} (\bibinfo {year} {2004})}\BibitemShut {NoStop}%
\bibitem [{\citenamefont {Ronga}(2006)}]{Francesco_Ronga_2006}%
  \BibitemOpen
  \bibfield  {author} {\bibinfo {author} {\bibfnamefont {F.}~\bibnamefont
  {Ronga}},\ }\bibfield  {title} {\bibinfo {title} {Detection of gravitational
  waves with resonant antennas},\ }\href
  {https://doi.org/10.1088/1742-6596/39/1/005} {\bibfield  {journal} {\bibinfo
  {journal} {Journal of Physics: Conference Series}\ }\textbf {\bibinfo
  {volume} {39}},\ \bibinfo {pages} {18} (\bibinfo {year} {2006})}\BibitemShut
  {NoStop}%
\bibitem [{\citenamefont {{Aguiar}}(2011)}]{2011RAA....11....1A}%
  \BibitemOpen
  \bibfield  {author} {\bibinfo {author} {\bibfnamefont {O.~D.}\ \bibnamefont
  {{Aguiar}}},\ }\bibfield  {title} {\bibinfo {title} {{Past, present and
  future of the Resonant-Mass gravitational wave detectors}},\ }\href
  {https://doi.org/10.1088/1674-4527/11/1/001} {\bibfield  {journal} {\bibinfo
  {journal} {Research in Astronomy and Astrophysics}\ }\textbf {\bibinfo
  {volume} {11}},\ \bibinfo {pages} {1} (\bibinfo {year} {2011})},\ \Eprint
  {https://arxiv.org/abs/1009.1138} {arXiv:1009.1138 [astro-ph.IM]}
  \BibitemShut {NoStop}%
\bibitem [{\citenamefont {{de Waard}}\ \emph {et~al.}(2003)\citenamefont {{de
  Waard}}, \citenamefont {{Gottardi}}, \citenamefont {{van Houwelingen}},
  \citenamefont {{Shumack}},\ and\ \citenamefont
  {{Frossati}}}]{2003CQGra..20S.143D}%
  \BibitemOpen
  \bibfield  {author} {\bibinfo {author} {\bibfnamefont {A.}~\bibnamefont {{de
  Waard}}}, \bibinfo {author} {\bibfnamefont {L.}~\bibnamefont {{Gottardi}}},
  \bibinfo {author} {\bibfnamefont {J.}~\bibnamefont {{van Houwelingen}}},
  \bibinfo {author} {\bibfnamefont {A.}~\bibnamefont {{Shumack}}},\ and\
  \bibinfo {author} {\bibfnamefont {G.}~\bibnamefont {{Frossati}}},\ }\bibfield
   {title} {\bibinfo {title} {{MiniGRAIL, the first spherical detector}},\
  }\href {https://doi.org/10.1088/0264-9381/20/10/317} {\bibfield  {journal}
  {\bibinfo  {journal} {Classical and Quantum Gravity}\ }\textbf {\bibinfo
  {volume} {20}},\ \bibinfo {pages} {S143} (\bibinfo {year}
  {2003})}\BibitemShut {NoStop}%
\bibitem [{\citenamefont {{Aguiar}}\ \emph {et~al.}(2002)\citenamefont
  {{Aguiar}}, \citenamefont {{Andrade}}, \citenamefont {{Camargo Filho}},
  \citenamefont {{Costa}}, \citenamefont {{de Ara{\'u}jo}}, \citenamefont {{de
  Rey Neto}}, \citenamefont {{de Souza}}, \citenamefont {{Fauth}},
  \citenamefont {{Frajuca}}, \citenamefont {{Frossati}}, \citenamefont
  {{Furtado}}, \citenamefont {{Furtado}}, \citenamefont {{Magalh{\~a}es}},
  \citenamefont {{Marinho}}, \citenamefont {{Matos}}, \citenamefont
  {{Meliani}}, \citenamefont {{Melo}}, \citenamefont {{Miranda}}, \citenamefont
  {{Oliveira}}, \citenamefont {{Ribeiro}}, \citenamefont {{Salles}},
  \citenamefont {{Stellati}},\ and\ \citenamefont
  {{Velloso}}}]{2002CQGra..19.1949A}%
  \BibitemOpen
  \bibfield  {author} {\bibinfo {author} {\bibfnamefont {O.~D.}\ \bibnamefont
  {{Aguiar}}}, \bibinfo {author} {\bibfnamefont {L.~A.}\ \bibnamefont
  {{Andrade}}}, \bibinfo {author} {\bibfnamefont {L.}~\bibnamefont {{Camargo
  Filho}}}, \bibinfo {author} {\bibfnamefont {C.~A.}\ \bibnamefont {{Costa}}},
  \bibinfo {author} {\bibfnamefont {J.~C.~N.}\ \bibnamefont {{de Ara{\'u}jo}}},
  \bibinfo {author} {\bibfnamefont {E.~C.}\ \bibnamefont {{de Rey Neto}}},
  \bibinfo {author} {\bibfnamefont {S.~T.}\ \bibnamefont {{de Souza}}},
  \bibinfo {author} {\bibfnamefont {A.~C.}\ \bibnamefont {{Fauth}}}, \bibinfo
  {author} {\bibfnamefont {C.}~\bibnamefont {{Frajuca}}}, \bibinfo {author}
  {\bibfnamefont {G.}~\bibnamefont {{Frossati}}}, \bibinfo {author}
  {\bibfnamefont {S.~R.}\ \bibnamefont {{Furtado}}}, \bibinfo {author}
  {\bibfnamefont {V.~G.~S.}\ \bibnamefont {{Furtado}}}, \bibinfo {author}
  {\bibfnamefont {N.~S.}\ \bibnamefont {{Magalh{\~a}es}}}, \bibinfo {author}
  {\bibfnamefont {J.}~\bibnamefont {{Marinho}}, \bibfnamefont {R.~M.}},
  \bibinfo {author} {\bibfnamefont {E.~S.}\ \bibnamefont {{Matos}}}, \bibinfo
  {author} {\bibfnamefont {M.~T.}\ \bibnamefont {{Meliani}}}, \bibinfo {author}
  {\bibfnamefont {J.~L.}\ \bibnamefont {{Melo}}}, \bibinfo {author}
  {\bibfnamefont {O.~D.}\ \bibnamefont {{Miranda}}}, \bibinfo {author}
  {\bibfnamefont {J.}~\bibnamefont {{Oliveira}}, \bibfnamefont {N.~F.}},
  \bibinfo {author} {\bibfnamefont {K.~L.}\ \bibnamefont {{Ribeiro}}}, \bibinfo
  {author} {\bibfnamefont {K.~B.~M.}\ \bibnamefont {{Salles}}}, \bibinfo
  {author} {\bibfnamefont {C.}~\bibnamefont {{Stellati}}},\ and\ \bibinfo
  {author} {\bibfnamefont {J.}~\bibnamefont {{Velloso}}, \bibfnamefont
  {W.~F.}},\ }\bibfield  {title} {\bibinfo {title} {{The status of the
  Brazilian spherical detector}},\ }\href
  {https://doi.org/10.1088/0264-9381/19/7/397} {\bibfield  {journal} {\bibinfo
  {journal} {Classical and Quantum Gravity}\ }\textbf {\bibinfo {volume}
  {19}},\ \bibinfo {pages} {1949} (\bibinfo {year} {2002})}\BibitemShut
  {NoStop}%
\bibitem [{Hil(2017)}]{Hill2017}%
  \BibitemOpen
  \bibfield  {title} {\bibinfo {title} {{How the green light was given for
  gravitational wave search}},\ }\href@noop {} {\bibfield  {journal} {\bibinfo
  {journal} {Notices of the American Mathematical Society}\ }\textbf {\bibinfo
  {volume} {64}},\ \bibinfo {pages} {686} (\bibinfo {year} {2017})}\BibitemShut
  {NoStop}%
\bibitem [{\citenamefont {Uchiyama}\ \emph {et~al.}(2006)\citenamefont
  {Uchiyama}, \citenamefont {Miyoki}, \citenamefont {Ohashi}, \citenamefont
  {Kuroda}, \citenamefont {Yamamoto}, \citenamefont {Tokunari}, \citenamefont
  {Akutsu}, \citenamefont {Kamagasako}, \citenamefont {Nakagawa}, \citenamefont
  {Kirihara}, \citenamefont {Agatsuma}, \citenamefont {Ishitsuka},
  \citenamefont {Tatsumi}, \citenamefont {Telada}, \citenamefont {Ando},
  \citenamefont {Tomaru}, \citenamefont {Suzuki}, \citenamefont {Sato},
  \citenamefont {Haruyama}, \citenamefont {Yamamoto},\ and\ \citenamefont
  {Shintomi}}]{Takashi_Uchiyama_2006}%
  \BibitemOpen
  \bibfield  {author} {\bibinfo {author} {\bibfnamefont {T.}~\bibnamefont
  {Uchiyama}}, \bibinfo {author} {\bibfnamefont {S.}~\bibnamefont {Miyoki}},
  \bibinfo {author} {\bibfnamefont {M.}~\bibnamefont {Ohashi}}, \bibinfo
  {author} {\bibfnamefont {K.}~\bibnamefont {Kuroda}}, \bibinfo {author}
  {\bibfnamefont {K.}~\bibnamefont {Yamamoto}}, \bibinfo {author}
  {\bibfnamefont {M.}~\bibnamefont {Tokunari}}, \bibinfo {author}
  {\bibfnamefont {T.}~\bibnamefont {Akutsu}}, \bibinfo {author} {\bibfnamefont
  {S.}~\bibnamefont {Kamagasako}}, \bibinfo {author} {\bibfnamefont
  {N.}~\bibnamefont {Nakagawa}}, \bibinfo {author} {\bibfnamefont
  {H.}~\bibnamefont {Kirihara}}, \bibinfo {author} {\bibfnamefont
  {K.}~\bibnamefont {Agatsuma}}, \bibinfo {author} {\bibfnamefont
  {H.}~\bibnamefont {Ishitsuka}}, \bibinfo {author} {\bibfnamefont
  {D.}~\bibnamefont {Tatsumi}}, \bibinfo {author} {\bibfnamefont
  {S.}~\bibnamefont {Telada}}, \bibinfo {author} {\bibfnamefont
  {M.}~\bibnamefont {Ando}}, \bibinfo {author} {\bibfnamefont {T.}~\bibnamefont
  {Tomaru}}, \bibinfo {author} {\bibfnamefont {T.}~\bibnamefont {Suzuki}},
  \bibinfo {author} {\bibfnamefont {N.}~\bibnamefont {Sato}}, \bibinfo {author}
  {\bibfnamefont {T.}~\bibnamefont {Haruyama}}, \bibinfo {author}
  {\bibfnamefont {A.}~\bibnamefont {Yamamoto}},\ and\ \bibinfo {author}
  {\bibfnamefont {T.}~\bibnamefont {Shintomi}},\ }\bibfield  {title} {\bibinfo
  {title} {Cryogenic systems of the cryogenic laser interferometer
  observatory},\ }\href {https://doi.org/10.1088/1742-6596/32/1/038} {\bibfield
   {journal} {\bibinfo  {journal} {Journal of Physics: Conference Series}\
  }\textbf {\bibinfo {volume} {32}},\ \bibinfo {pages} {259} (\bibinfo {year}
  {2006})}\BibitemShut {NoStop}%
\bibitem [{\citenamefont {Yamamoto}\ \emph {et~al.}(2008)\citenamefont
  {Yamamoto}, \citenamefont {Uchiyama}, \citenamefont {Miyoki}, \citenamefont
  {Ohashi}, \citenamefont {Kuroda}, \citenamefont {Ishitsuka}, \citenamefont
  {Akutsu}, \citenamefont {Telada}, \citenamefont {Tomaru}, \citenamefont
  {Suzuki}, \citenamefont {Sato}, \citenamefont {Saito}, \citenamefont
  {Higashi}, \citenamefont {Haruyama}, \citenamefont {Yamamoto}, \citenamefont
  {Shintomi}, \citenamefont {Tatsumi}, \citenamefont {Ando}, \citenamefont
  {Tagoshi}, \citenamefont {Kanda}, \citenamefont {Awaya}, \citenamefont
  {Yamagishi}, \citenamefont {Takahashi}, \citenamefont {Araya}, \citenamefont
  {Takamori}, \citenamefont {Takemoto}, \citenamefont {Higashi}, \citenamefont
  {Hayakawa}, \citenamefont {Morii},\ and\ \citenamefont
  {Akamatsu}}]{Yamamoto_2008}%
  \BibitemOpen
  \bibfield  {author} {\bibinfo {author} {\bibfnamefont {K.}~\bibnamefont
  {Yamamoto}}, \bibinfo {author} {\bibfnamefont {T.}~\bibnamefont {Uchiyama}},
  \bibinfo {author} {\bibfnamefont {S.}~\bibnamefont {Miyoki}}, \bibinfo
  {author} {\bibfnamefont {M.}~\bibnamefont {Ohashi}}, \bibinfo {author}
  {\bibfnamefont {K.}~\bibnamefont {Kuroda}}, \bibinfo {author} {\bibfnamefont
  {H.}~\bibnamefont {Ishitsuka}}, \bibinfo {author} {\bibfnamefont
  {T.}~\bibnamefont {Akutsu}}, \bibinfo {author} {\bibfnamefont
  {S.}~\bibnamefont {Telada}}, \bibinfo {author} {\bibfnamefont
  {T.}~\bibnamefont {Tomaru}}, \bibinfo {author} {\bibfnamefont
  {T.}~\bibnamefont {Suzuki}}, \bibinfo {author} {\bibfnamefont
  {N.}~\bibnamefont {Sato}}, \bibinfo {author} {\bibfnamefont {Y.}~\bibnamefont
  {Saito}}, \bibinfo {author} {\bibfnamefont {Y.}~\bibnamefont {Higashi}},
  \bibinfo {author} {\bibfnamefont {T.}~\bibnamefont {Haruyama}}, \bibinfo
  {author} {\bibfnamefont {A.}~\bibnamefont {Yamamoto}}, \bibinfo {author}
  {\bibfnamefont {T.}~\bibnamefont {Shintomi}}, \bibinfo {author}
  {\bibfnamefont {D.}~\bibnamefont {Tatsumi}}, \bibinfo {author} {\bibfnamefont
  {M.}~\bibnamefont {Ando}}, \bibinfo {author} {\bibfnamefont {H.}~\bibnamefont
  {Tagoshi}}, \bibinfo {author} {\bibfnamefont {N.}~\bibnamefont {Kanda}},
  \bibinfo {author} {\bibfnamefont {N.}~\bibnamefont {Awaya}}, \bibinfo
  {author} {\bibfnamefont {S.}~\bibnamefont {Yamagishi}}, \bibinfo {author}
  {\bibfnamefont {H.}~\bibnamefont {Takahashi}}, \bibinfo {author}
  {\bibfnamefont {A.}~\bibnamefont {Araya}}, \bibinfo {author} {\bibfnamefont
  {A.}~\bibnamefont {Takamori}}, \bibinfo {author} {\bibfnamefont
  {S.}~\bibnamefont {Takemoto}}, \bibinfo {author} {\bibfnamefont
  {T.}~\bibnamefont {Higashi}}, \bibinfo {author} {\bibfnamefont
  {H.}~\bibnamefont {Hayakawa}}, \bibinfo {author} {\bibfnamefont
  {W.}~\bibnamefont {Morii}},\ and\ \bibinfo {author} {\bibfnamefont
  {J.}~\bibnamefont {Akamatsu}},\ }\bibfield  {title} {\bibinfo {title}
  {Current status of the {CLIO} project},\ }\href
  {https://doi.org/10.1088/1742-6596/122/1/012002} {\bibfield  {journal}
  {\bibinfo  {journal} {Journal of Physics: Conference Series}\ }\textbf
  {\bibinfo {volume} {122}},\ \bibinfo {pages} {012002} (\bibinfo {year}
  {2008})}\BibitemShut {NoStop}%
\bibitem [{\citenamefont {Willke}\ \emph {et~al.}(2002)\citenamefont {Willke},
  \citenamefont {Aufmuth}, \citenamefont {Aulbert}, \citenamefont {Babak},
  \citenamefont {Balasubramanian}, \citenamefont {Barr}, \citenamefont
  {Berukoff}, \citenamefont {Bose}, \citenamefont {Cagnoli}, \citenamefont
  {Casey}, \citenamefont {Churches}, \citenamefont {Clubley}, \citenamefont
  {Colacino}, \citenamefont {Crooks}, \citenamefont {Cutler}, \citenamefont
  {Danzmann}, \citenamefont {Davies}, \citenamefont {Dupuis}, \citenamefont
  {Elliffe}, \citenamefont {Fallnich}, \citenamefont {Freise}, \citenamefont
  {Goßler}, \citenamefont {Grant}, \citenamefont {Grote}, \citenamefont
  {Heinzel}, \citenamefont {Heptonstall}, \citenamefont {Heurs}, \citenamefont
  {Hewitson}, \citenamefont {Hough}, \citenamefont {Jennrich}, \citenamefont
  {Kawabe}, \citenamefont {Kötter}, \citenamefont {Leonhardt}, \citenamefont
  {Lück}, \citenamefont {Malec}, \citenamefont {McNamara}, \citenamefont
  {McIntosh}, \citenamefont {Mossavi}, \citenamefont {Mohanty}, \citenamefont
  {Mukherjee}, \citenamefont {Nagano}, \citenamefont {Newton}, \citenamefont
  {Owen}, \citenamefont {Palmer}, \citenamefont {Papa}, \citenamefont {Plissi},
  \citenamefont {Quetschke}, \citenamefont {Robertson}, \citenamefont
  {Robertson}, \citenamefont {Rowan}, \citenamefont {Rüdiger}, \citenamefont
  {Sathyaprakash}, \citenamefont {Schilling}, \citenamefont {Schutz},
  \citenamefont {Senior}, \citenamefont {Sintes}, \citenamefont {Skeldon},
  \citenamefont {Sneddon}, \citenamefont {Stief}, \citenamefont {Strain},
  \citenamefont {Taylor}, \citenamefont {Torrie}, \citenamefont {Vecchio},
  \citenamefont {Ward}, \citenamefont {Weiland}, \citenamefont {Welling},
  \citenamefont {Williams}, \citenamefont {Winkler}, \citenamefont {Woan},\
  and\ \citenamefont {Zawischa}}]{B_Willke_2002}%
  \BibitemOpen
  \bibfield  {author} {\bibinfo {author} {\bibfnamefont {B.}~\bibnamefont
  {Willke}}, \bibinfo {author} {\bibfnamefont {P.}~\bibnamefont {Aufmuth}},
  \bibinfo {author} {\bibfnamefont {C.}~\bibnamefont {Aulbert}}, \bibinfo
  {author} {\bibfnamefont {S.}~\bibnamefont {Babak}}, \bibinfo {author}
  {\bibfnamefont {R.}~\bibnamefont {Balasubramanian}}, \bibinfo {author}
  {\bibfnamefont {B.~W.}\ \bibnamefont {Barr}}, \bibinfo {author}
  {\bibfnamefont {S.}~\bibnamefont {Berukoff}}, \bibinfo {author}
  {\bibfnamefont {S.}~\bibnamefont {Bose}}, \bibinfo {author} {\bibfnamefont
  {G.}~\bibnamefont {Cagnoli}}, \bibinfo {author} {\bibfnamefont {M.~M.}\
  \bibnamefont {Casey}}, \bibinfo {author} {\bibfnamefont {D.}~\bibnamefont
  {Churches}}, \bibinfo {author} {\bibfnamefont {D.}~\bibnamefont {Clubley}},
  \bibinfo {author} {\bibfnamefont {C.~N.}\ \bibnamefont {Colacino}}, \bibinfo
  {author} {\bibfnamefont {D.~R.~M.}\ \bibnamefont {Crooks}}, \bibinfo {author}
  {\bibfnamefont {C.}~\bibnamefont {Cutler}}, \bibinfo {author} {\bibfnamefont
  {K.}~\bibnamefont {Danzmann}}, \bibinfo {author} {\bibfnamefont
  {R.}~\bibnamefont {Davies}}, \bibinfo {author} {\bibfnamefont
  {R.}~\bibnamefont {Dupuis}}, \bibinfo {author} {\bibfnamefont
  {E.}~\bibnamefont {Elliffe}}, \bibinfo {author} {\bibfnamefont
  {C.}~\bibnamefont {Fallnich}}, \bibinfo {author} {\bibfnamefont
  {A.}~\bibnamefont {Freise}}, \bibinfo {author} {\bibfnamefont
  {S.}~\bibnamefont {Goßler}}, \bibinfo {author} {\bibfnamefont
  {A.}~\bibnamefont {Grant}}, \bibinfo {author} {\bibfnamefont
  {H.}~\bibnamefont {Grote}}, \bibinfo {author} {\bibfnamefont
  {G.}~\bibnamefont {Heinzel}}, \bibinfo {author} {\bibfnamefont
  {A.}~\bibnamefont {Heptonstall}}, \bibinfo {author} {\bibfnamefont
  {M.}~\bibnamefont {Heurs}}, \bibinfo {author} {\bibfnamefont
  {M.}~\bibnamefont {Hewitson}}, \bibinfo {author} {\bibfnamefont
  {J.}~\bibnamefont {Hough}}, \bibinfo {author} {\bibfnamefont
  {O.}~\bibnamefont {Jennrich}}, \bibinfo {author} {\bibfnamefont
  {K.}~\bibnamefont {Kawabe}}, \bibinfo {author} {\bibfnamefont
  {K.}~\bibnamefont {Kötter}}, \bibinfo {author} {\bibfnamefont
  {V.}~\bibnamefont {Leonhardt}}, \bibinfo {author} {\bibfnamefont
  {H.}~\bibnamefont {Lück}}, \bibinfo {author} {\bibfnamefont
  {M.}~\bibnamefont {Malec}}, \bibinfo {author} {\bibfnamefont {P.~W.}\
  \bibnamefont {McNamara}}, \bibinfo {author} {\bibfnamefont {S.~A.}\
  \bibnamefont {McIntosh}}, \bibinfo {author} {\bibfnamefont {K.}~\bibnamefont
  {Mossavi}}, \bibinfo {author} {\bibfnamefont {S.}~\bibnamefont {Mohanty}},
  \bibinfo {author} {\bibfnamefont {S.}~\bibnamefont {Mukherjee}}, \bibinfo
  {author} {\bibfnamefont {S.}~\bibnamefont {Nagano}}, \bibinfo {author}
  {\bibfnamefont {G.~P.}\ \bibnamefont {Newton}}, \bibinfo {author}
  {\bibfnamefont {B.~J.}\ \bibnamefont {Owen}}, \bibinfo {author}
  {\bibfnamefont {D.}~\bibnamefont {Palmer}}, \bibinfo {author} {\bibfnamefont
  {M.~A.}\ \bibnamefont {Papa}}, \bibinfo {author} {\bibfnamefont {M.~V.}\
  \bibnamefont {Plissi}}, \bibinfo {author} {\bibfnamefont {V.}~\bibnamefont
  {Quetschke}}, \bibinfo {author} {\bibfnamefont {D.~I.}\ \bibnamefont
  {Robertson}}, \bibinfo {author} {\bibfnamefont {N.~A.}\ \bibnamefont
  {Robertson}}, \bibinfo {author} {\bibfnamefont {S.}~\bibnamefont {Rowan}},
  \bibinfo {author} {\bibfnamefont {A.}~\bibnamefont {Rüdiger}}, \bibinfo
  {author} {\bibfnamefont {B.~S.}\ \bibnamefont {Sathyaprakash}}, \bibinfo
  {author} {\bibfnamefont {R.}~\bibnamefont {Schilling}}, \bibinfo {author}
  {\bibfnamefont {B.~F.}\ \bibnamefont {Schutz}}, \bibinfo {author}
  {\bibfnamefont {R.}~\bibnamefont {Senior}}, \bibinfo {author} {\bibfnamefont
  {A.~M.}\ \bibnamefont {Sintes}}, \bibinfo {author} {\bibfnamefont {K.~D.}\
  \bibnamefont {Skeldon}}, \bibinfo {author} {\bibfnamefont {P.}~\bibnamefont
  {Sneddon}}, \bibinfo {author} {\bibfnamefont {F.}~\bibnamefont {Stief}},
  \bibinfo {author} {\bibfnamefont {K.~A.}\ \bibnamefont {Strain}}, \bibinfo
  {author} {\bibfnamefont {I.}~\bibnamefont {Taylor}}, \bibinfo {author}
  {\bibfnamefont {C.~I.}\ \bibnamefont {Torrie}}, \bibinfo {author}
  {\bibfnamefont {A.}~\bibnamefont {Vecchio}}, \bibinfo {author} {\bibfnamefont
  {H.}~\bibnamefont {Ward}}, \bibinfo {author} {\bibfnamefont {U.}~\bibnamefont
  {Weiland}}, \bibinfo {author} {\bibfnamefont {H.}~\bibnamefont {Welling}},
  \bibinfo {author} {\bibfnamefont {P.}~\bibnamefont {Williams}}, \bibinfo
  {author} {\bibfnamefont {W.}~\bibnamefont {Winkler}}, \bibinfo {author}
  {\bibfnamefont {G.}~\bibnamefont {Woan}},\ and\ \bibinfo {author}
  {\bibfnamefont {I.}~\bibnamefont {Zawischa}},\ }\bibfield  {title} {\bibinfo
  {title} {The geo 600 gravitational wave detector},\ }\href
  {https://doi.org/10.1088/0264-9381/19/7/321} {\bibfield  {journal} {\bibinfo
  {journal} {Classical and Quantum Gravity}\ }\textbf {\bibinfo {volume}
  {19}},\ \bibinfo {pages} {1377} (\bibinfo {year} {2002})}\BibitemShut
  {NoStop}%
\bibitem [{\citenamefont {Gossler}\ \emph {et~al.}(2002)\citenamefont
  {Gossler}, \citenamefont {Casey}, \citenamefont {Freise}, \citenamefont
  {Grote}, \citenamefont {Lück}, \citenamefont {McNamara}, \citenamefont
  {Plissi}, \citenamefont {Robertson}, \citenamefont {Robertson}, \citenamefont
  {Skeldon}, \citenamefont {Strain}, \citenamefont {Torrie}, \citenamefont
  {Ward}, \citenamefont {Willke}, \citenamefont {Hough},\ and\ \citenamefont
  {Danzmann}}]{S_Gossler_2002}%
  \BibitemOpen
  \bibfield  {author} {\bibinfo {author} {\bibfnamefont {S.}~\bibnamefont
  {Gossler}}, \bibinfo {author} {\bibfnamefont {M.~M.}\ \bibnamefont {Casey}},
  \bibinfo {author} {\bibfnamefont {A.}~\bibnamefont {Freise}}, \bibinfo
  {author} {\bibfnamefont {H.}~\bibnamefont {Grote}}, \bibinfo {author}
  {\bibfnamefont {H.}~\bibnamefont {Lück}}, \bibinfo {author} {\bibfnamefont
  {P.}~\bibnamefont {McNamara}}, \bibinfo {author} {\bibfnamefont {M.~V.}\
  \bibnamefont {Plissi}}, \bibinfo {author} {\bibfnamefont {D.~I.}\
  \bibnamefont {Robertson}}, \bibinfo {author} {\bibfnamefont {N.~A.}\
  \bibnamefont {Robertson}}, \bibinfo {author} {\bibfnamefont {K.}~\bibnamefont
  {Skeldon}}, \bibinfo {author} {\bibfnamefont {K.~A.}\ \bibnamefont {Strain}},
  \bibinfo {author} {\bibfnamefont {C.~I.}\ \bibnamefont {Torrie}}, \bibinfo
  {author} {\bibfnamefont {H.}~\bibnamefont {Ward}}, \bibinfo {author}
  {\bibfnamefont {B.}~\bibnamefont {Willke}}, \bibinfo {author} {\bibfnamefont
  {J.}~\bibnamefont {Hough}},\ and\ \bibinfo {author} {\bibfnamefont
  {K.}~\bibnamefont {Danzmann}},\ }\bibfield  {title} {\bibinfo {title} {The
  modecleaner system and suspension aspects of geo 600},\ }\href
  {https://doi.org/10.1088/0264-9381/19/7/382} {\bibfield  {journal} {\bibinfo
  {journal} {Classical and Quantum Gravity}\ }\textbf {\bibinfo {volume}
  {19}},\ \bibinfo {pages} {1835} (\bibinfo {year} {2002})}\BibitemShut
  {NoStop}%
\bibitem [{\citenamefont {{Abramovici}}\ \emph {et~al.}(1992)\citenamefont
  {{Abramovici}}, \citenamefont {{Althouse}},\ and\ \citenamefont
  {{Drever}}}]{1992Sci...256..325A}%
  \BibitemOpen
  \bibfield  {author} {\bibinfo {author} {\bibfnamefont {A.}~\bibnamefont
  {{Abramovici}}}, \bibinfo {author} {\bibfnamefont {W.~E.}\ \bibnamefont
  {{Althouse}}},\ and\ \bibinfo {author} {\bibfnamefont {R.~W.~P.~{\it et
  al.}.}\ \bibnamefont {{Drever}}},\ }\bibfield  {title} {\bibinfo {title}
  {{LIGO - The Laser Interferometer Gravitational-Wave Observatory}},\
  }\href@noop {} {\bibfield  {journal} {\bibinfo  {journal} {Science}\ }\textbf
  {\bibinfo {volume} {256}},\ \bibinfo {pages} {325} (\bibinfo {year}
  {1992})}\BibitemShut {NoStop}%
\bibitem [{\citenamefont {{Takahashi}}\ and\ \citenamefont {{TAMA
  Collaboration}}(2003)}]{2003CQGra..20S.593T}%
  \BibitemOpen
  \bibfield  {author} {\bibinfo {author} {\bibfnamefont {R.}~\bibnamefont
  {{Takahashi}}}\ and\ \bibinfo {author} {\bibnamefont {{TAMA
  Collaboration}}},\ }\bibfield  {title} {\bibinfo {title} {{Operational status
  of TAMA300}},\ }\href {https://doi.org/10.1088/0264-9381/20/17/302}
  {\bibfield  {journal} {\bibinfo  {journal} {Classical and Quantum Gravity}\
  }\textbf {\bibinfo {volume} {20}},\ \bibinfo {pages} {S593} (\bibinfo {year}
  {2003})}\BibitemShut {NoStop}%
\bibitem [{\citenamefont {Caron}\ \emph {et~al.}(1996)\citenamefont {Caron},
  \citenamefont {Dominjon}, \citenamefont {Drezen}, \citenamefont {Flaminio},
  \citenamefont {Grave}, \citenamefont {Marion}, \citenamefont {Massonnet},
  \citenamefont {Mehmel}, \citenamefont {Morand}, \citenamefont {Mours},
  \citenamefont {Yvert}, \citenamefont {Babusci}, \citenamefont {Giordano},
  \citenamefont {Matone}, \citenamefont {Mackowski}, \citenamefont
  {Napolitano}, \citenamefont {Pinard}, \citenamefont {Dognin}, \citenamefont
  {Barone}, \citenamefont {Calloni}, \citenamefont {{Di Fiore}}, \citenamefont
  {Flagiello}, \citenamefont {Grado}, \citenamefont {Longo}, \citenamefont
  {Lops}, \citenamefont {Marano}, \citenamefont {Milano}, \citenamefont
  {Russo}, \citenamefont {Solimeno}, \citenamefont {Acker}, \citenamefont
  {Brillet}, \citenamefont {Bondu}, \citenamefont {Brisson}, \citenamefont
  {Cavalier}, \citenamefont {Heitmann}, \citenamefont {Hello}, \citenamefont
  {Jacquemet}, \citenamefont {Latrach}, \citenamefont {{Le Diberder}},
  \citenamefont {Man}, \citenamefont {Manh}, \citenamefont {Taubmann},
  \citenamefont {Vinet}, \citenamefont {Boccara}, \citenamefont {Gleyzes},
  \citenamefont {Roger}, \citenamefont {Loriette}, \citenamefont {Cagnoli},
  \citenamefont {Gammaitoni}, \citenamefont {Kovalik}, \citenamefont
  {Marchesoni}, \citenamefont {Punturo}, \citenamefont {Barsuglia},
  \citenamefont {Bernardini}, \citenamefont {Braccini}, \citenamefont
  {Bradaschia}, \citenamefont {{Del Fabbro}}, \citenamefont {{De Salvo}},
  \citenamefont {{Di Virgilio}}, \citenamefont {Ferrante}, \citenamefont
  {Fidecaro}, \citenamefont {Giassi}, \citenamefont {Giazotto}, \citenamefont
  {Gorini}, \citenamefont {Holloway}, \citenamefont {Lami}, \citenamefont
  {Lapenna}, \citenamefont {Losurdo}, \citenamefont {Mancini}, \citenamefont
  {Morganti}, \citenamefont {Palla}, \citenamefont {Pan}, \citenamefont
  {Passuello}, \citenamefont {Poggiani}, \citenamefont {Torelli}, \citenamefont
  {Zhang}, \citenamefont {Majorana}, \citenamefont {Puppo}, \citenamefont
  {Rapagnani},\ and\ \citenamefont {Ricci}}]{CARON1996107}%
  \BibitemOpen
  \bibfield  {author} {\bibinfo {author} {\bibfnamefont {B.}~\bibnamefont
  {Caron}}, \bibinfo {author} {\bibfnamefont {A.}~\bibnamefont {Dominjon}},
  \bibinfo {author} {\bibfnamefont {C.}~\bibnamefont {Drezen}}, \bibinfo
  {author} {\bibfnamefont {R.}~\bibnamefont {Flaminio}}, \bibinfo {author}
  {\bibfnamefont {X.}~\bibnamefont {Grave}}, \bibinfo {author} {\bibfnamefont
  {F.}~\bibnamefont {Marion}}, \bibinfo {author} {\bibfnamefont
  {L.}~\bibnamefont {Massonnet}}, \bibinfo {author} {\bibfnamefont
  {C.}~\bibnamefont {Mehmel}}, \bibinfo {author} {\bibfnamefont
  {R.}~\bibnamefont {Morand}}, \bibinfo {author} {\bibfnamefont
  {B.}~\bibnamefont {Mours}}, \bibinfo {author} {\bibfnamefont
  {M.}~\bibnamefont {Yvert}}, \bibinfo {author} {\bibfnamefont
  {D.}~\bibnamefont {Babusci}}, \bibinfo {author} {\bibfnamefont
  {G.}~\bibnamefont {Giordano}}, \bibinfo {author} {\bibfnamefont
  {G.}~\bibnamefont {Matone}}, \bibinfo {author} {\bibfnamefont {J.-M.}\
  \bibnamefont {Mackowski}}, \bibinfo {author} {\bibfnamefont {M.}~\bibnamefont
  {Napolitano}}, \bibinfo {author} {\bibfnamefont {L.}~\bibnamefont {Pinard}},
  \bibinfo {author} {\bibfnamefont {L.}~\bibnamefont {Dognin}}, \bibinfo
  {author} {\bibfnamefont {F.}~\bibnamefont {Barone}}, \bibinfo {author}
  {\bibfnamefont {E.}~\bibnamefont {Calloni}}, \bibinfo {author} {\bibfnamefont
  {L.}~\bibnamefont {{Di Fiore}}}, \bibinfo {author} {\bibfnamefont
  {M.}~\bibnamefont {Flagiello}}, \bibinfo {author} {\bibfnamefont
  {A.}~\bibnamefont {Grado}}, \bibinfo {author} {\bibfnamefont
  {M.}~\bibnamefont {Longo}}, \bibinfo {author} {\bibfnamefont
  {M.}~\bibnamefont {Lops}}, \bibinfo {author} {\bibfnamefont {S.}~\bibnamefont
  {Marano}}, \bibinfo {author} {\bibfnamefont {L.}~\bibnamefont {Milano}},
  \bibinfo {author} {\bibfnamefont {G.}~\bibnamefont {Russo}}, \bibinfo
  {author} {\bibfnamefont {S.}~\bibnamefont {Solimeno}}, \bibinfo {author}
  {\bibfnamefont {Y.}~\bibnamefont {Acker}}, \bibinfo {author} {\bibfnamefont
  {A.}~\bibnamefont {Brillet}}, \bibinfo {author} {\bibfnamefont
  {F.}~\bibnamefont {Bondu}}, \bibinfo {author} {\bibfnamefont
  {V.}~\bibnamefont {Brisson}}, \bibinfo {author} {\bibfnamefont
  {F.}~\bibnamefont {Cavalier}}, \bibinfo {author} {\bibfnamefont {M.~D.~H.}\
  \bibnamefont {Heitmann}}, \bibinfo {author} {\bibfnamefont {P.}~\bibnamefont
  {Hello}}, \bibinfo {author} {\bibfnamefont {M.}~\bibnamefont {Jacquemet}},
  \bibinfo {author} {\bibfnamefont {L.}~\bibnamefont {Latrach}}, \bibinfo
  {author} {\bibfnamefont {F.}~\bibnamefont {{Le Diberder}}}, \bibinfo {author}
  {\bibfnamefont {C.}~\bibnamefont {Man}}, \bibinfo {author} {\bibfnamefont
  {P.}~\bibnamefont {Manh}}, \bibinfo {author} {\bibfnamefont {M.}~\bibnamefont
  {Taubmann}}, \bibinfo {author} {\bibfnamefont {J.-Y.}\ \bibnamefont {Vinet}},
  \bibinfo {author} {\bibfnamefont {C.}~\bibnamefont {Boccara}}, \bibinfo
  {author} {\bibfnamefont {P.}~\bibnamefont {Gleyzes}}, \bibinfo {author}
  {\bibfnamefont {J.-P.}\ \bibnamefont {Roger}}, \bibinfo {author}
  {\bibfnamefont {V.}~\bibnamefont {Loriette}}, \bibinfo {author}
  {\bibfnamefont {G.}~\bibnamefont {Cagnoli}}, \bibinfo {author} {\bibfnamefont
  {L.}~\bibnamefont {Gammaitoni}}, \bibinfo {author} {\bibfnamefont
  {J.}~\bibnamefont {Kovalik}}, \bibinfo {author} {\bibfnamefont
  {F.}~\bibnamefont {Marchesoni}}, \bibinfo {author} {\bibfnamefont
  {M.}~\bibnamefont {Punturo}}, \bibinfo {author} {\bibfnamefont
  {M.}~\bibnamefont {Barsuglia}}, \bibinfo {author} {\bibfnamefont
  {M.}~\bibnamefont {Bernardini}}, \bibinfo {author} {\bibfnamefont
  {S.}~\bibnamefont {Braccini}}, \bibinfo {author} {\bibfnamefont
  {C.}~\bibnamefont {Bradaschia}}, \bibinfo {author} {\bibfnamefont
  {R.}~\bibnamefont {{Del Fabbro}}}, \bibinfo {author} {\bibfnamefont
  {R.}~\bibnamefont {{De Salvo}}}, \bibinfo {author} {\bibfnamefont
  {A.}~\bibnamefont {{Di Virgilio}}}, \bibinfo {author} {\bibfnamefont
  {I.}~\bibnamefont {Ferrante}}, \bibinfo {author} {\bibfnamefont
  {F.}~\bibnamefont {Fidecaro}}, \bibinfo {author} {\bibfnamefont
  {A.}~\bibnamefont {Giassi}}, \bibinfo {author} {\bibfnamefont
  {A.}~\bibnamefont {Giazotto}}, \bibinfo {author} {\bibfnamefont
  {G.}~\bibnamefont {Gorini}}, \bibinfo {author} {\bibfnamefont
  {L.}~\bibnamefont {Holloway}}, \bibinfo {author} {\bibfnamefont
  {S.}~\bibnamefont {Lami}}, \bibinfo {author} {\bibfnamefont {P.}~\bibnamefont
  {Lapenna}}, \bibinfo {author} {\bibfnamefont {G.}~\bibnamefont {Losurdo}},
  \bibinfo {author} {\bibfnamefont {S.}~\bibnamefont {Mancini}}, \bibinfo
  {author} {\bibfnamefont {M.}~\bibnamefont {Morganti}}, \bibinfo {author}
  {\bibfnamefont {F.}~\bibnamefont {Palla}}, \bibinfo {author} {\bibfnamefont
  {H.-B.}\ \bibnamefont {Pan}}, \bibinfo {author} {\bibfnamefont
  {D.}~\bibnamefont {Passuello}}, \bibinfo {author} {\bibfnamefont
  {R.}~\bibnamefont {Poggiani}}, \bibinfo {author} {\bibfnamefont
  {G.}~\bibnamefont {Torelli}}, \bibinfo {author} {\bibfnamefont
  {Z.}~\bibnamefont {Zhang}}, \bibinfo {author} {\bibfnamefont
  {E.}~\bibnamefont {Majorana}}, \bibinfo {author} {\bibfnamefont
  {P.}~\bibnamefont {Puppo}}, \bibinfo {author} {\bibfnamefont
  {P.}~\bibnamefont {Rapagnani}},\ and\ \bibinfo {author} {\bibfnamefont
  {F.}~\bibnamefont {Ricci}},\ }\bibfield  {title} {\bibinfo {title} {Status of
  the virgo experiment},\ }\href
  {https://doi.org/https://doi.org/10.1016/0920-5632(96)00220-4} {\bibfield
  {journal} {\bibinfo  {journal} {Nuclear Physics B - Proceedings Supplements}\
  }\textbf {\bibinfo {volume} {48}},\ \bibinfo {pages} {107} (\bibinfo {year}
  {1996})},\ \bibinfo {note} {proceedings of the Fourth International Workshop
  on Theoretical and Phenomenological Aspects of Underground
  Physics}\BibitemShut {NoStop}%
\bibitem [{\citenamefont {Dupuis}\ and\ \citenamefont {(forthe LIGO
  Scientific~collaboration)}(2006)}]{Rejean_J_Dupuis_2006}%
  \BibitemOpen
  \bibfield  {author} {\bibinfo {author} {\bibfnamefont {R.~J.}\ \bibnamefont
  {Dupuis}}\ and\ \bibinfo {author} {\bibnamefont {(forthe LIGO
  Scientific~collaboration)}},\ }\bibfield  {title} {\bibinfo {title} {Targeted
  searches for gravitational waves from radio pulsars},\ }\href
  {https://doi.org/10.1088/1742-6596/32/1/009} {\bibfield  {journal} {\bibinfo
  {journal} {Journal of Physics: Conference Series}\ }\textbf {\bibinfo
  {volume} {32}},\ \bibinfo {pages} {52} (\bibinfo {year} {2006})}\BibitemShut
  {NoStop}%
\bibitem [{\citenamefont {{Sintes}}\ and\ \citenamefont {{LIGO Scientific
  Collaboration}}(2006)}]{2006JPhCS..39...36S}%
  \BibitemOpen
  \bibfield  {author} {\bibinfo {author} {\bibfnamefont {A.~M.}\ \bibnamefont
  {{Sintes}}}\ and\ \bibinfo {author} {\bibnamefont {{LIGO Scientific
  Collaboration}}},\ }\bibfield  {title} {\bibinfo {title} {{Recent results on
  the search for continuous sources with LIGO and GEO 600}},\ }in\ \href
  {https://doi.org/10.1088/1742-6596/39/1/008} {\emph {\bibinfo {booktitle}
  {Journal of Physics Conference Series}}},\ \bibinfo {series} {Journal of
  Physics Conference Series}, Vol.~\bibinfo {volume} {39}\ (\bibinfo {year}
  {2006})\ pp.\ \bibinfo {pages} {36--38},\ \Eprint
  {https://arxiv.org/abs/gr-qc/0511152} {arXiv:gr-qc/0511152 [gr-qc]}
  \BibitemShut {NoStop}%
\bibitem [{\citenamefont {Abbott}\ \emph {et~al.}(2007)\citenamefont {Abbott},
  \citenamefont {Abbott}, \citenamefont {Adhikari}, \citenamefont {Agresti},
  \citenamefont {Ajith}, \citenamefont {Allen}, \citenamefont {Amin},
  \citenamefont {Anderson}, \citenamefont {Anderson}, \citenamefont {Arain},
  \citenamefont {Araya}, \citenamefont {Armandula}, \citenamefont {Ashley},
  \citenamefont {Aston}, \citenamefont {Aufmuth}, \citenamefont {Aulbert},
  \citenamefont {Babak}, \citenamefont {Ballmer}, \citenamefont {Bantilan},
  \citenamefont {Barish}, \citenamefont {Barker}, \citenamefont {Barker},
  \citenamefont {Barr}, \citenamefont {Barriga}, \citenamefont {Barton},
  \citenamefont {Bayer}, \citenamefont {Belczynski}, \citenamefont
  {Betzwieser}, \citenamefont {Beyersdorf}, \citenamefont {Bhawal},
  \citenamefont {Bilenko}, \citenamefont {Billingsley}, \citenamefont {Biswas},
  \citenamefont {Black}, \citenamefont {Blackburn}, \citenamefont {Blackburn},
  \citenamefont {Blair}, \citenamefont {Bland}, \citenamefont {Bogenstahl},
  \citenamefont {Bogue}, \citenamefont {Bork}, \citenamefont {Boschi},
  \citenamefont {Bose}, \citenamefont {Brady}, \citenamefont {Braginsky},
  \citenamefont {Brau}, \citenamefont {Brinkmann}, \citenamefont {Brooks},
  \citenamefont {Brown}, \citenamefont {Bullington}, \citenamefont {Bunkowski},
  \citenamefont {Buonanno}, \citenamefont {Burmeister}, \citenamefont {Busby},
  \citenamefont {Butler}, \citenamefont {Byer}, \citenamefont {Cadonati},
  \citenamefont {Cagnoli}, \citenamefont {Camp}, \citenamefont {Cannizzo},
  \citenamefont {Cannon}, \citenamefont {Cantley}, \citenamefont {Cao},
  \citenamefont {Cardenas}, \citenamefont {Carter}, \citenamefont {Casey},
  \citenamefont {Castaldi}, \citenamefont {Cepeda}, \citenamefont {Chalkey},
  \citenamefont {Charlton}, \citenamefont {Chatterji}, \citenamefont
  {Chelkowski}, \citenamefont {Chen}, \citenamefont {Chiadini}, \citenamefont
  {Chin}, \citenamefont {Chin}, \citenamefont {Chow}, \citenamefont
  {Christensen}, \citenamefont {Clark}, \citenamefont {Cochrane}, \citenamefont
  {Cokelaer}, \citenamefont {Colacino}, \citenamefont {Coldwell}, \citenamefont
  {Conte}, \citenamefont {Cook}, \citenamefont {Corbitt}, \citenamefont
  {Coward}, \citenamefont {Coyne}, \citenamefont {Creighton}, \citenamefont
  {Creighton}, \citenamefont {Croce}, \citenamefont {Crooks}, \citenamefont
  {Cruise}, \citenamefont {Cumming}, \citenamefont {Dalrymple}, \citenamefont
  {D'Ambrosio}, \citenamefont {Danzmann}, \citenamefont {Davies}, \citenamefont
  {DeBra}, \citenamefont {Degallaix}, \citenamefont {Degree}, \citenamefont
  {Demma}, \citenamefont {Dergachev}, \citenamefont {Desai}, \citenamefont
  {DeSalvo}, \citenamefont {Dhurandhar}, \citenamefont {D\'{\i}az},
  \citenamefont {Dickson}, \citenamefont {Di~Credico}, \citenamefont
  {Diederichs}, \citenamefont {Dietz}, \citenamefont {Doomes}, \citenamefont
  {Drever}, \citenamefont {Dumas}, \citenamefont {Dupuis}, \citenamefont
  {Dwyer}, \citenamefont {Ehrens}, \citenamefont {Espinoza}, \citenamefont
  {Etzel}, \citenamefont {Evans}, \citenamefont {Evans}, \citenamefont
  {Fairhurst}, \citenamefont {Fan}, \citenamefont {Fazi}, \citenamefont
  {Fejer}, \citenamefont {Finn}, \citenamefont {Fiumara}, \citenamefont
  {Fotopoulos}, \citenamefont {Franzen}, \citenamefont {Franzen}, \citenamefont
  {Freise}, \citenamefont {Frey}, \citenamefont {Fricke}, \citenamefont
  {Fritschel}, \citenamefont {Frolov}, \citenamefont {Fyffe}, \citenamefont
  {Galdi}, \citenamefont {Ganezer}, \citenamefont {Garofoli}, \citenamefont
  {Gholami}, \citenamefont {Giaime}, \citenamefont {Giampanis}, \citenamefont
  {Giardina}, \citenamefont {Goda}, \citenamefont {Goetz}, \citenamefont
  {Goggin}, \citenamefont {Gonz\'alez}, \citenamefont {Gossler}, \citenamefont
  {Grant}, \citenamefont {Gras}, \citenamefont {Gray}, \citenamefont {Gray},
  \citenamefont {Greenhalgh}, \citenamefont {Gretarsson}, \citenamefont
  {Grosso}, \citenamefont {Grote}, \citenamefont {Grunewald}, \citenamefont
  {Guenther}, \citenamefont {Gustafson}, \citenamefont {Hage}, \citenamefont
  {Hammer}, \citenamefont {Hanna}, \citenamefont {Hanson}, \citenamefont
  {Harms}, \citenamefont {Harry}, \citenamefont {Harstad}, \citenamefont
  {Hayler}, \citenamefont {Heefner}, \citenamefont {Heng}, \citenamefont
  {Heptonstall}, \citenamefont {Heurs}, \citenamefont {Hewitson}, \citenamefont
  {Hild}, \citenamefont {Hirose}, \citenamefont {Hoak}, \citenamefont {Hosken},
  \citenamefont {Hough}, \citenamefont {Howell}, \citenamefont {Hoyland},
  \citenamefont {Huttner}, \citenamefont {Ingram}, \citenamefont {Innerhofer},
  \citenamefont {Ito}, \citenamefont {Itoh}, \citenamefont {Ivanov},
  \citenamefont {Jackrel}, \citenamefont {Johnson}, \citenamefont {Johnson},
  \citenamefont {Jones}, \citenamefont {Jones}, \citenamefont {Jones},
  \citenamefont {Ju}, \citenamefont {Kalmus}, \citenamefont {Kalogera},
  \citenamefont {Kasprzyk}, \citenamefont {Katsavounidis}, \citenamefont
  {Kawabe}, \citenamefont {Kawamura}, \citenamefont {Kawazoe}, \citenamefont
  {Kells}, \citenamefont {Keppel}, \citenamefont {Khalili}, \citenamefont
  {Kim}, \citenamefont {King}, \citenamefont {Kissel}, \citenamefont
  {Klimenko}, \citenamefont {Kokeyama}, \citenamefont {Kondrashov},
  \citenamefont {Kopparapu}, \citenamefont {Kozak}, \citenamefont {Krishnan},
  \citenamefont {Kwee}, \citenamefont {Lam}, \citenamefont {Landry},
  \citenamefont {Lantz}, \citenamefont {Lazzarini}, \citenamefont {Lee},
  \citenamefont {Lei}, \citenamefont {Leiner}, \citenamefont {Leonhardt},
  \citenamefont {Leonor}, \citenamefont {Libbrecht}, \citenamefont {Lindquist},
  \citenamefont {Lockerbie}, \citenamefont {Longo}, \citenamefont {Lormand},
  \citenamefont {Lubinski}, \citenamefont {L\"uck}, \citenamefont
  {Machenschalk}, \citenamefont {MacInnis}, \citenamefont {Mageswaran},
  \citenamefont {Mailand}, \citenamefont {Malec}, \citenamefont {Mandic},
  \citenamefont {Marano}, \citenamefont {M\'arka}, \citenamefont {Markowitz},
  \citenamefont {Maros}, \citenamefont {Martin}, \citenamefont {Marx},
  \citenamefont {Mason}, \citenamefont {Matone}, \citenamefont {Matta},
  \citenamefont {Mavalvala}, \citenamefont {McCarthy}, \citenamefont
  {McClelland}, \citenamefont {McGuire}, \citenamefont {McHugh}, \citenamefont
  {McKenzie}, \citenamefont {McNabb}, \citenamefont {McWilliams}, \citenamefont
  {Meier}, \citenamefont {Melissinos}, \citenamefont {Mendell}, \citenamefont
  {Mercer}, \citenamefont {Meshkov}, \citenamefont {Messaritaki}, \citenamefont
  {Messenger}, \citenamefont {Meyers}, \citenamefont {Mikhailov}, \citenamefont
  {Mitra}, \citenamefont {Mitrofanov}, \citenamefont {Mitselmakher},
  \citenamefont {Mittleman}, \citenamefont {Miyakawa}, \citenamefont {Mohanty},
  \citenamefont {Moreno}, \citenamefont {Mossavi}, \citenamefont {MowLowry},
  \citenamefont {Moylan}, \citenamefont {Mudge}, \citenamefont {Mueller},
  \citenamefont {Mukherjee}, \citenamefont {M\"uller-Ebhardt}, \citenamefont
  {Munch}, \citenamefont {Murray}, \citenamefont {Myers}, \citenamefont
  {Myers}, \citenamefont {Nash}, \citenamefont {Newton}, \citenamefont
  {Nishizawa}, \citenamefont {Nocera}, \citenamefont {Numata}, \citenamefont
  {O'Reilly}, \citenamefont {O'Shaughnessy}, \citenamefont {Ottaway},
  \citenamefont {Overmier}, \citenamefont {Owen}, \citenamefont {Pan},
  \citenamefont {Papa}, \citenamefont {Parameshwaraiah}, \citenamefont
  {Parameswariah}, \citenamefont {Patel}, \citenamefont {Pedraza},
  \citenamefont {Penn}, \citenamefont {Pierro}, \citenamefont {Pinto},
  \citenamefont {Pitkin}, \citenamefont {Pletsch}, \citenamefont {Plissi},
  \citenamefont {Postiglione}, \citenamefont {Prix}, \citenamefont {Quetschke},
  \citenamefont {Raab}, \citenamefont {Rabeling}, \citenamefont {Radkins},
  \citenamefont {Rahkola}, \citenamefont {Rainer}, \citenamefont {Rakhmanov},
  \citenamefont {Rawlins}, \citenamefont {Ray-Majumder}, \citenamefont {Re},
  \citenamefont {Regimbau}, \citenamefont {Rehbein}, \citenamefont {Reid},
  \citenamefont {Reitze}, \citenamefont {Ribichini}, \citenamefont {Riesen},
  \citenamefont {Riles}, \citenamefont {Rivera}, \citenamefont {Robertson},
  \citenamefont {Robinson}, \citenamefont {Robinson}, \citenamefont {Roddy},
  \citenamefont {Rodriguez}, \citenamefont {Rogan}, \citenamefont {Rollins},
  \citenamefont {Romano}, \citenamefont {Romie}, \citenamefont {Route},
  \citenamefont {Rowan}, \citenamefont {R\"udiger}, \citenamefont {Ruet},
  \citenamefont {Russell}, \citenamefont {Ryan}, \citenamefont {Sakata},
  \citenamefont {Samidi}, \citenamefont {de~la Jordana}, \citenamefont
  {Sandberg}, \citenamefont {Sanders}, \citenamefont {Sannibale}, \citenamefont
  {Saraf}, \citenamefont {Sarin}, \citenamefont {Sathyaprakash}, \citenamefont
  {Sato}, \citenamefont {Saulson}, \citenamefont {Savage}, \citenamefont
  {Savov}, \citenamefont {Sazonov}, \citenamefont {Schediwy}, \citenamefont
  {Schilling}, \citenamefont {Schnabel}, \citenamefont {Schofield},
  \citenamefont {Schutz}, \citenamefont {Schwinberg}, \citenamefont {Scott},
  \citenamefont {Searle}, \citenamefont {Sears}, \citenamefont {Seifert},
  \citenamefont {Sellers}, \citenamefont {Sengupta}, \citenamefont {Shawhan},
  \citenamefont {Shoemaker}, \citenamefont {Sibley}, \citenamefont {Sidles},
  \citenamefont {Siemens}, \citenamefont {Sigg}, \citenamefont {Sinha},
  \citenamefont {Sintes}, \citenamefont {Slagmolen}, \citenamefont {Slutsky},
  \citenamefont {Smith}, \citenamefont {Smith}, \citenamefont {Somiya},
  \citenamefont {Strain}, \citenamefont {Strom}, \citenamefont {Stuver},
  \citenamefont {Summerscales}, \citenamefont {Sun}, \citenamefont {Sung},
  \citenamefont {Sutton}, \citenamefont {Takahashi}, \citenamefont {Tanner},
  \citenamefont {Tarallo}, \citenamefont {Taylor}, \citenamefont {Taylor},
  \citenamefont {Thacker}, \citenamefont {Thorne}, \citenamefont {Thorne},
  \citenamefont {Th\"uring}, \citenamefont {Tokmakov}, \citenamefont {Torres},
  \citenamefont {Torrie}, \citenamefont {Traylor}, \citenamefont {Trias},
  \citenamefont {Tyler}, \citenamefont {Ugolini}, \citenamefont {Ungarelli},
  \citenamefont {Urbanek}, \citenamefont {Vahlbruch}, \citenamefont
  {Vallisneri}, \citenamefont {Van Den~Broeck}, \citenamefont {van Putten},
  \citenamefont {Varvella}, \citenamefont {Vass}, \citenamefont {Vecchio},
  \citenamefont {Veitch}, \citenamefont {Veitch}, \citenamefont {Villar},
  \citenamefont {Vorvick}, \citenamefont {Vyachanin}, \citenamefont {Waldman},
  \citenamefont {Wallace}, \citenamefont {Ward}, \citenamefont {Ward},
  \citenamefont {Watts}, \citenamefont {Webber}, \citenamefont {Weidner},
  \citenamefont {Weinert}, \citenamefont {Weinstein}, \citenamefont {Weiss},
  \citenamefont {Wen}, \citenamefont {Wette}, \citenamefont {Whelan},
  \citenamefont {Whitbeck}, \citenamefont {Whitcomb}, \citenamefont {Whiting},
  \citenamefont {Wiley}, \citenamefont {Wilkinson}, \citenamefont {Willems},
  \citenamefont {Williams}, \citenamefont {Willke}, \citenamefont {Wilmut},
  \citenamefont {Winkler}, \citenamefont {Wipf}, \citenamefont {Wise},
  \citenamefont {Wiseman}, \citenamefont {Woan}, \citenamefont {Woods},
  \citenamefont {Wooley}, \citenamefont {Worden}, \citenamefont {Wu},
  \citenamefont {Yakushin}, \citenamefont {Yamamoto}, \citenamefont {Yan},
  \citenamefont {Yoshida}, \citenamefont {Yunes}, \citenamefont {Zanolin},
  \citenamefont {Zhang}, \citenamefont {Zhang}, \citenamefont {Zhao},
  \citenamefont {Zotov}, \citenamefont {Zucker}, \citenamefont {zur M\"uhlen},
  \citenamefont {Zweizig}, \citenamefont {Kramer},\ and\ \citenamefont
  {Lyne}}]{PhysRevD.76.042001}%
  \BibitemOpen
  \bibfield  {author} {\bibinfo {author} {\bibfnamefont {B.}~\bibnamefont
  {Abbott}}, \bibinfo {author} {\bibfnamefont {R.}~\bibnamefont {Abbott}},
  \bibinfo {author} {\bibfnamefont {R.}~\bibnamefont {Adhikari}}, \bibinfo
  {author} {\bibfnamefont {J.}~\bibnamefont {Agresti}}, \bibinfo {author}
  {\bibfnamefont {P.}~\bibnamefont {Ajith}}, \bibinfo {author} {\bibfnamefont
  {B.}~\bibnamefont {Allen}}, \bibinfo {author} {\bibfnamefont
  {R.}~\bibnamefont {Amin}}, \bibinfo {author} {\bibfnamefont {S.~B.}\
  \bibnamefont {Anderson}}, \bibinfo {author} {\bibfnamefont {W.~G.}\
  \bibnamefont {Anderson}}, \bibinfo {author} {\bibfnamefont {M.}~\bibnamefont
  {Arain}}, \bibinfo {author} {\bibfnamefont {M.}~\bibnamefont {Araya}},
  \bibinfo {author} {\bibfnamefont {H.}~\bibnamefont {Armandula}}, \bibinfo
  {author} {\bibfnamefont {M.}~\bibnamefont {Ashley}}, \bibinfo {author}
  {\bibfnamefont {S.}~\bibnamefont {Aston}}, \bibinfo {author} {\bibfnamefont
  {P.}~\bibnamefont {Aufmuth}}, \bibinfo {author} {\bibfnamefont
  {C.}~\bibnamefont {Aulbert}}, \bibinfo {author} {\bibfnamefont
  {S.}~\bibnamefont {Babak}}, \bibinfo {author} {\bibfnamefont
  {S.}~\bibnamefont {Ballmer}}, \bibinfo {author} {\bibfnamefont
  {H.}~\bibnamefont {Bantilan}}, \bibinfo {author} {\bibfnamefont {B.~C.}\
  \bibnamefont {Barish}}, \bibinfo {author} {\bibfnamefont {C.}~\bibnamefont
  {Barker}}, \bibinfo {author} {\bibfnamefont {D.}~\bibnamefont {Barker}},
  \bibinfo {author} {\bibfnamefont {B.}~\bibnamefont {Barr}}, \bibinfo {author}
  {\bibfnamefont {P.}~\bibnamefont {Barriga}}, \bibinfo {author} {\bibfnamefont
  {M.~A.}\ \bibnamefont {Barton}}, \bibinfo {author} {\bibfnamefont
  {K.}~\bibnamefont {Bayer}}, \bibinfo {author} {\bibfnamefont
  {K.}~\bibnamefont {Belczynski}}, \bibinfo {author} {\bibfnamefont
  {J.}~\bibnamefont {Betzwieser}}, \bibinfo {author} {\bibfnamefont {P.~T.}\
  \bibnamefont {Beyersdorf}}, \bibinfo {author} {\bibfnamefont
  {B.}~\bibnamefont {Bhawal}}, \bibinfo {author} {\bibfnamefont {I.~A.}\
  \bibnamefont {Bilenko}}, \bibinfo {author} {\bibfnamefont {G.}~\bibnamefont
  {Billingsley}}, \bibinfo {author} {\bibfnamefont {R.}~\bibnamefont {Biswas}},
  \bibinfo {author} {\bibfnamefont {E.}~\bibnamefont {Black}}, \bibinfo
  {author} {\bibfnamefont {K.}~\bibnamefont {Blackburn}}, \bibinfo {author}
  {\bibfnamefont {L.}~\bibnamefont {Blackburn}}, \bibinfo {author}
  {\bibfnamefont {D.}~\bibnamefont {Blair}}, \bibinfo {author} {\bibfnamefont
  {B.}~\bibnamefont {Bland}}, \bibinfo {author} {\bibfnamefont
  {J.}~\bibnamefont {Bogenstahl}}, \bibinfo {author} {\bibfnamefont
  {L.}~\bibnamefont {Bogue}}, \bibinfo {author} {\bibfnamefont
  {R.}~\bibnamefont {Bork}}, \bibinfo {author} {\bibfnamefont {V.}~\bibnamefont
  {Boschi}}, \bibinfo {author} {\bibfnamefont {S.}~\bibnamefont {Bose}},
  \bibinfo {author} {\bibfnamefont {P.~R.}\ \bibnamefont {Brady}}, \bibinfo
  {author} {\bibfnamefont {V.~B.}\ \bibnamefont {Braginsky}}, \bibinfo {author}
  {\bibfnamefont {J.~E.}\ \bibnamefont {Brau}}, \bibinfo {author}
  {\bibfnamefont {M.}~\bibnamefont {Brinkmann}}, \bibinfo {author}
  {\bibfnamefont {A.}~\bibnamefont {Brooks}}, \bibinfo {author} {\bibfnamefont
  {D.~A.}\ \bibnamefont {Brown}}, \bibinfo {author} {\bibfnamefont
  {A.}~\bibnamefont {Bullington}}, \bibinfo {author} {\bibfnamefont
  {A.}~\bibnamefont {Bunkowski}}, \bibinfo {author} {\bibfnamefont
  {A.}~\bibnamefont {Buonanno}}, \bibinfo {author} {\bibfnamefont
  {O.}~\bibnamefont {Burmeister}}, \bibinfo {author} {\bibfnamefont
  {D.}~\bibnamefont {Busby}}, \bibinfo {author} {\bibfnamefont {W.~E.}\
  \bibnamefont {Butler}}, \bibinfo {author} {\bibfnamefont {R.~L.}\
  \bibnamefont {Byer}}, \bibinfo {author} {\bibfnamefont {L.}~\bibnamefont
  {Cadonati}}, \bibinfo {author} {\bibfnamefont {G.}~\bibnamefont {Cagnoli}},
  \bibinfo {author} {\bibfnamefont {J.~B.}\ \bibnamefont {Camp}}, \bibinfo
  {author} {\bibfnamefont {J.}~\bibnamefont {Cannizzo}}, \bibinfo {author}
  {\bibfnamefont {K.}~\bibnamefont {Cannon}}, \bibinfo {author} {\bibfnamefont
  {C.~A.}\ \bibnamefont {Cantley}}, \bibinfo {author} {\bibfnamefont
  {J.}~\bibnamefont {Cao}}, \bibinfo {author} {\bibfnamefont {L.}~\bibnamefont
  {Cardenas}}, \bibinfo {author} {\bibfnamefont {K.}~\bibnamefont {Carter}},
  \bibinfo {author} {\bibfnamefont {M.~M.}\ \bibnamefont {Casey}}, \bibinfo
  {author} {\bibfnamefont {G.}~\bibnamefont {Castaldi}}, \bibinfo {author}
  {\bibfnamefont {C.}~\bibnamefont {Cepeda}}, \bibinfo {author} {\bibfnamefont
  {E.}~\bibnamefont {Chalkey}}, \bibinfo {author} {\bibfnamefont
  {P.}~\bibnamefont {Charlton}}, \bibinfo {author} {\bibfnamefont
  {S.}~\bibnamefont {Chatterji}}, \bibinfo {author} {\bibfnamefont
  {S.}~\bibnamefont {Chelkowski}}, \bibinfo {author} {\bibfnamefont
  {Y.}~\bibnamefont {Chen}}, \bibinfo {author} {\bibfnamefont {F.}~\bibnamefont
  {Chiadini}}, \bibinfo {author} {\bibfnamefont {D.}~\bibnamefont {Chin}},
  \bibinfo {author} {\bibfnamefont {E.}~\bibnamefont {Chin}}, \bibinfo {author}
  {\bibfnamefont {J.}~\bibnamefont {Chow}}, \bibinfo {author} {\bibfnamefont
  {N.}~\bibnamefont {Christensen}}, \bibinfo {author} {\bibfnamefont
  {J.}~\bibnamefont {Clark}}, \bibinfo {author} {\bibfnamefont
  {P.}~\bibnamefont {Cochrane}}, \bibinfo {author} {\bibfnamefont
  {T.}~\bibnamefont {Cokelaer}}, \bibinfo {author} {\bibfnamefont {C.~N.}\
  \bibnamefont {Colacino}}, \bibinfo {author} {\bibfnamefont {R.}~\bibnamefont
  {Coldwell}}, \bibinfo {author} {\bibfnamefont {R.}~\bibnamefont {Conte}},
  \bibinfo {author} {\bibfnamefont {D.}~\bibnamefont {Cook}}, \bibinfo {author}
  {\bibfnamefont {T.}~\bibnamefont {Corbitt}}, \bibinfo {author} {\bibfnamefont
  {D.}~\bibnamefont {Coward}}, \bibinfo {author} {\bibfnamefont
  {D.}~\bibnamefont {Coyne}}, \bibinfo {author} {\bibfnamefont {J.~D.~E.}\
  \bibnamefont {Creighton}}, \bibinfo {author} {\bibfnamefont {T.~D.}\
  \bibnamefont {Creighton}}, \bibinfo {author} {\bibfnamefont {R.~P.}\
  \bibnamefont {Croce}}, \bibinfo {author} {\bibfnamefont {D.~R.~M.}\
  \bibnamefont {Crooks}}, \bibinfo {author} {\bibfnamefont {A.~M.}\
  \bibnamefont {Cruise}}, \bibinfo {author} {\bibfnamefont {A.}~\bibnamefont
  {Cumming}}, \bibinfo {author} {\bibfnamefont {J.}~\bibnamefont {Dalrymple}},
  \bibinfo {author} {\bibfnamefont {E.}~\bibnamefont {D'Ambrosio}}, \bibinfo
  {author} {\bibfnamefont {K.}~\bibnamefont {Danzmann}}, \bibinfo {author}
  {\bibfnamefont {G.}~\bibnamefont {Davies}}, \bibinfo {author} {\bibfnamefont
  {D.}~\bibnamefont {DeBra}}, \bibinfo {author} {\bibfnamefont
  {J.}~\bibnamefont {Degallaix}}, \bibinfo {author} {\bibfnamefont
  {M.}~\bibnamefont {Degree}}, \bibinfo {author} {\bibfnamefont
  {T.}~\bibnamefont {Demma}}, \bibinfo {author} {\bibfnamefont
  {V.}~\bibnamefont {Dergachev}}, \bibinfo {author} {\bibfnamefont
  {S.}~\bibnamefont {Desai}}, \bibinfo {author} {\bibfnamefont
  {R.}~\bibnamefont {DeSalvo}}, \bibinfo {author} {\bibfnamefont
  {S.}~\bibnamefont {Dhurandhar}}, \bibinfo {author} {\bibfnamefont
  {M.}~\bibnamefont {D\'{\i}az}}, \bibinfo {author} {\bibfnamefont
  {J.}~\bibnamefont {Dickson}}, \bibinfo {author} {\bibfnamefont
  {A.}~\bibnamefont {Di~Credico}}, \bibinfo {author} {\bibfnamefont
  {G.}~\bibnamefont {Diederichs}}, \bibinfo {author} {\bibfnamefont
  {A.}~\bibnamefont {Dietz}}, \bibinfo {author} {\bibfnamefont {E.~E.}\
  \bibnamefont {Doomes}}, \bibinfo {author} {\bibfnamefont {R.~W.~P.}\
  \bibnamefont {Drever}}, \bibinfo {author} {\bibfnamefont {J.-C.}\
  \bibnamefont {Dumas}}, \bibinfo {author} {\bibfnamefont {R.~J.}\ \bibnamefont
  {Dupuis}}, \bibinfo {author} {\bibfnamefont {J.~G.}\ \bibnamefont {Dwyer}},
  \bibinfo {author} {\bibfnamefont {P.}~\bibnamefont {Ehrens}}, \bibinfo
  {author} {\bibfnamefont {E.}~\bibnamefont {Espinoza}}, \bibinfo {author}
  {\bibfnamefont {T.}~\bibnamefont {Etzel}}, \bibinfo {author} {\bibfnamefont
  {M.}~\bibnamefont {Evans}}, \bibinfo {author} {\bibfnamefont
  {T.}~\bibnamefont {Evans}}, \bibinfo {author} {\bibfnamefont
  {S.}~\bibnamefont {Fairhurst}}, \bibinfo {author} {\bibfnamefont
  {Y.}~\bibnamefont {Fan}}, \bibinfo {author} {\bibfnamefont {D.}~\bibnamefont
  {Fazi}}, \bibinfo {author} {\bibfnamefont {M.~M.}\ \bibnamefont {Fejer}},
  \bibinfo {author} {\bibfnamefont {L.~S.}\ \bibnamefont {Finn}}, \bibinfo
  {author} {\bibfnamefont {V.}~\bibnamefont {Fiumara}}, \bibinfo {author}
  {\bibfnamefont {N.}~\bibnamefont {Fotopoulos}}, \bibinfo {author}
  {\bibfnamefont {A.}~\bibnamefont {Franzen}}, \bibinfo {author} {\bibfnamefont
  {K.~Y.}\ \bibnamefont {Franzen}}, \bibinfo {author} {\bibfnamefont
  {A.}~\bibnamefont {Freise}}, \bibinfo {author} {\bibfnamefont
  {R.}~\bibnamefont {Frey}}, \bibinfo {author} {\bibfnamefont {T.}~\bibnamefont
  {Fricke}}, \bibinfo {author} {\bibfnamefont {P.}~\bibnamefont {Fritschel}},
  \bibinfo {author} {\bibfnamefont {V.~V.}\ \bibnamefont {Frolov}}, \bibinfo
  {author} {\bibfnamefont {M.}~\bibnamefont {Fyffe}}, \bibinfo {author}
  {\bibfnamefont {V.}~\bibnamefont {Galdi}}, \bibinfo {author} {\bibfnamefont
  {K.~S.}\ \bibnamefont {Ganezer}}, \bibinfo {author} {\bibfnamefont
  {J.}~\bibnamefont {Garofoli}}, \bibinfo {author} {\bibfnamefont
  {I.}~\bibnamefont {Gholami}}, \bibinfo {author} {\bibfnamefont {J.~A.}\
  \bibnamefont {Giaime}}, \bibinfo {author} {\bibfnamefont {S.}~\bibnamefont
  {Giampanis}}, \bibinfo {author} {\bibfnamefont {K.~D.}\ \bibnamefont
  {Giardina}}, \bibinfo {author} {\bibfnamefont {K.}~\bibnamefont {Goda}},
  \bibinfo {author} {\bibfnamefont {E.}~\bibnamefont {Goetz}}, \bibinfo
  {author} {\bibfnamefont {L.}~\bibnamefont {Goggin}}, \bibinfo {author}
  {\bibfnamefont {G.}~\bibnamefont {Gonz\'alez}}, \bibinfo {author}
  {\bibfnamefont {S.}~\bibnamefont {Gossler}}, \bibinfo {author} {\bibfnamefont
  {A.}~\bibnamefont {Grant}}, \bibinfo {author} {\bibfnamefont
  {S.}~\bibnamefont {Gras}}, \bibinfo {author} {\bibfnamefont {C.}~\bibnamefont
  {Gray}}, \bibinfo {author} {\bibfnamefont {M.}~\bibnamefont {Gray}}, \bibinfo
  {author} {\bibfnamefont {J.}~\bibnamefont {Greenhalgh}}, \bibinfo {author}
  {\bibfnamefont {A.~M.}\ \bibnamefont {Gretarsson}}, \bibinfo {author}
  {\bibfnamefont {R.}~\bibnamefont {Grosso}}, \bibinfo {author} {\bibfnamefont
  {H.}~\bibnamefont {Grote}}, \bibinfo {author} {\bibfnamefont
  {S.}~\bibnamefont {Grunewald}}, \bibinfo {author} {\bibfnamefont
  {M.}~\bibnamefont {Guenther}}, \bibinfo {author} {\bibfnamefont
  {R.}~\bibnamefont {Gustafson}}, \bibinfo {author} {\bibfnamefont
  {B.}~\bibnamefont {Hage}}, \bibinfo {author} {\bibfnamefont {D.}~\bibnamefont
  {Hammer}}, \bibinfo {author} {\bibfnamefont {C.}~\bibnamefont {Hanna}},
  \bibinfo {author} {\bibfnamefont {J.}~\bibnamefont {Hanson}}, \bibinfo
  {author} {\bibfnamefont {J.}~\bibnamefont {Harms}}, \bibinfo {author}
  {\bibfnamefont {G.}~\bibnamefont {Harry}}, \bibinfo {author} {\bibfnamefont
  {E.}~\bibnamefont {Harstad}}, \bibinfo {author} {\bibfnamefont
  {T.}~\bibnamefont {Hayler}}, \bibinfo {author} {\bibfnamefont
  {J.}~\bibnamefont {Heefner}}, \bibinfo {author} {\bibfnamefont {I.~S.}\
  \bibnamefont {Heng}}, \bibinfo {author} {\bibfnamefont {A.}~\bibnamefont
  {Heptonstall}}, \bibinfo {author} {\bibfnamefont {M.}~\bibnamefont {Heurs}},
  \bibinfo {author} {\bibfnamefont {M.}~\bibnamefont {Hewitson}}, \bibinfo
  {author} {\bibfnamefont {S.}~\bibnamefont {Hild}}, \bibinfo {author}
  {\bibfnamefont {E.}~\bibnamefont {Hirose}}, \bibinfo {author} {\bibfnamefont
  {D.}~\bibnamefont {Hoak}}, \bibinfo {author} {\bibfnamefont {D.}~\bibnamefont
  {Hosken}}, \bibinfo {author} {\bibfnamefont {J.}~\bibnamefont {Hough}},
  \bibinfo {author} {\bibfnamefont {E.}~\bibnamefont {Howell}}, \bibinfo
  {author} {\bibfnamefont {D.}~\bibnamefont {Hoyland}}, \bibinfo {author}
  {\bibfnamefont {S.~H.}\ \bibnamefont {Huttner}}, \bibinfo {author}
  {\bibfnamefont {D.}~\bibnamefont {Ingram}}, \bibinfo {author} {\bibfnamefont
  {E.}~\bibnamefont {Innerhofer}}, \bibinfo {author} {\bibfnamefont
  {M.}~\bibnamefont {Ito}}, \bibinfo {author} {\bibfnamefont {Y.}~\bibnamefont
  {Itoh}}, \bibinfo {author} {\bibfnamefont {A.}~\bibnamefont {Ivanov}},
  \bibinfo {author} {\bibfnamefont {D.}~\bibnamefont {Jackrel}}, \bibinfo
  {author} {\bibfnamefont {B.}~\bibnamefont {Johnson}}, \bibinfo {author}
  {\bibfnamefont {W.~W.}\ \bibnamefont {Johnson}}, \bibinfo {author}
  {\bibfnamefont {D.~I.}\ \bibnamefont {Jones}}, \bibinfo {author}
  {\bibfnamefont {G.}~\bibnamefont {Jones}}, \bibinfo {author} {\bibfnamefont
  {R.}~\bibnamefont {Jones}}, \bibinfo {author} {\bibfnamefont
  {L.}~\bibnamefont {Ju}}, \bibinfo {author} {\bibfnamefont {P.}~\bibnamefont
  {Kalmus}}, \bibinfo {author} {\bibfnamefont {V.}~\bibnamefont {Kalogera}},
  \bibinfo {author} {\bibfnamefont {D.}~\bibnamefont {Kasprzyk}}, \bibinfo
  {author} {\bibfnamefont {E.}~\bibnamefont {Katsavounidis}}, \bibinfo {author}
  {\bibfnamefont {K.}~\bibnamefont {Kawabe}}, \bibinfo {author} {\bibfnamefont
  {S.}~\bibnamefont {Kawamura}}, \bibinfo {author} {\bibfnamefont
  {F.}~\bibnamefont {Kawazoe}}, \bibinfo {author} {\bibfnamefont
  {W.}~\bibnamefont {Kells}}, \bibinfo {author} {\bibfnamefont {D.~G.}\
  \bibnamefont {Keppel}}, \bibinfo {author} {\bibfnamefont {F.~Y.}\
  \bibnamefont {Khalili}}, \bibinfo {author} {\bibfnamefont {C.}~\bibnamefont
  {Kim}}, \bibinfo {author} {\bibfnamefont {P.}~\bibnamefont {King}}, \bibinfo
  {author} {\bibfnamefont {J.~S.}\ \bibnamefont {Kissel}}, \bibinfo {author}
  {\bibfnamefont {S.}~\bibnamefont {Klimenko}}, \bibinfo {author}
  {\bibfnamefont {K.}~\bibnamefont {Kokeyama}}, \bibinfo {author}
  {\bibfnamefont {V.}~\bibnamefont {Kondrashov}}, \bibinfo {author}
  {\bibfnamefont {R.~K.}\ \bibnamefont {Kopparapu}}, \bibinfo {author}
  {\bibfnamefont {D.}~\bibnamefont {Kozak}}, \bibinfo {author} {\bibfnamefont
  {B.}~\bibnamefont {Krishnan}}, \bibinfo {author} {\bibfnamefont
  {P.}~\bibnamefont {Kwee}}, \bibinfo {author} {\bibfnamefont {P.~K.}\
  \bibnamefont {Lam}}, \bibinfo {author} {\bibfnamefont {M.}~\bibnamefont
  {Landry}}, \bibinfo {author} {\bibfnamefont {B.}~\bibnamefont {Lantz}},
  \bibinfo {author} {\bibfnamefont {A.}~\bibnamefont {Lazzarini}}, \bibinfo
  {author} {\bibfnamefont {B.}~\bibnamefont {Lee}}, \bibinfo {author}
  {\bibfnamefont {M.}~\bibnamefont {Lei}}, \bibinfo {author} {\bibfnamefont
  {J.}~\bibnamefont {Leiner}}, \bibinfo {author} {\bibfnamefont
  {V.}~\bibnamefont {Leonhardt}}, \bibinfo {author} {\bibfnamefont
  {I.}~\bibnamefont {Leonor}}, \bibinfo {author} {\bibfnamefont
  {K.}~\bibnamefont {Libbrecht}}, \bibinfo {author} {\bibfnamefont
  {P.}~\bibnamefont {Lindquist}}, \bibinfo {author} {\bibfnamefont {N.~A.}\
  \bibnamefont {Lockerbie}}, \bibinfo {author} {\bibfnamefont {M.}~\bibnamefont
  {Longo}}, \bibinfo {author} {\bibfnamefont {M.}~\bibnamefont {Lormand}},
  \bibinfo {author} {\bibfnamefont {M.}~\bibnamefont {Lubinski}}, \bibinfo
  {author} {\bibfnamefont {H.}~\bibnamefont {L\"uck}}, \bibinfo {author}
  {\bibfnamefont {B.}~\bibnamefont {Machenschalk}}, \bibinfo {author}
  {\bibfnamefont {M.}~\bibnamefont {MacInnis}}, \bibinfo {author}
  {\bibfnamefont {M.}~\bibnamefont {Mageswaran}}, \bibinfo {author}
  {\bibfnamefont {K.}~\bibnamefont {Mailand}}, \bibinfo {author} {\bibfnamefont
  {M.}~\bibnamefont {Malec}}, \bibinfo {author} {\bibfnamefont
  {V.}~\bibnamefont {Mandic}}, \bibinfo {author} {\bibfnamefont
  {S.}~\bibnamefont {Marano}}, \bibinfo {author} {\bibfnamefont
  {S.}~\bibnamefont {M\'arka}}, \bibinfo {author} {\bibfnamefont
  {J.}~\bibnamefont {Markowitz}}, \bibinfo {author} {\bibfnamefont
  {E.}~\bibnamefont {Maros}}, \bibinfo {author} {\bibfnamefont
  {I.}~\bibnamefont {Martin}}, \bibinfo {author} {\bibfnamefont {J.~N.}\
  \bibnamefont {Marx}}, \bibinfo {author} {\bibfnamefont {K.}~\bibnamefont
  {Mason}}, \bibinfo {author} {\bibfnamefont {L.}~\bibnamefont {Matone}},
  \bibinfo {author} {\bibfnamefont {V.}~\bibnamefont {Matta}}, \bibinfo
  {author} {\bibfnamefont {N.}~\bibnamefont {Mavalvala}}, \bibinfo {author}
  {\bibfnamefont {R.}~\bibnamefont {McCarthy}}, \bibinfo {author}
  {\bibfnamefont {D.~E.}\ \bibnamefont {McClelland}}, \bibinfo {author}
  {\bibfnamefont {S.~C.}\ \bibnamefont {McGuire}}, \bibinfo {author}
  {\bibfnamefont {M.}~\bibnamefont {McHugh}}, \bibinfo {author} {\bibfnamefont
  {K.}~\bibnamefont {McKenzie}}, \bibinfo {author} {\bibfnamefont {J.~W.~C.}\
  \bibnamefont {McNabb}}, \bibinfo {author} {\bibfnamefont {S.}~\bibnamefont
  {McWilliams}}, \bibinfo {author} {\bibfnamefont {T.}~\bibnamefont {Meier}},
  \bibinfo {author} {\bibfnamefont {A.}~\bibnamefont {Melissinos}}, \bibinfo
  {author} {\bibfnamefont {G.}~\bibnamefont {Mendell}}, \bibinfo {author}
  {\bibfnamefont {R.~A.}\ \bibnamefont {Mercer}}, \bibinfo {author}
  {\bibfnamefont {S.}~\bibnamefont {Meshkov}}, \bibinfo {author} {\bibfnamefont
  {E.}~\bibnamefont {Messaritaki}}, \bibinfo {author} {\bibfnamefont {C.~J.}\
  \bibnamefont {Messenger}}, \bibinfo {author} {\bibfnamefont {D.}~\bibnamefont
  {Meyers}}, \bibinfo {author} {\bibfnamefont {E.}~\bibnamefont {Mikhailov}},
  \bibinfo {author} {\bibfnamefont {S.}~\bibnamefont {Mitra}}, \bibinfo
  {author} {\bibfnamefont {V.~P.}\ \bibnamefont {Mitrofanov}}, \bibinfo
  {author} {\bibfnamefont {G.}~\bibnamefont {Mitselmakher}}, \bibinfo {author}
  {\bibfnamefont {R.}~\bibnamefont {Mittleman}}, \bibinfo {author}
  {\bibfnamefont {O.}~\bibnamefont {Miyakawa}}, \bibinfo {author}
  {\bibfnamefont {S.}~\bibnamefont {Mohanty}}, \bibinfo {author} {\bibfnamefont
  {G.}~\bibnamefont {Moreno}}, \bibinfo {author} {\bibfnamefont
  {K.}~\bibnamefont {Mossavi}}, \bibinfo {author} {\bibfnamefont
  {C.}~\bibnamefont {MowLowry}}, \bibinfo {author} {\bibfnamefont
  {A.}~\bibnamefont {Moylan}}, \bibinfo {author} {\bibfnamefont
  {D.}~\bibnamefont {Mudge}}, \bibinfo {author} {\bibfnamefont
  {G.}~\bibnamefont {Mueller}}, \bibinfo {author} {\bibfnamefont
  {S.}~\bibnamefont {Mukherjee}}, \bibinfo {author} {\bibfnamefont
  {H.}~\bibnamefont {M\"uller-Ebhardt}}, \bibinfo {author} {\bibfnamefont
  {J.}~\bibnamefont {Munch}}, \bibinfo {author} {\bibfnamefont
  {P.}~\bibnamefont {Murray}}, \bibinfo {author} {\bibfnamefont
  {E.}~\bibnamefont {Myers}}, \bibinfo {author} {\bibfnamefont
  {J.}~\bibnamefont {Myers}}, \bibinfo {author} {\bibfnamefont
  {T.}~\bibnamefont {Nash}}, \bibinfo {author} {\bibfnamefont {G.}~\bibnamefont
  {Newton}}, \bibinfo {author} {\bibfnamefont {A.}~\bibnamefont {Nishizawa}},
  \bibinfo {author} {\bibfnamefont {F.}~\bibnamefont {Nocera}}, \bibinfo
  {author} {\bibfnamefont {K.}~\bibnamefont {Numata}}, \bibinfo {author}
  {\bibfnamefont {B.}~\bibnamefont {O'Reilly}}, \bibinfo {author}
  {\bibfnamefont {R.}~\bibnamefont {O'Shaughnessy}}, \bibinfo {author}
  {\bibfnamefont {D.~J.}\ \bibnamefont {Ottaway}}, \bibinfo {author}
  {\bibfnamefont {H.}~\bibnamefont {Overmier}}, \bibinfo {author}
  {\bibfnamefont {B.~J.}\ \bibnamefont {Owen}}, \bibinfo {author}
  {\bibfnamefont {Y.}~\bibnamefont {Pan}}, \bibinfo {author} {\bibfnamefont
  {M.~A.}\ \bibnamefont {Papa}}, \bibinfo {author} {\bibfnamefont
  {V.}~\bibnamefont {Parameshwaraiah}}, \bibinfo {author} {\bibfnamefont
  {C.}~\bibnamefont {Parameswariah}}, \bibinfo {author} {\bibfnamefont
  {P.}~\bibnamefont {Patel}}, \bibinfo {author} {\bibfnamefont
  {M.}~\bibnamefont {Pedraza}}, \bibinfo {author} {\bibfnamefont
  {S.}~\bibnamefont {Penn}}, \bibinfo {author} {\bibfnamefont {V.}~\bibnamefont
  {Pierro}}, \bibinfo {author} {\bibfnamefont {I.~M.}\ \bibnamefont {Pinto}},
  \bibinfo {author} {\bibfnamefont {M.}~\bibnamefont {Pitkin}}, \bibinfo
  {author} {\bibfnamefont {H.}~\bibnamefont {Pletsch}}, \bibinfo {author}
  {\bibfnamefont {M.~V.}\ \bibnamefont {Plissi}}, \bibinfo {author}
  {\bibfnamefont {F.}~\bibnamefont {Postiglione}}, \bibinfo {author}
  {\bibfnamefont {R.}~\bibnamefont {Prix}}, \bibinfo {author} {\bibfnamefont
  {V.}~\bibnamefont {Quetschke}}, \bibinfo {author} {\bibfnamefont
  {F.}~\bibnamefont {Raab}}, \bibinfo {author} {\bibfnamefont {D.}~\bibnamefont
  {Rabeling}}, \bibinfo {author} {\bibfnamefont {H.}~\bibnamefont {Radkins}},
  \bibinfo {author} {\bibfnamefont {R.}~\bibnamefont {Rahkola}}, \bibinfo
  {author} {\bibfnamefont {N.}~\bibnamefont {Rainer}}, \bibinfo {author}
  {\bibfnamefont {M.}~\bibnamefont {Rakhmanov}}, \bibinfo {author}
  {\bibfnamefont {K.}~\bibnamefont {Rawlins}}, \bibinfo {author} {\bibfnamefont
  {S.}~\bibnamefont {Ray-Majumder}}, \bibinfo {author} {\bibfnamefont
  {V.}~\bibnamefont {Re}}, \bibinfo {author} {\bibfnamefont {T.}~\bibnamefont
  {Regimbau}}, \bibinfo {author} {\bibfnamefont {H.}~\bibnamefont {Rehbein}},
  \bibinfo {author} {\bibfnamefont {S.}~\bibnamefont {Reid}}, \bibinfo {author}
  {\bibfnamefont {D.~H.}\ \bibnamefont {Reitze}}, \bibinfo {author}
  {\bibfnamefont {L.}~\bibnamefont {Ribichini}}, \bibinfo {author}
  {\bibfnamefont {R.}~\bibnamefont {Riesen}}, \bibinfo {author} {\bibfnamefont
  {K.}~\bibnamefont {Riles}}, \bibinfo {author} {\bibfnamefont
  {B.}~\bibnamefont {Rivera}}, \bibinfo {author} {\bibfnamefont {N.~A.}\
  \bibnamefont {Robertson}}, \bibinfo {author} {\bibfnamefont {C.}~\bibnamefont
  {Robinson}}, \bibinfo {author} {\bibfnamefont {E.~L.}\ \bibnamefont
  {Robinson}}, \bibinfo {author} {\bibfnamefont {S.}~\bibnamefont {Roddy}},
  \bibinfo {author} {\bibfnamefont {A.}~\bibnamefont {Rodriguez}}, \bibinfo
  {author} {\bibfnamefont {A.~M.}\ \bibnamefont {Rogan}}, \bibinfo {author}
  {\bibfnamefont {J.}~\bibnamefont {Rollins}}, \bibinfo {author} {\bibfnamefont
  {J.~D.}\ \bibnamefont {Romano}}, \bibinfo {author} {\bibfnamefont
  {J.}~\bibnamefont {Romie}}, \bibinfo {author} {\bibfnamefont
  {R.}~\bibnamefont {Route}}, \bibinfo {author} {\bibfnamefont
  {S.}~\bibnamefont {Rowan}}, \bibinfo {author} {\bibfnamefont
  {A.}~\bibnamefont {R\"udiger}}, \bibinfo {author} {\bibfnamefont
  {L.}~\bibnamefont {Ruet}}, \bibinfo {author} {\bibfnamefont {P.}~\bibnamefont
  {Russell}}, \bibinfo {author} {\bibfnamefont {K.}~\bibnamefont {Ryan}},
  \bibinfo {author} {\bibfnamefont {S.}~\bibnamefont {Sakata}}, \bibinfo
  {author} {\bibfnamefont {M.}~\bibnamefont {Samidi}}, \bibinfo {author}
  {\bibfnamefont {L.~S.}\ \bibnamefont {de~la Jordana}}, \bibinfo {author}
  {\bibfnamefont {V.}~\bibnamefont {Sandberg}}, \bibinfo {author}
  {\bibfnamefont {G.~H.}\ \bibnamefont {Sanders}}, \bibinfo {author}
  {\bibfnamefont {V.}~\bibnamefont {Sannibale}}, \bibinfo {author}
  {\bibfnamefont {S.}~\bibnamefont {Saraf}}, \bibinfo {author} {\bibfnamefont
  {P.}~\bibnamefont {Sarin}}, \bibinfo {author} {\bibfnamefont {B.~S.}\
  \bibnamefont {Sathyaprakash}}, \bibinfo {author} {\bibfnamefont
  {S.}~\bibnamefont {Sato}}, \bibinfo {author} {\bibfnamefont {P.~R.}\
  \bibnamefont {Saulson}}, \bibinfo {author} {\bibfnamefont {R.}~\bibnamefont
  {Savage}}, \bibinfo {author} {\bibfnamefont {P.}~\bibnamefont {Savov}},
  \bibinfo {author} {\bibfnamefont {A.}~\bibnamefont {Sazonov}}, \bibinfo
  {author} {\bibfnamefont {S.}~\bibnamefont {Schediwy}}, \bibinfo {author}
  {\bibfnamefont {R.}~\bibnamefont {Schilling}}, \bibinfo {author}
  {\bibfnamefont {R.}~\bibnamefont {Schnabel}}, \bibinfo {author}
  {\bibfnamefont {R.}~\bibnamefont {Schofield}}, \bibinfo {author}
  {\bibfnamefont {B.~F.}\ \bibnamefont {Schutz}}, \bibinfo {author}
  {\bibfnamefont {P.}~\bibnamefont {Schwinberg}}, \bibinfo {author}
  {\bibfnamefont {S.~M.}\ \bibnamefont {Scott}}, \bibinfo {author}
  {\bibfnamefont {A.~C.}\ \bibnamefont {Searle}}, \bibinfo {author}
  {\bibfnamefont {B.}~\bibnamefont {Sears}}, \bibinfo {author} {\bibfnamefont
  {F.}~\bibnamefont {Seifert}}, \bibinfo {author} {\bibfnamefont
  {D.}~\bibnamefont {Sellers}}, \bibinfo {author} {\bibfnamefont {A.~S.}\
  \bibnamefont {Sengupta}}, \bibinfo {author} {\bibfnamefont {P.}~\bibnamefont
  {Shawhan}}, \bibinfo {author} {\bibfnamefont {D.~H.}\ \bibnamefont
  {Shoemaker}}, \bibinfo {author} {\bibfnamefont {A.}~\bibnamefont {Sibley}},
  \bibinfo {author} {\bibfnamefont {J.~A.}\ \bibnamefont {Sidles}}, \bibinfo
  {author} {\bibfnamefont {X.}~\bibnamefont {Siemens}}, \bibinfo {author}
  {\bibfnamefont {D.}~\bibnamefont {Sigg}}, \bibinfo {author} {\bibfnamefont
  {S.}~\bibnamefont {Sinha}}, \bibinfo {author} {\bibfnamefont {A.~M.}\
  \bibnamefont {Sintes}}, \bibinfo {author} {\bibfnamefont {B.~J.~J.}\
  \bibnamefont {Slagmolen}}, \bibinfo {author} {\bibfnamefont {J.}~\bibnamefont
  {Slutsky}}, \bibinfo {author} {\bibfnamefont {J.~R.}\ \bibnamefont {Smith}},
  \bibinfo {author} {\bibfnamefont {M.~R.}\ \bibnamefont {Smith}}, \bibinfo
  {author} {\bibfnamefont {K.}~\bibnamefont {Somiya}}, \bibinfo {author}
  {\bibfnamefont {K.~A.}\ \bibnamefont {Strain}}, \bibinfo {author}
  {\bibfnamefont {D.~M.}\ \bibnamefont {Strom}}, \bibinfo {author}
  {\bibfnamefont {A.}~\bibnamefont {Stuver}}, \bibinfo {author} {\bibfnamefont
  {T.~Z.}\ \bibnamefont {Summerscales}}, \bibinfo {author} {\bibfnamefont
  {K.-X.}\ \bibnamefont {Sun}}, \bibinfo {author} {\bibfnamefont
  {M.}~\bibnamefont {Sung}}, \bibinfo {author} {\bibfnamefont {P.~J.}\
  \bibnamefont {Sutton}}, \bibinfo {author} {\bibfnamefont {H.}~\bibnamefont
  {Takahashi}}, \bibinfo {author} {\bibfnamefont {D.~B.}\ \bibnamefont
  {Tanner}}, \bibinfo {author} {\bibfnamefont {M.}~\bibnamefont {Tarallo}},
  \bibinfo {author} {\bibfnamefont {R.}~\bibnamefont {Taylor}}, \bibinfo
  {author} {\bibfnamefont {R.}~\bibnamefont {Taylor}}, \bibinfo {author}
  {\bibfnamefont {J.}~\bibnamefont {Thacker}}, \bibinfo {author} {\bibfnamefont
  {K.~A.}\ \bibnamefont {Thorne}}, \bibinfo {author} {\bibfnamefont {K.~S.}\
  \bibnamefont {Thorne}}, \bibinfo {author} {\bibfnamefont {A.}~\bibnamefont
  {Th\"uring}}, \bibinfo {author} {\bibfnamefont {K.~V.}\ \bibnamefont
  {Tokmakov}}, \bibinfo {author} {\bibfnamefont {C.}~\bibnamefont {Torres}},
  \bibinfo {author} {\bibfnamefont {C.}~\bibnamefont {Torrie}}, \bibinfo
  {author} {\bibfnamefont {G.}~\bibnamefont {Traylor}}, \bibinfo {author}
  {\bibfnamefont {M.}~\bibnamefont {Trias}}, \bibinfo {author} {\bibfnamefont
  {W.}~\bibnamefont {Tyler}}, \bibinfo {author} {\bibfnamefont
  {D.}~\bibnamefont {Ugolini}}, \bibinfo {author} {\bibfnamefont
  {C.}~\bibnamefont {Ungarelli}}, \bibinfo {author} {\bibfnamefont
  {K.}~\bibnamefont {Urbanek}}, \bibinfo {author} {\bibfnamefont
  {H.}~\bibnamefont {Vahlbruch}}, \bibinfo {author} {\bibfnamefont
  {M.}~\bibnamefont {Vallisneri}}, \bibinfo {author} {\bibfnamefont
  {C.}~\bibnamefont {Van Den~Broeck}}, \bibinfo {author} {\bibfnamefont
  {M.}~\bibnamefont {van Putten}}, \bibinfo {author} {\bibfnamefont
  {M.}~\bibnamefont {Varvella}}, \bibinfo {author} {\bibfnamefont
  {S.}~\bibnamefont {Vass}}, \bibinfo {author} {\bibfnamefont {A.}~\bibnamefont
  {Vecchio}}, \bibinfo {author} {\bibfnamefont {J.}~\bibnamefont {Veitch}},
  \bibinfo {author} {\bibfnamefont {P.}~\bibnamefont {Veitch}}, \bibinfo
  {author} {\bibfnamefont {A.}~\bibnamefont {Villar}}, \bibinfo {author}
  {\bibfnamefont {C.}~\bibnamefont {Vorvick}}, \bibinfo {author} {\bibfnamefont
  {S.~P.}\ \bibnamefont {Vyachanin}}, \bibinfo {author} {\bibfnamefont {S.~J.}\
  \bibnamefont {Waldman}}, \bibinfo {author} {\bibfnamefont {L.}~\bibnamefont
  {Wallace}}, \bibinfo {author} {\bibfnamefont {H.}~\bibnamefont {Ward}},
  \bibinfo {author} {\bibfnamefont {R.}~\bibnamefont {Ward}}, \bibinfo {author}
  {\bibfnamefont {K.}~\bibnamefont {Watts}}, \bibinfo {author} {\bibfnamefont
  {D.}~\bibnamefont {Webber}}, \bibinfo {author} {\bibfnamefont
  {A.}~\bibnamefont {Weidner}}, \bibinfo {author} {\bibfnamefont
  {M.}~\bibnamefont {Weinert}}, \bibinfo {author} {\bibfnamefont
  {A.}~\bibnamefont {Weinstein}}, \bibinfo {author} {\bibfnamefont
  {R.}~\bibnamefont {Weiss}}, \bibinfo {author} {\bibfnamefont
  {S.}~\bibnamefont {Wen}}, \bibinfo {author} {\bibfnamefont {K.}~\bibnamefont
  {Wette}}, \bibinfo {author} {\bibfnamefont {J.~T.}\ \bibnamefont {Whelan}},
  \bibinfo {author} {\bibfnamefont {D.~M.}\ \bibnamefont {Whitbeck}}, \bibinfo
  {author} {\bibfnamefont {S.~E.}\ \bibnamefont {Whitcomb}}, \bibinfo {author}
  {\bibfnamefont {B.~F.}\ \bibnamefont {Whiting}}, \bibinfo {author}
  {\bibfnamefont {S.}~\bibnamefont {Wiley}}, \bibinfo {author} {\bibfnamefont
  {C.}~\bibnamefont {Wilkinson}}, \bibinfo {author} {\bibfnamefont {P.~A.}\
  \bibnamefont {Willems}}, \bibinfo {author} {\bibfnamefont {L.}~\bibnamefont
  {Williams}}, \bibinfo {author} {\bibfnamefont {B.}~\bibnamefont {Willke}},
  \bibinfo {author} {\bibfnamefont {I.}~\bibnamefont {Wilmut}}, \bibinfo
  {author} {\bibfnamefont {W.}~\bibnamefont {Winkler}}, \bibinfo {author}
  {\bibfnamefont {C.~C.}\ \bibnamefont {Wipf}}, \bibinfo {author}
  {\bibfnamefont {S.}~\bibnamefont {Wise}}, \bibinfo {author} {\bibfnamefont
  {A.~G.}\ \bibnamefont {Wiseman}}, \bibinfo {author} {\bibfnamefont
  {G.}~\bibnamefont {Woan}}, \bibinfo {author} {\bibfnamefont {D.}~\bibnamefont
  {Woods}}, \bibinfo {author} {\bibfnamefont {R.}~\bibnamefont {Wooley}},
  \bibinfo {author} {\bibfnamefont {J.}~\bibnamefont {Worden}}, \bibinfo
  {author} {\bibfnamefont {W.}~\bibnamefont {Wu}}, \bibinfo {author}
  {\bibfnamefont {I.}~\bibnamefont {Yakushin}}, \bibinfo {author}
  {\bibfnamefont {H.}~\bibnamefont {Yamamoto}}, \bibinfo {author}
  {\bibfnamefont {Z.}~\bibnamefont {Yan}}, \bibinfo {author} {\bibfnamefont
  {S.}~\bibnamefont {Yoshida}}, \bibinfo {author} {\bibfnamefont
  {N.}~\bibnamefont {Yunes}}, \bibinfo {author} {\bibfnamefont
  {M.}~\bibnamefont {Zanolin}}, \bibinfo {author} {\bibfnamefont
  {J.}~\bibnamefont {Zhang}}, \bibinfo {author} {\bibfnamefont
  {L.}~\bibnamefont {Zhang}}, \bibinfo {author} {\bibfnamefont
  {C.}~\bibnamefont {Zhao}}, \bibinfo {author} {\bibfnamefont {N.}~\bibnamefont
  {Zotov}}, \bibinfo {author} {\bibfnamefont {M.}~\bibnamefont {Zucker}},
  \bibinfo {author} {\bibfnamefont {H.}~\bibnamefont {zur M\"uhlen}}, \bibinfo
  {author} {\bibfnamefont {J.}~\bibnamefont {Zweizig}}, \bibinfo {author}
  {\bibfnamefont {M.}~\bibnamefont {Kramer}},\ and\ \bibinfo {author}
  {\bibfnamefont {A.~G.}\ \bibnamefont {Lyne}} (\bibinfo {collaboration} {LIGO
  Scientific Collaboration}),\ }\bibfield  {title} {\bibinfo {title} {Upper
  limits on gravitational wave emission from 78 radio pulsars},\ }\href
  {https://doi.org/10.1103/PhysRevD.76.042001} {\bibfield  {journal} {\bibinfo
  {journal} {Phys. Rev. D}\ }\textbf {\bibinfo {volume} {76}},\ \bibinfo
  {pages} {042001} (\bibinfo {year} {2007})}\BibitemShut {NoStop}%
\bibitem [{\citenamefont {Abbott}\ \emph {et~al.}(2008)\citenamefont {Abbott},
  \citenamefont {Abbott}, \citenamefont {Adhikari}, \citenamefont {Ajith},
  \citenamefont {Allen}, \citenamefont {Allen}, \citenamefont {Amin},
  \citenamefont {Anderson}, \citenamefont {Anderson}, \citenamefont {Arain},
  \citenamefont {Araya}, \citenamefont {Armandula}, \citenamefont {Armor},
  \citenamefont {Aso}, \citenamefont {Aston}, \citenamefont {Aufmuth},
  \citenamefont {Aulbert}, \citenamefont {Babak}, \citenamefont {Ballmer},
  \citenamefont {Bantilan}, \citenamefont {Barish}, \citenamefont {Barker},
  \citenamefont {Barker}, \citenamefont {Barr}, \citenamefont {Barriga},
  \citenamefont {Barton}, \citenamefont {Bartos}, \citenamefont {Bastarrika},
  \citenamefont {Bayer}, \citenamefont {Betzwieser}, \citenamefont
  {Beyersdorf}, \citenamefont {Bilenko}, \citenamefont {Billingsley},
  \citenamefont {Biswas}, \citenamefont {Black}, \citenamefont {Blackburn},
  \citenamefont {Blackburn}, \citenamefont {Blair}, \citenamefont {Bland},
  \citenamefont {Bodiya}, \citenamefont {Bogue}, \citenamefont {Bork},
  \citenamefont {Boschi}, \citenamefont {Bose}, \citenamefont {Brady},
  \citenamefont {Braginsky}, \citenamefont {Brau}, \citenamefont {Brinkmann},
  \citenamefont {Brooks}, \citenamefont {Brown}, \citenamefont {Brunet},
  \citenamefont {Bullington}, \citenamefont {Buonanno}, \citenamefont
  {Burmeister}, \citenamefont {Byer}, \citenamefont {Cadonati}, \citenamefont
  {Cagnoli}, \citenamefont {Camp}, \citenamefont {Cannizzo}, \citenamefont
  {Cannon}, \citenamefont {Cao}, \citenamefont {Cardenas}, \citenamefont
  {Casebolt}, \citenamefont {Castaldi}, \citenamefont {Cepeda}, \citenamefont
  {Chalkley}, \citenamefont {Charlton}, \citenamefont {Chatterji},
  \citenamefont {Chelkowski}, \citenamefont {Chen}, \citenamefont
  {Christensen}, \citenamefont {Clark}, \citenamefont {Clark}, \citenamefont
  {Cokelaer}, \citenamefont {Conte}, \citenamefont {Cook}, \citenamefont
  {Corbitt}, \citenamefont {Coyne}, \citenamefont {Creighton}, \citenamefont
  {Cumming}, \citenamefont {Cunningham}, \citenamefont {Cutler}, \citenamefont
  {Dalrymple}, \citenamefont {Danzmann}, \citenamefont {Davies}, \citenamefont
  {DeBra}, \citenamefont {Degallaix}, \citenamefont {Degree}, \citenamefont
  {Dergachev}, \citenamefont {Desai}, \citenamefont {DeSalvo}, \citenamefont
  {Dhurandhar}, \citenamefont {Díaz}, \citenamefont {Dickson}, \citenamefont
  {Credico}, \citenamefont {Dietz}, \citenamefont {Donovan}, \citenamefont
  {Dooley}, \citenamefont {Doomes}, \citenamefont {Drever}, \citenamefont
  {Duke}, \citenamefont {Dumas}, \citenamefont {Dupuis}, \citenamefont {Dwyer},
  \citenamefont {Echols}, \citenamefont {Effler}, \citenamefont {Ehrens},
  \citenamefont {Espinoza}, \citenamefont {Etzel}, \citenamefont {Evans},
  \citenamefont {Fairhurst}, \citenamefont {Fan}, \citenamefont {Fazi},
  \citenamefont {Fehrmann}, \citenamefont {Fejer}, \citenamefont {Finn},
  \citenamefont {Flasch}, \citenamefont {Fotopoulos}, \citenamefont {Freise},
  \citenamefont {Frey}, \citenamefont {Fricke}, \citenamefont {Fritschel},
  \citenamefont {Frolov}, \citenamefont {Fyffe}, \citenamefont {Garofoli},
  \citenamefont {Gholami}, \citenamefont {Giaime}, \citenamefont {Giampanis},
  \citenamefont {Giardina}, \citenamefont {Goda}, \citenamefont {Goetz},
  \citenamefont {Goggin}, \citenamefont {González}, \citenamefont {Gossler},
  \citenamefont {Gouaty}, \citenamefont {Grant}, \citenamefont {Gras},
  \citenamefont {Gray}, \citenamefont {Gray}, \citenamefont {Greenhalgh},
  \citenamefont {Gretarsson}, \citenamefont {Grimaldi}, \citenamefont {Grosso},
  \citenamefont {Grote}, \citenamefont {Grunewald}, \citenamefont {Guenther},
  \citenamefont {Gustafson}, \citenamefont {Gustafson}, \citenamefont {Hage},
  \citenamefont {Hallam}, \citenamefont {Hammer}, \citenamefont {Hanna},
  \citenamefont {Hanson}, \citenamefont {Harms}, \citenamefont {Harry},
  \citenamefont {Harstad}, \citenamefont {Hayama}, \citenamefont {Hayler},
  \citenamefont {Heefner}, \citenamefont {Heng}, \citenamefont {Hennessy},
  \citenamefont {Heptonstall}, \citenamefont {Hewitson}, \citenamefont {Hild},
  \citenamefont {Hirose}, \citenamefont {Hoak}, \citenamefont {Hosken},
  \citenamefont {Hough}, \citenamefont {Hughey}, \citenamefont {Huttner},
  \citenamefont {Ingram}, \citenamefont {Ito}, \citenamefont {Ivanov},
  \citenamefont {Johnson}, \citenamefont {Johnson}, \citenamefont {Jones},
  \citenamefont {Jones}, \citenamefont {Jones}, \citenamefont {Ju},
  \citenamefont {Kalmus}, \citenamefont {Kalogera}, \citenamefont {Kamat},
  \citenamefont {Kanner}, \citenamefont {Kasprzyk}, \citenamefont
  {Katsavounidis}, \citenamefont {Kawabe}, \citenamefont {Kawamura},
  \citenamefont {Kawazoe}, \citenamefont {Kells}, \citenamefont {Keppel},
  \citenamefont {Khalili}, \citenamefont {Khan}, \citenamefont {Khazanov},
  \citenamefont {Kim}, \citenamefont {King}, \citenamefont {Kissel},
  \citenamefont {Klimenko}, \citenamefont {Kokeyama}, \citenamefont
  {Kondrashov}, \citenamefont {Kopparapu}, \citenamefont {Kozak}, \citenamefont
  {Kozhevatov}, \citenamefont {Krishnan}, \citenamefont {Kwee}, \citenamefont
  {Lam}, \citenamefont {Landry}, \citenamefont {Lang}, \citenamefont {Lantz},
  \citenamefont {Lazzarini}, \citenamefont {Lei}, \citenamefont {Leindecker},
  \citenamefont {Leonhardt}, \citenamefont {Leonor}, \citenamefont {Libbrecht},
  \citenamefont {Lin}, \citenamefont {Lindquist}, \citenamefont {Lockerbie},
  \citenamefont {Lodhia}, \citenamefont {Lormand}, \citenamefont {Lu},
  \citenamefont {Lubinski}, \citenamefont {Lucianetti}, \citenamefont {Lück},
  \citenamefont {Machenschalk}, \citenamefont {MacInnis}, \citenamefont
  {Mageswaran}, \citenamefont {Mailand}, \citenamefont {Mandic}, \citenamefont
  {Márka}, \citenamefont {Márka}, \citenamefont {Markosyan}, \citenamefont
  {Markowitz}, \citenamefont {Maros}, \citenamefont {Martin}, \citenamefont
  {Martin}, \citenamefont {Marx}, \citenamefont {Mason}, \citenamefont
  {Matichard}, \citenamefont {Matone}, \citenamefont {Matzner}, \citenamefont
  {Mavalvala}, \citenamefont {McCarthy}, \citenamefont {McClelland},
  \citenamefont {McGuire}, \citenamefont {McHugh}, \citenamefont {McIntyre},
  \citenamefont {McIvor}, \citenamefont {McKechan}, \citenamefont {McKenzie},
  \citenamefont {Meier}, \citenamefont {Melissinos}, \citenamefont {Mendell},
  \citenamefont {Mercer}, \citenamefont {Meshkov}, \citenamefont {Messenger},
  \citenamefont {Meyers}, \citenamefont {Miller}, \citenamefont {Minelli},
  \citenamefont {Mitra}, \citenamefont {Mitrofanov}, \citenamefont
  {Mitselmakher}, \citenamefont {Mittleman}, \citenamefont {Miyakawa},
  \citenamefont {Moe}, \citenamefont {Mohanty}, \citenamefont {Moreno},
  \citenamefont {Mossavi}, \citenamefont {MowLowry}, \citenamefont {Mueller},
  \citenamefont {Mukherjee}, \citenamefont {Mukhopadhyay}, \citenamefont
  {Müller-Ebhardt}, \citenamefont {Munch}, \citenamefont {Murray},
  \citenamefont {Myers}, \citenamefont {Myers}, \citenamefont {Nash},
  \citenamefont {Nelson}, \citenamefont {Newton}, \citenamefont {Nishizawa},
  \citenamefont {Numata}, \citenamefont {O'Dell}, \citenamefont {Ogin},
  \citenamefont {O'Reilly}, \citenamefont {O'Shaughnessy}, \citenamefont
  {Ottaway}, \citenamefont {Ottens}, \citenamefont {Overmier}, \citenamefont
  {Owen}, \citenamefont {Pan}, \citenamefont {Pankow}, \citenamefont {Papa},
  \citenamefont {Parameshwaraiah}, \citenamefont {Patel}, \citenamefont
  {Pedraza}, \citenamefont {Penn}, \citenamefont {Perreca}, \citenamefont
  {Petrie}, \citenamefont {Pinto}, \citenamefont {Pitkin}, \citenamefont
  {Pletsch}, \citenamefont {Plissi}, \citenamefont {Postiglione}, \citenamefont
  {Principe}, \citenamefont {Prix}, \citenamefont {Quetschke}, \citenamefont
  {Raab}, \citenamefont {Rabeling}, \citenamefont {Radkins}, \citenamefont
  {Rainer}, \citenamefont {Rakhmanov}, \citenamefont {Ramsunder}, \citenamefont
  {Rehbein}, \citenamefont {Reid}, \citenamefont {Reitze}, \citenamefont
  {Riesen}, \citenamefont {Riles}, \citenamefont {Rivera}, \citenamefont
  {Robertson}, \citenamefont {Robinson}, \citenamefont {Robinson},
  \citenamefont {Roddy}, \citenamefont {Rodriguez}, \citenamefont {Rogan},
  \citenamefont {Rollins}, \citenamefont {Romano}, \citenamefont {Romie},
  \citenamefont {Route}, \citenamefont {Rowan}, \citenamefont {Rüdiger},
  \citenamefont {Ruet}, \citenamefont {Russell}, \citenamefont {Ryan},
  \citenamefont {Sakata}, \citenamefont {Samidi}, \citenamefont {de~la
  Jordana}, \citenamefont {Sandberg}, \citenamefont {Sannibale}, \citenamefont
  {Saraf}, \citenamefont {Sarin}, \citenamefont {Sathyaprakash}, \citenamefont
  {Sato}, \citenamefont {Saulson}, \citenamefont {Savage}, \citenamefont
  {Savov}, \citenamefont {Schediwy}, \citenamefont {Schilling}, \citenamefont
  {Schnabel}, \citenamefont {Schofield}, \citenamefont {Schutz}, \citenamefont
  {Schwinberg}, \citenamefont {Scott}, \citenamefont {Searle}, \citenamefont
  {Sears}, \citenamefont {Seifert}, \citenamefont {Sellers}, \citenamefont
  {Sengupta}, \citenamefont {Shawhan}, \citenamefont {Shoemaker}, \citenamefont
  {Sibley}, \citenamefont {Siemens}, \citenamefont {Sigg}, \citenamefont
  {Sinha}, \citenamefont {Sintes}, \citenamefont {Slagmolen}, \citenamefont
  {Slutsky}, \citenamefont {Smith}, \citenamefont {Smith}, \citenamefont
  {Smith}, \citenamefont {Somiya}, \citenamefont {Sorazu}, \citenamefont
  {Stein}, \citenamefont {Stochino}, \citenamefont {Stone}, \citenamefont
  {Strain}, \citenamefont {Strom}, \citenamefont {Stuver}, \citenamefont
  {Summerscales}, \citenamefont {Sun}, \citenamefont {Sung}, \citenamefont
  {Sutton}, \citenamefont {Takahashi}, \citenamefont {Tanner}, \citenamefont
  {Taylor}, \citenamefont {Taylor}, \citenamefont {Thacker}, \citenamefont
  {Thorne}, \citenamefont {Thorne}, \citenamefont {Thüring}, \citenamefont
  {Tinto}, \citenamefont {Tokmakov}, \citenamefont {Torres}, \citenamefont
  {Torrie}, \citenamefont {Traylor}, \citenamefont {Trias}, \citenamefont
  {Tyler}, \citenamefont {Ugolini}, \citenamefont {Ulmen}, \citenamefont
  {Urbanek}, \citenamefont {Vahlbruch}, \citenamefont {Broeck}, \citenamefont
  {van~der Sluys}, \citenamefont {Vass}, \citenamefont {Vaulin}, \citenamefont
  {Vecchio}, \citenamefont {Veitch}, \citenamefont {Veitch}, \citenamefont
  {Villar}, \citenamefont {Vorvick}, \citenamefont {Vyachanin}, \citenamefont
  {Waldman}, \citenamefont {Wallace}, \citenamefont {Ward}, \citenamefont
  {Ward}, \citenamefont {Weinert}, \citenamefont {Weinstein}, \citenamefont
  {Weiss}, \citenamefont {Wen}, \citenamefont {Wette}, \citenamefont {Whelan},
  \citenamefont {Whitcomb}, \citenamefont {Whiting}, \citenamefont {Wilkinson},
  \citenamefont {Willems}, \citenamefont {Williams}, \citenamefont {Williams},
  \citenamefont {Willke}, \citenamefont {Wilmut}, \citenamefont {Winkler},
  \citenamefont {Wipf}, \citenamefont {Wiseman}, \citenamefont {Woan},
  \citenamefont {Wooley}, \citenamefont {Worden}, \citenamefont {Wu},
  \citenamefont {Yakushin}, \citenamefont {Yamamoto}, \citenamefont {Yan},
  \citenamefont {Yoshida}, \citenamefont {Zanolin}, \citenamefont {Zhang},
  \citenamefont {Zhang}, \citenamefont {Zhao}, \citenamefont {Zotov},
  \citenamefont {Zucker}, \citenamefont {Zweizig},\ and\ \citenamefont
  {Collaboration)}}]{Abbott_2008}%
  \BibitemOpen
  \bibfield  {author} {\bibinfo {author} {\bibfnamefont {B.}~\bibnamefont
  {Abbott}}, \bibinfo {author} {\bibfnamefont {R.}~\bibnamefont {Abbott}},
  \bibinfo {author} {\bibfnamefont {R.}~\bibnamefont {Adhikari}}, \bibinfo
  {author} {\bibfnamefont {P.}~\bibnamefont {Ajith}}, \bibinfo {author}
  {\bibfnamefont {B.}~\bibnamefont {Allen}}, \bibinfo {author} {\bibfnamefont
  {G.}~\bibnamefont {Allen}}, \bibinfo {author} {\bibfnamefont
  {R.}~\bibnamefont {Amin}}, \bibinfo {author} {\bibfnamefont {S.~B.}\
  \bibnamefont {Anderson}}, \bibinfo {author} {\bibfnamefont {W.~G.}\
  \bibnamefont {Anderson}}, \bibinfo {author} {\bibfnamefont {M.~A.}\
  \bibnamefont {Arain}}, \bibinfo {author} {\bibfnamefont {M.}~\bibnamefont
  {Araya}}, \bibinfo {author} {\bibfnamefont {H.}~\bibnamefont {Armandula}},
  \bibinfo {author} {\bibfnamefont {P.}~\bibnamefont {Armor}}, \bibinfo
  {author} {\bibfnamefont {Y.}~\bibnamefont {Aso}}, \bibinfo {author}
  {\bibfnamefont {S.}~\bibnamefont {Aston}}, \bibinfo {author} {\bibfnamefont
  {P.}~\bibnamefont {Aufmuth}}, \bibinfo {author} {\bibfnamefont
  {C.}~\bibnamefont {Aulbert}}, \bibinfo {author} {\bibfnamefont
  {S.}~\bibnamefont {Babak}}, \bibinfo {author} {\bibfnamefont
  {S.}~\bibnamefont {Ballmer}}, \bibinfo {author} {\bibfnamefont
  {H.}~\bibnamefont {Bantilan}}, \bibinfo {author} {\bibfnamefont {B.~C.}\
  \bibnamefont {Barish}}, \bibinfo {author} {\bibfnamefont {C.}~\bibnamefont
  {Barker}}, \bibinfo {author} {\bibfnamefont {D.}~\bibnamefont {Barker}},
  \bibinfo {author} {\bibfnamefont {B.}~\bibnamefont {Barr}}, \bibinfo {author}
  {\bibfnamefont {P.}~\bibnamefont {Barriga}}, \bibinfo {author} {\bibfnamefont
  {M.~A.}\ \bibnamefont {Barton}}, \bibinfo {author} {\bibfnamefont
  {I.}~\bibnamefont {Bartos}}, \bibinfo {author} {\bibfnamefont
  {M.}~\bibnamefont {Bastarrika}}, \bibinfo {author} {\bibfnamefont
  {K.}~\bibnamefont {Bayer}}, \bibinfo {author} {\bibfnamefont
  {J.}~\bibnamefont {Betzwieser}}, \bibinfo {author} {\bibfnamefont {P.~T.}\
  \bibnamefont {Beyersdorf}}, \bibinfo {author} {\bibfnamefont {I.~A.}\
  \bibnamefont {Bilenko}}, \bibinfo {author} {\bibfnamefont {G.}~\bibnamefont
  {Billingsley}}, \bibinfo {author} {\bibfnamefont {R.}~\bibnamefont {Biswas}},
  \bibinfo {author} {\bibfnamefont {E.}~\bibnamefont {Black}}, \bibinfo
  {author} {\bibfnamefont {K.}~\bibnamefont {Blackburn}}, \bibinfo {author}
  {\bibfnamefont {L.}~\bibnamefont {Blackburn}}, \bibinfo {author}
  {\bibfnamefont {D.}~\bibnamefont {Blair}}, \bibinfo {author} {\bibfnamefont
  {B.}~\bibnamefont {Bland}}, \bibinfo {author} {\bibfnamefont {T.~P.}\
  \bibnamefont {Bodiya}}, \bibinfo {author} {\bibfnamefont {L.}~\bibnamefont
  {Bogue}}, \bibinfo {author} {\bibfnamefont {R.}~\bibnamefont {Bork}},
  \bibinfo {author} {\bibfnamefont {V.}~\bibnamefont {Boschi}}, \bibinfo
  {author} {\bibfnamefont {S.}~\bibnamefont {Bose}}, \bibinfo {author}
  {\bibfnamefont {P.~R.}\ \bibnamefont {Brady}}, \bibinfo {author}
  {\bibfnamefont {V.~B.}\ \bibnamefont {Braginsky}}, \bibinfo {author}
  {\bibfnamefont {J.~E.}\ \bibnamefont {Brau}}, \bibinfo {author}
  {\bibfnamefont {M.}~\bibnamefont {Brinkmann}}, \bibinfo {author}
  {\bibfnamefont {A.}~\bibnamefont {Brooks}}, \bibinfo {author} {\bibfnamefont
  {D.~A.}\ \bibnamefont {Brown}}, \bibinfo {author} {\bibfnamefont
  {G.}~\bibnamefont {Brunet}}, \bibinfo {author} {\bibfnamefont
  {A.}~\bibnamefont {Bullington}}, \bibinfo {author} {\bibfnamefont
  {A.}~\bibnamefont {Buonanno}}, \bibinfo {author} {\bibfnamefont
  {O.}~\bibnamefont {Burmeister}}, \bibinfo {author} {\bibfnamefont {R.~L.}\
  \bibnamefont {Byer}}, \bibinfo {author} {\bibfnamefont {L.}~\bibnamefont
  {Cadonati}}, \bibinfo {author} {\bibfnamefont {G.}~\bibnamefont {Cagnoli}},
  \bibinfo {author} {\bibfnamefont {J.~B.}\ \bibnamefont {Camp}}, \bibinfo
  {author} {\bibfnamefont {J.}~\bibnamefont {Cannizzo}}, \bibinfo {author}
  {\bibfnamefont {K.}~\bibnamefont {Cannon}}, \bibinfo {author} {\bibfnamefont
  {J.}~\bibnamefont {Cao}}, \bibinfo {author} {\bibfnamefont {L.}~\bibnamefont
  {Cardenas}}, \bibinfo {author} {\bibfnamefont {T.}~\bibnamefont {Casebolt}},
  \bibinfo {author} {\bibfnamefont {G.}~\bibnamefont {Castaldi}}, \bibinfo
  {author} {\bibfnamefont {C.}~\bibnamefont {Cepeda}}, \bibinfo {author}
  {\bibfnamefont {E.}~\bibnamefont {Chalkley}}, \bibinfo {author}
  {\bibfnamefont {P.}~\bibnamefont {Charlton}}, \bibinfo {author}
  {\bibfnamefont {S.}~\bibnamefont {Chatterji}}, \bibinfo {author}
  {\bibfnamefont {S.}~\bibnamefont {Chelkowski}}, \bibinfo {author}
  {\bibfnamefont {Y.}~\bibnamefont {Chen}}, \bibinfo {author} {\bibfnamefont
  {N.}~\bibnamefont {Christensen}}, \bibinfo {author} {\bibfnamefont
  {D.}~\bibnamefont {Clark}}, \bibinfo {author} {\bibfnamefont
  {J.}~\bibnamefont {Clark}}, \bibinfo {author} {\bibfnamefont
  {T.}~\bibnamefont {Cokelaer}}, \bibinfo {author} {\bibfnamefont
  {R.}~\bibnamefont {Conte}}, \bibinfo {author} {\bibfnamefont
  {D.}~\bibnamefont {Cook}}, \bibinfo {author} {\bibfnamefont {T.}~\bibnamefont
  {Corbitt}}, \bibinfo {author} {\bibfnamefont {D.}~\bibnamefont {Coyne}},
  \bibinfo {author} {\bibfnamefont {J.~D.~E.}\ \bibnamefont {Creighton}},
  \bibinfo {author} {\bibfnamefont {A.}~\bibnamefont {Cumming}}, \bibinfo
  {author} {\bibfnamefont {L.}~\bibnamefont {Cunningham}}, \bibinfo {author}
  {\bibfnamefont {R.~M.}\ \bibnamefont {Cutler}}, \bibinfo {author}
  {\bibfnamefont {J.}~\bibnamefont {Dalrymple}}, \bibinfo {author}
  {\bibfnamefont {K.}~\bibnamefont {Danzmann}}, \bibinfo {author}
  {\bibfnamefont {G.}~\bibnamefont {Davies}}, \bibinfo {author} {\bibfnamefont
  {D.}~\bibnamefont {DeBra}}, \bibinfo {author} {\bibfnamefont
  {J.}~\bibnamefont {Degallaix}}, \bibinfo {author} {\bibfnamefont
  {M.}~\bibnamefont {Degree}}, \bibinfo {author} {\bibfnamefont
  {V.}~\bibnamefont {Dergachev}}, \bibinfo {author} {\bibfnamefont
  {S.}~\bibnamefont {Desai}}, \bibinfo {author} {\bibfnamefont
  {R.}~\bibnamefont {DeSalvo}}, \bibinfo {author} {\bibfnamefont
  {S.}~\bibnamefont {Dhurandhar}}, \bibinfo {author} {\bibfnamefont
  {M.}~\bibnamefont {Díaz}}, \bibinfo {author} {\bibfnamefont
  {J.}~\bibnamefont {Dickson}}, \bibinfo {author} {\bibfnamefont {A.~D.}\
  \bibnamefont {Credico}}, \bibinfo {author} {\bibfnamefont {A.}~\bibnamefont
  {Dietz}}, \bibinfo {author} {\bibfnamefont {F.}~\bibnamefont {Donovan}},
  \bibinfo {author} {\bibfnamefont {K.~L.}\ \bibnamefont {Dooley}}, \bibinfo
  {author} {\bibfnamefont {E.~E.}\ \bibnamefont {Doomes}}, \bibinfo {author}
  {\bibfnamefont {R.~W.~P.}\ \bibnamefont {Drever}}, \bibinfo {author}
  {\bibfnamefont {I.}~\bibnamefont {Duke}}, \bibinfo {author} {\bibfnamefont
  {J.-C.}\ \bibnamefont {Dumas}}, \bibinfo {author} {\bibfnamefont {R.~J.}\
  \bibnamefont {Dupuis}}, \bibinfo {author} {\bibfnamefont {J.~G.}\
  \bibnamefont {Dwyer}}, \bibinfo {author} {\bibfnamefont {C.}~\bibnamefont
  {Echols}}, \bibinfo {author} {\bibfnamefont {A.}~\bibnamefont {Effler}},
  \bibinfo {author} {\bibfnamefont {P.}~\bibnamefont {Ehrens}}, \bibinfo
  {author} {\bibfnamefont {E.}~\bibnamefont {Espinoza}}, \bibinfo {author}
  {\bibfnamefont {T.}~\bibnamefont {Etzel}}, \bibinfo {author} {\bibfnamefont
  {T.}~\bibnamefont {Evans}}, \bibinfo {author} {\bibfnamefont
  {S.}~\bibnamefont {Fairhurst}}, \bibinfo {author} {\bibfnamefont
  {Y.}~\bibnamefont {Fan}}, \bibinfo {author} {\bibfnamefont {D.}~\bibnamefont
  {Fazi}}, \bibinfo {author} {\bibfnamefont {H.}~\bibnamefont {Fehrmann}},
  \bibinfo {author} {\bibfnamefont {M.~M.}\ \bibnamefont {Fejer}}, \bibinfo
  {author} {\bibfnamefont {L.~S.}\ \bibnamefont {Finn}}, \bibinfo {author}
  {\bibfnamefont {K.}~\bibnamefont {Flasch}}, \bibinfo {author} {\bibfnamefont
  {N.}~\bibnamefont {Fotopoulos}}, \bibinfo {author} {\bibfnamefont
  {A.}~\bibnamefont {Freise}}, \bibinfo {author} {\bibfnamefont
  {R.}~\bibnamefont {Frey}}, \bibinfo {author} {\bibfnamefont {T.}~\bibnamefont
  {Fricke}}, \bibinfo {author} {\bibfnamefont {P.}~\bibnamefont {Fritschel}},
  \bibinfo {author} {\bibfnamefont {V.~V.}\ \bibnamefont {Frolov}}, \bibinfo
  {author} {\bibfnamefont {M.}~\bibnamefont {Fyffe}}, \bibinfo {author}
  {\bibfnamefont {J.}~\bibnamefont {Garofoli}}, \bibinfo {author}
  {\bibfnamefont {I.}~\bibnamefont {Gholami}}, \bibinfo {author} {\bibfnamefont
  {J.~A.}\ \bibnamefont {Giaime}}, \bibinfo {author} {\bibfnamefont
  {S.}~\bibnamefont {Giampanis}}, \bibinfo {author} {\bibfnamefont {K.~D.}\
  \bibnamefont {Giardina}}, \bibinfo {author} {\bibfnamefont {K.}~\bibnamefont
  {Goda}}, \bibinfo {author} {\bibfnamefont {E.}~\bibnamefont {Goetz}},
  \bibinfo {author} {\bibfnamefont {L.}~\bibnamefont {Goggin}}, \bibinfo
  {author} {\bibfnamefont {G.}~\bibnamefont {González}}, \bibinfo {author}
  {\bibfnamefont {S.}~\bibnamefont {Gossler}}, \bibinfo {author} {\bibfnamefont
  {R.}~\bibnamefont {Gouaty}}, \bibinfo {author} {\bibfnamefont
  {A.}~\bibnamefont {Grant}}, \bibinfo {author} {\bibfnamefont
  {S.}~\bibnamefont {Gras}}, \bibinfo {author} {\bibfnamefont {C.}~\bibnamefont
  {Gray}}, \bibinfo {author} {\bibfnamefont {M.}~\bibnamefont {Gray}}, \bibinfo
  {author} {\bibfnamefont {R.~J.~S.}\ \bibnamefont {Greenhalgh}}, \bibinfo
  {author} {\bibfnamefont {A.~M.}\ \bibnamefont {Gretarsson}}, \bibinfo
  {author} {\bibfnamefont {F.}~\bibnamefont {Grimaldi}}, \bibinfo {author}
  {\bibfnamefont {R.}~\bibnamefont {Grosso}}, \bibinfo {author} {\bibfnamefont
  {H.}~\bibnamefont {Grote}}, \bibinfo {author} {\bibfnamefont
  {S.}~\bibnamefont {Grunewald}}, \bibinfo {author} {\bibfnamefont
  {M.}~\bibnamefont {Guenther}}, \bibinfo {author} {\bibfnamefont {E.~K.}\
  \bibnamefont {Gustafson}}, \bibinfo {author} {\bibfnamefont {R.}~\bibnamefont
  {Gustafson}}, \bibinfo {author} {\bibfnamefont {B.}~\bibnamefont {Hage}},
  \bibinfo {author} {\bibfnamefont {J.~M.}\ \bibnamefont {Hallam}}, \bibinfo
  {author} {\bibfnamefont {D.}~\bibnamefont {Hammer}}, \bibinfo {author}
  {\bibfnamefont {C.}~\bibnamefont {Hanna}}, \bibinfo {author} {\bibfnamefont
  {J.}~\bibnamefont {Hanson}}, \bibinfo {author} {\bibfnamefont
  {J.}~\bibnamefont {Harms}}, \bibinfo {author} {\bibfnamefont
  {G.}~\bibnamefont {Harry}}, \bibinfo {author} {\bibfnamefont
  {E.}~\bibnamefont {Harstad}}, \bibinfo {author} {\bibfnamefont
  {K.}~\bibnamefont {Hayama}}, \bibinfo {author} {\bibfnamefont
  {T.}~\bibnamefont {Hayler}}, \bibinfo {author} {\bibfnamefont
  {J.}~\bibnamefont {Heefner}}, \bibinfo {author} {\bibfnamefont {I.~S.}\
  \bibnamefont {Heng}}, \bibinfo {author} {\bibfnamefont {M.}~\bibnamefont
  {Hennessy}}, \bibinfo {author} {\bibfnamefont {A.}~\bibnamefont
  {Heptonstall}}, \bibinfo {author} {\bibfnamefont {M.}~\bibnamefont
  {Hewitson}}, \bibinfo {author} {\bibfnamefont {S.}~\bibnamefont {Hild}},
  \bibinfo {author} {\bibfnamefont {E.}~\bibnamefont {Hirose}}, \bibinfo
  {author} {\bibfnamefont {D.}~\bibnamefont {Hoak}}, \bibinfo {author}
  {\bibfnamefont {D.}~\bibnamefont {Hosken}}, \bibinfo {author} {\bibfnamefont
  {J.}~\bibnamefont {Hough}}, \bibinfo {author} {\bibfnamefont
  {B.}~\bibnamefont {Hughey}}, \bibinfo {author} {\bibfnamefont {S.~H.}\
  \bibnamefont {Huttner}}, \bibinfo {author} {\bibfnamefont {D.}~\bibnamefont
  {Ingram}}, \bibinfo {author} {\bibfnamefont {M.}~\bibnamefont {Ito}},
  \bibinfo {author} {\bibfnamefont {A.}~\bibnamefont {Ivanov}}, \bibinfo
  {author} {\bibfnamefont {B.}~\bibnamefont {Johnson}}, \bibinfo {author}
  {\bibfnamefont {W.~W.}\ \bibnamefont {Johnson}}, \bibinfo {author}
  {\bibfnamefont {D.~I.}\ \bibnamefont {Jones}}, \bibinfo {author}
  {\bibfnamefont {G.}~\bibnamefont {Jones}}, \bibinfo {author} {\bibfnamefont
  {R.}~\bibnamefont {Jones}}, \bibinfo {author} {\bibfnamefont
  {L.}~\bibnamefont {Ju}}, \bibinfo {author} {\bibfnamefont {P.}~\bibnamefont
  {Kalmus}}, \bibinfo {author} {\bibfnamefont {V.}~\bibnamefont {Kalogera}},
  \bibinfo {author} {\bibfnamefont {S.}~\bibnamefont {Kamat}}, \bibinfo
  {author} {\bibfnamefont {J.}~\bibnamefont {Kanner}}, \bibinfo {author}
  {\bibfnamefont {D.}~\bibnamefont {Kasprzyk}}, \bibinfo {author}
  {\bibfnamefont {E.}~\bibnamefont {Katsavounidis}}, \bibinfo {author}
  {\bibfnamefont {K.}~\bibnamefont {Kawabe}}, \bibinfo {author} {\bibfnamefont
  {S.}~\bibnamefont {Kawamura}}, \bibinfo {author} {\bibfnamefont
  {F.}~\bibnamefont {Kawazoe}}, \bibinfo {author} {\bibfnamefont
  {W.}~\bibnamefont {Kells}}, \bibinfo {author} {\bibfnamefont {D.~G.}\
  \bibnamefont {Keppel}}, \bibinfo {author} {\bibfnamefont {F.~Y.}\
  \bibnamefont {Khalili}}, \bibinfo {author} {\bibfnamefont {R.}~\bibnamefont
  {Khan}}, \bibinfo {author} {\bibfnamefont {E.}~\bibnamefont {Khazanov}},
  \bibinfo {author} {\bibfnamefont {C.}~\bibnamefont {Kim}}, \bibinfo {author}
  {\bibfnamefont {P.}~\bibnamefont {King}}, \bibinfo {author} {\bibfnamefont
  {J.~S.}\ \bibnamefont {Kissel}}, \bibinfo {author} {\bibfnamefont
  {S.}~\bibnamefont {Klimenko}}, \bibinfo {author} {\bibfnamefont
  {K.}~\bibnamefont {Kokeyama}}, \bibinfo {author} {\bibfnamefont
  {V.}~\bibnamefont {Kondrashov}}, \bibinfo {author} {\bibfnamefont {R.~K.}\
  \bibnamefont {Kopparapu}}, \bibinfo {author} {\bibfnamefont {D.}~\bibnamefont
  {Kozak}}, \bibinfo {author} {\bibfnamefont {I.}~\bibnamefont {Kozhevatov}},
  \bibinfo {author} {\bibfnamefont {B.}~\bibnamefont {Krishnan}}, \bibinfo
  {author} {\bibfnamefont {P.}~\bibnamefont {Kwee}}, \bibinfo {author}
  {\bibfnamefont {P.~K.}\ \bibnamefont {Lam}}, \bibinfo {author} {\bibfnamefont
  {M.}~\bibnamefont {Landry}}, \bibinfo {author} {\bibfnamefont {M.~M.}\
  \bibnamefont {Lang}}, \bibinfo {author} {\bibfnamefont {B.}~\bibnamefont
  {Lantz}}, \bibinfo {author} {\bibfnamefont {A.}~\bibnamefont {Lazzarini}},
  \bibinfo {author} {\bibfnamefont {M.}~\bibnamefont {Lei}}, \bibinfo {author}
  {\bibfnamefont {N.}~\bibnamefont {Leindecker}}, \bibinfo {author}
  {\bibfnamefont {V.}~\bibnamefont {Leonhardt}}, \bibinfo {author}
  {\bibfnamefont {I.}~\bibnamefont {Leonor}}, \bibinfo {author} {\bibfnamefont
  {K.}~\bibnamefont {Libbrecht}}, \bibinfo {author} {\bibfnamefont
  {H.}~\bibnamefont {Lin}}, \bibinfo {author} {\bibfnamefont {P.}~\bibnamefont
  {Lindquist}}, \bibinfo {author} {\bibfnamefont {N.~A.}\ \bibnamefont
  {Lockerbie}}, \bibinfo {author} {\bibfnamefont {D.}~\bibnamefont {Lodhia}},
  \bibinfo {author} {\bibfnamefont {M.}~\bibnamefont {Lormand}}, \bibinfo
  {author} {\bibfnamefont {P.}~\bibnamefont {Lu}}, \bibinfo {author}
  {\bibfnamefont {M.}~\bibnamefont {Lubinski}}, \bibinfo {author}
  {\bibfnamefont {A.}~\bibnamefont {Lucianetti}}, \bibinfo {author}
  {\bibfnamefont {H.}~\bibnamefont {Lück}}, \bibinfo {author} {\bibfnamefont
  {B.}~\bibnamefont {Machenschalk}}, \bibinfo {author} {\bibfnamefont
  {M.}~\bibnamefont {MacInnis}}, \bibinfo {author} {\bibfnamefont
  {M.}~\bibnamefont {Mageswaran}}, \bibinfo {author} {\bibfnamefont
  {K.}~\bibnamefont {Mailand}}, \bibinfo {author} {\bibfnamefont
  {V.}~\bibnamefont {Mandic}}, \bibinfo {author} {\bibfnamefont
  {S.}~\bibnamefont {Márka}}, \bibinfo {author} {\bibfnamefont
  {Z.}~\bibnamefont {Márka}}, \bibinfo {author} {\bibfnamefont
  {A.}~\bibnamefont {Markosyan}}, \bibinfo {author} {\bibfnamefont
  {J.}~\bibnamefont {Markowitz}}, \bibinfo {author} {\bibfnamefont
  {E.}~\bibnamefont {Maros}}, \bibinfo {author} {\bibfnamefont
  {I.}~\bibnamefont {Martin}}, \bibinfo {author} {\bibfnamefont {R.~M.}\
  \bibnamefont {Martin}}, \bibinfo {author} {\bibfnamefont {J.~N.}\
  \bibnamefont {Marx}}, \bibinfo {author} {\bibfnamefont {K.}~\bibnamefont
  {Mason}}, \bibinfo {author} {\bibfnamefont {F.}~\bibnamefont {Matichard}},
  \bibinfo {author} {\bibfnamefont {L.}~\bibnamefont {Matone}}, \bibinfo
  {author} {\bibfnamefont {R.}~\bibnamefont {Matzner}}, \bibinfo {author}
  {\bibfnamefont {N.}~\bibnamefont {Mavalvala}}, \bibinfo {author}
  {\bibfnamefont {R.}~\bibnamefont {McCarthy}}, \bibinfo {author}
  {\bibfnamefont {D.~E.}\ \bibnamefont {McClelland}}, \bibinfo {author}
  {\bibfnamefont {S.~C.}\ \bibnamefont {McGuire}}, \bibinfo {author}
  {\bibfnamefont {M.}~\bibnamefont {McHugh}}, \bibinfo {author} {\bibfnamefont
  {G.}~\bibnamefont {McIntyre}}, \bibinfo {author} {\bibfnamefont
  {G.}~\bibnamefont {McIvor}}, \bibinfo {author} {\bibfnamefont
  {D.}~\bibnamefont {McKechan}}, \bibinfo {author} {\bibfnamefont
  {K.}~\bibnamefont {McKenzie}}, \bibinfo {author} {\bibfnamefont
  {T.}~\bibnamefont {Meier}}, \bibinfo {author} {\bibfnamefont
  {A.}~\bibnamefont {Melissinos}}, \bibinfo {author} {\bibfnamefont
  {G.}~\bibnamefont {Mendell}}, \bibinfo {author} {\bibfnamefont {R.~A.}\
  \bibnamefont {Mercer}}, \bibinfo {author} {\bibfnamefont {S.}~\bibnamefont
  {Meshkov}}, \bibinfo {author} {\bibfnamefont {C.~J.}\ \bibnamefont
  {Messenger}}, \bibinfo {author} {\bibfnamefont {D.}~\bibnamefont {Meyers}},
  \bibinfo {author} {\bibfnamefont {J.}~\bibnamefont {Miller}}, \bibinfo
  {author} {\bibfnamefont {J.}~\bibnamefont {Minelli}}, \bibinfo {author}
  {\bibfnamefont {S.}~\bibnamefont {Mitra}}, \bibinfo {author} {\bibfnamefont
  {V.~P.}\ \bibnamefont {Mitrofanov}}, \bibinfo {author} {\bibfnamefont
  {G.}~\bibnamefont {Mitselmakher}}, \bibinfo {author} {\bibfnamefont
  {R.}~\bibnamefont {Mittleman}}, \bibinfo {author} {\bibfnamefont
  {O.}~\bibnamefont {Miyakawa}}, \bibinfo {author} {\bibfnamefont
  {B.}~\bibnamefont {Moe}}, \bibinfo {author} {\bibfnamefont {S.}~\bibnamefont
  {Mohanty}}, \bibinfo {author} {\bibfnamefont {G.}~\bibnamefont {Moreno}},
  \bibinfo {author} {\bibfnamefont {K.}~\bibnamefont {Mossavi}}, \bibinfo
  {author} {\bibfnamefont {C.}~\bibnamefont {MowLowry}}, \bibinfo {author}
  {\bibfnamefont {G.}~\bibnamefont {Mueller}}, \bibinfo {author} {\bibfnamefont
  {S.}~\bibnamefont {Mukherjee}}, \bibinfo {author} {\bibfnamefont
  {H.}~\bibnamefont {Mukhopadhyay}}, \bibinfo {author} {\bibfnamefont
  {H.}~\bibnamefont {Müller-Ebhardt}}, \bibinfo {author} {\bibfnamefont
  {J.}~\bibnamefont {Munch}}, \bibinfo {author} {\bibfnamefont
  {P.}~\bibnamefont {Murray}}, \bibinfo {author} {\bibfnamefont
  {E.}~\bibnamefont {Myers}}, \bibinfo {author} {\bibfnamefont
  {J.}~\bibnamefont {Myers}}, \bibinfo {author} {\bibfnamefont
  {T.}~\bibnamefont {Nash}}, \bibinfo {author} {\bibfnamefont {J.}~\bibnamefont
  {Nelson}}, \bibinfo {author} {\bibfnamefont {G.}~\bibnamefont {Newton}},
  \bibinfo {author} {\bibfnamefont {A.}~\bibnamefont {Nishizawa}}, \bibinfo
  {author} {\bibfnamefont {K.}~\bibnamefont {Numata}}, \bibinfo {author}
  {\bibfnamefont {J.}~\bibnamefont {O'Dell}}, \bibinfo {author} {\bibfnamefont
  {G.}~\bibnamefont {Ogin}}, \bibinfo {author} {\bibfnamefont {B.}~\bibnamefont
  {O'Reilly}}, \bibinfo {author} {\bibfnamefont {R.}~\bibnamefont
  {O'Shaughnessy}}, \bibinfo {author} {\bibfnamefont {D.~J.}\ \bibnamefont
  {Ottaway}}, \bibinfo {author} {\bibfnamefont {R.~S.}\ \bibnamefont {Ottens}},
  \bibinfo {author} {\bibfnamefont {H.}~\bibnamefont {Overmier}}, \bibinfo
  {author} {\bibfnamefont {B.~J.}\ \bibnamefont {Owen}}, \bibinfo {author}
  {\bibfnamefont {Y.}~\bibnamefont {Pan}}, \bibinfo {author} {\bibfnamefont
  {C.}~\bibnamefont {Pankow}}, \bibinfo {author} {\bibfnamefont {M.~A.}\
  \bibnamefont {Papa}}, \bibinfo {author} {\bibfnamefont {V.}~\bibnamefont
  {Parameshwaraiah}}, \bibinfo {author} {\bibfnamefont {P.}~\bibnamefont
  {Patel}}, \bibinfo {author} {\bibfnamefont {M.}~\bibnamefont {Pedraza}},
  \bibinfo {author} {\bibfnamefont {S.}~\bibnamefont {Penn}}, \bibinfo {author}
  {\bibfnamefont {A.}~\bibnamefont {Perreca}}, \bibinfo {author} {\bibfnamefont
  {T.}~\bibnamefont {Petrie}}, \bibinfo {author} {\bibfnamefont {I.~M.}\
  \bibnamefont {Pinto}}, \bibinfo {author} {\bibfnamefont {M.}~\bibnamefont
  {Pitkin}}, \bibinfo {author} {\bibfnamefont {H.~J.}\ \bibnamefont {Pletsch}},
  \bibinfo {author} {\bibfnamefont {M.~V.}\ \bibnamefont {Plissi}}, \bibinfo
  {author} {\bibfnamefont {F.}~\bibnamefont {Postiglione}}, \bibinfo {author}
  {\bibfnamefont {M.}~\bibnamefont {Principe}}, \bibinfo {author}
  {\bibfnamefont {R.}~\bibnamefont {Prix}}, \bibinfo {author} {\bibfnamefont
  {V.}~\bibnamefont {Quetschke}}, \bibinfo {author} {\bibfnamefont
  {F.}~\bibnamefont {Raab}}, \bibinfo {author} {\bibfnamefont {D.~S.}\
  \bibnamefont {Rabeling}}, \bibinfo {author} {\bibfnamefont {H.}~\bibnamefont
  {Radkins}}, \bibinfo {author} {\bibfnamefont {N.}~\bibnamefont {Rainer}},
  \bibinfo {author} {\bibfnamefont {M.}~\bibnamefont {Rakhmanov}}, \bibinfo
  {author} {\bibfnamefont {M.}~\bibnamefont {Ramsunder}}, \bibinfo {author}
  {\bibfnamefont {H.}~\bibnamefont {Rehbein}}, \bibinfo {author} {\bibfnamefont
  {S.}~\bibnamefont {Reid}}, \bibinfo {author} {\bibfnamefont {D.~H.}\
  \bibnamefont {Reitze}}, \bibinfo {author} {\bibfnamefont {R.}~\bibnamefont
  {Riesen}}, \bibinfo {author} {\bibfnamefont {K.}~\bibnamefont {Riles}},
  \bibinfo {author} {\bibfnamefont {B.}~\bibnamefont {Rivera}}, \bibinfo
  {author} {\bibfnamefont {N.~A.}\ \bibnamefont {Robertson}}, \bibinfo {author}
  {\bibfnamefont {C.}~\bibnamefont {Robinson}}, \bibinfo {author}
  {\bibfnamefont {E.~L.}\ \bibnamefont {Robinson}}, \bibinfo {author}
  {\bibfnamefont {S.}~\bibnamefont {Roddy}}, \bibinfo {author} {\bibfnamefont
  {A.}~\bibnamefont {Rodriguez}}, \bibinfo {author} {\bibfnamefont {A.~M.}\
  \bibnamefont {Rogan}}, \bibinfo {author} {\bibfnamefont {J.}~\bibnamefont
  {Rollins}}, \bibinfo {author} {\bibfnamefont {J.~D.}\ \bibnamefont {Romano}},
  \bibinfo {author} {\bibfnamefont {J.}~\bibnamefont {Romie}}, \bibinfo
  {author} {\bibfnamefont {R.}~\bibnamefont {Route}}, \bibinfo {author}
  {\bibfnamefont {S.}~\bibnamefont {Rowan}}, \bibinfo {author} {\bibfnamefont
  {A.}~\bibnamefont {Rüdiger}}, \bibinfo {author} {\bibfnamefont
  {L.}~\bibnamefont {Ruet}}, \bibinfo {author} {\bibfnamefont {P.}~\bibnamefont
  {Russell}}, \bibinfo {author} {\bibfnamefont {K.}~\bibnamefont {Ryan}},
  \bibinfo {author} {\bibfnamefont {S.}~\bibnamefont {Sakata}}, \bibinfo
  {author} {\bibfnamefont {M.}~\bibnamefont {Samidi}}, \bibinfo {author}
  {\bibfnamefont {L.~S.}\ \bibnamefont {de~la Jordana}}, \bibinfo {author}
  {\bibfnamefont {V.}~\bibnamefont {Sandberg}}, \bibinfo {author}
  {\bibfnamefont {V.}~\bibnamefont {Sannibale}}, \bibinfo {author}
  {\bibfnamefont {S.}~\bibnamefont {Saraf}}, \bibinfo {author} {\bibfnamefont
  {P.}~\bibnamefont {Sarin}}, \bibinfo {author} {\bibfnamefont {B.~S.}\
  \bibnamefont {Sathyaprakash}}, \bibinfo {author} {\bibfnamefont
  {S.}~\bibnamefont {Sato}}, \bibinfo {author} {\bibfnamefont {P.~R.}\
  \bibnamefont {Saulson}}, \bibinfo {author} {\bibfnamefont {R.}~\bibnamefont
  {Savage}}, \bibinfo {author} {\bibfnamefont {P.}~\bibnamefont {Savov}},
  \bibinfo {author} {\bibfnamefont {S.~W.}\ \bibnamefont {Schediwy}}, \bibinfo
  {author} {\bibfnamefont {R.}~\bibnamefont {Schilling}}, \bibinfo {author}
  {\bibfnamefont {R.}~\bibnamefont {Schnabel}}, \bibinfo {author}
  {\bibfnamefont {R.}~\bibnamefont {Schofield}}, \bibinfo {author}
  {\bibfnamefont {B.~F.}\ \bibnamefont {Schutz}}, \bibinfo {author}
  {\bibfnamefont {P.}~\bibnamefont {Schwinberg}}, \bibinfo {author}
  {\bibfnamefont {S.~M.}\ \bibnamefont {Scott}}, \bibinfo {author}
  {\bibfnamefont {A.~C.}\ \bibnamefont {Searle}}, \bibinfo {author}
  {\bibfnamefont {B.}~\bibnamefont {Sears}}, \bibinfo {author} {\bibfnamefont
  {F.}~\bibnamefont {Seifert}}, \bibinfo {author} {\bibfnamefont
  {D.}~\bibnamefont {Sellers}}, \bibinfo {author} {\bibfnamefont {A.~S.}\
  \bibnamefont {Sengupta}}, \bibinfo {author} {\bibfnamefont {P.}~\bibnamefont
  {Shawhan}}, \bibinfo {author} {\bibfnamefont {D.~H.}\ \bibnamefont
  {Shoemaker}}, \bibinfo {author} {\bibfnamefont {A.}~\bibnamefont {Sibley}},
  \bibinfo {author} {\bibfnamefont {X.}~\bibnamefont {Siemens}}, \bibinfo
  {author} {\bibfnamefont {D.}~\bibnamefont {Sigg}}, \bibinfo {author}
  {\bibfnamefont {S.}~\bibnamefont {Sinha}}, \bibinfo {author} {\bibfnamefont
  {A.~M.}\ \bibnamefont {Sintes}}, \bibinfo {author} {\bibfnamefont {B.~J.~J.}\
  \bibnamefont {Slagmolen}}, \bibinfo {author} {\bibfnamefont {J.}~\bibnamefont
  {Slutsky}}, \bibinfo {author} {\bibfnamefont {J.~R.}\ \bibnamefont {Smith}},
  \bibinfo {author} {\bibfnamefont {M.~R.}\ \bibnamefont {Smith}}, \bibinfo
  {author} {\bibfnamefont {N.~D.}\ \bibnamefont {Smith}}, \bibinfo {author}
  {\bibfnamefont {K.}~\bibnamefont {Somiya}}, \bibinfo {author} {\bibfnamefont
  {B.}~\bibnamefont {Sorazu}}, \bibinfo {author} {\bibfnamefont {L.~C.}\
  \bibnamefont {Stein}}, \bibinfo {author} {\bibfnamefont {A.}~\bibnamefont
  {Stochino}}, \bibinfo {author} {\bibfnamefont {R.}~\bibnamefont {Stone}},
  \bibinfo {author} {\bibfnamefont {K.~A.}\ \bibnamefont {Strain}}, \bibinfo
  {author} {\bibfnamefont {D.~M.}\ \bibnamefont {Strom}}, \bibinfo {author}
  {\bibfnamefont {A.}~\bibnamefont {Stuver}}, \bibinfo {author} {\bibfnamefont
  {T.~Z.}\ \bibnamefont {Summerscales}}, \bibinfo {author} {\bibfnamefont
  {K.-X.}\ \bibnamefont {Sun}}, \bibinfo {author} {\bibfnamefont
  {M.}~\bibnamefont {Sung}}, \bibinfo {author} {\bibfnamefont {P.~J.}\
  \bibnamefont {Sutton}}, \bibinfo {author} {\bibfnamefont {H.}~\bibnamefont
  {Takahashi}}, \bibinfo {author} {\bibfnamefont {D.~B.}\ \bibnamefont
  {Tanner}}, \bibinfo {author} {\bibfnamefont {R.}~\bibnamefont {Taylor}},
  \bibinfo {author} {\bibfnamefont {R.}~\bibnamefont {Taylor}}, \bibinfo
  {author} {\bibfnamefont {J.}~\bibnamefont {Thacker}}, \bibinfo {author}
  {\bibfnamefont {K.~A.}\ \bibnamefont {Thorne}}, \bibinfo {author}
  {\bibfnamefont {K.~S.}\ \bibnamefont {Thorne}}, \bibinfo {author}
  {\bibfnamefont {A.}~\bibnamefont {Thüring}}, \bibinfo {author}
  {\bibfnamefont {M.}~\bibnamefont {Tinto}}, \bibinfo {author} {\bibfnamefont
  {K.~V.}\ \bibnamefont {Tokmakov}}, \bibinfo {author} {\bibfnamefont
  {C.}~\bibnamefont {Torres}}, \bibinfo {author} {\bibfnamefont
  {C.}~\bibnamefont {Torrie}}, \bibinfo {author} {\bibfnamefont
  {G.}~\bibnamefont {Traylor}}, \bibinfo {author} {\bibfnamefont
  {M.}~\bibnamefont {Trias}}, \bibinfo {author} {\bibfnamefont
  {W.}~\bibnamefont {Tyler}}, \bibinfo {author} {\bibfnamefont
  {D.}~\bibnamefont {Ugolini}}, \bibinfo {author} {\bibfnamefont
  {J.}~\bibnamefont {Ulmen}}, \bibinfo {author} {\bibfnamefont
  {K.}~\bibnamefont {Urbanek}}, \bibinfo {author} {\bibfnamefont
  {H.}~\bibnamefont {Vahlbruch}}, \bibinfo {author} {\bibfnamefont {C.~V.~D.}\
  \bibnamefont {Broeck}}, \bibinfo {author} {\bibfnamefont {M.}~\bibnamefont
  {van~der Sluys}}, \bibinfo {author} {\bibfnamefont {S.}~\bibnamefont {Vass}},
  \bibinfo {author} {\bibfnamefont {R.}~\bibnamefont {Vaulin}}, \bibinfo
  {author} {\bibfnamefont {A.}~\bibnamefont {Vecchio}}, \bibinfo {author}
  {\bibfnamefont {J.}~\bibnamefont {Veitch}}, \bibinfo {author} {\bibfnamefont
  {P.}~\bibnamefont {Veitch}}, \bibinfo {author} {\bibfnamefont
  {A.}~\bibnamefont {Villar}}, \bibinfo {author} {\bibfnamefont
  {C.}~\bibnamefont {Vorvick}}, \bibinfo {author} {\bibfnamefont {S.~P.}\
  \bibnamefont {Vyachanin}}, \bibinfo {author} {\bibfnamefont {S.~J.}\
  \bibnamefont {Waldman}}, \bibinfo {author} {\bibfnamefont {L.}~\bibnamefont
  {Wallace}}, \bibinfo {author} {\bibfnamefont {H.}~\bibnamefont {Ward}},
  \bibinfo {author} {\bibfnamefont {R.}~\bibnamefont {Ward}}, \bibinfo {author}
  {\bibfnamefont {M.}~\bibnamefont {Weinert}}, \bibinfo {author} {\bibfnamefont
  {A.}~\bibnamefont {Weinstein}}, \bibinfo {author} {\bibfnamefont
  {R.}~\bibnamefont {Weiss}}, \bibinfo {author} {\bibfnamefont
  {S.}~\bibnamefont {Wen}}, \bibinfo {author} {\bibfnamefont {K.}~\bibnamefont
  {Wette}}, \bibinfo {author} {\bibfnamefont {J.~T.}\ \bibnamefont {Whelan}},
  \bibinfo {author} {\bibfnamefont {S.~E.}\ \bibnamefont {Whitcomb}}, \bibinfo
  {author} {\bibfnamefont {B.~F.}\ \bibnamefont {Whiting}}, \bibinfo {author}
  {\bibfnamefont {C.}~\bibnamefont {Wilkinson}}, \bibinfo {author}
  {\bibfnamefont {P.~A.}\ \bibnamefont {Willems}}, \bibinfo {author}
  {\bibfnamefont {H.~R.}\ \bibnamefont {Williams}}, \bibinfo {author}
  {\bibfnamefont {L.}~\bibnamefont {Williams}}, \bibinfo {author}
  {\bibfnamefont {B.}~\bibnamefont {Willke}}, \bibinfo {author} {\bibfnamefont
  {I.}~\bibnamefont {Wilmut}}, \bibinfo {author} {\bibfnamefont
  {W.}~\bibnamefont {Winkler}}, \bibinfo {author} {\bibfnamefont {C.~C.}\
  \bibnamefont {Wipf}}, \bibinfo {author} {\bibfnamefont {A.~G.}\ \bibnamefont
  {Wiseman}}, \bibinfo {author} {\bibfnamefont {G.}~\bibnamefont {Woan}},
  \bibinfo {author} {\bibfnamefont {R.}~\bibnamefont {Wooley}}, \bibinfo
  {author} {\bibfnamefont {J.}~\bibnamefont {Worden}}, \bibinfo {author}
  {\bibfnamefont {W.}~\bibnamefont {Wu}}, \bibinfo {author} {\bibfnamefont
  {I.}~\bibnamefont {Yakushin}}, \bibinfo {author} {\bibfnamefont
  {H.}~\bibnamefont {Yamamoto}}, \bibinfo {author} {\bibfnamefont
  {Z.}~\bibnamefont {Yan}}, \bibinfo {author} {\bibfnamefont {S.}~\bibnamefont
  {Yoshida}}, \bibinfo {author} {\bibfnamefont {M.}~\bibnamefont {Zanolin}},
  \bibinfo {author} {\bibfnamefont {J.}~\bibnamefont {Zhang}}, \bibinfo
  {author} {\bibfnamefont {L.}~\bibnamefont {Zhang}}, \bibinfo {author}
  {\bibfnamefont {C.}~\bibnamefont {Zhao}}, \bibinfo {author} {\bibfnamefont
  {N.}~\bibnamefont {Zotov}}, \bibinfo {author} {\bibfnamefont
  {M.}~\bibnamefont {Zucker}}, \bibinfo {author} {\bibfnamefont
  {J.}~\bibnamefont {Zweizig}},\ and\ \bibinfo {author} {\bibfnamefont {L.~S.}\
  \bibnamefont {Collaboration)}},\ }\bibfield  {title} {\bibinfo {title} {First
  joint search for gravitational-wave bursts in ligo and geo 600 data},\ }\href
  {https://doi.org/10.1088/0264-9381/25/24/245008} {\bibfield  {journal}
  {\bibinfo  {journal} {Classical and Quantum Gravity}\ }\textbf {\bibinfo
  {volume} {25}},\ \bibinfo {pages} {245008} (\bibinfo {year}
  {2008})}\BibitemShut {NoStop}%
\bibitem [{\citenamefont {Abadie}\ \emph {et~al.}(2010)\citenamefont {Abadie},
  \citenamefont {Abbott}, \citenamefont {Abbott}, \citenamefont {Accadia},
  \citenamefont {Acernese}, \citenamefont {Adhikari}, \citenamefont {Ajith},
  \citenamefont {Allen}, \citenamefont {Allen}, \citenamefont {Amador~Ceron},
  \citenamefont {Amin}, \citenamefont {Anderson}, \citenamefont {Anderson},
  \citenamefont {Antonucci}, \citenamefont {Arain}, \citenamefont {Araya},
  \citenamefont {Arun}, \citenamefont {Aso}, \citenamefont {Aston},
  \citenamefont {Astone}, \citenamefont {Aufmuth}, \citenamefont {Aulbert},
  \citenamefont {Babak}, \citenamefont {Baker}, \citenamefont {Ballardin},
  \citenamefont {Ballmer}, \citenamefont {Barker}, \citenamefont {Barone},
  \citenamefont {Barr}, \citenamefont {Barriga}, \citenamefont {Barsotti},
  \citenamefont {Barsuglia}, \citenamefont {Barton}, \citenamefont {Bartos},
  \citenamefont {Bassiri}, \citenamefont {Bastarrika}, \citenamefont {Bauer},
  \citenamefont {Behnke}, \citenamefont {Beker}, \citenamefont {Belletoile},
  \citenamefont {Benacquista}, \citenamefont {Betzwieser}, \citenamefont
  {Beyersdorf}, \citenamefont {Bigotta}, \citenamefont {Bilenko}, \citenamefont
  {Billingsley}, \citenamefont {Birindelli}, \citenamefont {Biswas},
  \citenamefont {Bizouard}, \citenamefont {Black}, \citenamefont {Blackburn},
  \citenamefont {Blackburn}, \citenamefont {Blair}, \citenamefont {Bland},
  \citenamefont {Blom}, \citenamefont {Boccara}, \citenamefont {Bock},
  \citenamefont {Bodiya}, \citenamefont {Bondarescu}, \citenamefont {Bondu},
  \citenamefont {Bonelli}, \citenamefont {Bonnand}, \citenamefont {Bork},
  \citenamefont {Born}, \citenamefont {Bose}, \citenamefont {Bosi},
  \citenamefont {Bouhou}, \citenamefont {Braccini}, \citenamefont {Bradaschia},
  \citenamefont {Brady}, \citenamefont {Braginsky}, \citenamefont {Brau},
  \citenamefont {Breyer}, \citenamefont {Bridges}, \citenamefont {Brillet},
  \citenamefont {Brinkmann}, \citenamefont {Brisson}, \citenamefont {Britzger},
  \citenamefont {Brooks}, \citenamefont {Brown}, \citenamefont
  {Budzy\ifmmode~\acute{n}\else \'{n}\fi{}ski}, \citenamefont {Bulik},
  \citenamefont {Bullington}, \citenamefont {Bulten}, \citenamefont {Buonanno},
  \citenamefont {Burmeister}, \citenamefont {Buskulic}, \citenamefont {Buy},
  \citenamefont {Byer}, \citenamefont {Cadonati}, \citenamefont {Cagnoli},
  \citenamefont {Cain}, \citenamefont {Calloni}, \citenamefont {Camp},
  \citenamefont {Campagna}, \citenamefont {Cannizzo}, \citenamefont {Cannon},
  \citenamefont {Canuel}, \citenamefont {Cao}, \citenamefont {Capano},
  \citenamefont {Carbognani}, \citenamefont {Cardenas}, \citenamefont
  {Caudill}, \citenamefont {Cavagli\`a}, \citenamefont {Cavalier},
  \citenamefont {Cavalieri}, \citenamefont {Cella}, \citenamefont {Cepeda},
  \citenamefont {Cesarini}, \citenamefont {Chalermsongsak}, \citenamefont
  {Chalkley}, \citenamefont {Charlton}, \citenamefont {Chassande-Mottin},
  \citenamefont {Chatterji}, \citenamefont {Chelkowski}, \citenamefont {Chen},
  \citenamefont {Chincarini}, \citenamefont {Christensen}, \citenamefont
  {Chua}, \citenamefont {Chung}, \citenamefont {Clark}, \citenamefont {Clark},
  \citenamefont {Clayton}, \citenamefont {Cleva}, \citenamefont {Coccia},
  \citenamefont {Colacino}, \citenamefont {Colas}, \citenamefont {Colla},
  \citenamefont {Colombini}, \citenamefont {Conte}, \citenamefont {Cook},
  \citenamefont {Corbitt}, \citenamefont {Cornish}, \citenamefont {Corsi},
  \citenamefont {Coulon}, \citenamefont {Coward}, \citenamefont {Coyne},
  \citenamefont {Creighton}, \citenamefont {Creighton}, \citenamefont {Cruise},
  \citenamefont {Culter}, \citenamefont {Cumming}, \citenamefont {Cunningham},
  \citenamefont {Cuoco}, \citenamefont {Dahl}, \citenamefont {Danilishin},
  \citenamefont {D'Antonio}, \citenamefont {Danzmann}, \citenamefont {Dattilo},
  \citenamefont {Daudert}, \citenamefont {Davier}, \citenamefont {Davies},
  \citenamefont {Daw}, \citenamefont {Day}, \citenamefont {Dayanga},
  \citenamefont {De~Rosa}, \citenamefont {DeBra}, \citenamefont {Degallaix},
  \citenamefont {del Prete}, \citenamefont {Dergachev}, \citenamefont
  {DeSalvo}, \citenamefont {Dhurandhar}, \citenamefont {Di~Fiore},
  \citenamefont {Di~Lieto}, \citenamefont {Di~Paolo~Emilio}, \citenamefont
  {Di~Virgilio}, \citenamefont {D\'{\i}az}, \citenamefont {Dietz},
  \citenamefont {Donovan}, \citenamefont {Dooley}, \citenamefont {Doomes},
  \citenamefont {Drago}, \citenamefont {Drever}, \citenamefont {Driggers},
  \citenamefont {Dueck}, \citenamefont {Duke}, \citenamefont {Dumas},
  \citenamefont {Dwyer}, \citenamefont {Edgar}, \citenamefont {Edwards},
  \citenamefont {Effler}, \citenamefont {Ehrens}, \citenamefont {Etzel},
  \citenamefont {Evans}, \citenamefont {Evans}, \citenamefont {Fafone},
  \citenamefont {Fairhurst}, \citenamefont {Faltas}, \citenamefont {Fan},
  \citenamefont {Fazi}, \citenamefont {Fehrmann}, \citenamefont {Ferrante},
  \citenamefont {Fidecaro}, \citenamefont {Finn}, \citenamefont {Fiori},
  \citenamefont {Flaminio}, \citenamefont {Flasch}, \citenamefont {Foley},
  \citenamefont {Forrest}, \citenamefont {Fotopoulos}, \citenamefont
  {Fournier}, \citenamefont {Franc}, \citenamefont {Frasca}, \citenamefont
  {Frasconi}, \citenamefont {Frede}, \citenamefont {Frei}, \citenamefont
  {Frei}, \citenamefont {Freise}, \citenamefont {Frey}, \citenamefont {Fricke},
  \citenamefont {Friedrich}, \citenamefont {Fritschel}, \citenamefont {Frolov},
  \citenamefont {Fulda}, \citenamefont {Fyffe}, \citenamefont {Galimberti},
  \citenamefont {Gammaitoni}, \citenamefont {Garofoli}, \citenamefont {Garufi},
  \citenamefont {Gemme}, \citenamefont {Genin}, \citenamefont {Gennai},
  \citenamefont {Ghosh}, \citenamefont {Giaime}, \citenamefont {Giampanis},
  \citenamefont {Giardina}, \citenamefont {Giazotto}, \citenamefont {Goetz},
  \citenamefont {Goggin}, \citenamefont {Gonz\'alez}, \citenamefont
  {Go\ss{}ler}, \citenamefont {Gouaty}, \citenamefont {Granata}, \citenamefont
  {Grant}, \citenamefont {Gras}, \citenamefont {Gray}, \citenamefont
  {Greenhalgh}, \citenamefont {Gretarsson}, \citenamefont {Greverie},
  \citenamefont {Grosso}, \citenamefont {Grote}, \citenamefont {Grunewald},
  \citenamefont {Guidi}, \citenamefont {Gustafson}, \citenamefont {Gustafson},
  \citenamefont {Hage}, \citenamefont {Hallam}, \citenamefont {Hammer},
  \citenamefont {Hammond}, \citenamefont {Hanna}, \citenamefont {Hanson},
  \citenamefont {Harms}, \citenamefont {Harry}, \citenamefont {Harry},
  \citenamefont {Harstad}, \citenamefont {Haughian}, \citenamefont {Hayama},
  \citenamefont {Hayau}, \citenamefont {Hayler}, \citenamefont {Heefner},
  \citenamefont {Heitmann}, \citenamefont {Hello}, \citenamefont {Heng},
  \citenamefont {Heptonstall}, \citenamefont {Hewitson}, \citenamefont {Hild},
  \citenamefont {Hirose}, \citenamefont {Hoak}, \citenamefont {Hodge},
  \citenamefont {Holt}, \citenamefont {Hosken}, \citenamefont {Hough},
  \citenamefont {Howell}, \citenamefont {Hoyland}, \citenamefont {Huet},
  \citenamefont {Hughey}, \citenamefont {Husa}, \citenamefont {Huttner},
  \citenamefont {Ingram}, \citenamefont {Isogai}, \citenamefont {Ivanov},
  \citenamefont {Jaranowski}, \citenamefont {Johnson}, \citenamefont {Jones},
  \citenamefont {Jones}, \citenamefont {Jones}, \citenamefont {Ju},
  \citenamefont {Kalmus}, \citenamefont {Kalogera}, \citenamefont {Kandhasamy},
  \citenamefont {Kanner}, \citenamefont {Katsavounidis}, \citenamefont
  {Kawabe}, \citenamefont {Kawamura}, \citenamefont {Kawazoe}, \citenamefont
  {Kells}, \citenamefont {Keppel}, \citenamefont {Khalaidovski}, \citenamefont
  {Khalili}, \citenamefont {Khan}, \citenamefont {Khazanov}, \citenamefont
  {Kim}, \citenamefont {King}, \citenamefont {Kissel}, \citenamefont
  {Klimenko}, \citenamefont {Kokeyama}, \citenamefont {Kondrashov},
  \citenamefont {Kopparapu}, \citenamefont {Koranda}, \citenamefont {Kowalska},
  \citenamefont {Kozak}, \citenamefont {Kringel}, \citenamefont {Krishnan},
  \citenamefont {Kr\'olak}, \citenamefont {Kuehn}, \citenamefont {Kullman},
  \citenamefont {Kumar}, \citenamefont {Kwee}, \citenamefont {Lam},
  \citenamefont {Landry}, \citenamefont {Lang}, \citenamefont {Lantz},
  \citenamefont {Lastzka}, \citenamefont {Lazzarini}, \citenamefont {Leaci},
  \citenamefont {Lei}, \citenamefont {Leindecker}, \citenamefont {Leonor},
  \citenamefont {Leroy}, \citenamefont {Letendre}, \citenamefont {Li},
  \citenamefont {Lin}, \citenamefont {Lindquist}, \citenamefont {Littenberg},
  \citenamefont {Lockerbie}, \citenamefont {Lodhia}, \citenamefont {Lorenzini},
  \citenamefont {Loriette}, \citenamefont {Lormand}, \citenamefont {Losurdo},
  \citenamefont {Lu}, \citenamefont {Lubinski}, \citenamefont {Lucianetti},
  \citenamefont {L\"uck}, \citenamefont {Lundgren}, \citenamefont
  {Machenschalk}, \citenamefont {MacInnis}, \citenamefont {Mageswaran},
  \citenamefont {Mailand}, \citenamefont {Majorana}, \citenamefont {Mak},
  \citenamefont {Maksimovic}, \citenamefont {Man}, \citenamefont {Mandel},
  \citenamefont {Mandic}, \citenamefont {Mantovani}, \citenamefont
  {Marchesoni}, \citenamefont {Marion}, \citenamefont {M\'arka}, \citenamefont
  {M\'arka}, \citenamefont {Markosyan}, \citenamefont {Markowitz},
  \citenamefont {Maros}, \citenamefont {Marque}, \citenamefont {Martelli},
  \citenamefont {Martin}, \citenamefont {Martin}, \citenamefont {Marx},
  \citenamefont {Mason}, \citenamefont {Masserot}, \citenamefont {Matichard},
  \citenamefont {Matone}, \citenamefont {Matzner}, \citenamefont {Mavalvala},
  \citenamefont {McCarthy}, \citenamefont {McClelland}, \citenamefont
  {McGuire}, \citenamefont {McIntyre}, \citenamefont {McKechan}, \citenamefont
  {Mehmet}, \citenamefont {Melatos}, \citenamefont {Melissinos}, \citenamefont
  {Mendell}, \citenamefont {Men\'endez}, \citenamefont {Mercer}, \citenamefont
  {Merill}, \citenamefont {Meshkov}, \citenamefont {Messenger}, \citenamefont
  {Meyer}, \citenamefont {Miao}, \citenamefont {Michel}, \citenamefont
  {Milano}, \citenamefont {Miller}, \citenamefont {Minenkov}, \citenamefont
  {Mino}, \citenamefont {Mitra}, \citenamefont {Mitrofanov}, \citenamefont
  {Mitselmakher}, \citenamefont {Mittleman}, \citenamefont {Miyakawa},
  \citenamefont {Moe}, \citenamefont {Mohan}, \citenamefont {Mohanty},
  \citenamefont {Mohapatra}, \citenamefont {Moreau}, \citenamefont {Moreno},
  \citenamefont {Morgado}, \citenamefont {Morgia}, \citenamefont {Mors},
  \citenamefont {Mosca}, \citenamefont {Moscatelli}, \citenamefont {Mossavi},
  \citenamefont {Mours}, \citenamefont {MowLowry}, \citenamefont {Mueller},
  \citenamefont {Mukherjee}, \citenamefont {Mullavey}, \citenamefont
  {M\"uller-Ebhardt}, \citenamefont {Munch}, \citenamefont {Murray},
  \citenamefont {Nash}, \citenamefont {Nawrodt}, \citenamefont {Nelson},
  \citenamefont {Neri}, \citenamefont {Newton}, \citenamefont {Nishida},
  \citenamefont {Nishizawa}, \citenamefont {Nocera}, \citenamefont {Ochsner},
  \citenamefont {O'Dell}, \citenamefont {Ogin}, \citenamefont {Oldenburg},
  \citenamefont {O'Reilly}, \citenamefont {O'Shaughnessy}, \citenamefont
  {Ottaway}, \citenamefont {Ottens}, \citenamefont {Overmier}, \citenamefont
  {Owen}, \citenamefont {Page}, \citenamefont {Pagliaroli}, \citenamefont
  {Palladino}, \citenamefont {Palomba}, \citenamefont {Pan}, \citenamefont
  {Pankow}, \citenamefont {Paoletti}, \citenamefont {Papa}, \citenamefont
  {Pardi}, \citenamefont {Parisi}, \citenamefont {Pasqualetti}, \citenamefont
  {Passaquieti}, \citenamefont {Passuello}, \citenamefont {Patel},
  \citenamefont {Pathak}, \citenamefont {Pedraza}, \citenamefont {Pekowsky},
  \citenamefont {Penn}, \citenamefont {Peralta}, \citenamefont {Perreca},
  \citenamefont {Persichetti}, \citenamefont {Pichot}, \citenamefont
  {Pickenpack}, \citenamefont {Piergiovanni}, \citenamefont {Pietka},
  \citenamefont {Pinard}, \citenamefont {Pinto}, \citenamefont {Pitkin},
  \citenamefont {Pletsch}, \citenamefont {Plissi}, \citenamefont {Poggiani},
  \citenamefont {Postiglione}, \citenamefont {Prato}, \citenamefont {Principe},
  \citenamefont {Prix}, \citenamefont {Prodi}, \citenamefont {Prokhorov},
  \citenamefont {Puncken}, \citenamefont {Punturo}, \citenamefont {Puppo},
  \citenamefont {Quetschke}, \citenamefont {Raab}, \citenamefont {Rabeling},
  \citenamefont {Rabeling}, \citenamefont {Radkins}, \citenamefont {Raffai},
  \citenamefont {Raics}, \citenamefont {Rakhmanov}, \citenamefont {Rapagnani},
  \citenamefont {Raymond}, \citenamefont {Re}, \citenamefont {Reed},
  \citenamefont {Reed}, \citenamefont {Regimbau}, \citenamefont {Rehbein},
  \citenamefont {Reid}, \citenamefont {Reitze}, \citenamefont {Ricci},
  \citenamefont {Riesen}, \citenamefont {Riles}, \citenamefont {Roberts},
  \citenamefont {Robertson}, \citenamefont {Robinet}, \citenamefont {Robinson},
  \citenamefont {Robinson}, \citenamefont {Rocchi}, \citenamefont {Roddy},
  \citenamefont {R\"over}, \citenamefont {Rolland}, \citenamefont {Rollins},
  \citenamefont {Romano}, \citenamefont {Romano}, \citenamefont {Romie},
  \citenamefont {Rosi\ifmmode~\acute{n}\else \'{n}\fi{}ska}, \citenamefont
  {Rowan}, \citenamefont {R\"udiger}, \citenamefont {Ruggi}, \citenamefont
  {Ryan}, \citenamefont {Sakata}, \citenamefont {Salemi}, \citenamefont
  {Sammut}, \citenamefont {Sancho de~la Jordana}, \citenamefont {Sandberg},
  \citenamefont {Sannibale}, \citenamefont {Santamar\'{\i}a}, \citenamefont
  {Santostasi}, \citenamefont {Saraf}, \citenamefont {Sarin}, \citenamefont
  {Sassolas}, \citenamefont {Sathyaprakash}, \citenamefont {Sato},
  \citenamefont {Satterthwaite}, \citenamefont {Saulson}, \citenamefont
  {Savage}, \citenamefont {Schilling}, \citenamefont {Schnabel}, \citenamefont
  {Schofield}, \citenamefont {Schulz}, \citenamefont {Schutz}, \citenamefont
  {Schwinberg}, \citenamefont {Scott}, \citenamefont {Scott}, \citenamefont
  {Searle}, \citenamefont {Seifert}, \citenamefont {Sellers}, \citenamefont
  {Sengupta}, \citenamefont {Sentenac}, \citenamefont {Sergeev}, \citenamefont
  {Shapiro}, \citenamefont {Shawhan}, \citenamefont {Shoemaker}, \citenamefont
  {Sibley}, \citenamefont {Siemens}, \citenamefont {Sigg}, \citenamefont
  {Sintes}, \citenamefont {Skelton}, \citenamefont {Slagmolen}, \citenamefont
  {Slutsky}, \citenamefont {Smith}, \citenamefont {Smith}, \citenamefont
  {Smith}, \citenamefont {Somiya}, \citenamefont {Sorazu}, \citenamefont
  {Sperandio}, \citenamefont {Stein}, \citenamefont {Stein}, \citenamefont
  {Steplewski}, \citenamefont {Stochino}, \citenamefont {Stone}, \citenamefont
  {Strain}, \citenamefont {Strigin}, \citenamefont {Stroeer}, \citenamefont
  {Sturani}, \citenamefont {Stuver}, \citenamefont {Summerscales},
  \citenamefont {Sung}, \citenamefont {Susmithan}, \citenamefont {Sutton},
  \citenamefont {Swinkels}, \citenamefont {Szokoly}, \citenamefont {Talukder},
  \citenamefont {Tanner}, \citenamefont {Tarabrin}, \citenamefont {Taylor},
  \citenamefont {Taylor}, \citenamefont {Thorne}, \citenamefont {Thorne},
  \citenamefont {Th\"uring}, \citenamefont {Titsler}, \citenamefont {Tokmakov},
  \citenamefont {Toncelli}, \citenamefont {Tonelli}, \citenamefont {Torres},
  \citenamefont {Torrie}, \citenamefont {Tournefier}, \citenamefont {Travasso},
  \citenamefont {Traylor}, \citenamefont {Trias}, \citenamefont {Trummer},
  \citenamefont {Turner}, \citenamefont {Ugolini}, \citenamefont {Urbanek},
  \citenamefont {Vahlbruch}, \citenamefont {Vajente}, \citenamefont
  {Vallisneri}, \citenamefont {van~den Brand}, \citenamefont {Van Den~Broeck},
  \citenamefont {van~der Putten}, \citenamefont {van~der Sluys}, \citenamefont
  {Vass}, \citenamefont {Vaulin}, \citenamefont {Vavoulidis}, \citenamefont
  {Vecchio}, \citenamefont {Vedovato}, \citenamefont {van Veggel},
  \citenamefont {Veitch}, \citenamefont {Veitch}, \citenamefont {Veltkamp},
  \citenamefont {Verkindt}, \citenamefont {Vetrano}, \citenamefont {Vicer\'e},
  \citenamefont {Villar}, \citenamefont {Vinet}, \citenamefont {Vocca},
  \citenamefont {Vorvick}, \citenamefont {Vyachanin}, \citenamefont {Waldman},
  \citenamefont {Wallace}, \citenamefont {Wanner}, \citenamefont {Ward},
  \citenamefont {Was}, \citenamefont {Wei}, \citenamefont {Weinert},
  \citenamefont {Weinstein}, \citenamefont {Weiss}, \citenamefont {Wen},
  \citenamefont {Wen}, \citenamefont {Wessels}, \citenamefont {West},
  \citenamefont {Westphal}, \citenamefont {Wette}, \citenamefont {Whelan},
  \citenamefont {Whitcomb}, \citenamefont {Whiting}, \citenamefont {Wilkinson},
  \citenamefont {Willems}, \citenamefont {Williams}, \citenamefont {Williams},
  \citenamefont {Willke}, \citenamefont {Wilmut}, \citenamefont {Winkelmann},
  \citenamefont {Winkler}, \citenamefont {Wipf}, \citenamefont {Wiseman},
  \citenamefont {Woan}, \citenamefont {Wooley}, \citenamefont {Worden},
  \citenamefont {Yakushin}, \citenamefont {Yamamoto}, \citenamefont {Yamamoto},
  \citenamefont {Yeaton-Massey}, \citenamefont {Yoshida}, \citenamefont
  {Yvert}, \citenamefont {Zanolin}, \citenamefont {Zhang}, \citenamefont
  {Zhang}, \citenamefont {Zhao}, \citenamefont {Zotov}, \citenamefont
  {Zucker},\ and\ \citenamefont {Zweizig}}]{PhysRevD.81.102001}%
  \BibitemOpen
  \bibfield  {author} {\bibinfo {author} {\bibfnamefont {J.}~\bibnamefont
  {Abadie}}, \bibinfo {author} {\bibfnamefont {B.~P.}\ \bibnamefont {Abbott}},
  \bibinfo {author} {\bibfnamefont {R.}~\bibnamefont {Abbott}}, \bibinfo
  {author} {\bibfnamefont {T.}~\bibnamefont {Accadia}}, \bibinfo {author}
  {\bibfnamefont {F.}~\bibnamefont {Acernese}}, \bibinfo {author}
  {\bibfnamefont {R.}~\bibnamefont {Adhikari}}, \bibinfo {author}
  {\bibfnamefont {P.}~\bibnamefont {Ajith}}, \bibinfo {author} {\bibfnamefont
  {B.}~\bibnamefont {Allen}}, \bibinfo {author} {\bibfnamefont
  {G.}~\bibnamefont {Allen}}, \bibinfo {author} {\bibfnamefont
  {E.}~\bibnamefont {Amador~Ceron}}, \bibinfo {author} {\bibfnamefont {R.~S.}\
  \bibnamefont {Amin}}, \bibinfo {author} {\bibfnamefont {S.~B.}\ \bibnamefont
  {Anderson}}, \bibinfo {author} {\bibfnamefont {W.~G.}\ \bibnamefont
  {Anderson}}, \bibinfo {author} {\bibfnamefont {F.}~\bibnamefont {Antonucci}},
  \bibinfo {author} {\bibfnamefont {M.~A.}\ \bibnamefont {Arain}}, \bibinfo
  {author} {\bibfnamefont {M.}~\bibnamefont {Araya}}, \bibinfo {author}
  {\bibfnamefont {K.~G.}\ \bibnamefont {Arun}}, \bibinfo {author}
  {\bibfnamefont {Y.}~\bibnamefont {Aso}}, \bibinfo {author} {\bibfnamefont
  {S.}~\bibnamefont {Aston}}, \bibinfo {author} {\bibfnamefont
  {P.}~\bibnamefont {Astone}}, \bibinfo {author} {\bibfnamefont
  {P.}~\bibnamefont {Aufmuth}}, \bibinfo {author} {\bibfnamefont
  {C.}~\bibnamefont {Aulbert}}, \bibinfo {author} {\bibfnamefont
  {S.}~\bibnamefont {Babak}}, \bibinfo {author} {\bibfnamefont
  {P.}~\bibnamefont {Baker}}, \bibinfo {author} {\bibfnamefont
  {G.}~\bibnamefont {Ballardin}}, \bibinfo {author} {\bibfnamefont
  {S.}~\bibnamefont {Ballmer}}, \bibinfo {author} {\bibfnamefont
  {D.}~\bibnamefont {Barker}}, \bibinfo {author} {\bibfnamefont
  {F.}~\bibnamefont {Barone}}, \bibinfo {author} {\bibfnamefont
  {B.}~\bibnamefont {Barr}}, \bibinfo {author} {\bibfnamefont {P.}~\bibnamefont
  {Barriga}}, \bibinfo {author} {\bibfnamefont {L.}~\bibnamefont {Barsotti}},
  \bibinfo {author} {\bibfnamefont {M.}~\bibnamefont {Barsuglia}}, \bibinfo
  {author} {\bibfnamefont {M.~A.}\ \bibnamefont {Barton}}, \bibinfo {author}
  {\bibfnamefont {I.}~\bibnamefont {Bartos}}, \bibinfo {author} {\bibfnamefont
  {R.}~\bibnamefont {Bassiri}}, \bibinfo {author} {\bibfnamefont
  {M.}~\bibnamefont {Bastarrika}}, \bibinfo {author} {\bibfnamefont {T.~S.}\
  \bibnamefont {Bauer}}, \bibinfo {author} {\bibfnamefont {B.}~\bibnamefont
  {Behnke}}, \bibinfo {author} {\bibfnamefont {M.~G.}\ \bibnamefont {Beker}},
  \bibinfo {author} {\bibfnamefont {A.}~\bibnamefont {Belletoile}}, \bibinfo
  {author} {\bibfnamefont {M.}~\bibnamefont {Benacquista}}, \bibinfo {author}
  {\bibfnamefont {J.}~\bibnamefont {Betzwieser}}, \bibinfo {author}
  {\bibfnamefont {P.~T.}\ \bibnamefont {Beyersdorf}}, \bibinfo {author}
  {\bibfnamefont {S.}~\bibnamefont {Bigotta}}, \bibinfo {author} {\bibfnamefont
  {I.~A.}\ \bibnamefont {Bilenko}}, \bibinfo {author} {\bibfnamefont
  {G.}~\bibnamefont {Billingsley}}, \bibinfo {author} {\bibfnamefont
  {S.}~\bibnamefont {Birindelli}}, \bibinfo {author} {\bibfnamefont
  {R.}~\bibnamefont {Biswas}}, \bibinfo {author} {\bibfnamefont {M.~A.}\
  \bibnamefont {Bizouard}}, \bibinfo {author} {\bibfnamefont {E.}~\bibnamefont
  {Black}}, \bibinfo {author} {\bibfnamefont {J.~K.}\ \bibnamefont
  {Blackburn}}, \bibinfo {author} {\bibfnamefont {L.}~\bibnamefont
  {Blackburn}}, \bibinfo {author} {\bibfnamefont {D.}~\bibnamefont {Blair}},
  \bibinfo {author} {\bibfnamefont {B.}~\bibnamefont {Bland}}, \bibinfo
  {author} {\bibfnamefont {M.}~\bibnamefont {Blom}}, \bibinfo {author}
  {\bibfnamefont {C.}~\bibnamefont {Boccara}}, \bibinfo {author} {\bibfnamefont
  {O.}~\bibnamefont {Bock}}, \bibinfo {author} {\bibfnamefont {T.~P.}\
  \bibnamefont {Bodiya}}, \bibinfo {author} {\bibfnamefont {R.}~\bibnamefont
  {Bondarescu}}, \bibinfo {author} {\bibfnamefont {F.}~\bibnamefont {Bondu}},
  \bibinfo {author} {\bibfnamefont {L.}~\bibnamefont {Bonelli}}, \bibinfo
  {author} {\bibfnamefont {R.}~\bibnamefont {Bonnand}}, \bibinfo {author}
  {\bibfnamefont {R.}~\bibnamefont {Bork}}, \bibinfo {author} {\bibfnamefont
  {M.}~\bibnamefont {Born}}, \bibinfo {author} {\bibfnamefont {S.}~\bibnamefont
  {Bose}}, \bibinfo {author} {\bibfnamefont {L.}~\bibnamefont {Bosi}}, \bibinfo
  {author} {\bibfnamefont {B.}~\bibnamefont {Bouhou}}, \bibinfo {author}
  {\bibfnamefont {S.}~\bibnamefont {Braccini}}, \bibinfo {author}
  {\bibfnamefont {C.}~\bibnamefont {Bradaschia}}, \bibinfo {author}
  {\bibfnamefont {P.~R.}\ \bibnamefont {Brady}}, \bibinfo {author}
  {\bibfnamefont {V.~B.}\ \bibnamefont {Braginsky}}, \bibinfo {author}
  {\bibfnamefont {J.~E.}\ \bibnamefont {Brau}}, \bibinfo {author}
  {\bibfnamefont {J.}~\bibnamefont {Breyer}}, \bibinfo {author} {\bibfnamefont
  {D.~O.}\ \bibnamefont {Bridges}}, \bibinfo {author} {\bibfnamefont
  {A.}~\bibnamefont {Brillet}}, \bibinfo {author} {\bibfnamefont
  {M.}~\bibnamefont {Brinkmann}}, \bibinfo {author} {\bibfnamefont
  {V.}~\bibnamefont {Brisson}}, \bibinfo {author} {\bibfnamefont
  {M.}~\bibnamefont {Britzger}}, \bibinfo {author} {\bibfnamefont {A.~F.}\
  \bibnamefont {Brooks}}, \bibinfo {author} {\bibfnamefont {D.~A.}\
  \bibnamefont {Brown}}, \bibinfo {author} {\bibfnamefont {R.}~\bibnamefont
  {Budzy\ifmmode~\acute{n}\else \'{n}\fi{}ski}}, \bibinfo {author}
  {\bibfnamefont {T.}~\bibnamefont {Bulik}}, \bibinfo {author} {\bibfnamefont
  {A.}~\bibnamefont {Bullington}}, \bibinfo {author} {\bibfnamefont {H.~J.}\
  \bibnamefont {Bulten}}, \bibinfo {author} {\bibfnamefont {A.}~\bibnamefont
  {Buonanno}}, \bibinfo {author} {\bibfnamefont {O.}~\bibnamefont
  {Burmeister}}, \bibinfo {author} {\bibfnamefont {D.}~\bibnamefont
  {Buskulic}}, \bibinfo {author} {\bibfnamefont {C.}~\bibnamefont {Buy}},
  \bibinfo {author} {\bibfnamefont {R.~L.}\ \bibnamefont {Byer}}, \bibinfo
  {author} {\bibfnamefont {L.}~\bibnamefont {Cadonati}}, \bibinfo {author}
  {\bibfnamefont {G.}~\bibnamefont {Cagnoli}}, \bibinfo {author} {\bibfnamefont
  {J.}~\bibnamefont {Cain}}, \bibinfo {author} {\bibfnamefont {E.}~\bibnamefont
  {Calloni}}, \bibinfo {author} {\bibfnamefont {J.~B.}\ \bibnamefont {Camp}},
  \bibinfo {author} {\bibfnamefont {E.}~\bibnamefont {Campagna}}, \bibinfo
  {author} {\bibfnamefont {J.}~\bibnamefont {Cannizzo}}, \bibinfo {author}
  {\bibfnamefont {K.~C.}\ \bibnamefont {Cannon}}, \bibinfo {author}
  {\bibfnamefont {B.}~\bibnamefont {Canuel}}, \bibinfo {author} {\bibfnamefont
  {J.}~\bibnamefont {Cao}}, \bibinfo {author} {\bibfnamefont {C.~D.}\
  \bibnamefont {Capano}}, \bibinfo {author} {\bibfnamefont {F.}~\bibnamefont
  {Carbognani}}, \bibinfo {author} {\bibfnamefont {L.}~\bibnamefont
  {Cardenas}}, \bibinfo {author} {\bibfnamefont {S.}~\bibnamefont {Caudill}},
  \bibinfo {author} {\bibfnamefont {M.}~\bibnamefont {Cavagli\`a}}, \bibinfo
  {author} {\bibfnamefont {F.}~\bibnamefont {Cavalier}}, \bibinfo {author}
  {\bibfnamefont {R.}~\bibnamefont {Cavalieri}}, \bibinfo {author}
  {\bibfnamefont {G.}~\bibnamefont {Cella}}, \bibinfo {author} {\bibfnamefont
  {C.}~\bibnamefont {Cepeda}}, \bibinfo {author} {\bibfnamefont
  {E.}~\bibnamefont {Cesarini}}, \bibinfo {author} {\bibfnamefont
  {T.}~\bibnamefont {Chalermsongsak}}, \bibinfo {author} {\bibfnamefont
  {E.}~\bibnamefont {Chalkley}}, \bibinfo {author} {\bibfnamefont
  {P.}~\bibnamefont {Charlton}}, \bibinfo {author} {\bibfnamefont
  {E.}~\bibnamefont {Chassande-Mottin}}, \bibinfo {author} {\bibfnamefont
  {S.}~\bibnamefont {Chatterji}}, \bibinfo {author} {\bibfnamefont
  {S.}~\bibnamefont {Chelkowski}}, \bibinfo {author} {\bibfnamefont
  {Y.}~\bibnamefont {Chen}}, \bibinfo {author} {\bibfnamefont {A.}~\bibnamefont
  {Chincarini}}, \bibinfo {author} {\bibfnamefont {N.}~\bibnamefont
  {Christensen}}, \bibinfo {author} {\bibfnamefont {S.~S.~Y.}\ \bibnamefont
  {Chua}}, \bibinfo {author} {\bibfnamefont {C.~T.~Y.}\ \bibnamefont {Chung}},
  \bibinfo {author} {\bibfnamefont {D.}~\bibnamefont {Clark}}, \bibinfo
  {author} {\bibfnamefont {J.}~\bibnamefont {Clark}}, \bibinfo {author}
  {\bibfnamefont {J.~H.}\ \bibnamefont {Clayton}}, \bibinfo {author}
  {\bibfnamefont {F.}~\bibnamefont {Cleva}}, \bibinfo {author} {\bibfnamefont
  {E.}~\bibnamefont {Coccia}}, \bibinfo {author} {\bibfnamefont {C.~N.}\
  \bibnamefont {Colacino}}, \bibinfo {author} {\bibfnamefont {J.}~\bibnamefont
  {Colas}}, \bibinfo {author} {\bibfnamefont {A.}~\bibnamefont {Colla}},
  \bibinfo {author} {\bibfnamefont {M.}~\bibnamefont {Colombini}}, \bibinfo
  {author} {\bibfnamefont {R.}~\bibnamefont {Conte}}, \bibinfo {author}
  {\bibfnamefont {D.}~\bibnamefont {Cook}}, \bibinfo {author} {\bibfnamefont
  {T.~R.~C.}\ \bibnamefont {Corbitt}}, \bibinfo {author} {\bibfnamefont
  {N.}~\bibnamefont {Cornish}}, \bibinfo {author} {\bibfnamefont
  {A.}~\bibnamefont {Corsi}}, \bibinfo {author} {\bibfnamefont {J.-P.}\
  \bibnamefont {Coulon}}, \bibinfo {author} {\bibfnamefont {D.}~\bibnamefont
  {Coward}}, \bibinfo {author} {\bibfnamefont {D.~C.}\ \bibnamefont {Coyne}},
  \bibinfo {author} {\bibfnamefont {J.~D.~E.}\ \bibnamefont {Creighton}},
  \bibinfo {author} {\bibfnamefont {T.~D.}\ \bibnamefont {Creighton}}, \bibinfo
  {author} {\bibfnamefont {A.~M.}\ \bibnamefont {Cruise}}, \bibinfo {author}
  {\bibfnamefont {R.~M.}\ \bibnamefont {Culter}}, \bibinfo {author}
  {\bibfnamefont {A.}~\bibnamefont {Cumming}}, \bibinfo {author} {\bibfnamefont
  {L.}~\bibnamefont {Cunningham}}, \bibinfo {author} {\bibfnamefont
  {E.}~\bibnamefont {Cuoco}}, \bibinfo {author} {\bibfnamefont
  {K.}~\bibnamefont {Dahl}}, \bibinfo {author} {\bibfnamefont {S.~L.}\
  \bibnamefont {Danilishin}}, \bibinfo {author} {\bibfnamefont
  {S.}~\bibnamefont {D'Antonio}}, \bibinfo {author} {\bibfnamefont
  {K.}~\bibnamefont {Danzmann}}, \bibinfo {author} {\bibfnamefont
  {V.}~\bibnamefont {Dattilo}}, \bibinfo {author} {\bibfnamefont
  {B.}~\bibnamefont {Daudert}}, \bibinfo {author} {\bibfnamefont
  {M.}~\bibnamefont {Davier}}, \bibinfo {author} {\bibfnamefont
  {G.}~\bibnamefont {Davies}}, \bibinfo {author} {\bibfnamefont {E.~J.}\
  \bibnamefont {Daw}}, \bibinfo {author} {\bibfnamefont {R.}~\bibnamefont
  {Day}}, \bibinfo {author} {\bibfnamefont {T.}~\bibnamefont {Dayanga}},
  \bibinfo {author} {\bibfnamefont {R.}~\bibnamefont {De~Rosa}}, \bibinfo
  {author} {\bibfnamefont {D.}~\bibnamefont {DeBra}}, \bibinfo {author}
  {\bibfnamefont {J.}~\bibnamefont {Degallaix}}, \bibinfo {author}
  {\bibfnamefont {M.}~\bibnamefont {del Prete}}, \bibinfo {author}
  {\bibfnamefont {V.}~\bibnamefont {Dergachev}}, \bibinfo {author}
  {\bibfnamefont {R.}~\bibnamefont {DeSalvo}}, \bibinfo {author} {\bibfnamefont
  {S.}~\bibnamefont {Dhurandhar}}, \bibinfo {author} {\bibfnamefont
  {L.}~\bibnamefont {Di~Fiore}}, \bibinfo {author} {\bibfnamefont
  {A.}~\bibnamefont {Di~Lieto}}, \bibinfo {author} {\bibfnamefont
  {M.}~\bibnamefont {Di~Paolo~Emilio}}, \bibinfo {author} {\bibfnamefont
  {A.}~\bibnamefont {Di~Virgilio}}, \bibinfo {author} {\bibfnamefont
  {M.}~\bibnamefont {D\'{\i}az}}, \bibinfo {author} {\bibfnamefont
  {A.}~\bibnamefont {Dietz}}, \bibinfo {author} {\bibfnamefont
  {F.}~\bibnamefont {Donovan}}, \bibinfo {author} {\bibfnamefont {K.~L.}\
  \bibnamefont {Dooley}}, \bibinfo {author} {\bibfnamefont {E.~E.}\
  \bibnamefont {Doomes}}, \bibinfo {author} {\bibfnamefont {M.}~\bibnamefont
  {Drago}}, \bibinfo {author} {\bibfnamefont {R.~W.~P.}\ \bibnamefont
  {Drever}}, \bibinfo {author} {\bibfnamefont {J.}~\bibnamefont {Driggers}},
  \bibinfo {author} {\bibfnamefont {J.}~\bibnamefont {Dueck}}, \bibinfo
  {author} {\bibfnamefont {I.}~\bibnamefont {Duke}}, \bibinfo {author}
  {\bibfnamefont {J.-C.}\ \bibnamefont {Dumas}}, \bibinfo {author}
  {\bibfnamefont {S.}~\bibnamefont {Dwyer}}, \bibinfo {author} {\bibfnamefont
  {M.}~\bibnamefont {Edgar}}, \bibinfo {author} {\bibfnamefont
  {M.}~\bibnamefont {Edwards}}, \bibinfo {author} {\bibfnamefont
  {A.}~\bibnamefont {Effler}}, \bibinfo {author} {\bibfnamefont
  {P.}~\bibnamefont {Ehrens}}, \bibinfo {author} {\bibfnamefont
  {T.}~\bibnamefont {Etzel}}, \bibinfo {author} {\bibfnamefont
  {M.}~\bibnamefont {Evans}}, \bibinfo {author} {\bibfnamefont
  {T.}~\bibnamefont {Evans}}, \bibinfo {author} {\bibfnamefont
  {V.}~\bibnamefont {Fafone}}, \bibinfo {author} {\bibfnamefont
  {S.}~\bibnamefont {Fairhurst}}, \bibinfo {author} {\bibfnamefont
  {Y.}~\bibnamefont {Faltas}}, \bibinfo {author} {\bibfnamefont
  {Y.}~\bibnamefont {Fan}}, \bibinfo {author} {\bibfnamefont {D.}~\bibnamefont
  {Fazi}}, \bibinfo {author} {\bibfnamefont {H.}~\bibnamefont {Fehrmann}},
  \bibinfo {author} {\bibfnamefont {I.}~\bibnamefont {Ferrante}}, \bibinfo
  {author} {\bibfnamefont {F.}~\bibnamefont {Fidecaro}}, \bibinfo {author}
  {\bibfnamefont {L.~S.}\ \bibnamefont {Finn}}, \bibinfo {author}
  {\bibfnamefont {I.}~\bibnamefont {Fiori}}, \bibinfo {author} {\bibfnamefont
  {R.}~\bibnamefont {Flaminio}}, \bibinfo {author} {\bibfnamefont
  {K.}~\bibnamefont {Flasch}}, \bibinfo {author} {\bibfnamefont
  {S.}~\bibnamefont {Foley}}, \bibinfo {author} {\bibfnamefont
  {C.}~\bibnamefont {Forrest}}, \bibinfo {author} {\bibfnamefont
  {N.}~\bibnamefont {Fotopoulos}}, \bibinfo {author} {\bibfnamefont {J.-D.}\
  \bibnamefont {Fournier}}, \bibinfo {author} {\bibfnamefont {J.}~\bibnamefont
  {Franc}}, \bibinfo {author} {\bibfnamefont {S.}~\bibnamefont {Frasca}},
  \bibinfo {author} {\bibfnamefont {F.}~\bibnamefont {Frasconi}}, \bibinfo
  {author} {\bibfnamefont {M.}~\bibnamefont {Frede}}, \bibinfo {author}
  {\bibfnamefont {M.}~\bibnamefont {Frei}}, \bibinfo {author} {\bibfnamefont
  {Z.}~\bibnamefont {Frei}}, \bibinfo {author} {\bibfnamefont {A.}~\bibnamefont
  {Freise}}, \bibinfo {author} {\bibfnamefont {R.}~\bibnamefont {Frey}},
  \bibinfo {author} {\bibfnamefont {T.~T.}\ \bibnamefont {Fricke}}, \bibinfo
  {author} {\bibfnamefont {D.}~\bibnamefont {Friedrich}}, \bibinfo {author}
  {\bibfnamefont {P.}~\bibnamefont {Fritschel}}, \bibinfo {author}
  {\bibfnamefont {V.~V.}\ \bibnamefont {Frolov}}, \bibinfo {author}
  {\bibfnamefont {P.}~\bibnamefont {Fulda}}, \bibinfo {author} {\bibfnamefont
  {M.}~\bibnamefont {Fyffe}}, \bibinfo {author} {\bibfnamefont
  {M.}~\bibnamefont {Galimberti}}, \bibinfo {author} {\bibfnamefont
  {L.}~\bibnamefont {Gammaitoni}}, \bibinfo {author} {\bibfnamefont {J.~A.}\
  \bibnamefont {Garofoli}}, \bibinfo {author} {\bibfnamefont {F.}~\bibnamefont
  {Garufi}}, \bibinfo {author} {\bibfnamefont {G.}~\bibnamefont {Gemme}},
  \bibinfo {author} {\bibfnamefont {E.}~\bibnamefont {Genin}}, \bibinfo
  {author} {\bibfnamefont {A.}~\bibnamefont {Gennai}}, \bibinfo {author}
  {\bibfnamefont {S.}~\bibnamefont {Ghosh}}, \bibinfo {author} {\bibfnamefont
  {J.~A.}\ \bibnamefont {Giaime}}, \bibinfo {author} {\bibfnamefont
  {S.}~\bibnamefont {Giampanis}}, \bibinfo {author} {\bibfnamefont {K.~D.}\
  \bibnamefont {Giardina}}, \bibinfo {author} {\bibfnamefont {A.}~\bibnamefont
  {Giazotto}}, \bibinfo {author} {\bibfnamefont {E.}~\bibnamefont {Goetz}},
  \bibinfo {author} {\bibfnamefont {L.~M.}\ \bibnamefont {Goggin}}, \bibinfo
  {author} {\bibfnamefont {G.}~\bibnamefont {Gonz\'alez}}, \bibinfo {author}
  {\bibfnamefont {S.}~\bibnamefont {Go\ss{}ler}}, \bibinfo {author}
  {\bibfnamefont {R.}~\bibnamefont {Gouaty}}, \bibinfo {author} {\bibfnamefont
  {M.}~\bibnamefont {Granata}}, \bibinfo {author} {\bibfnamefont
  {A.}~\bibnamefont {Grant}}, \bibinfo {author} {\bibfnamefont
  {S.}~\bibnamefont {Gras}}, \bibinfo {author} {\bibfnamefont {C.}~\bibnamefont
  {Gray}}, \bibinfo {author} {\bibfnamefont {R.~J.~S.}\ \bibnamefont
  {Greenhalgh}}, \bibinfo {author} {\bibfnamefont {A.~M.}\ \bibnamefont
  {Gretarsson}}, \bibinfo {author} {\bibfnamefont {C.}~\bibnamefont
  {Greverie}}, \bibinfo {author} {\bibfnamefont {R.}~\bibnamefont {Grosso}},
  \bibinfo {author} {\bibfnamefont {H.}~\bibnamefont {Grote}}, \bibinfo
  {author} {\bibfnamefont {S.}~\bibnamefont {Grunewald}}, \bibinfo {author}
  {\bibfnamefont {G.~M.}\ \bibnamefont {Guidi}}, \bibinfo {author}
  {\bibfnamefont {E.~K.}\ \bibnamefont {Gustafson}}, \bibinfo {author}
  {\bibfnamefont {R.}~\bibnamefont {Gustafson}}, \bibinfo {author}
  {\bibfnamefont {B.}~\bibnamefont {Hage}}, \bibinfo {author} {\bibfnamefont
  {J.~M.}\ \bibnamefont {Hallam}}, \bibinfo {author} {\bibfnamefont
  {D.}~\bibnamefont {Hammer}}, \bibinfo {author} {\bibfnamefont {G.~D.}\
  \bibnamefont {Hammond}}, \bibinfo {author} {\bibfnamefont {C.}~\bibnamefont
  {Hanna}}, \bibinfo {author} {\bibfnamefont {J.}~\bibnamefont {Hanson}},
  \bibinfo {author} {\bibfnamefont {J.}~\bibnamefont {Harms}}, \bibinfo
  {author} {\bibfnamefont {G.~M.}\ \bibnamefont {Harry}}, \bibinfo {author}
  {\bibfnamefont {I.~W.}\ \bibnamefont {Harry}}, \bibinfo {author}
  {\bibfnamefont {E.~D.}\ \bibnamefont {Harstad}}, \bibinfo {author}
  {\bibfnamefont {K.}~\bibnamefont {Haughian}}, \bibinfo {author}
  {\bibfnamefont {K.}~\bibnamefont {Hayama}}, \bibinfo {author} {\bibfnamefont
  {J.-F.}\ \bibnamefont {Hayau}}, \bibinfo {author} {\bibfnamefont
  {T.}~\bibnamefont {Hayler}}, \bibinfo {author} {\bibfnamefont
  {J.}~\bibnamefont {Heefner}}, \bibinfo {author} {\bibfnamefont
  {H.}~\bibnamefont {Heitmann}}, \bibinfo {author} {\bibfnamefont
  {P.}~\bibnamefont {Hello}}, \bibinfo {author} {\bibfnamefont {I.~S.}\
  \bibnamefont {Heng}}, \bibinfo {author} {\bibfnamefont {A.}~\bibnamefont
  {Heptonstall}}, \bibinfo {author} {\bibfnamefont {M.}~\bibnamefont
  {Hewitson}}, \bibinfo {author} {\bibfnamefont {S.}~\bibnamefont {Hild}},
  \bibinfo {author} {\bibfnamefont {E.}~\bibnamefont {Hirose}}, \bibinfo
  {author} {\bibfnamefont {D.}~\bibnamefont {Hoak}}, \bibinfo {author}
  {\bibfnamefont {K.~A.}\ \bibnamefont {Hodge}}, \bibinfo {author}
  {\bibfnamefont {K.}~\bibnamefont {Holt}}, \bibinfo {author} {\bibfnamefont
  {D.~J.}\ \bibnamefont {Hosken}}, \bibinfo {author} {\bibfnamefont
  {J.}~\bibnamefont {Hough}}, \bibinfo {author} {\bibfnamefont
  {E.}~\bibnamefont {Howell}}, \bibinfo {author} {\bibfnamefont
  {D.}~\bibnamefont {Hoyland}}, \bibinfo {author} {\bibfnamefont
  {D.}~\bibnamefont {Huet}}, \bibinfo {author} {\bibfnamefont {B.}~\bibnamefont
  {Hughey}}, \bibinfo {author} {\bibfnamefont {S.}~\bibnamefont {Husa}},
  \bibinfo {author} {\bibfnamefont {S.~H.}\ \bibnamefont {Huttner}}, \bibinfo
  {author} {\bibfnamefont {D.~R.}\ \bibnamefont {Ingram}}, \bibinfo {author}
  {\bibfnamefont {T.}~\bibnamefont {Isogai}}, \bibinfo {author} {\bibfnamefont
  {A.}~\bibnamefont {Ivanov}}, \bibinfo {author} {\bibfnamefont
  {P.}~\bibnamefont {Jaranowski}}, \bibinfo {author} {\bibfnamefont {W.~W.}\
  \bibnamefont {Johnson}}, \bibinfo {author} {\bibfnamefont {D.~I.}\
  \bibnamefont {Jones}}, \bibinfo {author} {\bibfnamefont {G.}~\bibnamefont
  {Jones}}, \bibinfo {author} {\bibfnamefont {R.}~\bibnamefont {Jones}},
  \bibinfo {author} {\bibfnamefont {L.}~\bibnamefont {Ju}}, \bibinfo {author}
  {\bibfnamefont {P.}~\bibnamefont {Kalmus}}, \bibinfo {author} {\bibfnamefont
  {V.}~\bibnamefont {Kalogera}}, \bibinfo {author} {\bibfnamefont
  {S.}~\bibnamefont {Kandhasamy}}, \bibinfo {author} {\bibfnamefont
  {J.}~\bibnamefont {Kanner}}, \bibinfo {author} {\bibfnamefont
  {E.}~\bibnamefont {Katsavounidis}}, \bibinfo {author} {\bibfnamefont
  {K.}~\bibnamefont {Kawabe}}, \bibinfo {author} {\bibfnamefont
  {S.}~\bibnamefont {Kawamura}}, \bibinfo {author} {\bibfnamefont
  {F.}~\bibnamefont {Kawazoe}}, \bibinfo {author} {\bibfnamefont
  {W.}~\bibnamefont {Kells}}, \bibinfo {author} {\bibfnamefont {D.~G.}\
  \bibnamefont {Keppel}}, \bibinfo {author} {\bibfnamefont {A.}~\bibnamefont
  {Khalaidovski}}, \bibinfo {author} {\bibfnamefont {F.~Y.}\ \bibnamefont
  {Khalili}}, \bibinfo {author} {\bibfnamefont {R.}~\bibnamefont {Khan}},
  \bibinfo {author} {\bibfnamefont {E.}~\bibnamefont {Khazanov}}, \bibinfo
  {author} {\bibfnamefont {H.}~\bibnamefont {Kim}}, \bibinfo {author}
  {\bibfnamefont {P.~J.}\ \bibnamefont {King}}, \bibinfo {author}
  {\bibfnamefont {J.~S.}\ \bibnamefont {Kissel}}, \bibinfo {author}
  {\bibfnamefont {S.}~\bibnamefont {Klimenko}}, \bibinfo {author}
  {\bibfnamefont {K.}~\bibnamefont {Kokeyama}}, \bibinfo {author}
  {\bibfnamefont {V.}~\bibnamefont {Kondrashov}}, \bibinfo {author}
  {\bibfnamefont {R.}~\bibnamefont {Kopparapu}}, \bibinfo {author}
  {\bibfnamefont {S.}~\bibnamefont {Koranda}}, \bibinfo {author} {\bibfnamefont
  {I.}~\bibnamefont {Kowalska}}, \bibinfo {author} {\bibfnamefont
  {D.}~\bibnamefont {Kozak}}, \bibinfo {author} {\bibfnamefont
  {V.}~\bibnamefont {Kringel}}, \bibinfo {author} {\bibfnamefont
  {B.}~\bibnamefont {Krishnan}}, \bibinfo {author} {\bibfnamefont
  {A.}~\bibnamefont {Kr\'olak}}, \bibinfo {author} {\bibfnamefont
  {G.}~\bibnamefont {Kuehn}}, \bibinfo {author} {\bibfnamefont
  {J.}~\bibnamefont {Kullman}}, \bibinfo {author} {\bibfnamefont
  {R.}~\bibnamefont {Kumar}}, \bibinfo {author} {\bibfnamefont
  {P.}~\bibnamefont {Kwee}}, \bibinfo {author} {\bibfnamefont {P.~K.}\
  \bibnamefont {Lam}}, \bibinfo {author} {\bibfnamefont {M.}~\bibnamefont
  {Landry}}, \bibinfo {author} {\bibfnamefont {M.}~\bibnamefont {Lang}},
  \bibinfo {author} {\bibfnamefont {B.}~\bibnamefont {Lantz}}, \bibinfo
  {author} {\bibfnamefont {N.}~\bibnamefont {Lastzka}}, \bibinfo {author}
  {\bibfnamefont {A.}~\bibnamefont {Lazzarini}}, \bibinfo {author}
  {\bibfnamefont {P.}~\bibnamefont {Leaci}}, \bibinfo {author} {\bibfnamefont
  {M.}~\bibnamefont {Lei}}, \bibinfo {author} {\bibfnamefont {N.}~\bibnamefont
  {Leindecker}}, \bibinfo {author} {\bibfnamefont {I.}~\bibnamefont {Leonor}},
  \bibinfo {author} {\bibfnamefont {N.}~\bibnamefont {Leroy}}, \bibinfo
  {author} {\bibfnamefont {N.}~\bibnamefont {Letendre}}, \bibinfo {author}
  {\bibfnamefont {T.~G.~F.}\ \bibnamefont {Li}}, \bibinfo {author}
  {\bibfnamefont {H.}~\bibnamefont {Lin}}, \bibinfo {author} {\bibfnamefont
  {P.~E.}\ \bibnamefont {Lindquist}}, \bibinfo {author} {\bibfnamefont {T.~B.}\
  \bibnamefont {Littenberg}}, \bibinfo {author} {\bibfnamefont {N.~A.}\
  \bibnamefont {Lockerbie}}, \bibinfo {author} {\bibfnamefont {D.}~\bibnamefont
  {Lodhia}}, \bibinfo {author} {\bibfnamefont {M.}~\bibnamefont {Lorenzini}},
  \bibinfo {author} {\bibfnamefont {V.}~\bibnamefont {Loriette}}, \bibinfo
  {author} {\bibfnamefont {M.}~\bibnamefont {Lormand}}, \bibinfo {author}
  {\bibfnamefont {G.}~\bibnamefont {Losurdo}}, \bibinfo {author} {\bibfnamefont
  {P.}~\bibnamefont {Lu}}, \bibinfo {author} {\bibfnamefont {M.}~\bibnamefont
  {Lubinski}}, \bibinfo {author} {\bibfnamefont {A.}~\bibnamefont
  {Lucianetti}}, \bibinfo {author} {\bibfnamefont {H.}~\bibnamefont {L\"uck}},
  \bibinfo {author} {\bibfnamefont {A.}~\bibnamefont {Lundgren}}, \bibinfo
  {author} {\bibfnamefont {B.}~\bibnamefont {Machenschalk}}, \bibinfo {author}
  {\bibfnamefont {M.}~\bibnamefont {MacInnis}}, \bibinfo {author}
  {\bibfnamefont {M.}~\bibnamefont {Mageswaran}}, \bibinfo {author}
  {\bibfnamefont {K.}~\bibnamefont {Mailand}}, \bibinfo {author} {\bibfnamefont
  {E.}~\bibnamefont {Majorana}}, \bibinfo {author} {\bibfnamefont
  {C.}~\bibnamefont {Mak}}, \bibinfo {author} {\bibfnamefont {I.}~\bibnamefont
  {Maksimovic}}, \bibinfo {author} {\bibfnamefont {N.}~\bibnamefont {Man}},
  \bibinfo {author} {\bibfnamefont {I.}~\bibnamefont {Mandel}}, \bibinfo
  {author} {\bibfnamefont {V.}~\bibnamefont {Mandic}}, \bibinfo {author}
  {\bibfnamefont {M.}~\bibnamefont {Mantovani}}, \bibinfo {author}
  {\bibfnamefont {F.}~\bibnamefont {Marchesoni}}, \bibinfo {author}
  {\bibfnamefont {F.}~\bibnamefont {Marion}}, \bibinfo {author} {\bibfnamefont
  {S.}~\bibnamefont {M\'arka}}, \bibinfo {author} {\bibfnamefont
  {Z.}~\bibnamefont {M\'arka}}, \bibinfo {author} {\bibfnamefont
  {A.}~\bibnamefont {Markosyan}}, \bibinfo {author} {\bibfnamefont
  {J.}~\bibnamefont {Markowitz}}, \bibinfo {author} {\bibfnamefont
  {E.}~\bibnamefont {Maros}}, \bibinfo {author} {\bibfnamefont
  {J.}~\bibnamefont {Marque}}, \bibinfo {author} {\bibfnamefont
  {F.}~\bibnamefont {Martelli}}, \bibinfo {author} {\bibfnamefont {I.~W.}\
  \bibnamefont {Martin}}, \bibinfo {author} {\bibfnamefont {R.~M.}\
  \bibnamefont {Martin}}, \bibinfo {author} {\bibfnamefont {J.~N.}\
  \bibnamefont {Marx}}, \bibinfo {author} {\bibfnamefont {K.}~\bibnamefont
  {Mason}}, \bibinfo {author} {\bibfnamefont {A.}~\bibnamefont {Masserot}},
  \bibinfo {author} {\bibfnamefont {F.}~\bibnamefont {Matichard}}, \bibinfo
  {author} {\bibfnamefont {L.}~\bibnamefont {Matone}}, \bibinfo {author}
  {\bibfnamefont {R.~A.}\ \bibnamefont {Matzner}}, \bibinfo {author}
  {\bibfnamefont {N.}~\bibnamefont {Mavalvala}}, \bibinfo {author}
  {\bibfnamefont {R.}~\bibnamefont {McCarthy}}, \bibinfo {author}
  {\bibfnamefont {D.~E.}\ \bibnamefont {McClelland}}, \bibinfo {author}
  {\bibfnamefont {S.~C.}\ \bibnamefont {McGuire}}, \bibinfo {author}
  {\bibfnamefont {G.}~\bibnamefont {McIntyre}}, \bibinfo {author}
  {\bibfnamefont {D.~J.~A.}\ \bibnamefont {McKechan}}, \bibinfo {author}
  {\bibfnamefont {M.}~\bibnamefont {Mehmet}}, \bibinfo {author} {\bibfnamefont
  {A.}~\bibnamefont {Melatos}}, \bibinfo {author} {\bibfnamefont {A.~C.}\
  \bibnamefont {Melissinos}}, \bibinfo {author} {\bibfnamefont
  {G.}~\bibnamefont {Mendell}}, \bibinfo {author} {\bibfnamefont {D.~F.}\
  \bibnamefont {Men\'endez}}, \bibinfo {author} {\bibfnamefont {R.~A.}\
  \bibnamefont {Mercer}}, \bibinfo {author} {\bibfnamefont {L.}~\bibnamefont
  {Merill}}, \bibinfo {author} {\bibfnamefont {S.}~\bibnamefont {Meshkov}},
  \bibinfo {author} {\bibfnamefont {C.}~\bibnamefont {Messenger}}, \bibinfo
  {author} {\bibfnamefont {M.~S.}\ \bibnamefont {Meyer}}, \bibinfo {author}
  {\bibfnamefont {H.}~\bibnamefont {Miao}}, \bibinfo {author} {\bibfnamefont
  {C.}~\bibnamefont {Michel}}, \bibinfo {author} {\bibfnamefont
  {L.}~\bibnamefont {Milano}}, \bibinfo {author} {\bibfnamefont
  {J.}~\bibnamefont {Miller}}, \bibinfo {author} {\bibfnamefont
  {Y.}~\bibnamefont {Minenkov}}, \bibinfo {author} {\bibfnamefont
  {Y.}~\bibnamefont {Mino}}, \bibinfo {author} {\bibfnamefont {S.}~\bibnamefont
  {Mitra}}, \bibinfo {author} {\bibfnamefont {V.~P.}\ \bibnamefont
  {Mitrofanov}}, \bibinfo {author} {\bibfnamefont {G.}~\bibnamefont
  {Mitselmakher}}, \bibinfo {author} {\bibfnamefont {R.}~\bibnamefont
  {Mittleman}}, \bibinfo {author} {\bibfnamefont {O.}~\bibnamefont {Miyakawa}},
  \bibinfo {author} {\bibfnamefont {B.}~\bibnamefont {Moe}}, \bibinfo {author}
  {\bibfnamefont {M.}~\bibnamefont {Mohan}}, \bibinfo {author} {\bibfnamefont
  {S.~D.}\ \bibnamefont {Mohanty}}, \bibinfo {author} {\bibfnamefont
  {S.~R.~P.}\ \bibnamefont {Mohapatra}}, \bibinfo {author} {\bibfnamefont
  {J.}~\bibnamefont {Moreau}}, \bibinfo {author} {\bibfnamefont
  {G.}~\bibnamefont {Moreno}}, \bibinfo {author} {\bibfnamefont
  {N.}~\bibnamefont {Morgado}}, \bibinfo {author} {\bibfnamefont
  {A.}~\bibnamefont {Morgia}}, \bibinfo {author} {\bibfnamefont
  {K.}~\bibnamefont {Mors}}, \bibinfo {author} {\bibfnamefont {S.}~\bibnamefont
  {Mosca}}, \bibinfo {author} {\bibfnamefont {V.}~\bibnamefont {Moscatelli}},
  \bibinfo {author} {\bibfnamefont {K.}~\bibnamefont {Mossavi}}, \bibinfo
  {author} {\bibfnamefont {B.}~\bibnamefont {Mours}}, \bibinfo {author}
  {\bibfnamefont {C.}~\bibnamefont {MowLowry}}, \bibinfo {author}
  {\bibfnamefont {G.}~\bibnamefont {Mueller}}, \bibinfo {author} {\bibfnamefont
  {S.}~\bibnamefont {Mukherjee}}, \bibinfo {author} {\bibfnamefont
  {A.}~\bibnamefont {Mullavey}}, \bibinfo {author} {\bibfnamefont
  {H.}~\bibnamefont {M\"uller-Ebhardt}}, \bibinfo {author} {\bibfnamefont
  {J.}~\bibnamefont {Munch}}, \bibinfo {author} {\bibfnamefont {P.~G.}\
  \bibnamefont {Murray}}, \bibinfo {author} {\bibfnamefont {T.}~\bibnamefont
  {Nash}}, \bibinfo {author} {\bibfnamefont {R.}~\bibnamefont {Nawrodt}},
  \bibinfo {author} {\bibfnamefont {J.}~\bibnamefont {Nelson}}, \bibinfo
  {author} {\bibfnamefont {I.}~\bibnamefont {Neri}}, \bibinfo {author}
  {\bibfnamefont {G.}~\bibnamefont {Newton}}, \bibinfo {author} {\bibfnamefont
  {E.}~\bibnamefont {Nishida}}, \bibinfo {author} {\bibfnamefont
  {A.}~\bibnamefont {Nishizawa}}, \bibinfo {author} {\bibfnamefont
  {F.}~\bibnamefont {Nocera}}, \bibinfo {author} {\bibfnamefont
  {E.}~\bibnamefont {Ochsner}}, \bibinfo {author} {\bibfnamefont
  {J.}~\bibnamefont {O'Dell}}, \bibinfo {author} {\bibfnamefont {G.~H.}\
  \bibnamefont {Ogin}}, \bibinfo {author} {\bibfnamefont {R.}~\bibnamefont
  {Oldenburg}}, \bibinfo {author} {\bibfnamefont {B.}~\bibnamefont {O'Reilly}},
  \bibinfo {author} {\bibfnamefont {R.}~\bibnamefont {O'Shaughnessy}}, \bibinfo
  {author} {\bibfnamefont {D.~J.}\ \bibnamefont {Ottaway}}, \bibinfo {author}
  {\bibfnamefont {R.~S.}\ \bibnamefont {Ottens}}, \bibinfo {author}
  {\bibfnamefont {H.}~\bibnamefont {Overmier}}, \bibinfo {author}
  {\bibfnamefont {B.~J.}\ \bibnamefont {Owen}}, \bibinfo {author}
  {\bibfnamefont {A.}~\bibnamefont {Page}}, \bibinfo {author} {\bibfnamefont
  {G.}~\bibnamefont {Pagliaroli}}, \bibinfo {author} {\bibfnamefont
  {L.}~\bibnamefont {Palladino}}, \bibinfo {author} {\bibfnamefont
  {C.}~\bibnamefont {Palomba}}, \bibinfo {author} {\bibfnamefont
  {Y.}~\bibnamefont {Pan}}, \bibinfo {author} {\bibfnamefont {C.}~\bibnamefont
  {Pankow}}, \bibinfo {author} {\bibfnamefont {F.}~\bibnamefont {Paoletti}},
  \bibinfo {author} {\bibfnamefont {M.~A.}\ \bibnamefont {Papa}}, \bibinfo
  {author} {\bibfnamefont {S.}~\bibnamefont {Pardi}}, \bibinfo {author}
  {\bibfnamefont {M.}~\bibnamefont {Parisi}}, \bibinfo {author} {\bibfnamefont
  {A.}~\bibnamefont {Pasqualetti}}, \bibinfo {author} {\bibfnamefont
  {R.}~\bibnamefont {Passaquieti}}, \bibinfo {author} {\bibfnamefont
  {D.}~\bibnamefont {Passuello}}, \bibinfo {author} {\bibfnamefont
  {P.}~\bibnamefont {Patel}}, \bibinfo {author} {\bibfnamefont
  {D.}~\bibnamefont {Pathak}}, \bibinfo {author} {\bibfnamefont
  {M.}~\bibnamefont {Pedraza}}, \bibinfo {author} {\bibfnamefont
  {L.}~\bibnamefont {Pekowsky}}, \bibinfo {author} {\bibfnamefont
  {S.}~\bibnamefont {Penn}}, \bibinfo {author} {\bibfnamefont {C.}~\bibnamefont
  {Peralta}}, \bibinfo {author} {\bibfnamefont {A.}~\bibnamefont {Perreca}},
  \bibinfo {author} {\bibfnamefont {G.}~\bibnamefont {Persichetti}}, \bibinfo
  {author} {\bibfnamefont {M.}~\bibnamefont {Pichot}}, \bibinfo {author}
  {\bibfnamefont {M.}~\bibnamefont {Pickenpack}}, \bibinfo {author}
  {\bibfnamefont {F.}~\bibnamefont {Piergiovanni}}, \bibinfo {author}
  {\bibfnamefont {M.}~\bibnamefont {Pietka}}, \bibinfo {author} {\bibfnamefont
  {L.}~\bibnamefont {Pinard}}, \bibinfo {author} {\bibfnamefont {I.~M.}\
  \bibnamefont {Pinto}}, \bibinfo {author} {\bibfnamefont {M.}~\bibnamefont
  {Pitkin}}, \bibinfo {author} {\bibfnamefont {H.~J.}\ \bibnamefont {Pletsch}},
  \bibinfo {author} {\bibfnamefont {M.~V.}\ \bibnamefont {Plissi}}, \bibinfo
  {author} {\bibfnamefont {R.}~\bibnamefont {Poggiani}}, \bibinfo {author}
  {\bibfnamefont {F.}~\bibnamefont {Postiglione}}, \bibinfo {author}
  {\bibfnamefont {M.}~\bibnamefont {Prato}}, \bibinfo {author} {\bibfnamefont
  {M.}~\bibnamefont {Principe}}, \bibinfo {author} {\bibfnamefont
  {R.}~\bibnamefont {Prix}}, \bibinfo {author} {\bibfnamefont {G.~A.}\
  \bibnamefont {Prodi}}, \bibinfo {author} {\bibfnamefont {L.}~\bibnamefont
  {Prokhorov}}, \bibinfo {author} {\bibfnamefont {O.}~\bibnamefont {Puncken}},
  \bibinfo {author} {\bibfnamefont {M.}~\bibnamefont {Punturo}}, \bibinfo
  {author} {\bibfnamefont {P.}~\bibnamefont {Puppo}}, \bibinfo {author}
  {\bibfnamefont {V.}~\bibnamefont {Quetschke}}, \bibinfo {author}
  {\bibfnamefont {F.~J.}\ \bibnamefont {Raab}}, \bibinfo {author}
  {\bibfnamefont {D.~S.}\ \bibnamefont {Rabeling}}, \bibinfo {author}
  {\bibfnamefont {D.~S.}\ \bibnamefont {Rabeling}}, \bibinfo {author}
  {\bibfnamefont {H.}~\bibnamefont {Radkins}}, \bibinfo {author} {\bibfnamefont
  {P.}~\bibnamefont {Raffai}}, \bibinfo {author} {\bibfnamefont
  {Z.}~\bibnamefont {Raics}}, \bibinfo {author} {\bibfnamefont
  {M.}~\bibnamefont {Rakhmanov}}, \bibinfo {author} {\bibfnamefont
  {P.}~\bibnamefont {Rapagnani}}, \bibinfo {author} {\bibfnamefont
  {V.}~\bibnamefont {Raymond}}, \bibinfo {author} {\bibfnamefont
  {V.}~\bibnamefont {Re}}, \bibinfo {author} {\bibfnamefont {C.~M.}\
  \bibnamefont {Reed}}, \bibinfo {author} {\bibfnamefont {T.}~\bibnamefont
  {Reed}}, \bibinfo {author} {\bibfnamefont {T.}~\bibnamefont {Regimbau}},
  \bibinfo {author} {\bibfnamefont {H.}~\bibnamefont {Rehbein}}, \bibinfo
  {author} {\bibfnamefont {S.}~\bibnamefont {Reid}}, \bibinfo {author}
  {\bibfnamefont {D.~H.}\ \bibnamefont {Reitze}}, \bibinfo {author}
  {\bibfnamefont {F.}~\bibnamefont {Ricci}}, \bibinfo {author} {\bibfnamefont
  {R.}~\bibnamefont {Riesen}}, \bibinfo {author} {\bibfnamefont
  {K.}~\bibnamefont {Riles}}, \bibinfo {author} {\bibfnamefont
  {P.}~\bibnamefont {Roberts}}, \bibinfo {author} {\bibfnamefont {N.~A.}\
  \bibnamefont {Robertson}}, \bibinfo {author} {\bibfnamefont {F.}~\bibnamefont
  {Robinet}}, \bibinfo {author} {\bibfnamefont {C.}~\bibnamefont {Robinson}},
  \bibinfo {author} {\bibfnamefont {E.~L.}\ \bibnamefont {Robinson}}, \bibinfo
  {author} {\bibfnamefont {A.}~\bibnamefont {Rocchi}}, \bibinfo {author}
  {\bibfnamefont {S.}~\bibnamefont {Roddy}}, \bibinfo {author} {\bibfnamefont
  {C.}~\bibnamefont {R\"over}}, \bibinfo {author} {\bibfnamefont
  {L.}~\bibnamefont {Rolland}}, \bibinfo {author} {\bibfnamefont
  {J.}~\bibnamefont {Rollins}}, \bibinfo {author} {\bibfnamefont {J.~D.}\
  \bibnamefont {Romano}}, \bibinfo {author} {\bibfnamefont {R.}~\bibnamefont
  {Romano}}, \bibinfo {author} {\bibfnamefont {J.~H.}\ \bibnamefont {Romie}},
  \bibinfo {author} {\bibfnamefont {D.}~\bibnamefont
  {Rosi\ifmmode~\acute{n}\else \'{n}\fi{}ska}}, \bibinfo {author}
  {\bibfnamefont {S.}~\bibnamefont {Rowan}}, \bibinfo {author} {\bibfnamefont
  {A.}~\bibnamefont {R\"udiger}}, \bibinfo {author} {\bibfnamefont
  {P.}~\bibnamefont {Ruggi}}, \bibinfo {author} {\bibfnamefont
  {K.}~\bibnamefont {Ryan}}, \bibinfo {author} {\bibfnamefont {S.}~\bibnamefont
  {Sakata}}, \bibinfo {author} {\bibfnamefont {F.}~\bibnamefont {Salemi}},
  \bibinfo {author} {\bibfnamefont {L.}~\bibnamefont {Sammut}}, \bibinfo
  {author} {\bibfnamefont {L.}~\bibnamefont {Sancho de~la Jordana}}, \bibinfo
  {author} {\bibfnamefont {V.}~\bibnamefont {Sandberg}}, \bibinfo {author}
  {\bibfnamefont {V.}~\bibnamefont {Sannibale}}, \bibinfo {author}
  {\bibfnamefont {L.}~\bibnamefont {Santamar\'{\i}a}}, \bibinfo {author}
  {\bibfnamefont {G.}~\bibnamefont {Santostasi}}, \bibinfo {author}
  {\bibfnamefont {S.}~\bibnamefont {Saraf}}, \bibinfo {author} {\bibfnamefont
  {P.}~\bibnamefont {Sarin}}, \bibinfo {author} {\bibfnamefont
  {B.}~\bibnamefont {Sassolas}}, \bibinfo {author} {\bibfnamefont {B.~S.}\
  \bibnamefont {Sathyaprakash}}, \bibinfo {author} {\bibfnamefont
  {S.}~\bibnamefont {Sato}}, \bibinfo {author} {\bibfnamefont {M.}~\bibnamefont
  {Satterthwaite}}, \bibinfo {author} {\bibfnamefont {P.~R.}\ \bibnamefont
  {Saulson}}, \bibinfo {author} {\bibfnamefont {R.}~\bibnamefont {Savage}},
  \bibinfo {author} {\bibfnamefont {R.}~\bibnamefont {Schilling}}, \bibinfo
  {author} {\bibfnamefont {R.}~\bibnamefont {Schnabel}}, \bibinfo {author}
  {\bibfnamefont {R.}~\bibnamefont {Schofield}}, \bibinfo {author}
  {\bibfnamefont {B.}~\bibnamefont {Schulz}}, \bibinfo {author} {\bibfnamefont
  {B.~F.}\ \bibnamefont {Schutz}}, \bibinfo {author} {\bibfnamefont
  {P.}~\bibnamefont {Schwinberg}}, \bibinfo {author} {\bibfnamefont
  {J.}~\bibnamefont {Scott}}, \bibinfo {author} {\bibfnamefont {S.~M.}\
  \bibnamefont {Scott}}, \bibinfo {author} {\bibfnamefont {A.~C.}\ \bibnamefont
  {Searle}}, \bibinfo {author} {\bibfnamefont {F.}~\bibnamefont {Seifert}},
  \bibinfo {author} {\bibfnamefont {D.}~\bibnamefont {Sellers}}, \bibinfo
  {author} {\bibfnamefont {A.~S.}\ \bibnamefont {Sengupta}}, \bibinfo {author}
  {\bibfnamefont {D.}~\bibnamefont {Sentenac}}, \bibinfo {author}
  {\bibfnamefont {A.}~\bibnamefont {Sergeev}}, \bibinfo {author} {\bibfnamefont
  {B.}~\bibnamefont {Shapiro}}, \bibinfo {author} {\bibfnamefont
  {P.}~\bibnamefont {Shawhan}}, \bibinfo {author} {\bibfnamefont {D.~H.}\
  \bibnamefont {Shoemaker}}, \bibinfo {author} {\bibfnamefont {A.}~\bibnamefont
  {Sibley}}, \bibinfo {author} {\bibfnamefont {X.}~\bibnamefont {Siemens}},
  \bibinfo {author} {\bibfnamefont {D.}~\bibnamefont {Sigg}}, \bibinfo {author}
  {\bibfnamefont {A.~M.}\ \bibnamefont {Sintes}}, \bibinfo {author}
  {\bibfnamefont {G.}~\bibnamefont {Skelton}}, \bibinfo {author} {\bibfnamefont
  {B.~J.~J.}\ \bibnamefont {Slagmolen}}, \bibinfo {author} {\bibfnamefont
  {J.}~\bibnamefont {Slutsky}}, \bibinfo {author} {\bibfnamefont {J.~R.}\
  \bibnamefont {Smith}}, \bibinfo {author} {\bibfnamefont {M.~R.}\ \bibnamefont
  {Smith}}, \bibinfo {author} {\bibfnamefont {N.~D.}\ \bibnamefont {Smith}},
  \bibinfo {author} {\bibfnamefont {K.}~\bibnamefont {Somiya}}, \bibinfo
  {author} {\bibfnamefont {B.}~\bibnamefont {Sorazu}}, \bibinfo {author}
  {\bibfnamefont {L.}~\bibnamefont {Sperandio}}, \bibinfo {author}
  {\bibfnamefont {A.~J.}\ \bibnamefont {Stein}}, \bibinfo {author}
  {\bibfnamefont {L.~C.}\ \bibnamefont {Stein}}, \bibinfo {author}
  {\bibfnamefont {S.}~\bibnamefont {Steplewski}}, \bibinfo {author}
  {\bibfnamefont {A.}~\bibnamefont {Stochino}}, \bibinfo {author}
  {\bibfnamefont {R.}~\bibnamefont {Stone}}, \bibinfo {author} {\bibfnamefont
  {K.~A.}\ \bibnamefont {Strain}}, \bibinfo {author} {\bibfnamefont
  {S.}~\bibnamefont {Strigin}}, \bibinfo {author} {\bibfnamefont
  {A.}~\bibnamefont {Stroeer}}, \bibinfo {author} {\bibfnamefont
  {R.}~\bibnamefont {Sturani}}, \bibinfo {author} {\bibfnamefont {A.~L.}\
  \bibnamefont {Stuver}}, \bibinfo {author} {\bibfnamefont {T.~Z.}\
  \bibnamefont {Summerscales}}, \bibinfo {author} {\bibfnamefont
  {M.}~\bibnamefont {Sung}}, \bibinfo {author} {\bibfnamefont {S.}~\bibnamefont
  {Susmithan}}, \bibinfo {author} {\bibfnamefont {P.~J.}\ \bibnamefont
  {Sutton}}, \bibinfo {author} {\bibfnamefont {B.}~\bibnamefont {Swinkels}},
  \bibinfo {author} {\bibfnamefont {G.~P.}\ \bibnamefont {Szokoly}}, \bibinfo
  {author} {\bibfnamefont {D.}~\bibnamefont {Talukder}}, \bibinfo {author}
  {\bibfnamefont {D.~B.}\ \bibnamefont {Tanner}}, \bibinfo {author}
  {\bibfnamefont {S.~P.}\ \bibnamefont {Tarabrin}}, \bibinfo {author}
  {\bibfnamefont {J.~R.}\ \bibnamefont {Taylor}}, \bibinfo {author}
  {\bibfnamefont {R.}~\bibnamefont {Taylor}}, \bibinfo {author} {\bibfnamefont
  {K.~A.}\ \bibnamefont {Thorne}}, \bibinfo {author} {\bibfnamefont {K.~S.}\
  \bibnamefont {Thorne}}, \bibinfo {author} {\bibfnamefont {A.}~\bibnamefont
  {Th\"uring}}, \bibinfo {author} {\bibfnamefont {C.}~\bibnamefont {Titsler}},
  \bibinfo {author} {\bibfnamefont {K.~V.}\ \bibnamefont {Tokmakov}}, \bibinfo
  {author} {\bibfnamefont {A.}~\bibnamefont {Toncelli}}, \bibinfo {author}
  {\bibfnamefont {M.}~\bibnamefont {Tonelli}}, \bibinfo {author} {\bibfnamefont
  {C.}~\bibnamefont {Torres}}, \bibinfo {author} {\bibfnamefont {C.~I.}\
  \bibnamefont {Torrie}}, \bibinfo {author} {\bibfnamefont {E.}~\bibnamefont
  {Tournefier}}, \bibinfo {author} {\bibfnamefont {F.}~\bibnamefont
  {Travasso}}, \bibinfo {author} {\bibfnamefont {G.}~\bibnamefont {Traylor}},
  \bibinfo {author} {\bibfnamefont {M.}~\bibnamefont {Trias}}, \bibinfo
  {author} {\bibfnamefont {J.}~\bibnamefont {Trummer}}, \bibinfo {author}
  {\bibfnamefont {L.}~\bibnamefont {Turner}}, \bibinfo {author} {\bibfnamefont
  {D.}~\bibnamefont {Ugolini}}, \bibinfo {author} {\bibfnamefont
  {K.}~\bibnamefont {Urbanek}}, \bibinfo {author} {\bibfnamefont
  {H.}~\bibnamefont {Vahlbruch}}, \bibinfo {author} {\bibfnamefont
  {G.}~\bibnamefont {Vajente}}, \bibinfo {author} {\bibfnamefont
  {M.}~\bibnamefont {Vallisneri}}, \bibinfo {author} {\bibfnamefont {J.~F.~J.}\
  \bibnamefont {van~den Brand}}, \bibinfo {author} {\bibfnamefont
  {C.}~\bibnamefont {Van Den~Broeck}}, \bibinfo {author} {\bibfnamefont
  {S.}~\bibnamefont {van~der Putten}}, \bibinfo {author} {\bibfnamefont
  {M.~V.}\ \bibnamefont {van~der Sluys}}, \bibinfo {author} {\bibfnamefont
  {S.}~\bibnamefont {Vass}}, \bibinfo {author} {\bibfnamefont {R.}~\bibnamefont
  {Vaulin}}, \bibinfo {author} {\bibfnamefont {M.}~\bibnamefont {Vavoulidis}},
  \bibinfo {author} {\bibfnamefont {A.}~\bibnamefont {Vecchio}}, \bibinfo
  {author} {\bibfnamefont {G.}~\bibnamefont {Vedovato}}, \bibinfo {author}
  {\bibfnamefont {A.~A.}\ \bibnamefont {van Veggel}}, \bibinfo {author}
  {\bibfnamefont {J.}~\bibnamefont {Veitch}}, \bibinfo {author} {\bibfnamefont
  {P.~J.}\ \bibnamefont {Veitch}}, \bibinfo {author} {\bibfnamefont
  {C.}~\bibnamefont {Veltkamp}}, \bibinfo {author} {\bibfnamefont
  {D.}~\bibnamefont {Verkindt}}, \bibinfo {author} {\bibfnamefont
  {F.}~\bibnamefont {Vetrano}}, \bibinfo {author} {\bibfnamefont
  {A.}~\bibnamefont {Vicer\'e}}, \bibinfo {author} {\bibfnamefont
  {A.}~\bibnamefont {Villar}}, \bibinfo {author} {\bibfnamefont {J.-Y.}\
  \bibnamefont {Vinet}}, \bibinfo {author} {\bibfnamefont {H.}~\bibnamefont
  {Vocca}}, \bibinfo {author} {\bibfnamefont {C.}~\bibnamefont {Vorvick}},
  \bibinfo {author} {\bibfnamefont {S.~P.}\ \bibnamefont {Vyachanin}}, \bibinfo
  {author} {\bibfnamefont {S.~J.}\ \bibnamefont {Waldman}}, \bibinfo {author}
  {\bibfnamefont {L.}~\bibnamefont {Wallace}}, \bibinfo {author} {\bibfnamefont
  {A.}~\bibnamefont {Wanner}}, \bibinfo {author} {\bibfnamefont {R.~L.}\
  \bibnamefont {Ward}}, \bibinfo {author} {\bibfnamefont {M.}~\bibnamefont
  {Was}}, \bibinfo {author} {\bibfnamefont {P.}~\bibnamefont {Wei}}, \bibinfo
  {author} {\bibfnamefont {M.}~\bibnamefont {Weinert}}, \bibinfo {author}
  {\bibfnamefont {A.~J.}\ \bibnamefont {Weinstein}}, \bibinfo {author}
  {\bibfnamefont {R.}~\bibnamefont {Weiss}}, \bibinfo {author} {\bibfnamefont
  {L.}~\bibnamefont {Wen}}, \bibinfo {author} {\bibfnamefont {S.}~\bibnamefont
  {Wen}}, \bibinfo {author} {\bibfnamefont {P.}~\bibnamefont {Wessels}},
  \bibinfo {author} {\bibfnamefont {M.}~\bibnamefont {West}}, \bibinfo {author}
  {\bibfnamefont {T.}~\bibnamefont {Westphal}}, \bibinfo {author}
  {\bibfnamefont {K.}~\bibnamefont {Wette}}, \bibinfo {author} {\bibfnamefont
  {J.~T.}\ \bibnamefont {Whelan}}, \bibinfo {author} {\bibfnamefont {S.~E.}\
  \bibnamefont {Whitcomb}}, \bibinfo {author} {\bibfnamefont {B.~F.}\
  \bibnamefont {Whiting}}, \bibinfo {author} {\bibfnamefont {C.}~\bibnamefont
  {Wilkinson}}, \bibinfo {author} {\bibfnamefont {P.~A.}\ \bibnamefont
  {Willems}}, \bibinfo {author} {\bibfnamefont {H.~R.}\ \bibnamefont
  {Williams}}, \bibinfo {author} {\bibfnamefont {L.}~\bibnamefont {Williams}},
  \bibinfo {author} {\bibfnamefont {B.}~\bibnamefont {Willke}}, \bibinfo
  {author} {\bibfnamefont {I.}~\bibnamefont {Wilmut}}, \bibinfo {author}
  {\bibfnamefont {L.}~\bibnamefont {Winkelmann}}, \bibinfo {author}
  {\bibfnamefont {W.}~\bibnamefont {Winkler}}, \bibinfo {author} {\bibfnamefont
  {C.~C.}\ \bibnamefont {Wipf}}, \bibinfo {author} {\bibfnamefont {A.~G.}\
  \bibnamefont {Wiseman}}, \bibinfo {author} {\bibfnamefont {G.}~\bibnamefont
  {Woan}}, \bibinfo {author} {\bibfnamefont {R.}~\bibnamefont {Wooley}},
  \bibinfo {author} {\bibfnamefont {J.}~\bibnamefont {Worden}}, \bibinfo
  {author} {\bibfnamefont {I.}~\bibnamefont {Yakushin}}, \bibinfo {author}
  {\bibfnamefont {H.}~\bibnamefont {Yamamoto}}, \bibinfo {author}
  {\bibfnamefont {K.}~\bibnamefont {Yamamoto}}, \bibinfo {author}
  {\bibfnamefont {D.}~\bibnamefont {Yeaton-Massey}}, \bibinfo {author}
  {\bibfnamefont {S.}~\bibnamefont {Yoshida}}, \bibinfo {author} {\bibfnamefont
  {M.}~\bibnamefont {Yvert}}, \bibinfo {author} {\bibfnamefont
  {M.}~\bibnamefont {Zanolin}}, \bibinfo {author} {\bibfnamefont
  {L.}~\bibnamefont {Zhang}}, \bibinfo {author} {\bibfnamefont
  {Z.}~\bibnamefont {Zhang}}, \bibinfo {author} {\bibfnamefont
  {C.}~\bibnamefont {Zhao}}, \bibinfo {author} {\bibfnamefont {N.}~\bibnamefont
  {Zotov}}, \bibinfo {author} {\bibfnamefont {M.~E.}\ \bibnamefont {Zucker}},\
  and\ \bibinfo {author} {\bibfnamefont {J.}~\bibnamefont {Zweizig}} (\bibinfo
  {collaboration} {The LIGO Scientific Collaboration and The Virgo
  Collaboration}),\ }\bibfield  {title} {\bibinfo {title} {All-sky search for
  gravitational-wave bursts in the first joint ligo-geo-virgo run},\ }\href
  {https://doi.org/10.1103/PhysRevD.81.102001} {\bibfield  {journal} {\bibinfo
  {journal} {Phys. Rev. D}\ }\textbf {\bibinfo {volume} {81}},\ \bibinfo
  {pages} {102001} (\bibinfo {year} {2010})}\BibitemShut {NoStop}%
\bibitem [{\citenamefont {Aasi}\ \emph {et~al.}(2014)\citenamefont {Aasi},
  \citenamefont {Abbott}, \citenamefont {Abbott}, \citenamefont {Abbott},
  \citenamefont {Abernathy}, \citenamefont {Acernese}, \citenamefont {Ackley},
  \citenamefont {Adams}, \citenamefont {Adams}, \citenamefont {Addesso},
  \citenamefont {Adhikari}, \citenamefont {Affeldt}, \citenamefont {Agathos},
  \citenamefont {Aggarwal}, \citenamefont {Aguiar}, \citenamefont {Ajith},
  \citenamefont {Alemic}, \citenamefont {Allen}, \citenamefont {Allocca},
  \citenamefont {Amariutei}, \citenamefont {Andersen}, \citenamefont
  {Anderson}, \citenamefont {Anderson}, \citenamefont {Anderson}, \citenamefont
  {Arai}, \citenamefont {Araya}, \citenamefont {Arceneaux}, \citenamefont
  {Areeda}, \citenamefont {Ast}, \citenamefont {Aston}, \citenamefont {Astone},
  \citenamefont {Aufmuth}, \citenamefont {Augustus}, \citenamefont {Aulbert},
  \citenamefont {Aylott}, \citenamefont {Babak}, \citenamefont {Baker},
  \citenamefont {Ballardin}, \citenamefont {Ballmer}, \citenamefont {Barayoga},
  \citenamefont {Barbet}, \citenamefont {Barish}, \citenamefont {Barker},
  \citenamefont {Barone}, \citenamefont {Barr}, \citenamefont {Barsotti},
  \citenamefont {Barsuglia}, \citenamefont {Barton}, \citenamefont {Bartos},
  \citenamefont {Bassiri}, \citenamefont {Basti}, \citenamefont {Batch},
  \citenamefont {Bauchrowitz}, \citenamefont {Bauer}, \citenamefont {Baune},
  \citenamefont {Bavigadda}, \citenamefont {Behnke}, \citenamefont {Bejger},
  \citenamefont {Beker}, \citenamefont {Belczynski}, \citenamefont {Bell},
  \citenamefont {Bell}, \citenamefont {Bergmann}, \citenamefont {Bersanetti},
  \citenamefont {Bertolini}, \citenamefont {Betzwieser}, \citenamefont
  {Bilenko}, \citenamefont {Billingsley}, \citenamefont {Birch}, \citenamefont
  {Biscans}, \citenamefont {Bitossi}, \citenamefont {Biwer}, \citenamefont
  {Bizouard}, \citenamefont {Black}, \citenamefont {Blackburn}, \citenamefont
  {Blackburn}, \citenamefont {Blair}, \citenamefont {Bloemen}, \citenamefont
  {Bock}, \citenamefont {Bodiya}, \citenamefont {Boer}, \citenamefont
  {Bogaert}, \citenamefont {Bogan}, \citenamefont {Bond}, \citenamefont
  {Bondu}, \citenamefont {Bonelli}, \citenamefont {Bonnand}, \citenamefont
  {Bork}, \citenamefont {Born}, \citenamefont {Boschi}, \citenamefont {Bose},
  \citenamefont {Bosi}, \citenamefont {Bradaschia}, \citenamefont {Brady},
  \citenamefont {Braginsky}, \citenamefont {Branchesi}, \citenamefont {Brau},
  \citenamefont {Briant}, \citenamefont {Bridges}, \citenamefont {Brillet},
  \citenamefont {Brinkmann}, \citenamefont {Brisson}, \citenamefont {Brooks},
  \citenamefont {Brown}, \citenamefont {Brown}, \citenamefont {Br\"uckner},
  \citenamefont {Buchman}, \citenamefont {Buikema}, \citenamefont {Bulik},
  \citenamefont {Bulten}, \citenamefont {Buonanno}, \citenamefont {Burman},
  \citenamefont {Buskulic}, \citenamefont {Buy}, \citenamefont {Cadonati},
  \citenamefont {Cagnoli}, \citenamefont {Cain}, \citenamefont
  {Calder\'on~Bustillo}, \citenamefont {Calloni}, \citenamefont {Camp},
  \citenamefont {Campsie}, \citenamefont {Cannon}, \citenamefont {Canuel},
  \citenamefont {Cao}, \citenamefont {Capano}, \citenamefont {Carbognani},
  \citenamefont {Carbone}, \citenamefont {Caride}, \citenamefont {Castaldi},
  \citenamefont {Caudill}, \citenamefont {Cavagli\`a}, \citenamefont
  {Cavalier}, \citenamefont {Cavalieri}, \citenamefont {Celerier},
  \citenamefont {Cella}, \citenamefont {Cepeda}, \citenamefont {Cesarini},
  \citenamefont {Chakraborty}, \citenamefont {Chalermsongsak}, \citenamefont
  {Chamberlin}, \citenamefont {Chao}, \citenamefont {Charlton}, \citenamefont
  {Chassande-Mottin}, \citenamefont {Chen}, \citenamefont {Chen}, \citenamefont
  {Chincarini}, \citenamefont {Chiummo}, \citenamefont {Cho}, \citenamefont
  {Cho}, \citenamefont {Chow}, \citenamefont {Christensen}, \citenamefont
  {Chu}, \citenamefont {Chua}, \citenamefont {Chung}, \citenamefont {Ciani},
  \citenamefont {Clara}, \citenamefont {Clark}, \citenamefont {Clark},
  \citenamefont {Clayton}, \citenamefont {Cleva}, \citenamefont {Coccia},
  \citenamefont {Cohadon}, \citenamefont {Colla}, \citenamefont {Collette},
  \citenamefont {Colombini}, \citenamefont {Cominsky}, \citenamefont
  {Constancio}, \citenamefont {Conte}, \citenamefont {Cook}, \citenamefont
  {Corbitt}, \citenamefont {Cornish}, \citenamefont {Corsi}, \citenamefont
  {Costa}, \citenamefont {Coughlin}, \citenamefont {Coulon}, \citenamefont
  {Countryman}, \citenamefont {Couvares}, \citenamefont {Coward}, \citenamefont
  {Cowart}, \citenamefont {Coyne}, \citenamefont {Coyne}, \citenamefont
  {Craig}, \citenamefont {Creighton}, \citenamefont {Croce}, \citenamefont
  {Crowder}, \citenamefont {Cumming}, \citenamefont {Cunningham}, \citenamefont
  {Cuoco}, \citenamefont {Cutler}, \citenamefont {Dahl}, \citenamefont
  {Dal~Canton}, \citenamefont {Damjanic}, \citenamefont {Danilishin},
  \citenamefont {D'Antonio}, \citenamefont {Danzmann}, \citenamefont {Dattilo},
  \citenamefont {Daveloza}, \citenamefont {Davier}, \citenamefont {Davies},
  \citenamefont {Daw}, \citenamefont {Day}, \citenamefont {Dayanga},
  \citenamefont {DeBra}, \citenamefont {Debreczeni}, \citenamefont {Degallaix},
  \citenamefont {Del\'eglise}, \citenamefont {Del~Pozzo}, \citenamefont
  {Del~Pozzo}, \citenamefont {Denker}, \citenamefont {Dent}, \citenamefont
  {Dereli}, \citenamefont {Dergachev}, \citenamefont {De~Rosa}, \citenamefont
  {DeRosa}, \citenamefont {DeSalvo}, \citenamefont {Dhurandhar}, \citenamefont
  {D\'{\i}az}, \citenamefont {Dickson}, \citenamefont {Di~Fiore}, \citenamefont
  {Di~Lieto}, \citenamefont {Di~Palma}, \citenamefont {Di~Virgilio},
  \citenamefont {Dolique}, \citenamefont {Dominguez}, \citenamefont {Donovan},
  \citenamefont {Dooley}, \citenamefont {Doravari}, \citenamefont {Douglas},
  \citenamefont {Downes}, \citenamefont {Drago}, \citenamefont {Drever},
  \citenamefont {Driggers}, \citenamefont {Du}, \citenamefont {Ducrot},
  \citenamefont {Dwyer}, \citenamefont {Eberle}, \citenamefont {Edo},
  \citenamefont {Edwards}, \citenamefont {Effler}, \citenamefont {Eggenstein},
  \citenamefont {Ehrens}, \citenamefont {Eichholz}, \citenamefont {Eikenberry},
  \citenamefont {Endr\ifmmode~\mbox{\H{o}}\else \H{o}\fi{}czi}, \citenamefont
  {Essick}, \citenamefont {Etzel}, \citenamefont {Evans}, \citenamefont
  {Evans}, \citenamefont {Factourovich}, \citenamefont {Fafone}, \citenamefont
  {Fairhurst}, \citenamefont {Fan}, \citenamefont {Fang}, \citenamefont
  {Farinon}, \citenamefont {Farr}, \citenamefont {Farr}, \citenamefont
  {Favata}, \citenamefont {Fazi}, \citenamefont {Fehrmann}, \citenamefont
  {Fejer}, \citenamefont {Feldbaum}, \citenamefont {Feroz}, \citenamefont
  {Ferrante}, \citenamefont {Ferreira}, \citenamefont {Ferrini}, \citenamefont
  {Fidecaro}, \citenamefont {Finn}, \citenamefont {Fiori}, \citenamefont
  {Fisher}, \citenamefont {Flaminio}, \citenamefont {Fotopoulos}, \citenamefont
  {Fournier}, \citenamefont {Franco}, \citenamefont {Frasca}, \citenamefont
  {Frasconi}, \citenamefont {Frede}, \citenamefont {Frei}, \citenamefont
  {Freise}, \citenamefont {Frey}, \citenamefont {Fricke}, \citenamefont
  {Fritschel}, \citenamefont {Frolov}, \citenamefont {Fulda}, \citenamefont
  {Fyffe}, \citenamefont {Gair}, \citenamefont {Gammaitoni}, \citenamefont
  {Gaonkar}, \citenamefont {Garufi}, \citenamefont {Gehrels}, \citenamefont
  {Gemme}, \citenamefont {Gendre}, \citenamefont {Genin}, \citenamefont
  {Gennai}, \citenamefont {Ghosh}, \citenamefont {Giaime}, \citenamefont
  {Giardina}, \citenamefont {Giazotto}, \citenamefont {Gill}, \citenamefont
  {Gleason}, \citenamefont {Goetz}, \citenamefont {Goetz}, \citenamefont
  {Gondan}, \citenamefont {Gonz\'alez}, \citenamefont {Gordon}, \citenamefont
  {Gorodetsky}, \citenamefont {Gossan}, \citenamefont {Go\ss{}ler},
  \citenamefont {Gouaty}, \citenamefont {Gr\"af}, \citenamefont {Graff},
  \citenamefont {Granata}, \citenamefont {Grant}, \citenamefont {Gras},
  \citenamefont {Gray}, \citenamefont {Greenhalgh}, \citenamefont {Gretarsson},
  \citenamefont {Groot}, \citenamefont {Grote}, \citenamefont {Grover},
  \citenamefont {Grunewald}, \citenamefont {Guidi}, \citenamefont {Guido},
  \citenamefont {Gushwa}, \citenamefont {Gustafson}, \citenamefont {Gustafson},
  \citenamefont {Ha}, \citenamefont {Hall}, \citenamefont {Hamilton},
  \citenamefont {Hammer}, \citenamefont {Hammond}, \citenamefont {Hanke},
  \citenamefont {Hanks}, \citenamefont {Hanna}, \citenamefont {Hannam},
  \citenamefont {Hanson}, \citenamefont {Haris}, \citenamefont {Harms},
  \citenamefont {Harry}, \citenamefont {Harry}, \citenamefont {Harstad},
  \citenamefont {Hart}, \citenamefont {Hartman}, \citenamefont {Haster},
  \citenamefont {Haughian}, \citenamefont {Heidmann}, \citenamefont {Heintze},
  \citenamefont {Heitmann}, \citenamefont {Hello}, \citenamefont {Hemming},
  \citenamefont {Hendry}, \citenamefont {Heng}, \citenamefont {Heptonstall},
  \citenamefont {Heurs}, \citenamefont {Hewitson}, \citenamefont {Hild},
  \citenamefont {Hoak}, \citenamefont {Hodge}, \citenamefont {Hofman},
  \citenamefont {Holt}, \citenamefont {Hopkins}, \citenamefont {Horrom},
  \citenamefont {Hoske}, \citenamefont {Hosken}, \citenamefont {Hough},
  \citenamefont {Howell}, \citenamefont {Hu}, \citenamefont {Huerta},
  \citenamefont {Hughey}, \citenamefont {Husa}, \citenamefont {Huttner},
  \citenamefont {Huynh}, \citenamefont {Huynh-Dinh}, \citenamefont {Idrisy},
  \citenamefont {Ingram}, \citenamefont {Inta}, \citenamefont {Islas},
  \citenamefont {Isogai}, \citenamefont {Ivanov}, \citenamefont {Iyer},
  \citenamefont {Izumi}, \citenamefont {Jacobson}, \citenamefont {Jang},
  \citenamefont {Jaranowski}, \citenamefont {Ji}, \citenamefont
  {Jim\'enez-Forteza}, \citenamefont {Johnson}, \citenamefont {Jones},
  \citenamefont {Jones}, \citenamefont {Jones}, \citenamefont {Jonker},
  \citenamefont {Ju}, \citenamefont {Kalmus}, \citenamefont {Kalogera},
  \citenamefont {Kandhasamy}, \citenamefont {Kang}, \citenamefont {Kanner},
  \citenamefont {Karlen}, \citenamefont {Kasprzack}, \citenamefont
  {Katsavounidis}, \citenamefont {Katzman}, \citenamefont {Kaufer},
  \citenamefont {Kaufer}, \citenamefont {Kaur}, \citenamefont {Kawabe},
  \citenamefont {Kawazoe}, \citenamefont {K\'ef\'elian}, \citenamefont
  {Keiser}, \citenamefont {Keitel}, \citenamefont {Kelley}, \citenamefont
  {Kells}, \citenamefont {Keppel}, \citenamefont {Khalaidovski}, \citenamefont
  {Khalili}, \citenamefont {Khazanov}, \citenamefont {Kim}, \citenamefont
  {Kim}, \citenamefont {Kim}, \citenamefont {Kim}, \citenamefont {Kim},
  \citenamefont {Kim}, \citenamefont {King}, \citenamefont {King},
  \citenamefont {Kinzel}, \citenamefont {Kissel}, \citenamefont {Klimenko},
  \citenamefont {Kline}, \citenamefont {Koehlenbeck}, \citenamefont {Kokeyama},
  \citenamefont {Kondrashov}, \citenamefont {Koranda}, \citenamefont {Korth},
  \citenamefont {Kowalska}, \citenamefont {Kozak}, \citenamefont {Kringel},
  \citenamefont {Krishnan}, \citenamefont {Kr\'olak}, \citenamefont {Kuehn},
  \citenamefont {Kumar}, \citenamefont {Kumar}, \citenamefont {Kumar},
  \citenamefont {Kumar}, \citenamefont {Kuo}, \citenamefont {Kutynia},
  \citenamefont {Lam}, \citenamefont {Landry}, \citenamefont {Lantz},
  \citenamefont {Larson}, \citenamefont {Lasky}, \citenamefont {Lazzaro},
  \citenamefont {Leaci}, \citenamefont {Leavey}, \citenamefont {Lebigot},
  \citenamefont {Lee}, \citenamefont {Lee}, \citenamefont {Lee}, \citenamefont
  {Lee}, \citenamefont {Lee}, \citenamefont {Leonardi}, \citenamefont {Leong},
  \citenamefont {Le~Roux}, \citenamefont {Leroy}, \citenamefont {Letendre},
  \citenamefont {Levin}, \citenamefont {Levine}, \citenamefont {Lewis},
  \citenamefont {Li}, \citenamefont {Libbrecht}, \citenamefont {Libson},
  \citenamefont {Lin}, \citenamefont {Littenberg}, \citenamefont {Lockerbie},
  \citenamefont {Lockett}, \citenamefont {Lodhia}, \citenamefont {Loew},
  \citenamefont {Logue}, \citenamefont {Lombardi}, \citenamefont {Lopez},
  \citenamefont {Lorenzini}, \citenamefont {Loriette}, \citenamefont {Lormand},
  \citenamefont {Losurdo}, \citenamefont {Lough}, \citenamefont {Lubinski},
  \citenamefont {L\"uck}, \citenamefont {Lundgren}, \citenamefont {Ma},
  \citenamefont {Macdonald}, \citenamefont {MacDonald}, \citenamefont
  {Machenschalk}, \citenamefont {MacInnis}, \citenamefont {Macleod},
  \citenamefont {Maga\~na Sandoval}, \citenamefont {Magee}, \citenamefont
  {Mageswaran}, \citenamefont {Maglione}, \citenamefont {Mailand},
  \citenamefont {Majorana}, \citenamefont {Maksimovic}, \citenamefont
  {Malvezzi}, \citenamefont {Man}, \citenamefont {Manca}, \citenamefont
  {Mandel}, \citenamefont {Mandic}, \citenamefont {Mangano}, \citenamefont
  {Mangini}, \citenamefont {Mansell}, \citenamefont {Mantovani}, \citenamefont
  {Marchesoni}, \citenamefont {Marion}, \citenamefont {M\'arka}, \citenamefont
  {M\'arka}, \citenamefont {Markosyan}, \citenamefont {Maros}, \citenamefont
  {Marque}, \citenamefont {Martelli}, \citenamefont {Martin}, \citenamefont
  {Martin}, \citenamefont {Martinelli}, \citenamefont {Martynov}, \citenamefont
  {Marx}, \citenamefont {Mason}, \citenamefont {Masserot}, \citenamefont
  {Massinger}, \citenamefont {Matichard}, \citenamefont {Matone}, \citenamefont
  {Mavalvala}, \citenamefont {May}, \citenamefont {Mazumder}, \citenamefont
  {Mazzolo}, \citenamefont {McCarthy}, \citenamefont {McClelland},
  \citenamefont {McGuire}, \citenamefont {McIntyre}, \citenamefont {McIver},
  \citenamefont {McLin}, \citenamefont {Meacher}, \citenamefont {Meadors},
  \citenamefont {Mehmet}, \citenamefont {Meidam}, \citenamefont {Meinders},
  \citenamefont {Melatos}, \citenamefont {Mendell}, \citenamefont {Mercer},
  \citenamefont {Meshkov}, \citenamefont {Messenger}, \citenamefont {Meyer},
  \citenamefont {Meyer}, \citenamefont {Meyers}, \citenamefont {Mezzani},
  \citenamefont {Miao}, \citenamefont {Michel}, \citenamefont {Mikhailov},
  \citenamefont {Milano}, \citenamefont {Miller}, \citenamefont {Minenkov},
  \citenamefont {Mingarelli}, \citenamefont {Mishra}, \citenamefont {Mitra},
  \citenamefont {Mitrofanov}, \citenamefont {Mitselmakher}, \citenamefont
  {Mittleman}, \citenamefont {Moe}, \citenamefont {Moggi}, \citenamefont
  {Mohan}, \citenamefont {Mohapatra}, \citenamefont {Moraru}, \citenamefont
  {Moreno}, \citenamefont {Morgado}, \citenamefont {Morriss}, \citenamefont
  {Mossavi}, \citenamefont {Mours}, \citenamefont {Mow-Lowry}, \citenamefont
  {Mueller}, \citenamefont {Mueller}, \citenamefont {Mukherjee}, \citenamefont
  {Mullavey}, \citenamefont {Munch}, \citenamefont {Murphy}, \citenamefont
  {Murray}, \citenamefont {Mytidis}, \citenamefont {Nagy}, \citenamefont
  {Nardecchia}, \citenamefont {Naticchioni}, \citenamefont {Nayak},
  \citenamefont {Necula}, \citenamefont {Nelemans}, \citenamefont {Neri},
  \citenamefont {Neri}, \citenamefont {Newton}, \citenamefont {Nguyen},
  \citenamefont {Nielsen}, \citenamefont {Nissanke}, \citenamefont {Nitz},
  \citenamefont {Nocera}, \citenamefont {Nolting}, \citenamefont {Normandin},
  \citenamefont {Nuttall}, \citenamefont {Ochsner}, \citenamefont {O'Dell},
  \citenamefont {Oelker}, \citenamefont {Oh}, \citenamefont {Oh}, \citenamefont
  {Ohme}, \citenamefont {Omar}, \citenamefont {Oppermann}, \citenamefont
  {Oram}, \citenamefont {O'Reilly}, \citenamefont {Ortega}, \citenamefont
  {O'Shaughnessy}, \citenamefont {Osthelder}, \citenamefont {Ottaway},
  \citenamefont {Ottens}, \citenamefont {Overmier}, \citenamefont {Owen},
  \citenamefont {Padilla}, \citenamefont {Pai}, \citenamefont {Palashov},
  \citenamefont {Palomba}, \citenamefont {Pan}, \citenamefont {Pan},
  \citenamefont {Pankow}, \citenamefont {Paoletti}, \citenamefont {Papa},
  \citenamefont {Paris}, \citenamefont {Pasqualetti}, \citenamefont
  {Passaquieti}, \citenamefont {Passuello}, \citenamefont {Patel},
  \citenamefont {Pedraza}, \citenamefont {Pele}, \citenamefont {Penn},
  \citenamefont {Perreca}, \citenamefont {Phelps}, \citenamefont {Pichot},
  \citenamefont {Pickenpack}, \citenamefont {Piergiovanni}, \citenamefont
  {Pierro}, \citenamefont {Pinard}, \citenamefont {Pinto}, \citenamefont
  {Pitkin}, \citenamefont {Poeld}, \citenamefont {Poggiani}, \citenamefont
  {Poteomkin}, \citenamefont {Powell}, \citenamefont {Prasad}, \citenamefont
  {Predoi}, \citenamefont {Premachandra}, \citenamefont {Prestegard},
  \citenamefont {Price}, \citenamefont {Prijatelj}, \citenamefont {Privitera},
  \citenamefont {Prodi}, \citenamefont {Prokhorov}, \citenamefont {Puncken},
  \citenamefont {Punturo}, \citenamefont {Puppo}, \citenamefont {P\"urrer},
  \citenamefont {Qin}, \citenamefont {Quetschke}, \citenamefont {Quintero},
  \citenamefont {Quitzow-James}, \citenamefont {Raab}, \citenamefont
  {Rabeling}, \citenamefont {R\'acz}, \citenamefont {Radkins}, \citenamefont
  {Raffai}, \citenamefont {Raja}, \citenamefont {Rajalakshmi}, \citenamefont
  {Rakhmanov}, \citenamefont {Ramet}, \citenamefont {Ramirez}, \citenamefont
  {Rapagnani}, \citenamefont {Raymond}, \citenamefont {Razzano}, \citenamefont
  {Re}, \citenamefont {Recchia}, \citenamefont {Reed}, \citenamefont
  {Regimbau}, \citenamefont {Reid}, \citenamefont {Reitze}, \citenamefont
  {Reula}, \citenamefont {Rhoades}, \citenamefont {Ricci}, \citenamefont
  {Riesen}, \citenamefont {Riles}, \citenamefont {Robertson}, \citenamefont
  {Robinet}, \citenamefont {Rocchi}, \citenamefont {Roddy}, \citenamefont
  {Rogstad}, \citenamefont {Rolland}, \citenamefont {Rollins}, \citenamefont
  {Romano}, \citenamefont {Romanov}, \citenamefont {Romie}, \citenamefont
  {Rosi\ifmmode~\acute{n}\else \'{n}\fi{}ska}, \citenamefont {Rowan},
  \citenamefont {R\"udiger}, \citenamefont {Ruggi}, \citenamefont {Ryan},
  \citenamefont {Salemi}, \citenamefont {Sammut}, \citenamefont {Sandberg},
  \citenamefont {Sanders}, \citenamefont {Sankar}, \citenamefont {Sannibale},
  \citenamefont {Santiago-Prieto}, \citenamefont {Saracco}, \citenamefont
  {Sassolas}, \citenamefont {Sathyaprakash}, \citenamefont {Saulson},
  \citenamefont {Savage}, \citenamefont {Scheuer}, \citenamefont {Schilling},
  \citenamefont {Schilman}, \citenamefont {Schmidt}, \citenamefont {Schnabel},
  \citenamefont {Schofield}, \citenamefont {Schreiber}, \citenamefont
  {Schuette}, \citenamefont {Schutz}, \citenamefont {Scott}, \citenamefont
  {Scott}, \citenamefont {Sellers}, \citenamefont {Sengupta}, \citenamefont
  {Sentenac}, \citenamefont {Sequino}, \citenamefont {Sergeev}, \citenamefont
  {Shaddock}, \citenamefont {Shah}, \citenamefont {Shahriar}, \citenamefont
  {Shaltev}, \citenamefont {Shao}, \citenamefont {Shapiro}, \citenamefont
  {Shawhan}, \citenamefont {Shoemaker}, \citenamefont {Sidery}, \citenamefont
  {Siellez}, \citenamefont {Siemens}, \citenamefont {Sigg}, \citenamefont
  {Simakov}, \citenamefont {Singer}, \citenamefont {Singer}, \citenamefont
  {Singh}, \citenamefont {Sintes}, \citenamefont {Slagmolen}, \citenamefont
  {Slutsky}, \citenamefont {Smith}, \citenamefont {Smith}, \citenamefont
  {Smith}, \citenamefont {Smith-Lefebvre}, \citenamefont {Son}, \citenamefont
  {Sorazu}, \citenamefont {Souradeep}, \citenamefont {Staley}, \citenamefont
  {Stebbins}, \citenamefont {Steinke}, \citenamefont {Steinlechner},
  \citenamefont {Steinlechner}, \citenamefont {Stephens}, \citenamefont
  {Steplewski}, \citenamefont {Stevenson}, \citenamefont {Stone}, \citenamefont
  {Stops}, \citenamefont {Strain}, \citenamefont {Straniero}, \citenamefont
  {Strigin}, \citenamefont {Sturani}, \citenamefont {Stuver}, \citenamefont
  {Summerscales}, \citenamefont {Susmithan}, \citenamefont {Sutton},
  \citenamefont {Swinkels}, \citenamefont {Tacca}, \citenamefont {Talukder},
  \citenamefont {Tanner}, \citenamefont {Tao}, \citenamefont {Tarabrin},
  \citenamefont {Taylor}, \citenamefont {Tellez}, \citenamefont
  {Thirugnanasambandam}, \citenamefont {Thomas}, \citenamefont {Thomas},
  \citenamefont {Thorne}, \citenamefont {Thorne}, \citenamefont {Thrane},
  \citenamefont {Tiwari}, \citenamefont {Tokmakov}, \citenamefont {Tomlinson},
  \citenamefont {Tonelli}, \citenamefont {Torres}, \citenamefont {Torrie},
  \citenamefont {Travasso}, \citenamefont {Traylor}, \citenamefont {Trias},
  \citenamefont {Tse}, \citenamefont {Tshilumba}, \citenamefont {Tuennermann},
  \citenamefont {Ugolini}, \citenamefont {Unnikrishnan}, \citenamefont {Urban},
  \citenamefont {Usman}, \citenamefont {Vahlbruch}, \citenamefont {Vajente},
  \citenamefont {Valdes}, \citenamefont {Vallisneri}, \citenamefont {van
  Beuzekom}, \citenamefont {van~den Brand}, \citenamefont {Van Den~Broeck},
  \citenamefont {van~der Sluys}, \citenamefont {van Heijningen}, \citenamefont
  {van Veggel}, \citenamefont {Vass}, \citenamefont {Vas\'uth}, \citenamefont
  {Vaulin}, \citenamefont {Vecchio}, \citenamefont {Vedovato}, \citenamefont
  {Veitch}, \citenamefont {Veitch}, \citenamefont {Venkateswara}, \citenamefont
  {Verkindt}, \citenamefont {Vetrano}, \citenamefont {Vicer\'e}, \citenamefont
  {Vincent-Finley}, \citenamefont {Vinet}, \citenamefont {Vitale},
  \citenamefont {Vo}, \citenamefont {Vocca}, \citenamefont {Vorvick},
  \citenamefont {Vousden}, \citenamefont {Vyachanin}, \citenamefont {Wade},
  \citenamefont {Wade}, \citenamefont {Wade}, \citenamefont {Walker},
  \citenamefont {Wallace}, \citenamefont {Walsh}, \citenamefont {Wang},
  \citenamefont {Wang}, \citenamefont {Ward}, \citenamefont {Was},
  \citenamefont {Weaver}, \citenamefont {Wei}, \citenamefont {Weinert},
  \citenamefont {Weinstein}, \citenamefont {Weiss}, \citenamefont {Welborn},
  \citenamefont {Wen}, \citenamefont {Wessels}, \citenamefont {West},
  \citenamefont {Westphal}, \citenamefont {Wette}, \citenamefont {Whelan},
  \citenamefont {White}, \citenamefont {Whiting}, \citenamefont {Wiesner},
  \citenamefont {Wilkinson}, \citenamefont {Williams}, \citenamefont
  {Williams}, \citenamefont {Williams}, \citenamefont {Williams}, \citenamefont
  {Williamson}, \citenamefont {Willis}, \citenamefont {Willke}, \citenamefont
  {Wimmer}, \citenamefont {Winkler}, \citenamefont {Wipf}, \citenamefont
  {Wiseman}, \citenamefont {Wittel}, \citenamefont {Woan}, \citenamefont
  {Wolovick}, \citenamefont {Worden}, \citenamefont {Wu}, \citenamefont
  {Yablon}, \citenamefont {Yakushin}, \citenamefont {Yam}, \citenamefont
  {Yamamoto}, \citenamefont {Yancey}, \citenamefont {Yang}, \citenamefont
  {Yoshida}, \citenamefont {Yvert}, \citenamefont {Zadro\ifmmode~\dot{z}\else
  \.{z}\fi{}ny}, \citenamefont {Zanolin}, \citenamefont {Zendri}, \citenamefont
  {Zhang}, \citenamefont {Zhang}, \citenamefont {Zhao}, \citenamefont {Zhu},
  \citenamefont {Zhu}, \citenamefont {Zucker}, \citenamefont {Zuraw},\ and\
  \citenamefont {Zweizig}}]{PhysRevD.89.122004}%
  \BibitemOpen
  \bibfield  {author} {\bibinfo {author} {\bibfnamefont {J.}~\bibnamefont
  {Aasi}}, \bibinfo {author} {\bibfnamefont {B.~P.}\ \bibnamefont {Abbott}},
  \bibinfo {author} {\bibfnamefont {R.}~\bibnamefont {Abbott}}, \bibinfo
  {author} {\bibfnamefont {T.}~\bibnamefont {Abbott}}, \bibinfo {author}
  {\bibfnamefont {M.~R.}\ \bibnamefont {Abernathy}}, \bibinfo {author}
  {\bibfnamefont {F.}~\bibnamefont {Acernese}}, \bibinfo {author}
  {\bibfnamefont {K.}~\bibnamefont {Ackley}}, \bibinfo {author} {\bibfnamefont
  {C.}~\bibnamefont {Adams}}, \bibinfo {author} {\bibfnamefont
  {T.}~\bibnamefont {Adams}}, \bibinfo {author} {\bibfnamefont
  {P.}~\bibnamefont {Addesso}}, \bibinfo {author} {\bibfnamefont {R.~X.}\
  \bibnamefont {Adhikari}}, \bibinfo {author} {\bibfnamefont {C.}~\bibnamefont
  {Affeldt}}, \bibinfo {author} {\bibfnamefont {M.}~\bibnamefont {Agathos}},
  \bibinfo {author} {\bibfnamefont {N.}~\bibnamefont {Aggarwal}}, \bibinfo
  {author} {\bibfnamefont {O.~D.}\ \bibnamefont {Aguiar}}, \bibinfo {author}
  {\bibfnamefont {P.}~\bibnamefont {Ajith}}, \bibinfo {author} {\bibfnamefont
  {A.}~\bibnamefont {Alemic}}, \bibinfo {author} {\bibfnamefont
  {B.}~\bibnamefont {Allen}}, \bibinfo {author} {\bibfnamefont
  {A.}~\bibnamefont {Allocca}}, \bibinfo {author} {\bibfnamefont
  {D.}~\bibnamefont {Amariutei}}, \bibinfo {author} {\bibfnamefont
  {M.}~\bibnamefont {Andersen}}, \bibinfo {author} {\bibfnamefont {R.~A.}\
  \bibnamefont {Anderson}}, \bibinfo {author} {\bibfnamefont {S.~B.}\
  \bibnamefont {Anderson}}, \bibinfo {author} {\bibfnamefont {W.~G.}\
  \bibnamefont {Anderson}}, \bibinfo {author} {\bibfnamefont {K.}~\bibnamefont
  {Arai}}, \bibinfo {author} {\bibfnamefont {M.~C.}\ \bibnamefont {Araya}},
  \bibinfo {author} {\bibfnamefont {C.}~\bibnamefont {Arceneaux}}, \bibinfo
  {author} {\bibfnamefont {J.~S.}\ \bibnamefont {Areeda}}, \bibinfo {author}
  {\bibfnamefont {S.}~\bibnamefont {Ast}}, \bibinfo {author} {\bibfnamefont
  {S.~M.}\ \bibnamefont {Aston}}, \bibinfo {author} {\bibfnamefont
  {P.}~\bibnamefont {Astone}}, \bibinfo {author} {\bibfnamefont
  {P.}~\bibnamefont {Aufmuth}}, \bibinfo {author} {\bibfnamefont
  {H.}~\bibnamefont {Augustus}}, \bibinfo {author} {\bibfnamefont
  {C.}~\bibnamefont {Aulbert}}, \bibinfo {author} {\bibfnamefont {B.~E.}\
  \bibnamefont {Aylott}}, \bibinfo {author} {\bibfnamefont {S.}~\bibnamefont
  {Babak}}, \bibinfo {author} {\bibfnamefont {P.~T.}\ \bibnamefont {Baker}},
  \bibinfo {author} {\bibfnamefont {G.}~\bibnamefont {Ballardin}}, \bibinfo
  {author} {\bibfnamefont {S.~W.}\ \bibnamefont {Ballmer}}, \bibinfo {author}
  {\bibfnamefont {J.~C.}\ \bibnamefont {Barayoga}}, \bibinfo {author}
  {\bibfnamefont {M.}~\bibnamefont {Barbet}}, \bibinfo {author} {\bibfnamefont
  {B.~C.}\ \bibnamefont {Barish}}, \bibinfo {author} {\bibfnamefont
  {D.}~\bibnamefont {Barker}}, \bibinfo {author} {\bibfnamefont
  {F.}~\bibnamefont {Barone}}, \bibinfo {author} {\bibfnamefont
  {B.}~\bibnamefont {Barr}}, \bibinfo {author} {\bibfnamefont {L.}~\bibnamefont
  {Barsotti}}, \bibinfo {author} {\bibfnamefont {M.}~\bibnamefont {Barsuglia}},
  \bibinfo {author} {\bibfnamefont {M.~A.}\ \bibnamefont {Barton}}, \bibinfo
  {author} {\bibfnamefont {I.}~\bibnamefont {Bartos}}, \bibinfo {author}
  {\bibfnamefont {R.}~\bibnamefont {Bassiri}}, \bibinfo {author} {\bibfnamefont
  {A.}~\bibnamefont {Basti}}, \bibinfo {author} {\bibfnamefont {J.~C.}\
  \bibnamefont {Batch}}, \bibinfo {author} {\bibfnamefont {J.}~\bibnamefont
  {Bauchrowitz}}, \bibinfo {author} {\bibfnamefont {T.~S.}\ \bibnamefont
  {Bauer}}, \bibinfo {author} {\bibfnamefont {C.}~\bibnamefont {Baune}},
  \bibinfo {author} {\bibfnamefont {V.}~\bibnamefont {Bavigadda}}, \bibinfo
  {author} {\bibfnamefont {B.}~\bibnamefont {Behnke}}, \bibinfo {author}
  {\bibfnamefont {M.}~\bibnamefont {Bejger}}, \bibinfo {author} {\bibfnamefont
  {M.~G.}\ \bibnamefont {Beker}}, \bibinfo {author} {\bibfnamefont
  {C.}~\bibnamefont {Belczynski}}, \bibinfo {author} {\bibfnamefont {A.~S.}\
  \bibnamefont {Bell}}, \bibinfo {author} {\bibfnamefont {C.}~\bibnamefont
  {Bell}}, \bibinfo {author} {\bibfnamefont {G.}~\bibnamefont {Bergmann}},
  \bibinfo {author} {\bibfnamefont {D.}~\bibnamefont {Bersanetti}}, \bibinfo
  {author} {\bibfnamefont {A.}~\bibnamefont {Bertolini}}, \bibinfo {author}
  {\bibfnamefont {J.}~\bibnamefont {Betzwieser}}, \bibinfo {author}
  {\bibfnamefont {I.~A.}\ \bibnamefont {Bilenko}}, \bibinfo {author}
  {\bibfnamefont {G.}~\bibnamefont {Billingsley}}, \bibinfo {author}
  {\bibfnamefont {J.}~\bibnamefont {Birch}}, \bibinfo {author} {\bibfnamefont
  {S.}~\bibnamefont {Biscans}}, \bibinfo {author} {\bibfnamefont
  {M.}~\bibnamefont {Bitossi}}, \bibinfo {author} {\bibfnamefont
  {C.}~\bibnamefont {Biwer}}, \bibinfo {author} {\bibfnamefont {M.~A.}\
  \bibnamefont {Bizouard}}, \bibinfo {author} {\bibfnamefont {E.}~\bibnamefont
  {Black}}, \bibinfo {author} {\bibfnamefont {J.~K.}\ \bibnamefont
  {Blackburn}}, \bibinfo {author} {\bibfnamefont {L.}~\bibnamefont
  {Blackburn}}, \bibinfo {author} {\bibfnamefont {D.}~\bibnamefont {Blair}},
  \bibinfo {author} {\bibfnamefont {S.}~\bibnamefont {Bloemen}}, \bibinfo
  {author} {\bibfnamefont {O.}~\bibnamefont {Bock}}, \bibinfo {author}
  {\bibfnamefont {T.~P.}\ \bibnamefont {Bodiya}}, \bibinfo {author}
  {\bibfnamefont {M.}~\bibnamefont {Boer}}, \bibinfo {author} {\bibfnamefont
  {G.}~\bibnamefont {Bogaert}}, \bibinfo {author} {\bibfnamefont
  {C.}~\bibnamefont {Bogan}}, \bibinfo {author} {\bibfnamefont
  {C.}~\bibnamefont {Bond}}, \bibinfo {author} {\bibfnamefont {F.}~\bibnamefont
  {Bondu}}, \bibinfo {author} {\bibfnamefont {L.}~\bibnamefont {Bonelli}},
  \bibinfo {author} {\bibfnamefont {R.}~\bibnamefont {Bonnand}}, \bibinfo
  {author} {\bibfnamefont {R.}~\bibnamefont {Bork}}, \bibinfo {author}
  {\bibfnamefont {M.}~\bibnamefont {Born}}, \bibinfo {author} {\bibfnamefont
  {V.}~\bibnamefont {Boschi}}, \bibinfo {author} {\bibfnamefont
  {S.}~\bibnamefont {Bose}}, \bibinfo {author} {\bibfnamefont {L.}~\bibnamefont
  {Bosi}}, \bibinfo {author} {\bibfnamefont {C.}~\bibnamefont {Bradaschia}},
  \bibinfo {author} {\bibfnamefont {P.~R.}\ \bibnamefont {Brady}}, \bibinfo
  {author} {\bibfnamefont {V.~B.}\ \bibnamefont {Braginsky}}, \bibinfo {author}
  {\bibfnamefont {M.}~\bibnamefont {Branchesi}}, \bibinfo {author}
  {\bibfnamefont {J.~E.}\ \bibnamefont {Brau}}, \bibinfo {author}
  {\bibfnamefont {T.}~\bibnamefont {Briant}}, \bibinfo {author} {\bibfnamefont
  {D.~O.}\ \bibnamefont {Bridges}}, \bibinfo {author} {\bibfnamefont
  {A.}~\bibnamefont {Brillet}}, \bibinfo {author} {\bibfnamefont
  {M.}~\bibnamefont {Brinkmann}}, \bibinfo {author} {\bibfnamefont
  {V.}~\bibnamefont {Brisson}}, \bibinfo {author} {\bibfnamefont {A.~F.}\
  \bibnamefont {Brooks}}, \bibinfo {author} {\bibfnamefont {D.~A.}\
  \bibnamefont {Brown}}, \bibinfo {author} {\bibfnamefont {D.~D.}\ \bibnamefont
  {Brown}}, \bibinfo {author} {\bibfnamefont {F.}~\bibnamefont {Br\"uckner}},
  \bibinfo {author} {\bibfnamefont {S.}~\bibnamefont {Buchman}}, \bibinfo
  {author} {\bibfnamefont {A.}~\bibnamefont {Buikema}}, \bibinfo {author}
  {\bibfnamefont {T.}~\bibnamefont {Bulik}}, \bibinfo {author} {\bibfnamefont
  {H.~J.}\ \bibnamefont {Bulten}}, \bibinfo {author} {\bibfnamefont
  {A.}~\bibnamefont {Buonanno}}, \bibinfo {author} {\bibfnamefont
  {R.}~\bibnamefont {Burman}}, \bibinfo {author} {\bibfnamefont
  {D.}~\bibnamefont {Buskulic}}, \bibinfo {author} {\bibfnamefont
  {C.}~\bibnamefont {Buy}}, \bibinfo {author} {\bibfnamefont {L.}~\bibnamefont
  {Cadonati}}, \bibinfo {author} {\bibfnamefont {G.}~\bibnamefont {Cagnoli}},
  \bibinfo {author} {\bibfnamefont {J.}~\bibnamefont {Cain}}, \bibinfo {author}
  {\bibfnamefont {J.}~\bibnamefont {Calder\'on~Bustillo}}, \bibinfo {author}
  {\bibfnamefont {E.}~\bibnamefont {Calloni}}, \bibinfo {author} {\bibfnamefont
  {J.~B.}\ \bibnamefont {Camp}}, \bibinfo {author} {\bibfnamefont
  {P.}~\bibnamefont {Campsie}}, \bibinfo {author} {\bibfnamefont {K.~C.}\
  \bibnamefont {Cannon}}, \bibinfo {author} {\bibfnamefont {B.}~\bibnamefont
  {Canuel}}, \bibinfo {author} {\bibfnamefont {J.}~\bibnamefont {Cao}},
  \bibinfo {author} {\bibfnamefont {C.~D.}\ \bibnamefont {Capano}}, \bibinfo
  {author} {\bibfnamefont {F.}~\bibnamefont {Carbognani}}, \bibinfo {author}
  {\bibfnamefont {L.}~\bibnamefont {Carbone}}, \bibinfo {author} {\bibfnamefont
  {S.}~\bibnamefont {Caride}}, \bibinfo {author} {\bibfnamefont
  {G.}~\bibnamefont {Castaldi}}, \bibinfo {author} {\bibfnamefont
  {S.}~\bibnamefont {Caudill}}, \bibinfo {author} {\bibfnamefont
  {M.}~\bibnamefont {Cavagli\`a}}, \bibinfo {author} {\bibfnamefont
  {F.}~\bibnamefont {Cavalier}}, \bibinfo {author} {\bibfnamefont
  {R.}~\bibnamefont {Cavalieri}}, \bibinfo {author} {\bibfnamefont
  {C.}~\bibnamefont {Celerier}}, \bibinfo {author} {\bibfnamefont
  {G.}~\bibnamefont {Cella}}, \bibinfo {author} {\bibfnamefont
  {C.}~\bibnamefont {Cepeda}}, \bibinfo {author} {\bibfnamefont
  {E.}~\bibnamefont {Cesarini}}, \bibinfo {author} {\bibfnamefont
  {R.}~\bibnamefont {Chakraborty}}, \bibinfo {author} {\bibfnamefont
  {T.}~\bibnamefont {Chalermsongsak}}, \bibinfo {author} {\bibfnamefont
  {S.~J.}\ \bibnamefont {Chamberlin}}, \bibinfo {author} {\bibfnamefont
  {S.}~\bibnamefont {Chao}}, \bibinfo {author} {\bibfnamefont {P.}~\bibnamefont
  {Charlton}}, \bibinfo {author} {\bibfnamefont {E.}~\bibnamefont
  {Chassande-Mottin}}, \bibinfo {author} {\bibfnamefont {X.}~\bibnamefont
  {Chen}}, \bibinfo {author} {\bibfnamefont {Y.}~\bibnamefont {Chen}}, \bibinfo
  {author} {\bibfnamefont {A.}~\bibnamefont {Chincarini}}, \bibinfo {author}
  {\bibfnamefont {A.}~\bibnamefont {Chiummo}}, \bibinfo {author} {\bibfnamefont
  {H.~S.}\ \bibnamefont {Cho}}, \bibinfo {author} {\bibfnamefont
  {M.}~\bibnamefont {Cho}}, \bibinfo {author} {\bibfnamefont {J.~H.}\
  \bibnamefont {Chow}}, \bibinfo {author} {\bibfnamefont {N.}~\bibnamefont
  {Christensen}}, \bibinfo {author} {\bibfnamefont {Q.}~\bibnamefont {Chu}},
  \bibinfo {author} {\bibfnamefont {S.~S.~Y.}\ \bibnamefont {Chua}}, \bibinfo
  {author} {\bibfnamefont {S.}~\bibnamefont {Chung}}, \bibinfo {author}
  {\bibfnamefont {G.}~\bibnamefont {Ciani}}, \bibinfo {author} {\bibfnamefont
  {F.}~\bibnamefont {Clara}}, \bibinfo {author} {\bibfnamefont {D.~E.}\
  \bibnamefont {Clark}}, \bibinfo {author} {\bibfnamefont {J.~A.}\ \bibnamefont
  {Clark}}, \bibinfo {author} {\bibfnamefont {J.~H.}\ \bibnamefont {Clayton}},
  \bibinfo {author} {\bibfnamefont {F.}~\bibnamefont {Cleva}}, \bibinfo
  {author} {\bibfnamefont {E.}~\bibnamefont {Coccia}}, \bibinfo {author}
  {\bibfnamefont {P.-F.}\ \bibnamefont {Cohadon}}, \bibinfo {author}
  {\bibfnamefont {A.}~\bibnamefont {Colla}}, \bibinfo {author} {\bibfnamefont
  {C.}~\bibnamefont {Collette}}, \bibinfo {author} {\bibfnamefont
  {M.}~\bibnamefont {Colombini}}, \bibinfo {author} {\bibfnamefont
  {L.}~\bibnamefont {Cominsky}}, \bibinfo {author} {\bibfnamefont
  {M.}~\bibnamefont {Constancio}}, \bibinfo {author} {\bibfnamefont
  {A.}~\bibnamefont {Conte}}, \bibinfo {author} {\bibfnamefont
  {D.}~\bibnamefont {Cook}}, \bibinfo {author} {\bibfnamefont {T.~R.}\
  \bibnamefont {Corbitt}}, \bibinfo {author} {\bibfnamefont {N.}~\bibnamefont
  {Cornish}}, \bibinfo {author} {\bibfnamefont {A.}~\bibnamefont {Corsi}},
  \bibinfo {author} {\bibfnamefont {C.~A.}\ \bibnamefont {Costa}}, \bibinfo
  {author} {\bibfnamefont {M.~W.}\ \bibnamefont {Coughlin}}, \bibinfo {author}
  {\bibfnamefont {J.-P.}\ \bibnamefont {Coulon}}, \bibinfo {author}
  {\bibfnamefont {S.}~\bibnamefont {Countryman}}, \bibinfo {author}
  {\bibfnamefont {P.}~\bibnamefont {Couvares}}, \bibinfo {author}
  {\bibfnamefont {D.~M.}\ \bibnamefont {Coward}}, \bibinfo {author}
  {\bibfnamefont {M.~J.}\ \bibnamefont {Cowart}}, \bibinfo {author}
  {\bibfnamefont {D.~C.}\ \bibnamefont {Coyne}}, \bibinfo {author}
  {\bibfnamefont {R.}~\bibnamefont {Coyne}}, \bibinfo {author} {\bibfnamefont
  {K.}~\bibnamefont {Craig}}, \bibinfo {author} {\bibfnamefont {J.~D.~E.}\
  \bibnamefont {Creighton}}, \bibinfo {author} {\bibfnamefont {R.~P.}\
  \bibnamefont {Croce}}, \bibinfo {author} {\bibfnamefont {S.~G.}\ \bibnamefont
  {Crowder}}, \bibinfo {author} {\bibfnamefont {A.}~\bibnamefont {Cumming}},
  \bibinfo {author} {\bibfnamefont {L.}~\bibnamefont {Cunningham}}, \bibinfo
  {author} {\bibfnamefont {E.}~\bibnamefont {Cuoco}}, \bibinfo {author}
  {\bibfnamefont {C.}~\bibnamefont {Cutler}}, \bibinfo {author} {\bibfnamefont
  {K.}~\bibnamefont {Dahl}}, \bibinfo {author} {\bibfnamefont {T.}~\bibnamefont
  {Dal~Canton}}, \bibinfo {author} {\bibfnamefont {M.}~\bibnamefont
  {Damjanic}}, \bibinfo {author} {\bibfnamefont {S.~L.}\ \bibnamefont
  {Danilishin}}, \bibinfo {author} {\bibfnamefont {S.}~\bibnamefont
  {D'Antonio}}, \bibinfo {author} {\bibfnamefont {K.}~\bibnamefont {Danzmann}},
  \bibinfo {author} {\bibfnamefont {V.}~\bibnamefont {Dattilo}}, \bibinfo
  {author} {\bibfnamefont {H.}~\bibnamefont {Daveloza}}, \bibinfo {author}
  {\bibfnamefont {M.}~\bibnamefont {Davier}}, \bibinfo {author} {\bibfnamefont
  {G.~S.}\ \bibnamefont {Davies}}, \bibinfo {author} {\bibfnamefont {E.~J.}\
  \bibnamefont {Daw}}, \bibinfo {author} {\bibfnamefont {R.}~\bibnamefont
  {Day}}, \bibinfo {author} {\bibfnamefont {T.}~\bibnamefont {Dayanga}},
  \bibinfo {author} {\bibfnamefont {D.}~\bibnamefont {DeBra}}, \bibinfo
  {author} {\bibfnamefont {G.}~\bibnamefont {Debreczeni}}, \bibinfo {author}
  {\bibfnamefont {J.}~\bibnamefont {Degallaix}}, \bibinfo {author}
  {\bibfnamefont {S.}~\bibnamefont {Del\'eglise}}, \bibinfo {author}
  {\bibfnamefont {W.}~\bibnamefont {Del~Pozzo}}, \bibinfo {author}
  {\bibfnamefont {W.}~\bibnamefont {Del~Pozzo}}, \bibinfo {author}
  {\bibfnamefont {T.}~\bibnamefont {Denker}}, \bibinfo {author} {\bibfnamefont
  {T.}~\bibnamefont {Dent}}, \bibinfo {author} {\bibfnamefont {H.}~\bibnamefont
  {Dereli}}, \bibinfo {author} {\bibfnamefont {V.}~\bibnamefont {Dergachev}},
  \bibinfo {author} {\bibfnamefont {R.}~\bibnamefont {De~Rosa}}, \bibinfo
  {author} {\bibfnamefont {R.~T.}\ \bibnamefont {DeRosa}}, \bibinfo {author}
  {\bibfnamefont {R.}~\bibnamefont {DeSalvo}}, \bibinfo {author} {\bibfnamefont
  {S.}~\bibnamefont {Dhurandhar}}, \bibinfo {author} {\bibfnamefont
  {M.}~\bibnamefont {D\'{\i}az}}, \bibinfo {author} {\bibfnamefont
  {J.}~\bibnamefont {Dickson}}, \bibinfo {author} {\bibfnamefont
  {L.}~\bibnamefont {Di~Fiore}}, \bibinfo {author} {\bibfnamefont
  {A.}~\bibnamefont {Di~Lieto}}, \bibinfo {author} {\bibfnamefont
  {I.}~\bibnamefont {Di~Palma}}, \bibinfo {author} {\bibfnamefont
  {A.}~\bibnamefont {Di~Virgilio}}, \bibinfo {author} {\bibfnamefont
  {V.}~\bibnamefont {Dolique}}, \bibinfo {author} {\bibfnamefont
  {E.}~\bibnamefont {Dominguez}}, \bibinfo {author} {\bibfnamefont
  {F.}~\bibnamefont {Donovan}}, \bibinfo {author} {\bibfnamefont {K.~L.}\
  \bibnamefont {Dooley}}, \bibinfo {author} {\bibfnamefont {S.}~\bibnamefont
  {Doravari}}, \bibinfo {author} {\bibfnamefont {R.}~\bibnamefont {Douglas}},
  \bibinfo {author} {\bibfnamefont {T.~P.}\ \bibnamefont {Downes}}, \bibinfo
  {author} {\bibfnamefont {M.}~\bibnamefont {Drago}}, \bibinfo {author}
  {\bibfnamefont {R.~W.~P.}\ \bibnamefont {Drever}}, \bibinfo {author}
  {\bibfnamefont {J.~C.}\ \bibnamefont {Driggers}}, \bibinfo {author}
  {\bibfnamefont {Z.}~\bibnamefont {Du}}, \bibinfo {author} {\bibfnamefont
  {M.}~\bibnamefont {Ducrot}}, \bibinfo {author} {\bibfnamefont
  {S.}~\bibnamefont {Dwyer}}, \bibinfo {author} {\bibfnamefont
  {T.}~\bibnamefont {Eberle}}, \bibinfo {author} {\bibfnamefont
  {T.}~\bibnamefont {Edo}}, \bibinfo {author} {\bibfnamefont {M.}~\bibnamefont
  {Edwards}}, \bibinfo {author} {\bibfnamefont {A.}~\bibnamefont {Effler}},
  \bibinfo {author} {\bibfnamefont {H.-B.}\ \bibnamefont {Eggenstein}},
  \bibinfo {author} {\bibfnamefont {P.}~\bibnamefont {Ehrens}}, \bibinfo
  {author} {\bibfnamefont {J.}~\bibnamefont {Eichholz}}, \bibinfo {author}
  {\bibfnamefont {S.~S.}\ \bibnamefont {Eikenberry}}, \bibinfo {author}
  {\bibfnamefont {G.}~\bibnamefont {Endr\ifmmode~\mbox{\H{o}}\else
  \H{o}\fi{}czi}}, \bibinfo {author} {\bibfnamefont {R.}~\bibnamefont
  {Essick}}, \bibinfo {author} {\bibfnamefont {T.}~\bibnamefont {Etzel}},
  \bibinfo {author} {\bibfnamefont {M.}~\bibnamefont {Evans}}, \bibinfo
  {author} {\bibfnamefont {T.}~\bibnamefont {Evans}}, \bibinfo {author}
  {\bibfnamefont {M.}~\bibnamefont {Factourovich}}, \bibinfo {author}
  {\bibfnamefont {V.}~\bibnamefont {Fafone}}, \bibinfo {author} {\bibfnamefont
  {S.}~\bibnamefont {Fairhurst}}, \bibinfo {author} {\bibfnamefont
  {X.}~\bibnamefont {Fan}}, \bibinfo {author} {\bibfnamefont {Q.}~\bibnamefont
  {Fang}}, \bibinfo {author} {\bibfnamefont {S.}~\bibnamefont {Farinon}},
  \bibinfo {author} {\bibfnamefont {B.}~\bibnamefont {Farr}}, \bibinfo {author}
  {\bibfnamefont {W.~M.}\ \bibnamefont {Farr}}, \bibinfo {author}
  {\bibfnamefont {M.}~\bibnamefont {Favata}}, \bibinfo {author} {\bibfnamefont
  {D.}~\bibnamefont {Fazi}}, \bibinfo {author} {\bibfnamefont {H.}~\bibnamefont
  {Fehrmann}}, \bibinfo {author} {\bibfnamefont {M.~M.}\ \bibnamefont {Fejer}},
  \bibinfo {author} {\bibfnamefont {D.}~\bibnamefont {Feldbaum}}, \bibinfo
  {author} {\bibfnamefont {F.}~\bibnamefont {Feroz}}, \bibinfo {author}
  {\bibfnamefont {I.}~\bibnamefont {Ferrante}}, \bibinfo {author}
  {\bibfnamefont {E.~C.}\ \bibnamefont {Ferreira}}, \bibinfo {author}
  {\bibfnamefont {F.}~\bibnamefont {Ferrini}}, \bibinfo {author} {\bibfnamefont
  {F.}~\bibnamefont {Fidecaro}}, \bibinfo {author} {\bibfnamefont {L.~S.}\
  \bibnamefont {Finn}}, \bibinfo {author} {\bibfnamefont {I.}~\bibnamefont
  {Fiori}}, \bibinfo {author} {\bibfnamefont {R.~P.}\ \bibnamefont {Fisher}},
  \bibinfo {author} {\bibfnamefont {R.}~\bibnamefont {Flaminio}}, \bibinfo
  {author} {\bibfnamefont {N.}~\bibnamefont {Fotopoulos}}, \bibinfo {author}
  {\bibfnamefont {J.-D.}\ \bibnamefont {Fournier}}, \bibinfo {author}
  {\bibfnamefont {S.}~\bibnamefont {Franco}}, \bibinfo {author} {\bibfnamefont
  {S.}~\bibnamefont {Frasca}}, \bibinfo {author} {\bibfnamefont
  {F.}~\bibnamefont {Frasconi}}, \bibinfo {author} {\bibfnamefont
  {M.}~\bibnamefont {Frede}}, \bibinfo {author} {\bibfnamefont
  {Z.}~\bibnamefont {Frei}}, \bibinfo {author} {\bibfnamefont {A.}~\bibnamefont
  {Freise}}, \bibinfo {author} {\bibfnamefont {R.}~\bibnamefont {Frey}},
  \bibinfo {author} {\bibfnamefont {T.~T.}\ \bibnamefont {Fricke}}, \bibinfo
  {author} {\bibfnamefont {P.}~\bibnamefont {Fritschel}}, \bibinfo {author}
  {\bibfnamefont {V.~V.}\ \bibnamefont {Frolov}}, \bibinfo {author}
  {\bibfnamefont {P.}~\bibnamefont {Fulda}}, \bibinfo {author} {\bibfnamefont
  {M.}~\bibnamefont {Fyffe}}, \bibinfo {author} {\bibfnamefont {J.~R.}\
  \bibnamefont {Gair}}, \bibinfo {author} {\bibfnamefont {L.}~\bibnamefont
  {Gammaitoni}}, \bibinfo {author} {\bibfnamefont {S.}~\bibnamefont {Gaonkar}},
  \bibinfo {author} {\bibfnamefont {F.}~\bibnamefont {Garufi}}, \bibinfo
  {author} {\bibfnamefont {N.}~\bibnamefont {Gehrels}}, \bibinfo {author}
  {\bibfnamefont {G.}~\bibnamefont {Gemme}}, \bibinfo {author} {\bibfnamefont
  {B.}~\bibnamefont {Gendre}}, \bibinfo {author} {\bibfnamefont
  {E.}~\bibnamefont {Genin}}, \bibinfo {author} {\bibfnamefont
  {A.}~\bibnamefont {Gennai}}, \bibinfo {author} {\bibfnamefont
  {S.}~\bibnamefont {Ghosh}}, \bibinfo {author} {\bibfnamefont {J.~A.}\
  \bibnamefont {Giaime}}, \bibinfo {author} {\bibfnamefont {K.~D.}\
  \bibnamefont {Giardina}}, \bibinfo {author} {\bibfnamefont {A.}~\bibnamefont
  {Giazotto}}, \bibinfo {author} {\bibfnamefont {C.}~\bibnamefont {Gill}},
  \bibinfo {author} {\bibfnamefont {J.}~\bibnamefont {Gleason}}, \bibinfo
  {author} {\bibfnamefont {E.}~\bibnamefont {Goetz}}, \bibinfo {author}
  {\bibfnamefont {R.}~\bibnamefont {Goetz}}, \bibinfo {author} {\bibfnamefont
  {L.}~\bibnamefont {Gondan}}, \bibinfo {author} {\bibfnamefont
  {G.}~\bibnamefont {Gonz\'alez}}, \bibinfo {author} {\bibfnamefont
  {N.}~\bibnamefont {Gordon}}, \bibinfo {author} {\bibfnamefont {M.~L.}\
  \bibnamefont {Gorodetsky}}, \bibinfo {author} {\bibfnamefont
  {S.}~\bibnamefont {Gossan}}, \bibinfo {author} {\bibfnamefont
  {S.}~\bibnamefont {Go\ss{}ler}}, \bibinfo {author} {\bibfnamefont
  {R.}~\bibnamefont {Gouaty}}, \bibinfo {author} {\bibfnamefont
  {C.}~\bibnamefont {Gr\"af}}, \bibinfo {author} {\bibfnamefont {P.~B.}\
  \bibnamefont {Graff}}, \bibinfo {author} {\bibfnamefont {M.}~\bibnamefont
  {Granata}}, \bibinfo {author} {\bibfnamefont {A.}~\bibnamefont {Grant}},
  \bibinfo {author} {\bibfnamefont {S.}~\bibnamefont {Gras}}, \bibinfo {author}
  {\bibfnamefont {C.}~\bibnamefont {Gray}}, \bibinfo {author} {\bibfnamefont
  {R.~J.~S.}\ \bibnamefont {Greenhalgh}}, \bibinfo {author} {\bibfnamefont
  {A.~M.}\ \bibnamefont {Gretarsson}}, \bibinfo {author} {\bibfnamefont
  {P.}~\bibnamefont {Groot}}, \bibinfo {author} {\bibfnamefont
  {H.}~\bibnamefont {Grote}}, \bibinfo {author} {\bibfnamefont
  {K.}~\bibnamefont {Grover}}, \bibinfo {author} {\bibfnamefont
  {S.}~\bibnamefont {Grunewald}}, \bibinfo {author} {\bibfnamefont {G.~M.}\
  \bibnamefont {Guidi}}, \bibinfo {author} {\bibfnamefont {C.~J.}\ \bibnamefont
  {Guido}}, \bibinfo {author} {\bibfnamefont {K.}~\bibnamefont {Gushwa}},
  \bibinfo {author} {\bibfnamefont {E.~K.}\ \bibnamefont {Gustafson}}, \bibinfo
  {author} {\bibfnamefont {R.}~\bibnamefont {Gustafson}}, \bibinfo {author}
  {\bibfnamefont {J.}~\bibnamefont {Ha}}, \bibinfo {author} {\bibfnamefont
  {E.~D.}\ \bibnamefont {Hall}}, \bibinfo {author} {\bibfnamefont
  {W.}~\bibnamefont {Hamilton}}, \bibinfo {author} {\bibfnamefont
  {D.}~\bibnamefont {Hammer}}, \bibinfo {author} {\bibfnamefont
  {G.}~\bibnamefont {Hammond}}, \bibinfo {author} {\bibfnamefont
  {M.}~\bibnamefont {Hanke}}, \bibinfo {author} {\bibfnamefont
  {J.}~\bibnamefont {Hanks}}, \bibinfo {author} {\bibfnamefont
  {C.}~\bibnamefont {Hanna}}, \bibinfo {author} {\bibfnamefont {M.~D.}\
  \bibnamefont {Hannam}}, \bibinfo {author} {\bibfnamefont {J.}~\bibnamefont
  {Hanson}}, \bibinfo {author} {\bibfnamefont {K.}~\bibnamefont {Haris}},
  \bibinfo {author} {\bibfnamefont {J.}~\bibnamefont {Harms}}, \bibinfo
  {author} {\bibfnamefont {G.~M.}\ \bibnamefont {Harry}}, \bibinfo {author}
  {\bibfnamefont {I.~W.}\ \bibnamefont {Harry}}, \bibinfo {author}
  {\bibfnamefont {E.~D.}\ \bibnamefont {Harstad}}, \bibinfo {author}
  {\bibfnamefont {M.}~\bibnamefont {Hart}}, \bibinfo {author} {\bibfnamefont
  {M.~T.}\ \bibnamefont {Hartman}}, \bibinfo {author} {\bibfnamefont {C.-J.}\
  \bibnamefont {Haster}}, \bibinfo {author} {\bibfnamefont {K.}~\bibnamefont
  {Haughian}}, \bibinfo {author} {\bibfnamefont {A.}~\bibnamefont {Heidmann}},
  \bibinfo {author} {\bibfnamefont {M.}~\bibnamefont {Heintze}}, \bibinfo
  {author} {\bibfnamefont {H.}~\bibnamefont {Heitmann}}, \bibinfo {author}
  {\bibfnamefont {P.}~\bibnamefont {Hello}}, \bibinfo {author} {\bibfnamefont
  {G.}~\bibnamefont {Hemming}}, \bibinfo {author} {\bibfnamefont
  {M.}~\bibnamefont {Hendry}}, \bibinfo {author} {\bibfnamefont {I.~S.}\
  \bibnamefont {Heng}}, \bibinfo {author} {\bibfnamefont {A.~W.}\ \bibnamefont
  {Heptonstall}}, \bibinfo {author} {\bibfnamefont {M.}~\bibnamefont {Heurs}},
  \bibinfo {author} {\bibfnamefont {M.}~\bibnamefont {Hewitson}}, \bibinfo
  {author} {\bibfnamefont {S.}~\bibnamefont {Hild}}, \bibinfo {author}
  {\bibfnamefont {D.}~\bibnamefont {Hoak}}, \bibinfo {author} {\bibfnamefont
  {K.~A.}\ \bibnamefont {Hodge}}, \bibinfo {author} {\bibfnamefont
  {D.}~\bibnamefont {Hofman}}, \bibinfo {author} {\bibfnamefont
  {K.}~\bibnamefont {Holt}}, \bibinfo {author} {\bibfnamefont {P.}~\bibnamefont
  {Hopkins}}, \bibinfo {author} {\bibfnamefont {T.}~\bibnamefont {Horrom}},
  \bibinfo {author} {\bibfnamefont {D.}~\bibnamefont {Hoske}}, \bibinfo
  {author} {\bibfnamefont {D.~J.}\ \bibnamefont {Hosken}}, \bibinfo {author}
  {\bibfnamefont {J.}~\bibnamefont {Hough}}, \bibinfo {author} {\bibfnamefont
  {E.~J.}\ \bibnamefont {Howell}}, \bibinfo {author} {\bibfnamefont
  {Y.}~\bibnamefont {Hu}}, \bibinfo {author} {\bibfnamefont {E.}~\bibnamefont
  {Huerta}}, \bibinfo {author} {\bibfnamefont {B.}~\bibnamefont {Hughey}},
  \bibinfo {author} {\bibfnamefont {S.}~\bibnamefont {Husa}}, \bibinfo {author}
  {\bibfnamefont {S.~H.}\ \bibnamefont {Huttner}}, \bibinfo {author}
  {\bibfnamefont {M.}~\bibnamefont {Huynh}}, \bibinfo {author} {\bibfnamefont
  {T.}~\bibnamefont {Huynh-Dinh}}, \bibinfo {author} {\bibfnamefont
  {A.}~\bibnamefont {Idrisy}}, \bibinfo {author} {\bibfnamefont {D.~R.}\
  \bibnamefont {Ingram}}, \bibinfo {author} {\bibfnamefont {R.}~\bibnamefont
  {Inta}}, \bibinfo {author} {\bibfnamefont {G.}~\bibnamefont {Islas}},
  \bibinfo {author} {\bibfnamefont {T.}~\bibnamefont {Isogai}}, \bibinfo
  {author} {\bibfnamefont {A.}~\bibnamefont {Ivanov}}, \bibinfo {author}
  {\bibfnamefont {B.~R.}\ \bibnamefont {Iyer}}, \bibinfo {author}
  {\bibfnamefont {K.}~\bibnamefont {Izumi}}, \bibinfo {author} {\bibfnamefont
  {M.}~\bibnamefont {Jacobson}}, \bibinfo {author} {\bibfnamefont
  {H.}~\bibnamefont {Jang}}, \bibinfo {author} {\bibfnamefont {P.}~\bibnamefont
  {Jaranowski}}, \bibinfo {author} {\bibfnamefont {Y.}~\bibnamefont {Ji}},
  \bibinfo {author} {\bibfnamefont {F.}~\bibnamefont {Jim\'enez-Forteza}},
  \bibinfo {author} {\bibfnamefont {W.~W.}\ \bibnamefont {Johnson}}, \bibinfo
  {author} {\bibfnamefont {D.~I.}\ \bibnamefont {Jones}}, \bibinfo {author}
  {\bibfnamefont {G.}~\bibnamefont {Jones}}, \bibinfo {author} {\bibfnamefont
  {R.}~\bibnamefont {Jones}}, \bibinfo {author} {\bibfnamefont {R.~J.~G.}\
  \bibnamefont {Jonker}}, \bibinfo {author} {\bibfnamefont {L.}~\bibnamefont
  {Ju}}, \bibinfo {author} {\bibfnamefont {P.}~\bibnamefont {Kalmus}}, \bibinfo
  {author} {\bibfnamefont {V.}~\bibnamefont {Kalogera}}, \bibinfo {author}
  {\bibfnamefont {S.}~\bibnamefont {Kandhasamy}}, \bibinfo {author}
  {\bibfnamefont {G.}~\bibnamefont {Kang}}, \bibinfo {author} {\bibfnamefont
  {J.~B.}\ \bibnamefont {Kanner}}, \bibinfo {author} {\bibfnamefont
  {J.}~\bibnamefont {Karlen}}, \bibinfo {author} {\bibfnamefont
  {M.}~\bibnamefont {Kasprzack}}, \bibinfo {author} {\bibfnamefont
  {E.}~\bibnamefont {Katsavounidis}}, \bibinfo {author} {\bibfnamefont
  {W.}~\bibnamefont {Katzman}}, \bibinfo {author} {\bibfnamefont
  {H.}~\bibnamefont {Kaufer}}, \bibinfo {author} {\bibfnamefont
  {S.}~\bibnamefont {Kaufer}}, \bibinfo {author} {\bibfnamefont
  {T.}~\bibnamefont {Kaur}}, \bibinfo {author} {\bibfnamefont {K.}~\bibnamefont
  {Kawabe}}, \bibinfo {author} {\bibfnamefont {F.}~\bibnamefont {Kawazoe}},
  \bibinfo {author} {\bibfnamefont {F.}~\bibnamefont {K\'ef\'elian}}, \bibinfo
  {author} {\bibfnamefont {G.~M.}\ \bibnamefont {Keiser}}, \bibinfo {author}
  {\bibfnamefont {D.}~\bibnamefont {Keitel}}, \bibinfo {author} {\bibfnamefont
  {D.~B.}\ \bibnamefont {Kelley}}, \bibinfo {author} {\bibfnamefont
  {W.}~\bibnamefont {Kells}}, \bibinfo {author} {\bibfnamefont {D.~G.}\
  \bibnamefont {Keppel}}, \bibinfo {author} {\bibfnamefont {A.}~\bibnamefont
  {Khalaidovski}}, \bibinfo {author} {\bibfnamefont {F.~Y.}\ \bibnamefont
  {Khalili}}, \bibinfo {author} {\bibfnamefont {E.~A.}\ \bibnamefont
  {Khazanov}}, \bibinfo {author} {\bibfnamefont {C.}~\bibnamefont {Kim}},
  \bibinfo {author} {\bibfnamefont {K.}~\bibnamefont {Kim}}, \bibinfo {author}
  {\bibfnamefont {N.~G.}\ \bibnamefont {Kim}}, \bibinfo {author} {\bibfnamefont
  {N.}~\bibnamefont {Kim}}, \bibinfo {author} {\bibfnamefont {S.}~\bibnamefont
  {Kim}}, \bibinfo {author} {\bibfnamefont {Y.-M.}\ \bibnamefont {Kim}},
  \bibinfo {author} {\bibfnamefont {E.~J.}\ \bibnamefont {King}}, \bibinfo
  {author} {\bibfnamefont {P.~J.}\ \bibnamefont {King}}, \bibinfo {author}
  {\bibfnamefont {D.~L.}\ \bibnamefont {Kinzel}}, \bibinfo {author}
  {\bibfnamefont {J.~S.}\ \bibnamefont {Kissel}}, \bibinfo {author}
  {\bibfnamefont {S.}~\bibnamefont {Klimenko}}, \bibinfo {author}
  {\bibfnamefont {J.}~\bibnamefont {Kline}}, \bibinfo {author} {\bibfnamefont
  {S.}~\bibnamefont {Koehlenbeck}}, \bibinfo {author} {\bibfnamefont
  {K.}~\bibnamefont {Kokeyama}}, \bibinfo {author} {\bibfnamefont
  {V.}~\bibnamefont {Kondrashov}}, \bibinfo {author} {\bibfnamefont
  {S.}~\bibnamefont {Koranda}}, \bibinfo {author} {\bibfnamefont {W.~Z.}\
  \bibnamefont {Korth}}, \bibinfo {author} {\bibfnamefont {I.}~\bibnamefont
  {Kowalska}}, \bibinfo {author} {\bibfnamefont {D.~B.}\ \bibnamefont {Kozak}},
  \bibinfo {author} {\bibfnamefont {V.}~\bibnamefont {Kringel}}, \bibinfo
  {author} {\bibfnamefont {B.}~\bibnamefont {Krishnan}}, \bibinfo {author}
  {\bibfnamefont {A.}~\bibnamefont {Kr\'olak}}, \bibinfo {author}
  {\bibfnamefont {G.}~\bibnamefont {Kuehn}}, \bibinfo {author} {\bibfnamefont
  {A.}~\bibnamefont {Kumar}}, \bibinfo {author} {\bibfnamefont {D.~N.}\
  \bibnamefont {Kumar}}, \bibinfo {author} {\bibfnamefont {P.}~\bibnamefont
  {Kumar}}, \bibinfo {author} {\bibfnamefont {R.}~\bibnamefont {Kumar}},
  \bibinfo {author} {\bibfnamefont {L.}~\bibnamefont {Kuo}}, \bibinfo {author}
  {\bibfnamefont {A.}~\bibnamefont {Kutynia}}, \bibinfo {author} {\bibfnamefont
  {P.~K.}\ \bibnamefont {Lam}}, \bibinfo {author} {\bibfnamefont
  {M.}~\bibnamefont {Landry}}, \bibinfo {author} {\bibfnamefont
  {B.}~\bibnamefont {Lantz}}, \bibinfo {author} {\bibfnamefont
  {S.}~\bibnamefont {Larson}}, \bibinfo {author} {\bibfnamefont {P.~D.}\
  \bibnamefont {Lasky}}, \bibinfo {author} {\bibfnamefont {C.}~\bibnamefont
  {Lazzaro}}, \bibinfo {author} {\bibfnamefont {P.}~\bibnamefont {Leaci}},
  \bibinfo {author} {\bibfnamefont {S.}~\bibnamefont {Leavey}}, \bibinfo
  {author} {\bibfnamefont {E.~O.}\ \bibnamefont {Lebigot}}, \bibinfo {author}
  {\bibfnamefont {C.~H.}\ \bibnamefont {Lee}}, \bibinfo {author} {\bibfnamefont
  {H.~K.}\ \bibnamefont {Lee}}, \bibinfo {author} {\bibfnamefont {H.~M.}\
  \bibnamefont {Lee}}, \bibinfo {author} {\bibfnamefont {J.}~\bibnamefont
  {Lee}}, \bibinfo {author} {\bibfnamefont {P.~J.}\ \bibnamefont {Lee}},
  \bibinfo {author} {\bibfnamefont {M.}~\bibnamefont {Leonardi}}, \bibinfo
  {author} {\bibfnamefont {J.~R.}\ \bibnamefont {Leong}}, \bibinfo {author}
  {\bibfnamefont {A.}~\bibnamefont {Le~Roux}}, \bibinfo {author} {\bibfnamefont
  {N.}~\bibnamefont {Leroy}}, \bibinfo {author} {\bibfnamefont
  {N.}~\bibnamefont {Letendre}}, \bibinfo {author} {\bibfnamefont
  {Y.}~\bibnamefont {Levin}}, \bibinfo {author} {\bibfnamefont
  {B.}~\bibnamefont {Levine}}, \bibinfo {author} {\bibfnamefont
  {J.}~\bibnamefont {Lewis}}, \bibinfo {author} {\bibfnamefont {T.~G.~F.}\
  \bibnamefont {Li}}, \bibinfo {author} {\bibfnamefont {K.}~\bibnamefont
  {Libbrecht}}, \bibinfo {author} {\bibfnamefont {A.}~\bibnamefont {Libson}},
  \bibinfo {author} {\bibfnamefont {A.~C.}\ \bibnamefont {Lin}}, \bibinfo
  {author} {\bibfnamefont {T.~B.}\ \bibnamefont {Littenberg}}, \bibinfo
  {author} {\bibfnamefont {N.~A.}\ \bibnamefont {Lockerbie}}, \bibinfo {author}
  {\bibfnamefont {V.}~\bibnamefont {Lockett}}, \bibinfo {author} {\bibfnamefont
  {D.}~\bibnamefont {Lodhia}}, \bibinfo {author} {\bibfnamefont
  {K.}~\bibnamefont {Loew}}, \bibinfo {author} {\bibfnamefont {J.}~\bibnamefont
  {Logue}}, \bibinfo {author} {\bibfnamefont {A.~L.}\ \bibnamefont {Lombardi}},
  \bibinfo {author} {\bibfnamefont {E.}~\bibnamefont {Lopez}}, \bibinfo
  {author} {\bibfnamefont {M.}~\bibnamefont {Lorenzini}}, \bibinfo {author}
  {\bibfnamefont {V.}~\bibnamefont {Loriette}}, \bibinfo {author}
  {\bibfnamefont {M.}~\bibnamefont {Lormand}}, \bibinfo {author} {\bibfnamefont
  {G.}~\bibnamefont {Losurdo}}, \bibinfo {author} {\bibfnamefont
  {J.}~\bibnamefont {Lough}}, \bibinfo {author} {\bibfnamefont {M.~J.}\
  \bibnamefont {Lubinski}}, \bibinfo {author} {\bibfnamefont {H.}~\bibnamefont
  {L\"uck}}, \bibinfo {author} {\bibfnamefont {A.~P.}\ \bibnamefont
  {Lundgren}}, \bibinfo {author} {\bibfnamefont {Y.}~\bibnamefont {Ma}},
  \bibinfo {author} {\bibfnamefont {E.~P.}\ \bibnamefont {Macdonald}}, \bibinfo
  {author} {\bibfnamefont {T.}~\bibnamefont {MacDonald}}, \bibinfo {author}
  {\bibfnamefont {B.}~\bibnamefont {Machenschalk}}, \bibinfo {author}
  {\bibfnamefont {M.}~\bibnamefont {MacInnis}}, \bibinfo {author}
  {\bibfnamefont {D.~M.}\ \bibnamefont {Macleod}}, \bibinfo {author}
  {\bibfnamefont {F.}~\bibnamefont {Maga\~na Sandoval}}, \bibinfo {author}
  {\bibfnamefont {R.}~\bibnamefont {Magee}}, \bibinfo {author} {\bibfnamefont
  {M.}~\bibnamefont {Mageswaran}}, \bibinfo {author} {\bibfnamefont
  {C.}~\bibnamefont {Maglione}}, \bibinfo {author} {\bibfnamefont
  {K.}~\bibnamefont {Mailand}}, \bibinfo {author} {\bibfnamefont
  {E.}~\bibnamefont {Majorana}}, \bibinfo {author} {\bibfnamefont
  {I.}~\bibnamefont {Maksimovic}}, \bibinfo {author} {\bibfnamefont
  {V.}~\bibnamefont {Malvezzi}}, \bibinfo {author} {\bibfnamefont
  {N.}~\bibnamefont {Man}}, \bibinfo {author} {\bibfnamefont {G.~M.}\
  \bibnamefont {Manca}}, \bibinfo {author} {\bibfnamefont {I.}~\bibnamefont
  {Mandel}}, \bibinfo {author} {\bibfnamefont {V.}~\bibnamefont {Mandic}},
  \bibinfo {author} {\bibfnamefont {V.}~\bibnamefont {Mangano}}, \bibinfo
  {author} {\bibfnamefont {N.~M.}\ \bibnamefont {Mangini}}, \bibinfo {author}
  {\bibfnamefont {G.}~\bibnamefont {Mansell}}, \bibinfo {author} {\bibfnamefont
  {M.}~\bibnamefont {Mantovani}}, \bibinfo {author} {\bibfnamefont
  {F.}~\bibnamefont {Marchesoni}}, \bibinfo {author} {\bibfnamefont
  {F.}~\bibnamefont {Marion}}, \bibinfo {author} {\bibfnamefont
  {S.}~\bibnamefont {M\'arka}}, \bibinfo {author} {\bibfnamefont
  {Z.}~\bibnamefont {M\'arka}}, \bibinfo {author} {\bibfnamefont
  {A.}~\bibnamefont {Markosyan}}, \bibinfo {author} {\bibfnamefont
  {E.}~\bibnamefont {Maros}}, \bibinfo {author} {\bibfnamefont
  {J.}~\bibnamefont {Marque}}, \bibinfo {author} {\bibfnamefont
  {F.}~\bibnamefont {Martelli}}, \bibinfo {author} {\bibfnamefont {I.~W.}\
  \bibnamefont {Martin}}, \bibinfo {author} {\bibfnamefont {R.~M.}\
  \bibnamefont {Martin}}, \bibinfo {author} {\bibfnamefont {L.}~\bibnamefont
  {Martinelli}}, \bibinfo {author} {\bibfnamefont {D.}~\bibnamefont
  {Martynov}}, \bibinfo {author} {\bibfnamefont {J.~N.}\ \bibnamefont {Marx}},
  \bibinfo {author} {\bibfnamefont {K.}~\bibnamefont {Mason}}, \bibinfo
  {author} {\bibfnamefont {A.}~\bibnamefont {Masserot}}, \bibinfo {author}
  {\bibfnamefont {T.~J.}\ \bibnamefont {Massinger}}, \bibinfo {author}
  {\bibfnamefont {F.}~\bibnamefont {Matichard}}, \bibinfo {author}
  {\bibfnamefont {L.}~\bibnamefont {Matone}}, \bibinfo {author} {\bibfnamefont
  {N.}~\bibnamefont {Mavalvala}}, \bibinfo {author} {\bibfnamefont
  {G.}~\bibnamefont {May}}, \bibinfo {author} {\bibfnamefont {N.}~\bibnamefont
  {Mazumder}}, \bibinfo {author} {\bibfnamefont {G.}~\bibnamefont {Mazzolo}},
  \bibinfo {author} {\bibfnamefont {R.}~\bibnamefont {McCarthy}}, \bibinfo
  {author} {\bibfnamefont {D.~E.}\ \bibnamefont {McClelland}}, \bibinfo
  {author} {\bibfnamefont {S.~C.}\ \bibnamefont {McGuire}}, \bibinfo {author}
  {\bibfnamefont {G.}~\bibnamefont {McIntyre}}, \bibinfo {author}
  {\bibfnamefont {J.}~\bibnamefont {McIver}}, \bibinfo {author} {\bibfnamefont
  {K.}~\bibnamefont {McLin}}, \bibinfo {author} {\bibfnamefont
  {D.}~\bibnamefont {Meacher}}, \bibinfo {author} {\bibfnamefont {G.~D.}\
  \bibnamefont {Meadors}}, \bibinfo {author} {\bibfnamefont {M.}~\bibnamefont
  {Mehmet}}, \bibinfo {author} {\bibfnamefont {J.}~\bibnamefont {Meidam}},
  \bibinfo {author} {\bibfnamefont {M.}~\bibnamefont {Meinders}}, \bibinfo
  {author} {\bibfnamefont {A.}~\bibnamefont {Melatos}}, \bibinfo {author}
  {\bibfnamefont {G.}~\bibnamefont {Mendell}}, \bibinfo {author} {\bibfnamefont
  {R.~A.}\ \bibnamefont {Mercer}}, \bibinfo {author} {\bibfnamefont
  {S.}~\bibnamefont {Meshkov}}, \bibinfo {author} {\bibfnamefont
  {C.}~\bibnamefont {Messenger}}, \bibinfo {author} {\bibfnamefont
  {A.}~\bibnamefont {Meyer}}, \bibinfo {author} {\bibfnamefont {M.~S.}\
  \bibnamefont {Meyer}}, \bibinfo {author} {\bibfnamefont {P.~M.}\ \bibnamefont
  {Meyers}}, \bibinfo {author} {\bibfnamefont {F.}~\bibnamefont {Mezzani}},
  \bibinfo {author} {\bibfnamefont {H.}~\bibnamefont {Miao}}, \bibinfo {author}
  {\bibfnamefont {C.}~\bibnamefont {Michel}}, \bibinfo {author} {\bibfnamefont
  {E.~E.}\ \bibnamefont {Mikhailov}}, \bibinfo {author} {\bibfnamefont
  {L.}~\bibnamefont {Milano}}, \bibinfo {author} {\bibfnamefont
  {J.}~\bibnamefont {Miller}}, \bibinfo {author} {\bibfnamefont
  {Y.}~\bibnamefont {Minenkov}}, \bibinfo {author} {\bibfnamefont {C.~M.~F.}\
  \bibnamefont {Mingarelli}}, \bibinfo {author} {\bibfnamefont
  {C.}~\bibnamefont {Mishra}}, \bibinfo {author} {\bibfnamefont
  {S.}~\bibnamefont {Mitra}}, \bibinfo {author} {\bibfnamefont {V.~P.}\
  \bibnamefont {Mitrofanov}}, \bibinfo {author} {\bibfnamefont
  {G.}~\bibnamefont {Mitselmakher}}, \bibinfo {author} {\bibfnamefont
  {R.}~\bibnamefont {Mittleman}}, \bibinfo {author} {\bibfnamefont
  {B.}~\bibnamefont {Moe}}, \bibinfo {author} {\bibfnamefont {A.}~\bibnamefont
  {Moggi}}, \bibinfo {author} {\bibfnamefont {M.}~\bibnamefont {Mohan}},
  \bibinfo {author} {\bibfnamefont {S.~R.~P.}\ \bibnamefont {Mohapatra}},
  \bibinfo {author} {\bibfnamefont {D.}~\bibnamefont {Moraru}}, \bibinfo
  {author} {\bibfnamefont {G.}~\bibnamefont {Moreno}}, \bibinfo {author}
  {\bibfnamefont {N.}~\bibnamefont {Morgado}}, \bibinfo {author} {\bibfnamefont
  {S.~R.}\ \bibnamefont {Morriss}}, \bibinfo {author} {\bibfnamefont
  {K.}~\bibnamefont {Mossavi}}, \bibinfo {author} {\bibfnamefont
  {B.}~\bibnamefont {Mours}}, \bibinfo {author} {\bibfnamefont {C.~M.}\
  \bibnamefont {Mow-Lowry}}, \bibinfo {author} {\bibfnamefont {C.~L.}\
  \bibnamefont {Mueller}}, \bibinfo {author} {\bibfnamefont {G.}~\bibnamefont
  {Mueller}}, \bibinfo {author} {\bibfnamefont {S.}~\bibnamefont {Mukherjee}},
  \bibinfo {author} {\bibfnamefont {A.}~\bibnamefont {Mullavey}}, \bibinfo
  {author} {\bibfnamefont {J.}~\bibnamefont {Munch}}, \bibinfo {author}
  {\bibfnamefont {D.}~\bibnamefont {Murphy}}, \bibinfo {author} {\bibfnamefont
  {P.~G.}\ \bibnamefont {Murray}}, \bibinfo {author} {\bibfnamefont
  {A.}~\bibnamefont {Mytidis}}, \bibinfo {author} {\bibfnamefont {M.~F.}\
  \bibnamefont {Nagy}}, \bibinfo {author} {\bibfnamefont {I.}~\bibnamefont
  {Nardecchia}}, \bibinfo {author} {\bibfnamefont {L.}~\bibnamefont
  {Naticchioni}}, \bibinfo {author} {\bibfnamefont {R.~K.}\ \bibnamefont
  {Nayak}}, \bibinfo {author} {\bibfnamefont {V.}~\bibnamefont {Necula}},
  \bibinfo {author} {\bibfnamefont {G.}~\bibnamefont {Nelemans}}, \bibinfo
  {author} {\bibfnamefont {I.}~\bibnamefont {Neri}}, \bibinfo {author}
  {\bibfnamefont {M.}~\bibnamefont {Neri}}, \bibinfo {author} {\bibfnamefont
  {G.}~\bibnamefont {Newton}}, \bibinfo {author} {\bibfnamefont
  {T.}~\bibnamefont {Nguyen}}, \bibinfo {author} {\bibfnamefont {A.~B.}\
  \bibnamefont {Nielsen}}, \bibinfo {author} {\bibfnamefont {S.}~\bibnamefont
  {Nissanke}}, \bibinfo {author} {\bibfnamefont {A.~H.}\ \bibnamefont {Nitz}},
  \bibinfo {author} {\bibfnamefont {F.}~\bibnamefont {Nocera}}, \bibinfo
  {author} {\bibfnamefont {D.}~\bibnamefont {Nolting}}, \bibinfo {author}
  {\bibfnamefont {M.~E.~N.}\ \bibnamefont {Normandin}}, \bibinfo {author}
  {\bibfnamefont {L.~K.}\ \bibnamefont {Nuttall}}, \bibinfo {author}
  {\bibfnamefont {E.}~\bibnamefont {Ochsner}}, \bibinfo {author} {\bibfnamefont
  {J.}~\bibnamefont {O'Dell}}, \bibinfo {author} {\bibfnamefont
  {E.}~\bibnamefont {Oelker}}, \bibinfo {author} {\bibfnamefont {J.~J.}\
  \bibnamefont {Oh}}, \bibinfo {author} {\bibfnamefont {S.~H.}\ \bibnamefont
  {Oh}}, \bibinfo {author} {\bibfnamefont {F.}~\bibnamefont {Ohme}}, \bibinfo
  {author} {\bibfnamefont {S.}~\bibnamefont {Omar}}, \bibinfo {author}
  {\bibfnamefont {P.}~\bibnamefont {Oppermann}}, \bibinfo {author}
  {\bibfnamefont {R.}~\bibnamefont {Oram}}, \bibinfo {author} {\bibfnamefont
  {B.}~\bibnamefont {O'Reilly}}, \bibinfo {author} {\bibfnamefont
  {W.}~\bibnamefont {Ortega}}, \bibinfo {author} {\bibfnamefont
  {R.}~\bibnamefont {O'Shaughnessy}}, \bibinfo {author} {\bibfnamefont
  {C.}~\bibnamefont {Osthelder}}, \bibinfo {author} {\bibfnamefont {D.~J.}\
  \bibnamefont {Ottaway}}, \bibinfo {author} {\bibfnamefont {R.~S.}\
  \bibnamefont {Ottens}}, \bibinfo {author} {\bibfnamefont {H.}~\bibnamefont
  {Overmier}}, \bibinfo {author} {\bibfnamefont {B.~J.}\ \bibnamefont {Owen}},
  \bibinfo {author} {\bibfnamefont {C.}~\bibnamefont {Padilla}}, \bibinfo
  {author} {\bibfnamefont {A.}~\bibnamefont {Pai}}, \bibinfo {author}
  {\bibfnamefont {O.}~\bibnamefont {Palashov}}, \bibinfo {author}
  {\bibfnamefont {C.}~\bibnamefont {Palomba}}, \bibinfo {author} {\bibfnamefont
  {H.}~\bibnamefont {Pan}}, \bibinfo {author} {\bibfnamefont {Y.}~\bibnamefont
  {Pan}}, \bibinfo {author} {\bibfnamefont {C.}~\bibnamefont {Pankow}},
  \bibinfo {author} {\bibfnamefont {F.}~\bibnamefont {Paoletti}}, \bibinfo
  {author} {\bibfnamefont {M.~A.}\ \bibnamefont {Papa}}, \bibinfo {author}
  {\bibfnamefont {H.}~\bibnamefont {Paris}}, \bibinfo {author} {\bibfnamefont
  {A.}~\bibnamefont {Pasqualetti}}, \bibinfo {author} {\bibfnamefont
  {R.}~\bibnamefont {Passaquieti}}, \bibinfo {author} {\bibfnamefont
  {D.}~\bibnamefont {Passuello}}, \bibinfo {author} {\bibfnamefont
  {P.}~\bibnamefont {Patel}}, \bibinfo {author} {\bibfnamefont
  {M.}~\bibnamefont {Pedraza}}, \bibinfo {author} {\bibfnamefont
  {A.}~\bibnamefont {Pele}}, \bibinfo {author} {\bibfnamefont {S.}~\bibnamefont
  {Penn}}, \bibinfo {author} {\bibfnamefont {A.}~\bibnamefont {Perreca}},
  \bibinfo {author} {\bibfnamefont {M.}~\bibnamefont {Phelps}}, \bibinfo
  {author} {\bibfnamefont {M.}~\bibnamefont {Pichot}}, \bibinfo {author}
  {\bibfnamefont {M.}~\bibnamefont {Pickenpack}}, \bibinfo {author}
  {\bibfnamefont {F.}~\bibnamefont {Piergiovanni}}, \bibinfo {author}
  {\bibfnamefont {V.}~\bibnamefont {Pierro}}, \bibinfo {author} {\bibfnamefont
  {L.}~\bibnamefont {Pinard}}, \bibinfo {author} {\bibfnamefont {I.~M.}\
  \bibnamefont {Pinto}}, \bibinfo {author} {\bibfnamefont {M.}~\bibnamefont
  {Pitkin}}, \bibinfo {author} {\bibfnamefont {J.}~\bibnamefont {Poeld}},
  \bibinfo {author} {\bibfnamefont {R.}~\bibnamefont {Poggiani}}, \bibinfo
  {author} {\bibfnamefont {A.}~\bibnamefont {Poteomkin}}, \bibinfo {author}
  {\bibfnamefont {J.}~\bibnamefont {Powell}}, \bibinfo {author} {\bibfnamefont
  {J.}~\bibnamefont {Prasad}}, \bibinfo {author} {\bibfnamefont
  {V.}~\bibnamefont {Predoi}}, \bibinfo {author} {\bibfnamefont
  {S.}~\bibnamefont {Premachandra}}, \bibinfo {author} {\bibfnamefont
  {T.}~\bibnamefont {Prestegard}}, \bibinfo {author} {\bibfnamefont {L.~R.}\
  \bibnamefont {Price}}, \bibinfo {author} {\bibfnamefont {M.}~\bibnamefont
  {Prijatelj}}, \bibinfo {author} {\bibfnamefont {S.}~\bibnamefont
  {Privitera}}, \bibinfo {author} {\bibfnamefont {G.~A.}\ \bibnamefont
  {Prodi}}, \bibinfo {author} {\bibfnamefont {L.}~\bibnamefont {Prokhorov}},
  \bibinfo {author} {\bibfnamefont {O.}~\bibnamefont {Puncken}}, \bibinfo
  {author} {\bibfnamefont {M.}~\bibnamefont {Punturo}}, \bibinfo {author}
  {\bibfnamefont {P.}~\bibnamefont {Puppo}}, \bibinfo {author} {\bibfnamefont
  {M.}~\bibnamefont {P\"urrer}}, \bibinfo {author} {\bibfnamefont
  {J.}~\bibnamefont {Qin}}, \bibinfo {author} {\bibfnamefont {V.}~\bibnamefont
  {Quetschke}}, \bibinfo {author} {\bibfnamefont {E.}~\bibnamefont {Quintero}},
  \bibinfo {author} {\bibfnamefont {R.}~\bibnamefont {Quitzow-James}}, \bibinfo
  {author} {\bibfnamefont {F.~J.}\ \bibnamefont {Raab}}, \bibinfo {author}
  {\bibfnamefont {D.~S.}\ \bibnamefont {Rabeling}}, \bibinfo {author}
  {\bibfnamefont {I.}~\bibnamefont {R\'acz}}, \bibinfo {author} {\bibfnamefont
  {H.}~\bibnamefont {Radkins}}, \bibinfo {author} {\bibfnamefont
  {P.}~\bibnamefont {Raffai}}, \bibinfo {author} {\bibfnamefont
  {S.}~\bibnamefont {Raja}}, \bibinfo {author} {\bibfnamefont {G.}~\bibnamefont
  {Rajalakshmi}}, \bibinfo {author} {\bibfnamefont {M.}~\bibnamefont
  {Rakhmanov}}, \bibinfo {author} {\bibfnamefont {C.}~\bibnamefont {Ramet}},
  \bibinfo {author} {\bibfnamefont {K.}~\bibnamefont {Ramirez}}, \bibinfo
  {author} {\bibfnamefont {P.}~\bibnamefont {Rapagnani}}, \bibinfo {author}
  {\bibfnamefont {V.}~\bibnamefont {Raymond}}, \bibinfo {author} {\bibfnamefont
  {M.}~\bibnamefont {Razzano}}, \bibinfo {author} {\bibfnamefont
  {V.}~\bibnamefont {Re}}, \bibinfo {author} {\bibfnamefont {S.}~\bibnamefont
  {Recchia}}, \bibinfo {author} {\bibfnamefont {C.~M.}\ \bibnamefont {Reed}},
  \bibinfo {author} {\bibfnamefont {T.}~\bibnamefont {Regimbau}}, \bibinfo
  {author} {\bibfnamefont {S.}~\bibnamefont {Reid}}, \bibinfo {author}
  {\bibfnamefont {D.~H.}\ \bibnamefont {Reitze}}, \bibinfo {author}
  {\bibfnamefont {O.}~\bibnamefont {Reula}}, \bibinfo {author} {\bibfnamefont
  {E.}~\bibnamefont {Rhoades}}, \bibinfo {author} {\bibfnamefont
  {F.}~\bibnamefont {Ricci}}, \bibinfo {author} {\bibfnamefont
  {R.}~\bibnamefont {Riesen}}, \bibinfo {author} {\bibfnamefont
  {K.}~\bibnamefont {Riles}}, \bibinfo {author} {\bibfnamefont {N.~A.}\
  \bibnamefont {Robertson}}, \bibinfo {author} {\bibfnamefont {F.}~\bibnamefont
  {Robinet}}, \bibinfo {author} {\bibfnamefont {A.}~\bibnamefont {Rocchi}},
  \bibinfo {author} {\bibfnamefont {S.~B.}\ \bibnamefont {Roddy}}, \bibinfo
  {author} {\bibfnamefont {S.}~\bibnamefont {Rogstad}}, \bibinfo {author}
  {\bibfnamefont {L.}~\bibnamefont {Rolland}}, \bibinfo {author} {\bibfnamefont
  {J.~G.}\ \bibnamefont {Rollins}}, \bibinfo {author} {\bibfnamefont
  {R.}~\bibnamefont {Romano}}, \bibinfo {author} {\bibfnamefont
  {G.}~\bibnamefont {Romanov}}, \bibinfo {author} {\bibfnamefont {J.~H.}\
  \bibnamefont {Romie}}, \bibinfo {author} {\bibfnamefont {D.}~\bibnamefont
  {Rosi\ifmmode~\acute{n}\else \'{n}\fi{}ska}}, \bibinfo {author}
  {\bibfnamefont {S.}~\bibnamefont {Rowan}}, \bibinfo {author} {\bibfnamefont
  {A.}~\bibnamefont {R\"udiger}}, \bibinfo {author} {\bibfnamefont
  {P.}~\bibnamefont {Ruggi}}, \bibinfo {author} {\bibfnamefont
  {K.}~\bibnamefont {Ryan}}, \bibinfo {author} {\bibfnamefont {F.}~\bibnamefont
  {Salemi}}, \bibinfo {author} {\bibfnamefont {L.}~\bibnamefont {Sammut}},
  \bibinfo {author} {\bibfnamefont {V.}~\bibnamefont {Sandberg}}, \bibinfo
  {author} {\bibfnamefont {J.~R.}\ \bibnamefont {Sanders}}, \bibinfo {author}
  {\bibfnamefont {S.}~\bibnamefont {Sankar}}, \bibinfo {author} {\bibfnamefont
  {V.}~\bibnamefont {Sannibale}}, \bibinfo {author} {\bibfnamefont
  {I.}~\bibnamefont {Santiago-Prieto}}, \bibinfo {author} {\bibfnamefont
  {E.}~\bibnamefont {Saracco}}, \bibinfo {author} {\bibfnamefont
  {B.}~\bibnamefont {Sassolas}}, \bibinfo {author} {\bibfnamefont {B.~S.}\
  \bibnamefont {Sathyaprakash}}, \bibinfo {author} {\bibfnamefont {P.~R.}\
  \bibnamefont {Saulson}}, \bibinfo {author} {\bibfnamefont {R.}~\bibnamefont
  {Savage}}, \bibinfo {author} {\bibfnamefont {J.}~\bibnamefont {Scheuer}},
  \bibinfo {author} {\bibfnamefont {R.}~\bibnamefont {Schilling}}, \bibinfo
  {author} {\bibfnamefont {M.}~\bibnamefont {Schilman}}, \bibinfo {author}
  {\bibfnamefont {P.}~\bibnamefont {Schmidt}}, \bibinfo {author} {\bibfnamefont
  {R.}~\bibnamefont {Schnabel}}, \bibinfo {author} {\bibfnamefont {R.~M.~S.}\
  \bibnamefont {Schofield}}, \bibinfo {author} {\bibfnamefont {E.}~\bibnamefont
  {Schreiber}}, \bibinfo {author} {\bibfnamefont {D.}~\bibnamefont {Schuette}},
  \bibinfo {author} {\bibfnamefont {B.~F.}\ \bibnamefont {Schutz}}, \bibinfo
  {author} {\bibfnamefont {J.}~\bibnamefont {Scott}}, \bibinfo {author}
  {\bibfnamefont {S.~M.}\ \bibnamefont {Scott}}, \bibinfo {author}
  {\bibfnamefont {D.}~\bibnamefont {Sellers}}, \bibinfo {author} {\bibfnamefont
  {A.~S.}\ \bibnamefont {Sengupta}}, \bibinfo {author} {\bibfnamefont
  {D.}~\bibnamefont {Sentenac}}, \bibinfo {author} {\bibfnamefont
  {V.}~\bibnamefont {Sequino}}, \bibinfo {author} {\bibfnamefont
  {A.}~\bibnamefont {Sergeev}}, \bibinfo {author} {\bibfnamefont {D.~A.}\
  \bibnamefont {Shaddock}}, \bibinfo {author} {\bibfnamefont {S.}~\bibnamefont
  {Shah}}, \bibinfo {author} {\bibfnamefont {M.~S.}\ \bibnamefont {Shahriar}},
  \bibinfo {author} {\bibfnamefont {M.}~\bibnamefont {Shaltev}}, \bibinfo
  {author} {\bibfnamefont {Z.}~\bibnamefont {Shao}}, \bibinfo {author}
  {\bibfnamefont {B.}~\bibnamefont {Shapiro}}, \bibinfo {author} {\bibfnamefont
  {P.}~\bibnamefont {Shawhan}}, \bibinfo {author} {\bibfnamefont {D.~H.}\
  \bibnamefont {Shoemaker}}, \bibinfo {author} {\bibfnamefont {T.~L.}\
  \bibnamefont {Sidery}}, \bibinfo {author} {\bibfnamefont {K.}~\bibnamefont
  {Siellez}}, \bibinfo {author} {\bibfnamefont {X.}~\bibnamefont {Siemens}},
  \bibinfo {author} {\bibfnamefont {D.}~\bibnamefont {Sigg}}, \bibinfo {author}
  {\bibfnamefont {D.}~\bibnamefont {Simakov}}, \bibinfo {author} {\bibfnamefont
  {A.}~\bibnamefont {Singer}}, \bibinfo {author} {\bibfnamefont
  {L.}~\bibnamefont {Singer}}, \bibinfo {author} {\bibfnamefont
  {R.}~\bibnamefont {Singh}}, \bibinfo {author} {\bibfnamefont {A.~M.}\
  \bibnamefont {Sintes}}, \bibinfo {author} {\bibfnamefont {B.~J.~J.}\
  \bibnamefont {Slagmolen}}, \bibinfo {author} {\bibfnamefont {J.}~\bibnamefont
  {Slutsky}}, \bibinfo {author} {\bibfnamefont {J.~R.}\ \bibnamefont {Smith}},
  \bibinfo {author} {\bibfnamefont {M.~R.}\ \bibnamefont {Smith}}, \bibinfo
  {author} {\bibfnamefont {R.~J.~E.}\ \bibnamefont {Smith}}, \bibinfo {author}
  {\bibfnamefont {N.~D.}\ \bibnamefont {Smith-Lefebvre}}, \bibinfo {author}
  {\bibfnamefont {E.~J.}\ \bibnamefont {Son}}, \bibinfo {author} {\bibfnamefont
  {B.}~\bibnamefont {Sorazu}}, \bibinfo {author} {\bibfnamefont
  {T.}~\bibnamefont {Souradeep}}, \bibinfo {author} {\bibfnamefont
  {A.}~\bibnamefont {Staley}}, \bibinfo {author} {\bibfnamefont
  {J.}~\bibnamefont {Stebbins}}, \bibinfo {author} {\bibfnamefont
  {M.}~\bibnamefont {Steinke}}, \bibinfo {author} {\bibfnamefont
  {J.}~\bibnamefont {Steinlechner}}, \bibinfo {author} {\bibfnamefont
  {S.}~\bibnamefont {Steinlechner}}, \bibinfo {author} {\bibfnamefont {B.~C.}\
  \bibnamefont {Stephens}}, \bibinfo {author} {\bibfnamefont {S.}~\bibnamefont
  {Steplewski}}, \bibinfo {author} {\bibfnamefont {S.}~\bibnamefont
  {Stevenson}}, \bibinfo {author} {\bibfnamefont {R.}~\bibnamefont {Stone}},
  \bibinfo {author} {\bibfnamefont {D.}~\bibnamefont {Stops}}, \bibinfo
  {author} {\bibfnamefont {K.~A.}\ \bibnamefont {Strain}}, \bibinfo {author}
  {\bibfnamefont {N.}~\bibnamefont {Straniero}}, \bibinfo {author}
  {\bibfnamefont {S.}~\bibnamefont {Strigin}}, \bibinfo {author} {\bibfnamefont
  {R.}~\bibnamefont {Sturani}}, \bibinfo {author} {\bibfnamefont {A.~L.}\
  \bibnamefont {Stuver}}, \bibinfo {author} {\bibfnamefont {T.~Z.}\
  \bibnamefont {Summerscales}}, \bibinfo {author} {\bibfnamefont
  {S.}~\bibnamefont {Susmithan}}, \bibinfo {author} {\bibfnamefont {P.~J.}\
  \bibnamefont {Sutton}}, \bibinfo {author} {\bibfnamefont {B.}~\bibnamefont
  {Swinkels}}, \bibinfo {author} {\bibfnamefont {M.}~\bibnamefont {Tacca}},
  \bibinfo {author} {\bibfnamefont {D.}~\bibnamefont {Talukder}}, \bibinfo
  {author} {\bibfnamefont {D.~B.}\ \bibnamefont {Tanner}}, \bibinfo {author}
  {\bibfnamefont {J.}~\bibnamefont {Tao}}, \bibinfo {author} {\bibfnamefont
  {S.~P.}\ \bibnamefont {Tarabrin}}, \bibinfo {author} {\bibfnamefont
  {R.}~\bibnamefont {Taylor}}, \bibinfo {author} {\bibfnamefont
  {G.}~\bibnamefont {Tellez}}, \bibinfo {author} {\bibfnamefont {M.~P.}\
  \bibnamefont {Thirugnanasambandam}}, \bibinfo {author} {\bibfnamefont
  {M.}~\bibnamefont {Thomas}}, \bibinfo {author} {\bibfnamefont
  {P.}~\bibnamefont {Thomas}}, \bibinfo {author} {\bibfnamefont {K.~A.}\
  \bibnamefont {Thorne}}, \bibinfo {author} {\bibfnamefont {K.~S.}\
  \bibnamefont {Thorne}}, \bibinfo {author} {\bibfnamefont {E.}~\bibnamefont
  {Thrane}}, \bibinfo {author} {\bibfnamefont {V.}~\bibnamefont {Tiwari}},
  \bibinfo {author} {\bibfnamefont {K.~V.}\ \bibnamefont {Tokmakov}}, \bibinfo
  {author} {\bibfnamefont {C.}~\bibnamefont {Tomlinson}}, \bibinfo {author}
  {\bibfnamefont {M.}~\bibnamefont {Tonelli}}, \bibinfo {author} {\bibfnamefont
  {C.~V.}\ \bibnamefont {Torres}}, \bibinfo {author} {\bibfnamefont {C.~I.}\
  \bibnamefont {Torrie}}, \bibinfo {author} {\bibfnamefont {F.}~\bibnamefont
  {Travasso}}, \bibinfo {author} {\bibfnamefont {G.}~\bibnamefont {Traylor}},
  \bibinfo {author} {\bibfnamefont {M.}~\bibnamefont {Trias}}, \bibinfo
  {author} {\bibfnamefont {M.}~\bibnamefont {Tse}}, \bibinfo {author}
  {\bibfnamefont {D.}~\bibnamefont {Tshilumba}}, \bibinfo {author}
  {\bibfnamefont {H.}~\bibnamefont {Tuennermann}}, \bibinfo {author}
  {\bibfnamefont {D.}~\bibnamefont {Ugolini}}, \bibinfo {author} {\bibfnamefont
  {C.~S.}\ \bibnamefont {Unnikrishnan}}, \bibinfo {author} {\bibfnamefont
  {A.~L.}\ \bibnamefont {Urban}}, \bibinfo {author} {\bibfnamefont {S.~A.}\
  \bibnamefont {Usman}}, \bibinfo {author} {\bibfnamefont {H.}~\bibnamefont
  {Vahlbruch}}, \bibinfo {author} {\bibfnamefont {G.}~\bibnamefont {Vajente}},
  \bibinfo {author} {\bibfnamefont {G.}~\bibnamefont {Valdes}}, \bibinfo
  {author} {\bibfnamefont {M.}~\bibnamefont {Vallisneri}}, \bibinfo {author}
  {\bibfnamefont {M.}~\bibnamefont {van Beuzekom}}, \bibinfo {author}
  {\bibfnamefont {J.~F.~J.}\ \bibnamefont {van~den Brand}}, \bibinfo {author}
  {\bibfnamefont {C.}~\bibnamefont {Van Den~Broeck}}, \bibinfo {author}
  {\bibfnamefont {M.~V.}\ \bibnamefont {van~der Sluys}}, \bibinfo {author}
  {\bibfnamefont {J.}~\bibnamefont {van Heijningen}}, \bibinfo {author}
  {\bibfnamefont {A.~A.}\ \bibnamefont {van Veggel}}, \bibinfo {author}
  {\bibfnamefont {S.}~\bibnamefont {Vass}}, \bibinfo {author} {\bibfnamefont
  {M.}~\bibnamefont {Vas\'uth}}, \bibinfo {author} {\bibfnamefont
  {R.}~\bibnamefont {Vaulin}}, \bibinfo {author} {\bibfnamefont
  {A.}~\bibnamefont {Vecchio}}, \bibinfo {author} {\bibfnamefont
  {G.}~\bibnamefont {Vedovato}}, \bibinfo {author} {\bibfnamefont
  {J.}~\bibnamefont {Veitch}}, \bibinfo {author} {\bibfnamefont {P.~J.}\
  \bibnamefont {Veitch}}, \bibinfo {author} {\bibfnamefont {K.}~\bibnamefont
  {Venkateswara}}, \bibinfo {author} {\bibfnamefont {D.}~\bibnamefont
  {Verkindt}}, \bibinfo {author} {\bibfnamefont {F.}~\bibnamefont {Vetrano}},
  \bibinfo {author} {\bibfnamefont {A.}~\bibnamefont {Vicer\'e}}, \bibinfo
  {author} {\bibfnamefont {R.}~\bibnamefont {Vincent-Finley}}, \bibinfo
  {author} {\bibfnamefont {J.-Y.}\ \bibnamefont {Vinet}}, \bibinfo {author}
  {\bibfnamefont {S.}~\bibnamefont {Vitale}}, \bibinfo {author} {\bibfnamefont
  {T.}~\bibnamefont {Vo}}, \bibinfo {author} {\bibfnamefont {H.}~\bibnamefont
  {Vocca}}, \bibinfo {author} {\bibfnamefont {C.}~\bibnamefont {Vorvick}},
  \bibinfo {author} {\bibfnamefont {W.~D.}\ \bibnamefont {Vousden}}, \bibinfo
  {author} {\bibfnamefont {S.~P.}\ \bibnamefont {Vyachanin}}, \bibinfo {author}
  {\bibfnamefont {A.~R.}\ \bibnamefont {Wade}}, \bibinfo {author}
  {\bibfnamefont {L.}~\bibnamefont {Wade}}, \bibinfo {author} {\bibfnamefont
  {M.}~\bibnamefont {Wade}}, \bibinfo {author} {\bibfnamefont {M.}~\bibnamefont
  {Walker}}, \bibinfo {author} {\bibfnamefont {L.}~\bibnamefont {Wallace}},
  \bibinfo {author} {\bibfnamefont {S.}~\bibnamefont {Walsh}}, \bibinfo
  {author} {\bibfnamefont {M.}~\bibnamefont {Wang}}, \bibinfo {author}
  {\bibfnamefont {X.}~\bibnamefont {Wang}}, \bibinfo {author} {\bibfnamefont
  {R.~L.}\ \bibnamefont {Ward}}, \bibinfo {author} {\bibfnamefont
  {M.}~\bibnamefont {Was}}, \bibinfo {author} {\bibfnamefont {B.}~\bibnamefont
  {Weaver}}, \bibinfo {author} {\bibfnamefont {L.-W.}\ \bibnamefont {Wei}},
  \bibinfo {author} {\bibfnamefont {M.}~\bibnamefont {Weinert}}, \bibinfo
  {author} {\bibfnamefont {A.~J.}\ \bibnamefont {Weinstein}}, \bibinfo {author}
  {\bibfnamefont {R.}~\bibnamefont {Weiss}}, \bibinfo {author} {\bibfnamefont
  {T.}~\bibnamefont {Welborn}}, \bibinfo {author} {\bibfnamefont
  {L.}~\bibnamefont {Wen}}, \bibinfo {author} {\bibfnamefont {P.}~\bibnamefont
  {Wessels}}, \bibinfo {author} {\bibfnamefont {M.}~\bibnamefont {West}},
  \bibinfo {author} {\bibfnamefont {T.}~\bibnamefont {Westphal}}, \bibinfo
  {author} {\bibfnamefont {K.}~\bibnamefont {Wette}}, \bibinfo {author}
  {\bibfnamefont {J.~T.}\ \bibnamefont {Whelan}}, \bibinfo {author}
  {\bibfnamefont {D.~J.}\ \bibnamefont {White}}, \bibinfo {author}
  {\bibfnamefont {B.~F.}\ \bibnamefont {Whiting}}, \bibinfo {author}
  {\bibfnamefont {K.}~\bibnamefont {Wiesner}}, \bibinfo {author} {\bibfnamefont
  {C.}~\bibnamefont {Wilkinson}}, \bibinfo {author} {\bibfnamefont
  {K.}~\bibnamefont {Williams}}, \bibinfo {author} {\bibfnamefont
  {L.}~\bibnamefont {Williams}}, \bibinfo {author} {\bibfnamefont
  {R.}~\bibnamefont {Williams}}, \bibinfo {author} {\bibfnamefont {T.~D.}\
  \bibnamefont {Williams}}, \bibinfo {author} {\bibfnamefont {A.~R.}\
  \bibnamefont {Williamson}}, \bibinfo {author} {\bibfnamefont {J.~L.}\
  \bibnamefont {Willis}}, \bibinfo {author} {\bibfnamefont {B.}~\bibnamefont
  {Willke}}, \bibinfo {author} {\bibfnamefont {M.}~\bibnamefont {Wimmer}},
  \bibinfo {author} {\bibfnamefont {W.}~\bibnamefont {Winkler}}, \bibinfo
  {author} {\bibfnamefont {C.~C.}\ \bibnamefont {Wipf}}, \bibinfo {author}
  {\bibfnamefont {A.~G.}\ \bibnamefont {Wiseman}}, \bibinfo {author}
  {\bibfnamefont {H.}~\bibnamefont {Wittel}}, \bibinfo {author} {\bibfnamefont
  {G.}~\bibnamefont {Woan}}, \bibinfo {author} {\bibfnamefont {N.}~\bibnamefont
  {Wolovick}}, \bibinfo {author} {\bibfnamefont {J.}~\bibnamefont {Worden}},
  \bibinfo {author} {\bibfnamefont {Y.}~\bibnamefont {Wu}}, \bibinfo {author}
  {\bibfnamefont {J.}~\bibnamefont {Yablon}}, \bibinfo {author} {\bibfnamefont
  {I.}~\bibnamefont {Yakushin}}, \bibinfo {author} {\bibfnamefont
  {W.}~\bibnamefont {Yam}}, \bibinfo {author} {\bibfnamefont {H.}~\bibnamefont
  {Yamamoto}}, \bibinfo {author} {\bibfnamefont {C.~C.}\ \bibnamefont
  {Yancey}}, \bibinfo {author} {\bibfnamefont {H.}~\bibnamefont {Yang}},
  \bibinfo {author} {\bibfnamefont {S.}~\bibnamefont {Yoshida}}, \bibinfo
  {author} {\bibfnamefont {M.}~\bibnamefont {Yvert}}, \bibinfo {author}
  {\bibfnamefont {A.}~\bibnamefont {Zadro\ifmmode~\dot{z}\else \.{z}\fi{}ny}},
  \bibinfo {author} {\bibfnamefont {M.}~\bibnamefont {Zanolin}}, \bibinfo
  {author} {\bibfnamefont {J.-P.}\ \bibnamefont {Zendri}}, \bibinfo {author}
  {\bibfnamefont {F.}~\bibnamefont {Zhang}}, \bibinfo {author} {\bibfnamefont
  {L.}~\bibnamefont {Zhang}}, \bibinfo {author} {\bibfnamefont
  {C.}~\bibnamefont {Zhao}}, \bibinfo {author} {\bibfnamefont {H.}~\bibnamefont
  {Zhu}}, \bibinfo {author} {\bibfnamefont {X.~J.}\ \bibnamefont {Zhu}},
  \bibinfo {author} {\bibfnamefont {M.~E.}\ \bibnamefont {Zucker}}, \bibinfo
  {author} {\bibfnamefont {S.}~\bibnamefont {Zuraw}},\ and\ \bibinfo {author}
  {\bibfnamefont {J.}~\bibnamefont {Zweizig}} (\bibinfo {collaboration} {LIGO
  Scientific Collaboration and Virgo Collaboration}),\ }\bibfield  {title}
  {\bibinfo {title} {Methods and results of a search for gravitational waves
  associated with gamma-ray bursts using the geo 600, ligo, and virgo
  detectors},\ }\href {https://doi.org/10.1103/PhysRevD.89.122004} {\bibfield
  {journal} {\bibinfo  {journal} {Phys. Rev. D}\ }\textbf {\bibinfo {volume}
  {89}},\ \bibinfo {pages} {122004} (\bibinfo {year} {2014})}\BibitemShut
  {NoStop}%
\bibitem [{\citenamefont {{Moore}}\ \emph {et~al.}(2015)\citenamefont
  {{Moore}}, \citenamefont {{Cole}},\ and\ \citenamefont
  {{Berry}}}]{Moore_2015}%
  \BibitemOpen
  \bibfield  {author} {\bibinfo {author} {\bibfnamefont {C.~J.}\ \bibnamefont
  {{Moore}}}, \bibinfo {author} {\bibfnamefont {R.~H.}\ \bibnamefont
  {{Cole}}},\ and\ \bibinfo {author} {\bibfnamefont {C.~P.~L.}\ \bibnamefont
  {{Berry}}},\ }\bibfield  {title} {\bibinfo {title} {{Gravitational-wave
  sensitivity curves}},\ }\href {https://doi.org/10.1088/0264-9381/32/1/015014}
  {\bibfield  {journal} {\bibinfo  {journal} {Classical and Quantum Gravity}\
  }\textbf {\bibinfo {volume} {32}},\ \bibinfo {eid} {015014} (\bibinfo {year}
  {2015})},\ \Eprint {https://arxiv.org/abs/1408.0740} {arXiv:1408.0740
  [gr-qc]} \BibitemShut {NoStop}%
\bibitem [{\citenamefont {{Harry}}(2012)}]{2012mgm..conf..628H}%
  \BibitemOpen
  \bibfield  {author} {\bibinfo {author} {\bibfnamefont {G.~M.}\ \bibnamefont
  {{Harry}}},\ }\bibfield  {title} {\bibinfo {title} {{Second Generation
  Gravitational Wave Detectors}},\ }in\ \href
  {https://doi.org/10.1142/9789814374552_0032} {\emph {\bibinfo {booktitle}
  {Twelfth Marcel Grossmann Meeting on General Relativity}}},\ \bibinfo
  {editor} {edited by\ \bibinfo {editor} {\bibfnamefont {A.~H.}\ \bibnamefont
  {{Chamseddine}}}}\ (\bibinfo {year} {2012})\ pp.\ \bibinfo {pages}
  {628--644}\BibitemShut {NoStop}%
\bibitem [{\citenamefont {{Dooley}}\ \emph {et~al.}(2016)\citenamefont
  {{Dooley}}, \citenamefont {{Leong}}, \citenamefont {{Adams}}, \citenamefont
  {{Affeldt}}, \citenamefont {{Bisht}}, \citenamefont {{Bogan}}, \citenamefont
  {{Degallaix}}, \citenamefont {{Gr{\"a}f}}, \citenamefont {{Hild}},
  \citenamefont {{Hough}}, \citenamefont {{Khalaidovski}}, \citenamefont
  {{Lastzka}}, \citenamefont {{Lough}}, \citenamefont {{L{\"u}ck}},
  \citenamefont {{Macleod}}, \citenamefont {{Nuttall}}, \citenamefont
  {{Prijatelj}}, \citenamefont {{Schnabel}}, \citenamefont {{Schreiber}},
  \citenamefont {{Slutsky}}, \citenamefont {{Sorazu}}, \citenamefont
  {{Strain}}, \citenamefont {{Vahlbruch}}, \citenamefont {{W{\k{a}}s}},
  \citenamefont {{Willke}}, \citenamefont {{Wittel}}, \citenamefont
  {{Danzmann}},\ and\ \citenamefont {{Grote}}}]{2016CQGra..33g5009D}%
  \BibitemOpen
  \bibfield  {author} {\bibinfo {author} {\bibfnamefont {K.~L.}\ \bibnamefont
  {{Dooley}}}, \bibinfo {author} {\bibfnamefont {J.~R.}\ \bibnamefont
  {{Leong}}}, \bibinfo {author} {\bibfnamefont {T.}~\bibnamefont {{Adams}}},
  \bibinfo {author} {\bibfnamefont {C.}~\bibnamefont {{Affeldt}}}, \bibinfo
  {author} {\bibfnamefont {A.}~\bibnamefont {{Bisht}}}, \bibinfo {author}
  {\bibfnamefont {C.}~\bibnamefont {{Bogan}}}, \bibinfo {author} {\bibfnamefont
  {J.}~\bibnamefont {{Degallaix}}}, \bibinfo {author} {\bibfnamefont
  {C.}~\bibnamefont {{Gr{\"a}f}}}, \bibinfo {author} {\bibfnamefont
  {S.}~\bibnamefont {{Hild}}}, \bibinfo {author} {\bibfnamefont
  {J.}~\bibnamefont {{Hough}}}, \bibinfo {author} {\bibfnamefont
  {A.}~\bibnamefont {{Khalaidovski}}}, \bibinfo {author} {\bibfnamefont
  {N.}~\bibnamefont {{Lastzka}}}, \bibinfo {author} {\bibfnamefont
  {J.}~\bibnamefont {{Lough}}}, \bibinfo {author} {\bibfnamefont
  {H.}~\bibnamefont {{L{\"u}ck}}}, \bibinfo {author} {\bibfnamefont
  {D.}~\bibnamefont {{Macleod}}}, \bibinfo {author} {\bibfnamefont
  {L.}~\bibnamefont {{Nuttall}}}, \bibinfo {author} {\bibfnamefont
  {M.}~\bibnamefont {{Prijatelj}}}, \bibinfo {author} {\bibfnamefont
  {R.}~\bibnamefont {{Schnabel}}}, \bibinfo {author} {\bibfnamefont
  {E.}~\bibnamefont {{Schreiber}}}, \bibinfo {author} {\bibfnamefont
  {J.}~\bibnamefont {{Slutsky}}}, \bibinfo {author} {\bibfnamefont
  {B.}~\bibnamefont {{Sorazu}}}, \bibinfo {author} {\bibfnamefont {K.~A.}\
  \bibnamefont {{Strain}}}, \bibinfo {author} {\bibfnamefont {H.}~\bibnamefont
  {{Vahlbruch}}}, \bibinfo {author} {\bibfnamefont {M.}~\bibnamefont
  {{W{\k{a}}s}}}, \bibinfo {author} {\bibfnamefont {B.}~\bibnamefont
  {{Willke}}}, \bibinfo {author} {\bibfnamefont {H.}~\bibnamefont {{Wittel}}},
  \bibinfo {author} {\bibfnamefont {K.}~\bibnamefont {{Danzmann}}},\ and\
  \bibinfo {author} {\bibfnamefont {H.}~\bibnamefont {{Grote}}},\ }\bibfield
  {title} {\bibinfo {title} {{GEO 600 and the GEO-HF upgrade program: successes
  and challenges}},\ }\href {https://doi.org/10.1088/0264-9381/33/7/075009}
  {\bibfield  {journal} {\bibinfo  {journal} {Classical and Quantum Gravity}\
  }\textbf {\bibinfo {volume} {33}},\ \bibinfo {eid} {075009} (\bibinfo {year}
  {2016})},\ \Eprint {https://arxiv.org/abs/1510.00317} {arXiv:1510.00317
  [physics.ins-det]} \BibitemShut {NoStop}%
\bibitem [{\citenamefont {{LIGO Scientific Collaboration}}\ \emph
  {et~al.}(2015)\citenamefont {{LIGO Scientific Collaboration}}, \citenamefont
  {{Aasi}}, \citenamefont {{Abbott}}, \citenamefont {{Abbott}}, \citenamefont
  {{Abbott}}, \citenamefont {{Abernathy}}, \citenamefont {{Ackley}},
  \citenamefont {{Adams}}, \citenamefont {{Adams}}, \citenamefont {{Addesso}},
  \citenamefont {{Adhikari}}, \citenamefont {{Adya}}, \citenamefont
  {{Affeldt}}, \citenamefont {{Aggarwal}}, \citenamefont {{Aguiar}},
  \citenamefont {{Ain}}, \citenamefont {{Ajith}}, \citenamefont {{Alemic}},
  \citenamefont {{Allen}}, \citenamefont {{Amariutei}}, \citenamefont
  {{Anderson}}, \citenamefont {{Anderson}}, \citenamefont {{Arai}},
  \citenamefont {{Araya}}, \citenamefont {{Arceneaux}}, \citenamefont
  {{Areeda}}, \citenamefont {{Ashton}}, \citenamefont {{Ast}}, \citenamefont
  {{Aston}}, \citenamefont {{Aufmuth}}, \citenamefont {{Aulbert}},
  \citenamefont {{Aylott}}, \citenamefont {{Babak}}, \citenamefont {{Baker}},
  \citenamefont {{Ballmer}}, \citenamefont {{Barayoga}}, \citenamefont
  {{Barbet}}, \citenamefont {{Barclay}}, \citenamefont {{Barish}},
  \citenamefont {{Barker}}, \citenamefont {{Barr}}, \citenamefont {{Barsotti}},
  \citenamefont {{Bartlett}}, \citenamefont {{Barton}}, \citenamefont
  {{Bartos}}, \citenamefont {{Bassiri}}, \citenamefont {{Batch}}, \citenamefont
  {{Baune}}, \citenamefont {{Behnke}}, \citenamefont {{Bell}}, \citenamefont
  {{Bell}}, \citenamefont {{Benacquista}}, \citenamefont {{Bergman}},
  \citenamefont {{Bergmann}}, \citenamefont {{Berry}}, \citenamefont
  {{Betzwieser}}, \citenamefont {{Bhagwat}}, \citenamefont {{Bhandare}},
  \citenamefont {{Bilenko}}, \citenamefont {{Billingsley}}, \citenamefont
  {{Birch}}, \citenamefont {{Biscans}}, \citenamefont {{Biwer}}, \citenamefont
  {{Blackburn}}, \citenamefont {{Blackburn}}, \citenamefont {{Blair}},
  \citenamefont {{Blair}}, \citenamefont {{Bock}}, \citenamefont {{Bodiya}},
  \citenamefont {{Bojtos}}, \citenamefont {{Bond}}, \citenamefont {{Bork}},
  \citenamefont {{Born}}, \citenamefont {{Bose}}, \citenamefont {{Brady}},
  \citenamefont {{Braginsky}}, \citenamefont {{Brau}}, \citenamefont
  {{Bridges}}, \citenamefont {{Brinkmann}}, \citenamefont {{Brooks}},
  \citenamefont {{Brown}}, \citenamefont {{Brown}}, \citenamefont {{Brown}},
  \citenamefont {{Buchman}}, \citenamefont {{Buikema}}, \citenamefont
  {{Buonanno}}, \citenamefont {{Cadonati}}, \citenamefont {{Calder{\'o}n
  Bustillo}}, \citenamefont {{Camp}}, \citenamefont {{Cannon}}, \citenamefont
  {{Cao}}, \citenamefont {{Capano}}, \citenamefont {{Caride}}, \citenamefont
  {{Caudill}}, \citenamefont {{Cavagli{\`a}}}, \citenamefont {{Cepeda}},
  \citenamefont {{Chakraborty}}, \citenamefont {{Chalermsongsak}},
  \citenamefont {{Chamberlin}}, \citenamefont {{Chao}}, \citenamefont
  {{Charlton}}, \citenamefont {{Chen}}, \citenamefont {{Cho}}, \citenamefont
  {{Cho}}, \citenamefont {{Chow}}, \citenamefont {{Christensen}}, \citenamefont
  {{Chu}}, \citenamefont {{Chung}}, \citenamefont {{Ciani}}, \citenamefont
  {{Clara}}, \citenamefont {{Clark}}, \citenamefont {{Collette}}, \citenamefont
  {{Cominsky}}, \citenamefont {{Constancio}}, \citenamefont {{Cook}},
  \citenamefont {{Corbitt}}, \citenamefont {{Cornish}}, \citenamefont
  {{Corsi}}, \citenamefont {{Costa}}, \citenamefont {{Coughlin}}, \citenamefont
  {{Countryman}}, \citenamefont {{Couvares}}, \citenamefont {{Coward}},
  \citenamefont {{Cowart}}, \citenamefont {{Coyne}}, \citenamefont {{Coyne}},
  \citenamefont {{Craig}}, \citenamefont {{Creighton}}, \citenamefont
  {{Creighton}}, \citenamefont {{Cripe}}, \citenamefont {{Crowder}},
  \citenamefont {{Cumming}}, \citenamefont {{Cunningham}}, \citenamefont
  {{Cutler}}, \citenamefont {{Dahl}}, \citenamefont {{Dal Canton}},
  \citenamefont {{Damjanic}}, \citenamefont {{Danilishin}}, \citenamefont
  {{Danzmann}}, \citenamefont {{Dartez}}, \citenamefont {{Dave}}, \citenamefont
  {{Daveloza}}, \citenamefont {{Davies}}, \citenamefont {{Daw}}, \citenamefont
  {{DeBra}}, \citenamefont {{Del Pozzo}}, \citenamefont {{Denker}},
  \citenamefont {{Dent}}, \citenamefont {{Dergachev}}, \citenamefont
  {{DeRosa}}, \citenamefont {{DeSalvo}}, \citenamefont {{Dhurandhar}},
  \citenamefont {{D{\textasciiacute}{\i}az}}, \citenamefont {{Di Palma}},
  \citenamefont {{Dojcinoski}}, \citenamefont {{Dominguez}}, \citenamefont
  {{Donovan}}, \citenamefont {{Dooley}}, \citenamefont {{Doravari}},
  \citenamefont {{Douglas}}, \citenamefont {{Downes}}, \citenamefont
  {{Driggers}}, \citenamefont {{Du}}, \citenamefont {{Dwyer}}, \citenamefont
  {{Eberle}}, \citenamefont {{Edo}}, \citenamefont {{Edwards}}, \citenamefont
  {{Edwards}}, \citenamefont {{Effler}}, \citenamefont {{Eggenstein}},
  \citenamefont {{Ehrens}}, \citenamefont {{Eichholz}}, \citenamefont
  {{Eikenberry}}, \citenamefont {{Essick}}, \citenamefont {{Etzel}},
  \citenamefont {{Evans}}, \citenamefont {{Evans}}, \citenamefont
  {{Factourovich}}, \citenamefont {{Fairhurst}}, \citenamefont {{Fan}},
  \citenamefont {{Fang}}, \citenamefont {{Farr}}, \citenamefont {{Farr}},
  \citenamefont {{Favata}}, \citenamefont {{Fays}}, \citenamefont {{Fehrmann}},
  \citenamefont {{Fejer}}, \citenamefont {{Feldbaum}}, \citenamefont
  {{Ferreira}}, \citenamefont {{Fisher}}, \citenamefont {{Frei}}, \citenamefont
  {{Freise}}, \citenamefont {{Frey}}, \citenamefont {{Fricke}}, \citenamefont
  {{Fritschel}}, \citenamefont {{Frolov}}, \citenamefont {{Fuentes-Tapia}},
  \citenamefont {{Fulda}}, \citenamefont {{Fyffe}}, \citenamefont {{Gair}},
  \citenamefont {{Gaonkar}}, \citenamefont {{Gehrels}}, \citenamefont
  {{Gergely}}, \citenamefont {{Giaime}}, \citenamefont {{Giardina}},
  \citenamefont {{Gleason}}, \citenamefont {{Goetz}}, \citenamefont {{Goetz}},
  \citenamefont {{Gondan}}, \citenamefont {{Gonz{\'a}lez}}, \citenamefont
  {{Gordon}}, \citenamefont {{Gorodetsky}}, \citenamefont {{Gossan}},
  \citenamefont {{Go{\ss}ler}}, \citenamefont {{Gr{\"a}f}}, \citenamefont
  {{Graff}}, \citenamefont {{Grant}}, \citenamefont {{Gras}}, \citenamefont
  {{Gray}}, \citenamefont {{Greenhalgh}}, \citenamefont {{Gretarsson}},
  \citenamefont {{Grote}}, \citenamefont {{Grunewald}}, \citenamefont
  {{Guido}}, \citenamefont {{Guo}}, \citenamefont {{Gushwa}}, \citenamefont
  {{Gustafson}}, \citenamefont {{Gustafson}}, \citenamefont {{Hacker}},
  \citenamefont {{Hall}}, \citenamefont {{Hammond}}, \citenamefont {{Hanke}},
  \citenamefont {{Hanks}}, \citenamefont {{Hanna}}, \citenamefont {{Hannam}},
  \citenamefont {{Hanson}}, \citenamefont {{Hardwick}}, \citenamefont
  {{Harry}}, \citenamefont {{Harry}}, \citenamefont {{Hart}}, \citenamefont
  {{Hartman}}, \citenamefont {{Haster}}, \citenamefont {{Haughian}},
  \citenamefont {{Hee}}, \citenamefont {{Heintze}}, \citenamefont {{Heinzel}},
  \citenamefont {{Hendry}}, \citenamefont {{Heng}}, \citenamefont
  {{Heptonstall}}, \citenamefont {{Heurs}}, \citenamefont {{Hewitson}},
  \citenamefont {{Hild}}, \citenamefont {{Hoak}}, \citenamefont {{Hodge}},
  \citenamefont {{Hollitt}}, \citenamefont {{Holt}}, \citenamefont {{Hopkins}},
  \citenamefont {{Hosken}}, \citenamefont {{Hough}}, \citenamefont {{Houston}},
  \citenamefont {{Howell}}, \citenamefont {{Hu}}, \citenamefont {{Huerta}},
  \citenamefont {{Hughey}}, \citenamefont {{Husa}}, \citenamefont {{Huttner}},
  \citenamefont {{Huynh}}, \citenamefont {{Huynh-Dinh}}, \citenamefont
  {{Idrisy}}, \citenamefont {{Indik}}, \citenamefont {{Ingram}}, \citenamefont
  {{Inta}}, \citenamefont {{Islas}}, \citenamefont {{Isler}}, \citenamefont
  {{Isogai}}, \citenamefont {{Iyer}}, \citenamefont {{Izumi}}, \citenamefont
  {{Jacobson}}, \citenamefont {{Jang}}, \citenamefont {{Jawahar}},
  \citenamefont {{Ji}}, \citenamefont {{Jim{\'e}nez-Forteza}}, \citenamefont
  {{Johnson}}, \citenamefont {{Jones}}, \citenamefont {{Jones}}, \citenamefont
  {{Ju}}, \citenamefont {{Haris}}, \citenamefont {{Kalogera}}, \citenamefont
  {{Kandhasamy}}, \citenamefont {{Kang}}, \citenamefont {{Kanner}},
  \citenamefont {{Katsavounidis}}, \citenamefont {{Katzman}}, \citenamefont
  {{Kaufer}}, \citenamefont {{Kaufer}}, \citenamefont {{Kaur}}, \citenamefont
  {{Kawabe}}, \citenamefont {{Kawazoe}}, \citenamefont {{Keiser}},
  \citenamefont {{Keitel}}, \citenamefont {{Kelley}}, \citenamefont {{Kells}},
  \citenamefont {{Keppel}}, \citenamefont {{Key}}, \citenamefont
  {{Khalaidovski}}, \citenamefont {{Khalili}}, \citenamefont {{Khazanov}},
  \citenamefont {{Kim}}, \citenamefont {{Kim}}, \citenamefont {{Kim}},
  \citenamefont {{Kim}}, \citenamefont {{Kim}}, \citenamefont {{King}},
  \citenamefont {{King}}, \citenamefont {{Kinzel}}, \citenamefont {{Kissel}},
  \citenamefont {{Klimenko}}, \citenamefont {{Kline}}, \citenamefont
  {{Koehlenbeck}}, \citenamefont {{Kokeyama}}, \citenamefont {{Kondrashov}},
  \citenamefont {{Korobko}}, \citenamefont {{Korth}}, \citenamefont {{Kozak}},
  \citenamefont {{Kringel}}, \citenamefont {{Krishnan}}, \citenamefont
  {{Krueger}}, \citenamefont {{Kuehn}}, \citenamefont {{Kumar}}, \citenamefont
  {{Kumar}}, \citenamefont {{Kuo}}, \citenamefont {{Landry}}, \citenamefont
  {{Lantz}}, \citenamefont {{Larson}}, \citenamefont {{Lasky}}, \citenamefont
  {{Lazzarini}}, \citenamefont {{Lazzaro}}, \citenamefont {{Le}}, \citenamefont
  {{Leaci}}, \citenamefont {{Leavey}}, \citenamefont {{Lebigot}}, \citenamefont
  {{Lee}}, \citenamefont {{Lee}}, \citenamefont {{Lee}}, \citenamefont
  {{Leong}}, \citenamefont {{Levin}}, \citenamefont {{Levine}}, \citenamefont
  {{Lewis}}, \citenamefont {{Li}}, \citenamefont {{Libbrecht}}, \citenamefont
  {{Libson}}, \citenamefont {{Lin}}, \citenamefont {{Littenberg}},
  \citenamefont {{Lockerbie}}, \citenamefont {{Lockett}}, \citenamefont
  {{Logue}}, \citenamefont {{Lombardi}}, \citenamefont {{Lormand}},
  \citenamefont {{Lough}}, \citenamefont {{Lubinski}}, \citenamefont
  {{L{\"u}ck}}, \citenamefont {{Lundgren}}, \citenamefont {{Lynch}},
  \citenamefont {{Ma}}, \citenamefont {{Macarthur}}, \citenamefont
  {{MacDonald}}, \citenamefont {{Machenschalk}}, \citenamefont {{MacInnis}},
  \citenamefont {{Macleod}}, \citenamefont {{Maga{\~n}a-Sandoval}},
  \citenamefont {{Magee}}, \citenamefont {{Mageswaran}}, \citenamefont
  {{Maglione}}, \citenamefont {{Mailand}}, \citenamefont {{Mandel}},
  \citenamefont {{Mandic}}, \citenamefont {{Mangano}}, \citenamefont
  {{Mansell}}, \citenamefont {{M{\'a}rka}}, \citenamefont {{M{\'a}rka}},
  \citenamefont {{Markosyan}}, \citenamefont {{Maros}}, \citenamefont
  {{Martin}}, \citenamefont {{Martin}}, \citenamefont {{Martynov}},
  \citenamefont {{Marx}}, \citenamefont {{Mason}}, \citenamefont {{Massinger}},
  \citenamefont {{Matichard}}, \citenamefont {{Matone}}, \citenamefont
  {{Mavalvala}}, \citenamefont {{Mazumder}}, \citenamefont {{Mazzolo}},
  \citenamefont {{McCarthy}}, \citenamefont {{McClelland}}, \citenamefont
  {{McCormick}}, \citenamefont {{McGuire}}, \citenamefont {{McIntyre}},
  \citenamefont {{McIver}}, \citenamefont {{McLin}}, \citenamefont
  {{McWilliams}}, \citenamefont {{Meadors}}, \citenamefont {{Meinders}},
  \citenamefont {{Melatos}}, \citenamefont {{Mendell}}, \citenamefont
  {{Mercer}}, \citenamefont {{Meshkov}}, \citenamefont {{Messenger}},
  \citenamefont {{Meyers}}, \citenamefont {{Miao}}, \citenamefont
  {{Middleton}}, \citenamefont {{Mikhailov}}, \citenamefont {{Miller}},
  \citenamefont {{Miller}}, \citenamefont {{Millhouse}}, \citenamefont
  {{Ming}}, \citenamefont {{Mirshekari}}, \citenamefont {{Mishra}},
  \citenamefont {{Mitra}}, \citenamefont {{Mitrofanov}}, \citenamefont
  {{Mitselmakher}}, \citenamefont {{Mittleman}}, \citenamefont {{Moe}},
  \citenamefont {{Mohanty}}, \citenamefont {{Mohapatra}}, \citenamefont
  {{Moore}}, \citenamefont {{Moraru}}, \citenamefont {{Moreno}}, \citenamefont
  {{Morriss}}, \citenamefont {{Mossavi}}, \citenamefont {{Mow-Lowry}},
  \citenamefont {{Mueller}}, \citenamefont {{Mueller}}, \citenamefont
  {{Mukherjee}}, \citenamefont {{Mullavey}}, \citenamefont {{Munch}},
  \citenamefont {{Murphy}}, \citenamefont {{Murray}}, \citenamefont
  {{Mytidis}}, \citenamefont {{Nash}}, \citenamefont {{Nayak}}, \citenamefont
  {{Necula}}, \citenamefont {{Nedkova}}, \citenamefont {{Newton}},
  \citenamefont {{Nguyen}}, \citenamefont {{Nielsen}}, \citenamefont
  {{Nissanke}}, \citenamefont {{Nitz}}, \citenamefont {{Nolting}},
  \citenamefont {{Normandin}}, \citenamefont {{Nuttall}}, \citenamefont
  {{Ochsner}}, \citenamefont {{O'Dell}}, \citenamefont {{Oelker}},
  \citenamefont {{Ogin}}, \citenamefont {{Oh}}, \citenamefont {{Oh}},
  \citenamefont {{Ohme}}, \citenamefont {{Oppermann}}, \citenamefont {{Oram}},
  \citenamefont {{O'Reilly}}, \citenamefont {{Ortega}}, \citenamefont
  {{O'Shaughnessy}}, \citenamefont {{Osthelder}}, \citenamefont {{Ott}},
  \citenamefont {{Ottaway}}, \citenamefont {{Ottens}}, \citenamefont
  {{Overmier}}, \citenamefont {{Owen}}, \citenamefont {{Padilla}},
  \citenamefont {{Pai}}, \citenamefont {{Pai}}, \citenamefont {{Palashov}},
  \citenamefont {{Pal-Singh}}, \citenamefont {{Pan}}, \citenamefont {{Pankow}},
  \citenamefont {{Pannarale}}, \citenamefont {{Pant}}, \citenamefont {{Papa}},
  \citenamefont {{Paris}}, \citenamefont {{Patrick}}, \citenamefont
  {{Pedraza}}, \citenamefont {{Pekowsky}}, \citenamefont {{Pele}},
  \citenamefont {{Penn}}, \citenamefont {{Perreca}}, \citenamefont {{Phelps}},
  \citenamefont {{Pierro}}, \citenamefont {{Pinto}}, \citenamefont {{Pitkin}},
  \citenamefont {{Poeld}}, \citenamefont {{Post}}, \citenamefont {{Poteomkin}},
  \citenamefont {{Powell}}, \citenamefont {{Prasad}}, \citenamefont {{Predoi}},
  \citenamefont {{Premachandra}}, \citenamefont {{Prestegard}}, \citenamefont
  {{Price}}, \citenamefont {{Principe}}, \citenamefont {{Privitera}},
  \citenamefont {{Prix}}, \citenamefont {{Prokhorov}}, \citenamefont
  {{Puncken}}, \citenamefont {{P{\"u}rrer}}, \citenamefont {{Qin}},
  \citenamefont {{Quetschke}}, \citenamefont {{Quintero}}, \citenamefont
  {{Quiroga}}, \citenamefont {{Quitzow-James}}, \citenamefont {{Raab}},
  \citenamefont {{Rabeling}}, \citenamefont {{Radkins}}, \citenamefont
  {{Raffai}}, \citenamefont {{Raja}}, \citenamefont {{Rajalakshmi}},
  \citenamefont {{Rakhmanov}}, \citenamefont {{Ramirez}}, \citenamefont
  {{Raymond}}, \citenamefont {{Reed}}, \citenamefont {{Reid}}, \citenamefont
  {{Reitze}}, \citenamefont {{Reula}}, \citenamefont {{Riles}}, \citenamefont
  {{Robertson}}, \citenamefont {{Robie}}, \citenamefont {{Rollins}},
  \citenamefont {{Roma}}, \citenamefont {{Romano}}, \citenamefont {{Romanov}},
  \citenamefont {{Romie}}, \citenamefont {{Rowan}}, \citenamefont
  {{R{\"u}diger}}, \citenamefont {{Ryan}}, \citenamefont {{Sachdev}},
  \citenamefont {{Sadecki}}, \citenamefont {{Sadeghian}}, \citenamefont
  {{Saleem}}, \citenamefont {{Salemi}}, \citenamefont {{Sammut}}, \citenamefont
  {{Sandberg}}, \citenamefont {{Sanders}}, \citenamefont {{Sannibale}},
  \citenamefont {{Santiago-Prieto}}, \citenamefont {{Sathyaprakash}},
  \citenamefont {{Saulson}}, \citenamefont {{Savage}}, \citenamefont
  {{Sawadsky}}, \citenamefont {{Scheuer}}, \citenamefont {{Schilling}},
  \citenamefont {{Schmidt}}, \citenamefont {{Schnabel}}, \citenamefont
  {{Schofield}}, \citenamefont {{Schreiber}}, \citenamefont {{Schuette}},
  \citenamefont {{Schutz}}, \citenamefont {{Scott}}, \citenamefont {{Scott}},
  \citenamefont {{Sellers}}, \citenamefont {{Sengupta}}, \citenamefont
  {{Sergeev}}, \citenamefont {{Serna}}, \citenamefont {{Sevigny}},
  \citenamefont {{Shaddock}}, \citenamefont {{Shahriar}}, \citenamefont
  {{Shaltev}}, \citenamefont {{Shao}}, \citenamefont {{Shapiro}}, \citenamefont
  {{Shawhan}}, \citenamefont {{Shoemaker}}, \citenamefont {{Sidery}},
  \citenamefont {{Siemens}}, \citenamefont {{Sigg}}, \citenamefont {{Silva}},
  \citenamefont {{Simakov}}, \citenamefont {{Singer}}, \citenamefont
  {{Singer}}, \citenamefont {{Singh}}, \citenamefont {{Sintes}}, \citenamefont
  {{Slagmolen}}, \citenamefont {{Smith}}, \citenamefont {{Smith}},
  \citenamefont {{Smith}}, \citenamefont {{Smith-Lefebvre}}, \citenamefont
  {{Son}}, \citenamefont {{Sorazu}}, \citenamefont {{Souradeep}}, \citenamefont
  {{Staley}}, \citenamefont {{Stebbins}}, \citenamefont {{Steinke}},
  \citenamefont {{Steinlechner}}, \citenamefont {{Steinlechner}}, \citenamefont
  {{Steinmeyer}}, \citenamefont {{Stephens}}, \citenamefont {{Steplewski}},
  \citenamefont {{Stevenson}}, \citenamefont {{Stone}}, \citenamefont
  {{Strain}}, \citenamefont {{Strigin}}, \citenamefont {{Sturani}},
  \citenamefont {{Stuver}}, \citenamefont {{Summerscales}}, \citenamefont
  {{Sutton}}, \citenamefont {{Szczepanczyk}}, \citenamefont {{Szeifert}},
  \citenamefont {{Talukder}}, \citenamefont {{Tanner}}, \citenamefont
  {{T{\'a}pai}}, \citenamefont {{Tarabrin}}, \citenamefont {{Taracchini}},
  \citenamefont {{Taylor}}, \citenamefont {{Tellez}}, \citenamefont {{Theeg}},
  \citenamefont {{Thirugnanasambandam}}, \citenamefont {{Thomas}},
  \citenamefont {{Thomas}}, \citenamefont {{Thorne}}, \citenamefont {{Thorne}},
  \citenamefont {{Thrane}}, \citenamefont {{Tiwari}}, \citenamefont
  {{Tomlinson}}, \citenamefont {{Torres}}, \citenamefont {{Torrie}},
  \citenamefont {{Traylor}}, \citenamefont {{Tse}}, \citenamefont
  {{Tshilumba}}, \citenamefont {{Ugolini}}, \citenamefont {{Unnikrishnan}},
  \citenamefont {{Urban}}, \citenamefont {{Usman}}, \citenamefont
  {{Vahlbruch}}, \citenamefont {{Vajente}}, \citenamefont {{Valdes}},
  \citenamefont {{Vallisneri}}, \citenamefont {{van Veggel}}, \citenamefont
  {{Vass}}, \citenamefont {{Vaulin}}, \citenamefont {{Vecchio}}, \citenamefont
  {{Veitch}}, \citenamefont {{Veitch}}, \citenamefont {{Venkateswara}},
  \citenamefont {{Vincent-Finley}}, \citenamefont {{Vitale}}, \citenamefont
  {{Vo}}, \citenamefont {{Vorvick}}, \citenamefont {{Vousden}}, \citenamefont
  {{Vyatchanin}}, \citenamefont {{Wade}}, \citenamefont {{Wade}}, \citenamefont
  {{Wade}}, \citenamefont {{Walker}}, \citenamefont {{Wallace}}, \citenamefont
  {{Walsh}}, \citenamefont {{Wang}}, \citenamefont {{Wang}}, \citenamefont
  {{Wang}}, \citenamefont {{Ward}}, \citenamefont {{Warner}}, \citenamefont
  {{Was}}, \citenamefont {{Weaver}}, \citenamefont {{Weinert}}, \citenamefont
  {{Weinstein}}, \citenamefont {{Weiss}}, \citenamefont {{Welborn}},
  \citenamefont {{Wen}}, \citenamefont {{Wessels}}, \citenamefont {{Westphal}},
  \citenamefont {{Wette}}, \citenamefont {{Whelan}}, \citenamefont
  {{Whitcomb}}, \citenamefont {{White}}, \citenamefont {{Whiting}},
  \citenamefont {{Wilkinson}}, \citenamefont {{Williams}}, \citenamefont
  {{Williams}}, \citenamefont {{Williamson}}, \citenamefont {{Willis}},
  \citenamefont {{Willke}}, \citenamefont {{Wimmer}}, \citenamefont
  {{Winkler}}, \citenamefont {{Wipf}}, \citenamefont {{Wittel}}, \citenamefont
  {{Woan}}, \citenamefont {{Worden}}, \citenamefont {{Xie}}, \citenamefont
  {{Yablon}}, \citenamefont {{Yakushin}}, \citenamefont {{Yam}}, \citenamefont
  {{Yamamoto}}, \citenamefont {{Yancey}}, \citenamefont {{Yang}}, \citenamefont
  {{Zanolin}}, \citenamefont {{Zhang}}, \citenamefont {{Zhang}}, \citenamefont
  {{Zhang}}, \citenamefont {{Zhang}}, \citenamefont {{Zhao}}, \citenamefont
  {{Zhou}}, \citenamefont {{Zhu}}, \citenamefont {{Zucker}}, \citenamefont
  {{Zuraw}},\ and\ \citenamefont {{Zweizig}}}]{LIGO_2015}%
  \BibitemOpen
  \bibfield  {author} {\bibinfo {author} {\bibnamefont {{LIGO Scientific
  Collaboration}}}, \bibinfo {author} {\bibfnamefont {J.}~\bibnamefont
  {{Aasi}}}, \bibinfo {author} {\bibfnamefont {B.~P.}\ \bibnamefont
  {{Abbott}}}, \bibinfo {author} {\bibfnamefont {R.}~\bibnamefont {{Abbott}}},
  \bibinfo {author} {\bibfnamefont {T.}~\bibnamefont {{Abbott}}}, \bibinfo
  {author} {\bibfnamefont {M.~R.}\ \bibnamefont {{Abernathy}}}, \bibinfo
  {author} {\bibfnamefont {K.}~\bibnamefont {{Ackley}}}, \bibinfo {author}
  {\bibfnamefont {C.}~\bibnamefont {{Adams}}}, \bibinfo {author} {\bibfnamefont
  {T.}~\bibnamefont {{Adams}}}, \bibinfo {author} {\bibfnamefont
  {P.}~\bibnamefont {{Addesso}}}, \bibinfo {author} {\bibfnamefont {R.~X.}\
  \bibnamefont {{Adhikari}}}, \bibinfo {author} {\bibfnamefont
  {V.}~\bibnamefont {{Adya}}}, \bibinfo {author} {\bibfnamefont
  {C.}~\bibnamefont {{Affeldt}}}, \bibinfo {author} {\bibfnamefont
  {N.}~\bibnamefont {{Aggarwal}}}, \bibinfo {author} {\bibfnamefont {O.~D.}\
  \bibnamefont {{Aguiar}}}, \bibinfo {author} {\bibfnamefont {A.}~\bibnamefont
  {{Ain}}}, \bibinfo {author} {\bibfnamefont {P.}~\bibnamefont {{Ajith}}},
  \bibinfo {author} {\bibfnamefont {A.}~\bibnamefont {{Alemic}}}, \bibinfo
  {author} {\bibfnamefont {B.}~\bibnamefont {{Allen}}}, \bibinfo {author}
  {\bibfnamefont {D.}~\bibnamefont {{Amariutei}}}, \bibinfo {author}
  {\bibfnamefont {S.~B.}\ \bibnamefont {{Anderson}}}, \bibinfo {author}
  {\bibfnamefont {W.~G.}\ \bibnamefont {{Anderson}}}, \bibinfo {author}
  {\bibfnamefont {K.}~\bibnamefont {{Arai}}}, \bibinfo {author} {\bibfnamefont
  {M.~C.}\ \bibnamefont {{Araya}}}, \bibinfo {author} {\bibfnamefont
  {C.}~\bibnamefont {{Arceneaux}}}, \bibinfo {author} {\bibfnamefont {J.~S.}\
  \bibnamefont {{Areeda}}}, \bibinfo {author} {\bibfnamefont {G.}~\bibnamefont
  {{Ashton}}}, \bibinfo {author} {\bibfnamefont {S.}~\bibnamefont {{Ast}}},
  \bibinfo {author} {\bibfnamefont {S.~M.}\ \bibnamefont {{Aston}}}, \bibinfo
  {author} {\bibfnamefont {P.}~\bibnamefont {{Aufmuth}}}, \bibinfo {author}
  {\bibfnamefont {C.}~\bibnamefont {{Aulbert}}}, \bibinfo {author}
  {\bibfnamefont {B.~E.}\ \bibnamefont {{Aylott}}}, \bibinfo {author}
  {\bibfnamefont {S.}~\bibnamefont {{Babak}}}, \bibinfo {author} {\bibfnamefont
  {P.~T.}\ \bibnamefont {{Baker}}}, \bibinfo {author} {\bibfnamefont {S.~W.}\
  \bibnamefont {{Ballmer}}}, \bibinfo {author} {\bibfnamefont {J.~C.}\
  \bibnamefont {{Barayoga}}}, \bibinfo {author} {\bibfnamefont
  {M.}~\bibnamefont {{Barbet}}}, \bibinfo {author} {\bibfnamefont
  {S.}~\bibnamefont {{Barclay}}}, \bibinfo {author} {\bibfnamefont {B.~C.}\
  \bibnamefont {{Barish}}}, \bibinfo {author} {\bibfnamefont {D.}~\bibnamefont
  {{Barker}}}, \bibinfo {author} {\bibfnamefont {B.}~\bibnamefont {{Barr}}},
  \bibinfo {author} {\bibfnamefont {L.}~\bibnamefont {{Barsotti}}}, \bibinfo
  {author} {\bibfnamefont {J.}~\bibnamefont {{Bartlett}}}, \bibinfo {author}
  {\bibfnamefont {M.~A.}\ \bibnamefont {{Barton}}}, \bibinfo {author}
  {\bibfnamefont {I.}~\bibnamefont {{Bartos}}}, \bibinfo {author}
  {\bibfnamefont {R.}~\bibnamefont {{Bassiri}}}, \bibinfo {author}
  {\bibfnamefont {J.~C.}\ \bibnamefont {{Batch}}}, \bibinfo {author}
  {\bibfnamefont {C.}~\bibnamefont {{Baune}}}, \bibinfo {author} {\bibfnamefont
  {B.}~\bibnamefont {{Behnke}}}, \bibinfo {author} {\bibfnamefont {A.~S.}\
  \bibnamefont {{Bell}}}, \bibinfo {author} {\bibfnamefont {C.}~\bibnamefont
  {{Bell}}}, \bibinfo {author} {\bibfnamefont {M.}~\bibnamefont
  {{Benacquista}}}, \bibinfo {author} {\bibfnamefont {J.}~\bibnamefont
  {{Bergman}}}, \bibinfo {author} {\bibfnamefont {G.}~\bibnamefont
  {{Bergmann}}}, \bibinfo {author} {\bibfnamefont {C.~P.~L.}\ \bibnamefont
  {{Berry}}}, \bibinfo {author} {\bibfnamefont {J.}~\bibnamefont
  {{Betzwieser}}}, \bibinfo {author} {\bibfnamefont {S.}~\bibnamefont
  {{Bhagwat}}}, \bibinfo {author} {\bibfnamefont {R.}~\bibnamefont
  {{Bhandare}}}, \bibinfo {author} {\bibfnamefont {I.~A.}\ \bibnamefont
  {{Bilenko}}}, \bibinfo {author} {\bibfnamefont {G.}~\bibnamefont
  {{Billingsley}}}, \bibinfo {author} {\bibfnamefont {J.}~\bibnamefont
  {{Birch}}}, \bibinfo {author} {\bibfnamefont {S.}~\bibnamefont {{Biscans}}},
  \bibinfo {author} {\bibfnamefont {C.}~\bibnamefont {{Biwer}}}, \bibinfo
  {author} {\bibfnamefont {J.~K.}\ \bibnamefont {{Blackburn}}}, \bibinfo
  {author} {\bibfnamefont {L.}~\bibnamefont {{Blackburn}}}, \bibinfo {author}
  {\bibfnamefont {C.~D.}\ \bibnamefont {{Blair}}}, \bibinfo {author}
  {\bibfnamefont {D.}~\bibnamefont {{Blair}}}, \bibinfo {author} {\bibfnamefont
  {O.}~\bibnamefont {{Bock}}}, \bibinfo {author} {\bibfnamefont {T.~P.}\
  \bibnamefont {{Bodiya}}}, \bibinfo {author} {\bibfnamefont {P.}~\bibnamefont
  {{Bojtos}}}, \bibinfo {author} {\bibfnamefont {C.}~\bibnamefont {{Bond}}},
  \bibinfo {author} {\bibfnamefont {R.}~\bibnamefont {{Bork}}}, \bibinfo
  {author} {\bibfnamefont {M.}~\bibnamefont {{Born}}}, \bibinfo {author}
  {\bibfnamefont {S.}~\bibnamefont {{Bose}}}, \bibinfo {author} {\bibfnamefont
  {P.~R.}\ \bibnamefont {{Brady}}}, \bibinfo {author} {\bibfnamefont {V.~B.}\
  \bibnamefont {{Braginsky}}}, \bibinfo {author} {\bibfnamefont {J.~E.}\
  \bibnamefont {{Brau}}}, \bibinfo {author} {\bibfnamefont {D.~O.}\
  \bibnamefont {{Bridges}}}, \bibinfo {author} {\bibfnamefont {M.}~\bibnamefont
  {{Brinkmann}}}, \bibinfo {author} {\bibfnamefont {A.~F.}\ \bibnamefont
  {{Brooks}}}, \bibinfo {author} {\bibfnamefont {D.~A.}\ \bibnamefont
  {{Brown}}}, \bibinfo {author} {\bibfnamefont {D.~D.}\ \bibnamefont
  {{Brown}}}, \bibinfo {author} {\bibfnamefont {N.~M.}\ \bibnamefont
  {{Brown}}}, \bibinfo {author} {\bibfnamefont {S.}~\bibnamefont {{Buchman}}},
  \bibinfo {author} {\bibfnamefont {A.}~\bibnamefont {{Buikema}}}, \bibinfo
  {author} {\bibfnamefont {A.}~\bibnamefont {{Buonanno}}}, \bibinfo {author}
  {\bibfnamefont {L.}~\bibnamefont {{Cadonati}}}, \bibinfo {author}
  {\bibfnamefont {J.}~\bibnamefont {{Calder{\'o}n Bustillo}}}, \bibinfo
  {author} {\bibfnamefont {J.~B.}\ \bibnamefont {{Camp}}}, \bibinfo {author}
  {\bibfnamefont {K.~C.}\ \bibnamefont {{Cannon}}}, \bibinfo {author}
  {\bibfnamefont {J.}~\bibnamefont {{Cao}}}, \bibinfo {author} {\bibfnamefont
  {C.~D.}\ \bibnamefont {{Capano}}}, \bibinfo {author} {\bibfnamefont
  {S.}~\bibnamefont {{Caride}}}, \bibinfo {author} {\bibfnamefont
  {S.}~\bibnamefont {{Caudill}}}, \bibinfo {author} {\bibfnamefont
  {M.}~\bibnamefont {{Cavagli{\`a}}}}, \bibinfo {author} {\bibfnamefont
  {C.}~\bibnamefont {{Cepeda}}}, \bibinfo {author} {\bibfnamefont
  {R.}~\bibnamefont {{Chakraborty}}}, \bibinfo {author} {\bibfnamefont
  {T.}~\bibnamefont {{Chalermsongsak}}}, \bibinfo {author} {\bibfnamefont
  {S.~J.}\ \bibnamefont {{Chamberlin}}}, \bibinfo {author} {\bibfnamefont
  {S.}~\bibnamefont {{Chao}}}, \bibinfo {author} {\bibfnamefont
  {P.}~\bibnamefont {{Charlton}}}, \bibinfo {author} {\bibfnamefont
  {Y.}~\bibnamefont {{Chen}}}, \bibinfo {author} {\bibfnamefont {H.~S.}\
  \bibnamefont {{Cho}}}, \bibinfo {author} {\bibfnamefont {M.}~\bibnamefont
  {{Cho}}}, \bibinfo {author} {\bibfnamefont {J.~H.}\ \bibnamefont {{Chow}}},
  \bibinfo {author} {\bibfnamefont {N.}~\bibnamefont {{Christensen}}}, \bibinfo
  {author} {\bibfnamefont {Q.}~\bibnamefont {{Chu}}}, \bibinfo {author}
  {\bibfnamefont {S.}~\bibnamefont {{Chung}}}, \bibinfo {author} {\bibfnamefont
  {G.}~\bibnamefont {{Ciani}}}, \bibinfo {author} {\bibfnamefont
  {F.}~\bibnamefont {{Clara}}}, \bibinfo {author} {\bibfnamefont {J.~A.}\
  \bibnamefont {{Clark}}}, \bibinfo {author} {\bibfnamefont {C.}~\bibnamefont
  {{Collette}}}, \bibinfo {author} {\bibfnamefont {L.}~\bibnamefont
  {{Cominsky}}}, \bibinfo {author} {\bibfnamefont {J.}~\bibnamefont
  {{Constancio}}, \bibfnamefont {M.}}, \bibinfo {author} {\bibfnamefont
  {D.}~\bibnamefont {{Cook}}}, \bibinfo {author} {\bibfnamefont {T.~R.}\
  \bibnamefont {{Corbitt}}}, \bibinfo {author} {\bibfnamefont {N.}~\bibnamefont
  {{Cornish}}}, \bibinfo {author} {\bibfnamefont {A.}~\bibnamefont {{Corsi}}},
  \bibinfo {author} {\bibfnamefont {C.~A.}\ \bibnamefont {{Costa}}}, \bibinfo
  {author} {\bibfnamefont {M.~W.}\ \bibnamefont {{Coughlin}}}, \bibinfo
  {author} {\bibfnamefont {S.}~\bibnamefont {{Countryman}}}, \bibinfo {author}
  {\bibfnamefont {P.}~\bibnamefont {{Couvares}}}, \bibinfo {author}
  {\bibfnamefont {D.~M.}\ \bibnamefont {{Coward}}}, \bibinfo {author}
  {\bibfnamefont {M.~J.}\ \bibnamefont {{Cowart}}}, \bibinfo {author}
  {\bibfnamefont {D.~C.}\ \bibnamefont {{Coyne}}}, \bibinfo {author}
  {\bibfnamefont {R.}~\bibnamefont {{Coyne}}}, \bibinfo {author} {\bibfnamefont
  {K.}~\bibnamefont {{Craig}}}, \bibinfo {author} {\bibfnamefont {J.~D.~E.}\
  \bibnamefont {{Creighton}}}, \bibinfo {author} {\bibfnamefont {T.~D.}\
  \bibnamefont {{Creighton}}}, \bibinfo {author} {\bibfnamefont
  {J.}~\bibnamefont {{Cripe}}}, \bibinfo {author} {\bibfnamefont {S.~G.}\
  \bibnamefont {{Crowder}}}, \bibinfo {author} {\bibfnamefont {A.}~\bibnamefont
  {{Cumming}}}, \bibinfo {author} {\bibfnamefont {L.}~\bibnamefont
  {{Cunningham}}}, \bibinfo {author} {\bibfnamefont {C.}~\bibnamefont
  {{Cutler}}}, \bibinfo {author} {\bibfnamefont {K.}~\bibnamefont {{Dahl}}},
  \bibinfo {author} {\bibfnamefont {T.}~\bibnamefont {{Dal Canton}}}, \bibinfo
  {author} {\bibfnamefont {M.}~\bibnamefont {{Damjanic}}}, \bibinfo {author}
  {\bibfnamefont {S.~L.}\ \bibnamefont {{Danilishin}}}, \bibinfo {author}
  {\bibfnamefont {K.}~\bibnamefont {{Danzmann}}}, \bibinfo {author}
  {\bibfnamefont {L.}~\bibnamefont {{Dartez}}}, \bibinfo {author}
  {\bibfnamefont {I.}~\bibnamefont {{Dave}}}, \bibinfo {author} {\bibfnamefont
  {H.}~\bibnamefont {{Daveloza}}}, \bibinfo {author} {\bibfnamefont {G.~S.}\
  \bibnamefont {{Davies}}}, \bibinfo {author} {\bibfnamefont {E.~J.}\
  \bibnamefont {{Daw}}}, \bibinfo {author} {\bibfnamefont {D.}~\bibnamefont
  {{DeBra}}}, \bibinfo {author} {\bibfnamefont {W.}~\bibnamefont {{Del
  Pozzo}}}, \bibinfo {author} {\bibfnamefont {T.}~\bibnamefont {{Denker}}},
  \bibinfo {author} {\bibfnamefont {T.}~\bibnamefont {{Dent}}}, \bibinfo
  {author} {\bibfnamefont {V.}~\bibnamefont {{Dergachev}}}, \bibinfo {author}
  {\bibfnamefont {R.~T.}\ \bibnamefont {{DeRosa}}}, \bibinfo {author}
  {\bibfnamefont {R.}~\bibnamefont {{DeSalvo}}}, \bibinfo {author}
  {\bibfnamefont {S.}~\bibnamefont {{Dhurandhar}}}, \bibinfo {author}
  {\bibfnamefont {M.}~\bibnamefont {{D{\textasciiacute}{\i}az}}}, \bibinfo
  {author} {\bibfnamefont {I.}~\bibnamefont {{Di Palma}}}, \bibinfo {author}
  {\bibfnamefont {G.}~\bibnamefont {{Dojcinoski}}}, \bibinfo {author}
  {\bibfnamefont {E.}~\bibnamefont {{Dominguez}}}, \bibinfo {author}
  {\bibfnamefont {F.}~\bibnamefont {{Donovan}}}, \bibinfo {author}
  {\bibfnamefont {K.~L.}\ \bibnamefont {{Dooley}}}, \bibinfo {author}
  {\bibfnamefont {S.}~\bibnamefont {{Doravari}}}, \bibinfo {author}
  {\bibfnamefont {R.}~\bibnamefont {{Douglas}}}, \bibinfo {author}
  {\bibfnamefont {T.~P.}\ \bibnamefont {{Downes}}}, \bibinfo {author}
  {\bibfnamefont {J.~C.}\ \bibnamefont {{Driggers}}}, \bibinfo {author}
  {\bibfnamefont {Z.}~\bibnamefont {{Du}}}, \bibinfo {author} {\bibfnamefont
  {S.}~\bibnamefont {{Dwyer}}}, \bibinfo {author} {\bibfnamefont
  {T.}~\bibnamefont {{Eberle}}}, \bibinfo {author} {\bibfnamefont
  {T.}~\bibnamefont {{Edo}}}, \bibinfo {author} {\bibfnamefont
  {M.}~\bibnamefont {{Edwards}}}, \bibinfo {author} {\bibfnamefont
  {M.}~\bibnamefont {{Edwards}}}, \bibinfo {author} {\bibfnamefont
  {A.}~\bibnamefont {{Effler}}}, \bibinfo {author} {\bibfnamefont {H.~B.}\
  \bibnamefont {{Eggenstein}}}, \bibinfo {author} {\bibfnamefont
  {P.}~\bibnamefont {{Ehrens}}}, \bibinfo {author} {\bibfnamefont
  {J.}~\bibnamefont {{Eichholz}}}, \bibinfo {author} {\bibfnamefont {S.~S.}\
  \bibnamefont {{Eikenberry}}}, \bibinfo {author} {\bibfnamefont
  {R.}~\bibnamefont {{Essick}}}, \bibinfo {author} {\bibfnamefont
  {T.}~\bibnamefont {{Etzel}}}, \bibinfo {author} {\bibfnamefont
  {M.}~\bibnamefont {{Evans}}}, \bibinfo {author} {\bibfnamefont
  {T.}~\bibnamefont {{Evans}}}, \bibinfo {author} {\bibfnamefont
  {M.}~\bibnamefont {{Factourovich}}}, \bibinfo {author} {\bibfnamefont
  {S.}~\bibnamefont {{Fairhurst}}}, \bibinfo {author} {\bibfnamefont
  {X.}~\bibnamefont {{Fan}}}, \bibinfo {author} {\bibfnamefont
  {Q.}~\bibnamefont {{Fang}}}, \bibinfo {author} {\bibfnamefont
  {B.}~\bibnamefont {{Farr}}}, \bibinfo {author} {\bibfnamefont {W.~M.}\
  \bibnamefont {{Farr}}}, \bibinfo {author} {\bibfnamefont {M.}~\bibnamefont
  {{Favata}}}, \bibinfo {author} {\bibfnamefont {M.}~\bibnamefont {{Fays}}},
  \bibinfo {author} {\bibfnamefont {H.}~\bibnamefont {{Fehrmann}}}, \bibinfo
  {author} {\bibfnamefont {M.~M.}\ \bibnamefont {{Fejer}}}, \bibinfo {author}
  {\bibfnamefont {D.}~\bibnamefont {{Feldbaum}}}, \bibinfo {author}
  {\bibfnamefont {E.~C.}\ \bibnamefont {{Ferreira}}}, \bibinfo {author}
  {\bibfnamefont {R.~P.}\ \bibnamefont {{Fisher}}}, \bibinfo {author}
  {\bibfnamefont {Z.}~\bibnamefont {{Frei}}}, \bibinfo {author} {\bibfnamefont
  {A.}~\bibnamefont {{Freise}}}, \bibinfo {author} {\bibfnamefont
  {R.}~\bibnamefont {{Frey}}}, \bibinfo {author} {\bibfnamefont {T.~T.}\
  \bibnamefont {{Fricke}}}, \bibinfo {author} {\bibfnamefont {P.}~\bibnamefont
  {{Fritschel}}}, \bibinfo {author} {\bibfnamefont {V.~V.}\ \bibnamefont
  {{Frolov}}}, \bibinfo {author} {\bibfnamefont {S.}~\bibnamefont
  {{Fuentes-Tapia}}}, \bibinfo {author} {\bibfnamefont {P.}~\bibnamefont
  {{Fulda}}}, \bibinfo {author} {\bibfnamefont {M.}~\bibnamefont {{Fyffe}}},
  \bibinfo {author} {\bibfnamefont {J.~R.}\ \bibnamefont {{Gair}}}, \bibinfo
  {author} {\bibfnamefont {S.}~\bibnamefont {{Gaonkar}}}, \bibinfo {author}
  {\bibfnamefont {N.}~\bibnamefont {{Gehrels}}}, \bibinfo {author}
  {\bibfnamefont {L.~{\'A}.}\ \bibnamefont {{Gergely}}}, \bibinfo {author}
  {\bibfnamefont {J.~A.}\ \bibnamefont {{Giaime}}}, \bibinfo {author}
  {\bibfnamefont {K.~D.}\ \bibnamefont {{Giardina}}}, \bibinfo {author}
  {\bibfnamefont {J.}~\bibnamefont {{Gleason}}}, \bibinfo {author}
  {\bibfnamefont {E.}~\bibnamefont {{Goetz}}}, \bibinfo {author} {\bibfnamefont
  {R.}~\bibnamefont {{Goetz}}}, \bibinfo {author} {\bibfnamefont
  {L.}~\bibnamefont {{Gondan}}}, \bibinfo {author} {\bibfnamefont
  {G.}~\bibnamefont {{Gonz{\'a}lez}}}, \bibinfo {author} {\bibfnamefont
  {N.}~\bibnamefont {{Gordon}}}, \bibinfo {author} {\bibfnamefont {M.~L.}\
  \bibnamefont {{Gorodetsky}}}, \bibinfo {author} {\bibfnamefont
  {S.}~\bibnamefont {{Gossan}}}, \bibinfo {author} {\bibfnamefont
  {S.}~\bibnamefont {{Go{\ss}ler}}}, \bibinfo {author} {\bibfnamefont
  {C.}~\bibnamefont {{Gr{\"a}f}}}, \bibinfo {author} {\bibfnamefont {P.~B.}\
  \bibnamefont {{Graff}}}, \bibinfo {author} {\bibfnamefont {A.}~\bibnamefont
  {{Grant}}}, \bibinfo {author} {\bibfnamefont {S.}~\bibnamefont {{Gras}}},
  \bibinfo {author} {\bibfnamefont {C.}~\bibnamefont {{Gray}}}, \bibinfo
  {author} {\bibfnamefont {R.~J.~S.}\ \bibnamefont {{Greenhalgh}}}, \bibinfo
  {author} {\bibfnamefont {A.~M.}\ \bibnamefont {{Gretarsson}}}, \bibinfo
  {author} {\bibfnamefont {H.}~\bibnamefont {{Grote}}}, \bibinfo {author}
  {\bibfnamefont {S.}~\bibnamefont {{Grunewald}}}, \bibinfo {author}
  {\bibfnamefont {C.~J.}\ \bibnamefont {{Guido}}}, \bibinfo {author}
  {\bibfnamefont {X.}~\bibnamefont {{Guo}}}, \bibinfo {author} {\bibfnamefont
  {K.}~\bibnamefont {{Gushwa}}}, \bibinfo {author} {\bibfnamefont {E.~K.}\
  \bibnamefont {{Gustafson}}}, \bibinfo {author} {\bibfnamefont
  {R.}~\bibnamefont {{Gustafson}}}, \bibinfo {author} {\bibfnamefont
  {J.}~\bibnamefont {{Hacker}}}, \bibinfo {author} {\bibfnamefont {E.~D.}\
  \bibnamefont {{Hall}}}, \bibinfo {author} {\bibfnamefont {G.}~\bibnamefont
  {{Hammond}}}, \bibinfo {author} {\bibfnamefont {M.}~\bibnamefont {{Hanke}}},
  \bibinfo {author} {\bibfnamefont {J.}~\bibnamefont {{Hanks}}}, \bibinfo
  {author} {\bibfnamefont {C.}~\bibnamefont {{Hanna}}}, \bibinfo {author}
  {\bibfnamefont {M.~D.}\ \bibnamefont {{Hannam}}}, \bibinfo {author}
  {\bibfnamefont {J.}~\bibnamefont {{Hanson}}}, \bibinfo {author}
  {\bibfnamefont {T.}~\bibnamefont {{Hardwick}}}, \bibinfo {author}
  {\bibfnamefont {G.~M.}\ \bibnamefont {{Harry}}}, \bibinfo {author}
  {\bibfnamefont {I.~W.}\ \bibnamefont {{Harry}}}, \bibinfo {author}
  {\bibfnamefont {M.}~\bibnamefont {{Hart}}}, \bibinfo {author} {\bibfnamefont
  {M.~T.}\ \bibnamefont {{Hartman}}}, \bibinfo {author} {\bibfnamefont {C.~J.}\
  \bibnamefont {{Haster}}}, \bibinfo {author} {\bibfnamefont {K.}~\bibnamefont
  {{Haughian}}}, \bibinfo {author} {\bibfnamefont {S.}~\bibnamefont {{Hee}}},
  \bibinfo {author} {\bibfnamefont {M.}~\bibnamefont {{Heintze}}}, \bibinfo
  {author} {\bibfnamefont {G.}~\bibnamefont {{Heinzel}}}, \bibinfo {author}
  {\bibfnamefont {M.}~\bibnamefont {{Hendry}}}, \bibinfo {author}
  {\bibfnamefont {I.~S.}\ \bibnamefont {{Heng}}}, \bibinfo {author}
  {\bibfnamefont {A.~W.}\ \bibnamefont {{Heptonstall}}}, \bibinfo {author}
  {\bibfnamefont {M.}~\bibnamefont {{Heurs}}}, \bibinfo {author} {\bibfnamefont
  {M.}~\bibnamefont {{Hewitson}}}, \bibinfo {author} {\bibfnamefont
  {S.}~\bibnamefont {{Hild}}}, \bibinfo {author} {\bibfnamefont
  {D.}~\bibnamefont {{Hoak}}}, \bibinfo {author} {\bibfnamefont {K.~A.}\
  \bibnamefont {{Hodge}}}, \bibinfo {author} {\bibfnamefont {S.~E.}\
  \bibnamefont {{Hollitt}}}, \bibinfo {author} {\bibfnamefont {K.}~\bibnamefont
  {{Holt}}}, \bibinfo {author} {\bibfnamefont {P.}~\bibnamefont {{Hopkins}}},
  \bibinfo {author} {\bibfnamefont {D.~J.}\ \bibnamefont {{Hosken}}}, \bibinfo
  {author} {\bibfnamefont {J.}~\bibnamefont {{Hough}}}, \bibinfo {author}
  {\bibfnamefont {E.}~\bibnamefont {{Houston}}}, \bibinfo {author}
  {\bibfnamefont {E.~J.}\ \bibnamefont {{Howell}}}, \bibinfo {author}
  {\bibfnamefont {Y.~M.}\ \bibnamefont {{Hu}}}, \bibinfo {author}
  {\bibfnamefont {E.}~\bibnamefont {{Huerta}}}, \bibinfo {author}
  {\bibfnamefont {B.}~\bibnamefont {{Hughey}}}, \bibinfo {author}
  {\bibfnamefont {S.}~\bibnamefont {{Husa}}}, \bibinfo {author} {\bibfnamefont
  {S.~H.}\ \bibnamefont {{Huttner}}}, \bibinfo {author} {\bibfnamefont
  {M.}~\bibnamefont {{Huynh}}}, \bibinfo {author} {\bibfnamefont
  {T.}~\bibnamefont {{Huynh-Dinh}}}, \bibinfo {author} {\bibfnamefont
  {A.}~\bibnamefont {{Idrisy}}}, \bibinfo {author} {\bibfnamefont
  {N.}~\bibnamefont {{Indik}}}, \bibinfo {author} {\bibfnamefont {D.~R.}\
  \bibnamefont {{Ingram}}}, \bibinfo {author} {\bibfnamefont {R.}~\bibnamefont
  {{Inta}}}, \bibinfo {author} {\bibfnamefont {G.}~\bibnamefont {{Islas}}},
  \bibinfo {author} {\bibfnamefont {J.~C.}\ \bibnamefont {{Isler}}}, \bibinfo
  {author} {\bibfnamefont {T.}~\bibnamefont {{Isogai}}}, \bibinfo {author}
  {\bibfnamefont {B.~R.}\ \bibnamefont {{Iyer}}}, \bibinfo {author}
  {\bibfnamefont {K.}~\bibnamefont {{Izumi}}}, \bibinfo {author} {\bibfnamefont
  {M.}~\bibnamefont {{Jacobson}}}, \bibinfo {author} {\bibfnamefont
  {H.}~\bibnamefont {{Jang}}}, \bibinfo {author} {\bibfnamefont
  {S.}~\bibnamefont {{Jawahar}}}, \bibinfo {author} {\bibfnamefont
  {Y.}~\bibnamefont {{Ji}}}, \bibinfo {author} {\bibfnamefont {F.}~\bibnamefont
  {{Jim{\'e}nez-Forteza}}}, \bibinfo {author} {\bibfnamefont {W.~W.}\
  \bibnamefont {{Johnson}}}, \bibinfo {author} {\bibfnamefont {D.~I.}\
  \bibnamefont {{Jones}}}, \bibinfo {author} {\bibfnamefont {R.}~\bibnamefont
  {{Jones}}}, \bibinfo {author} {\bibfnamefont {L.}~\bibnamefont {{Ju}}},
  \bibinfo {author} {\bibfnamefont {K.}~\bibnamefont {{Haris}}}, \bibinfo
  {author} {\bibfnamefont {V.}~\bibnamefont {{Kalogera}}}, \bibinfo {author}
  {\bibfnamefont {S.}~\bibnamefont {{Kandhasamy}}}, \bibinfo {author}
  {\bibfnamefont {G.}~\bibnamefont {{Kang}}}, \bibinfo {author} {\bibfnamefont
  {J.~B.}\ \bibnamefont {{Kanner}}}, \bibinfo {author} {\bibfnamefont
  {E.}~\bibnamefont {{Katsavounidis}}}, \bibinfo {author} {\bibfnamefont
  {W.}~\bibnamefont {{Katzman}}}, \bibinfo {author} {\bibfnamefont
  {H.}~\bibnamefont {{Kaufer}}}, \bibinfo {author} {\bibfnamefont
  {S.}~\bibnamefont {{Kaufer}}}, \bibinfo {author} {\bibfnamefont
  {T.}~\bibnamefont {{Kaur}}}, \bibinfo {author} {\bibfnamefont
  {K.}~\bibnamefont {{Kawabe}}}, \bibinfo {author} {\bibfnamefont
  {F.}~\bibnamefont {{Kawazoe}}}, \bibinfo {author} {\bibfnamefont {G.~M.}\
  \bibnamefont {{Keiser}}}, \bibinfo {author} {\bibfnamefont {D.}~\bibnamefont
  {{Keitel}}}, \bibinfo {author} {\bibfnamefont {D.~B.}\ \bibnamefont
  {{Kelley}}}, \bibinfo {author} {\bibfnamefont {W.}~\bibnamefont {{Kells}}},
  \bibinfo {author} {\bibfnamefont {D.~G.}\ \bibnamefont {{Keppel}}}, \bibinfo
  {author} {\bibfnamefont {J.~S.}\ \bibnamefont {{Key}}}, \bibinfo {author}
  {\bibfnamefont {A.}~\bibnamefont {{Khalaidovski}}}, \bibinfo {author}
  {\bibfnamefont {F.~Y.}\ \bibnamefont {{Khalili}}}, \bibinfo {author}
  {\bibfnamefont {E.~A.}\ \bibnamefont {{Khazanov}}}, \bibinfo {author}
  {\bibfnamefont {C.}~\bibnamefont {{Kim}}}, \bibinfo {author} {\bibfnamefont
  {K.}~\bibnamefont {{Kim}}}, \bibinfo {author} {\bibfnamefont {N.~G.}\
  \bibnamefont {{Kim}}}, \bibinfo {author} {\bibfnamefont {N.}~\bibnamefont
  {{Kim}}}, \bibinfo {author} {\bibfnamefont {Y.~M.}\ \bibnamefont {{Kim}}},
  \bibinfo {author} {\bibfnamefont {E.~J.}\ \bibnamefont {{King}}}, \bibinfo
  {author} {\bibfnamefont {P.~J.}\ \bibnamefont {{King}}}, \bibinfo {author}
  {\bibfnamefont {D.~L.}\ \bibnamefont {{Kinzel}}}, \bibinfo {author}
  {\bibfnamefont {J.~S.}\ \bibnamefont {{Kissel}}}, \bibinfo {author}
  {\bibfnamefont {S.}~\bibnamefont {{Klimenko}}}, \bibinfo {author}
  {\bibfnamefont {J.}~\bibnamefont {{Kline}}}, \bibinfo {author} {\bibfnamefont
  {S.}~\bibnamefont {{Koehlenbeck}}}, \bibinfo {author} {\bibfnamefont
  {K.}~\bibnamefont {{Kokeyama}}}, \bibinfo {author} {\bibfnamefont
  {V.}~\bibnamefont {{Kondrashov}}}, \bibinfo {author} {\bibfnamefont
  {M.}~\bibnamefont {{Korobko}}}, \bibinfo {author} {\bibfnamefont {W.~Z.}\
  \bibnamefont {{Korth}}}, \bibinfo {author} {\bibfnamefont {D.~B.}\
  \bibnamefont {{Kozak}}}, \bibinfo {author} {\bibfnamefont {V.}~\bibnamefont
  {{Kringel}}}, \bibinfo {author} {\bibfnamefont {B.}~\bibnamefont
  {{Krishnan}}}, \bibinfo {author} {\bibfnamefont {C.}~\bibnamefont
  {{Krueger}}}, \bibinfo {author} {\bibfnamefont {G.}~\bibnamefont {{Kuehn}}},
  \bibinfo {author} {\bibfnamefont {A.}~\bibnamefont {{Kumar}}}, \bibinfo
  {author} {\bibfnamefont {P.}~\bibnamefont {{Kumar}}}, \bibinfo {author}
  {\bibfnamefont {L.}~\bibnamefont {{Kuo}}}, \bibinfo {author} {\bibfnamefont
  {M.}~\bibnamefont {{Landry}}}, \bibinfo {author} {\bibfnamefont
  {B.}~\bibnamefont {{Lantz}}}, \bibinfo {author} {\bibfnamefont
  {S.}~\bibnamefont {{Larson}}}, \bibinfo {author} {\bibfnamefont {P.~D.}\
  \bibnamefont {{Lasky}}}, \bibinfo {author} {\bibfnamefont {A.}~\bibnamefont
  {{Lazzarini}}}, \bibinfo {author} {\bibfnamefont {C.}~\bibnamefont
  {{Lazzaro}}}, \bibinfo {author} {\bibfnamefont {J.}~\bibnamefont {{Le}}},
  \bibinfo {author} {\bibfnamefont {P.}~\bibnamefont {{Leaci}}}, \bibinfo
  {author} {\bibfnamefont {S.}~\bibnamefont {{Leavey}}}, \bibinfo {author}
  {\bibfnamefont {E.~O.}\ \bibnamefont {{Lebigot}}}, \bibinfo {author}
  {\bibfnamefont {C.~H.}\ \bibnamefont {{Lee}}}, \bibinfo {author}
  {\bibfnamefont {H.~K.}\ \bibnamefont {{Lee}}}, \bibinfo {author}
  {\bibfnamefont {H.~M.}\ \bibnamefont {{Lee}}}, \bibinfo {author}
  {\bibfnamefont {J.~R.}\ \bibnamefont {{Leong}}}, \bibinfo {author}
  {\bibfnamefont {Y.}~\bibnamefont {{Levin}}}, \bibinfo {author} {\bibfnamefont
  {B.}~\bibnamefont {{Levine}}}, \bibinfo {author} {\bibfnamefont
  {J.}~\bibnamefont {{Lewis}}}, \bibinfo {author} {\bibfnamefont {T.~G.~F.}\
  \bibnamefont {{Li}}}, \bibinfo {author} {\bibfnamefont {K.}~\bibnamefont
  {{Libbrecht}}}, \bibinfo {author} {\bibfnamefont {A.}~\bibnamefont
  {{Libson}}}, \bibinfo {author} {\bibfnamefont {A.~C.}\ \bibnamefont {{Lin}}},
  \bibinfo {author} {\bibfnamefont {T.~B.}\ \bibnamefont {{Littenberg}}},
  \bibinfo {author} {\bibfnamefont {N.~A.}\ \bibnamefont {{Lockerbie}}},
  \bibinfo {author} {\bibfnamefont {V.}~\bibnamefont {{Lockett}}}, \bibinfo
  {author} {\bibfnamefont {J.}~\bibnamefont {{Logue}}}, \bibinfo {author}
  {\bibfnamefont {A.~L.}\ \bibnamefont {{Lombardi}}}, \bibinfo {author}
  {\bibfnamefont {M.}~\bibnamefont {{Lormand}}}, \bibinfo {author}
  {\bibfnamefont {J.}~\bibnamefont {{Lough}}}, \bibinfo {author} {\bibfnamefont
  {M.~J.}\ \bibnamefont {{Lubinski}}}, \bibinfo {author} {\bibfnamefont
  {H.}~\bibnamefont {{L{\"u}ck}}}, \bibinfo {author} {\bibfnamefont {A.~P.}\
  \bibnamefont {{Lundgren}}}, \bibinfo {author} {\bibfnamefont
  {R.}~\bibnamefont {{Lynch}}}, \bibinfo {author} {\bibfnamefont
  {Y.}~\bibnamefont {{Ma}}}, \bibinfo {author} {\bibfnamefont {J.}~\bibnamefont
  {{Macarthur}}}, \bibinfo {author} {\bibfnamefont {T.}~\bibnamefont
  {{MacDonald}}}, \bibinfo {author} {\bibfnamefont {B.}~\bibnamefont
  {{Machenschalk}}}, \bibinfo {author} {\bibfnamefont {M.}~\bibnamefont
  {{MacInnis}}}, \bibinfo {author} {\bibfnamefont {D.~M.}\ \bibnamefont
  {{Macleod}}}, \bibinfo {author} {\bibfnamefont {F.}~\bibnamefont
  {{Maga{\~n}a-Sandoval}}}, \bibinfo {author} {\bibfnamefont {R.}~\bibnamefont
  {{Magee}}}, \bibinfo {author} {\bibfnamefont {M.}~\bibnamefont
  {{Mageswaran}}}, \bibinfo {author} {\bibfnamefont {C.}~\bibnamefont
  {{Maglione}}}, \bibinfo {author} {\bibfnamefont {K.}~\bibnamefont
  {{Mailand}}}, \bibinfo {author} {\bibfnamefont {I.}~\bibnamefont {{Mandel}}},
  \bibinfo {author} {\bibfnamefont {V.}~\bibnamefont {{Mandic}}}, \bibinfo
  {author} {\bibfnamefont {V.}~\bibnamefont {{Mangano}}}, \bibinfo {author}
  {\bibfnamefont {G.~L.}\ \bibnamefont {{Mansell}}}, \bibinfo {author}
  {\bibfnamefont {S.}~\bibnamefont {{M{\'a}rka}}}, \bibinfo {author}
  {\bibfnamefont {Z.}~\bibnamefont {{M{\'a}rka}}}, \bibinfo {author}
  {\bibfnamefont {A.}~\bibnamefont {{Markosyan}}}, \bibinfo {author}
  {\bibfnamefont {E.}~\bibnamefont {{Maros}}}, \bibinfo {author} {\bibfnamefont
  {I.~W.}\ \bibnamefont {{Martin}}}, \bibinfo {author} {\bibfnamefont {R.~M.}\
  \bibnamefont {{Martin}}}, \bibinfo {author} {\bibfnamefont {D.}~\bibnamefont
  {{Martynov}}}, \bibinfo {author} {\bibfnamefont {J.~N.}\ \bibnamefont
  {{Marx}}}, \bibinfo {author} {\bibfnamefont {K.}~\bibnamefont {{Mason}}},
  \bibinfo {author} {\bibfnamefont {T.~J.}\ \bibnamefont {{Massinger}}},
  \bibinfo {author} {\bibfnamefont {F.}~\bibnamefont {{Matichard}}}, \bibinfo
  {author} {\bibfnamefont {L.}~\bibnamefont {{Matone}}}, \bibinfo {author}
  {\bibfnamefont {N.}~\bibnamefont {{Mavalvala}}}, \bibinfo {author}
  {\bibfnamefont {N.}~\bibnamefont {{Mazumder}}}, \bibinfo {author}
  {\bibfnamefont {G.}~\bibnamefont {{Mazzolo}}}, \bibinfo {author}
  {\bibfnamefont {R.}~\bibnamefont {{McCarthy}}}, \bibinfo {author}
  {\bibfnamefont {D.~E.}\ \bibnamefont {{McClelland}}}, \bibinfo {author}
  {\bibfnamefont {S.}~\bibnamefont {{McCormick}}}, \bibinfo {author}
  {\bibfnamefont {S.~C.}\ \bibnamefont {{McGuire}}}, \bibinfo {author}
  {\bibfnamefont {G.}~\bibnamefont {{McIntyre}}}, \bibinfo {author}
  {\bibfnamefont {J.}~\bibnamefont {{McIver}}}, \bibinfo {author}
  {\bibfnamefont {K.}~\bibnamefont {{McLin}}}, \bibinfo {author} {\bibfnamefont
  {S.}~\bibnamefont {{McWilliams}}}, \bibinfo {author} {\bibfnamefont {G.~D.}\
  \bibnamefont {{Meadors}}}, \bibinfo {author} {\bibfnamefont {M.}~\bibnamefont
  {{Meinders}}}, \bibinfo {author} {\bibfnamefont {A.}~\bibnamefont
  {{Melatos}}}, \bibinfo {author} {\bibfnamefont {G.}~\bibnamefont
  {{Mendell}}}, \bibinfo {author} {\bibfnamefont {R.~A.}\ \bibnamefont
  {{Mercer}}}, \bibinfo {author} {\bibfnamefont {S.}~\bibnamefont {{Meshkov}}},
  \bibinfo {author} {\bibfnamefont {C.}~\bibnamefont {{Messenger}}}, \bibinfo
  {author} {\bibfnamefont {P.~M.}\ \bibnamefont {{Meyers}}}, \bibinfo {author}
  {\bibfnamefont {H.}~\bibnamefont {{Miao}}}, \bibinfo {author} {\bibfnamefont
  {H.}~\bibnamefont {{Middleton}}}, \bibinfo {author} {\bibfnamefont {E.~E.}\
  \bibnamefont {{Mikhailov}}}, \bibinfo {author} {\bibfnamefont
  {A.}~\bibnamefont {{Miller}}}, \bibinfo {author} {\bibfnamefont
  {J.}~\bibnamefont {{Miller}}}, \bibinfo {author} {\bibfnamefont
  {M.}~\bibnamefont {{Millhouse}}}, \bibinfo {author} {\bibfnamefont
  {J.}~\bibnamefont {{Ming}}}, \bibinfo {author} {\bibfnamefont
  {S.}~\bibnamefont {{Mirshekari}}}, \bibinfo {author} {\bibfnamefont
  {C.}~\bibnamefont {{Mishra}}}, \bibinfo {author} {\bibfnamefont
  {S.}~\bibnamefont {{Mitra}}}, \bibinfo {author} {\bibfnamefont {V.~P.}\
  \bibnamefont {{Mitrofanov}}}, \bibinfo {author} {\bibfnamefont
  {G.}~\bibnamefont {{Mitselmakher}}}, \bibinfo {author} {\bibfnamefont
  {R.}~\bibnamefont {{Mittleman}}}, \bibinfo {author} {\bibfnamefont
  {B.}~\bibnamefont {{Moe}}}, \bibinfo {author} {\bibfnamefont {S.~D.}\
  \bibnamefont {{Mohanty}}}, \bibinfo {author} {\bibfnamefont {S.~R.~P.}\
  \bibnamefont {{Mohapatra}}}, \bibinfo {author} {\bibfnamefont
  {B.}~\bibnamefont {{Moore}}}, \bibinfo {author} {\bibfnamefont
  {D.}~\bibnamefont {{Moraru}}}, \bibinfo {author} {\bibfnamefont
  {G.}~\bibnamefont {{Moreno}}}, \bibinfo {author} {\bibfnamefont {S.~R.}\
  \bibnamefont {{Morriss}}}, \bibinfo {author} {\bibfnamefont {K.}~\bibnamefont
  {{Mossavi}}}, \bibinfo {author} {\bibfnamefont {C.~M.}\ \bibnamefont
  {{Mow-Lowry}}}, \bibinfo {author} {\bibfnamefont {C.~L.}\ \bibnamefont
  {{Mueller}}}, \bibinfo {author} {\bibfnamefont {G.}~\bibnamefont
  {{Mueller}}}, \bibinfo {author} {\bibfnamefont {S.}~\bibnamefont
  {{Mukherjee}}}, \bibinfo {author} {\bibfnamefont {A.}~\bibnamefont
  {{Mullavey}}}, \bibinfo {author} {\bibfnamefont {J.}~\bibnamefont {{Munch}}},
  \bibinfo {author} {\bibfnamefont {D.}~\bibnamefont {{Murphy}}}, \bibinfo
  {author} {\bibfnamefont {P.~G.}\ \bibnamefont {{Murray}}}, \bibinfo {author}
  {\bibfnamefont {A.}~\bibnamefont {{Mytidis}}}, \bibinfo {author}
  {\bibfnamefont {T.}~\bibnamefont {{Nash}}}, \bibinfo {author} {\bibfnamefont
  {R.~K.}\ \bibnamefont {{Nayak}}}, \bibinfo {author} {\bibfnamefont
  {V.}~\bibnamefont {{Necula}}}, \bibinfo {author} {\bibfnamefont
  {K.}~\bibnamefont {{Nedkova}}}, \bibinfo {author} {\bibfnamefont
  {G.}~\bibnamefont {{Newton}}}, \bibinfo {author} {\bibfnamefont
  {T.}~\bibnamefont {{Nguyen}}}, \bibinfo {author} {\bibfnamefont {A.~B.}\
  \bibnamefont {{Nielsen}}}, \bibinfo {author} {\bibfnamefont {S.}~\bibnamefont
  {{Nissanke}}}, \bibinfo {author} {\bibfnamefont {A.~H.}\ \bibnamefont
  {{Nitz}}}, \bibinfo {author} {\bibfnamefont {D.}~\bibnamefont {{Nolting}}},
  \bibinfo {author} {\bibfnamefont {M.~E.~N.}\ \bibnamefont {{Normandin}}},
  \bibinfo {author} {\bibfnamefont {L.~K.}\ \bibnamefont {{Nuttall}}}, \bibinfo
  {author} {\bibfnamefont {E.}~\bibnamefont {{Ochsner}}}, \bibinfo {author}
  {\bibfnamefont {J.}~\bibnamefont {{O'Dell}}}, \bibinfo {author}
  {\bibfnamefont {E.}~\bibnamefont {{Oelker}}}, \bibinfo {author}
  {\bibfnamefont {G.~H.}\ \bibnamefont {{Ogin}}}, \bibinfo {author}
  {\bibfnamefont {J.~J.}\ \bibnamefont {{Oh}}}, \bibinfo {author}
  {\bibfnamefont {S.~H.}\ \bibnamefont {{Oh}}}, \bibinfo {author}
  {\bibfnamefont {F.}~\bibnamefont {{Ohme}}}, \bibinfo {author} {\bibfnamefont
  {P.}~\bibnamefont {{Oppermann}}}, \bibinfo {author} {\bibfnamefont
  {R.}~\bibnamefont {{Oram}}}, \bibinfo {author} {\bibfnamefont
  {B.}~\bibnamefont {{O'Reilly}}}, \bibinfo {author} {\bibfnamefont
  {W.}~\bibnamefont {{Ortega}}}, \bibinfo {author} {\bibfnamefont
  {R.}~\bibnamefont {{O'Shaughnessy}}}, \bibinfo {author} {\bibfnamefont
  {C.}~\bibnamefont {{Osthelder}}}, \bibinfo {author} {\bibfnamefont {C.~D.}\
  \bibnamefont {{Ott}}}, \bibinfo {author} {\bibfnamefont {D.~J.}\ \bibnamefont
  {{Ottaway}}}, \bibinfo {author} {\bibfnamefont {R.~S.}\ \bibnamefont
  {{Ottens}}}, \bibinfo {author} {\bibfnamefont {H.}~\bibnamefont
  {{Overmier}}}, \bibinfo {author} {\bibfnamefont {B.~J.}\ \bibnamefont
  {{Owen}}}, \bibinfo {author} {\bibfnamefont {C.}~\bibnamefont {{Padilla}}},
  \bibinfo {author} {\bibfnamefont {A.}~\bibnamefont {{Pai}}}, \bibinfo
  {author} {\bibfnamefont {S.}~\bibnamefont {{Pai}}}, \bibinfo {author}
  {\bibfnamefont {O.}~\bibnamefont {{Palashov}}}, \bibinfo {author}
  {\bibfnamefont {A.}~\bibnamefont {{Pal-Singh}}}, \bibinfo {author}
  {\bibfnamefont {H.}~\bibnamefont {{Pan}}}, \bibinfo {author} {\bibfnamefont
  {C.}~\bibnamefont {{Pankow}}}, \bibinfo {author} {\bibfnamefont
  {F.}~\bibnamefont {{Pannarale}}}, \bibinfo {author} {\bibfnamefont {B.~C.}\
  \bibnamefont {{Pant}}}, \bibinfo {author} {\bibfnamefont {M.~A.}\
  \bibnamefont {{Papa}}}, \bibinfo {author} {\bibfnamefont {H.}~\bibnamefont
  {{Paris}}}, \bibinfo {author} {\bibfnamefont {Z.}~\bibnamefont {{Patrick}}},
  \bibinfo {author} {\bibfnamefont {M.}~\bibnamefont {{Pedraza}}}, \bibinfo
  {author} {\bibfnamefont {L.}~\bibnamefont {{Pekowsky}}}, \bibinfo {author}
  {\bibfnamefont {A.}~\bibnamefont {{Pele}}}, \bibinfo {author} {\bibfnamefont
  {S.}~\bibnamefont {{Penn}}}, \bibinfo {author} {\bibfnamefont
  {A.}~\bibnamefont {{Perreca}}}, \bibinfo {author} {\bibfnamefont
  {M.}~\bibnamefont {{Phelps}}}, \bibinfo {author} {\bibfnamefont
  {V.}~\bibnamefont {{Pierro}}}, \bibinfo {author} {\bibfnamefont {I.~M.}\
  \bibnamefont {{Pinto}}}, \bibinfo {author} {\bibfnamefont {M.}~\bibnamefont
  {{Pitkin}}}, \bibinfo {author} {\bibfnamefont {J.}~\bibnamefont {{Poeld}}},
  \bibinfo {author} {\bibfnamefont {A.}~\bibnamefont {{Post}}}, \bibinfo
  {author} {\bibfnamefont {A.}~\bibnamefont {{Poteomkin}}}, \bibinfo {author}
  {\bibfnamefont {J.}~\bibnamefont {{Powell}}}, \bibinfo {author}
  {\bibfnamefont {J.}~\bibnamefont {{Prasad}}}, \bibinfo {author}
  {\bibfnamefont {V.}~\bibnamefont {{Predoi}}}, \bibinfo {author}
  {\bibfnamefont {S.}~\bibnamefont {{Premachandra}}}, \bibinfo {author}
  {\bibfnamefont {T.}~\bibnamefont {{Prestegard}}}, \bibinfo {author}
  {\bibfnamefont {L.~R.}\ \bibnamefont {{Price}}}, \bibinfo {author}
  {\bibfnamefont {M.}~\bibnamefont {{Principe}}}, \bibinfo {author}
  {\bibfnamefont {S.}~\bibnamefont {{Privitera}}}, \bibinfo {author}
  {\bibfnamefont {R.}~\bibnamefont {{Prix}}}, \bibinfo {author} {\bibfnamefont
  {L.}~\bibnamefont {{Prokhorov}}}, \bibinfo {author} {\bibfnamefont
  {O.}~\bibnamefont {{Puncken}}}, \bibinfo {author} {\bibfnamefont
  {M.}~\bibnamefont {{P{\"u}rrer}}}, \bibinfo {author} {\bibfnamefont
  {J.}~\bibnamefont {{Qin}}}, \bibinfo {author} {\bibfnamefont
  {V.}~\bibnamefont {{Quetschke}}}, \bibinfo {author} {\bibfnamefont
  {E.}~\bibnamefont {{Quintero}}}, \bibinfo {author} {\bibfnamefont
  {G.}~\bibnamefont {{Quiroga}}}, \bibinfo {author} {\bibfnamefont
  {R.}~\bibnamefont {{Quitzow-James}}}, \bibinfo {author} {\bibfnamefont
  {F.~J.}\ \bibnamefont {{Raab}}}, \bibinfo {author} {\bibfnamefont {D.~S.}\
  \bibnamefont {{Rabeling}}}, \bibinfo {author} {\bibfnamefont
  {H.}~\bibnamefont {{Radkins}}}, \bibinfo {author} {\bibfnamefont
  {P.}~\bibnamefont {{Raffai}}}, \bibinfo {author} {\bibfnamefont
  {S.}~\bibnamefont {{Raja}}}, \bibinfo {author} {\bibfnamefont
  {G.}~\bibnamefont {{Rajalakshmi}}}, \bibinfo {author} {\bibfnamefont
  {M.}~\bibnamefont {{Rakhmanov}}}, \bibinfo {author} {\bibfnamefont
  {K.}~\bibnamefont {{Ramirez}}}, \bibinfo {author} {\bibfnamefont
  {V.}~\bibnamefont {{Raymond}}}, \bibinfo {author} {\bibfnamefont {C.~M.}\
  \bibnamefont {{Reed}}}, \bibinfo {author} {\bibfnamefont {S.}~\bibnamefont
  {{Reid}}}, \bibinfo {author} {\bibfnamefont {D.~H.}\ \bibnamefont
  {{Reitze}}}, \bibinfo {author} {\bibfnamefont {O.}~\bibnamefont {{Reula}}},
  \bibinfo {author} {\bibfnamefont {K.}~\bibnamefont {{Riles}}}, \bibinfo
  {author} {\bibfnamefont {N.~A.}\ \bibnamefont {{Robertson}}}, \bibinfo
  {author} {\bibfnamefont {R.}~\bibnamefont {{Robie}}}, \bibinfo {author}
  {\bibfnamefont {J.~G.}\ \bibnamefont {{Rollins}}}, \bibinfo {author}
  {\bibfnamefont {V.}~\bibnamefont {{Roma}}}, \bibinfo {author} {\bibfnamefont
  {J.~D.}\ \bibnamefont {{Romano}}}, \bibinfo {author} {\bibfnamefont
  {G.}~\bibnamefont {{Romanov}}}, \bibinfo {author} {\bibfnamefont {J.~H.}\
  \bibnamefont {{Romie}}}, \bibinfo {author} {\bibfnamefont {S.}~\bibnamefont
  {{Rowan}}}, \bibinfo {author} {\bibfnamefont {A.}~\bibnamefont
  {{R{\"u}diger}}}, \bibinfo {author} {\bibfnamefont {K.}~\bibnamefont
  {{Ryan}}}, \bibinfo {author} {\bibfnamefont {S.}~\bibnamefont {{Sachdev}}},
  \bibinfo {author} {\bibfnamefont {T.}~\bibnamefont {{Sadecki}}}, \bibinfo
  {author} {\bibfnamefont {L.}~\bibnamefont {{Sadeghian}}}, \bibinfo {author}
  {\bibfnamefont {M.}~\bibnamefont {{Saleem}}}, \bibinfo {author}
  {\bibfnamefont {F.}~\bibnamefont {{Salemi}}}, \bibinfo {author}
  {\bibfnamefont {L.}~\bibnamefont {{Sammut}}}, \bibinfo {author}
  {\bibfnamefont {V.}~\bibnamefont {{Sandberg}}}, \bibinfo {author}
  {\bibfnamefont {J.~R.}\ \bibnamefont {{Sanders}}}, \bibinfo {author}
  {\bibfnamefont {V.}~\bibnamefont {{Sannibale}}}, \bibinfo {author}
  {\bibfnamefont {I.}~\bibnamefont {{Santiago-Prieto}}}, \bibinfo {author}
  {\bibfnamefont {B.~S.}\ \bibnamefont {{Sathyaprakash}}}, \bibinfo {author}
  {\bibfnamefont {P.~R.}\ \bibnamefont {{Saulson}}}, \bibinfo {author}
  {\bibfnamefont {R.}~\bibnamefont {{Savage}}}, \bibinfo {author}
  {\bibfnamefont {A.}~\bibnamefont {{Sawadsky}}}, \bibinfo {author}
  {\bibfnamefont {J.}~\bibnamefont {{Scheuer}}}, \bibinfo {author}
  {\bibfnamefont {R.}~\bibnamefont {{Schilling}}}, \bibinfo {author}
  {\bibfnamefont {P.}~\bibnamefont {{Schmidt}}}, \bibinfo {author}
  {\bibfnamefont {R.}~\bibnamefont {{Schnabel}}}, \bibinfo {author}
  {\bibfnamefont {R.~M.~S.}\ \bibnamefont {{Schofield}}}, \bibinfo {author}
  {\bibfnamefont {E.}~\bibnamefont {{Schreiber}}}, \bibinfo {author}
  {\bibfnamefont {D.}~\bibnamefont {{Schuette}}}, \bibinfo {author}
  {\bibfnamefont {B.~F.}\ \bibnamefont {{Schutz}}}, \bibinfo {author}
  {\bibfnamefont {J.}~\bibnamefont {{Scott}}}, \bibinfo {author} {\bibfnamefont
  {S.~M.}\ \bibnamefont {{Scott}}}, \bibinfo {author} {\bibfnamefont
  {D.}~\bibnamefont {{Sellers}}}, \bibinfo {author} {\bibfnamefont {A.~S.}\
  \bibnamefont {{Sengupta}}}, \bibinfo {author} {\bibfnamefont
  {A.}~\bibnamefont {{Sergeev}}}, \bibinfo {author} {\bibfnamefont
  {G.}~\bibnamefont {{Serna}}}, \bibinfo {author} {\bibfnamefont
  {A.}~\bibnamefont {{Sevigny}}}, \bibinfo {author} {\bibfnamefont {D.~A.}\
  \bibnamefont {{Shaddock}}}, \bibinfo {author} {\bibfnamefont {M.~S.}\
  \bibnamefont {{Shahriar}}}, \bibinfo {author} {\bibfnamefont
  {M.}~\bibnamefont {{Shaltev}}}, \bibinfo {author} {\bibfnamefont
  {Z.}~\bibnamefont {{Shao}}}, \bibinfo {author} {\bibfnamefont
  {B.}~\bibnamefont {{Shapiro}}}, \bibinfo {author} {\bibfnamefont
  {P.}~\bibnamefont {{Shawhan}}}, \bibinfo {author} {\bibfnamefont {D.~H.}\
  \bibnamefont {{Shoemaker}}}, \bibinfo {author} {\bibfnamefont {T.~L.}\
  \bibnamefont {{Sidery}}}, \bibinfo {author} {\bibfnamefont {X.}~\bibnamefont
  {{Siemens}}}, \bibinfo {author} {\bibfnamefont {D.}~\bibnamefont {{Sigg}}},
  \bibinfo {author} {\bibfnamefont {A.~D.}\ \bibnamefont {{Silva}}}, \bibinfo
  {author} {\bibfnamefont {D.}~\bibnamefont {{Simakov}}}, \bibinfo {author}
  {\bibfnamefont {A.}~\bibnamefont {{Singer}}}, \bibinfo {author}
  {\bibfnamefont {L.}~\bibnamefont {{Singer}}}, \bibinfo {author}
  {\bibfnamefont {R.}~\bibnamefont {{Singh}}}, \bibinfo {author} {\bibfnamefont
  {A.~M.}\ \bibnamefont {{Sintes}}}, \bibinfo {author} {\bibfnamefont
  {B.~J.~J.}\ \bibnamefont {{Slagmolen}}}, \bibinfo {author} {\bibfnamefont
  {J.~R.}\ \bibnamefont {{Smith}}}, \bibinfo {author} {\bibfnamefont {M.~R.}\
  \bibnamefont {{Smith}}}, \bibinfo {author} {\bibfnamefont {R.~J.~E.}\
  \bibnamefont {{Smith}}}, \bibinfo {author} {\bibfnamefont {N.~D.}\
  \bibnamefont {{Smith-Lefebvre}}}, \bibinfo {author} {\bibfnamefont {E.~J.}\
  \bibnamefont {{Son}}}, \bibinfo {author} {\bibfnamefont {B.}~\bibnamefont
  {{Sorazu}}}, \bibinfo {author} {\bibfnamefont {T.}~\bibnamefont
  {{Souradeep}}}, \bibinfo {author} {\bibfnamefont {A.}~\bibnamefont
  {{Staley}}}, \bibinfo {author} {\bibfnamefont {J.}~\bibnamefont
  {{Stebbins}}}, \bibinfo {author} {\bibfnamefont {M.}~\bibnamefont
  {{Steinke}}}, \bibinfo {author} {\bibfnamefont {J.}~\bibnamefont
  {{Steinlechner}}}, \bibinfo {author} {\bibfnamefont {S.}~\bibnamefont
  {{Steinlechner}}}, \bibinfo {author} {\bibfnamefont {D.}~\bibnamefont
  {{Steinmeyer}}}, \bibinfo {author} {\bibfnamefont {B.~C.}\ \bibnamefont
  {{Stephens}}}, \bibinfo {author} {\bibfnamefont {S.}~\bibnamefont
  {{Steplewski}}}, \bibinfo {author} {\bibfnamefont {S.}~\bibnamefont
  {{Stevenson}}}, \bibinfo {author} {\bibfnamefont {R.}~\bibnamefont
  {{Stone}}}, \bibinfo {author} {\bibfnamefont {K.~A.}\ \bibnamefont
  {{Strain}}}, \bibinfo {author} {\bibfnamefont {S.}~\bibnamefont {{Strigin}}},
  \bibinfo {author} {\bibfnamefont {R.}~\bibnamefont {{Sturani}}}, \bibinfo
  {author} {\bibfnamefont {A.~L.}\ \bibnamefont {{Stuver}}}, \bibinfo {author}
  {\bibfnamefont {T.~Z.}\ \bibnamefont {{Summerscales}}}, \bibinfo {author}
  {\bibfnamefont {P.~J.}\ \bibnamefont {{Sutton}}}, \bibinfo {author}
  {\bibfnamefont {M.}~\bibnamefont {{Szczepanczyk}}}, \bibinfo {author}
  {\bibfnamefont {G.}~\bibnamefont {{Szeifert}}}, \bibinfo {author}
  {\bibfnamefont {D.}~\bibnamefont {{Talukder}}}, \bibinfo {author}
  {\bibfnamefont {D.~B.}\ \bibnamefont {{Tanner}}}, \bibinfo {author}
  {\bibfnamefont {M.}~\bibnamefont {{T{\'a}pai}}}, \bibinfo {author}
  {\bibfnamefont {S.~P.}\ \bibnamefont {{Tarabrin}}}, \bibinfo {author}
  {\bibfnamefont {A.}~\bibnamefont {{Taracchini}}}, \bibinfo {author}
  {\bibfnamefont {R.}~\bibnamefont {{Taylor}}}, \bibinfo {author}
  {\bibfnamefont {G.}~\bibnamefont {{Tellez}}}, \bibinfo {author}
  {\bibfnamefont {T.}~\bibnamefont {{Theeg}}}, \bibinfo {author} {\bibfnamefont
  {M.~P.}\ \bibnamefont {{Thirugnanasambandam}}}, \bibinfo {author}
  {\bibfnamefont {M.}~\bibnamefont {{Thomas}}}, \bibinfo {author}
  {\bibfnamefont {P.}~\bibnamefont {{Thomas}}}, \bibinfo {author}
  {\bibfnamefont {K.~A.}\ \bibnamefont {{Thorne}}}, \bibinfo {author}
  {\bibfnamefont {K.~S.}\ \bibnamefont {{Thorne}}}, \bibinfo {author}
  {\bibfnamefont {E.}~\bibnamefont {{Thrane}}}, \bibinfo {author}
  {\bibfnamefont {V.}~\bibnamefont {{Tiwari}}}, \bibinfo {author}
  {\bibfnamefont {C.}~\bibnamefont {{Tomlinson}}}, \bibinfo {author}
  {\bibfnamefont {C.~V.}\ \bibnamefont {{Torres}}}, \bibinfo {author}
  {\bibfnamefont {C.~I.}\ \bibnamefont {{Torrie}}}, \bibinfo {author}
  {\bibfnamefont {G.}~\bibnamefont {{Traylor}}}, \bibinfo {author}
  {\bibfnamefont {M.}~\bibnamefont {{Tse}}}, \bibinfo {author} {\bibfnamefont
  {D.}~\bibnamefont {{Tshilumba}}}, \bibinfo {author} {\bibfnamefont
  {D.}~\bibnamefont {{Ugolini}}}, \bibinfo {author} {\bibfnamefont {C.~S.}\
  \bibnamefont {{Unnikrishnan}}}, \bibinfo {author} {\bibfnamefont {A.~L.}\
  \bibnamefont {{Urban}}}, \bibinfo {author} {\bibfnamefont {S.~A.}\
  \bibnamefont {{Usman}}}, \bibinfo {author} {\bibfnamefont {H.}~\bibnamefont
  {{Vahlbruch}}}, \bibinfo {author} {\bibfnamefont {G.}~\bibnamefont
  {{Vajente}}}, \bibinfo {author} {\bibfnamefont {G.}~\bibnamefont {{Valdes}}},
  \bibinfo {author} {\bibfnamefont {M.}~\bibnamefont {{Vallisneri}}}, \bibinfo
  {author} {\bibfnamefont {A.~A.}\ \bibnamefont {{van Veggel}}}, \bibinfo
  {author} {\bibfnamefont {S.}~\bibnamefont {{Vass}}}, \bibinfo {author}
  {\bibfnamefont {R.}~\bibnamefont {{Vaulin}}}, \bibinfo {author}
  {\bibfnamefont {A.}~\bibnamefont {{Vecchio}}}, \bibinfo {author}
  {\bibfnamefont {J.}~\bibnamefont {{Veitch}}}, \bibinfo {author}
  {\bibfnamefont {P.~J.}\ \bibnamefont {{Veitch}}}, \bibinfo {author}
  {\bibfnamefont {K.}~\bibnamefont {{Venkateswara}}}, \bibinfo {author}
  {\bibfnamefont {R.}~\bibnamefont {{Vincent-Finley}}}, \bibinfo {author}
  {\bibfnamefont {S.}~\bibnamefont {{Vitale}}}, \bibinfo {author}
  {\bibfnamefont {T.}~\bibnamefont {{Vo}}}, \bibinfo {author} {\bibfnamefont
  {C.}~\bibnamefont {{Vorvick}}}, \bibinfo {author} {\bibfnamefont {W.~D.}\
  \bibnamefont {{Vousden}}}, \bibinfo {author} {\bibfnamefont {S.~P.}\
  \bibnamefont {{Vyatchanin}}}, \bibinfo {author} {\bibfnamefont {A.~R.}\
  \bibnamefont {{Wade}}}, \bibinfo {author} {\bibfnamefont {L.}~\bibnamefont
  {{Wade}}}, \bibinfo {author} {\bibfnamefont {M.}~\bibnamefont {{Wade}}},
  \bibinfo {author} {\bibfnamefont {M.}~\bibnamefont {{Walker}}}, \bibinfo
  {author} {\bibfnamefont {L.}~\bibnamefont {{Wallace}}}, \bibinfo {author}
  {\bibfnamefont {S.}~\bibnamefont {{Walsh}}}, \bibinfo {author} {\bibfnamefont
  {H.}~\bibnamefont {{Wang}}}, \bibinfo {author} {\bibfnamefont
  {M.}~\bibnamefont {{Wang}}}, \bibinfo {author} {\bibfnamefont
  {X.}~\bibnamefont {{Wang}}}, \bibinfo {author} {\bibfnamefont {R.~L.}\
  \bibnamefont {{Ward}}}, \bibinfo {author} {\bibfnamefont {J.}~\bibnamefont
  {{Warner}}}, \bibinfo {author} {\bibfnamefont {M.}~\bibnamefont {{Was}}},
  \bibinfo {author} {\bibfnamefont {B.}~\bibnamefont {{Weaver}}}, \bibinfo
  {author} {\bibfnamefont {M.}~\bibnamefont {{Weinert}}}, \bibinfo {author}
  {\bibfnamefont {A.~J.}\ \bibnamefont {{Weinstein}}}, \bibinfo {author}
  {\bibfnamefont {R.}~\bibnamefont {{Weiss}}}, \bibinfo {author} {\bibfnamefont
  {T.}~\bibnamefont {{Welborn}}}, \bibinfo {author} {\bibfnamefont
  {L.}~\bibnamefont {{Wen}}}, \bibinfo {author} {\bibfnamefont
  {P.}~\bibnamefont {{Wessels}}}, \bibinfo {author} {\bibfnamefont
  {T.}~\bibnamefont {{Westphal}}}, \bibinfo {author} {\bibfnamefont
  {K.}~\bibnamefont {{Wette}}}, \bibinfo {author} {\bibfnamefont {J.~T.}\
  \bibnamefont {{Whelan}}}, \bibinfo {author} {\bibfnamefont {S.~E.}\
  \bibnamefont {{Whitcomb}}}, \bibinfo {author} {\bibfnamefont {D.~J.}\
  \bibnamefont {{White}}}, \bibinfo {author} {\bibfnamefont {B.~F.}\
  \bibnamefont {{Whiting}}}, \bibinfo {author} {\bibfnamefont {C.}~\bibnamefont
  {{Wilkinson}}}, \bibinfo {author} {\bibfnamefont {L.}~\bibnamefont
  {{Williams}}}, \bibinfo {author} {\bibfnamefont {R.}~\bibnamefont
  {{Williams}}}, \bibinfo {author} {\bibfnamefont {A.~R.}\ \bibnamefont
  {{Williamson}}}, \bibinfo {author} {\bibfnamefont {J.~L.}\ \bibnamefont
  {{Willis}}}, \bibinfo {author} {\bibfnamefont {B.}~\bibnamefont {{Willke}}},
  \bibinfo {author} {\bibfnamefont {M.}~\bibnamefont {{Wimmer}}}, \bibinfo
  {author} {\bibfnamefont {W.}~\bibnamefont {{Winkler}}}, \bibinfo {author}
  {\bibfnamefont {C.~C.}\ \bibnamefont {{Wipf}}}, \bibinfo {author}
  {\bibfnamefont {H.}~\bibnamefont {{Wittel}}}, \bibinfo {author}
  {\bibfnamefont {G.}~\bibnamefont {{Woan}}}, \bibinfo {author} {\bibfnamefont
  {J.}~\bibnamefont {{Worden}}}, \bibinfo {author} {\bibfnamefont
  {S.}~\bibnamefont {{Xie}}}, \bibinfo {author} {\bibfnamefont
  {J.}~\bibnamefont {{Yablon}}}, \bibinfo {author} {\bibfnamefont
  {I.}~\bibnamefont {{Yakushin}}}, \bibinfo {author} {\bibfnamefont
  {W.}~\bibnamefont {{Yam}}}, \bibinfo {author} {\bibfnamefont
  {H.}~\bibnamefont {{Yamamoto}}}, \bibinfo {author} {\bibfnamefont {C.~C.}\
  \bibnamefont {{Yancey}}}, \bibinfo {author} {\bibfnamefont {Q.}~\bibnamefont
  {{Yang}}}, \bibinfo {author} {\bibfnamefont {M.}~\bibnamefont {{Zanolin}}},
  \bibinfo {author} {\bibfnamefont {F.}~\bibnamefont {{Zhang}}}, \bibinfo
  {author} {\bibfnamefont {L.}~\bibnamefont {{Zhang}}}, \bibinfo {author}
  {\bibfnamefont {M.}~\bibnamefont {{Zhang}}}, \bibinfo {author} {\bibfnamefont
  {Y.}~\bibnamefont {{Zhang}}}, \bibinfo {author} {\bibfnamefont
  {C.}~\bibnamefont {{Zhao}}}, \bibinfo {author} {\bibfnamefont
  {M.}~\bibnamefont {{Zhou}}}, \bibinfo {author} {\bibfnamefont {X.~J.}\
  \bibnamefont {{Zhu}}}, \bibinfo {author} {\bibfnamefont {M.~E.}\ \bibnamefont
  {{Zucker}}}, \bibinfo {author} {\bibfnamefont {S.}~\bibnamefont {{Zuraw}}},\
  and\ \bibinfo {author} {\bibfnamefont {J.}~\bibnamefont {{Zweizig}}},\
  }\bibfield  {title} {\bibinfo {title} {{Advanced LIGO}},\ }\href
  {https://doi.org/10.1088/0264-9381/32/7/074001} {\bibfield  {journal}
  {\bibinfo  {journal} {Classical and Quantum Gravity}\ }\textbf {\bibinfo
  {volume} {32}},\ \bibinfo {eid} {074001} (\bibinfo {year} {2015})},\ \Eprint
  {https://arxiv.org/abs/1411.4547} {arXiv:1411.4547 [gr-qc]} \BibitemShut
  {NoStop}%
\bibitem [{\citenamefont {Acernese}\ \emph {et~al.}(2014)\citenamefont
  {Acernese}, \citenamefont {Agathos}, \citenamefont {Agatsuma}, \citenamefont
  {Aisa}, \citenamefont {Allemandou}, \citenamefont {Allocca}, \citenamefont
  {Amarni}, \citenamefont {Astone}, \citenamefont {Balestri}, \citenamefont
  {Ballardin}, \citenamefont {Barone}, \citenamefont {Baronick}, \citenamefont
  {Barsuglia}, \citenamefont {Basti}, \citenamefont {Basti}, \citenamefont
  {Bauer}, \citenamefont {Bavigadda}, \citenamefont {Bejger}, \citenamefont
  {Beker}, \citenamefont {Belczynski}, \citenamefont {Bersanetti},
  \citenamefont {Bertolini}, \citenamefont {Bitossi}, \citenamefont {Bizouard},
  \citenamefont {Bloemen}, \citenamefont {Blom}, \citenamefont {Boer},
  \citenamefont {Bogaert}, \citenamefont {Bondi}, \citenamefont {Bondu},
  \citenamefont {Bonelli}, \citenamefont {Bonnand}, \citenamefont {Boschi},
  \citenamefont {Bosi}, \citenamefont {Bouedo}, \citenamefont {Bradaschia},
  \citenamefont {Branchesi}, \citenamefont {Briant}, \citenamefont {Brillet},
  \citenamefont {Brisson}, \citenamefont {Bulik}, \citenamefont {Bulten},
  \citenamefont {Buskulic}, \citenamefont {Buy}, \citenamefont {Cagnoli},
  \citenamefont {Calloni}, \citenamefont {Campeggi}, \citenamefont {Canuel},
  \citenamefont {Carbognani}, \citenamefont {Cavalier}, \citenamefont
  {Cavalieri}, \citenamefont {Cella}, \citenamefont {Cesarini}, \citenamefont
  {Chassande-Mottin}, \citenamefont {Chincarini}, \citenamefont {Chiummo},
  \citenamefont {Chua}, \citenamefont {Cleva}, \citenamefont {Coccia},
  \citenamefont {Cohadon}, \citenamefont {Colla}, \citenamefont {Colombini},
  \citenamefont {Conte}, \citenamefont {Coulon}, \citenamefont {Cuoco},
  \citenamefont {Dalmaz}, \citenamefont {D'Antonio}, \citenamefont {Dattilo},
  \citenamefont {Davier}, \citenamefont {Day}, \citenamefont {Debreczeni},
  \citenamefont {Degallaix}, \citenamefont {Del{\'{e} }glise}, \citenamefont
  {Pozzo}, \citenamefont {Dereli}, \citenamefont {Rosa}, \citenamefont {Fiore},
  \citenamefont {Lieto}, \citenamefont {Virgilio}, \citenamefont {Doets},
  \citenamefont {Dolique}, \citenamefont {Drago}, \citenamefont {Ducrot},
  \citenamefont {Endr{\H{o}}czi}, \citenamefont {Fafone}, \citenamefont
  {Farinon}, \citenamefont {Ferrante}, \citenamefont {Ferrini}, \citenamefont
  {Fidecaro}, \citenamefont {Fiori}, \citenamefont {Flaminio}, \citenamefont
  {Fournier}, \citenamefont {Franco}, \citenamefont {Frasca}, \citenamefont
  {Frasconi}, \citenamefont {Gammaitoni}, \citenamefont {Garufi}, \citenamefont
  {Gaspard}, \citenamefont {Gatto}, \citenamefont {Gemme}, \citenamefont
  {Gendre}, \citenamefont {Genin}, \citenamefont {Gennai}, \citenamefont
  {Ghosh}, \citenamefont {Giacobone}, \citenamefont {Giazotto}, \citenamefont
  {Gouaty}, \citenamefont {Granata}, \citenamefont {Greco}, \citenamefont
  {Groot}, \citenamefont {Guidi}, \citenamefont {Harms}, \citenamefont
  {Heidmann}, \citenamefont {Heitmann}, \citenamefont {Hello}, \citenamefont
  {Hemming}, \citenamefont {Hennes}, \citenamefont {Hofman}, \citenamefont
  {Jaranowski}, \citenamefont {Jonker}, \citenamefont {Kasprzack},
  \citenamefont {K{\'{e}}f{\'{e}}lian}, \citenamefont {Kowalska}, \citenamefont
  {Kraan}, \citenamefont {Kr{\'{o}}lak}, \citenamefont {Kutynia}, \citenamefont
  {Lazzaro}, \citenamefont {Leonardi}, \citenamefont {Leroy}, \citenamefont
  {Letendre}, \citenamefont {Li}, \citenamefont {Lieunard}, \citenamefont
  {Lorenzini}, \citenamefont {Loriette}, \citenamefont {Losurdo}, \citenamefont
  {Magazz{\`{u}}}, \citenamefont {Majorana}, \citenamefont {Maksimovic},
  \citenamefont {Malvezzi}, \citenamefont {Man}, \citenamefont {Mangano},
  \citenamefont {Mantovani}, \citenamefont {Marchesoni}, \citenamefont
  {Marion}, \citenamefont {Marque}, \citenamefont {Martelli}, \citenamefont
  {Martellini}, \citenamefont {Masserot}, \citenamefont {Meacher},
  \citenamefont {Meidam}, \citenamefont {Mezzani}, \citenamefont {Michel},
  \citenamefont {Milano}, \citenamefont {Minenkov}, \citenamefont {Moggi},
  \citenamefont {Mohan}, \citenamefont {Montani}, \citenamefont {Morgado},
  \citenamefont {Mours}, \citenamefont {Mul}, \citenamefont {Nagy},
  \citenamefont {Nardecchia}, \citenamefont {Naticchioni}, \citenamefont
  {Nelemans}, \citenamefont {Neri}, \citenamefont {Neri}, \citenamefont
  {Nocera}, \citenamefont {Pacaud}, \citenamefont {Palomba}, \citenamefont
  {Paoletti}, \citenamefont {Paoli}, \citenamefont {Pasqualetti}, \citenamefont
  {Passaquieti}, \citenamefont {Passuello}, \citenamefont {Perciballi},
  \citenamefont {Petit}, \citenamefont {Pichot}, \citenamefont {Piergiovanni},
  \citenamefont {Pillant}, \citenamefont {Piluso}, \citenamefont {Pinard},
  \citenamefont {Poggiani}, \citenamefont {Prijatelj}, \citenamefont {Prodi},
  \citenamefont {Punturo}, \citenamefont {Puppo}, \citenamefont {Rabeling},
  \citenamefont {R{\'{a}}cz}, \citenamefont {Rapagnani}, \citenamefont
  {Razzano}, \citenamefont {Re}, \citenamefont {Regimbau}, \citenamefont
  {Ricci}, \citenamefont {Robinet}, \citenamefont {Rocchi}, \citenamefont
  {Rolland}, \citenamefont {Romano}, \citenamefont {Rosi{\'{n}}ska},
  \citenamefont {Ruggi}, \citenamefont {Saracco}, \citenamefont {Sassolas},
  \citenamefont {Schimmel}, \citenamefont {Sentenac}, \citenamefont {Sequino},
  \citenamefont {Shah}, \citenamefont {Siellez}, \citenamefont {Straniero},
  \citenamefont {Swinkels}, \citenamefont {Tacca}, \citenamefont {Tonelli},
  \citenamefont {Travasso}, \citenamefont {Turconi}, \citenamefont {Vajente},
  \citenamefont {van Bakel}, \citenamefont {van Beuzekom}, \citenamefont
  {van~den Brand}, \citenamefont {Broeck}, \citenamefont {van~der Sluys},
  \citenamefont {van Heijningen}, \citenamefont {Vas{\'{u}}th}, \citenamefont
  {Vedovato}, \citenamefont {Veitch}, \citenamefont {Verkindt}, \citenamefont
  {Vetrano}, \citenamefont {Vicer{\'{e}}}, \citenamefont {Vinet}, \citenamefont
  {Visser}, \citenamefont {Vocca}, \citenamefont {Ward}, \citenamefont {Was},
  \citenamefont {Wei}, \citenamefont {Yvert}, \citenamefont {{\.{z}}ny},\ and\
  \citenamefont {Zendri}}]{Acernese_2014}%
  \BibitemOpen
  \bibfield  {author} {\bibinfo {author} {\bibfnamefont {F.}~\bibnamefont
  {Acernese}}, \bibinfo {author} {\bibfnamefont {M.}~\bibnamefont {Agathos}},
  \bibinfo {author} {\bibfnamefont {K.}~\bibnamefont {Agatsuma}}, \bibinfo
  {author} {\bibfnamefont {D.}~\bibnamefont {Aisa}}, \bibinfo {author}
  {\bibfnamefont {N.}~\bibnamefont {Allemandou}}, \bibinfo {author}
  {\bibfnamefont {A.}~\bibnamefont {Allocca}}, \bibinfo {author} {\bibfnamefont
  {J.}~\bibnamefont {Amarni}}, \bibinfo {author} {\bibfnamefont
  {P.}~\bibnamefont {Astone}}, \bibinfo {author} {\bibfnamefont
  {G.}~\bibnamefont {Balestri}}, \bibinfo {author} {\bibfnamefont
  {G.}~\bibnamefont {Ballardin}}, \bibinfo {author} {\bibfnamefont
  {F.}~\bibnamefont {Barone}}, \bibinfo {author} {\bibfnamefont {J.-P.}\
  \bibnamefont {Baronick}}, \bibinfo {author} {\bibfnamefont {M.}~\bibnamefont
  {Barsuglia}}, \bibinfo {author} {\bibfnamefont {A.}~\bibnamefont {Basti}},
  \bibinfo {author} {\bibfnamefont {F.}~\bibnamefont {Basti}}, \bibinfo
  {author} {\bibfnamefont {T.~S.}\ \bibnamefont {Bauer}}, \bibinfo {author}
  {\bibfnamefont {V.}~\bibnamefont {Bavigadda}}, \bibinfo {author}
  {\bibfnamefont {M.}~\bibnamefont {Bejger}}, \bibinfo {author} {\bibfnamefont
  {M.~G.}\ \bibnamefont {Beker}}, \bibinfo {author} {\bibfnamefont
  {C.}~\bibnamefont {Belczynski}}, \bibinfo {author} {\bibfnamefont
  {D.}~\bibnamefont {Bersanetti}}, \bibinfo {author} {\bibfnamefont
  {A.}~\bibnamefont {Bertolini}}, \bibinfo {author} {\bibfnamefont
  {M.}~\bibnamefont {Bitossi}}, \bibinfo {author} {\bibfnamefont {M.~A.}\
  \bibnamefont {Bizouard}}, \bibinfo {author} {\bibfnamefont {S.}~\bibnamefont
  {Bloemen}}, \bibinfo {author} {\bibfnamefont {M.}~\bibnamefont {Blom}},
  \bibinfo {author} {\bibfnamefont {M.}~\bibnamefont {Boer}}, \bibinfo {author}
  {\bibfnamefont {G.}~\bibnamefont {Bogaert}}, \bibinfo {author} {\bibfnamefont
  {D.}~\bibnamefont {Bondi}}, \bibinfo {author} {\bibfnamefont
  {F.}~\bibnamefont {Bondu}}, \bibinfo {author} {\bibfnamefont
  {L.}~\bibnamefont {Bonelli}}, \bibinfo {author} {\bibfnamefont
  {R.}~\bibnamefont {Bonnand}}, \bibinfo {author} {\bibfnamefont
  {V.}~\bibnamefont {Boschi}}, \bibinfo {author} {\bibfnamefont
  {L.}~\bibnamefont {Bosi}}, \bibinfo {author} {\bibfnamefont {T.}~\bibnamefont
  {Bouedo}}, \bibinfo {author} {\bibfnamefont {C.}~\bibnamefont {Bradaschia}},
  \bibinfo {author} {\bibfnamefont {M.}~\bibnamefont {Branchesi}}, \bibinfo
  {author} {\bibfnamefont {T.}~\bibnamefont {Briant}}, \bibinfo {author}
  {\bibfnamefont {A.}~\bibnamefont {Brillet}}, \bibinfo {author} {\bibfnamefont
  {V.}~\bibnamefont {Brisson}}, \bibinfo {author} {\bibfnamefont
  {T.}~\bibnamefont {Bulik}}, \bibinfo {author} {\bibfnamefont {H.~J.}\
  \bibnamefont {Bulten}}, \bibinfo {author} {\bibfnamefont {D.}~\bibnamefont
  {Buskulic}}, \bibinfo {author} {\bibfnamefont {C.}~\bibnamefont {Buy}},
  \bibinfo {author} {\bibfnamefont {G.}~\bibnamefont {Cagnoli}}, \bibinfo
  {author} {\bibfnamefont {E.}~\bibnamefont {Calloni}}, \bibinfo {author}
  {\bibfnamefont {C.}~\bibnamefont {Campeggi}}, \bibinfo {author}
  {\bibfnamefont {B.}~\bibnamefont {Canuel}}, \bibinfo {author} {\bibfnamefont
  {F.}~\bibnamefont {Carbognani}}, \bibinfo {author} {\bibfnamefont
  {F.}~\bibnamefont {Cavalier}}, \bibinfo {author} {\bibfnamefont
  {R.}~\bibnamefont {Cavalieri}}, \bibinfo {author} {\bibfnamefont
  {G.}~\bibnamefont {Cella}}, \bibinfo {author} {\bibfnamefont
  {E.}~\bibnamefont {Cesarini}}, \bibinfo {author} {\bibfnamefont
  {E.}~\bibnamefont {Chassande-Mottin}}, \bibinfo {author} {\bibfnamefont
  {A.}~\bibnamefont {Chincarini}}, \bibinfo {author} {\bibfnamefont
  {A.}~\bibnamefont {Chiummo}}, \bibinfo {author} {\bibfnamefont
  {S.}~\bibnamefont {Chua}}, \bibinfo {author} {\bibfnamefont {F.}~\bibnamefont
  {Cleva}}, \bibinfo {author} {\bibfnamefont {E.}~\bibnamefont {Coccia}},
  \bibinfo {author} {\bibfnamefont {P.-F.}\ \bibnamefont {Cohadon}}, \bibinfo
  {author} {\bibfnamefont {A.}~\bibnamefont {Colla}}, \bibinfo {author}
  {\bibfnamefont {M.}~\bibnamefont {Colombini}}, \bibinfo {author}
  {\bibfnamefont {A.}~\bibnamefont {Conte}}, \bibinfo {author} {\bibfnamefont
  {J.-P.}\ \bibnamefont {Coulon}}, \bibinfo {author} {\bibfnamefont
  {E.}~\bibnamefont {Cuoco}}, \bibinfo {author} {\bibfnamefont
  {A.}~\bibnamefont {Dalmaz}}, \bibinfo {author} {\bibfnamefont
  {S.}~\bibnamefont {D'Antonio}}, \bibinfo {author} {\bibfnamefont
  {V.}~\bibnamefont {Dattilo}}, \bibinfo {author} {\bibfnamefont
  {M.}~\bibnamefont {Davier}}, \bibinfo {author} {\bibfnamefont
  {R.}~\bibnamefont {Day}}, \bibinfo {author} {\bibfnamefont {G.}~\bibnamefont
  {Debreczeni}}, \bibinfo {author} {\bibfnamefont {J.}~\bibnamefont
  {Degallaix}}, \bibinfo {author} {\bibfnamefont {S.}~\bibnamefont {Del{\'{e}
  }glise}}, \bibinfo {author} {\bibfnamefont {W.~D.}\ \bibnamefont {Pozzo}},
  \bibinfo {author} {\bibfnamefont {H.}~\bibnamefont {Dereli}}, \bibinfo
  {author} {\bibfnamefont {R.~D.}\ \bibnamefont {Rosa}}, \bibinfo {author}
  {\bibfnamefont {L.~D.}\ \bibnamefont {Fiore}}, \bibinfo {author}
  {\bibfnamefont {A.~D.}\ \bibnamefont {Lieto}}, \bibinfo {author}
  {\bibfnamefont {A.~D.}\ \bibnamefont {Virgilio}}, \bibinfo {author}
  {\bibfnamefont {M.}~\bibnamefont {Doets}}, \bibinfo {author} {\bibfnamefont
  {V.}~\bibnamefont {Dolique}}, \bibinfo {author} {\bibfnamefont
  {M.}~\bibnamefont {Drago}}, \bibinfo {author} {\bibfnamefont
  {M.}~\bibnamefont {Ducrot}}, \bibinfo {author} {\bibfnamefont
  {G.}~\bibnamefont {Endr{\H{o}}czi}}, \bibinfo {author} {\bibfnamefont
  {V.}~\bibnamefont {Fafone}}, \bibinfo {author} {\bibfnamefont
  {S.}~\bibnamefont {Farinon}}, \bibinfo {author} {\bibfnamefont
  {I.}~\bibnamefont {Ferrante}}, \bibinfo {author} {\bibfnamefont
  {F.}~\bibnamefont {Ferrini}}, \bibinfo {author} {\bibfnamefont
  {F.}~\bibnamefont {Fidecaro}}, \bibinfo {author} {\bibfnamefont
  {I.}~\bibnamefont {Fiori}}, \bibinfo {author} {\bibfnamefont
  {R.}~\bibnamefont {Flaminio}}, \bibinfo {author} {\bibfnamefont {J.-D.}\
  \bibnamefont {Fournier}}, \bibinfo {author} {\bibfnamefont {S.}~\bibnamefont
  {Franco}}, \bibinfo {author} {\bibfnamefont {S.}~\bibnamefont {Frasca}},
  \bibinfo {author} {\bibfnamefont {F.}~\bibnamefont {Frasconi}}, \bibinfo
  {author} {\bibfnamefont {L.}~\bibnamefont {Gammaitoni}}, \bibinfo {author}
  {\bibfnamefont {F.}~\bibnamefont {Garufi}}, \bibinfo {author} {\bibfnamefont
  {M.}~\bibnamefont {Gaspard}}, \bibinfo {author} {\bibfnamefont
  {A.}~\bibnamefont {Gatto}}, \bibinfo {author} {\bibfnamefont
  {G.}~\bibnamefont {Gemme}}, \bibinfo {author} {\bibfnamefont
  {B.}~\bibnamefont {Gendre}}, \bibinfo {author} {\bibfnamefont
  {E.}~\bibnamefont {Genin}}, \bibinfo {author} {\bibfnamefont
  {A.}~\bibnamefont {Gennai}}, \bibinfo {author} {\bibfnamefont
  {S.}~\bibnamefont {Ghosh}}, \bibinfo {author} {\bibfnamefont
  {L.}~\bibnamefont {Giacobone}}, \bibinfo {author} {\bibfnamefont
  {A.}~\bibnamefont {Giazotto}}, \bibinfo {author} {\bibfnamefont
  {R.}~\bibnamefont {Gouaty}}, \bibinfo {author} {\bibfnamefont
  {M.}~\bibnamefont {Granata}}, \bibinfo {author} {\bibfnamefont
  {G.}~\bibnamefont {Greco}}, \bibinfo {author} {\bibfnamefont
  {P.}~\bibnamefont {Groot}}, \bibinfo {author} {\bibfnamefont {G.~M.}\
  \bibnamefont {Guidi}}, \bibinfo {author} {\bibfnamefont {J.}~\bibnamefont
  {Harms}}, \bibinfo {author} {\bibfnamefont {A.}~\bibnamefont {Heidmann}},
  \bibinfo {author} {\bibfnamefont {H.}~\bibnamefont {Heitmann}}, \bibinfo
  {author} {\bibfnamefont {P.}~\bibnamefont {Hello}}, \bibinfo {author}
  {\bibfnamefont {G.}~\bibnamefont {Hemming}}, \bibinfo {author} {\bibfnamefont
  {E.}~\bibnamefont {Hennes}}, \bibinfo {author} {\bibfnamefont
  {D.}~\bibnamefont {Hofman}}, \bibinfo {author} {\bibfnamefont
  {P.}~\bibnamefont {Jaranowski}}, \bibinfo {author} {\bibfnamefont {R.~J.~G.}\
  \bibnamefont {Jonker}}, \bibinfo {author} {\bibfnamefont {M.}~\bibnamefont
  {Kasprzack}}, \bibinfo {author} {\bibfnamefont {F.}~\bibnamefont
  {K{\'{e}}f{\'{e}}lian}}, \bibinfo {author} {\bibfnamefont {I.}~\bibnamefont
  {Kowalska}}, \bibinfo {author} {\bibfnamefont {M.}~\bibnamefont {Kraan}},
  \bibinfo {author} {\bibfnamefont {A.}~\bibnamefont {Kr{\'{o}}lak}}, \bibinfo
  {author} {\bibfnamefont {A.}~\bibnamefont {Kutynia}}, \bibinfo {author}
  {\bibfnamefont {C.}~\bibnamefont {Lazzaro}}, \bibinfo {author} {\bibfnamefont
  {M.}~\bibnamefont {Leonardi}}, \bibinfo {author} {\bibfnamefont
  {N.}~\bibnamefont {Leroy}}, \bibinfo {author} {\bibfnamefont
  {N.}~\bibnamefont {Letendre}}, \bibinfo {author} {\bibfnamefont {T.~G.~F.}\
  \bibnamefont {Li}}, \bibinfo {author} {\bibfnamefont {B.}~\bibnamefont
  {Lieunard}}, \bibinfo {author} {\bibfnamefont {M.}~\bibnamefont {Lorenzini}},
  \bibinfo {author} {\bibfnamefont {V.}~\bibnamefont {Loriette}}, \bibinfo
  {author} {\bibfnamefont {G.}~\bibnamefont {Losurdo}}, \bibinfo {author}
  {\bibfnamefont {C.}~\bibnamefont {Magazz{\`{u}}}}, \bibinfo {author}
  {\bibfnamefont {E.}~\bibnamefont {Majorana}}, \bibinfo {author}
  {\bibfnamefont {I.}~\bibnamefont {Maksimovic}}, \bibinfo {author}
  {\bibfnamefont {V.}~\bibnamefont {Malvezzi}}, \bibinfo {author}
  {\bibfnamefont {N.}~\bibnamefont {Man}}, \bibinfo {author} {\bibfnamefont
  {V.}~\bibnamefont {Mangano}}, \bibinfo {author} {\bibfnamefont
  {M.}~\bibnamefont {Mantovani}}, \bibinfo {author} {\bibfnamefont
  {F.}~\bibnamefont {Marchesoni}}, \bibinfo {author} {\bibfnamefont
  {F.}~\bibnamefont {Marion}}, \bibinfo {author} {\bibfnamefont
  {J.}~\bibnamefont {Marque}}, \bibinfo {author} {\bibfnamefont
  {F.}~\bibnamefont {Martelli}}, \bibinfo {author} {\bibfnamefont
  {L.}~\bibnamefont {Martellini}}, \bibinfo {author} {\bibfnamefont
  {A.}~\bibnamefont {Masserot}}, \bibinfo {author} {\bibfnamefont
  {D.}~\bibnamefont {Meacher}}, \bibinfo {author} {\bibfnamefont
  {J.}~\bibnamefont {Meidam}}, \bibinfo {author} {\bibfnamefont
  {F.}~\bibnamefont {Mezzani}}, \bibinfo {author} {\bibfnamefont
  {C.}~\bibnamefont {Michel}}, \bibinfo {author} {\bibfnamefont
  {L.}~\bibnamefont {Milano}}, \bibinfo {author} {\bibfnamefont
  {Y.}~\bibnamefont {Minenkov}}, \bibinfo {author} {\bibfnamefont
  {A.}~\bibnamefont {Moggi}}, \bibinfo {author} {\bibfnamefont
  {M.}~\bibnamefont {Mohan}}, \bibinfo {author} {\bibfnamefont
  {M.}~\bibnamefont {Montani}}, \bibinfo {author} {\bibfnamefont
  {N.}~\bibnamefont {Morgado}}, \bibinfo {author} {\bibfnamefont
  {B.}~\bibnamefont {Mours}}, \bibinfo {author} {\bibfnamefont
  {F.}~\bibnamefont {Mul}}, \bibinfo {author} {\bibfnamefont {M.~F.}\
  \bibnamefont {Nagy}}, \bibinfo {author} {\bibfnamefont {I.}~\bibnamefont
  {Nardecchia}}, \bibinfo {author} {\bibfnamefont {L.}~\bibnamefont
  {Naticchioni}}, \bibinfo {author} {\bibfnamefont {G.}~\bibnamefont
  {Nelemans}}, \bibinfo {author} {\bibfnamefont {I.}~\bibnamefont {Neri}},
  \bibinfo {author} {\bibfnamefont {M.}~\bibnamefont {Neri}}, \bibinfo {author}
  {\bibfnamefont {F.}~\bibnamefont {Nocera}}, \bibinfo {author} {\bibfnamefont
  {E.}~\bibnamefont {Pacaud}}, \bibinfo {author} {\bibfnamefont
  {C.}~\bibnamefont {Palomba}}, \bibinfo {author} {\bibfnamefont
  {F.}~\bibnamefont {Paoletti}}, \bibinfo {author} {\bibfnamefont
  {A.}~\bibnamefont {Paoli}}, \bibinfo {author} {\bibfnamefont
  {A.}~\bibnamefont {Pasqualetti}}, \bibinfo {author} {\bibfnamefont
  {R.}~\bibnamefont {Passaquieti}}, \bibinfo {author} {\bibfnamefont
  {D.}~\bibnamefont {Passuello}}, \bibinfo {author} {\bibfnamefont
  {M.}~\bibnamefont {Perciballi}}, \bibinfo {author} {\bibfnamefont
  {S.}~\bibnamefont {Petit}}, \bibinfo {author} {\bibfnamefont
  {M.}~\bibnamefont {Pichot}}, \bibinfo {author} {\bibfnamefont
  {F.}~\bibnamefont {Piergiovanni}}, \bibinfo {author} {\bibfnamefont
  {G.}~\bibnamefont {Pillant}}, \bibinfo {author} {\bibfnamefont
  {A.}~\bibnamefont {Piluso}}, \bibinfo {author} {\bibfnamefont
  {L.}~\bibnamefont {Pinard}}, \bibinfo {author} {\bibfnamefont
  {R.}~\bibnamefont {Poggiani}}, \bibinfo {author} {\bibfnamefont
  {M.}~\bibnamefont {Prijatelj}}, \bibinfo {author} {\bibfnamefont {G.~A.}\
  \bibnamefont {Prodi}}, \bibinfo {author} {\bibfnamefont {M.}~\bibnamefont
  {Punturo}}, \bibinfo {author} {\bibfnamefont {P.}~\bibnamefont {Puppo}},
  \bibinfo {author} {\bibfnamefont {D.~S.}\ \bibnamefont {Rabeling}}, \bibinfo
  {author} {\bibfnamefont {I.}~\bibnamefont {R{\'{a}}cz}}, \bibinfo {author}
  {\bibfnamefont {P.}~\bibnamefont {Rapagnani}}, \bibinfo {author}
  {\bibfnamefont {M.}~\bibnamefont {Razzano}}, \bibinfo {author} {\bibfnamefont
  {V.}~\bibnamefont {Re}}, \bibinfo {author} {\bibfnamefont {T.}~\bibnamefont
  {Regimbau}}, \bibinfo {author} {\bibfnamefont {F.}~\bibnamefont {Ricci}},
  \bibinfo {author} {\bibfnamefont {F.}~\bibnamefont {Robinet}}, \bibinfo
  {author} {\bibfnamefont {A.}~\bibnamefont {Rocchi}}, \bibinfo {author}
  {\bibfnamefont {L.}~\bibnamefont {Rolland}}, \bibinfo {author} {\bibfnamefont
  {R.}~\bibnamefont {Romano}}, \bibinfo {author} {\bibfnamefont
  {D.}~\bibnamefont {Rosi{\'{n}}ska}}, \bibinfo {author} {\bibfnamefont
  {P.}~\bibnamefont {Ruggi}}, \bibinfo {author} {\bibfnamefont
  {E.}~\bibnamefont {Saracco}}, \bibinfo {author} {\bibfnamefont
  {B.}~\bibnamefont {Sassolas}}, \bibinfo {author} {\bibfnamefont
  {F.}~\bibnamefont {Schimmel}}, \bibinfo {author} {\bibfnamefont
  {D.}~\bibnamefont {Sentenac}}, \bibinfo {author} {\bibfnamefont
  {V.}~\bibnamefont {Sequino}}, \bibinfo {author} {\bibfnamefont
  {S.}~\bibnamefont {Shah}}, \bibinfo {author} {\bibfnamefont {K.}~\bibnamefont
  {Siellez}}, \bibinfo {author} {\bibfnamefont {N.}~\bibnamefont {Straniero}},
  \bibinfo {author} {\bibfnamefont {B.}~\bibnamefont {Swinkels}}, \bibinfo
  {author} {\bibfnamefont {M.}~\bibnamefont {Tacca}}, \bibinfo {author}
  {\bibfnamefont {M.}~\bibnamefont {Tonelli}}, \bibinfo {author} {\bibfnamefont
  {F.}~\bibnamefont {Travasso}}, \bibinfo {author} {\bibfnamefont
  {M.}~\bibnamefont {Turconi}}, \bibinfo {author} {\bibfnamefont
  {G.}~\bibnamefont {Vajente}}, \bibinfo {author} {\bibfnamefont
  {N.}~\bibnamefont {van Bakel}}, \bibinfo {author} {\bibfnamefont
  {M.}~\bibnamefont {van Beuzekom}}, \bibinfo {author} {\bibfnamefont
  {J.~F.~J.}\ \bibnamefont {van~den Brand}}, \bibinfo {author} {\bibfnamefont
  {C.~V.~D.}\ \bibnamefont {Broeck}}, \bibinfo {author} {\bibfnamefont {M.~V.}\
  \bibnamefont {van~der Sluys}}, \bibinfo {author} {\bibfnamefont
  {J.}~\bibnamefont {van Heijningen}}, \bibinfo {author} {\bibfnamefont
  {M.}~\bibnamefont {Vas{\'{u}}th}}, \bibinfo {author} {\bibfnamefont
  {G.}~\bibnamefont {Vedovato}}, \bibinfo {author} {\bibfnamefont
  {J.}~\bibnamefont {Veitch}}, \bibinfo {author} {\bibfnamefont
  {D.}~\bibnamefont {Verkindt}}, \bibinfo {author} {\bibfnamefont
  {F.}~\bibnamefont {Vetrano}}, \bibinfo {author} {\bibfnamefont
  {A.}~\bibnamefont {Vicer{\'{e}}}}, \bibinfo {author} {\bibfnamefont {J.-Y.}\
  \bibnamefont {Vinet}}, \bibinfo {author} {\bibfnamefont {G.}~\bibnamefont
  {Visser}}, \bibinfo {author} {\bibfnamefont {H.}~\bibnamefont {Vocca}},
  \bibinfo {author} {\bibfnamefont {R.}~\bibnamefont {Ward}}, \bibinfo {author}
  {\bibfnamefont {M.}~\bibnamefont {Was}}, \bibinfo {author} {\bibfnamefont
  {L.-W.}\ \bibnamefont {Wei}}, \bibinfo {author} {\bibfnamefont
  {M.}~\bibnamefont {Yvert}}, \bibinfo {author} {\bibfnamefont {A.~Z.}\
  \bibnamefont {{\.{z}}ny}},\ and\ \bibinfo {author} {\bibfnamefont {J.-P.}\
  \bibnamefont {Zendri}},\ }\bibfield  {title} {\bibinfo {title} {Advanced
  virgo: a second-generation interferometric gravitational wave detector},\
  }\href {https://doi.org/10.1088/0264-9381/32/2/024001} {\bibfield  {journal}
  {\bibinfo  {journal} {Classical and Quantum Gravity}\ }\textbf {\bibinfo
  {volume} {32}},\ \bibinfo {pages} {024001} (\bibinfo {year}
  {2014})}\BibitemShut {NoStop}%
\bibitem [{\citenamefont {Abe}\ \emph {et~al.}(2022)\citenamefont {Abe},
  \citenamefont {Akutsu}, \citenamefont {Ando}, \citenamefont {Araya},
  \citenamefont {Aritomi}, \citenamefont {Asada}, \citenamefont {Aso},
  \citenamefont {Bae}, \citenamefont {Bajpai}, \citenamefont {Cannon},
  \citenamefont {Cao}, \citenamefont {Capocasa}, \citenamefont {Chan},
  \citenamefont {Chen}, \citenamefont {Chen}, \citenamefont {Eisenmann},
  \citenamefont {Flaminio}, \citenamefont {Fong}, \citenamefont {Fujikawa},
  \citenamefont {Fujimoto}, \citenamefont {Hadiputrawan}, \citenamefont
  {Haino}, \citenamefont {Han}, \citenamefont {Hayama}, \citenamefont
  {Himemoto}, \citenamefont {Hirata}, \citenamefont {Hirose}, \citenamefont
  {Ho}, \citenamefont {Hsieh}, \citenamefont {Hsieh}, \citenamefont {Hsiung},
  \citenamefont {Huang}, \citenamefont {Huang}, \citenamefont {Huang},
  \citenamefont {Huang}, \citenamefont {Hui}, \citenamefont {Inayoshi},
  \citenamefont {Inoue}, \citenamefont {Itoh}, \citenamefont {Jung},
  \citenamefont {Kajita}, \citenamefont {Kamiizumi}, \citenamefont {Kanda},
  \citenamefont {Kato}, \citenamefont {Kim}, \citenamefont {Kim}, \citenamefont
  {Kim}, \citenamefont {Kobayashi}, \citenamefont {Kohri}, \citenamefont
  {Kokeyama}, \citenamefont {Kong}, \citenamefont {Koyama}, \citenamefont
  {Kozakai}, \citenamefont {Kume}, \citenamefont {Kuroyanagi}, \citenamefont
  {Kwak}, \citenamefont {Lee}, \citenamefont {Lee}, \citenamefont {Lee},
  \citenamefont {Leonardi}, \citenamefont {Li}, \citenamefont {Li},
  \citenamefont {Lin}, \citenamefont {Lin}, \citenamefont {Lin}, \citenamefont
  {Lin}, \citenamefont {Liu}, \citenamefont {Luo}, \citenamefont {Ma’arif},
  \citenamefont {Michimura}, \citenamefont {Mio}, \citenamefont {Miyakawa},
  \citenamefont {Miyo}, \citenamefont {Miyoki}, \citenamefont {Morisue},
  \citenamefont {Nakamura}, \citenamefont {Nakano}, \citenamefont {Nakano},
  \citenamefont {Narikawa}, \citenamefont {Quynh}, \citenamefont {Nishimoto},
  \citenamefont {Nishizawa}, \citenamefont {Obayashi}, \citenamefont {Oh},
  \citenamefont {Ohashi}, \citenamefont {Ohashi}, \citenamefont {Ohkawa},
  \citenamefont {Okutani}, \citenamefont {Oohara}, \citenamefont {Oshino},
  \citenamefont {Pan}, \citenamefont {Parisi}, \citenamefont {Park},
  \citenamefont {Arellano}, \citenamefont {Saha}, \citenamefont {Sakai},
  \citenamefont {Sawada}, \citenamefont {Sekiguchi}, \citenamefont {Shao},
  \citenamefont {Shikano}, \citenamefont {Shimizu}, \citenamefont {Shimode},
  \citenamefont {Shinkai}, \citenamefont {Shoda}, \citenamefont {Somiya},
  \citenamefont {Song}, \citenamefont {Sugimoto}, \citenamefont {Suresh},
  \citenamefont {Suzuki}, \citenamefont {Suzuki}, \citenamefont {Suzuki},
  \citenamefont {Tagoshi}, \citenamefont {Takahashi}, \citenamefont
  {Takahashi}, \citenamefont {Takeda}, \citenamefont {Takeda}, \citenamefont
  {Taruya}, \citenamefont {Tomaru}, \citenamefont {Tomura}, \citenamefont
  {Trozzo}, \citenamefont {Tsang}, \citenamefont {Tsuchida}, \citenamefont
  {Tsutsui}, \citenamefont {Tuyenbayev}, \citenamefont {Uchikata},
  \citenamefont {Uchiyama}, \citenamefont {Uehara}, \citenamefont {Ueno},
  \citenamefont {Ushiba}, \citenamefont {Putten}, \citenamefont {Washimi},
  \citenamefont {Wu}, \citenamefont {Wu}, \citenamefont {Yamada}, \citenamefont
  {Yamamoto}, \citenamefont {Yamamoto}, \citenamefont {Yamazaki}, \citenamefont
  {Yeh}, \citenamefont {Yokoyama}, \citenamefont {Yokozawa}, \citenamefont
  {Yuzurihara}, \citenamefont {Zeidler},\ and\ \citenamefont
  {Zhao}}]{galaxies10030063}%
  \BibitemOpen
  \bibfield  {author} {\bibinfo {author} {\bibfnamefont {H.}~\bibnamefont
  {Abe}}, \bibinfo {author} {\bibfnamefont {T.}~\bibnamefont {Akutsu}},
  \bibinfo {author} {\bibfnamefont {M.}~\bibnamefont {Ando}}, \bibinfo {author}
  {\bibfnamefont {A.}~\bibnamefont {Araya}}, \bibinfo {author} {\bibfnamefont
  {N.}~\bibnamefont {Aritomi}}, \bibinfo {author} {\bibfnamefont
  {H.}~\bibnamefont {Asada}}, \bibinfo {author} {\bibfnamefont
  {Y.}~\bibnamefont {Aso}}, \bibinfo {author} {\bibfnamefont {S.}~\bibnamefont
  {Bae}}, \bibinfo {author} {\bibfnamefont {R.}~\bibnamefont {Bajpai}},
  \bibinfo {author} {\bibfnamefont {K.}~\bibnamefont {Cannon}}, \bibinfo
  {author} {\bibfnamefont {Z.}~\bibnamefont {Cao}}, \bibinfo {author}
  {\bibfnamefont {E.}~\bibnamefont {Capocasa}}, \bibinfo {author}
  {\bibfnamefont {M.~L.}\ \bibnamefont {Chan}}, \bibinfo {author}
  {\bibfnamefont {D.}~\bibnamefont {Chen}}, \bibinfo {author} {\bibfnamefont
  {Y.-R.}\ \bibnamefont {Chen}}, \bibinfo {author} {\bibfnamefont
  {M.}~\bibnamefont {Eisenmann}}, \bibinfo {author} {\bibfnamefont
  {R.}~\bibnamefont {Flaminio}}, \bibinfo {author} {\bibfnamefont {H.~K.}\
  \bibnamefont {Fong}}, \bibinfo {author} {\bibfnamefont {Y.}~\bibnamefont
  {Fujikawa}}, \bibinfo {author} {\bibfnamefont {Y.}~\bibnamefont {Fujimoto}},
  \bibinfo {author} {\bibfnamefont {I.~P.~W.}\ \bibnamefont {Hadiputrawan}},
  \bibinfo {author} {\bibfnamefont {S.}~\bibnamefont {Haino}}, \bibinfo
  {author} {\bibfnamefont {W.}~\bibnamefont {Han}}, \bibinfo {author}
  {\bibfnamefont {K.}~\bibnamefont {Hayama}}, \bibinfo {author} {\bibfnamefont
  {Y.}~\bibnamefont {Himemoto}}, \bibinfo {author} {\bibfnamefont
  {N.}~\bibnamefont {Hirata}}, \bibinfo {author} {\bibfnamefont
  {C.}~\bibnamefont {Hirose}}, \bibinfo {author} {\bibfnamefont {T.-C.}\
  \bibnamefont {Ho}}, \bibinfo {author} {\bibfnamefont {B.-H.}\ \bibnamefont
  {Hsieh}}, \bibinfo {author} {\bibfnamefont {H.-F.}\ \bibnamefont {Hsieh}},
  \bibinfo {author} {\bibfnamefont {C.-H.}\ \bibnamefont {Hsiung}}, \bibinfo
  {author} {\bibfnamefont {H.-Y.}\ \bibnamefont {Huang}}, \bibinfo {author}
  {\bibfnamefont {P.}~\bibnamefont {Huang}}, \bibinfo {author} {\bibfnamefont
  {Y.-C.}\ \bibnamefont {Huang}}, \bibinfo {author} {\bibfnamefont {Y.-J.}\
  \bibnamefont {Huang}}, \bibinfo {author} {\bibfnamefont {D.~C.~Y.}\
  \bibnamefont {Hui}}, \bibinfo {author} {\bibfnamefont {K.}~\bibnamefont
  {Inayoshi}}, \bibinfo {author} {\bibfnamefont {Y.}~\bibnamefont {Inoue}},
  \bibinfo {author} {\bibfnamefont {Y.}~\bibnamefont {Itoh}}, \bibinfo {author}
  {\bibfnamefont {P.-J.}\ \bibnamefont {Jung}}, \bibinfo {author}
  {\bibfnamefont {T.}~\bibnamefont {Kajita}}, \bibinfo {author} {\bibfnamefont
  {M.}~\bibnamefont {Kamiizumi}}, \bibinfo {author} {\bibfnamefont
  {N.}~\bibnamefont {Kanda}}, \bibinfo {author} {\bibfnamefont
  {T.}~\bibnamefont {Kato}}, \bibinfo {author} {\bibfnamefont {C.}~\bibnamefont
  {Kim}}, \bibinfo {author} {\bibfnamefont {J.}~\bibnamefont {Kim}}, \bibinfo
  {author} {\bibfnamefont {Y.-M.}\ \bibnamefont {Kim}}, \bibinfo {author}
  {\bibfnamefont {Y.}~\bibnamefont {Kobayashi}}, \bibinfo {author}
  {\bibfnamefont {K.}~\bibnamefont {Kohri}}, \bibinfo {author} {\bibfnamefont
  {K.}~\bibnamefont {Kokeyama}}, \bibinfo {author} {\bibfnamefont {A.~K.~H.}\
  \bibnamefont {Kong}}, \bibinfo {author} {\bibfnamefont {N.}~\bibnamefont
  {Koyama}}, \bibinfo {author} {\bibfnamefont {C.}~\bibnamefont {Kozakai}},
  \bibinfo {author} {\bibfnamefont {J.}~\bibnamefont {Kume}}, \bibinfo {author}
  {\bibfnamefont {S.}~\bibnamefont {Kuroyanagi}}, \bibinfo {author}
  {\bibfnamefont {K.}~\bibnamefont {Kwak}}, \bibinfo {author} {\bibfnamefont
  {E.}~\bibnamefont {Lee}}, \bibinfo {author} {\bibfnamefont {H.~W.}\
  \bibnamefont {Lee}}, \bibinfo {author} {\bibfnamefont {R.-K.}\ \bibnamefont
  {Lee}}, \bibinfo {author} {\bibfnamefont {M.}~\bibnamefont {Leonardi}},
  \bibinfo {author} {\bibfnamefont {K.-L.}\ \bibnamefont {Li}}, \bibinfo
  {author} {\bibfnamefont {P.}~\bibnamefont {Li}}, \bibinfo {author}
  {\bibfnamefont {L.~C.-C.}\ \bibnamefont {Lin}}, \bibinfo {author}
  {\bibfnamefont {C.-Y.}\ \bibnamefont {Lin}}, \bibinfo {author} {\bibfnamefont
  {E.-T.}\ \bibnamefont {Lin}}, \bibinfo {author} {\bibfnamefont {H.-L.}\
  \bibnamefont {Lin}}, \bibinfo {author} {\bibfnamefont {G.-C.}\ \bibnamefont
  {Liu}}, \bibinfo {author} {\bibfnamefont {L.-W.}\ \bibnamefont {Luo}},
  \bibinfo {author} {\bibfnamefont {M.}~\bibnamefont {Ma’arif}}, \bibinfo
  {author} {\bibfnamefont {Y.}~\bibnamefont {Michimura}}, \bibinfo {author}
  {\bibfnamefont {N.}~\bibnamefont {Mio}}, \bibinfo {author} {\bibfnamefont
  {O.}~\bibnamefont {Miyakawa}}, \bibinfo {author} {\bibfnamefont
  {K.}~\bibnamefont {Miyo}}, \bibinfo {author} {\bibfnamefont {S.}~\bibnamefont
  {Miyoki}}, \bibinfo {author} {\bibfnamefont {N.}~\bibnamefont {Morisue}},
  \bibinfo {author} {\bibfnamefont {K.}~\bibnamefont {Nakamura}}, \bibinfo
  {author} {\bibfnamefont {H.}~\bibnamefont {Nakano}}, \bibinfo {author}
  {\bibfnamefont {M.}~\bibnamefont {Nakano}}, \bibinfo {author} {\bibfnamefont
  {T.}~\bibnamefont {Narikawa}}, \bibinfo {author} {\bibfnamefont {L.~N.}\
  \bibnamefont {Quynh}}, \bibinfo {author} {\bibfnamefont {T.}~\bibnamefont
  {Nishimoto}}, \bibinfo {author} {\bibfnamefont {A.}~\bibnamefont
  {Nishizawa}}, \bibinfo {author} {\bibfnamefont {Y.}~\bibnamefont {Obayashi}},
  \bibinfo {author} {\bibfnamefont {K.}~\bibnamefont {Oh}}, \bibinfo {author}
  {\bibfnamefont {M.}~\bibnamefont {Ohashi}}, \bibinfo {author} {\bibfnamefont
  {T.}~\bibnamefont {Ohashi}}, \bibinfo {author} {\bibfnamefont
  {M.}~\bibnamefont {Ohkawa}}, \bibinfo {author} {\bibfnamefont
  {Y.}~\bibnamefont {Okutani}}, \bibinfo {author} {\bibfnamefont {K.-i.}\
  \bibnamefont {Oohara}}, \bibinfo {author} {\bibfnamefont {S.}~\bibnamefont
  {Oshino}}, \bibinfo {author} {\bibfnamefont {K.-C.}\ \bibnamefont {Pan}},
  \bibinfo {author} {\bibfnamefont {A.}~\bibnamefont {Parisi}}, \bibinfo
  {author} {\bibfnamefont {J.~G.}\ \bibnamefont {Park}}, \bibinfo {author}
  {\bibfnamefont {F.~E.~P.}\ \bibnamefont {Arellano}}, \bibinfo {author}
  {\bibfnamefont {S.}~\bibnamefont {Saha}}, \bibinfo {author} {\bibfnamefont
  {K.}~\bibnamefont {Sakai}}, \bibinfo {author} {\bibfnamefont
  {T.}~\bibnamefont {Sawada}}, \bibinfo {author} {\bibfnamefont
  {Y.}~\bibnamefont {Sekiguchi}}, \bibinfo {author} {\bibfnamefont
  {L.}~\bibnamefont {Shao}}, \bibinfo {author} {\bibfnamefont {Y.}~\bibnamefont
  {Shikano}}, \bibinfo {author} {\bibfnamefont {H.}~\bibnamefont {Shimizu}},
  \bibinfo {author} {\bibfnamefont {K.}~\bibnamefont {Shimode}}, \bibinfo
  {author} {\bibfnamefont {H.}~\bibnamefont {Shinkai}}, \bibinfo {author}
  {\bibfnamefont {A.}~\bibnamefont {Shoda}}, \bibinfo {author} {\bibfnamefont
  {K.}~\bibnamefont {Somiya}}, \bibinfo {author} {\bibfnamefont
  {I.}~\bibnamefont {Song}}, \bibinfo {author} {\bibfnamefont {R.}~\bibnamefont
  {Sugimoto}}, \bibinfo {author} {\bibfnamefont {J.}~\bibnamefont {Suresh}},
  \bibinfo {author} {\bibfnamefont {T.}~\bibnamefont {Suzuki}}, \bibinfo
  {author} {\bibfnamefont {T.}~\bibnamefont {Suzuki}}, \bibinfo {author}
  {\bibfnamefont {T.}~\bibnamefont {Suzuki}}, \bibinfo {author} {\bibfnamefont
  {H.}~\bibnamefont {Tagoshi}}, \bibinfo {author} {\bibfnamefont
  {H.}~\bibnamefont {Takahashi}}, \bibinfo {author} {\bibfnamefont
  {R.}~\bibnamefont {Takahashi}}, \bibinfo {author} {\bibfnamefont
  {H.}~\bibnamefont {Takeda}}, \bibinfo {author} {\bibfnamefont
  {M.}~\bibnamefont {Takeda}}, \bibinfo {author} {\bibfnamefont
  {A.}~\bibnamefont {Taruya}}, \bibinfo {author} {\bibfnamefont
  {T.}~\bibnamefont {Tomaru}}, \bibinfo {author} {\bibfnamefont
  {T.}~\bibnamefont {Tomura}}, \bibinfo {author} {\bibfnamefont
  {L.}~\bibnamefont {Trozzo}}, \bibinfo {author} {\bibfnamefont {T.~T.~L.}\
  \bibnamefont {Tsang}}, \bibinfo {author} {\bibfnamefont {S.}~\bibnamefont
  {Tsuchida}}, \bibinfo {author} {\bibfnamefont {T.}~\bibnamefont {Tsutsui}},
  \bibinfo {author} {\bibfnamefont {D.}~\bibnamefont {Tuyenbayev}}, \bibinfo
  {author} {\bibfnamefont {N.}~\bibnamefont {Uchikata}}, \bibinfo {author}
  {\bibfnamefont {T.}~\bibnamefont {Uchiyama}}, \bibinfo {author}
  {\bibfnamefont {T.}~\bibnamefont {Uehara}}, \bibinfo {author} {\bibfnamefont
  {K.}~\bibnamefont {Ueno}}, \bibinfo {author} {\bibfnamefont {T.}~\bibnamefont
  {Ushiba}}, \bibinfo {author} {\bibfnamefont {M.~H. P. M.~v.}\ \bibnamefont
  {Putten}}, \bibinfo {author} {\bibfnamefont {T.}~\bibnamefont {Washimi}},
  \bibinfo {author} {\bibfnamefont {C.-M.}\ \bibnamefont {Wu}}, \bibinfo
  {author} {\bibfnamefont {H.-C.}\ \bibnamefont {Wu}}, \bibinfo {author}
  {\bibfnamefont {T.}~\bibnamefont {Yamada}}, \bibinfo {author} {\bibfnamefont
  {K.}~\bibnamefont {Yamamoto}}, \bibinfo {author} {\bibfnamefont
  {T.}~\bibnamefont {Yamamoto}}, \bibinfo {author} {\bibfnamefont
  {R.}~\bibnamefont {Yamazaki}}, \bibinfo {author} {\bibfnamefont {S.-W.}\
  \bibnamefont {Yeh}}, \bibinfo {author} {\bibfnamefont {J.}~\bibnamefont
  {Yokoyama}}, \bibinfo {author} {\bibfnamefont {T.}~\bibnamefont {Yokozawa}},
  \bibinfo {author} {\bibfnamefont {H.}~\bibnamefont {Yuzurihara}}, \bibinfo
  {author} {\bibfnamefont {S.}~\bibnamefont {Zeidler}},\ and\ \bibinfo {author}
  {\bibfnamefont {Y.}~\bibnamefont {Zhao}},\ }\bibfield  {title} {\bibinfo
  {title} {The current status and future prospects of kagra, the large-scale
  cryogenic gravitational wave telescope built in the kamioka underground},\
  }\bibfield  {journal} {\bibinfo  {journal} {Galaxies}\ }\textbf {\bibinfo
  {volume} {10}},\ \href {https://doi.org/10.3390/galaxies10030063}
  {10.3390/galaxies10030063} (\bibinfo {year} {2022})\BibitemShut {NoStop}%
\bibitem [{\citenamefont {{Unnikrishnan}}(2013)}]{2013IJMPD..2241010U}%
  \BibitemOpen
  \bibfield  {author} {\bibinfo {author} {\bibfnamefont {C.~S.}\ \bibnamefont
  {{Unnikrishnan}}},\ }\bibfield  {title} {\bibinfo {title} {{IndIGO and
  Ligo-India Scope and Plans for Gravitational Wave Research and Precision
  Metrology in India}},\ }\href {https://doi.org/10.1142/S0218271813410101}
  {\bibfield  {journal} {\bibinfo  {journal} {International Journal of Modern
  Physics D}\ }\textbf {\bibinfo {volume} {22}},\ \bibinfo {eid} {1341010}
  (\bibinfo {year} {2013})},\ \Eprint {https://arxiv.org/abs/1510.06059}
  {arXiv:1510.06059 [physics.ins-det]} \BibitemShut {NoStop}%
\bibitem [{\citenamefont {Punturo}\ \emph {et~al.}(2010)\citenamefont
  {Punturo}, \citenamefont {Abernathy}, \citenamefont {Acernese}, \citenamefont
  {Allen}, \citenamefont {Andersson}, \citenamefont {Arun}, \citenamefont
  {Barone}, \citenamefont {Barr}, \citenamefont {Barsuglia}, \citenamefont
  {Beker}, \citenamefont {Beveridge}, \citenamefont {Birindelli}, \citenamefont
  {Bose}, \citenamefont {Bosi}, \citenamefont {Braccini}, \citenamefont
  {Bradaschia}, \citenamefont {Bulik}, \citenamefont {Calloni}, \citenamefont
  {Cella},\ and\ \citenamefont {Yamamoto}}]{The_Einstein_Telescope}%
  \BibitemOpen
  \bibfield  {author} {\bibinfo {author} {\bibfnamefont {M.}~\bibnamefont
  {Punturo}}, \bibinfo {author} {\bibfnamefont {M.}~\bibnamefont {Abernathy}},
  \bibinfo {author} {\bibfnamefont {F.}~\bibnamefont {Acernese}}, \bibinfo
  {author} {\bibfnamefont {B.}~\bibnamefont {Allen}}, \bibinfo {author}
  {\bibfnamefont {N.}~\bibnamefont {Andersson}}, \bibinfo {author}
  {\bibfnamefont {K.}~\bibnamefont {Arun}}, \bibinfo {author} {\bibfnamefont
  {F.}~\bibnamefont {Barone}}, \bibinfo {author} {\bibfnamefont
  {B.}~\bibnamefont {Barr}}, \bibinfo {author} {\bibfnamefont {M.}~\bibnamefont
  {Barsuglia}}, \bibinfo {author} {\bibfnamefont {M.}~\bibnamefont {Beker}},
  \bibinfo {author} {\bibfnamefont {N.}~\bibnamefont {Beveridge}}, \bibinfo
  {author} {\bibfnamefont {S.}~\bibnamefont {Birindelli}}, \bibinfo {author}
  {\bibfnamefont {S.}~\bibnamefont {Bose}}, \bibinfo {author} {\bibfnamefont
  {L.}~\bibnamefont {Bosi}}, \bibinfo {author} {\bibfnamefont {S.}~\bibnamefont
  {Braccini}}, \bibinfo {author} {\bibfnamefont {C.}~\bibnamefont
  {Bradaschia}}, \bibinfo {author} {\bibfnamefont {T.}~\bibnamefont {Bulik}},
  \bibinfo {author} {\bibfnamefont {E.}~\bibnamefont {Calloni}}, \bibinfo
  {author} {\bibfnamefont {G.}~\bibnamefont {Cella}},\ and\ \bibinfo {author}
  {\bibfnamefont {K.}~\bibnamefont {Yamamoto}},\ }\bibfield  {title} {\bibinfo
  {title} {The einstein telescope: A third-generation gravitational wave
  observatory},\ }\href {https://doi.org/10.1088/0264-9381/27/19/194002}
  {\bibfield  {journal} {\bibinfo  {journal} {Classical and Quantum Gravity}\
  }\textbf {\bibinfo {volume} {27}} (\bibinfo {year} {2010})}\BibitemShut
  {NoStop}%
\bibitem [{\citenamefont {Reitze}\ \emph {et~al.}(2019)\citenamefont {Reitze},
  \citenamefont {Adhikari}, \citenamefont {Ballmer}, \citenamefont {Barish},
  \citenamefont {Barsotti}, \citenamefont {Billingsley}, \citenamefont {Brown},
  \citenamefont {Chen}, \citenamefont {Coyne}, \citenamefont {Eisenstein},
  \citenamefont {Evans}, \citenamefont {Fritschel}, \citenamefont {Hall},
  \citenamefont {Lazzarini}, \citenamefont {Lovelace}, \citenamefont {Read},
  \citenamefont {Sathyaprakash}, \citenamefont {Shoemaker}, \citenamefont
  {Smith}, \citenamefont {Torrie}, \citenamefont {Vitale}, \citenamefont
  {Weiss}, \citenamefont {Wipf},\ and\ \citenamefont
  {Zucker}}]{reitze2019cosmic}%
  \BibitemOpen
  \bibfield  {author} {\bibinfo {author} {\bibfnamefont {D.}~\bibnamefont
  {Reitze}}, \bibinfo {author} {\bibfnamefont {R.~X.}\ \bibnamefont
  {Adhikari}}, \bibinfo {author} {\bibfnamefont {S.}~\bibnamefont {Ballmer}},
  \bibinfo {author} {\bibfnamefont {B.}~\bibnamefont {Barish}}, \bibinfo
  {author} {\bibfnamefont {L.}~\bibnamefont {Barsotti}}, \bibinfo {author}
  {\bibfnamefont {G.}~\bibnamefont {Billingsley}}, \bibinfo {author}
  {\bibfnamefont {D.~A.}\ \bibnamefont {Brown}}, \bibinfo {author}
  {\bibfnamefont {Y.}~\bibnamefont {Chen}}, \bibinfo {author} {\bibfnamefont
  {D.}~\bibnamefont {Coyne}}, \bibinfo {author} {\bibfnamefont
  {R.}~\bibnamefont {Eisenstein}}, \bibinfo {author} {\bibfnamefont
  {M.}~\bibnamefont {Evans}}, \bibinfo {author} {\bibfnamefont
  {P.}~\bibnamefont {Fritschel}}, \bibinfo {author} {\bibfnamefont {E.~D.}\
  \bibnamefont {Hall}}, \bibinfo {author} {\bibfnamefont {A.}~\bibnamefont
  {Lazzarini}}, \bibinfo {author} {\bibfnamefont {G.}~\bibnamefont {Lovelace}},
  \bibinfo {author} {\bibfnamefont {J.}~\bibnamefont {Read}}, \bibinfo {author}
  {\bibfnamefont {B.~S.}\ \bibnamefont {Sathyaprakash}}, \bibinfo {author}
  {\bibfnamefont {D.}~\bibnamefont {Shoemaker}}, \bibinfo {author}
  {\bibfnamefont {J.}~\bibnamefont {Smith}}, \bibinfo {author} {\bibfnamefont
  {C.}~\bibnamefont {Torrie}}, \bibinfo {author} {\bibfnamefont
  {S.}~\bibnamefont {Vitale}}, \bibinfo {author} {\bibfnamefont
  {R.}~\bibnamefont {Weiss}}, \bibinfo {author} {\bibfnamefont
  {C.}~\bibnamefont {Wipf}},\ and\ \bibinfo {author} {\bibfnamefont
  {M.}~\bibnamefont {Zucker}},\ }\href@noop {} {\bibinfo {title} {Cosmic
  explorer: The u.s. contribution to gravitational-wave astronomy beyond ligo}}
  (\bibinfo {year} {2019}),\ \Eprint {https://arxiv.org/abs/1907.04833}
  {arXiv:1907.04833 [astro-ph.IM]} \BibitemShut {NoStop}%
\bibitem [{\citenamefont {Kawamura}\ \emph {et~al.}(2021)\citenamefont
  {Kawamura}, \citenamefont {Ando}, \citenamefont {Seto}, \citenamefont {Sato},
  \citenamefont {Musha}, \citenamefont {Kawano}, \citenamefont {Yokoyama},
  \citenamefont {Tanaka}, \citenamefont {Ioka}, \citenamefont {Akutsu},
  \citenamefont {Takashima}, \citenamefont {Agatsuma}, \citenamefont {Araya},
  \citenamefont {Aritomi}, \citenamefont {Asada}, \citenamefont {Chiba},
  \citenamefont {Eguchi}, \citenamefont {Enoki}, \citenamefont {Fujimoto},
  \citenamefont {Fujita}, \citenamefont {Futamase}, \citenamefont {Harada},
  \citenamefont {Hayama}, \citenamefont {Himemoto}, \citenamefont {Hiramatsu},
  \citenamefont {Hong}, \citenamefont {Hosokawa}, \citenamefont {Ichiki},
  \citenamefont {Ikari}, \citenamefont {Ishihara}, \citenamefont {Ishikawa},
  \citenamefont {Itoh}, \citenamefont {Ito}, \citenamefont {Iwaguchi},
  \citenamefont {Izumi}, \citenamefont {Kanda}, \citenamefont {Kanemura},
  \citenamefont {Kawazoe}, \citenamefont {Kobayashi}, \citenamefont {Kohri},
  \citenamefont {Kojima}, \citenamefont {Kokeyama}, \citenamefont {Kotake},
  \citenamefont {Kuroyanagi}, \citenamefont {Maeda}, \citenamefont
  {Matsushita}, \citenamefont {Michimura}, \citenamefont {Morimoto},
  \citenamefont {Mukohyama}, \citenamefont {Nagano}, \citenamefont {Nagano},
  \citenamefont {Naito}, \citenamefont {Nakamura}, \citenamefont {Nakamura},
  \citenamefont {Nakano}, \citenamefont {Nakao}, \citenamefont {Nakasuka},
  \citenamefont {Nakayama}, \citenamefont {Nakazawa}, \citenamefont
  {Nishizawa}, \citenamefont {Ohkawa}, \citenamefont {Oohara}, \citenamefont
  {Sago}, \citenamefont {Saijo}, \citenamefont {Sakagami}, \citenamefont
  {Sakai}, \citenamefont {Sato}, \citenamefont {Shibata}, \citenamefont
  {Shinkai}, \citenamefont {Shoda}, \citenamefont {Somiya}, \citenamefont
  {Sotani}, \citenamefont {Takahashi}, \citenamefont {Takahashi}, \citenamefont
  {Akiteru}, \citenamefont {Taniguchi}, \citenamefont {Taruya}, \citenamefont
  {Tsubono}, \citenamefont {Tsujikawa}, \citenamefont {Ueda}, \citenamefont
  {Ueda}, \citenamefont {Watanabe}, \citenamefont {Yagi}, \citenamefont
  {Yamada}, \citenamefont {Yokoyama}, \citenamefont {Yoo},\ and\ \citenamefont
  {Zhu}}]{10.1093_ptep_ptab019}%
  \BibitemOpen
  \bibfield  {author} {\bibinfo {author} {\bibfnamefont {S.}~\bibnamefont
  {Kawamura}}, \bibinfo {author} {\bibfnamefont {M.}~\bibnamefont {Ando}},
  \bibinfo {author} {\bibfnamefont {N.}~\bibnamefont {Seto}}, \bibinfo {author}
  {\bibfnamefont {S.}~\bibnamefont {Sato}}, \bibinfo {author} {\bibfnamefont
  {M.}~\bibnamefont {Musha}}, \bibinfo {author} {\bibfnamefont
  {I.}~\bibnamefont {Kawano}}, \bibinfo {author} {\bibfnamefont
  {J.}~\bibnamefont {Yokoyama}}, \bibinfo {author} {\bibfnamefont
  {T.}~\bibnamefont {Tanaka}}, \bibinfo {author} {\bibfnamefont
  {K.}~\bibnamefont {Ioka}}, \bibinfo {author} {\bibfnamefont {T.}~\bibnamefont
  {Akutsu}}, \bibinfo {author} {\bibfnamefont {T.}~\bibnamefont {Takashima}},
  \bibinfo {author} {\bibfnamefont {K.}~\bibnamefont {Agatsuma}}, \bibinfo
  {author} {\bibfnamefont {A.}~\bibnamefont {Araya}}, \bibinfo {author}
  {\bibfnamefont {N.}~\bibnamefont {Aritomi}}, \bibinfo {author} {\bibfnamefont
  {H.}~\bibnamefont {Asada}}, \bibinfo {author} {\bibfnamefont
  {T.}~\bibnamefont {Chiba}}, \bibinfo {author} {\bibfnamefont
  {S.}~\bibnamefont {Eguchi}}, \bibinfo {author} {\bibfnamefont
  {M.}~\bibnamefont {Enoki}}, \bibinfo {author} {\bibfnamefont {M.-K.}\
  \bibnamefont {Fujimoto}}, \bibinfo {author} {\bibfnamefont {R.}~\bibnamefont
  {Fujita}}, \bibinfo {author} {\bibfnamefont {T.}~\bibnamefont {Futamase}},
  \bibinfo {author} {\bibfnamefont {T.}~\bibnamefont {Harada}}, \bibinfo
  {author} {\bibfnamefont {K.}~\bibnamefont {Hayama}}, \bibinfo {author}
  {\bibfnamefont {Y.}~\bibnamefont {Himemoto}}, \bibinfo {author}
  {\bibfnamefont {T.}~\bibnamefont {Hiramatsu}}, \bibinfo {author}
  {\bibfnamefont {F.-L.}\ \bibnamefont {Hong}}, \bibinfo {author}
  {\bibfnamefont {M.}~\bibnamefont {Hosokawa}}, \bibinfo {author}
  {\bibfnamefont {K.}~\bibnamefont {Ichiki}}, \bibinfo {author} {\bibfnamefont
  {S.}~\bibnamefont {Ikari}}, \bibinfo {author} {\bibfnamefont
  {H.}~\bibnamefont {Ishihara}}, \bibinfo {author} {\bibfnamefont
  {T.}~\bibnamefont {Ishikawa}}, \bibinfo {author} {\bibfnamefont
  {Y.}~\bibnamefont {Itoh}}, \bibinfo {author} {\bibfnamefont {T.}~\bibnamefont
  {Ito}}, \bibinfo {author} {\bibfnamefont {S.}~\bibnamefont {Iwaguchi}},
  \bibinfo {author} {\bibfnamefont {K.}~\bibnamefont {Izumi}}, \bibinfo
  {author} {\bibfnamefont {N.}~\bibnamefont {Kanda}}, \bibinfo {author}
  {\bibfnamefont {S.}~\bibnamefont {Kanemura}}, \bibinfo {author}
  {\bibfnamefont {F.}~\bibnamefont {Kawazoe}}, \bibinfo {author} {\bibfnamefont
  {S.}~\bibnamefont {Kobayashi}}, \bibinfo {author} {\bibfnamefont
  {K.}~\bibnamefont {Kohri}}, \bibinfo {author} {\bibfnamefont
  {Y.}~\bibnamefont {Kojima}}, \bibinfo {author} {\bibfnamefont
  {K.}~\bibnamefont {Kokeyama}}, \bibinfo {author} {\bibfnamefont
  {K.}~\bibnamefont {Kotake}}, \bibinfo {author} {\bibfnamefont
  {S.}~\bibnamefont {Kuroyanagi}}, \bibinfo {author} {\bibfnamefont {K.-i.}\
  \bibnamefont {Maeda}}, \bibinfo {author} {\bibfnamefont {S.}~\bibnamefont
  {Matsushita}}, \bibinfo {author} {\bibfnamefont {Y.}~\bibnamefont
  {Michimura}}, \bibinfo {author} {\bibfnamefont {T.}~\bibnamefont {Morimoto}},
  \bibinfo {author} {\bibfnamefont {S.}~\bibnamefont {Mukohyama}}, \bibinfo
  {author} {\bibfnamefont {K.}~\bibnamefont {Nagano}}, \bibinfo {author}
  {\bibfnamefont {S.}~\bibnamefont {Nagano}}, \bibinfo {author} {\bibfnamefont
  {T.}~\bibnamefont {Naito}}, \bibinfo {author} {\bibfnamefont
  {K.}~\bibnamefont {Nakamura}}, \bibinfo {author} {\bibfnamefont
  {T.}~\bibnamefont {Nakamura}}, \bibinfo {author} {\bibfnamefont
  {H.}~\bibnamefont {Nakano}}, \bibinfo {author} {\bibfnamefont
  {K.}~\bibnamefont {Nakao}}, \bibinfo {author} {\bibfnamefont
  {S.}~\bibnamefont {Nakasuka}}, \bibinfo {author} {\bibfnamefont
  {Y.}~\bibnamefont {Nakayama}}, \bibinfo {author} {\bibfnamefont
  {K.}~\bibnamefont {Nakazawa}}, \bibinfo {author} {\bibfnamefont
  {A.}~\bibnamefont {Nishizawa}}, \bibinfo {author} {\bibfnamefont
  {M.}~\bibnamefont {Ohkawa}}, \bibinfo {author} {\bibfnamefont
  {K.}~\bibnamefont {Oohara}}, \bibinfo {author} {\bibfnamefont
  {N.}~\bibnamefont {Sago}}, \bibinfo {author} {\bibfnamefont {M.}~\bibnamefont
  {Saijo}}, \bibinfo {author} {\bibfnamefont {M.}~\bibnamefont {Sakagami}},
  \bibinfo {author} {\bibfnamefont {S.-i.}\ \bibnamefont {Sakai}}, \bibinfo
  {author} {\bibfnamefont {T.}~\bibnamefont {Sato}}, \bibinfo {author}
  {\bibfnamefont {M.}~\bibnamefont {Shibata}}, \bibinfo {author} {\bibfnamefont
  {H.}~\bibnamefont {Shinkai}}, \bibinfo {author} {\bibfnamefont
  {A.}~\bibnamefont {Shoda}}, \bibinfo {author} {\bibfnamefont
  {K.}~\bibnamefont {Somiya}}, \bibinfo {author} {\bibfnamefont
  {H.}~\bibnamefont {Sotani}}, \bibinfo {author} {\bibfnamefont
  {R.}~\bibnamefont {Takahashi}}, \bibinfo {author} {\bibfnamefont
  {H.}~\bibnamefont {Takahashi}}, \bibinfo {author} {\bibfnamefont
  {T.}~\bibnamefont {Akiteru}}, \bibinfo {author} {\bibfnamefont
  {K.}~\bibnamefont {Taniguchi}}, \bibinfo {author} {\bibfnamefont
  {A.}~\bibnamefont {Taruya}}, \bibinfo {author} {\bibfnamefont
  {K.}~\bibnamefont {Tsubono}}, \bibinfo {author} {\bibfnamefont
  {S.}~\bibnamefont {Tsujikawa}}, \bibinfo {author} {\bibfnamefont
  {A.}~\bibnamefont {Ueda}}, \bibinfo {author} {\bibfnamefont {K.-i.}\
  \bibnamefont {Ueda}}, \bibinfo {author} {\bibfnamefont {I.}~\bibnamefont
  {Watanabe}}, \bibinfo {author} {\bibfnamefont {K.}~\bibnamefont {Yagi}},
  \bibinfo {author} {\bibfnamefont {R.}~\bibnamefont {Yamada}}, \bibinfo
  {author} {\bibfnamefont {S.}~\bibnamefont {Yokoyama}}, \bibinfo {author}
  {\bibfnamefont {C.-M.}\ \bibnamefont {Yoo}},\ and\ \bibinfo {author}
  {\bibfnamefont {Z.-H.}\ \bibnamefont {Zhu}},\ }\bibfield  {title} {\bibinfo
  {title} {{Current status of space gravitational wave antenna DECIGO and
  B-DECIGO}},\ }\href {https://doi.org/10.1093/ptep/ptab019} {\bibfield
  {journal} {\bibinfo  {journal} {Progress of Theoretical and Experimental
  Physics}\ }\textbf {\bibinfo {volume} {2021}},\ \bibinfo {pages} {05A105}
  (\bibinfo {year} {2021})},\ \Eprint
  {https://arxiv.org/abs/https://academic.oup.com/ptep/article-pdf/2021/5/05A105/38109685/ptab019.pdf}
  {https://academic.oup.com/ptep/article-pdf/2021/5/05A105/38109685/ptab019.pdf}
  \BibitemShut {NoStop}%
\bibitem [{\citenamefont {Audley}\ \emph {et~al.}(2017)\citenamefont {Audley},
  \citenamefont {Babak}, \citenamefont {Baker}, \citenamefont {Barausse},
  \citenamefont {Bender}, \citenamefont {Berti}, \citenamefont {Binetruy},
  \citenamefont {Born}, \citenamefont {Bortoluzzi}, \citenamefont {Camp},
  \citenamefont {Caprini}, \citenamefont {Cardoso}, \citenamefont {Colpi},
  \citenamefont {Conklin}, \citenamefont {Cornish}, \citenamefont {Cutler},
  \citenamefont {Danzmann}, \citenamefont {Dolesi}, \citenamefont {Ferraioli},\
  and\ \citenamefont {F~Sopuerta}}]{Lisa_2017}%
  \BibitemOpen
  \bibfield  {author} {\bibinfo {author} {\bibfnamefont {H.}~\bibnamefont
  {Audley}}, \bibinfo {author} {\bibfnamefont {S.}~\bibnamefont {Babak}},
  \bibinfo {author} {\bibfnamefont {J.}~\bibnamefont {Baker}}, \bibinfo
  {author} {\bibfnamefont {E.}~\bibnamefont {Barausse}}, \bibinfo {author}
  {\bibfnamefont {P.}~\bibnamefont {Bender}}, \bibinfo {author} {\bibfnamefont
  {E.}~\bibnamefont {Berti}}, \bibinfo {author} {\bibfnamefont
  {P.}~\bibnamefont {Binetruy}}, \bibinfo {author} {\bibfnamefont
  {M.}~\bibnamefont {Born}}, \bibinfo {author} {\bibfnamefont {D.}~\bibnamefont
  {Bortoluzzi}}, \bibinfo {author} {\bibfnamefont {J.}~\bibnamefont {Camp}},
  \bibinfo {author} {\bibfnamefont {C.}~\bibnamefont {Caprini}}, \bibinfo
  {author} {\bibfnamefont {V.}~\bibnamefont {Cardoso}}, \bibinfo {author}
  {\bibfnamefont {M.}~\bibnamefont {Colpi}}, \bibinfo {author} {\bibfnamefont
  {J.}~\bibnamefont {Conklin}}, \bibinfo {author} {\bibfnamefont
  {N.}~\bibnamefont {Cornish}}, \bibinfo {author} {\bibfnamefont
  {C.}~\bibnamefont {Cutler}}, \bibinfo {author} {\bibfnamefont
  {K.}~\bibnamefont {Danzmann}}, \bibinfo {author} {\bibfnamefont
  {R.}~\bibnamefont {Dolesi}}, \bibinfo {author} {\bibfnamefont
  {L.}~\bibnamefont {Ferraioli}},\ and\ \bibinfo {author} {\bibfnamefont
  {C.}~\bibnamefont {F~Sopuerta}},\ }\bibfield  {title} {\bibinfo {title}
  {Laser interferometer space antenna},\ }\href@noop {} {\  (\bibinfo {year}
  {2017})}\BibitemShut {NoStop}%
\bibitem [{\citenamefont {Ruan}\ \emph {et~al.}(2020)\citenamefont {Ruan},
  \citenamefont {Guo}, \citenamefont {Cai},\ and\ \citenamefont
  {Zhang}}]{ruan2020taiji}%
  \BibitemOpen
  \bibfield  {author} {\bibinfo {author} {\bibfnamefont {W.-H.}\ \bibnamefont
  {Ruan}}, \bibinfo {author} {\bibfnamefont {Z.-K.}\ \bibnamefont {Guo}},
  \bibinfo {author} {\bibfnamefont {R.-G.}\ \bibnamefont {Cai}},\ and\ \bibinfo
  {author} {\bibfnamefont {Y.-Z.}\ \bibnamefont {Zhang}},\ }\href@noop {}
  {\bibinfo {title} {Taiji program: Gravitational-wave sources}} (\bibinfo
  {year} {2020}),\ \Eprint {https://arxiv.org/abs/1807.09495} {arXiv:1807.09495
  [gr-qc]} \BibitemShut {NoStop}%
\bibitem [{\citenamefont {{Luo}}\ \emph {et~al.}(2016)\citenamefont {{Luo}},
  \citenamefont {{Chen}}, \citenamefont {{Duan}}, \citenamefont {{Gong}},
  \citenamefont {{Hu}}, \citenamefont {{Ji}}, \citenamefont {{Liu}},
  \citenamefont {{Mei}}, \citenamefont {{Milyukov}}, \citenamefont {{Sazhin}},\
  and\ \citenamefont {et~al.}}]{2016CQGra..33c5010L}%
  \BibitemOpen
  \bibfield  {author} {\bibinfo {author} {\bibfnamefont {J.}~\bibnamefont
  {{Luo}}}, \bibinfo {author} {\bibfnamefont {L.-S.}\ \bibnamefont {{Chen}}},
  \bibinfo {author} {\bibfnamefont {H.-Z.}\ \bibnamefont {{Duan}}}, \bibinfo
  {author} {\bibfnamefont {Y.-G.}\ \bibnamefont {{Gong}}}, \bibinfo {author}
  {\bibfnamefont {S.}~\bibnamefont {{Hu}}}, \bibinfo {author} {\bibfnamefont
  {J.}~\bibnamefont {{Ji}}}, \bibinfo {author} {\bibfnamefont {Q.}~\bibnamefont
  {{Liu}}}, \bibinfo {author} {\bibfnamefont {J.}~\bibnamefont {{Mei}}},
  \bibinfo {author} {\bibfnamefont {V.}~\bibnamefont {{Milyukov}}}, \bibinfo
  {author} {\bibfnamefont {M.}~\bibnamefont {{Sazhin}}},\ and\ \bibinfo
  {author} {\bibnamefont {et~al.}},\ }\bibfield  {title} {\bibinfo {title}
  {{TianQin: a space-borne gravitational wave detector}},\ }\href
  {https://doi.org/10.1088/0264-9381/33/3/035010} {\bibfield  {journal}
  {\bibinfo  {journal} {Classical and Quantum Gravity}\ }\textbf {\bibinfo
  {volume} {33}},\ \bibinfo {eid} {035010} (\bibinfo {year} {2016})},\ \Eprint
  {https://arxiv.org/abs/1512.02076} {arXiv:1512.02076 [astro-ph.IM]}
  \BibitemShut {NoStop}%
\bibitem [{\citenamefont {Ballantini}\ \emph {et~al.}(2003)\citenamefont
  {Ballantini}, \citenamefont {Bernard}, \citenamefont {Chiaveri},
  \citenamefont {Chincarini}, \citenamefont {Gemme}, \citenamefont {Losito},
  \citenamefont {Parodi},\ and\ \citenamefont {Picasso}}]{R_Ballantini_2003}%
  \BibitemOpen
  \bibfield  {author} {\bibinfo {author} {\bibfnamefont {R.}~\bibnamefont
  {Ballantini}}, \bibinfo {author} {\bibfnamefont {P.}~\bibnamefont {Bernard}},
  \bibinfo {author} {\bibfnamefont {E.}~\bibnamefont {Chiaveri}}, \bibinfo
  {author} {\bibfnamefont {A.}~\bibnamefont {Chincarini}}, \bibinfo {author}
  {\bibfnamefont {G.}~\bibnamefont {Gemme}}, \bibinfo {author} {\bibfnamefont
  {R.}~\bibnamefont {Losito}}, \bibinfo {author} {\bibfnamefont
  {R.}~\bibnamefont {Parodi}},\ and\ \bibinfo {author} {\bibfnamefont
  {E.}~\bibnamefont {Picasso}},\ }\bibfield  {title} {\bibinfo {title} {A
  detector of high frequency gravitational waves based on coupled microwave
  cavities},\ }\href {https://doi.org/10.1088/0264-9381/20/15/316} {\bibfield
  {journal} {\bibinfo  {journal} {Classical and Quantum Gravity}\ }\textbf
  {\bibinfo {volume} {20}},\ \bibinfo {pages} {3505} (\bibinfo {year}
  {2003})}\BibitemShut {NoStop}%
\bibitem [{\citenamefont {Cruise}\ and\ \citenamefont
  {Ingley}(2005)}]{Cruise_2005}%
  \BibitemOpen
  \bibfield  {author} {\bibinfo {author} {\bibfnamefont {A.~M.}\ \bibnamefont
  {Cruise}}\ and\ \bibinfo {author} {\bibfnamefont {R.~M.~J.}\ \bibnamefont
  {Ingley}},\ }\bibfield  {title} {\bibinfo {title} {A correlation detector for
  very high frequency gravitational waves},\ }\href
  {https://doi.org/10.1088/0264-9381/22/10/046} {\bibfield  {journal} {\bibinfo
   {journal} {Classical and Quantum Gravity}\ }\textbf {\bibinfo {volume}
  {22}},\ \bibinfo {pages} {S479} (\bibinfo {year} {2005})}\BibitemShut
  {NoStop}%
\bibitem [{\citenamefont {Canuel}\ \emph {et~al.}(2018)\citenamefont {Canuel},
  \citenamefont {Bertoldi}, \citenamefont {Amand}, \citenamefont {di~Borgo},
  \citenamefont {Chantrait}, \citenamefont {Danquigny}, \citenamefont
  {{\'{A}}lvarez}, \citenamefont {Fang}, \citenamefont {Freise}, \citenamefont
  {Geiger}, \citenamefont {Gillot}, \citenamefont {Henry}, \citenamefont
  {Hinderer}, \citenamefont {Holleville}, \citenamefont {Junca}, \citenamefont
  {Lef{\`{e}}vre}, \citenamefont {Merzougui}, \citenamefont {Mielec},
  \citenamefont {Monfret}, \citenamefont {Pelisson}, \citenamefont
  {Prevedelli}, \citenamefont {Reynaud}, \citenamefont {Riou}, \citenamefont
  {Rogister}, \citenamefont {Rosat}, \citenamefont {Cormier}, \citenamefont
  {Landragin}, \citenamefont {Chaibi}, \citenamefont {Gaffet},\ and\
  \citenamefont {Bouyer}}]{Canuel_2018}%
  \BibitemOpen
  \bibfield  {author} {\bibinfo {author} {\bibfnamefont {B.}~\bibnamefont
  {Canuel}}, \bibinfo {author} {\bibfnamefont {A.}~\bibnamefont {Bertoldi}},
  \bibinfo {author} {\bibfnamefont {L.}~\bibnamefont {Amand}}, \bibinfo
  {author} {\bibfnamefont {E.~P.}\ \bibnamefont {di~Borgo}}, \bibinfo {author}
  {\bibfnamefont {T.}~\bibnamefont {Chantrait}}, \bibinfo {author}
  {\bibfnamefont {C.}~\bibnamefont {Danquigny}}, \bibinfo {author}
  {\bibfnamefont {M.~D.}\ \bibnamefont {{\'{A}}lvarez}}, \bibinfo {author}
  {\bibfnamefont {B.}~\bibnamefont {Fang}}, \bibinfo {author} {\bibfnamefont
  {A.}~\bibnamefont {Freise}}, \bibinfo {author} {\bibfnamefont
  {R.}~\bibnamefont {Geiger}}, \bibinfo {author} {\bibfnamefont
  {J.}~\bibnamefont {Gillot}}, \bibinfo {author} {\bibfnamefont
  {S.}~\bibnamefont {Henry}}, \bibinfo {author} {\bibfnamefont
  {J.}~\bibnamefont {Hinderer}}, \bibinfo {author} {\bibfnamefont
  {D.}~\bibnamefont {Holleville}}, \bibinfo {author} {\bibfnamefont
  {J.}~\bibnamefont {Junca}}, \bibinfo {author} {\bibfnamefont
  {G.}~\bibnamefont {Lef{\`{e}}vre}}, \bibinfo {author} {\bibfnamefont
  {M.}~\bibnamefont {Merzougui}}, \bibinfo {author} {\bibfnamefont
  {N.}~\bibnamefont {Mielec}}, \bibinfo {author} {\bibfnamefont
  {T.}~\bibnamefont {Monfret}}, \bibinfo {author} {\bibfnamefont
  {S.}~\bibnamefont {Pelisson}}, \bibinfo {author} {\bibfnamefont
  {M.}~\bibnamefont {Prevedelli}}, \bibinfo {author} {\bibfnamefont
  {S.}~\bibnamefont {Reynaud}}, \bibinfo {author} {\bibfnamefont
  {I.}~\bibnamefont {Riou}}, \bibinfo {author} {\bibfnamefont {Y.}~\bibnamefont
  {Rogister}}, \bibinfo {author} {\bibfnamefont {S.}~\bibnamefont {Rosat}},
  \bibinfo {author} {\bibfnamefont {E.}~\bibnamefont {Cormier}}, \bibinfo
  {author} {\bibfnamefont {A.}~\bibnamefont {Landragin}}, \bibinfo {author}
  {\bibfnamefont {W.}~\bibnamefont {Chaibi}}, \bibinfo {author} {\bibfnamefont
  {S.}~\bibnamefont {Gaffet}},\ and\ \bibinfo {author} {\bibfnamefont
  {P.}~\bibnamefont {Bouyer}},\ }\bibfield  {title} {\bibinfo {title}
  {Exploring gravity with the {MIGA} large scale atom interferometer},\
  }\bibfield  {journal} {\bibinfo  {journal} {Scientific Reports}\ }\textbf
  {\bibinfo {volume} {8}},\ \href {https://doi.org/10.1038/s41598-018-32165-z}
  {10.1038/s41598-018-32165-z} (\bibinfo {year} {2018})\BibitemShut {NoStop}%
\bibitem [{\citenamefont {Beckwith}\ \emph {et~al.}(2020)\citenamefont
  {Beckwith}, \citenamefont {Baker}, \citenamefont {Access}, \citenamefont
  {Beckwith},\ and\ \citenamefont {Baker}}]{Li-Baker_HFRGW_Detector}%
  \BibitemOpen
  \bibfield  {author} {\bibinfo {author} {\bibfnamefont {A.}~\bibnamefont
  {Beckwith}}, \bibinfo {author} {\bibfnamefont {R.}~\bibnamefont {Baker}},
  \bibinfo {author} {\bibfnamefont {O.}~\bibnamefont {Access}}, \bibinfo
  {author} {\bibfnamefont {A.}~\bibnamefont {Beckwith}},\ and\ \bibinfo
  {author} {\bibfnamefont {R.}~\bibnamefont {Baker}},\ }\bibfield  {title}
  {\bibinfo {title} {Value of high-frequency relic gravitational wave (hfrgw)
  detection to astrophysics and fabrication and utilization of the li-baker
  hfrgw detector},\ }\href {https://doi.org/10.4236/jhepgc.2020.61010}
  {\bibfield  {journal} {\bibinfo  {journal} {Journal of High Energy Physics,
  Gravitation and Cosmology}\ }\textbf {\bibinfo {volume} {6}},\ \bibinfo
  {pages} {103} (\bibinfo {year} {2020})}\BibitemShut {NoStop}%
\bibitem [{\citenamefont {{Taylor}}\ and\ \citenamefont
  {{Weisberg}}(1989)}]{1989ApJ...345..434T}%
  \BibitemOpen
  \bibfield  {author} {\bibinfo {author} {\bibfnamefont {J.~H.}\ \bibnamefont
  {{Taylor}}}\ and\ \bibinfo {author} {\bibfnamefont {J.~M.}\ \bibnamefont
  {{Weisberg}}},\ }\bibfield  {title} {\bibinfo {title} {{Further Experimental
  Tests of Relativistic Gravity Using the Binary Pulsar PSR 1913+16}},\ }\href
  {https://doi.org/10.1086/167917} {\bibfield  {journal} {\bibinfo  {journal}
  {\apj}\ }\textbf {\bibinfo {volume} {345}},\ \bibinfo {pages} {434} (\bibinfo
  {year} {1989})}\BibitemShut {NoStop}%
\bibitem [{\citenamefont {{Weisberg}}\ \emph {et~al.}(2010)\citenamefont
  {{Weisberg}}, \citenamefont {{Nice}},\ and\ \citenamefont
  {{Taylor}}}]{2010ApJ...722.1030W}%
  \BibitemOpen
  \bibfield  {author} {\bibinfo {author} {\bibfnamefont {J.~M.}\ \bibnamefont
  {{Weisberg}}}, \bibinfo {author} {\bibfnamefont {D.~J.}\ \bibnamefont
  {{Nice}}},\ and\ \bibinfo {author} {\bibfnamefont {J.~H.}\ \bibnamefont
  {{Taylor}}},\ }\bibfield  {title} {\bibinfo {title} {{Timing Measurements of
  the Relativistic Binary Pulsar PSR B1913+16}},\ }\href
  {https://doi.org/10.1088/0004-637X/722/2/1030} {\bibfield  {journal}
  {\bibinfo  {journal} {\apj}\ }\textbf {\bibinfo {volume} {722}},\ \bibinfo
  {pages} {1030} (\bibinfo {year} {2010})},\ \Eprint
  {https://arxiv.org/abs/1011.0718} {arXiv:1011.0718 [astro-ph.GA]}
  \BibitemShut {NoStop}%
\bibitem [{\citenamefont {{Abbott}}\ \emph
  {et~al.}(2019{\natexlab{b}})\citenamefont {{Abbott}}, \citenamefont
  {{Abbott}}, \citenamefont {{Abbott}}, \citenamefont {{Abraham}},
  \citenamefont {{Acernese}}, \citenamefont {{Ackley}}, \citenamefont
  {{Adams}}, \citenamefont {{Adhikari}}, \citenamefont {{Adya}}, \citenamefont
  {{Affeldt}}, \citenamefont {{Agathos}}, \citenamefont {{Agatsuma}},
  \citenamefont {{Aggarwal}}, \citenamefont {{Aguiar}}, \citenamefont
  {{Aiello}}, \citenamefont {{Ain}}, \citenamefont {{Ajith}}, \citenamefont
  {{Allen}}, \citenamefont {{Allocca}}, \citenamefont {{Aloy}}, \citenamefont
  {{Altin}}, \citenamefont {{Amato}}, \citenamefont {{Ananyeva}}, \citenamefont
  {{Anderson}}, \citenamefont {{Anderson}}, \citenamefont {{Angelova}},
  \citenamefont {{Antier}}, \citenamefont {{Appert}}, \citenamefont {{Arai}},
  \citenamefont {{Araya}}, \citenamefont {{Areeda}}, \citenamefont
  {{Ar{\`e}ne}}, \citenamefont {{Arnaud}}, \citenamefont {{Arun}},
  \citenamefont {{Ascenzi}}, \citenamefont {{Ashton}}, \citenamefont {{Aston}},
  \citenamefont {{Astone}}, \citenamefont {{Aubin}}, \citenamefont {{Aufmuth}},
  \citenamefont {{AultONeal}}, \citenamefont {{Austin}}, \citenamefont
  {{Avendano}}, \citenamefont {{Avila-Alvarez}}, \citenamefont {{Babak}},
  \citenamefont {{Bacon}}, \citenamefont {{Badaracco}}, \citenamefont
  {{Bader}}, \citenamefont {{Bae}}, \citenamefont {{Baker}}, \citenamefont
  {{Baldaccini}}, \citenamefont {{Ballardin}}, \citenamefont {{Ballmer}},
  \citenamefont {{Banagiri}}, \citenamefont {{Barayoga}}, \citenamefont
  {{Barclay}}, \citenamefont {{Barish}}, \citenamefont {{Barker}},
  \citenamefont {{Barkett}}, \citenamefont {{Barnum}}, \citenamefont
  {{Barone}}, \citenamefont {{Barr}}, \citenamefont {{Barsotti}}, \citenamefont
  {{Barsuglia}}, \citenamefont {{Barta}}, \citenamefont {{Bartlett}},
  \citenamefont {{Bartos}}, \citenamefont {{Bassiri}}, \citenamefont {{Basti}},
  \citenamefont {{Bawaj}}, \citenamefont {{Bayley}}, \citenamefont {{Bazzan}},
  \citenamefont {{B{\'e}csy}}, \citenamefont {{Bejger}}, \citenamefont
  {{Belahcene}}, \citenamefont {{Bell}}, \citenamefont {{Beniwal}},
  \citenamefont {{Berger}}, \citenamefont {{Bergmann}}, \citenamefont
  {{Bernuzzi}}, \citenamefont {{Bero}}, \citenamefont {{Berry}}, \citenamefont
  {{Bersanetti}}, \citenamefont {{Bertolini}}, \citenamefont {{Betzwieser}},
  \citenamefont {{Bhand are}}, \citenamefont {{Bidler}}, \citenamefont
  {{Bilenko}}, \citenamefont {{Bilgili}}, \citenamefont {{Billingsley}},
  \citenamefont {{Birch}}, \citenamefont {{Birney}}, \citenamefont
  {{Birnholtz}}, \citenamefont {{Biscans}}, \citenamefont {{Biscoveanu}},
  \citenamefont {{Bisht}}, \citenamefont {{Bitossi}}, \citenamefont
  {{Bizouard}}, \citenamefont {{Blackburn}}, \citenamefont {{Blair}},
  \citenamefont {{Blair}}, \citenamefont {{Blair}}, \citenamefont {{Bloemen}},
  \citenamefont {{Bode}}, \citenamefont {{Boer}}, \citenamefont {{Boetzel}},
  \citenamefont {{Bogaert}}, \citenamefont {{Bondu}}, \citenamefont
  {{Bonilla}}, \citenamefont {{Bonnand}}, \citenamefont {{Booker}},
  \citenamefont {{Boom}}, \citenamefont {{Booth}}, \citenamefont {{Bork}},
  \citenamefont {{Boschi}}, \citenamefont {{Bose}}, \citenamefont {{Bossie}},
  \citenamefont {{Bossilkov}}, \citenamefont {{Bosveld}}, \citenamefont
  {{Bouffanais}}, \citenamefont {{Bozzi}}, \citenamefont {{Bradaschia}},
  \citenamefont {{Brady}}, \citenamefont {{Bramley}}, \citenamefont
  {{Branchesi}}, \citenamefont {{Brau}}, \citenamefont {{Briant}},
  \citenamefont {{Briggs}}, \citenamefont {{Brighenti}}, \citenamefont
  {{Brillet}}, \citenamefont {{Brinkmann}}, \citenamefont {{Brisson}},
  \citenamefont {{Brockill}}, \citenamefont {{Brooks}}, \citenamefont
  {{Brown}}, \citenamefont {{Brunett}}, \citenamefont {{Buikema}},
  \citenamefont {{Bulik}}, \citenamefont {{Bulten}}, \citenamefont
  {{Buonanno}}, \citenamefont {{Buscicchio}}, \citenamefont {{Buskulic}},
  \citenamefont {{Buy}}, \citenamefont {{Byer}}, \citenamefont {{Cabero}},
  \citenamefont {{Cadonati}}, \citenamefont {{Cagnoli}}, \citenamefont
  {{Cahillane}}, \citenamefont {{Calder{\'o}n Bustillo}}, \citenamefont
  {{Callister}}, \citenamefont {{Calloni}}, \citenamefont {{Camp}},
  \citenamefont {{Campbell}}, \citenamefont {{Canepa}}, \citenamefont
  {{Cannon}}, \citenamefont {{Cao}}, \citenamefont {{Cao}}, \citenamefont
  {{Capocasa}}, \citenamefont {{Carbognani}}, \citenamefont {{Caride}},
  \citenamefont {{Carney}}, \citenamefont {{Carullo}}, \citenamefont
  {{Casanueva Diaz}}, \citenamefont {{Casentini}}, \citenamefont {{Caudill}},
  \citenamefont {{Cavagli{\`a}}}, \citenamefont {{Cavalier}}, \citenamefont
  {{Cavalieri}}, \citenamefont {{Cella}}, \citenamefont
  {{Cerd{\'a}-Dur{\'a}n}}, \citenamefont {{Cerretani}}, \citenamefont
  {{Cesarini}}, \citenamefont {{Chaibi}}, \citenamefont {{Chakravarti}},
  \citenamefont {{Chamberlin}}, \citenamefont {{Chan}}, \citenamefont {{Chao}},
  \citenamefont {{Charlton}}, \citenamefont {{Chase}}, \citenamefont {{Chassand
  e-Mottin}}, \citenamefont {{Chatterjee}}, \citenamefont {{Chaturvedi}},
  \citenamefont {{LIGO Scientific Collaboration}},\ and\ \citenamefont {{Virgo
  Collaboration}}}]{2019ApJ...882L..24A}%
  \BibitemOpen
  \bibfield  {author} {\bibinfo {author} {\bibfnamefont {B.~P.}\ \bibnamefont
  {{Abbott}}}, \bibinfo {author} {\bibfnamefont {R.}~\bibnamefont {{Abbott}}},
  \bibinfo {author} {\bibfnamefont {T.~D.}\ \bibnamefont {{Abbott}}}, \bibinfo
  {author} {\bibfnamefont {S.}~\bibnamefont {{Abraham}}}, \bibinfo {author}
  {\bibfnamefont {F.}~\bibnamefont {{Acernese}}}, \bibinfo {author}
  {\bibfnamefont {K.}~\bibnamefont {{Ackley}}}, \bibinfo {author}
  {\bibfnamefont {C.}~\bibnamefont {{Adams}}}, \bibinfo {author} {\bibfnamefont
  {R.~X.}\ \bibnamefont {{Adhikari}}}, \bibinfo {author} {\bibfnamefont
  {V.~B.}\ \bibnamefont {{Adya}}}, \bibinfo {author} {\bibfnamefont
  {C.}~\bibnamefont {{Affeldt}}}, \bibinfo {author} {\bibfnamefont
  {M.}~\bibnamefont {{Agathos}}}, \bibinfo {author} {\bibfnamefont
  {K.}~\bibnamefont {{Agatsuma}}}, \bibinfo {author} {\bibfnamefont
  {N.}~\bibnamefont {{Aggarwal}}}, \bibinfo {author} {\bibfnamefont {O.~D.}\
  \bibnamefont {{Aguiar}}}, \bibinfo {author} {\bibfnamefont {L.}~\bibnamefont
  {{Aiello}}}, \bibinfo {author} {\bibfnamefont {A.}~\bibnamefont {{Ain}}},
  \bibinfo {author} {\bibfnamefont {P.}~\bibnamefont {{Ajith}}}, \bibinfo
  {author} {\bibfnamefont {G.}~\bibnamefont {{Allen}}}, \bibinfo {author}
  {\bibfnamefont {A.}~\bibnamefont {{Allocca}}}, \bibinfo {author}
  {\bibfnamefont {M.~A.}\ \bibnamefont {{Aloy}}}, \bibinfo {author}
  {\bibfnamefont {P.~A.}\ \bibnamefont {{Altin}}}, \bibinfo {author}
  {\bibfnamefont {A.}~\bibnamefont {{Amato}}}, \bibinfo {author} {\bibfnamefont
  {A.}~\bibnamefont {{Ananyeva}}}, \bibinfo {author} {\bibfnamefont {S.~B.}\
  \bibnamefont {{Anderson}}}, \bibinfo {author} {\bibfnamefont {W.~G.}\
  \bibnamefont {{Anderson}}}, \bibinfo {author} {\bibfnamefont {S.~V.}\
  \bibnamefont {{Angelova}}}, \bibinfo {author} {\bibfnamefont
  {S.}~\bibnamefont {{Antier}}}, \bibinfo {author} {\bibfnamefont
  {S.}~\bibnamefont {{Appert}}}, \bibinfo {author} {\bibfnamefont
  {K.}~\bibnamefont {{Arai}}}, \bibinfo {author} {\bibfnamefont {M.~C.}\
  \bibnamefont {{Araya}}}, \bibinfo {author} {\bibfnamefont {J.~S.}\
  \bibnamefont {{Areeda}}}, \bibinfo {author} {\bibfnamefont {M.}~\bibnamefont
  {{Ar{\`e}ne}}}, \bibinfo {author} {\bibfnamefont {N.}~\bibnamefont
  {{Arnaud}}}, \bibinfo {author} {\bibfnamefont {K.~G.}\ \bibnamefont
  {{Arun}}}, \bibinfo {author} {\bibfnamefont {S.}~\bibnamefont {{Ascenzi}}},
  \bibinfo {author} {\bibfnamefont {G.}~\bibnamefont {{Ashton}}}, \bibinfo
  {author} {\bibfnamefont {S.~M.}\ \bibnamefont {{Aston}}}, \bibinfo {author}
  {\bibfnamefont {P.}~\bibnamefont {{Astone}}}, \bibinfo {author}
  {\bibfnamefont {F.}~\bibnamefont {{Aubin}}}, \bibinfo {author} {\bibfnamefont
  {P.}~\bibnamefont {{Aufmuth}}}, \bibinfo {author} {\bibfnamefont
  {K.}~\bibnamefont {{AultONeal}}}, \bibinfo {author} {\bibfnamefont
  {C.}~\bibnamefont {{Austin}}}, \bibinfo {author} {\bibfnamefont
  {V.}~\bibnamefont {{Avendano}}}, \bibinfo {author} {\bibfnamefont
  {A.}~\bibnamefont {{Avila-Alvarez}}}, \bibinfo {author} {\bibfnamefont
  {S.}~\bibnamefont {{Babak}}}, \bibinfo {author} {\bibfnamefont
  {P.}~\bibnamefont {{Bacon}}}, \bibinfo {author} {\bibfnamefont
  {F.}~\bibnamefont {{Badaracco}}}, \bibinfo {author} {\bibfnamefont
  {M.~K.~M.}\ \bibnamefont {{Bader}}}, \bibinfo {author} {\bibfnamefont
  {S.}~\bibnamefont {{Bae}}}, \bibinfo {author} {\bibfnamefont {P.~T.}\
  \bibnamefont {{Baker}}}, \bibinfo {author} {\bibfnamefont {F.}~\bibnamefont
  {{Baldaccini}}}, \bibinfo {author} {\bibfnamefont {G.}~\bibnamefont
  {{Ballardin}}}, \bibinfo {author} {\bibfnamefont {S.~W.}\ \bibnamefont
  {{Ballmer}}}, \bibinfo {author} {\bibfnamefont {S.}~\bibnamefont
  {{Banagiri}}}, \bibinfo {author} {\bibfnamefont {J.~C.}\ \bibnamefont
  {{Barayoga}}}, \bibinfo {author} {\bibfnamefont {S.~E.}\ \bibnamefont
  {{Barclay}}}, \bibinfo {author} {\bibfnamefont {B.~C.}\ \bibnamefont
  {{Barish}}}, \bibinfo {author} {\bibfnamefont {D.}~\bibnamefont {{Barker}}},
  \bibinfo {author} {\bibfnamefont {K.}~\bibnamefont {{Barkett}}}, \bibinfo
  {author} {\bibfnamefont {S.}~\bibnamefont {{Barnum}}}, \bibinfo {author}
  {\bibfnamefont {F.}~\bibnamefont {{Barone}}}, \bibinfo {author}
  {\bibfnamefont {B.}~\bibnamefont {{Barr}}}, \bibinfo {author} {\bibfnamefont
  {L.}~\bibnamefont {{Barsotti}}}, \bibinfo {author} {\bibfnamefont
  {M.}~\bibnamefont {{Barsuglia}}}, \bibinfo {author} {\bibfnamefont
  {D.}~\bibnamefont {{Barta}}}, \bibinfo {author} {\bibfnamefont
  {J.}~\bibnamefont {{Bartlett}}}, \bibinfo {author} {\bibfnamefont
  {I.}~\bibnamefont {{Bartos}}}, \bibinfo {author} {\bibfnamefont
  {R.}~\bibnamefont {{Bassiri}}}, \bibinfo {author} {\bibfnamefont
  {A.}~\bibnamefont {{Basti}}}, \bibinfo {author} {\bibfnamefont
  {M.}~\bibnamefont {{Bawaj}}}, \bibinfo {author} {\bibfnamefont {J.~C.}\
  \bibnamefont {{Bayley}}}, \bibinfo {author} {\bibfnamefont {M.}~\bibnamefont
  {{Bazzan}}}, \bibinfo {author} {\bibfnamefont {B.}~\bibnamefont
  {{B{\'e}csy}}}, \bibinfo {author} {\bibfnamefont {M.}~\bibnamefont
  {{Bejger}}}, \bibinfo {author} {\bibfnamefont {I.}~\bibnamefont
  {{Belahcene}}}, \bibinfo {author} {\bibfnamefont {A.~S.}\ \bibnamefont
  {{Bell}}}, \bibinfo {author} {\bibfnamefont {D.}~\bibnamefont {{Beniwal}}},
  \bibinfo {author} {\bibfnamefont {B.~K.}\ \bibnamefont {{Berger}}}, \bibinfo
  {author} {\bibfnamefont {G.}~\bibnamefont {{Bergmann}}}, \bibinfo {author}
  {\bibfnamefont {S.}~\bibnamefont {{Bernuzzi}}}, \bibinfo {author}
  {\bibfnamefont {J.~J.}\ \bibnamefont {{Bero}}}, \bibinfo {author}
  {\bibfnamefont {C.~P.~L.}\ \bibnamefont {{Berry}}}, \bibinfo {author}
  {\bibfnamefont {D.}~\bibnamefont {{Bersanetti}}}, \bibinfo {author}
  {\bibfnamefont {A.}~\bibnamefont {{Bertolini}}}, \bibinfo {author}
  {\bibfnamefont {J.}~\bibnamefont {{Betzwieser}}}, \bibinfo {author}
  {\bibfnamefont {R.}~\bibnamefont {{Bhand are}}}, \bibinfo {author}
  {\bibfnamefont {J.}~\bibnamefont {{Bidler}}}, \bibinfo {author}
  {\bibfnamefont {I.~A.}\ \bibnamefont {{Bilenko}}}, \bibinfo {author}
  {\bibfnamefont {S.~A.}\ \bibnamefont {{Bilgili}}}, \bibinfo {author}
  {\bibfnamefont {G.}~\bibnamefont {{Billingsley}}}, \bibinfo {author}
  {\bibfnamefont {J.}~\bibnamefont {{Birch}}}, \bibinfo {author} {\bibfnamefont
  {R.}~\bibnamefont {{Birney}}}, \bibinfo {author} {\bibfnamefont
  {O.}~\bibnamefont {{Birnholtz}}}, \bibinfo {author} {\bibfnamefont
  {S.}~\bibnamefont {{Biscans}}}, \bibinfo {author} {\bibfnamefont
  {S.}~\bibnamefont {{Biscoveanu}}}, \bibinfo {author} {\bibfnamefont
  {A.}~\bibnamefont {{Bisht}}}, \bibinfo {author} {\bibfnamefont
  {M.}~\bibnamefont {{Bitossi}}}, \bibinfo {author} {\bibfnamefont {M.~A.}\
  \bibnamefont {{Bizouard}}}, \bibinfo {author} {\bibfnamefont {J.~K.}\
  \bibnamefont {{Blackburn}}}, \bibinfo {author} {\bibfnamefont {C.~D.}\
  \bibnamefont {{Blair}}}, \bibinfo {author} {\bibfnamefont {D.~G.}\
  \bibnamefont {{Blair}}}, \bibinfo {author} {\bibfnamefont {R.~M.}\
  \bibnamefont {{Blair}}}, \bibinfo {author} {\bibfnamefont {S.}~\bibnamefont
  {{Bloemen}}}, \bibinfo {author} {\bibfnamefont {N.}~\bibnamefont {{Bode}}},
  \bibinfo {author} {\bibfnamefont {M.}~\bibnamefont {{Boer}}}, \bibinfo
  {author} {\bibfnamefont {Y.}~\bibnamefont {{Boetzel}}}, \bibinfo {author}
  {\bibfnamefont {G.}~\bibnamefont {{Bogaert}}}, \bibinfo {author}
  {\bibfnamefont {F.}~\bibnamefont {{Bondu}}}, \bibinfo {author} {\bibfnamefont
  {E.}~\bibnamefont {{Bonilla}}}, \bibinfo {author} {\bibfnamefont
  {R.}~\bibnamefont {{Bonnand}}}, \bibinfo {author} {\bibfnamefont
  {P.}~\bibnamefont {{Booker}}}, \bibinfo {author} {\bibfnamefont {B.~A.}\
  \bibnamefont {{Boom}}}, \bibinfo {author} {\bibfnamefont {C.~D.}\
  \bibnamefont {{Booth}}}, \bibinfo {author} {\bibfnamefont {R.}~\bibnamefont
  {{Bork}}}, \bibinfo {author} {\bibfnamefont {V.}~\bibnamefont {{Boschi}}},
  \bibinfo {author} {\bibfnamefont {S.}~\bibnamefont {{Bose}}}, \bibinfo
  {author} {\bibfnamefont {K.}~\bibnamefont {{Bossie}}}, \bibinfo {author}
  {\bibfnamefont {V.}~\bibnamefont {{Bossilkov}}}, \bibinfo {author}
  {\bibfnamefont {J.}~\bibnamefont {{Bosveld}}}, \bibinfo {author}
  {\bibfnamefont {Y.}~\bibnamefont {{Bouffanais}}}, \bibinfo {author}
  {\bibfnamefont {A.}~\bibnamefont {{Bozzi}}}, \bibinfo {author} {\bibfnamefont
  {C.}~\bibnamefont {{Bradaschia}}}, \bibinfo {author} {\bibfnamefont {P.~R.}\
  \bibnamefont {{Brady}}}, \bibinfo {author} {\bibfnamefont {A.}~\bibnamefont
  {{Bramley}}}, \bibinfo {author} {\bibfnamefont {M.}~\bibnamefont
  {{Branchesi}}}, \bibinfo {author} {\bibfnamefont {J.~E.}\ \bibnamefont
  {{Brau}}}, \bibinfo {author} {\bibfnamefont {T.}~\bibnamefont {{Briant}}},
  \bibinfo {author} {\bibfnamefont {J.~H.}\ \bibnamefont {{Briggs}}}, \bibinfo
  {author} {\bibfnamefont {F.}~\bibnamefont {{Brighenti}}}, \bibinfo {author}
  {\bibfnamefont {A.}~\bibnamefont {{Brillet}}}, \bibinfo {author}
  {\bibfnamefont {M.}~\bibnamefont {{Brinkmann}}}, \bibinfo {author}
  {\bibfnamefont {V.}~\bibnamefont {{Brisson}}}, \bibinfo {author}
  {\bibfnamefont {P.}~\bibnamefont {{Brockill}}}, \bibinfo {author}
  {\bibfnamefont {A.~F.}\ \bibnamefont {{Brooks}}}, \bibinfo {author}
  {\bibfnamefont {D.~D.}\ \bibnamefont {{Brown}}}, \bibinfo {author}
  {\bibfnamefont {S.}~\bibnamefont {{Brunett}}}, \bibinfo {author}
  {\bibfnamefont {A.}~\bibnamefont {{Buikema}}}, \bibinfo {author}
  {\bibfnamefont {T.}~\bibnamefont {{Bulik}}}, \bibinfo {author} {\bibfnamefont
  {H.~J.}\ \bibnamefont {{Bulten}}}, \bibinfo {author} {\bibfnamefont
  {A.}~\bibnamefont {{Buonanno}}}, \bibinfo {author} {\bibfnamefont
  {R.}~\bibnamefont {{Buscicchio}}}, \bibinfo {author} {\bibfnamefont
  {D.}~\bibnamefont {{Buskulic}}}, \bibinfo {author} {\bibfnamefont
  {C.}~\bibnamefont {{Buy}}}, \bibinfo {author} {\bibfnamefont {R.~L.}\
  \bibnamefont {{Byer}}}, \bibinfo {author} {\bibfnamefont {M.}~\bibnamefont
  {{Cabero}}}, \bibinfo {author} {\bibfnamefont {L.}~\bibnamefont
  {{Cadonati}}}, \bibinfo {author} {\bibfnamefont {G.}~\bibnamefont
  {{Cagnoli}}}, \bibinfo {author} {\bibfnamefont {C.}~\bibnamefont
  {{Cahillane}}}, \bibinfo {author} {\bibfnamefont {J.}~\bibnamefont
  {{Calder{\'o}n Bustillo}}}, \bibinfo {author} {\bibfnamefont {T.~A.}\
  \bibnamefont {{Callister}}}, \bibinfo {author} {\bibfnamefont
  {E.}~\bibnamefont {{Calloni}}}, \bibinfo {author} {\bibfnamefont {J.~B.}\
  \bibnamefont {{Camp}}}, \bibinfo {author} {\bibfnamefont {W.~A.}\
  \bibnamefont {{Campbell}}}, \bibinfo {author} {\bibfnamefont
  {M.}~\bibnamefont {{Canepa}}}, \bibinfo {author} {\bibfnamefont {K.~C.}\
  \bibnamefont {{Cannon}}}, \bibinfo {author} {\bibfnamefont {H.}~\bibnamefont
  {{Cao}}}, \bibinfo {author} {\bibfnamefont {J.}~\bibnamefont {{Cao}}},
  \bibinfo {author} {\bibfnamefont {E.}~\bibnamefont {{Capocasa}}}, \bibinfo
  {author} {\bibfnamefont {F.}~\bibnamefont {{Carbognani}}}, \bibinfo {author}
  {\bibfnamefont {S.}~\bibnamefont {{Caride}}}, \bibinfo {author}
  {\bibfnamefont {M.~F.}\ \bibnamefont {{Carney}}}, \bibinfo {author}
  {\bibfnamefont {G.}~\bibnamefont {{Carullo}}}, \bibinfo {author}
  {\bibfnamefont {J.}~\bibnamefont {{Casanueva Diaz}}}, \bibinfo {author}
  {\bibfnamefont {C.}~\bibnamefont {{Casentini}}}, \bibinfo {author}
  {\bibfnamefont {S.}~\bibnamefont {{Caudill}}}, \bibinfo {author}
  {\bibfnamefont {M.}~\bibnamefont {{Cavagli{\`a}}}}, \bibinfo {author}
  {\bibfnamefont {F.}~\bibnamefont {{Cavalier}}}, \bibinfo {author}
  {\bibfnamefont {R.}~\bibnamefont {{Cavalieri}}}, \bibinfo {author}
  {\bibfnamefont {G.}~\bibnamefont {{Cella}}}, \bibinfo {author} {\bibfnamefont
  {P.}~\bibnamefont {{Cerd{\'a}-Dur{\'a}n}}}, \bibinfo {author} {\bibfnamefont
  {G.}~\bibnamefont {{Cerretani}}}, \bibinfo {author} {\bibfnamefont
  {E.}~\bibnamefont {{Cesarini}}}, \bibinfo {author} {\bibfnamefont
  {O.}~\bibnamefont {{Chaibi}}}, \bibinfo {author} {\bibfnamefont
  {K.}~\bibnamefont {{Chakravarti}}}, \bibinfo {author} {\bibfnamefont {S.~J.}\
  \bibnamefont {{Chamberlin}}}, \bibinfo {author} {\bibfnamefont
  {M.}~\bibnamefont {{Chan}}}, \bibinfo {author} {\bibfnamefont
  {S.}~\bibnamefont {{Chao}}}, \bibinfo {author} {\bibfnamefont
  {P.}~\bibnamefont {{Charlton}}}, \bibinfo {author} {\bibfnamefont {E.~A.}\
  \bibnamefont {{Chase}}}, \bibinfo {author} {\bibfnamefont {E.}~\bibnamefont
  {{Chassand e-Mottin}}}, \bibinfo {author} {\bibfnamefont {D.}~\bibnamefont
  {{Chatterjee}}}, \bibinfo {author} {\bibfnamefont {M.}~\bibnamefont
  {{Chaturvedi}}}, \bibinfo {author} {\bibnamefont {{LIGO Scientific
  Collaboration}}},\ and\ \bibinfo {author} {\bibnamefont {{Virgo
  Collaboration}}},\ }\bibfield  {title} {\bibinfo {title} {{Binary Black Hole
  Population Properties Inferred from the First and Second Observing Runs of
  Advanced LIGO and Advanced Virgo}},\ }\href
  {https://doi.org/10.3847/2041-8213/ab3800} {\bibfield  {journal} {\bibinfo
  {journal} {\apjl}\ }\textbf {\bibinfo {volume} {882}},\ \bibinfo {eid} {L24}
  (\bibinfo {year} {2019}{\natexlab{b}})},\ \Eprint
  {https://arxiv.org/abs/1811.12940} {arXiv:1811.12940 [astro-ph.HE]}
  \BibitemShut {NoStop}%
\bibitem [{\citenamefont {{Abbott}}\ \emph
  {et~al.}(2021{\natexlab{c}})\citenamefont {{Abbott}}, \citenamefont
  {{Abbott}}, \citenamefont {{Abraham}}, \citenamefont {{Acernese}},
  \citenamefont {{Ackley}}, \citenamefont {{Adams}}, \citenamefont {{Adams}},
  \citenamefont {{Adhikari}}, \citenamefont {{Adya}}, \citenamefont
  {{Affeldt}}, \citenamefont {{Agarwal}}, \citenamefont {{Agathos}},
  \citenamefont {{Agatsuma}}, \citenamefont {{Aggarwal}}, \citenamefont
  {{Aguiar}}, \citenamefont {{Aiello}}, \citenamefont {{Ain}}, \citenamefont
  {{Ajith}}, \citenamefont {{Akutsu}}, \citenamefont {{Aleman}}, \citenamefont
  {{Allen}}, \citenamefont {{Allocca}}, \citenamefont {{Altin}}, \citenamefont
  {{Amato}}, \citenamefont {{Anand}}, \citenamefont {{Ananyeva}}, \citenamefont
  {{Zlochower}}, \citenamefont {{Zucker}}, \citenamefont {{Zweizig}},
  \citenamefont {{Ligo Scientific Collaboration}}, \citenamefont {{VIRGO
  Collaboration}},\ and\ \citenamefont {{KAGRA
  Collaboration}}}]{Abbott:2021-first-NSBH}%
  \BibitemOpen
  \bibfield  {author} {\bibinfo {author} {\bibfnamefont {R.}~\bibnamefont
  {{Abbott}}}, \bibinfo {author} {\bibfnamefont {T.~D.}\ \bibnamefont
  {{Abbott}}}, \bibinfo {author} {\bibfnamefont {S.}~\bibnamefont {{Abraham}}},
  \bibinfo {author} {\bibfnamefont {F.}~\bibnamefont {{Acernese}}}, \bibinfo
  {author} {\bibfnamefont {K.}~\bibnamefont {{Ackley}}}, \bibinfo {author}
  {\bibfnamefont {A.}~\bibnamefont {{Adams}}}, \bibinfo {author} {\bibfnamefont
  {C.}~\bibnamefont {{Adams}}}, \bibinfo {author} {\bibfnamefont {R.~X.}\
  \bibnamefont {{Adhikari}}}, \bibinfo {author} {\bibfnamefont {V.~B.}\
  \bibnamefont {{Adya}}}, \bibinfo {author} {\bibfnamefont {C.}~\bibnamefont
  {{Affeldt}}}, \bibinfo {author} {\bibfnamefont {D.}~\bibnamefont
  {{Agarwal}}}, \bibinfo {author} {\bibfnamefont {M.}~\bibnamefont
  {{Agathos}}}, \bibinfo {author} {\bibfnamefont {K.}~\bibnamefont
  {{Agatsuma}}}, \bibinfo {author} {\bibfnamefont {N.}~\bibnamefont
  {{Aggarwal}}}, \bibinfo {author} {\bibfnamefont {O.~D.}\ \bibnamefont
  {{Aguiar}}}, \bibinfo {author} {\bibfnamefont {L.}~\bibnamefont {{Aiello}}},
  \bibinfo {author} {\bibfnamefont {A.}~\bibnamefont {{Ain}}}, \bibinfo
  {author} {\bibfnamefont {P.}~\bibnamefont {{Ajith}}}, \bibinfo {author}
  {\bibfnamefont {T.}~\bibnamefont {{Akutsu}}}, \bibinfo {author}
  {\bibfnamefont {K.~M.}\ \bibnamefont {{Aleman}}}, \bibinfo {author}
  {\bibfnamefont {G.}~\bibnamefont {{Allen}}}, \bibinfo {author} {\bibfnamefont
  {A.}~\bibnamefont {{Allocca}}}, \bibinfo {author} {\bibfnamefont {P.~A.}\
  \bibnamefont {{Altin}}}, \bibinfo {author} {\bibfnamefont {A.}~\bibnamefont
  {{Amato}}}, \bibinfo {author} {\bibfnamefont {S.}~\bibnamefont {{Anand}}},
  \bibinfo {author} {\bibfnamefont {A.}~\bibnamefont {{Ananyeva}}}, \bibinfo
  {author} {\bibfnamefont {Y.}~\bibnamefont {{Zlochower}}}, \bibinfo {author}
  {\bibfnamefont {M.~E.}\ \bibnamefont {{Zucker}}}, \bibinfo {author}
  {\bibfnamefont {J.}~\bibnamefont {{Zweizig}}}, \bibinfo {author}
  {\bibnamefont {{Ligo Scientific Collaboration}}}, \bibinfo {author}
  {\bibnamefont {{VIRGO Collaboration}}},\ and\ \bibinfo {author} {\bibnamefont
  {{KAGRA Collaboration}}},\ }\bibfield  {title} {\bibinfo {title}
  {{Observation of Gravitational Waves from Two Neutron Star-Black Hole
  Coalescences}},\ }\href {https://doi.org/10.3847/2041-8213/ac082e} {\bibfield
   {journal} {\bibinfo  {journal} {\apjl}\ }\textbf {\bibinfo {volume} {915}},\
  \bibinfo {eid} {L5} (\bibinfo {year} {2021}{\natexlab{c}})},\ \Eprint
  {https://arxiv.org/abs/2106.15163} {arXiv:2106.15163 [astro-ph.HE]}
  \BibitemShut {NoStop}%
\end{thebibliography}%

\end{document}